\documentclass
[aps,prb,amsfonts,amssymb,twocolumn,amsmath,preprintnumbers,nofootinbib,floatfix,superscriptaddress,notitlepage]{revtex4-1}%
\usepackage[dvips]{graphics}
\usepackage{graphicx}
\usepackage{bm}
\usepackage{amsmath}
\usepackage{amsfonts}
\usepackage{amssymb}
\usepackage{xcolor}
\usepackage{subfigure}
\usepackage{tabularx}
\usepackage{multirow}
\usepackage[colorlinks=true, pdfstartview=FitV,  linkcolor=blue, citecolor=blue, urlcolor=blue]{hyperref}
\usepackage{hypcap}
\usepackage{braket}
\usepackage{commath}%
\graphicspath{ {figures/} }
\usepackage{chemformula}
\definecolor{colorhhy}{rgb}{0.9, 0.17, 0.31}
\newcommand{\hhy}[1]{\textcolor{colorhhy}{HH:#1}}
\newcommand{\DC}[1]{\textcolor{blue}{DC: #1}}
\newcommand{\YJ}[1]{\textcolor{blue}{YJ: #1}}

\usepackage{ulem}

\makeatother

\begin{document}
\title{ Catalogue of Phonon Instabilities in Symmetry Group 191 Kagome \ch{MT6Z6} Materials }

\author{X. Feng}
\email{xiaolong.feng@cpfs.mpg.de}
\affiliation{Max Planck Institute for Chemical Physics of Solids, 01187 Dresden, Germany}

\author{Y. Jiang}
\affiliation{Donostia International Physics Center (DIPC), Paseo Manuel de Lardizábal. 20018, San Sebastián, Spain}

\author{H. Hu}
\affiliation{Donostia International Physics Center (DIPC), Paseo Manuel de Lardizábal. 20018, San Sebastián, Spain}

\author{D. C\u{a}lug\u{a}ru}
\affiliation{Department of Physics, Princeton University, Princeton, NJ 08544, USA}

\author{N. Regnault}
\affiliation{Laboratoire de Physique de l’Ecole normale supérieure,
ENS, Université PSL, CNRS, Sorbonne Université,
Université Paris-Diderot, Sorbonne Paris Cité, 75005 Paris, France}
\affiliation{Department of Physics, Princeton University, Princeton, NJ 08544, USA}

\author{M. G. Vergniory}
\affiliation{Donostia International Physics Center (DIPC), Paseo Manuel de Lardizábal. 20018, San Sebastián, Spain}
\affiliation{Max Planck Institute for Chemical Physics of Solids, 01187 Dresden, Germany}

\author{C. Felser}
\email{Claudia.Felser@cpfs.mpg.de}
\affiliation{Max Planck Institute for Chemical Physics of Solids, 01187 Dresden, Germany}

\author{S. Blanco-Canosa}
\affiliation{Donostia International Physics Center (DIPC), Paseo Manuel de Lardizábal. 20018, San Sebastián, Spain}
\affiliation{IKERBASQUE, Basque Foundation for Science, 48013 Bilbao, Spain}

\author{B. Andrei Bernevig}
\email{bernevig@princeton.edu}
\affiliation{Donostia International Physics Center (DIPC), Paseo Manuel de Lardizábal. 20018, San Sebastián, Spain}
\affiliation{Department of Physics, Princeton University, Princeton, NJ 08544, USA}
\affiliation{IKERBASQUE, Basque Foundation for Science, 48013 Bilbao, Spain}

\begin{abstract}
Kagome materials manifest rich physical properties due to the emergence of abundant electronic phases. 
Here, we carry out a high-throughput first-principles study of the kagome 1:6:6 family \ch{MT6Z6} materials in space group 191, focusing on their phonon instability and electronic flat bands. 
Different \ch{MT6Z6} kagome candidates reveal a remarkable variety of kagome flat bands ranging from unfilled, partially filled, to fully filled. Notably, the Mn/Fe-166 compounds exhibit partially filled flat bands with a pronounced sharp peak in the density of states near the Fermi level, leading to magnetic orders that polarize the bands and stabilize the otherwise unstable phonon. 
When the flat bands are located away from the Fermi level, we find a large number of phonon instabilities, which can be classified into three types, based on the phonon dispersion and vibrational modes. 
Type-I instabilities involve the in-plane distortion of kagome nets, while type-II and type-III present out-of-plane distortion of trigonal M and Z atoms. We take \ch{MgNi6Ge6} and \ch{HfNi6In6} as examples to illustrate the possible CDW structures derived from the emergent type-I and type-II instabilities. The type-I instability in \ch{MgNi6Ge6} suggests a nematic phase transition, governed by the local twisting of kagome nets. The type-II instability in \ch{HfNi6In6} may result in a hexagonal-to-orthorhombic transition, offering insight into the formation of \ch{MT6Z6} in other space groups. Additionally, the predicted \ch{ScNb6Sn6} is analyzed as an example of the type-III instability.
Our predictions suggest a vast kagome family with rich properties induced by the flat bands, possible CDW transitions, and their interplay with magnetism.
\end{abstract}

\maketitle

\section{\label{sec:intro} Introduction}

Kagome materials, composed of corner-sharing triangles, provide a unique opportunity to investigate nontrivial topology and correlation effects~\cite{mielke_ferromagnetic_1991,obrien_strongly_2010,norman_colloquium_2016}. These materials have been shown to exhibit a range of fascinating physical properties, including Dirac/Weyl semimetals~\cite{ye_massive_2018,kang_dirac_2020,TbMnSn2020}, charge density waves (CDWs)~\cite{kiesel_unconventional_2013,wang_competing_2013,christensen_theory_2021,jiang_unconventional_2021,tan_charge_2021,li_observation_2021,subedi_hexagonal--base-centered-orthorhombic_2022}, quantum spin liquids~\cite{yan_spin-liquid_2011,han_fractionalized_2012,fu_evidence_2015,norman_colloquium_2016}, and unconventional superconductors~\cite{kohn_new_1965,ko_doped_2009,chen_roton_2021,ortiz_superconductivity_2021,wu_nature_2021}. The underlying physical mechanisms behind these properties can be linked to the kagome lattices, which exhibit a variety of intriguing electronic states at different fillings~\cite{yu_chiral_2012,wang_competing_2013,kang_dirac_2020,kang_topological_2020,liu_orbital-selective_2020,cho_emergence_2021,li_discovery_2022,kang_twofold_2022,kang_charge_2023,pokharel_frustrated_2023}, including flat bands, van Hove singularities (vHSs), and Dirac nodes. The flat bands and van Hove singularities show distinct density of states (DOS) and electron susceptibility, indicating an enhanced electron-electron correlation and the potential for Stoner flat band/Fermi surface instability~\cite{obrien_strongly_2010,zhao_electronic_2021,zhao_cascade_2021,ye_structural_2022,guo_correlated_2023}. 

The recent kagome materials 1:3:5 class \ch{AV3Sb5} ($A$=K, Rb, and Cs), with a perfect V kagome lattice exhibit competing charge orders and superconductivity.~\cite{ortiz_new_2019,ortiz_superconductivity_2021,liang_three-dimensional_2021,wang_competition_2021,yu_concurrence_2021,jiang_unconventional_2021,yu_unusual_2021,chen_double_2021,wu_nature_2021,kang_twofold_2022,zheng_emergent_2022,zhu_double-dome_2022}. Superconducting transitions are reported from 0.92 K to 2.5 K and occur at lower temperature than the charge density wave (CDW) transitions with $T_{CDW}$ ranging from 78 to 102 K~\cite{ortiz_csv3sb5_2020,ortiz_superconductivity_2021,yin_superconductivity_2021,li_observation_2021,xu_multiband_2021,chen_roton_2021,yin_topological_2022,jiang_kagome_2023}. 
Interestingly, a giant anomalous Hall effect was observed below $T_{CDW}$, denoting the possible time-reversal-symmetry breaking, which still remains controversial~\cite{yang_giant_2020,yu_concurrence_2021,mielke_time-reversal_2022,hu_time-reversal_2023, xing2023optical}. %

Recently, a CDW was also observed in kagome magnet FeGe below $T_{CDW}=100K$ with wavevectors identical to that of \ch{AV3Sb5}~\cite{teng_discovery_2022,yin_discovery_2022,setty_electron_2022,zhao_photoemission_2023,wu_annealing_2023,chen_charge_2023,shi_disordered_2023,wu_symmetry_2023}. While the kagome Fe lattice with partially filled $3d$ orbitals exhibits flat bands close to the Fermi level in the paramagnetic phase, the strong correlation first induces magnetic order and leads to the collinear A-type antiferromagnetic order below $T_{N}=410K$~\cite{ohoyama_new_1963, haggstrom_mossbauer_1975,forsyth_low-temperature_1978,bernhard_neutron_1984,wu_symmetry_2023,teng_magnetism_2023,wang_enhanced_2023,ma_theory_2023,chen_competing_2023}. In contrast to the usual CDW materials which exhibit imaginary modes in their pristine phase, FeGe in the AFM phase shows only positive modes, indicating a  distinct origin for the CDW~\cite{miao_charge_2022, JiangYi_kagomeI_2023, wu_novel_2023, zhou_magnetic_2023}.

Another interesting kagome family is the 1:6:6 class \ch{MT6Z6} (M = metallic elements and rare-earth [RE] elements; $T$ = transition metals; Z = main group elements)~\cite{CoFeNiGe1981,venturini_filling_2006,fredrickson_origins_2008,VENTURINI2008,baranov_magnetism_2011, ghimire2020competing}, which hosts two kagome layers of transition metals in the unit cell. Recently, CDW transitions have also been observed in \ch{ScV6Sn6} with two V kagome layers, presenting a competition between two phases at the CDW wavevector $\bar{K} = \frac{1}{3}\frac{1}{3}\frac{1}{3}$ and a different softening wavevector $H=\frac{1}{3}\frac{1}{3}\frac{1}{2}$~\cite{ScVSn2022,hu_kagome_2023,cao_competing_2023,zhang_destabilization_2022,cheng_nanoscale_2023,pokharel_frustrated_2023,destefano_pseudogap_2023}. 
{Anomalous Hall effect %
is also reported in the CDW phase, suggesting the possible breaking of time-reversal-symmetry in the kagome material}~\cite{guguchia_hidden_2023,yi_charge_2023,mozaffari_universal_2023}. The phonon softening in \ch{ScV6Sn6} is mainly driven by the electron-phonon coupling, and can be well captured by the the out-of-plane vibrations of weakly coupled 1D chains formed by the trigonal Sn atoms ~\cite{korshunov_softening_2023,hu_kagome_2023,cao_competing_2023,tan_abundant_2023,lee_nature_2023,hu_phonon_2023,gu_phonon_2023}. 

In this manuscript, we report on our high-throughput first-principle study of electronic band structures and phonon spectrum of  293 materials in the 1:6:6 family \ch{MT6Z6} of hexagonal \ch{HfFe6Ge6}-type structures in space group (SG) 191, revealing a wide range of possible CDW and nematic transitions, as well as correlations between these and the position of the flat bands relative to the Fermi level in the electronic band structure. 
We start with an introduction to the crystal structure of \ch{MT6Z6}, which can be understood as the intercalation of M atoms in 1:1 family \ch{TZ} parent structures. The electronic band structures of the 1:6:6 family show similar kagome band features with varying fillings, which are primarily determined by the valence of $d$ electrons of T atoms, and are further changed by different Z and M atoms. When the flat bands are close to half-filling, the compounds present sharp DOS peaks close to the Fermi level, leading to instability and magnetic orders. When the flat bands are away from the Fermi level $E_f$, vHSs could appear near $E_f$. A large number of phonon instabilities are found and categorized into three types: type-I is contributed by the in-plane displacements from kagome T atoms, and type-II and -III are mainly from the out-of-plane displacements of trigonal M-Z-Z-M chains. {We use \ch{MgNi6Ge6}, \ch{HfNi6In6}, and \ch{ScNb6Sn6} to demonstrate the properties of type-I, II and III instabilities, respectively.} 

\section{\label{sec:struct} Crystal structures of 1:6:6 family}

\begin{figure}[t]
\includegraphics[width=0.9\linewidth]{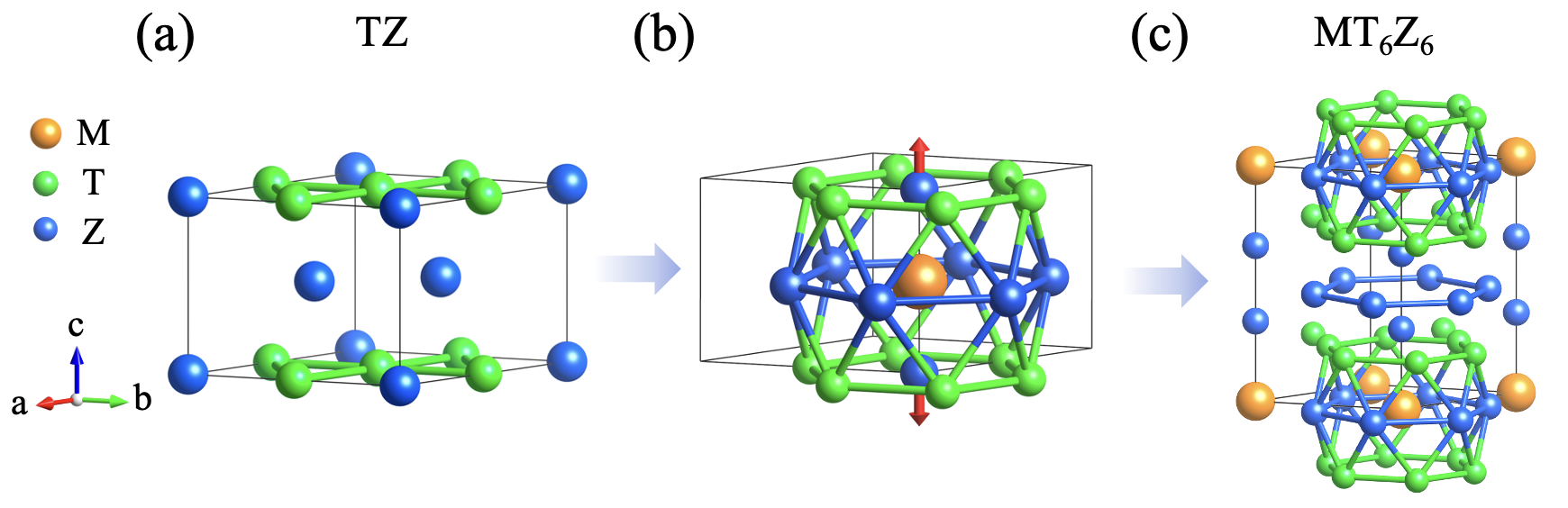}
\caption{Crystal structures of kagome 1:1 family (CoSn-type) TZ and 1:6:6 family \ch{MT6Z6} in space group 191 $P6/mmm$. (a) Kagome 1:1 family \ch{TZ} structure consists of two layers, one is the hexagonal Z atoms (blue), and the other is the kagome T atoms (green) and trigonal Z atoms at the same plane (blue). (b) The intercalation of M atoms (brown) in the CoSn-type structure drives trigonal Z atoms away from the kagome plane, resulting in a doubled structure along the hexagonal $c$ direction to form the \ch{MT6Z6} structure in (c). 
}
\label{fig:struct}
\end{figure}
We start from the crystal structure of the 1:6:6 family \ch{MT6Z6}. The 1:6:6 family can be understood as the intercalation of M atoms in the 1:1 \ch{TZ} structures. %
As shown in Fig.~\ref{fig:struct}(a), \ch{TZ} materials are formed by the simple alternating stacking of the kagome/trigonal layer and honeycomb layer, with representative 0:3:3 materials FeGe, FeSn, and CoSn. In the \ch{TZ} materials, the trigonal Z1 lattice is at Wyckoff position $1a=(0,0,0)$, the honeycomb Z2 lattice at $2d=(\frac{1}{3},\frac{2}{3},\frac{1}{2}), (\frac{2}{3},\frac{1}{3},\frac{1}{2})$, and the kagome T lattice at $3f=(\frac{1}{2},0,0), (\frac{1}{2},\frac{1}{2},0), (0, \frac{1}{2},0)$ site. The coordinates are written under the hexagonal bases $\bm{a}_1=(a,0,0), \bm{a}_2=(-\frac{a}{2},\frac{\sqrt{3}a}{2},0), \bm{a}_3=(0,0,c)$, with $a,c$ being the lattice constants. The two inequivalent Z sublattices form a large vacancy available for M atoms. The intercalation of M atoms drives the trigonal Z1 atoms away from the kagome plane, forming a compact M-Z chain through the hexagonal tunnel of the kagome/honeycomb lattice. 
The varying arrangements of M atoms in the M-Z chains, surrounded by perfect or distorted kagome layers, allow for the existence of 1:6:6 family across a range of space groups, including $P6/mmm$, $Immm$, and $Cmcm$~\cite{venturini_filling_2006,fredrickson_origins_2008,baranov_magnetism_2011}, encompassing small unit cells to large superstructures. Representative structures are illustrated in Supplementary Note II~\cite{supp}.
This study focuses on the high-symmetry 1:6:6 structures crystallizing in SG 191 $P6/mmm$, with an enlarged unit cell doubled along the $c$-direction compared to TZ structures, as presented in Fig.~\ref{fig:struct}(c).
\begin{table*}[htbp]
\global\long\def\arraystretch{1.12}
\caption{Collection of phonon stability with irreps for \ch{MT6Z6} kagome family, obtained by non-spin-polarized calculations without SOC. The first column denotes the $TZ$ ingredient, and the second column denotes the results for the CoSn-type $TZ$ structures, while the columns present the results for \ch{MT6Z6}. IRREPs denote the irreducible representations of the lowest imaginary phonon at high-symmetry points. "S" denotes the stable phonon at low-T. "-" are not included in this survey. "*" denote that they have been experimentally observed in either \ch{HfFe6Ge6}- or \ch{YCo6Ge6}-type hexagonal structure. Here, the finite temperature is considered at the harmonic level via the smearing approach. High-T and low-T refer to a Fermi-Dirac smearing of 0.4 eV and 0.05 eV, respectively. Red (blue) color denotes the unstable (stable) phonon at high-T. Note that in NiIn-166 the lowest modes obtained from $2\times2\times2$ supercells could be located at generic $\bm{q}$ instead of high-symmetry points, such as \ch{HfNi6In6} and \ch{LuNi6In6}. {Experimentally, instabilities have been reported in \ch{ScV6Sn6}~\cite{ScVSn2022}, \ch{MgCo6Ge6}~\cite{MgCoGe2021}, and \ch{YbCo6Ge6}~\cite{YbCoGe2022}.} The specific phonon spectra and notations are described in the Supplementary Note IV~\cite{supp}.}
\label{Tab:phonon}
\setlength{\tabcolsep}{0.6mm}{
\begin{tabular}{c|c|c|c|c|c|c|c|c|c|c|c|c|c|c|c|c|c|c|c|c|c|c|c}
\hline
\hline
 & TZ & Li & Mg & Ca & Sc & Y & Lu & Ti & Zr & Hf & Nb & Ta & Pr & Nd & Sm & Gd & Tb & Dy & Ho & Er & Tm & Yb & U  \\
\hline
VGe & S & S & {\color{red}${H_4}$} & {\color{blue}${H_4}$} & S & S & S & S & S & S & S & S & S & S & S & S & S & S & S & S & S & S & -  \\ 
VSn & S & S & {\color{red}${H_4}$} & {\color{blue}${H_4}$} & {\color{blue}${H_4}$$^*$} & S$^*$ & S$^*$ & S & S & S & S & S & S & S & S$^*$ & S$^*$ & S$^*$ & S$^*$ & S$^*$ & S$^*$ & S$^*$ & {\color{blue}${H_1}$$^*$} & -  \\ 
NbSn & S & {\color{red}${A_5^-}$} & {\color{red}${H_4}$} & {\color{blue}${H_4}$} & {\color{blue}${H_4}$} & S$^*$ & -  & S & S & S & -  & -  & S & S & S & S & S$^*$ & S & {\color{blue}${H_4}$} & {\color{blue}${H_4}$} & {\color{blue}${H_4}$} & {\color{blue}${H_4}$} & -  \\ 
CrGe & S & S & S & S & S$^*$ & S$^*$ & S & S & S & S & S & S & S & S & S & S$^*$ & S$^*$ & S$^*$ & S$^*$ & S$^*$ & S$^*$ & S$^*$ & -  \\ 
MnGe & {\color{blue}${\Gamma_5^-}$} & {\color{blue}${\Gamma_5^-}$} & {\color{blue}${\Gamma_5^-}$$^*$} & {\color{blue}${\Gamma_5^-}$} & S$^*$ & S$^*$ & S$^*$ & S & S$^*$ & S$^*$ & S & S & S & S$^*$ & S$^*$ & S$^*$ & S$^*$ & S$^*$ & S$^*$ & S$^*$ & S$^*$ & {\color{blue}${\Gamma_5^-}$$^*$} & -  \\
\hline
\hline
\end{tabular}}
\end{table*} 

\begin{figure}[htbp]
\includegraphics[width=0.45\textwidth]{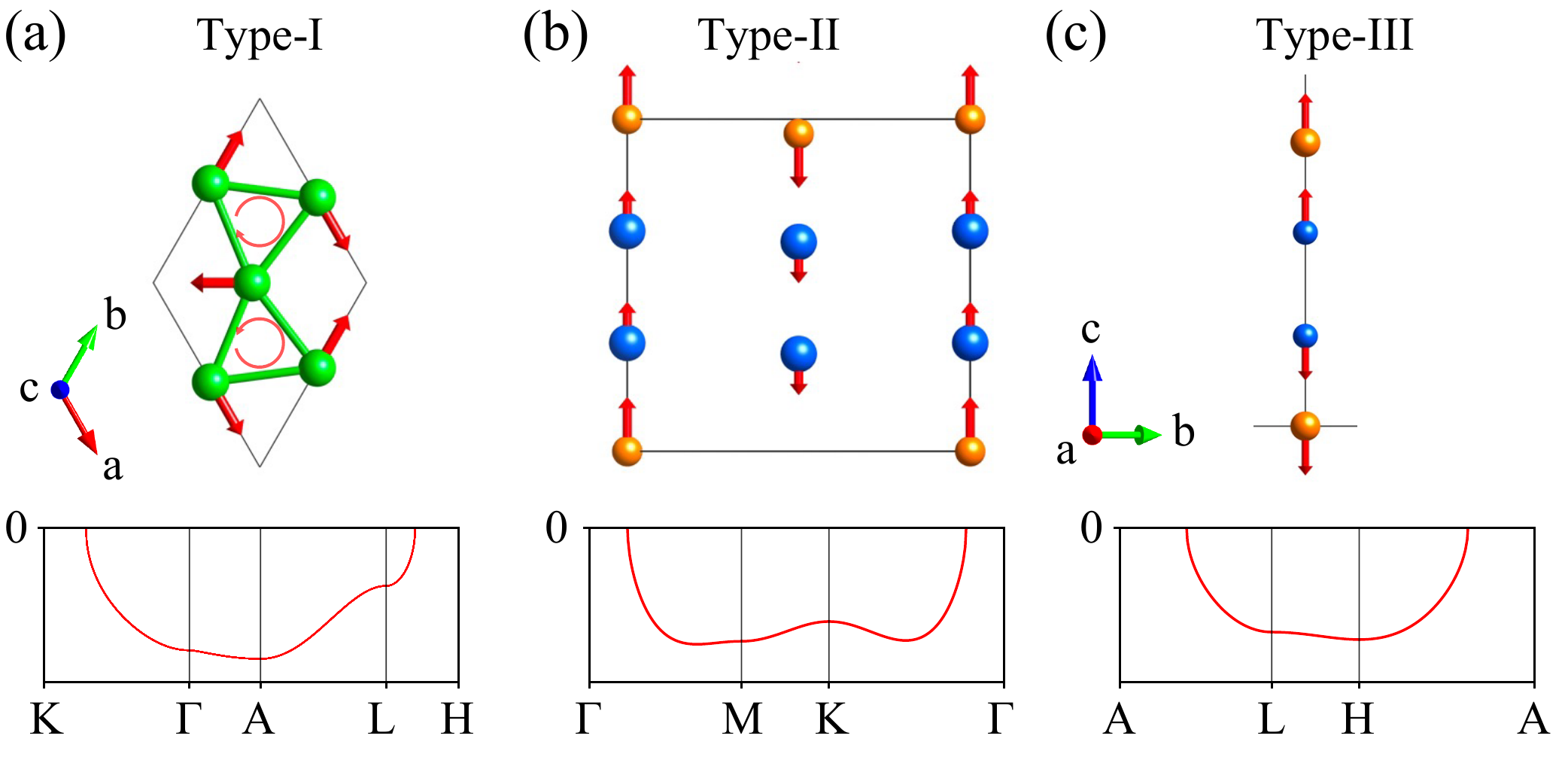}
\caption{Dominant vibration modes with corresponding imaginary instabilities in \ch{MT6Z6} compounds. (a) Type-I instabilities occurring along the $\Gamma-A$ path are primarily characterized by in-plane vibrations of the kagome T atoms, which occur in CoGe-166 and NiSi/Ge-166. The top view of distorted kagome nets shows the twisting of kagome T nets, resulting in the relative rotations of small triangles. (b) Type-II instabilities, which occur on the $k_z=0$ plane, are dominated by out-of-plane vibrations in the M-Z chains in the case of NiIn-166. The displacements illustrate the out-of-plane movements of M and Z atoms in the same direction within the M-Z chain but in the opposite direction in neighbor chains. (c) Type -III instabilities, observed on the $k_z=\pi$ plane, dominated by out-of-plane vibrations in the $M-Z$ chains at the edges of VGe/Sn-166 and NbSn-166. {One notable example of such a material is \ch{ScV6Sn6}, which exhibits competing charge orders at lower temperatures~\cite{ScVSn2022,korshunov_softening_2023,cao_competing_2023}.} 
}
\label{fig:modes}
\end{figure}

\section{\label{sec:result} High-throughput computation results} 

In this section, we introduce the main findings of the high-throughput calculations of the electronic and phonon properties of the 1:6:6 family. Our results reveal a wide range of phonon instabilities in \ch{MT6Z6} family as presented in Sec.~\ref{ssec:phonon}. Their different electronic properties are then discussed in Sec.~\ref{ssec:flatband}. In the 1:6:6 family, the band structures at the Fermi level are mainly determined by the $d$ orbitals of T atoms at kagome sites, which can be derived from their TZ parent structures. When the kagome flat bands are at the Fermi level and partially filled, e.g. in the paramagnetic (PM) phase of T=Mn/Fe 1:6:6 classes, the flat bands exhibit a sharp DOS peak near the Fermi level, which gives rise to magnetic instabilities. The magnetic order splits the sharp peak into two peaks below and above the Fermi level, thereby stabilizing the soft phonon that exists in these systems in the PM phase. In Sec.~\ref{ssec:mag}, we discuss in detail the magnetic orders and their effects on the electron and phonon spectrum in the Mn/Fe 1:6:6 family.

When the flat bands are situated at a distance from the Fermi level, vHSs could appear in proximity to the Fermi level, which may account for phonon instability and CDW transitions. Based on the dispersion of soft phonons and their vibration modes, we further categorize these instabilities into three different types, as shown in Fig.~\ref{fig:modes}. 
Type-I instabilities occur along the $\Gamma-A$ line, governed by the in-plane vibrations of the kagome lattice, which breaks $C_6/C_2$ symmetry. 
Type-II instabilities present a nearly flat imaginary phonon on the $k_z=0$ plane, which suggests the atomic movement in the same direction within M-Z chains. The arrangement of M-Z chains of opposite directions then results in the breaking of rotational symmetry. 
Type-III instabilities present nearly flat imaginary phonons on the $k_z=\pi$ plane, characterized by the atomic movement in opposite directions within one M-Z chain. %
{In the following section, taking \ch{MgNi6Ge6}, \ch{HfNi6In6} and \ch{ScNb6Sn6} as examples,}
we characterize each instability by the representations of the soft phonon modes and show that a clear pattern emerges.

\subsection{\label{ssec:phonon} Classification of CDW instabilities}

We now summarize the phonon stability in the 1:6:6 family. Among the 293 \ch{MT6Z6} structures, 166 of them are found to be stable, and 103 of them only present instabilities at low-T, while the remaining 24 present instabilities at both low-T ($0.05$ eV) and high-T ($0.4$ eV) (T represents here the smearing, and not necessarily the instability, temperature).
To further characterize the instabilities, we analyze the irreducible representations (IRREPs) at high-symmetry points and project their atomic weights as described in Supplementary Note IV~\cite{supp}. The IRREPs follow the convention of \textit{Bilbao Crystallographic Server}\cite{aroyo2006bilbao1, aroyo2006bilbao2}. 
A complete collection of the instabilities in the 1:6:6 family is presented in Tab.~\ref{Tab:phonon}, marked by the IRREPs of the leading instability at high-symmetry points. %
From the results, we classify these instabilities into three different types, as shown in Fig.~\ref{fig:modes}. A small group of materials with Type-I or III instability has been observed in experiments\cite{ScVSn2022, MgCoGe2021, YbCoGe2022}. 
We provide their theoretical understanding and enumerate new materials that can also realize these two types of CDWs. On the other hand, type-II is a new class of CDW that we here predict. In the following, we first introduce the distinct features of these three types of instabilities and then give more detailed investigations on type-I and II with realistic materials as examples. 

The three types of phonon instabilities have the following main features:
\begin{itemize}
    \item Type-I has leading instabilities along the $\Gamma-A$ line, dominated by the in-plane displacements of kagome nets, which exist in CoGe-166, NiSi-166, and NiGe-166 families. Though the modes at M and L points may also go soft, the soft modes at $\Gamma$ or A point are at much lower energy. The displacement patterns reveal the breaking of rotational symmetry $C_6/C_2$ in the kagome plane. The soft modes at the $A$ points also break the translational symmetry, leading to a $1\times1\times2$ order doubled along $c$ direction. Instead, the soft modes at $\Gamma$ modes do not break translational symmetry, suggesting a nematic phase transition. Type-I instability has been reported in CoGe-166, including \ch{MgCo6Ge6}~\cite{MgCoGe2021} and \ch{YbCo6Ge6}~\cite{YbCoGe2022}, where in-plane kagome displacement patterns and a cell doubling along the $c$ direction are observed, suggesting the distortion derived from the soft mode at $A$. 

    \item Type-II presents nearly flat instabilities on the $k_z=0$ plane, which occurs in many compounds within the NiIn-166 family, including \ch{HfNi6In6} and \ch{LuNi6In6}. Different from the reported type-III in \ch{ScV6Sn6} where the trigonal sites move in opposite directions within the M-Z chain, the type-II instability suggests the trigonal sites move in the same direction. The leading instability could be located at generic $\bm{q}$ points, rather than high-symmetry points, suggesting the possible incommensurability. The displacement pattern of M-Z chains derived from a single M mode reveals a hexagonal-to-orthorhombic transition, which provides insight into the formation of 1:6:6 compounds in other space groups~\cite{venturini_filling_2006}. To date, no type-II CDW transitions have been observed in experiments yet.  

    \item Type-III contains the nearly flat imaginary optical phonon on the $k_z=\pi$ plane, with a leading imaginary mode at $H$ point of IRREP $H_4$, which exists in the VGe/VSn/NbSn-166 families. One representative material is \ch{ScV6Sn6}~\cite{ScVSn2022}, which has recently gained significant attention. 
    {Most of the other compounds with type-III instability are predicted, and further experimental verification is required.} 
\end{itemize}
The instabilities in Mn/Fe-166 compounds in the PM phase are attributed to the partial filling of flat bands near the Fermi level and will be stabilized after the inclusion of magnetic orders, which are discussed in Sec.~\ref{ssec:mag}. 
To elaborate on the type-I and type-II instabilities, we choose Ni-based 1:6:6 as the representatives, as both instabilities are presented. 
For the type-III instability, the predicted \ch{ScNb6Sn6} is introduced, instead of the widely discussed \ch{ScV6Sn6}. %

In the Ni-based 1:6:6 family \ch{MNi6Z6}, we consider group III-A element In, and group IVA elements Si/Ge/Sn at Z sites. In the four 1:1 family building blocks \ch{NiZ} (Z=In, Si, Ge, and Sn), only NiIn has been experimentally reported and presents stable phonons here. The other three \ch{NiZ} present instabilities at both high-T and low-T, with the same leading IRREP $A_2^-$ as listed in Tab.~\ref{Tab:phonon}. With the insertion of M elements, NiIn-166 exhibits a nearly flat imaginary phonon on the $k_z=0$ plane while NiSi/Ge-166 presents leading instabilities along the $\Gamma-A$ path. One exception is the stable NiSn-166 class, which presents no instabilities.

\begin{figure*}[htbp]
\includegraphics[width=0.8\textwidth]{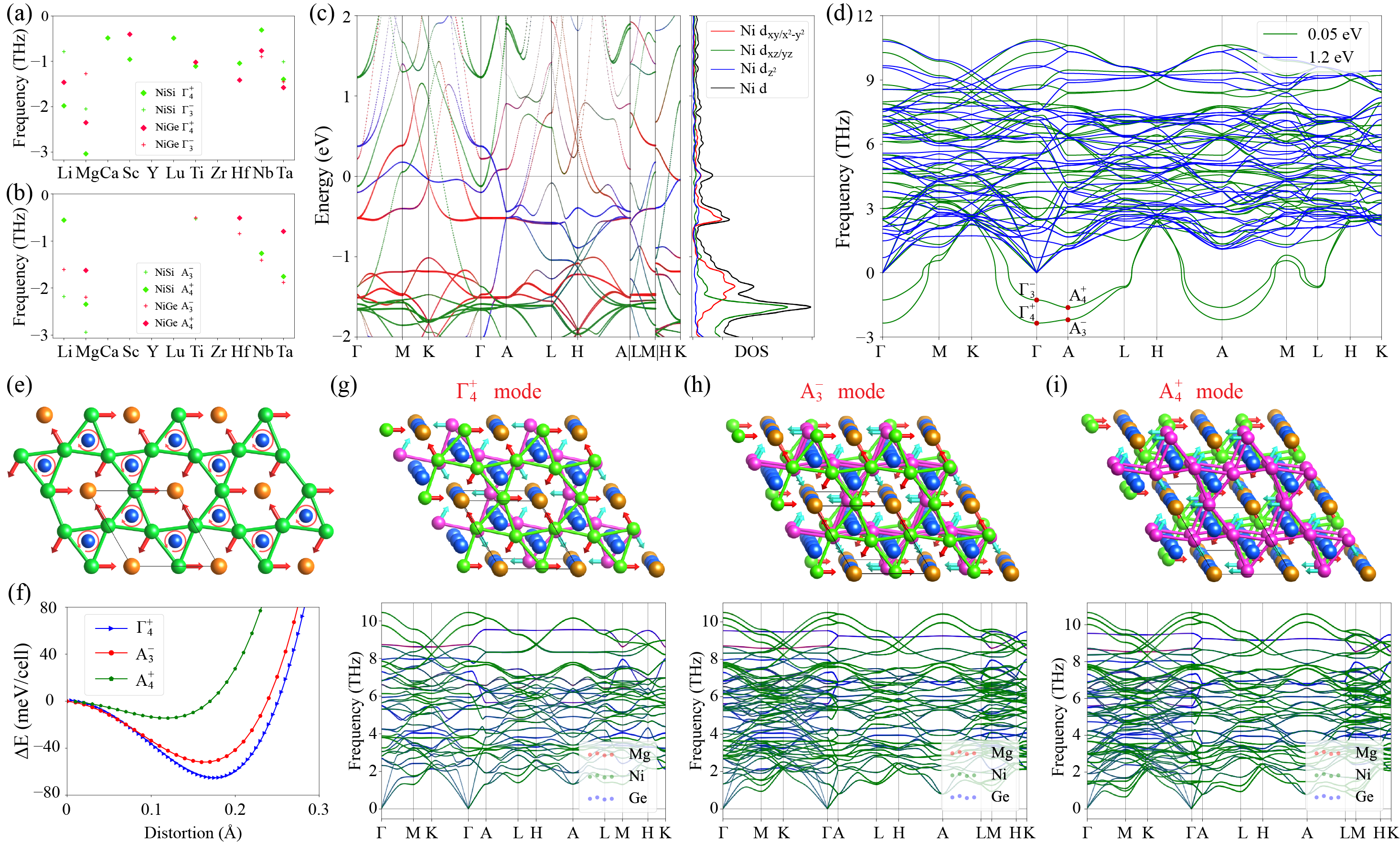}
\caption{Type-I phonon instabilities and possible CDWs in NiSi/Ge-166. (a-b) Representations of imaginary modes at high symmetry points, $\Gamma$ and A, in NiSi/Ge-166 classes. The NiSi/Ge-166 compounds share similar representations, while their position on the frequency scale may vary. The leading imaginary modes are located at either $\Gamma$ or $A$ points, similar to those observed in CoGe-166. (c-d) Band structure and phonon spectrum of \ch{MgNi6Ge6}. The electronic band structure, weighted by Ni $3d$ orbitals, reveals vHS from $d_{xz/yz}$ orbital. Flat band segments composed of $d_{xy/x^2-y^2}$ and $d_{z^2}$ orbitals are observed along $\Gamma-K$ path close to the Fermi level, accompanied by a peaked DOS at the Fermi level. The phonon calculation at low temperature reveals two soft branches along the $\Gamma-A$ path. Atomic projections suggest the imaginary branches are governed by kagome Ni atoms, which can be reproduced from motions in the bilayer kagome lattice. A stable phonon with a smearing value of $1.2$ eV is also demonstrated. (e) Displacement pattern of Ni atoms in single kagome layer, indicating a breaking of $C_6/C_2$ symmetry. This displacement pattern serves as the building block for $\Gamma$ and A imaginary modes. (f) The total energy of CDW structures as a function of distortion of Ni atoms. The energy is given per pristine unit cell and the pristine phase is set to zero energy. 
The nematic phase derived from the $\Gamma_4^+$ mode exhibits the lowest energy among the CDW structures. (g-i) Possible stable CDW structures and phonon spectra derived from $\Gamma_4^+$, $A_3^-$, and $A_4^+$ modes, respectively. The green and magenta kagome layer, marked as A and B, denotes the two opposite distortions with corresponding displacements indicated by the red and cyan arrows. The CDW structures present distinct stacking of distorted kagome layers, i.e., MBAM for $\Gamma_4^+$, {MAAM for $\Gamma_3^-$}, MABMBAM for $A_4^-$, and MAAMBBM for $A_4^+$, where M denotes the M atom layer. Similar CDW structures for \ch{MgCo6Ge6} can be found in Supplementary Note IIIA~\cite{supp}. %
}
\label{Fig:pho_NiGe}
\end{figure*}

\textbf{Type-I instabilities.} For the type-I phonon instability in the NiSi/Ge-166 family, the collection of their imaginary frequency and IRREPs at low-T is shown in Fig.~\ref{Fig:pho_NiGe}(a)-(b). The symmetry analysis reveals similar IRREPs for their leading instabilities governed by kagome nets. This similarity suggests that they undergo similar distortions. Meanwhile, the existence of two equivalent kagome layers in one unit cell also contributes to two nearly degenerate kagome phonon bands, which are also prominent in CoGe-166, as presented in the Supplementary Note IIIA. 
The weak interactions between kagome lattices and other atoms enable the two-dimensional nature of kagome lattices, namely, the flatness along the $\Gamma-A$ path. %
While Si/Ge/Sn are isovalent, their atomic radii increase from Si to Ge and Sn, leading to the increase of the lattice parameters in Ge and Sn compared to Si. 
Compared to the phonon instabilities in CoGe-166 only at low-T, the instabilities in NiSi/Ge-166 are presented at both low-T and high-T, which may indicate higher temperatures to stabilize the phonon.

\begin{table}[htbp]
\global\long\def\arraystretch{1.12}
\caption{Possible subgroups of SG 191 $P6/mmm$ driven by imaginary phonon modes in \ch{MgNi6Ge6}.}
\label{Tab:type_I_subgroup_irrep}
\setlength{\tabcolsep}{2.5mm}{
\begin{tabular}{c|c|c}
\hline
\hline
 $f$(THz) & IRREP & Subgroup   \\
\hline
$-2.353$ & $\Gamma_4^+$ & 162~$P\bar{3}1m$   \\
$-1.274$ & $\Gamma_3^-$ & 189~$P\bar{6}2m$ \\
$-2.195$ & $A_3^-$ & 193~$P6_3/mcm$  \\
$-1.625$ & $A_4^+$ & 193~$P6_3/mcm$  \\
\hline
\hline
\end{tabular}}
\end{table}

Taking \ch{MgNi6Ge6} as an example, from the band structure in fig.~\ref{Fig:pho_NiGe}(c), one can observe van Hove singularity contributed by $d_{xz/yz}$ orbital, and flat band segments contributed by $d_{z^2}$ and $d_{xy/x^2-y^2}$ orbitals near the Fermi level, accompanied by a weak peak in the DOS. In Fig.~\ref{Fig:pho_NiGe}(d), two phonon spectra with smearing values of 0.05 eV and 1.2 eV are displayed. At the lower temperature regime, two widespread (in $k$-space) imaginary branches appear over $\Gamma$, A, and L points, which can be stabilized with a higher smearing value, such as 1.2 eV in Fig.~\ref{Fig:pho_NiGe}(d).
The lower four modes are identified as $\Gamma_4^+$, $\Gamma_3^-$, $A_3^-$, and $A_4^+$ modes at $\Gamma$ and A points. According to the IRREPs, one can obtain the corresponding distortion pattern and possible subgroups of the resulting phase, as listed in Tab.~\ref{Tab:type_I_subgroup_irrep}.  %

While the $\Gamma$ modes preserve translation symmetry with broken discrete rotational symmetry, the $A$ modes would break translation symmetry, resulting in a doubled unit cell along the $c$ direction. %
Atomic projections indicate that all the imaginary phonon modes are dominated by the in-plane vibrations of the kagome Ni atoms. The displacements of these four leading modes also show similar patterns in the kagome layers. 
The typical displacement pattern in a single kagome layer is shown in Fig.~\ref{Fig:pho_NiGe}(e). 
From the top view, the triangles surrounding the honeycomb Ge atoms (blue) are rotated in either a clockwise or counterclockwise direction. The hexagons of six Ni atoms (green) atoms surrounding Mg {(orange) and trigonal Ge (superimposed by Mg)} atoms are distorted, leading to three close and three distant Ni atoms, indicating the $C_6/C_2$ symmetry breaking. %
With $C_3$ symmetry retained, the Dirac points at $K$ are preserved.

Following the displacements, the total energy as a function of distortion of the Ni atoms can be obtained as shown in Fig~\ref{Fig:pho_NiGe}(f). Here, three stable CDW structures are obtained for $\Gamma_4^+$, $A_3^-$, and $A_4^+$ IRREPs, which are 65 meV/cell, 51 meV/cell, and 14 meV/cell lower than the pristine structure, respectively %
. For the $\Gamma_4^+$ mode, the neighboring kagome layers present the same displacement but in an opposite phase with inversion symmetry preserved, which can be denoted as MBAM stacking as shown in Fig.~\ref{Fig:pho_NiGe}(g) (we use M to denote the layer formed by the M atoms, and A, B the two kagome layers of opposite displacements.). The structures derived from the $A_i$-mode are doubled along the $c$ direction with different phase arrangements, which can be referred to as MABMBAM structures for $A_3^-$ and MAAMBBM structures for $A_4^+$, as shown in Fig.~\ref{Fig:pho_NiGe}(h) and (i), respectively. %
Remark that although these two structures have different displacement stackings along the $c$ direction, they correspond to the same SG 193 $P6_3/mcm$ as shown in Tab.~\ref{Tab:type_I_subgroup_irrep}. %

Similar arguments can be applied to {phase transitions} in the CoGe-166 family. For example, among the four lower imaginary modes in \ch{MgCo6Ge6}, the $A_3^-$ mode presents the lowest frequency, denoting an MABMBAM distortion. Meanwhile, $A_3^-$ modes are more energetically favored as presented in the {Supplementary Note IIIA}~\cite{supp}, which agrees well with the experimental observation~\cite{MgCoGe2021}. 
{A phase transition is also reported in \ch{YbCo6Ge6} with the in-plane distortion of kagome nets. Note that structural disorder is reported within trigonal Y-Ge chains, leading to Yb$_{0.5}$Co$_3$Ge$_3$ structure (See Supplementary Note II~\cite{supp} and~\cite{YbCoGe2020,YbCoGe2022}).}
The structure is doubled along $c$ direction at a lower temperature, featured by a distinct peak near 95 K in the resistivity measurement with twisted kagome sublattices~\cite{YbCoGe2022}.

\begin{figure}[htbp]
\includegraphics[width=0.48\textwidth]{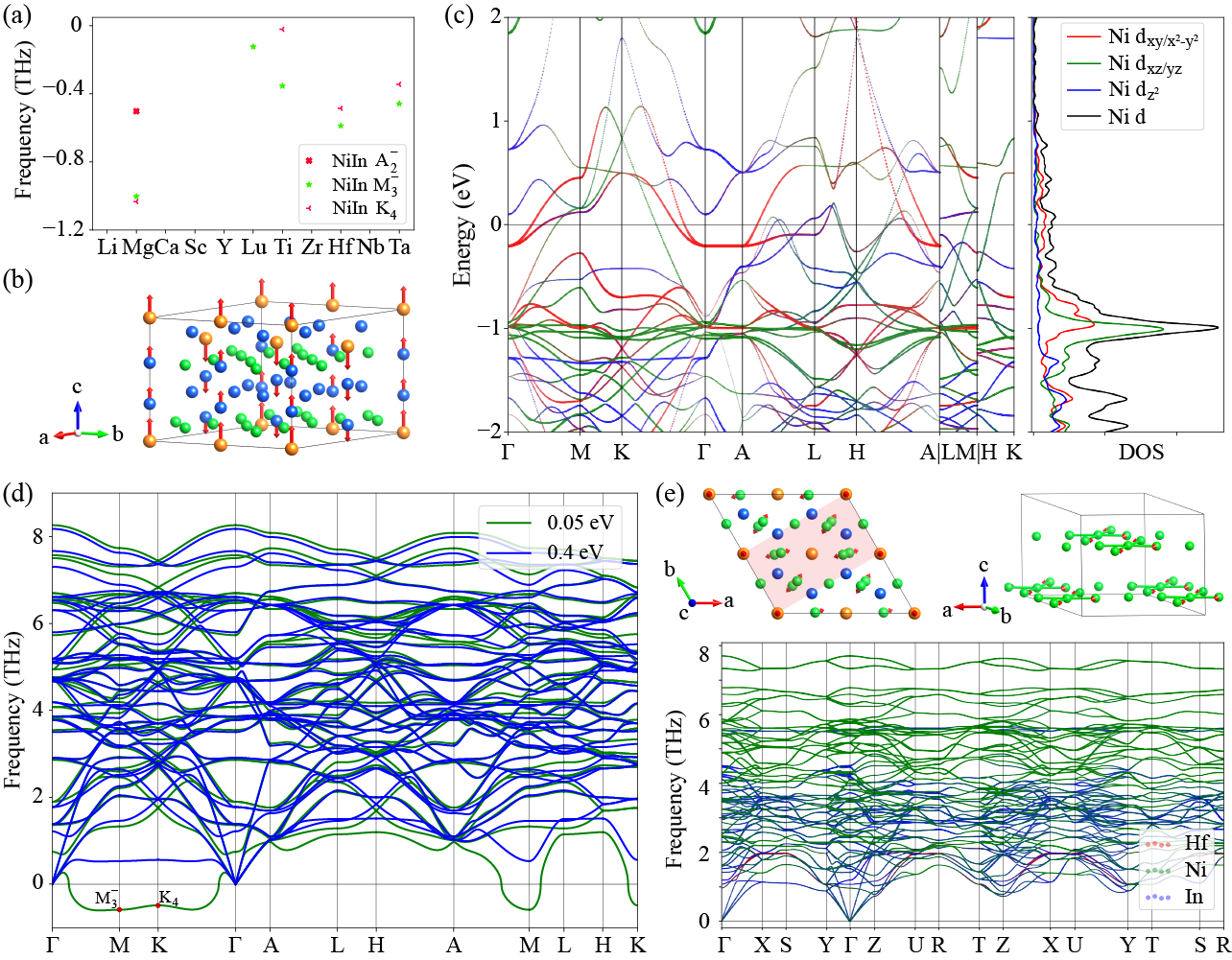}
\caption{Type-II phonon instabilities and possible CDW in NiIn-166 compounds. (a) Representations of imaginary modes at high symmetry points, M and K, in NiIn-166 compounds. Note that the leading imaginary modes in NiIn-166 compounds may be located at generic $q$ points instead of high-symmetry points. (b) Typical distorted structure derived from single M mode, illustrating the dominant out-of-plane displacements of trigonal M and Z atoms. (c-d) Band structure and phonon spectrum of \ch{HfNi6In6}. The band projection of Ni $3d$ orbital reveals flat band segments composed of $d_{xy/x^2-y^2}$ and $d_{z^2}$ orbitals near the Fermi level, accompanied by a moderate DOS peak. The phonon spectra are obtained using a $2\times2\times2$ supercell at smearing values of $0.05$ eV and $0.4$ eV. The low-T phonon shows the lowest imaginary mode at generic $\bm{q}$ rather than high-symmetry points, suggesting the possibility of incommensurability. (e) Possible stable commensurate CDW structure and corresponding stable phonon spectrum derived from a single M mode. The shaded area represents the unit cell of distorted CDW unit cell in SG $Pmmn$ (No. 59). In addition to the out-of-plane displacements of trigonal Hf and In atoms, in-plane distortion of kagome Ni nets is also presented.%
}
\label{Fig:pho_NiIn}
\end{figure}

\textbf{Type-II instabilities.} For the type-II instability, nearly flat imaginary phonon modes are found on the $k_z=0$ plane in the predicted NiIn-166 family, {such as \ch{HfNi6In6} in Fig.~\ref{Fig:pho_NiIn}.} 
The IRREPs and corresponding frequencies at the M and $K$ points are collected in Fig.~\ref{Fig:pho_NiIn}(a). Generally, the M point shows a lower frequency compared to the $K$ point. A typical displacement pattern derived from a single M mode is shown in Fig.~\ref{Fig:pho_NiIn}(b). The atoms within each M-Z chain exhibit coordinated motion. However, the motion alternates along the $\bm{a}$ direction while remaining uniform along the $\bm{b}$ direction %
, indicating the breaking of the $C_3$ symmetry. 

Taking \ch{HfNi6In6} as an example, its electronic band structure and phonon spectrum are presented in Fig.~\ref{Fig:pho_NiIn}(c-d). There are vHSs and flat band segments close to the Fermi level, contributed by Ni $3d$ orbitals. The phonon spectrum shows flat band instabilities on the $k_z=0$ plane at low-T, with the lowest frequency at a generic $\bm{q}$, but very close in energy to the frequency at the high-symmetry M or $K$ points, as shown in Fig.~\eqref{Fig:pho_NiIn}(d). The projections suggest that the imaginary modes are dominated by the out-of-plane vibrations of trigonal Hf and In atoms, accompanied by the distortions of kagome nets. %
Note that in \ch{ScV6Sn6} the imaginary mode {at the $k_z=\pi$ plane} shows the opposite motions of Sc and Sn atoms within each Sc-Sn chain, {corresponding to a mirror even phonon mode}. In contrast, the Hf and In atoms in the Hf-In chain exhibit movements in the same direction, {representing a mirror odd feature.} %
Fig.~\eqref{Fig:pho_NiIn}(b) illustrates a typical distorted structure obtained from a single M mode $M_3^-$ at $(\frac{1}{2},0,0)$, which exhibits alternating out-of-plane motions of Hf-In chains along $\bm{a}$ direction. Derived from the single M mode, a stable phonon of CDW structure in SG $Pmmn$ (No. 59) is obtained as shown in Fig.~\eqref{Fig:pho_NiIn}(e). 
Besides the stable structure from a single M mode, it is also possible to obtain CDW structures from the combination of multiple $\bm{q}$ modes, which requires further investigation. 
It is worth noting that, in contrast to the slight shift between the M-Z chains observed in this case, 
the 1:6:6 compounds in SG $Immm$ (No. 71) display a similar but half-hexagonal-$c$ shift in the adjacent M-Z chain, surrounded by the distorted kagome lattice. %

\begin{figure}[htbp]
\includegraphics[width=0.48\textwidth]{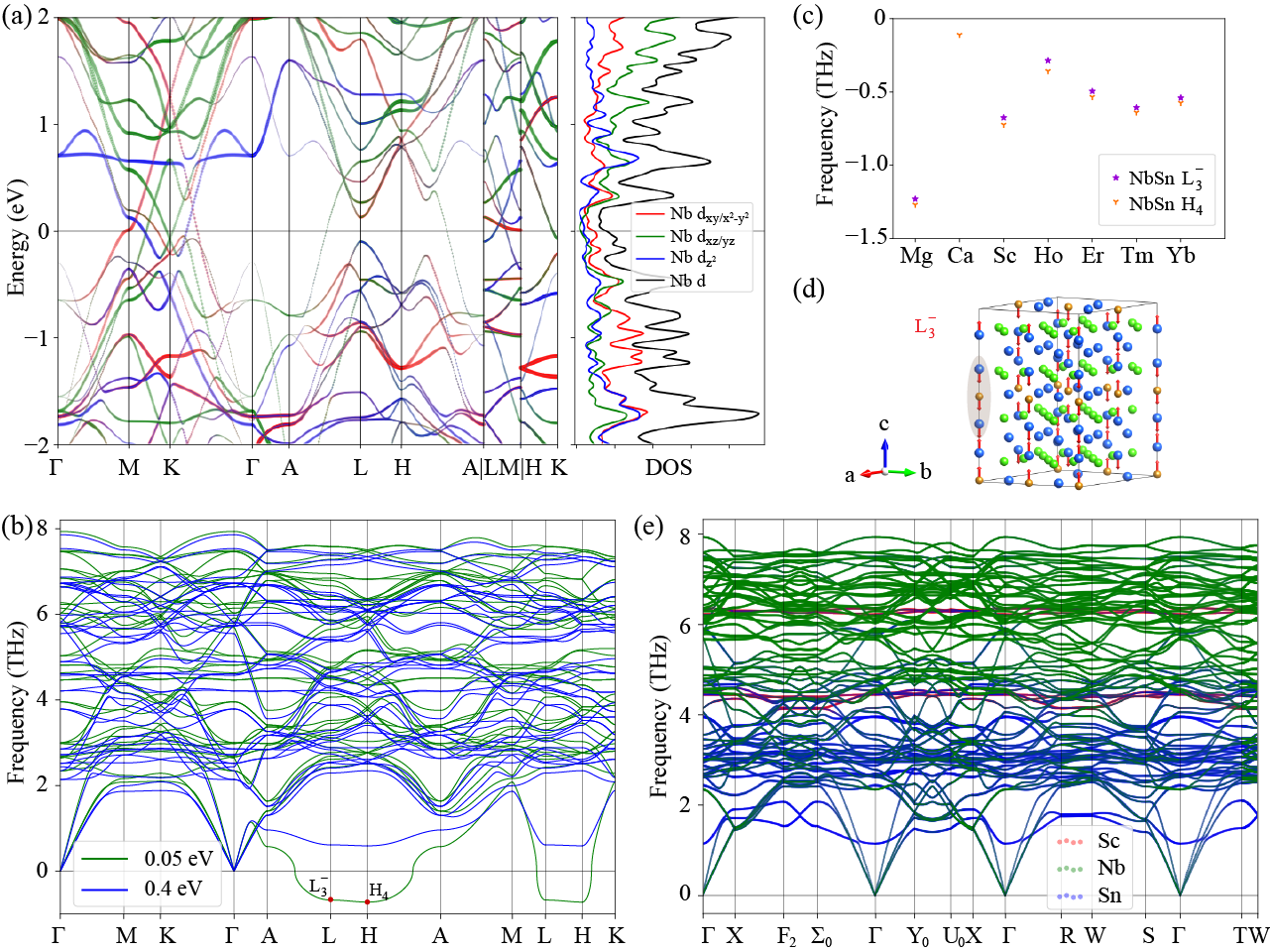}
\caption{Type-III phonon instabilities and possible CDW in NbSn-166 compounds. (a-b) Band structure and phonon spectrum of \ch{ScNb6Sn6}. The band projection presents vHSs from Nb $d_{xy/x^2-y^2}$ and $d_{z^2}$ orbitals close to the Fermi level. The phonon spectra are obtained with a $2\times2\times2$ supercell at smearing values of $0.05$ eV and $0.4$ eV. At low temperatures, the phonon spectrum exhibits the lowest imaginary mode at $H$, resembling the phonon behavior reported in \ch{ScV6Sn6}. (c) Representations of imaginary modes at $L$ and $H$ points in NbSn-166 compounds. (d-e) Typical distorted structure and corresponding stable phonon spectrum derived from a single $L_3^-$ mode. These distortions primarily involve the out-of-plane displacements of trigonal M and Sn atoms. Unlike the mirror odd mode in type-II phonon instabilities, the imaginary mode here corresponds to a mirror even mode, exhibiting opposite movement of Sc and Sn atoms within the Sc-Sn-Sn-Sc chain. Specifically, the Sn-Sn distances are significantly modulated below CDW transitions, while the Sn-Sc bonding exhibits negligible differences.}
\label{Fig:pho_NbSn}
\end{figure}

\textbf{Type-III instabilities.} For type-III instability, the nearly flat phonon band is present at $k_z=\pi$ plane in the VGe/VSn/NbSn-166 families. Here, we take the predicted \ch{ScNb6Sn6} as an example, as shown in Fig.~\ref{Fig:pho_NbSn}. Obviously, the isovalent Nb $4d^3$ has the same electron fillings as V $3d^3$. The Fermi level is dominated by Nb $4d$ electrons, with vHSs from $d_{xy/x^2-y^2}$ and $d_z^2$ orbitals close to the Fermi level. However, the heavier Nb also indicates a larger bandwidth and relatively stronger SOC strength. One direct consequence is that the unfilled flat bands are located at higher energy compared to those in V $3d$ kagome compounds, as shown in the later Fig.~\ref{Fig:kgm}. Meanwhile, One can obtain the nearly flat imaginary phonon mode at low temperatures, dominated by the out-of-plane vibrations of trigonal M and Sn atoms as shown in Fig.~\ref{Fig:pho_NbSn}(b). Similar phonon instabilities are also found in other NbSn-166 candidates, as shown in Fig.~\ref{Fig:pho_NbSn}(c). Here, one CDW structure derived from the single $L_3^-$ mode is illustrated in Fig.~\ref{Fig:pho_NbSn}(d), with the stable phonon spectrum in Fig.~\ref{Fig:pho_NbSn}(e). The Sn-Sn bond distance experiences a significant modulation while the Sc-Sn bond shows a negligible difference.

\subsection{\label{ssec:flatband}Flat bands from kagome lattice}

\begin{figure}[htbp]
\includegraphics[width=0.5\textwidth]{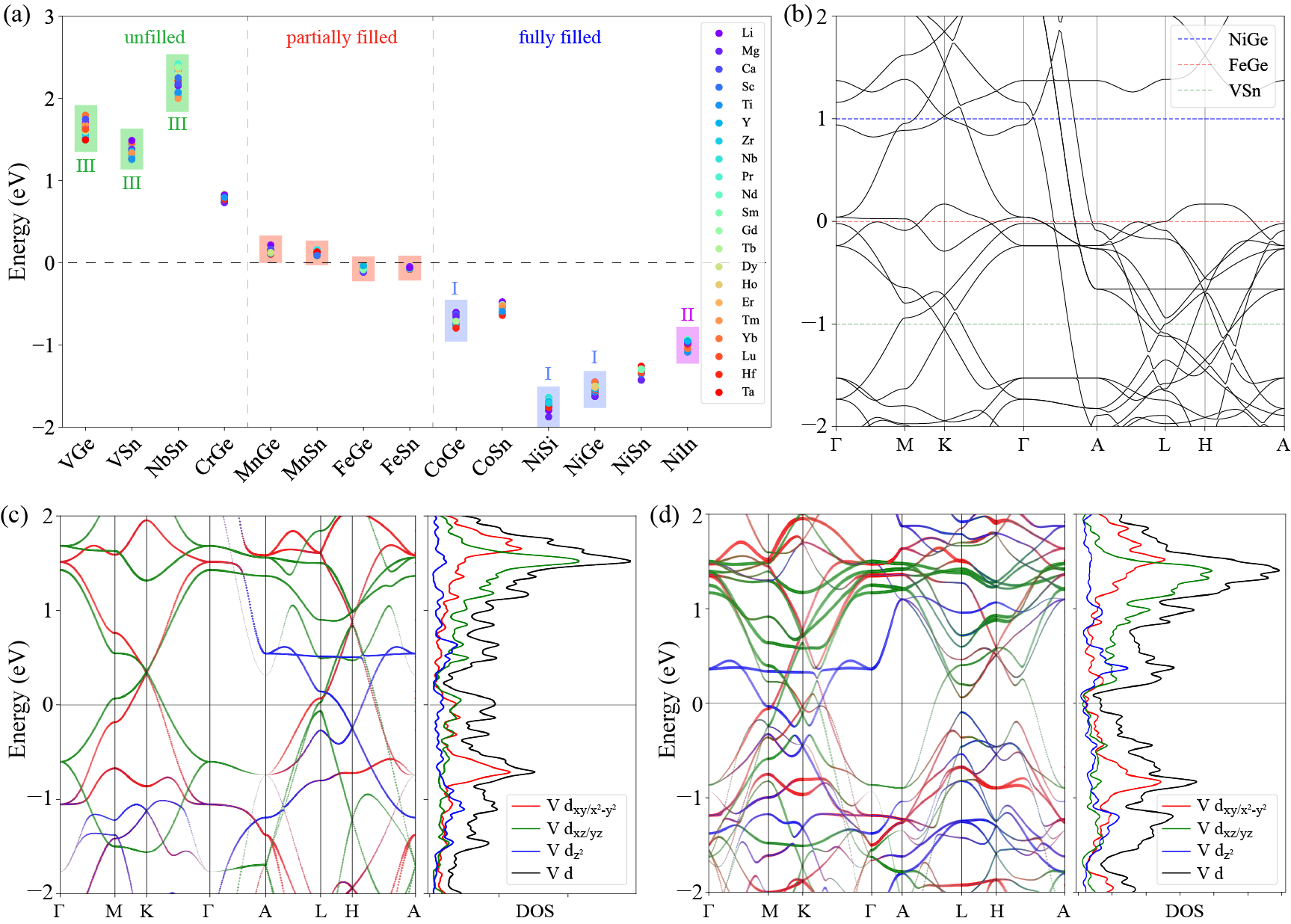}
\caption{Flat bands in 1:6:6 compounds. (a) Peaks of DOS in 1:6:6 compounds obtained from non-spin-polarized calculations without SOC. 
The flat bands that contribute to the DOS peak are classified into three classes, i.e., unfilled, partially filled, and fully filled. The green blocks denote the presence of type-III instabilities in VGe/VSn/NbSn-166 with unfilled kagome-derived bands, while the blue and purple blocks denote the presence of type-I instabilities in CoGe/NiSi/NiGe-166 and Type-II instabilities in NiIn-166, respectively. The red block signifies the presence of instabilities associated with the sharp DOS near the Fermi level obtained from the non-spin-polarized calculations. %
(b) Band structure of the minimal tight-binding model for FeGe from \cite{JiangYi_kagomeI_2023}, where there are two quasi-flat bands near the Fermi level contributed by $d_{xy/x^2-y^2}$ and $d_{xz/yz}$ orbitals of Fe. The approximated Fermi levels of VSn, FeGe, and NiGe families are illustrated to present the unfilled, partially filled, and fully filled flat bands. 
(c) Band structure of predicted 1:1 family VSn structures with the projection of V $3d$ orbitals. The kagome bands from $d_{xy/x^2-y^2}$ and $d_{xz/yz}$ orbitals are located at the same energy scale while the kagome bands from $d_{z^2}$ sit at lower energy. %
(d) Band structure of \ch{LiV6Sn6}. The intercalation of Li drives triangular Sn away from the kagome plane, leading to a doubled unit cell along the $c$ direction. The bands on $k_z=0,\pi$ planes in the \ch{VSn} BZ are folded together to $k_z=0$ plane in the BZ of \ch{LiV6Sn6}. The Fermi level is slightly shifted upwards due to the additional valence electron contributed by Li. %
}
\label{Fig:kgm}
\end{figure}

In this section, we discuss the electronic properties of the 1:6:6 family \ch{MT6Z6}. We {confirm} that the bands near the Fermi level are predominantly given by the $d$ orbitals of kagome T atoms, with the Fermi level mainly determined by the number of valence $d$ electrons in T. 
{The position of the highest peak in the DOS given by the flat bands is related to the type of the CDW or magnetic instabilities.}

We begin with a brief review of the ideal $s$-orbital kagome bands. 
The kagome lattice intrinsically hosts interesting electronic states at different fillings, including flat bands, Dirac points, and van Hove singularities. For the ideal three-band nearest neighbor $s$-orbital kagome model (assuming the flat band is at the top), there is one Dirac point located at the $\frac{4}{12}$ filling, and two vHSs at the $\frac{3}{12}$ and $\frac{5}{12}$ fillings, respectively. 
When the vHS is located at the Fermi level, there is a perfect Fermi surface nesting with a peak of the density of states that leads to a divergent electronic susceptibility, which is crucial for understanding distinct charge/spin orders, and has been widely discussed in the $A$V$_3$Sb$_5$ family~\cite{kiesel_sublattice_2012,kiesel_unconventional_2013,wu_nature_2021}. 
Furthermore, the flat bands would be located at the Fermi level when the filling is larger than $\frac{8}{12}$, which is featured by a large peak in the DOS. 
The flat bands near the Fermi level introduce strong fluctuations that indicate the instability, such as the AFM instability, of the PM phase at low temperatures.

In the 1:1 family TZ and the 1:6:6 family \ch{MT6Z6}, the kagome lattices are composed of $3d$ transition metals from V to Ni (together with $4d$ Nb) with partially filled $d$ orbitals, which contribute most to the bands near the Fermi level. The presence of Z atoms at triangular and honeycomb sites prevents the existence of an ideal kagome flat band. 
In Ref.\cite{JiangYi_kagomeI_2023}, we successfully decompose the spaghetti-like band structure of the 1:1 family FeGe into three decoupled groups. The decomposition is based on both symmetry and chemical analysis, where 
the $d$ orbitals on kagome sites are split into three groups under SG 191 symmetries, i.e., the in-plane $d_{xy/x^2-y^2}$, out-of-plane $d_{xz/yz}$ and $d_{z^2}$ orbitals. 
These three groups of $d$ orbitals are then coupled to specific $p$ orbitals of trigonal and honeycomb Z to form bipartite crystalline lattice\cite{calugaru_general_2022, regnault_catalogue_2022} which gives a faithful explanation of the emergent flat bands near the Fermi level. 
The two flat bands from $d_{xy/x^2-y^2}$ and $d_{xz/yz}$ groups are located closely in energy, which contribute to the largest peak in the DOS. The band structure of the minimal model is shown in Fig.~\eqref{Fig:kgm}(b). 
We also use FeGe as a fundamental ``LEGO'' building block to construct the band structures of the whole 1:6:6 family, which can be seen as the doubled and perturbed version of bands in the 1:1 family. 

Fig.~\eqref{Fig:kgm}(a) shows the collection of the energies of the highest DOS peak in the entire 1:6:6 family, obtained from non-spin-polarized calculations without SOC. The location of the DOS peak is mainly determined by the TZ constituents, primarily from the transition element T. When the $d$ orbitals are close to half-filled, the flat bands are located close to the Fermi level with a sharp DOS peak, indicating possible instabilities of the Fermi surface and enhanced correlations. This is the case for Mn($3d^5$)- and Fe($3d^6$)-166, in which magnetic orders have been extensively reported. Fewer valence $d$ electrons will lift the flat bands above the Fermi level in V/Nb/Cr-166, and more valence $d$ electrons will push the flat bands below the Fermi level in Co/Ni-166. 
In CrGe-166 the flat bands from Cr $3d_{z^2}$ orbitals are located close to the Fermi level, {which contributes to a tiny peak in DOS, while the highest peak from $d_{xy/x^2-y^2}, d_{xz/yz}$ is completely unoccupied at about 1 eV. The relatively small DOS at $E_f$ could account for the absence of } long-range magnetic orders in CrGe-166, although small moments are obtained from the magnetization measurements \cite{RCrGe1994, YCrGe2013}. %
{In Fig.~\eqref{Fig:kgm}(b), we show the band structure and DOS from the minimal model of FeGe in Ref.\cite{JiangYi_kagomeI_2023}. The approximated Fermi levels for VSn-, FeGe-, and NiGe-166 are marked in the plot at $E=-1, 0, 1$ eV, respectively. These three Fermi levels in the minimal model give an average filling of 3.42, 6.63, and 7.95 $d$ electrons per kagome site, which are close to the number of valence $d$ electrons in V ($3d^3$), Fe ($3d^6$), and Ni ($3d^8$), showing the faithfulness of the model. %
}

For the Z sites, the isovalent Ge/Sn present no significant difference when $d$ orbitals are close to half-filling as in Mn($3d^5$) and Fe($3d^6$). %
Instead, the substitution from Ge to Sn shows opposite trends when the $d$ orbitals deviate from half-filling. For V($3d^3$) with smaller $d$-fillings, the substitution of Ge by Sn will fill the kagome bands more, while for Co($3d^7$) and Ni($3d^8$) with larger $d$-fillings, the {substitution of Ge by Sn} will fill the kagome bands less. %

At last, although the intercalation of M sites pushes the triangular Z atoms away from the kagome plane, the M atoms have minor effects on the band structures {besides a rigid shift of the Fermi level.}
With additional electron filling from the M site, the electronic properties of \ch{MT6Z6} can be derived from their \ch{TZ} parent compounds~\cite{JiangYi_kagomeI_2023}. 
Taking VSn-166 as an example, the band structure for the predicted 1:1 VSn is shown in Fig.~\ref{Fig:kgm}(c), which has stable phonon at both low-T and high-T. 
The flat bands from $d_{xy/x^2-y^2}$ and $d_{xz/yz}$ orbitals, which contribute to the highest DOS peak, are located at $\sim 1.5$ eV, with the lower vHS near the Fermi level. The flat bands from $d_{z^2}$ on the $k_z=\pi$ plane are located at $\sim 0.5$ eV, with the higher vHS near the Fermi level. In \ch{LiV6Sn6} with two V kagome layers, the flat band from $d_{z^2}$ orbitals on the $k_z=\pi$ plane is folded onto the $k_z=0$ plane. Meanwhile, all the kagome bands are preserved at almost the same energy scale as those in VSn, with a slight downward shift due to the additional electron filling contributed by Li.%

\subsection{\label{ssec:mag}Magnetism}

\begin{figure*}[htbp]
\includegraphics[width=0.95\textwidth]{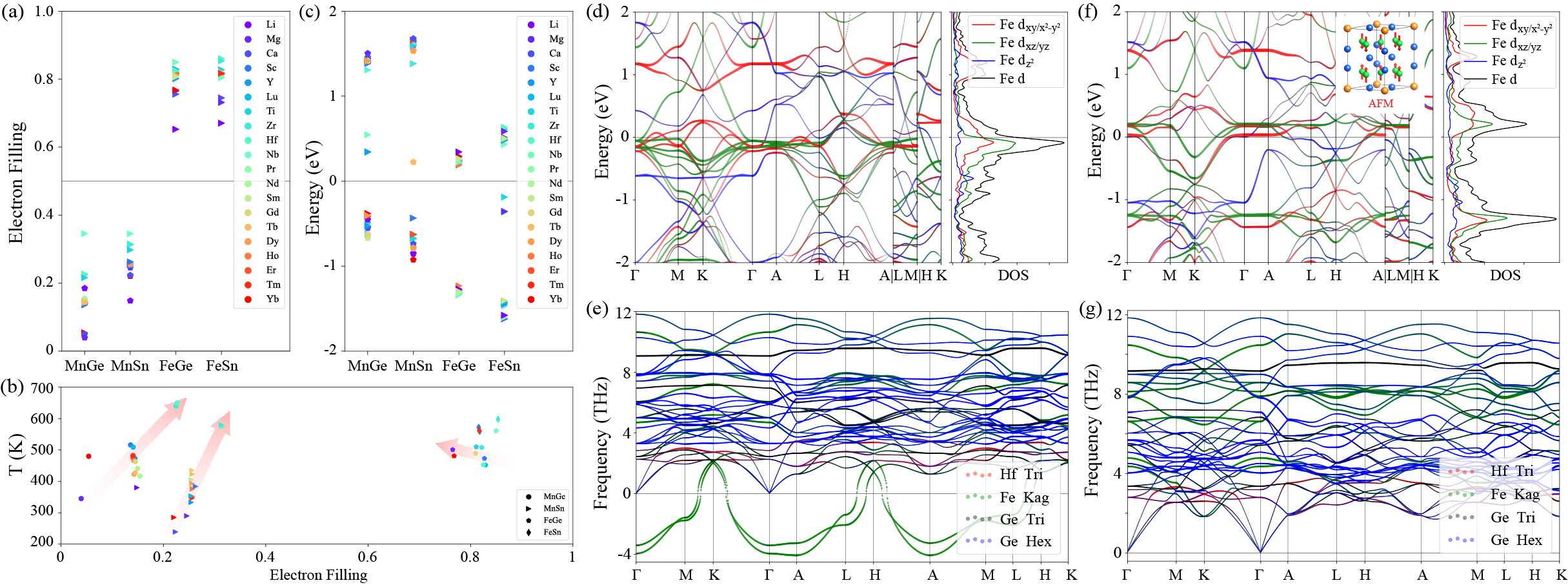}
\caption{Splitting of flat bands in \ch{MT6Z6} under magnetic orders. (a) Electron filling of flat bands close to the Fermi level obtained from non-spin-polarized calculations. The filling of the flat bands is determined by setting a density of states (DOS) cutoff. Here, the triangles (pentagons) denote antiferromagnetic (ferromagnetic) orders for spin-polarized band structures and phonon calculations. Note that Mn/Fe-166 tends to exhibit AFM %
orders, {with FM intralayer interactions and AFM interlayer interactions,} except for several MnGe/Sn-166 compounds with lower fillings, which exhibit FM orders, {such as \ch{AMn6Sn6}(A=Li,Mg, and Ca)~\cite{LiMnSn2006,LiMnSn2021,MgCaMnSn1998} and \ch{MgMn6Ge6}~\cite{MgMnGe2008}.} (b) The evolution of magnetic transition temperature relative to the electron fillings of flat bands. When the fillings approach half filling, the compounds in general exhibit higher transition temperatures $T_{N/C}$ within each 1:6:6 class. For example, for MnSn-166, \ch{MgMn6Sn6} with a lower filling has an FM $T_C=290K$, while \ch{ZrMn6Sn6} with a higher filling has an AFM $T_N=580K$.  %
(c) Collection of DOS peaks in Mn/Fe-166 obtained from spin-polarized calculations, considering collinear magnetic orders without Hubbard U. The sharp peak observed in the PM phase is split into two moderate sharp peaks below and above the Fermi level when magnetic order is present. (d,f) Band structures and (e,g)phonon spectrum for \ch{HfFe6Ge6} without and with magnetic order. In the PM phase, the kagome bands present a sharp-peaked DOS in proximity to the Fermi level, indicating an unstable PM state characterized by imaginary branches. Upon introducing magnetic order, the Fermi surface undergoes modification, displaying the splitting of the DOS peak into two moderate peaks positioned below and above the Fermi level, accompanied by stable phonons.}
\label{Fig:mag_HfFeGe}
\end{figure*}

In this section, we discuss the electron fillings of flat bands in Mn/Fe-166, which give rise to the instability in the PM phase and lead to magnetic orders. This instability is signaled by imaginary phonon frequencies. 

The Mn/Fe-166 compounds exhibit pronounced DOS peaks close to the Fermi level, attributed to the near half-filling of the $3d$ orbitals, as shown in Fig.~\ref{Fig:mag_HfFeGe}(a).
Specifically, Mn-166 exhibits less than half-filling while Fe-166 exhibits more than half-filling, which is consistent with their DOS peaks above and below the Fermi level in Fig.~\ref{Fig:kgm}(a). 
The sharp DOS peak (or, equivalently speaking, the presence of flat bands) near the Fermi level indicates the possible instabilities of the system at low temperatures, which can also be supported by the observations of soft phonons in our non-spin-polarized calculations as shown in Tab.~\ref{Tab:phonon}. 
Indeed, experimentally, most Mn/Fe-166 compounds exhibit (magnetic) instabilities in their PM phases at lower temperatures.

Although the PM phases host unstable phonons, the phonon instabilities are eliminated with the presence of magnetic orders. Experimentally, all existing Mn/Fe-166 are reported to exhibit strong magnetic orders~\cite{baranov_magnetism_2011}. Essentially, the tendency toward magnetic order is enhanced by the flat bands that appear near the Fermi energy and provide a large density of state. 
As shown in Fig.~\ref{Fig:mag_HfFeGe}(a), most of Mn/Fe-166 tends to form antiferromagnetic (AFM) order similar to their 1:1 counterparts FeGe and FeSn, with FM intralayer interactions but AFM interlayer interactions.
{The Fe kagome lattices in Fe-166 compounds usually exhibit easy-axis anisotropy} 
along the { hexagonal} $c$ direction (out-of-plane)~\cite{ScTmLuZrFeSn2000,RFeGeSn2001} {while in 1:1 FeSn the iron moments exhibit easy-plane anisotropy inplane~\cite{sales_electronic_2019}}. 

{In contrast, Mn-166 compounds are found to exhibit more diverse magnetic structures, such as easy-plane FM \ch{LiMn6Sn6}~\cite{LiMnSn2021}, easy-plane AFM \ch{HfMn6Sn6}~\cite{MgZrHfMnSn2002}, flat spiral \ch{TbMn6Ge6}~\cite{zhou_metamagnetic_2023}, and easy-axis AFM \ch{LuMn6Ge6}~\cite{zhou_metamagnetic_2023}. Competing magnetic orders or metamagnetic transitions are also observed with decreasing temperature when rare-earth elements are involved
~\cite{ghimire2020competing,zhou_metamagnetic_2023}}. Here, we estimate the filling of flat bands from $d_{xy/x^2-y^2}$ and $d_{xz/yz}$ orbitals from DOS, as presented in Fig.~\ref{Fig:mag_HfFeGe}. Further details are given in the Supplementary Note IIIB~\cite{supp}. When the fillings approach half-filling of the flat bands, %
Mn-166 exhibits AFM orders, including M=Zr and Hf. At lower fillings ($<\sim$0.2), Mn-166 compounds prefer FM orders, e.g., in \ch{MgMn6Sn6} and \ch{MgMn6Ge6}. However, %
Mn-166 compounds could still exhibit complicated magnetic orders, including spiral magnetic structures~\cite{baranov_magnetism_2011,zhou_metamagnetic_2023,Lee_interplay_2023}, due to the complex magnetic interactions. 
In particular, when rare-earth elements with localized $4f$ orbitals are involved, their magnetic anisotropies and M-T interactions, as well as the strong SOC, would also contribute to diverse magnetic structures~\cite{ghimire2020competing,zhou_metamagnetic_2023,Lee_interplay_2023}. 

The evolution of flat band fillings is also accompanied by the magnetic transition temperature. As shown in Fig.~\ref{Fig:mag_HfFeGe}(b), the experimental magnetic transition temperatures $T_{N/C}$ for AFM/FM are collected with respect to the filling of flat bands. In general, the transition temperatures exhibit an increasing trend as the filling of the flat bands approaches half-filling in each Mn/Fe-1:6:6 class.
For example, \ch{MgMn6Sn6} exhibits an FM transition at $T_c=290K$ with a filling of 0.22~\cite{MgCaMnSn1998,MgZrHfMnSn2002}, %
while \ch{ZrMn6Sn6} exhibits an AFM transition at $T_N=580$ with a filling of 0.31. Although MnSn-166 has higher fillings compared to MnGe-166 %
, the enlarged cell weakens the interactions, resulting in lower transition temperatures and more diverse magnetic structures. %

With the magnetic orders, additional calculations are carried out to check the band structures and phonon stability of Mn/Fe-166. Here, collinear magnetic orders are considered without additional Hubbard U. Electronically, the presence of magnetism splits the bands and may pull vHSs and Dirac points close to the Fermi level. Spin-polarized calculations reveal that the phonon spectra for Mn/Fe-166 are stabilized without imaginary modes with the presence of magnetic orders, as discussed in {the Supplementary Note III}~\cite{supp}. Note that a first-order CDW transition is still observed in kagome 1:1 FeGe, without imaginary phonons, when magnetic order is present with vHS closed to the Fermi level~\cite{teng_discovery_2022,yin_discovery_2022, teng_magnetism_2023}, indicating the significant interplay between spin, phonon, vHSs, and charge density waves in correlated systems.

We focus on \ch{HfFe6Ge6} as an example to demonstrate the effect of magnetic order on the phonon spectrum. \ch{HfFe6Ge6} exhibits A-type AFM order, with FM intralayer interactions and AFM interlayer interactions. As shown in Fig.~\eqref{Fig:mag_HfFeGe}(d), the non-spin-polarized band structure presents {two} partially filled flat bands. Correspondingly, a sharp DOS peak appears close to the Fermi level contributed by Fe $d$ orbitals. The phonon spectrum in the PM phase shows instabilities dominated by kagome Fe atoms as shown in Fig.~\eqref{Fig:mag_HfFeGe}(e). In the magnetic phase, the flat bands are split into two distant groups, with two DOS peaks away from the Fermi level, as shown in Fig.~\eqref{Fig:mag_HfFeGe}(f). Correspondingly, stable phonons are then obtained in the AFM phase as shown in Fig.~\eqref{Fig:mag_HfFeGe}(g). 

\section{Discussion \label{sec:dis}}

Several basic aspects of the above calculations should be clarified. First, when $4f$ elements are involved in 1:6:6 compounds, the local $4f$ moments are reported to exhibit weak magnetic order at lower temperatures, corresponding to additional magnetic transitions at lower temperatures, even in the absence of magnetic kagome lattice~\cite{RCrGe1994,YbVSn2022}. Particularly, when $4f$ electrons are involved with the presence of magnetic kagome T lattice, additional M-T interactions give rise to complex magnetic orders in experiments~\cite{YbFeGe2005,zhou_metamagnetic_2023,Lee_interplay_2023}. Here, we take $4f$ electrons as the core electrons in our calculations of phonon spectrum and electronic bands. 
Second, the finite temperature phonon was considered at the harmonic level via the (electronic) smearing approach. 
Unless otherwise stated, the high-T and low-T refer to a Fermi-Dirac smearing of 0.4 eV and 0.05 eV, respectively. Therefore, the imaginary phonon at high-T (0.4 eV) may be stabilized at even higher temperatures/smearing values. For example, we demonstrate a stable phonon with a smearing of 1.2 eV for \ch{MgNi6Ge6} in~\ref{Fig:pho_NiGe}(d), which suggests the stabilization at higher temperatures. 

For the \ch{MT6Z6} family, the element substitutions can give rise to a large compound diversity, such as the group-IVA elements Si/Ge/Sn at Z. Since the elements have the same valence electrons, the band structures show similar features. Instead, the substitution may significantly modify the lattice parameters, $a$ and $c$, namely the interlayer and intralayer distances of kagome layers as shown in the Supplementary Note. For example, the substitution from Ge to Sn can increase the distances and weaken the interactions in magnetic kagome lattices, leading to lower magnetic transition temperatures, especially in MnGe/Sn-166 compounds. As for the phonon instabilities, although the substitution from Ge to Sn with heavier mass,  {the larger size helps stabilize the phonon}, which occurs in Co/Ni-166 classes. %

The M atoms, ranging from metal elements, and transition metals, to rare-earth metals, contribute to the diverse properties in each TZ-166 class. One significant property is the involvement of $4f$ orbitals from rare-earth elements, {which may exhibit magnetism at lower temperatures~\cite{YVSnGdVSn2021,YbVSn2022,SmVSn2023}}. In Mn/Fe-166, it also leads to diverse magnetic structures due to the presence of its distinct anisotropy and M-T interactions. %
The key element {in understanding \ch{MT6Z6}} is the presence of the kagome T lattice, which plays a dominant role in shaping the main characteristics at the Fermi level. 
As T varies from V to Ni, the fillings of the kagome bands vary considerably, leading to distinct states at the Fermi level, such as flat bands, Dirac points, and vHSs. In contrast, the Z and M sites exert a minor influence on the fillings as shown in Fig.~\ref{Fig:kgm} and Fig.~\ref{Fig:mag_HfFeGe}. %
The kagome bands can be further extended to a wider range when $4d$ and $5d$ elements are included as shown for NbSn-166 in Fig~\eqref{Fig:pho_NbSn}. Although Nb has the same number of valence electrons as V, the DOS peaks of NbSn-166 show larger variations compared to those of VSn-166 due to the increased bandwidth of the $4d$ orbitals. %

Meanwhile, the substitution of elements can also modify the location of flat bands as depicted in Fig.~\ref{Fig:kgm}. Particularly, one can expect to control the presence of magnetism by replacing Z elements beyond Mn/Fe-166. One proposal is to introduce Ga/In in Co-166. According to the filling in Fig.~\ref{Fig:kgm}(a), one may expect the presence of a sharp DOS peak close to the Fermi level in CoGa/In-166 compounds, which have the same number of valence electrons as FeGe/Sn-166. 

\section{summary \label{sec:sum}}

In summary, we have performed a first-principles study of the electronic band structures and phonon spectrum for 1:6:6 \ch{MT6Z6} compounds. The electronic band structures reveal the multitude of fillings of kagome flat bands derived from the kagome lattice. 
Kagome bands can be considered as three different groups from the in-plane $d_{xy/x^2-y^2}$, out-of-plane $d_{xz/yz}$ and $d_{z^2}$ orbitals. %
Their locations present a strong dependence on the TZ constituent, especially the kagome T. When the T $d$ orbitals are close to half-filled, a pronounced DOS peak is presented close to the Fermi level, leading to the presence of instabilities in the PM phase and magnetic orders. Meanwhile, the magnetic order drives the splitting of the sharp DOS peak into two distant DOS peaks positioned below and above the Fermi level, resulting in the stabilization of the phonon spectrum in the magnetic phase.

When the flat bands are away from the Fermi level, a range of instabilities are also demonstrated. Instead of kagome flat bands, one can find van Hove singularities or flat band segments close to the Fermi level. Based on their vibrational modes, these instabilities are classified into three types. 
Type-I instabilities occur along the $\Gamma-A$ path, and are characterized by the locally twisted distortions of kagome nets, breaking the $C_6/C2$ symmetry. 
We take \ch{MgNi6Ge6} as an example to illustrate the possible CDW structures, which energetically favor a nematic phase. Experimentally, such distortions have been reported in \ch{MgCo6Ge6}~\cite{MgCoGe2021} and \ch{YbCo6Ge6}~\cite{YbCoGe2022}.
Type-II instabilities primarily occur on the $k_z=0$ plane, and are characterized by out-of-plane vibrations of trigonal M and Z atoms in the M-Z chain. A typical example is the NiIn-166 compounds. From a single M mode, a stable CDW structure is demonstrated, but others can be possible. The hexagonal-to-orthorhombic transition suggests the possible formation of 1:6:6 in other space groups~\cite{venturini_filling_2006}. 
Type-III instabilities occur on the $k_z=\pi$ plane, dominated by out-of-plane vibrations of trigonal M and Z atoms in the M-Z chain. A recent example of this is \ch{ScV6Sn6}, where CDW transitions have been observed at low temperatures. Similar vibration modes are also predicted in the Nb-166 compounds, such as \ch{ScNb6Sn6}.

Our study uncovers the intricate interplay between the kagome bands, magnetism, and instability. 
The extensive \ch{MT6Z6} family with varying fillings provides a playground to explore the appealing properties embedded within the kagome bands.

\section{Acknowledgments}
We would like to thank Titus Neupert, Laura Classen, Chandra Shekhar, Subhajit Roychowdhury, Changjiang Yi, Shogo Yamashita %
for discussions.

Y.J. and H.H. were supported by the European Research Council (ERC) under the European Union’s Horizon 2020 research and innovation program (Grant Agreement No. 101020833) as well as by IKUR Strategy. 
D.C\u{a}l. acknowledges the hospitality of the Donostia International Physics Center, at which this work was carried out. 
D.C\u{a}l.  was supported by the European Research Council (ERC) under the European Union’s Horizon 2020 research and innovation program (grant agreement no. 101020833) and by the Simons Investigator Grant No. 404513. B.A.B was supported by the Gordon and Betty Moore Foundation through Grant No.GBMF8685 towards the Princeton theory program, the Gordon and Betty Moore Foundation’s EPiQS Initiative (Grant No. GBMF11070), Office of Naval Research (ONR Grant No. N00014-20-1-2303), Global Collaborative Network Grant at Princeton University, BSF Israel US foundation No. 2018226, NSF-MERSEC (Grant No. MERSEC DMR 2011750). 
C.F. acknowledges financial support from Deutsche Forschungsgemeinschaft (DFG) under SFB1143 (Project No. 247310070); Würzburg-Dresden Cluster of Excellence on Complexity and Topology in Quantum Matter—ct.qmat (EXC 2147, project no. 390858490).
S.B-C. acknowledges financial support from the MINECO of Spain through the project PID2021-122609NB-C21 and by MCIN and by the European Union Next Generation EU/PRTR-C17.I1, as well as by IKUR Strategy under the collaboration agreement between Ikerbasque Foundation and DIPC on behalf of the Department of Education of the Basque Government.
M.G.V. acknowledges support to the Spanish Ministerio de Ciencia e Innovacion (grant PID2022-142008NB-I00), partial support from European Research Council (ERC) grant agreement no. 101020833, the European Union NextGenerationEU/PRTR-C17.I1, by the IKUR Strategy under the collaboration agreement between Ikerbasque Foundation and DIPC on behalf of the Department of Education of the Basque Government and the Ministry for Digital Transformation and of Civil Service of the Spanish Government through the QUANTUM ENIA project call - Quantum Spain project, and by the European Union through the Recovery, Transformation and Resilience Plan - NextGenerationEU within the framework of the Digital Spain 2026 Agenda. 
M.G.V. and C.F. acknowledge funding from the Deutsche Forschungsgemeinschaft (DFG, German Research Foundation) for the project FOR 5249 (QUAST).

\bibliography{ref}

\end{document}


\title{ Supplemental Material for "Catalogue of Phonon Instabilities in Symmetry Group 191 Kagome \ch{MT6Z6} Materials" }

\author{X. Feng}
\email{xiaolong.feng@cpfs.mpg.de}
\affiliation{Max Planck Institute for Chemical Physics of Solids, 01187 Dresden, Germany}

\author{Y. Jiang}
\affiliation{Donostia International Physics Center (DIPC), Paseo Manuel de Lardizábal. 20018, San Sebastián, Spain}

\author{H. Hu}
\affiliation{Donostia International Physics Center (DIPC), Paseo Manuel de Lardizábal. 20018, San Sebastián, Spain}

\author{D. C\u{a}lug\u{a}ru}
\affiliation{Department of Physics, Princeton University, Princeton, NJ 08544, USA}

\author{N. Regnault}
\affiliation{Laboratoire de Physique de l’Ecole normale supérieure,
ENS, Université PSL, CNRS, Sorbonne Université,
Université Paris-Diderot, Sorbonne Paris Cité, 75005 Paris, France}
\affiliation{Department of Physics, Princeton University, Princeton, NJ 08544, USA}

\author{M. G. Vergniory}
\affiliation{Donostia International Physics Center (DIPC), Paseo Manuel de Lardizábal. 20018, San Sebastián, Spain}
\affiliation{Max Planck Institute for Chemical Physics of Solids, 01187 Dresden, Germany}

\author{C. Felser}
\email{Claudia.Felser@cpfs.mpg.de}
\affiliation{Max Planck Institute for Chemical Physics of Solids, 01187 Dresden, Germany}

\author{S. Blanco-Canosa}
\affiliation{Donostia International Physics Center (DIPC), Paseo Manuel de Lardizábal. 20018, San Sebastián, Spain}
\affiliation{IKERBASQUE, Basque Foundation for Science, 48013 Bilbao, Spain}

\author{B. Andrei Bernevig}
\email{bernevig@princeton.edu}
\affiliation{Donostia International Physics Center (DIPC), Paseo Manuel de Lardizábal. 20018, San Sebastián, Spain}
\affiliation{Department of Physics, Princeton University, Princeton, NJ 08544, USA}
\affiliation{IKERBASQUE, Basque Foundation for Science, 48013 Bilbao, Spain}

\maketitle

\tableofcontents

\newpage
\label{smsec:col}
The supplementary material is organized as follows. 
\begin{itemize}
    \item Sec.~\ref{smsec:method} describes the first-principle methodology applied to obtain the phonon spectrum and band structures. 
    \item Sec.~\ref{smssec:structure} gives a more detailed description of structures of \ch{MT6Z6} compounds. A brief collection of their theoretically or experimentally reported structures is shown in Tab.~\ref{tab:166_1} and~\ref{tab:166_2}, where we include the space groups and references. 
    \item Sec.~\ref{smssec:stat} briefly summarizes the electronic band features and phonon stabilities for each 1:6:6 class. The phonon instabilities are summarized in Tab.~\ref{Tab:SM_phononcollection}. 
    
    \item Sec.~\ref{smsec:allcompounds} collects all results for each 1:1 and 1:6:6 compound in this study. 
    \begin{itemize}
        \item In Sec.~\ref{smssec:TZ}, the phonon spectrum and band structures for TZ parent structures are presented with atomic projections. 
        \item From Sec.~\ref{smssec:VGe} to Sec.~\ref{smssec:NbSn}, the phonon spectrum and band structures for all 166 compounds in each TZ-class are presented. For all compounds, both high-T and low-T phonon spectra are presented at the harmonic level via the smearing approach.
        \item Particularly, for Mn/Fe-166 compounds, additional results with the presence of collinear magnetic orderings are also included without Hubbard U, which show no instability. 
    \end{itemize}
\end{itemize}

\section{Calculation methods \label{smsec:method}}

\subsection{First-principle calculation}
The structural relaxation is performed using Quantum Espresso with a force convergence criterion of 10$^{-6}$ Ry~\cite{giannozzi_quantum_2009}. The standard solid-state pseudopotentials (SSSP) library is adopted to describe of functional~\cite{prandini_precision_2018}. The kinetic energy cutoff for wave functions and charge density are set to be 90 Ry and 700 Ry, respectively. With the obtained relaxed structure, the phonon spectra and electronic band structures are calculated by VASP~\cite{kresse_ab_1994,kresse_efficiency_1996,kresse_efficient_1996,kresse_ultrasoft_1999}. 
The phonon spectra and force constants are extracted with the PHONOPY\cite{togo_first_2015, togo2023first} code based on density functional perturbation theory (DFPT), using a $2\times2\times2$ supercell. The energy cutoff is 450 eV. The generalized gradient approximation (GGA) with the Perbew-Burke-Ernzerhof (PBE) scheme is adopted for the exchange-correlation functional~\cite{perdew_generalized_1996}. Without special notations, all the calculations are performed without SOC based on the non-spin-polarized description.

\subsection{Structural distortions}

{The equilibrium geometry is obtained when the forces acting on atoms vanish~\cite{2001phonons,togo_first_2015,pallikara2022physical}}
\begin{align}
    F=-\frac{\partial E(\bm{R})}{\partial \bm{R}}=0,
\end{align}
where $E(\bm{R})$ is the total energy as a function of the atomic positions $\bm{R}$. The atomic vibrations are determined by second-order derivatives of energy within harmonic approximation. The energy can be expanded with respect to the displacements as
\begin{align}
    E(\bm{u}) = E_{\bm{R}_0} + \sum_{j,j'}\sum_{\alpha,\beta}\Phi^{jj'}_{\alpha\beta}u_{\alpha}^j u_{\beta}^{j'},
\end{align}
where $j(j')$ denotes the atoms, $\alpha (\beta)$ denotes the Cartesian directions, $\bm{u}$ are the displacements from their equilibrium positions $\bm{R_0}$, and $\Psi_{\alpha\beta}^{jj'}=\frac{\partial^2 E}{\partial u_{\alpha}^{j}\partial u_{\beta}^{j'}}$ is the second-order force constant. The Fourier transform of force constants gives the dynamic matrix as
\begin{align}
    \bm{D}(\bm{q})_{\alpha\beta}^{jj'} = \frac{1}{\sqrt{m_j m_{j'}}} \sum_{l'} \Psi_{\alpha\beta}^{j0,j'l'} \exp[i \bm{q}\cdot (\bm{R_{j'l'}}-\bm{R_{j0}})],
\end{align}
where $\bm{q}$ is the phonon wave vector, $m_{j}$ is the mass of atom $j$, and $l'$ denotes the unit cells of atom $j'$. Therefore, the motion of atoms can be obtained by solving the equation
\begin{align}
    \bm{D}(\bm{q}) \bm{e}(\bm{q}) = \omega^2(\bm{q}) \bm{e}(\bm{q}).
\end{align}
The eigenvalues are the squared frequencies $\omega(\bm{q})$. The displacements of atom $j$ can then be obtained from the eigenvectors $\bm{e}(\bm{q})$ as
\begin{align}
    \bm{u}^j_{\bm{q}} = \frac{A}{\sqrt{m_j}} \text{Re}[\bm{e}_j(\bm{q}) \exp{(i\phi + i\bm{q}\cdot\bm{R}_j)}],
\end{align}
where $\phi$ is a phase factor and $A$ is the displacement amplitude.

For a phonon mode $\bm{q}$, the corresponding harmonic energy is given by $E(\bm{q})=\frac{1}{2}\omega^2(\bm{q})A^2$.
One can obtain $E(\bm{q})<0$ when $\omega^2(\bm{q})<0$. Therefore, an imaginary phonon at $\bm{q}$ suggests that the system may undergo a phase transition with structural distortions to minimize the energy, which means the instability of pristine structure.

\section{Structures}\label{smssec:structure}

\begin{figure*}[ht]
\begin{center}
\includegraphics[width=0.9\linewidth]{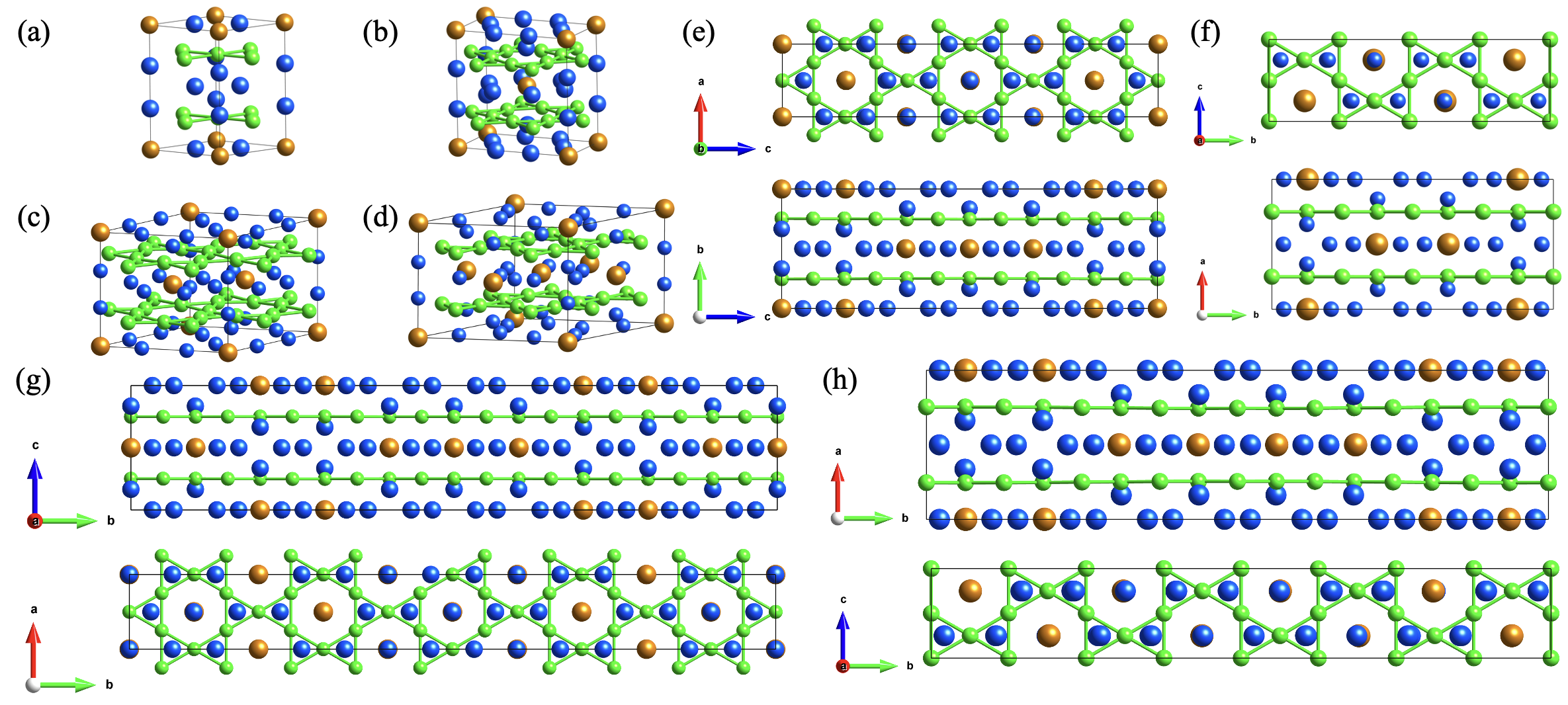}
\caption{Typical structures of 166 \ch{MT6Z6} compounds crystallized in various space groups, including SG 191 $P6/mmm$ (a) \ch{HfFe6Ge6}-type, (c) \ch{LiFe6Ge6}-type, and (d) \ch{ScNi6Ge6}-type, SG 71 $Immm$ (b) \ch{ScFe6Ga6}-type, (e) \ch{YFe6Sn6}-type, SG 63 $Cmcm$ (f) \ch{TbFe6Sn6}-type, and (h) \ch{ErFe6Sn6}-type, and SG 65 $Cmmm$ (g) \ch{DyFe6Sn6}-type.}
\label{Fig:166}
\end{center}
\end{figure*}

In the main text, we introduced the crystal structure of the hexagonal \ch{MT6Z6} family, which can be understood as the intercalation of M atoms in 1:1 $TZ$ nets. However, in the literature, there are more complicated superstructures in the 1:6:6 family, exhibiting different space groups. These superstructures are induced by variations in the M sites with potentially distorted kagome nets, as illustrated in Fig.~\ref{Fig:166}~\cite{venturini_filling_2006,baranov_magnetism_2011}. Experimental characterization of these structures is typically performed using X-ray diffraction. The obtained diffraction patterns are then used to solve and refine the structures, allowing for the determination of lattice parameters, atomic coordinates, and occupations. Ideally, structures without disorder can be obtained from high-quality single crystals. 
The typical unit cell of hexagonal \ch{MT6Z6} compounds consists of two kagome layers located at $6i(\frac{1}{2},0,\frac{1}{4}+\gamma)$, separated by honeycomb Z3/Z4 lattices at sites $2c(\frac{1}{3},\frac{2}{3},0)$ and $2d(\frac{1}{3},\frac{2}{3},\frac{1}{2})$. The sandwiched honeycomb-kagome structures are surrounded by the M-Z chains located at trigonal sites, with M at $1a(0,0,0)$ and Z1 at $2e(0,0,z_1)$, respectively. The presence of M atoms causes the trigonal Z1 atoms to deviate from the kagome plane. Additionally, the kagome layer is also positioned away from the $\frac{c}{4}$ plane, resulting in two distinct interlayer distances between the kagome lattices.

\begin{figure*}[ht]
\begin{center}
\includegraphics[width=0.65\linewidth]{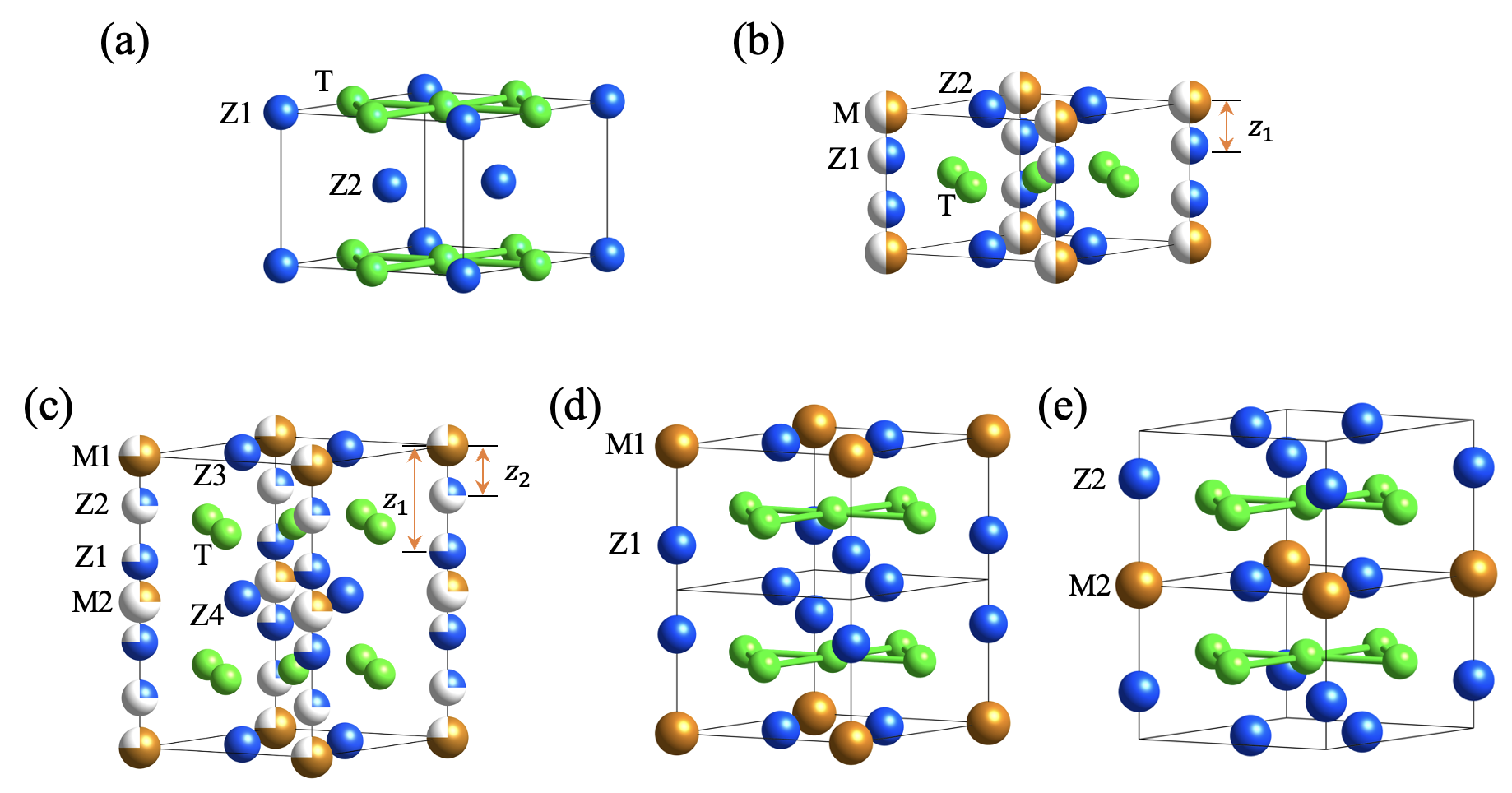}
\caption{%
Disordered and order structures for hexagonal \ch{MT6Z6} materials. (a) CoSn-type parent structure, including two inequivalent trigonal Z1 and honeycomb Z2 sites and one T kagome site. (b) \ch{YCo6Ge6}-type disordered structure. In XRD measurement, the disorder in the M-Z chain would suggest a reduced 166 structure like the CoSn-type structure, which is also denoted as Y$_{0.5}$Co$_3$Ge$_3$ in the reduced cell. (c) \ch{HfFe6Ge6}-type disordered structure. M can be partially assigned to two different trigonal sites, M1 and M2, in one unit cell, along with two trigonal Z sites, Z1 and Z2, which can be considered as two different sets of ordered structures as shown in (d) and (e). Note that only when tha kagome nets are located at $\frac{1}{4}c$ and $\frac{3}{4}c$ plane, namely $\gamma=0$, the two sets are related by a half-$c$ translation. Despite the difference in the distance between the kagome layer and trigonal M layer, the band calculations reveal no discernible distinctions between the two structures. %
}
\label{Fig:166_order}
\end{center}
\end{figure*}

However, in experimental studies, two common disordered hexagonal \ch{MT6Z6} structures, \ch{YCo6Ge6}-type~\cite{venturini_filling_2006,YbCoGe2020,YbCoGe2022} and \ch{HfFe6Ge6}-type~\cite{SmVSn2023}, are reported as shown in Fig.~\ref{Fig:166_order}. Their lattice parameters are given in Tab.~\ref{Tab:unitcell}. It is worth noting that in both structures the disorder arises from the arrangement of trigonal M and Z sites, which may depend on the thermal history during synthesis~\cite{venturini_filling_2006,YbCoGe2020,CaFeCoGe2023}. 
For \ch{YCo6Ge6}-type disordered structures, the disorder leads to the half-$c$ unit cell, as shown in Fig.~\ref{Fig:166_order}(b), which has similar size as the parent CoSn-type TZ structure. For the TZ structure, the T atoms are positioned at $3f(0,\frac{1}{2},0)$ sites forming kagome nets, and the Z atoms are located at two different sites $1a(0,0,0)$ and $2d(\frac{1}{3},\frac{2}{3},0)$, forming trigonal and honeycomb lattices. And the trigonal Z also lies on the kagome plane. However, in the \ch{YCo6Ge6}-type structures, the intercalation of M atoms at the trigonal sites $1a(0,0,0)$ drives the trigonal Z atoms away from the kagome plane into the Z1 site at $2e(0,0,z_1)$. It is also important to note that both the M and Z1 sites in such ordered structures are usually close to half occupied.

For hexagonal \ch{HfFe6Ge6}-type disordered structures, M atoms may partially sit at two trigonal sites, M1 at $1a(0,0,0)$ and M2 at $1b(0,0,\frac{1}{2})$, accompanied by the corresponding trigonal Z1 at $2e(0,0,z_1)$ and Z2 at $2e(0,0,z_2)$, as shown in Fig.~\ref{Fig:166_order}(c) and Tab.~\ref{Tab:unitcell}. %
Meanwhile, the kagome T will be located at $6i(\frac{1}{2},0,\frac{1}{4}+\gamma)$. With a nonzero $\gamma$, the two kagome planes are no longer mirror planes as in the CoSn-type and \ch{YCo6Ge6}-type disordered structure, leading to two distinct interlayer distances. But they are still related to each other by the mirrors at the honeycomb planes. %

It is worth noting that some of these disordered structures can be further refined to ordered superstructures via additional synthesis procedures~\cite{CaFeCoGe2023}. Fig.~\ref{Fig:166} illustrates typical ordered \ch{MT6Z6} structures in different space groups, including SG 191 $P6/mmm$ (a) \ch{HfFe6Ge6}-type, (c) \ch{LiFe6Ge6}-type, and (d) \ch{ScNi6Ge6}-type, SG 71 $Immm$ (b) \ch{ScFe6Ga6}-type , (e) \ch{YFe6Sn6}-type, SG 63 $Cmcm$ (f) \ch{TbFe6Sn6}-type, and (h) \ch{ErFe6Sn6}-type, and SG 65 $Cmmm$ (g) \ch{DyFe6Sn6}-type.

In our calculations, the disordered structures were initially treated as two configurations with a regular $c$ lattice, where the two sets of partial occupations were considered separately, as shown in Fig.~\ref{Fig:166_order}(d) and~\ref{Fig:166_order}(e). Generally, both configurations yield similar results upon relaxation. Hence, only one of them will be presented. %
By performing element substitution and structural relaxation, we successfully obtained the ordered structures within the hexagonal \ch{HfFe6Ge6}-type unit cell for this particular study.

\begin{table}[h]
\global\long\def\arraystretch{1.12}
\caption{Typical unit cell of hexagonal CoSn-type, \ch{YCo6Ge6}-type and \ch{HfFe6Ge6}-type structure in space group $P6/mmm$, including the info of occupations. The structures are presented in Fig.~\ref{Fig:166_order}.}
\label{Tab:unitcell} 
\setlength{\tabcolsep}{2mm}{
\begin{tabular}{c|c|c||c|c|c||c|c|c}
\hline
\hline
Co-type & WP & Occup & Y-type & WP & Occup & Hf-type & WP & Occup  \\
\hline
& & & M & $1a~(0,0,0)$ &  0.5  & M1 & $1a~(0,0,0)$ &  $1-\delta$   \\
& & &  & & & M2 & $1b~(0,0,\frac{1}{2})$ &  $\delta$   \\
\hline
T & $3f(0,\frac{1}{2},0)$ & 1.0 & T & $3g~(\frac{1}{2},0,\frac{1}{2})$ &  1.0  & T & $6i~(\frac{1}{2},0,\frac{1}{4}+\gamma)$ &  1.0   \\
\hline
Z1 & $1a(0,0,0)$ & 1.0 & Z1 & $2e~(0,0,z_1)$ &  0.5  & Z1 & $2e~(0,0,z_1)$ &  $1-\delta$    \\
& & & & & & Z2 & $2e~(0,0,z_2)$ &  $\delta$   \\
\hline
Z2 & $2d(\frac{1}{3},\frac{2}{3},0)$ & 1.0 & Z2 & $2c~(\frac{1}{3},\frac{2}{3},0)$ &  1.0    & Z3 & $2c~(\frac{1}{3},\frac{2}{3},0)$ &  1.0  \\
& & & & & &  Z4 & $2d~(\frac{1}{3},\frac{2}{3},\frac{1}{2})$ &  1.0   \\
\hline
\hline
\end{tabular}}
\end{table}

\begin{figure*}
\begin{center}
\subfloat[]{
\label{fig:struct_a}
\includegraphics[width=0.44\textwidth]{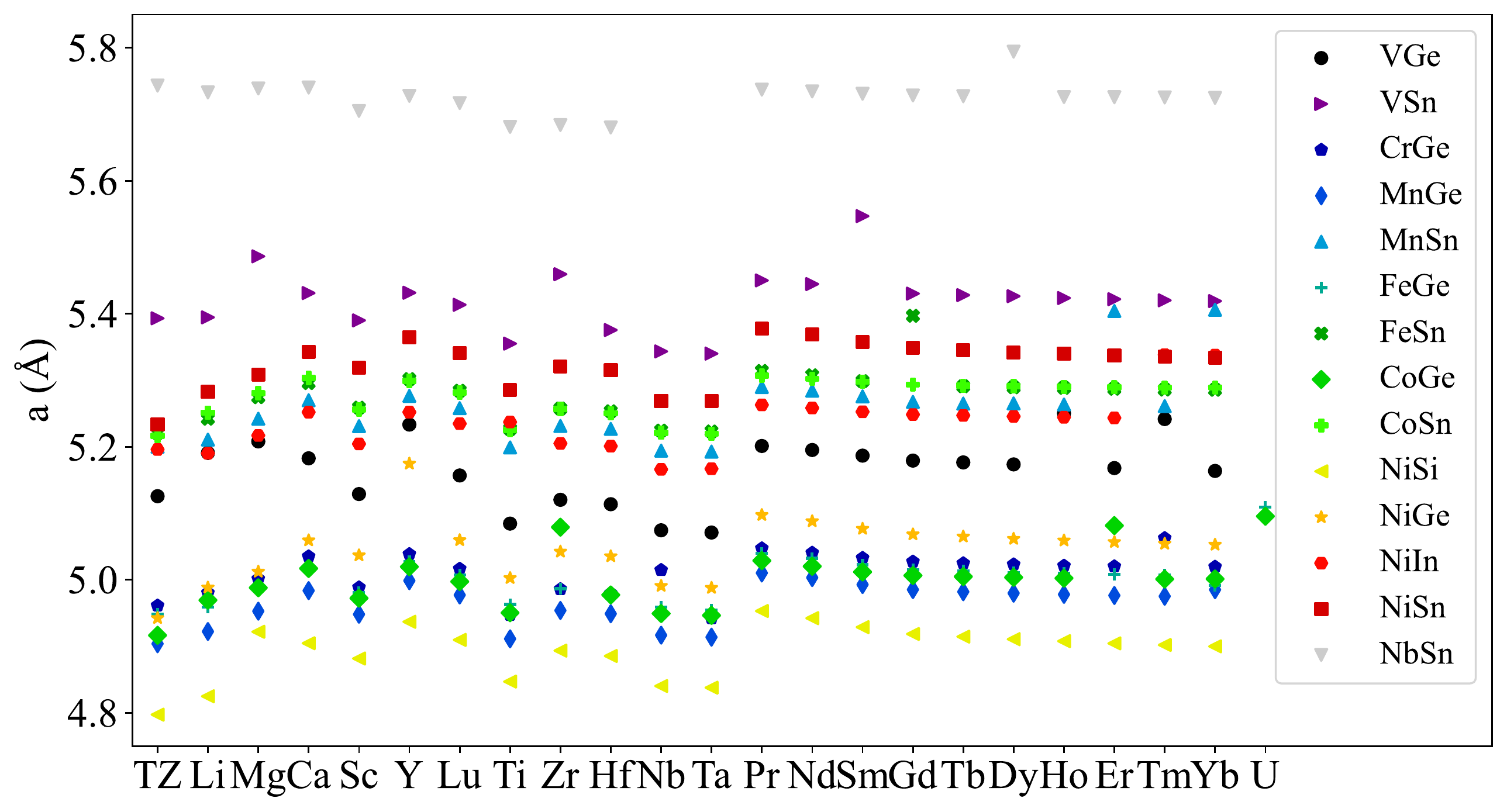}
}
\hspace{2pt}\subfloat[]{
\label{fig:struct_c}
\includegraphics[width=0.44\textwidth]{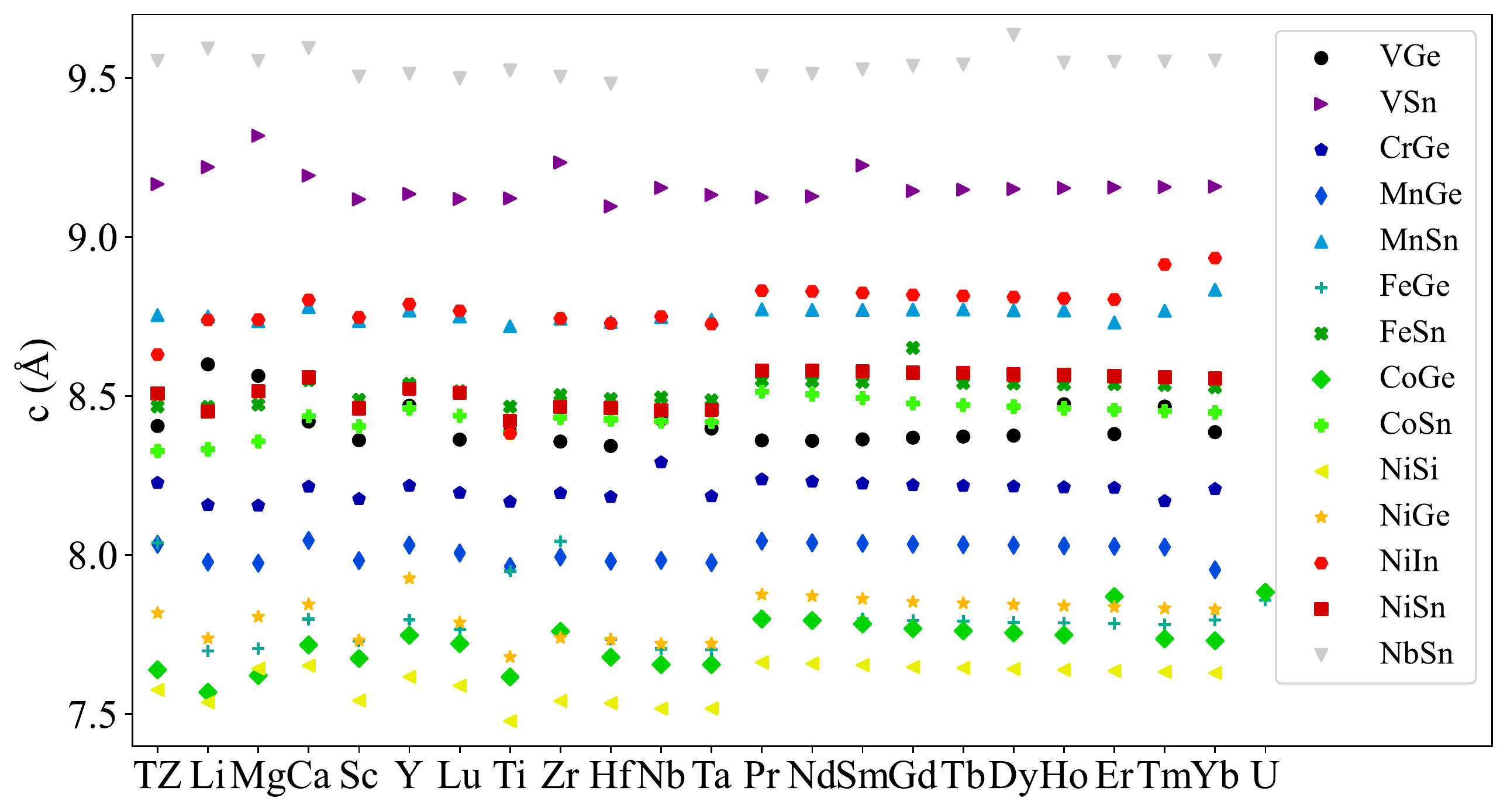}
}
\\
\subfloat[]{
\label{fig:struct_dt2}
\includegraphics[width=0.44\textwidth]{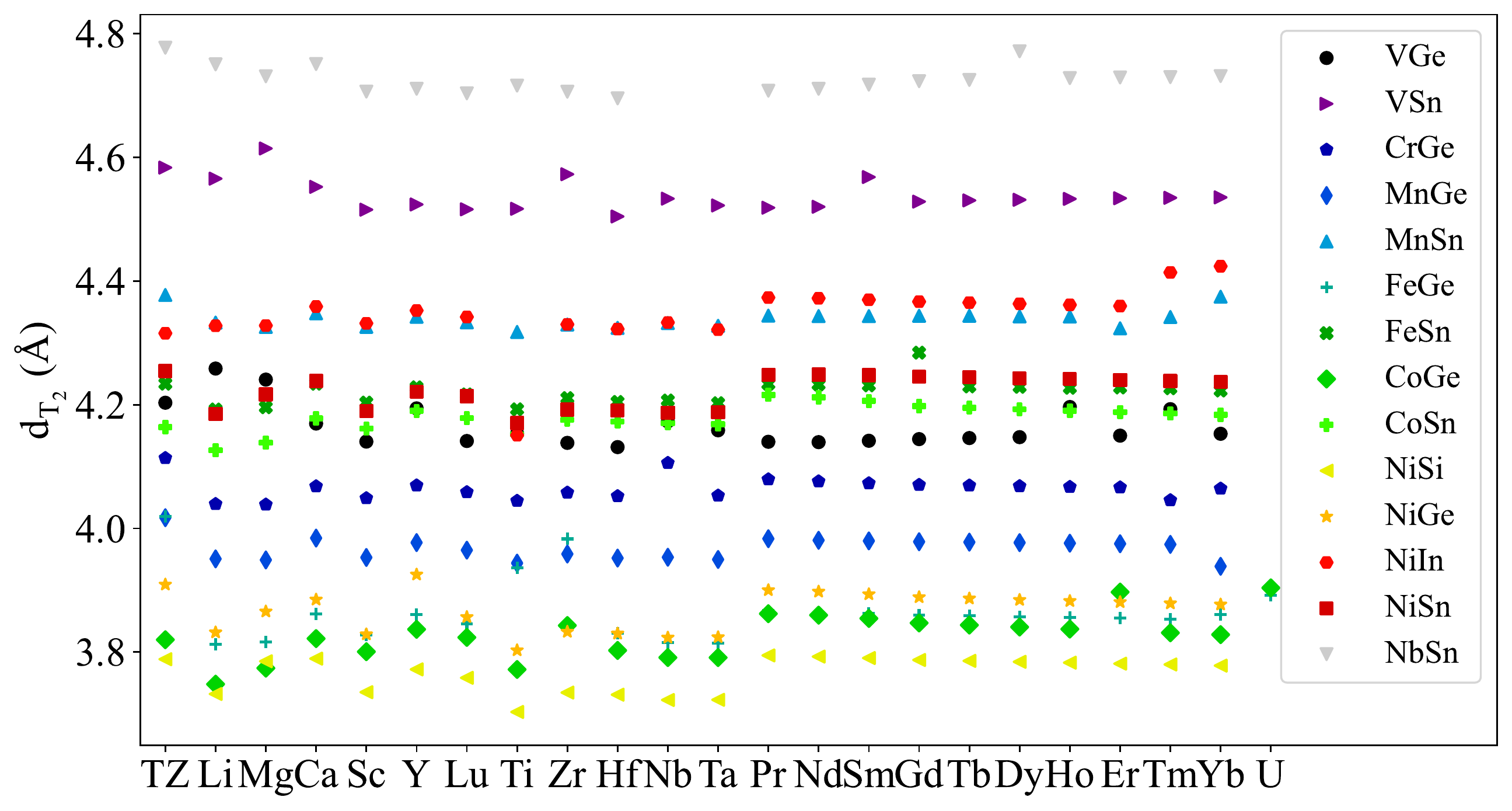}
}
\hspace{2pt}\subfloat[]{
\label{fig:struct_dzz}
\includegraphics[width=0.44\textwidth]{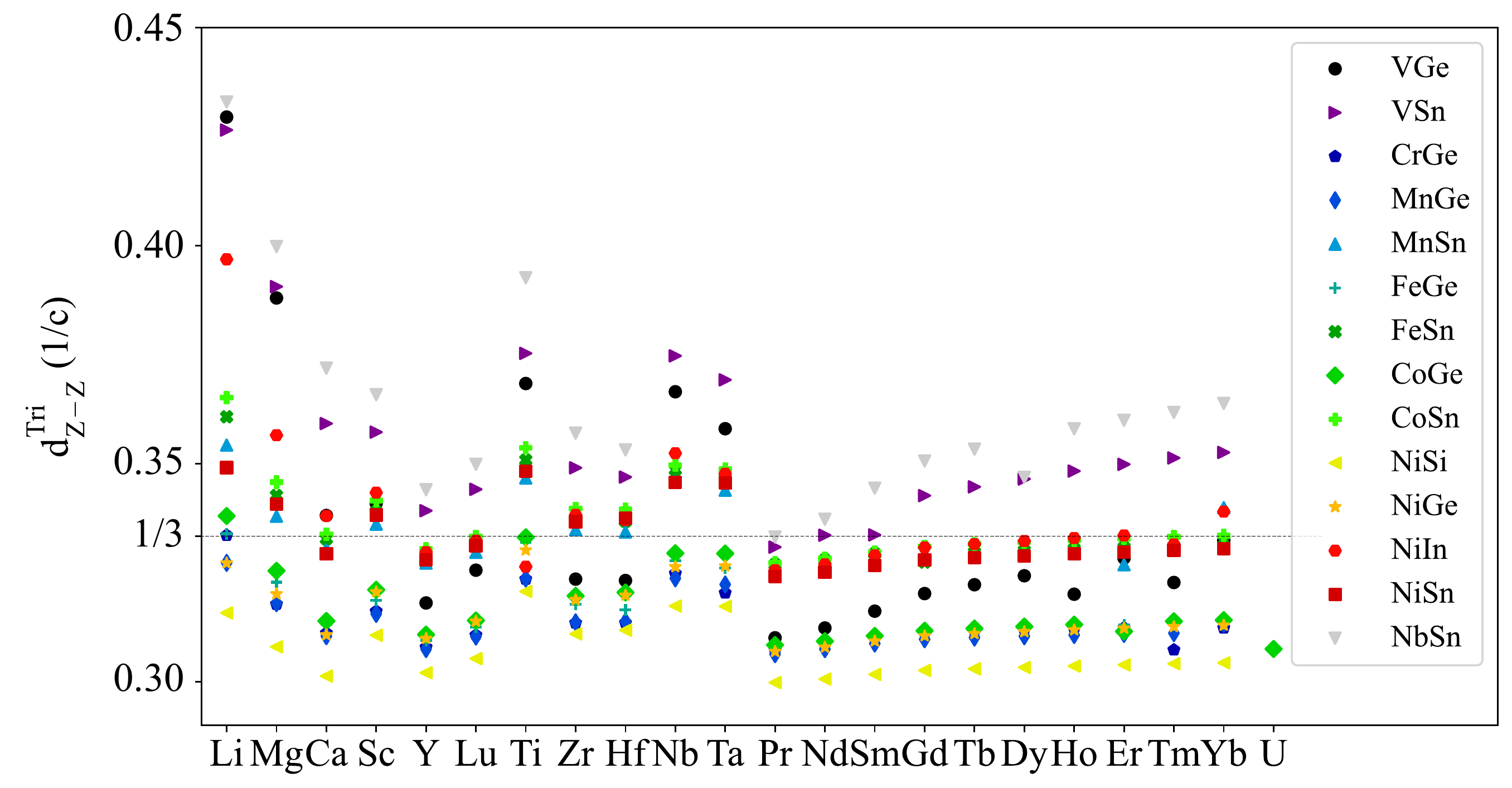}
}
\caption{Structural parameters for 166 \ch{MT6Z6} compounds, including lattice parameters (a)$a$, and (b)$c$, (c) nearest interlayer distance of kagome layer $d_{T_2}$, and (d) bond distance between Z atoms within trigonal M-Z-Z-M chains $d_{\text{Z}^\text{T}-\text{Z}^\text{T}}$. Note that for type-III instability, such as in \ch{ScV6Sn6}~\cite{ScVSn2022,pokharel_frustrated_2023}, the significant modulation is the bonding distance between Z atoms within trigonal M-Z-Z-M chains. %
The first 'TZ' denotes the relaxed parameters for $TZ$ parent structure and $c$ is doubled for comparison.}
\label{Fig:parameter}
\end{center}
\end{figure*}

Fig.~\ref{Fig:parameter} provides the structural parameters for relaxed 166 compounds in SG 191, including lattice parameters $a$ and $c$, the nearest interlayer distance of kagome lattice $d_{T2}$, and the nearest distance between two trigonal Z atoms $d_{\text{Z}^\text{T}-\text{Z}^\text{T}}$. Additionally, the nearest intralayer distance in kagome lattice $d_{T1}=\frac{a}{2}$ is included in the table of lattice parameters for each compound. It is evident that the lattice parameters exhibit variations among different compositions, primarily influenced by the atomic sizes, particularly those of the TZ nets. For example, in the case of Ni-166 with the kagome Ni lattices, the lattice parameters, $a$ and $c$, increase as the Z atoms change from Si to Ge to Sn due to the increased atomic sizes. %

Tab.~\ref{tab:166_1} and~\ref{tab:166_2} present a concise collection of reported \ch{MT6Z6} structures found in previous literature, whether through experimental studies or theoretical databases. The tables include information regarding the atomic occupations in these structures. In particular, an occupation of 0.5 indicates the presence of \ch{YCo6Ge6}-type disorder. It is important to note that the disorder observed in early reports may have been resolved as a result of advancements in experimental instrumentation and synthesis techniques, leading to the characterization of ordered structures or superstructures in these materials.

\begin{table*}[ht]\label{tab:SM_struct1}
\global\long\def\arraystretch{1.1}
\caption{Brief collection of $MT_6Z_6$ reported in experiments or databases. $MT_6Z_6$ can crystallize in space groups 191 $P6/mmm$, 71 $Immm$, 65 $Cmmm$, 63 $Cmcm$, 62 $Pnma$. Typical structures can be found in Fig.~\ref{Fig:166}.}
\centering
\setlength{\tabcolsep}{1mm}{
\begin{tabular}{rccccccc} 
  \hline
  \hline

M(TZ)$_6$ & VSn & VGe & CrGe & MnSn & MnGe & FeGa & FeGe \\
\hline
Li [$2s^1$]  & 191$^a$ & 65$^a$ &  & 191~\cite{LiMnSn2006,LiMnSn2021} &  &  & $^1$191~\cite{CoFeNiGe1981}   \\ 
\hline
Mg [$3s^2$]  & 191$^a$ & 191$^a$ &  & 191~\cite{MgZrHfMnSn2002} & {\color{black} 191}~\cite{MgMnGe2008} &  & $^1${\color{black} 191}~\cite{CoFeNiGe1981}  \\ 
Ca [$4s^2$]  &  &  &  & 191~\cite{MgCaMnSn1998} &  &  & $^{0.5}${\color{black} 191}~\cite{CaFeCoGe2023}, 63~\cite{CaFeGe2022,CaFeCoGe2023}   \\ 
\hline
Sc [$3d^1$] & $^1$191~\cite{ScVSn2022} & 191$^a$ & $^1${\color{black}191}~\cite{ScCrGe1988} & $^1${\color{black} 191}~\cite{RMnSn1991} & $^1${\color{black} 191}~\cite{RTGe1992,ZrLuScMnGe1999} & 71~\cite{RFeGa1990} & $^1${\color{black} 191}~\cite{RTGe1992}   \\ 
Y [$4d^1$] & $^1${\color{black}191}~\cite{RVSn2011,GdVSn2021,YVSnGdVSn2021} & 191$^a$ & $^1${\color{black}191}~\cite{YCrGe2013,YCrGe2022}  & $^1$191~\cite{RMnSn1991,YMnSn2021} & $^1$191~\cite{RTGe1992,YMnGe-LuMnSn1993} & 71~\cite{RFeGa1990} & $^{0.5}${\color{black} 191}$^{c}$~\cite{RFeGe1994}, $^{0.76}$63~\cite{RTGe1992,RFeGe1994}  \\ 
Lu [$5d^1$] & $^{0.962}${\color{black}191}~\cite{RVSn2011} &  &  & $^1${\color{black} 191}~\cite{RMnSn1991,YMnGe-LuMnSn1993,RMnSn1999} & $^1${\color{black} 191}~\cite{RTGe1992,ZrLuScMnGe1999} & 71~\cite{RFeGa1990} & $^{0.5}${\color{black} 191}$^{c}$~\cite{RTGe1992,YLuHfFeGe1998,RFeGeSn2001}   \\ 
\hline
Ti [$3d^2$] & 191$^a$ &  &  &  &  &  & $^1$191~\cite{ScTiZrHfNb2000}   \\
\hline
\hline
\end{tabular}}
\begin{tablenotes}
\item $^a$ Structures obtained from Open Quantum Materials Database (OQMD)~\cite{OQMD2013};
\item $^b$ Structures obtained from Materials Project~\cite{MaterialsProject2013};
\item $^c$ occupation info obtained from ICSD, denoting the occupation of M atom at $1a(0,0,0)$ position. $0.5$ will lead to type-I disorder structure.
\end{tablenotes}
\label{tab:166_1}
\end{table*}

\begin{table*}[ht]\label{tab:SM_struct2}
\global\long\def\arraystretch{1.1}
\caption{Brief collection of MT$_6$Z$_6$ with space group info. Generally at room temperature, they can crystallize in space group 191 $P6/mmm$, 71 $Immm$, 65 $Cmmm$, 63 $Cmcm$, 62 $Pnma$.}
\centering
\setlength{\tabcolsep}{1mm}{
\begin{tabular}{rcccccccc} 
  \hline
  \hline
M(TZ)$_6$ & FeSn & FeAl & CoGa & CoGe & CoSn & NiSi & NiGe & NbSn \\
\hline
Li [$2s^1$]  &  &  &  & $^1${\color{black} 191}~\cite{CoFeNiGe1981} & 191$^a$ & $^1$191~\cite{CoFeNiGe1981} & $^1$191~\cite{CoFeNiGe1981} &   \\ 
\hline
Mg [$3s^2$]  & 191$^a$ &  &  & $^1${\color{black} 191}~\cite{CoFeNiGe1981,MgCoGe2006,MgCoGe2021} & 65$^a$ & $^{0.5}${\color{black} 191}~\cite{CoFeNiGe1981} & $^{0.5}${\color{black} 191}~\cite{CoFeNiGe1981} &   \\ 
Ca [$4s^2$]  &  &  &  & $^{0.5}${\color{black} 191}~\cite{CaFeCoGe2023}, 71~\cite{CaFeCoGe2023} &  &  &  &   \\ 
\hline
Sc [$3d^1$] & $^{0.85}${\color{black} 191}~\cite{RMnSnScFeSn1988} & 71$^b$ & 71$^b$ & $^1${\color{black} 191}~\cite{CoFeNiGe1981} & 65$^a$ & $^{0.5}${\color{black} 191}$^{c}$ & $^1$191~\cite{CoFeNiGe1981} &   \\ 
Y [$4d^1$] & $^{0.5}${\color{black} 191}$^{c}$, 71~\cite{RFeSn1991,RFeGeSn2001} & 71$^b$ & 71$^b$ & $^{0.5}${\color{black} 191}$^{c}$~\cite{CoFeNiGe1981} & $^{0.5}${\color{black} 191}$^{c}$~\cite{RCoSn1984,YCoSn2017} &  & 191$^a$ & $^1${\color{black} 191}~\cite{YNbSn2012}  \\ 
Lu [$5d^1$] & $^{0.5}${\color{black} 191}$^{c}$~\cite{RFeGeSn2001} &  &  & $^{0.5}${\color{black} 191}~\cite{CoFeNiGe1981} & $^{0.5}${\color{black} 191}$^{c}$~\cite{RCoSn1984} &  &  &   \\ 
\hline
Ti [$3d^2$] & 191$^a$ &  &  & $^1${\color{black} 191}~\cite{TiCoGe2019} & 65$^a$ &  & 191$^a$ &   \\
\hline
\hline
\end{tabular}}
\label{tab:166_2}
\end{table*}

\section{brief overview of results in each 1:6:6 family}\label{smssec:stat}

In this paper, we carry out a high-throughput first-principle study of 293 \ch{MT6Z6} compounds and 14 parent TZ structures in total, focusing on their phonon instability and band structures. Generally, the phonon spectrum provide valuable insights into the possible structural phase transitions that occur at low temperatures, often associated with the breaking of specific symmetries. Therefore, we expect the pristine structures of these compounds to exhibit a high-symmetry phase at higher temperatures and undergo a phase transition into a low-symmetry phase as temperature is lowered, accompanied by the breaking of certain symmetries, such as translational or rotational. The phonon spectra can help identify these temperature-induced phase transitions. It is worth noting that there have been recent reports of time-reversal-symmetry breaking below the CDW temperature, which is still under debate\cite{yang_giant_2020,yu_concurrence_2021,mielke_time-reversal_2022,hu_time-reversal_2023, xing2023optical}. 
Regarding the phonon instability, we conduct the phonon calculation at two smearing values to account for finite temperature effects, 0.4 eV for higher temperature and 0.05 eV for lower temperature. It is crucial to acknowledge that in this context, the temperature effect here primarily arises from the electronic contribution using smearing methods, which involve applying a Fermi-Dirac distribution. However, the actual temperature effect should also consider the contribution of ions. The thermal motion of ions can significantly impact the overall stability and behavior of the system at different temperatures. Such anharmonic effects have been included in some packages like SSCHA\cite{monacelli2021stochastic}, SCAILD\cite{souvatzis2008entropy}, and TDEP\cite{hellman2013temperature}, which require significant computation resources and are beyond the scope of this study.

\begin{table}[htbp]
\global\long\def\arraystretch{1.12}
\captionsetup{width=.95\textwidth}
\caption{Brief collection of phonon results for \ch{MT6Z6} compounds. HLS denotes stable phonon at both low-T and high-T, HSLU denotes stable phonon at high-T but unstable phonon at low-T, and HLU denotes unstable phonon at both high-T and low-T. Here, the finite temperature phonon are calculated at harmonic level via smearing approach. High-T and low-T denote a Fermi-Dirac smearing value of 0.4 eV and 0.05 eV, respectively.}
\label{Tab:SM_phononcollection}
\setlength{\tabcolsep}{1.mm}{
\begin{tabular}{c|c|l|c|l}
\hline
\hline
 & Count & Materials & Count & Materials  \\
\hline
\hline
\end{tabular}}
\end{table}
 
\subsection{Non-spin polarized calculation}\label{smssec:nonsp}

\begin{figure*}[ht]
\begin{center}
\includegraphics[width=0.9\linewidth]{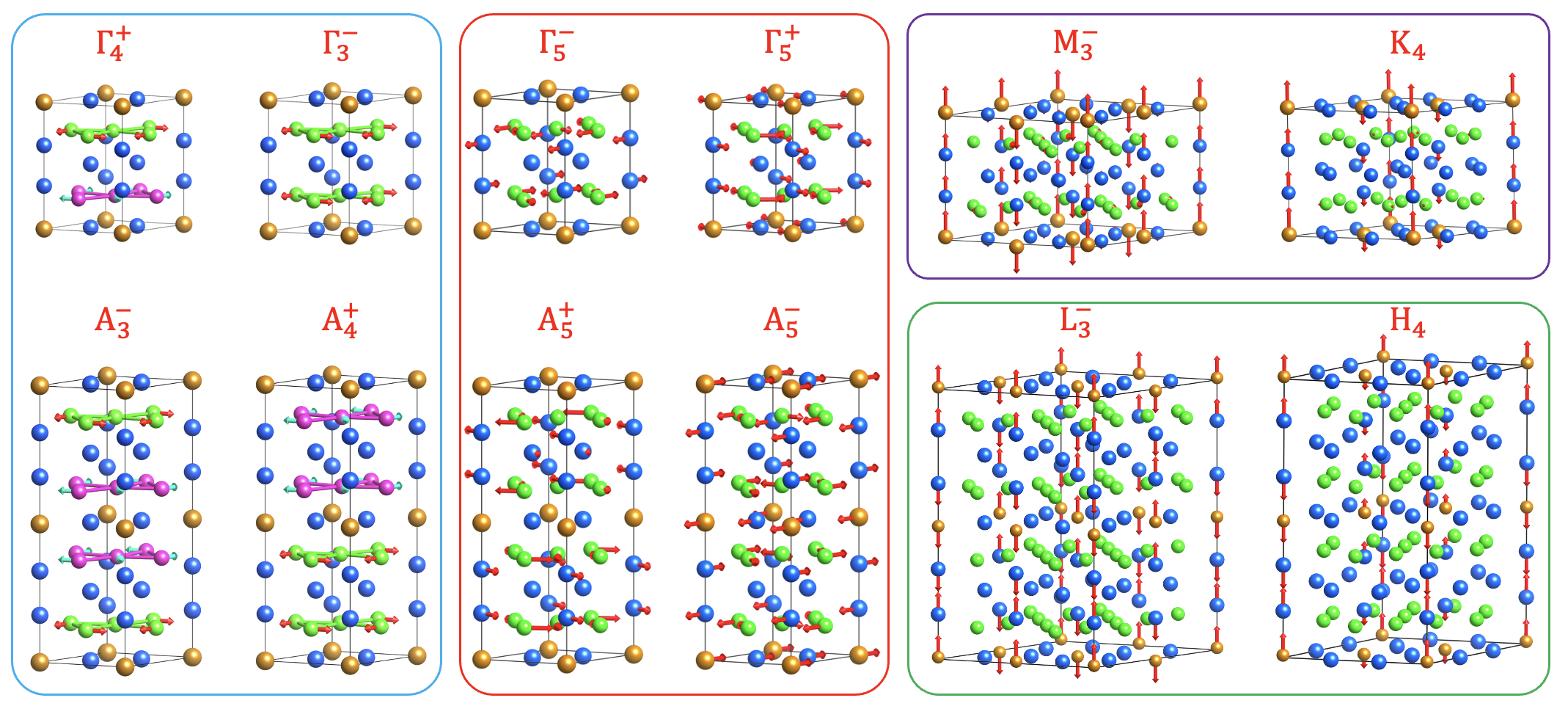}
\caption{Displacements corresponding to the representative imaginary phonons in \ch{MT6Z6} kagome compounds. Blue, purple and green blocks present the characteristic distortions observed in type-I, type-II, and type-III phonon instabilities at high symmetry points. Red block shows the possible distortions corresponding to the imaginary phonon of Mn/Fe-based 1:6:6 kagome compounds in their paramagnetic phase. However, these instabilities would be eliminated in the presence of magnetic orders.}
\label{Fig:dist_irrep}
\end{center}
\end{figure*}

Here, we briefly summarize the results, regarding the electronic features and phonon instabilities, for non-spin polarized calculations. 
A summary of phonon instabilities is listed in Tab.~\ref{Tab:SM_phononcollection}, obtained from non-spin-polarized calculations. Out of the 293 \ch{MT6Z6} compounds, 166 of them are found to be stable at both high-T and low-T, 24 are unstable at both high-T and low-T. The remaining 103 are stable at high-T but became unstable at low-T. As previously mentioned, the finite temperature effect via the smearing method does not correspond to the actual temperature. The instability of the 24 unstable at high-T compounds may potentially be stabilized at higher smearing temperatures. For instance, the main text presents an example of \ch{MgNi6Ge6} at higher smearing of 1.2 eV. In Fig.~\ref{Fig:dist_irrep}, typical distortions obtained from the imaginary phonons are presented. For type-I instability, the displacements corresponding to the imaginary modes at $\Gamma$ and $A$ are presented. These instabilities arise primarily from the in-plane movements of the kagome sites. For type-II instability, we display the displacements for $M_3^-$ and $K_4$ modes, which involve the out-of-plane movements in the trigonal M-Z-Z-M chains. Regarding type-III instability, we provide the displacements for $L_3^-$ and $H_4$ modes. In addition, the distorted suggested from the imaginary modes, $\Gamma_5^{\pm}$ and $A_5^{\pm}$, in Mn/Fe-166 are also presented. However, these distortions would be completely eliminated with the inclusion of magnetic order as demonstrated in Sec.~\ref{smssec:sp_MnFe}. The imaginary modes for Mn/Fe-166 primarily involve the in-plane vibrations of all sites. 

Throughout this SM, we use `kagome flat bands' to refer to the flat bands that exhibit the largest DOS peak. Remark that there can be additional flat segments of bands in the band structures that contribute to smaller peaks in DOS.

\textbf{\hyperref[smssec:VGe]{VGe}/\hyperref[smssec:VSn]{VSn}/\hyperref[smssec:NbSn]{NbSn}-166}
In \hyperref[smssec:VGe]{VGe}-166 class, the electronic kagome flat bands exists at $\sim 1.8$ eV above the Fermi level and contribute to the prominent DOS peak. Correspondingly, van Hove singularities (vHSs) may appear close to the Fermi level. For example,%
in \ch{LiV6Ge6}, multiple vHSs are observed in the vicinity of the Fermi level, primarily originating from V $3d$ orbitals. The position of the vHS related to the Fermi level could be modulated by adjusting the filling of the system through the choice of the M element. For example, in \ch{ScV6Ge6} and \ch{ZrV6Ge6}, the presence of a vHS precisely at the Fermi level can be achieved by selecting Sc and Zr as the M element, respectivel. 
According to the phonon calculations in VGe-166, the compounds \ch{MgV6Ge6} and \ch{CaV6Ge6} are the only ones found to be unstable. These instabilities occur specifically on the $k_z=\pi$ plane and belong to type-III instability, which is primarily characterized by the out-of-plane movements of trigonal Mg/Ca and Ge atoms. In the case of \ch{CaV6Ge6}, the unstable phonon mode shows a tendency to harden at high temperatures, indicating a potential stabilization of the instability with increasing temperature.On the other hand, for \ch{MgV6Ge6}, the observed instability may require even higher temperatures in order to stabilize. %

In \hyperref[smssec:VSn]{VSn}-166 class, the flat bands are typically found to be situated around $1.4$ eV above the Fermi level, resulting in vHSs close to the Fermi level. This particular class has attracted significant attention due to the recent discovery of charge density wave in \ch{ScV6Sn6}, characterized by the out-of-plane movement within the trigonal Sc-Sn chains. There have been reports suggesting that the distinct atomic size at the M sites %
may play a role in the emergence of charge density waves. Notably, doping with either Y or Lu has been shown to suppress the CDW in \ch{ScV6Sn6}~\cite{meier_tiny_2023}. Additionally, external pressure can also be applied to suppress the CDW in doped compounds~\cite{meier_tiny_2023}. %

\begin{figure}[htbp]
\begin{center}
\includegraphics[width=0.45\textwidth]{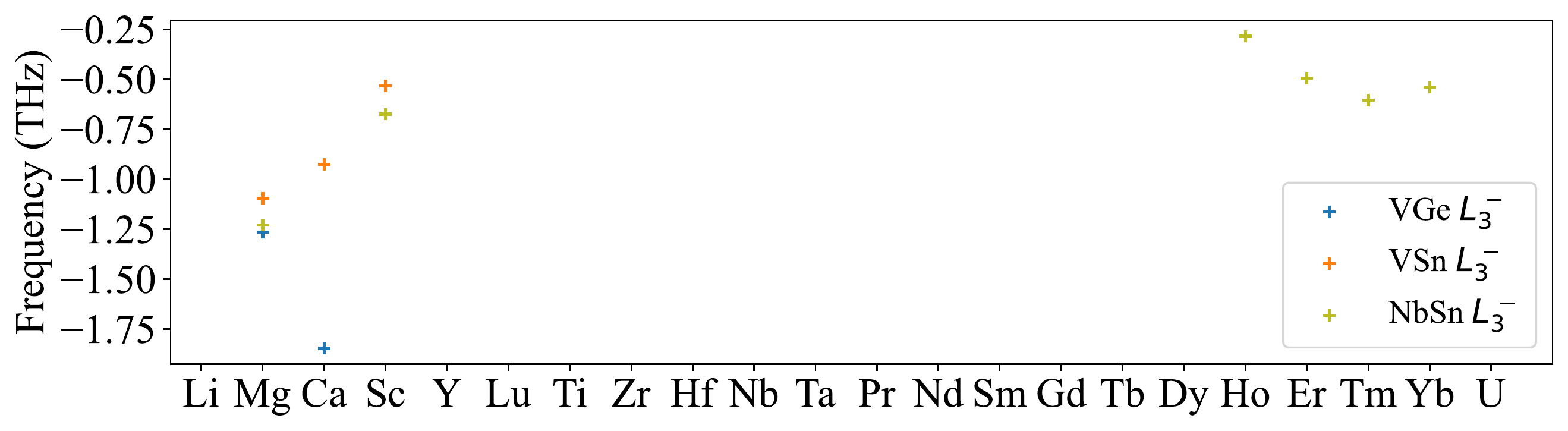}
\includegraphics[width=0.45\textwidth]{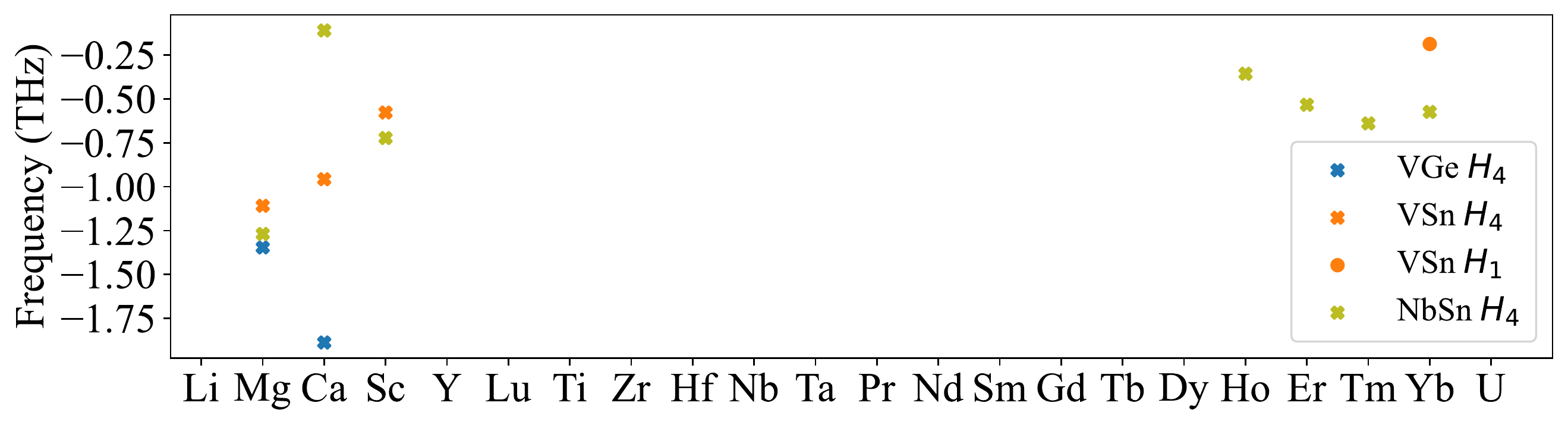}
\caption{Imaginary phonon at $L$ and $H$ points in VGe/VSn/NbSn-166.}
\label{Fig:ImagPho_VNb}
\end{center}
\end{figure}

In \hyperref[smssec:NbSn]{NbSn}-166 class, Nb $4d^3$ forms the kagome nets, instead of V $3d^3$. With the same valence electrons, Nb-based kagome compounds present similar band fillings and multiple vHSs near the Fermi level as V-based kagome. However, due to the larger bandwidth of $4d$ electrons, the unfilled kagome flat bands are located at approximately $2.1$ eV above the Fermi level compared to around $1.4$ eV for V-based kagome compounds. %
The phonon instability on the $k_z=\pi$ plane indicates similar distortion patterns as observed in V-166 compounds, indicating comparable behavior in terms of structural instability. Besides, additional degenerate instabilities are also presented at the $\Gamma-A$ path when M=Li and Mg. These instabilities are dominated by the in-plane vibrations of M atoms, which are shown in Fig.~\ref{Fig:dist_irrep} for $\Gamma_5^-$ and $A_5^-$ modes. %

\textbf{\hyperref[smssec:CrGe]{CrGe-166}}
In this class, the kagome flat bands are found to be situated at around $0.8$ eV above the Fermi level. However, there is an additional flat band from $d_{z^2}$ orbital located near the Fermi level. For example, in \ch{LiCr6Ge6} the flat band from $d_{z^2}$ is located exactly at the Fermi level, resulting in a small DOS peak. Nevertheless, the phonon spectrum does not show any instabilities. Unlike Mn/Fe-166, which features a larger DOS peak that leads to magnetism, this small DOS peak in \ch{LiCr6Ge6} does not give rise to magnetism. 

\textbf{\hyperref[smssec:CoGe]{CoGe}/\hyperref[smssec:CoSn]{CoSn}-166}
Many of the \hyperref[smssec:CoGe]{CoGe}-166 cell is reported to present disorder at the trigonal M-Ge chain, resulting in a half-$c$ unit cell. Thus, they are also referred to as M$_{0.5}$Co$_3$Ge$_3$ in the reduced unit cell, such as Y$_{0.5}$\ch{Co3Ge3} and Yb$_{0.5}$\ch{Co3Ge3}. In CoGe-166, the kagome flat bands are fully occupied and located approximately $-0.8$ eV below the Fermi level. However, near the Fermi level on the $k_z=\pi$ plane, one can still observe a van Hove singularity and a segment of flat bands. 
Meanwhile, two Dirac cones can be found at the H point in close proximity to the Fermi level. %
Phonon calculations indicate the widespread presence of instabilities among CoGe-166 compounds, with imaginary modes occurring along the $\Gamma-A$ path.  Atomic projections suggest that the imaginary modes are dominated by in-plane vibrations within the Co kagome nets. 

For \hyperref[smssec:CoSn]{CoSn}-166, the replacement of Ge by isovalent Sn leads to the expansion of unit cell due to larger atomic size, as indicated by the lattice parameters shown in Fig.~\ref{Fig:parameter}. However, no instabilities are observed in CoSn-166. %
Nevertheless, one can still identify in-plane vibration modes in kagome nets, which lie just above the acoustic modes, such as $\Gamma_4^+$ mode at 1.710 Hz in \ch{YCo6Sn6}. Such a mode corresponds to the local rotation of Co triangles, which governs the distortions in CoGe-166. %

\begin{figure}[htbp]
\begin{center}
\includegraphics[width=0.45\textwidth]{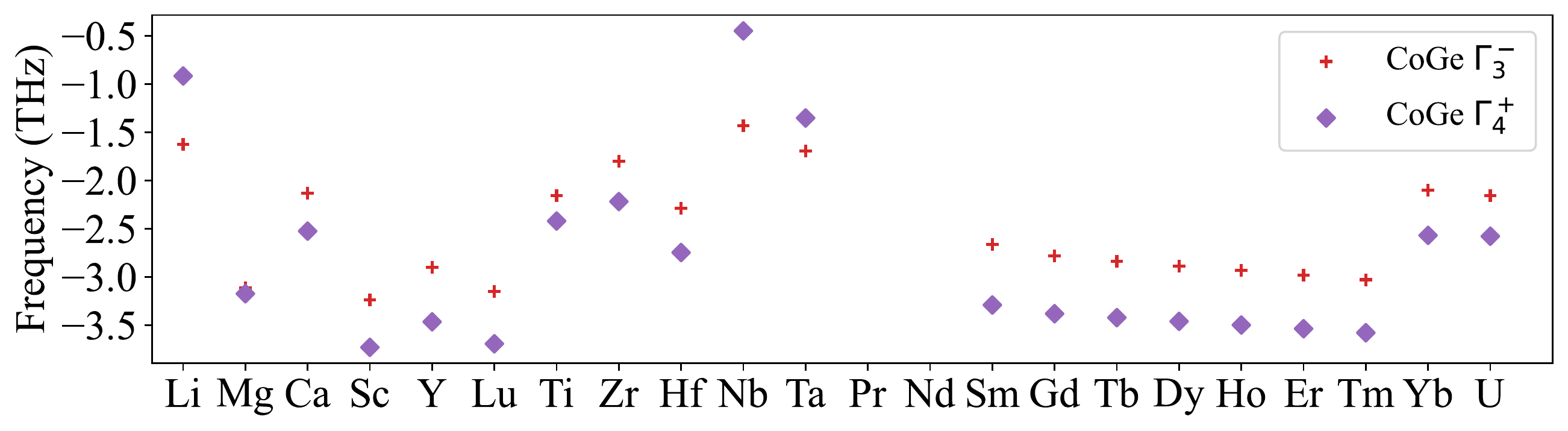}
\includegraphics[width=0.45\textwidth]{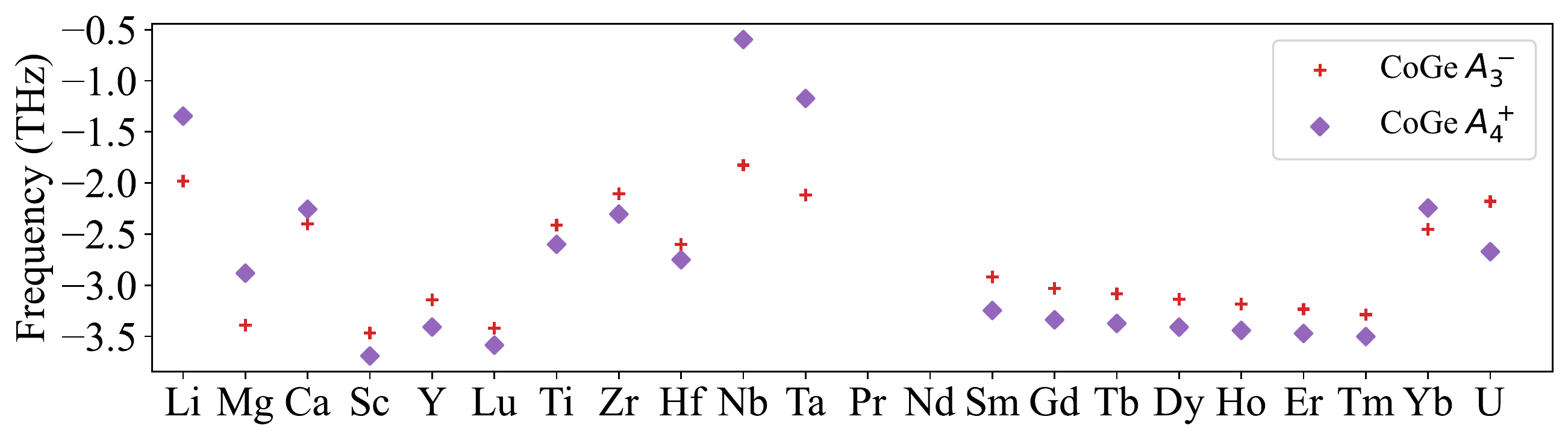}
\caption{Imaginary phonon at $\Gamma$ and $A$ points in CoGe-166. Generally, one can observe two close imaginary modes at $\Gamma$ and $A$ points, corresponding to two close imaginary branches in the phonon spectra.}
\label{Fig:ImagPho_CoGe}
\end{center}
\end{figure}

\begin{figure}[htbp]
\begin{center}
\includegraphics[width=0.85\textwidth]{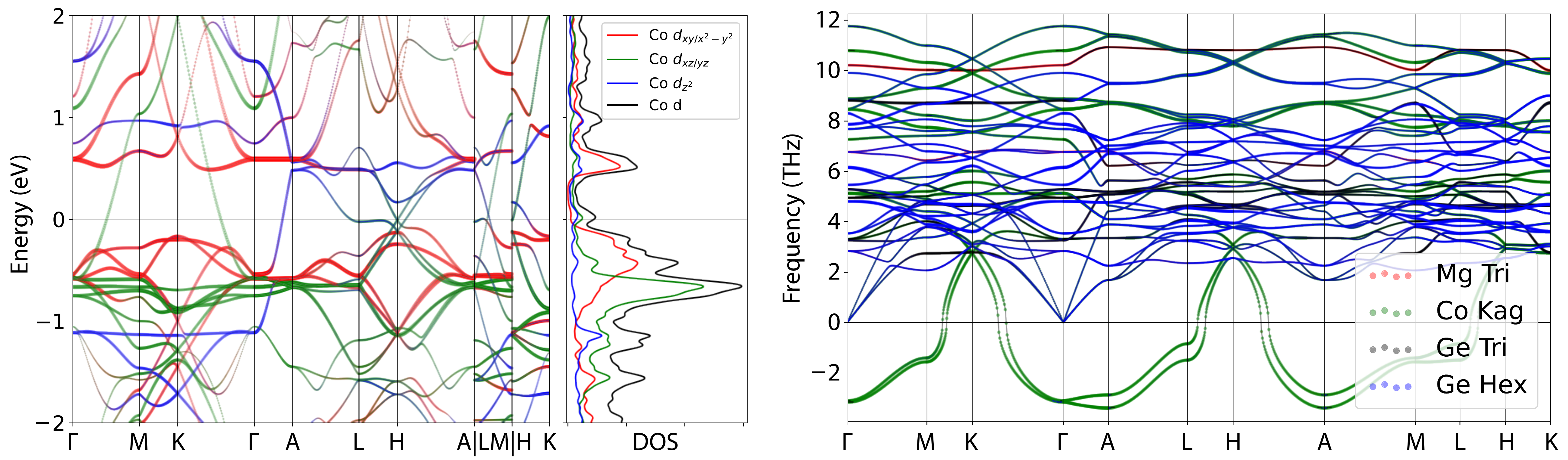}
\caption{Band structure and low-T phonon spectrum for \ch{MgCo6Ge6}. The band structure reveals a pronounced peak at $\sim-0.7$ eV below the Fermi surface, contributed by Co $d_{xy/x^2-y^2}$ and $d_{xz/yz}$ orbitals. Near the Fermi level, one can find vHS at M point and Dirac points at K points near the Fermi level. Meanwhile, \ch{MgCo6Ge6} exhibit two widespread imaginary branches in the phonon spectrum, governed by kagome Co atoms. }
\label{Fig:MgCoGe}
\end{center}
\end{figure}

Taking \ch{MgCo6Ge6} as an example, the band structure and phonon spectrum are presented in Fig.~\ref{Fig:MgCoGe}. Notably, two significant imaginary bands are observed along the high-symmetry paths connecting the $\Gamma$, A, M, and L points. 
At $\Gamma$ and A points, there are lower four modes identified as $\Gamma_4^+$, $\Gamma_3^-$, $A_3^-$, and $A_4^+$.
Based on the IRREPs, one can deduce the potential distorted subgroups associated with these modes, as listed in Tab.~\ref{Tab:subgroup_irrep}. %

\begin{table*}[htbp]
\global\long\def\arraystretch{1.12}
\captionsetup{width=.75\textwidth}
\caption{Possible subgroups of 191 $P6/mmm$ driven by imaginary phonon modes in \ch{MgCo6Ge6}.}
\label{Tab:subgroup_irrep}
\setlength{\tabcolsep}{0.5mm}{
\begin{tabular}{c|c|c||c|c|c}
\hline
\hline
 $f$(THz) & IRREP & Subgroup & $f$(THz) & IRREP & Subgroup   \\
\hline
$-3.171$ & $\Gamma_4^+$ & 162~$P\bar{3}1m$  & $-1.573$ & $M_1^-$ & 50~$Pban$ \\
 $-3.112$ & $\Gamma_3^-$ & 189~$P\bar{6}2m$ & $-1.414$ & $M_3^+$ & 53~$Pmna$ \\
$-3.393$ & $A_3^-$ & 193~$P6_3/mcm$ & $-1.503$ & $L_3^+$ & 74~$Imma$ \\
$-2.282$ & $A_4^+$ & 193~$P6_3/mcm$ & $-0.856$ & $L_1^-$ & 72~$Ibam$ \\
\hline
\hline
\end{tabular}}
\end{table*}

\begin{figure}[htbp]
\begin{center}
\includegraphics[width=0.85\textwidth]{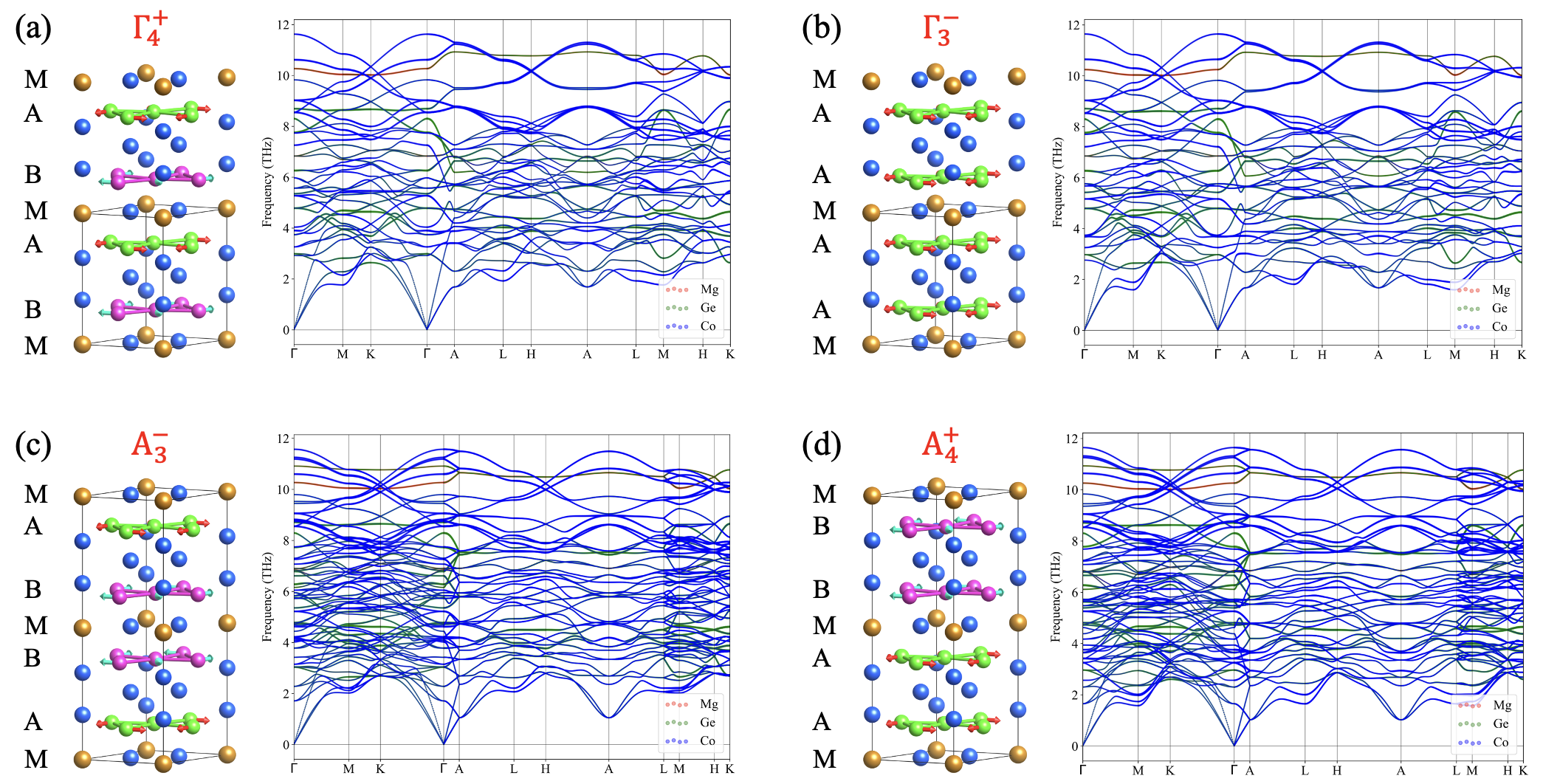}
\caption{Possible stable CDW structures derived by imaginary phonon at $\Gamma$ and A points in \ch{MgCo6Ge6}. As presented in the main text, the distortions are mainly dominated by the in-plane local rotation of the small triangles in the kagome nets. Marking the kagome layer of one rotating direction as A, the kagome layer of the opposite direction as B, the distorted structures can then be marked as MBAM, MAAM, MABMBAM, and MAAMBBM, derived from $\Gamma_4^+$, $\Gamma_3^-$, $A_3^-$, and $A_4^+$, respectively, with M for the plane of M site. The solid lines denote the unit cell. Here, the $\Gamma_i$ modes don't break the translational symmetry, preserving the unit cell, while the $A_i$ modes break the translational symmetry, leading to the doubling along $c$ direction.}
\label{Fig:CDW_MgCoGe}
\end{center}
\end{figure}

Based on the given four unstable IRREPs, the corresponding displaced structures can be obtained, as depicted in Fig~\ref{Fig:CDW_MgCoGe}. %
While the $\Gamma$ modes preserve the unit cell, the A modes would double the unit cell along $c$ direction, leading to the $1\times1\times2$ order. Atomic projections indicate that all the imaginary phonon modes primarily arise from the vibrations of the kagome Co atoms. Further displacement analysis shows that all these imaginary modes primarily involve the in-plane vibrations of the kagome Co layer. The in-plane displacement induce distortions in the kagome Co sublattices, breaking the $C_6/C_2$ symmetry while the small triangles are locally twisted as presented in the main text. %
The structure derived from $\Gamma_3^-$ mode corresponds to a simple stacking of the distorted kagome layer along the $c$ direction, as presented in Fig.~\ref{Fig:CDW_MgCoGe}, which can be denoted as MAAM stacking with A representing the distorted kagome layer and M representing the plane of trigonal M sublattice. In the case of the $\Gamma_4^+$-derived structure, the distortion in the two kagome layers are in opposite directions, and this can be denoted as MBAM stacking. Similarly, the structures derived from A modes are denoted as MAAMBBM and MABMBAM for $A_4^+$ and $A_3^-$, respectively, which exhibit a doubling along the $c$ direction due to the broken translational symmetry. %

\begin{figure}[htbp]
\begin{center}
\includegraphics[width=0.4\textwidth]{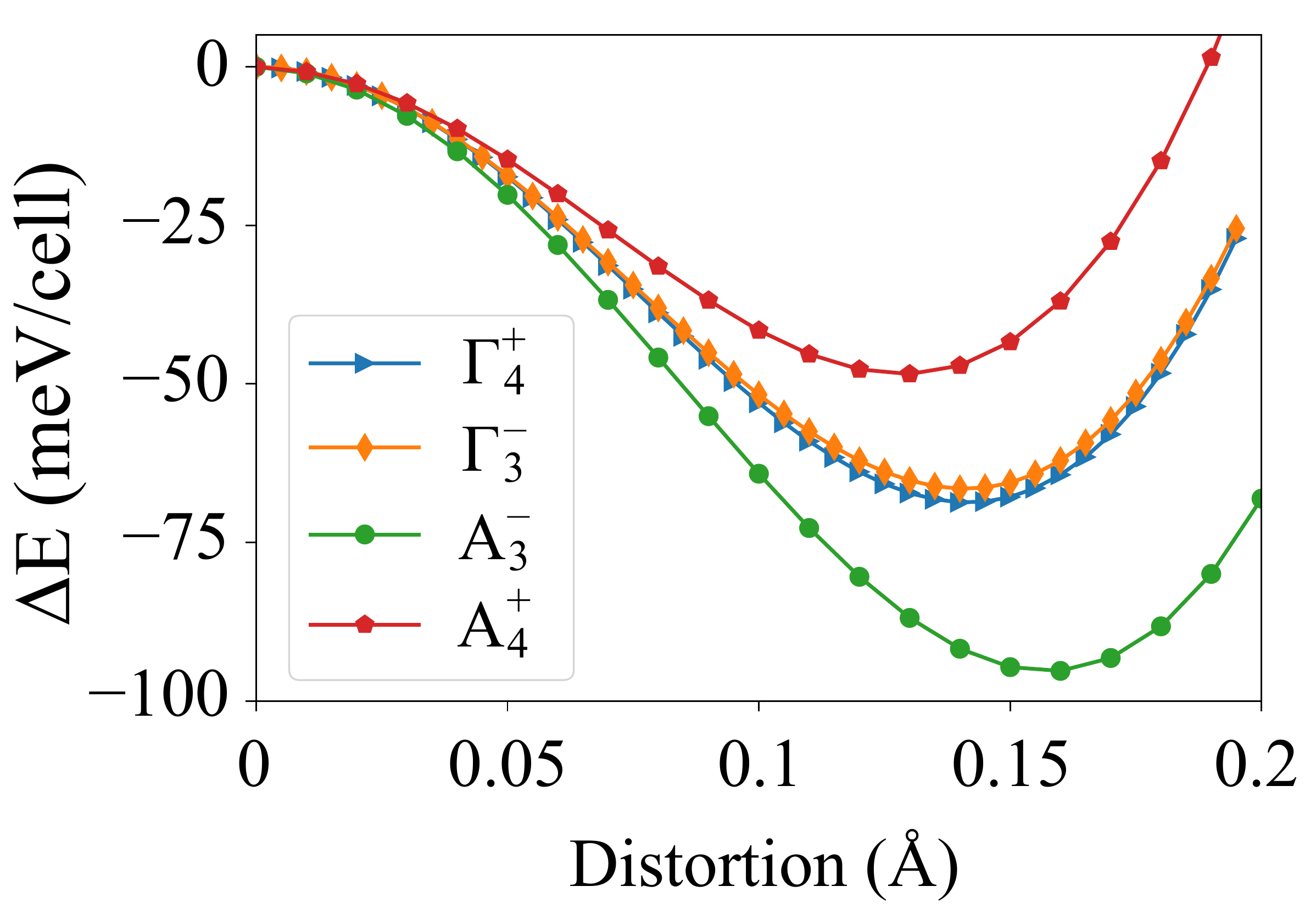}
\caption{Energy evolution of derived CDW structures as a function of the displacement of Co atoms, referred to the energy of the pristine unit cell. Results favor the CDW structure derived from $A_3^-$ mode, which matches the experimental report~\cite{MgCoGe2021}.}
\label{Fig:E_cdw_CoGe}
\end{center}
\end{figure}

The energy evolution reveals that the $A_3^-$ structure is the energetically favored structure, as illustrated in Fig.~\ref{Fig:E_cdw_CoGe}. This finding aligns with the reported phase transition in \ch{MgCo6Ge6}~\cite{MgCoGe2021}, where a locally twisted structure with a double-$c$ supercell is presented. However, it is worth noting that no clear signals of this phase transition are observed in their heat and resistivity measurements. %

\textbf{\hyperref[smssec:NiSi]{NiSi}/\hyperref[smssec:NiGe]{NiGe}/\hyperref[smssec:NiSn]{NiSn}-166}
In \hyperref[smssec:NiSi]{NiSi}/\hyperref[smssec:NiGe]{NiGe}/\hyperref[smssec:NiSn]{NiSn} class, the flat bands are fully occupied and positioned below the Fermi level. However, despite being filled, vHSs and segments of flat bands can still be observed near the Fermi level. For example, in \ch{YNi6Si6}, vHSs arising from the $d_{xy/x^2-y^2}$ and $d_{xz/yz}$ orbitals are present in proximity to the Fermi level, accompanied by Dirac points located at $K$ points.

\begin{figure}[htbp]
\begin{center}
\includegraphics[width=0.45\textwidth]{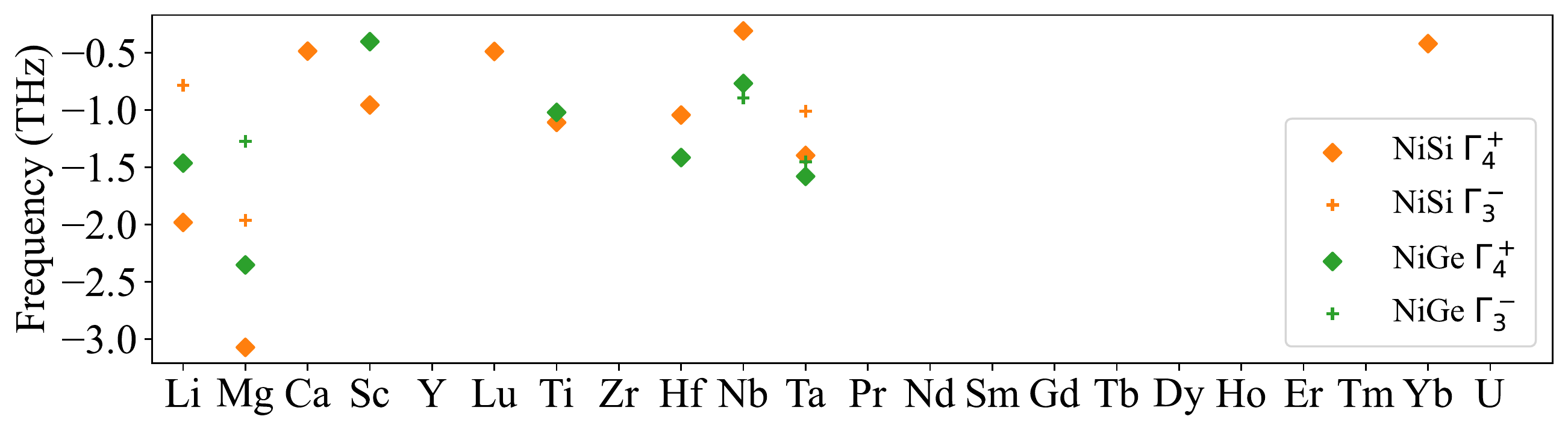}
\includegraphics[width=0.45\textwidth]{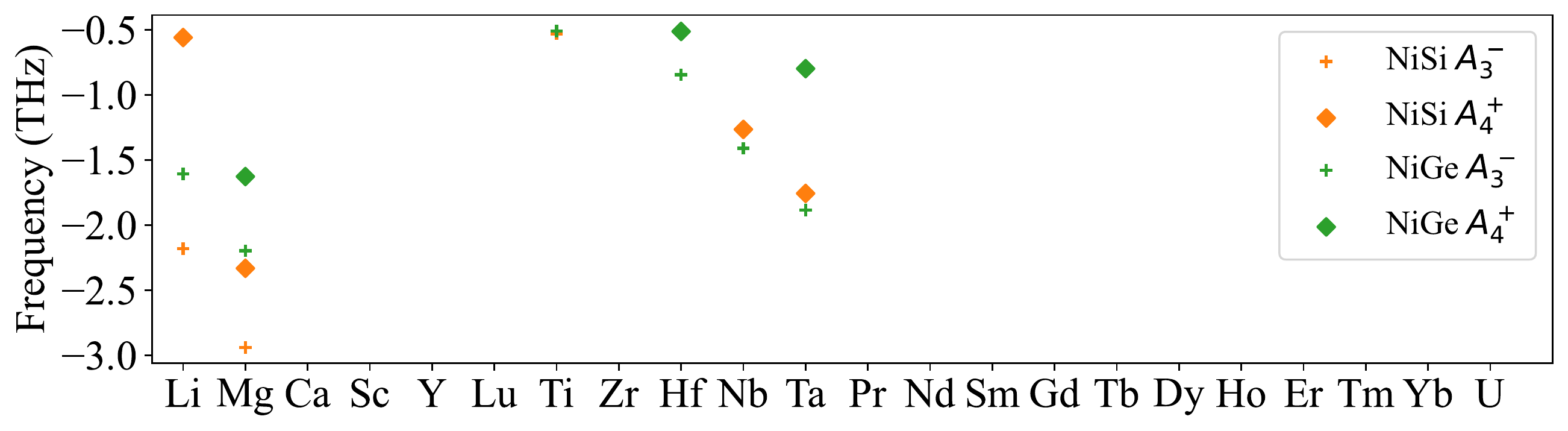}
\caption{Imaginary phonon at $\Gamma$ and $A$ points in NiSi/NiGe-166 while NiSn-166 compounds show no instabilities.}
\label{Fig:ImagPho_Ni}
\end{center}
\end{figure}

While these compounds exhibit similar band features near the Fermi level, phonon instabilities are only found in NiSi/NiGe-166 compounds. %
Fig.~\ref{Fig:ImagPho_Ni} provides an overview of the instabilities observed in these compounds. One can observe the hardening %
trend of imaginary phonon from Si to Ge, and their eventual disappearance in Sn ones.

\textbf{\hyperref[smssec:NiIn]{NiIn-166}}
In \hyperref[smssec:NiIn]{NiIn} class, similar electronic structures are observed, with all flat bands below the Fermi level being fully occupied. Similarly, vHSs can be found near the Fermi level, along with segments of flat bands. Though these electronic features resemble those seen in other Ni-166 compounds with one less valence electron, the instabilities exhibit distinct characteristics, occurring specifically on the $k_z=0$ plane. %
Atomic projections also indicate the dominance of trigonal M and In atoms in band weight, rather than the kagome Ni atoms. The vibration modes observed in these compounds suggest coordinated movements of the M and In atoms in the same direction within the M-In chain. The out-of-phase movement of neighboring chains can result in the breaking of rotational symmetry. Notably, in some compounds, the leading instability may appear at a generic $\bm{q}$ point instead of high symmetry points. This highlights the importance of considering the full Brillouin zone and the possibility of unconventional instabilities occurring at non-high-symmetry points.

\begin{figure}[htbp]
\begin{center}
\includegraphics[width=0.45\textwidth]{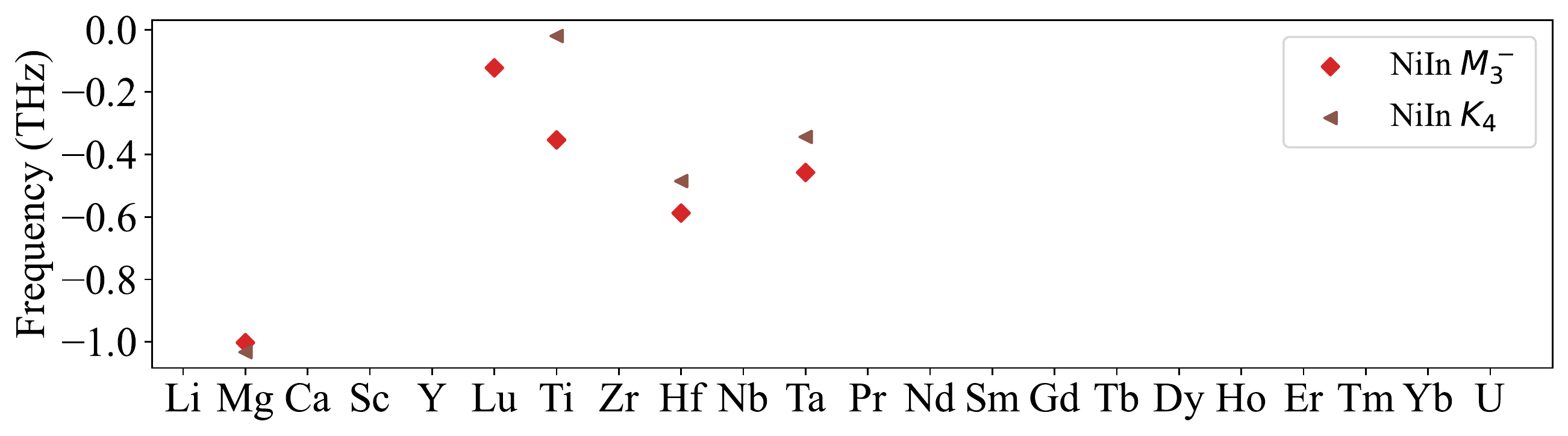}
\caption{Imaginary phonons at M and $K$ points in NiIn-166.}
\label{Fig:ImagPho_NiIn}
\end{center}
\end{figure}

\subsection{Spin-polarized calculations}\label{smssec:sp_MnFe}

In this section, we summarize the results for Mn/Fe-166 compounds both without and with magnetic orderings. However, there are several important points that need clarification regarding our calculations with magnetic ordering. Firstly, when considering rare-earth elements, we do not include the $4f$ electrons in our calculations. Therefore, our focus will be solely on the magnetic ordering of the kagome $d$ orbitals. Additionally, experimental measurements have also reported the presence of diverse magnetic structures even in the absence of $4f$ elements at lower temperatures, such as the noncollinear double cones and flat spiral magnetic structures~\cite{baranov_magnetism_2011,zhou_metamagnetic_2023}. One particular magnetic arrangement is the collinear ($+--+$) AFM order, which doubles the unit cell along the $c$ direction to incorporate four kagome layers in the magnetic unit cell. However, for the sake of simplicity, we only consider the magnetic orderings ($+-$) instead of the more complex ($+--+$) arrangement. %
No extra Hubbard U is considered in current calculations.

\begin{figure}[htbp]
\begin{center}
\includegraphics[width=0.6\textwidth]{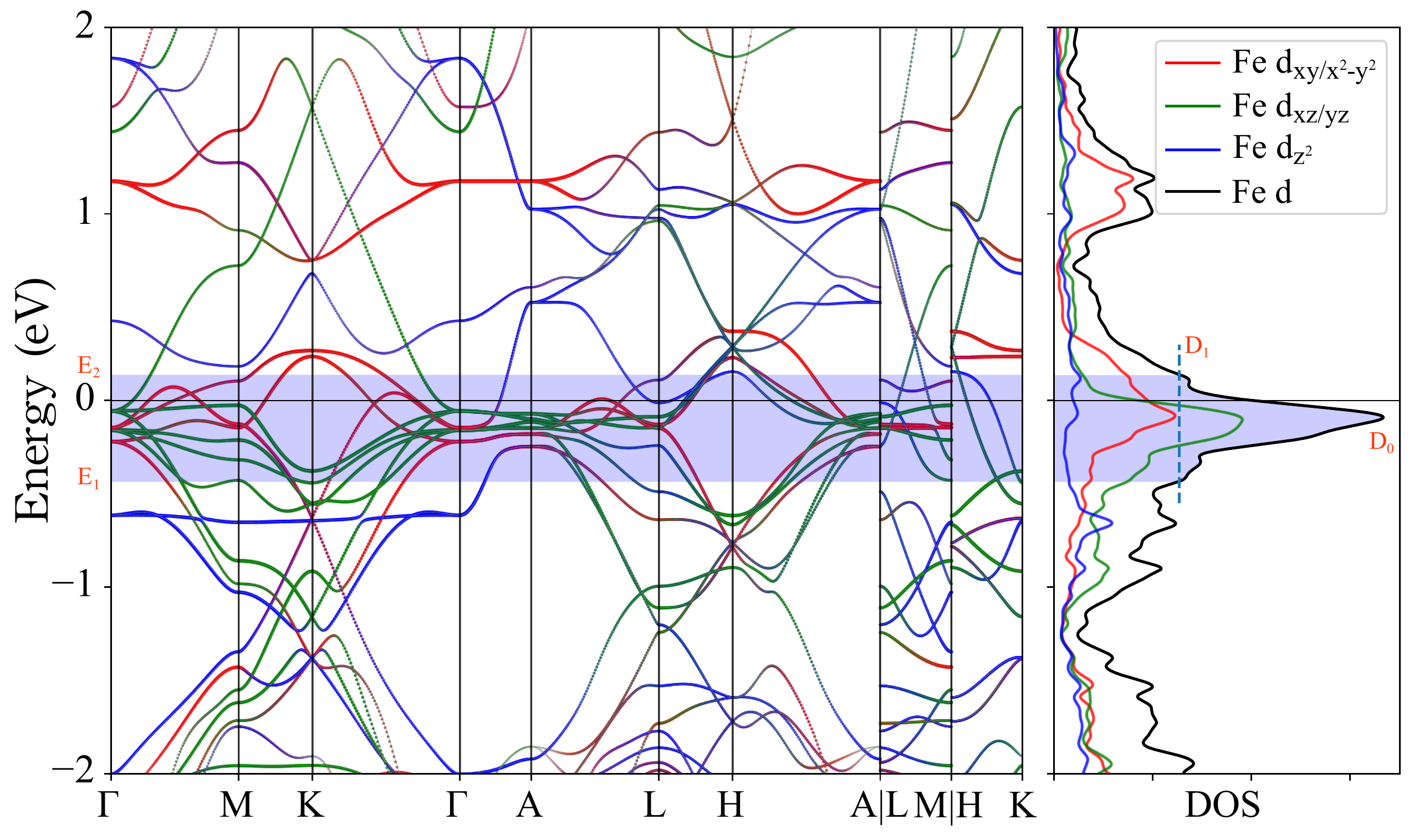}
\caption{Fillings of flat bands estimated from DOS by taking a DOS cutoff at $D_1 = \beta D_0$ in the energy range from $E_1$ to $E_2$, where $D_0$ represents the peaked DOS. The filling of flat bands can be estimated from the portion below the Fermi level.}
\label{Fig:filling}
\end{center}
\end{figure}

\begin{figure}[htbp]
\begin{center}
\includegraphics[width=0.45\textwidth]{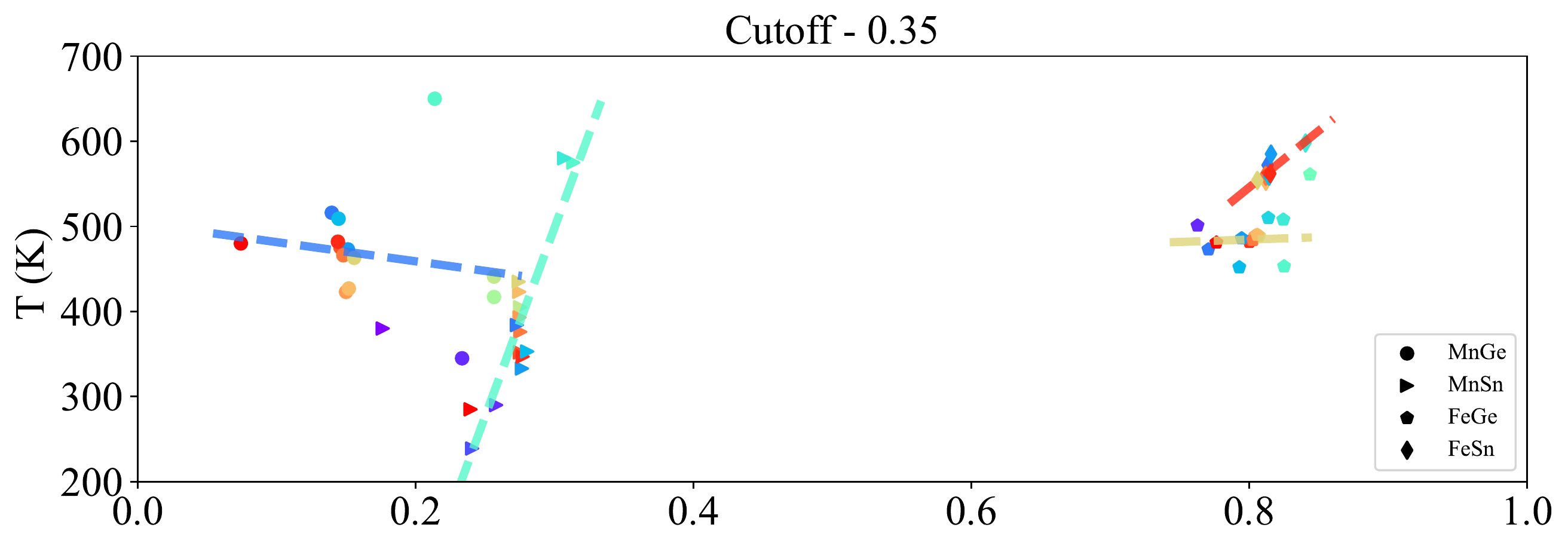}
\includegraphics[width=0.45\textwidth]{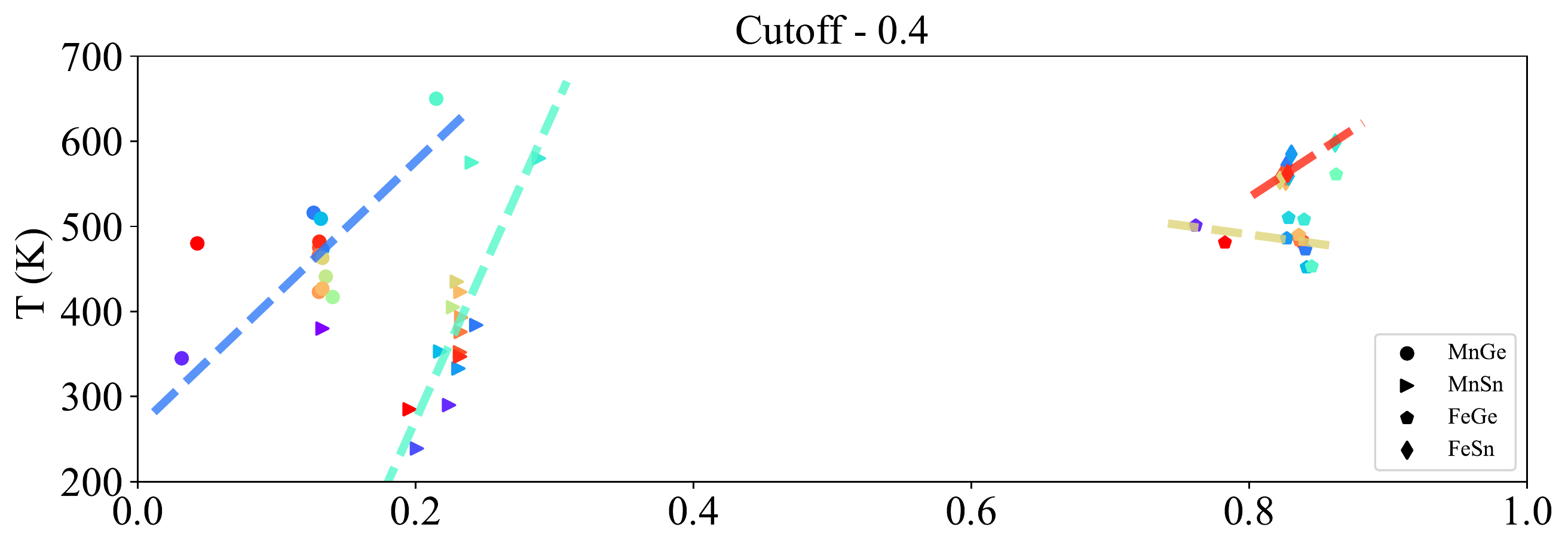}\\
\includegraphics[width=0.45\textwidth]{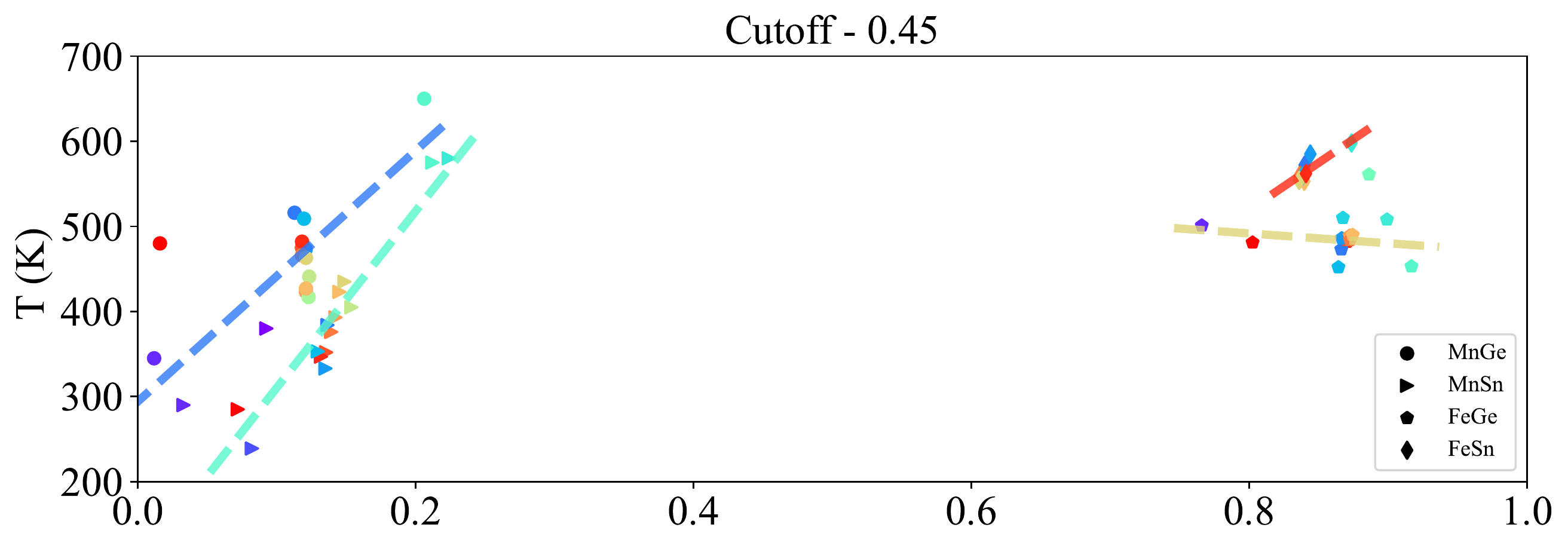}
\includegraphics[width=0.45\textwidth]{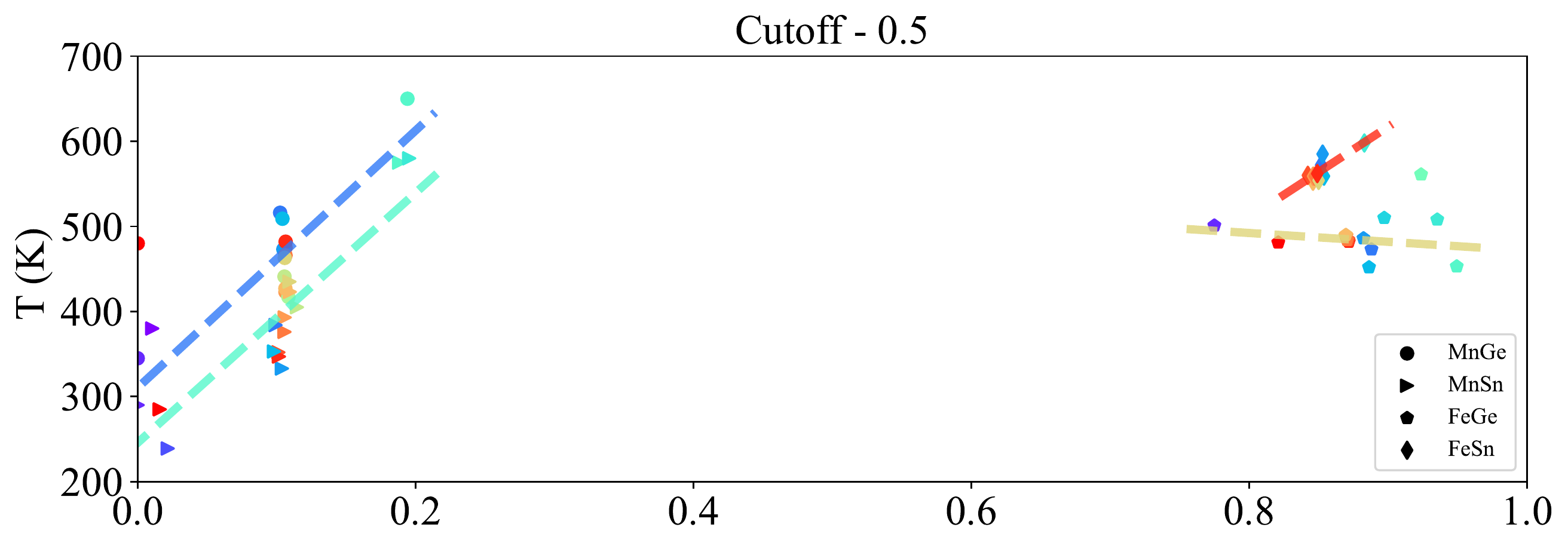}
\caption{Fillings of flat bands for different cutoff with $\gamma=0.35,0.5,0.45,0.5$. Linear fittings are also presented as dashed lines for each class. With the exception of FeSn-166, all other three classes suggest a higher transition temperature when they approach half filling. The deviation in FeSn-166 could be attributed to the limited data sample available for analysis. Meanwhile, these compounds may also crystallize in superstructures with space groups other than SG 191, which can also contribute to the deviation. Note that \ch{NbFe6Ge6} is not included in the fitting process for FeGe-166.}
\label{Fig:ff}
\end{center}
\end{figure}

\begin{figure}[htbp]
\begin{center}
\includegraphics[width=0.45\textwidth]{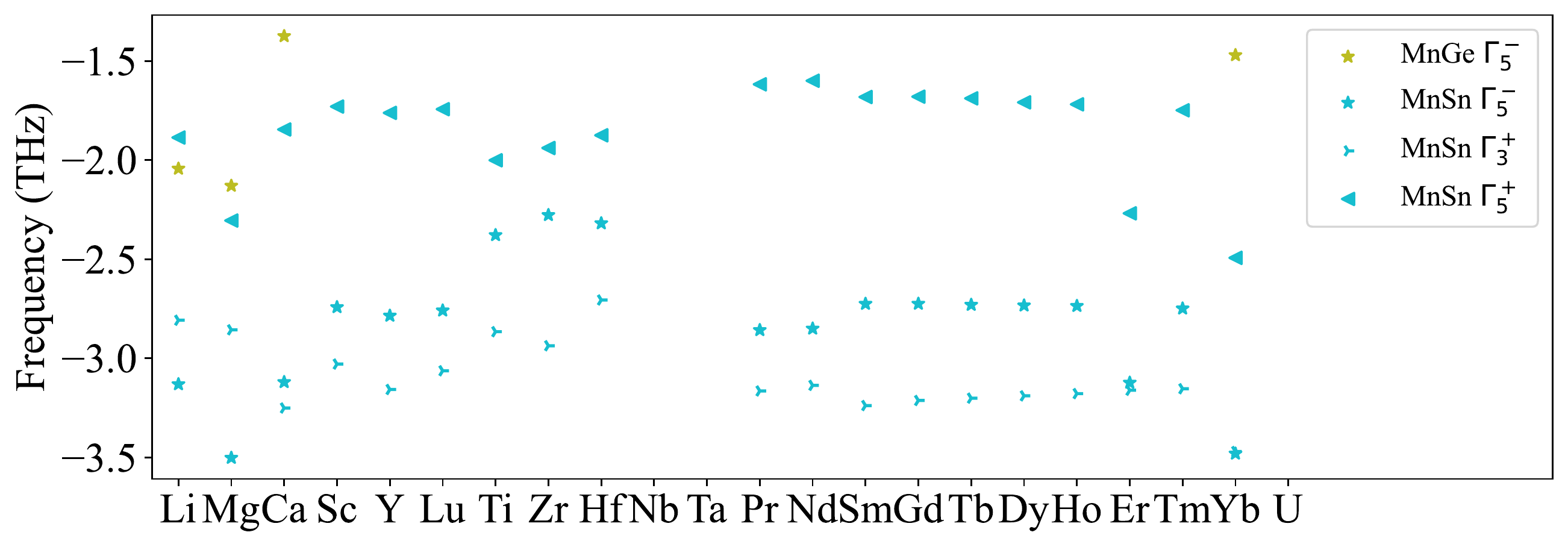}
\includegraphics[width=0.45\textwidth]{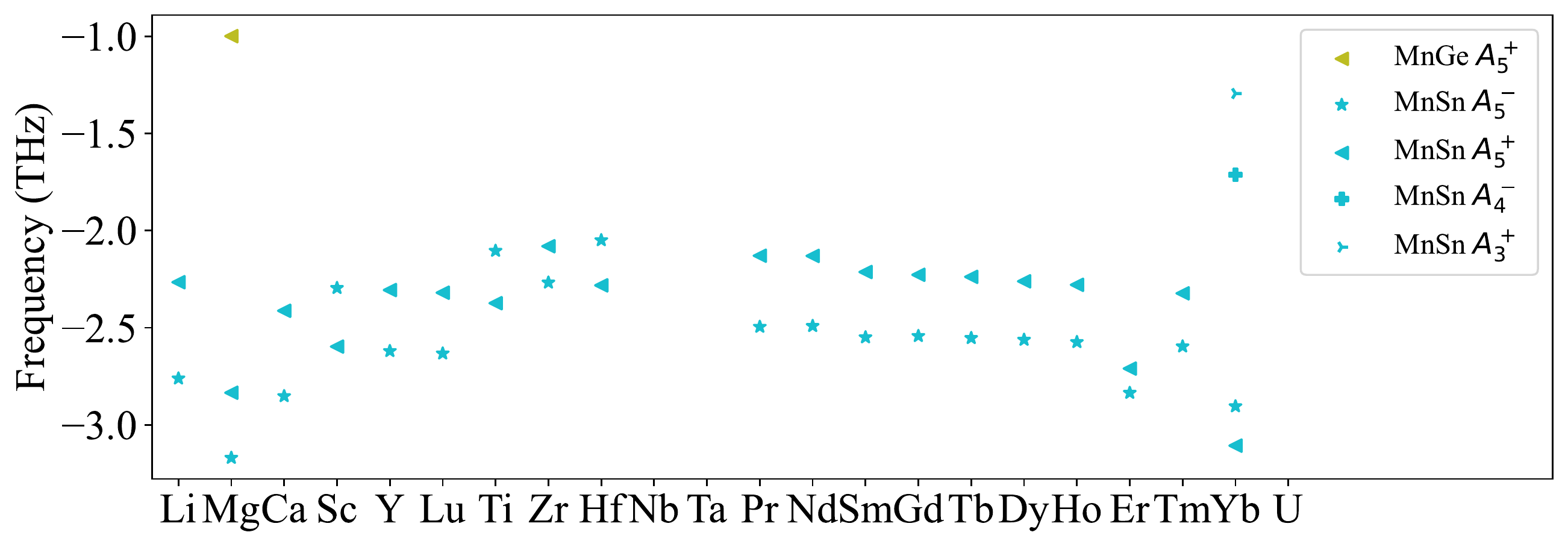}\\
\includegraphics[width=0.45\textwidth]{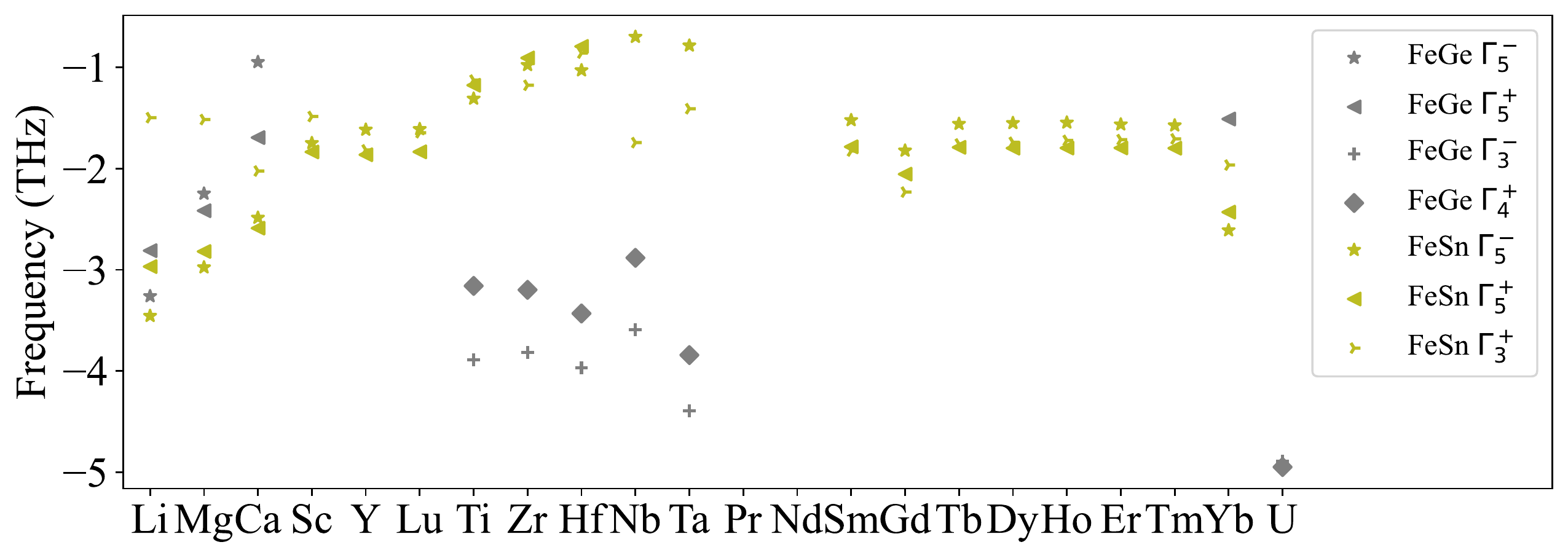}
\includegraphics[width=0.45\textwidth]{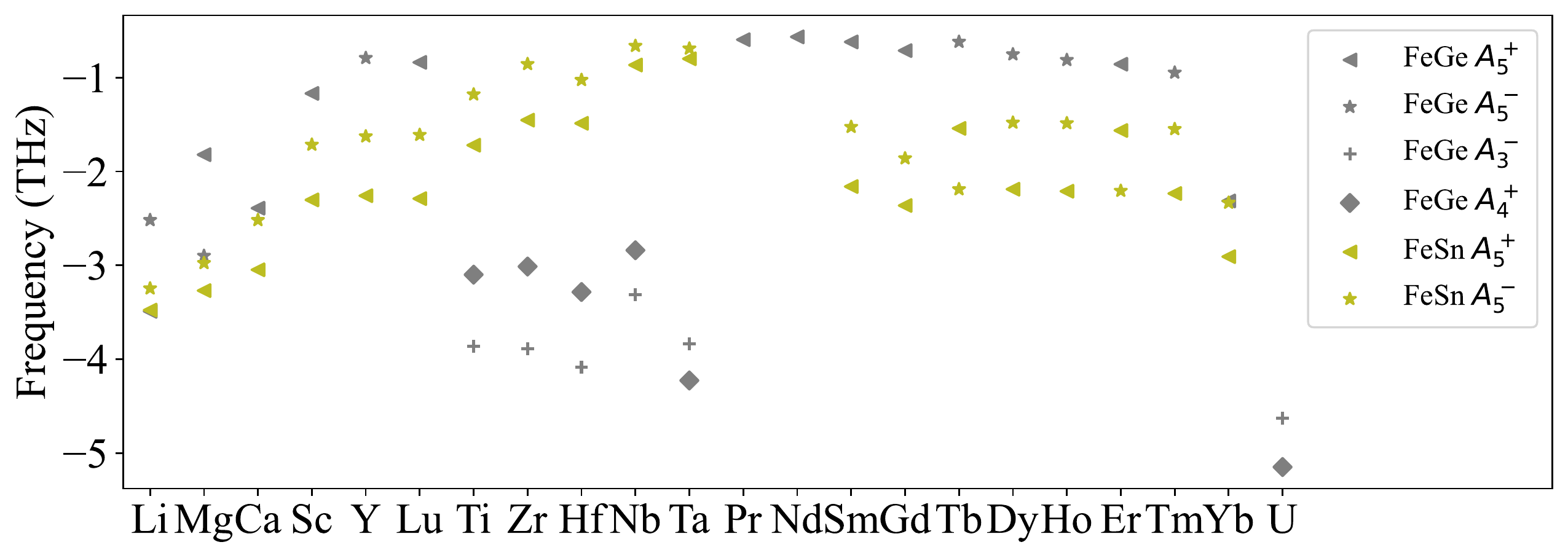}
\caption{Imaginary phonon at $\Gamma$ and $A$ points in Mn/Fe-166.}
\label{Fig:ImagPho_MnFe}
\end{center}
\end{figure}

For Mn/Fe-166, a notable feature is the presence of partially filled kagome flat bands near the Fermi level in the paramagnetic phase, characterized by the pronounced DOS peak near the Fermi level, which is mainly contributed by $d_{xy/x^2-y^2}$ and $d_{xz/yz}$ orbitals, as shown in Fig.~\ref{Fig:filling}. The fillings of the flat bands can be estimated from the DOS by introducing a cutoff at $D_1 = \beta D_0$, where $D_0$ represents the peaked DOS. The distribution of the flat bands from the $d_{xy/x^2-y^2}$ and $d_{xz/yz}$ orbitals is mainly confined between energies $E_1$ and $E_2$, with a small portion of dispersive bands. Then the filling of flat bands from $d$ can be estimated from the portion below the Fermi level. Note that the flat bands here are not perfectly flat with zero bandwidth as the ideal flat band in the three-band kagome model consists solely of $s$-orbital with nearest neighbor hopping  or longer range very fine-tuned hoppings. Additional hoppings, such as next-nearest hoppings and interlayer hoppings, can easily destroy the perfect flatness. Furthermore, the $d_{z^2}$ orbital has a minor influence on the fillings due to its significantly smaller DOS in the considered energy range. To compare the fillings for different compounds, fillings for various values of $\gamma$ (e.g., $\gamma=0.35, 0.4, 0.45, 0.5$) are presented in Fig.~\ref{Fig:ff}. Except for a small cutoff at $\gamma=0.35$, all other results exhibit a consistent trend in the temperature-filling dependence. With the exception of FeSn-166, the other three families demonstrate higher transition temperatures when their fillings approach half-filling. %
The deviation observed in FeSn-166 could be attributed to the limited number of measured kagome compounds within that family. %
Additionally, those compounds may crystallize in superstructures with space groups other than SG 191, which also contributes to the deviation. Such high DOS near the Fermi level suggests the instability in the paramagnetic phase, giving rise to the magnetic orderings.
As shown in Fig.~\ref{Fig:ImagPho_MnFe}, theoretical instabilities exist widely in Mn/Fe-166, with the leading imaginary phonons located along the $\Gamma-A$ path.

In comparison to the non-spin-polarized calculation, the inclusion of collinear AFM/FM magnetic order eliminates all of the phonon instabilities, as shown in Tab.~\ref{Tab:phonon_mag}, where A and F denote AFM and FM orders, respectively. Here, $*$ denotes that the magnetic orderings are suggested from the experimental measurements. The results indicate that the introduction of magnetic orderings stabilizes the lattice and removes the lattice instabilities observed in the non-magnetic case. %

\begin{table*}[htbp]
\global\long\def\arraystretch{1.12}
\caption{Collections of phonon with magnetism included. '*' denotes the experimental observation of specific magnetic orderings of kagome sites. 'S/U' stands for stable/unstable at low-T, and 'A/F' for the suggested AFM/FM order in the calculations. Here, all phonon spectra are stabilized.}
\label{Tab:phonon_mag}
\setlength{\tabcolsep}{0.8mm}{
\begin{tabular}{c|c|c|c|c|c|c|c|c|c|c|c|c|c|c|c|c|c|c|c|c|c}
\hline
\hline
   & Li & Mg & Ca & Sc & Y & Lu & Ti & Zr & Hf & Nb & Pr & Nd & Sm & Gd & Tb & Dy & Ho & Er & Tm & Yb & U \\ 
\hline
MnGe &  $S_{F}$  &  $S_{F}^{*}$  &  $S_{F}$  &  $S_{A}^{*}$  &  $S_{A}^{*}$  &  $S_{A}^{*}$  &  $S_{A}$  &  $S_{A}^{*}$  &  $S_{A}^{*}$  &  $S_{A}^{*}$  &  $S_{F}$  &  $S_{F}^{*}$  &  $-_{F}^{*}$  &  $S_{F}^{*}$  &  $S_{A}^{*}$  &  $S_{A}^{*}$  &  $S_{A}^{*}$  &  $S_{A}^{*}$  &  $S_{A}^{*}$  &  $S_{A}^{*}$  &  $-$  \\ 
MnSn &  $S_{F}^{*}$  &  $S_{F}^{*}$  &  $S_{F}^{*}$  &  $S_{A}^{*}$  &  $S_{A}^{*}$  &  $S_{A}^{*}$  &  $S_{A}$  &  $S_{A}^{*}$  &  $S_{A}^{*}$  &  $S_{A}^{*}$  &  $S_{F}^{*}$  &  $S_{F}^{*}$  &  $S_{F}^{*}$  &  $S_{F}^{*}$  &  $S_{F}^{*}$  &  $S_{F}^{*}$  &  $S_{F}^{*}$  &  $S_{A}^{*}$  &  $S_{A}^{*}$  &  $S_{F}^{*}$  &  $-$  \\ 
FeGe &  $S_{A}^{*}$  &  $S_{A}^{*}$  &  $S_{A}^{*}$  &  $S_{A}^{*}$  &  $S_{A}^{*}$  &  $S_{A}^{*}$  &  $S_{A}^{*}$  &  $S_{A}^{*}$  &  $S_{A}^{*}$  &  $S_{A}^{*}$  &  $S_{A}$  &  $S_{A}$  &  $S_{A}$  &  $S_{A}^{*}$  &  $S_{A}^{*}$  &  $S_{A}^{*}$  &  $S_{A}^{*}$  &  $S_{A}^{*}$  &  $S_{A}^{*}$  &  $S_{A}^{*}$  &  $S_{A}^{*}$  \\ 
FeSn &  $S_{A}$  &  $S_{A}^{*}$  &  $S_{A}$  &  $S_{A}^{*}$  &  $S_{A}^{*}$  &  $S_{A}$  &  $S_{A}^{*}$  &  $S_{A}^{*}$  &  $S_{A}^{*}$  &  $S_{A}$  &  $-$  &  $-$  &  $S_{A}$  &  $S_{A}^{*}$  &  $S_{A}^{*}$  &  $S_{A}^{*}$  &  $S_{A}^{*}$  &  $S_{A}^{*}$  &  $S_{A}^{*}$  &  $S_{A}$  &  $-$ \\ 
\hline
\hline
\end{tabular}}
\end{table*}

\section{Results}\label{smsec:allcompounds}

In this section, we list all results for TZ parent structures and \ch{MT6Z6} compounds, including their band structures and phonon spectra, as well as the projection of the dominant kagome $d$ orbitals in the electronic bands. These orbitals are categorized into three groups, i.e., in-plane $d_{xy/x^2-y^2}$, out-of-plane $d_{xz/yz}$ and $d_{z^2}$ orbitals. The phonon spectra of \ch{MT6Z6} are projected regarding their sites, namely, trigonal M, kagome T, trigonal Z1, and honeycomb Z2. The phonon spectra are further projected onto in-plane and out-of-plane vibrations, providing insights into the vibrational modes associated with these sites. Meanwhile, the IRREPs of phonon spectra at high-symmetry points are provided to track the evolution of imaginary modes within the \ch{MT6Z6} compounds. Overall, these results offer a detailed understanding of the electronic and vibrational properties of both the TZ parent structures and the \ch{MT6Z6} compounds.

\subsection{\hyperref[smsec:col]{TZ parent structures}}\label{smssec:TZ}

\subsubsection{\label{sms2sec:VGe}\hyperref[smsec:col]{\texorpdfstring{VGe}{}}~{\color{green} (high-T stable, low-T stable)}}

\begin{figure}[htbp]
\centering
\includegraphics[width=0.5\textwidth]{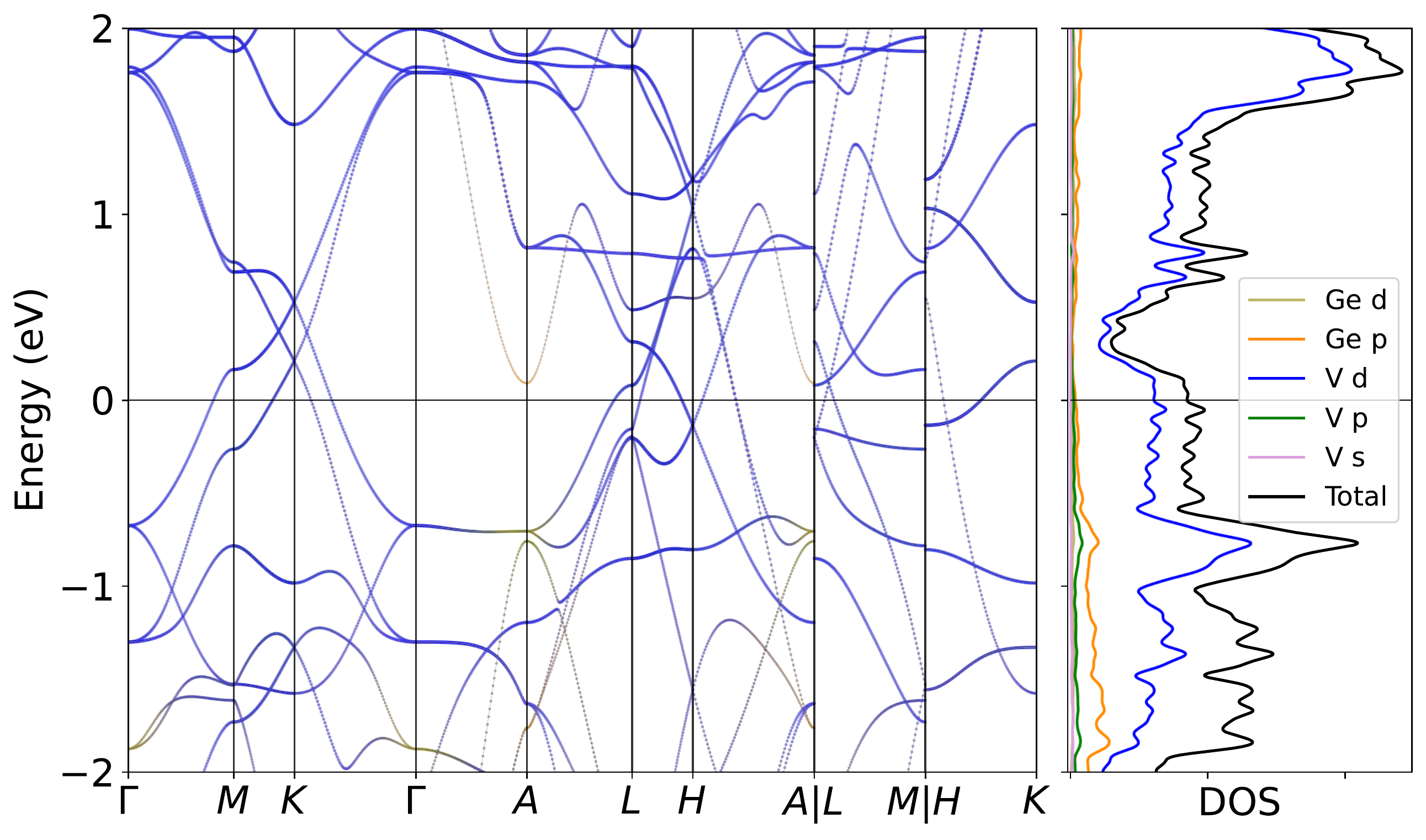}
\caption{Band structure for VGe without SOC. }
\label{fig:VGeband}
\end{figure}

\begin{figure}[H]
\centering
\subfloat[Relaxed phonon at low-T.]{
\label{fig:VGe_relaxed_Low}
\includegraphics[width=0.45\textwidth]{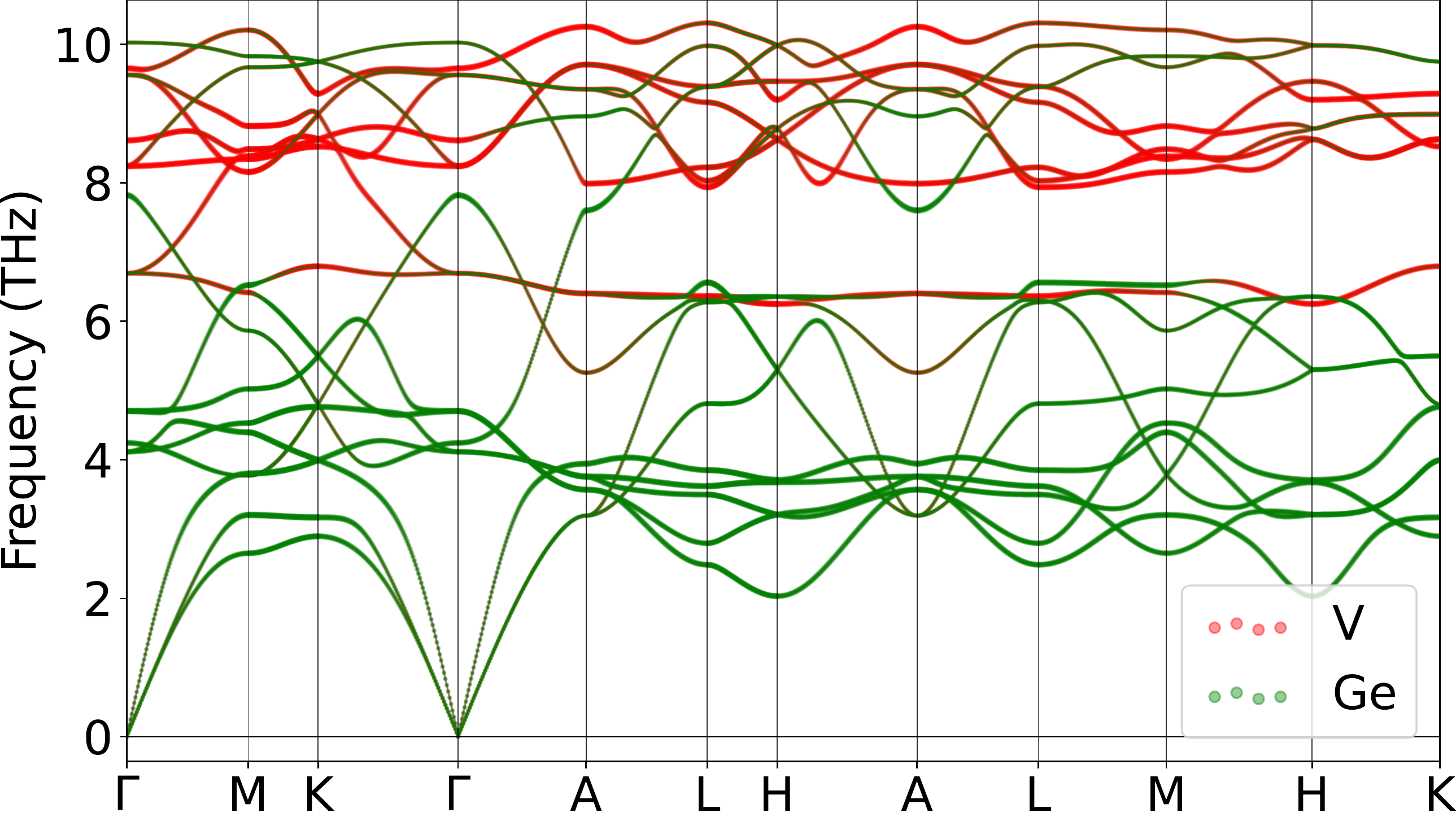}
}
\hspace{5pt}\subfloat[Relaxed phonon at high-T.]{
\label{fig:VGe_relaxed_High}
\includegraphics[width=0.45\textwidth]{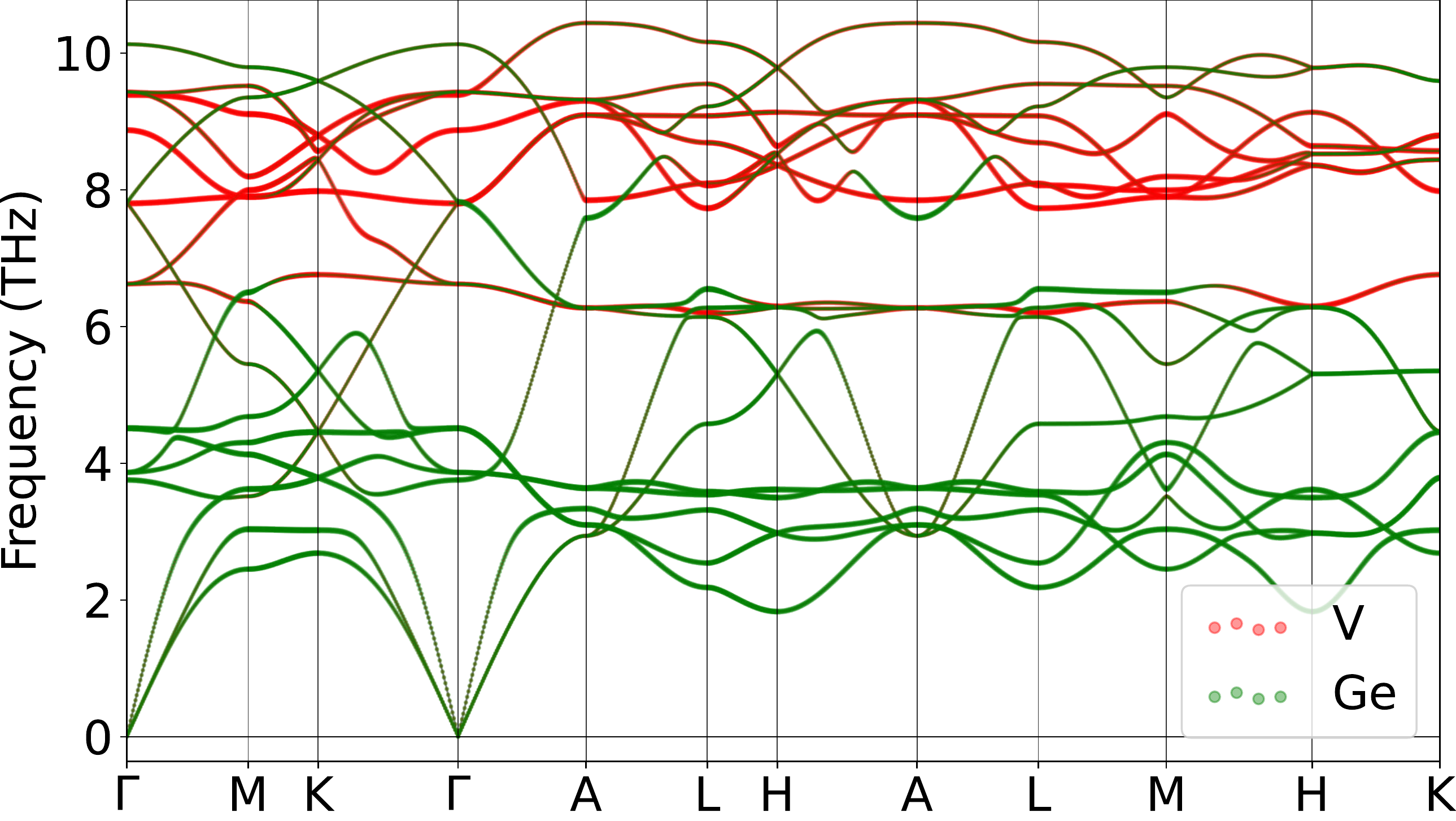}
}
\caption{Atomically projected phonon spectrum for VGe.}
\label{fig:VGepho}
\end{figure}

\subsubsection{\label{sms2sec:VSn}\hyperref[smsec:col]{\texorpdfstring{VSn}{}}~{\color{green} (high-T stable, low-T stable)}}

\begin{figure}[htbp]
\centering
\includegraphics[width=0.5\textwidth]{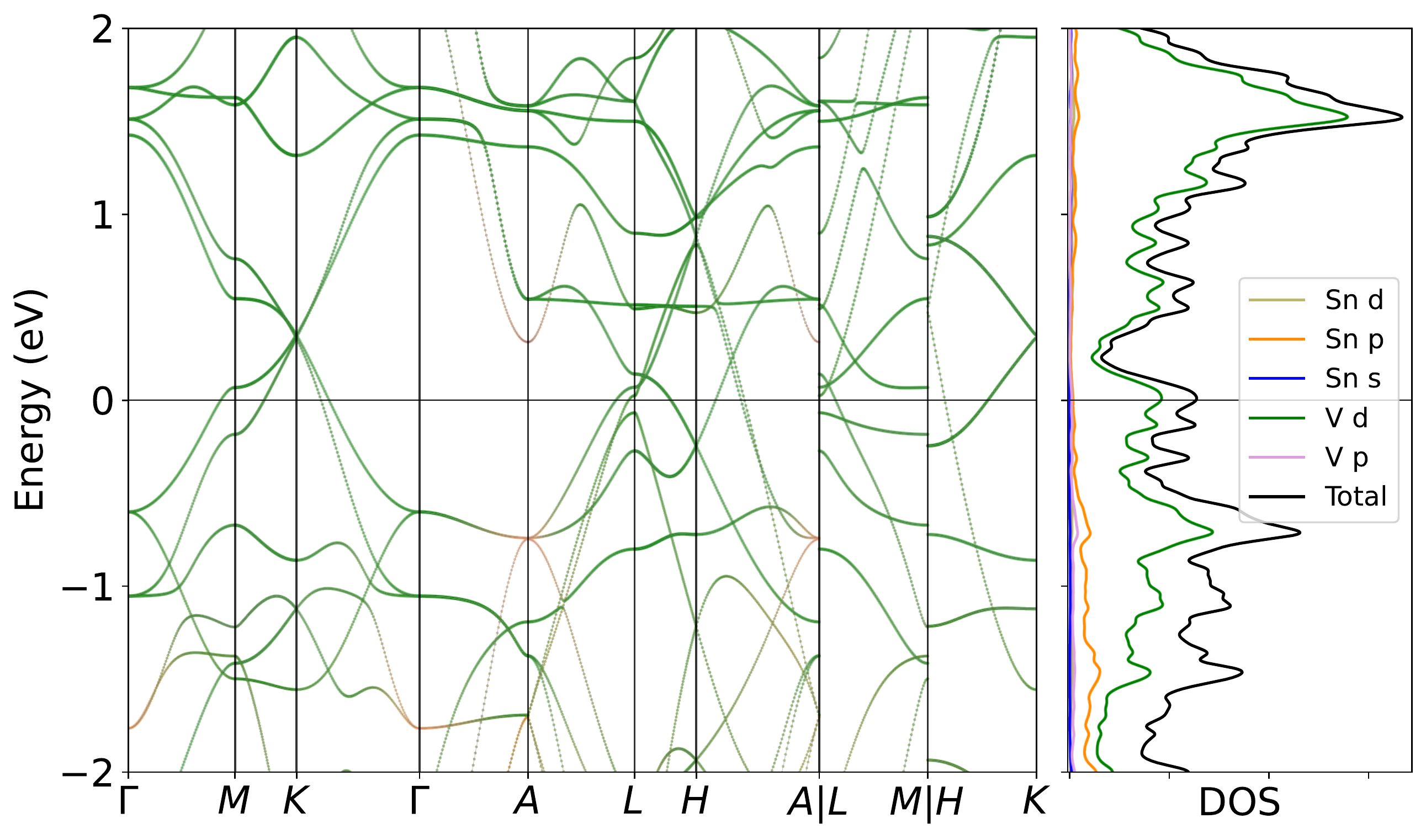}
\caption{Band structure for VSn without SOC. }
\label{fig:VSnband}
\end{figure}

\begin{figure}[H]
\centering
\subfloat[Relaxed phonon at low-T.]{
\label{fig:VSn_relaxed_Low}
\includegraphics[width=0.45\textwidth]{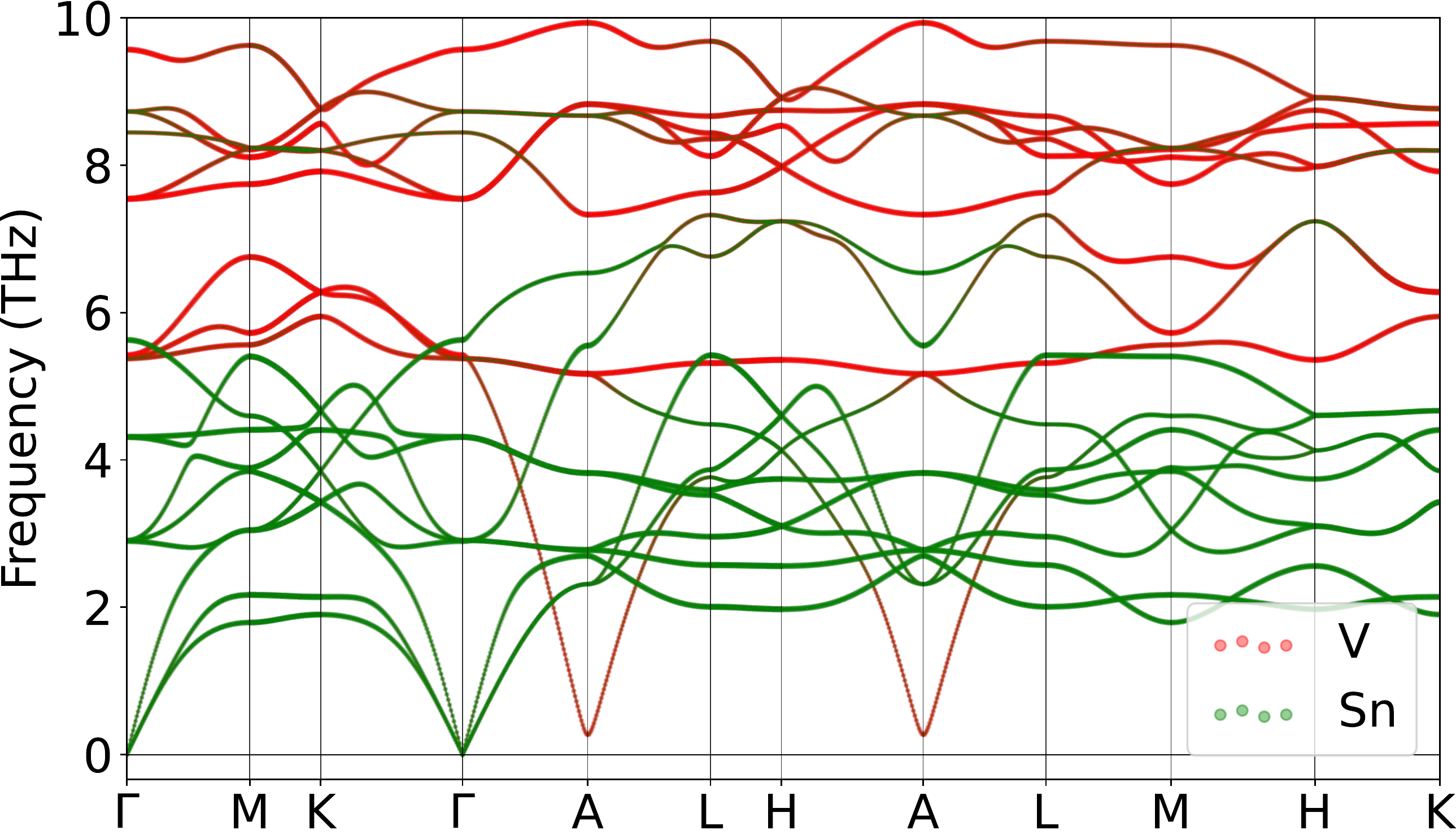}
}
\hspace{5pt}\subfloat[Relaxed phonon at high-T.]{
\label{fig:VSn_relaxed_High}
\includegraphics[width=0.45\textwidth]{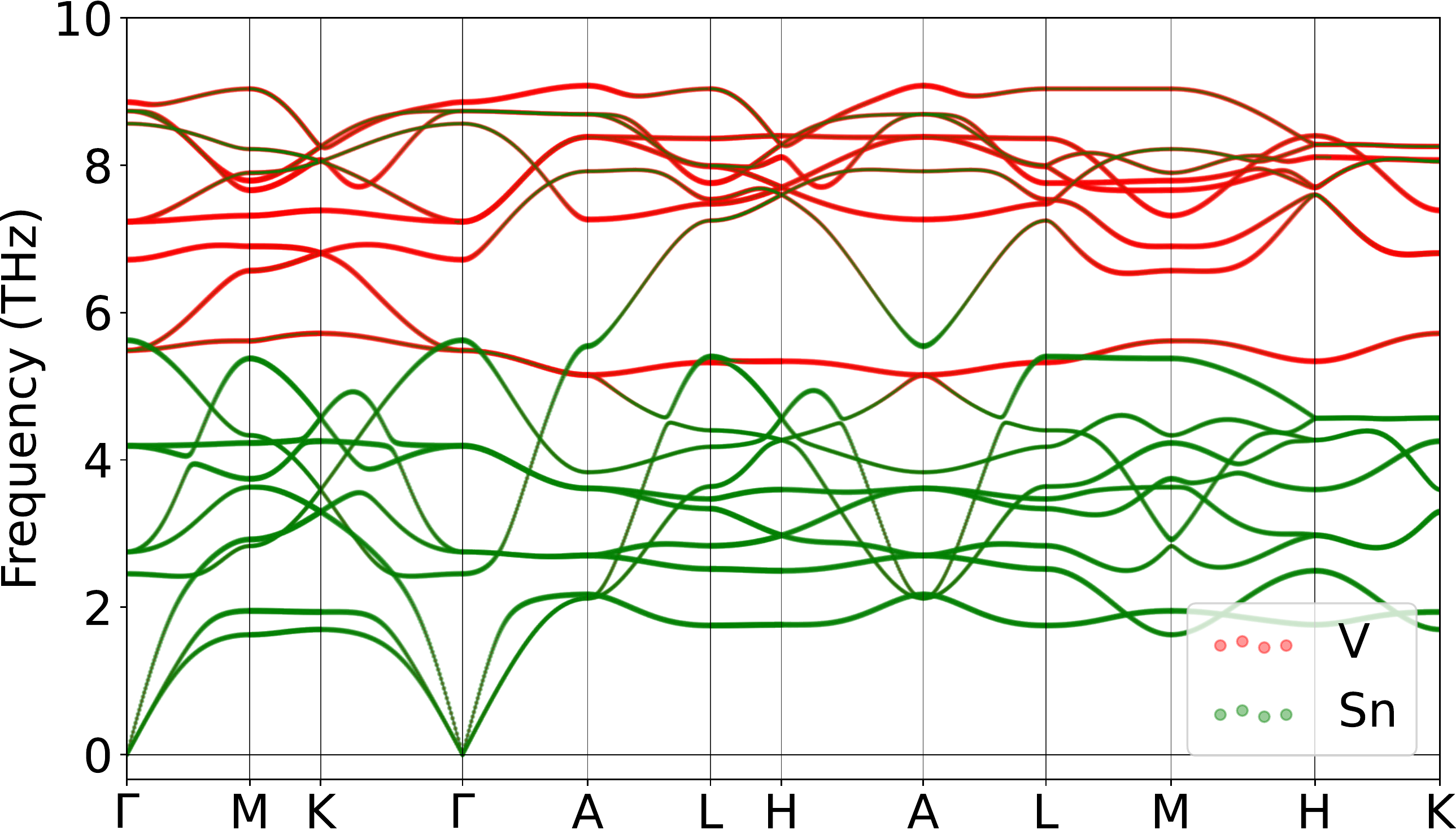}
}
\caption{Atomically projected phonon spectrum for VSn.}
\label{fig:VSnpho}
\end{figure}

\subsubsection{\label{sms2sec:CrGe}\hyperref[smsec:col]{\texorpdfstring{CrGe}{}}~{\color{green} (high-T stable, low-T stable)}}

\begin{figure}[htbp]
\centering
\includegraphics[width=0.5\textwidth]{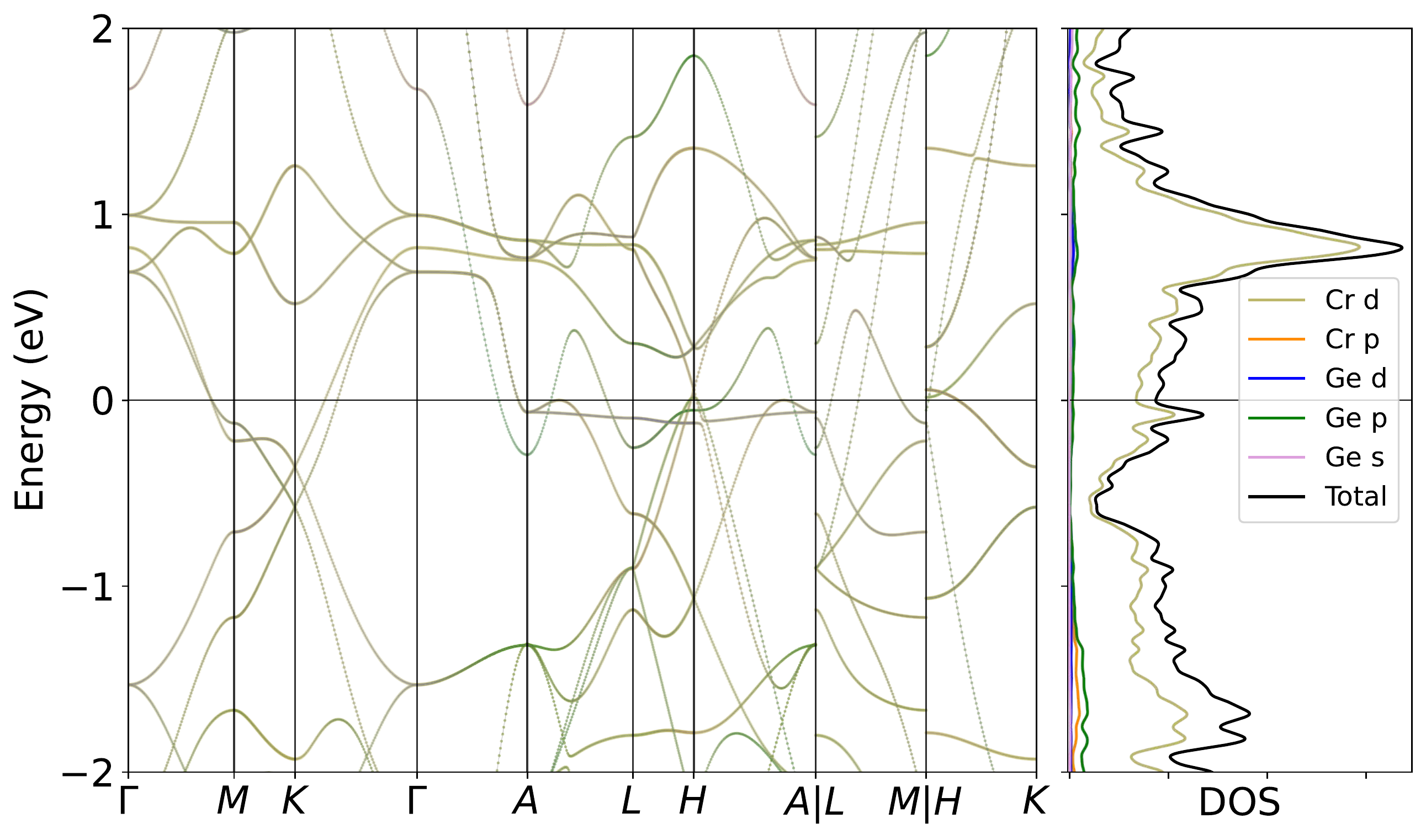}
\caption{Band structure for CrGe without SOC. }
\label{fig:CrGeband}
\end{figure}

\begin{figure}[H]
\centering
\subfloat[Relaxed phonon at low-T.]{
\label{fig:CrGe_relaxed_Low}
\includegraphics[width=0.45\textwidth]{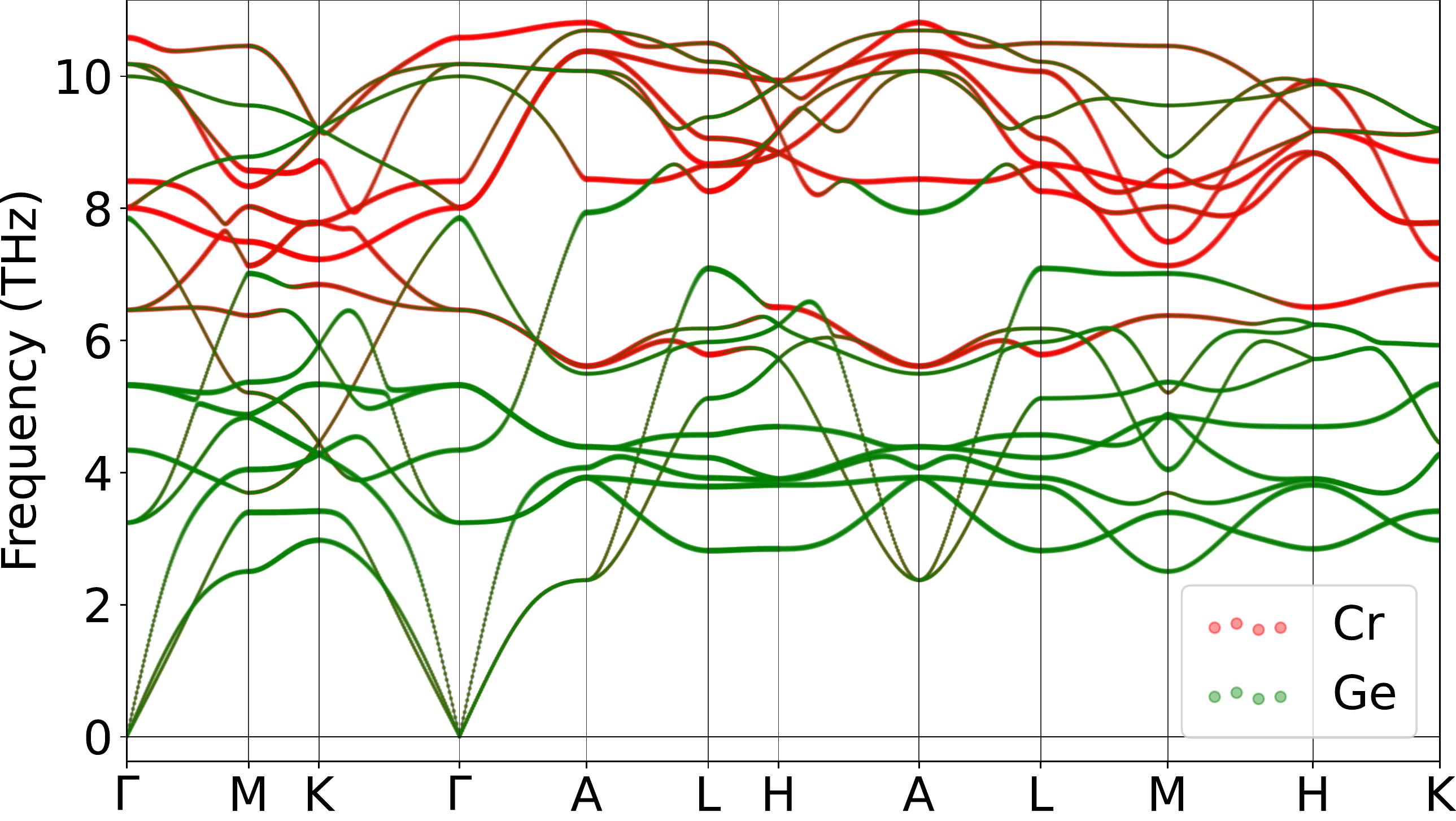}
}
\hspace{5pt}\subfloat[Relaxed phonon at high-T.]{
\label{fig:CrGe_relaxed_High}
\includegraphics[width=0.45\textwidth]{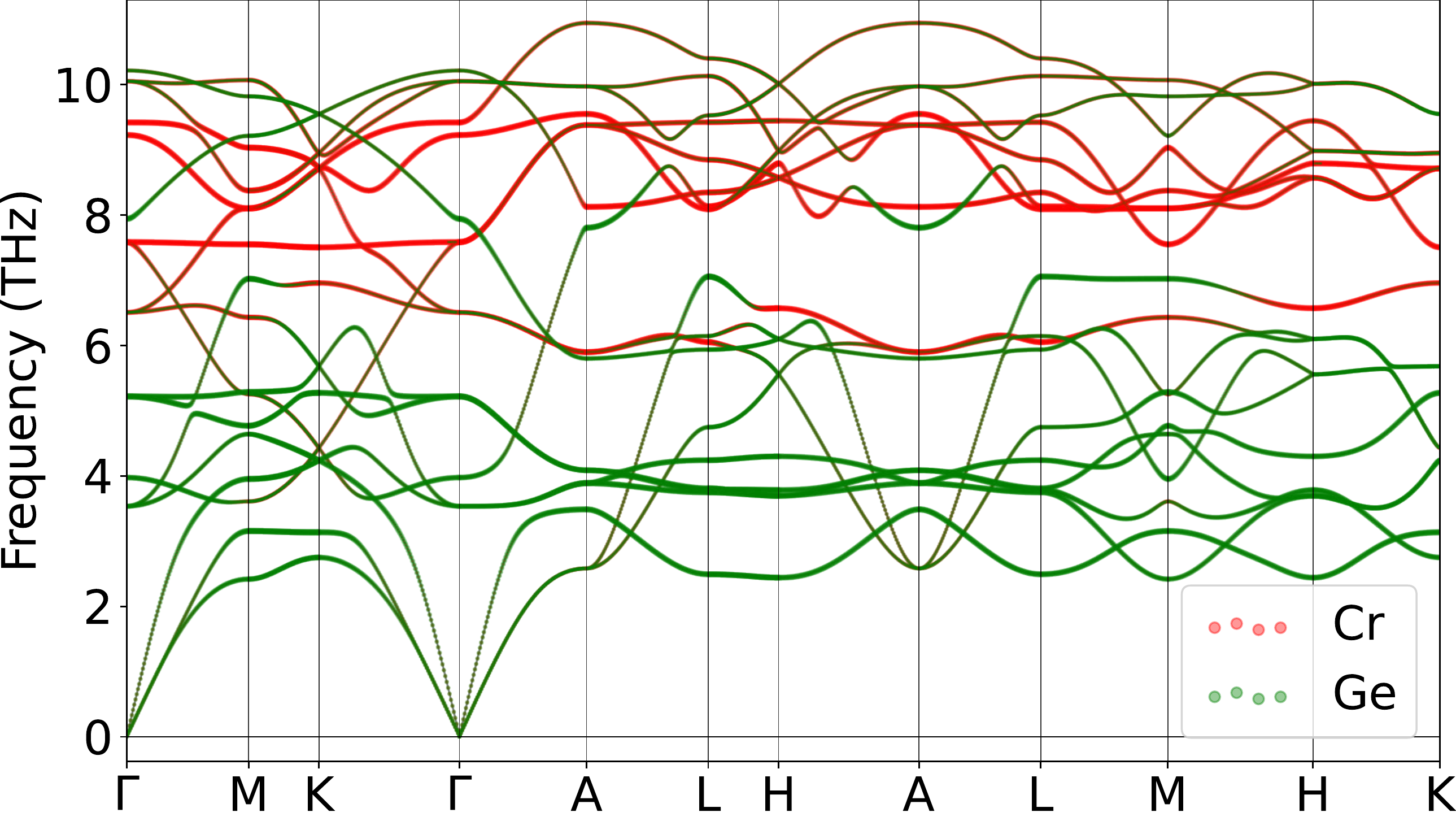}
}
\caption{Atomically projected phonon spectrum for CrGe.}
\label{fig:CrGepho}
\end{figure}

\subsubsection{\label{sms2sec:MnGe}\hyperref[smsec:col]{\texorpdfstring{MnGe}{}}~{\color{blue} (high-T stable, low-T unstable)}}

\begin{figure}[htbp]
\centering
\includegraphics[width=0.5\textwidth]{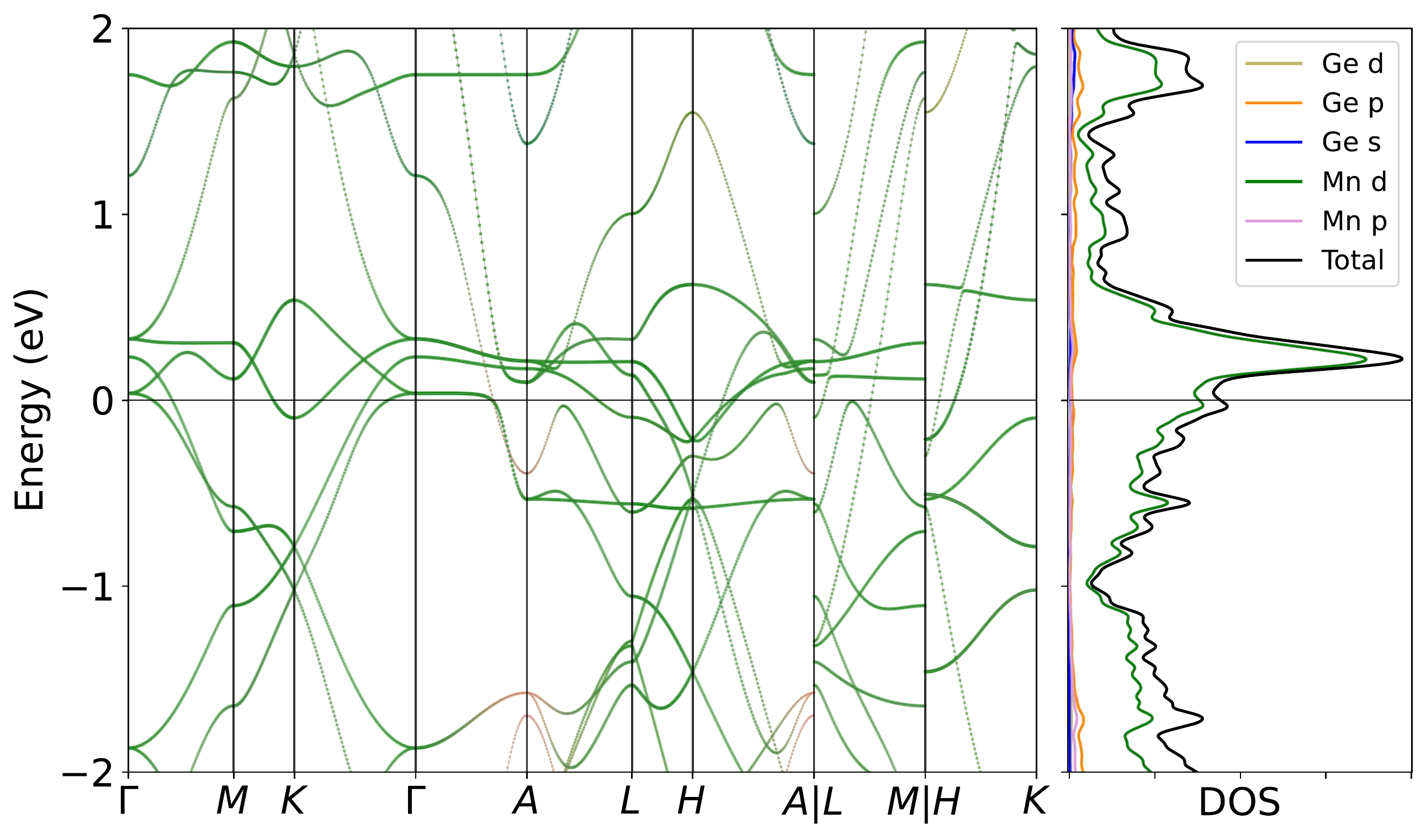}
\caption{Band structure for MnGe without SOC. }
\label{fig:MnGeband}
\end{figure}

\begin{figure}[H]
\centering
\subfloat[Relaxed phonon at low-T.]{
\label{fig:MnGe_relaxed_Low}
\includegraphics[width=0.45\textwidth]{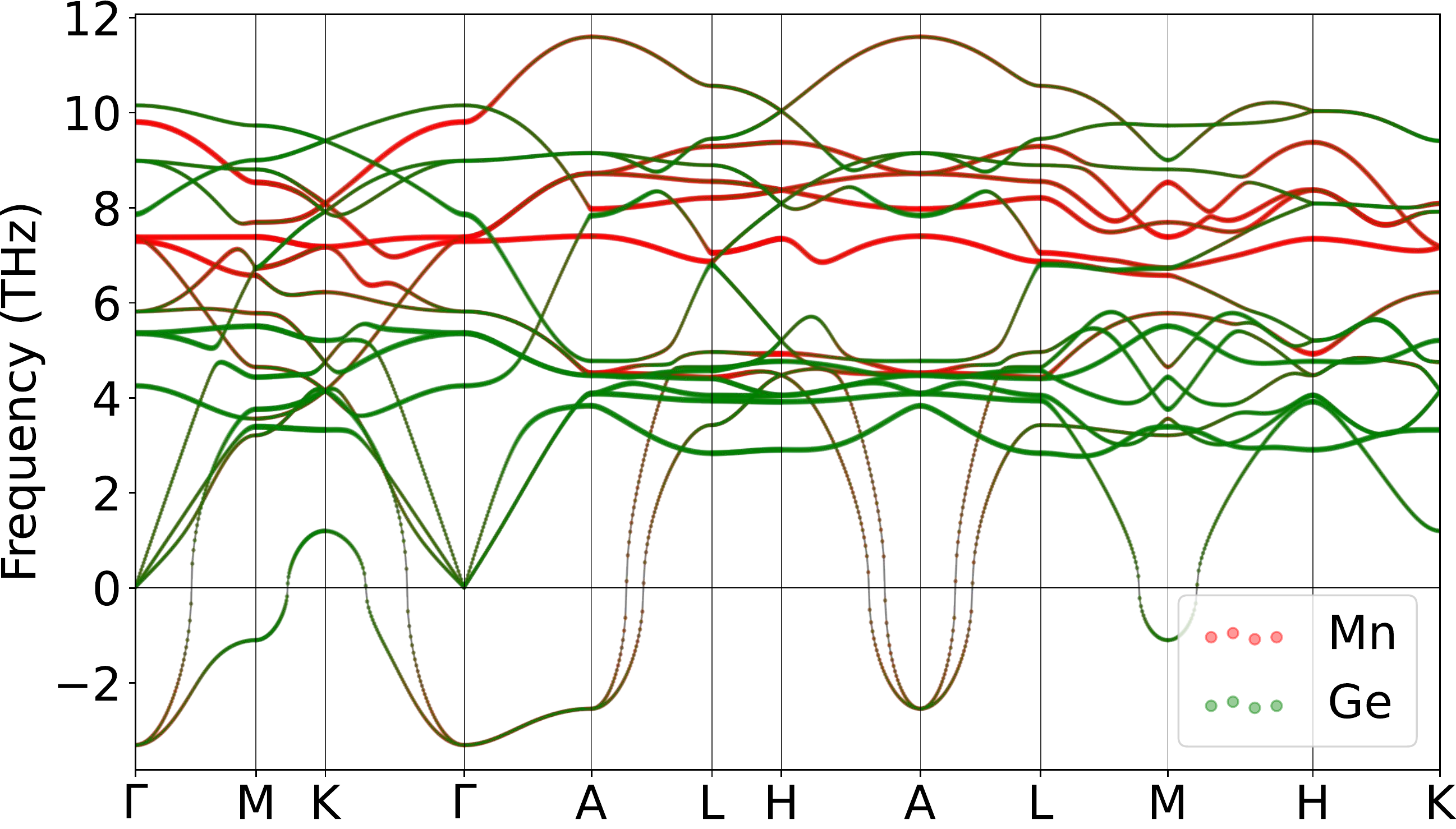}
}
\hspace{5pt}\subfloat[Relaxed phonon at high-T.]{
\label{fig:MnGe_relaxed_High}
\includegraphics[width=0.45\textwidth]{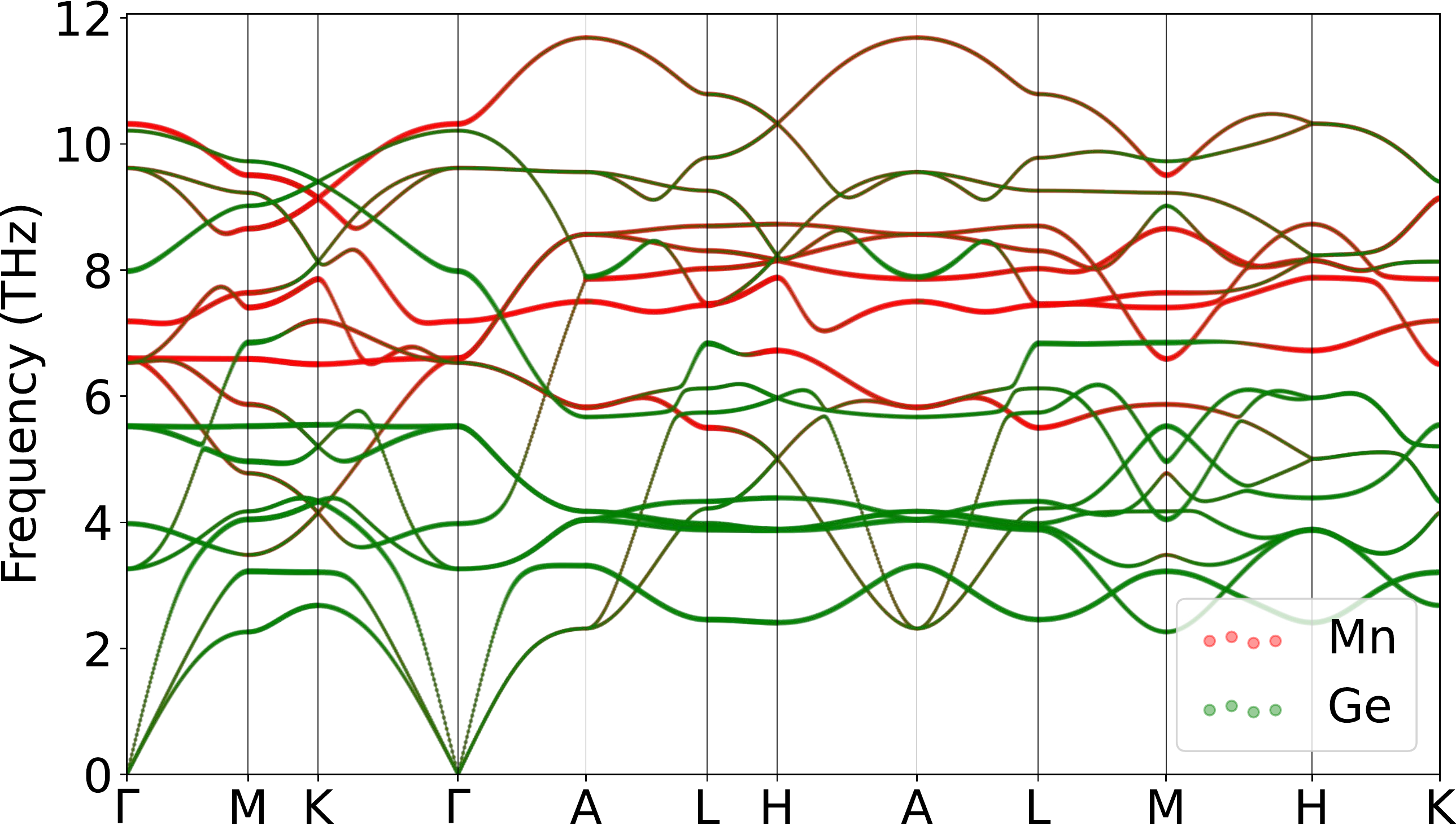}
}
\caption{Atomically projected phonon spectrum for MnGe.}
\label{fig:MnGepho}
\end{figure}

\subsubsection{\label{sms2sec:MnSn}\hyperref[smsec:col]{\texorpdfstring{MnSn}{}}~{\color{blue} (high-T stable, low-T unstable)}}

\begin{figure}[htbp]
\centering
\includegraphics[width=0.5\textwidth]{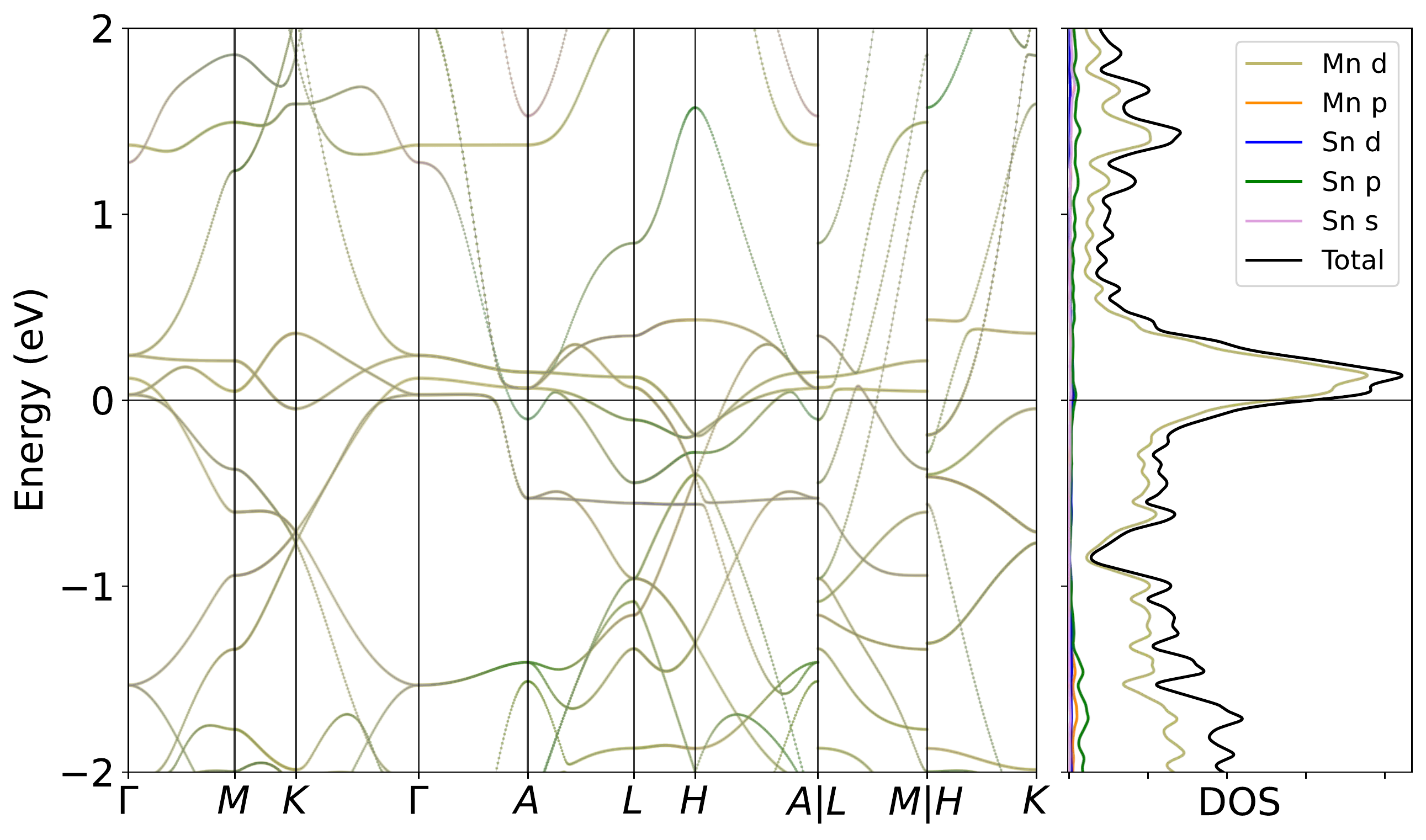}
\caption{Band structure for MnSn without SOC. }
\label{fig:MnSnband}
\end{figure}

\begin{figure}[H]
\centering
\subfloat[Relaxed phonon at low-T.]{
\label{fig:MnSn_relaxed_Low}
\includegraphics[width=0.45\textwidth]{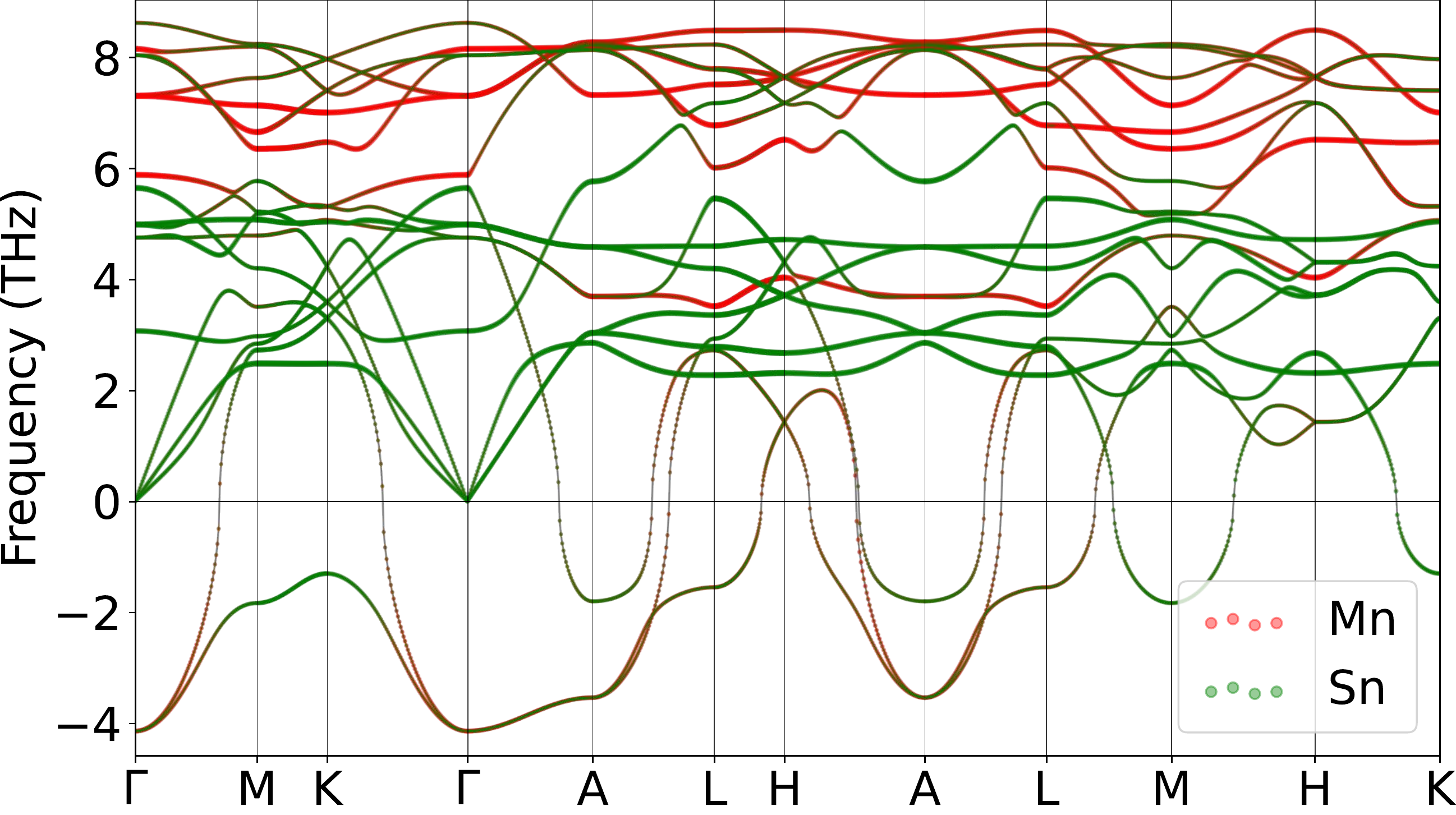}
}
\hspace{5pt}\subfloat[Relaxed phonon at high-T.]{
\label{fig:MnSn_relaxed_High}
\includegraphics[width=0.45\textwidth]{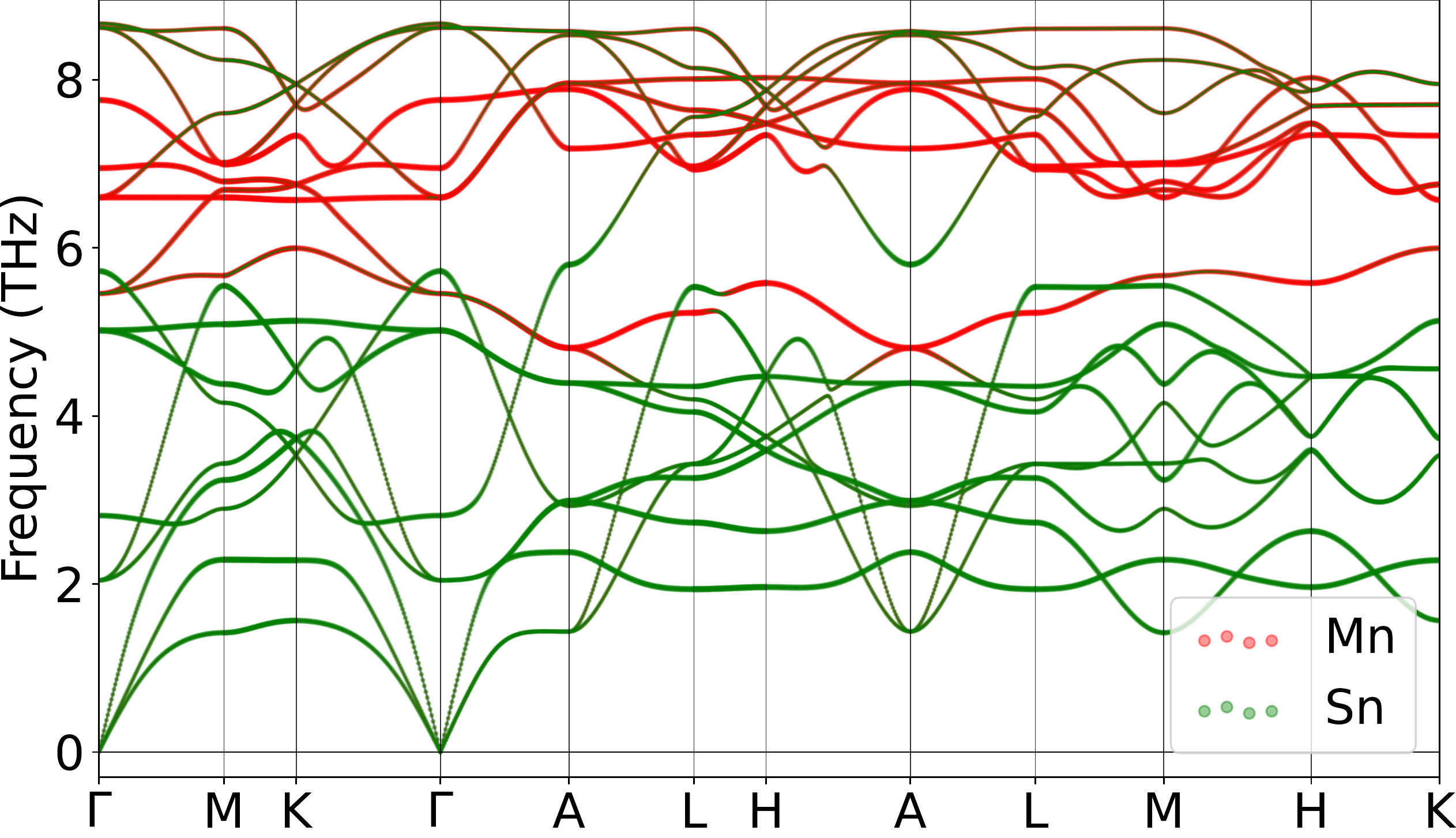}
}
\caption{Atomically projected phonon spectrum for MnSn.}
\label{fig:MnSnpho}
\end{figure}

\subsubsection{\label{sms2sec:FeGe}\hyperref[smsec:col]{\texorpdfstring{FeGe}{}}~{\color{blue} (high-T stable, low-T unstable)}}

\begin{figure}[htbp]
\centering
\includegraphics[width=0.5\textwidth]{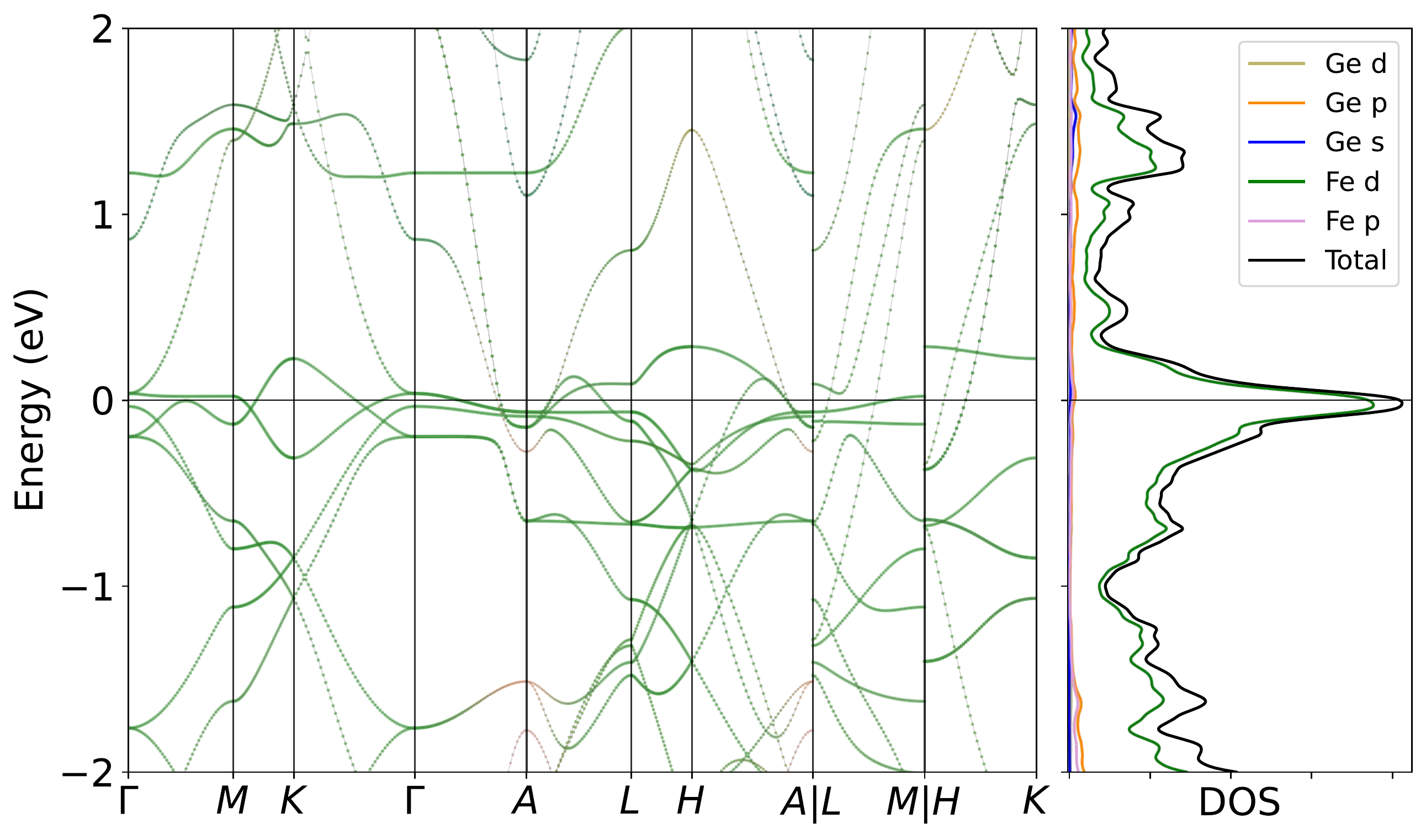}
\caption{Band structure for FeGe without SOC. }
\label{fig:FeGeband}
\end{figure}

\begin{figure}[H]
\centering
\subfloat[Relaxed phonon at low-T.]{
\label{fig:FeGe_relaxed_Low}
\includegraphics[width=0.45\textwidth]{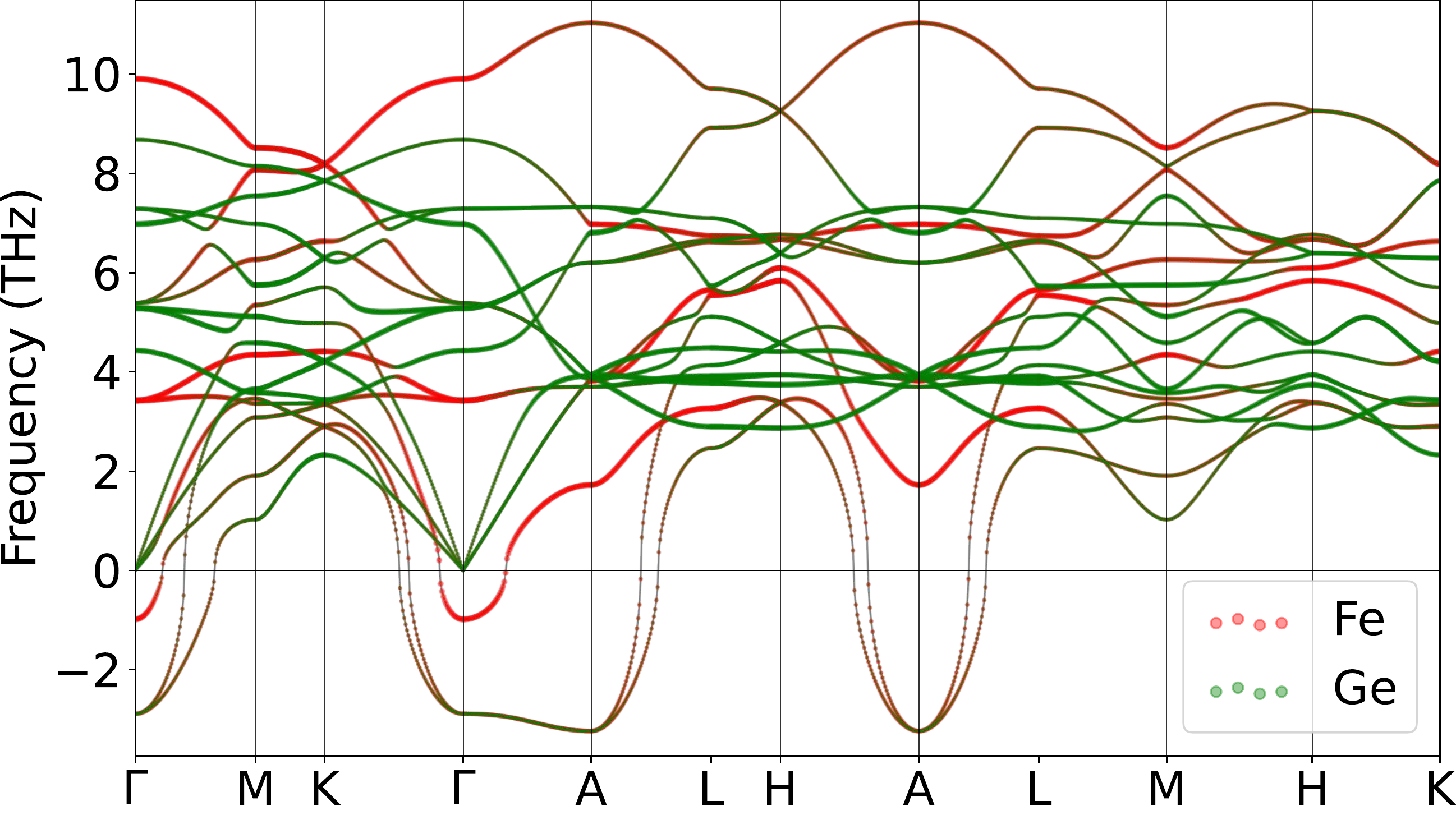}
}
\hspace{5pt}\subfloat[Relaxed phonon at high-T.]{
\label{fig:FeGe_relaxed_High}
\includegraphics[width=0.45\textwidth]{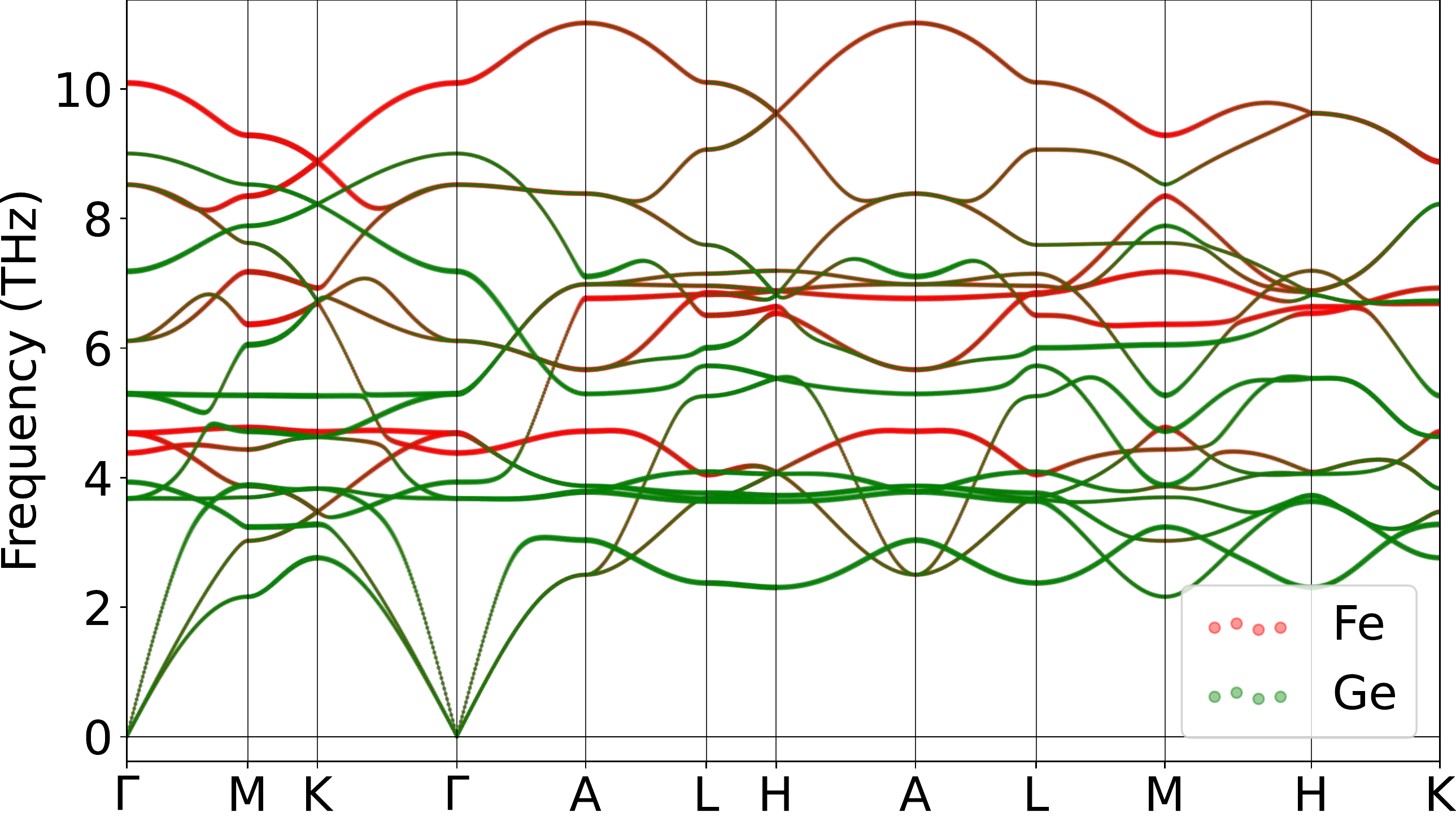}
}
\caption{Atomically projected phonon spectrum for FeGe.}
\label{fig:FeGepho}
\end{figure}

\subsubsection{\label{sms2sec:FeSn}\hyperref[smsec:col]{\texorpdfstring{FeSn}{}}~{\color{blue} (high-T stable, low-T unstable)}}

\begin{figure}[htbp]
\centering
\includegraphics[width=0.5\textwidth]{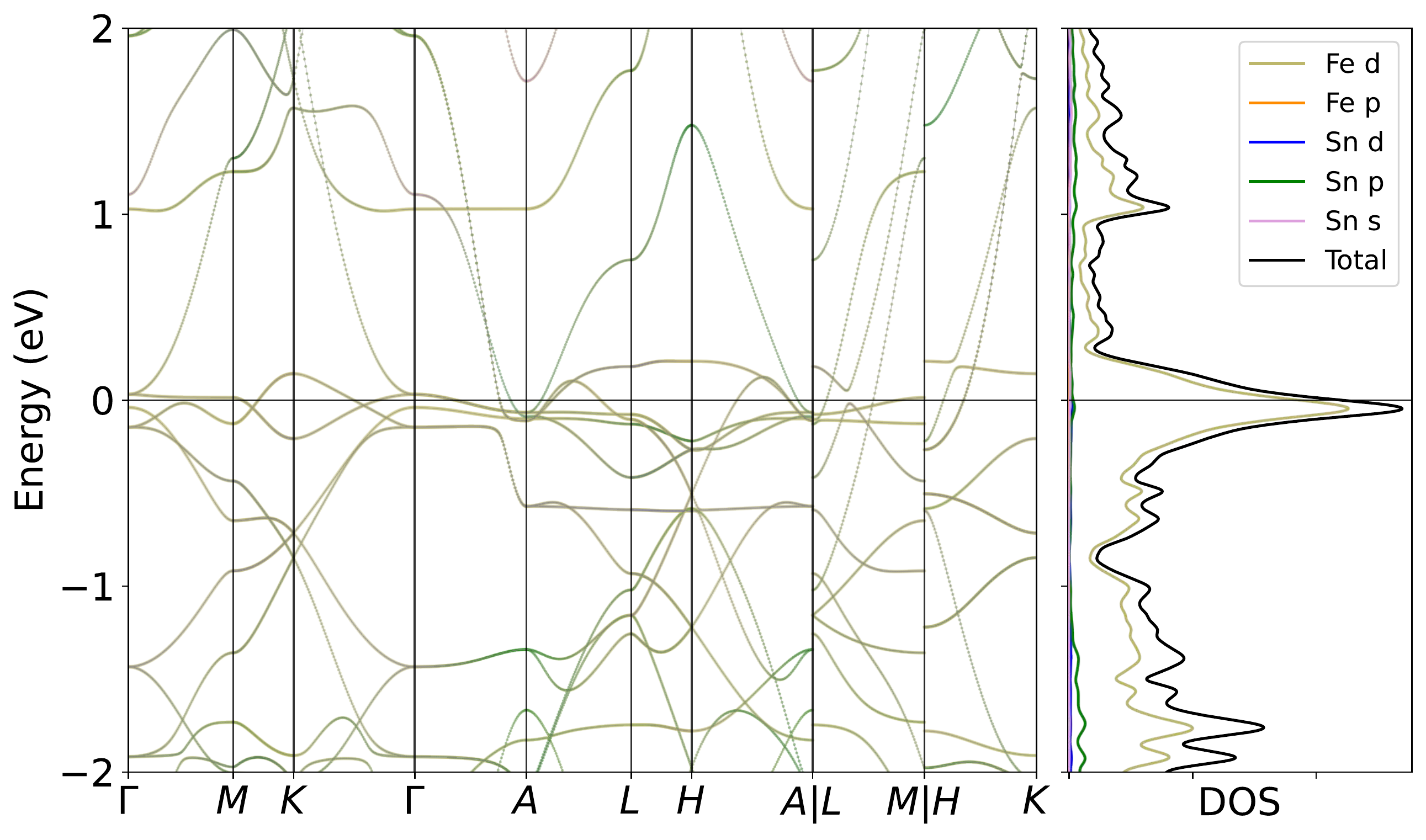}
\caption{Band structure for FeSn without SOC. }
\label{fig:FeSnband}
\end{figure}

\begin{figure}[H]
\centering
\subfloat[Relaxed phonon at low-T.]{
\label{fig:FeSn_relaxed_Low}
\includegraphics[width=0.45\textwidth]{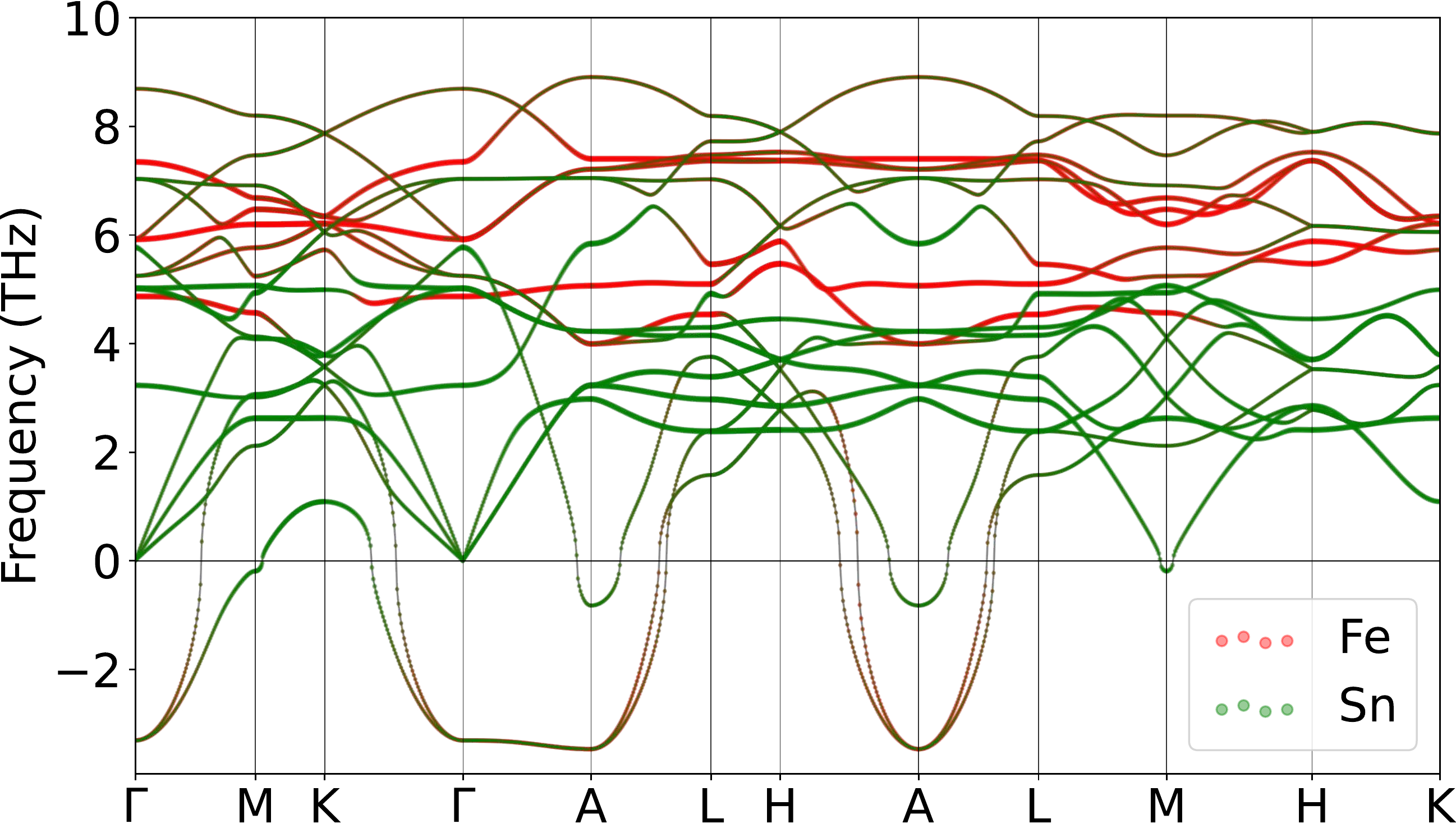}
}
\hspace{5pt}\subfloat[Relaxed phonon at high-T.]{
\label{fig:FeSn_relaxed_High}
\includegraphics[width=0.45\textwidth]{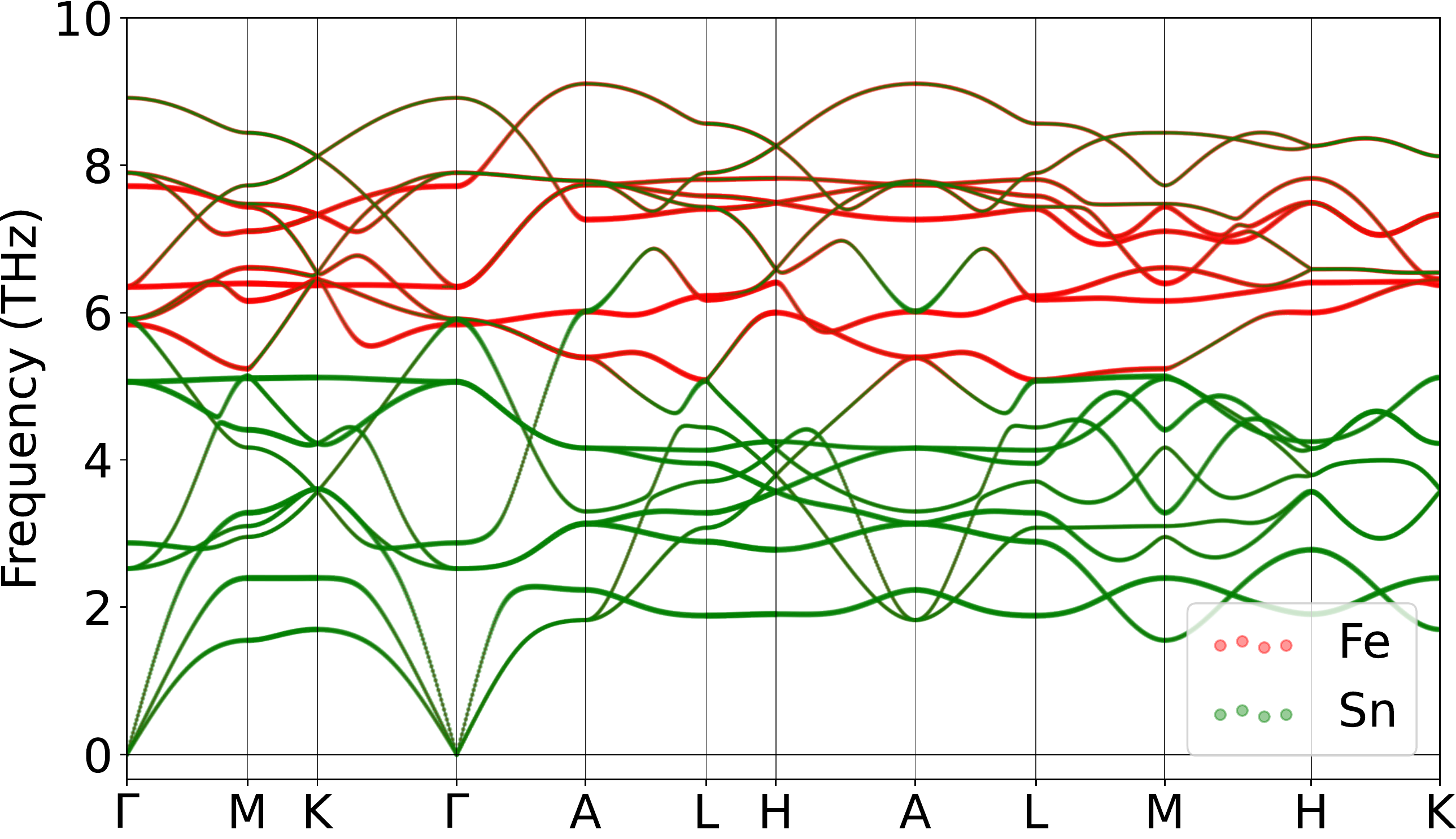}
}
\caption{Atomically projected phonon spectrum for FeSn.}
\label{fig:FeSnpho}
\end{figure}

\subsubsection{\label{sms2sec:CoGe}\hyperref[smsec:col]{\texorpdfstring{CoGe}{}}~{\color{blue} (high-T stable, low-T unstable)}}

\begin{figure}[htbp]
\centering
\includegraphics[width=0.5\textwidth]{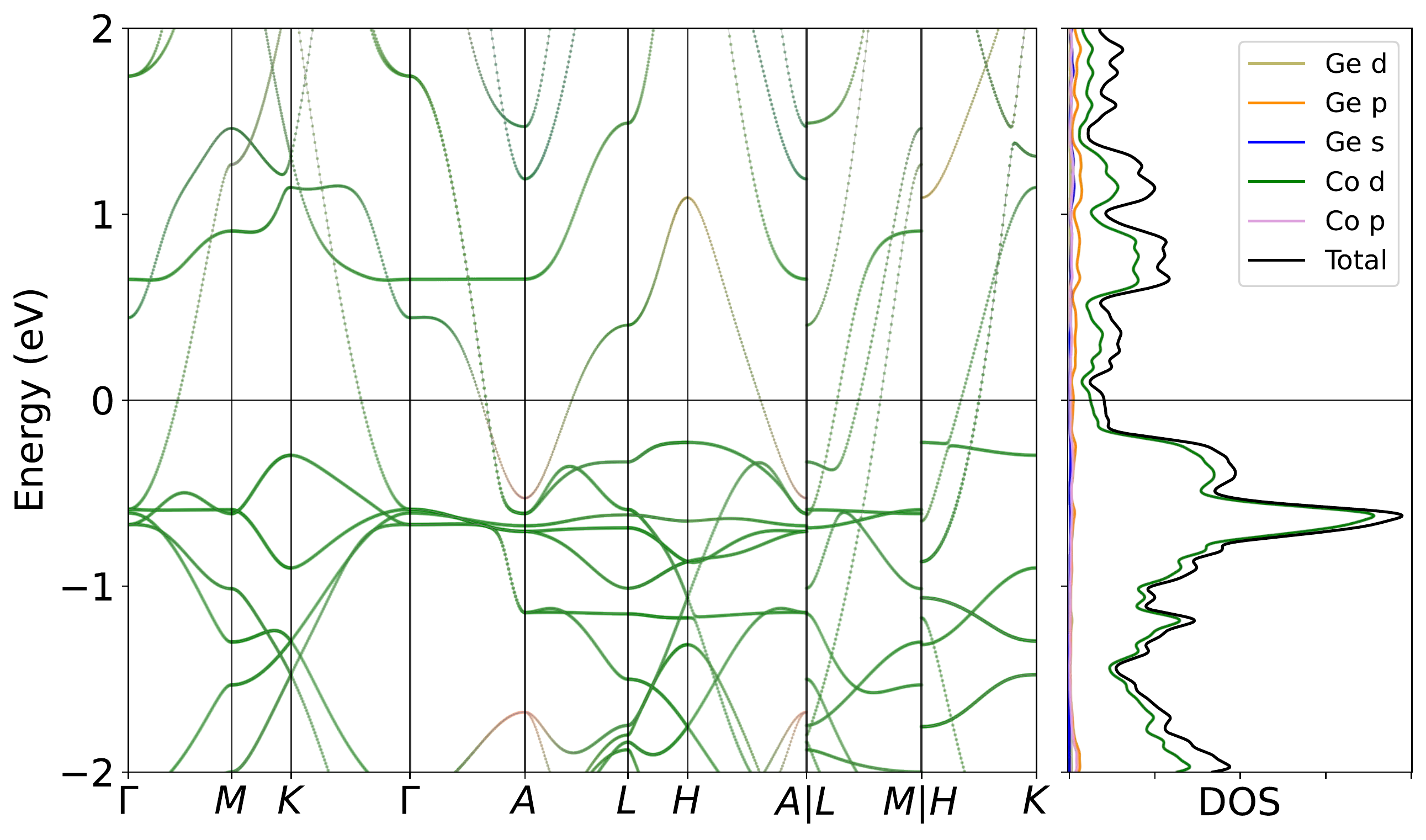}
\caption{Band structure for CoGe without SOC. }
\label{fig:CoGeband}
\end{figure}

\begin{figure}[H]
\centering
\subfloat[Relaxed phonon at low-T.]{
\label{fig:CoGe_relaxed_Low}
\includegraphics[width=0.45\textwidth]{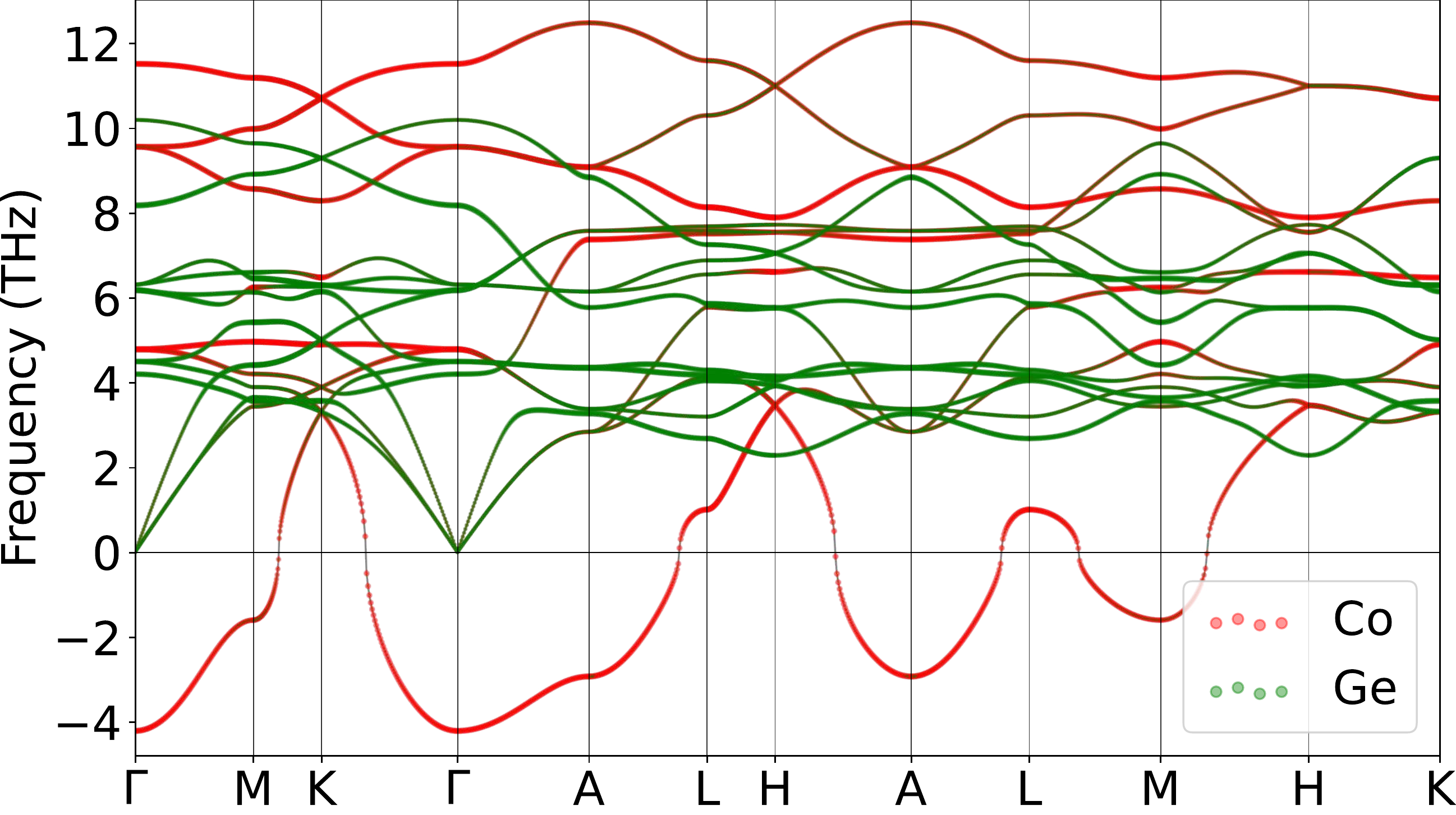}
}
\hspace{5pt}\subfloat[Relaxed phonon at high-T.]{
\label{fig:CoGe_relaxed_High}
\includegraphics[width=0.45\textwidth]{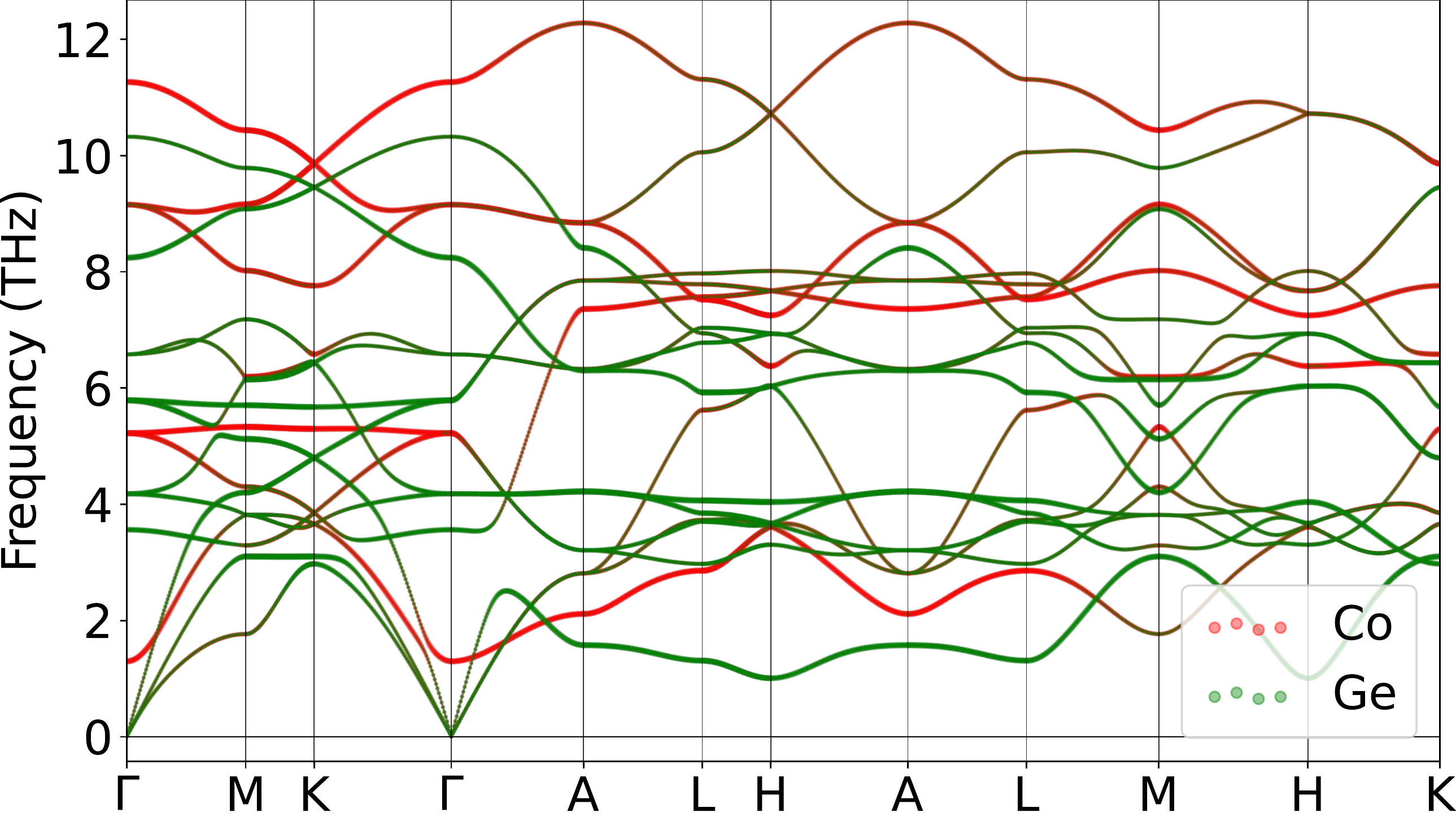}
}
\caption{Atomically projected phonon spectrum for CoGe.}
\label{fig:CoGepho}
\end{figure}

\subsubsection{\label{sms2sec:CoSn}\hyperref[smsec:col]{\texorpdfstring{CoSn}{}}~{\color{green} (high-T stable, low-T stable)}}

\begin{figure}[htbp]
\centering
\includegraphics[width=0.5\textwidth]{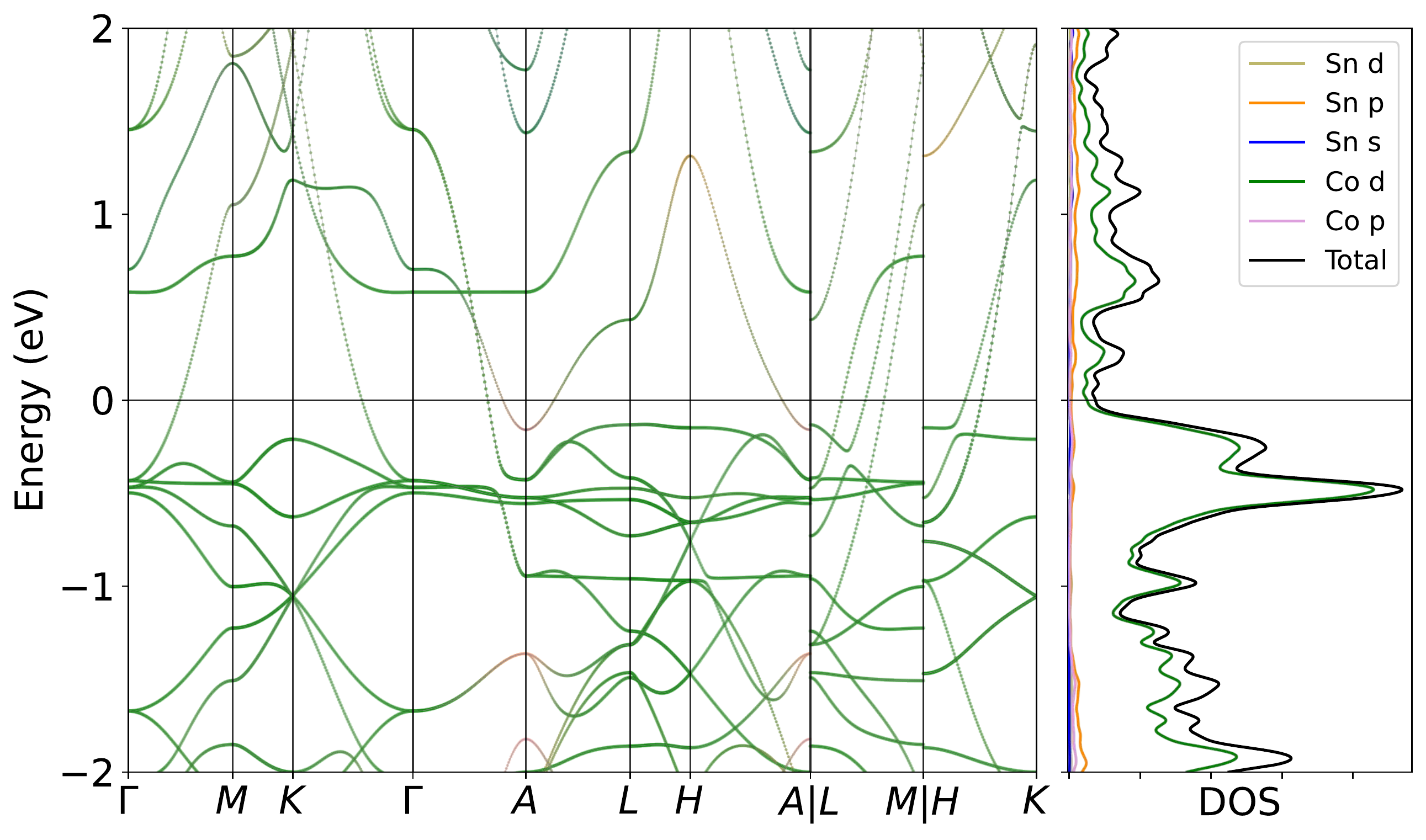}
\caption{Band structure for CoSn without SOC. }
\label{fig:CoSnband}
\end{figure}

\begin{figure}[H]
\centering
\subfloat[Relaxed phonon at low-T.]{
\label{fig:CoSn_relaxed_Low}
\includegraphics[width=0.45\textwidth]{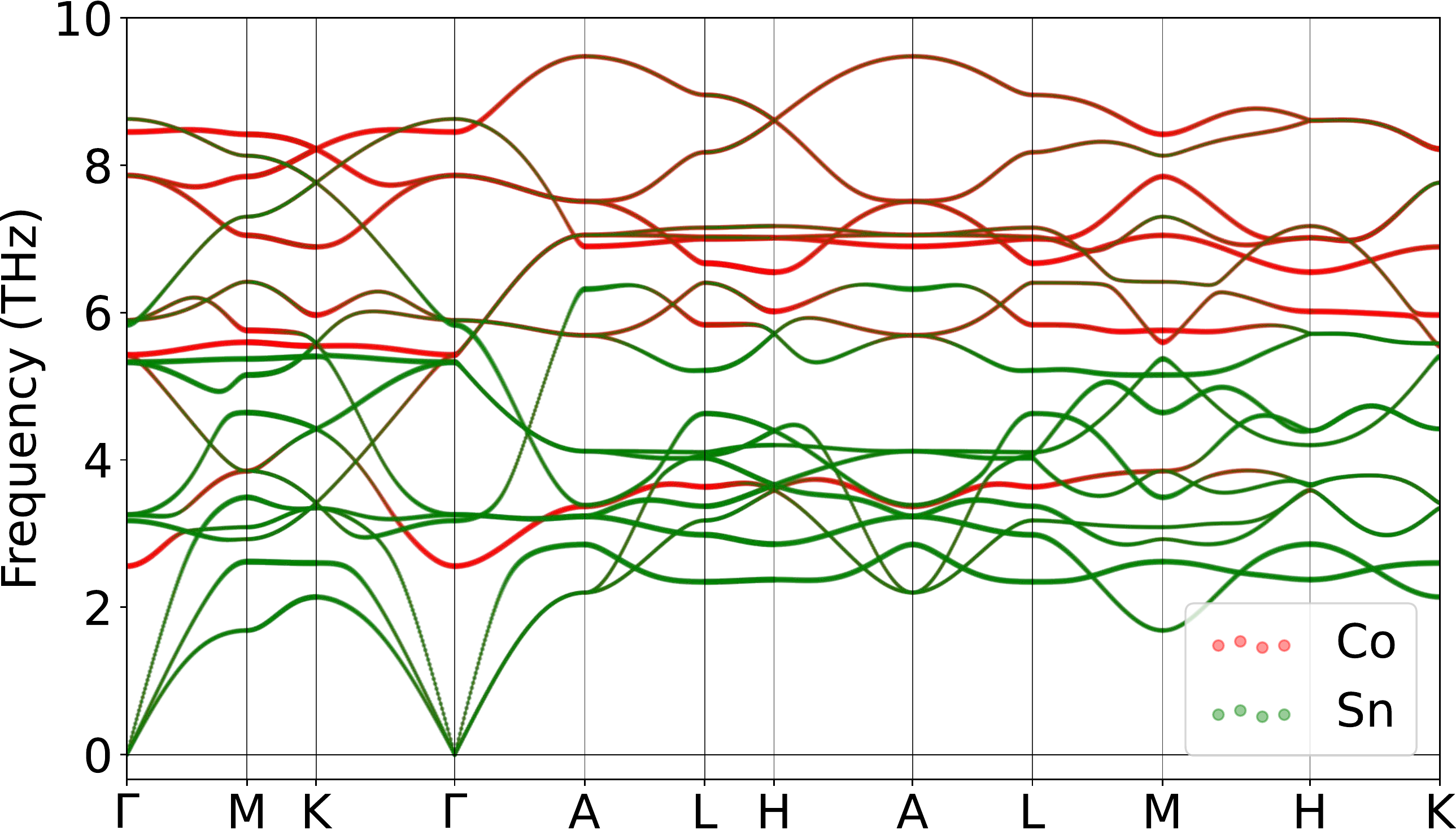}
}
\hspace{5pt}\subfloat[Relaxed phonon at high-T.]{
\label{fig:CoSn_relaxed_High}
\includegraphics[width=0.45\textwidth]{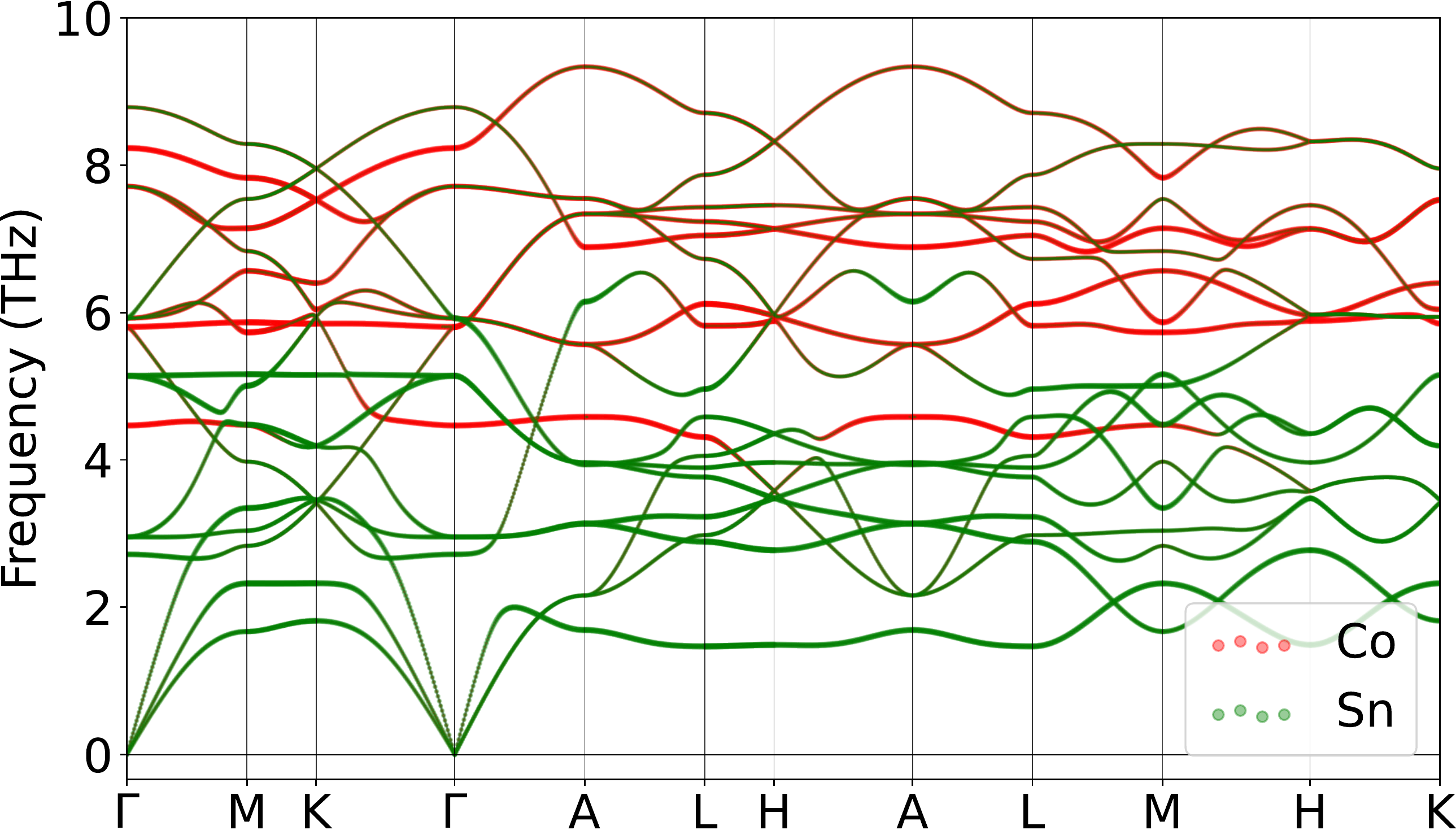}
}
\caption{Atomically projected phonon spectrum for CoSn.}
\label{fig:CoSnpho}
\end{figure}

\subsubsection{\label{sms2sec:NiSi}\hyperref[smsec:col]{\texorpdfstring{NiSi}{}}~{\color{red} (high-T unstable, low-T unstable)}}

\begin{figure}[htbp]
\centering
\includegraphics[width=0.5\textwidth]{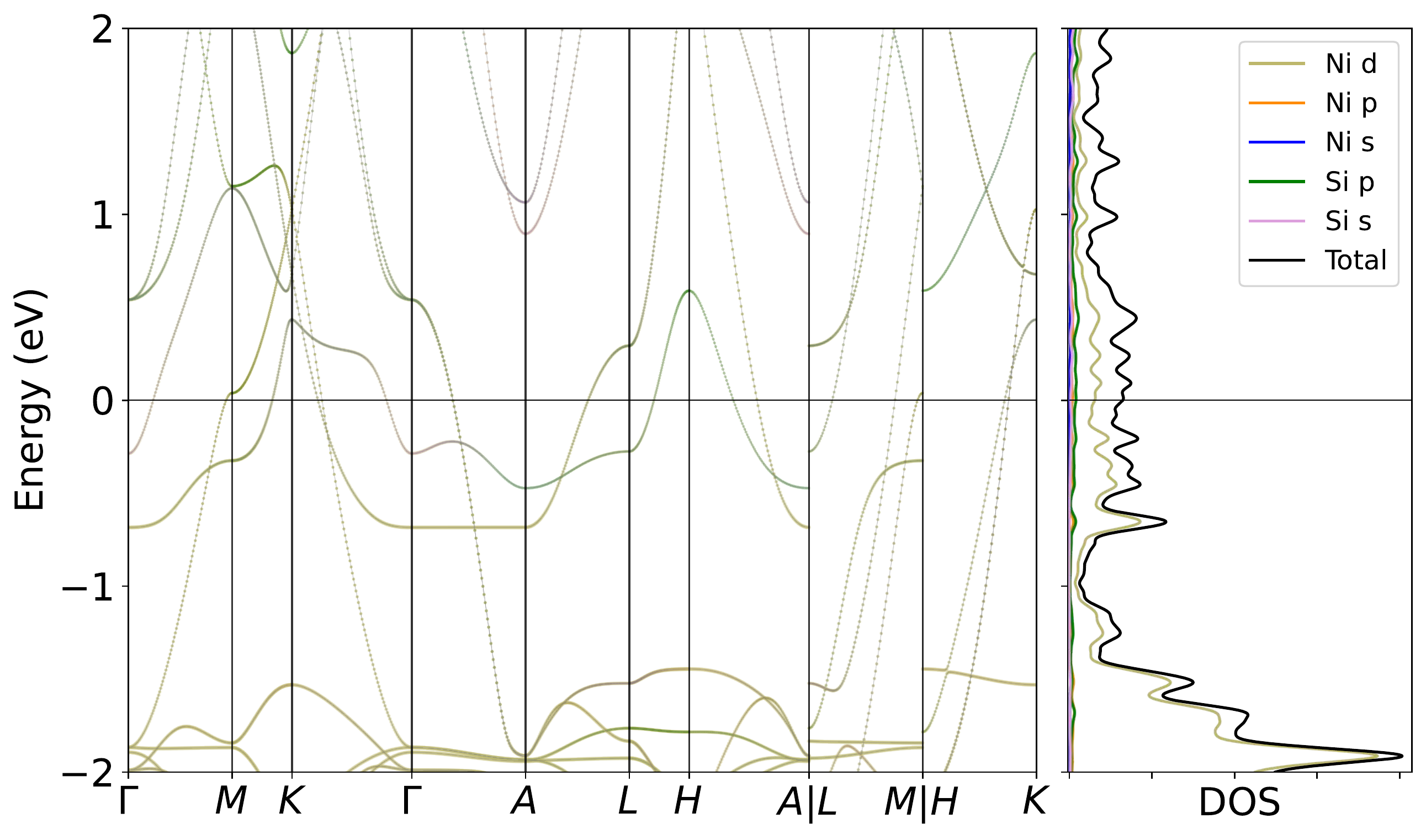}
\caption{Band structure for NiSi without SOC. }
\label{fig:NiSiband}
\end{figure}

\begin{figure}[H]
\centering
\subfloat[Relaxed phonon at low-T.]{
\label{fig:NiSi_relaxed_Low}
\includegraphics[width=0.45\textwidth]{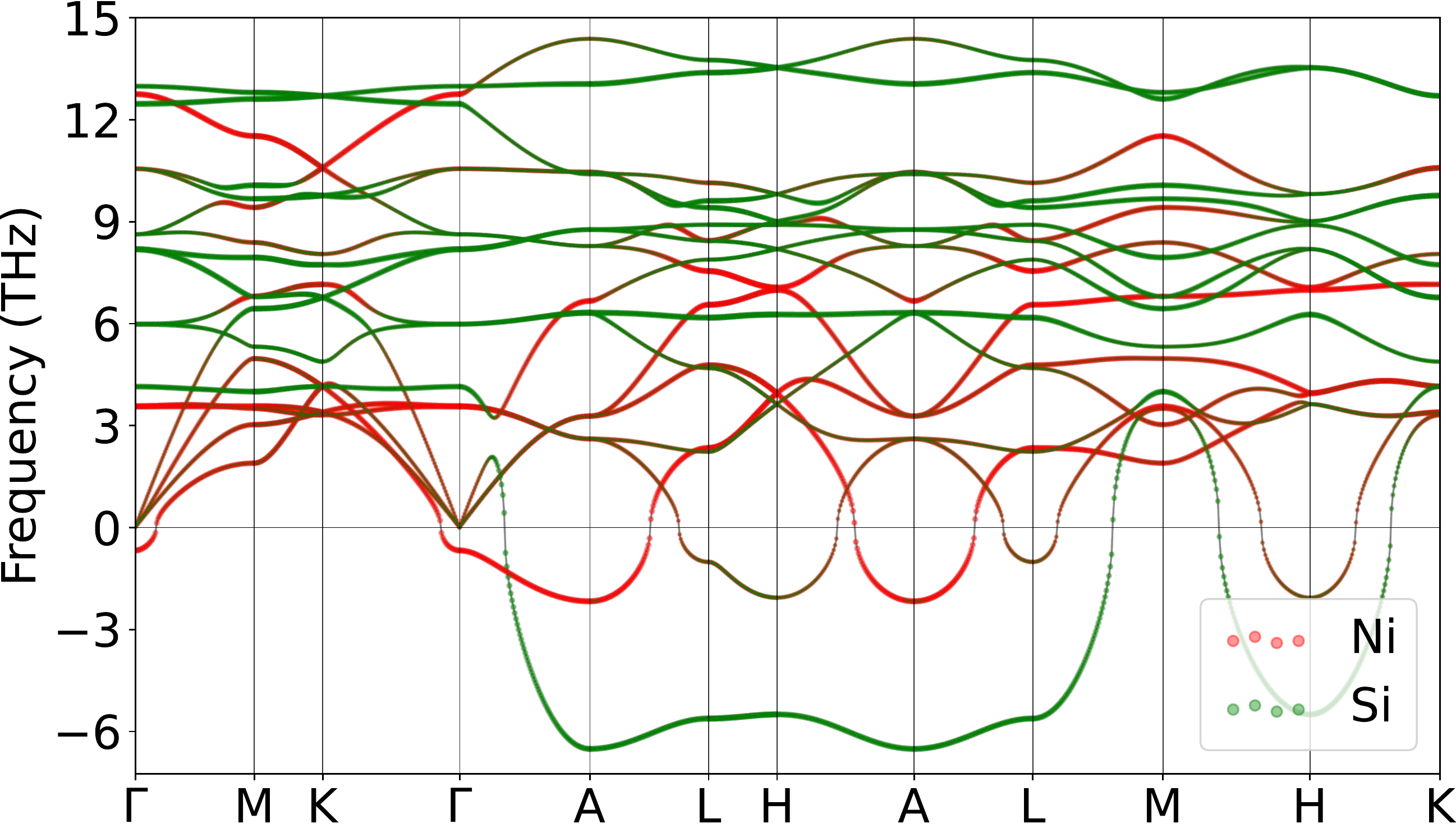}
}
\hspace{5pt}\subfloat[Relaxed phonon at high-T.]{
\label{fig:NiSi_relaxed_High}
\includegraphics[width=0.45\textwidth]{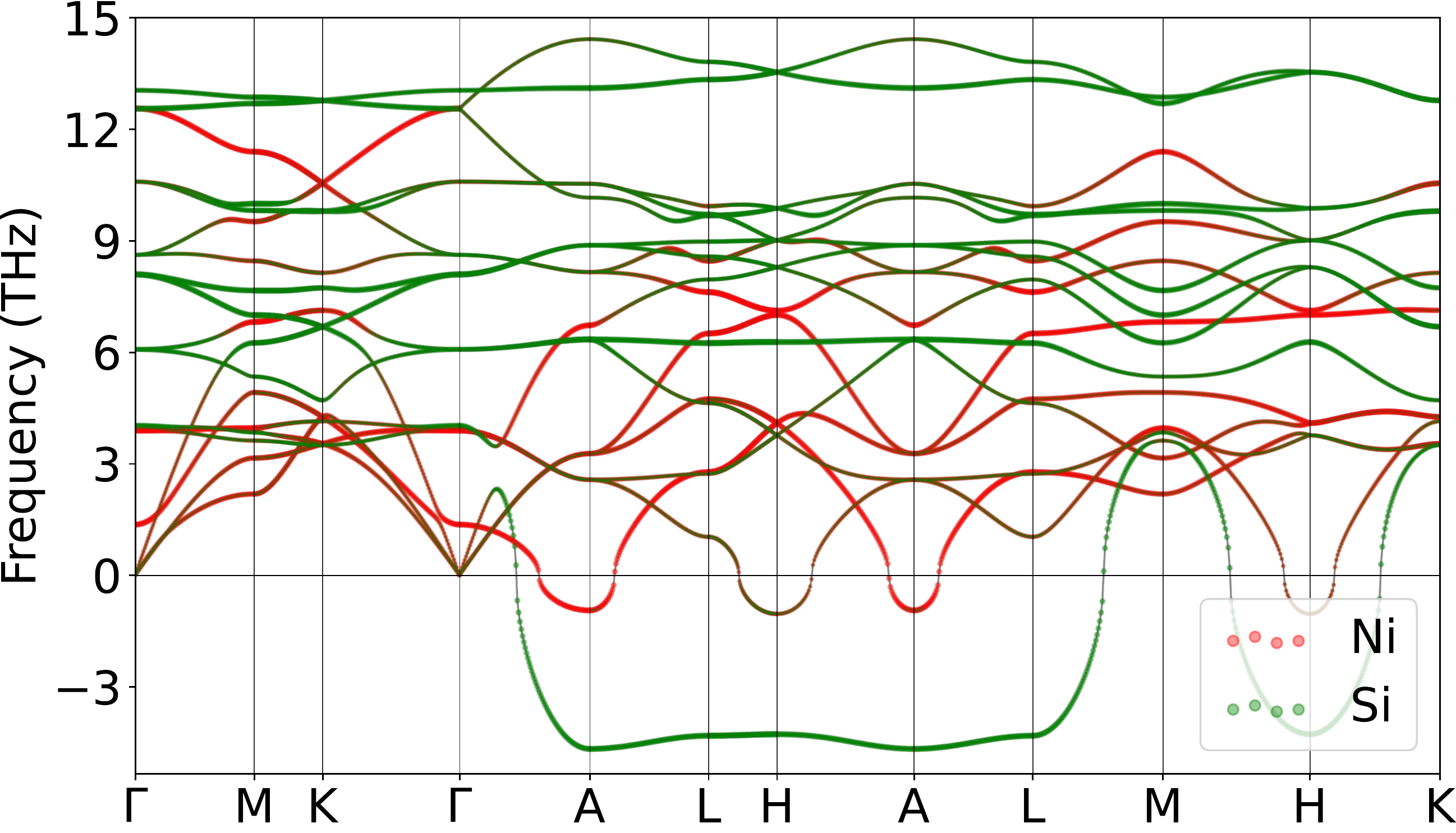}
}
\caption{Atomically projected phonon spectrum for NiSi.}
\label{fig:NiSipho}
\end{figure}

\subsubsection{\label{sms2sec:NiGe}\hyperref[smsec:col]{\texorpdfstring{NiGe}{}}~{\color{red} (high-T unstable, low-T unstable)}}

\begin{figure}[htbp]
\centering
\includegraphics[width=0.5\textwidth]{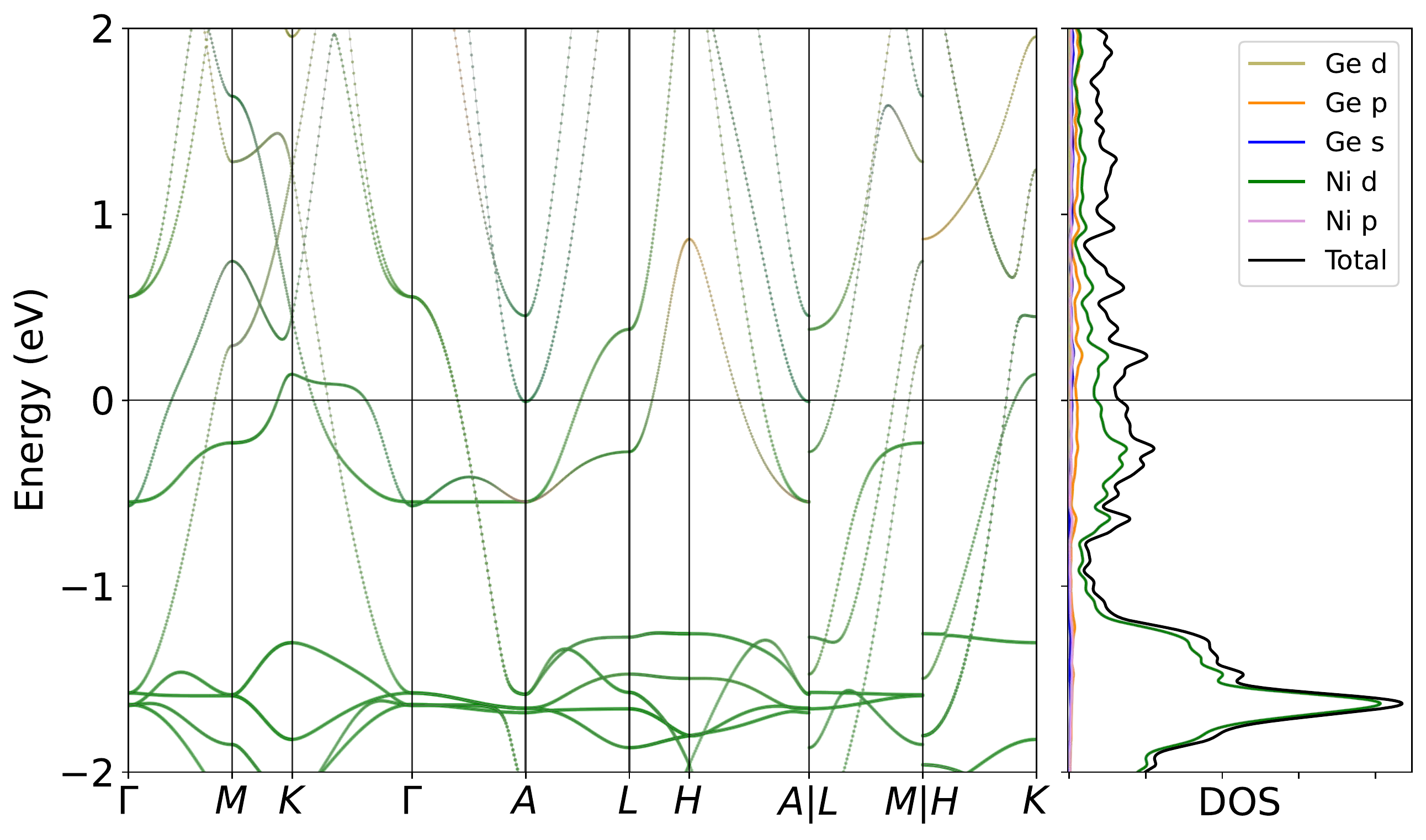}
\caption{Band structure for NiGe without SOC. }
\label{fig:NiGeband}
\end{figure}

\begin{figure}[H]
\centering
\subfloat[Relaxed phonon at low-T.]{
\label{fig:NiGe_relaxed_Low}
\includegraphics[width=0.45\textwidth]{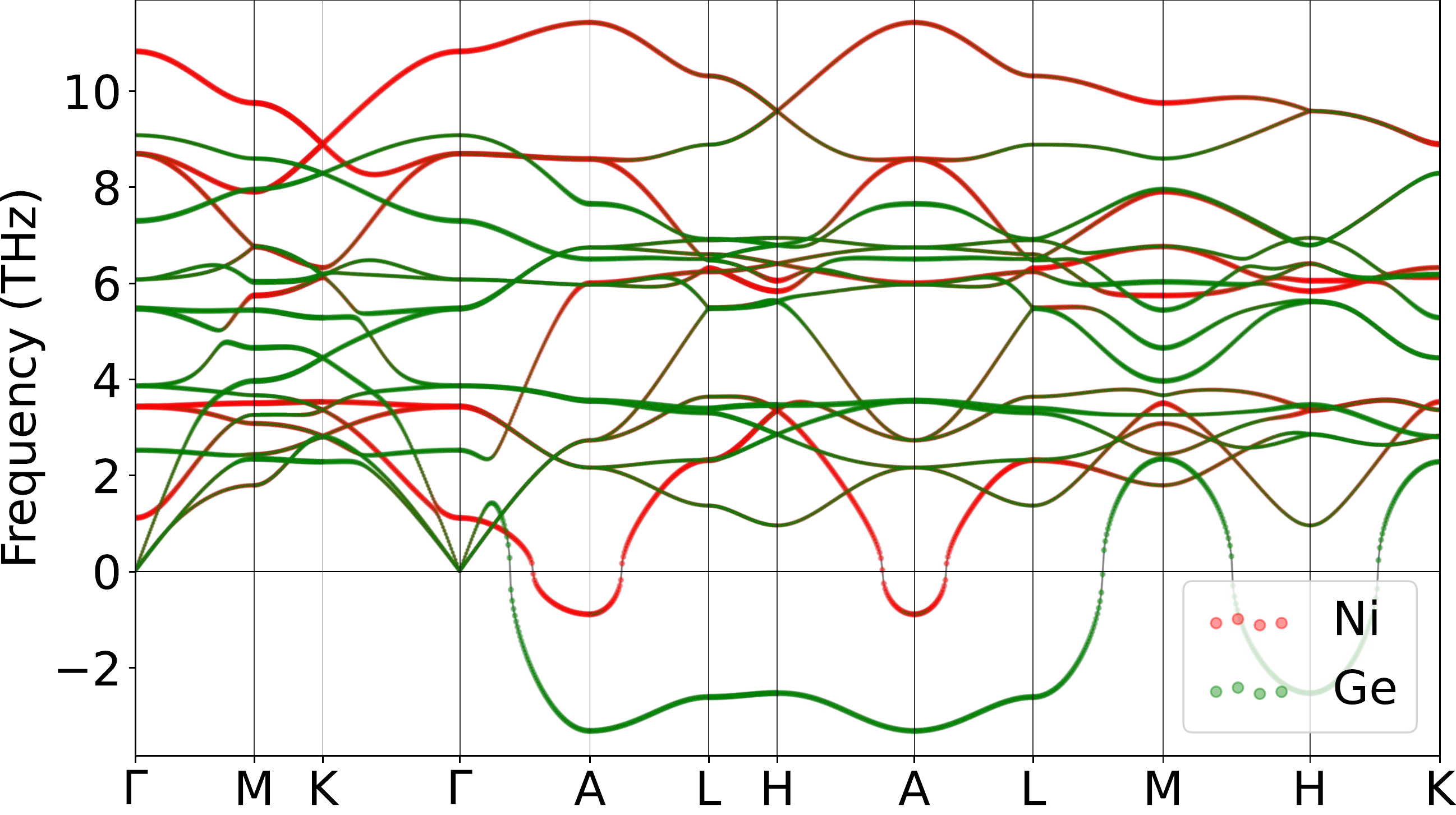}
}
\hspace{5pt}\subfloat[Relaxed phonon at high-T.]{
\label{fig:NiGe_relaxed_High}
\includegraphics[width=0.45\textwidth]{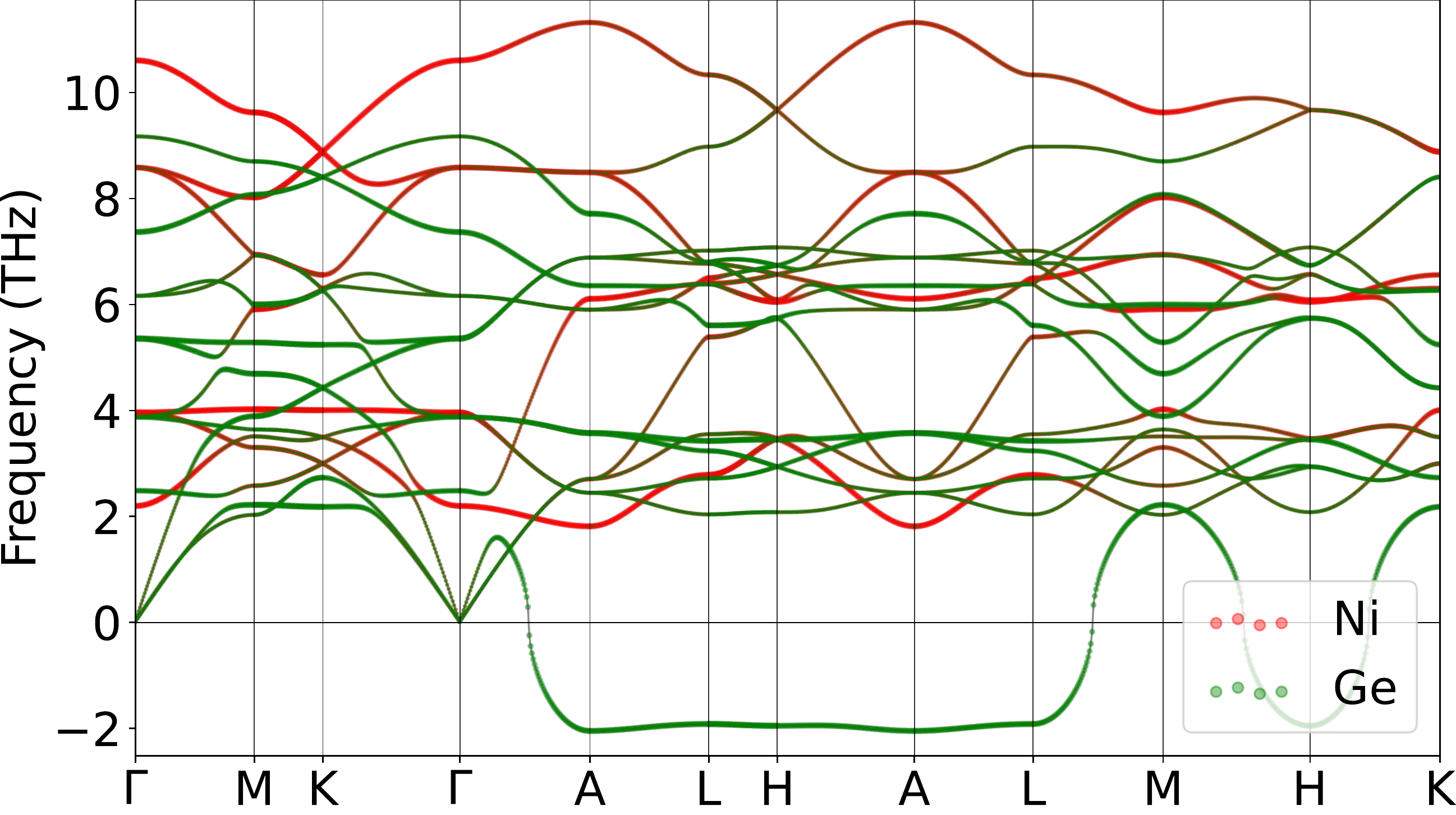}
}
\caption{Atomically projected phonon spectrum for NiGe.}
\label{fig:NiGepho}
\end{figure}

\subsubsection{\label{sms2sec:NiIn}\hyperref[smsec:col]{\texorpdfstring{NiIn}{}}~{\color{green} (high-T stable, low-T stable)}}

\begin{figure}[htbp]
\centering
\includegraphics[width=0.5\textwidth]{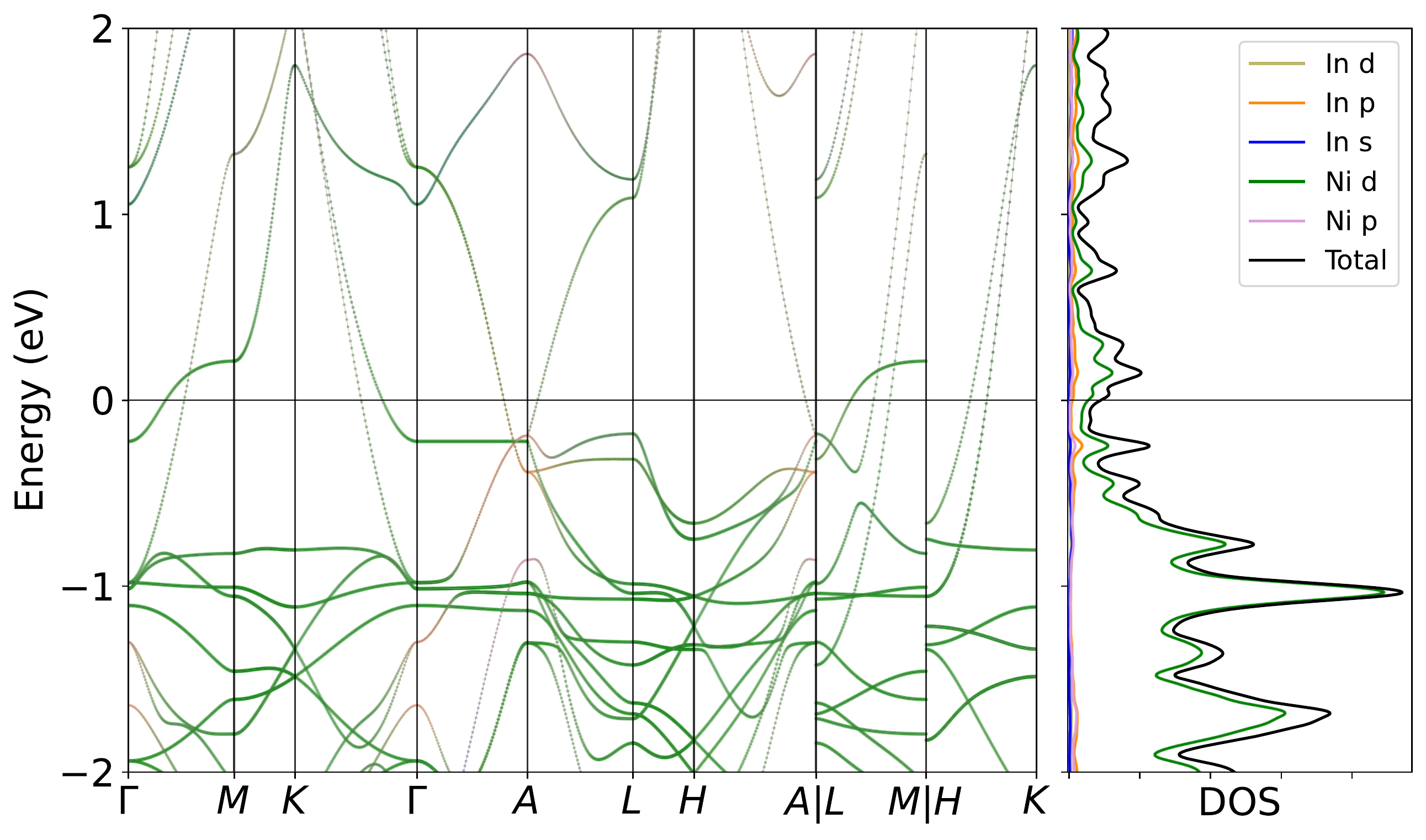}
\caption{Band structure for NiIn without SOC. }
\label{fig:NiInband}
\end{figure}

\begin{figure}[H]
\centering
\subfloat[Relaxed phonon at low-T.]{
\label{fig:NiIn_relaxed_Low}
\includegraphics[width=0.45\textwidth]{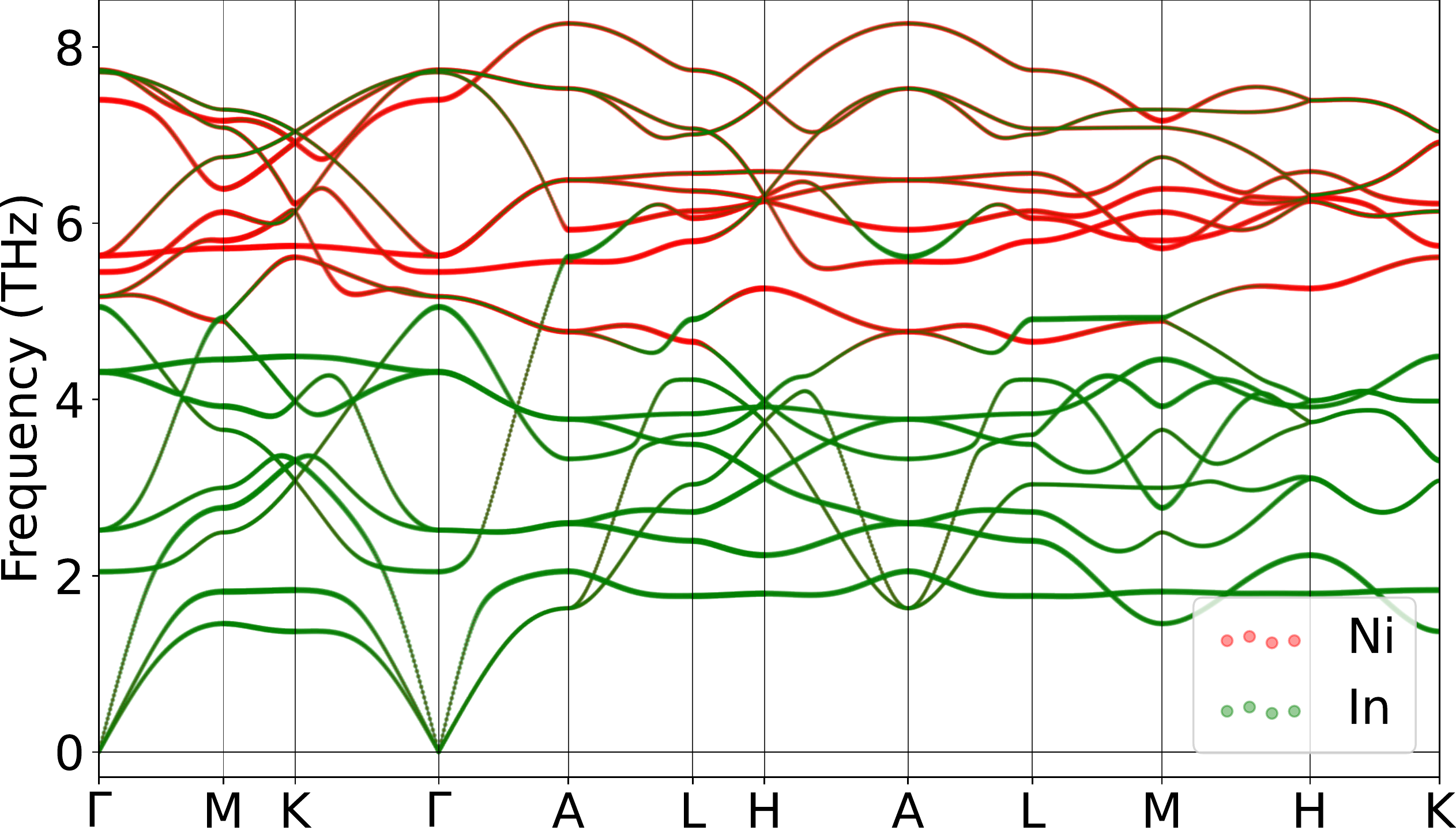}
}
\hspace{5pt}\subfloat[Relaxed phonon at high-T.]{
\label{fig:NiIn_relaxed_High}
\includegraphics[width=0.45\textwidth]{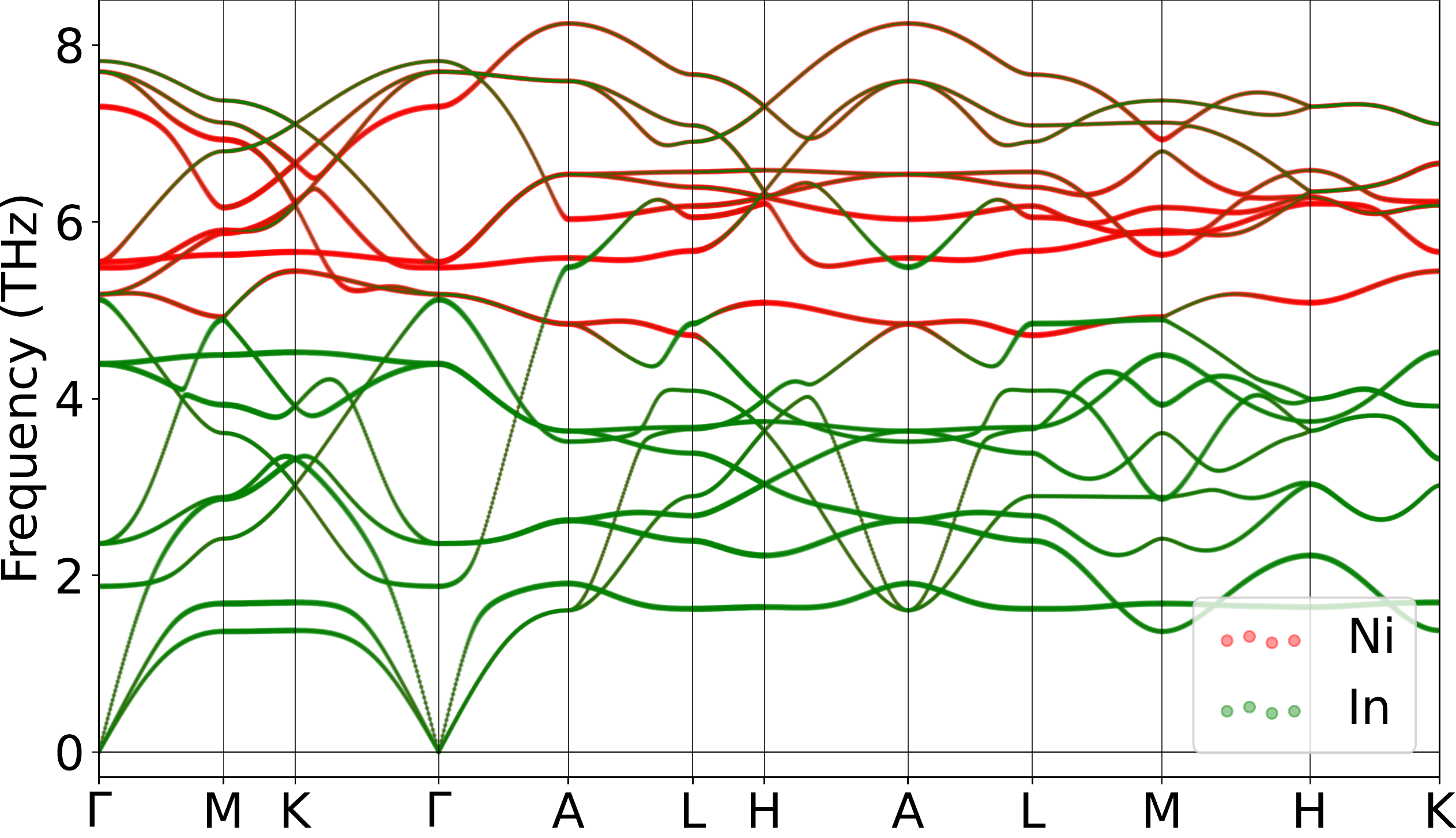}
}
\caption{Atomically projected phonon spectrum for NiIn.}
\label{fig:NiInpho}
\end{figure}

\subsubsection{\label{sms2sec:NiSn}\hyperref[smsec:col]{\texorpdfstring{NiSn}{}}~{\color{red} (high-T unstable, low-T unstable)}}

\begin{figure}[htbp]
\centering
\includegraphics[width=0.5\textwidth]{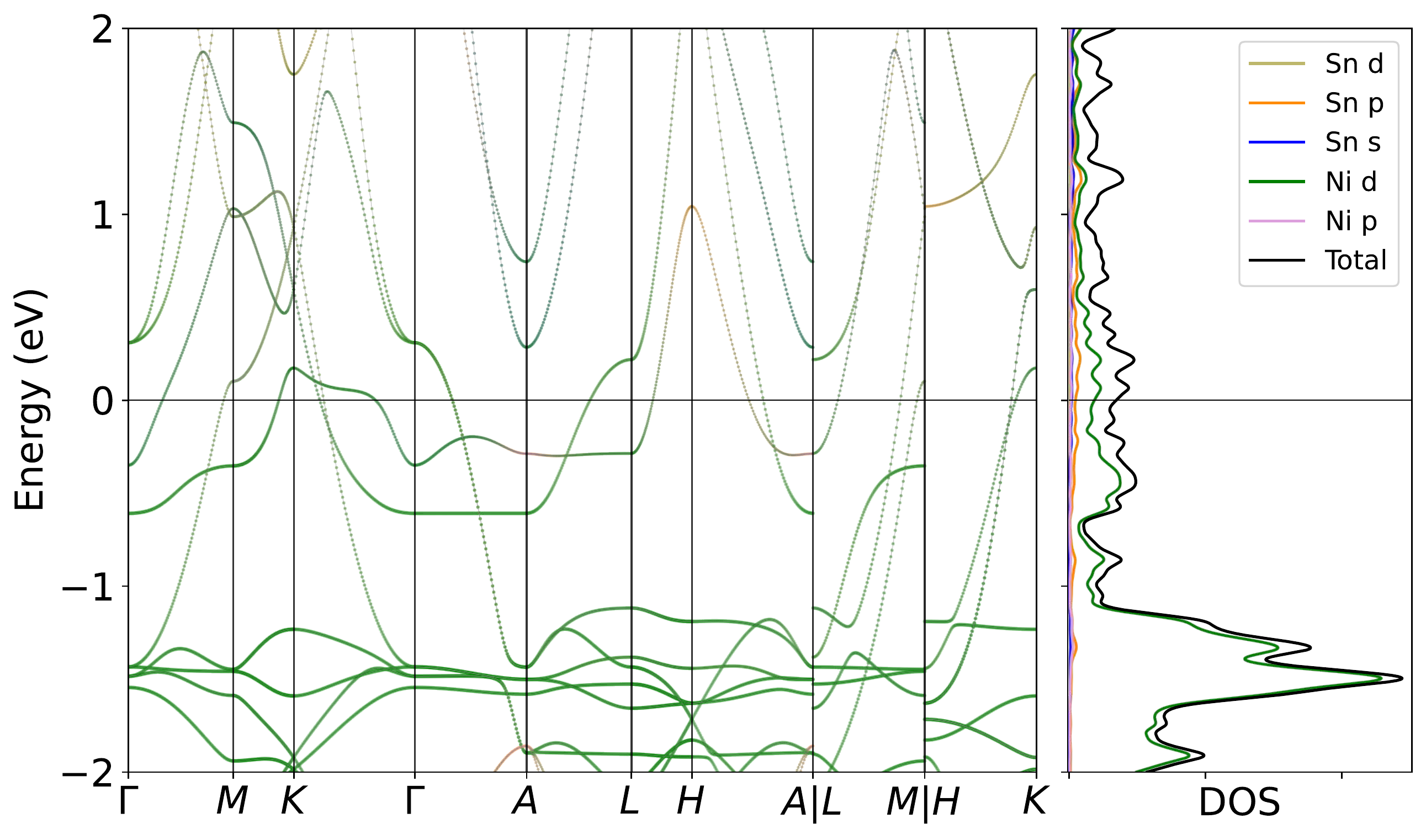}
\caption{Band structure for NiSn without SOC. }
\label{fig:NiSnband}
\end{figure}

\begin{figure}[H]
\centering
\subfloat[Relaxed phonon at low-T.]{
\label{fig:NiSn_relaxed_Low}
\includegraphics[width=0.45\textwidth]{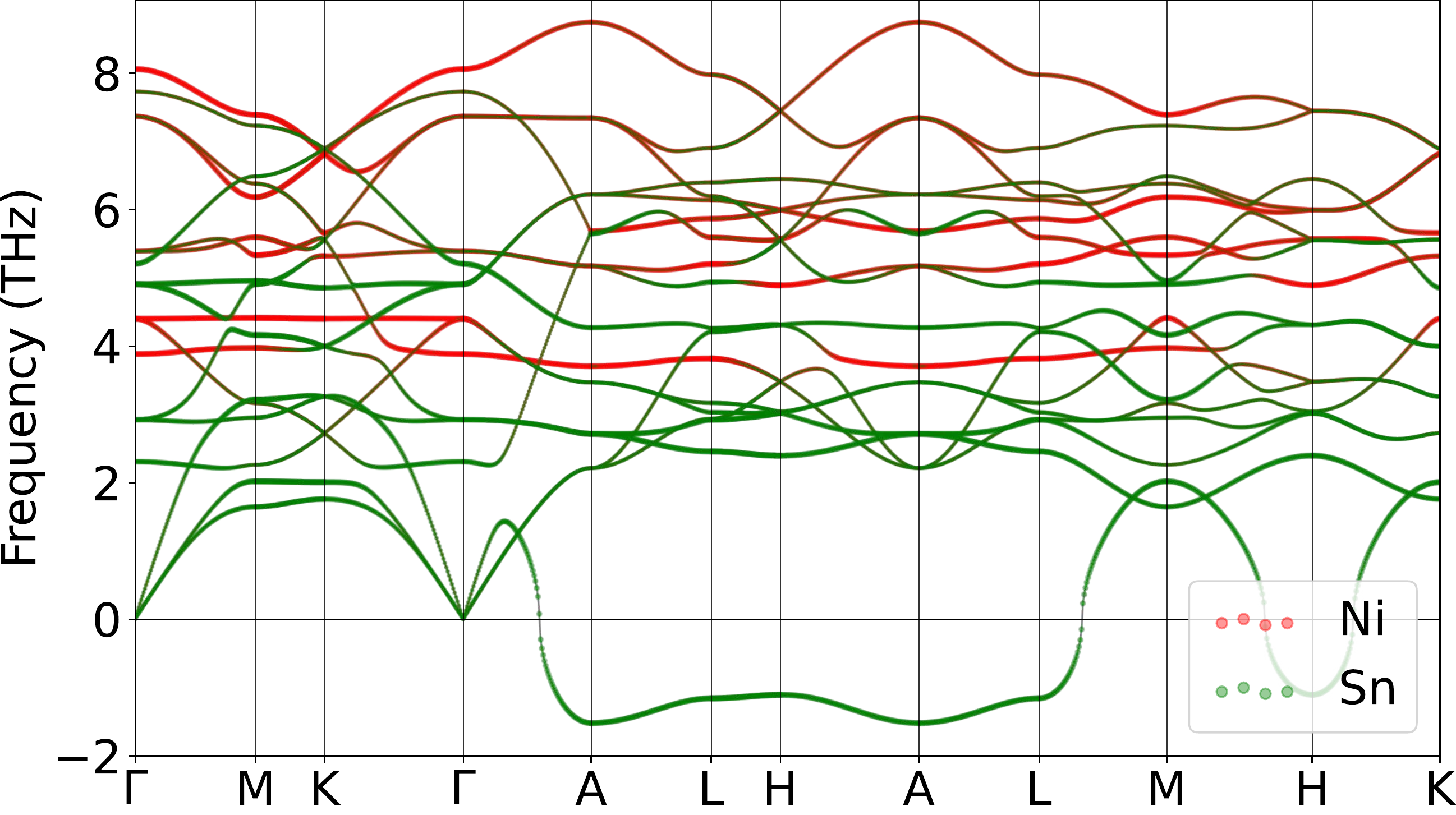}
}
\hspace{5pt}\subfloat[Relaxed phonon at high-T.]{
\label{fig:NiSn_relaxed_High}
\includegraphics[width=0.45\textwidth]{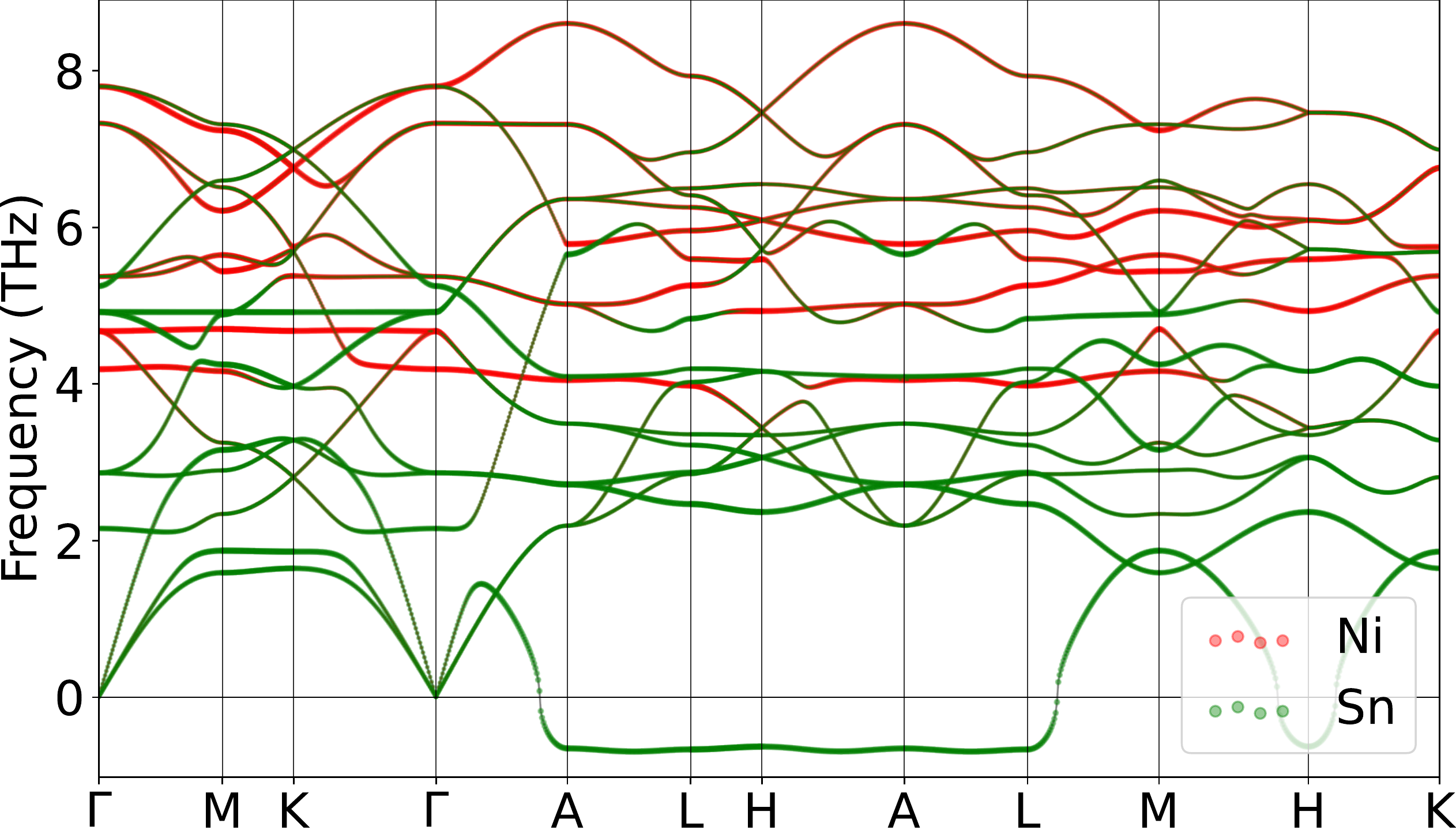}
}
\caption{Atomically projected phonon spectrum for NiSn.}
\label{fig:NiSnpho}
\end{figure}

\subsubsection{\label{sms2sec:NbSn}\hyperref[smsec:col]{\texorpdfstring{NbSn}{}}~{\color{green} (high-T stable, low-T stable)}}

\begin{figure}[htbp]
\centering
\includegraphics[width=0.5\textwidth]{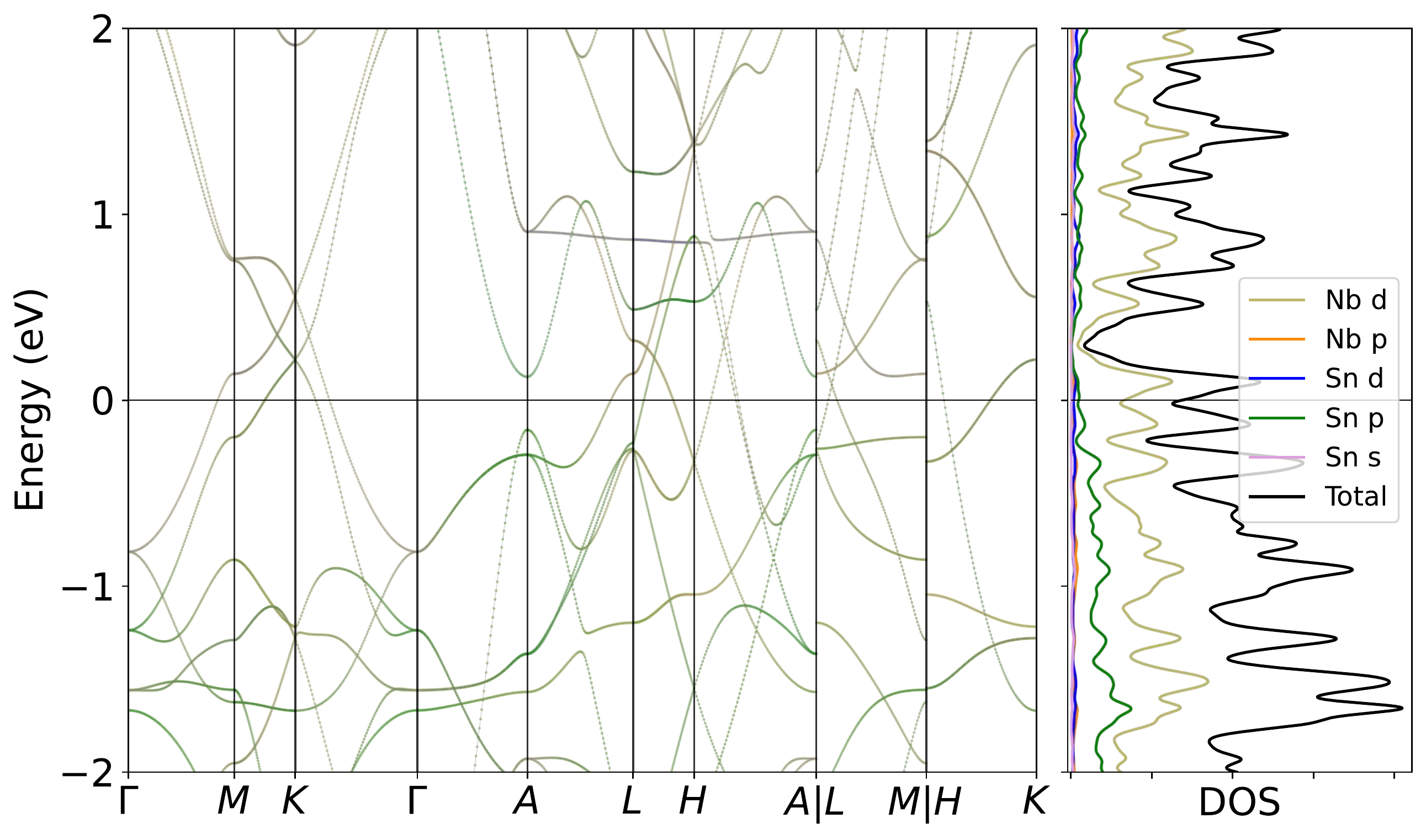}
\caption{Band structure for NbSn without SOC. }
\label{fig:NbSnband}
\end{figure}

\begin{figure}[H]
\centering
\subfloat[Relaxed phonon at low-T.]{
\label{fig:NbSn_relaxed_Low}
\includegraphics[width=0.45\textwidth]{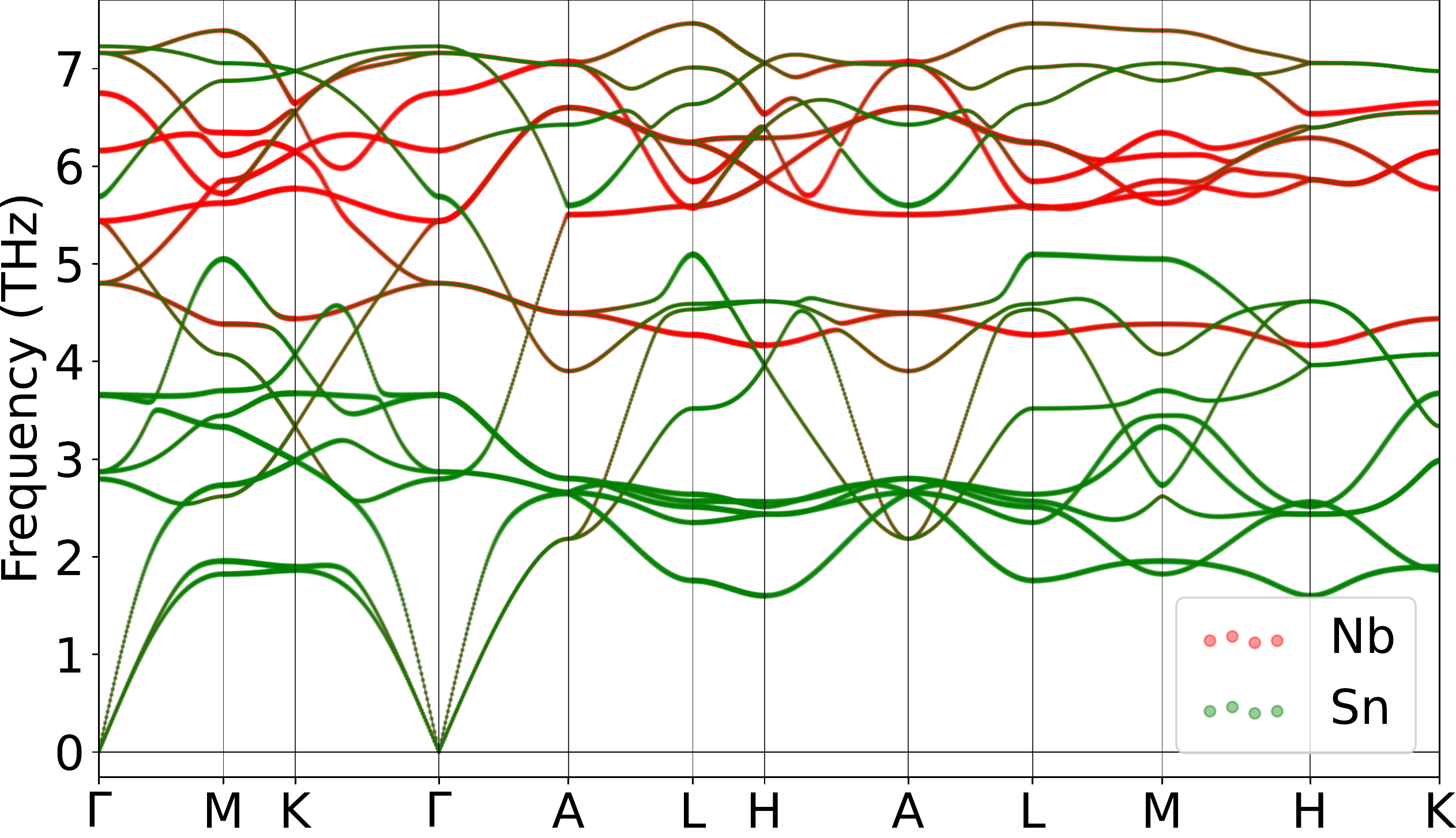}
}
\hspace{5pt}\subfloat[Relaxed phonon at high-T.]{
\label{fig:NbSn_relaxed_High}
\includegraphics[width=0.45\textwidth]{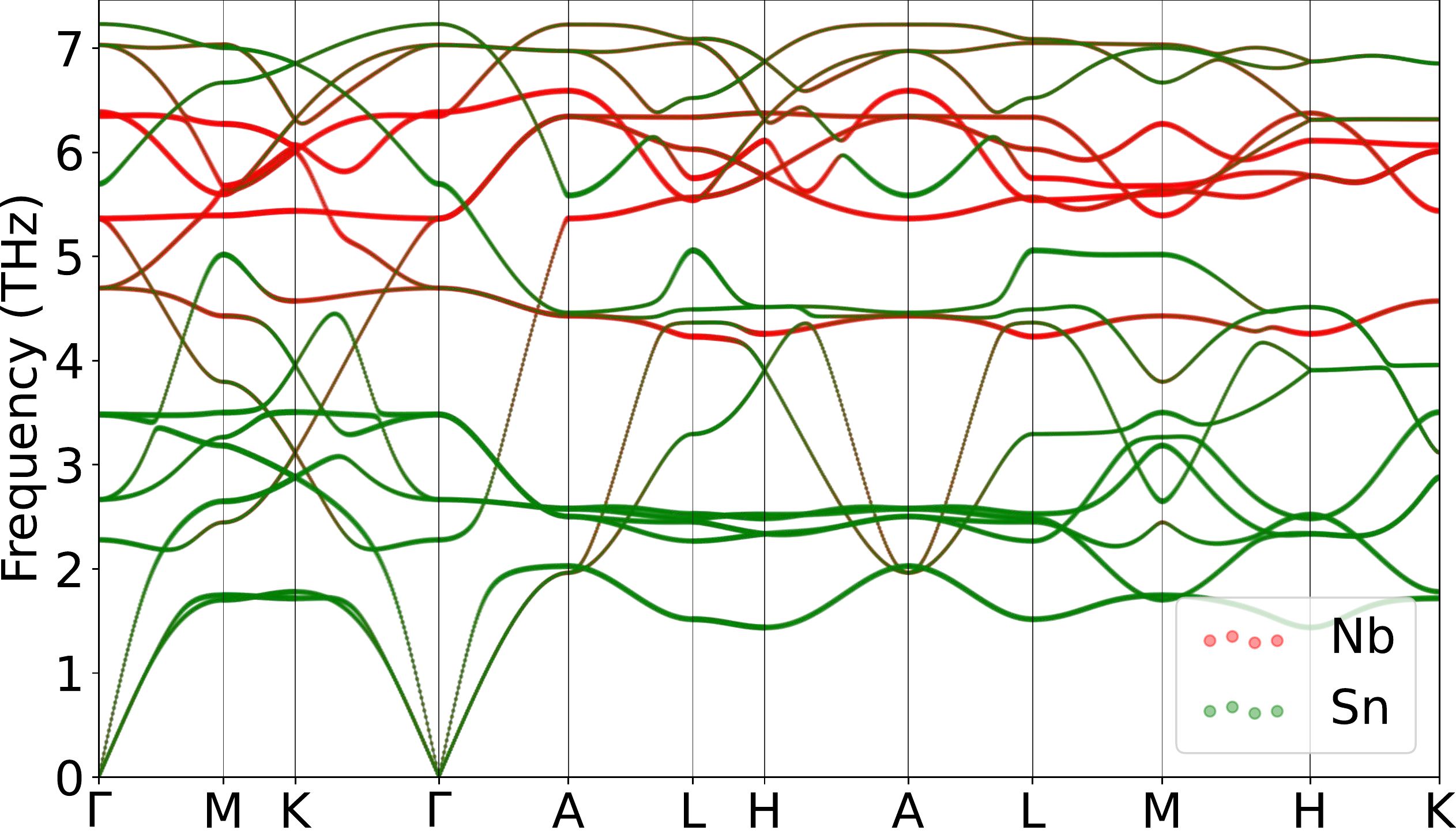}
}
\caption{Atomically projected phonon spectrum for NbSn.}
\label{fig:NbSnpho}
\end{figure}

\subsection{\hyperref[smsec:col]{VGe}}~\label{smssec:VGe}

\subsubsection{\label{sms2sec:LiV6Ge6}\hyperref[smsec:col]{\texorpdfstring{LiV$_6$Ge$_6$}{}}~{\color{green}  (High-T stable, Low-T stable) }}

\begin{table}[htbp]
\global\long\def\arraystretch{1.12}
\captionsetup{width=.85\textwidth}
\caption{Structure parameters of LiV$_6$Ge$_6$ after relaxation. It contains 333 electrons. }
\label{tab:LiV6Ge6}
\setlength{\tabcolsep}{4mm}{
\begin{tabular}{c|c||c|c}
\hline
\hline
 Atom & Coordinates & \multicolumn{2}{c}{Parameters~($\AA$)} \\
\hline
Li ~ 1a & ($0,0,0$) & $a$ & $5.1909$ \\
V ~ 6i & ($\frac{1}{2},0,0.2505$) & $c$ & $8.6002$ \\
Ge1 ~ 2c & ($\frac{1}{3},\frac{2}{3},0$) & $d^{Tri}_{Z-Z}$ & $3.6932$ \\
Ge2 ~ 2d & ($\frac{1}{3},\frac{2}{3},\frac{1}{2}$) & $d_{T_{1}}$ & $2.5954$ \\
Ge3 ~ 2e & ($0,0,0.2853$) & $d_{T_{2}}$ & $4.2921$ \\
\hline
\hline
\end{tabular}}
\end{table}

\begin{figure}[htbp]
\centering
\includegraphics[width=0.85\textwidth]{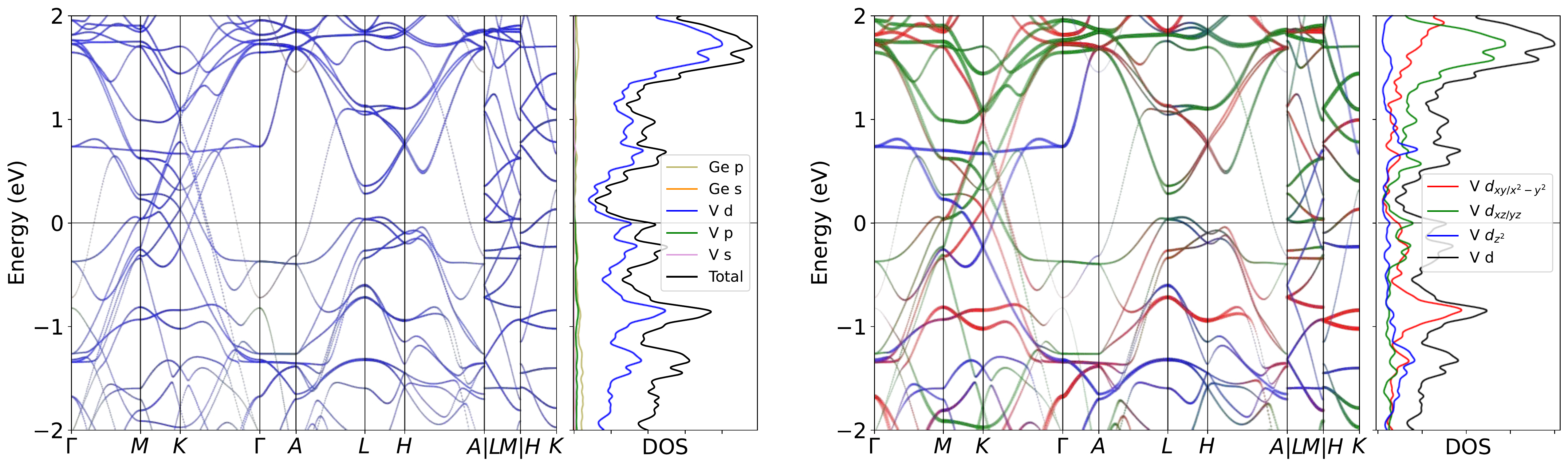}
\caption{Band structure for LiV$_6$Ge$_6$ without SOC. The projection of V $3d$ orbitals is also presented. The band structures clearly reveal the dominating of V $3d$ orbitals  near the Fermi level, as well as the pronounced peak.}
\label{fig:LiV6Ge6band}
\end{figure}

\begin{figure}[H]
\centering
\subfloat[Relaxed phonon at low-T.]{
\label{fig:LiV6Ge6_relaxed_Low}
\includegraphics[width=0.45\textwidth]{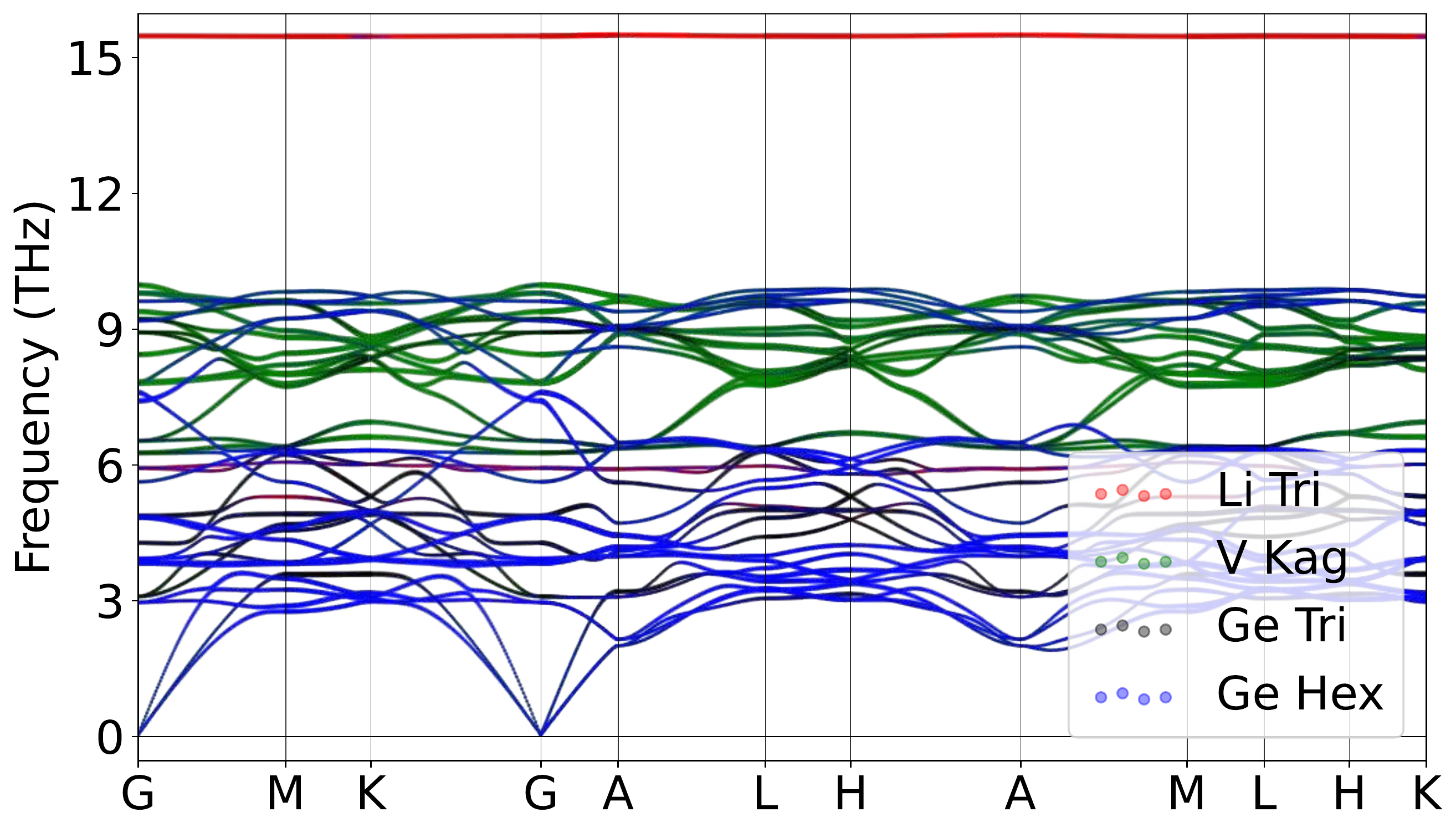}
}
\hspace{5pt}\subfloat[Relaxed phonon at high-T.]{
\label{fig:LiV6Ge6_relaxed_High}
\includegraphics[width=0.45\textwidth]{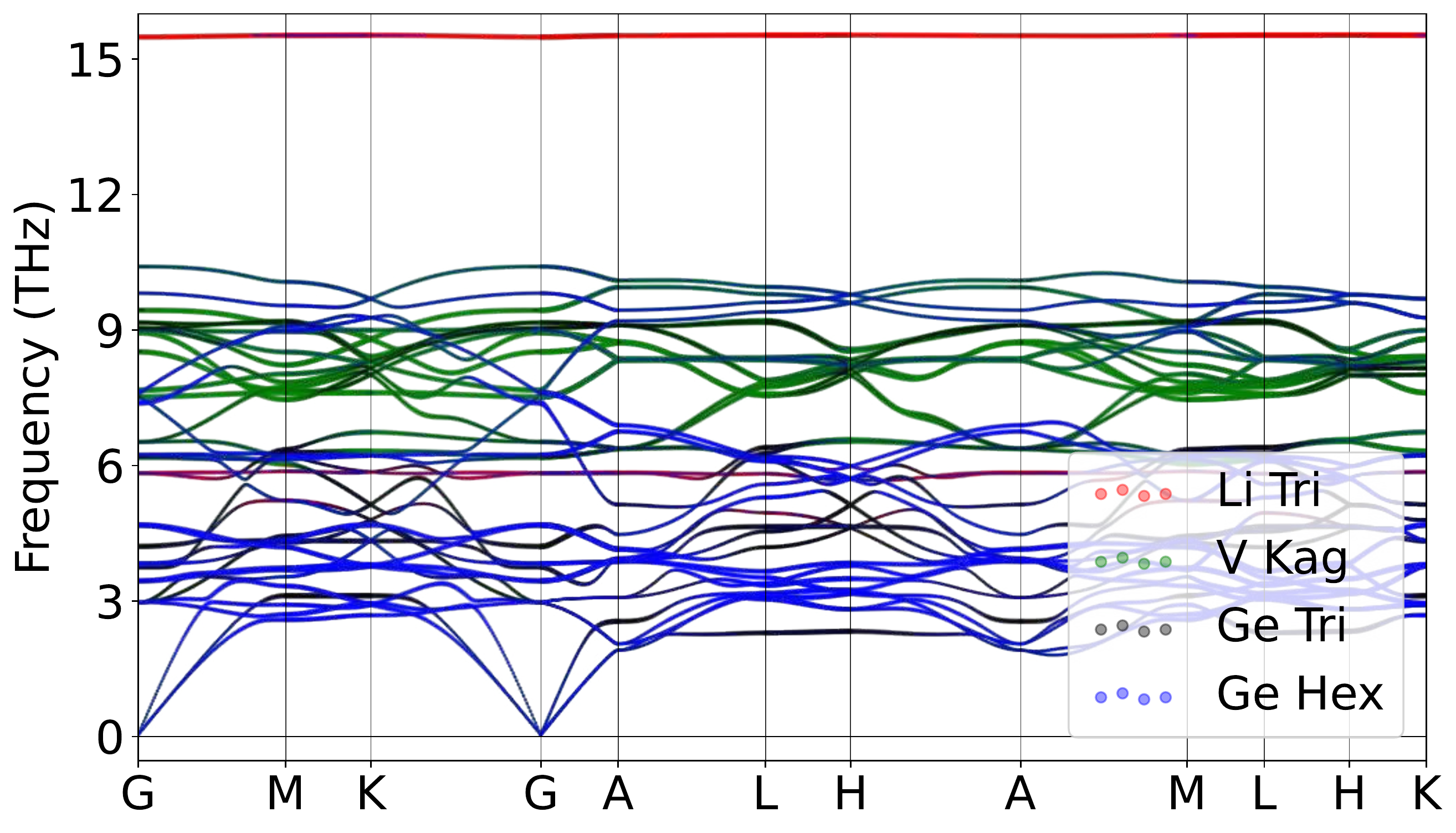}
}
\caption{Atomically projected phonon spectrum for LiV$_6$Ge$_6$.}
\label{fig:LiV6Ge6pho}
\end{figure}

\begin{figure}[H]
\centering
\label{fig:LiV6Ge6_relaxed_Low_xz}
\includegraphics[width=0.85\textwidth]{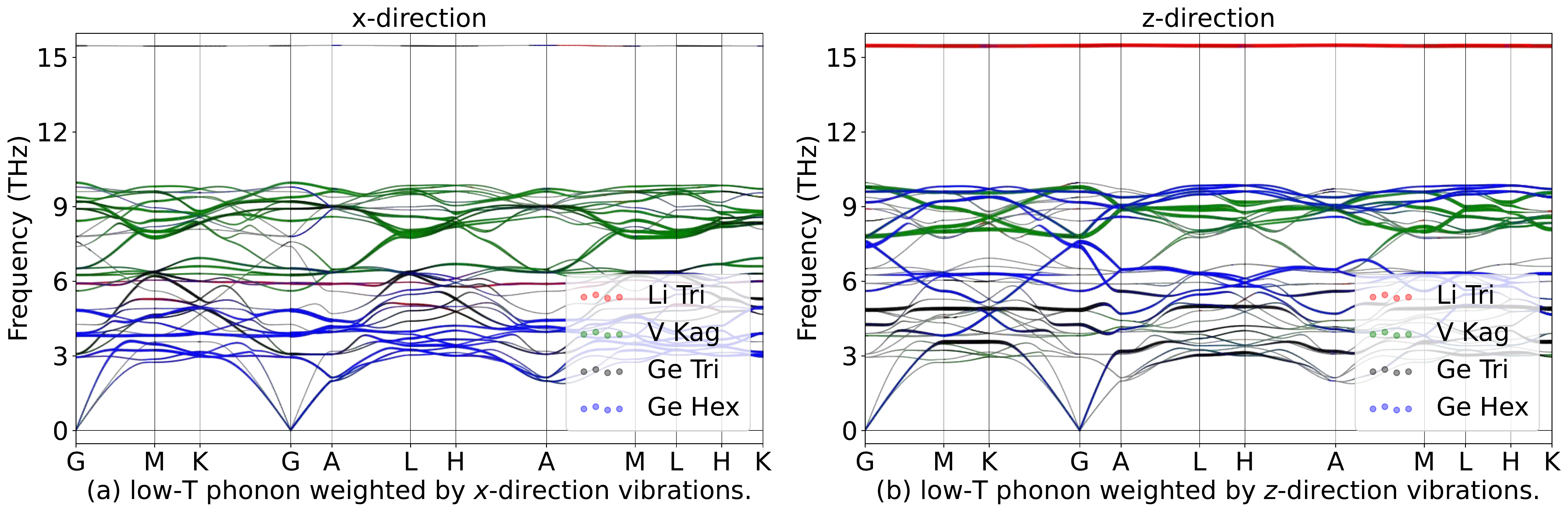}
\hspace{5pt}\label{fig:LiV6Ge6_relaxed_High_xz}
\includegraphics[width=0.85\textwidth]{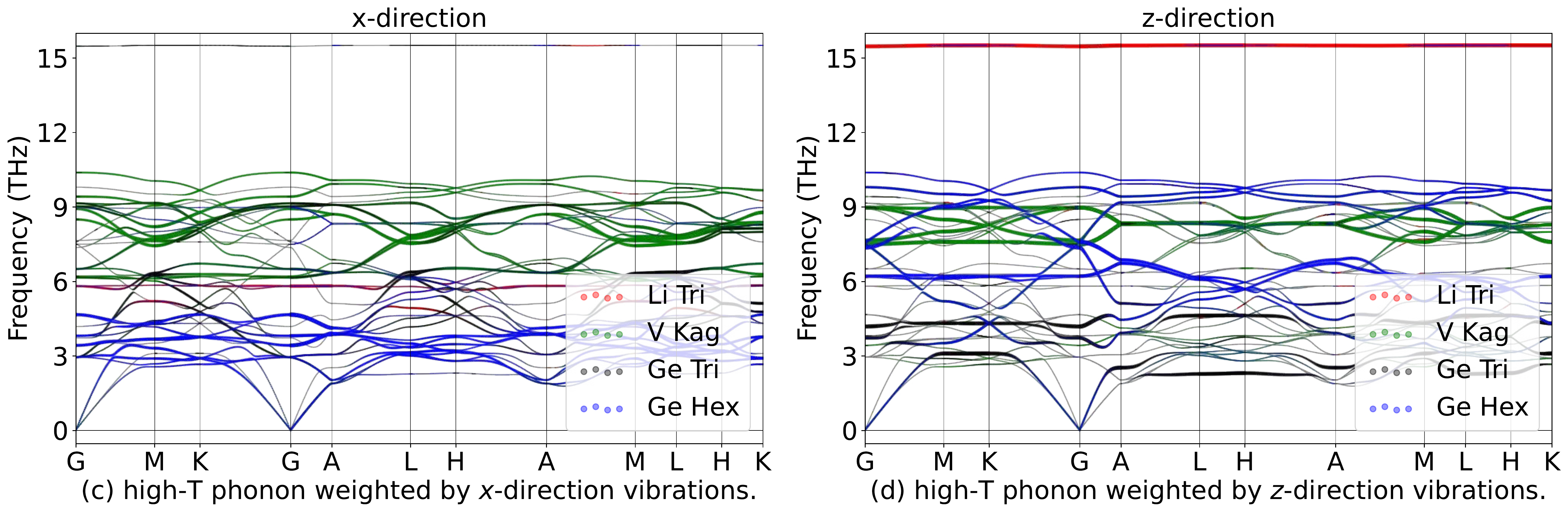}
\caption{Phonon spectrum for LiV$_6$Ge$_6$in in-plane ($x$) and out-of-plane ($z$) directions.}
\label{fig:LiV6Ge6phoxyz}
\end{figure}

\begin{table}[htbp]
\global\long\def\arraystretch{1.12}
\captionsetup{width=.95\textwidth}
\caption{IRREPs at high-symmetry points for LiV$_6$Ge$_6$ phonon spectrum at Low-T.}
\label{tab:191_LiV6Ge6_Low}
\setlength{\tabcolsep}{1.mm}{
\begin{tabular}{c|cc|cc|cc|cc|cc|cc}
\hline
\hline
 \multirow{2}{*}{Band}  & \multicolumn{2}{c}{$\Gamma$ $(0,0,0)$}  & \multicolumn{2}{c}{$A$ $(0,0,\frac{1}{2})$}  & \multicolumn{2}{c}{$M$ $(\frac{1}{2},0,0)$}  & \multicolumn{2}{c}{$L$ $(\frac{1}{2},0,\frac{1}{2})$}  & \multicolumn{2}{c}{$K$ $(\frac{1}{3},\frac{1}{3},0)$}  & \multicolumn{2}{c}{$H$ $(\frac{1}{3},\frac{1}{3},\frac{1}{2})$}   \\
\cline{2-13}
 & $f$ (THz) & Reps & $f$ (THz) & Reps & $f$ (THz) & Reps & $f$ (THz) & Reps & $f$ (THz) & Reps & $f$ (THz) & Reps  \\
\hline
1 & \multirow{3}{*}{  -0.000} & \multirow{3}{*}{$-$$|$$-$} & \multirow{2}{*}{   1.864} & \multirow{2}{*}{$E_{1g}$$|$${A_5^+}$} &    2.531 & $B_{2u}$$|$${M_2^-}$ &    2.908 & $B_{2u}$$|$${L_2^-}$ & \multirow{2}{*}{   2.769} & \multirow{2}{*}{$E'$$|$${K_6}$} & \multirow{2}{*}{   2.721} & \multirow{2}{*}{$E'$$|$${H_6}$} \\
\cline{1-1}\cline{6-7}\cline{8-9}
2 &   &   &   &   &    2.737 & $B_{1g}$$|$${M_3^+}$ &    2.954 & $B_{1u}$$|$${L_3^-}$ &   &   &   &   \\
\cline{1-1}\cline{4-5}\cline{6-7}\cline{8-9}\cline{10-11}\cline{12-13}
3 &   &   & \multirow{2}{*}{   1.959} & \multirow{2}{*}{$E_{1u}$$|$${A_5^-}$} &    3.094 & $A_{g}$$|$${M_1^+}$ &    2.979 & $B_{3u}$$|$${L_4^-}$ &    2.888 & $A_{1}'$$|$${K_1}$ &    3.067 & $A_{2}''$$|$${H_4}$ \\
\cline{1-1}\cline{2-3}\cline{6-7}\cline{8-9}\cline{10-11}\cline{12-13}
4 & \multirow{2}{*}{   2.544} & \multirow{2}{*}{$E_{1u}$$|$${\Gamma_5^-}$} &   &   &    3.177 & $B_{3u}$$|$${M_4^-}$ &    3.036 & $A_{u}$$|$${L_1^-}$ &    3.027 & $A_{2}'$$|$${K_2}$ &    3.196 & $A_{1}''$$|$${H_3}$ \\
\cline{1-1}\cline{4-5}\cline{6-7}\cline{8-9}\cline{10-11}\cline{12-13}
5 &   &   & \multirow{2}{*}{   2.709} & \multirow{2}{*}{$E_{1u}$$|$${A_5^-}$} &    3.478 & $B_{1u}$$|$${M_3^-}$ &    3.272 & $B_{2g}$$|$${L_2^+}$ &    3.493 & $A_{2}''$$|$${K_4}$ & \multirow{2}{*}{   3.259} & \multirow{2}{*}{$E''$$|$${H_5}$} \\
\hline
\hline
\end{tabular}}
\end{table}

\begin{table}[htbp]
\global\long\def\arraystretch{1.12}
\captionsetup{width=.95\textwidth}
\caption{IRREPs at high-symmetry points for LiV$_6$Ge$_6$ phonon spectrum at High-T.}
\label{tab:191_LiV6Ge6_High}
\setlength{\tabcolsep}{1.mm}{
\begin{tabular}{c|cc|cc|cc|cc|cc|cc}
\hline
\hline
 \multirow{2}{*}{Band}  & \multicolumn{2}{c}{$\Gamma$ $(0,0,0)$}  & \multicolumn{2}{c}{$A$ $(0,0,\frac{1}{2})$}  & \multicolumn{2}{c}{$M$ $(\frac{1}{2},0,0)$}  & \multicolumn{2}{c}{$L$ $(\frac{1}{2},0,\frac{1}{2})$}  & \multicolumn{2}{c}{$K$ $(\frac{1}{3},\frac{1}{3},0)$}  & \multicolumn{2}{c}{$H$ $(\frac{1}{3},\frac{1}{3},\frac{1}{2})$}   \\
\cline{2-13}
 & $f$ (THz) & Reps & $f$ (THz) & Reps & $f$ (THz) & Reps & $f$ (THz) & Reps & $f$ (THz) & Reps & $f$ (THz) & Reps  \\
\hline
1 & \multirow{3}{*}{  -0.000} & \multirow{3}{*}{$-$$|$$-$} & \multirow{2}{*}{   1.750} & \multirow{2}{*}{$E_{1g}$$|$${A_5^+}$} &    2.499 & $B_{2u}$$|$${M_2^-}$ &    2.166 & $B_{1u}$$|$${L_3^-}$ &    2.606 & $A_{1}'$$|$${K_1}$ &    2.204 & $A_{2}''$$|$${H_4}$ \\
\cline{1-1}\cline{6-7}\cline{8-9}\cline{10-11}\cline{12-13}
2 &   &   &   &   &    2.585 & $B_{1g}$$|$${M_3^+}$ &    2.877 & $B_{3u}$$|$${L_4^-}$ & \multirow{2}{*}{   2.756} & \multirow{2}{*}{$E'$$|$${K_6}$} & \multirow{2}{*}{   2.668} & \multirow{2}{*}{$E'$$|$${H_6}$} \\
\cline{1-1}\cline{4-5}\cline{6-7}\cline{8-9}
3 &   &   & \multirow{2}{*}{   1.923} & \multirow{2}{*}{$E_{1u}$$|$${A_5^-}$} &    2.790 & $A_{g}$$|$${M_1^+}$ &    2.914 & $A_{u}$$|$${L_1^-}$ &   &   &   &   \\
\cline{1-1}\cline{2-3}\cline{6-7}\cline{8-9}\cline{10-11}\cline{12-13}
4 & \multirow{2}{*}{   2.709} & \multirow{2}{*}{$E_{1u}$$|$${\Gamma_5^-}$} &   &   &    2.951 & $B_{1u}$$|$${M_3^-}$ &    2.988 & $B_{2u}$$|$${L_2^-}$ &    2.833 & $A_{2}'$$|$${K_2}$ & \multirow{2}{*}{   2.989} & \multirow{2}{*}{$E''$$|$${H_5}$} \\
\cline{1-1}\cline{4-5}\cline{6-7}\cline{8-9}\cline{10-11}
5 &   &   &    2.373 & $A_{2u}$$|$${A_2^-}$ &    3.133 & $B_{3u}$$|$${M_4^-}$ &    3.019 & $B_{2g}$$|$${L_2^+}$ &    2.951 & $A_{2}''$$|$${K_4}$ &   &   \\
\hline
\hline
\end{tabular}}
\end{table}
\subsubsection{\label{sms2sec:MgV6Ge6}\hyperref[smsec:col]{\texorpdfstring{MgV$_6$Ge$_6$}{}}~{\color{red}  (High-T unstable, Low-T unstable) }}

\begin{table}[htbp]
\global\long\def\arraystretch{1.12}
\captionsetup{width=.85\textwidth}
\caption{Structure parameters of MgV$_6$Ge$_6$ after relaxation. It contains 342 electrons. }
\label{tab:MgV6Ge6}
\setlength{\tabcolsep}{4mm}{
\begin{tabular}{c|c||c|c}
\hline
\hline
 Atom & Coordinates & \multicolumn{2}{c}{Parameters~($\AA$)} \\
\hline
Mg ~ 1a & ($0,0,0$) & $a$ & $5.2082$ \\
V ~ 6i & ($\frac{1}{2},0,0.2503$) & $c$ & $8.5642$ \\
Ge1 ~ 2c & ($\frac{1}{3},\frac{2}{3},0$) & $d^{Tri}_{Z-Z}$ & $3.3225$ \\
Ge2 ~ 2d & ($\frac{1}{3},\frac{2}{3},\frac{1}{2}$) & $d_{T_{1}}$ & $2.6041$ \\
Ge3 ~ 2e & ($0,0,0.3060$) & $d_{T_{2}}$ & $4.2769$ \\
\hline
\hline
\end{tabular}}
\end{table}

\begin{figure}[htbp]
\centering
\includegraphics[width=0.85\textwidth]{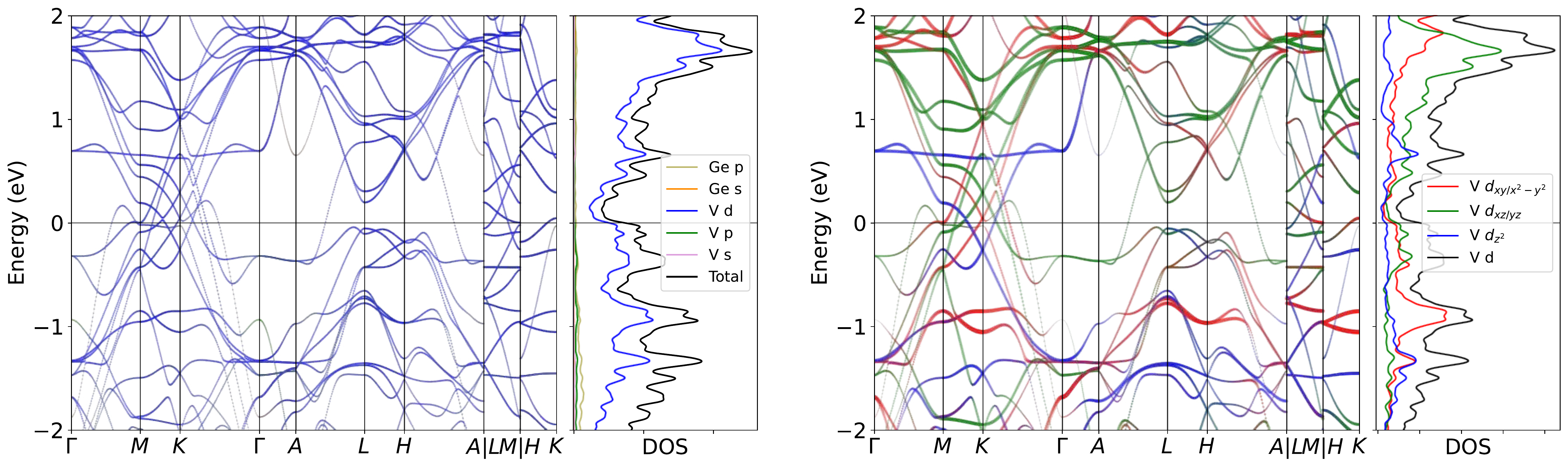}
\caption{Band structure for MgV$_6$Ge$_6$ without SOC. The projection of V $3d$ orbitals is also presented. The band structures clearly reveal the dominating of V $3d$ orbitals  near the Fermi level, as well as the pronounced peak.}
\label{fig:MgV6Ge6band}
\end{figure}

\begin{figure}[H]
\centering
\subfloat[Relaxed phonon at low-T.]{
\label{fig:MgV6Ge6_relaxed_Low}
\includegraphics[width=0.45\textwidth]{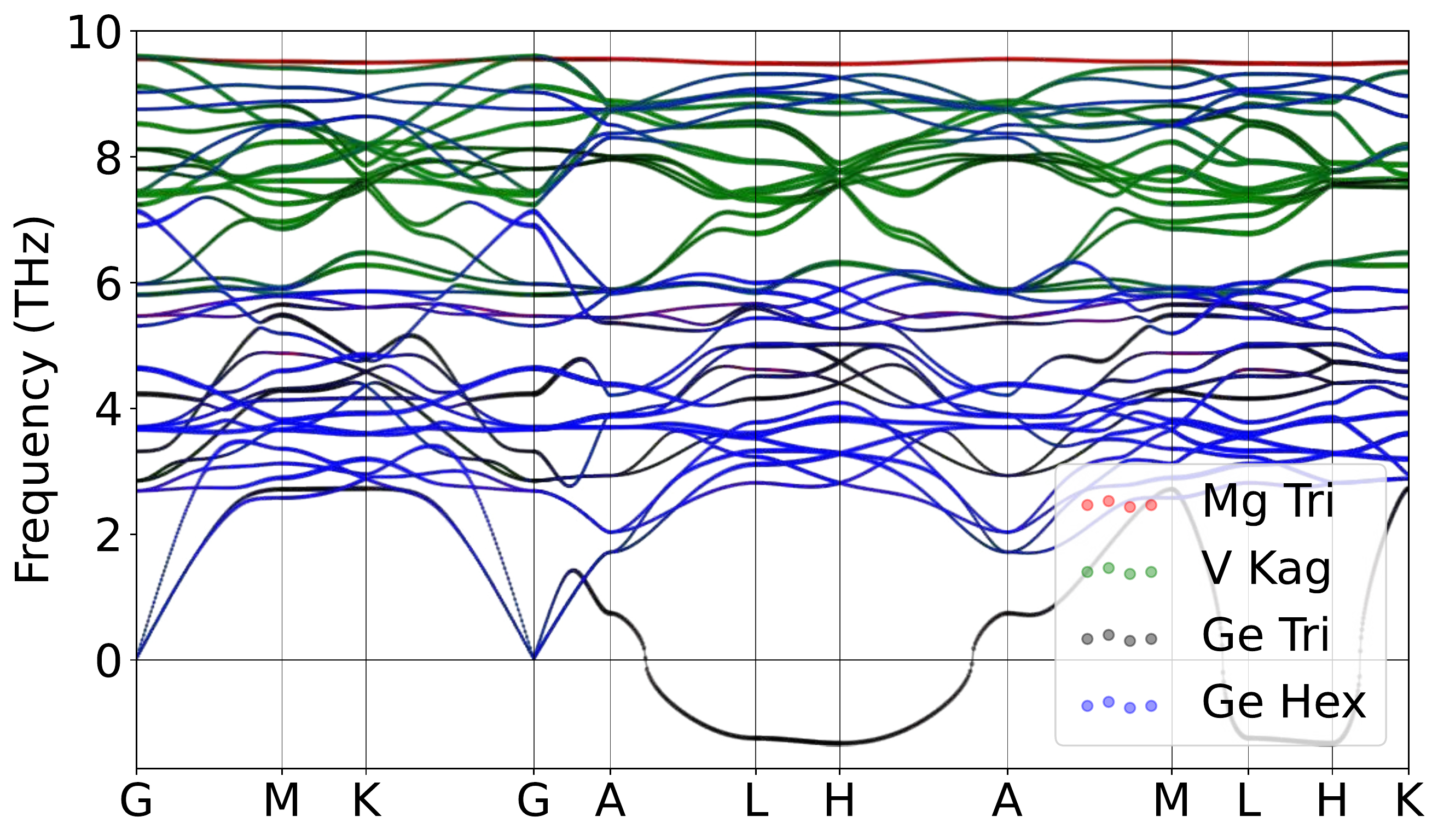}
}
\hspace{5pt}\subfloat[Relaxed phonon at high-T.]{
\label{fig:MgV6Ge6_relaxed_High}
\includegraphics[width=0.45\textwidth]{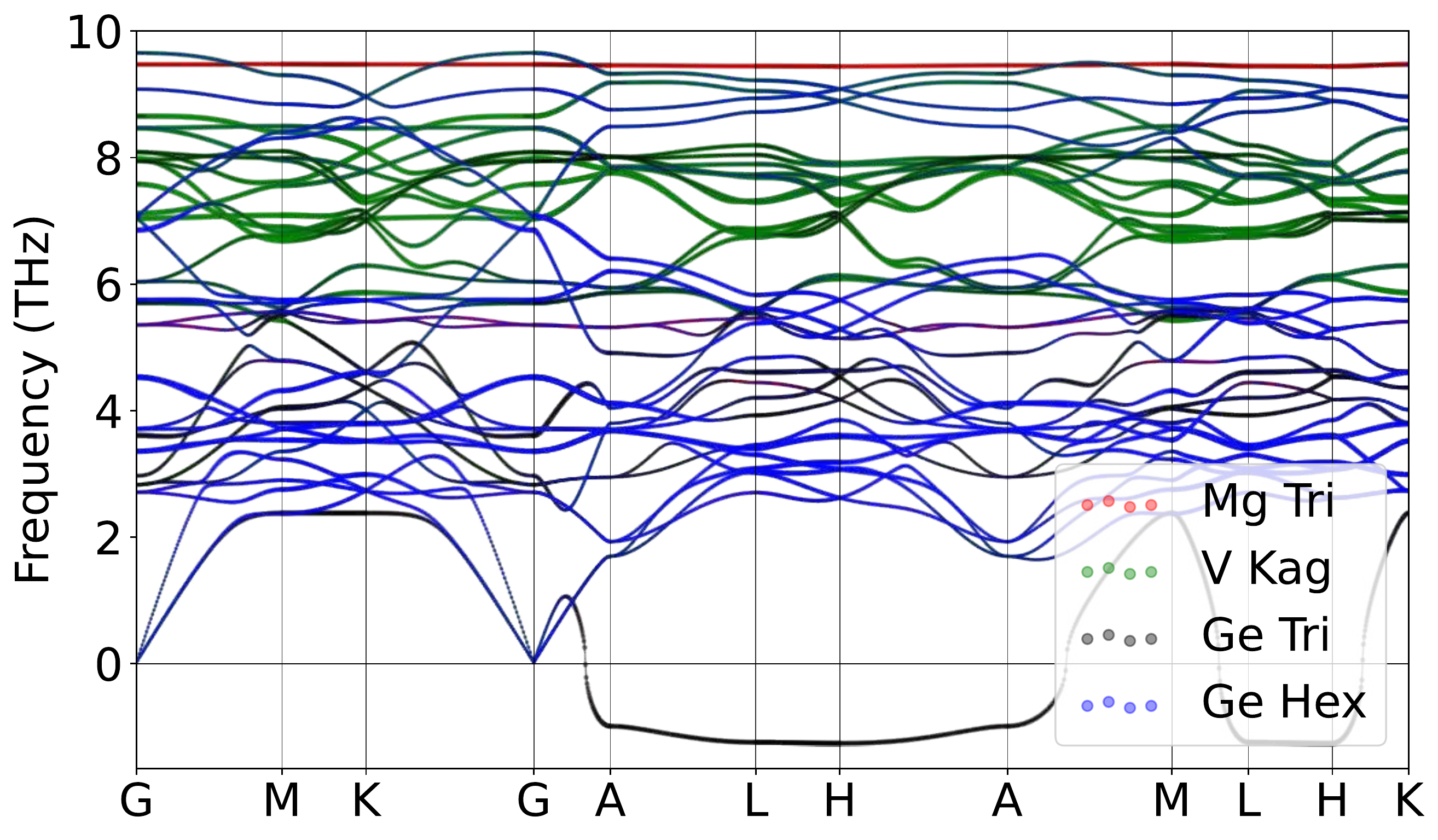}
}
\caption{Atomically projected phonon spectrum for MgV$_6$Ge$_6$.}
\label{fig:MgV6Ge6pho}
\end{figure}

\begin{figure}[H]
\centering
\label{fig:MgV6Ge6_relaxed_Low_xz}
\includegraphics[width=0.85\textwidth]{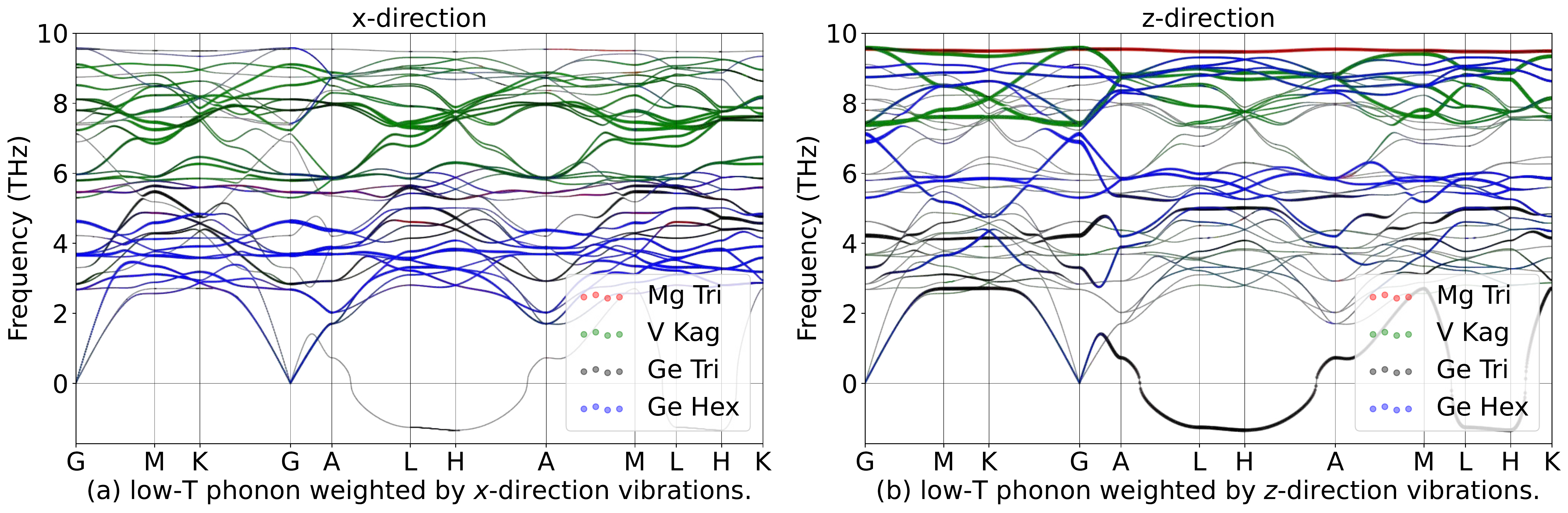}
\hspace{5pt}\label{fig:MgV6Ge6_relaxed_High_xz}
\includegraphics[width=0.85\textwidth]{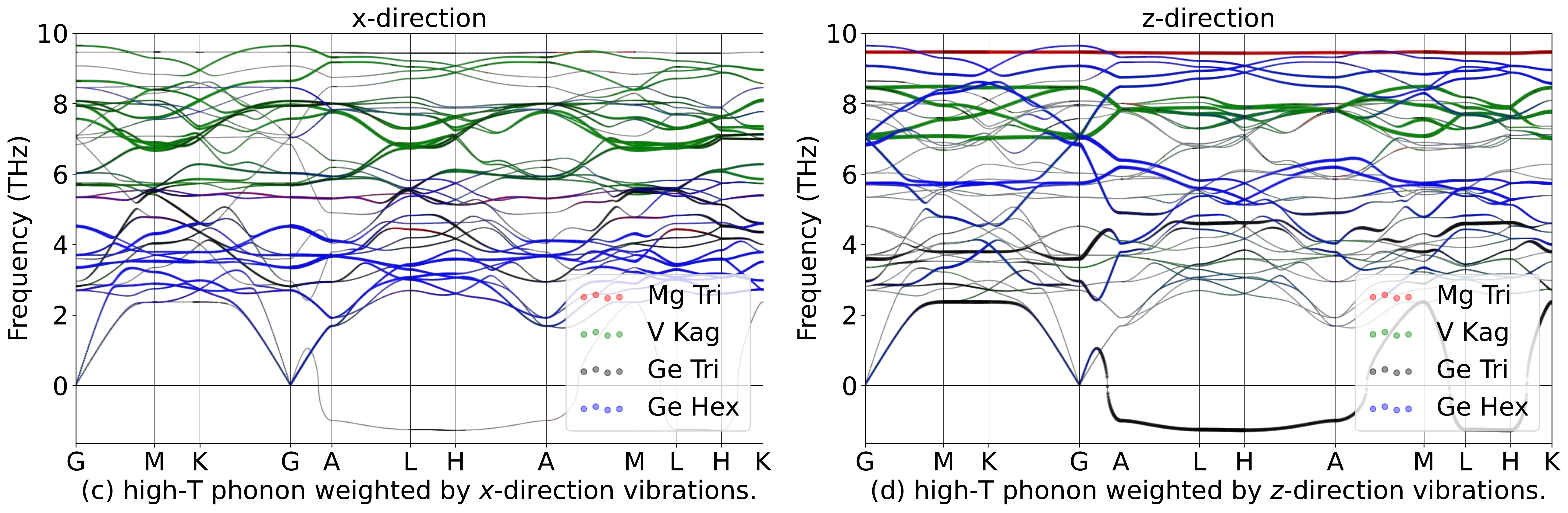}
\caption{Phonon spectrum for MgV$_6$Ge$_6$in in-plane ($x$) and out-of-plane ($z$) directions.}
\label{fig:MgV6Ge6phoxyz}
\end{figure}

\begin{table}[htbp]
\global\long\def\arraystretch{1.12}
\captionsetup{width=.95\textwidth}
\caption{IRREPs at high-symmetry points for MgV$_6$Ge$_6$ phonon spectrum at Low-T.}
\label{tab:191_MgV6Ge6_Low}
\setlength{\tabcolsep}{1.mm}{
\begin{tabular}{c|cc|cc|cc|cc|cc|cc}
\hline
\hline
 \multirow{2}{*}{Band}  & \multicolumn{2}{c}{$\Gamma$ $(0,0,0)$}  & \multicolumn{2}{c}{$A$ $(0,0,\frac{1}{2})$}  & \multicolumn{2}{c}{$M$ $(\frac{1}{2},0,0)$}  & \multicolumn{2}{c}{$L$ $(\frac{1}{2},0,\frac{1}{2})$}  & \multicolumn{2}{c}{$K$ $(\frac{1}{3},\frac{1}{3},0)$}  & \multicolumn{2}{c}{$H$ $(\frac{1}{3},\frac{1}{3},\frac{1}{2})$}   \\
\cline{2-13}
 & $f$ (THz) & Reps & $f$ (THz) & Reps & $f$ (THz) & Reps & $f$ (THz) & Reps & $f$ (THz) & Reps & $f$ (THz) & Reps  \\
\hline
1 & \multirow{3}{*}{  -0.000} & \multirow{3}{*}{$-$$|$$-$} &    0.721 & $A_{2u}$$|$${A_2^-}$ &    2.553 & $B_{2u}$$|$${M_2^-}$ &   -1.265 & $B_{1u}$$|$${L_3^-}$ &    2.702 & $A_{2}''$$|$${K_4}$ &   -1.346 & $A_{2}''$$|$${H_4}$ \\
\cline{1-1}\cline{4-5}\cline{6-7}\cline{8-9}\cline{10-11}\cline{12-13}
2 &   &   & \multirow{2}{*}{   1.691} & \multirow{2}{*}{$E_{1g}$$|$${A_5^+}$} &    2.692 & $B_{1u}$$|$${M_3^-}$ &    2.794 & $B_{2u}$$|$${L_2^-}$ & \multirow{2}{*}{   2.859} & \multirow{2}{*}{$E'$$|$${K_6}$} & \multirow{2}{*}{   2.791} & \multirow{2}{*}{$E'$$|$${H_6}$} \\
\cline{1-1}\cline{6-7}\cline{8-9}
3 &   &   &   &   &    2.873 & $B_{1g}$$|$${M_3^+}$ &    3.087 & $A_{u}$$|$${L_1^-}$ &   &   &   &   \\
\cline{1-1}\cline{2-3}\cline{4-5}\cline{6-7}\cline{8-9}\cline{10-11}\cline{12-13}
4 & \multirow{2}{*}{   2.667} & \multirow{2}{*}{$E_{1u}$$|$${\Gamma_5^-}$} & \multirow{2}{*}{   2.010} & \multirow{2}{*}{$E_{1u}$$|$${A_5^-}$} &    3.102 & $A_{g}$$|$${M_1^+}$ &    3.207 & $B_{3u}$$|$${L_4^-}$ &    2.932 & $A_{1}'$$|$${K_1}$ &    3.258 & $A_{1}''$$|$${H_3}$ \\
\cline{1-1}\cline{6-7}\cline{8-9}\cline{10-11}\cline{12-13}
5 &   &   &   &   &    3.335 & $B_{3u}$$|$${M_4^-}$ &    3.301 & $B_{2g}$$|$${L_2^+}$ &    3.172 & $A_{2}'$$|$${K_2}$ & \multirow{2}{*}{   3.259} & \multirow{2}{*}{$E''$$|$${H_5}$} \\
\hline
\hline
\end{tabular}}
\end{table}

\begin{table}[htbp]
\global\long\def\arraystretch{1.12}
\captionsetup{width=.95\textwidth}
\caption{IRREPs at high-symmetry points for MgV$_6$Ge$_6$ phonon spectrum at High-T.}
\label{tab:191_MgV6Ge6_High}
\setlength{\tabcolsep}{1.mm}{
\begin{tabular}{c|cc|cc|cc|cc|cc|cc}
\hline
\hline
 \multirow{2}{*}{Band}  & \multicolumn{2}{c}{$\Gamma$ $(0,0,0)$}  & \multicolumn{2}{c}{$A$ $(0,0,\frac{1}{2})$}  & \multicolumn{2}{c}{$M$ $(\frac{1}{2},0,0)$}  & \multicolumn{2}{c}{$L$ $(\frac{1}{2},0,\frac{1}{2})$}  & \multicolumn{2}{c}{$K$ $(\frac{1}{3},\frac{1}{3},0)$}  & \multicolumn{2}{c}{$H$ $(\frac{1}{3},\frac{1}{3},\frac{1}{2})$}   \\
\cline{2-13}
 & $f$ (THz) & Reps & $f$ (THz) & Reps & $f$ (THz) & Reps & $f$ (THz) & Reps & $f$ (THz) & Reps & $f$ (THz) & Reps  \\
\hline
1 & \multirow{3}{*}{   0.000} & \multirow{3}{*}{$-$$|$$-$} &   -1.008 & $A_{2u}$$|$${A_2^-}$ &    2.347 & $B_{2u}$$|$${M_2^-}$ &   -1.263 & $B_{1u}$$|$${L_3^-}$ &    2.359 & $A_{2}''$$|$${K_4}$ &   -1.279 & $A_{2}''$$|$${H_4}$ \\
\cline{1-1}\cline{4-5}\cline{6-7}\cline{8-9}\cline{10-11}\cline{12-13}
2 &   &   & \multirow{2}{*}{   1.676} & \multirow{2}{*}{$E_{1g}$$|$${A_5^+}$} &    2.360 & $B_{1u}$$|$${M_3^-}$ &    2.682 & $B_{2u}$$|$${L_2^-}$ &    2.704 & $A_{1}'$$|$${K_1}$ & \multirow{2}{*}{   2.599} & \multirow{2}{*}{$E'$$|$${H_6}$} \\
\cline{1-1}\cline{6-7}\cline{8-9}\cline{10-11}
3 &   &   &   &   &    2.731 & $B_{1g}$$|$${M_3^+}$ &    2.991 & $B_{3u}$$|$${L_4^-}$ & \multirow{2}{*}{   2.718} & \multirow{2}{*}{$E'$$|$${K_6}$} &   &   \\
\cline{1-1}\cline{2-3}\cline{4-5}\cline{6-7}\cline{8-9}\cline{12-13}
4 & \multirow{2}{*}{   2.689} & \multirow{2}{*}{$E_{1u}$$|$${\Gamma_5^-}$} & \multirow{2}{*}{   1.908} & \multirow{2}{*}{$E_{1u}$$|$${A_5^-}$} &    2.879 & $A_{g}$$|$${M_1^+}$ &    3.022 & $A_{u}$$|$${L_1^-}$ &   &   & \multirow{2}{*}{   3.050} & \multirow{2}{*}{$E''$$|$${H_5}$} \\
\cline{1-1}\cline{6-7}\cline{8-9}\cline{10-11}
5 &   &   &   &   &    3.206 & $B_{3u}$$|$${M_4^-}$ &    3.064 & $B_{2g}$$|$${L_2^+}$ &    2.969 & $A_{2}'$$|$${K_2}$ &   &   \\
\hline
\hline
\end{tabular}}
\end{table}
\subsubsection{\label{sms2sec:CaV6Ge6}\hyperref[smsec:col]{\texorpdfstring{CaV$_6$Ge$_6$}{}}~{\color{blue}  (High-T stable, Low-T unstable) }}

\begin{table}[htbp]
\global\long\def\arraystretch{1.12}
\captionsetup{width=.85\textwidth}
\caption{Structure parameters of CaV$_6$Ge$_6$ after relaxation. It contains 350 electrons. }
\label{tab:CaV6Ge6}
\setlength{\tabcolsep}{4mm}{
\begin{tabular}{c|c||c|c}
\hline
\hline
 Atom & Coordinates & \multicolumn{2}{c}{Parameters~($\AA$)} \\
\hline
Ca ~ 1a & ($0,0,0$) & $a$ & $5.1828$ \\
V ~ 6i & ($\frac{1}{2},0,0.2522$) & $c$ & $8.4206$ \\
Ge1 ~ 2c & ($\frac{1}{3},\frac{2}{3},0$) & $d^{Tri}_{Z-Z}$ & $2.8469$ \\
Ge2 ~ 2d & ($\frac{1}{3},\frac{2}{3},\frac{1}{2}$) & $d_{T_{1}}$ & $2.5914$ \\
Ge3 ~ 2e & ($0,0,0.3310$) & $d_{T_{2}}$ & $4.1734$ \\
\hline
\hline
\end{tabular}}
\end{table}

\begin{figure}[htbp]
\centering
\includegraphics[width=0.85\textwidth]{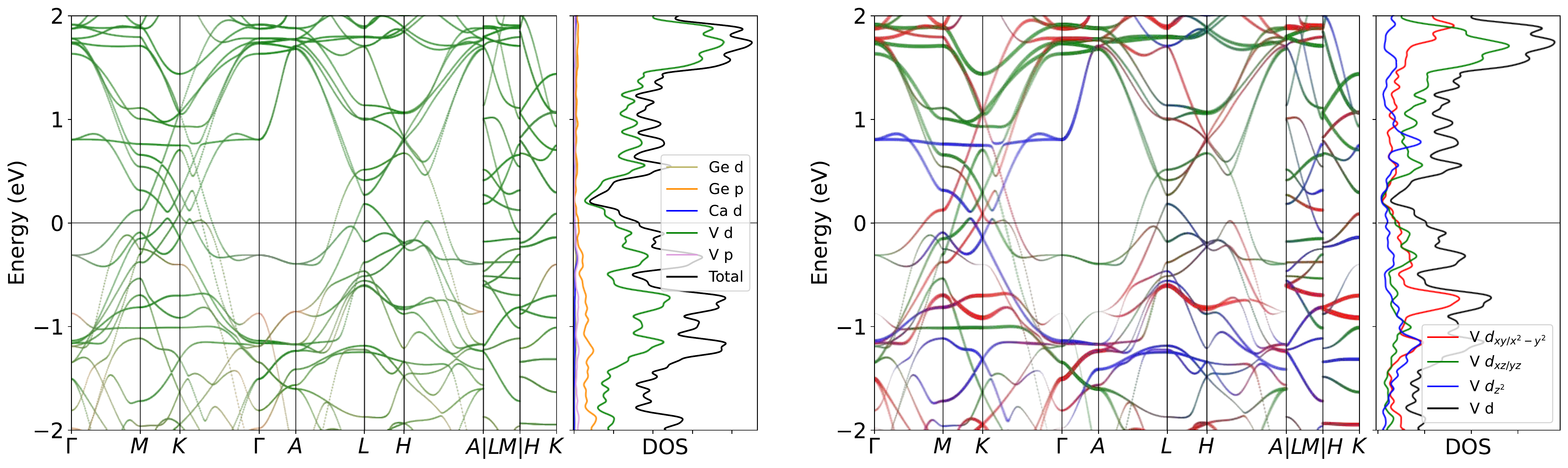}
\caption{Band structure for CaV$_6$Ge$_6$ without SOC. The projection of V $3d$ orbitals is also presented. The band structures clearly reveal the dominating of V $3d$ orbitals  near the Fermi level, as well as the pronounced peak.}
\label{fig:CaV6Ge6band}
\end{figure}

\begin{figure}[H]
\centering
\subfloat[Relaxed phonon at low-T.]{
\label{fig:CaV6Ge6_relaxed_Low}
\includegraphics[width=0.45\textwidth]{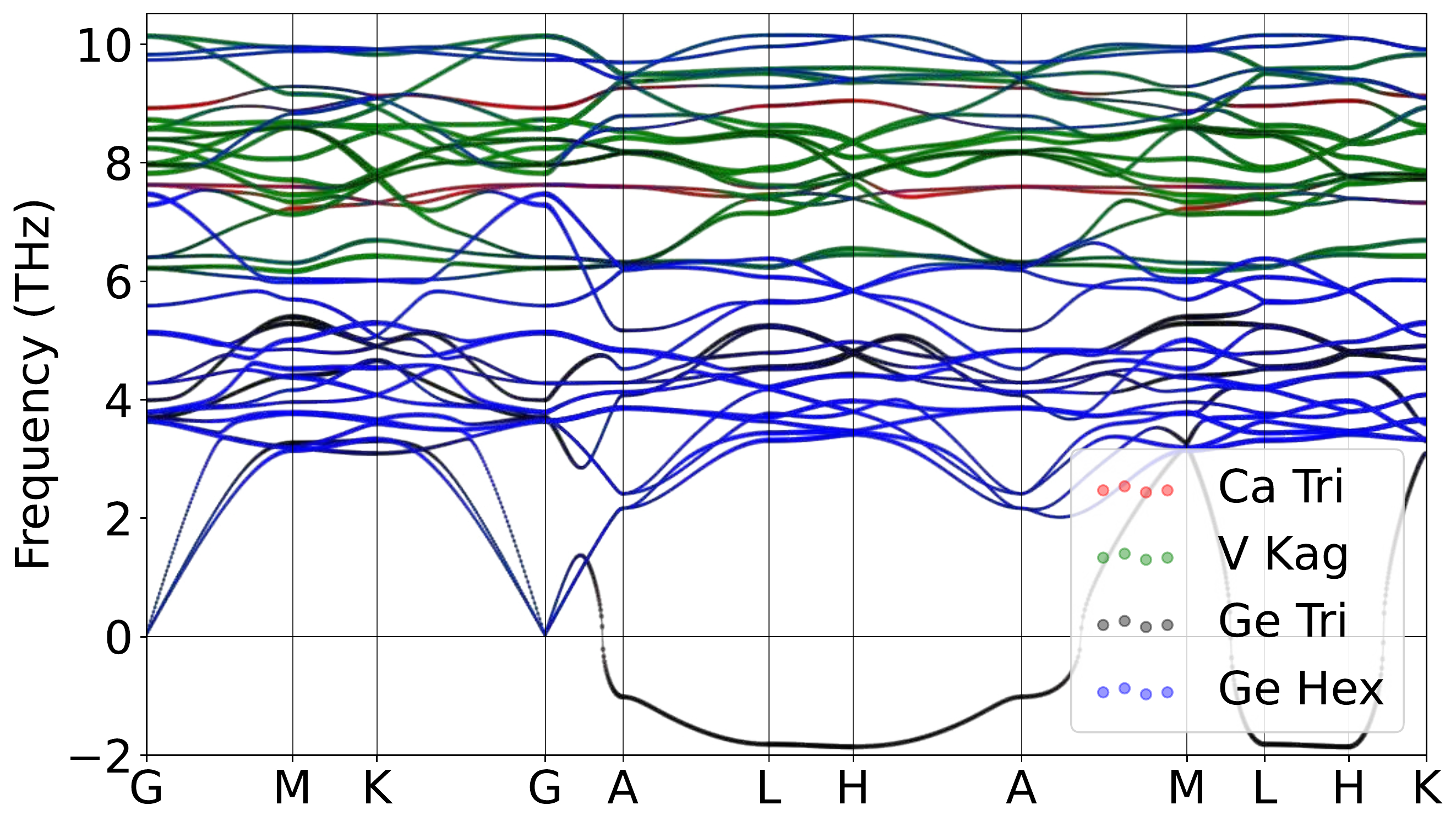}
}
\hspace{5pt}\subfloat[Relaxed phonon at high-T.]{
\label{fig:CaV6Ge6_relaxed_High}
\includegraphics[width=0.45\textwidth]{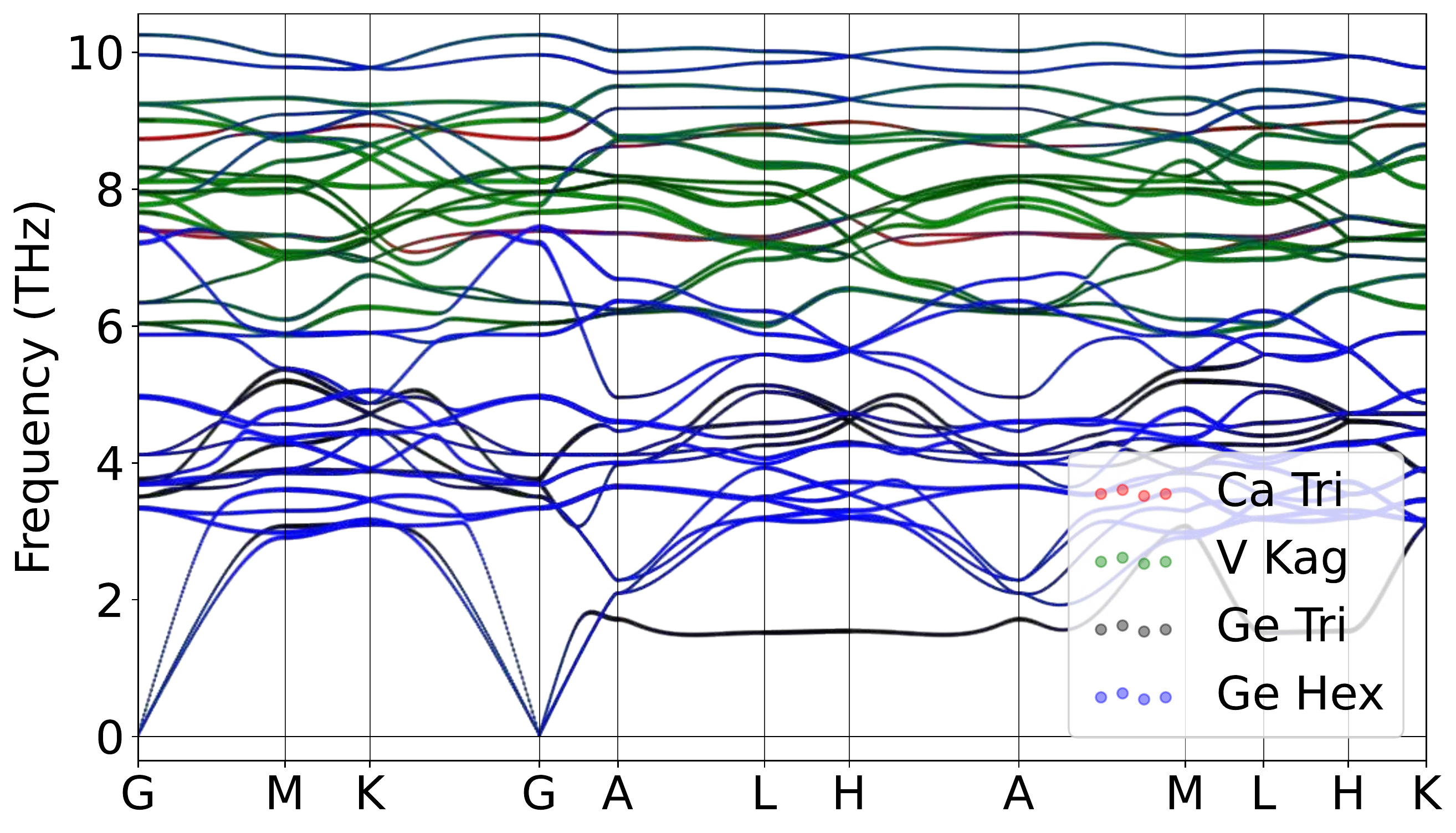}
}
\caption{Atomically projected phonon spectrum for CaV$_6$Ge$_6$.}
\label{fig:CaV6Ge6pho}
\end{figure}

\begin{figure}[H]
\centering
\label{fig:CaV6Ge6_relaxed_Low_xz}
\includegraphics[width=0.85\textwidth]{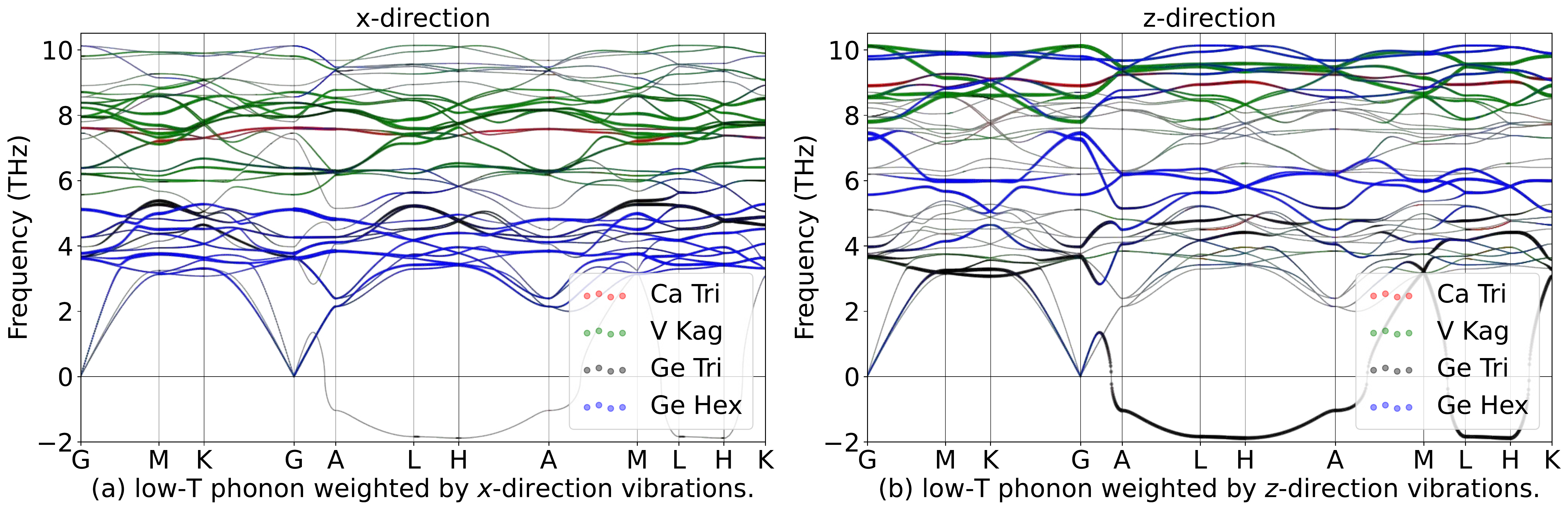}
\hspace{5pt}\label{fig:CaV6Ge6_relaxed_High_xz}
\includegraphics[width=0.85\textwidth]{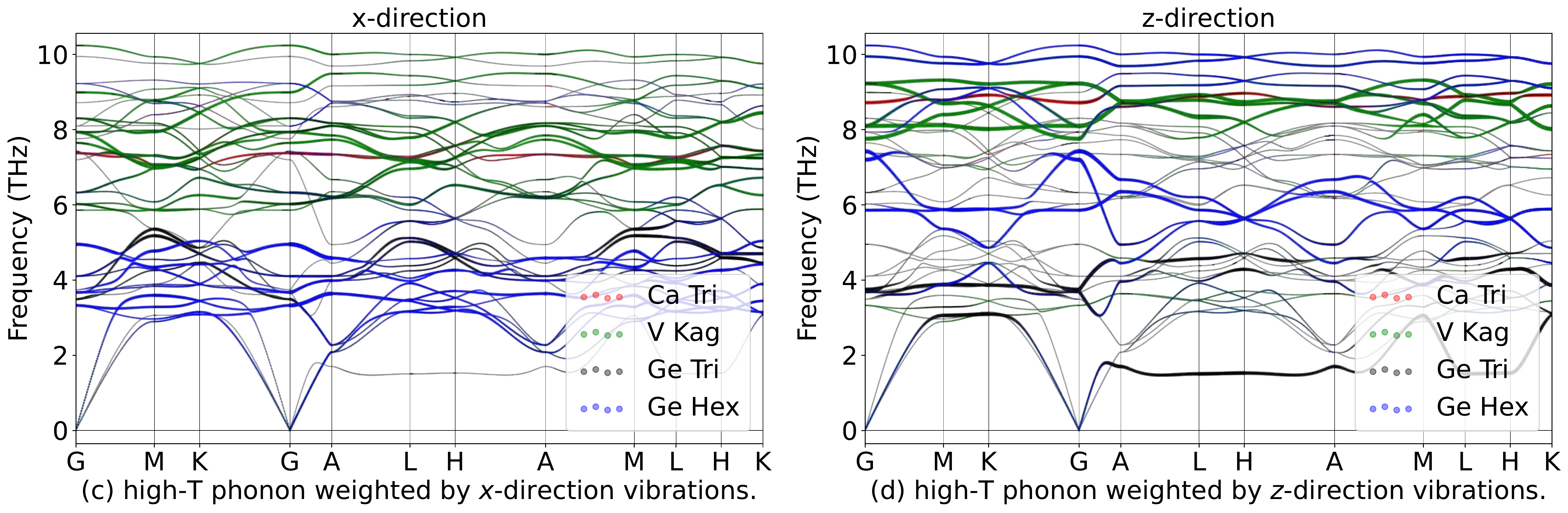}
\caption{Phonon spectrum for CaV$_6$Ge$_6$in in-plane ($x$) and out-of-plane ($z$) directions.}
\label{fig:CaV6Ge6phoxyz}
\end{figure}

\begin{table}[htbp]
\global\long\def\arraystretch{1.12}
\captionsetup{width=.95\textwidth}
\caption{IRREPs at high-symmetry points for CaV$_6$Ge$_6$ phonon spectrum at Low-T.}
\label{tab:191_CaV6Ge6_Low}
\setlength{\tabcolsep}{1.mm}{
\begin{tabular}{c|cc|cc|cc|cc|cc|cc}
\hline
\hline
 \multirow{2}{*}{Band}  & \multicolumn{2}{c}{$\Gamma$ $(0,0,0)$}  & \multicolumn{2}{c}{$A$ $(0,0,\frac{1}{2})$}  & \multicolumn{2}{c}{$M$ $(\frac{1}{2},0,0)$}  & \multicolumn{2}{c}{$L$ $(\frac{1}{2},0,\frac{1}{2})$}  & \multicolumn{2}{c}{$K$ $(\frac{1}{3},\frac{1}{3},0)$}  & \multicolumn{2}{c}{$H$ $(\frac{1}{3},\frac{1}{3},\frac{1}{2})$}   \\
\cline{2-13}
 & $f$ (THz) & Reps & $f$ (THz) & Reps & $f$ (THz) & Reps & $f$ (THz) & Reps & $f$ (THz) & Reps & $f$ (THz) & Reps  \\
\hline
1 & \multirow{3}{*}{  -0.000} & \multirow{3}{*}{$-$$|$$-$} &   -1.046 & $A_{2u}$$|$${A_2^-}$ &    3.112 & $B_{1g}$$|$${M_3^+}$ &   -1.846 & $B_{1u}$$|$${L_3^-}$ &    3.060 & $A_{1}'$$|$${K_1}$ &   -1.887 & $A_{2}''$$|$${H_4}$ \\
\cline{1-1}\cline{4-5}\cline{6-7}\cline{8-9}\cline{10-11}\cline{12-13}
2 &   &   & \multirow{2}{*}{   2.133} & \multirow{2}{*}{$E_{1g}$$|$${A_5^+}$} &    3.157 & $B_{2u}$$|$${M_2^-}$ &    3.281 & $A_{u}$$|$${L_1^-}$ &    3.272 & $A_{2}''$$|$${K_4}$ &    3.385 & $A_{1}''$$|$${H_3}$ \\
\cline{1-1}\cline{6-7}\cline{8-9}\cline{10-11}\cline{12-13}
3 &   &   &   &   &    3.172 & $A_{g}$$|$${M_1^+}$ &    3.405 & $B_{2g}$$|$${L_2^+}$ &    3.300 & $A_{2}'$$|$${K_2}$ & \multirow{2}{*}{   3.433} & \multirow{2}{*}{$E''$$|$${H_5}$} \\
\cline{1-1}\cline{2-3}\cline{4-5}\cline{6-7}\cline{8-9}\cline{10-11}
4 & \multirow{2}{*}{   3.603} & \multirow{2}{*}{$E_{2g}$$|$${\Gamma_6^+}$} & \multirow{2}{*}{   2.381} & \multirow{2}{*}{$E_{1u}$$|$${A_5^-}$} &    3.236 & $B_{1u}$$|$${M_3^-}$ &    3.598 & $B_{3g}$$|$${L_4^+}$ &    3.554 & $A_{1}'$$|$${K_1}$ &   &   \\
\cline{1-1}\cline{6-7}\cline{8-9}\cline{10-11}\cline{12-13}
5 &   &   &   &   &    3.730 & $A_{g}$$|$${M_1^+}$ &    3.675 & $B_{1u}$$|$${L_3^-}$ & \multirow{2}{*}{   3.623} & \multirow{2}{*}{$E'$$|$${K_6}$} & \multirow{2}{*}{   3.765} & \multirow{2}{*}{$E'$$|$${H_6}$} \\
\hline
\hline
\end{tabular}}
\end{table}

\begin{table}[htbp]
\global\long\def\arraystretch{1.12}
\captionsetup{width=.95\textwidth}
\caption{IRREPs at high-symmetry points for CaV$_6$Ge$_6$ phonon spectrum at High-T.}
\label{tab:191_CaV6Ge6_High}
\setlength{\tabcolsep}{1.mm}{
\begin{tabular}{c|cc|cc|cc|cc|cc|cc}
\hline
\hline
 \multirow{2}{*}{Band}  & \multicolumn{2}{c}{$\Gamma$ $(0,0,0)$}  & \multicolumn{2}{c}{$A$ $(0,0,\frac{1}{2})$}  & \multicolumn{2}{c}{$M$ $(\frac{1}{2},0,0)$}  & \multicolumn{2}{c}{$L$ $(\frac{1}{2},0,\frac{1}{2})$}  & \multicolumn{2}{c}{$K$ $(\frac{1}{3},\frac{1}{3},0)$}  & \multicolumn{2}{c}{$H$ $(\frac{1}{3},\frac{1}{3},\frac{1}{2})$}   \\
\cline{2-13}
 & $f$ (THz) & Reps & $f$ (THz) & Reps & $f$ (THz) & Reps & $f$ (THz) & Reps & $f$ (THz) & Reps & $f$ (THz) & Reps  \\
\hline
1 & \multirow{3}{*}{  -0.000} & \multirow{3}{*}{$-$$|$$-$} &    1.688 & $A_{2u}$$|$${A_2^-}$ &    2.882 & $B_{2u}$$|$${M_2^-}$ &    1.492 & $B_{1u}$$|$${L_3^-}$ &    3.069 & $A_{1}'$$|$${K_1}$ &    1.514 & $A_{2}''$$|$${H_4}$ \\
\cline{1-1}\cline{4-5}\cline{6-7}\cline{8-9}\cline{10-11}\cline{12-13}
2 &   &   & \multirow{2}{*}{   2.068} & \multirow{2}{*}{$E_{1g}$$|$${A_5^+}$} &    2.953 & $B_{1g}$$|$${M_3^+}$ &    3.147 & $B_{2g}$$|$${L_2^+}$ &    3.080 & $A_{2}''$$|$${K_4}$ & \multirow{2}{*}{   3.174} & \multirow{2}{*}{$E''$$|$${H_5}$} \\
\cline{1-1}\cline{6-7}\cline{8-9}\cline{10-11}
3 &   &   &   &   &    3.045 & $B_{1u}$$|$${M_3^-}$ &    3.166 & $A_{u}$$|$${L_1^-}$ &    3.137 & $A_{2}'$$|$${K_2}$ &   &   \\
\cline{1-1}\cline{2-3}\cline{4-5}\cline{6-7}\cline{8-9}\cline{10-11}\cline{12-13}
4 & \multirow{2}{*}{   3.312} & \multirow{2}{*}{$E_{2g}$$|$${\Gamma_6^+}$} & \multirow{2}{*}{   2.258} & \multirow{2}{*}{$E_{1u}$$|$${A_5^-}$} &    3.271 & $A_{g}$$|$${M_1^+}$ &    3.435 & $B_{1u}$$|$${L_3^-}$ & \multirow{2}{*}{   3.429} & \multirow{2}{*}{$E'$$|$${K_6}$} &    3.271 & $A_{1}''$$|$${H_3}$ \\
\cline{1-1}\cline{6-7}\cline{8-9}\cline{12-13}
5 &   &   &   &   &    3.579 & $B_{3u}$$|$${M_4^-}$ &    3.449 & $B_{3g}$$|$${L_4^+}$ &   &   & \multirow{2}{*}{   3.512} & \multirow{2}{*}{$E'$$|$${H_6}$} \\
\hline
\hline
\end{tabular}}
\end{table}
\subsubsection{\label{sms2sec:ScV6Ge6}\hyperref[smsec:col]{\texorpdfstring{ScV$_6$Ge$_6$}{}}~{\color{green}  (High-T stable, Low-T stable) }}

\begin{table}[htbp]
\global\long\def\arraystretch{1.12}
\captionsetup{width=.85\textwidth}
\caption{Structure parameters of ScV$_6$Ge$_6$ after relaxation. It contains 351 electrons. }
\label{tab:ScV6Ge6}
\setlength{\tabcolsep}{4mm}{
\begin{tabular}{c|c||c|c}
\hline
\hline
 Atom & Coordinates & \multicolumn{2}{c}{Parameters~($\AA$)} \\
\hline
Sc ~ 1a & ($0,0,0$) & $a$ & $5.1290$ \\
V ~ 6i & ($\frac{1}{2},0,0.2500$) & $c$ & $8.3616$ \\
Ge1 ~ 2c & ($\frac{1}{3},\frac{2}{3},0$) & $d^{Tri}_{Z-Z}$ & $2.8504$ \\
Ge2 ~ 2d & ($\frac{1}{3},\frac{2}{3},\frac{1}{2}$) & $d_{T_{1}}$ & $2.5645$ \\
Ge3 ~ 2e & ($0,0,0.3296$) & $d_{T_{2}}$ & $4.1804$ \\
\hline
\hline
\end{tabular}}
\end{table}

\begin{figure}[htbp]
\centering
\includegraphics[width=0.85\textwidth]{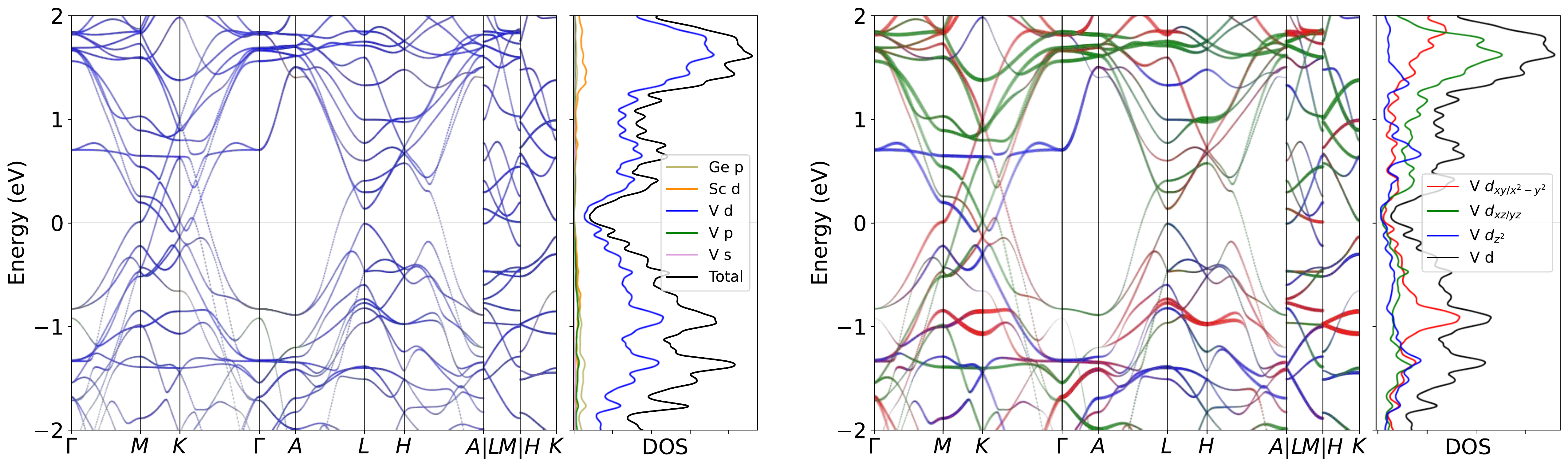}
\caption{Band structure for ScV$_6$Ge$_6$ without SOC. The projection of V $3d$ orbitals is also presented. The band structures clearly reveal the dominating of V $3d$ orbitals  near the Fermi level, as well as the pronounced peak.}
\label{fig:ScV6Ge6band}
\end{figure}

\begin{figure}[H]
\centering
\subfloat[Relaxed phonon at low-T.]{
\label{fig:ScV6Ge6_relaxed_Low}
\includegraphics[width=0.45\textwidth]{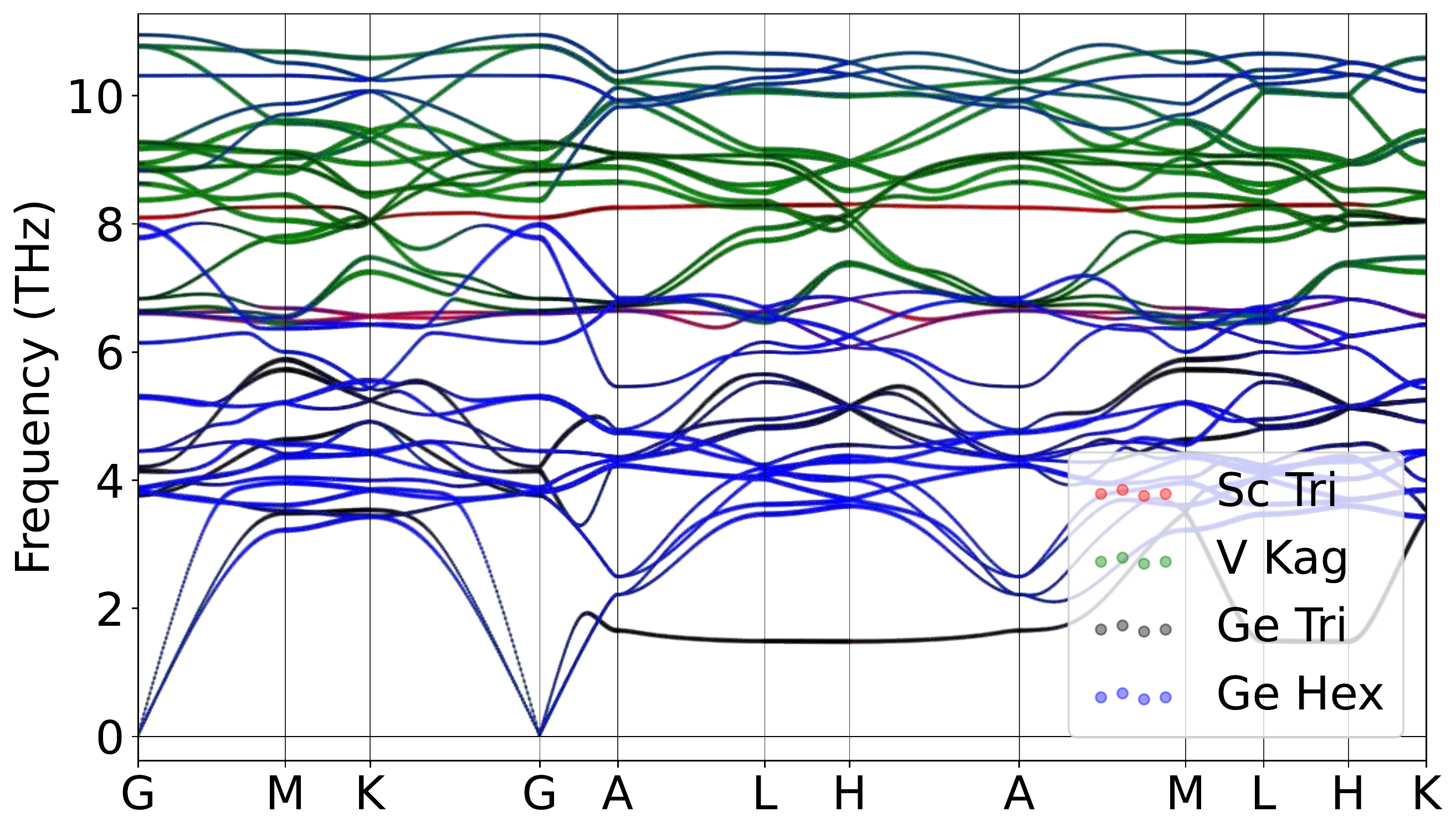}
}
\hspace{5pt}\subfloat[Relaxed phonon at high-T.]{
\label{fig:ScV6Ge6_relaxed_High}
\includegraphics[width=0.45\textwidth]{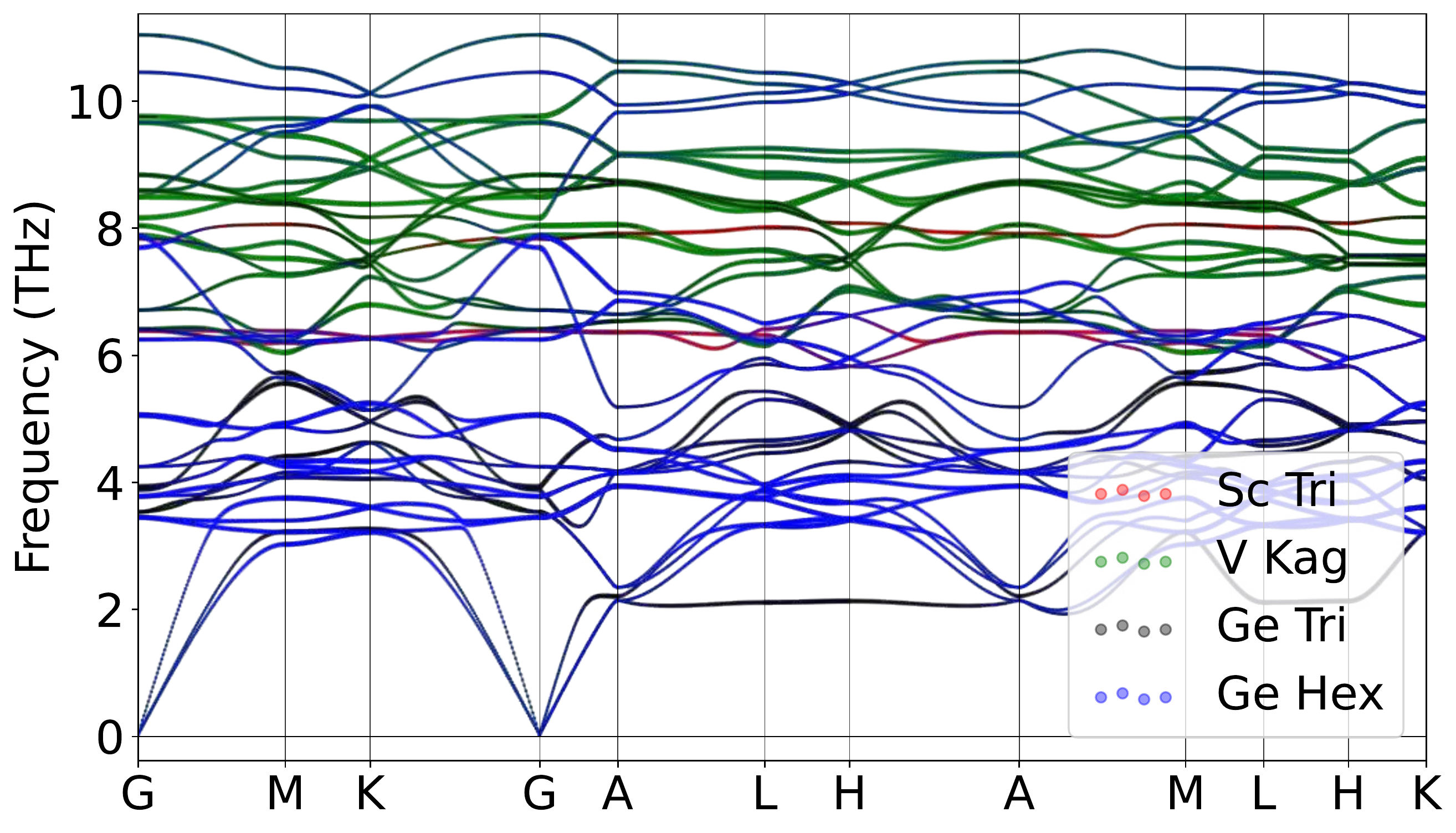}
}
\caption{Atomically projected phonon spectrum for ScV$_6$Ge$_6$.}
\label{fig:ScV6Ge6pho}
\end{figure}

\begin{figure}[H]
\centering
\label{fig:ScV6Ge6_relaxed_Low_xz}
\includegraphics[width=0.85\textwidth]{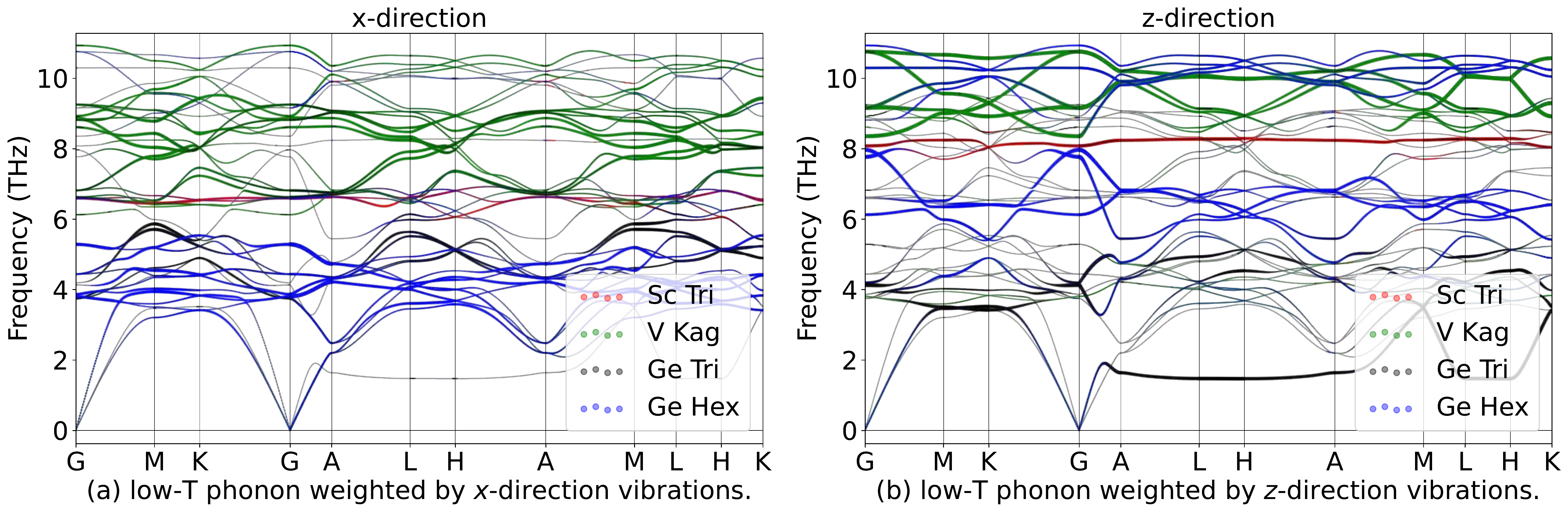}
\hspace{5pt}\label{fig:ScV6Ge6_relaxed_High_xz}
\includegraphics[width=0.85\textwidth]{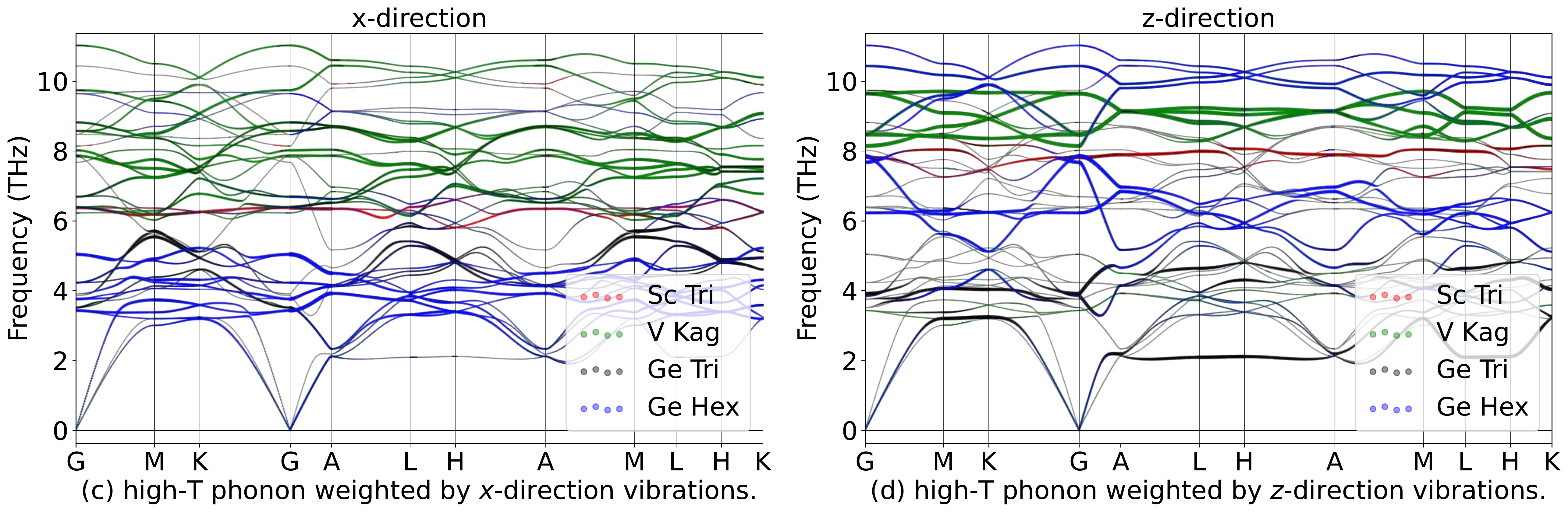}
\caption{Phonon spectrum for ScV$_6$Ge$_6$in in-plane ($x$) and out-of-plane ($z$) directions.}
\label{fig:ScV6Ge6phoxyz}
\end{figure}

\begin{table}[htbp]
\global\long\def\arraystretch{1.12}
\captionsetup{width=.95\textwidth}
\caption{IRREPs at high-symmetry points for ScV$_6$Ge$_6$ phonon spectrum at Low-T.}
\label{tab:191_ScV6Ge6_Low}
\setlength{\tabcolsep}{1.mm}{
\begin{tabular}{c|cc|cc|cc|cc|cc|cc}
\hline
\hline
 \multirow{2}{*}{Band}  & \multicolumn{2}{c}{$\Gamma$ $(0,0,0)$}  & \multicolumn{2}{c}{$A$ $(0,0,\frac{1}{2})$}  & \multicolumn{2}{c}{$M$ $(\frac{1}{2},0,0)$}  & \multicolumn{2}{c}{$L$ $(\frac{1}{2},0,\frac{1}{2})$}  & \multicolumn{2}{c}{$K$ $(\frac{1}{3},\frac{1}{3},0)$}  & \multicolumn{2}{c}{$H$ $(\frac{1}{3},\frac{1}{3},\frac{1}{2})$}   \\
\cline{2-13}
 & $f$ (THz) & Reps & $f$ (THz) & Reps & $f$ (THz) & Reps & $f$ (THz) & Reps & $f$ (THz) & Reps & $f$ (THz) & Reps  \\
\hline
1 & \multirow{3}{*}{  -0.000} & \multirow{3}{*}{$-$$|$$-$} &    1.625 & $A_{2u}$$|$${A_2^-}$ &    3.192 & $B_{1g}$$|$${M_3^+}$ &    1.460 & $B_{1u}$$|$${L_3^-}$ &    3.401 & $A_{2}'$$|$${K_2}$ &    1.456 & $A_{2}''$$|$${H_4}$ \\
\cline{1-1}\cline{4-5}\cline{6-7}\cline{8-9}\cline{10-11}\cline{12-13}
2 &   &   & \multirow{2}{*}{   2.187} & \multirow{2}{*}{$E_{1g}$$|$${A_5^+}$} &    3.454 & $B_{1u}$$|$${M_3^-}$ &    3.438 & $A_{u}$$|$${L_1^-}$ &    3.406 & $A_{1}'$$|$${K_1}$ &    3.568 & $A_{1}''$$|$${H_3}$ \\
\cline{1-1}\cline{6-7}\cline{8-9}\cline{10-11}\cline{12-13}
3 &   &   &   &   &    3.491 & $A_{g}$$|$${M_1^+}$ &    3.593 & $B_{2g}$$|$${L_2^+}$ &    3.505 & $A_{2}''$$|$${K_4}$ & \multirow{2}{*}{   3.668} & \multirow{2}{*}{$E''$$|$${H_5}$} \\
\cline{1-1}\cline{2-3}\cline{4-5}\cline{6-7}\cline{8-9}\cline{10-11}
4 & \multirow{2}{*}{   3.731} & \multirow{2}{*}{$E_{1g}$$|$${\Gamma_5^+}$} & \multirow{2}{*}{   2.466} & \multirow{2}{*}{$E_{1u}$$|$${A_5^-}$} &    3.581 & $B_{2u}$$|$${M_2^-}$ &    3.988 & $B_{1u}$$|$${L_3^-}$ & \multirow{2}{*}{   3.816} & \multirow{2}{*}{$E'$$|$${K_6}$} &   &   \\
\cline{1-1}\cline{6-7}\cline{8-9}\cline{12-13}
5 &   &   &   &   &    3.930 & $B_{3u}$$|$${M_4^-}$ &    4.032 & $B_{3g}$$|$${L_4^+}$ &   &   & \multirow{2}{*}{   3.988} & \multirow{2}{*}{$E'$$|$${H_6}$} \\
\hline
\hline
\end{tabular}}
\end{table}

\begin{table}[htbp]
\global\long\def\arraystretch{1.12}
\captionsetup{width=.95\textwidth}
\caption{IRREPs at high-symmetry points for ScV$_6$Ge$_6$ phonon spectrum at High-T.}
\label{tab:191_ScV6Ge6_High}
\setlength{\tabcolsep}{1.mm}{
\begin{tabular}{c|cc|cc|cc|cc|cc|cc}
\hline
\hline
 \multirow{2}{*}{Band}  & \multicolumn{2}{c}{$\Gamma$ $(0,0,0)$}  & \multicolumn{2}{c}{$A$ $(0,0,\frac{1}{2})$}  & \multicolumn{2}{c}{$M$ $(\frac{1}{2},0,0)$}  & \multicolumn{2}{c}{$L$ $(\frac{1}{2},0,\frac{1}{2})$}  & \multicolumn{2}{c}{$K$ $(\frac{1}{3},\frac{1}{3},0)$}  & \multicolumn{2}{c}{$H$ $(\frac{1}{3},\frac{1}{3},\frac{1}{2})$}   \\
\cline{2-13}
 & $f$ (THz) & Reps & $f$ (THz) & Reps & $f$ (THz) & Reps & $f$ (THz) & Reps & $f$ (THz) & Reps & $f$ (THz) & Reps  \\
\hline
1 & \multirow{3}{*}{  -0.000} & \multirow{3}{*}{$-$$|$$-$} & \multirow{2}{*}{   2.109} & \multirow{2}{*}{$E_{1g}$$|$${A_5^+}$} &    2.995 & $B_{1g}$$|$${M_3^+}$ &    2.081 & $B_{1u}$$|$${L_3^-}$ &    3.182 & $A_{2}'$$|$${K_2}$ &    2.103 & $A_{2}''$$|$${H_4}$ \\
\cline{1-1}\cline{6-7}\cline{8-9}\cline{10-11}\cline{12-13}
2 &   &   &   &   &    3.186 & $B_{2u}$$|$${M_2^-}$ &    3.290 & $A_{u}$$|$${L_1^-}$ &    3.201 & $A_{1}'$$|$${K_1}$ & \multirow{2}{*}{   3.375} & \multirow{2}{*}{$E''$$|$${H_5}$} \\
\cline{1-1}\cline{4-5}\cline{6-7}\cline{8-9}\cline{10-11}
3 &   &   &    2.177 & $A_{2u}$$|$${A_2^-}$ &    3.194 & $B_{1u}$$|$${M_3^-}$ &    3.306 & $B_{2g}$$|$${L_2^+}$ &    3.237 & $A_{2}''$$|$${K_4}$ &   &   \\
\cline{1-1}\cline{2-3}\cline{4-5}\cline{6-7}\cline{8-9}\cline{10-11}\cline{12-13}
4 & \multirow{2}{*}{   3.419} & \multirow{2}{*}{$E_{2g}$$|$${\Gamma_6^+}$} & \multirow{2}{*}{   2.320} & \multirow{2}{*}{$E_{1u}$$|$${A_5^-}$} &    3.368 & $A_{g}$$|$${M_1^+}$ &    3.702 & $B_{1u}$$|$${L_3^-}$ & \multirow{2}{*}{   3.577} & \multirow{2}{*}{$E'$$|$${K_6}$} &    3.396 & $A_{1}''$$|$${H_3}$ \\
\cline{1-1}\cline{6-7}\cline{8-9}\cline{12-13}
5 &   &   &   &   &    3.723 & $B_{3u}$$|$${M_4^-}$ &    3.737 & $B_{3g}$$|$${L_4^+}$ &   &   & \multirow{2}{*}{   3.652} & \multirow{2}{*}{$E'$$|$${H_6}$} \\
\hline
\hline
\end{tabular}}
\end{table}
\subsubsection{\label{sms2sec:TiV6Ge6}\hyperref[smsec:col]{\texorpdfstring{TiV$_6$Ge$_6$}{}}~{\color{green}  (High-T stable, Low-T stable) }}

\begin{table}[htbp]
\global\long\def\arraystretch{1.12}
\captionsetup{width=.85\textwidth}
\caption{Structure parameters of TiV$_6$Ge$_6$ after relaxation. It contains 352 electrons. }
\label{tab:TiV6Ge6}
\setlength{\tabcolsep}{4mm}{
\begin{tabular}{c|c||c|c}
\hline
\hline
 Atom & Coordinates & \multicolumn{2}{c}{Parameters~($\AA$)} \\
\hline
Ti ~ 1a & ($0,0,0$) & $a$ & $5.0846$ \\
V ~ 6i & ($\frac{1}{2},0,0.2491$) & $c$ & $8.3877$ \\
Ge1 ~ 2c & ($\frac{1}{3},\frac{2}{3},0$) & $d^{Tri}_{Z-Z}$ & $3.0895$ \\
Ge2 ~ 2d & ($\frac{1}{3},\frac{2}{3},\frac{1}{2}$) & $d_{T_{1}}$ & $2.5423$ \\
Ge3 ~ 2e & ($0,0,0.3158$) & $d_{T_{2}}$ & $4.1785$ \\
\hline
\hline
\end{tabular}}
\end{table}

\begin{figure}[htbp]
\centering
\includegraphics[width=0.85\textwidth]{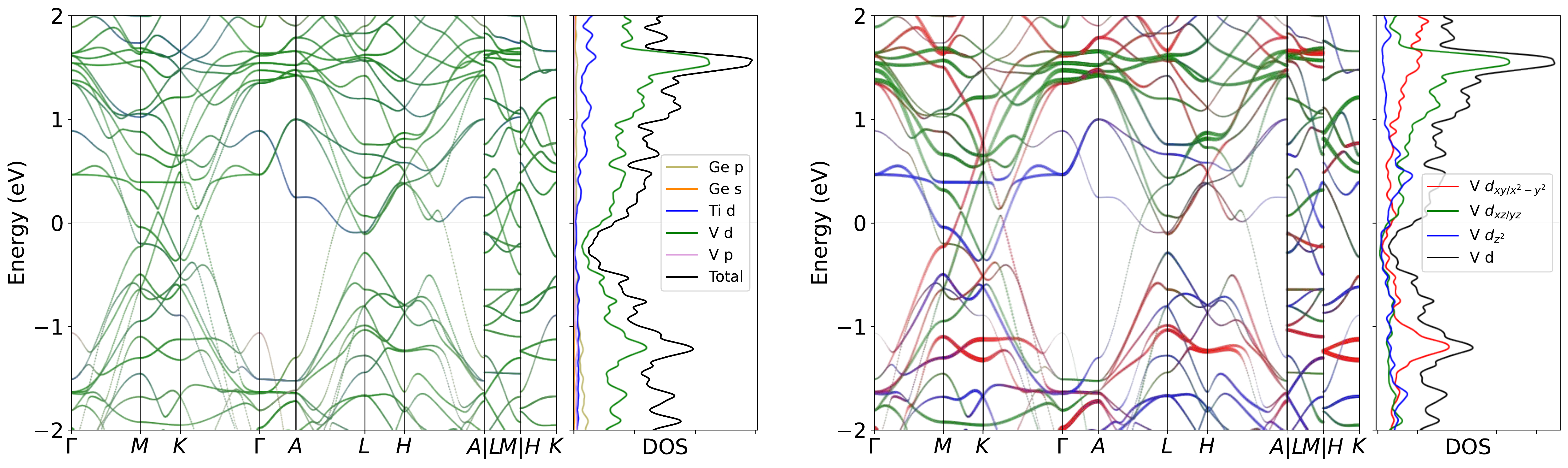}
\caption{Band structure for TiV$_6$Ge$_6$ without SOC. The projection of V $3d$ orbitals is also presented. The band structures clearly reveal the dominating of V $3d$ orbitals  near the Fermi level, as well as the pronounced peak.}
\label{fig:TiV6Ge6band}
\end{figure}

\begin{figure}[H]
\centering
\subfloat[Relaxed phonon at low-T.]{
\label{fig:TiV6Ge6_relaxed_Low}
\includegraphics[width=0.45\textwidth]{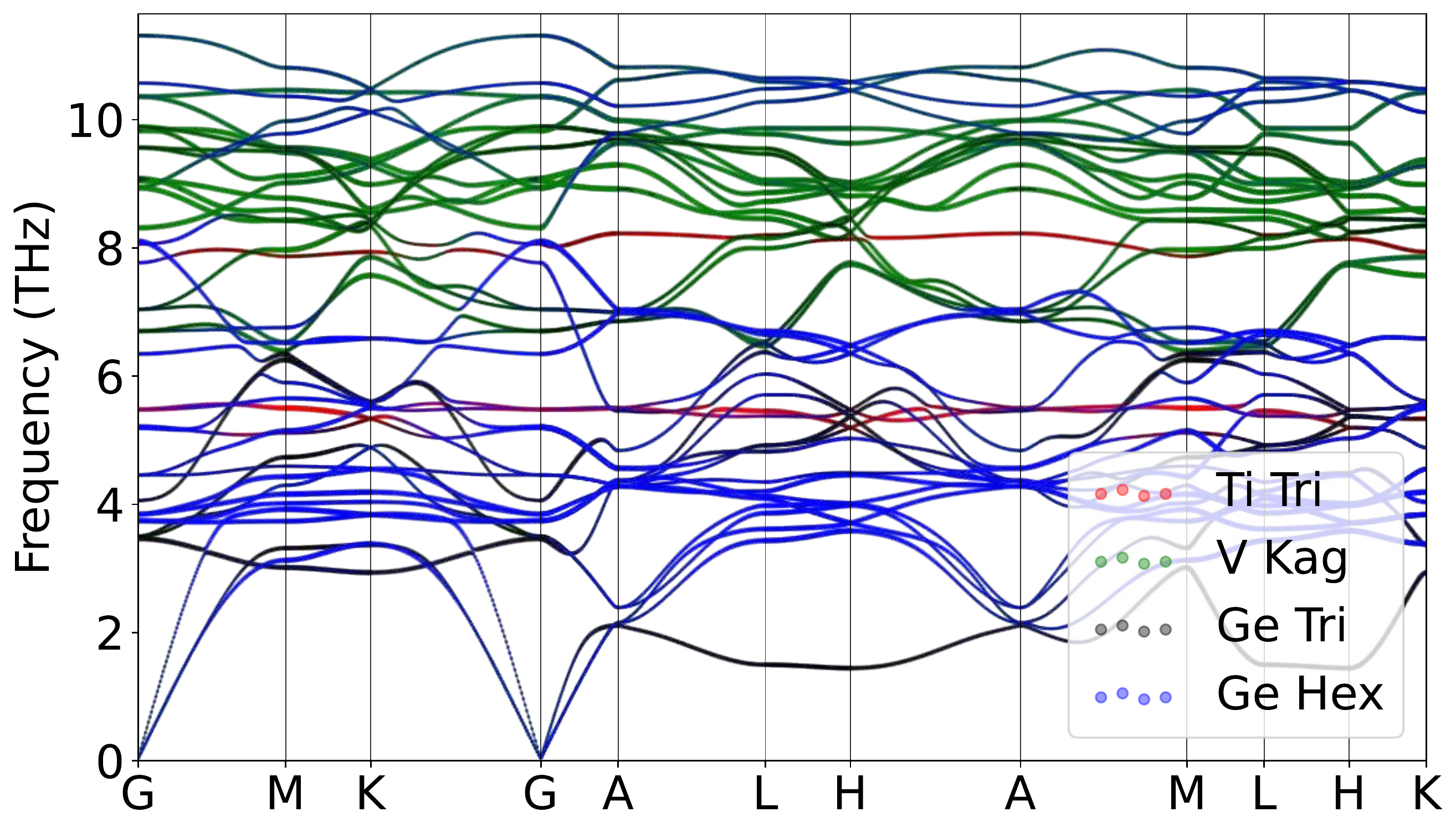}
}
\hspace{5pt}\subfloat[Relaxed phonon at high-T.]{
\label{fig:TiV6Ge6_relaxed_High}
\includegraphics[width=0.45\textwidth]{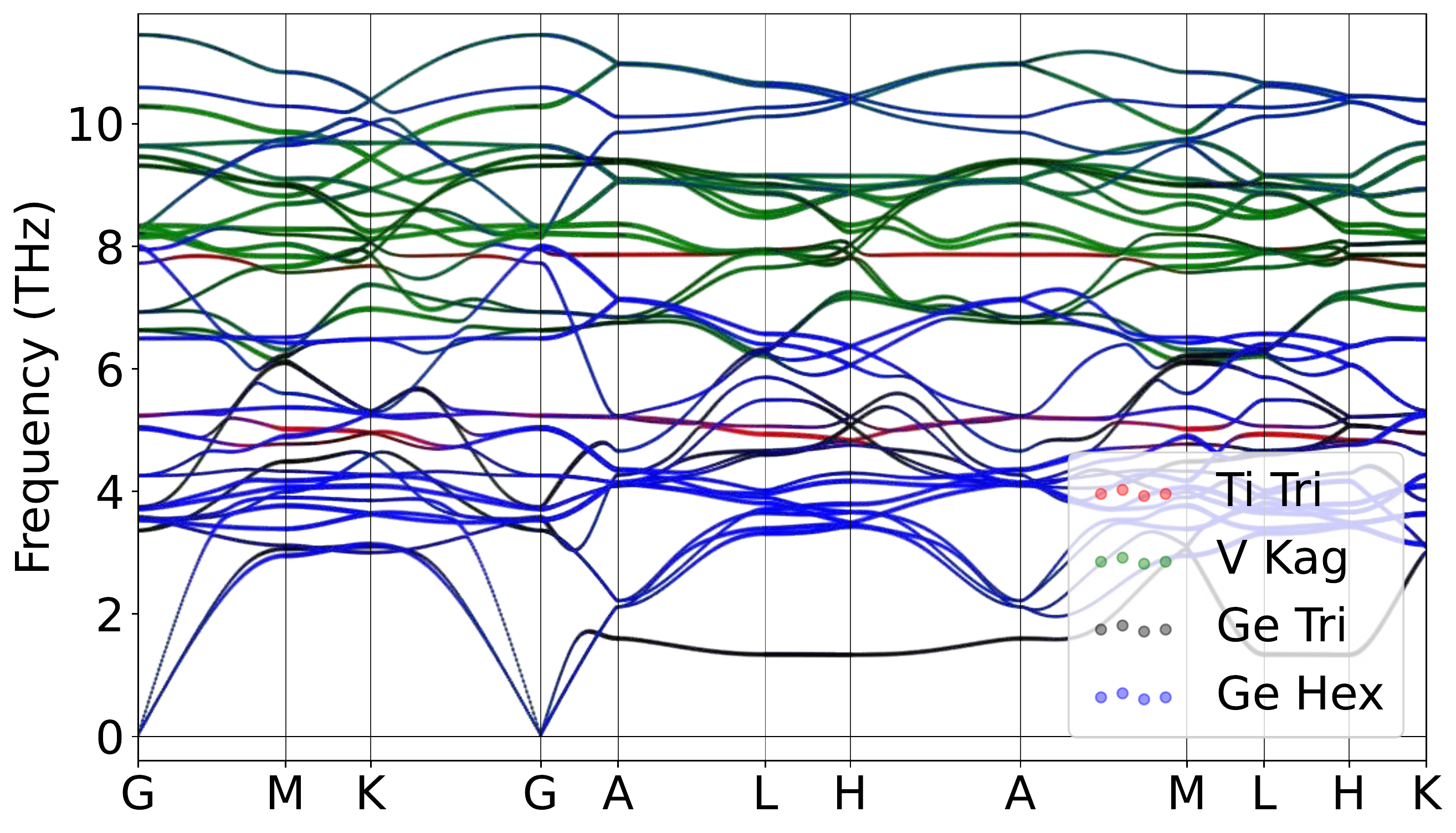}
}
\caption{Atomically projected phonon spectrum for TiV$_6$Ge$_6$.}
\label{fig:TiV6Ge6pho}
\end{figure}

\begin{figure}[H]
\centering
\label{fig:TiV6Ge6_relaxed_Low_xz}
\includegraphics[width=0.85\textwidth]{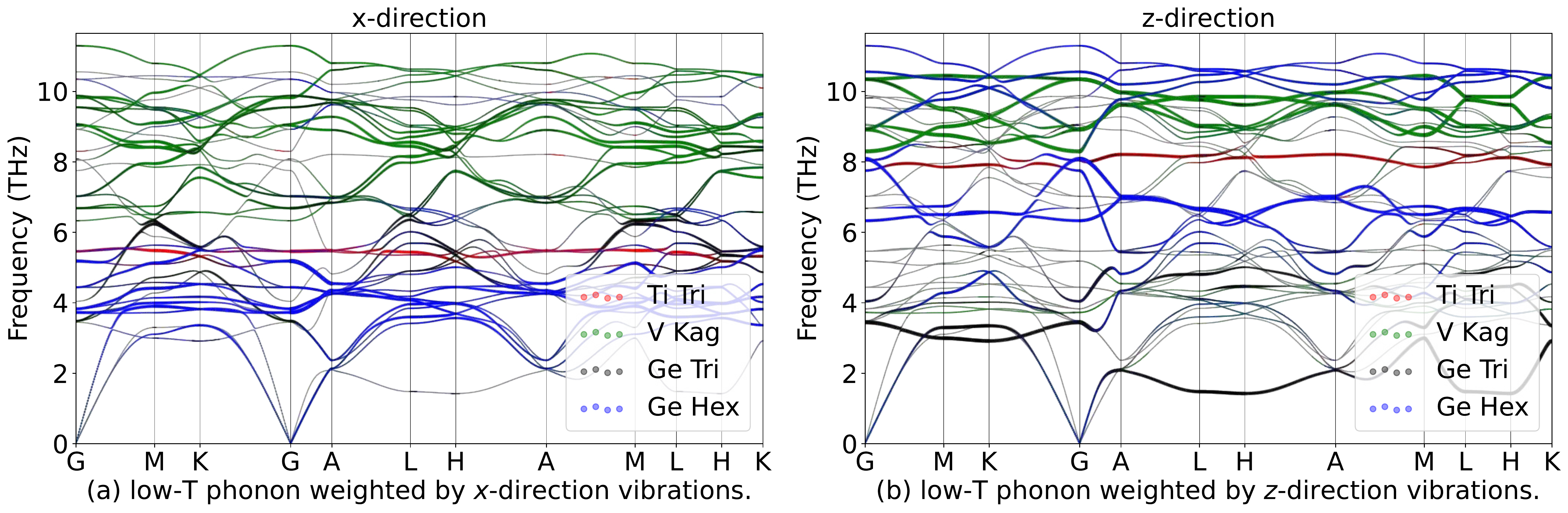}
\hspace{5pt}\label{fig:TiV6Ge6_relaxed_High_xz}
\includegraphics[width=0.85\textwidth]{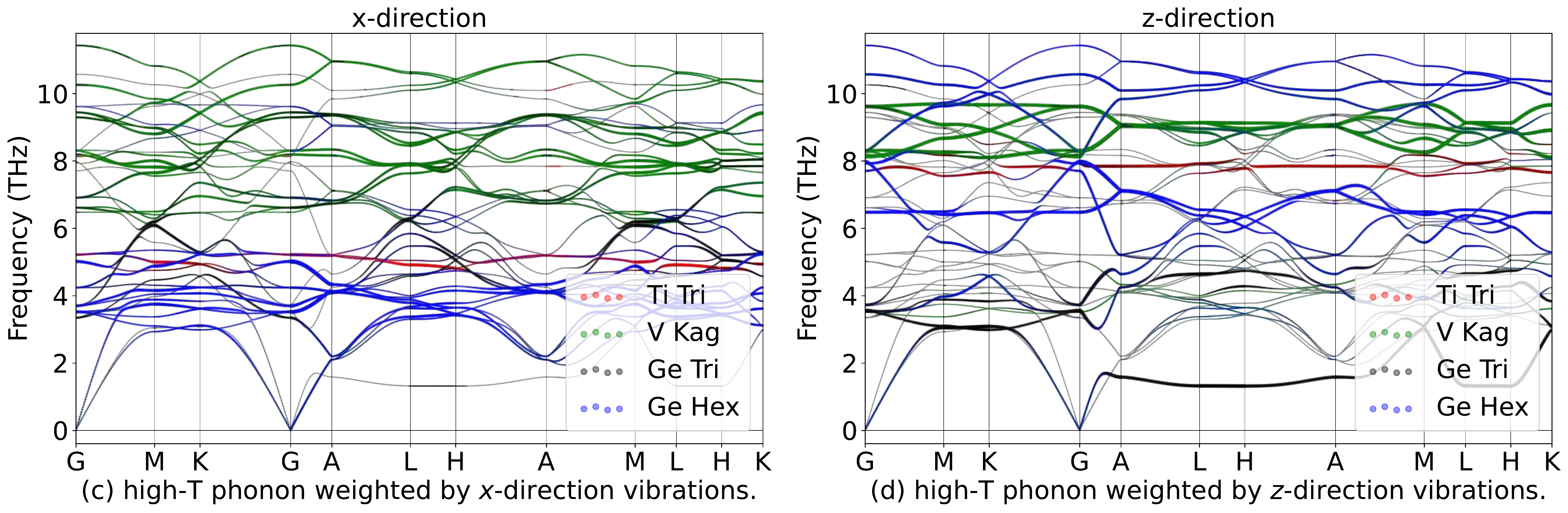}
\caption{Phonon spectrum for TiV$_6$Ge$_6$in in-plane ($x$) and out-of-plane ($z$) directions.}
\label{fig:TiV6Ge6phoxyz}
\end{figure}

\begin{table}[htbp]
\global\long\def\arraystretch{1.12}
\captionsetup{width=.95\textwidth}
\caption{IRREPs at high-symmetry points for TiV$_6$Ge$_6$ phonon spectrum at Low-T.}
\label{tab:191_TiV6Ge6_Low}
\setlength{\tabcolsep}{1.mm}{
\begin{tabular}{c|cc|cc|cc|cc|cc|cc}
\hline
\hline
 \multirow{2}{*}{Band}  & \multicolumn{2}{c}{$\Gamma$ $(0,0,0)$}  & \multicolumn{2}{c}{$A$ $(0,0,\frac{1}{2})$}  & \multicolumn{2}{c}{$M$ $(\frac{1}{2},0,0)$}  & \multicolumn{2}{c}{$L$ $(\frac{1}{2},0,\frac{1}{2})$}  & \multicolumn{2}{c}{$K$ $(\frac{1}{3},\frac{1}{3},0)$}  & \multicolumn{2}{c}{$H$ $(\frac{1}{3},\frac{1}{3},\frac{1}{2})$}   \\
\cline{2-13}
 & $f$ (THz) & Reps & $f$ (THz) & Reps & $f$ (THz) & Reps & $f$ (THz) & Reps & $f$ (THz) & Reps & $f$ (THz) & Reps  \\
\hline
1 & \multirow{3}{*}{  -0.000} & \multirow{3}{*}{$-$$|$$-$} &    2.079 & $A_{2u}$$|$${A_2^-}$ &    2.983 & $A_{g}$$|$${M_1^+}$ &    1.468 & $B_{1u}$$|$${L_3^-}$ &    2.905 & $A_{1}'$$|$${K_1}$ &    1.415 & $A_{2}''$$|$${H_4}$ \\
\cline{1-1}\cline{4-5}\cline{6-7}\cline{8-9}\cline{10-11}\cline{12-13}
2 &   &   & \multirow{2}{*}{   2.121} & \multirow{2}{*}{$E_{1g}$$|$${A_5^+}$} &    3.097 & $B_{1g}$$|$${M_3^+}$ &    3.403 & $A_{u}$$|$${L_1^-}$ &    3.335 & $A_{2}''$$|$${K_4}$ &    3.554 & $A_{1}''$$|$${H_3}$ \\
\cline{1-1}\cline{6-7}\cline{8-9}\cline{10-11}\cline{12-13}
3 &   &   &   &   &    3.286 & $B_{1u}$$|$${M_3^-}$ &    3.584 & $B_{2g}$$|$${L_2^+}$ &    3.352 & $A_{2}'$$|$${K_2}$ & \multirow{2}{*}{   3.678} & \multirow{2}{*}{$E''$$|$${H_5}$} \\
\cline{1-1}\cline{2-3}\cline{4-5}\cline{6-7}\cline{8-9}\cline{10-11}
4 &    3.435 & $A_{1g}$$|$${\Gamma_1^+}$ & \multirow{2}{*}{   2.358} & \multirow{2}{*}{$E_{1u}$$|$${A_5^-}$} &    3.705 & $B_{2u}$$|$${M_2^-}$ &    3.831 & $B_{1g}$$|$${L_3^+}$ & \multirow{2}{*}{   3.807} & \multirow{2}{*}{$E'$$|$${K_6}$} &   &   \\
\cline{1-1}\cline{2-3}\cline{6-7}\cline{8-9}\cline{12-13}
5 & \multirow{2}{*}{   3.471} & \multirow{2}{*}{$E_{1g}$$|$${\Gamma_5^+}$} &   &   &    3.890 & $B_{3u}$$|$${M_4^-}$ &    3.948 & $B_{3u}$$|$${L_4^-}$ &   &   &    3.953 & $A_{2}'$$|$${H_2}$ \\
\hline
\hline
\end{tabular}}
\end{table}

\begin{table}[htbp]
\global\long\def\arraystretch{1.12}
\captionsetup{width=.95\textwidth}
\caption{IRREPs at high-symmetry points for TiV$_6$Ge$_6$ phonon spectrum at High-T.}
\label{tab:191_TiV6Ge6_High}
\setlength{\tabcolsep}{1.mm}{
\begin{tabular}{c|cc|cc|cc|cc|cc|cc}
\hline
\hline
 \multirow{2}{*}{Band}  & \multicolumn{2}{c}{$\Gamma$ $(0,0,0)$}  & \multicolumn{2}{c}{$A$ $(0,0,\frac{1}{2})$}  & \multicolumn{2}{c}{$M$ $(\frac{1}{2},0,0)$}  & \multicolumn{2}{c}{$L$ $(\frac{1}{2},0,\frac{1}{2})$}  & \multicolumn{2}{c}{$K$ $(\frac{1}{3},\frac{1}{3},0)$}  & \multicolumn{2}{c}{$H$ $(\frac{1}{3},\frac{1}{3},\frac{1}{2})$}   \\
\cline{2-13}
 & $f$ (THz) & Reps & $f$ (THz) & Reps & $f$ (THz) & Reps & $f$ (THz) & Reps & $f$ (THz) & Reps & $f$ (THz) & Reps  \\
\hline
1 & \multirow{3}{*}{  -0.000} & \multirow{3}{*}{$-$$|$$-$} &    1.570 & $A_{2u}$$|$${A_2^-}$ &    2.920 & $B_{1g}$$|$${M_3^+}$ &    1.309 & $B_{1u}$$|$${L_3^-}$ &    2.971 & $A_{1}'$$|$${K_1}$ &    1.304 & $A_{2}''$$|$${H_4}$ \\
\cline{1-1}\cline{4-5}\cline{6-7}\cline{8-9}\cline{10-11}\cline{12-13}
2 &   &   & \multirow{2}{*}{   2.088} & \multirow{2}{*}{$E_{1g}$$|$${A_5^+}$} &    3.033 & $B_{1u}$$|$${M_3^-}$ &    3.287 & $A_{u}$$|$${L_1^-}$ &    3.072 & $A_{2}''$$|$${K_4}$ &    3.400 & $A_{1}''$$|$${H_3}$ \\
\cline{1-1}\cline{6-7}\cline{8-9}\cline{10-11}\cline{12-13}
3 &   &   &   &   &    3.087 & $A_{g}$$|$${M_1^+}$ &    3.362 & $B_{2g}$$|$${L_2^+}$ &    3.105 & $A_{2}'$$|$${K_2}$ & \multirow{2}{*}{   3.450} & \multirow{2}{*}{$E''$$|$${H_5}$} \\
\cline{1-1}\cline{2-3}\cline{4-5}\cline{6-7}\cline{8-9}\cline{10-11}
4 & \multirow{2}{*}{   3.331} & \multirow{2}{*}{$E_{1g}$$|$${\Gamma_5^+}$} & \multirow{2}{*}{   2.186} & \multirow{2}{*}{$E_{1u}$$|$${A_5^-}$} &    3.359 & $B_{2u}$$|$${M_2^-}$ &    3.619 & $B_{3u}$$|$${L_4^-}$ & \multirow{2}{*}{   3.602} & \multirow{2}{*}{$E'$$|$${K_6}$} &   &   \\
\cline{1-1}\cline{6-7}\cline{8-9}\cline{12-13}
5 &   &   &   &   &    3.734 & $B_{3u}$$|$${M_4^-}$ &    3.674 & $B_{1g}$$|$${L_3^+}$ &   &   & \multirow{2}{*}{   3.630} & \multirow{2}{*}{$E'$$|$${H_6}$} \\
\hline
\hline
\end{tabular}}
\end{table}
\subsubsection{\label{sms2sec:YV6Ge6}\hyperref[smsec:col]{\texorpdfstring{YV$_6$Ge$_6$}{}}~{\color{green}  (High-T stable, Low-T stable) }}

\begin{table}[htbp]
\global\long\def\arraystretch{1.12}
\captionsetup{width=.85\textwidth}
\caption{Structure parameters of YV$_6$Ge$_6$ after relaxation. It contains 369 electrons. }
\label{tab:YV6Ge6}
\setlength{\tabcolsep}{4mm}{
\begin{tabular}{c|c||c|c}
\hline
\hline
 Atom & Coordinates & \multicolumn{2}{c}{Parameters~($\AA$)} \\
\hline
Y ~ 1a & ($0,0,0$) & $a$ & $5.2335$ \\
V ~ 6i & ($\frac{1}{2},0,0.2513$) & $c$ & $8.4702$ \\
Ge1 ~ 2c & ($\frac{1}{3},\frac{2}{3},0$) & $d^{Tri}_{Z-Z}$ & $2.6937$ \\
Ge2 ~ 2d & ($\frac{1}{3},\frac{2}{3},\frac{1}{2}$) & $d_{T_{1}}$ & $2.6167$ \\
Ge3 ~ 2e & ($0,0,0.3410$) & $d_{T_{2}}$ & $4.2126$ \\
\hline
\hline
\end{tabular}}
\end{table}

\begin{figure}[htbp]
\centering
\includegraphics[width=0.85\textwidth]{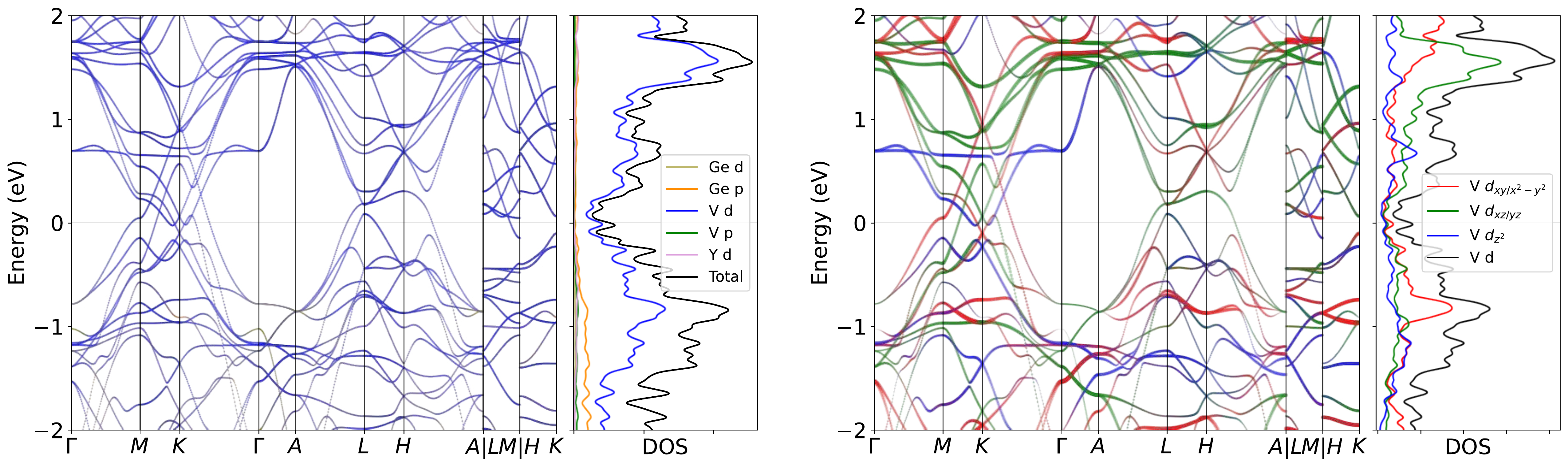}
\caption{Band structure for YV$_6$Ge$_6$ without SOC. The projection of V $3d$ orbitals is also presented. The band structures clearly reveal the dominating of V $3d$ orbitals  near the Fermi level, as well as the pronounced peak.}
\label{fig:YV6Ge6band}
\end{figure}

\begin{figure}[H]
\centering
\subfloat[Relaxed phonon at low-T.]{
\label{fig:YV6Ge6_relaxed_Low}
\includegraphics[width=0.45\textwidth]{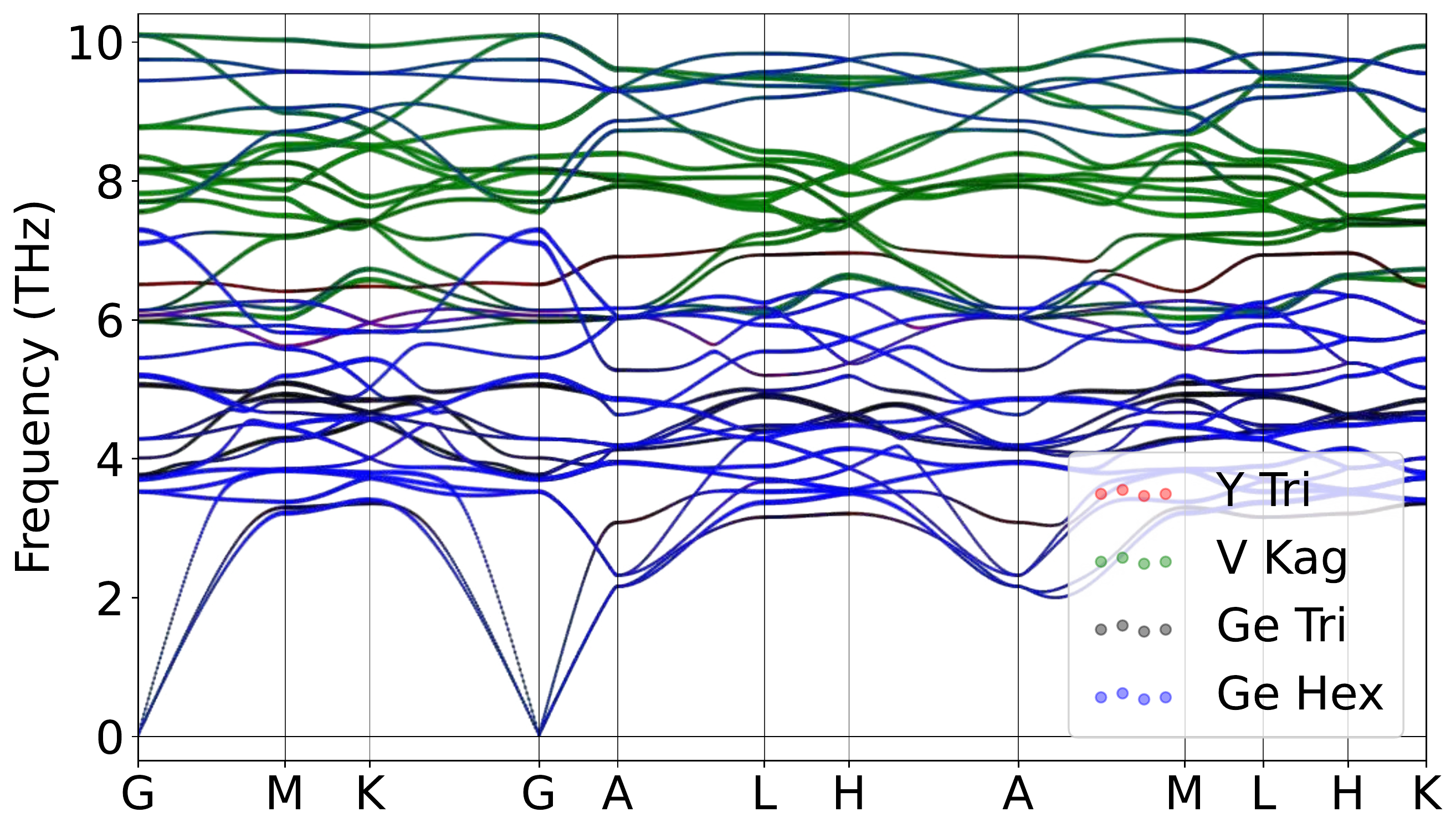}
}
\hspace{5pt}\subfloat[Relaxed phonon at high-T.]{
\label{fig:YV6Ge6_relaxed_High}
\includegraphics[width=0.45\textwidth]{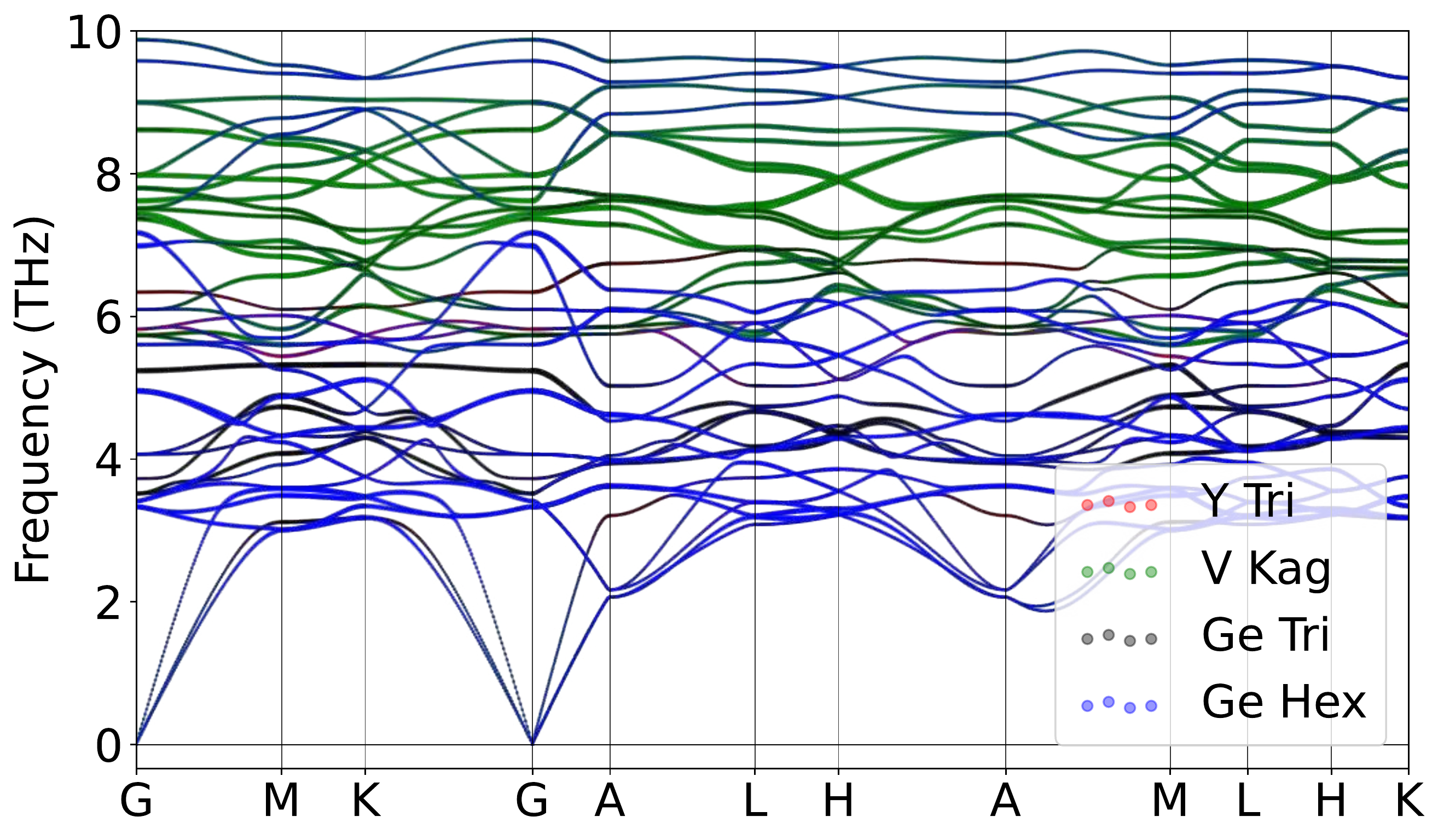}
}
\caption{Atomically projected phonon spectrum for YV$_6$Ge$_6$.}
\label{fig:YV6Ge6pho}
\end{figure}

\begin{figure}[H]
\centering
\label{fig:YV6Ge6_relaxed_Low_xz}
\includegraphics[width=0.85\textwidth]{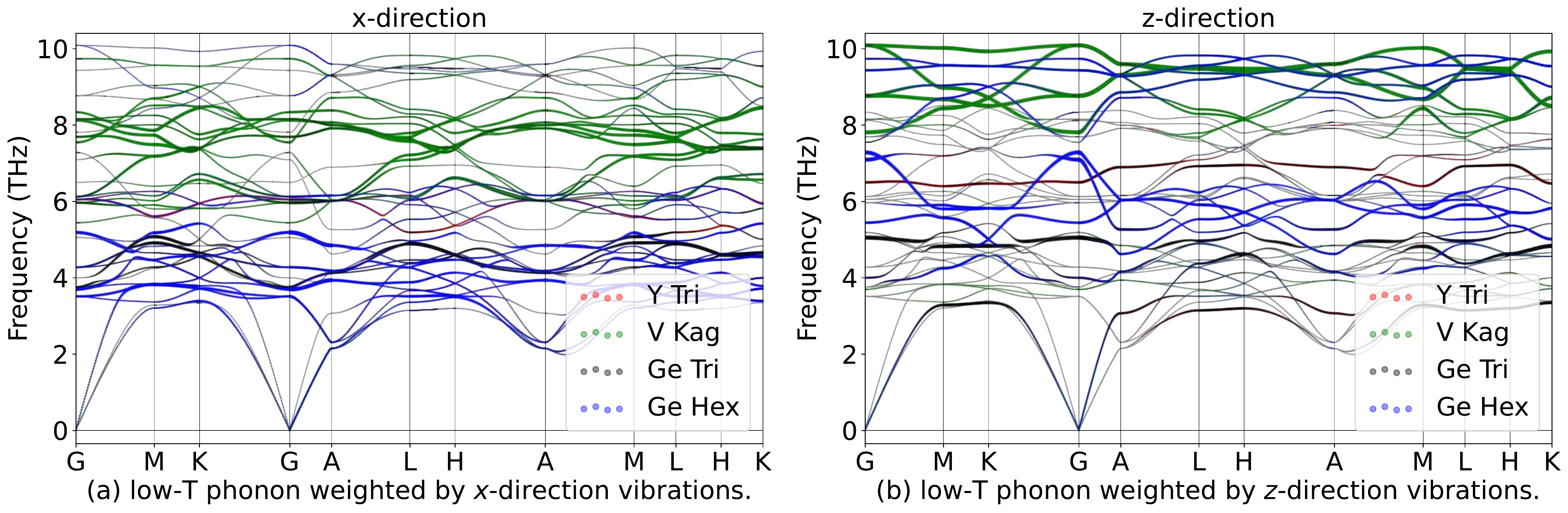}
\hspace{5pt}\label{fig:YV6Ge6_relaxed_High_xz}
\includegraphics[width=0.85\textwidth]{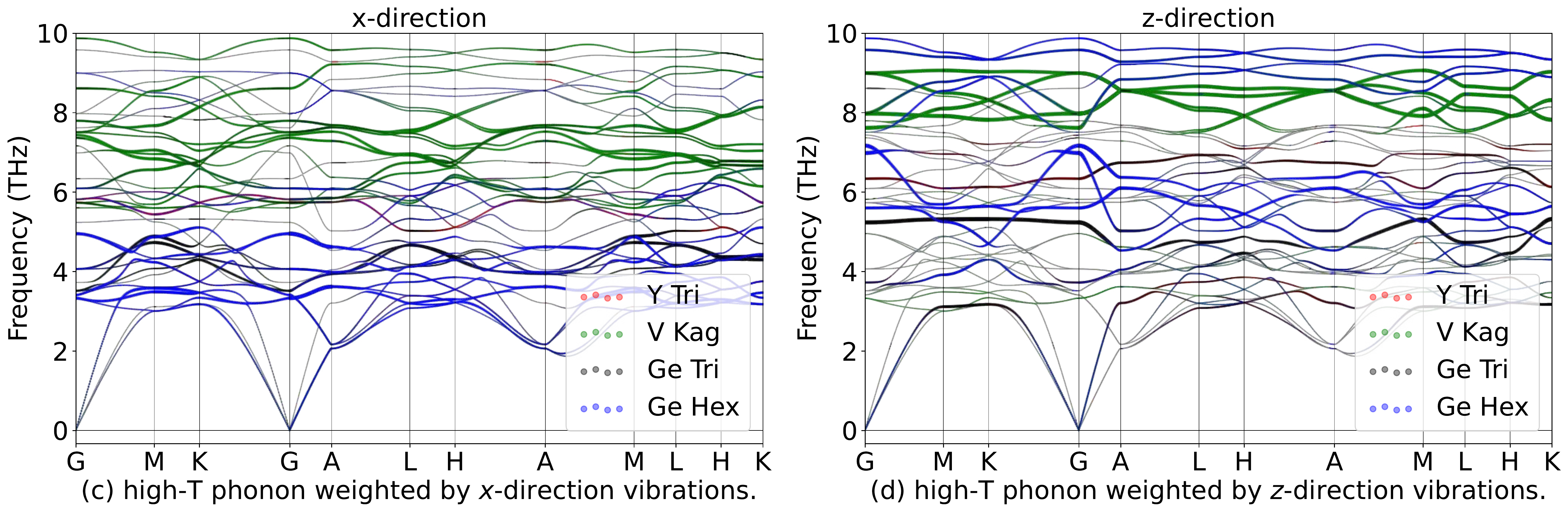}
\caption{Phonon spectrum for YV$_6$Ge$_6$in in-plane ($x$) and out-of-plane ($z$) directions.}
\label{fig:YV6Ge6phoxyz}
\end{figure}

\begin{table}[htbp]
\global\long\def\arraystretch{1.12}
\captionsetup{width=.95\textwidth}
\caption{IRREPs at high-symmetry points for YV$_6$Ge$_6$ phonon spectrum at Low-T.}
\label{tab:191_YV6Ge6_Low}
\setlength{\tabcolsep}{1.mm}{
\begin{tabular}{c|cc|cc|cc|cc|cc|cc}
\hline
\hline
 \multirow{2}{*}{Band}  & \multicolumn{2}{c}{$\Gamma$ $(0,0,0)$}  & \multicolumn{2}{c}{$A$ $(0,0,\frac{1}{2})$}  & \multicolumn{2}{c}{$M$ $(\frac{1}{2},0,0)$}  & \multicolumn{2}{c}{$L$ $(\frac{1}{2},0,\frac{1}{2})$}  & \multicolumn{2}{c}{$K$ $(\frac{1}{3},\frac{1}{3},0)$}  & \multicolumn{2}{c}{$H$ $(\frac{1}{3},\frac{1}{3},\frac{1}{2})$}   \\
\cline{2-13}
 & $f$ (THz) & Reps & $f$ (THz) & Reps & $f$ (THz) & Reps & $f$ (THz) & Reps & $f$ (THz) & Reps & $f$ (THz) & Reps  \\
\hline
1 & \multirow{3}{*}{  -0.000} & \multirow{3}{*}{$-$$|$$-$} & \multirow{2}{*}{   2.135} & \multirow{2}{*}{$E_{1g}$$|$${A_5^+}$} &    3.193 & $B_{1g}$$|$${M_3^+}$ &    3.132 & $B_{1u}$$|$${L_3^-}$ &    3.331 & $A_{2}''$$|$${K_4}$ &    3.183 & $A_{2}''$$|$${H_4}$ \\
\cline{1-1}\cline{6-7}\cline{8-9}\cline{10-11}\cline{12-13}
2 &   &   &   &   &    3.272 & $B_{1u}$$|$${M_3^-}$ &    3.344 & $A_{u}$$|$${L_1^-}$ &    3.378 & $A_{2}'$$|$${K_2}$ &    3.477 & $A_{1}''$$|$${H_3}$ \\
\cline{1-1}\cline{4-5}\cline{6-7}\cline{8-9}\cline{10-11}\cline{12-13}
3 &   &   & \multirow{2}{*}{   2.295} & \multirow{2}{*}{$E_{1u}$$|$${A_5^-}$} &    3.351 & $B_{2u}$$|$${M_2^-}$ &    3.499 & $B_{2g}$$|$${L_2^+}$ & \multirow{2}{*}{   3.699} & \multirow{2}{*}{$E'$$|$${K_6}$} & \multirow{2}{*}{   3.521} & \multirow{2}{*}{$E''$$|$${H_5}$} \\
\cline{1-1}\cline{2-3}\cline{6-7}\cline{8-9}
4 & \multirow{2}{*}{   3.498} & \multirow{2}{*}{$E_{1u}$$|$${\Gamma_5^-}$} &   &   &    3.805 & $B_{3u}$$|$${M_4^-}$ &    3.644 & $B_{2u}$$|$${L_2^-}$ &   &   &   &   \\
\cline{1-1}\cline{4-5}\cline{6-7}\cline{8-9}\cline{10-11}\cline{12-13}
5 &   &   &    3.056 & $A_{2u}$$|$${A_2^-}$ &    3.818 & $A_{g}$$|$${M_1^+}$ &    3.682 & $B_{3g}$$|$${L_4^+}$ &    3.771 & $A_{1}'$$|$${K_1}$ & \multirow{2}{*}{   3.840} & \multirow{2}{*}{$E'$$|$${H_6}$} \\
\hline
\hline
\end{tabular}}
\end{table}

\begin{table}[htbp]
\global\long\def\arraystretch{1.12}
\captionsetup{width=.95\textwidth}
\caption{IRREPs at high-symmetry points for YV$_6$Ge$_6$ phonon spectrum at High-T.}
\label{tab:191_YV6Ge6_High}
\setlength{\tabcolsep}{1.mm}{
\begin{tabular}{c|cc|cc|cc|cc|cc|cc}
\hline
\hline
 \multirow{2}{*}{Band}  & \multicolumn{2}{c}{$\Gamma$ $(0,0,0)$}  & \multicolumn{2}{c}{$A$ $(0,0,\frac{1}{2})$}  & \multicolumn{2}{c}{$M$ $(\frac{1}{2},0,0)$}  & \multicolumn{2}{c}{$L$ $(\frac{1}{2},0,\frac{1}{2})$}  & \multicolumn{2}{c}{$K$ $(\frac{1}{3},\frac{1}{3},0)$}  & \multicolumn{2}{c}{$H$ $(\frac{1}{3},\frac{1}{3},\frac{1}{2})$}   \\
\cline{2-13}
 & $f$ (THz) & Reps & $f$ (THz) & Reps & $f$ (THz) & Reps & $f$ (THz) & Reps & $f$ (THz) & Reps & $f$ (THz) & Reps  \\
\hline
1 & \multirow{3}{*}{  -0.000} & \multirow{3}{*}{$-$$|$$-$} & \multirow{2}{*}{   2.049} & \multirow{2}{*}{$E_{1g}$$|$${A_5^+}$} &    2.989 & $B_{2u}$$|$${M_2^-}$ &    3.067 & $B_{1u}$$|$${L_3^-}$ &    3.160 & $A_{2}''$$|$${K_4}$ &    3.201 & $A_{2}''$$|$${H_4}$ \\
\cline{1-1}\cline{6-7}\cline{8-9}\cline{10-11}\cline{12-13}
2 &   &   &   &   &    2.995 & $B_{1g}$$|$${M_3^+}$ &    3.174 & $A_{u}$$|$${L_1^-}$ &    3.165 & $A_{2}'$$|$${K_2}$ & \multirow{2}{*}{   3.211} & \multirow{2}{*}{$E''$$|$${H_5}$} \\
\cline{1-1}\cline{4-5}\cline{6-7}\cline{8-9}\cline{10-11}
3 &   &   & \multirow{2}{*}{   2.153} & \multirow{2}{*}{$E_{1u}$$|$${A_5^-}$} &    3.101 & $B_{1u}$$|$${M_3^-}$ &    3.190 & $B_{2g}$$|$${L_2^+}$ &    3.327 & $A_{1}'$$|$${K_1}$ &   &   \\
\cline{1-1}\cline{2-3}\cline{6-7}\cline{8-9}\cline{10-11}\cline{12-13}
4 & \multirow{2}{*}{   3.316} & \multirow{2}{*}{$E_{2g}$$|$${\Gamma_6^+}$} &   &   &    3.474 & $A_{g}$$|$${M_1^+}$ &    3.375 & $B_{2u}$$|$${L_2^-}$ & \multirow{2}{*}{   3.455} & \multirow{2}{*}{$E'$$|$${K_6}$} &    3.287 & $A_{1}''$$|$${H_3}$ \\
\cline{1-1}\cline{4-5}\cline{6-7}\cline{8-9}\cline{12-13}
5 &   &   &    3.193 & $A_{2u}$$|$${A_2^-}$ &    3.573 & $B_{3u}$$|$${M_4^-}$ &    3.382 & $B_{3g}$$|$${L_4^+}$ &   &   & \multirow{2}{*}{   3.536} & \multirow{2}{*}{$E'$$|$${H_6}$} \\
\hline
\hline
\end{tabular}}
\end{table}
\subsubsection{\label{sms2sec:ZrV6Ge6}\hyperref[smsec:col]{\texorpdfstring{ZrV$_6$Ge$_6$}{}}~{\color{green}  (High-T stable, Low-T stable) }}

\begin{table}[htbp]
\global\long\def\arraystretch{1.12}
\captionsetup{width=.85\textwidth}
\caption{Structure parameters of ZrV$_6$Ge$_6$ after relaxation. It contains 370 electrons. }
\label{tab:ZrV6Ge6}
\setlength{\tabcolsep}{4mm}{
\begin{tabular}{c|c||c|c}
\hline
\hline
 Atom & Coordinates & \multicolumn{2}{c}{Parameters~($\AA$)} \\
\hline
Zr ~ 1a & ($0,0,0$) & $a$ & $5.1204$ \\
V ~ 6i & ($\frac{1}{2},0,0.2493$) & $c$ & $8.3575$ \\
Ge1 ~ 2c & ($\frac{1}{3},\frac{2}{3},0$) & $d^{Tri}_{Z-Z}$ & $2.7036$ \\
Ge2 ~ 2d & ($\frac{1}{3},\frac{2}{3},\frac{1}{2}$) & $d_{T_{1}}$ & $2.5602$ \\
Ge3 ~ 2e & ($0,0,0.3383$) & $d_{T_{2}}$ & $4.1671$ \\
\hline
\hline
\end{tabular}}
\end{table}

\begin{figure}[htbp]
\centering
\includegraphics[width=0.85\textwidth]{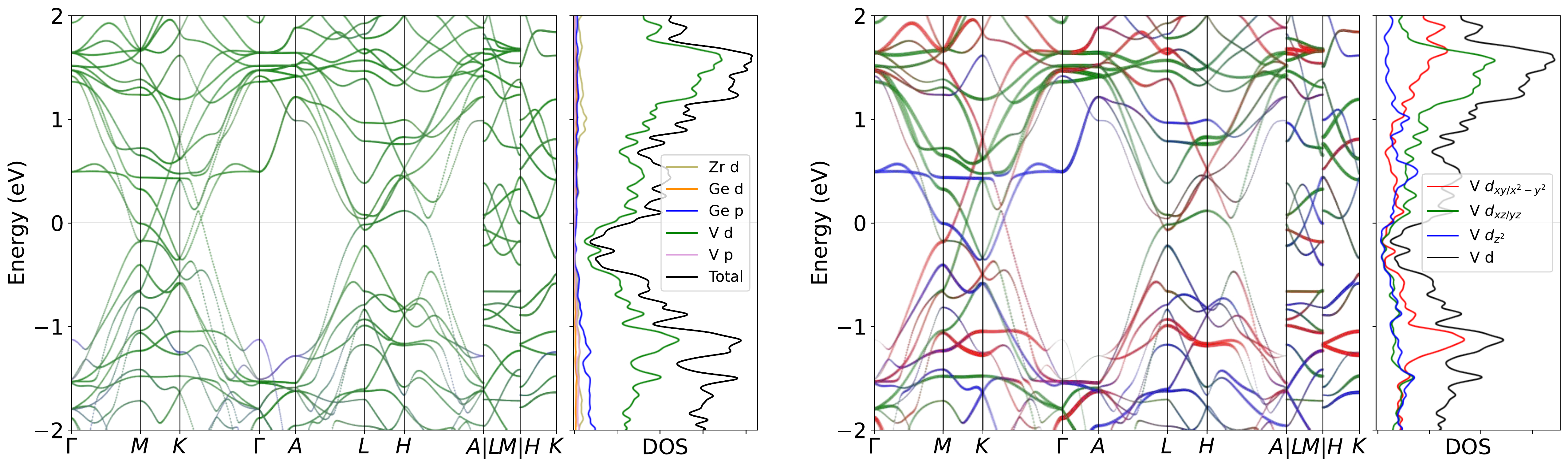}
\caption{Band structure for ZrV$_6$Ge$_6$ without SOC. The projection of V $3d$ orbitals is also presented. The band structures clearly reveal the dominating of V $3d$ orbitals  near the Fermi level, as well as the pronounced peak.}
\label{fig:ZrV6Ge6band}
\end{figure}

\begin{figure}[H]
\centering
\subfloat[Relaxed phonon at low-T.]{
\label{fig:ZrV6Ge6_relaxed_Low}
\includegraphics[width=0.45\textwidth]{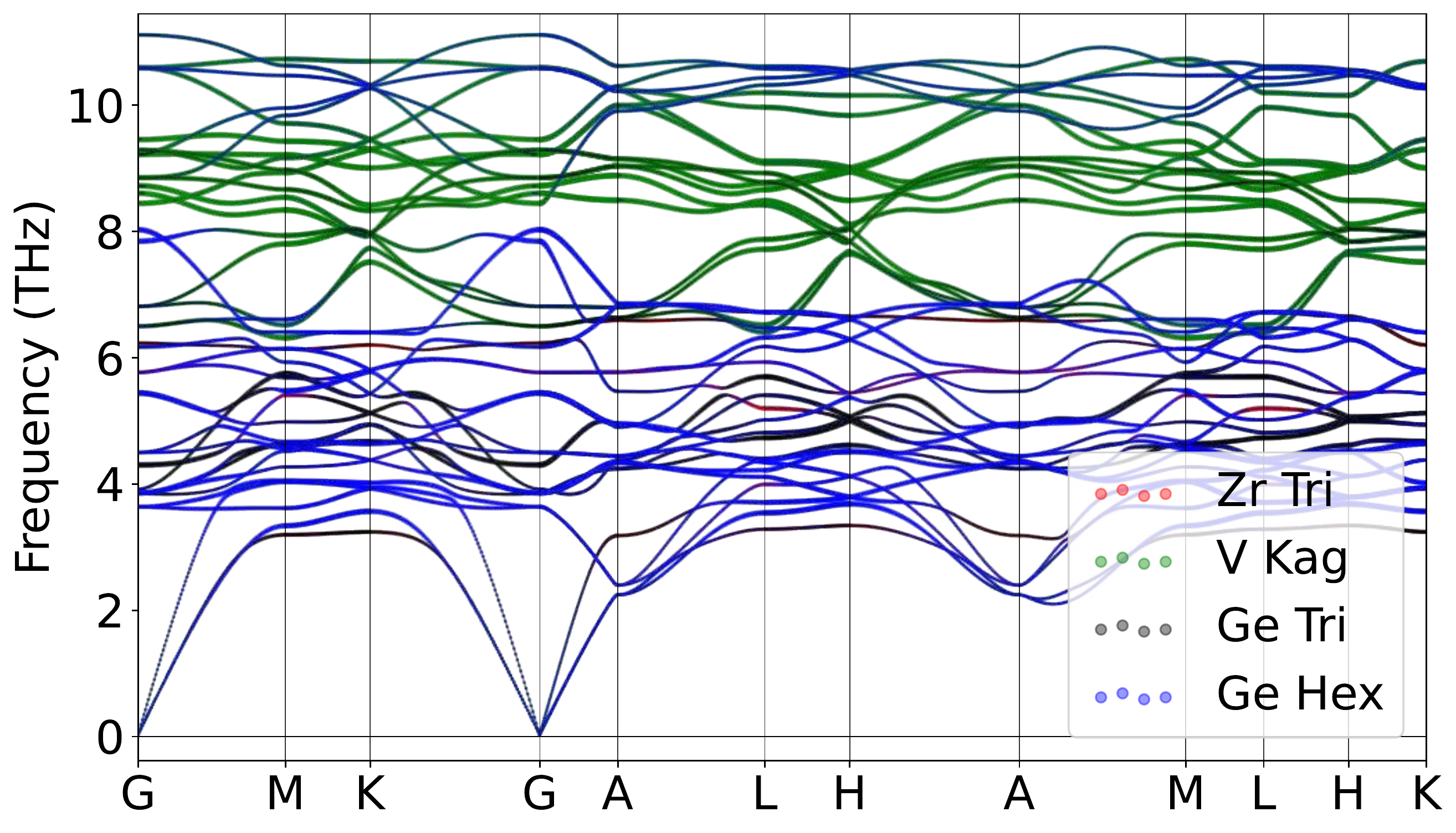}
}
\hspace{5pt}\subfloat[Relaxed phonon at high-T.]{
\label{fig:ZrV6Ge6_relaxed_High}
\includegraphics[width=0.45\textwidth]{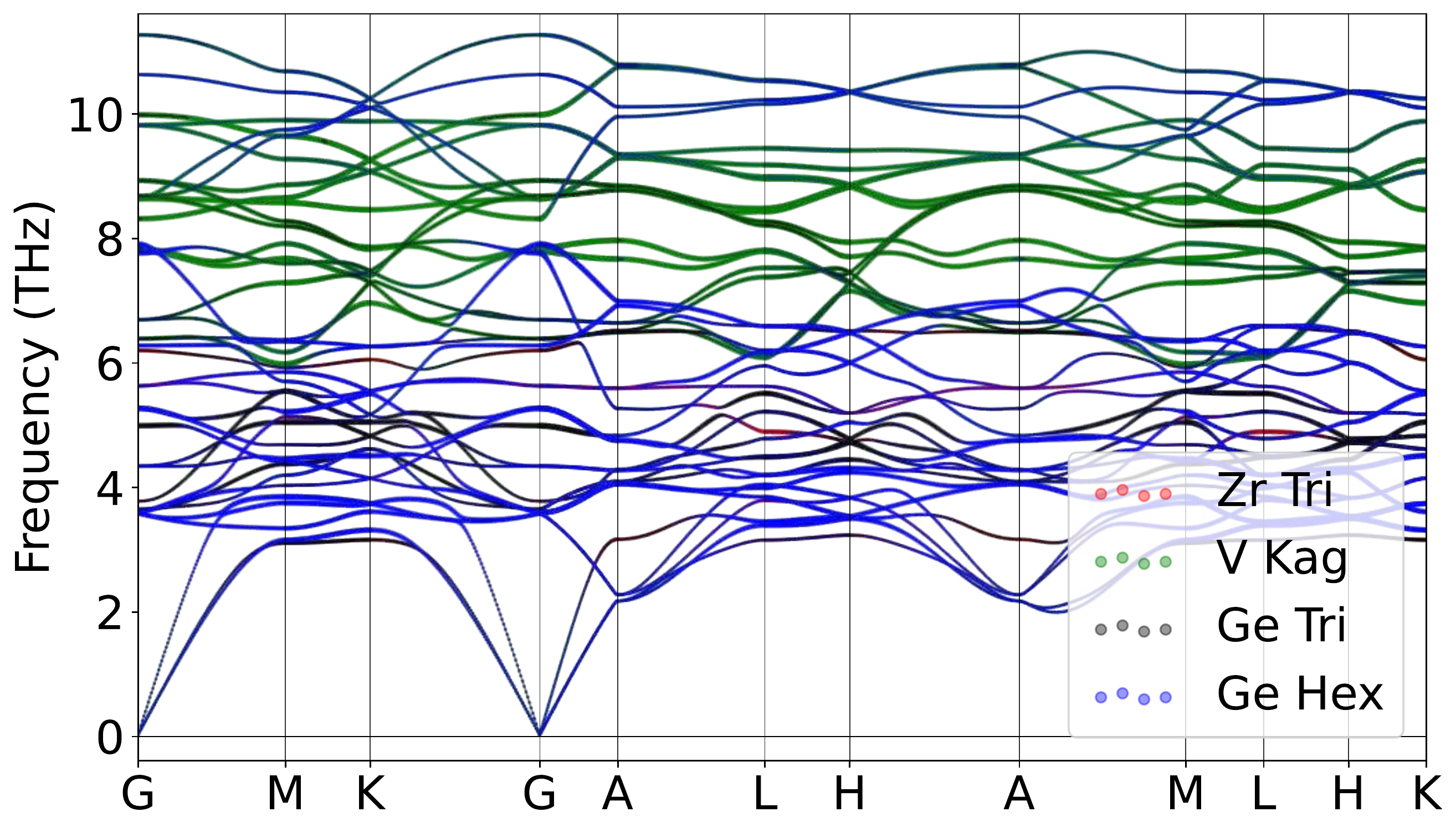}
}
\caption{Atomically projected phonon spectrum for ZrV$_6$Ge$_6$.}
\label{fig:ZrV6Ge6pho}
\end{figure}

\begin{figure}[H]
\centering
\label{fig:ZrV6Ge6_relaxed_Low_xz}
\includegraphics[width=0.85\textwidth]{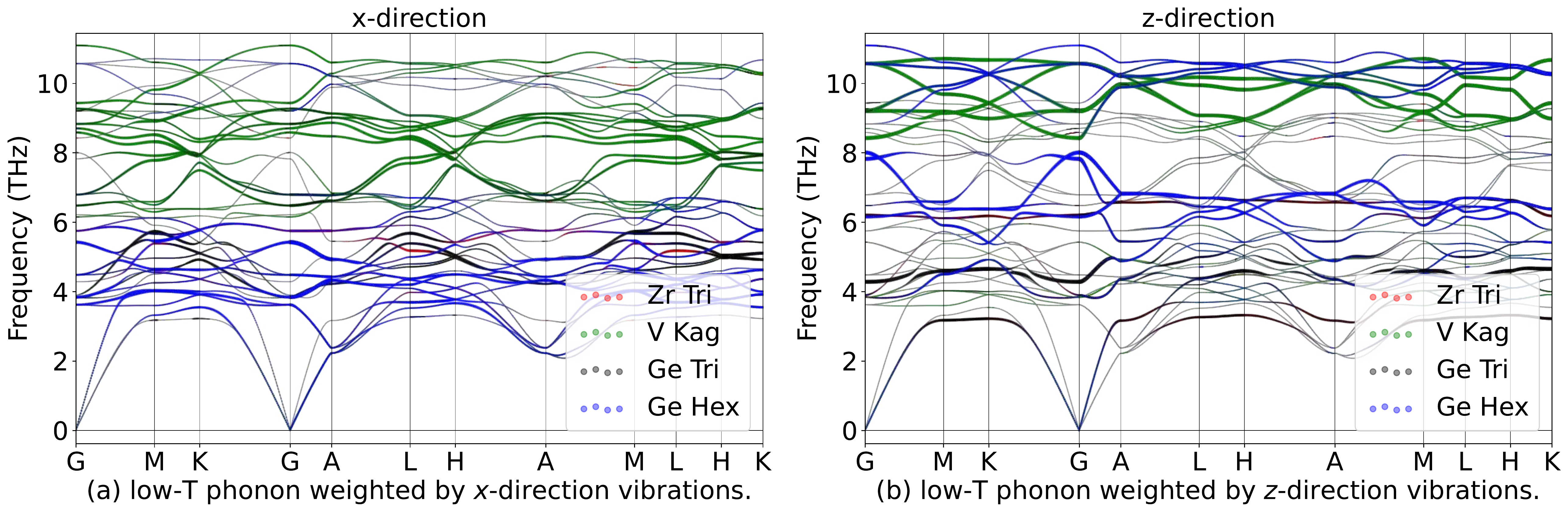}
\hspace{5pt}\label{fig:ZrV6Ge6_relaxed_High_xz}
\includegraphics[width=0.85\textwidth]{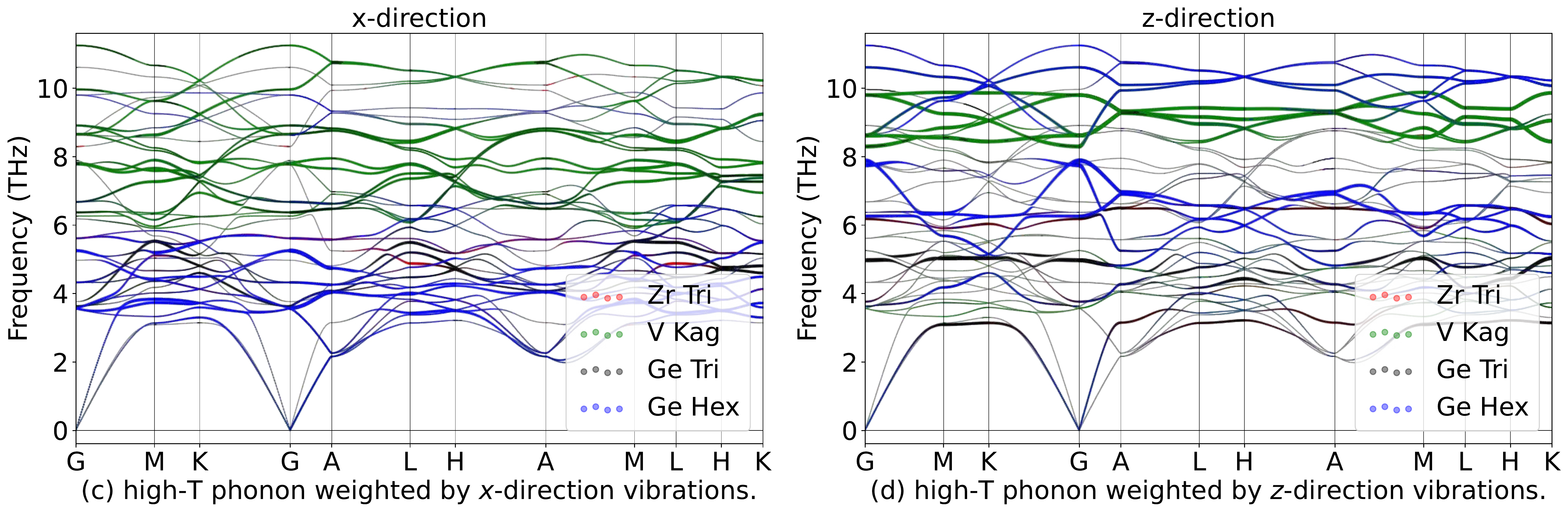}
\caption{Phonon spectrum for ZrV$_6$Ge$_6$in in-plane ($x$) and out-of-plane ($z$) directions.}
\label{fig:ZrV6Ge6phoxyz}
\end{figure}

\begin{table}[htbp]
\global\long\def\arraystretch{1.12}
\captionsetup{width=.95\textwidth}
\caption{IRREPs at high-symmetry points for ZrV$_6$Ge$_6$ phonon spectrum at Low-T.}
\label{tab:191_ZrV6Ge6_Low}
\setlength{\tabcolsep}{1.mm}{
\begin{tabular}{c|cc|cc|cc|cc|cc|cc}
\hline
\hline
 \multirow{2}{*}{Band}  & \multicolumn{2}{c}{$\Gamma$ $(0,0,0)$}  & \multicolumn{2}{c}{$A$ $(0,0,\frac{1}{2})$}  & \multicolumn{2}{c}{$M$ $(\frac{1}{2},0,0)$}  & \multicolumn{2}{c}{$L$ $(\frac{1}{2},0,\frac{1}{2})$}  & \multicolumn{2}{c}{$K$ $(\frac{1}{3},\frac{1}{3},0)$}  & \multicolumn{2}{c}{$H$ $(\frac{1}{3},\frac{1}{3},\frac{1}{2})$}   \\
\cline{2-13}
 & $f$ (THz) & Reps & $f$ (THz) & Reps & $f$ (THz) & Reps & $f$ (THz) & Reps & $f$ (THz) & Reps & $f$ (THz) & Reps  \\
\hline
1 & \multirow{3}{*}{   0.000} & \multirow{3}{*}{$-$$|$$-$} & \multirow{2}{*}{   2.214} & \multirow{2}{*}{$E_{1g}$$|$${A_5^+}$} &    3.165 & $B_{1u}$$|$${M_3^-}$ &    3.254 & $B_{1u}$$|$${L_3^-}$ &    3.210 & $A_{2}''$$|$${K_4}$ &    3.312 & $A_{2}''$$|$${H_4}$ \\
\cline{1-1}\cline{6-7}\cline{8-9}\cline{10-11}\cline{12-13}
2 &   &   &   &   &    3.308 & $B_{1g}$$|$${M_3^+}$ &    3.512 & $A_{u}$$|$${L_1^-}$ &    3.536 & $A_{2}'$$|$${K_2}$ &    3.653 & $A_{1}''$$|$${H_3}$ \\
\cline{1-1}\cline{4-5}\cline{6-7}\cline{8-9}\cline{10-11}\cline{12-13}
3 &   &   & \multirow{2}{*}{   2.367} & \multirow{2}{*}{$E_{1u}$$|$${A_5^-}$} &    3.586 & $B_{2u}$$|$${M_2^-}$ &    3.678 & $B_{2g}$$|$${L_2^+}$ & \multirow{2}{*}{   3.903} & \multirow{2}{*}{$E'$$|$${K_6}$} & \multirow{2}{*}{   3.762} & \multirow{2}{*}{$E''$$|$${H_5}$} \\
\cline{1-1}\cline{2-3}\cline{6-7}\cline{8-9}
4 & \multirow{2}{*}{   3.609} & \multirow{2}{*}{$E_{1u}$$|$${\Gamma_5^-}$} &   &   &    4.010 & $B_{3u}$$|$${M_4^-}$ &    3.963 & $B_{2u}$$|$${L_2^-}$ &   &   &   &   \\
\cline{1-1}\cline{4-5}\cline{6-7}\cline{8-9}\cline{10-11}\cline{12-13}
5 &   &   &    3.153 & $A_{2u}$$|$${A_2^-}$ &    4.015 & $A_{g}$$|$${M_1^+}$ &    4.082 & $B_{3g}$$|$${L_4^+}$ &    3.988 & $A_{1}'$$|$${K_1}$ & \multirow{2}{*}{   4.090} & \multirow{2}{*}{$E'$$|$${H_6}$} \\
\hline
\hline
\end{tabular}}
\end{table}

\begin{table}[htbp]
\global\long\def\arraystretch{1.12}
\captionsetup{width=.95\textwidth}
\caption{IRREPs at high-symmetry points for ZrV$_6$Ge$_6$ phonon spectrum at High-T.}
\label{tab:191_ZrV6Ge6_High}
\setlength{\tabcolsep}{1.mm}{
\begin{tabular}{c|cc|cc|cc|cc|cc|cc}
\hline
\hline
 \multirow{2}{*}{Band}  & \multicolumn{2}{c}{$\Gamma$ $(0,0,0)$}  & \multicolumn{2}{c}{$A$ $(0,0,\frac{1}{2})$}  & \multicolumn{2}{c}{$M$ $(\frac{1}{2},0,0)$}  & \multicolumn{2}{c}{$L$ $(\frac{1}{2},0,\frac{1}{2})$}  & \multicolumn{2}{c}{$K$ $(\frac{1}{3},\frac{1}{3},0)$}  & \multicolumn{2}{c}{$H$ $(\frac{1}{3},\frac{1}{3},\frac{1}{2})$}   \\
\cline{2-13}
 & $f$ (THz) & Reps & $f$ (THz) & Reps & $f$ (THz) & Reps & $f$ (THz) & Reps & $f$ (THz) & Reps & $f$ (THz) & Reps  \\
\hline
1 & \multirow{3}{*}{  -0.000} & \multirow{3}{*}{$-$$|$$-$} & \multirow{2}{*}{   2.147} & \multirow{2}{*}{$E_{1g}$$|$${A_5^+}$} &    3.077 & $B_{1u}$$|$${M_3^-}$ &    3.125 & $B_{1u}$$|$${L_3^-}$ &    3.129 & $A_{2}''$$|$${K_4}$ &    3.204 & $A_{2}''$$|$${H_4}$ \\
\cline{1-1}\cline{6-7}\cline{8-9}\cline{10-11}\cline{12-13}
2 &   &   &   &   &    3.125 & $B_{1g}$$|$${M_3^+}$ &    3.365 & $A_{u}$$|$${L_1^-}$ &    3.286 & $A_{2}'$$|$${K_2}$ &    3.473 & $A_{1}''$$|$${H_3}$ \\
\cline{1-1}\cline{4-5}\cline{6-7}\cline{8-9}\cline{10-11}\cline{12-13}
3 &   &   & \multirow{2}{*}{   2.246} & \multirow{2}{*}{$E_{1u}$$|$${A_5^-}$} &    3.313 & $B_{2u}$$|$${M_2^-}$ &    3.418 & $B_{2g}$$|$${L_2^+}$ &    3.580 & $A_{1}'$$|$${K_1}$ & \multirow{2}{*}{   3.501} & \multirow{2}{*}{$E''$$|$${H_5}$} \\
\cline{1-1}\cline{2-3}\cline{6-7}\cline{8-9}\cline{10-11}
4 & \multirow{2}{*}{   3.547} & \multirow{2}{*}{$E_{1u}$$|$${\Gamma_5^-}$} &   &   &    3.720 & $A_{g}$$|$${M_1^+}$ &    3.760 & $B_{2u}$$|$${L_2^-}$ & \multirow{2}{*}{   3.709} & \multirow{2}{*}{$E'$$|$${K_6}$} &   &   \\
\cline{1-1}\cline{4-5}\cline{6-7}\cline{8-9}\cline{12-13}
5 &   &   &    3.136 & $A_{2u}$$|$${A_2^-}$ &    3.820 & $B_{3u}$$|$${M_4^-}$ &    3.817 & $B_{3g}$$|$${L_4^+}$ &   &   & \multirow{2}{*}{   3.802} & \multirow{2}{*}{$E'$$|$${H_6}$} \\
\hline
\hline
\end{tabular}}
\end{table}
\subsubsection{\label{sms2sec:NbV6Ge6}\hyperref[smsec:col]{\texorpdfstring{NbV$_6$Ge$_6$}{}}~{\color{green}  (High-T stable, Low-T stable) }}

\begin{table}[htbp]
\global\long\def\arraystretch{1.12}
\captionsetup{width=.85\textwidth}
\caption{Structure parameters of NbV$_6$Ge$_6$ after relaxation. It contains 371 electrons. }
\label{tab:NbV6Ge6}
\setlength{\tabcolsep}{4mm}{
\begin{tabular}{c|c||c|c}
\hline
\hline
 Atom & Coordinates & \multicolumn{2}{c}{Parameters~($\AA$)} \\
\hline
Nb ~ 1a & ($0,0,0$) & $a$ & $5.0745$ \\
V ~ 6i & ($\frac{1}{2},0,0.2495$) & $c$ & $8.4275$ \\
Ge1 ~ 2c & ($\frac{1}{3},\frac{2}{3},0$) & $d^{Tri}_{Z-Z}$ & $3.0883$ \\
Ge2 ~ 2d & ($\frac{1}{3},\frac{2}{3},\frac{1}{2}$) & $d_{T_{1}}$ & $2.5373$ \\
Ge3 ~ 2e & ($0,0,0.3168$) & $d_{T_{2}}$ & $4.2053$ \\
\hline
\hline
\end{tabular}}
\end{table}

\begin{figure}[htbp]
\centering
\includegraphics[width=0.85\textwidth]{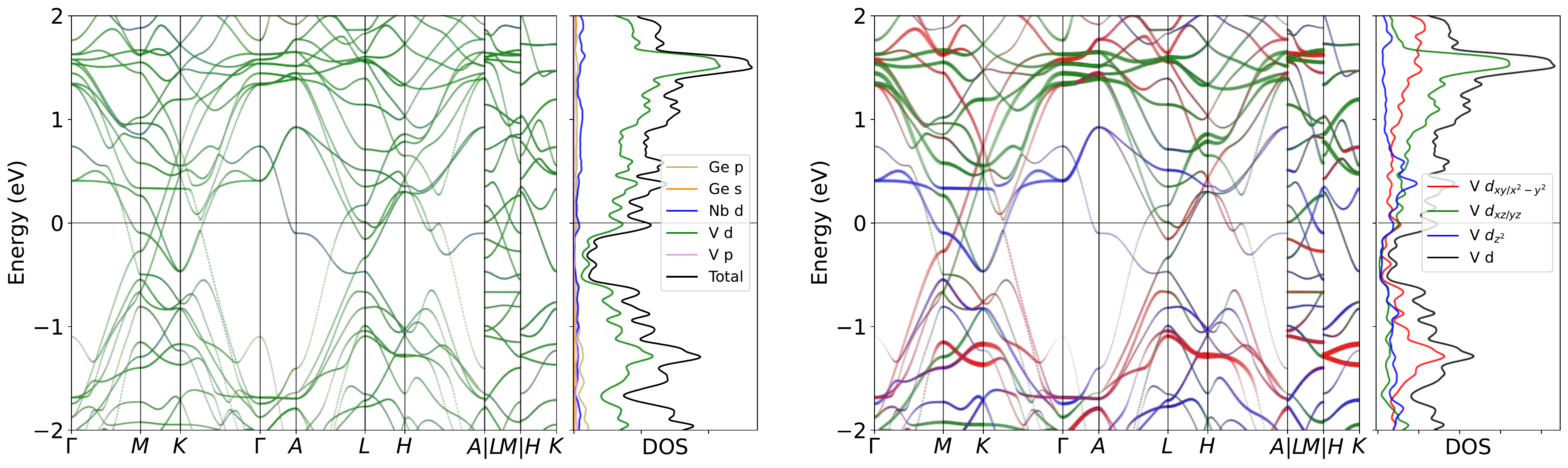}
\caption{Band structure for NbV$_6$Ge$_6$ without SOC. The projection of V $3d$ orbitals is also presented. The band structures clearly reveal the dominating of V $3d$ orbitals  near the Fermi level, as well as the pronounced peak.}
\label{fig:NbV6Ge6band}
\end{figure}

\begin{figure}[H]
\centering
\subfloat[Relaxed phonon at low-T.]{
\label{fig:NbV6Ge6_relaxed_Low}
\includegraphics[width=0.45\textwidth]{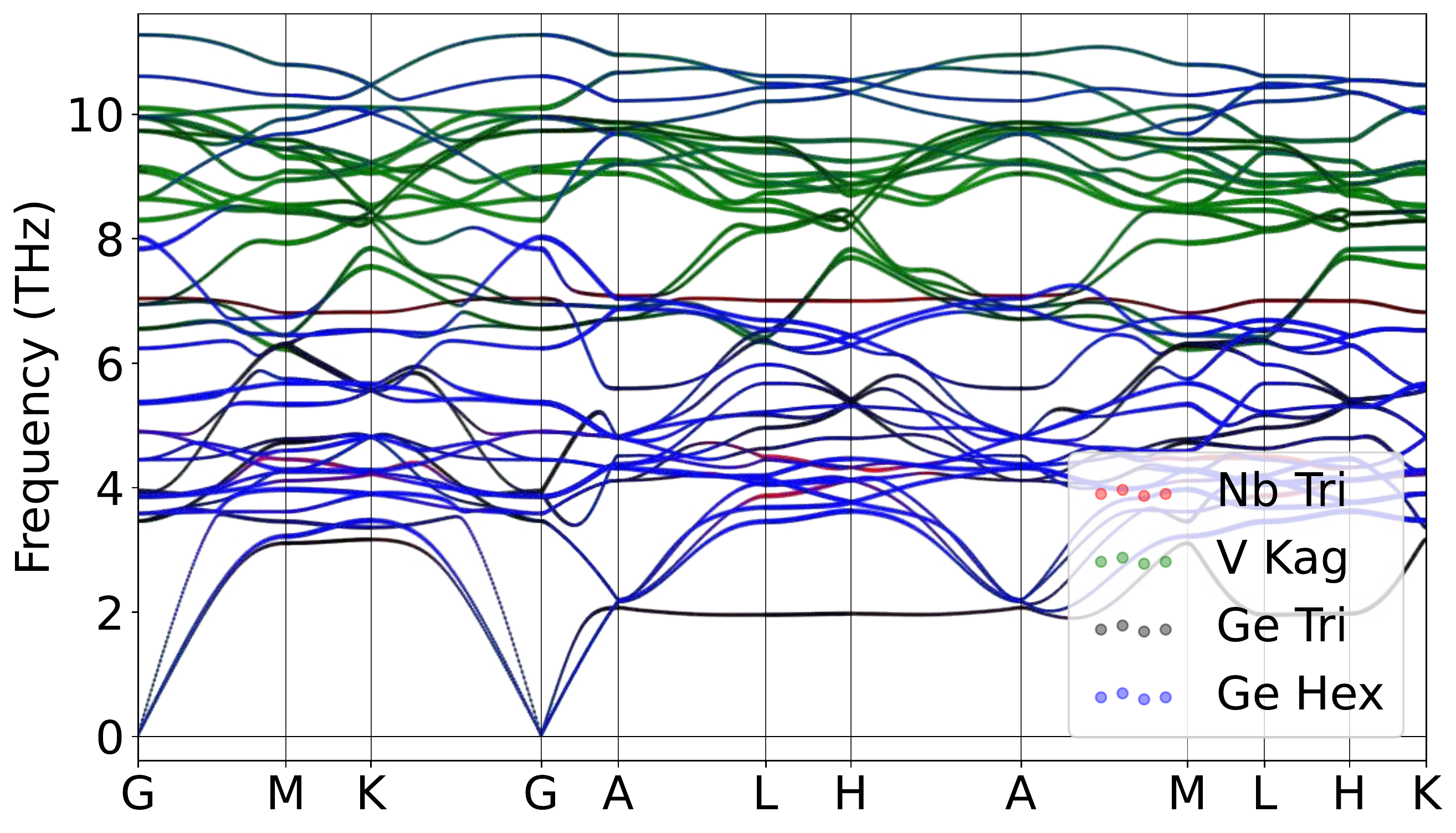}
}
\hspace{5pt}\subfloat[Relaxed phonon at high-T.]{
\label{fig:NbV6Ge6_relaxed_High}
\includegraphics[width=0.45\textwidth]{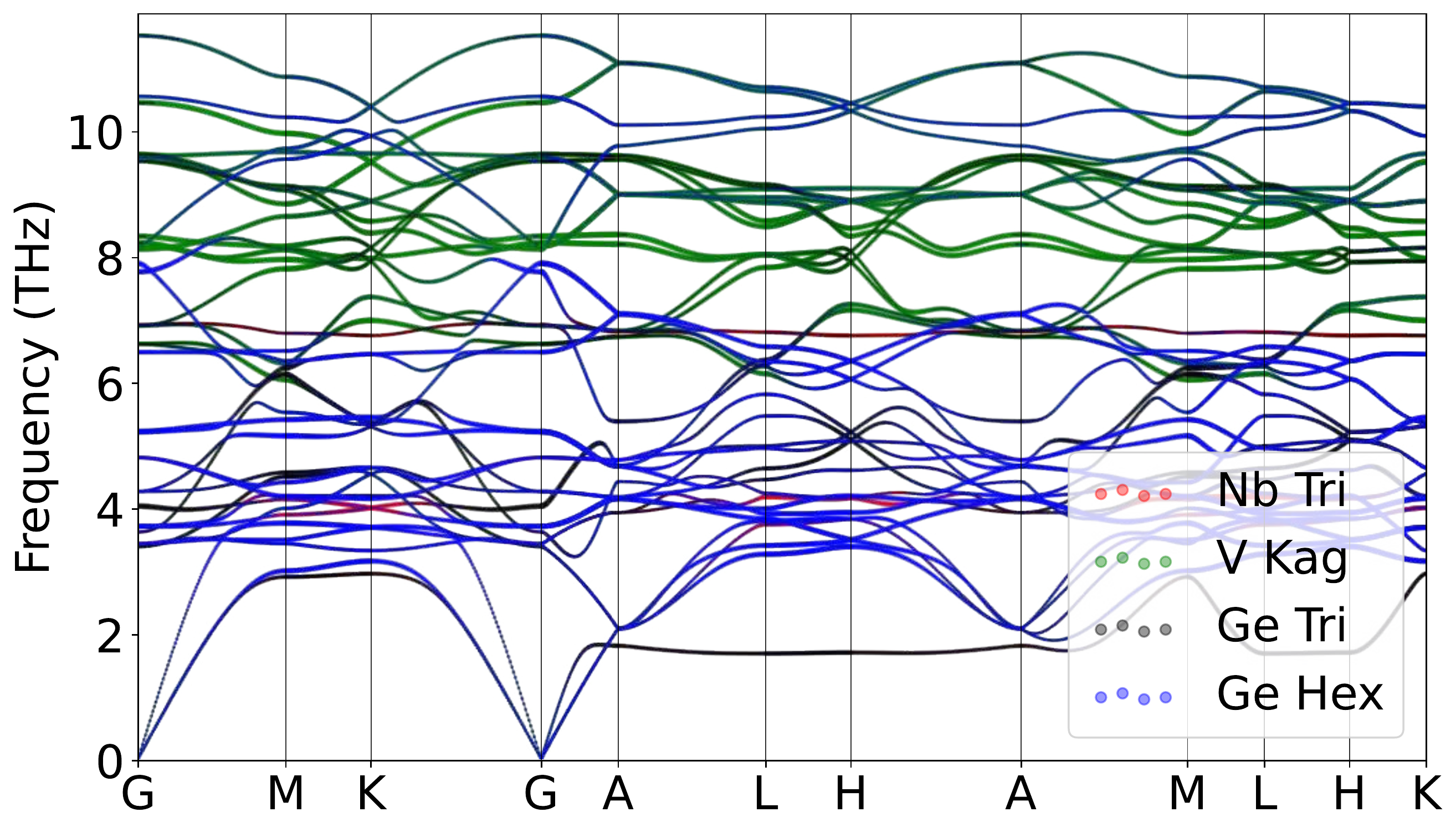}
}
\caption{Atomically projected phonon spectrum for NbV$_6$Ge$_6$.}
\label{fig:NbV6Ge6pho}
\end{figure}

\begin{figure}[H]
\centering
\label{fig:NbV6Ge6_relaxed_Low_xz}
\includegraphics[width=0.85\textwidth]{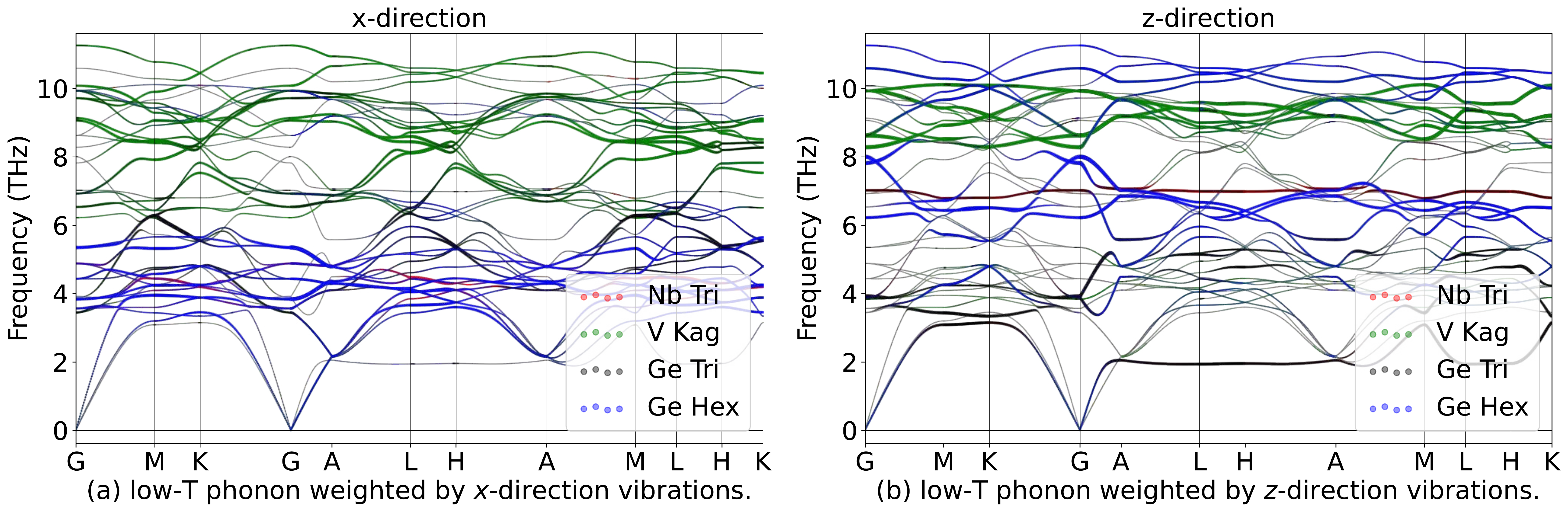}
\hspace{5pt}\label{fig:NbV6Ge6_relaxed_High_xz}
\includegraphics[width=0.85\textwidth]{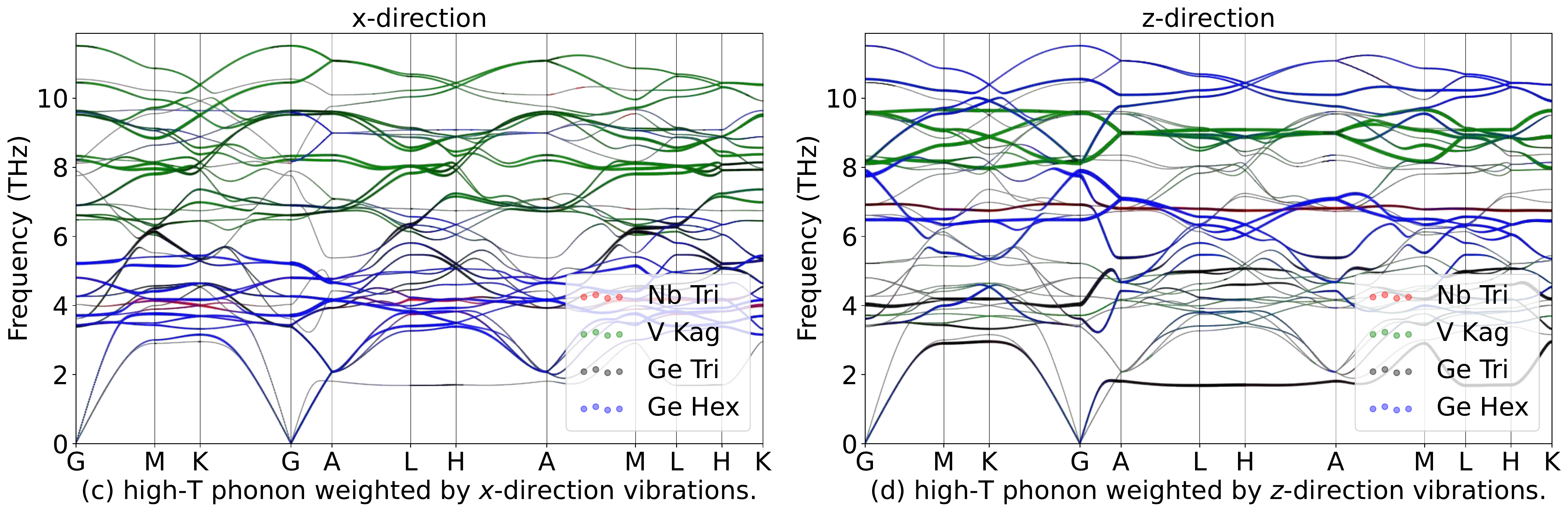}
\caption{Phonon spectrum for NbV$_6$Ge$_6$in in-plane ($x$) and out-of-plane ($z$) directions.}
\label{fig:NbV6Ge6phoxyz}
\end{figure}

\begin{table}[htbp]
\global\long\def\arraystretch{1.12}
\captionsetup{width=.95\textwidth}
\caption{IRREPs at high-symmetry points for NbV$_6$Ge$_6$ phonon spectrum at Low-T.}
\label{tab:191_NbV6Ge6_Low}
\setlength{\tabcolsep}{1.mm}{
\begin{tabular}{c|cc|cc|cc|cc|cc|cc}
\hline
\hline
 \multirow{2}{*}{Band}  & \multicolumn{2}{c}{$\Gamma$ $(0,0,0)$}  & \multicolumn{2}{c}{$A$ $(0,0,\frac{1}{2})$}  & \multicolumn{2}{c}{$M$ $(\frac{1}{2},0,0)$}  & \multicolumn{2}{c}{$L$ $(\frac{1}{2},0,\frac{1}{2})$}  & \multicolumn{2}{c}{$K$ $(\frac{1}{3},\frac{1}{3},0)$}  & \multicolumn{2}{c}{$H$ $(\frac{1}{3},\frac{1}{3},\frac{1}{2})$}   \\
\cline{2-13}
 & $f$ (THz) & Reps & $f$ (THz) & Reps & $f$ (THz) & Reps & $f$ (THz) & Reps & $f$ (THz) & Reps & $f$ (THz) & Reps  \\
\hline
1 & \multirow{3}{*}{  -0.000} & \multirow{3}{*}{$-$$|$$-$} &    2.042 & $A_{2u}$$|$${A_2^-}$ &    3.077 & $B_{1u}$$|$${M_3^-}$ &    1.926 & $B_{1u}$$|$${L_3^-}$ &    3.137 & $A_{2}''$$|$${K_4}$ &    1.944 & $A_{2}''$$|$${H_4}$ \\
\cline{1-1}\cline{4-5}\cline{6-7}\cline{8-9}\cline{10-11}\cline{12-13}
2 &   &   & \multirow{2}{*}{   2.135} & \multirow{2}{*}{$E_{1g}$$|$${A_5^+}$} &    3.188 & $B_{1g}$$|$${M_3^+}$ &    3.424 & $A_{u}$$|$${L_1^-}$ &    3.329 & $A_{1}'$$|$${K_1}$ &    3.591 & $A_{1}''$$|$${H_3}$ \\
\cline{1-1}\cline{6-7}\cline{8-9}\cline{10-11}\cline{12-13}
3 &   &   &   &   &    3.424 & $A_{g}$$|$${M_1^+}$ &    3.650 & $B_{2g}$$|$${L_2^+}$ &    3.441 & $A_{2}'$$|$${K_2}$ & \multirow{2}{*}{   3.732} & \multirow{2}{*}{$E''$$|$${H_5}$} \\
\cline{1-1}\cline{2-3}\cline{4-5}\cline{6-7}\cline{8-9}\cline{10-11}
4 & \multirow{2}{*}{   3.432} & \multirow{2}{*}{$E_{1g}$$|$${\Gamma_5^+}$} & \multirow{2}{*}{   2.173} & \multirow{2}{*}{$E_{1u}$$|$${A_5^-}$} &    3.583 & $B_{2u}$$|$${M_2^-}$ &    3.836 & $B_{2u}$$|$${L_2^-}$ & \multirow{2}{*}{   3.869} & \multirow{2}{*}{$E'$$|$${K_6}$} &   &   \\
\cline{1-1}\cline{6-7}\cline{8-9}\cline{12-13}
5 &   &   &   &   &    3.939 & $B_{3u}$$|$${M_4^-}$ &    4.020 & $B_{1g}$$|$${L_3^+}$ &   &   &    4.089 & $A_{2}'$$|$${H_2}$ \\
\hline
\hline
\end{tabular}}
\end{table}

\begin{table}[htbp]
\global\long\def\arraystretch{1.12}
\captionsetup{width=.95\textwidth}
\caption{IRREPs at high-symmetry points for NbV$_6$Ge$_6$ phonon spectrum at High-T.}
\label{tab:191_NbV6Ge6_High}
\setlength{\tabcolsep}{1.mm}{
\begin{tabular}{c|cc|cc|cc|cc|cc|cc}
\hline
\hline
 \multirow{2}{*}{Band}  & \multicolumn{2}{c}{$\Gamma$ $(0,0,0)$}  & \multicolumn{2}{c}{$A$ $(0,0,\frac{1}{2})$}  & \multicolumn{2}{c}{$M$ $(\frac{1}{2},0,0)$}  & \multicolumn{2}{c}{$L$ $(\frac{1}{2},0,\frac{1}{2})$}  & \multicolumn{2}{c}{$K$ $(\frac{1}{3},\frac{1}{3},0)$}  & \multicolumn{2}{c}{$H$ $(\frac{1}{3},\frac{1}{3},\frac{1}{2})$}   \\
\cline{2-13}
 & $f$ (THz) & Reps & $f$ (THz) & Reps & $f$ (THz) & Reps & $f$ (THz) & Reps & $f$ (THz) & Reps & $f$ (THz) & Reps  \\
\hline
1 & \multirow{3}{*}{   0.000} & \multirow{3}{*}{$-$$|$$-$} &    1.798 & $A_{2u}$$|$${A_2^-}$ &    2.893 & $B_{1u}$$|$${M_3^-}$ &    1.675 & $B_{1u}$$|$${L_3^-}$ &    2.943 & $A_{2}''$$|$${K_4}$ &    1.690 & $A_{2}''$$|$${H_4}$ \\
\cline{1-1}\cline{4-5}\cline{6-7}\cline{8-9}\cline{10-11}\cline{12-13}
2 &   &   & \multirow{2}{*}{   2.059} & \multirow{2}{*}{$E_{1u}$$|$${A_5^-}$} &    2.990 & $B_{1g}$$|$${M_3^+}$ &    3.253 & $A_{u}$$|$${L_1^-}$ &    3.144 & $A_{2}'$$|$${K_2}$ &    3.375 & $A_{1}''$$|$${H_3}$ \\
\cline{1-1}\cline{6-7}\cline{8-9}\cline{10-11}\cline{12-13}
3 &   &   &   &   &    3.422 & $A_{g}$$|$${M_1^+}$ &    3.393 & $B_{2g}$$|$${L_2^+}$ &    3.310 & $A_{1}'$$|$${K_1}$ & \multirow{2}{*}{   3.484} & \multirow{2}{*}{$E''$$|$${H_5}$} \\
\cline{1-1}\cline{2-3}\cline{4-5}\cline{6-7}\cline{8-9}\cline{10-11}
4 & \multirow{2}{*}{   3.376} & \multirow{2}{*}{$E_{1g}$$|$${\Gamma_5^+}$} & \multirow{2}{*}{   2.079} & \multirow{2}{*}{$E_{1g}$$|$${A_5^+}$} &    3.483 & $B_{2u}$$|$${M_2^-}$ &    3.728 & $B_{2u}$$|$${L_2^-}$ & \multirow{2}{*}{   3.680} & \multirow{2}{*}{$E'$$|$${K_6}$} &   &   \\
\cline{1-1}\cline{6-7}\cline{8-9}\cline{12-13}
5 &   &   &   &   &    3.747 & $B_{3u}$$|$${M_4^-}$ &    3.780 & $B_{3u}$$|$${L_4^-}$ &   &   & \multirow{2}{*}{   3.813} & \multirow{2}{*}{$E'$$|$${H_6}$} \\
\hline
\hline
\end{tabular}}
\end{table}
\subsubsection{\label{sms2sec:PrV6Ge6}\hyperref[smsec:col]{\texorpdfstring{PrV$_6$Ge$_6$}{}}~{\color{green}  (High-T stable, Low-T stable) }}

\begin{table}[htbp]
\global\long\def\arraystretch{1.12}
\captionsetup{width=.85\textwidth}
\caption{Structure parameters of PrV$_6$Ge$_6$ after relaxation. It contains 389 electrons. }
\label{tab:PrV6Ge6}
\setlength{\tabcolsep}{4mm}{
\begin{tabular}{c|c||c|c}
\hline
\hline
 Atom & Coordinates & \multicolumn{2}{c}{Parameters~($\AA$)} \\
\hline
Pr ~ 1a & ($0,0,0$) & $a$ & $5.2013$ \\
V ~ 6i & ($\frac{1}{2},0,0.2505$) & $c$ & $8.3611$ \\
Ge1 ~ 2c & ($\frac{1}{3},\frac{2}{3},0$) & $d^{Tri}_{Z-Z}$ & $2.5922$ \\
Ge2 ~ 2d & ($\frac{1}{3},\frac{2}{3},\frac{1}{2}$) & $d_{T_{1}}$ & $2.6006$ \\
Ge3 ~ 2e & ($0,0,0.3450$) & $d_{T_{2}}$ & $4.1728$ \\
\hline
\hline
\end{tabular}}
\end{table}

\begin{figure}[htbp]
\centering
\includegraphics[width=0.85\textwidth]{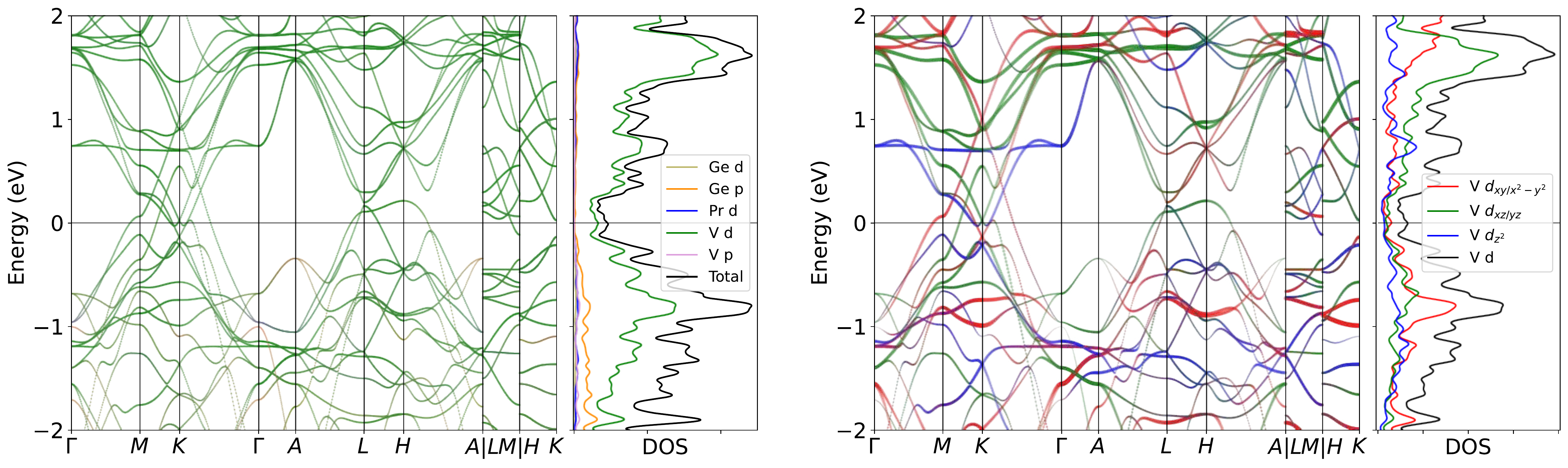}
\caption{Band structure for PrV$_6$Ge$_6$ without SOC. The projection of V $3d$ orbitals is also presented. The band structures clearly reveal the dominating of V $3d$ orbitals  near the Fermi level, as well as the pronounced peak.}
\label{fig:PrV6Ge6band}
\end{figure}

\begin{figure}[H]
\centering
\subfloat[Relaxed phonon at low-T.]{
\label{fig:PrV6Ge6_relaxed_Low}
\includegraphics[width=0.45\textwidth]{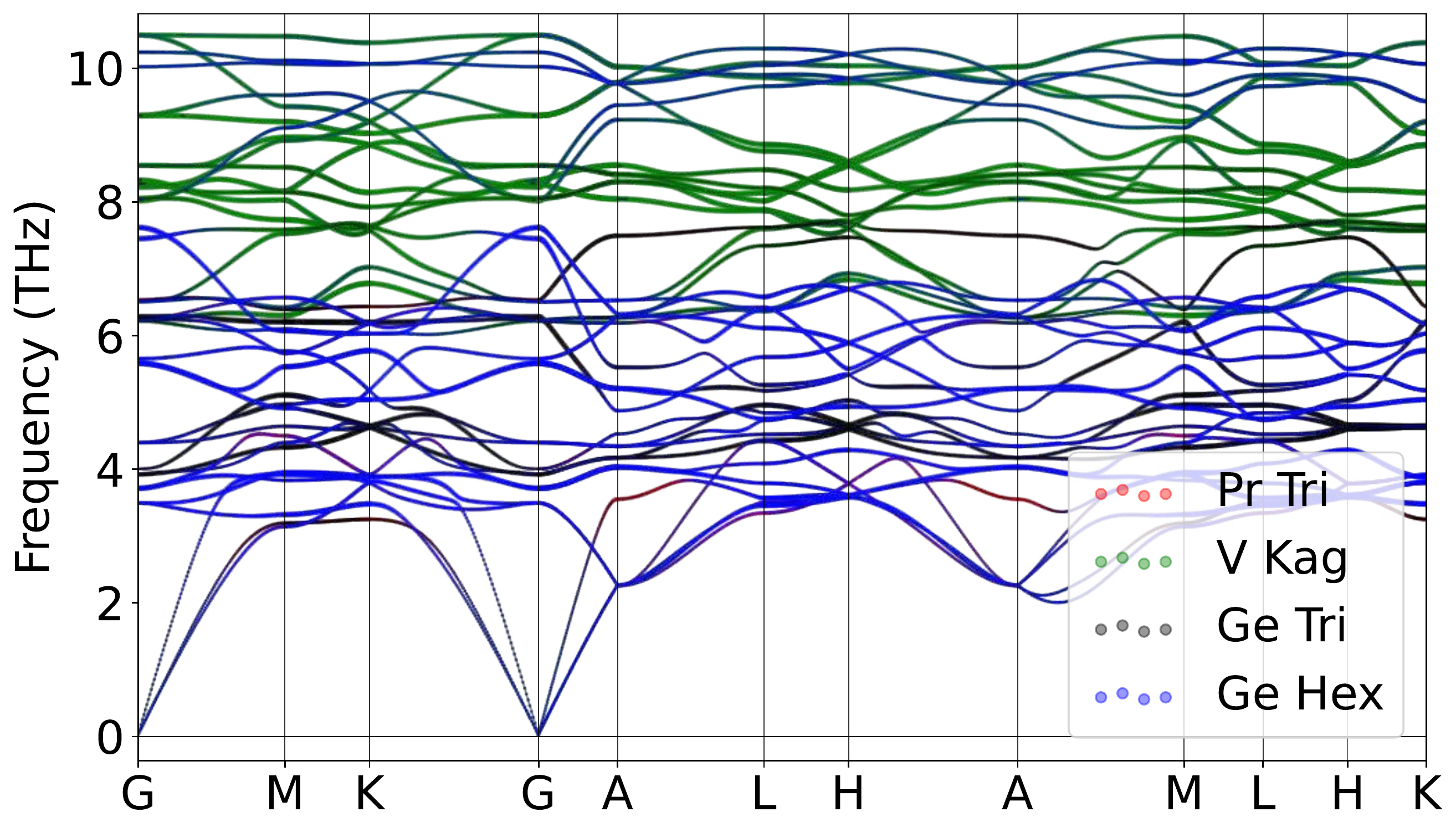}
}
\hspace{5pt}\subfloat[Relaxed phonon at high-T.]{
\label{fig:PrV6Ge6_relaxed_High}
\includegraphics[width=0.45\textwidth]{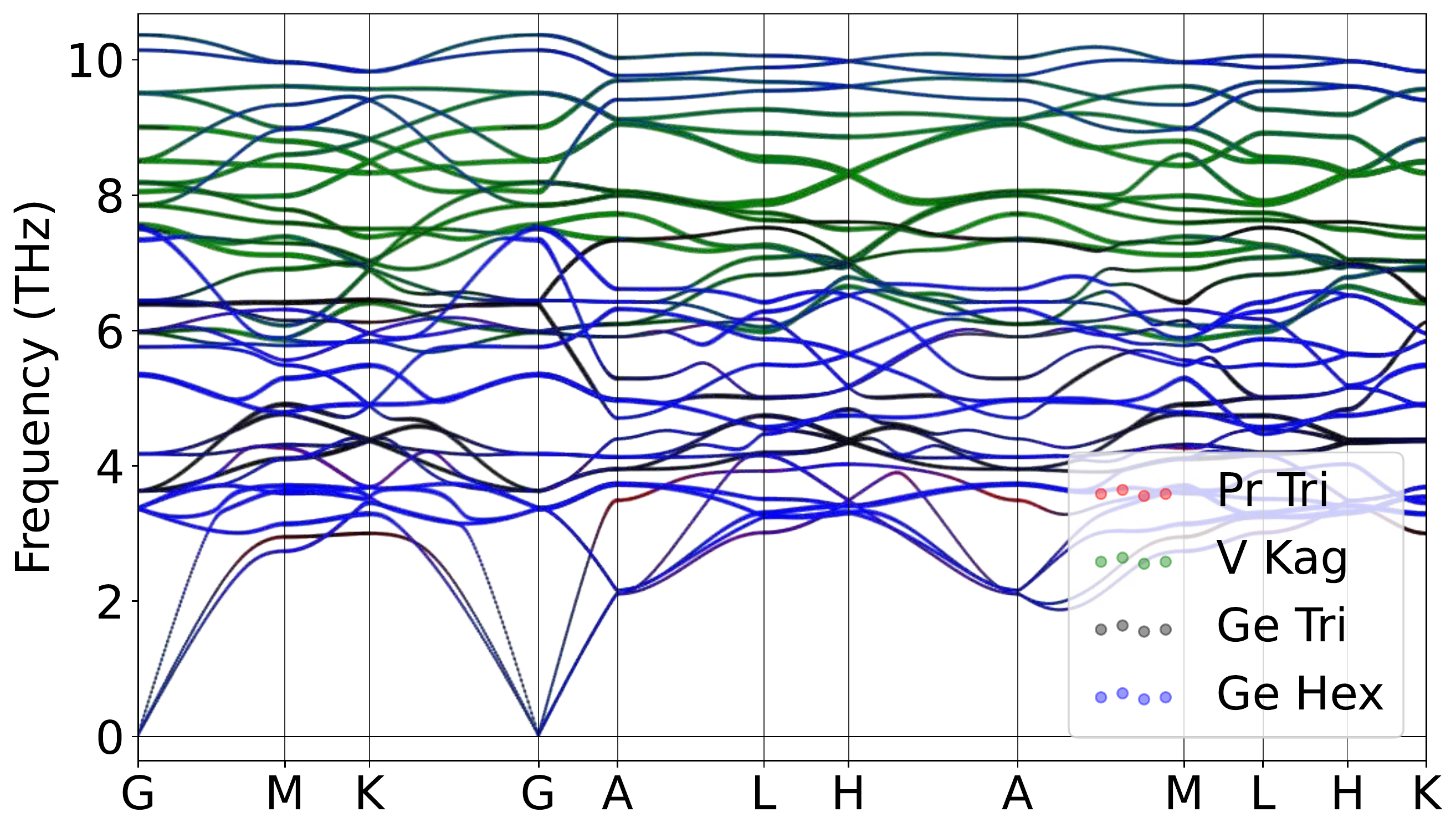}
}
\caption{Atomically projected phonon spectrum for PrV$_6$Ge$_6$.}
\label{fig:PrV6Ge6pho}
\end{figure}

\begin{figure}[H]
\centering
\label{fig:PrV6Ge6_relaxed_Low_xz}
\includegraphics[width=0.85\textwidth]{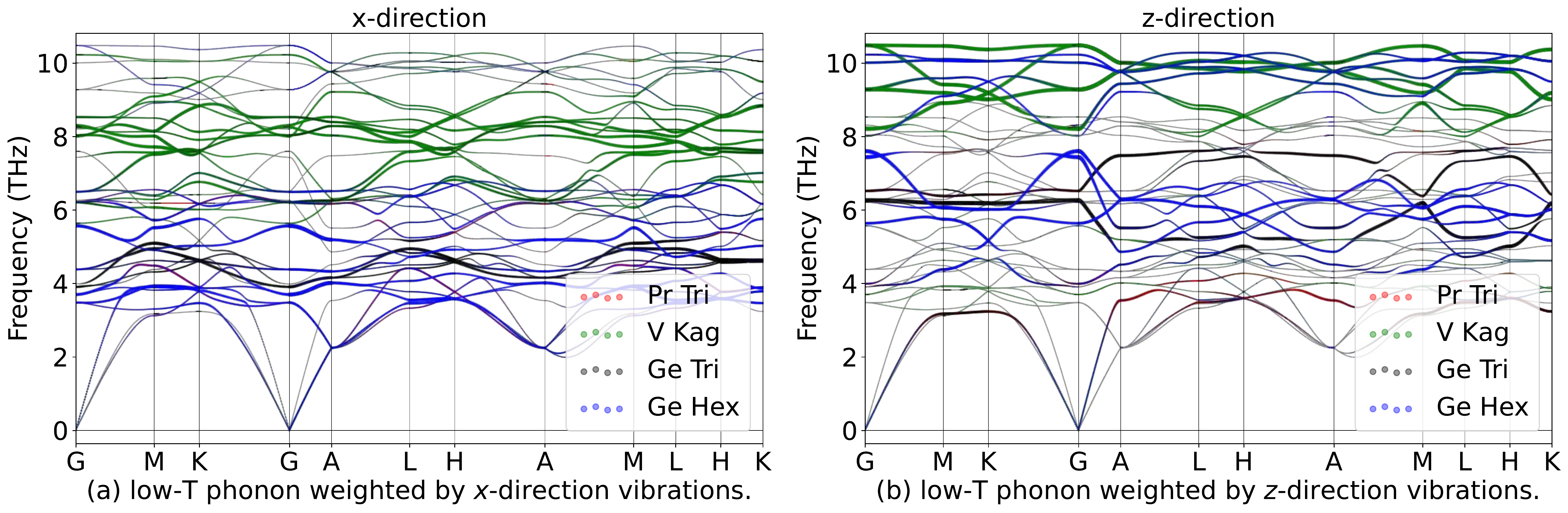}
\hspace{5pt}\label{fig:PrV6Ge6_relaxed_High_xz}
\includegraphics[width=0.85\textwidth]{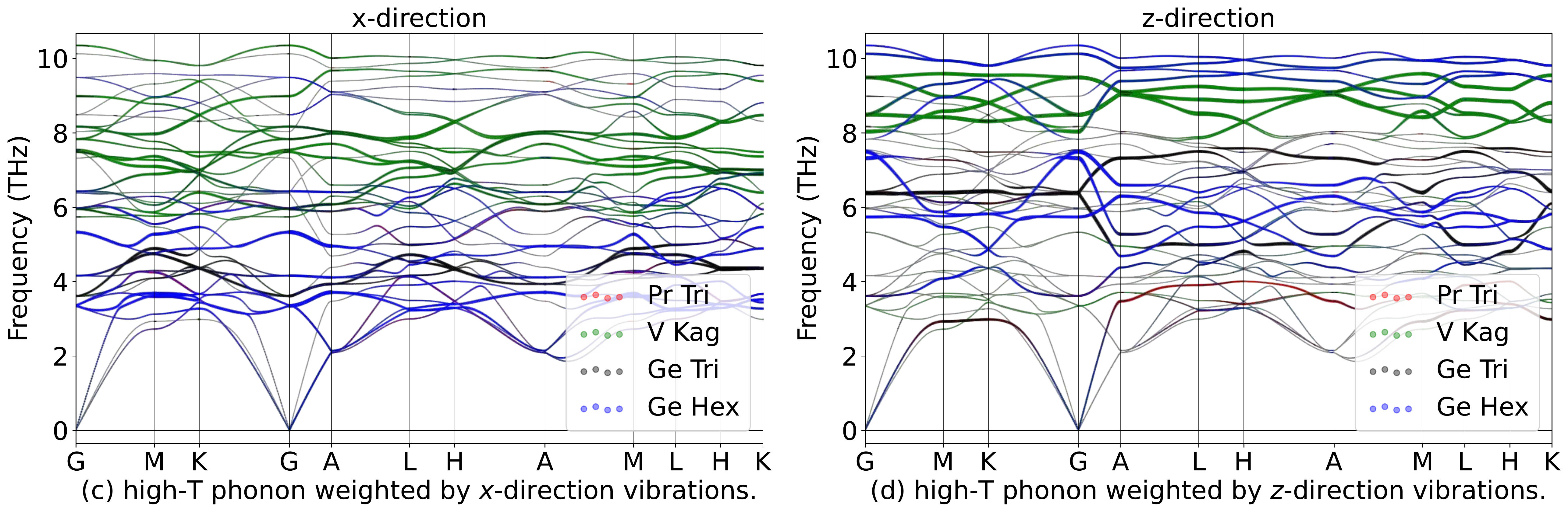}
\caption{Phonon spectrum for PrV$_6$Ge$_6$in in-plane ($x$) and out-of-plane ($z$) directions.}
\label{fig:PrV6Ge6phoxyz}
\end{figure}

\begin{table}[htbp]
\global\long\def\arraystretch{1.12}
\captionsetup{width=.95\textwidth}
\caption{IRREPs at high-symmetry points for PrV$_6$Ge$_6$ phonon spectrum at Low-T.}
\label{tab:191_PrV6Ge6_Low}
\setlength{\tabcolsep}{1.mm}{
\begin{tabular}{c|cc|cc|cc|cc|cc|cc}
\hline
\hline
 \multirow{2}{*}{Band}  & \multicolumn{2}{c}{$\Gamma$ $(0,0,0)$}  & \multicolumn{2}{c}{$A$ $(0,0,\frac{1}{2})$}  & \multicolumn{2}{c}{$M$ $(\frac{1}{2},0,0)$}  & \multicolumn{2}{c}{$L$ $(\frac{1}{2},0,\frac{1}{2})$}  & \multicolumn{2}{c}{$K$ $(\frac{1}{3},\frac{1}{3},0)$}  & \multicolumn{2}{c}{$H$ $(\frac{1}{3},\frac{1}{3},\frac{1}{2})$}   \\
\cline{2-13}
 & $f$ (THz) & Reps & $f$ (THz) & Reps & $f$ (THz) & Reps & $f$ (THz) & Reps & $f$ (THz) & Reps & $f$ (THz) & Reps  \\
\hline
1 & \multirow{3}{*}{   0.000} & \multirow{3}{*}{$-$$|$$-$} & \multirow{2}{*}{   2.225} & \multirow{2}{*}{$E_{1u}$$|$${A_5^-}$} &    3.113 & $B_{2u}$$|$${M_2^-}$ &    3.317 & $B_{2u}$$|$${L_2^-}$ &    3.223 & $A_{2}''$$|$${K_4}$ &    3.557 & $A_{1}''$$|$${H_3}$ \\
\cline{1-1}\cline{6-7}\cline{8-9}\cline{10-11}\cline{12-13}
2 &   &   &   &   &    3.165 & $B_{1u}$$|$${M_3^-}$ &    3.433 & $A_{u}$$|$${L_1^-}$ &    3.453 & $A_{2}'$$|$${K_2}$ & \multirow{2}{*}{   3.583} & \multirow{2}{*}{$E''$$|$${H_5}$} \\
\cline{1-1}\cline{4-5}\cline{6-7}\cline{8-9}\cline{10-11}
3 &   &   & \multirow{2}{*}{   2.244} & \multirow{2}{*}{$E_{1g}$$|$${A_5^+}$} &    3.293 & $B_{1g}$$|$${M_3^+}$ &    3.476 & $B_{1u}$$|$${L_3^-}$ & \multirow{2}{*}{   3.771} & \multirow{2}{*}{$E'$$|$${K_6}$} &   &   \\
\cline{1-1}\cline{2-3}\cline{6-7}\cline{8-9}\cline{12-13}
4 & \multirow{2}{*}{   3.467} & \multirow{2}{*}{$E_{1u}$$|$${\Gamma_5^-}$} &   &   &    3.805 & $B_{2u}$$|$${M_2^-}$ &    3.539 & $B_{2g}$$|$${L_2^+}$ &   &   &    3.589 & $A_{2}''$$|$${H_4}$ \\
\cline{1-1}\cline{4-5}\cline{6-7}\cline{8-9}\cline{10-11}\cline{12-13}
5 &   &   &    3.524 & $A_{2u}$$|$${A_2^-}$ &    3.898 & $B_{3u}$$|$${M_4^-}$ &    3.763 & $B_{3g}$$|$${L_4^+}$ &    3.849 & $A_{1}'$$|$${K_1}$ & \multirow{2}{*}{   3.758} & \multirow{2}{*}{$E'$$|$${H_6}$} \\
\hline
\hline
\end{tabular}}
\end{table}

\begin{table}[htbp]
\global\long\def\arraystretch{1.12}
\captionsetup{width=.95\textwidth}
\caption{IRREPs at high-symmetry points for PrV$_6$Ge$_6$ phonon spectrum at High-T.}
\label{tab:191_PrV6Ge6_High}
\setlength{\tabcolsep}{1.mm}{
\begin{tabular}{c|cc|cc|cc|cc|cc|cc}
\hline
\hline
 \multirow{2}{*}{Band}  & \multicolumn{2}{c}{$\Gamma$ $(0,0,0)$}  & \multicolumn{2}{c}{$A$ $(0,0,\frac{1}{2})$}  & \multicolumn{2}{c}{$M$ $(\frac{1}{2},0,0)$}  & \multicolumn{2}{c}{$L$ $(\frac{1}{2},0,\frac{1}{2})$}  & \multicolumn{2}{c}{$K$ $(\frac{1}{3},\frac{1}{3},0)$}  & \multicolumn{2}{c}{$H$ $(\frac{1}{3},\frac{1}{3},\frac{1}{2})$}   \\
\cline{2-13}
 & $f$ (THz) & Reps & $f$ (THz) & Reps & $f$ (THz) & Reps & $f$ (THz) & Reps & $f$ (THz) & Reps & $f$ (THz) & Reps  \\
\hline
1 & \multirow{3}{*}{  -0.000} & \multirow{3}{*}{$-$$|$$-$} & \multirow{2}{*}{   2.078} & \multirow{2}{*}{$E_{1u}$$|$${A_5^-}$} &    2.711 & $B_{2u}$$|$${M_2^-}$ &    2.985 & $B_{2u}$$|$${L_2^-}$ &    2.974 & $A_{2}''$$|$${K_4}$ & \multirow{2}{*}{   3.280} & \multirow{2}{*}{$E''$$|$${H_5}$} \\
\cline{1-1}\cline{6-7}\cline{8-9}\cline{10-11}
2 &   &   &   &   &    2.922 & $B_{1u}$$|$${M_3^-}$ &    3.209 & $B_{1u}$$|$${L_3^-}$ &    3.262 & $A_{2}'$$|$${K_2}$ &   &   \\
\cline{1-1}\cline{4-5}\cline{6-7}\cline{8-9}\cline{10-11}\cline{12-13}
3 &   &   & \multirow{2}{*}{   2.136} & \multirow{2}{*}{$E_{1g}$$|$${A_5^+}$} &    3.114 & $B_{1g}$$|$${M_3^+}$ &    3.234 & $B_{2g}$$|$${L_2^+}$ &    3.428 & $A_{1}'$$|$${K_1}$ &    3.382 & $A_{1}''$$|$${H_3}$ \\
\cline{1-1}\cline{2-3}\cline{6-7}\cline{8-9}\cline{10-11}\cline{12-13}
4 & \multirow{2}{*}{   3.338} & \multirow{2}{*}{$E_{2g}$$|$${\Gamma_6^+}$} &   &   &    3.560 & $B_{2u}$$|$${M_2^-}$ &    3.277 & $A_{u}$$|$${L_1^-}$ & \multirow{2}{*}{   3.514} & \multirow{2}{*}{$E'$$|$${K_6}$} &    3.385 & $A_{2}''$$|$${H_4}$ \\
\cline{1-1}\cline{4-5}\cline{6-7}\cline{8-9}\cline{12-13}
5 &   &   &    3.462 & $A_{2u}$$|$${A_2^-}$ &    3.604 & $A_{g}$$|$${M_1^+}$ &    3.481 & $B_{3g}$$|$${L_4^+}$ &   &   & \multirow{2}{*}{   3.460} & \multirow{2}{*}{$E'$$|$${H_6}$} \\
\hline
\hline
\end{tabular}}
\end{table}
\subsubsection{\label{sms2sec:NdV6Ge6}\hyperref[smsec:col]{\texorpdfstring{NdV$_6$Ge$_6$}{}}~{\color{green}  (High-T stable, Low-T stable) }}

\begin{table}[htbp]
\global\long\def\arraystretch{1.12}
\captionsetup{width=.85\textwidth}
\caption{Structure parameters of NdV$_6$Ge$_6$ after relaxation. It contains 390 electrons. }
\label{tab:NdV6Ge6}
\setlength{\tabcolsep}{4mm}{
\begin{tabular}{c|c||c|c}
\hline
\hline
 Atom & Coordinates & \multicolumn{2}{c}{Parameters~($\AA$)} \\
\hline
Nd ~ 1a & ($0,0,0$) & $a$ & $5.1953$ \\
V ~ 6i & ($\frac{1}{2},0,0.2505$) & $c$ & $8.3599$ \\
Ge1 ~ 2c & ($\frac{1}{3},\frac{2}{3},0$) & $d^{Tri}_{Z-Z}$ & $2.6107$ \\
Ge2 ~ 2d & ($\frac{1}{3},\frac{2}{3},\frac{1}{2}$) & $d_{T_{1}}$ & $2.5976$ \\
Ge3 ~ 2e & ($0,0,0.3439$) & $d_{T_{2}}$ & $4.1723$ \\
\hline
\hline
\end{tabular}}
\end{table}

\begin{figure}[htbp]
\centering
\includegraphics[width=0.85\textwidth]{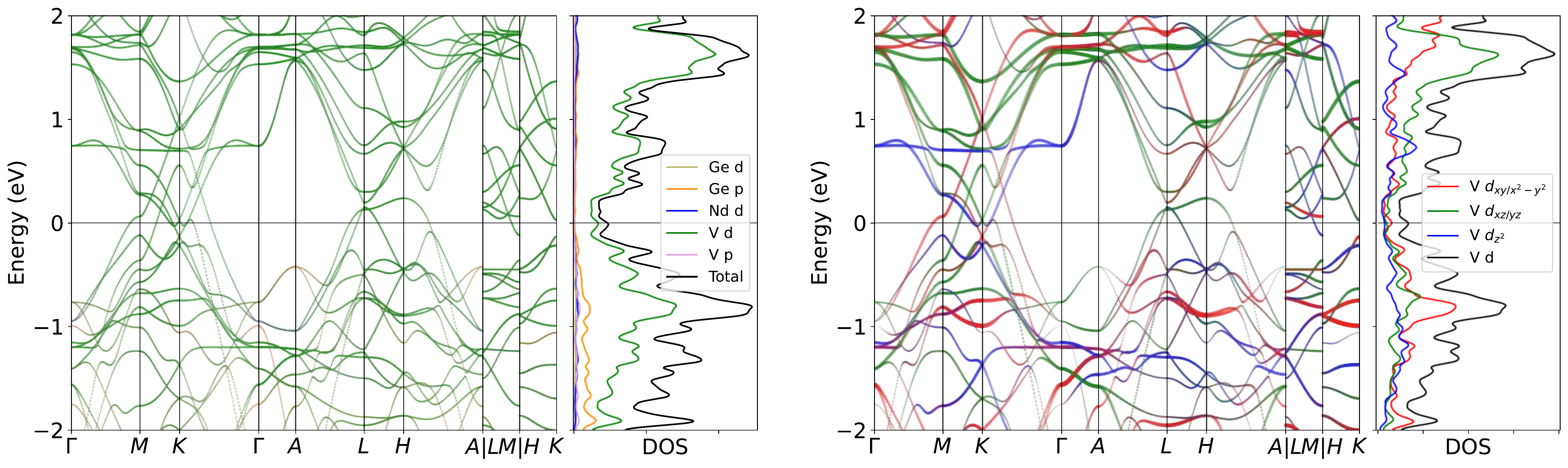}
\caption{Band structure for NdV$_6$Ge$_6$ without SOC. The projection of V $3d$ orbitals is also presented. The band structures clearly reveal the dominating of V $3d$ orbitals  near the Fermi level, as well as the pronounced peak.}
\label{fig:NdV6Ge6band}
\end{figure}

\begin{figure}[H]
\centering
\subfloat[Relaxed phonon at low-T.]{
\label{fig:NdV6Ge6_relaxed_Low}
\includegraphics[width=0.45\textwidth]{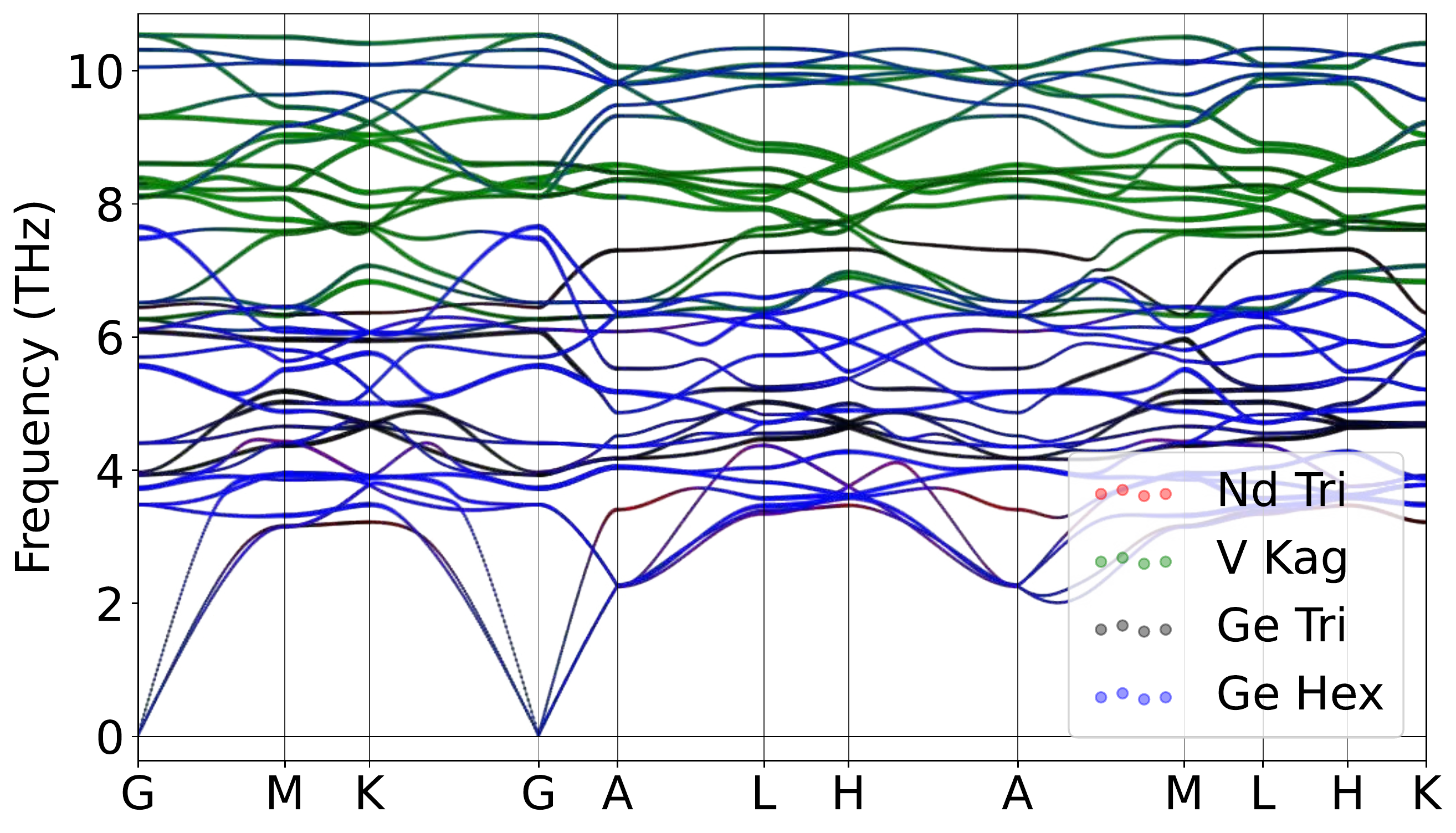}
}
\hspace{5pt}\subfloat[Relaxed phonon at high-T.]{
\label{fig:NdV6Ge6_relaxed_High}
\includegraphics[width=0.45\textwidth]{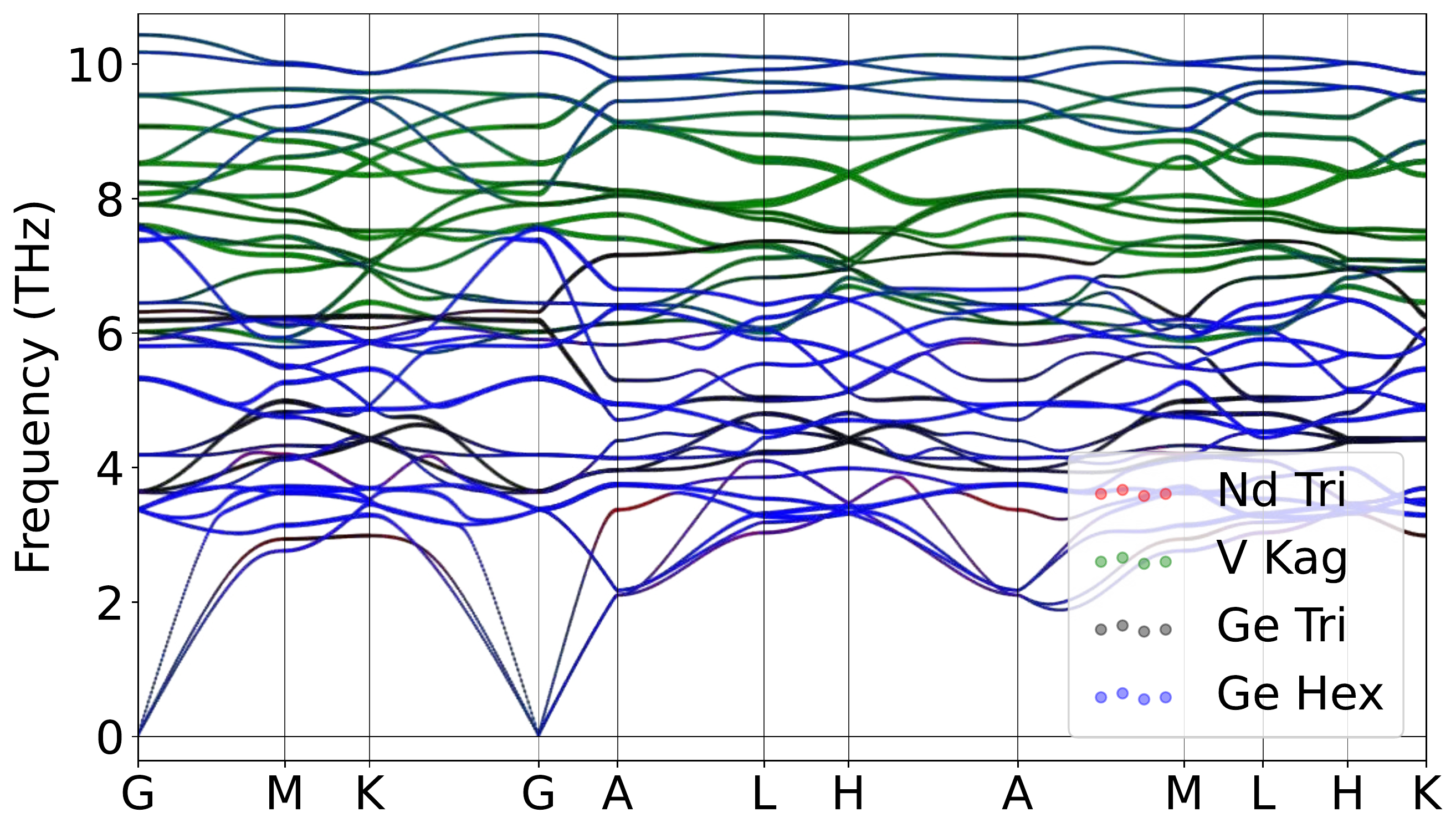}
}
\caption{Atomically projected phonon spectrum for NdV$_6$Ge$_6$.}
\label{fig:NdV6Ge6pho}
\end{figure}

\begin{figure}[H]
\centering
\label{fig:NdV6Ge6_relaxed_Low_xz}
\includegraphics[width=0.85\textwidth]{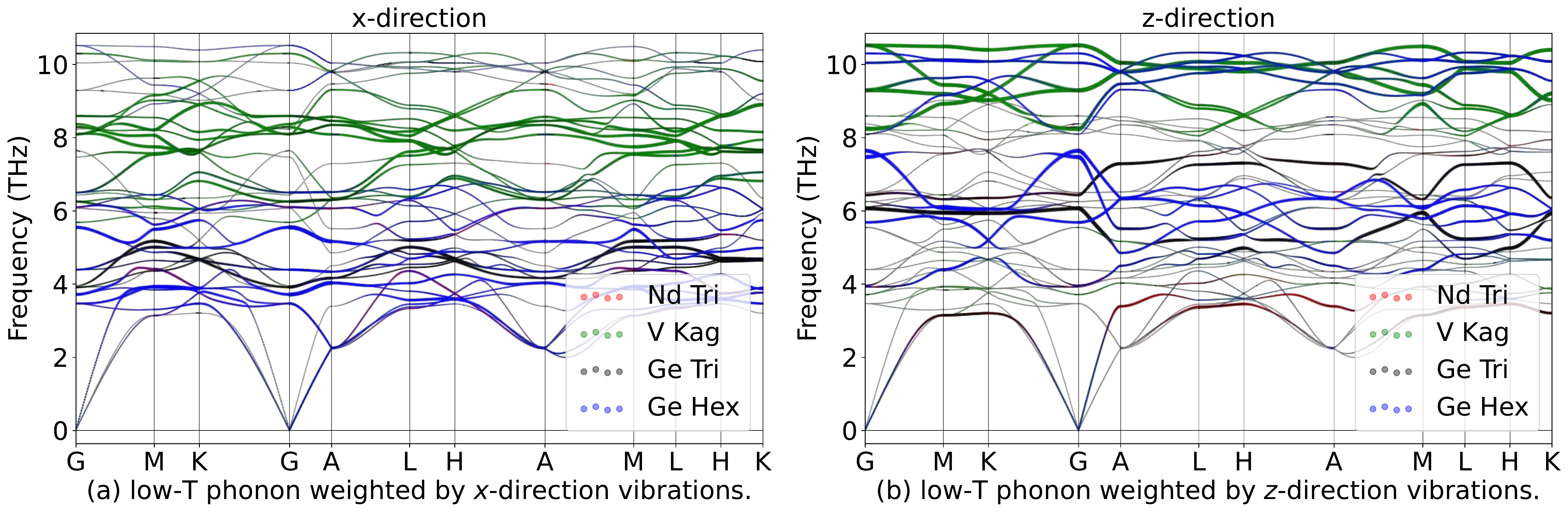}
\hspace{5pt}\label{fig:NdV6Ge6_relaxed_High_xz}
\includegraphics[width=0.85\textwidth]{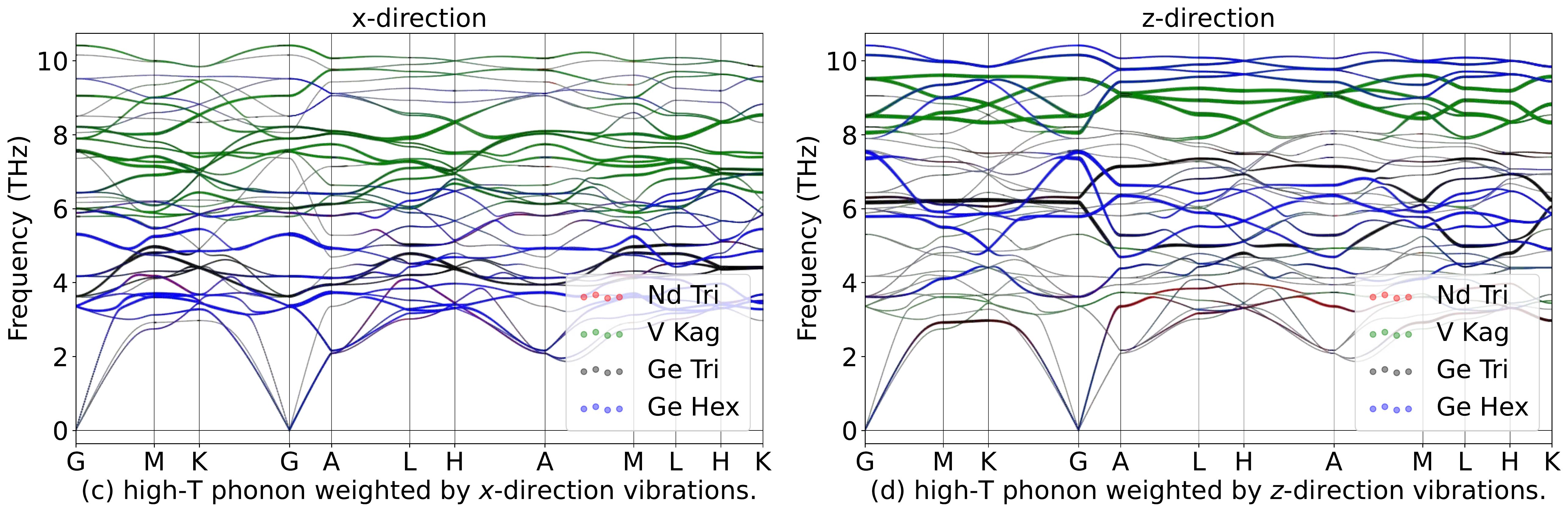}
\caption{Phonon spectrum for NdV$_6$Ge$_6$in in-plane ($x$) and out-of-plane ($z$) directions.}
\label{fig:NdV6Ge6phoxyz}
\end{figure}

\begin{table}[htbp]
\global\long\def\arraystretch{1.12}
\captionsetup{width=.95\textwidth}
\caption{IRREPs at high-symmetry points for NdV$_6$Ge$_6$ phonon spectrum at Low-T.}
\label{tab:191_NdV6Ge6_Low}
\setlength{\tabcolsep}{1.mm}{
\begin{tabular}{c|cc|cc|cc|cc|cc|cc}
\hline
\hline
 \multirow{2}{*}{Band}  & \multicolumn{2}{c}{$\Gamma$ $(0,0,0)$}  & \multicolumn{2}{c}{$A$ $(0,0,\frac{1}{2})$}  & \multicolumn{2}{c}{$M$ $(\frac{1}{2},0,0)$}  & \multicolumn{2}{c}{$L$ $(\frac{1}{2},0,\frac{1}{2})$}  & \multicolumn{2}{c}{$K$ $(\frac{1}{3},\frac{1}{3},0)$}  & \multicolumn{2}{c}{$H$ $(\frac{1}{3},\frac{1}{3},\frac{1}{2})$}   \\
\cline{2-13}
 & $f$ (THz) & Reps & $f$ (THz) & Reps & $f$ (THz) & Reps & $f$ (THz) & Reps & $f$ (THz) & Reps & $f$ (THz) & Reps  \\
\hline
1 & \multirow{3}{*}{  -0.000} & \multirow{3}{*}{$-$$|$$-$} & \multirow{2}{*}{   2.222} & \multirow{2}{*}{$E_{1u}$$|$${A_5^-}$} &    3.121 & $B_{2u}$$|$${M_2^-}$ &    3.318 & $B_{2u}$$|$${L_2^-}$ &    3.192 & $A_{2}''$$|$${K_4}$ &    3.444 & $A_{2}''$$|$${H_4}$ \\
\cline{1-1}\cline{6-7}\cline{8-9}\cline{10-11}\cline{12-13}
2 &   &   &   &   &    3.136 & $B_{1u}$$|$${M_3^-}$ &    3.365 & $B_{1u}$$|$${L_3^-}$ &    3.455 & $A_{2}'$$|$${K_2}$ &    3.562 & $A_{1}''$$|$${H_3}$ \\
\cline{1-1}\cline{4-5}\cline{6-7}\cline{8-9}\cline{10-11}\cline{12-13}
3 &   &   & \multirow{2}{*}{   2.251} & \multirow{2}{*}{$E_{1g}$$|$${A_5^+}$} &    3.292 & $B_{1g}$$|$${M_3^+}$ &    3.436 & $A_{u}$$|$${L_1^-}$ & \multirow{2}{*}{   3.751} & \multirow{2}{*}{$E'$$|$${K_6}$} & \multirow{2}{*}{   3.593} & \multirow{2}{*}{$E''$$|$${H_5}$} \\
\cline{1-1}\cline{2-3}\cline{6-7}\cline{8-9}
4 & \multirow{2}{*}{   3.456} & \multirow{2}{*}{$E_{1u}$$|$${\Gamma_5^-}$} &   &   &    3.827 & $B_{2u}$$|$${M_2^-}$ &    3.548 & $B_{2g}$$|$${L_2^+}$ &   &   &   &   \\
\cline{1-1}\cline{4-5}\cline{6-7}\cline{8-9}\cline{10-11}\cline{12-13}
5 &   &   &    3.379 & $A_{2u}$$|$${A_2^-}$ &    3.905 & $B_{3u}$$|$${M_4^-}$ &    3.792 & $B_{3g}$$|$${L_4^+}$ &    3.851 & $A_{1}'$$|$${K_1}$ & \multirow{2}{*}{   3.729} & \multirow{2}{*}{$E'$$|$${H_6}$} \\
\hline
\hline
\end{tabular}}
\end{table}

\begin{table}[htbp]
\global\long\def\arraystretch{1.12}
\captionsetup{width=.95\textwidth}
\caption{IRREPs at high-symmetry points for NdV$_6$Ge$_6$ phonon spectrum at High-T.}
\label{tab:191_NdV6Ge6_High}
\setlength{\tabcolsep}{1.mm}{
\begin{tabular}{c|cc|cc|cc|cc|cc|cc}
\hline
\hline
 \multirow{2}{*}{Band}  & \multicolumn{2}{c}{$\Gamma$ $(0,0,0)$}  & \multicolumn{2}{c}{$A$ $(0,0,\frac{1}{2})$}  & \multicolumn{2}{c}{$M$ $(\frac{1}{2},0,0)$}  & \multicolumn{2}{c}{$L$ $(\frac{1}{2},0,\frac{1}{2})$}  & \multicolumn{2}{c}{$K$ $(\frac{1}{3},\frac{1}{3},0)$}  & \multicolumn{2}{c}{$H$ $(\frac{1}{3},\frac{1}{3},\frac{1}{2})$}   \\
\cline{2-13}
 & $f$ (THz) & Reps & $f$ (THz) & Reps & $f$ (THz) & Reps & $f$ (THz) & Reps & $f$ (THz) & Reps & $f$ (THz) & Reps  \\
\hline
1 & \multirow{3}{*}{  -0.000} & \multirow{3}{*}{$-$$|$$-$} & \multirow{2}{*}{   2.075} & \multirow{2}{*}{$E_{1u}$$|$${A_5^-}$} &    2.734 & $B_{2u}$$|$${M_2^-}$ &    3.003 & $B_{2u}$$|$${L_2^-}$ &    2.959 & $A_{2}''$$|$${K_4}$ & \multirow{2}{*}{   3.294} & \multirow{2}{*}{$E''$$|$${H_5}$} \\
\cline{1-1}\cline{6-7}\cline{8-9}\cline{10-11}
2 &   &   &   &   &    2.909 & $B_{1u}$$|$${M_3^-}$ &    3.156 & $B_{1u}$$|$${L_3^-}$ &    3.265 & $A_{2}'$$|$${K_2}$ &   &   \\
\cline{1-1}\cline{4-5}\cline{6-7}\cline{8-9}\cline{10-11}\cline{12-13}
3 &   &   & \multirow{2}{*}{   2.150} & \multirow{2}{*}{$E_{1g}$$|$${A_5^+}$} &    3.114 & $B_{1g}$$|$${M_3^+}$ &    3.247 & $B_{2g}$$|$${L_2^+}$ &    3.428 & $A_{1}'$$|$${K_1}$ &    3.302 & $A_{2}''$$|$${H_4}$ \\
\cline{1-1}\cline{2-3}\cline{6-7}\cline{8-9}\cline{10-11}\cline{12-13}
4 & \multirow{2}{*}{   3.343} & \multirow{2}{*}{$E_{1u}$$|$${\Gamma_5^-}$} &   &   &    3.583 & $B_{2u}$$|$${M_2^-}$ &    3.283 & $A_{u}$$|$${L_1^-}$ & \multirow{2}{*}{   3.493} & \multirow{2}{*}{$E'$$|$${K_6}$} &    3.389 & $A_{1}''$$|$${H_3}$ \\
\cline{1-1}\cline{4-5}\cline{6-7}\cline{8-9}\cline{12-13}
5 &   &   &    3.343 & $A_{2u}$$|$${A_2^-}$ &    3.607 & $A_{g}$$|$${M_1^+}$ &    3.508 & $B_{3g}$$|$${L_4^+}$ &   &   & \multirow{2}{*}{   3.437} & \multirow{2}{*}{$E'$$|$${H_6}$} \\
\hline
\hline
\end{tabular}}
\end{table}
\subsubsection{\label{sms2sec:SmV6Ge6}\hyperref[smsec:col]{\texorpdfstring{SmV$_6$Ge$_6$}{}}~{\color{green}  (High-T stable, Low-T stable) }}

\begin{table}[htbp]
\global\long\def\arraystretch{1.12}
\captionsetup{width=.85\textwidth}
\caption{Structure parameters of SmV$_6$Ge$_6$ after relaxation. It contains 392 electrons. }
\label{tab:SmV6Ge6}
\setlength{\tabcolsep}{4mm}{
\begin{tabular}{c|c||c|c}
\hline
\hline
 Atom & Coordinates & \multicolumn{2}{c}{Parameters~($\AA$)} \\
\hline
Sm ~ 1a & ($0,0,0$) & $a$ & $5.1868$ \\
V ~ 6i & ($\frac{1}{2},0,0.2506$) & $c$ & $8.3644$ \\
Ge1 ~ 2c & ($\frac{1}{3},\frac{2}{3},0$) & $d^{Tri}_{Z-Z}$ & $2.6443$ \\
Ge2 ~ 2d & ($\frac{1}{3},\frac{2}{3},\frac{1}{2}$) & $d_{T_{1}}$ & $2.5934$ \\
Ge3 ~ 2e & ($0,0,0.3419$) & $d_{T_{2}}$ & $4.1717$ \\
\hline
\hline
\end{tabular}}
\end{table}

\begin{figure}[htbp]
\centering
\includegraphics[width=0.85\textwidth]{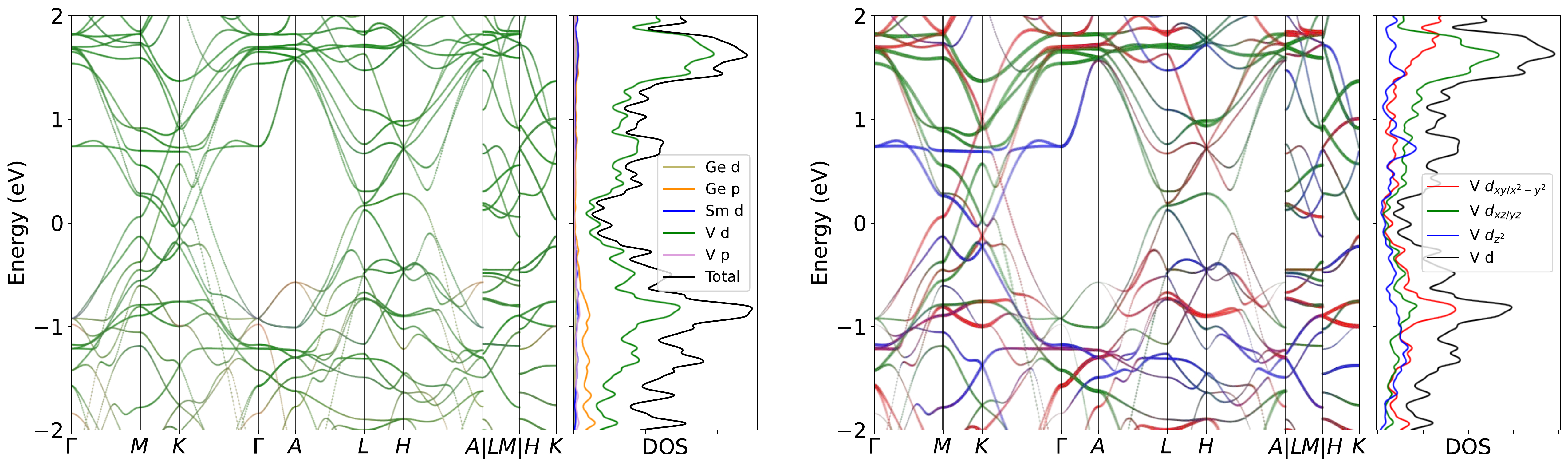}
\caption{Band structure for SmV$_6$Ge$_6$ without SOC. The projection of V $3d$ orbitals is also presented. The band structures clearly reveal the dominating of V $3d$ orbitals  near the Fermi level, as well as the pronounced peak.}
\label{fig:SmV6Ge6band}
\end{figure}

\begin{figure}[H]
\centering
\subfloat[Relaxed phonon at low-T.]{
\label{fig:SmV6Ge6_relaxed_Low}
\includegraphics[width=0.45\textwidth]{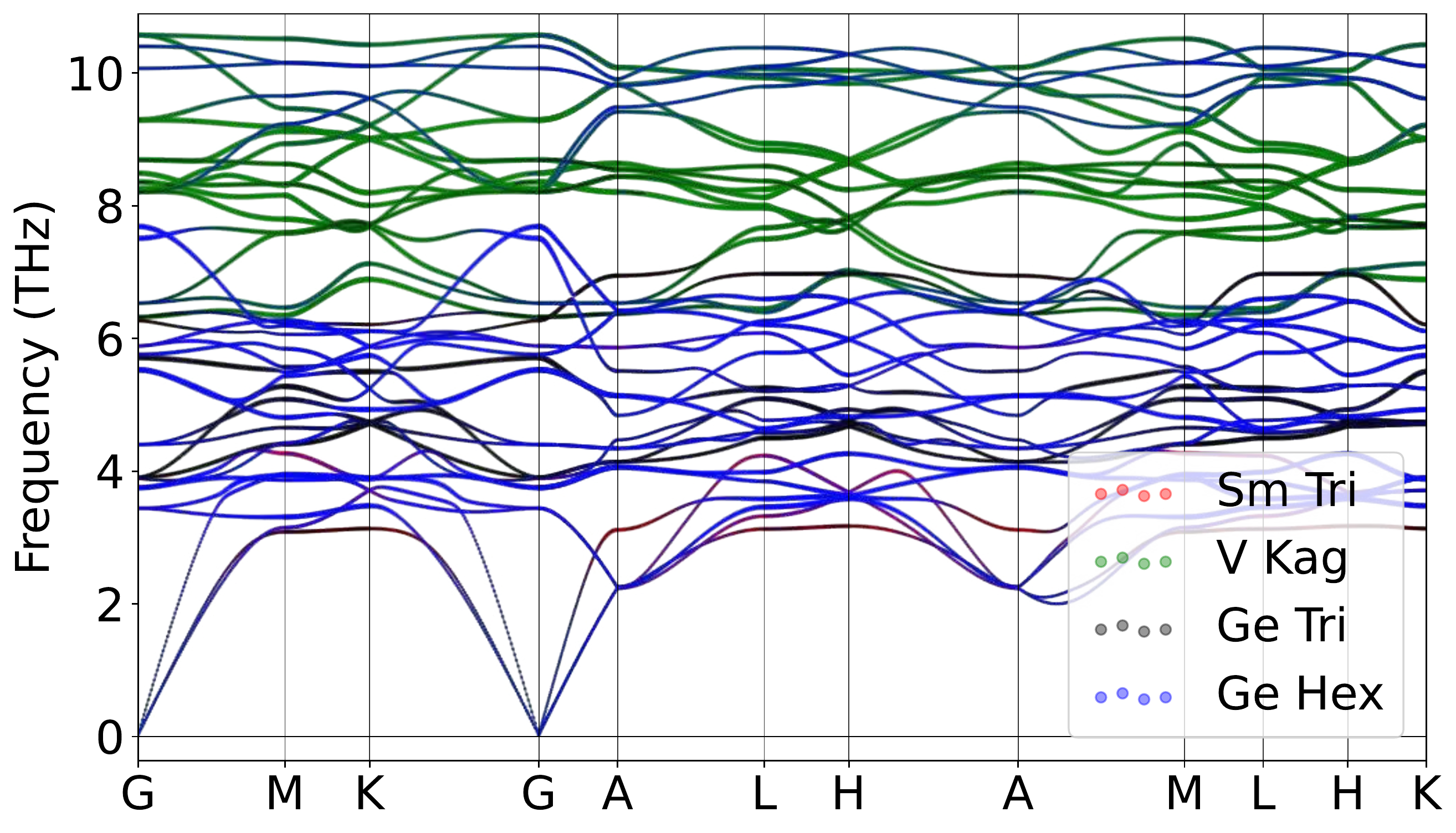}
}
\hspace{5pt}\subfloat[Relaxed phonon at high-T.]{
\label{fig:SmV6Ge6_relaxed_High}
\includegraphics[width=0.45\textwidth]{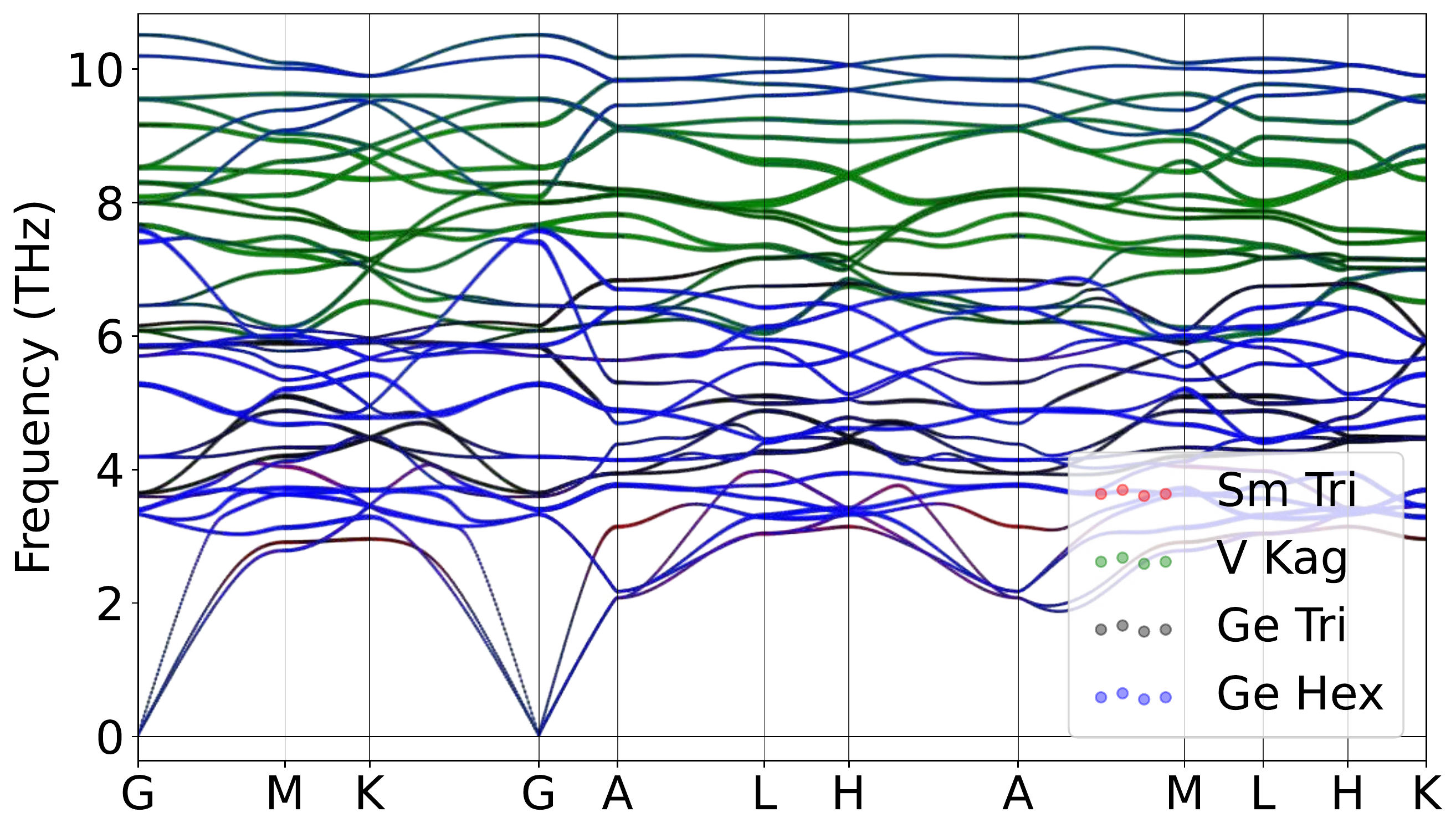}
}
\caption{Atomically projected phonon spectrum for SmV$_6$Ge$_6$.}
\label{fig:SmV6Ge6pho}
\end{figure}

\begin{figure}[H]
\centering
\label{fig:SmV6Ge6_relaxed_Low_xz}
\includegraphics[width=0.85\textwidth]{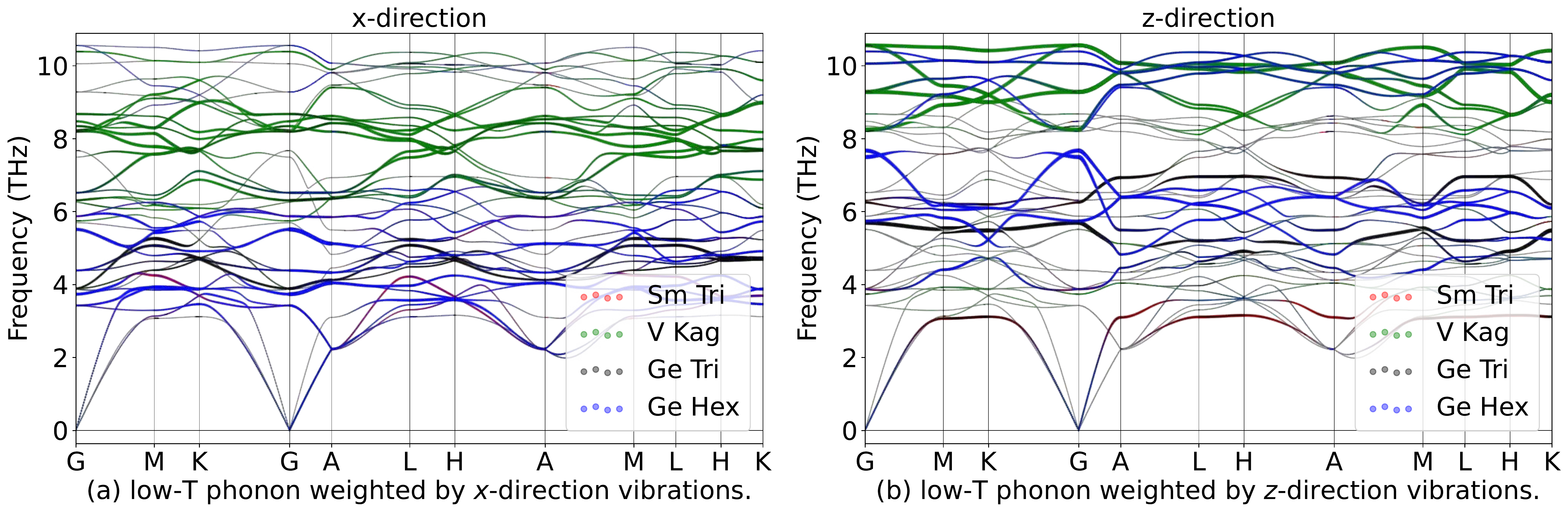}
\hspace{5pt}\label{fig:SmV6Ge6_relaxed_High_xz}
\includegraphics[width=0.85\textwidth]{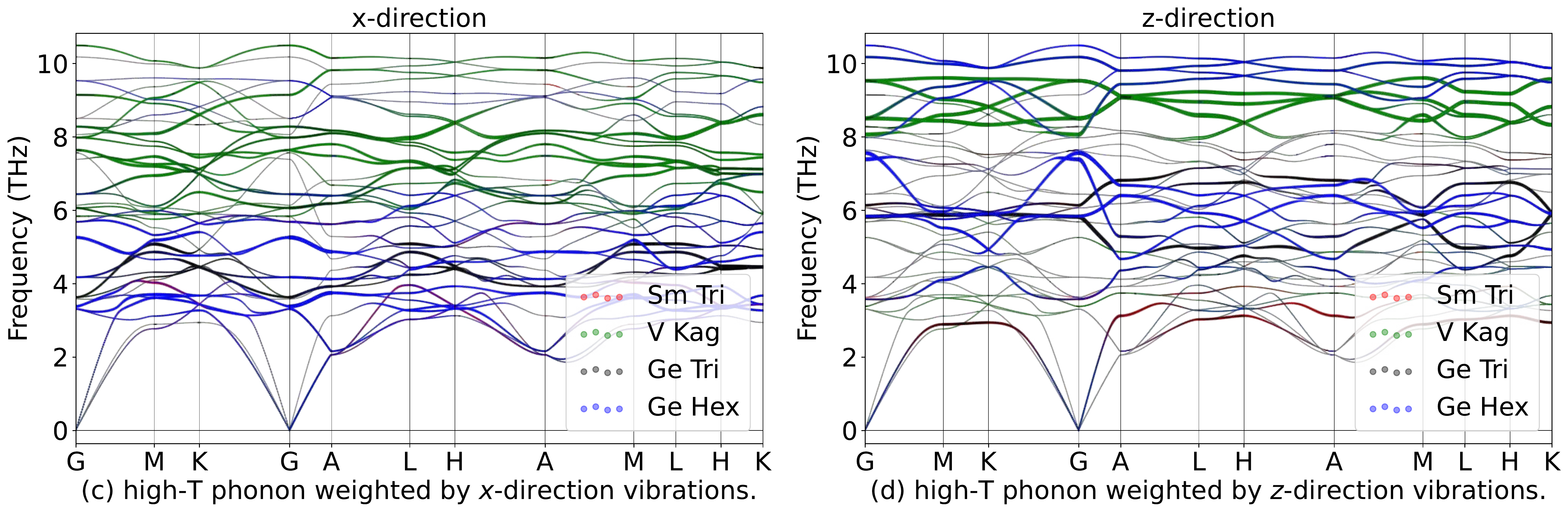}
\caption{Phonon spectrum for SmV$_6$Ge$_6$in in-plane ($x$) and out-of-plane ($z$) directions.}
\label{fig:SmV6Ge6phoxyz}
\end{figure}

\begin{table}[htbp]
\global\long\def\arraystretch{1.12}
\captionsetup{width=.95\textwidth}
\caption{IRREPs at high-symmetry points for SmV$_6$Ge$_6$ phonon spectrum at Low-T.}
\label{tab:191_SmV6Ge6_Low}
\setlength{\tabcolsep}{1.mm}{
\begin{tabular}{c|cc|cc|cc|cc|cc|cc}
\hline
\hline
 \multirow{2}{*}{Band}  & \multicolumn{2}{c}{$\Gamma$ $(0,0,0)$}  & \multicolumn{2}{c}{$A$ $(0,0,\frac{1}{2})$}  & \multicolumn{2}{c}{$M$ $(\frac{1}{2},0,0)$}  & \multicolumn{2}{c}{$L$ $(\frac{1}{2},0,\frac{1}{2})$}  & \multicolumn{2}{c}{$K$ $(\frac{1}{3},\frac{1}{3},0)$}  & \multicolumn{2}{c}{$H$ $(\frac{1}{3},\frac{1}{3},\frac{1}{2})$}   \\
\cline{2-13}
 & $f$ (THz) & Reps & $f$ (THz) & Reps & $f$ (THz) & Reps & $f$ (THz) & Reps & $f$ (THz) & Reps & $f$ (THz) & Reps  \\
\hline
1 & \multirow{3}{*}{  -0.000} & \multirow{3}{*}{$-$$|$$-$} & \multirow{2}{*}{   2.203} & \multirow{2}{*}{$E_{1u}$$|$${A_5^-}$} &    3.057 & $B_{1u}$$|$${M_3^-}$ &    3.100 & $B_{1u}$$|$${L_3^-}$ &    3.107 & $A_{2}''$$|$${K_4}$ &    3.145 & $A_{2}''$$|$${H_4}$ \\
\cline{1-1}\cline{6-7}\cline{8-9}\cline{10-11}\cline{12-13}
2 &   &   &   &   &    3.121 & $B_{2u}$$|$${M_2^-}$ &    3.295 & $B_{2u}$$|$${L_2^-}$ &    3.450 & $A_{2}'$$|$${K_2}$ &    3.559 & $A_{1}''$$|$${H_3}$ \\
\cline{1-1}\cline{4-5}\cline{6-7}\cline{8-9}\cline{10-11}\cline{12-13}
3 &   &   & \multirow{2}{*}{   2.231} & \multirow{2}{*}{$E_{1g}$$|$${A_5^+}$} &    3.281 & $B_{1g}$$|$${M_3^+}$ &    3.432 & $A_{u}$$|$${L_1^-}$ & \multirow{2}{*}{   3.682} & \multirow{2}{*}{$E'$$|$${K_6}$} & \multirow{2}{*}{   3.602} & \multirow{2}{*}{$E''$$|$${H_5}$} \\
\cline{1-1}\cline{2-3}\cline{6-7}\cline{8-9}
4 & \multirow{2}{*}{   3.413} & \multirow{2}{*}{$E_{1u}$$|$${\Gamma_5^-}$} &   &   &    3.835 & $B_{2u}$$|$${M_2^-}$ &    3.554 & $B_{2g}$$|$${L_2^+}$ &   &   &   &   \\
\cline{1-1}\cline{4-5}\cline{6-7}\cline{8-9}\cline{10-11}\cline{12-13}
5 &   &   &    3.088 & $A_{2u}$$|$${A_2^-}$ &    3.906 & $B_{3u}$$|$${M_4^-}$ &    3.818 & $B_{3g}$$|$${L_4^+}$ &    3.851 & $A_{1}'$$|$${K_1}$ & \multirow{2}{*}{   3.652} & \multirow{2}{*}{$E'$$|$${H_6}$} \\
\hline
\hline
\end{tabular}}
\end{table}

\begin{table}[htbp]
\global\long\def\arraystretch{1.12}
\captionsetup{width=.95\textwidth}
\caption{IRREPs at high-symmetry points for SmV$_6$Ge$_6$ phonon spectrum at High-T.}
\label{tab:191_SmV6Ge6_High}
\setlength{\tabcolsep}{1.mm}{
%
}
\end{table}
\subsubsection{\label{sms2sec:GdV6Ge6}\hyperref[smsec:col]{\texorpdfstring{GdV$_6$Ge$_6$}{}}~{\color{green}  (High-T stable, Low-T stable) }}

\begin{table}[htbp]
\global\long\def\arraystretch{1.12}
\captionsetup{width=.85\textwidth}
\caption{Structure parameters of GdV$_6$Ge$_6$ after relaxation. It contains 394 electrons. }
\label{tab:GdV6Ge6}
\setlength{\tabcolsep}{4mm}{
\begin{tabular}{c|c||c|c}
\hline
\hline
 Atom & Coordinates & \multicolumn{2}{c}{Parameters~($\AA$)} \\
\hline
Gd ~ 1a & ($0,0,0$) & $a$ & $5.1793$ \\
V ~ 6i & ($\frac{1}{2},0,0.2508$) & $c$ & $8.3698$ \\
Ge1 ~ 2c & ($\frac{1}{3},\frac{2}{3},0$) & $d^{Tri}_{Z-Z}$ & $2.6798$ \\
Ge2 ~ 2d & ($\frac{1}{3},\frac{2}{3},\frac{1}{2}$) & $d_{T_{1}}$ & $2.5897$ \\
Ge3 ~ 2e & ($0,0,0.3399$) & $d_{T_{2}}$ & $4.1719$ \\
\hline
\hline
\end{tabular}}
\end{table}

\begin{figure}[htbp]
\centering
\includegraphics[width=0.85\textwidth]{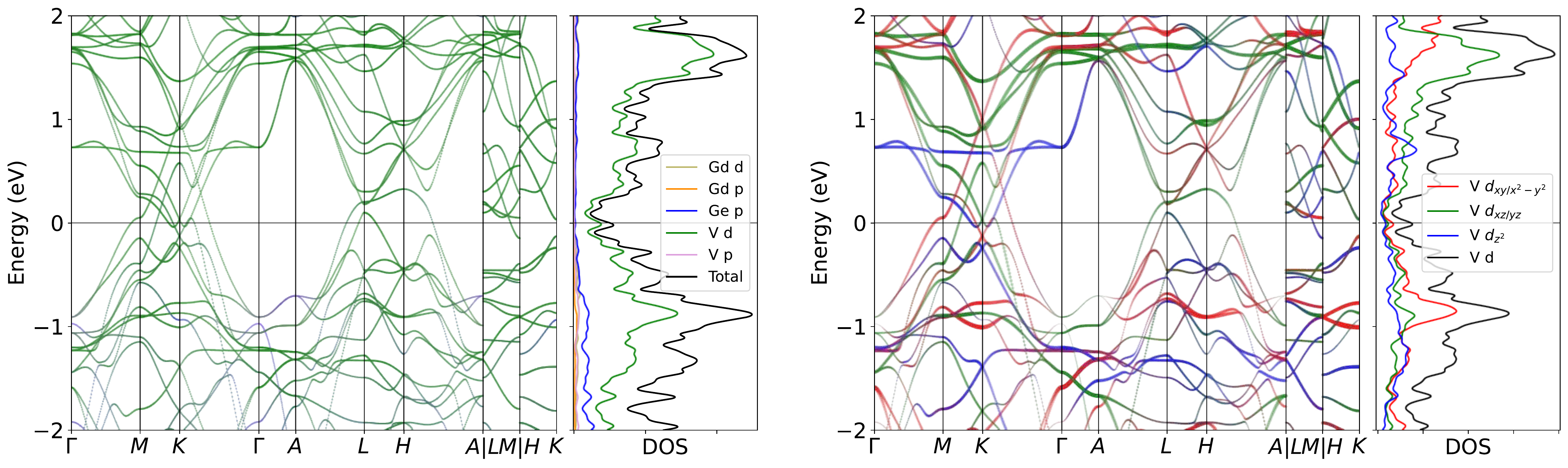}
\caption{Band structure for GdV$_6$Ge$_6$ without SOC. The projection of V $3d$ orbitals is also presented. The band structures clearly reveal the dominating of V $3d$ orbitals  near the Fermi level, as well as the pronounced peak.}
\label{fig:GdV6Ge6band}
\end{figure}

\begin{figure}[H]
\centering
\subfloat[Relaxed phonon at low-T.]{
\label{fig:GdV6Ge6_relaxed_Low}
\includegraphics[width=0.45\textwidth]{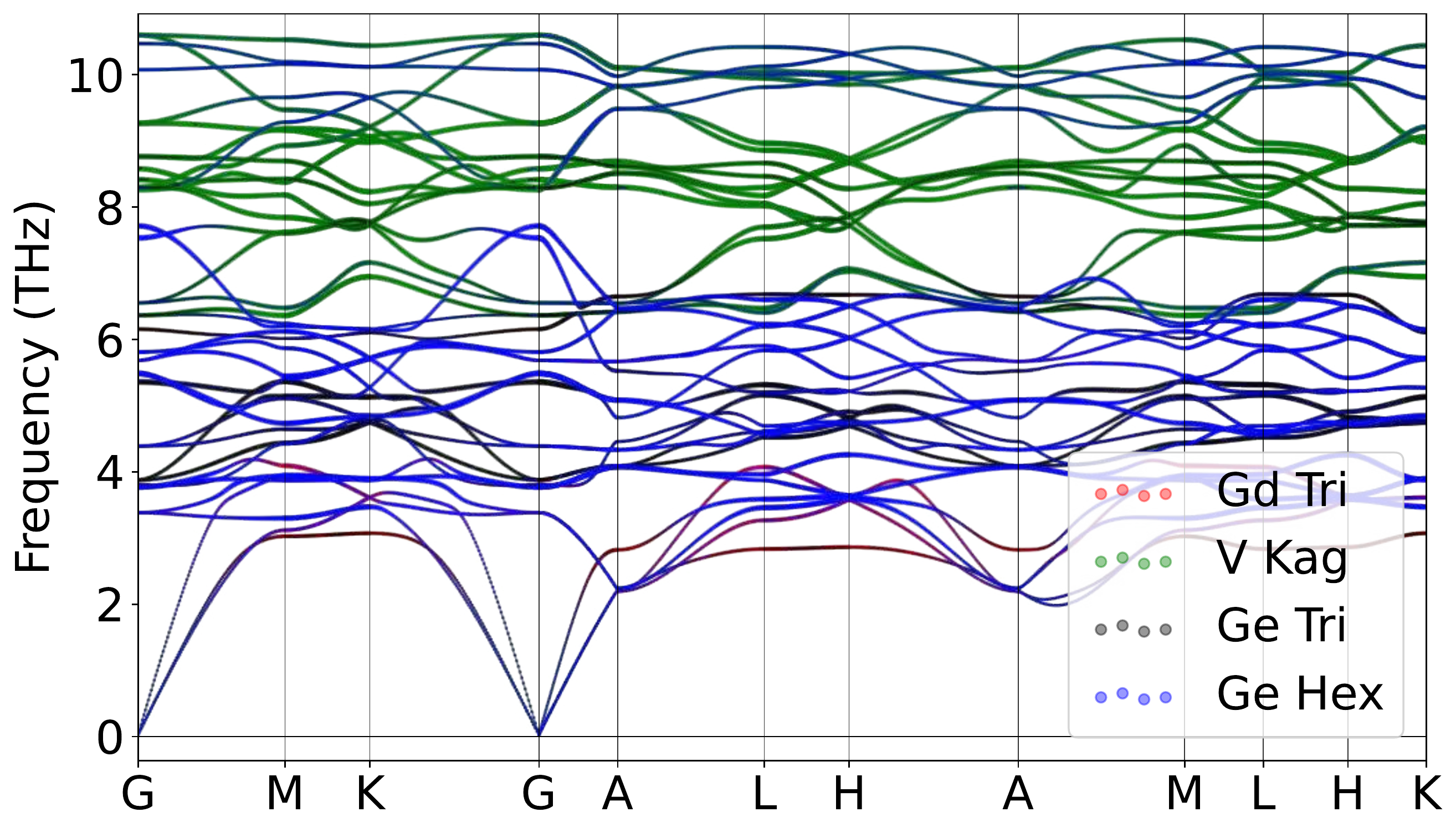}
}
\hspace{5pt}\subfloat[Relaxed phonon at high-T.]{
\label{fig:GdV6Ge6_relaxed_High}
\includegraphics[width=0.45\textwidth]{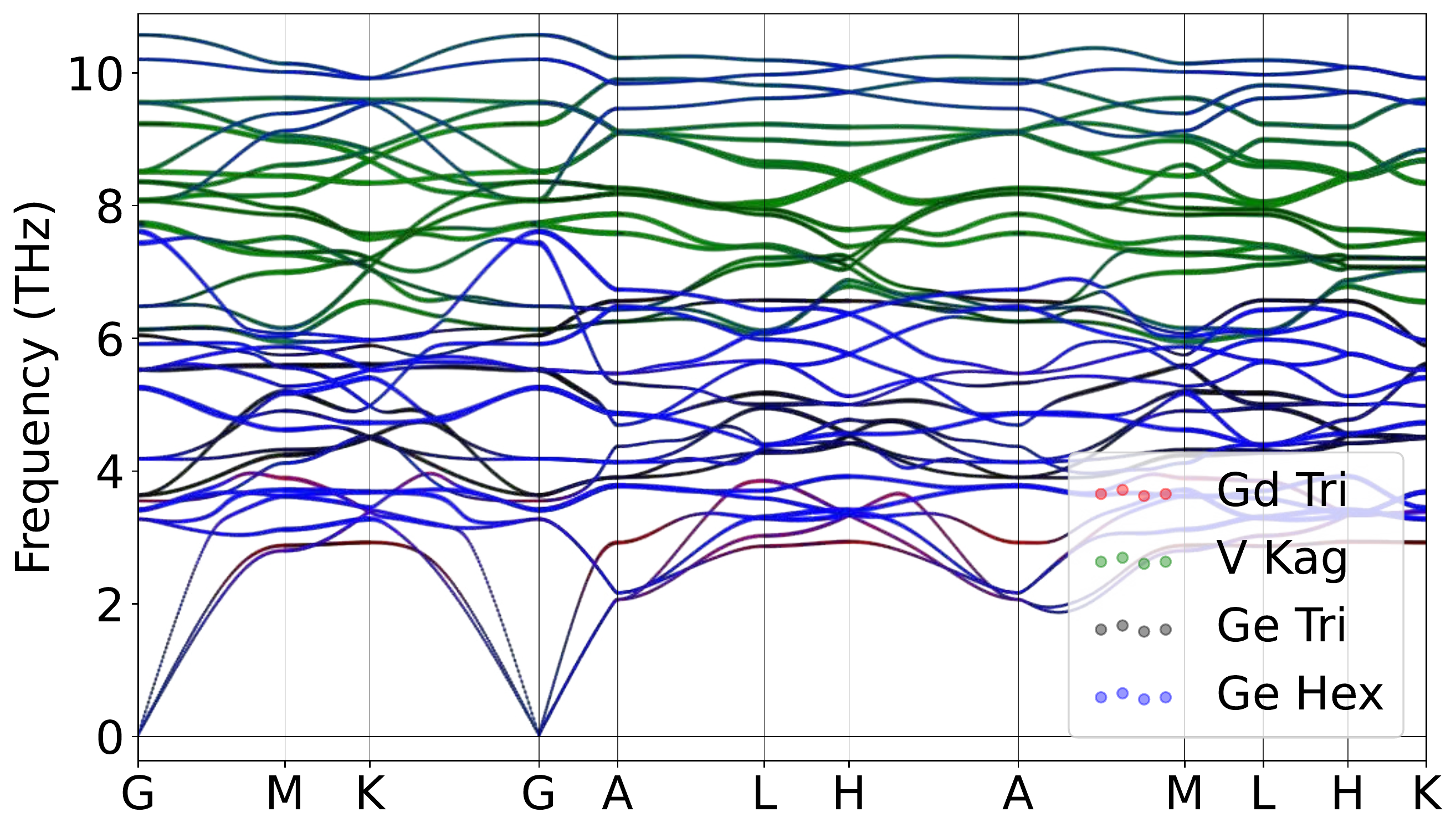}
}
\caption{Atomically projected phonon spectrum for GdV$_6$Ge$_6$.}
\label{fig:GdV6Ge6pho}
\end{figure}

\begin{figure}[H]
\centering
\label{fig:GdV6Ge6_relaxed_Low_xz}
\includegraphics[width=0.85\textwidth]{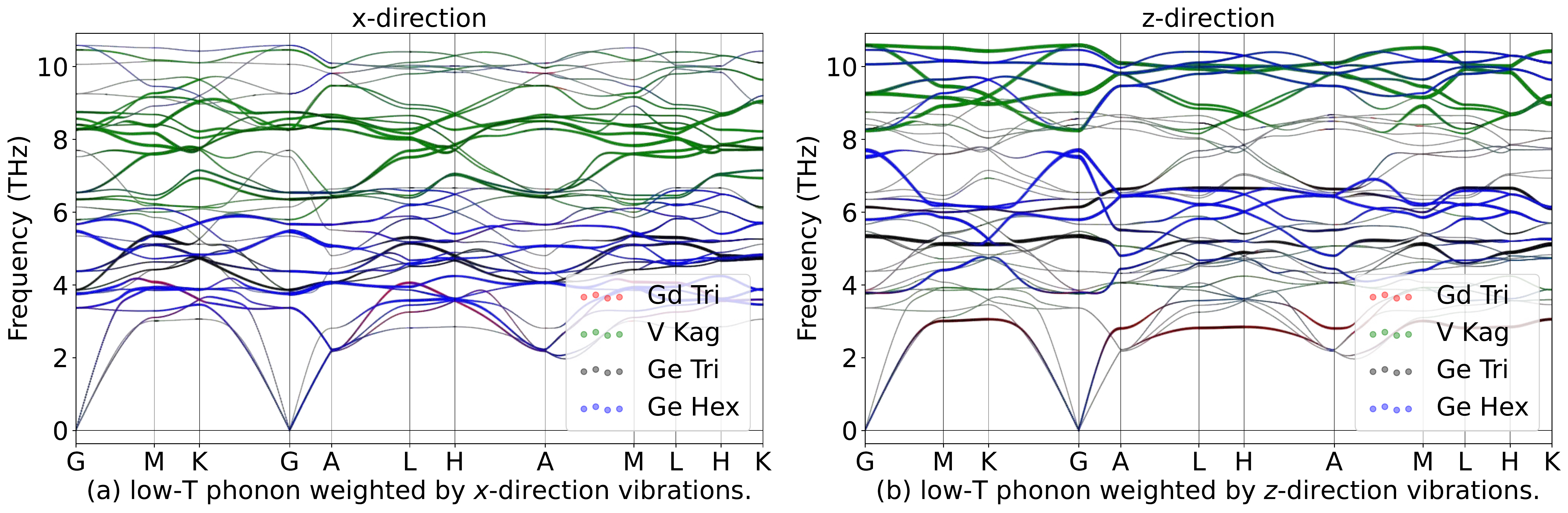}
\hspace{5pt}\label{fig:GdV6Ge6_relaxed_High_xz}
\includegraphics[width=0.85\textwidth]{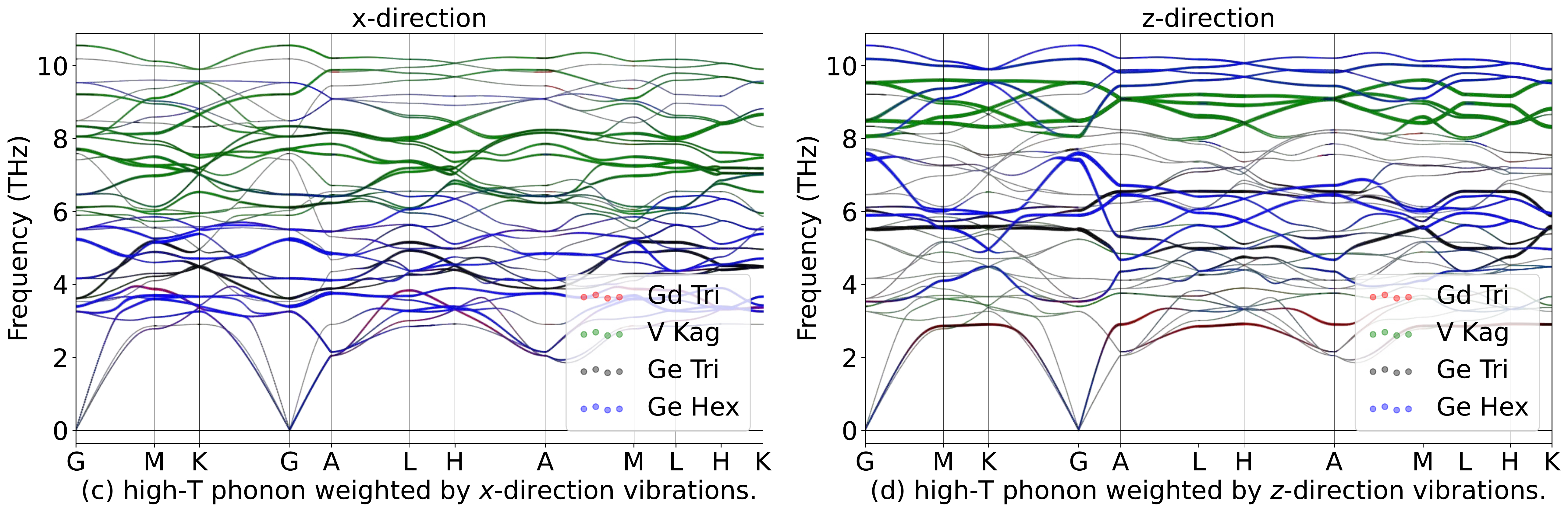}
\caption{Phonon spectrum for GdV$_6$Ge$_6$in in-plane ($x$) and out-of-plane ($z$) directions.}
\label{fig:GdV6Ge6phoxyz}
\end{figure}

\begin{table}[htbp]
\global\long\def\arraystretch{1.12}
\captionsetup{width=.95\textwidth}
\caption{IRREPs at high-symmetry points for GdV$_6$Ge$_6$ phonon spectrum at Low-T.}
\label{tab:191_GdV6Ge6_Low}
\setlength{\tabcolsep}{1.mm}{
}
\end{table}

\begin{table}[htbp]
\global\long\def\arraystretch{1.12}
\captionsetup{width=.95\textwidth}
\caption{IRREPs at high-symmetry points for GdV$_6$Ge$_6$ phonon spectrum at High-T.}
\label{tab:191_GdV6Ge6_High}
\setlength{\tabcolsep}{1.mm}{
%
}
\end{table}
\subsubsection{\label{sms2sec:TbV6Ge6}\hyperref[smsec:col]{\texorpdfstring{TbV$_6$Ge$_6$}{}}~{\color{green}  (High-T stable, Low-T stable) }}

\begin{table}[htbp]
\global\long\def\arraystretch{1.12}
\captionsetup{width=.85\textwidth}
\caption{Structure parameters of TbV$_6$Ge$_6$ after relaxation. It contains 395 electrons. }
\label{tab:TbV6Ge6}
\setlength{\tabcolsep}{4mm}{
\begin{tabular}{c|c||c|c}
\hline
\hline
 Atom & Coordinates & \multicolumn{2}{c}{Parameters~($\AA$)} \\
\hline
Tb ~ 1a & ($0,0,0$) & $a$ & $5.1766$ \\
V ~ 6i & ($\frac{1}{2},0,0.2509$) & $c$ & $8.3732$ \\
Ge1 ~ 2c & ($\frac{1}{3},\frac{2}{3},0$) & $d^{Tri}_{Z-Z}$ & $2.6979$ \\
Ge2 ~ 2d & ($\frac{1}{3},\frac{2}{3},\frac{1}{2}$) & $d_{T_{1}}$ & $2.5883$ \\
Ge3 ~ 2e & ($0,0,0.3389$) & $d_{T_{2}}$ & $4.1723$ \\
\hline
\hline
\end{tabular}}
\end{table}

\begin{figure}[htbp]
\centering
\includegraphics[width=0.85\textwidth]{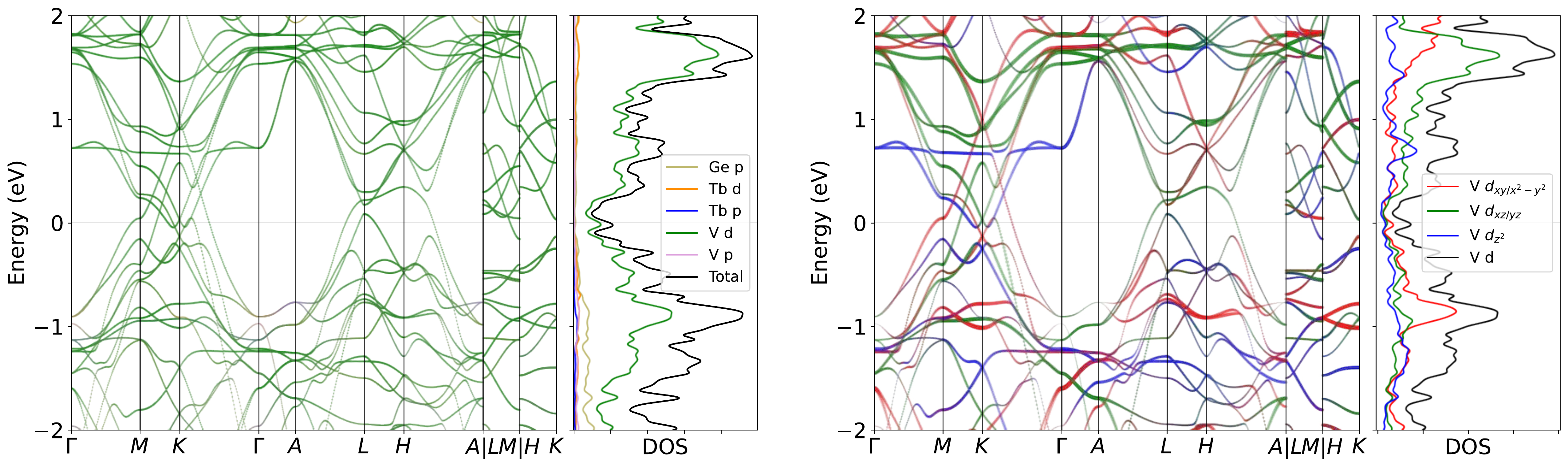}
\caption{Band structure for TbV$_6$Ge$_6$ without SOC. The projection of V $3d$ orbitals is also presented. The band structures clearly reveal the dominating of V $3d$ orbitals  near the Fermi level, as well as the pronounced peak.}
\label{fig:TbV6Ge6band}
\end{figure}

\begin{figure}[H]
\centering
\subfloat[Relaxed phonon at low-T.]{
\label{fig:TbV6Ge6_relaxed_Low}
\includegraphics[width=0.45\textwidth]{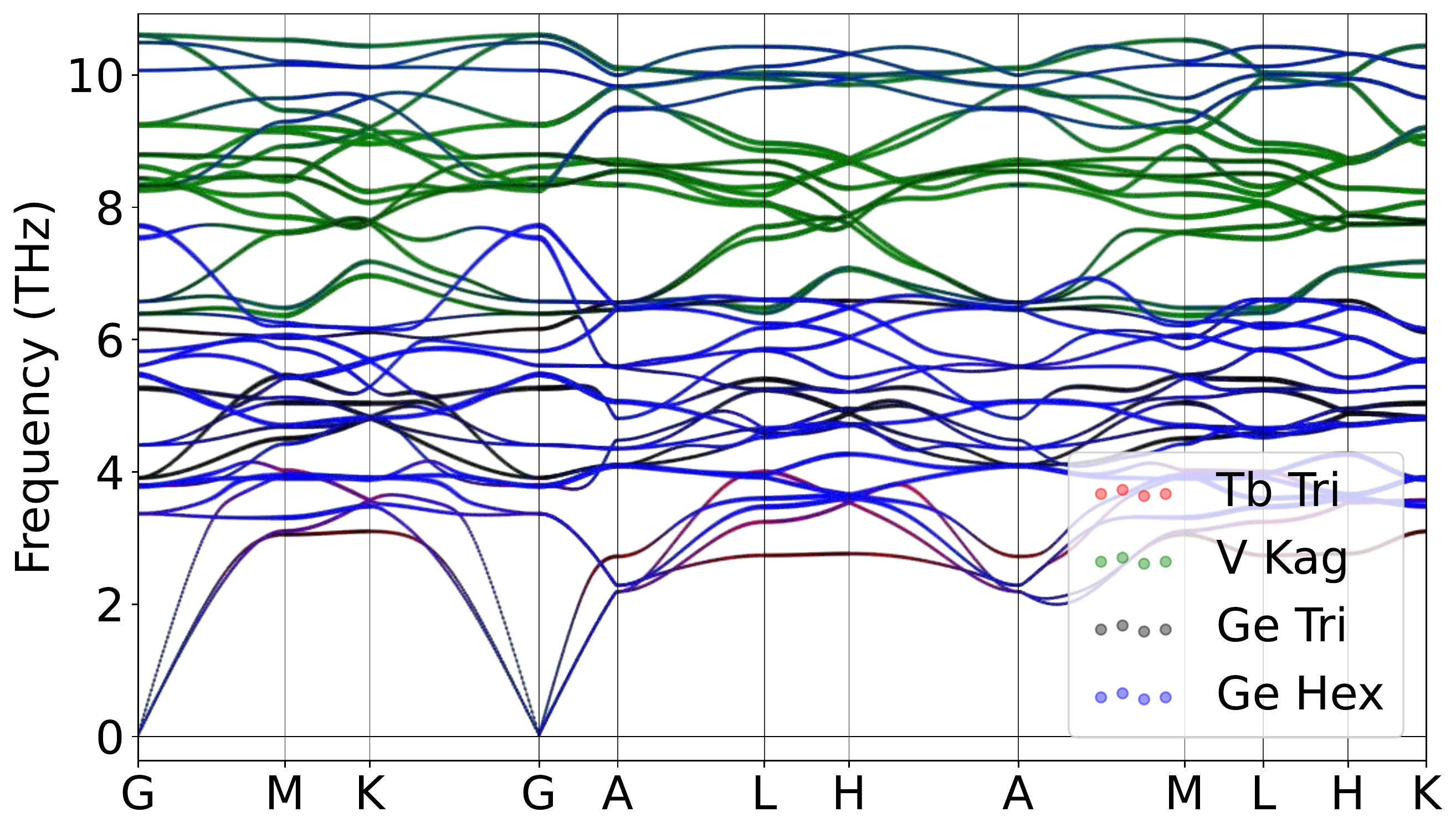}
}
\hspace{5pt}\subfloat[Relaxed phonon at high-T.]{
\label{fig:TbV6Ge6_relaxed_High}
\includegraphics[width=0.45\textwidth]{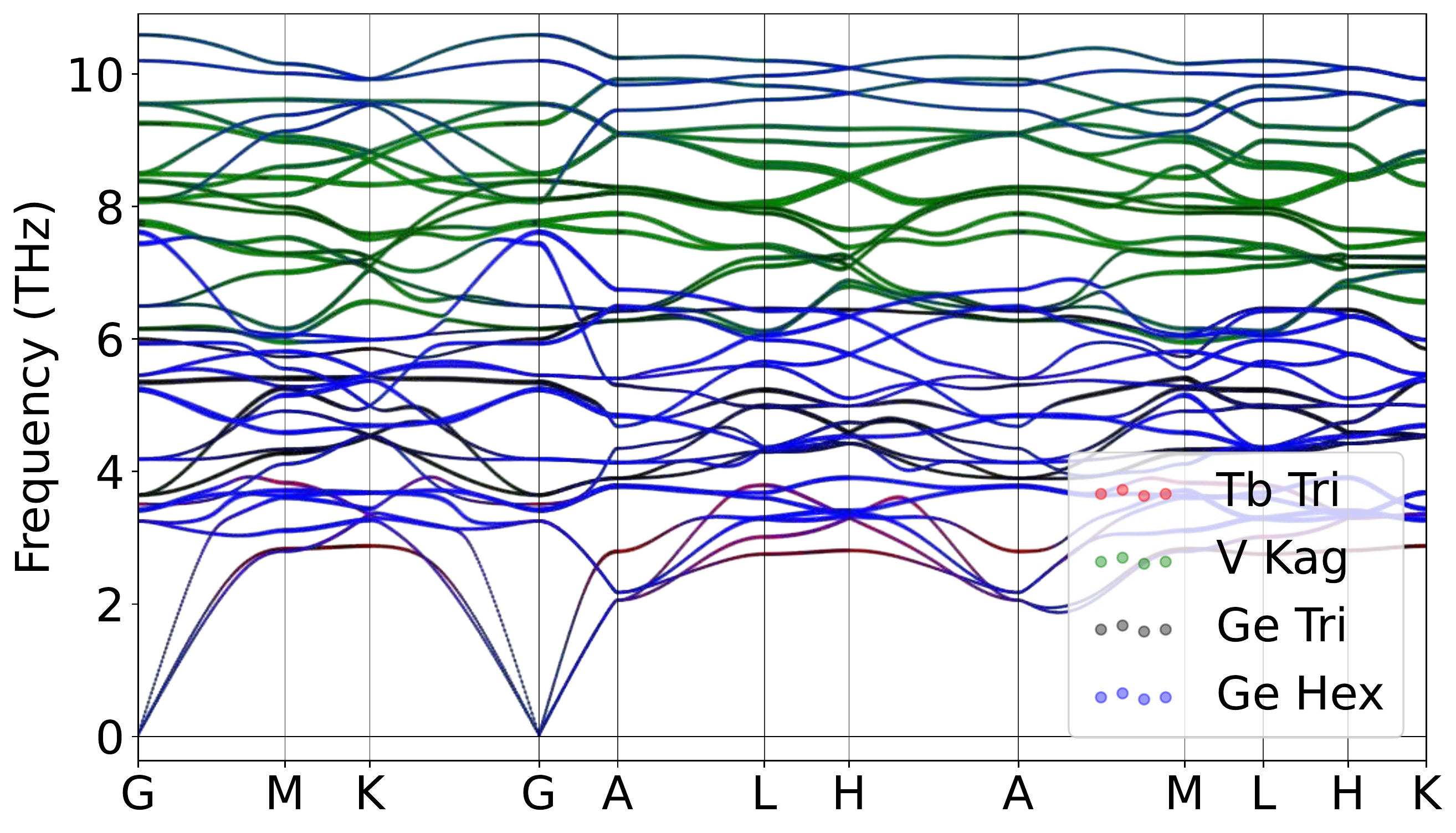}
}
\caption{Atomically projected phonon spectrum for TbV$_6$Ge$_6$.}
\label{fig:TbV6Ge6pho}
\end{figure}

\begin{figure}[H]
\centering
\label{fig:TbV6Ge6_relaxed_Low_xz}
\includegraphics[width=0.85\textwidth]{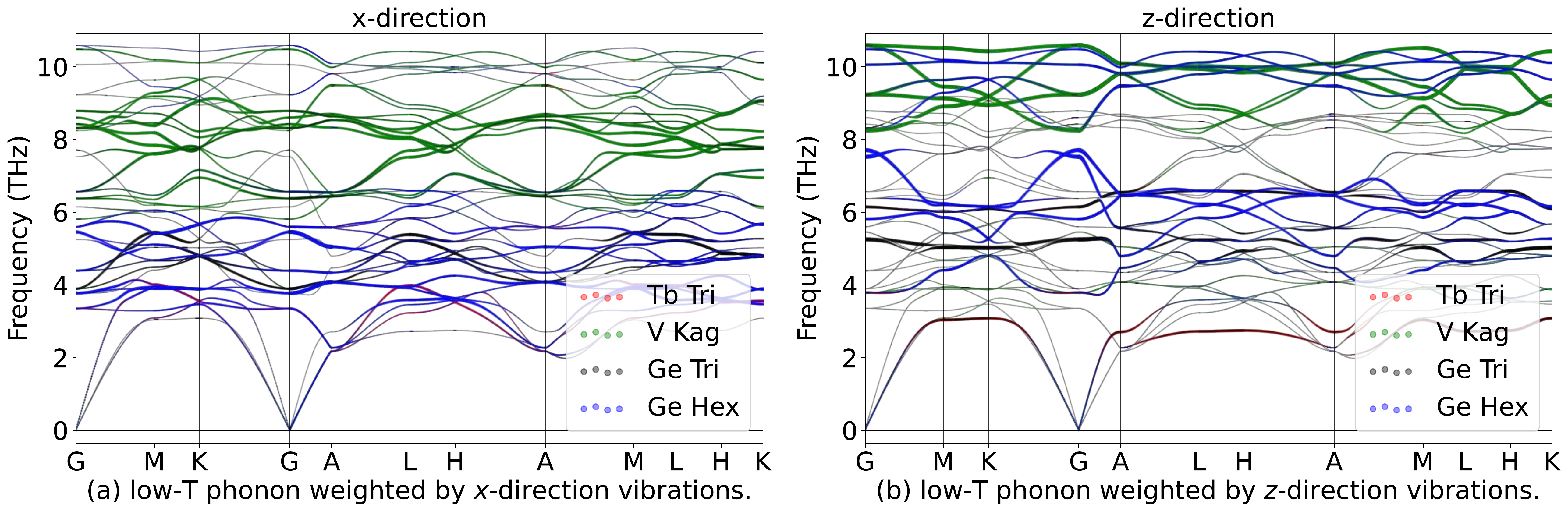}
\hspace{5pt}\label{fig:TbV6Ge6_relaxed_High_xz}
\includegraphics[width=0.85\textwidth]{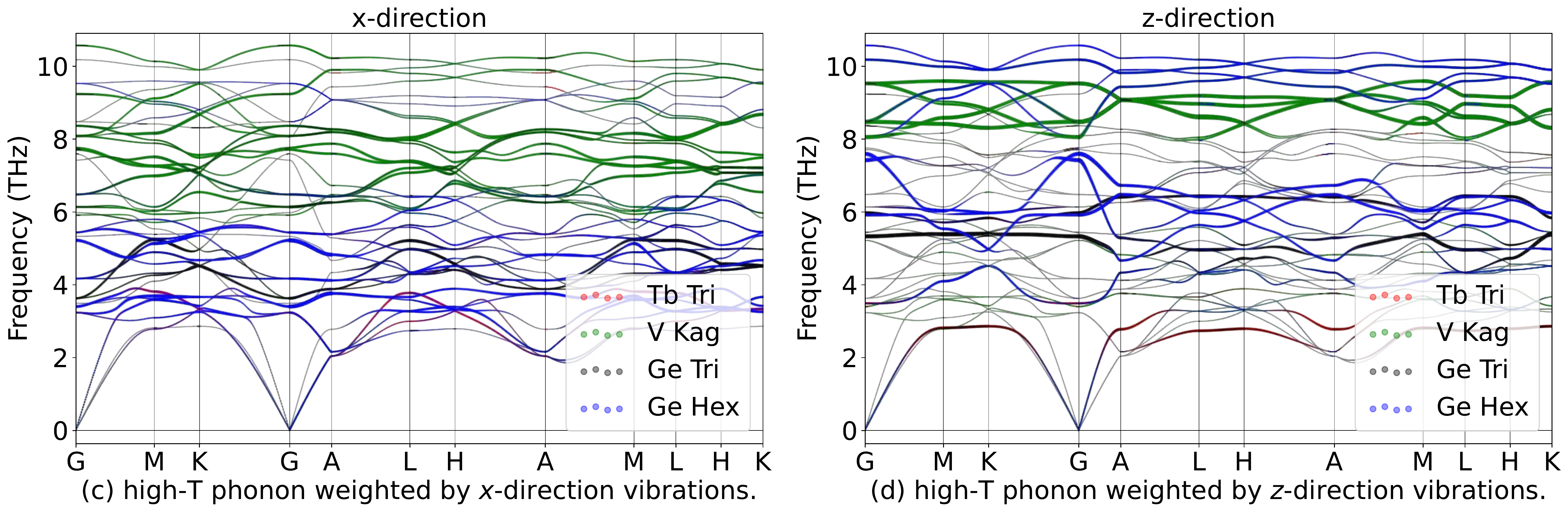}
\caption{Phonon spectrum for TbV$_6$Ge$_6$in in-plane ($x$) and out-of-plane ($z$) directions.}
\label{fig:TbV6Ge6phoxyz}
\end{figure}

\begin{table}[htbp]
\global\long\def\arraystretch{1.12}
\captionsetup{width=.95\textwidth}
\caption{IRREPs at high-symmetry points for TbV$_6$Ge$_6$ phonon spectrum at Low-T.}
\label{tab:191_TbV6Ge6_Low}
\setlength{\tabcolsep}{1.mm}{
}
\end{table}

\begin{table}[htbp]
\global\long\def\arraystretch{1.12}
\captionsetup{width=.95\textwidth}
\caption{IRREPs at high-symmetry points for TbV$_6$Ge$_6$ phonon spectrum at High-T.}
\label{tab:191_TbV6Ge6_High}
\setlength{\tabcolsep}{1.mm}{
%
}
\end{table}
\subsubsection{\label{sms2sec:DyV6Ge6}\hyperref[smsec:col]{\texorpdfstring{DyV$_6$Ge$_6$}{}}~{\color{green}  (High-T stable, Low-T stable) }}

\begin{table}[htbp]
\global\long\def\arraystretch{1.12}
\captionsetup{width=.85\textwidth}
\caption{Structure parameters of DyV$_6$Ge$_6$ after relaxation. It contains 396 electrons. }
\label{tab:DyV6Ge6}
\setlength{\tabcolsep}{4mm}{
\begin{tabular}{c|c||c|c}
\hline
\hline
 Atom & Coordinates & \multicolumn{2}{c}{Parameters~($\AA$)} \\
\hline
Dy ~ 1a & ($0,0,0$) & $a$ & $5.1736$ \\
V ~ 6i & ($\frac{1}{2},0,0.2509$) & $c$ & $8.3762$ \\
Ge1 ~ 2c & ($\frac{1}{3},\frac{2}{3},0$) & $d^{Tri}_{Z-Z}$ & $2.7161$ \\
Ge2 ~ 2d & ($\frac{1}{3},\frac{2}{3},\frac{1}{2}$) & $d_{T_{1}}$ & $2.5868$ \\
Ge3 ~ 2e & ($0,0,0.3379$) & $d_{T_{2}}$ & $4.1728$ \\
\hline
\hline
\end{tabular}}
\end{table}

\begin{figure}[htbp]
\centering
\includegraphics[width=0.85\textwidth]{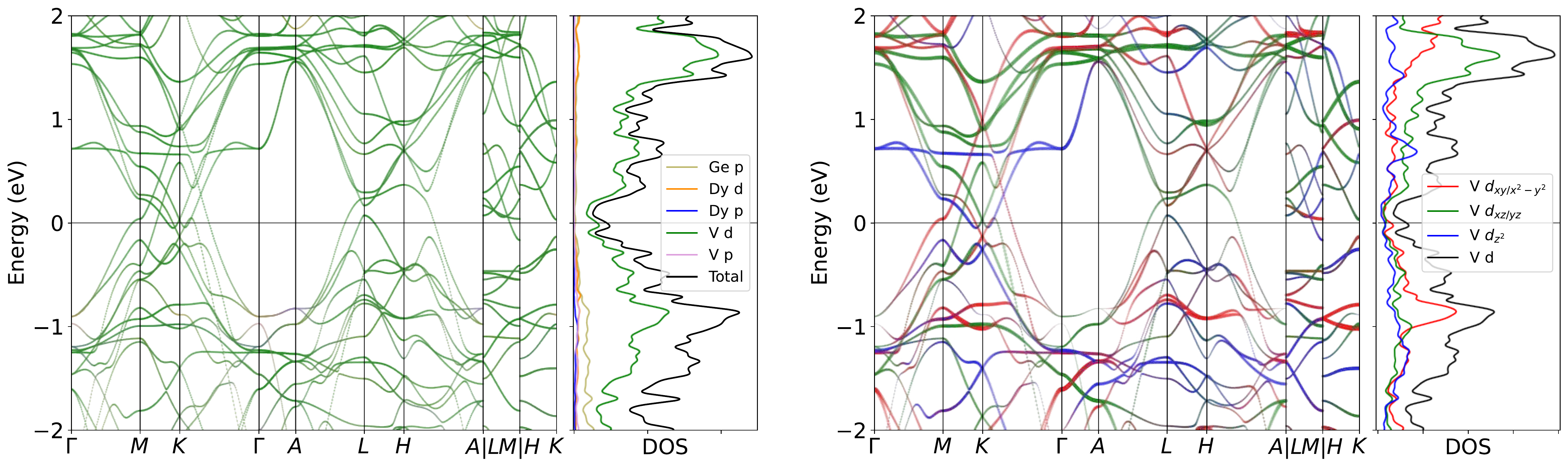}
\caption{Band structure for DyV$_6$Ge$_6$ without SOC. The projection of V $3d$ orbitals is also presented. The band structures clearly reveal the dominating of V $3d$ orbitals  near the Fermi level, as well as the pronounced peak.}
\label{fig:DyV6Ge6band}
\end{figure}

\begin{figure}[H]
\centering
\subfloat[Relaxed phonon at low-T.]{
\label{fig:DyV6Ge6_relaxed_Low}
\includegraphics[width=0.45\textwidth]{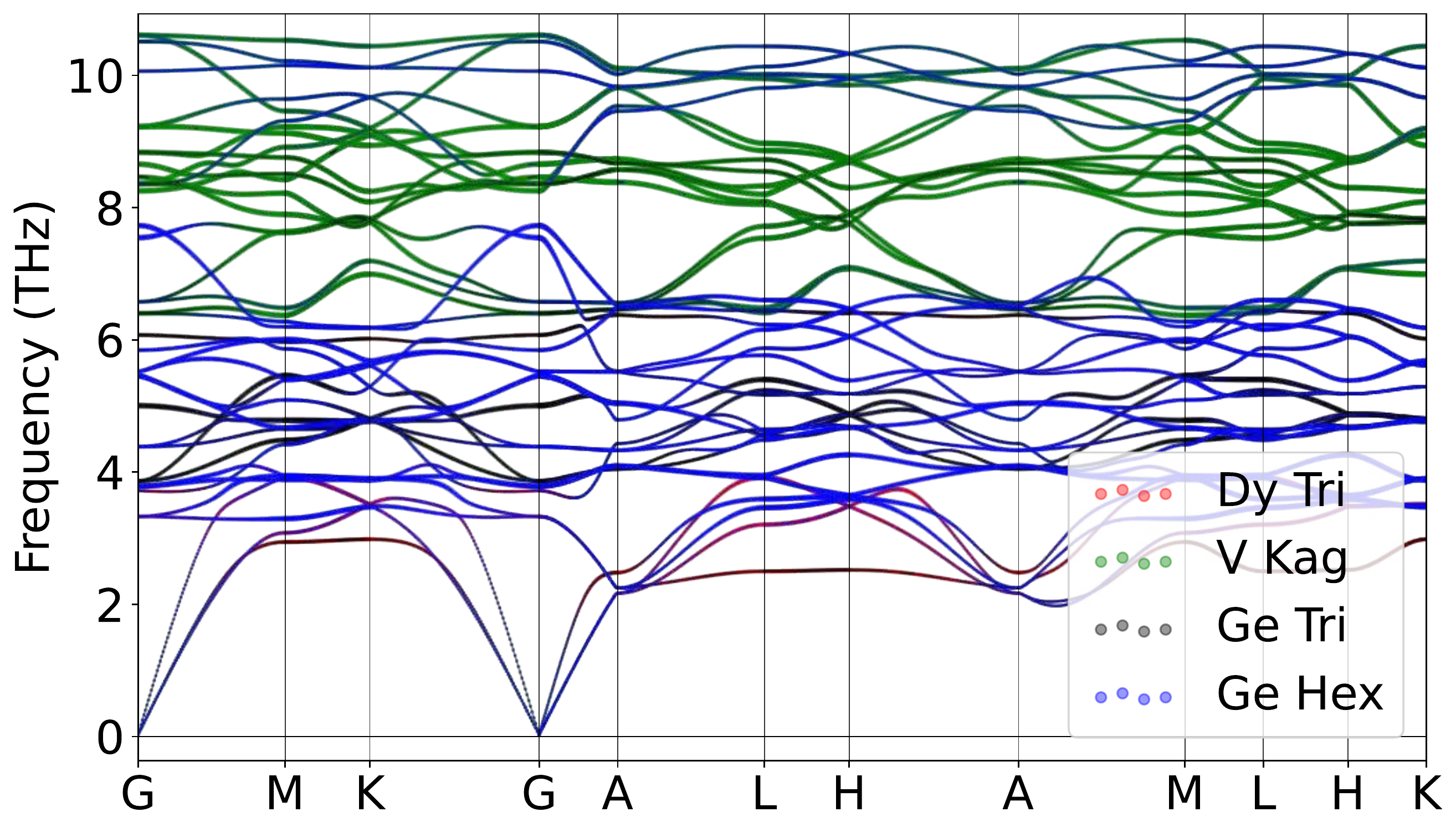}
}
\hspace{5pt}\subfloat[Relaxed phonon at high-T.]{
\label{fig:DyV6Ge6_relaxed_High}
\includegraphics[width=0.45\textwidth]{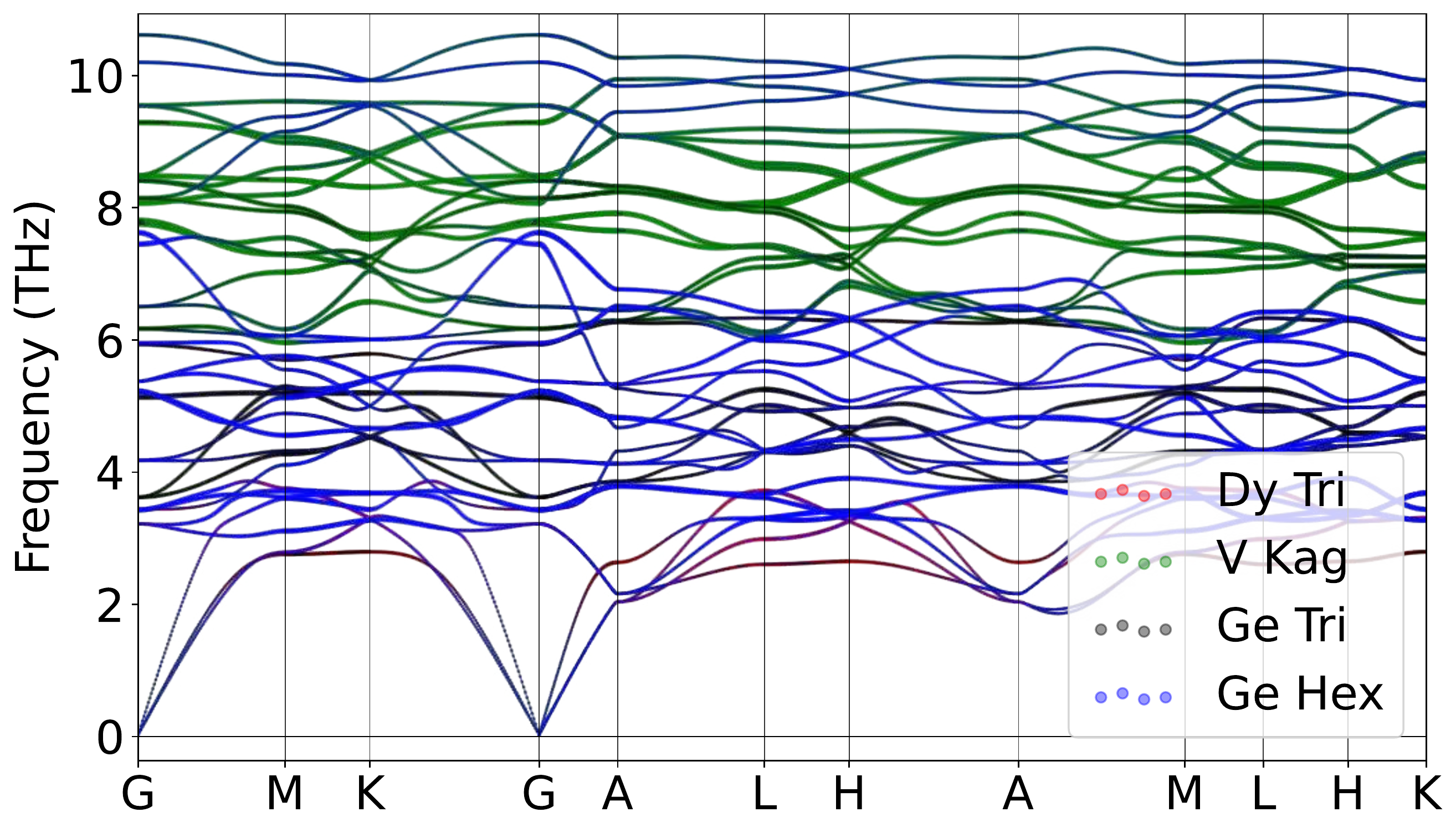}
}
\caption{Atomically projected phonon spectrum for DyV$_6$Ge$_6$.}
\label{fig:DyV6Ge6pho}
\end{figure}

\begin{figure}[H]
\centering
\label{fig:DyV6Ge6_relaxed_Low_xz}
\includegraphics[width=0.85\textwidth]{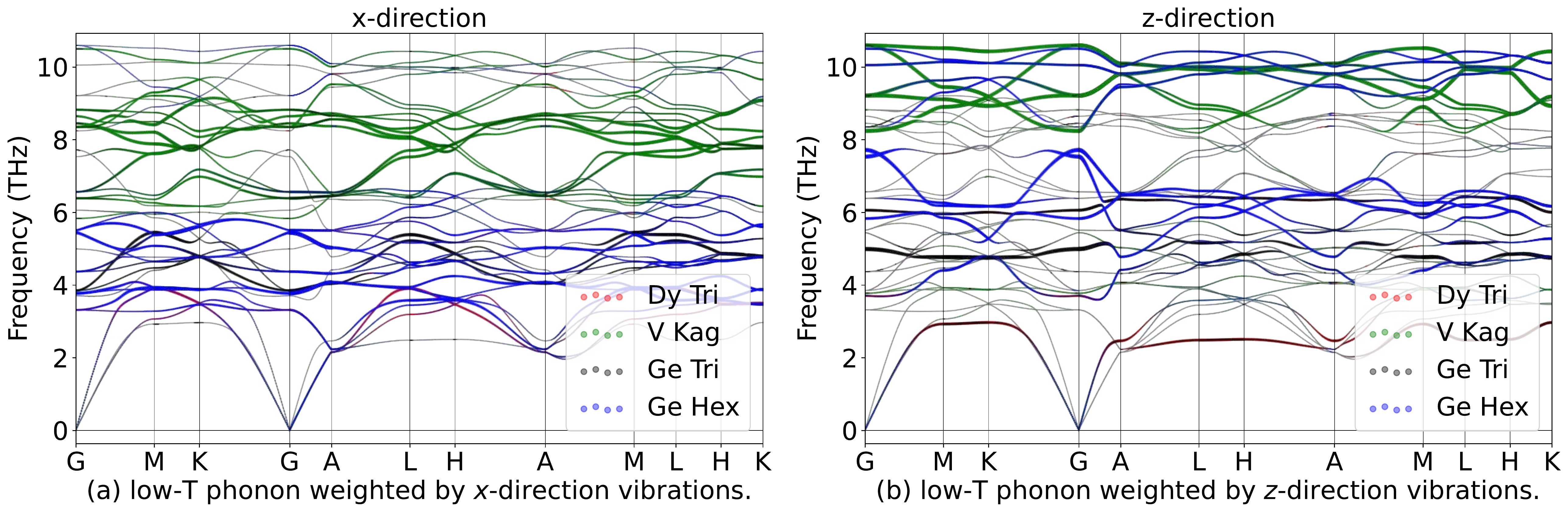}
\hspace{5pt}\label{fig:DyV6Ge6_relaxed_High_xz}
\includegraphics[width=0.85\textwidth]{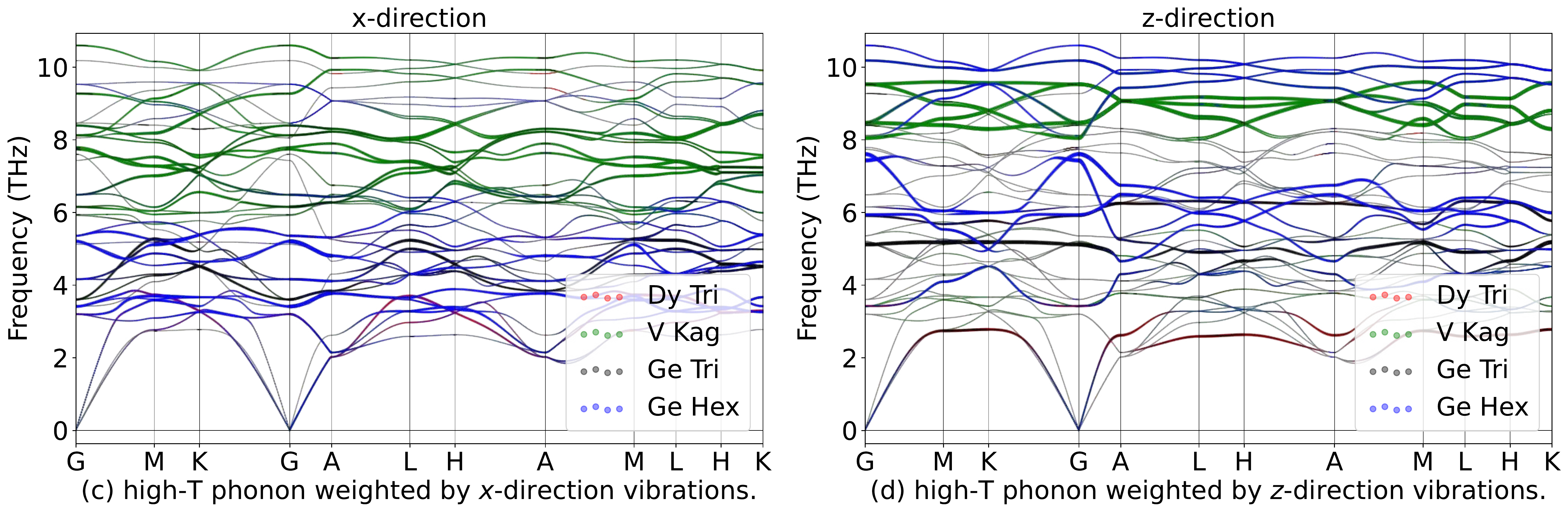}
\caption{Phonon spectrum for DyV$_6$Ge$_6$in in-plane ($x$) and out-of-plane ($z$) directions.}
\label{fig:DyV6Ge6phoxyz}
\end{figure}

\begin{table}[htbp]
\global\long\def\arraystretch{1.12}
\captionsetup{width=.95\textwidth}
\caption{IRREPs at high-symmetry points for DyV$_6$Ge$_6$ phonon spectrum at Low-T.}
\label{tab:191_DyV6Ge6_Low}
\setlength{\tabcolsep}{1.mm}{
}
\end{table}

\begin{table}[htbp]
\global\long\def\arraystretch{1.12}
\captionsetup{width=.95\textwidth}
\caption{IRREPs at high-symmetry points for DyV$_6$Ge$_6$ phonon spectrum at High-T.}
\label{tab:191_DyV6Ge6_High}
\setlength{\tabcolsep}{1.mm}{
%
}
\end{table}
\subsubsection{\label{sms2sec:HoV6Ge6}\hyperref[smsec:col]{\texorpdfstring{HoV$_6$Ge$_6$}{}}~{\color{green}  (High-T stable, Low-T stable) }}

\begin{table}[htbp]
\global\long\def\arraystretch{1.12}
\captionsetup{width=.85\textwidth}
\caption{Structure parameters of HoV$_6$Ge$_6$ after relaxation. It contains 397 electrons. }
\label{tab:HoV6Ge6}
\setlength{\tabcolsep}{4mm}{
\begin{tabular}{c|c||c|c}
\hline
\hline
 Atom & Coordinates & \multicolumn{2}{c}{Parameters~($\AA$)} \\
\hline
Ho ~ 1a & ($0,0,0$) & $a$ & $5.2507$ \\
V ~ 6i & ($\frac{1}{2},0,0.2511$) & $c$ & $8.4746$ \\
Ge1 ~ 2c & ($\frac{1}{3},\frac{2}{3},0$) & $d^{Tri}_{Z-Z}$ & $2.7121$ \\
Ge2 ~ 2d & ($\frac{1}{3},\frac{2}{3},\frac{1}{2}$) & $d_{T_{1}}$ & $2.6253$ \\
Ge3 ~ 2e & ($0,0,0.3400$) & $d_{T_{2}}$ & $4.2180$ \\
\hline
\hline
\end{tabular}}
\end{table}

\begin{figure}[htbp]
\centering
\includegraphics[width=0.85\textwidth]{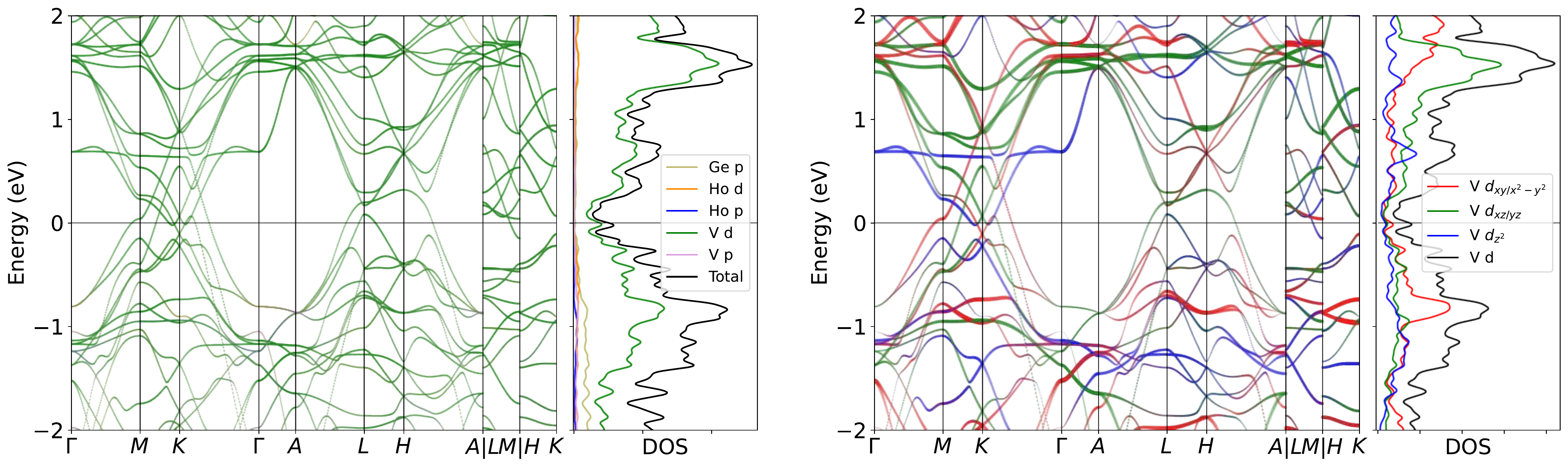}
\caption{Band structure for HoV$_6$Ge$_6$ without SOC. The projection of V $3d$ orbitals is also presented. The band structures clearly reveal the dominating of V $3d$ orbitals  near the Fermi level, as well as the pronounced peak.}
\label{fig:HoV6Ge6band}
\end{figure}

\begin{figure}[H]
\centering
\subfloat[Relaxed phonon at low-T.]{
\label{fig:HoV6Ge6_relaxed_Low}
\includegraphics[width=0.45\textwidth]{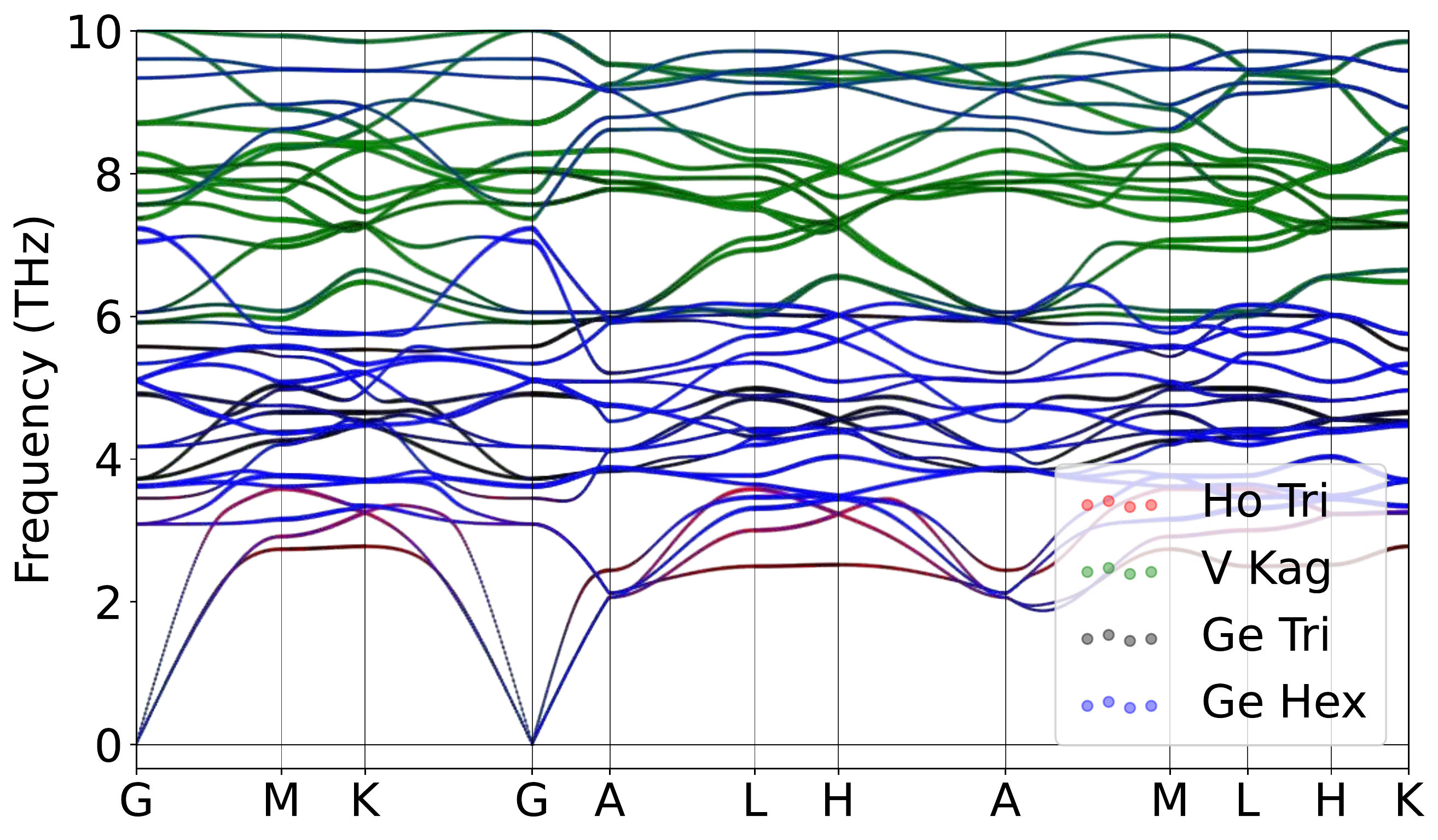}
}
\hspace{5pt}\subfloat[Relaxed phonon at high-T.]{
\label{fig:HoV6Ge6_relaxed_High}
\includegraphics[width=0.45\textwidth]{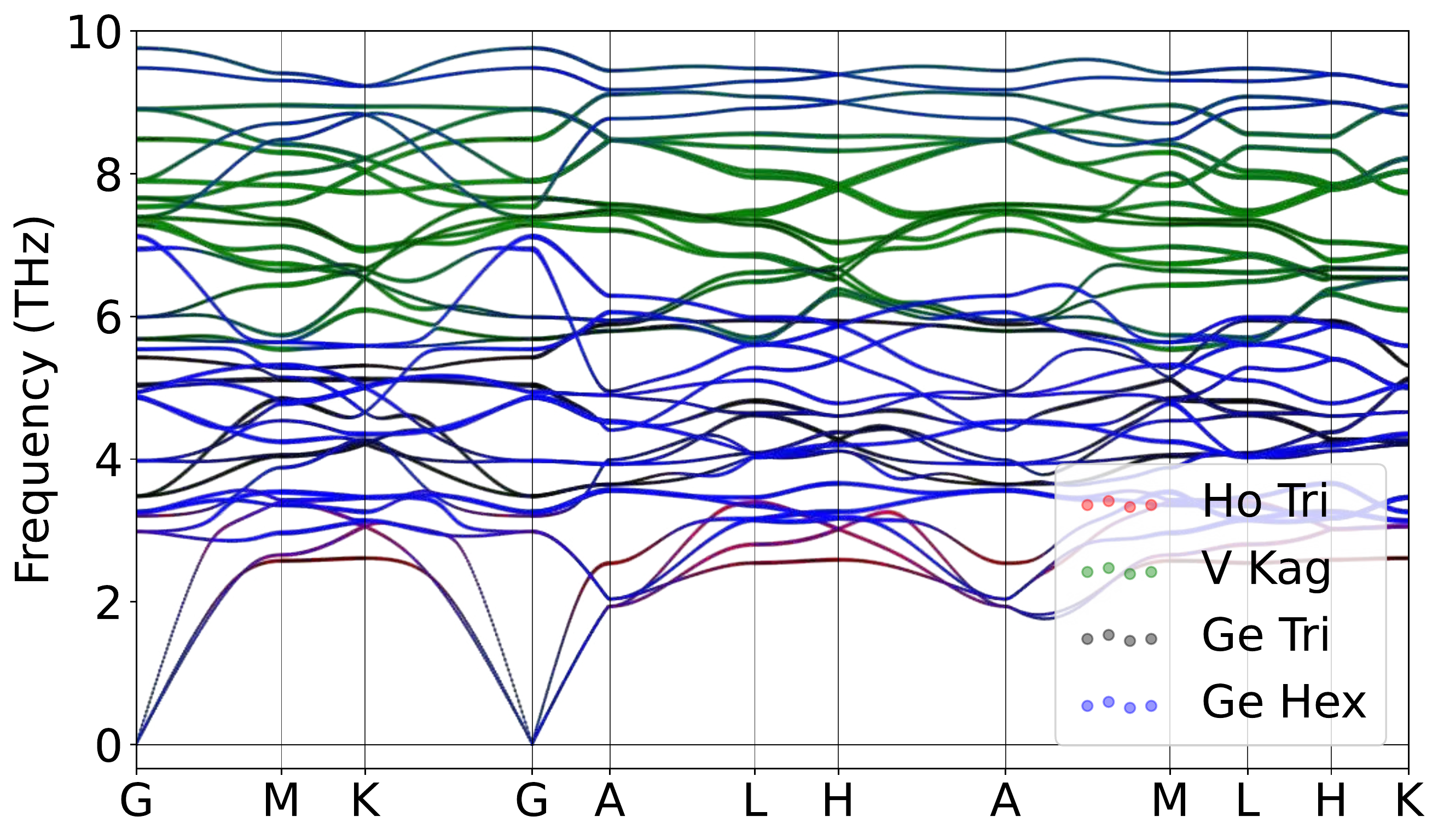}
}
\caption{Atomically projected phonon spectrum for HoV$_6$Ge$_6$.}
\label{fig:HoV6Ge6pho}
\end{figure}

\begin{figure}[H]
\centering
\label{fig:HoV6Ge6_relaxed_Low_xz}
\includegraphics[width=0.85\textwidth]{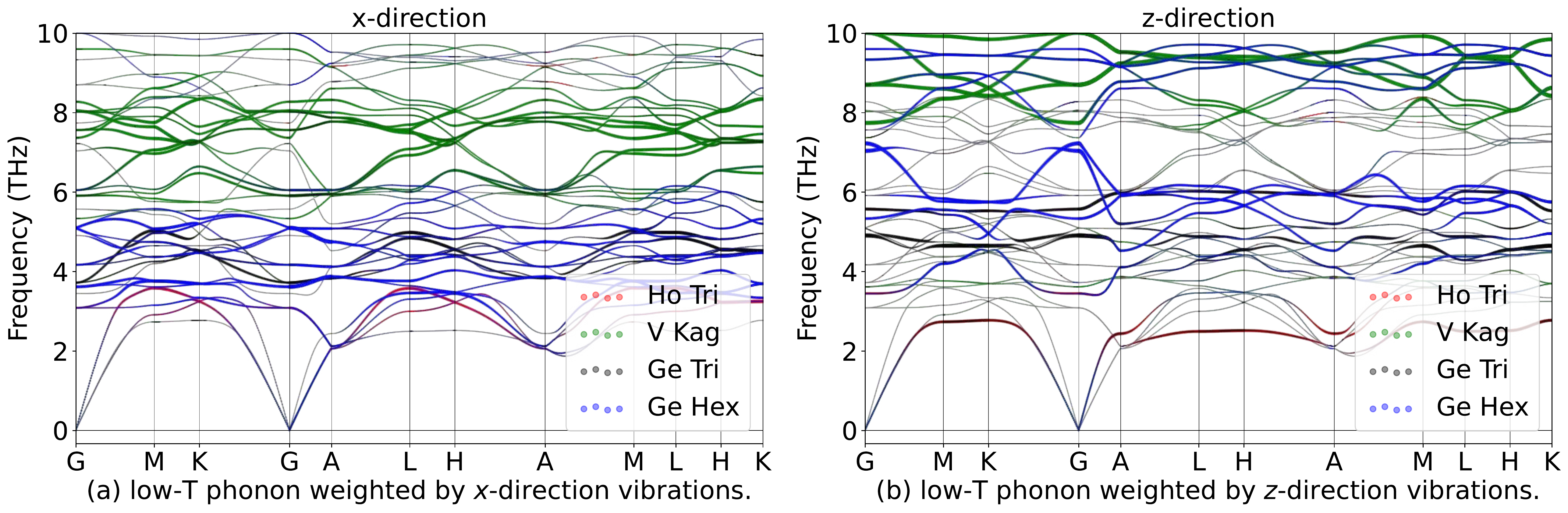}
\hspace{5pt}\label{fig:HoV6Ge6_relaxed_High_xz}
\includegraphics[width=0.85\textwidth]{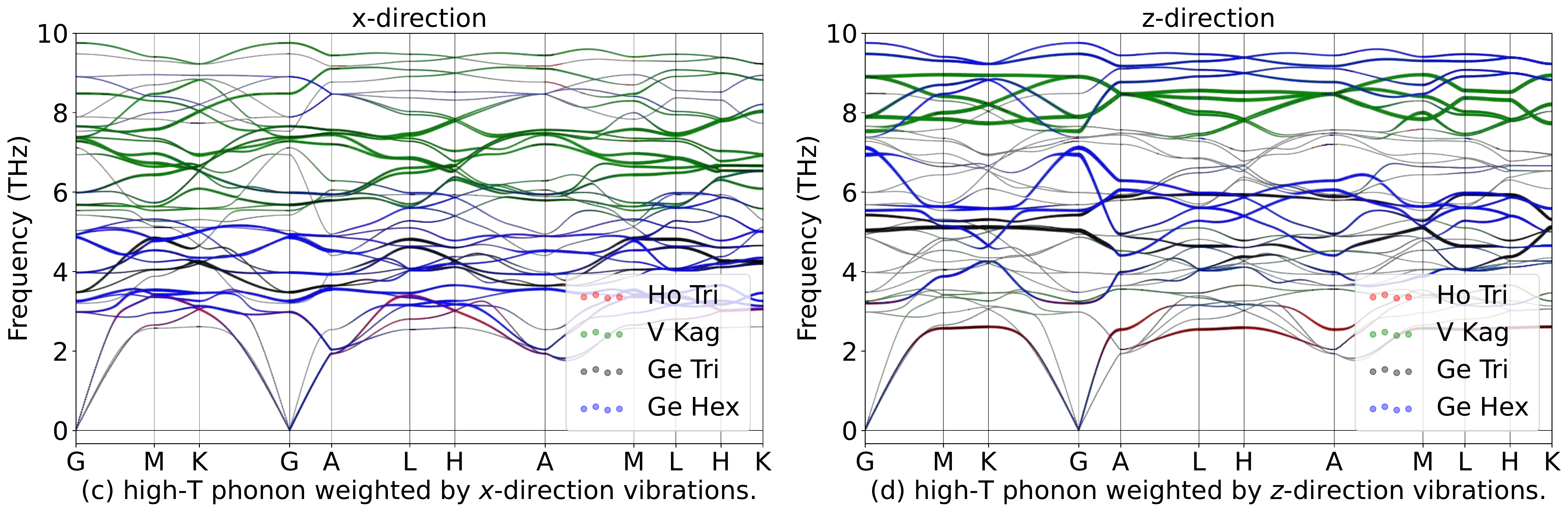}
\caption{Phonon spectrum for HoV$_6$Ge$_6$in in-plane ($x$) and out-of-plane ($z$) directions.}
\label{fig:HoV6Ge6phoxyz}
\end{figure}

\begin{table}[htbp]
\global\long\def\arraystretch{1.12}
\captionsetup{width=.95\textwidth}
\caption{IRREPs at high-symmetry points for HoV$_6$Ge$_6$ phonon spectrum at Low-T.}
\label{tab:191_HoV6Ge6_Low}
\setlength{\tabcolsep}{1.mm}{
}
\end{table}

\begin{table}[htbp]
\global\long\def\arraystretch{1.12}
\captionsetup{width=.95\textwidth}
\caption{IRREPs at high-symmetry points for HoV$_6$Ge$_6$ phonon spectrum at High-T.}
\label{tab:191_HoV6Ge6_High}
\setlength{\tabcolsep}{1.mm}{
%
}
\end{table}
\subsubsection{\label{sms2sec:ErV6Ge6}\hyperref[smsec:col]{\texorpdfstring{ErV$_6$Ge$_6$}{}}~{\color{green}  (High-T stable, Low-T stable) }}

\begin{table}[htbp]
\global\long\def\arraystretch{1.12}
\captionsetup{width=.85\textwidth}
\caption{Structure parameters of ErV$_6$Ge$_6$ after relaxation. It contains 398 electrons. }
\label{tab:ErV6Ge6}
\setlength{\tabcolsep}{4mm}{
%
}
\end{table}

\begin{figure}[htbp]
\centering
\includegraphics[width=0.85\textwidth]{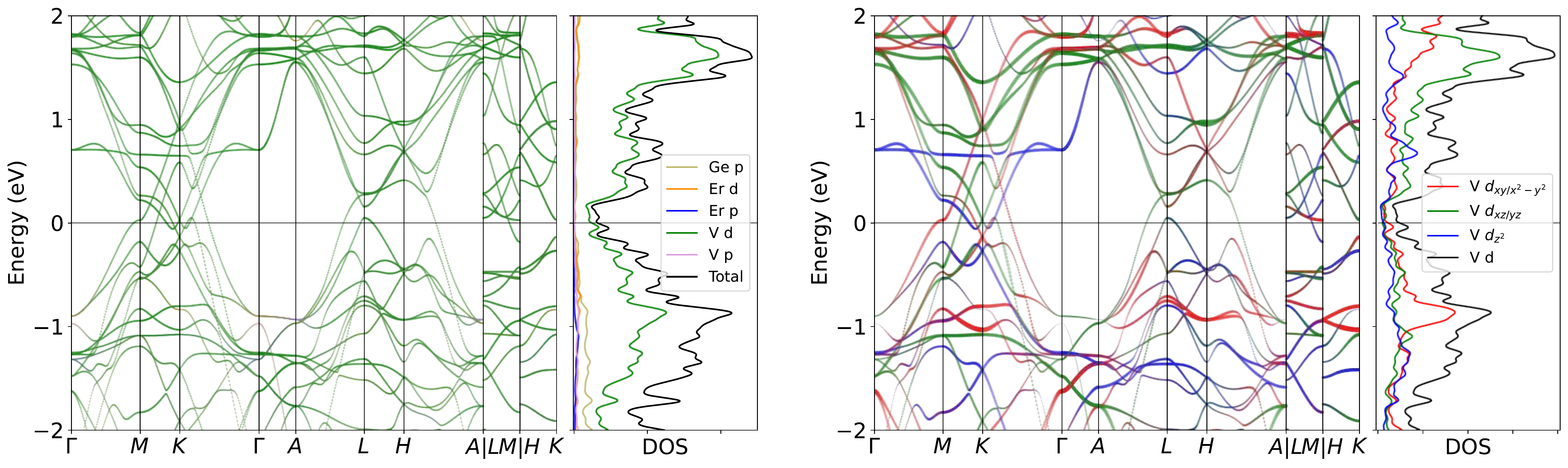}
\caption{Band structure for ErV$_6$Ge$_6$ without SOC. The projection of V $3d$ orbitals is also presented. The band structures clearly reveal the dominating of V $3d$ orbitals  near the Fermi level, as well as the pronounced peak.}
\label{fig:ErV6Ge6band}
\end{figure}

\begin{figure}[H]
\centering
\subfloat[Relaxed phonon at low-T.]{
\label{fig:ErV6Ge6_relaxed_Low}
\includegraphics[width=0.45\textwidth]{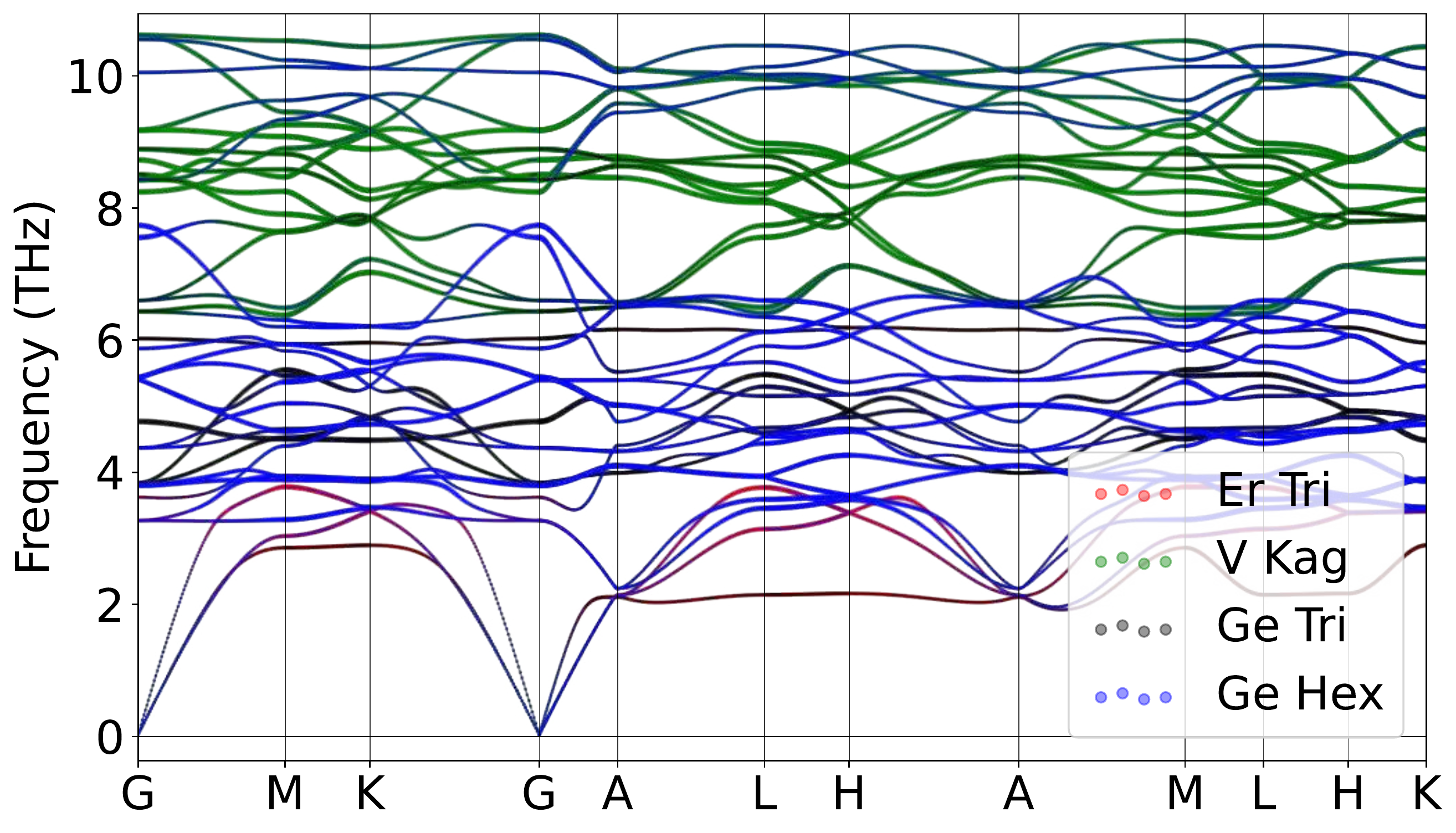}
}
\hspace{5pt}\subfloat[Relaxed phonon at high-T.]{
\label{fig:ErV6Ge6_relaxed_High}
\includegraphics[width=0.45\textwidth]{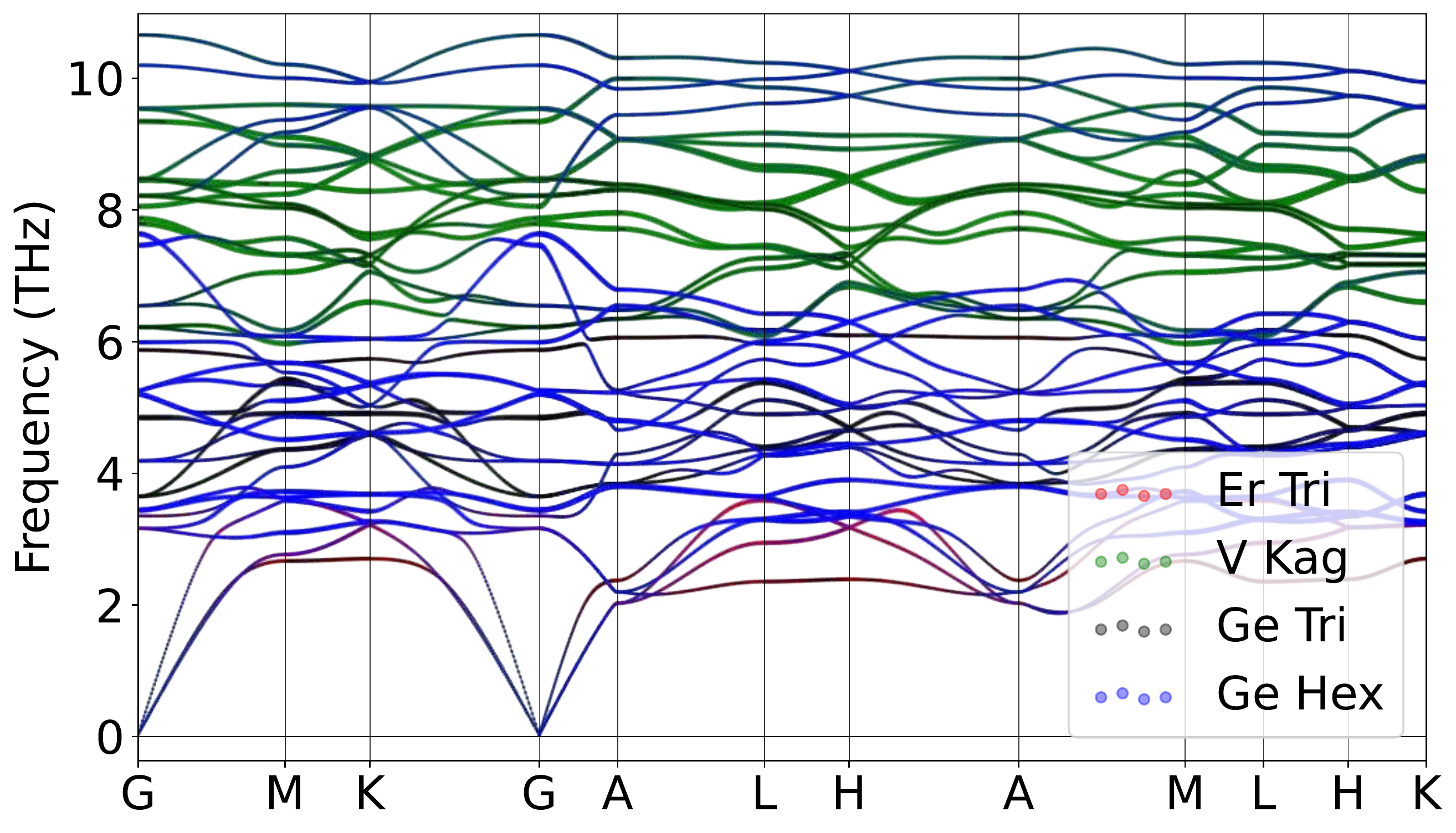}
}
\caption{Atomically projected phonon spectrum for ErV$_6$Ge$_6$.}
\label{fig:ErV6Ge6pho}
\end{figure}

\begin{figure}[H]
\centering
\label{fig:ErV6Ge6_relaxed_Low_xz}
\includegraphics[width=0.85\textwidth]{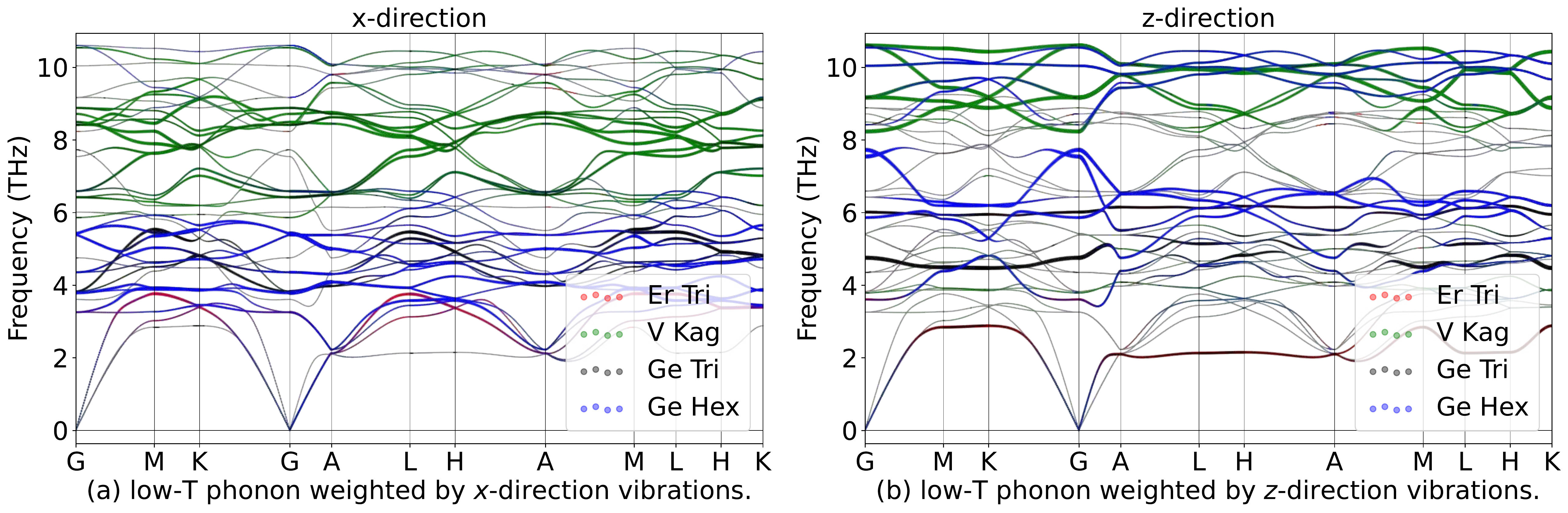}
\hspace{5pt}\label{fig:ErV6Ge6_relaxed_High_xz}
\includegraphics[width=0.85\textwidth]{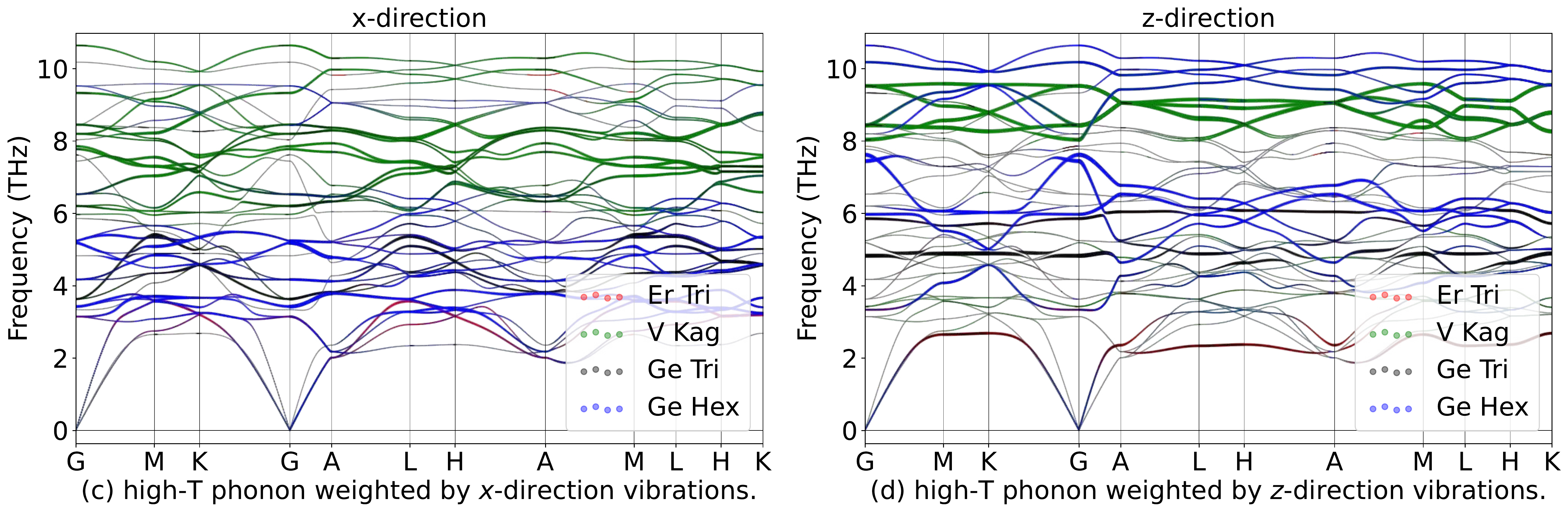}
\caption{Phonon spectrum for ErV$_6$Ge$_6$in in-plane ($x$) and out-of-plane ($z$) directions.}
\label{fig:ErV6Ge6phoxyz}
\end{figure}

\begin{table}[htbp]
\global\long\def\arraystretch{1.12}
\captionsetup{width=.95\textwidth}
\caption{IRREPs at high-symmetry points for ErV$_6$Ge$_6$ phonon spectrum at Low-T.}
\label{tab:191_ErV6Ge6_Low}
\setlength{\tabcolsep}{1.mm}{
}
\end{table}

\begin{table}[htbp]
\global\long\def\arraystretch{1.12}
\captionsetup{width=.95\textwidth}
\caption{IRREPs at high-symmetry points for ErV$_6$Ge$_6$ phonon spectrum at High-T.}
\label{tab:191_ErV6Ge6_High}
\setlength{\tabcolsep}{1.mm}{
%
}
\end{table}
\subsubsection{\label{sms2sec:TmV6Ge6}\hyperref[smsec:col]{\texorpdfstring{TmV$_6$Ge$_6$}{}}~{\color{green}  (High-T stable, Low-T stable) }}

\begin{table}[htbp]
\global\long\def\arraystretch{1.12}
\captionsetup{width=.85\textwidth}
\caption{Structure parameters of TmV$_6$Ge$_6$ after relaxation. It contains 399 electrons. }
\label{tab:TmV6Ge6}
\setlength{\tabcolsep}{4mm}{
%
}
\end{table}

\begin{figure}[htbp]
\centering
\includegraphics[width=0.85\textwidth]{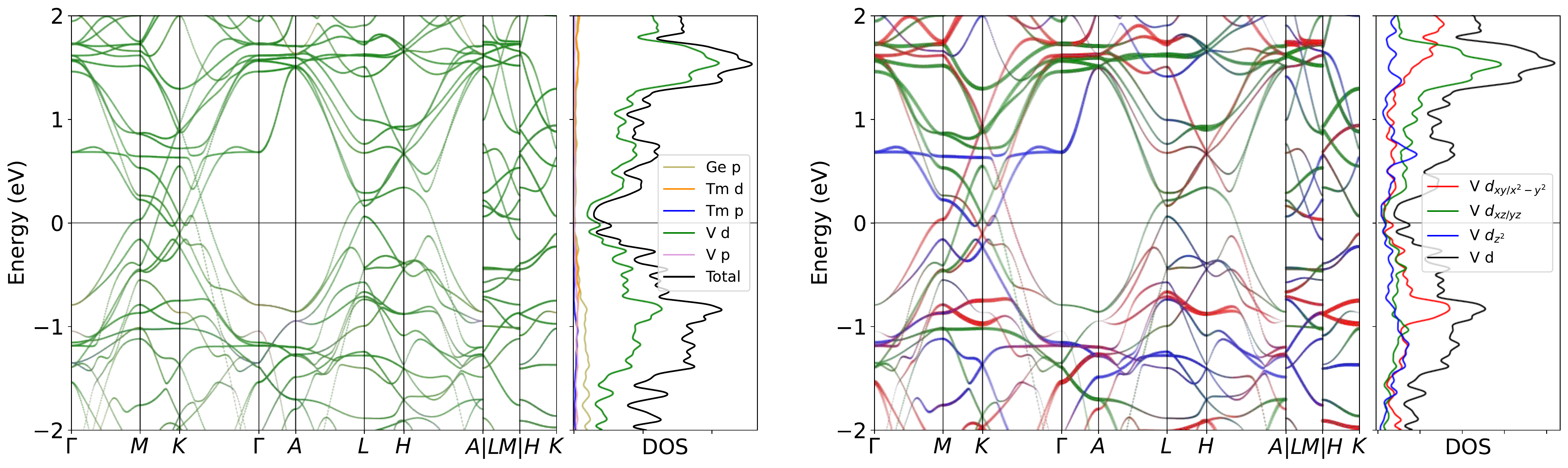}
\caption{Band structure for TmV$_6$Ge$_6$ without SOC. The projection of V $3d$ orbitals is also presented. The band structures clearly reveal the dominating of V $3d$ orbitals  near the Fermi level, as well as the pronounced peak.}
\label{fig:TmV6Ge6band}
\end{figure}

\begin{figure}[H]
\centering
\subfloat[Relaxed phonon at low-T.]{
\label{fig:TmV6Ge6_relaxed_Low}
\includegraphics[width=0.45\textwidth]{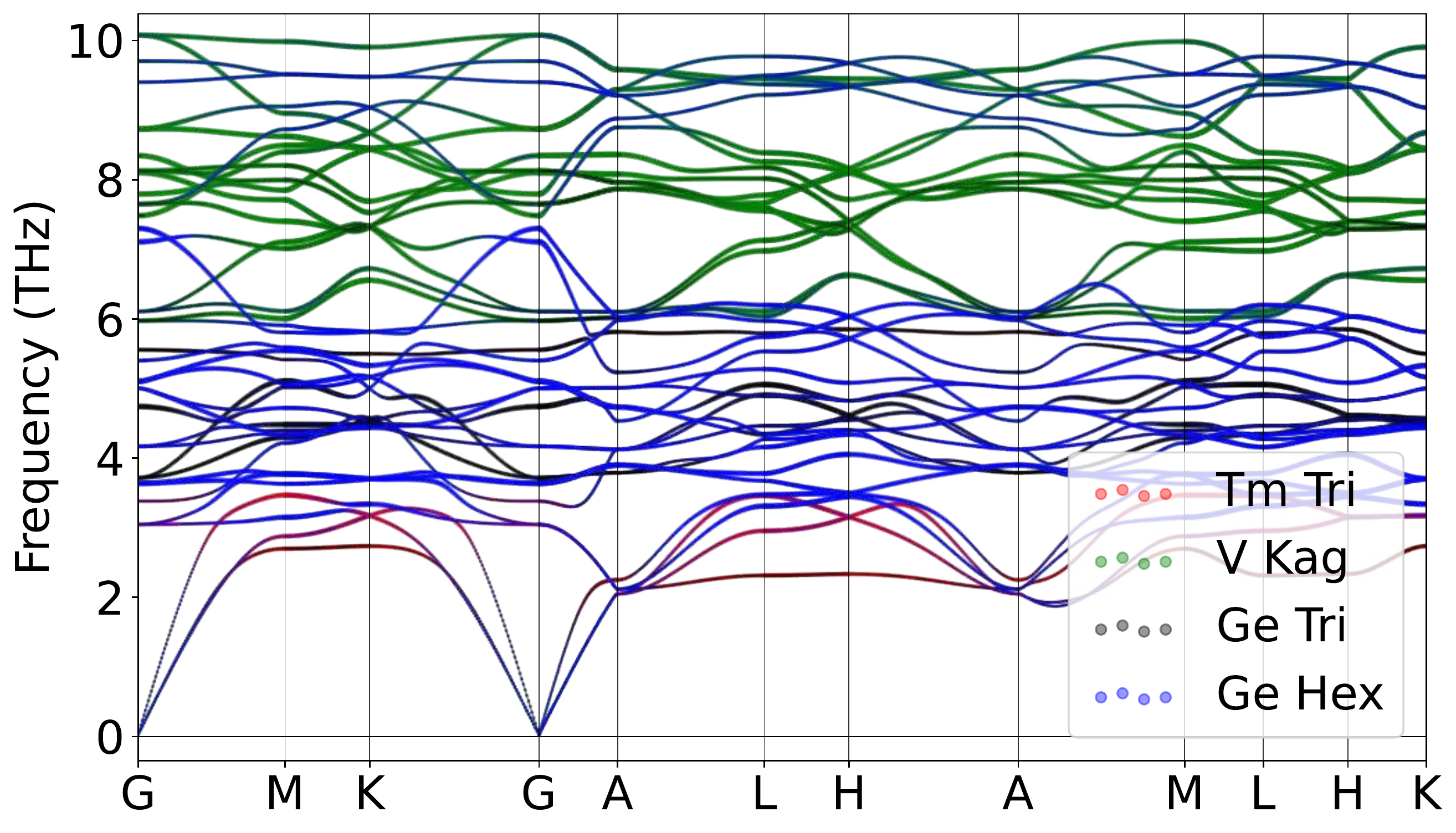}
}
\hspace{5pt}\subfloat[Relaxed phonon at high-T.]{
\label{fig:TmV6Ge6_relaxed_High}
\includegraphics[width=0.45\textwidth]{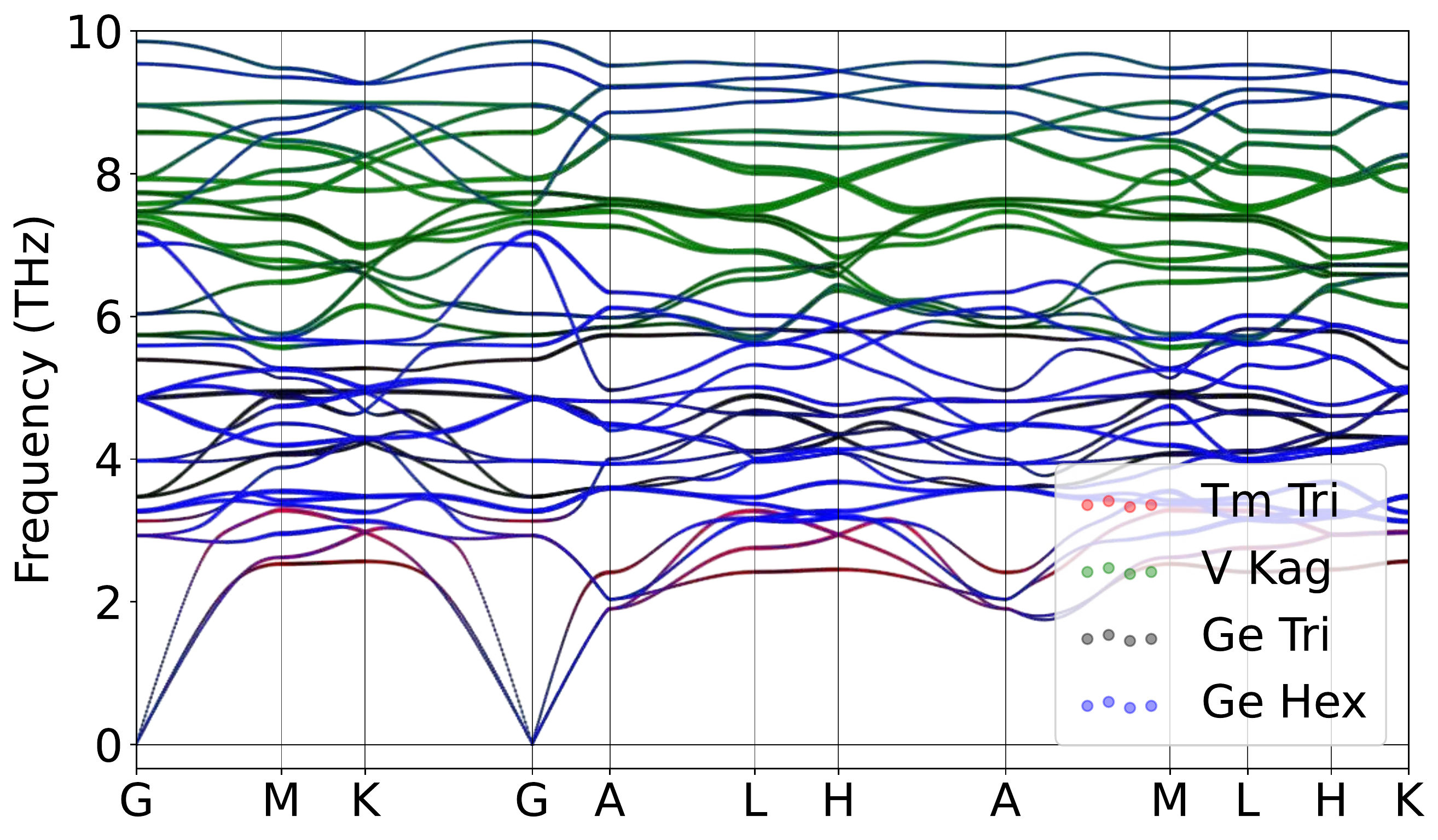}
}
\caption{Atomically projected phonon spectrum for TmV$_6$Ge$_6$.}
\label{fig:TmV6Ge6pho}
\end{figure}

\begin{figure}[H]
\centering
\label{fig:TmV6Ge6_relaxed_Low_xz}
\includegraphics[width=0.85\textwidth]{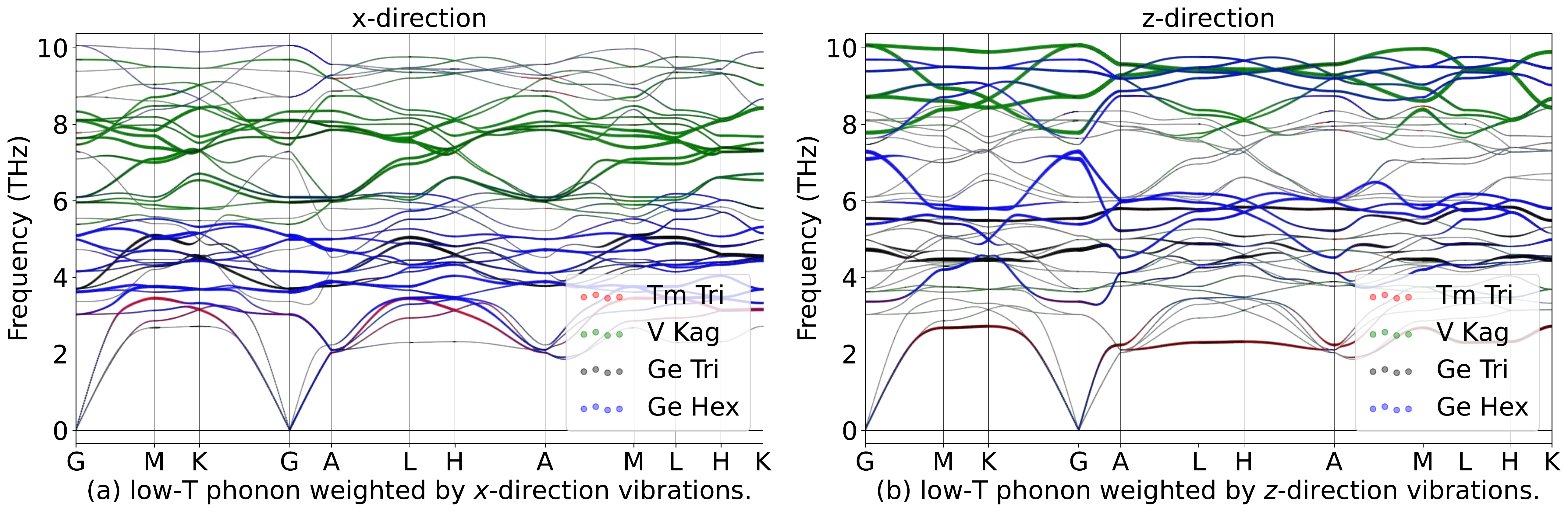}
\hspace{5pt}\label{fig:TmV6Ge6_relaxed_High_xz}
\includegraphics[width=0.85\textwidth]{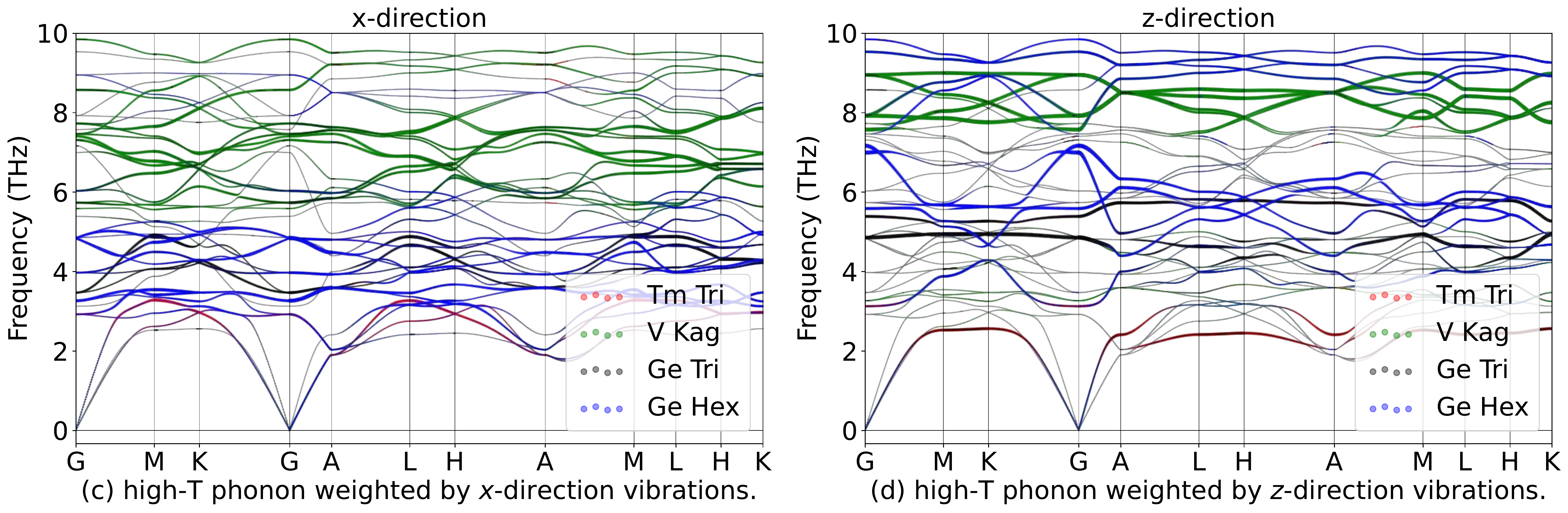}
\caption{Phonon spectrum for TmV$_6$Ge$_6$in in-plane ($x$) and out-of-plane ($z$) directions.}
\label{fig:TmV6Ge6phoxyz}
\end{figure}

\begin{table}[htbp]
\global\long\def\arraystretch{1.12}
\captionsetup{width=.95\textwidth}
\caption{IRREPs at high-symmetry points for TmV$_6$Ge$_6$ phonon spectrum at Low-T.}
\label{tab:191_TmV6Ge6_Low}
\setlength{\tabcolsep}{1.mm}{
}
\end{table}

\begin{table}[htbp]
\global\long\def\arraystretch{1.12}
\captionsetup{width=.95\textwidth}
\caption{IRREPs at high-symmetry points for TmV$_6$Ge$_6$ phonon spectrum at High-T.}
\label{tab:191_TmV6Ge6_High}
\setlength{\tabcolsep}{1.mm}{
%
}
\end{table}
\subsubsection{\label{sms2sec:YbV6Ge6}\hyperref[smsec:col]{\texorpdfstring{YbV$_6$Ge$_6$}{}}~{\color{green}  (High-T stable, Low-T stable) }}

\begin{table}[htbp]
\global\long\def\arraystretch{1.12}
\captionsetup{width=.85\textwidth}
\caption{Structure parameters of YbV$_6$Ge$_6$ after relaxation. It contains 400 electrons. }
\label{tab:YbV6Ge6}
\setlength{\tabcolsep}{4mm}{
%
}
\end{table}

\begin{figure}[htbp]
\centering
\includegraphics[width=0.85\textwidth]{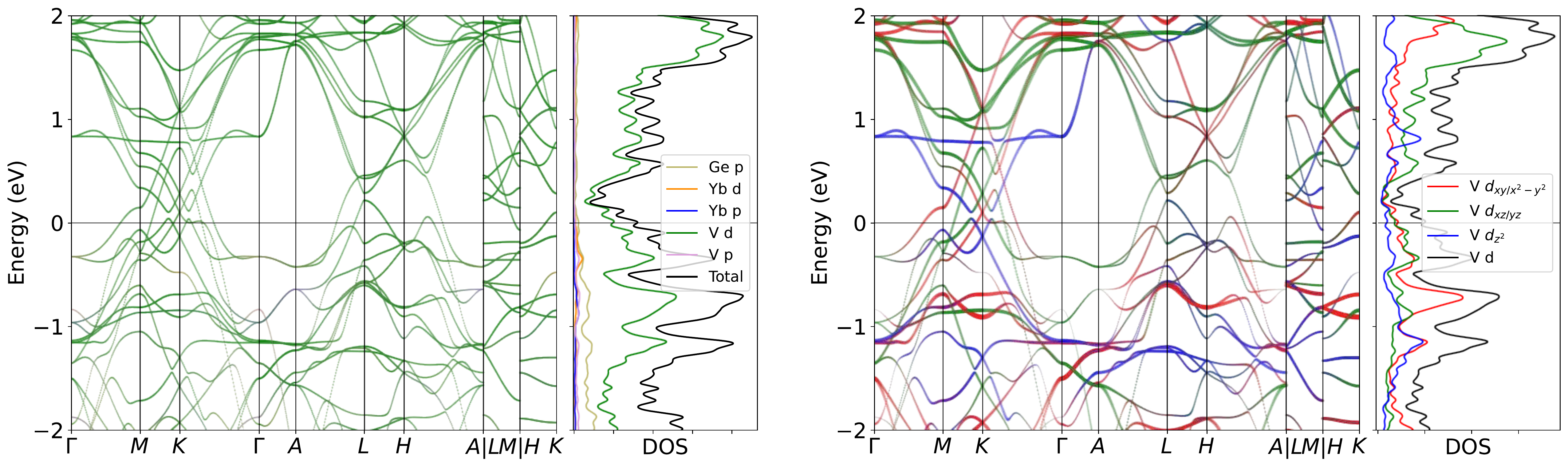}
\caption{Band structure for YbV$_6$Ge$_6$ without SOC. The projection of V $3d$ orbitals is also presented. The band structures clearly reveal the dominating of V $3d$ orbitals  near the Fermi level, as well as the pronounced peak.}
\label{fig:YbV6Ge6band}
\end{figure}

\begin{figure}[H]
\centering
\subfloat[Relaxed phonon at low-T.]{
\label{fig:YbV6Ge6_relaxed_Low}
\includegraphics[width=0.45\textwidth]{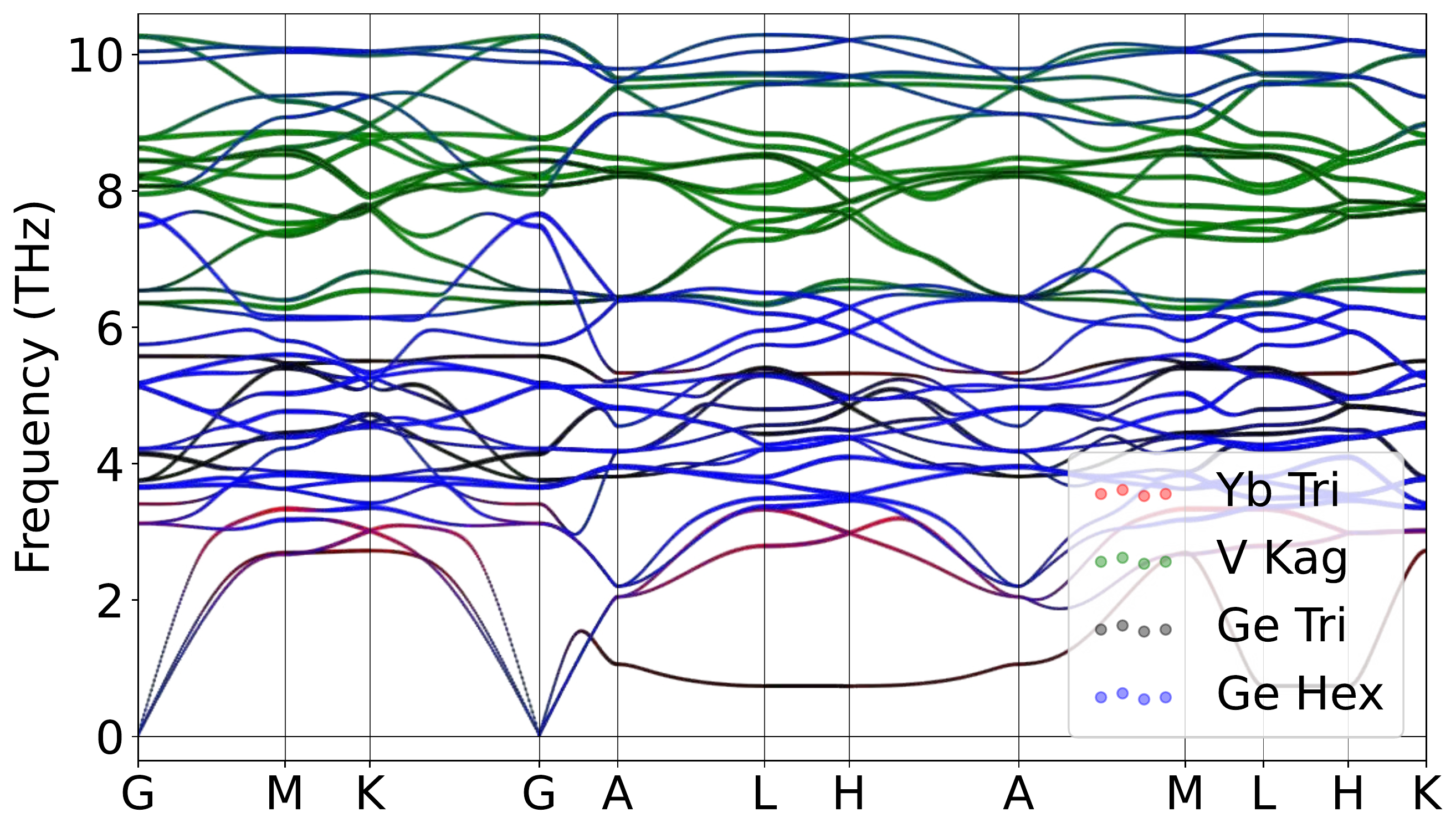}
}
\hspace{5pt}\subfloat[Relaxed phonon at high-T.]{
\label{fig:YbV6Ge6_relaxed_High}
\includegraphics[width=0.45\textwidth]{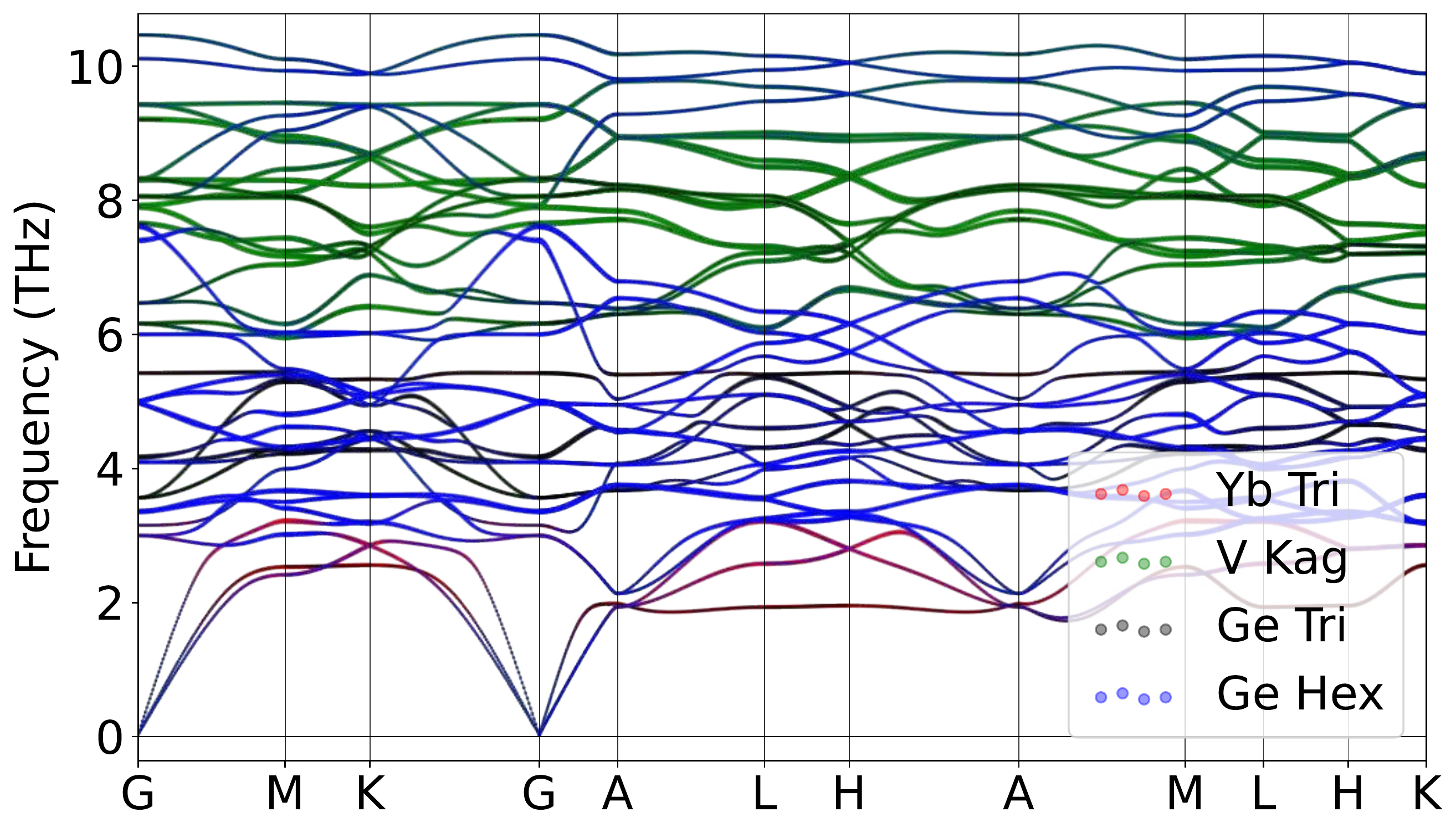}
}
\caption{Atomically projected phonon spectrum for YbV$_6$Ge$_6$.}
\label{fig:YbV6Ge6pho}
\end{figure}

\begin{figure}[H]
\centering
\label{fig:YbV6Ge6_relaxed_Low_xz}
\includegraphics[width=0.85\textwidth]{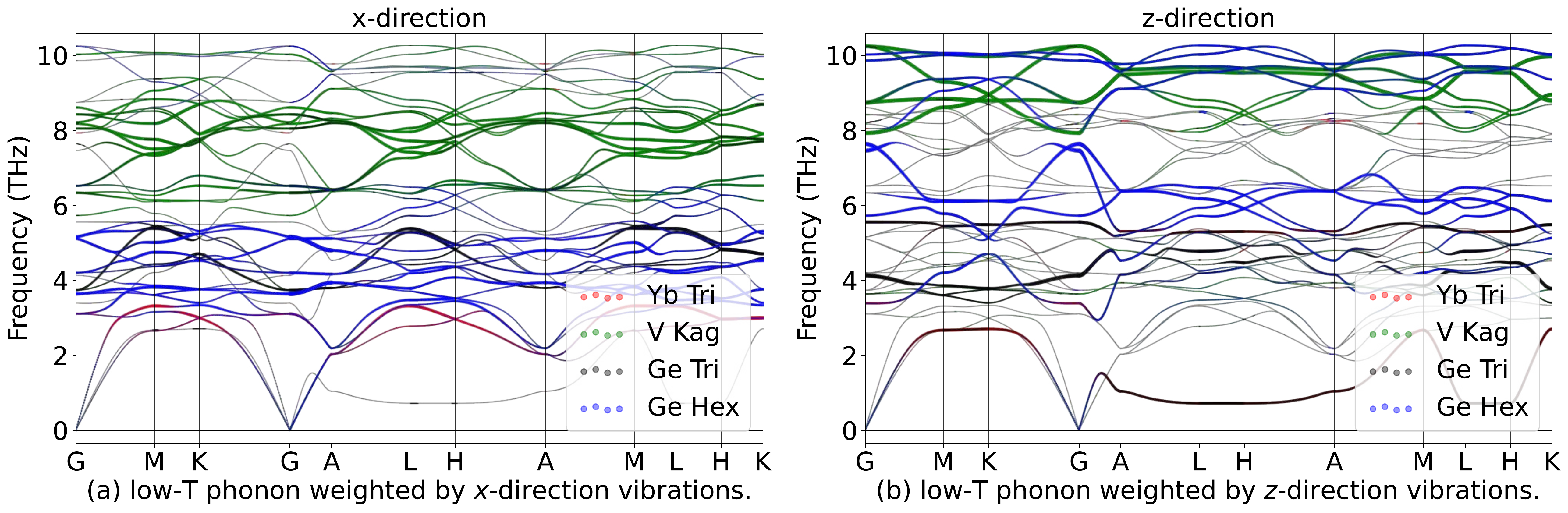}
\hspace{5pt}\label{fig:YbV6Ge6_relaxed_High_xz}
\includegraphics[width=0.85\textwidth]{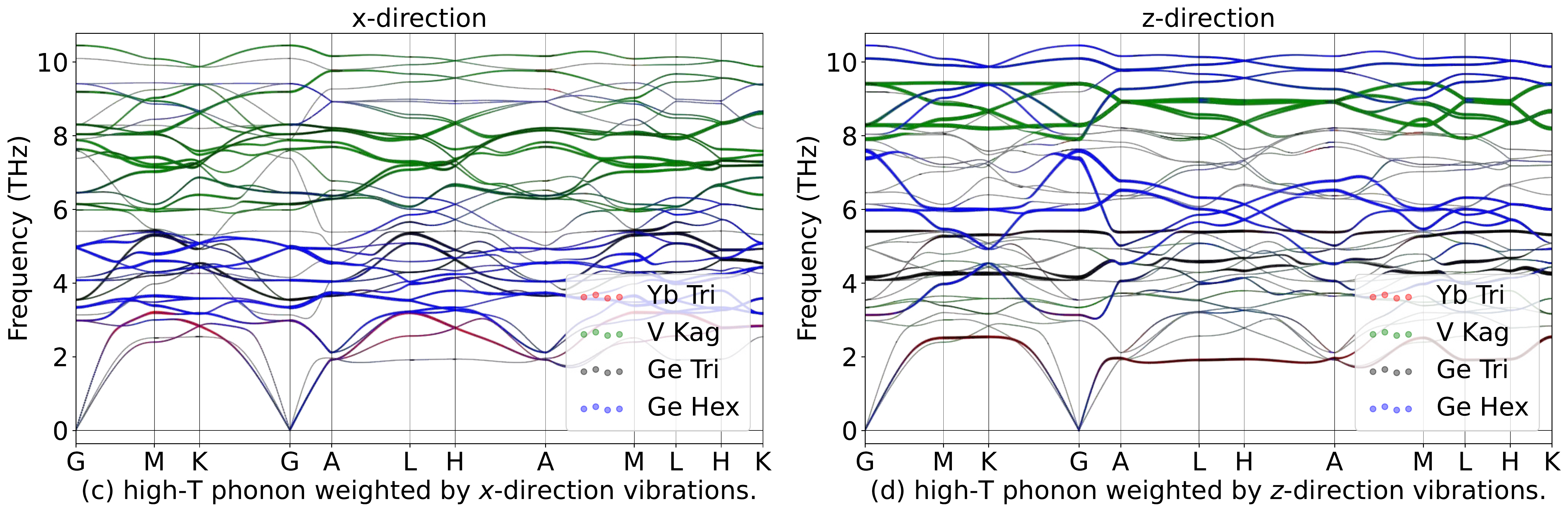}
\caption{Phonon spectrum for YbV$_6$Ge$_6$in in-plane ($x$) and out-of-plane ($z$) directions.}
\label{fig:YbV6Ge6phoxyz}
\end{figure}

\begin{table}[htbp]
\global\long\def\arraystretch{1.12}
\captionsetup{width=.95\textwidth}
\caption{IRREPs at high-symmetry points for YbV$_6$Ge$_6$ phonon spectrum at Low-T.}
\label{tab:191_YbV6Ge6_Low}
\setlength{\tabcolsep}{1.mm}{
}
\end{table}

\begin{table}[htbp]
\global\long\def\arraystretch{1.12}
\captionsetup{width=.95\textwidth}
\caption{IRREPs at high-symmetry points for YbV$_6$Ge$_6$ phonon spectrum at High-T.}
\label{tab:191_YbV6Ge6_High}
\setlength{\tabcolsep}{1.mm}{
%
}
\end{table}
\subsubsection{\label{sms2sec:LuV6Ge6}\hyperref[smsec:col]{\texorpdfstring{LuV$_6$Ge$_6$}{}}~{\color{green}  (High-T stable, Low-T stable) }}

\begin{table}[htbp]
\global\long\def\arraystretch{1.12}
\captionsetup{width=.85\textwidth}
\caption{Structure parameters of LuV$_6$Ge$_6$ after relaxation. It contains 401 electrons. }
\label{tab:LuV6Ge6}
\setlength{\tabcolsep}{4mm}{
%
}
\end{table}

\begin{figure}[htbp]
\centering
\includegraphics[width=0.85\textwidth]{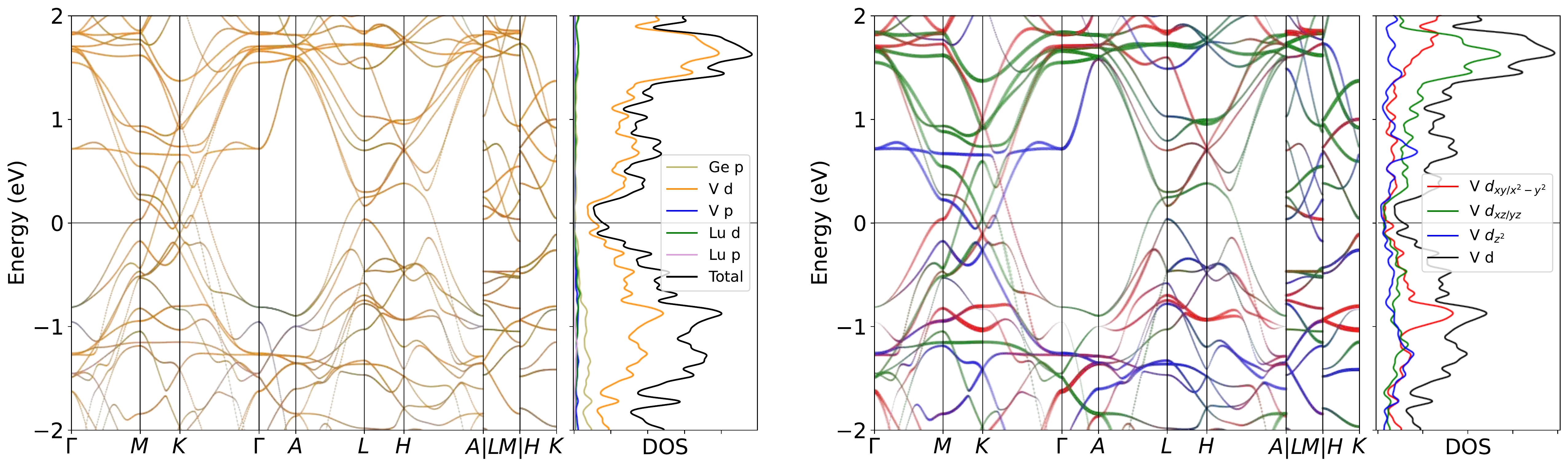}
\caption{Band structure for LuV$_6$Ge$_6$ without SOC. The projection of V $3d$ orbitals is also presented. The band structures clearly reveal the dominating of V $3d$ orbitals  near the Fermi level, as well as the pronounced peak.}
\label{fig:LuV6Ge6band}
\end{figure}

\begin{figure}[H]
\centering
\subfloat[Relaxed phonon at low-T.]{
\label{fig:LuV6Ge6_relaxed_Low}
\includegraphics[width=0.45\textwidth]{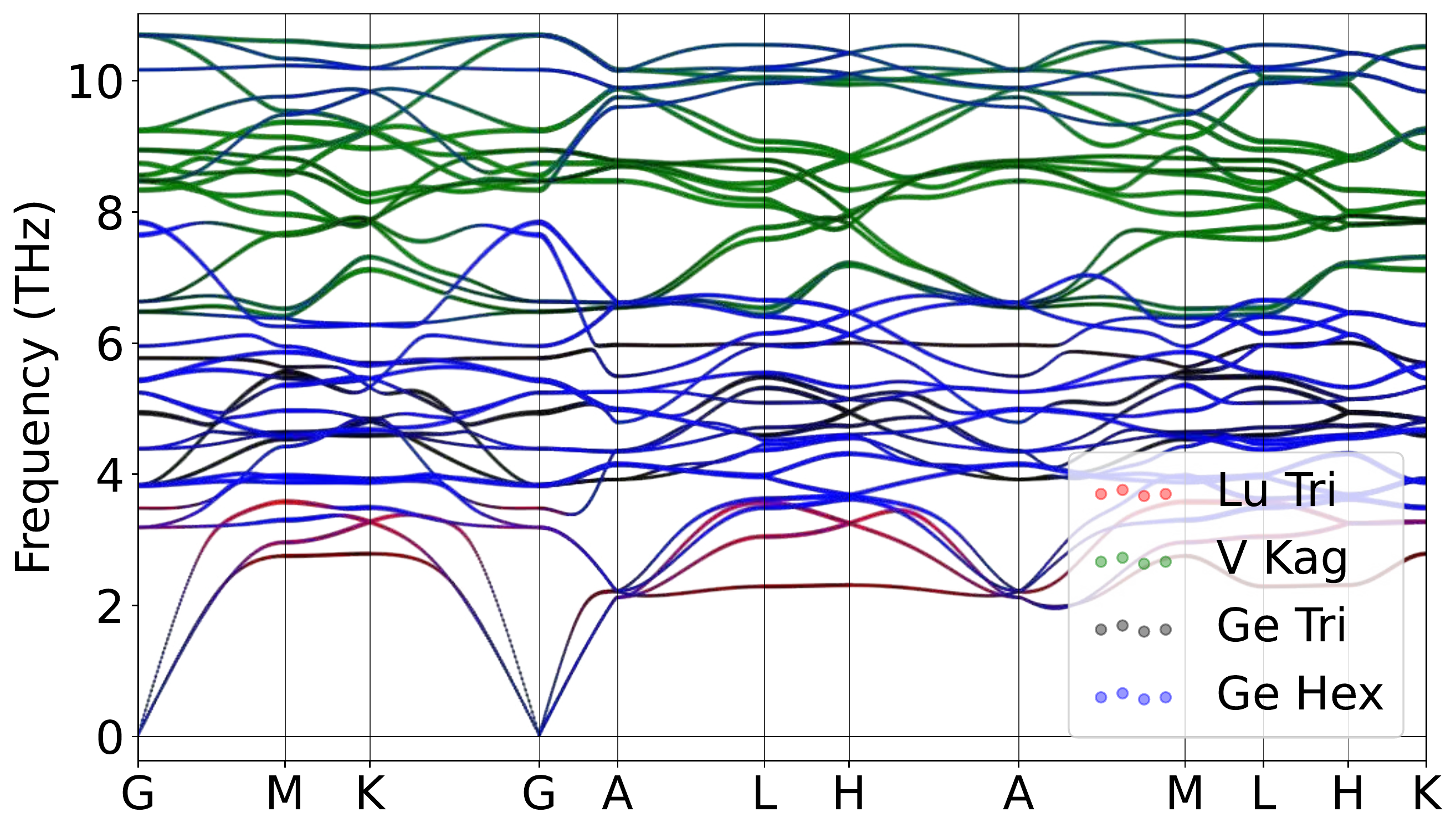}
}
\hspace{5pt}\subfloat[Relaxed phonon at high-T.]{
\label{fig:LuV6Ge6_relaxed_High}
\includegraphics[width=0.45\textwidth]{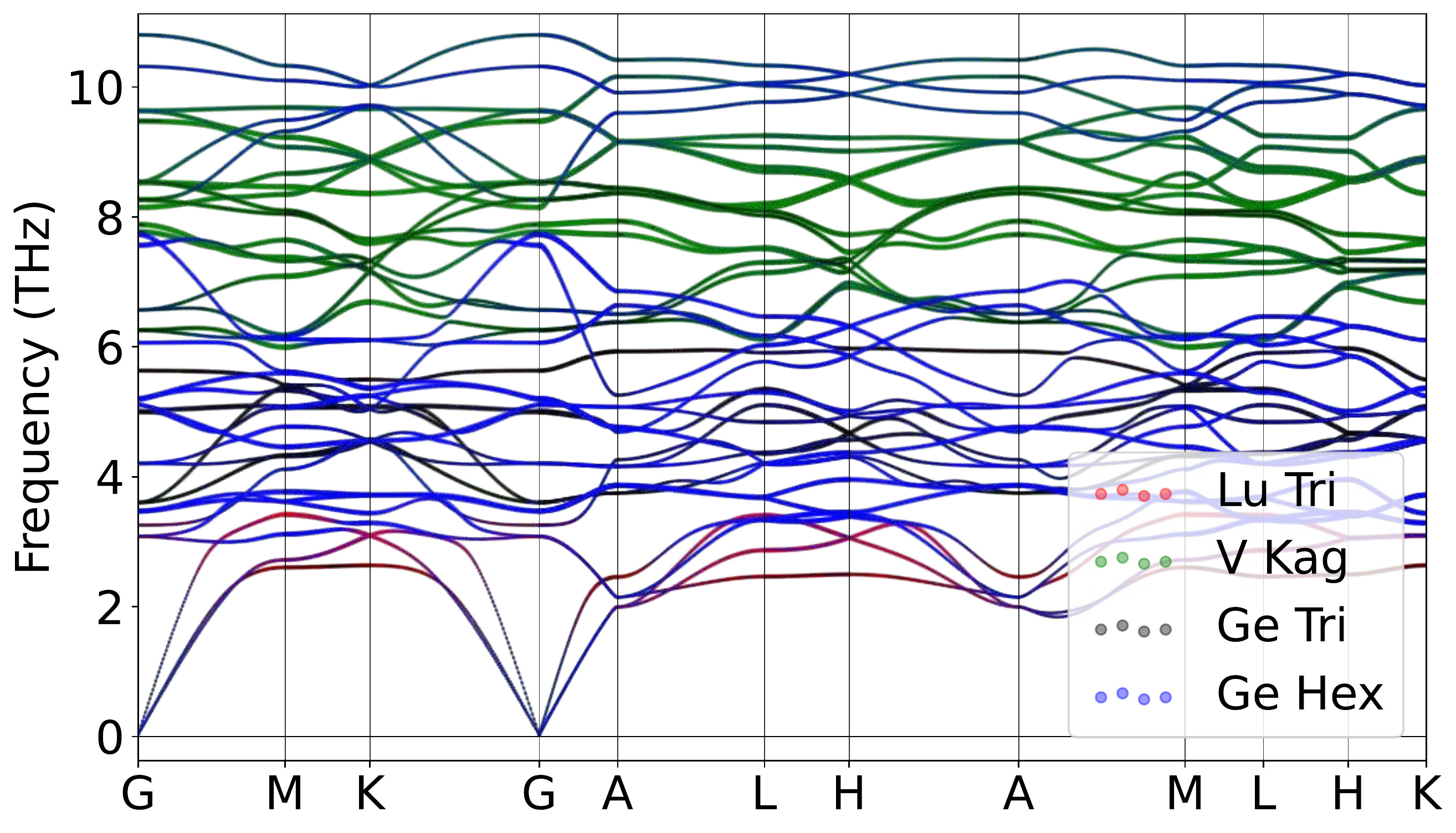}
}
\caption{Atomically projected phonon spectrum for LuV$_6$Ge$_6$.}
\label{fig:LuV6Ge6pho}
\end{figure}

\begin{figure}[H]
\centering
\label{fig:LuV6Ge6_relaxed_Low_xz}
\includegraphics[width=0.85\textwidth]{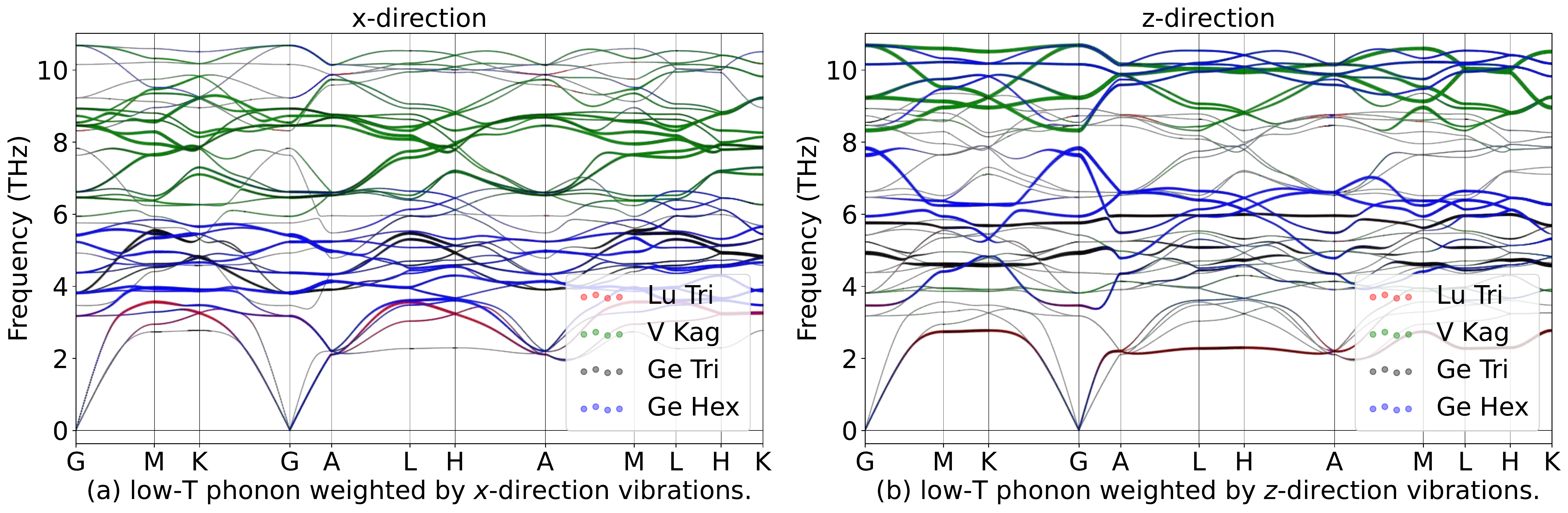}
\hspace{5pt}\label{fig:LuV6Ge6_relaxed_High_xz}
\includegraphics[width=0.85\textwidth]{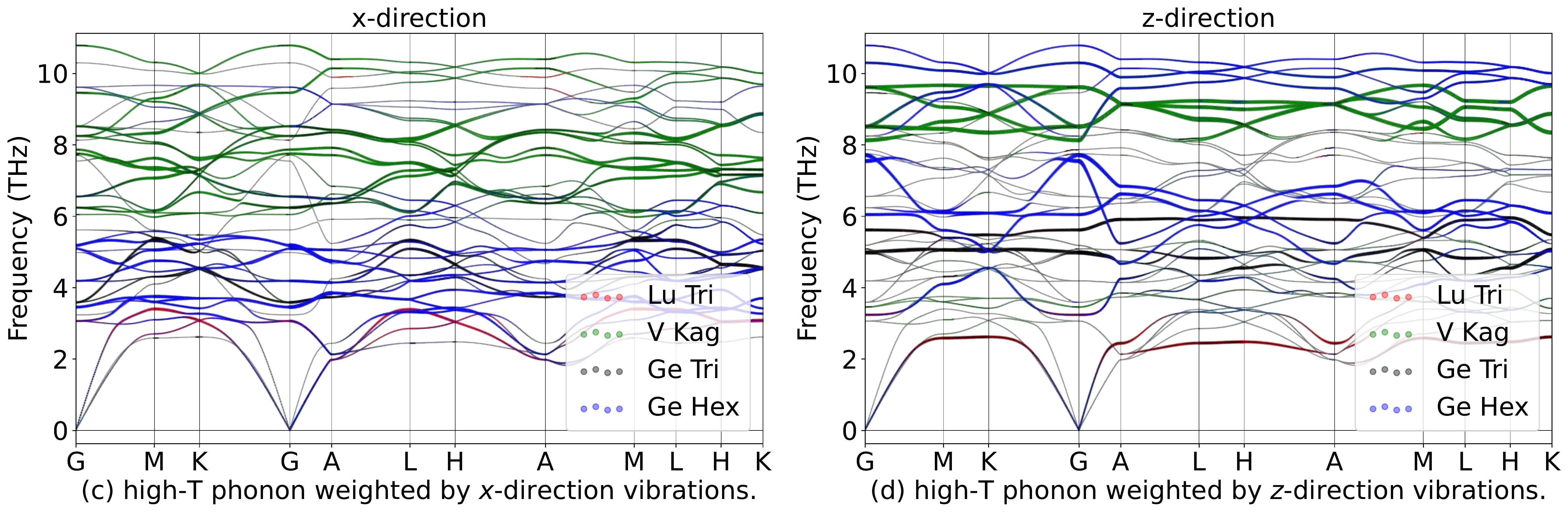}
\caption{Phonon spectrum for LuV$_6$Ge$_6$in in-plane ($x$) and out-of-plane ($z$) directions.}
\label{fig:LuV6Ge6phoxyz}
\end{figure}

\begin{table}[htbp]
\global\long\def\arraystretch{1.12}
\captionsetup{width=.95\textwidth}
\caption{IRREPs at high-symmetry points for LuV$_6$Ge$_6$ phonon spectrum at Low-T.}
\label{tab:191_LuV6Ge6_Low}
\setlength{\tabcolsep}{1.mm}{
}
\end{table}

\begin{table}[htbp]
\global\long\def\arraystretch{1.12}
\captionsetup{width=.95\textwidth}
\caption{IRREPs at high-symmetry points for LuV$_6$Ge$_6$ phonon spectrum at High-T.}
\label{tab:191_LuV6Ge6_High}
\setlength{\tabcolsep}{1.mm}{
%
}
\end{table}
\subsubsection{\label{sms2sec:HfV6Ge6}\hyperref[smsec:col]{\texorpdfstring{HfV$_6$Ge$_6$}{}}~{\color{green}  (High-T stable, Low-T stable) }}

\begin{table}[htbp]
\global\long\def\arraystretch{1.12}
\captionsetup{width=.85\textwidth}
\caption{Structure parameters of HfV$_6$Ge$_6$ after relaxation. It contains 402 electrons. }
\label{tab:HfV6Ge6}
\setlength{\tabcolsep}{4mm}{
%
}
\end{table}

\begin{figure}[htbp]
\centering
\includegraphics[width=0.85\textwidth]{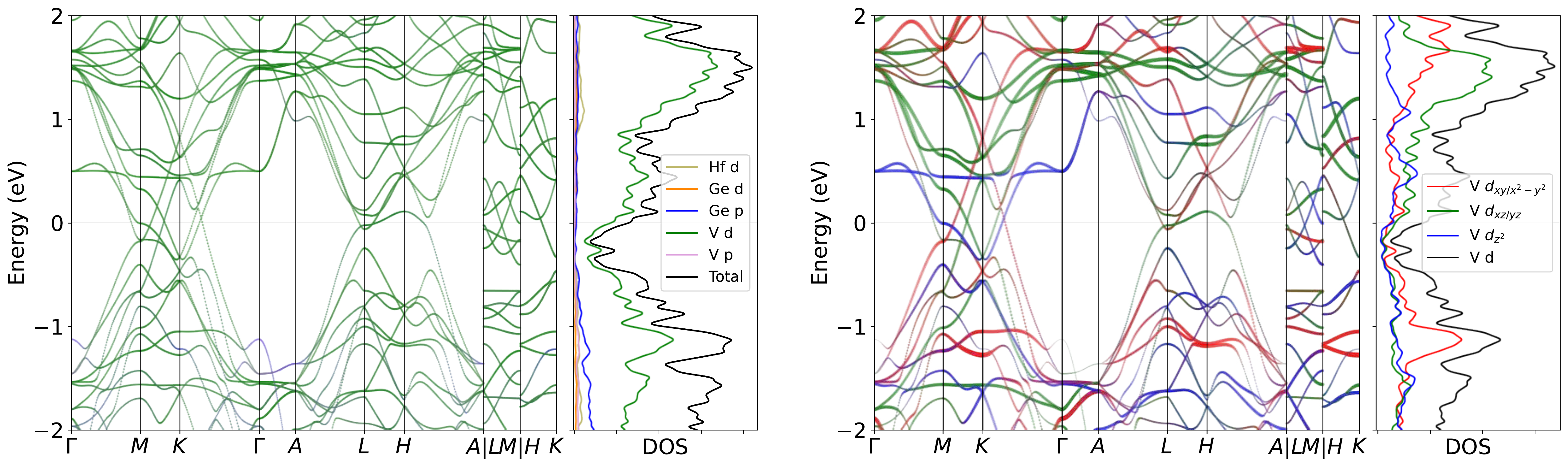}
\caption{Band structure for HfV$_6$Ge$_6$ without SOC. The projection of V $3d$ orbitals is also presented. The band structures clearly reveal the dominating of V $3d$ orbitals  near the Fermi level, as well as the pronounced peak.}
\label{fig:HfV6Ge6band}
\end{figure}

\begin{figure}[H]
\centering
\subfloat[Relaxed phonon at low-T.]{
\label{fig:HfV6Ge6_relaxed_Low}
\includegraphics[width=0.45\textwidth]{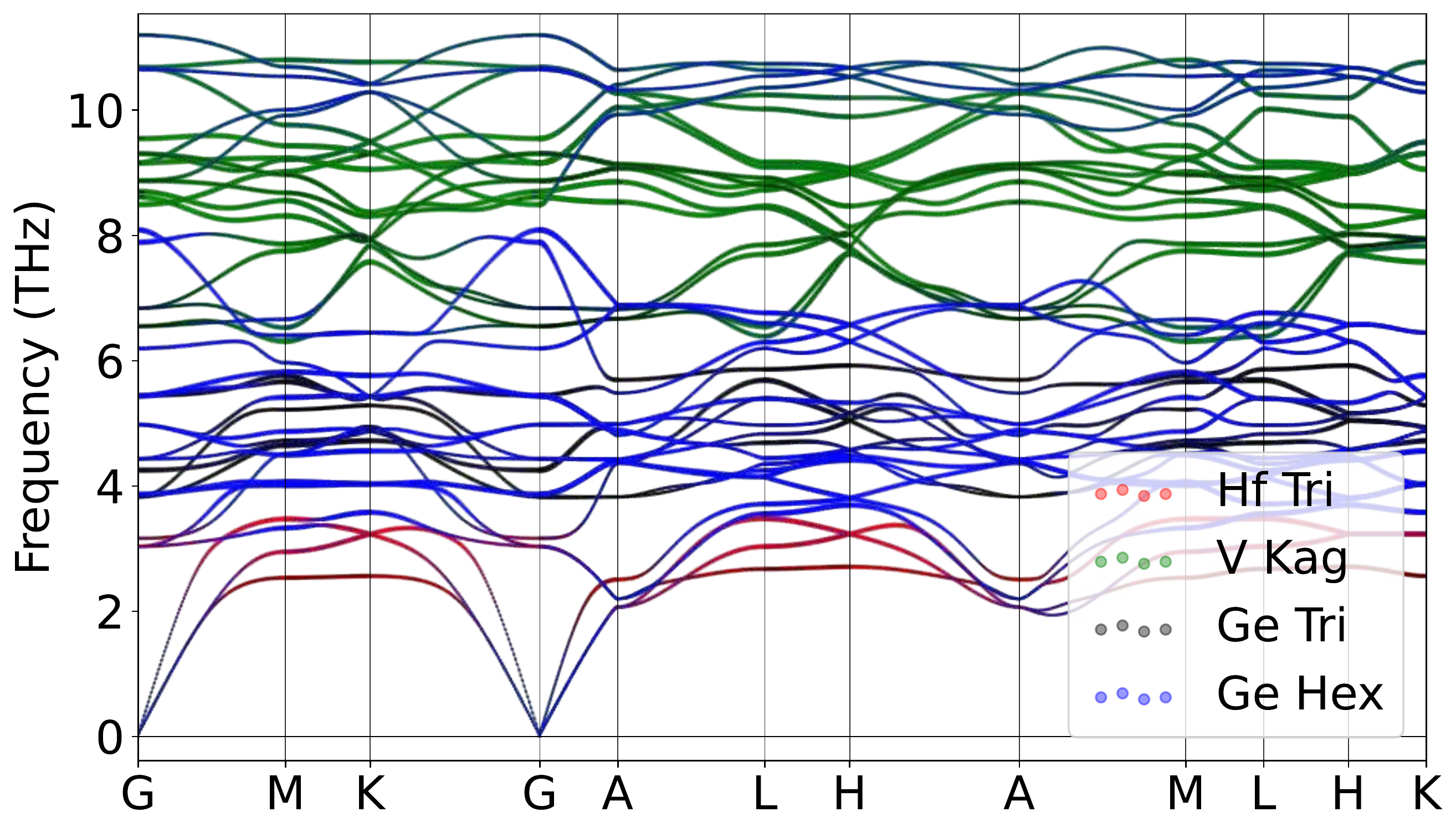}
}
\hspace{5pt}\subfloat[Relaxed phonon at high-T.]{
\label{fig:HfV6Ge6_relaxed_High}
\includegraphics[width=0.45\textwidth]{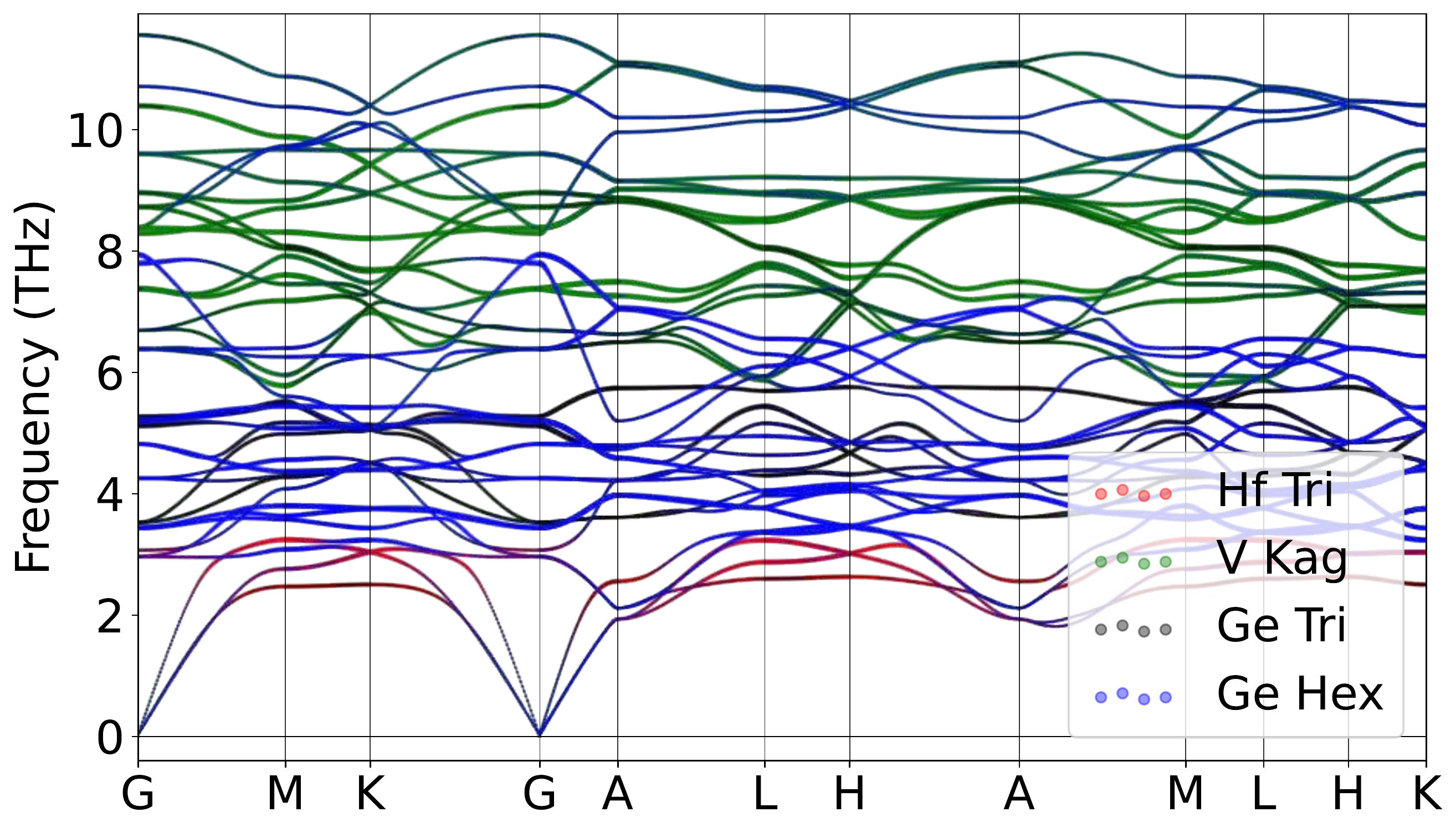}
}
\caption{Atomically projected phonon spectrum for HfV$_6$Ge$_6$.}
\label{fig:HfV6Ge6pho}
\end{figure}

\begin{figure}[H]
\centering
\label{fig:HfV6Ge6_relaxed_Low_xz}
\includegraphics[width=0.85\textwidth]{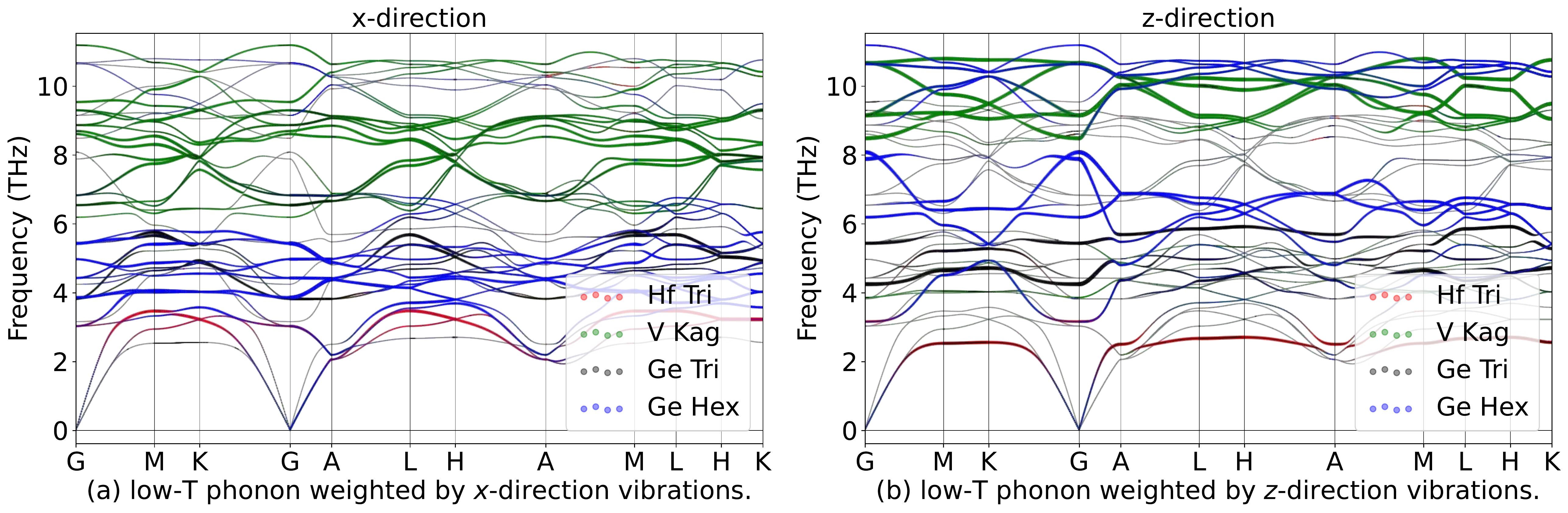}
\hspace{5pt}\label{fig:HfV6Ge6_relaxed_High_xz}
\includegraphics[width=0.85\textwidth]{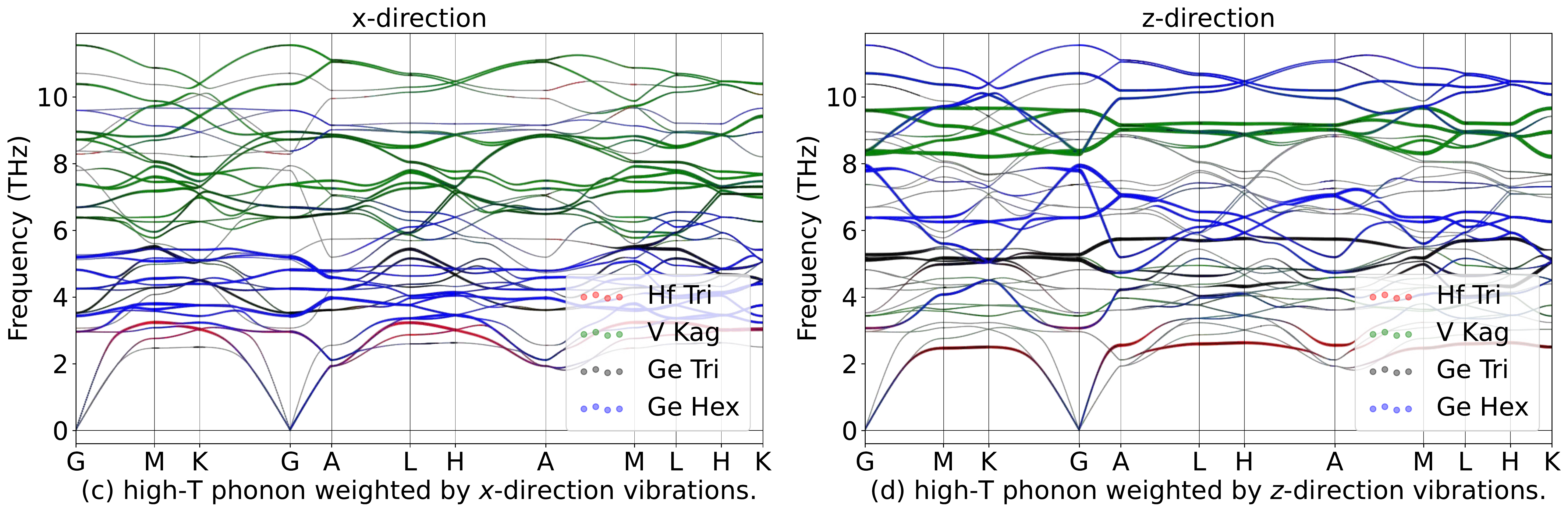}
\caption{Phonon spectrum for HfV$_6$Ge$_6$in in-plane ($x$) and out-of-plane ($z$) directions.}
\label{fig:HfV6Ge6phoxyz}
\end{figure}

\begin{table}[htbp]
\global\long\def\arraystretch{1.12}
\captionsetup{width=.95\textwidth}
\caption{IRREPs at high-symmetry points for HfV$_6$Ge$_6$ phonon spectrum at Low-T.}
\label{tab:191_HfV6Ge6_Low}
\setlength{\tabcolsep}{1.mm}{
}
\end{table}

\begin{table}[htbp]
\global\long\def\arraystretch{1.12}
\captionsetup{width=.95\textwidth}
\caption{IRREPs at high-symmetry points for HfV$_6$Ge$_6$ phonon spectrum at High-T.}
\label{tab:191_HfV6Ge6_High}
\setlength{\tabcolsep}{1.mm}{
%
}
\end{table}
\subsubsection{\label{sms2sec:TaV6Ge6}\hyperref[smsec:col]{\texorpdfstring{TaV$_6$Ge$_6$}{}}~{\color{green}  (High-T stable, Low-T stable) }}

\begin{table}[htbp]
\global\long\def\arraystretch{1.12}
\captionsetup{width=.85\textwidth}
\caption{Structure parameters of TaV$_6$Ge$_6$ after relaxation. It contains 403 electrons. }
\label{tab:TaV6Ge6}
\setlength{\tabcolsep}{4mm}{
%
}
\end{table}

\begin{figure}[htbp]
\centering
\includegraphics[width=0.85\textwidth]{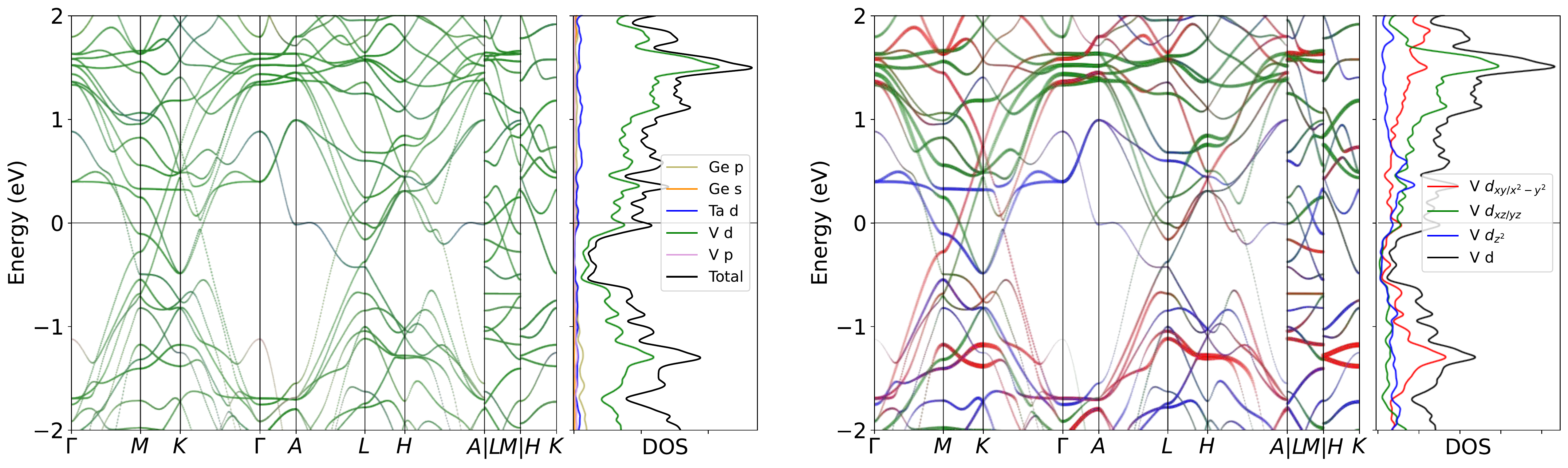}
\caption{Band structure for TaV$_6$Ge$_6$ without SOC. The projection of V $3d$ orbitals is also presented. The band structures clearly reveal the dominating of V $3d$ orbitals  near the Fermi level, as well as the pronounced peak.}
\label{fig:TaV6Ge6band}
\end{figure}

\begin{figure}[H]
\centering
\subfloat[Relaxed phonon at low-T.]{
\label{fig:TaV6Ge6_relaxed_Low}
\includegraphics[width=0.45\textwidth]{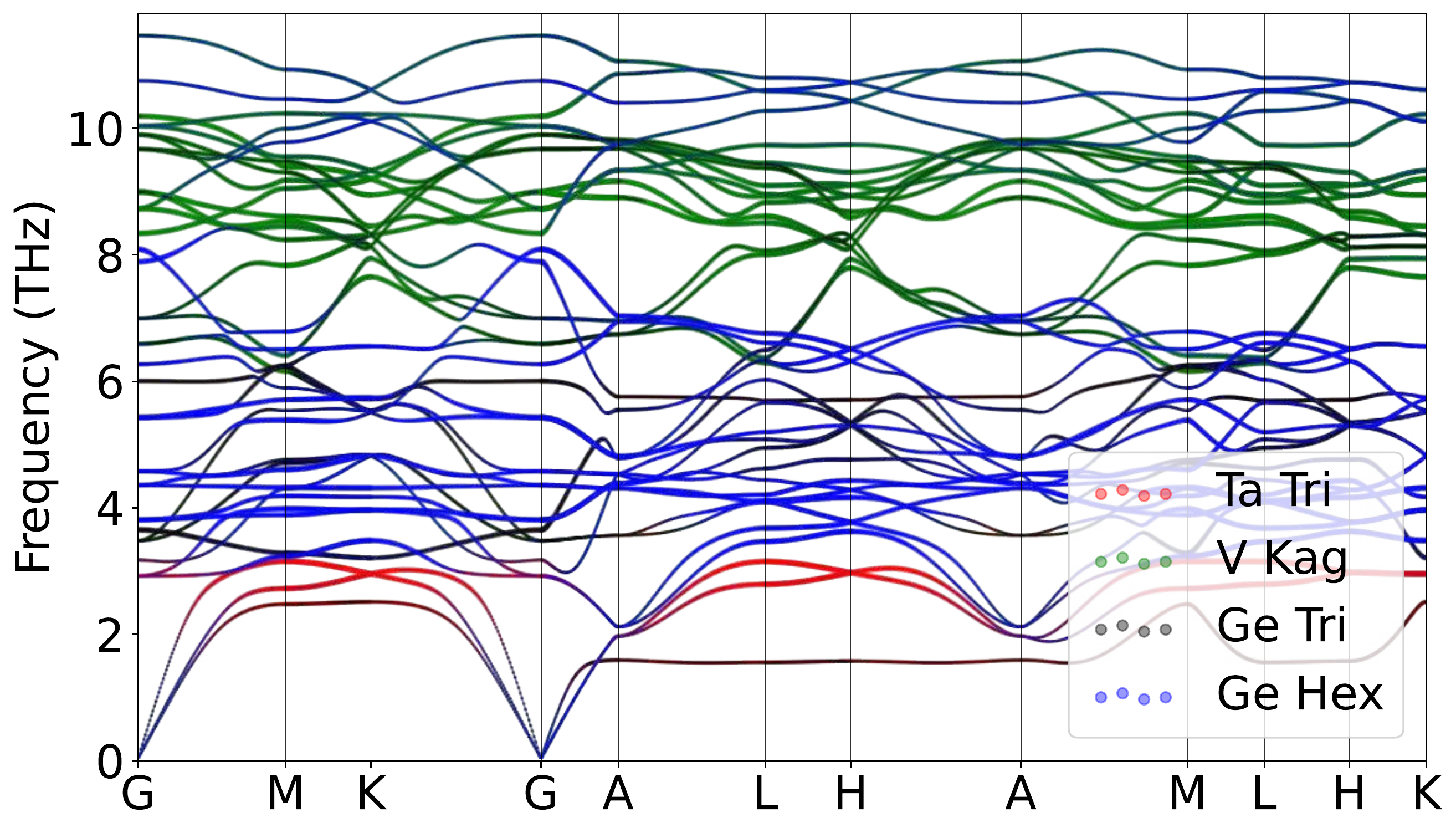}
}
\hspace{5pt}\subfloat[Relaxed phonon at high-T.]{
\label{fig:TaV6Ge6_relaxed_High}
\includegraphics[width=0.45\textwidth]{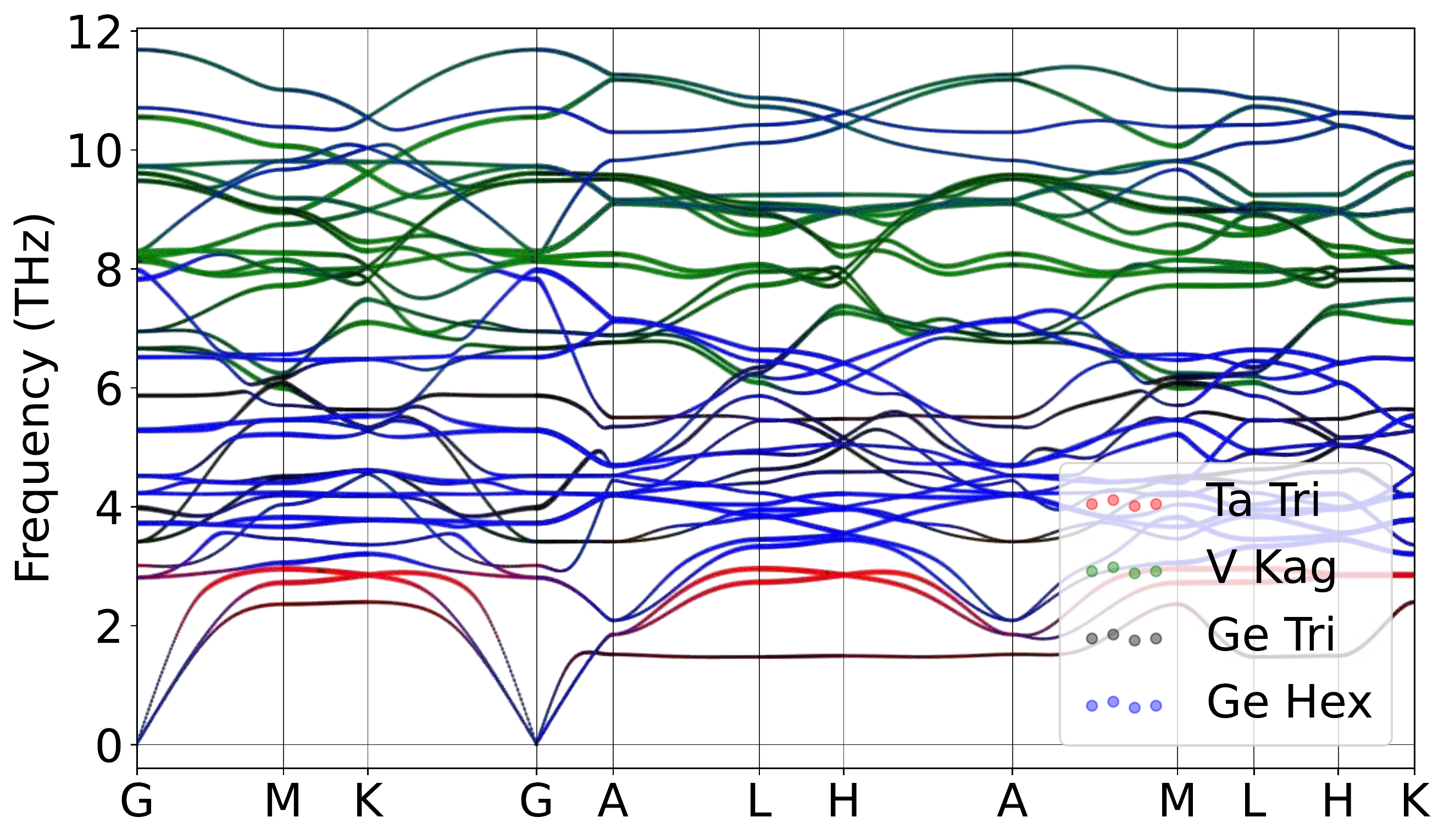}
}
\caption{Atomically projected phonon spectrum for TaV$_6$Ge$_6$.}
\label{fig:TaV6Ge6pho}
\end{figure}

\begin{figure}[H]
\centering
\label{fig:TaV6Ge6_relaxed_Low_xz}
\includegraphics[width=0.85\textwidth]{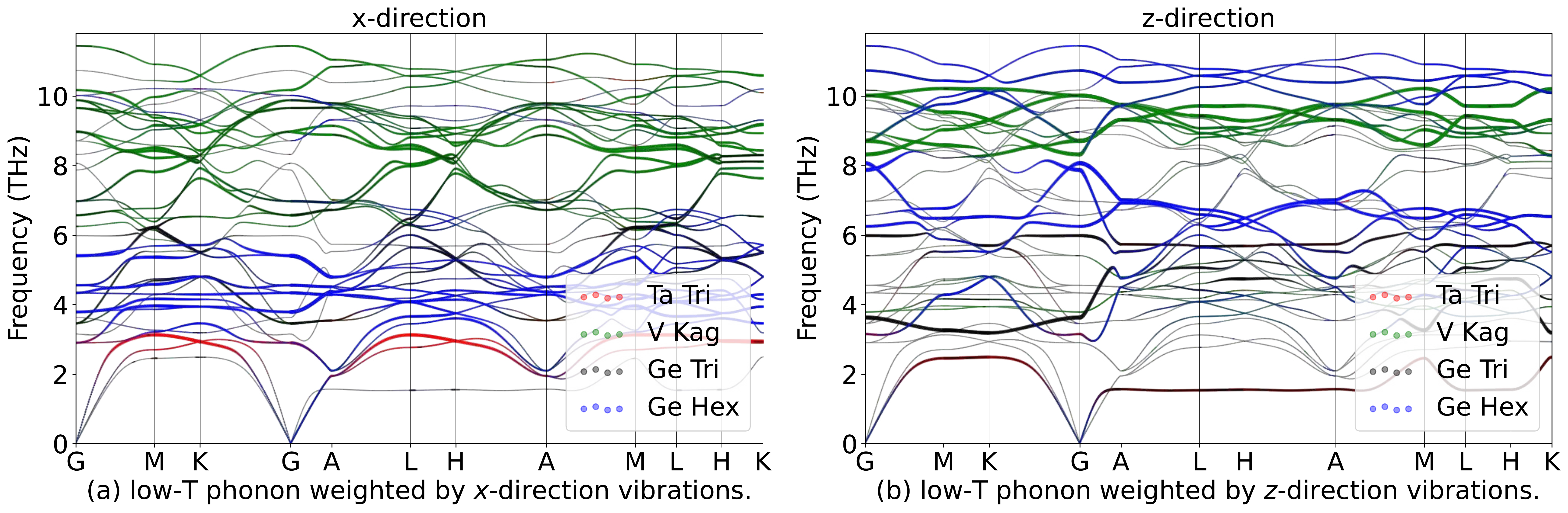}
\hspace{5pt}\label{fig:TaV6Ge6_relaxed_High_xz}
\includegraphics[width=0.85\textwidth]{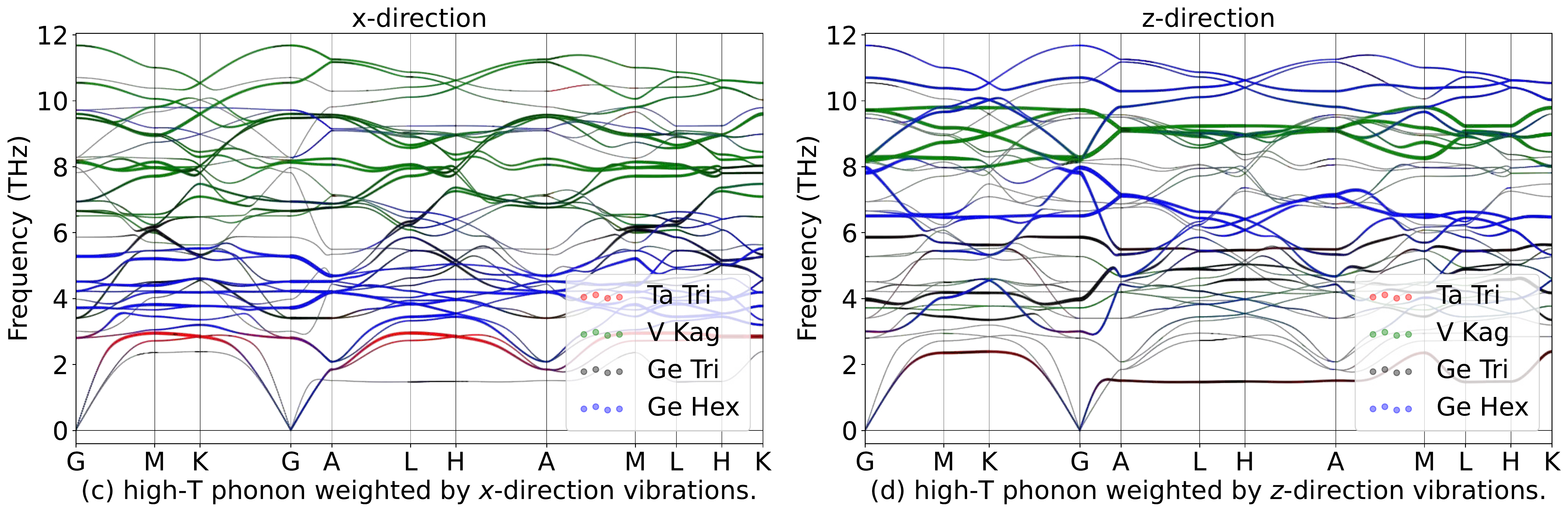}
\caption{Phonon spectrum for TaV$_6$Ge$_6$in in-plane ($x$) and out-of-plane ($z$) directions.}
\label{fig:TaV6Ge6phoxyz}
\end{figure}

\begin{table}[htbp]
\global\long\def\arraystretch{1.12}
\captionsetup{width=.95\textwidth}
\caption{IRREPs at high-symmetry points for TaV$_6$Ge$_6$ phonon spectrum at Low-T.}
\label{tab:191_TaV6Ge6_Low}
\setlength{\tabcolsep}{1.mm}{
}
\end{table}

\begin{table}[htbp]
\global\long\def\arraystretch{1.12}
\captionsetup{width=.95\textwidth}
\caption{IRREPs at high-symmetry points for TaV$_6$Ge$_6$ phonon spectrum at High-T.}
\label{tab:191_TaV6Ge6_High}
\setlength{\tabcolsep}{1.mm}{
%
}
\end{table}
\subsection{\hyperref[smsec:col]{VSn}}~\label{smssec:VSn}

\subsubsection{\label{sms2sec:LiV6Sn6}\hyperref[smsec:col]{\texorpdfstring{LiV$_6$Sn$_6$}{}}~{\color{green}  (High-T stable, Low-T stable) }}

\begin{table}[htbp]
\global\long\def\arraystretch{1.12}
\captionsetup{width=.85\textwidth}
\caption{Structure parameters of LiV$_6$Sn$_6$ after relaxation. It contains 441 electrons. }
\label{tab:LiV6Sn6}
\setlength{\tabcolsep}{4mm}{
%
}
\end{table}

\begin{figure}[htbp]
\centering
\includegraphics[width=0.85\textwidth]{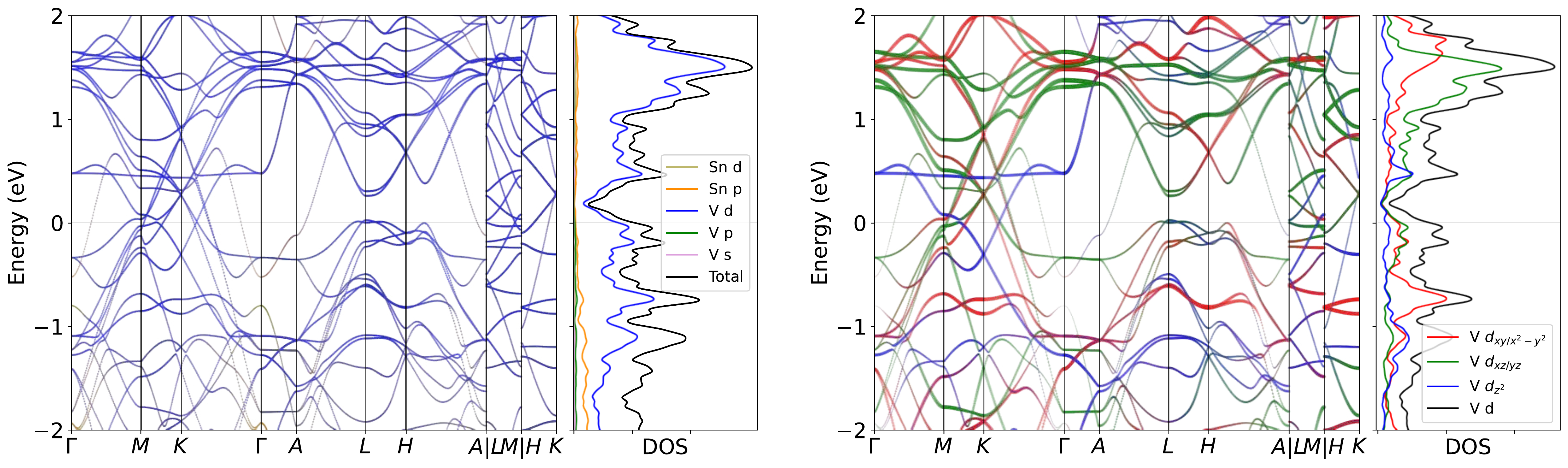}
\caption{Band structure for LiV$_6$Sn$_6$ without SOC. The projection of V $3d$ orbitals is also presented. The band structures clearly reveal the dominating of V $3d$ orbitals  near the Fermi level, as well as the pronounced peak.}
\label{fig:LiV6Sn6band}
\end{figure}

\begin{figure}[H]
\centering
\subfloat[Relaxed phonon at low-T.]{
\label{fig:LiV6Sn6_relaxed_Low}
\includegraphics[width=0.45\textwidth]{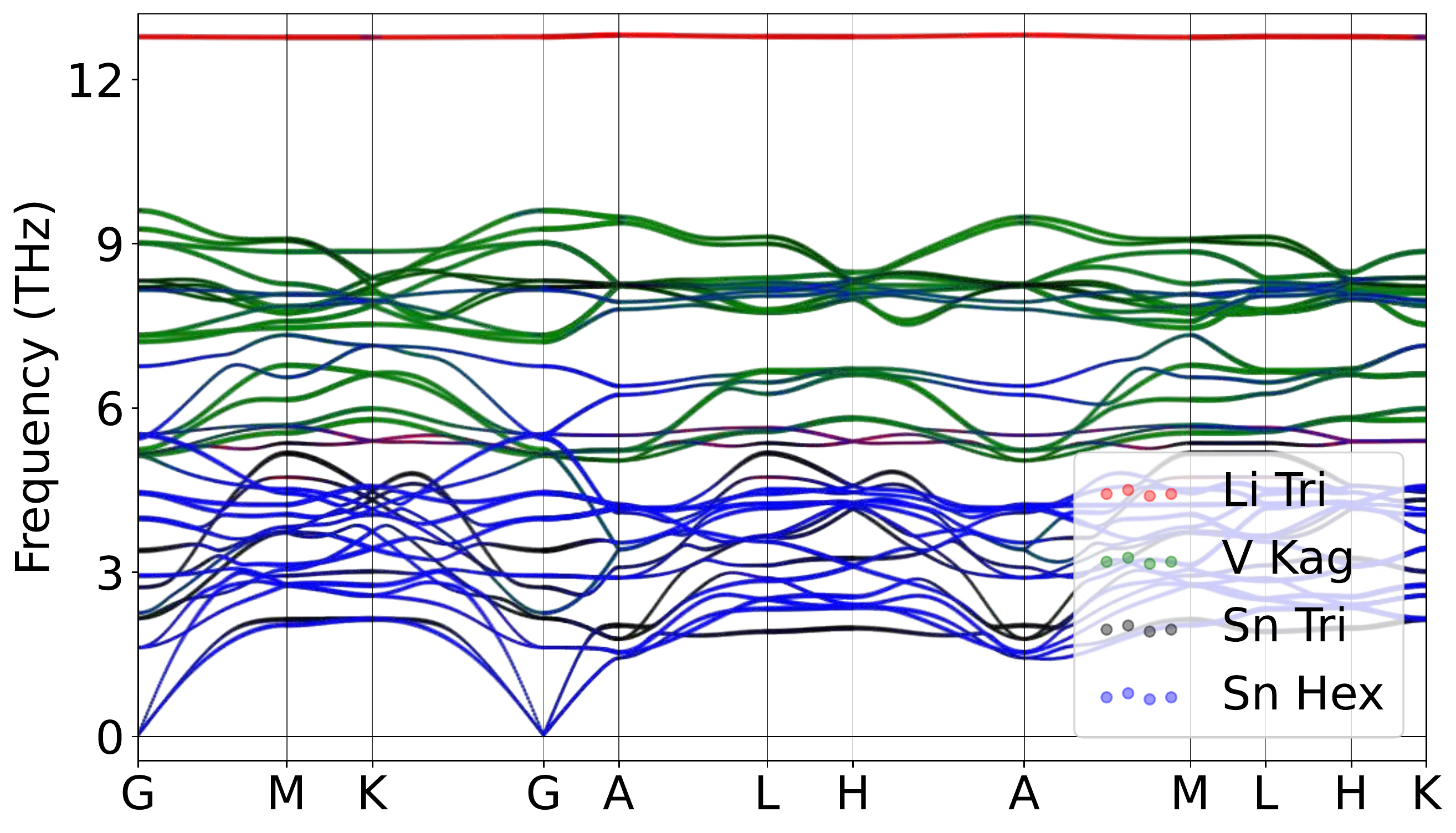}
}
\hspace{5pt}\subfloat[Relaxed phonon at high-T.]{
\label{fig:LiV6Sn6_relaxed_High}
\includegraphics[width=0.45\textwidth]{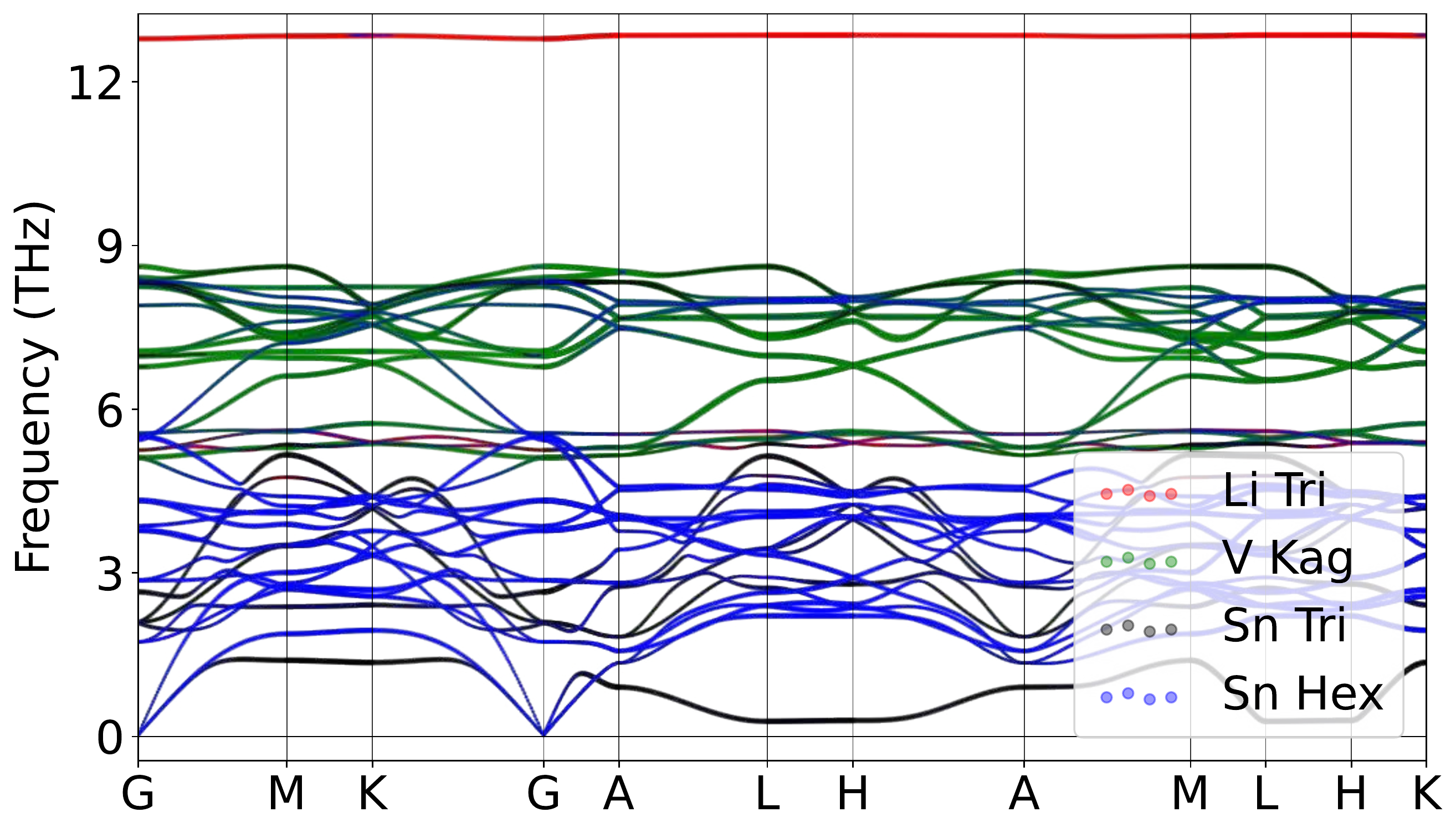}
}
\caption{Atomically projected phonon spectrum for LiV$_6$Sn$_6$.}
\label{fig:LiV6Sn6pho}
\end{figure}

\begin{figure}[H]
\centering
\label{fig:LiV6Sn6_relaxed_Low_xz}
\includegraphics[width=0.85\textwidth]{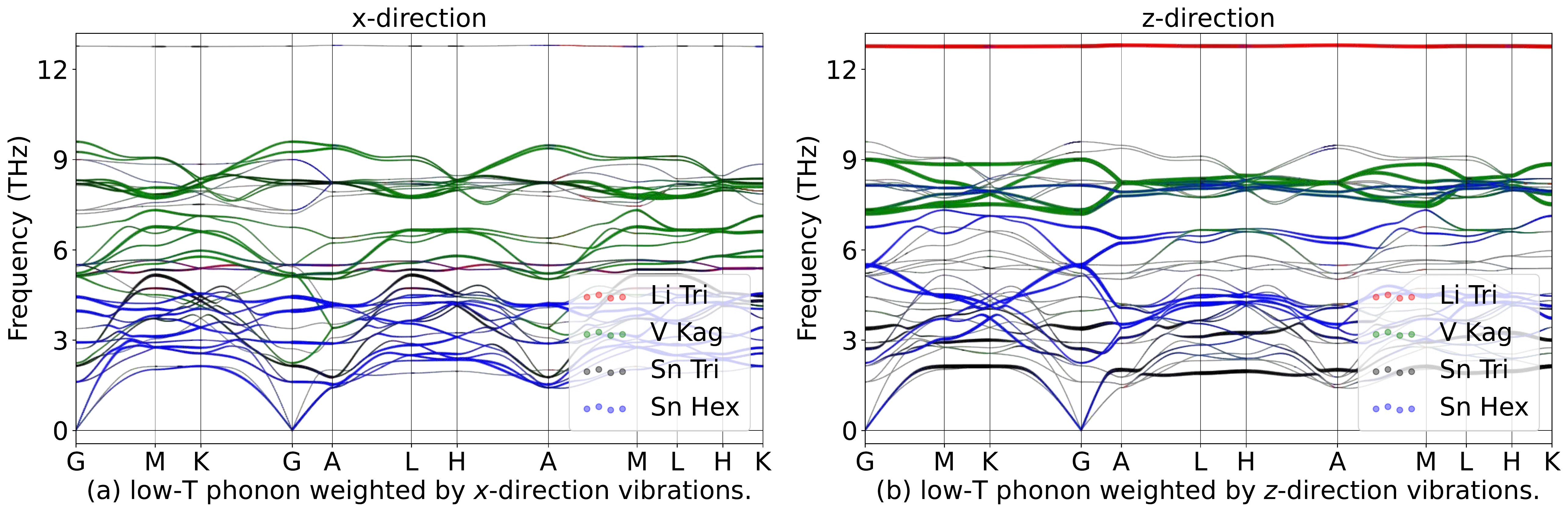}
\hspace{5pt}\label{fig:LiV6Sn6_relaxed_High_xz}
\includegraphics[width=0.85\textwidth]{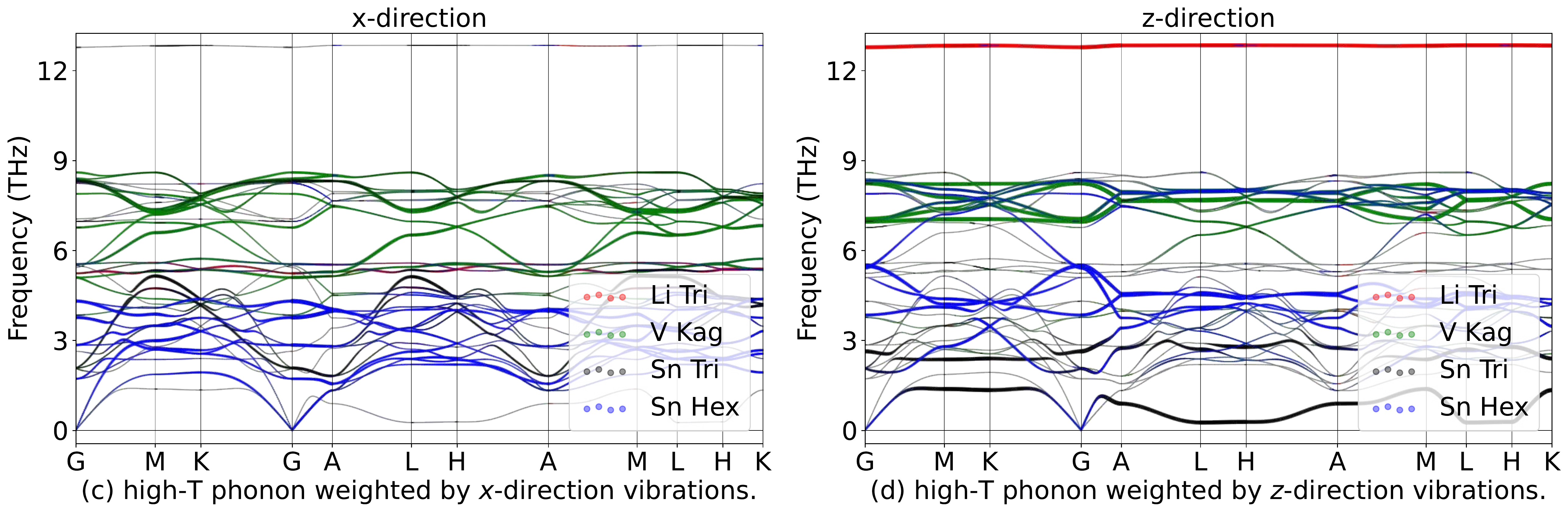}
\caption{Phonon spectrum for LiV$_6$Sn$_6$in in-plane ($x$) and out-of-plane ($z$) directions.}
\label{fig:LiV6Sn6phoxyz}
\end{figure}

\begin{table}[htbp]
\global\long\def\arraystretch{1.12}
\captionsetup{width=.95\textwidth}
\caption{IRREPs at high-symmetry points for LiV$_6$Sn$_6$ phonon spectrum at Low-T.}
\label{tab:191_LiV6Sn6_Low}
\setlength{\tabcolsep}{1.mm}{
}
\end{table}

\begin{table}[htbp]
\global\long\def\arraystretch{1.12}
\captionsetup{width=.95\textwidth}
\caption{IRREPs at high-symmetry points for LiV$_6$Sn$_6$ phonon spectrum at High-T.}
\label{tab:191_LiV6Sn6_High}
\setlength{\tabcolsep}{1.mm}{
%
}
\end{table}
\subsubsection{\label{sms2sec:MgV6Sn6}\hyperref[smsec:col]{\texorpdfstring{MgV$_6$Sn$_6$}{}}~{\color{red}  (High-T unstable, Low-T unstable) }}

\begin{table}[htbp]
\global\long\def\arraystretch{1.12}
\captionsetup{width=.85\textwidth}
\caption{Structure parameters of MgV$_6$Sn$_6$ after relaxation. It contains 450 electrons. }
\label{tab:MgV6Sn6}
\setlength{\tabcolsep}{4mm}{
%
}
\end{table}

\begin{figure}[htbp]
\centering
\includegraphics[width=0.85\textwidth]{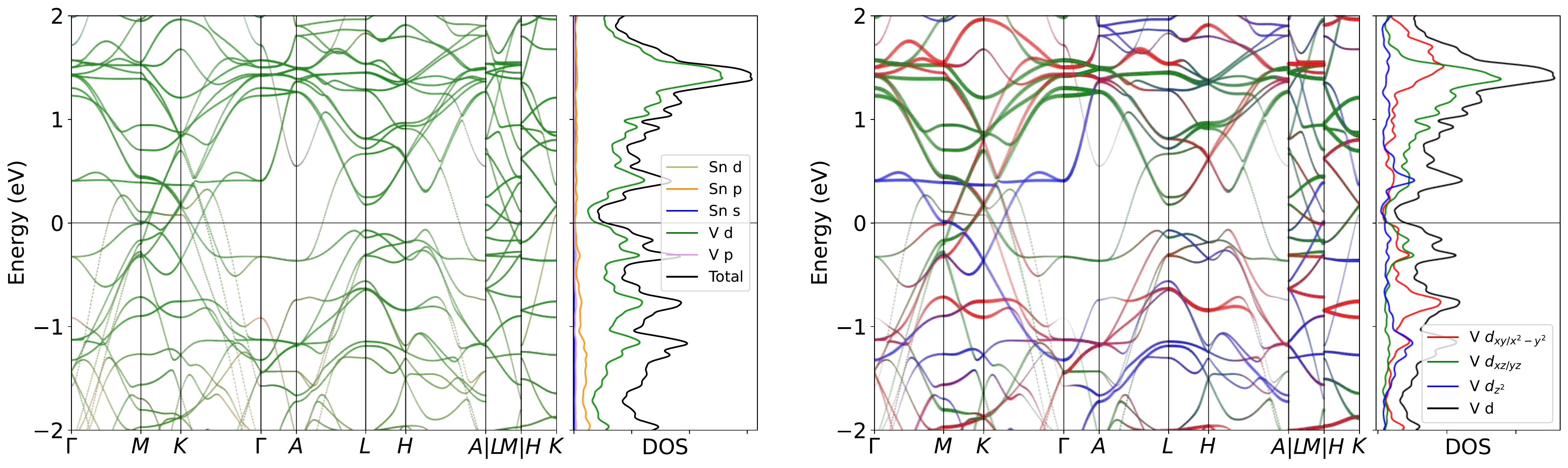}
\caption{Band structure for MgV$_6$Sn$_6$ without SOC. The projection of V $3d$ orbitals is also presented. The band structures clearly reveal the dominating of V $3d$ orbitals  near the Fermi level, as well as the pronounced peak.}
\label{fig:MgV6Sn6band}
\end{figure}

\begin{figure}[H]
\centering
\subfloat[Relaxed phonon at low-T.]{
\label{fig:MgV6Sn6_relaxed_Low}
\includegraphics[width=0.45\textwidth]{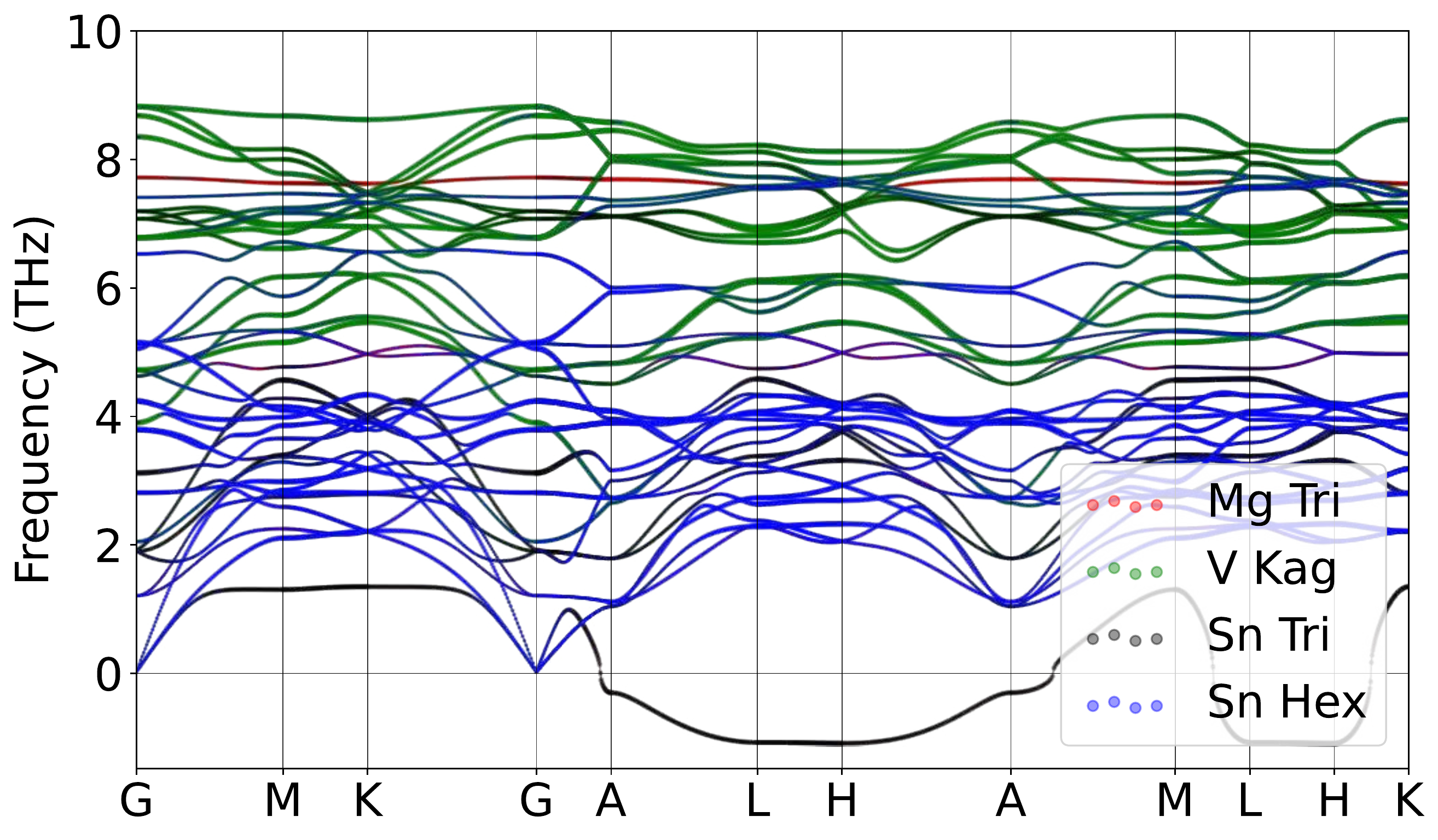}
}
\hspace{5pt}\subfloat[Relaxed phonon at high-T.]{
\label{fig:MgV6Sn6_relaxed_High}
\includegraphics[width=0.45\textwidth]{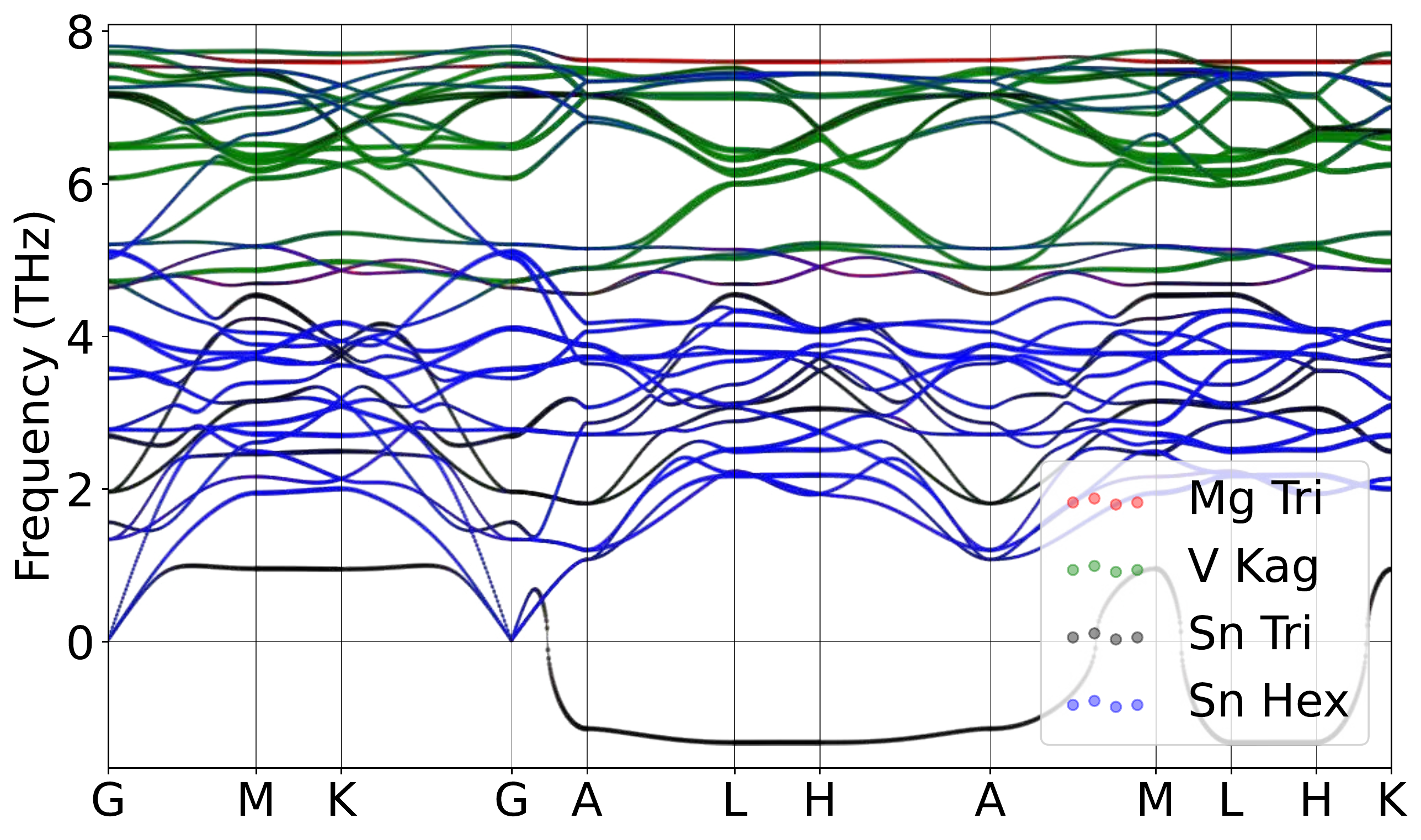}
}
\caption{Atomically projected phonon spectrum for MgV$_6$Sn$_6$.}
\label{fig:MgV6Sn6pho}
\end{figure}

\begin{figure}[H]
\centering
\label{fig:MgV6Sn6_relaxed_Low_xz}
\includegraphics[width=0.85\textwidth]{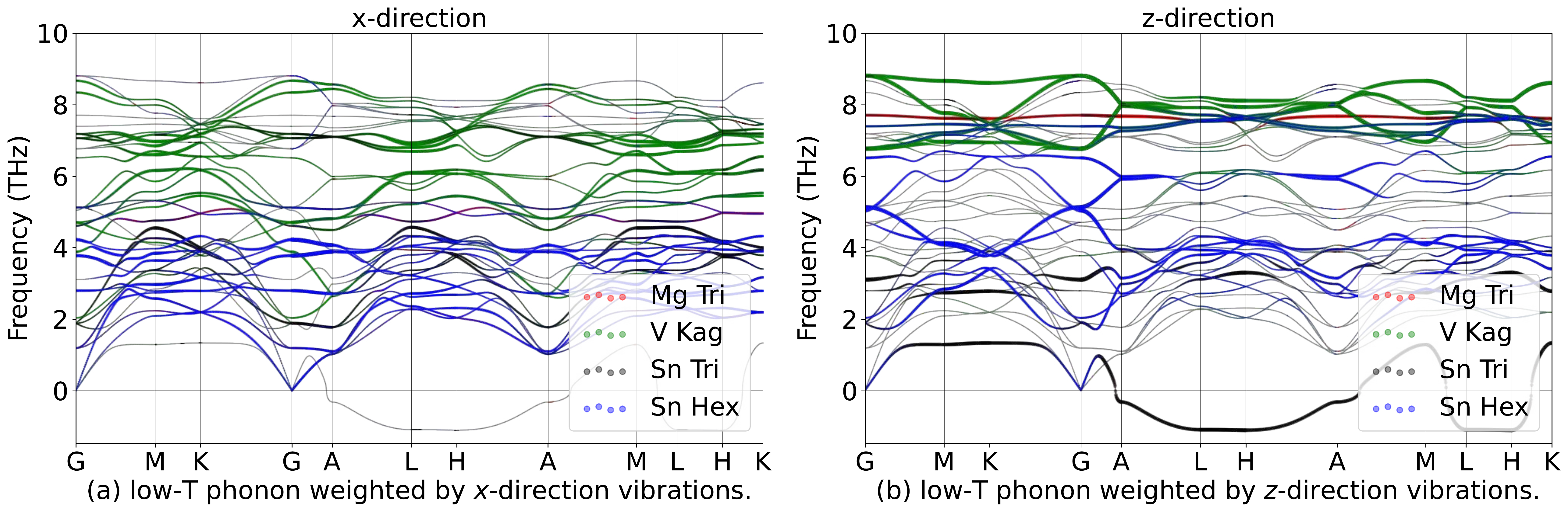}
\hspace{5pt}\label{fig:MgV6Sn6_relaxed_High_xz}
\includegraphics[width=0.85\textwidth]{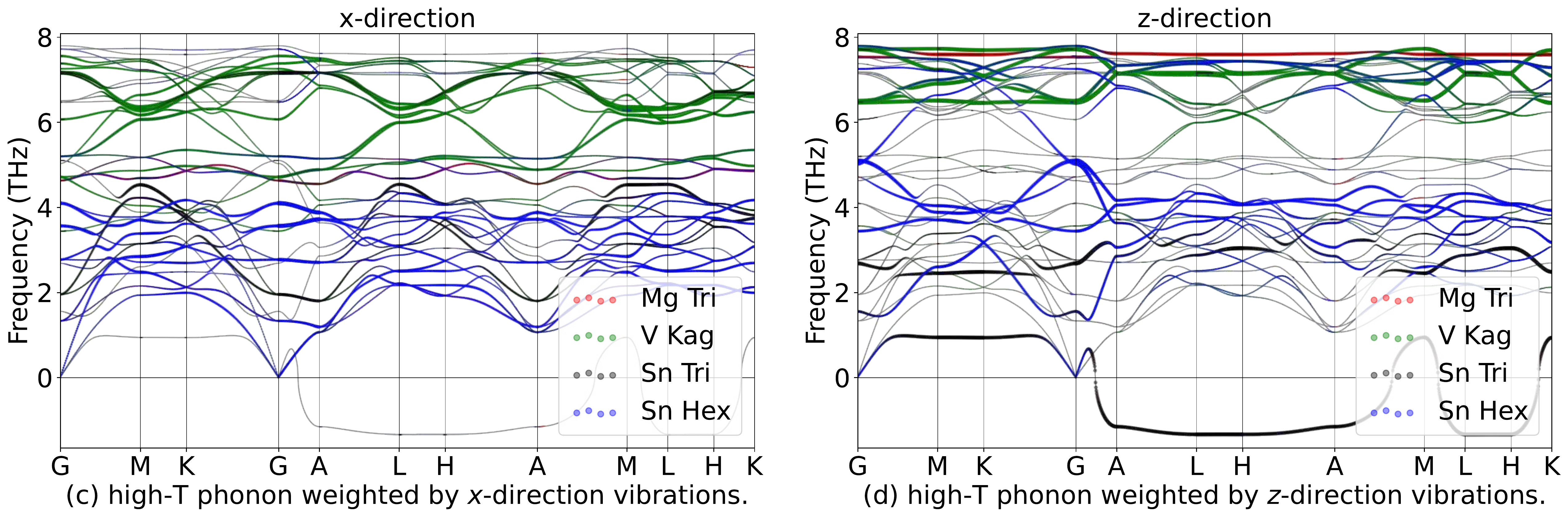}
\caption{Phonon spectrum for MgV$_6$Sn$_6$in in-plane ($x$) and out-of-plane ($z$) directions.}
\label{fig:MgV6Sn6phoxyz}
\end{figure}

\begin{table}[htbp]
\global\long\def\arraystretch{1.12}
\captionsetup{width=.95\textwidth}
\caption{IRREPs at high-symmetry points for MgV$_6$Sn$_6$ phonon spectrum at Low-T.}
\label{tab:191_MgV6Sn6_Low}
\setlength{\tabcolsep}{1.mm}{
}
\end{table}

\begin{table}[htbp]
\global\long\def\arraystretch{1.12}
\captionsetup{width=.95\textwidth}
\caption{IRREPs at high-symmetry points for MgV$_6$Sn$_6$ phonon spectrum at High-T.}
\label{tab:191_MgV6Sn6_High}
\setlength{\tabcolsep}{1.mm}{
%
}
\end{table}
\subsubsection{\label{sms2sec:CaV6Sn6}\hyperref[smsec:col]{\texorpdfstring{CaV$_6$Sn$_6$}{}}~{\color{blue}  (High-T stable, Low-T unstable) }}

\begin{table}[htbp]
\global\long\def\arraystretch{1.12}
\captionsetup{width=.85\textwidth}
\caption{Structure parameters of CaV$_6$Sn$_6$ after relaxation. It contains 458 electrons. }
\label{tab:CaV6Sn6}
\setlength{\tabcolsep}{4mm}{
%
}
\end{table}

\begin{figure}[htbp]
\centering
\includegraphics[width=0.85\textwidth]{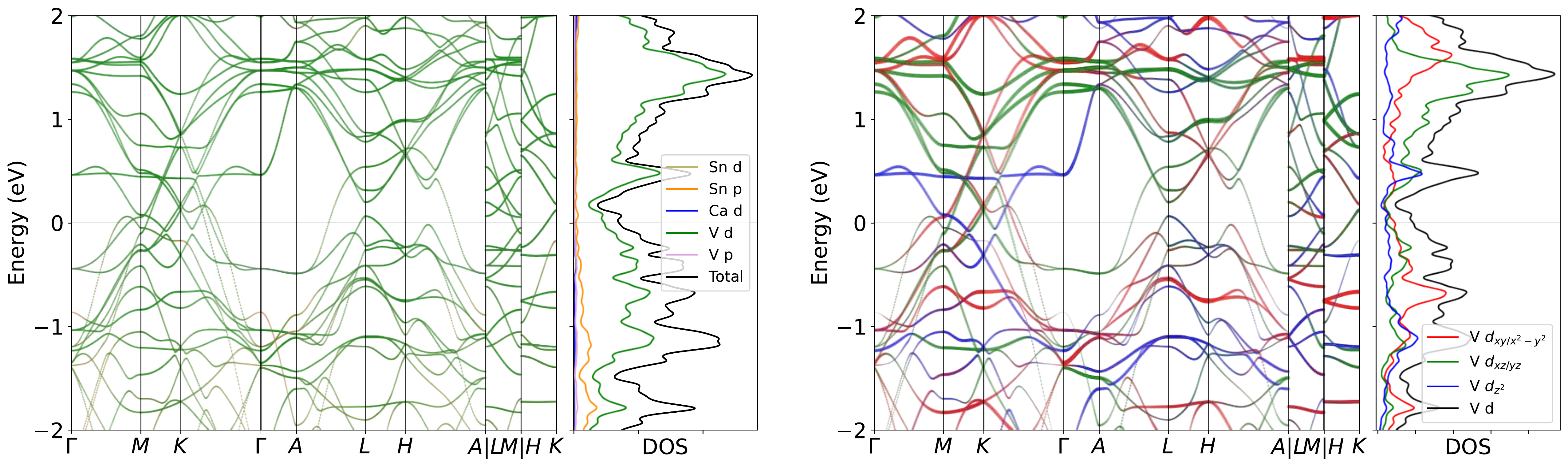}
\caption{Band structure for CaV$_6$Sn$_6$ without SOC. The projection of V $3d$ orbitals is also presented. The band structures clearly reveal the dominating of V $3d$ orbitals  near the Fermi level, as well as the pronounced peak.}
\label{fig:CaV6Sn6band}
\end{figure}

\begin{figure}[H]
\centering
\subfloat[Relaxed phonon at low-T.]{
\label{fig:CaV6Sn6_relaxed_Low}
\includegraphics[width=0.45\textwidth]{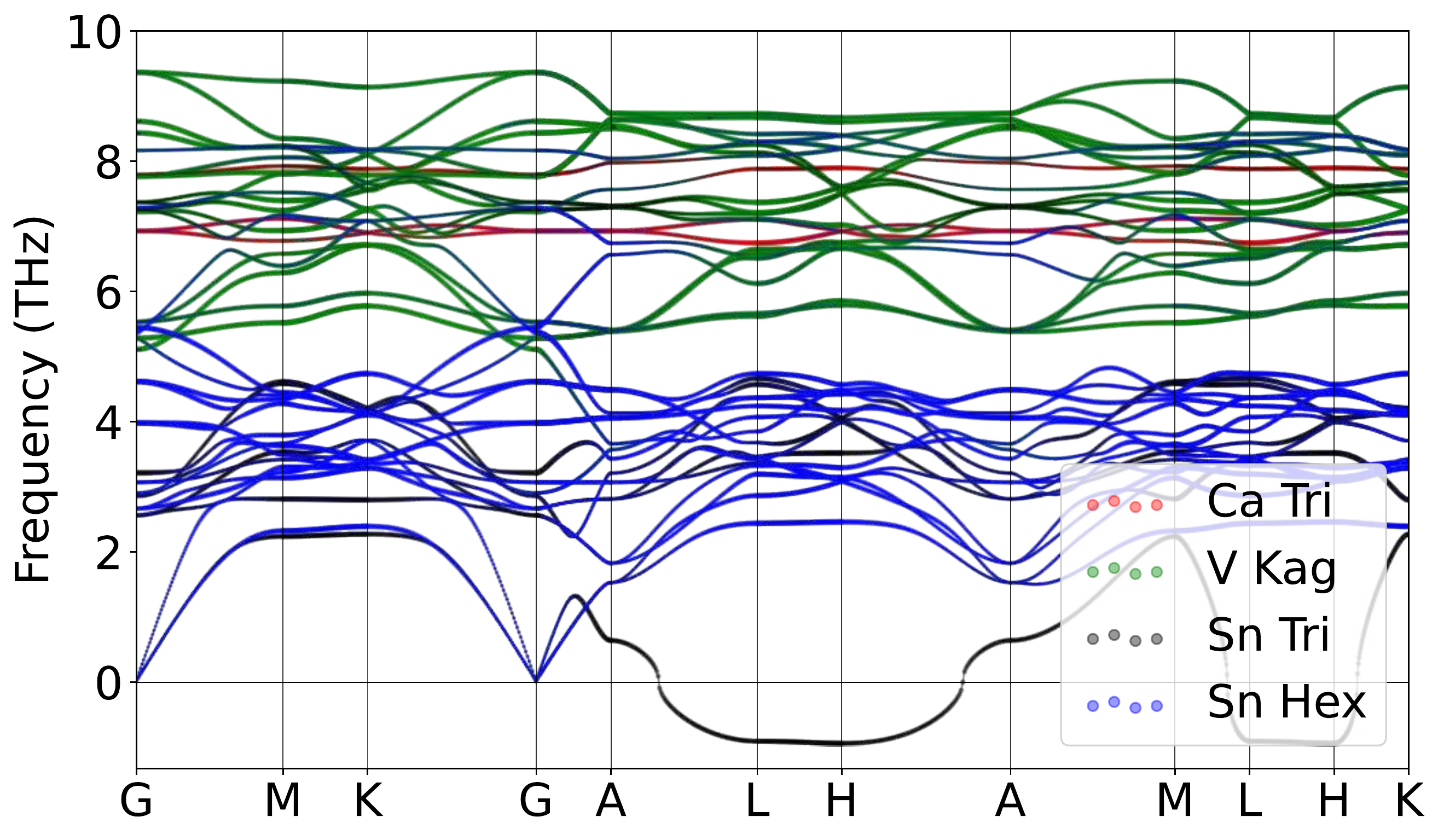}
}
\hspace{5pt}\subfloat[Relaxed phonon at high-T.]{
\label{fig:CaV6Sn6_relaxed_High}
\includegraphics[width=0.45\textwidth]{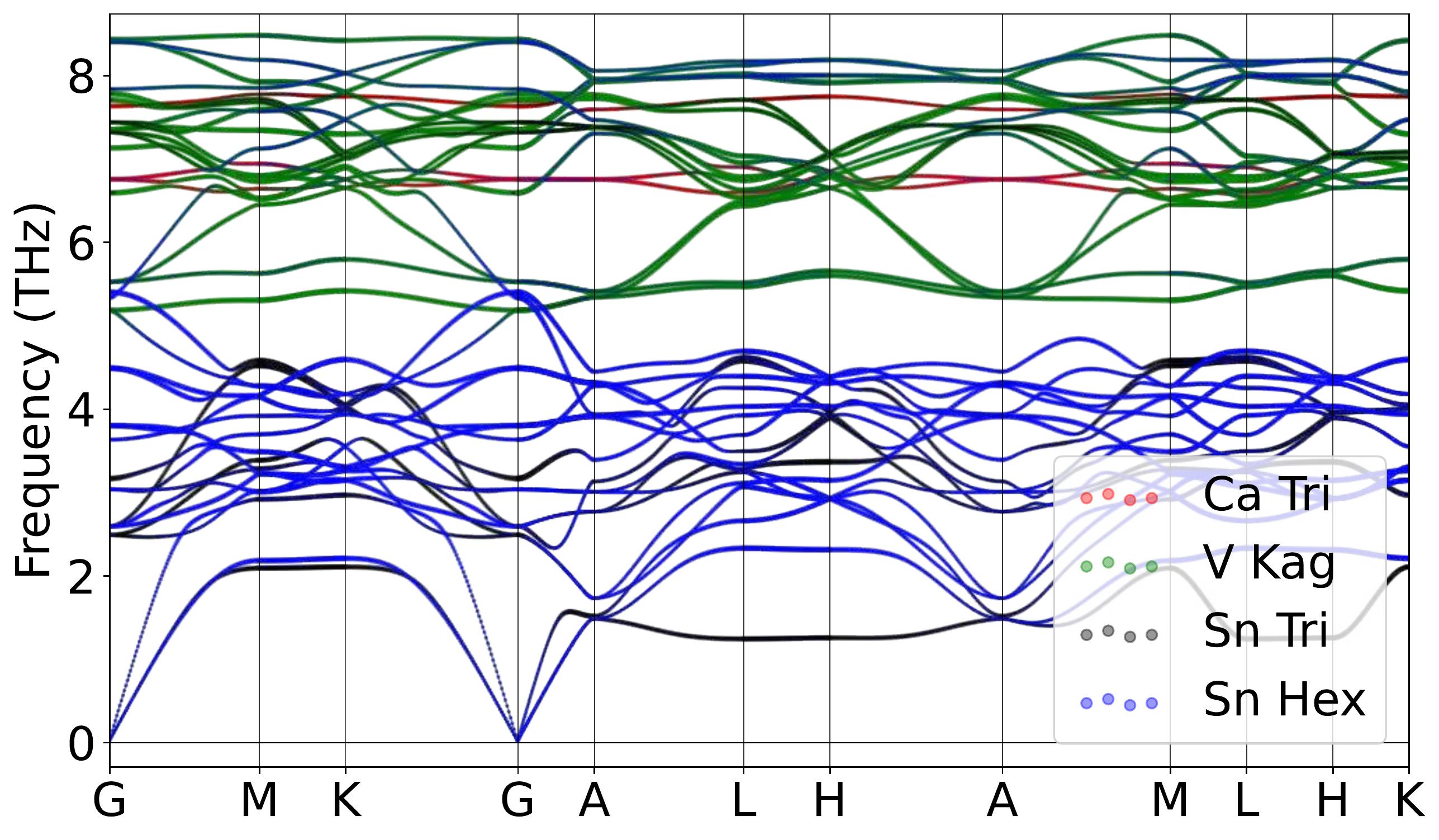}
}
\caption{Atomically projected phonon spectrum for CaV$_6$Sn$_6$.}
\label{fig:CaV6Sn6pho}
\end{figure}

\begin{figure}[H]
\centering
\label{fig:CaV6Sn6_relaxed_Low_xz}
\includegraphics[width=0.85\textwidth]{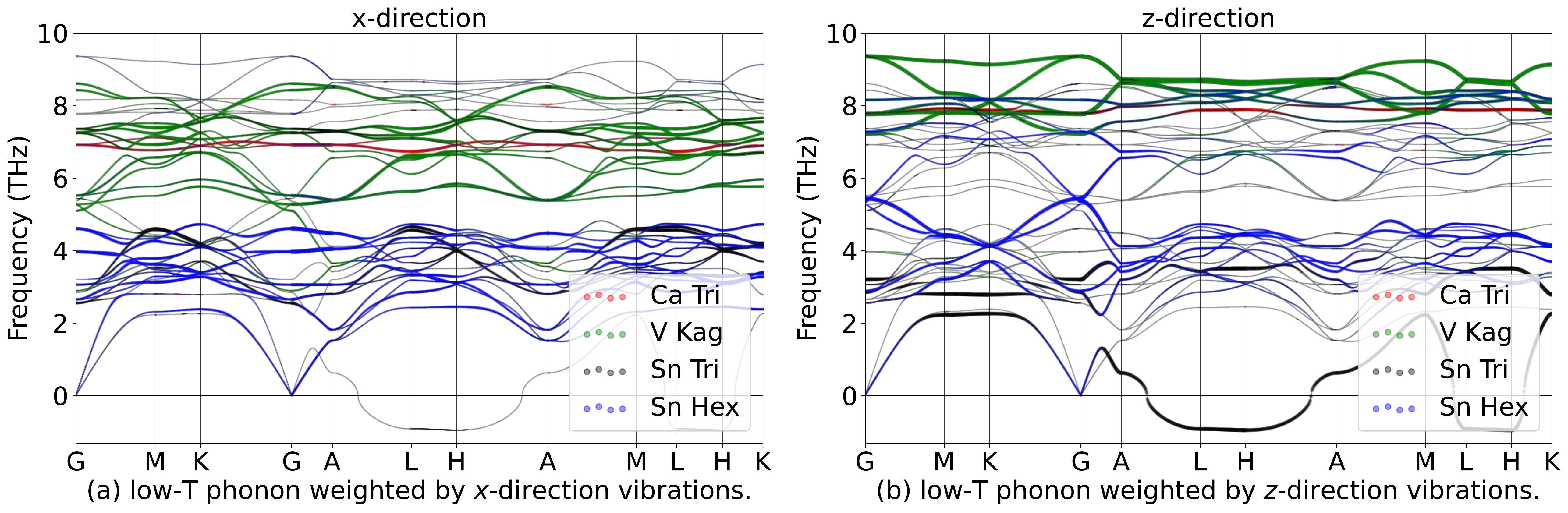}
\hspace{5pt}\label{fig:CaV6Sn6_relaxed_High_xz}
\includegraphics[width=0.85\textwidth]{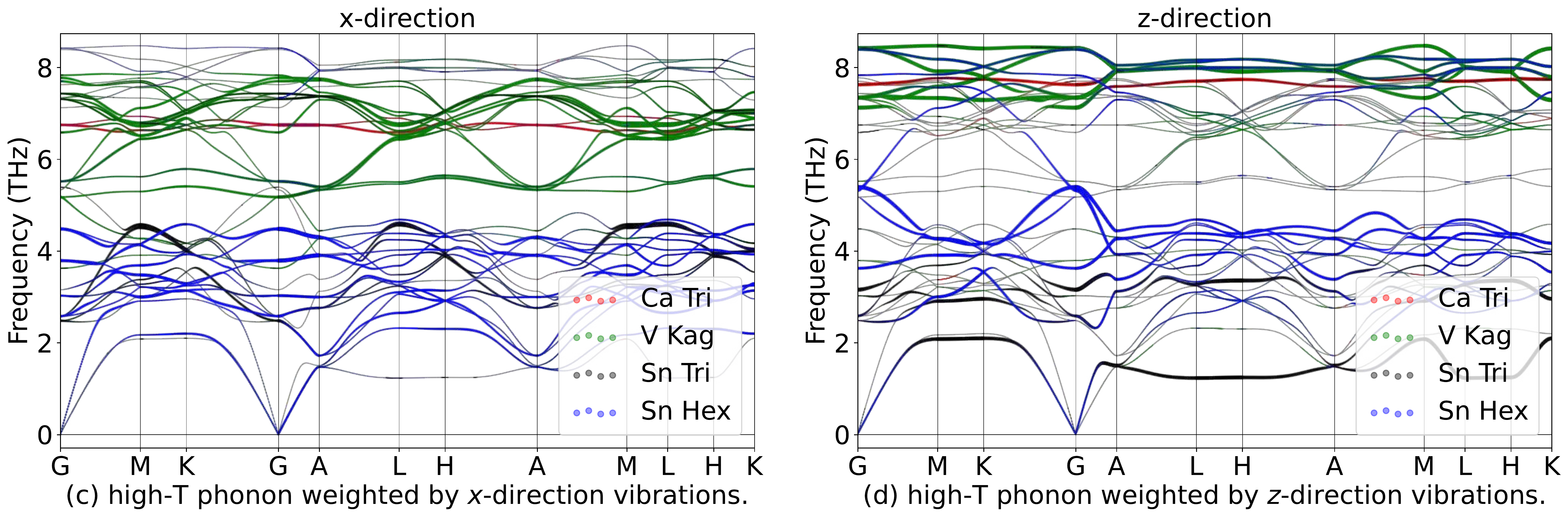}
\caption{Phonon spectrum for CaV$_6$Sn$_6$in in-plane ($x$) and out-of-plane ($z$) directions.}
\label{fig:CaV6Sn6phoxyz}
\end{figure}

\begin{table}[htbp]
\global\long\def\arraystretch{1.12}
\captionsetup{width=.95\textwidth}
\caption{IRREPs at high-symmetry points for CaV$_6$Sn$_6$ phonon spectrum at Low-T.}
\label{tab:191_CaV6Sn6_Low}
\setlength{\tabcolsep}{1.mm}{
}
\end{table}

\begin{table}[htbp]
\global\long\def\arraystretch{1.12}
\captionsetup{width=.95\textwidth}
\caption{IRREPs at high-symmetry points for CaV$_6$Sn$_6$ phonon spectrum at High-T.}
\label{tab:191_CaV6Sn6_High}
\setlength{\tabcolsep}{1.mm}{
%
}
\end{table}
\subsubsection{\label{sms2sec:ScV6Sn6}\hyperref[smsec:col]{\texorpdfstring{ScV$_6$Sn$_6$}{}}~{\color{blue}  (High-T stable, Low-T unstable) }}

\begin{table}[htbp]
\global\long\def\arraystretch{1.12}
\captionsetup{width=.85\textwidth}
\caption{Structure parameters of ScV$_6$Sn$_6$ after relaxation. It contains 459 electrons. }
\label{tab:ScV6Sn6}
\setlength{\tabcolsep}{4mm}{
%
}
\end{table}

\begin{figure}[htbp]
\centering
\includegraphics[width=0.85\textwidth]{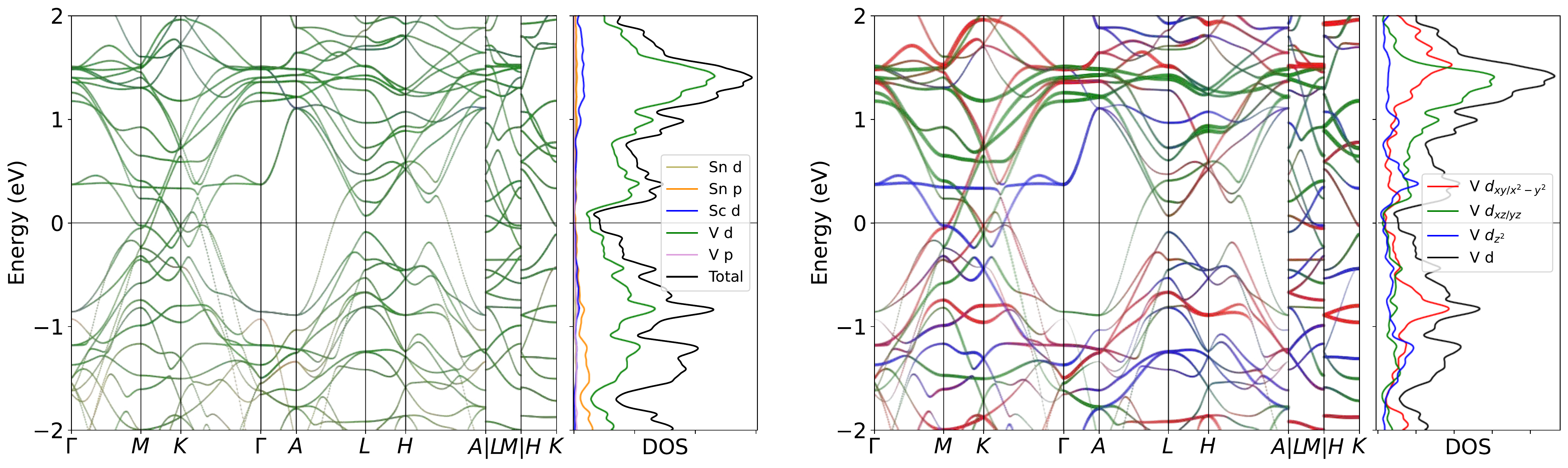}
\caption{Band structure for ScV$_6$Sn$_6$ without SOC. The projection of V $3d$ orbitals is also presented. The band structures clearly reveal the dominating of V $3d$ orbitals  near the Fermi level, as well as the pronounced peak.}
\label{fig:ScV6Sn6band}
\end{figure}

\begin{figure}[H]
\centering
\subfloat[Relaxed phonon at low-T.]{
\label{fig:ScV6Sn6_relaxed_Low}
\includegraphics[width=0.45\textwidth]{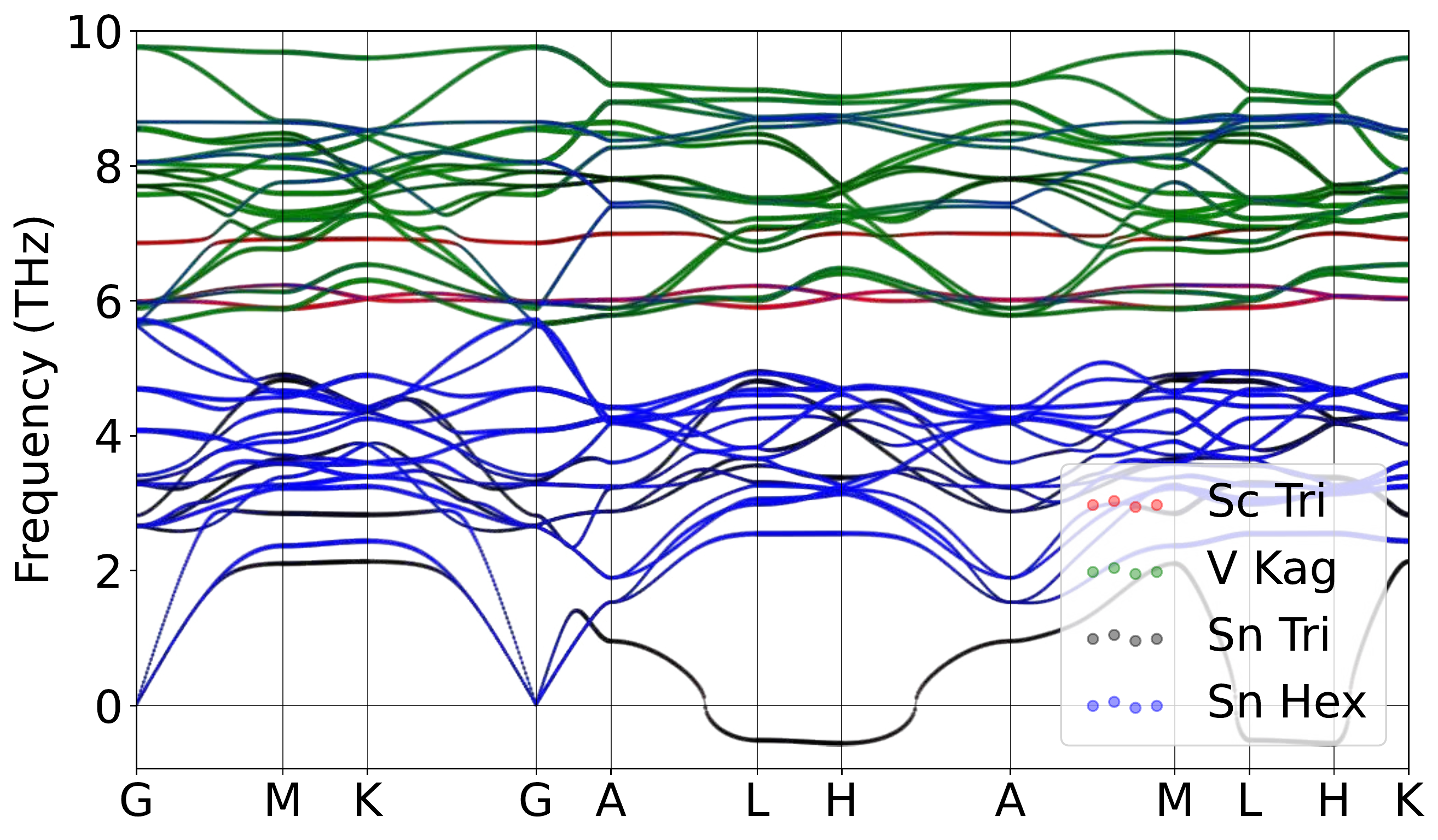}
}
\hspace{5pt}\subfloat[Relaxed phonon at high-T.]{
\label{fig:ScV6Sn6_relaxed_High}
\includegraphics[width=0.45\textwidth]{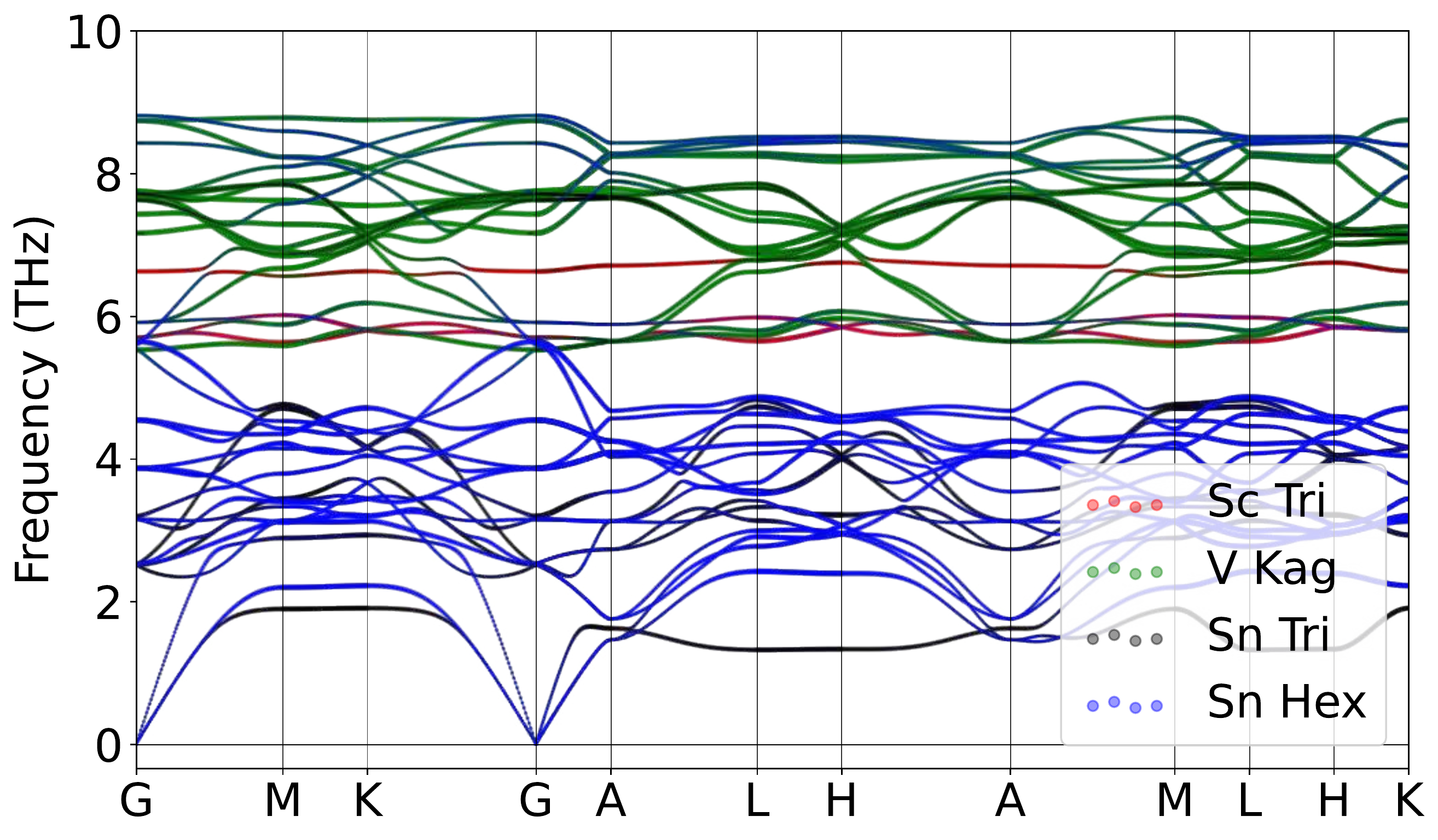}
}
\caption{Atomically projected phonon spectrum for ScV$_6$Sn$_6$.}
\label{fig:ScV6Sn6pho}
\end{figure}

\begin{figure}[H]
\centering
\label{fig:ScV6Sn6_relaxed_Low_xz}
\includegraphics[width=0.85\textwidth]{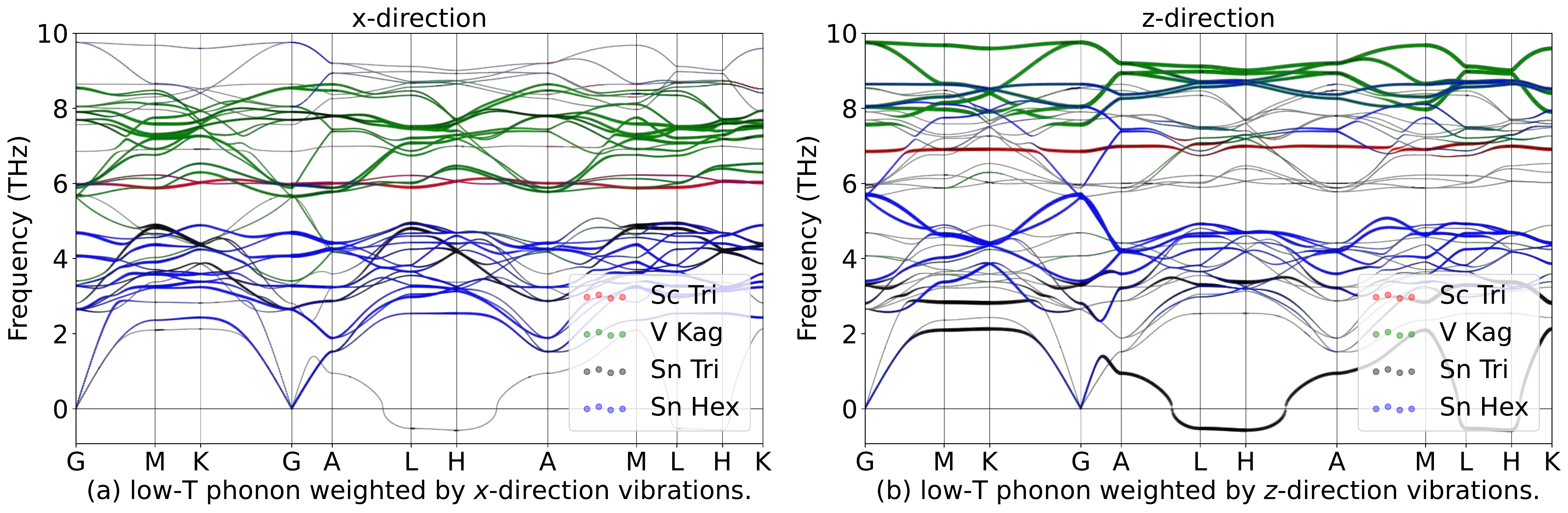}
\hspace{5pt}\label{fig:ScV6Sn6_relaxed_High_xz}
\includegraphics[width=0.85\textwidth]{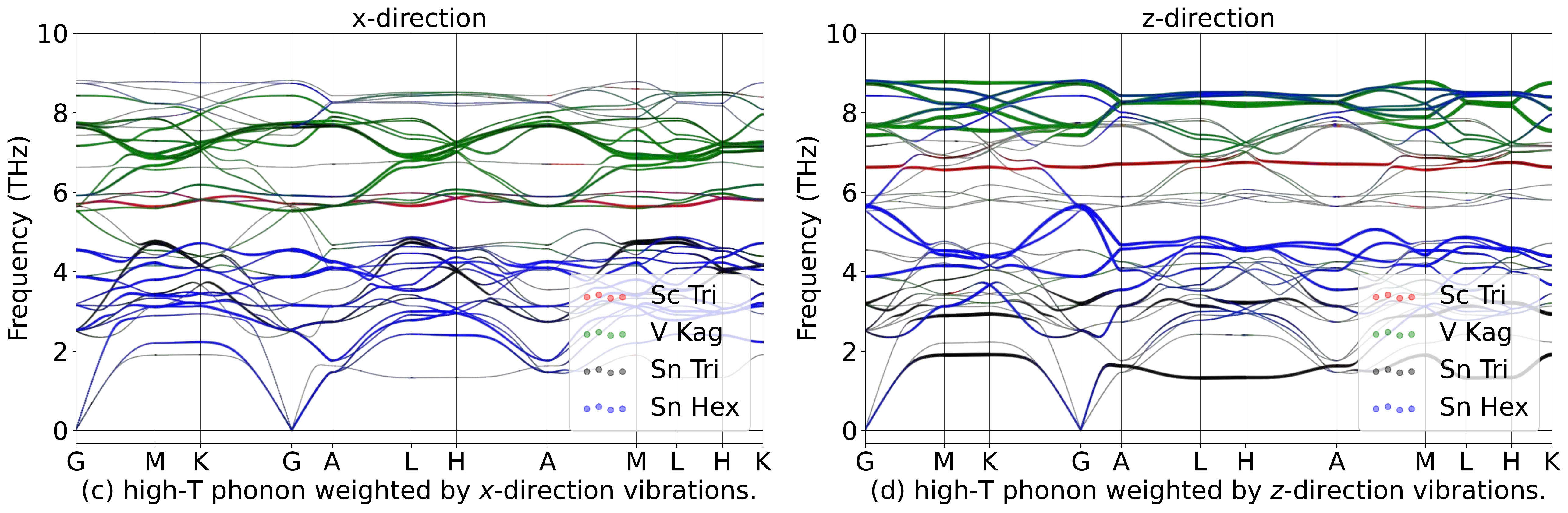}
\caption{Phonon spectrum for ScV$_6$Sn$_6$in in-plane ($x$) and out-of-plane ($z$) directions.}
\label{fig:ScV6Sn6phoxyz}
\end{figure}

\begin{table}[htbp]
\global\long\def\arraystretch{1.12}
\captionsetup{width=.95\textwidth}
\caption{IRREPs at high-symmetry points for ScV$_6$Sn$_6$ phonon spectrum at Low-T.}
\label{tab:191_ScV6Sn6_Low}
\setlength{\tabcolsep}{1.mm}{
}
\end{table}

\begin{table}[htbp]
\global\long\def\arraystretch{1.12}
\captionsetup{width=.95\textwidth}
\caption{IRREPs at high-symmetry points for ScV$_6$Sn$_6$ phonon spectrum at High-T.}
\label{tab:191_ScV6Sn6_High}
\setlength{\tabcolsep}{1.mm}{
%
}
\end{table}
\subsubsection{\label{sms2sec:TiV6Sn6}\hyperref[smsec:col]{\texorpdfstring{TiV$_6$Sn$_6$}{}}~{\color{green}  (High-T stable, Low-T stable) }}

\begin{table}[htbp]
\global\long\def\arraystretch{1.12}
\captionsetup{width=.85\textwidth}
\caption{Structure parameters of TiV$_6$Sn$_6$ after relaxation. It contains 460 electrons. }
\label{tab:TiV6Sn6}
\setlength{\tabcolsep}{4mm}{
%
}
\end{table}

\begin{figure}[htbp]
\centering
\includegraphics[width=0.85\textwidth]{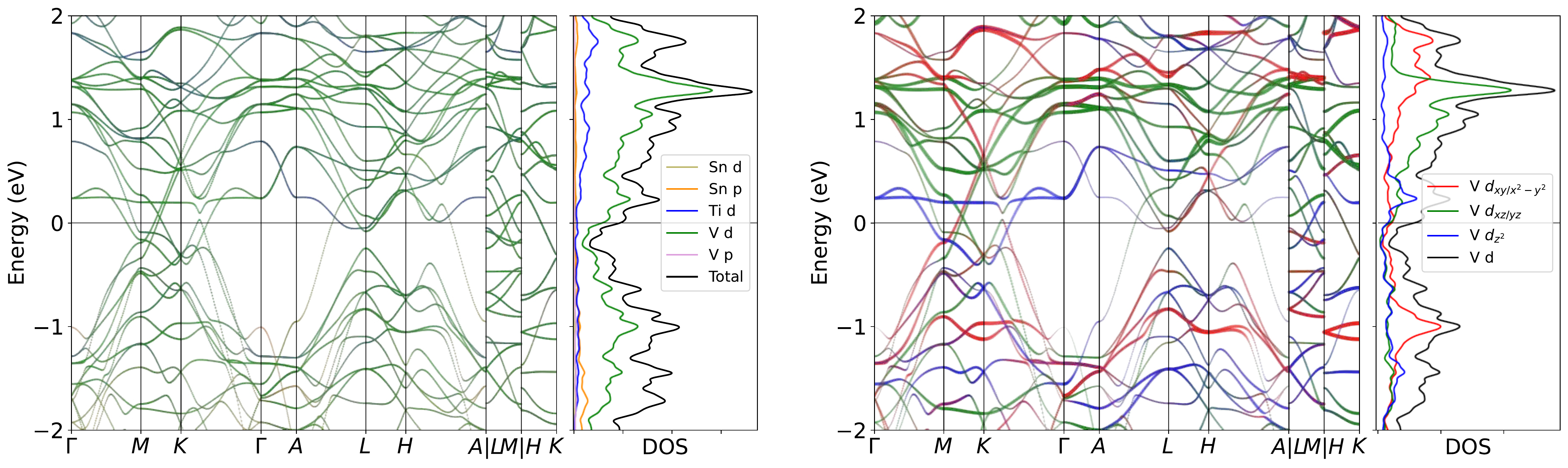}
\caption{Band structure for TiV$_6$Sn$_6$ without SOC. The projection of V $3d$ orbitals is also presented. The band structures clearly reveal the dominating of V $3d$ orbitals  near the Fermi level, as well as the pronounced peak.}
\label{fig:TiV6Sn6band}
\end{figure}

\begin{figure}[H]
\centering
\subfloat[Relaxed phonon at low-T.]{
\label{fig:TiV6Sn6_relaxed_Low}
\includegraphics[width=0.45\textwidth]{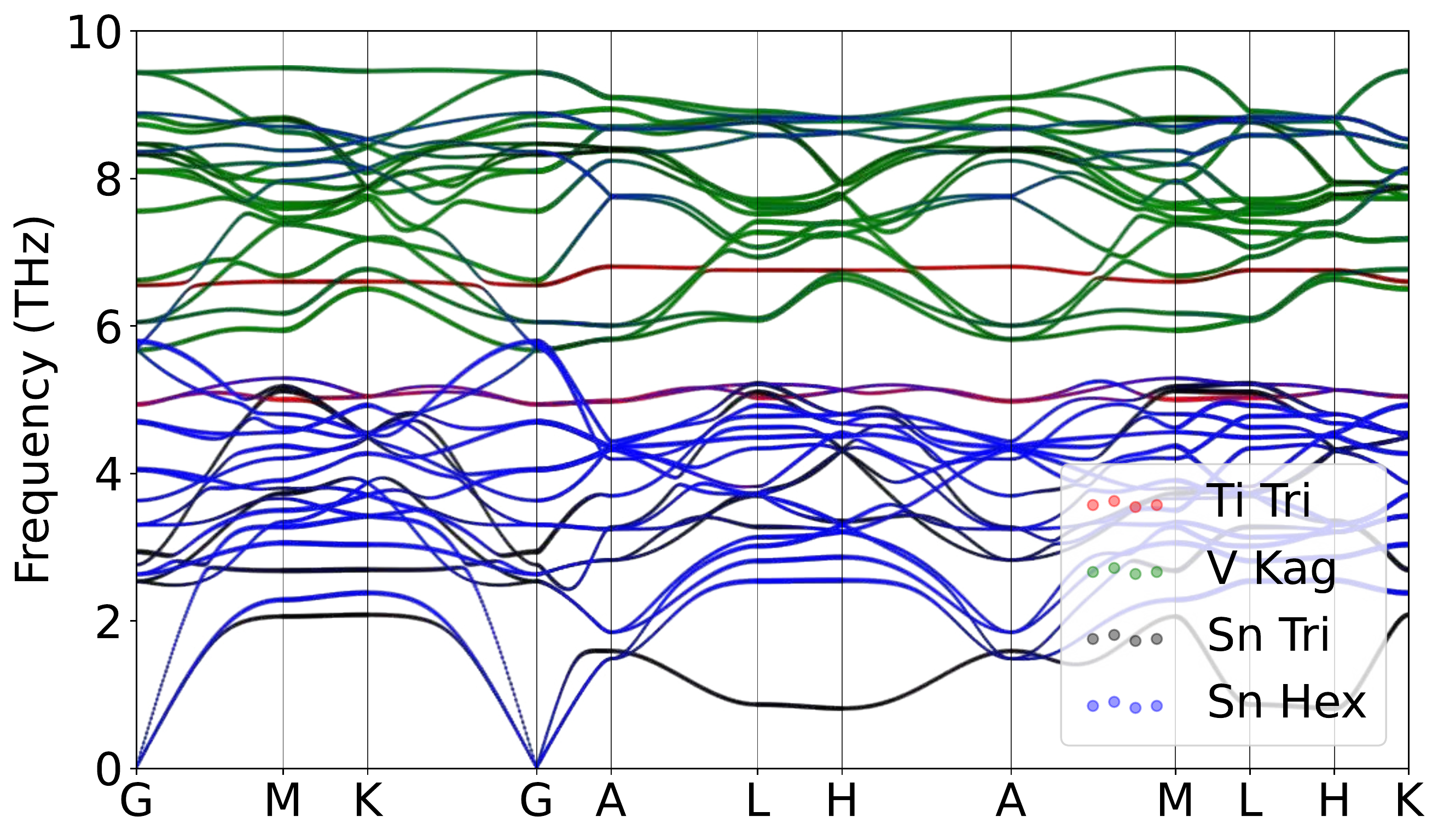}
}
\hspace{5pt}\subfloat[Relaxed phonon at high-T.]{
\label{fig:TiV6Sn6_relaxed_High}
\includegraphics[width=0.45\textwidth]{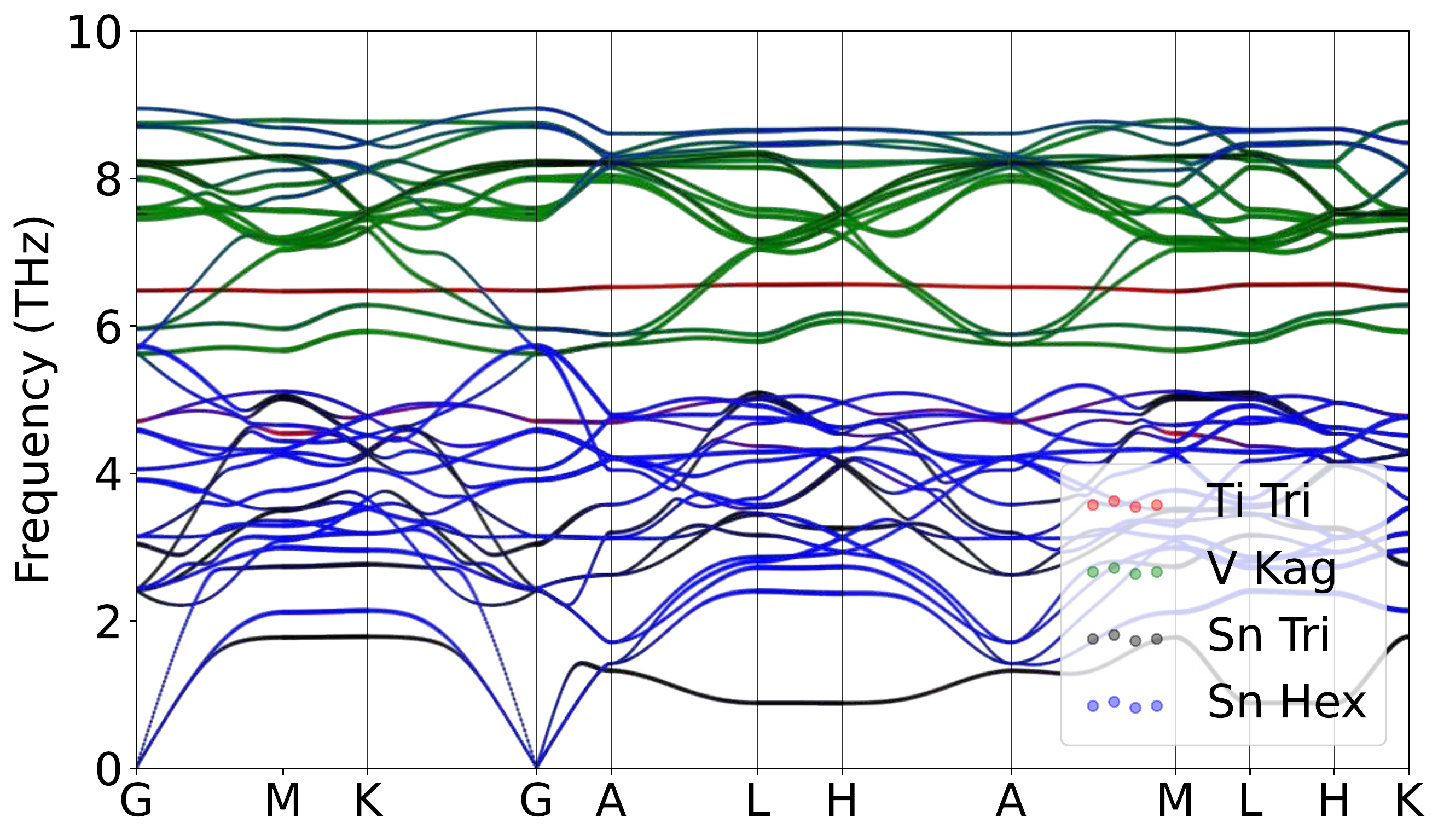}
}
\caption{Atomically projected phonon spectrum for TiV$_6$Sn$_6$.}
\label{fig:TiV6Sn6pho}
\end{figure}

\begin{figure}[H]
\centering
\label{fig:TiV6Sn6_relaxed_Low_xz}
\includegraphics[width=0.85\textwidth]{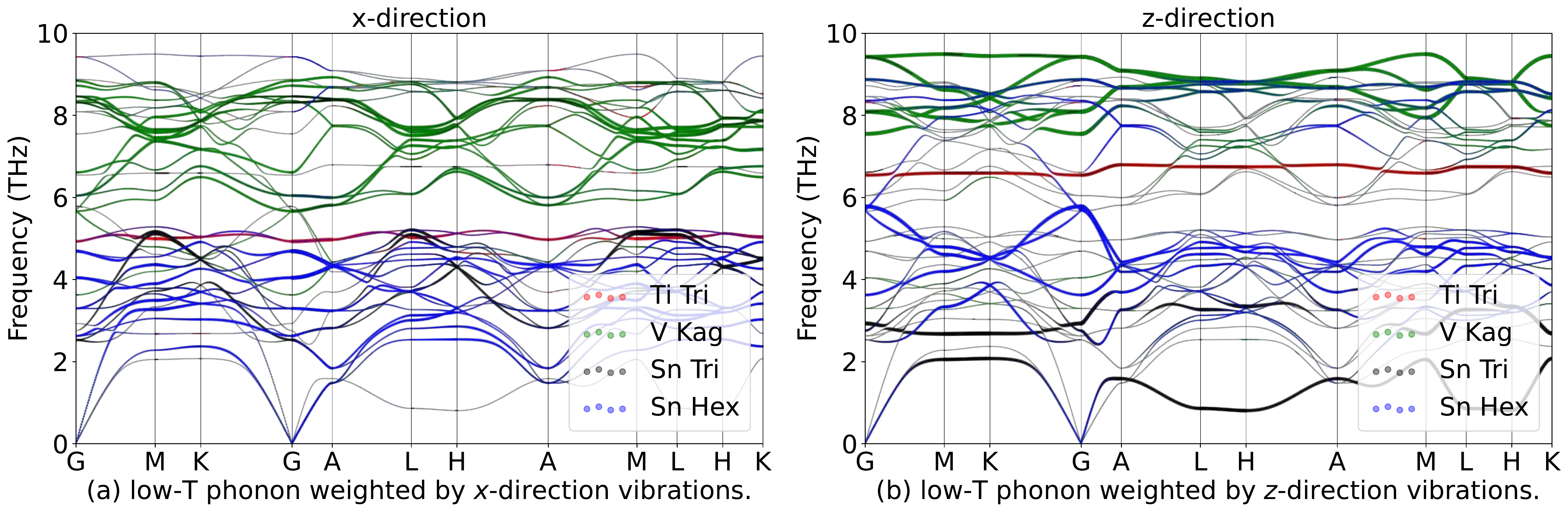}
\hspace{5pt}\label{fig:TiV6Sn6_relaxed_High_xz}
\includegraphics[width=0.85\textwidth]{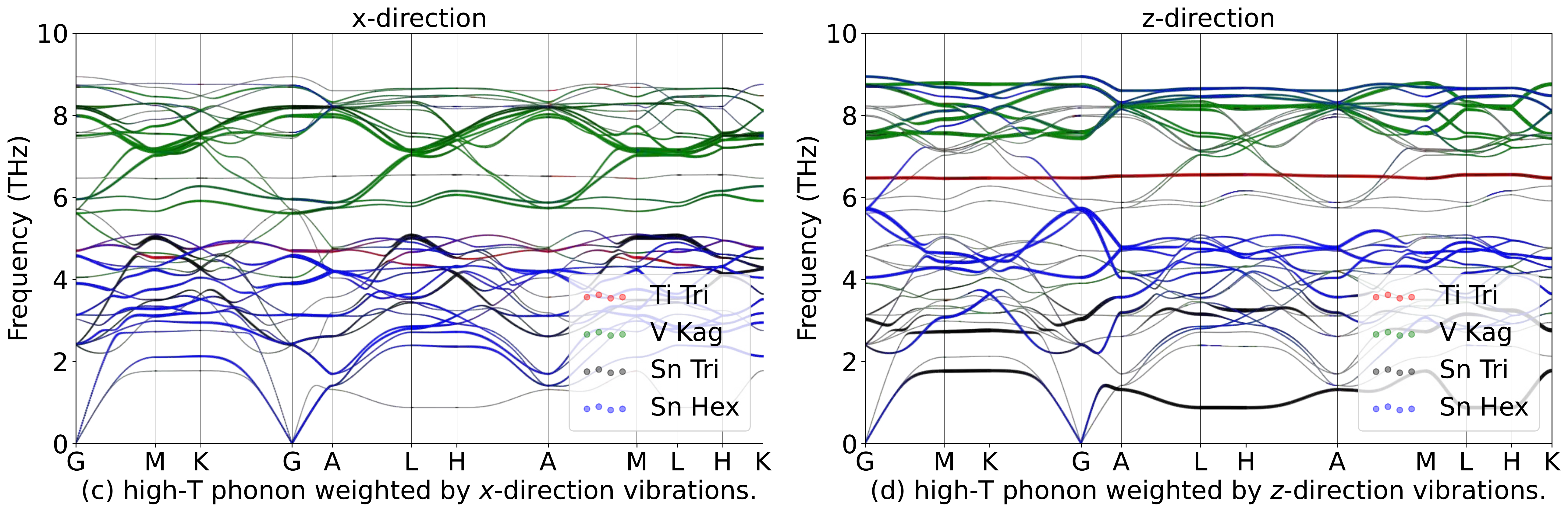}
\caption{Phonon spectrum for TiV$_6$Sn$_6$in in-plane ($x$) and out-of-plane ($z$) directions.}
\label{fig:TiV6Sn6phoxyz}
\end{figure}

\begin{table}[htbp]
\global\long\def\arraystretch{1.12}
\captionsetup{width=.95\textwidth}
\caption{IRREPs at high-symmetry points for TiV$_6$Sn$_6$ phonon spectrum at Low-T.}
\label{tab:191_TiV6Sn6_Low}
\setlength{\tabcolsep}{1.mm}{
}
\end{table}

\begin{table}[htbp]
\global\long\def\arraystretch{1.12}
\captionsetup{width=.95\textwidth}
\caption{IRREPs at high-symmetry points for TiV$_6$Sn$_6$ phonon spectrum at High-T.}
\label{tab:191_TiV6Sn6_High}
\setlength{\tabcolsep}{1.mm}{
%
}
\end{table}
\subsubsection{\label{sms2sec:YV6Sn6}\hyperref[smsec:col]{\texorpdfstring{YV$_6$Sn$_6$}{}}~{\color{green}  (High-T stable, Low-T stable) }}

\begin{table}[htbp]
\global\long\def\arraystretch{1.12}
\captionsetup{width=.85\textwidth}
\caption{Structure parameters of YV$_6$Sn$_6$ after relaxation. It contains 477 electrons. }
\label{tab:YV6Sn6}
\setlength{\tabcolsep}{4mm}{
%
}
\end{table}

\begin{figure}[htbp]
\centering
\includegraphics[width=0.85\textwidth]{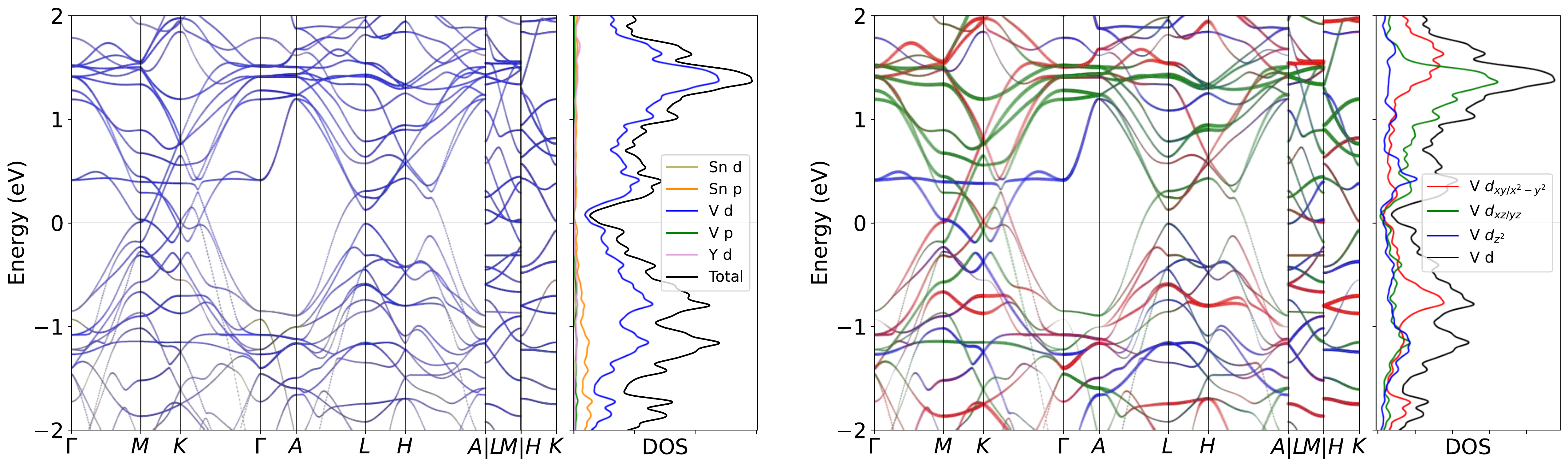}
\caption{Band structure for YV$_6$Sn$_6$ without SOC. The projection of V $3d$ orbitals is also presented. The band structures clearly reveal the dominating of V $3d$ orbitals  near the Fermi level, as well as the pronounced peak.}
\label{fig:YV6Sn6band}
\end{figure}

\begin{figure}[H]
\centering
\subfloat[Relaxed phonon at low-T.]{
\label{fig:YV6Sn6_relaxed_Low}
\includegraphics[width=0.45\textwidth]{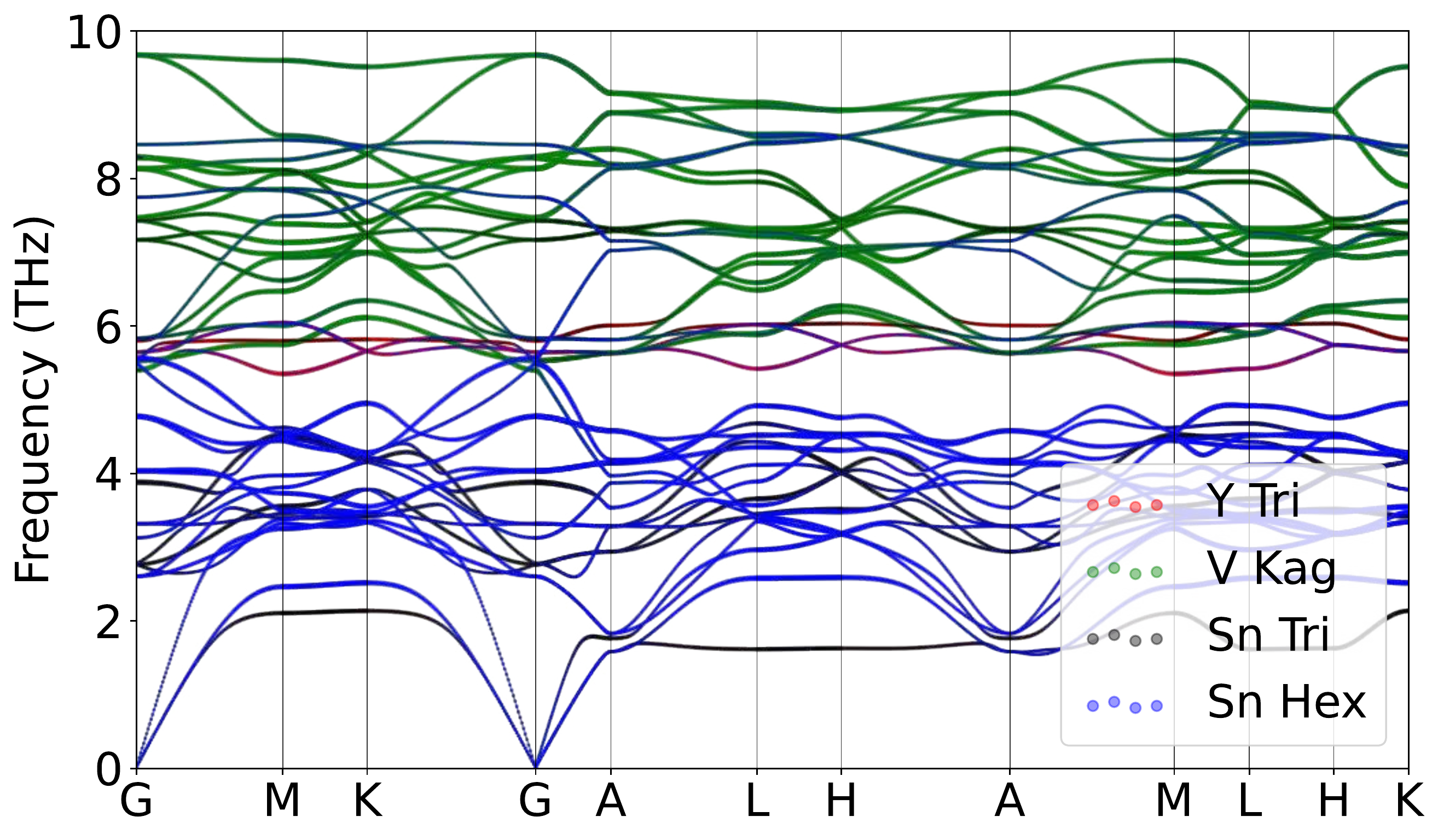}
}
\hspace{5pt}\subfloat[Relaxed phonon at high-T.]{
\label{fig:YV6Sn6_relaxed_High}
\includegraphics[width=0.45\textwidth]{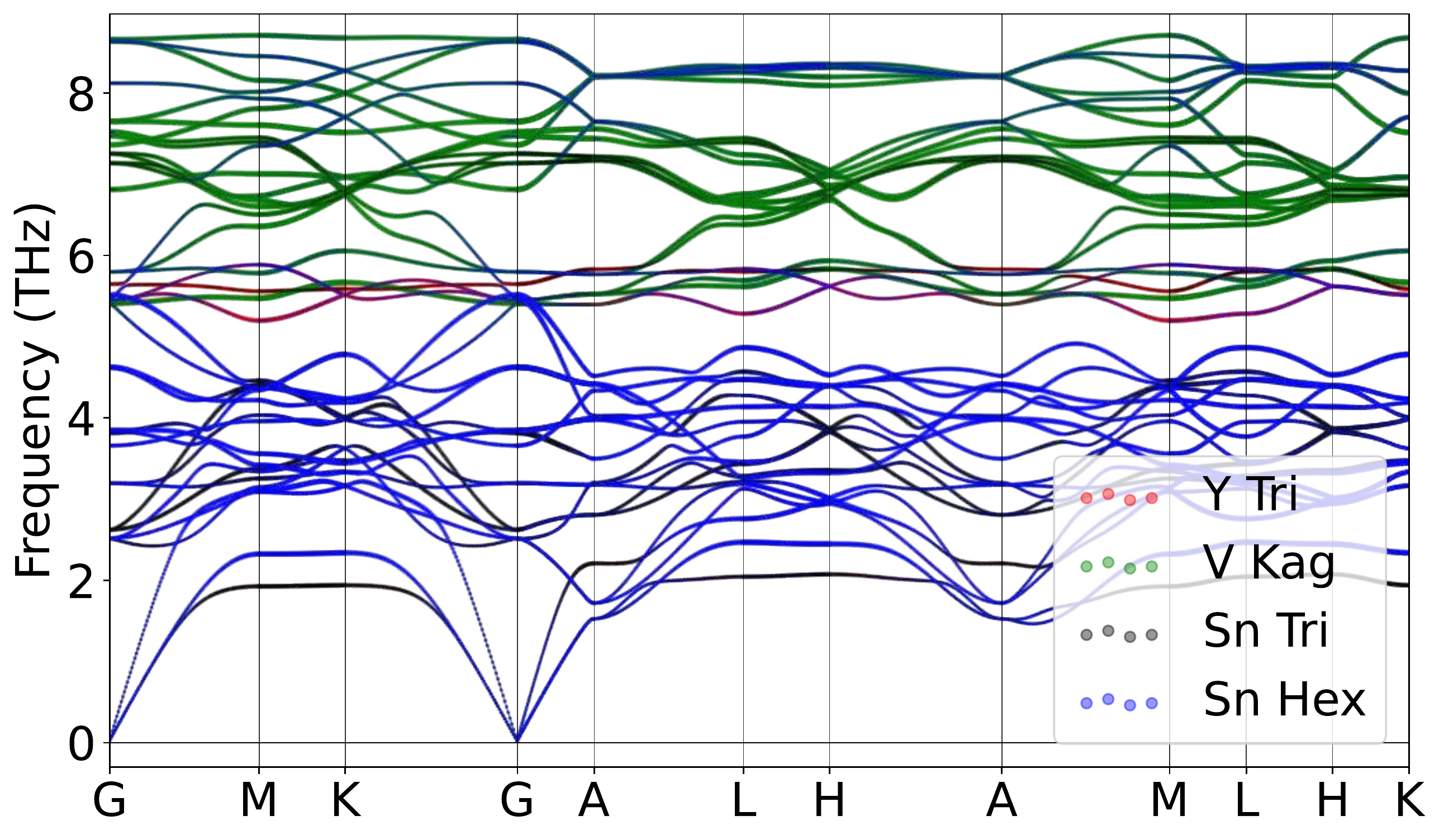}
}
\caption{Atomically projected phonon spectrum for YV$_6$Sn$_6$.}
\label{fig:YV6Sn6pho}
\end{figure}

\begin{figure}[H]
\centering
\label{fig:YV6Sn6_relaxed_Low_xz}
\includegraphics[width=0.85\textwidth]{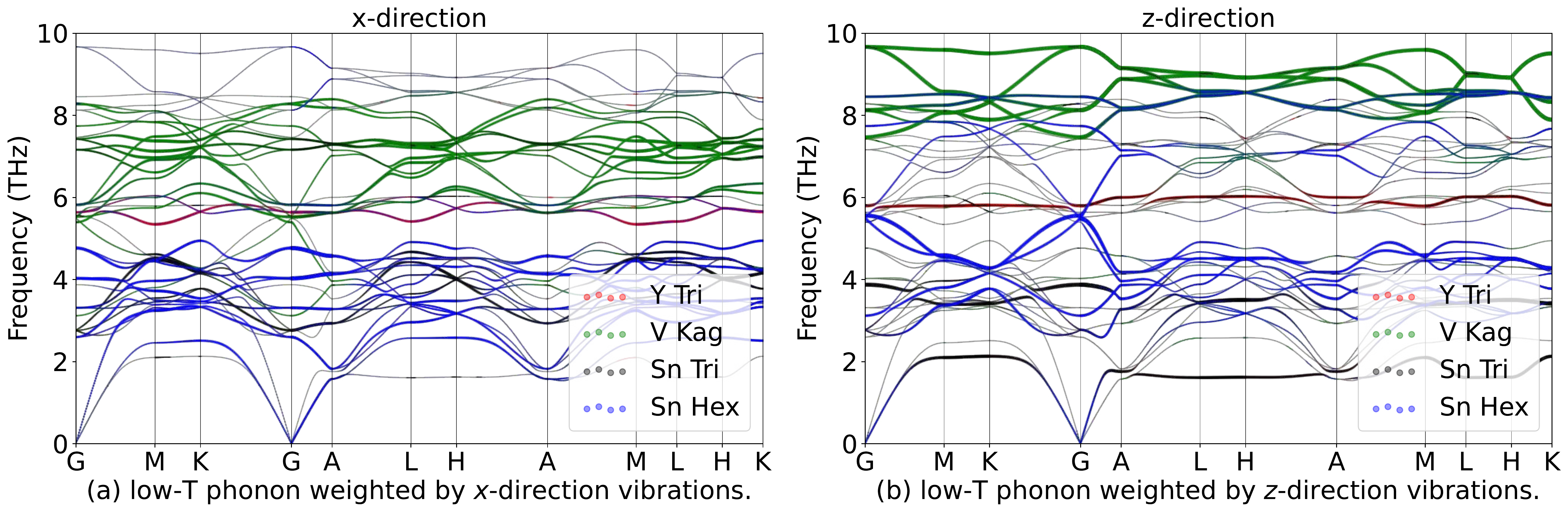}
\hspace{5pt}\label{fig:YV6Sn6_relaxed_High_xz}
\includegraphics[width=0.85\textwidth]{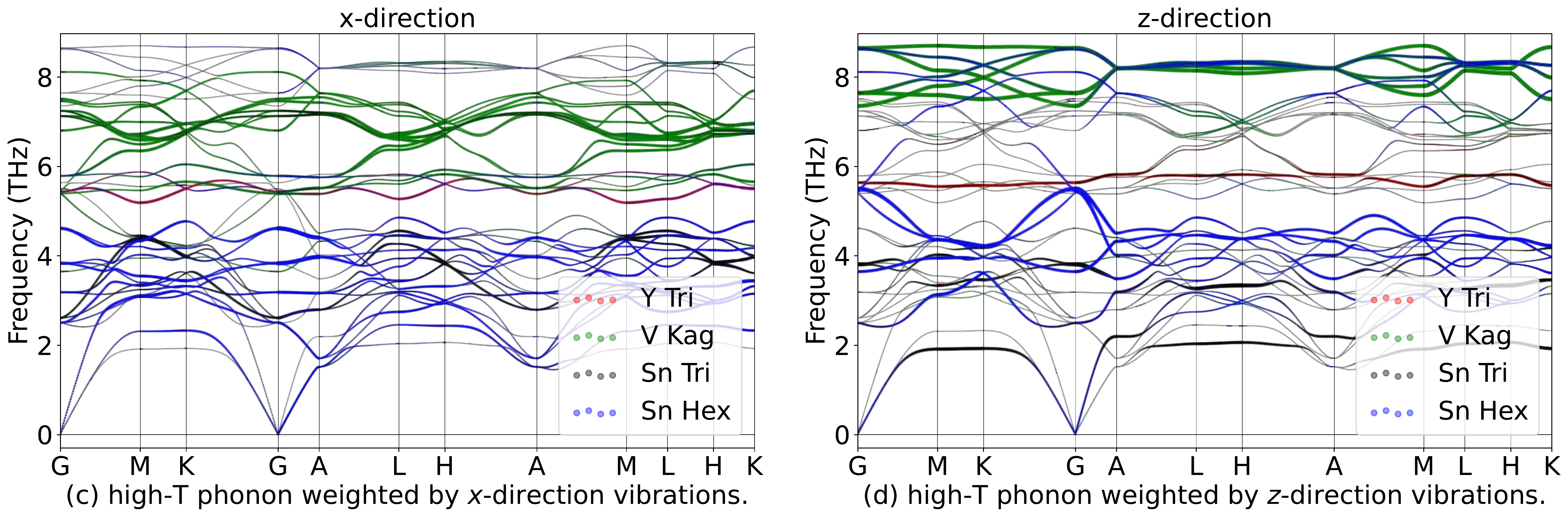}
\caption{Phonon spectrum for YV$_6$Sn$_6$in in-plane ($x$) and out-of-plane ($z$) directions.}
\label{fig:YV6Sn6phoxyz}
\end{figure}

\begin{table}[htbp]
\global\long\def\arraystretch{1.12}
\captionsetup{width=.95\textwidth}
\caption{IRREPs at high-symmetry points for YV$_6$Sn$_6$ phonon spectrum at Low-T.}
\label{tab:191_YV6Sn6_Low}
\setlength{\tabcolsep}{1.mm}{
}
\end{table}

\begin{table}[htbp]
\global\long\def\arraystretch{1.12}
\captionsetup{width=.95\textwidth}
\caption{IRREPs at high-symmetry points for YV$_6$Sn$_6$ phonon spectrum at High-T.}
\label{tab:191_YV6Sn6_High}
\setlength{\tabcolsep}{1.mm}{
%
}
\end{table}
\subsubsection{\label{sms2sec:ZrV6Sn6}\hyperref[smsec:col]{\texorpdfstring{ZrV$_6$Sn$_6$}{}}~{\color{green}  (High-T stable, Low-T stable) }}

\begin{table}[htbp]
\global\long\def\arraystretch{1.12}
\captionsetup{width=.85\textwidth}
\caption{Structure parameters of ZrV$_6$Sn$_6$ after relaxation. It contains 478 electrons. }
\label{tab:ZrV6Sn6}
\setlength{\tabcolsep}{4mm}{
%
}
\end{table}

\begin{figure}[htbp]
\centering
\includegraphics[width=0.85\textwidth]{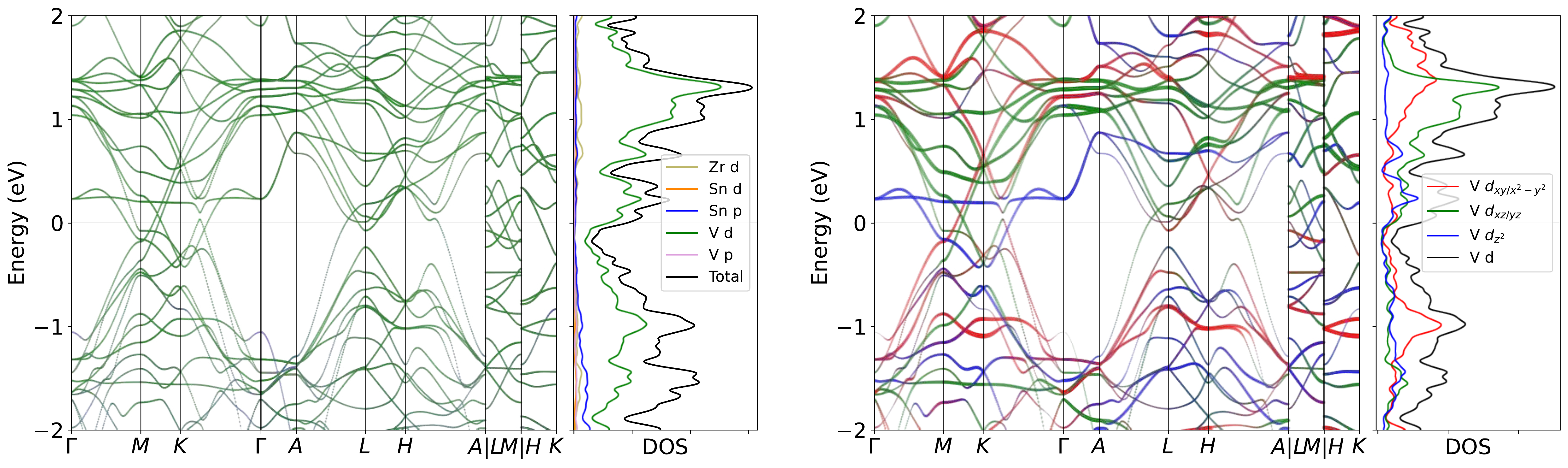}
\caption{Band structure for ZrV$_6$Sn$_6$ without SOC. The projection of V $3d$ orbitals is also presented. The band structures clearly reveal the dominating of V $3d$ orbitals  near the Fermi level, as well as the pronounced peak.}
\label{fig:ZrV6Sn6band}
\end{figure}

\begin{figure}[H]
\centering
\subfloat[Relaxed phonon at low-T.]{
\label{fig:ZrV6Sn6_relaxed_Low}
\includegraphics[width=0.45\textwidth]{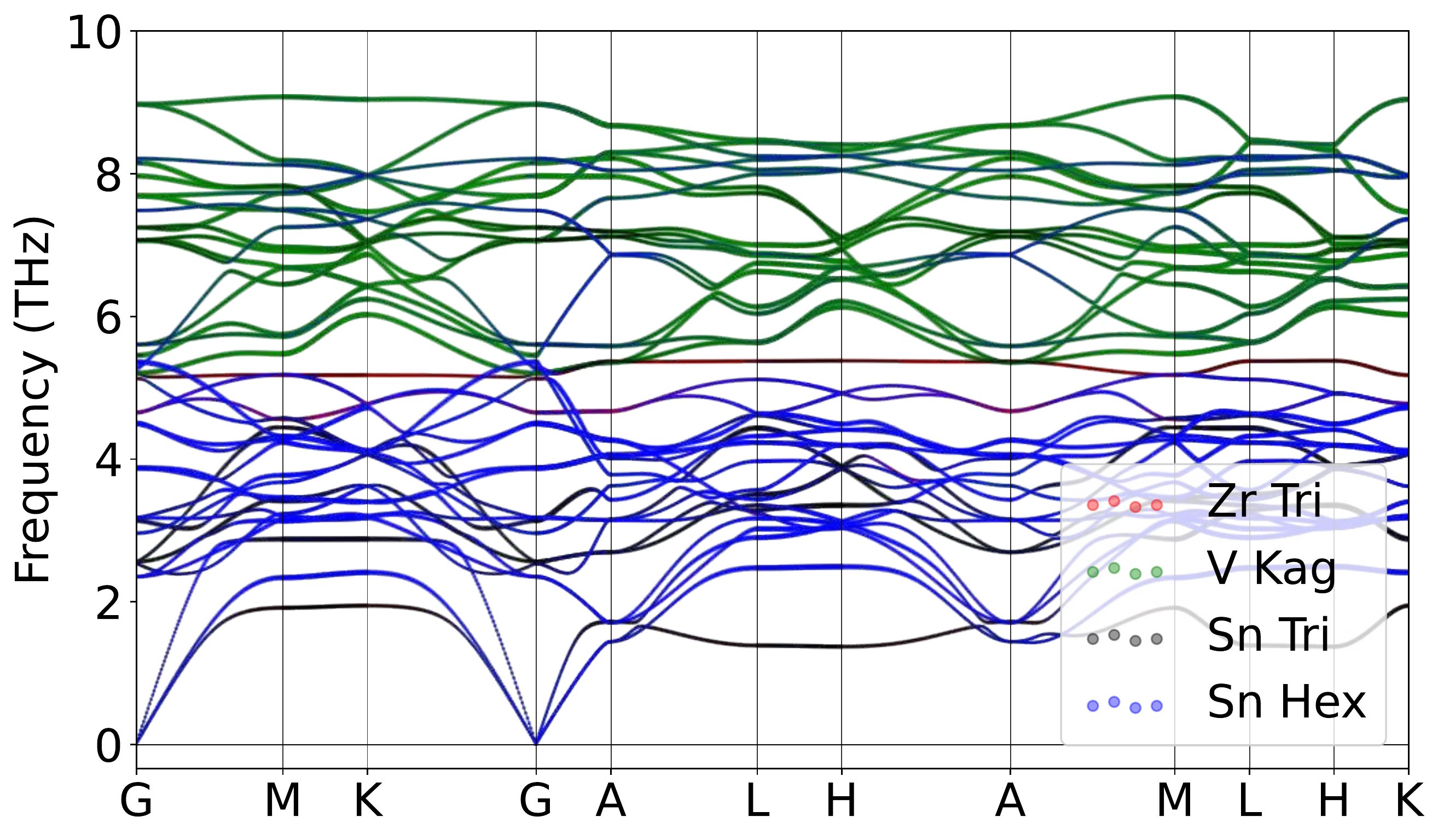}
}
\hspace{5pt}\subfloat[Relaxed phonon at high-T.]{
\label{fig:ZrV6Sn6_relaxed_High}
\includegraphics[width=0.45\textwidth]{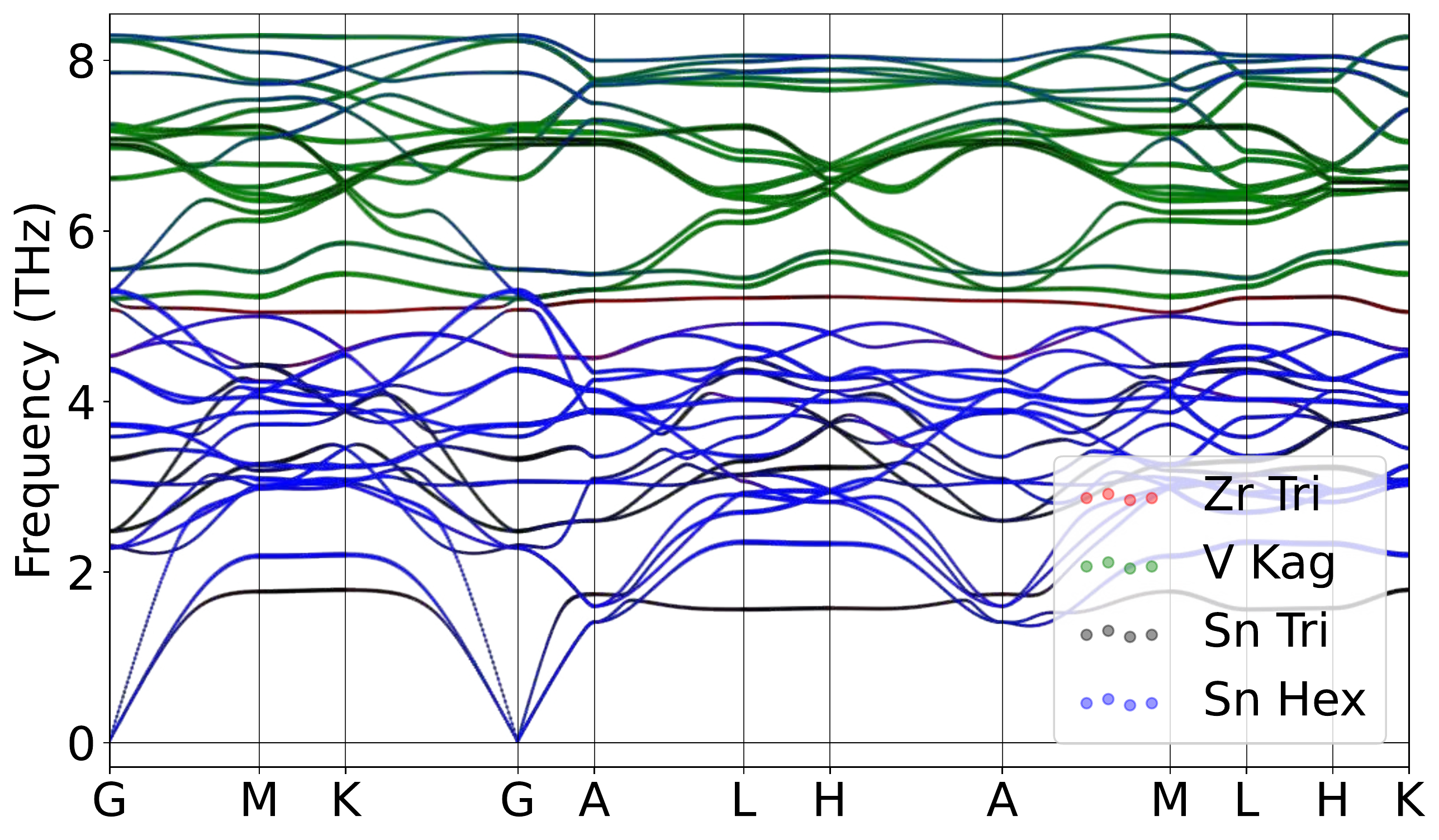}
}
\caption{Atomically projected phonon spectrum for ZrV$_6$Sn$_6$.}
\label{fig:ZrV6Sn6pho}
\end{figure}

\begin{figure}[H]
\centering
\label{fig:ZrV6Sn6_relaxed_Low_xz}
\includegraphics[width=0.85\textwidth]{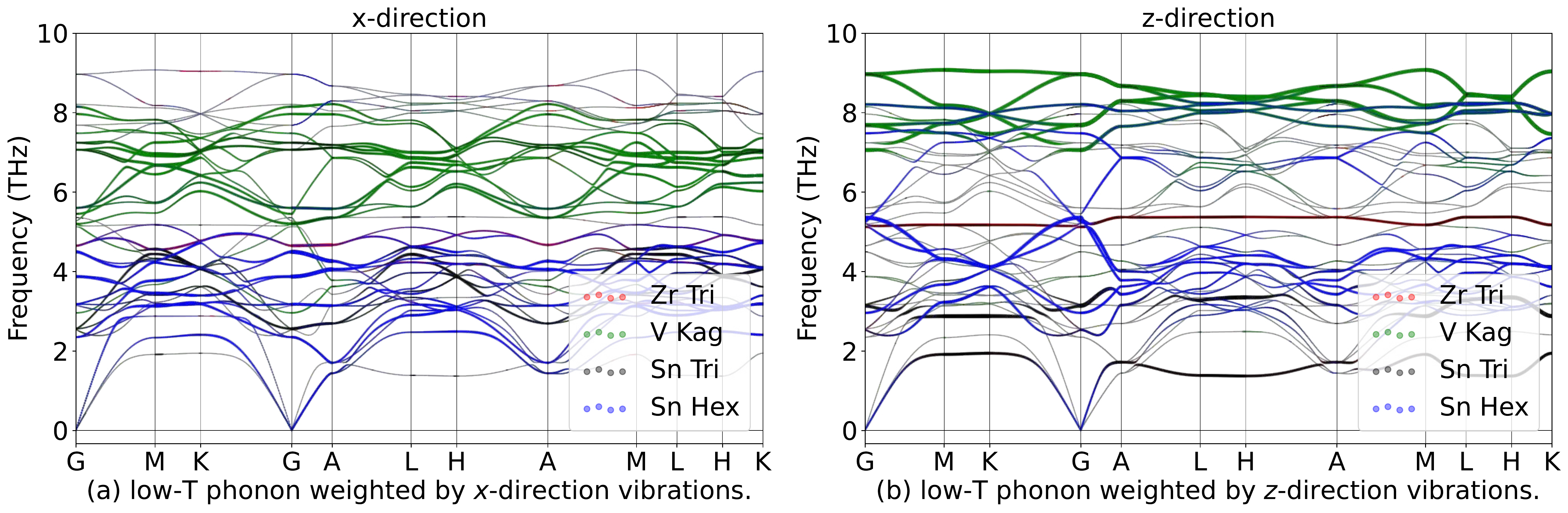}
\hspace{5pt}\label{fig:ZrV6Sn6_relaxed_High_xz}
\includegraphics[width=0.85\textwidth]{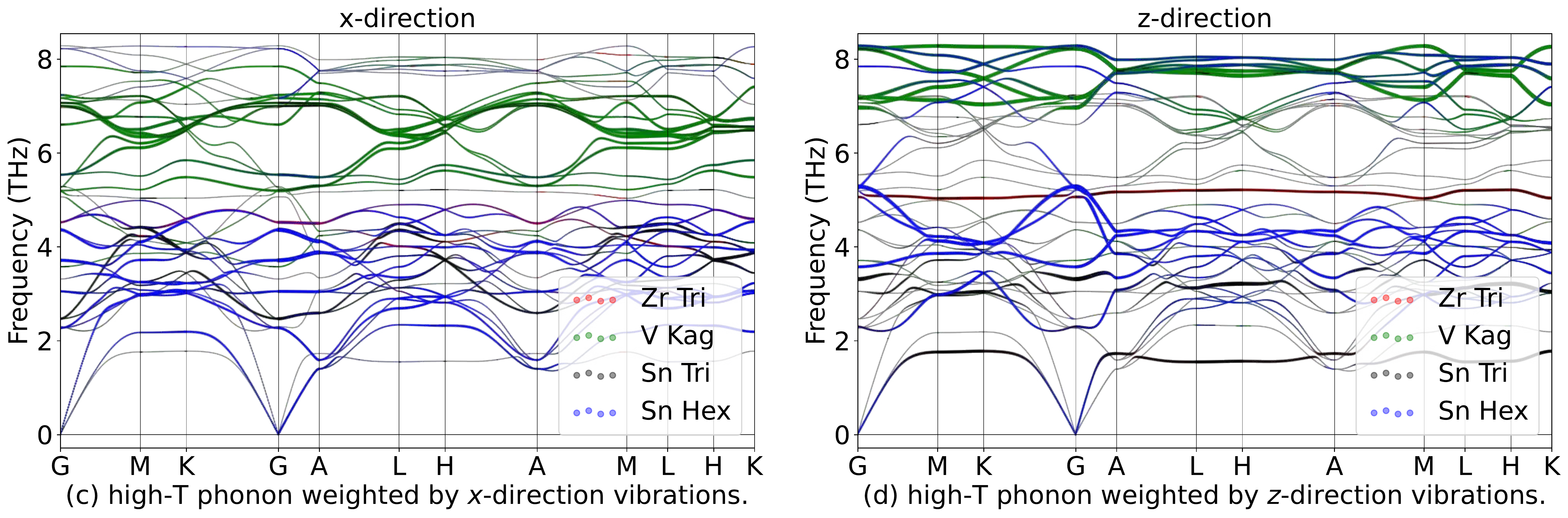}
\caption{Phonon spectrum for ZrV$_6$Sn$_6$in in-plane ($x$) and out-of-plane ($z$) directions.}
\label{fig:ZrV6Sn6phoxyz}
\end{figure}

\begin{table}[htbp]
\global\long\def\arraystretch{1.12}
\captionsetup{width=.95\textwidth}
\caption{IRREPs at high-symmetry points for ZrV$_6$Sn$_6$ phonon spectrum at Low-T.}
\label{tab:191_ZrV6Sn6_Low}
\setlength{\tabcolsep}{1.mm}{
}
\end{table}

\begin{table}[htbp]
\global\long\def\arraystretch{1.12}
\captionsetup{width=.95\textwidth}
\caption{IRREPs at high-symmetry points for ZrV$_6$Sn$_6$ phonon spectrum at High-T.}
\label{tab:191_ZrV6Sn6_High}
\setlength{\tabcolsep}{1.mm}{
%
}
\end{table}
\subsubsection{\label{sms2sec:NbV6Sn6}\hyperref[smsec:col]{\texorpdfstring{NbV$_6$Sn$_6$}{}}~{\color{green}  (High-T stable, Low-T stable) }}

\begin{table}[htbp]
\global\long\def\arraystretch{1.12}
\captionsetup{width=.85\textwidth}
\caption{Structure parameters of NbV$_6$Sn$_6$ after relaxation. It contains 479 electrons. }
\label{tab:NbV6Sn6}
\setlength{\tabcolsep}{4mm}{
%
}
\end{table}

\begin{figure}[htbp]
\centering
\includegraphics[width=0.85\textwidth]{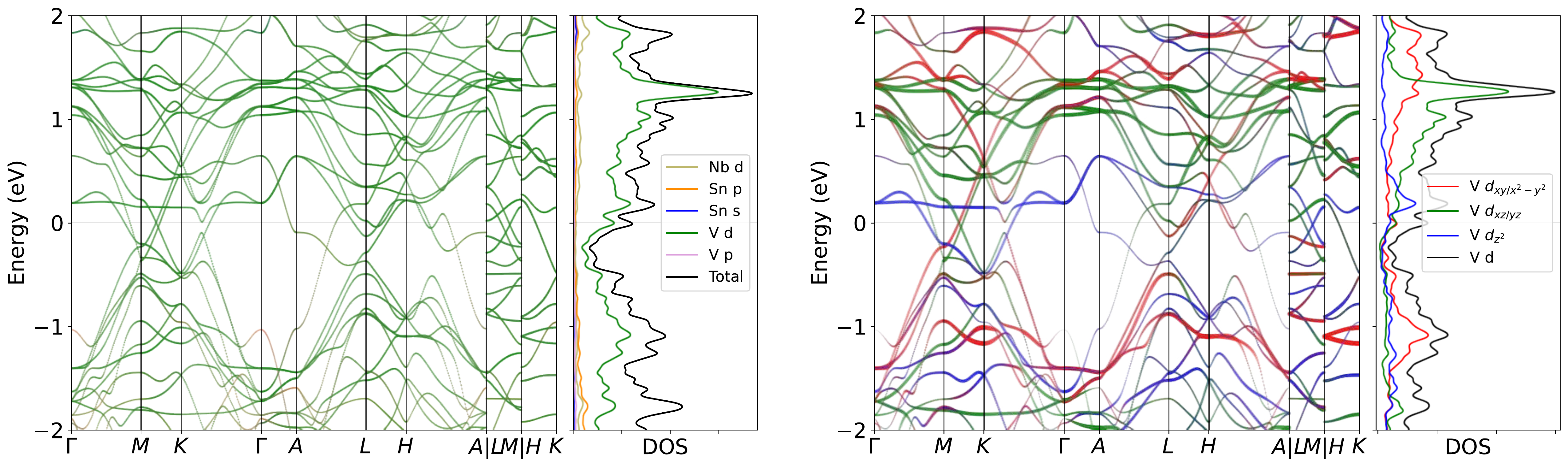}
\caption{Band structure for NbV$_6$Sn$_6$ without SOC. The projection of V $3d$ orbitals is also presented. The band structures clearly reveal the dominating of V $3d$ orbitals  near the Fermi level, as well as the pronounced peak.}
\label{fig:NbV6Sn6band}
\end{figure}

\begin{figure}[H]
\centering
\subfloat[Relaxed phonon at low-T.]{
\label{fig:NbV6Sn6_relaxed_Low}
\includegraphics[width=0.45\textwidth]{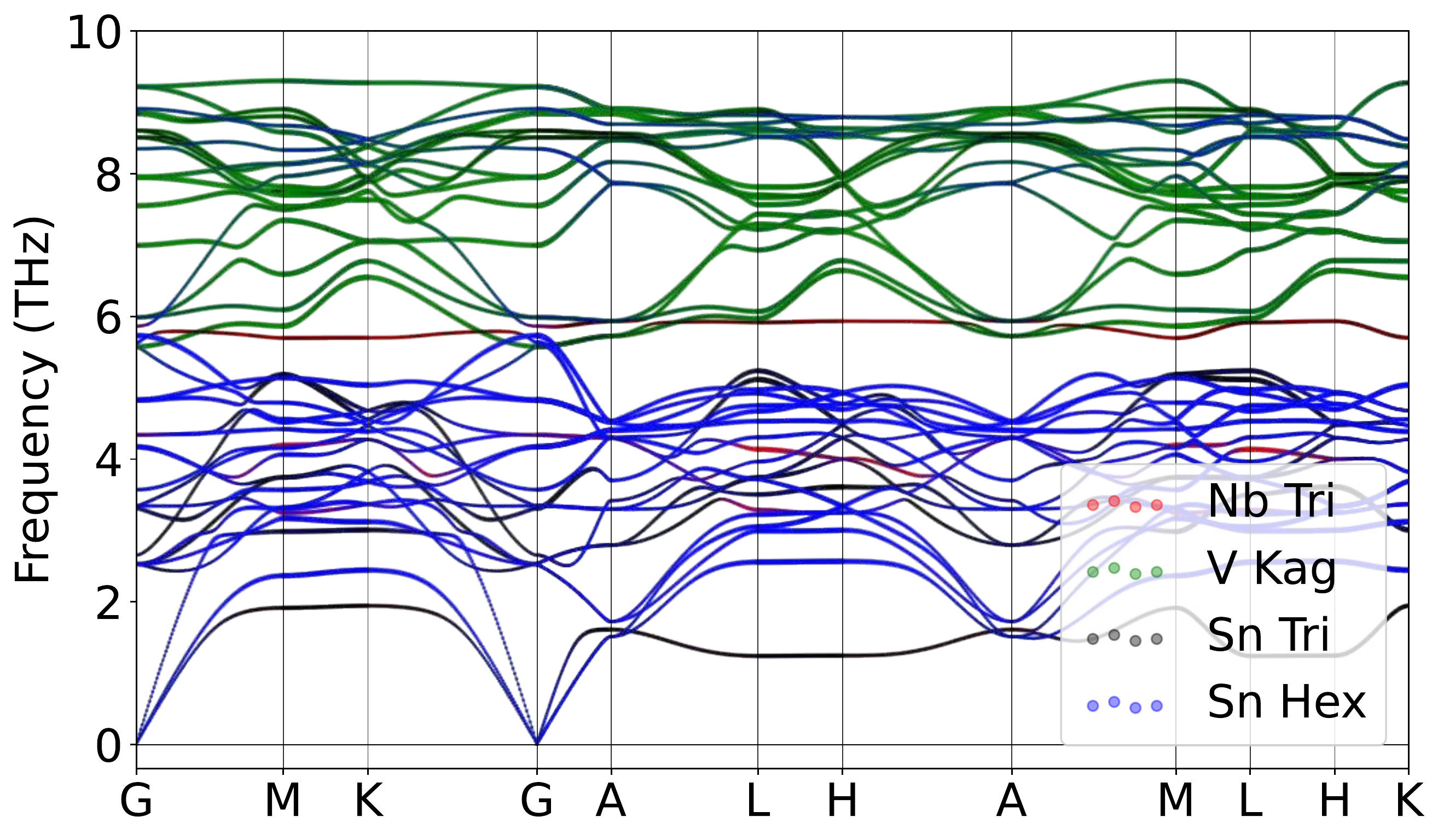}
}
\hspace{5pt}\subfloat[Relaxed phonon at high-T.]{
\label{fig:NbV6Sn6_relaxed_High}
\includegraphics[width=0.45\textwidth]{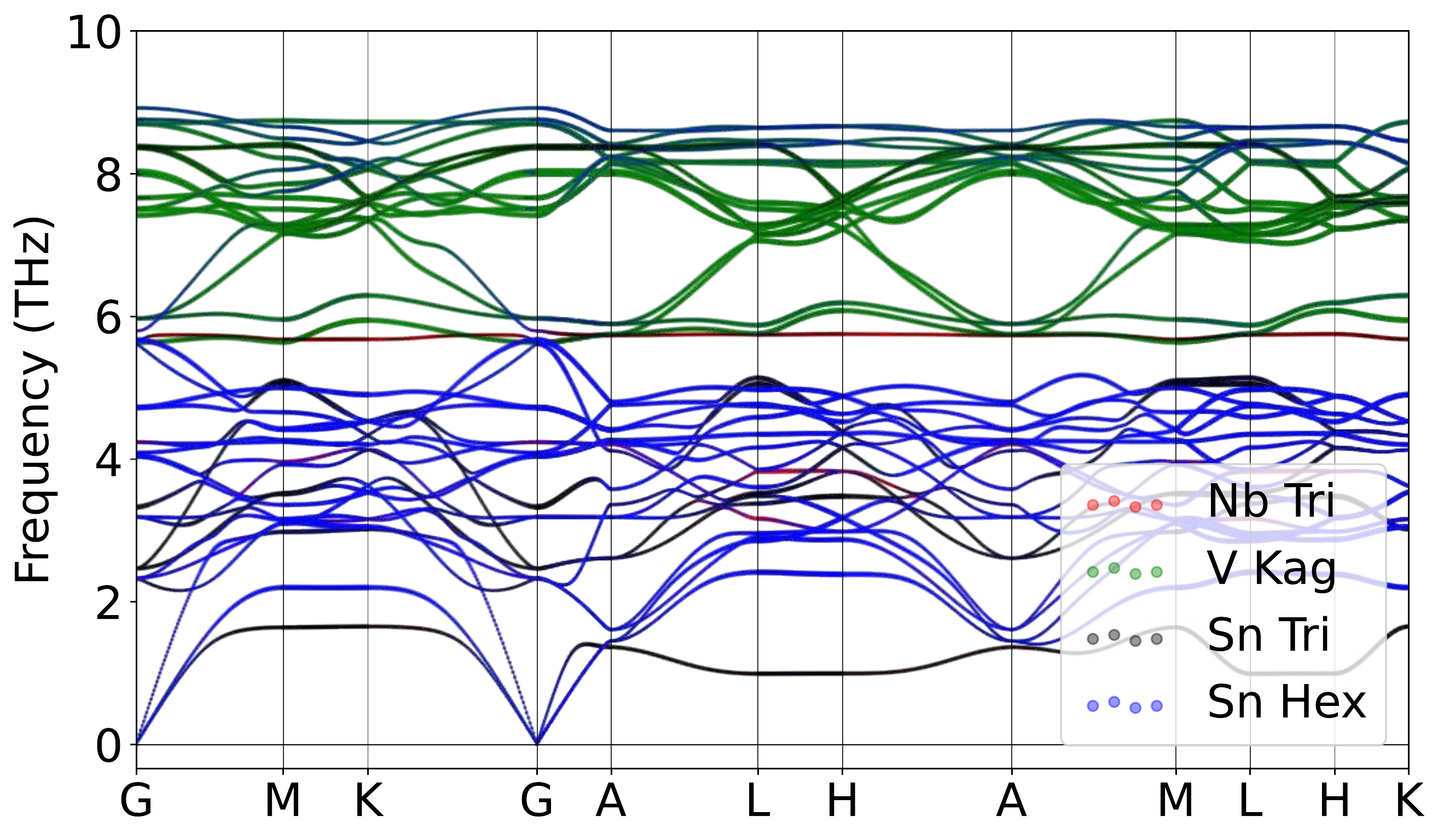}
}
\caption{Atomically projected phonon spectrum for NbV$_6$Sn$_6$.}
\label{fig:NbV6Sn6pho}
\end{figure}

\begin{figure}[H]
\centering
\label{fig:NbV6Sn6_relaxed_Low_xz}
\includegraphics[width=0.85\textwidth]{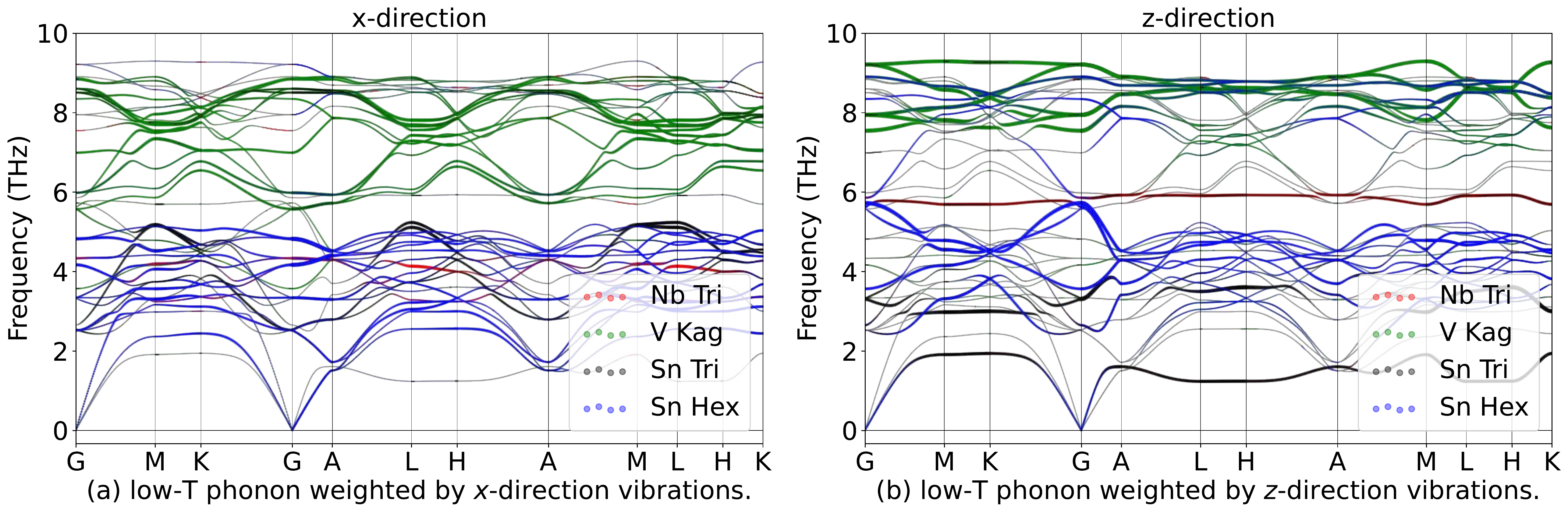}
\hspace{5pt}\label{fig:NbV6Sn6_relaxed_High_xz}
\includegraphics[width=0.85\textwidth]{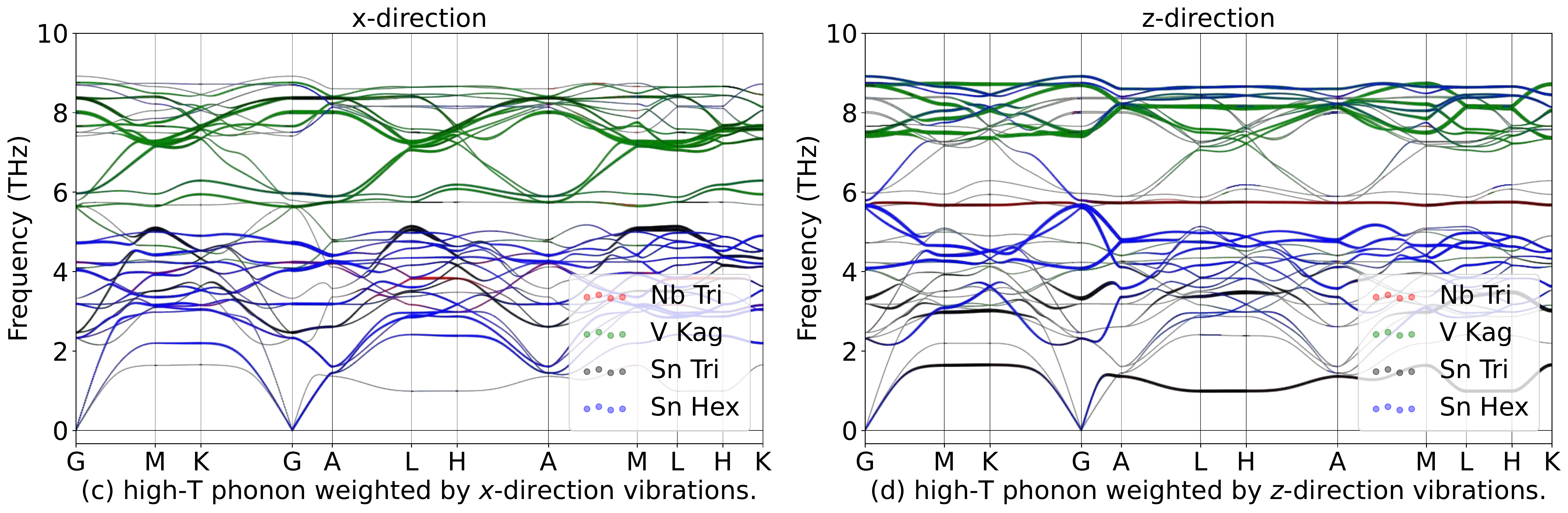}
\caption{Phonon spectrum for NbV$_6$Sn$_6$in in-plane ($x$) and out-of-plane ($z$) directions.}
\label{fig:NbV6Sn6phoxyz}
\end{figure}

\begin{table}[htbp]
\global\long\def\arraystretch{1.12}
\captionsetup{width=.95\textwidth}
\caption{IRREPs at high-symmetry points for NbV$_6$Sn$_6$ phonon spectrum at Low-T.}
\label{tab:191_NbV6Sn6_Low}
\setlength{\tabcolsep}{1.mm}{
}
\end{table}

\begin{table}[htbp]
\global\long\def\arraystretch{1.12}
\captionsetup{width=.95\textwidth}
\caption{IRREPs at high-symmetry points for NbV$_6$Sn$_6$ phonon spectrum at High-T.}
\label{tab:191_NbV6Sn6_High}
\setlength{\tabcolsep}{1.mm}{
%
}
\end{table}
\subsubsection{\label{sms2sec:PrV6Sn6}\hyperref[smsec:col]{\texorpdfstring{PrV$_6$Sn$_6$}{}}~{\color{green}  (High-T stable, Low-T stable) }}

\begin{table}[htbp]
\global\long\def\arraystretch{1.12}
\captionsetup{width=.85\textwidth}
\caption{Structure parameters of PrV$_6$Sn$_6$ after relaxation. It contains 497 electrons. }
\label{tab:PrV6Sn6}
\setlength{\tabcolsep}{4mm}{
%
}
\end{table}

\begin{figure}[htbp]
\centering
\includegraphics[width=0.85\textwidth]{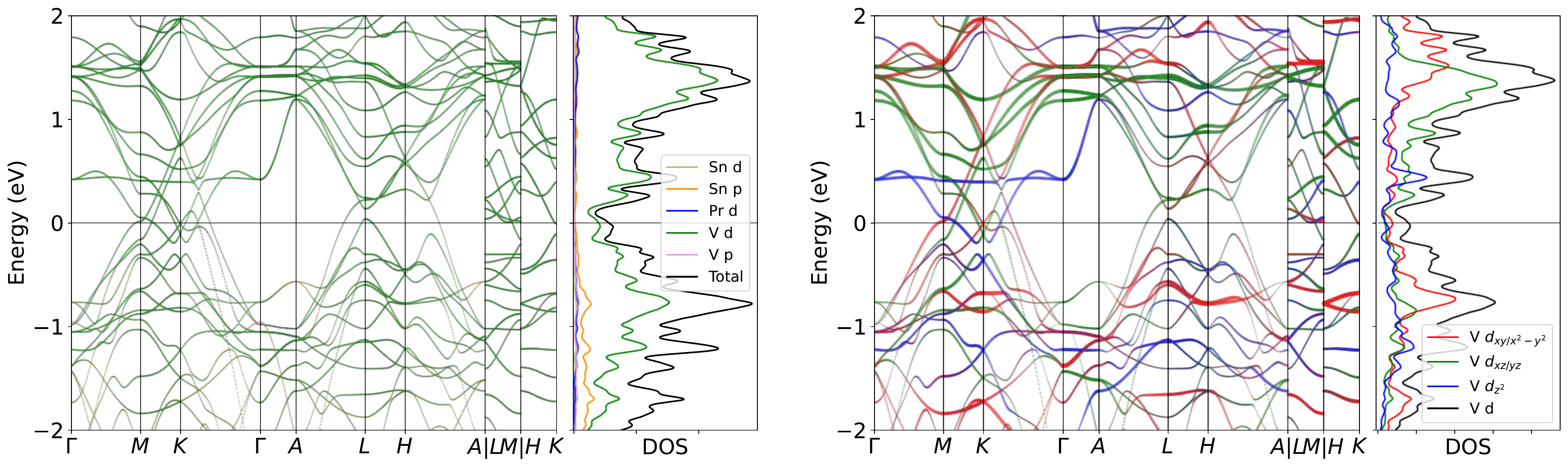}
\caption{Band structure for PrV$_6$Sn$_6$ without SOC. The projection of V $3d$ orbitals is also presented. The band structures clearly reveal the dominating of V $3d$ orbitals  near the Fermi level, as well as the pronounced peak.}
\label{fig:PrV6Sn6band}
\end{figure}

\begin{figure}[H]
\centering
\subfloat[Relaxed phonon at low-T.]{
\label{fig:PrV6Sn6_relaxed_Low}
\includegraphics[width=0.45\textwidth]{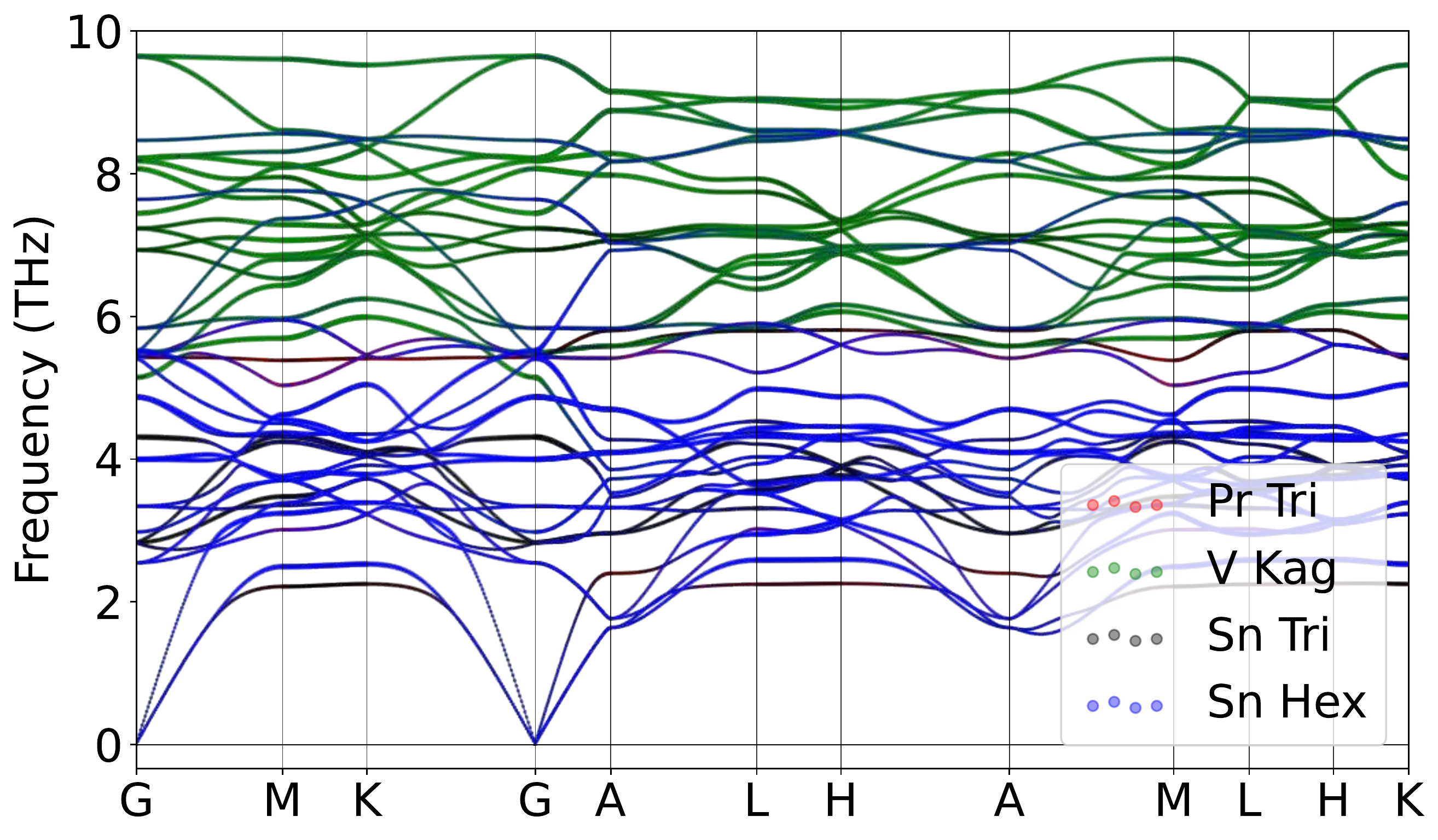}
}
\hspace{5pt}\subfloat[Relaxed phonon at high-T.]{
\label{fig:PrV6Sn6_relaxed_High}
\includegraphics[width=0.45\textwidth]{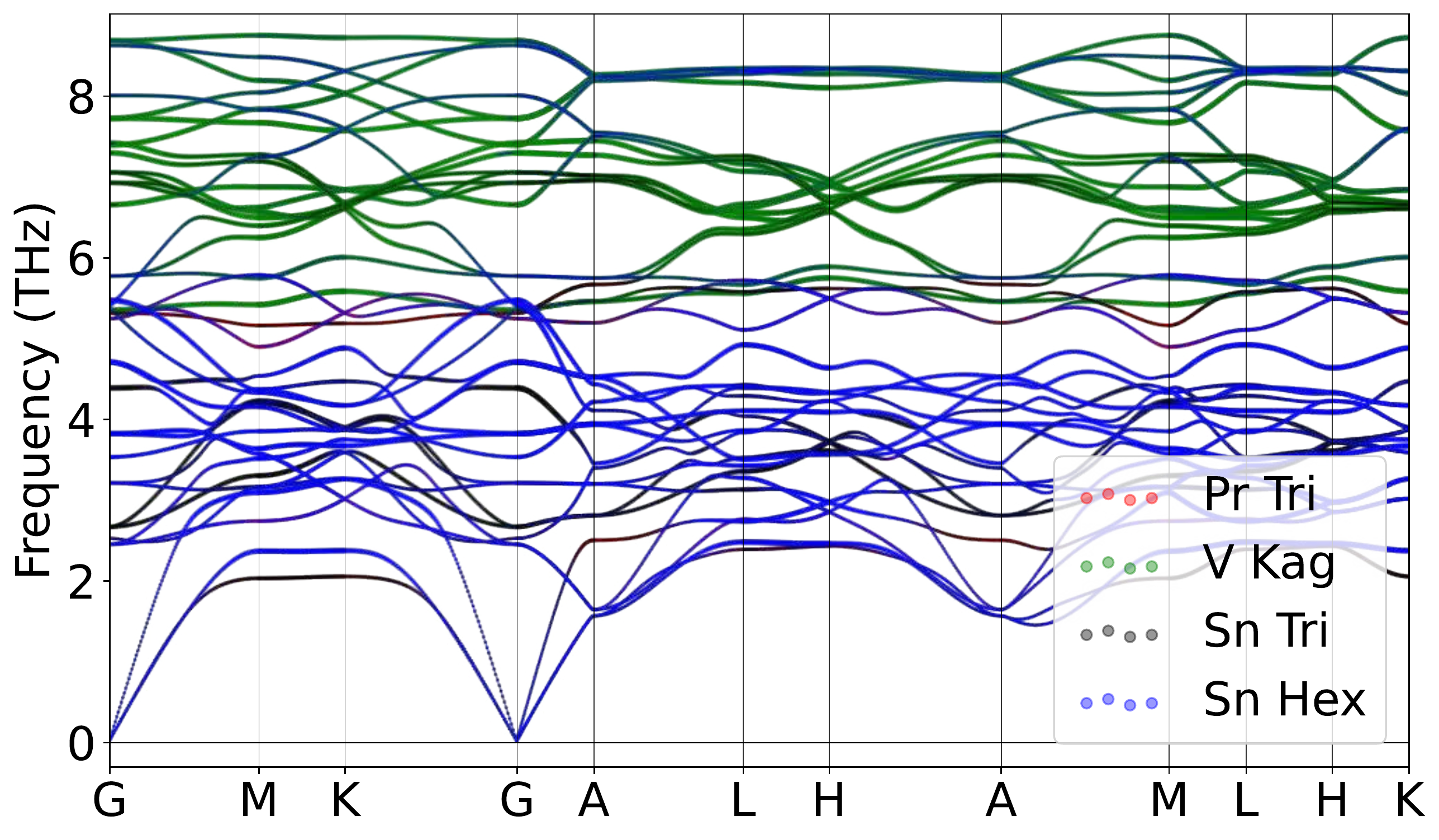}
}
\caption{Atomically projected phonon spectrum for PrV$_6$Sn$_6$.}
\label{fig:PrV6Sn6pho}
\end{figure}

\begin{figure}[H]
\centering
\label{fig:PrV6Sn6_relaxed_Low_xz}
\includegraphics[width=0.85\textwidth]{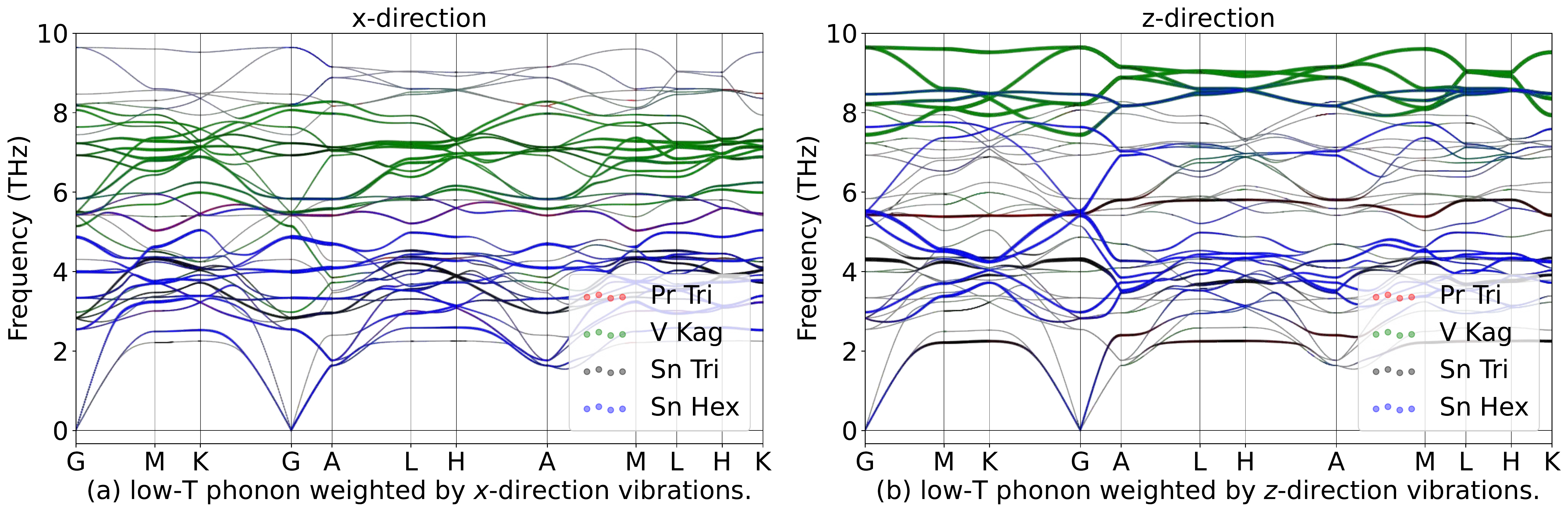}
\hspace{5pt}\label{fig:PrV6Sn6_relaxed_High_xz}
\includegraphics[width=0.85\textwidth]{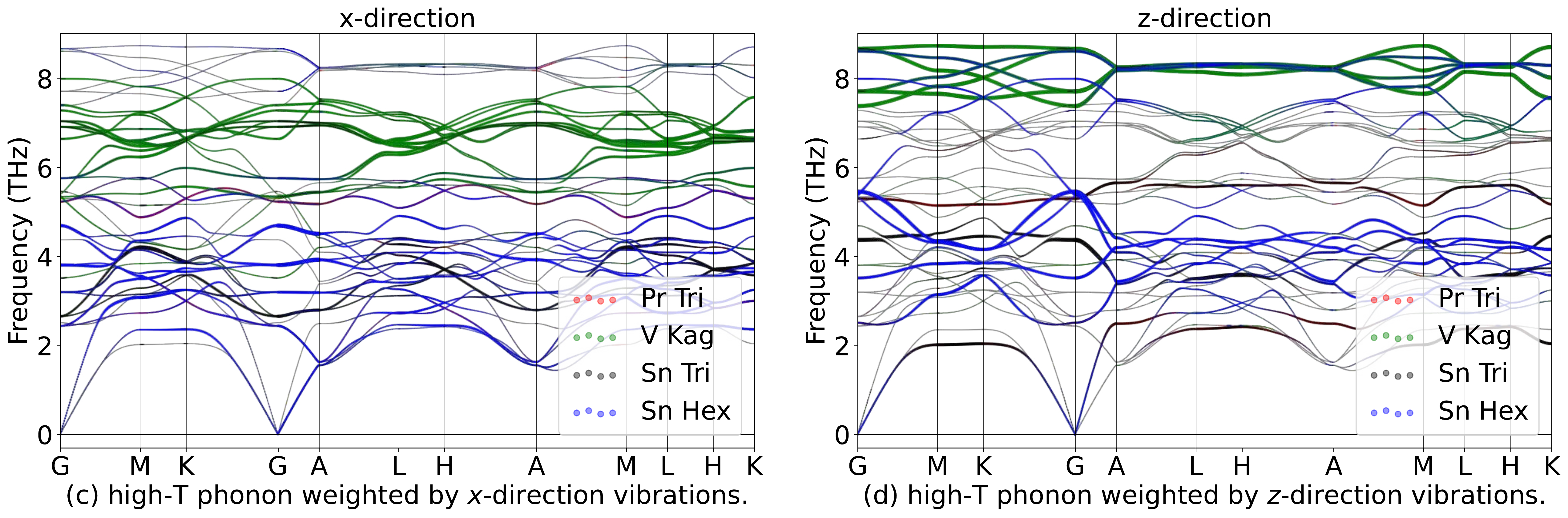}
\caption{Phonon spectrum for PrV$_6$Sn$_6$in in-plane ($x$) and out-of-plane ($z$) directions.}
\label{fig:PrV6Sn6phoxyz}
\end{figure}

\begin{table}[htbp]
\global\long\def\arraystretch{1.12}
\captionsetup{width=.95\textwidth}
\caption{IRREPs at high-symmetry points for PrV$_6$Sn$_6$ phonon spectrum at Low-T.}
\label{tab:191_PrV6Sn6_Low}
\setlength{\tabcolsep}{1.mm}{
}
\end{table}

\begin{table}[htbp]
\global\long\def\arraystretch{1.12}
\captionsetup{width=.95\textwidth}
\caption{IRREPs at high-symmetry points for PrV$_6$Sn$_6$ phonon spectrum at High-T.}
\label{tab:191_PrV6Sn6_High}
\setlength{\tabcolsep}{1.mm}{
%
}
\end{table}
\subsubsection{\label{sms2sec:NdV6Sn6}\hyperref[smsec:col]{\texorpdfstring{NdV$_6$Sn$_6$}{}}~{\color{green}  (High-T stable, Low-T stable) }}

\begin{table}[htbp]
\global\long\def\arraystretch{1.12}
\captionsetup{width=.85\textwidth}
\caption{Structure parameters of NdV$_6$Sn$_6$ after relaxation. It contains 498 electrons. }
\label{tab:NdV6Sn6}
\setlength{\tabcolsep}{4mm}{
%
}
\end{table}

\begin{figure}[htbp]
\centering
\includegraphics[width=0.85\textwidth]{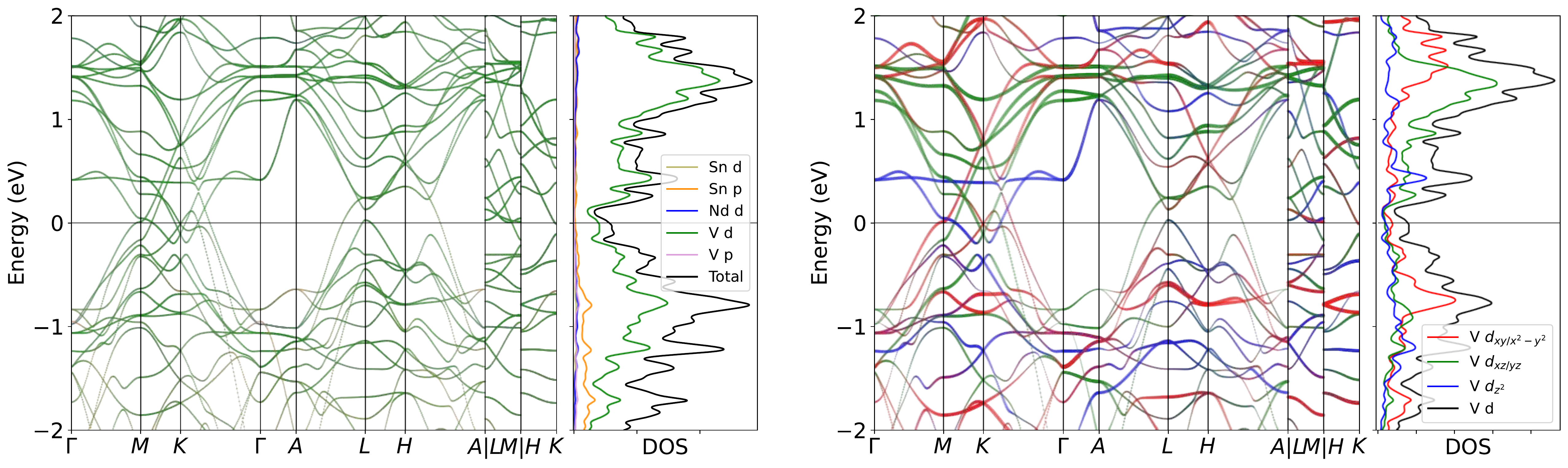}
\caption{Band structure for NdV$_6$Sn$_6$ without SOC. The projection of V $3d$ orbitals is also presented. The band structures clearly reveal the dominating of V $3d$ orbitals  near the Fermi level, as well as the pronounced peak.}
\label{fig:NdV6Sn6band}
\end{figure}

\begin{figure}[H]
\centering
\subfloat[Relaxed phonon at low-T.]{
\label{fig:NdV6Sn6_relaxed_Low}
\includegraphics[width=0.45\textwidth]{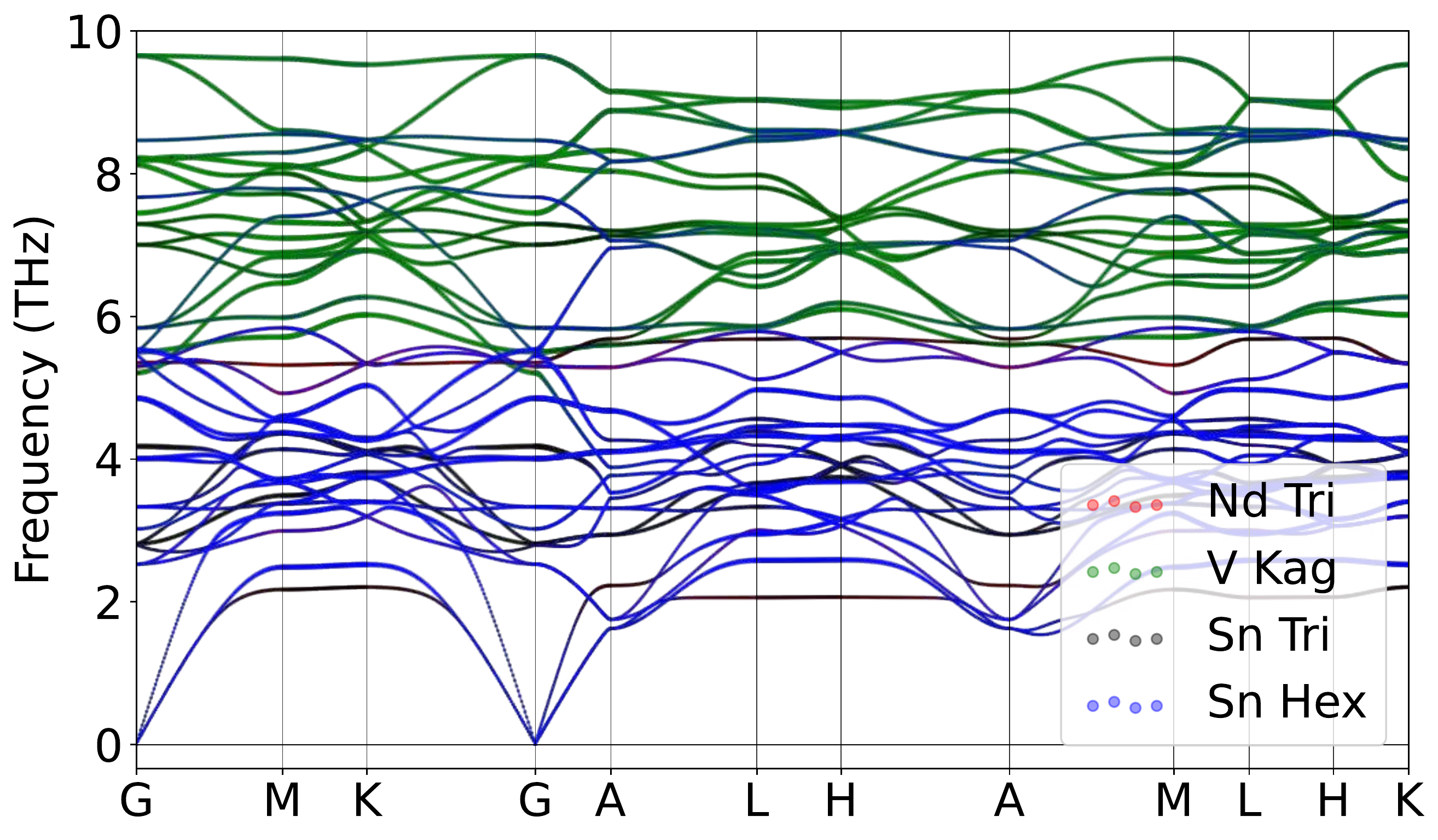}
}
\hspace{5pt}\subfloat[Relaxed phonon at high-T.]{
\label{fig:NdV6Sn6_relaxed_High}
\includegraphics[width=0.45\textwidth]{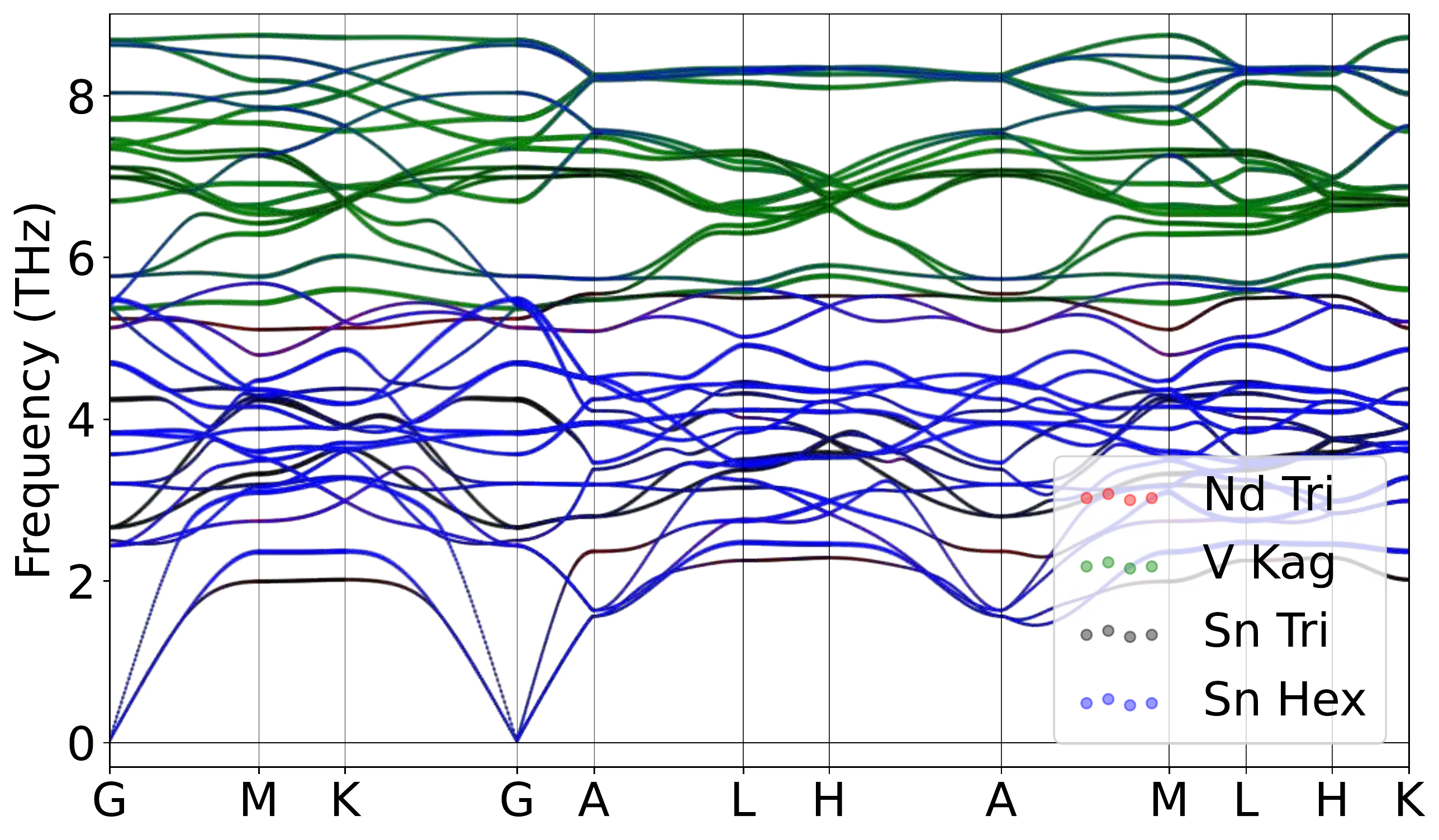}
}
\caption{Atomically projected phonon spectrum for NdV$_6$Sn$_6$.}
\label{fig:NdV6Sn6pho}
\end{figure}

\begin{figure}[H]
\centering
\label{fig:NdV6Sn6_relaxed_Low_xz}
\includegraphics[width=0.85\textwidth]{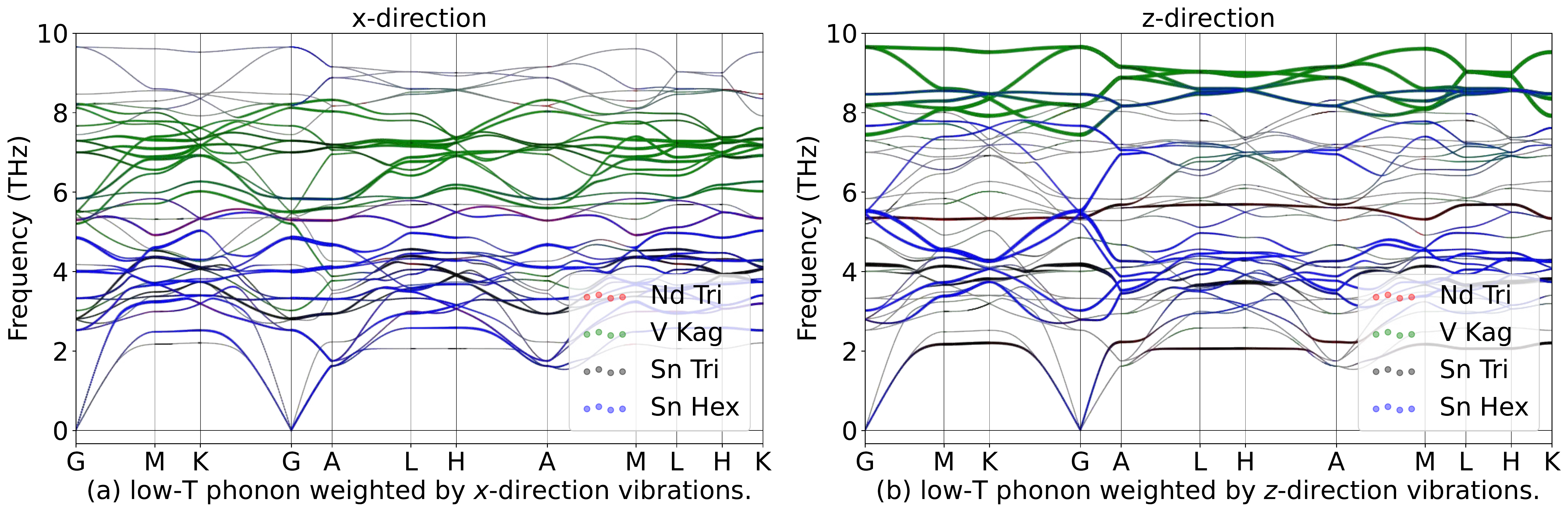}
\hspace{5pt}\label{fig:NdV6Sn6_relaxed_High_xz}
\includegraphics[width=0.85\textwidth]{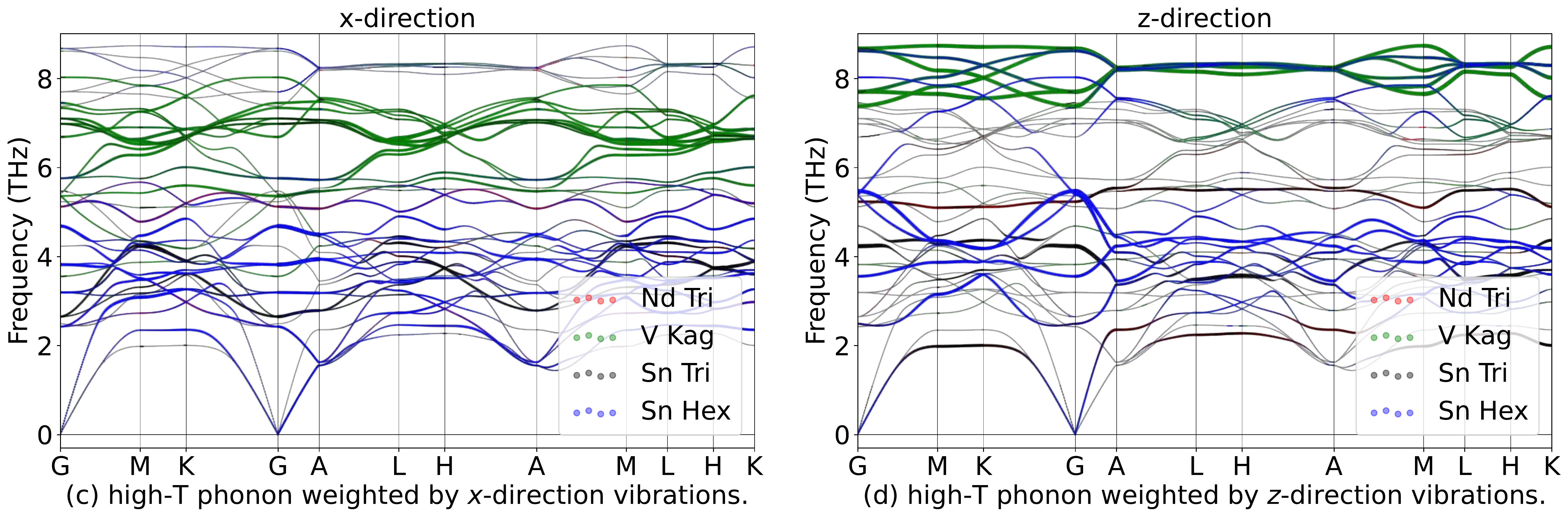}
\caption{Phonon spectrum for NdV$_6$Sn$_6$in in-plane ($x$) and out-of-plane ($z$) directions.}
\label{fig:NdV6Sn6phoxyz}
\end{figure}

\begin{table}[htbp]
\global\long\def\arraystretch{1.12}
\captionsetup{width=.95\textwidth}
\caption{IRREPs at high-symmetry points for NdV$_6$Sn$_6$ phonon spectrum at Low-T.}
\label{tab:191_NdV6Sn6_Low}
\setlength{\tabcolsep}{1.mm}{
}
\end{table}

\begin{table}[htbp]
\global\long\def\arraystretch{1.12}
\captionsetup{width=.95\textwidth}
\caption{IRREPs at high-symmetry points for NdV$_6$Sn$_6$ phonon spectrum at High-T.}
\label{tab:191_NdV6Sn6_High}
\setlength{\tabcolsep}{1.mm}{
%
}
\end{table}
\subsubsection{\label{sms2sec:SmV6Sn6}\hyperref[smsec:col]{\texorpdfstring{SmV$_6$Sn$_6$}{}}~{\color{green}  (High-T stable, Low-T stable) }}

\begin{table}[htbp]
\global\long\def\arraystretch{1.12}
\captionsetup{width=.85\textwidth}
\caption{Structure parameters of SmV$_6$Sn$_6$ after relaxation. It contains 500 electrons. }
\label{tab:SmV6Sn6}
\setlength{\tabcolsep}{4mm}{
%
}
\end{table}

\begin{figure}[htbp]
\centering
\includegraphics[width=0.85\textwidth]{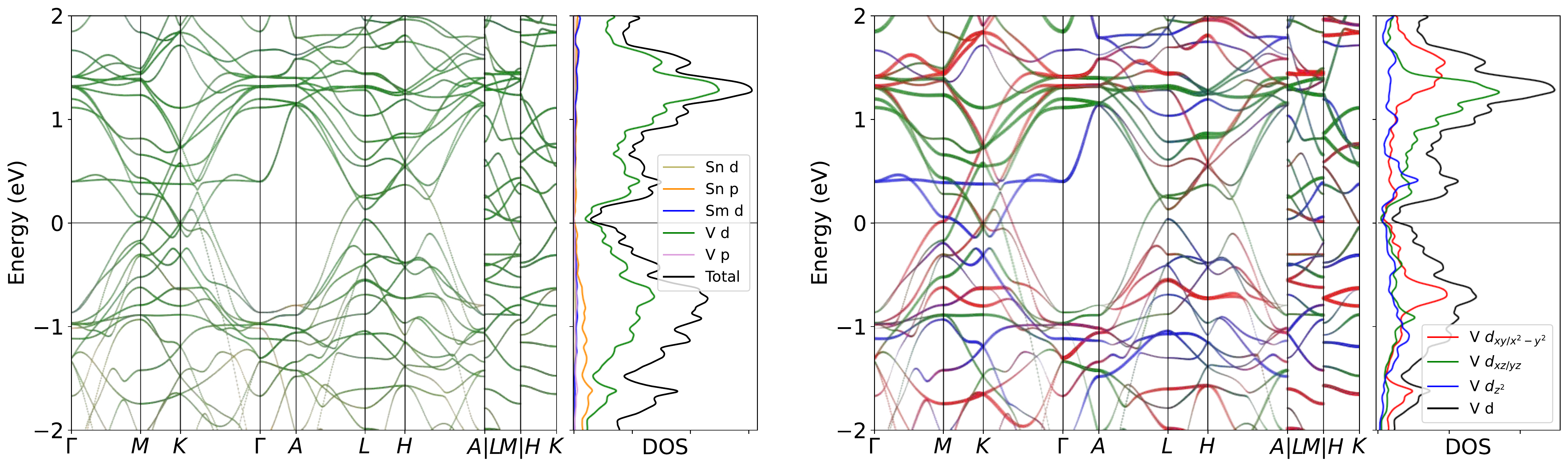}
\caption{Band structure for SmV$_6$Sn$_6$ without SOC. The projection of V $3d$ orbitals is also presented. The band structures clearly reveal the dominating of V $3d$ orbitals  near the Fermi level, as well as the pronounced peak.}
\label{fig:SmV6Sn6band}
\end{figure}

\begin{figure}[H]
\centering
\subfloat[Relaxed phonon at low-T.]{
\label{fig:SmV6Sn6_relaxed_Low}
\includegraphics[width=0.45\textwidth]{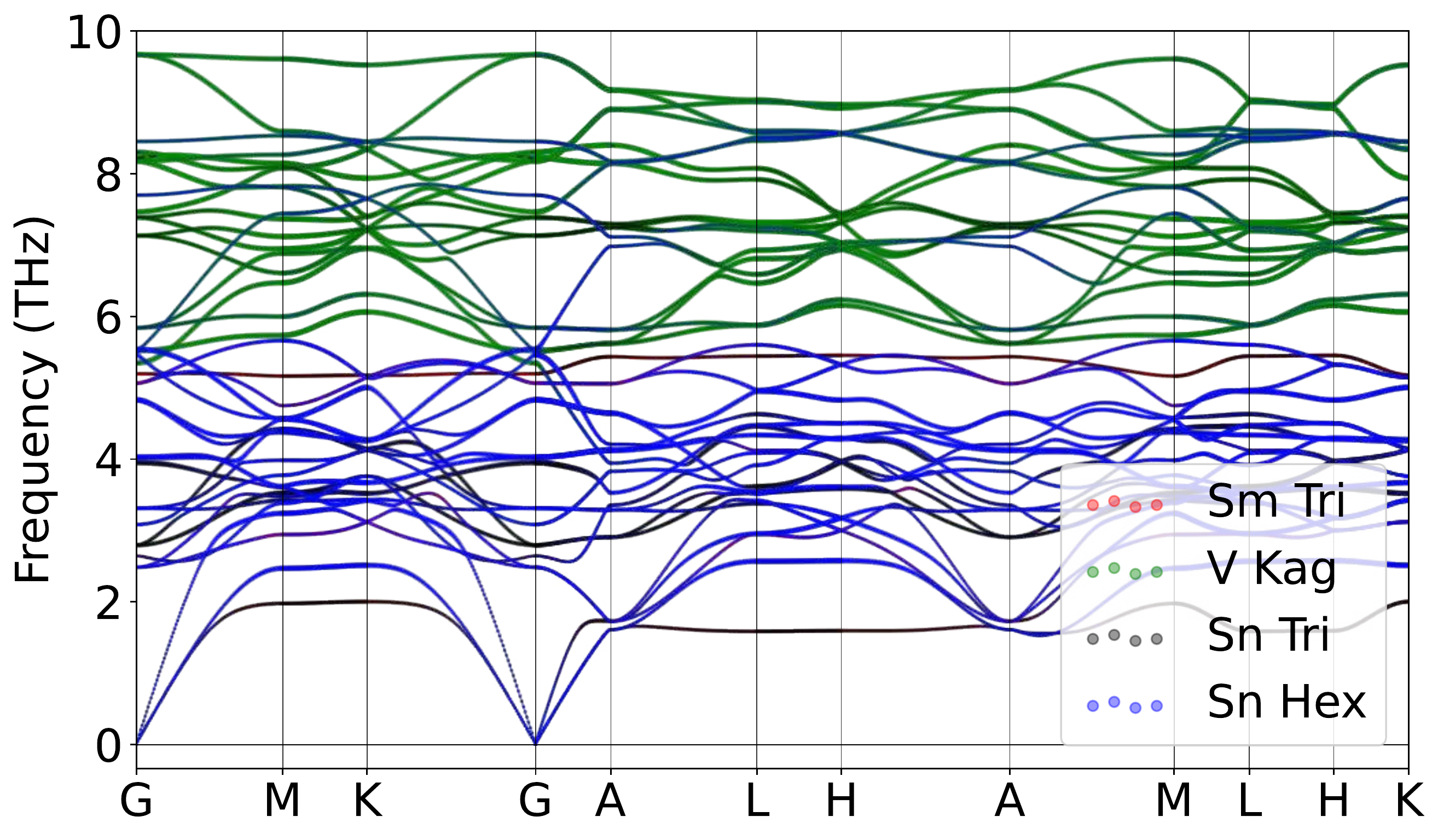}
}
\hspace{5pt}\subfloat[Relaxed phonon at high-T.]{
\label{fig:SmV6Sn6_relaxed_High}
\includegraphics[width=0.45\textwidth]{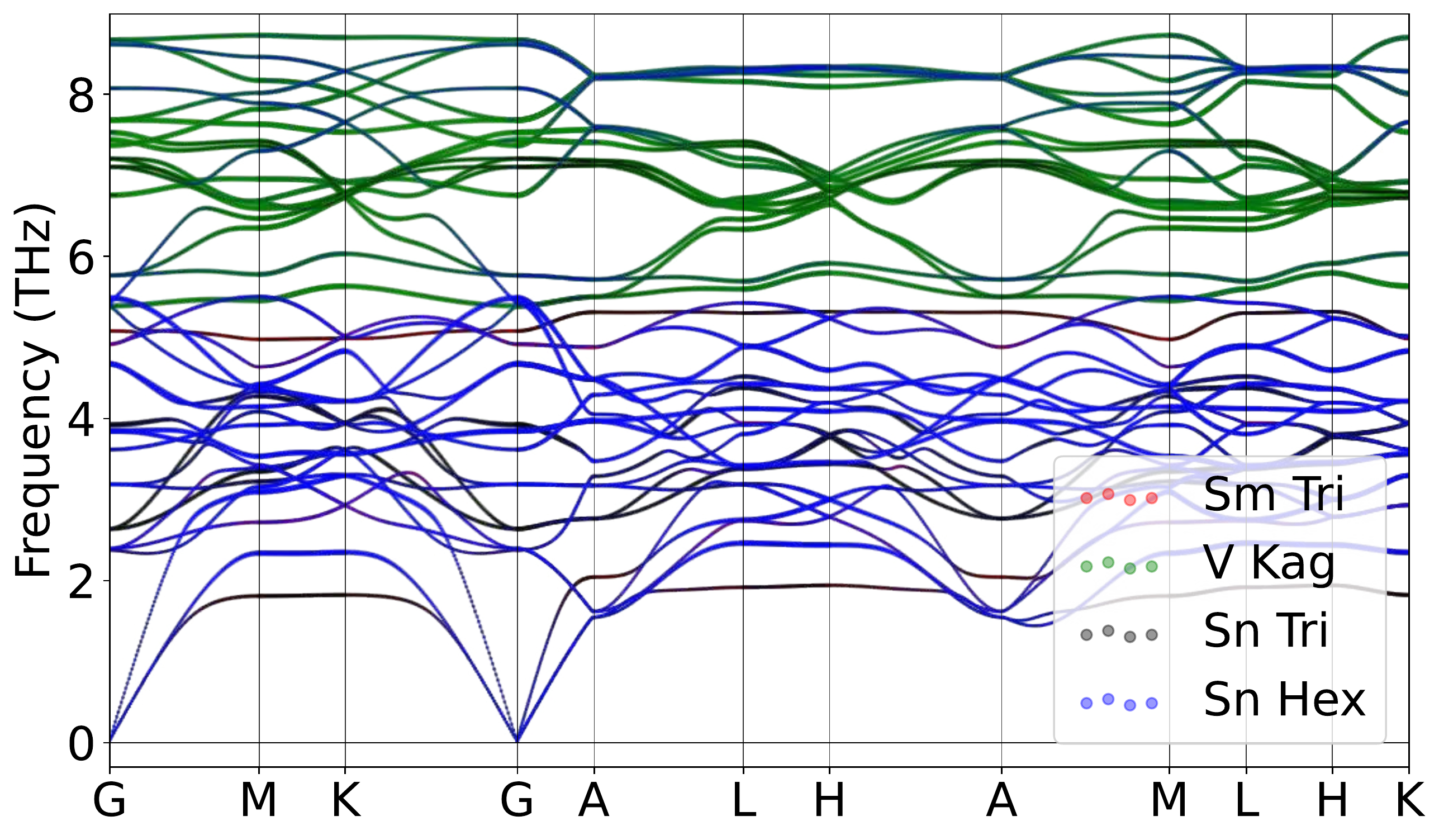}
}
\caption{Atomically projected phonon spectrum for SmV$_6$Sn$_6$.}
\label{fig:SmV6Sn6pho}
\end{figure}

\begin{figure}[H]
\centering
\label{fig:SmV6Sn6_relaxed_Low_xz}
\includegraphics[width=0.85\textwidth]{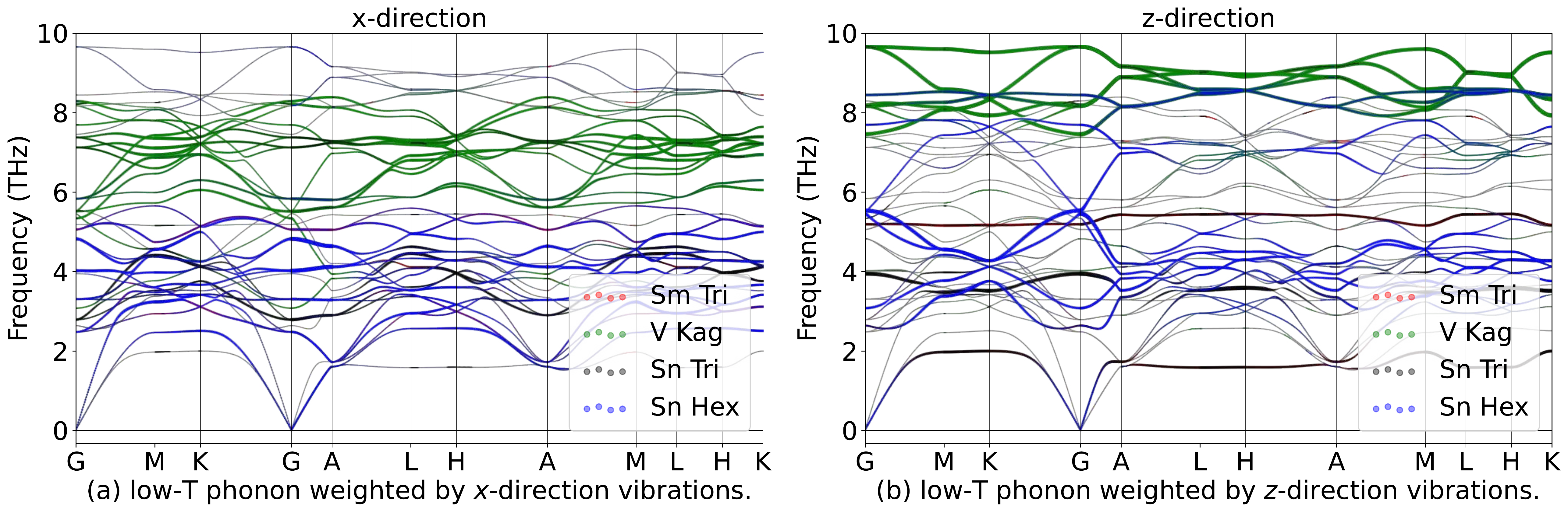}
\hspace{5pt}\label{fig:SmV6Sn6_relaxed_High_xz}
\includegraphics[width=0.85\textwidth]{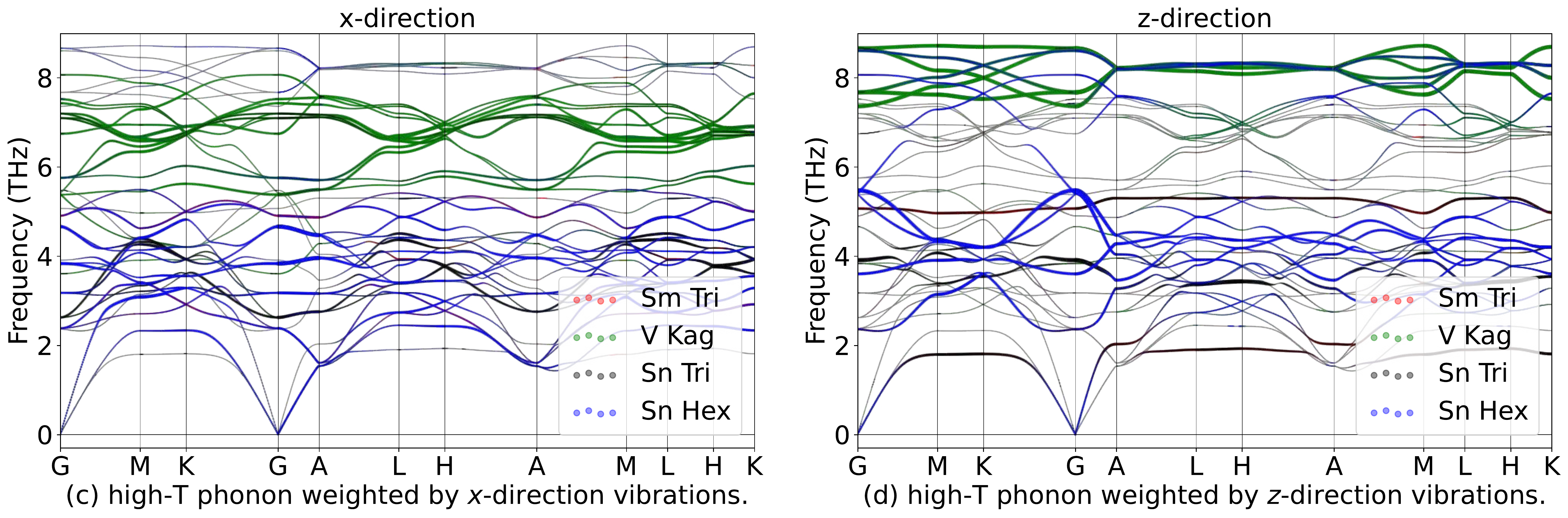}
\caption{Phonon spectrum for SmV$_6$Sn$_6$in in-plane ($x$) and out-of-plane ($z$) directions.}
\label{fig:SmV6Sn6phoxyz}
\end{figure}

\begin{table}[htbp]
\global\long\def\arraystretch{1.12}
\captionsetup{width=.95\textwidth}
\caption{IRREPs at high-symmetry points for SmV$_6$Sn$_6$ phonon spectrum at Low-T.}
\label{tab:191_SmV6Sn6_Low}
\setlength{\tabcolsep}{1.mm}{
}
\end{table}

\begin{table}[htbp]
\global\long\def\arraystretch{1.12}
\captionsetup{width=.95\textwidth}
\caption{IRREPs at high-symmetry points for SmV$_6$Sn$_6$ phonon spectrum at High-T.}
\label{tab:191_SmV6Sn6_High}
\setlength{\tabcolsep}{1.mm}{
%
}
\end{table}
\subsubsection{\label{sms2sec:GdV6Sn6}\hyperref[smsec:col]{\texorpdfstring{GdV$_6$Sn$_6$}{}}~{\color{green}  (High-T stable, Low-T stable) }}

\begin{table}[htbp]
\global\long\def\arraystretch{1.12}
\captionsetup{width=.85\textwidth}
\caption{Structure parameters of GdV$_6$Sn$_6$ after relaxation. It contains 502 electrons. }
\label{tab:GdV6Sn6}
\setlength{\tabcolsep}{4mm}{
%
}
\end{table}

\begin{figure}[htbp]
\centering
\includegraphics[width=0.85\textwidth]{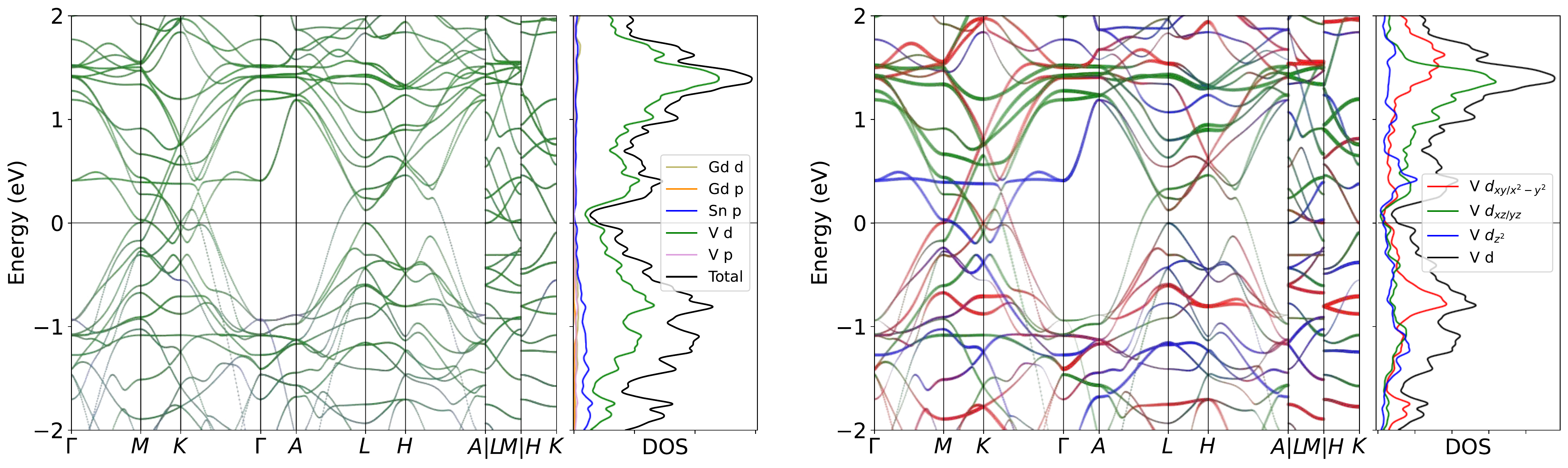}
\caption{Band structure for GdV$_6$Sn$_6$ without SOC. The projection of V $3d$ orbitals is also presented. The band structures clearly reveal the dominating of V $3d$ orbitals  near the Fermi level, as well as the pronounced peak.}
\label{fig:GdV6Sn6band}
\end{figure}

\begin{figure}[H]
\centering
\subfloat[Relaxed phonon at low-T.]{
\label{fig:GdV6Sn6_relaxed_Low}
\includegraphics[width=0.45\textwidth]{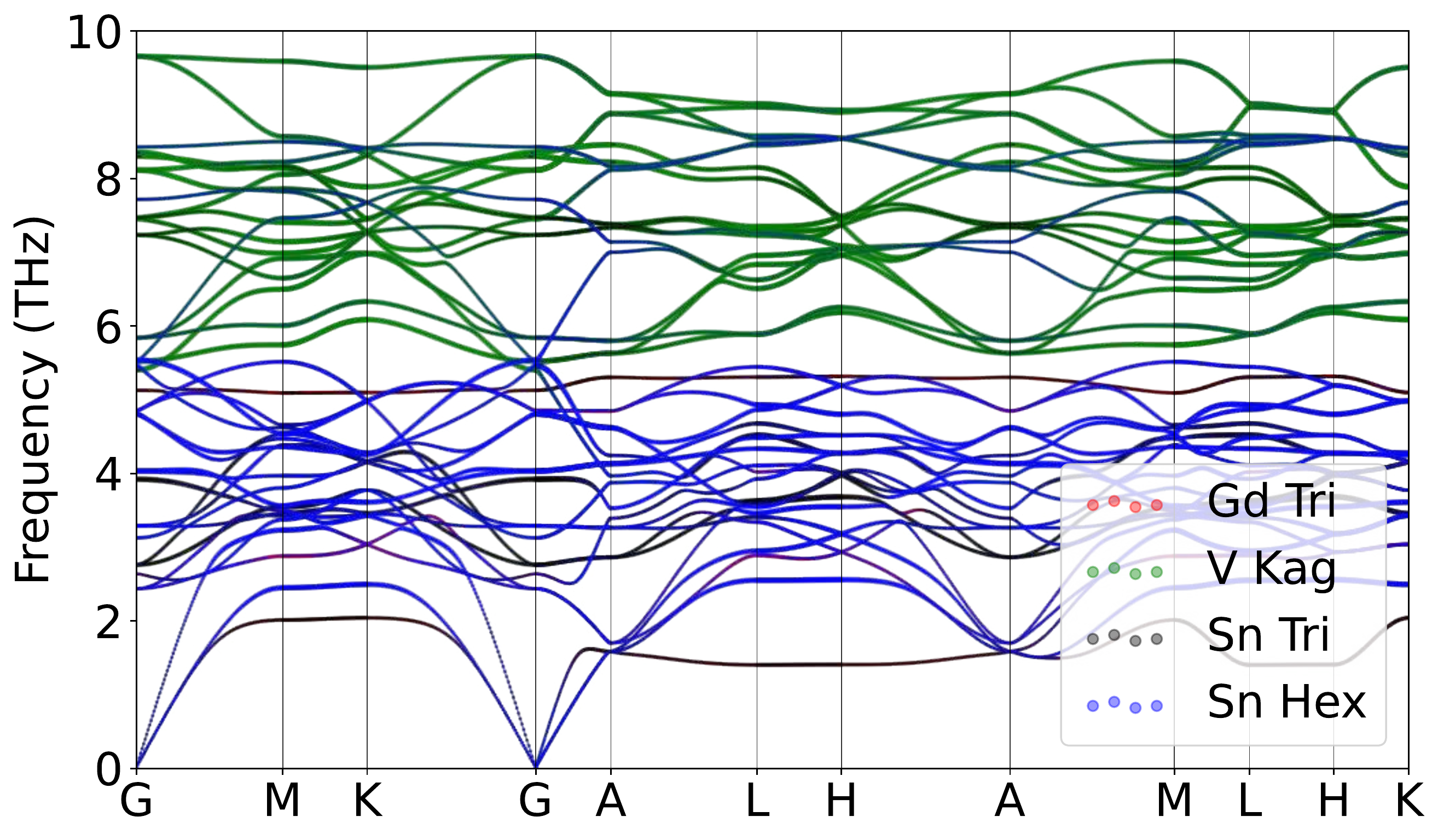}
}
\hspace{5pt}\subfloat[Relaxed phonon at high-T.]{
\label{fig:GdV6Sn6_relaxed_High}
\includegraphics[width=0.45\textwidth]{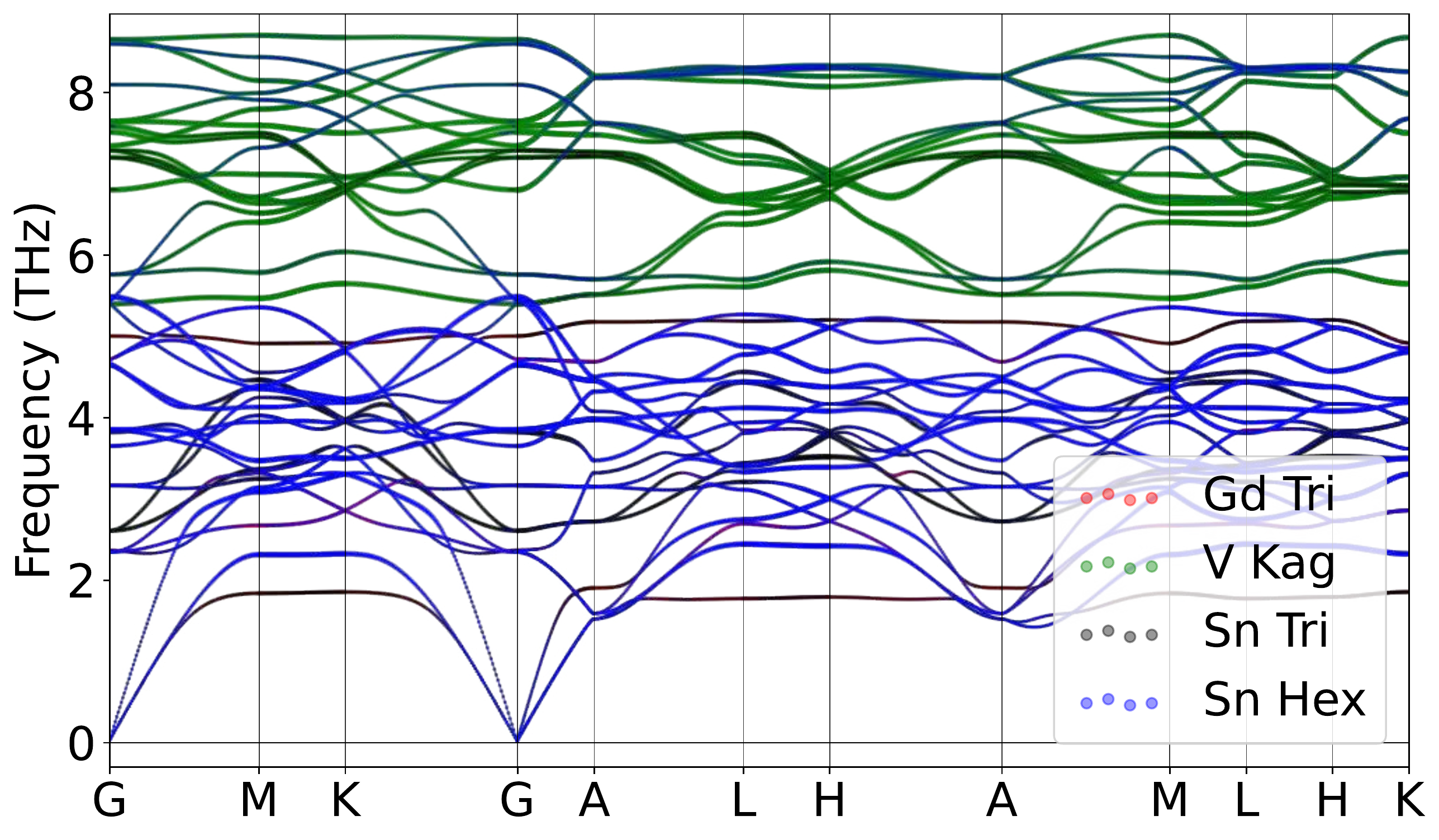}
}
\caption{Atomically projected phonon spectrum for GdV$_6$Sn$_6$.}
\label{fig:GdV6Sn6pho}
\end{figure}

\begin{figure}[H]
\centering
\label{fig:GdV6Sn6_relaxed_Low_xz}
\includegraphics[width=0.85\textwidth]{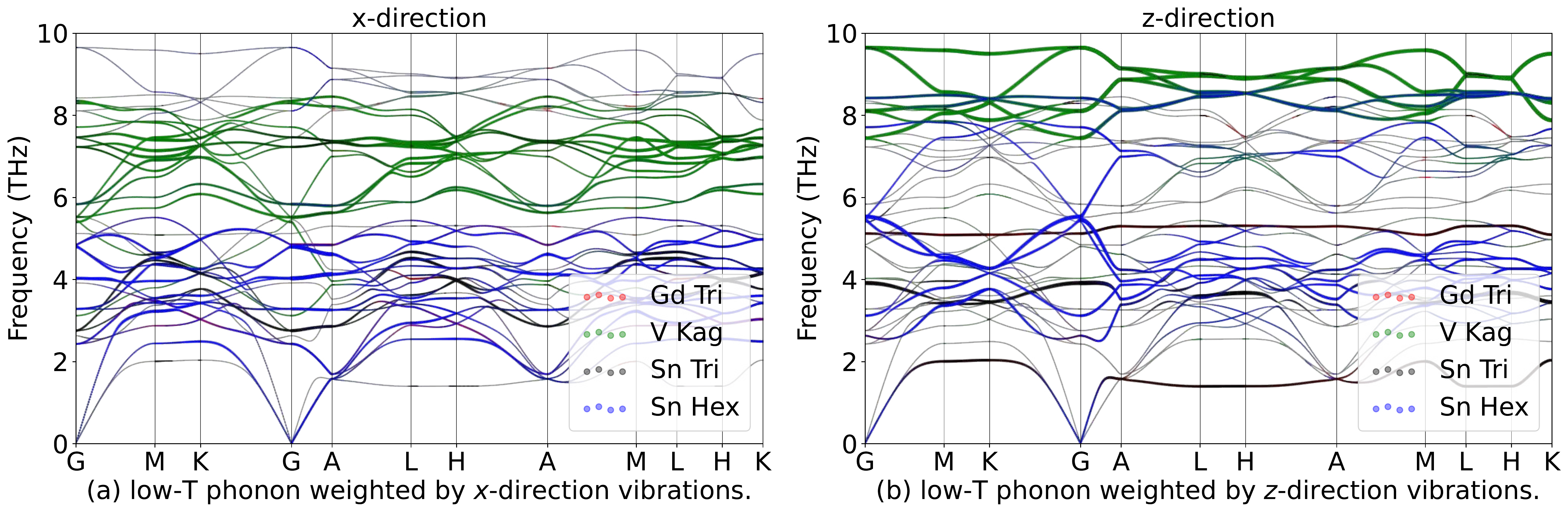}
\hspace{5pt}\label{fig:GdV6Sn6_relaxed_High_xz}
\includegraphics[width=0.85\textwidth]{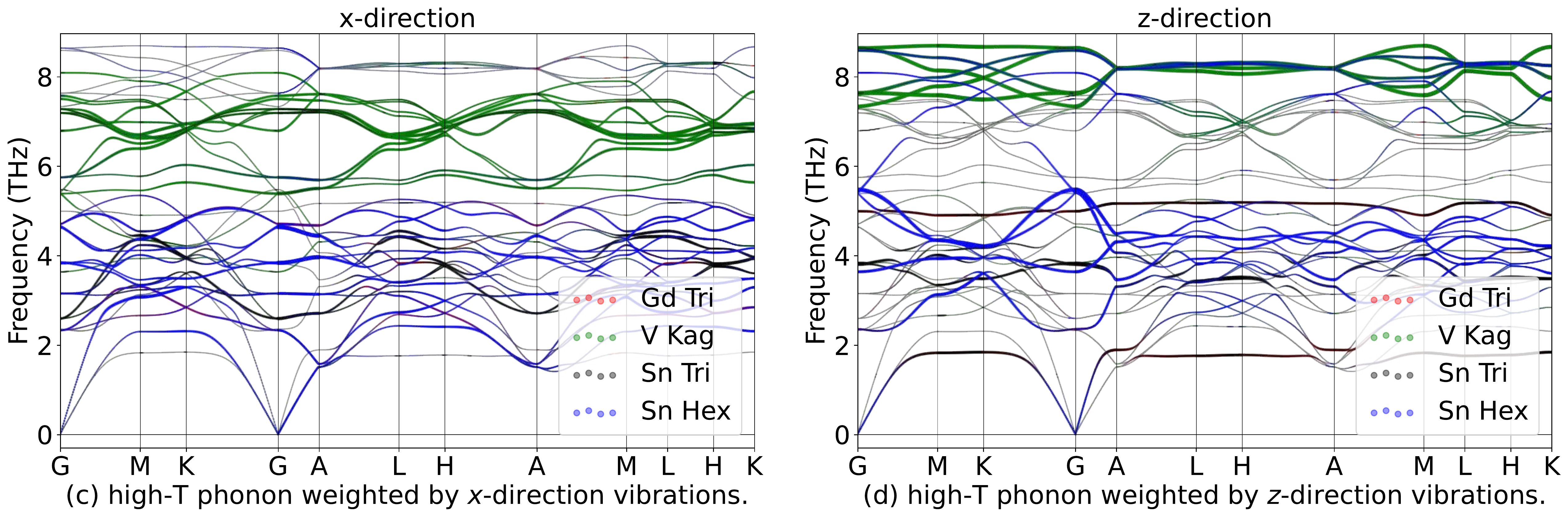}
\caption{Phonon spectrum for GdV$_6$Sn$_6$in in-plane ($x$) and out-of-plane ($z$) directions.}
\label{fig:GdV6Sn6phoxyz}
\end{figure}

\begin{table}[htbp]
\global\long\def\arraystretch{1.12}
\captionsetup{width=.95\textwidth}
\caption{IRREPs at high-symmetry points for GdV$_6$Sn$_6$ phonon spectrum at Low-T.}
\label{tab:191_GdV6Sn6_Low}
\setlength{\tabcolsep}{1.mm}{
}
\end{table}

\begin{table}[htbp]
\global\long\def\arraystretch{1.12}
\captionsetup{width=.95\textwidth}
\caption{IRREPs at high-symmetry points for GdV$_6$Sn$_6$ phonon spectrum at High-T.}
\label{tab:191_GdV6Sn6_High}
\setlength{\tabcolsep}{1.mm}{
%
}
\end{table}
\subsubsection{\label{sms2sec:TbV6Sn6}\hyperref[smsec:col]{\texorpdfstring{TbV$_6$Sn$_6$}{}}~{\color{green}  (High-T stable, Low-T stable) }}

\begin{table}[htbp]
\global\long\def\arraystretch{1.12}
\captionsetup{width=.85\textwidth}
\caption{Structure parameters of TbV$_6$Sn$_6$ after relaxation. It contains 503 electrons. }
\label{tab:TbV6Sn6}
\setlength{\tabcolsep}{4mm}{
%
}
\end{table}

\begin{figure}[htbp]
\centering
\includegraphics[width=0.85\textwidth]{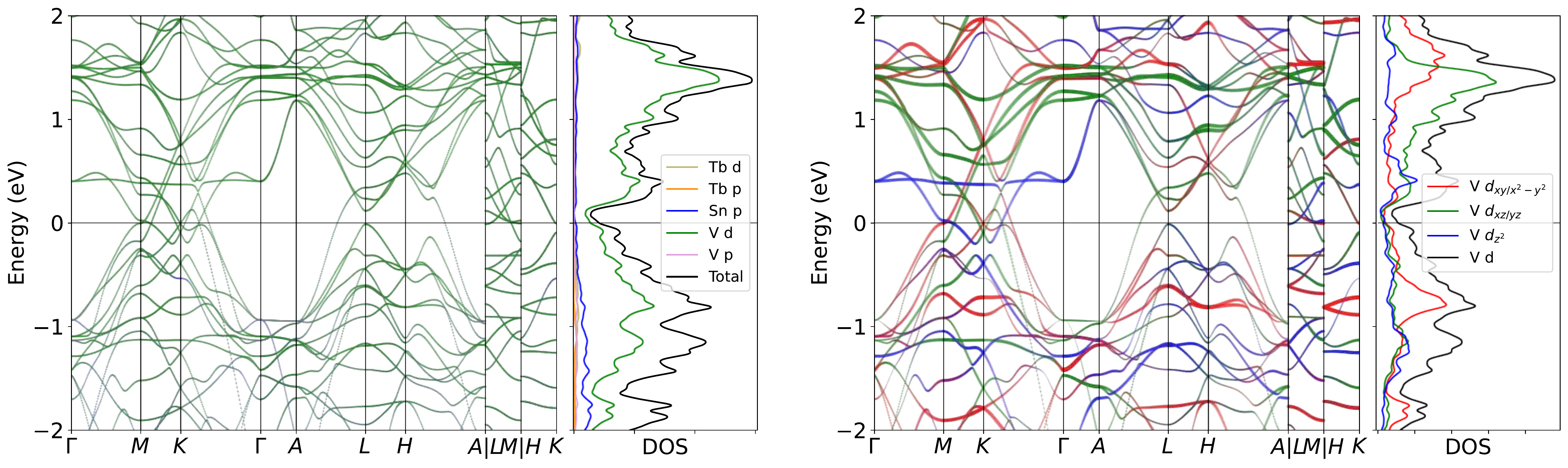}
\caption{Band structure for TbV$_6$Sn$_6$ without SOC. The projection of V $3d$ orbitals is also presented. The band structures clearly reveal the dominating of V $3d$ orbitals  near the Fermi level, as well as the pronounced peak.}
\label{fig:TbV6Sn6band}
\end{figure}

\begin{figure}[H]
\centering
\subfloat[Relaxed phonon at low-T.]{
\label{fig:TbV6Sn6_relaxed_Low}
\includegraphics[width=0.45\textwidth]{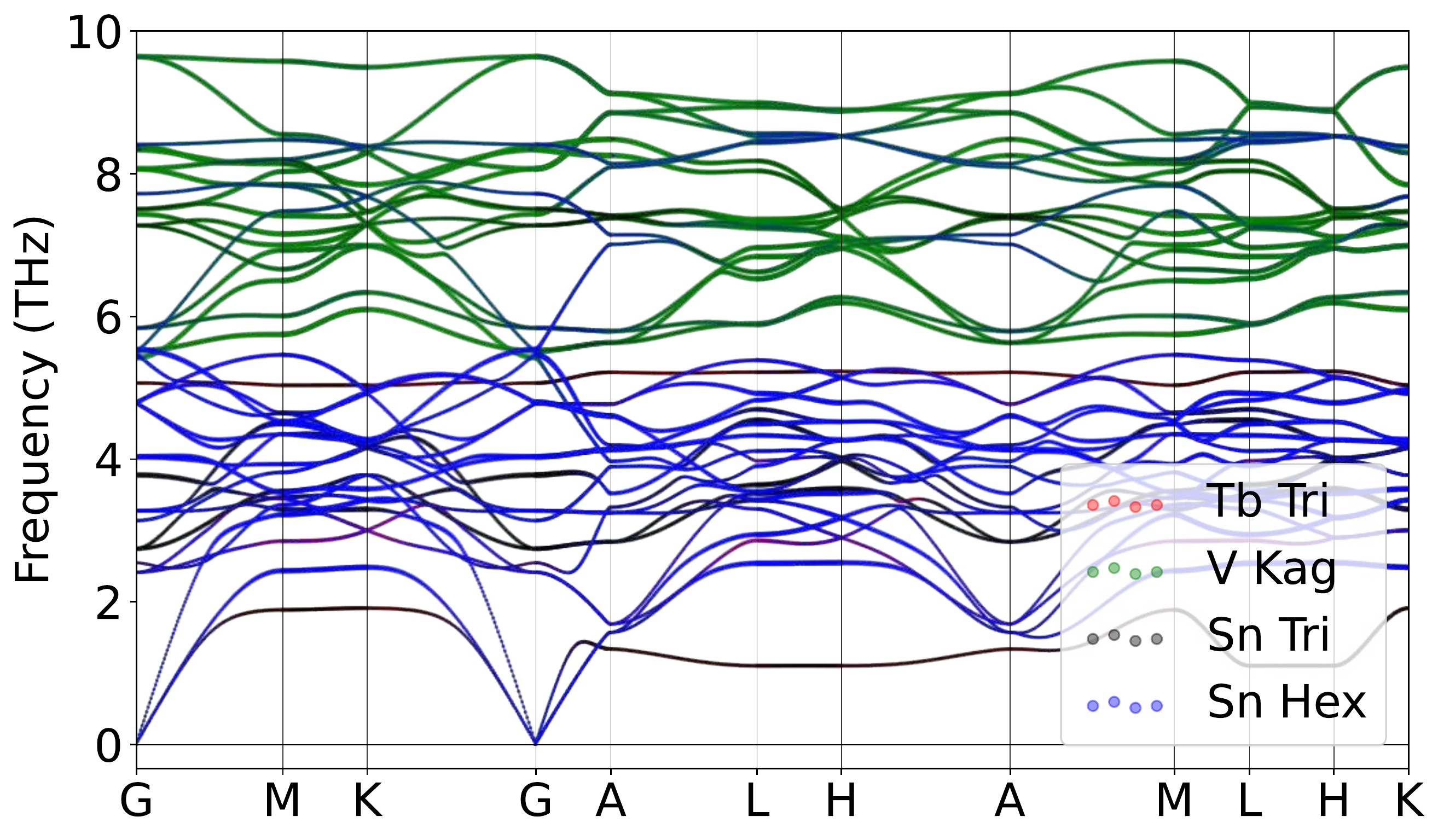}
}
\hspace{5pt}\subfloat[Relaxed phonon at high-T.]{
\label{fig:TbV6Sn6_relaxed_High}
\includegraphics[width=0.45\textwidth]{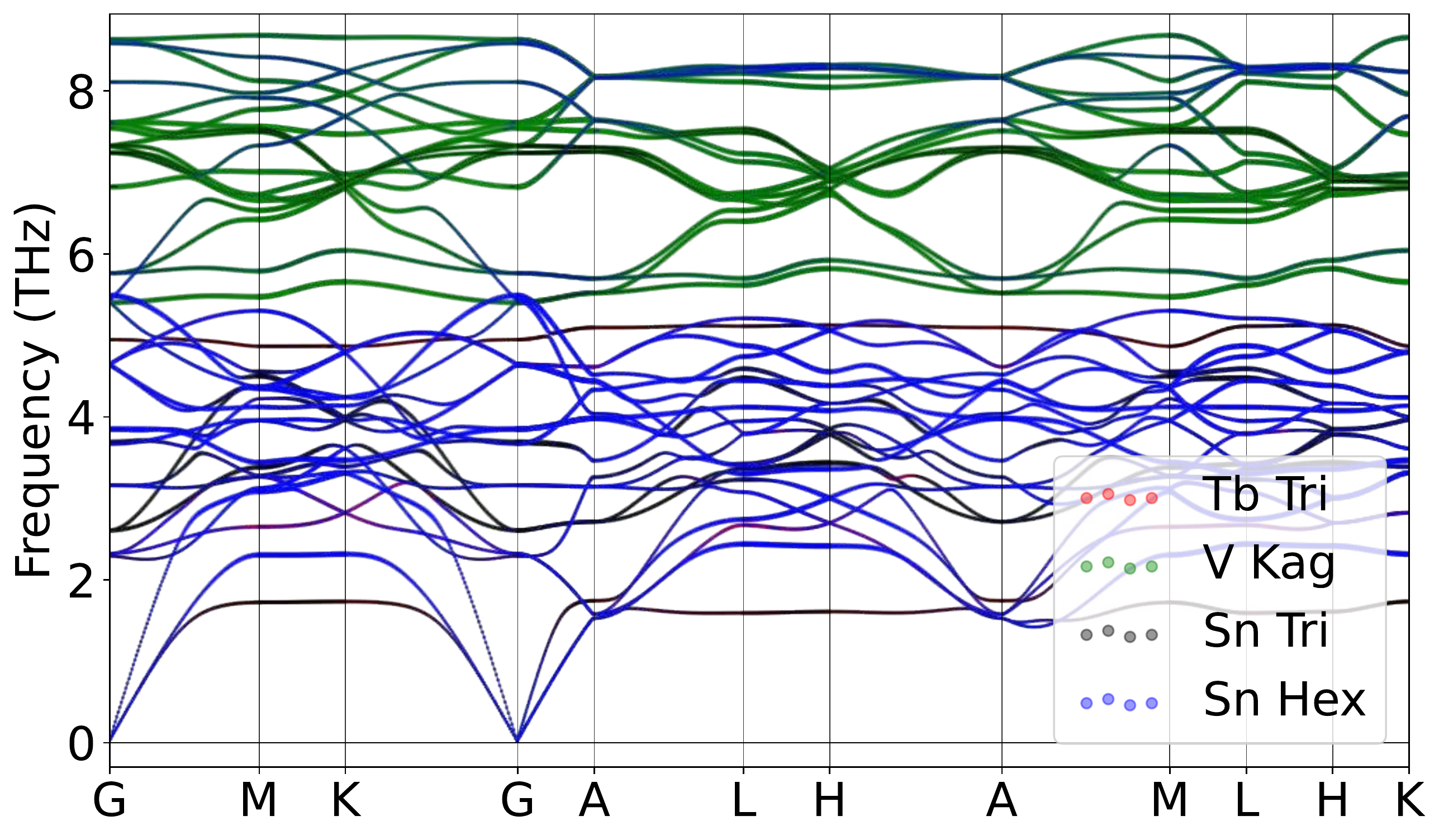}
}
\caption{Atomically projected phonon spectrum for TbV$_6$Sn$_6$.}
\label{fig:TbV6Sn6pho}
\end{figure}

\begin{figure}[H]
\centering
\label{fig:TbV6Sn6_relaxed_Low_xz}
\includegraphics[width=0.85\textwidth]{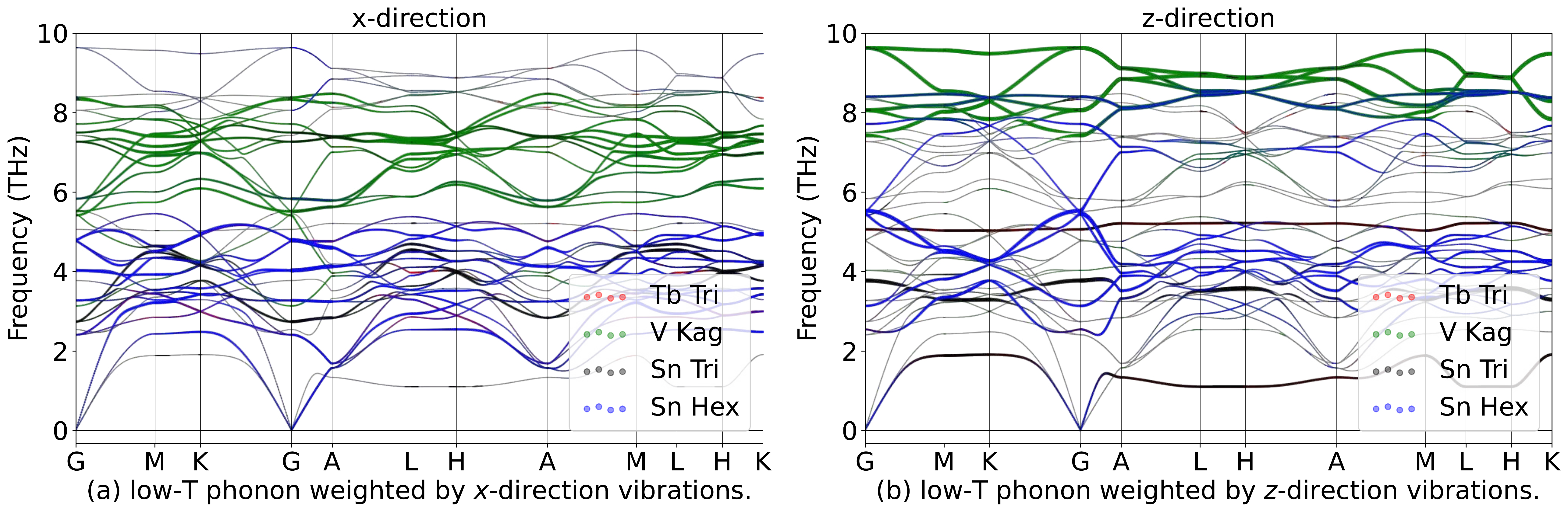}
\hspace{5pt}\label{fig:TbV6Sn6_relaxed_High_xz}
\includegraphics[width=0.85\textwidth]{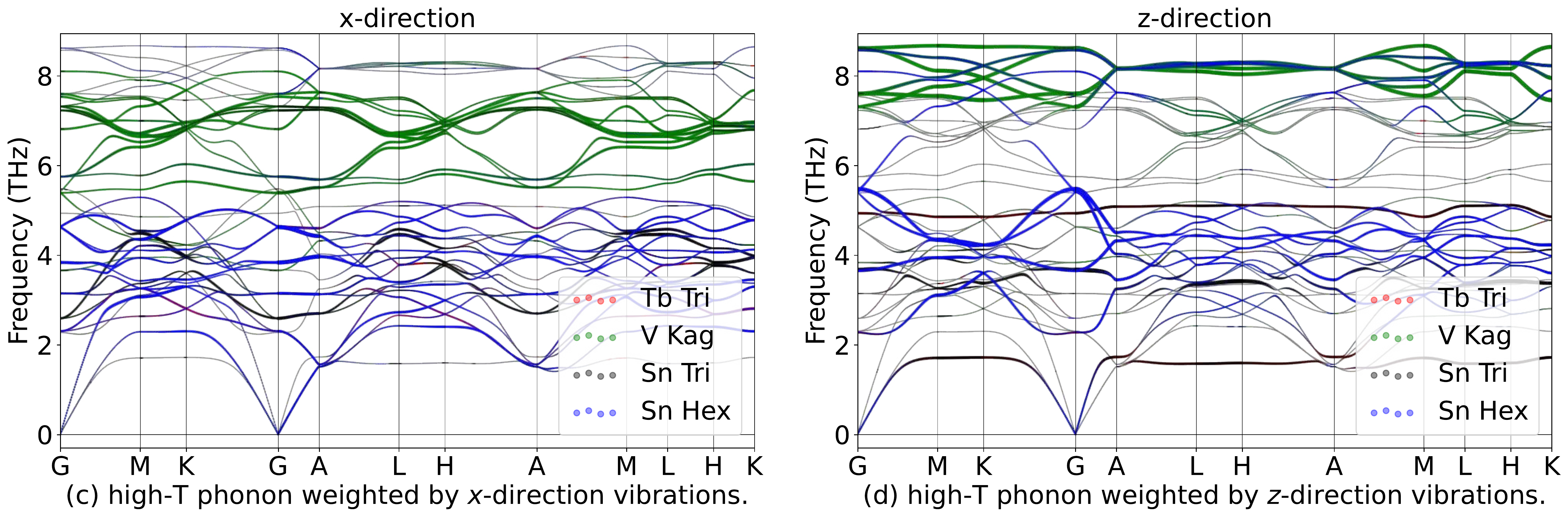}
\caption{Phonon spectrum for TbV$_6$Sn$_6$in in-plane ($x$) and out-of-plane ($z$) directions.}
\label{fig:TbV6Sn6phoxyz}
\end{figure}

\begin{table}[htbp]
\global\long\def\arraystretch{1.12}
\captionsetup{width=.95\textwidth}
\caption{IRREPs at high-symmetry points for TbV$_6$Sn$_6$ phonon spectrum at Low-T.}
\label{tab:191_TbV6Sn6_Low}
\setlength{\tabcolsep}{1.mm}{
}
\end{table}

\begin{table}[htbp]
\global\long\def\arraystretch{1.12}
\captionsetup{width=.95\textwidth}
\caption{IRREPs at high-symmetry points for TbV$_6$Sn$_6$ phonon spectrum at High-T.}
\label{tab:191_TbV6Sn6_High}
\setlength{\tabcolsep}{1.mm}{
%
}
\end{table}
\subsubsection{\label{sms2sec:DyV6Sn6}\hyperref[smsec:col]{\texorpdfstring{DyV$_6$Sn$_6$}{}}~{\color{green}  (High-T stable, Low-T stable) }}

\begin{table}[htbp]
\global\long\def\arraystretch{1.12}
\captionsetup{width=.85\textwidth}
\caption{Structure parameters of DyV$_6$Sn$_6$ after relaxation. It contains 504 electrons. }
\label{tab:DyV6Sn6}
\setlength{\tabcolsep}{4mm}{
%
}
\end{table}

\begin{figure}[htbp]
\centering
\includegraphics[width=0.85\textwidth]{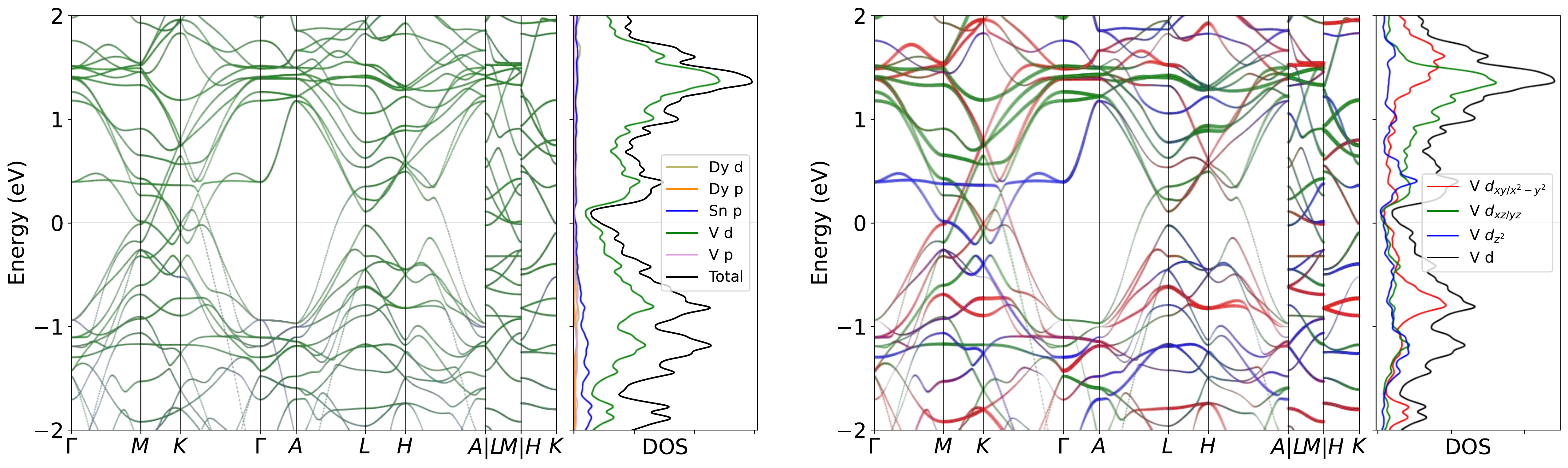}
\caption{Band structure for DyV$_6$Sn$_6$ without SOC. The projection of V $3d$ orbitals is also presented. The band structures clearly reveal the dominating of V $3d$ orbitals  near the Fermi level, as well as the pronounced peak.}
\label{fig:DyV6Sn6band}
\end{figure}

\begin{figure}[H]
\centering
\subfloat[Relaxed phonon at low-T.]{
\label{fig:DyV6Sn6_relaxed_Low}
\includegraphics[width=0.45\textwidth]{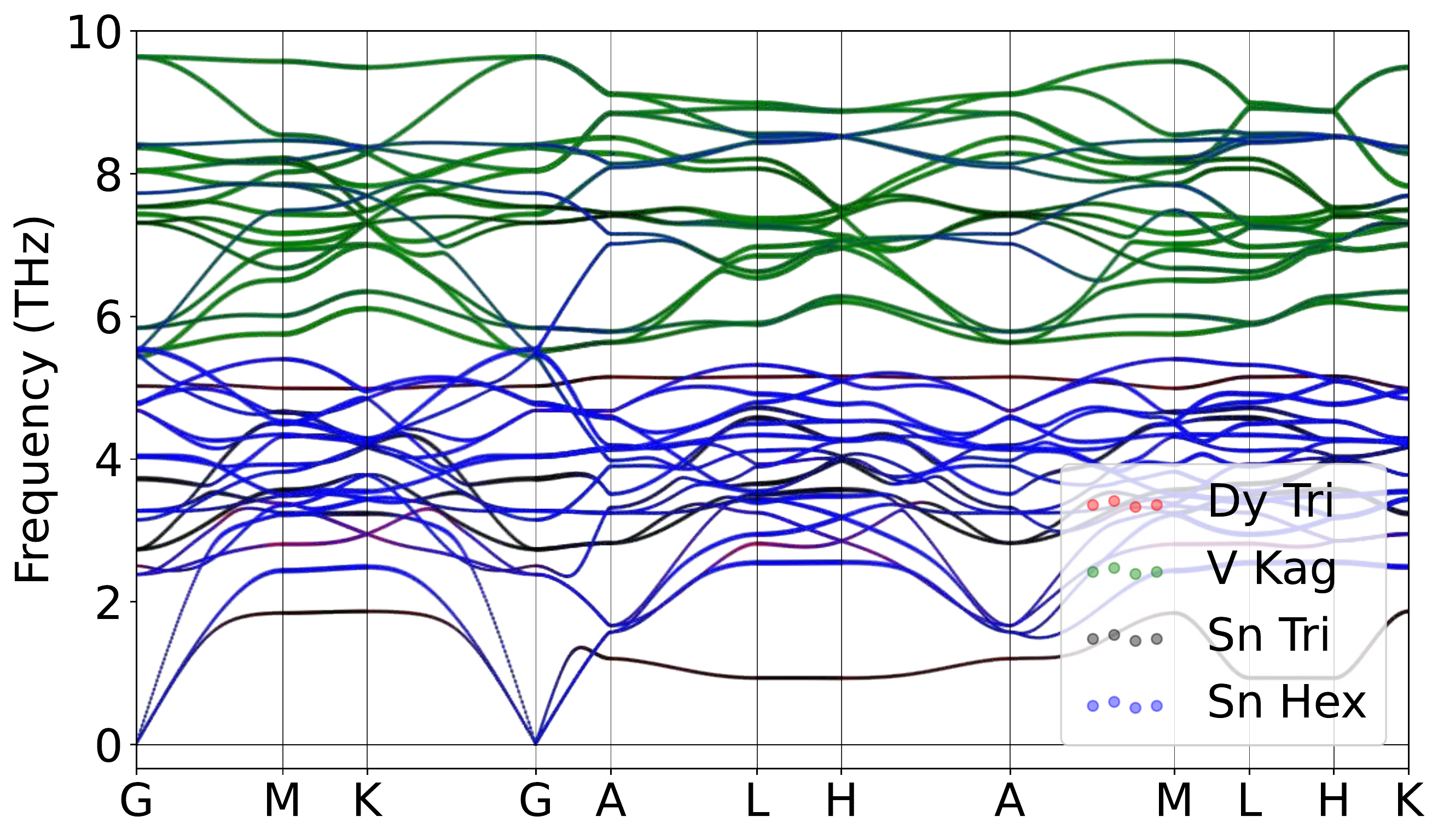}
}
\hspace{5pt}\subfloat[Relaxed phonon at high-T.]{
\label{fig:DyV6Sn6_relaxed_High}
\includegraphics[width=0.45\textwidth]{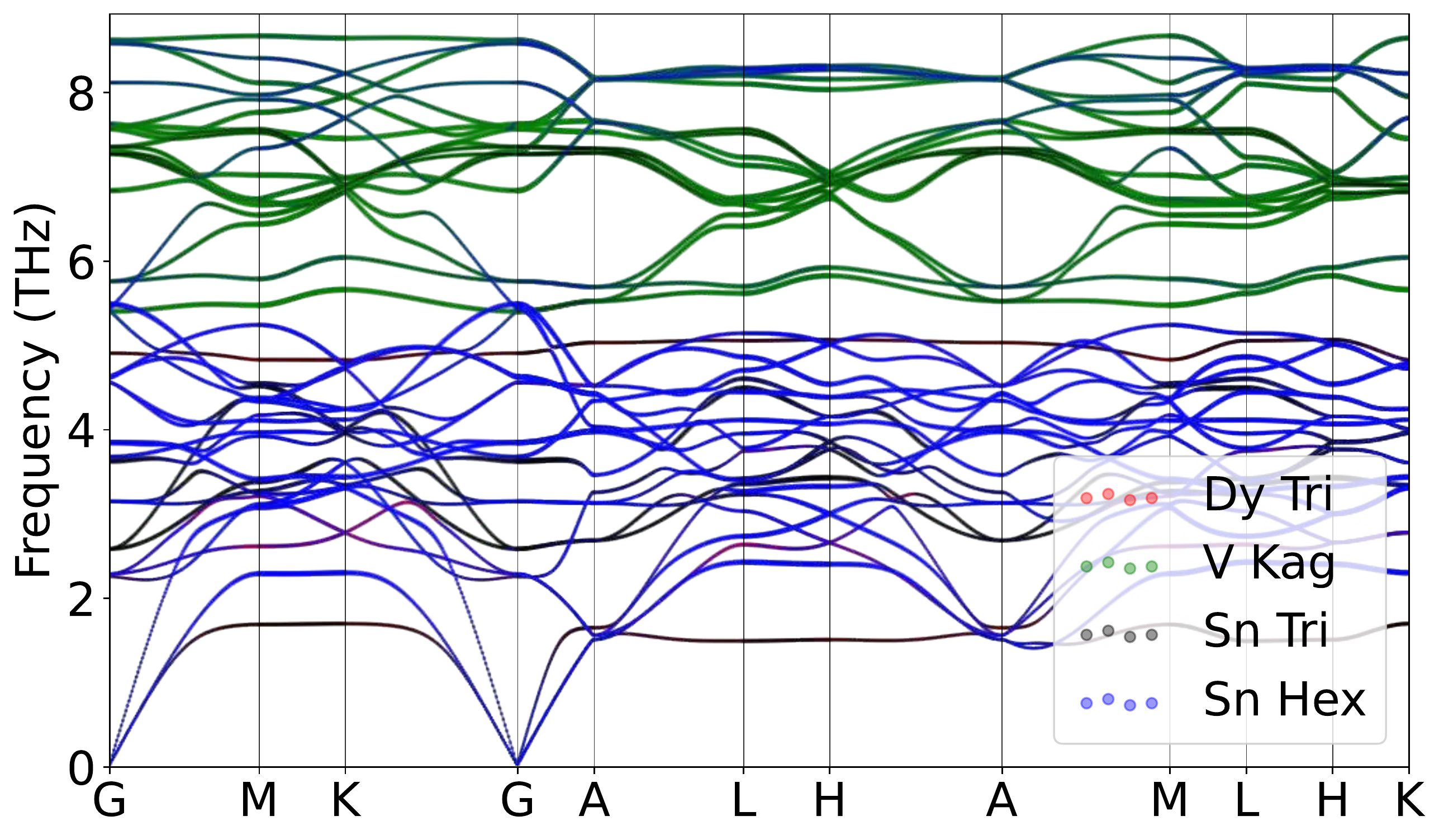}
}
\caption{Atomically projected phonon spectrum for DyV$_6$Sn$_6$.}
\label{fig:DyV6Sn6pho}
\end{figure}

\begin{figure}[H]
\centering
\label{fig:DyV6Sn6_relaxed_Low_xz}
\includegraphics[width=0.85\textwidth]{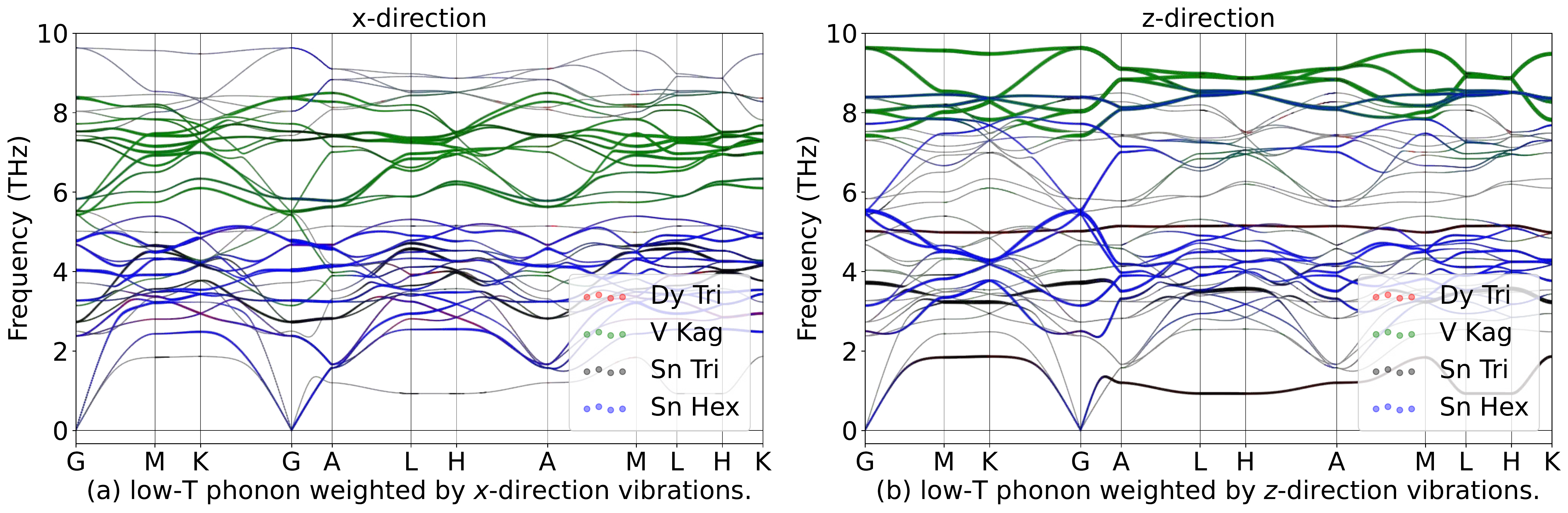}
\hspace{5pt}\label{fig:DyV6Sn6_relaxed_High_xz}
\includegraphics[width=0.85\textwidth]{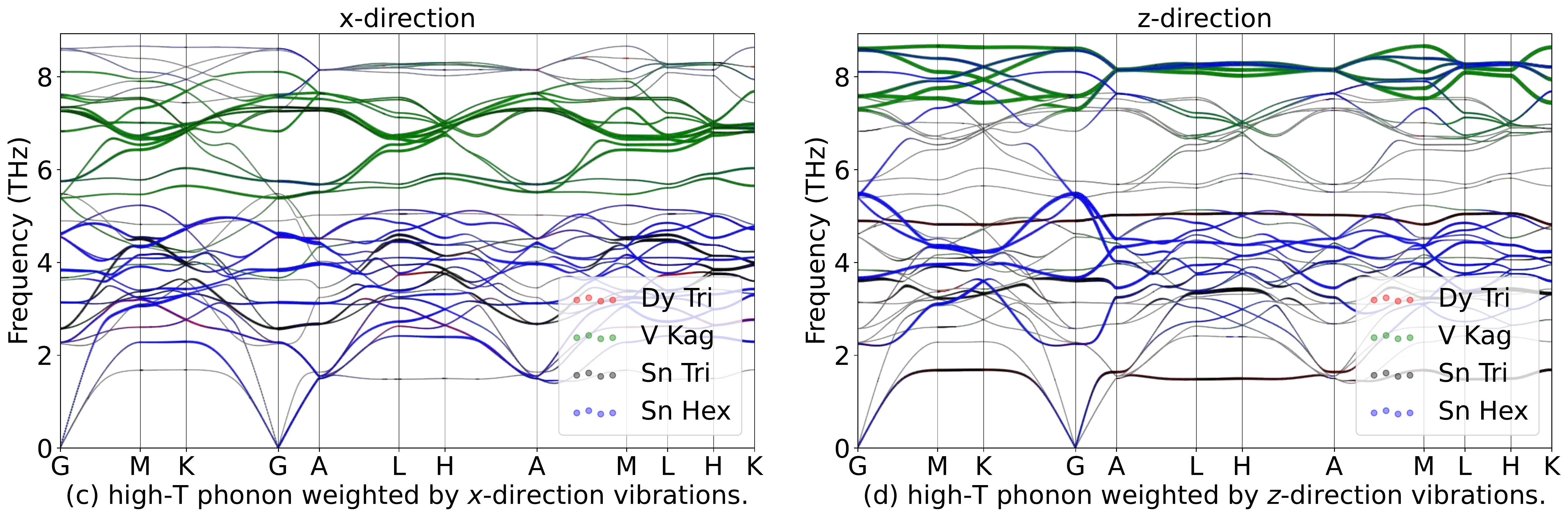}
\caption{Phonon spectrum for DyV$_6$Sn$_6$in in-plane ($x$) and out-of-plane ($z$) directions.}
\label{fig:DyV6Sn6phoxyz}
\end{figure}

\begin{table}[htbp]
\global\long\def\arraystretch{1.12}
\captionsetup{width=.95\textwidth}
\caption{IRREPs at high-symmetry points for DyV$_6$Sn$_6$ phonon spectrum at Low-T.}
\label{tab:191_DyV6Sn6_Low}
\setlength{\tabcolsep}{1.mm}{
}
\end{table}

\begin{table}[htbp]
\global\long\def\arraystretch{1.12}
\captionsetup{width=.95\textwidth}
\caption{IRREPs at high-symmetry points for DyV$_6$Sn$_6$ phonon spectrum at High-T.}
\label{tab:191_DyV6Sn6_High}
\setlength{\tabcolsep}{1.mm}{
%
}
\end{table}
\subsubsection{\label{sms2sec:HoV6Sn6}\hyperref[smsec:col]{\texorpdfstring{HoV$_6$Sn$_6$}{}}~{\color{green}  (High-T stable, Low-T stable) }}

\begin{table}[htbp]
\global\long\def\arraystretch{1.12}
\captionsetup{width=.85\textwidth}
\caption{Structure parameters of HoV$_6$Sn$_6$ after relaxation. It contains 505 electrons. }
\label{tab:HoV6Sn6}
\setlength{\tabcolsep}{4mm}{
%
}
\end{table}

\begin{figure}[htbp]
\centering
\includegraphics[width=0.85\textwidth]{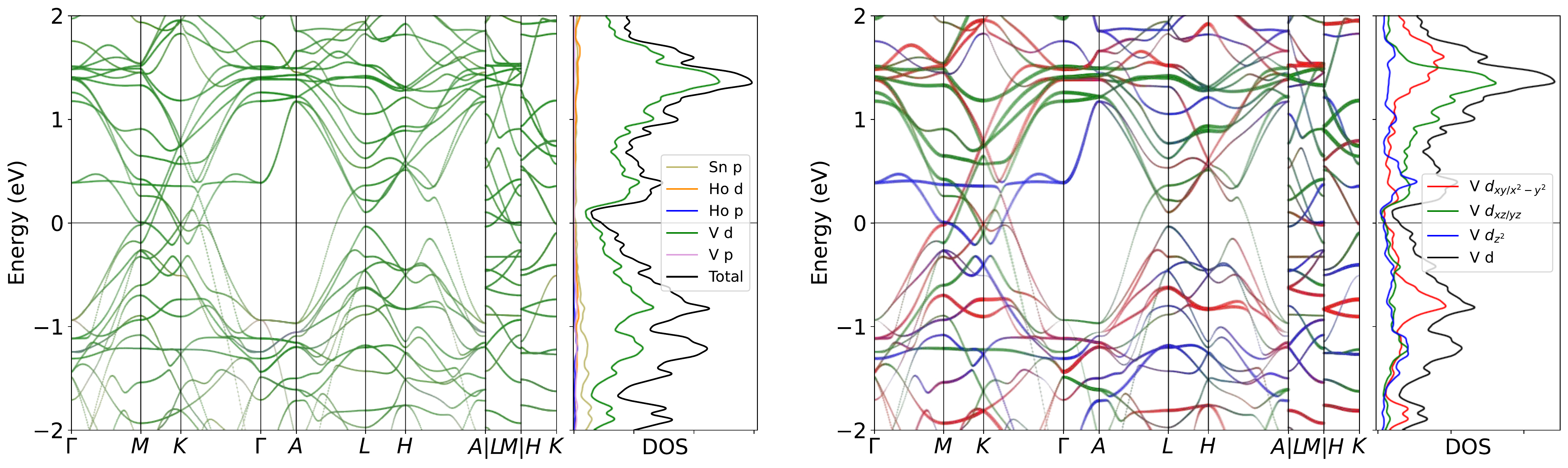}
\caption{Band structure for HoV$_6$Sn$_6$ without SOC. The projection of V $3d$ orbitals is also presented. The band structures clearly reveal the dominating of V $3d$ orbitals  near the Fermi level, as well as the pronounced peak.}
\label{fig:HoV6Sn6band}
\end{figure}

\begin{figure}[H]
\centering
\subfloat[Relaxed phonon at low-T.]{
\label{fig:HoV6Sn6_relaxed_Low}
\includegraphics[width=0.45\textwidth]{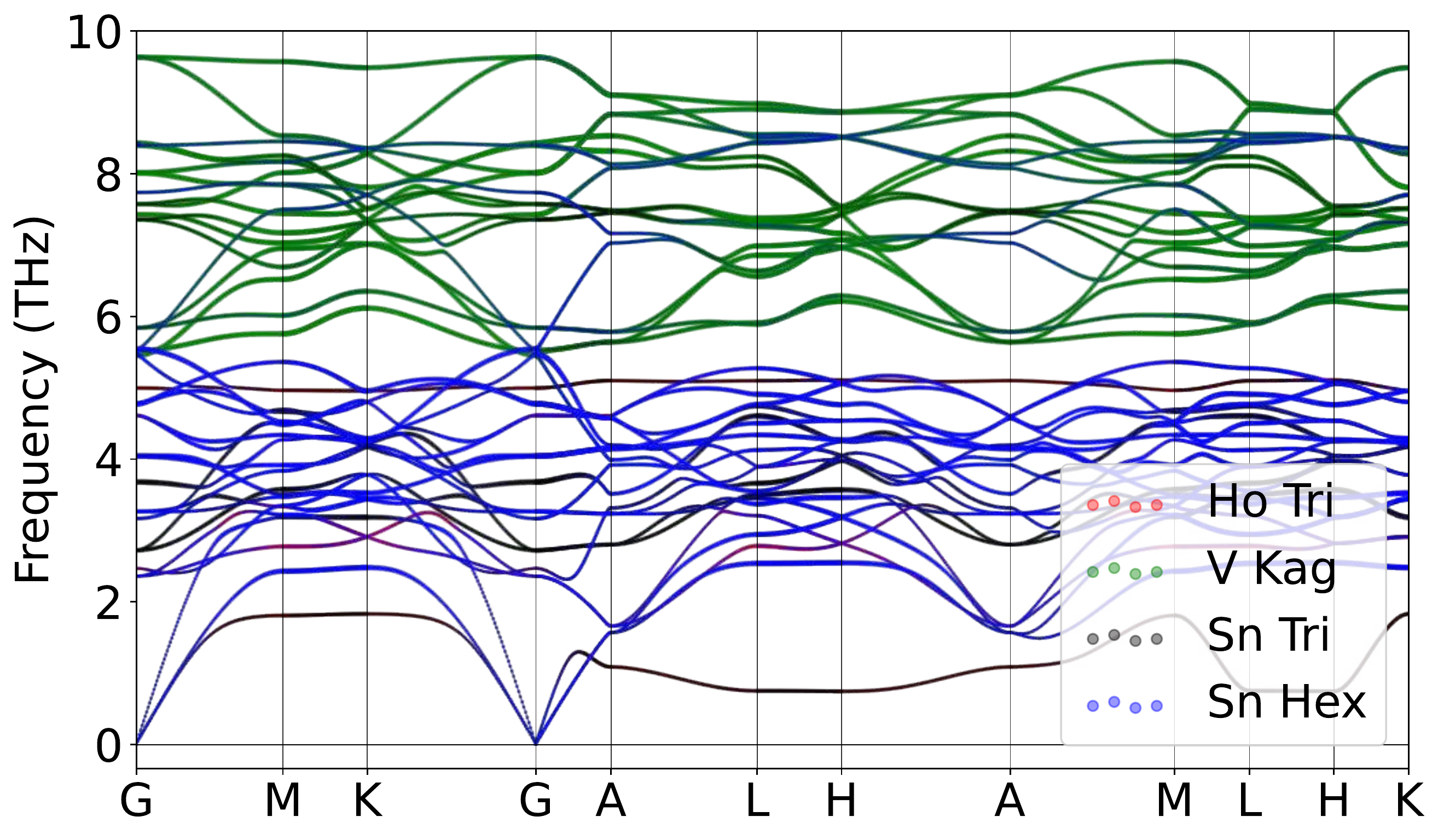}
}
\hspace{5pt}\subfloat[Relaxed phonon at high-T.]{
\label{fig:HoV6Sn6_relaxed_High}
\includegraphics[width=0.45\textwidth]{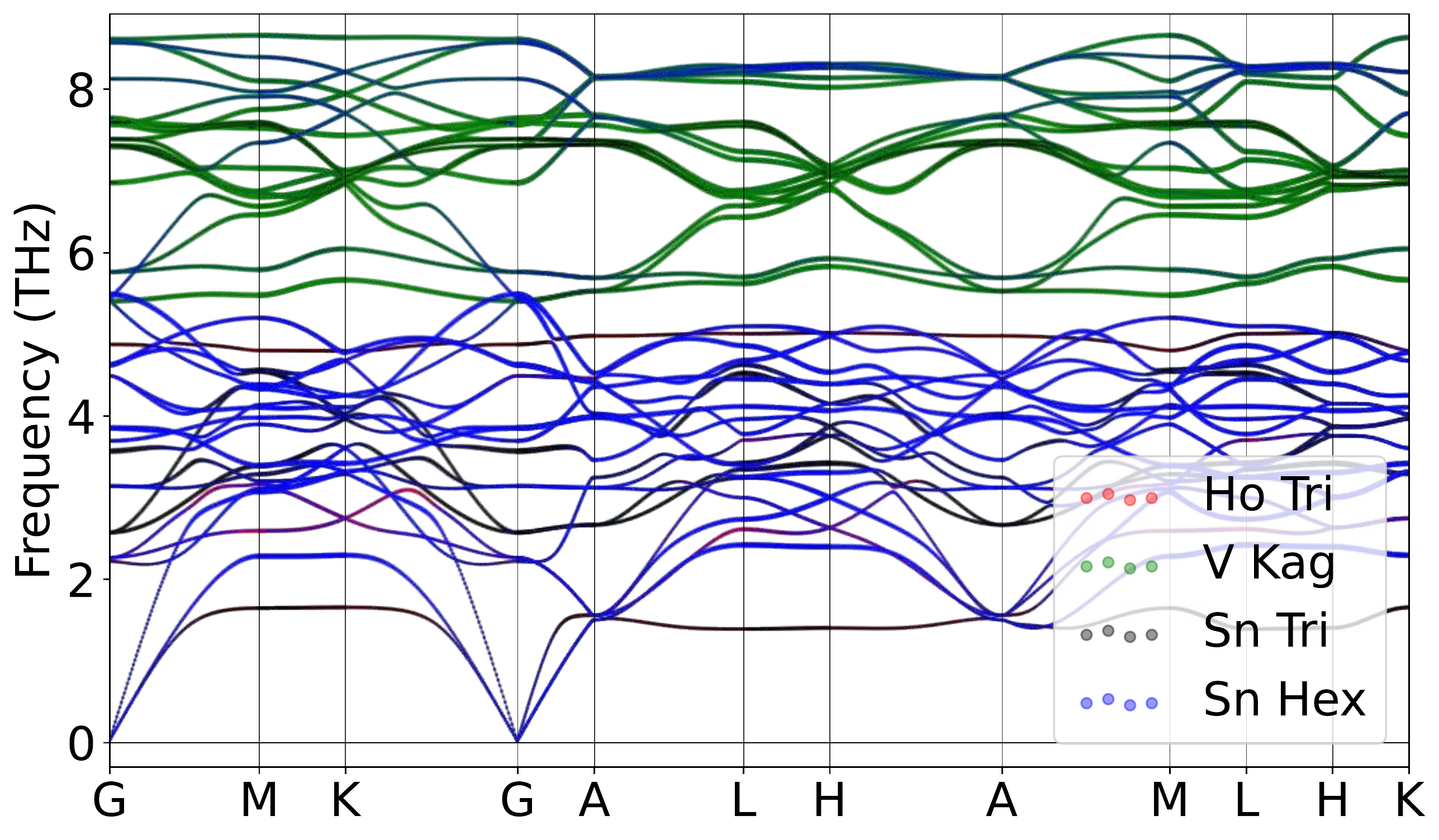}
}
\caption{Atomically projected phonon spectrum for HoV$_6$Sn$_6$.}
\label{fig:HoV6Sn6pho}
\end{figure}

\begin{figure}[H]
\centering
\label{fig:HoV6Sn6_relaxed_Low_xz}
\includegraphics[width=0.85\textwidth]{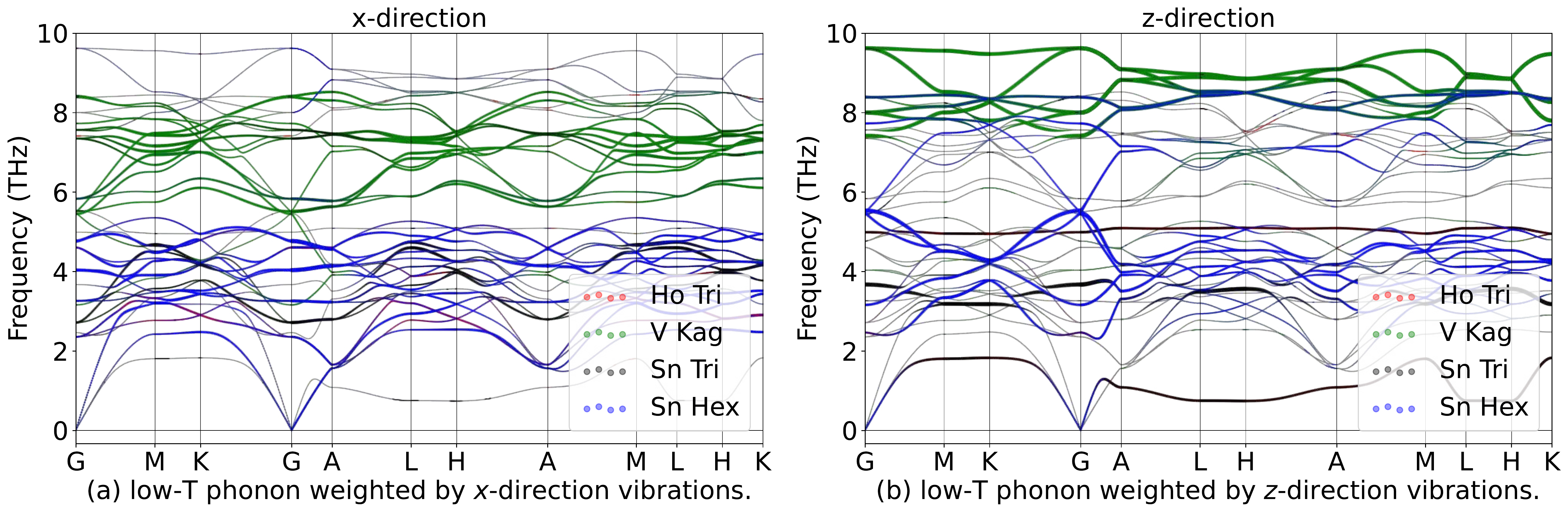}
\hspace{5pt}\label{fig:HoV6Sn6_relaxed_High_xz}
\includegraphics[width=0.85\textwidth]{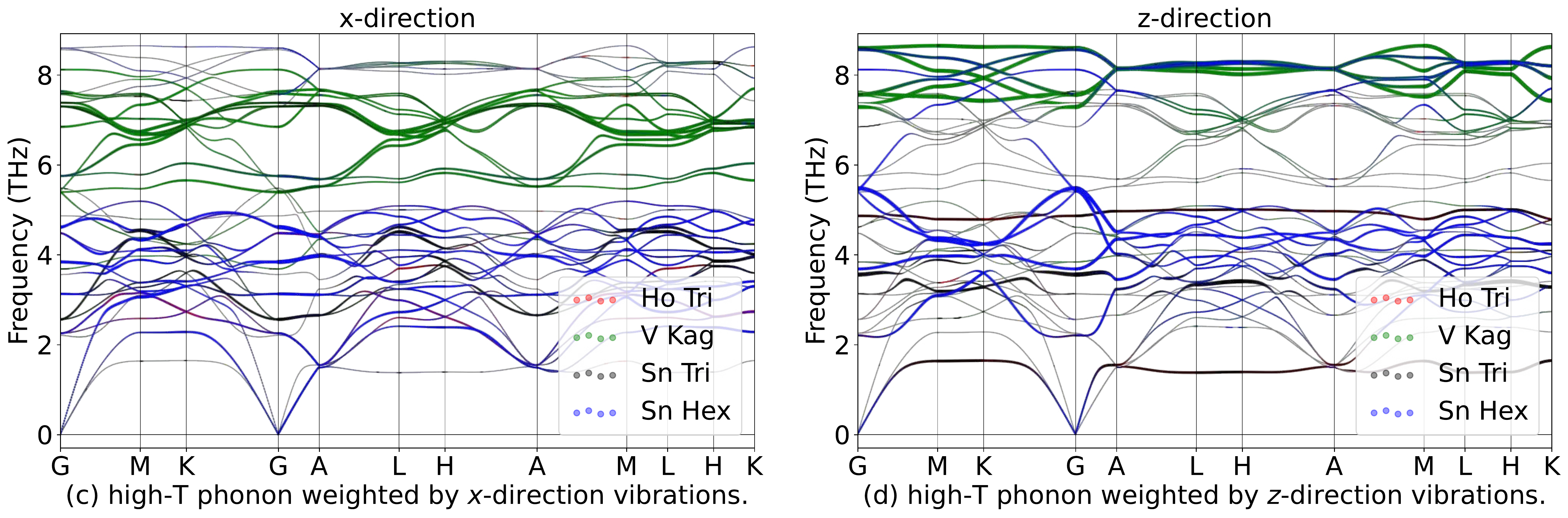}
\caption{Phonon spectrum for HoV$_6$Sn$_6$in in-plane ($x$) and out-of-plane ($z$) directions.}
\label{fig:HoV6Sn6phoxyz}
\end{figure}

\begin{table}[htbp]
\global\long\def\arraystretch{1.12}
\captionsetup{width=.95\textwidth}
\caption{IRREPs at high-symmetry points for HoV$_6$Sn$_6$ phonon spectrum at Low-T.}
\label{tab:191_HoV6Sn6_Low}
\setlength{\tabcolsep}{1.mm}{
}
\end{table}

\begin{table}[htbp]
\global\long\def\arraystretch{1.12}
\captionsetup{width=.95\textwidth}
\caption{IRREPs at high-symmetry points for HoV$_6$Sn$_6$ phonon spectrum at High-T.}
\label{tab:191_HoV6Sn6_High}
\setlength{\tabcolsep}{1.mm}{
%
}
\end{table}
\subsubsection{\label{sms2sec:ErV6Sn6}\hyperref[smsec:col]{\texorpdfstring{ErV$_6$Sn$_6$}{}}~{\color{green}  (High-T stable, Low-T stable) }}

\begin{table}[htbp]
\global\long\def\arraystretch{1.12}
\captionsetup{width=.85\textwidth}
\caption{Structure parameters of ErV$_6$Sn$_6$ after relaxation. It contains 506 electrons. }
\label{tab:ErV6Sn6}
\setlength{\tabcolsep}{4mm}{
%
}
\end{table}

\begin{figure}[htbp]
\centering
\includegraphics[width=0.85\textwidth]{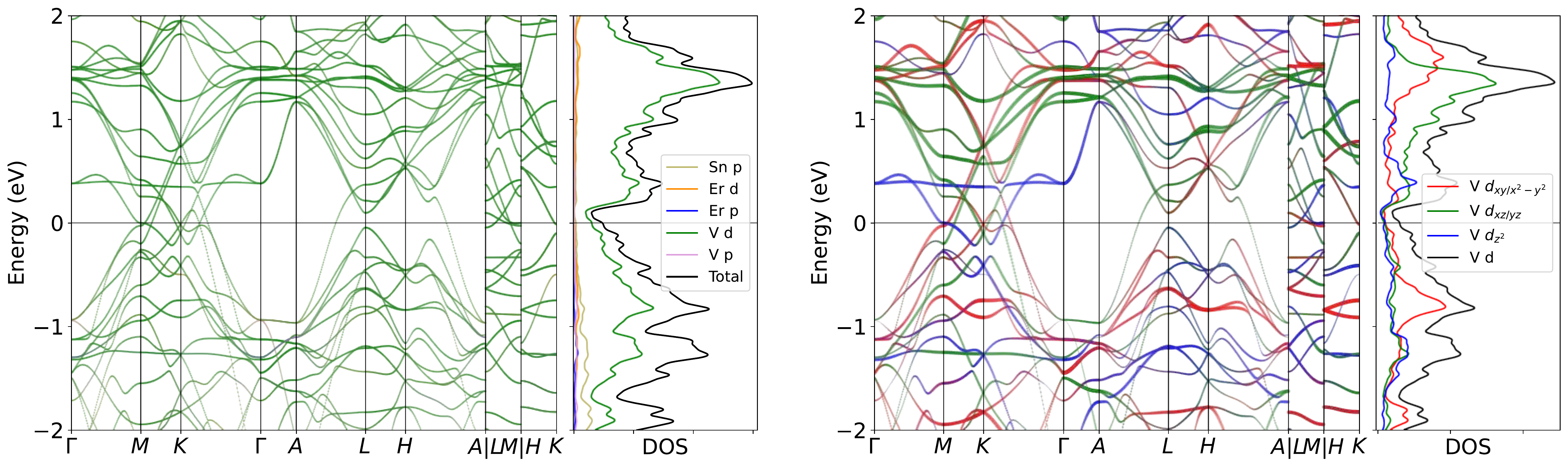}
\caption{Band structure for ErV$_6$Sn$_6$ without SOC. The projection of V $3d$ orbitals is also presented. The band structures clearly reveal the dominating of V $3d$ orbitals  near the Fermi level, as well as the pronounced peak.}
\label{fig:ErV6Sn6band}
\end{figure}

\begin{figure}[H]
\centering
\subfloat[Relaxed phonon at low-T.]{
\label{fig:ErV6Sn6_relaxed_Low}
\includegraphics[width=0.45\textwidth]{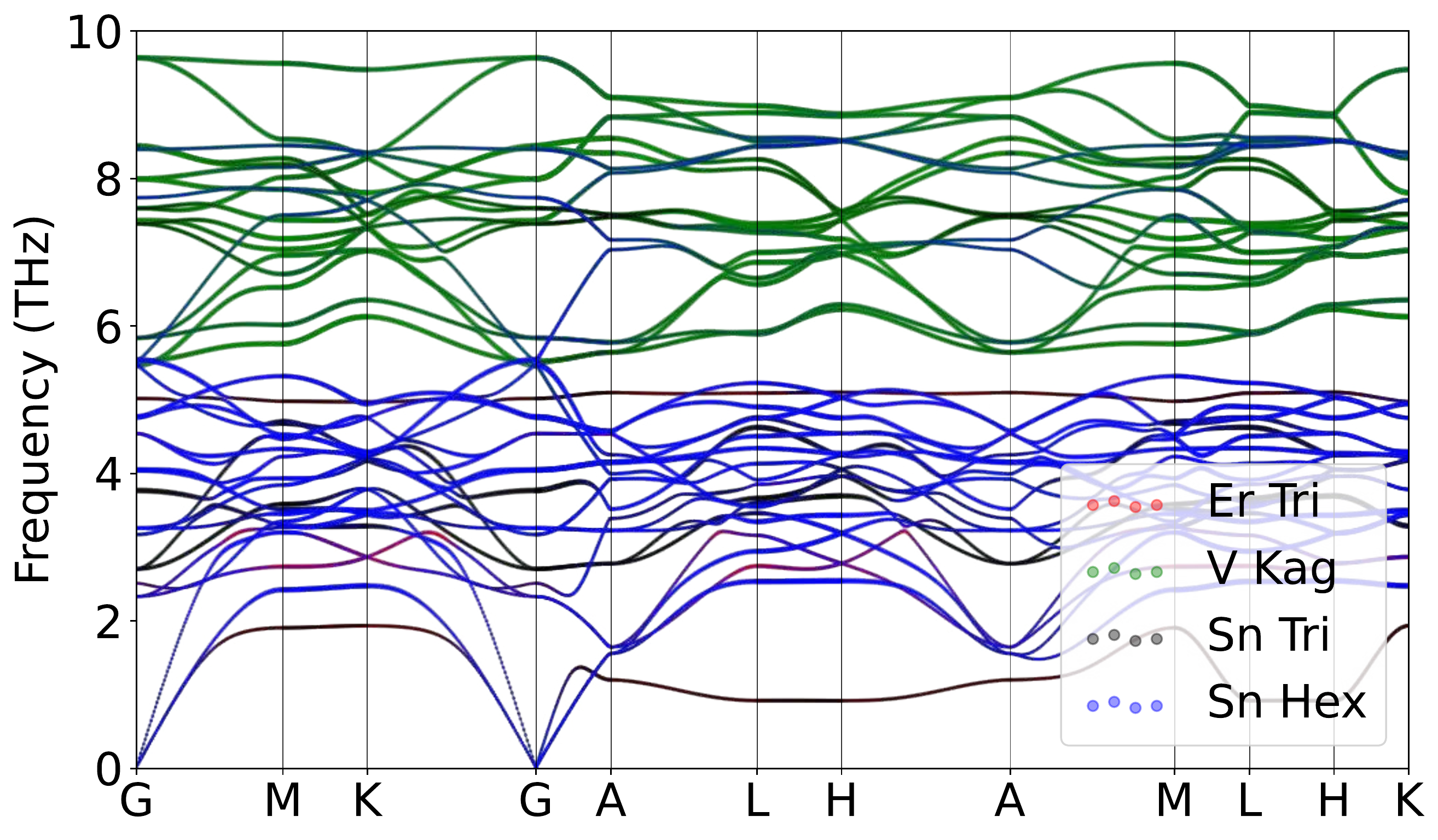}
}
\hspace{5pt}\subfloat[Relaxed phonon at high-T.]{
\label{fig:ErV6Sn6_relaxed_High}
\includegraphics[width=0.45\textwidth]{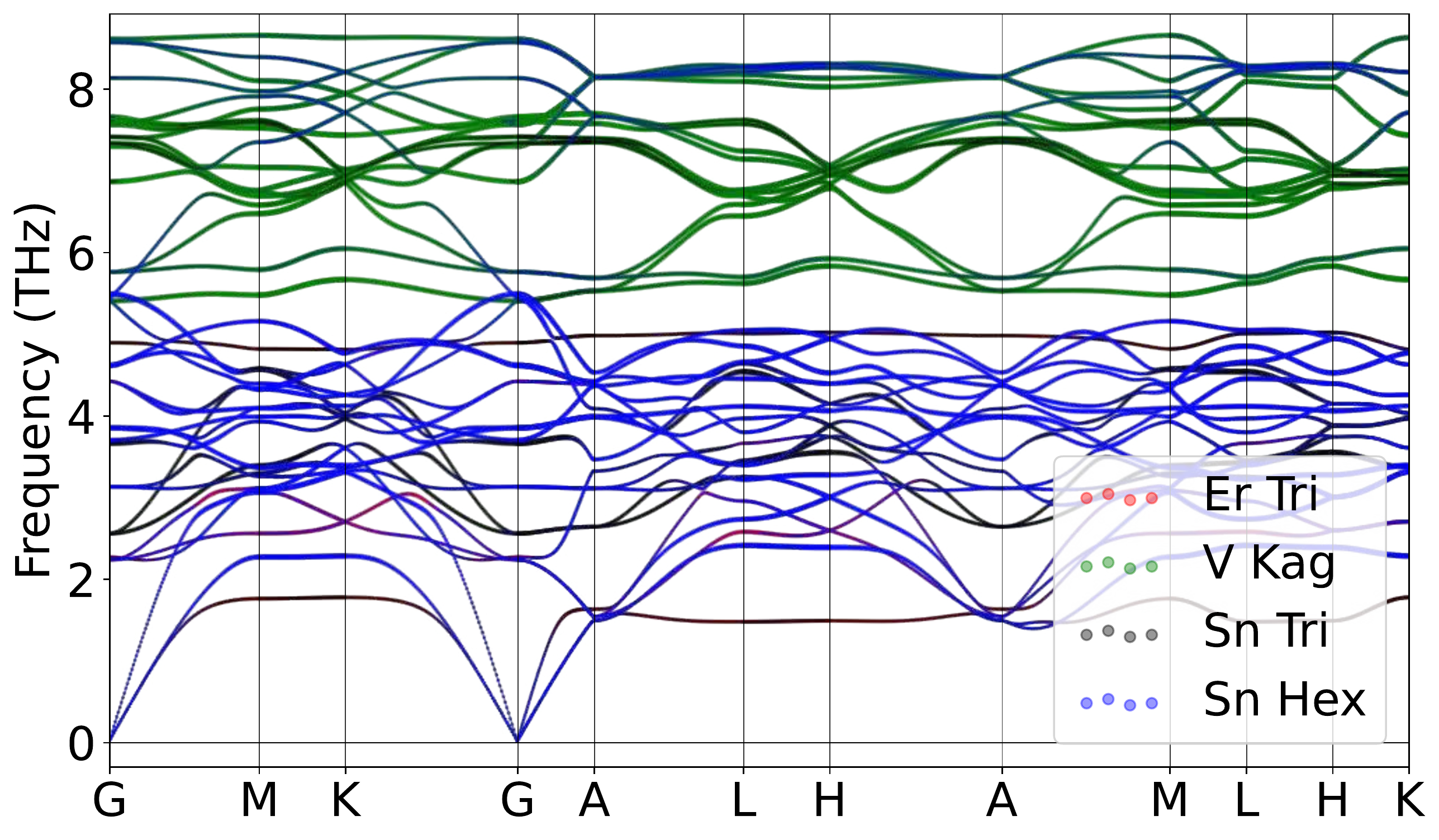}
}
\caption{Atomically projected phonon spectrum for ErV$_6$Sn$_6$.}
\label{fig:ErV6Sn6pho}
\end{figure}

\begin{figure}[H]
\centering
\label{fig:ErV6Sn6_relaxed_Low_xz}
\includegraphics[width=0.85\textwidth]{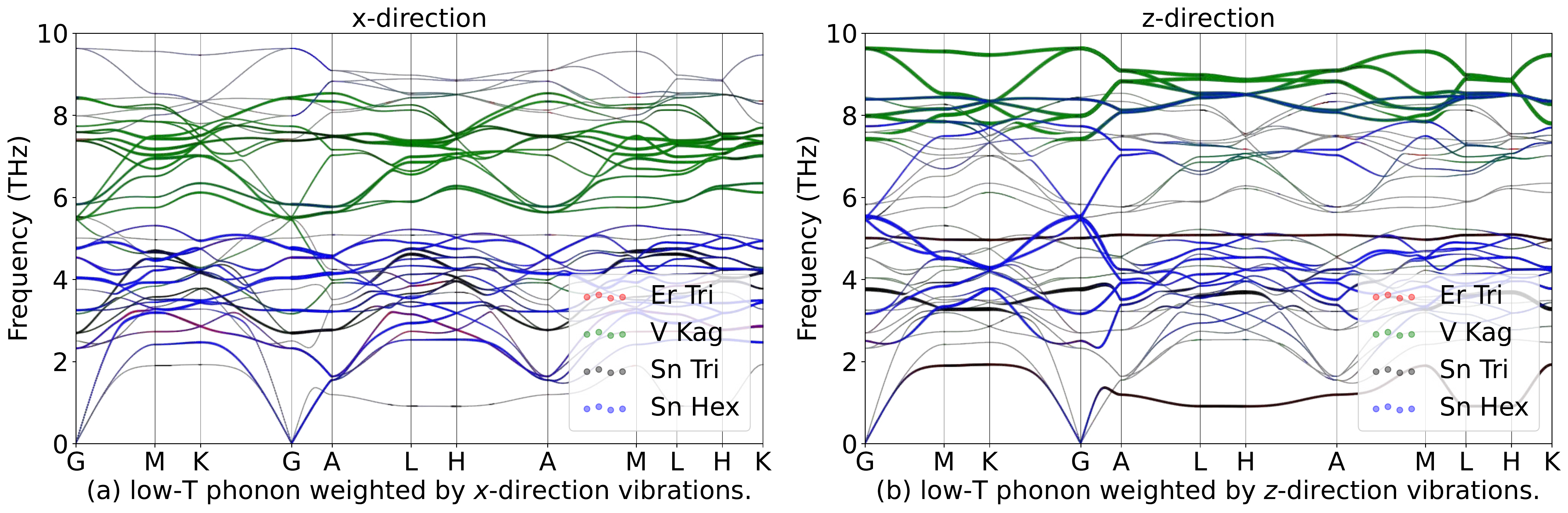}
\hspace{5pt}\label{fig:ErV6Sn6_relaxed_High_xz}
\includegraphics[width=0.85\textwidth]{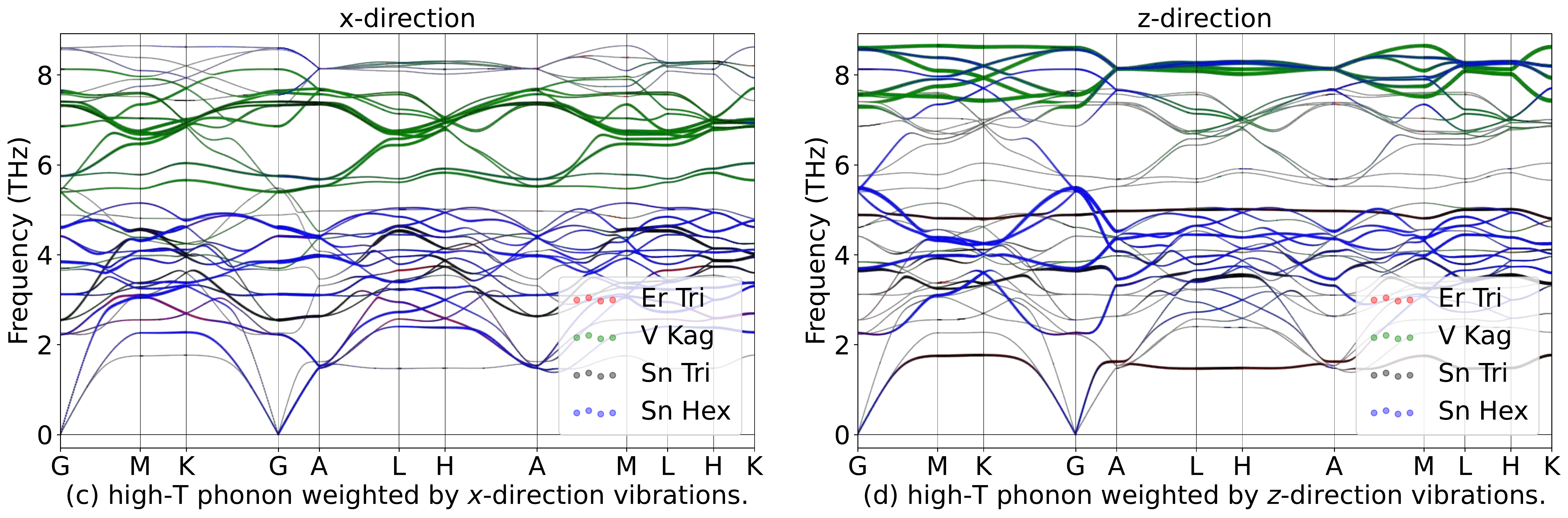}
\caption{Phonon spectrum for ErV$_6$Sn$_6$in in-plane ($x$) and out-of-plane ($z$) directions.}
\label{fig:ErV6Sn6phoxyz}
\end{figure}

\begin{table}[htbp]
\global\long\def\arraystretch{1.12}
\captionsetup{width=.95\textwidth}
\caption{IRREPs at high-symmetry points for ErV$_6$Sn$_6$ phonon spectrum at Low-T.}
\label{tab:191_ErV6Sn6_Low}
\setlength{\tabcolsep}{1.mm}{
}
\end{table}

\begin{table}[htbp]
\global\long\def\arraystretch{1.12}
\captionsetup{width=.95\textwidth}
\caption{IRREPs at high-symmetry points for ErV$_6$Sn$_6$ phonon spectrum at High-T.}
\label{tab:191_ErV6Sn6_High}
\setlength{\tabcolsep}{1.mm}{
%
}
\end{table}
\subsubsection{\label{sms2sec:TmV6Sn6}\hyperref[smsec:col]{\texorpdfstring{TmV$_6$Sn$_6$}{}}~{\color{green}  (High-T stable, Low-T stable) }}

\begin{table}[htbp]
\global\long\def\arraystretch{1.12}
\captionsetup{width=.85\textwidth}
\caption{Structure parameters of TmV$_6$Sn$_6$ after relaxation. It contains 507 electrons. }
\label{tab:TmV6Sn6}
\setlength{\tabcolsep}{4mm}{
%
}
\end{table}

\begin{figure}[htbp]
\centering
\includegraphics[width=0.85\textwidth]{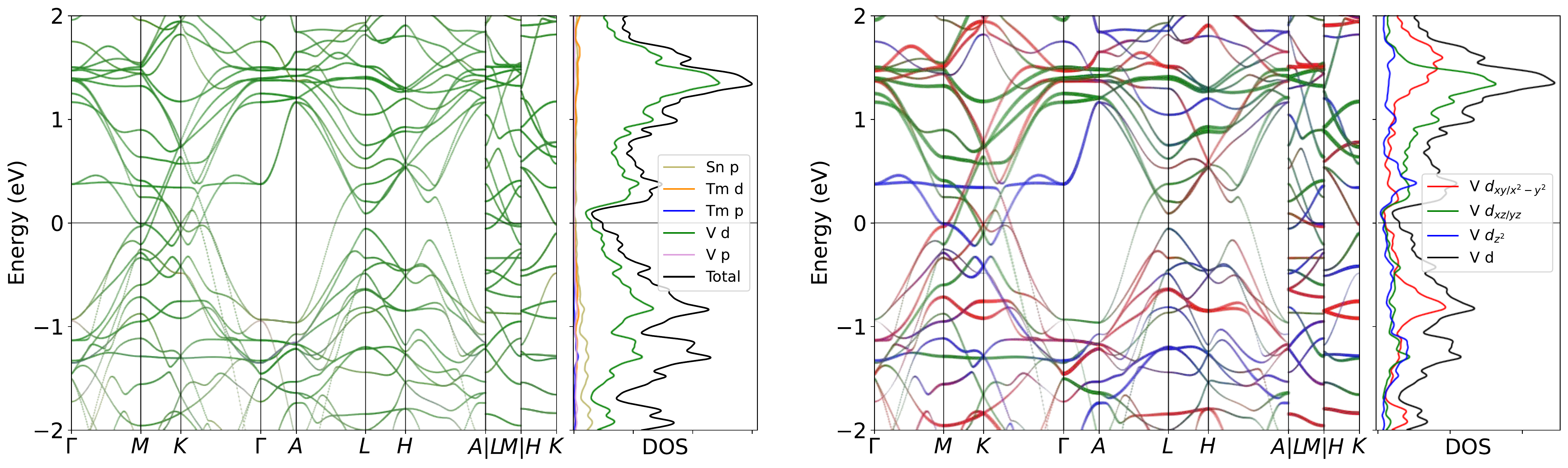}
\caption{Band structure for TmV$_6$Sn$_6$ without SOC. The projection of V $3d$ orbitals is also presented. The band structures clearly reveal the dominating of V $3d$ orbitals  near the Fermi level, as well as the pronounced peak.}
\label{fig:TmV6Sn6band}
\end{figure}

\begin{figure}[H]
\centering
\subfloat[Relaxed phonon at low-T.]{
\label{fig:TmV6Sn6_relaxed_Low}
\includegraphics[width=0.45\textwidth]{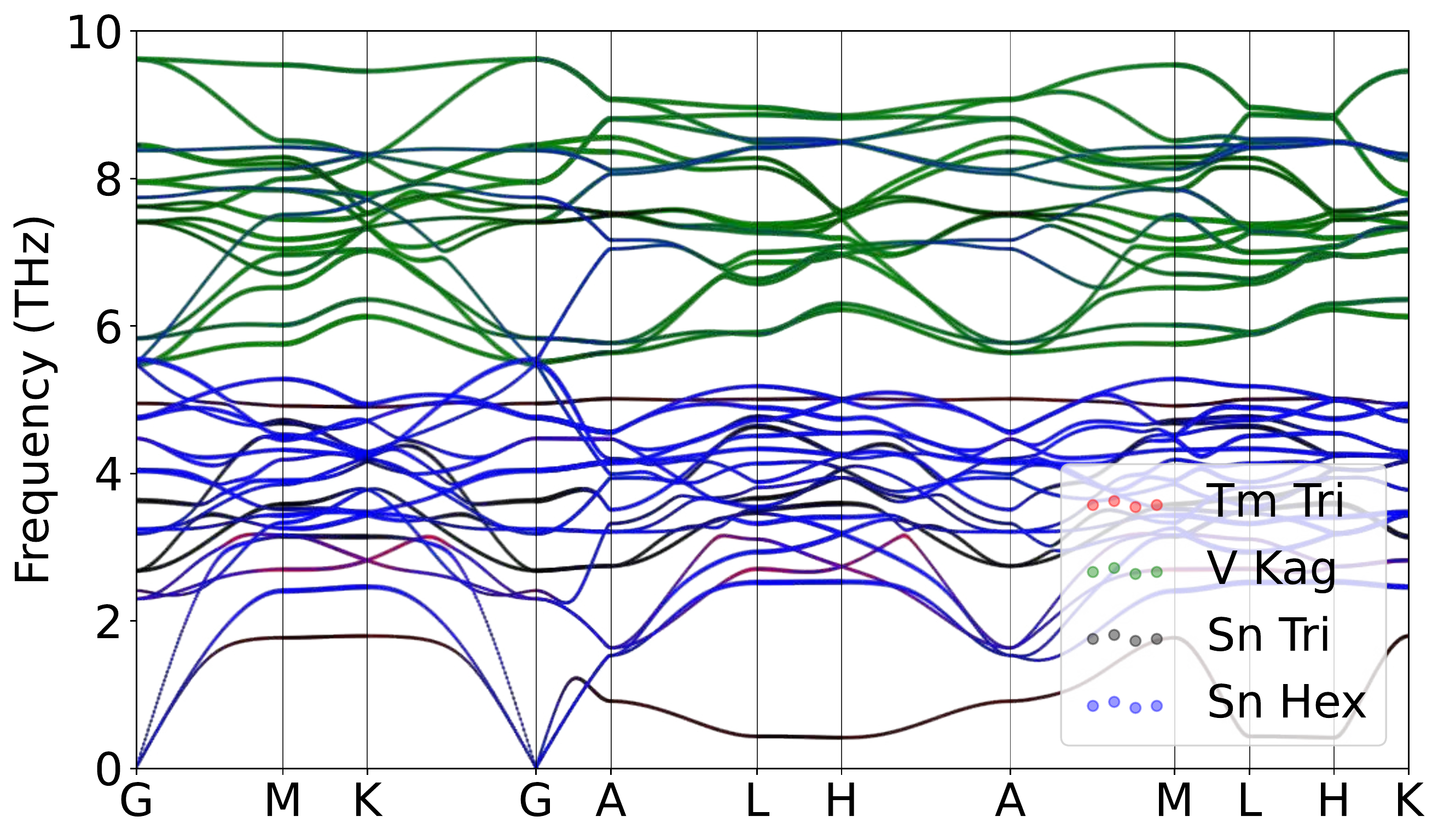}
}
\hspace{5pt}\subfloat[Relaxed phonon at high-T.]{
\label{fig:TmV6Sn6_relaxed_High}
\includegraphics[width=0.45\textwidth]{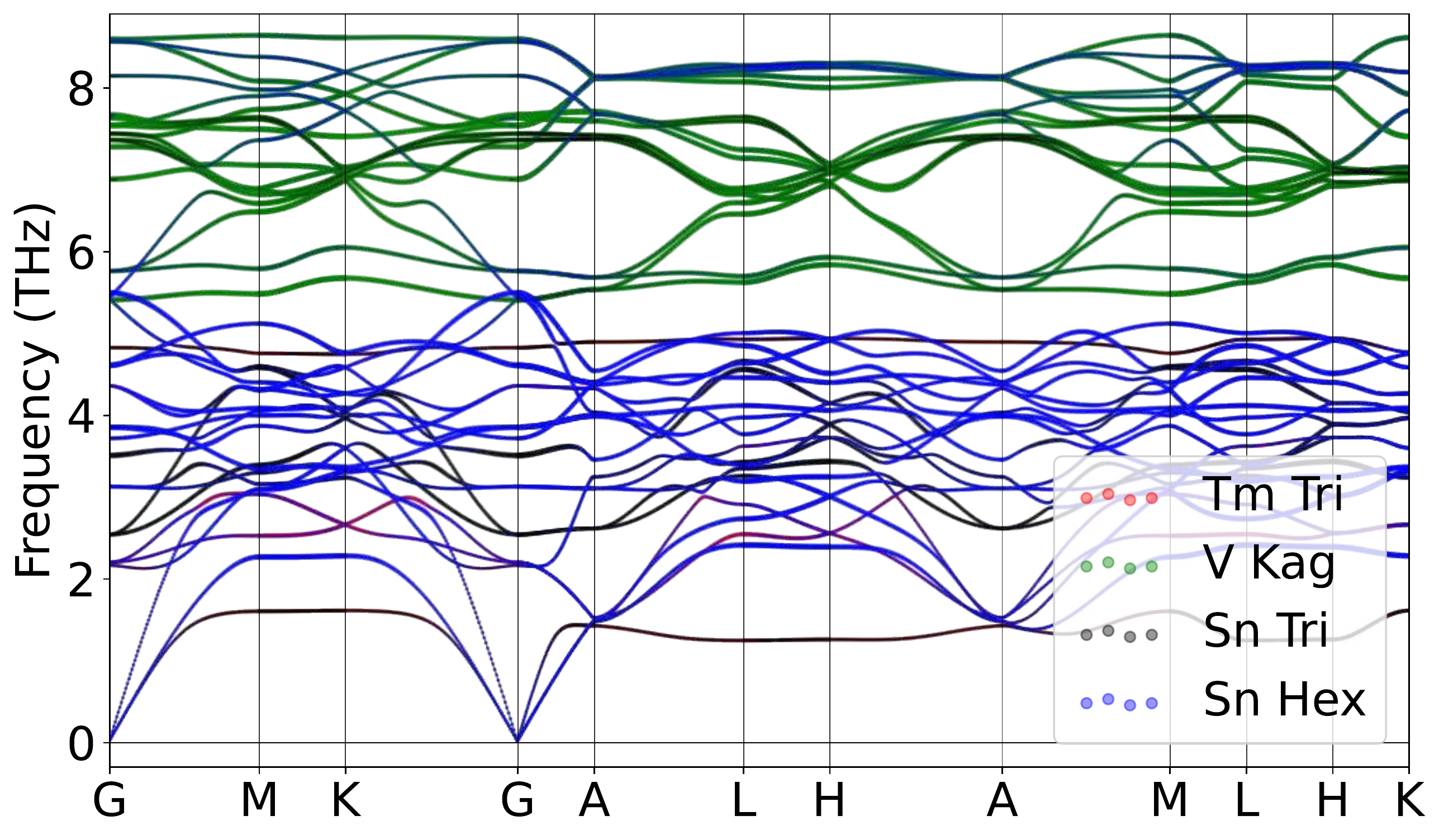}
}
\caption{Atomically projected phonon spectrum for TmV$_6$Sn$_6$.}
\label{fig:TmV6Sn6pho}
\end{figure}

\begin{figure}[H]
\centering
\label{fig:TmV6Sn6_relaxed_Low_xz}
\includegraphics[width=0.85\textwidth]{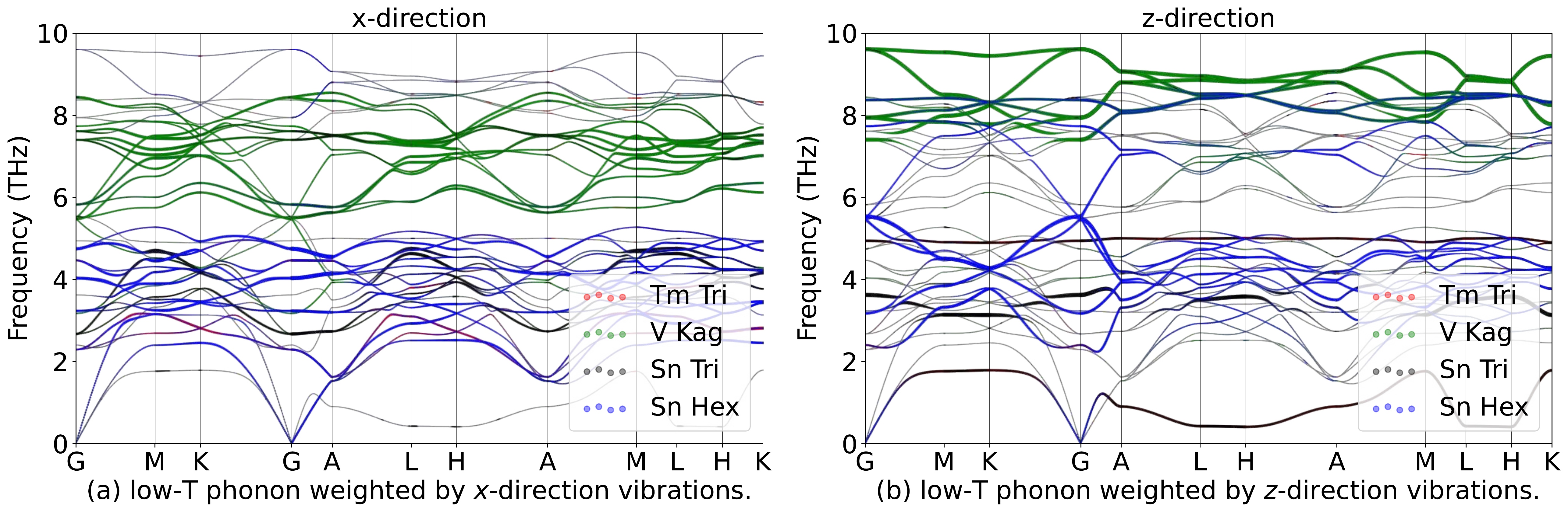}
\hspace{5pt}\label{fig:TmV6Sn6_relaxed_High_xz}
\includegraphics[width=0.85\textwidth]{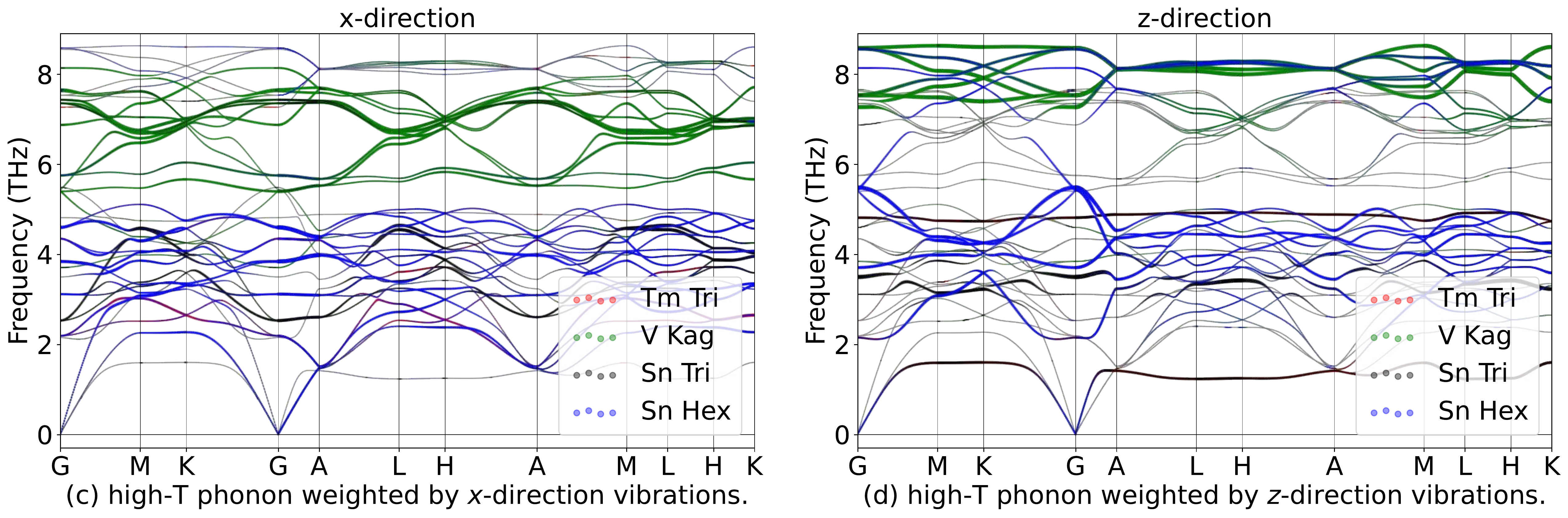}
\caption{Phonon spectrum for TmV$_6$Sn$_6$in in-plane ($x$) and out-of-plane ($z$) directions.}
\label{fig:TmV6Sn6phoxyz}
\end{figure}

\begin{table}[htbp]
\global\long\def\arraystretch{1.12}
\captionsetup{width=.95\textwidth}
\caption{IRREPs at high-symmetry points for TmV$_6$Sn$_6$ phonon spectrum at Low-T.}
\label{tab:191_TmV6Sn6_Low}
\setlength{\tabcolsep}{1.mm}{
}
\end{table}

\begin{table}[htbp]
\global\long\def\arraystretch{1.12}
\captionsetup{width=.95\textwidth}
\caption{IRREPs at high-symmetry points for TmV$_6$Sn$_6$ phonon spectrum at High-T.}
\label{tab:191_TmV6Sn6_High}
\setlength{\tabcolsep}{1.mm}{
%
}
\end{table}
\subsubsection{\label{sms2sec:YbV6Sn6}\hyperref[smsec:col]{\texorpdfstring{YbV$_6$Sn$_6$}{}}~{\color{blue}  (High-T stable, Low-T unstable) }}

\begin{table}[htbp]
\global\long\def\arraystretch{1.12}
\captionsetup{width=.85\textwidth}
\caption{Structure parameters of YbV$_6$Sn$_6$ after relaxation. It contains 508 electrons. }
\label{tab:YbV6Sn6}
\setlength{\tabcolsep}{4mm}{
%
}
\end{table}

\begin{figure}[htbp]
\centering
\includegraphics[width=0.85\textwidth]{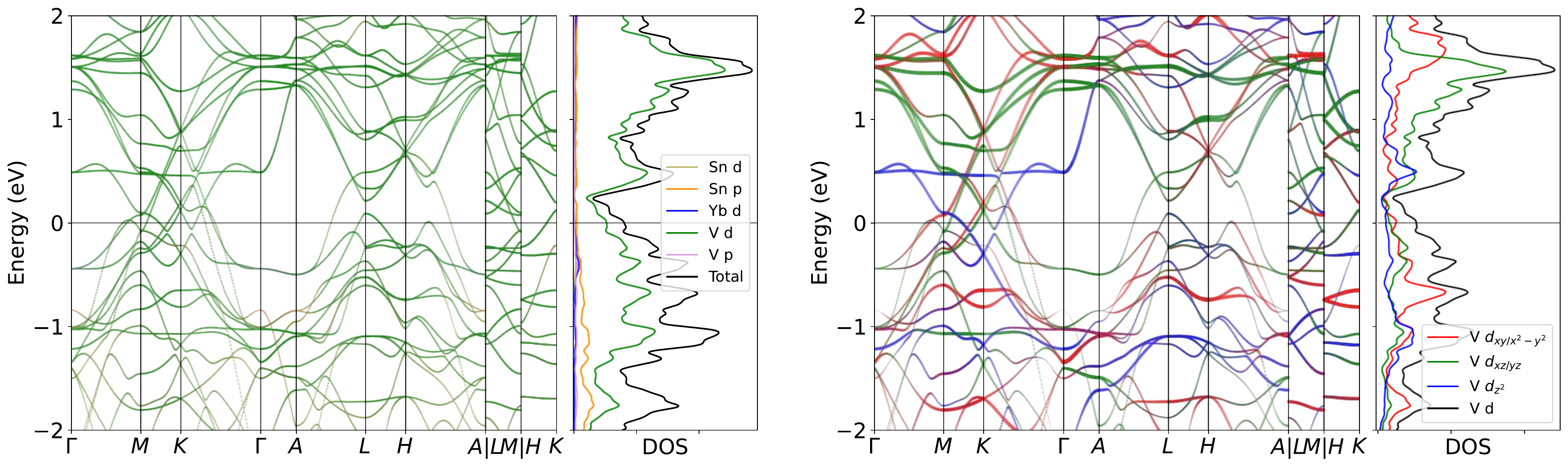}
\caption{Band structure for YbV$_6$Sn$_6$ without SOC. The projection of V $3d$ orbitals is also presented. The band structures clearly reveal the dominating of V $3d$ orbitals  near the Fermi level, as well as the pronounced peak.}
\label{fig:YbV6Sn6band}
\end{figure}

\begin{figure}[H]
\centering
\subfloat[Relaxed phonon at low-T.]{
\label{fig:YbV6Sn6_relaxed_Low}
\includegraphics[width=0.45\textwidth]{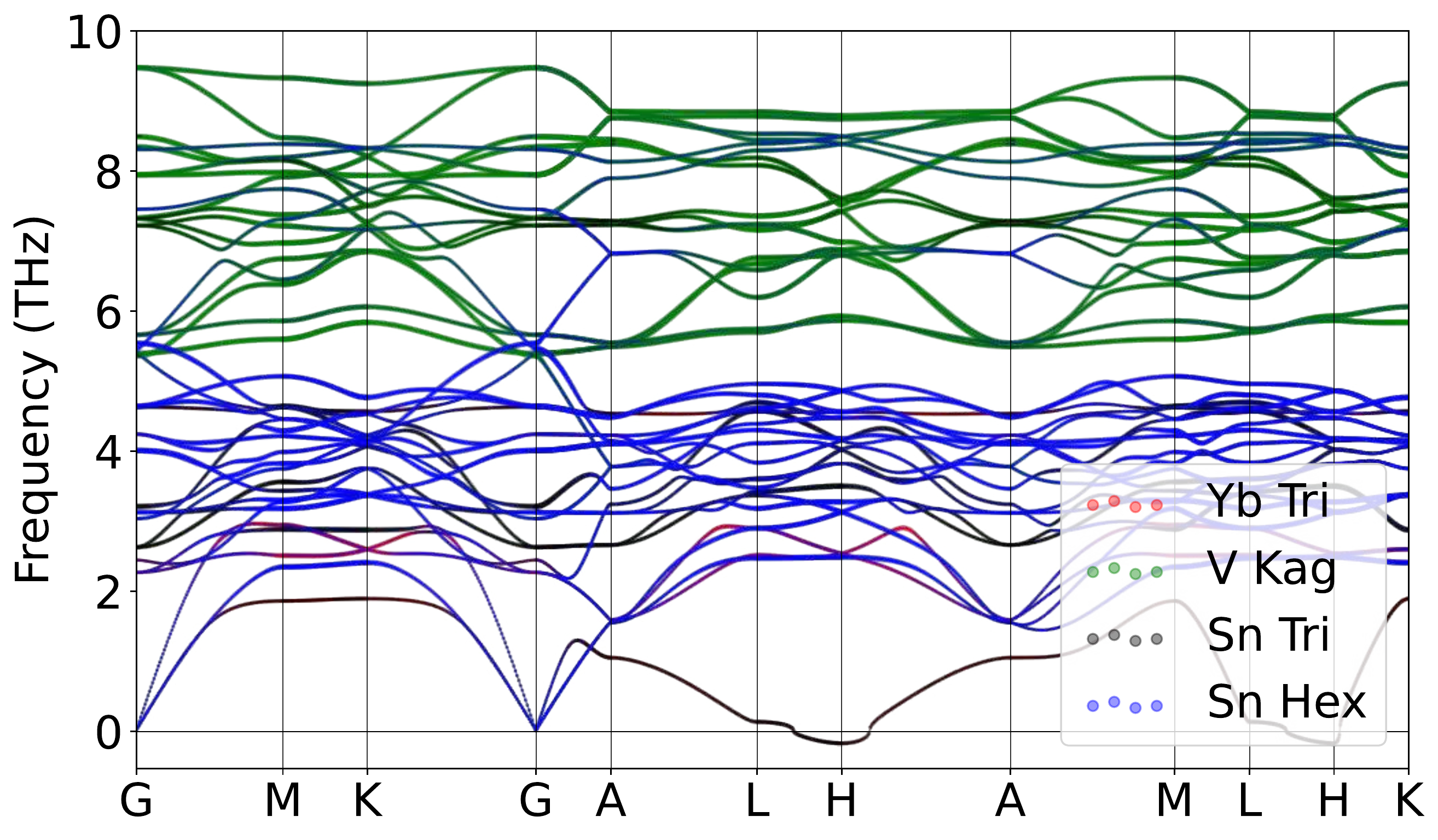}
}
\hspace{5pt}\subfloat[Relaxed phonon at high-T.]{
\label{fig:YbV6Sn6_relaxed_High}
\includegraphics[width=0.45\textwidth]{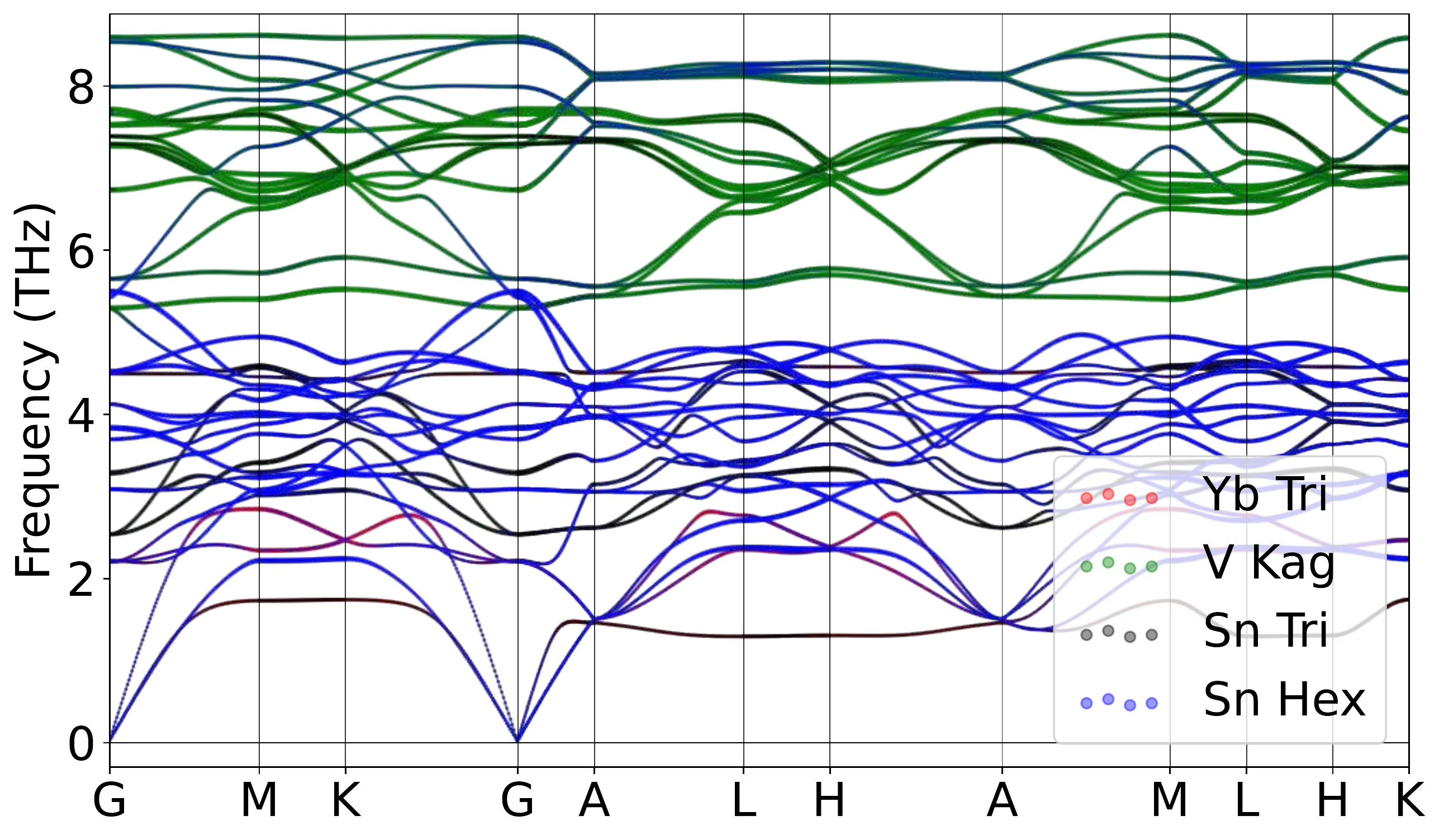}
}
\caption{Atomically projected phonon spectrum for YbV$_6$Sn$_6$.}
\label{fig:YbV6Sn6pho}
\end{figure}

\begin{figure}[H]
\centering
\label{fig:YbV6Sn6_relaxed_Low_xz}
\includegraphics[width=0.85\textwidth]{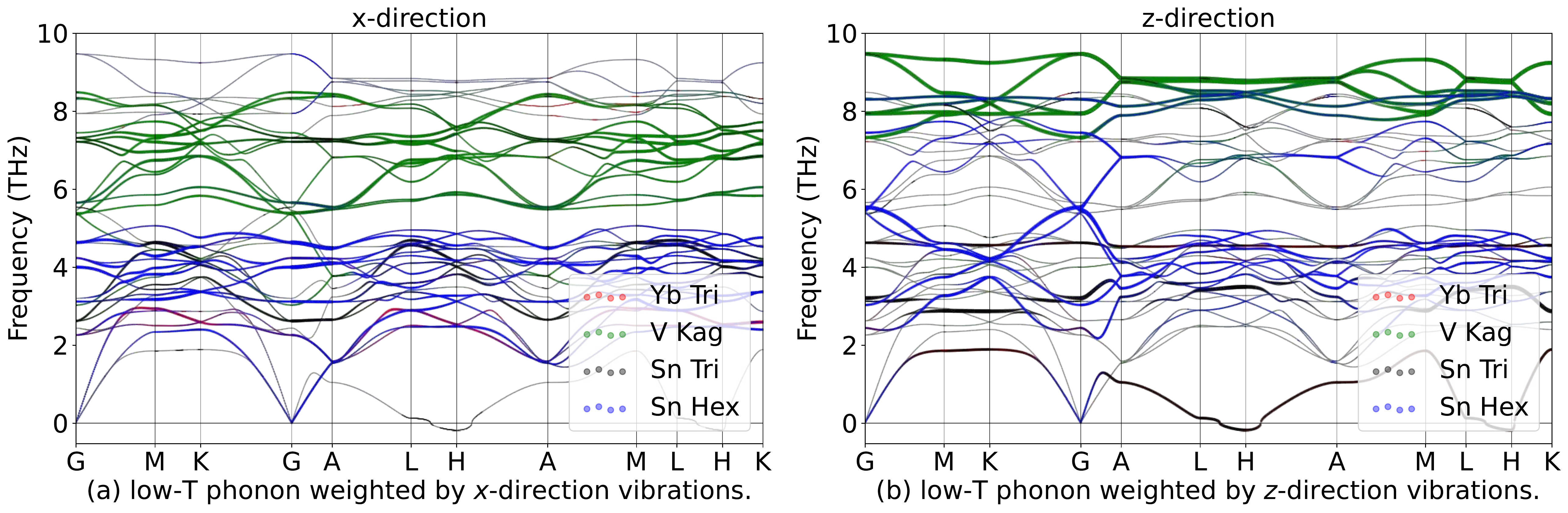}
\hspace{5pt}\label{fig:YbV6Sn6_relaxed_High_xz}
\includegraphics[width=0.85\textwidth]{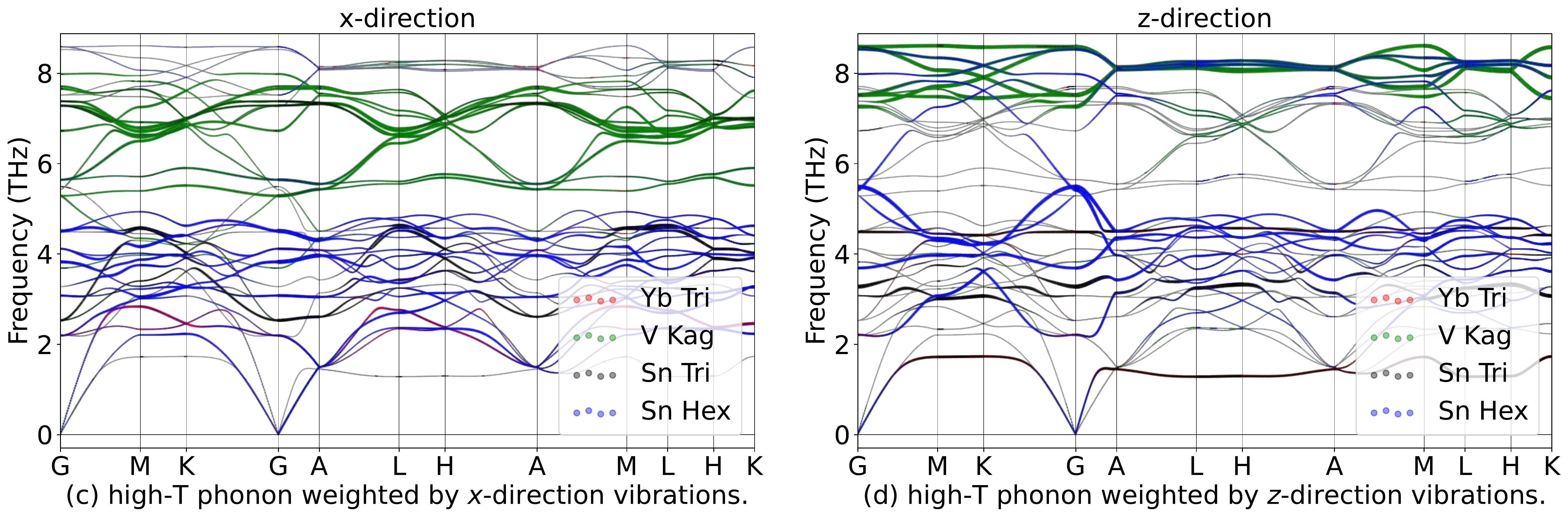}
\caption{Phonon spectrum for YbV$_6$Sn$_6$in in-plane ($x$) and out-of-plane ($z$) directions.}
\label{fig:YbV6Sn6phoxyz}
\end{figure}

\begin{table}[htbp]
\global\long\def\arraystretch{1.12}
\captionsetup{width=.95\textwidth}
\caption{IRREPs at high-symmetry points for YbV$_6$Sn$_6$ phonon spectrum at Low-T.}
\label{tab:191_YbV6Sn6_Low}
\setlength{\tabcolsep}{1.mm}{
}
\end{table}

\begin{table}[htbp]
\global\long\def\arraystretch{1.12}
\captionsetup{width=.95\textwidth}
\caption{IRREPs at high-symmetry points for YbV$_6$Sn$_6$ phonon spectrum at High-T.}
\label{tab:191_YbV6Sn6_High}
\setlength{\tabcolsep}{1.mm}{
%
}
\end{table}
\subsubsection{\label{sms2sec:LuV6Sn6}\hyperref[smsec:col]{\texorpdfstring{LuV$_6$Sn$_6$}{}}~{\color{green}  (High-T stable, Low-T stable) }}

\begin{table}[htbp]
\global\long\def\arraystretch{1.12}
\captionsetup{width=.85\textwidth}
\caption{Structure parameters of LuV$_6$Sn$_6$ after relaxation. It contains 509 electrons. }
\label{tab:LuV6Sn6}
\setlength{\tabcolsep}{4mm}{
%
}
\end{table}

\begin{figure}[htbp]
\centering
\includegraphics[width=0.85\textwidth]{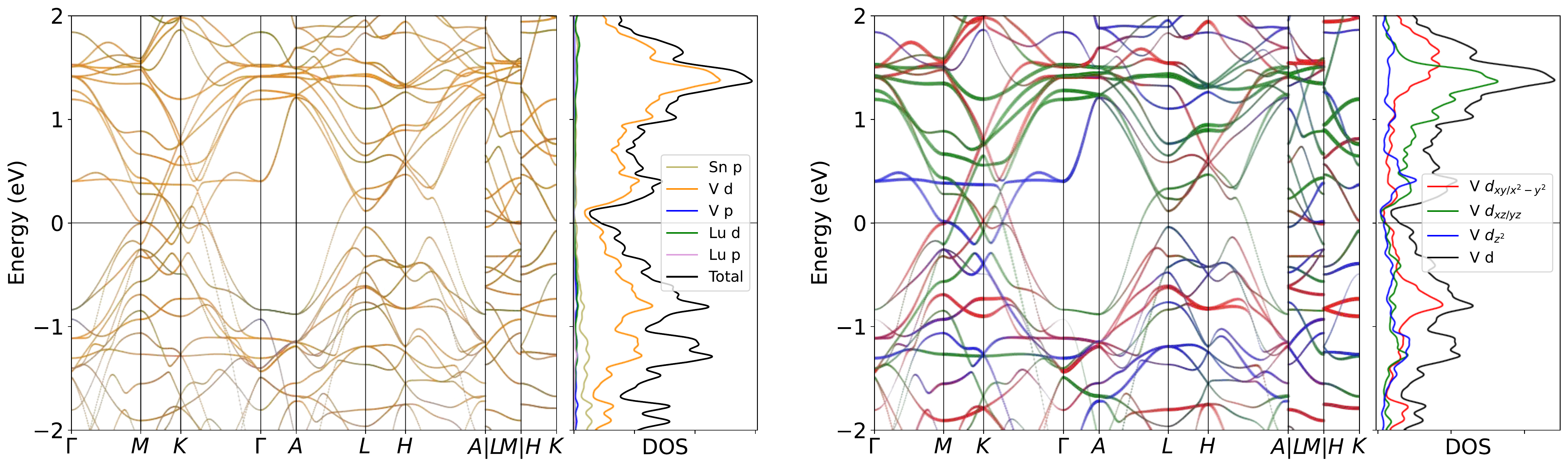}
\caption{Band structure for LuV$_6$Sn$_6$ without SOC. The projection of V $3d$ orbitals is also presented. The band structures clearly reveal the dominating of V $3d$ orbitals  near the Fermi level, as well as the pronounced peak.}
\label{fig:LuV6Sn6band}
\end{figure}

\begin{figure}[H]
\centering
\subfloat[Relaxed phonon at low-T.]{
\label{fig:LuV6Sn6_relaxed_Low}
\includegraphics[width=0.45\textwidth]{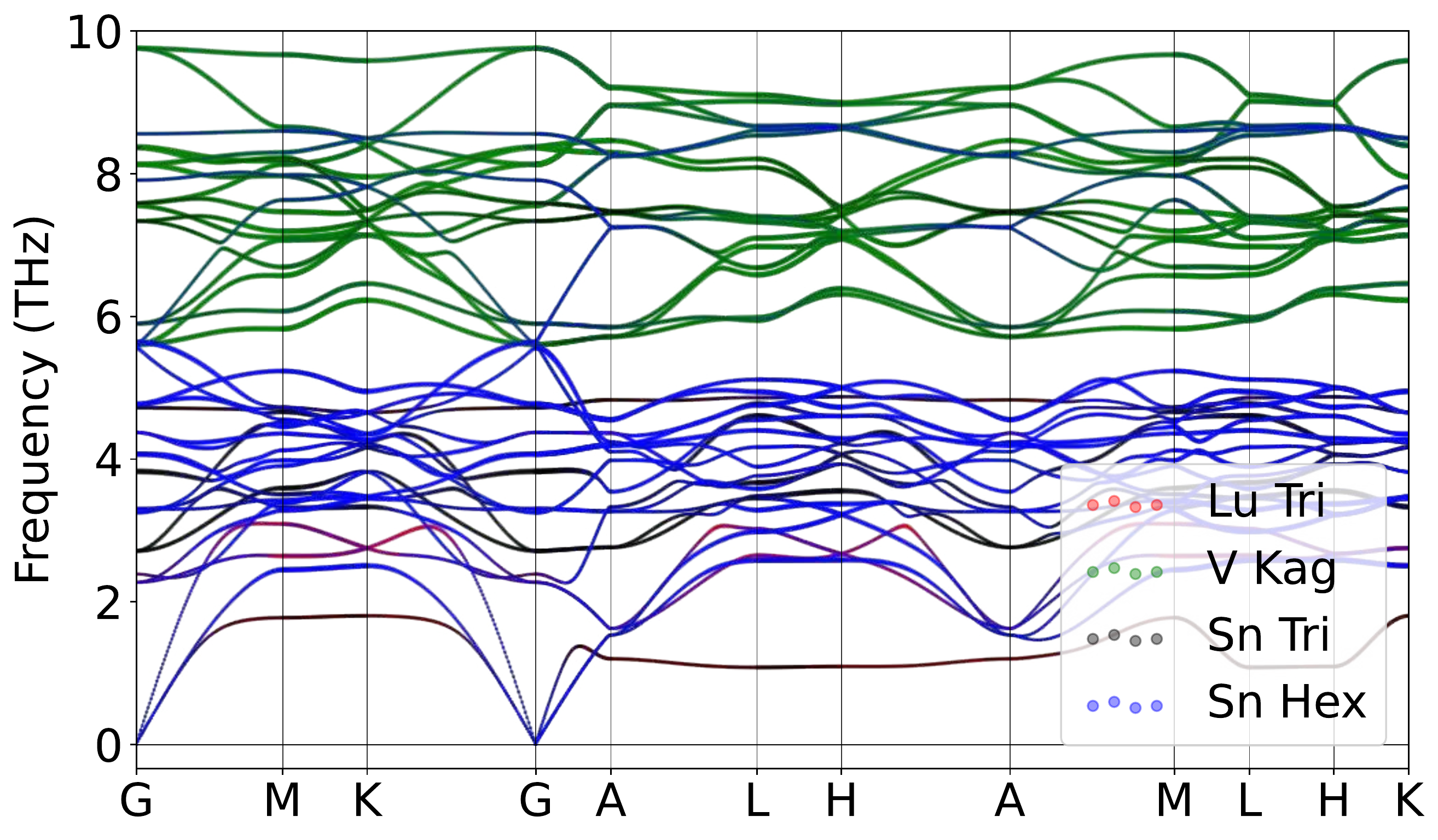}
}
\hspace{5pt}\subfloat[Relaxed phonon at high-T.]{
\label{fig:LuV6Sn6_relaxed_High}
\includegraphics[width=0.45\textwidth]{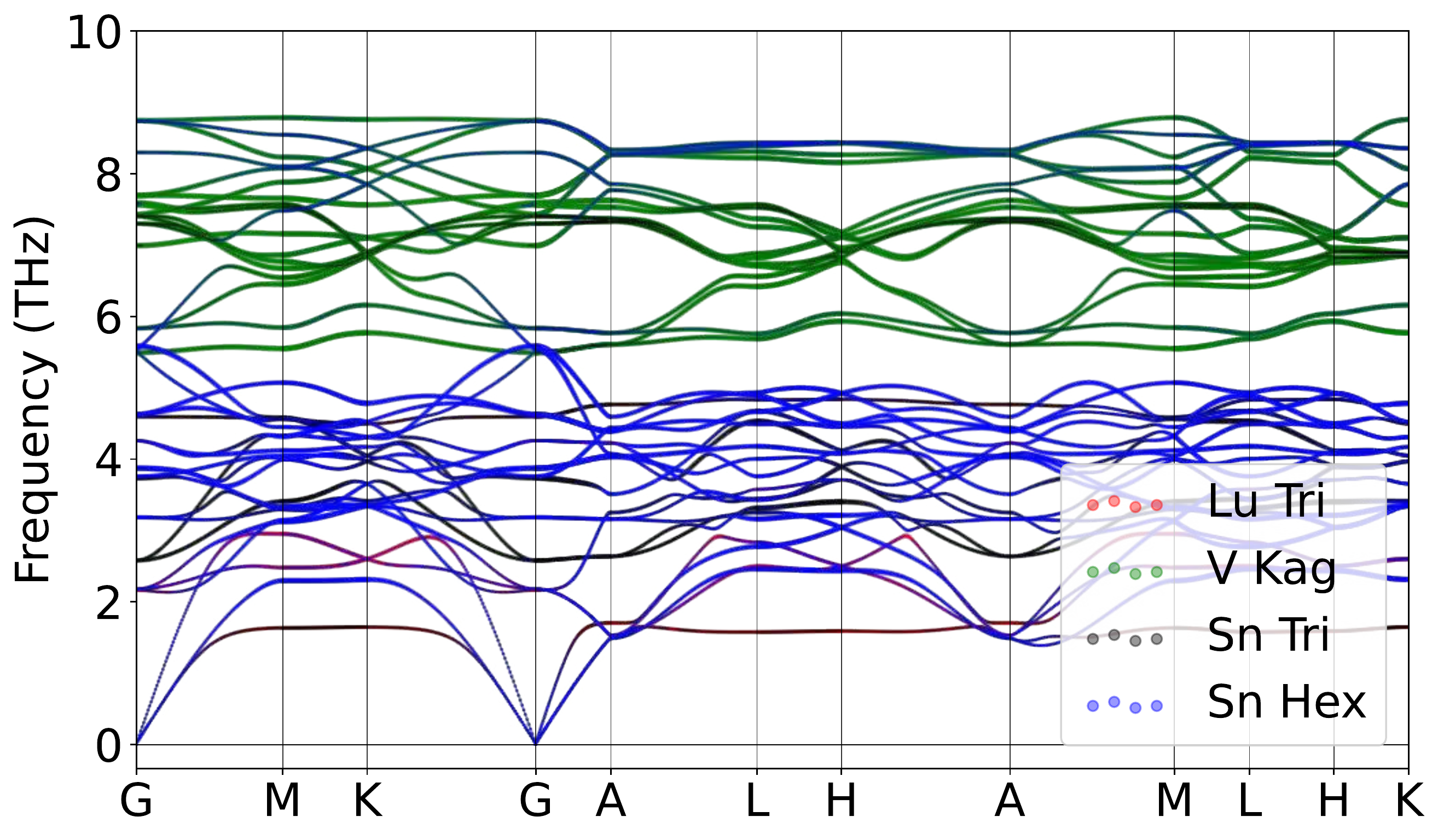}
}
\caption{Atomically projected phonon spectrum for LuV$_6$Sn$_6$.}
\label{fig:LuV6Sn6pho}
\end{figure}

\begin{figure}[H]
\centering
\label{fig:LuV6Sn6_relaxed_Low_xz}
\includegraphics[width=0.85\textwidth]{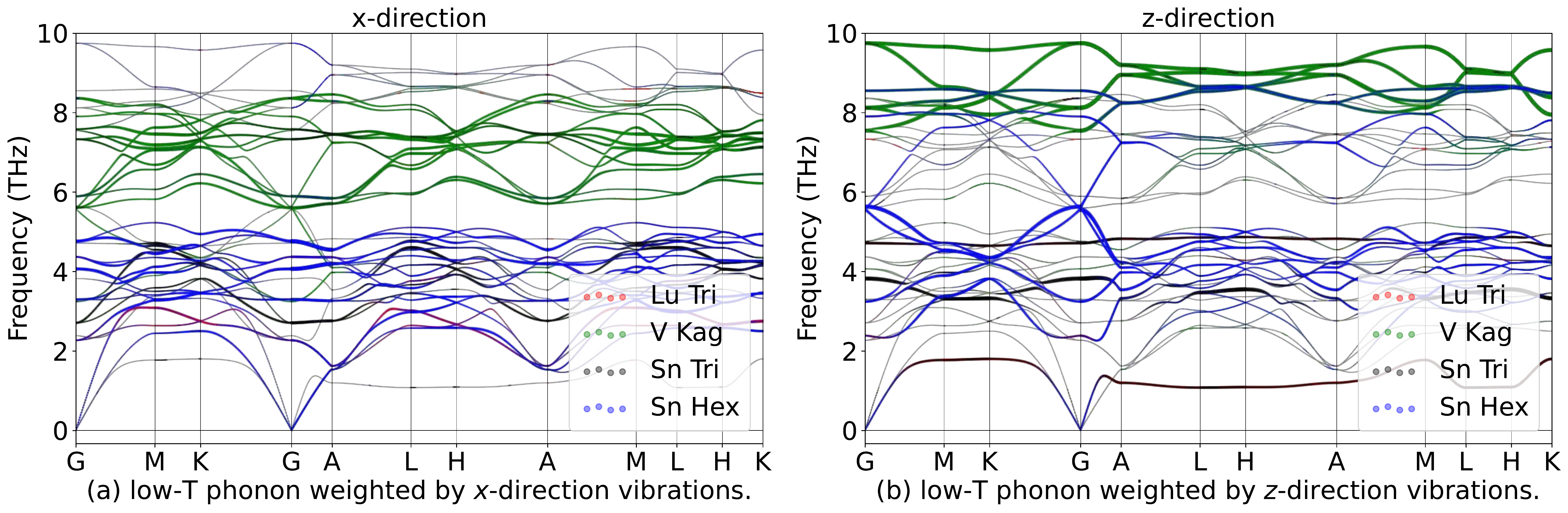}
\hspace{5pt}\label{fig:LuV6Sn6_relaxed_High_xz}
\includegraphics[width=0.85\textwidth]{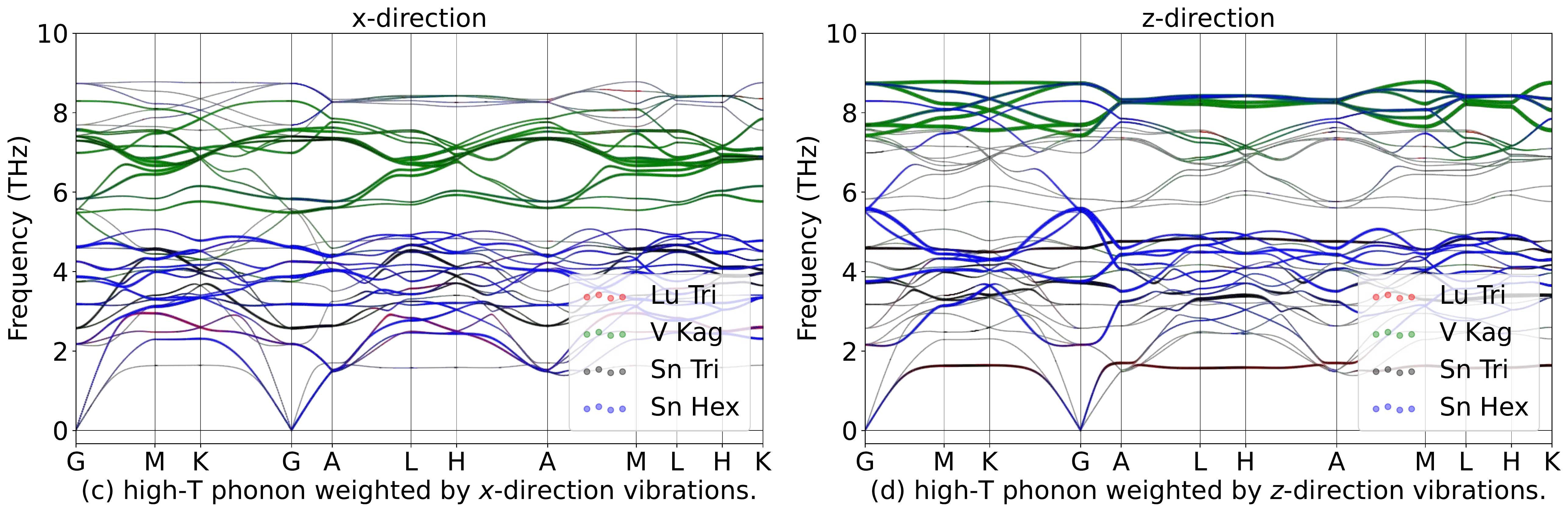}
\caption{Phonon spectrum for LuV$_6$Sn$_6$in in-plane ($x$) and out-of-plane ($z$) directions.}
\label{fig:LuV6Sn6phoxyz}
\end{figure}

\begin{table}[htbp]
\global\long\def\arraystretch{1.12}
\captionsetup{width=.95\textwidth}
\caption{IRREPs at high-symmetry points for LuV$_6$Sn$_6$ phonon spectrum at Low-T.}
\label{tab:191_LuV6Sn6_Low}
\setlength{\tabcolsep}{1.mm}{
}
\end{table}

\begin{table}[htbp]
\global\long\def\arraystretch{1.12}
\captionsetup{width=.95\textwidth}
\caption{IRREPs at high-symmetry points for LuV$_6$Sn$_6$ phonon spectrum at High-T.}
\label{tab:191_LuV6Sn6_High}
\setlength{\tabcolsep}{1.mm}{
%
}
\end{table}
\subsubsection{\label{sms2sec:HfV6Sn6}\hyperref[smsec:col]{\texorpdfstring{HfV$_6$Sn$_6$}{}}~{\color{green}  (High-T stable, Low-T stable) }}

\begin{table}[htbp]
\global\long\def\arraystretch{1.12}
\captionsetup{width=.85\textwidth}
\caption{Structure parameters of HfV$_6$Sn$_6$ after relaxation. It contains 510 electrons. }
\label{tab:HfV6Sn6}
\setlength{\tabcolsep}{4mm}{
%
}
\end{table}

\begin{figure}[htbp]
\centering
\includegraphics[width=0.85\textwidth]{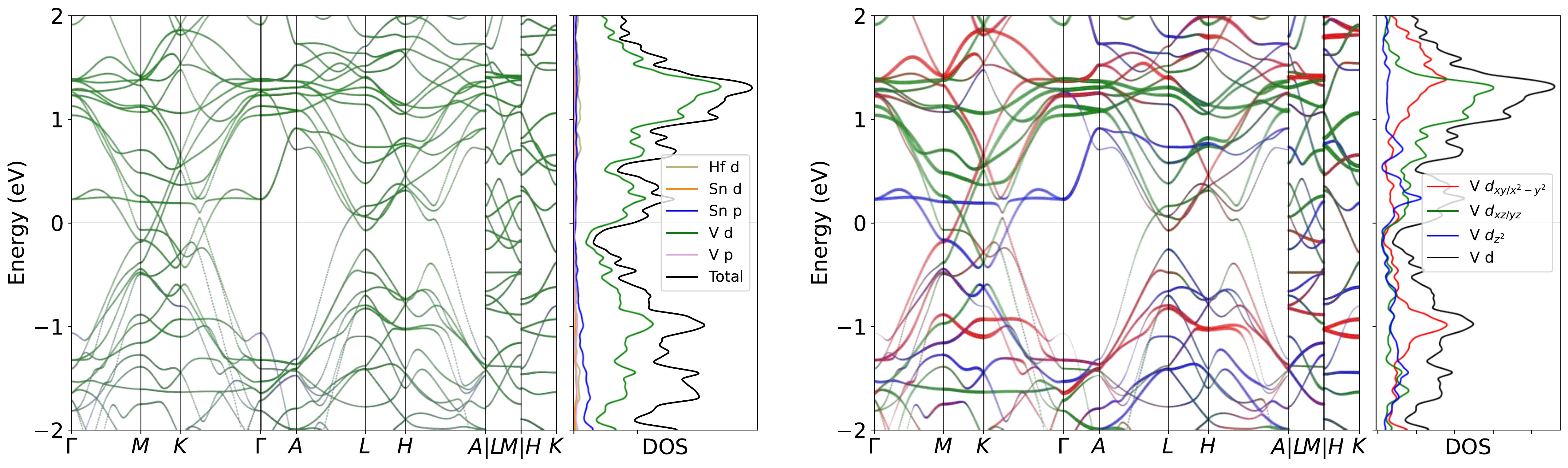}
\caption{Band structure for HfV$_6$Sn$_6$ without SOC. The projection of V $3d$ orbitals is also presented. The band structures clearly reveal the dominating of V $3d$ orbitals  near the Fermi level, as well as the pronounced peak.}
\label{fig:HfV6Sn6band}
\end{figure}

\begin{figure}[H]
\centering
\subfloat[Relaxed phonon at low-T.]{
\label{fig:HfV6Sn6_relaxed_Low}
\includegraphics[width=0.45\textwidth]{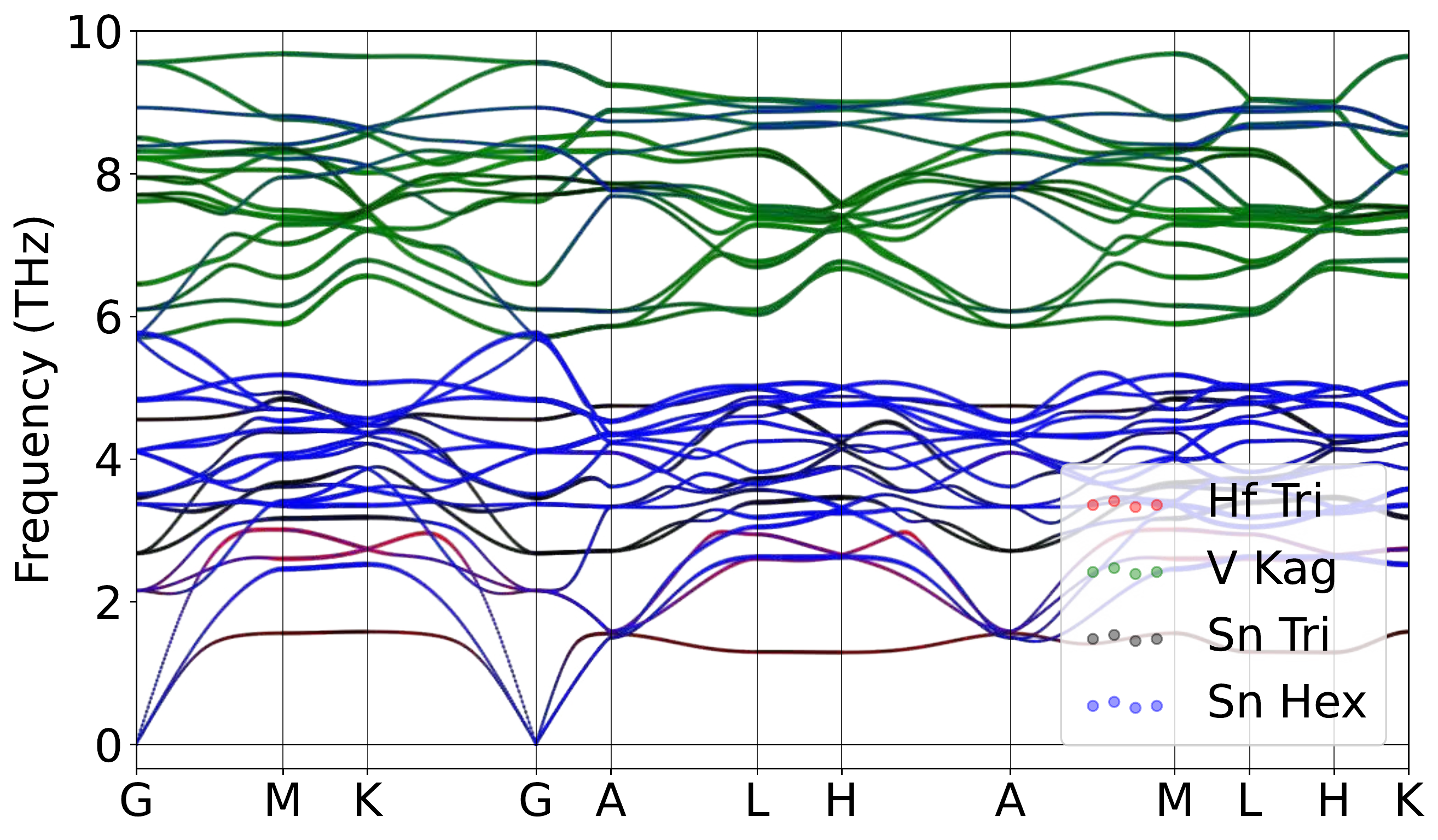}
}
\hspace{5pt}\subfloat[Relaxed phonon at high-T.]{
\label{fig:HfV6Sn6_relaxed_High}
\includegraphics[width=0.45\textwidth]{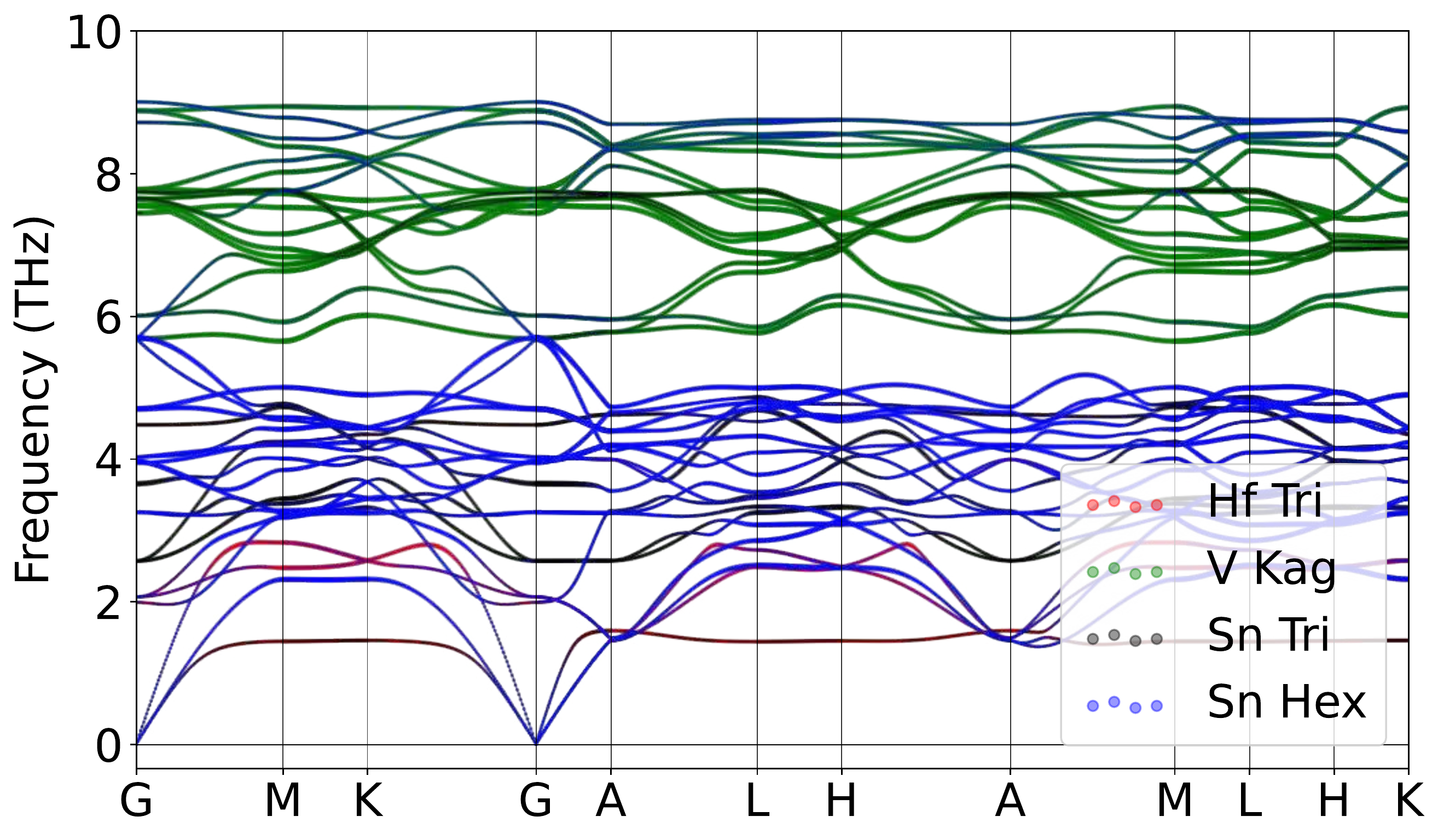}
}
\caption{Atomically projected phonon spectrum for HfV$_6$Sn$_6$.}
\label{fig:HfV6Sn6pho}
\end{figure}

\begin{figure}[H]
\centering
\label{fig:HfV6Sn6_relaxed_Low_xz}
\includegraphics[width=0.85\textwidth]{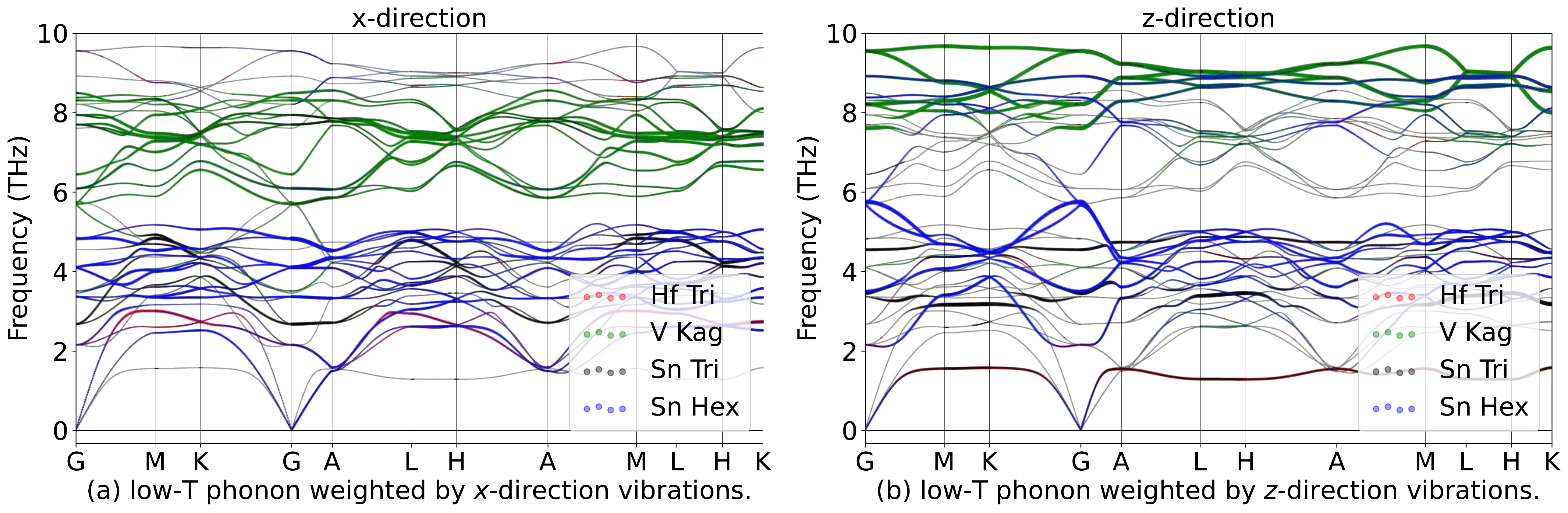}
\hspace{5pt}\label{fig:HfV6Sn6_relaxed_High_xz}
\includegraphics[width=0.85\textwidth]{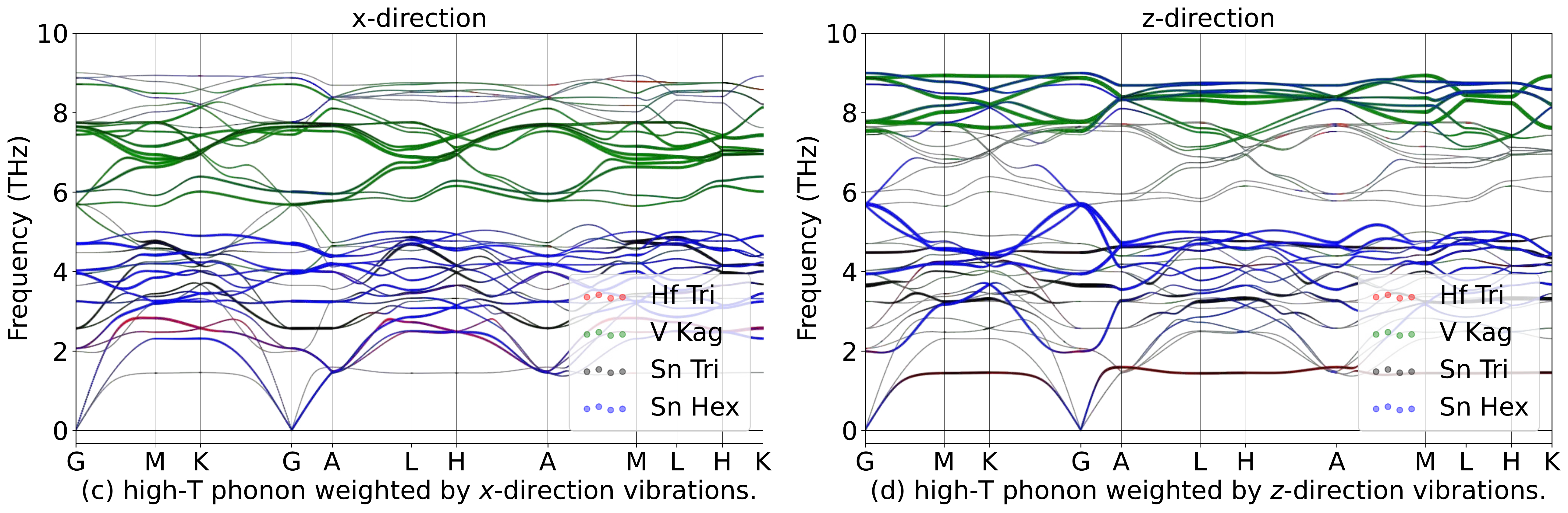}
\caption{Phonon spectrum for HfV$_6$Sn$_6$in in-plane ($x$) and out-of-plane ($z$) directions.}
\label{fig:HfV6Sn6phoxyz}
\end{figure}

\begin{table}[htbp]
\global\long\def\arraystretch{1.12}
\captionsetup{width=.95\textwidth}
\caption{IRREPs at high-symmetry points for HfV$_6$Sn$_6$ phonon spectrum at Low-T.}
\label{tab:191_HfV6Sn6_Low}
\setlength{\tabcolsep}{1.mm}{
}
\end{table}

\begin{table}[htbp]
\global\long\def\arraystretch{1.12}
\captionsetup{width=.95\textwidth}
\caption{IRREPs at high-symmetry points for HfV$_6$Sn$_6$ phonon spectrum at High-T.}
\label{tab:191_HfV6Sn6_High}
\setlength{\tabcolsep}{1.mm}{
%
}
\end{table}
\subsubsection{\label{sms2sec:TaV6Sn6}\hyperref[smsec:col]{\texorpdfstring{TaV$_6$Sn$_6$}{}}~{\color{green}  (High-T stable, Low-T stable) }}

\begin{table}[htbp]
\global\long\def\arraystretch{1.12}
\captionsetup{width=.85\textwidth}
\caption{Structure parameters of TaV$_6$Sn$_6$ after relaxation. It contains 511 electrons. }
\label{tab:TaV6Sn6}
\setlength{\tabcolsep}{4mm}{
%
}
\end{table}

\begin{figure}[htbp]
\centering
\includegraphics[width=0.85\textwidth]{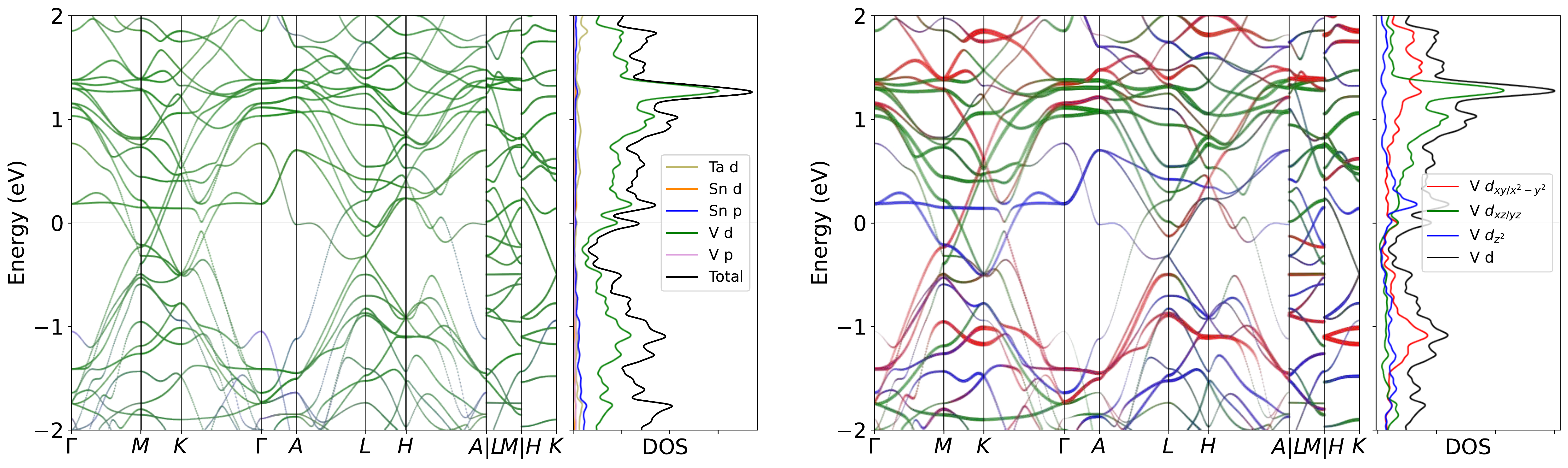}
\caption{Band structure for TaV$_6$Sn$_6$ without SOC. The projection of V $3d$ orbitals is also presented. The band structures clearly reveal the dominating of V $3d$ orbitals  near the Fermi level, as well as the pronounced peak.}
\label{fig:TaV6Sn6band}
\end{figure}

\begin{figure}[H]
\centering
\subfloat[Relaxed phonon at low-T.]{
\label{fig:TaV6Sn6_relaxed_Low}
\includegraphics[width=0.45\textwidth]{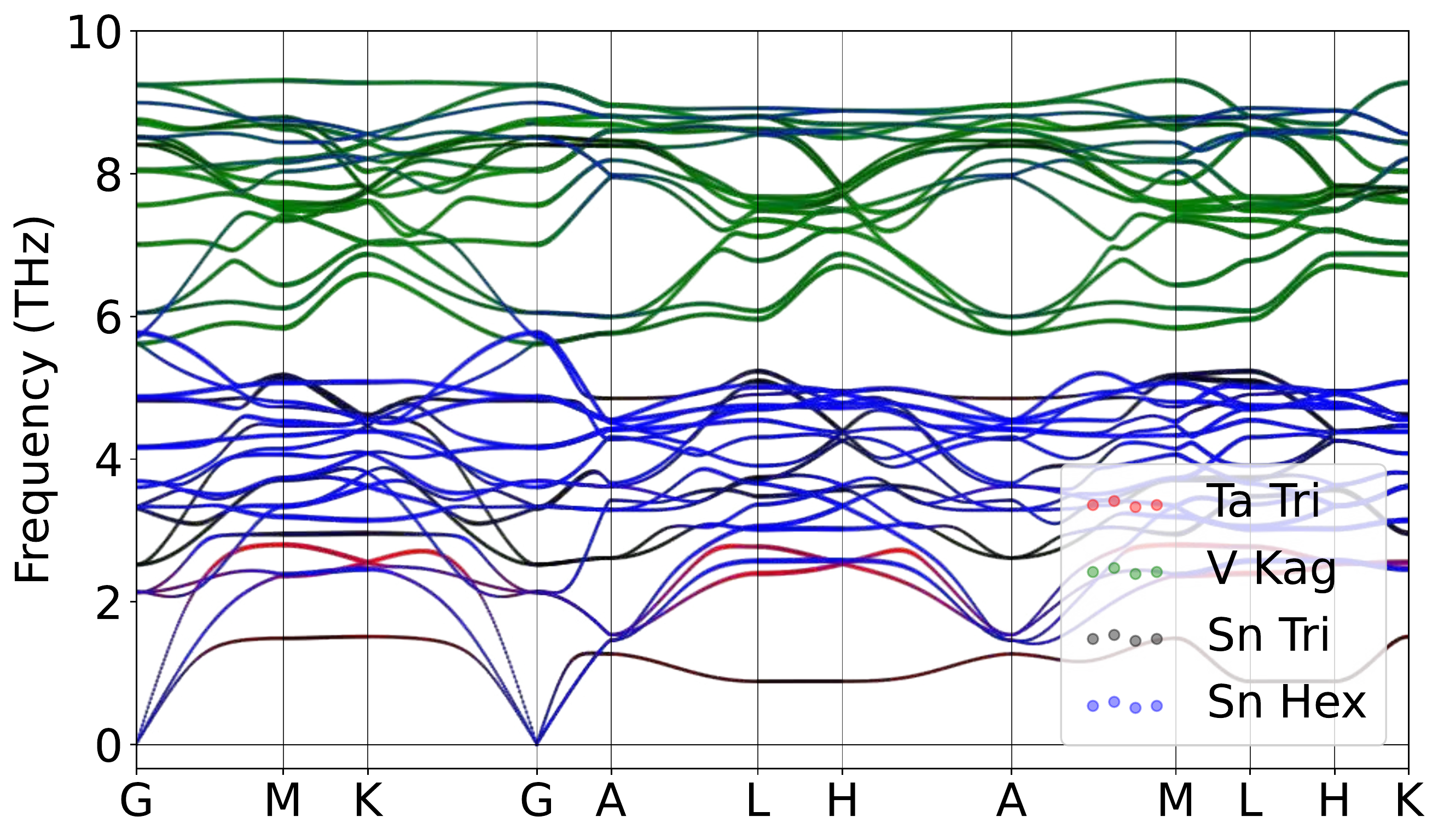}
}
\hspace{5pt}\subfloat[Relaxed phonon at high-T.]{
\label{fig:TaV6Sn6_relaxed_High}
\includegraphics[width=0.45\textwidth]{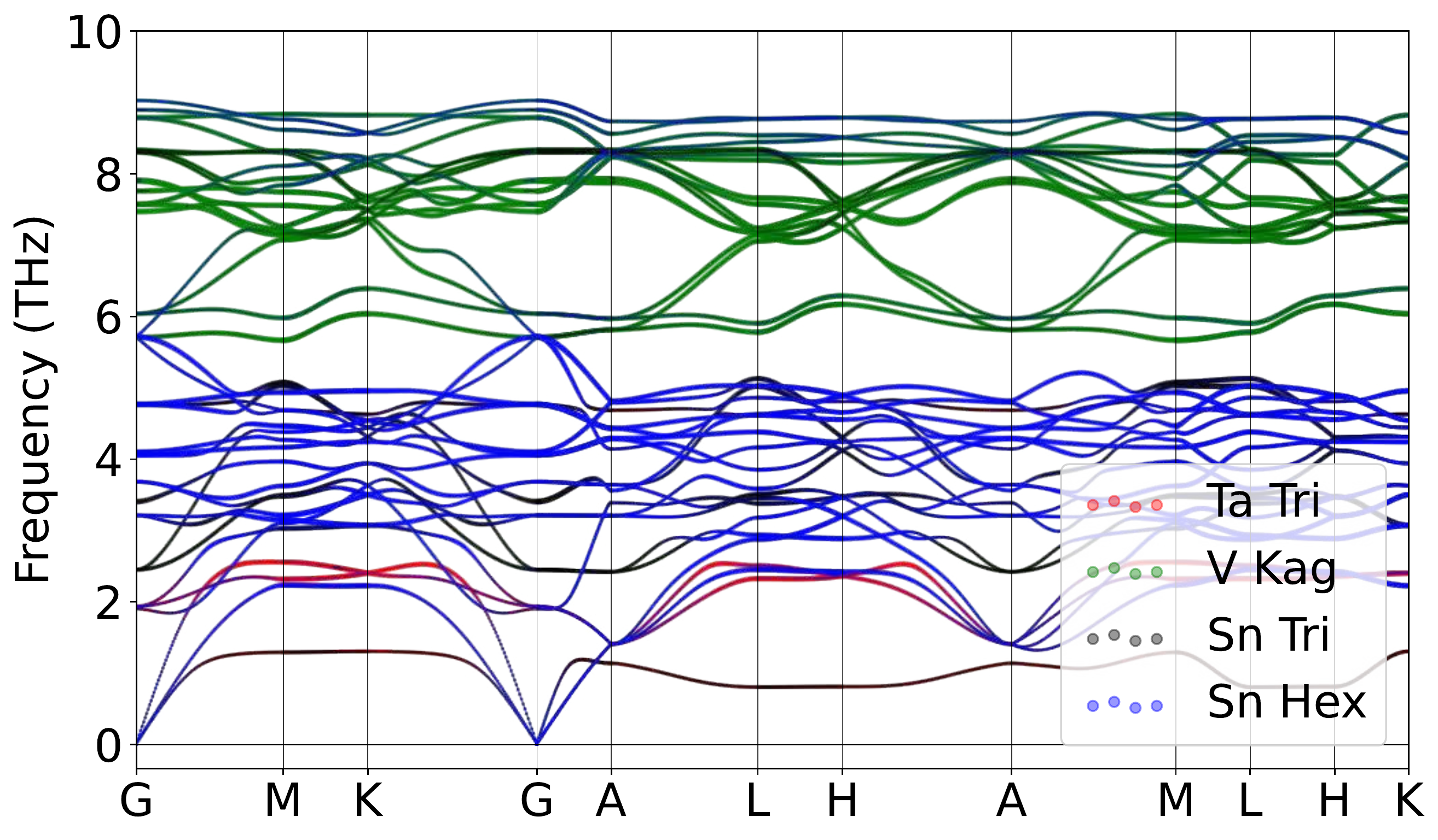}
}
\caption{Atomically projected phonon spectrum for TaV$_6$Sn$_6$.}
\label{fig:TaV6Sn6pho}
\end{figure}

\begin{figure}[H]
\centering
\label{fig:TaV6Sn6_relaxed_Low_xz}
\includegraphics[width=0.85\textwidth]{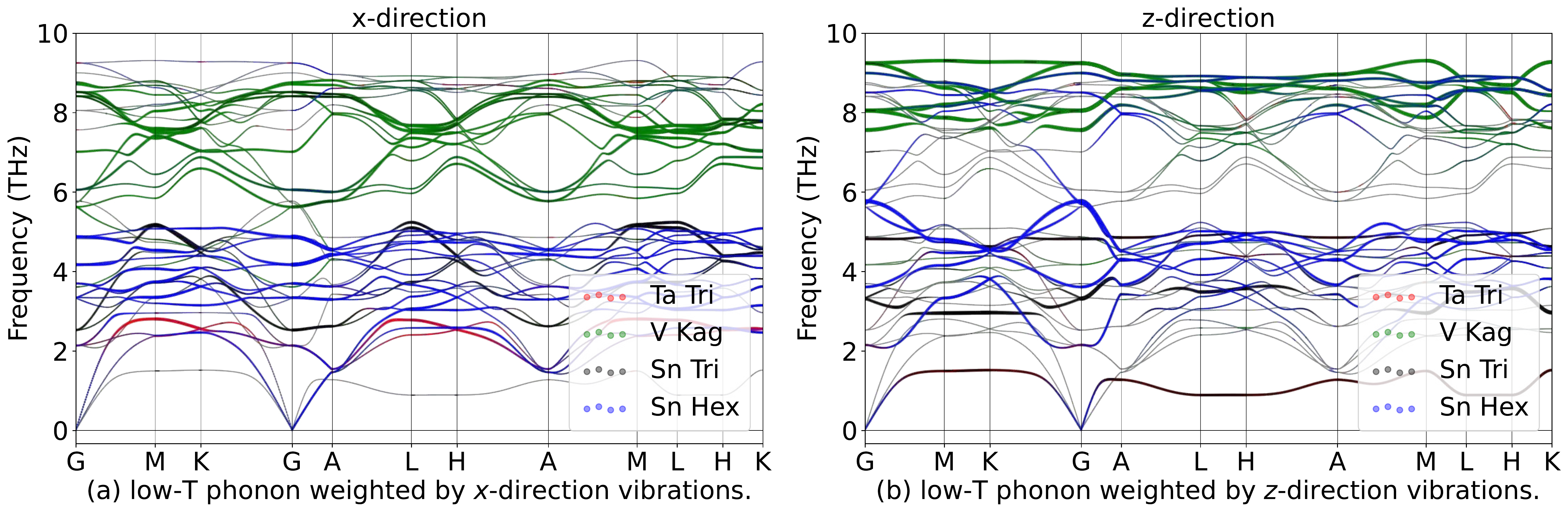}
\hspace{5pt}\label{fig:TaV6Sn6_relaxed_High_xz}
\includegraphics[width=0.85\textwidth]{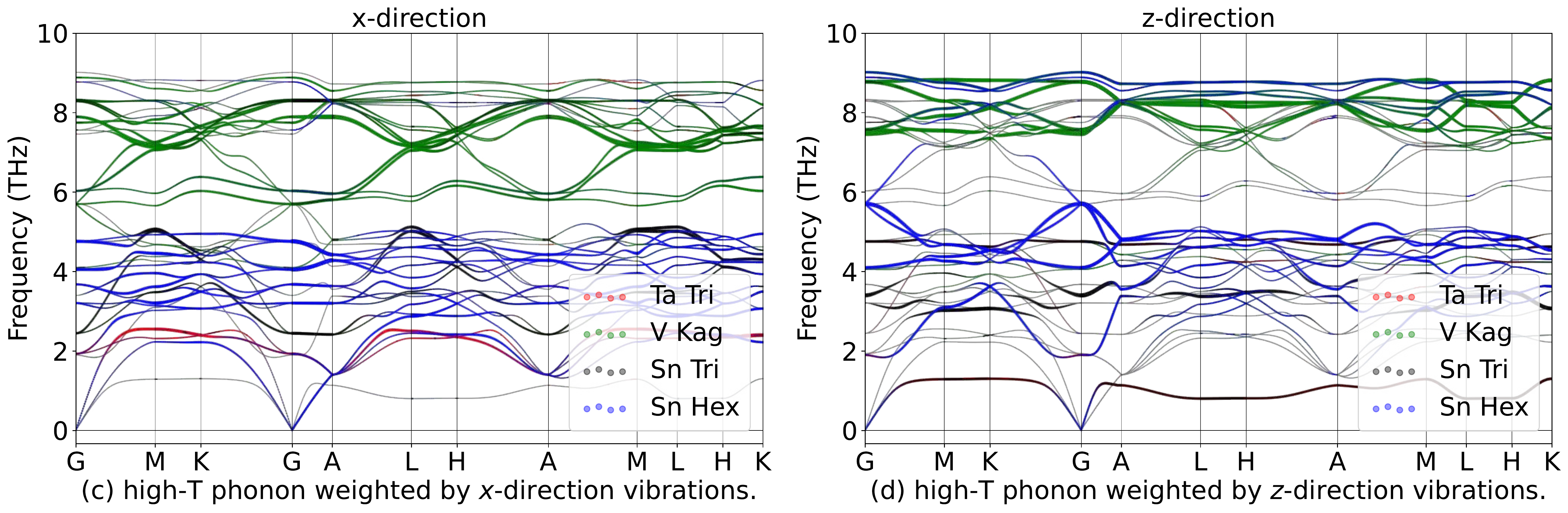}
\caption{Phonon spectrum for TaV$_6$Sn$_6$in in-plane ($x$) and out-of-plane ($z$) directions.}
\label{fig:TaV6Sn6phoxyz}
\end{figure}

\begin{table}[htbp]
\global\long\def\arraystretch{1.12}
\captionsetup{width=.95\textwidth}
\caption{IRREPs at high-symmetry points for TaV$_6$Sn$_6$ phonon spectrum at Low-T.}
\label{tab:191_TaV6Sn6_Low}
\setlength{\tabcolsep}{1.mm}{
}
\end{table}

\begin{table}[htbp]
\global\long\def\arraystretch{1.12}
\captionsetup{width=.95\textwidth}
\caption{IRREPs at high-symmetry points for TaV$_6$Sn$_6$ phonon spectrum at High-T.}
\label{tab:191_TaV6Sn6_High}
\setlength{\tabcolsep}{1.mm}{
%
}
\end{table}
\subsection{\hyperref[smsec:col]{CrGe}}~\label{smssec:CrGe}

\subsubsection{\label{sms2sec:LiCr6Ge6}\hyperref[smsec:col]{\texorpdfstring{LiCr$_6$Ge$_6$}{}}~{\color{green}  (High-T stable, Low-T stable) }}

\begin{table}[htbp]
\global\long\def\arraystretch{1.12}
\captionsetup{width=.85\textwidth}
\caption{Structure parameters of LiCr$_6$Ge$_6$ after relaxation. It contains 339 electrons. }
\label{tab:LiCr6Ge6}
\setlength{\tabcolsep}{4mm}{
%
}
\end{table}

\begin{figure}[htbp]
\centering
\includegraphics[width=0.85\textwidth]{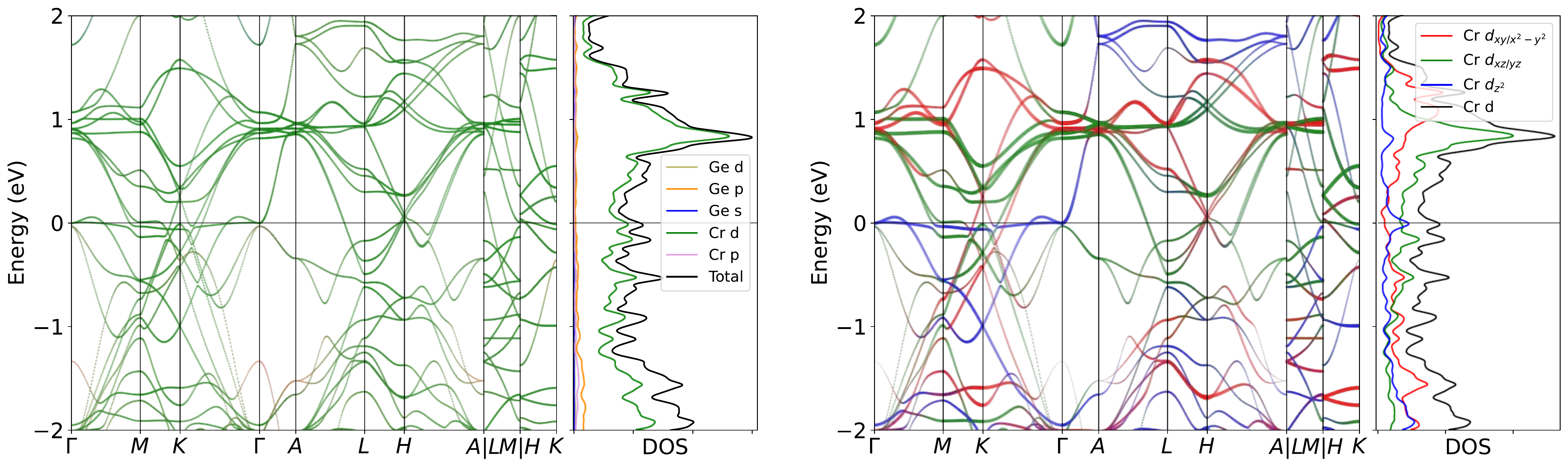}
\caption{Band structure for LiCr$_6$Ge$_6$ without SOC. The projection of Cr $3d$ orbitals is also presented. The band structures clearly reveal the dominating of Cr $3d$ orbitals  near the Fermi level, as well as the pronounced peak.}
\label{fig:LiCr6Ge6band}
\end{figure}

\begin{figure}[H]
\centering
\subfloat[Relaxed phonon at low-T.]{
\label{fig:LiCr6Ge6_relaxed_Low}
\includegraphics[width=0.45\textwidth]{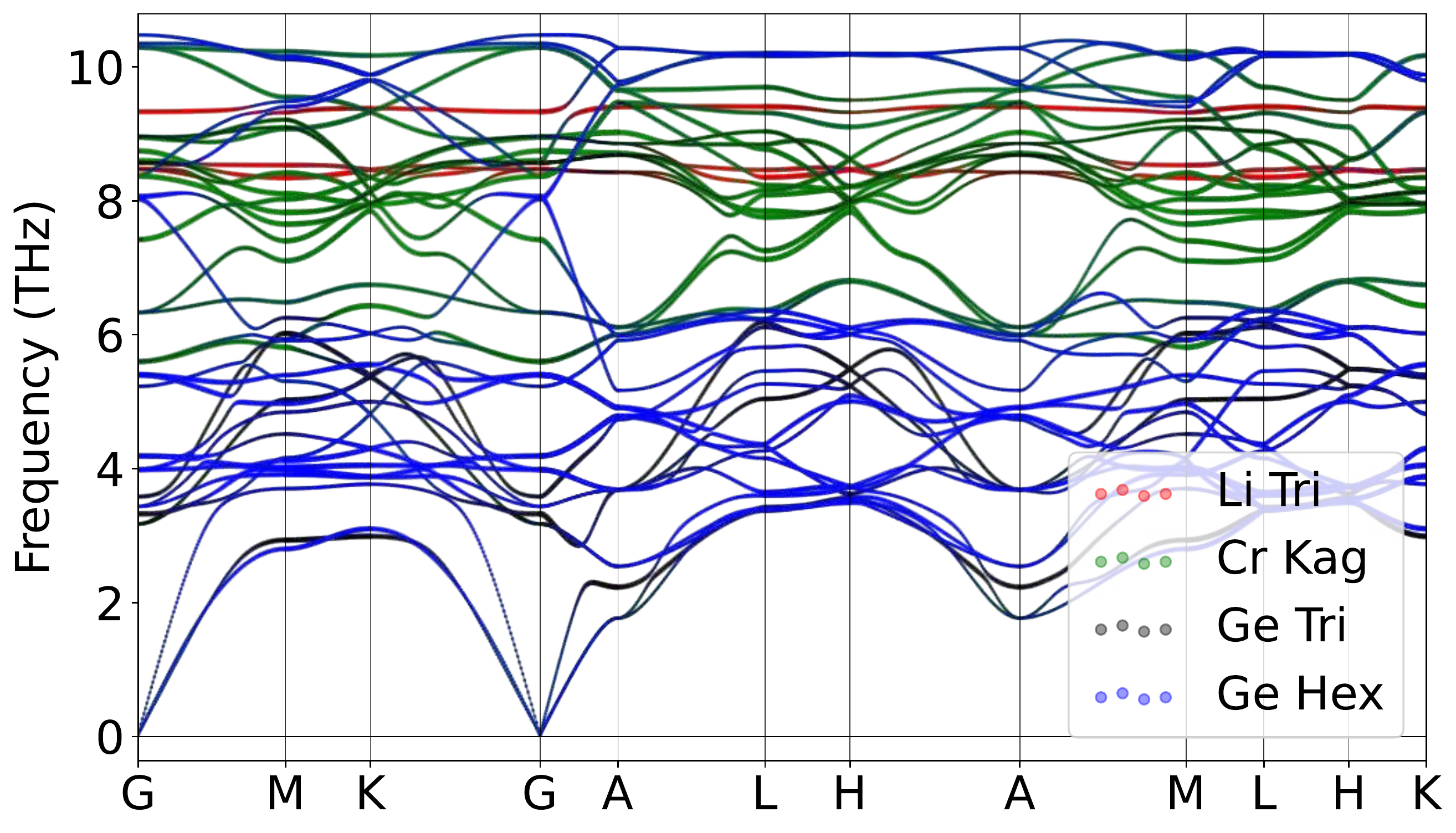}
}
\hspace{5pt}\subfloat[Relaxed phonon at high-T.]{
\label{fig:LiCr6Ge6_relaxed_High}
\includegraphics[width=0.45\textwidth]{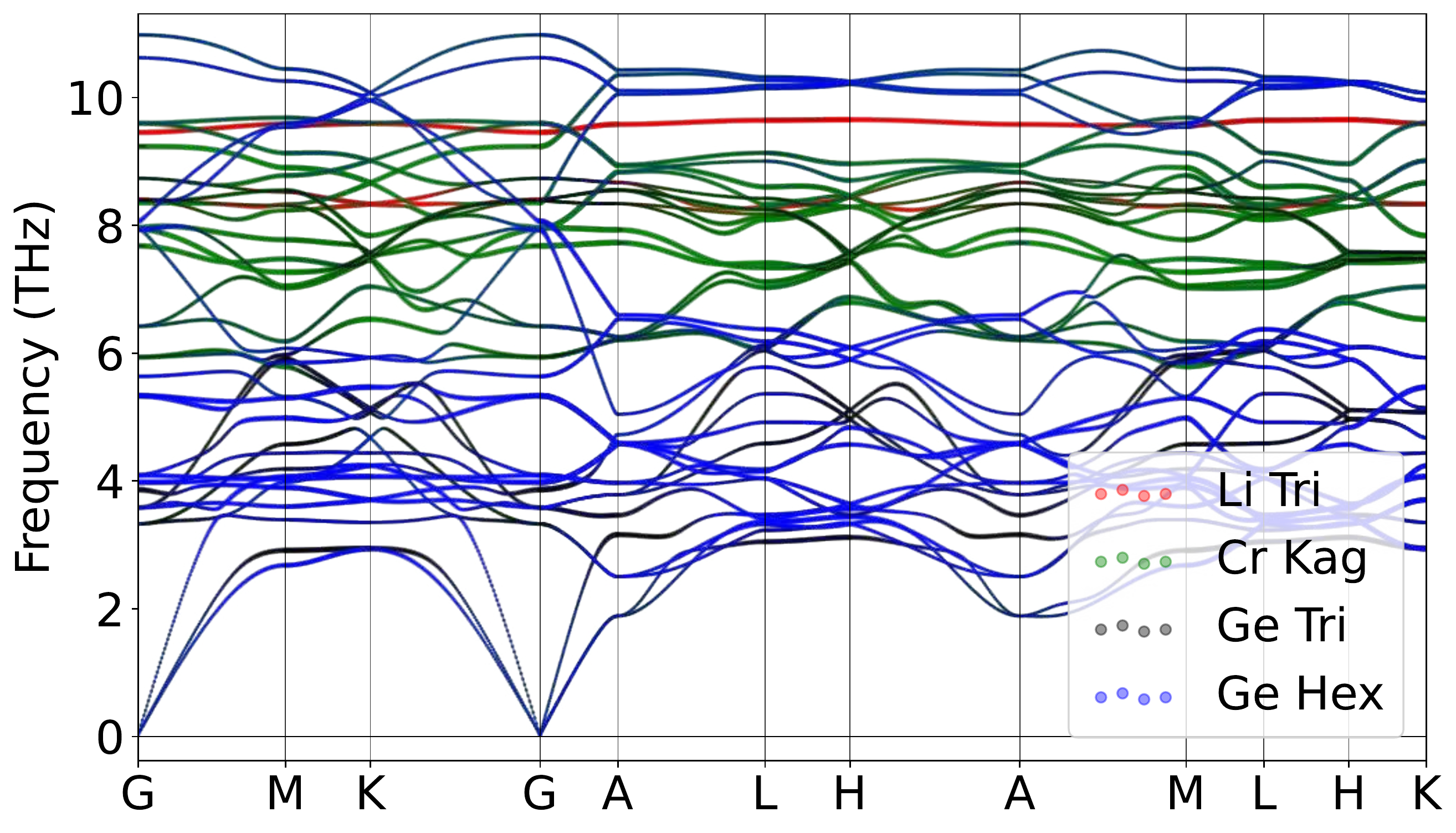}
}
\caption{Atomically projected phonon spectrum for LiCr$_6$Ge$_6$.}
\label{fig:LiCr6Ge6pho}
\end{figure}

\begin{figure}[H]
\centering
\label{fig:LiCr6Ge6_relaxed_Low_xz}
\includegraphics[width=0.85\textwidth]{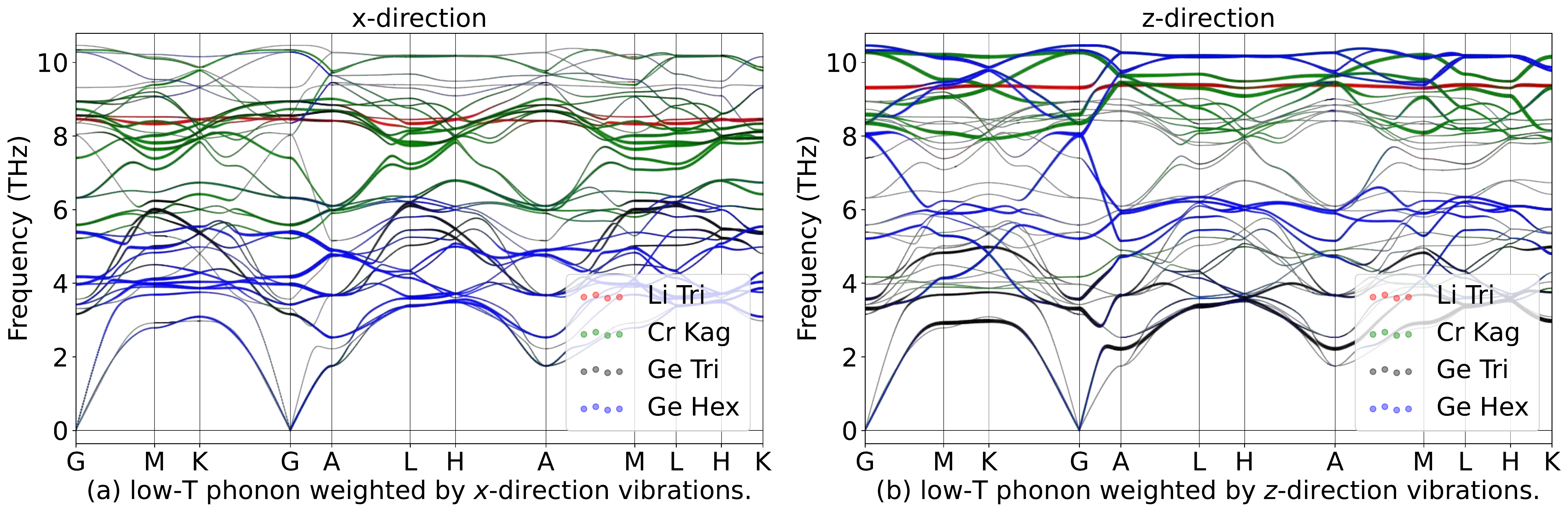}
\hspace{5pt}\label{fig:LiCr6Ge6_relaxed_High_xz}
\includegraphics[width=0.85\textwidth]{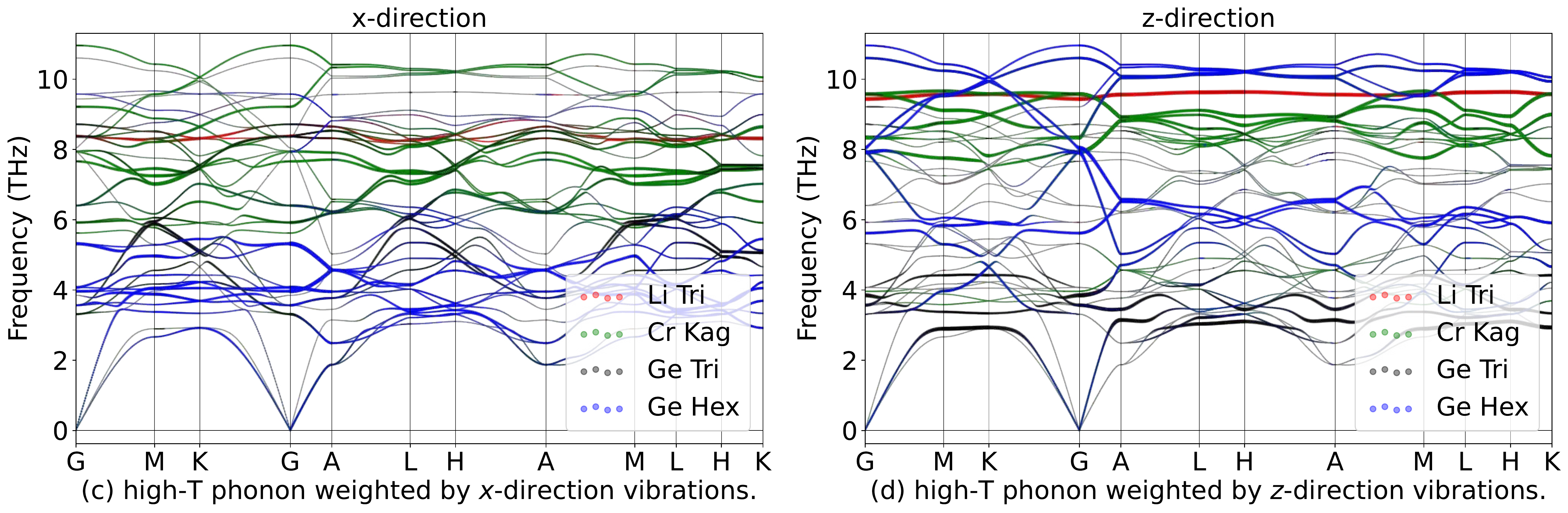}
\caption{Phonon spectrum for LiCr$_6$Ge$_6$in in-plane ($x$) and out-of-plane ($z$) directions.}
\label{fig:LiCr6Ge6phoxyz}
\end{figure}

\begin{table}[htbp]
\global\long\def\arraystretch{1.12}
\captionsetup{width=.95\textwidth}
\caption{IRREPs at high-symmetry points for LiCr$_6$Ge$_6$ phonon spectrum at Low-T.}
\label{tab:191_LiCr6Ge6_Low}
\setlength{\tabcolsep}{1.mm}{
}
\end{table}

\begin{table}[htbp]
\global\long\def\arraystretch{1.12}
\captionsetup{width=.95\textwidth}
\caption{IRREPs at high-symmetry points for LiCr$_6$Ge$_6$ phonon spectrum at High-T.}
\label{tab:191_LiCr6Ge6_High}
\setlength{\tabcolsep}{1.mm}{
%
}
\end{table}
\subsubsection{\label{sms2sec:MgCr6Ge6}\hyperref[smsec:col]{\texorpdfstring{MgCr$_6$Ge$_6$}{}}~{\color{green}  (High-T stable, Low-T stable) }}

\begin{table}[htbp]
\global\long\def\arraystretch{1.12}
\captionsetup{width=.85\textwidth}
\caption{Structure parameters of MgCr$_6$Ge$_6$ after relaxation. It contains 348 electrons. }
\label{tab:MgCr6Ge6}
\setlength{\tabcolsep}{4mm}{
%
}
\end{table}

\begin{figure}[htbp]
\centering
\includegraphics[width=0.85\textwidth]{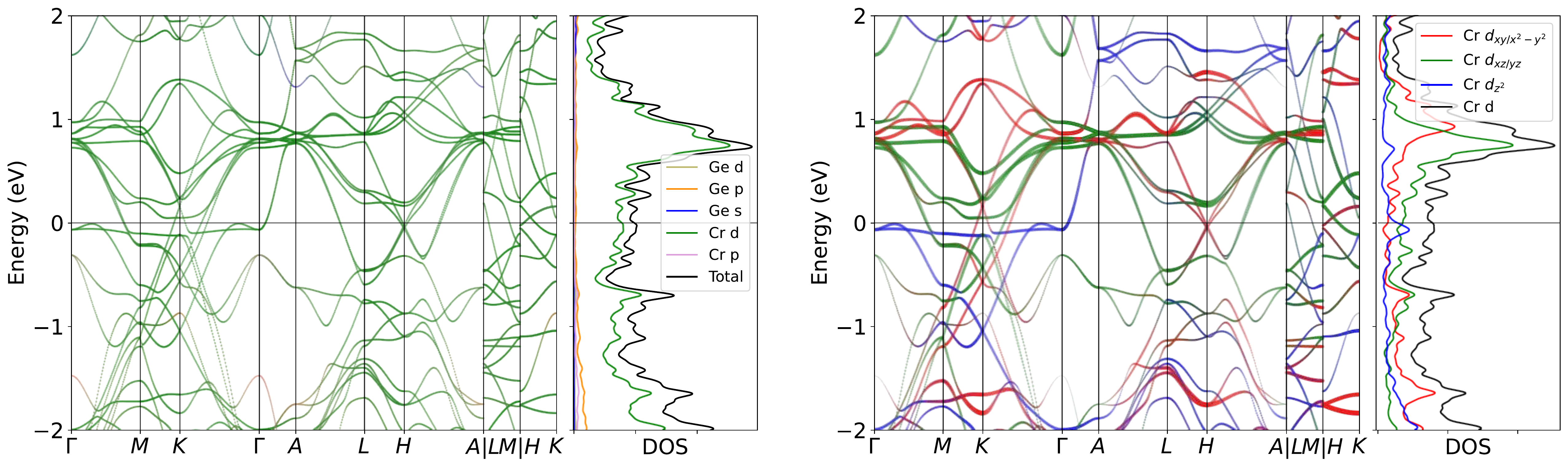}
\caption{Band structure for MgCr$_6$Ge$_6$ without SOC. The projection of Cr $3d$ orbitals is also presented. The band structures clearly reveal the dominating of Cr $3d$ orbitals  near the Fermi level, as well as the pronounced peak.}
\label{fig:MgCr6Ge6band}
\end{figure}

\begin{figure}[H]
\centering
\subfloat[Relaxed phonon at low-T.]{
\label{fig:MgCr6Ge6_relaxed_Low}
\includegraphics[width=0.45\textwidth]{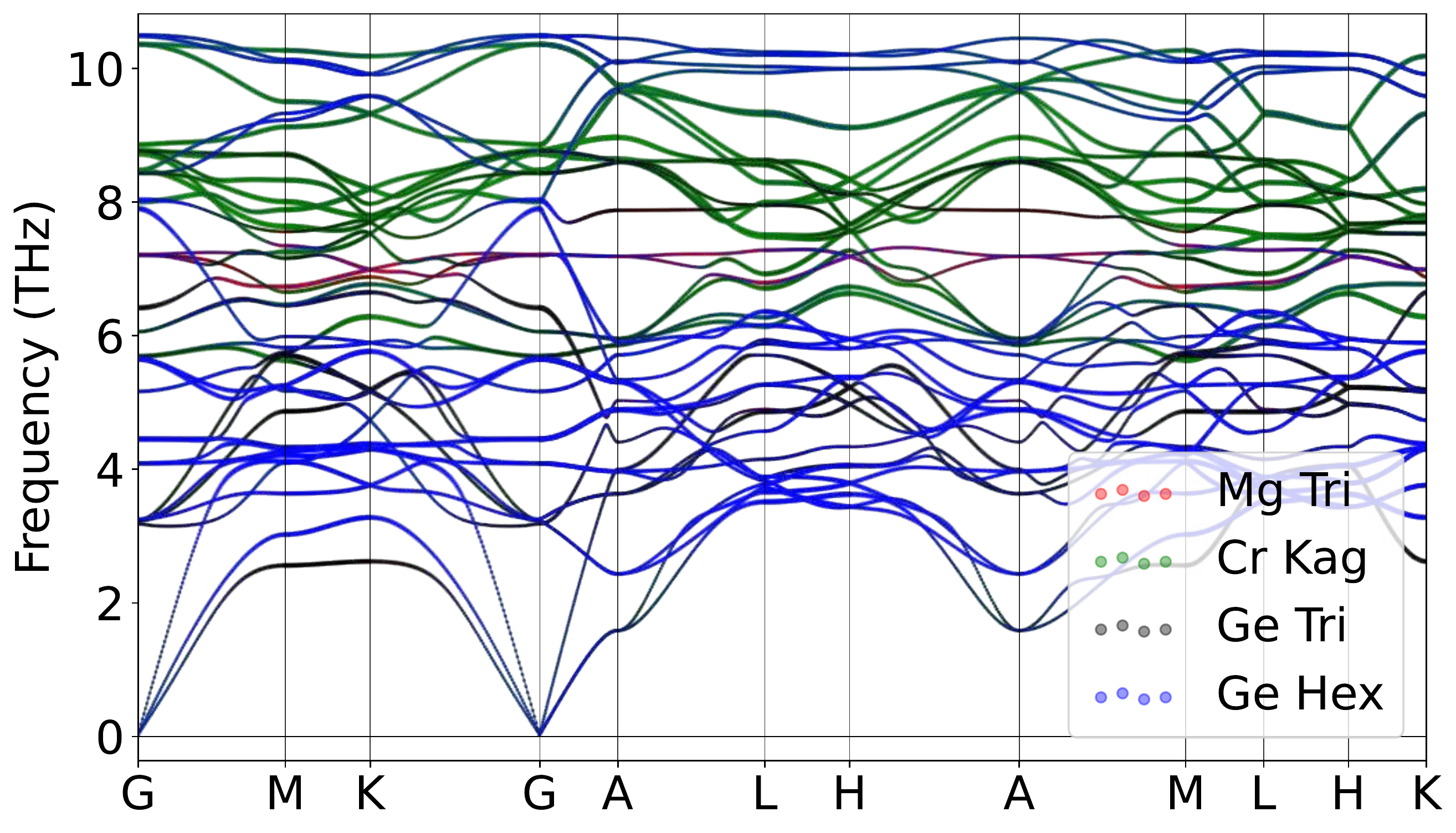}
}
\hspace{5pt}\subfloat[Relaxed phonon at high-T.]{
\label{fig:MgCr6Ge6_relaxed_High}
\includegraphics[width=0.45\textwidth]{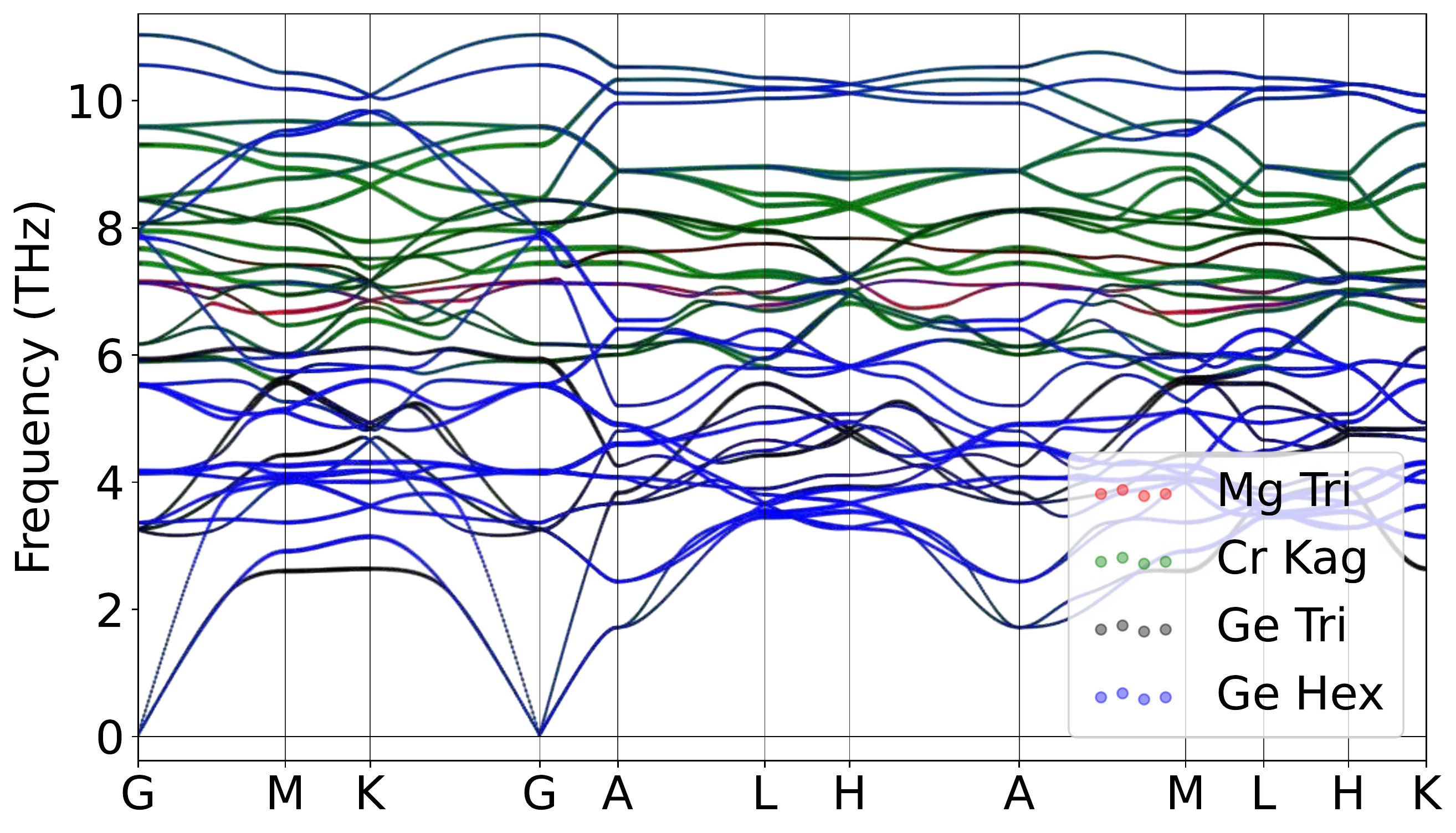}
}
\caption{Atomically projected phonon spectrum for MgCr$_6$Ge$_6$.}
\label{fig:MgCr6Ge6pho}
\end{figure}

\begin{figure}[H]
\centering
\label{fig:MgCr6Ge6_relaxed_Low_xz}
\includegraphics[width=0.85\textwidth]{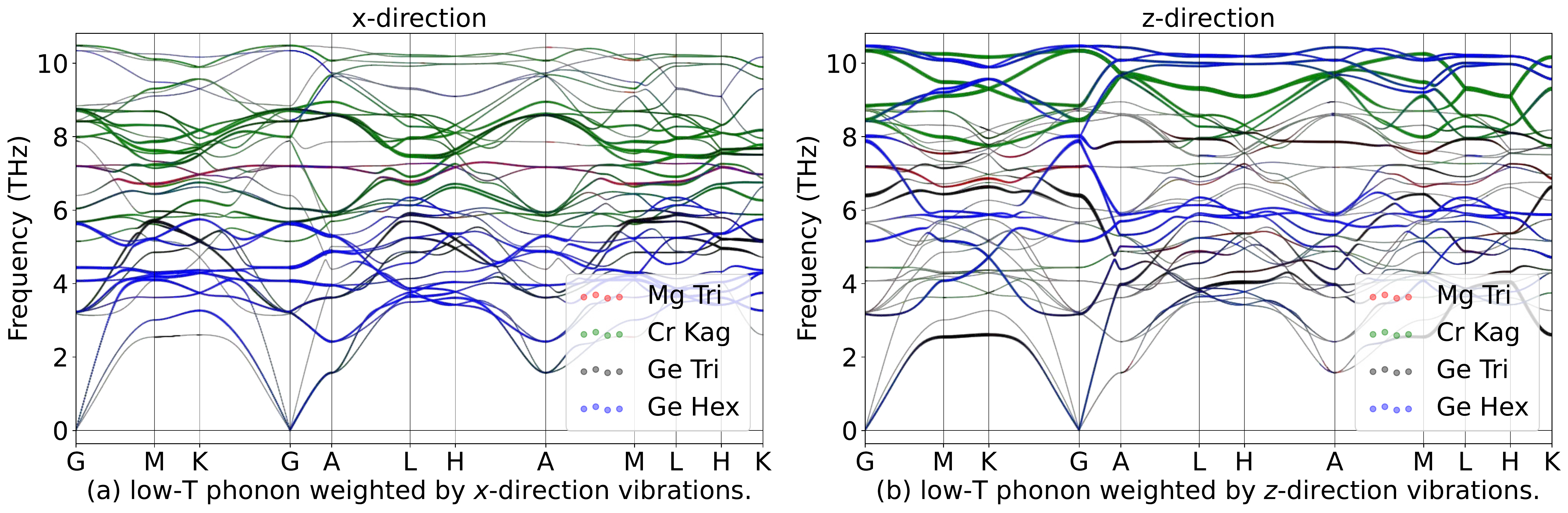}
\hspace{5pt}\label{fig:MgCr6Ge6_relaxed_High_xz}
\includegraphics[width=0.85\textwidth]{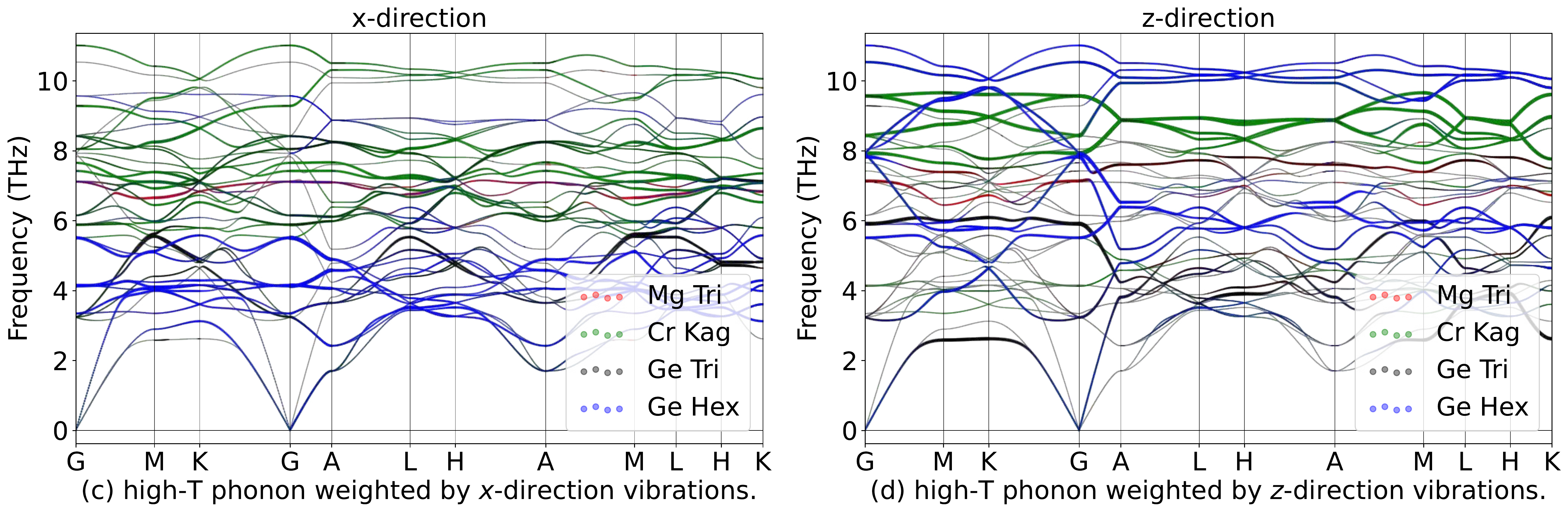}
\caption{Phonon spectrum for MgCr$_6$Ge$_6$in in-plane ($x$) and out-of-plane ($z$) directions.}
\label{fig:MgCr6Ge6phoxyz}
\end{figure}

\begin{table}[htbp]
\global\long\def\arraystretch{1.12}
\captionsetup{width=.95\textwidth}
\caption{IRREPs at high-symmetry points for MgCr$_6$Ge$_6$ phonon spectrum at Low-T.}
\label{tab:191_MgCr6Ge6_Low}
\setlength{\tabcolsep}{1.mm}{
}
\end{table}

\begin{table}[htbp]
\global\long\def\arraystretch{1.12}
\captionsetup{width=.95\textwidth}
\caption{IRREPs at high-symmetry points for MgCr$_6$Ge$_6$ phonon spectrum at High-T.}
\label{tab:191_MgCr6Ge6_High}
\setlength{\tabcolsep}{1.mm}{
%
}
\end{table}
\subsubsection{\label{sms2sec:CaCr6Ge6}\hyperref[smsec:col]{\texorpdfstring{CaCr$_6$Ge$_6$}{}}~{\color{green}  (High-T stable, Low-T stable) }}

\begin{table}[htbp]
\global\long\def\arraystretch{1.12}
\captionsetup{width=.85\textwidth}
\caption{Structure parameters of CaCr$_6$Ge$_6$ after relaxation. It contains 356 electrons. }
\label{tab:CaCr6Ge6}
\setlength{\tabcolsep}{4mm}{
%
}
\end{table}

\begin{figure}[htbp]
\centering
\includegraphics[width=0.85\textwidth]{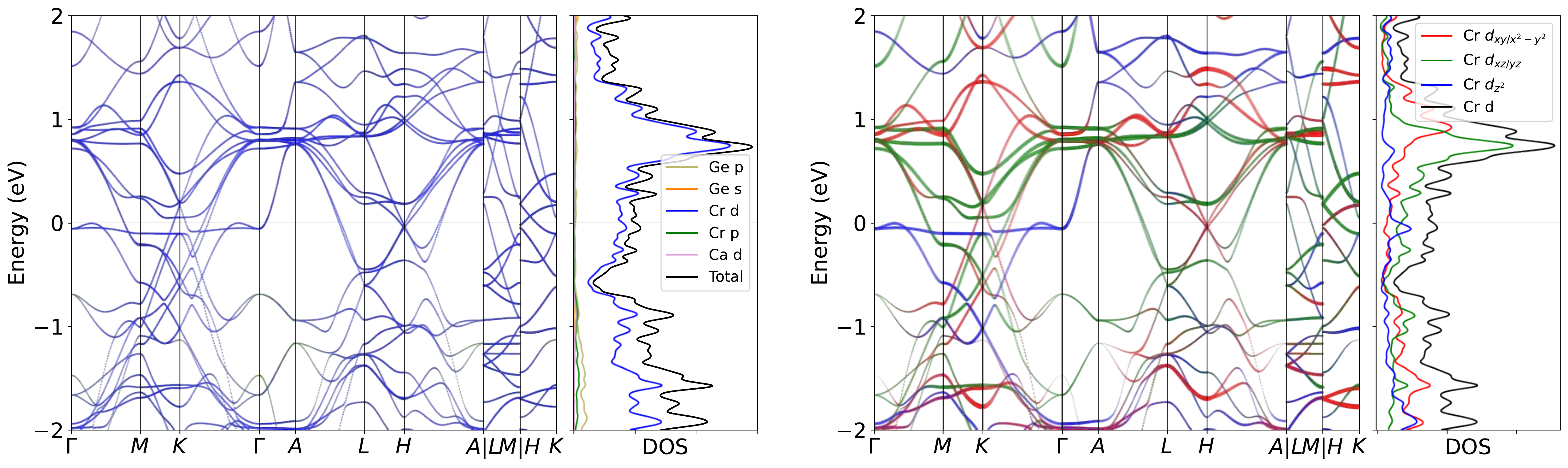}
\caption{Band structure for CaCr$_6$Ge$_6$ without SOC. The projection of Cr $3d$ orbitals is also presented. The band structures clearly reveal the dominating of Cr $3d$ orbitals  near the Fermi level, as well as the pronounced peak.}
\label{fig:CaCr6Ge6band}
\end{figure}

\begin{figure}[H]
\centering
\subfloat[Relaxed phonon at low-T.]{
\label{fig:CaCr6Ge6_relaxed_Low}
\includegraphics[width=0.45\textwidth]{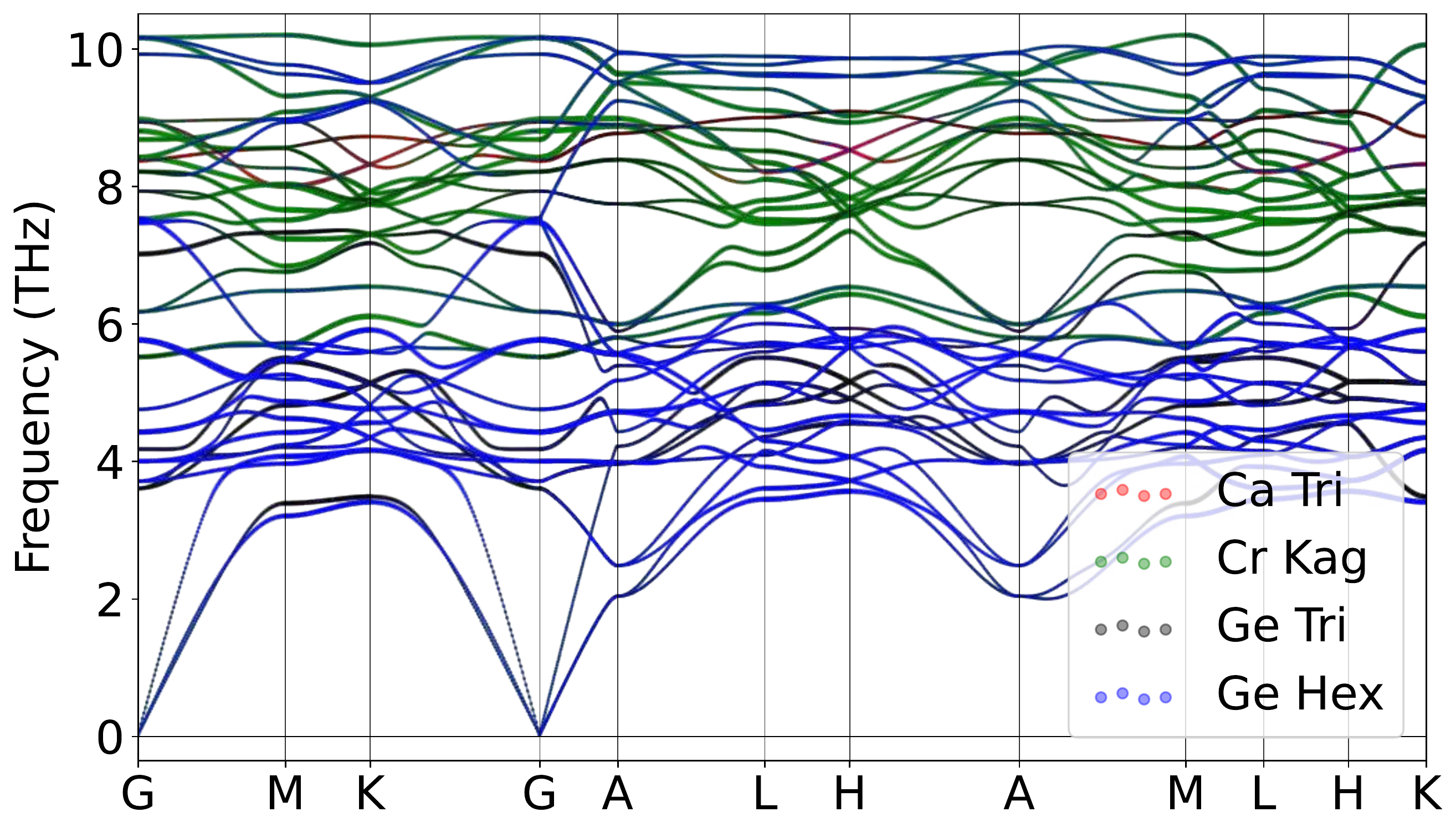}
}
\hspace{5pt}\subfloat[Relaxed phonon at high-T.]{
\label{fig:CaCr6Ge6_relaxed_High}
\includegraphics[width=0.45\textwidth]{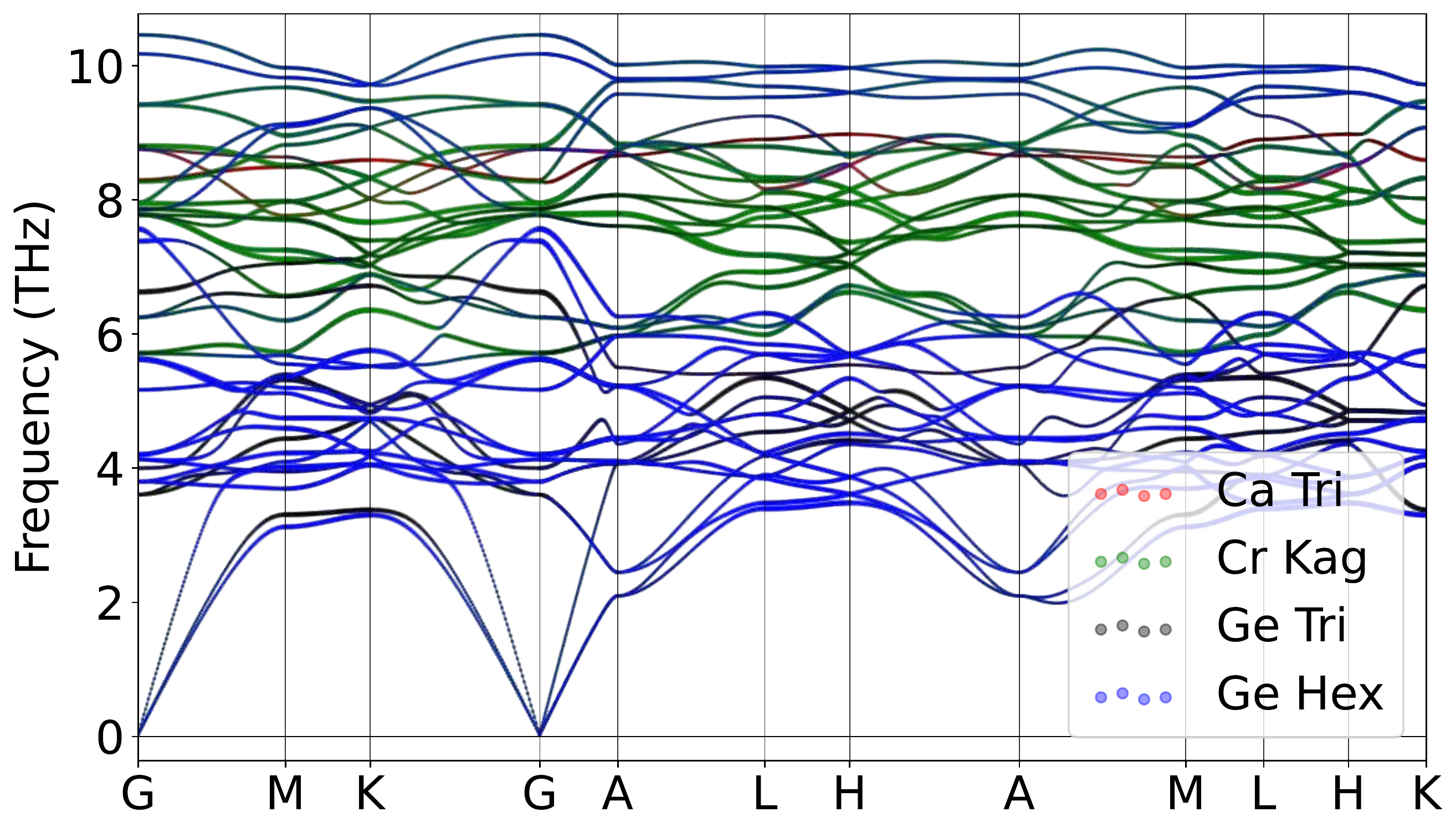}
}
\caption{Atomically projected phonon spectrum for CaCr$_6$Ge$_6$.}
\label{fig:CaCr6Ge6pho}
\end{figure}

\begin{figure}[H]
\centering
\label{fig:CaCr6Ge6_relaxed_Low_xz}
\includegraphics[width=0.85\textwidth]{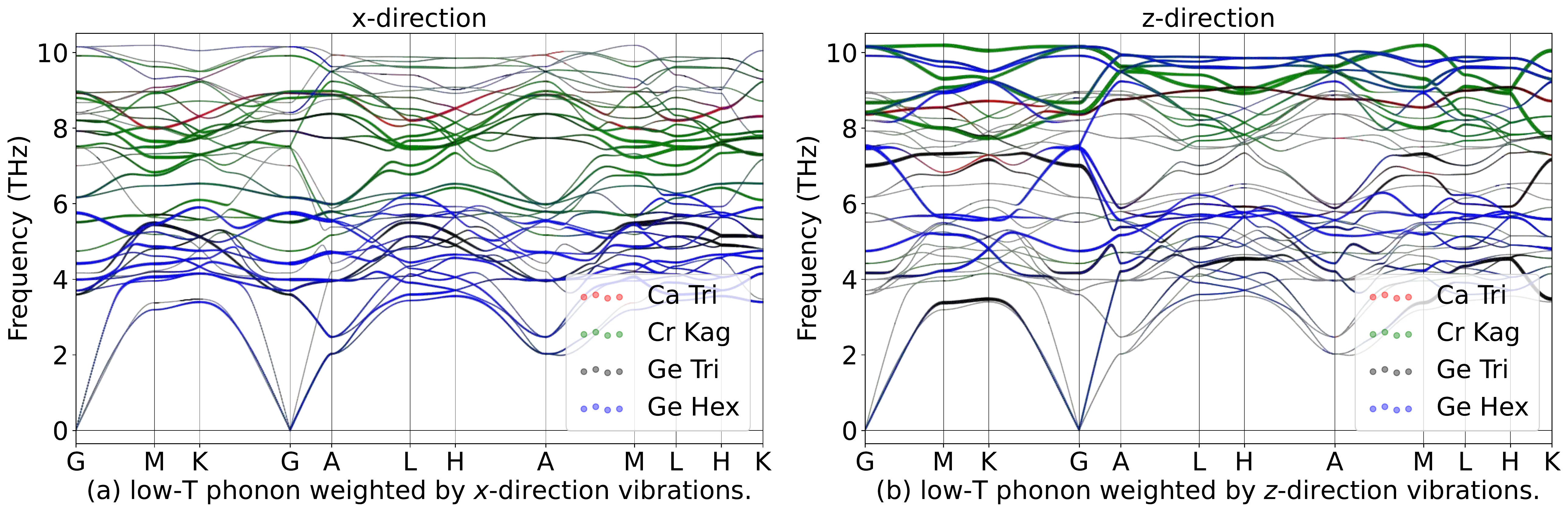}
\hspace{5pt}\label{fig:CaCr6Ge6_relaxed_High_xz}
\includegraphics[width=0.85\textwidth]{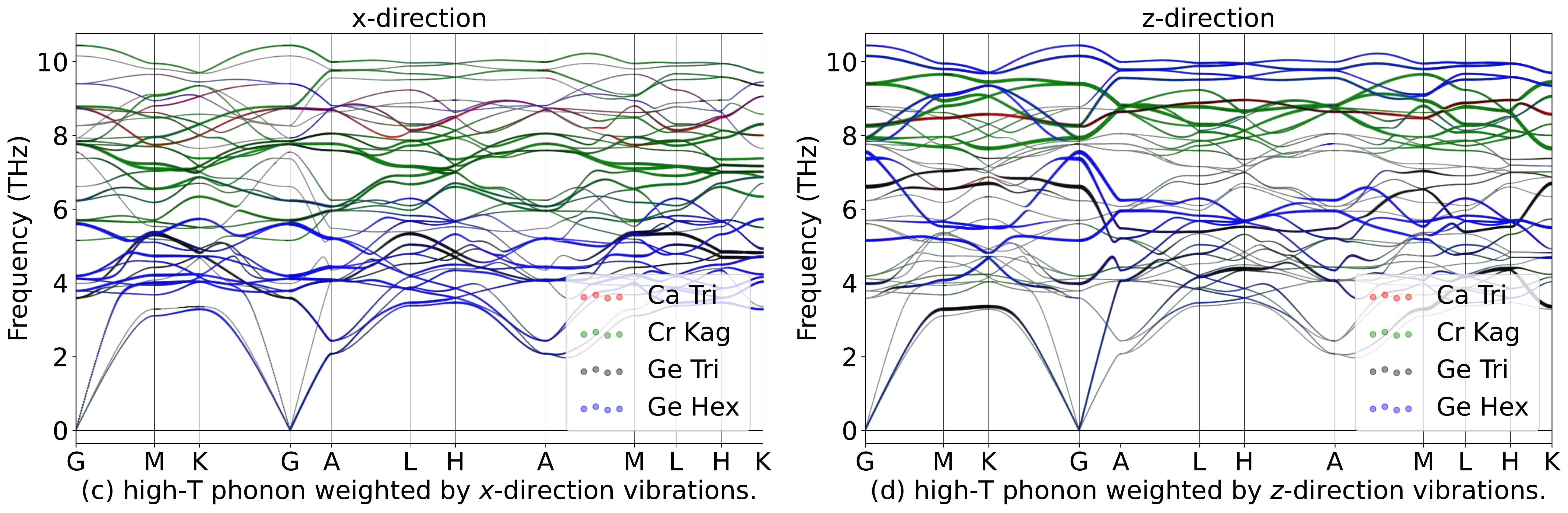}
\caption{Phonon spectrum for CaCr$_6$Ge$_6$in in-plane ($x$) and out-of-plane ($z$) directions.}
\label{fig:CaCr6Ge6phoxyz}
\end{figure}

\begin{table}[htbp]
\global\long\def\arraystretch{1.12}
\captionsetup{width=.95\textwidth}
\caption{IRREPs at high-symmetry points for CaCr$_6$Ge$_6$ phonon spectrum at Low-T.}
\label{tab:191_CaCr6Ge6_Low}
\setlength{\tabcolsep}{1.mm}{
}
\end{table}

\begin{table}[htbp]
\global\long\def\arraystretch{1.12}
\captionsetup{width=.95\textwidth}
\caption{IRREPs at high-symmetry points for CaCr$_6$Ge$_6$ phonon spectrum at High-T.}
\label{tab:191_CaCr6Ge6_High}
\setlength{\tabcolsep}{1.mm}{
%
}
\end{table}
\subsubsection{\label{sms2sec:ScCr6Ge6}\hyperref[smsec:col]{\texorpdfstring{ScCr$_6$Ge$_6$}{}}~{\color{green}  (High-T stable, Low-T stable) }}

\begin{table}[htbp]
\global\long\def\arraystretch{1.12}
\captionsetup{width=.85\textwidth}
\caption{Structure parameters of ScCr$_6$Ge$_6$ after relaxation. It contains 357 electrons. }
\label{tab:ScCr6Ge6}
\setlength{\tabcolsep}{4mm}{
%
}
\end{table}

\begin{figure}[htbp]
\centering
\includegraphics[width=0.85\textwidth]{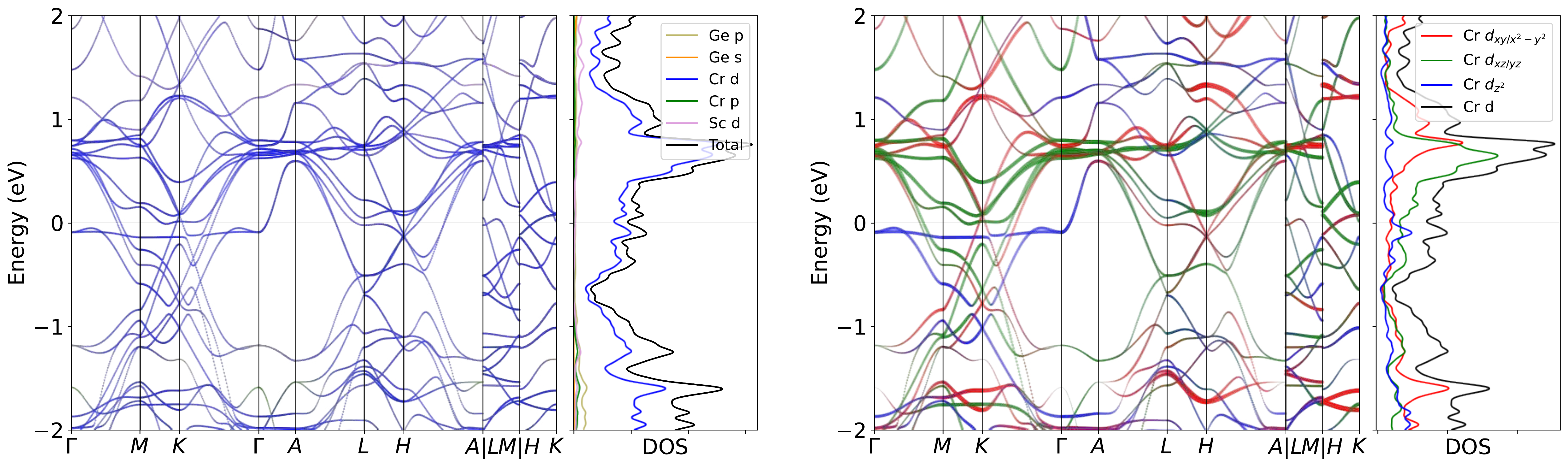}
\caption{Band structure for ScCr$_6$Ge$_6$ without SOC. The projection of Cr $3d$ orbitals is also presented. The band structures clearly reveal the dominating of Cr $3d$ orbitals  near the Fermi level, as well as the pronounced peak.}
\label{fig:ScCr6Ge6band}
\end{figure}

\begin{figure}[H]
\centering
\subfloat[Relaxed phonon at low-T.]{
\label{fig:ScCr6Ge6_relaxed_Low}
\includegraphics[width=0.45\textwidth]{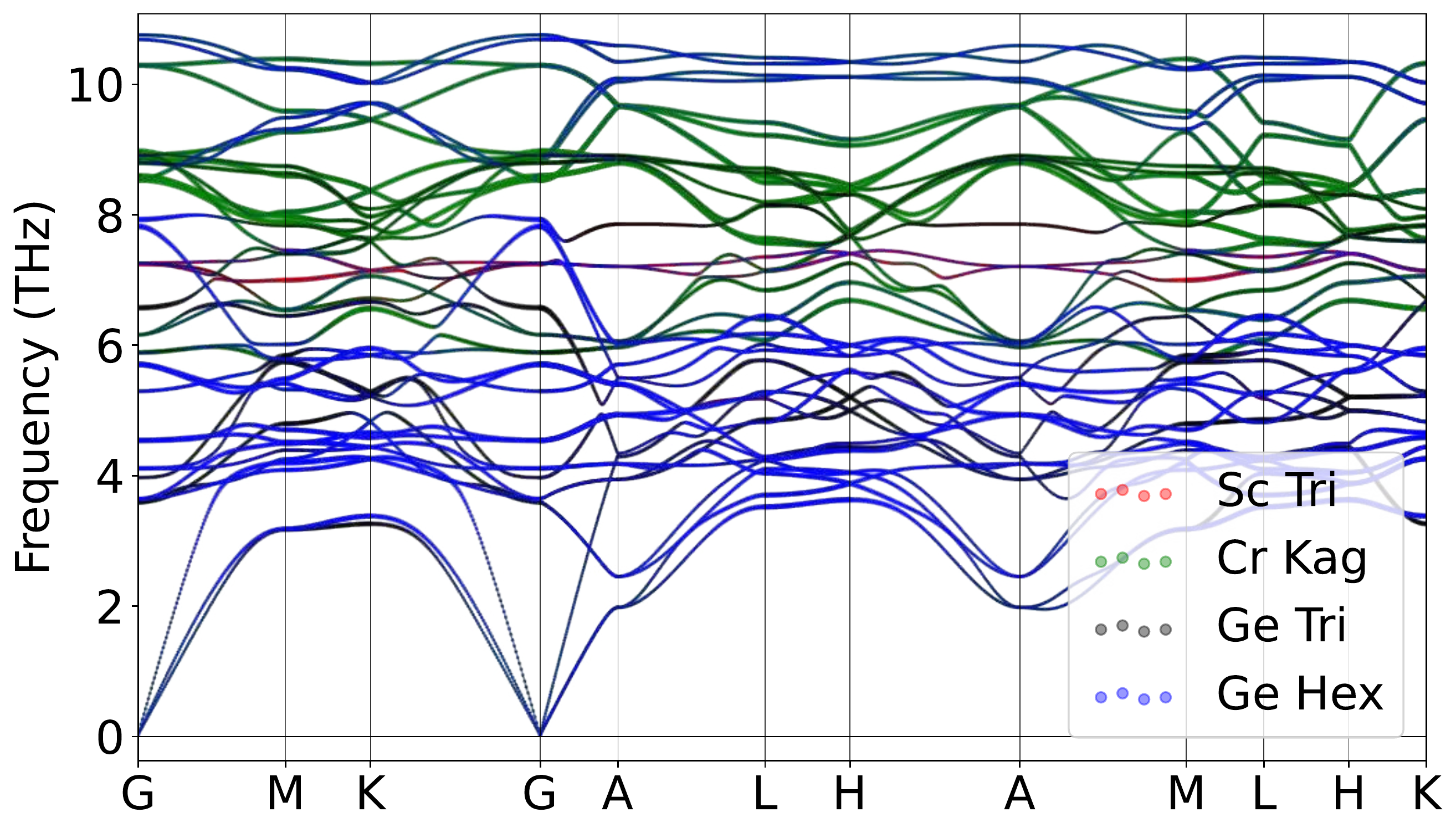}
}
\hspace{5pt}\subfloat[Relaxed phonon at high-T.]{
\label{fig:ScCr6Ge6_relaxed_High}
\includegraphics[width=0.45\textwidth]{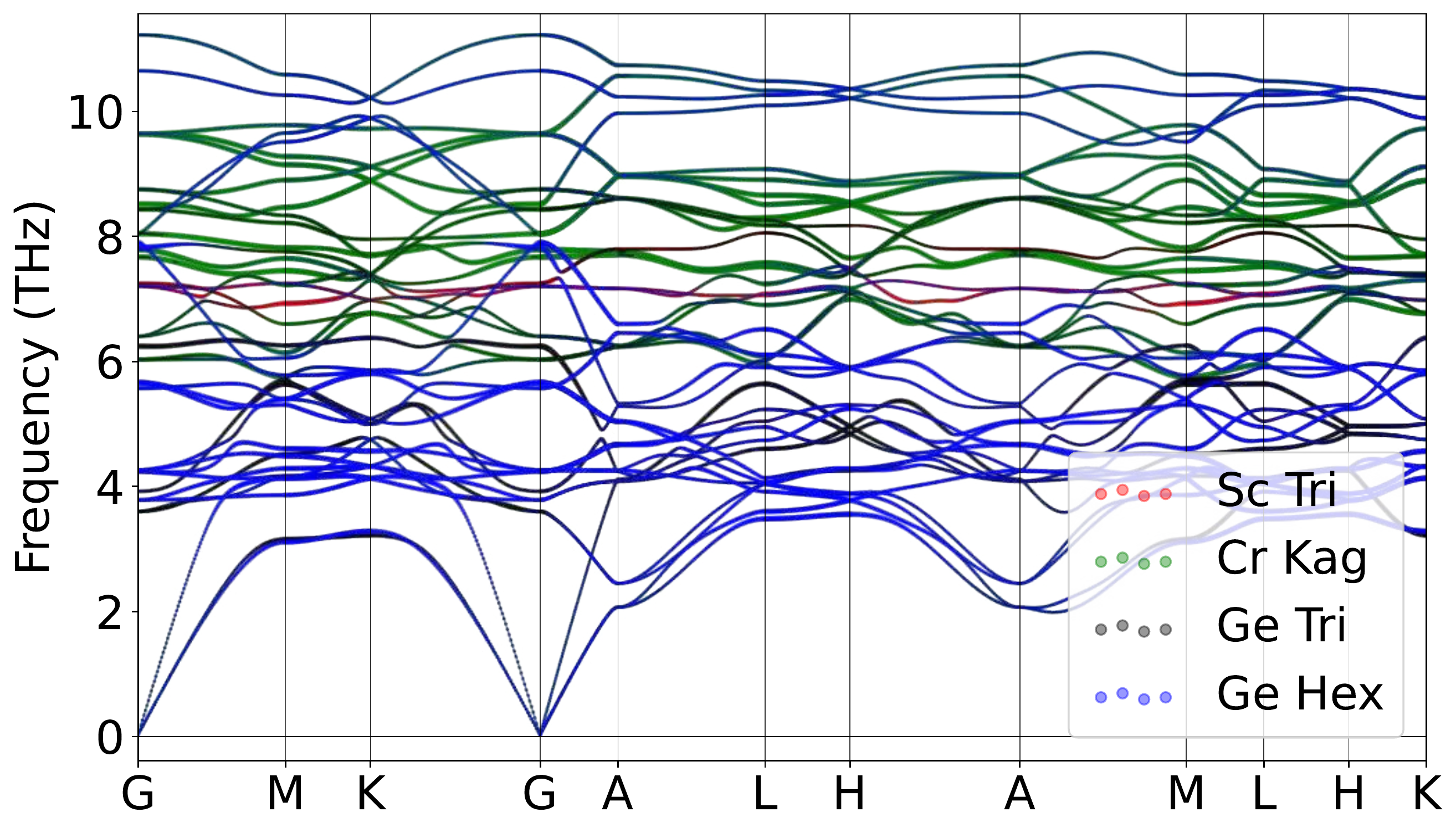}
}
\caption{Atomically projected phonon spectrum for ScCr$_6$Ge$_6$.}
\label{fig:ScCr6Ge6pho}
\end{figure}

\begin{figure}[H]
\centering
\label{fig:ScCr6Ge6_relaxed_Low_xz}
\includegraphics[width=0.85\textwidth]{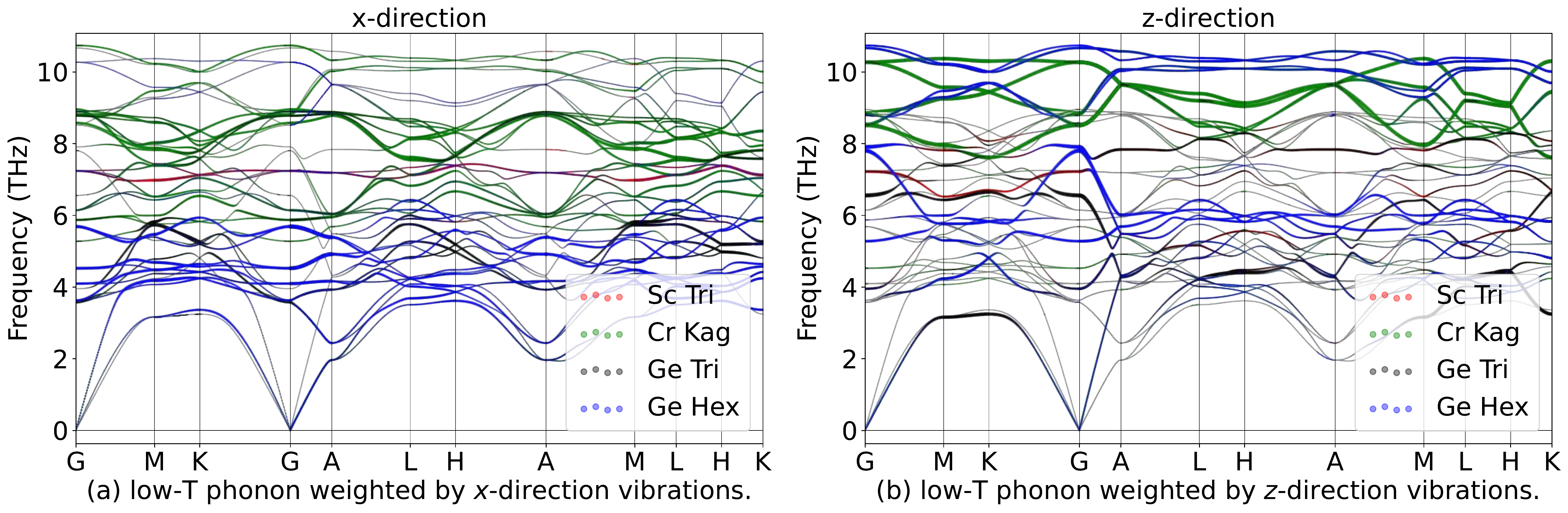}
\hspace{5pt}\label{fig:ScCr6Ge6_relaxed_High_xz}
\includegraphics[width=0.85\textwidth]{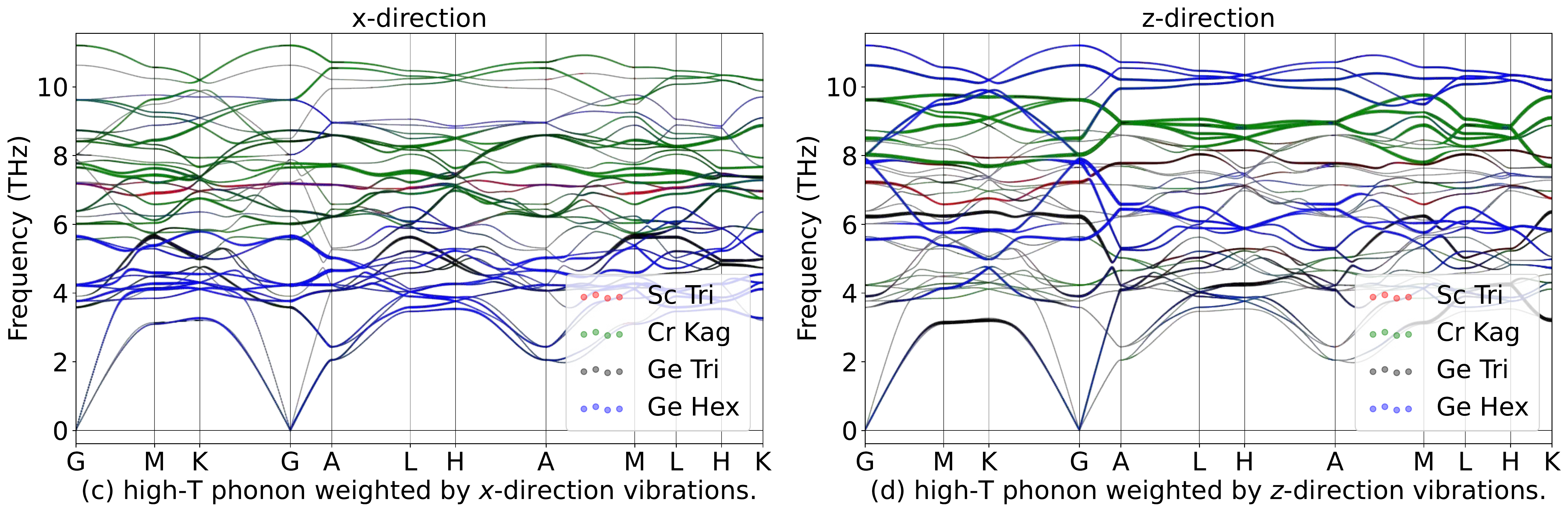}
\caption{Phonon spectrum for ScCr$_6$Ge$_6$in in-plane ($x$) and out-of-plane ($z$) directions.}
\label{fig:ScCr6Ge6phoxyz}
\end{figure}

\begin{table}[htbp]
\global\long\def\arraystretch{1.12}
\captionsetup{width=.95\textwidth}
\caption{IRREPs at high-symmetry points for ScCr$_6$Ge$_6$ phonon spectrum at Low-T.}
\label{tab:191_ScCr6Ge6_Low}
\setlength{\tabcolsep}{1.mm}{
}
\end{table}

\begin{table}[htbp]
\global\long\def\arraystretch{1.12}
\captionsetup{width=.95\textwidth}
\caption{IRREPs at high-symmetry points for ScCr$_6$Ge$_6$ phonon spectrum at High-T.}
\label{tab:191_ScCr6Ge6_High}
\setlength{\tabcolsep}{1.mm}{
%
}
\end{table}
\subsubsection{\label{sms2sec:TiCr6Ge6}\hyperref[smsec:col]{\texorpdfstring{TiCr$_6$Ge$_6$}{}}~{\color{green}  (High-T stable, Low-T stable) }}

\begin{table}[htbp]
\global\long\def\arraystretch{1.12}
\captionsetup{width=.85\textwidth}
\caption{Structure parameters of TiCr$_6$Ge$_6$ after relaxation. It contains 358 electrons. }
\label{tab:TiCr6Ge6}
\setlength{\tabcolsep}{4mm}{
%
}
\end{table}

\begin{figure}[htbp]
\centering
\includegraphics[width=0.85\textwidth]{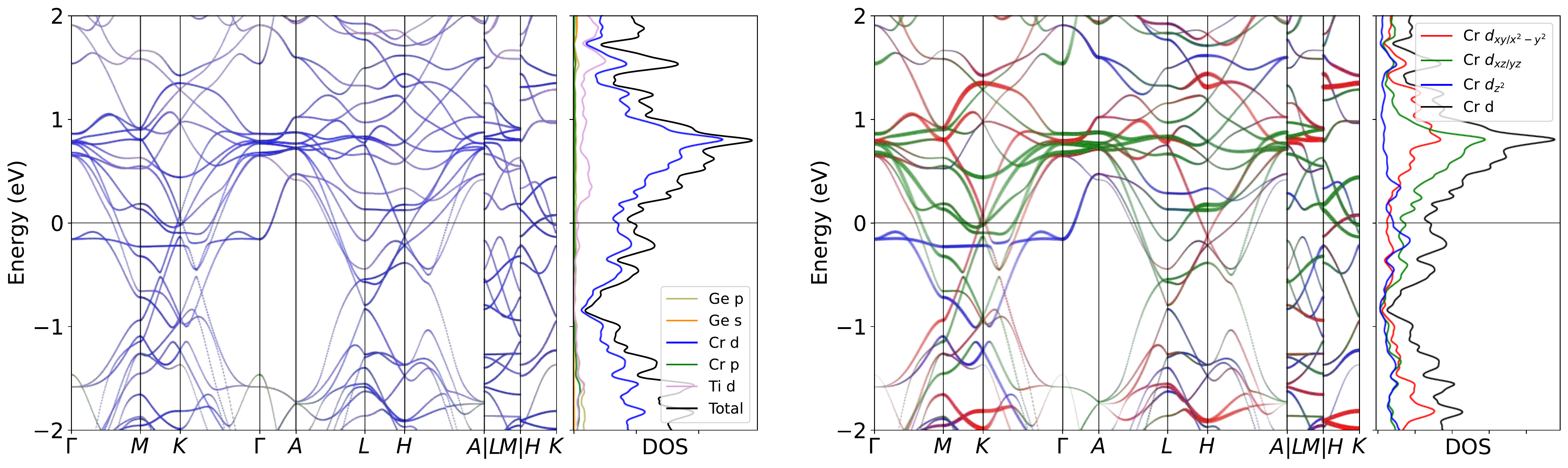}
\caption{Band structure for TiCr$_6$Ge$_6$ without SOC. The projection of Cr $3d$ orbitals is also presented. The band structures clearly reveal the dominating of Cr $3d$ orbitals  near the Fermi level, as well as the pronounced peak.}
\label{fig:TiCr6Ge6band}
\end{figure}

\begin{figure}[H]
\centering
\subfloat[Relaxed phonon at low-T.]{
\label{fig:TiCr6Ge6_relaxed_Low}
\includegraphics[width=0.45\textwidth]{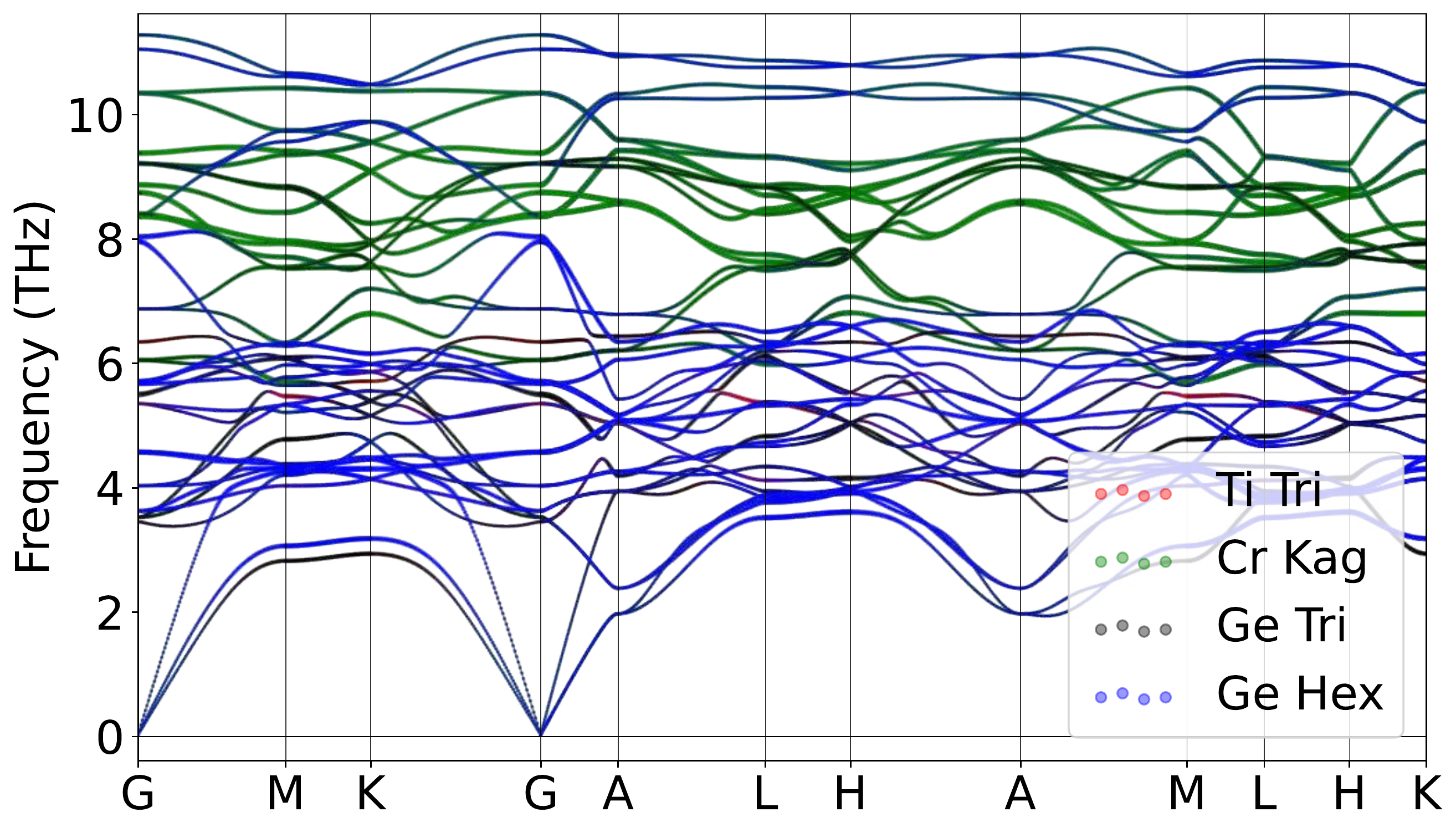}
}
\hspace{5pt}\subfloat[Relaxed phonon at high-T.]{
\label{fig:TiCr6Ge6_relaxed_High}
\includegraphics[width=0.45\textwidth]{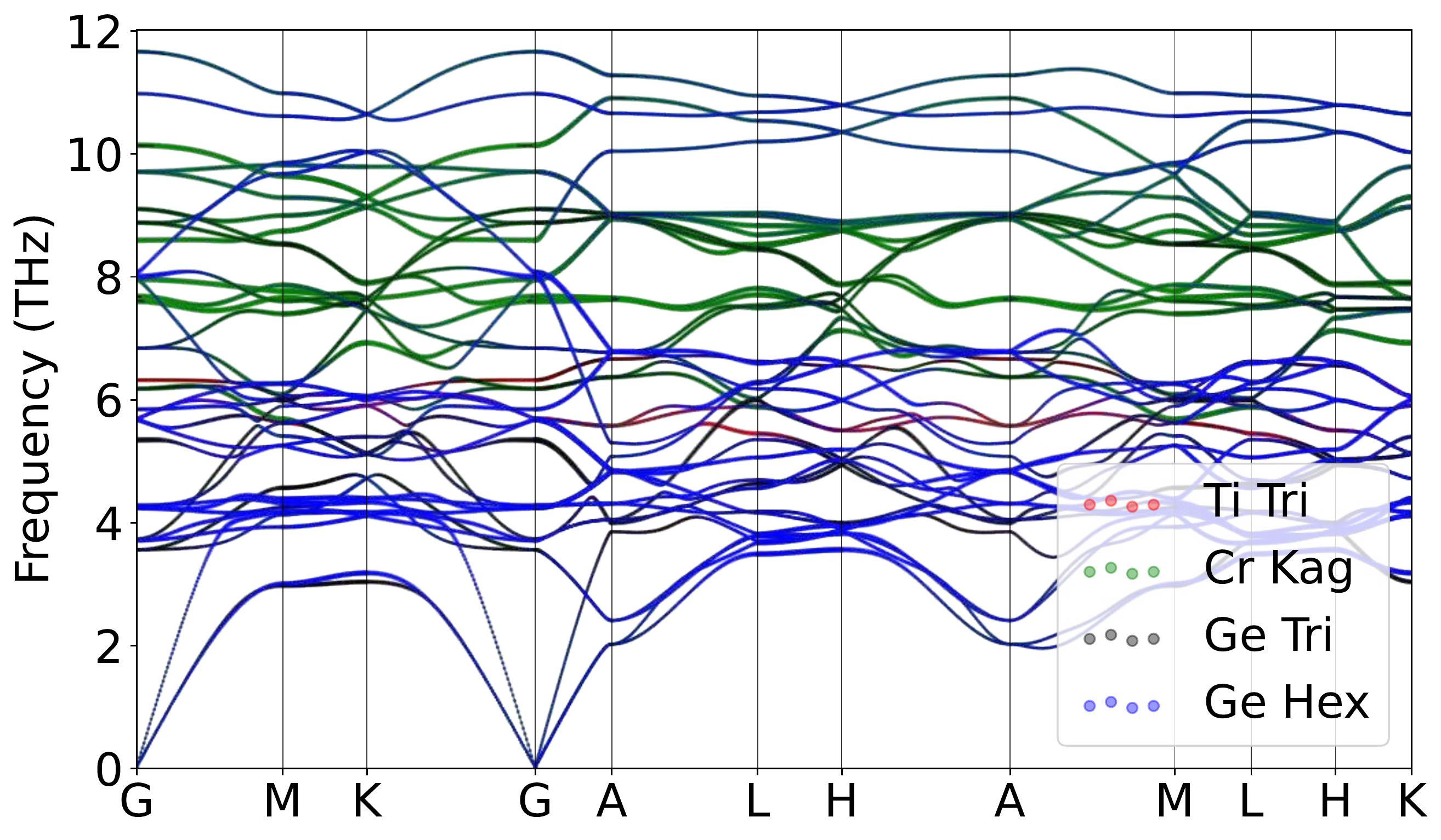}
}
\caption{Atomically projected phonon spectrum for TiCr$_6$Ge$_6$.}
\label{fig:TiCr6Ge6pho}
\end{figure}

\begin{figure}[H]
\centering
\label{fig:TiCr6Ge6_relaxed_Low_xz}
\includegraphics[width=0.85\textwidth]{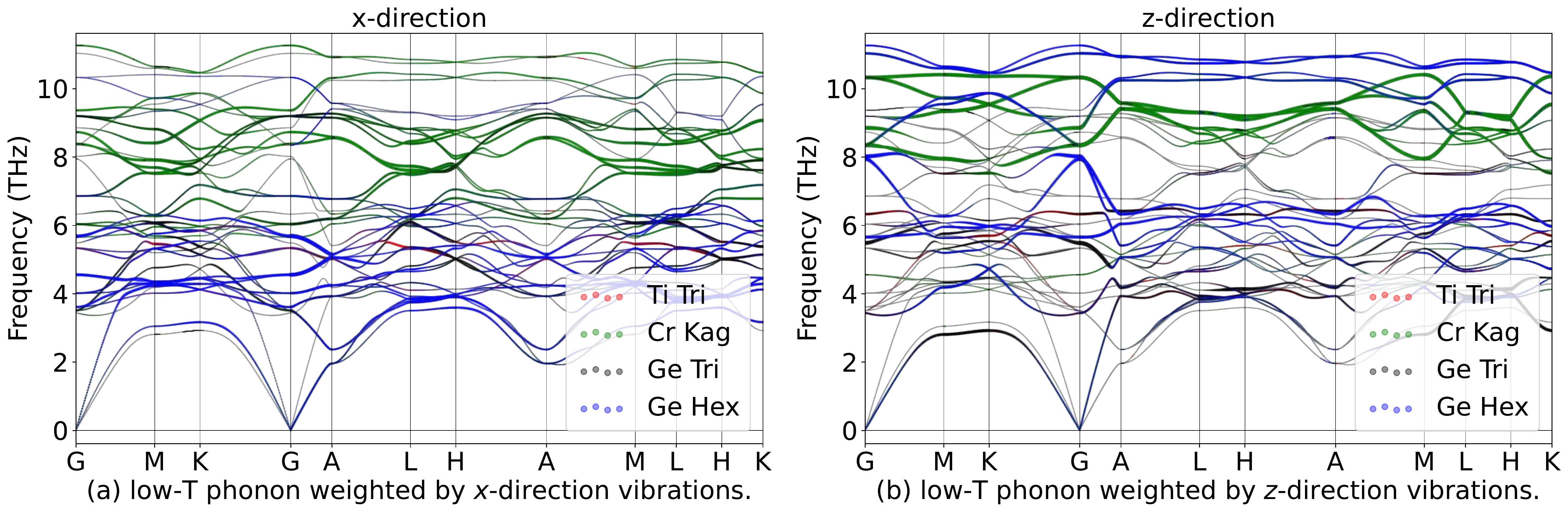}
\hspace{5pt}\label{fig:TiCr6Ge6_relaxed_High_xz}
\includegraphics[width=0.85\textwidth]{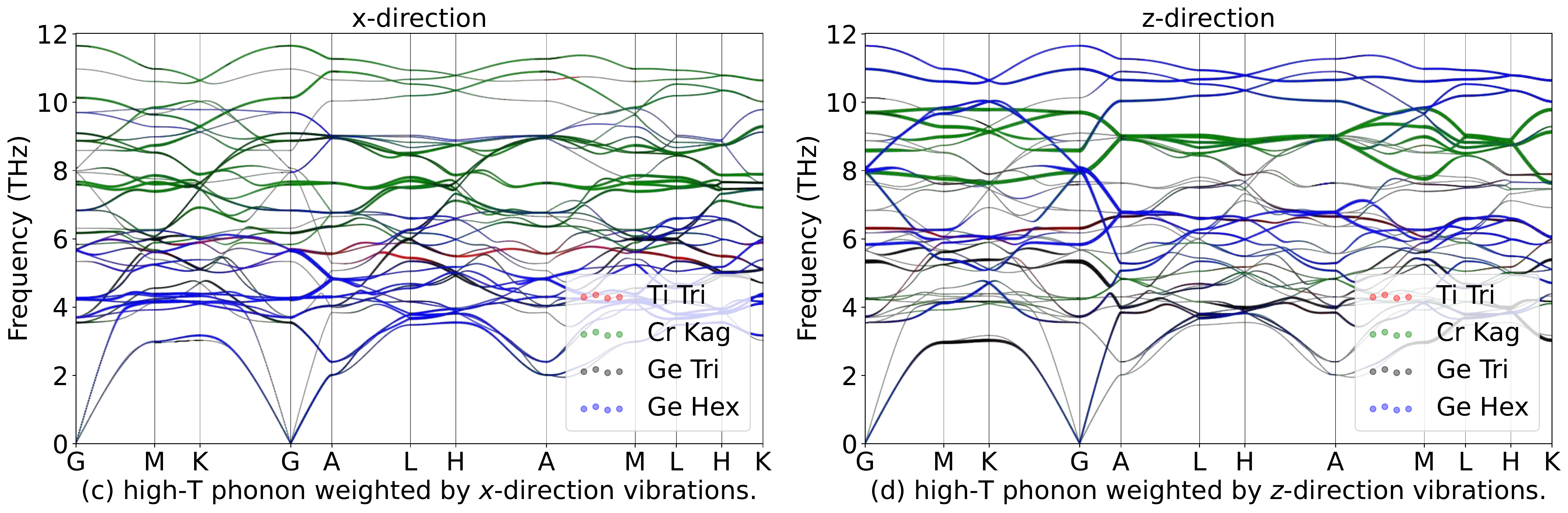}
\caption{Phonon spectrum for TiCr$_6$Ge$_6$in in-plane ($x$) and out-of-plane ($z$) directions.}
\label{fig:TiCr6Ge6phoxyz}
\end{figure}

\begin{table}[htbp]
\global\long\def\arraystretch{1.12}
\captionsetup{width=.95\textwidth}
\caption{IRREPs at high-symmetry points for TiCr$_6$Ge$_6$ phonon spectrum at Low-T.}
\label{tab:191_TiCr6Ge6_Low}
\setlength{\tabcolsep}{1.mm}{
}
\end{table}

\begin{table}[htbp]
\global\long\def\arraystretch{1.12}
\captionsetup{width=.95\textwidth}
\caption{IRREPs at high-symmetry points for TiCr$_6$Ge$_6$ phonon spectrum at High-T.}
\label{tab:191_TiCr6Ge6_High}
\setlength{\tabcolsep}{1.mm}{
%
}
\end{table}
\subsubsection{\label{sms2sec:YCr6Ge6}\hyperref[smsec:col]{\texorpdfstring{YCr$_6$Ge$_6$}{}}~{\color{green}  (High-T stable, Low-T stable) }}

\begin{table}[htbp]
\global\long\def\arraystretch{1.12}
\captionsetup{width=.85\textwidth}
\caption{Structure parameters of YCr$_6$Ge$_6$ after relaxation. It contains 375 electrons. }
\label{tab:YCr6Ge6}
\setlength{\tabcolsep}{4mm}{
%
}
\end{table}

\begin{figure}[htbp]
\centering
\includegraphics[width=0.85\textwidth]{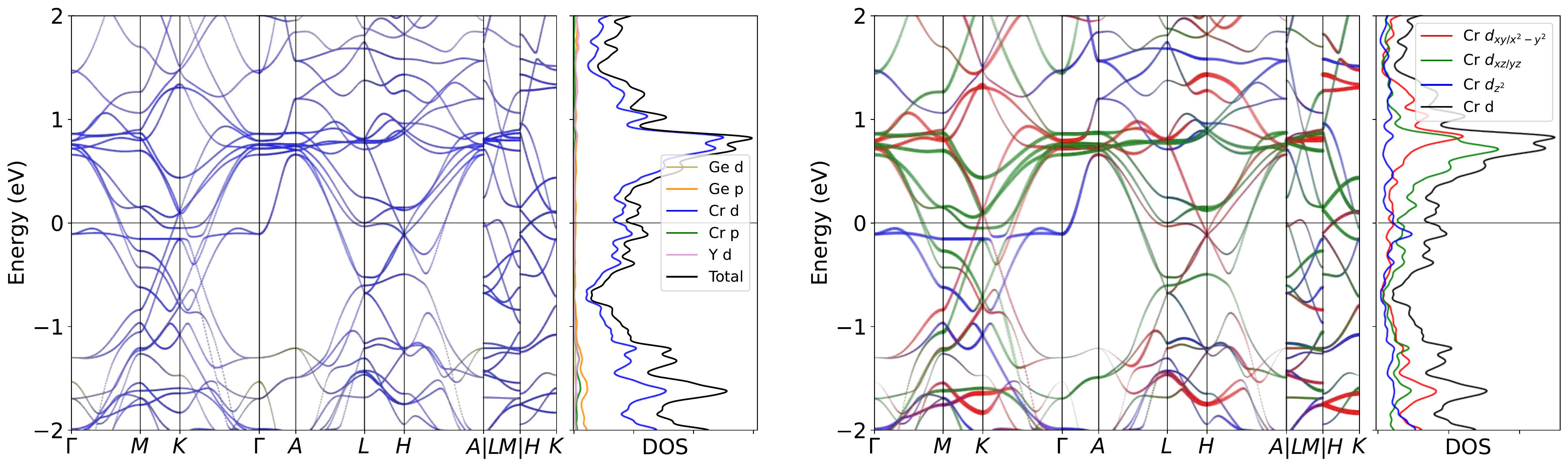}
\caption{Band structure for YCr$_6$Ge$_6$ without SOC. The projection of Cr $3d$ orbitals is also presented. The band structures clearly reveal the dominating of Cr $3d$ orbitals  near the Fermi level, as well as the pronounced peak.}
\label{fig:YCr6Ge6band}
\end{figure}

\begin{figure}[H]
\centering
\subfloat[Relaxed phonon at low-T.]{
\label{fig:YCr6Ge6_relaxed_Low}
\includegraphics[width=0.45\textwidth]{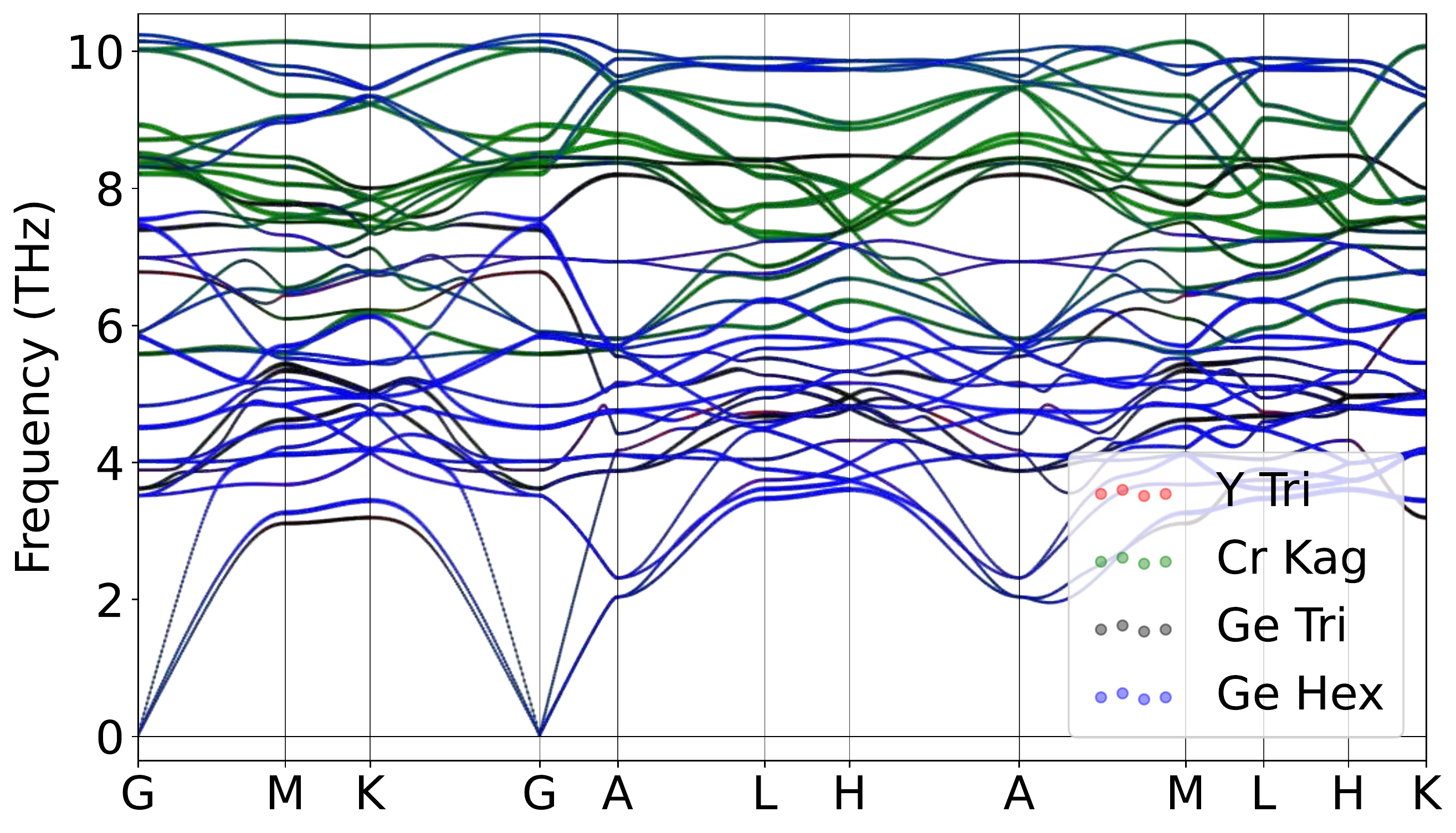}
}
\hspace{5pt}\subfloat[Relaxed phonon at high-T.]{
\label{fig:YCr6Ge6_relaxed_High}
\includegraphics[width=0.45\textwidth]{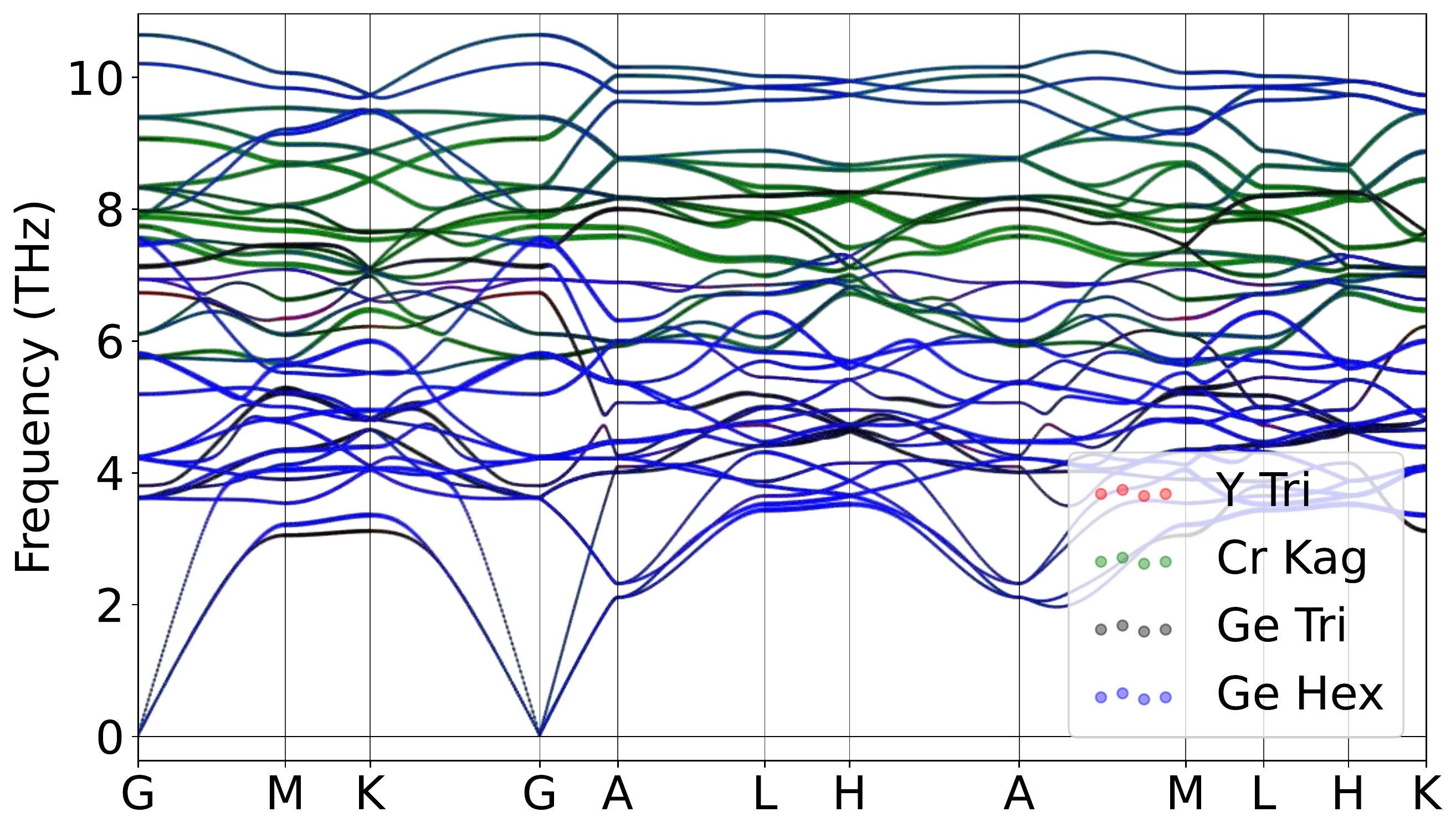}
}
\caption{Atomically projected phonon spectrum for YCr$_6$Ge$_6$.}
\label{fig:YCr6Ge6pho}
\end{figure}

\begin{figure}[H]
\centering
\label{fig:YCr6Ge6_relaxed_Low_xz}
\includegraphics[width=0.85\textwidth]{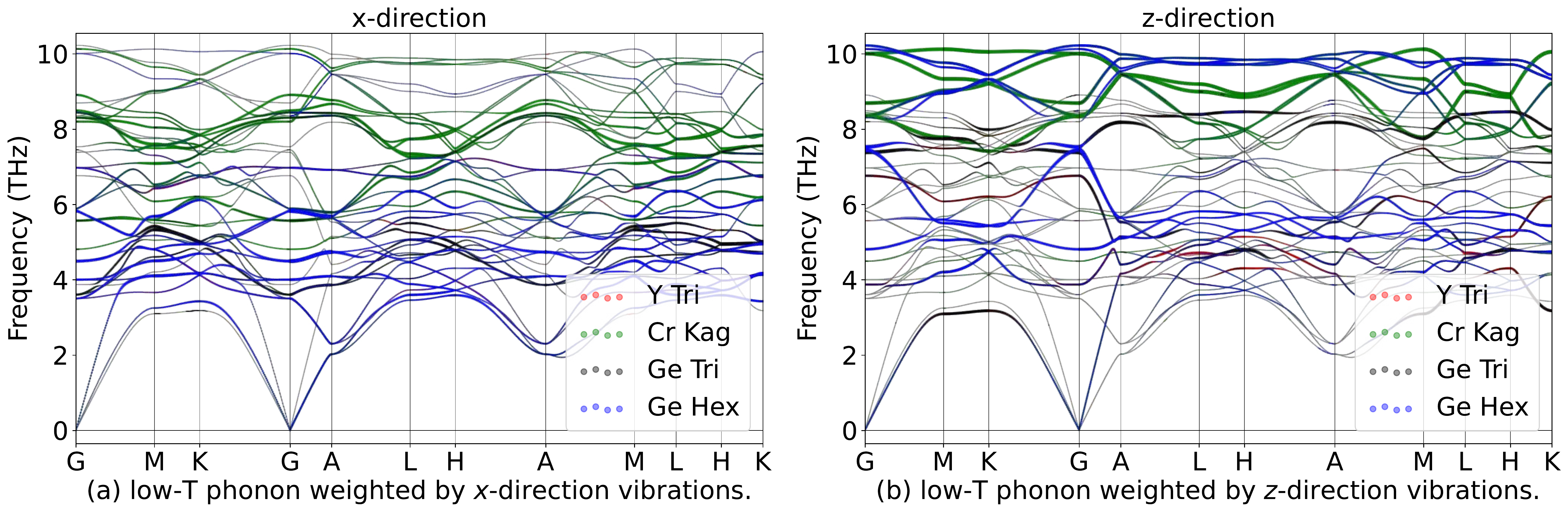}
\hspace{5pt}\label{fig:YCr6Ge6_relaxed_High_xz}
\includegraphics[width=0.85\textwidth]{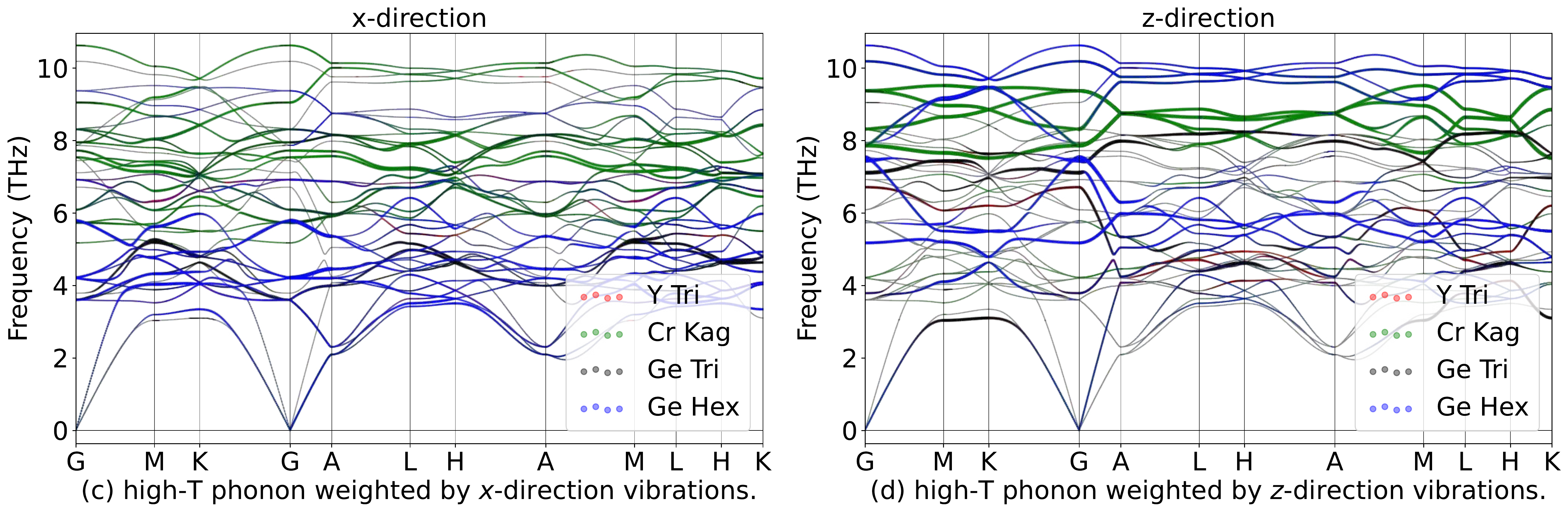}
\caption{Phonon spectrum for YCr$_6$Ge$_6$in in-plane ($x$) and out-of-plane ($z$) directions.}
\label{fig:YCr6Ge6phoxyz}
\end{figure}

\begin{table}[htbp]
\global\long\def\arraystretch{1.12}
\captionsetup{width=.95\textwidth}
\caption{IRREPs at high-symmetry points for YCr$_6$Ge$_6$ phonon spectrum at Low-T.}
\label{tab:191_YCr6Ge6_Low}
\setlength{\tabcolsep}{1.mm}{
}
\end{table}

\begin{table}[htbp]
\global\long\def\arraystretch{1.12}
\captionsetup{width=.95\textwidth}
\caption{IRREPs at high-symmetry points for YCr$_6$Ge$_6$ phonon spectrum at High-T.}
\label{tab:191_YCr6Ge6_High}
\setlength{\tabcolsep}{1.mm}{
%
}
\end{table}
\subsubsection{\label{sms2sec:ZrCr6Ge6}\hyperref[smsec:col]{\texorpdfstring{ZrCr$_6$Ge$_6$}{}}~{\color{green}  (High-T stable, Low-T stable) }}

\begin{table}[htbp]
\global\long\def\arraystretch{1.12}
\captionsetup{width=.85\textwidth}
\caption{Structure parameters of ZrCr$_6$Ge$_6$ after relaxation. It contains 376 electrons. }
\label{tab:ZrCr6Ge6}
\setlength{\tabcolsep}{4mm}{
%
}
\end{table}

\begin{figure}[htbp]
\centering
\includegraphics[width=0.85\textwidth]{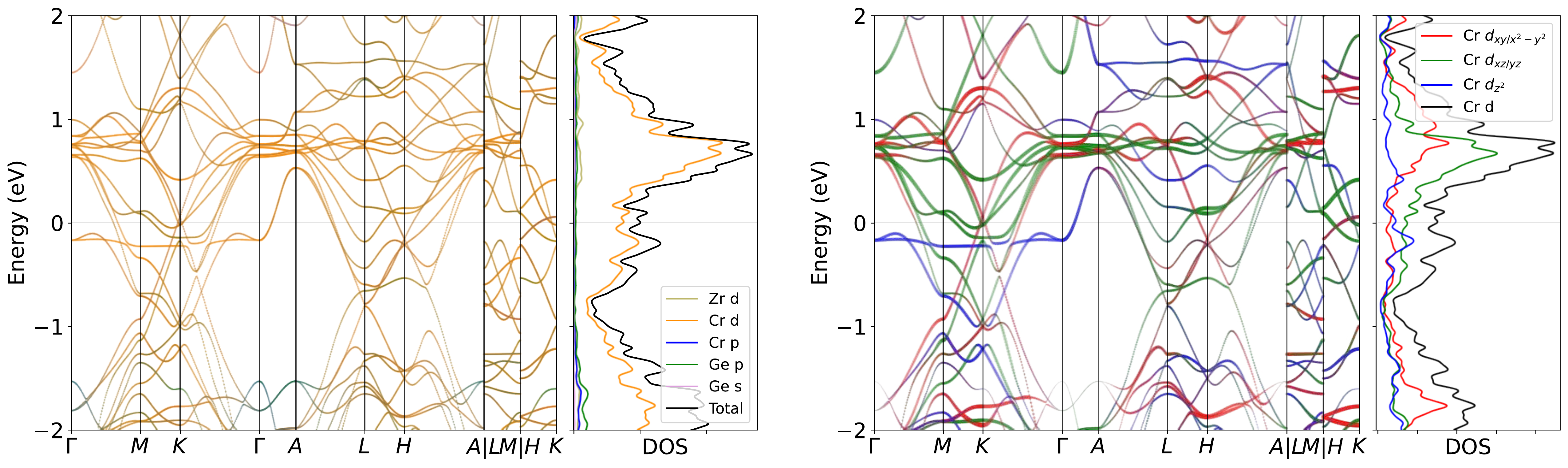}
\caption{Band structure for ZrCr$_6$Ge$_6$ without SOC. The projection of Cr $3d$ orbitals is also presented. The band structures clearly reveal the dominating of Cr $3d$ orbitals  near the Fermi level, as well as the pronounced peak.}
\label{fig:ZrCr6Ge6band}
\end{figure}

\begin{figure}[H]
\centering
\subfloat[Relaxed phonon at low-T.]{
\label{fig:ZrCr6Ge6_relaxed_Low}
\includegraphics[width=0.45\textwidth]{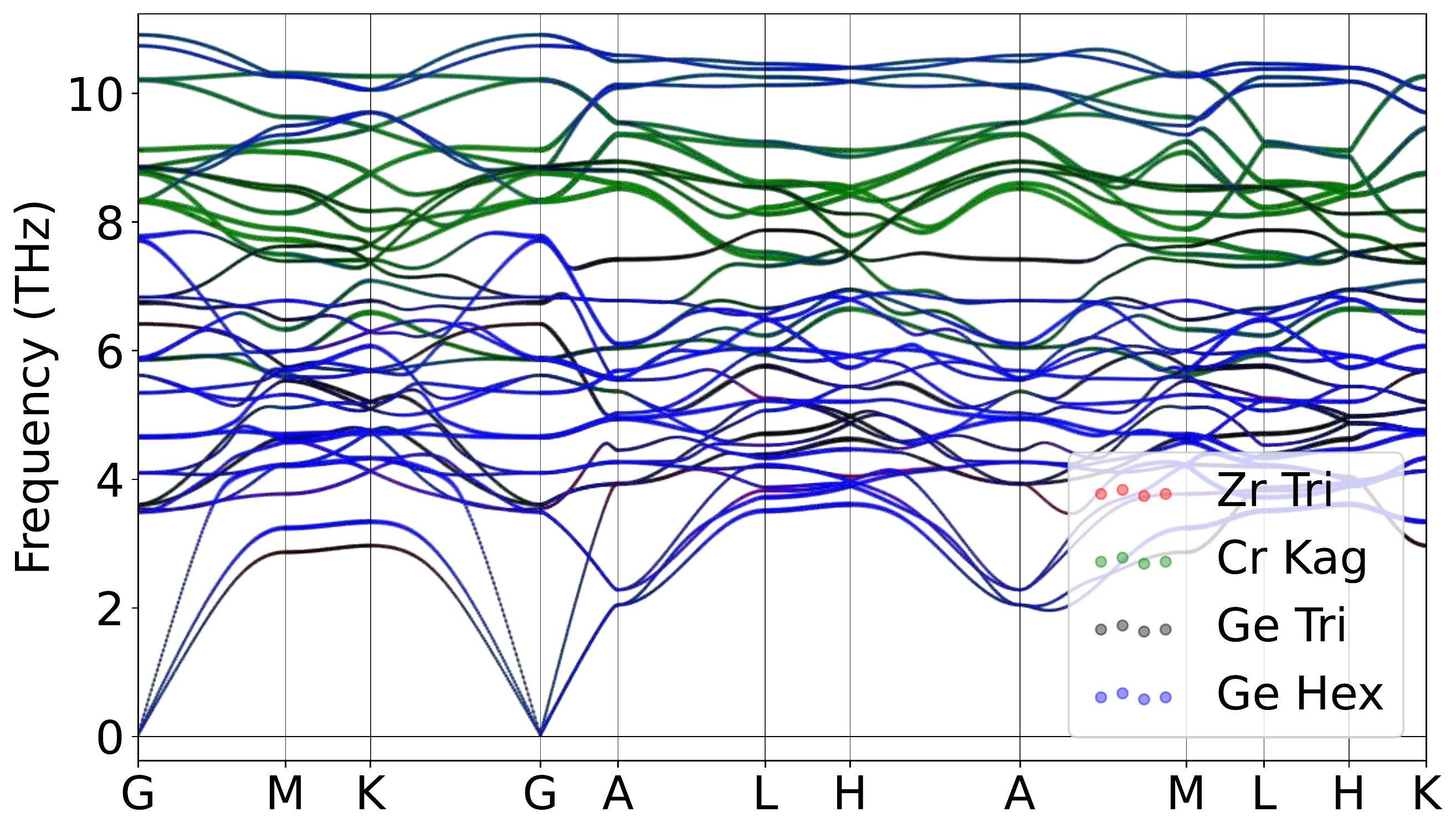}
}
\hspace{5pt}\subfloat[Relaxed phonon at high-T.]{
\label{fig:ZrCr6Ge6_relaxed_High}
\includegraphics[width=0.45\textwidth]{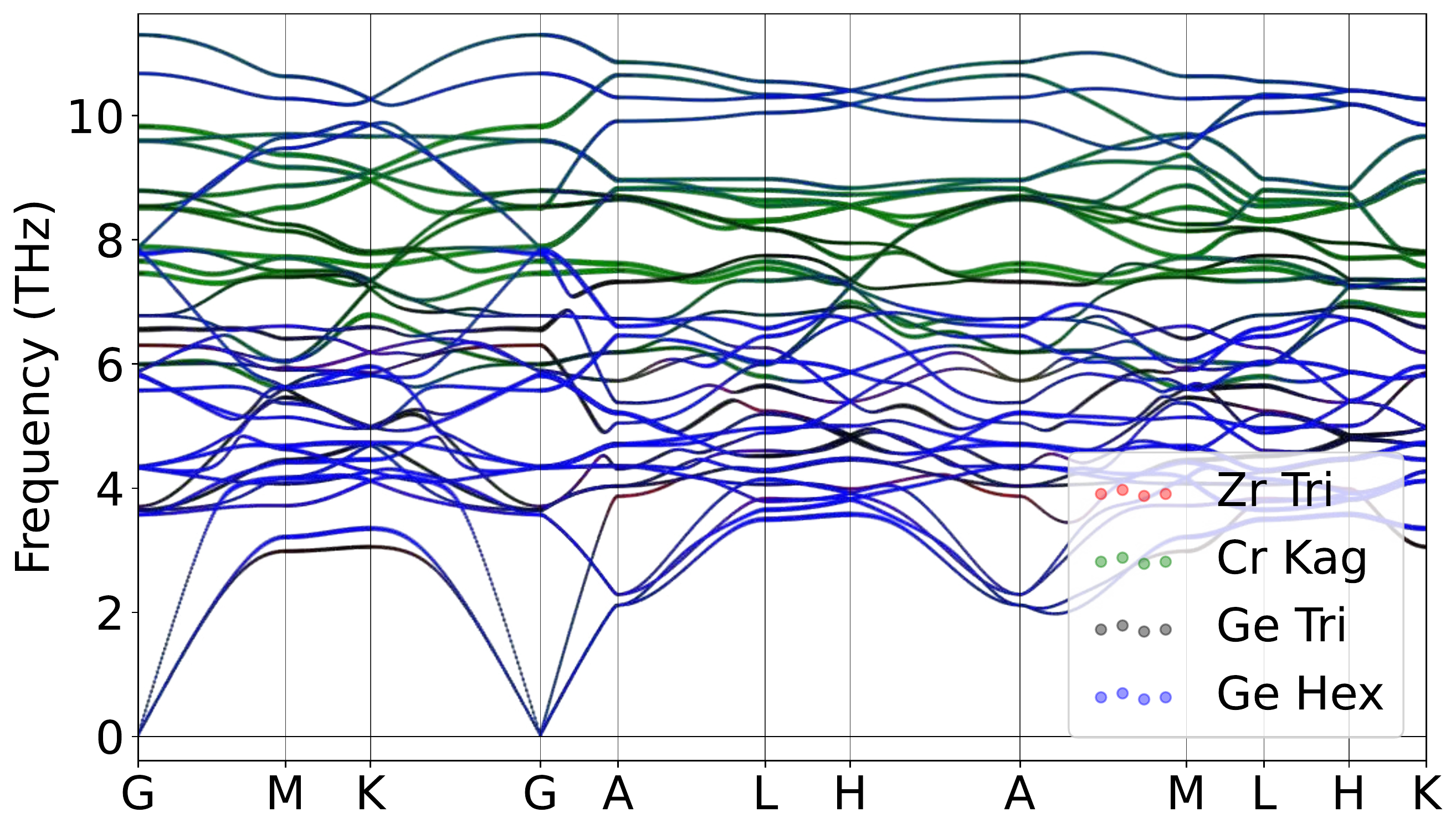}
}
\caption{Atomically projected phonon spectrum for ZrCr$_6$Ge$_6$.}
\label{fig:ZrCr6Ge6pho}
\end{figure}

\begin{figure}[H]
\centering
\label{fig:ZrCr6Ge6_relaxed_Low_xz}
\includegraphics[width=0.85\textwidth]{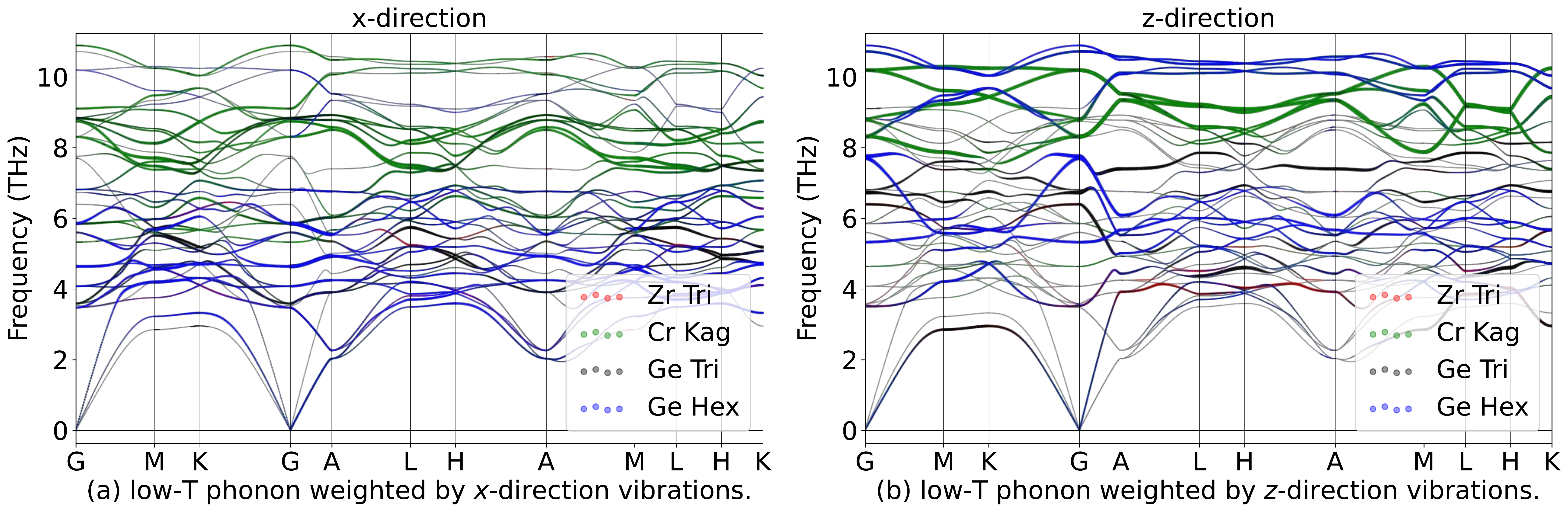}
\hspace{5pt}\label{fig:ZrCr6Ge6_relaxed_High_xz}
\includegraphics[width=0.85\textwidth]{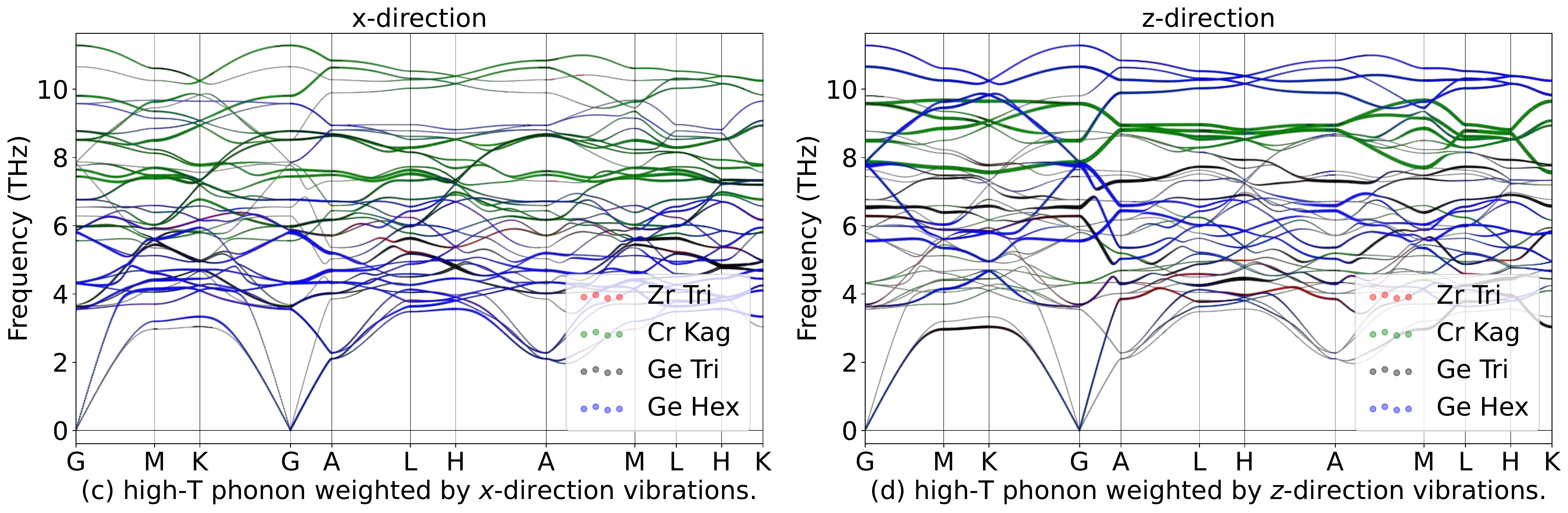}
\caption{Phonon spectrum for ZrCr$_6$Ge$_6$in in-plane ($x$) and out-of-plane ($z$) directions.}
\label{fig:ZrCr6Ge6phoxyz}
\end{figure}

\begin{table}[htbp]
\global\long\def\arraystretch{1.12}
\captionsetup{width=.95\textwidth}
\caption{IRREPs at high-symmetry points for ZrCr$_6$Ge$_6$ phonon spectrum at Low-T.}
\label{tab:191_ZrCr6Ge6_Low}
\setlength{\tabcolsep}{1.mm}{
}
\end{table}

\begin{table}[htbp]
\global\long\def\arraystretch{1.12}
\captionsetup{width=.95\textwidth}
\caption{IRREPs at high-symmetry points for ZrCr$_6$Ge$_6$ phonon spectrum at High-T.}
\label{tab:191_ZrCr6Ge6_High}
\setlength{\tabcolsep}{1.mm}{
%
}
\end{table}
\subsubsection{\label{sms2sec:NbCr6Ge6}\hyperref[smsec:col]{\texorpdfstring{NbCr$_6$Ge$_6$}{}}~{\color{green}  (High-T stable, Low-T stable) }}

\begin{table}[htbp]
\global\long\def\arraystretch{1.12}
\captionsetup{width=.85\textwidth}
\caption{Structure parameters of NbCr$_6$Ge$_6$ after relaxation. It contains 377 electrons. }
\label{tab:NbCr6Ge6}
\setlength{\tabcolsep}{4mm}{
%
}
\end{table}

\begin{figure}[htbp]
\centering
\includegraphics[width=0.85\textwidth]{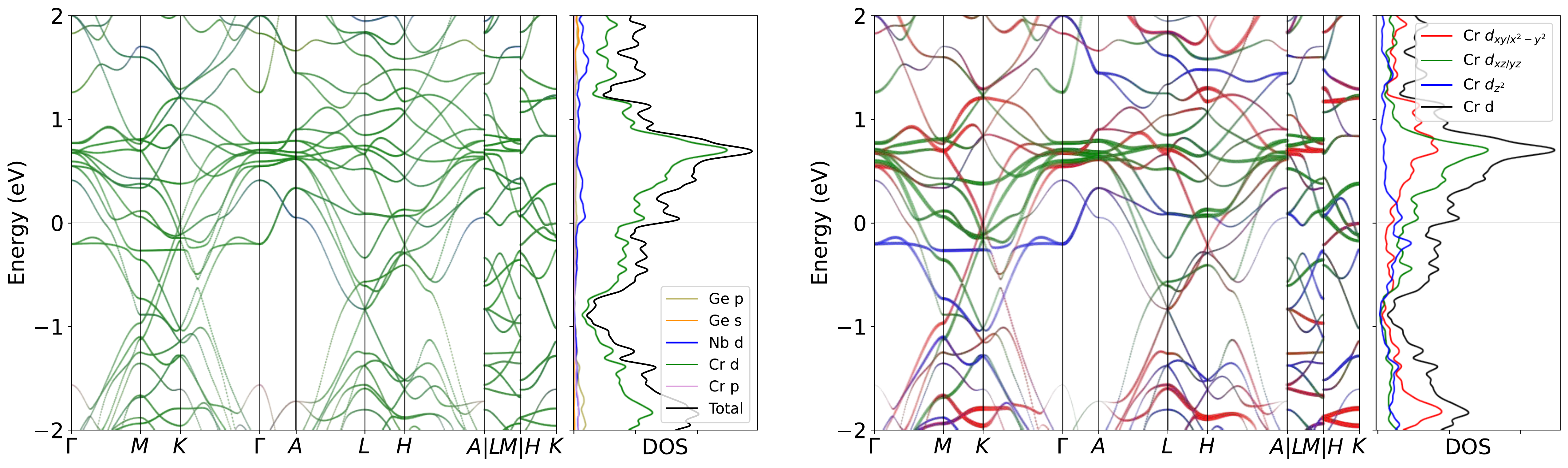}
\caption{Band structure for NbCr$_6$Ge$_6$ without SOC. The projection of Cr $3d$ orbitals is also presented. The band structures clearly reveal the dominating of Cr $3d$ orbitals  near the Fermi level, as well as the pronounced peak.}
\label{fig:NbCr6Ge6band}
\end{figure}

\begin{figure}[H]
\centering
\subfloat[Relaxed phonon at low-T.]{
\label{fig:NbCr6Ge6_relaxed_Low}
\includegraphics[width=0.45\textwidth]{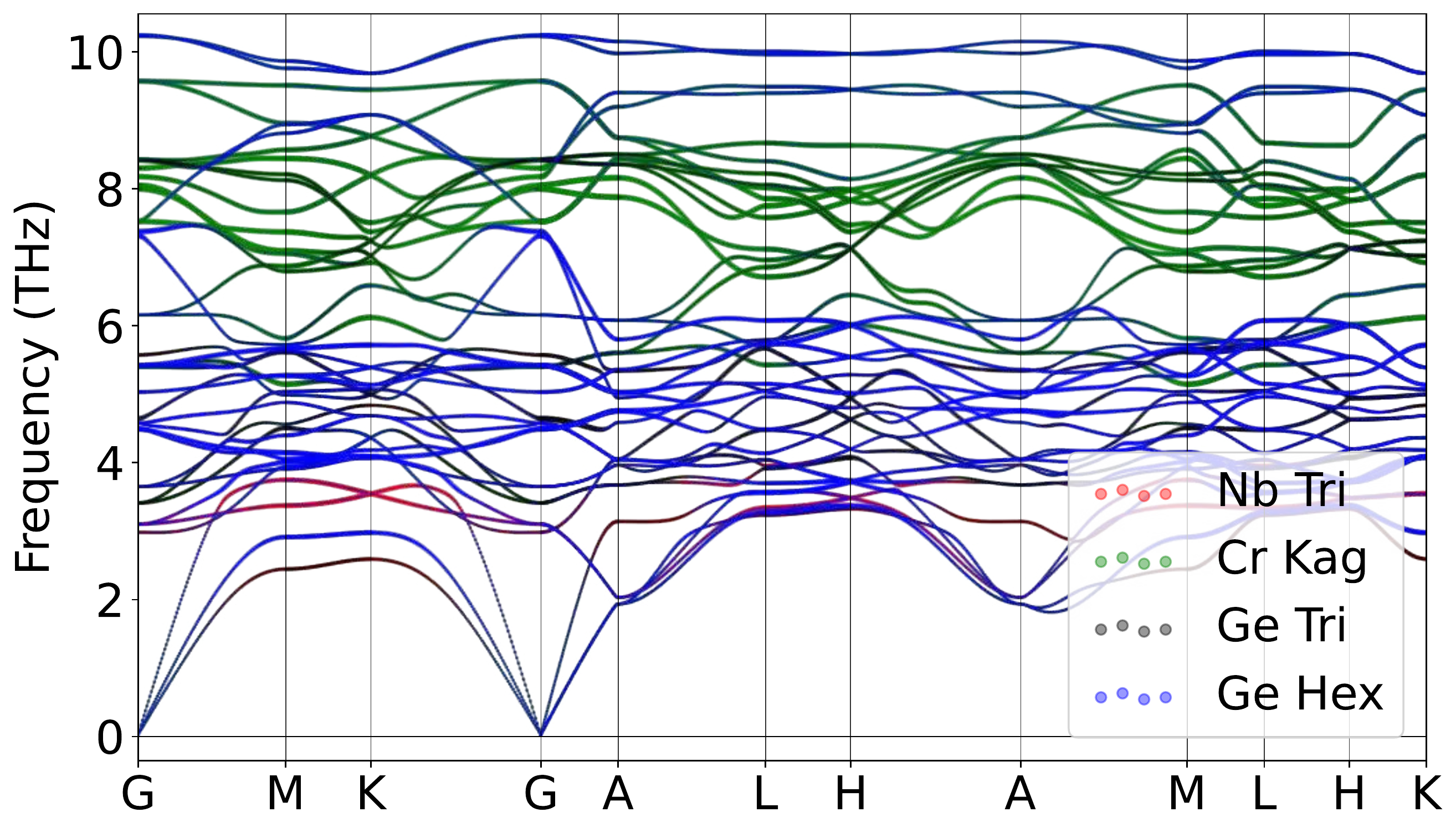}
}
\hspace{5pt}\subfloat[Relaxed phonon at high-T.]{
\label{fig:NbCr6Ge6_relaxed_High}
\includegraphics[width=0.45\textwidth]{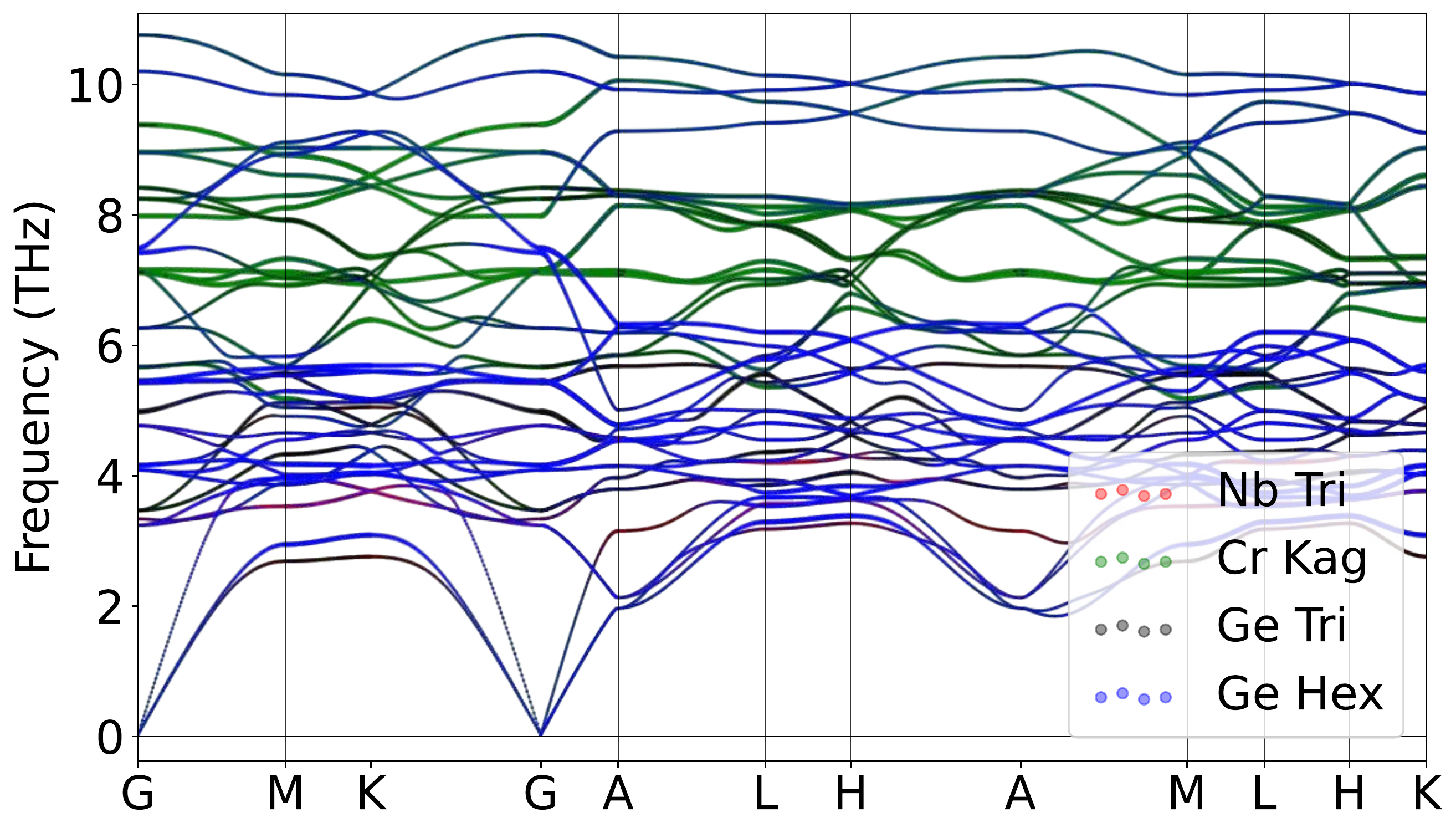}
}
\caption{Atomically projected phonon spectrum for NbCr$_6$Ge$_6$.}
\label{fig:NbCr6Ge6pho}
\end{figure}

\begin{figure}[H]
\centering
\label{fig:NbCr6Ge6_relaxed_Low_xz}
\includegraphics[width=0.85\textwidth]{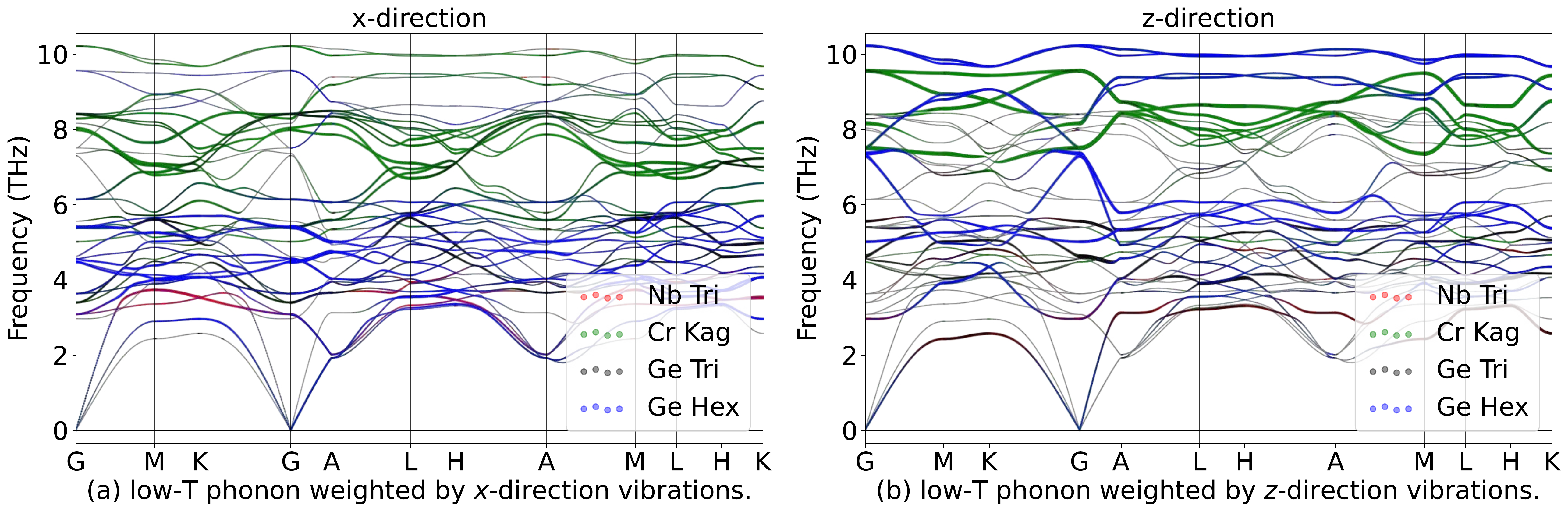}
\hspace{5pt}\label{fig:NbCr6Ge6_relaxed_High_xz}
\includegraphics[width=0.85\textwidth]{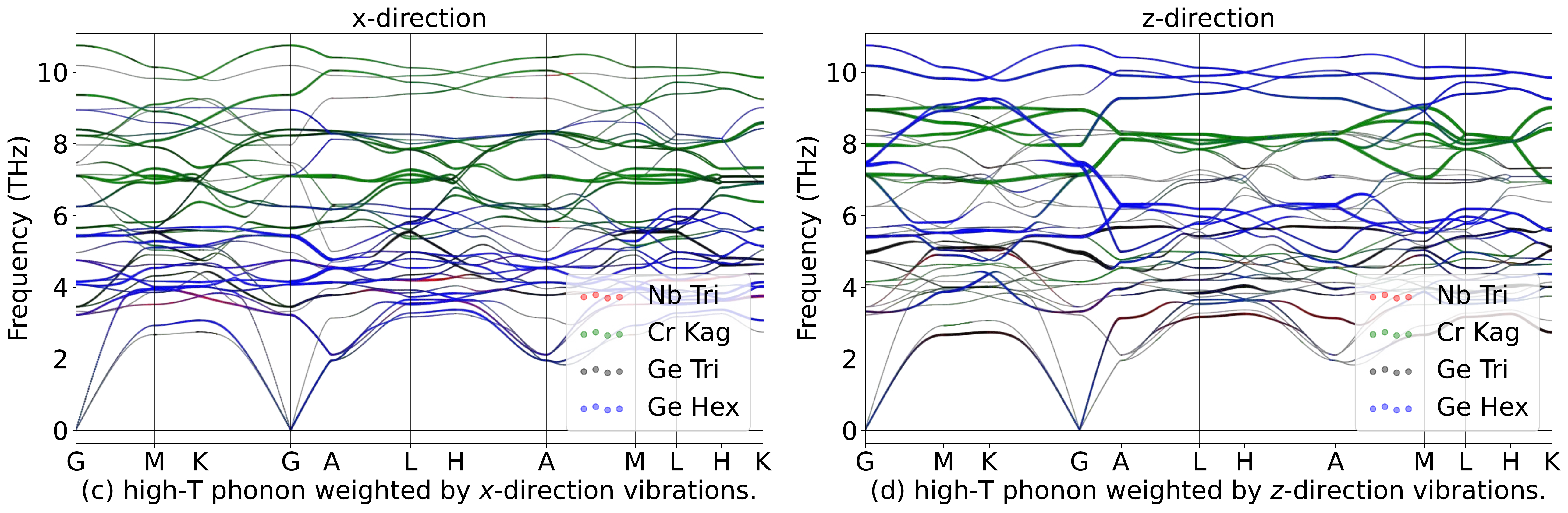}
\caption{Phonon spectrum for NbCr$_6$Ge$_6$in in-plane ($x$) and out-of-plane ($z$) directions.}
\label{fig:NbCr6Ge6phoxyz}
\end{figure}

\begin{table}[htbp]
\global\long\def\arraystretch{1.12}
\captionsetup{width=.95\textwidth}
\caption{IRREPs at high-symmetry points for NbCr$_6$Ge$_6$ phonon spectrum at Low-T.}
\label{tab:191_NbCr6Ge6_Low}
\setlength{\tabcolsep}{1.mm}{
}
\end{table}

\begin{table}[htbp]
\global\long\def\arraystretch{1.12}
\captionsetup{width=.95\textwidth}
\caption{IRREPs at high-symmetry points for NbCr$_6$Ge$_6$ phonon spectrum at High-T.}
\label{tab:191_NbCr6Ge6_High}
\setlength{\tabcolsep}{1.mm}{
%
}
\end{table}
\subsubsection{\label{sms2sec:PrCr6Ge6}\hyperref[smsec:col]{\texorpdfstring{PrCr$_6$Ge$_6$}{}}~{\color{green}  (High-T stable, Low-T stable) }}

\begin{table}[htbp]
\global\long\def\arraystretch{1.12}
\captionsetup{width=.85\textwidth}
\caption{Structure parameters of PrCr$_6$Ge$_6$ after relaxation. It contains 395 electrons. }
\label{tab:PrCr6Ge6}
\setlength{\tabcolsep}{4mm}{
%
}
\end{table}

\begin{figure}[htbp]
\centering
\includegraphics[width=0.85\textwidth]{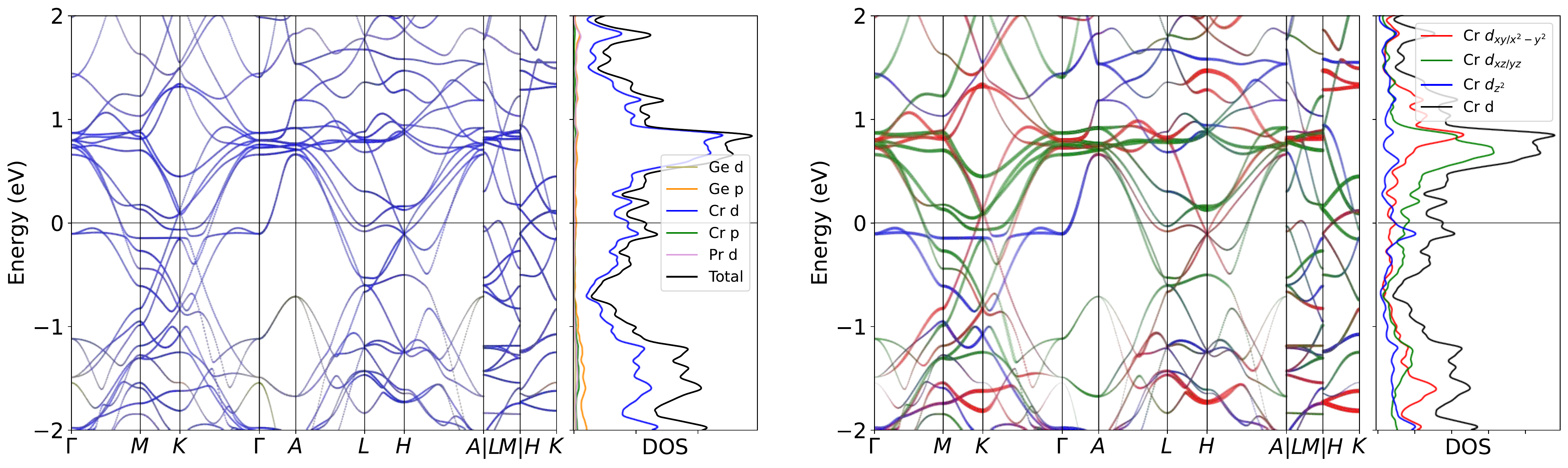}
\caption{Band structure for PrCr$_6$Ge$_6$ without SOC. The projection of Cr $3d$ orbitals is also presented. The band structures clearly reveal the dominating of Cr $3d$ orbitals  near the Fermi level, as well as the pronounced peak.}
\label{fig:PrCr6Ge6band}
\end{figure}

\begin{figure}[H]
\centering
\subfloat[Relaxed phonon at low-T.]{
\label{fig:PrCr6Ge6_relaxed_Low}
\includegraphics[width=0.45\textwidth]{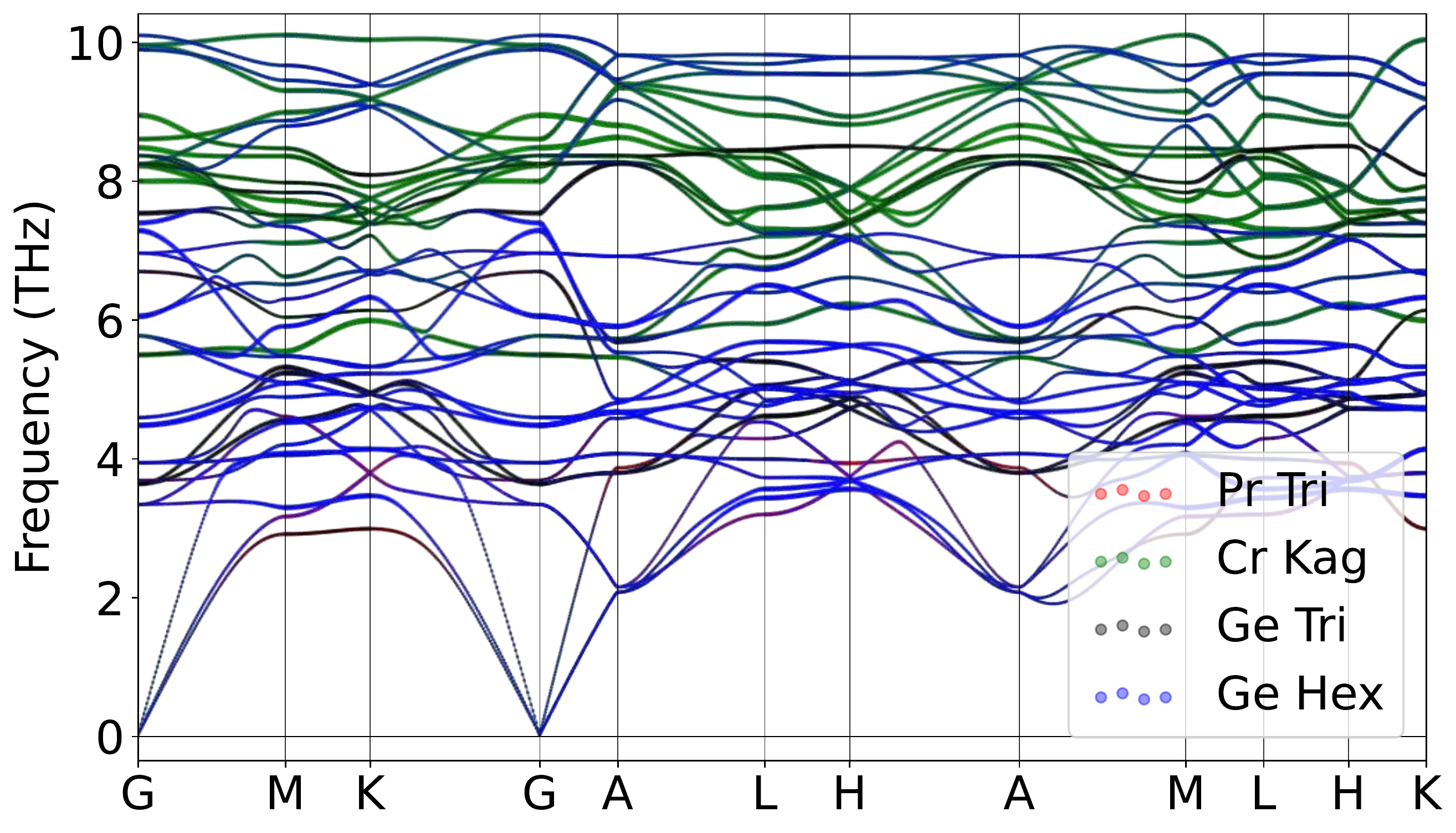}
}
\hspace{5pt}\subfloat[Relaxed phonon at high-T.]{
\label{fig:PrCr6Ge6_relaxed_High}
\includegraphics[width=0.45\textwidth]{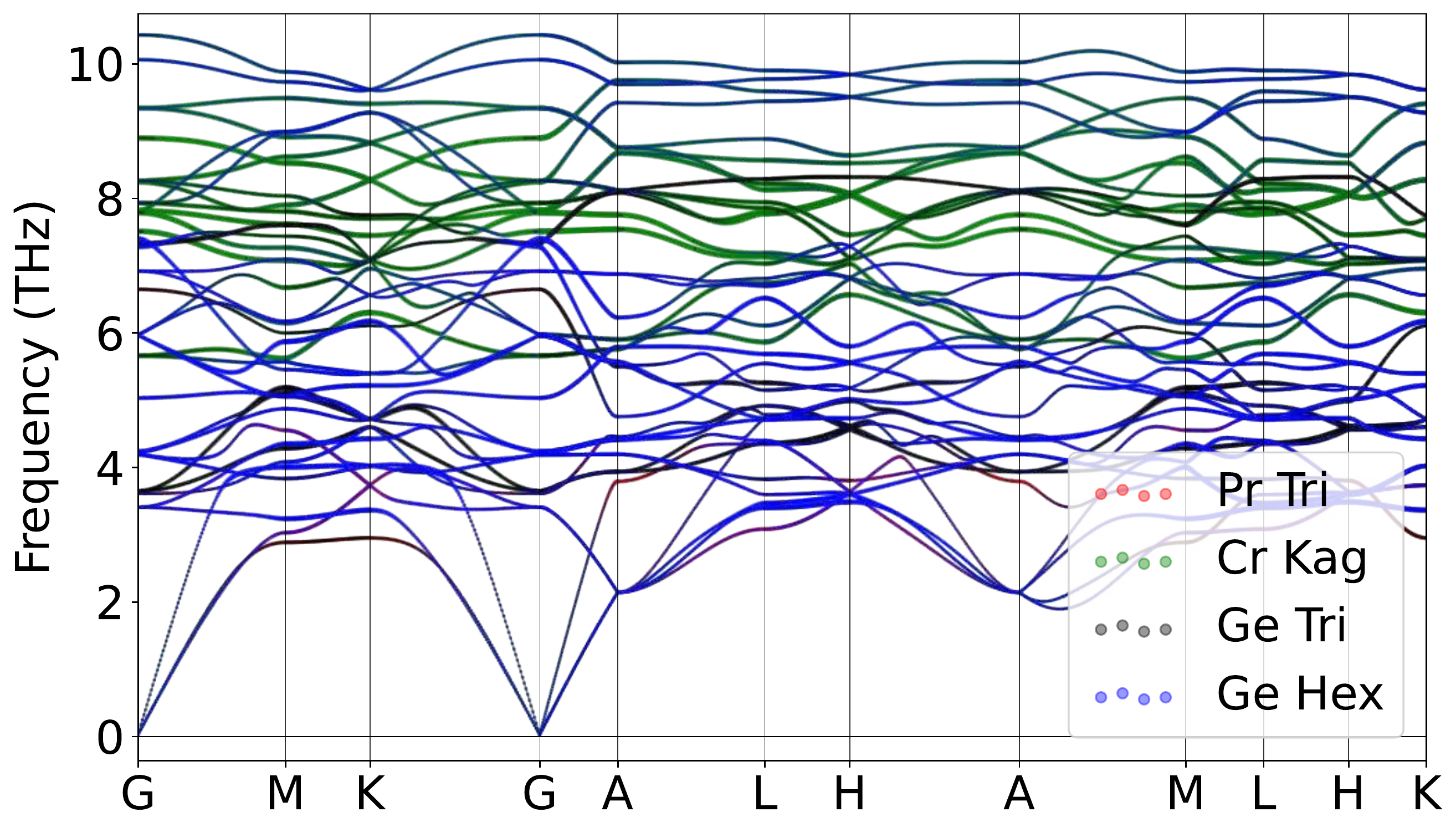}
}
\caption{Atomically projected phonon spectrum for PrCr$_6$Ge$_6$.}
\label{fig:PrCr6Ge6pho}
\end{figure}

\begin{figure}[H]
\centering
\label{fig:PrCr6Ge6_relaxed_Low_xz}
\includegraphics[width=0.85\textwidth]{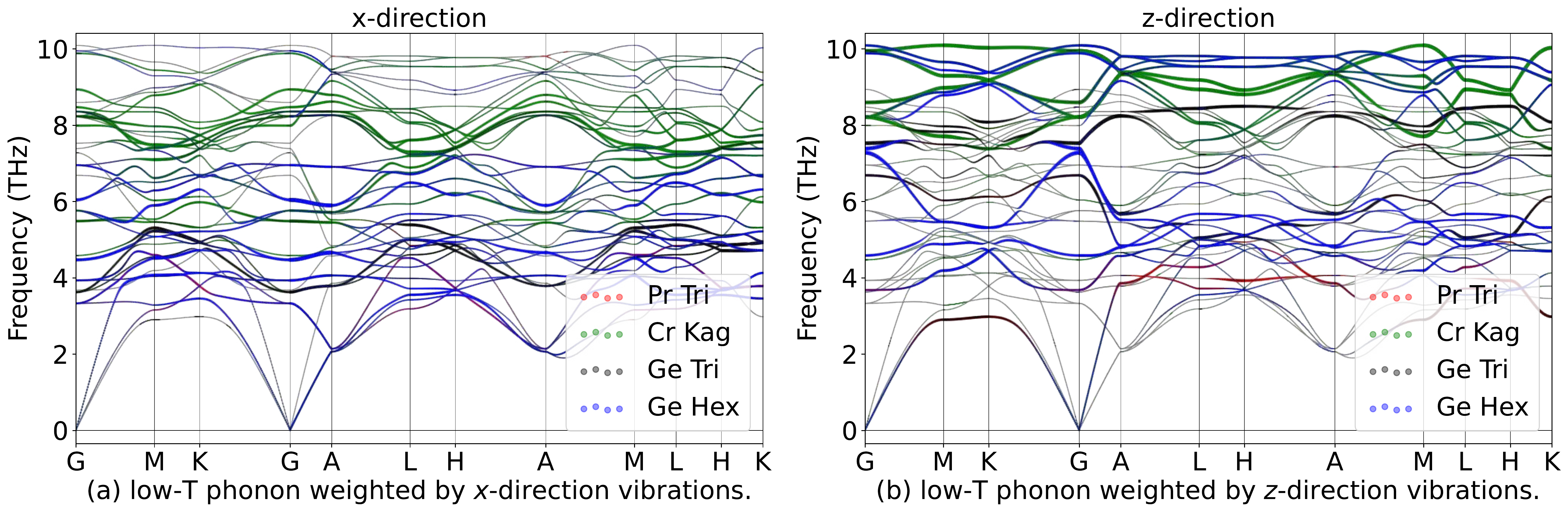}
\hspace{5pt}\label{fig:PrCr6Ge6_relaxed_High_xz}
\includegraphics[width=0.85\textwidth]{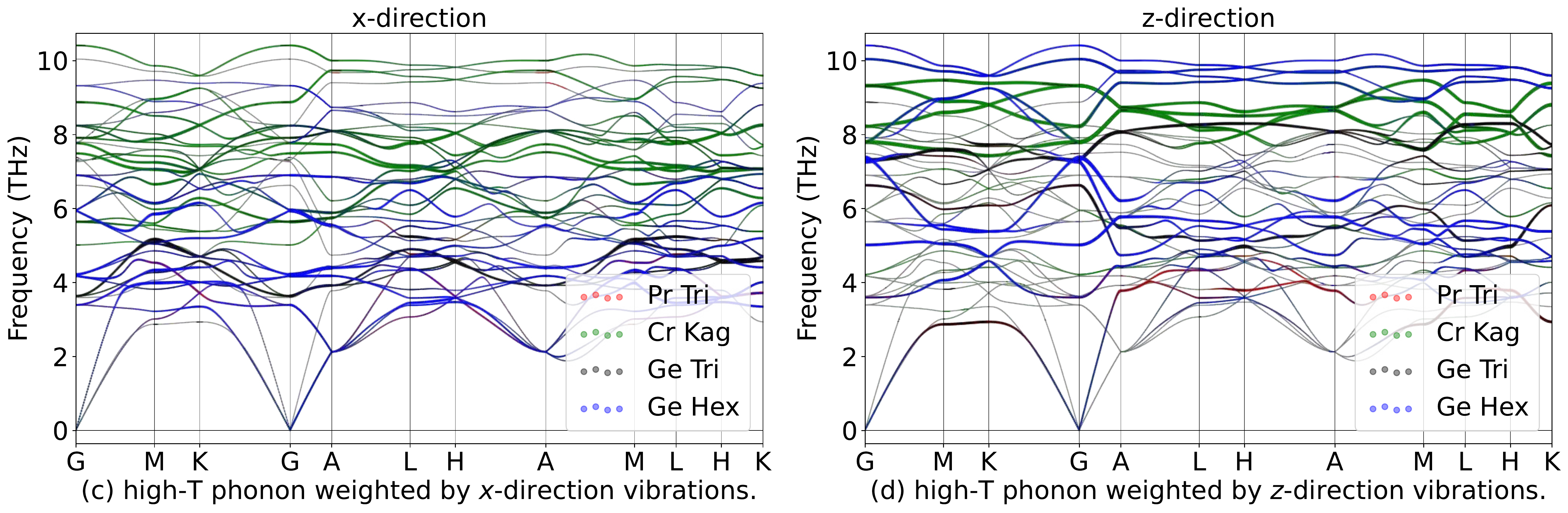}
\caption{Phonon spectrum for PrCr$_6$Ge$_6$in in-plane ($x$) and out-of-plane ($z$) directions.}
\label{fig:PrCr6Ge6phoxyz}
\end{figure}

\begin{table}[htbp]
\global\long\def\arraystretch{1.12}
\captionsetup{width=.95\textwidth}
\caption{IRREPs at high-symmetry points for PrCr$_6$Ge$_6$ phonon spectrum at Low-T.}
\label{tab:191_PrCr6Ge6_Low}
\setlength{\tabcolsep}{1.mm}{
}
\end{table}

\begin{table}[htbp]
\global\long\def\arraystretch{1.12}
\captionsetup{width=.95\textwidth}
\caption{IRREPs at high-symmetry points for PrCr$_6$Ge$_6$ phonon spectrum at High-T.}
\label{tab:191_PrCr6Ge6_High}
\setlength{\tabcolsep}{1.mm}{
%
}
\end{table}
\subsubsection{\label{sms2sec:NdCr6Ge6}\hyperref[smsec:col]{\texorpdfstring{NdCr$_6$Ge$_6$}{}}~{\color{green}  (High-T stable, Low-T stable) }}

\begin{table}[htbp]
\global\long\def\arraystretch{1.12}
\captionsetup{width=.85\textwidth}
\caption{Structure parameters of NdCr$_6$Ge$_6$ after relaxation. It contains 396 electrons. }
\label{tab:NdCr6Ge6}
\setlength{\tabcolsep}{4mm}{
%
}
\end{table}

\begin{figure}[htbp]
\centering
\includegraphics[width=0.85\textwidth]{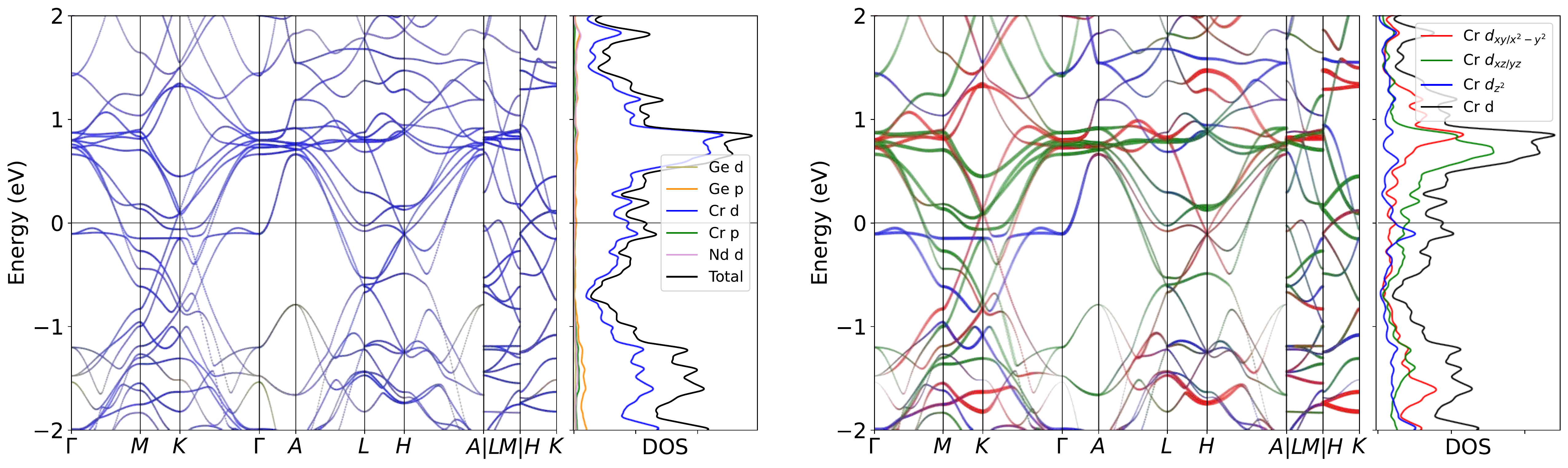}
\caption{Band structure for NdCr$_6$Ge$_6$ without SOC. The projection of Cr $3d$ orbitals is also presented. The band structures clearly reveal the dominating of Cr $3d$ orbitals  near the Fermi level, as well as the pronounced peak.}
\label{fig:NdCr6Ge6band}
\end{figure}

\begin{figure}[H]
\centering
\subfloat[Relaxed phonon at low-T.]{
\label{fig:NdCr6Ge6_relaxed_Low}
\includegraphics[width=0.45\textwidth]{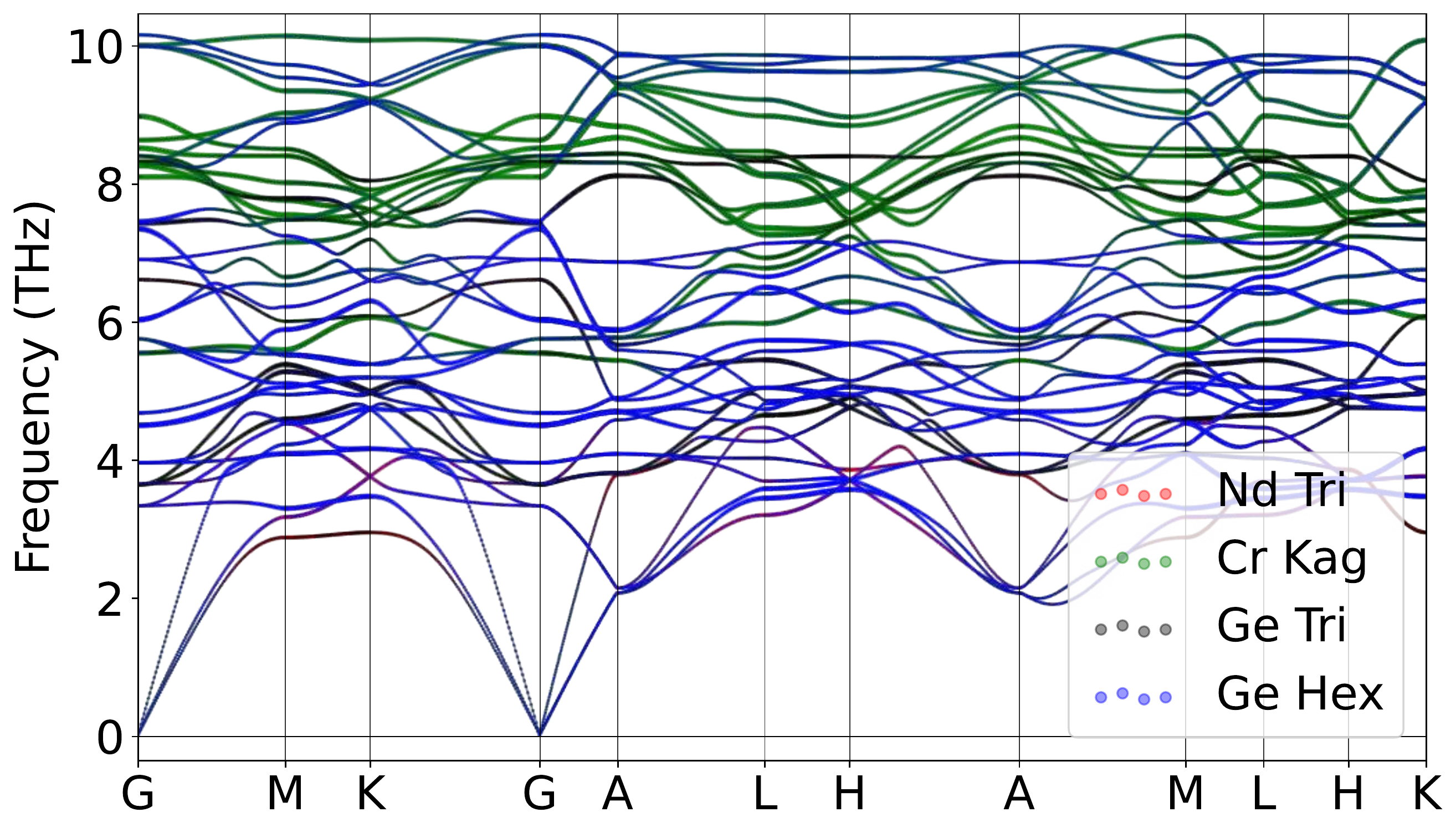}
}
\hspace{5pt}\subfloat[Relaxed phonon at high-T.]{
\label{fig:NdCr6Ge6_relaxed_High}
\includegraphics[width=0.45\textwidth]{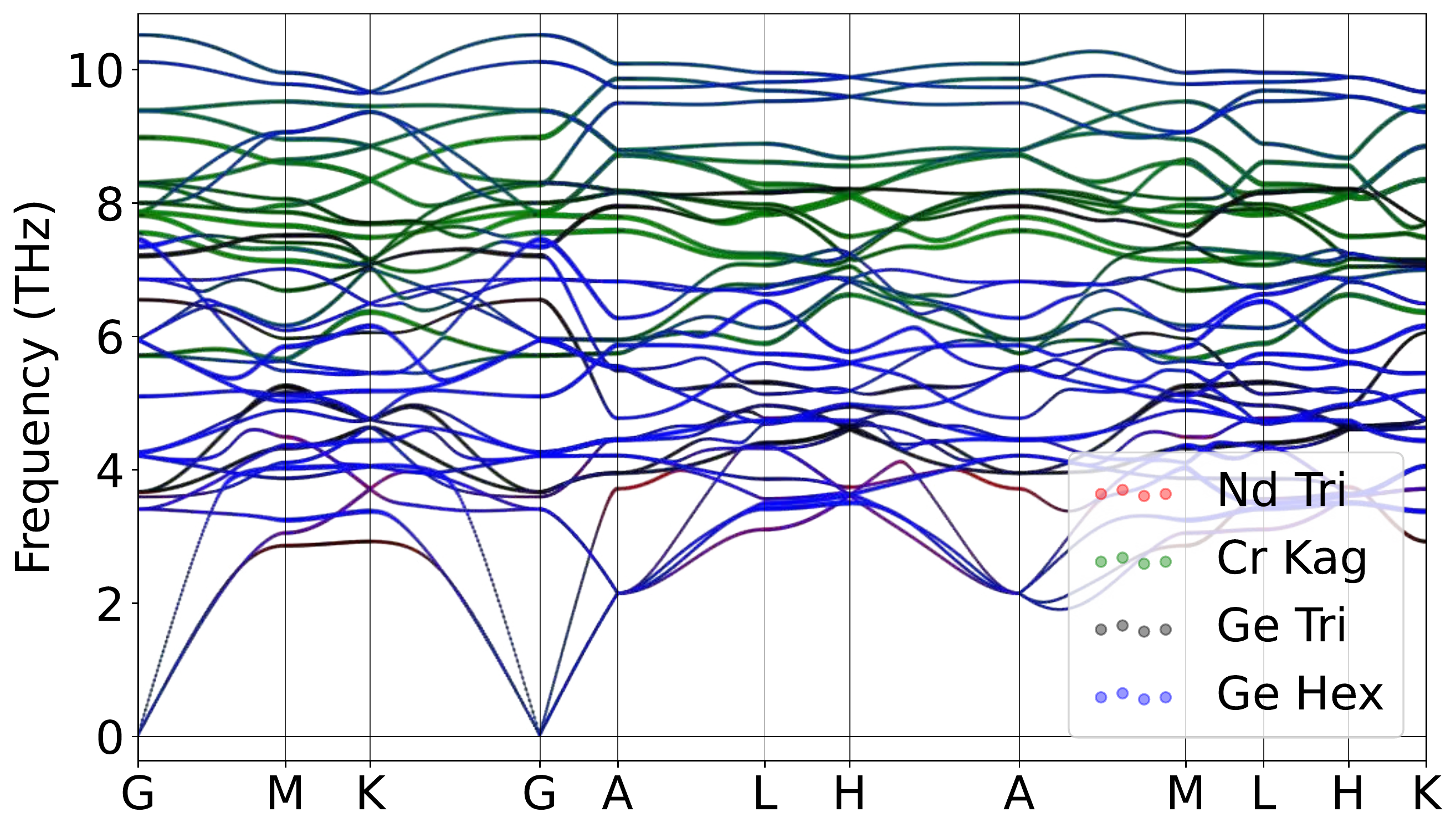}
}
\caption{Atomically projected phonon spectrum for NdCr$_6$Ge$_6$.}
\label{fig:NdCr6Ge6pho}
\end{figure}

\begin{figure}[H]
\centering
\label{fig:NdCr6Ge6_relaxed_Low_xz}
\includegraphics[width=0.85\textwidth]{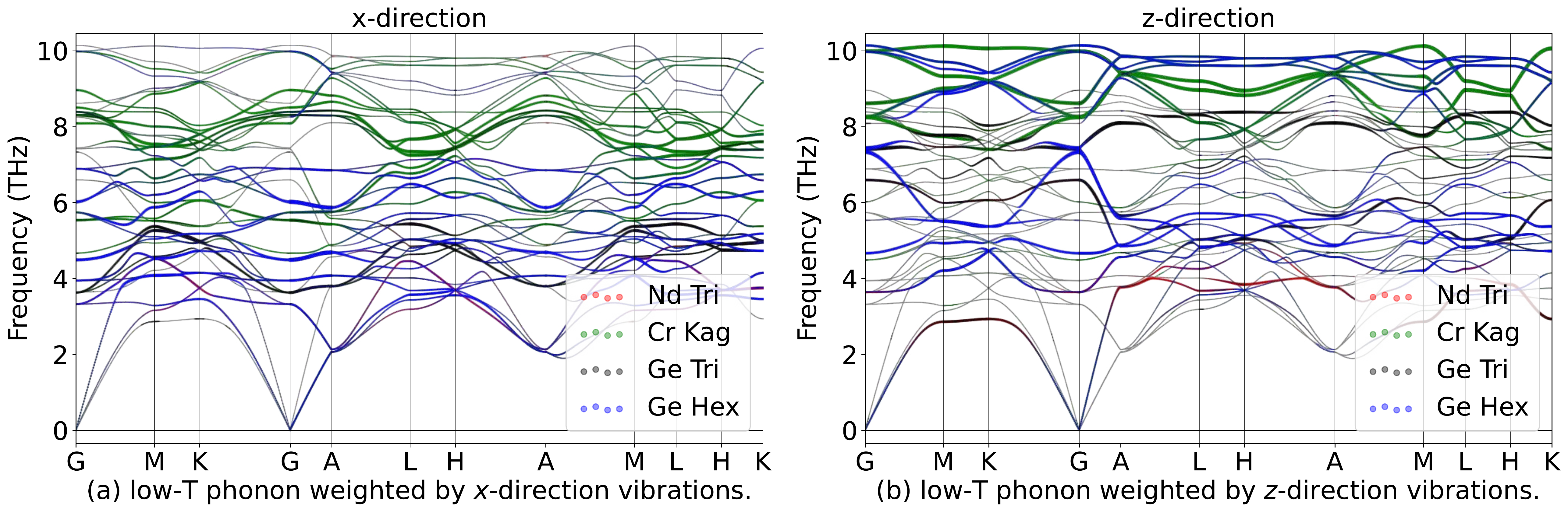}
\hspace{5pt}\label{fig:NdCr6Ge6_relaxed_High_xz}
\includegraphics[width=0.85\textwidth]{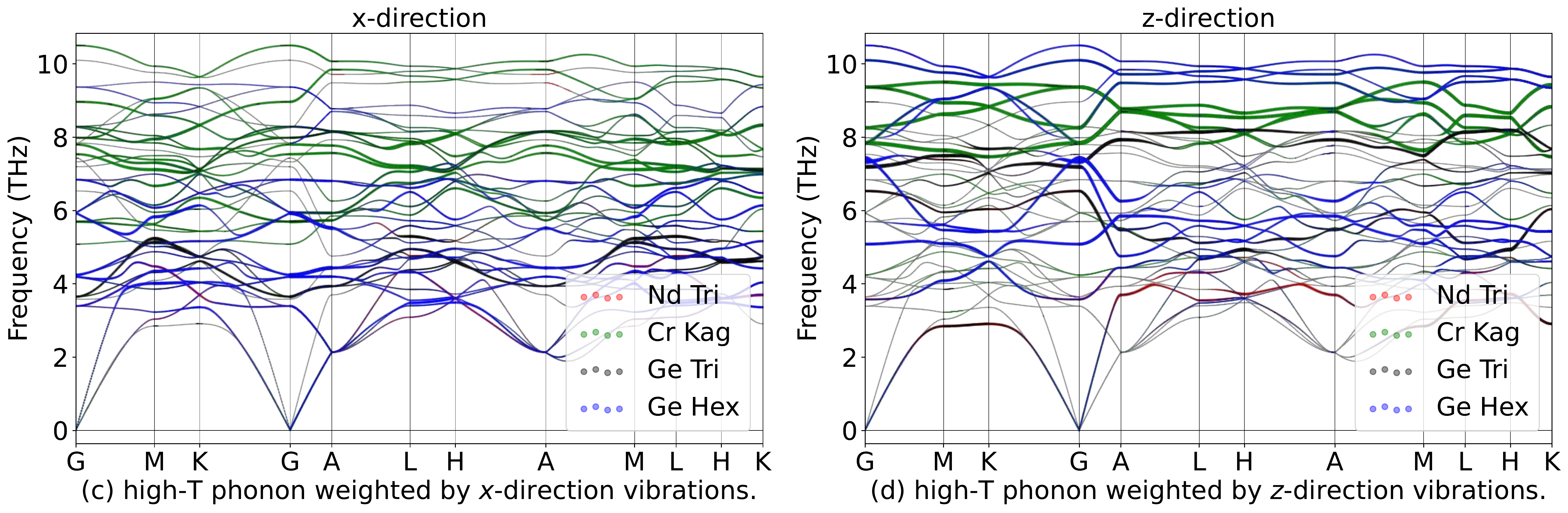}
\caption{Phonon spectrum for NdCr$_6$Ge$_6$in in-plane ($x$) and out-of-plane ($z$) directions.}
\label{fig:NdCr6Ge6phoxyz}
\end{figure}

\begin{table}[htbp]
\global\long\def\arraystretch{1.12}
\captionsetup{width=.95\textwidth}
\caption{IRREPs at high-symmetry points for NdCr$_6$Ge$_6$ phonon spectrum at Low-T.}
\label{tab:191_NdCr6Ge6_Low}
\setlength{\tabcolsep}{1.mm}{
}
\end{table}

\begin{table}[htbp]
\global\long\def\arraystretch{1.12}
\captionsetup{width=.95\textwidth}
\caption{IRREPs at high-symmetry points for NdCr$_6$Ge$_6$ phonon spectrum at High-T.}
\label{tab:191_NdCr6Ge6_High}
\setlength{\tabcolsep}{1.mm}{
%
}
\end{table}
\subsubsection{\label{sms2sec:SmCr6Ge6}\hyperref[smsec:col]{\texorpdfstring{SmCr$_6$Ge$_6$}{}}~{\color{green}  (High-T stable, Low-T stable) }}

\begin{table}[htbp]
\global\long\def\arraystretch{1.12}
\captionsetup{width=.85\textwidth}
\caption{Structure parameters of SmCr$_6$Ge$_6$ after relaxation. It contains 398 electrons. }
\label{tab:SmCr6Ge6}
\setlength{\tabcolsep}{4mm}{
%
}
\end{table}

\begin{figure}[htbp]
\centering
\includegraphics[width=0.85\textwidth]{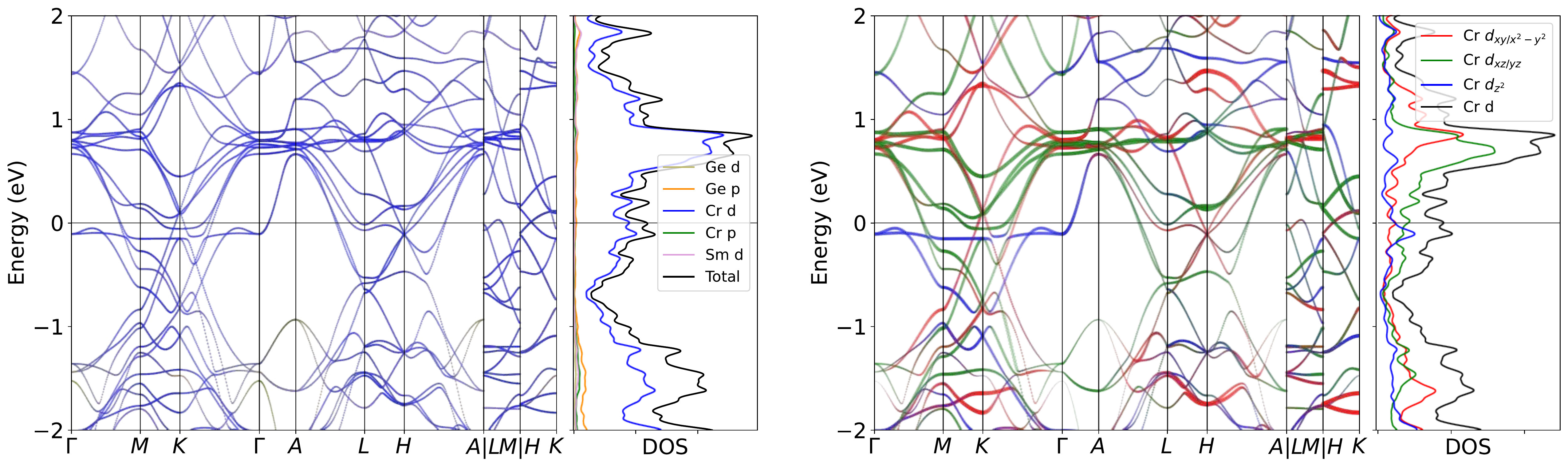}
\caption{Band structure for SmCr$_6$Ge$_6$ without SOC. The projection of Cr $3d$ orbitals is also presented. The band structures clearly reveal the dominating of Cr $3d$ orbitals  near the Fermi level, as well as the pronounced peak.}
\label{fig:SmCr6Ge6band}
\end{figure}

\begin{figure}[H]
\centering
\subfloat[Relaxed phonon at low-T.]{
\label{fig:SmCr6Ge6_relaxed_Low}
\includegraphics[width=0.45\textwidth]{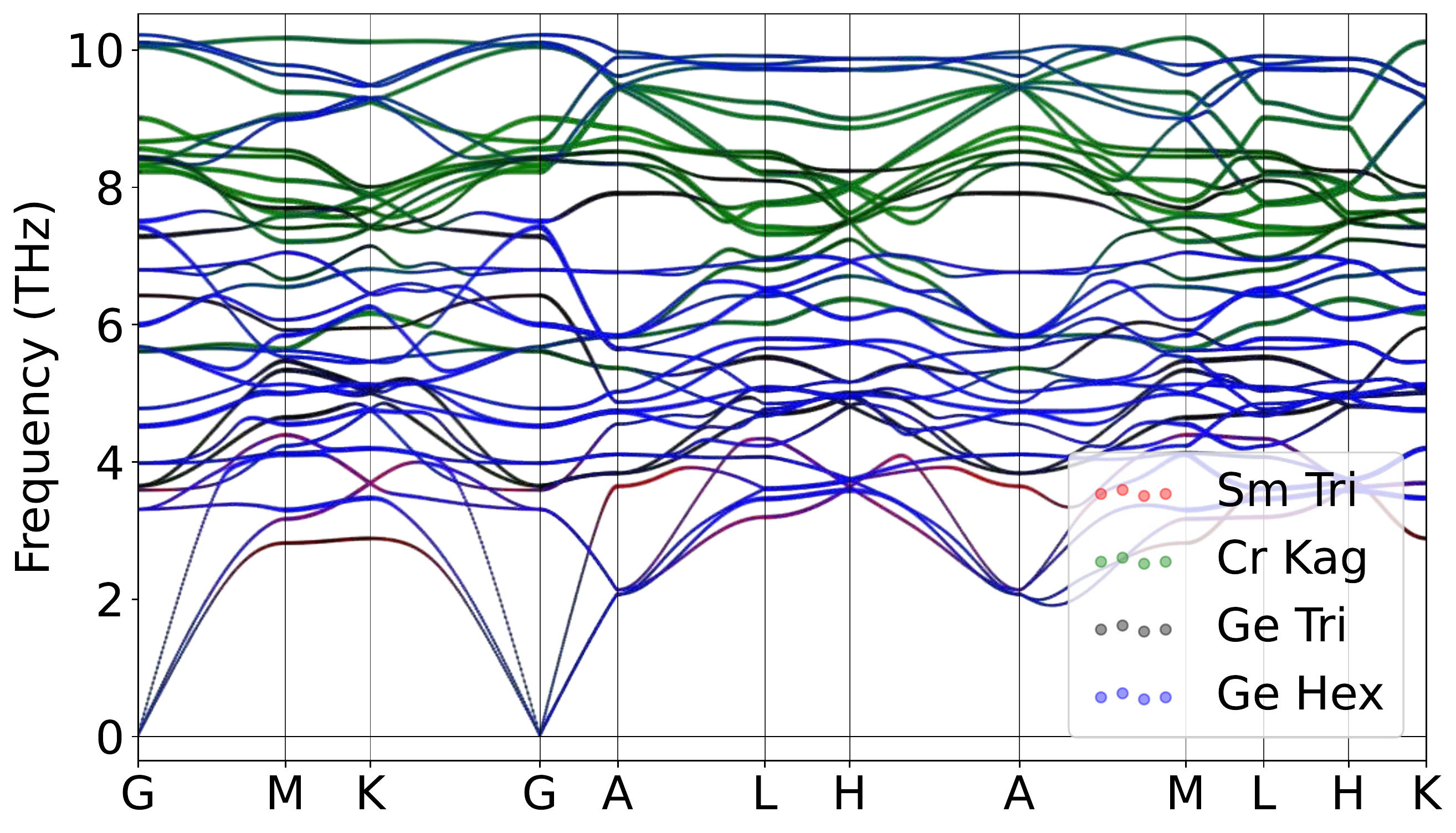}
}
\hspace{5pt}\subfloat[Relaxed phonon at high-T.]{
\label{fig:SmCr6Ge6_relaxed_High}
\includegraphics[width=0.45\textwidth]{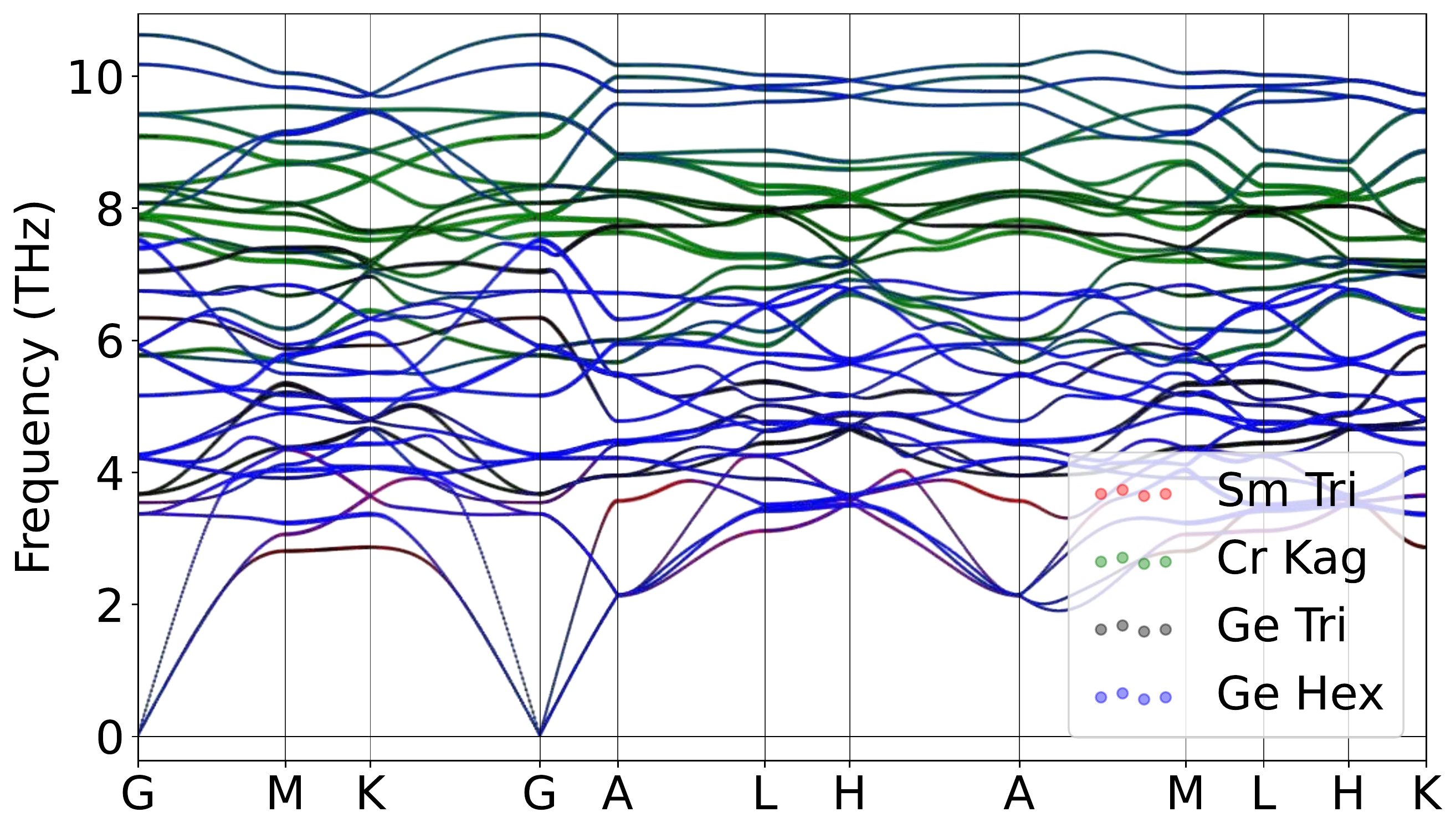}
}
\caption{Atomically projected phonon spectrum for SmCr$_6$Ge$_6$.}
\label{fig:SmCr6Ge6pho}
\end{figure}

\begin{figure}[H]
\centering
\label{fig:SmCr6Ge6_relaxed_Low_xz}
\includegraphics[width=0.85\textwidth]{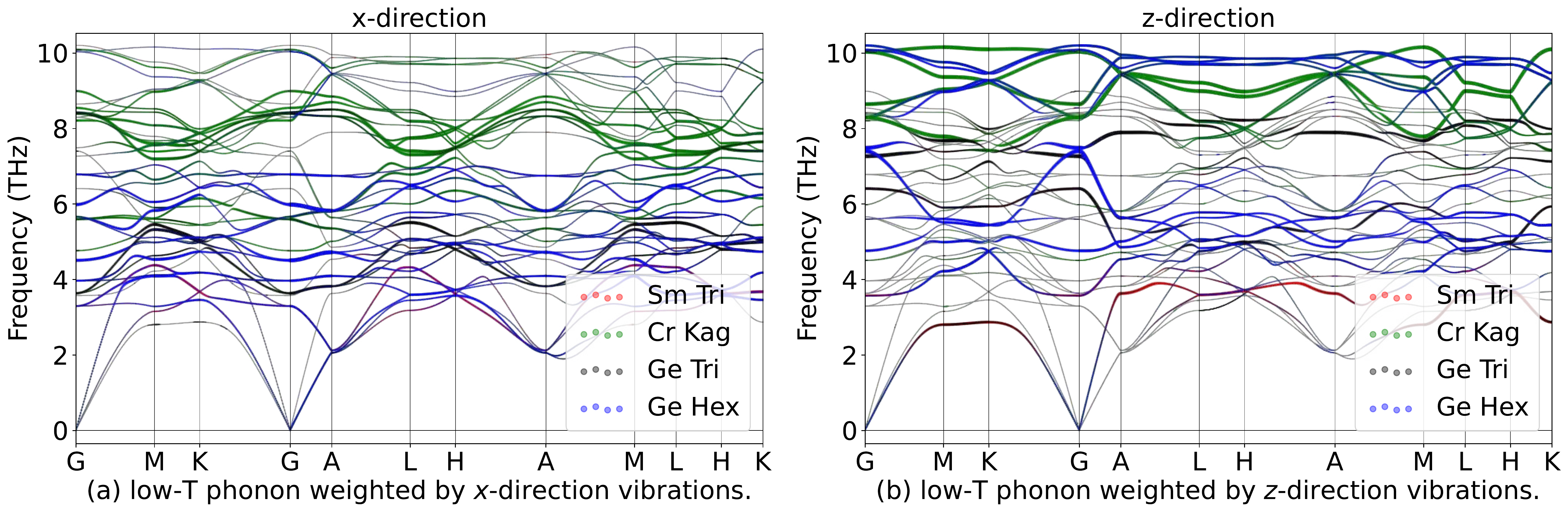}
\hspace{5pt}\label{fig:SmCr6Ge6_relaxed_High_xz}
\includegraphics[width=0.85\textwidth]{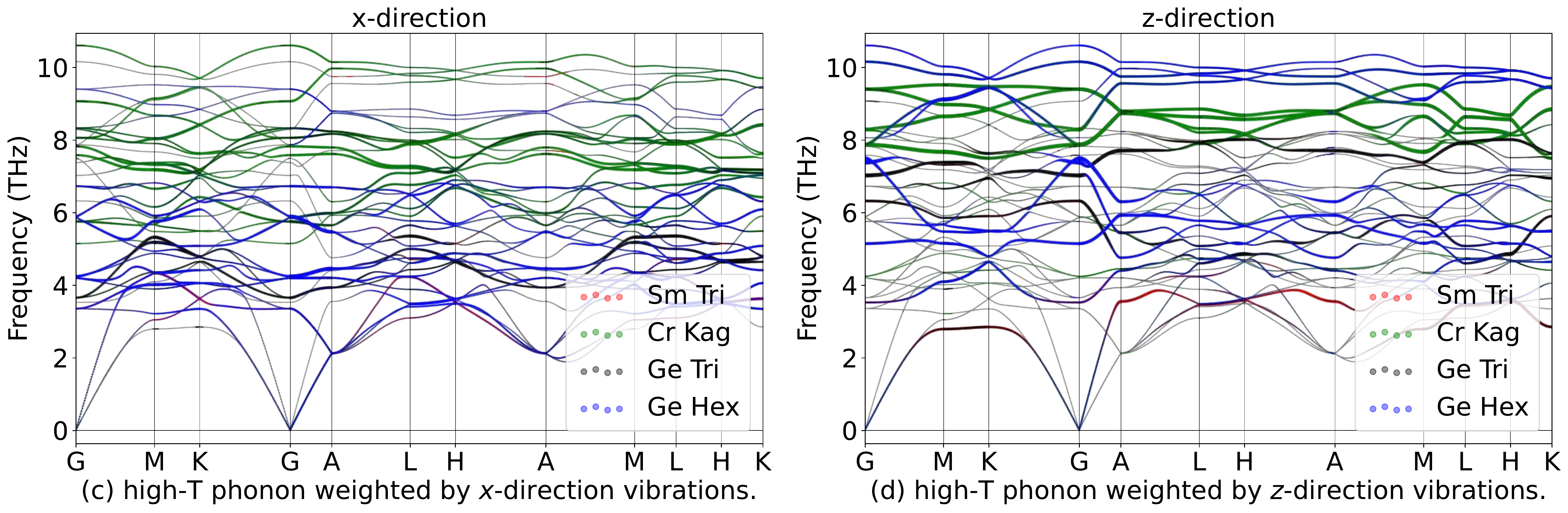}
\caption{Phonon spectrum for SmCr$_6$Ge$_6$in in-plane ($x$) and out-of-plane ($z$) directions.}
\label{fig:SmCr6Ge6phoxyz}
\end{figure}

\begin{table}[htbp]
\global\long\def\arraystretch{1.12}
\captionsetup{width=.95\textwidth}
\caption{IRREPs at high-symmetry points for SmCr$_6$Ge$_6$ phonon spectrum at Low-T.}
\label{tab:191_SmCr6Ge6_Low}
\setlength{\tabcolsep}{1.mm}{
}
\end{table}

\begin{table}[htbp]
\global\long\def\arraystretch{1.12}
\captionsetup{width=.95\textwidth}
\caption{IRREPs at high-symmetry points for SmCr$_6$Ge$_6$ phonon spectrum at High-T.}
\label{tab:191_SmCr6Ge6_High}
\setlength{\tabcolsep}{1.mm}{
%
}
\end{table}
\subsubsection{\label{sms2sec:GdCr6Ge6}\hyperref[smsec:col]{\texorpdfstring{GdCr$_6$Ge$_6$}{}}~{\color{green}  (High-T stable, Low-T stable) }}

\begin{table}[htbp]
\global\long\def\arraystretch{1.12}
\captionsetup{width=.85\textwidth}
\caption{Structure parameters of GdCr$_6$Ge$_6$ after relaxation. It contains 400 electrons. }
\label{tab:GdCr6Ge6}
\setlength{\tabcolsep}{4mm}{
%
}
\end{table}

\begin{figure}[htbp]
\centering
\includegraphics[width=0.85\textwidth]{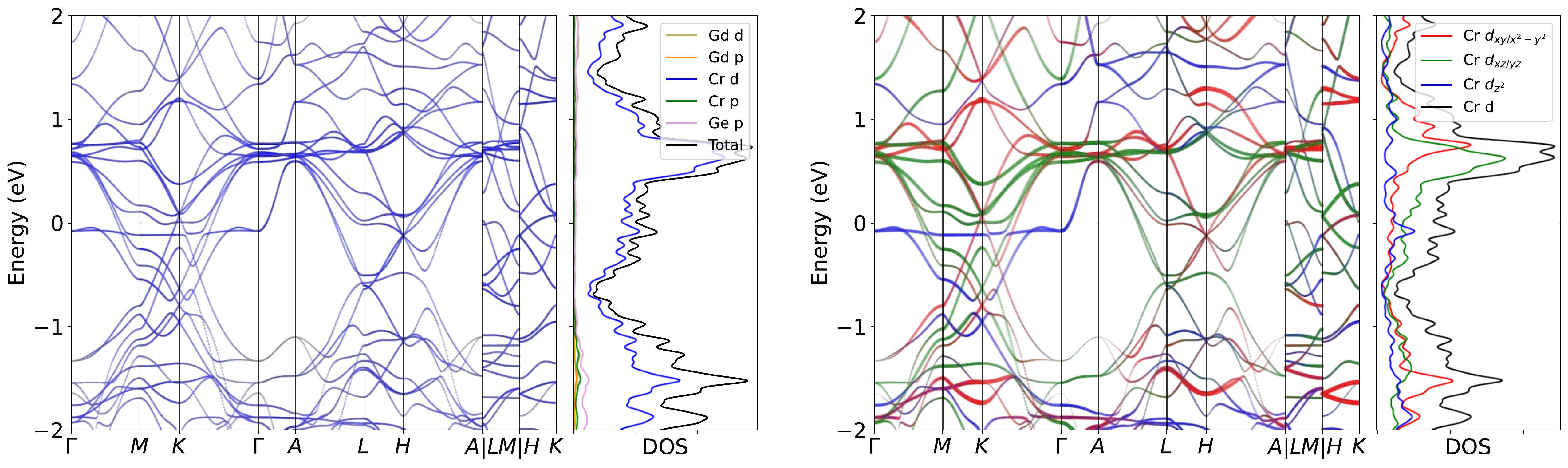}
\caption{Band structure for GdCr$_6$Ge$_6$ without SOC. The projection of Cr $3d$ orbitals is also presented. The band structures clearly reveal the dominating of Cr $3d$ orbitals  near the Fermi level, as well as the pronounced peak.}
\label{fig:GdCr6Ge6band}
\end{figure}

\begin{figure}[H]
\centering
\subfloat[Relaxed phonon at low-T.]{
\label{fig:GdCr6Ge6_relaxed_Low}
\includegraphics[width=0.45\textwidth]{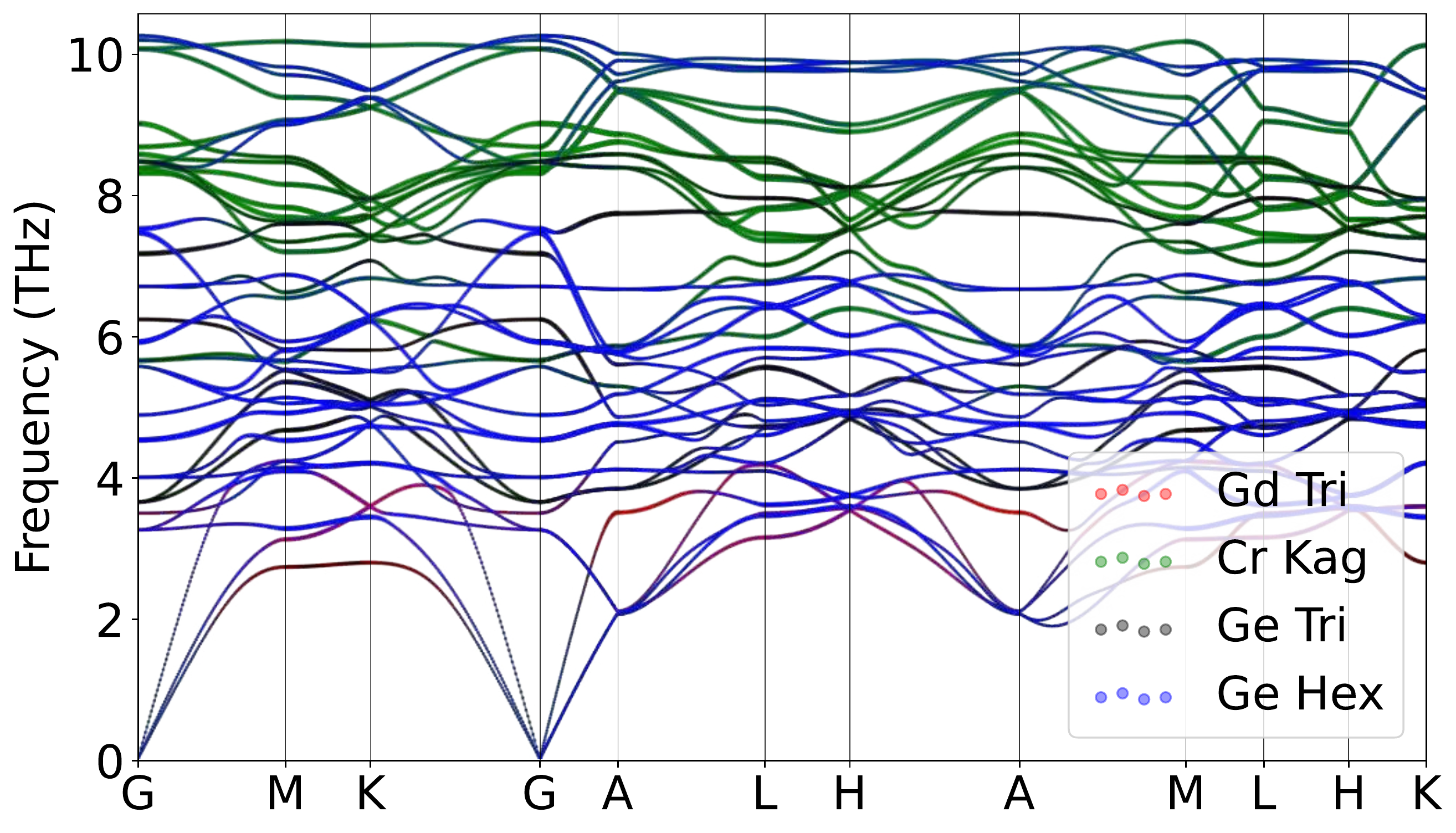}
}
\hspace{5pt}\subfloat[Relaxed phonon at high-T.]{
\label{fig:GdCr6Ge6_relaxed_High}
\includegraphics[width=0.45\textwidth]{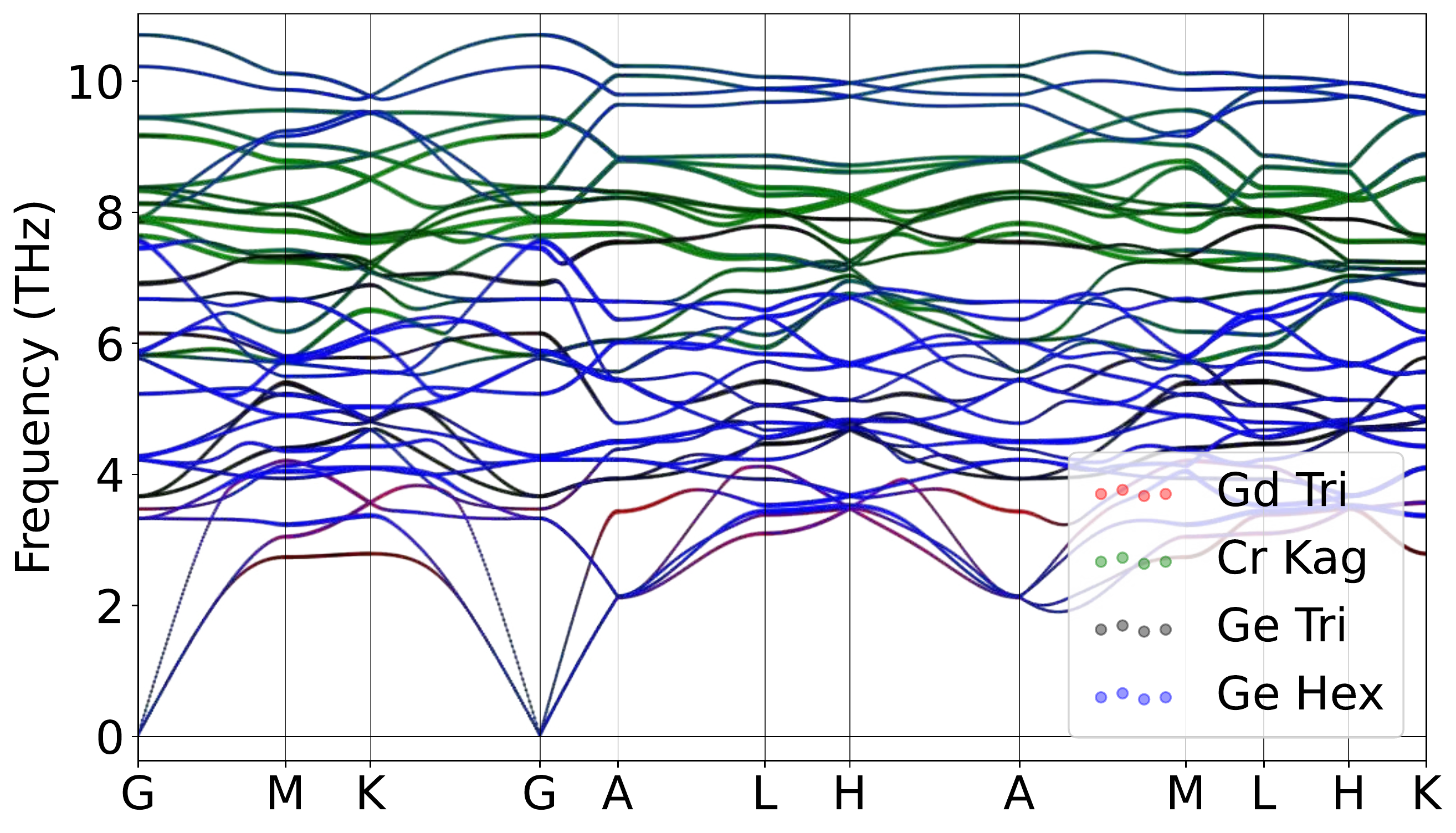}
}
\caption{Atomically projected phonon spectrum for GdCr$_6$Ge$_6$.}
\label{fig:GdCr6Ge6pho}
\end{figure}

\begin{figure}[H]
\centering
\label{fig:GdCr6Ge6_relaxed_Low_xz}
\includegraphics[width=0.85\textwidth]{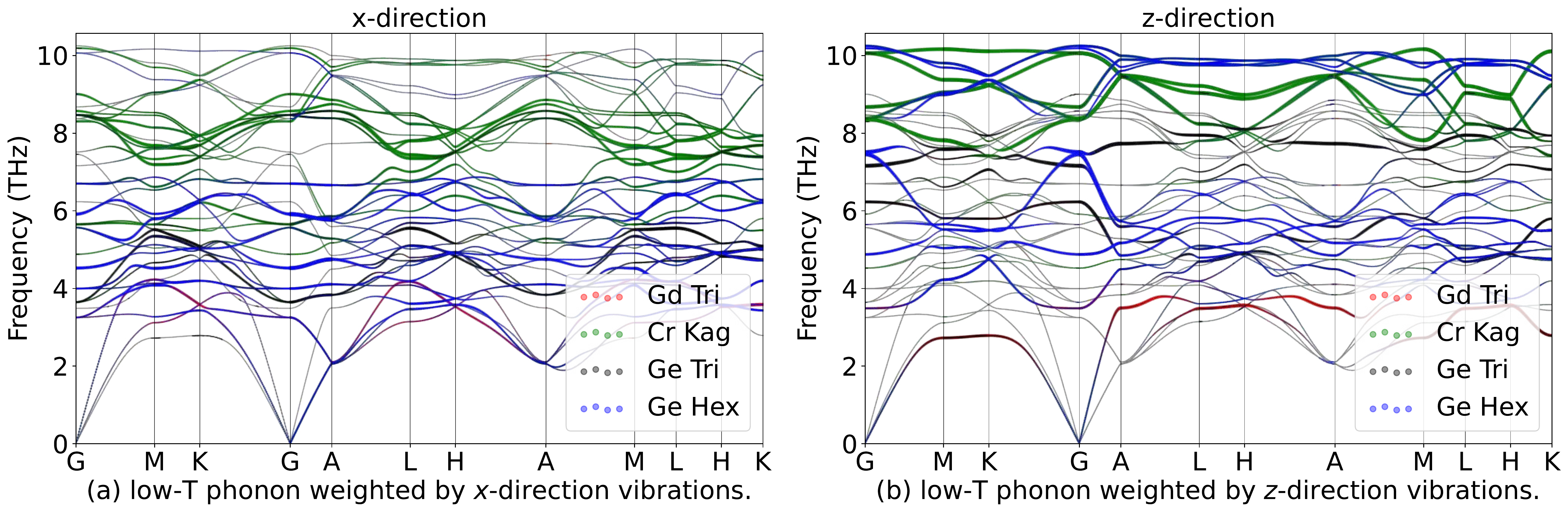}
\hspace{5pt}\label{fig:GdCr6Ge6_relaxed_High_xz}
\includegraphics[width=0.85\textwidth]{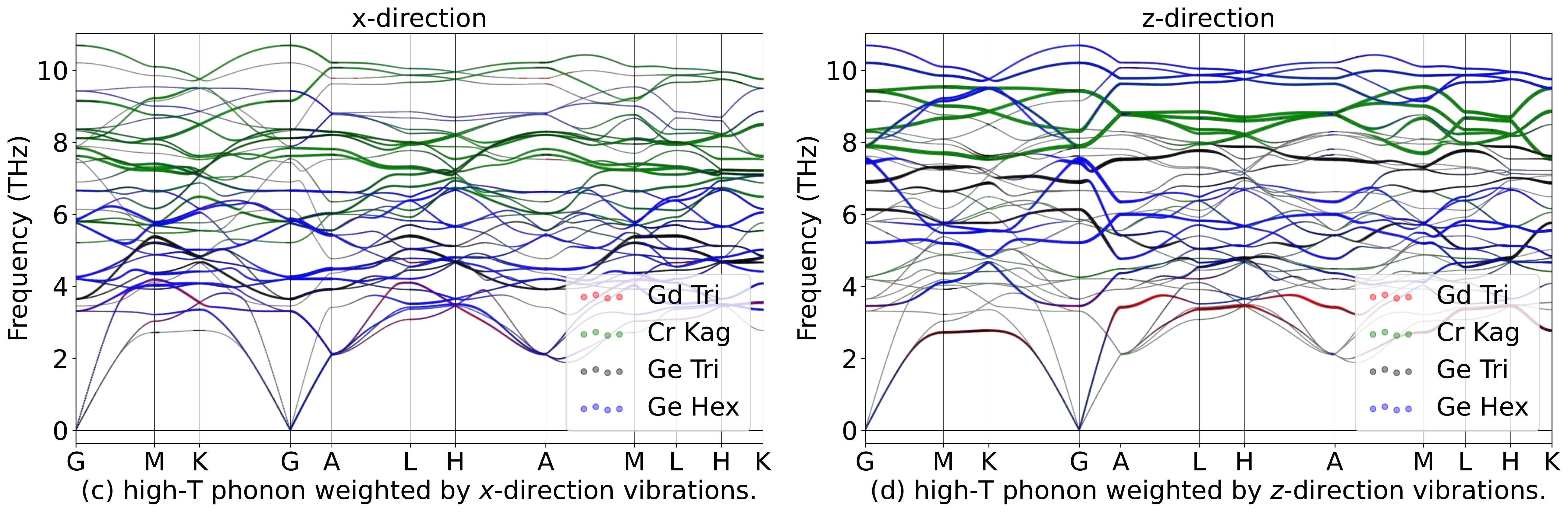}
\caption{Phonon spectrum for GdCr$_6$Ge$_6$in in-plane ($x$) and out-of-plane ($z$) directions.}
\label{fig:GdCr6Ge6phoxyz}
\end{figure}

\begin{table}[htbp]
\global\long\def\arraystretch{1.12}
\captionsetup{width=.95\textwidth}
\caption{IRREPs at high-symmetry points for GdCr$_6$Ge$_6$ phonon spectrum at Low-T.}
\label{tab:191_GdCr6Ge6_Low}
\setlength{\tabcolsep}{1.mm}{
}
\end{table}

\begin{table}[htbp]
\global\long\def\arraystretch{1.12}
\captionsetup{width=.95\textwidth}
\caption{IRREPs at high-symmetry points for GdCr$_6$Ge$_6$ phonon spectrum at High-T.}
\label{tab:191_GdCr6Ge6_High}
\setlength{\tabcolsep}{1.mm}{
%
}
\end{table}
\subsubsection{\label{sms2sec:TbCr6Ge6}\hyperref[smsec:col]{\texorpdfstring{TbCr$_6$Ge$_6$}{}}~{\color{green}  (High-T stable, Low-T stable) }}

\begin{table}[htbp]
\global\long\def\arraystretch{1.12}
\captionsetup{width=.85\textwidth}
\caption{Structure parameters of TbCr$_6$Ge$_6$ after relaxation. It contains 401 electrons. }
\label{tab:TbCr6Ge6}
\setlength{\tabcolsep}{4mm}{
%
}
\end{table}

\begin{figure}[htbp]
\centering
\includegraphics[width=0.85\textwidth]{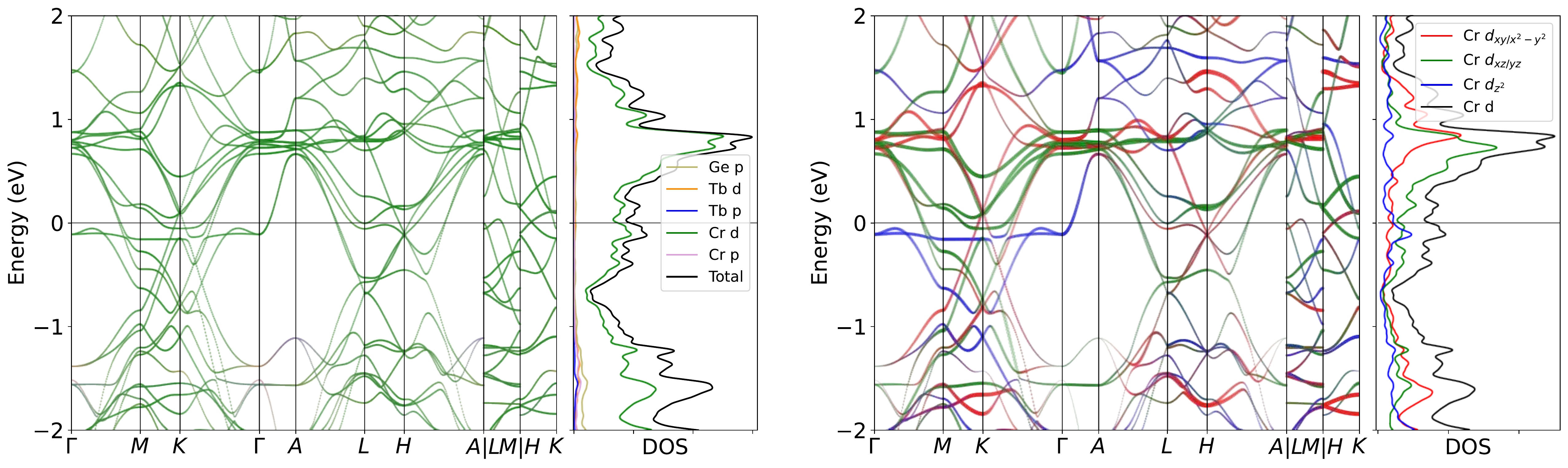}
\caption{Band structure for TbCr$_6$Ge$_6$ without SOC. The projection of Cr $3d$ orbitals is also presented. The band structures clearly reveal the dominating of Cr $3d$ orbitals  near the Fermi level, as well as the pronounced peak.}
\label{fig:TbCr6Ge6band}
\end{figure}

\begin{figure}[H]
\centering
\subfloat[Relaxed phonon at low-T.]{
\label{fig:TbCr6Ge6_relaxed_Low}
\includegraphics[width=0.45\textwidth]{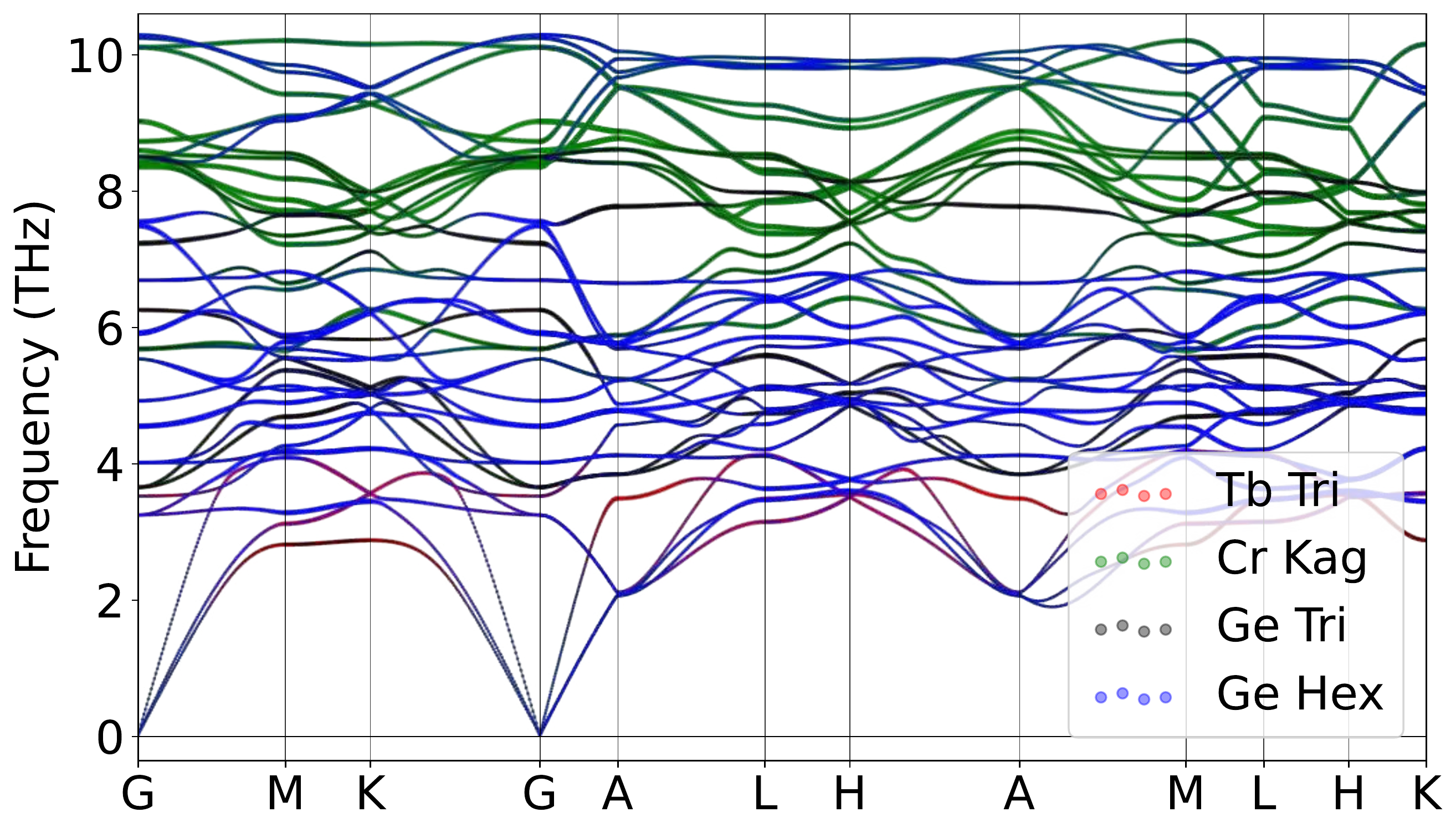}
}
\hspace{5pt}\subfloat[Relaxed phonon at high-T.]{
\label{fig:TbCr6Ge6_relaxed_High}
\includegraphics[width=0.45\textwidth]{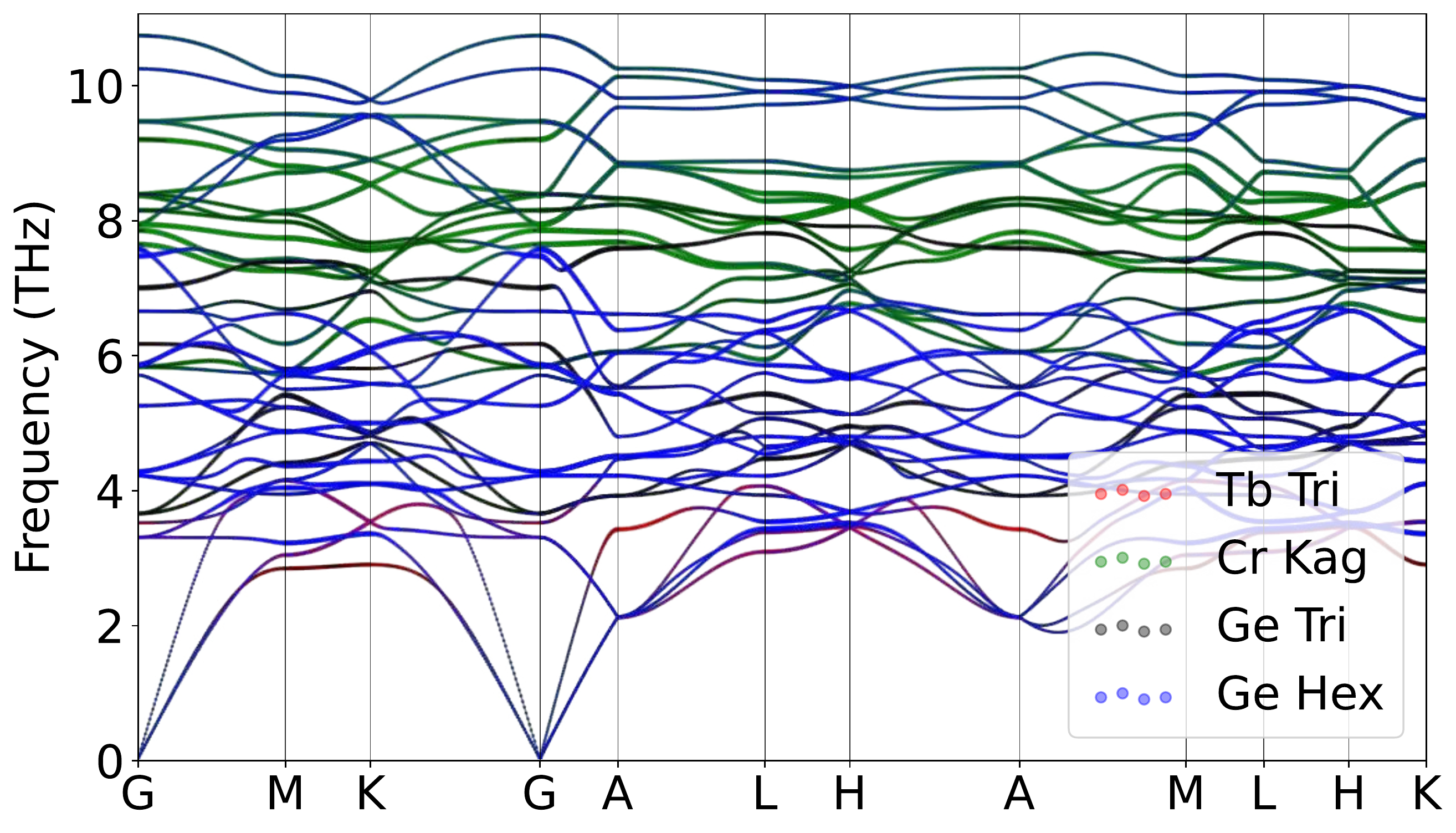}
}
\caption{Atomically projected phonon spectrum for TbCr$_6$Ge$_6$.}
\label{fig:TbCr6Ge6pho}
\end{figure}

\begin{figure}[H]
\centering
\label{fig:TbCr6Ge6_relaxed_Low_xz}
\includegraphics[width=0.85\textwidth]{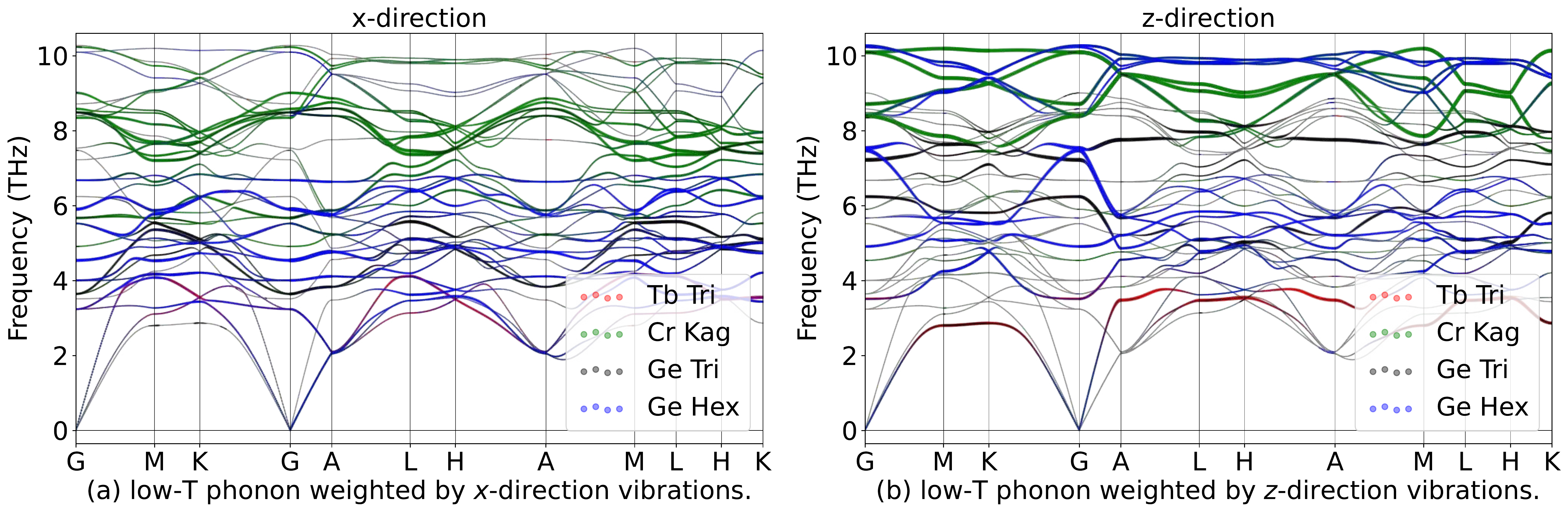}
\hspace{5pt}\label{fig:TbCr6Ge6_relaxed_High_xz}
\includegraphics[width=0.85\textwidth]{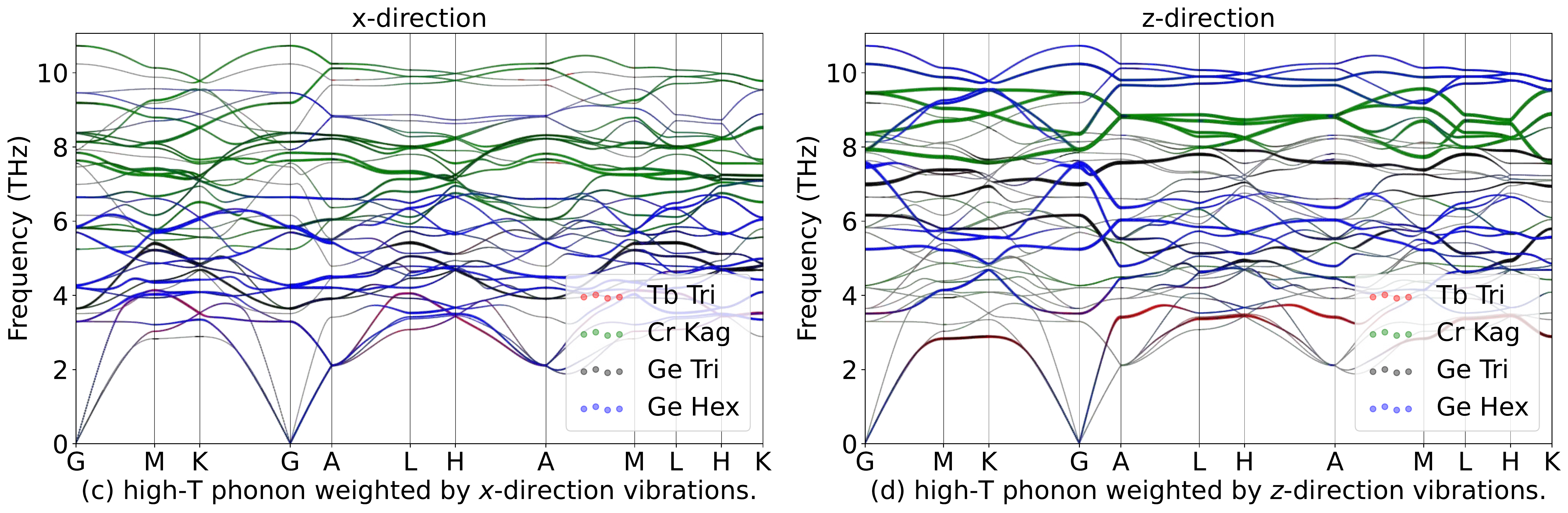}
\caption{Phonon spectrum for TbCr$_6$Ge$_6$in in-plane ($x$) and out-of-plane ($z$) directions.}
\label{fig:TbCr6Ge6phoxyz}
\end{figure}

\begin{table}[htbp]
\global\long\def\arraystretch{1.12}
\captionsetup{width=.95\textwidth}
\caption{IRREPs at high-symmetry points for TbCr$_6$Ge$_6$ phonon spectrum at Low-T.}
\label{tab:191_TbCr6Ge6_Low}
\setlength{\tabcolsep}{1.mm}{
}
\end{table}

\begin{table}[htbp]
\global\long\def\arraystretch{1.12}
\captionsetup{width=.95\textwidth}
\caption{IRREPs at high-symmetry points for TbCr$_6$Ge$_6$ phonon spectrum at High-T.}
\label{tab:191_TbCr6Ge6_High}
\setlength{\tabcolsep}{1.mm}{
%
}
\end{table}
\subsubsection{\label{sms2sec:DyCr6Ge6}\hyperref[smsec:col]{\texorpdfstring{DyCr$_6$Ge$_6$}{}}~{\color{green}  (High-T stable, Low-T stable) }}

\begin{table}[htbp]
\global\long\def\arraystretch{1.12}
\captionsetup{width=.85\textwidth}
\caption{Structure parameters of DyCr$_6$Ge$_6$ after relaxation. It contains 402 electrons. }
\label{tab:DyCr6Ge6}
\setlength{\tabcolsep}{4mm}{
%
}
\end{table}

\begin{figure}[htbp]
\centering
\includegraphics[width=0.85\textwidth]{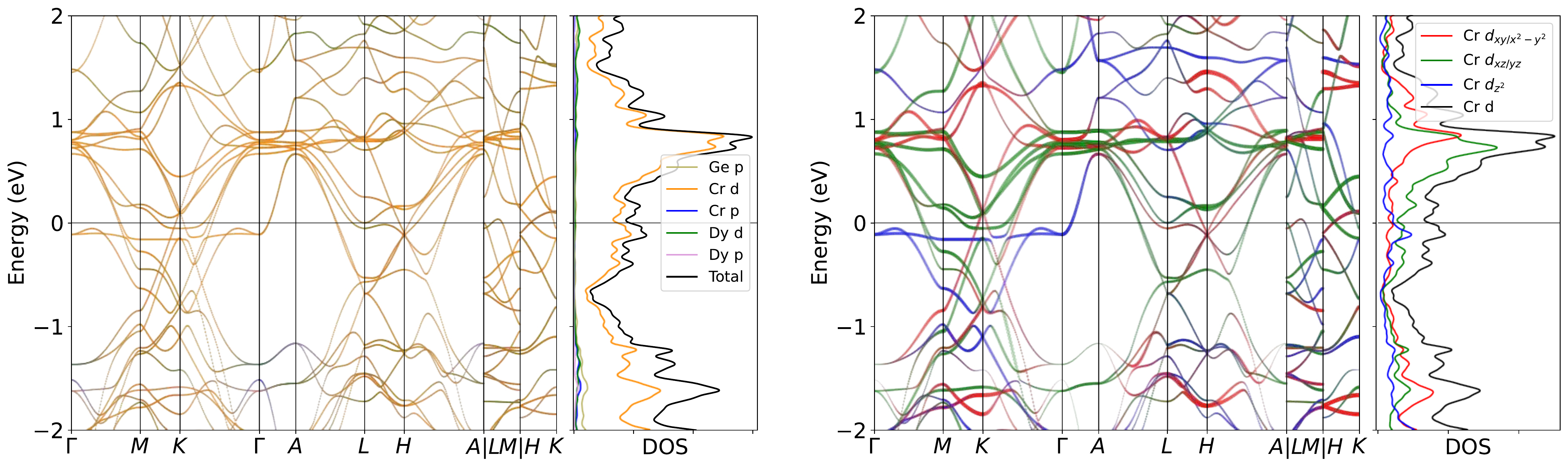}
\caption{Band structure for DyCr$_6$Ge$_6$ without SOC. The projection of Cr $3d$ orbitals is also presented. The band structures clearly reveal the dominating of Cr $3d$ orbitals  near the Fermi level, as well as the pronounced peak.}
\label{fig:DyCr6Ge6band}
\end{figure}

\begin{figure}[H]
\centering
\subfloat[Relaxed phonon at low-T.]{
\label{fig:DyCr6Ge6_relaxed_Low}
\includegraphics[width=0.45\textwidth]{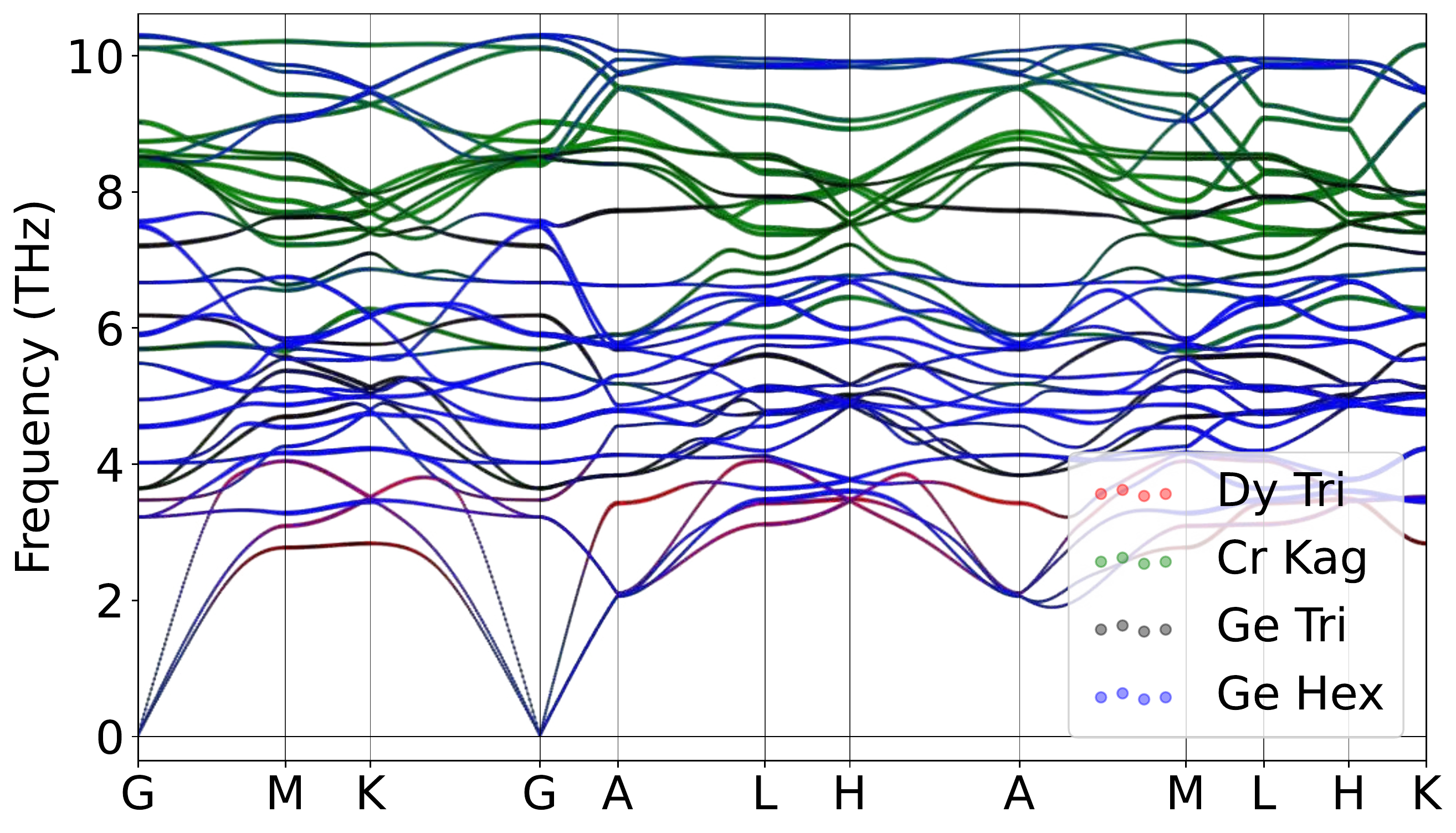}
}
\hspace{5pt}\subfloat[Relaxed phonon at high-T.]{
\label{fig:DyCr6Ge6_relaxed_High}
\includegraphics[width=0.45\textwidth]{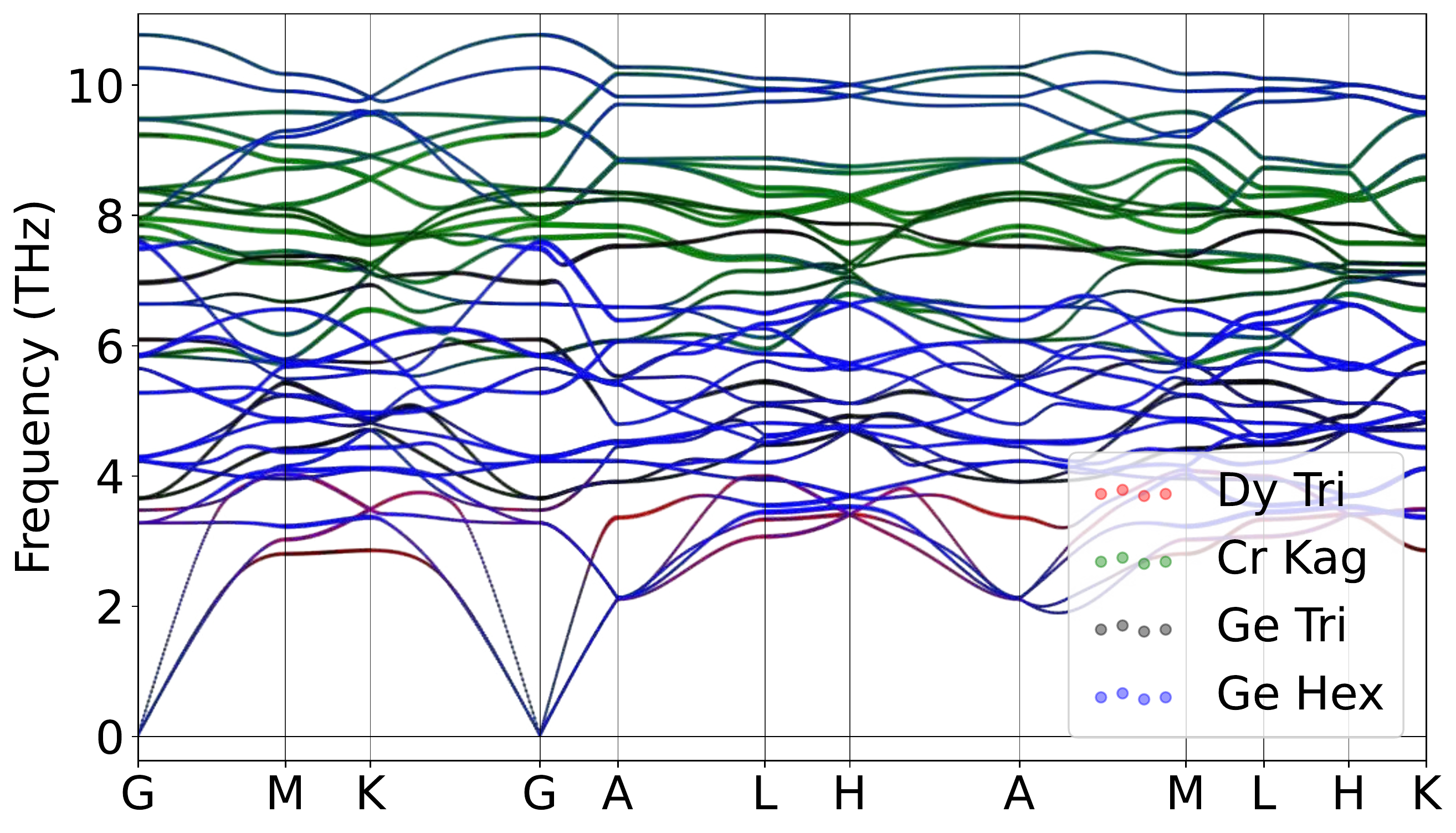}
}
\caption{Atomically projected phonon spectrum for DyCr$_6$Ge$_6$.}
\label{fig:DyCr6Ge6pho}
\end{figure}

\begin{figure}[H]
\centering
\label{fig:DyCr6Ge6_relaxed_Low_xz}
\includegraphics[width=0.85\textwidth]{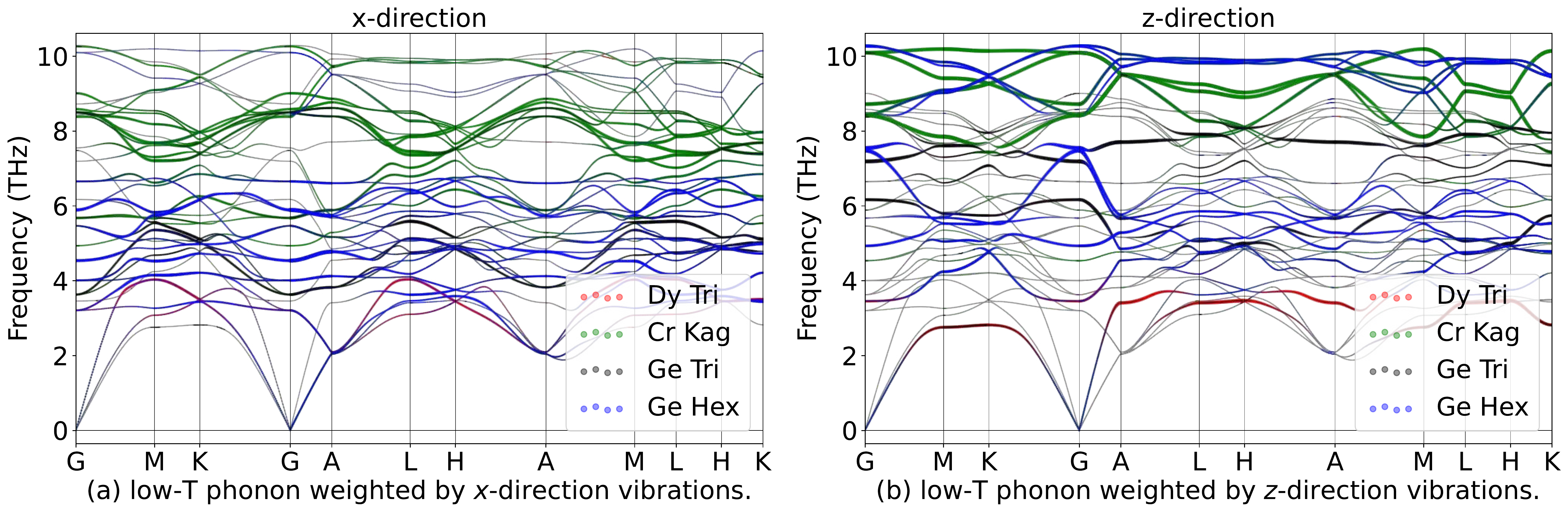}
\hspace{5pt}\label{fig:DyCr6Ge6_relaxed_High_xz}
\includegraphics[width=0.85\textwidth]{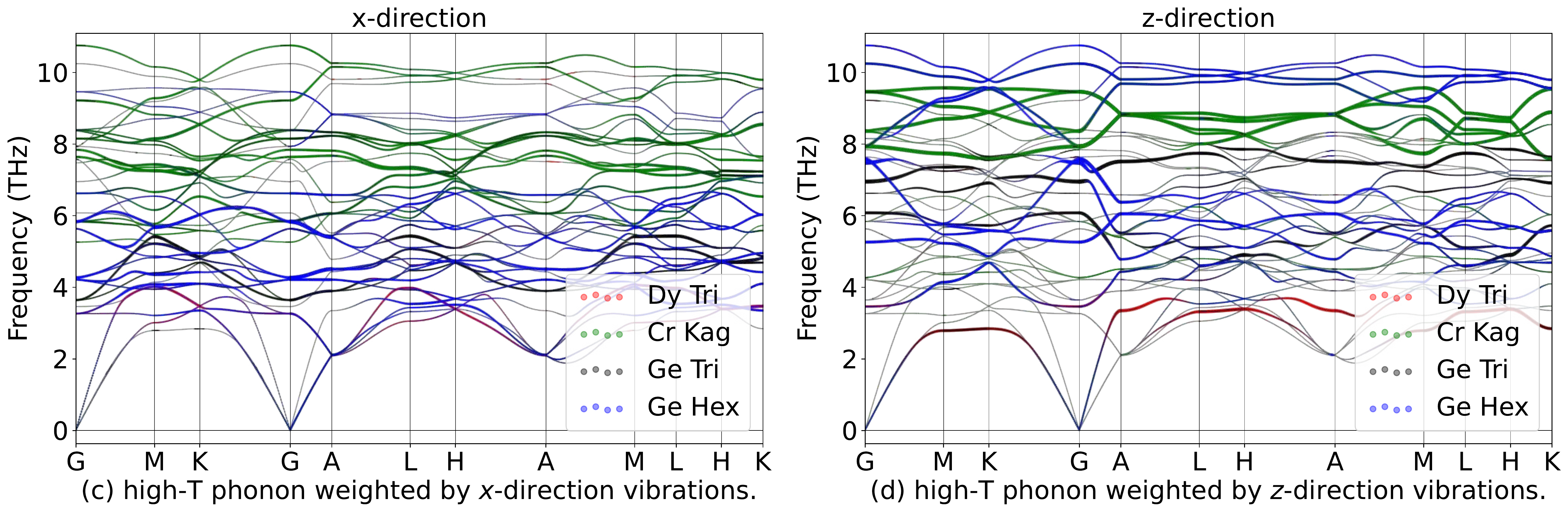}
\caption{Phonon spectrum for DyCr$_6$Ge$_6$in in-plane ($x$) and out-of-plane ($z$) directions.}
\label{fig:DyCr6Ge6phoxyz}
\end{figure}

\begin{table}[htbp]
\global\long\def\arraystretch{1.12}
\captionsetup{width=.95\textwidth}
\caption{IRREPs at high-symmetry points for DyCr$_6$Ge$_6$ phonon spectrum at Low-T.}
\label{tab:191_DyCr6Ge6_Low}
\setlength{\tabcolsep}{1.mm}{
}
\end{table}

\begin{table}[htbp]
\global\long\def\arraystretch{1.12}
\captionsetup{width=.95\textwidth}
\caption{IRREPs at high-symmetry points for DyCr$_6$Ge$_6$ phonon spectrum at High-T.}
\label{tab:191_DyCr6Ge6_High}
\setlength{\tabcolsep}{1.mm}{
%
}
\end{table}
\subsubsection{\label{sms2sec:HoCr6Ge6}\hyperref[smsec:col]{\texorpdfstring{HoCr$_6$Ge$_6$}{}}~{\color{green}  (High-T stable, Low-T stable) }}

\begin{table}[htbp]
\global\long\def\arraystretch{1.12}
\captionsetup{width=.85\textwidth}
\caption{Structure parameters of HoCr$_6$Ge$_6$ after relaxation. It contains 403 electrons. }
\label{tab:HoCr6Ge6}
\setlength{\tabcolsep}{4mm}{
%
}
\end{table}

\begin{figure}[htbp]
\centering
\includegraphics[width=0.85\textwidth]{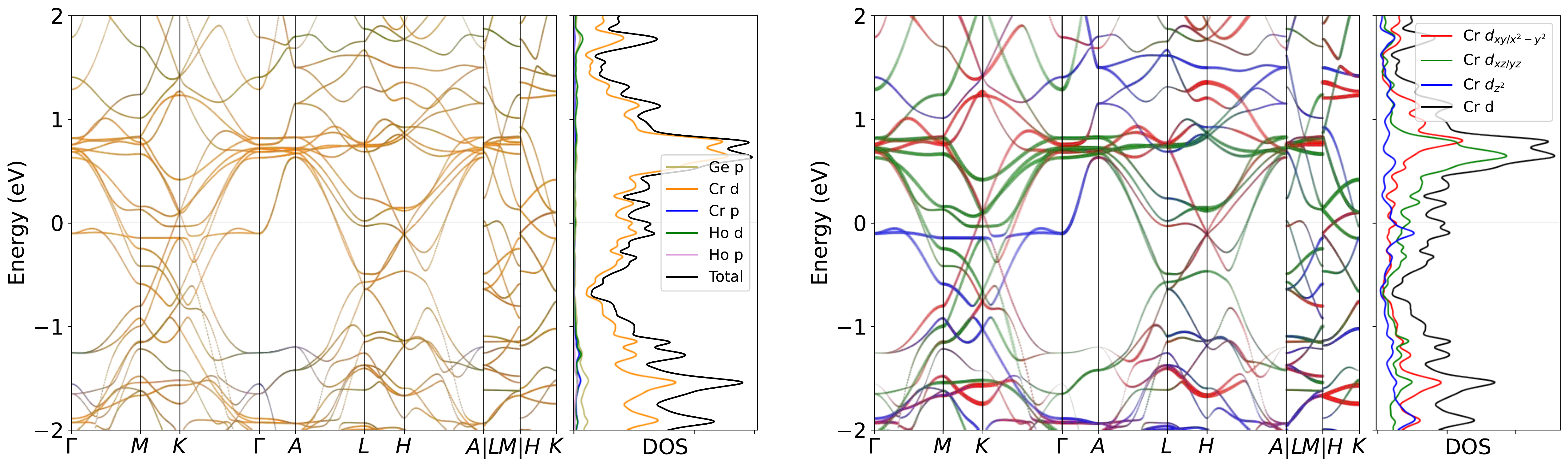}
\caption{Band structure for HoCr$_6$Ge$_6$ without SOC. The projection of Cr $3d$ orbitals is also presented. The band structures clearly reveal the dominating of Cr $3d$ orbitals  near the Fermi level, as well as the pronounced peak.}
\label{fig:HoCr6Ge6band}
\end{figure}

\begin{figure}[H]
\centering
\subfloat[Relaxed phonon at low-T.]{
\label{fig:HoCr6Ge6_relaxed_Low}
\includegraphics[width=0.45\textwidth]{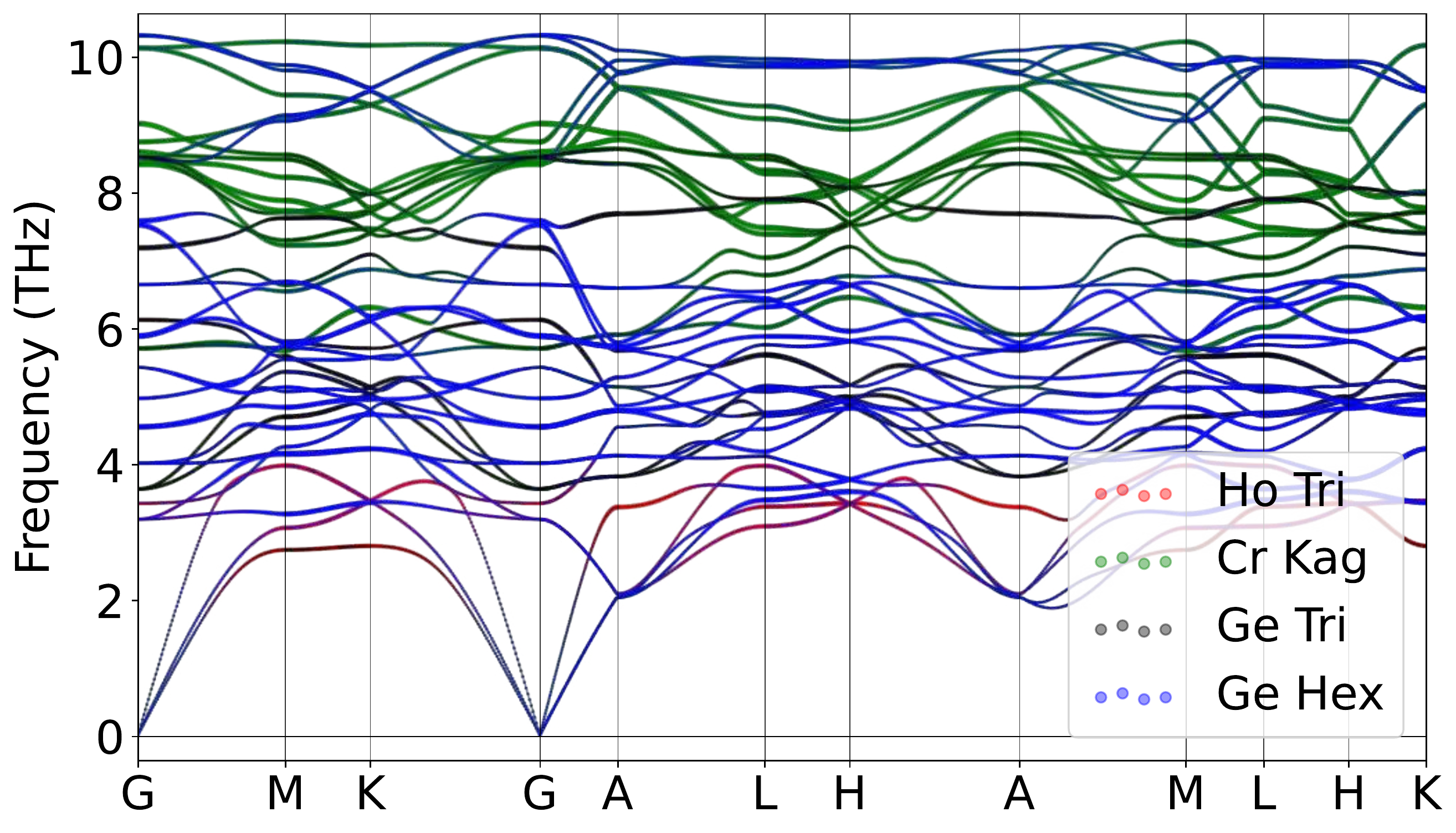}
}
\hspace{5pt}\subfloat[Relaxed phonon at high-T.]{
\label{fig:HoCr6Ge6_relaxed_High}
\includegraphics[width=0.45\textwidth]{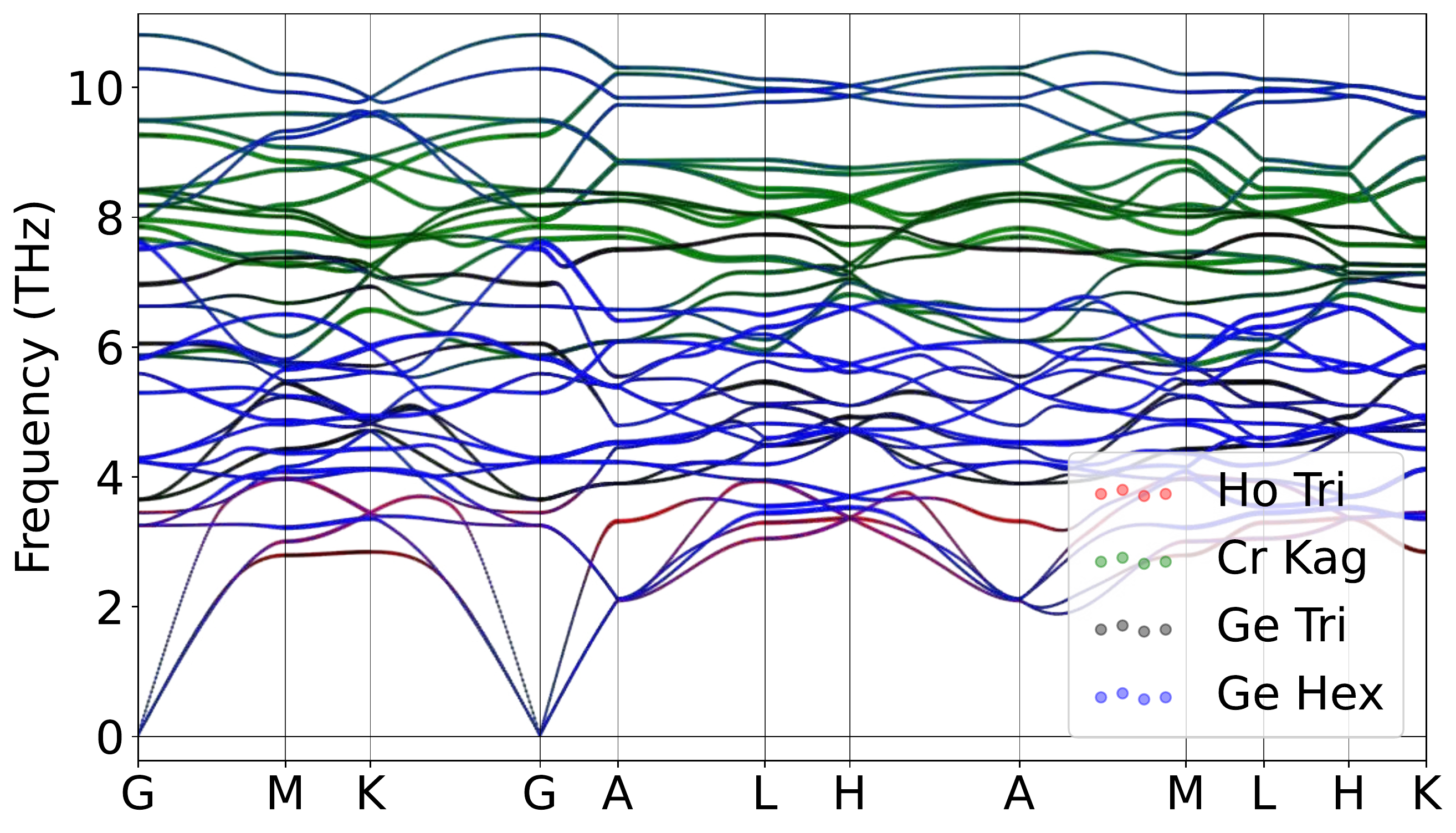}
}
\caption{Atomically projected phonon spectrum for HoCr$_6$Ge$_6$.}
\label{fig:HoCr6Ge6pho}
\end{figure}

\begin{figure}[H]
\centering
\label{fig:HoCr6Ge6_relaxed_Low_xz}
\includegraphics[width=0.85\textwidth]{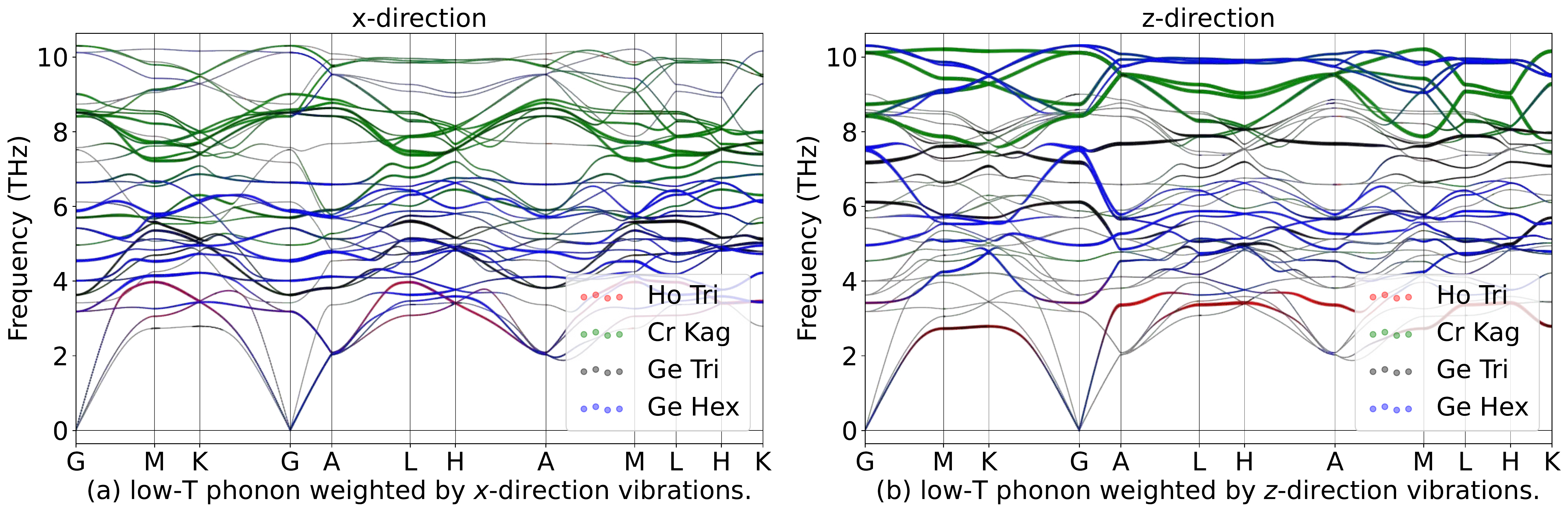}
\hspace{5pt}\label{fig:HoCr6Ge6_relaxed_High_xz}
\includegraphics[width=0.85\textwidth]{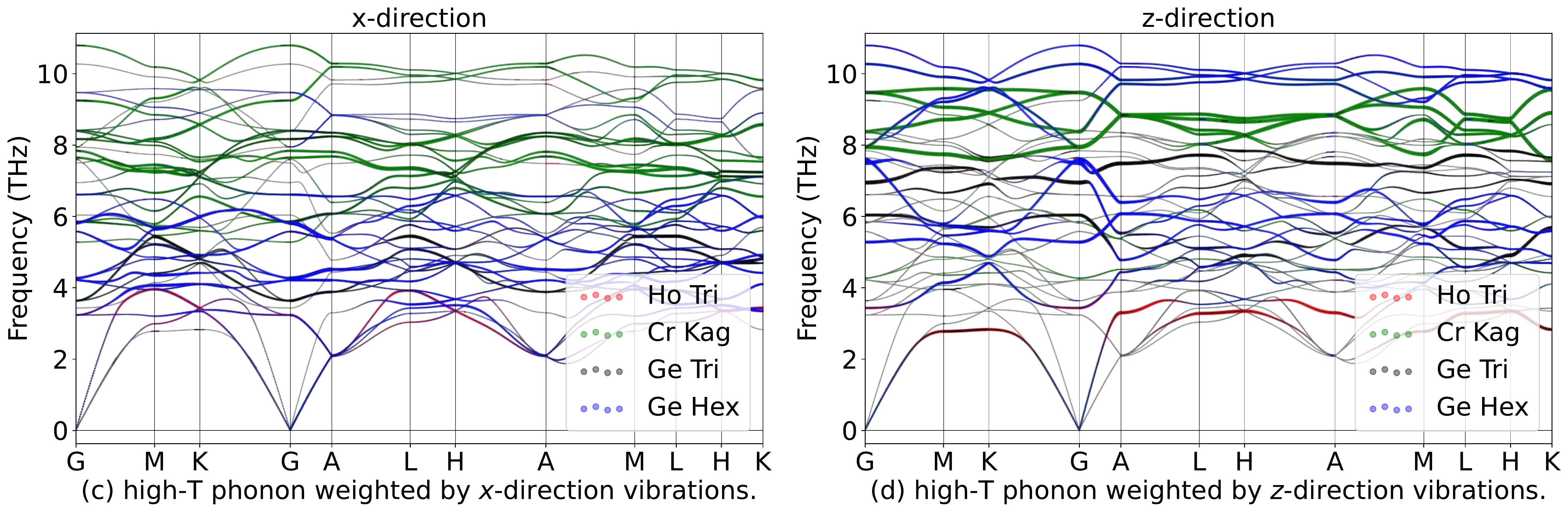}
\caption{Phonon spectrum for HoCr$_6$Ge$_6$in in-plane ($x$) and out-of-plane ($z$) directions.}
\label{fig:HoCr6Ge6phoxyz}
\end{figure}

\begin{table}[htbp]
\global\long\def\arraystretch{1.12}
\captionsetup{width=.95\textwidth}
\caption{IRREPs at high-symmetry points for HoCr$_6$Ge$_6$ phonon spectrum at Low-T.}
\label{tab:191_HoCr6Ge6_Low}
\setlength{\tabcolsep}{1.mm}{
}
\end{table}

\begin{table}[htbp]
\global\long\def\arraystretch{1.12}
\captionsetup{width=.95\textwidth}
\caption{IRREPs at high-symmetry points for HoCr$_6$Ge$_6$ phonon spectrum at High-T.}
\label{tab:191_HoCr6Ge6_High}
\setlength{\tabcolsep}{1.mm}{
%
}
\end{table}
\subsubsection{\label{sms2sec:ErCr6Ge6}\hyperref[smsec:col]{\texorpdfstring{ErCr$_6$Ge$_6$}{}}~{\color{green}  (High-T stable, Low-T stable) }}

\begin{table}[htbp]
\global\long\def\arraystretch{1.12}
\captionsetup{width=.85\textwidth}
\caption{Structure parameters of ErCr$_6$Ge$_6$ after relaxation. It contains 404 electrons. }
\label{tab:ErCr6Ge6}
\setlength{\tabcolsep}{4mm}{
%
}
\end{table}

\begin{figure}[htbp]
\centering
\includegraphics[width=0.85\textwidth]{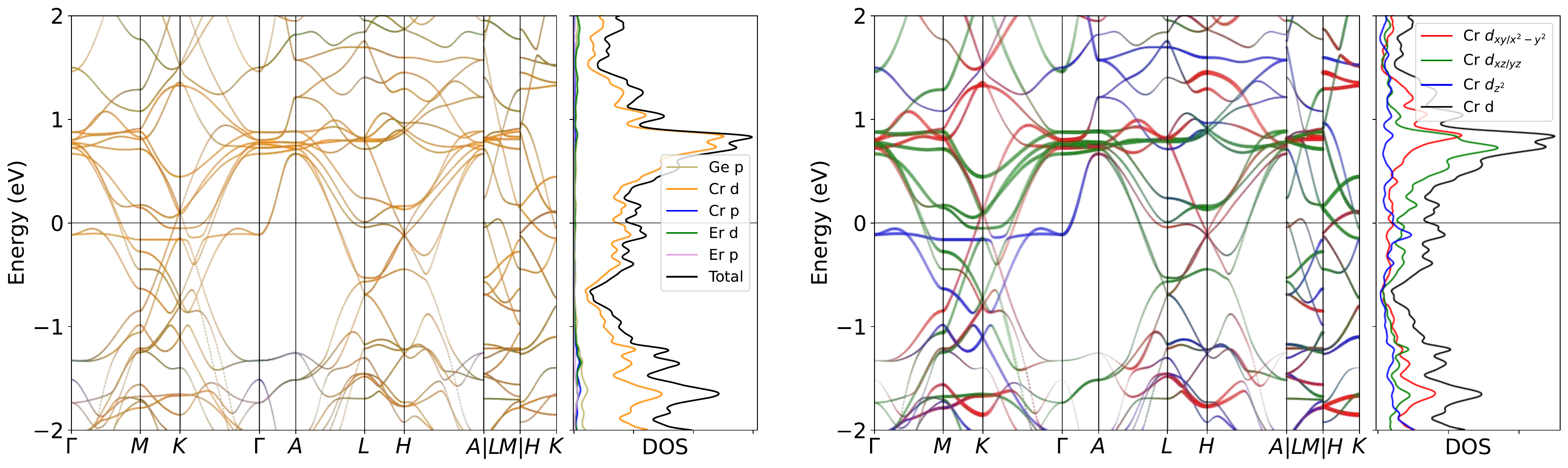}
\caption{Band structure for ErCr$_6$Ge$_6$ without SOC. The projection of Cr $3d$ orbitals is also presented. The band structures clearly reveal the dominating of Cr $3d$ orbitals  near the Fermi level, as well as the pronounced peak.}
\label{fig:ErCr6Ge6band}
\end{figure}

\begin{figure}[H]
\centering
\subfloat[Relaxed phonon at low-T.]{
\label{fig:ErCr6Ge6_relaxed_Low}
\includegraphics[width=0.45\textwidth]{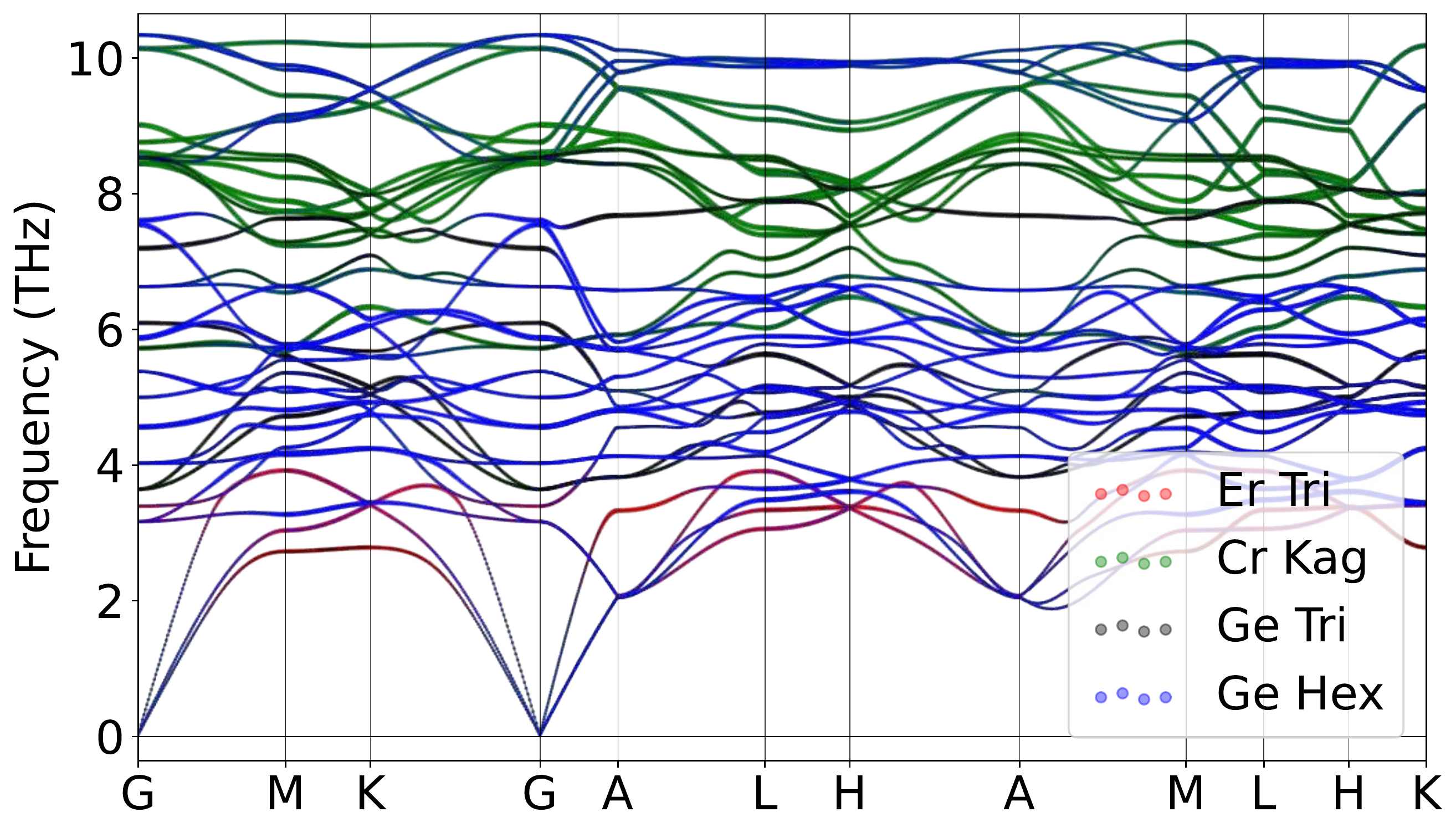}
}
\hspace{5pt}\subfloat[Relaxed phonon at high-T.]{
\label{fig:ErCr6Ge6_relaxed_High}
\includegraphics[width=0.45\textwidth]{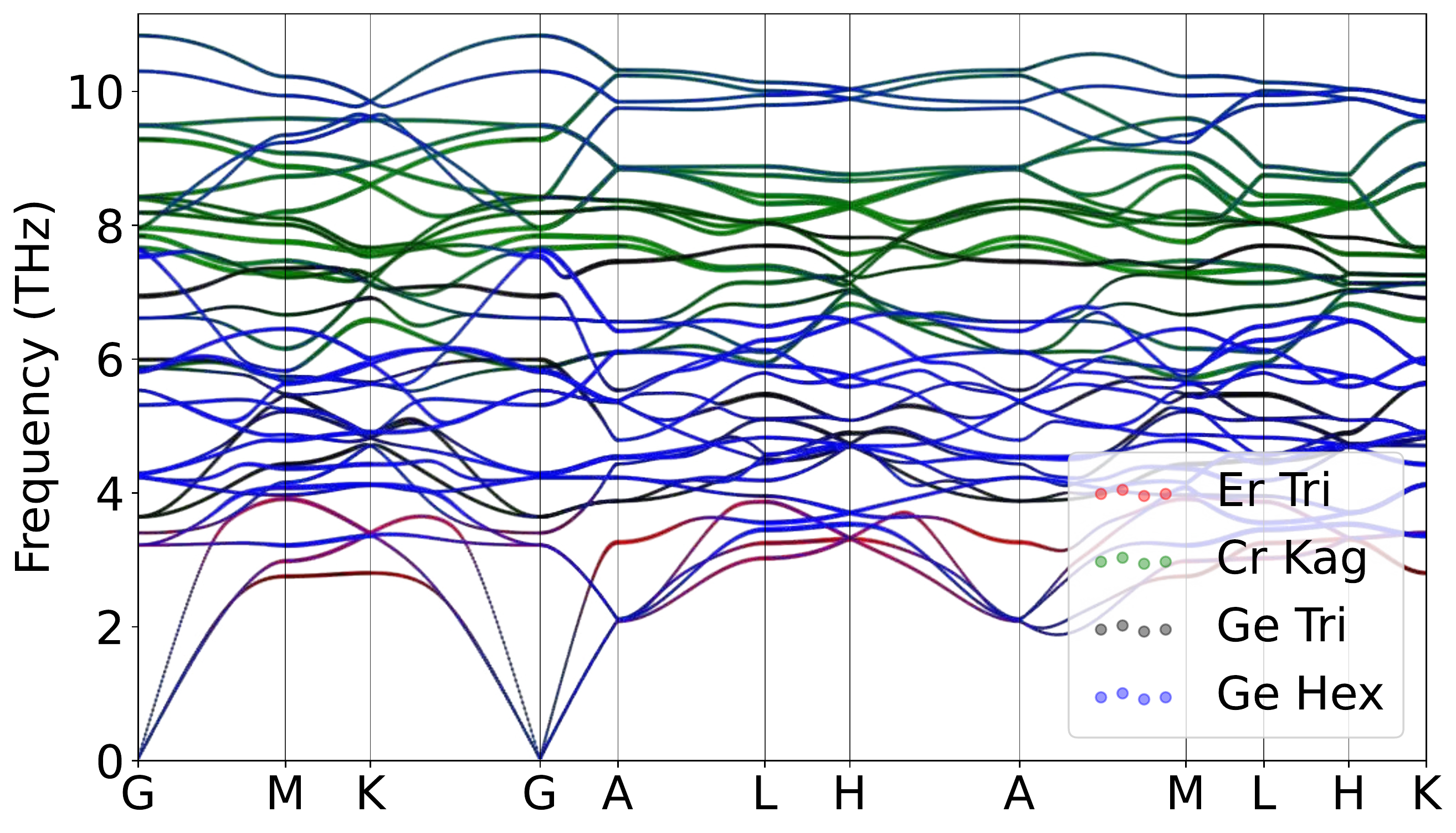}
}
\caption{Atomically projected phonon spectrum for ErCr$_6$Ge$_6$.}
\label{fig:ErCr6Ge6pho}
\end{figure}

\begin{figure}[H]
\centering
\label{fig:ErCr6Ge6_relaxed_Low_xz}
\includegraphics[width=0.85\textwidth]{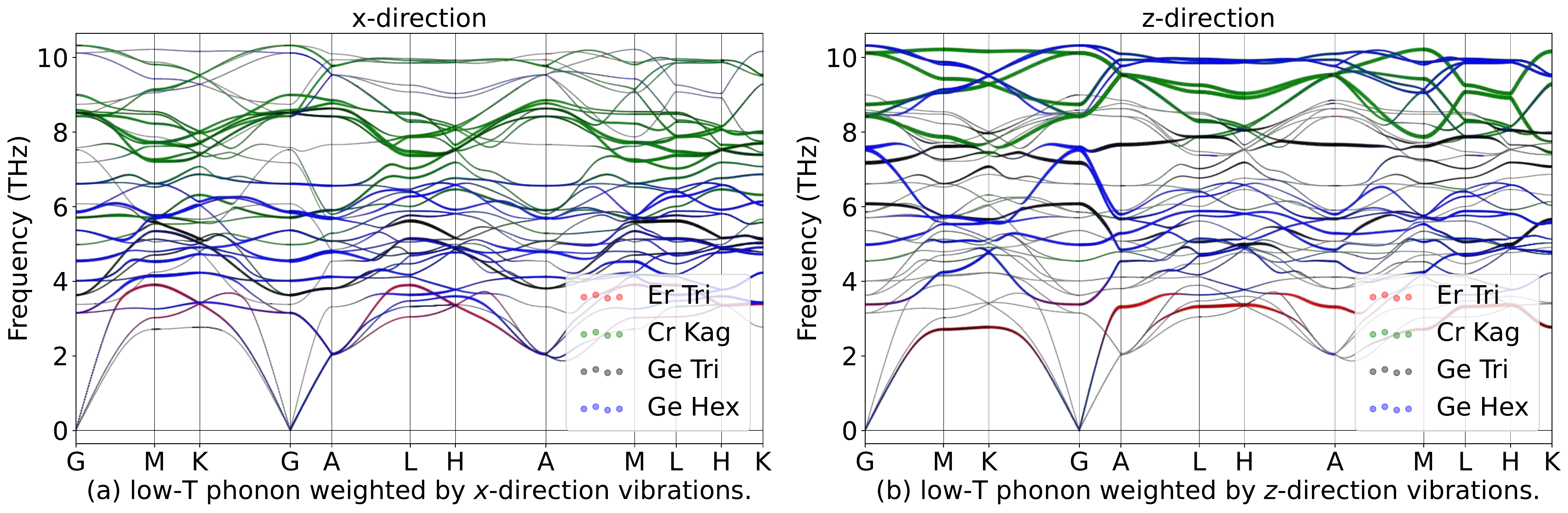}
\hspace{5pt}\label{fig:ErCr6Ge6_relaxed_High_xz}
\includegraphics[width=0.85\textwidth]{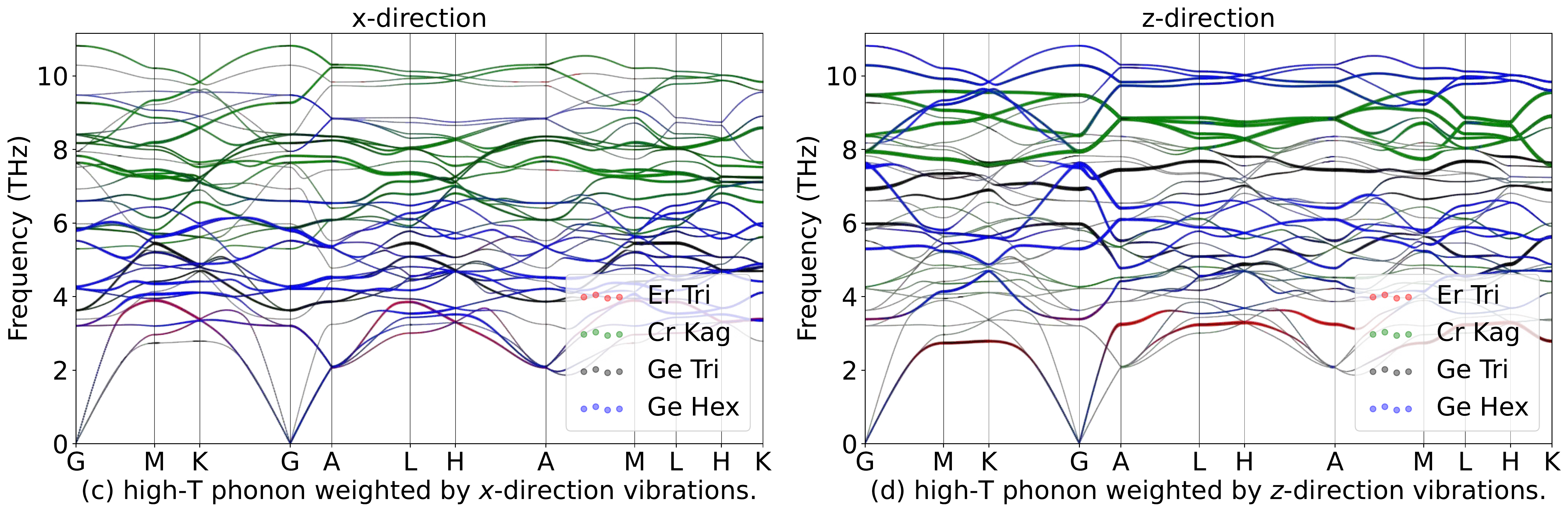}
\caption{Phonon spectrum for ErCr$_6$Ge$_6$in in-plane ($x$) and out-of-plane ($z$) directions.}
\label{fig:ErCr6Ge6phoxyz}
\end{figure}

\begin{table}[htbp]
\global\long\def\arraystretch{1.12}
\captionsetup{width=.95\textwidth}
\caption{IRREPs at high-symmetry points for ErCr$_6$Ge$_6$ phonon spectrum at Low-T.}
\label{tab:191_ErCr6Ge6_Low}
\setlength{\tabcolsep}{1.mm}{
}
\end{table}

\begin{table}[htbp]
\global\long\def\arraystretch{1.12}
\captionsetup{width=.95\textwidth}
\caption{IRREPs at high-symmetry points for ErCr$_6$Ge$_6$ phonon spectrum at High-T.}
\label{tab:191_ErCr6Ge6_High}
\setlength{\tabcolsep}{1.mm}{
%
}
\end{table}
\subsubsection{\label{sms2sec:TmCr6Ge6}\hyperref[smsec:col]{\texorpdfstring{TmCr$_6$Ge$_6$}{}}~{\color{green}  (High-T stable, Low-T stable) }}

\begin{table}[htbp]
\global\long\def\arraystretch{1.12}
\captionsetup{width=.85\textwidth}
\caption{Structure parameters of TmCr$_6$Ge$_6$ after relaxation. It contains 405 electrons. }
\label{tab:TmCr6Ge6}
\setlength{\tabcolsep}{4mm}{
%
}
\end{table}

\begin{figure}[htbp]
\centering
\includegraphics[width=0.85\textwidth]{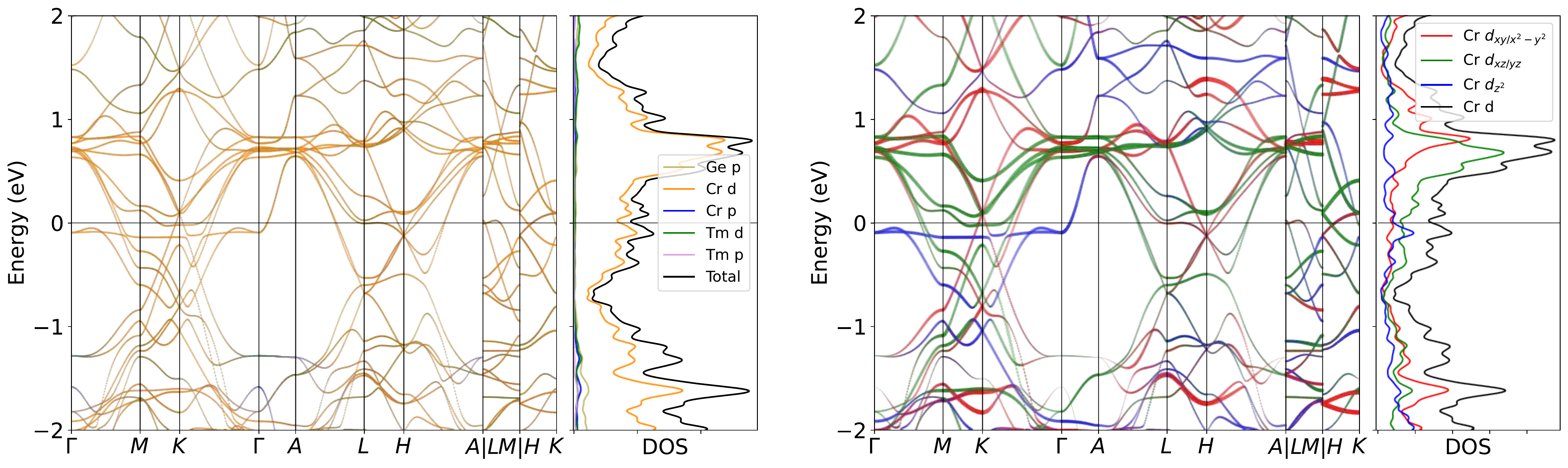}
\caption{Band structure for TmCr$_6$Ge$_6$ without SOC. The projection of Cr $3d$ orbitals is also presented. The band structures clearly reveal the dominating of Cr $3d$ orbitals  near the Fermi level, as well as the pronounced peak.}
\label{fig:TmCr6Ge6band}
\end{figure}

\begin{figure}[H]
\centering
\subfloat[Relaxed phonon at low-T.]{
\label{fig:TmCr6Ge6_relaxed_Low}
\includegraphics[width=0.45\textwidth]{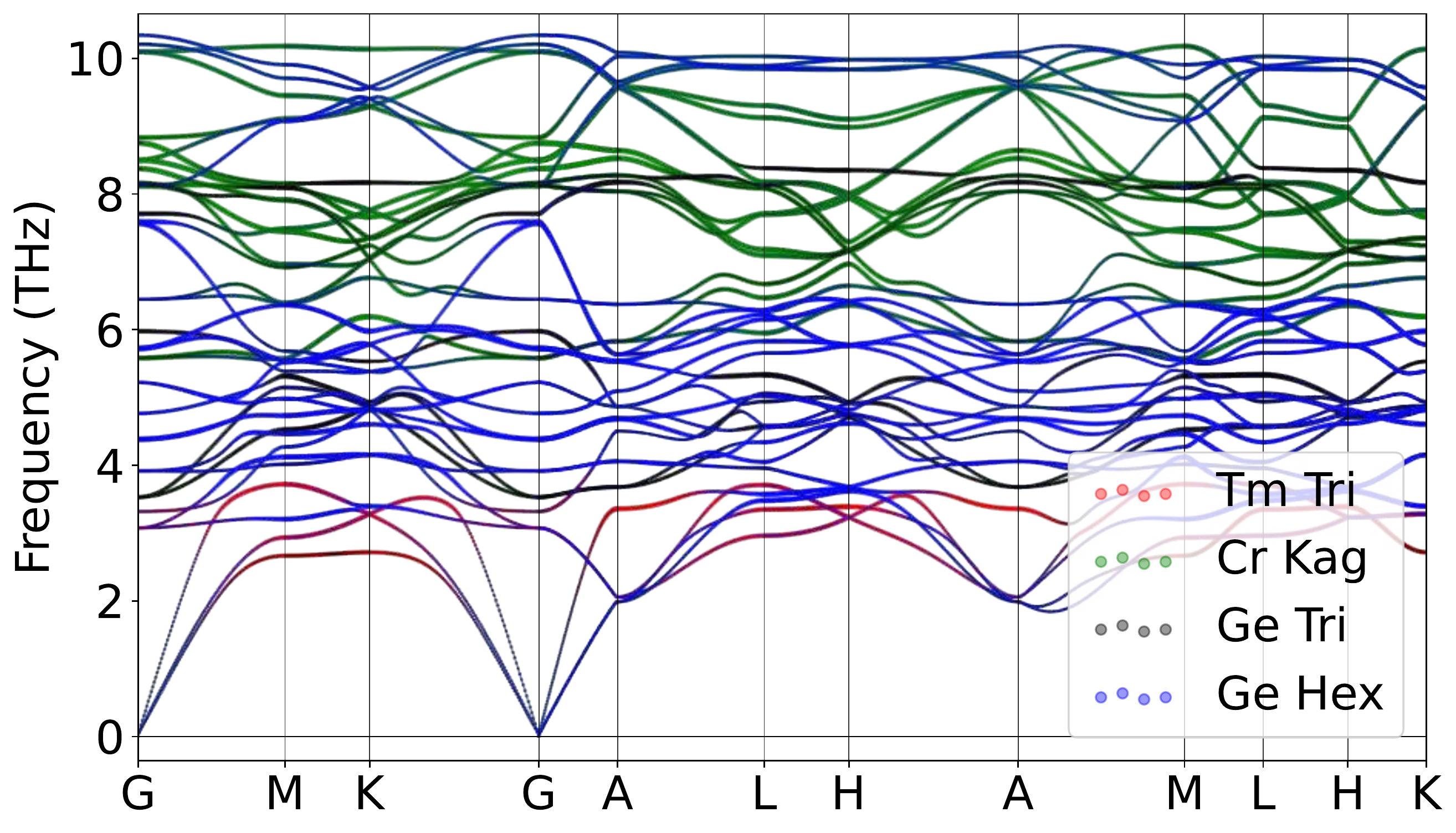}
}
\hspace{5pt}\subfloat[Relaxed phonon at high-T.]{
\label{fig:TmCr6Ge6_relaxed_High}
\includegraphics[width=0.45\textwidth]{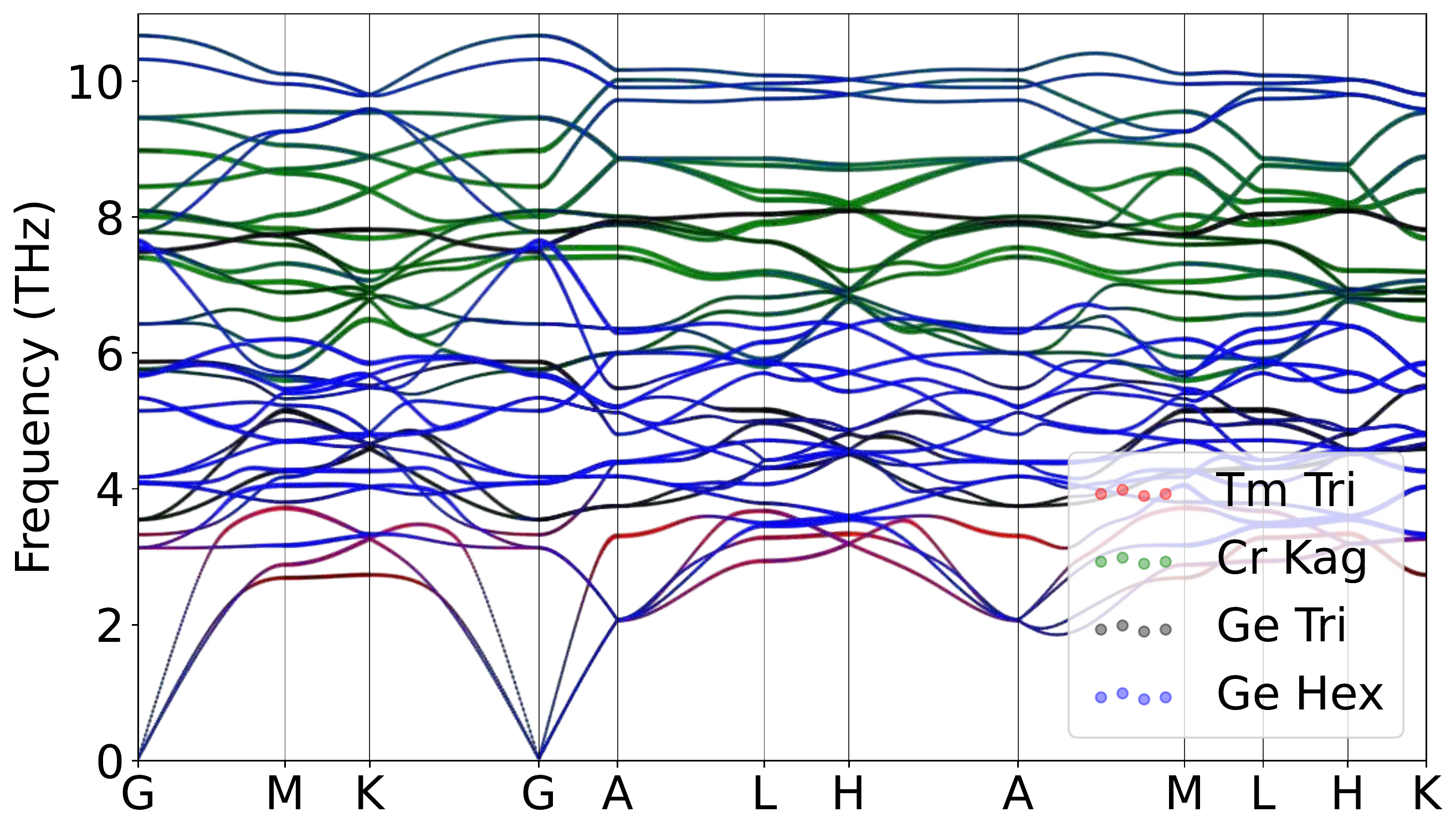}
}
\caption{Atomically projected phonon spectrum for TmCr$_6$Ge$_6$.}
\label{fig:TmCr6Ge6pho}
\end{figure}

\begin{figure}[H]
\centering
\label{fig:TmCr6Ge6_relaxed_Low_xz}
\includegraphics[width=0.85\textwidth]{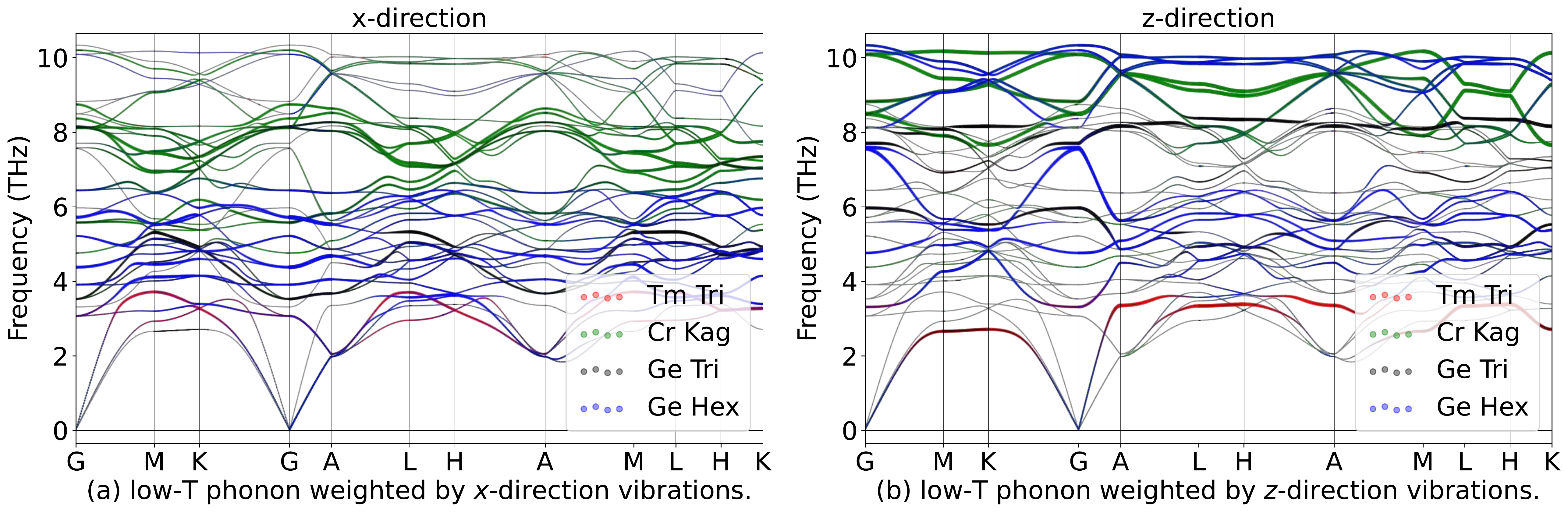}
\hspace{5pt}\label{fig:TmCr6Ge6_relaxed_High_xz}
\includegraphics[width=0.85\textwidth]{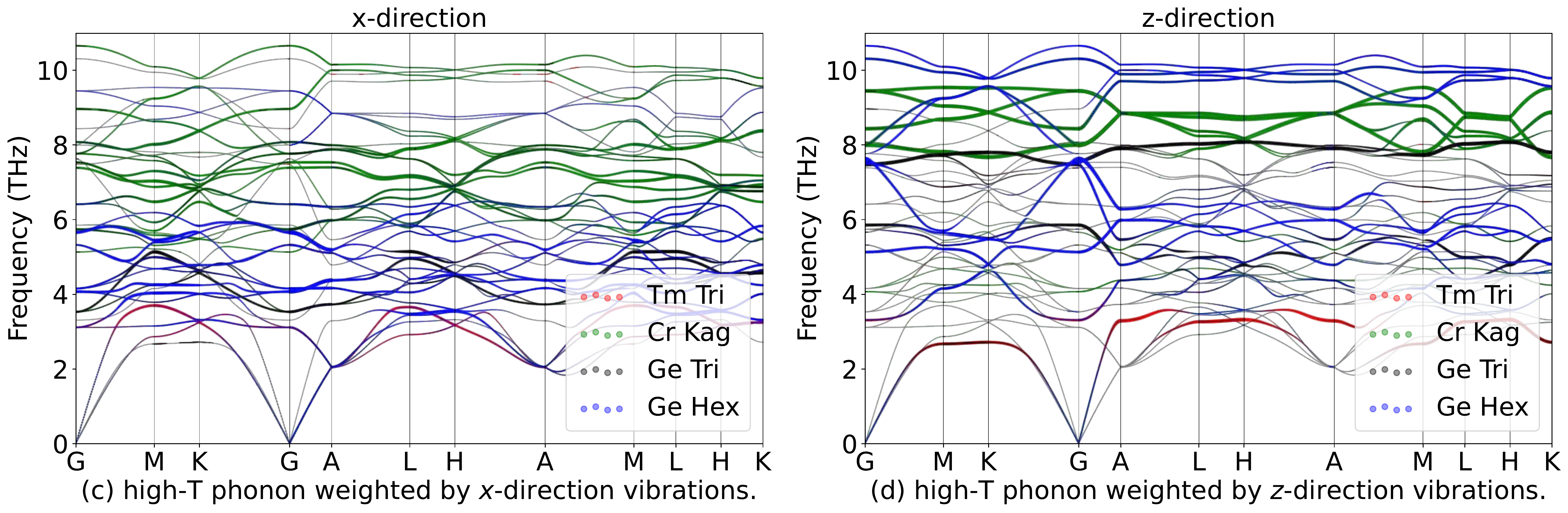}
\caption{Phonon spectrum for TmCr$_6$Ge$_6$in in-plane ($x$) and out-of-plane ($z$) directions.}
\label{fig:TmCr6Ge6phoxyz}
\end{figure}

\begin{table}[htbp]
\global\long\def\arraystretch{1.12}
\captionsetup{width=.95\textwidth}
\caption{IRREPs at high-symmetry points for TmCr$_6$Ge$_6$ phonon spectrum at Low-T.}
\label{tab:191_TmCr6Ge6_Low}
\setlength{\tabcolsep}{1.mm}{
}
\end{table}

\begin{table}[htbp]
\global\long\def\arraystretch{1.12}
\captionsetup{width=.95\textwidth}
\caption{IRREPs at high-symmetry points for TmCr$_6$Ge$_6$ phonon spectrum at High-T.}
\label{tab:191_TmCr6Ge6_High}
\setlength{\tabcolsep}{1.mm}{
%
}
\end{table}
\subsubsection{\label{sms2sec:YbCr6Ge6}\hyperref[smsec:col]{\texorpdfstring{YbCr$_6$Ge$_6$}{}}~{\color{green}  (High-T stable, Low-T stable) }}

\begin{table}[htbp]
\global\long\def\arraystretch{1.12}
\captionsetup{width=.85\textwidth}
\caption{Structure parameters of YbCr$_6$Ge$_6$ after relaxation. It contains 406 electrons. }
\label{tab:YbCr6Ge6}
\setlength{\tabcolsep}{4mm}{
%
}
\end{table}

\begin{figure}[htbp]
\centering
\includegraphics[width=0.85\textwidth]{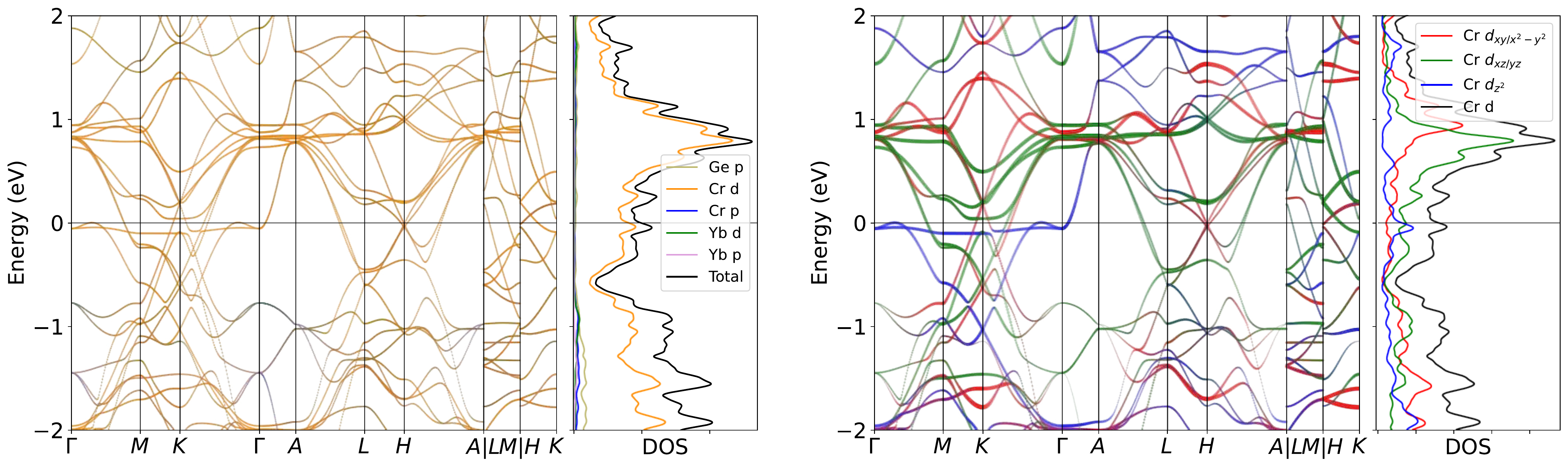}
\caption{Band structure for YbCr$_6$Ge$_6$ without SOC. The projection of Cr $3d$ orbitals is also presented. The band structures clearly reveal the dominating of Cr $3d$ orbitals  near the Fermi level, as well as the pronounced peak.}
\label{fig:YbCr6Ge6band}
\end{figure}

\begin{figure}[H]
\centering
\subfloat[Relaxed phonon at low-T.]{
\label{fig:YbCr6Ge6_relaxed_Low}
\includegraphics[width=0.45\textwidth]{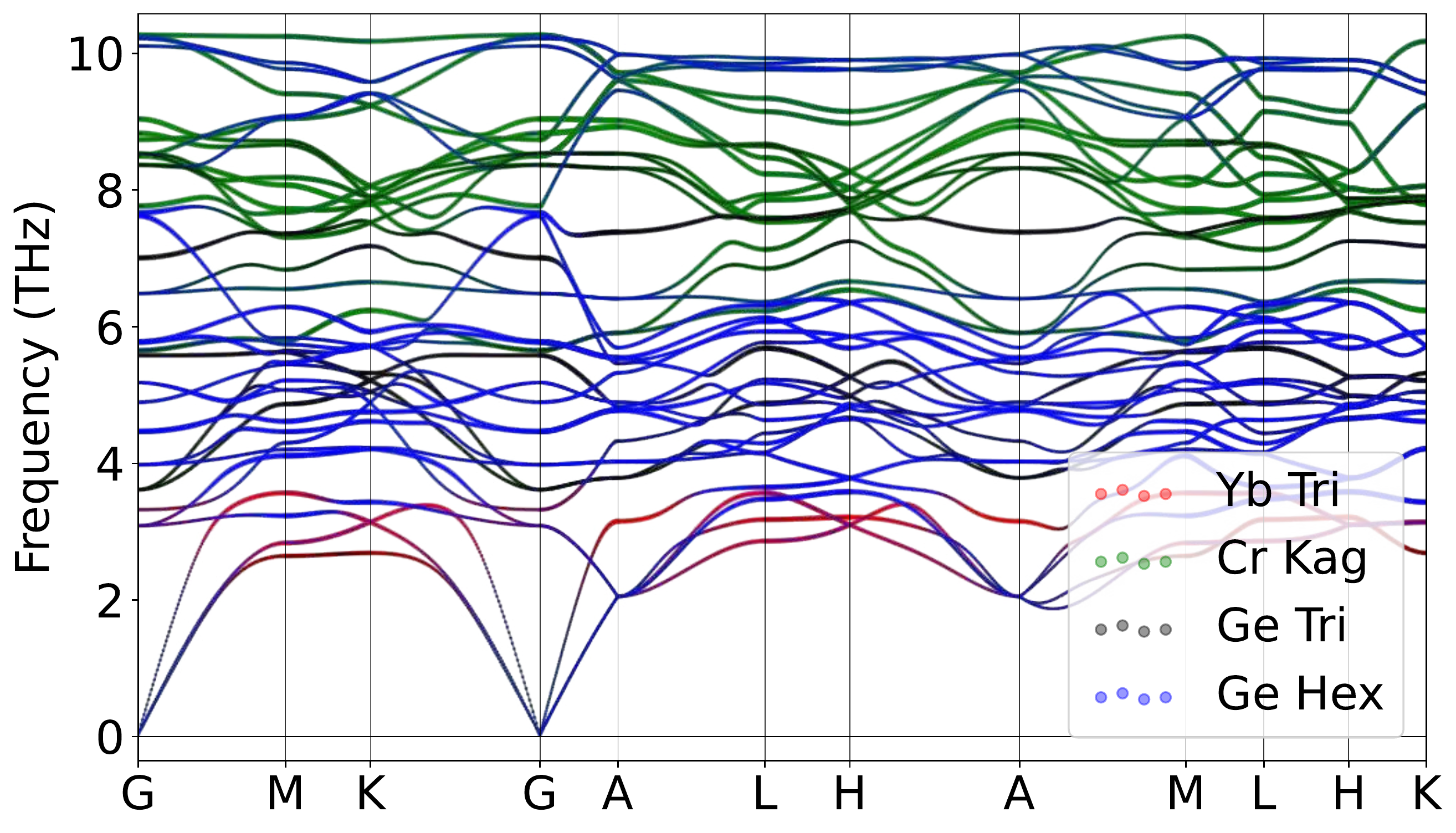}
}
\hspace{5pt}\subfloat[Relaxed phonon at high-T.]{
\label{fig:YbCr6Ge6_relaxed_High}
\includegraphics[width=0.45\textwidth]{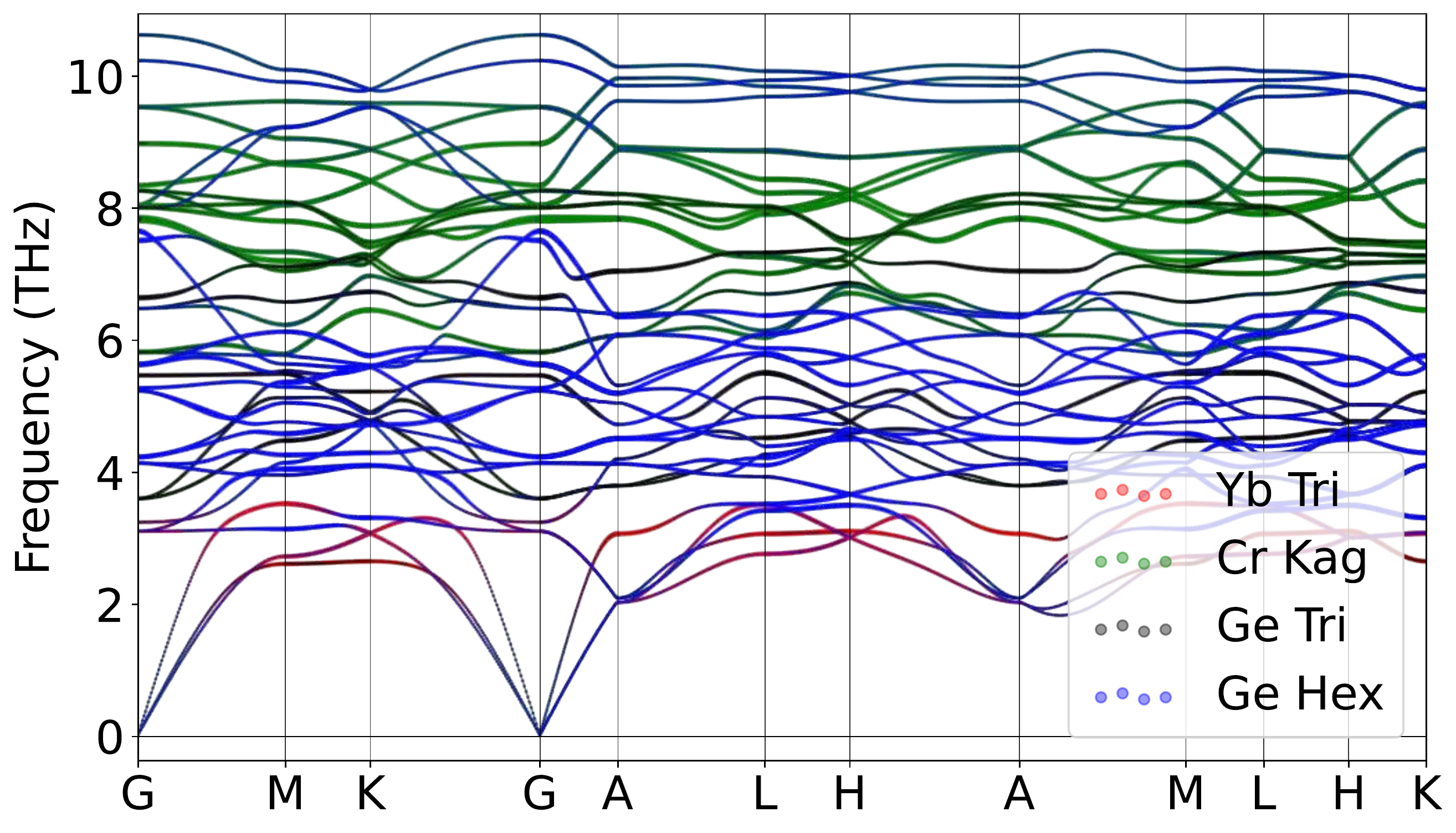}
}
\caption{Atomically projected phonon spectrum for YbCr$_6$Ge$_6$.}
\label{fig:YbCr6Ge6pho}
\end{figure}

\begin{figure}[H]
\centering
\label{fig:YbCr6Ge6_relaxed_Low_xz}
\includegraphics[width=0.85\textwidth]{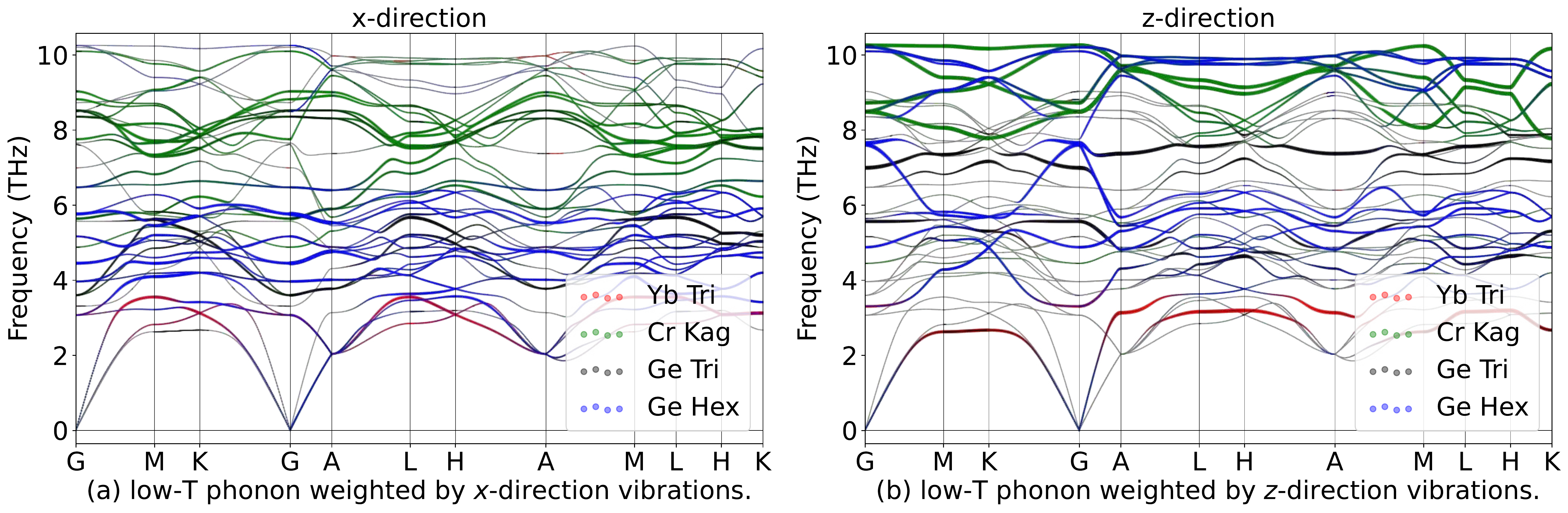}
\hspace{5pt}\label{fig:YbCr6Ge6_relaxed_High_xz}
\includegraphics[width=0.85\textwidth]{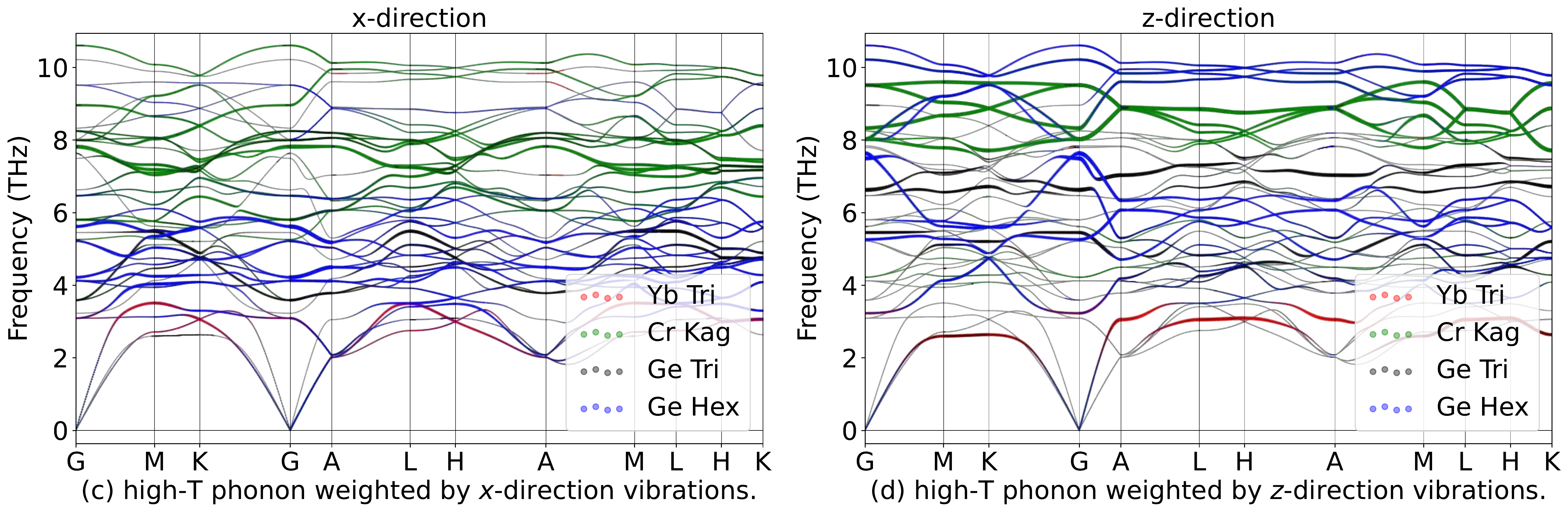}
\caption{Phonon spectrum for YbCr$_6$Ge$_6$in in-plane ($x$) and out-of-plane ($z$) directions.}
\label{fig:YbCr6Ge6phoxyz}
\end{figure}

\begin{table}[htbp]
\global\long\def\arraystretch{1.12}
\captionsetup{width=.95\textwidth}
\caption{IRREPs at high-symmetry points for YbCr$_6$Ge$_6$ phonon spectrum at Low-T.}
\label{tab:191_YbCr6Ge6_Low}
\setlength{\tabcolsep}{1.mm}{
}
\end{table}

\begin{table}[htbp]
\global\long\def\arraystretch{1.12}
\captionsetup{width=.95\textwidth}
\caption{IRREPs at high-symmetry points for YbCr$_6$Ge$_6$ phonon spectrum at High-T.}
\label{tab:191_YbCr6Ge6_High}
\setlength{\tabcolsep}{1.mm}{
%
}
\end{table}
\subsubsection{\label{sms2sec:LuCr6Ge6}\hyperref[smsec:col]{\texorpdfstring{LuCr$_6$Ge$_6$}{}}~{\color{green}  (High-T stable, Low-T stable) }}

\begin{table}[htbp]
\global\long\def\arraystretch{1.12}
\captionsetup{width=.85\textwidth}
\caption{Structure parameters of LuCr$_6$Ge$_6$ after relaxation. It contains 407 electrons. }
\label{tab:LuCr6Ge6}
\setlength{\tabcolsep}{4mm}{
%
}
\end{table}

\begin{figure}[htbp]
\centering
\includegraphics[width=0.85\textwidth]{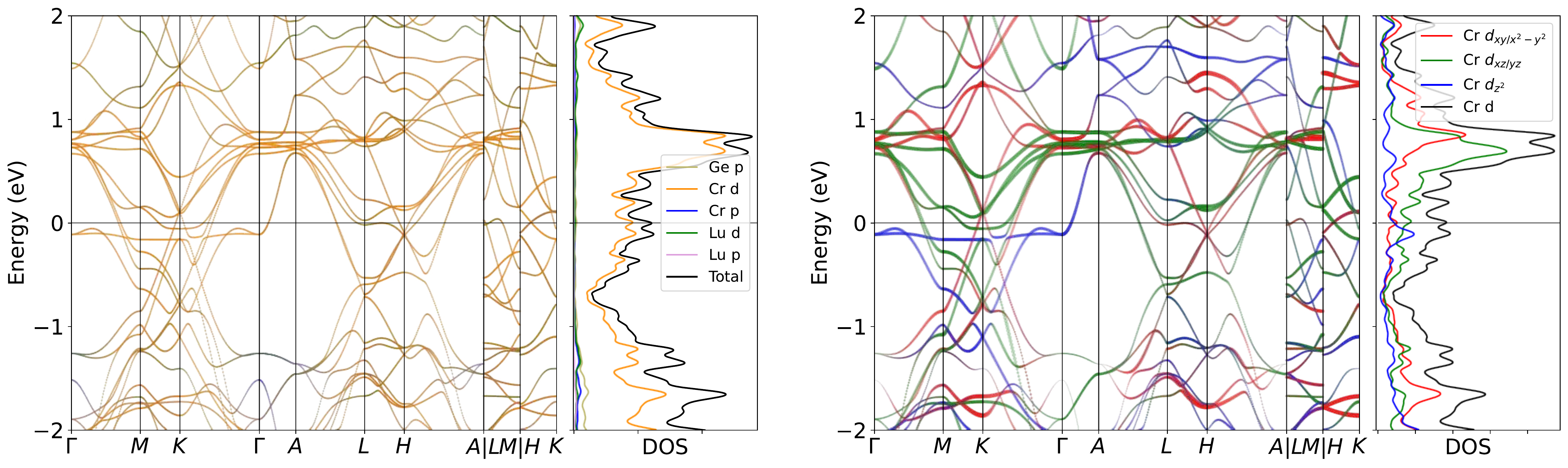}
\caption{Band structure for LuCr$_6$Ge$_6$ without SOC. The projection of Cr $3d$ orbitals is also presented. The band structures clearly reveal the dominating of Cr $3d$ orbitals  near the Fermi level, as well as the pronounced peak.}
\label{fig:LuCr6Ge6band}
\end{figure}

\begin{figure}[H]
\centering
\subfloat[Relaxed phonon at low-T.]{
\label{fig:LuCr6Ge6_relaxed_Low}
\includegraphics[width=0.45\textwidth]{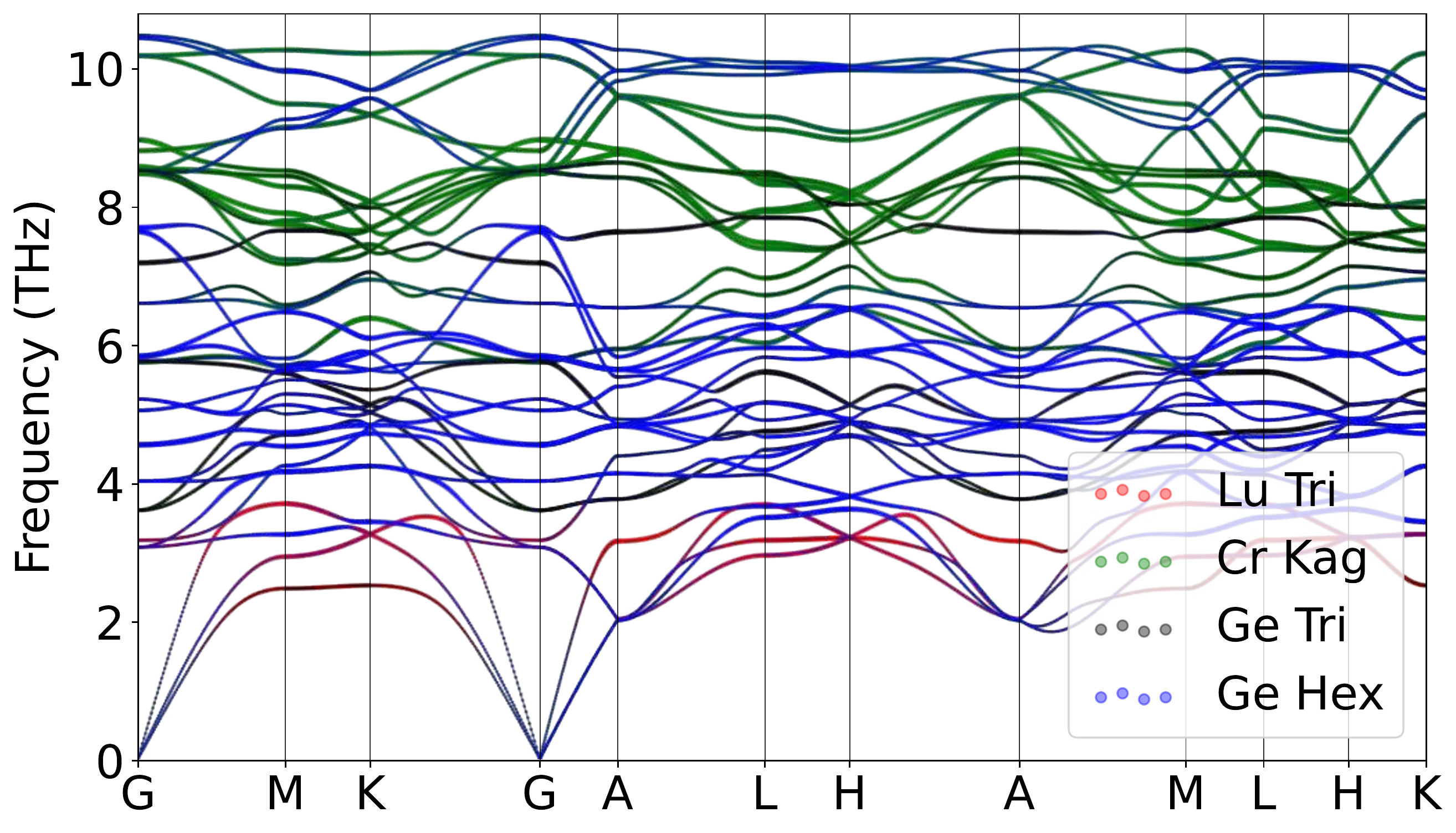}
}
\hspace{5pt}\subfloat[Relaxed phonon at high-T.]{
\label{fig:LuCr6Ge6_relaxed_High}
\includegraphics[width=0.45\textwidth]{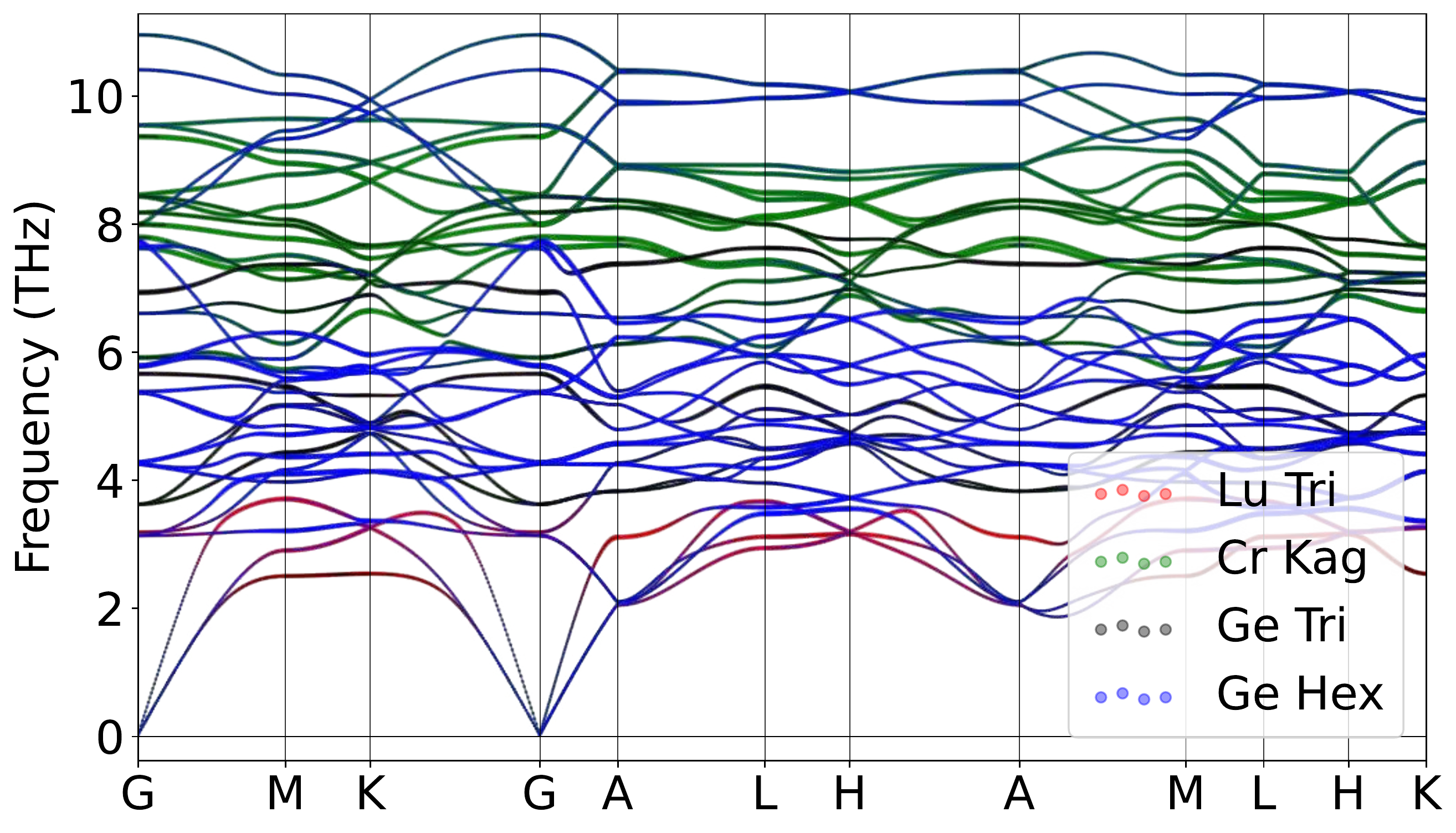}
}
\caption{Atomically projected phonon spectrum for LuCr$_6$Ge$_6$.}
\label{fig:LuCr6Ge6pho}
\end{figure}

\begin{figure}[H]
\centering
\label{fig:LuCr6Ge6_relaxed_Low_xz}
\includegraphics[width=0.85\textwidth]{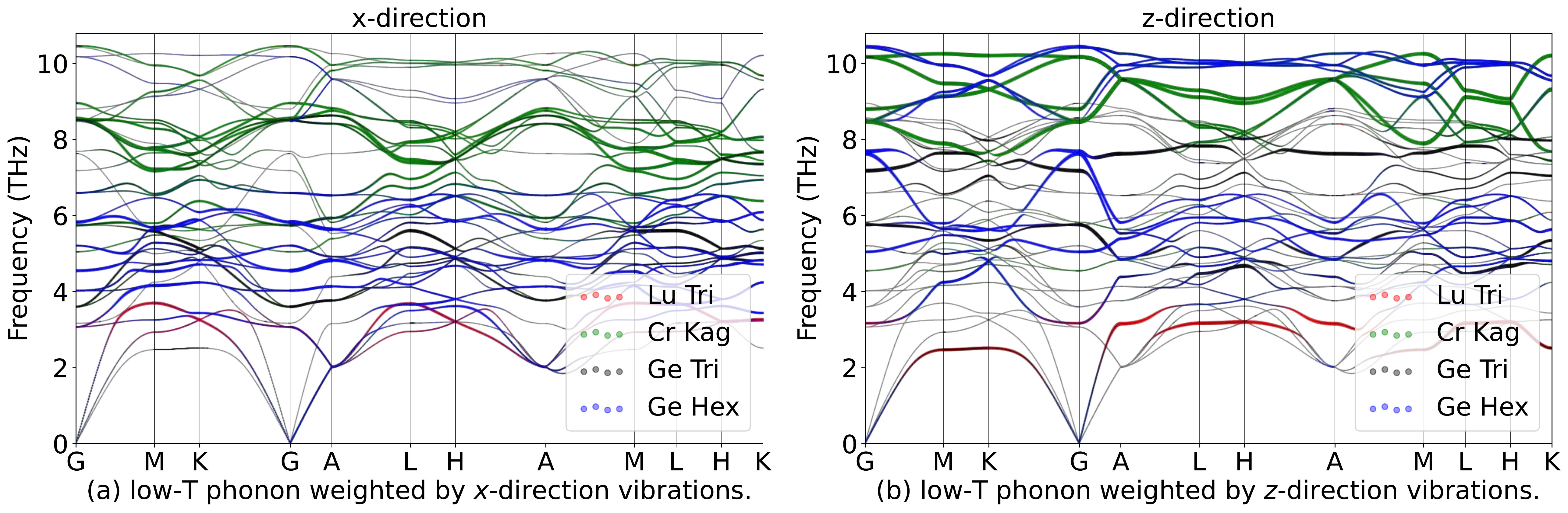}
\hspace{5pt}\label{fig:LuCr6Ge6_relaxed_High_xz}
\includegraphics[width=0.85\textwidth]{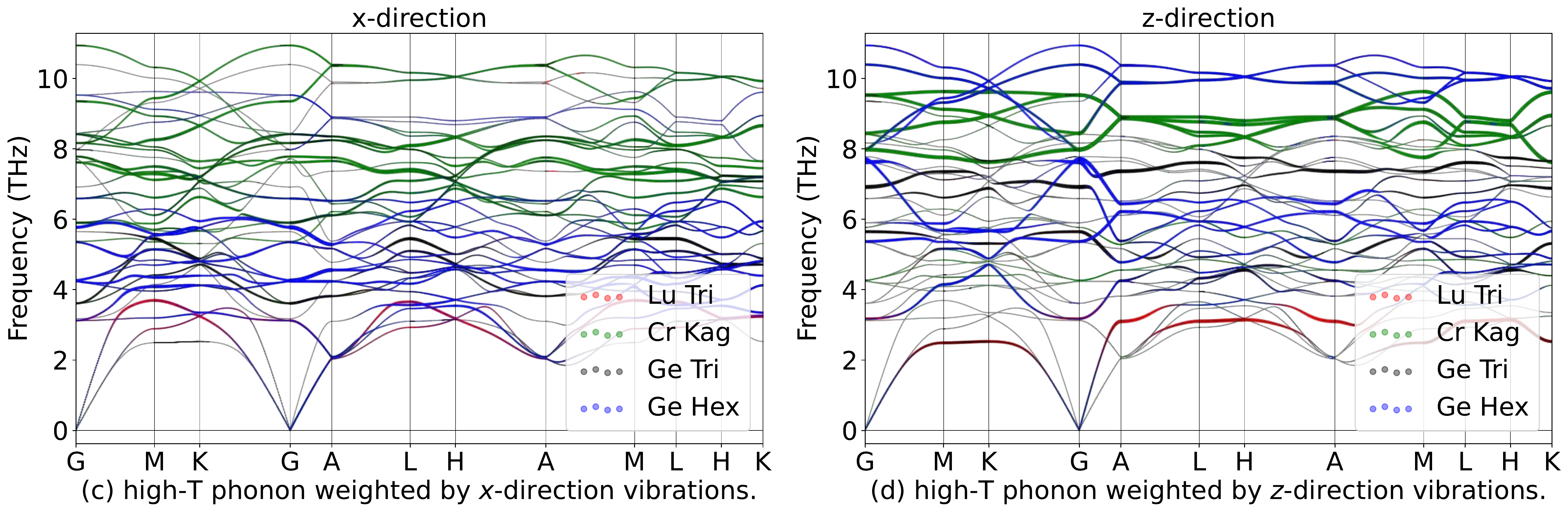}
\caption{Phonon spectrum for LuCr$_6$Ge$_6$in in-plane ($x$) and out-of-plane ($z$) directions.}
\label{fig:LuCr6Ge6phoxyz}
\end{figure}

\begin{table}[htbp]
\global\long\def\arraystretch{1.12}
\captionsetup{width=.95\textwidth}
\caption{IRREPs at high-symmetry points for LuCr$_6$Ge$_6$ phonon spectrum at Low-T.}
\label{tab:191_LuCr6Ge6_Low}
\setlength{\tabcolsep}{1.mm}{
}
\end{table}

\begin{table}[htbp]
\global\long\def\arraystretch{1.12}
\captionsetup{width=.95\textwidth}
\caption{IRREPs at high-symmetry points for LuCr$_6$Ge$_6$ phonon spectrum at High-T.}
\label{tab:191_LuCr6Ge6_High}
\setlength{\tabcolsep}{1.mm}{
%
}
\end{table}
\subsubsection{\label{sms2sec:HfCr6Ge6}\hyperref[smsec:col]{\texorpdfstring{HfCr$_6$Ge$_6$}{}}~{\color{green}  (High-T stable, Low-T stable) }}

\begin{table}[htbp]
\global\long\def\arraystretch{1.12}
\captionsetup{width=.85\textwidth}
\caption{Structure parameters of HfCr$_6$Ge$_6$ after relaxation. It contains 408 electrons. }
\label{tab:HfCr6Ge6}
\setlength{\tabcolsep}{4mm}{
%
}
\end{table}

\begin{figure}[htbp]
\centering
\includegraphics[width=0.85\textwidth]{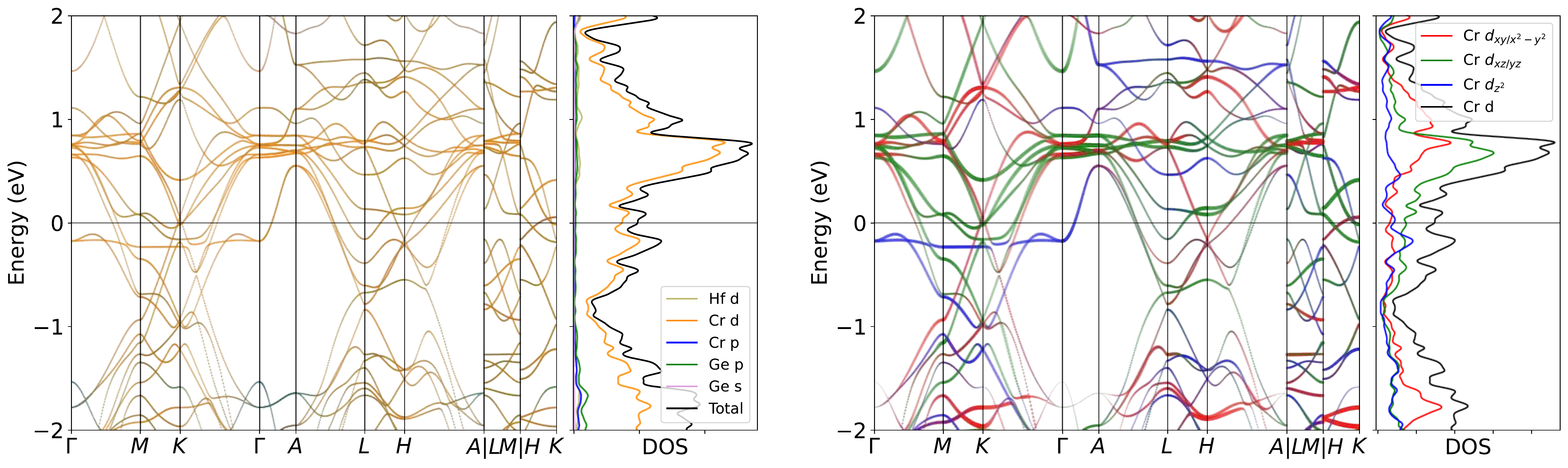}
\caption{Band structure for HfCr$_6$Ge$_6$ without SOC. The projection of Cr $3d$ orbitals is also presented. The band structures clearly reveal the dominating of Cr $3d$ orbitals  near the Fermi level, as well as the pronounced peak.}
\label{fig:HfCr6Ge6band}
\end{figure}

\begin{figure}[H]
\centering
\subfloat[Relaxed phonon at low-T.]{
\label{fig:HfCr6Ge6_relaxed_Low}
\includegraphics[width=0.45\textwidth]{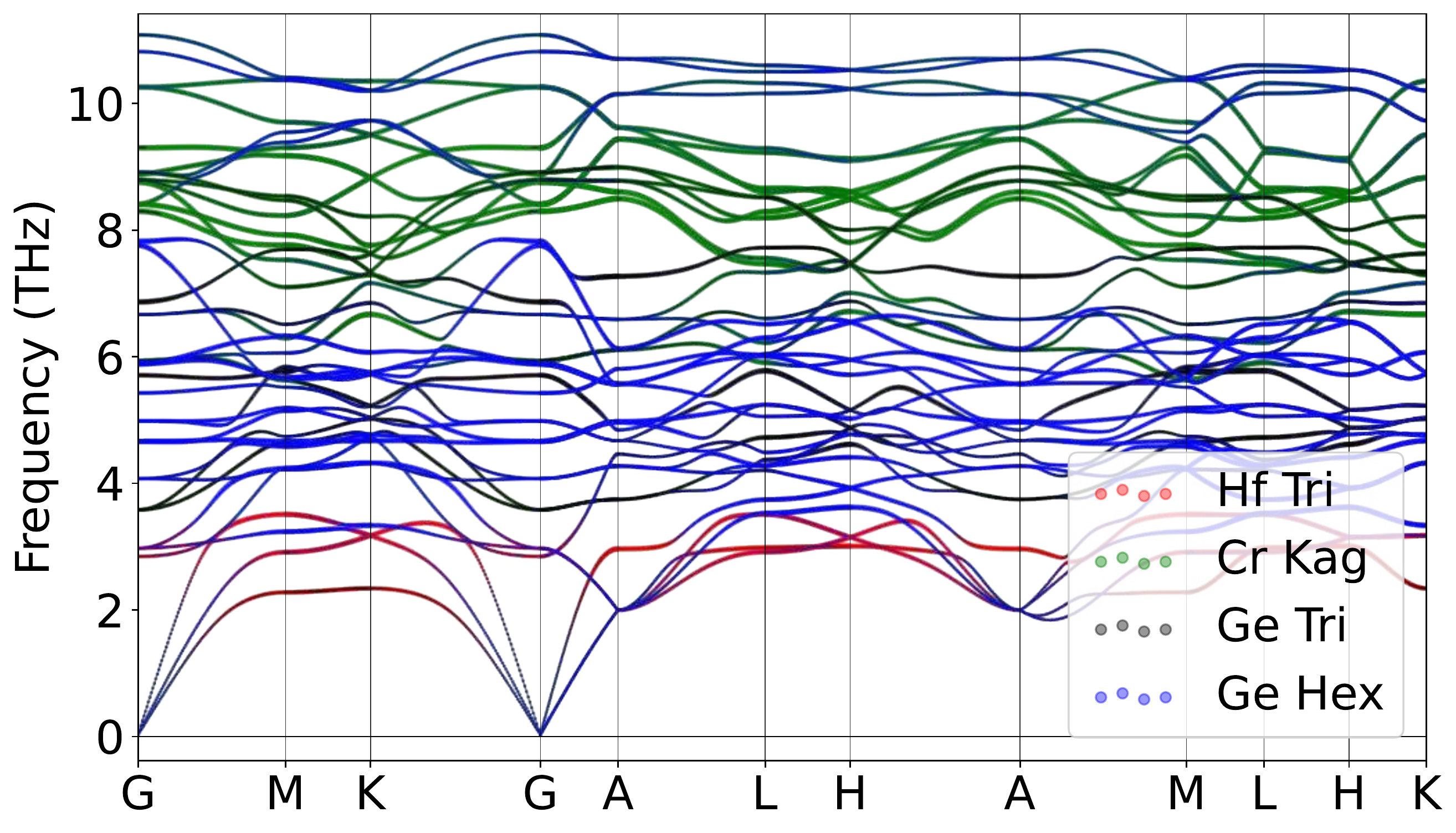}
}
\hspace{5pt}\subfloat[Relaxed phonon at high-T.]{
\label{fig:HfCr6Ge6_relaxed_High}
\includegraphics[width=0.45\textwidth]{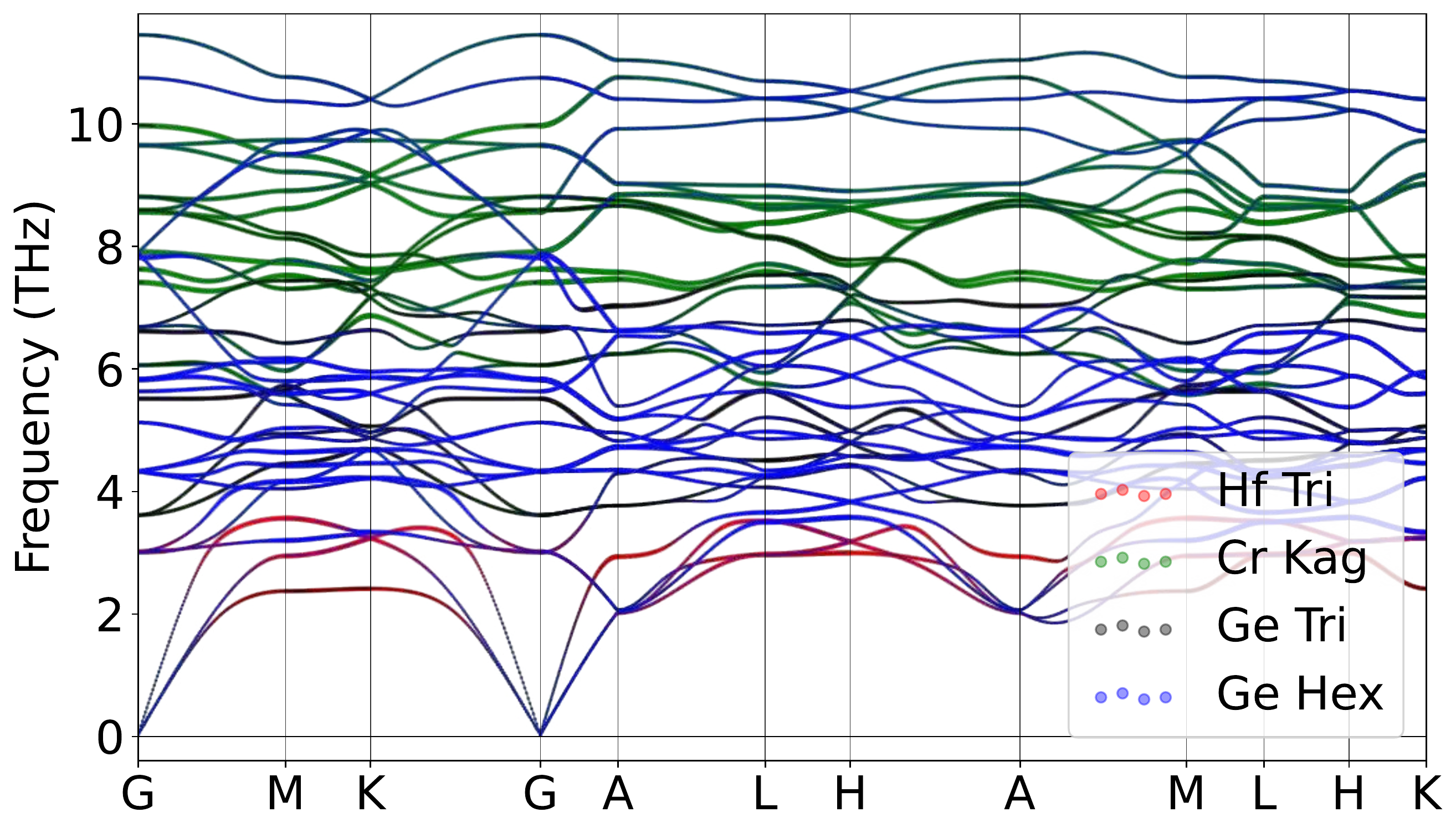}
}
\caption{Atomically projected phonon spectrum for HfCr$_6$Ge$_6$.}
\label{fig:HfCr6Ge6pho}
\end{figure}

\begin{figure}[H]
\centering
\label{fig:HfCr6Ge6_relaxed_Low_xz}
\includegraphics[width=0.85\textwidth]{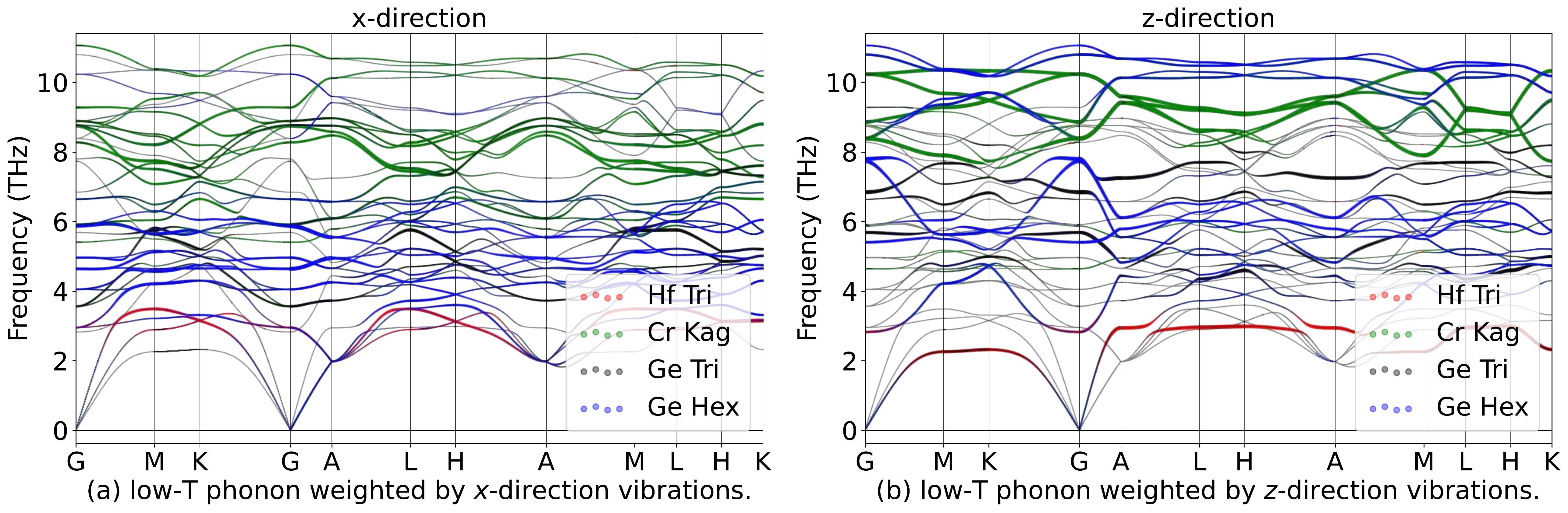}
\hspace{5pt}\label{fig:HfCr6Ge6_relaxed_High_xz}
\includegraphics[width=0.85\textwidth]{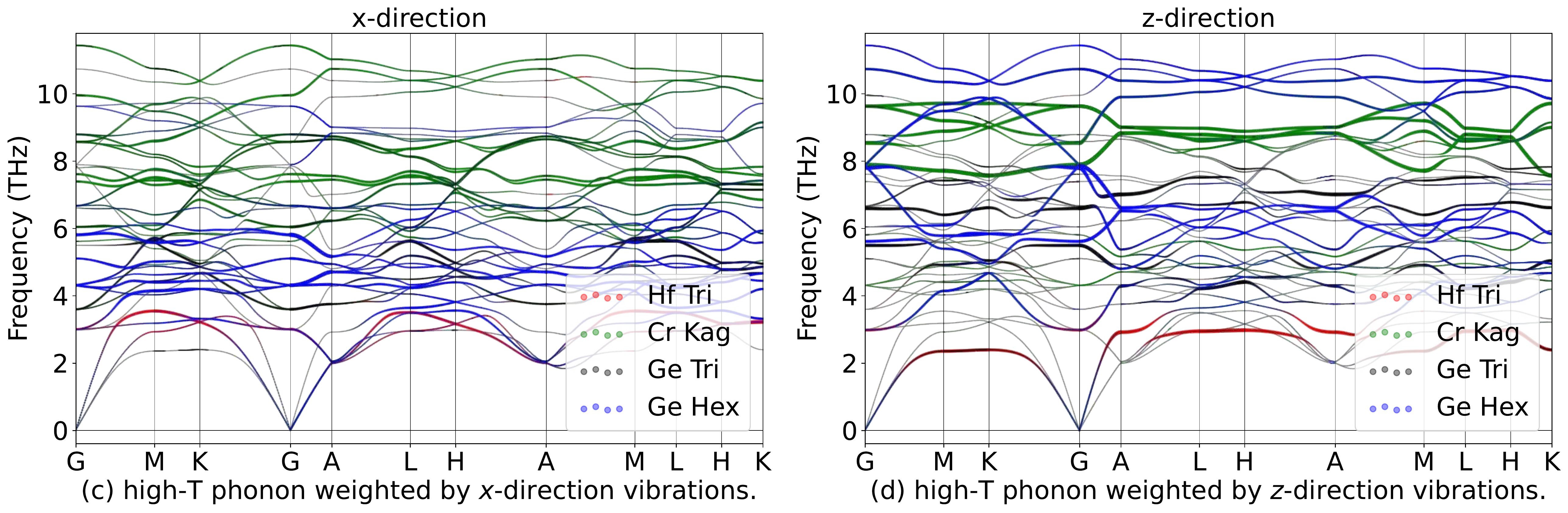}
\caption{Phonon spectrum for HfCr$_6$Ge$_6$in in-plane ($x$) and out-of-plane ($z$) directions.}
\label{fig:HfCr6Ge6phoxyz}
\end{figure}

\begin{table}[htbp]
\global\long\def\arraystretch{1.12}
\captionsetup{width=.95\textwidth}
\caption{IRREPs at high-symmetry points for HfCr$_6$Ge$_6$ phonon spectrum at Low-T.}
\label{tab:191_HfCr6Ge6_Low}
\setlength{\tabcolsep}{1.mm}{
}
\end{table}

\begin{table}[htbp]
\global\long\def\arraystretch{1.12}
\captionsetup{width=.95\textwidth}
\caption{IRREPs at high-symmetry points for HfCr$_6$Ge$_6$ phonon spectrum at High-T.}
\label{tab:191_HfCr6Ge6_High}
\setlength{\tabcolsep}{1.mm}{
%
}
\end{table}
\subsubsection{\label{sms2sec:TaCr6Ge6}\hyperref[smsec:col]{\texorpdfstring{TaCr$_6$Ge$_6$}{}}~{\color{green}  (High-T stable, Low-T stable) }}

\begin{table}[htbp]
\global\long\def\arraystretch{1.12}
\captionsetup{width=.85\textwidth}
\caption{Structure parameters of TaCr$_6$Ge$_6$ after relaxation. It contains 409 electrons. }
\label{tab:TaCr6Ge6}
\setlength{\tabcolsep}{4mm}{
%
}
\end{table}

\begin{figure}[htbp]
\centering
\includegraphics[width=0.85\textwidth]{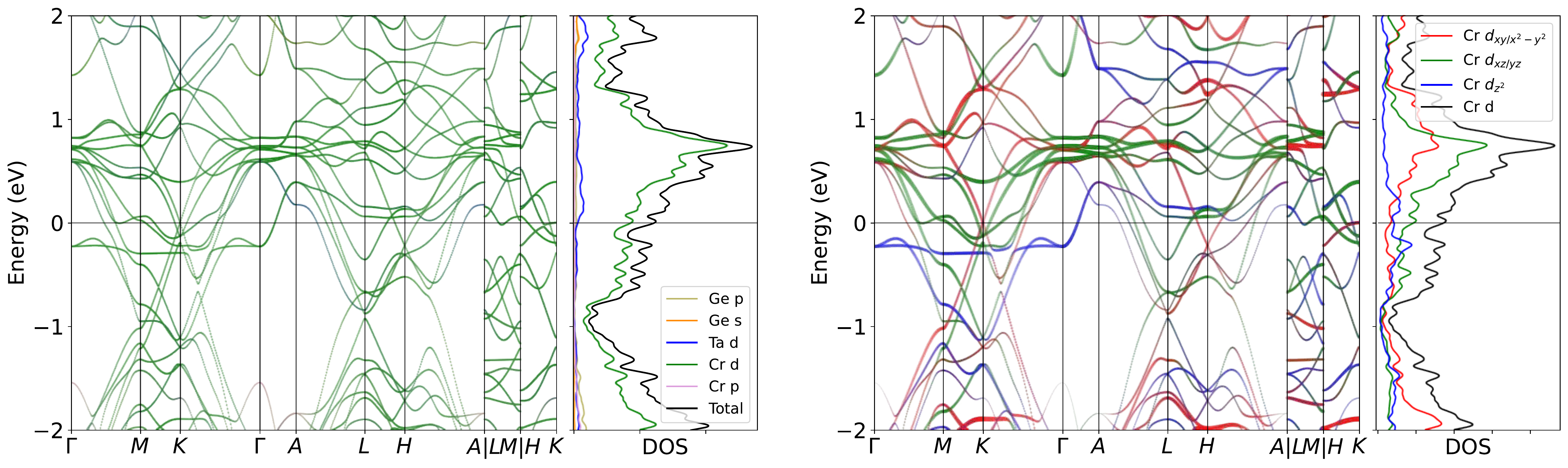}
\caption{Band structure for TaCr$_6$Ge$_6$ without SOC. The projection of Cr $3d$ orbitals is also presented. The band structures clearly reveal the dominating of Cr $3d$ orbitals  near the Fermi level, as well as the pronounced peak.}
\label{fig:TaCr6Ge6band}
\end{figure}

\begin{figure}[H]
\centering
\subfloat[Relaxed phonon at low-T.]{
\label{fig:TaCr6Ge6_relaxed_Low}
\includegraphics[width=0.45\textwidth]{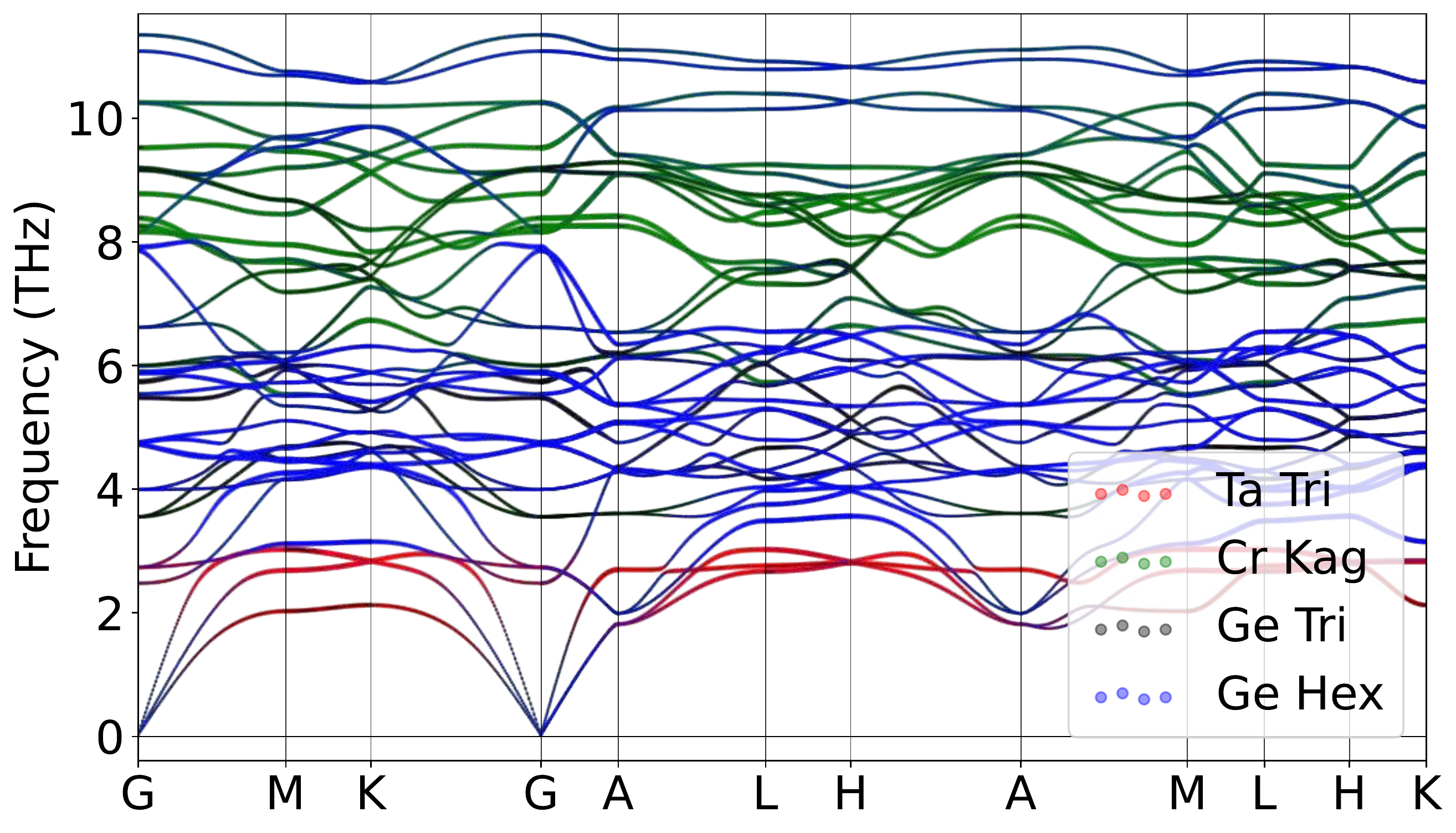}
}
\hspace{5pt}\subfloat[Relaxed phonon at high-T.]{
\label{fig:TaCr6Ge6_relaxed_High}
\includegraphics[width=0.45\textwidth]{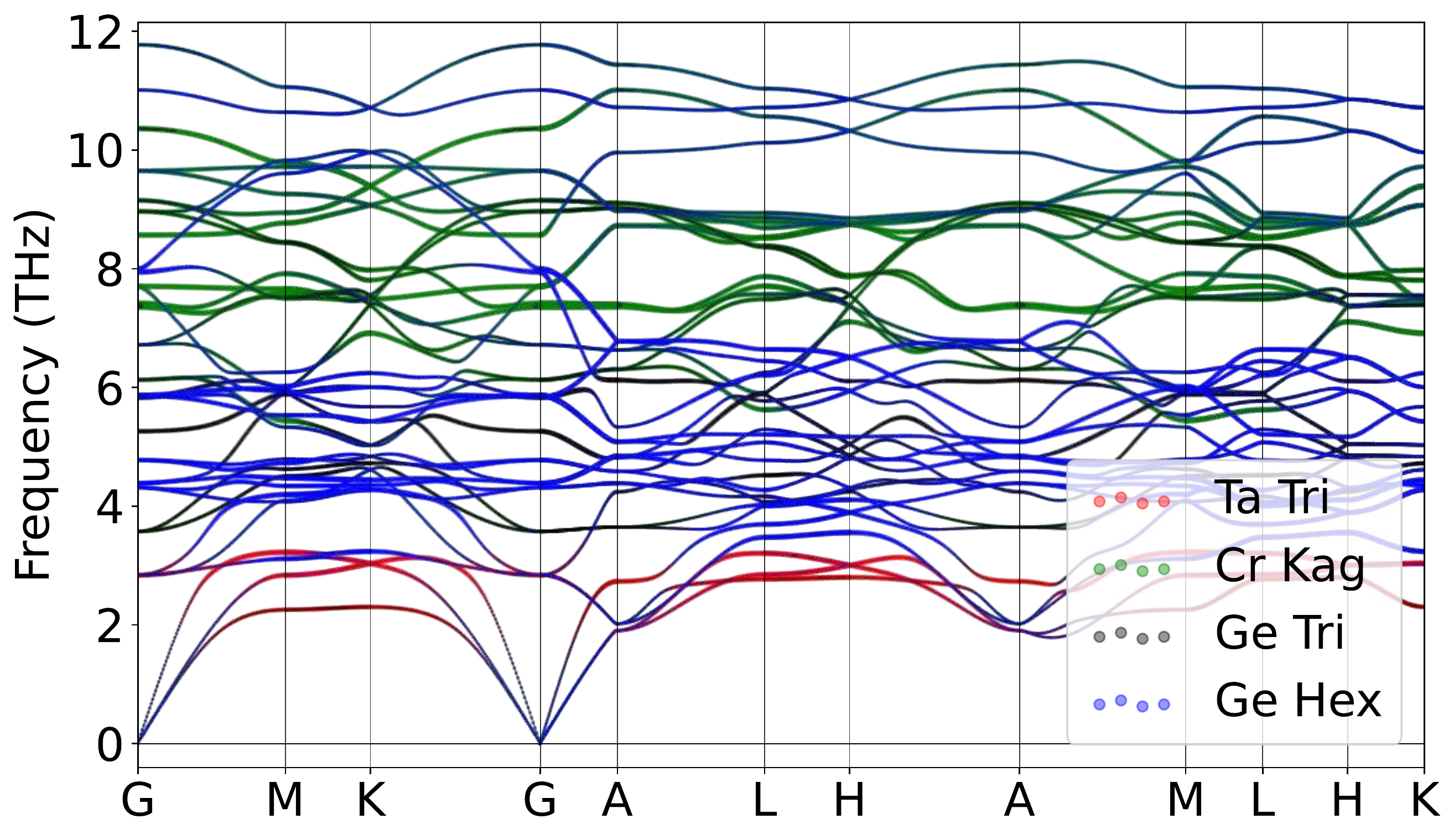}
}
\caption{Atomically projected phonon spectrum for TaCr$_6$Ge$_6$.}
\label{fig:TaCr6Ge6pho}
\end{figure}

\begin{figure}[H]
\centering
\label{fig:TaCr6Ge6_relaxed_Low_xz}
\includegraphics[width=0.85\textwidth]{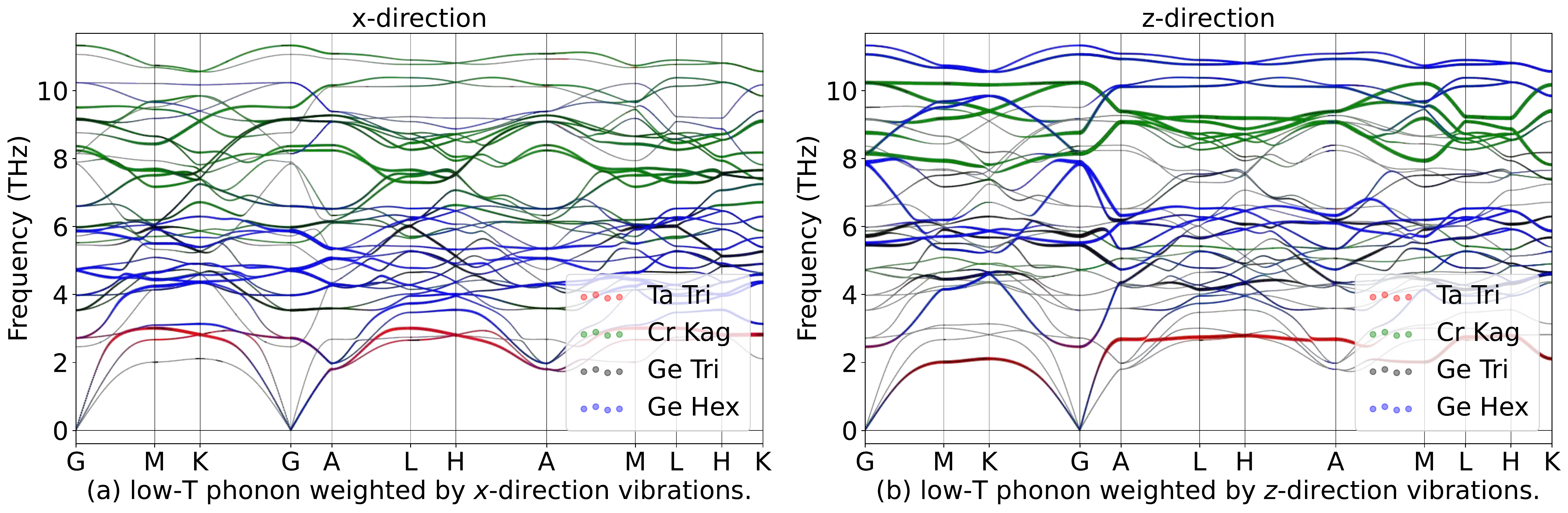}
\hspace{5pt}\label{fig:TaCr6Ge6_relaxed_High_xz}
\includegraphics[width=0.85\textwidth]{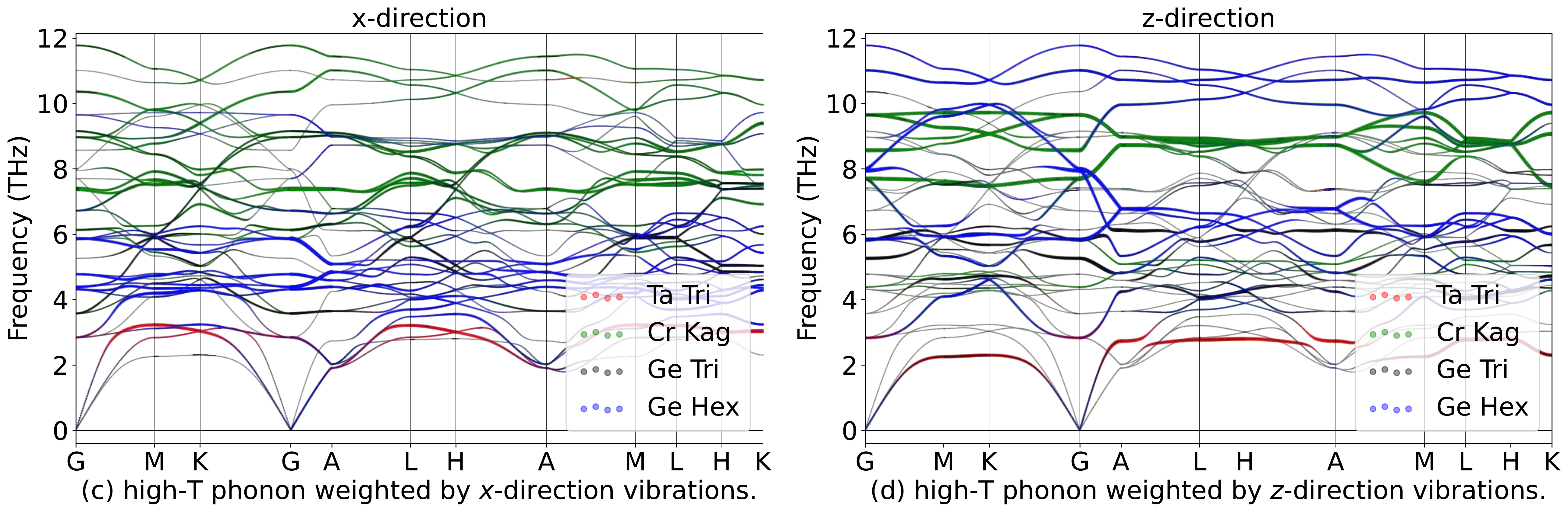}
\caption{Phonon spectrum for TaCr$_6$Ge$_6$in in-plane ($x$) and out-of-plane ($z$) directions.}
\label{fig:TaCr6Ge6phoxyz}
\end{figure}

\begin{table}[htbp]
\global\long\def\arraystretch{1.12}
\captionsetup{width=.95\textwidth}
\caption{IRREPs at high-symmetry points for TaCr$_6$Ge$_6$ phonon spectrum at Low-T.}
\label{tab:191_TaCr6Ge6_Low}
\setlength{\tabcolsep}{1.mm}{
}
\end{table}

\begin{table}[htbp]
\global\long\def\arraystretch{1.12}
\captionsetup{width=.95\textwidth}
\caption{IRREPs at high-symmetry points for TaCr$_6$Ge$_6$ phonon spectrum at High-T.}
\label{tab:191_TaCr6Ge6_High}
\setlength{\tabcolsep}{1.mm}{
%
}
\end{table}
\subsection{\hyperref[smsec:col]{MnGe}}~\label{smssec:MnGe}

\subsubsection{\label{sms2sec:LiMn6Ge6}\hyperref[smsec:col]{\texorpdfstring{LiMn$_6$Ge$_6$}{}}~{\color{blue}  (High-T stable, Low-T unstable) }}

\begin{table}[htbp]
\global\long\def\arraystretch{1.12}
\captionsetup{width=.85\textwidth}
\caption{Structure parameters of LiMn$_6$Ge$_6$ after relaxation. It contains 345 electrons. }
\label{tab:LiMn6Ge6}
\setlength{\tabcolsep}{4mm}{
%
}
\end{table}

\begin{figure}[htbp]
\centering
\includegraphics[width=0.85\textwidth]{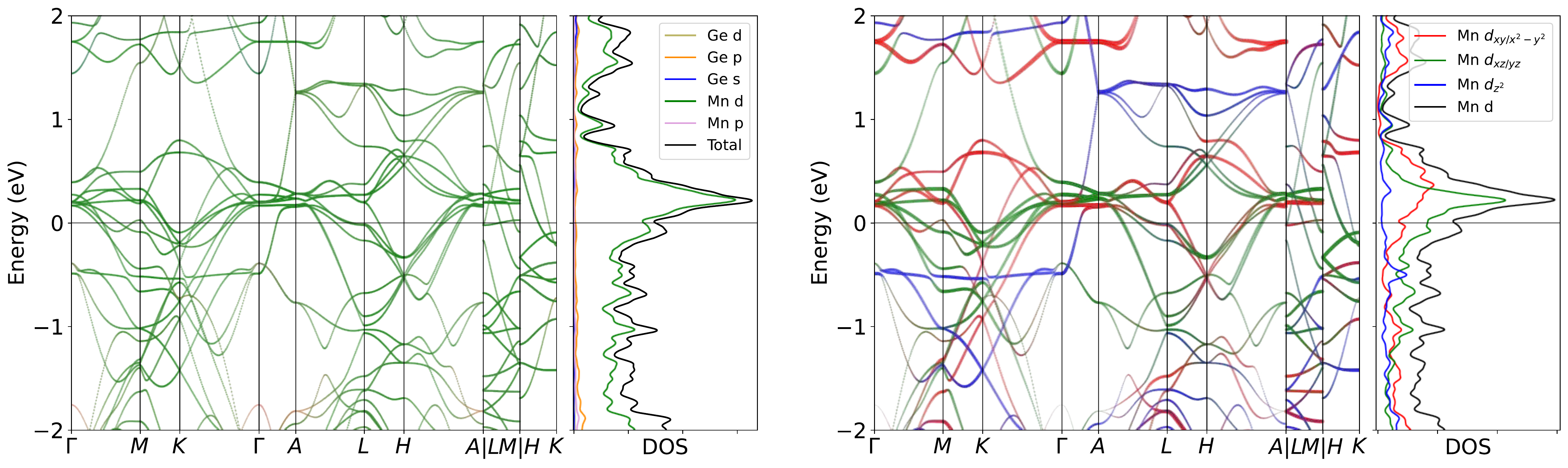}
\caption{Band structure for LiMn$_6$Ge$_6$ without SOC in paramagnetic phase. The projection of Mn $3d$ orbitals is also presented. The band structures clearly reveal the dominating of Mn $3d$ orbitals  near the Fermi level, as well as the pronounced peak.}
\label{fig:LiMn6Ge6band}
\end{figure}

\begin{figure}[H]
\centering
\subfloat[Relaxed phonon at low-T.]{
\label{fig:LiMn6Ge6_relaxed_Low}
\includegraphics[width=0.45\textwidth]{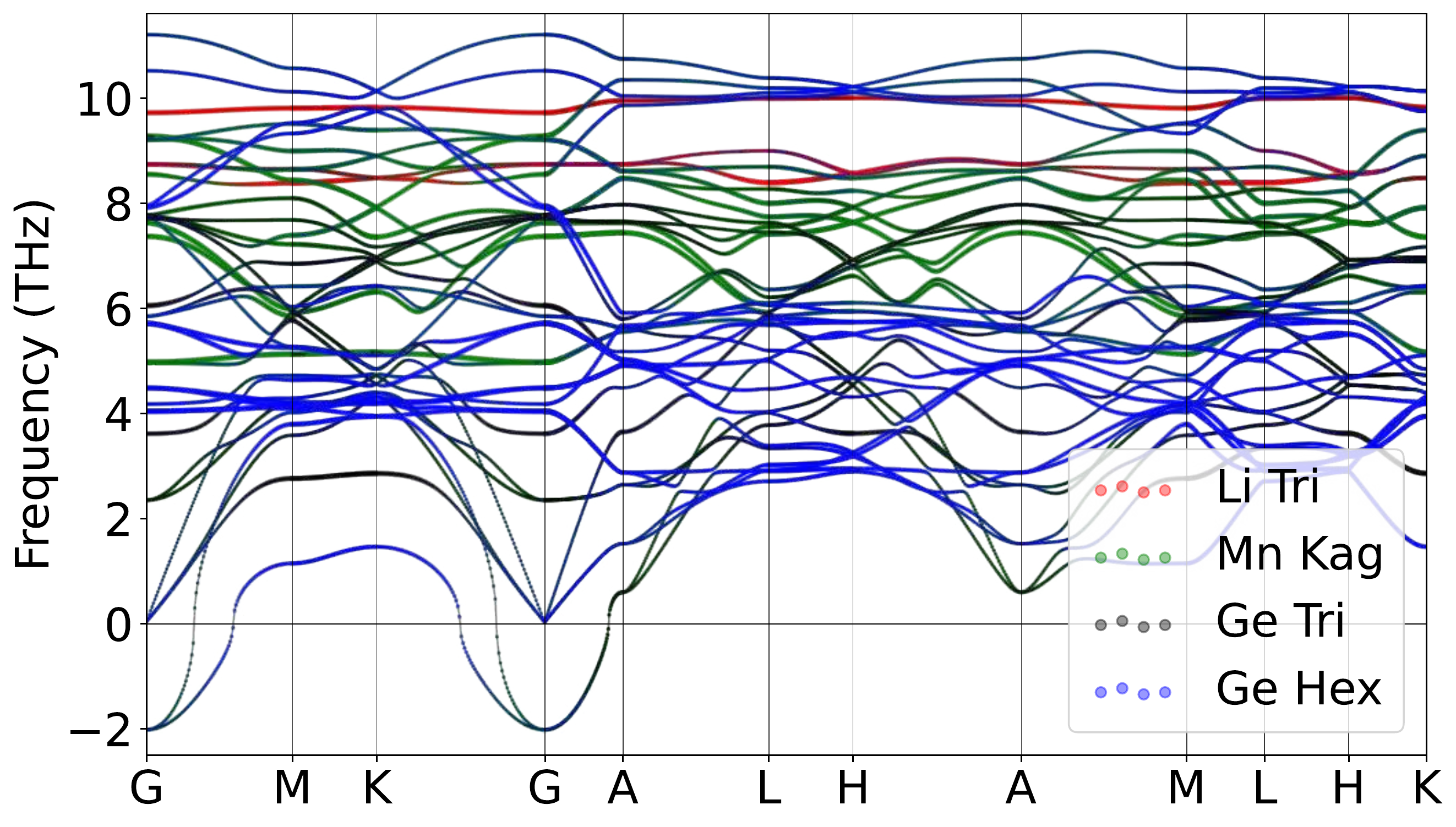}
}
\hspace{5pt}\subfloat[Relaxed phonon at high-T.]{
\label{fig:LiMn6Ge6_relaxed_High}
\includegraphics[width=0.45\textwidth]{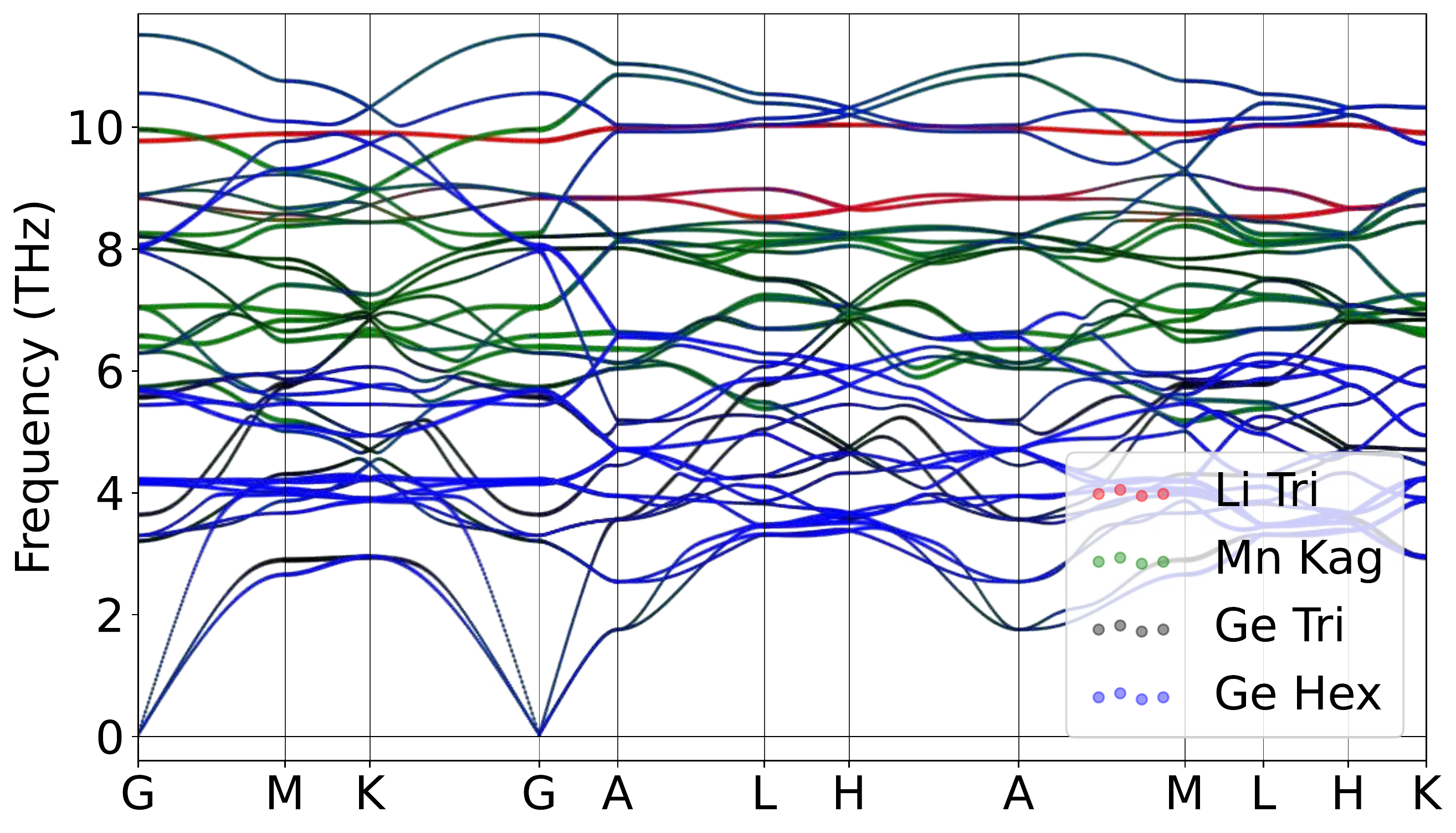}
}
\caption{Atomically projected phonon spectrum for LiMn$_6$Ge$_6$ in paramagnetic phase.}
\label{fig:LiMn6Ge6pho}
\end{figure}

\begin{figure}[htbp]
\centering
\includegraphics[width=0.32\textwidth]{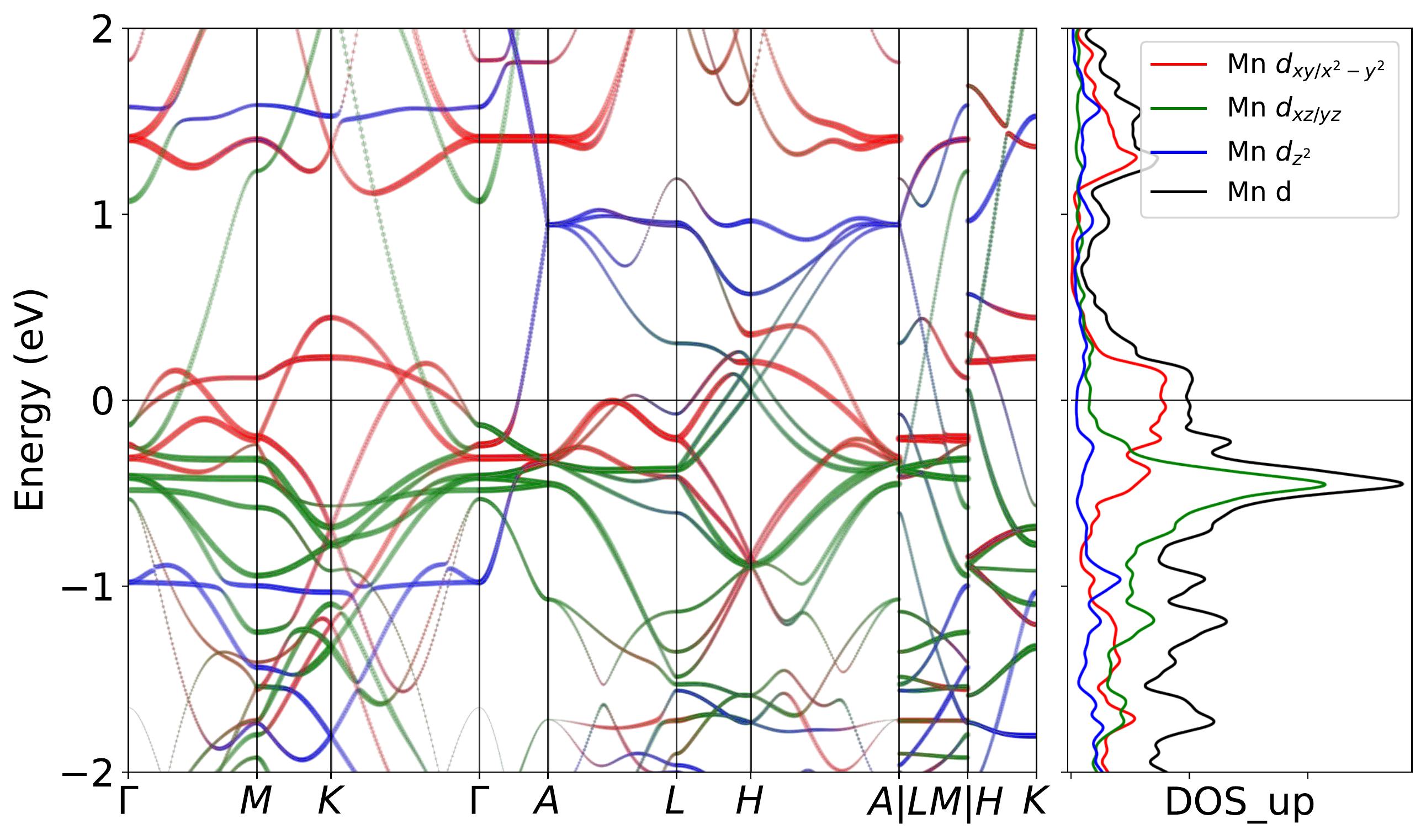}
\includegraphics[width=0.32\textwidth]{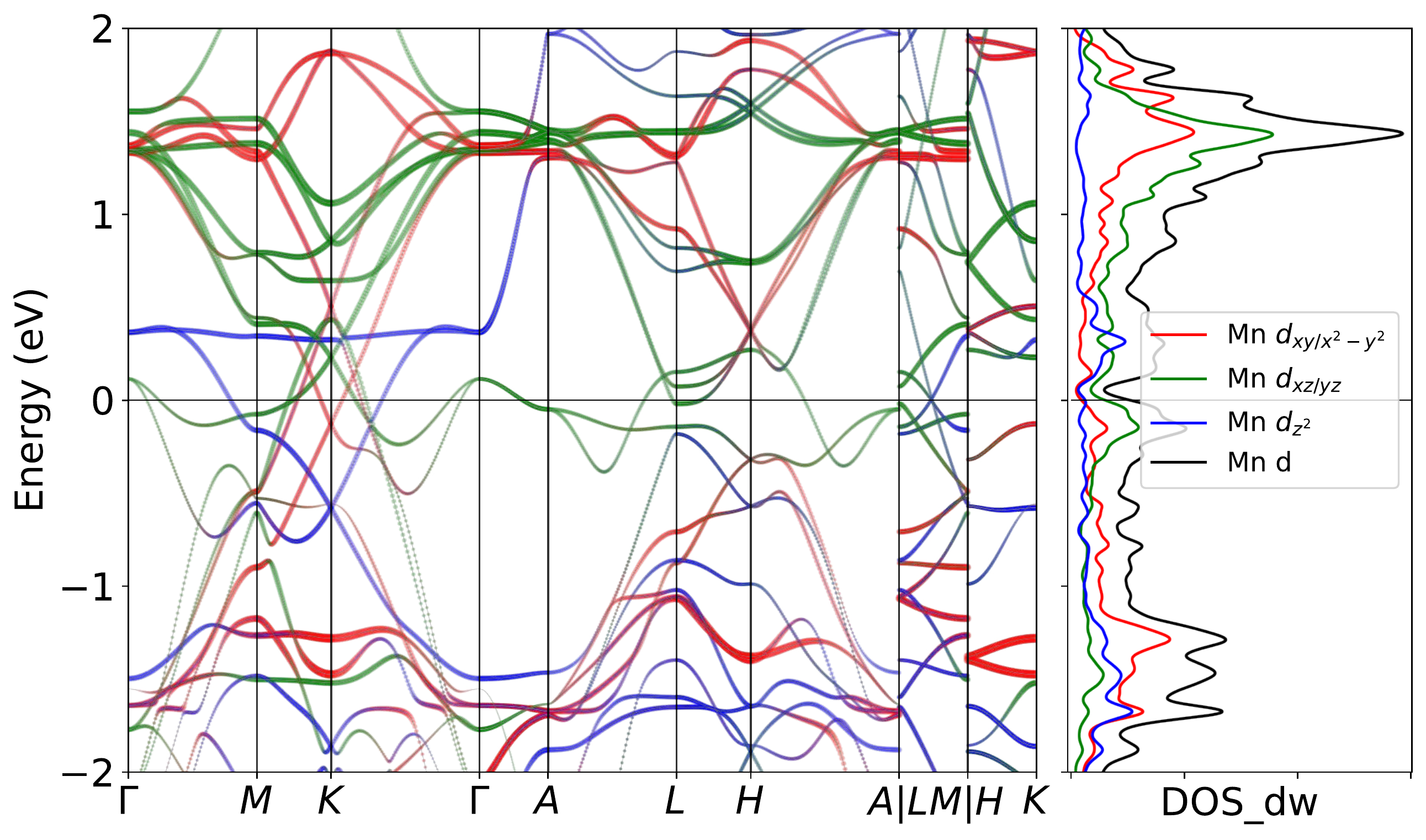}
\includegraphics[width=0.32\textwidth]{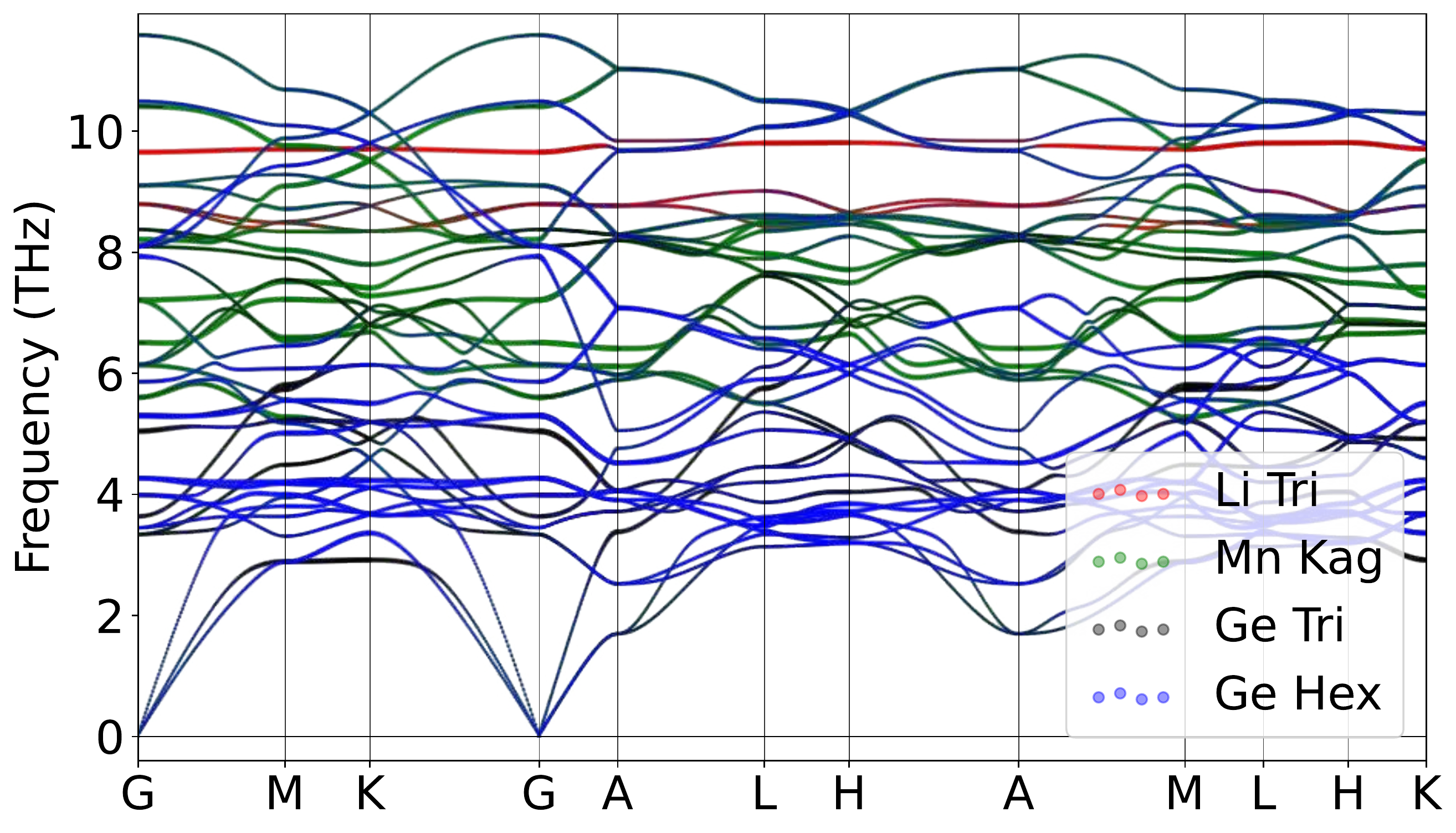}
\caption{Band structure for LiMn$_6$Ge$_6$ with FM order without SOC for spin (Left)up and (Middle)down channels. The projection of Mn $3d$ orbitals is also presented. (Right)Correspondingly, a stable phonon was demonstrated with the collinear magnetic order without Hubbard U at lower temperature regime, weighted by atomic projections.}
\label{fig:LiMn6Ge6mag}
\end{figure}

\begin{figure}[H]
\centering
\label{fig:LiMn6Ge6_relaxed_Low_xz}
\includegraphics[width=0.85\textwidth]{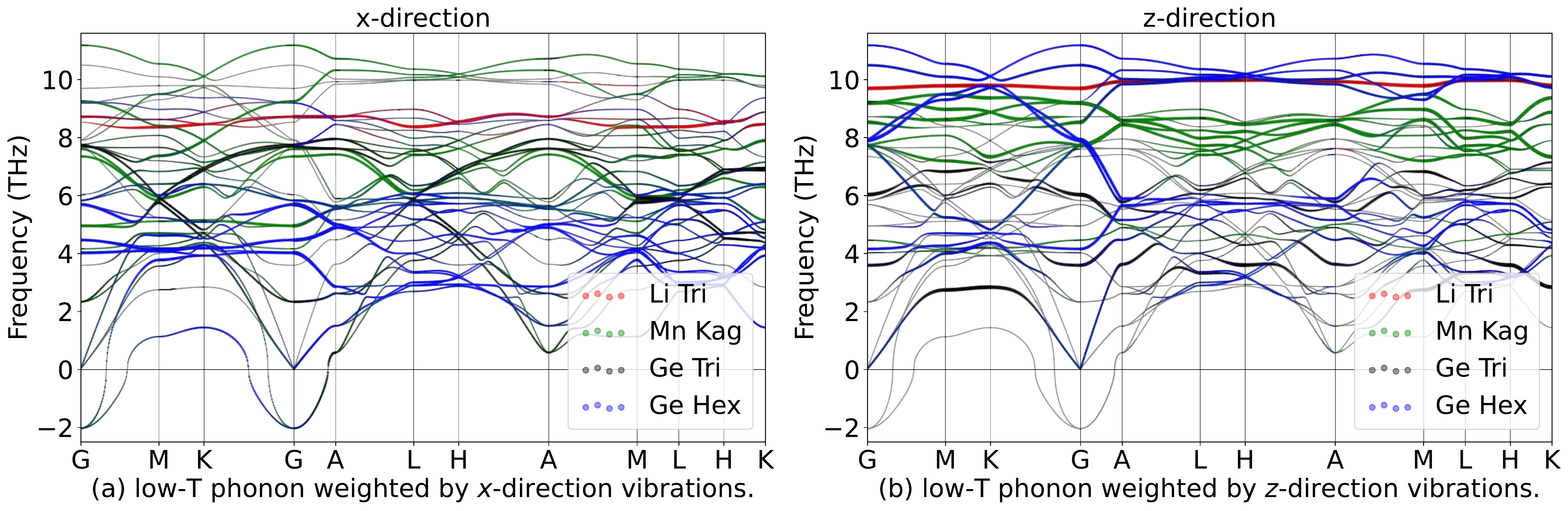}
\hspace{5pt}\label{fig:LiMn6Ge6_relaxed_High_xz}
\includegraphics[width=0.85\textwidth]{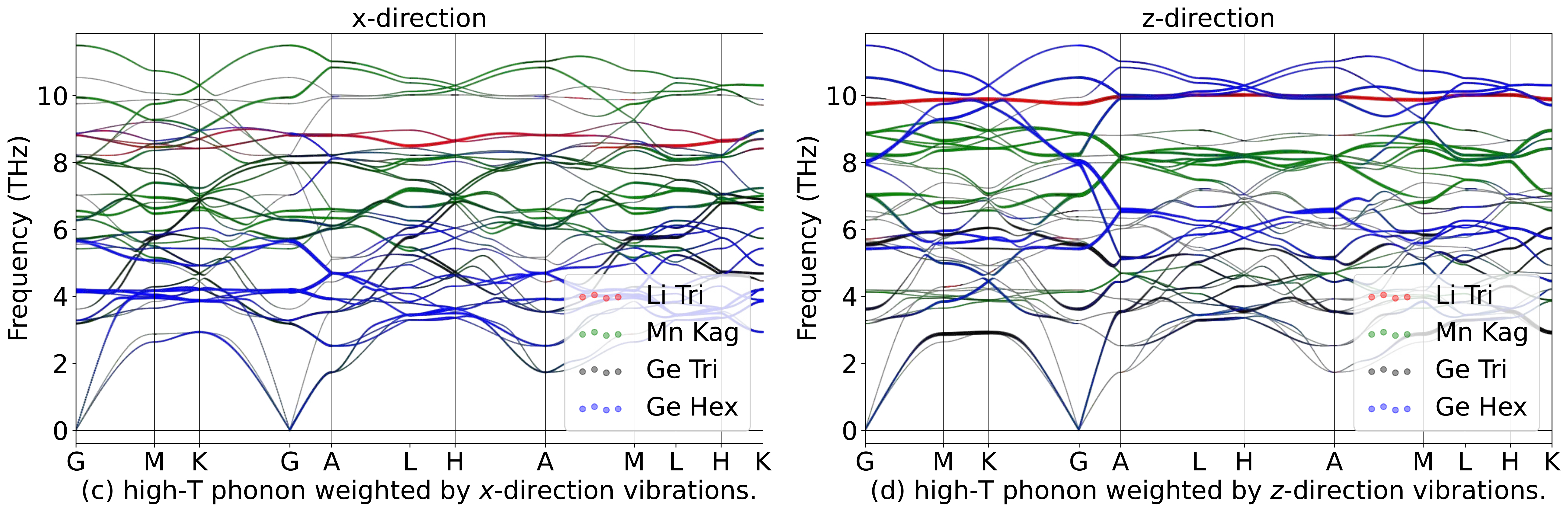}
\caption{Phonon spectrum for LiMn$_6$Ge$_6$in in-plane ($x$) and out-of-plane ($z$) directions in paramagnetic phase.}
\label{fig:LiMn6Ge6phoxyz}
\end{figure}

\begin{table}[htbp]
\global\long\def\arraystretch{1.12}
\captionsetup{width=.95\textwidth}
\caption{IRREPs at high-symmetry points for LiMn$_6$Ge$_6$ phonon spectrum at Low-T.}
\label{tab:191_LiMn6Ge6_Low}
\setlength{\tabcolsep}{1.mm}{
}
\end{table}

\begin{table}[htbp]
\global\long\def\arraystretch{1.12}
\captionsetup{width=.95\textwidth}
\caption{IRREPs at high-symmetry points for LiMn$_6$Ge$_6$ phonon spectrum at High-T.}
\label{tab:191_LiMn6Ge6_High}
\setlength{\tabcolsep}{1.mm}{
%
}
\end{table}
\subsubsection{\label{sms2sec:MgMn6Ge6}\hyperref[smsec:col]{\texorpdfstring{MgMn$_6$Ge$_6$}{}}~{\color{blue}  (High-T stable, Low-T unstable) }}

\begin{table}[htbp]
\global\long\def\arraystretch{1.12}
\captionsetup{width=.85\textwidth}
\caption{Structure parameters of MgMn$_6$Ge$_6$ after relaxation. It contains 354 electrons. }
\label{tab:MgMn6Ge6}
\setlength{\tabcolsep}{4mm}{
%
}
\end{table}

\begin{figure}[htbp]
\centering
\includegraphics[width=0.85\textwidth]{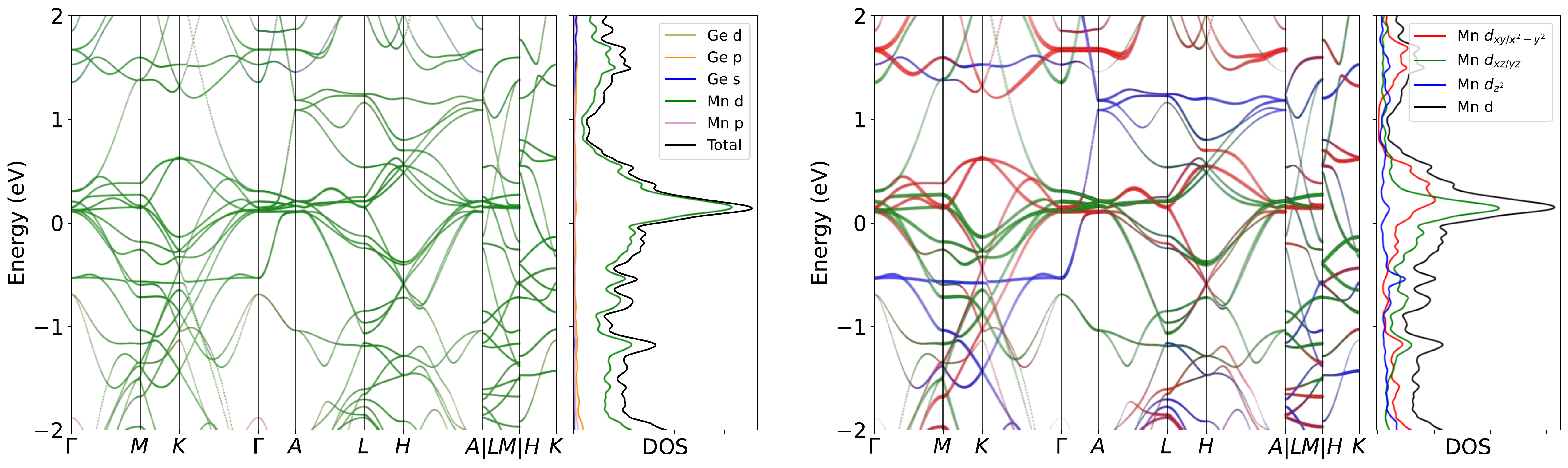}
\caption{Band structure for MgMn$_6$Ge$_6$ without SOC in paramagnetic phase. The projection of Mn $3d$ orbitals is also presented. The band structures clearly reveal the dominating of Mn $3d$ orbitals  near the Fermi level, as well as the pronounced peak.}
\label{fig:MgMn6Ge6band}
\end{figure}

\begin{figure}[H]
\centering
\subfloat[Relaxed phonon at low-T.]{
\label{fig:MgMn6Ge6_relaxed_Low}
\includegraphics[width=0.45\textwidth]{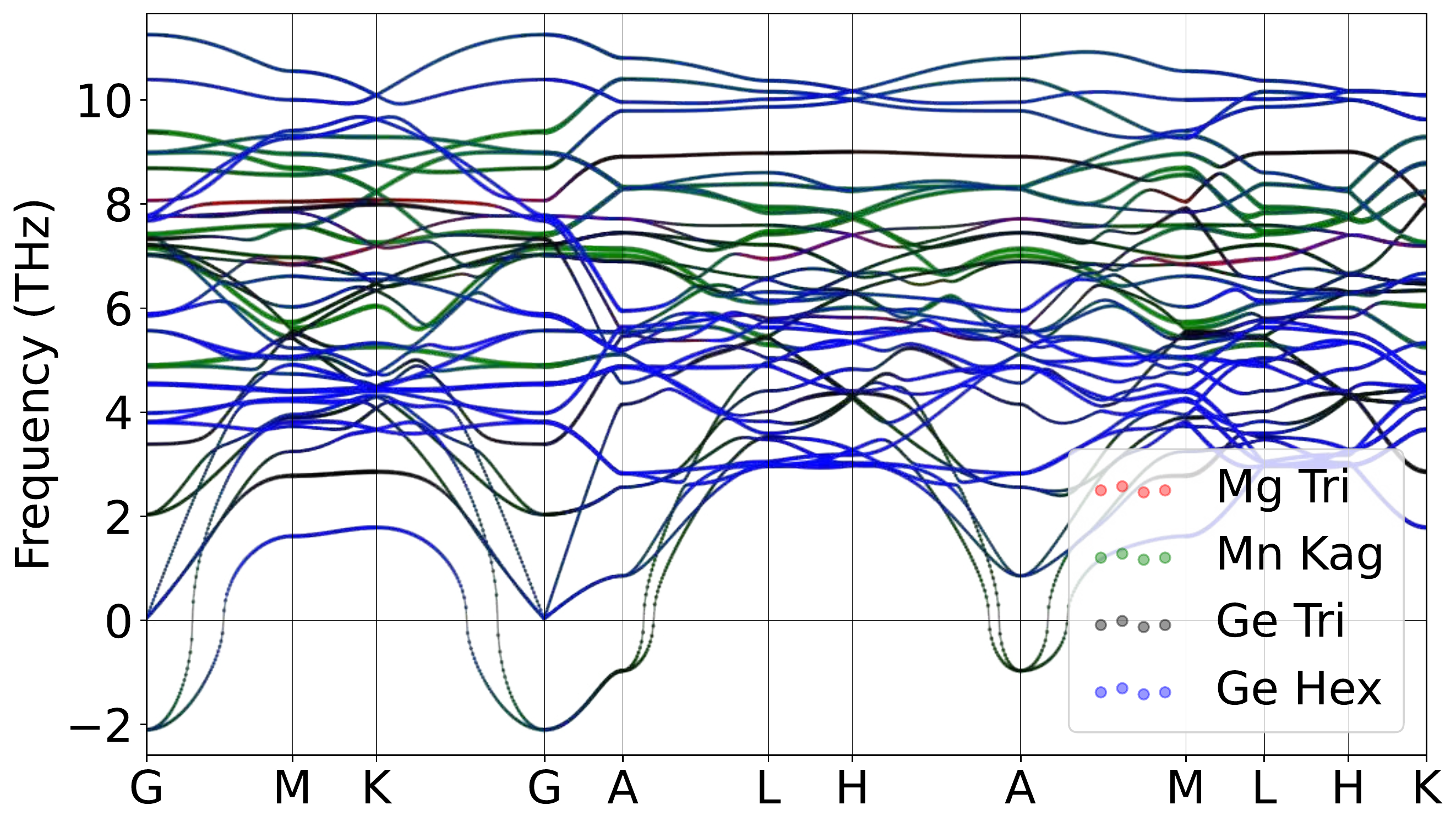}
}
\hspace{5pt}\subfloat[Relaxed phonon at high-T.]{
\label{fig:MgMn6Ge6_relaxed_High}
\includegraphics[width=0.45\textwidth]{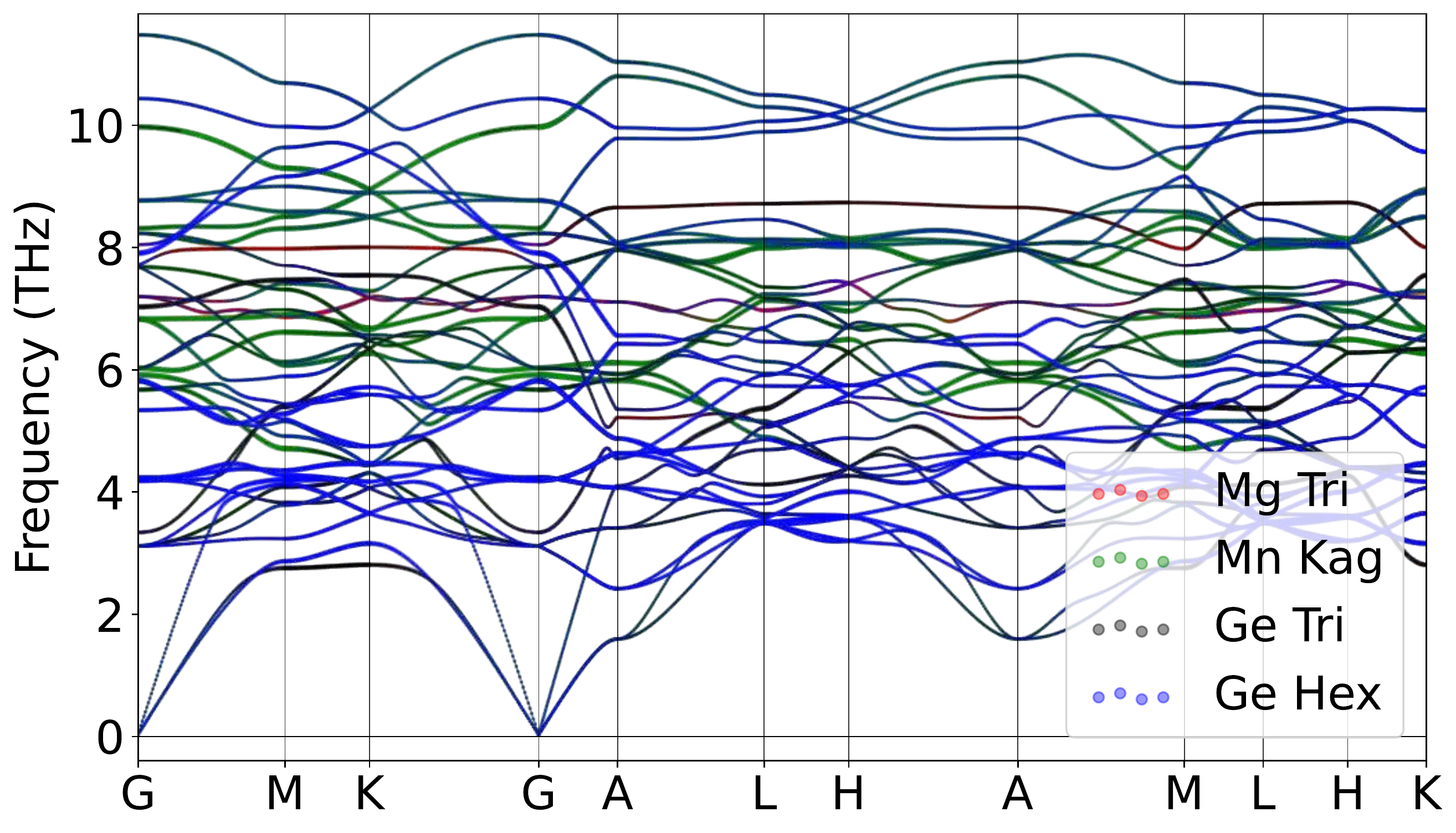}
}
\caption{Atomically projected phonon spectrum for MgMn$_6$Ge$_6$ in paramagnetic phase.}
\label{fig:MgMn6Ge6pho}
\end{figure}

\begin{figure}[htbp]
\centering
\includegraphics[width=0.32\textwidth]{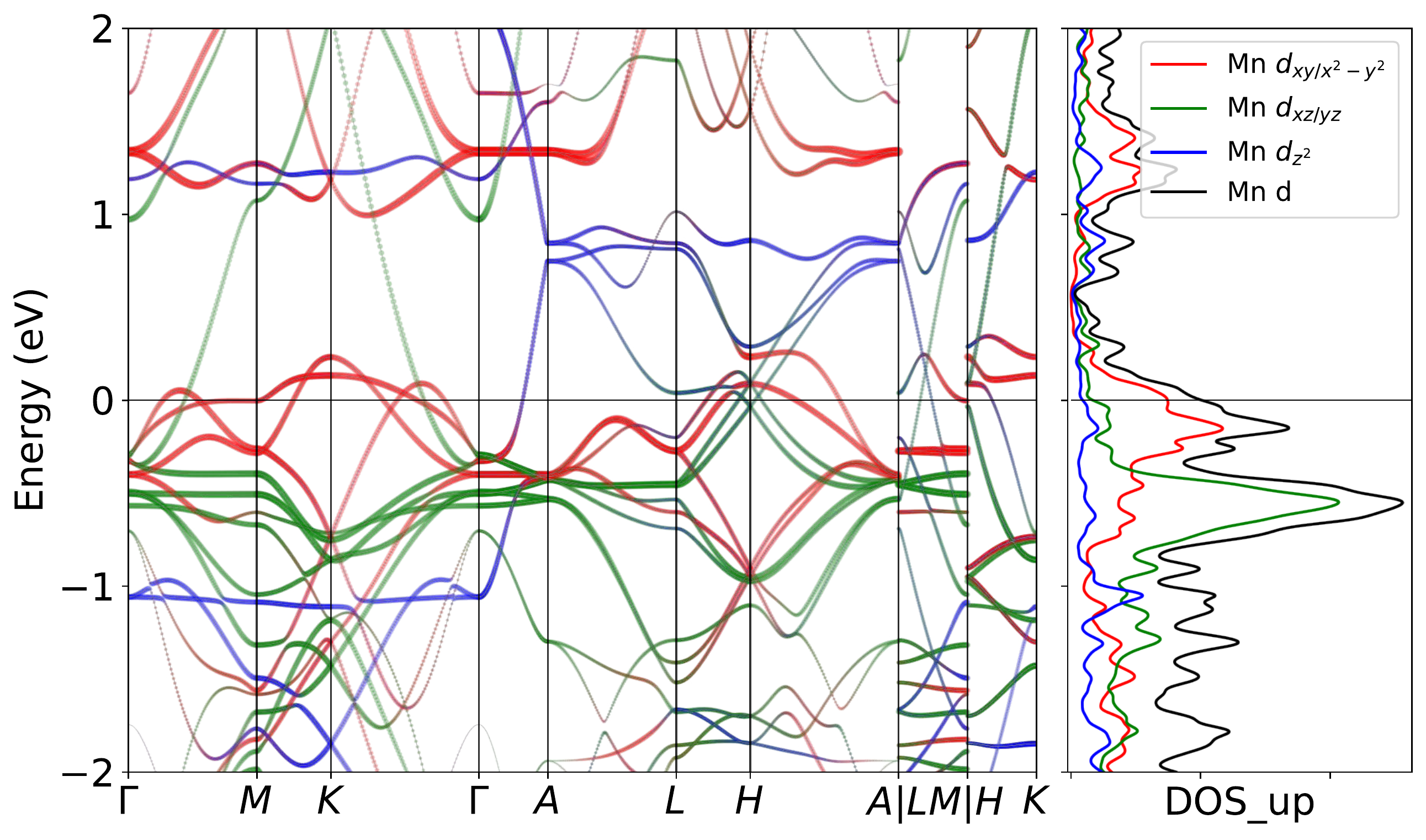}
\includegraphics[width=0.32\textwidth]{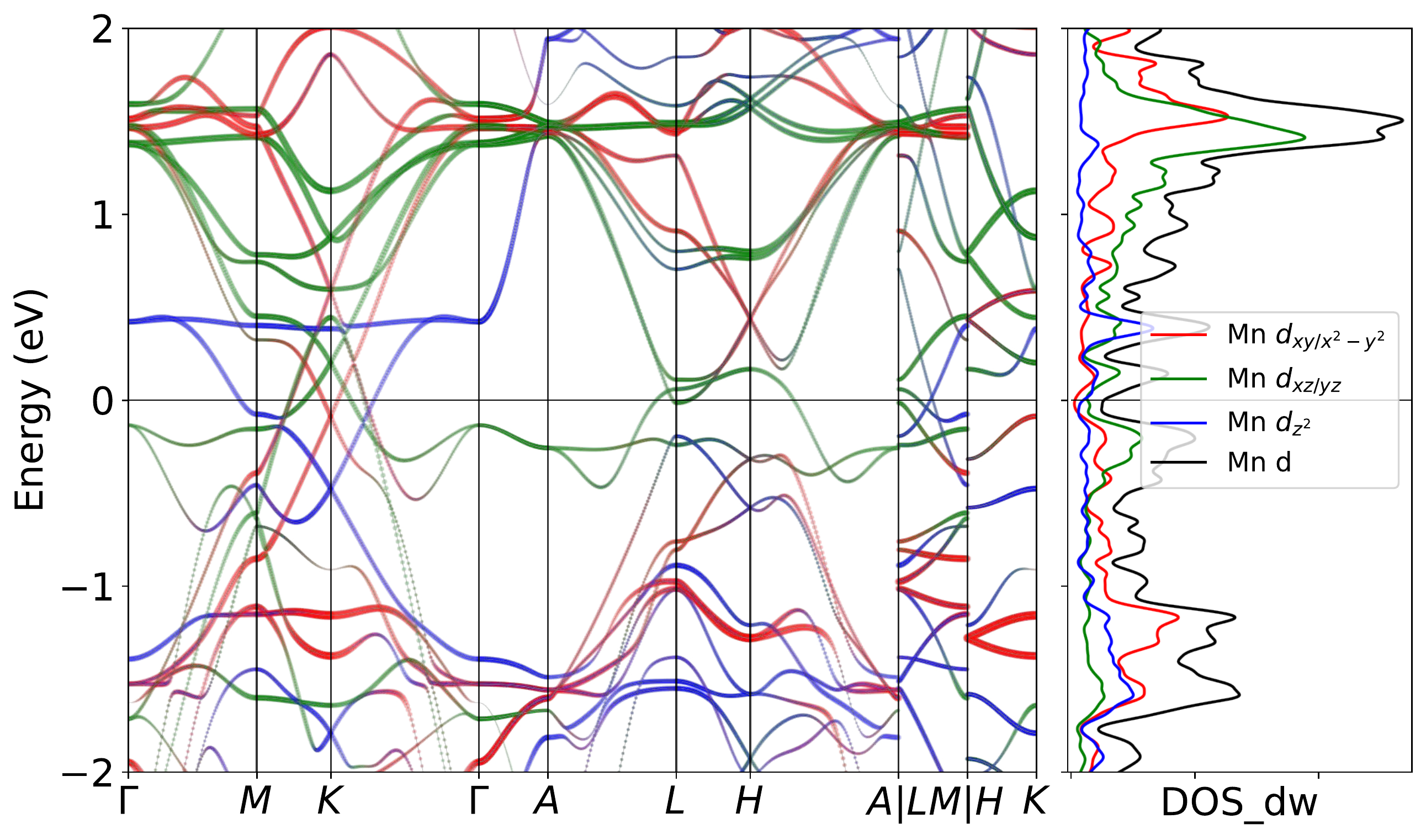}
\includegraphics[width=0.32\textwidth]{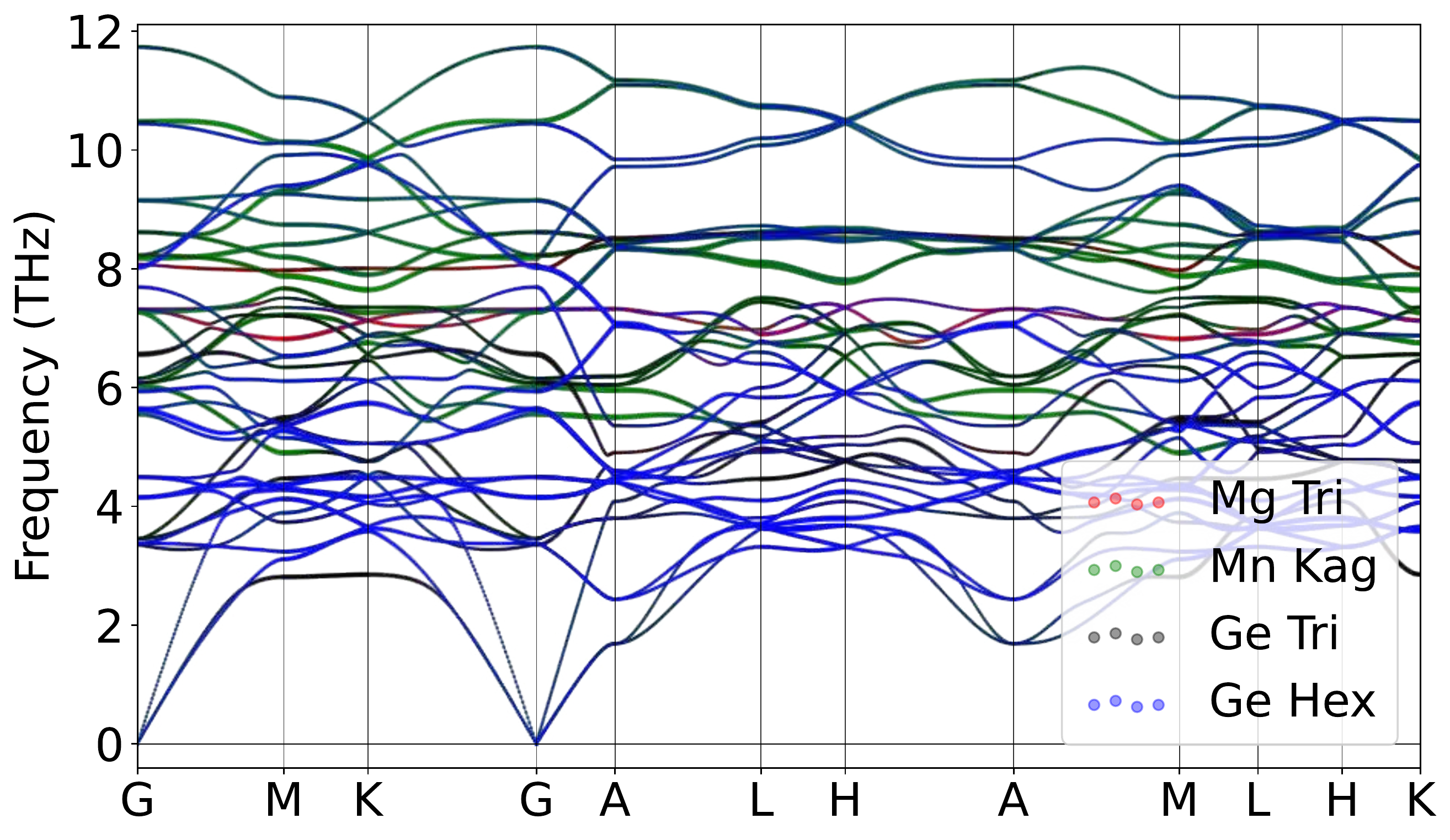}
\caption{Band structure for MgMn$_6$Ge$_6$ with FM order without SOC for spin (Left)up and (Middle)down channels. The projection of Mn $3d$ orbitals is also presented. (Right)Correspondingly, a stable phonon was demonstrated with the collinear magnetic order without Hubbard U at lower temperature regime, weighted by atomic projections.}
\label{fig:MgMn6Ge6mag}
\end{figure}

\begin{figure}[H]
\centering
\label{fig:MgMn6Ge6_relaxed_Low_xz}
\includegraphics[width=0.85\textwidth]{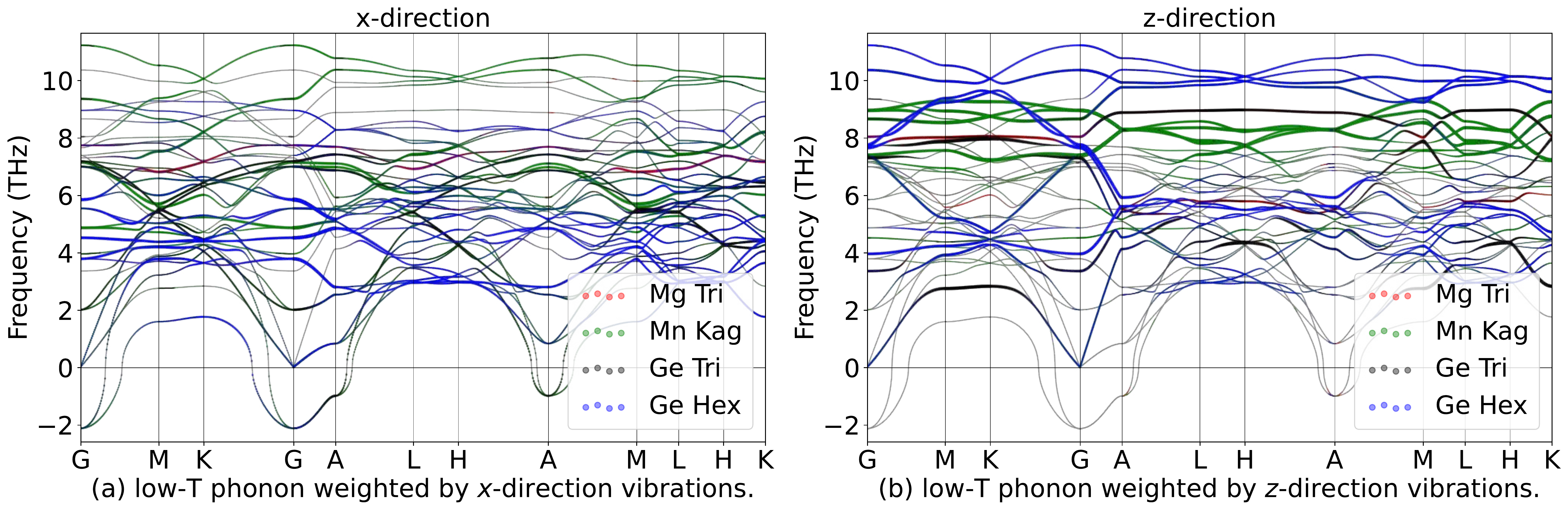}
\hspace{5pt}\label{fig:MgMn6Ge6_relaxed_High_xz}
\includegraphics[width=0.85\textwidth]{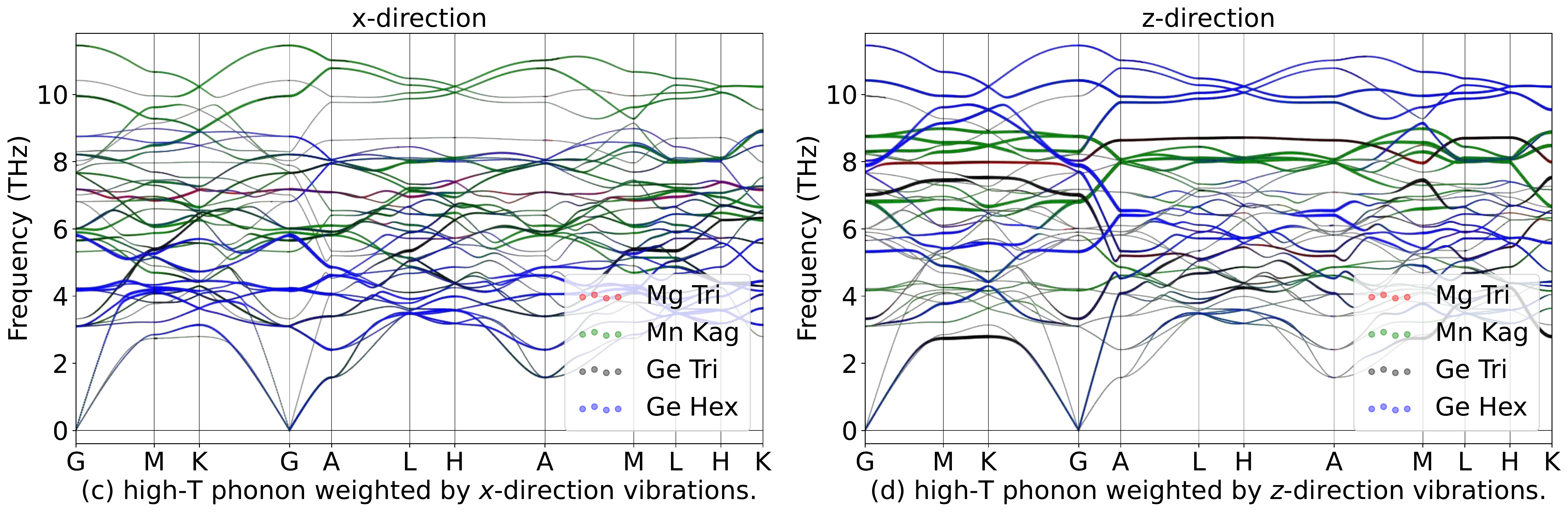}
\caption{Phonon spectrum for MgMn$_6$Ge$_6$in in-plane ($x$) and out-of-plane ($z$) directions in paramagnetic phase.}
\label{fig:MgMn6Ge6phoxyz}
\end{figure}

\begin{table}[htbp]
\global\long\def\arraystretch{1.12}
\captionsetup{width=.95\textwidth}
\caption{IRREPs at high-symmetry points for MgMn$_6$Ge$_6$ phonon spectrum at Low-T.}
\label{tab:191_MgMn6Ge6_Low}
\setlength{\tabcolsep}{1.mm}{
}
\end{table}

\begin{table}[htbp]
\global\long\def\arraystretch{1.12}
\captionsetup{width=.95\textwidth}
\caption{IRREPs at high-symmetry points for MgMn$_6$Ge$_6$ phonon spectrum at High-T.}
\label{tab:191_MgMn6Ge6_High}
\setlength{\tabcolsep}{1.mm}{
%
}
\end{table}
\subsubsection{\label{sms2sec:CaMn6Ge6}\hyperref[smsec:col]{\texorpdfstring{CaMn$_6$Ge$_6$}{}}~{\color{blue}  (High-T stable, Low-T unstable) }}

\begin{table}[htbp]
\global\long\def\arraystretch{1.12}
\captionsetup{width=.85\textwidth}
\caption{Structure parameters of CaMn$_6$Ge$_6$ after relaxation. It contains 362 electrons. }
\label{tab:CaMn6Ge6}
\setlength{\tabcolsep}{4mm}{
%
}
\end{table}

\begin{figure}[htbp]
\centering
\includegraphics[width=0.85\textwidth]{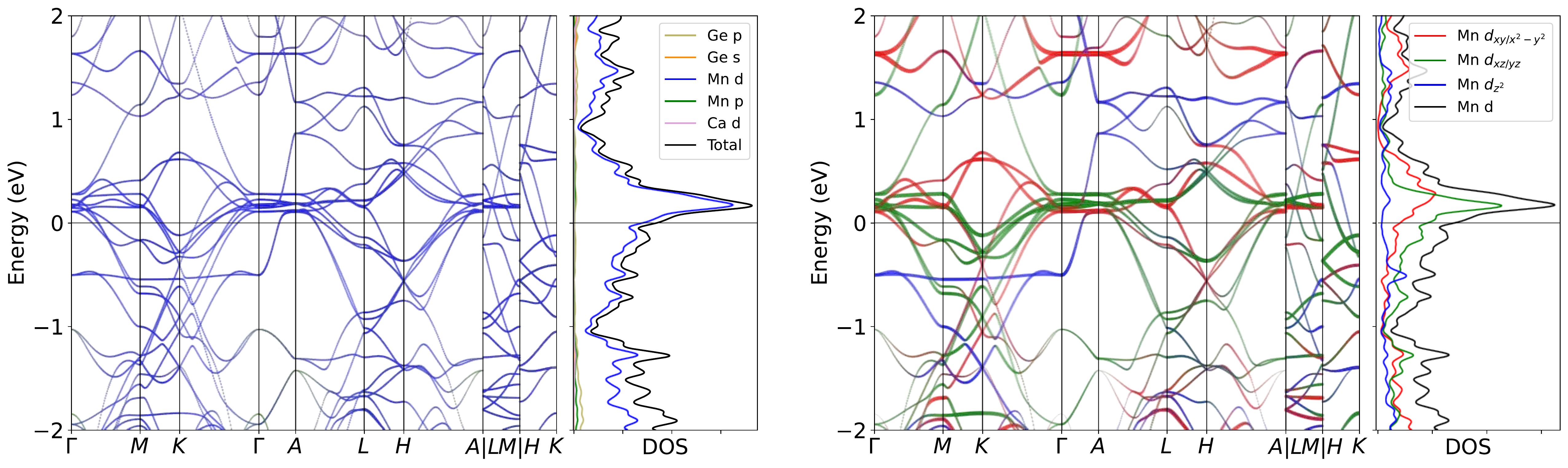}
\caption{Band structure for CaMn$_6$Ge$_6$ without SOC in paramagnetic phase. The projection of Mn $3d$ orbitals is also presented. The band structures clearly reveal the dominating of Mn $3d$ orbitals  near the Fermi level, as well as the pronounced peak.}
\label{fig:CaMn6Ge6band}
\end{figure}

\begin{figure}[H]
\centering
\subfloat[Relaxed phonon at low-T.]{
\label{fig:CaMn6Ge6_relaxed_Low}
\includegraphics[width=0.45\textwidth]{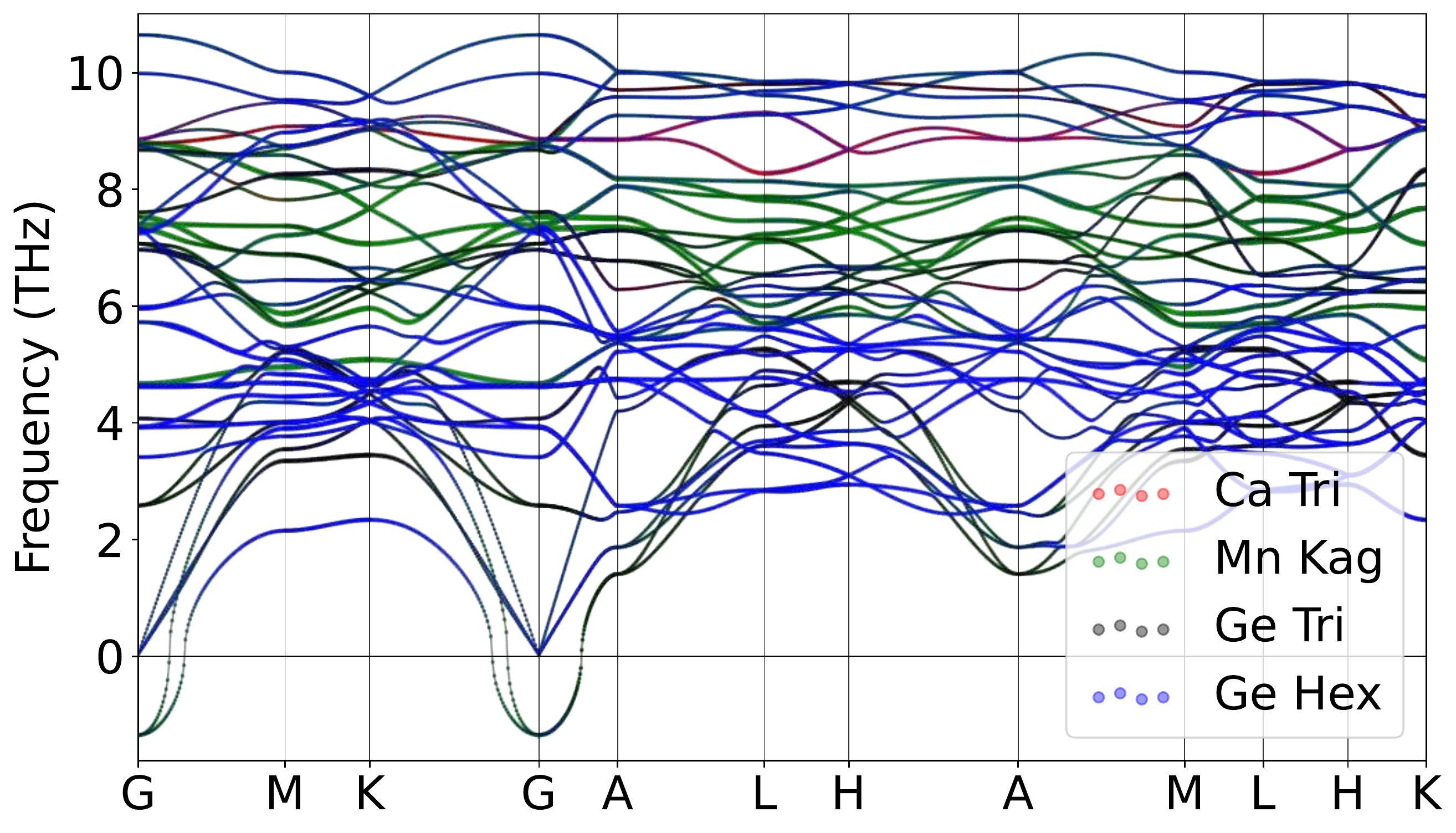}
}
\hspace{5pt}\subfloat[Relaxed phonon at high-T.]{
\label{fig:CaMn6Ge6_relaxed_High}
\includegraphics[width=0.45\textwidth]{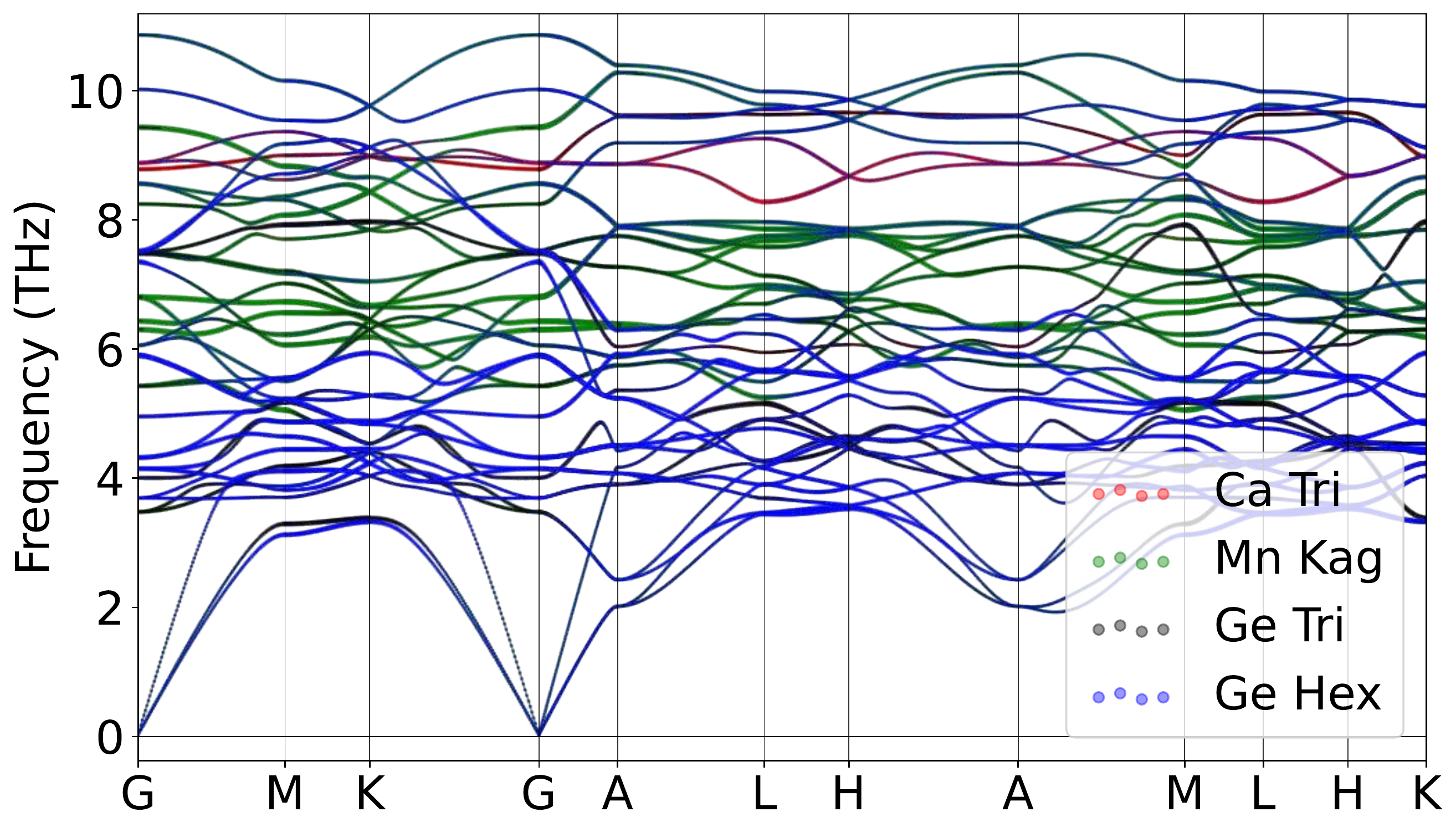}
}
\caption{Atomically projected phonon spectrum for CaMn$_6$Ge$_6$ in paramagnetic phase.}
\label{fig:CaMn6Ge6pho}
\end{figure}

\begin{figure}[htbp]
\centering
\includegraphics[width=0.32\textwidth]{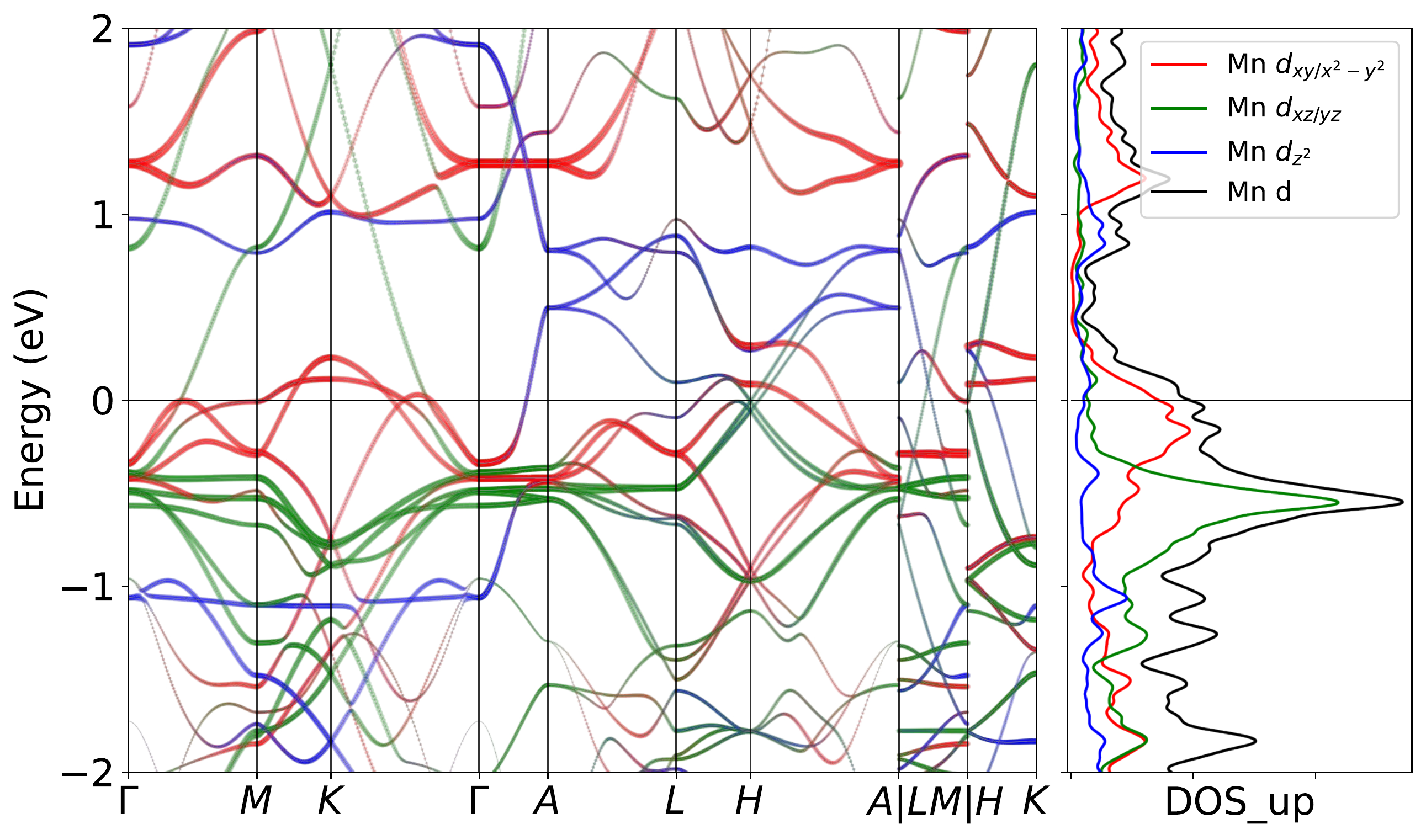}
\includegraphics[width=0.32\textwidth]{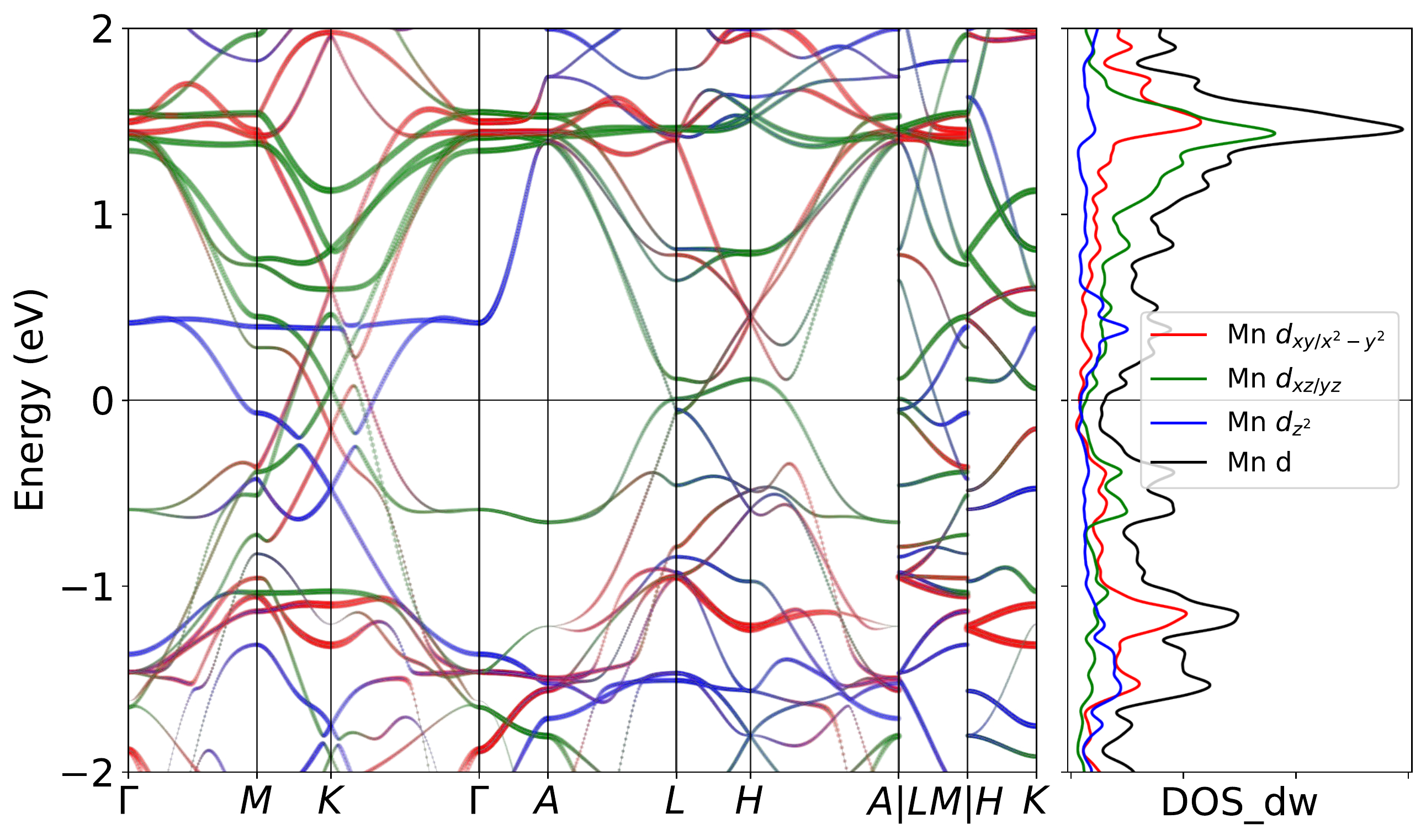}
\includegraphics[width=0.32\textwidth]{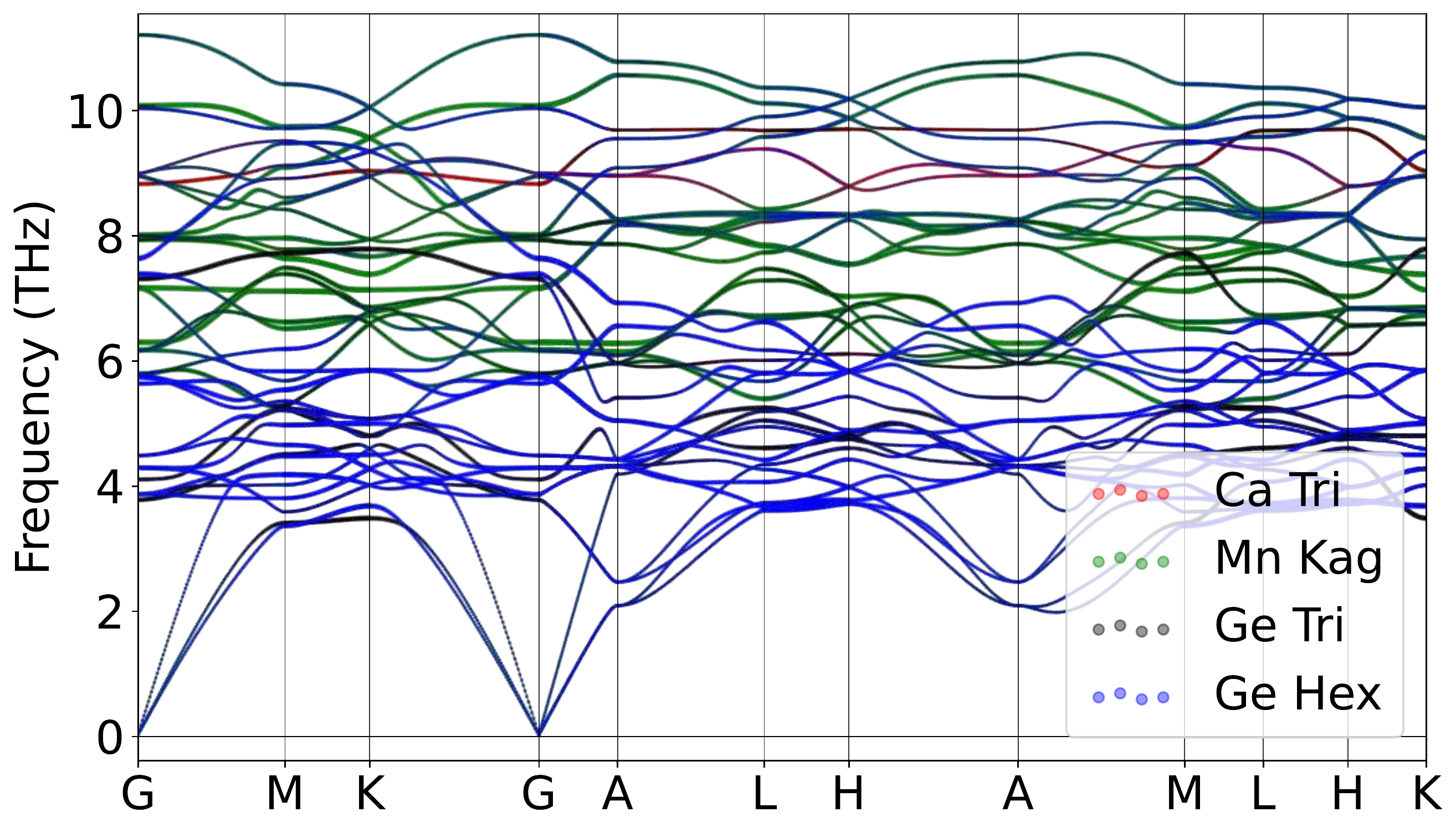}
\caption{Band structure for CaMn$_6$Ge$_6$ with FM order without SOC for spin (Left)up and (Middle)down channels. The projection of Mn $3d$ orbitals is also presented. (Right)Correspondingly, a stable phonon was demonstrated with the collinear magnetic order without Hubbard U at lower temperature regime, weighted by atomic projections.}
\label{fig:CaMn6Ge6mag}
\end{figure}

\begin{figure}[H]
\centering
\label{fig:CaMn6Ge6_relaxed_Low_xz}
\includegraphics[width=0.85\textwidth]{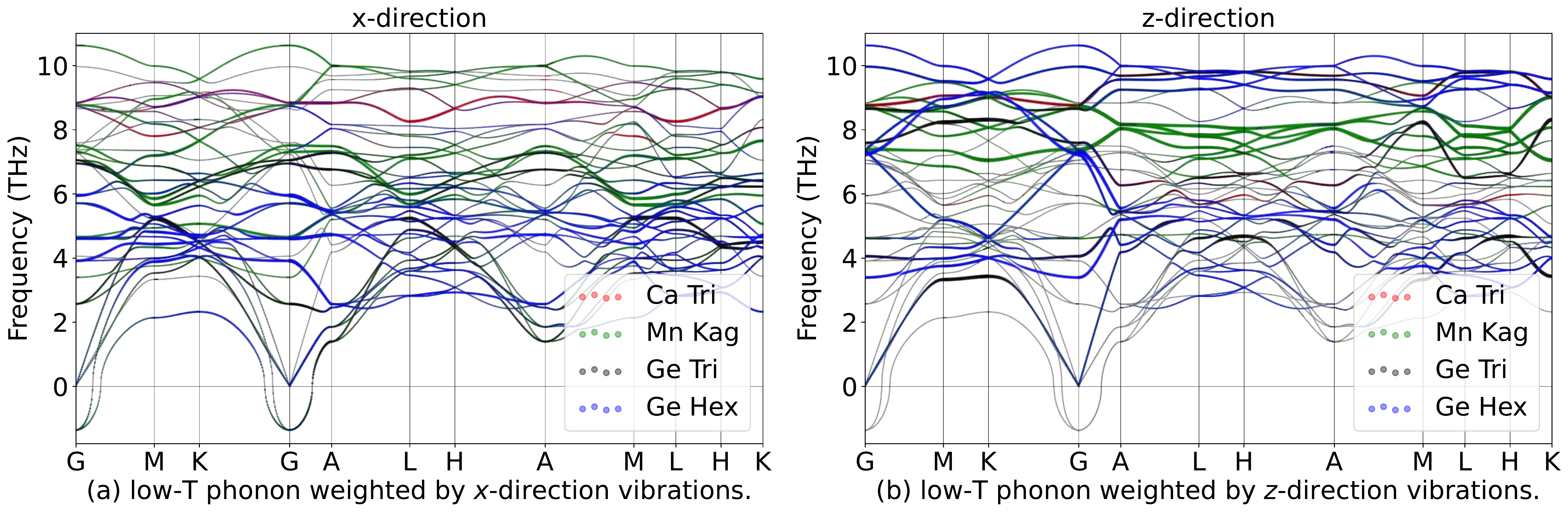}
\hspace{5pt}\label{fig:CaMn6Ge6_relaxed_High_xz}
\includegraphics[width=0.85\textwidth]{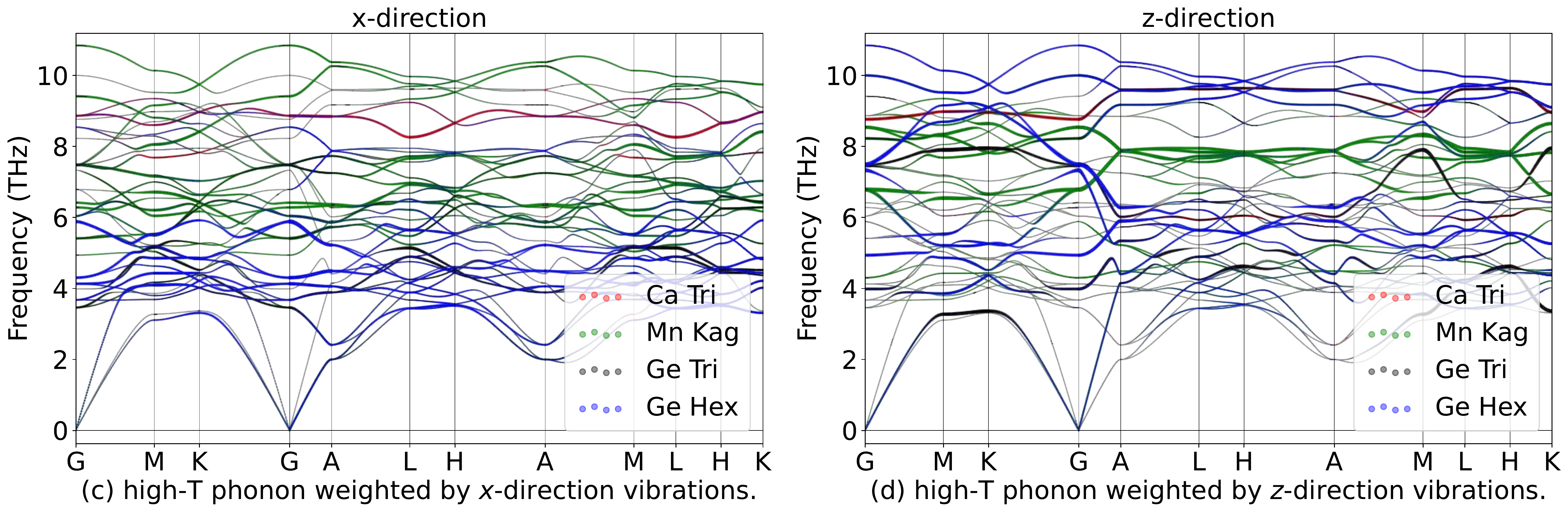}
\caption{Phonon spectrum for CaMn$_6$Ge$_6$in in-plane ($x$) and out-of-plane ($z$) directions in paramagnetic phase.}
\label{fig:CaMn6Ge6phoxyz}
\end{figure}

\begin{table}[htbp]
\global\long\def\arraystretch{1.12}
\captionsetup{width=.95\textwidth}
\caption{IRREPs at high-symmetry points for CaMn$_6$Ge$_6$ phonon spectrum at Low-T.}
\label{tab:191_CaMn6Ge6_Low}
\setlength{\tabcolsep}{1.mm}{
}
\end{table}

\begin{table}[htbp]
\global\long\def\arraystretch{1.12}
\captionsetup{width=.95\textwidth}
\caption{IRREPs at high-symmetry points for CaMn$_6$Ge$_6$ phonon spectrum at High-T.}
\label{tab:191_CaMn6Ge6_High}
\setlength{\tabcolsep}{1.mm}{
%
}
\end{table}
\subsubsection{\label{sms2sec:ScMn6Ge6}\hyperref[smsec:col]{\texorpdfstring{ScMn$_6$Ge$_6$}{}}~{\color{green}  (High-T stable, Low-T stable) }}

\begin{table}[htbp]
\global\long\def\arraystretch{1.12}
\captionsetup{width=.85\textwidth}
\caption{Structure parameters of ScMn$_6$Ge$_6$ after relaxation. It contains 363 electrons. }
\label{tab:ScMn6Ge6}
\setlength{\tabcolsep}{4mm}{
%
}
\end{table}

\begin{figure}[htbp]
\centering
\includegraphics[width=0.85\textwidth]{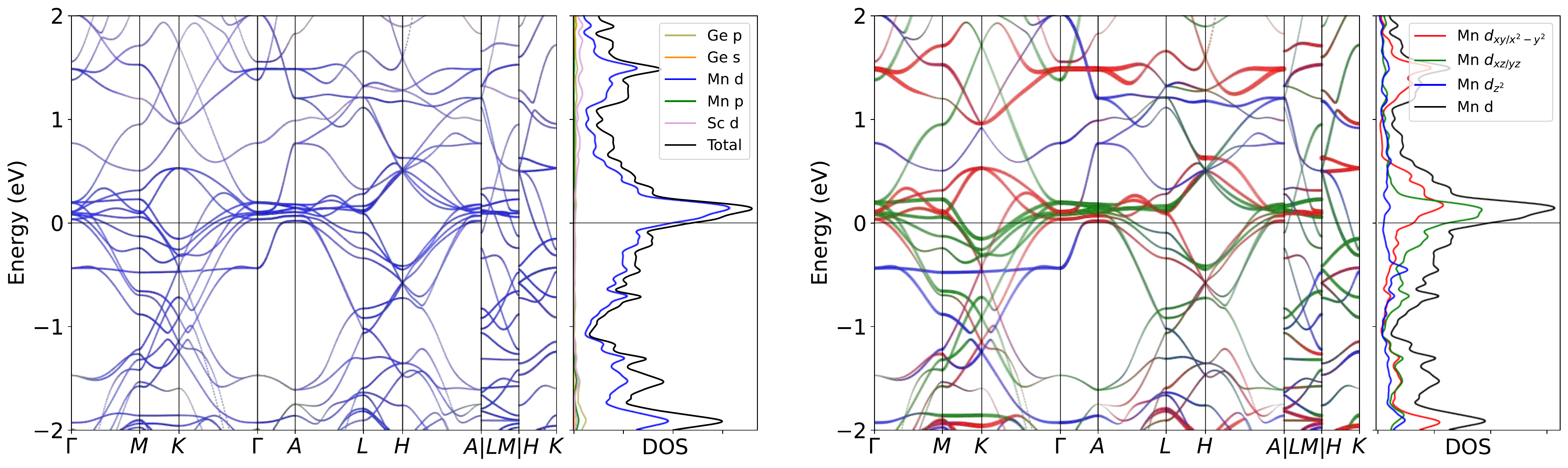}
\caption{Band structure for ScMn$_6$Ge$_6$ without SOC in paramagnetic phase. The projection of Mn $3d$ orbitals is also presented. The band structures clearly reveal the dominating of Mn $3d$ orbitals  near the Fermi level, as well as the pronounced peak.}
\label{fig:ScMn6Ge6band}
\end{figure}

\begin{figure}[H]
\centering
\subfloat[Relaxed phonon at low-T.]{
\label{fig:ScMn6Ge6_relaxed_Low}
\includegraphics[width=0.45\textwidth]{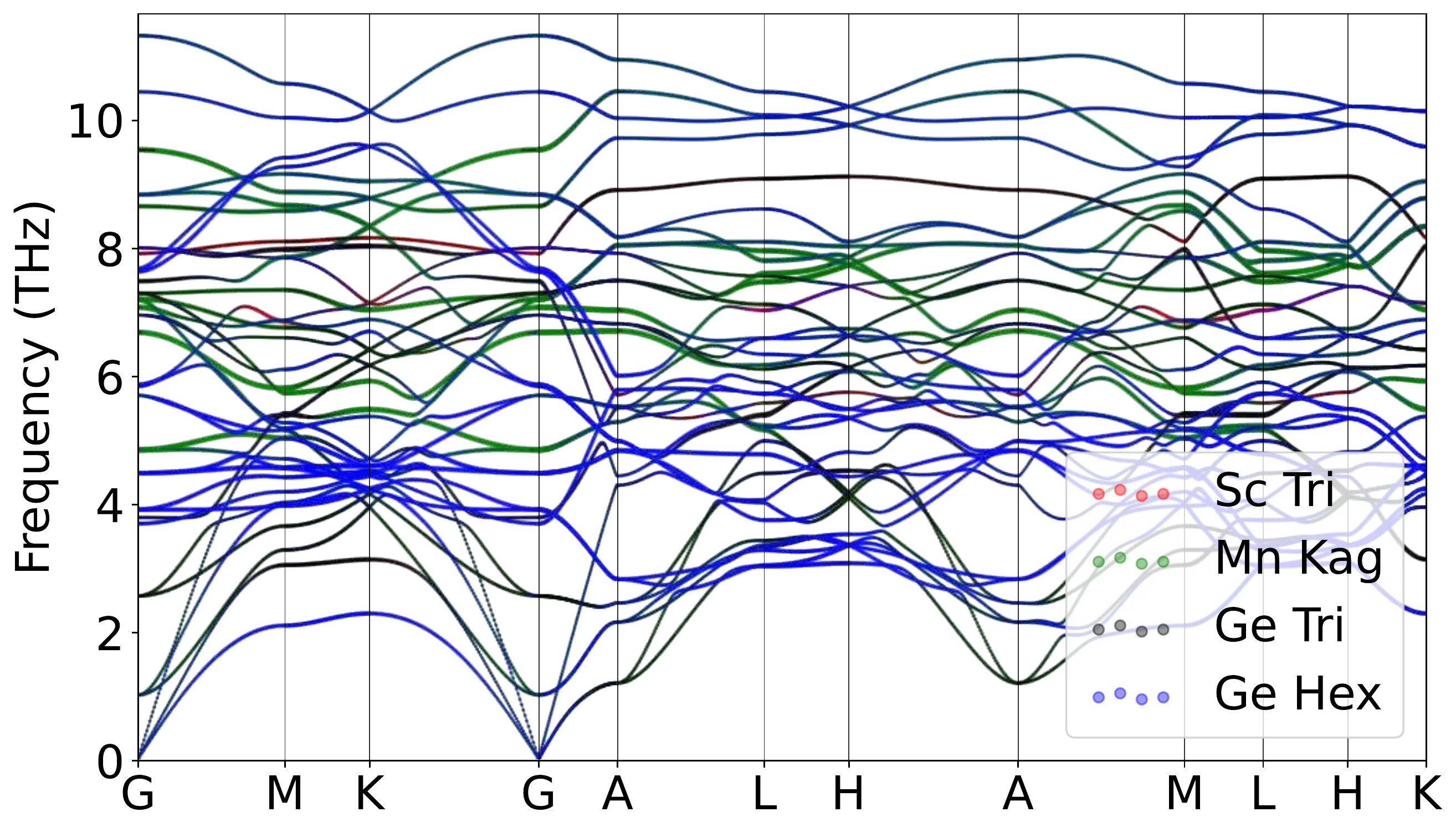}
}
\hspace{5pt}\subfloat[Relaxed phonon at high-T.]{
\label{fig:ScMn6Ge6_relaxed_High}
\includegraphics[width=0.45\textwidth]{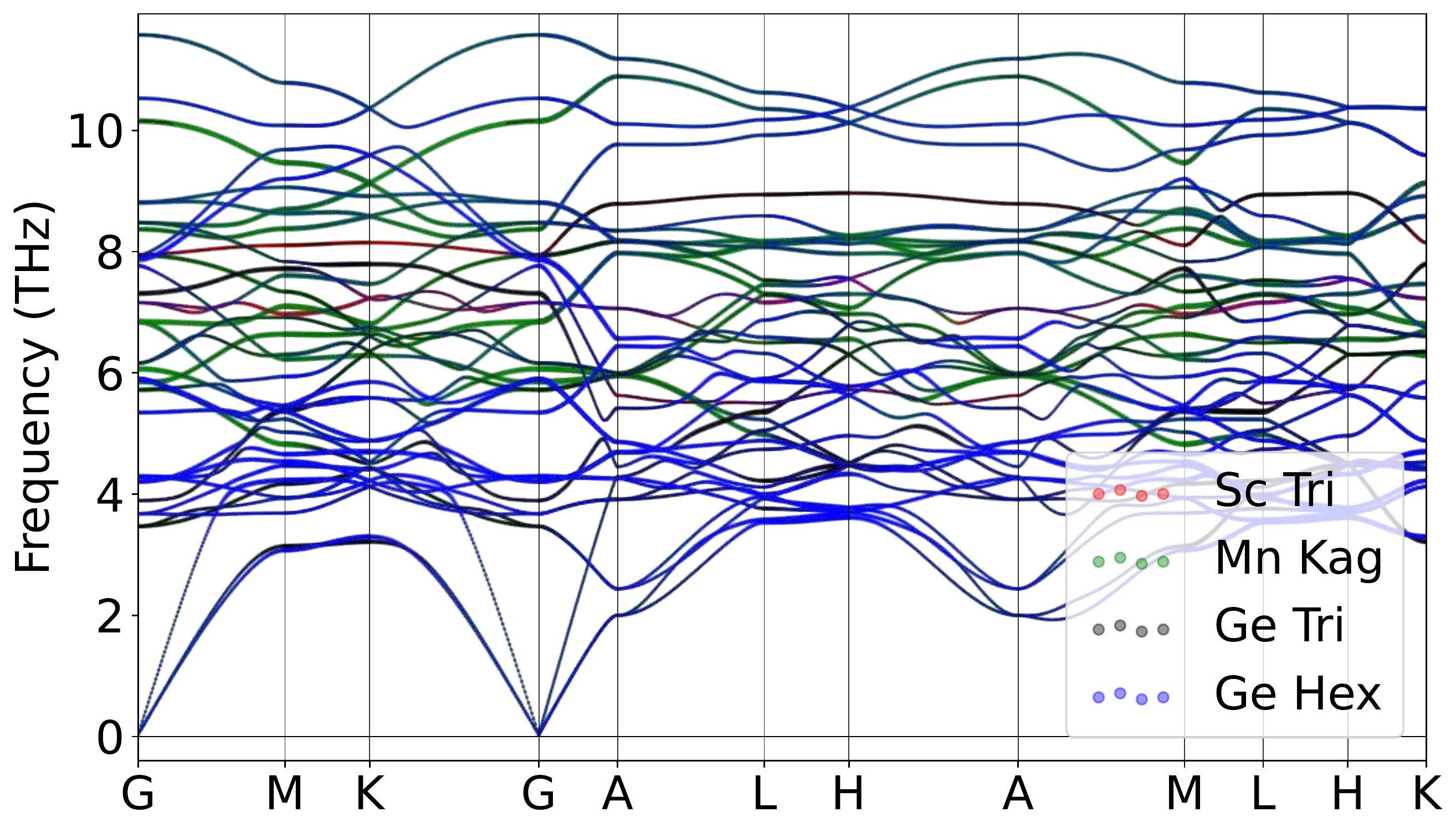}
}
\caption{Atomically projected phonon spectrum for ScMn$_6$Ge$_6$ in paramagnetic phase.}
\label{fig:ScMn6Ge6pho}
\end{figure}

\begin{figure}[htbp]
\centering
\includegraphics[width=0.45\textwidth]{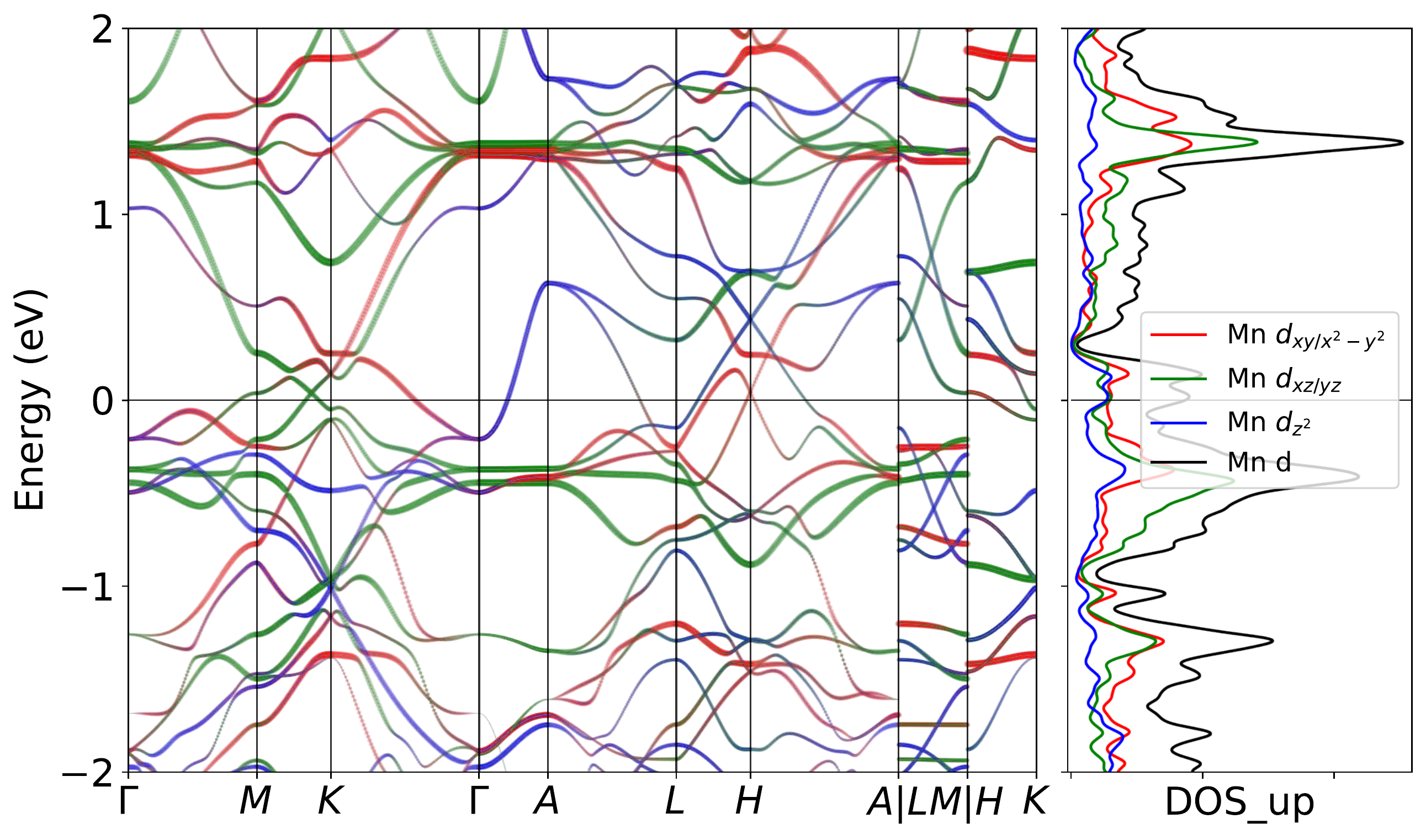}
\includegraphics[width=0.45\textwidth]{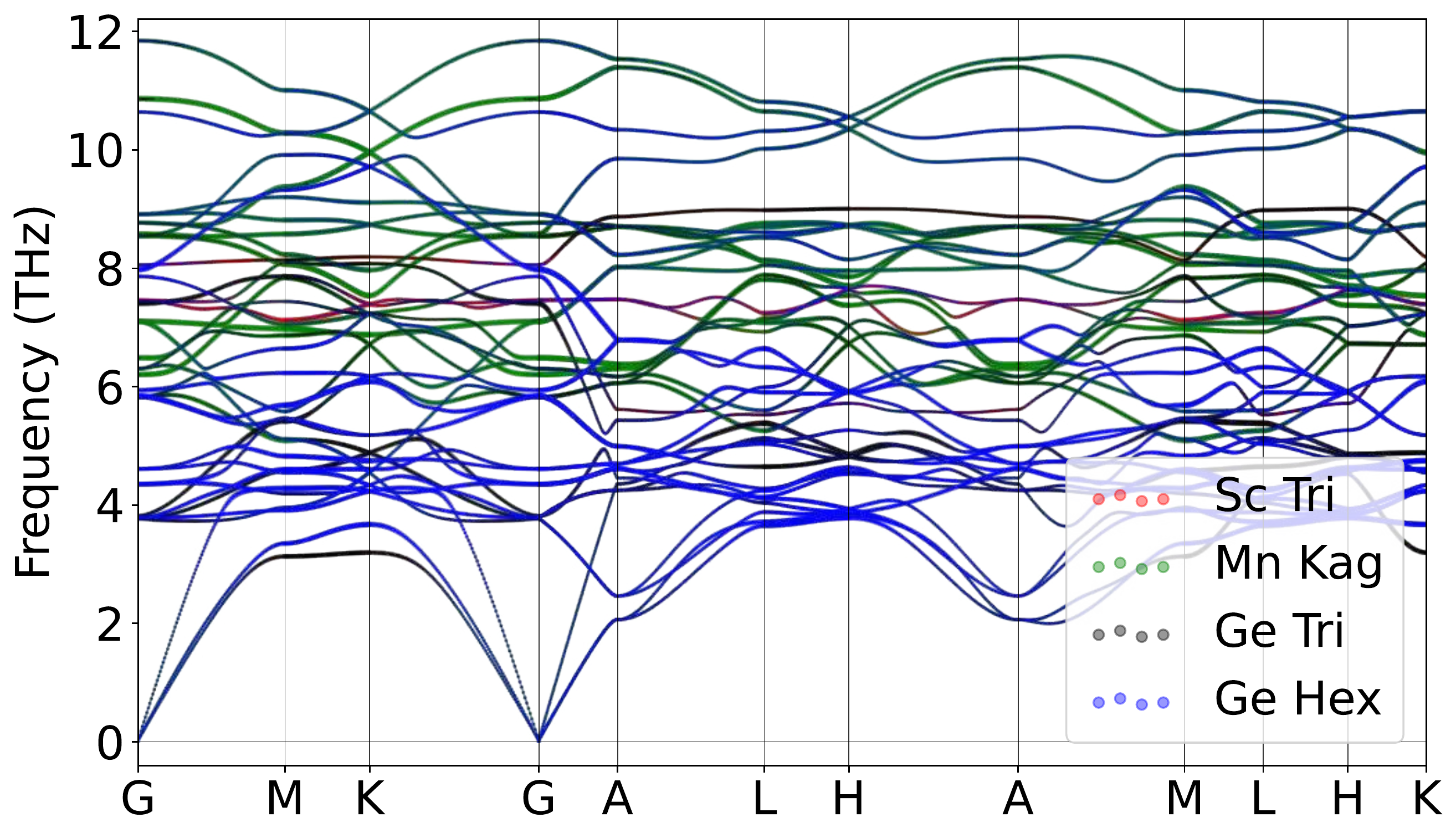}
\caption{Band structure for ScMn$_6$Ge$_6$ with AFM order without SOC for spin (Left)up channel. The projection of Mn $3d$ orbitals is also presented. (Right)Correspondingly, a stable phonon was demonstrated with the collinear magnetic order without Hubbard U at lower temperature regime, weighted by atomic projections.}
\label{fig:ScMn6Ge6mag}
\end{figure}

\begin{figure}[H]
\centering
\label{fig:ScMn6Ge6_relaxed_Low_xz}
\includegraphics[width=0.85\textwidth]{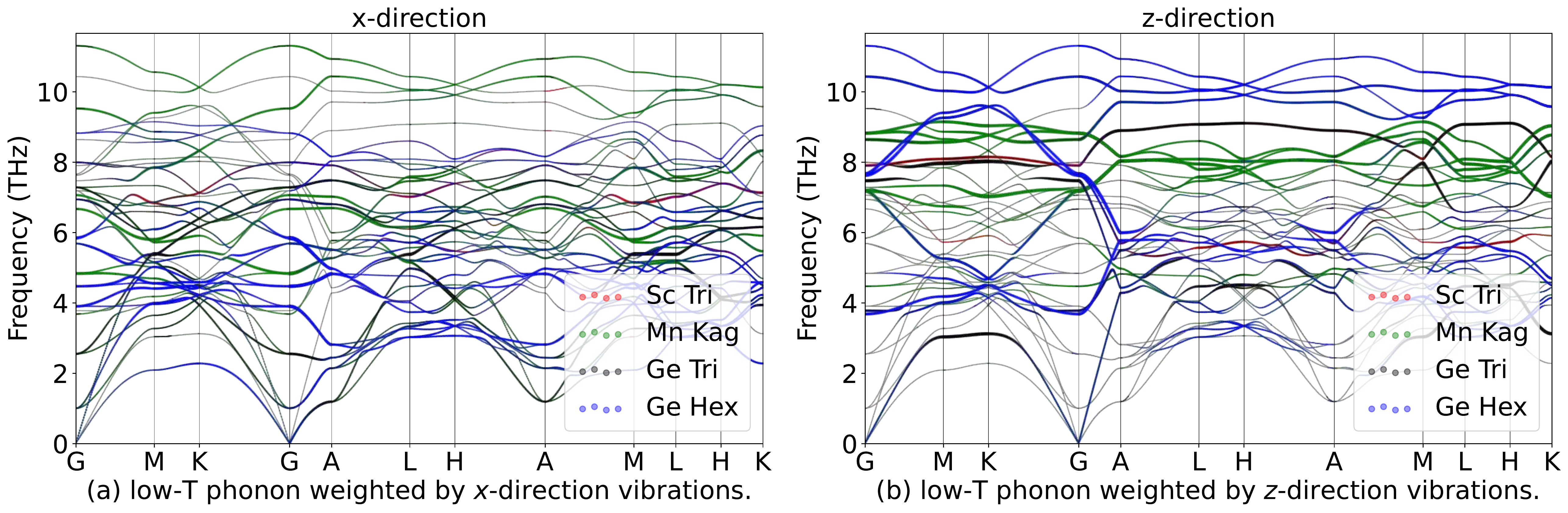}
\hspace{5pt}\label{fig:ScMn6Ge6_relaxed_High_xz}
\includegraphics[width=0.85\textwidth]{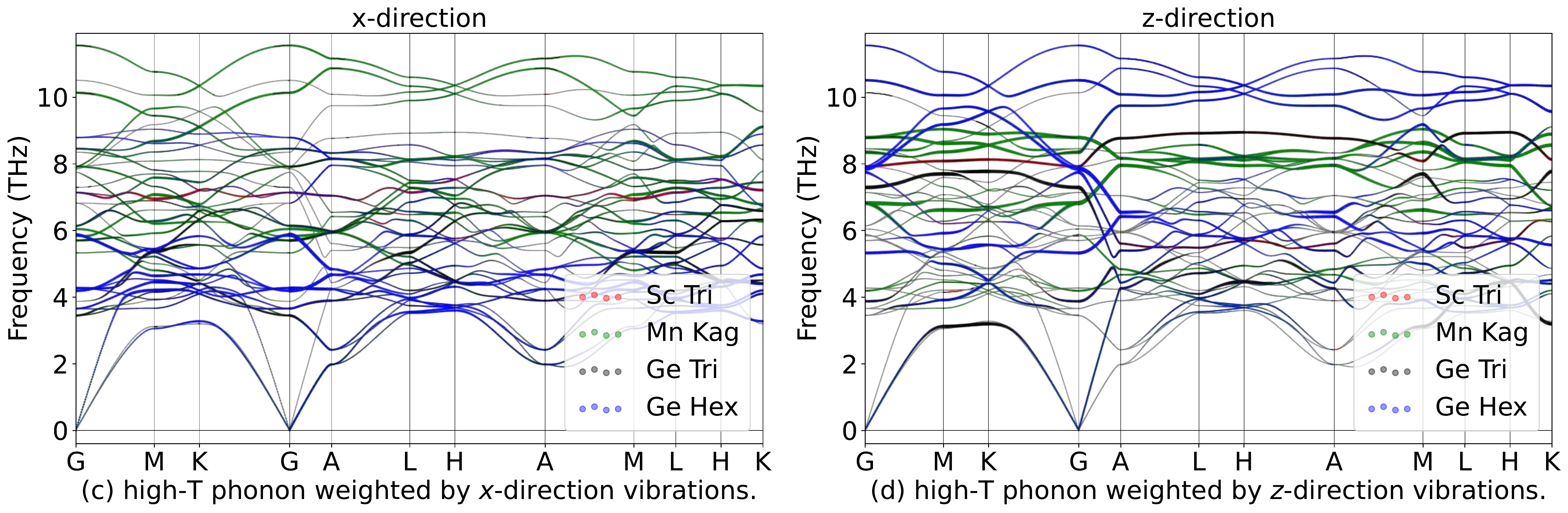}
\caption{Phonon spectrum for ScMn$_6$Ge$_6$in in-plane ($x$) and out-of-plane ($z$) directions in paramagnetic phase.}
\label{fig:ScMn6Ge6phoxyz}
\end{figure}

\begin{table}[htbp]
\global\long\def\arraystretch{1.12}
\captionsetup{width=.95\textwidth}
\caption{IRREPs at high-symmetry points for ScMn$_6$Ge$_6$ phonon spectrum at Low-T.}
\label{tab:191_ScMn6Ge6_Low}
\setlength{\tabcolsep}{1.mm}{
}
\end{table}

\begin{table}[htbp]
\global\long\def\arraystretch{1.12}
\captionsetup{width=.95\textwidth}
\caption{IRREPs at high-symmetry points for ScMn$_6$Ge$_6$ phonon spectrum at High-T.}
\label{tab:191_ScMn6Ge6_High}
\setlength{\tabcolsep}{1.mm}{
%
}
\end{table}
\subsubsection{\label{sms2sec:TiMn6Ge6}\hyperref[smsec:col]{\texorpdfstring{TiMn$_6$Ge$_6$}{}}~{\color{green}  (High-T stable, Low-T stable) }}

\begin{table}[htbp]
\global\long\def\arraystretch{1.12}
\captionsetup{width=.85\textwidth}
\caption{Structure parameters of TiMn$_6$Ge$_6$ after relaxation. It contains 364 electrons. }
\label{tab:TiMn6Ge6}
\setlength{\tabcolsep}{4mm}{
%
}
\end{table}

\begin{figure}[htbp]
\centering
\includegraphics[width=0.85\textwidth]{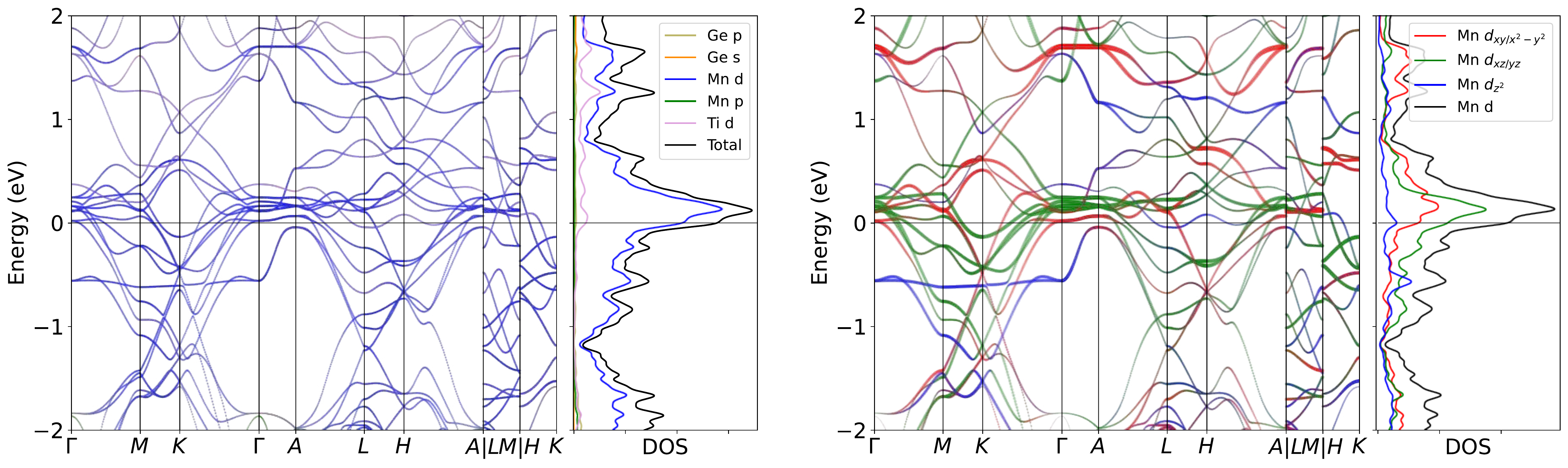}
\caption{Band structure for TiMn$_6$Ge$_6$ without SOC in paramagnetic phase. The projection of Mn $3d$ orbitals is also presented. The band structures clearly reveal the dominating of Mn $3d$ orbitals  near the Fermi level, as well as the pronounced peak.}
\label{fig:TiMn6Ge6band}
\end{figure}

\begin{figure}[H]
\centering
\subfloat[Relaxed phonon at low-T.]{
\label{fig:TiMn6Ge6_relaxed_Low}
\includegraphics[width=0.45\textwidth]{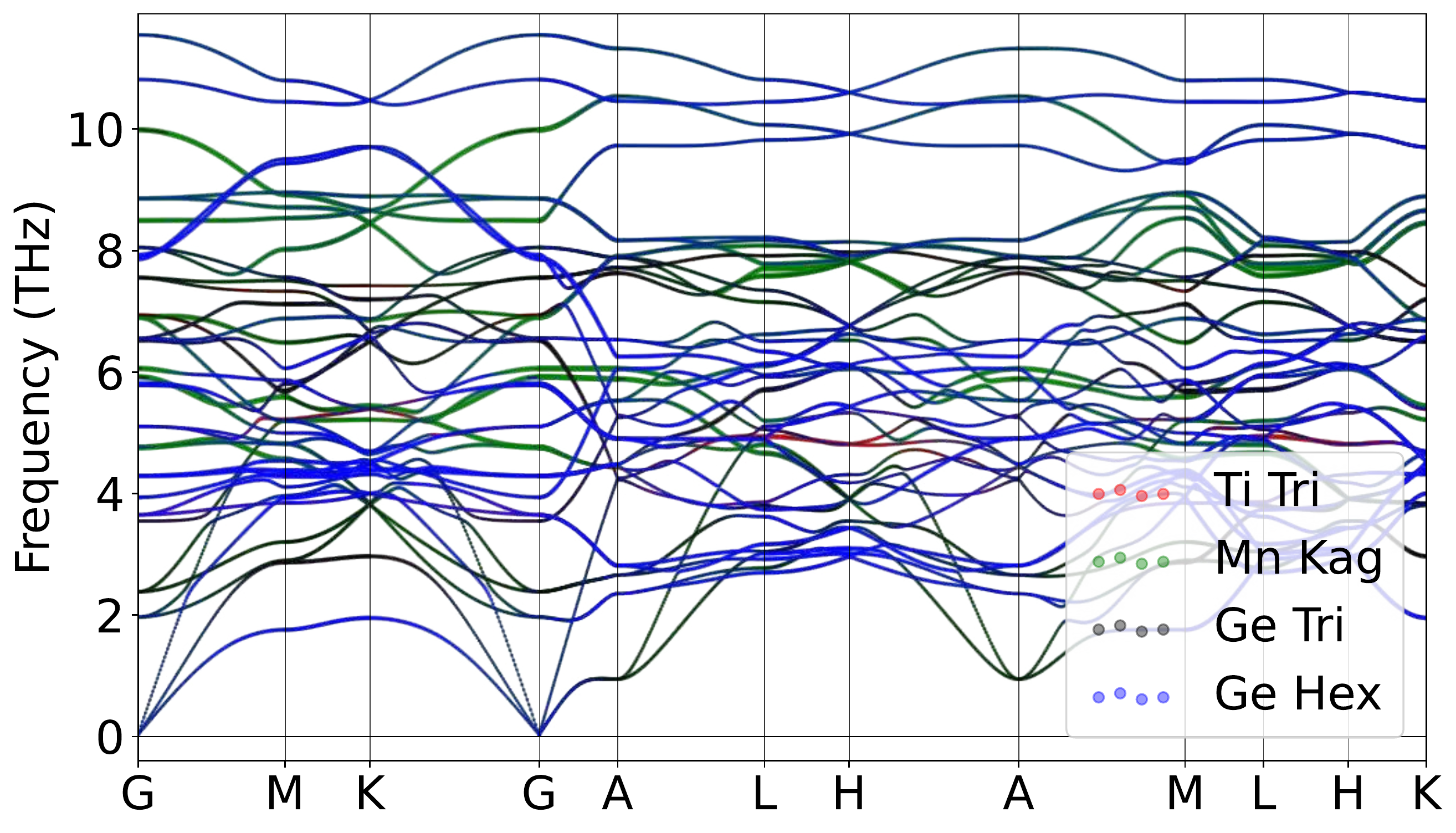}
}
\hspace{5pt}\subfloat[Relaxed phonon at high-T.]{
\label{fig:TiMn6Ge6_relaxed_High}
\includegraphics[width=0.45\textwidth]{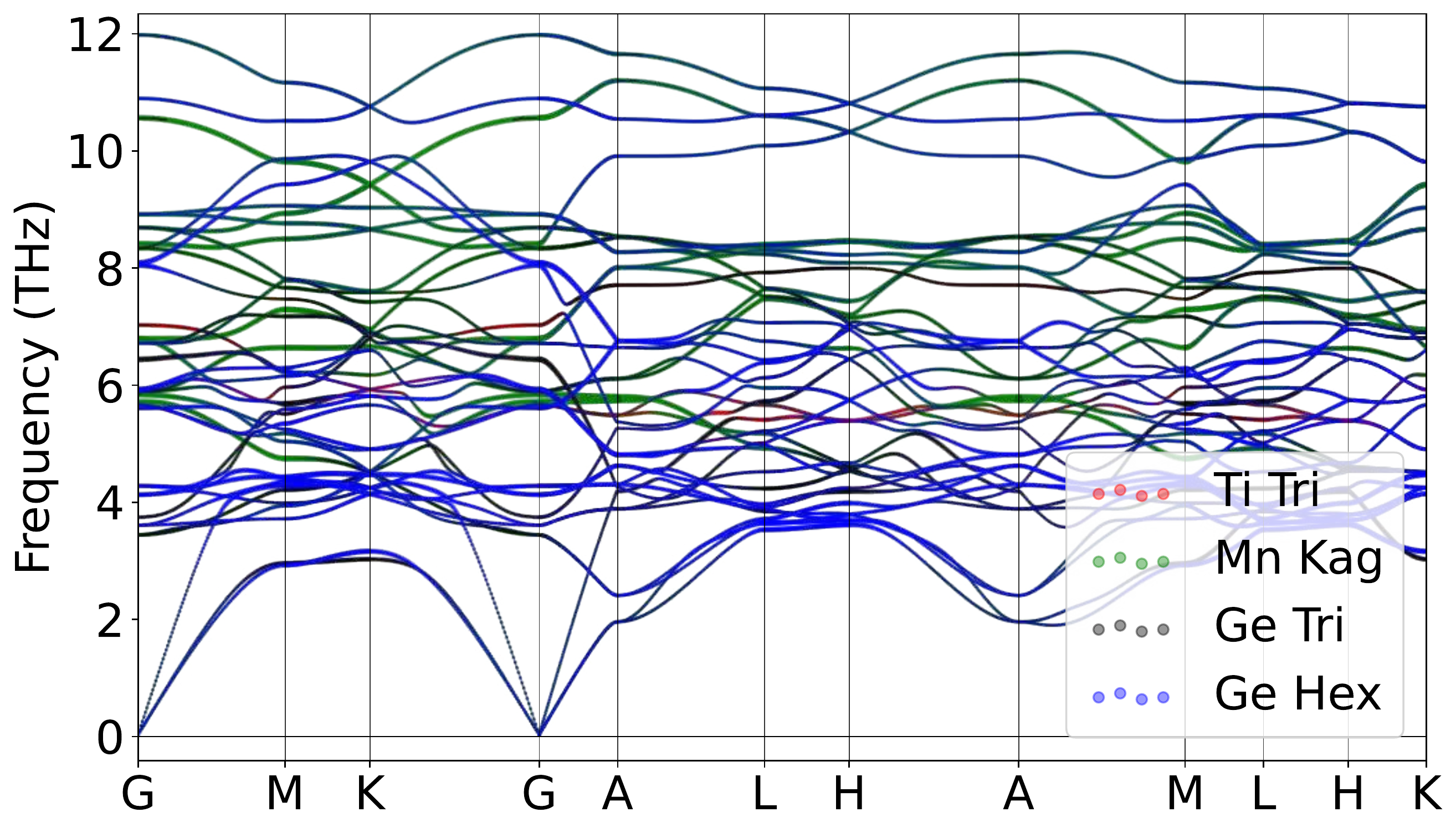}
}
\caption{Atomically projected phonon spectrum for TiMn$_6$Ge$_6$ in paramagnetic phase.}
\label{fig:TiMn6Ge6pho}
\end{figure}

\begin{figure}[htbp]
\centering
\includegraphics[width=0.45\textwidth]{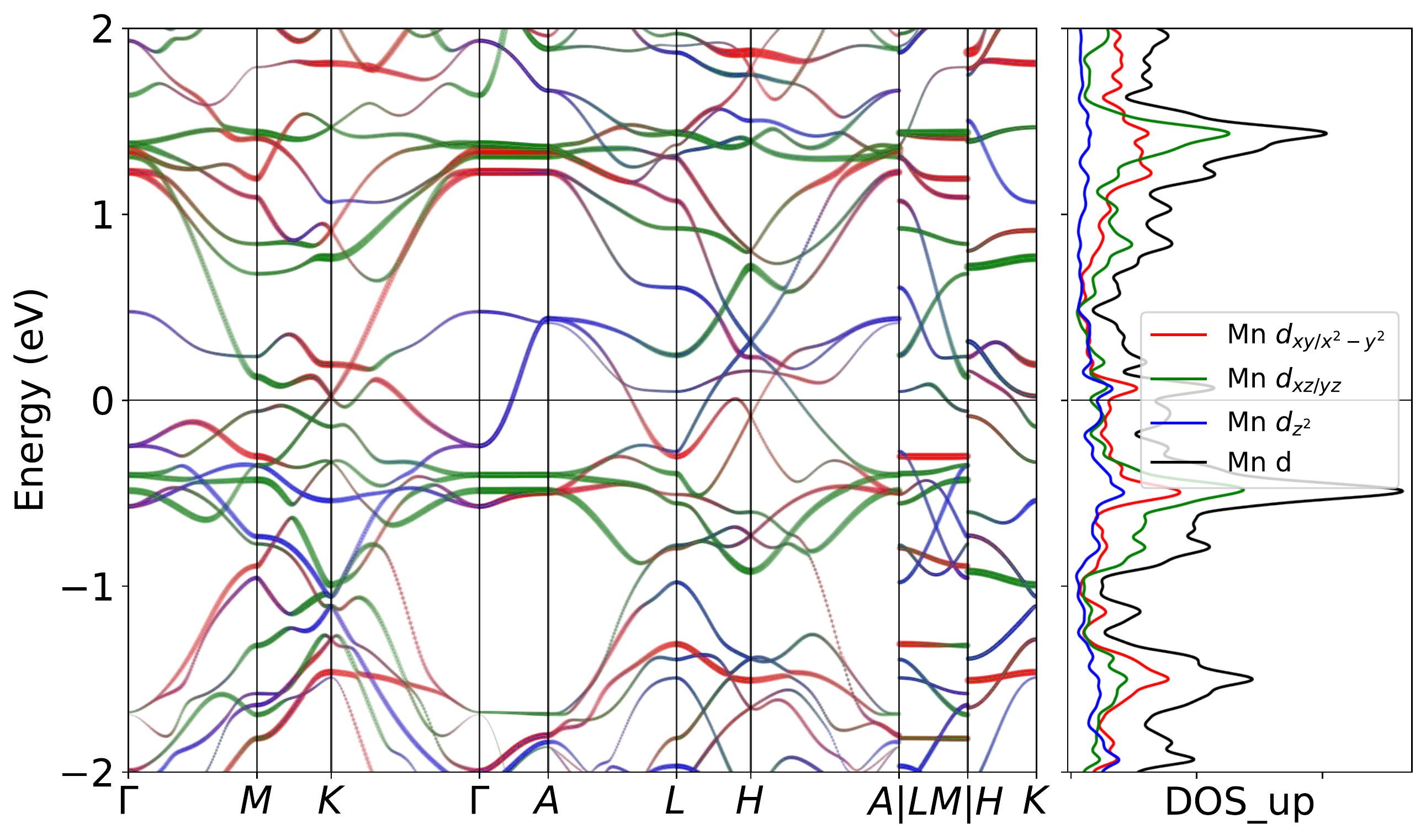}
\includegraphics[width=0.45\textwidth]{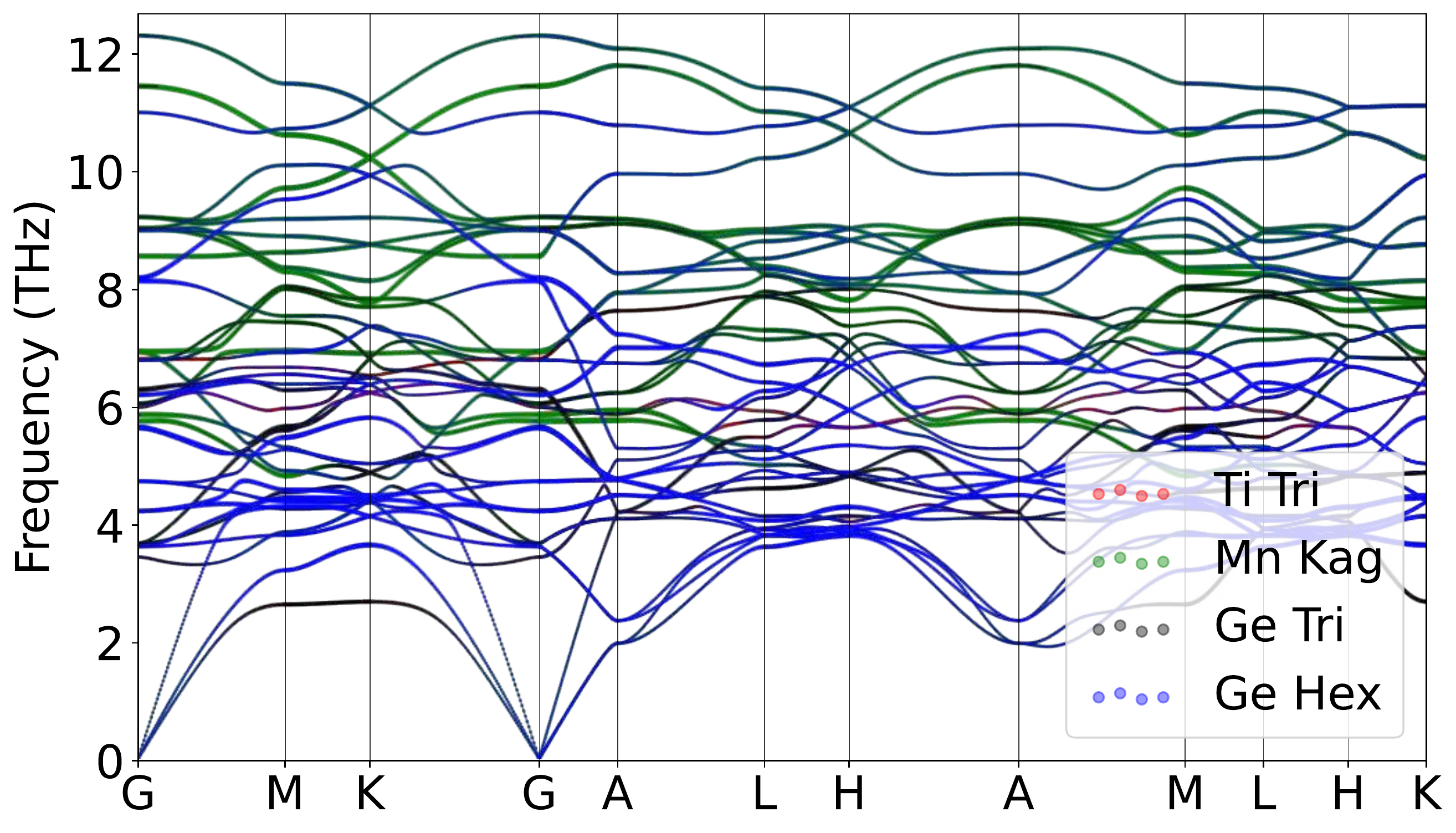}
\caption{Band structure for TiMn$_6$Ge$_6$ with AFM order without SOC for spin (Left)up channel. The projection of Mn $3d$ orbitals is also presented. (Right)Correspondingly, a stable phonon was demonstrated with the collinear magnetic order without Hubbard U at lower temperature regime, weighted by atomic projections.}
\label{fig:TiMn6Ge6mag}
\end{figure}

\begin{figure}[H]
\centering
\label{fig:TiMn6Ge6_relaxed_Low_xz}
\includegraphics[width=0.85\textwidth]{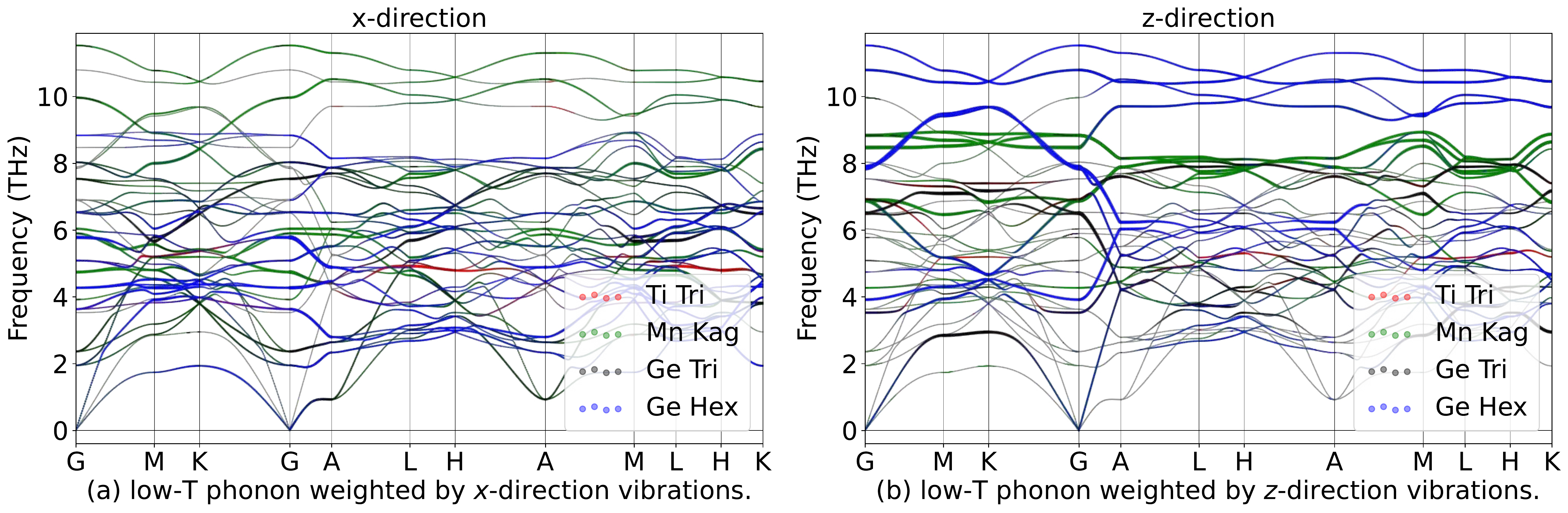}
\hspace{5pt}\label{fig:TiMn6Ge6_relaxed_High_xz}
\includegraphics[width=0.85\textwidth]{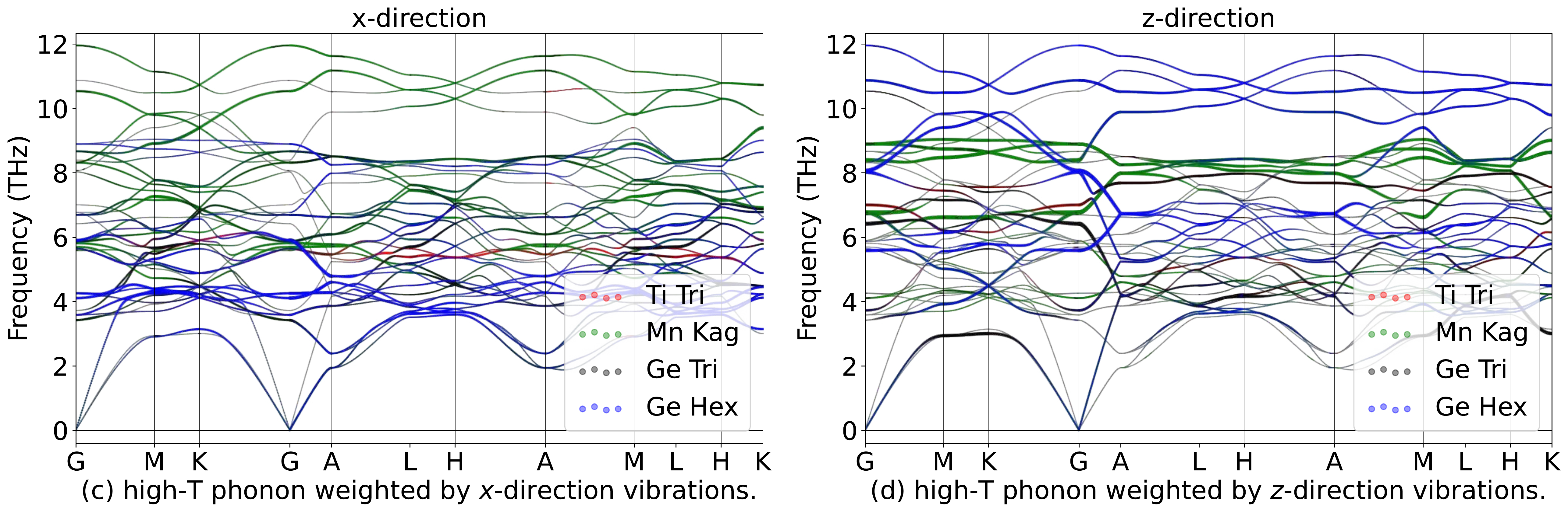}
\caption{Phonon spectrum for TiMn$_6$Ge$_6$in in-plane ($x$) and out-of-plane ($z$) directions in paramagnetic phase.}
\label{fig:TiMn6Ge6phoxyz}
\end{figure}

\begin{table}[htbp]
\global\long\def\arraystretch{1.12}
\captionsetup{width=.95\textwidth}
\caption{IRREPs at high-symmetry points for TiMn$_6$Ge$_6$ phonon spectrum at Low-T.}
\label{tab:191_TiMn6Ge6_Low}
\setlength{\tabcolsep}{1.mm}{
}
\end{table}

\begin{table}[htbp]
\global\long\def\arraystretch{1.12}
\captionsetup{width=.95\textwidth}
\caption{IRREPs at high-symmetry points for TiMn$_6$Ge$_6$ phonon spectrum at High-T.}
\label{tab:191_TiMn6Ge6_High}
\setlength{\tabcolsep}{1.mm}{
%
}
\end{table}
\subsubsection{\label{sms2sec:YMn6Ge6}\hyperref[smsec:col]{\texorpdfstring{YMn$_6$Ge$_6$}{}}~{\color{green}  (High-T stable, Low-T stable) }}

\begin{table}[htbp]
\global\long\def\arraystretch{1.12}
\captionsetup{width=.85\textwidth}
\caption{Structure parameters of YMn$_6$Ge$_6$ after relaxation. It contains 381 electrons. }
\label{tab:YMn6Ge6}
\setlength{\tabcolsep}{4mm}{
%
}
\end{table}

\begin{figure}[htbp]
\centering
\includegraphics[width=0.85\textwidth]{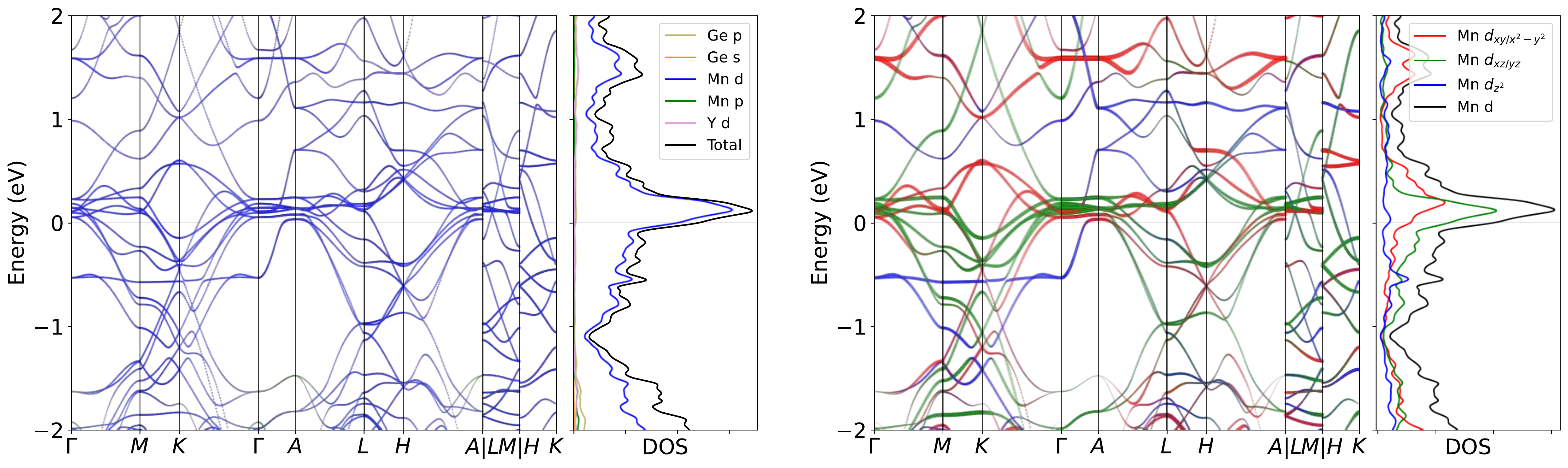}
\caption{Band structure for YMn$_6$Ge$_6$ without SOC in paramagnetic phase. The projection of Mn $3d$ orbitals is also presented. The band structures clearly reveal the dominating of Mn $3d$ orbitals  near the Fermi level, as well as the pronounced peak.}
\label{fig:YMn6Ge6band}
\end{figure}

\begin{figure}[H]
\centering
\subfloat[Relaxed phonon at low-T.]{
\label{fig:YMn6Ge6_relaxed_Low}
\includegraphics[width=0.45\textwidth]{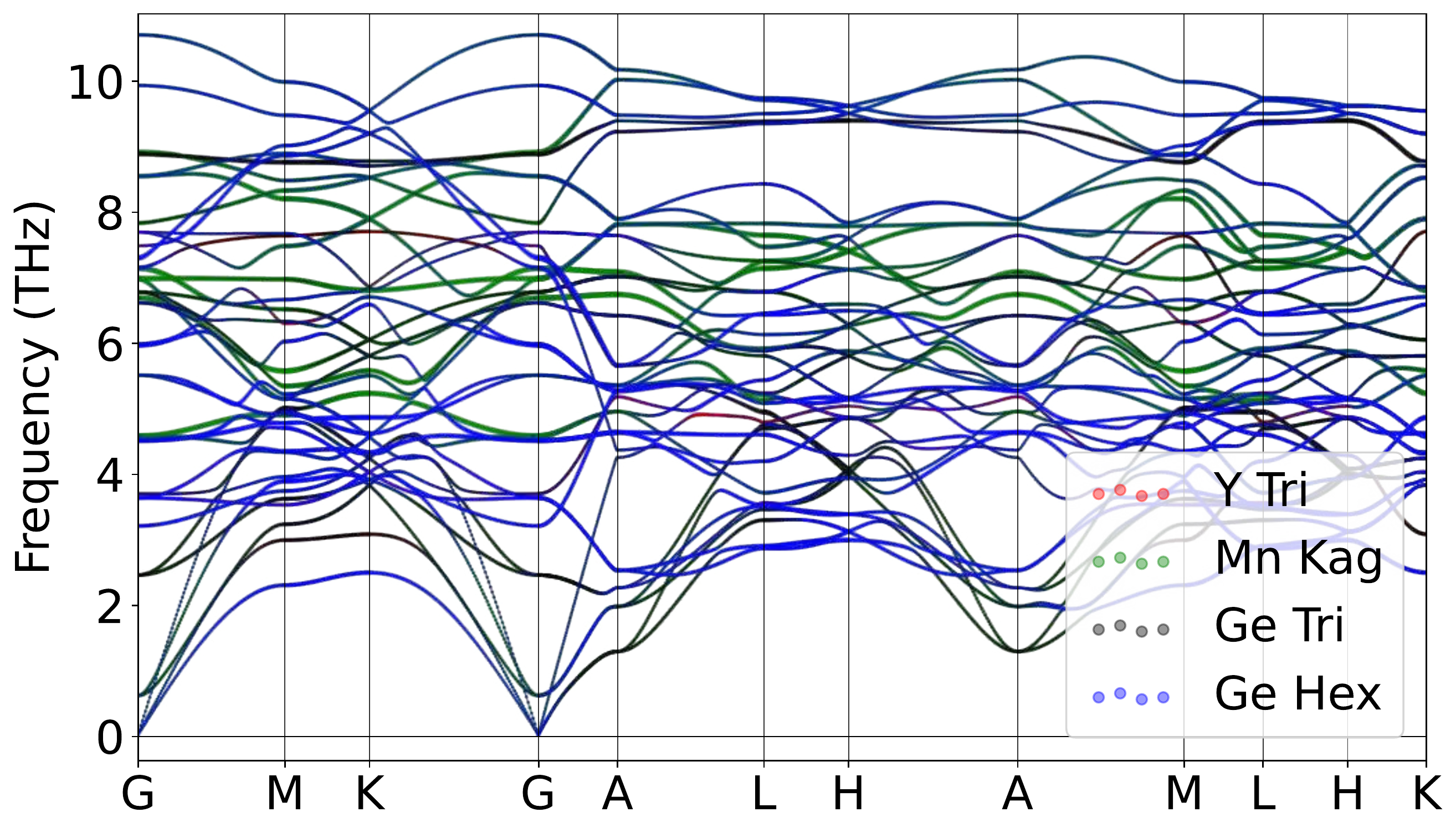}
}
\hspace{5pt}\subfloat[Relaxed phonon at high-T.]{
\label{fig:YMn6Ge6_relaxed_High}
\includegraphics[width=0.45\textwidth]{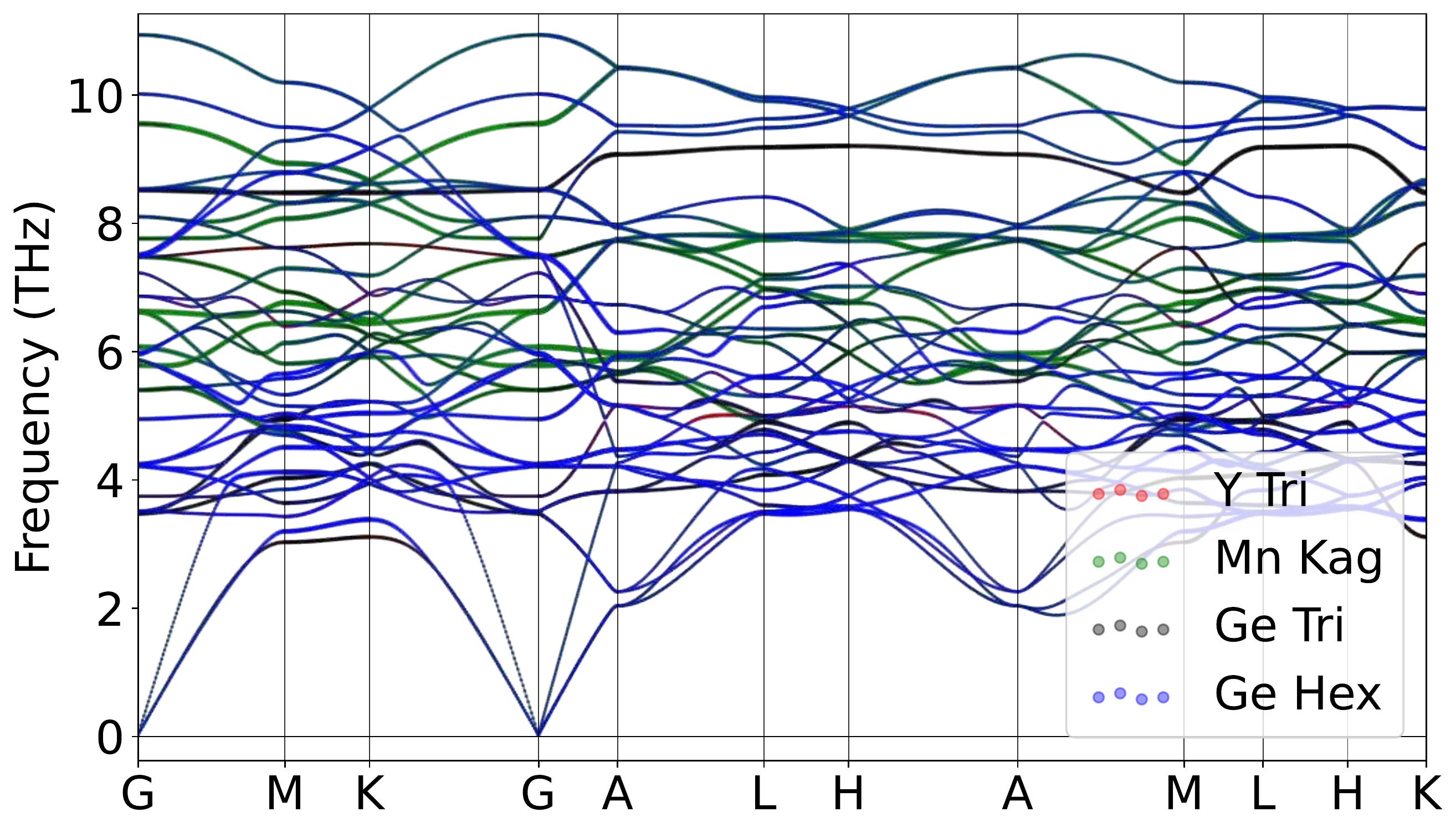}
}
\caption{Atomically projected phonon spectrum for YMn$_6$Ge$_6$ in paramagnetic phase.}
\label{fig:YMn6Ge6pho}
\end{figure}

\begin{figure}[htbp]
\centering
\includegraphics[width=0.45\textwidth]{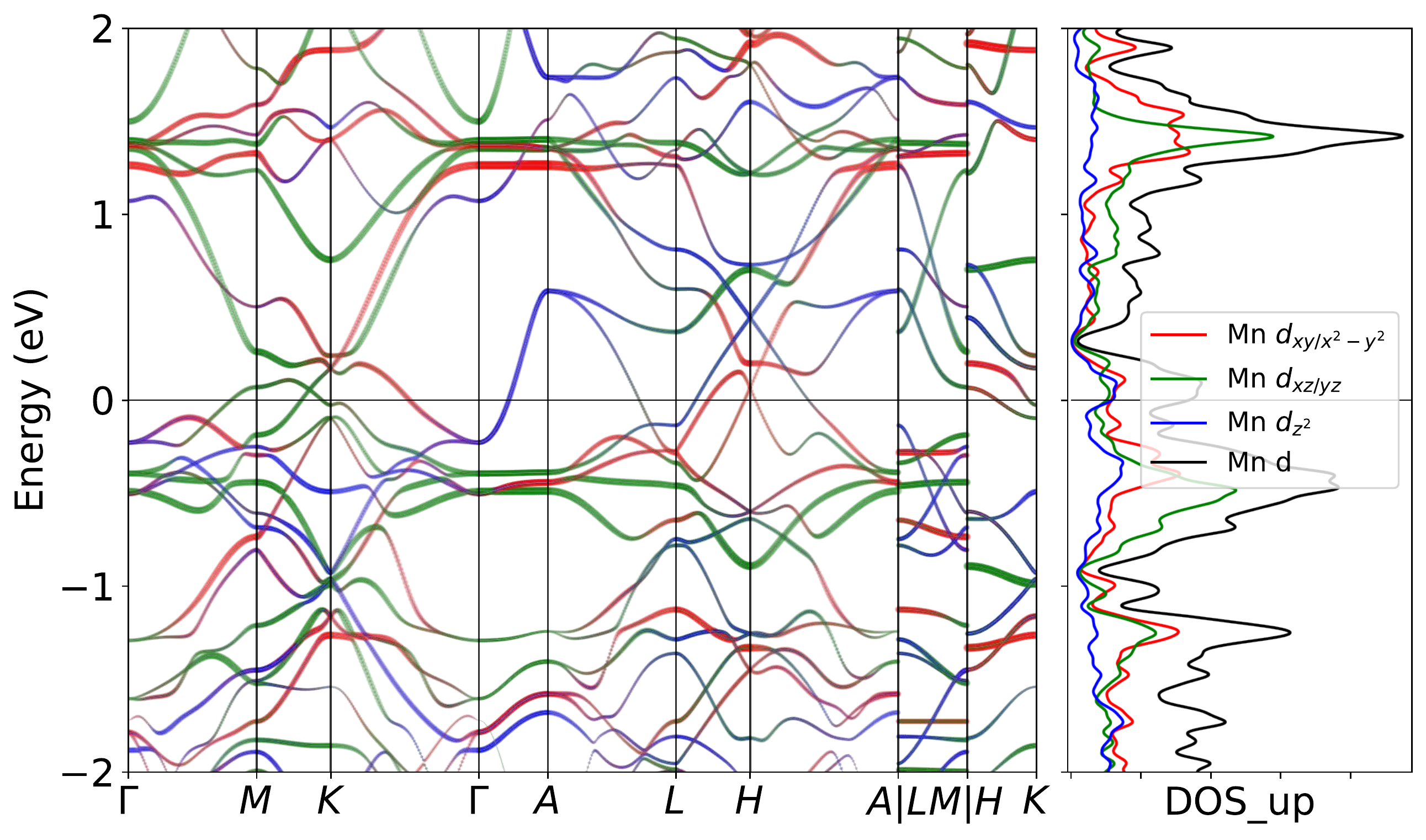}
\includegraphics[width=0.45\textwidth]{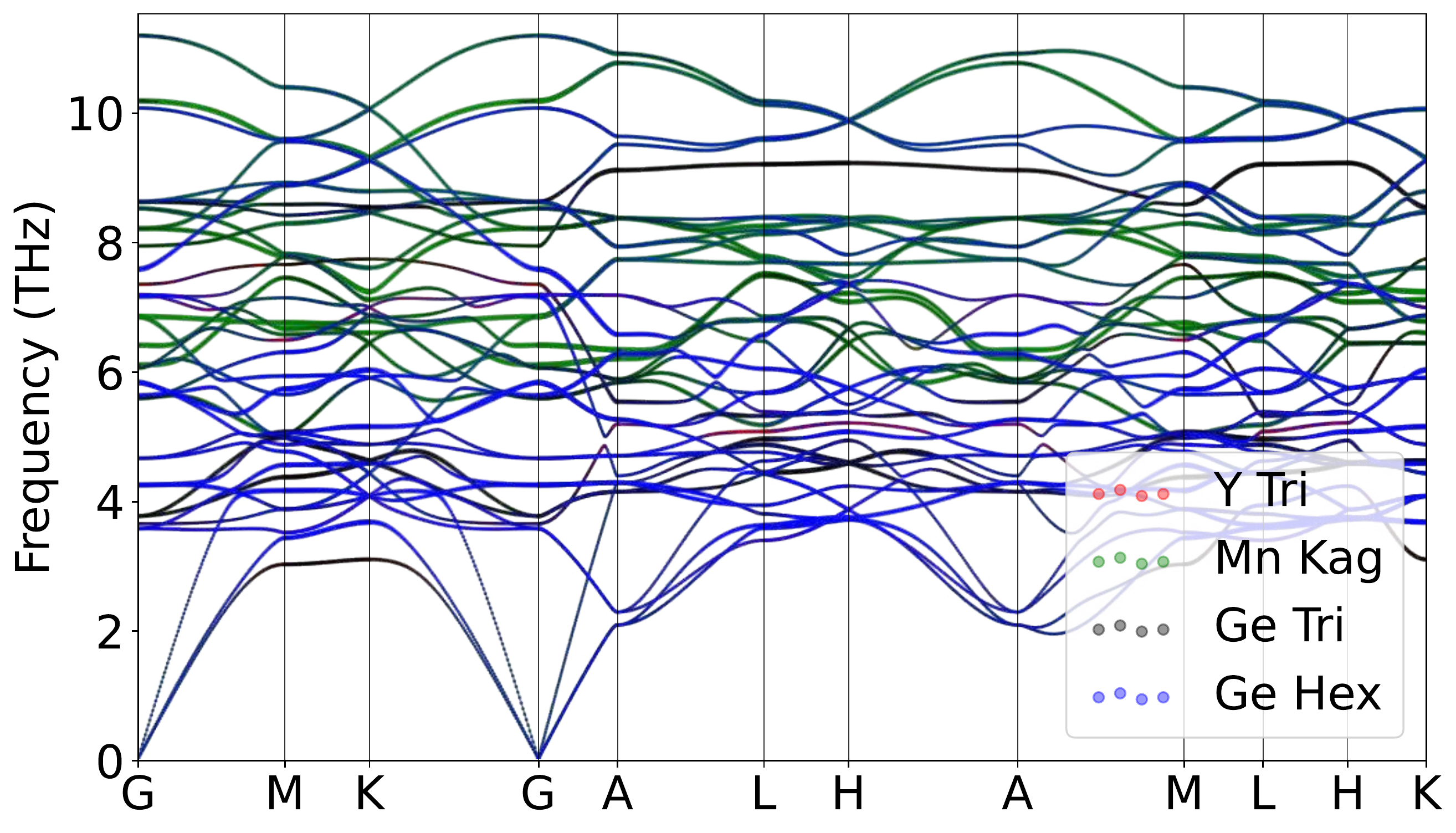}
\caption{Band structure for YMn$_6$Ge$_6$ with AFM order without SOC for spin (Left)up channel. The projection of Mn $3d$ orbitals is also presented. (Right)Correspondingly, a stable phonon was demonstrated with the collinear magnetic order without Hubbard U at lower temperature regime, weighted by atomic projections.}
\label{fig:YMn6Ge6mag}
\end{figure}

\begin{figure}[H]
\centering
\label{fig:YMn6Ge6_relaxed_Low_xz}
\includegraphics[width=0.85\textwidth]{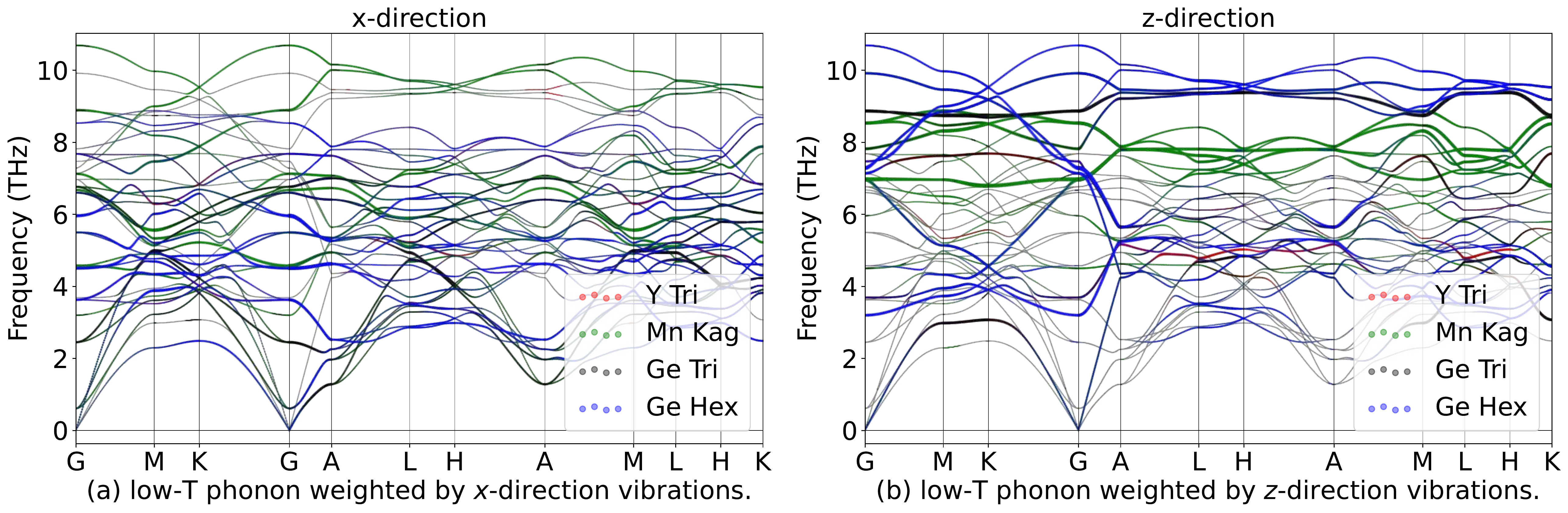}
\hspace{5pt}\label{fig:YMn6Ge6_relaxed_High_xz}
\includegraphics[width=0.85\textwidth]{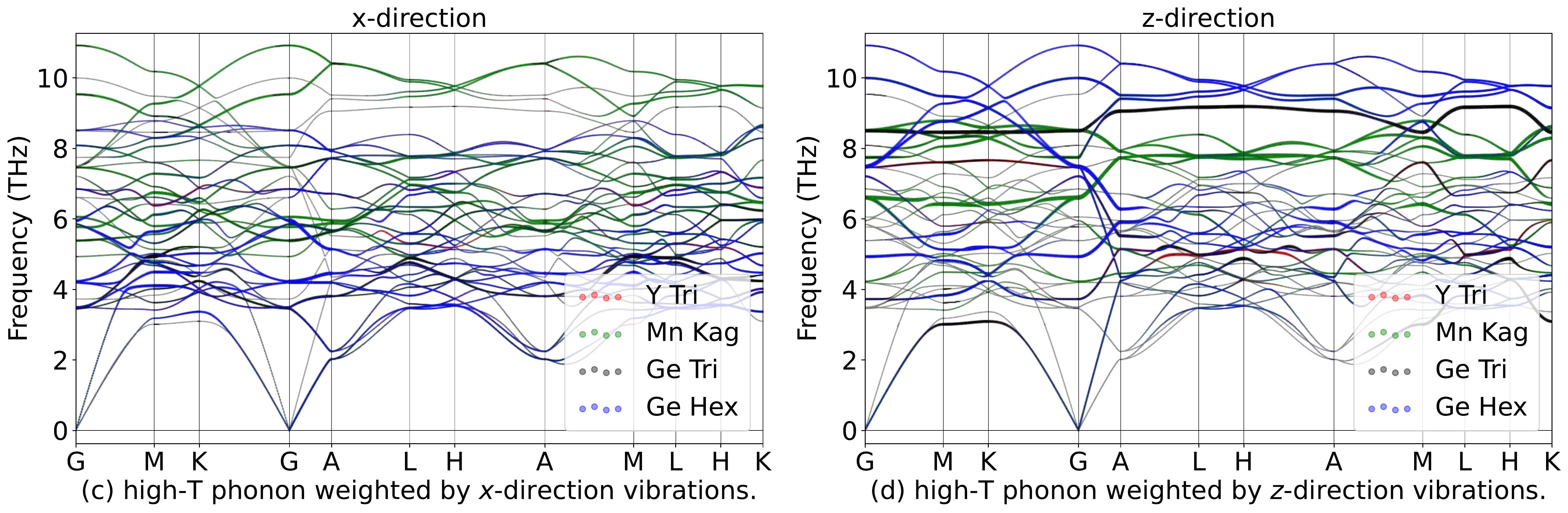}
\caption{Phonon spectrum for YMn$_6$Ge$_6$in in-plane ($x$) and out-of-plane ($z$) directions in paramagnetic phase.}
\label{fig:YMn6Ge6phoxyz}
\end{figure}

\begin{table}[htbp]
\global\long\def\arraystretch{1.12}
\captionsetup{width=.95\textwidth}
\caption{IRREPs at high-symmetry points for YMn$_6$Ge$_6$ phonon spectrum at Low-T.}
\label{tab:191_YMn6Ge6_Low}
\setlength{\tabcolsep}{1.mm}{
}
\end{table}

\begin{table}[htbp]
\global\long\def\arraystretch{1.12}
\captionsetup{width=.95\textwidth}
\caption{IRREPs at high-symmetry points for YMn$_6$Ge$_6$ phonon spectrum at High-T.}
\label{tab:191_YMn6Ge6_High}
\setlength{\tabcolsep}{1.mm}{
%
}
\end{table}
\subsubsection{\label{sms2sec:ZrMn6Ge6}\hyperref[smsec:col]{\texorpdfstring{ZrMn$_6$Ge$_6$}{}}~{\color{green}  (High-T stable, Low-T stable) }}

\begin{table}[htbp]
\global\long\def\arraystretch{1.12}
\captionsetup{width=.85\textwidth}
\caption{Structure parameters of ZrMn$_6$Ge$_6$ after relaxation. It contains 382 electrons. }
\label{tab:ZrMn6Ge6}
\setlength{\tabcolsep}{4mm}{
%
}
\end{table}

\begin{figure}[htbp]
\centering
\includegraphics[width=0.85\textwidth]{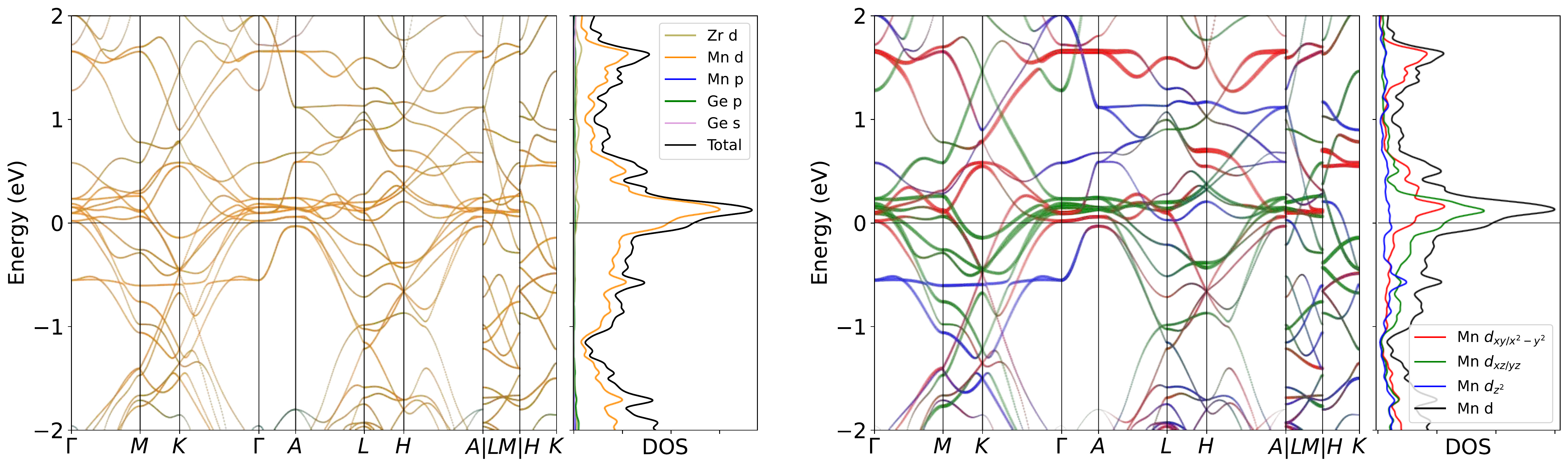}
\caption{Band structure for ZrMn$_6$Ge$_6$ without SOC in paramagnetic phase. The projection of Mn $3d$ orbitals is also presented. The band structures clearly reveal the dominating of Mn $3d$ orbitals  near the Fermi level, as well as the pronounced peak.}
\label{fig:ZrMn6Ge6band}
\end{figure}

\begin{figure}[H]
\centering
\subfloat[Relaxed phonon at low-T.]{
\label{fig:ZrMn6Ge6_relaxed_Low}
\includegraphics[width=0.45\textwidth]{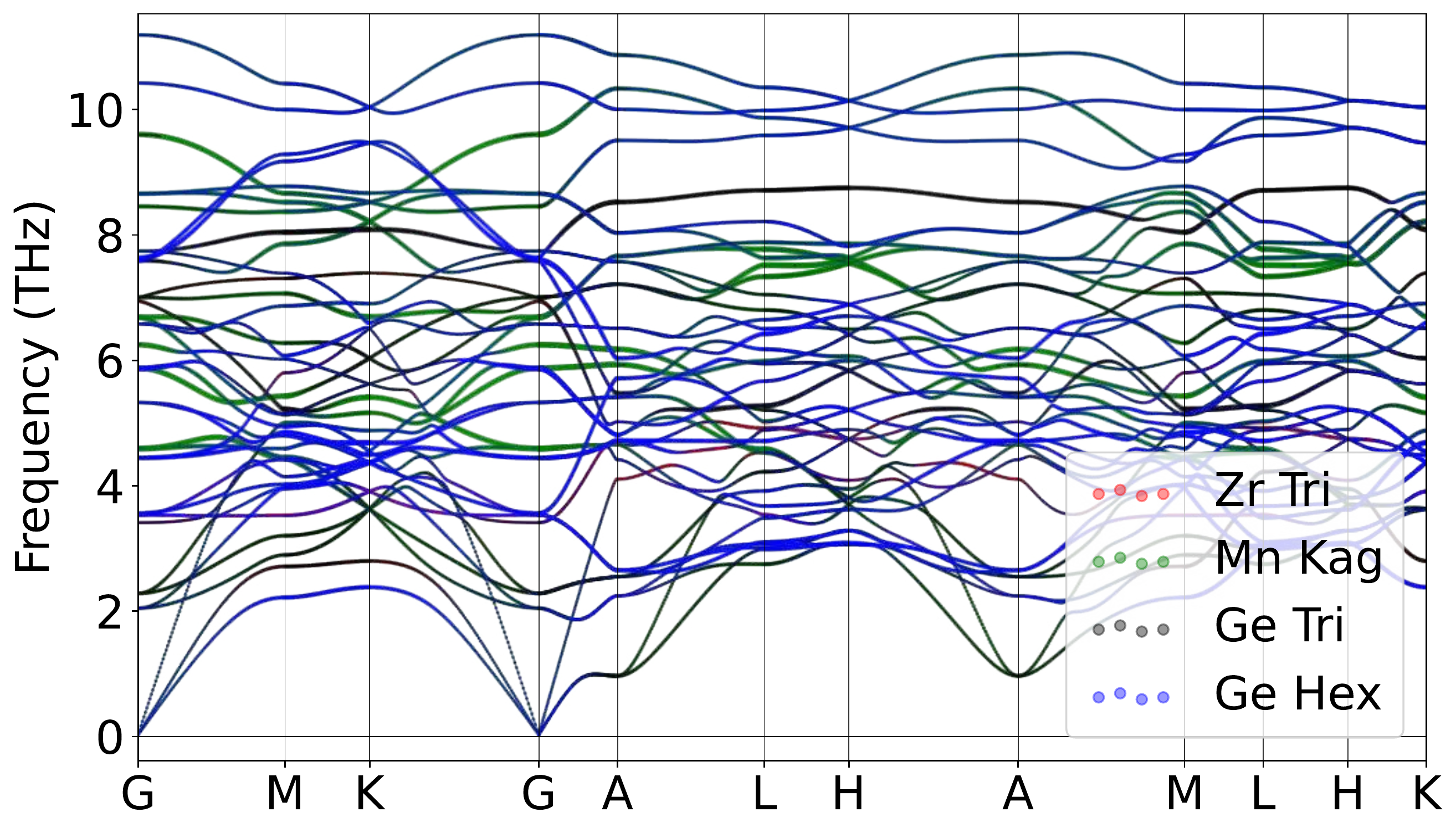}
}
\hspace{5pt}\subfloat[Relaxed phonon at high-T.]{
\label{fig:ZrMn6Ge6_relaxed_High}
\includegraphics[width=0.45\textwidth]{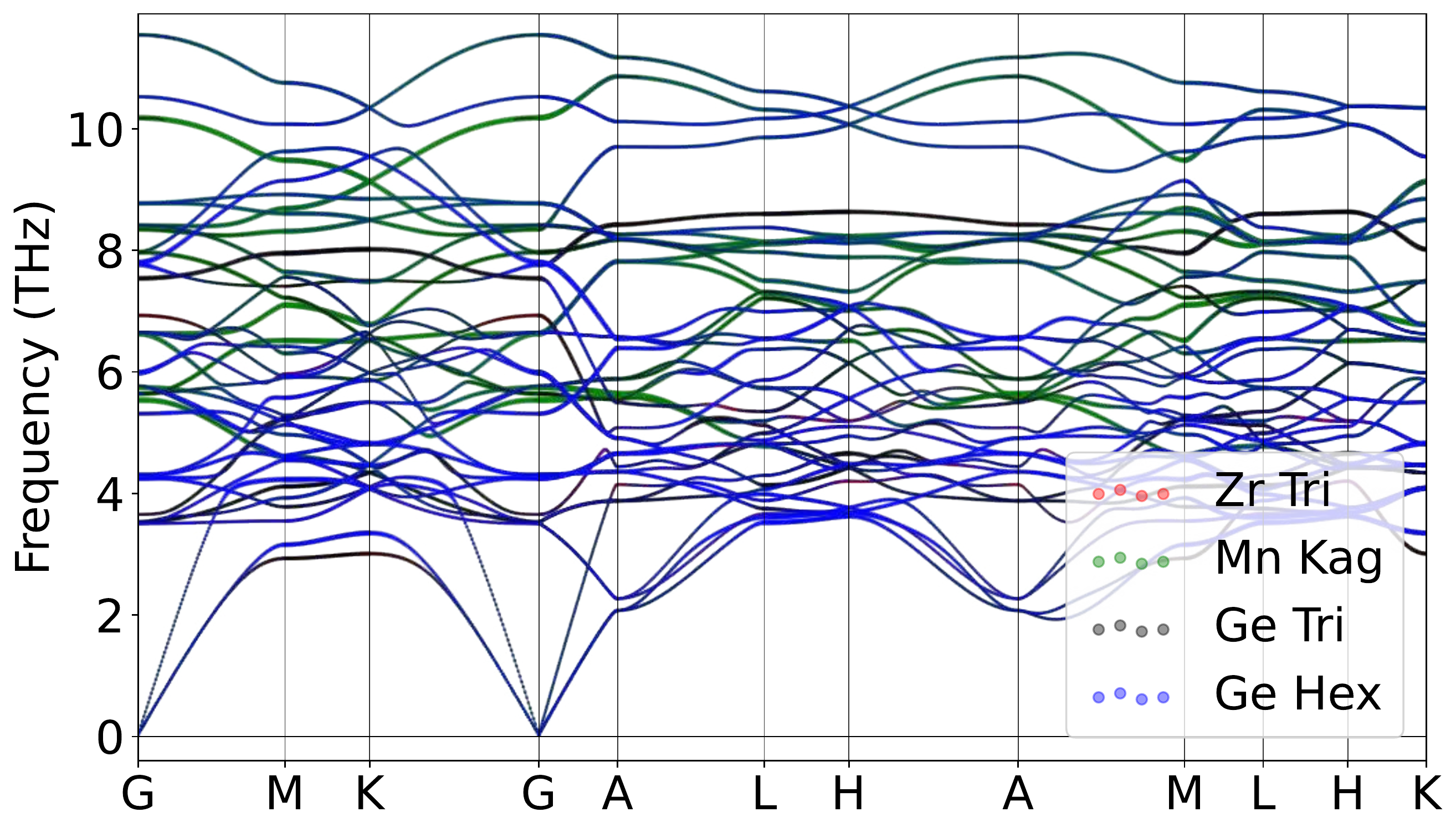}
}
\caption{Atomically projected phonon spectrum for ZrMn$_6$Ge$_6$ in paramagnetic phase.}
\label{fig:ZrMn6Ge6pho}
\end{figure}

\begin{figure}[htbp]
\centering
\includegraphics[width=0.45\textwidth]{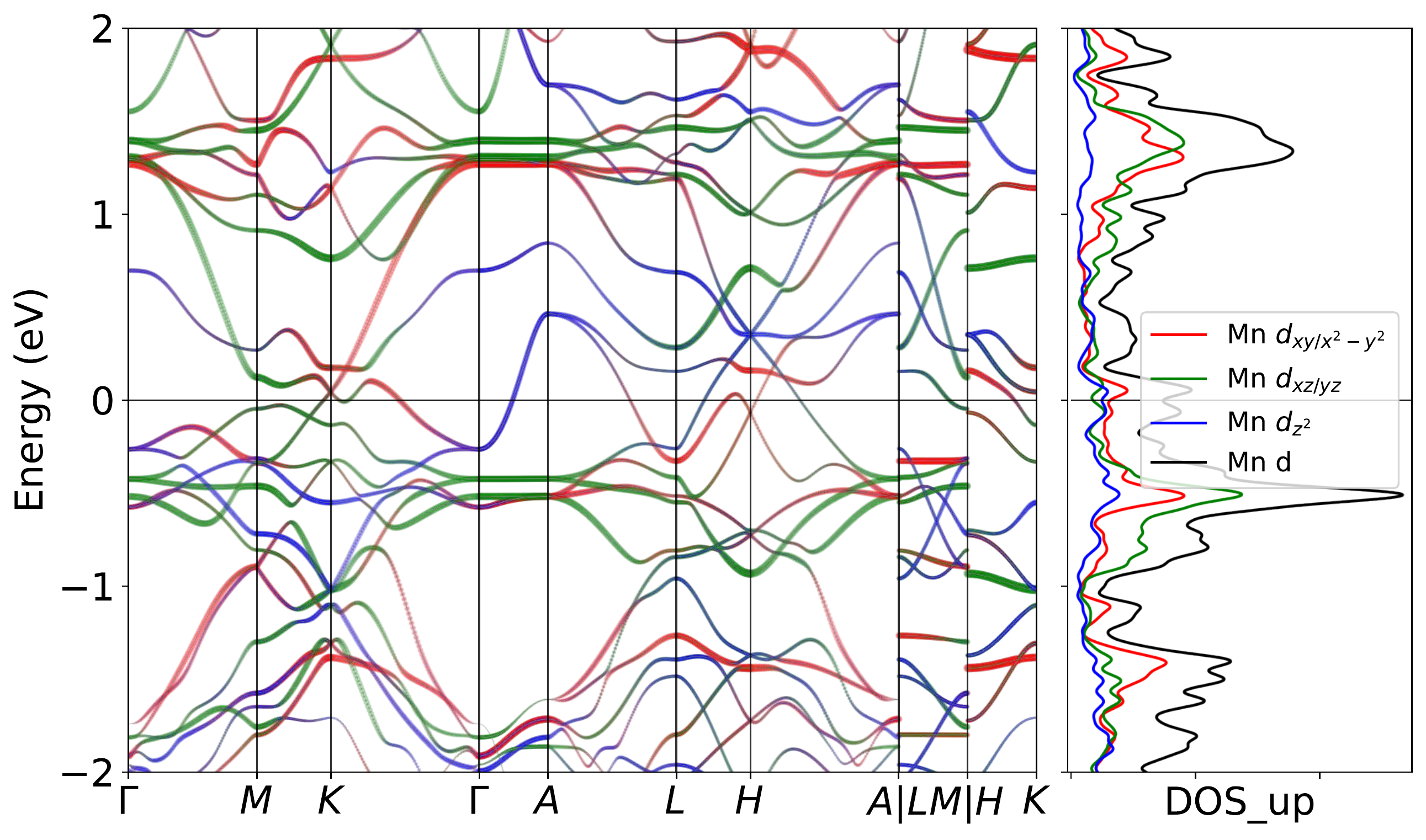}
\includegraphics[width=0.45\textwidth]{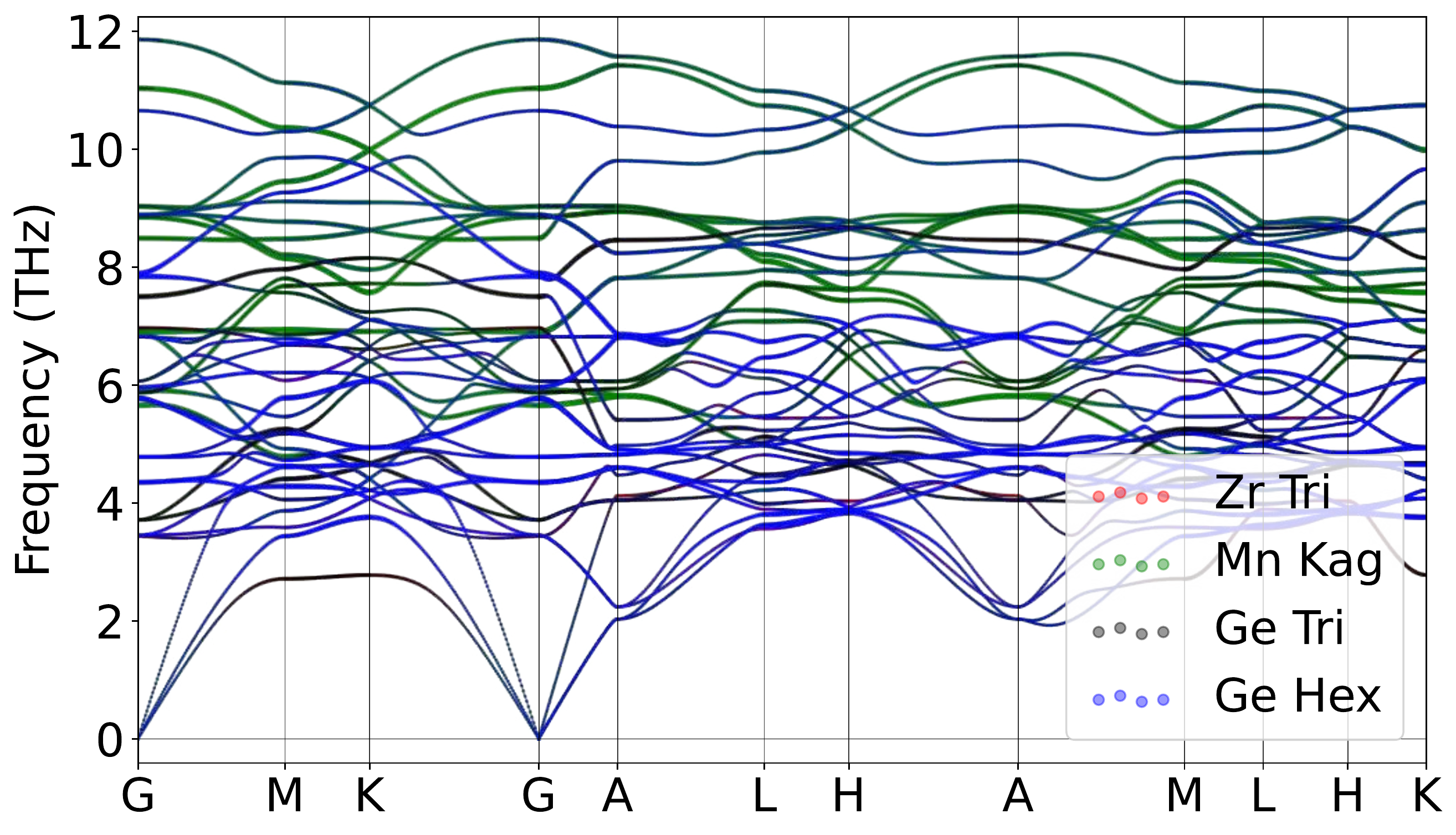}
\caption{Band structure for ZrMn$_6$Ge$_6$ with AFM order without SOC for spin (Left)up channel. The projection of Mn $3d$ orbitals is also presented. (Right)Correspondingly, a stable phonon was demonstrated with the collinear magnetic order without Hubbard U at lower temperature regime, weighted by atomic projections.}
\label{fig:ZrMn6Ge6mag}
\end{figure}

\begin{figure}[H]
\centering
\label{fig:ZrMn6Ge6_relaxed_Low_xz}
\includegraphics[width=0.85\textwidth]{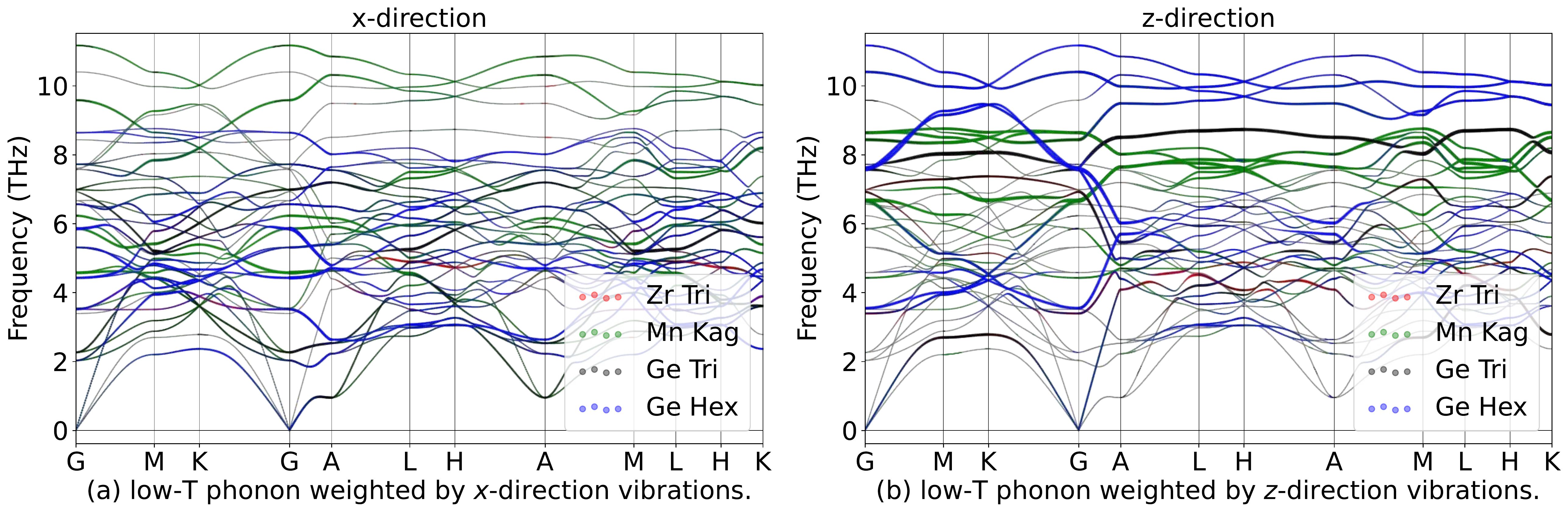}
\hspace{5pt}\label{fig:ZrMn6Ge6_relaxed_High_xz}
\includegraphics[width=0.85\textwidth]{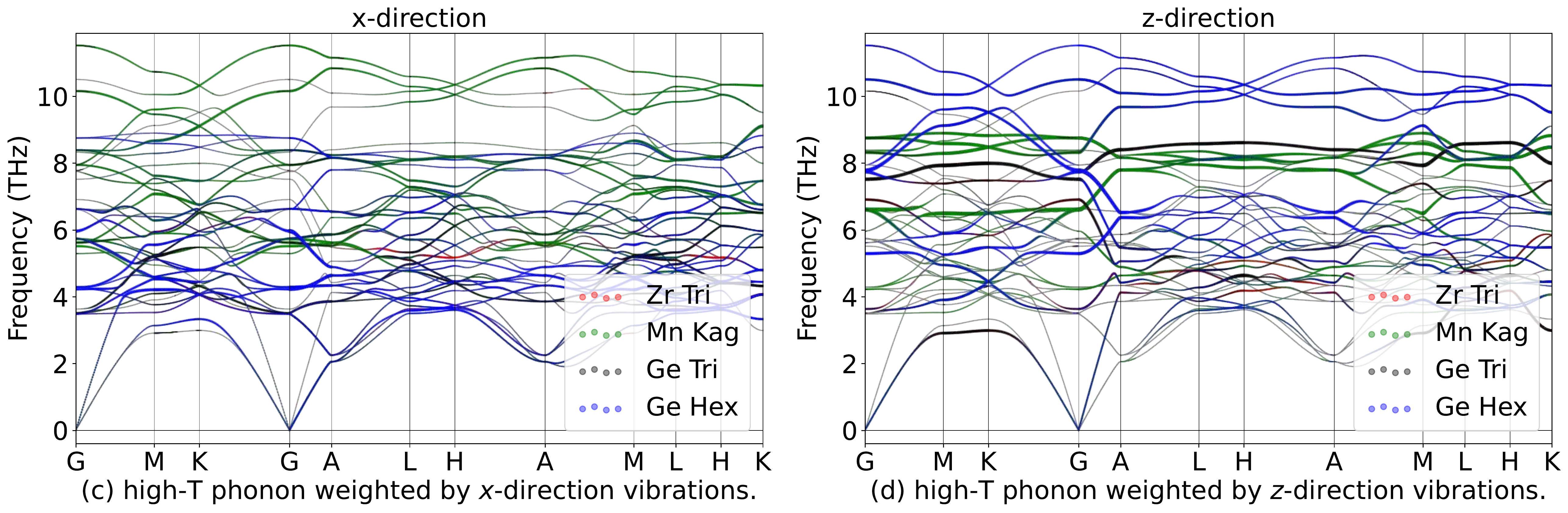}
\caption{Phonon spectrum for ZrMn$_6$Ge$_6$in in-plane ($x$) and out-of-plane ($z$) directions in paramagnetic phase.}
\label{fig:ZrMn6Ge6phoxyz}
\end{figure}

\begin{table}[htbp]
\global\long\def\arraystretch{1.12}
\captionsetup{width=.95\textwidth}
\caption{IRREPs at high-symmetry points for ZrMn$_6$Ge$_6$ phonon spectrum at Low-T.}
\label{tab:191_ZrMn6Ge6_Low}
\setlength{\tabcolsep}{1.mm}{
}
\end{table}

\begin{table}[htbp]
\global\long\def\arraystretch{1.12}
\captionsetup{width=.95\textwidth}
\caption{IRREPs at high-symmetry points for ZrMn$_6$Ge$_6$ phonon spectrum at High-T.}
\label{tab:191_ZrMn6Ge6_High}
\setlength{\tabcolsep}{1.mm}{
%
}
\end{table}
\subsubsection{\label{sms2sec:NbMn6Ge6}\hyperref[smsec:col]{\texorpdfstring{NbMn$_6$Ge$_6$}{}}~{\color{green}  (High-T stable, Low-T stable) }}

\begin{table}[htbp]
\global\long\def\arraystretch{1.12}
\captionsetup{width=.85\textwidth}
\caption{Structure parameters of NbMn$_6$Ge$_6$ after relaxation. It contains 383 electrons. }
\label{tab:NbMn6Ge6}
\setlength{\tabcolsep}{4mm}{
%
}
\end{table}

\begin{figure}[htbp]
\centering
\includegraphics[width=0.85\textwidth]{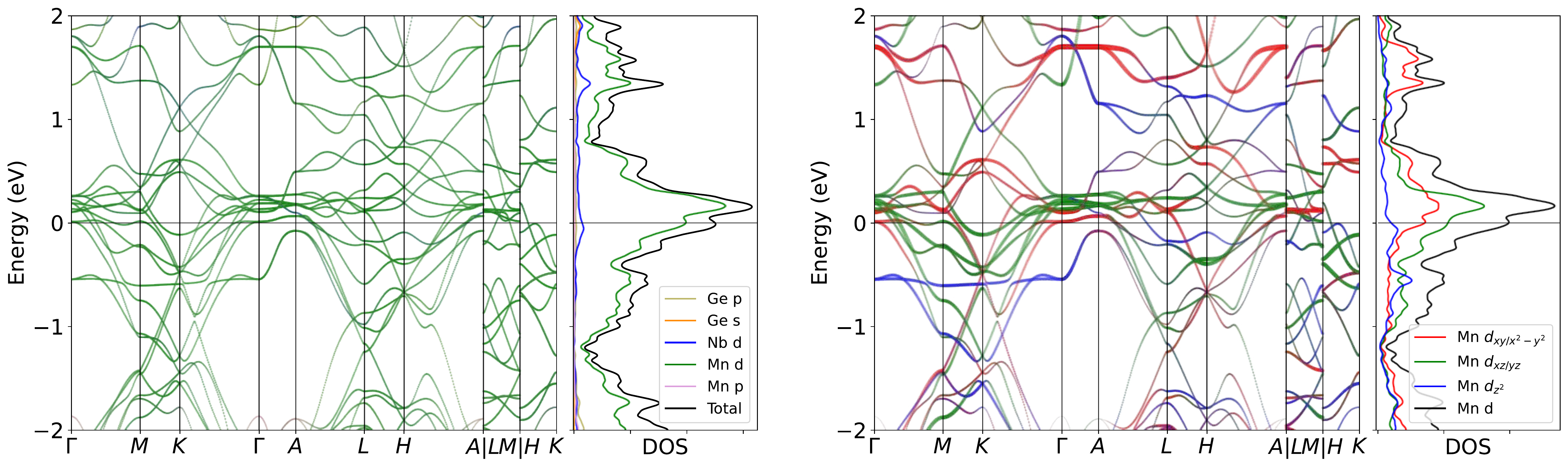}
\caption{Band structure for NbMn$_6$Ge$_6$ without SOC in paramagnetic phase. The projection of Mn $3d$ orbitals is also presented. The band structures clearly reveal the dominating of Mn $3d$ orbitals  near the Fermi level, as well as the pronounced peak.}
\label{fig:NbMn6Ge6band}
\end{figure}

\begin{figure}[H]
\centering
\subfloat[Relaxed phonon at low-T.]{
\label{fig:NbMn6Ge6_relaxed_Low}
\includegraphics[width=0.45\textwidth]{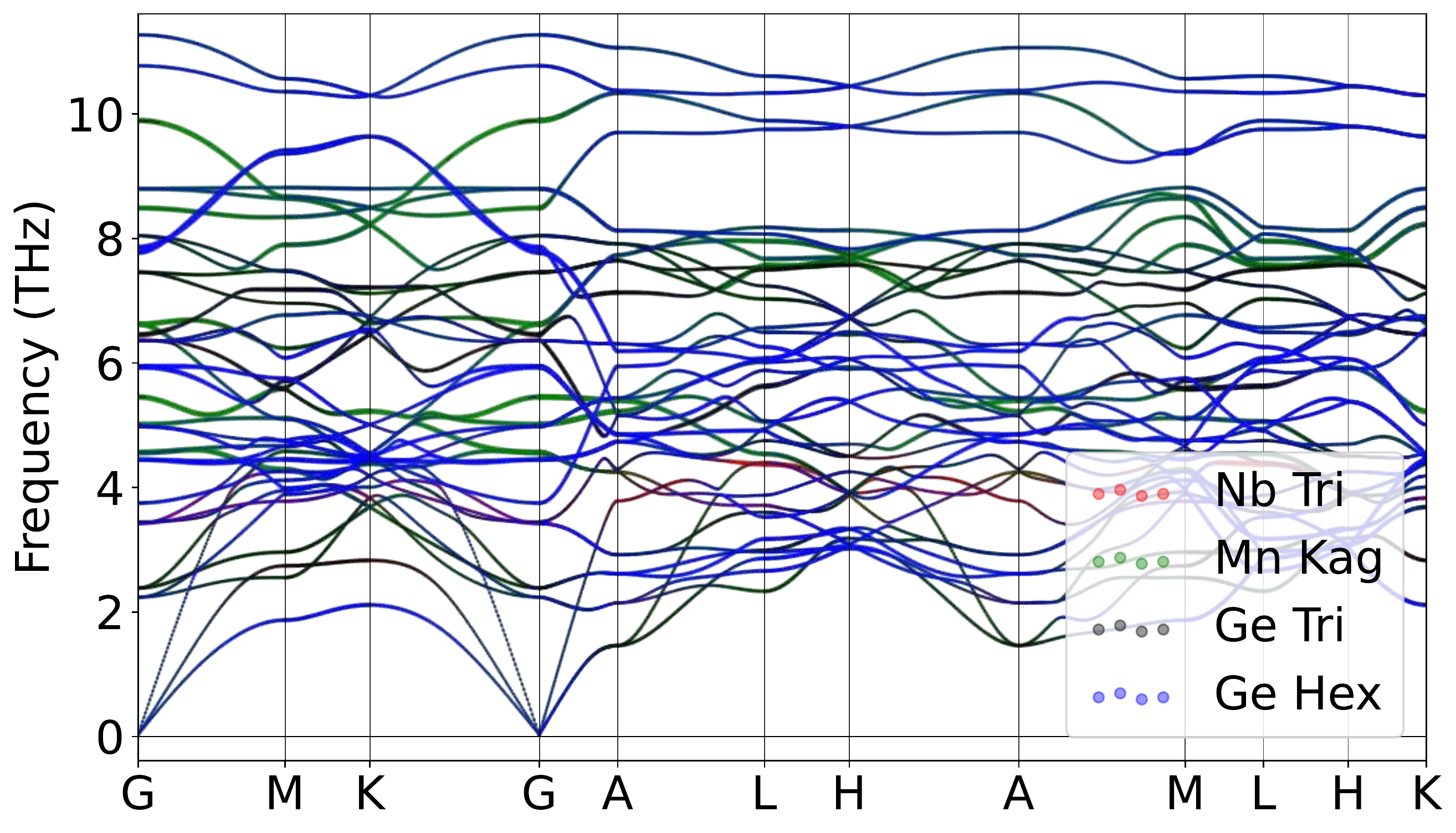}
}
\hspace{5pt}\subfloat[Relaxed phonon at high-T.]{
\label{fig:NbMn6Ge6_relaxed_High}
\includegraphics[width=0.45\textwidth]{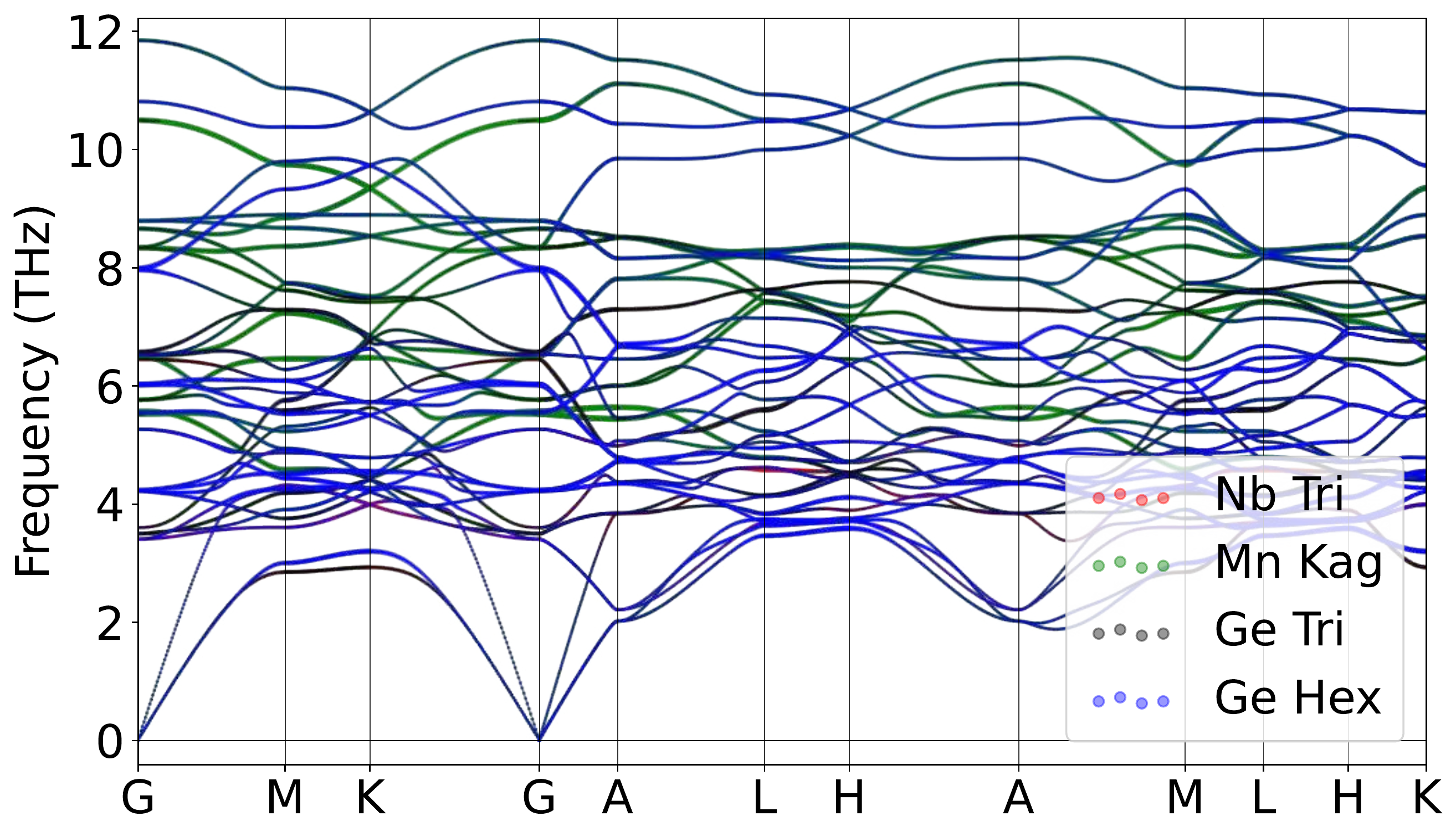}
}
\caption{Atomically projected phonon spectrum for NbMn$_6$Ge$_6$ in paramagnetic phase.}
\label{fig:NbMn6Ge6pho}
\end{figure}

\begin{figure}[htbp]
\centering
\includegraphics[width=0.45\textwidth]{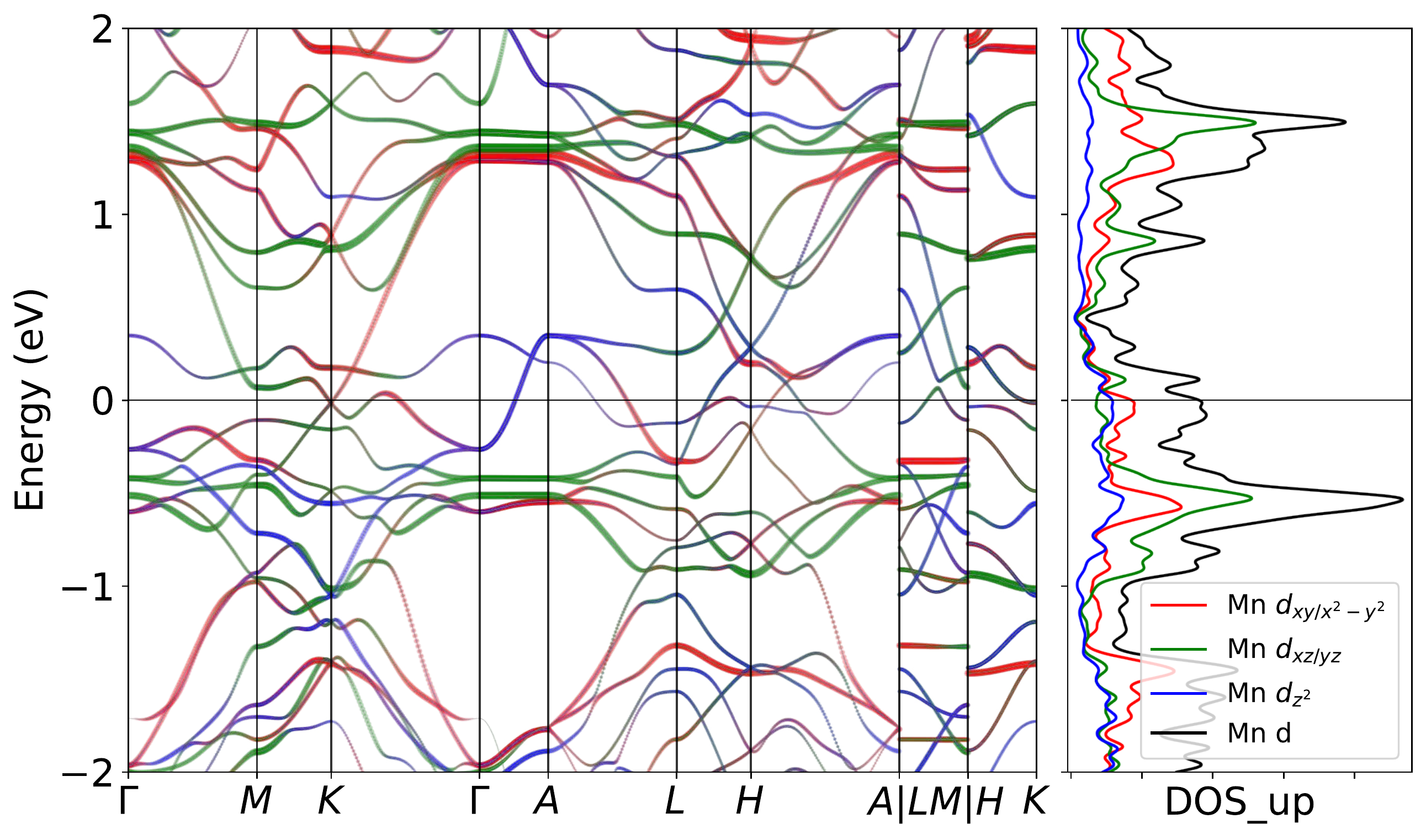}
\includegraphics[width=0.45\textwidth]{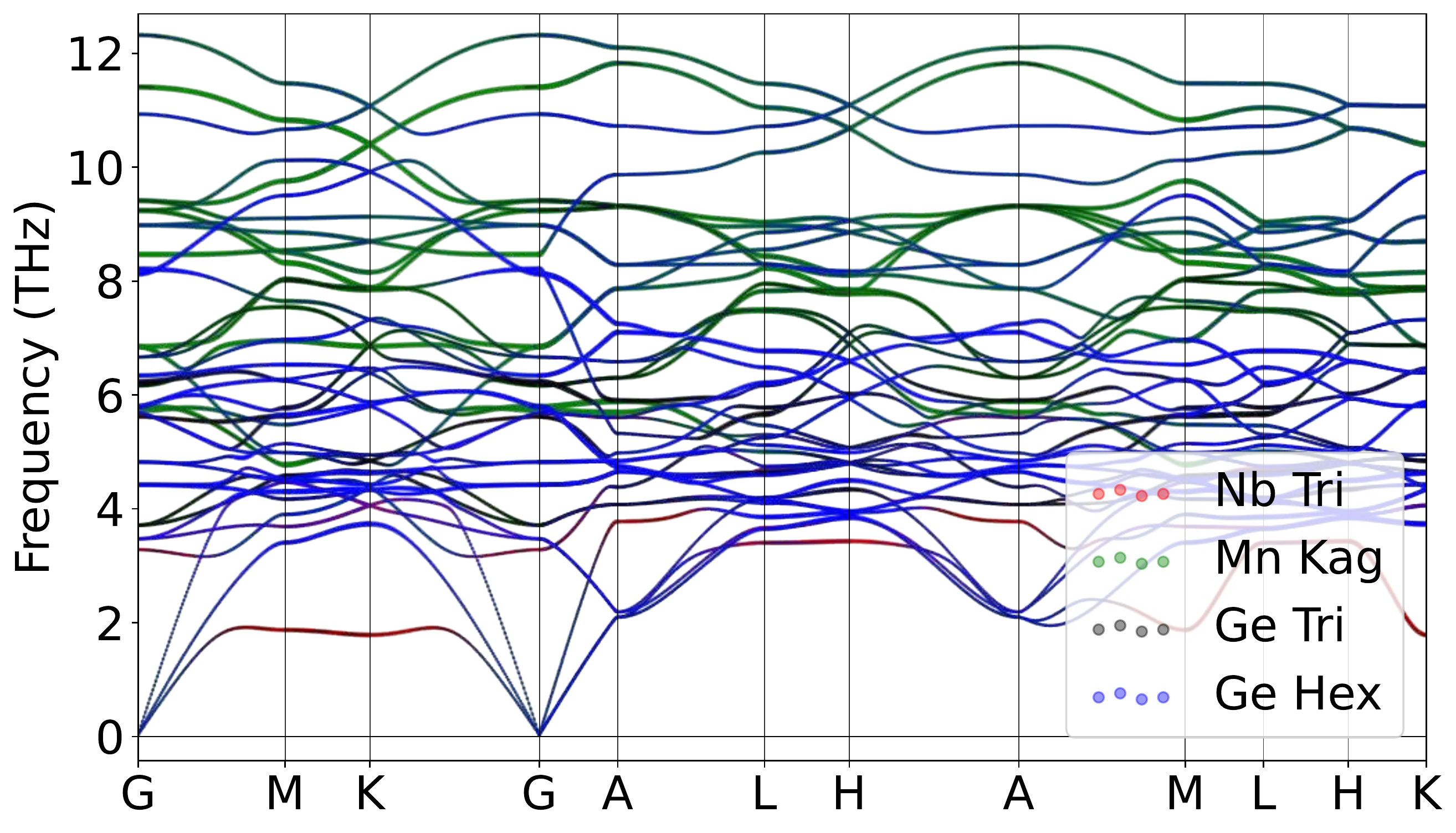}
\caption{Band structure for NbMn$_6$Ge$_6$ with AFM order without SOC for spin (Left)up channel. The projection of Mn $3d$ orbitals is also presented. (Right)Correspondingly, a stable phonon was demonstrated with the collinear magnetic order without Hubbard U at lower temperature regime, weighted by atomic projections.}
\label{fig:NbMn6Ge6mag}
\end{figure}

\begin{figure}[H]
\centering
\label{fig:NbMn6Ge6_relaxed_Low_xz}
\includegraphics[width=0.85\textwidth]{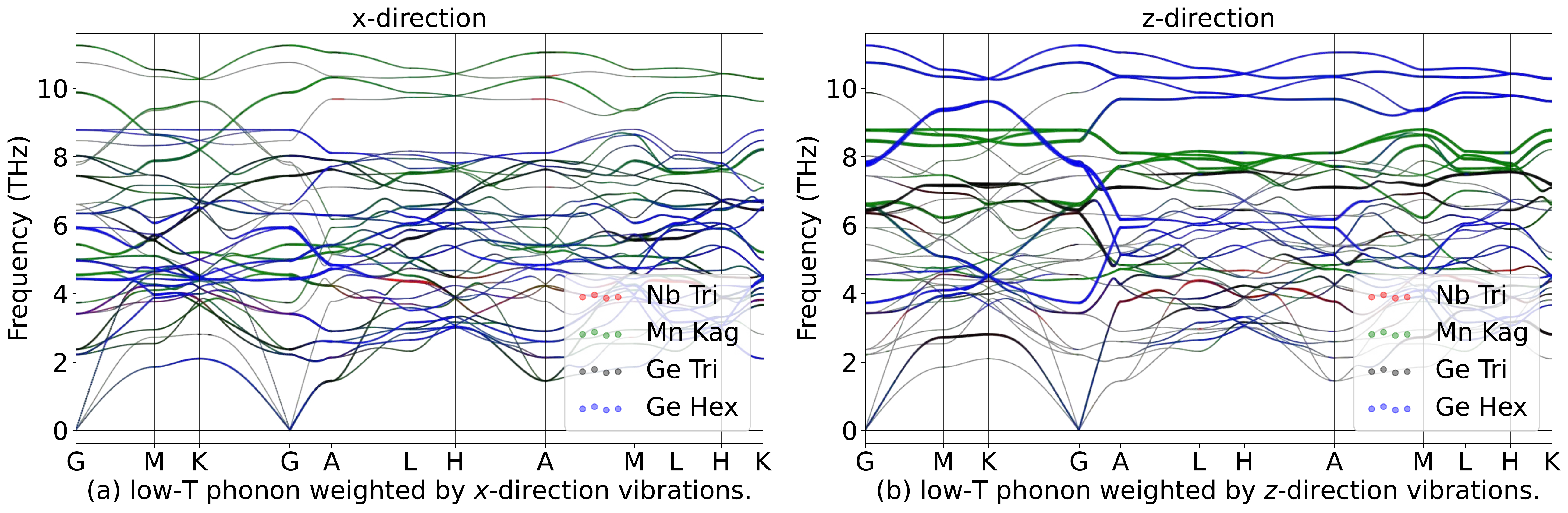}
\hspace{5pt}\label{fig:NbMn6Ge6_relaxed_High_xz}
\includegraphics[width=0.85\textwidth]{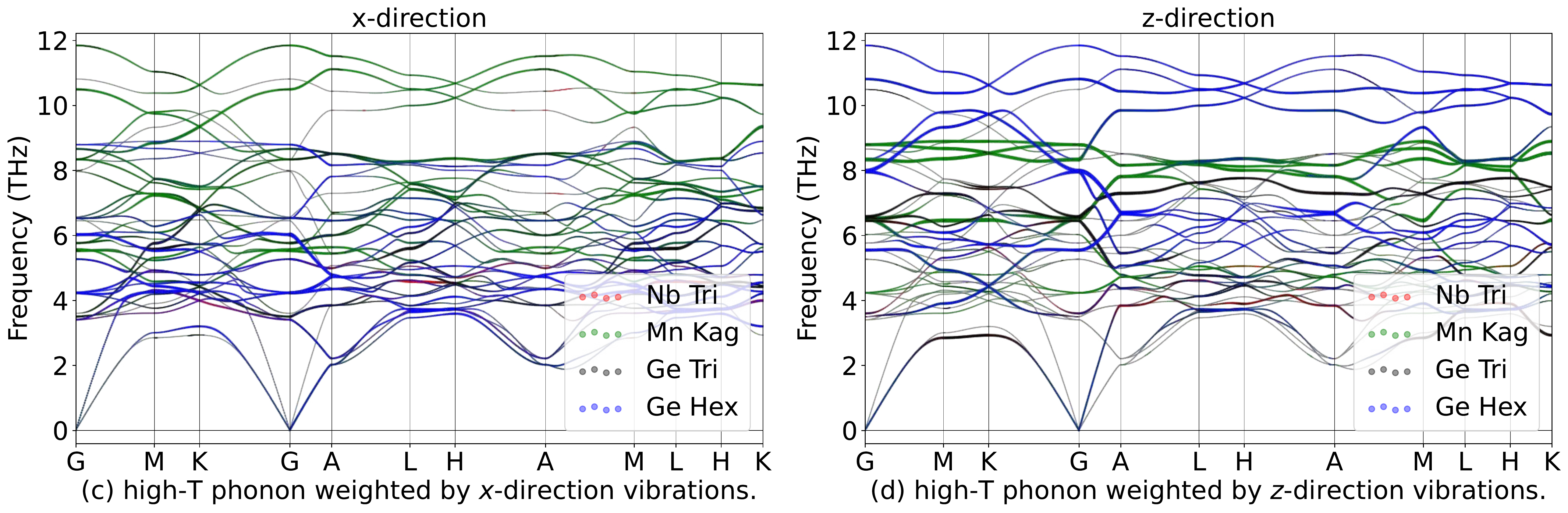}
\caption{Phonon spectrum for NbMn$_6$Ge$_6$in in-plane ($x$) and out-of-plane ($z$) directions in paramagnetic phase.}
\label{fig:NbMn6Ge6phoxyz}
\end{figure}

\begin{table}[htbp]
\global\long\def\arraystretch{1.12}
\captionsetup{width=.95\textwidth}
\caption{IRREPs at high-symmetry points for NbMn$_6$Ge$_6$ phonon spectrum at Low-T.}
\label{tab:191_NbMn6Ge6_Low}
\setlength{\tabcolsep}{1.mm}{
}
\end{table}

\begin{table}[htbp]
\global\long\def\arraystretch{1.12}
\captionsetup{width=.95\textwidth}
\caption{IRREPs at high-symmetry points for NbMn$_6$Ge$_6$ phonon spectrum at High-T.}
\label{tab:191_NbMn6Ge6_High}
\setlength{\tabcolsep}{1.mm}{
%
}
\end{table}
\subsubsection{\label{sms2sec:PrMn6Ge6}\hyperref[smsec:col]{\texorpdfstring{PrMn$_6$Ge$_6$}{}}~{\color{green}  (High-T stable, Low-T stable) }}

\begin{table}[htbp]
\global\long\def\arraystretch{1.12}
\captionsetup{width=.85\textwidth}
\caption{Structure parameters of PrMn$_6$Ge$_6$ after relaxation. It contains 401 electrons. }
\label{tab:PrMn6Ge6}
\setlength{\tabcolsep}{4mm}{
%
}
\end{table}

\begin{figure}[htbp]
\centering
\includegraphics[width=0.85\textwidth]{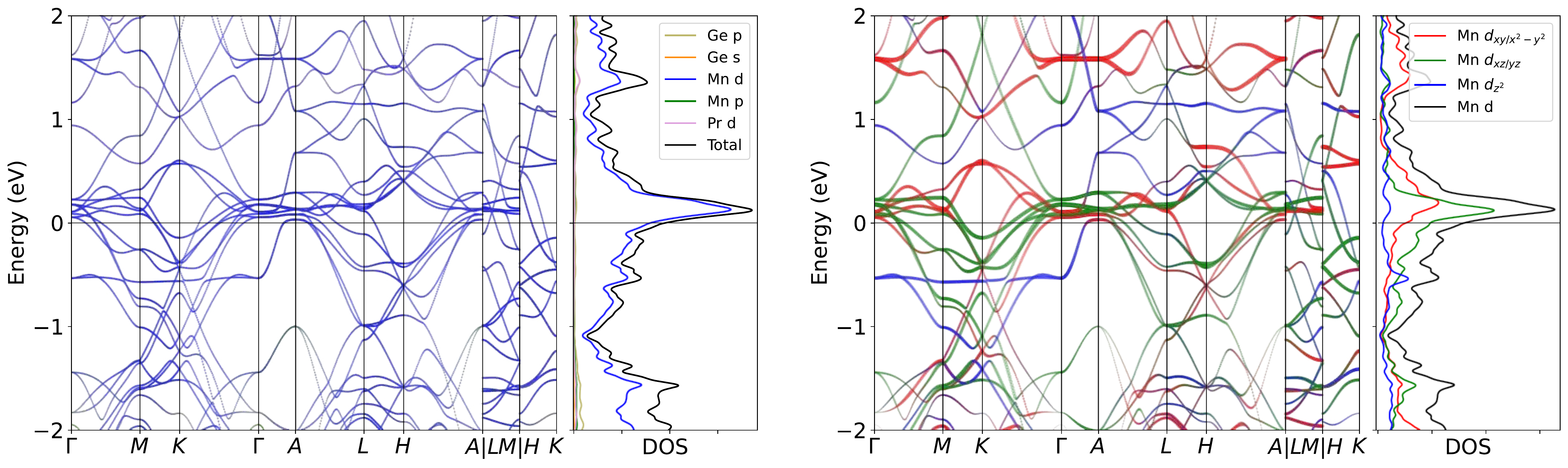}
\caption{Band structure for PrMn$_6$Ge$_6$ without SOC in paramagnetic phase. The projection of Mn $3d$ orbitals is also presented. The band structures clearly reveal the dominating of Mn $3d$ orbitals  near the Fermi level, as well as the pronounced peak.}
\label{fig:PrMn6Ge6band}
\end{figure}

\begin{figure}[H]
\centering
\subfloat[Relaxed phonon at low-T.]{
\label{fig:PrMn6Ge6_relaxed_Low}
\includegraphics[width=0.45\textwidth]{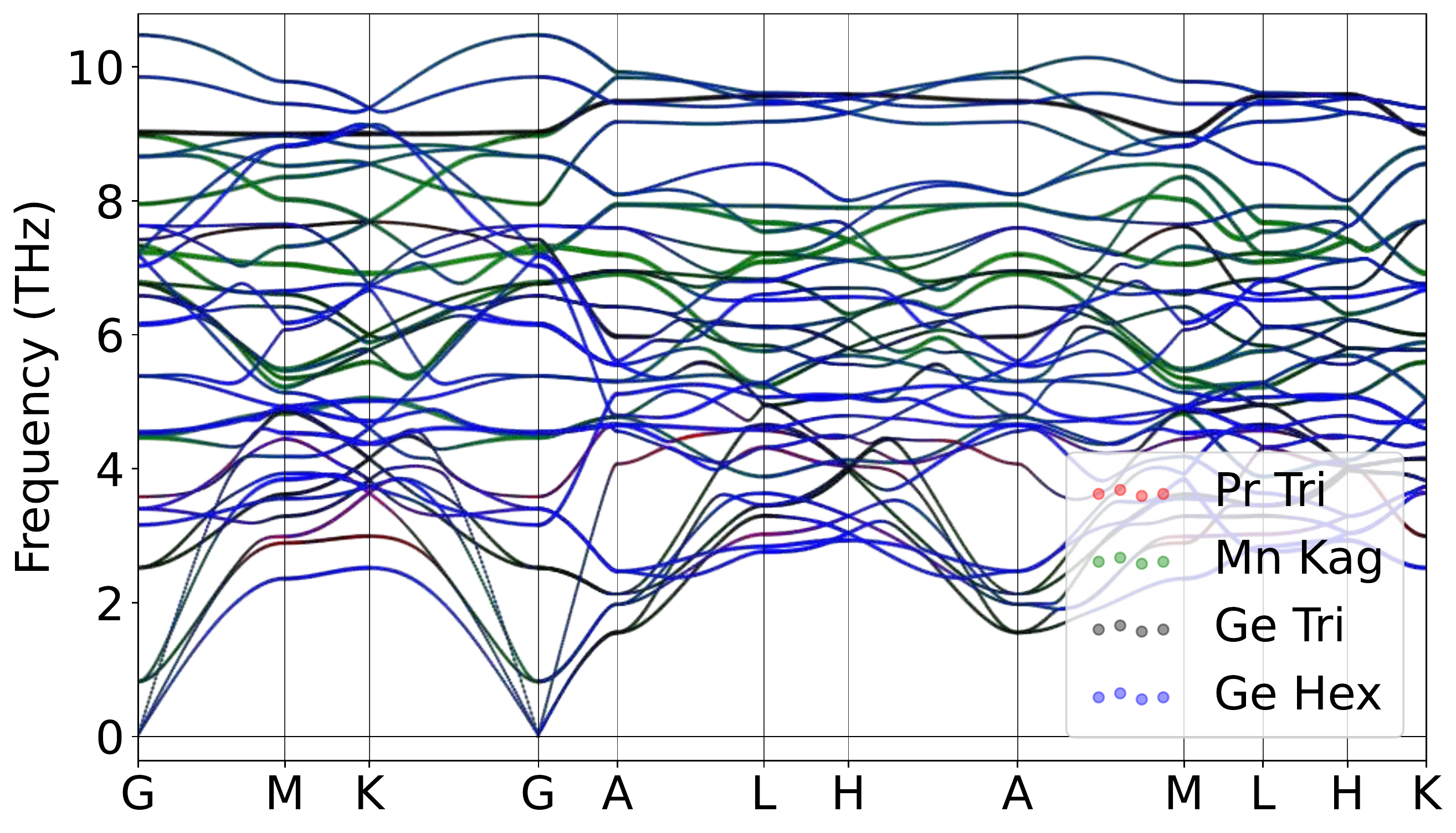}
}
\hspace{5pt}\subfloat[Relaxed phonon at high-T.]{
\label{fig:PrMn6Ge6_relaxed_High}
\includegraphics[width=0.45\textwidth]{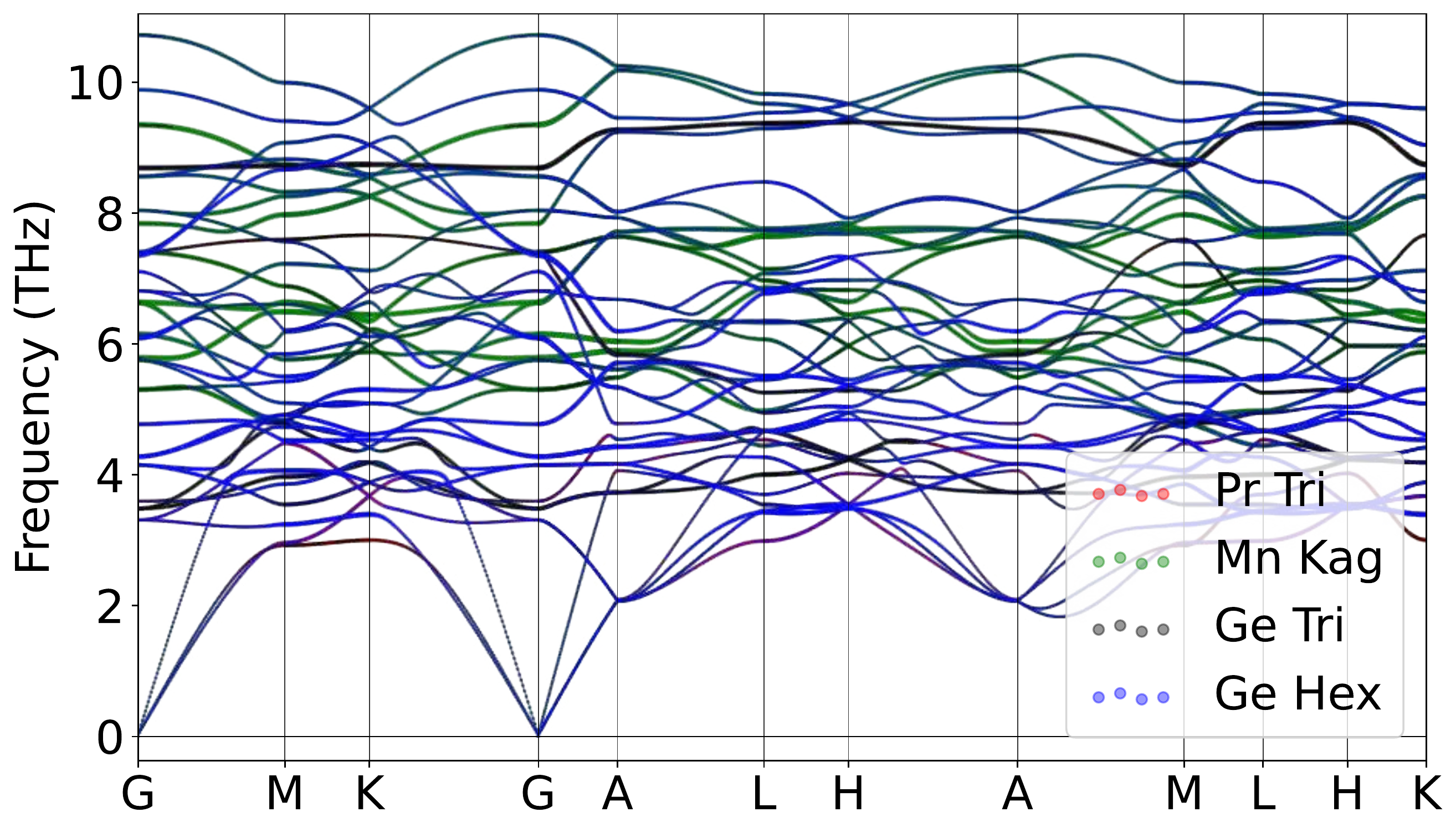}
}
\caption{Atomically projected phonon spectrum for PrMn$_6$Ge$_6$ in paramagnetic phase.}
\label{fig:PrMn6Ge6pho}
\end{figure}

\begin{figure}[htbp]
\centering
\includegraphics[width=0.32\textwidth]{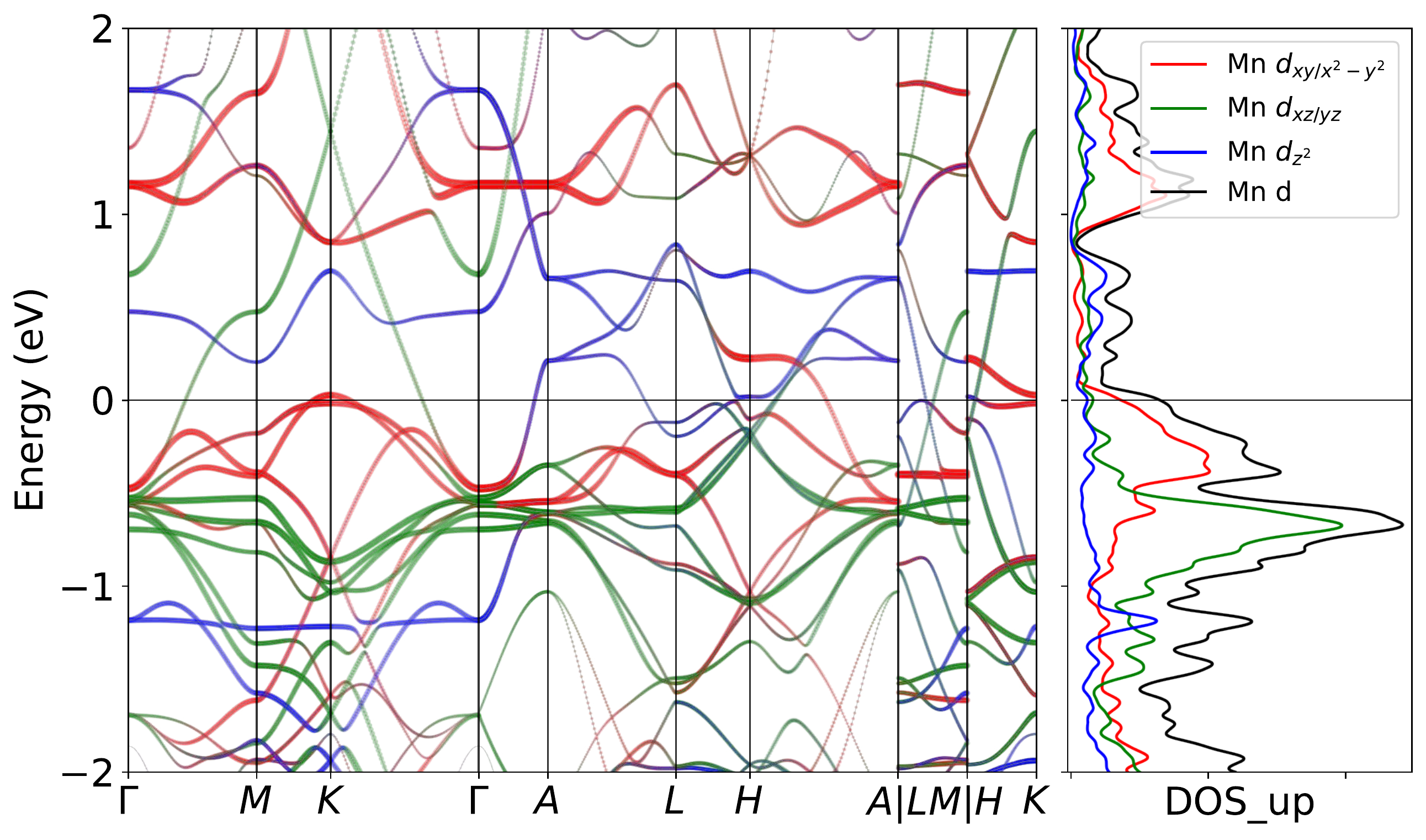}
\includegraphics[width=0.32\textwidth]{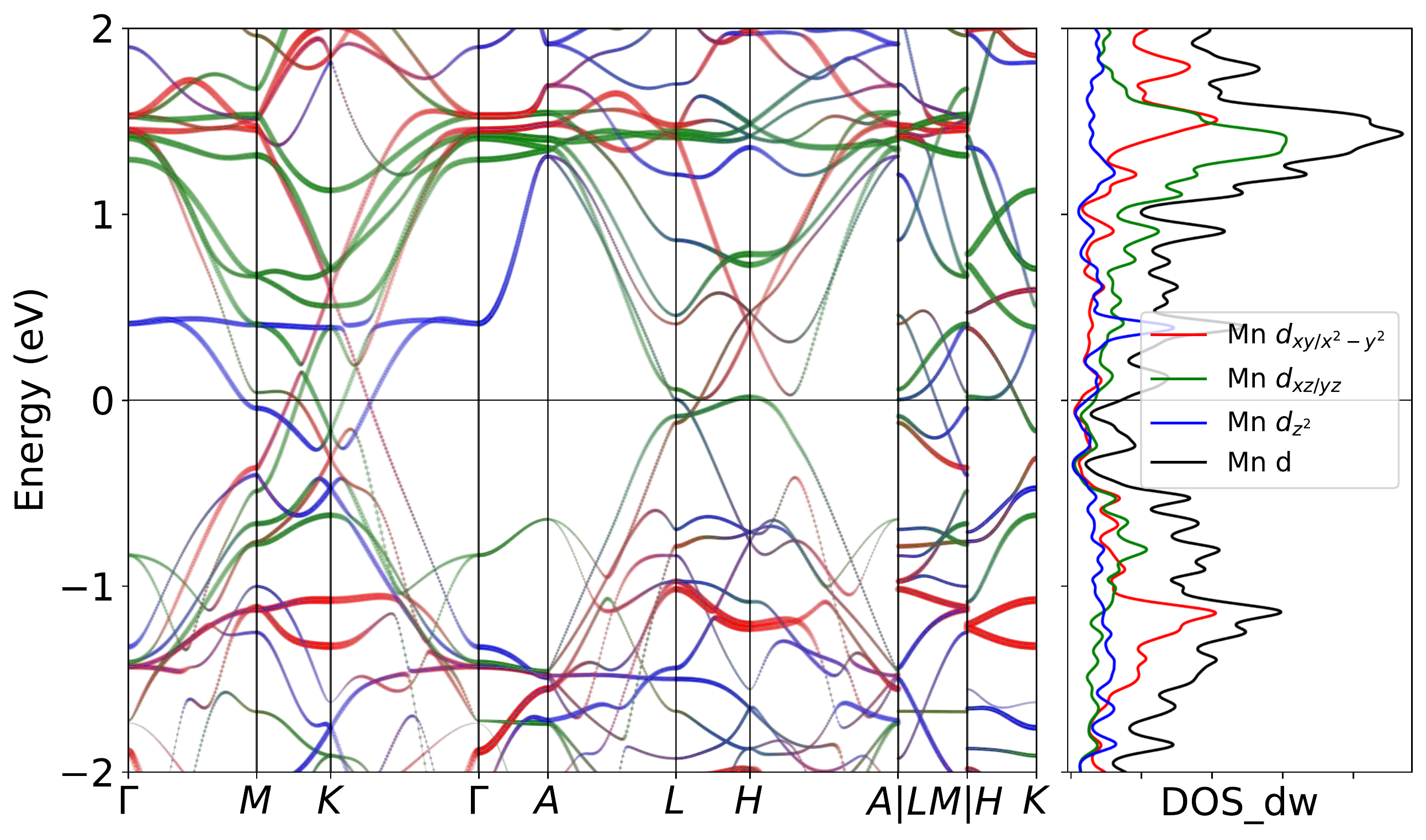}
\includegraphics[width=0.32\textwidth]{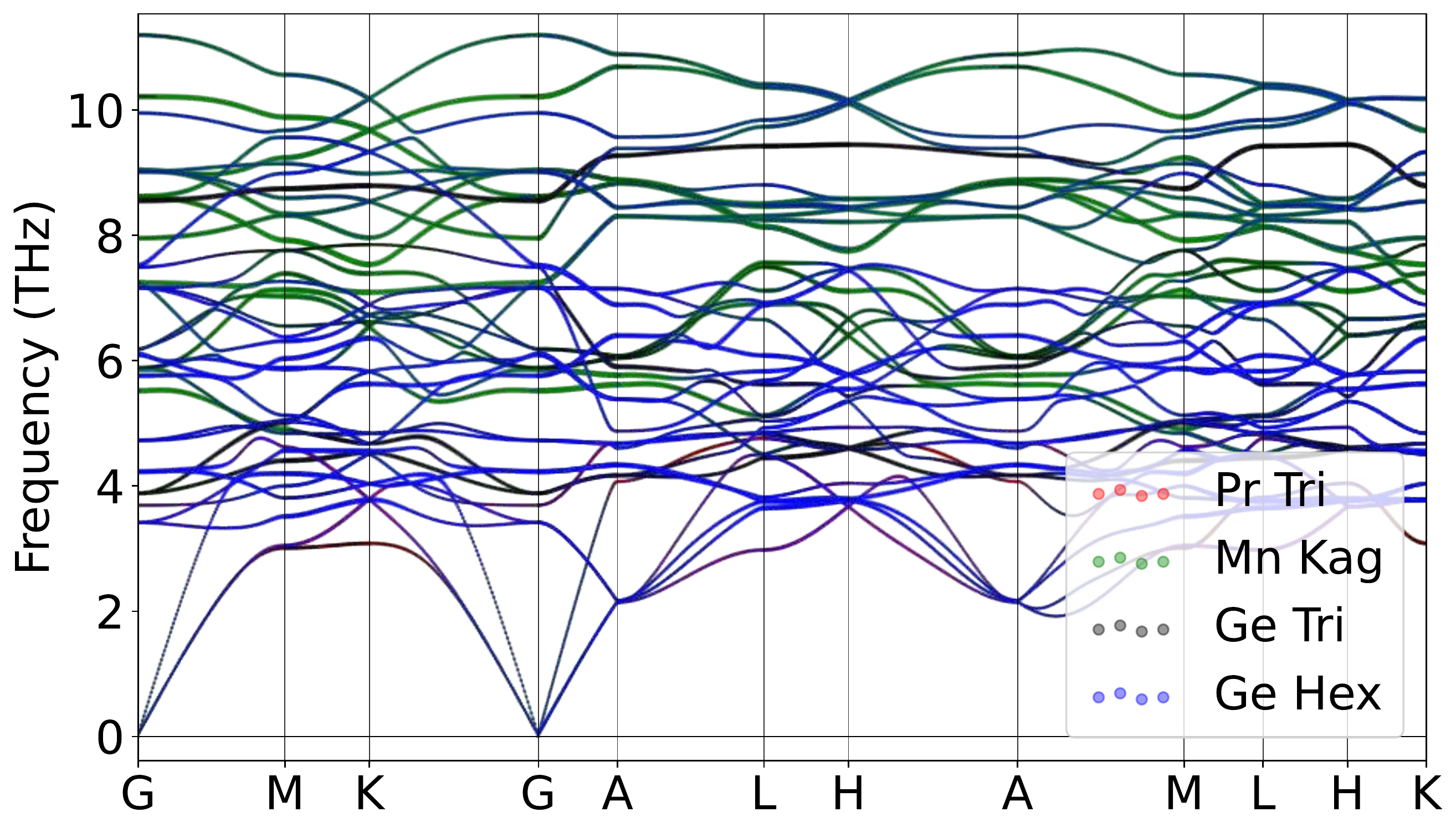}
\caption{Band structure for PrMn$_6$Ge$_6$ with FM order without SOC for spin (Left)up and (Middle)down channels. The projection of Mn $3d$ orbitals is also presented. (Right)Correspondingly, a stable phonon was demonstrated with the collinear magnetic order without Hubbard U at lower temperature regime, weighted by atomic projections.}
\label{fig:PrMn6Ge6mag}
\end{figure}

\begin{figure}[H]
\centering
\label{fig:PrMn6Ge6_relaxed_Low_xz}
\includegraphics[width=0.85\textwidth]{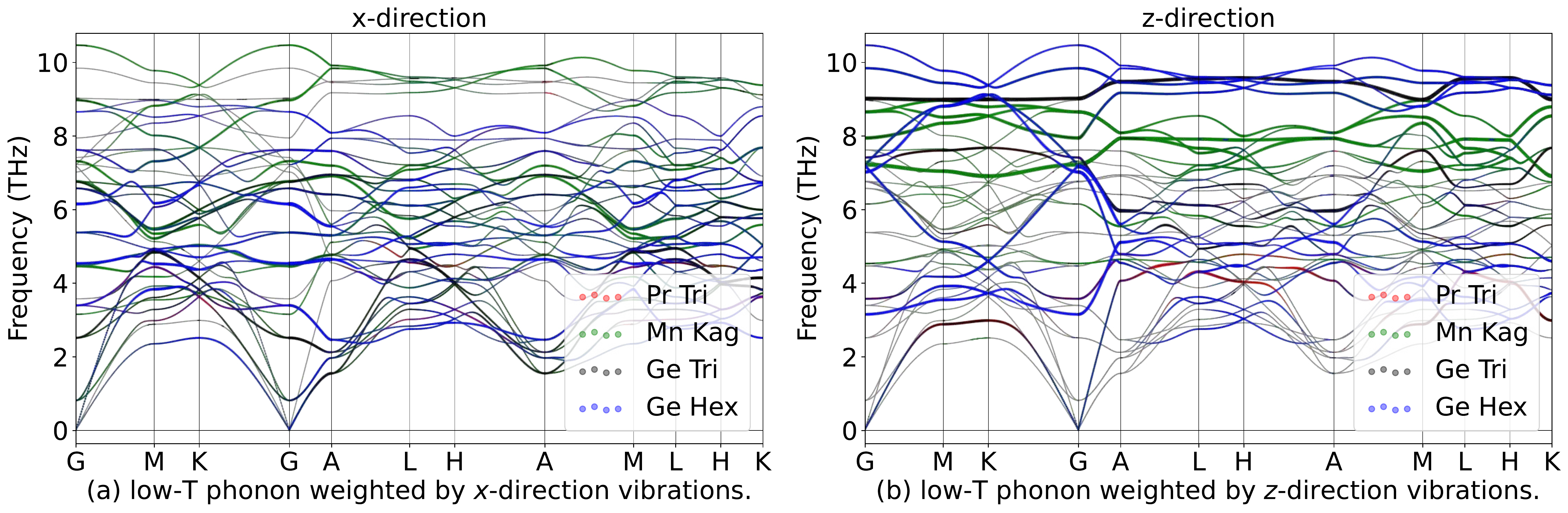}
\hspace{5pt}\label{fig:PrMn6Ge6_relaxed_High_xz}
\includegraphics[width=0.85\textwidth]{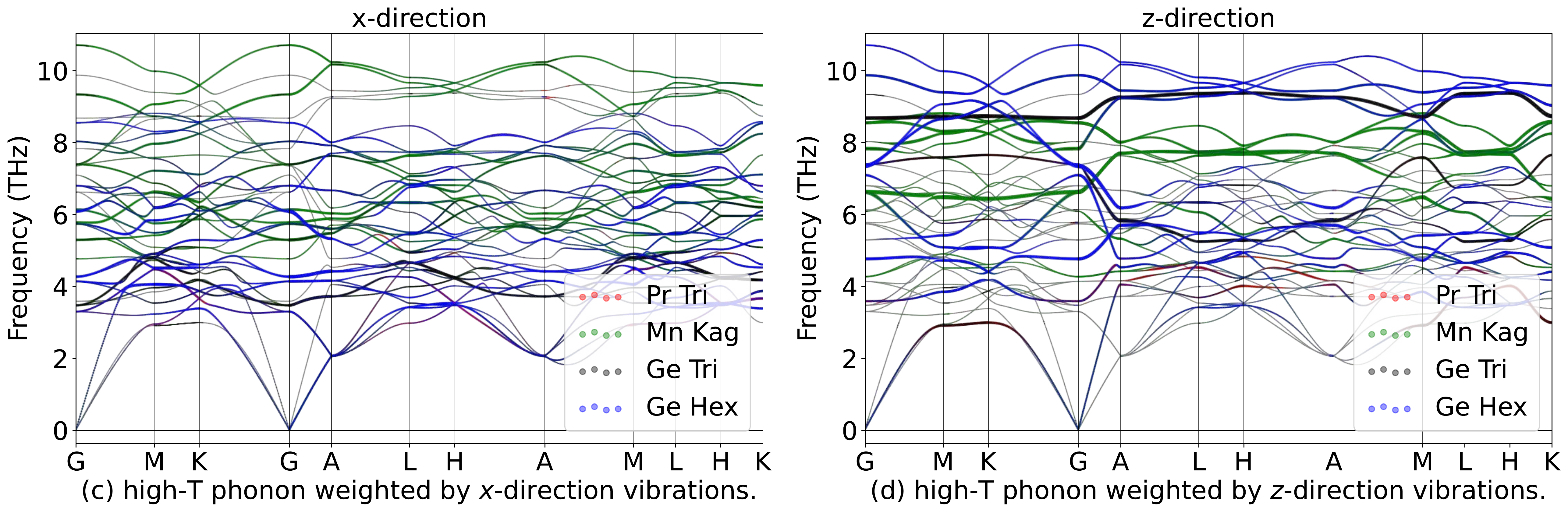}
\caption{Phonon spectrum for PrMn$_6$Ge$_6$in in-plane ($x$) and out-of-plane ($z$) directions in paramagnetic phase.}
\label{fig:PrMn6Ge6phoxyz}
\end{figure}

\begin{table}[htbp]
\global\long\def\arraystretch{1.12}
\captionsetup{width=.95\textwidth}
\caption{IRREPs at high-symmetry points for PrMn$_6$Ge$_6$ phonon spectrum at Low-T.}
\label{tab:191_PrMn6Ge6_Low}
\setlength{\tabcolsep}{1.mm}{
}
\end{table}

\begin{table}[htbp]
\global\long\def\arraystretch{1.12}
\captionsetup{width=.95\textwidth}
\caption{IRREPs at high-symmetry points for PrMn$_6$Ge$_6$ phonon spectrum at High-T.}
\label{tab:191_PrMn6Ge6_High}
\setlength{\tabcolsep}{1.mm}{
%
}
\end{table}
\subsubsection{\label{sms2sec:NdMn6Ge6}\hyperref[smsec:col]{\texorpdfstring{NdMn$_6$Ge$_6$}{}}~{\color{green}  (High-T stable, Low-T stable) }}

\begin{table}[htbp]
\global\long\def\arraystretch{1.12}
\captionsetup{width=.85\textwidth}
\caption{Structure parameters of NdMn$_6$Ge$_6$ after relaxation. It contains 402 electrons. }
\label{tab:NdMn6Ge6}
\setlength{\tabcolsep}{4mm}{
%
}
\end{table}

\begin{figure}[htbp]
\centering
\includegraphics[width=0.85\textwidth]{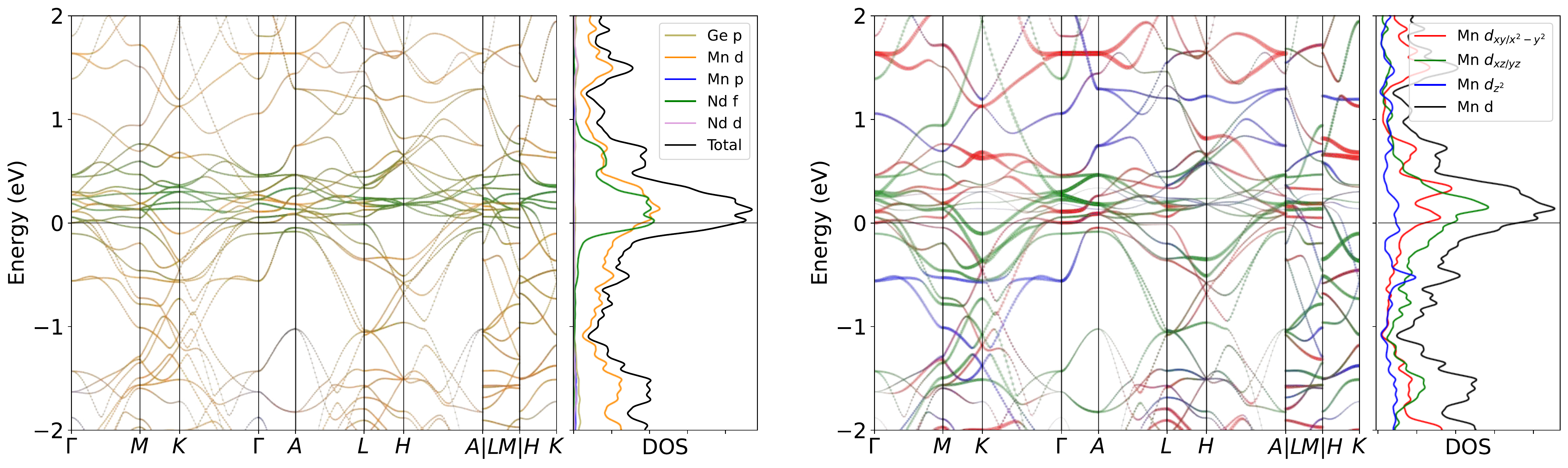}
\caption{Band structure for NdMn$_6$Ge$_6$ without SOC in paramagnetic phase. The projection of Mn $3d$ orbitals is also presented. The band structures clearly reveal the dominating of Mn $3d$ orbitals  near the Fermi level, as well as the pronounced peak.}
\label{fig:NdMn6Ge6band}
\end{figure}

\begin{figure}[H]
\centering
\subfloat[Relaxed phonon at low-T.]{
\label{fig:NdMn6Ge6_relaxed_Low}
\includegraphics[width=0.45\textwidth]{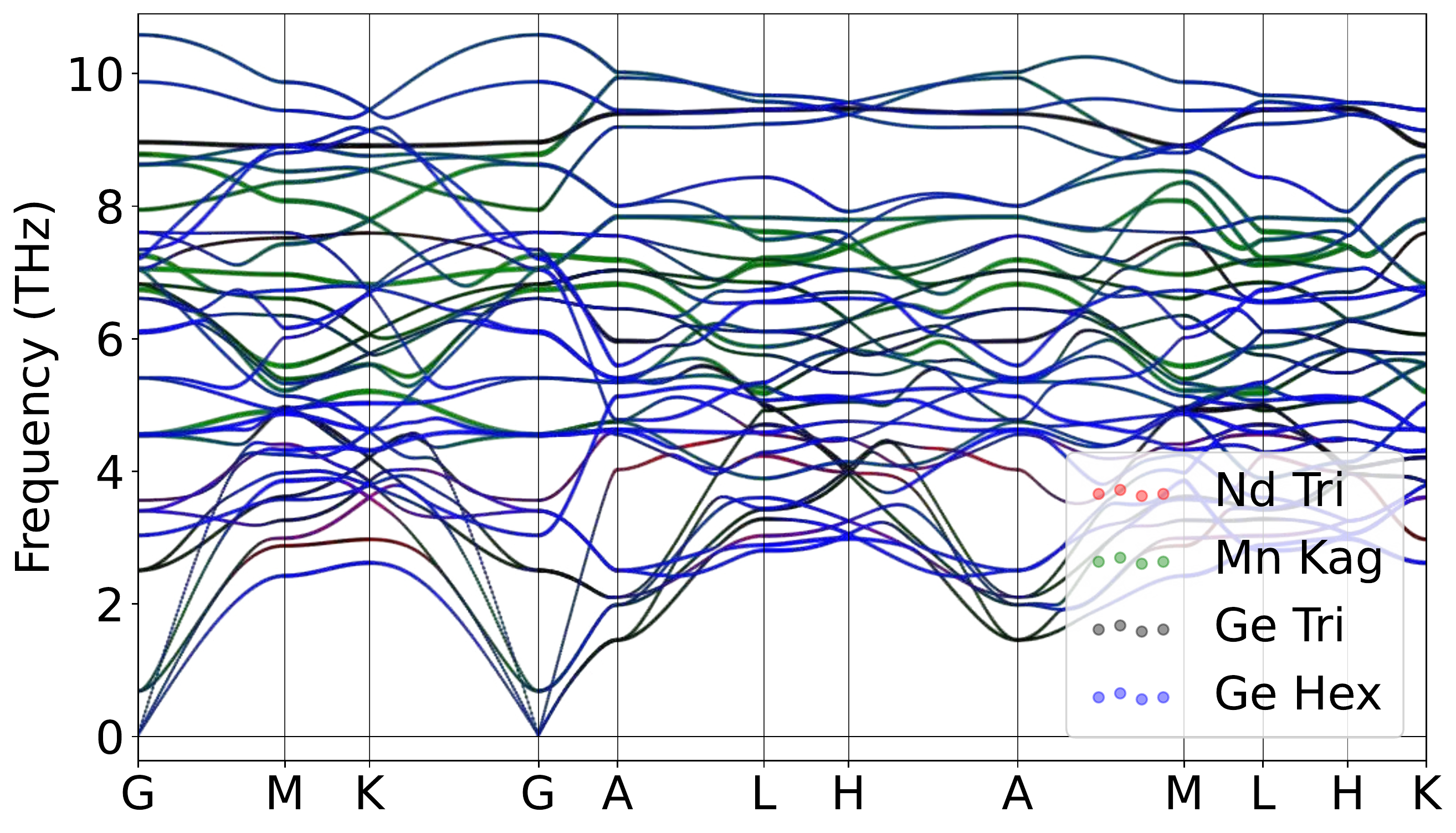}
}
\hspace{5pt}\subfloat[Relaxed phonon at high-T.]{
\label{fig:NdMn6Ge6_relaxed_High}
\includegraphics[width=0.45\textwidth]{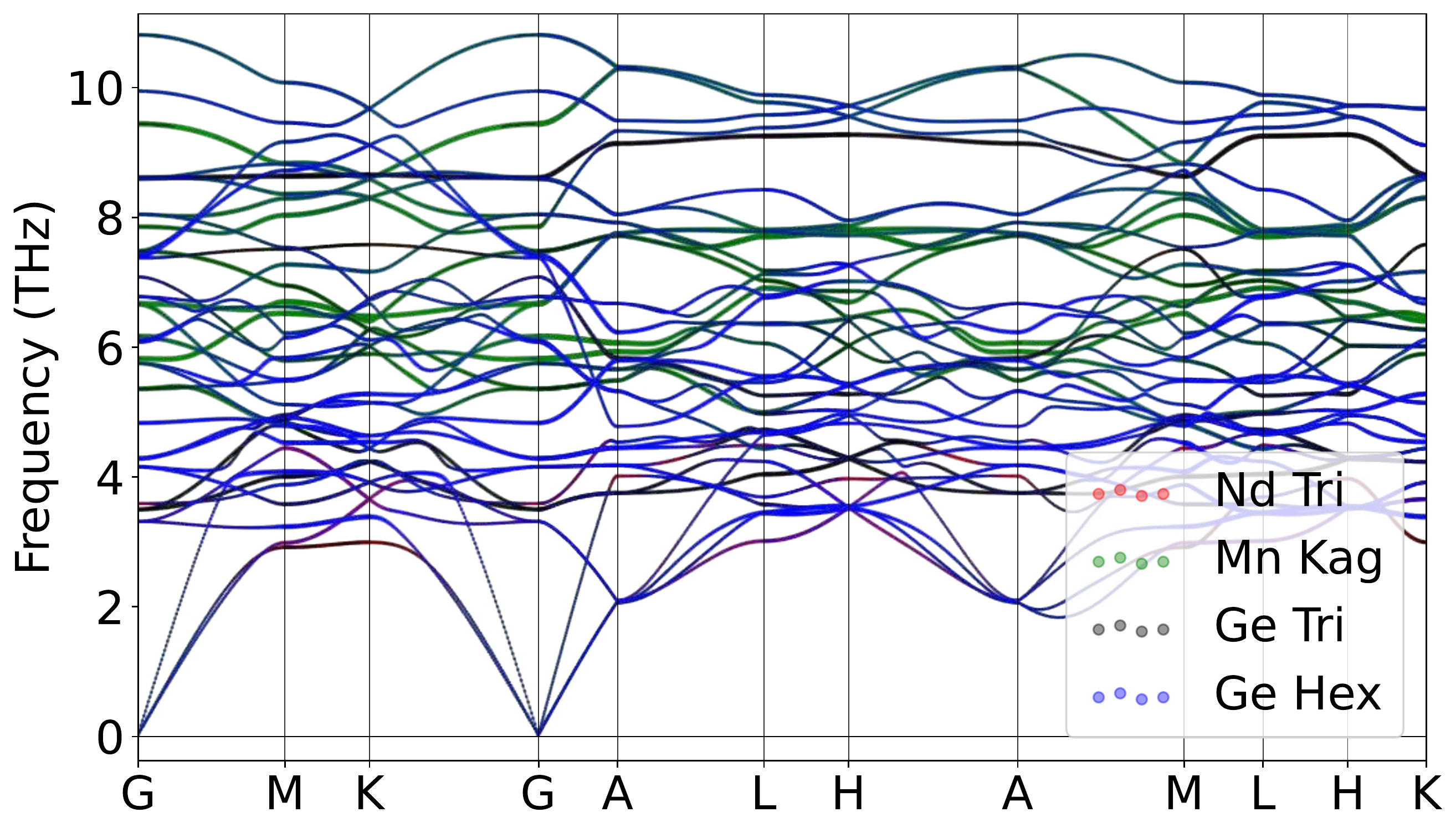}
}
\caption{Atomically projected phonon spectrum for NdMn$_6$Ge$_6$ in paramagnetic phase.}
\label{fig:NdMn6Ge6pho}
\end{figure}

\begin{figure}[htbp]
\centering
\includegraphics[width=0.32\textwidth]{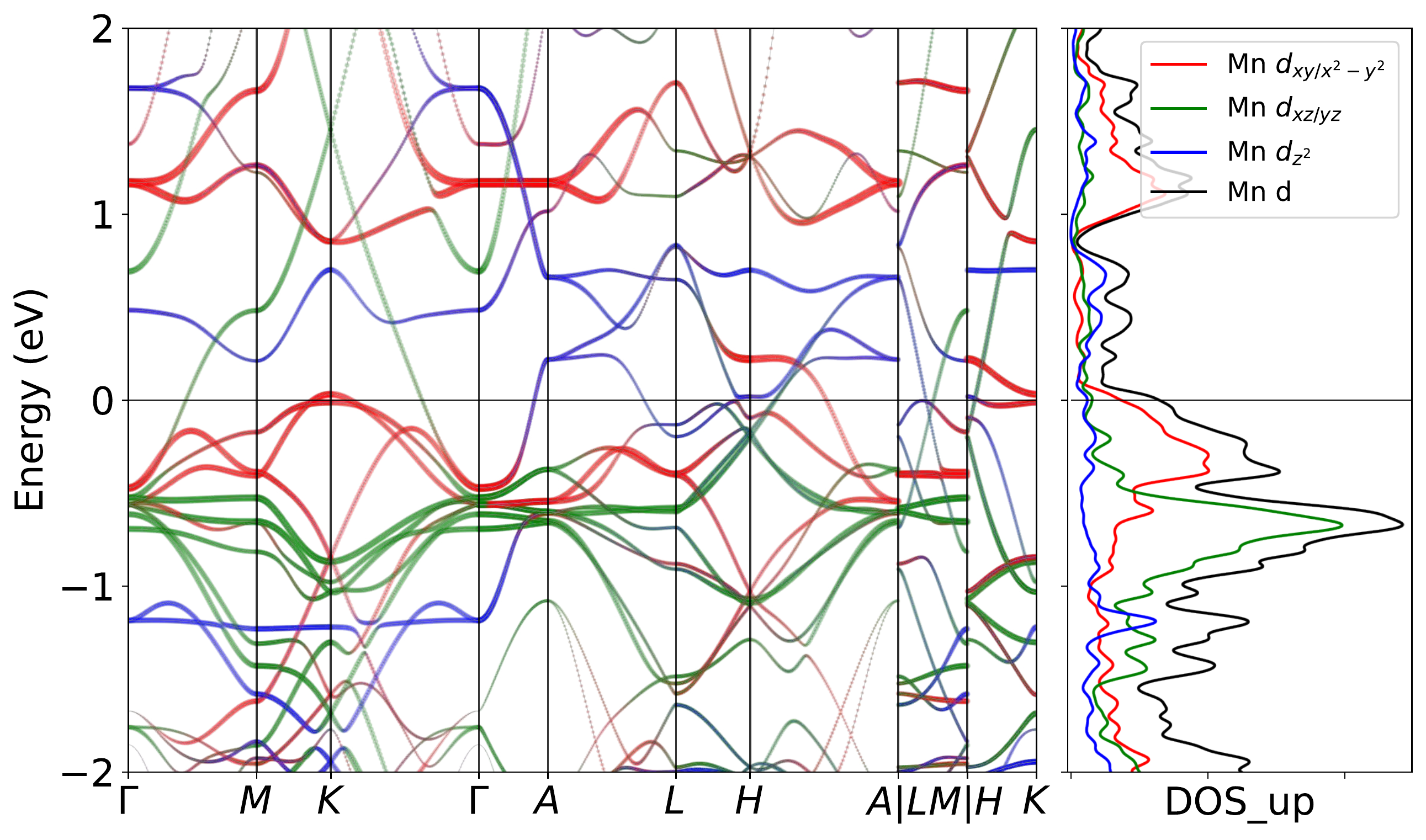}
\includegraphics[width=0.32\textwidth]{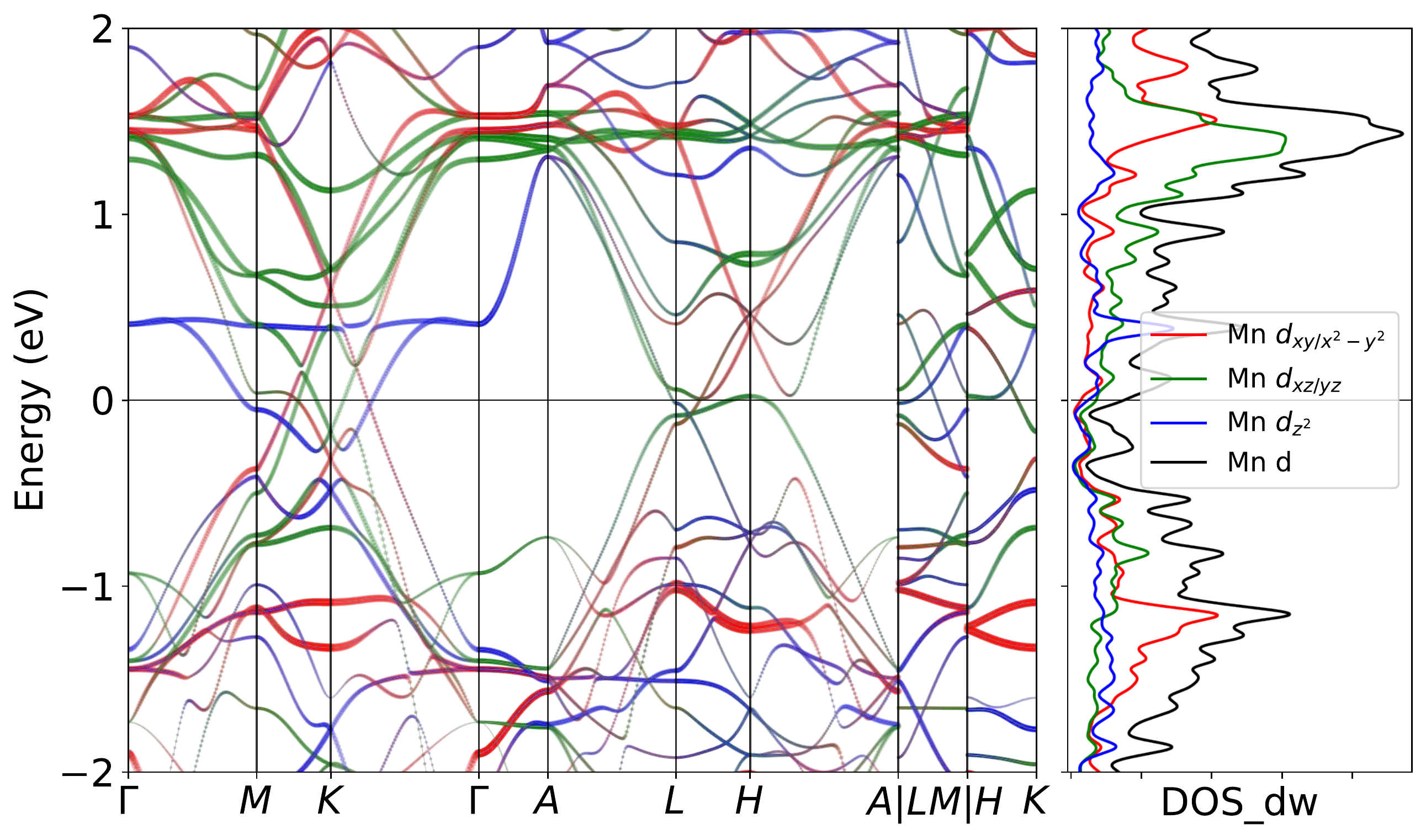}
\includegraphics[width=0.32\textwidth]{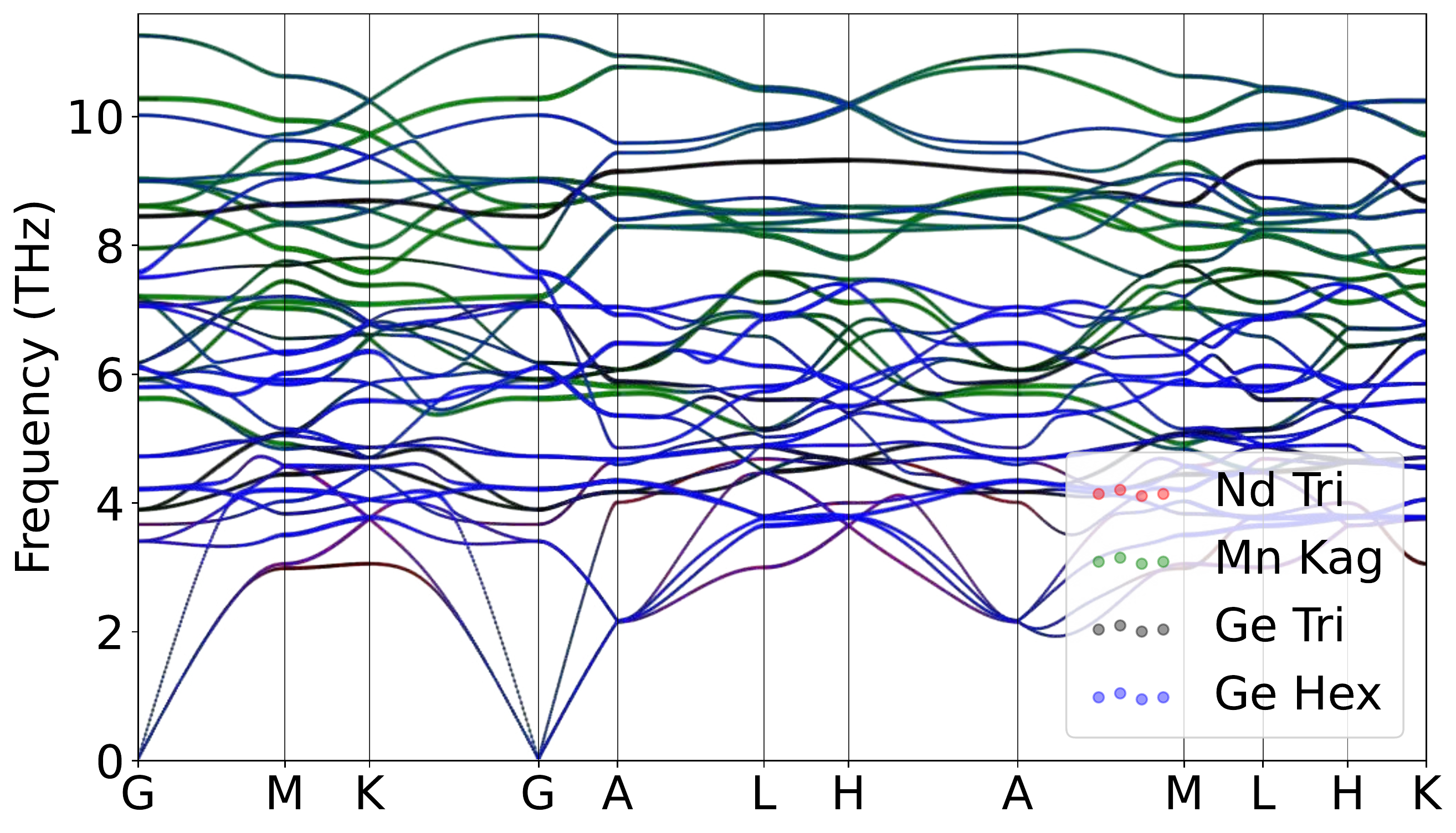}
\caption{Band structure for NdMn$_6$Ge$_6$ with FM order without SOC for spin (Left)up and (Middle)down channels. The projection of Mn $3d$ orbitals is also presented. (Right)Correspondingly, a stable phonon was demonstrated with the collinear magnetic order without Hubbard U at lower temperature regime, weighted by atomic projections.}
\label{fig:NdMn6Ge6mag}
\end{figure}

\begin{figure}[H]
\centering
\label{fig:NdMn6Ge6_relaxed_Low_xz}
\includegraphics[width=0.85\textwidth]{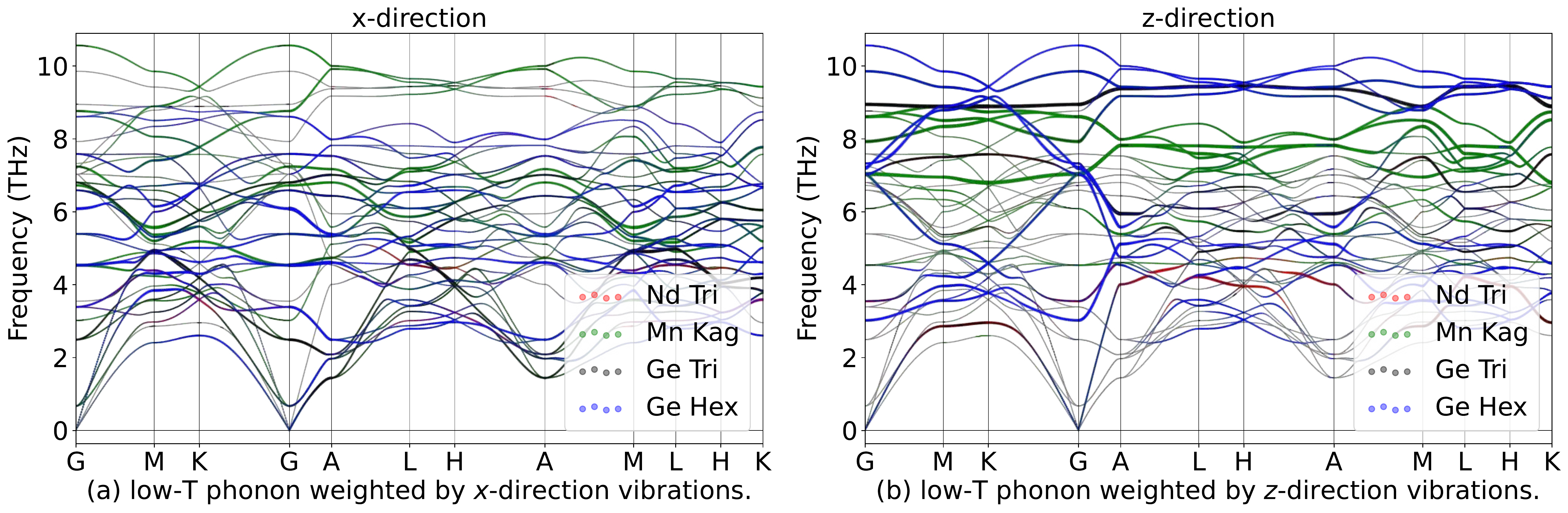}
\hspace{5pt}\label{fig:NdMn6Ge6_relaxed_High_xz}
\includegraphics[width=0.85\textwidth]{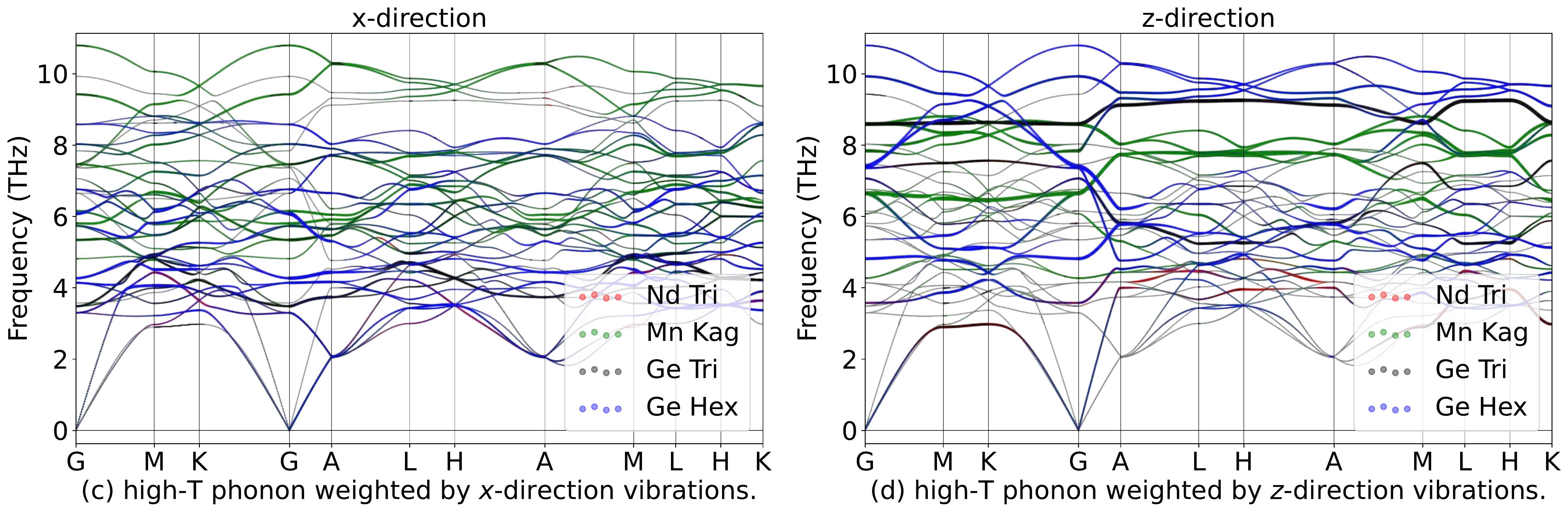}
\caption{Phonon spectrum for NdMn$_6$Ge$_6$in in-plane ($x$) and out-of-plane ($z$) directions in paramagnetic phase.}
\label{fig:NdMn6Ge6phoxyz}
\end{figure}

\begin{table}[htbp]
\global\long\def\arraystretch{1.12}
\captionsetup{width=.95\textwidth}
\caption{IRREPs at high-symmetry points for NdMn$_6$Ge$_6$ phonon spectrum at Low-T.}
\label{tab:191_NdMn6Ge6_Low}
\setlength{\tabcolsep}{1.mm}{
}
\end{table}

\begin{table}[htbp]
\global\long\def\arraystretch{1.12}
\captionsetup{width=.95\textwidth}
\caption{IRREPs at high-symmetry points for NdMn$_6$Ge$_6$ phonon spectrum at High-T.}
\label{tab:191_NdMn6Ge6_High}
\setlength{\tabcolsep}{1.mm}{
%
}
\end{table}
\subsubsection{\label{sms2sec:SmMn6Ge6}\hyperref[smsec:col]{\texorpdfstring{SmMn$_6$Ge$_6$}{}}~{\color{green}  (High-T stable, Low-T stable) }}

\begin{table}[htbp]
\global\long\def\arraystretch{1.12}
\captionsetup{width=.85\textwidth}
\caption{Structure parameters of SmMn$_6$Ge$_6$ after relaxation. It contains 404 electrons. }
\label{tab:SmMn6Ge6}
\setlength{\tabcolsep}{4mm}{
%
}
\end{table}

\begin{figure}[htbp]
\centering
\includegraphics[width=0.85\textwidth]{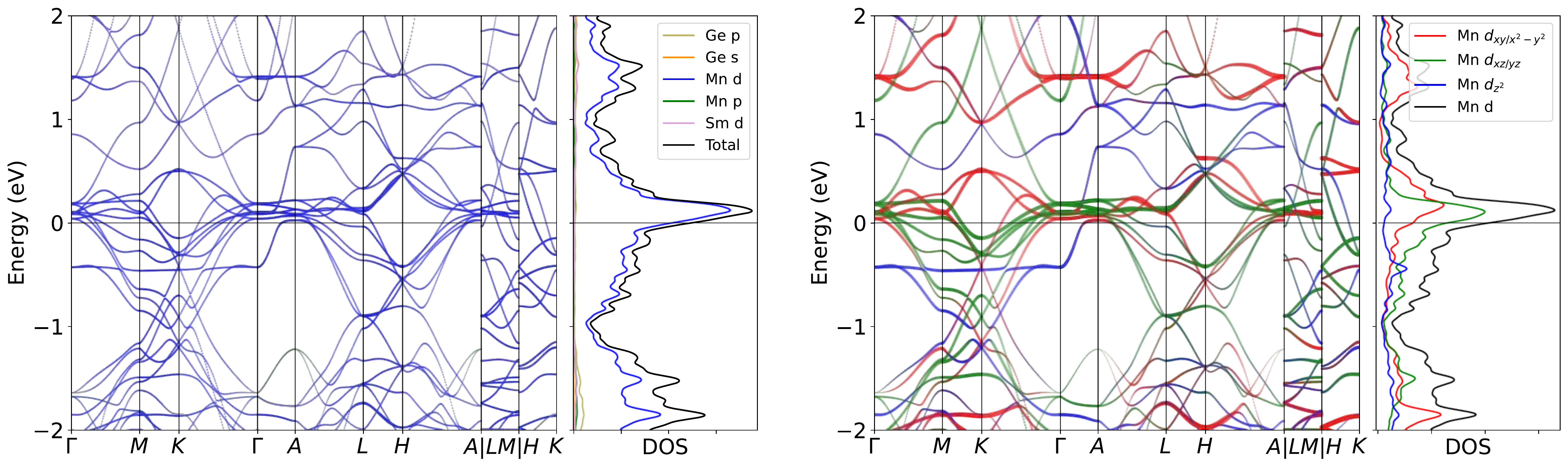}
\caption{Band structure for SmMn$_6$Ge$_6$ without SOC in paramagnetic phase. The projection of Mn $3d$ orbitals is also presented. The band structures clearly reveal the dominating of Mn $3d$ orbitals  near the Fermi level, as well as the pronounced peak.}
\label{fig:SmMn6Ge6band}
\end{figure}

\begin{figure}[H]
\centering
\subfloat[Relaxed phonon at low-T.]{
\label{fig:SmMn6Ge6_relaxed_Low}
\includegraphics[width=0.45\textwidth]{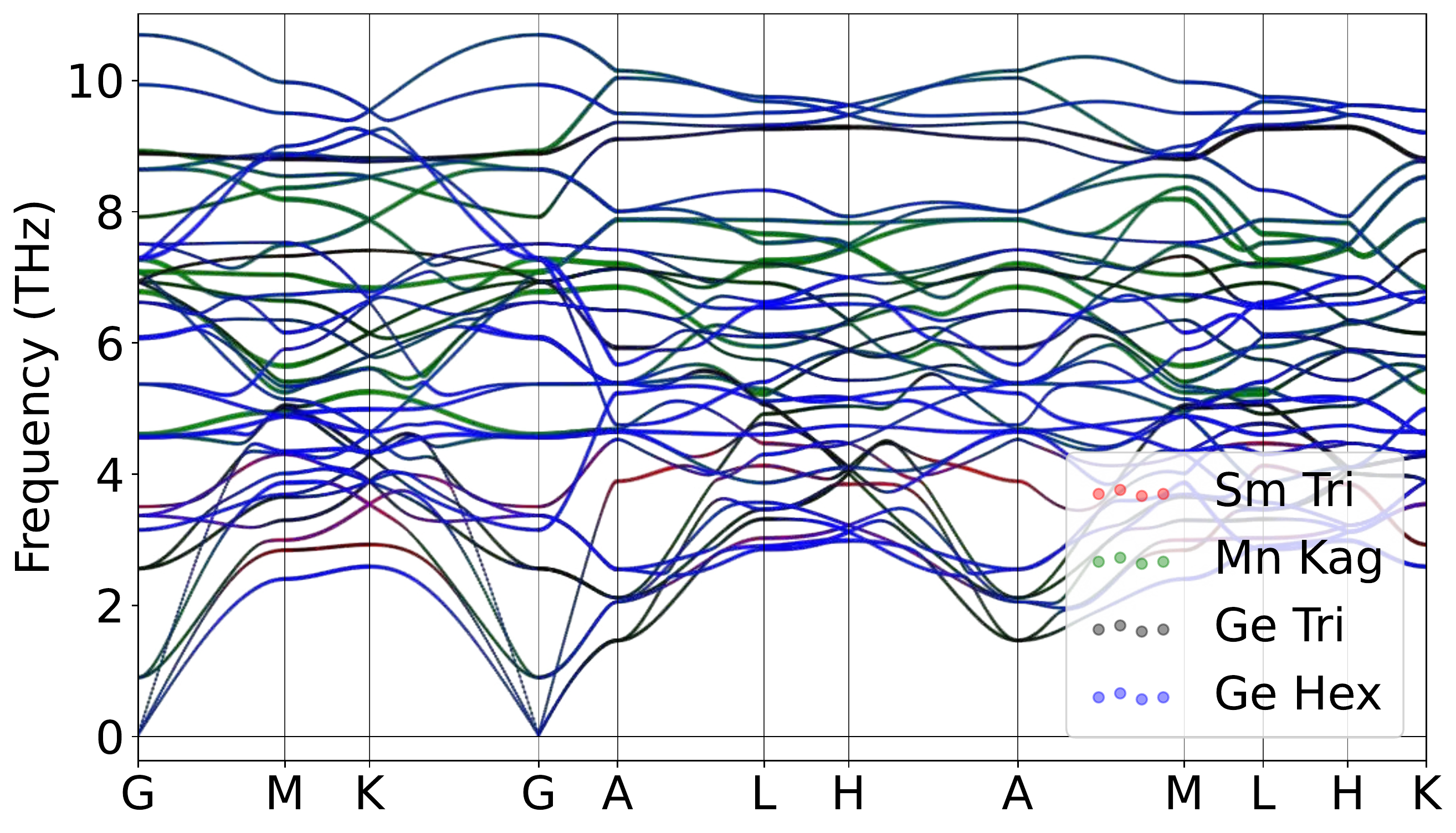}
}
\hspace{5pt}\subfloat[Relaxed phonon at high-T.]{
\label{fig:SmMn6Ge6_relaxed_High}
\includegraphics[width=0.45\textwidth]{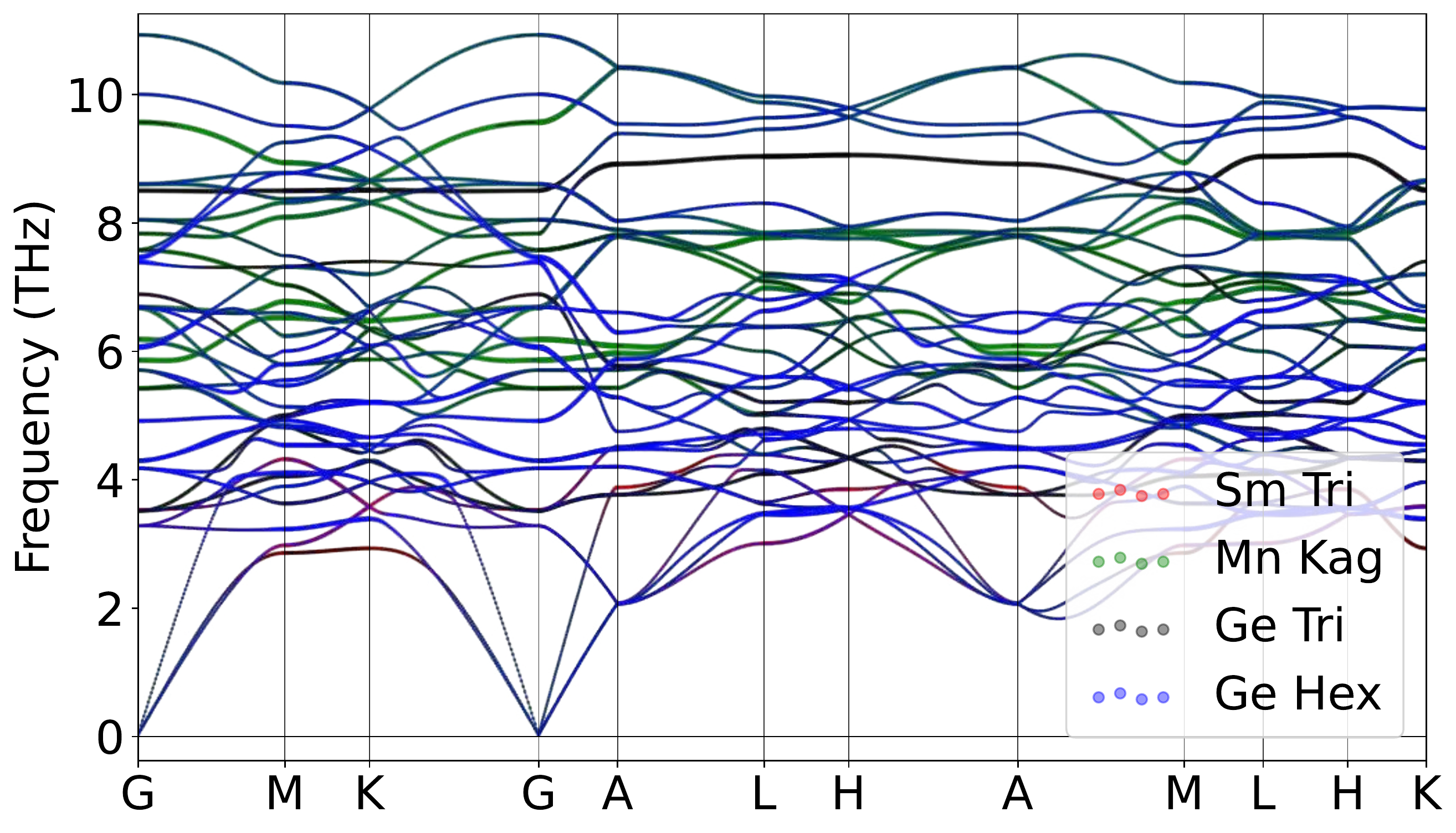}
}
\caption{Atomically projected phonon spectrum for SmMn$_6$Ge$_6$ in paramagnetic phase.}
\label{fig:SmMn6Ge6pho}
\end{figure}

\begin{figure}[htbp]
\centering
\includegraphics[width=0.32\textwidth]{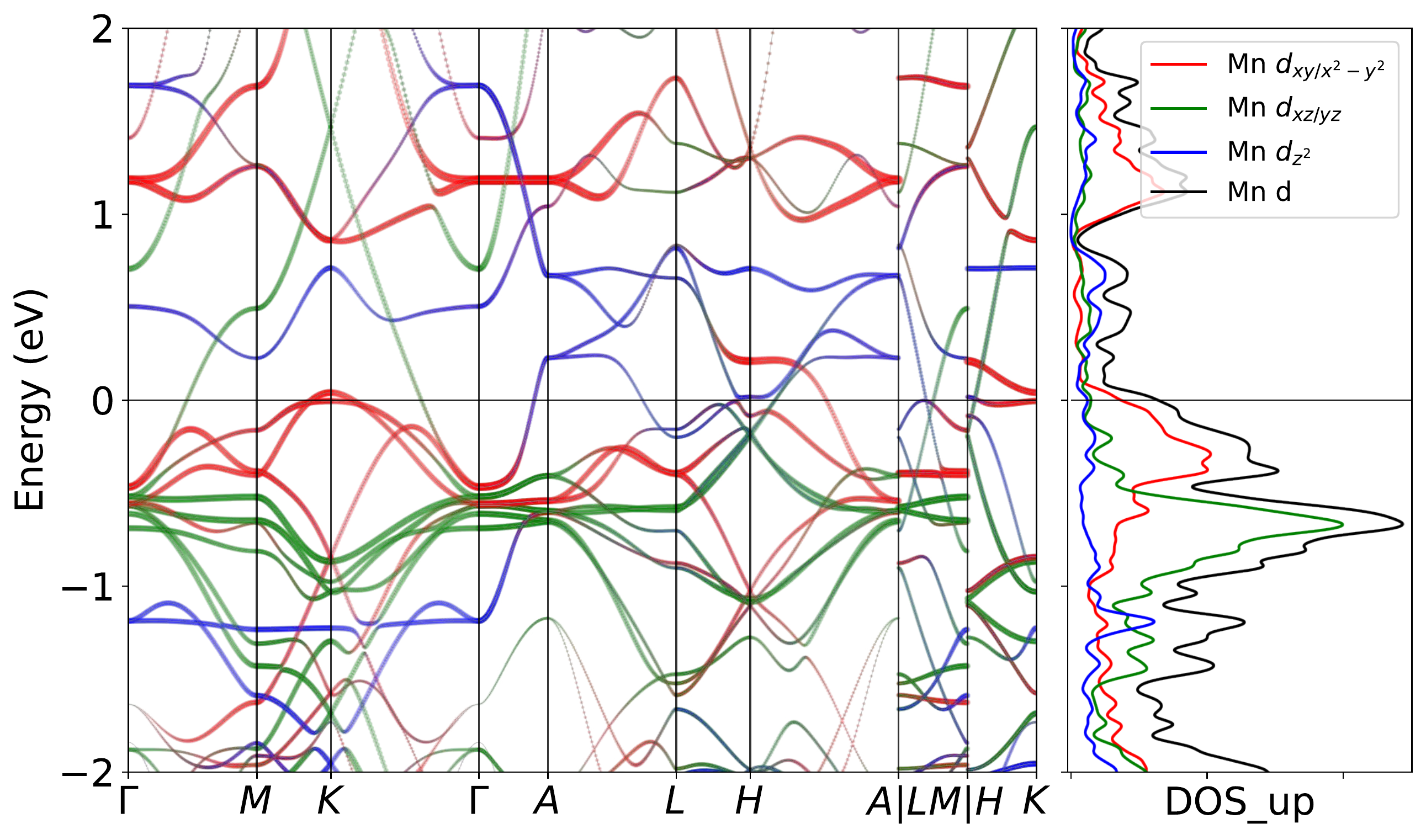}
\includegraphics[width=0.32\textwidth]{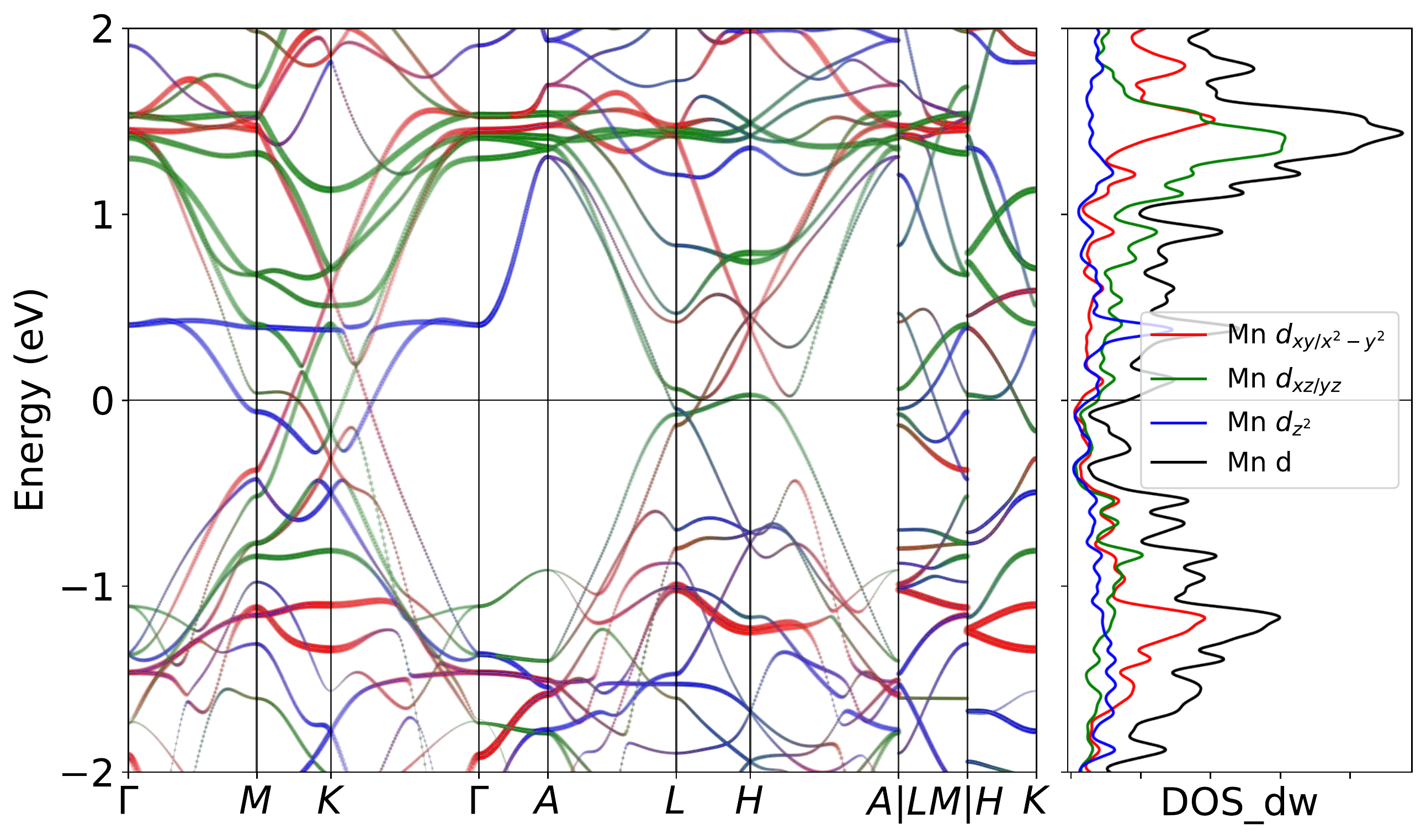}
\includegraphics[width=0.32\textwidth]{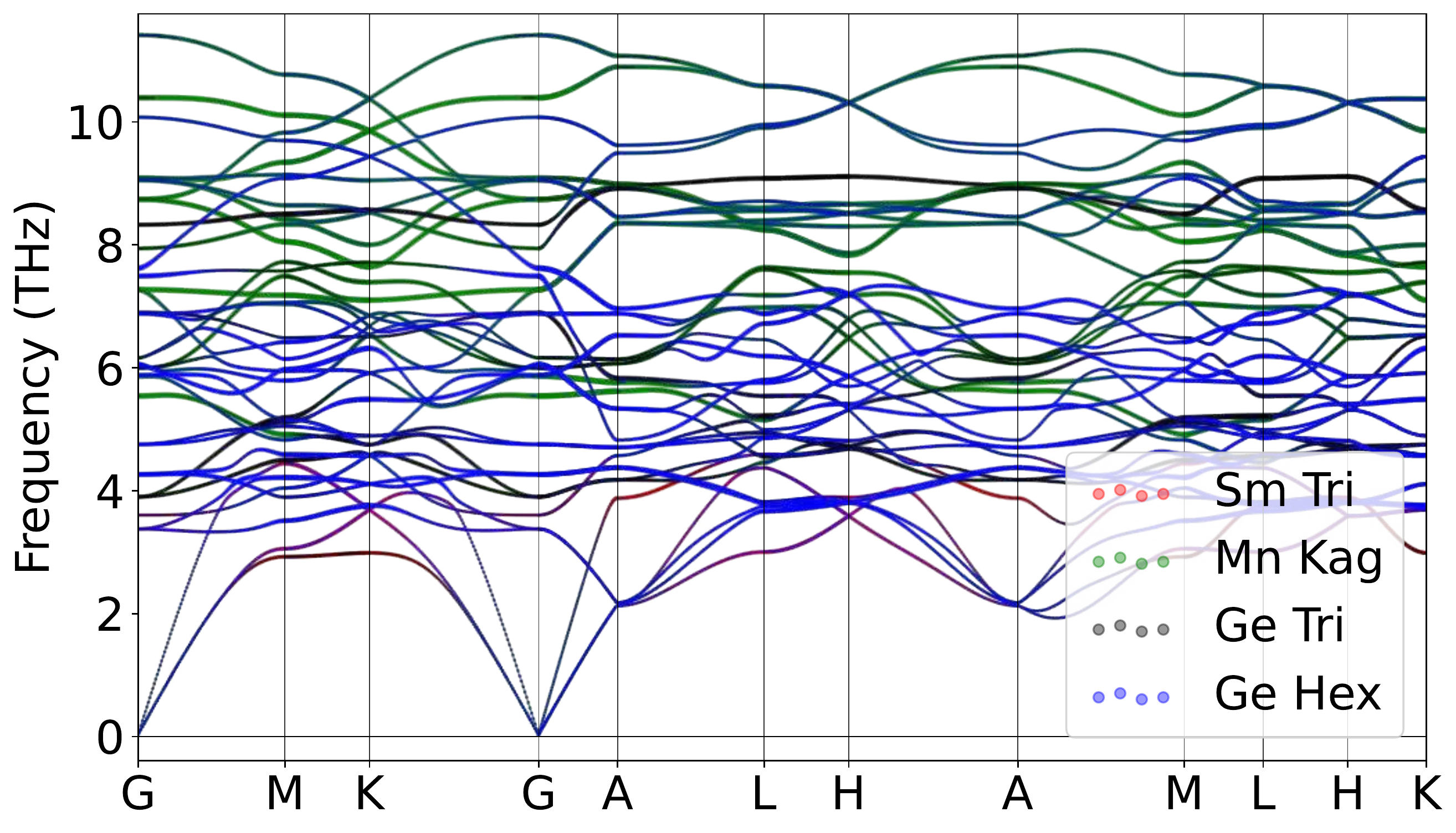}
\caption{Band structure for SmMn$_6$Ge$_6$ with FM order without SOC for spin (Left)up and (Middle)down channels. The projection of Mn $3d$ orbitals is also presented. (Right)Correspondingly, a stable phonon was demonstrated with the collinear magnetic order without Hubbard U at lower temperature regime, weighted by atomic projections.}
\label{fig:SmMn6Ge6mag}
\end{figure}

\begin{figure}[H]
\centering
\label{fig:SmMn6Ge6_relaxed_Low_xz}
\includegraphics[width=0.85\textwidth]{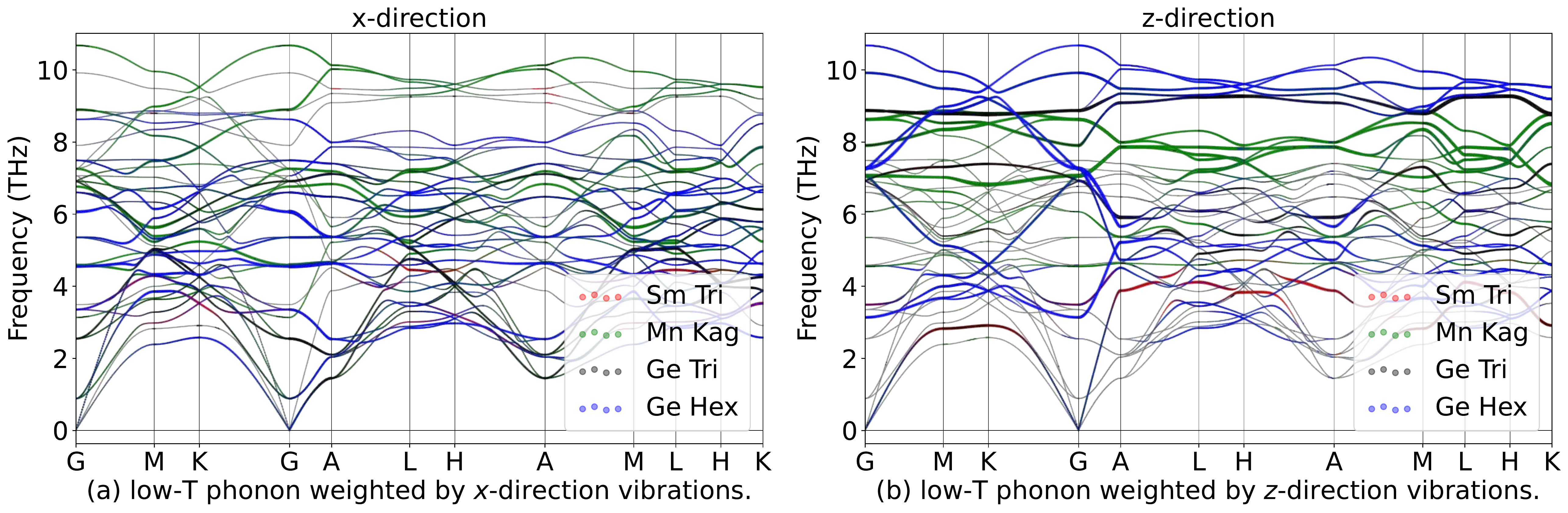}
\hspace{5pt}\label{fig:SmMn6Ge6_relaxed_High_xz}
\includegraphics[width=0.85\textwidth]{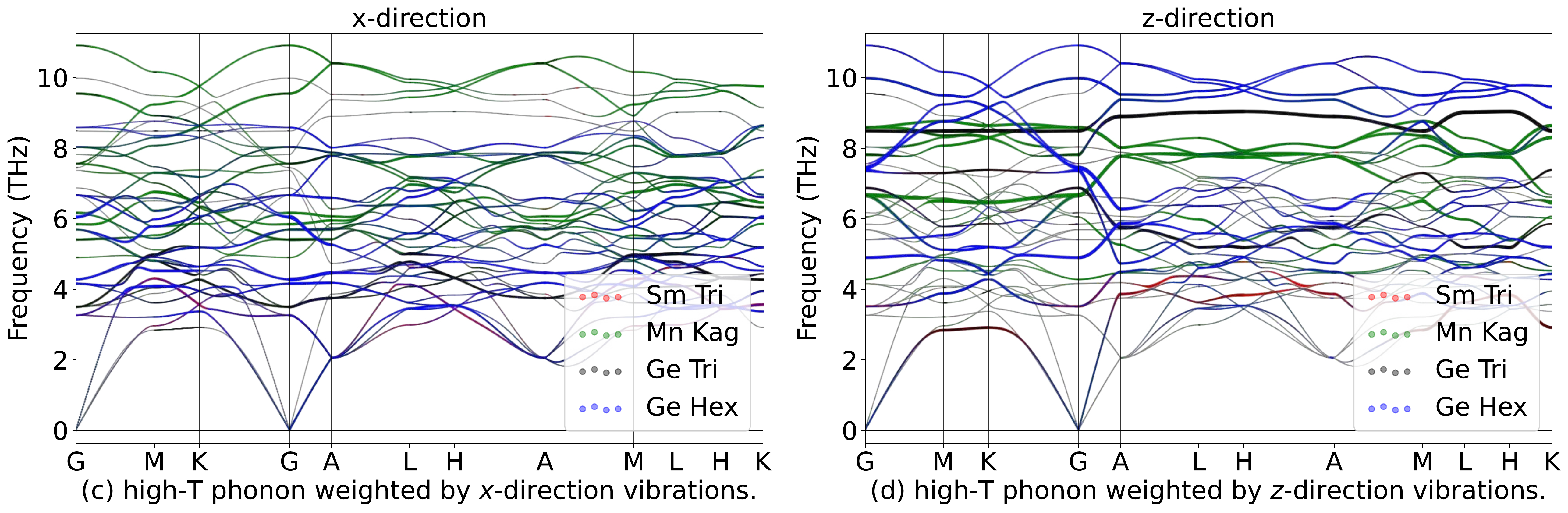}
\caption{Phonon spectrum for SmMn$_6$Ge$_6$in in-plane ($x$) and out-of-plane ($z$) directions in paramagnetic phase.}
\label{fig:SmMn6Ge6phoxyz}
\end{figure}

\begin{table}[htbp]
\global\long\def\arraystretch{1.12}
\captionsetup{width=.95\textwidth}
\caption{IRREPs at high-symmetry points for SmMn$_6$Ge$_6$ phonon spectrum at Low-T.}
\label{tab:191_SmMn6Ge6_Low}
\setlength{\tabcolsep}{1.mm}{
}
\end{table}

\begin{table}[htbp]
\global\long\def\arraystretch{1.12}
\captionsetup{width=.95\textwidth}
\caption{IRREPs at high-symmetry points for SmMn$_6$Ge$_6$ phonon spectrum at High-T.}
\label{tab:191_SmMn6Ge6_High}
\setlength{\tabcolsep}{1.mm}{
%
}
\end{table}
\subsubsection{\label{sms2sec:GdMn6Ge6}\hyperref[smsec:col]{\texorpdfstring{GdMn$_6$Ge$_6$}{}}~{\color{green}  (High-T stable, Low-T stable) }}

\begin{table}[htbp]
\global\long\def\arraystretch{1.12}
\captionsetup{width=.85\textwidth}
\caption{Structure parameters of GdMn$_6$Ge$_6$ after relaxation. It contains 406 electrons. }
\label{tab:GdMn6Ge6}
\setlength{\tabcolsep}{4mm}{
%
}
\end{table}

\begin{figure}[htbp]
\centering
\includegraphics[width=0.85\textwidth]{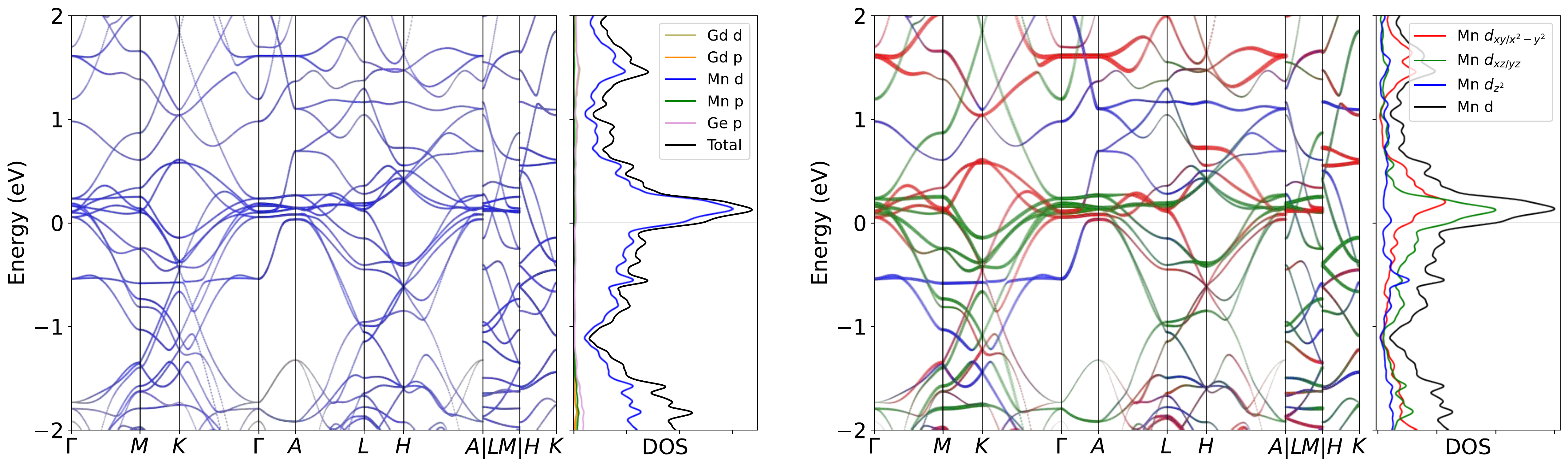}
\caption{Band structure for GdMn$_6$Ge$_6$ without SOC in paramagnetic phase. The projection of Mn $3d$ orbitals is also presented. The band structures clearly reveal the dominating of Mn $3d$ orbitals  near the Fermi level, as well as the pronounced peak.}
\label{fig:GdMn6Ge6band}
\end{figure}

\begin{figure}[H]
\centering
\subfloat[Relaxed phonon at low-T.]{
\label{fig:GdMn6Ge6_relaxed_Low}
\includegraphics[width=0.45\textwidth]{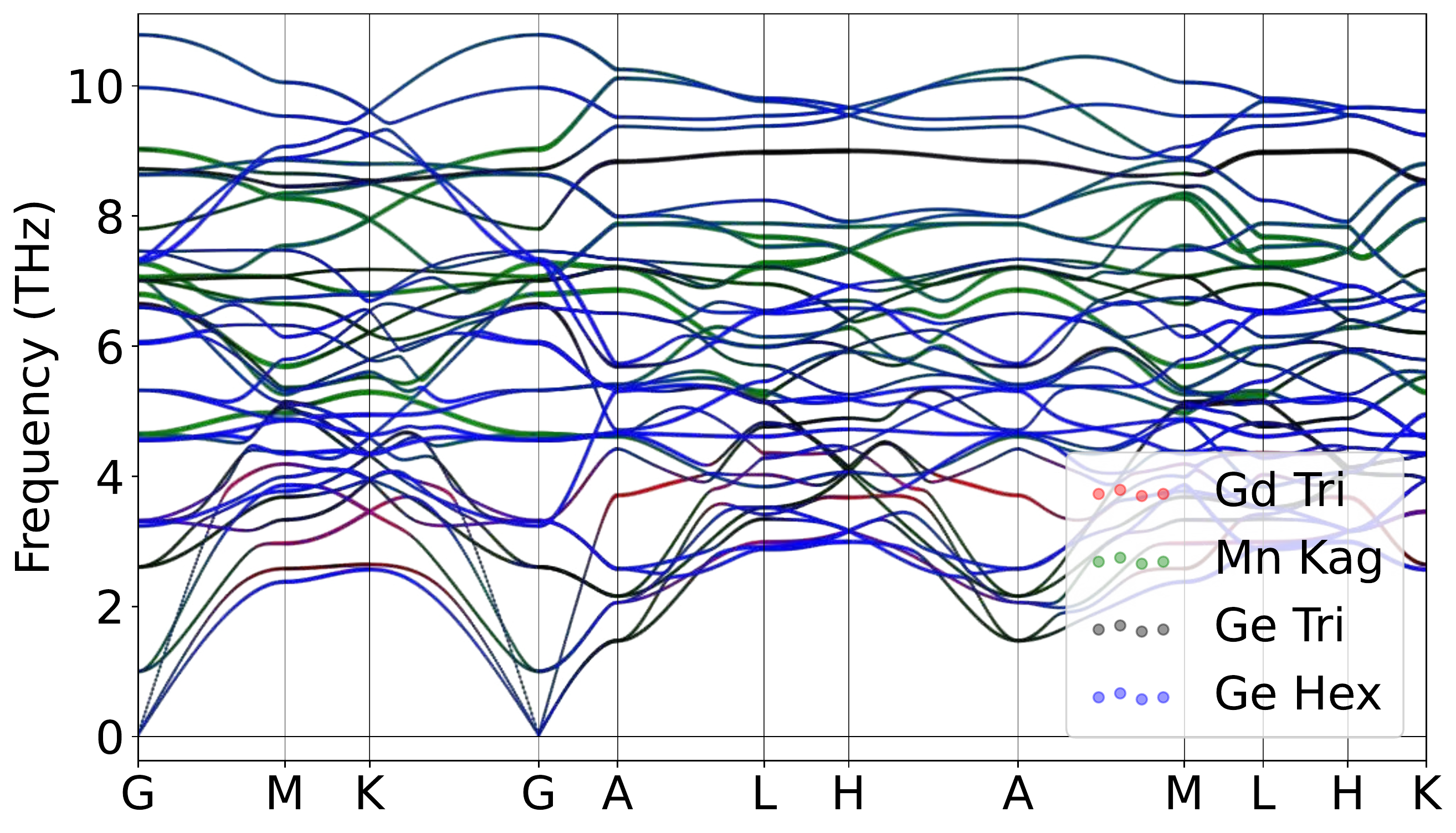}
}
\hspace{5pt}\subfloat[Relaxed phonon at high-T.]{
\label{fig:GdMn6Ge6_relaxed_High}
\includegraphics[width=0.45\textwidth]{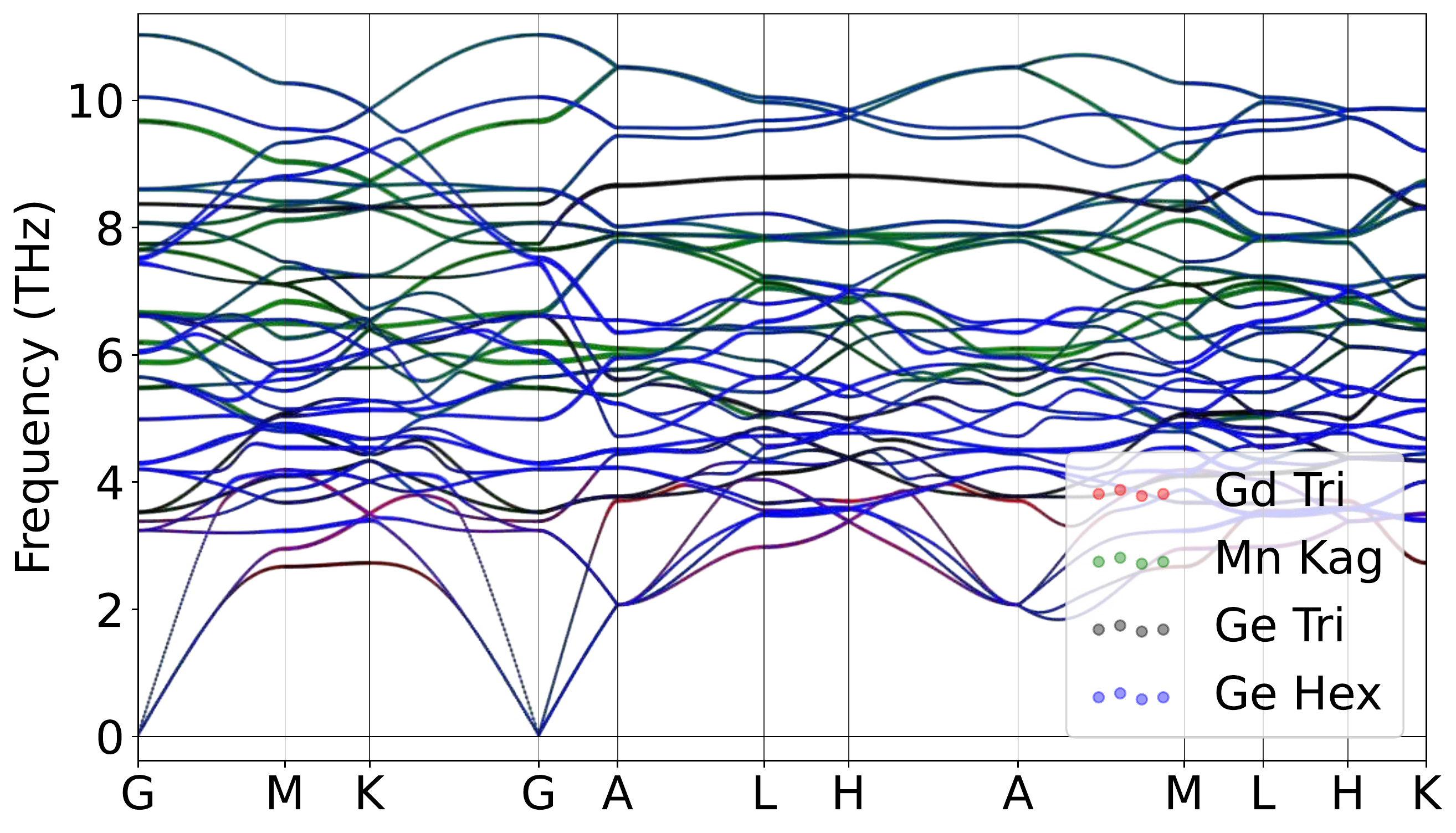}
}
\caption{Atomically projected phonon spectrum for GdMn$_6$Ge$_6$ in paramagnetic phase.}
\label{fig:GdMn6Ge6pho}
\end{figure}

\begin{figure}[htbp]
\centering
\includegraphics[width=0.32\textwidth]{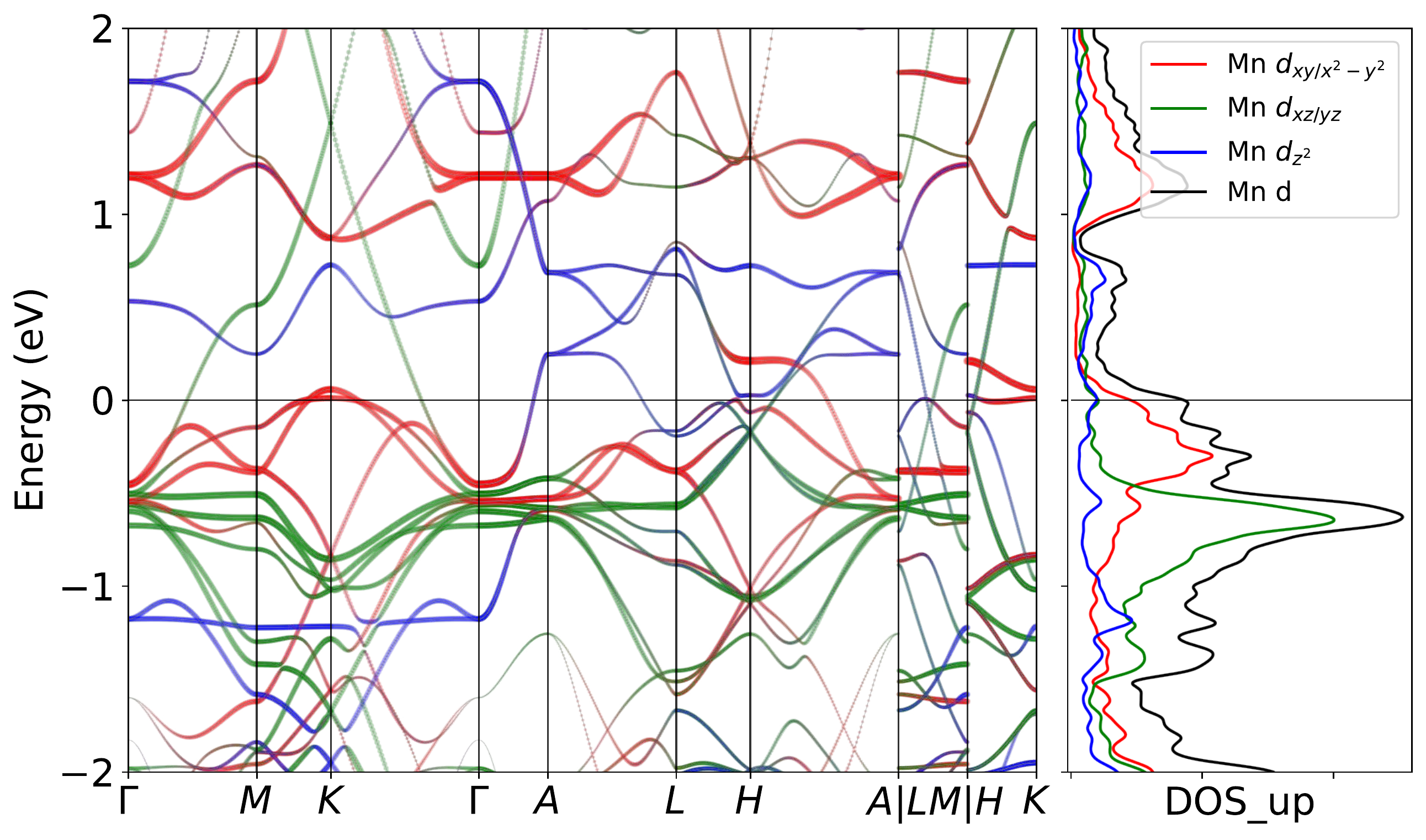}
\includegraphics[width=0.32\textwidth]{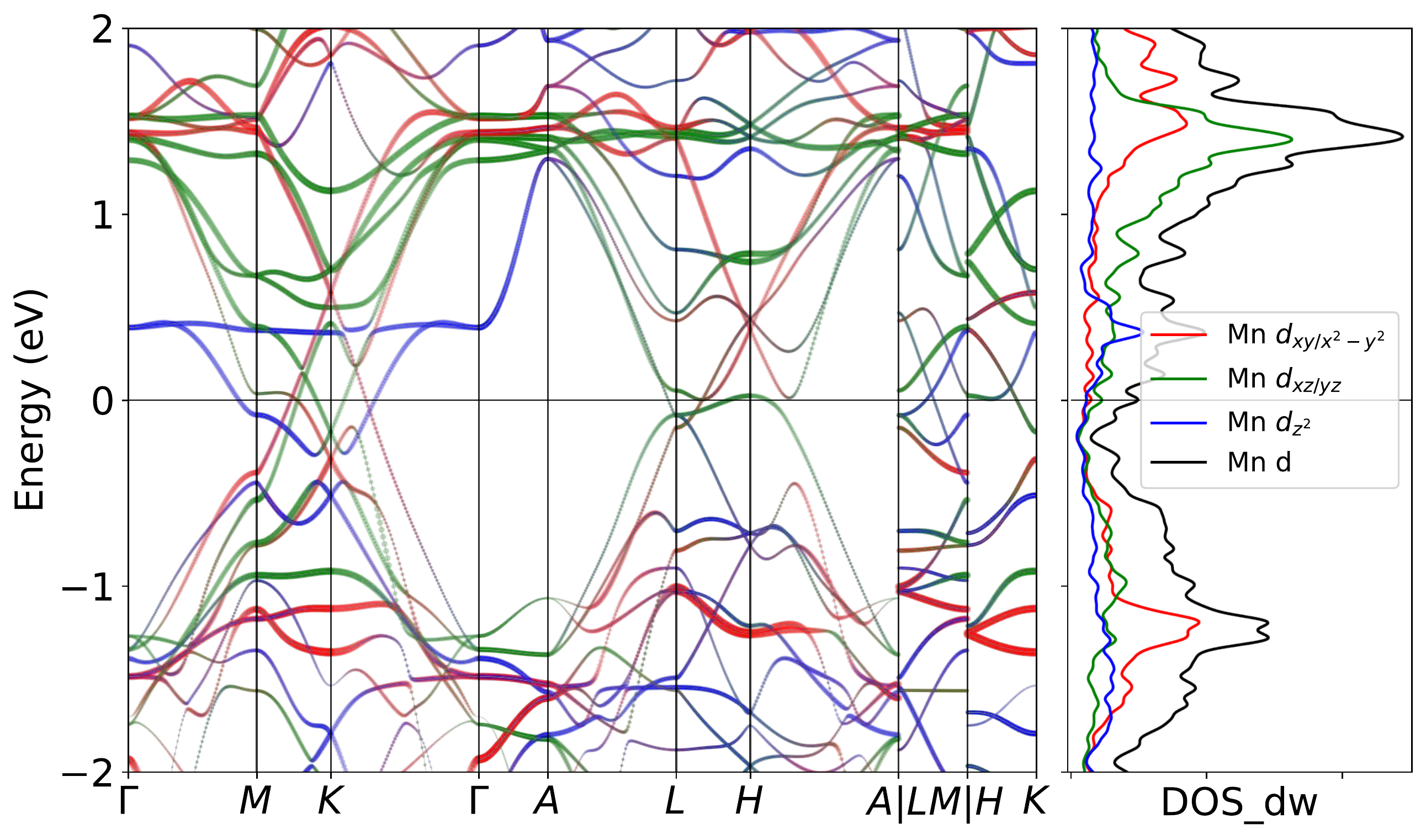}
\includegraphics[width=0.32\textwidth]{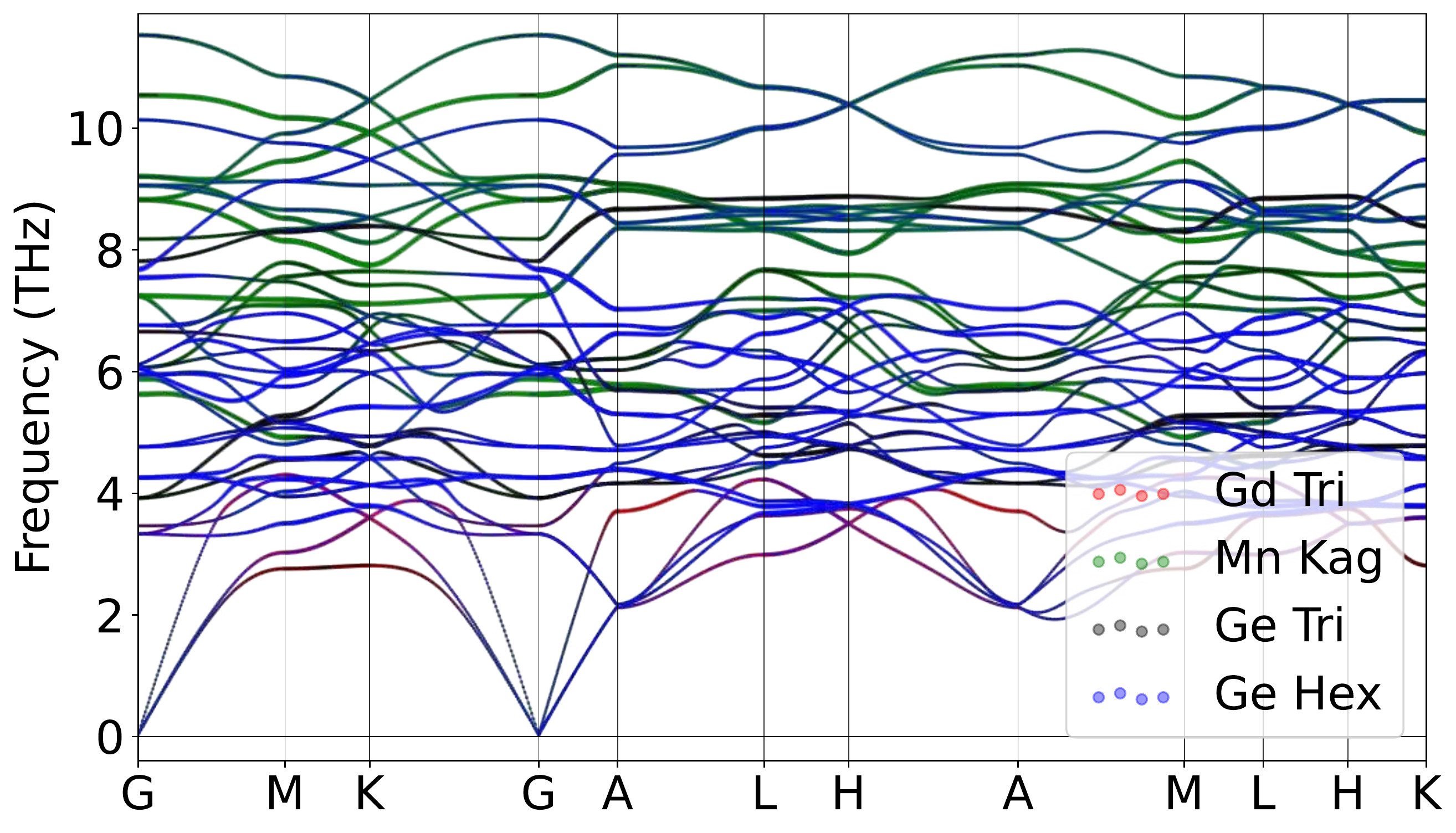}
\caption{Band structure for GdMn$_6$Ge$_6$ with FM order without SOC for spin (Left)up and (Middle)down channels. The projection of Mn $3d$ orbitals is also presented. (Right)Correspondingly, a stable phonon was demonstrated with the collinear magnetic order without Hubbard U at lower temperature regime, weighted by atomic projections.}
\label{fig:GdMn6Ge6mag}
\end{figure}

\begin{figure}[H]
\centering
\label{fig:GdMn6Ge6_relaxed_Low_xz}
\includegraphics[width=0.85\textwidth]{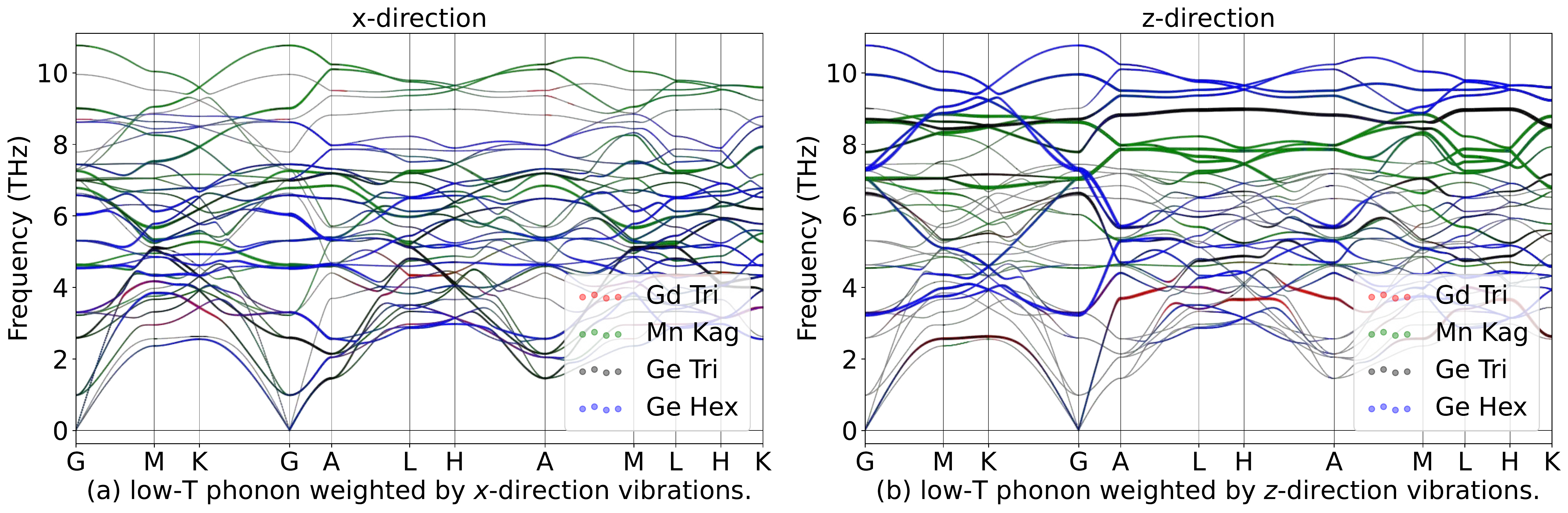}
\hspace{5pt}\label{fig:GdMn6Ge6_relaxed_High_xz}
\includegraphics[width=0.85\textwidth]{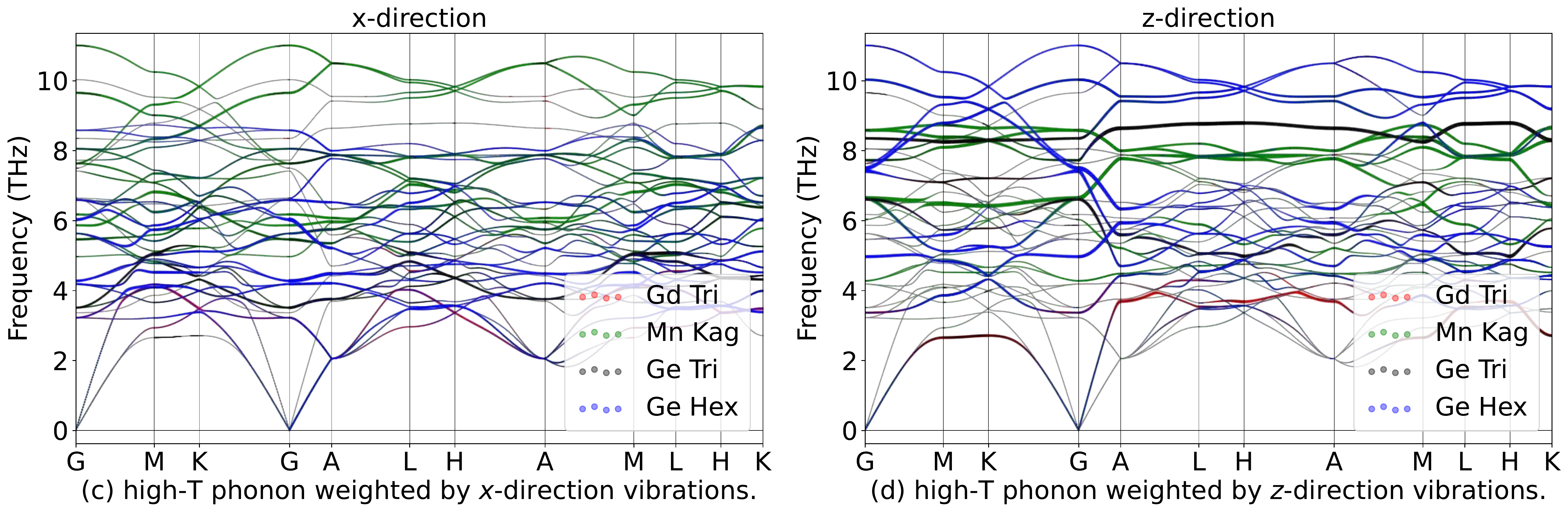}
\caption{Phonon spectrum for GdMn$_6$Ge$_6$in in-plane ($x$) and out-of-plane ($z$) directions in paramagnetic phase.}
\label{fig:GdMn6Ge6phoxyz}
\end{figure}

\begin{table}[htbp]
\global\long\def\arraystretch{1.12}
\captionsetup{width=.95\textwidth}
\caption{IRREPs at high-symmetry points for GdMn$_6$Ge$_6$ phonon spectrum at Low-T.}
\label{tab:191_GdMn6Ge6_Low}
\setlength{\tabcolsep}{1.mm}{
}
\end{table}

\begin{table}[htbp]
\global\long\def\arraystretch{1.12}
\captionsetup{width=.95\textwidth}
\caption{IRREPs at high-symmetry points for GdMn$_6$Ge$_6$ phonon spectrum at High-T.}
\label{tab:191_GdMn6Ge6_High}
\setlength{\tabcolsep}{1.mm}{
%
}
\end{table}
\subsubsection{\label{sms2sec:TbMn6Ge6}\hyperref[smsec:col]{\texorpdfstring{TbMn$_6$Ge$_6$}{}}~{\color{green}  (High-T stable, Low-T stable) }}

\begin{table}[htbp]
\global\long\def\arraystretch{1.12}
\captionsetup{width=.85\textwidth}
\caption{Structure parameters of TbMn$_6$Ge$_6$ after relaxation. It contains 407 electrons. }
\label{tab:TbMn6Ge6}
\setlength{\tabcolsep}{4mm}{
%
}
\end{table}

\begin{figure}[htbp]
\centering
\includegraphics[width=0.85\textwidth]{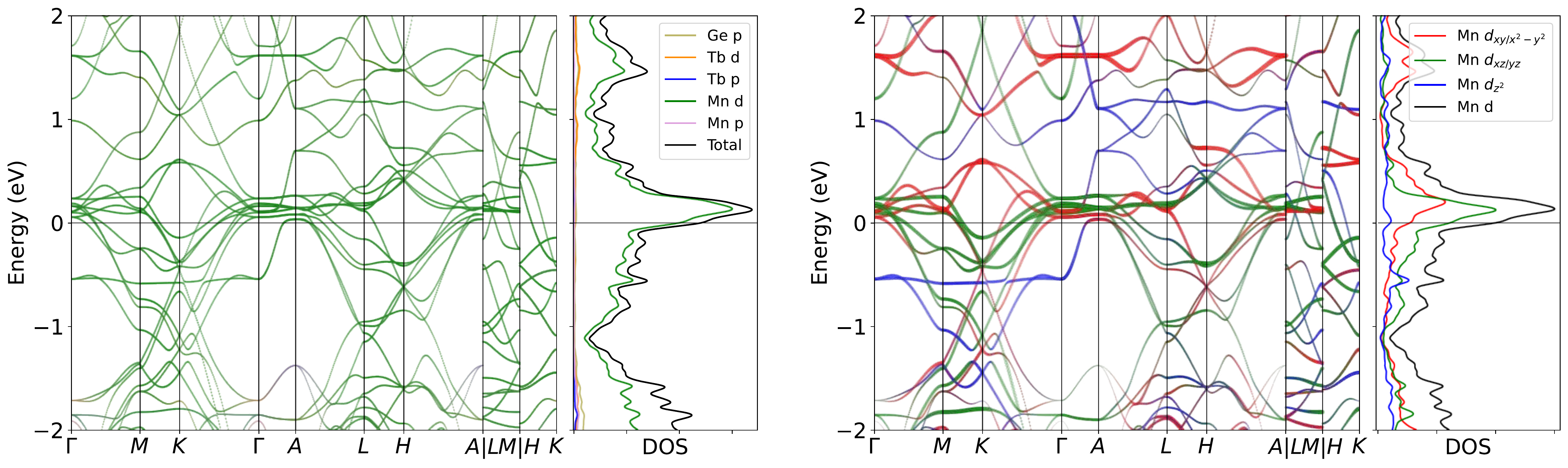}
\caption{Band structure for TbMn$_6$Ge$_6$ without SOC in paramagnetic phase. The projection of Mn $3d$ orbitals is also presented. The band structures clearly reveal the dominating of Mn $3d$ orbitals  near the Fermi level, as well as the pronounced peak.}
\label{fig:TbMn6Ge6band}
\end{figure}

\begin{figure}[H]
\centering
\subfloat[Relaxed phonon at low-T.]{
\label{fig:TbMn6Ge6_relaxed_Low}
\includegraphics[width=0.45\textwidth]{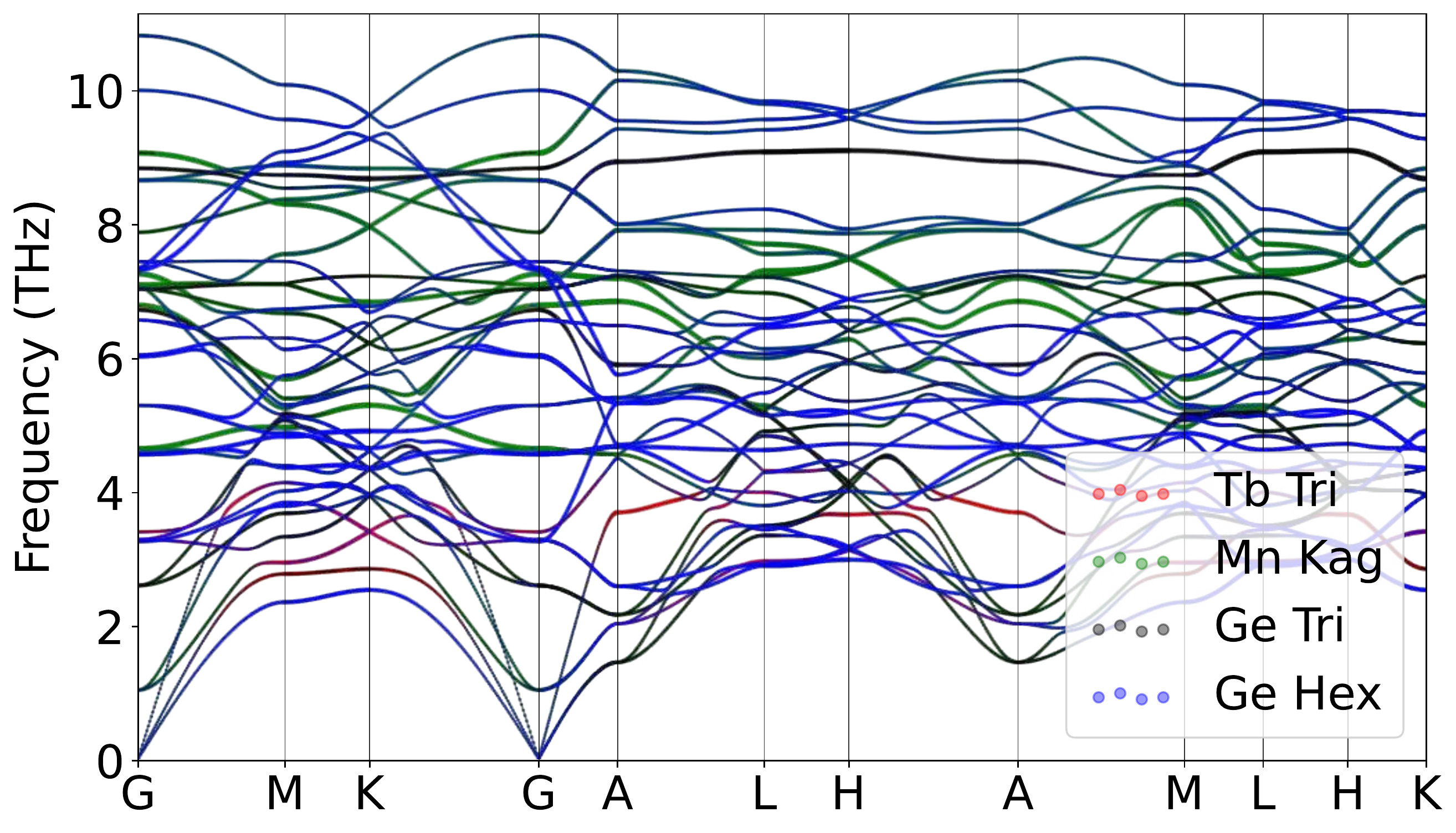}
}
\hspace{5pt}\subfloat[Relaxed phonon at high-T.]{
\label{fig:TbMn6Ge6_relaxed_High}
\includegraphics[width=0.45\textwidth]{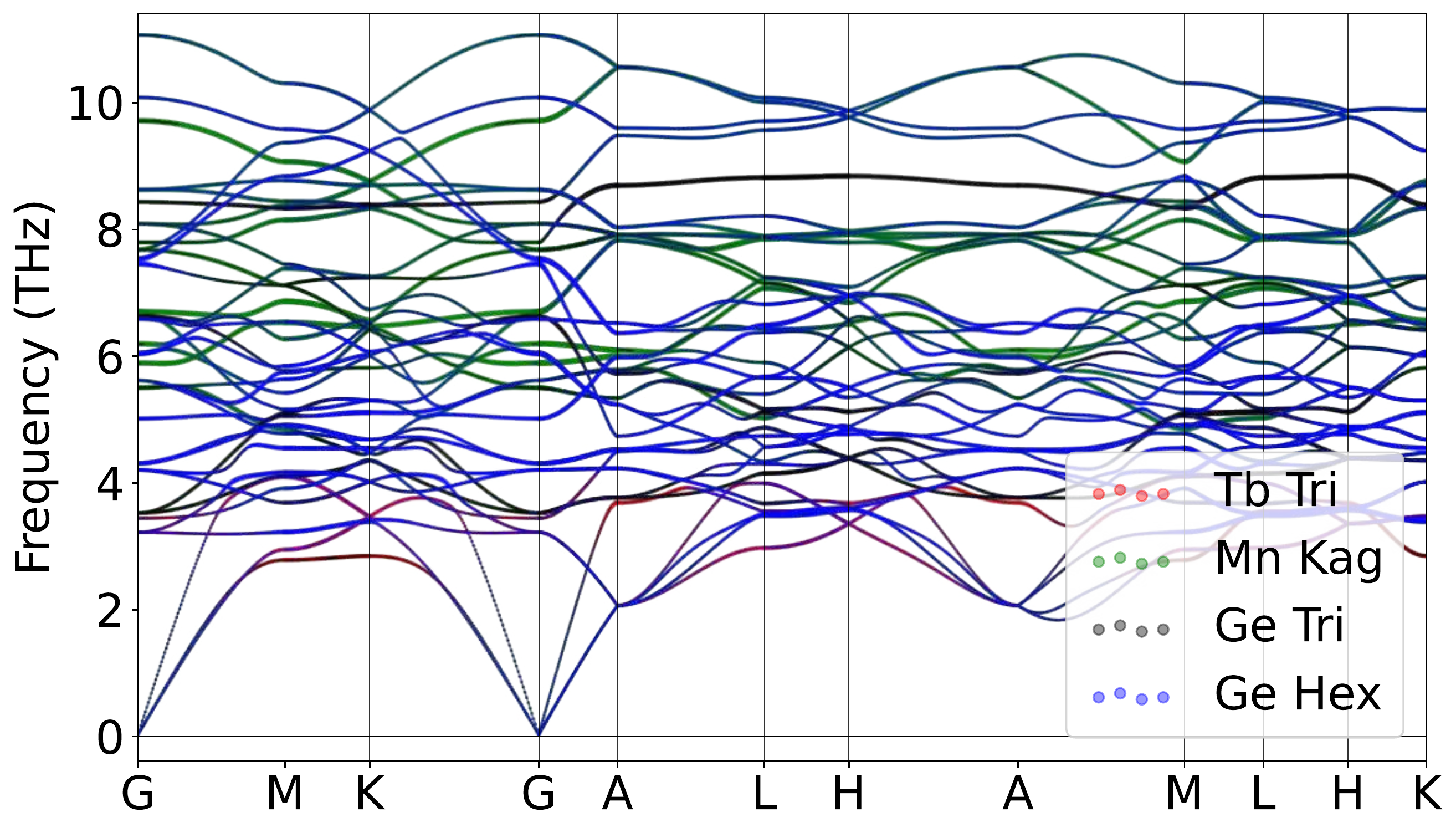}
}
\caption{Atomically projected phonon spectrum for TbMn$_6$Ge$_6$ in paramagnetic phase.}
\label{fig:TbMn6Ge6pho}
\end{figure}

\begin{figure}[htbp]
\centering
\includegraphics[width=0.45\textwidth]{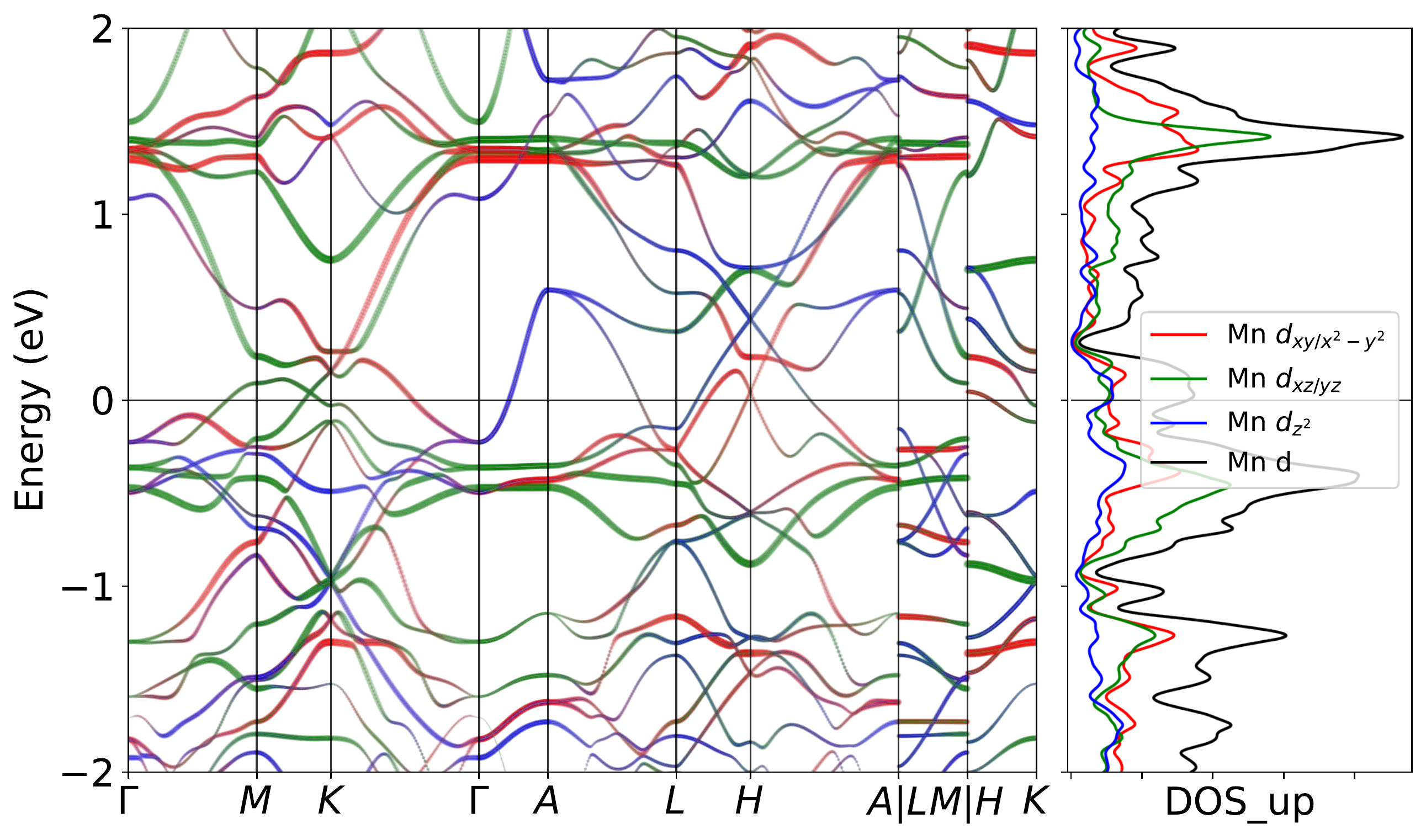}
\includegraphics[width=0.45\textwidth]{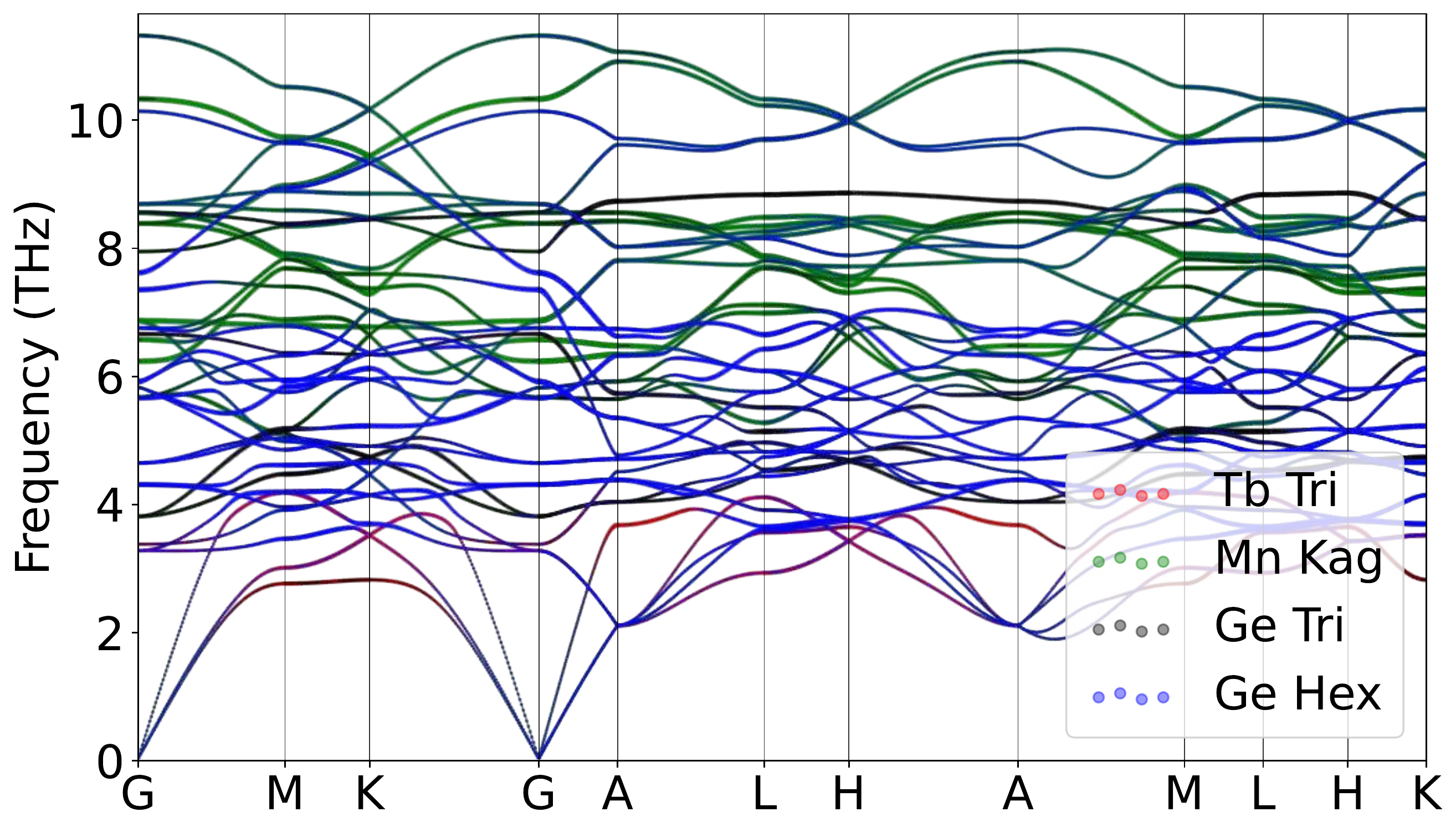}
\caption{Band structure for TbMn$_6$Ge$_6$ with AFM order without SOC for spin (Left)up channel. The projection of Mn $3d$ orbitals is also presented. (Right)Correspondingly, a stable phonon was demonstrated with the collinear magnetic order without Hubbard U at lower temperature regime, weighted by atomic projections.}
\label{fig:TbMn6Ge6mag}
\end{figure}

\begin{figure}[H]
\centering
\label{fig:TbMn6Ge6_relaxed_Low_xz}
\includegraphics[width=0.85\textwidth]{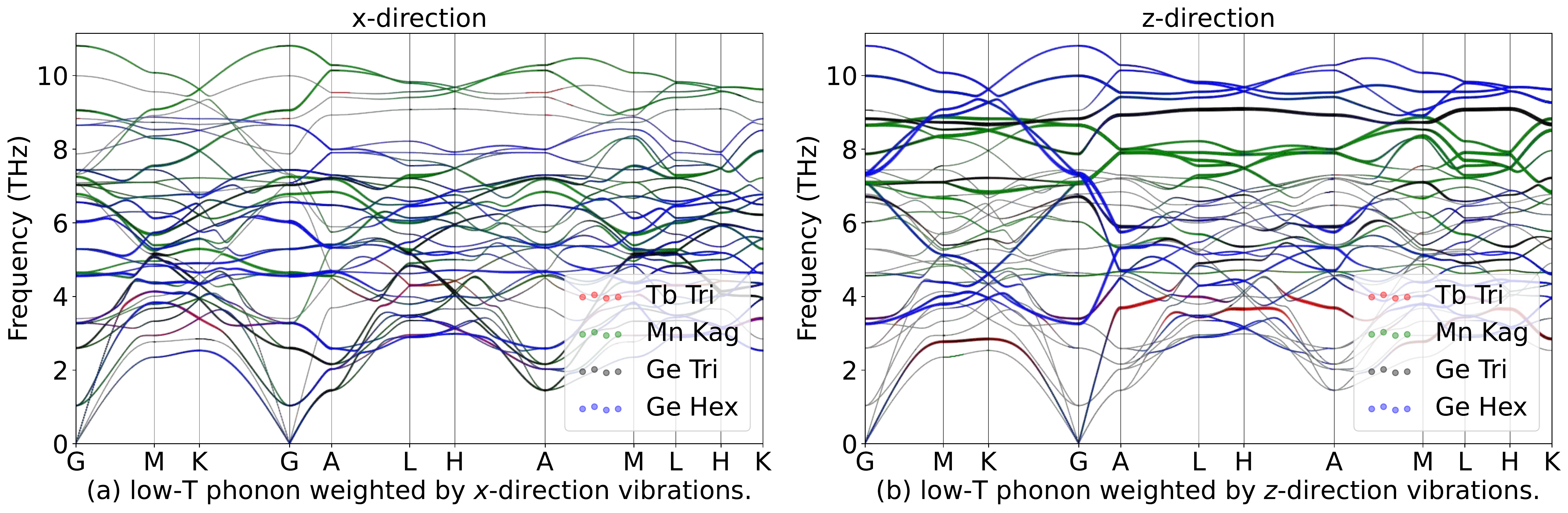}
\hspace{5pt}\label{fig:TbMn6Ge6_relaxed_High_xz}
\includegraphics[width=0.85\textwidth]{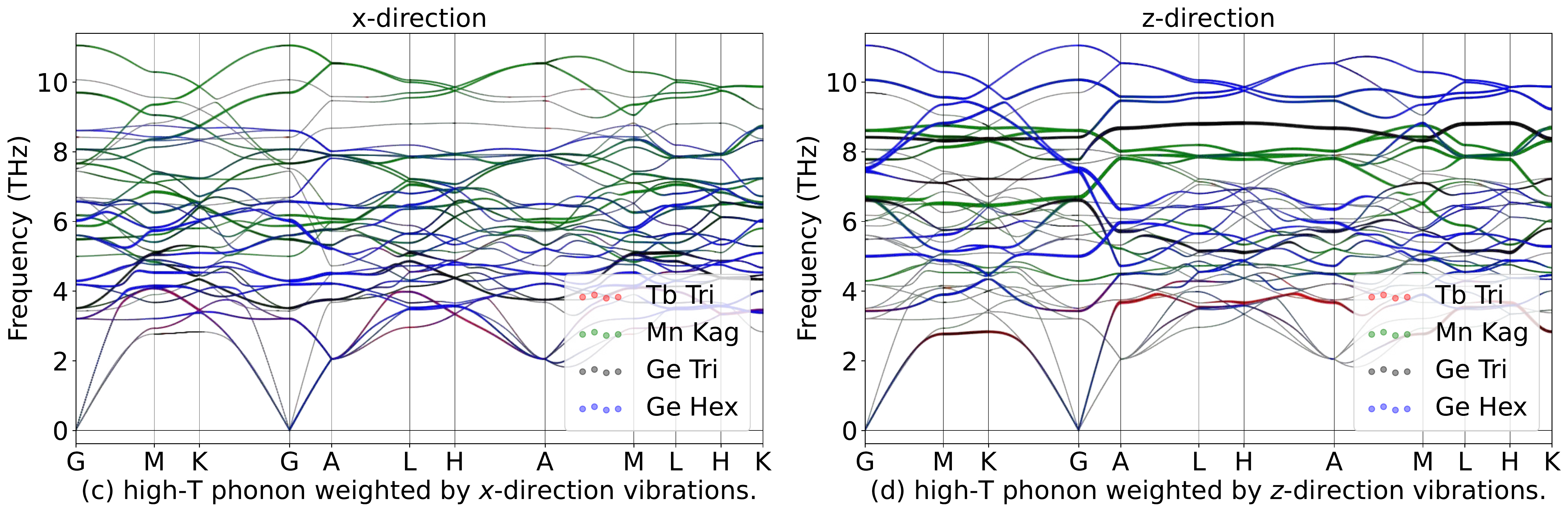}
\caption{Phonon spectrum for TbMn$_6$Ge$_6$in in-plane ($x$) and out-of-plane ($z$) directions in paramagnetic phase.}
\label{fig:TbMn6Ge6phoxyz}
\end{figure}

\begin{table}[htbp]
\global\long\def\arraystretch{1.12}
\captionsetup{width=.95\textwidth}
\caption{IRREPs at high-symmetry points for TbMn$_6$Ge$_6$ phonon spectrum at Low-T.}
\label{tab:191_TbMn6Ge6_Low}
\setlength{\tabcolsep}{1.mm}{
}
\end{table}

\begin{table}[htbp]
\global\long\def\arraystretch{1.12}
\captionsetup{width=.95\textwidth}
\caption{IRREPs at high-symmetry points for TbMn$_6$Ge$_6$ phonon spectrum at High-T.}
\label{tab:191_TbMn6Ge6_High}
\setlength{\tabcolsep}{1.mm}{
%
}
\end{table}
\subsubsection{\label{sms2sec:DyMn6Ge6}\hyperref[smsec:col]{\texorpdfstring{DyMn$_6$Ge$_6$}{}}~{\color{green}  (High-T stable, Low-T stable) }}

\begin{table}[htbp]
\global\long\def\arraystretch{1.12}
\captionsetup{width=.85\textwidth}
\caption{Structure parameters of DyMn$_6$Ge$_6$ after relaxation. It contains 408 electrons. }
\label{tab:DyMn6Ge6}
\setlength{\tabcolsep}{4mm}{
%
}
\end{table}

\begin{figure}[htbp]
\centering
\includegraphics[width=0.85\textwidth]{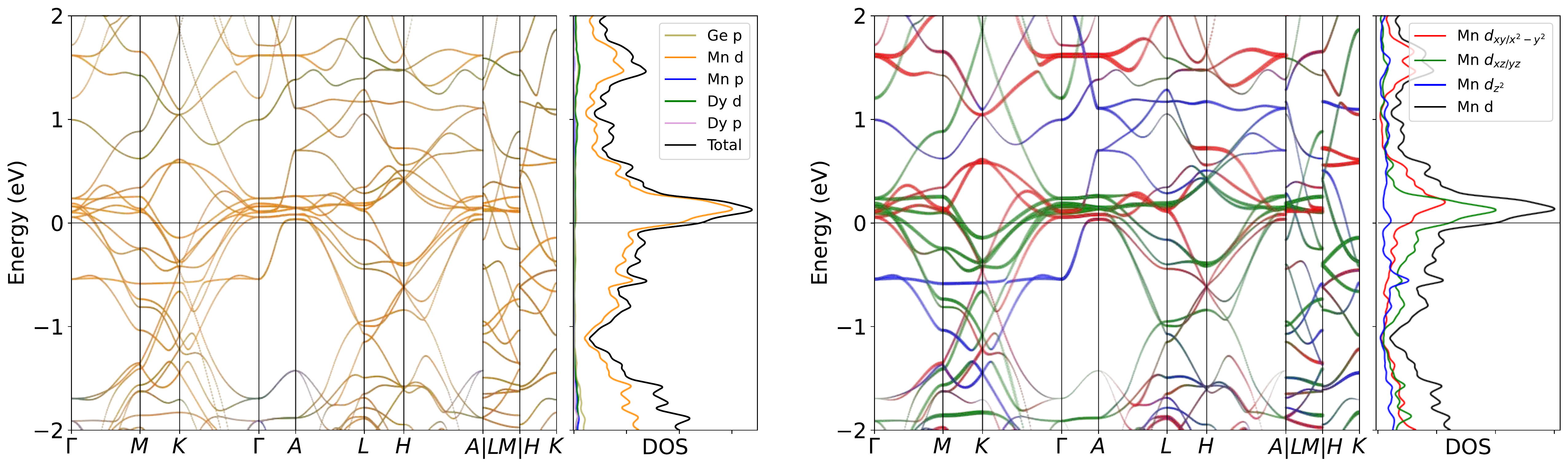}
\caption{Band structure for DyMn$_6$Ge$_6$ without SOC in paramagnetic phase. The projection of Mn $3d$ orbitals is also presented. The band structures clearly reveal the dominating of Mn $3d$ orbitals  near the Fermi level, as well as the pronounced peak.}
\label{fig:DyMn6Ge6band}
\end{figure}

\begin{figure}[H]
\centering
\subfloat[Relaxed phonon at low-T.]{
\label{fig:DyMn6Ge6_relaxed_Low}
\includegraphics[width=0.45\textwidth]{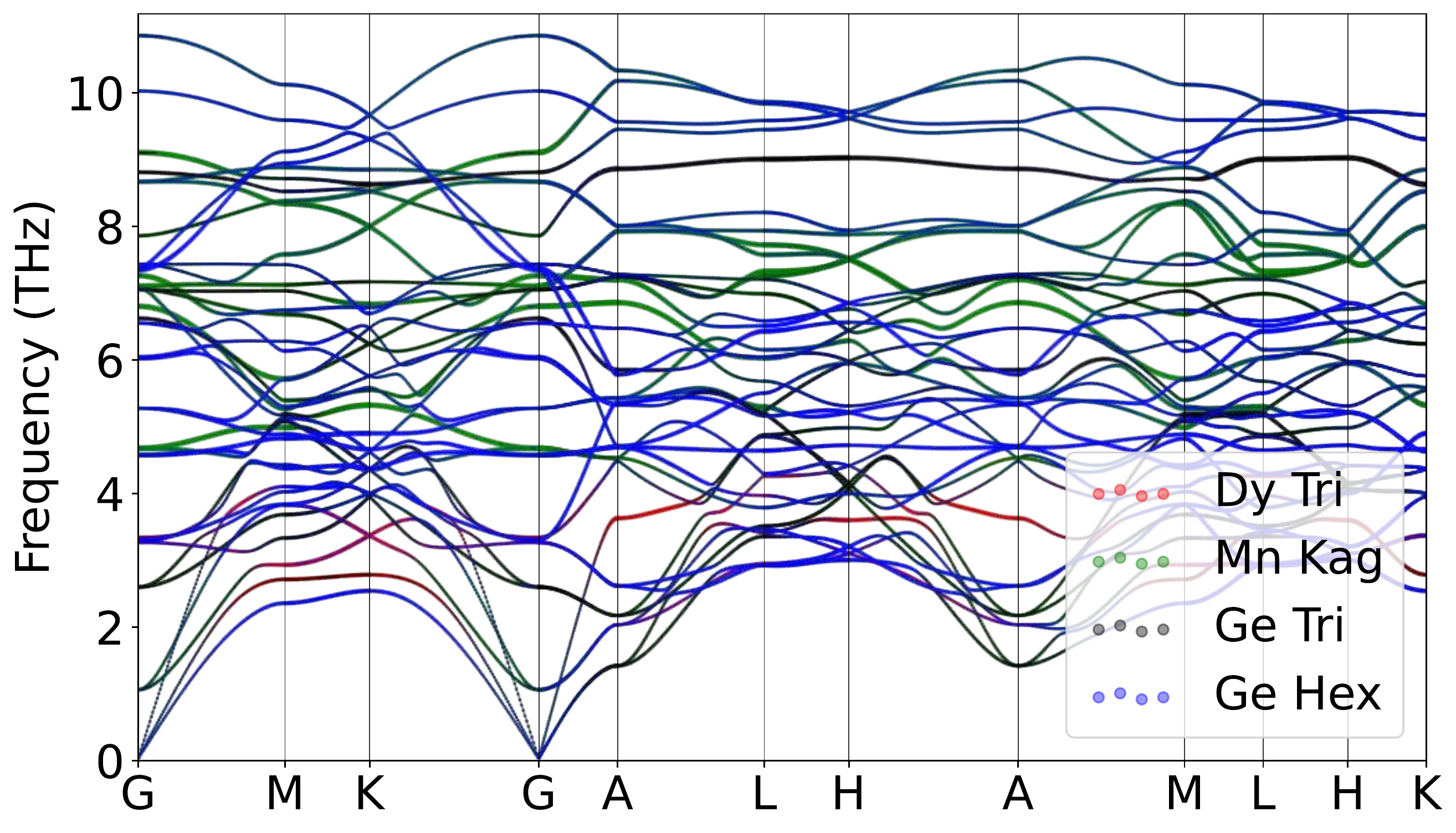}
}
\hspace{5pt}\subfloat[Relaxed phonon at high-T.]{
\label{fig:DyMn6Ge6_relaxed_High}
\includegraphics[width=0.45\textwidth]{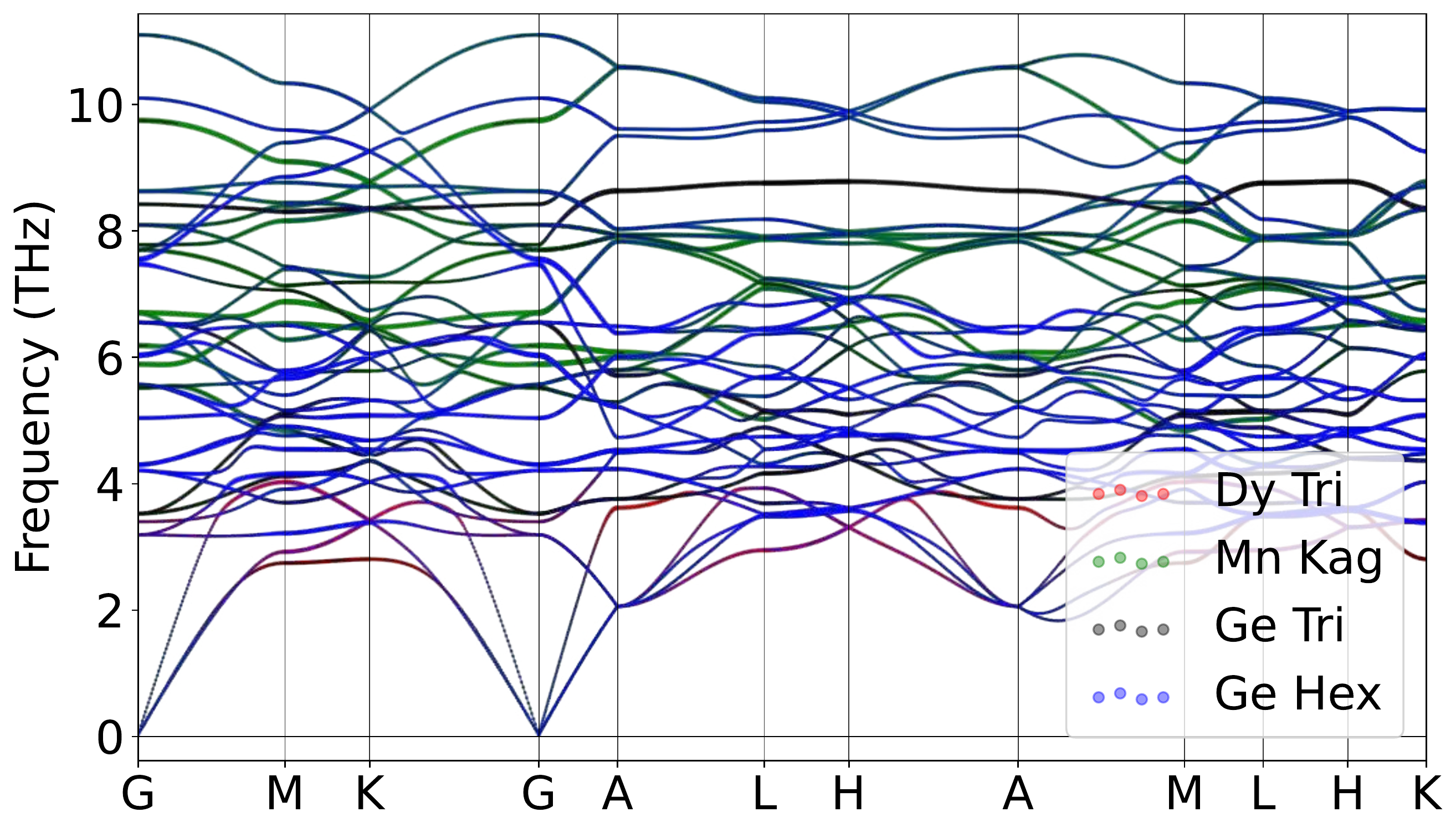}
}
\caption{Atomically projected phonon spectrum for DyMn$_6$Ge$_6$ in paramagnetic phase.}
\label{fig:DyMn6Ge6pho}
\end{figure}

\begin{figure}[htbp]
\centering
\includegraphics[width=0.45\textwidth]{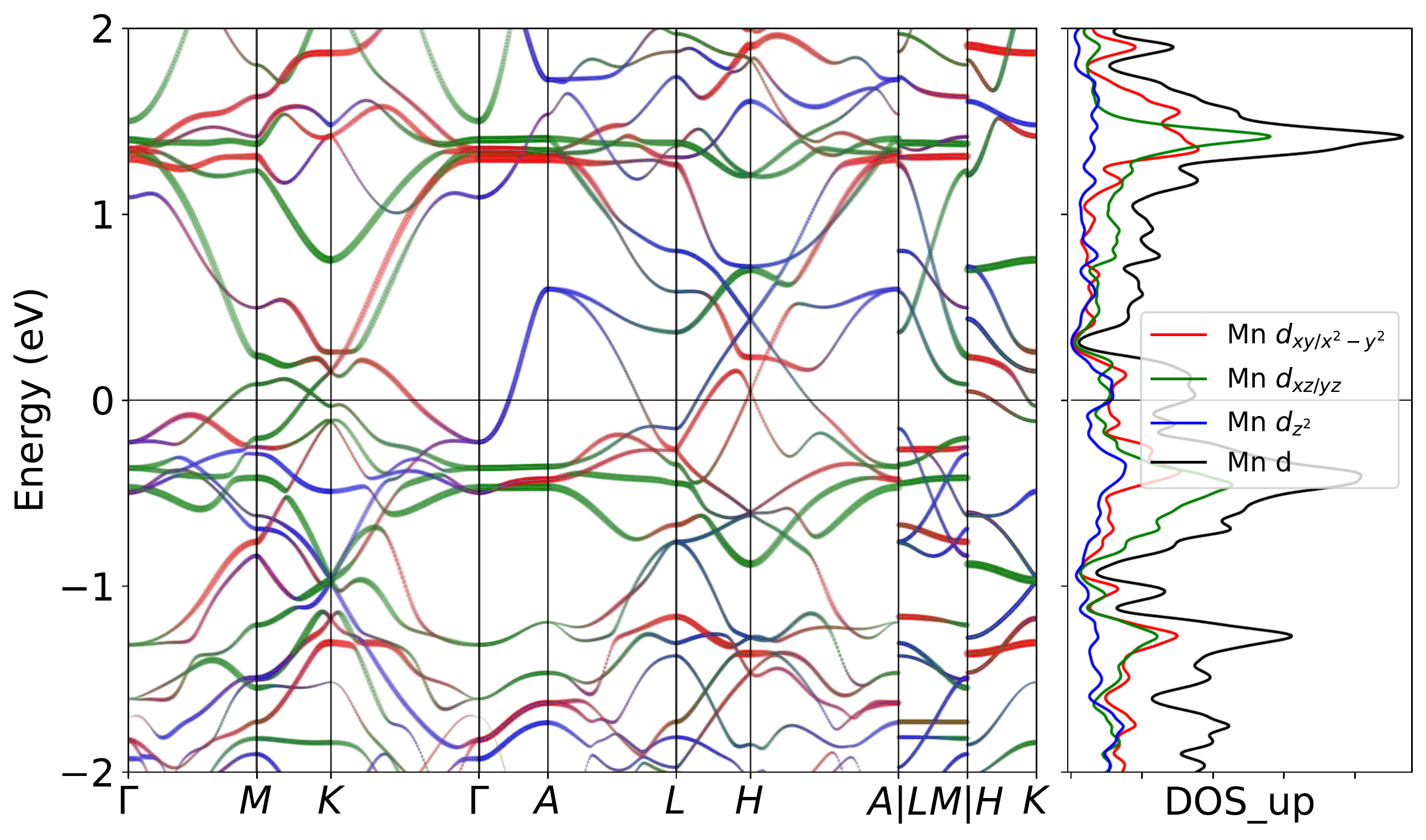}
\includegraphics[width=0.45\textwidth]{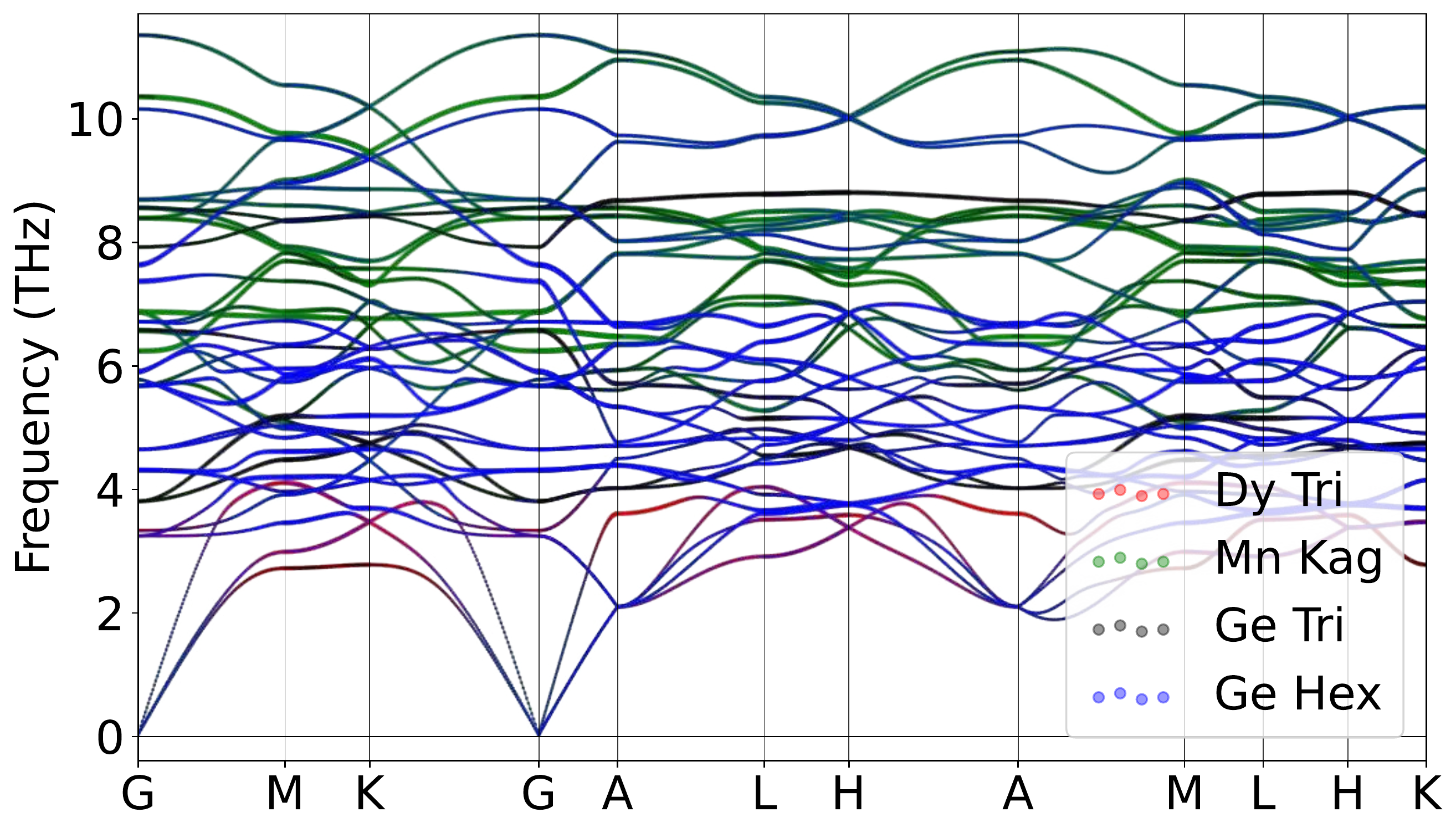}
\caption{Band structure for DyMn$_6$Ge$_6$ with AFM order without SOC for spin (Left)up channel. The projection of Mn $3d$ orbitals is also presented. (Right)Correspondingly, a stable phonon was demonstrated with the collinear magnetic order without Hubbard U at lower temperature regime, weighted by atomic projections.}
\label{fig:DyMn6Ge6mag}
\end{figure}

\begin{figure}[H]
\centering
\label{fig:DyMn6Ge6_relaxed_Low_xz}
\includegraphics[width=0.85\textwidth]{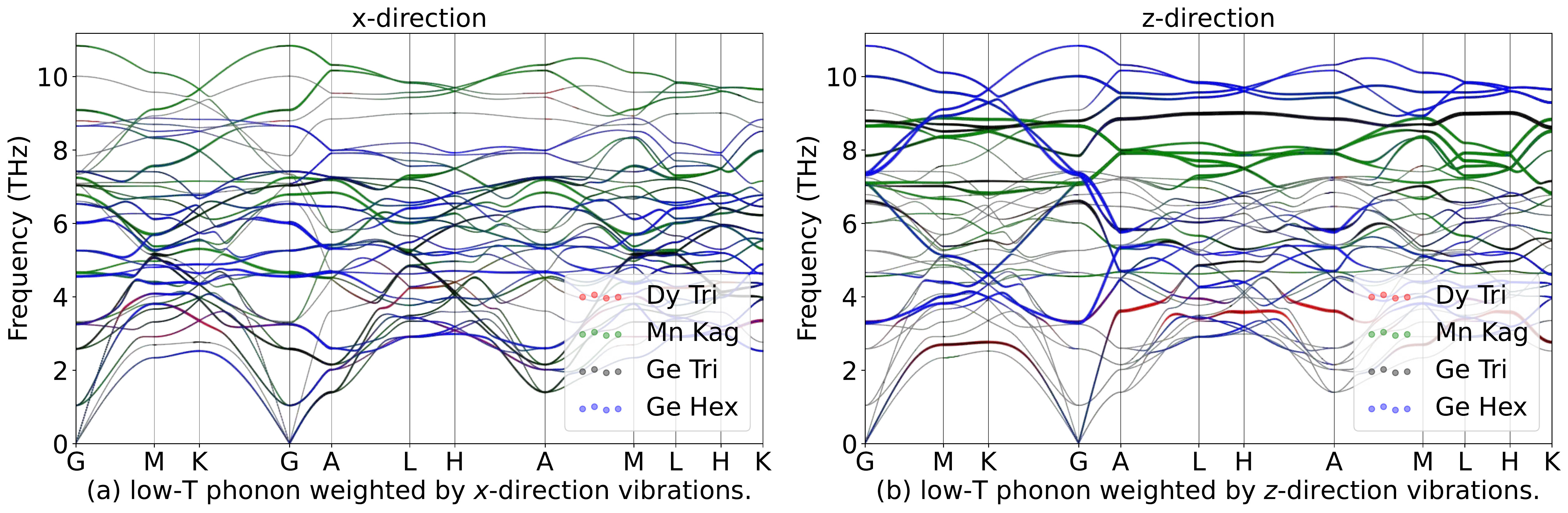}
\hspace{5pt}\label{fig:DyMn6Ge6_relaxed_High_xz}
\includegraphics[width=0.85\textwidth]{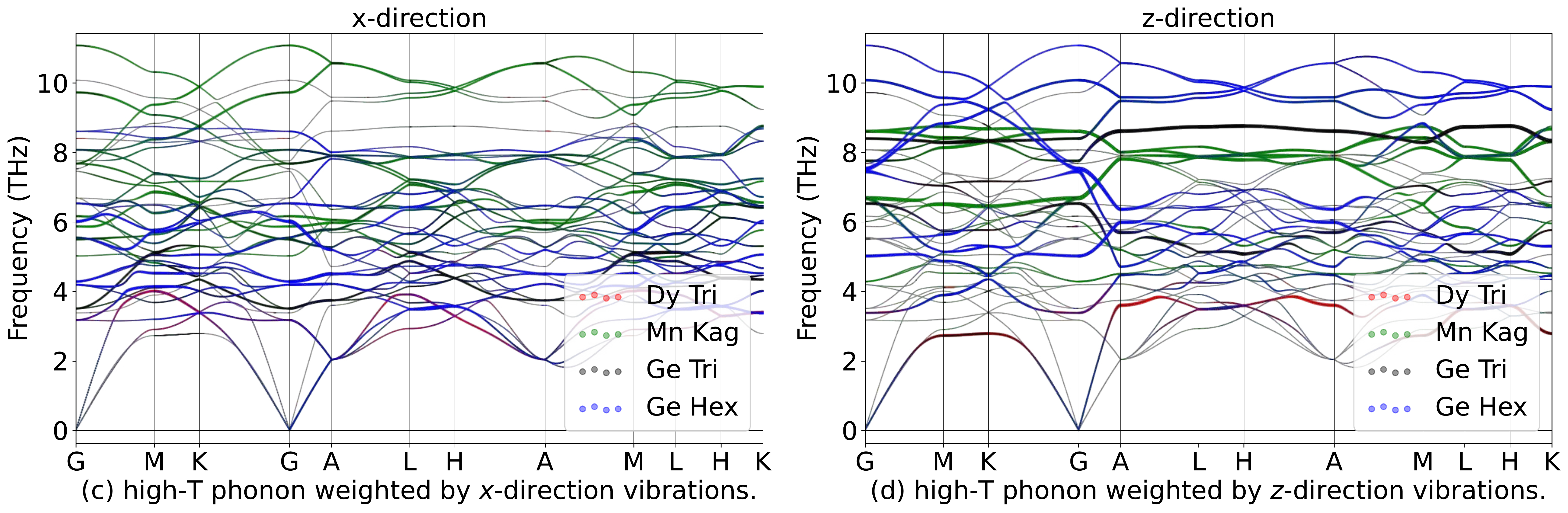}
\caption{Phonon spectrum for DyMn$_6$Ge$_6$in in-plane ($x$) and out-of-plane ($z$) directions in paramagnetic phase.}
\label{fig:DyMn6Ge6phoxyz}
\end{figure}

\begin{table}[htbp]
\global\long\def\arraystretch{1.12}
\captionsetup{width=.95\textwidth}
\caption{IRREPs at high-symmetry points for DyMn$_6$Ge$_6$ phonon spectrum at Low-T.}
\label{tab:191_DyMn6Ge6_Low}
\setlength{\tabcolsep}{1.mm}{
}
\end{table}

\begin{table}[htbp]
\global\long\def\arraystretch{1.12}
\captionsetup{width=.95\textwidth}
\caption{IRREPs at high-symmetry points for DyMn$_6$Ge$_6$ phonon spectrum at High-T.}
\label{tab:191_DyMn6Ge6_High}
\setlength{\tabcolsep}{1.mm}{
%
}
\end{table}
\subsubsection{\label{sms2sec:HoMn6Ge6}\hyperref[smsec:col]{\texorpdfstring{HoMn$_6$Ge$_6$}{}}~{\color{green}  (High-T stable, Low-T stable) }}

\begin{table}[htbp]
\global\long\def\arraystretch{1.12}
\captionsetup{width=.85\textwidth}
\caption{Structure parameters of HoMn$_6$Ge$_6$ after relaxation. It contains 409 electrons. }
\label{tab:HoMn6Ge6}
\setlength{\tabcolsep}{4mm}{
%
}
\end{table}

\begin{figure}[htbp]
\centering
\includegraphics[width=0.85\textwidth]{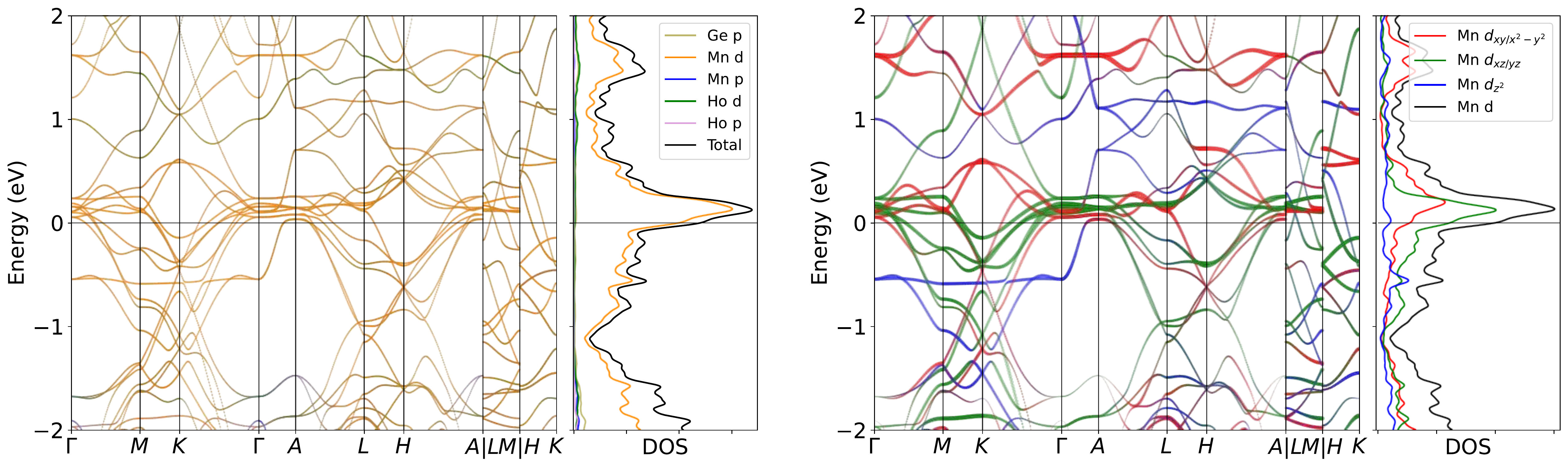}
\caption{Band structure for HoMn$_6$Ge$_6$ without SOC in paramagnetic phase. The projection of Mn $3d$ orbitals is also presented. The band structures clearly reveal the dominating of Mn $3d$ orbitals  near the Fermi level, as well as the pronounced peak.}
\label{fig:HoMn6Ge6band}
\end{figure}

\begin{figure}[H]
\centering
\subfloat[Relaxed phonon at low-T.]{
\label{fig:HoMn6Ge6_relaxed_Low}
\includegraphics[width=0.45\textwidth]{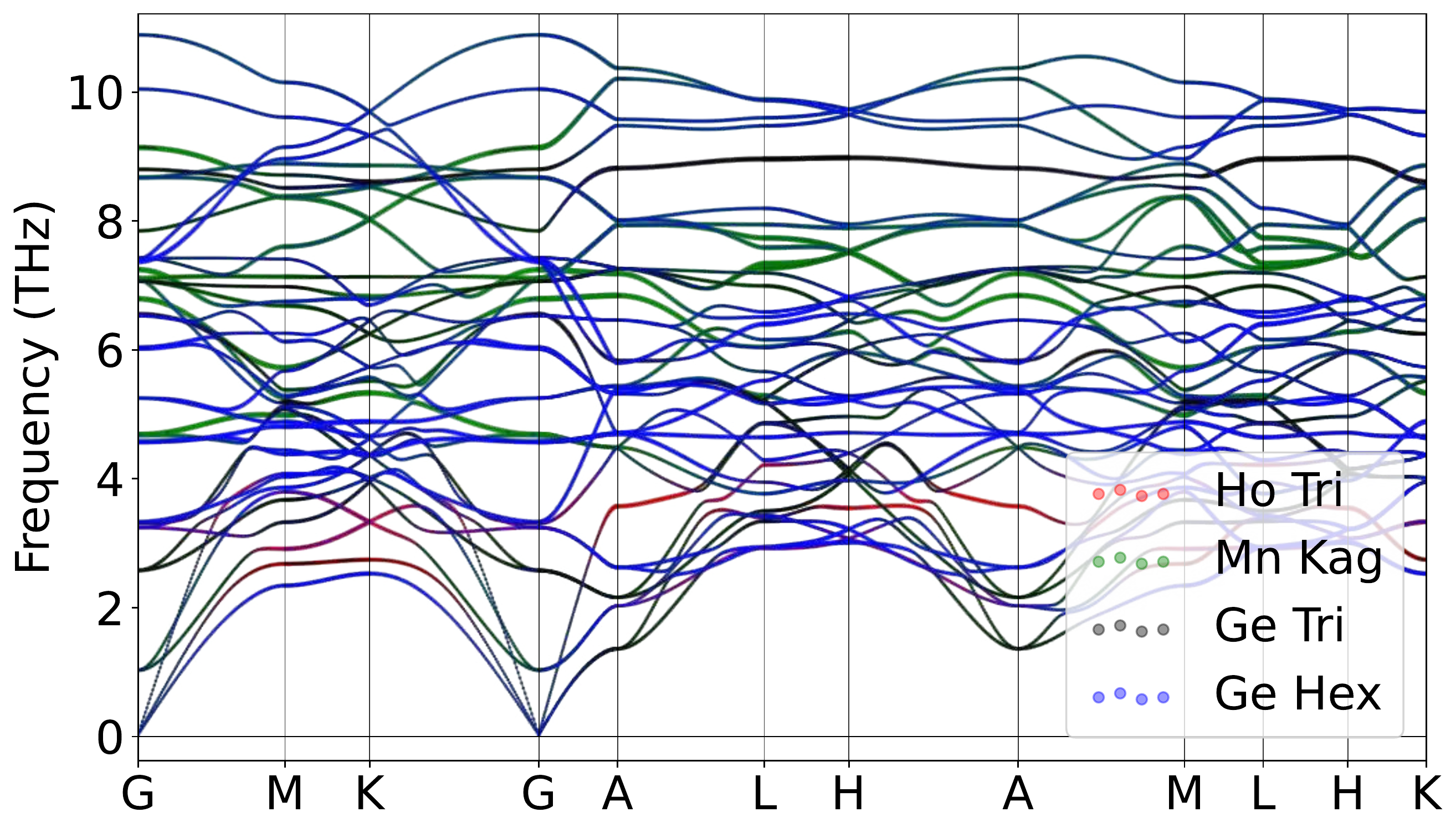}
}
\hspace{5pt}\subfloat[Relaxed phonon at high-T.]{
\label{fig:HoMn6Ge6_relaxed_High}
\includegraphics[width=0.45\textwidth]{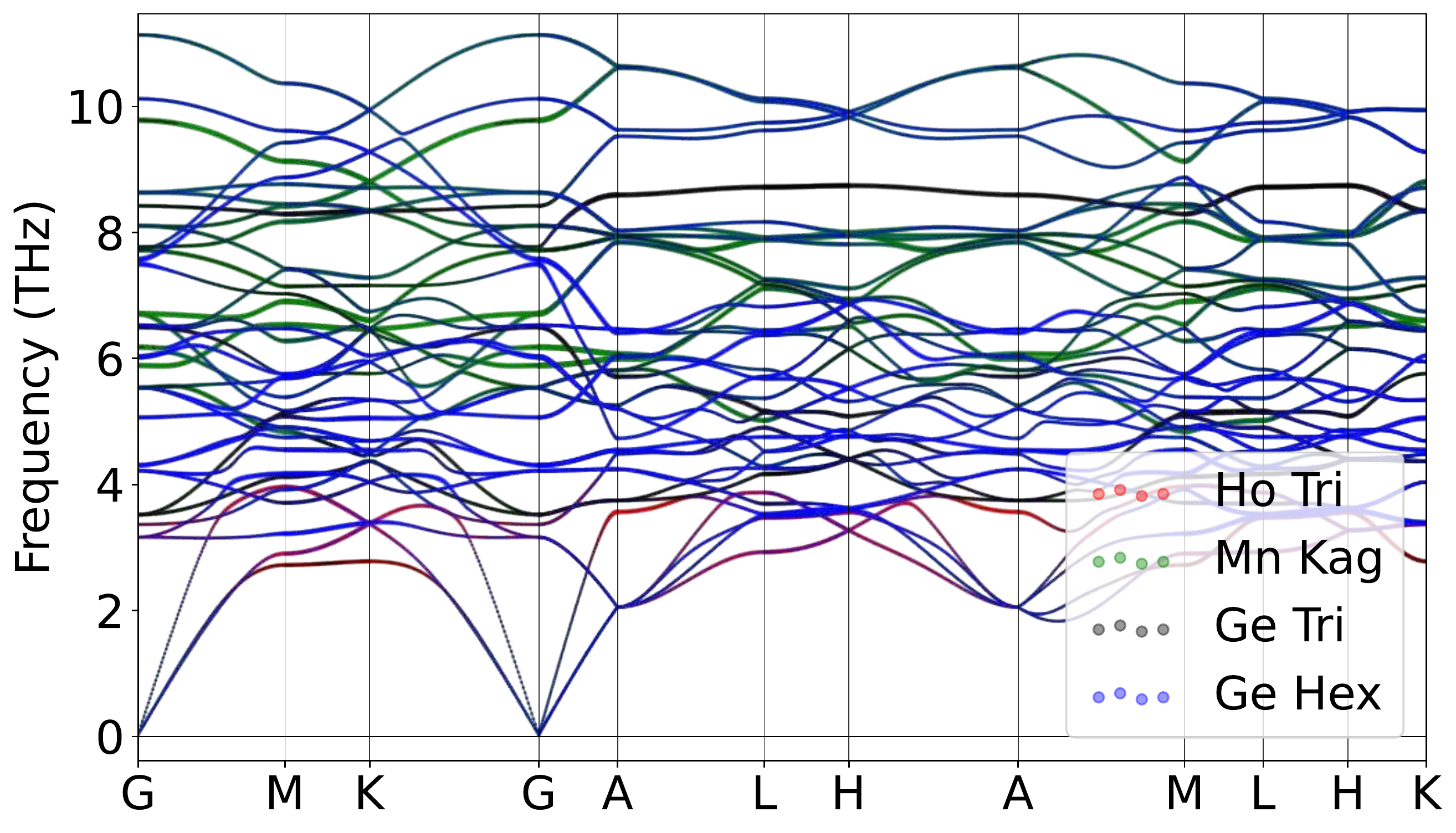}
}
\caption{Atomically projected phonon spectrum for HoMn$_6$Ge$_6$ in paramagnetic phase.}
\label{fig:HoMn6Ge6pho}
\end{figure}

\begin{figure}[htbp]
\centering
\includegraphics[width=0.45\textwidth]{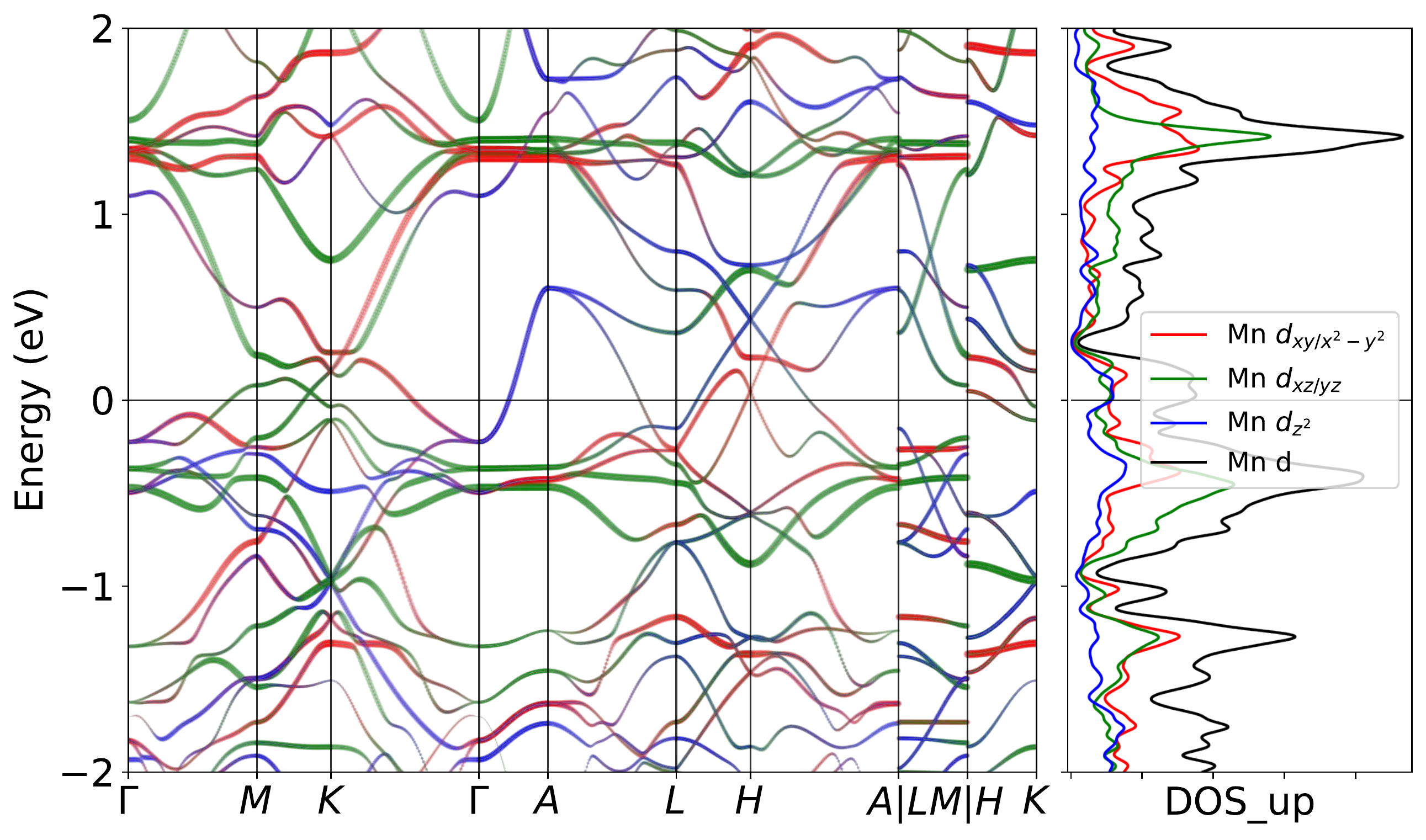}
\includegraphics[width=0.45\textwidth]{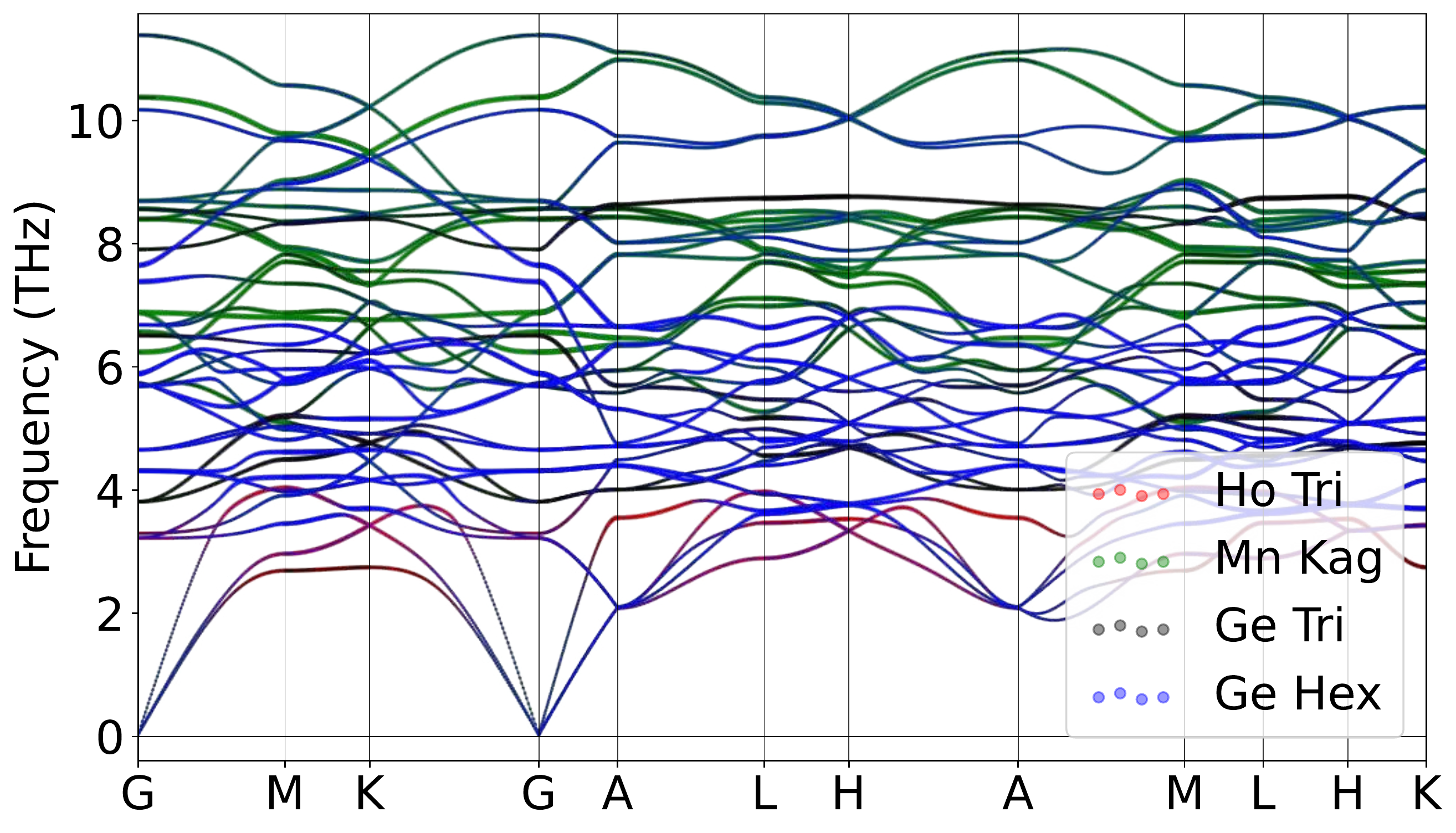}
\caption{Band structure for HoMn$_6$Ge$_6$ with AFM order without SOC for spin (Left)up channel. The projection of Mn $3d$ orbitals is also presented. (Right)Correspondingly, a stable phonon was demonstrated with the collinear magnetic order without Hubbard U at lower temperature regime, weighted by atomic projections.}
\label{fig:HoMn6Ge6mag}
\end{figure}

\begin{figure}[H]
\centering
\label{fig:HoMn6Ge6_relaxed_Low_xz}
\includegraphics[width=0.85\textwidth]{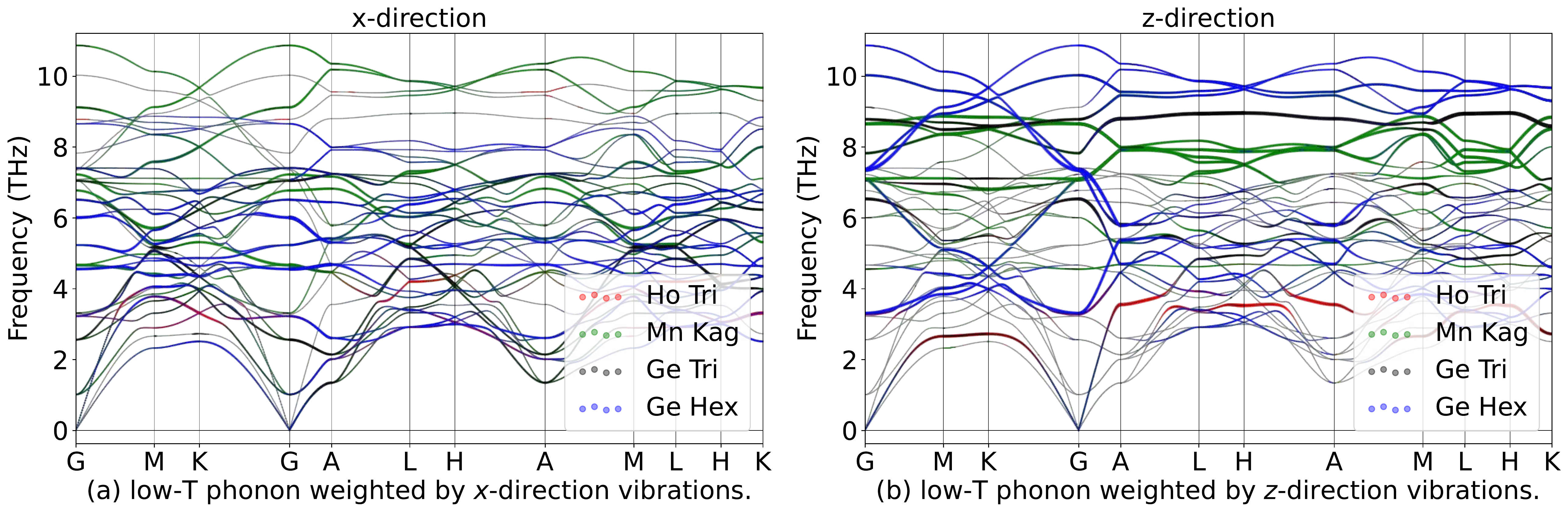}
\hspace{5pt}\label{fig:HoMn6Ge6_relaxed_High_xz}
\includegraphics[width=0.85\textwidth]{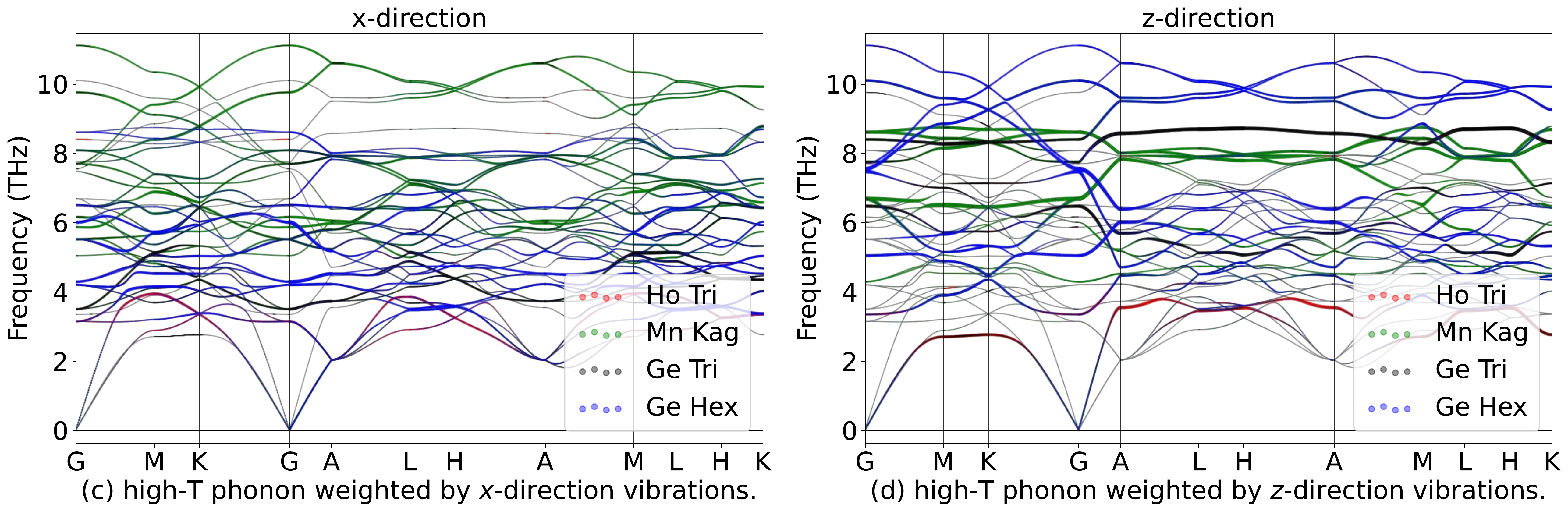}
\caption{Phonon spectrum for HoMn$_6$Ge$_6$in in-plane ($x$) and out-of-plane ($z$) directions in paramagnetic phase.}
\label{fig:HoMn6Ge6phoxyz}
\end{figure}

\begin{table}[htbp]
\global\long\def\arraystretch{1.12}
\captionsetup{width=.95\textwidth}
\caption{IRREPs at high-symmetry points for HoMn$_6$Ge$_6$ phonon spectrum at Low-T.}
\label{tab:191_HoMn6Ge6_Low}
\setlength{\tabcolsep}{1.mm}{
}
\end{table}

\begin{table}[htbp]
\global\long\def\arraystretch{1.12}
\captionsetup{width=.95\textwidth}
\caption{IRREPs at high-symmetry points for HoMn$_6$Ge$_6$ phonon spectrum at High-T.}
\label{tab:191_HoMn6Ge6_High}
\setlength{\tabcolsep}{1.mm}{
%
}
\end{table}
\subsubsection{\label{sms2sec:ErMn6Ge6}\hyperref[smsec:col]{\texorpdfstring{ErMn$_6$Ge$_6$}{}}~{\color{green}  (High-T stable, Low-T stable) }}

\begin{table}[htbp]
\global\long\def\arraystretch{1.12}
\captionsetup{width=.85\textwidth}
\caption{Structure parameters of ErMn$_6$Ge$_6$ after relaxation. It contains 410 electrons. }
\label{tab:ErMn6Ge6}
\setlength{\tabcolsep}{4mm}{
%
}
\end{table}

\begin{figure}[htbp]
\centering
\includegraphics[width=0.85\textwidth]{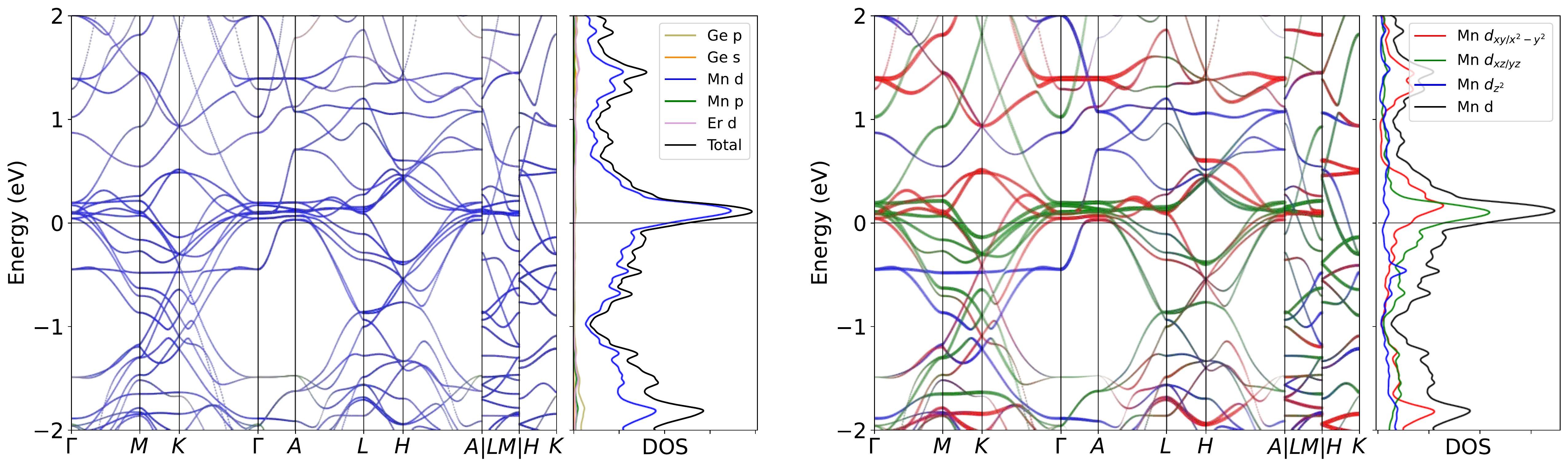}
\caption{Band structure for ErMn$_6$Ge$_6$ without SOC in paramagnetic phase. The projection of Mn $3d$ orbitals is also presented. The band structures clearly reveal the dominating of Mn $3d$ orbitals  near the Fermi level, as well as the pronounced peak.}
\label{fig:ErMn6Ge6band}
\end{figure}

\begin{figure}[H]
\centering
\subfloat[Relaxed phonon at low-T.]{
\label{fig:ErMn6Ge6_relaxed_Low}
\includegraphics[width=0.45\textwidth]{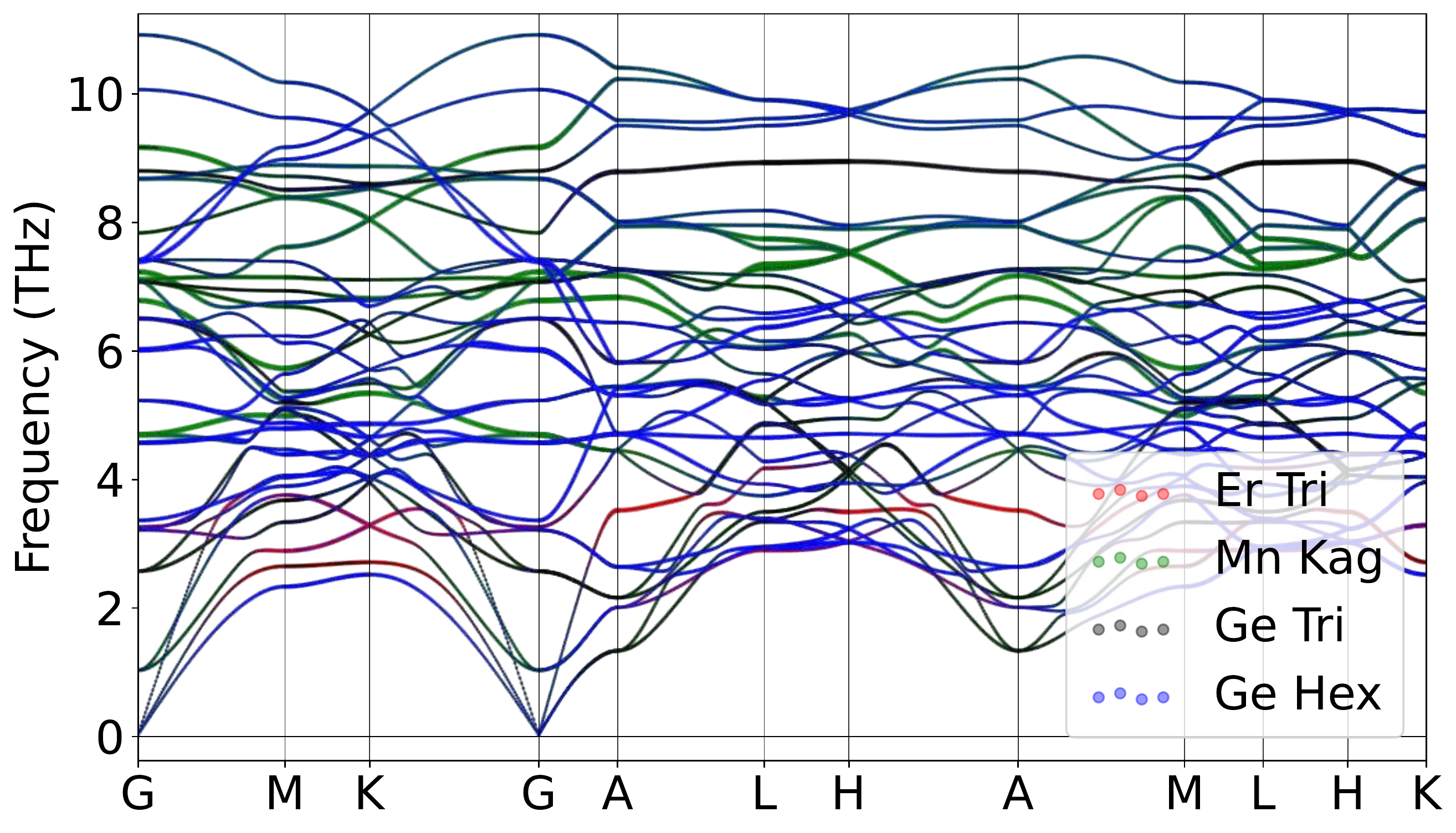}
}
\hspace{5pt}\subfloat[Relaxed phonon at high-T.]{
\label{fig:ErMn6Ge6_relaxed_High}
\includegraphics[width=0.45\textwidth]{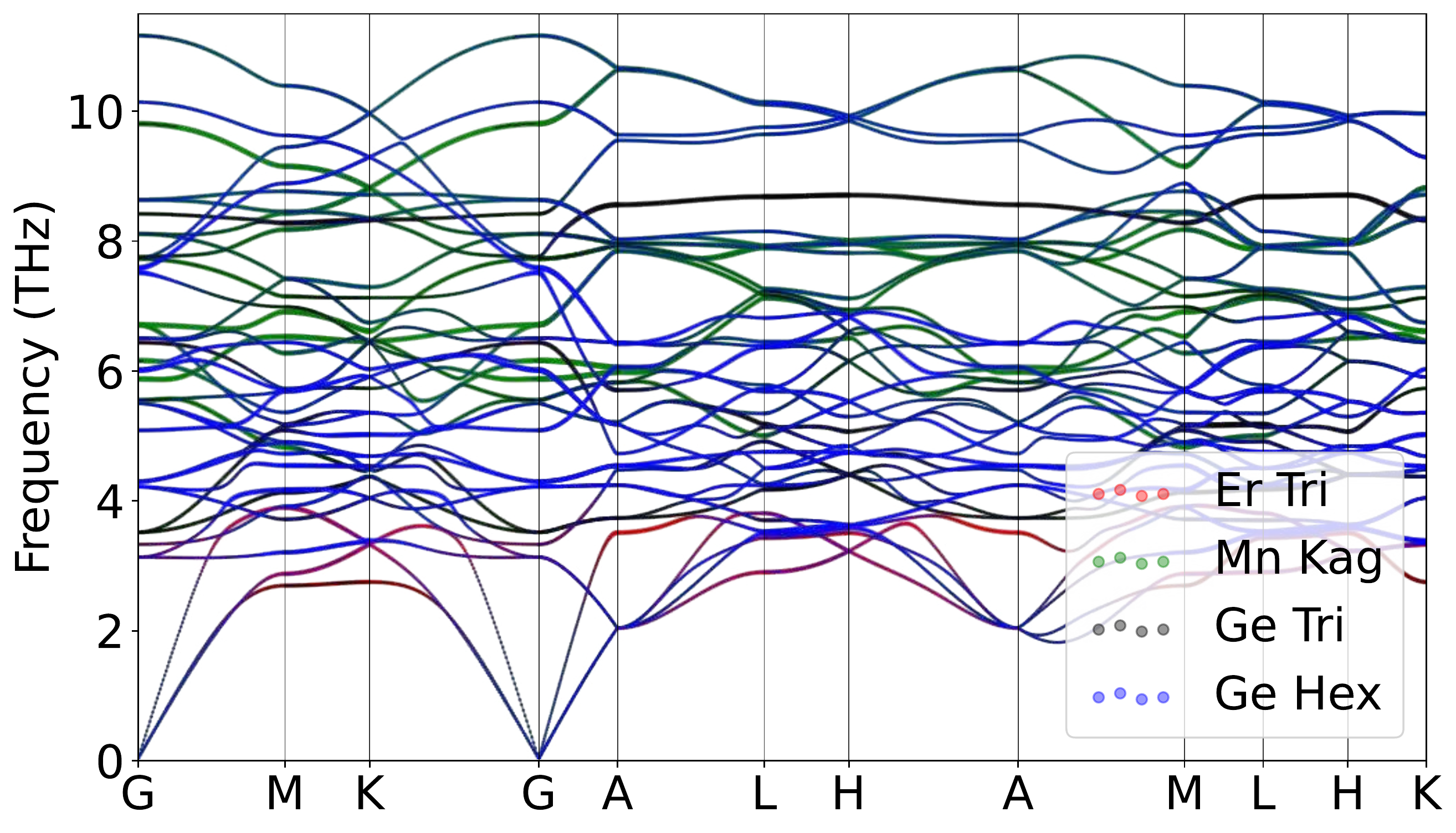}
}
\caption{Atomically projected phonon spectrum for ErMn$_6$Ge$_6$ in paramagnetic phase.}
\label{fig:ErMn6Ge6pho}
\end{figure}

\begin{figure}[htbp]
\centering
\includegraphics[width=0.45\textwidth]{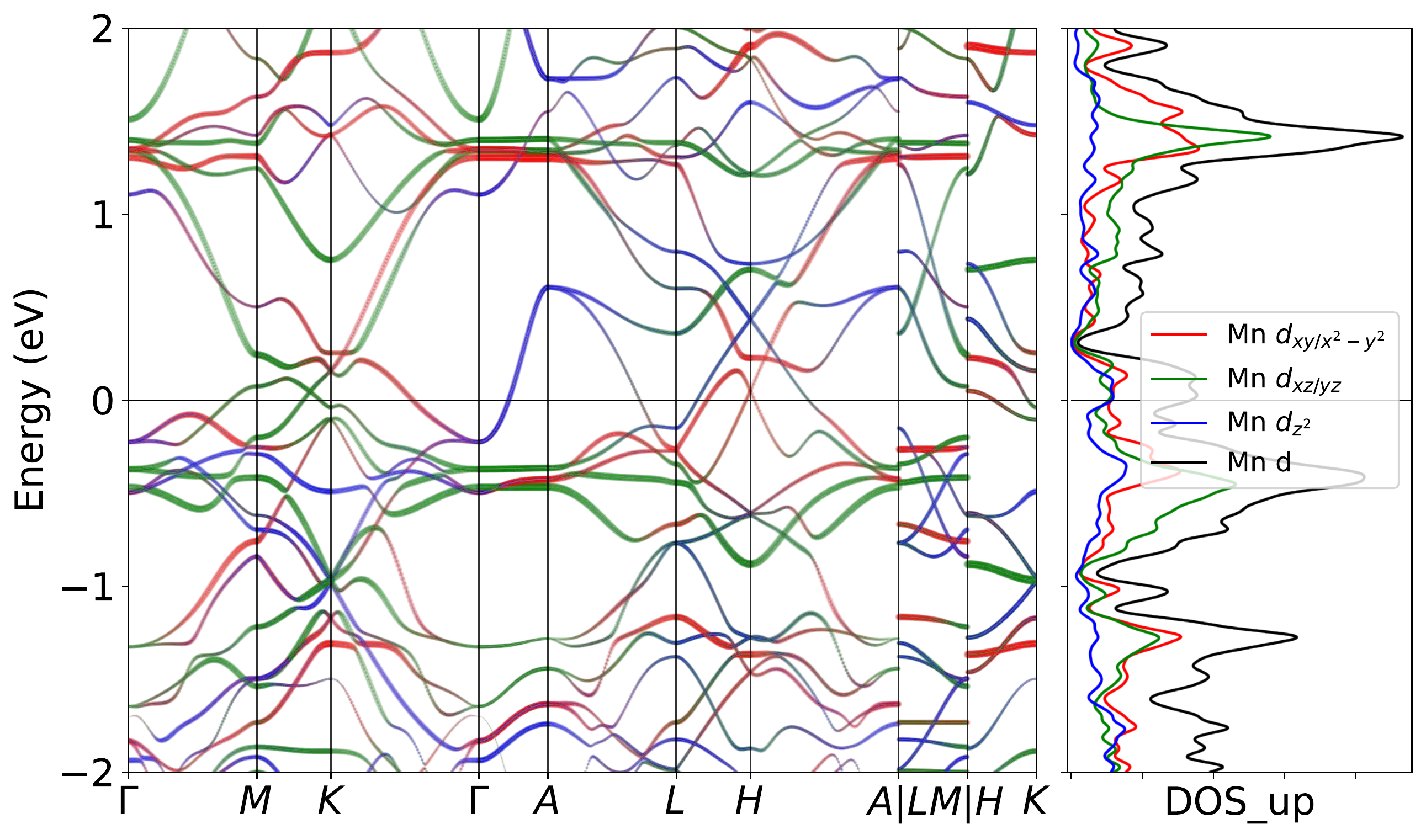}
\includegraphics[width=0.45\textwidth]{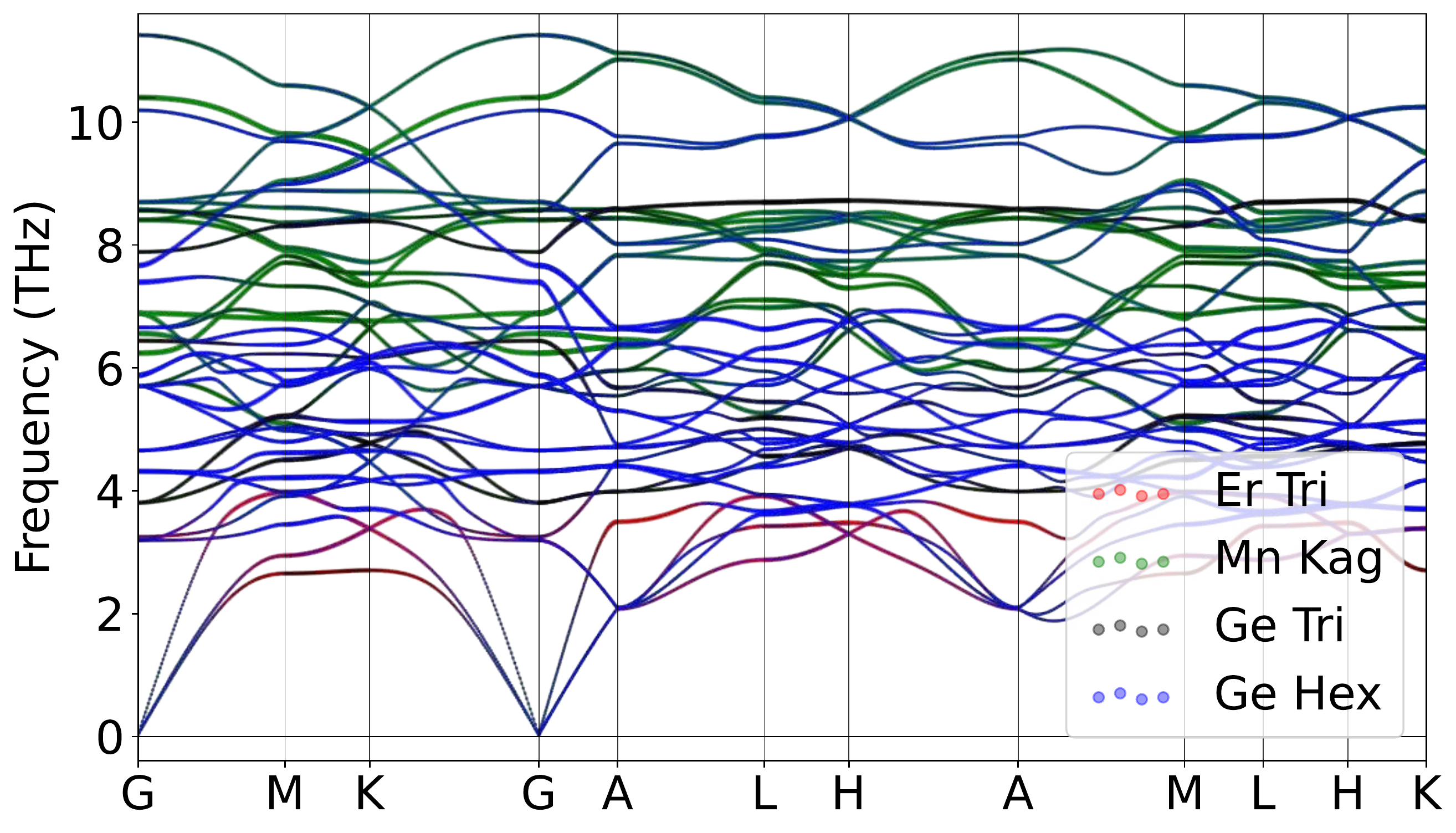}
\caption{Band structure for ErMn$_6$Ge$_6$ with AFM order without SOC for spin (Left)up channel. The projection of Mn $3d$ orbitals is also presented. (Right)Correspondingly, a stable phonon was demonstrated with the collinear magnetic order without Hubbard U at lower temperature regime, weighted by atomic projections.}
\label{fig:ErMn6Ge6mag}
\end{figure}

\begin{figure}[H]
\centering
\label{fig:ErMn6Ge6_relaxed_Low_xz}
\includegraphics[width=0.85\textwidth]{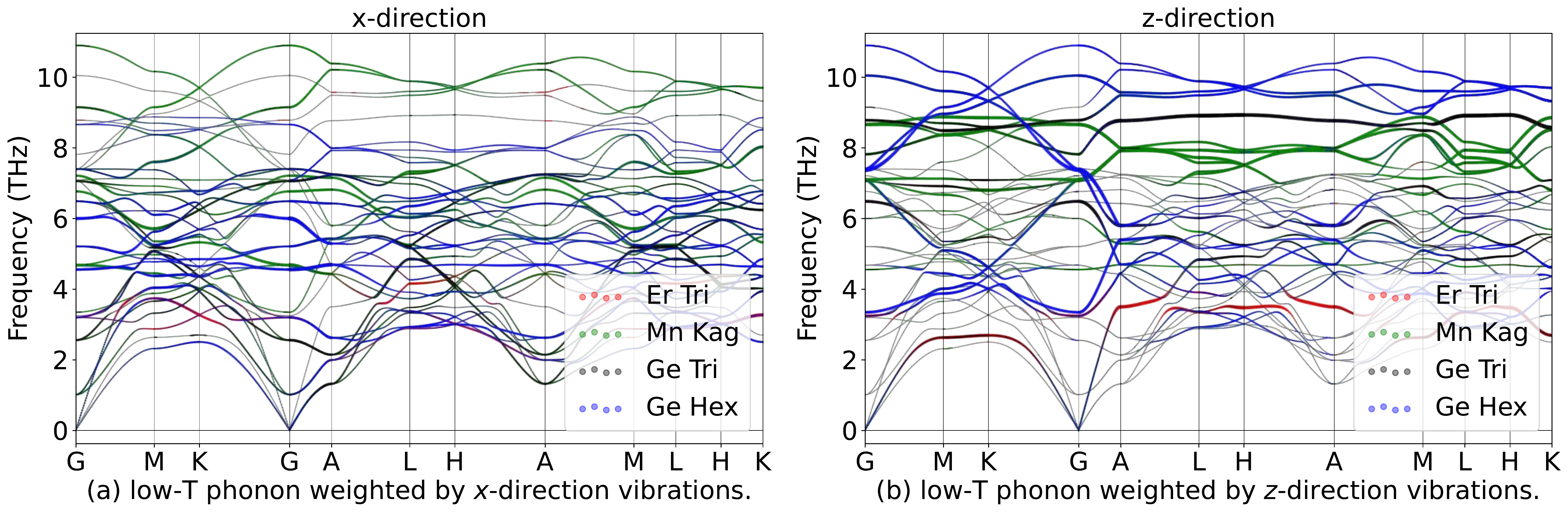}
\hspace{5pt}\label{fig:ErMn6Ge6_relaxed_High_xz}
\includegraphics[width=0.85\textwidth]{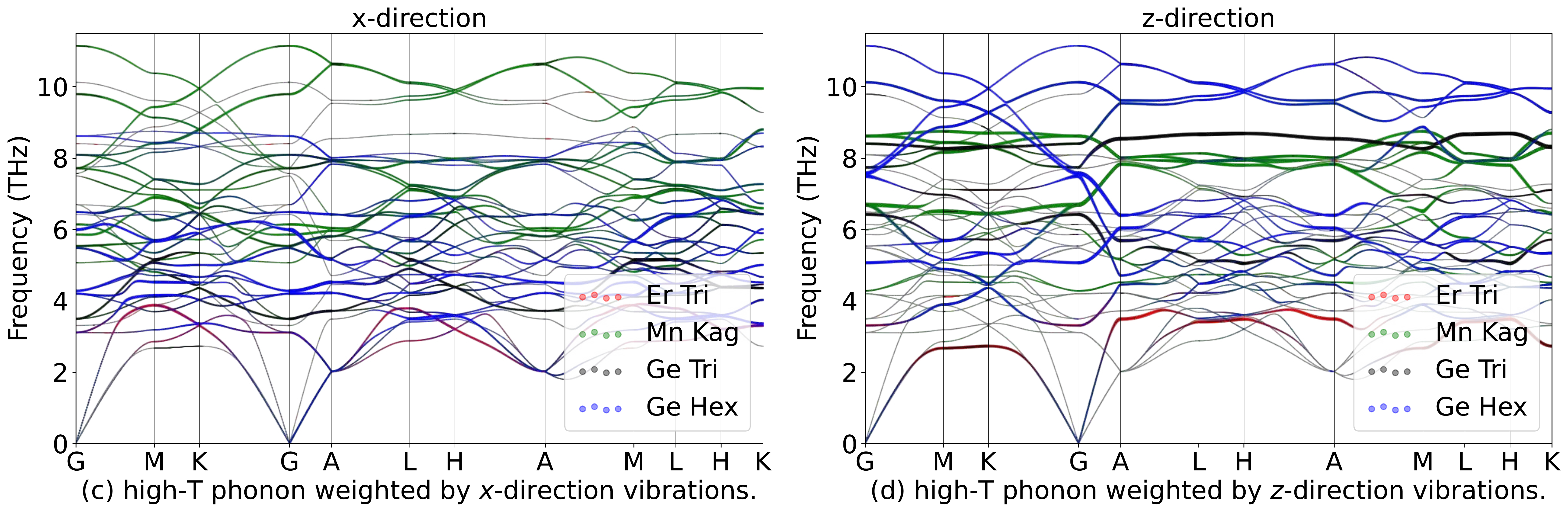}
\caption{Phonon spectrum for ErMn$_6$Ge$_6$in in-plane ($x$) and out-of-plane ($z$) directions in paramagnetic phase.}
\label{fig:ErMn6Ge6phoxyz}
\end{figure}

\begin{table}[htbp]
\global\long\def\arraystretch{1.12}
\captionsetup{width=.95\textwidth}
\caption{IRREPs at high-symmetry points for ErMn$_6$Ge$_6$ phonon spectrum at Low-T.}
\label{tab:191_ErMn6Ge6_Low}
\setlength{\tabcolsep}{1.mm}{
}
\end{table}

\begin{table}[htbp]
\global\long\def\arraystretch{1.12}
\captionsetup{width=.95\textwidth}
\caption{IRREPs at high-symmetry points for ErMn$_6$Ge$_6$ phonon spectrum at High-T.}
\label{tab:191_ErMn6Ge6_High}
\setlength{\tabcolsep}{1.mm}{
%
}
\end{table}
\subsubsection{\label{sms2sec:TmMn6Ge6}\hyperref[smsec:col]{\texorpdfstring{TmMn$_6$Ge$_6$}{}}~{\color{green}  (High-T stable, Low-T stable) }}

\begin{table}[htbp]
\global\long\def\arraystretch{1.12}
\captionsetup{width=.85\textwidth}
\caption{Structure parameters of TmMn$_6$Ge$_6$ after relaxation. It contains 411 electrons. }
\label{tab:TmMn6Ge6}
\setlength{\tabcolsep}{4mm}{
%
}
\end{table}

\begin{figure}[htbp]
\centering
\includegraphics[width=0.85\textwidth]{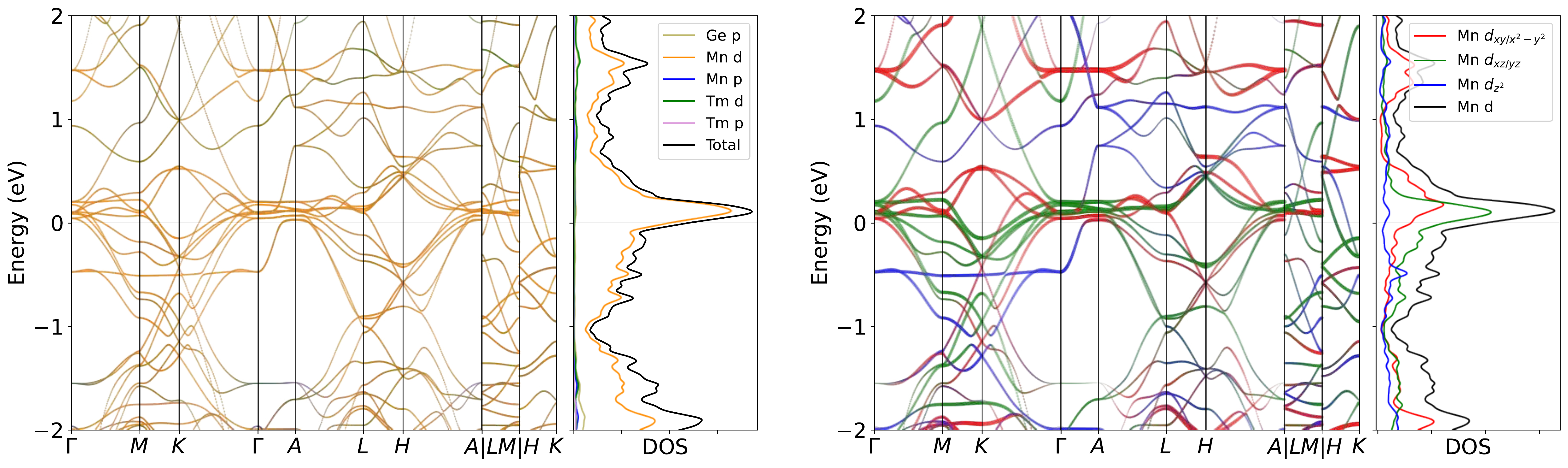}
\caption{Band structure for TmMn$_6$Ge$_6$ without SOC in paramagnetic phase. The projection of Mn $3d$ orbitals is also presented. The band structures clearly reveal the dominating of Mn $3d$ orbitals  near the Fermi level, as well as the pronounced peak.}
\label{fig:TmMn6Ge6band}
\end{figure}

\begin{figure}[H]
\centering
\subfloat[Relaxed phonon at low-T.]{
\label{fig:TmMn6Ge6_relaxed_Low}
\includegraphics[width=0.45\textwidth]{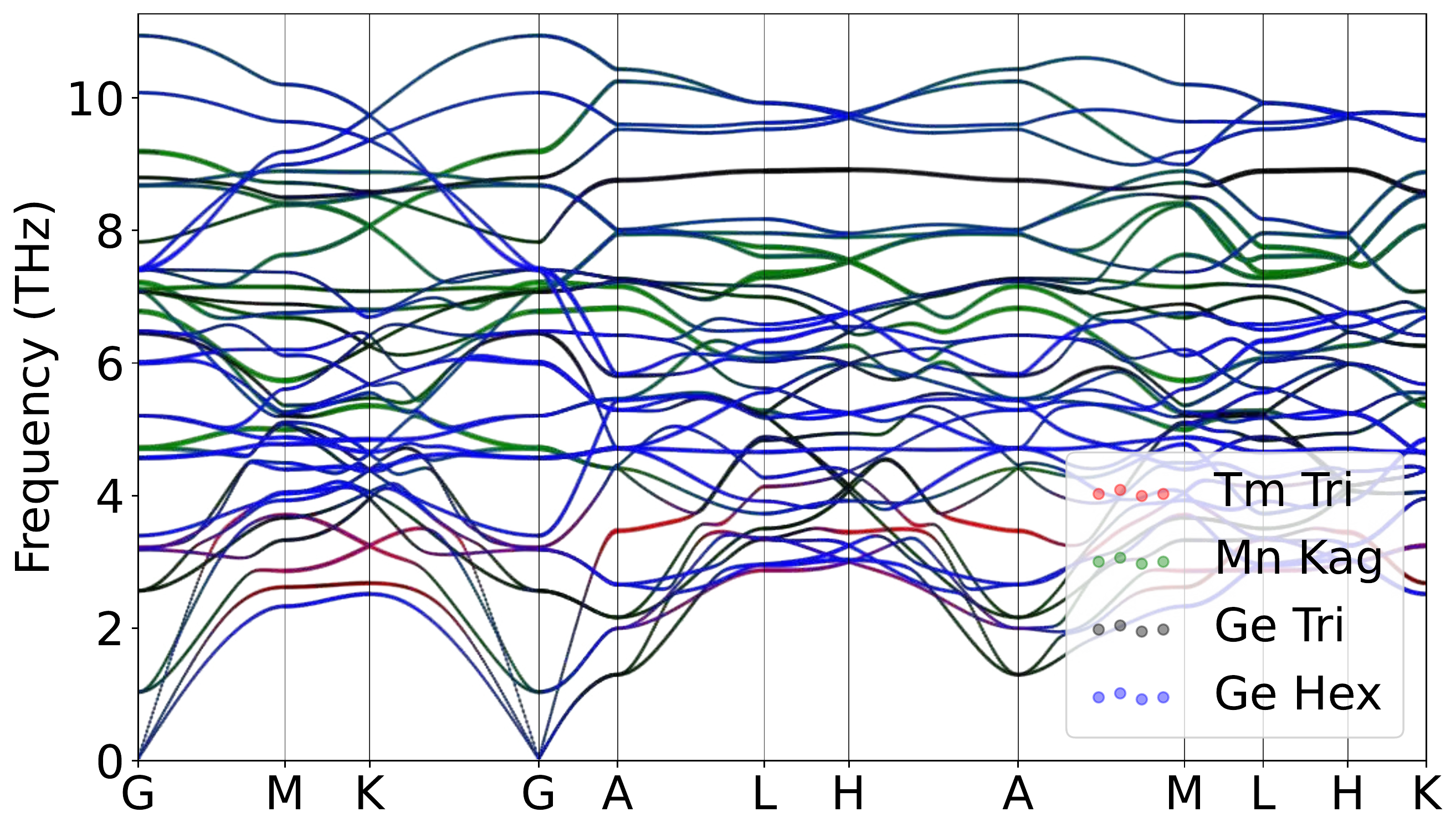}
}
\hspace{5pt}\subfloat[Relaxed phonon at high-T.]{
\label{fig:TmMn6Ge6_relaxed_High}
\includegraphics[width=0.45\textwidth]{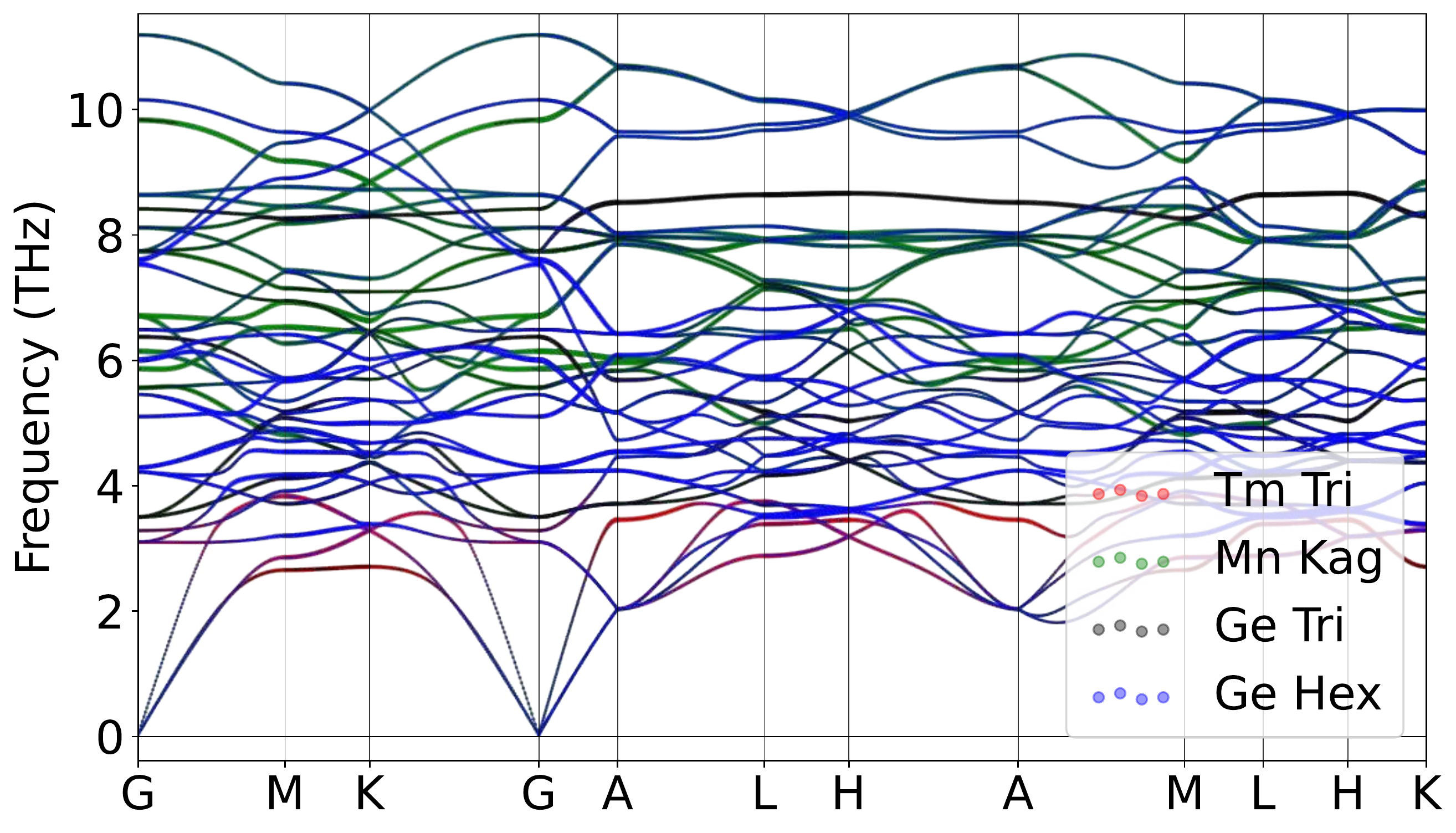}
}
\caption{Atomically projected phonon spectrum for TmMn$_6$Ge$_6$ in paramagnetic phase.}
\label{fig:TmMn6Ge6pho}
\end{figure}

\begin{figure}[htbp]
\centering
\includegraphics[width=0.45\textwidth]{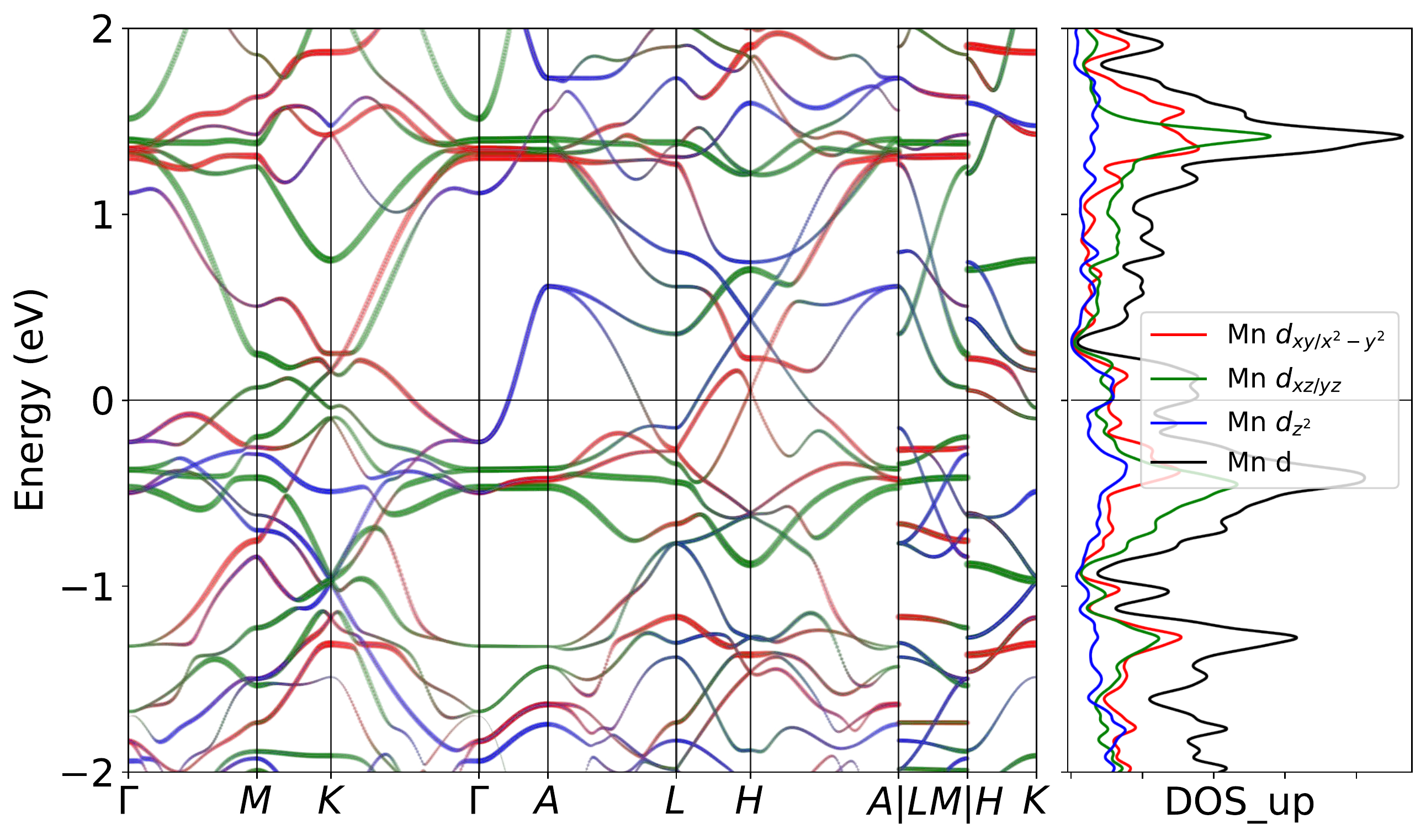}
\includegraphics[width=0.45\textwidth]{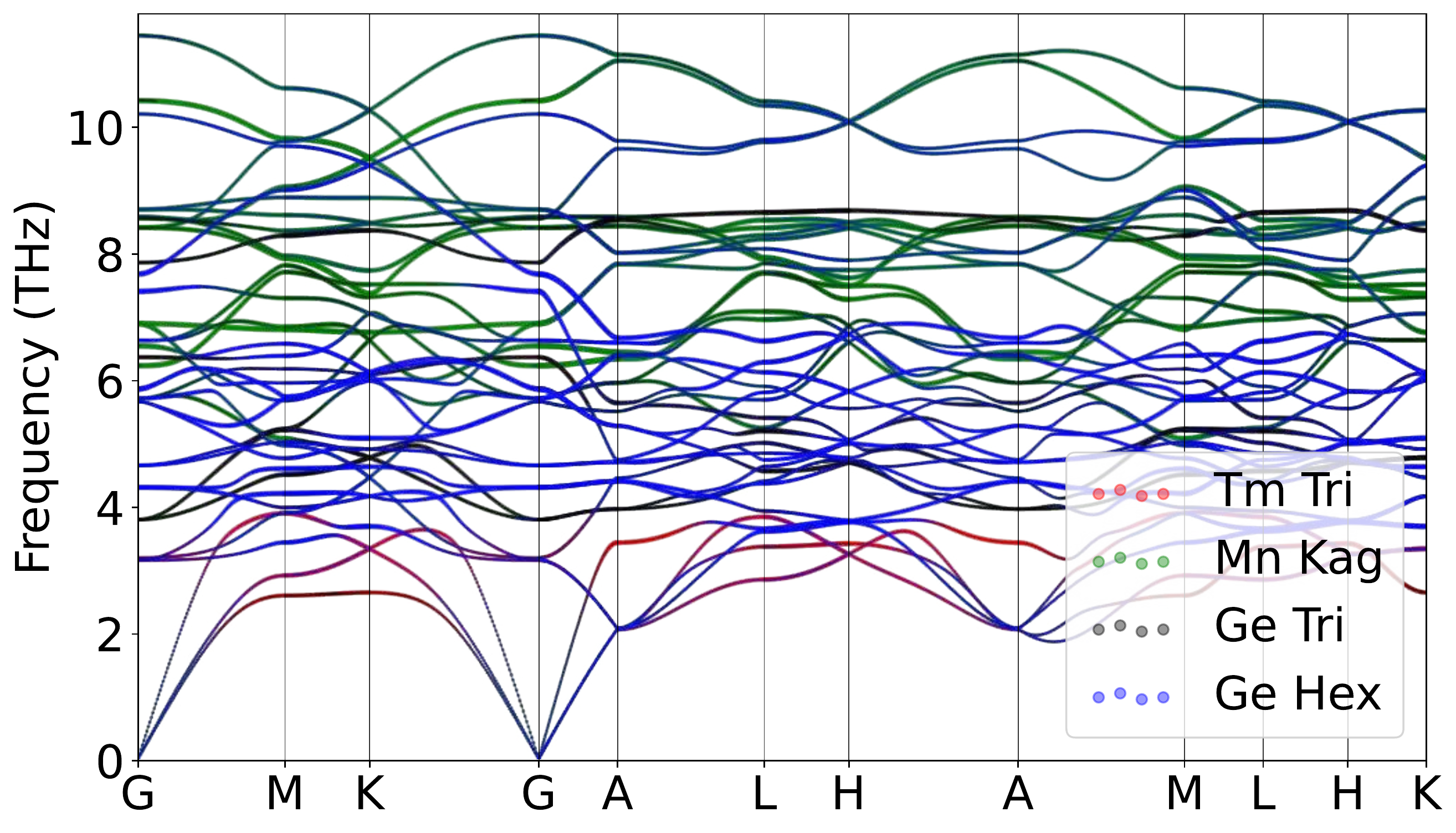}
\caption{Band structure for TmMn$_6$Ge$_6$ with AFM order without SOC for spin (Left)up channel. The projection of Mn $3d$ orbitals is also presented. (Right)Correspondingly, a stable phonon was demonstrated with the collinear magnetic order without Hubbard U at lower temperature regime, weighted by atomic projections.}
\label{fig:TmMn6Ge6mag}
\end{figure}

\begin{figure}[H]
\centering
\label{fig:TmMn6Ge6_relaxed_Low_xz}
\includegraphics[width=0.85\textwidth]{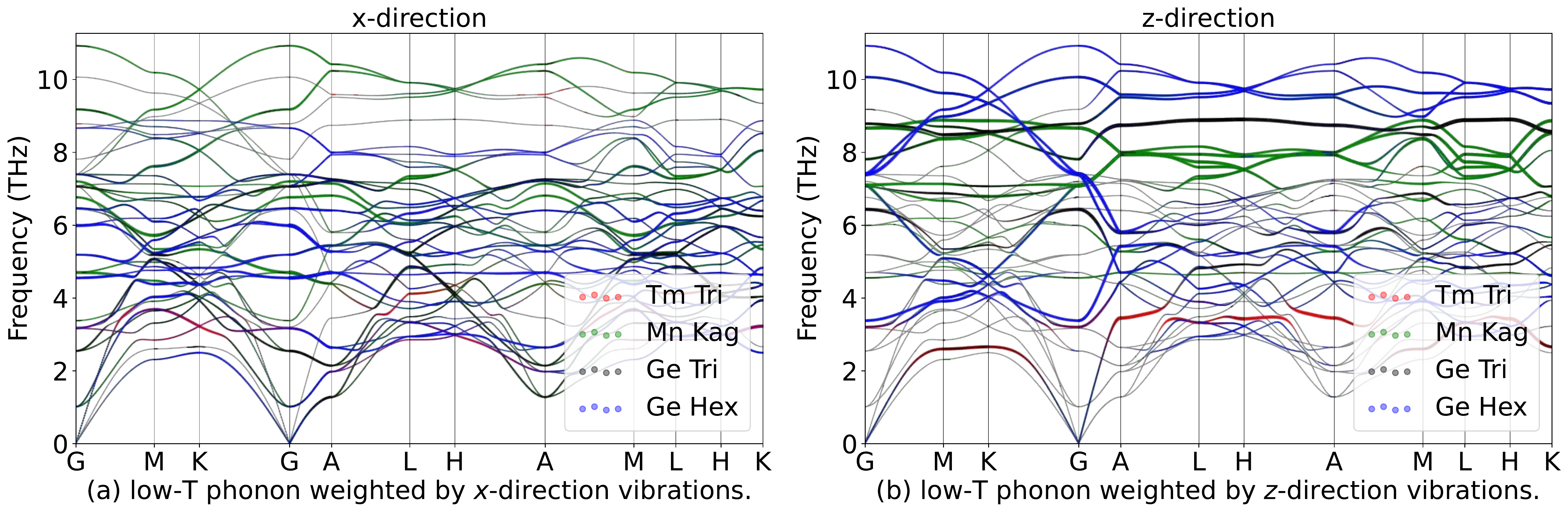}
\hspace{5pt}\label{fig:TmMn6Ge6_relaxed_High_xz}
\includegraphics[width=0.85\textwidth]{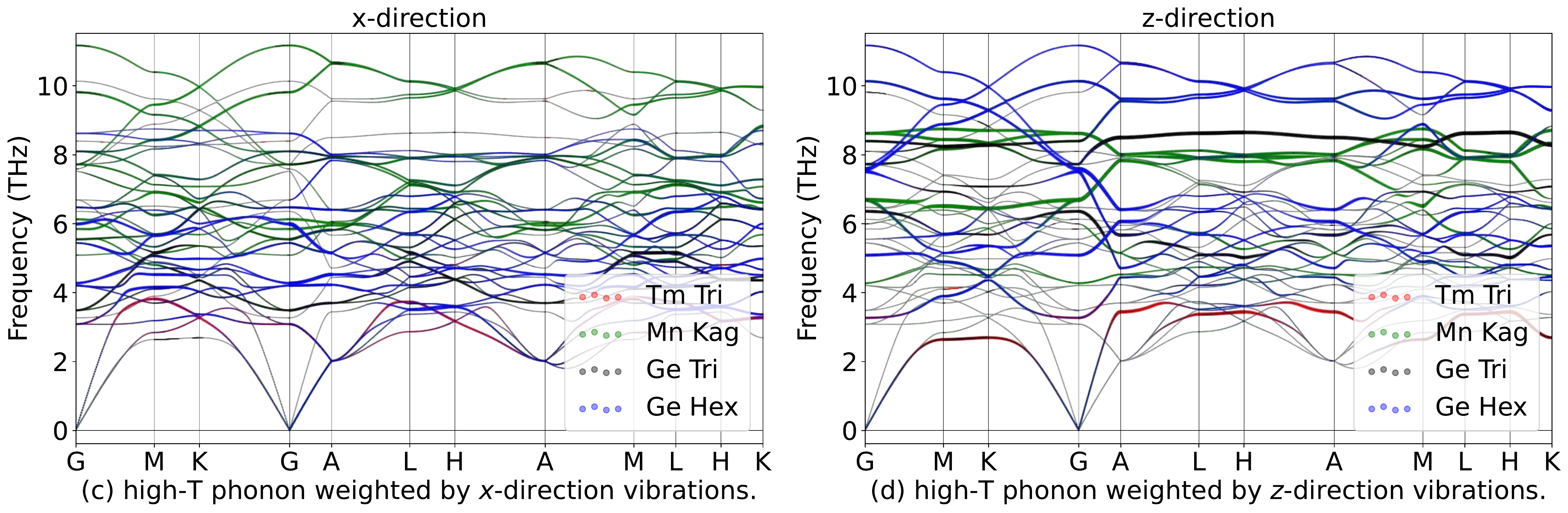}
\caption{Phonon spectrum for TmMn$_6$Ge$_6$in in-plane ($x$) and out-of-plane ($z$) directions in paramagnetic phase.}
\label{fig:TmMn6Ge6phoxyz}
\end{figure}

\begin{table}[htbp]
\global\long\def\arraystretch{1.12}
\captionsetup{width=.95\textwidth}
\caption{IRREPs at high-symmetry points for TmMn$_6$Ge$_6$ phonon spectrum at Low-T.}
\label{tab:191_TmMn6Ge6_Low}
\setlength{\tabcolsep}{1.mm}{
}
\end{table}

\begin{table}[htbp]
\global\long\def\arraystretch{1.12}
\captionsetup{width=.95\textwidth}
\caption{IRREPs at high-symmetry points for TmMn$_6$Ge$_6$ phonon spectrum at High-T.}
\label{tab:191_TmMn6Ge6_High}
\setlength{\tabcolsep}{1.mm}{
%
}
\end{table}
\subsubsection{\label{sms2sec:YbMn6Ge6}\hyperref[smsec:col]{\texorpdfstring{YbMn$_6$Ge$_6$}{}}~{\color{blue}  (High-T stable, Low-T unstable) }}

\begin{table}[htbp]
\global\long\def\arraystretch{1.12}
\captionsetup{width=.85\textwidth}
\caption{Structure parameters of YbMn$_6$Ge$_6$ after relaxation. It contains 412 electrons. }
\label{tab:YbMn6Ge6}
\setlength{\tabcolsep}{4mm}{
%
}
\end{table}

\begin{figure}[htbp]
\centering
\includegraphics[width=0.85\textwidth]{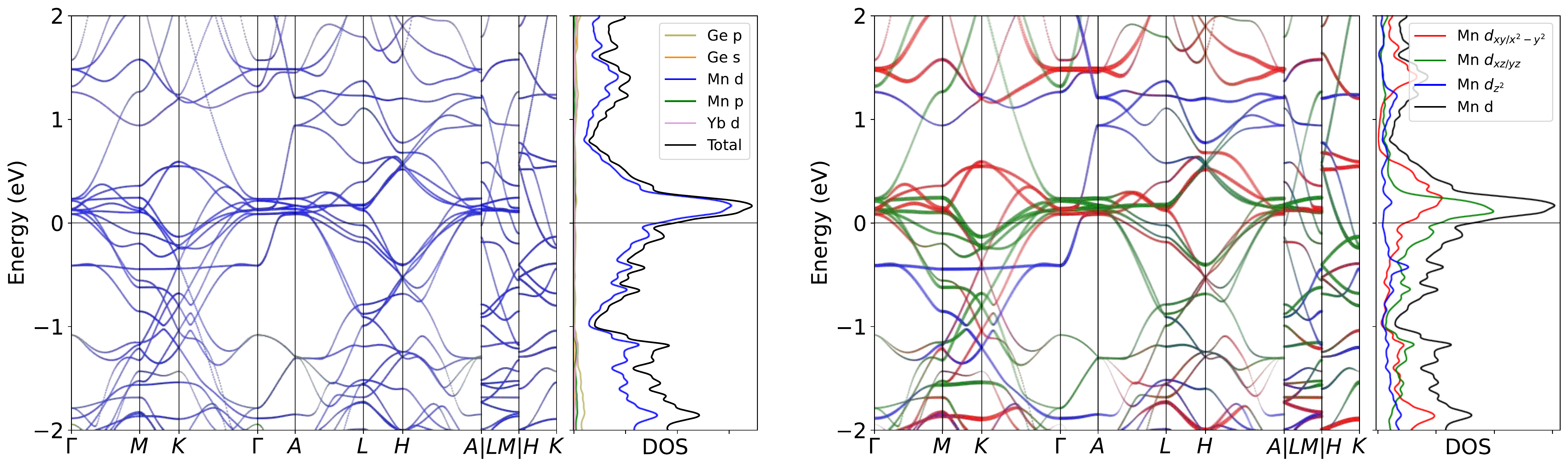}
\caption{Band structure for YbMn$_6$Ge$_6$ without SOC in paramagnetic phase. The projection of Mn $3d$ orbitals is also presented. The band structures clearly reveal the dominating of Mn $3d$ orbitals  near the Fermi level, as well as the pronounced peak.}
\label{fig:YbMn6Ge6band}
\end{figure}

\begin{figure}[H]
\centering
\subfloat[Relaxed phonon at low-T.]{
\label{fig:YbMn6Ge6_relaxed_Low}
\includegraphics[width=0.45\textwidth]{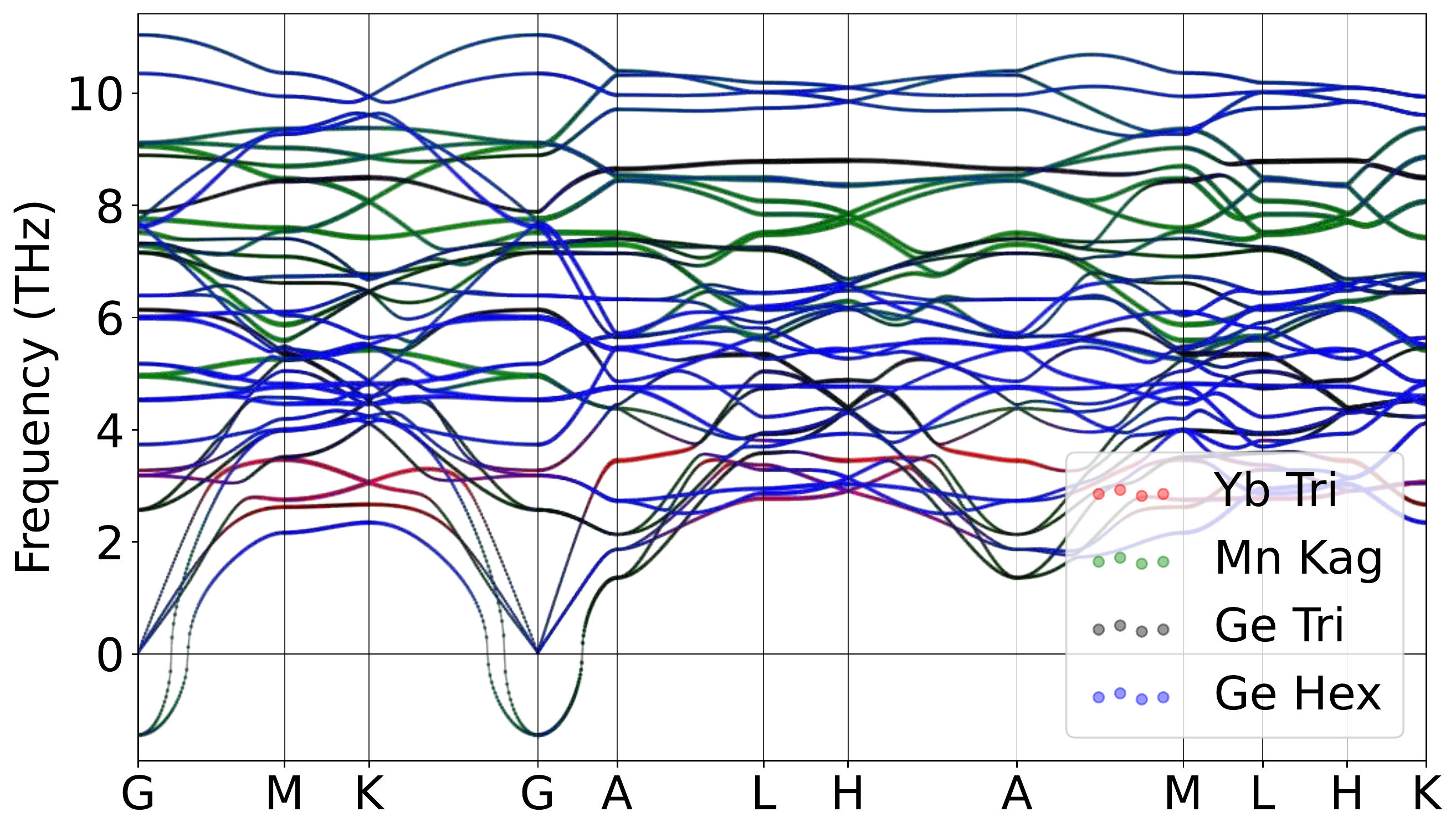}
}
\hspace{5pt}\subfloat[Relaxed phonon at high-T.]{
\label{fig:YbMn6Ge6_relaxed_High}
\includegraphics[width=0.45\textwidth]{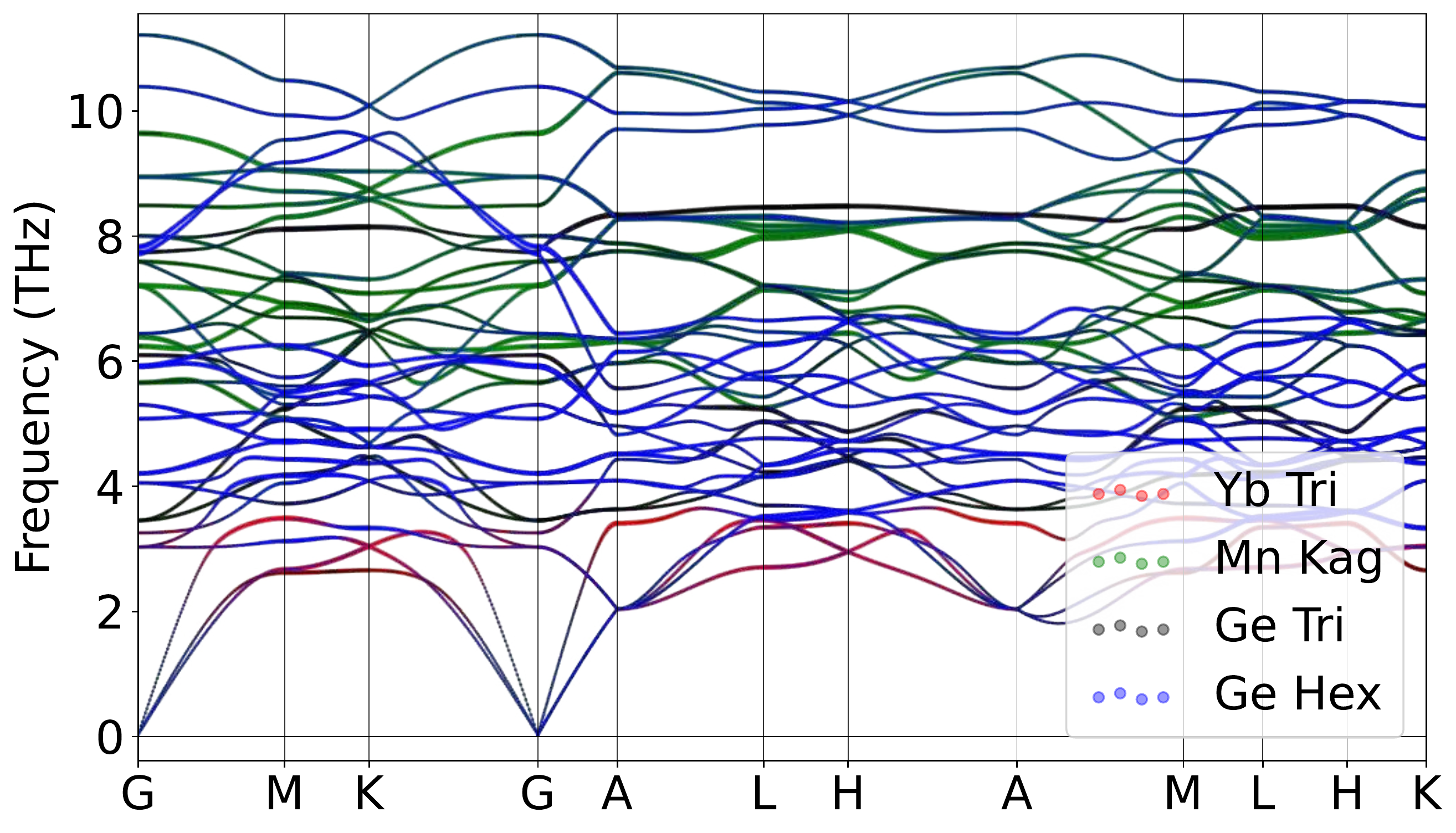}
}
\caption{Atomically projected phonon spectrum for YbMn$_6$Ge$_6$ in paramagnetic phase.}
\label{fig:YbMn6Ge6pho}
\end{figure}

\begin{figure}[htbp]
\centering
\includegraphics[width=0.45\textwidth]{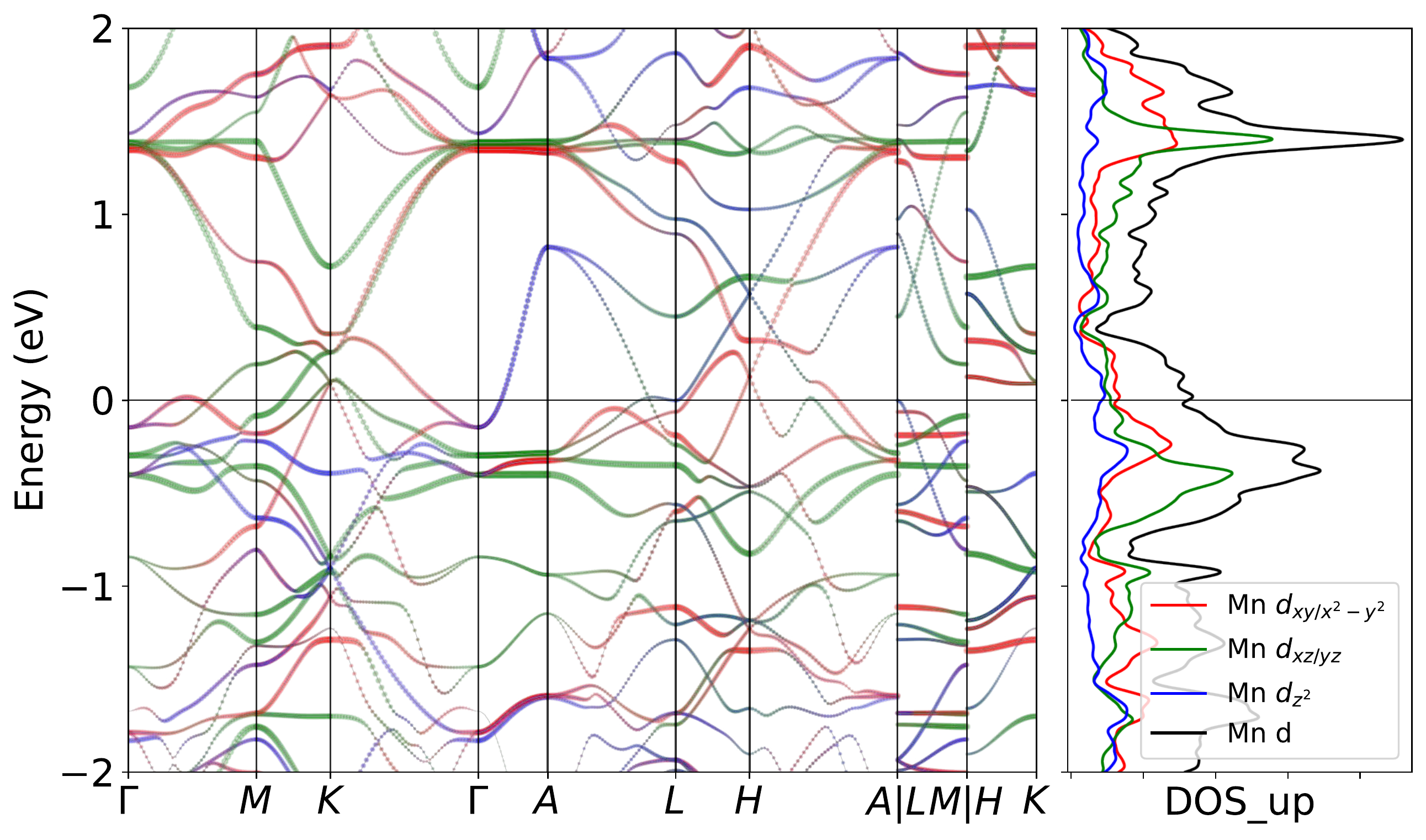}
\includegraphics[width=0.45\textwidth]{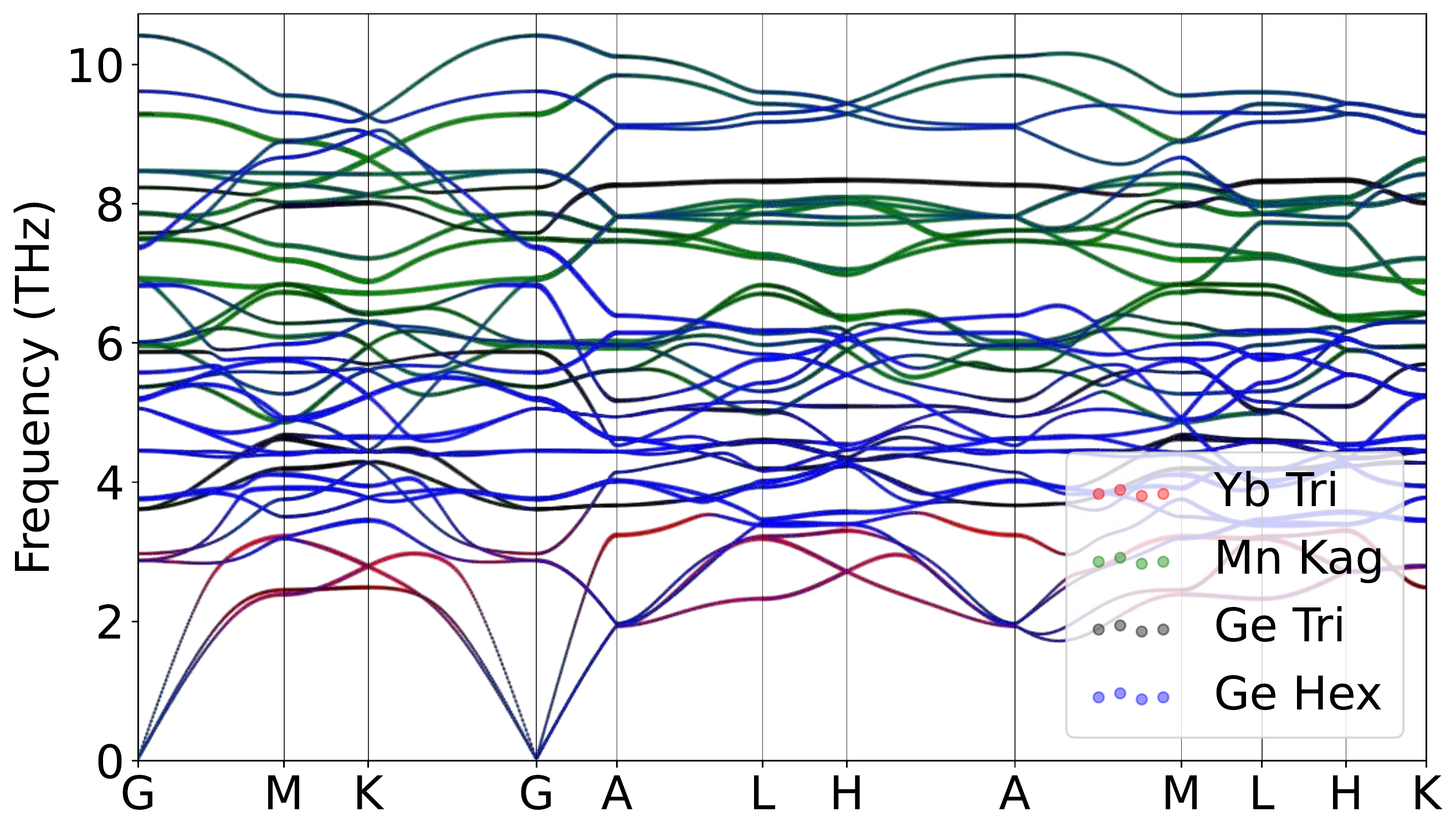}
\caption{Band structure for YbMn$_6$Ge$_6$ with AFM order without SOC for spin (Left)up channel. The projection of Mn $3d$ orbitals is also presented. (Right)Correspondingly, a stable phonon was demonstrated with the collinear magnetic order without Hubbard U at lower temperature regime, weighted by atomic projections.}
\label{fig:YbMn6Ge6mag}
\end{figure}

\begin{figure}[H]
\centering
\label{fig:YbMn6Ge6_relaxed_Low_xz}
\includegraphics[width=0.85\textwidth]{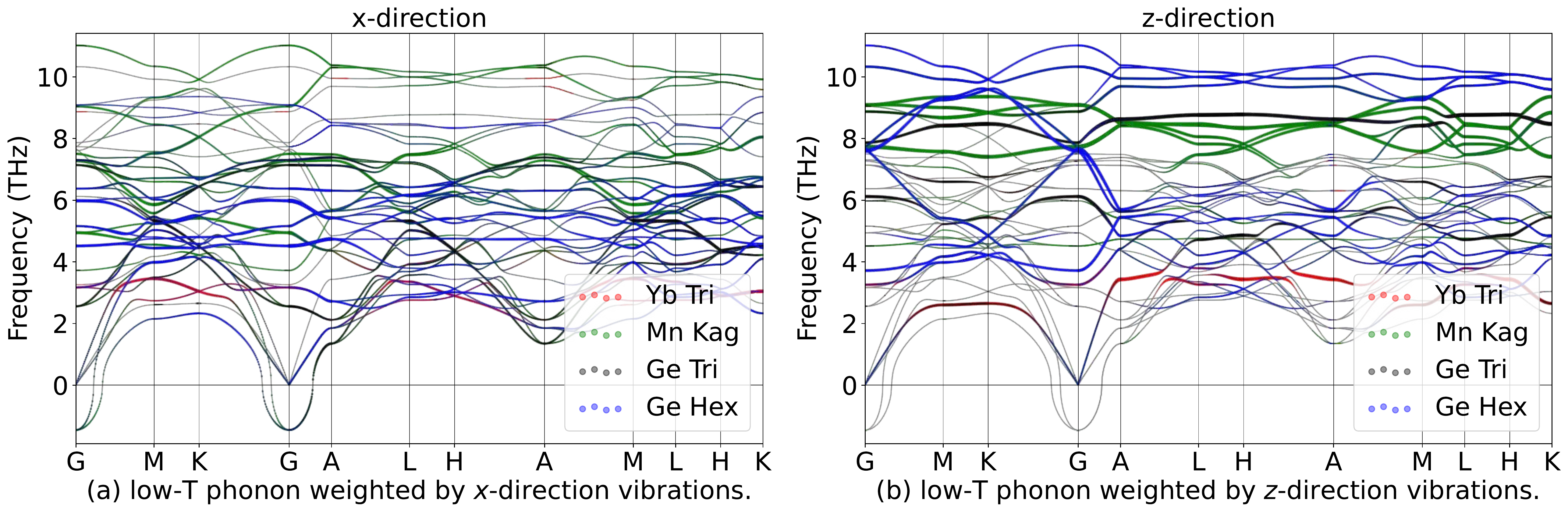}
\hspace{5pt}\label{fig:YbMn6Ge6_relaxed_High_xz}
\includegraphics[width=0.85\textwidth]{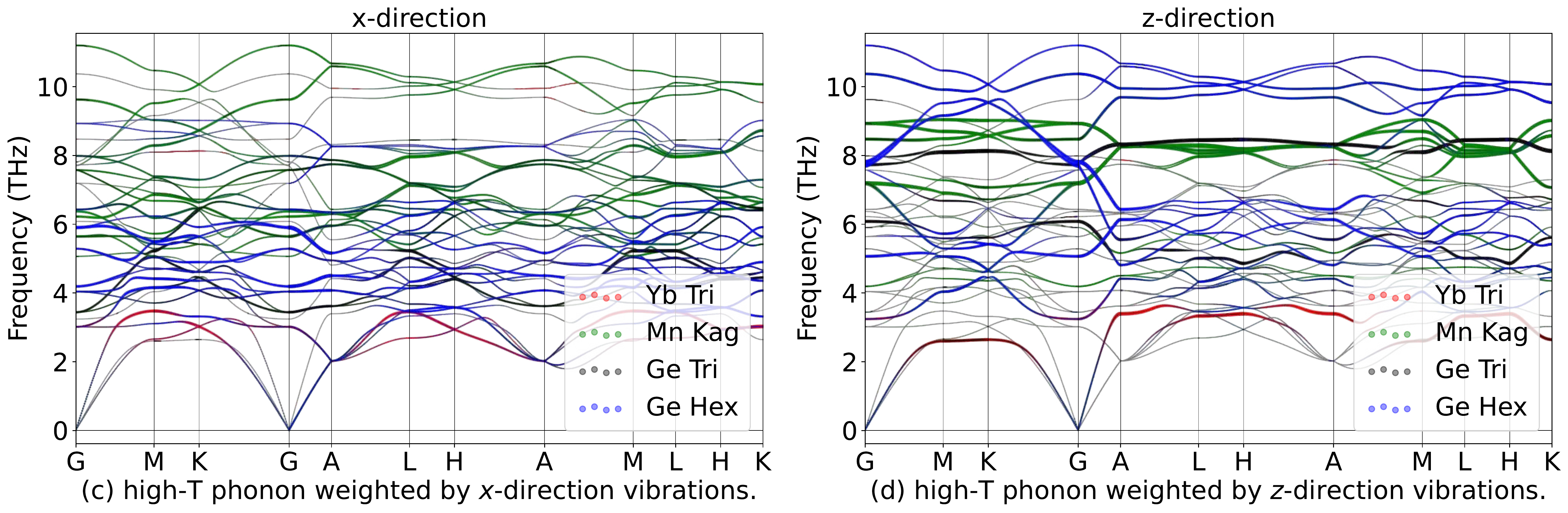}
\caption{Phonon spectrum for YbMn$_6$Ge$_6$in in-plane ($x$) and out-of-plane ($z$) directions in paramagnetic phase.}
\label{fig:YbMn6Ge6phoxyz}
\end{figure}

\begin{table}[htbp]
\global\long\def\arraystretch{1.12}
\captionsetup{width=.95\textwidth}
\caption{IRREPs at high-symmetry points for YbMn$_6$Ge$_6$ phonon spectrum at Low-T.}
\label{tab:191_YbMn6Ge6_Low}
\setlength{\tabcolsep}{1.mm}{
}
\end{table}

\begin{table}[htbp]
\global\long\def\arraystretch{1.12}
\captionsetup{width=.95\textwidth}
\caption{IRREPs at high-symmetry points for YbMn$_6$Ge$_6$ phonon spectrum at High-T.}
\label{tab:191_YbMn6Ge6_High}
\setlength{\tabcolsep}{1.mm}{
%
}
\end{table}
\subsubsection{\label{sms2sec:LuMn6Ge6}\hyperref[smsec:col]{\texorpdfstring{LuMn$_6$Ge$_6$}{}}~{\color{green}  (High-T stable, Low-T stable) }}

\begin{table}[htbp]
\global\long\def\arraystretch{1.12}
\captionsetup{width=.85\textwidth}
\caption{Structure parameters of LuMn$_6$Ge$_6$ after relaxation. It contains 413 electrons. }
\label{tab:LuMn6Ge6}
\setlength{\tabcolsep}{4mm}{
%
}
\end{table}

\begin{figure}[htbp]
\centering
\includegraphics[width=0.85\textwidth]{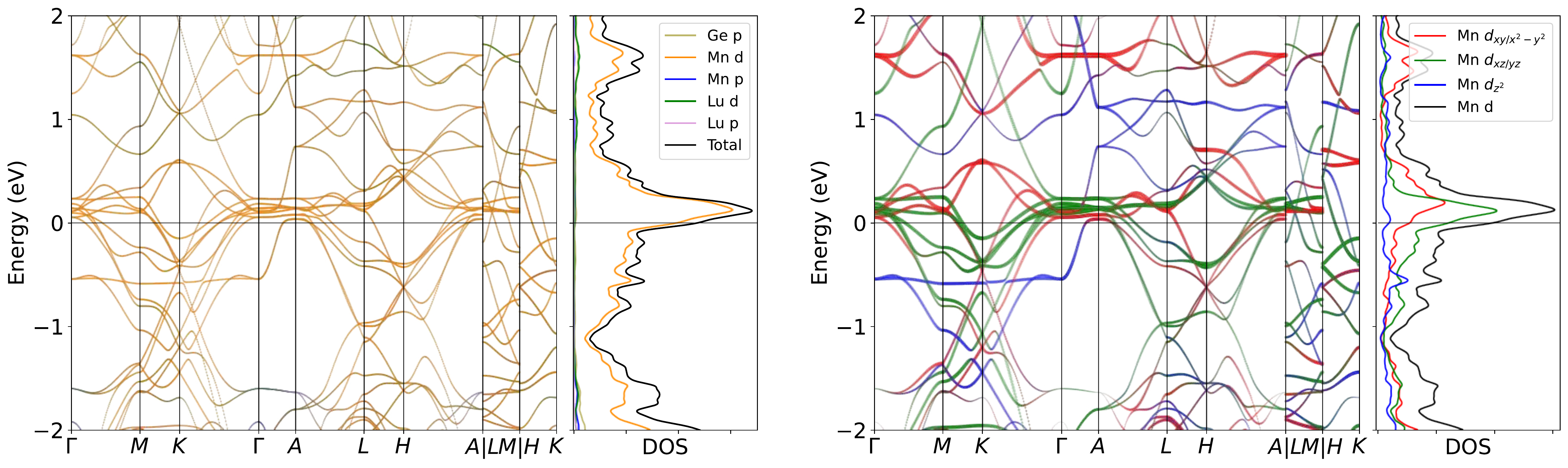}
\caption{Band structure for LuMn$_6$Ge$_6$ without SOC in paramagnetic phase. The projection of Mn $3d$ orbitals is also presented. The band structures clearly reveal the dominating of Mn $3d$ orbitals  near the Fermi level, as well as the pronounced peak.}
\label{fig:LuMn6Ge6band}
\end{figure}

\begin{figure}[H]
\centering
\subfloat[Relaxed phonon at low-T.]{
\label{fig:LuMn6Ge6_relaxed_Low}
\includegraphics[width=0.45\textwidth]{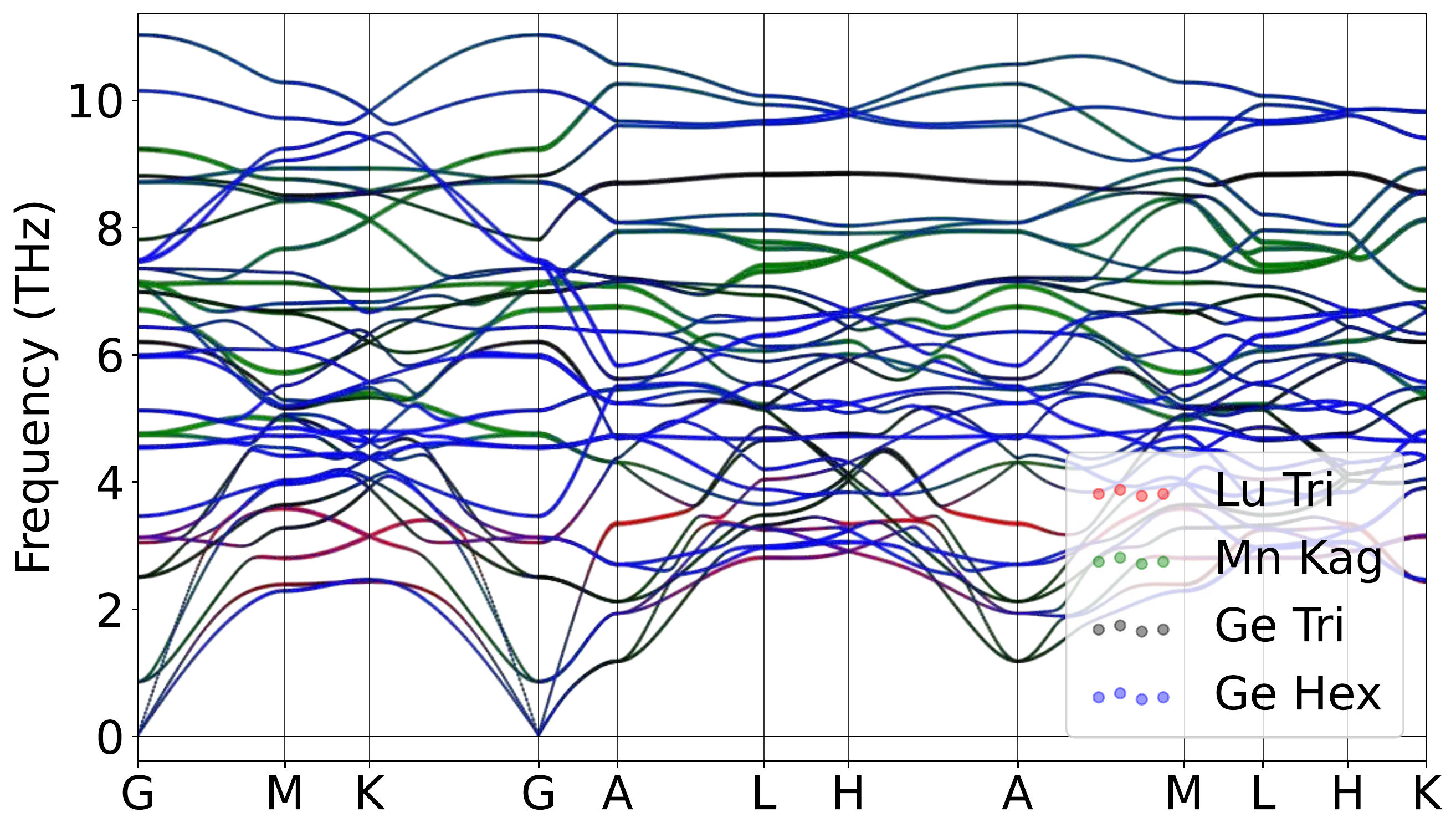}
}
\hspace{5pt}\subfloat[Relaxed phonon at high-T.]{
\label{fig:LuMn6Ge6_relaxed_High}
\includegraphics[width=0.45\textwidth]{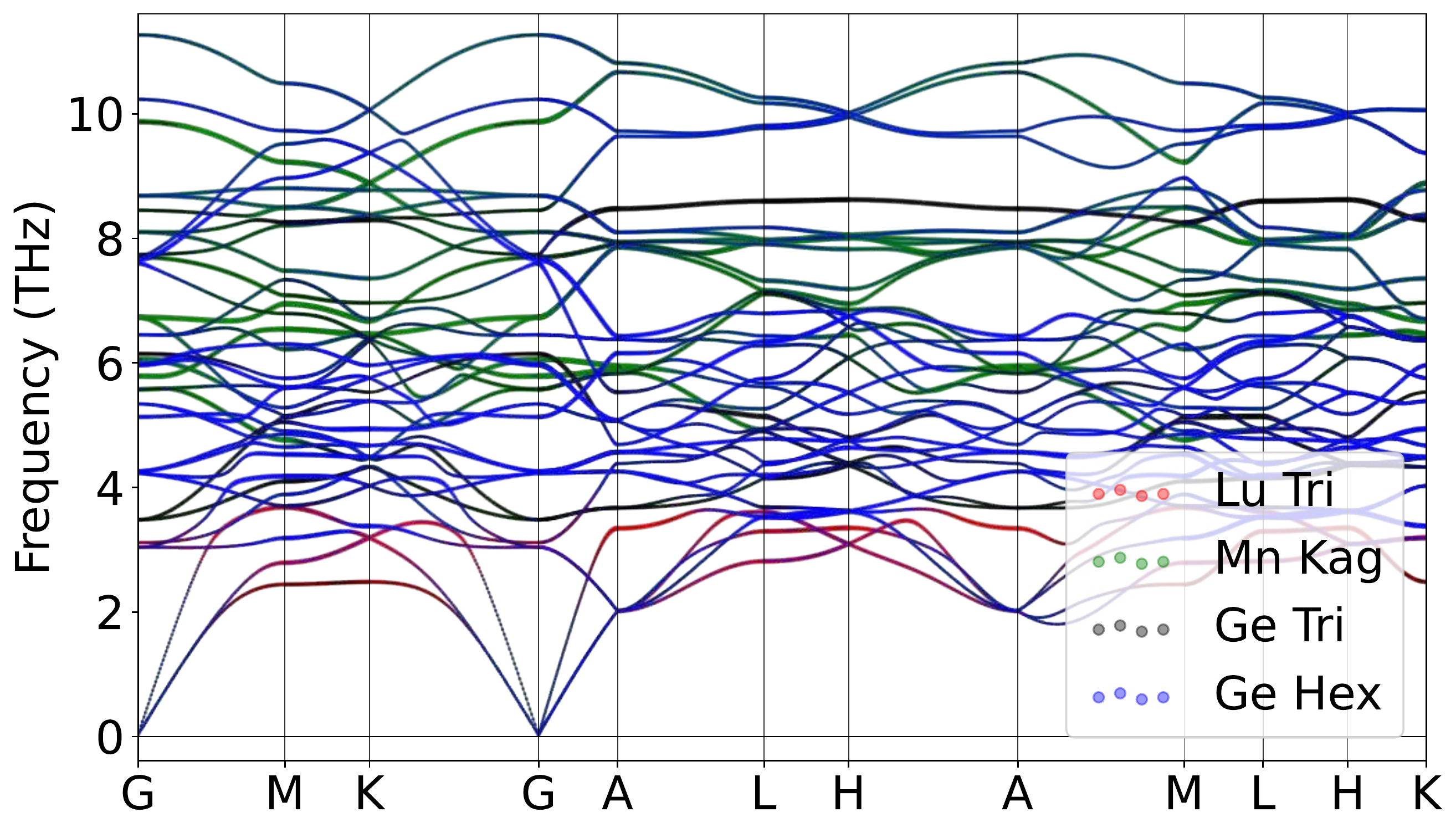}
}
\caption{Atomically projected phonon spectrum for LuMn$_6$Ge$_6$ in paramagnetic phase.}
\label{fig:LuMn6Ge6pho}
\end{figure}

\begin{figure}[htbp]
\centering
\includegraphics[width=0.45\textwidth]{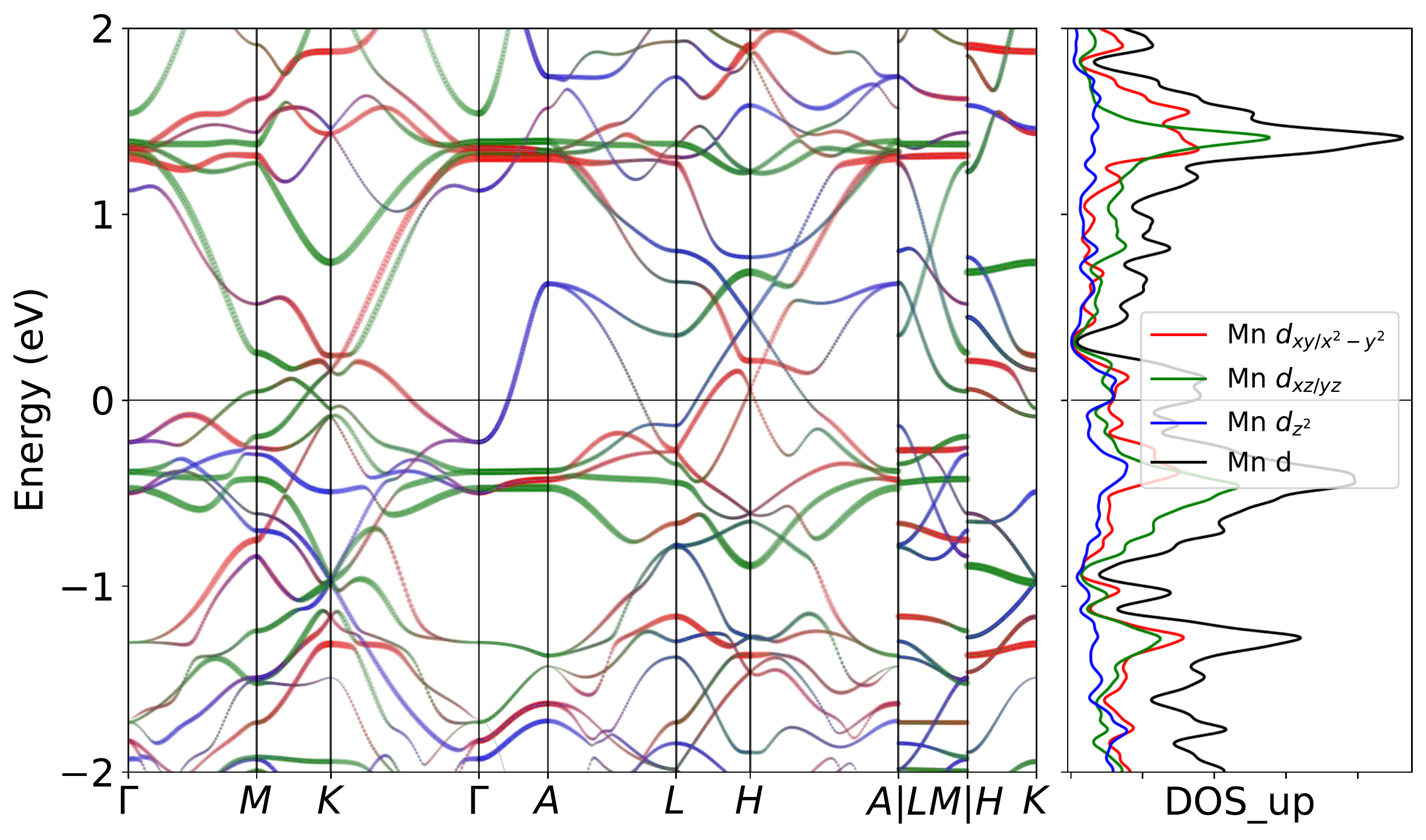}
\includegraphics[width=0.45\textwidth]{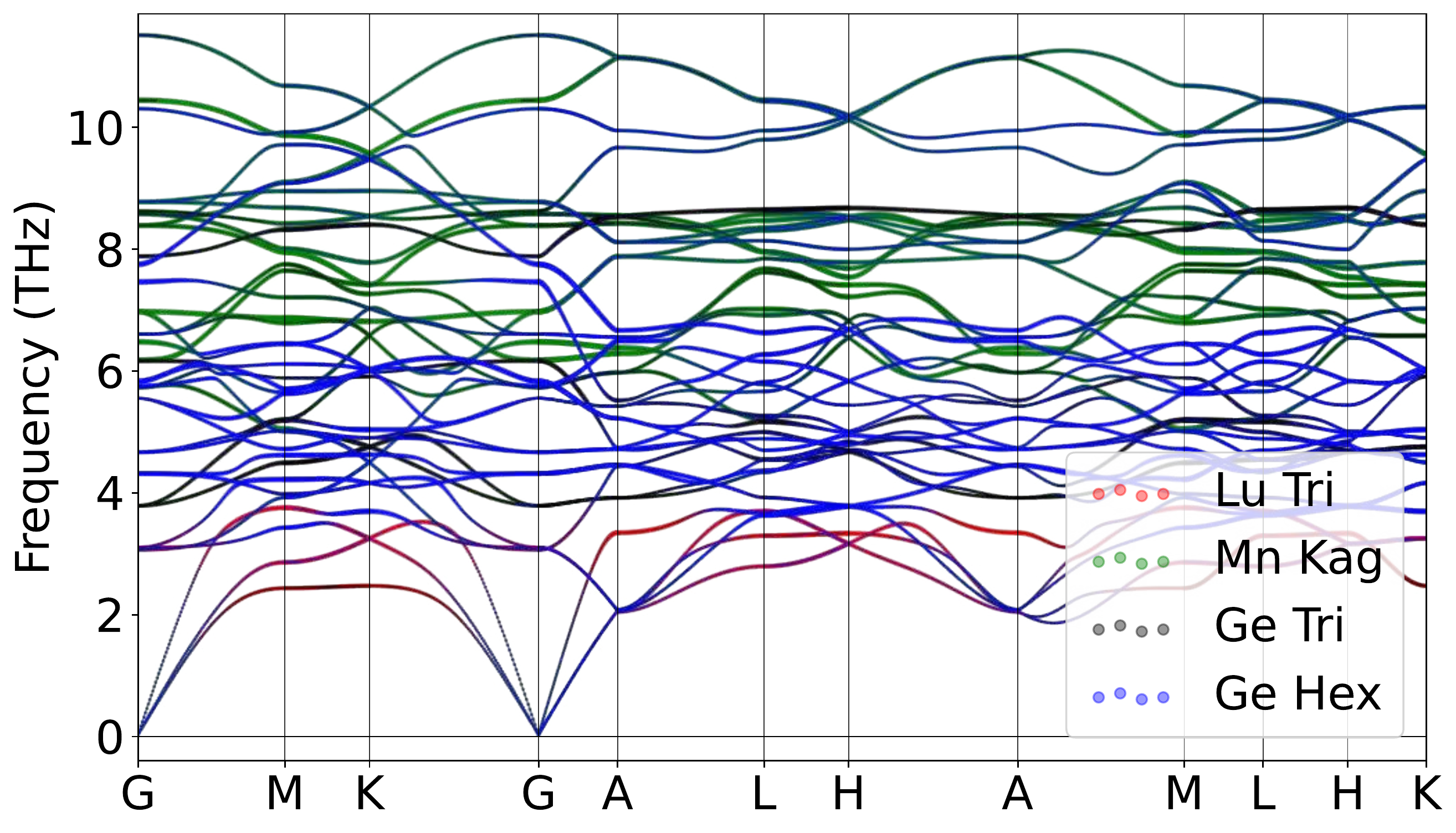}
\caption{Band structure for LuMn$_6$Ge$_6$ with AFM order without SOC for spin (Left)up channel. The projection of Mn $3d$ orbitals is also presented. (Right)Correspondingly, a stable phonon was demonstrated with the collinear magnetic order without Hubbard U at lower temperature regime, weighted by atomic projections.}
\label{fig:LuMn6Ge6mag}
\end{figure}

\begin{figure}[H]
\centering
\label{fig:LuMn6Ge6_relaxed_Low_xz}
\includegraphics[width=0.85\textwidth]{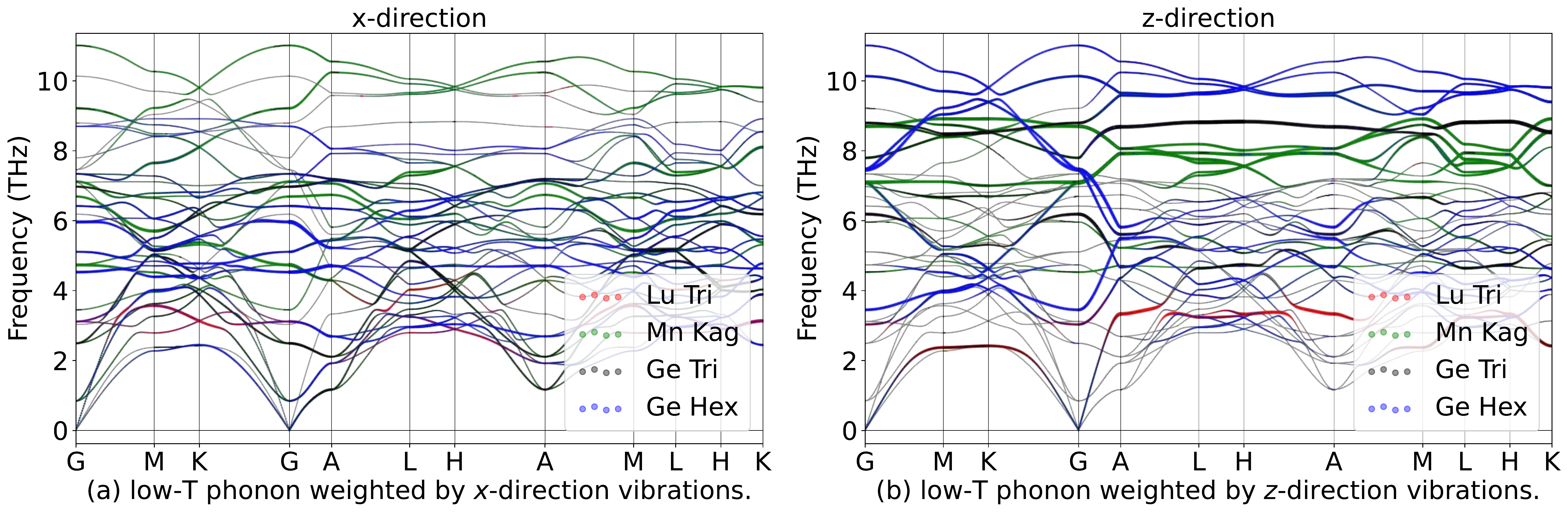}
\hspace{5pt}\label{fig:LuMn6Ge6_relaxed_High_xz}
\includegraphics[width=0.85\textwidth]{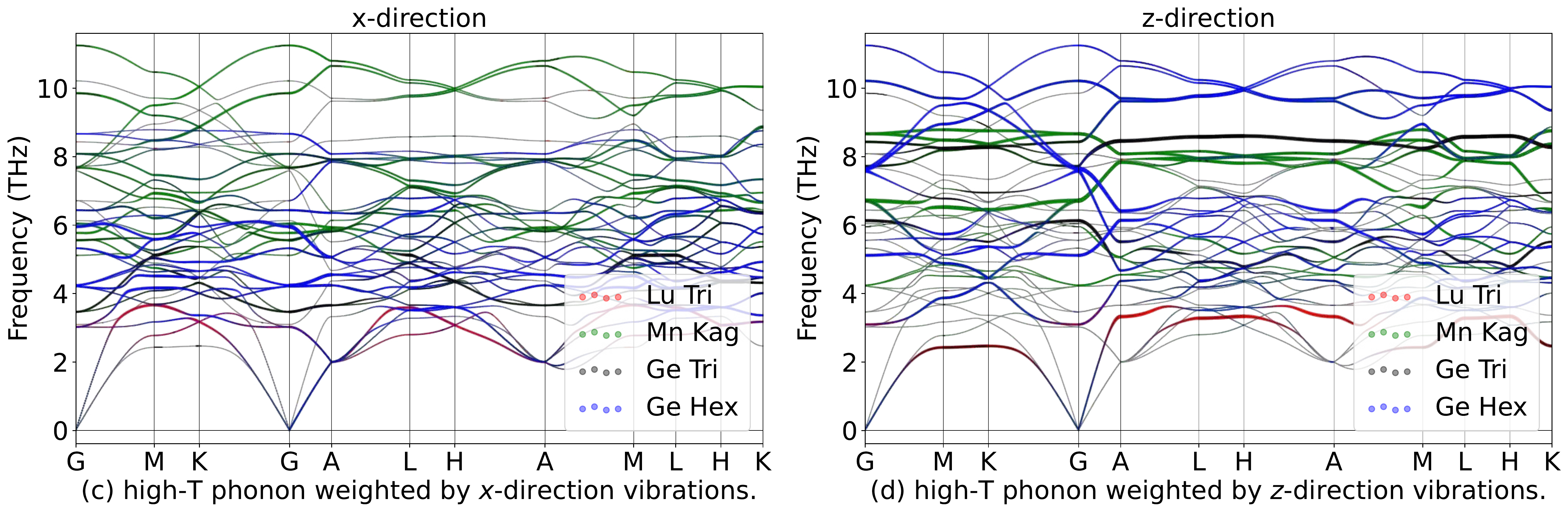}
\caption{Phonon spectrum for LuMn$_6$Ge$_6$in in-plane ($x$) and out-of-plane ($z$) directions in paramagnetic phase.}
\label{fig:LuMn6Ge6phoxyz}
\end{figure}

\begin{table}[htbp]
\global\long\def\arraystretch{1.12}
\captionsetup{width=.95\textwidth}
\caption{IRREPs at high-symmetry points for LuMn$_6$Ge$_6$ phonon spectrum at Low-T.}
\label{tab:191_LuMn6Ge6_Low}
\setlength{\tabcolsep}{1.mm}{
}
\end{table}

\begin{table}[htbp]
\global\long\def\arraystretch{1.12}
\captionsetup{width=.95\textwidth}
\caption{IRREPs at high-symmetry points for LuMn$_6$Ge$_6$ phonon spectrum at High-T.}
\label{tab:191_LuMn6Ge6_High}
\setlength{\tabcolsep}{1.mm}{
%
}
\end{table}
\subsubsection{\label{sms2sec:HfMn6Ge6}\hyperref[smsec:col]{\texorpdfstring{HfMn$_6$Ge$_6$}{}}~{\color{green}  (High-T stable, Low-T stable) }}

\begin{table}[htbp]
\global\long\def\arraystretch{1.12}
\captionsetup{width=.85\textwidth}
\caption{Structure parameters of HfMn$_6$Ge$_6$ after relaxation. It contains 414 electrons. }
\label{tab:HfMn6Ge6}
\setlength{\tabcolsep}{4mm}{
%
}
\end{table}

\begin{figure}[htbp]
\centering
\includegraphics[width=0.85\textwidth]{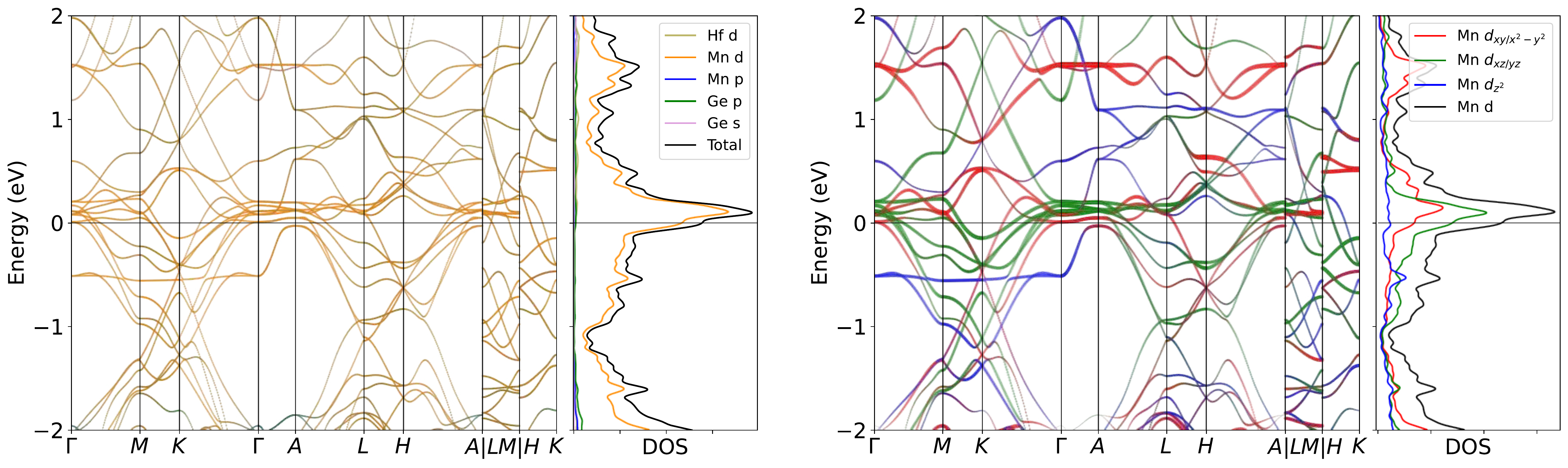}
\caption{Band structure for HfMn$_6$Ge$_6$ without SOC in paramagnetic phase. The projection of Mn $3d$ orbitals is also presented. The band structures clearly reveal the dominating of Mn $3d$ orbitals  near the Fermi level, as well as the pronounced peak.}
\label{fig:HfMn6Ge6band}
\end{figure}

\begin{figure}[H]
\centering
\subfloat[Relaxed phonon at low-T.]{
\label{fig:HfMn6Ge6_relaxed_Low}
\includegraphics[width=0.45\textwidth]{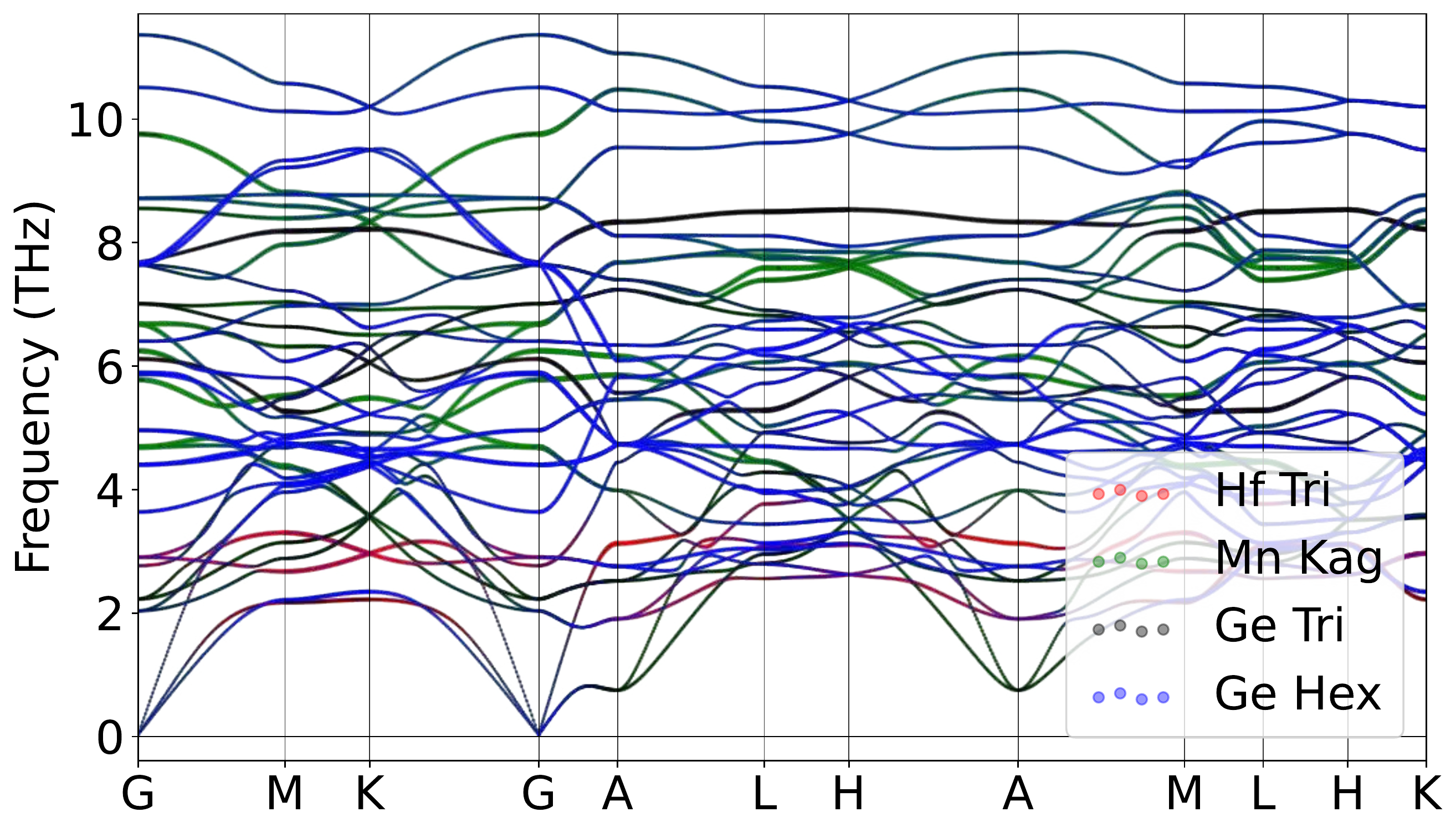}
}
\hspace{5pt}\subfloat[Relaxed phonon at high-T.]{
\label{fig:HfMn6Ge6_relaxed_High}
\includegraphics[width=0.45\textwidth]{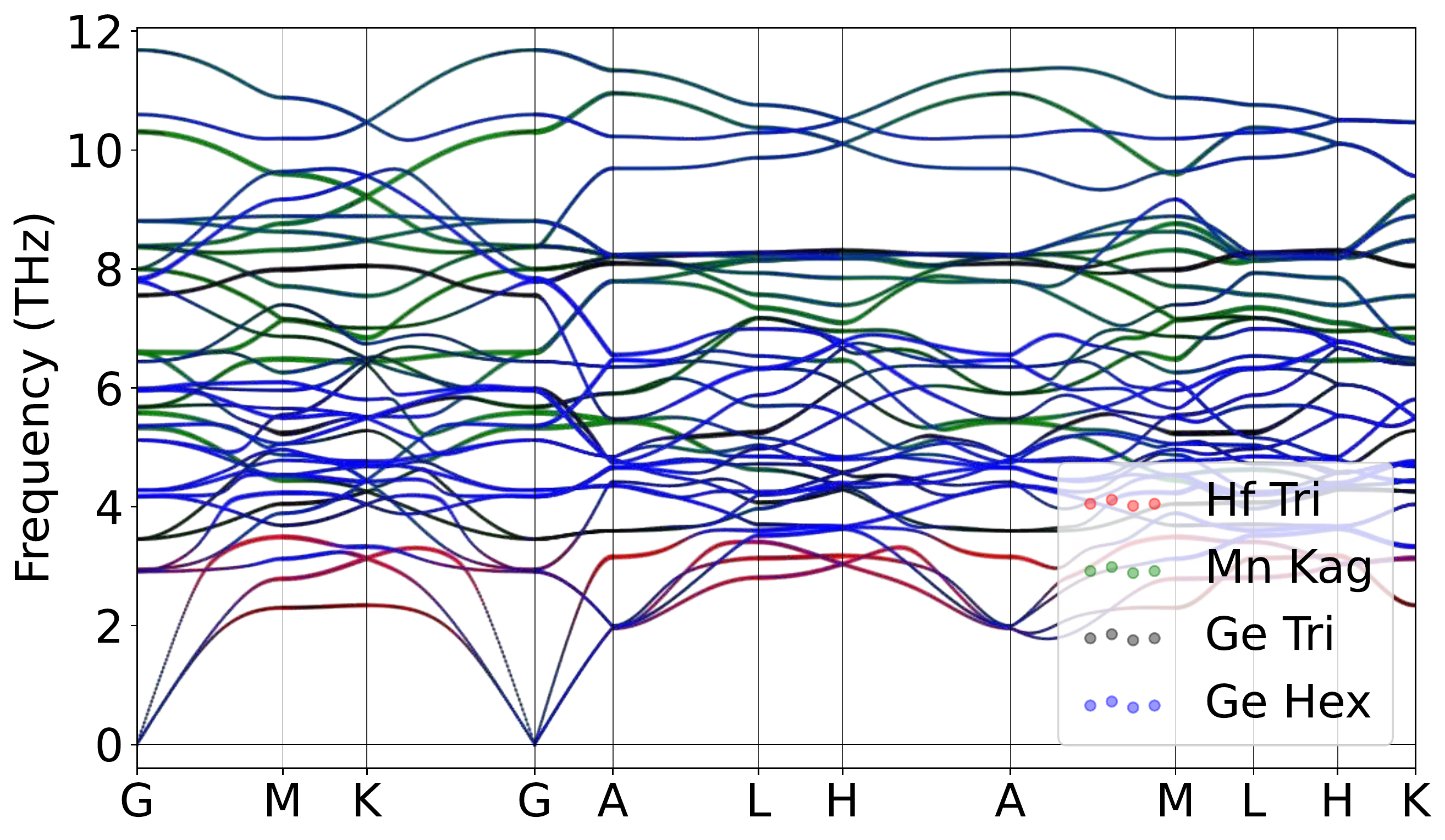}
}
\caption{Atomically projected phonon spectrum for HfMn$_6$Ge$_6$ in paramagnetic phase.}
\label{fig:HfMn6Ge6pho}
\end{figure}

\begin{figure}[htbp]
\centering
\includegraphics[width=0.45\textwidth]{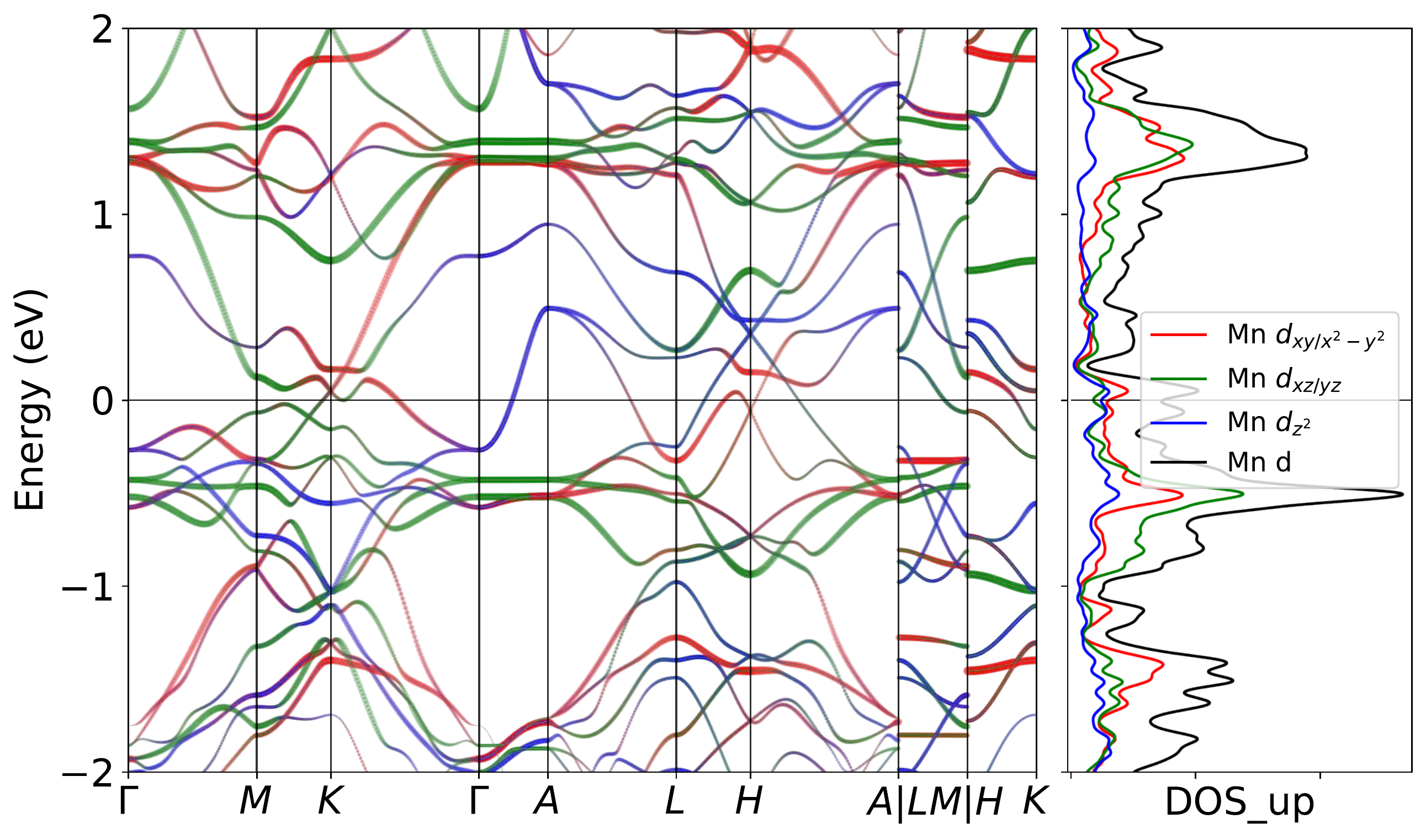}
\includegraphics[width=0.45\textwidth]{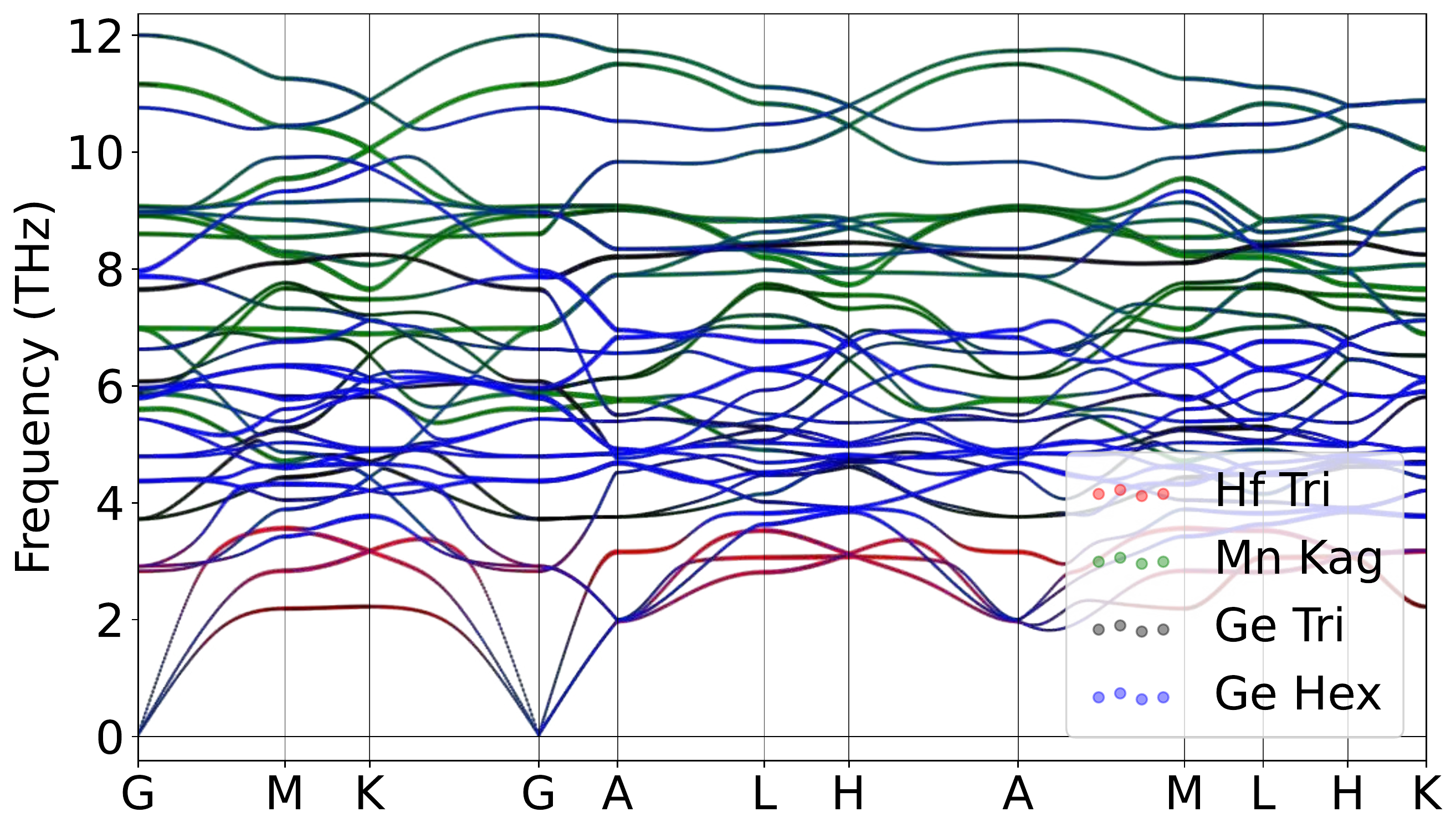}
\caption{Band structure for HfMn$_6$Ge$_6$ with AFM order without SOC for spin (Left)up channel. The projection of Mn $3d$ orbitals is also presented. (Right)Correspondingly, a stable phonon was demonstrated with the collinear magnetic order without Hubbard U at lower temperature regime, weighted by atomic projections.}
\label{fig:HfMn6Ge6mag}
\end{figure}

\begin{figure}[H]
\centering
\label{fig:HfMn6Ge6_relaxed_Low_xz}
\includegraphics[width=0.85\textwidth]{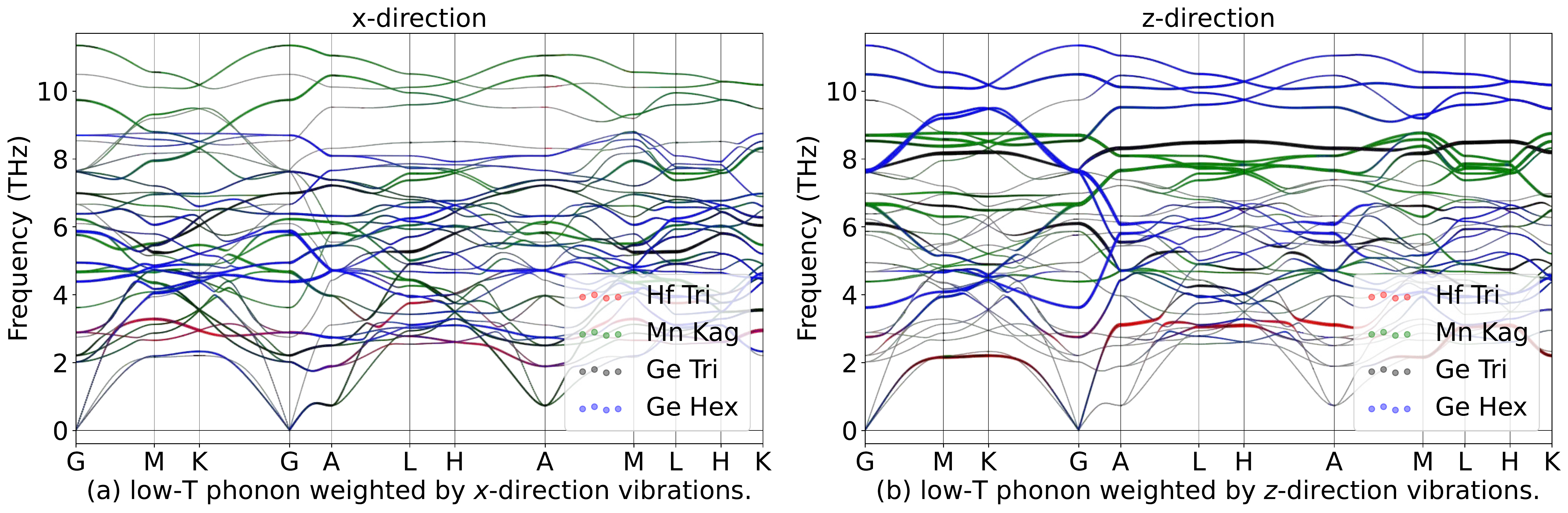}
\hspace{5pt}\label{fig:HfMn6Ge6_relaxed_High_xz}
\includegraphics[width=0.85\textwidth]{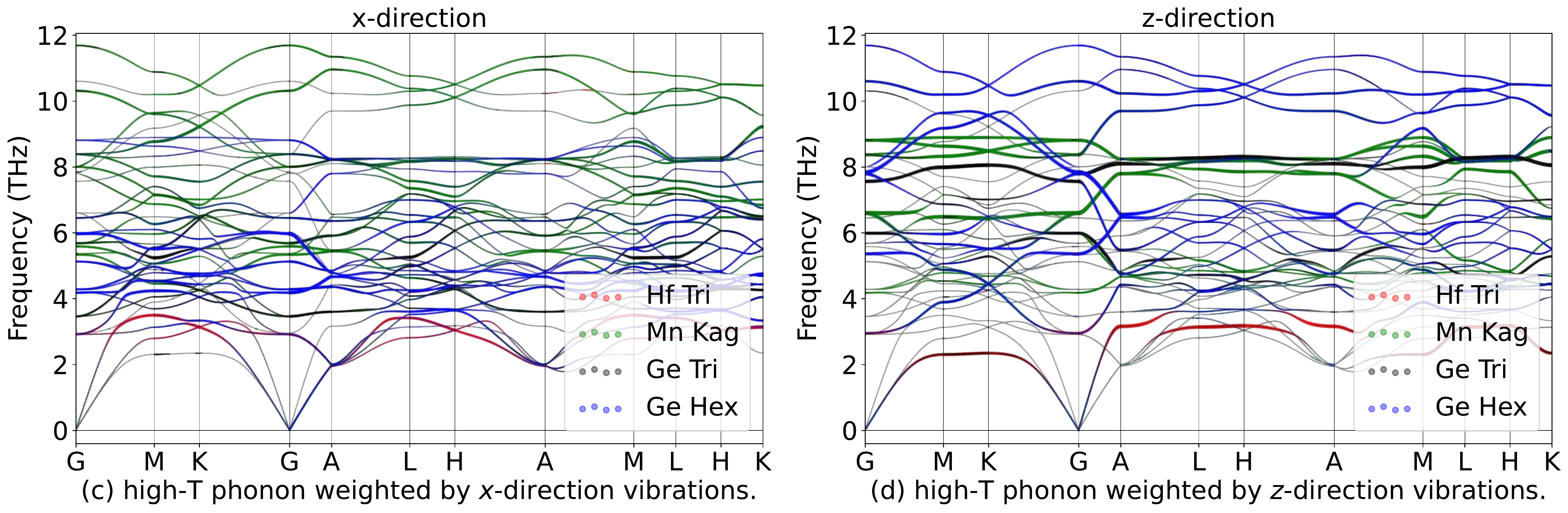}
\caption{Phonon spectrum for HfMn$_6$Ge$_6$in in-plane ($x$) and out-of-plane ($z$) directions in paramagnetic phase.}
\label{fig:HfMn6Ge6phoxyz}
\end{figure}

\begin{table}[htbp]
\global\long\def\arraystretch{1.12}
\captionsetup{width=.95\textwidth}
\caption{IRREPs at high-symmetry points for HfMn$_6$Ge$_6$ phonon spectrum at Low-T.}
\label{tab:191_HfMn6Ge6_Low}
\setlength{\tabcolsep}{1.mm}{
}
\end{table}

\begin{table}[htbp]
\global\long\def\arraystretch{1.12}
\captionsetup{width=.95\textwidth}
\caption{IRREPs at high-symmetry points for HfMn$_6$Ge$_6$ phonon spectrum at High-T.}
\label{tab:191_HfMn6Ge6_High}
\setlength{\tabcolsep}{1.mm}{
%
}
\end{table}
\subsubsection{\label{sms2sec:TaMn6Ge6}\hyperref[smsec:col]{\texorpdfstring{TaMn$_6$Ge$_6$}{}}~{\color{green}  (High-T stable, Low-T stable) }}

\begin{table}[htbp]
\global\long\def\arraystretch{1.12}
\captionsetup{width=.85\textwidth}
\caption{Structure parameters of TaMn$_6$Ge$_6$ after relaxation. It contains 415 electrons. }
\label{tab:TaMn6Ge6}
\setlength{\tabcolsep}{4mm}{
%
}
\end{table}

\begin{figure}[htbp]
\centering
\includegraphics[width=0.85\textwidth]{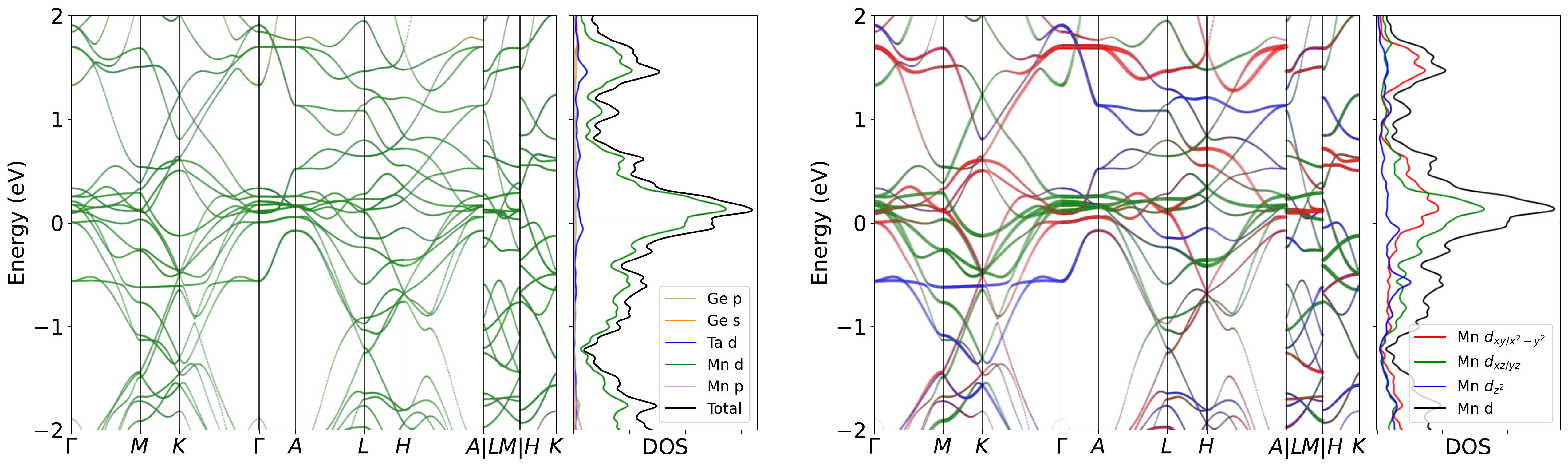}
\caption{Band structure for TaMn$_6$Ge$_6$ without SOC in paramagnetic phase. The projection of Mn $3d$ orbitals is also presented. The band structures clearly reveal the dominating of Mn $3d$ orbitals  near the Fermi level, as well as the pronounced peak.}
\label{fig:TaMn6Ge6band}
\end{figure}

\begin{figure}[H]
\centering
\subfloat[Relaxed phonon at low-T.]{
\label{fig:TaMn6Ge6_relaxed_Low}
\includegraphics[width=0.45\textwidth]{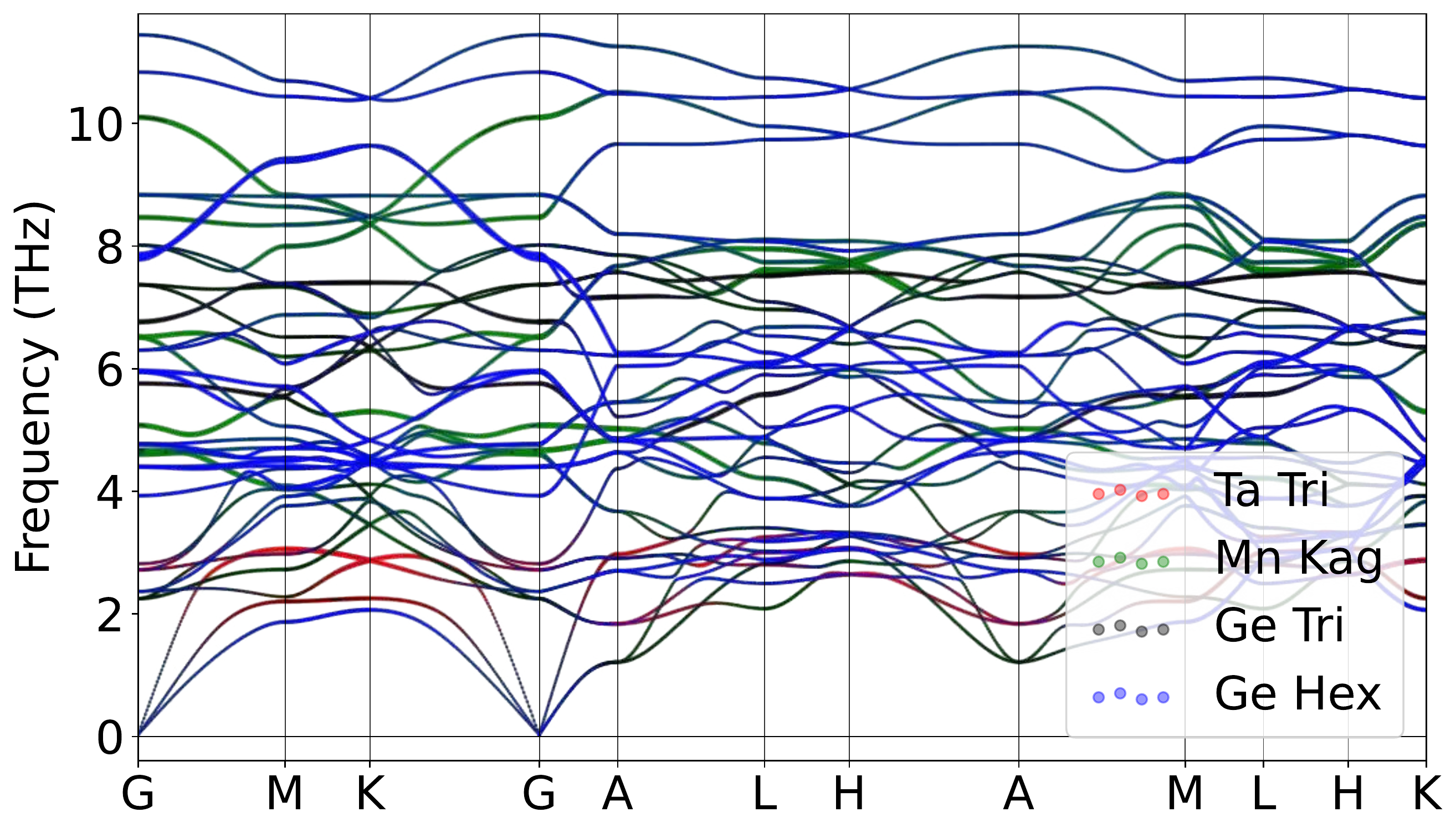}
}
\hspace{5pt}\subfloat[Relaxed phonon at high-T.]{
\label{fig:TaMn6Ge6_relaxed_High}
\includegraphics[width=0.45\textwidth]{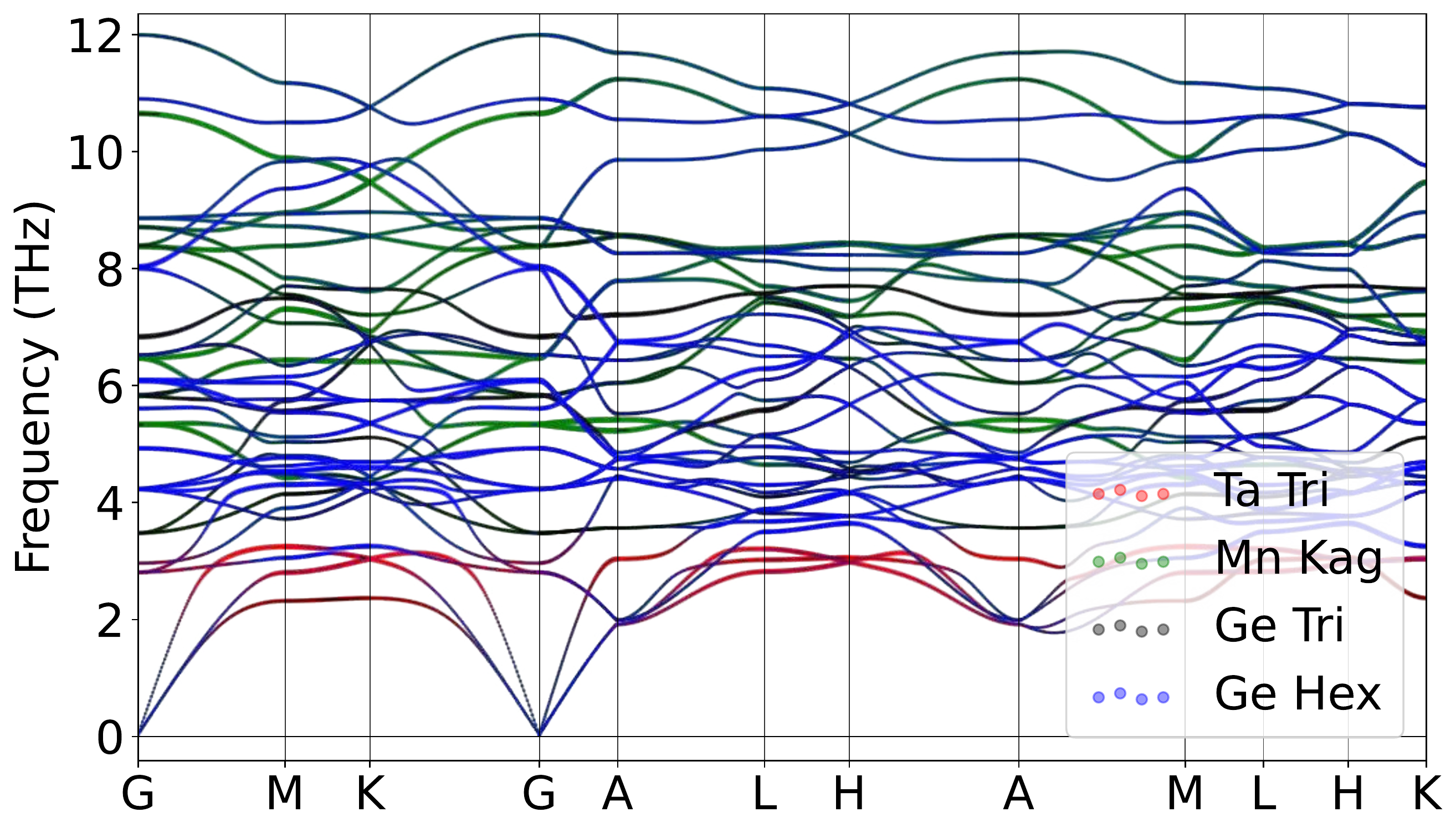}
}
\caption{Atomically projected phonon spectrum for TaMn$_6$Ge$_6$ in paramagnetic phase.}
\label{fig:TaMn6Ge6pho}
\end{figure}

\begin{figure}[htbp]
\centering
\includegraphics[width=0.45\textwidth]{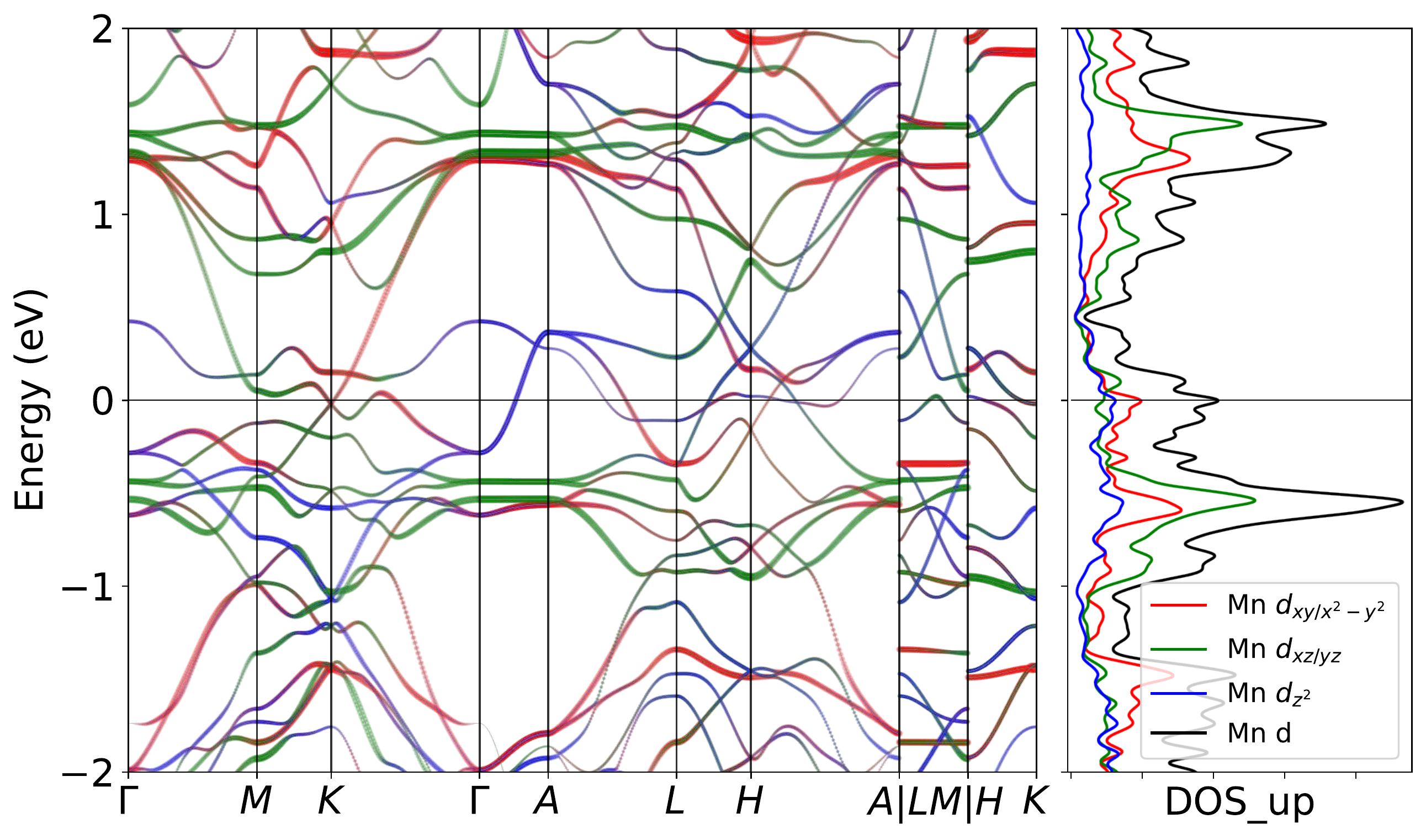}
\includegraphics[width=0.45\textwidth]{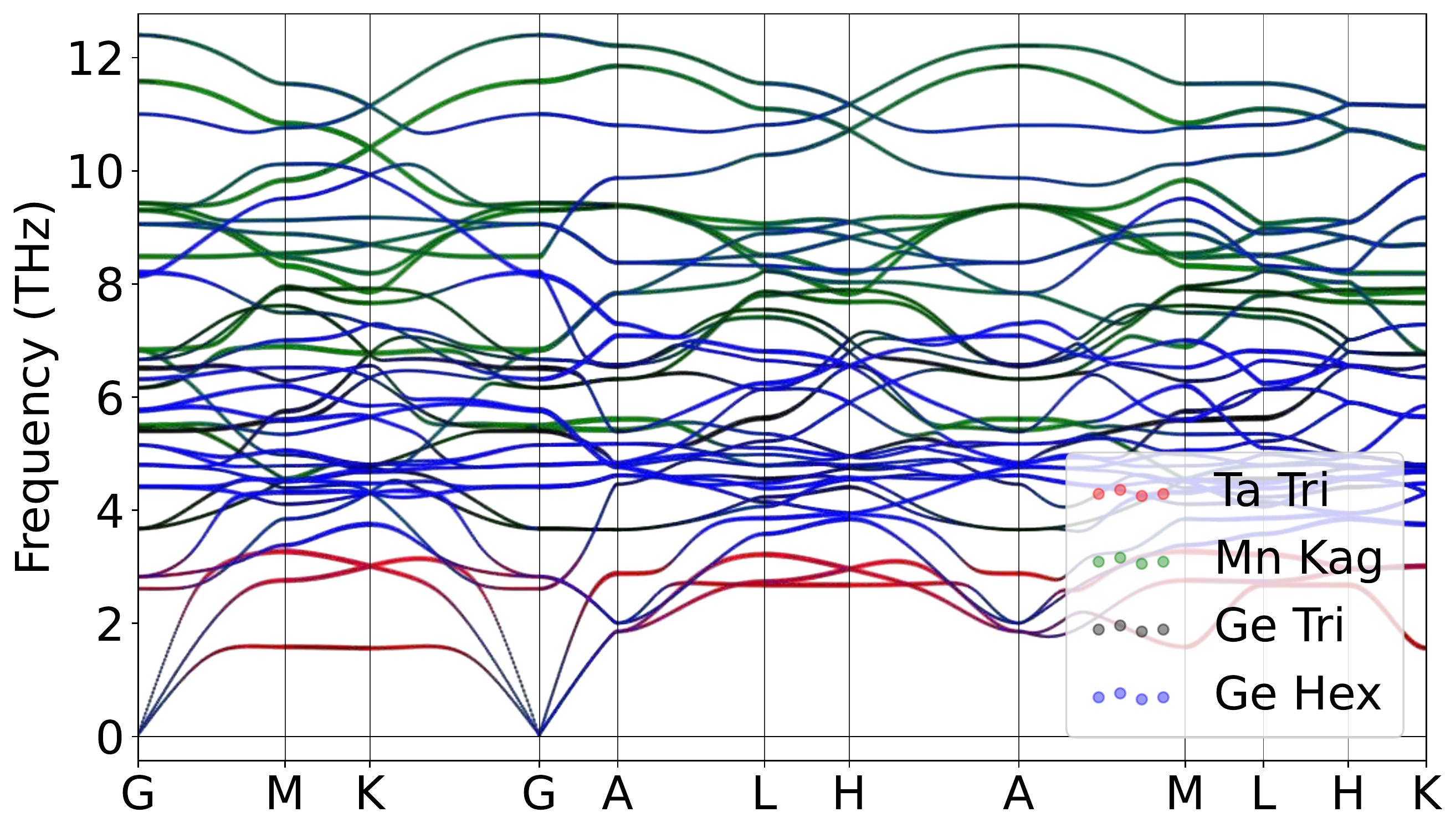}
\caption{Band structure for TaMn$_6$Ge$_6$ with AFM order without SOC for spin (Left)up channel. The projection of Mn $3d$ orbitals is also presented. (Right)Correspondingly, a stable phonon was demonstrated with the collinear magnetic order without Hubbard U at lower temperature regime, weighted by atomic projections.}
\label{fig:TaMn6Ge6mag}
\end{figure}

\begin{figure}[H]
\centering
\label{fig:TaMn6Ge6_relaxed_Low_xz}
\includegraphics[width=0.85\textwidth]{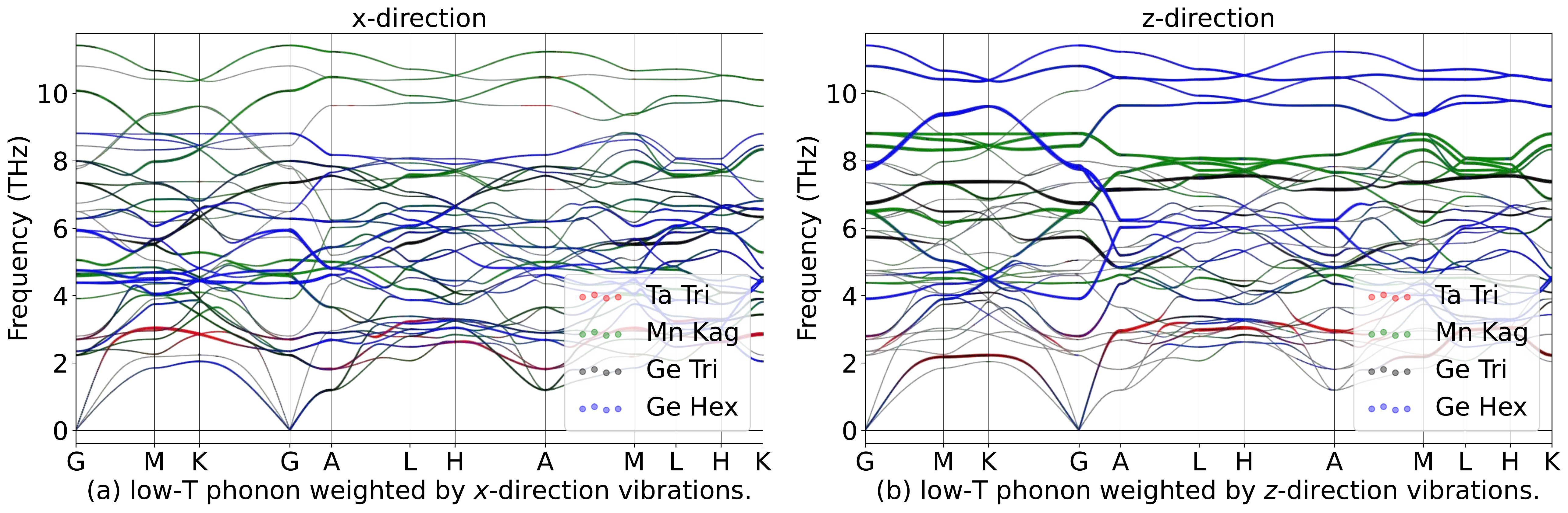}
\hspace{5pt}\label{fig:TaMn6Ge6_relaxed_High_xz}
\includegraphics[width=0.85\textwidth]{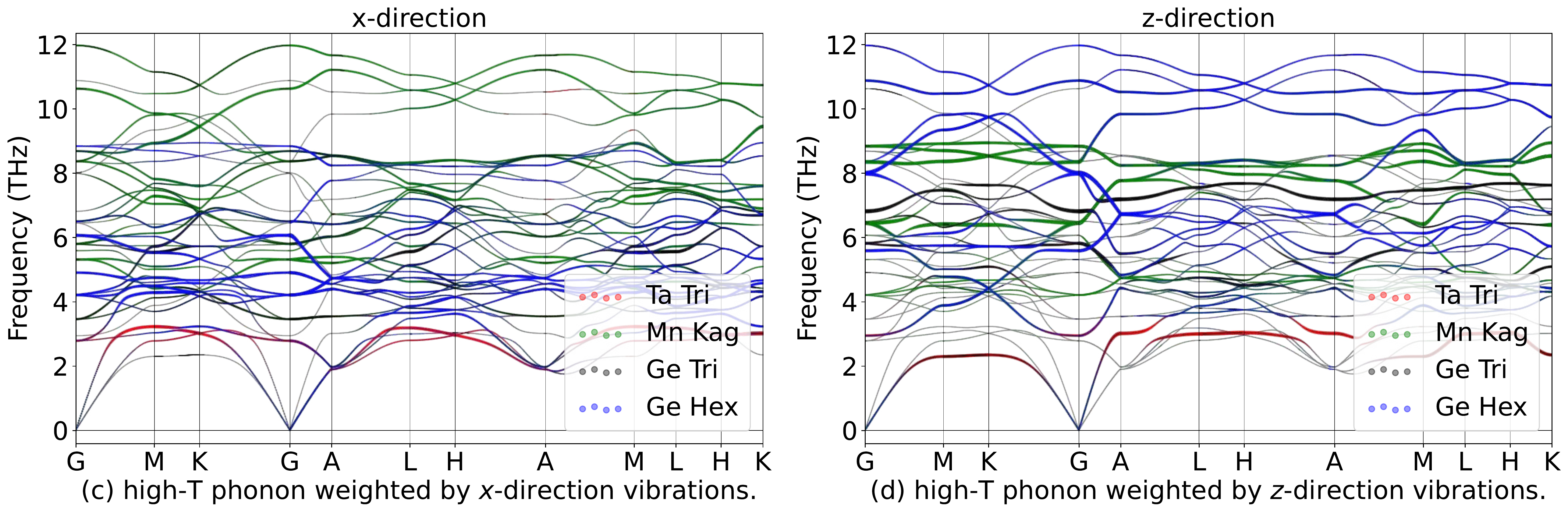}
\caption{Phonon spectrum for TaMn$_6$Ge$_6$in in-plane ($x$) and out-of-plane ($z$) directions in paramagnetic phase.}
\label{fig:TaMn6Ge6phoxyz}
\end{figure}

\begin{table}[htbp]
\global\long\def\arraystretch{1.12}
\captionsetup{width=.95\textwidth}
\caption{IRREPs at high-symmetry points for TaMn$_6$Ge$_6$ phonon spectrum at Low-T.}
\label{tab:191_TaMn6Ge6_Low}
\setlength{\tabcolsep}{1.mm}{
}
\end{table}

\begin{table}[htbp]
\global\long\def\arraystretch{1.12}
\captionsetup{width=.95\textwidth}
\caption{IRREPs at high-symmetry points for TaMn$_6$Ge$_6$ phonon spectrum at High-T.}
\label{tab:191_TaMn6Ge6_High}
\setlength{\tabcolsep}{1.mm}{
%
}
\end{table}
\subsection{\hyperref[smsec:col]{MnSn}}~\label{smssec:MnSn}

\subsubsection{\label{sms2sec:LiMn6Sn6}\hyperref[smsec:col]{\texorpdfstring{LiMn$_6$Sn$_6$}{}}~{\color{blue}  (High-T stable, Low-T unstable) }}

\begin{table}[htbp]
\global\long\def\arraystretch{1.12}
\captionsetup{width=.85\textwidth}
\caption{Structure parameters of LiMn$_6$Sn$_6$ after relaxation. It contains 453 electrons. }
\label{tab:LiMn6Sn6}
\setlength{\tabcolsep}{4mm}{
%
}
\end{table}

\begin{figure}[htbp]
\centering
\includegraphics[width=0.85\textwidth]{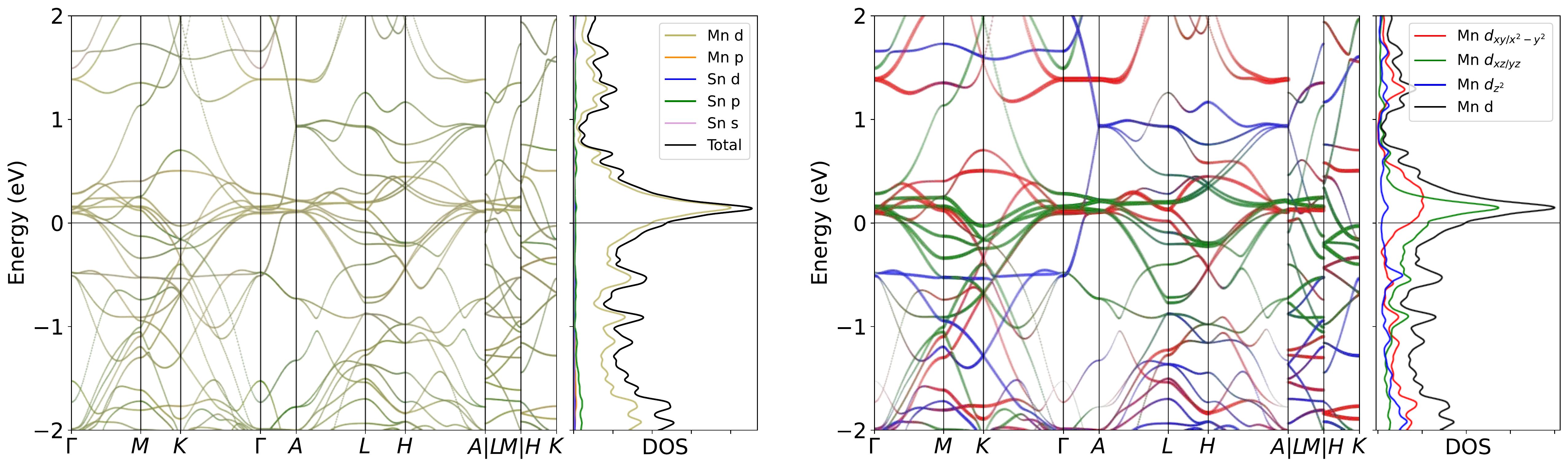}
\caption{Band structure for LiMn$_6$Sn$_6$ without SOC in paramagnetic phase. The projection of Mn $3d$ orbitals is also presented. The band structures clearly reveal the dominating of Mn $3d$ orbitals  near the Fermi level, as well as the pronounced peak.}
\label{fig:LiMn6Sn6band}
\end{figure}

\begin{figure}[H]
\centering
\subfloat[Relaxed phonon at low-T.]{
\label{fig:LiMn6Sn6_relaxed_Low}
\includegraphics[width=0.45\textwidth]{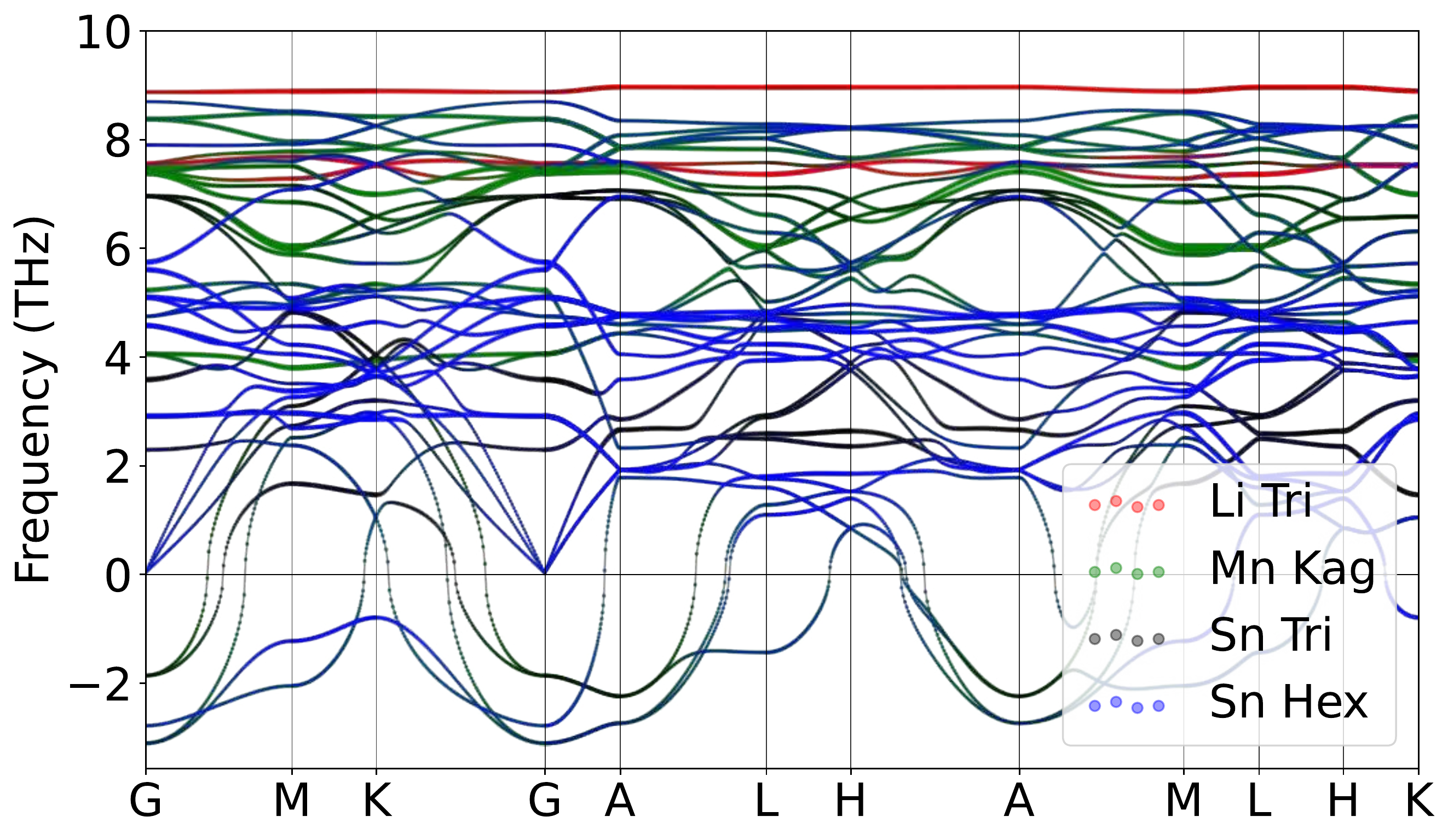}
}
\hspace{5pt}\subfloat[Relaxed phonon at high-T.]{
\label{fig:LiMn6Sn6_relaxed_High}
\includegraphics[width=0.45\textwidth]{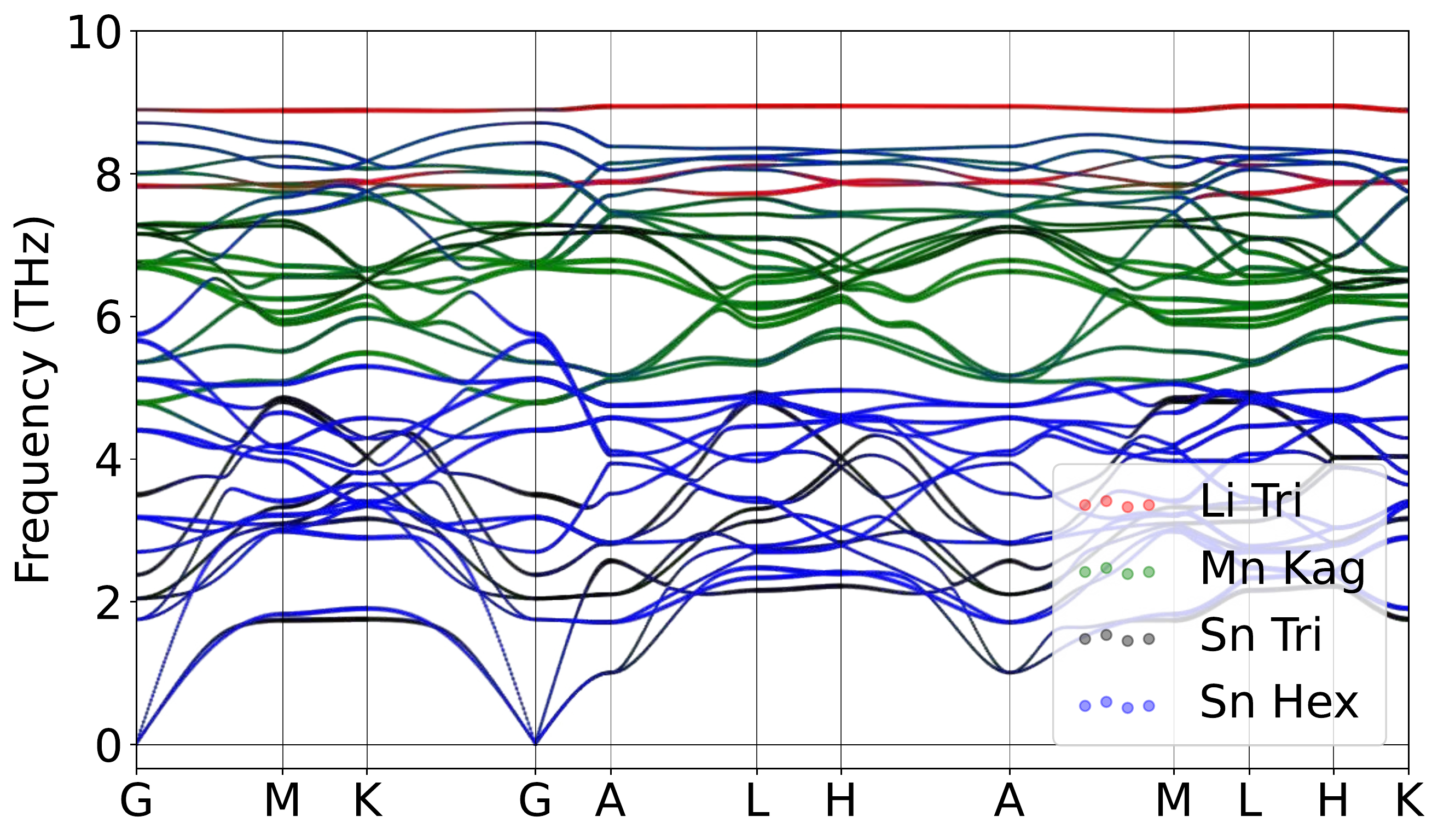}
}
\caption{Atomically projected phonon spectrum for LiMn$_6$Sn$_6$ in paramagnetic phase.}
\label{fig:LiMn6Sn6pho}
\end{figure}

\begin{figure}[htbp]
\centering
\includegraphics[width=0.32\textwidth]{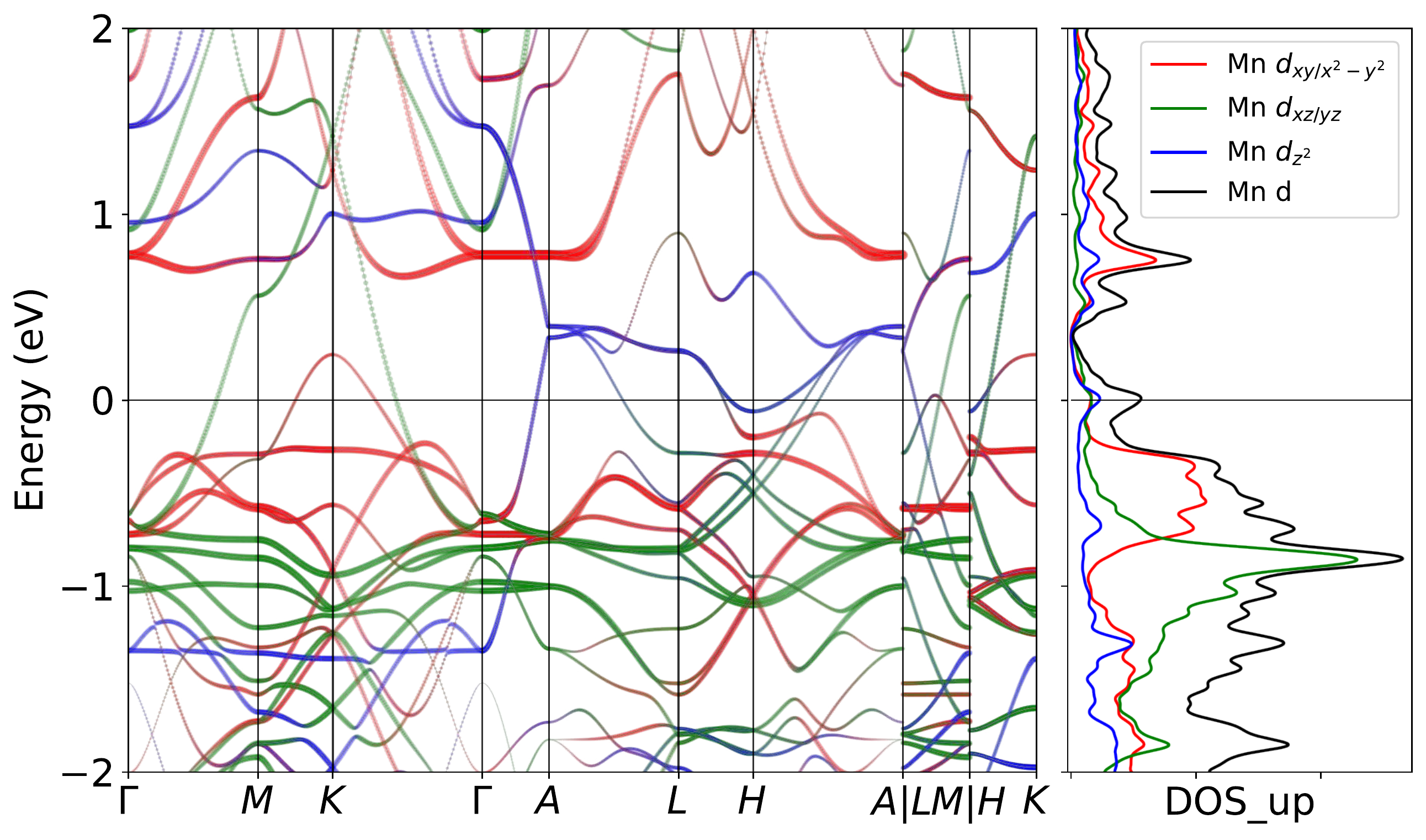}
\includegraphics[width=0.32\textwidth]{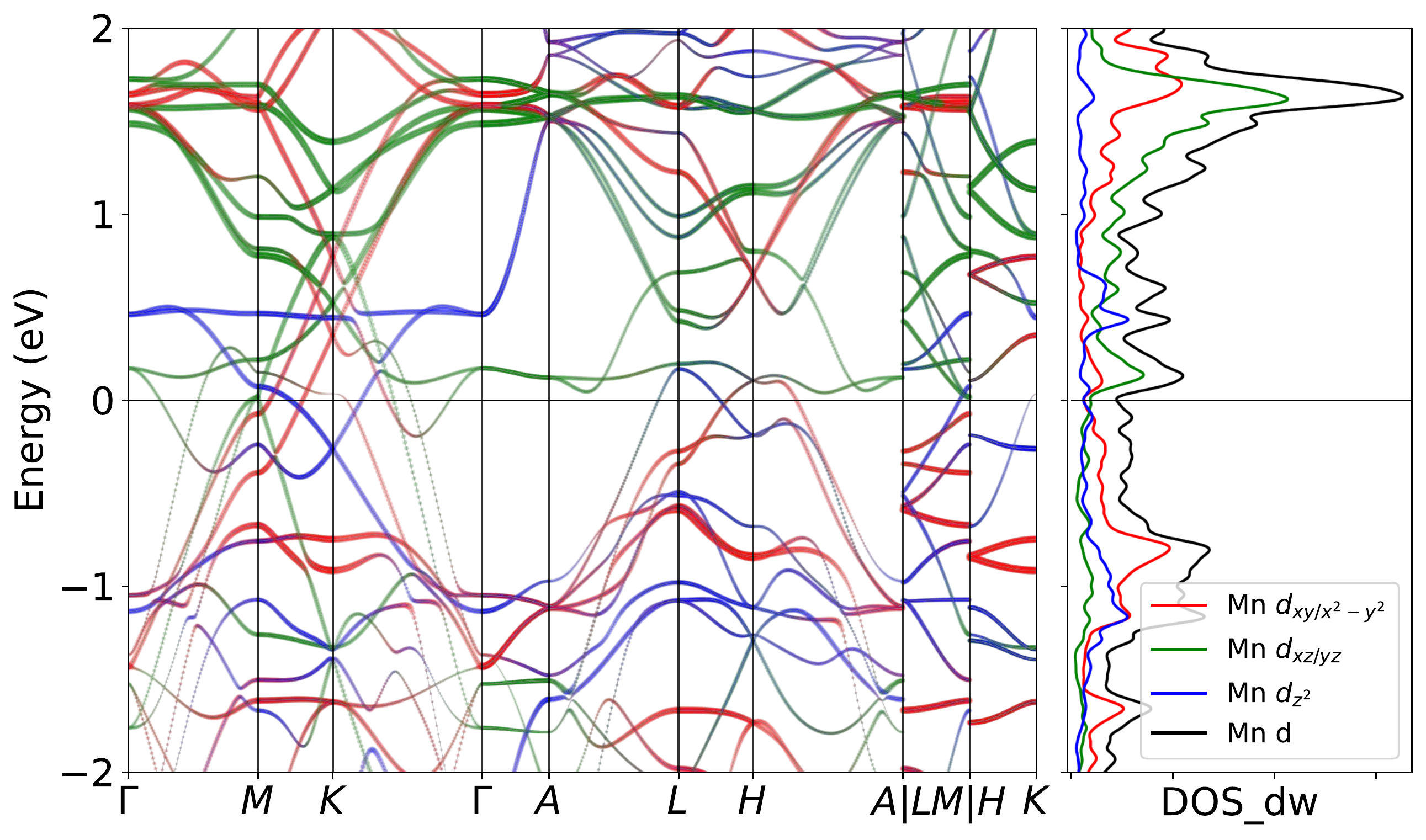}
\includegraphics[width=0.32\textwidth]{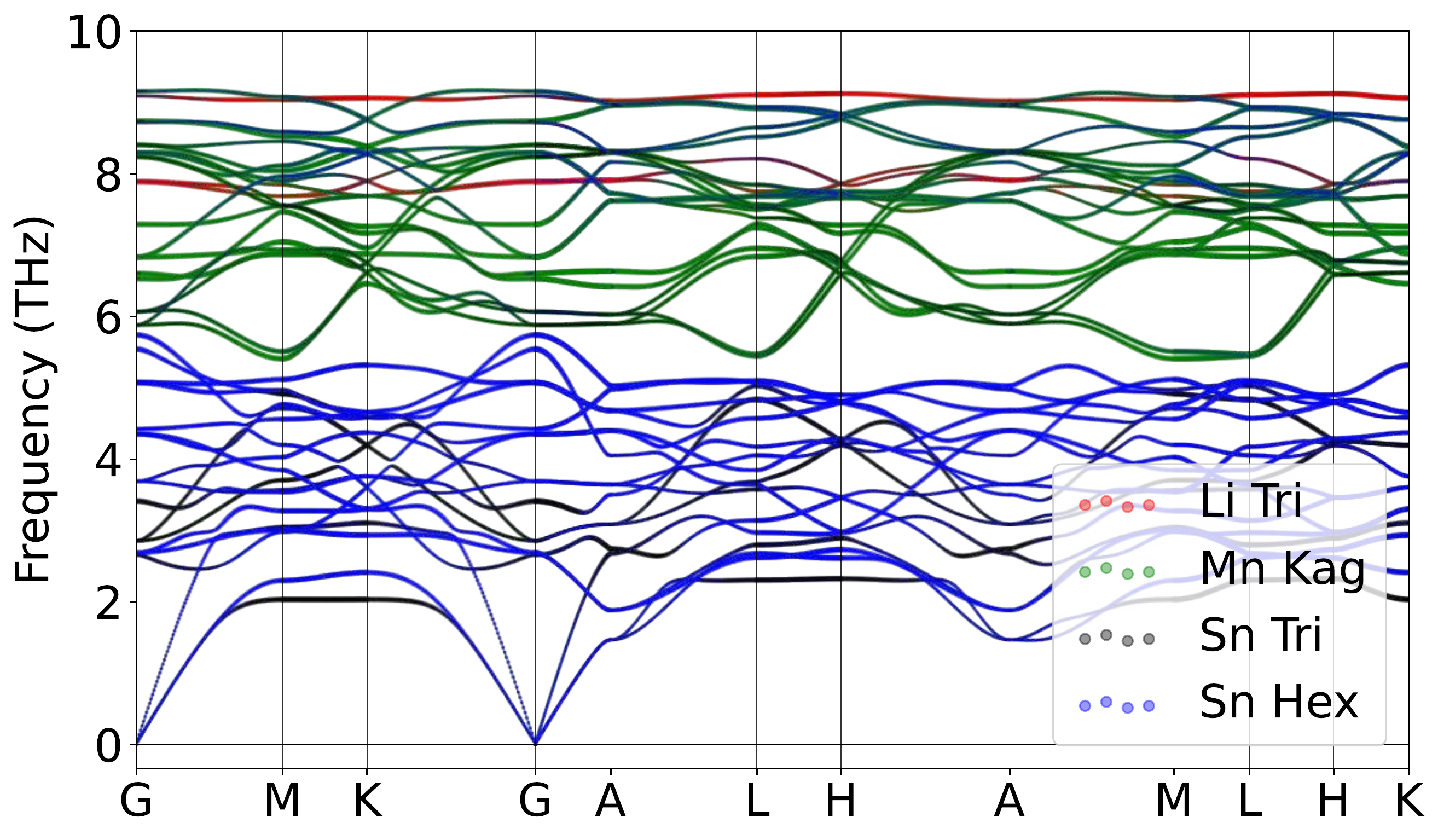}
\caption{Band structure for LiMn$_6$Sn$_6$ with FM order without SOC for spin (Left)up and (Middle)down channels. The projection of Mn $3d$ orbitals is also presented. (Right)Correspondingly, a stable phonon was demonstrated with the collinear magnetic order without Hubbard U at lower temperature regime, weighted by atomic projections.}
\label{fig:LiMn6Sn6mag}
\end{figure}

\begin{figure}[H]
\centering
\label{fig:LiMn6Sn6_relaxed_Low_xz}
\includegraphics[width=0.85\textwidth]{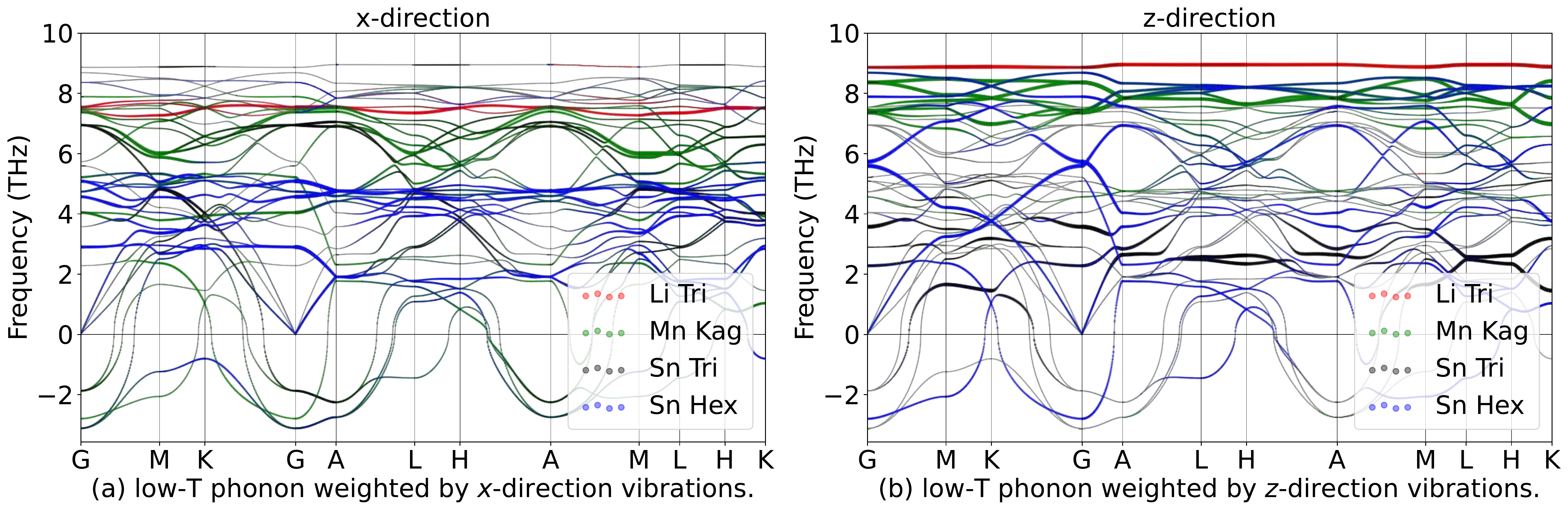}
\hspace{5pt}\label{fig:LiMn6Sn6_relaxed_High_xz}
\includegraphics[width=0.85\textwidth]{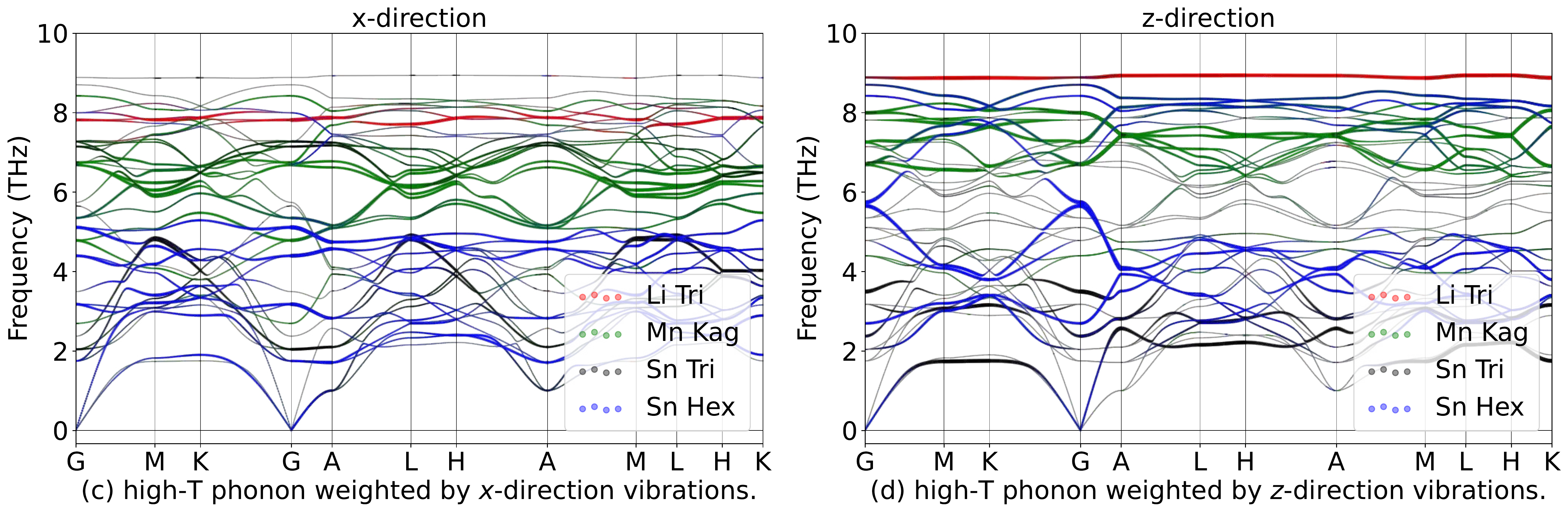}
\caption{Phonon spectrum for LiMn$_6$Sn$_6$in in-plane ($x$) and out-of-plane ($z$) directions in paramagnetic phase.}
\label{fig:LiMn6Sn6phoxyz}
\end{figure}

\begin{table}[htbp]
\global\long\def\arraystretch{1.12}
\captionsetup{width=.95\textwidth}
\caption{IRREPs at high-symmetry points for LiMn$_6$Sn$_6$ phonon spectrum at Low-T.}
\label{tab:191_LiMn6Sn6_Low}
\setlength{\tabcolsep}{1.mm}{
}
\end{table}

\begin{table}[htbp]
\global\long\def\arraystretch{1.12}
\captionsetup{width=.95\textwidth}
\caption{IRREPs at high-symmetry points for LiMn$_6$Sn$_6$ phonon spectrum at High-T.}
\label{tab:191_LiMn6Sn6_High}
\setlength{\tabcolsep}{1.mm}{
%
}
\end{table}
\subsubsection{\label{sms2sec:MgMn6Sn6}\hyperref[smsec:col]{\texorpdfstring{MgMn$_6$Sn$_6$}{}}~{\color{blue}  (High-T stable, Low-T unstable) }}

\begin{table}[htbp]
\global\long\def\arraystretch{1.12}
\captionsetup{width=.85\textwidth}
\caption{Structure parameters of MgMn$_6$Sn$_6$ after relaxation. It contains 462 electrons. }
\label{tab:MgMn6Sn6}
\setlength{\tabcolsep}{4mm}{
%
}
\end{table}

\begin{figure}[htbp]
\centering
\includegraphics[width=0.85\textwidth]{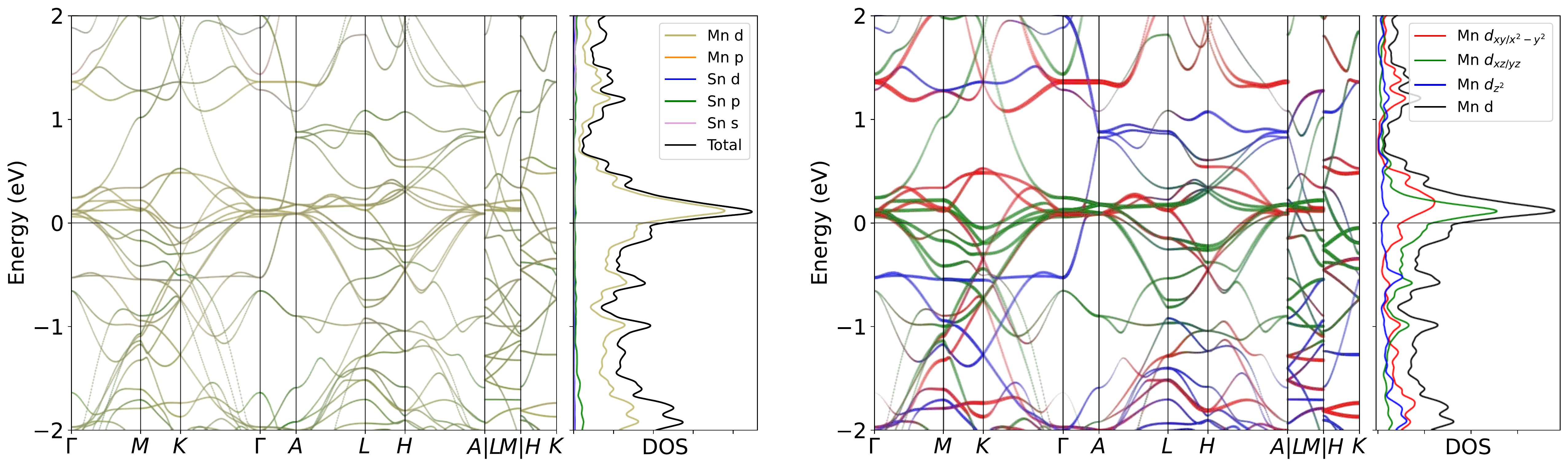}
\caption{Band structure for MgMn$_6$Sn$_6$ without SOC in paramagnetic phase. The projection of Mn $3d$ orbitals is also presented. The band structures clearly reveal the dominating of Mn $3d$ orbitals  near the Fermi level, as well as the pronounced peak.}
\label{fig:MgMn6Sn6band}
\end{figure}

\begin{figure}[H]
\centering
\subfloat[Relaxed phonon at low-T.]{
\label{fig:MgMn6Sn6_relaxed_Low}
\includegraphics[width=0.45\textwidth]{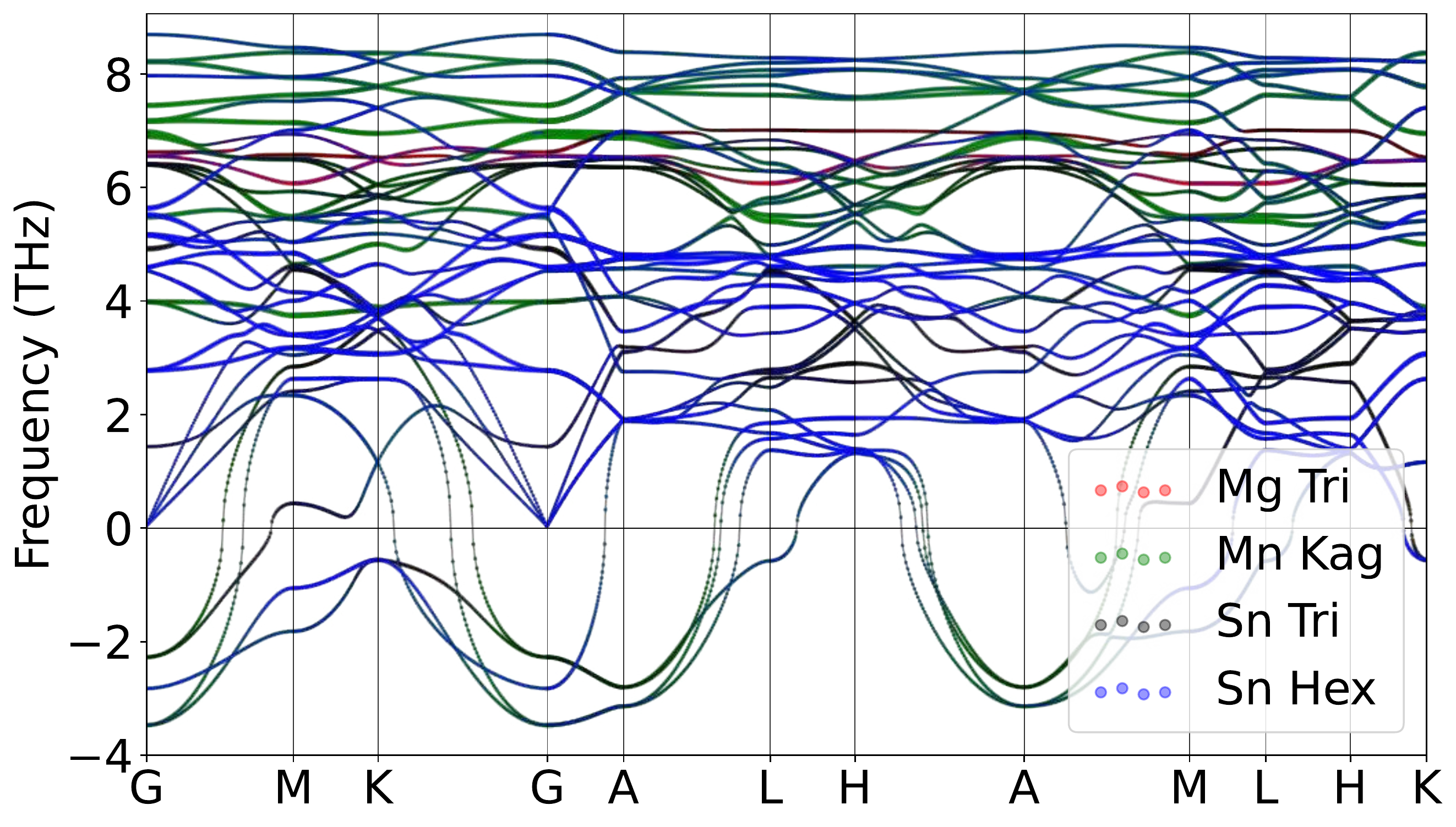}
}
\hspace{5pt}\subfloat[Relaxed phonon at high-T.]{
\label{fig:MgMn6Sn6_relaxed_High}
\includegraphics[width=0.45\textwidth]{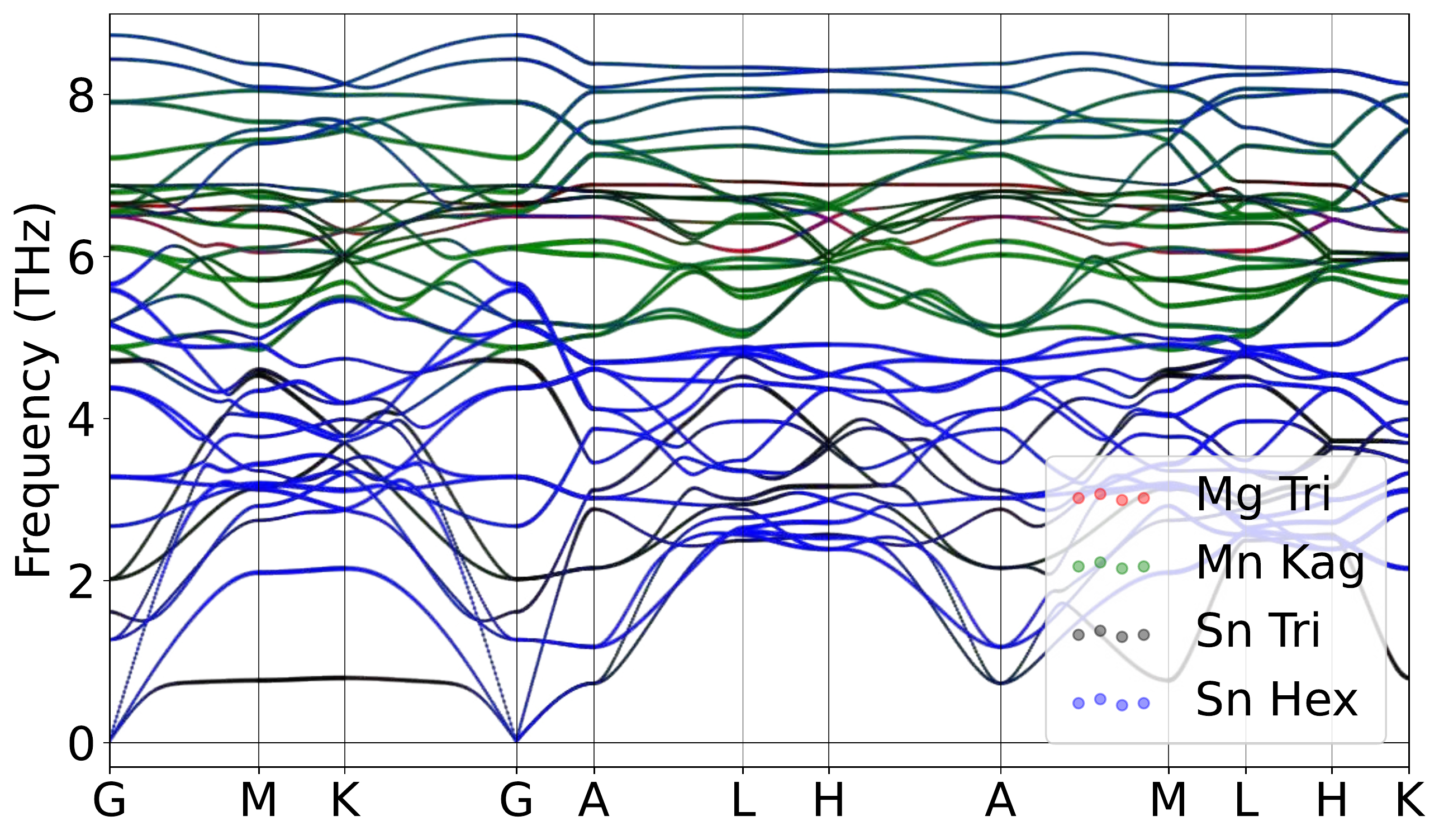}
}
\caption{Atomically projected phonon spectrum for MgMn$_6$Sn$_6$ in paramagnetic phase.}
\label{fig:MgMn6Sn6pho}
\end{figure}

\begin{figure}[htbp]
\centering
\includegraphics[width=0.32\textwidth]{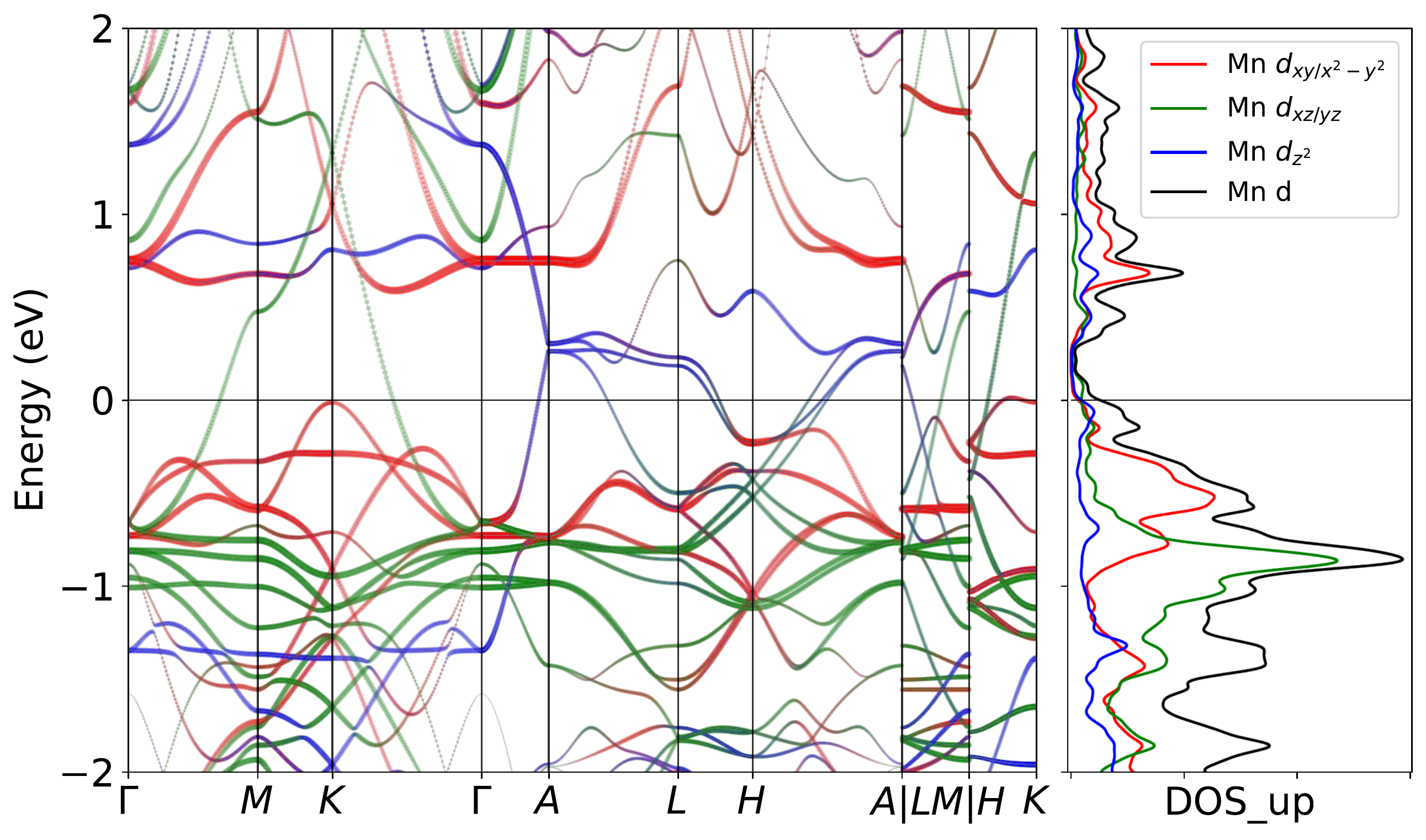}
\includegraphics[width=0.32\textwidth]{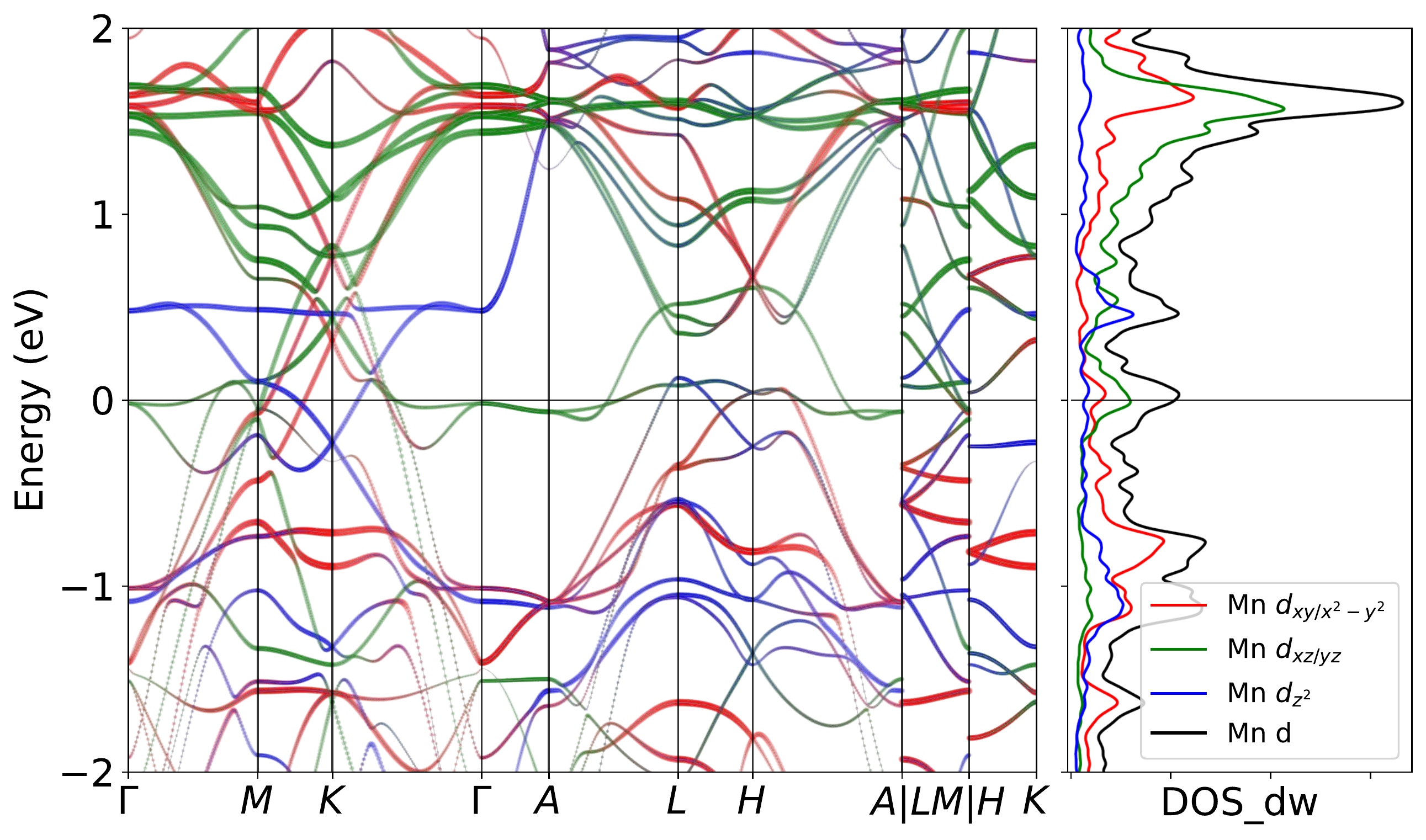}
\includegraphics[width=0.32\textwidth]{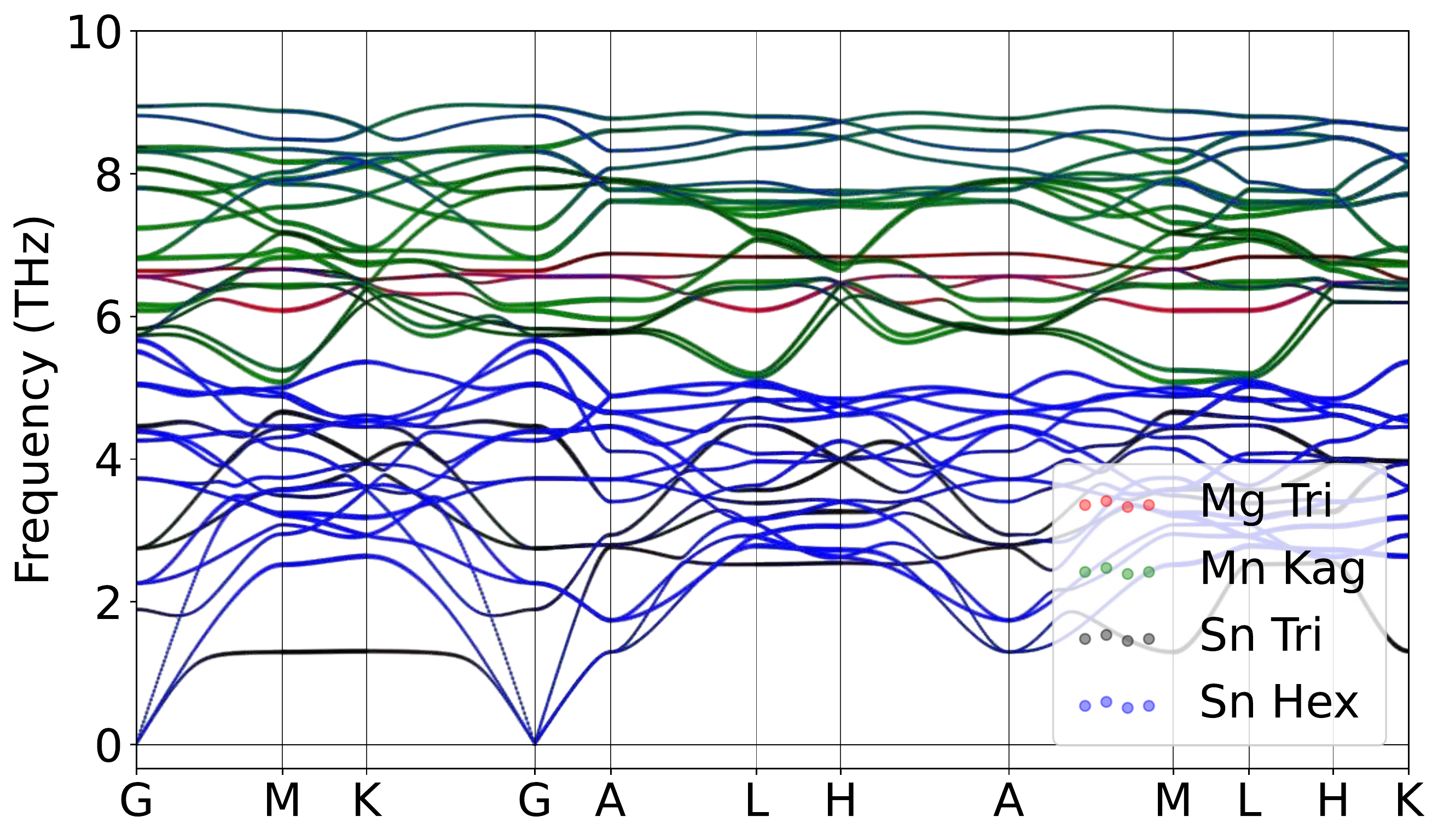}
\caption{Band structure for MgMn$_6$Sn$_6$ with FM order without SOC for spin (Left)up and (Middle)down channels. The projection of Mn $3d$ orbitals is also presented. (Right)Correspondingly, a stable phonon was demonstrated with the collinear magnetic order without Hubbard U at lower temperature regime, weighted by atomic projections.}
\label{fig:MgMn6Sn6mag}
\end{figure}

\begin{figure}[H]
\centering
\label{fig:MgMn6Sn6_relaxed_Low_xz}
\includegraphics[width=0.85\textwidth]{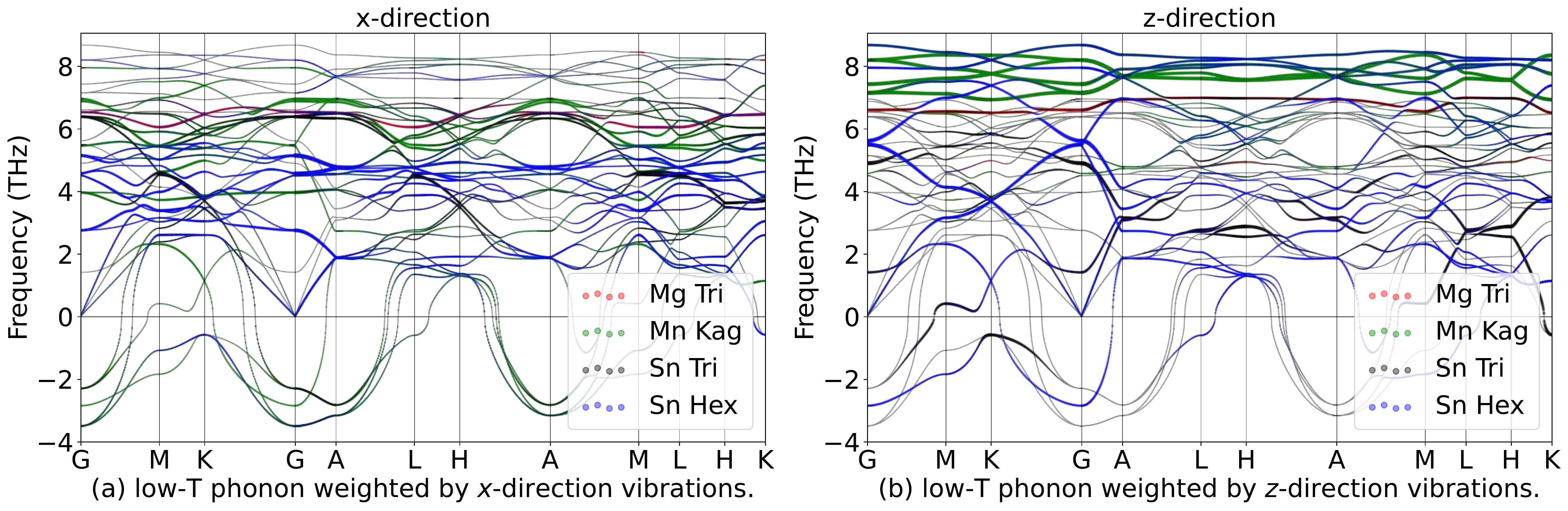}
\hspace{5pt}\label{fig:MgMn6Sn6_relaxed_High_xz}
\includegraphics[width=0.85\textwidth]{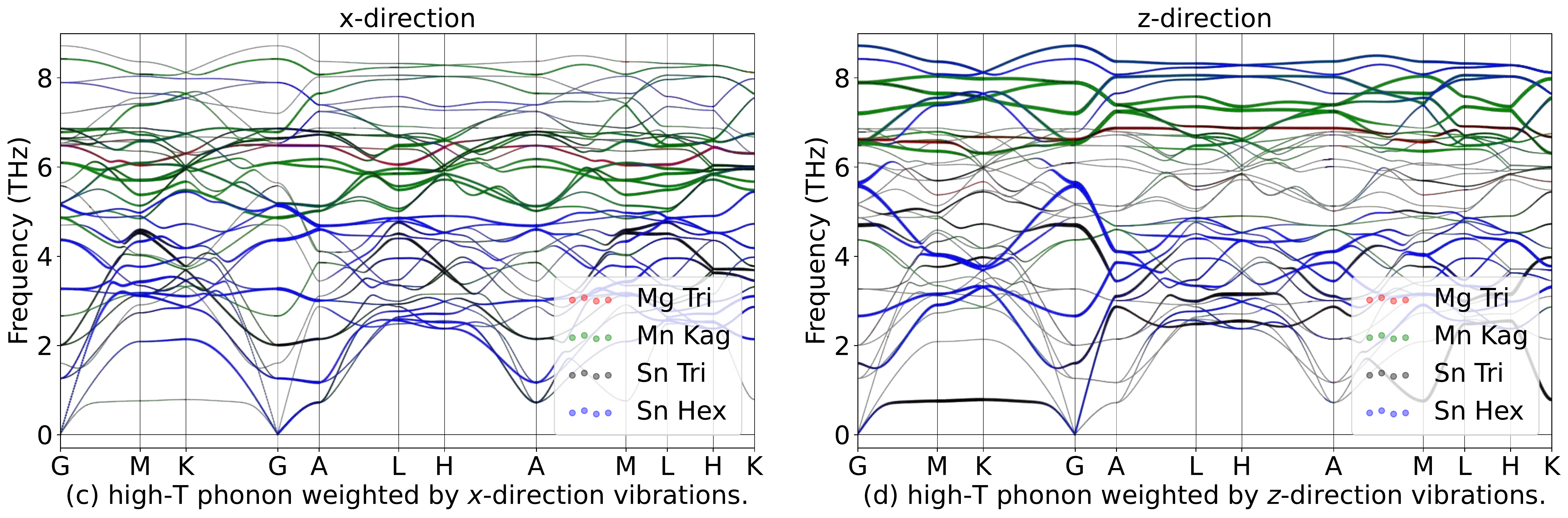}
\caption{Phonon spectrum for MgMn$_6$Sn$_6$in in-plane ($x$) and out-of-plane ($z$) directions in paramagnetic phase.}
\label{fig:MgMn6Sn6phoxyz}
\end{figure}

\begin{table}[htbp]
\global\long\def\arraystretch{1.12}
\captionsetup{width=.95\textwidth}
\caption{IRREPs at high-symmetry points for MgMn$_6$Sn$_6$ phonon spectrum at Low-T.}
\label{tab:191_MgMn6Sn6_Low}
\setlength{\tabcolsep}{1.mm}{
}
\end{table}

\begin{table}[htbp]
\global\long\def\arraystretch{1.12}
\captionsetup{width=.95\textwidth}
\caption{IRREPs at high-symmetry points for MgMn$_6$Sn$_6$ phonon spectrum at High-T.}
\label{tab:191_MgMn6Sn6_High}
\setlength{\tabcolsep}{1.mm}{
%
}
\end{table}
\subsubsection{\label{sms2sec:CaMn6Sn6}\hyperref[smsec:col]{\texorpdfstring{CaMn$_6$Sn$_6$}{}}~{\color{blue}  (High-T stable, Low-T unstable) }}

\begin{table}[htbp]
\global\long\def\arraystretch{1.12}
\captionsetup{width=.85\textwidth}
\caption{Structure parameters of CaMn$_6$Sn$_6$ after relaxation. It contains 470 electrons. }
\label{tab:CaMn6Sn6}
\setlength{\tabcolsep}{4mm}{
%
}
\end{table}

\begin{figure}[htbp]
\centering
\includegraphics[width=0.85\textwidth]{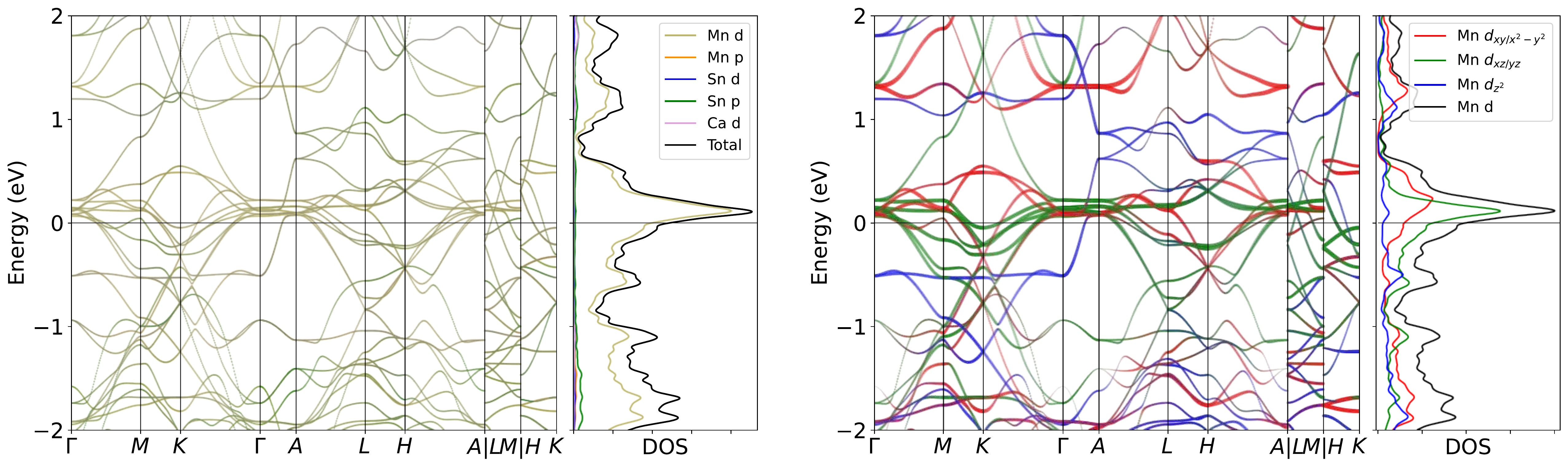}
\caption{Band structure for CaMn$_6$Sn$_6$ without SOC in paramagnetic phase. The projection of Mn $3d$ orbitals is also presented. The band structures clearly reveal the dominating of Mn $3d$ orbitals  near the Fermi level, as well as the pronounced peak.}
\label{fig:CaMn6Sn6band}
\end{figure}

\begin{figure}[H]
\centering
\subfloat[Relaxed phonon at low-T.]{
\label{fig:CaMn6Sn6_relaxed_Low}
\includegraphics[width=0.45\textwidth]{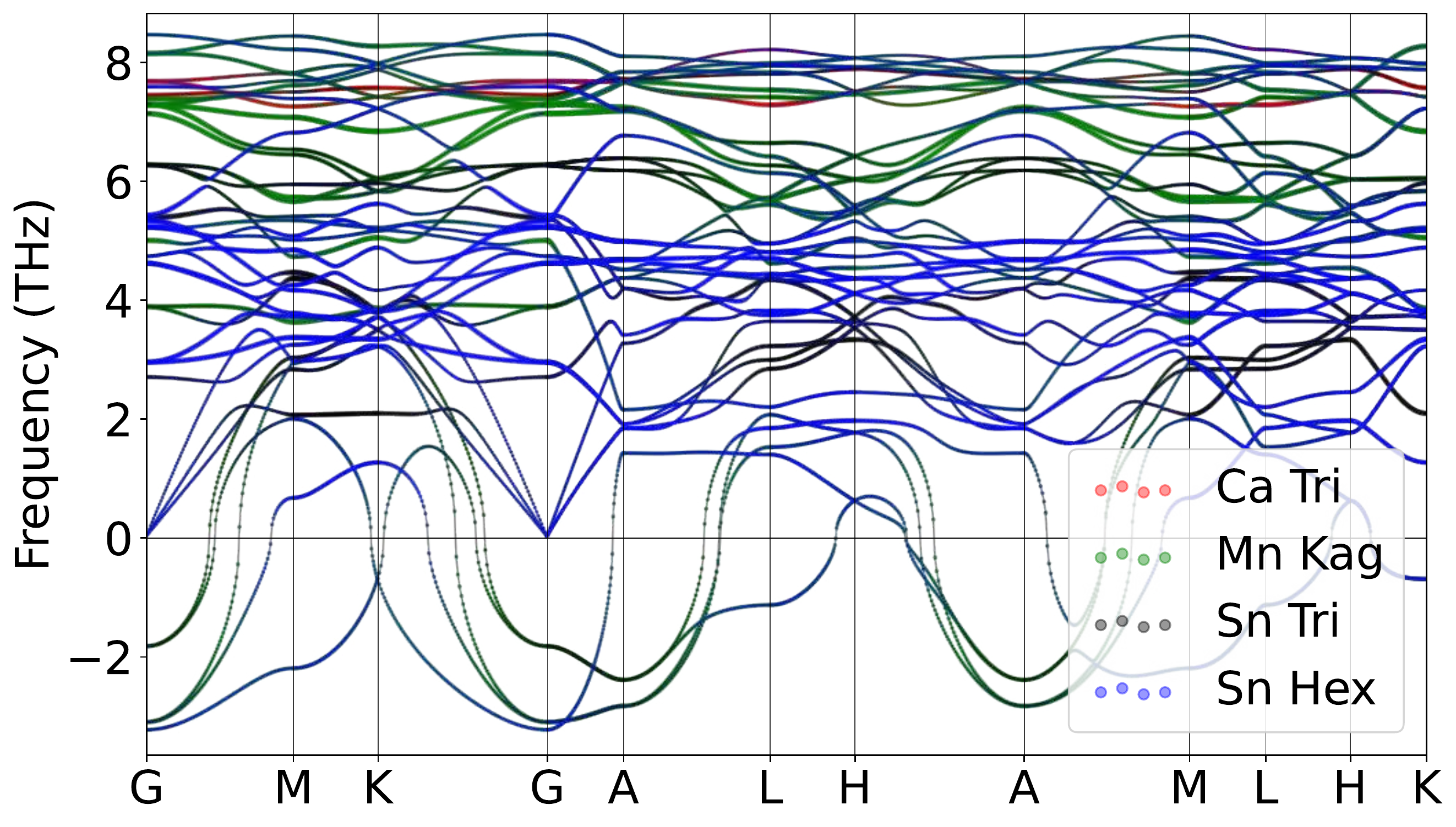}
}
\hspace{5pt}\subfloat[Relaxed phonon at high-T.]{
\label{fig:CaMn6Sn6_relaxed_High}
\includegraphics[width=0.45\textwidth]{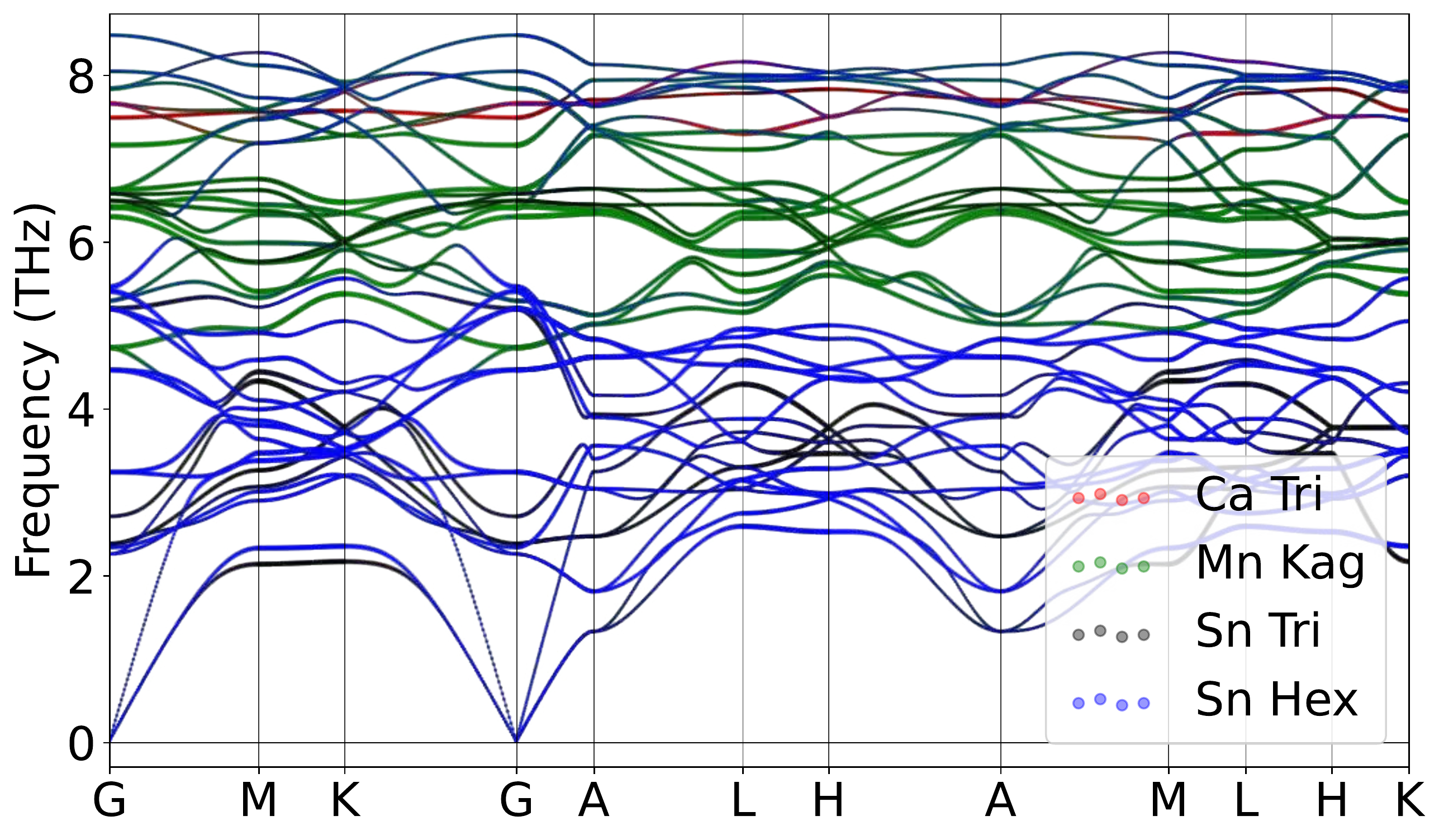}
}
\caption{Atomically projected phonon spectrum for CaMn$_6$Sn$_6$ in paramagnetic phase.}
\label{fig:CaMn6Sn6pho}
\end{figure}

\begin{figure}[htbp]
\centering
\includegraphics[width=0.32\textwidth]{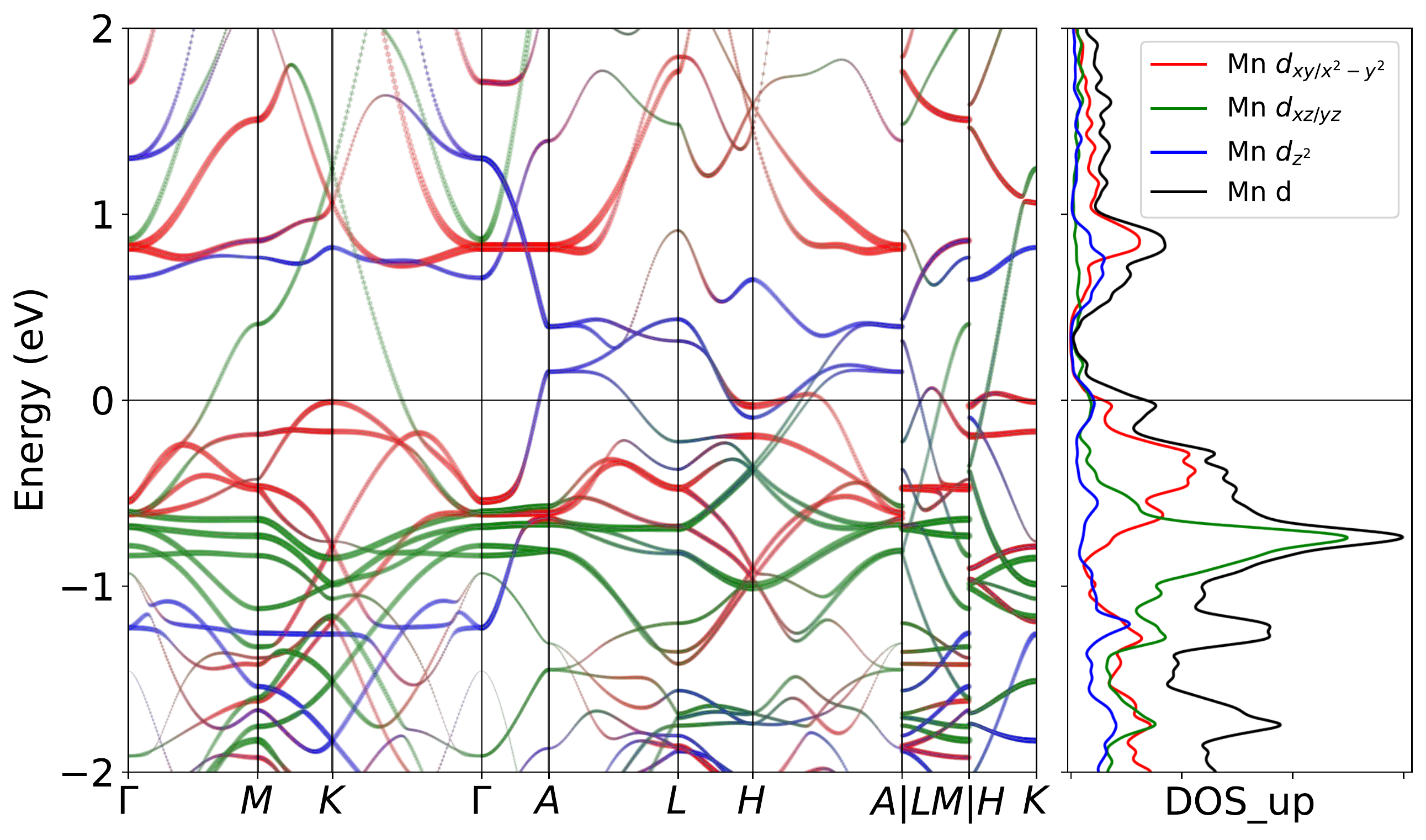}
\includegraphics[width=0.32\textwidth]{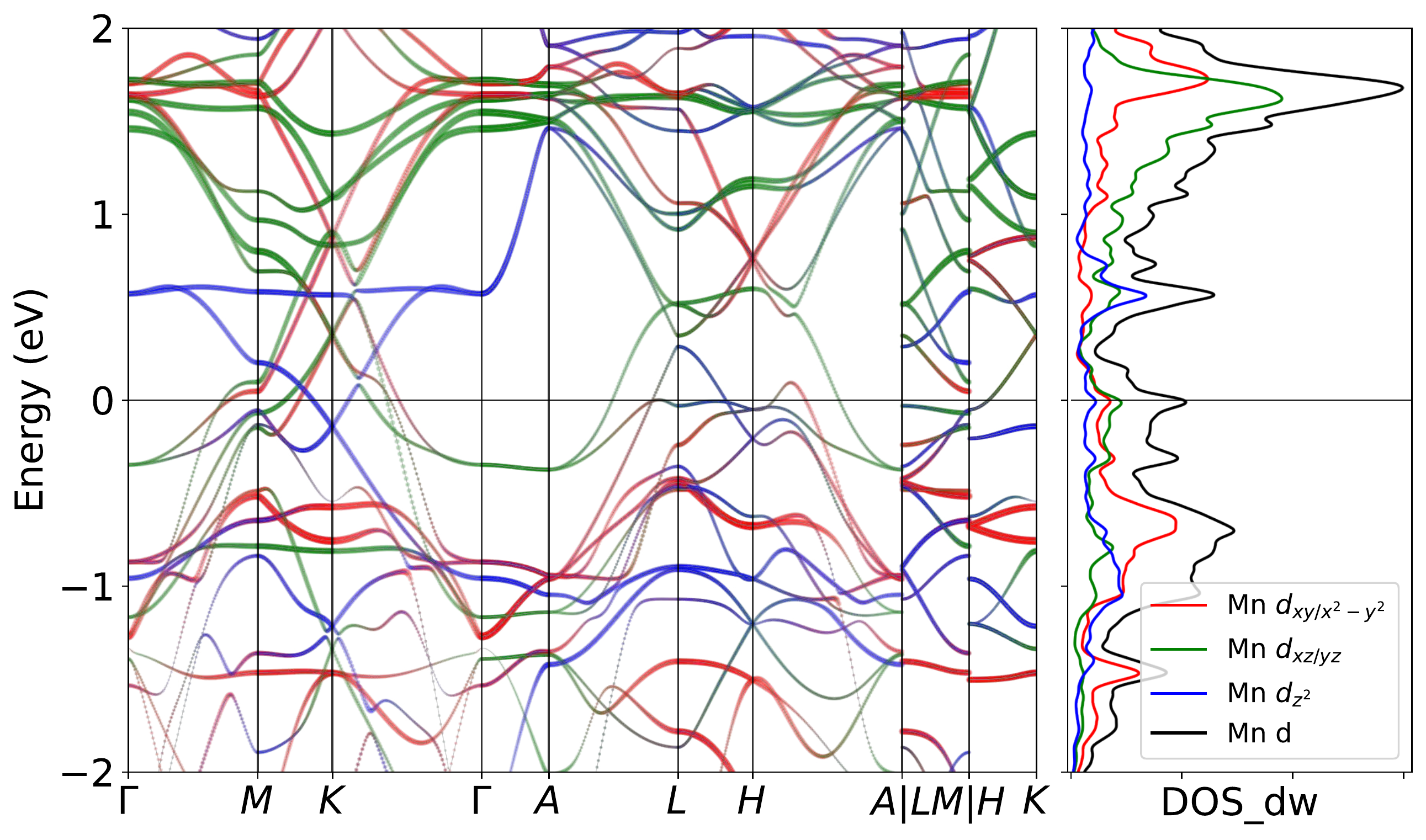}
\includegraphics[width=0.32\textwidth]{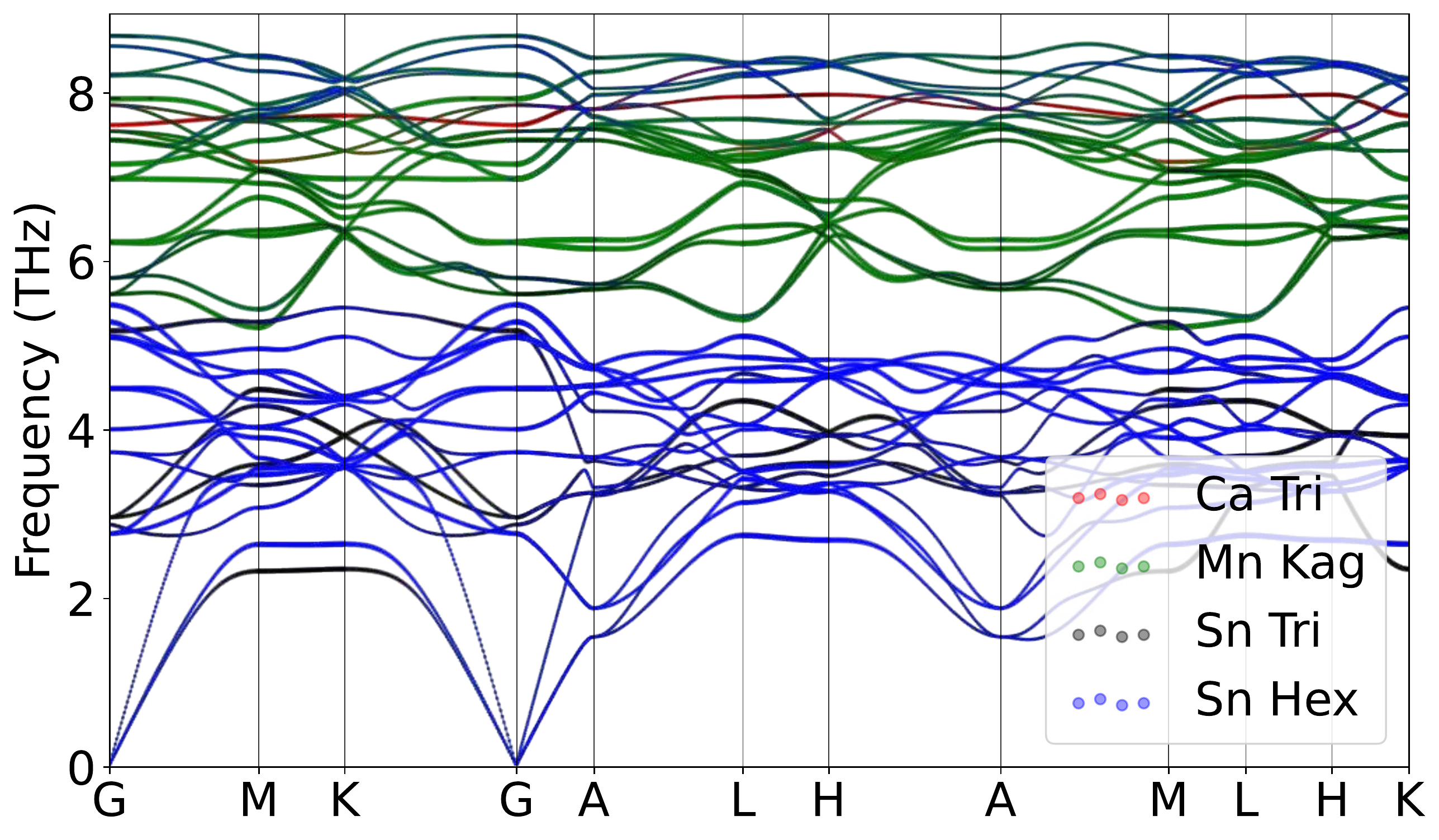}
\caption{Band structure for CaMn$_6$Sn$_6$ with FM order without SOC for spin (Left)up and (Middle)down channels. The projection of Mn $3d$ orbitals is also presented. (Right)Correspondingly, a stable phonon was demonstrated with the collinear magnetic order without Hubbard U at lower temperature regime, weighted by atomic projections.}
\label{fig:CaMn6Sn6mag}
\end{figure}

\begin{figure}[H]
\centering
\label{fig:CaMn6Sn6_relaxed_Low_xz}
\includegraphics[width=0.85\textwidth]{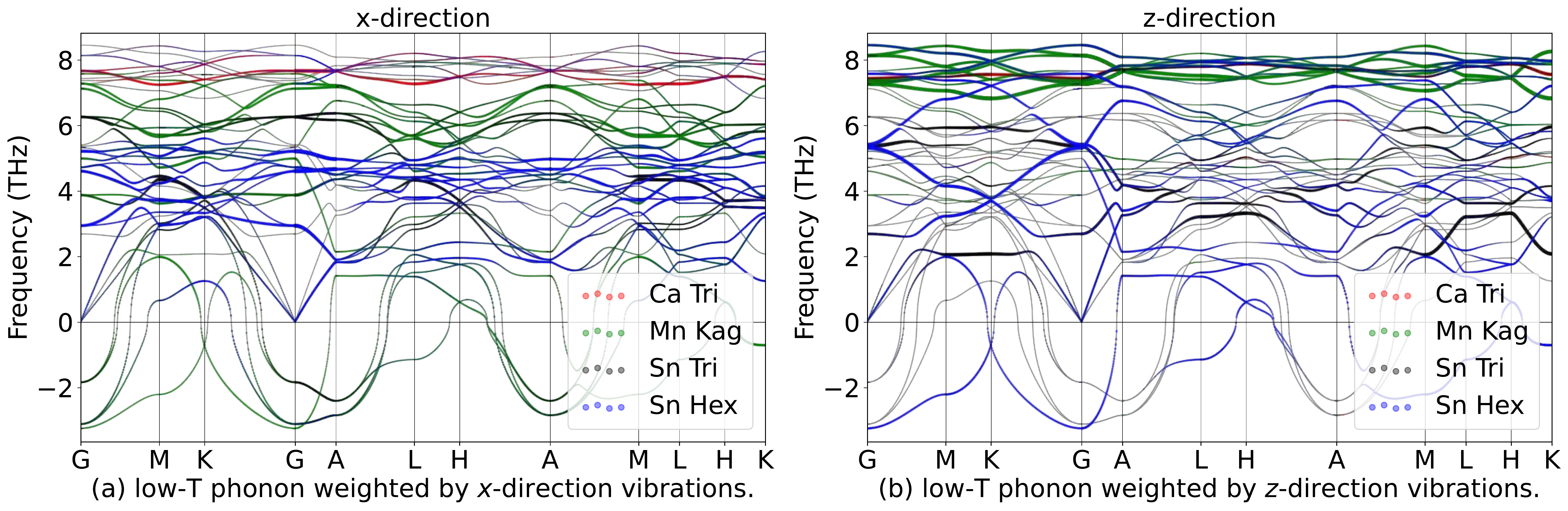}
\hspace{5pt}\label{fig:CaMn6Sn6_relaxed_High_xz}
\includegraphics[width=0.85\textwidth]{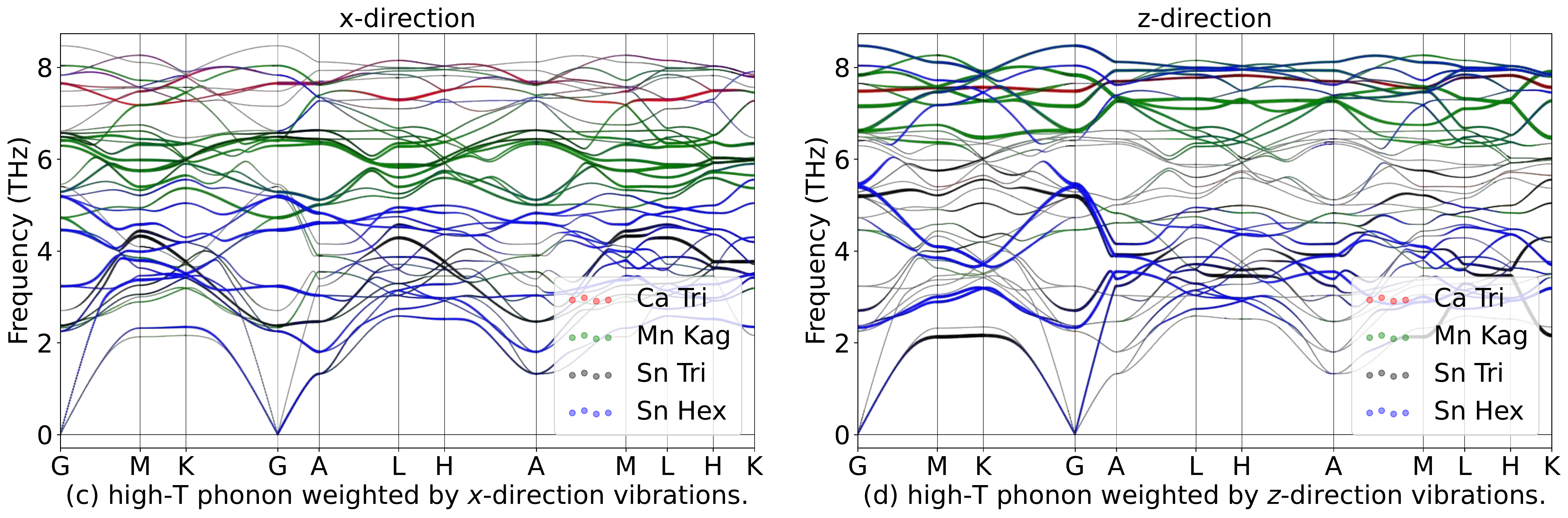}
\caption{Phonon spectrum for CaMn$_6$Sn$_6$in in-plane ($x$) and out-of-plane ($z$) directions in paramagnetic phase.}
\label{fig:CaMn6Sn6phoxyz}
\end{figure}

\begin{table}[htbp]
\global\long\def\arraystretch{1.12}
\captionsetup{width=.95\textwidth}
\caption{IRREPs at high-symmetry points for CaMn$_6$Sn$_6$ phonon spectrum at Low-T.}
\label{tab:191_CaMn6Sn6_Low}
\setlength{\tabcolsep}{1.mm}{
}
\end{table}

\begin{table}[htbp]
\global\long\def\arraystretch{1.12}
\captionsetup{width=.95\textwidth}
\caption{IRREPs at high-symmetry points for CaMn$_6$Sn$_6$ phonon spectrum at High-T.}
\label{tab:191_CaMn6Sn6_High}
\setlength{\tabcolsep}{1.mm}{
%
}
\end{table}
\subsubsection{\label{sms2sec:ScMn6Sn6}\hyperref[smsec:col]{\texorpdfstring{ScMn$_6$Sn$_6$}{}}~{\color{blue}  (High-T stable, Low-T unstable) }}

\begin{table}[htbp]
\global\long\def\arraystretch{1.12}
\captionsetup{width=.85\textwidth}
\caption{Structure parameters of ScMn$_6$Sn$_6$ after relaxation. It contains 471 electrons. }
\label{tab:ScMn6Sn6}
\setlength{\tabcolsep}{4mm}{
%
}
\end{table}

\begin{figure}[htbp]
\centering
\includegraphics[width=0.85\textwidth]{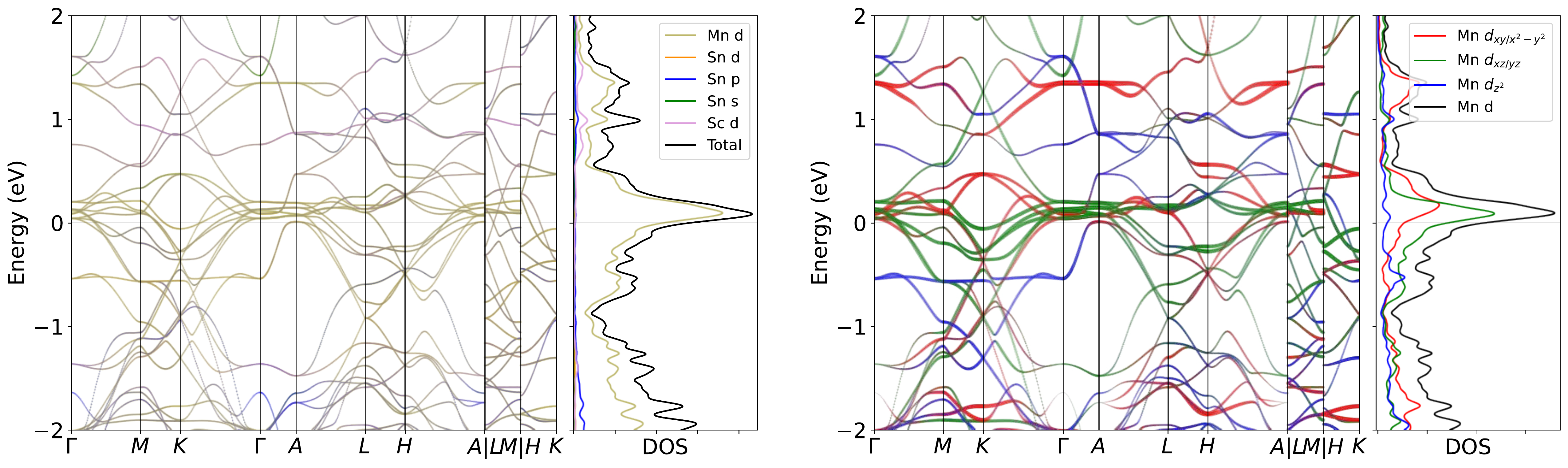}
\caption{Band structure for ScMn$_6$Sn$_6$ without SOC in paramagnetic phase. The projection of Mn $3d$ orbitals is also presented. The band structures clearly reveal the dominating of Mn $3d$ orbitals  near the Fermi level, as well as the pronounced peak.}
\label{fig:ScMn6Sn6band}
\end{figure}

\begin{figure}[H]
\centering
\subfloat[Relaxed phonon at low-T.]{
\label{fig:ScMn6Sn6_relaxed_Low}
\includegraphics[width=0.45\textwidth]{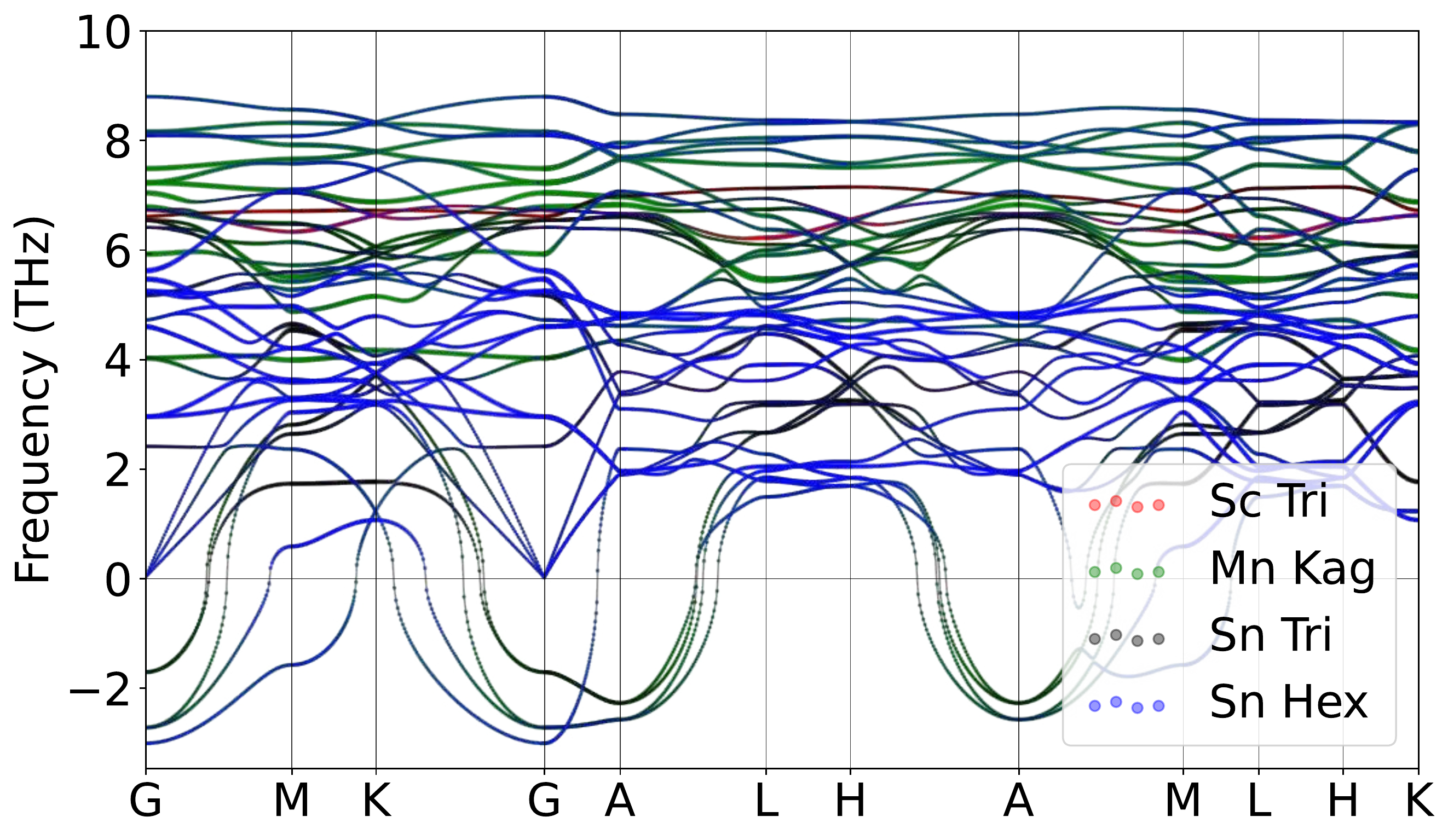}
}
\hspace{5pt}\subfloat[Relaxed phonon at high-T.]{
\label{fig:ScMn6Sn6_relaxed_High}
\includegraphics[width=0.45\textwidth]{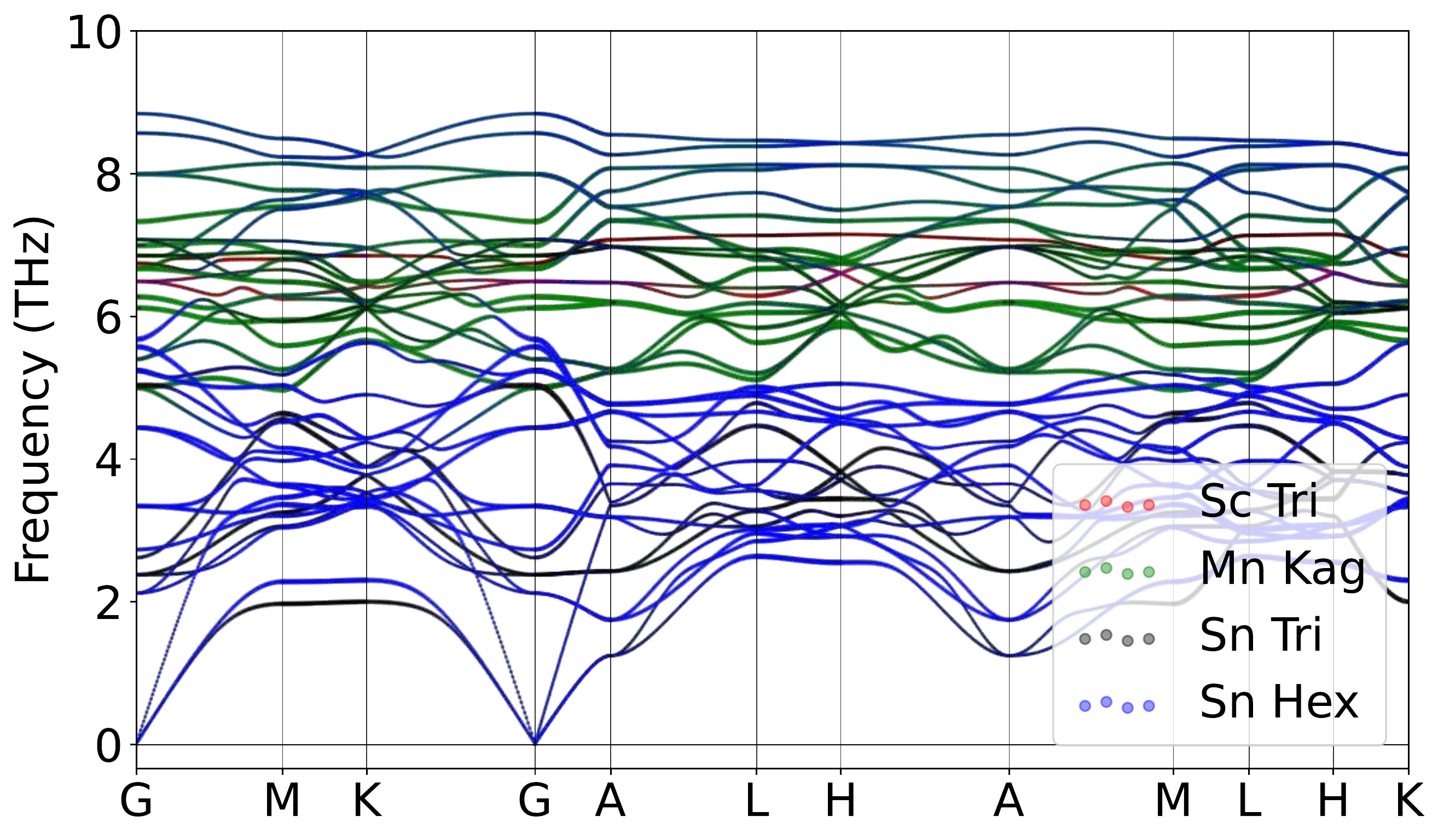}
}
\caption{Atomically projected phonon spectrum for ScMn$_6$Sn$_6$ in paramagnetic phase.}
\label{fig:ScMn6Sn6pho}
\end{figure}

\begin{figure}[htbp]
\centering
\includegraphics[width=0.45\textwidth]{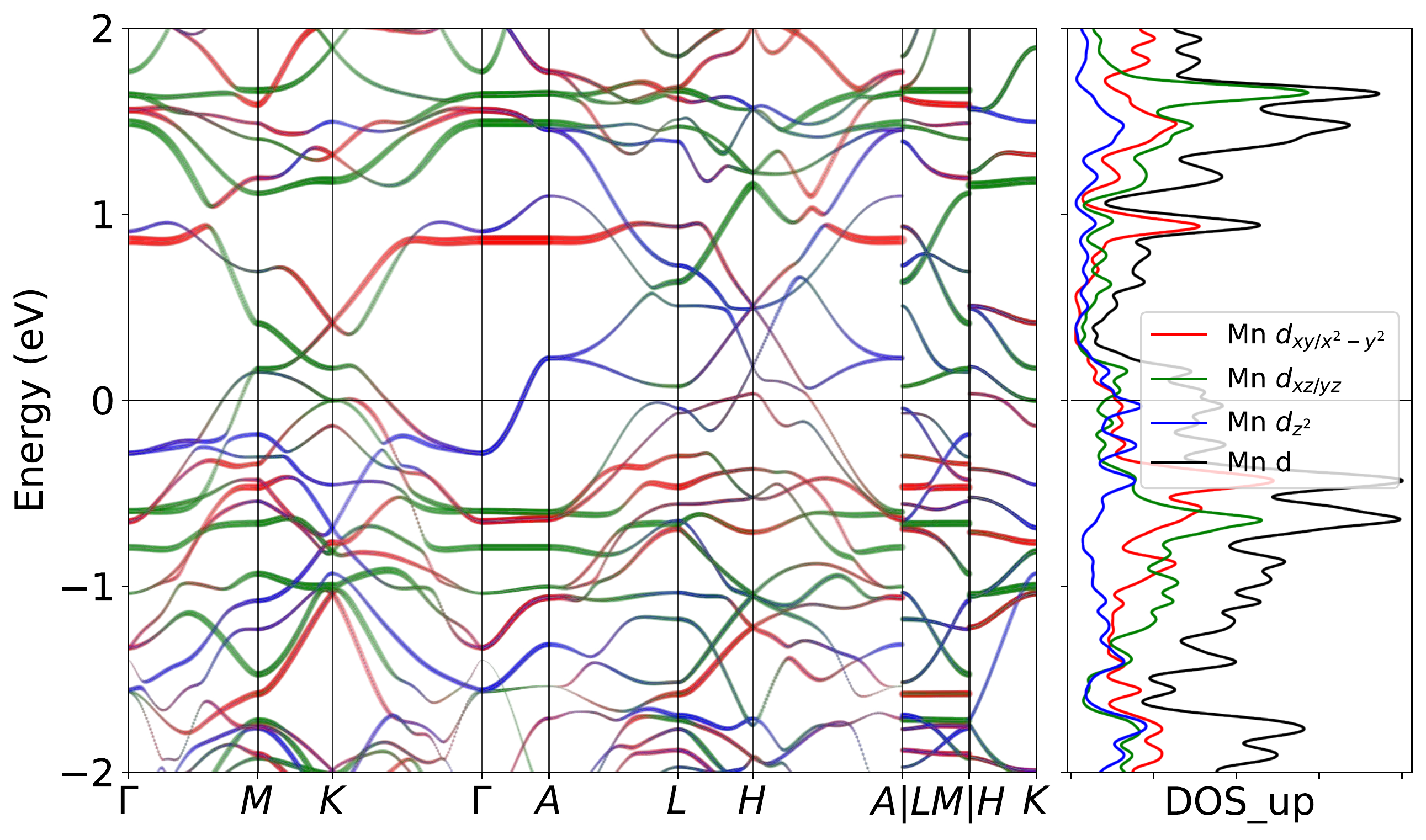}
\includegraphics[width=0.45\textwidth]{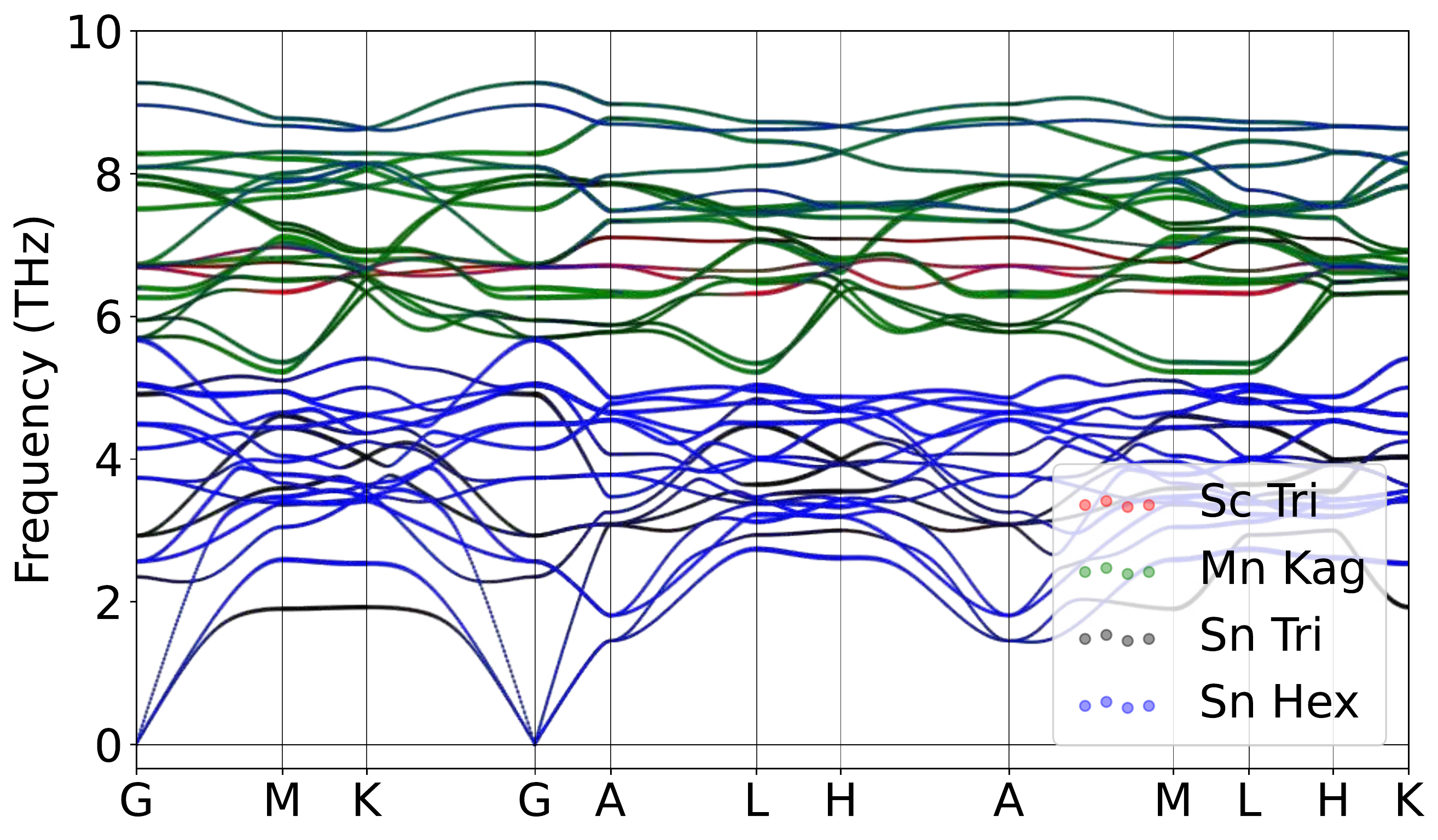}
\caption{Band structure for ScMn$_6$Sn$_6$ with AFM order without SOC for spin (Left)up channel. The projection of Mn $3d$ orbitals is also presented. (Right)Correspondingly, a stable phonon was demonstrated with the collinear magnetic order without Hubbard U at lower temperature regime, weighted by atomic projections.}
\label{fig:ScMn6Sn6mag}
\end{figure}

\begin{figure}[H]
\centering
\label{fig:ScMn6Sn6_relaxed_Low_xz}
\includegraphics[width=0.85\textwidth]{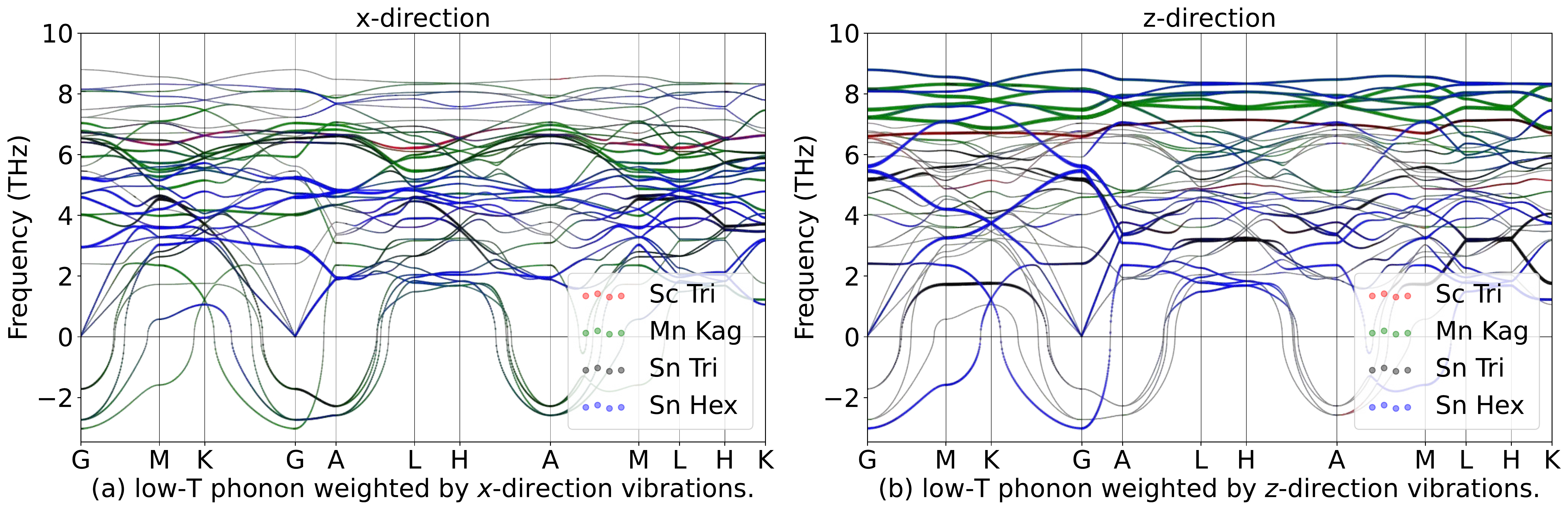}
\hspace{5pt}\label{fig:ScMn6Sn6_relaxed_High_xz}
\includegraphics[width=0.85\textwidth]{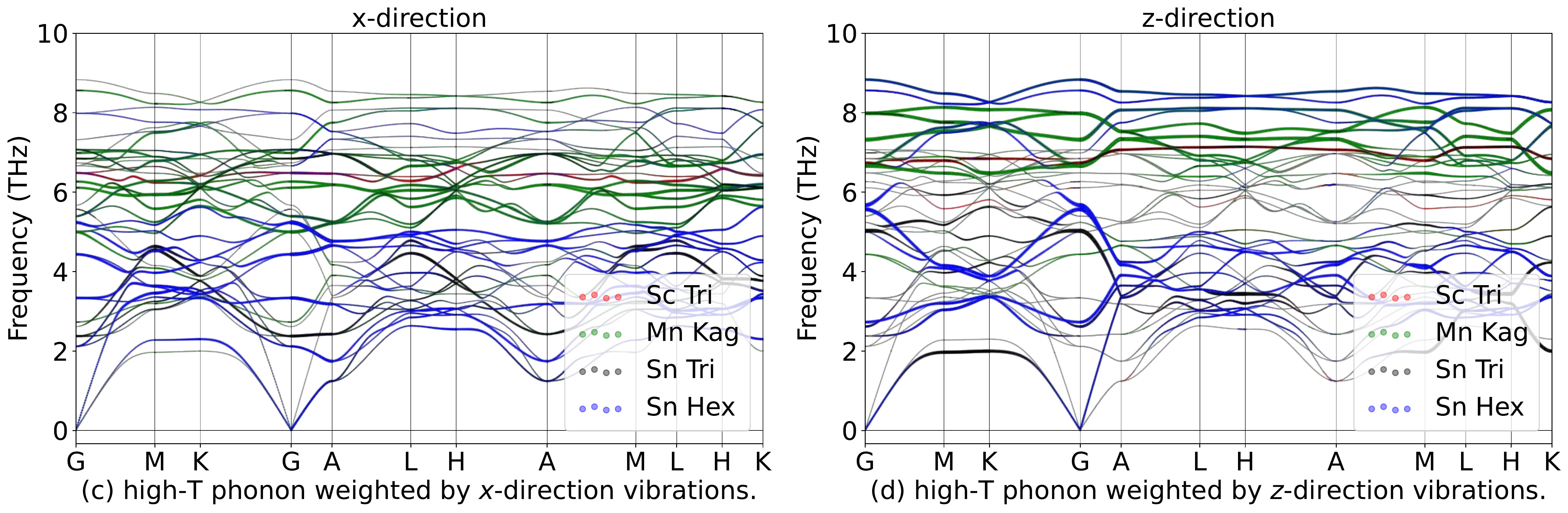}
\caption{Phonon spectrum for ScMn$_6$Sn$_6$in in-plane ($x$) and out-of-plane ($z$) directions in paramagnetic phase.}
\label{fig:ScMn6Sn6phoxyz}
\end{figure}

\begin{table}[htbp]
\global\long\def\arraystretch{1.12}
\captionsetup{width=.95\textwidth}
\caption{IRREPs at high-symmetry points for ScMn$_6$Sn$_6$ phonon spectrum at Low-T.}
\label{tab:191_ScMn6Sn6_Low}
\setlength{\tabcolsep}{1.mm}{
}
\end{table}

\begin{table}[htbp]
\global\long\def\arraystretch{1.12}
\captionsetup{width=.95\textwidth}
\caption{IRREPs at high-symmetry points for ScMn$_6$Sn$_6$ phonon spectrum at High-T.}
\label{tab:191_ScMn6Sn6_High}
\setlength{\tabcolsep}{1.mm}{
%
}
\end{table}
\subsubsection{\label{sms2sec:TiMn6Sn6}\hyperref[smsec:col]{\texorpdfstring{TiMn$_6$Sn$_6$}{}}~{\color{blue}  (High-T stable, Low-T unstable) }}

\begin{table}[htbp]
\global\long\def\arraystretch{1.12}
\captionsetup{width=.85\textwidth}
\caption{Structure parameters of TiMn$_6$Sn$_6$ after relaxation. It contains 472 electrons. }
\label{tab:TiMn6Sn6}
\setlength{\tabcolsep}{4mm}{
%
}
\end{table}

\begin{figure}[htbp]
\centering
\includegraphics[width=0.85\textwidth]{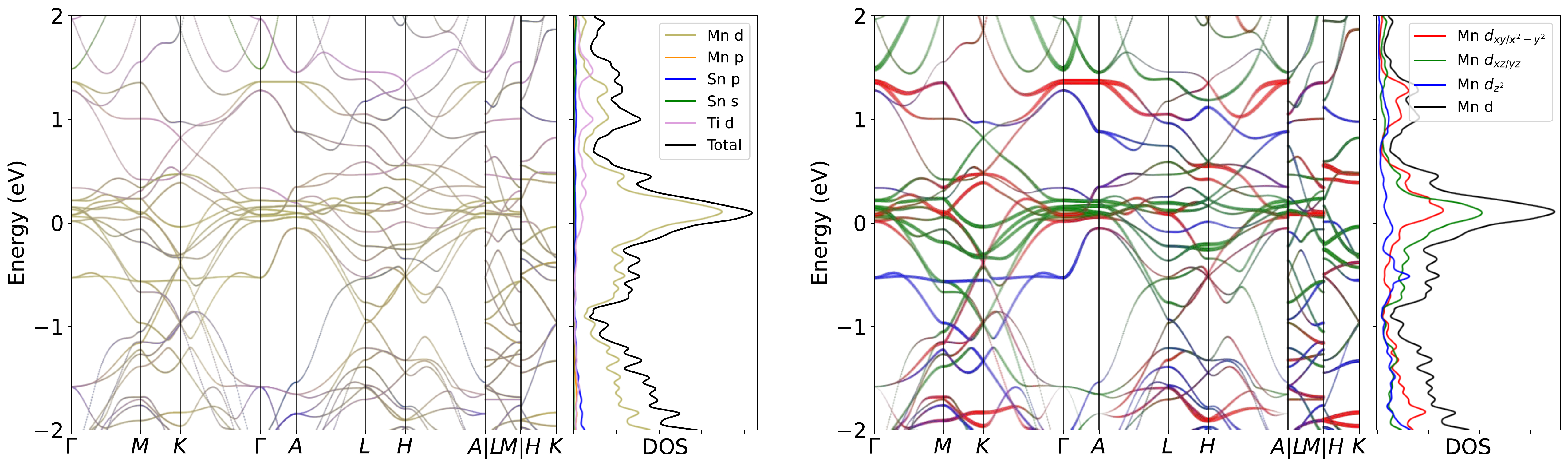}
\caption{Band structure for TiMn$_6$Sn$_6$ without SOC in paramagnetic phase. The projection of Mn $3d$ orbitals is also presented. The band structures clearly reveal the dominating of Mn $3d$ orbitals  near the Fermi level, as well as the pronounced peak.}
\label{fig:TiMn6Sn6band}
\end{figure}

\begin{figure}[H]
\centering
\subfloat[Relaxed phonon at low-T.]{
\label{fig:TiMn6Sn6_relaxed_Low}
\includegraphics[width=0.45\textwidth]{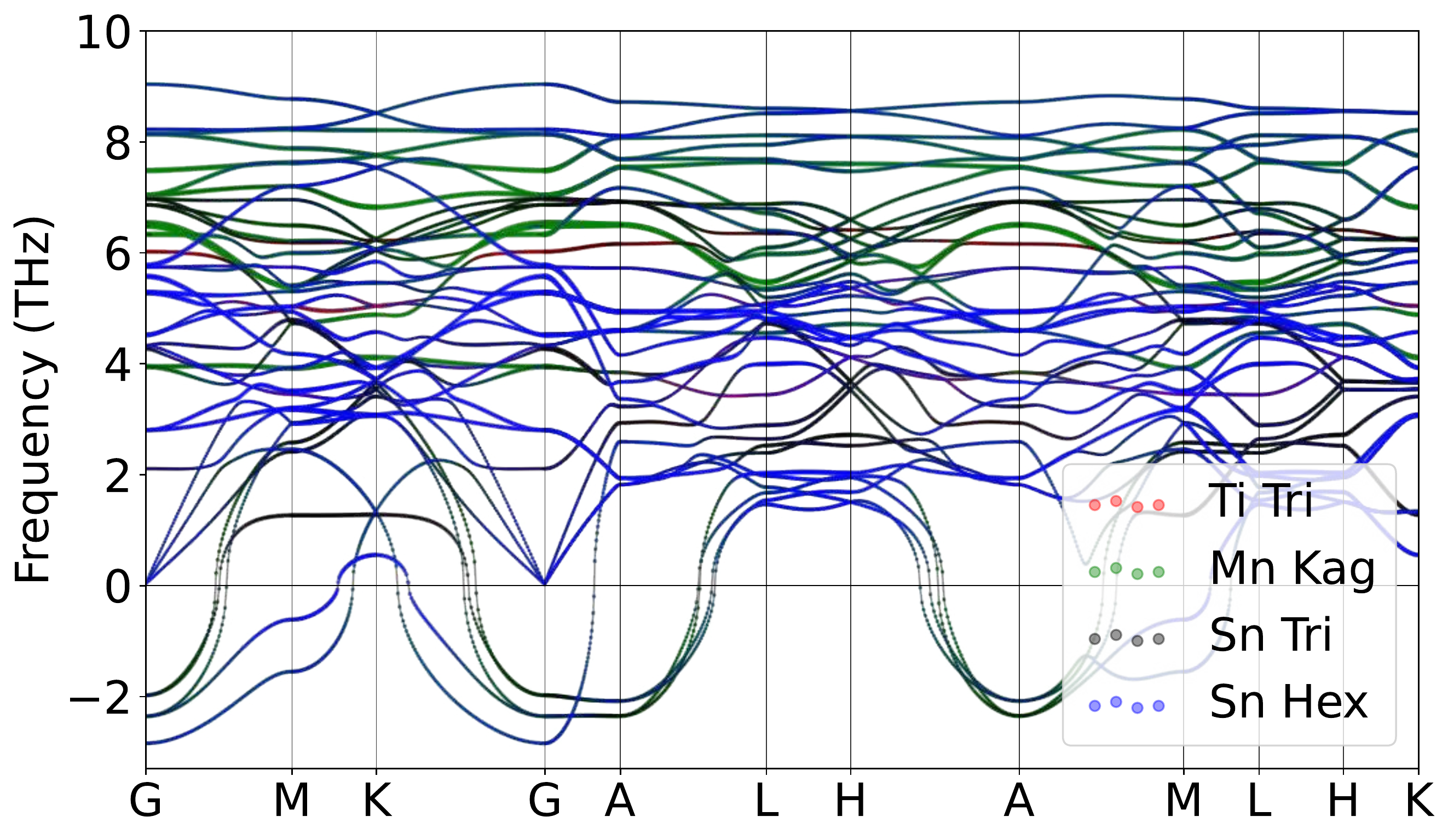}
}
\hspace{5pt}\subfloat[Relaxed phonon at high-T.]{
\label{fig:TiMn6Sn6_relaxed_High}
\includegraphics[width=0.45\textwidth]{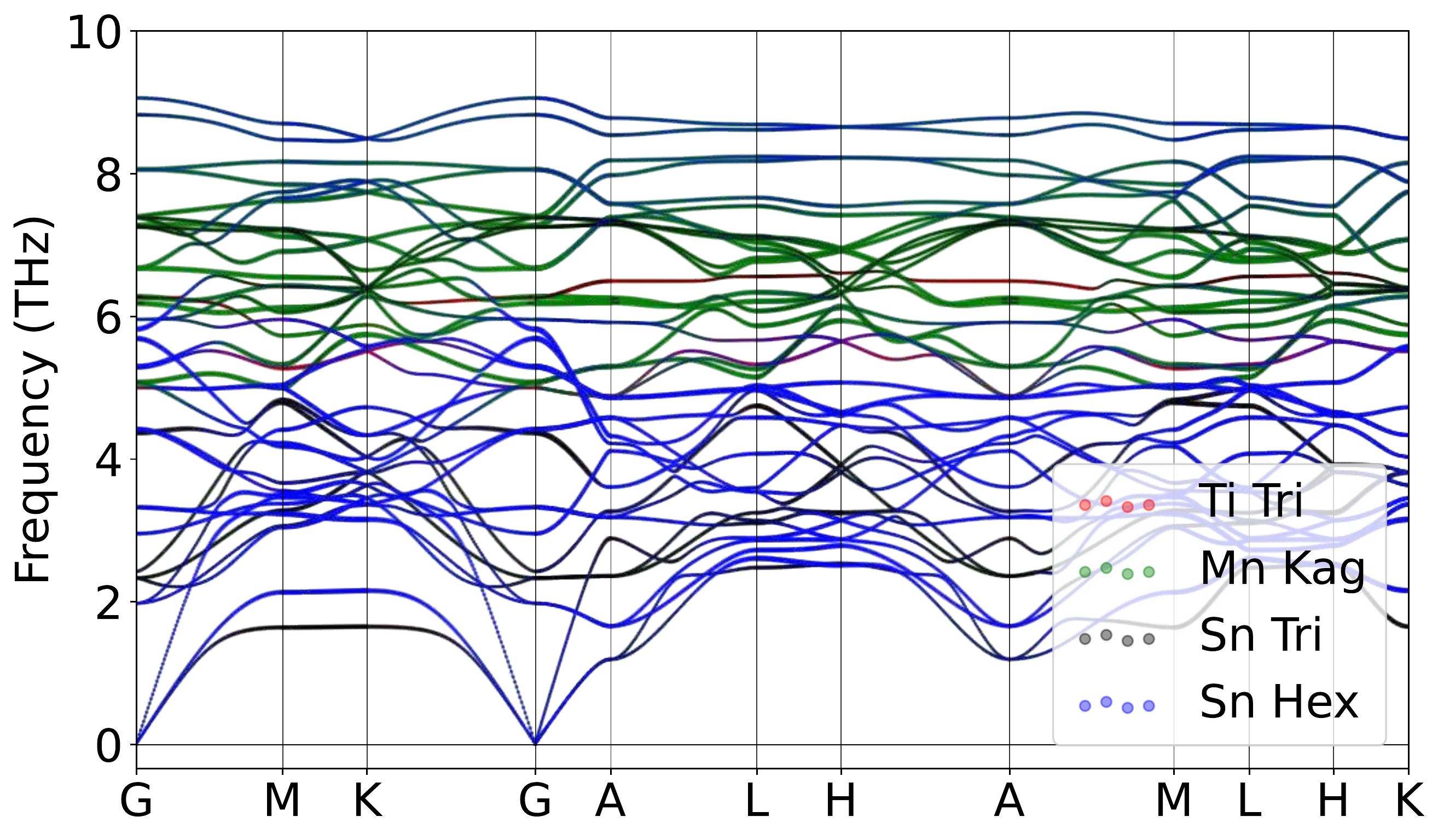}
}
\caption{Atomically projected phonon spectrum for TiMn$_6$Sn$_6$ in paramagnetic phase.}
\label{fig:TiMn6Sn6pho}
\end{figure}

\begin{figure}[htbp]
\centering
\includegraphics[width=0.45\textwidth]{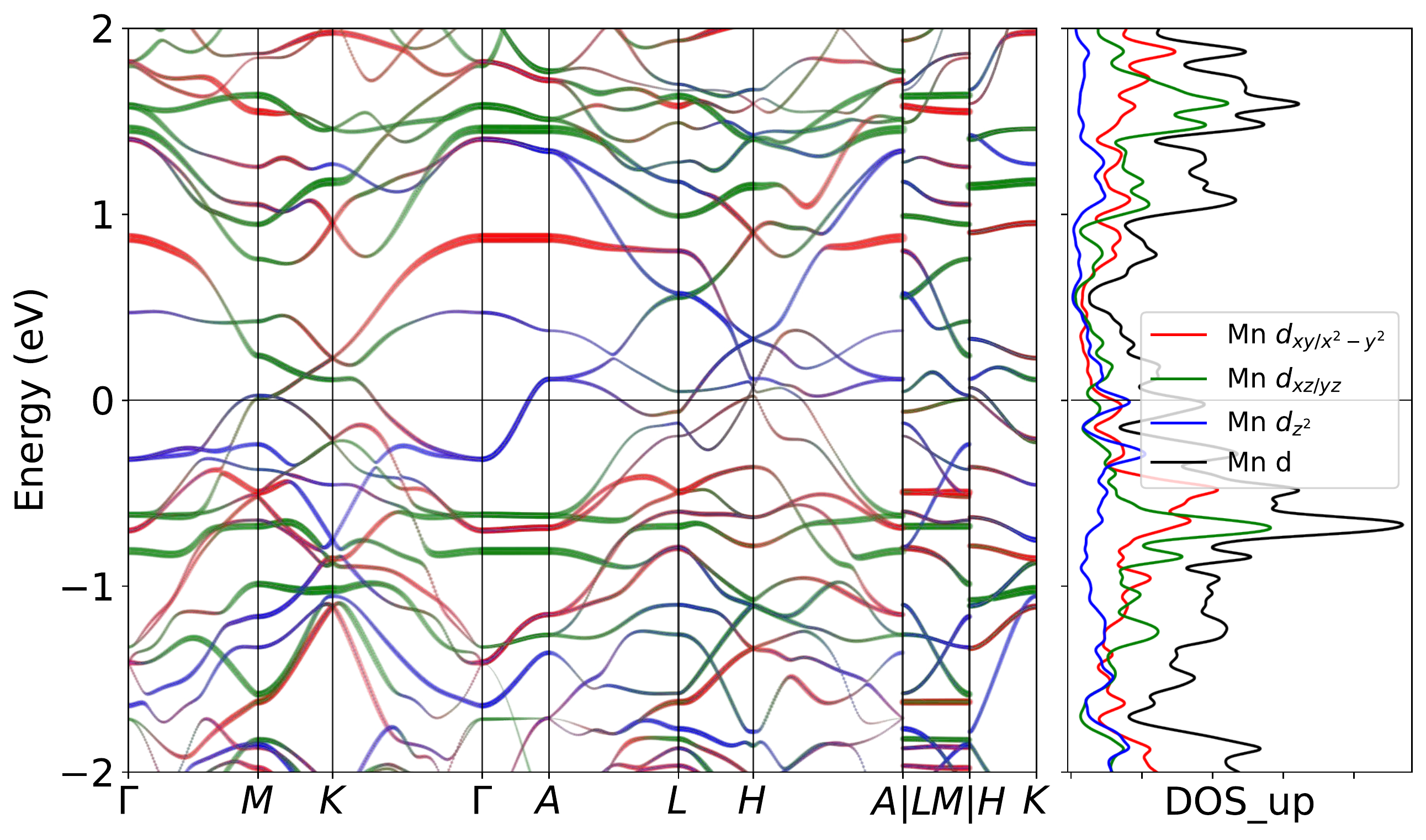}
\includegraphics[width=0.45\textwidth]{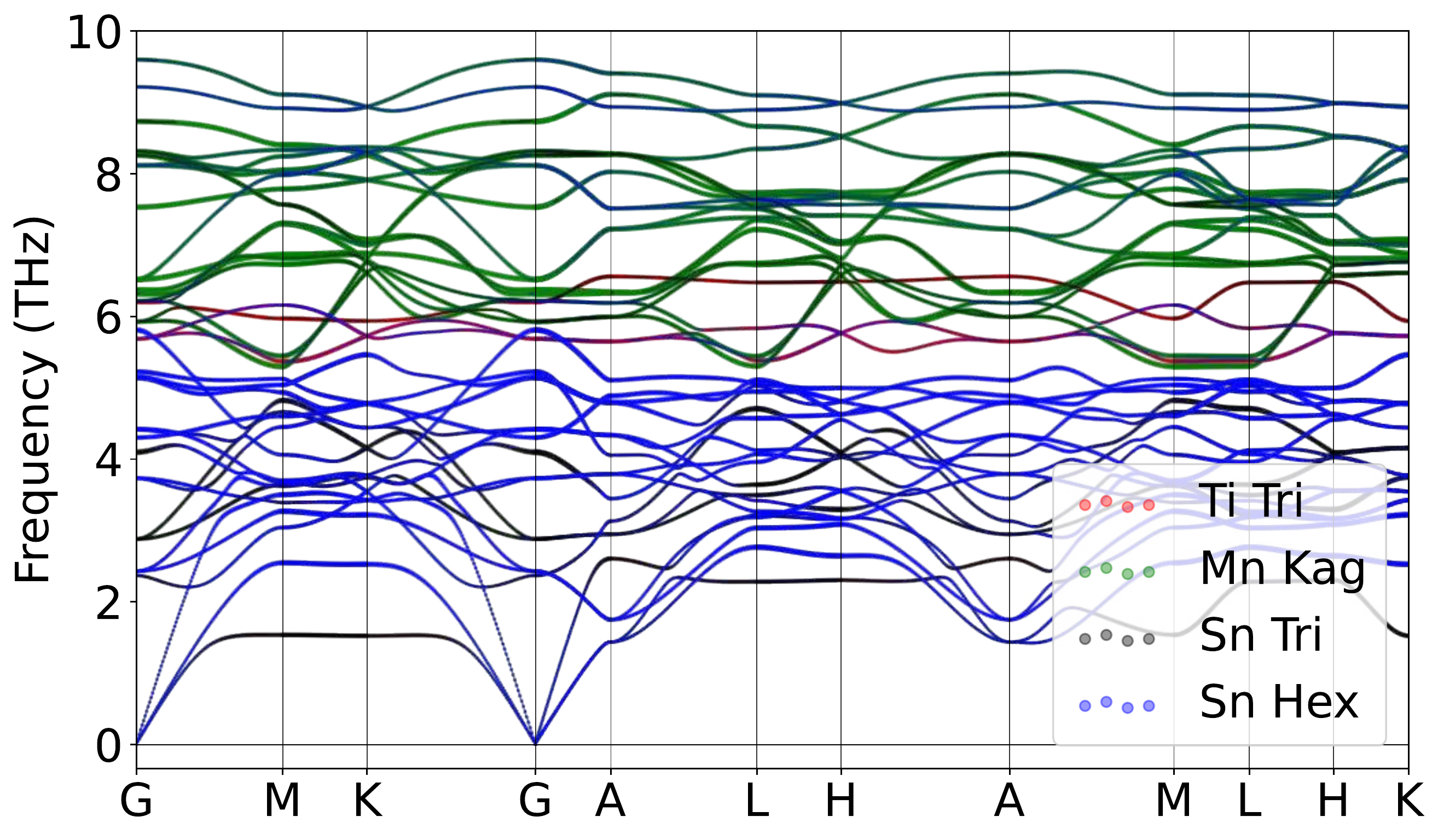}
\caption{Band structure for TiMn$_6$Sn$_6$ with AFM order without SOC for spin (Left)up channel. The projection of Mn $3d$ orbitals is also presented. (Right)Correspondingly, a stable phonon was demonstrated with the collinear magnetic order without Hubbard U at lower temperature regime, weighted by atomic projections.}
\label{fig:TiMn6Sn6mag}
\end{figure}

\begin{figure}[H]
\centering
\label{fig:TiMn6Sn6_relaxed_Low_xz}
\includegraphics[width=0.85\textwidth]{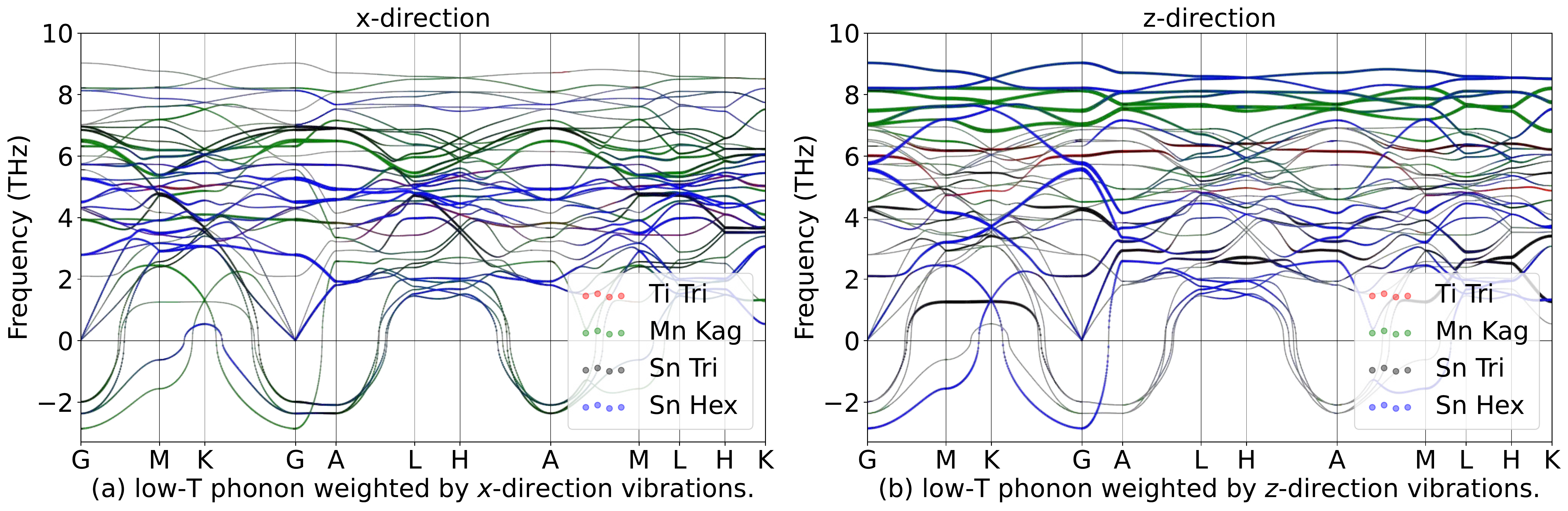}
\hspace{5pt}\label{fig:TiMn6Sn6_relaxed_High_xz}
\includegraphics[width=0.85\textwidth]{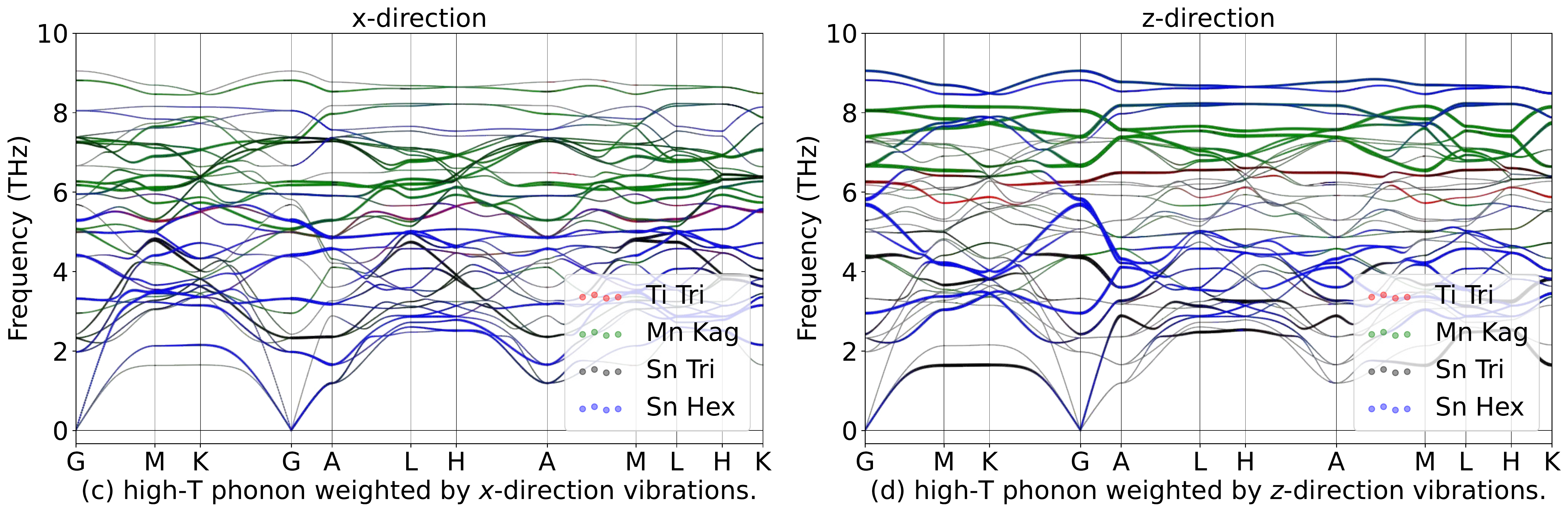}
\caption{Phonon spectrum for TiMn$_6$Sn$_6$in in-plane ($x$) and out-of-plane ($z$) directions in paramagnetic phase.}
\label{fig:TiMn6Sn6phoxyz}
\end{figure}

\begin{table}[htbp]
\global\long\def\arraystretch{1.12}
\captionsetup{width=.95\textwidth}
\caption{IRREPs at high-symmetry points for TiMn$_6$Sn$_6$ phonon spectrum at Low-T.}
\label{tab:191_TiMn6Sn6_Low}
\setlength{\tabcolsep}{1.mm}{
}
\end{table}

\begin{table}[htbp]
\global\long\def\arraystretch{1.12}
\captionsetup{width=.95\textwidth}
\caption{IRREPs at high-symmetry points for TiMn$_6$Sn$_6$ phonon spectrum at High-T.}
\label{tab:191_TiMn6Sn6_High}
\setlength{\tabcolsep}{1.mm}{
%
}
\end{table}
\subsubsection{\label{sms2sec:YMn6Sn6}\hyperref[smsec:col]{\texorpdfstring{YMn$_6$Sn$_6$}{}}~{\color{blue}  (High-T stable, Low-T unstable) }}

\begin{table}[htbp]
\global\long\def\arraystretch{1.12}
\captionsetup{width=.85\textwidth}
\caption{Structure parameters of YMn$_6$Sn$_6$ after relaxation. It contains 489 electrons. }
\label{tab:YMn6Sn6}
\setlength{\tabcolsep}{4mm}{
%
}
\end{table}

\begin{figure}[htbp]
\centering
\includegraphics[width=0.85\textwidth]{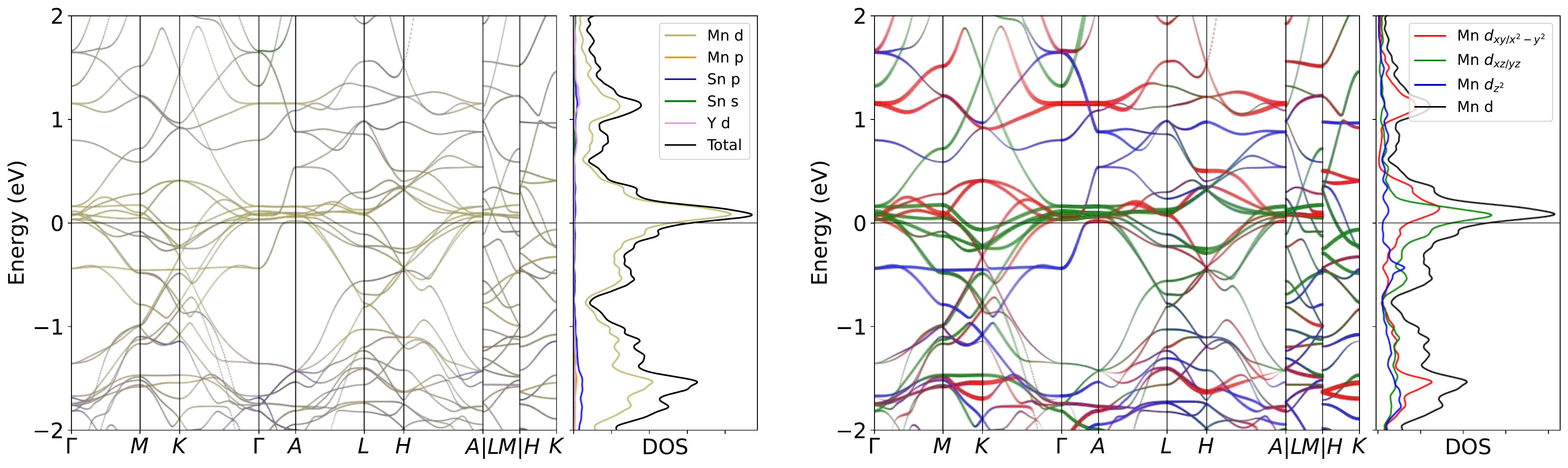}
\caption{Band structure for YMn$_6$Sn$_6$ without SOC in paramagnetic phase. The projection of Mn $3d$ orbitals is also presented. The band structures clearly reveal the dominating of Mn $3d$ orbitals  near the Fermi level, as well as the pronounced peak.}
\label{fig:YMn6Sn6band}
\end{figure}

\begin{figure}[H]
\centering
\subfloat[Relaxed phonon at low-T.]{
\label{fig:YMn6Sn6_relaxed_Low}
\includegraphics[width=0.45\textwidth]{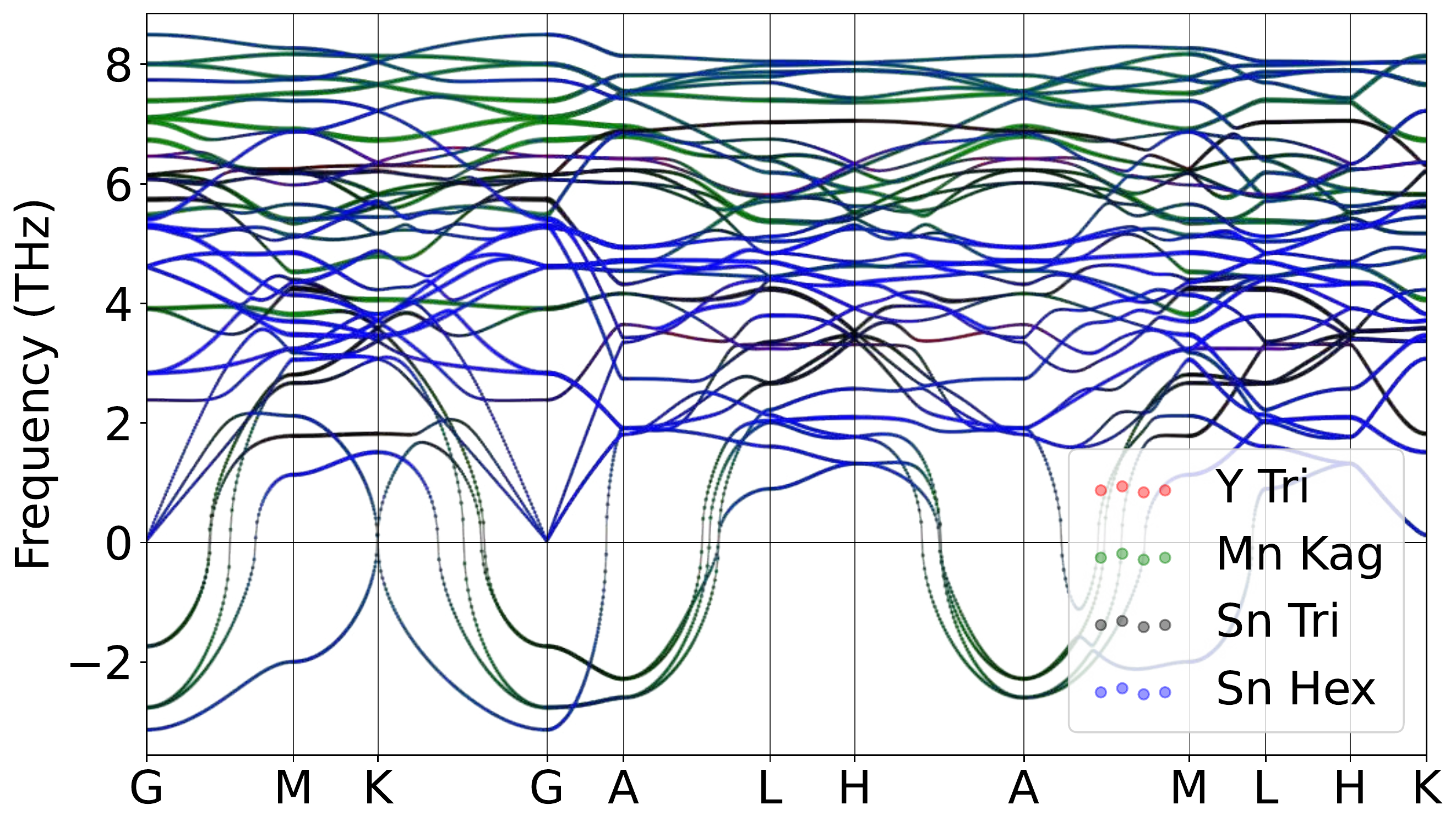}
}
\hspace{5pt}\subfloat[Relaxed phonon at high-T.]{
\label{fig:YMn6Sn6_relaxed_High}
\includegraphics[width=0.45\textwidth]{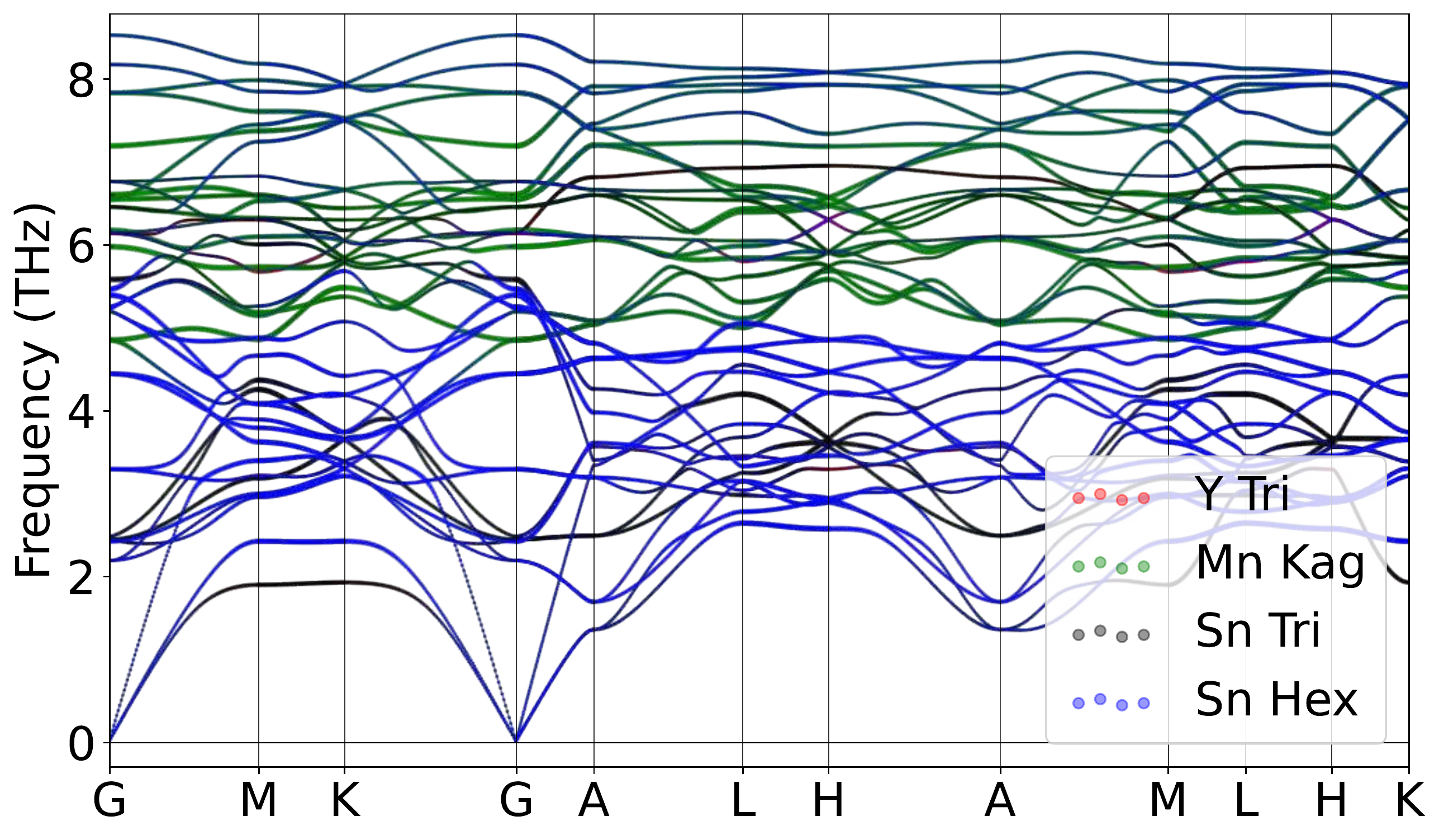}
}
\caption{Atomically projected phonon spectrum for YMn$_6$Sn$_6$ in paramagnetic phase.}
\label{fig:YMn6Sn6pho}
\end{figure}

\begin{figure}[htbp]
\centering
\includegraphics[width=0.45\textwidth]{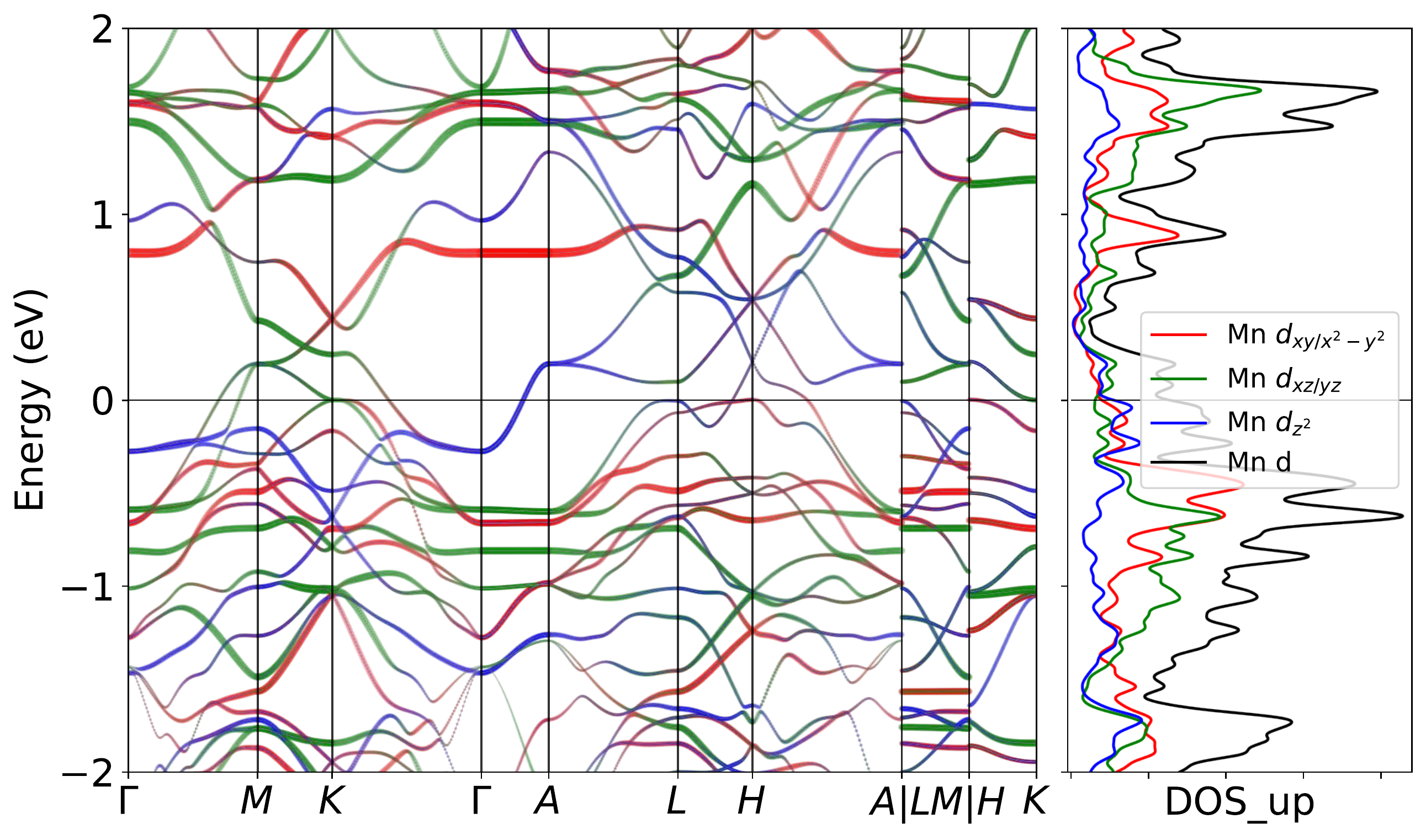}
\includegraphics[width=0.45\textwidth]{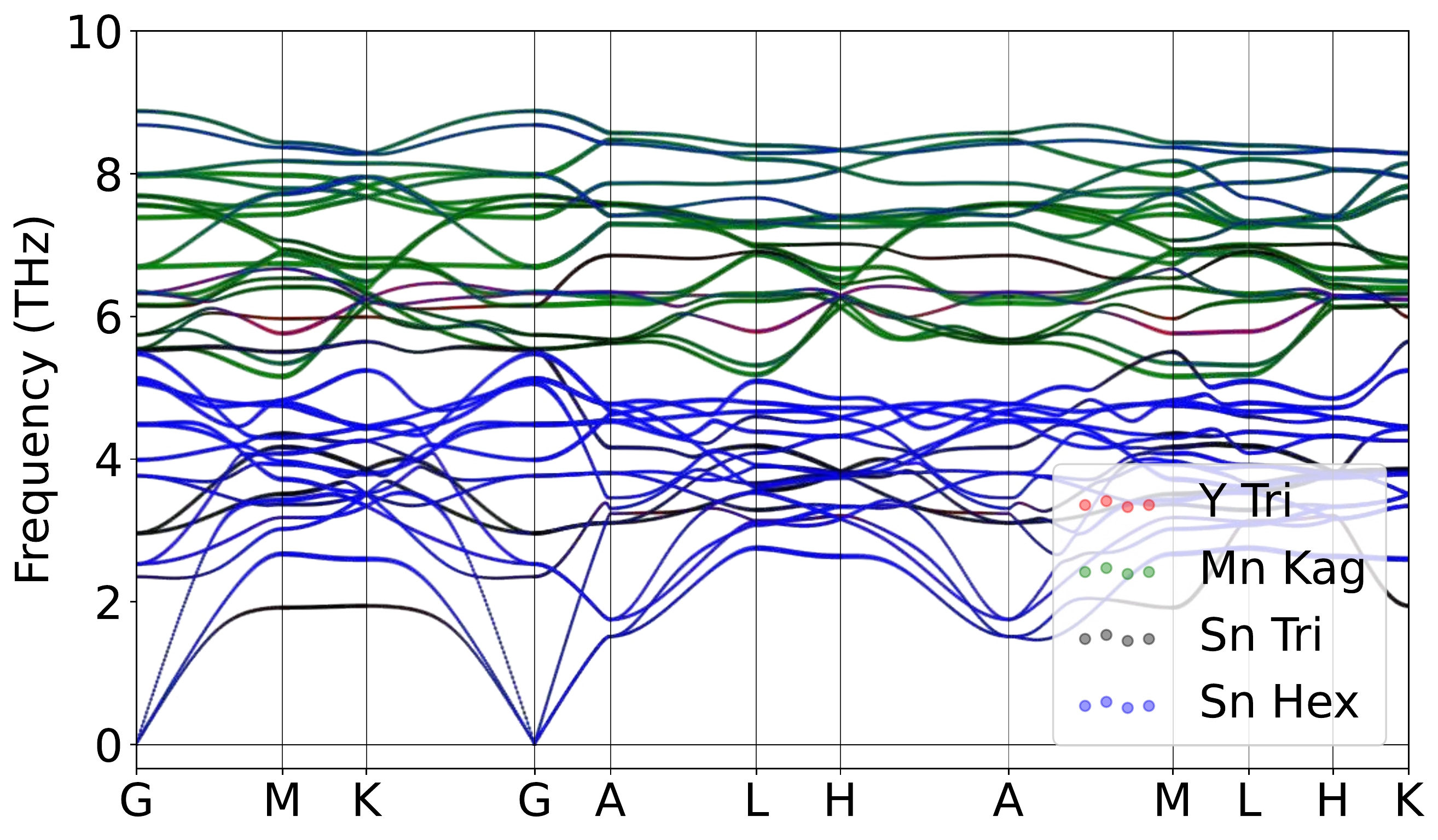}
\caption{Band structure for YMn$_6$Sn$_6$ with AFM order without SOC for spin (Left)up channel. The projection of Mn $3d$ orbitals is also presented. (Right)Correspondingly, a stable phonon was demonstrated with the collinear magnetic order without Hubbard U at lower temperature regime, weighted by atomic projections.}
\label{fig:YMn6Sn6mag}
\end{figure}

\begin{figure}[H]
\centering
\label{fig:YMn6Sn6_relaxed_Low_xz}
\includegraphics[width=0.85\textwidth]{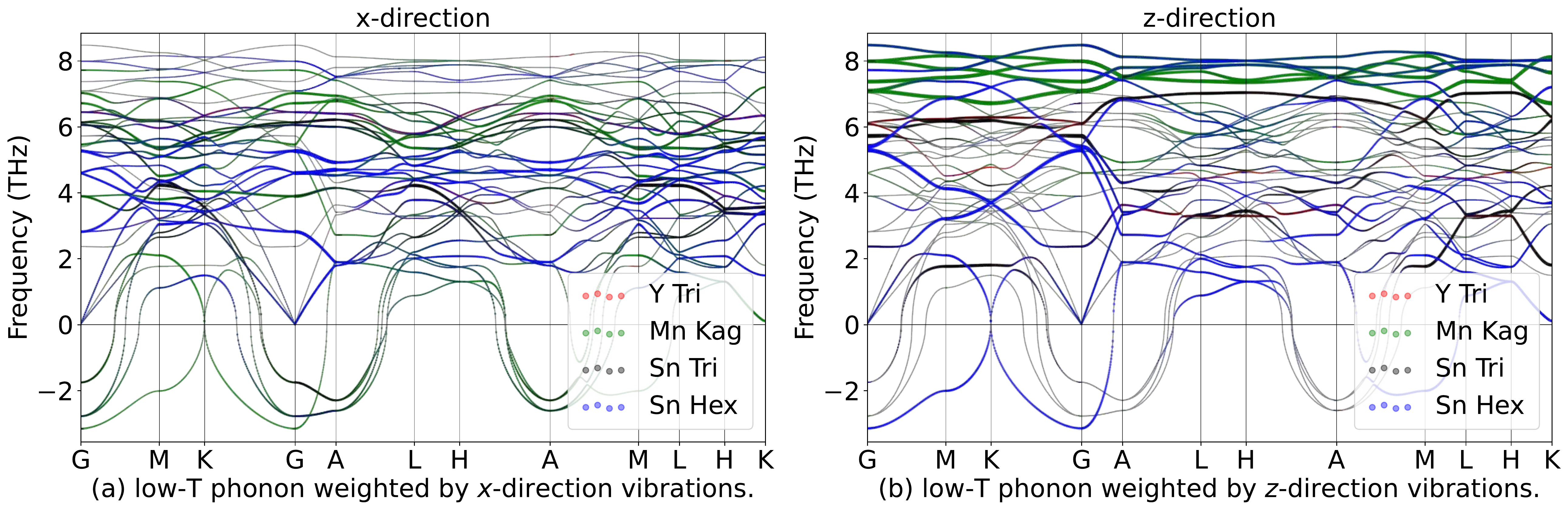}
\hspace{5pt}\label{fig:YMn6Sn6_relaxed_High_xz}
\includegraphics[width=0.85\textwidth]{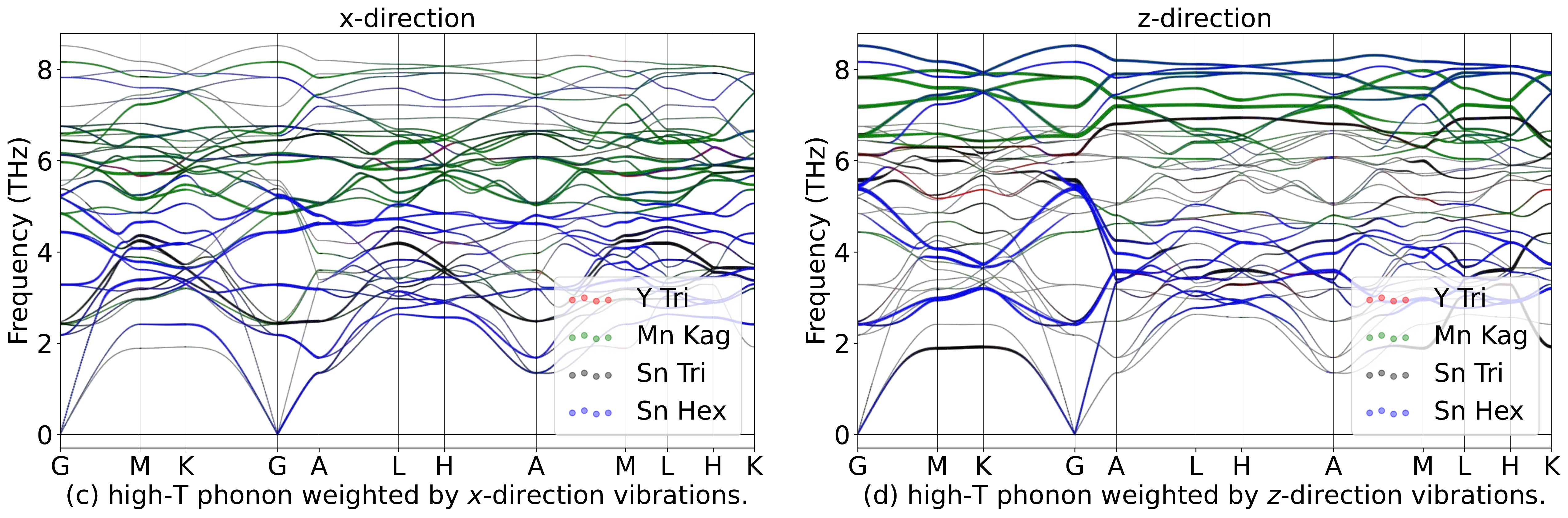}
\caption{Phonon spectrum for YMn$_6$Sn$_6$in in-plane ($x$) and out-of-plane ($z$) directions in paramagnetic phase.}
\label{fig:YMn6Sn6phoxyz}
\end{figure}

\begin{table}[htbp]
\global\long\def\arraystretch{1.12}
\captionsetup{width=.95\textwidth}
\caption{IRREPs at high-symmetry points for YMn$_6$Sn$_6$ phonon spectrum at Low-T.}
\label{tab:191_YMn6Sn6_Low}
\setlength{\tabcolsep}{1.mm}{
}
\end{table}

\begin{table}[htbp]
\global\long\def\arraystretch{1.12}
\captionsetup{width=.95\textwidth}
\caption{IRREPs at high-symmetry points for YMn$_6$Sn$_6$ phonon spectrum at High-T.}
\label{tab:191_YMn6Sn6_High}
\setlength{\tabcolsep}{1.mm}{
%
}
\end{table}
\subsubsection{\label{sms2sec:ZrMn6Sn6}\hyperref[smsec:col]{\texorpdfstring{ZrMn$_6$Sn$_6$}{}}~{\color{blue}  (High-T stable, Low-T unstable) }}

\begin{table}[htbp]
\global\long\def\arraystretch{1.12}
\captionsetup{width=.85\textwidth}
\caption{Structure parameters of ZrMn$_6$Sn$_6$ after relaxation. It contains 490 electrons. }
\label{tab:ZrMn6Sn6}
\setlength{\tabcolsep}{4mm}{
%
}
\end{table}

\begin{figure}[htbp]
\centering
\includegraphics[width=0.85\textwidth]{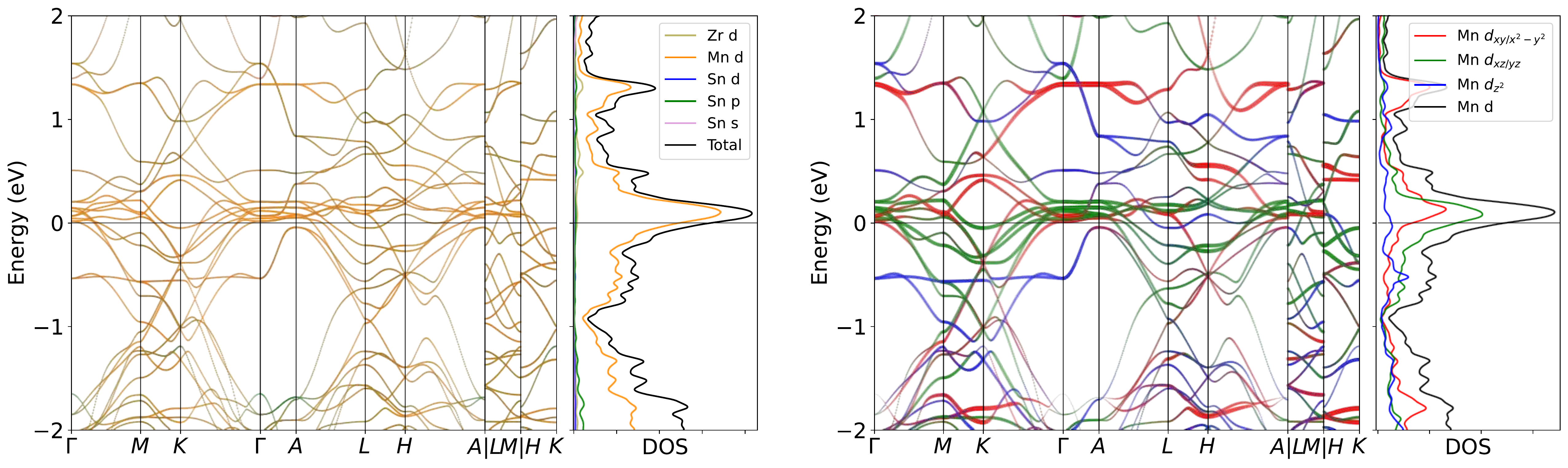}
\caption{Band structure for ZrMn$_6$Sn$_6$ without SOC in paramagnetic phase. The projection of Mn $3d$ orbitals is also presented. The band structures clearly reveal the dominating of Mn $3d$ orbitals  near the Fermi level, as well as the pronounced peak.}
\label{fig:ZrMn6Sn6band}
\end{figure}

\begin{figure}[H]
\centering
\subfloat[Relaxed phonon at low-T.]{
\label{fig:ZrMn6Sn6_relaxed_Low}
\includegraphics[width=0.45\textwidth]{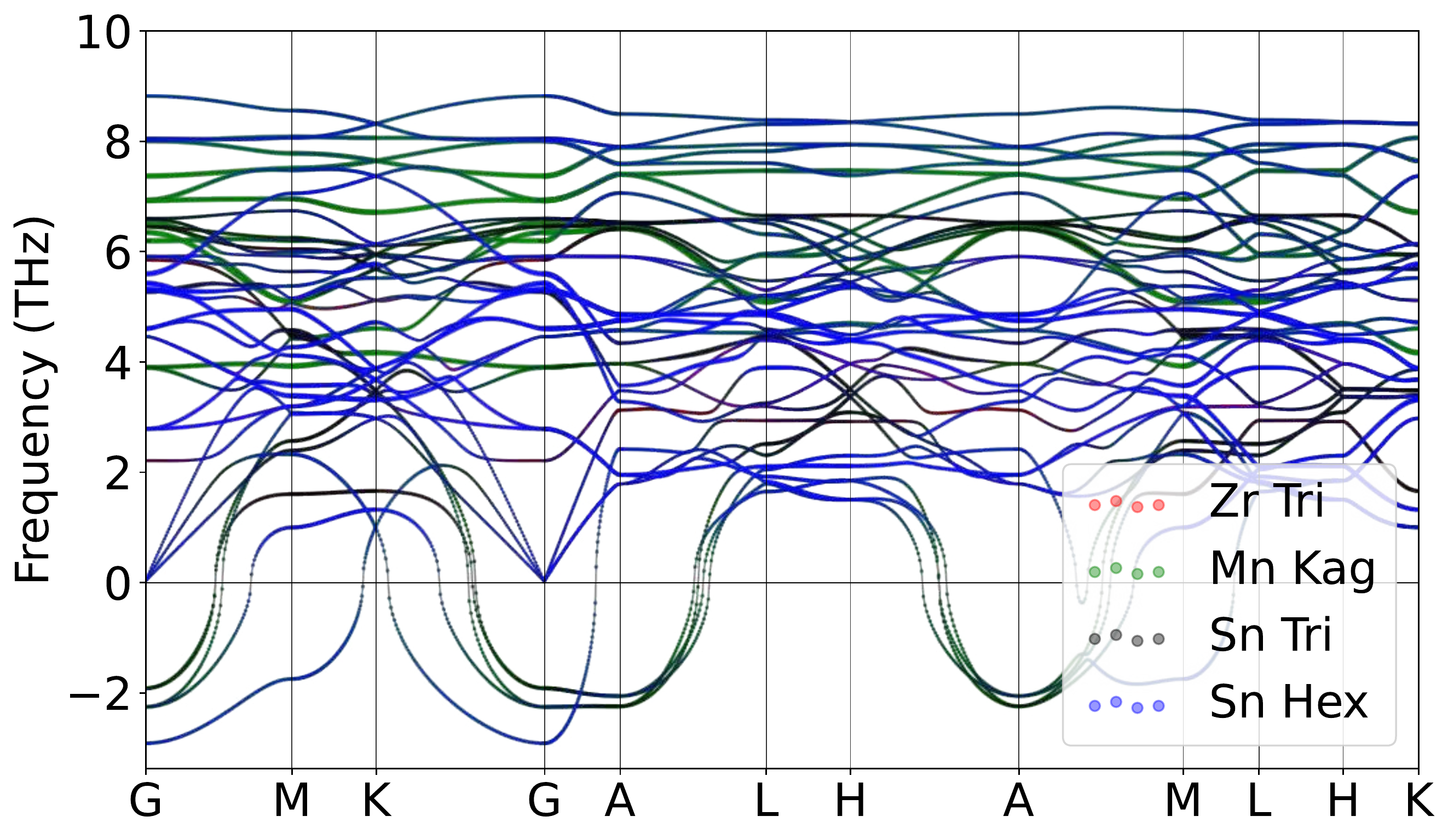}
}
\hspace{5pt}\subfloat[Relaxed phonon at high-T.]{
\label{fig:ZrMn6Sn6_relaxed_High}
\includegraphics[width=0.45\textwidth]{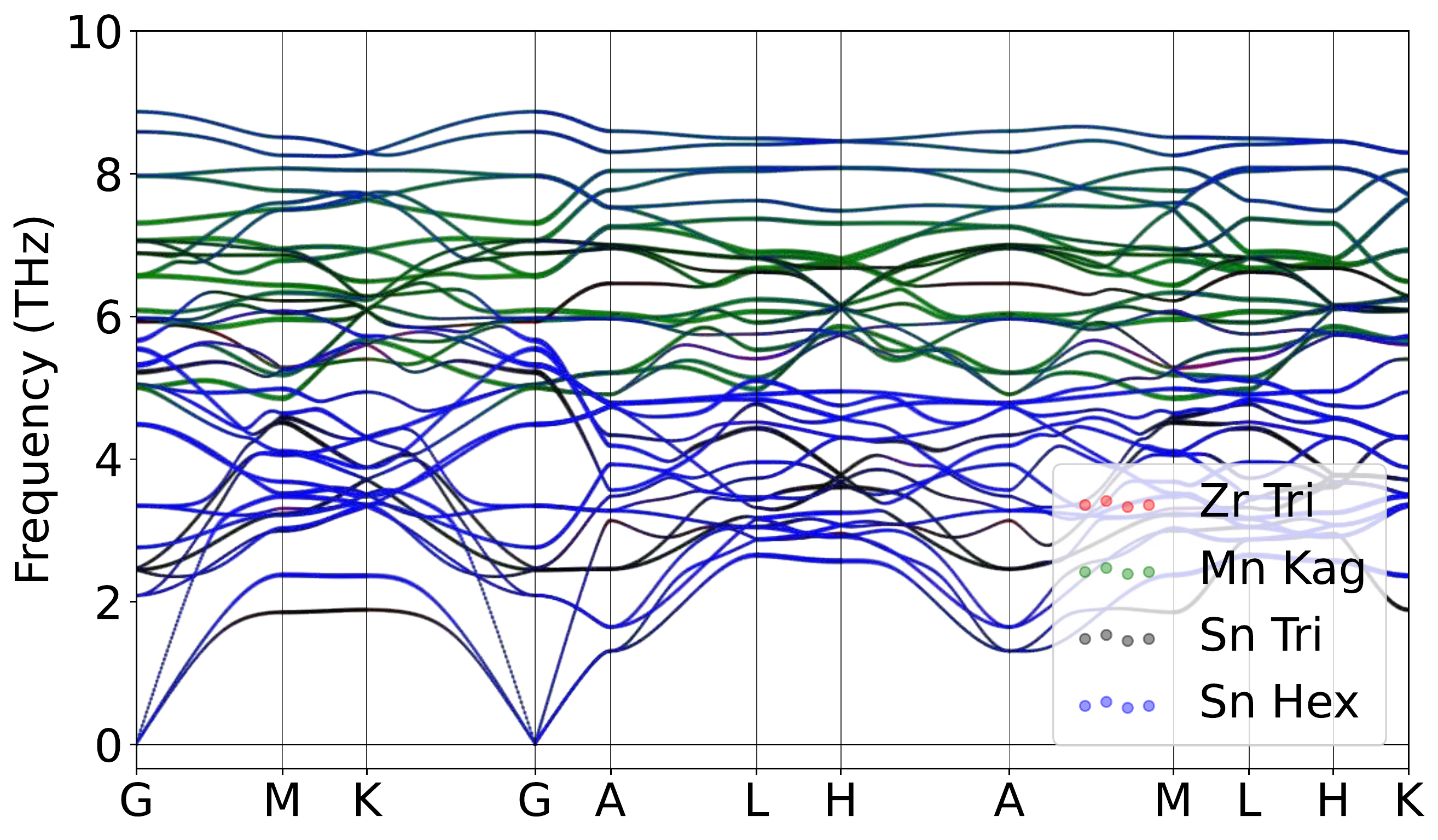}
}
\caption{Atomically projected phonon spectrum for ZrMn$_6$Sn$_6$ in paramagnetic phase.}
\label{fig:ZrMn6Sn6pho}
\end{figure}

\begin{figure}[htbp]
\centering
\includegraphics[width=0.45\textwidth]{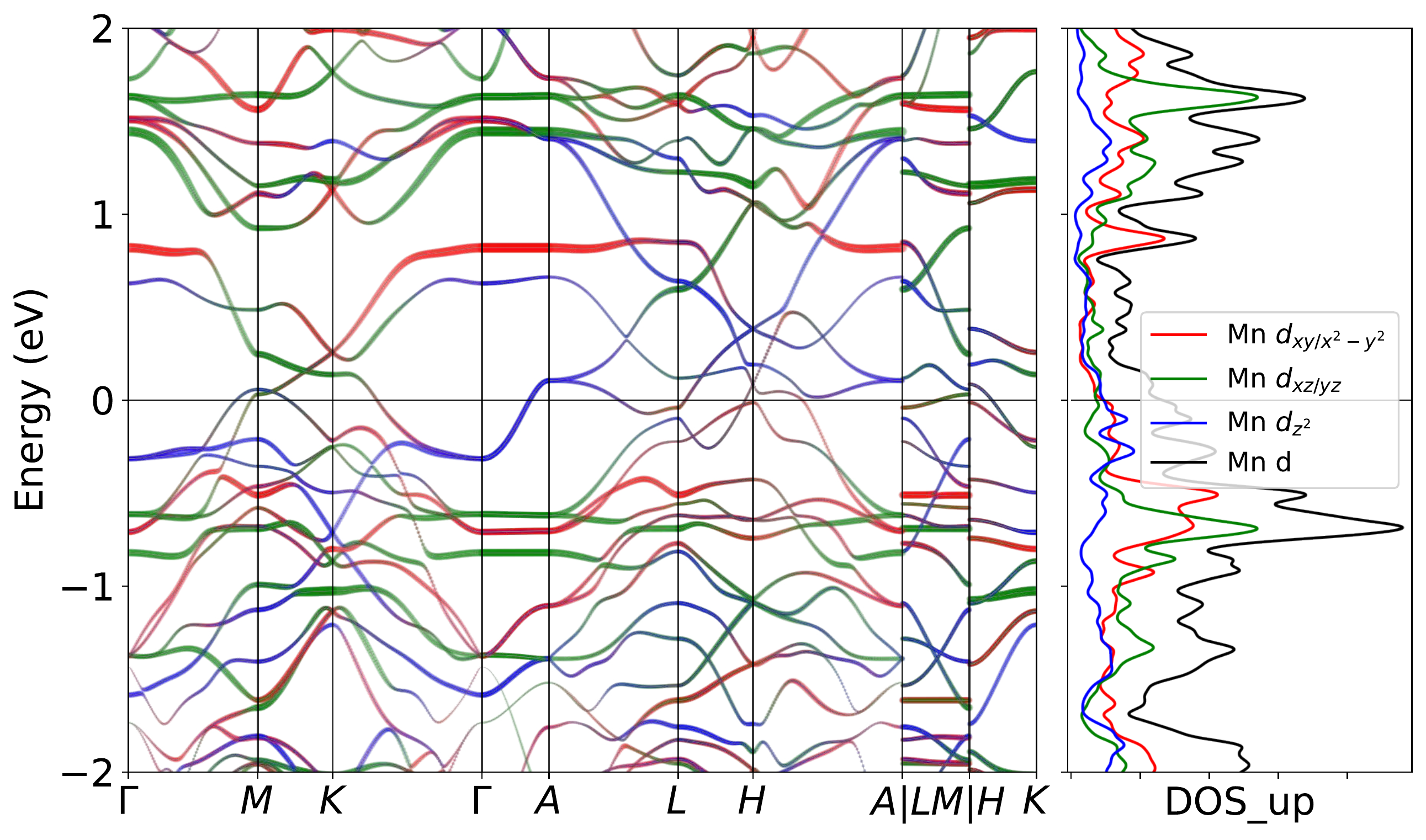}
\includegraphics[width=0.45\textwidth]{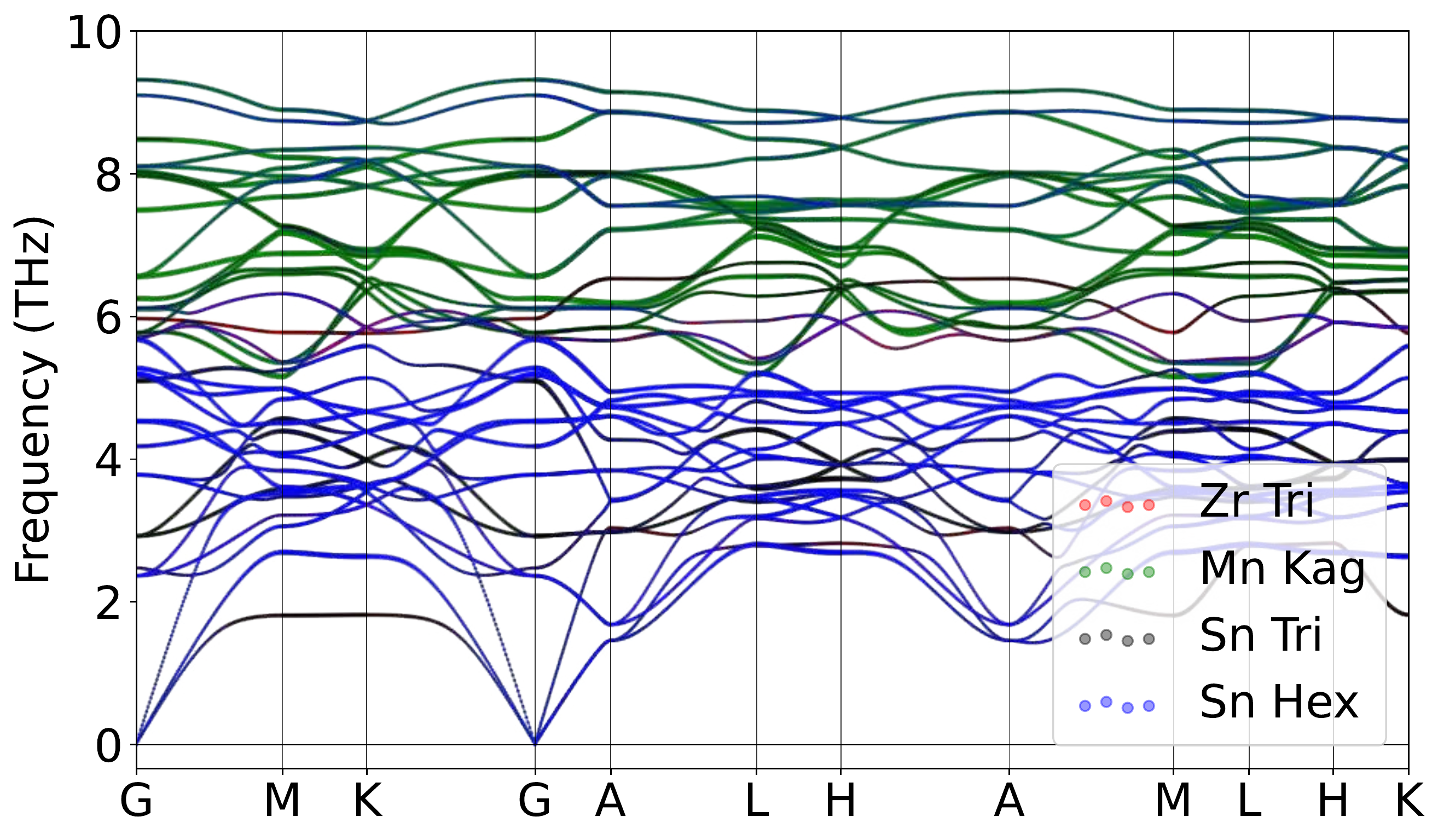}
\caption{Band structure for ZrMn$_6$Sn$_6$ with AFM order without SOC for spin (Left)up channel. The projection of Mn $3d$ orbitals is also presented. (Right)Correspondingly, a stable phonon was demonstrated with the collinear magnetic order without Hubbard U at lower temperature regime, weighted by atomic projections.}
\label{fig:ZrMn6Sn6mag}
\end{figure}

\begin{figure}[H]
\centering
\label{fig:ZrMn6Sn6_relaxed_Low_xz}
\includegraphics[width=0.85\textwidth]{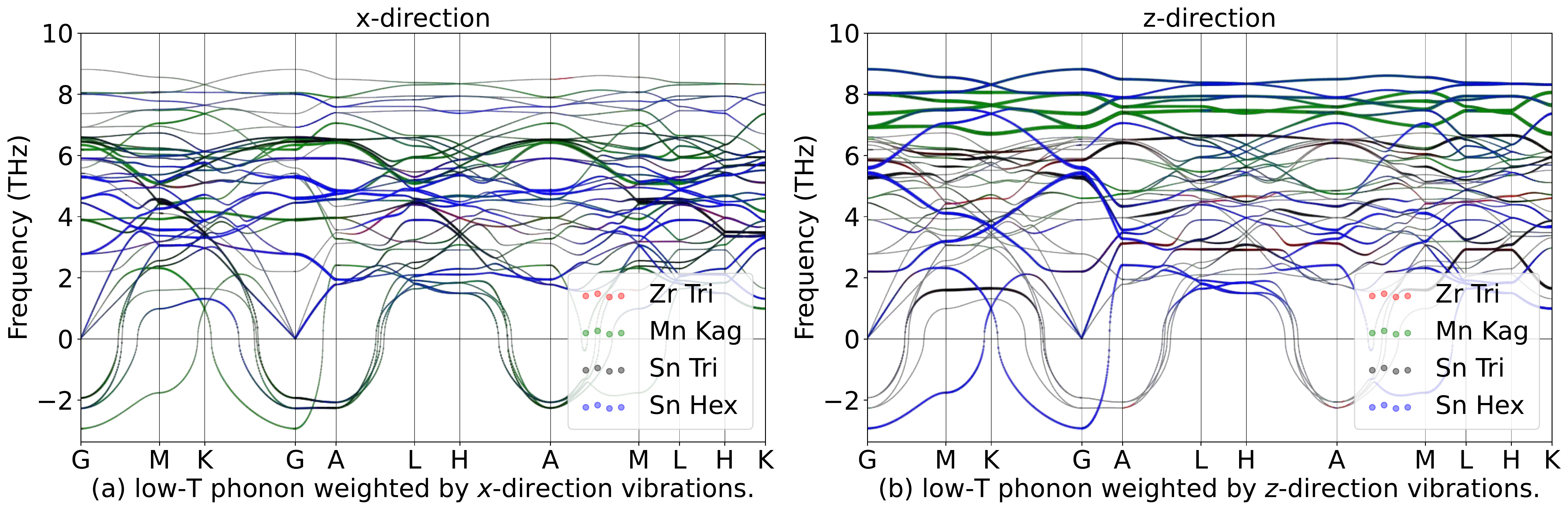}
\hspace{5pt}\label{fig:ZrMn6Sn6_relaxed_High_xz}
\includegraphics[width=0.85\textwidth]{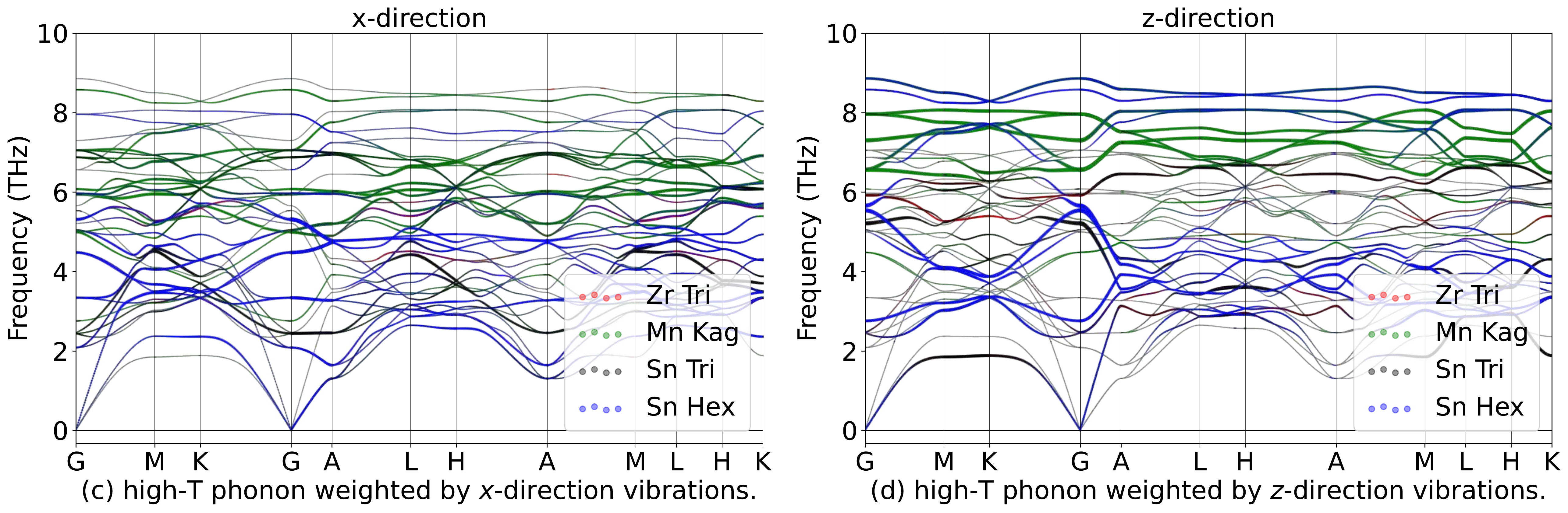}
\caption{Phonon spectrum for ZrMn$_6$Sn$_6$in in-plane ($x$) and out-of-plane ($z$) directions in paramagnetic phase.}
\label{fig:ZrMn6Sn6phoxyz}
\end{figure}

\begin{table}[htbp]
\global\long\def\arraystretch{1.12}
\captionsetup{width=.95\textwidth}
\caption{IRREPs at high-symmetry points for ZrMn$_6$Sn$_6$ phonon spectrum at Low-T.}
\label{tab:191_ZrMn6Sn6_Low}
\setlength{\tabcolsep}{1.mm}{
}
\end{table}

\begin{table}[htbp]
\global\long\def\arraystretch{1.12}
\captionsetup{width=.95\textwidth}
\caption{IRREPs at high-symmetry points for ZrMn$_6$Sn$_6$ phonon spectrum at High-T.}
\label{tab:191_ZrMn6Sn6_High}
\setlength{\tabcolsep}{1.mm}{
%
}
\end{table}
\subsubsection{\label{sms2sec:NbMn6Sn6}\hyperref[smsec:col]{\texorpdfstring{NbMn$_6$Sn$_6$}{}}~{\color{blue}  (High-T stable, Low-T unstable) }}

\begin{table}[htbp]
\global\long\def\arraystretch{1.12}
\captionsetup{width=.85\textwidth}
\caption{Structure parameters of NbMn$_6$Sn$_6$ after relaxation. It contains 491 electrons. }
\label{tab:NbMn6Sn6}
\setlength{\tabcolsep}{4mm}{
%
}
\end{table}

\begin{figure}[htbp]
\centering
\includegraphics[width=0.85\textwidth]{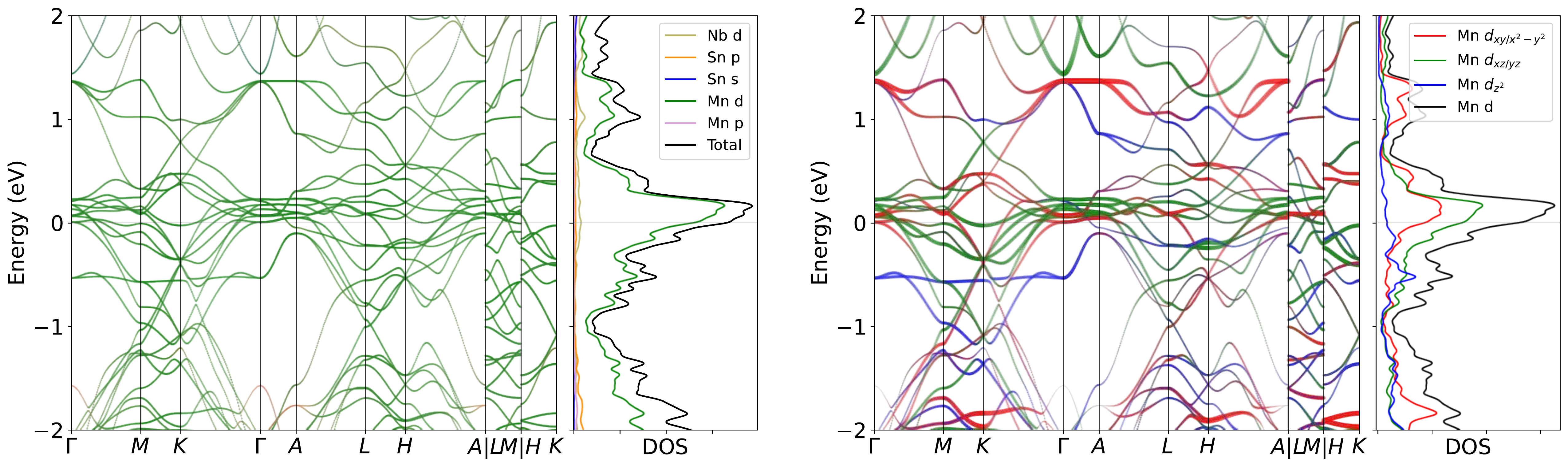}
\caption{Band structure for NbMn$_6$Sn$_6$ without SOC in paramagnetic phase. The projection of Mn $3d$ orbitals is also presented. The band structures clearly reveal the dominating of Mn $3d$ orbitals  near the Fermi level, as well as the pronounced peak.}
\label{fig:NbMn6Sn6band}
\end{figure}

\begin{figure}[H]
\centering
\subfloat[Relaxed phonon at low-T.]{
\label{fig:NbMn6Sn6_relaxed_Low}
\includegraphics[width=0.45\textwidth]{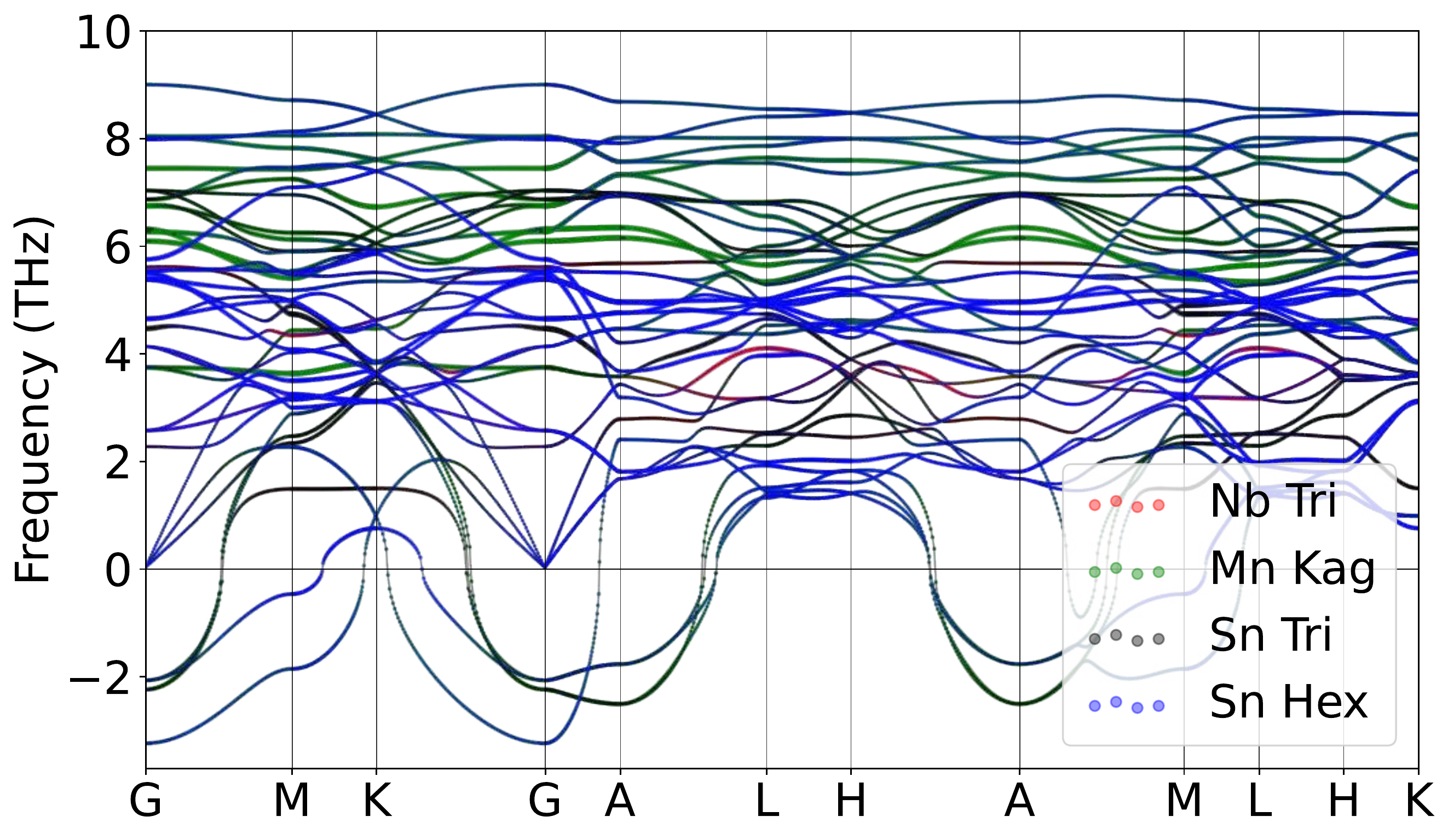}
}
\hspace{5pt}\subfloat[Relaxed phonon at high-T.]{
\label{fig:NbMn6Sn6_relaxed_High}
\includegraphics[width=0.45\textwidth]{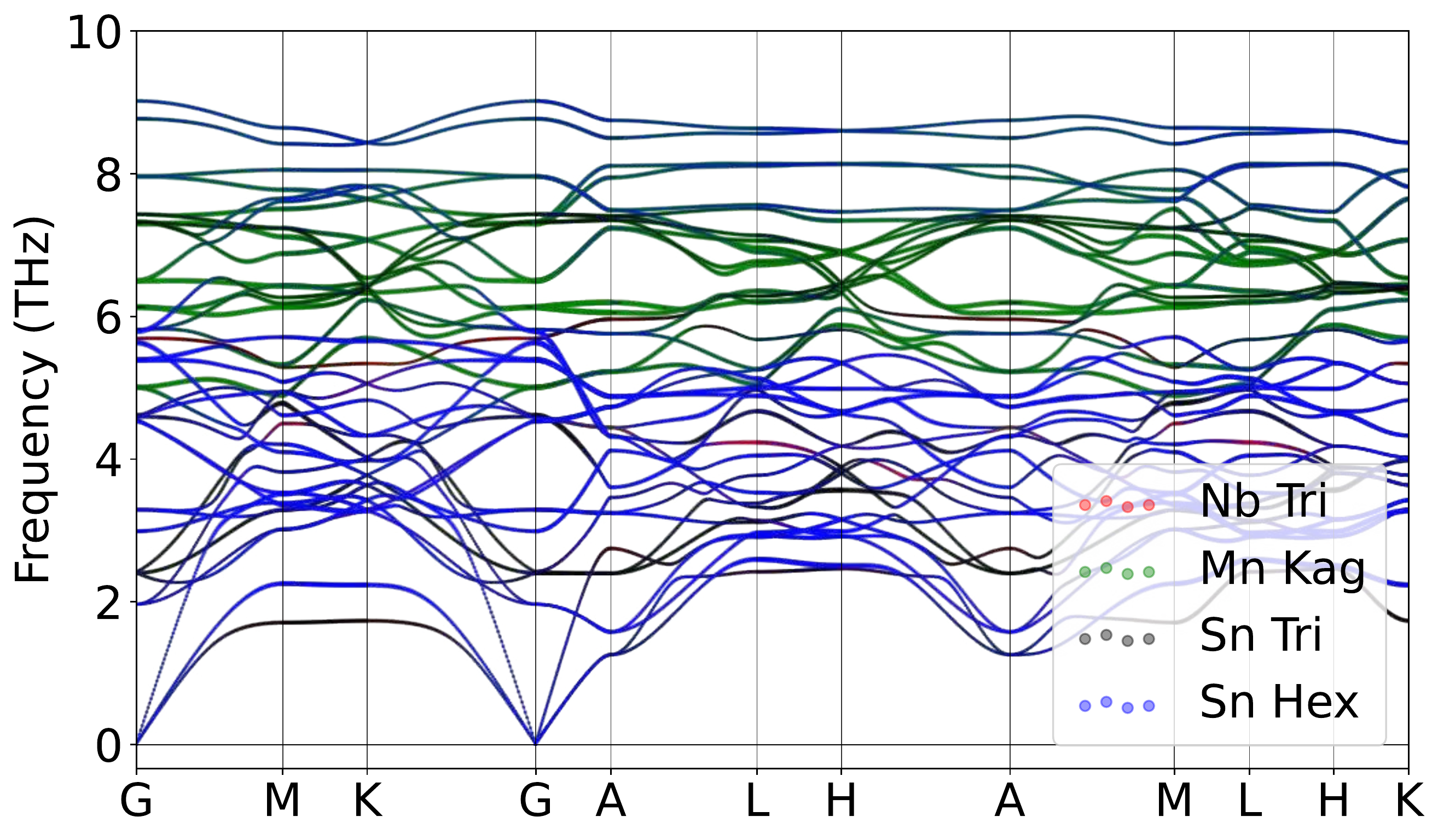}
}
\caption{Atomically projected phonon spectrum for NbMn$_6$Sn$_6$ in paramagnetic phase.}
\label{fig:NbMn6Sn6pho}
\end{figure}

\begin{figure}[htbp]
\centering
\includegraphics[width=0.45\textwidth]{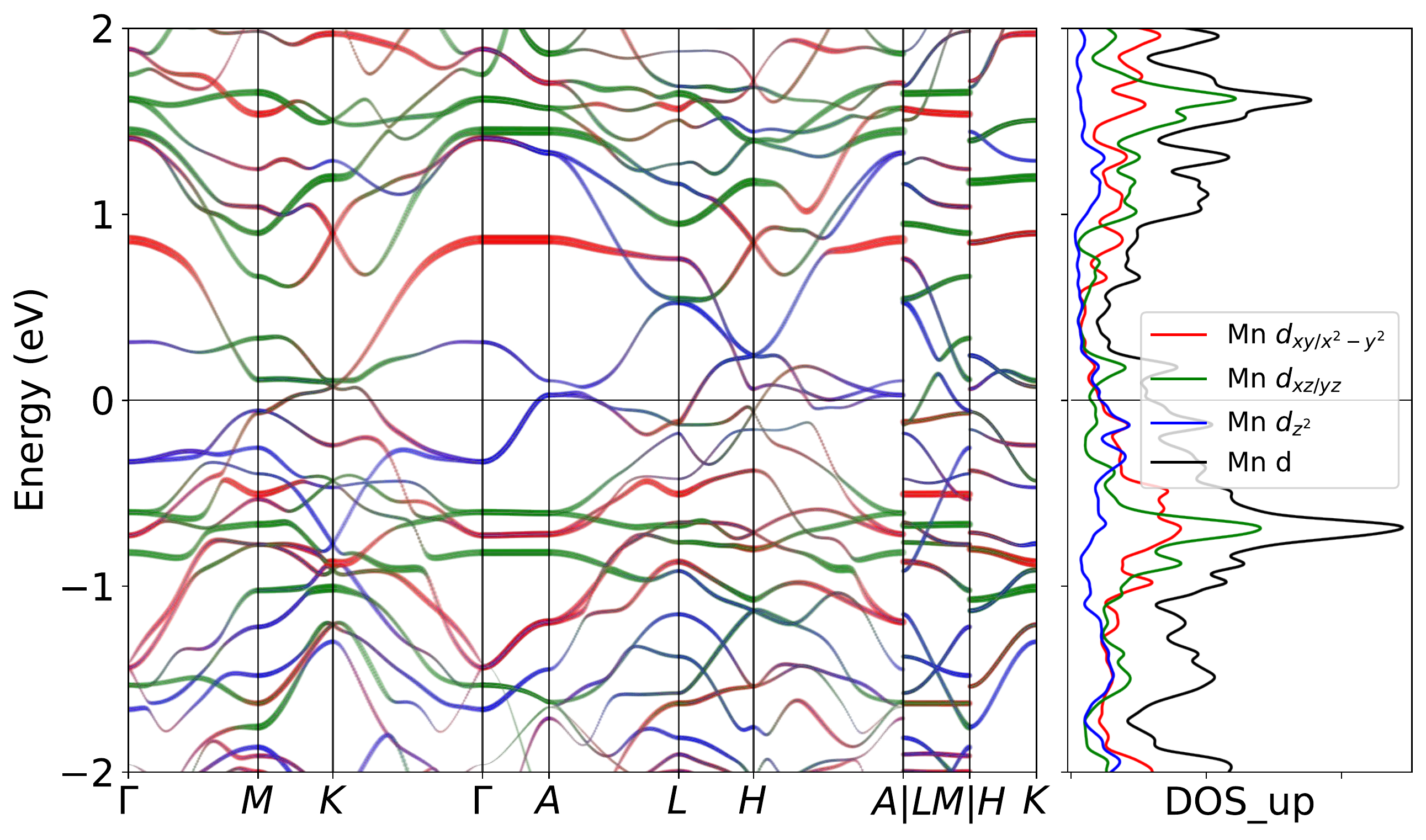}
\includegraphics[width=0.45\textwidth]{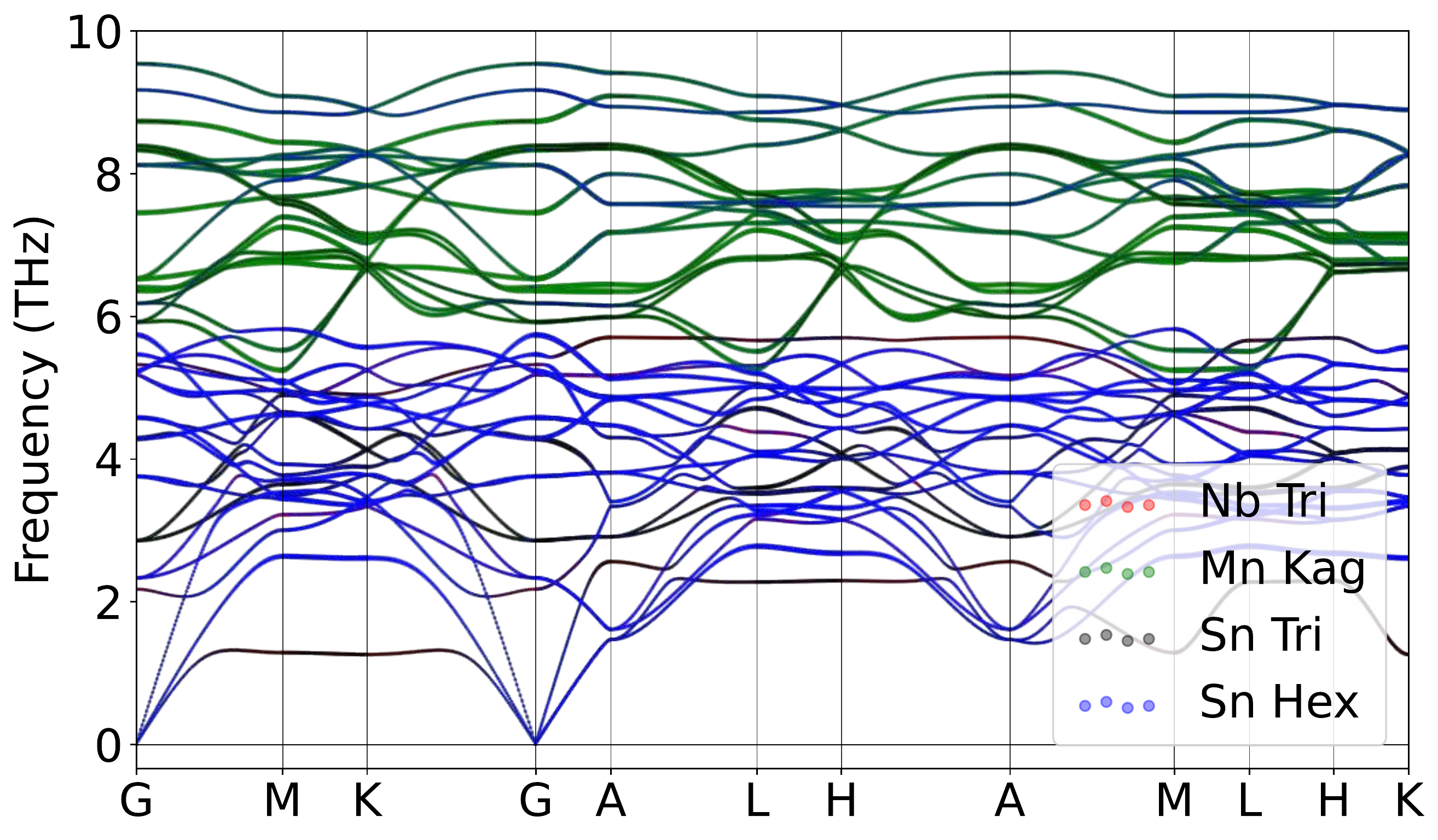}
\caption{Band structure for NbMn$_6$Sn$_6$ with AFM order without SOC for spin (Left)up channel. The projection of Mn $3d$ orbitals is also presented. (Right)Correspondingly, a stable phonon was demonstrated with the collinear magnetic order without Hubbard U at lower temperature regime, weighted by atomic projections.}
\label{fig:NbMn6Sn6mag}
\end{figure}

\begin{figure}[H]
\centering
\label{fig:NbMn6Sn6_relaxed_Low_xz}
\includegraphics[width=0.85\textwidth]{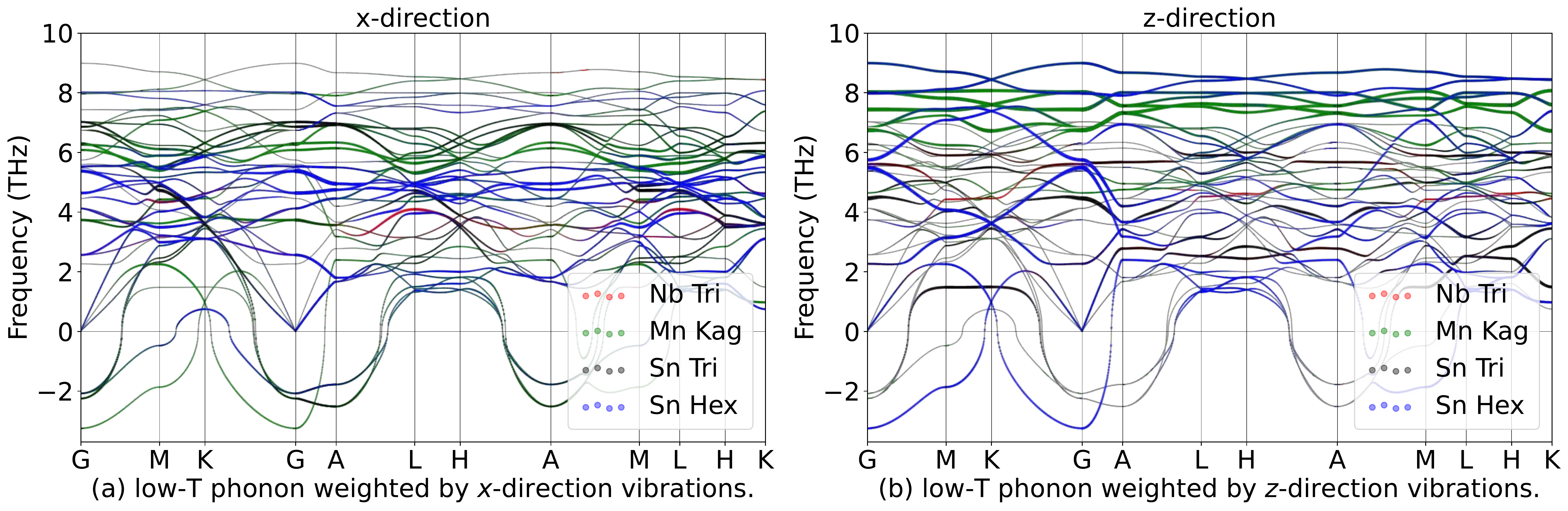}
\hspace{5pt}\label{fig:NbMn6Sn6_relaxed_High_xz}
\includegraphics[width=0.85\textwidth]{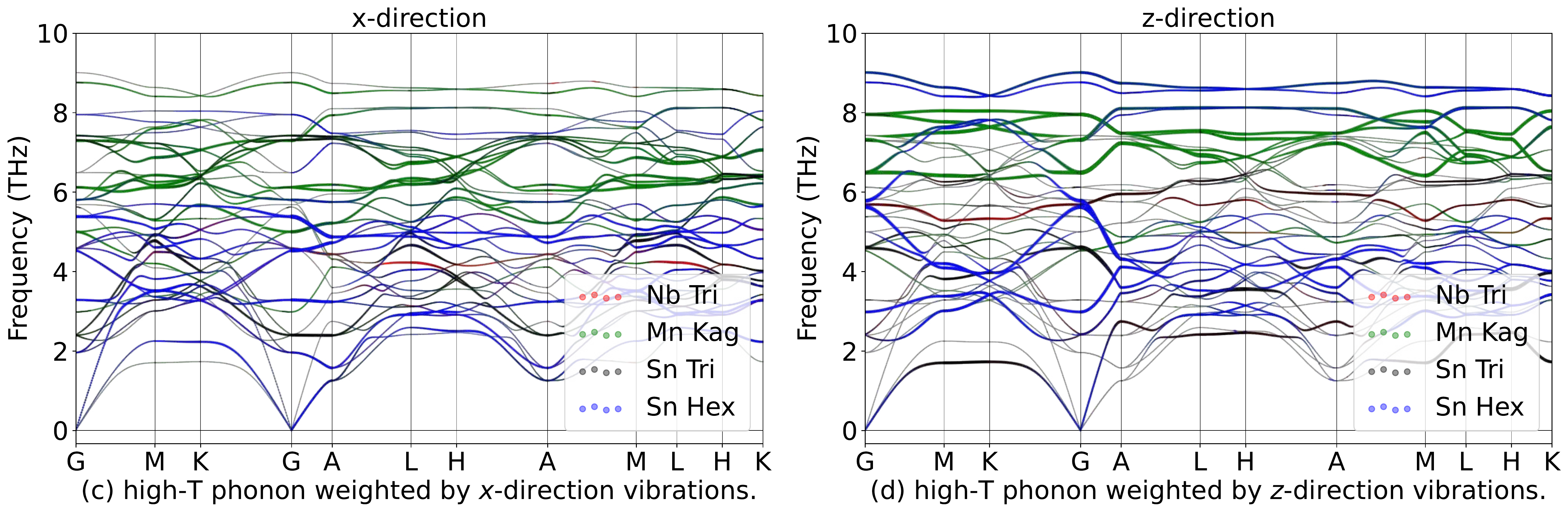}
\caption{Phonon spectrum for NbMn$_6$Sn$_6$in in-plane ($x$) and out-of-plane ($z$) directions in paramagnetic phase.}
\label{fig:NbMn6Sn6phoxyz}
\end{figure}

\begin{table}[htbp]
\global\long\def\arraystretch{1.12}
\captionsetup{width=.95\textwidth}
\caption{IRREPs at high-symmetry points for NbMn$_6$Sn$_6$ phonon spectrum at Low-T.}
\label{tab:191_NbMn6Sn6_Low}
\setlength{\tabcolsep}{1.mm}{
}
\end{table}

\begin{table}[htbp]
\global\long\def\arraystretch{1.12}
\captionsetup{width=.95\textwidth}
\caption{IRREPs at high-symmetry points for NbMn$_6$Sn$_6$ phonon spectrum at High-T.}
\label{tab:191_NbMn6Sn6_High}
\setlength{\tabcolsep}{1.mm}{
%
}
\end{table}
\subsubsection{\label{sms2sec:PrMn6Sn6}\hyperref[smsec:col]{\texorpdfstring{PrMn$_6$Sn$_6$}{}}~{\color{blue}  (High-T stable, Low-T unstable) }}

\begin{table}[htbp]
\global\long\def\arraystretch{1.12}
\captionsetup{width=.85\textwidth}
\caption{Structure parameters of PrMn$_6$Sn$_6$ after relaxation. It contains 509 electrons. }
\label{tab:PrMn6Sn6}
\setlength{\tabcolsep}{4mm}{
%
}
\end{table}

\begin{figure}[htbp]
\centering
\includegraphics[width=0.85\textwidth]{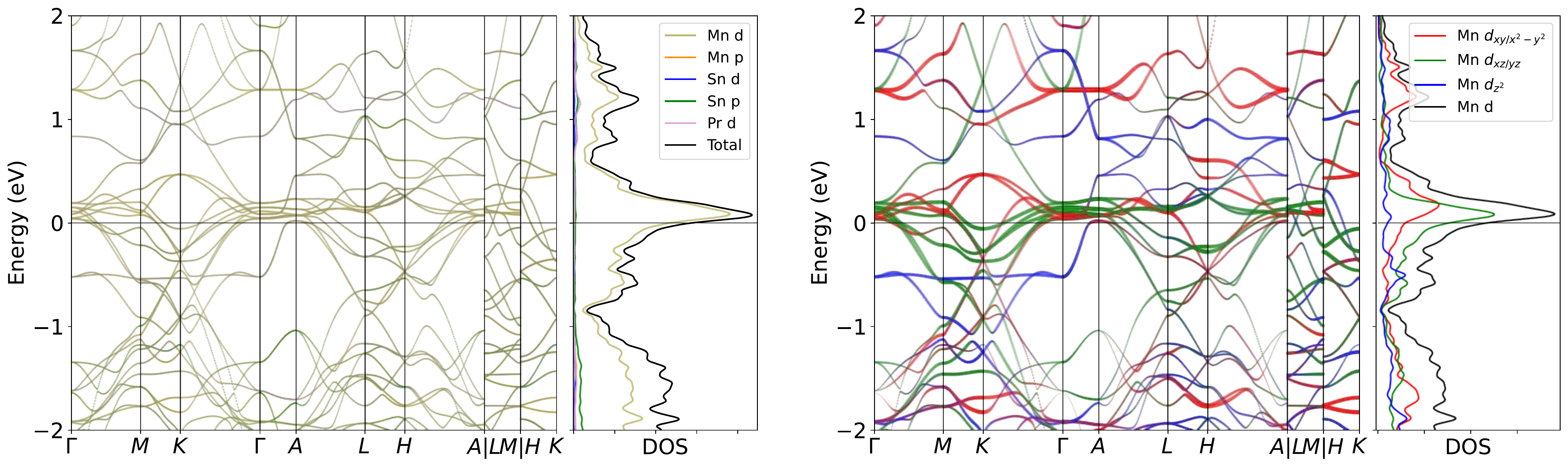}
\caption{Band structure for PrMn$_6$Sn$_6$ without SOC in paramagnetic phase. The projection of Mn $3d$ orbitals is also presented. The band structures clearly reveal the dominating of Mn $3d$ orbitals  near the Fermi level, as well as the pronounced peak.}
\label{fig:PrMn6Sn6band}
\end{figure}

\begin{figure}[H]
\centering
\subfloat[Relaxed phonon at low-T.]{
\label{fig:PrMn6Sn6_relaxed_Low}
\includegraphics[width=0.45\textwidth]{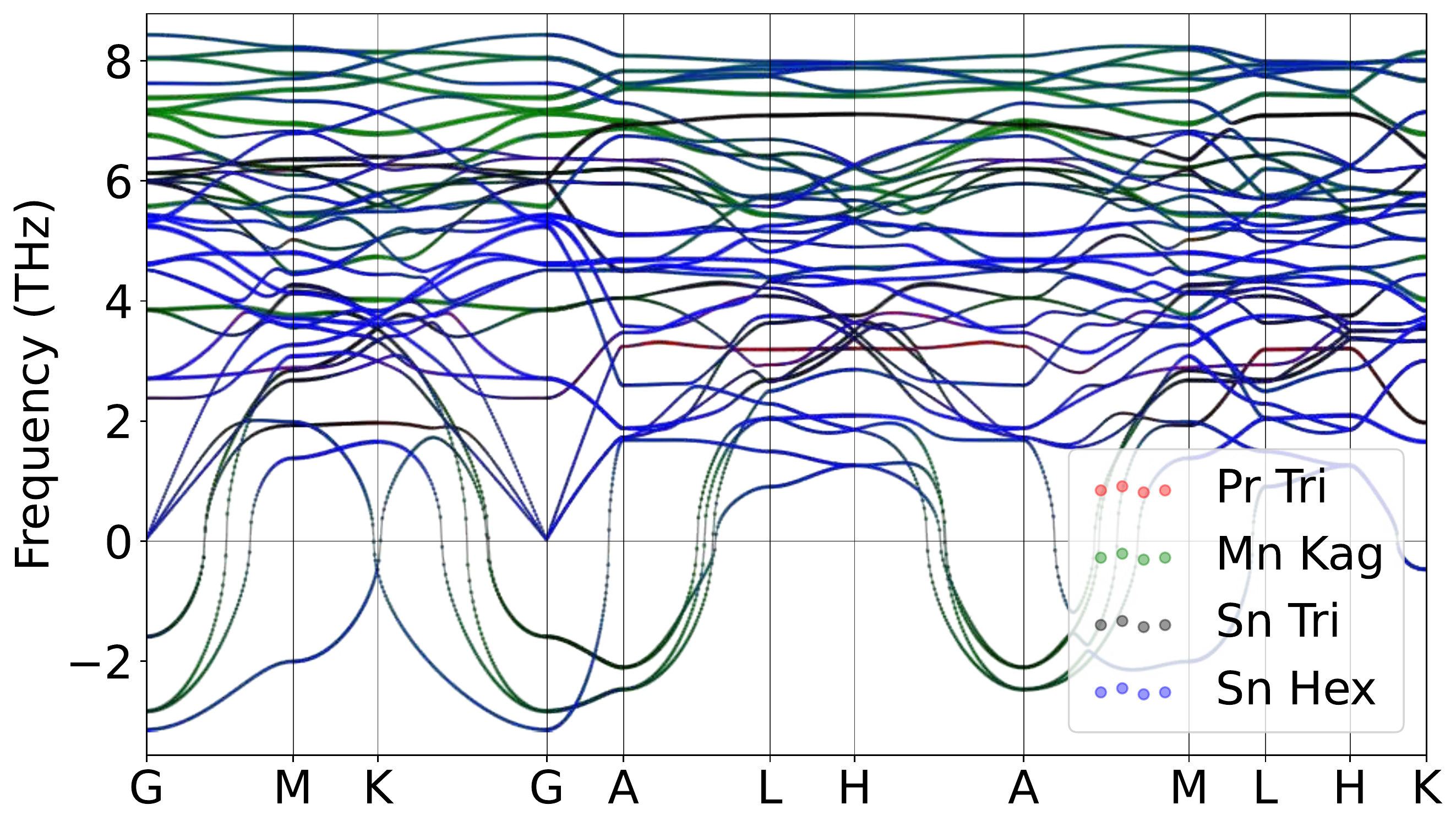}
}
\hspace{5pt}\subfloat[Relaxed phonon at high-T.]{
\label{fig:PrMn6Sn6_relaxed_High}
\includegraphics[width=0.45\textwidth]{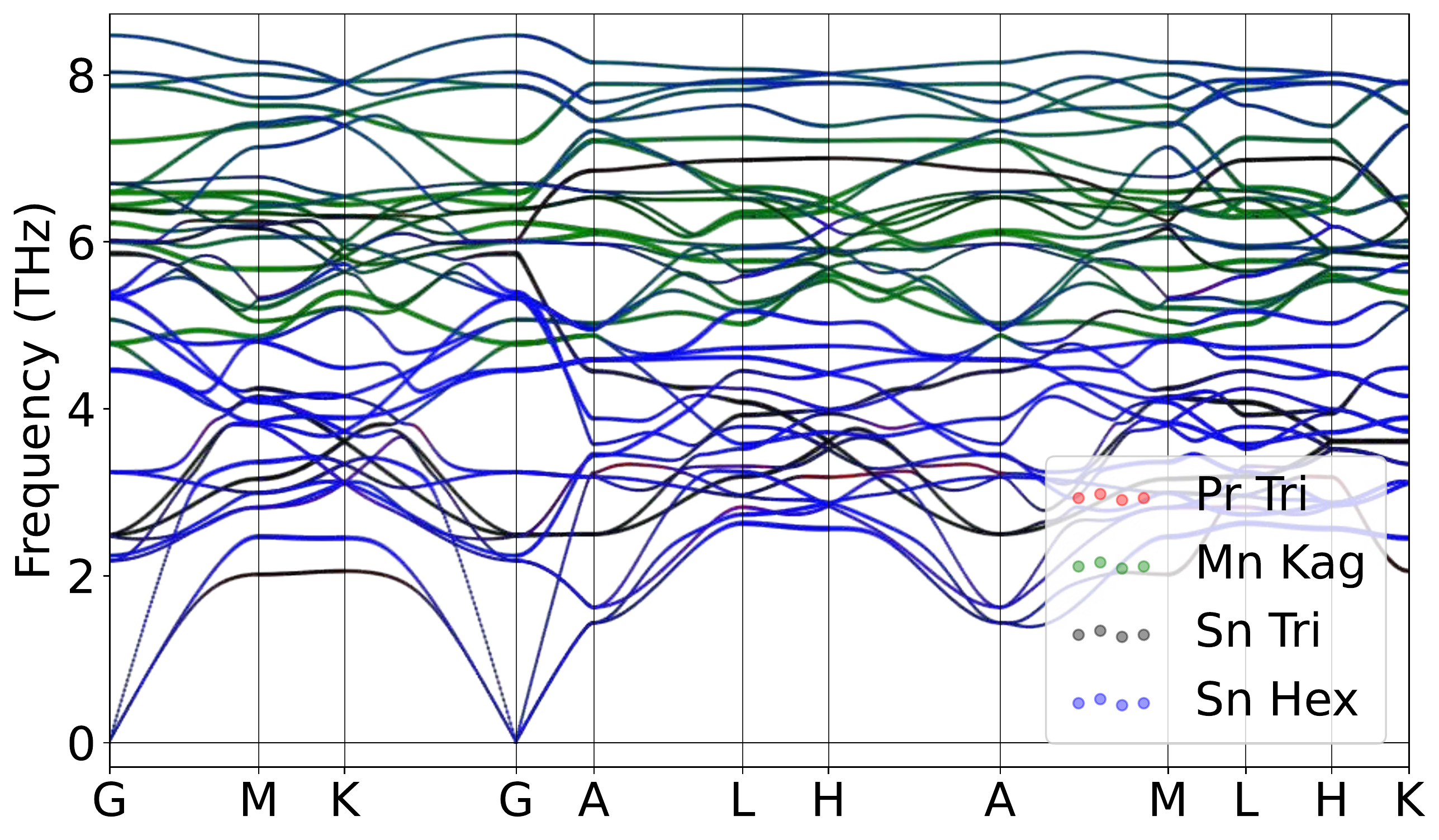}
}
\caption{Atomically projected phonon spectrum for PrMn$_6$Sn$_6$ in paramagnetic phase.}
\label{fig:PrMn6Sn6pho}
\end{figure}

\begin{figure}[htbp]
\centering
\includegraphics[width=0.32\textwidth]{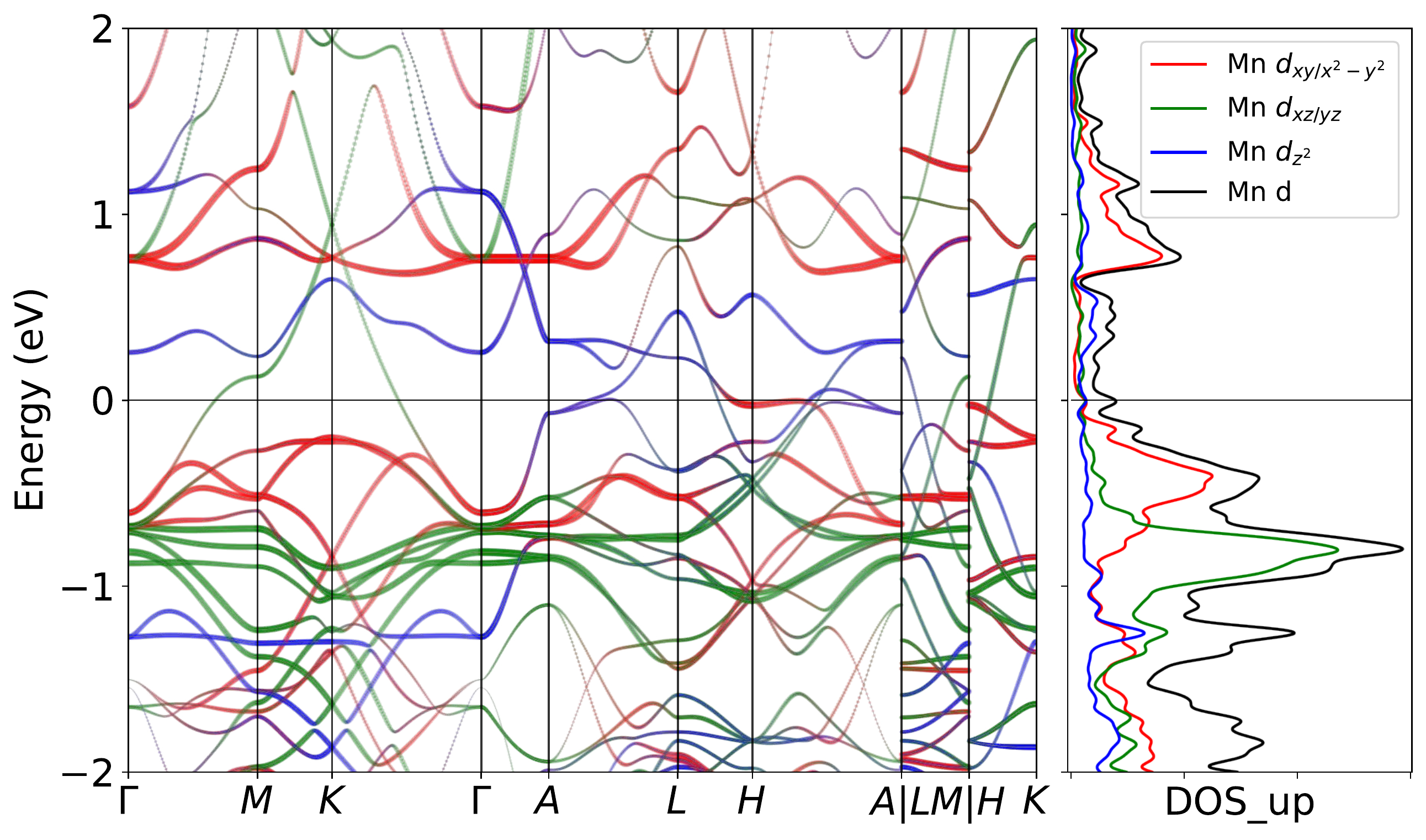}
\includegraphics[width=0.32\textwidth]{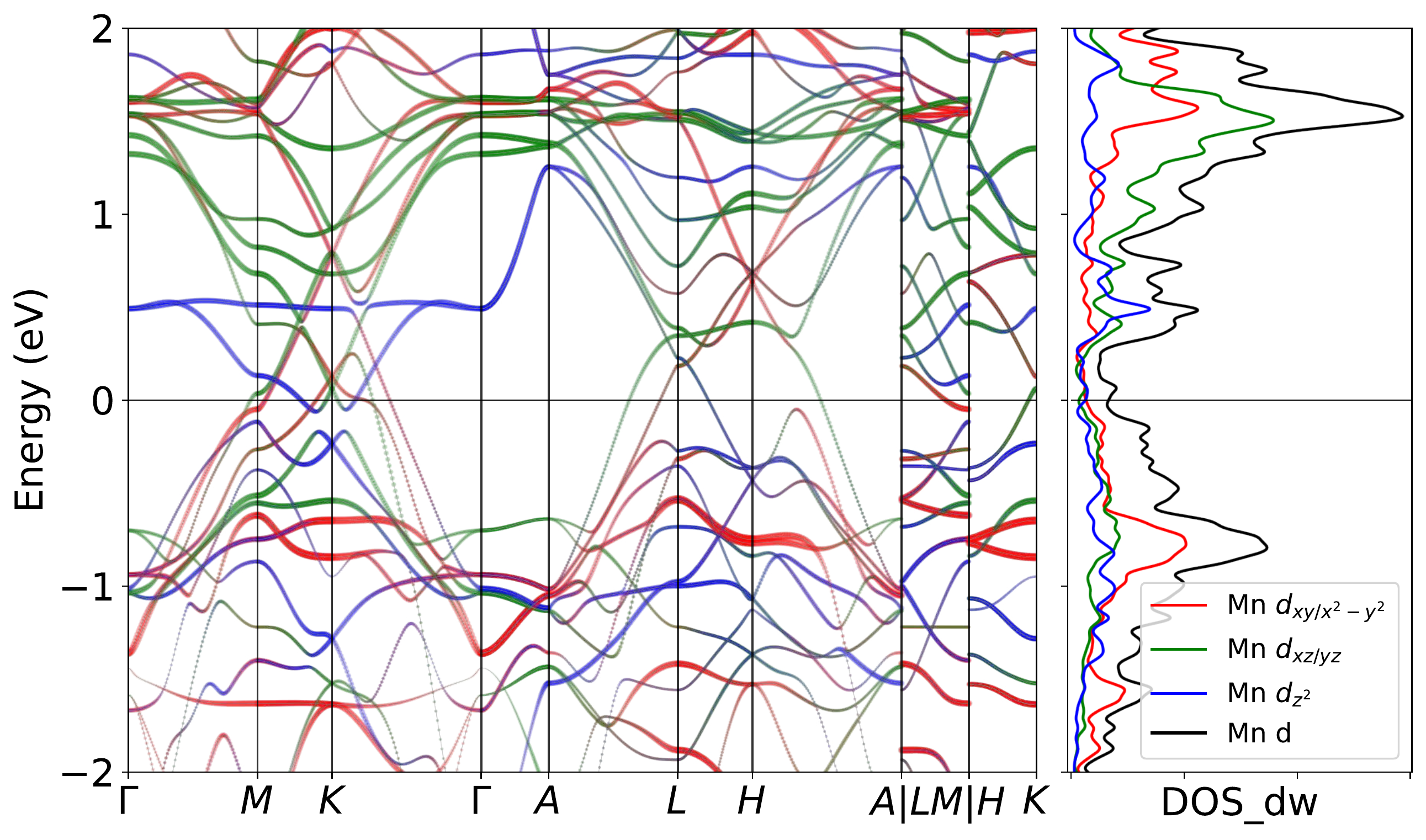}
\includegraphics[width=0.32\textwidth]{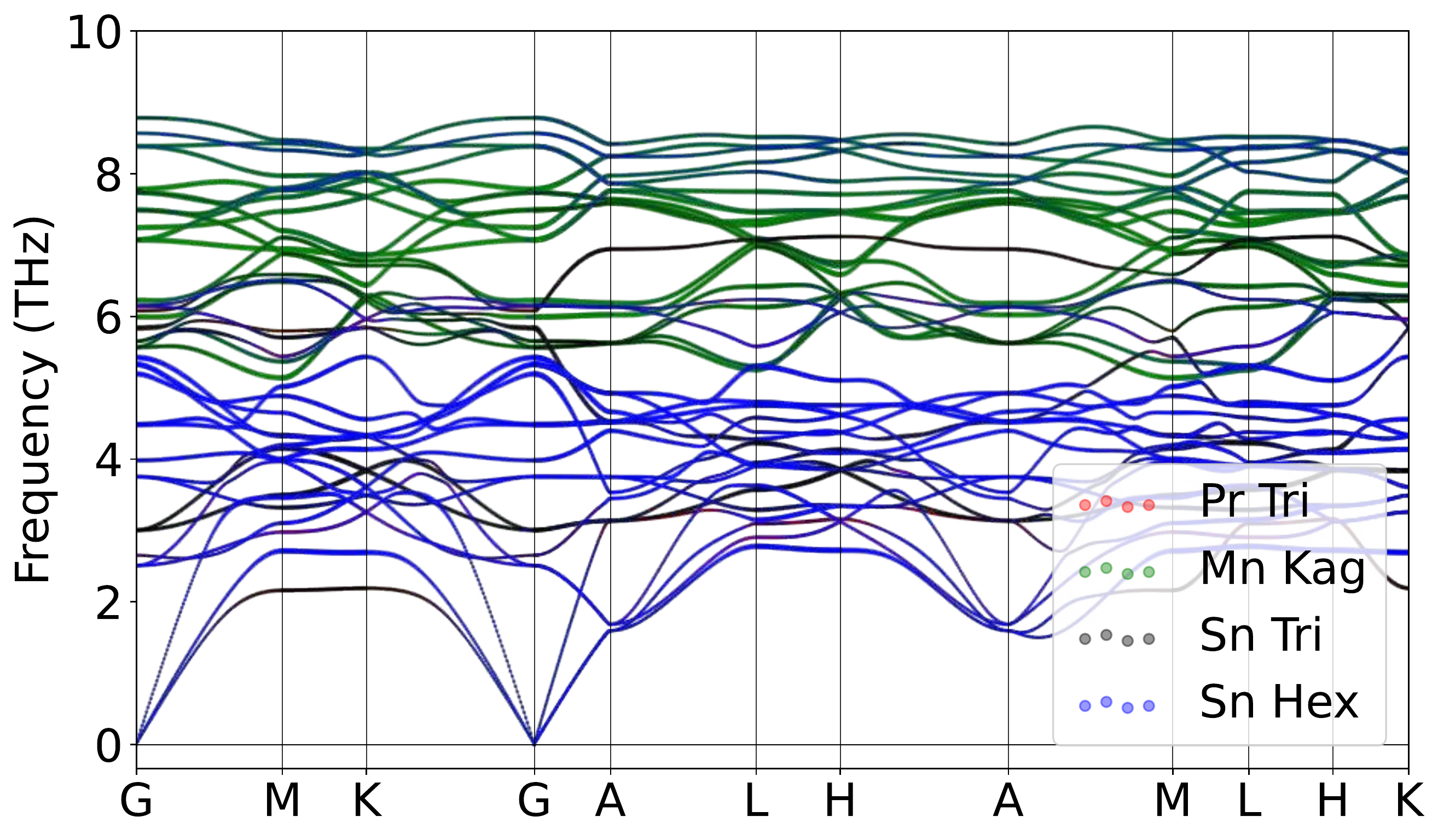}
\caption{Band structure for PrMn$_6$Sn$_6$ with FM order without SOC for spin (Left)up and (Middle)down channels. The projection of Mn $3d$ orbitals is also presented. (Right)Correspondingly, a stable phonon was demonstrated with the collinear magnetic order without Hubbard U at lower temperature regime, weighted by atomic projections.}
\label{fig:PrMn6Sn6mag}
\end{figure}

\begin{figure}[H]
\centering
\label{fig:PrMn6Sn6_relaxed_Low_xz}
\includegraphics[width=0.85\textwidth]{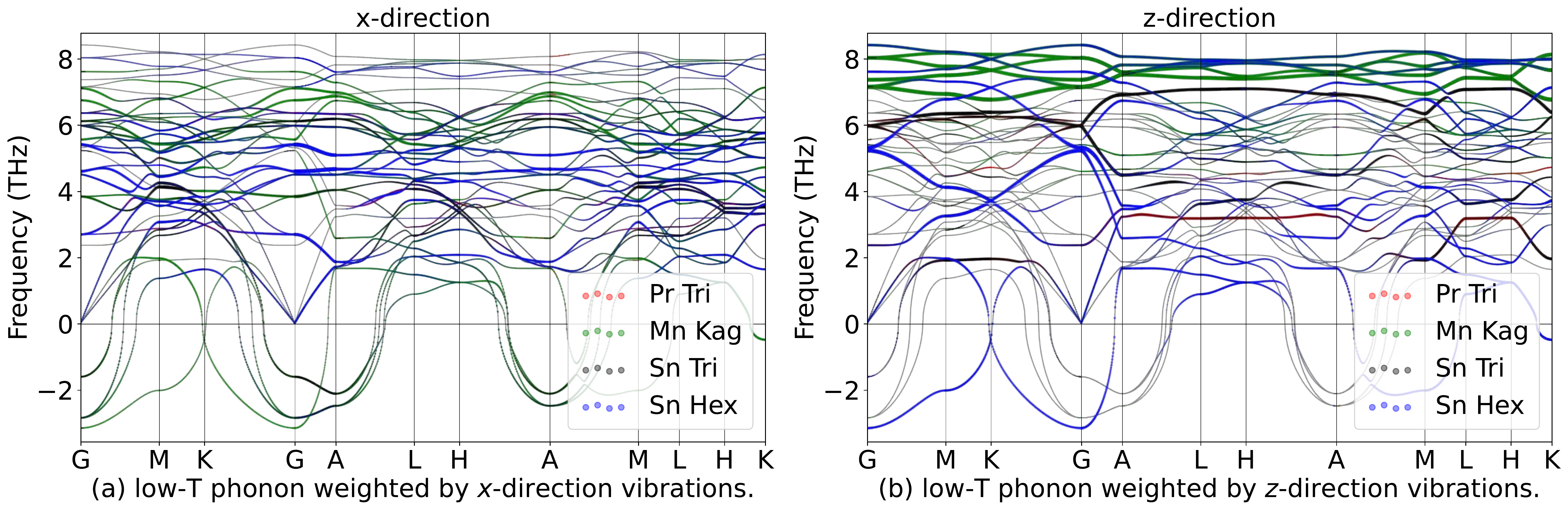}
\hspace{5pt}\label{fig:PrMn6Sn6_relaxed_High_xz}
\includegraphics[width=0.85\textwidth]{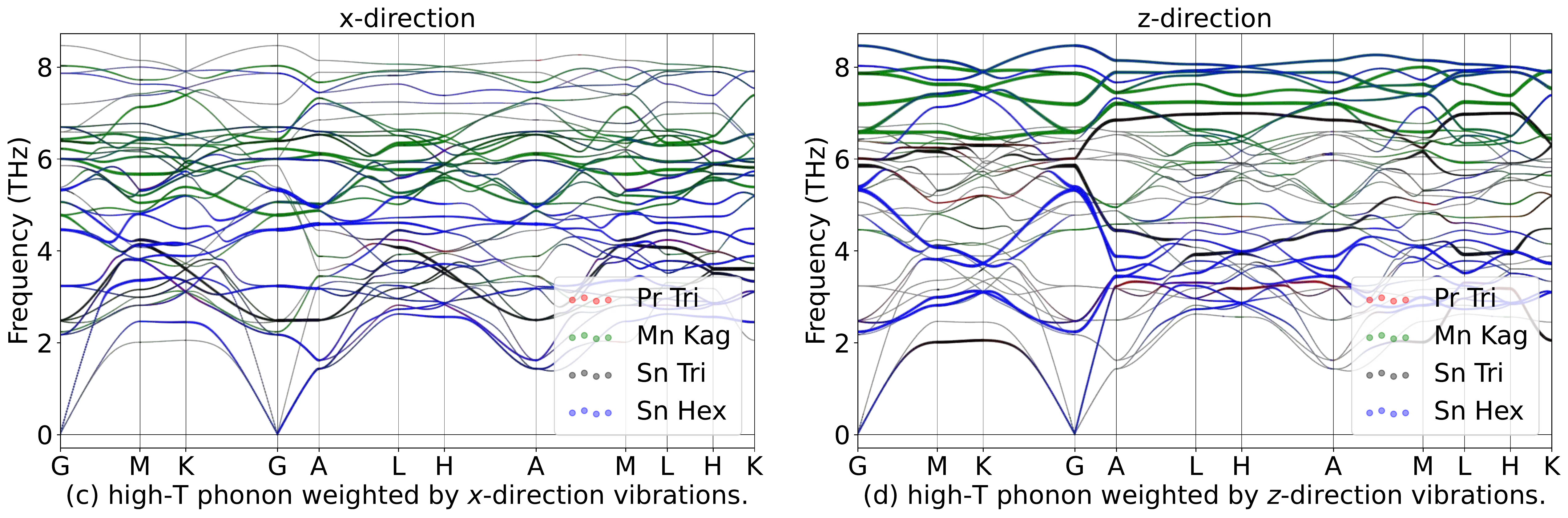}
\caption{Phonon spectrum for PrMn$_6$Sn$_6$in in-plane ($x$) and out-of-plane ($z$) directions in paramagnetic phase.}
\label{fig:PrMn6Sn6phoxyz}
\end{figure}

\begin{table}[htbp]
\global\long\def\arraystretch{1.12}
\captionsetup{width=.95\textwidth}
\caption{IRREPs at high-symmetry points for PrMn$_6$Sn$_6$ phonon spectrum at Low-T.}
\label{tab:191_PrMn6Sn6_Low}
\setlength{\tabcolsep}{1.mm}{
}
\end{table}

\begin{table}[htbp]
\global\long\def\arraystretch{1.12}
\captionsetup{width=.95\textwidth}
\caption{IRREPs at high-symmetry points for PrMn$_6$Sn$_6$ phonon spectrum at High-T.}
\label{tab:191_PrMn6Sn6_High}
\setlength{\tabcolsep}{1.mm}{
%
}
\end{table}
\subsubsection{\label{sms2sec:NdMn6Sn6}\hyperref[smsec:col]{\texorpdfstring{NdMn$_6$Sn$_6$}{}}~{\color{blue}  (High-T stable, Low-T unstable) }}

\begin{table}[htbp]
\global\long\def\arraystretch{1.12}
\captionsetup{width=.85\textwidth}
\caption{Structure parameters of NdMn$_6$Sn$_6$ after relaxation. It contains 510 electrons. }
\label{tab:NdMn6Sn6}
\setlength{\tabcolsep}{4mm}{
%
}
\end{table}

\begin{figure}[htbp]
\centering
\includegraphics[width=0.85\textwidth]{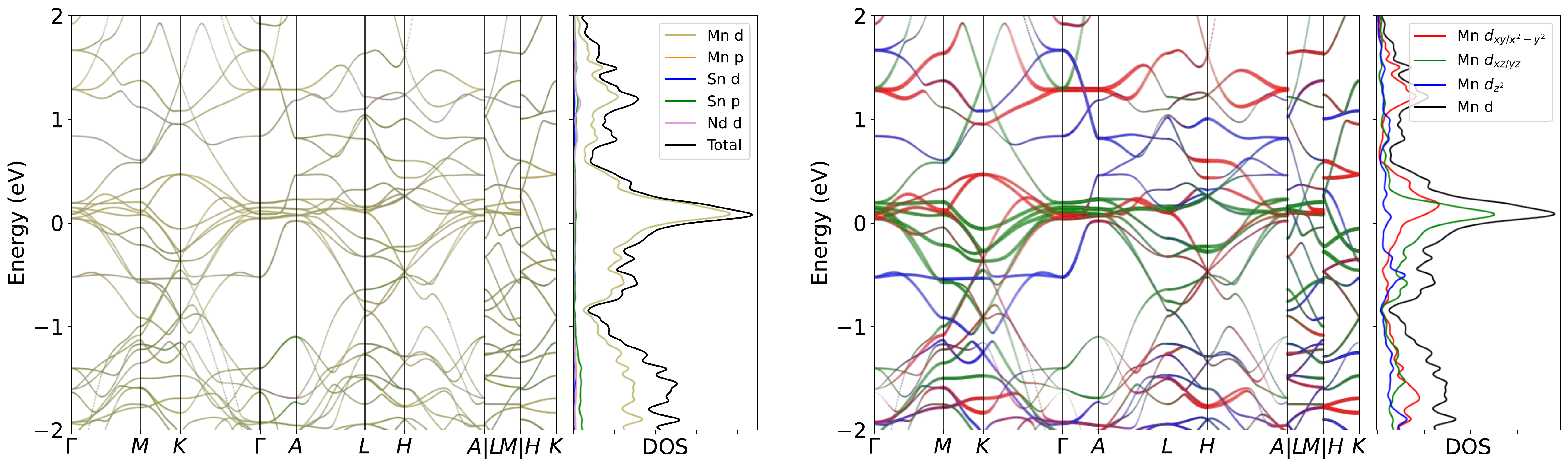}
\caption{Band structure for NdMn$_6$Sn$_6$ without SOC in paramagnetic phase. The projection of Mn $3d$ orbitals is also presented. The band structures clearly reveal the dominating of Mn $3d$ orbitals  near the Fermi level, as well as the pronounced peak.}
\label{fig:NdMn6Sn6band}
\end{figure}

\begin{figure}[H]
\centering
\subfloat[Relaxed phonon at low-T.]{
\label{fig:NdMn6Sn6_relaxed_Low}
\includegraphics[width=0.45\textwidth]{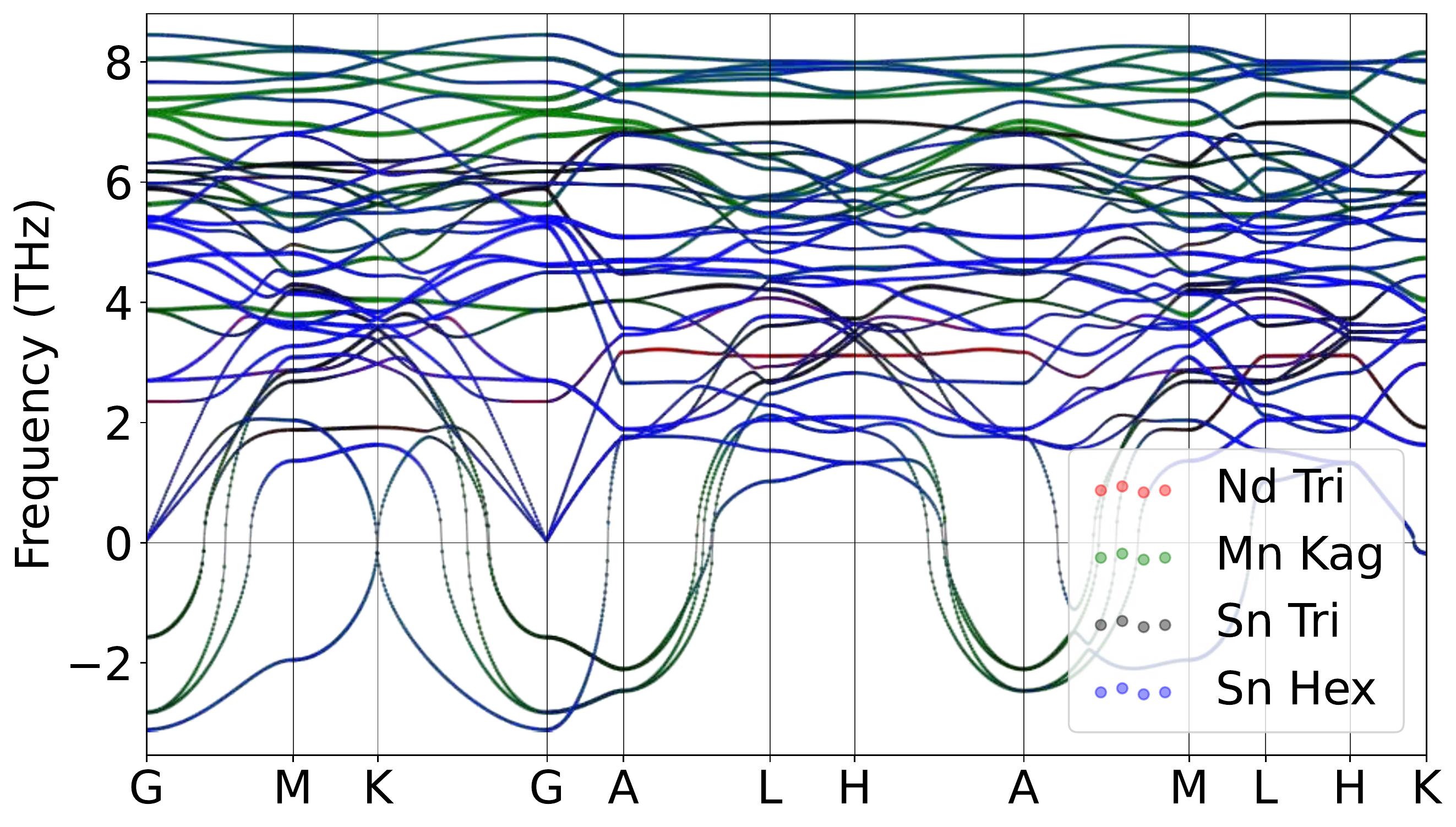}
}
\hspace{5pt}\subfloat[Relaxed phonon at high-T.]{
\label{fig:NdMn6Sn6_relaxed_High}
\includegraphics[width=0.45\textwidth]{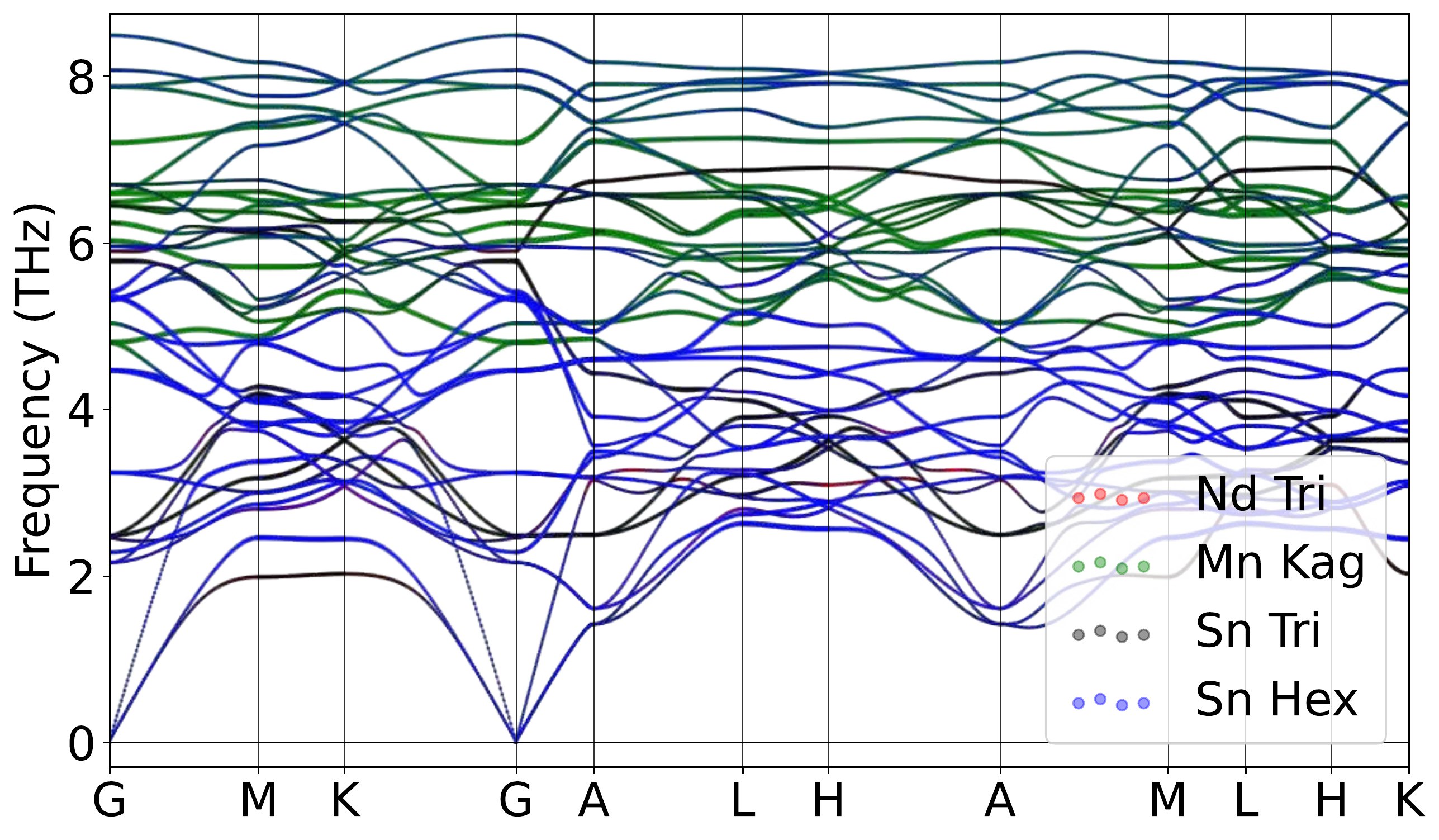}
}
\caption{Atomically projected phonon spectrum for NdMn$_6$Sn$_6$ in paramagnetic phase.}
\label{fig:NdMn6Sn6pho}
\end{figure}

\begin{figure}[htbp]
\centering
\includegraphics[width=0.32\textwidth]{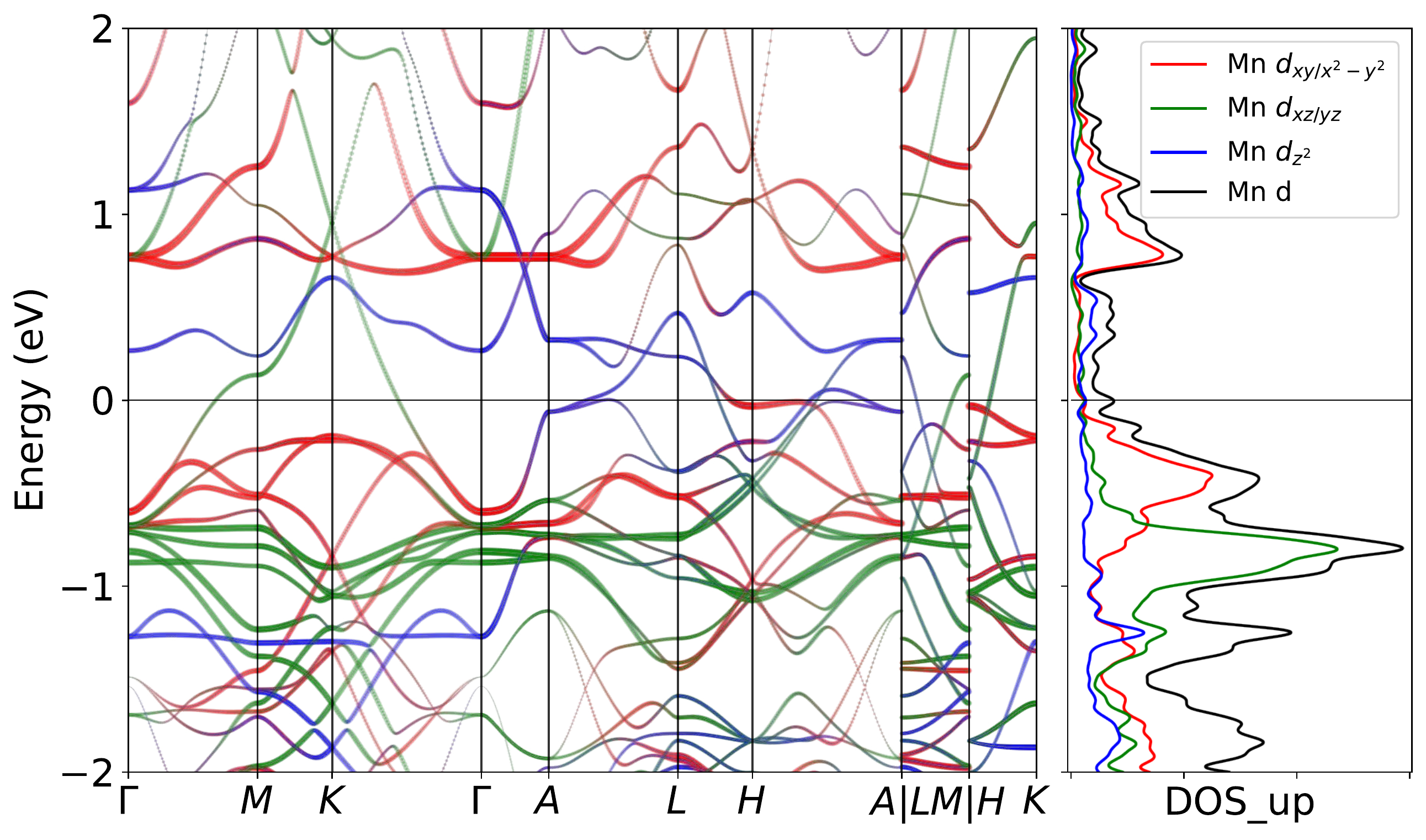}
\includegraphics[width=0.32\textwidth]{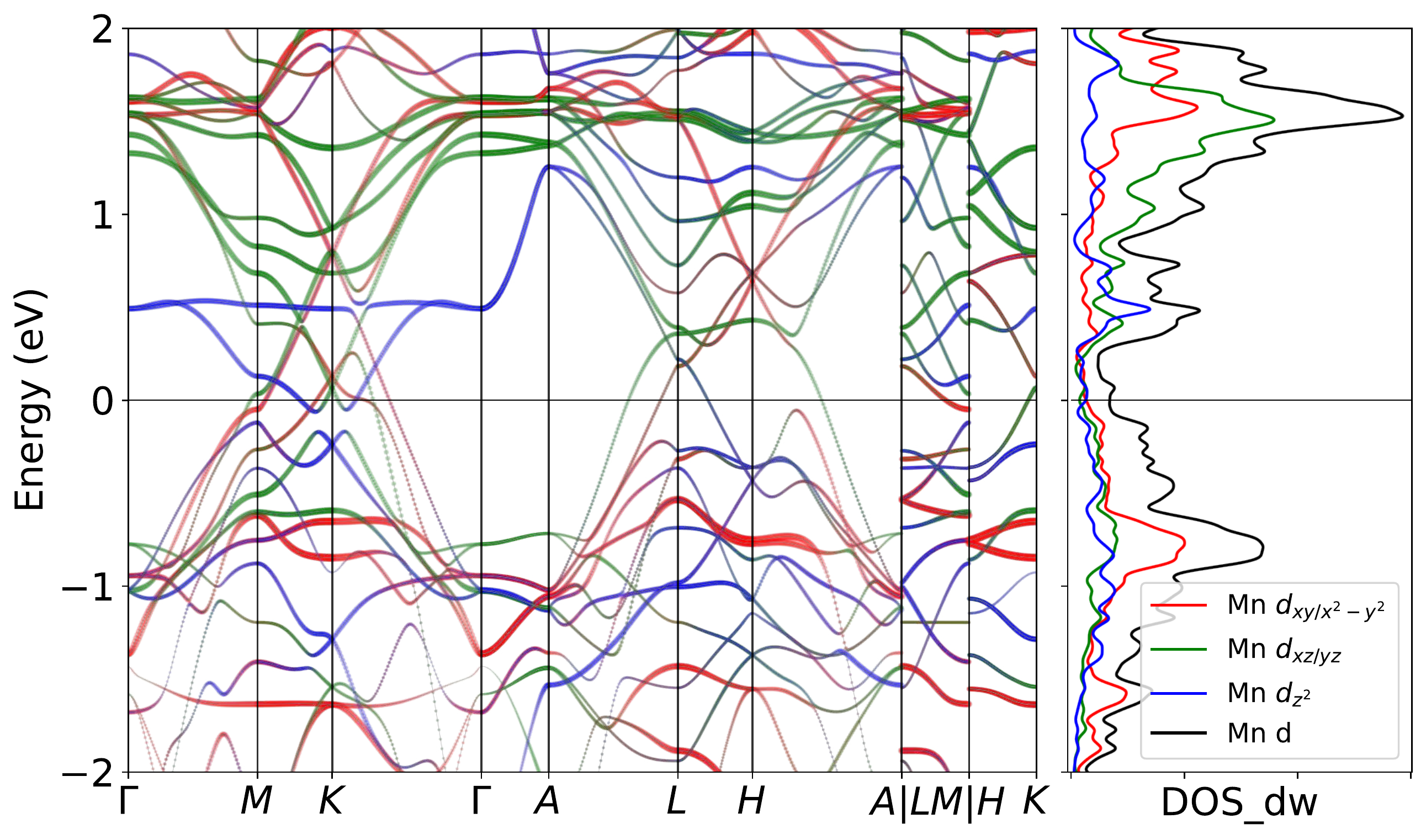}
\includegraphics[width=0.32\textwidth]{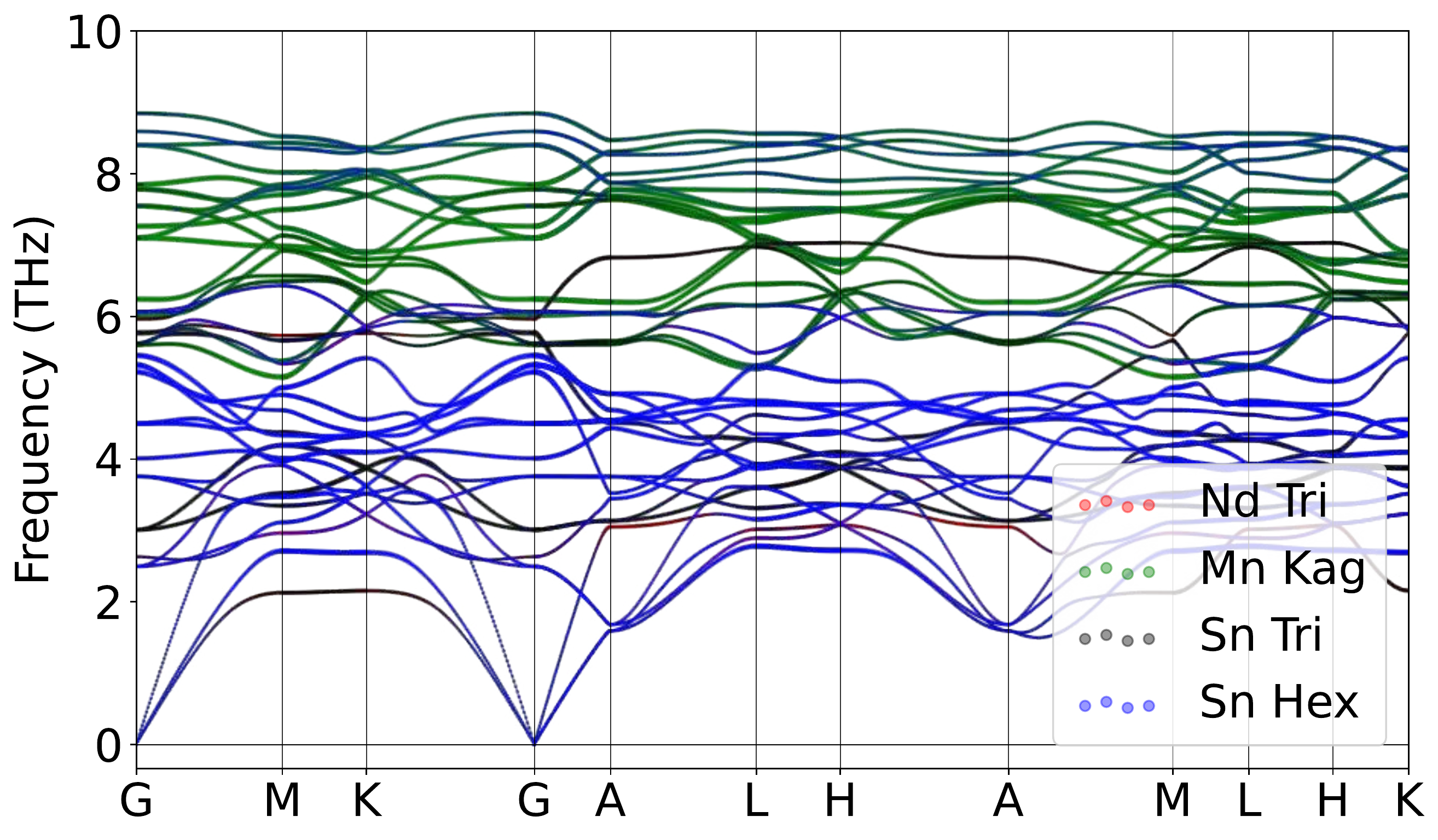}
\caption{Band structure for NdMn$_6$Sn$_6$ with FM order without SOC for spin (Left)up and (Middle)down channels. The projection of Mn $3d$ orbitals is also presented. (Right)Correspondingly, a stable phonon was demonstrated with the collinear magnetic order without Hubbard U at lower temperature regime, weighted by atomic projections.}
\label{fig:NdMn6Sn6mag}
\end{figure}

\begin{figure}[H]
\centering
\label{fig:NdMn6Sn6_relaxed_Low_xz}
\includegraphics[width=0.85\textwidth]{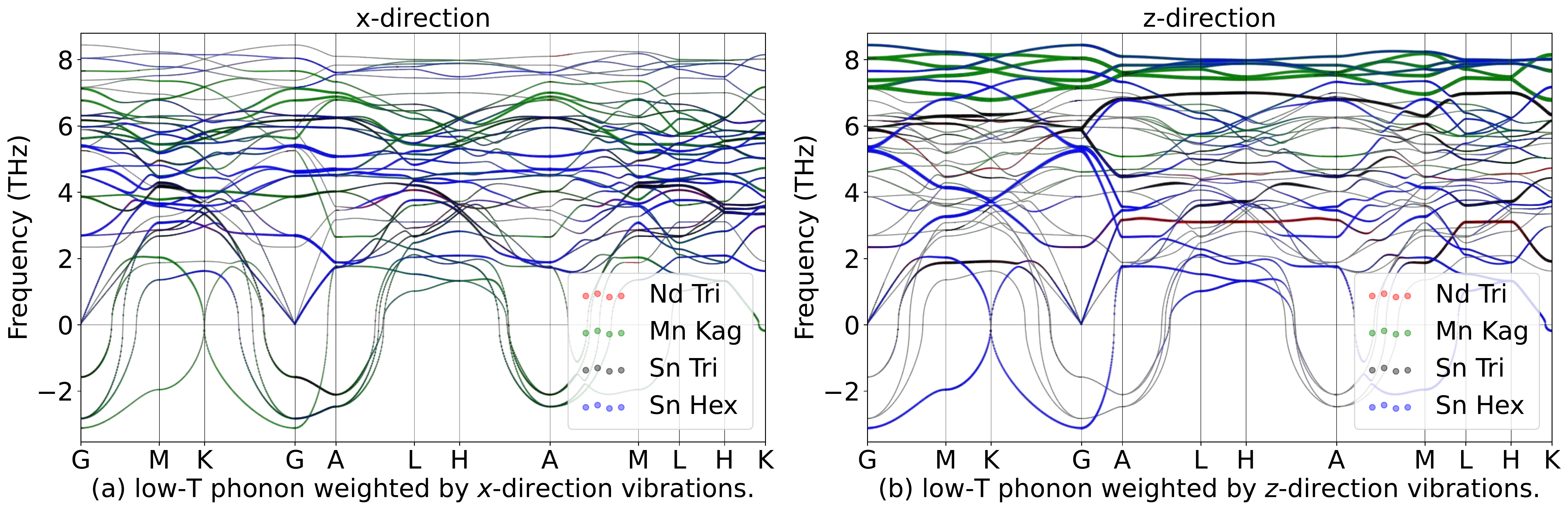}
\hspace{5pt}\label{fig:NdMn6Sn6_relaxed_High_xz}
\includegraphics[width=0.85\textwidth]{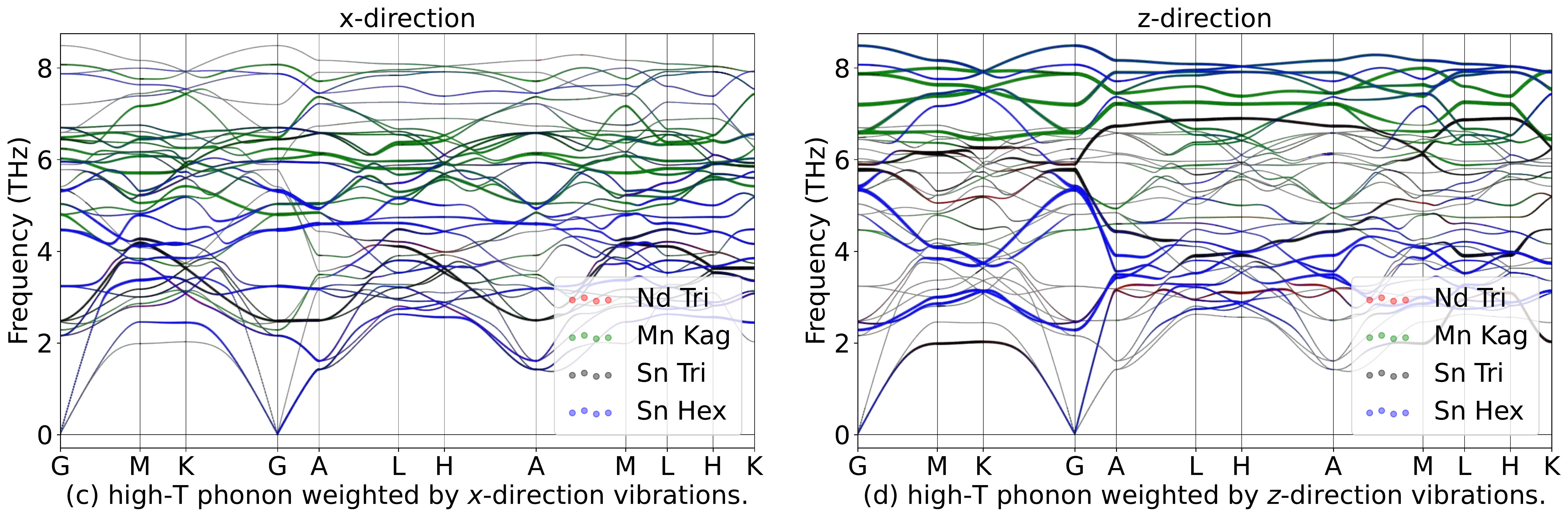}
\caption{Phonon spectrum for NdMn$_6$Sn$_6$in in-plane ($x$) and out-of-plane ($z$) directions in paramagnetic phase.}
\label{fig:NdMn6Sn6phoxyz}
\end{figure}

\begin{table}[htbp]
\global\long\def\arraystretch{1.12}
\captionsetup{width=.95\textwidth}
\caption{IRREPs at high-symmetry points for NdMn$_6$Sn$_6$ phonon spectrum at Low-T.}
\label{tab:191_NdMn6Sn6_Low}
\setlength{\tabcolsep}{1.mm}{
}
\end{table}

\begin{table}[htbp]
\global\long\def\arraystretch{1.12}
\captionsetup{width=.95\textwidth}
\caption{IRREPs at high-symmetry points for NdMn$_6$Sn$_6$ phonon spectrum at High-T.}
\label{tab:191_NdMn6Sn6_High}
\setlength{\tabcolsep}{1.mm}{
%
}
\end{table}
\subsubsection{\label{sms2sec:SmMn6Sn6}\hyperref[smsec:col]{\texorpdfstring{SmMn$_6$Sn$_6$}{}}~{\color{blue}  (High-T stable, Low-T unstable) }}

\begin{table}[htbp]
\global\long\def\arraystretch{1.12}
\captionsetup{width=.85\textwidth}
\caption{Structure parameters of SmMn$_6$Sn$_6$ after relaxation. It contains 512 electrons. }
\label{tab:SmMn6Sn6}
\setlength{\tabcolsep}{4mm}{
%
}
\end{table}

\begin{figure}[htbp]
\centering
\includegraphics[width=0.85\textwidth]{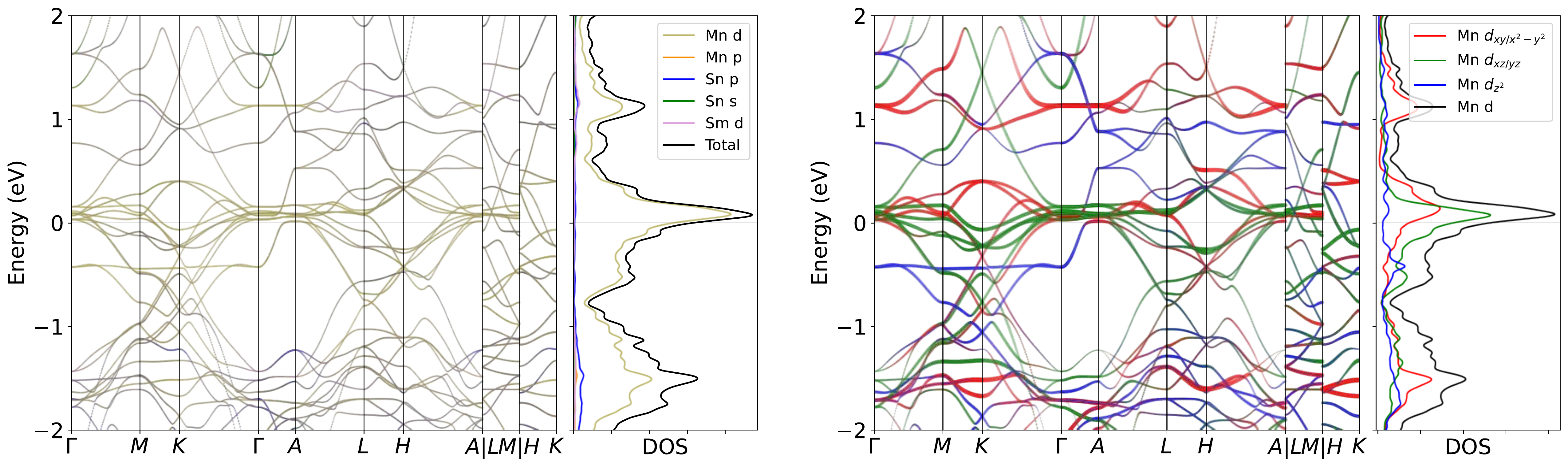}
\caption{Band structure for SmMn$_6$Sn$_6$ without SOC in paramagnetic phase. The projection of Mn $3d$ orbitals is also presented. The band structures clearly reveal the dominating of Mn $3d$ orbitals  near the Fermi level, as well as the pronounced peak.}
\label{fig:SmMn6Sn6band}
\end{figure}

\begin{figure}[H]
\centering
\subfloat[Relaxed phonon at low-T.]{
\label{fig:SmMn6Sn6_relaxed_Low}
\includegraphics[width=0.45\textwidth]{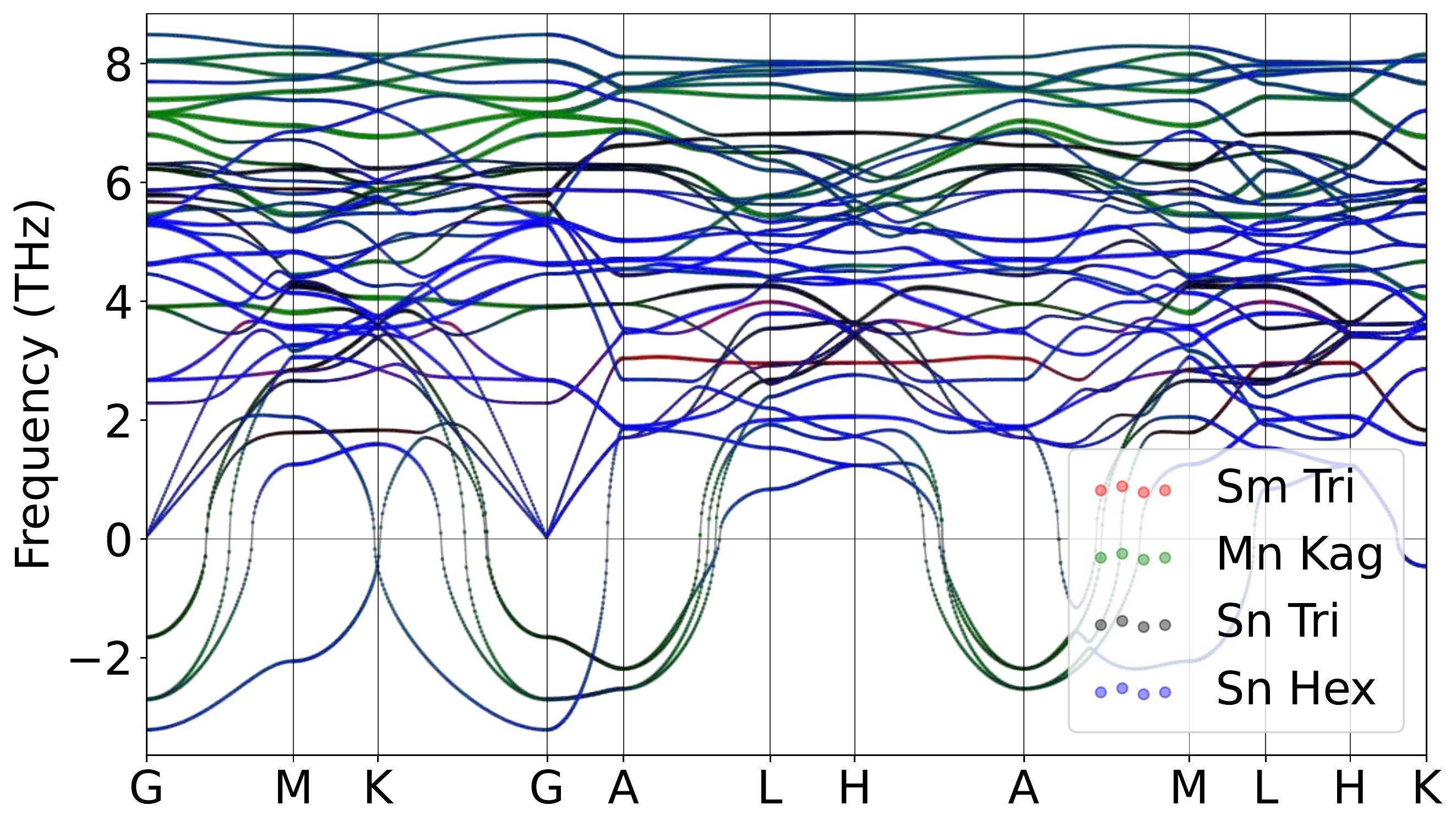}
}
\hspace{5pt}\subfloat[Relaxed phonon at high-T.]{
\label{fig:SmMn6Sn6_relaxed_High}
\includegraphics[width=0.45\textwidth]{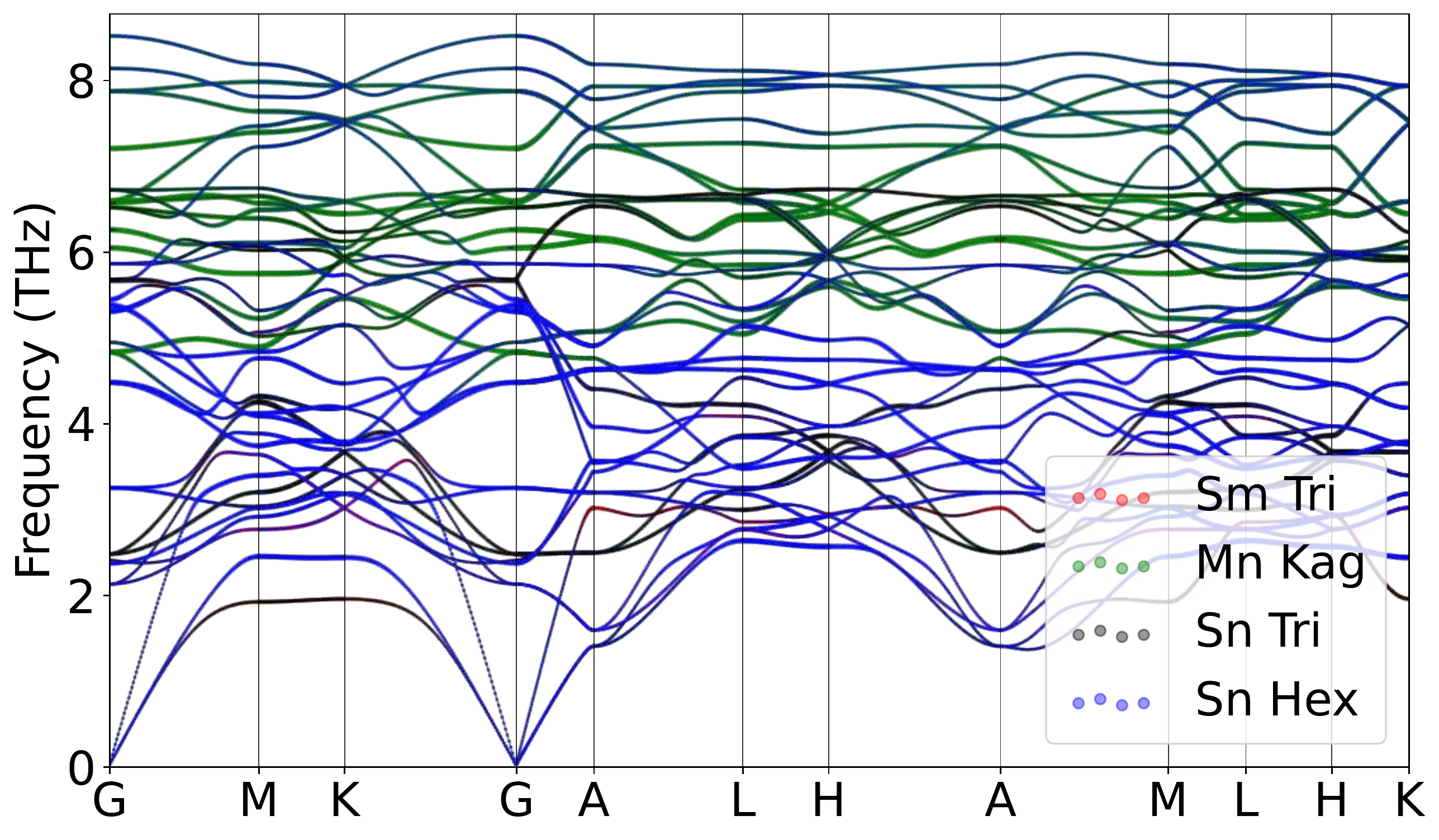}
}
\caption{Atomically projected phonon spectrum for SmMn$_6$Sn$_6$ in paramagnetic phase.}
\label{fig:SmMn6Sn6pho}
\end{figure}

\begin{figure}[htbp]
\centering
\includegraphics[width=0.32\textwidth]{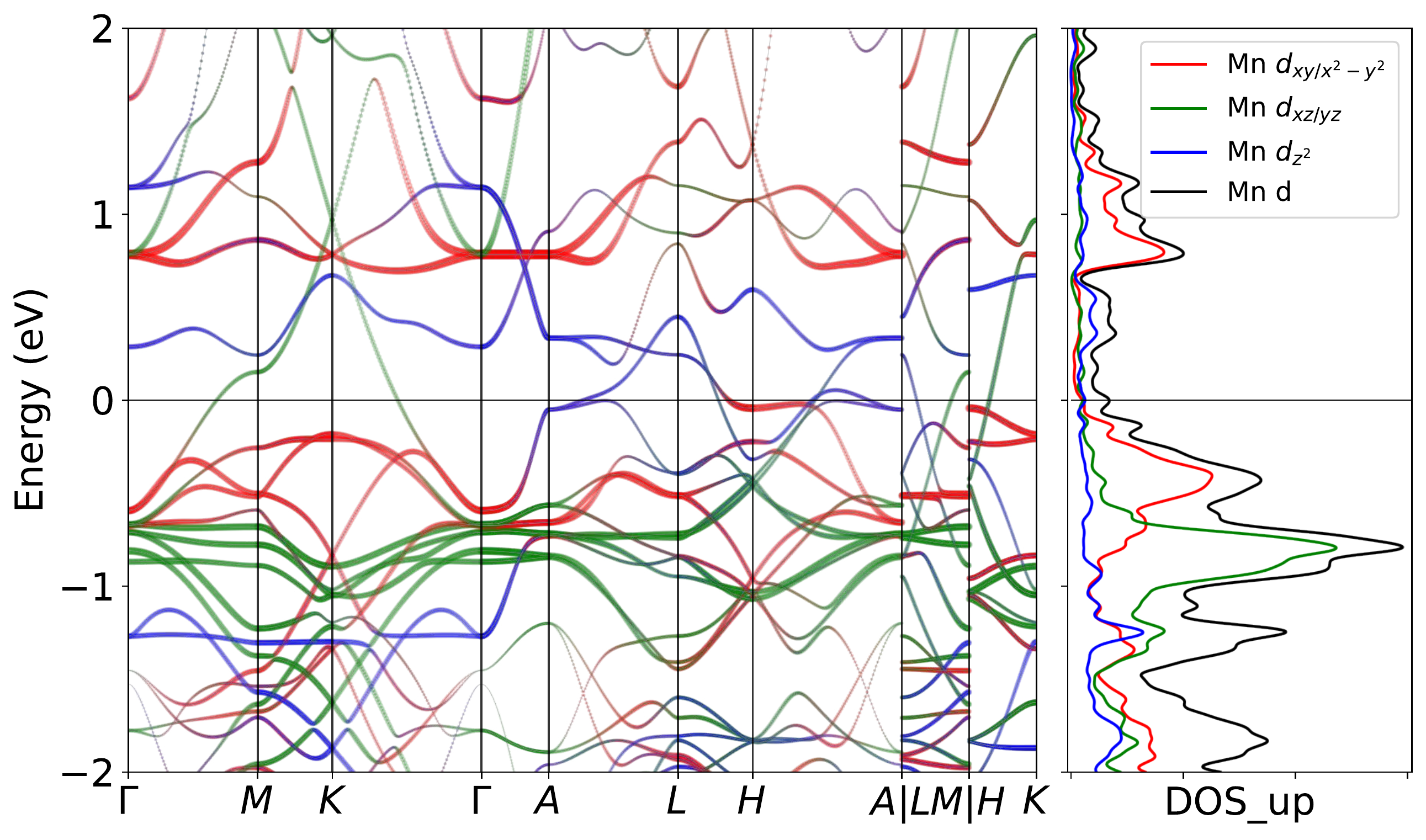}
\includegraphics[width=0.32\textwidth]{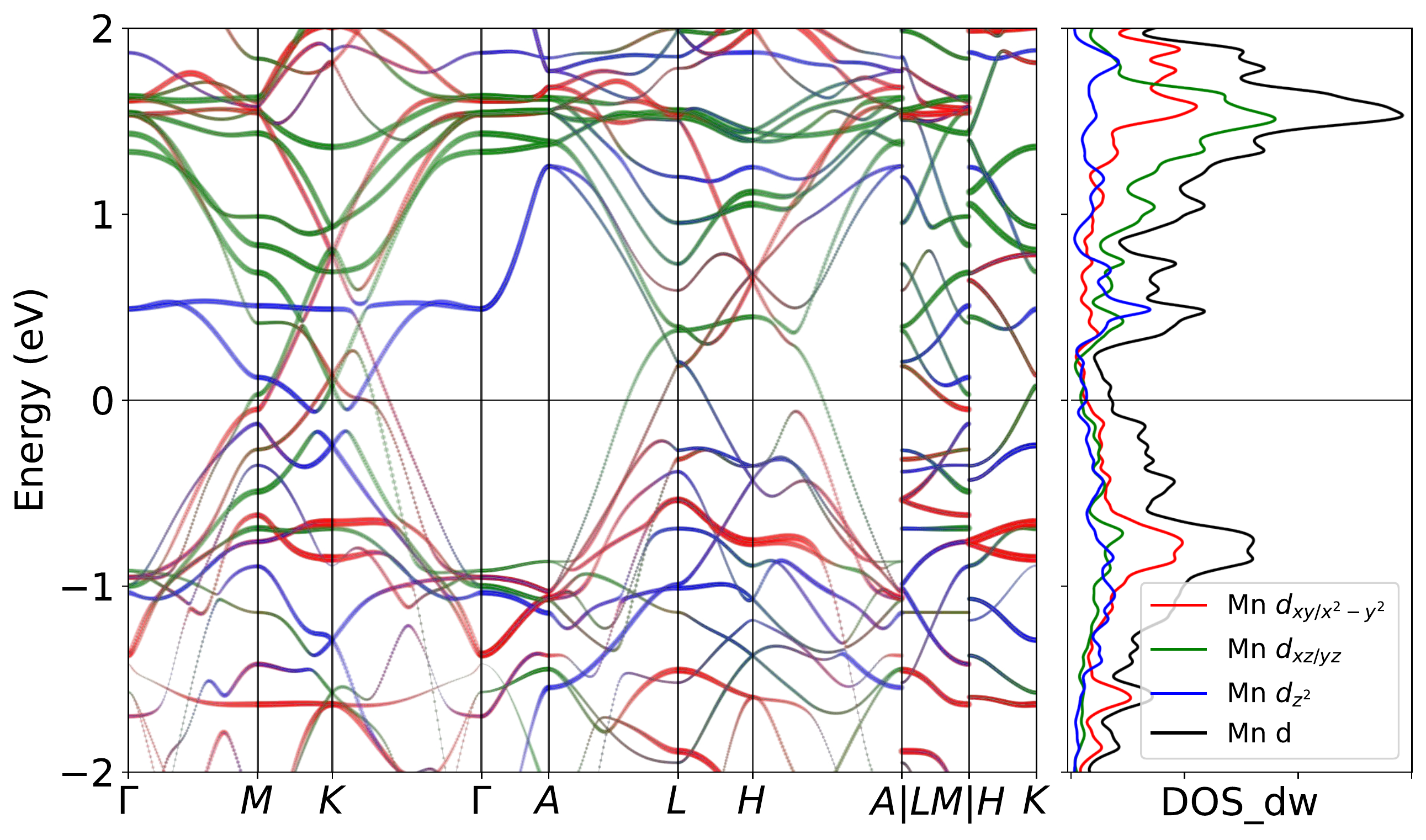}
\includegraphics[width=0.32\textwidth]{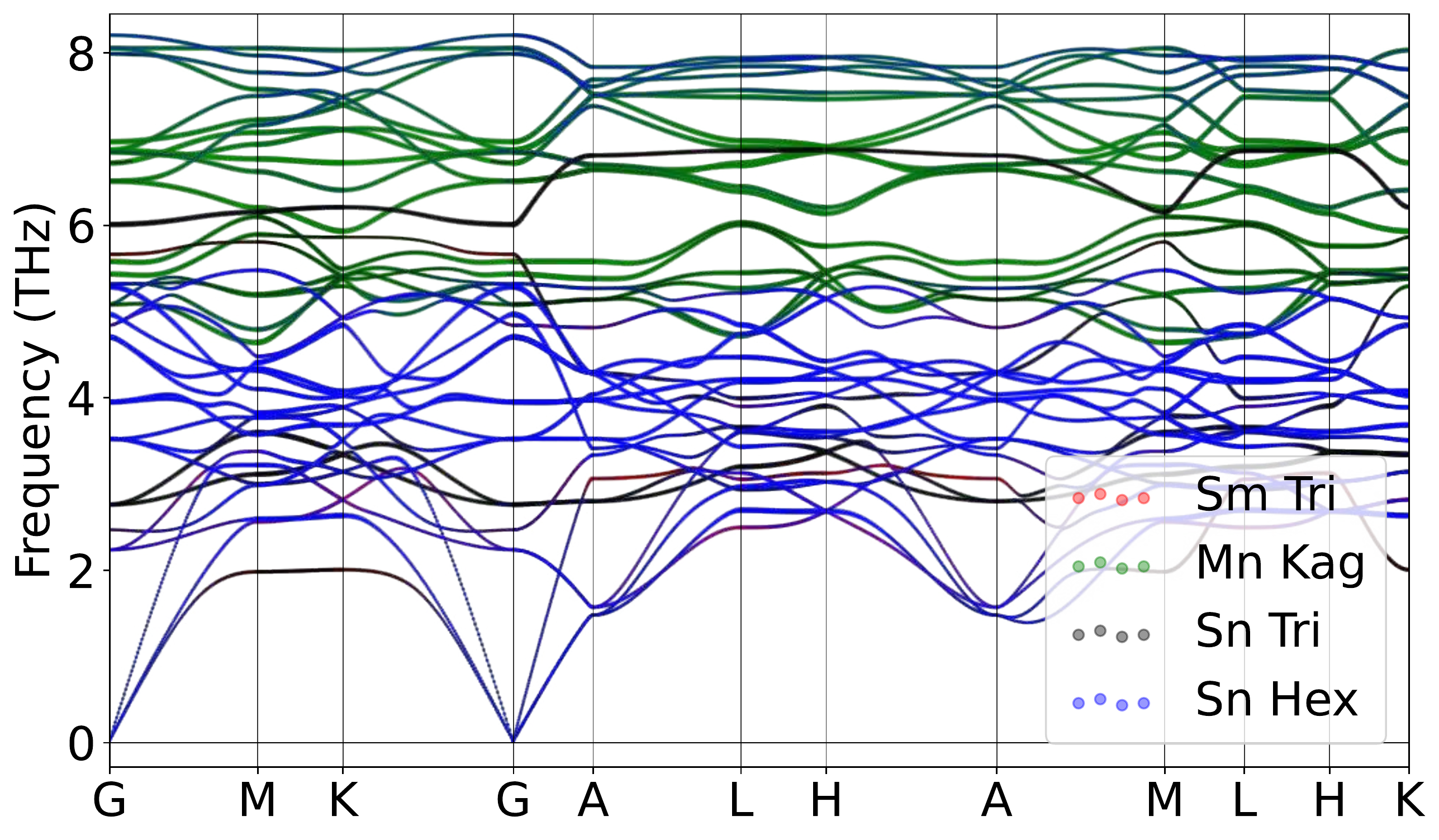}
\caption{Band structure for SmMn$_6$Sn$_6$ with FM order without SOC for spin (Left)up and (Middle)down channels. The projection of Mn $3d$ orbitals is also presented. (Right)Correspondingly, a stable phonon was demonstrated with the collinear magnetic order without Hubbard U at lower temperature regime, weighted by atomic projections.}
\label{fig:SmMn6Sn6mag}
\end{figure}

\begin{figure}[H]
\centering
\label{fig:SmMn6Sn6_relaxed_Low_xz}
\includegraphics[width=0.85\textwidth]{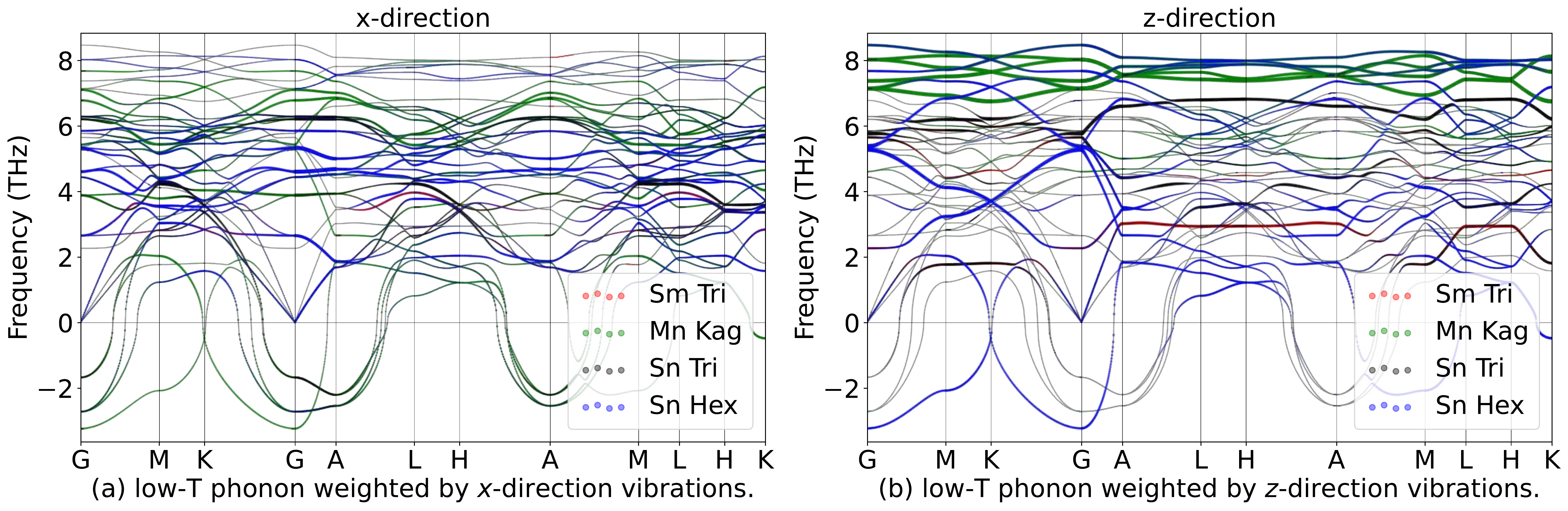}
\hspace{5pt}\label{fig:SmMn6Sn6_relaxed_High_xz}
\includegraphics[width=0.85\textwidth]{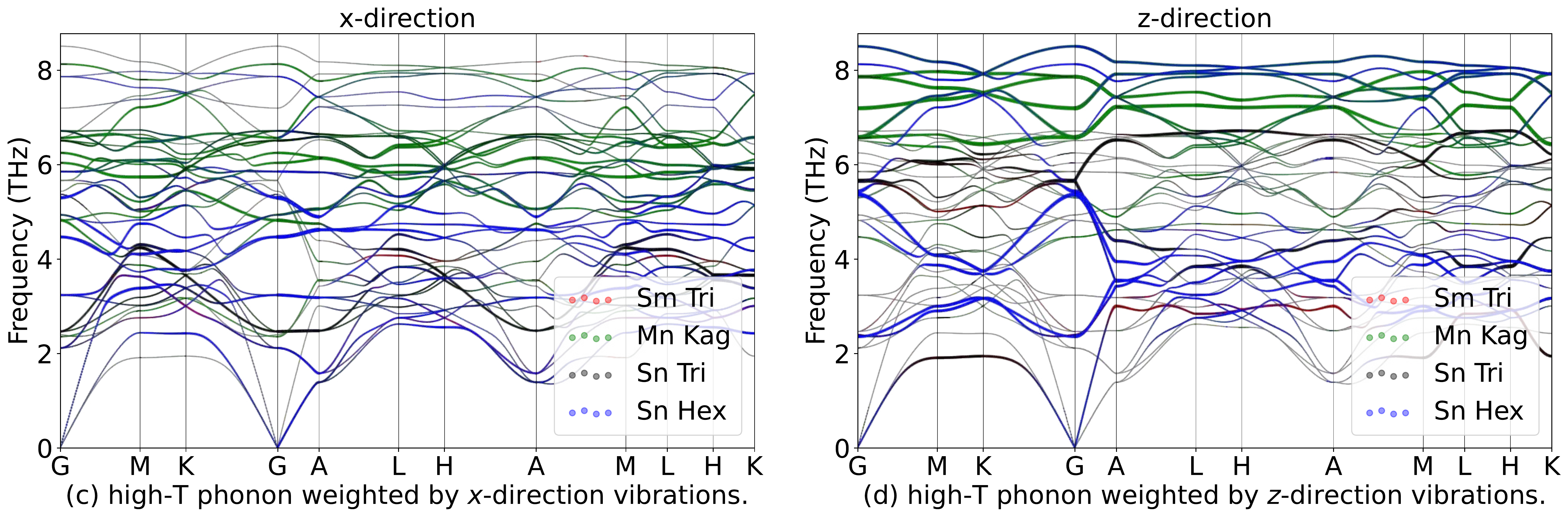}
\caption{Phonon spectrum for SmMn$_6$Sn$_6$in in-plane ($x$) and out-of-plane ($z$) directions in paramagnetic phase.}
\label{fig:SmMn6Sn6phoxyz}
\end{figure}

\begin{table}[htbp]
\global\long\def\arraystretch{1.12}
\captionsetup{width=.95\textwidth}
\caption{IRREPs at high-symmetry points for SmMn$_6$Sn$_6$ phonon spectrum at Low-T.}
\label{tab:191_SmMn6Sn6_Low}
\setlength{\tabcolsep}{1.mm}{
}
\end{table}

\begin{table}[htbp]
\global\long\def\arraystretch{1.12}
\captionsetup{width=.95\textwidth}
\caption{IRREPs at high-symmetry points for SmMn$_6$Sn$_6$ phonon spectrum at High-T.}
\label{tab:191_SmMn6Sn6_High}
\setlength{\tabcolsep}{1.mm}{
%
}
\end{table}
\subsubsection{\label{sms2sec:GdMn6Sn6}\hyperref[smsec:col]{\texorpdfstring{GdMn$_6$Sn$_6$}{}}~{\color{blue}  (High-T stable, Low-T unstable) }}

\begin{table}[htbp]
\global\long\def\arraystretch{1.12}
\captionsetup{width=.85\textwidth}
\caption{Structure parameters of GdMn$_6$Sn$_6$ after relaxation. It contains 514 electrons. }
\label{tab:GdMn6Sn6}
\setlength{\tabcolsep}{4mm}{
%
}
\end{table}

\begin{figure}[htbp]
\centering
\includegraphics[width=0.85\textwidth]{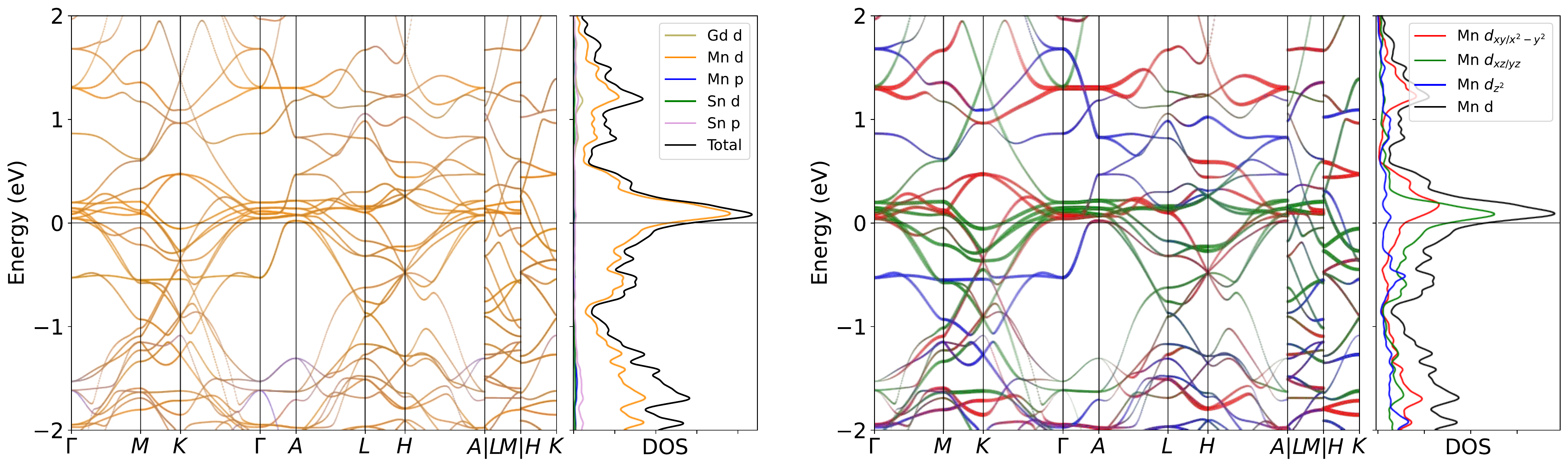}
\caption{Band structure for GdMn$_6$Sn$_6$ without SOC in paramagnetic phase. The projection of Mn $3d$ orbitals is also presented. The band structures clearly reveal the dominating of Mn $3d$ orbitals  near the Fermi level, as well as the pronounced peak.}
\label{fig:GdMn6Sn6band}
\end{figure}

\begin{figure}[H]
\centering
\subfloat[Relaxed phonon at low-T.]{
\label{fig:GdMn6Sn6_relaxed_Low}
\includegraphics[width=0.45\textwidth]{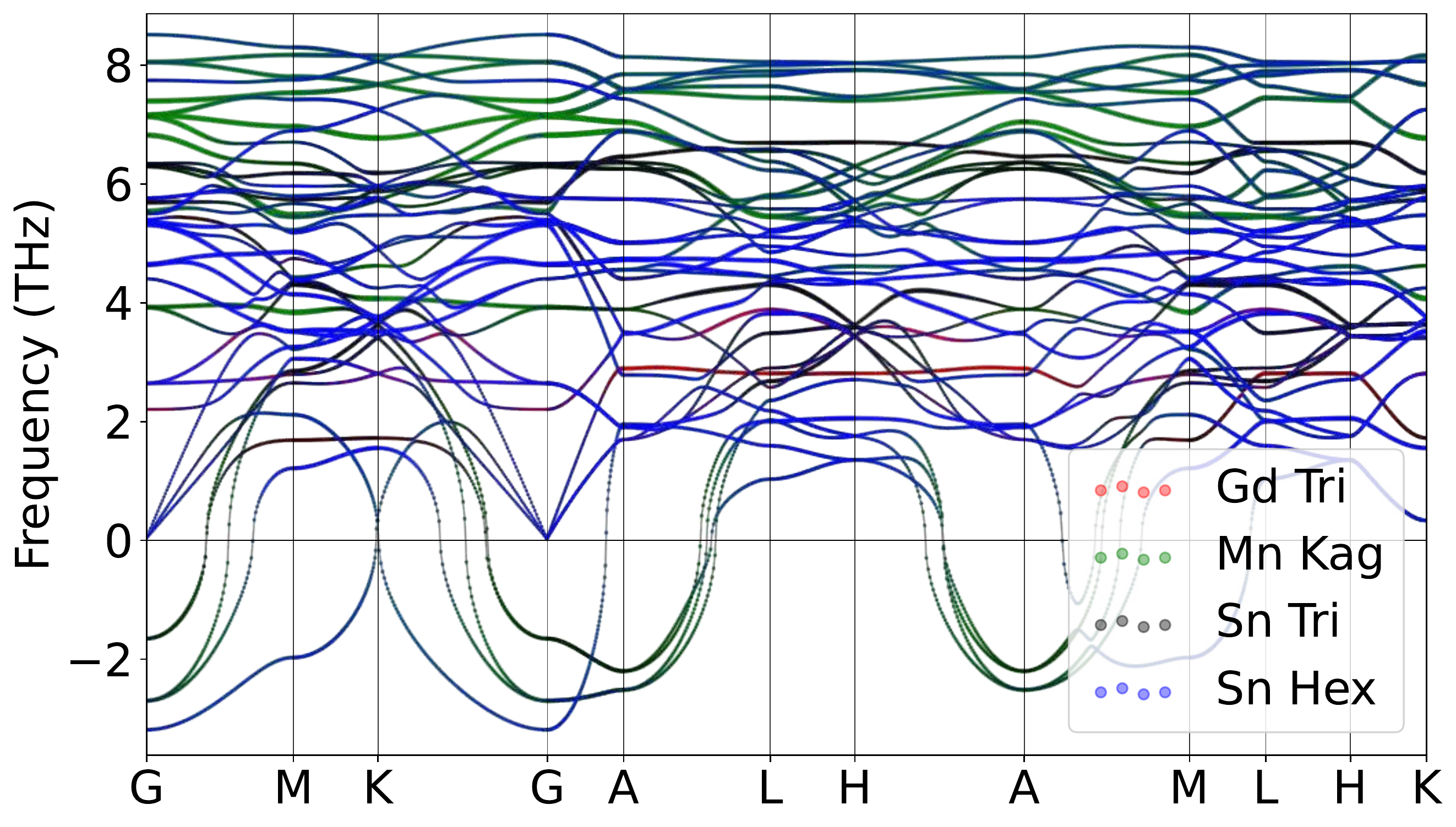}
}
\hspace{5pt}\subfloat[Relaxed phonon at high-T.]{
\label{fig:GdMn6Sn6_relaxed_High}
\includegraphics[width=0.45\textwidth]{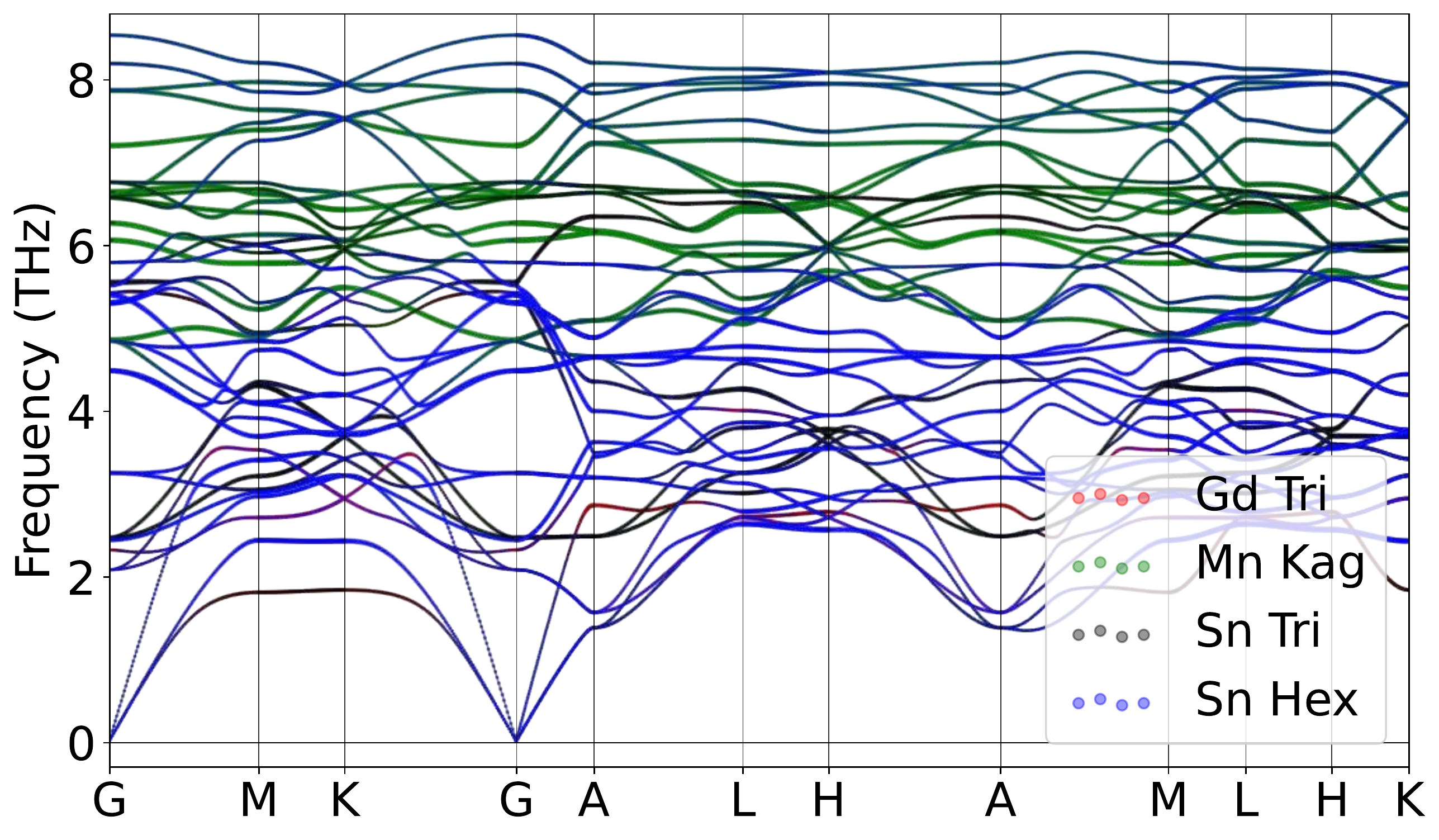}
}
\caption{Atomically projected phonon spectrum for GdMn$_6$Sn$_6$ in paramagnetic phase.}
\label{fig:GdMn6Sn6pho}
\end{figure}

\begin{figure}[htbp]
\centering
\includegraphics[width=0.32\textwidth]{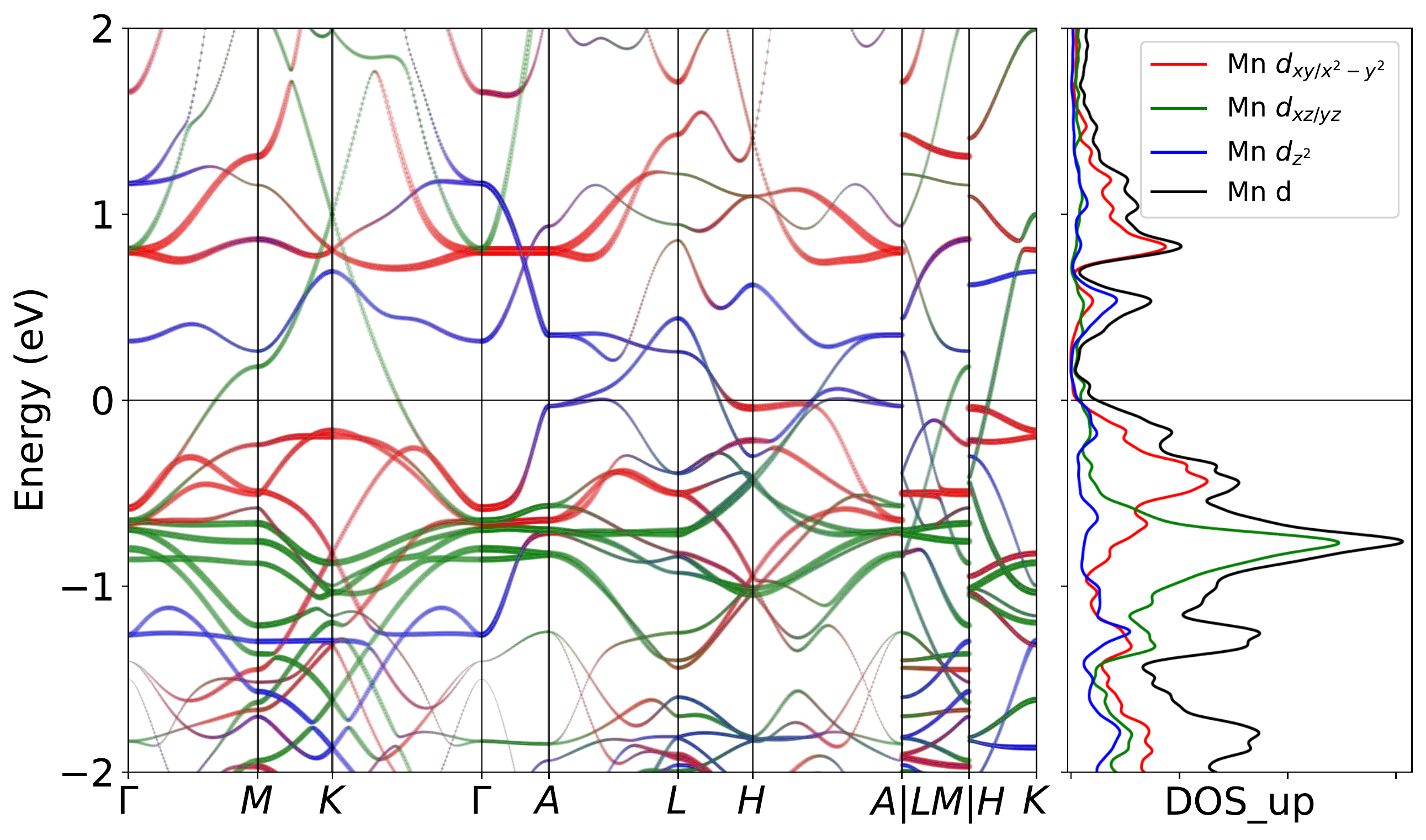}
\includegraphics[width=0.32\textwidth]{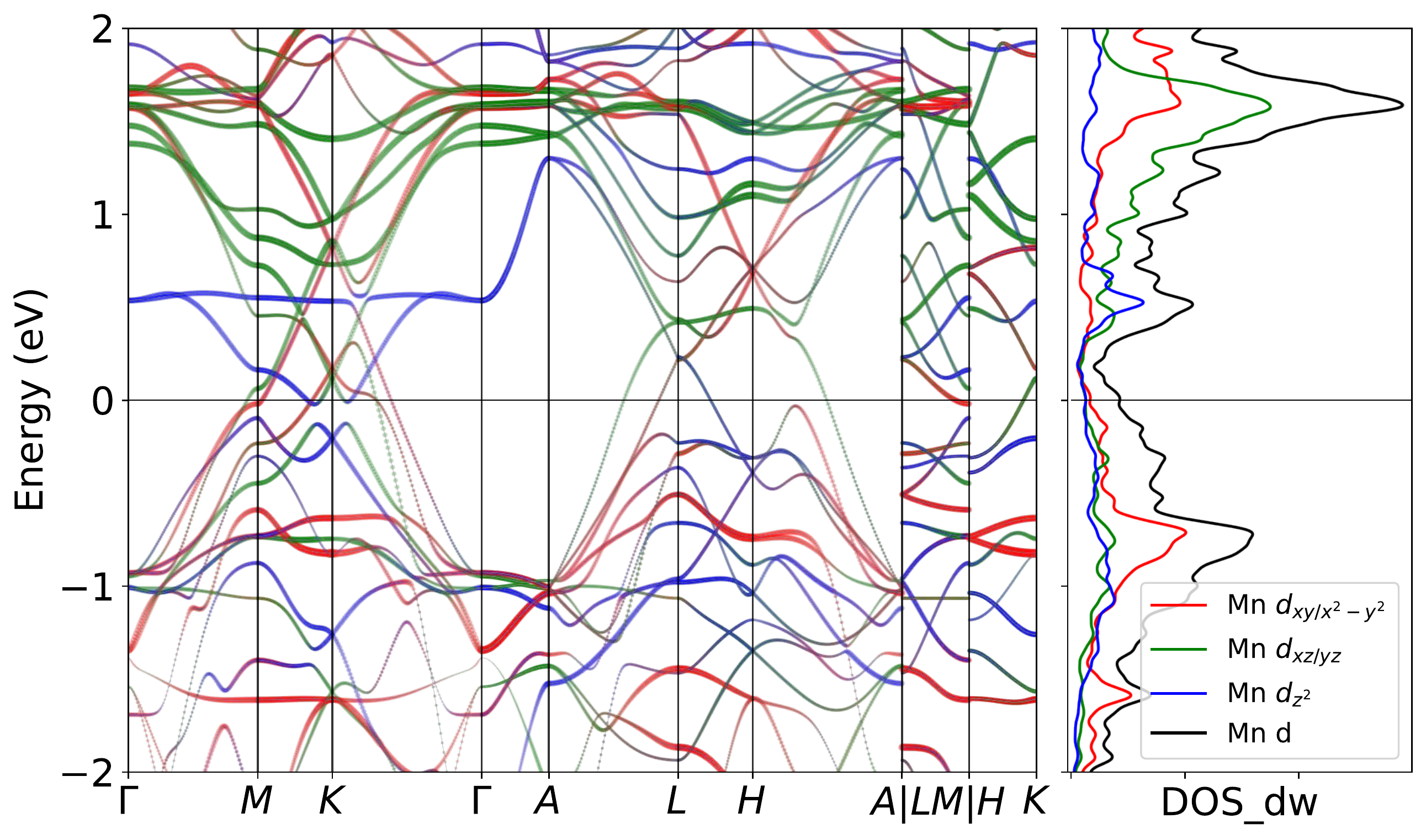}
\includegraphics[width=0.32\textwidth]{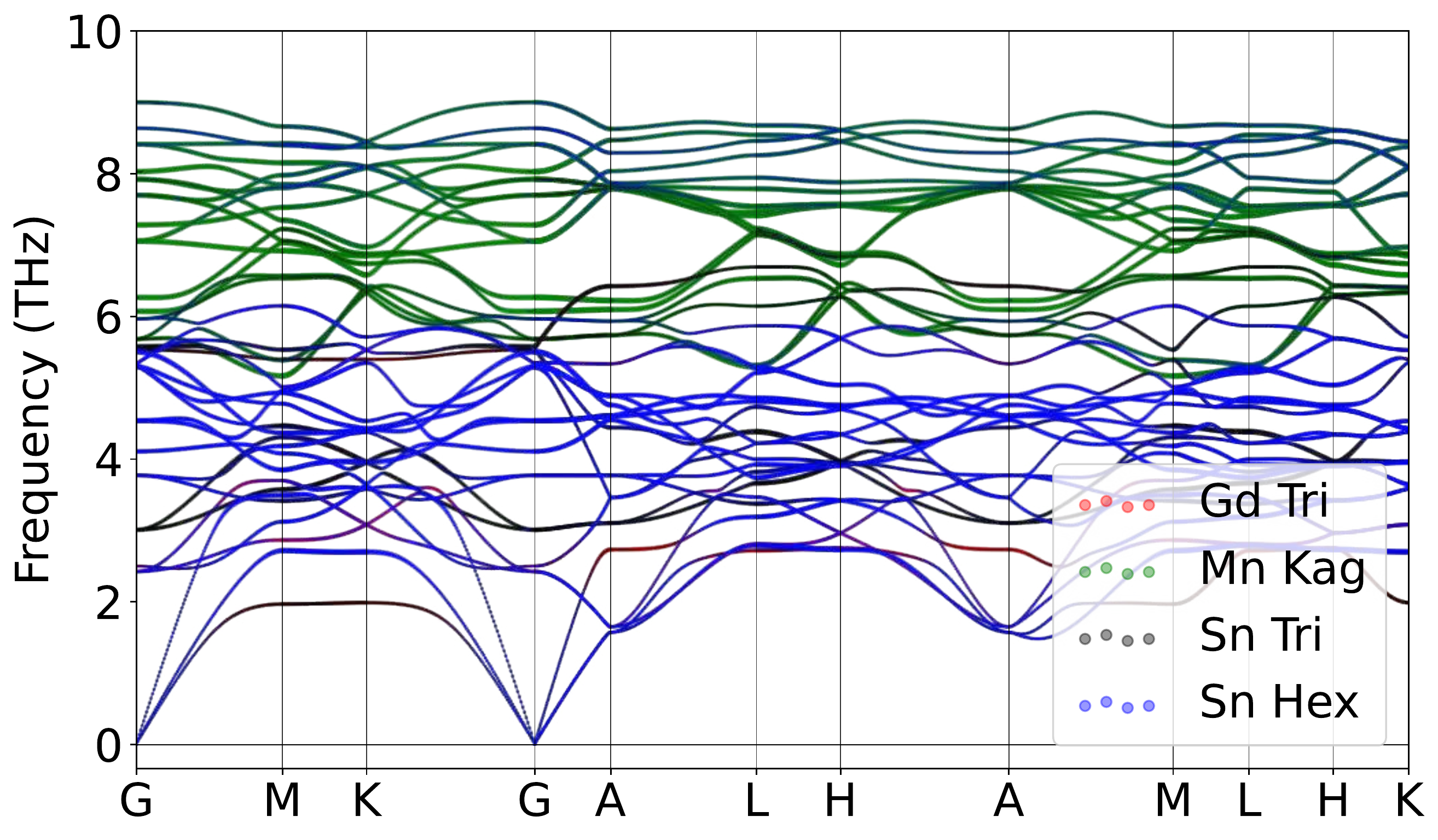}
\caption{Band structure for GdMn$_6$Sn$_6$ with FM order without SOC for spin (Left)up and (Middle)down channels. The projection of Mn $3d$ orbitals is also presented. (Right)Correspondingly, a stable phonon was demonstrated with the collinear magnetic order without Hubbard U at lower temperature regime, weighted by atomic projections.}
\label{fig:GdMn6Sn6mag}
\end{figure}

\begin{figure}[H]
\centering
\label{fig:GdMn6Sn6_relaxed_Low_xz}
\includegraphics[width=0.85\textwidth]{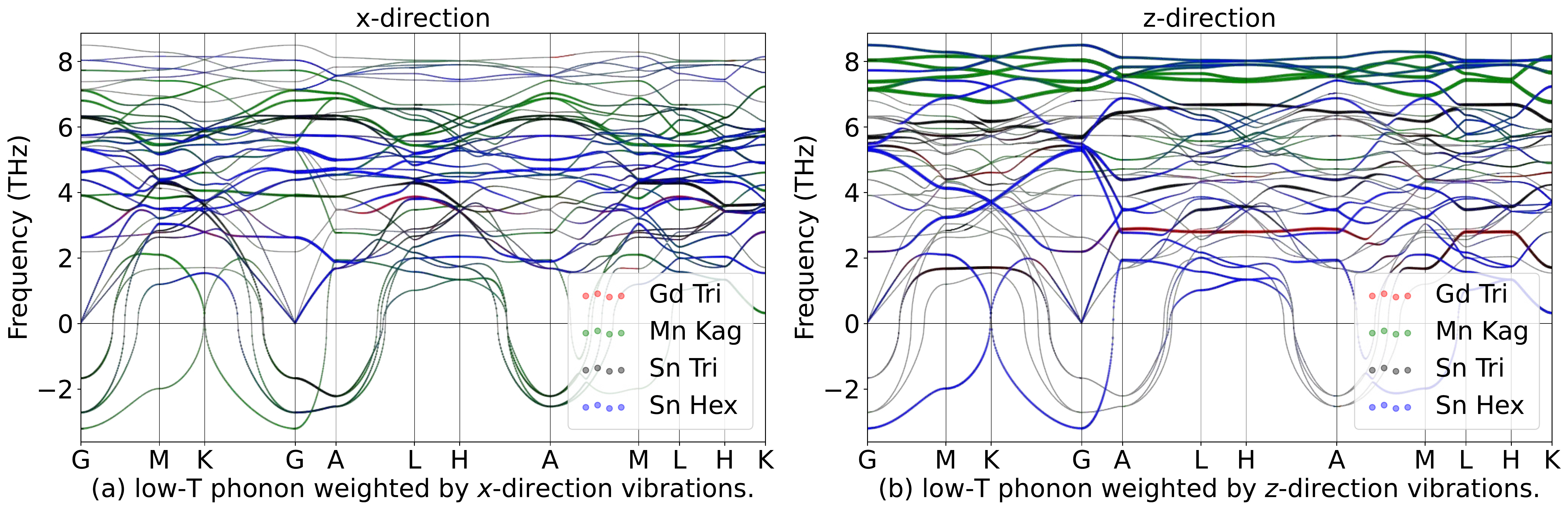}
\hspace{5pt}\label{fig:GdMn6Sn6_relaxed_High_xz}
\includegraphics[width=0.85\textwidth]{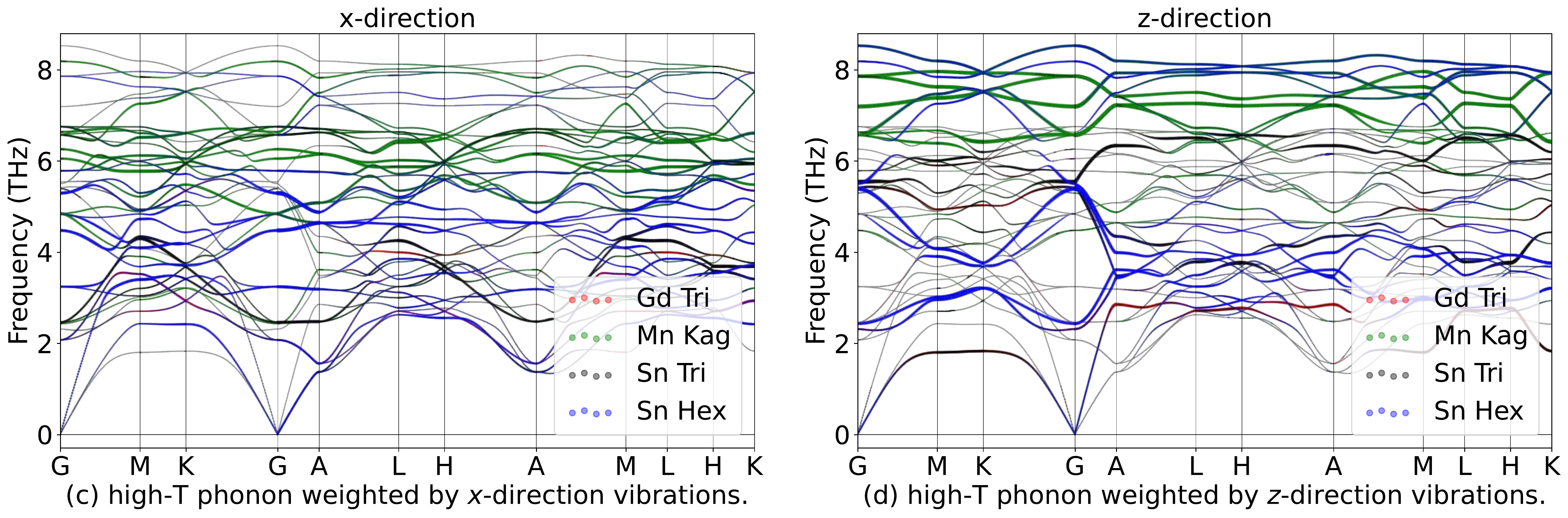}
\caption{Phonon spectrum for GdMn$_6$Sn$_6$in in-plane ($x$) and out-of-plane ($z$) directions in paramagnetic phase.}
\label{fig:GdMn6Sn6phoxyz}
\end{figure}

\begin{table}[htbp]
\global\long\def\arraystretch{1.12}
\captionsetup{width=.95\textwidth}
\caption{IRREPs at high-symmetry points for GdMn$_6$Sn$_6$ phonon spectrum at Low-T.}
\label{tab:191_GdMn6Sn6_Low}
\setlength{\tabcolsep}{1.mm}{
}
\end{table}

\begin{table}[htbp]
\global\long\def\arraystretch{1.12}
\captionsetup{width=.95\textwidth}
\caption{IRREPs at high-symmetry points for GdMn$_6$Sn$_6$ phonon spectrum at High-T.}
\label{tab:191_GdMn6Sn6_High}
\setlength{\tabcolsep}{1.mm}{
%
}
\end{table}
\subsubsection{\label{sms2sec:TbMn6Sn6}\hyperref[smsec:col]{\texorpdfstring{TbMn$_6$Sn$_6$}{}}~{\color{blue}  (High-T stable, Low-T unstable) }}

\begin{table}[htbp]
\global\long\def\arraystretch{1.12}
\captionsetup{width=.85\textwidth}
\caption{Structure parameters of TbMn$_6$Sn$_6$ after relaxation. It contains 515 electrons. }
\label{tab:TbMn6Sn6}
\setlength{\tabcolsep}{4mm}{
%
}
\end{table}

\begin{figure}[htbp]
\centering
\includegraphics[width=0.85\textwidth]{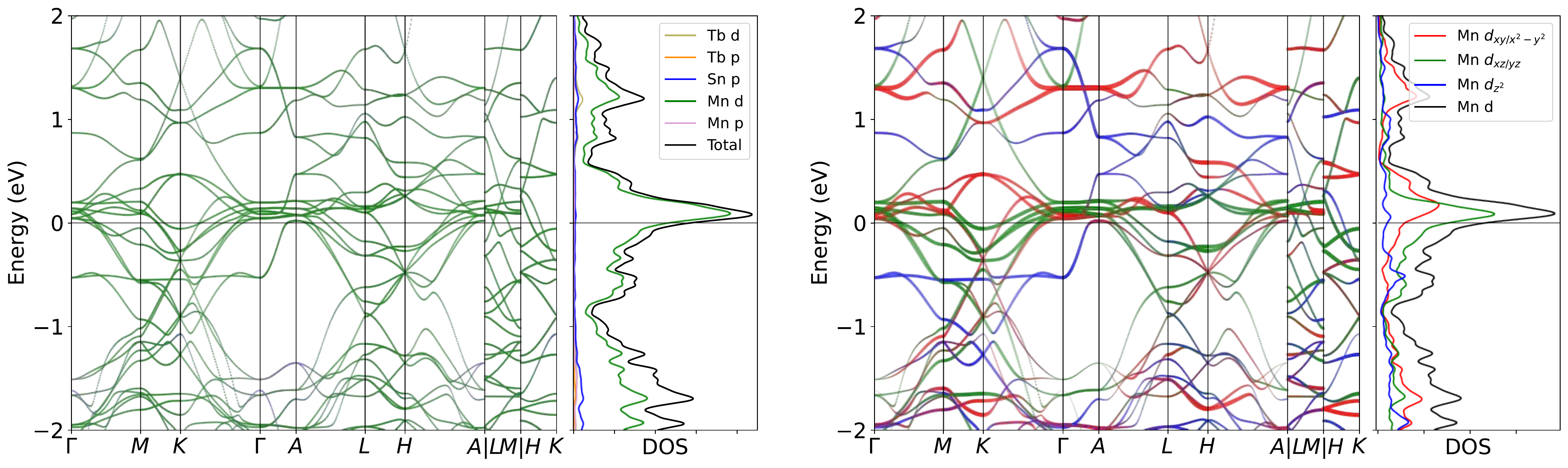}
\caption{Band structure for TbMn$_6$Sn$_6$ without SOC in paramagnetic phase. The projection of Mn $3d$ orbitals is also presented. The band structures clearly reveal the dominating of Mn $3d$ orbitals  near the Fermi level, as well as the pronounced peak.}
\label{fig:TbMn6Sn6band}
\end{figure}

\begin{figure}[H]
\centering
\subfloat[Relaxed phonon at low-T.]{
\label{fig:TbMn6Sn6_relaxed_Low}
\includegraphics[width=0.45\textwidth]{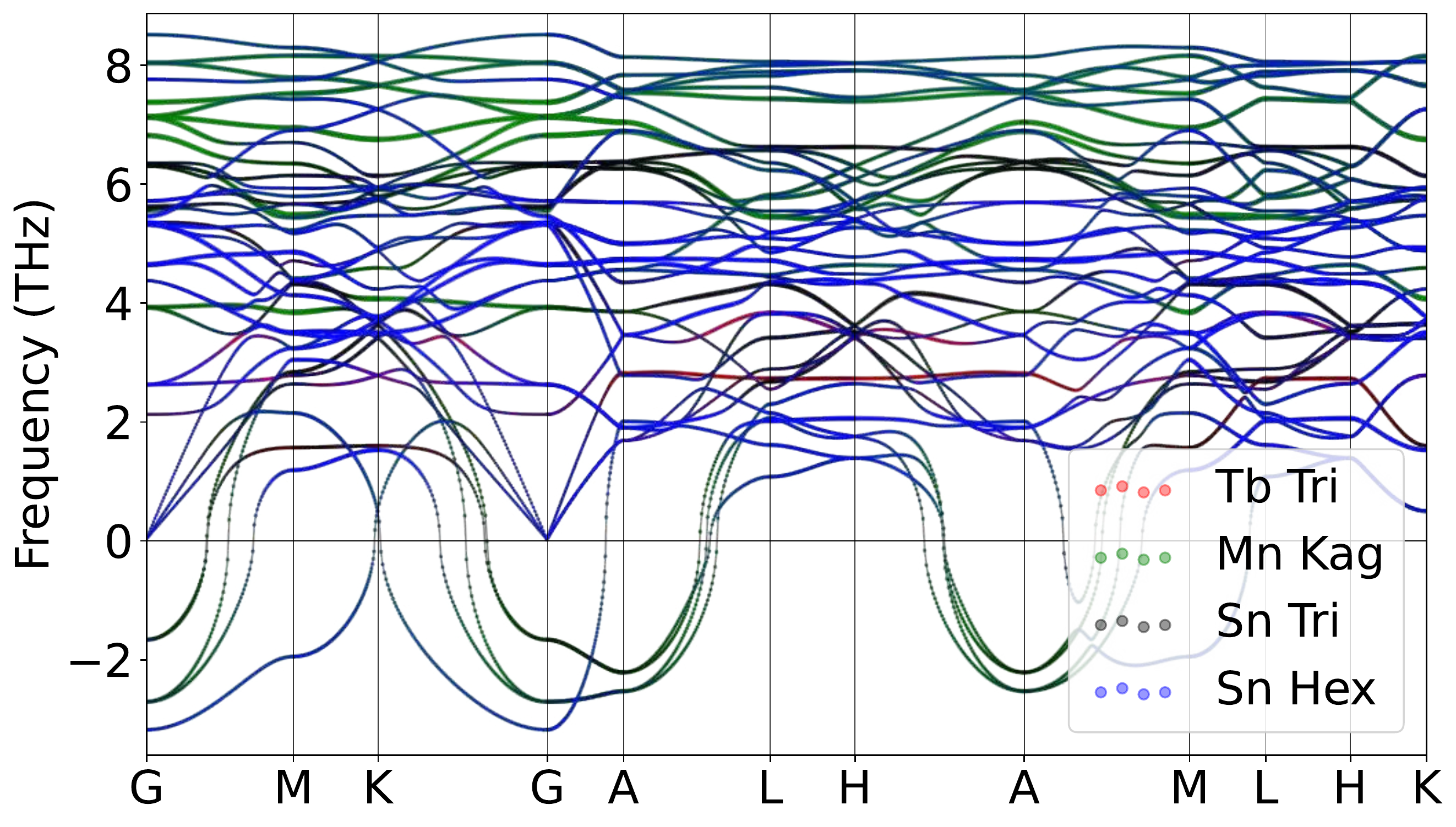}
}
\hspace{5pt}\subfloat[Relaxed phonon at high-T.]{
\label{fig:TbMn6Sn6_relaxed_High}
\includegraphics[width=0.45\textwidth]{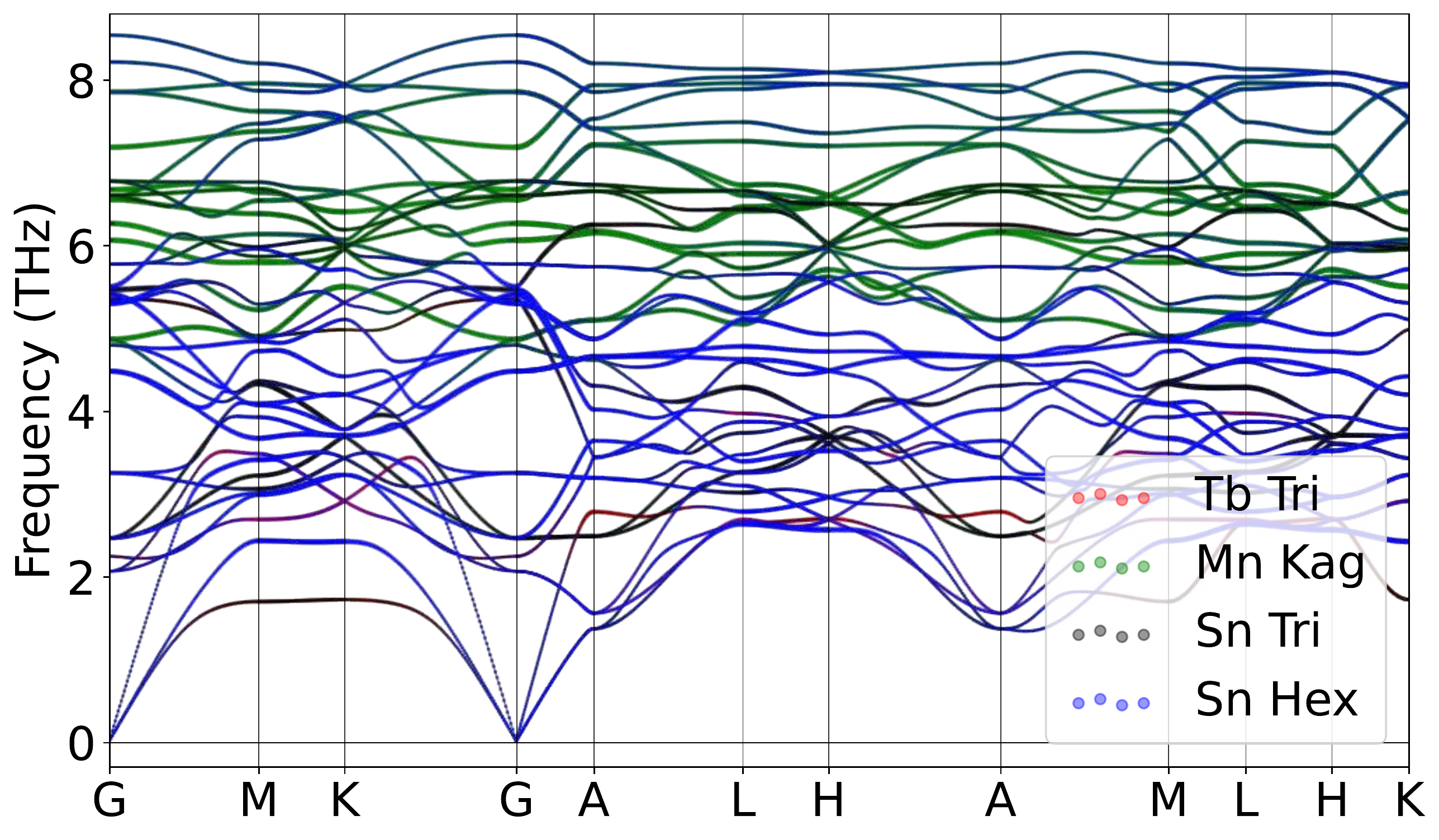}
}
\caption{Atomically projected phonon spectrum for TbMn$_6$Sn$_6$ in paramagnetic phase.}
\label{fig:TbMn6Sn6pho}
\end{figure}

\begin{figure}[htbp]
\centering
\includegraphics[width=0.32\textwidth]{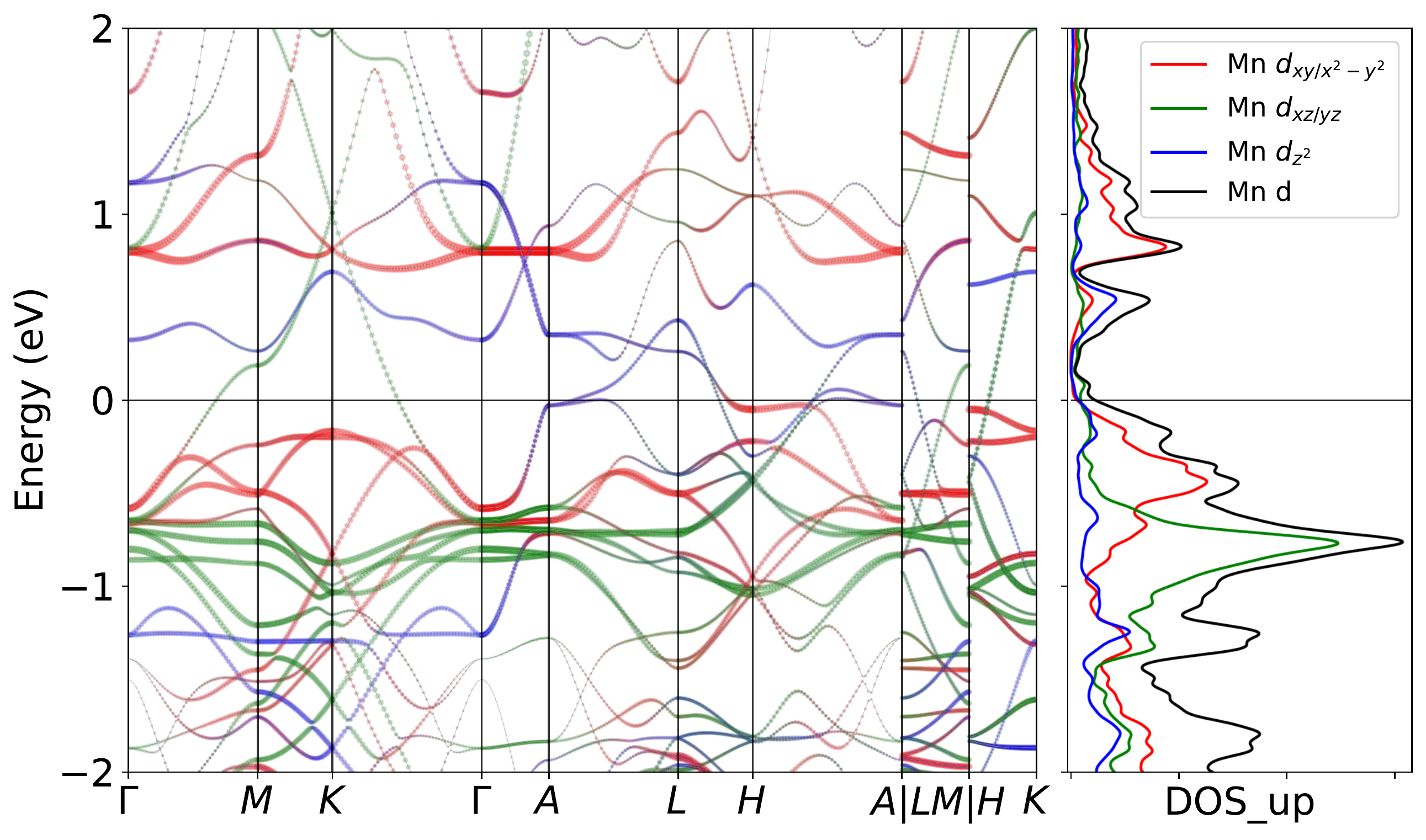}
\includegraphics[width=0.32\textwidth]{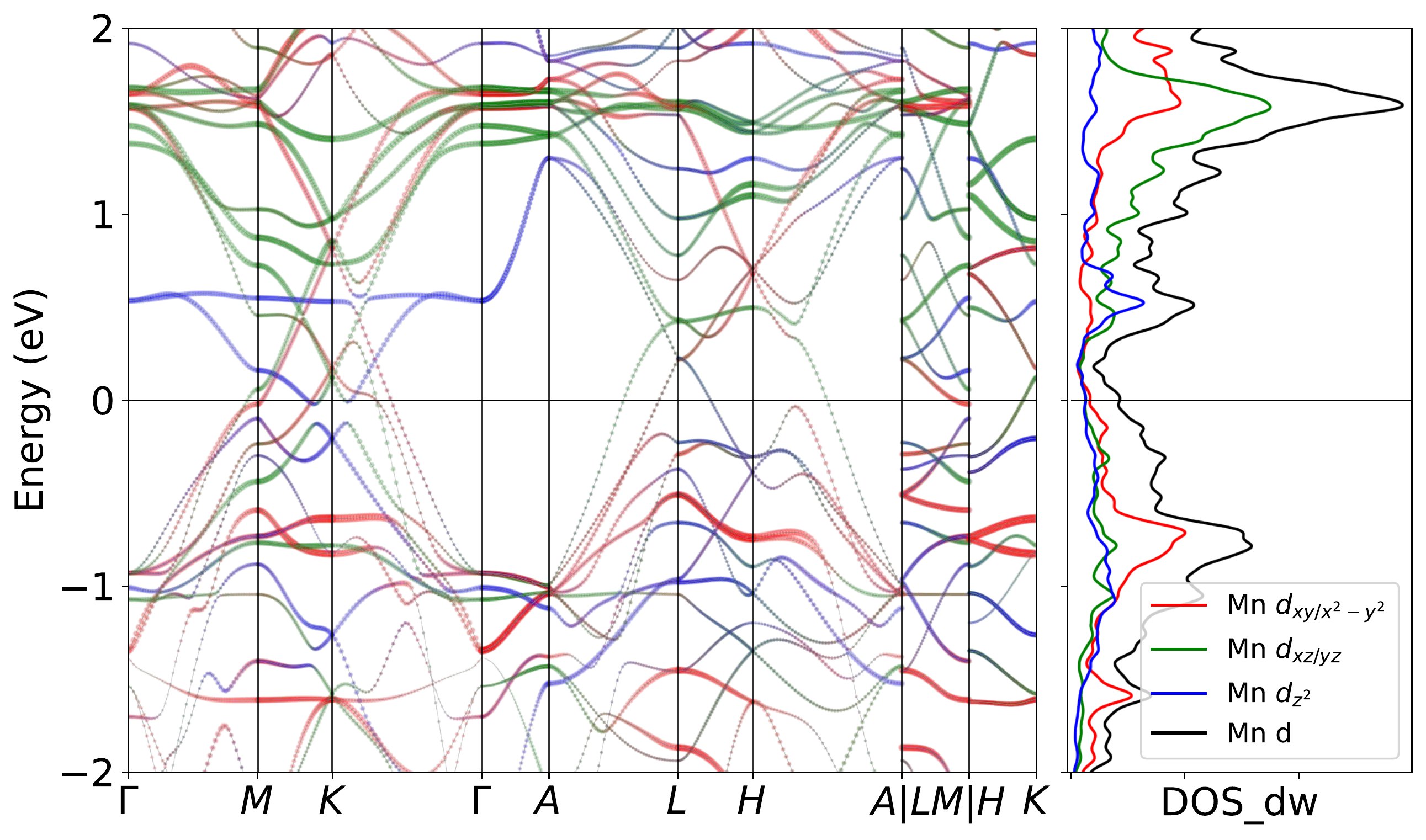}
\includegraphics[width=0.32\textwidth]{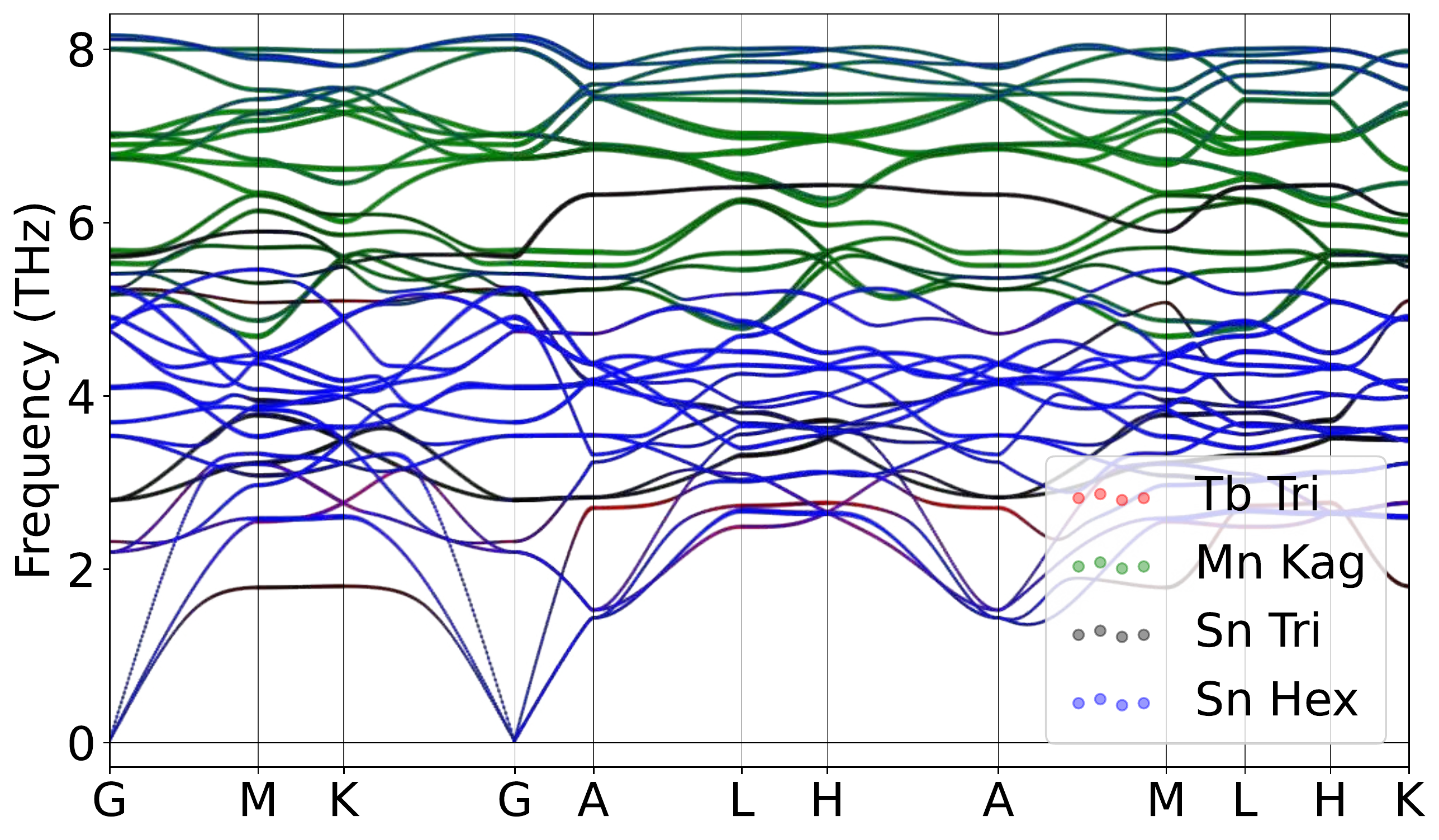}
\caption{Band structure for TbMn$_6$Sn$_6$ with FM order without SOC for spin (Left)up and (Middle)down channels. The projection of Mn $3d$ orbitals is also presented. (Right)Correspondingly, a stable phonon was demonstrated with the collinear magnetic order without Hubbard U at lower temperature regime, weighted by atomic projections.}
\label{fig:TbMn6Sn6mag}
\end{figure}

\begin{figure}[H]
\centering
\label{fig:TbMn6Sn6_relaxed_Low_xz}
\includegraphics[width=0.85\textwidth]{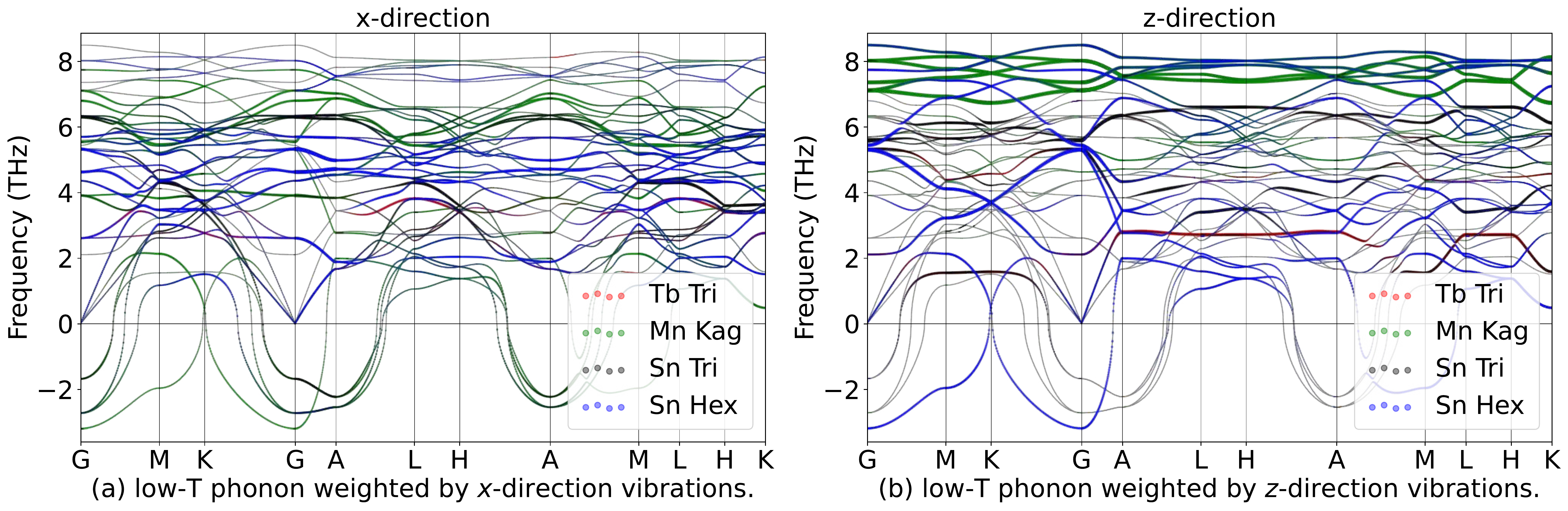}
\hspace{5pt}\label{fig:TbMn6Sn6_relaxed_High_xz}
\includegraphics[width=0.85\textwidth]{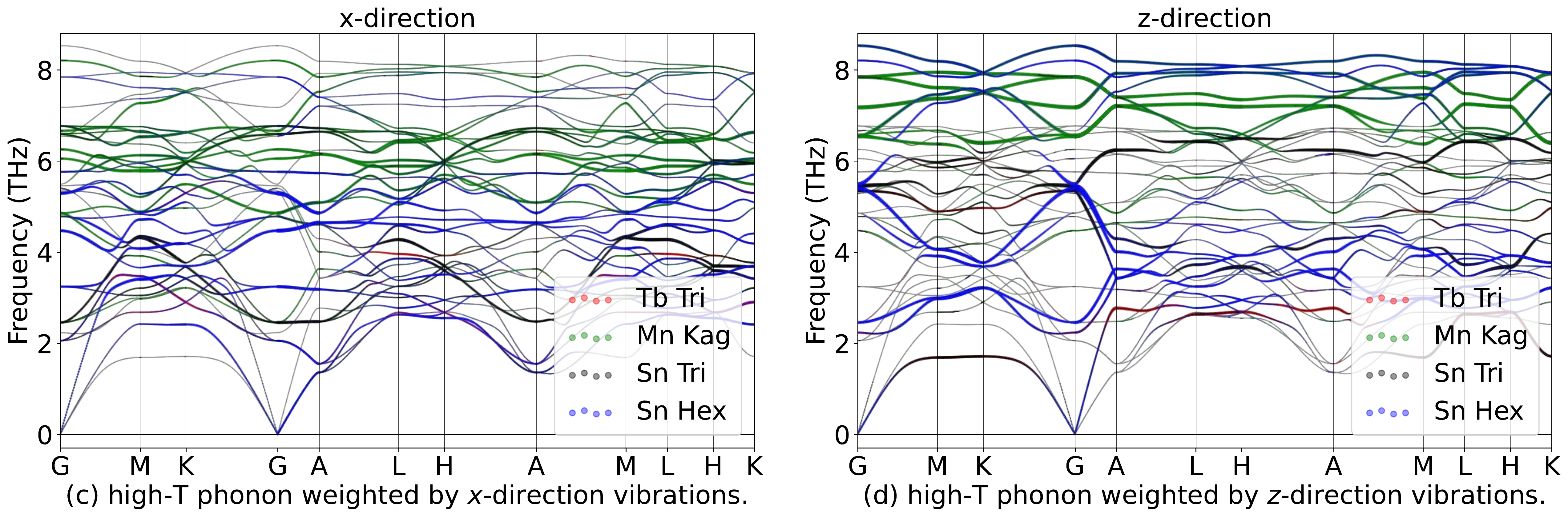}
\caption{Phonon spectrum for TbMn$_6$Sn$_6$in in-plane ($x$) and out-of-plane ($z$) directions in paramagnetic phase.}
\label{fig:TbMn6Sn6phoxyz}
\end{figure}

\begin{table}[htbp]
\global\long\def\arraystretch{1.12}
\captionsetup{width=.95\textwidth}
\caption{IRREPs at high-symmetry points for TbMn$_6$Sn$_6$ phonon spectrum at Low-T.}
\label{tab:191_TbMn6Sn6_Low}
\setlength{\tabcolsep}{1.mm}{
}
\end{table}

\begin{table}[htbp]
\global\long\def\arraystretch{1.12}
\captionsetup{width=.95\textwidth}
\caption{IRREPs at high-symmetry points for TbMn$_6$Sn$_6$ phonon spectrum at High-T.}
\label{tab:191_TbMn6Sn6_High}
\setlength{\tabcolsep}{1.mm}{
%
}
\end{table}
\subsubsection{\label{sms2sec:DyMn6Sn6}\hyperref[smsec:col]{\texorpdfstring{DyMn$_6$Sn$_6$}{}}~{\color{blue}  (High-T stable, Low-T unstable) }}

\begin{table}[htbp]
\global\long\def\arraystretch{1.12}
\captionsetup{width=.85\textwidth}
\caption{Structure parameters of DyMn$_6$Sn$_6$ after relaxation. It contains 516 electrons. }
\label{tab:DyMn6Sn6}
\setlength{\tabcolsep}{4mm}{
%
}
\end{table}

\begin{figure}[htbp]
\centering
\includegraphics[width=0.85\textwidth]{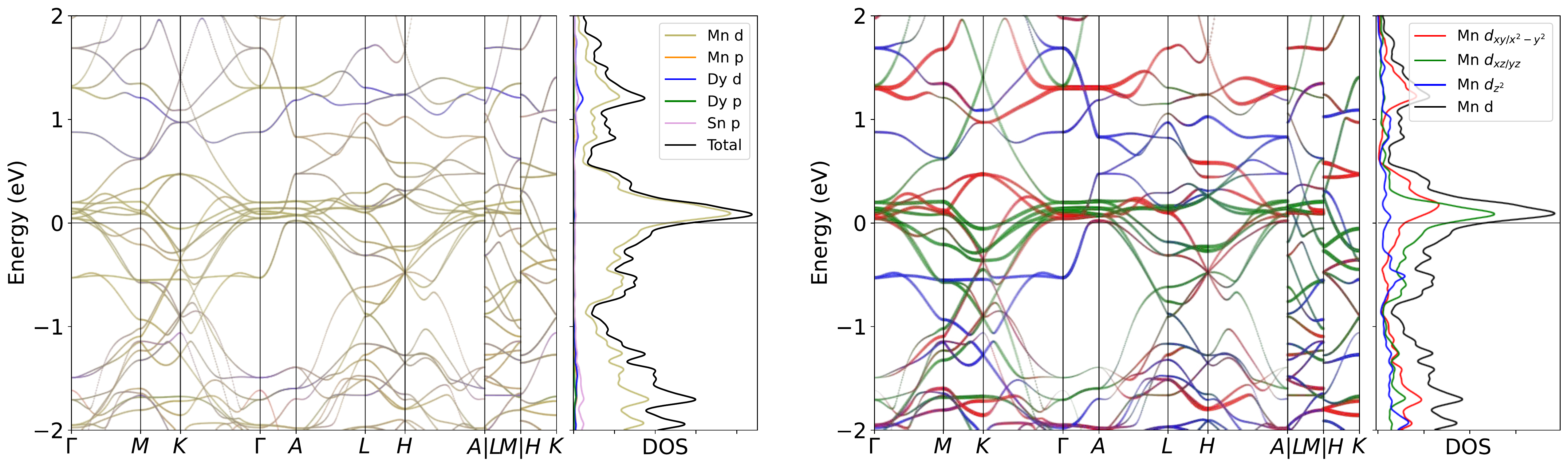}
\caption{Band structure for DyMn$_6$Sn$_6$ without SOC in paramagnetic phase. The projection of Mn $3d$ orbitals is also presented. The band structures clearly reveal the dominating of Mn $3d$ orbitals  near the Fermi level, as well as the pronounced peak.}
\label{fig:DyMn6Sn6band}
\end{figure}

\begin{figure}[H]
\centering
\subfloat[Relaxed phonon at low-T.]{
\label{fig:DyMn6Sn6_relaxed_Low}
\includegraphics[width=0.45\textwidth]{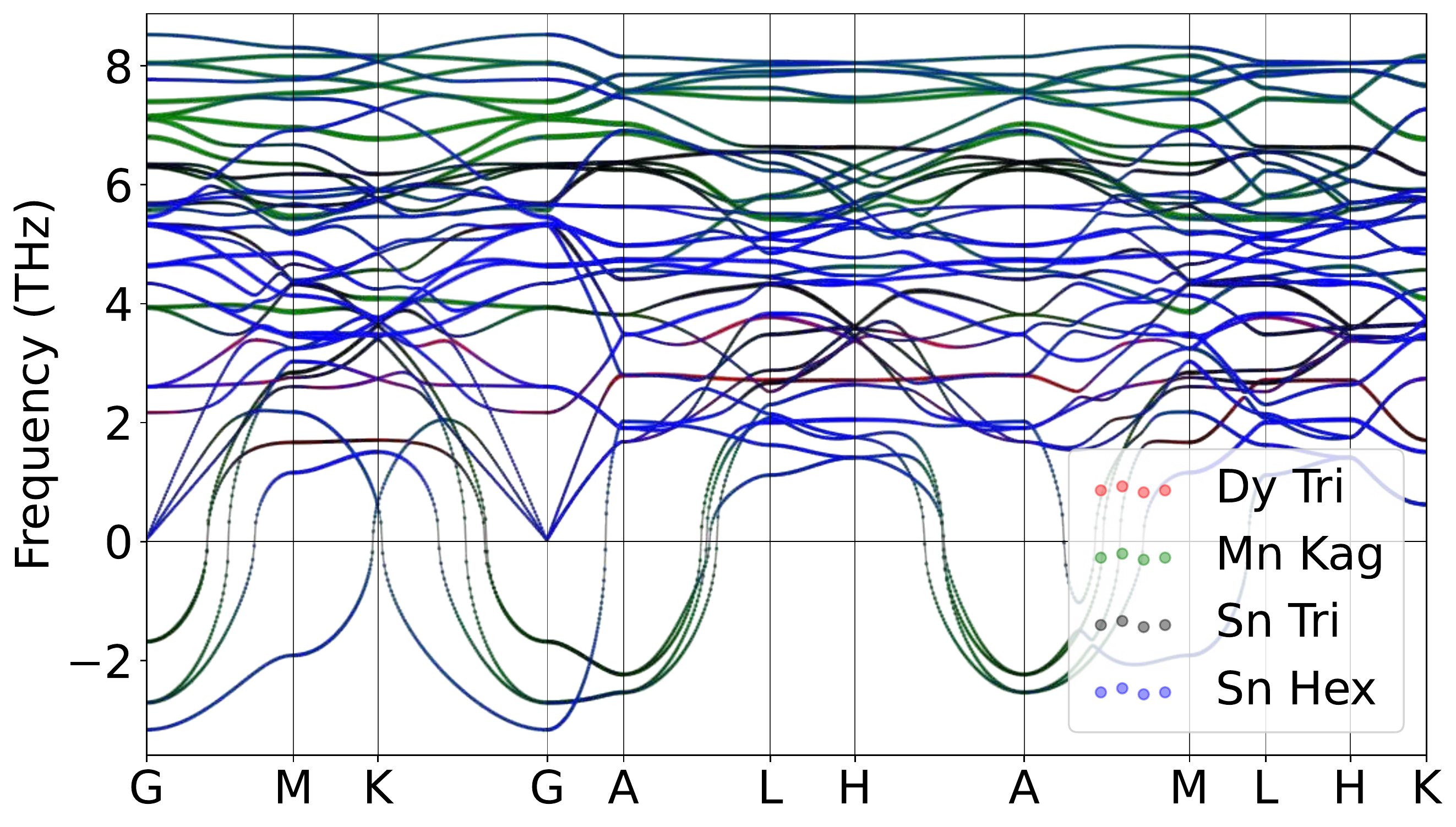}
}
\hspace{5pt}\subfloat[Relaxed phonon at high-T.]{
\label{fig:DyMn6Sn6_relaxed_High}
\includegraphics[width=0.45\textwidth]{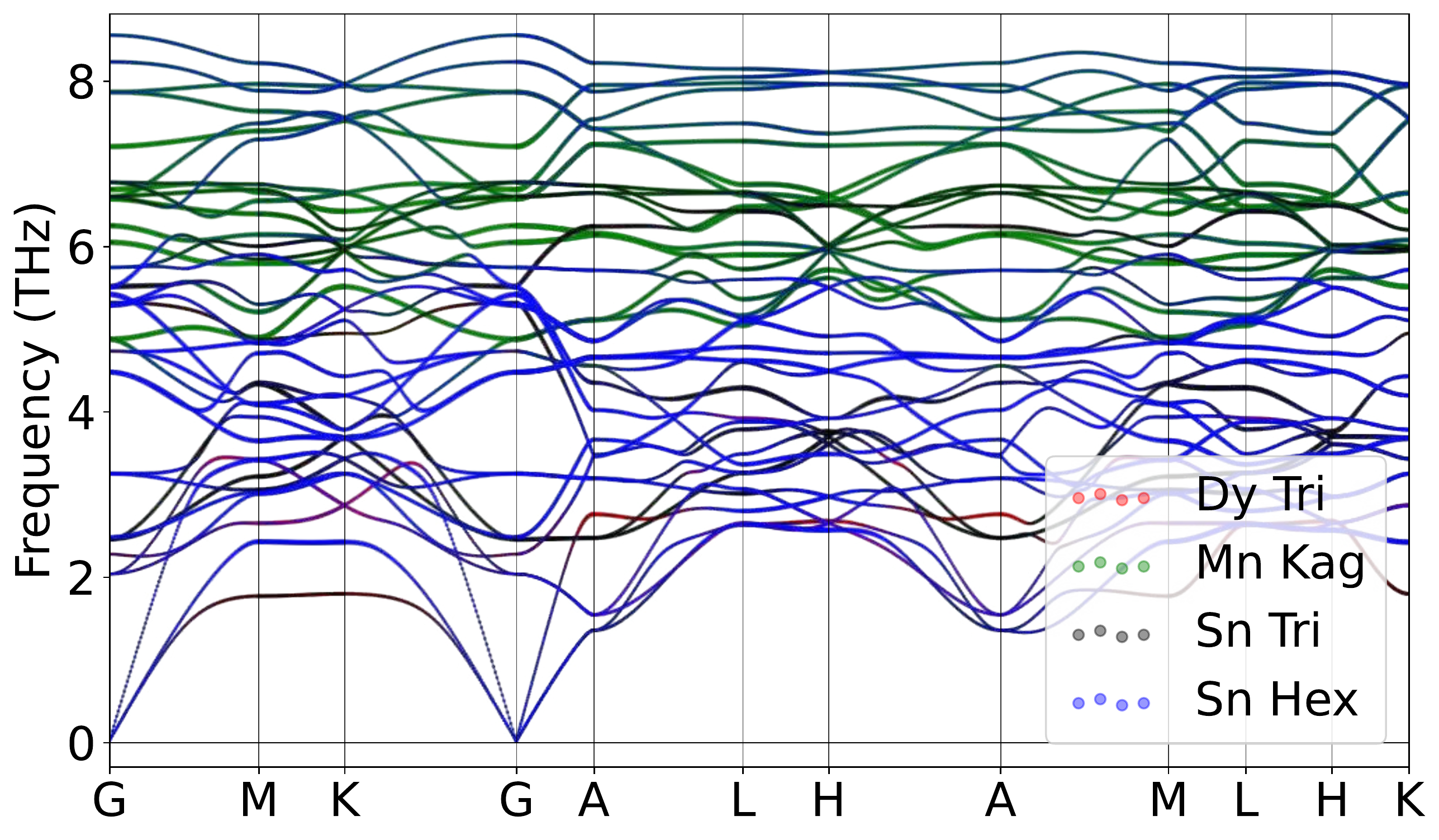}
}
\caption{Atomically projected phonon spectrum for DyMn$_6$Sn$_6$ in paramagnetic phase.}
\label{fig:DyMn6Sn6pho}
\end{figure}

\begin{figure}[htbp]
\centering
\includegraphics[width=0.32\textwidth]{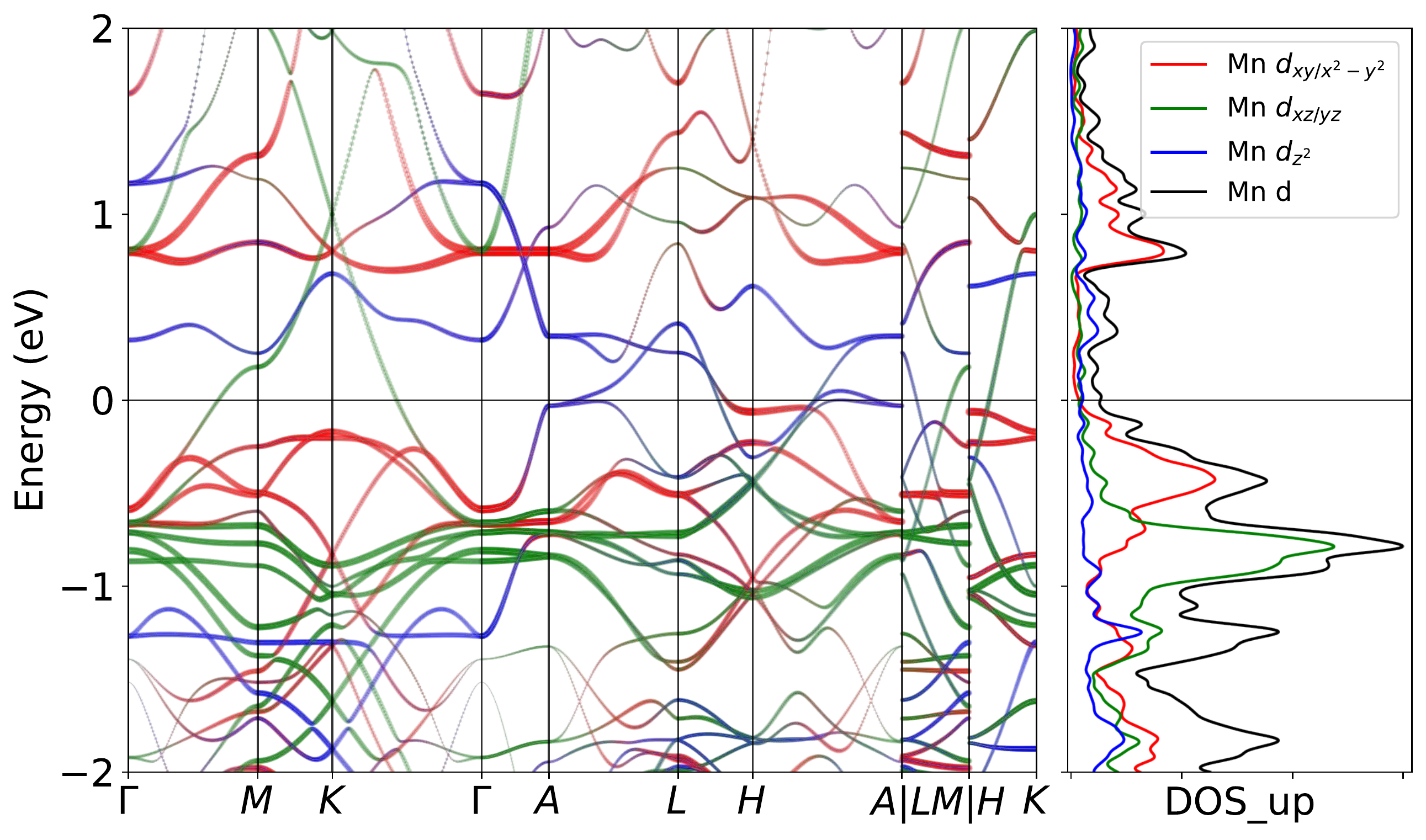}
\includegraphics[width=0.32\textwidth]{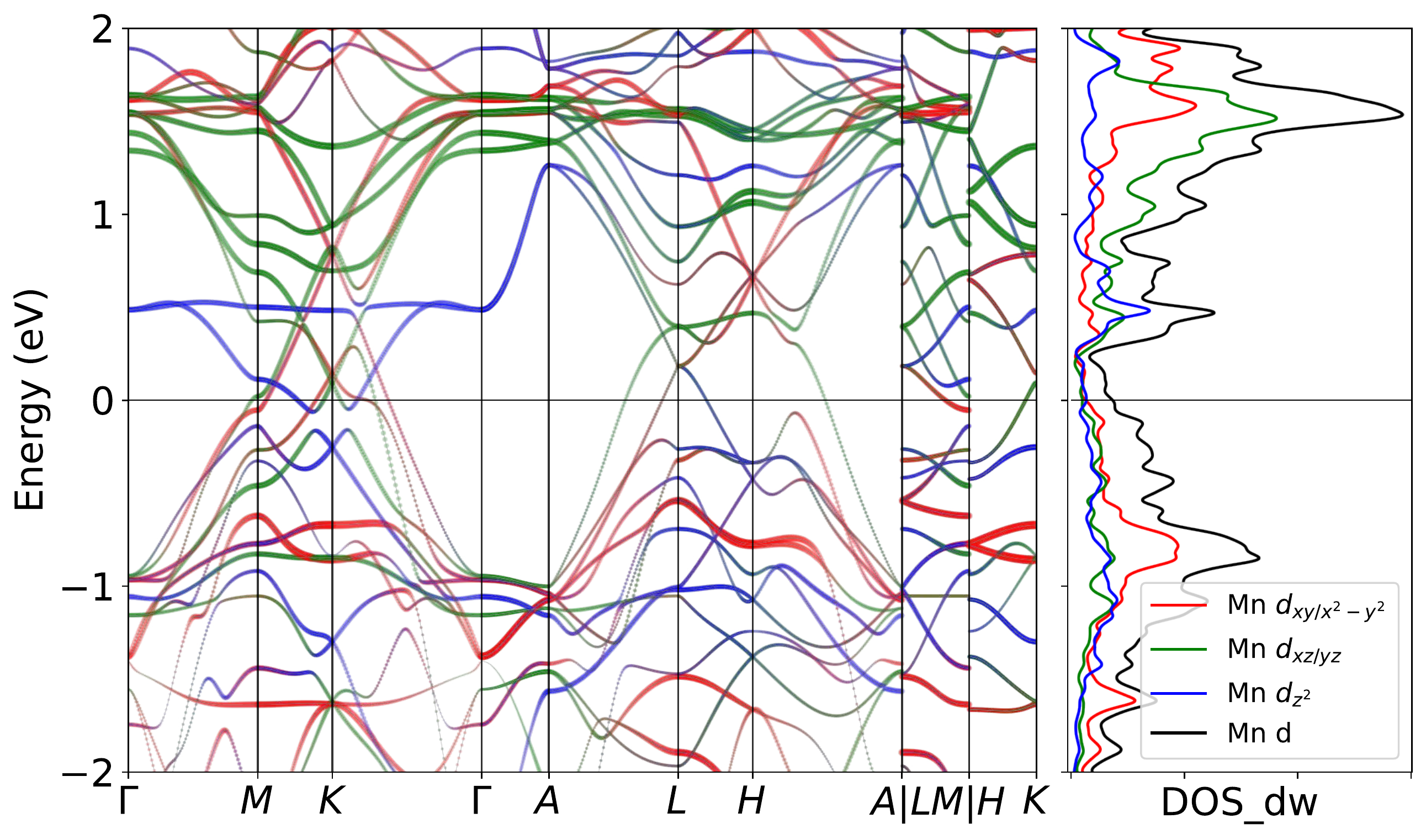}
\includegraphics[width=0.32\textwidth]{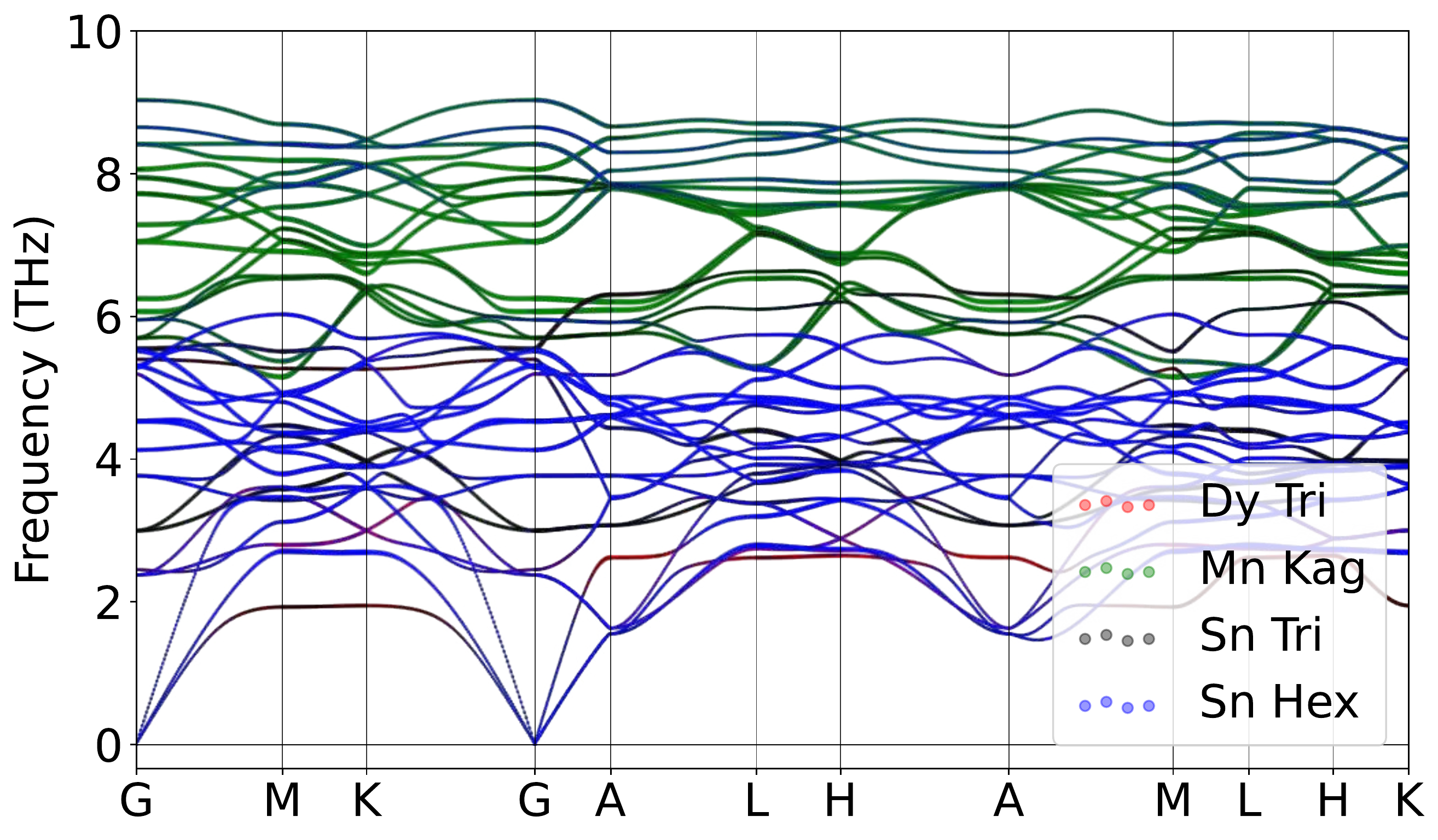}
\caption{Band structure for DyMn$_6$Sn$_6$ with FM order without SOC for spin (Left)up and (Middle)down channels. The projection of Mn $3d$ orbitals is also presented. (Right)Correspondingly, a stable phonon was demonstrated with the collinear magnetic order without Hubbard U at lower temperature regime, weighted by atomic projections.}
\label{fig:DyMn6Sn6mag}
\end{figure}

\begin{figure}[H]
\centering
\label{fig:DyMn6Sn6_relaxed_Low_xz}
\includegraphics[width=0.85\textwidth]{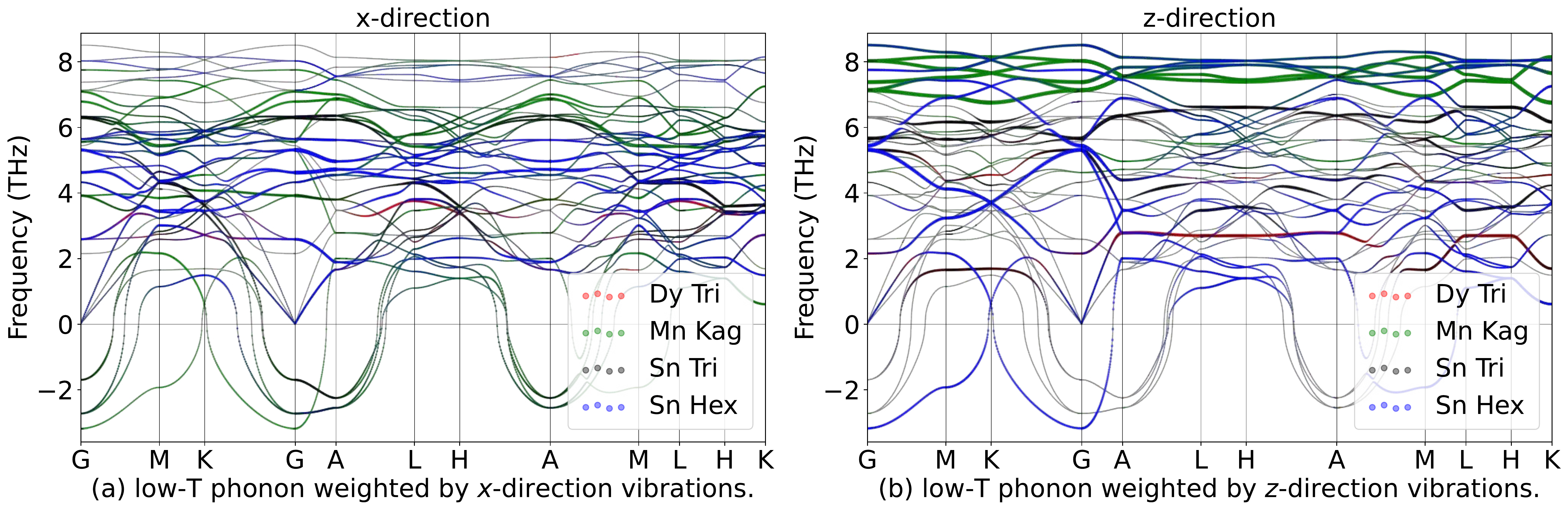}
\hspace{5pt}\label{fig:DyMn6Sn6_relaxed_High_xz}
\includegraphics[width=0.85\textwidth]{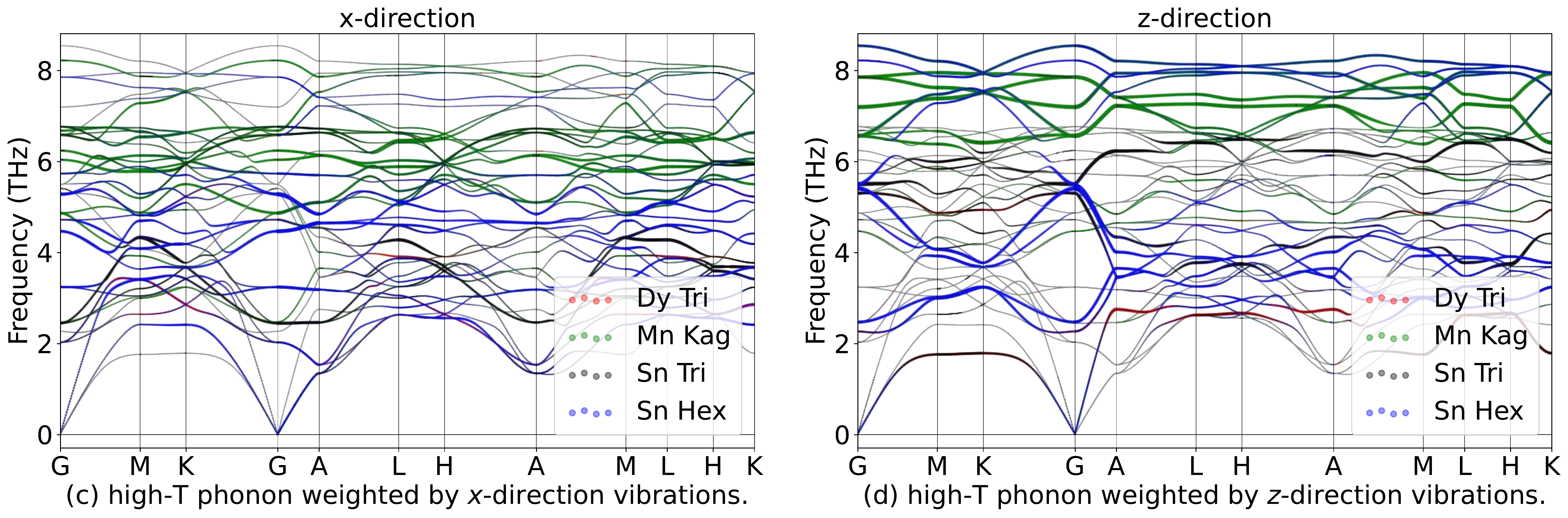}
\caption{Phonon spectrum for DyMn$_6$Sn$_6$in in-plane ($x$) and out-of-plane ($z$) directions in paramagnetic phase.}
\label{fig:DyMn6Sn6phoxyz}
\end{figure}

\begin{table}[htbp]
\global\long\def\arraystretch{1.12}
\captionsetup{width=.95\textwidth}
\caption{IRREPs at high-symmetry points for DyMn$_6$Sn$_6$ phonon spectrum at Low-T.}
\label{tab:191_DyMn6Sn6_Low}
\setlength{\tabcolsep}{1.mm}{
}
\end{table}

\begin{table}[htbp]
\global\long\def\arraystretch{1.12}
\captionsetup{width=.95\textwidth}
\caption{IRREPs at high-symmetry points for DyMn$_6$Sn$_6$ phonon spectrum at High-T.}
\label{tab:191_DyMn6Sn6_High}
\setlength{\tabcolsep}{1.mm}{
%
}
\end{table}
\subsubsection{\label{sms2sec:HoMn6Sn6}\hyperref[smsec:col]{\texorpdfstring{HoMn$_6$Sn$_6$}{}}~{\color{blue}  (High-T stable, Low-T unstable) }}

\begin{table}[htbp]
\global\long\def\arraystretch{1.12}
\captionsetup{width=.85\textwidth}
\caption{Structure parameters of HoMn$_6$Sn$_6$ after relaxation. It contains 517 electrons. }
\label{tab:HoMn6Sn6}
\setlength{\tabcolsep}{4mm}{
%
}
\end{table}

\begin{figure}[htbp]
\centering
\includegraphics[width=0.85\textwidth]{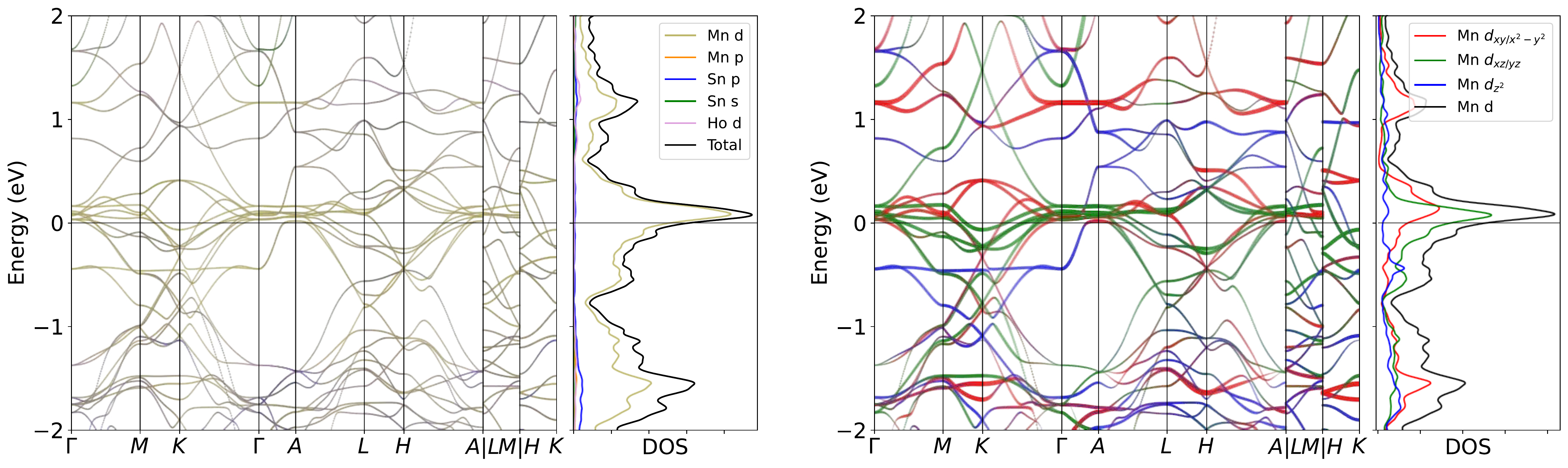}
\caption{Band structure for HoMn$_6$Sn$_6$ without SOC in paramagnetic phase. The projection of Mn $3d$ orbitals is also presented. The band structures clearly reveal the dominating of Mn $3d$ orbitals  near the Fermi level, as well as the pronounced peak.}
\label{fig:HoMn6Sn6band}
\end{figure}

\begin{figure}[H]
\centering
\subfloat[Relaxed phonon at low-T.]{
\label{fig:HoMn6Sn6_relaxed_Low}
\includegraphics[width=0.45\textwidth]{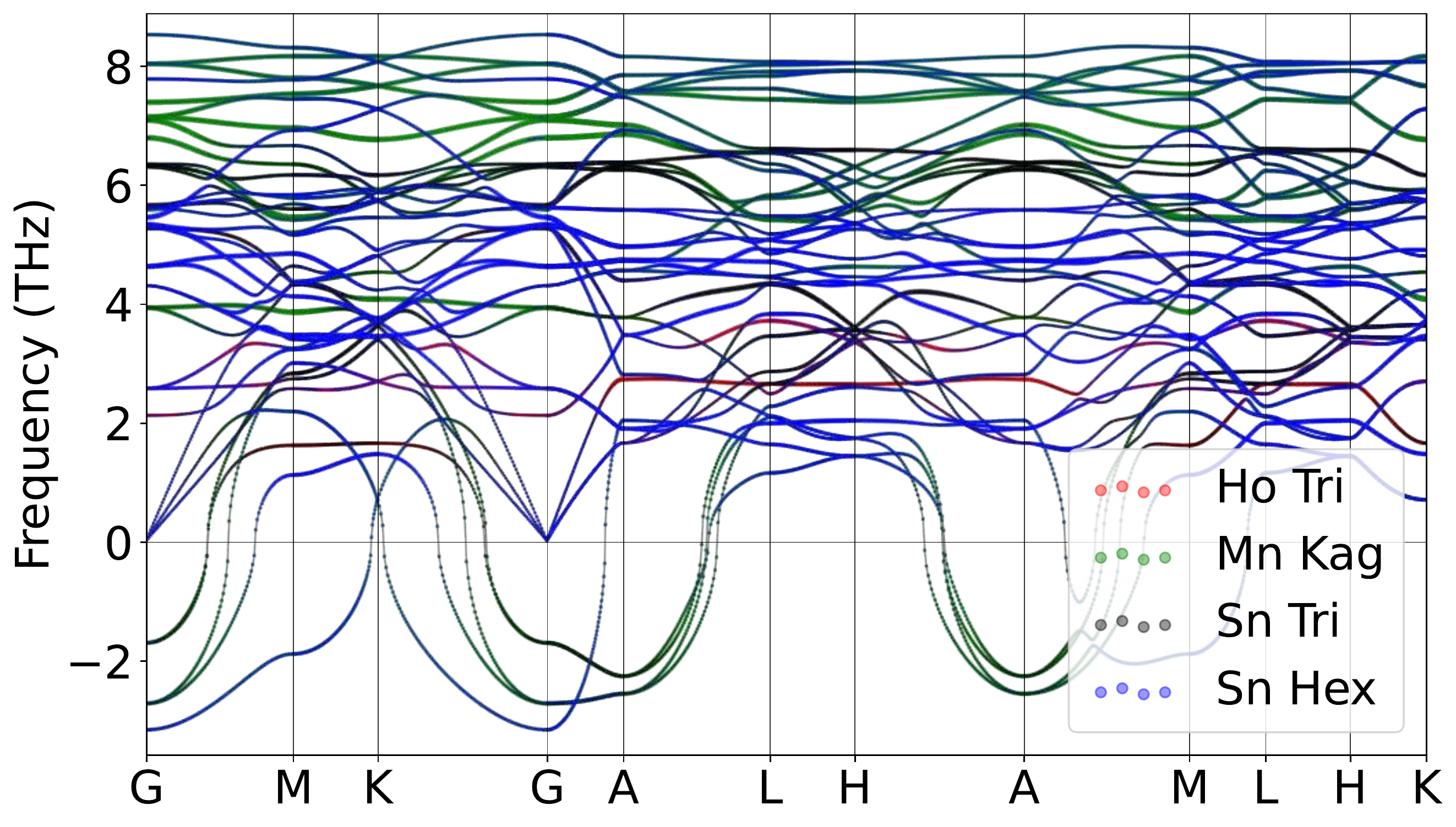}
}
\hspace{5pt}\subfloat[Relaxed phonon at high-T.]{
\label{fig:HoMn6Sn6_relaxed_High}
\includegraphics[width=0.45\textwidth]{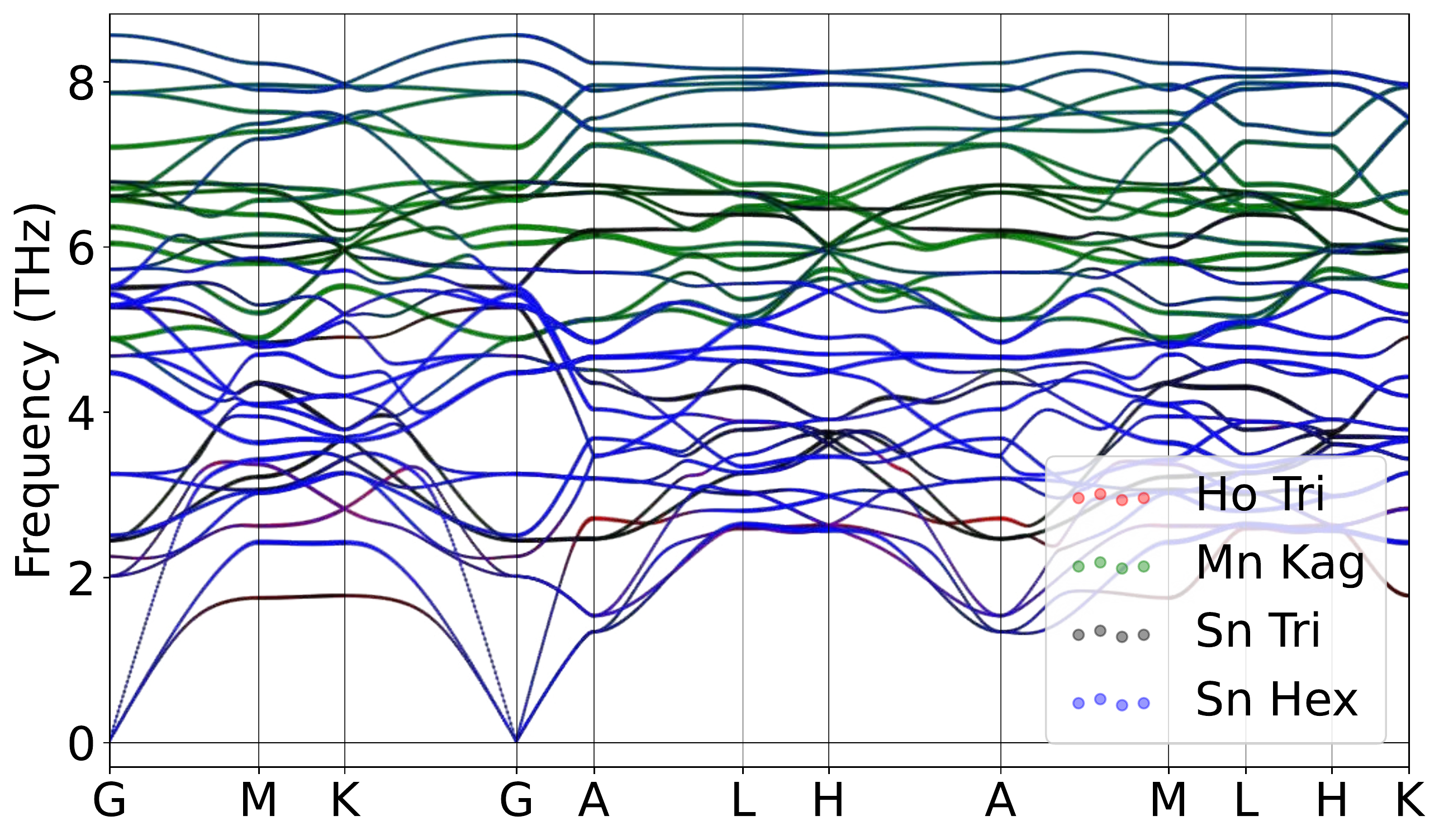}
}
\caption{Atomically projected phonon spectrum for HoMn$_6$Sn$_6$ in paramagnetic phase.}
\label{fig:HoMn6Sn6pho}
\end{figure}

\begin{figure}[htbp]
\centering
\includegraphics[width=0.32\textwidth]{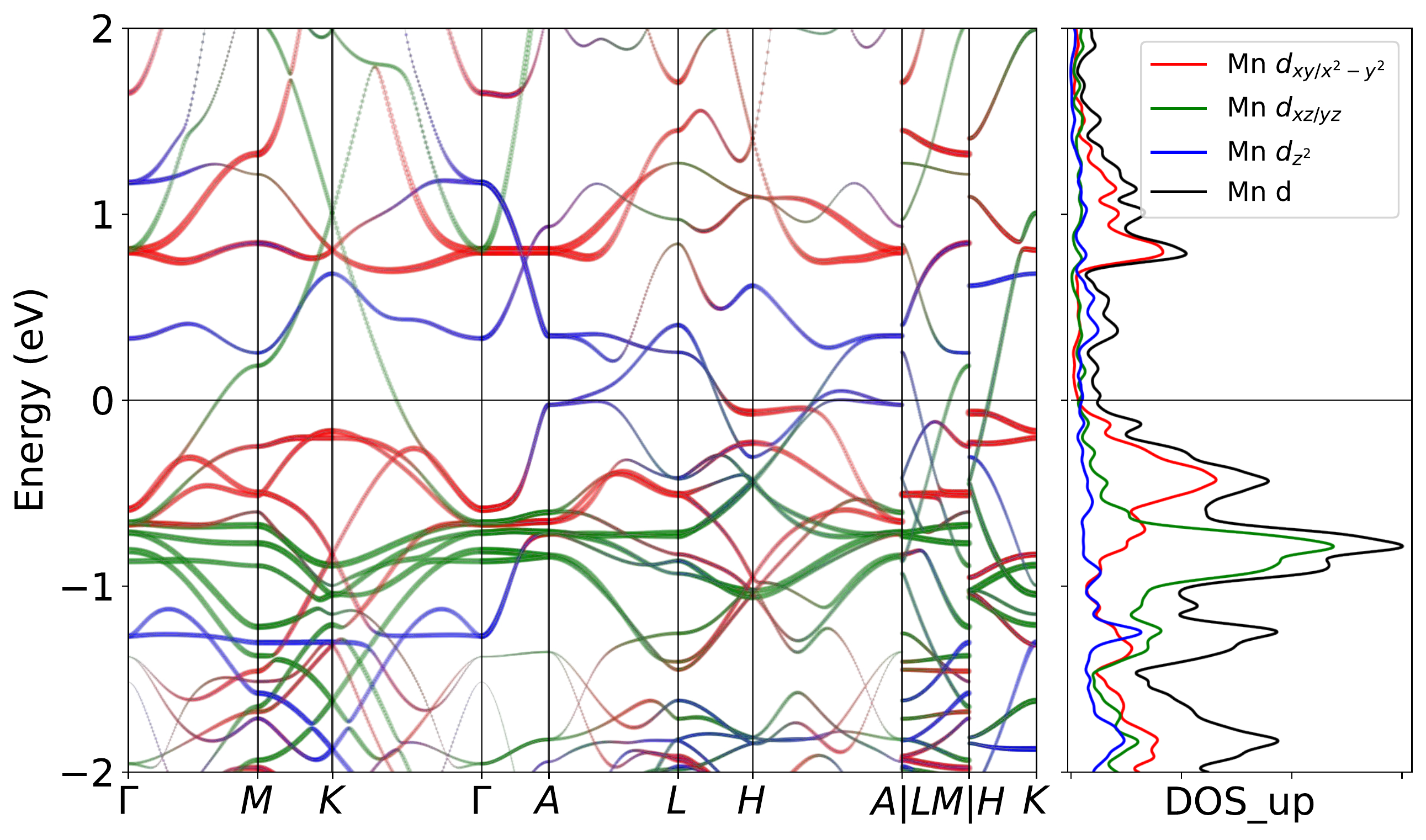}
\includegraphics[width=0.32\textwidth]{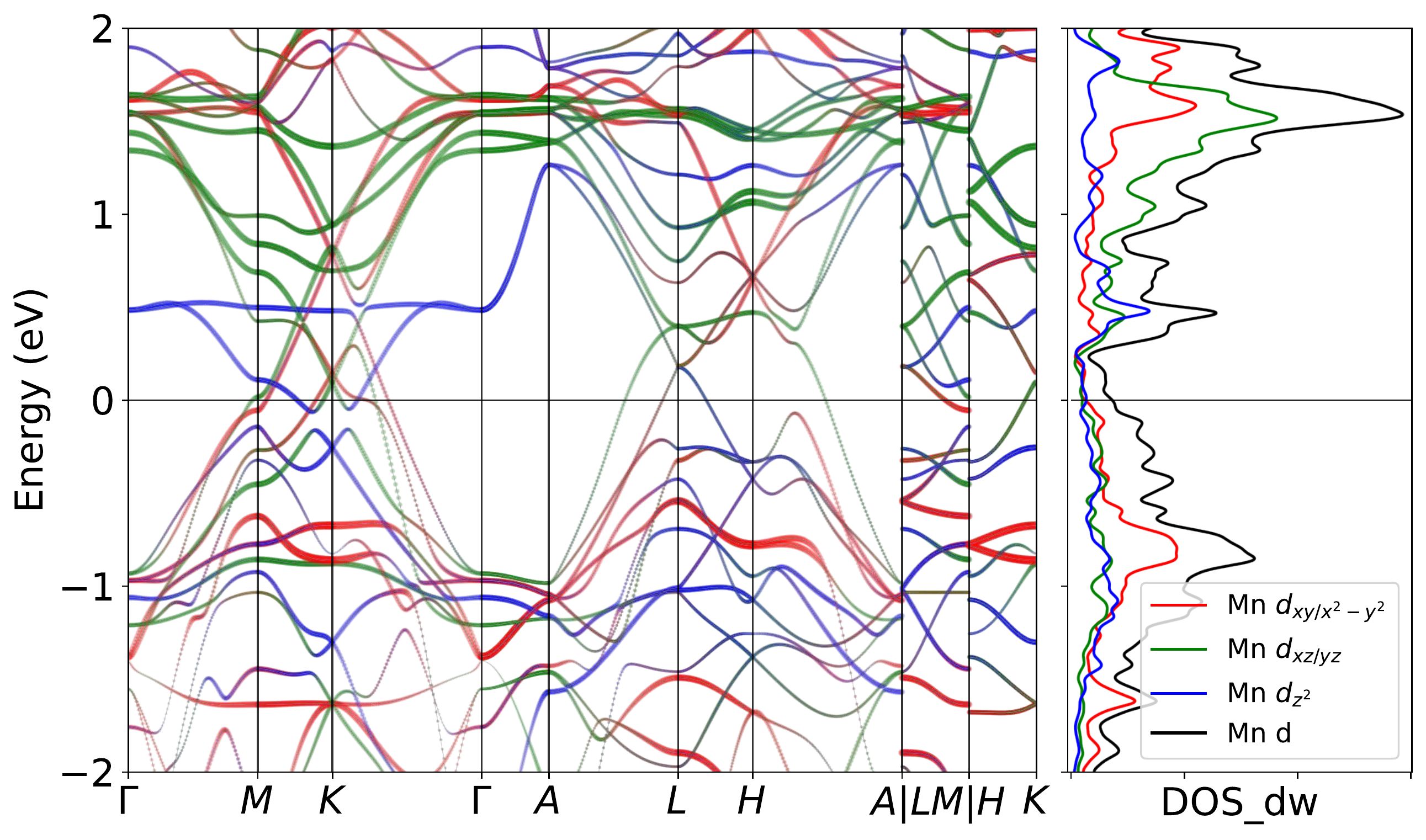}
\includegraphics[width=0.32\textwidth]{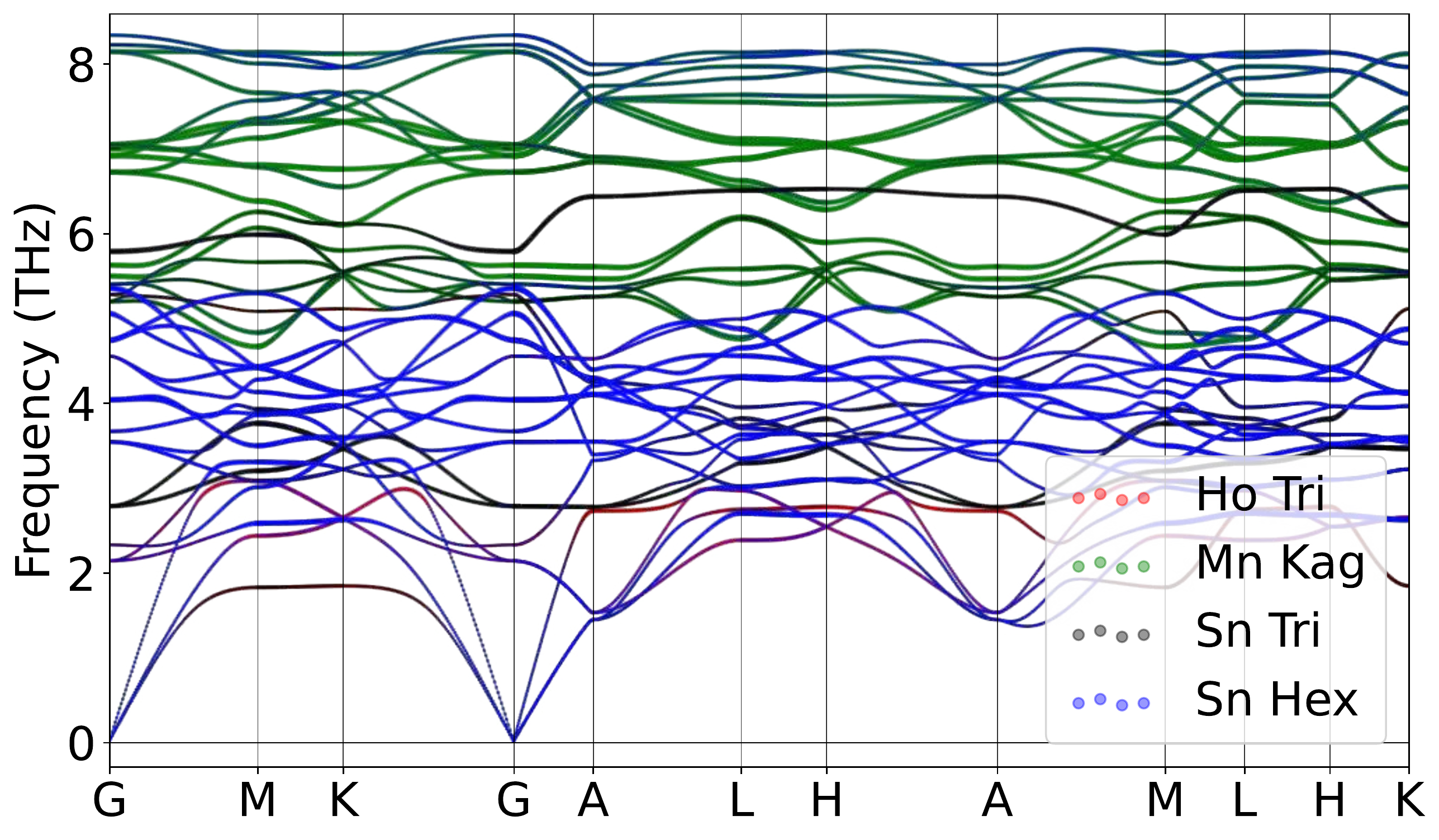}
\caption{Band structure for HoMn$_6$Sn$_6$ with FM order without SOC for spin (Left)up and (Middle)down channels. The projection of Mn $3d$ orbitals is also presented. (Right)Correspondingly, a stable phonon was demonstrated with the collinear magnetic order without Hubbard U at lower temperature regime, weighted by atomic projections.}
\label{fig:HoMn6Sn6mag}
\end{figure}

\begin{figure}[H]
\centering
\label{fig:HoMn6Sn6_relaxed_Low_xz}
\includegraphics[width=0.85\textwidth]{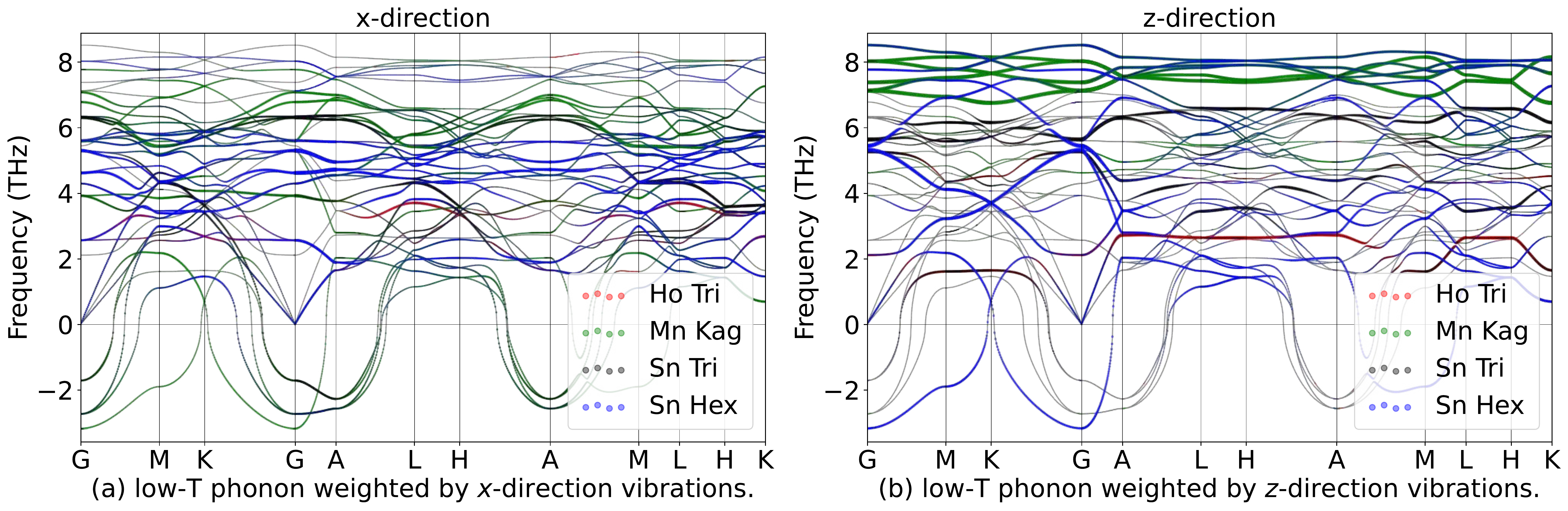}
\hspace{5pt}\label{fig:HoMn6Sn6_relaxed_High_xz}
\includegraphics[width=0.85\textwidth]{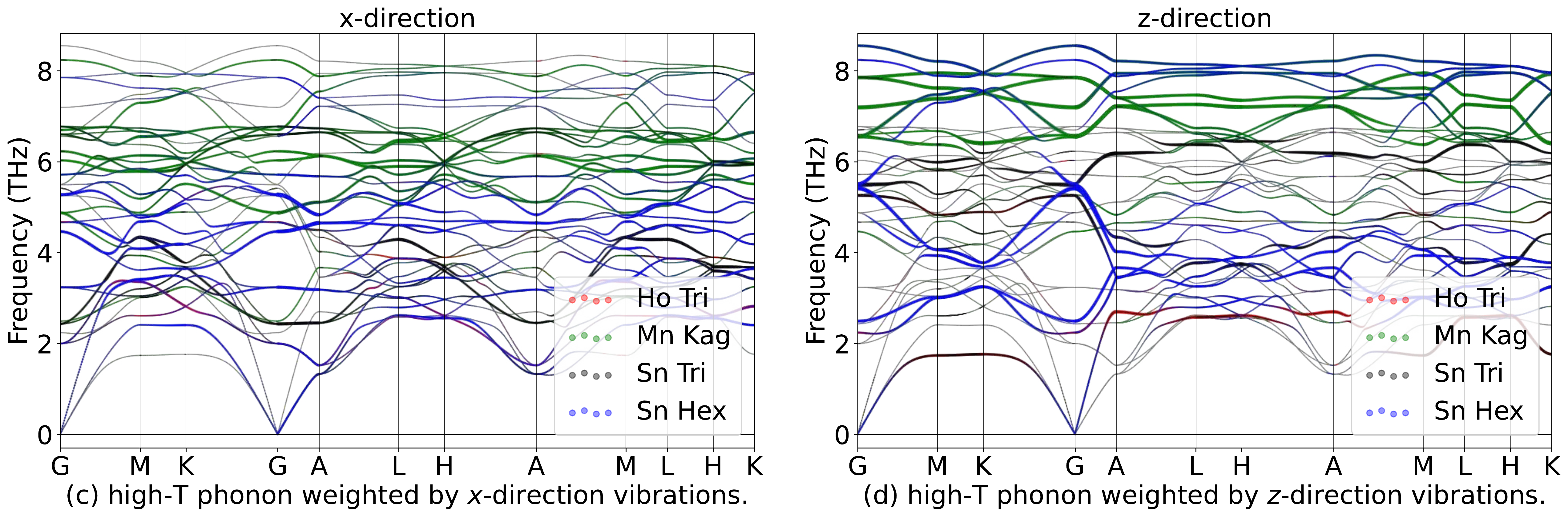}
\caption{Phonon spectrum for HoMn$_6$Sn$_6$in in-plane ($x$) and out-of-plane ($z$) directions in paramagnetic phase.}
\label{fig:HoMn6Sn6phoxyz}
\end{figure}

\begin{table}[htbp]
\global\long\def\arraystretch{1.12}
\captionsetup{width=.95\textwidth}
\caption{IRREPs at high-symmetry points for HoMn$_6$Sn$_6$ phonon spectrum at Low-T.}
\label{tab:191_HoMn6Sn6_Low}
\setlength{\tabcolsep}{1.mm}{
}
\end{table}

\begin{table}[htbp]
\global\long\def\arraystretch{1.12}
\captionsetup{width=.95\textwidth}
\caption{IRREPs at high-symmetry points for HoMn$_6$Sn$_6$ phonon spectrum at High-T.}
\label{tab:191_HoMn6Sn6_High}
\setlength{\tabcolsep}{1.mm}{
%
}
\end{table}
\subsubsection{\label{sms2sec:ErMn6Sn6}\hyperref[smsec:col]{\texorpdfstring{ErMn$_6$Sn$_6$}{}}~{\color{blue}  (High-T stable, Low-T unstable) }}

\begin{table}[htbp]
\global\long\def\arraystretch{1.12}
\captionsetup{width=.85\textwidth}
\caption{Structure parameters of ErMn$_6$Sn$_6$ after relaxation. It contains 518 electrons. }
\label{tab:ErMn6Sn6}
\setlength{\tabcolsep}{4mm}{
%
}
\end{table}

\begin{figure}[htbp]
\centering
\includegraphics[width=0.85\textwidth]{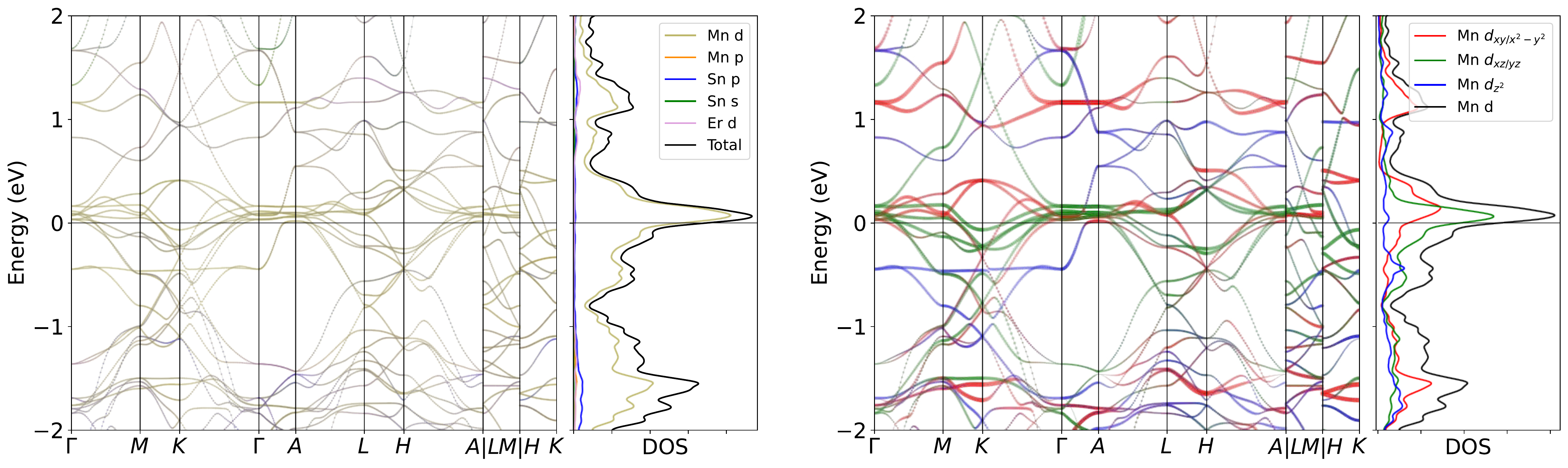}
\caption{Band structure for ErMn$_6$Sn$_6$ without SOC in paramagnetic phase. The projection of Mn $3d$ orbitals is also presented. The band structures clearly reveal the dominating of Mn $3d$ orbitals  near the Fermi level, as well as the pronounced peak.}
\label{fig:ErMn6Sn6band}
\end{figure}

\begin{figure}[H]
\centering
\subfloat[Relaxed phonon at low-T.]{
\label{fig:ErMn6Sn6_relaxed_Low}
\includegraphics[width=0.45\textwidth]{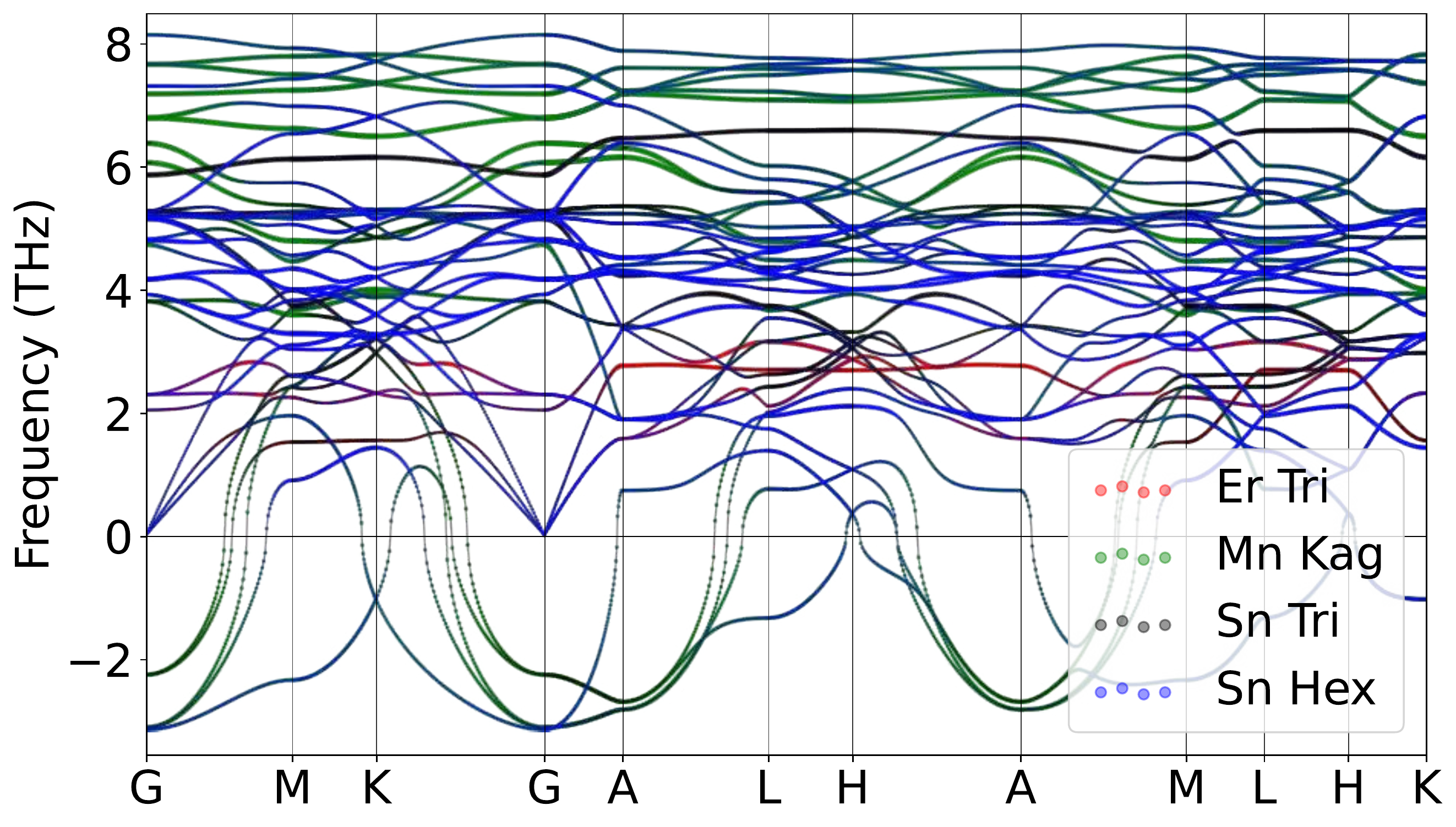}
}
\hspace{5pt}\subfloat[Relaxed phonon at high-T.]{
\label{fig:ErMn6Sn6_relaxed_High}
\includegraphics[width=0.45\textwidth]{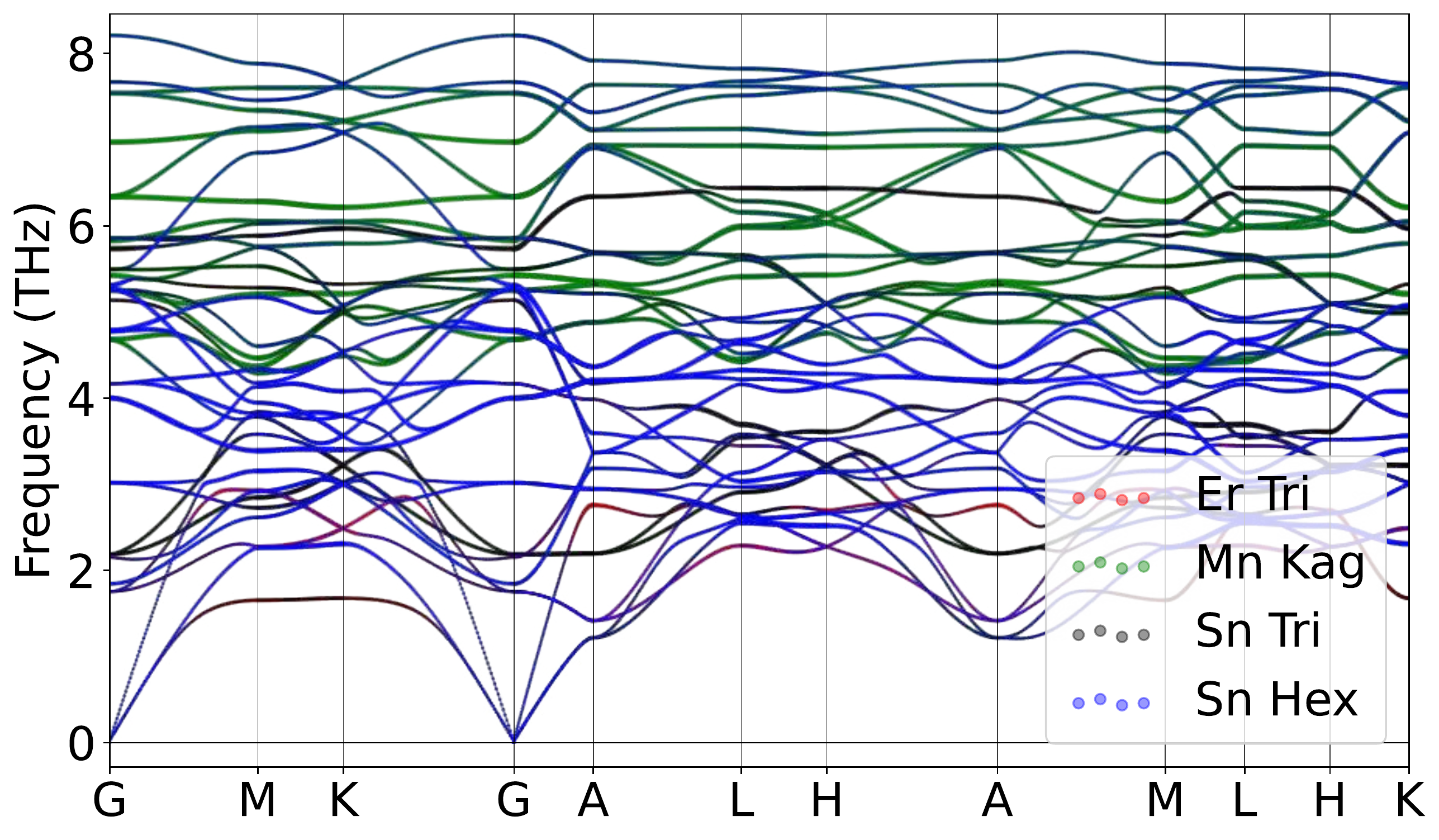}
}
\caption{Atomically projected phonon spectrum for ErMn$_6$Sn$_6$ in paramagnetic phase.}
\label{fig:ErMn6Sn6pho}
\end{figure}

\begin{figure}[htbp]
\centering
\includegraphics[width=0.45\textwidth]{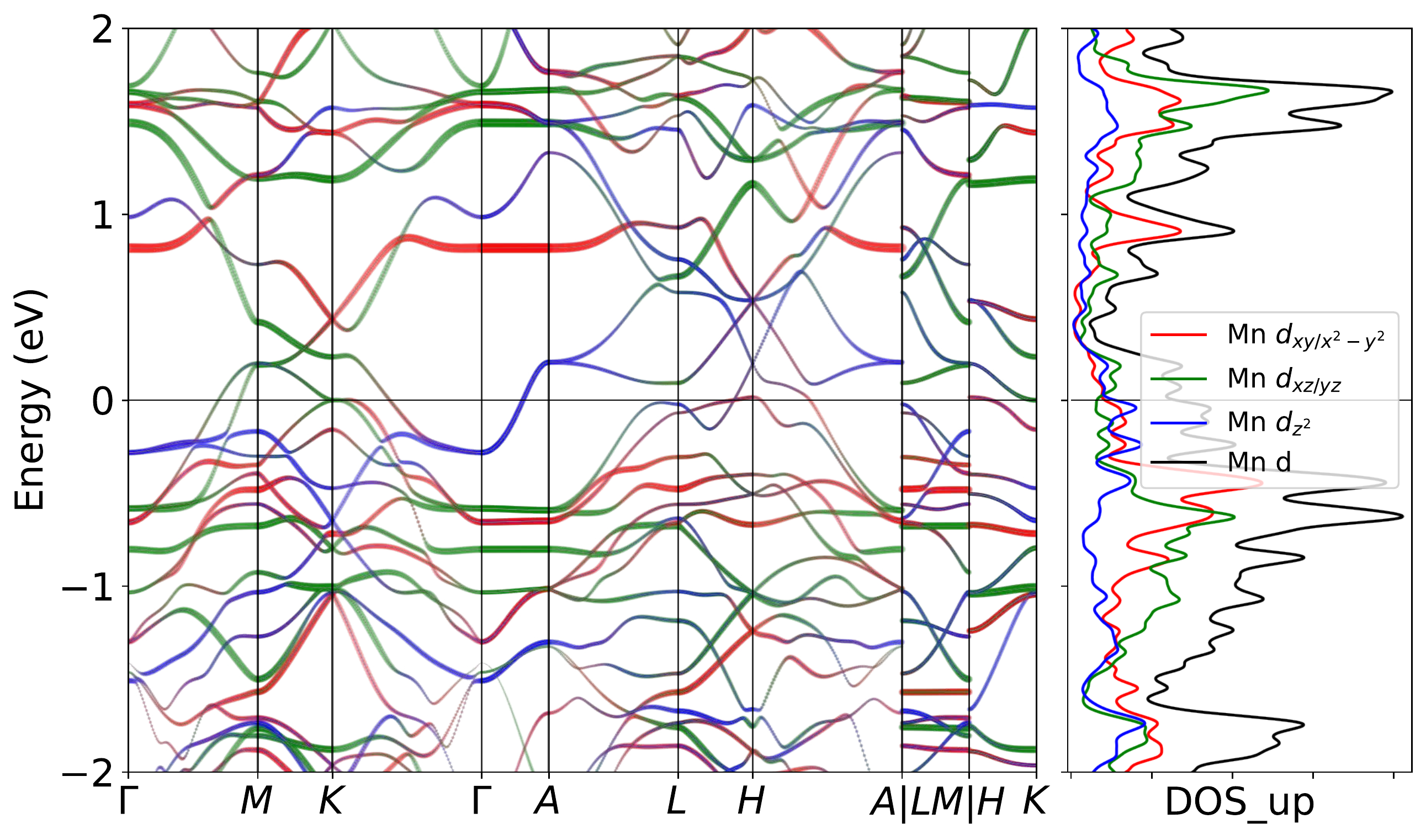}
\includegraphics[width=0.45\textwidth]{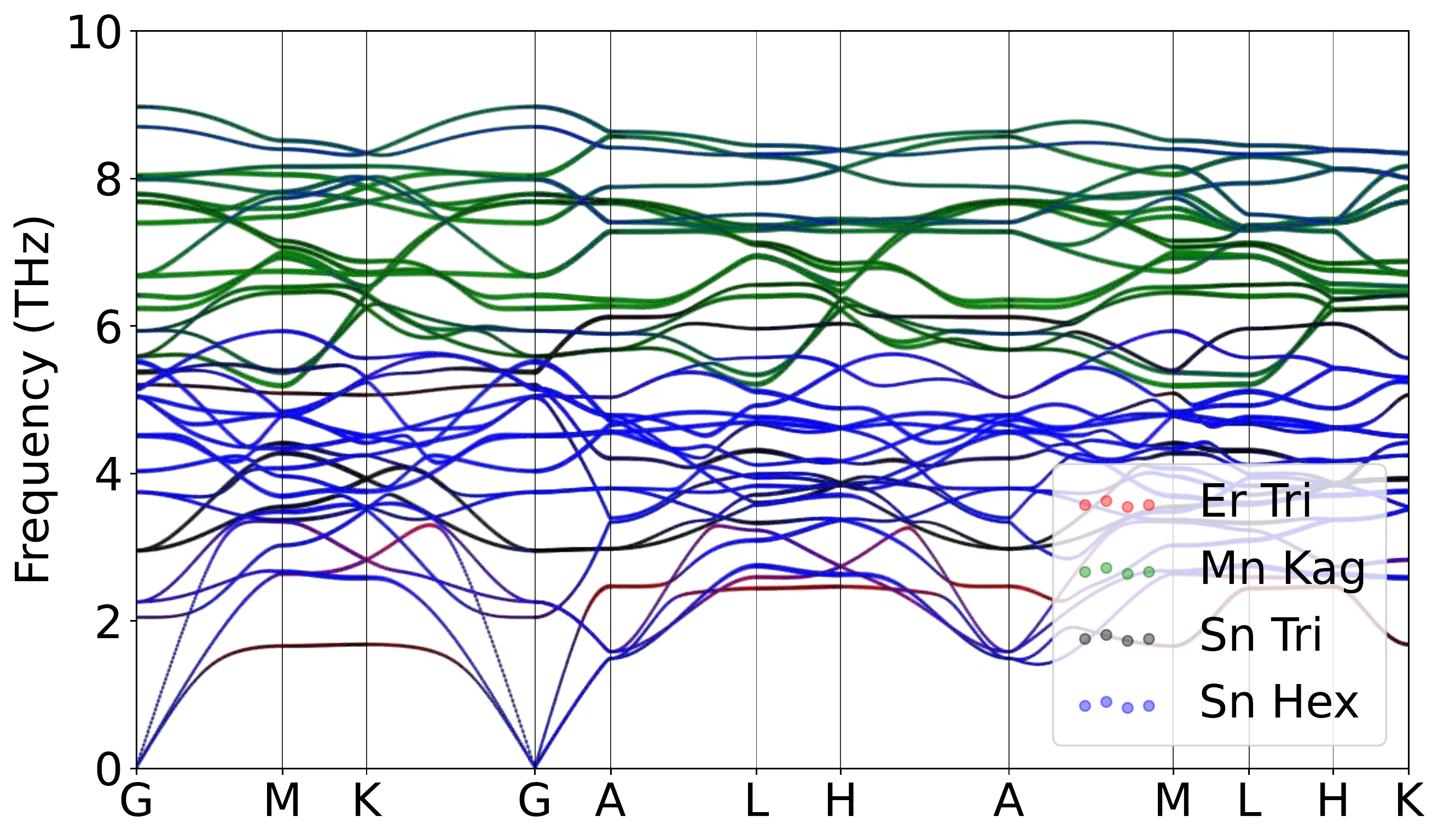}
\caption{Band structure for ErMn$_6$Sn$_6$ with AFM order without SOC for spin (Left)up channel. The projection of Mn $3d$ orbitals is also presented. (Right)Correspondingly, a stable phonon was demonstrated with the collinear magnetic order without Hubbard U at lower temperature regime, weighted by atomic projections.}
\label{fig:ErMn6Sn6mag}
\end{figure}

\begin{figure}[H]
\centering
\label{fig:ErMn6Sn6_relaxed_Low_xz}
\includegraphics[width=0.85\textwidth]{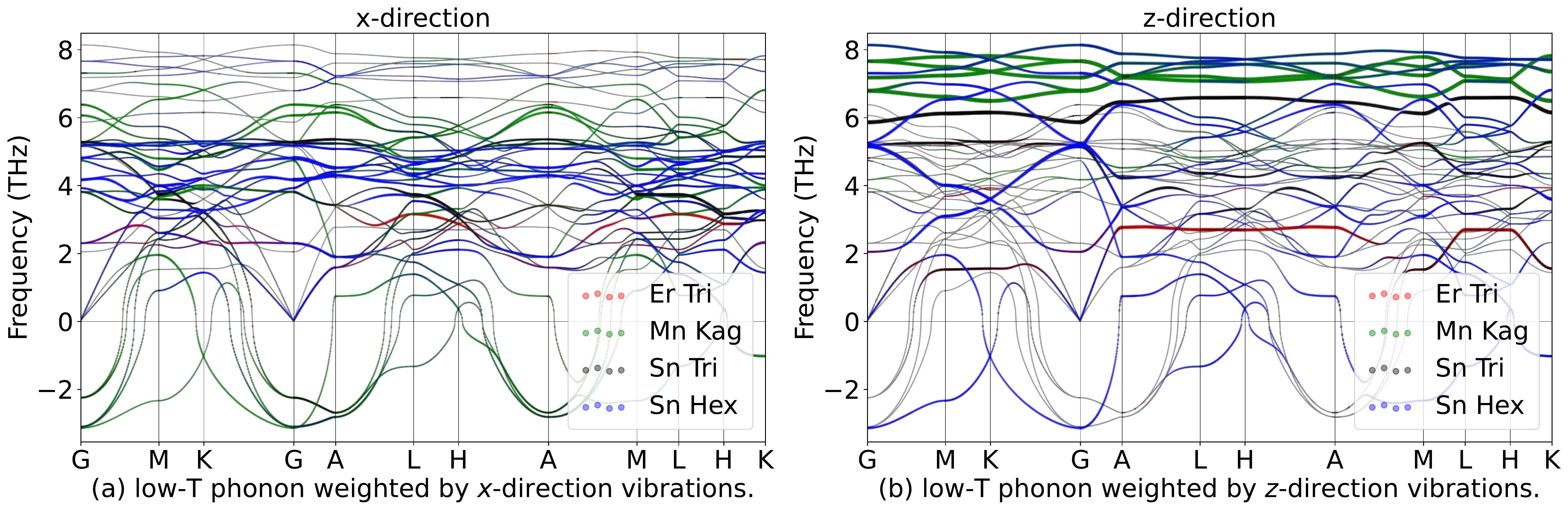}
\hspace{5pt}\label{fig:ErMn6Sn6_relaxed_High_xz}
\includegraphics[width=0.85\textwidth]{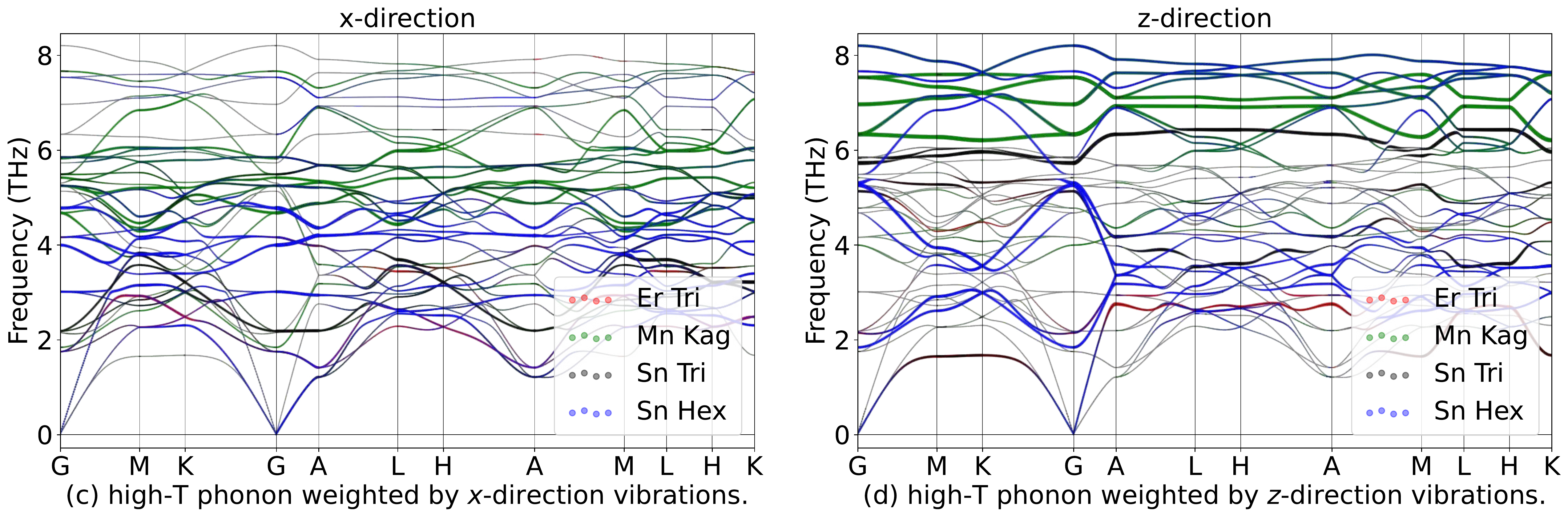}
\caption{Phonon spectrum for ErMn$_6$Sn$_6$in in-plane ($x$) and out-of-plane ($z$) directions in paramagnetic phase.}
\label{fig:ErMn6Sn6phoxyz}
\end{figure}

\begin{table}[htbp]
\global\long\def\arraystretch{1.12}
\captionsetup{width=.95\textwidth}
\caption{IRREPs at high-symmetry points for ErMn$_6$Sn$_6$ phonon spectrum at Low-T.}
\label{tab:191_ErMn6Sn6_Low}
\setlength{\tabcolsep}{1.mm}{
}
\end{table}

\begin{table}[htbp]
\global\long\def\arraystretch{1.12}
\captionsetup{width=.95\textwidth}
\caption{IRREPs at high-symmetry points for ErMn$_6$Sn$_6$ phonon spectrum at High-T.}
\label{tab:191_ErMn6Sn6_High}
\setlength{\tabcolsep}{1.mm}{
%
}
\end{table}
\subsubsection{\label{sms2sec:TmMn6Sn6}\hyperref[smsec:col]{\texorpdfstring{TmMn$_6$Sn$_6$}{}}~{\color{blue}  (High-T stable, Low-T unstable) }}

\begin{table}[htbp]
\global\long\def\arraystretch{1.12}
\captionsetup{width=.85\textwidth}
\caption{Structure parameters of TmMn$_6$Sn$_6$ after relaxation. It contains 519 electrons. }
\label{tab:TmMn6Sn6}
\setlength{\tabcolsep}{4mm}{
%
}
\end{table}

\begin{figure}[htbp]
\centering
\includegraphics[width=0.85\textwidth]{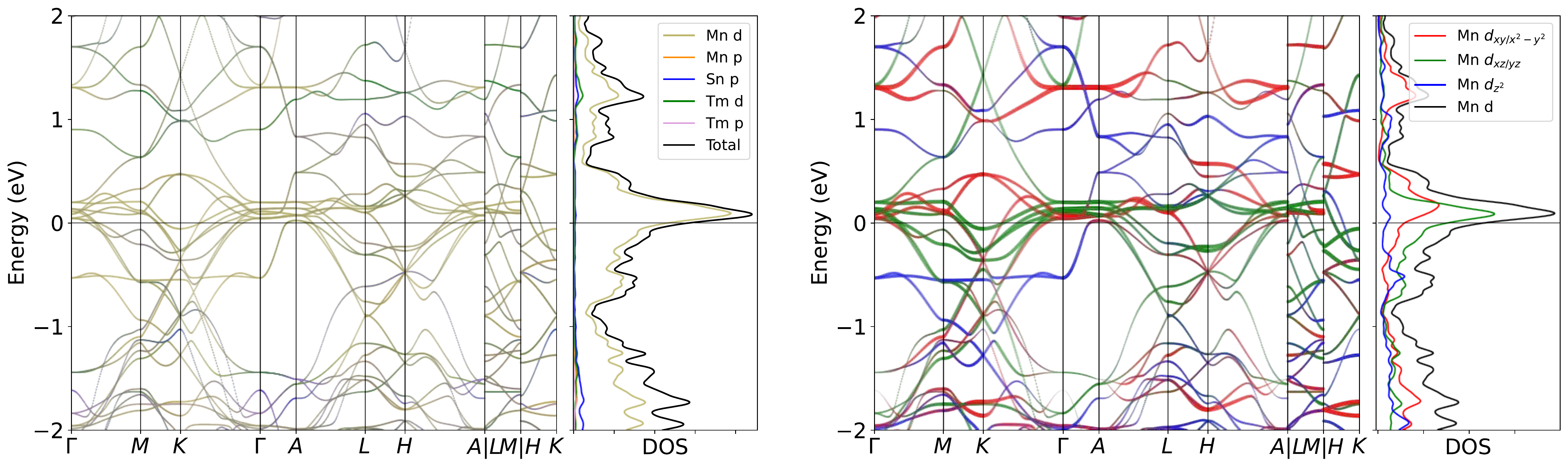}
\caption{Band structure for TmMn$_6$Sn$_6$ without SOC in paramagnetic phase. The projection of Mn $3d$ orbitals is also presented. The band structures clearly reveal the dominating of Mn $3d$ orbitals  near the Fermi level, as well as the pronounced peak.}
\label{fig:TmMn6Sn6band}
\end{figure}

\begin{figure}[H]
\centering
\subfloat[Relaxed phonon at low-T.]{
\label{fig:TmMn6Sn6_relaxed_Low}
\includegraphics[width=0.45\textwidth]{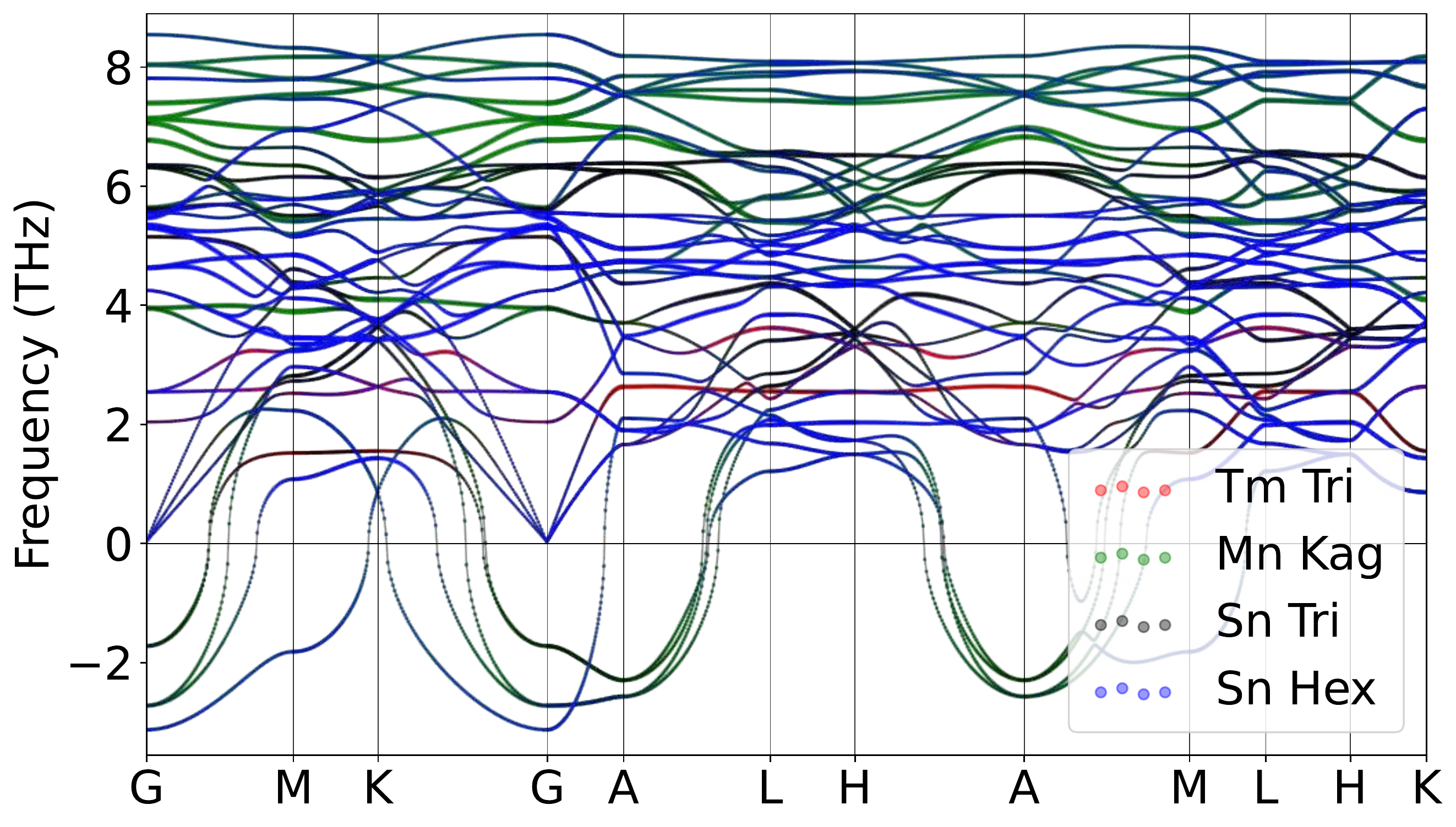}
}
\hspace{5pt}\subfloat[Relaxed phonon at high-T.]{
\label{fig:TmMn6Sn6_relaxed_High}
\includegraphics[width=0.45\textwidth]{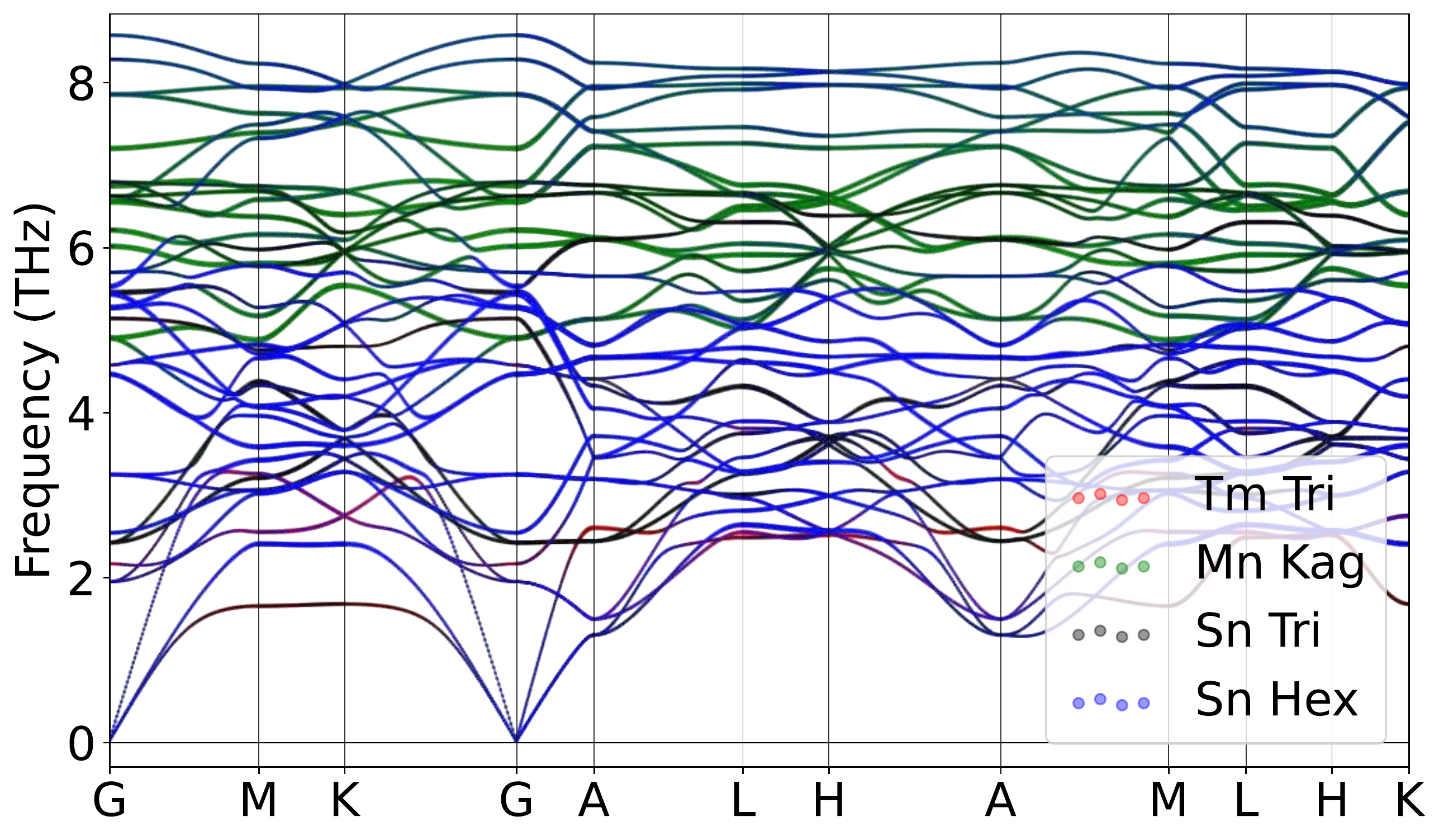}
}
\caption{Atomically projected phonon spectrum for TmMn$_6$Sn$_6$ in paramagnetic phase.}
\label{fig:TmMn6Sn6pho}
\end{figure}

\begin{figure}[htbp]
\centering
\includegraphics[width=0.45\textwidth]{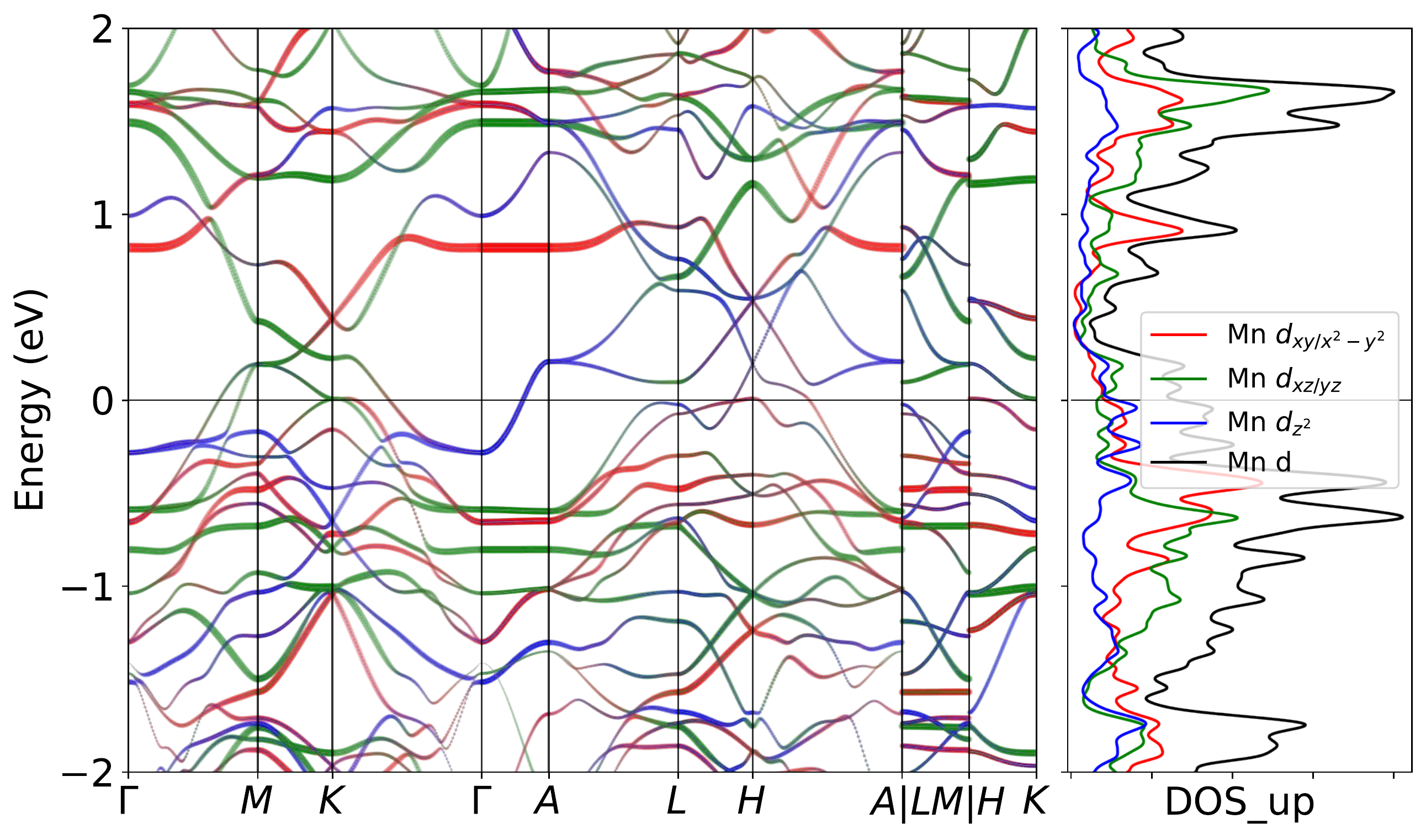}
\includegraphics[width=0.45\textwidth]{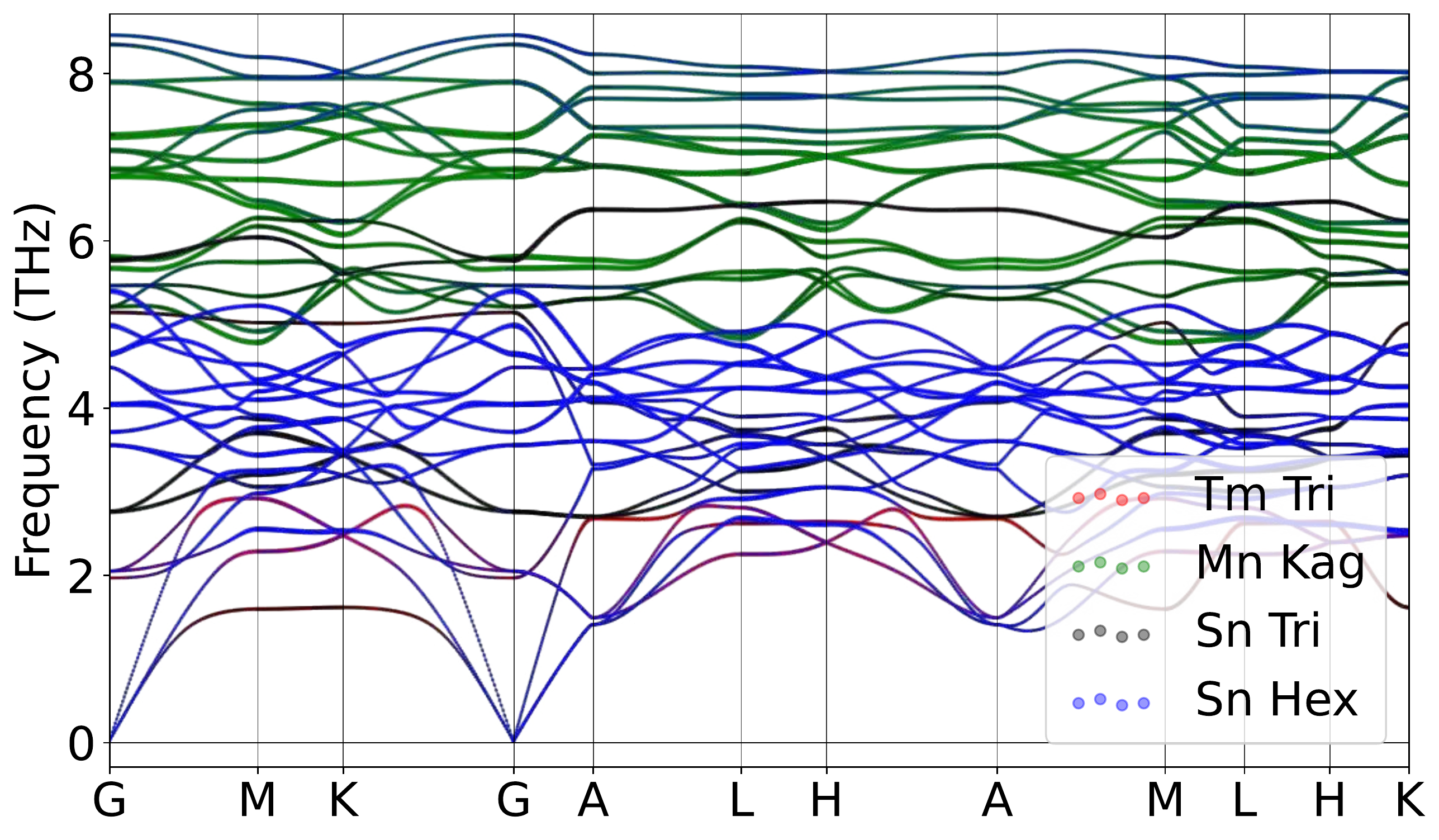}
\caption{Band structure for TmMn$_6$Sn$_6$ with AFM order without SOC for spin (Left)up channel. The projection of Mn $3d$ orbitals is also presented. (Right)Correspondingly, a stable phonon was demonstrated with the collinear magnetic order without Hubbard U at lower temperature regime, weighted by atomic projections.}
\label{fig:TmMn6Sn6mag}
\end{figure}

\begin{figure}[H]
\centering
\label{fig:TmMn6Sn6_relaxed_Low_xz}
\includegraphics[width=0.85\textwidth]{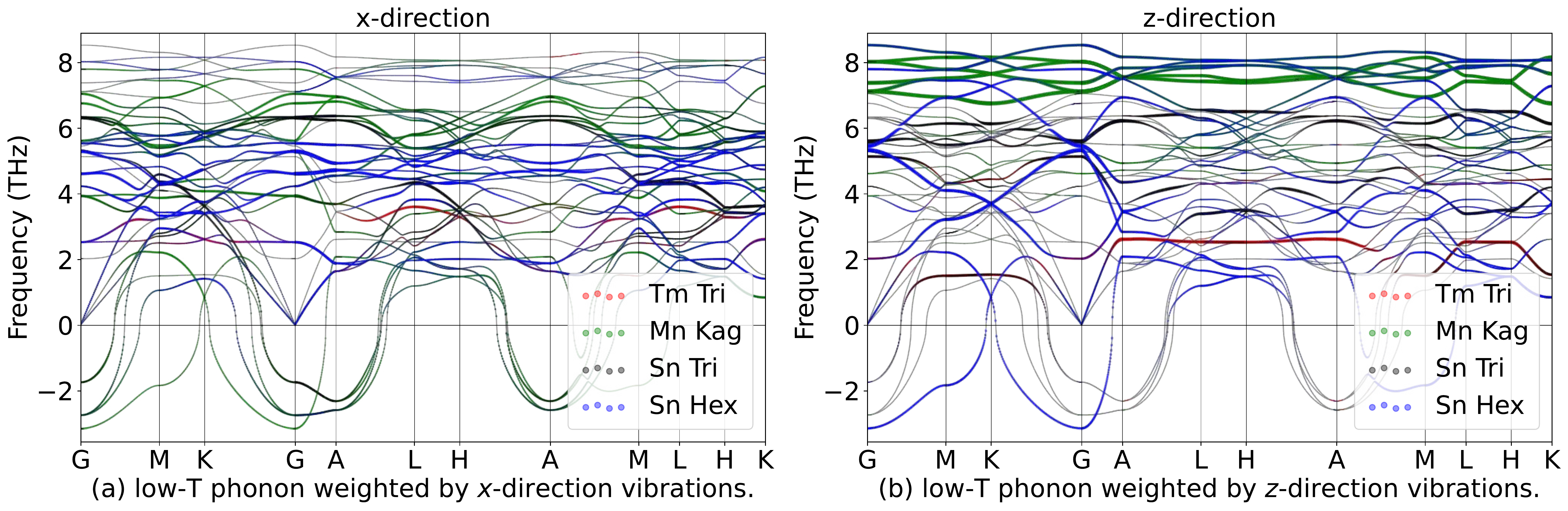}
\hspace{5pt}\label{fig:TmMn6Sn6_relaxed_High_xz}
\includegraphics[width=0.85\textwidth]{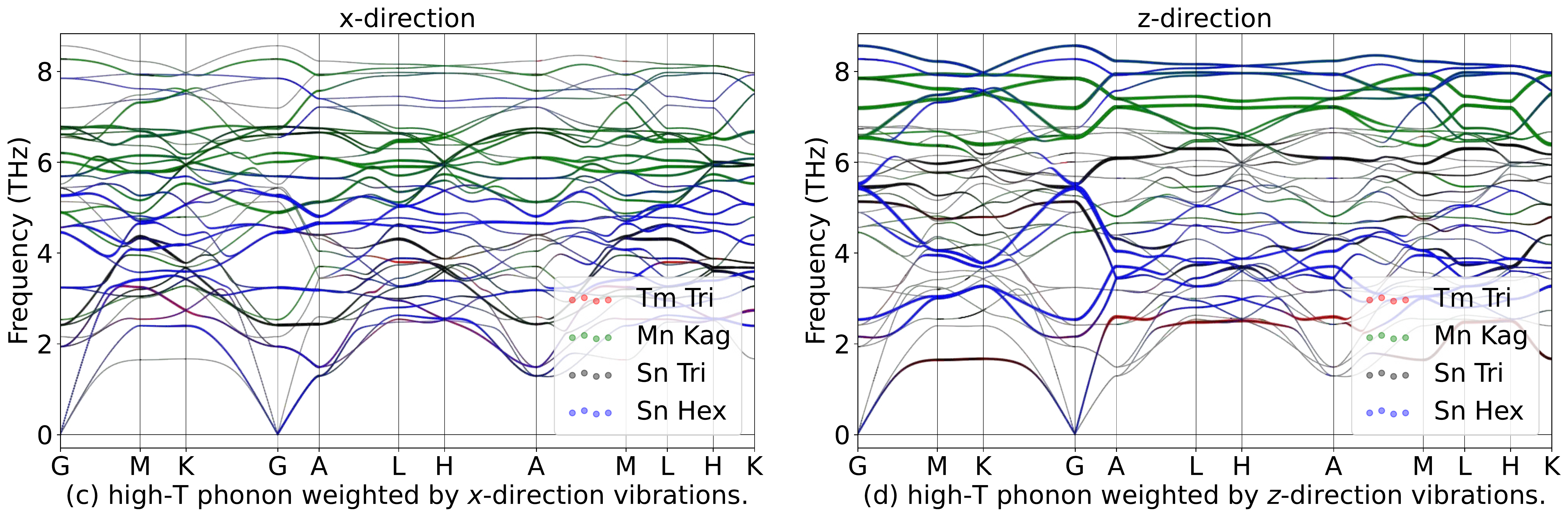}
\caption{Phonon spectrum for TmMn$_6$Sn$_6$in in-plane ($x$) and out-of-plane ($z$) directions in paramagnetic phase.}
\label{fig:TmMn6Sn6phoxyz}
\end{figure}

\begin{table}[htbp]
\global\long\def\arraystretch{1.12}
\captionsetup{width=.95\textwidth}
\caption{IRREPs at high-symmetry points for TmMn$_6$Sn$_6$ phonon spectrum at Low-T.}
\label{tab:191_TmMn6Sn6_Low}
\setlength{\tabcolsep}{1.mm}{
}
\end{table}

\begin{table}[htbp]
\global\long\def\arraystretch{1.12}
\captionsetup{width=.95\textwidth}
\caption{IRREPs at high-symmetry points for TmMn$_6$Sn$_6$ phonon spectrum at High-T.}
\label{tab:191_TmMn6Sn6_High}
\setlength{\tabcolsep}{1.mm}{
%
}
\end{table}
\subsubsection{\label{sms2sec:YbMn6Sn6}\hyperref[smsec:col]{\texorpdfstring{YbMn$_6$Sn$_6$}{}}~{\color{blue}  (High-T stable, Low-T unstable) }}

\begin{table}[htbp]
\global\long\def\arraystretch{1.12}
\captionsetup{width=.85\textwidth}
\caption{Structure parameters of YbMn$_6$Sn$_6$ after relaxation. It contains 520 electrons. }
\label{tab:YbMn6Sn6}
\setlength{\tabcolsep}{4mm}{
%
}
\end{table}

\begin{figure}[htbp]
\centering
\includegraphics[width=0.85\textwidth]{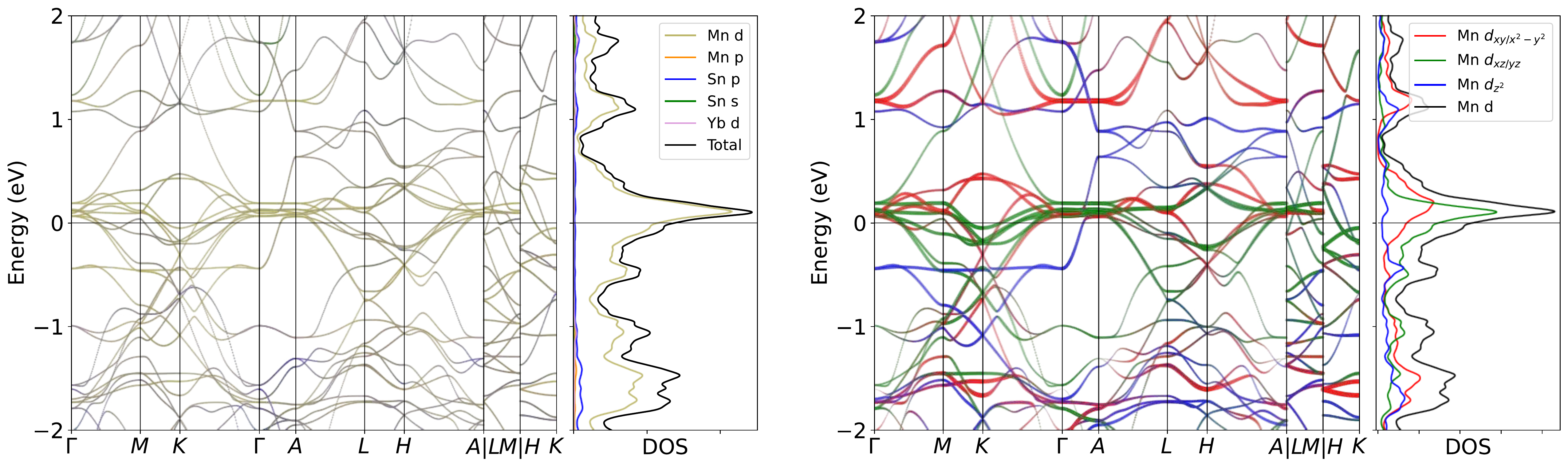}
\caption{Band structure for YbMn$_6$Sn$_6$ without SOC in paramagnetic phase. The projection of Mn $3d$ orbitals is also presented. The band structures clearly reveal the dominating of Mn $3d$ orbitals  near the Fermi level, as well as the pronounced peak.}
\label{fig:YbMn6Sn6band}
\end{figure}

\begin{figure}[H]
\centering
\subfloat[Relaxed phonon at low-T.]{
\label{fig:YbMn6Sn6_relaxed_Low}
\includegraphics[width=0.45\textwidth]{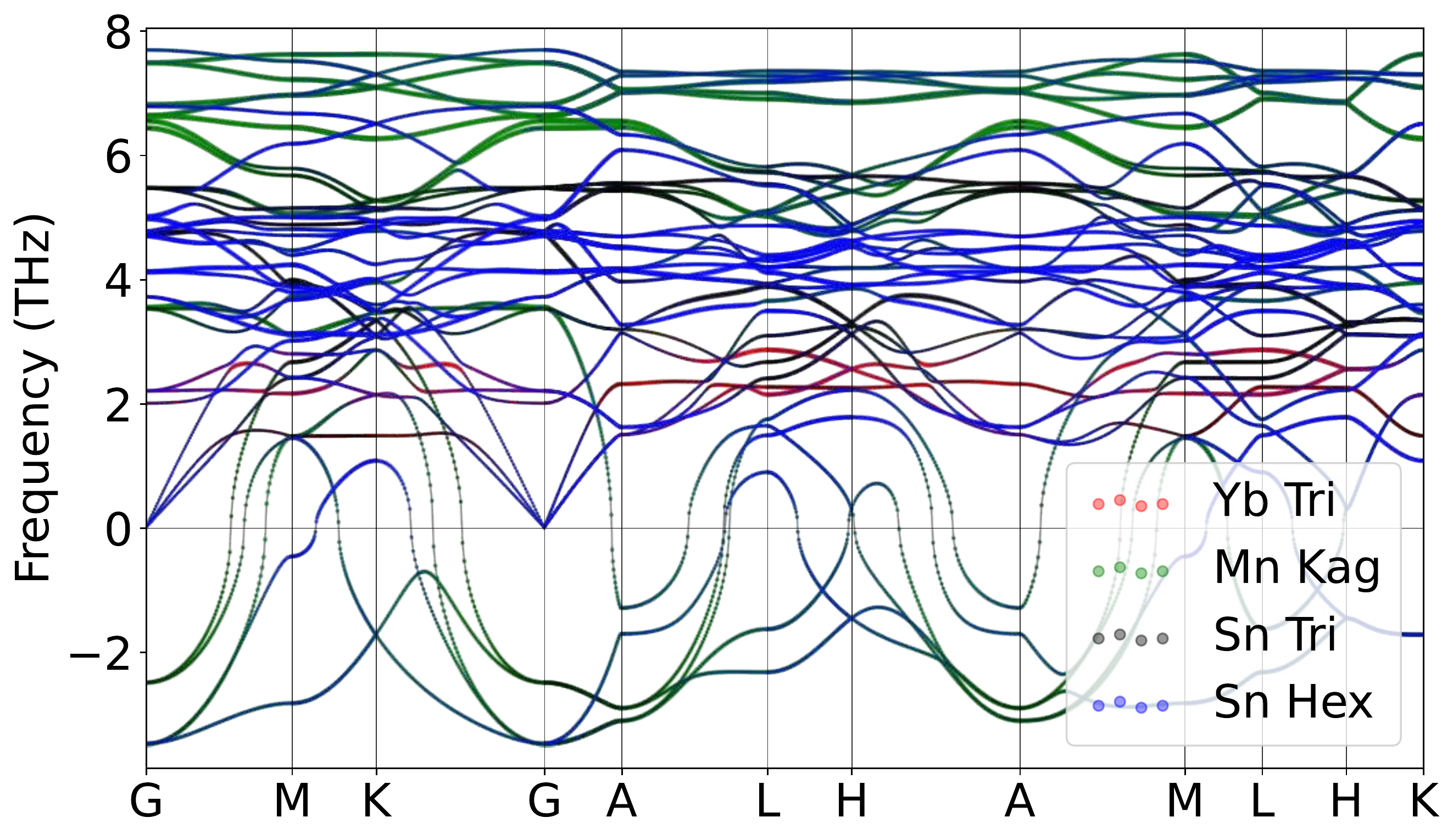}
}
\hspace{5pt}\subfloat[Relaxed phonon at high-T.]{
\label{fig:YbMn6Sn6_relaxed_High}
\includegraphics[width=0.45\textwidth]{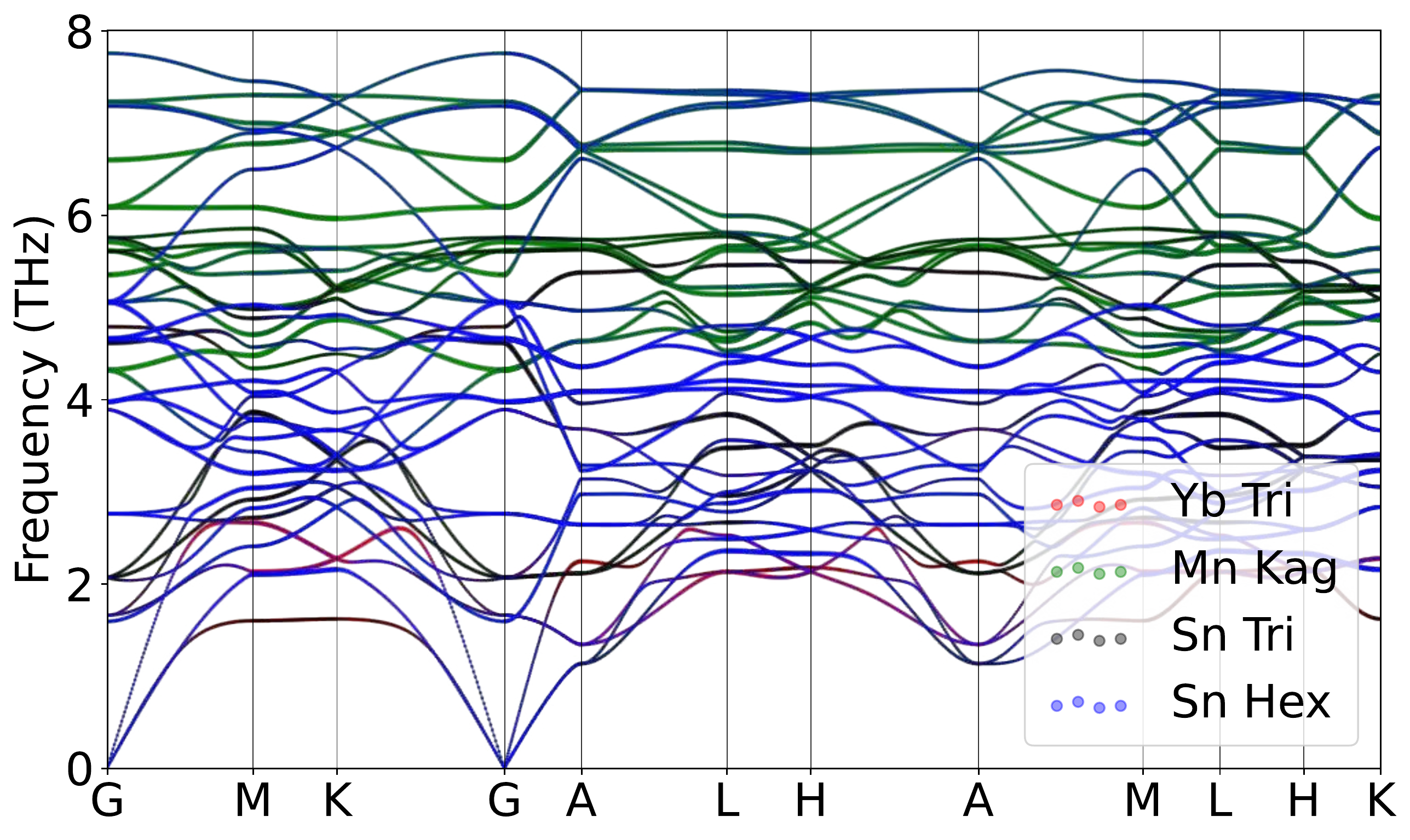}
}
\caption{Atomically projected phonon spectrum for YbMn$_6$Sn$_6$ in paramagnetic phase.}
\label{fig:YbMn6Sn6pho}
\end{figure}

\begin{figure}[htbp]
\centering
\includegraphics[width=0.32\textwidth]{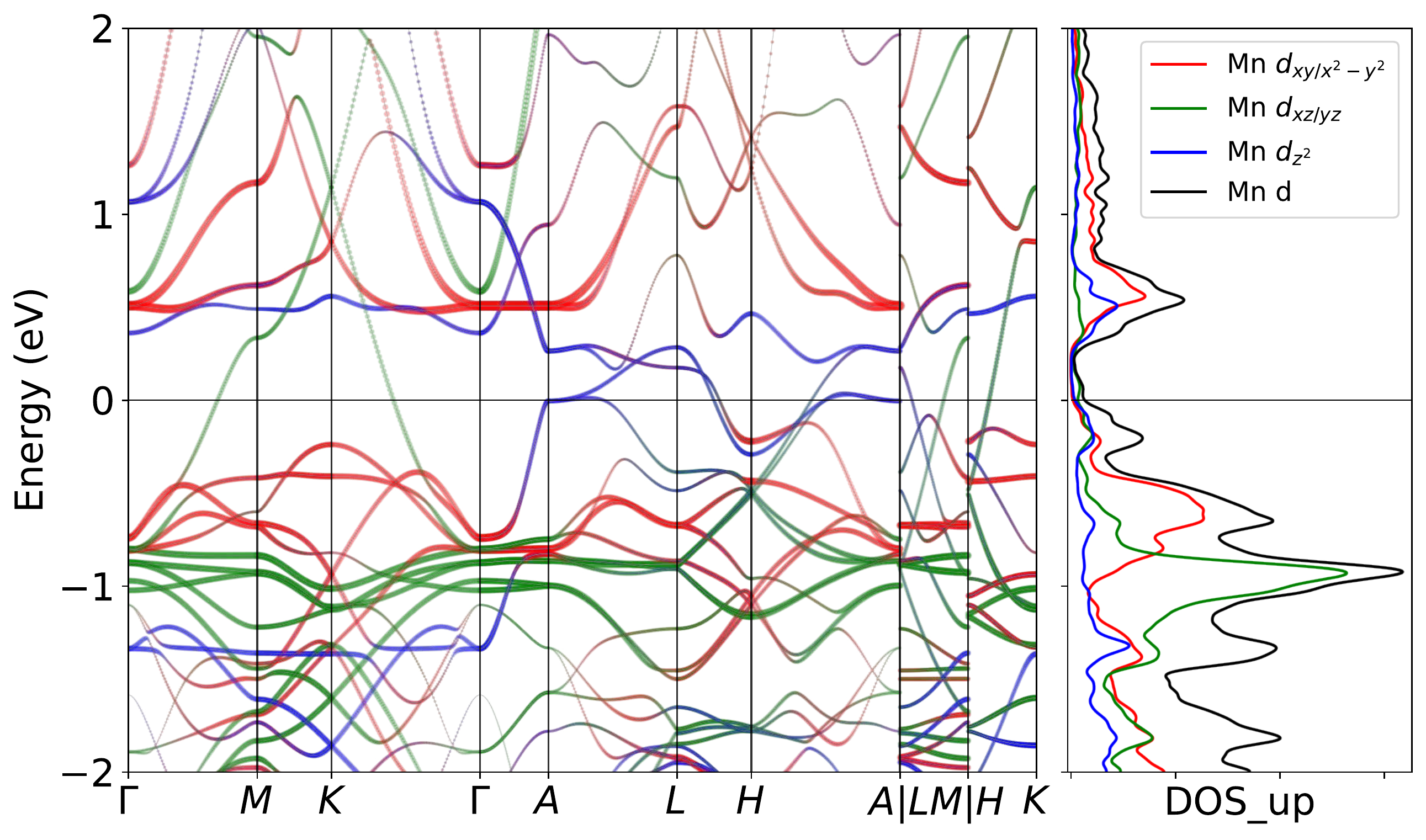}
\includegraphics[width=0.32\textwidth]{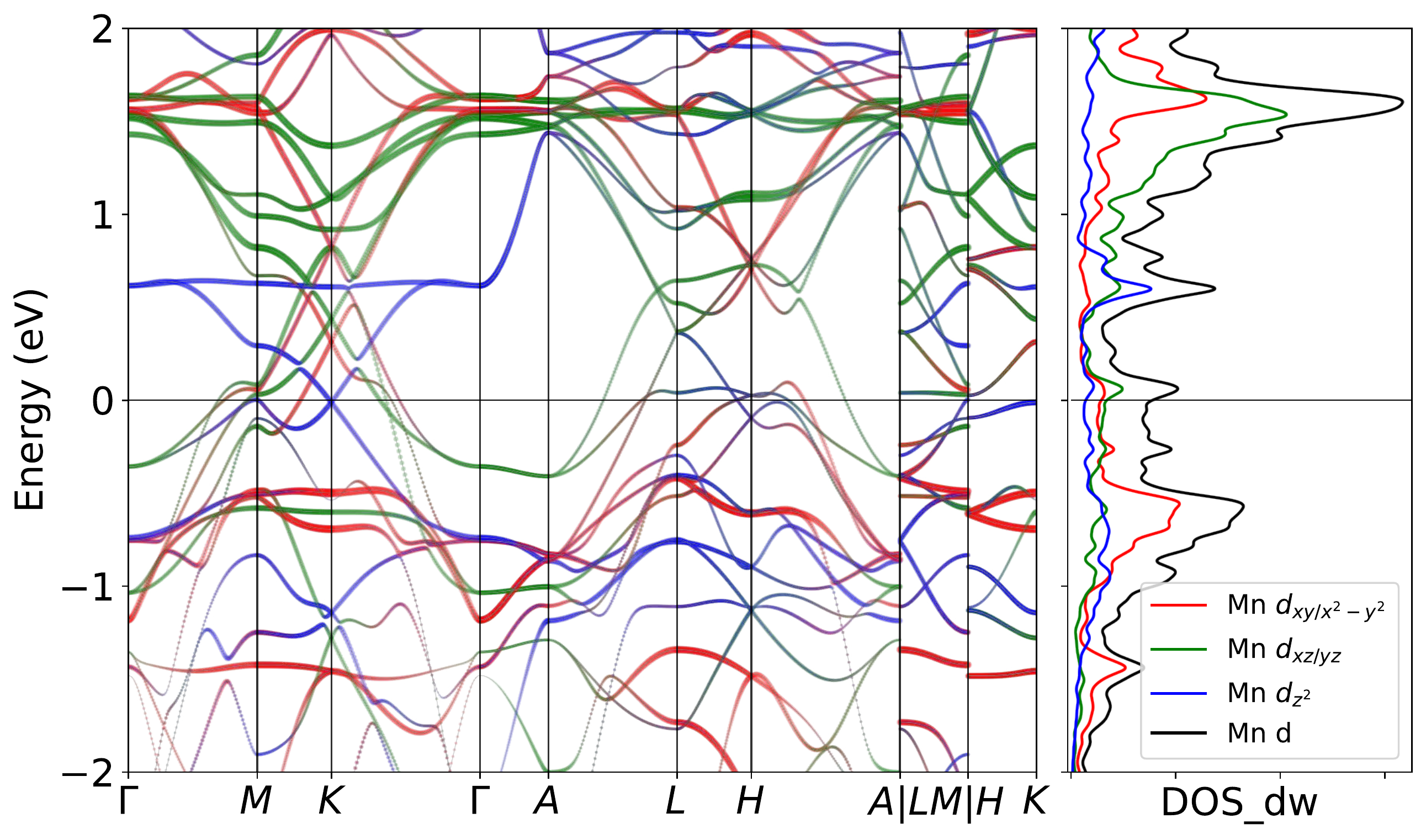}
\includegraphics[width=0.32\textwidth]{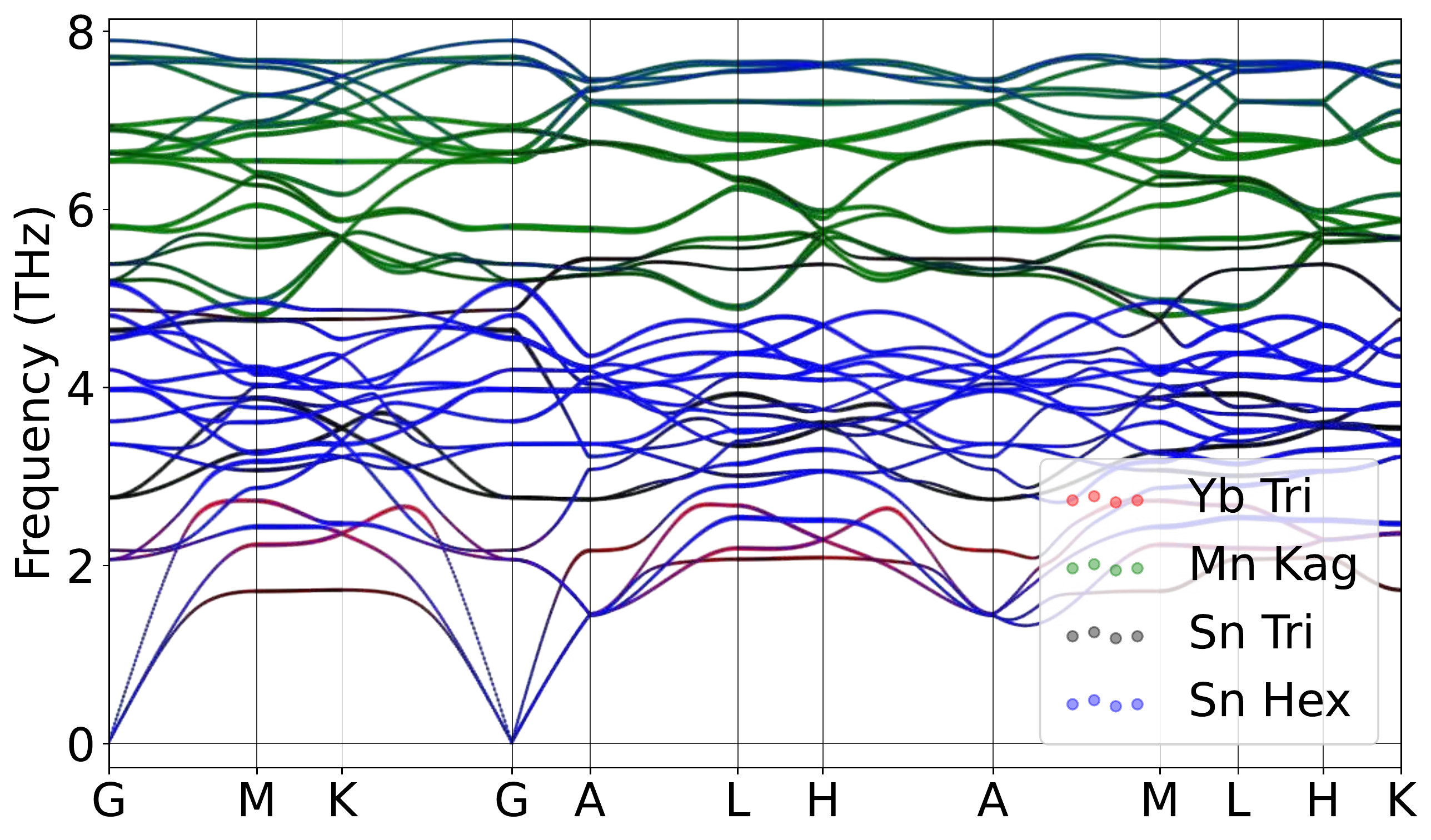}
\caption{Band structure for YbMn$_6$Sn$_6$ with FM order without SOC for spin (Left)up and (Middle)down channels. The projection of Mn $3d$ orbitals is also presented. (Right)Correspondingly, a stable phonon was demonstrated with the collinear magnetic order without Hubbard U at lower temperature regime, weighted by atomic projections.}
\label{fig:YbMn6Sn6mag}
\end{figure}

\begin{figure}[H]
\centering
\label{fig:YbMn6Sn6_relaxed_Low_xz}
\includegraphics[width=0.85\textwidth]{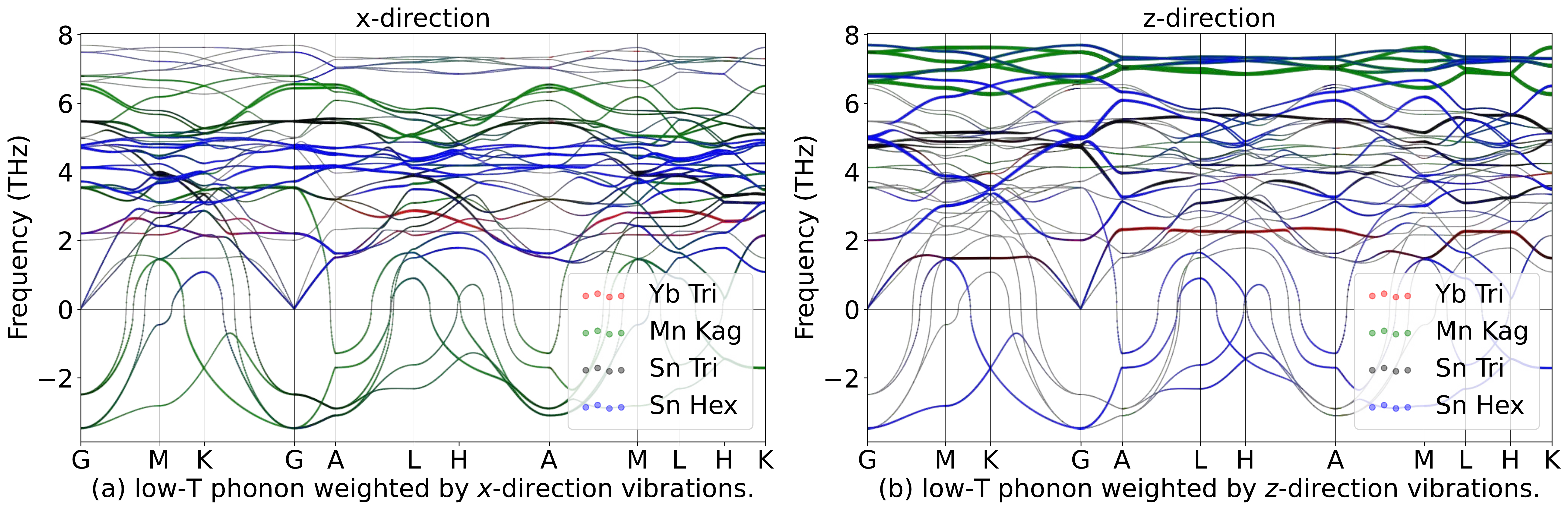}
\hspace{5pt}\label{fig:YbMn6Sn6_relaxed_High_xz}
\includegraphics[width=0.85\textwidth]{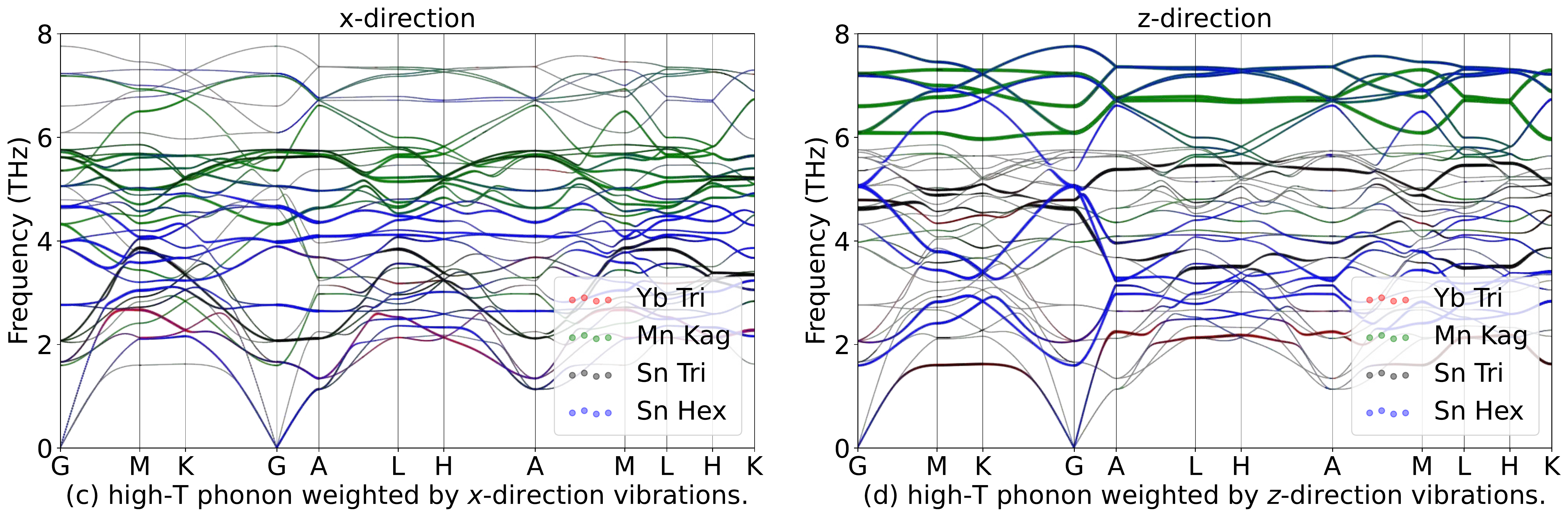}
\caption{Phonon spectrum for YbMn$_6$Sn$_6$in in-plane ($x$) and out-of-plane ($z$) directions in paramagnetic phase.}
\label{fig:YbMn6Sn6phoxyz}
\end{figure}

\begin{table}[htbp]
\global\long\def\arraystretch{1.12}
\captionsetup{width=.95\textwidth}
\caption{IRREPs at high-symmetry points for YbMn$_6$Sn$_6$ phonon spectrum at Low-T.}
\label{tab:191_YbMn6Sn6_Low}
\setlength{\tabcolsep}{1.mm}{
}
\end{table}

\begin{table}[htbp]
\global\long\def\arraystretch{1.12}
\captionsetup{width=.95\textwidth}
\caption{IRREPs at high-symmetry points for YbMn$_6$Sn$_6$ phonon spectrum at High-T.}
\label{tab:191_YbMn6Sn6_High}
\setlength{\tabcolsep}{1.mm}{
%
}
\end{table}
\subsubsection{\label{sms2sec:LuMn6Sn6}\hyperref[smsec:col]{\texorpdfstring{LuMn$_6$Sn$_6$}{}}~{\color{blue}  (High-T stable, Low-T unstable) }}

\begin{table}[htbp]
\global\long\def\arraystretch{1.12}
\captionsetup{width=.85\textwidth}
\caption{Structure parameters of LuMn$_6$Sn$_6$ after relaxation. It contains 521 electrons. }
\label{tab:LuMn6Sn6}
\setlength{\tabcolsep}{4mm}{
%
}
\end{table}

\begin{figure}[htbp]
\centering
\includegraphics[width=0.85\textwidth]{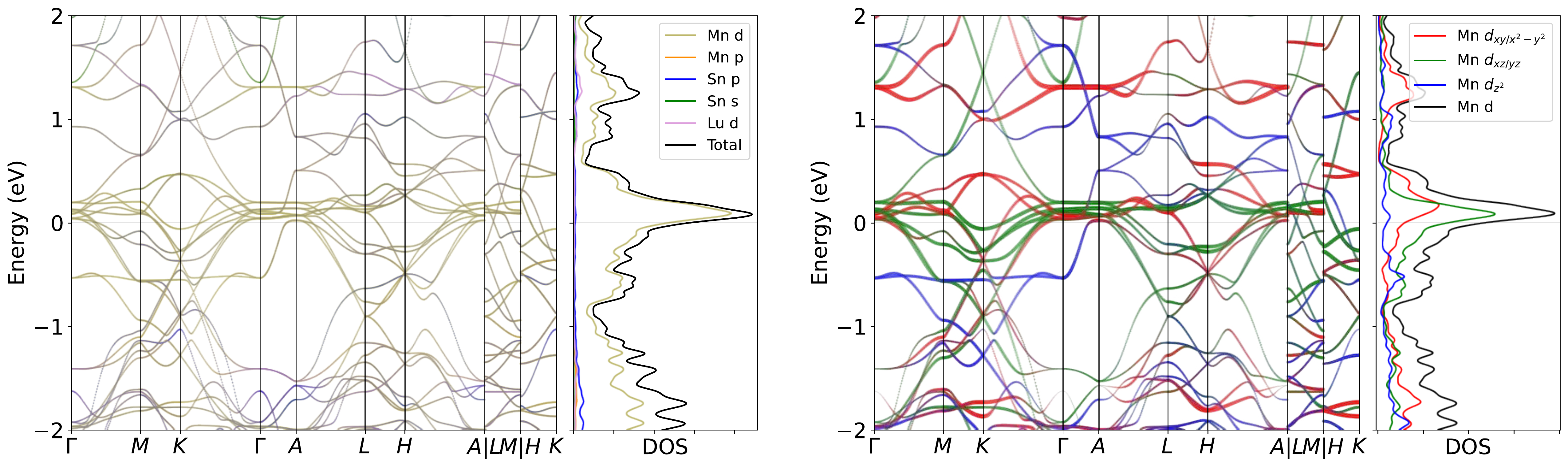}
\caption{Band structure for LuMn$_6$Sn$_6$ without SOC in paramagnetic phase. The projection of Mn $3d$ orbitals is also presented. The band structures clearly reveal the dominating of Mn $3d$ orbitals  near the Fermi level, as well as the pronounced peak.}
\label{fig:LuMn6Sn6band}
\end{figure}

\begin{figure}[H]
\centering
\subfloat[Relaxed phonon at low-T.]{
\label{fig:LuMn6Sn6_relaxed_Low}
\includegraphics[width=0.45\textwidth]{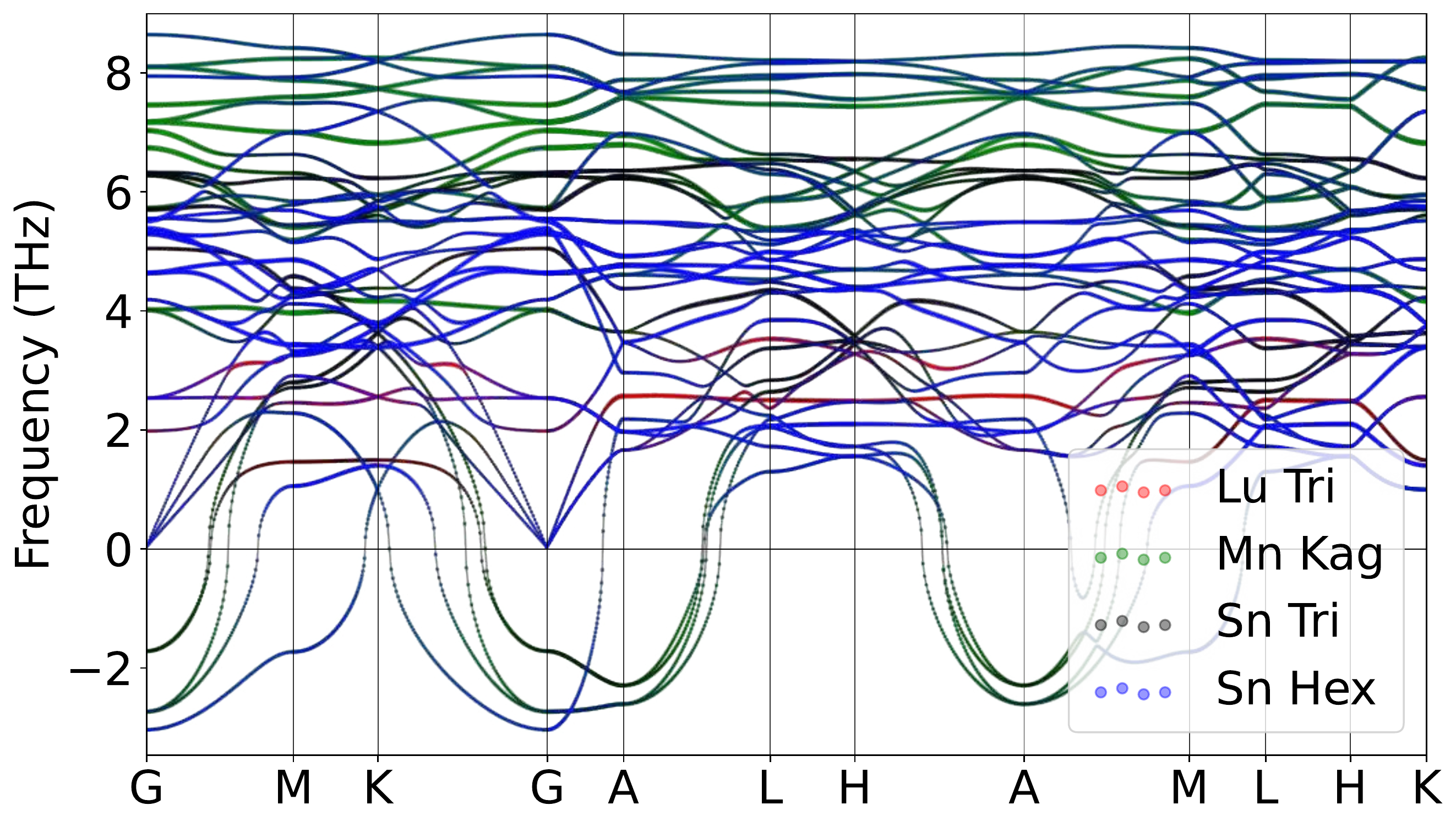}
}
\hspace{5pt}\subfloat[Relaxed phonon at high-T.]{
\label{fig:LuMn6Sn6_relaxed_High}
\includegraphics[width=0.45\textwidth]{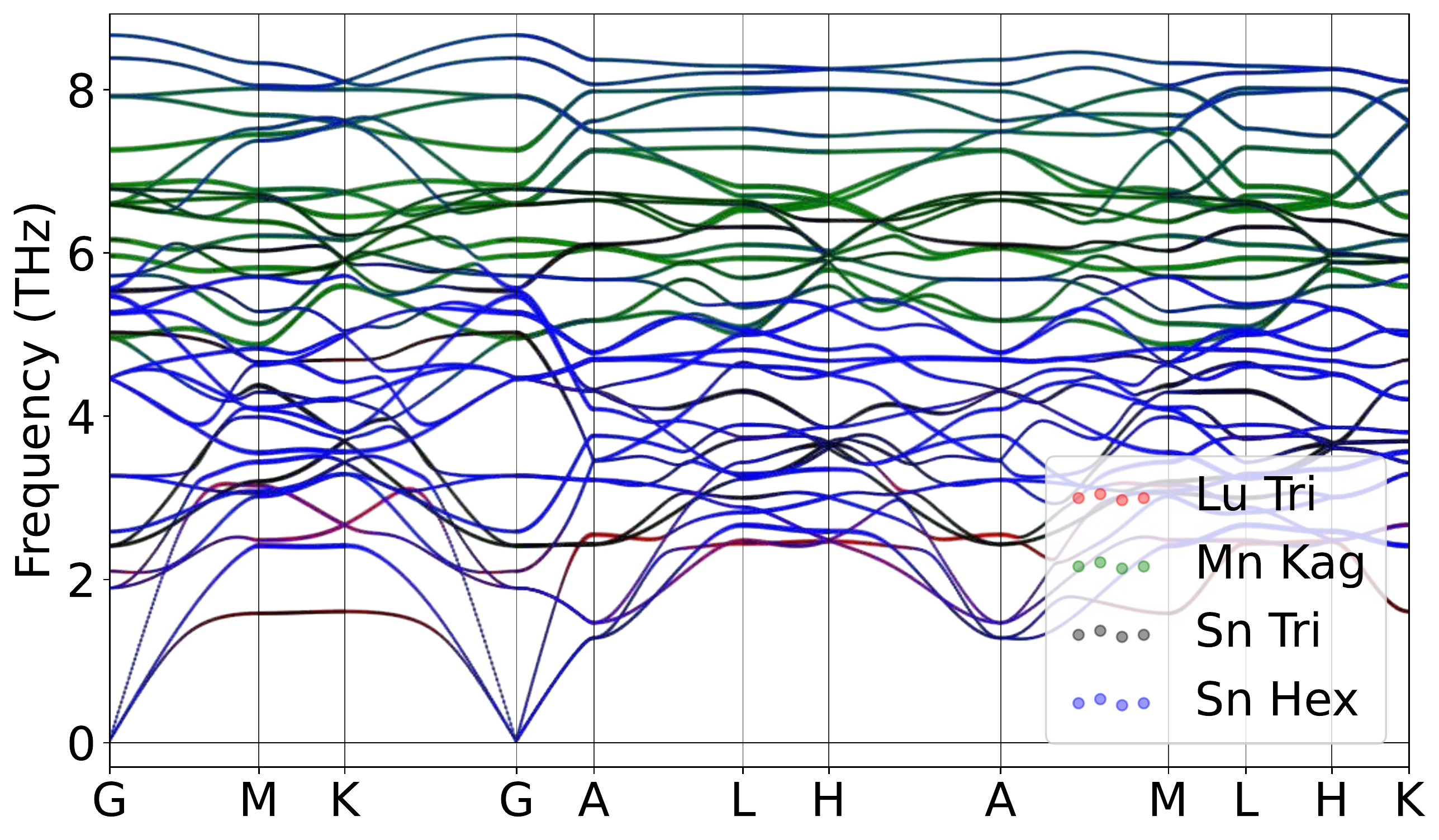}
}
\caption{Atomically projected phonon spectrum for LuMn$_6$Sn$_6$ in paramagnetic phase.}
\label{fig:LuMn6Sn6pho}
\end{figure}

\begin{figure}[htbp]
\centering
\includegraphics[width=0.45\textwidth]{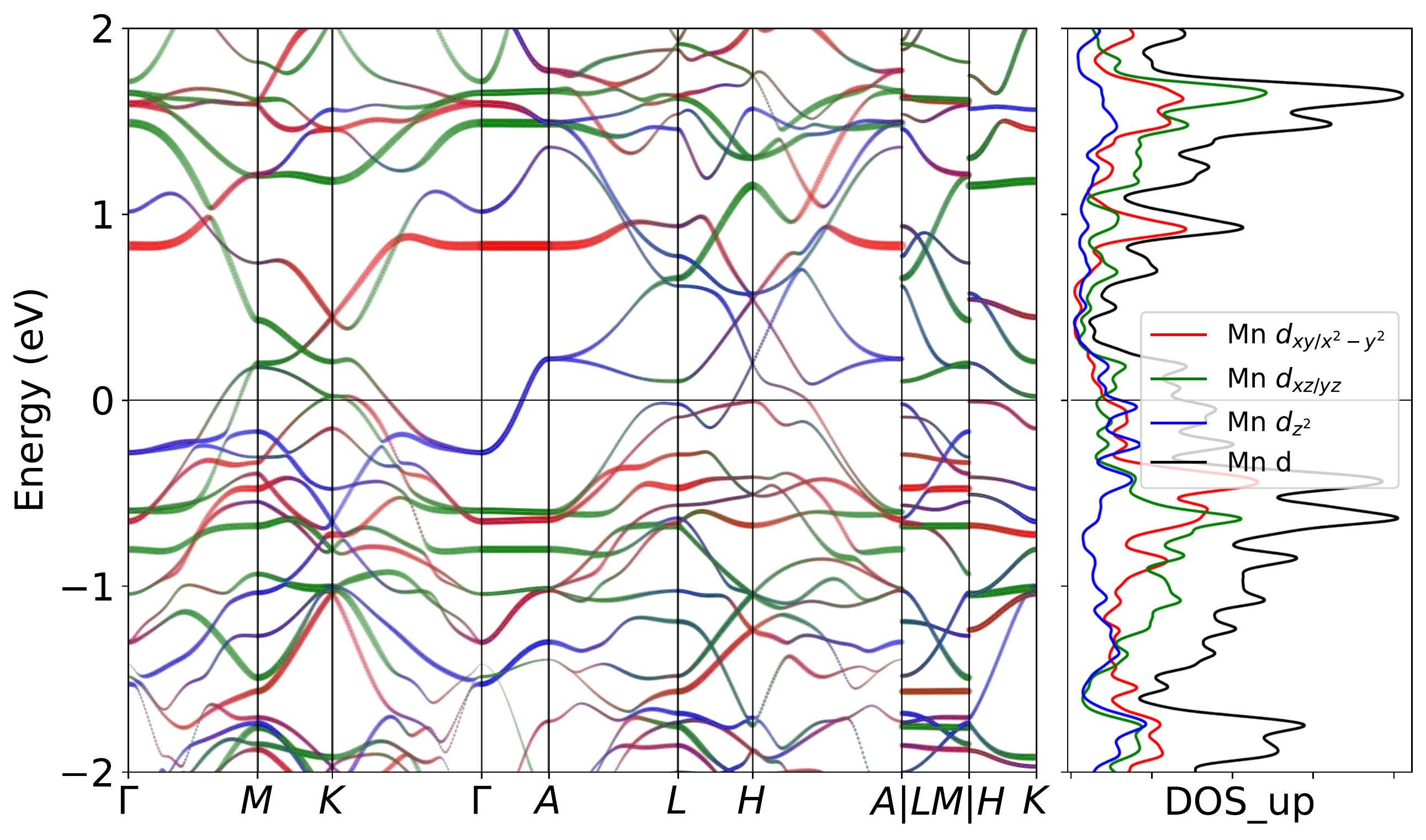}
\includegraphics[width=0.45\textwidth]{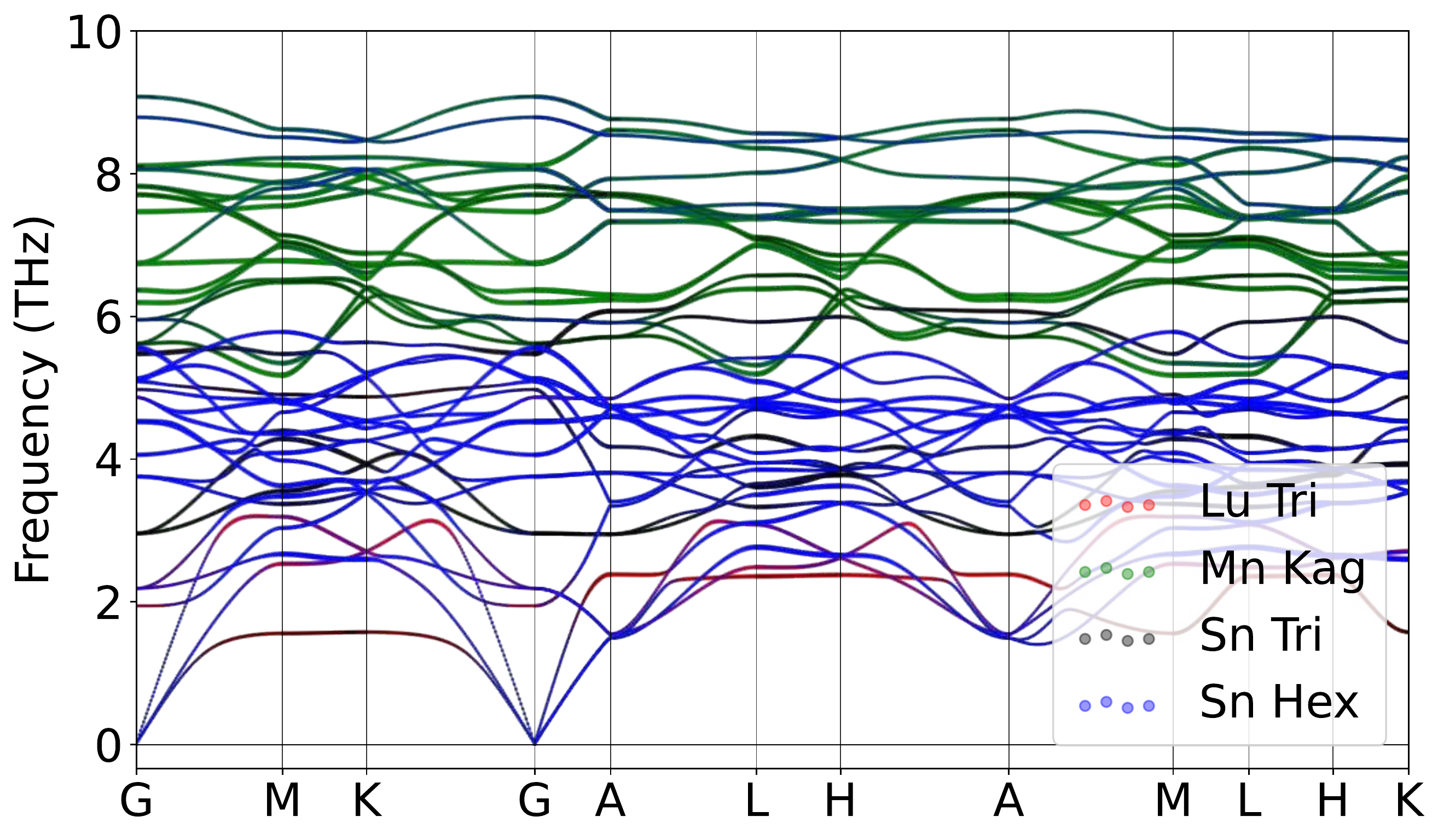}
\caption{Band structure for LuMn$_6$Sn$_6$ with AFM order without SOC for spin (Left)up channel. The projection of Mn $3d$ orbitals is also presented. (Right)Correspondingly, a stable phonon was demonstrated with the collinear magnetic order without Hubbard U at lower temperature regime, weighted by atomic projections.}
\label{fig:LuMn6Sn6mag}
\end{figure}

\begin{figure}[H]
\centering
\label{fig:LuMn6Sn6_relaxed_Low_xz}
\includegraphics[width=0.85\textwidth]{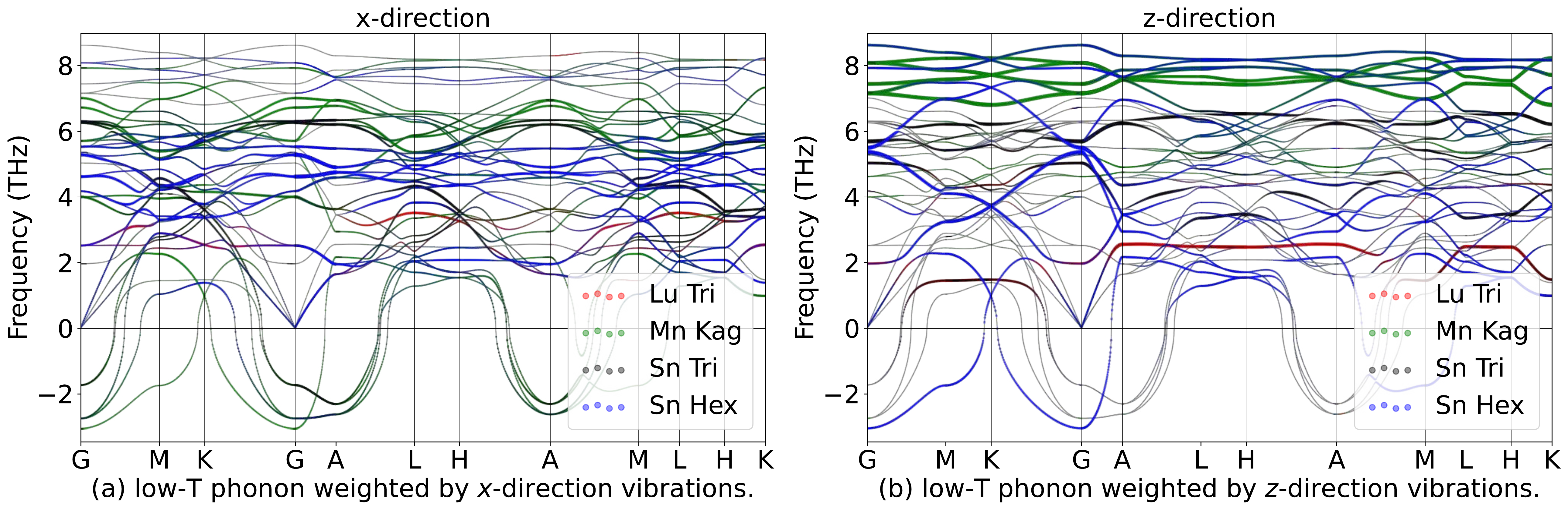}
\hspace{5pt}\label{fig:LuMn6Sn6_relaxed_High_xz}
\includegraphics[width=0.85\textwidth]{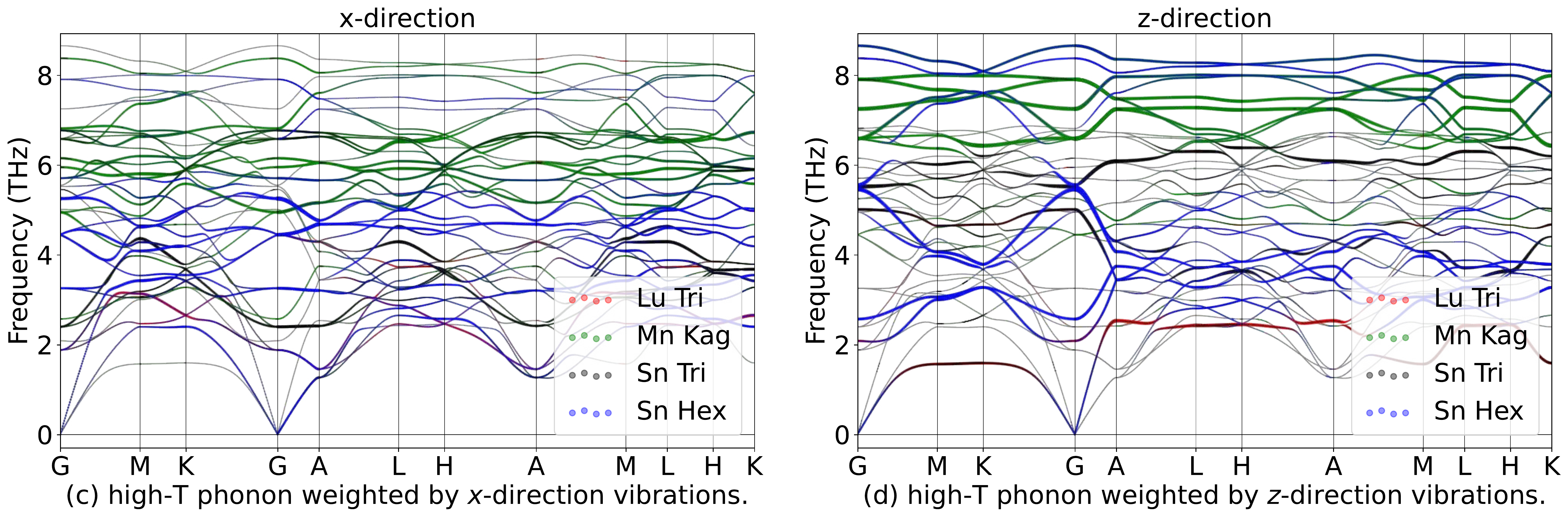}
\caption{Phonon spectrum for LuMn$_6$Sn$_6$in in-plane ($x$) and out-of-plane ($z$) directions in paramagnetic phase.}
\label{fig:LuMn6Sn6phoxyz}
\end{figure}

\begin{table}[htbp]
\global\long\def\arraystretch{1.12}
\captionsetup{width=.95\textwidth}
\caption{IRREPs at high-symmetry points for LuMn$_6$Sn$_6$ phonon spectrum at Low-T.}
\label{tab:191_LuMn6Sn6_Low}
\setlength{\tabcolsep}{1.mm}{
}
\end{table}

\begin{table}[htbp]
\global\long\def\arraystretch{1.12}
\captionsetup{width=.95\textwidth}
\caption{IRREPs at high-symmetry points for LuMn$_6$Sn$_6$ phonon spectrum at High-T.}
\label{tab:191_LuMn6Sn6_High}
\setlength{\tabcolsep}{1.mm}{
%
}
\end{table}
\subsubsection{\label{sms2sec:HfMn6Sn6}\hyperref[smsec:col]{\texorpdfstring{HfMn$_6$Sn$_6$}{}}~{\color{blue}  (High-T stable, Low-T unstable) }}

\begin{table}[htbp]
\global\long\def\arraystretch{1.12}
\captionsetup{width=.85\textwidth}
\caption{Structure parameters of HfMn$_6$Sn$_6$ after relaxation. It contains 522 electrons. }
\label{tab:HfMn6Sn6}
\setlength{\tabcolsep}{4mm}{
%
}
\end{table}

\begin{figure}[htbp]
\centering
\includegraphics[width=0.85\textwidth]{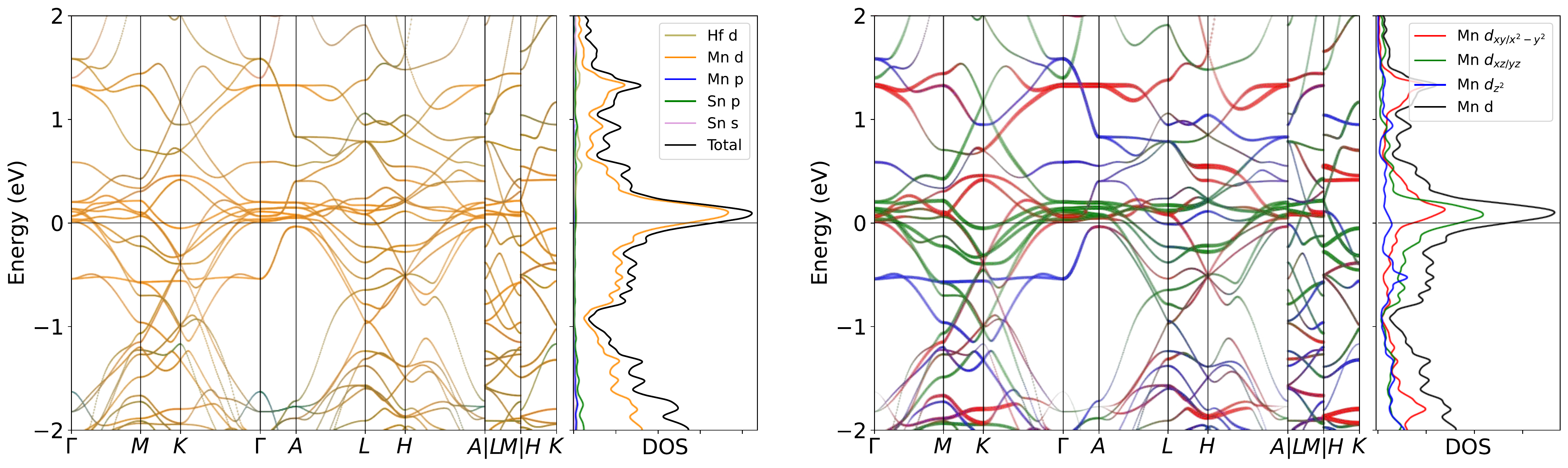}
\caption{Band structure for HfMn$_6$Sn$_6$ without SOC in paramagnetic phase. The projection of Mn $3d$ orbitals is also presented. The band structures clearly reveal the dominating of Mn $3d$ orbitals  near the Fermi level, as well as the pronounced peak.}
\label{fig:HfMn6Sn6band}
\end{figure}

\begin{figure}[H]
\centering
\subfloat[Relaxed phonon at low-T.]{
\label{fig:HfMn6Sn6_relaxed_Low}
\includegraphics[width=0.45\textwidth]{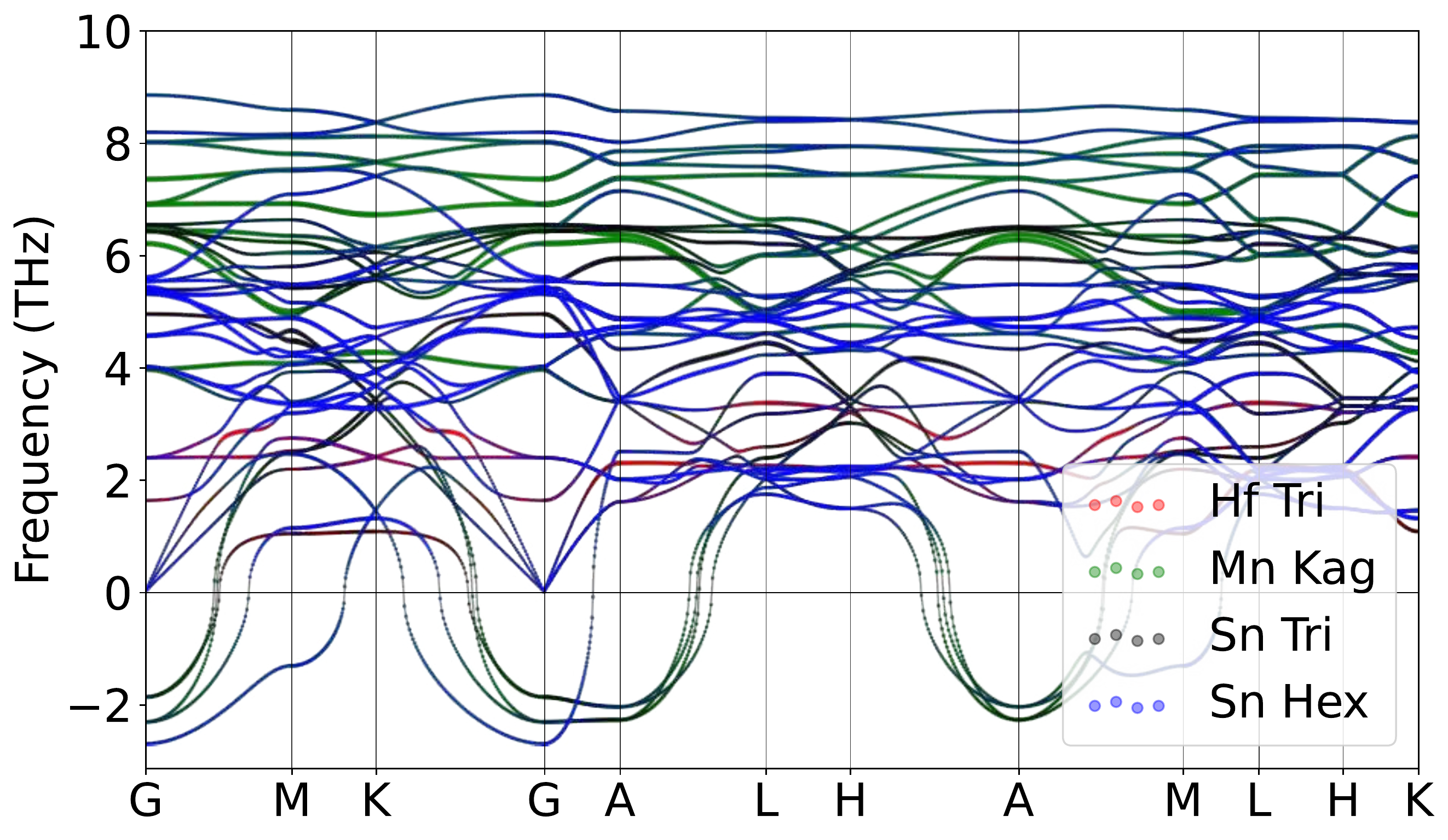}
}
\hspace{5pt}\subfloat[Relaxed phonon at high-T.]{
\label{fig:HfMn6Sn6_relaxed_High}
\includegraphics[width=0.45\textwidth]{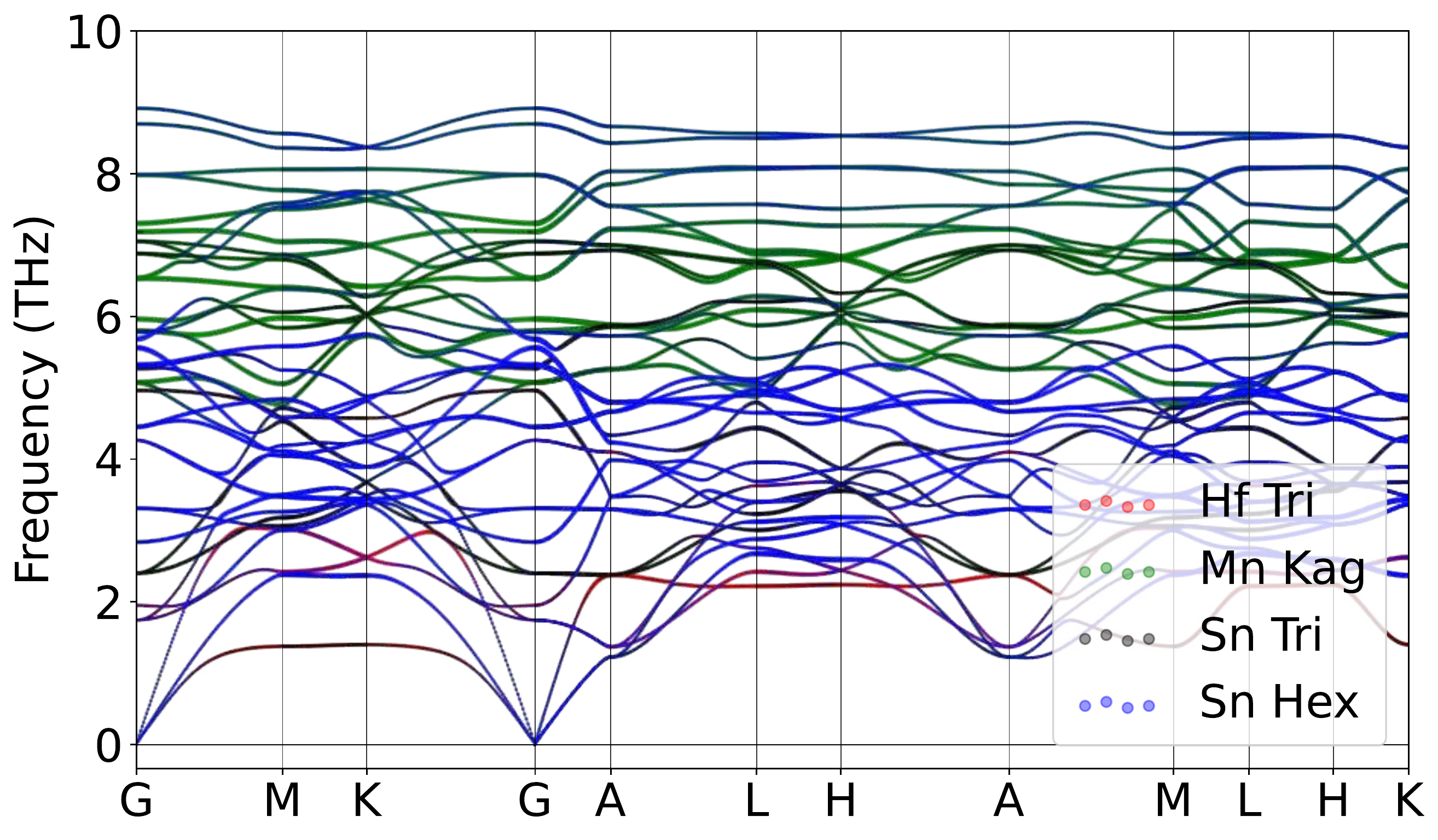}
}
\caption{Atomically projected phonon spectrum for HfMn$_6$Sn$_6$ in paramagnetic phase.}
\label{fig:HfMn6Sn6pho}
\end{figure}

\begin{figure}[htbp]
\centering
\includegraphics[width=0.45\textwidth]{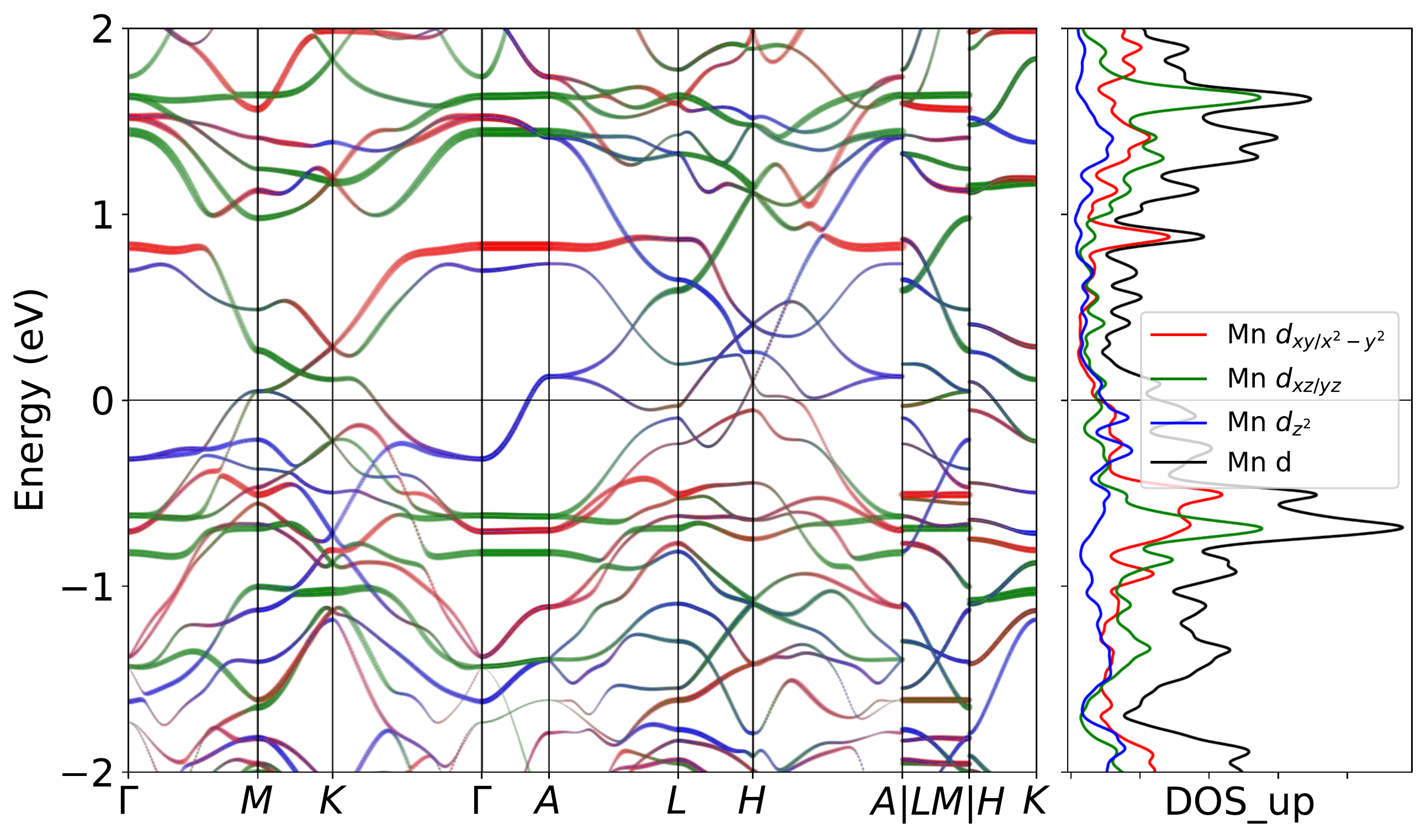}
\includegraphics[width=0.45\textwidth]{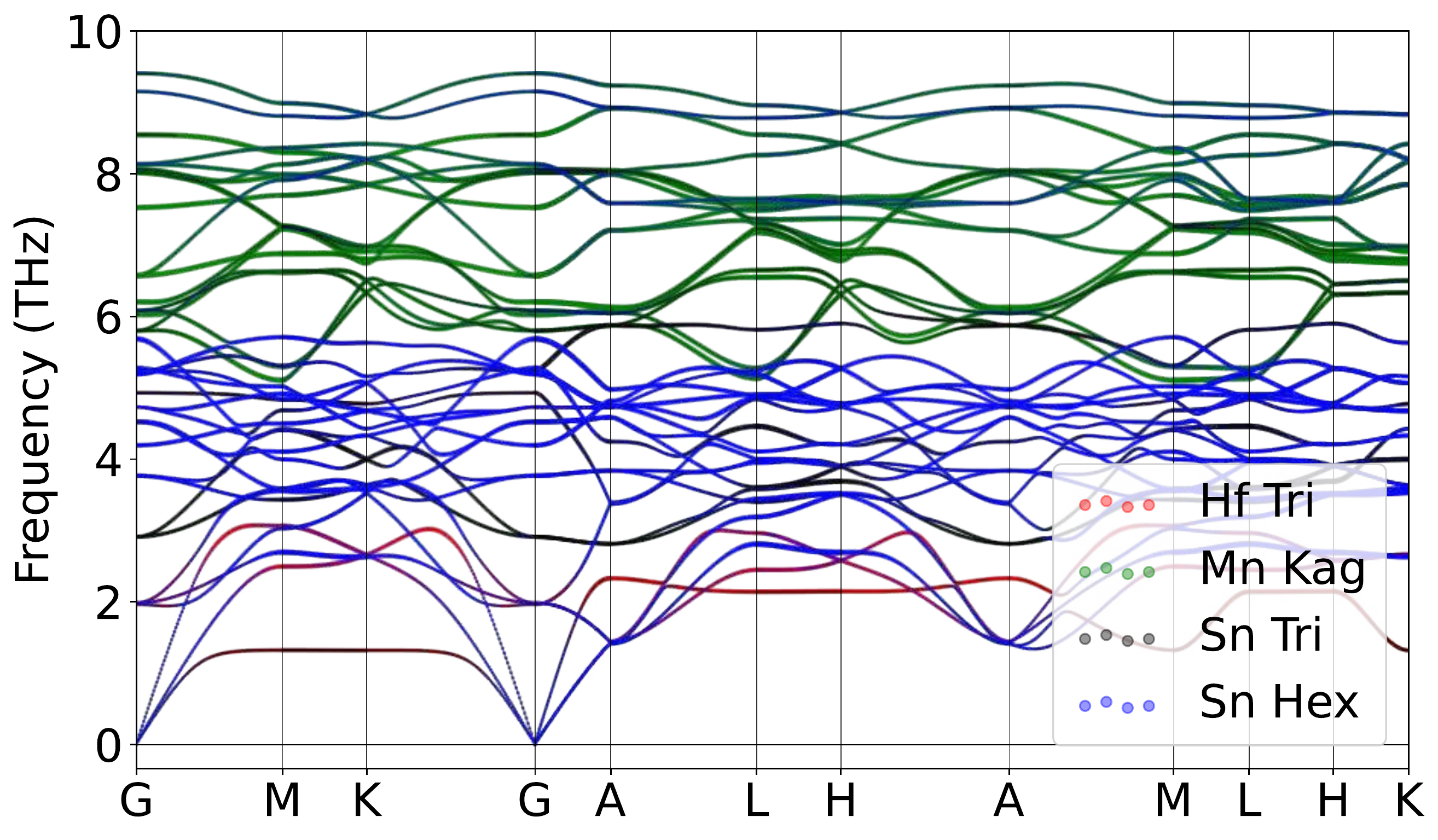}
\caption{Band structure for HfMn$_6$Sn$_6$ with AFM order without SOC for spin (Left)up channel. The projection of Mn $3d$ orbitals is also presented. (Right)Correspondingly, a stable phonon was demonstrated with the collinear magnetic order without Hubbard U at lower temperature regime, weighted by atomic projections.}
\label{fig:HfMn6Sn6mag}
\end{figure}

\begin{figure}[H]
\centering
\label{fig:HfMn6Sn6_relaxed_Low_xz}
\includegraphics[width=0.85\textwidth]{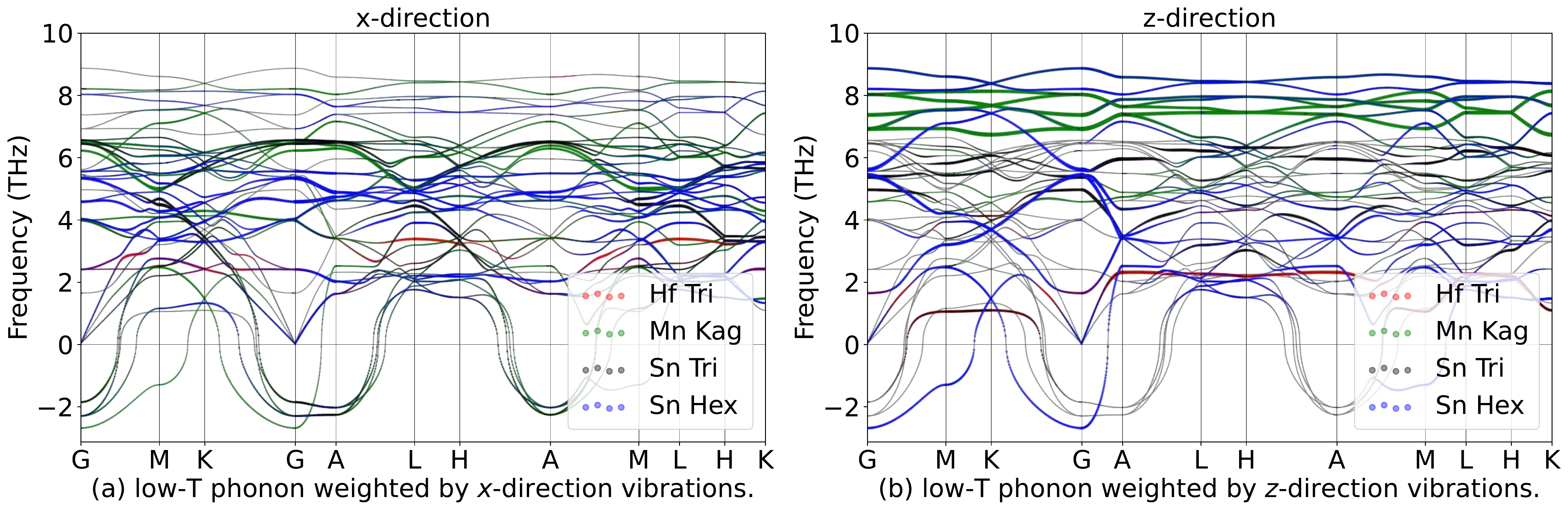}
\hspace{5pt}\label{fig:HfMn6Sn6_relaxed_High_xz}
\includegraphics[width=0.85\textwidth]{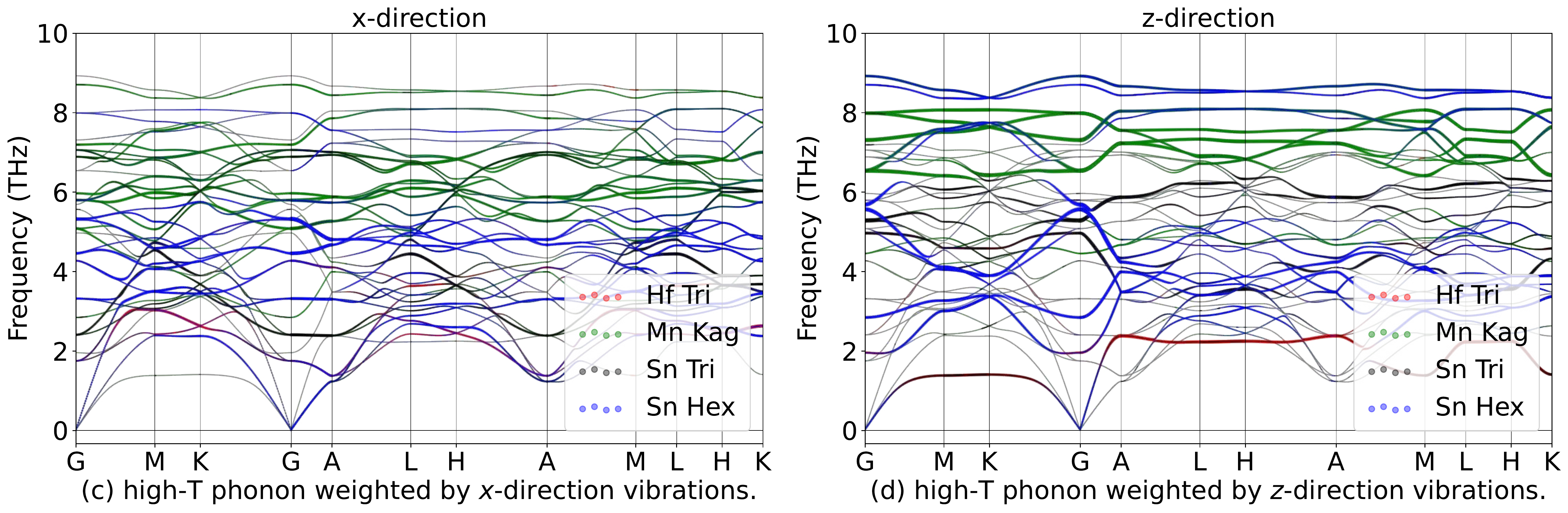}
\caption{Phonon spectrum for HfMn$_6$Sn$_6$in in-plane ($x$) and out-of-plane ($z$) directions in paramagnetic phase.}
\label{fig:HfMn6Sn6phoxyz}
\end{figure}

\begin{table}[htbp]
\global\long\def\arraystretch{1.12}
\captionsetup{width=.95\textwidth}
\caption{IRREPs at high-symmetry points for HfMn$_6$Sn$_6$ phonon spectrum at Low-T.}
\label{tab:191_HfMn6Sn6_Low}
\setlength{\tabcolsep}{1.mm}{
}
\end{table}

\begin{table}[htbp]
\global\long\def\arraystretch{1.12}
\captionsetup{width=.95\textwidth}
\caption{IRREPs at high-symmetry points for HfMn$_6$Sn$_6$ phonon spectrum at High-T.}
\label{tab:191_HfMn6Sn6_High}
\setlength{\tabcolsep}{1.mm}{
%
}
\end{table}
\subsubsection{\label{sms2sec:TaMn6Sn6}\hyperref[smsec:col]{\texorpdfstring{TaMn$_6$Sn$_6$}{}}~{\color{blue}  (High-T stable, Low-T unstable) }}

\begin{table}[htbp]
\global\long\def\arraystretch{1.12}
\captionsetup{width=.85\textwidth}
\caption{Structure parameters of TaMn$_6$Sn$_6$ after relaxation. It contains 523 electrons. }
\label{tab:TaMn6Sn6}
\setlength{\tabcolsep}{4mm}{
%
}
\end{table}

\begin{figure}[htbp]
\centering
\includegraphics[width=0.85\textwidth]{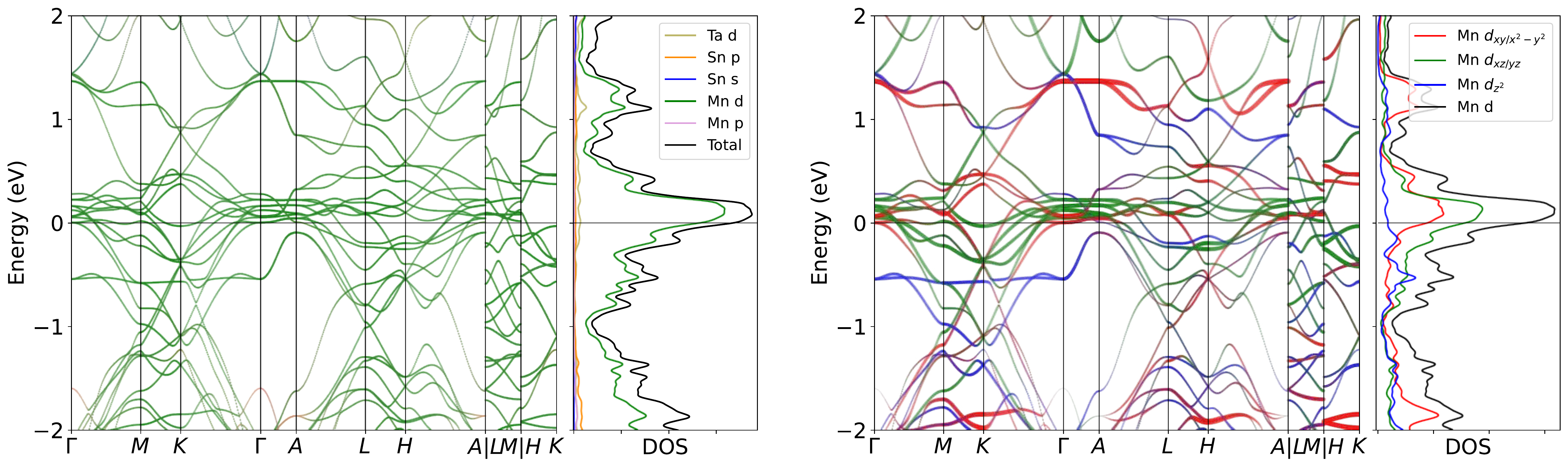}
\caption{Band structure for TaMn$_6$Sn$_6$ without SOC in paramagnetic phase. The projection of Mn $3d$ orbitals is also presented. The band structures clearly reveal the dominating of Mn $3d$ orbitals  near the Fermi level, as well as the pronounced peak.}
\label{fig:TaMn6Sn6band}
\end{figure}

\begin{figure}[H]
\centering
\subfloat[Relaxed phonon at low-T.]{
\label{fig:TaMn6Sn6_relaxed_Low}
\includegraphics[width=0.45\textwidth]{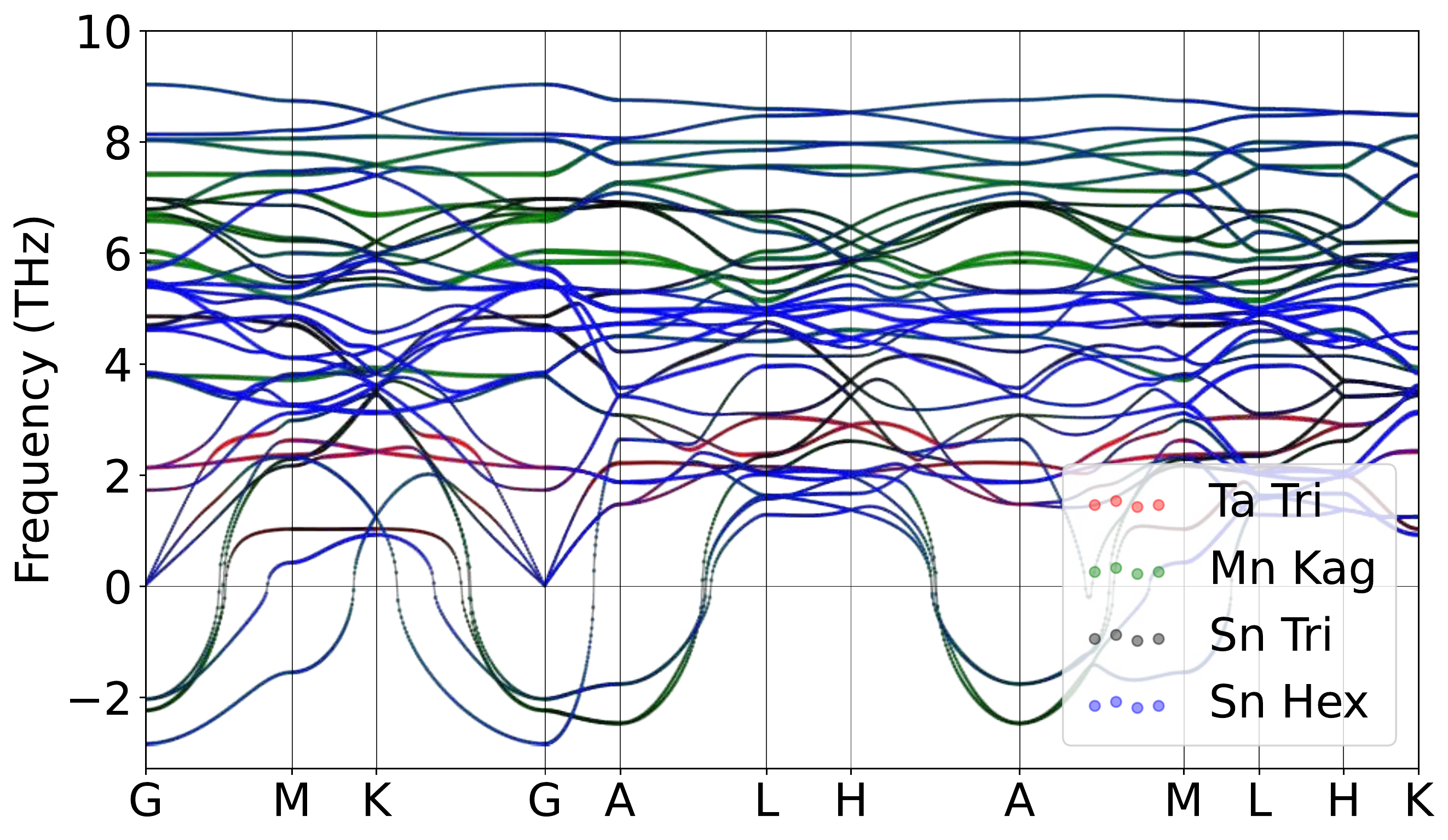}
}
\hspace{5pt}\subfloat[Relaxed phonon at high-T.]{
\label{fig:TaMn6Sn6_relaxed_High}
\includegraphics[width=0.45\textwidth]{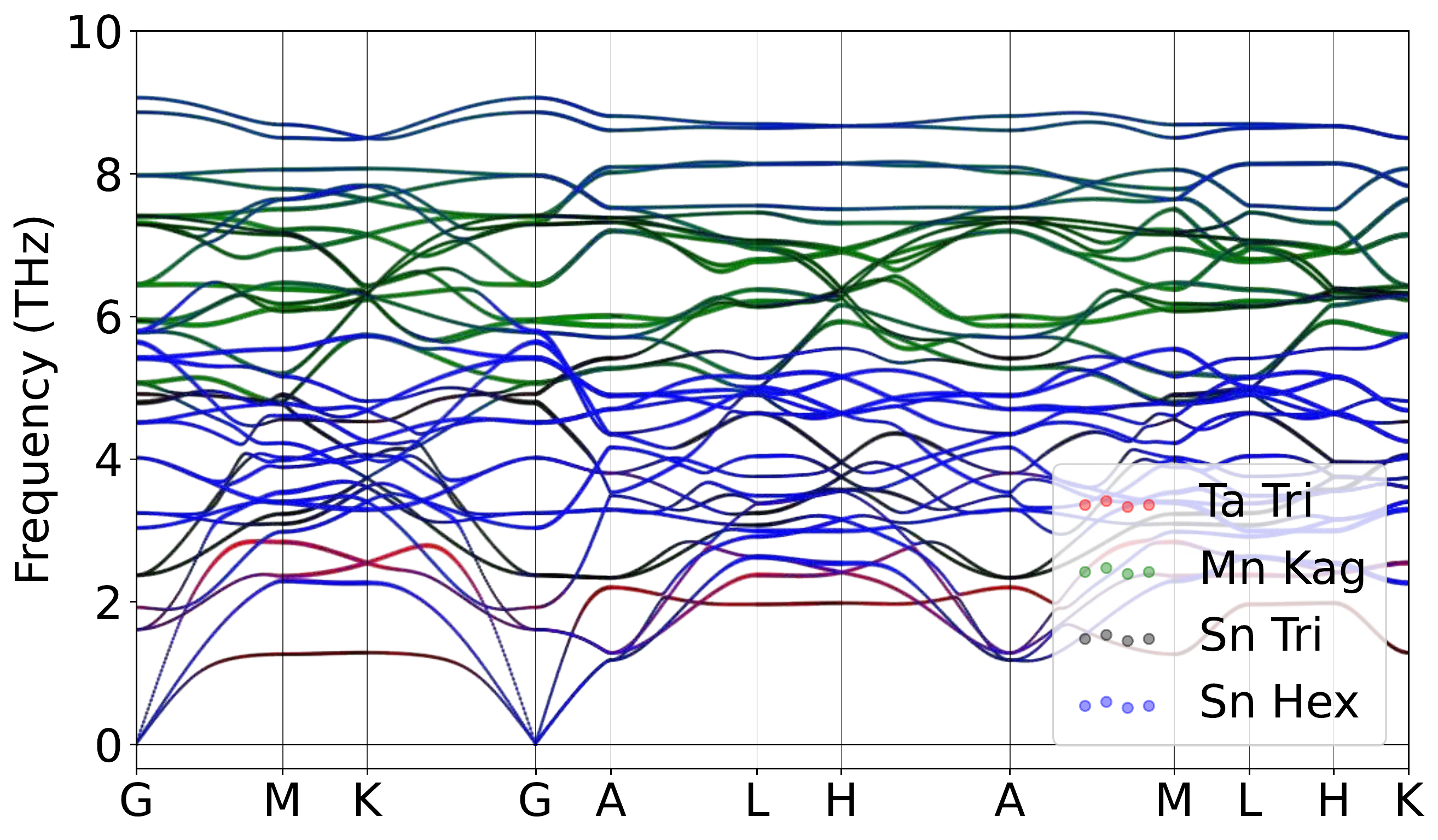}
}
\caption{Atomically projected phonon spectrum for TaMn$_6$Sn$_6$ in paramagnetic phase.}
\label{fig:TaMn6Sn6pho}
\end{figure}

\begin{figure}[htbp]
\centering
\includegraphics[width=0.45\textwidth]{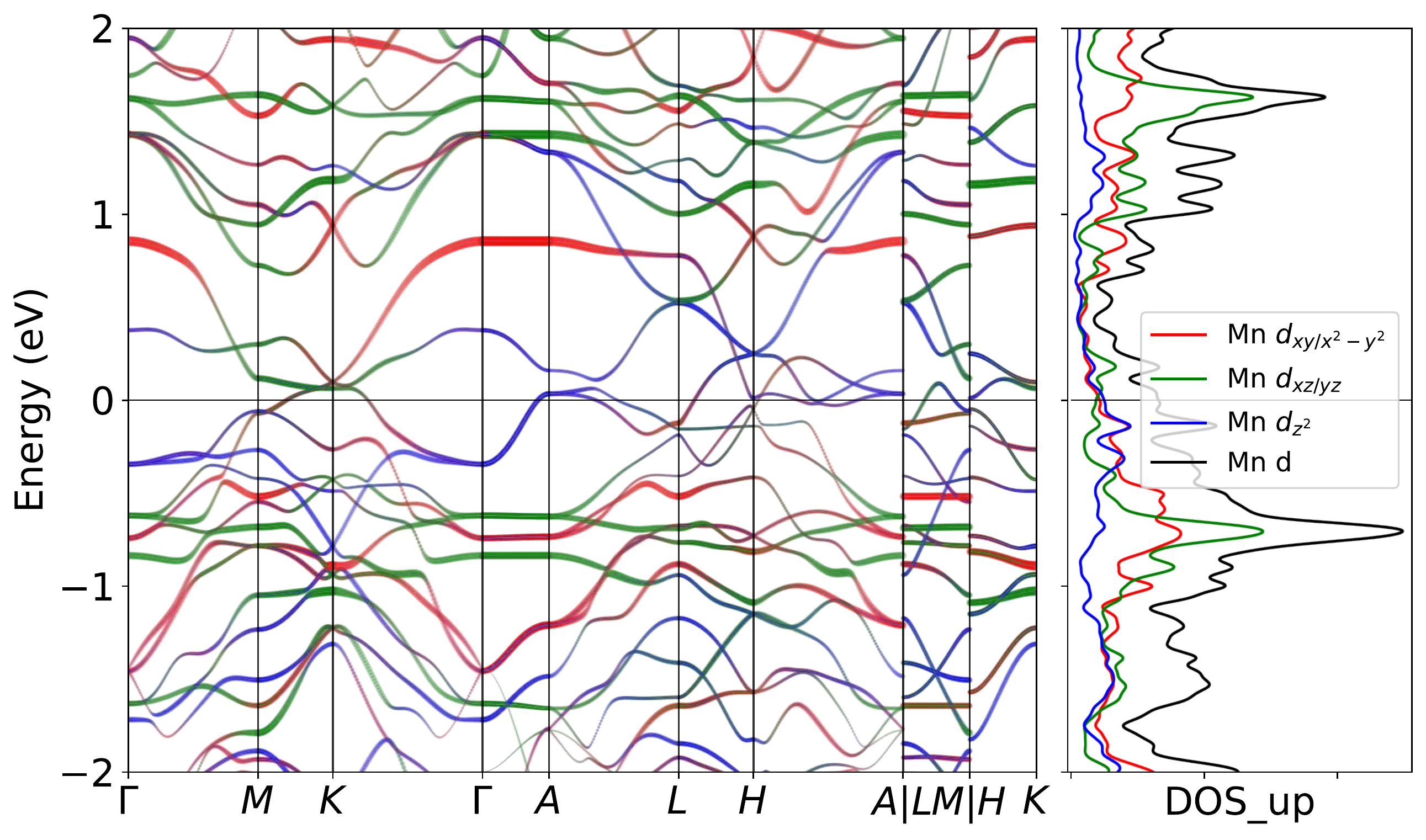}
\caption{Band structure for TaMn$_6$Sn$_6$ with AFM order without SOC for spin (Left)up channel. The projection of Mn $3d$ orbitals is also presented.}
\label{fig:TaMn6Sn6mag}
\end{figure}

\begin{figure}[H]
\centering
\label{fig:TaMn6Sn6_relaxed_Low_xz}
\includegraphics[width=0.85\textwidth]{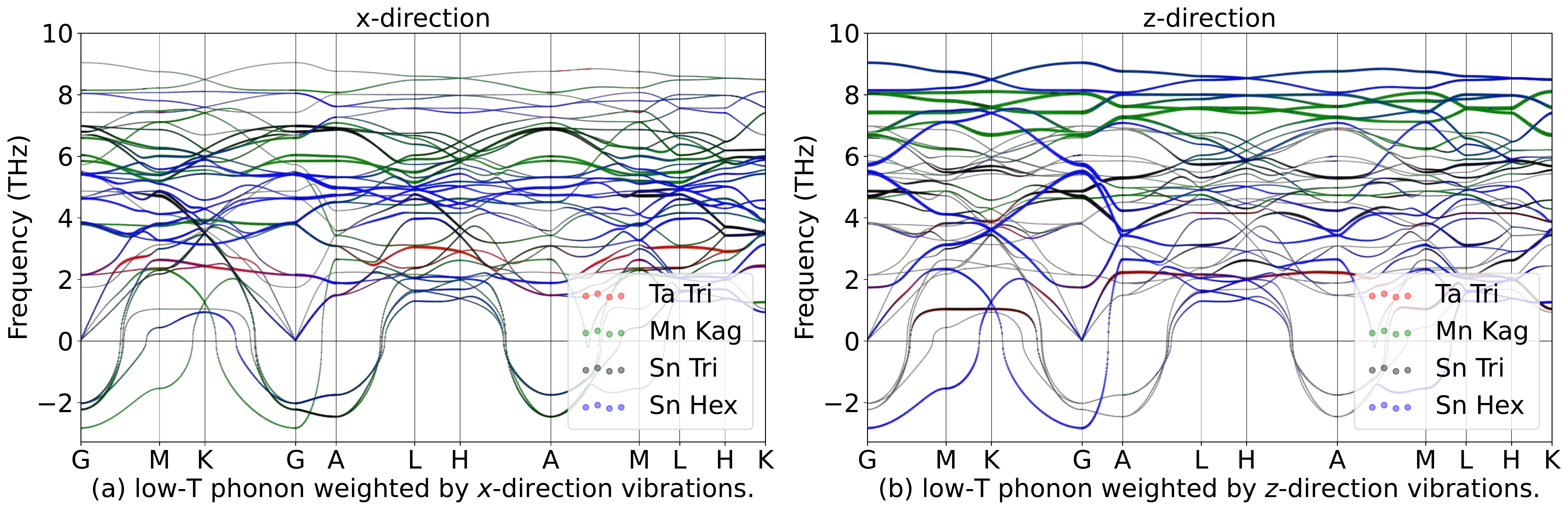}
\hspace{5pt}\label{fig:TaMn6Sn6_relaxed_High_xz}
\includegraphics[width=0.85\textwidth]{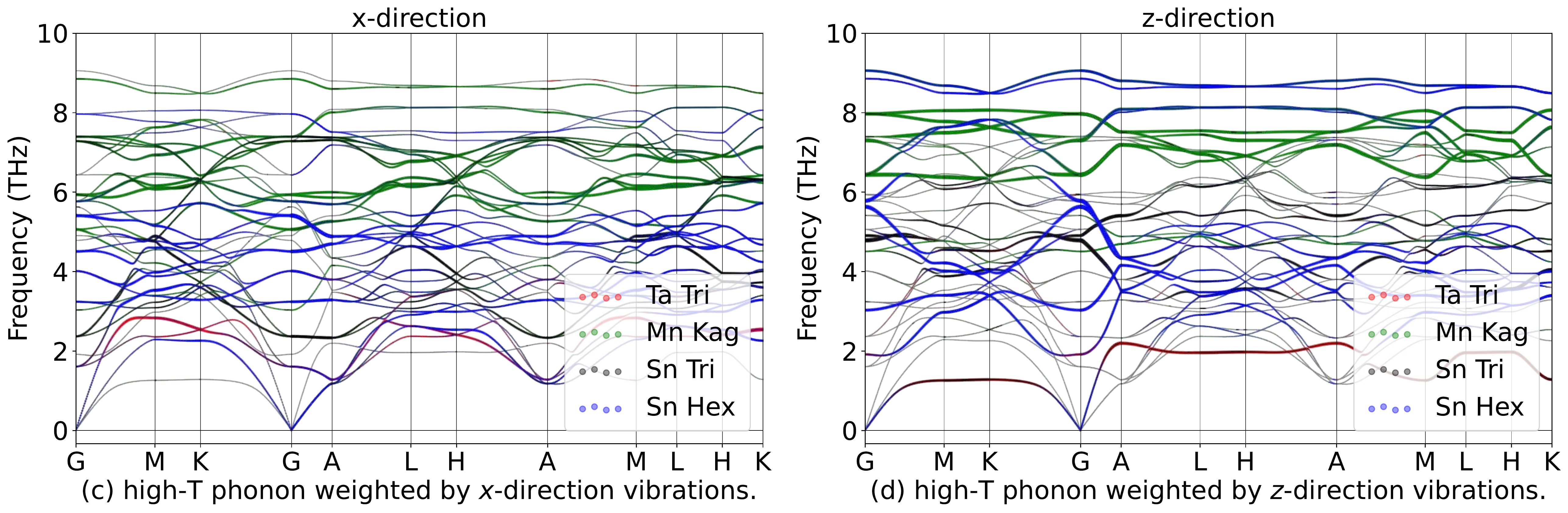}
\caption{Phonon spectrum for TaMn$_6$Sn$_6$in in-plane ($x$) and out-of-plane ($z$) directions in paramagnetic phase.}
\label{fig:TaMn6Sn6phoxyz}
\end{figure}

\begin{table}[htbp]
\global\long\def\arraystretch{1.12}
\captionsetup{width=.95\textwidth}
\caption{IRREPs at high-symmetry points for TaMn$_6$Sn$_6$ phonon spectrum at Low-T.}
\label{tab:191_TaMn6Sn6_Low}
\setlength{\tabcolsep}{1.mm}{
}
\end{table}

\begin{table}[htbp]
\global\long\def\arraystretch{1.12}
\captionsetup{width=.95\textwidth}
\caption{IRREPs at high-symmetry points for TaMn$_6$Sn$_6$ phonon spectrum at High-T.}
\label{tab:191_TaMn6Sn6_High}
\setlength{\tabcolsep}{1.mm}{
%
}
\end{table}
\subsection{\hyperref[smsec:col]{FeGe}}~\label{smssec:FeGe}

\subsubsection{\label{sms2sec:LiFe6Ge6}\hyperref[smsec:col]{\texorpdfstring{LiFe$_6$Ge$_6$}{}}~{\color{blue}  (High-T stable, Low-T unstable) }}

\begin{table}[htbp]
\global\long\def\arraystretch{1.12}
\captionsetup{width=.85\textwidth}
\caption{Structure parameters of LiFe$_6$Ge$_6$ after relaxation. It contains 351 electrons. }
\label{tab:LiFe6Ge6}
\setlength{\tabcolsep}{4mm}{
%
}
\end{table}

\begin{figure}[htbp]
\centering
\includegraphics[width=0.85\textwidth]{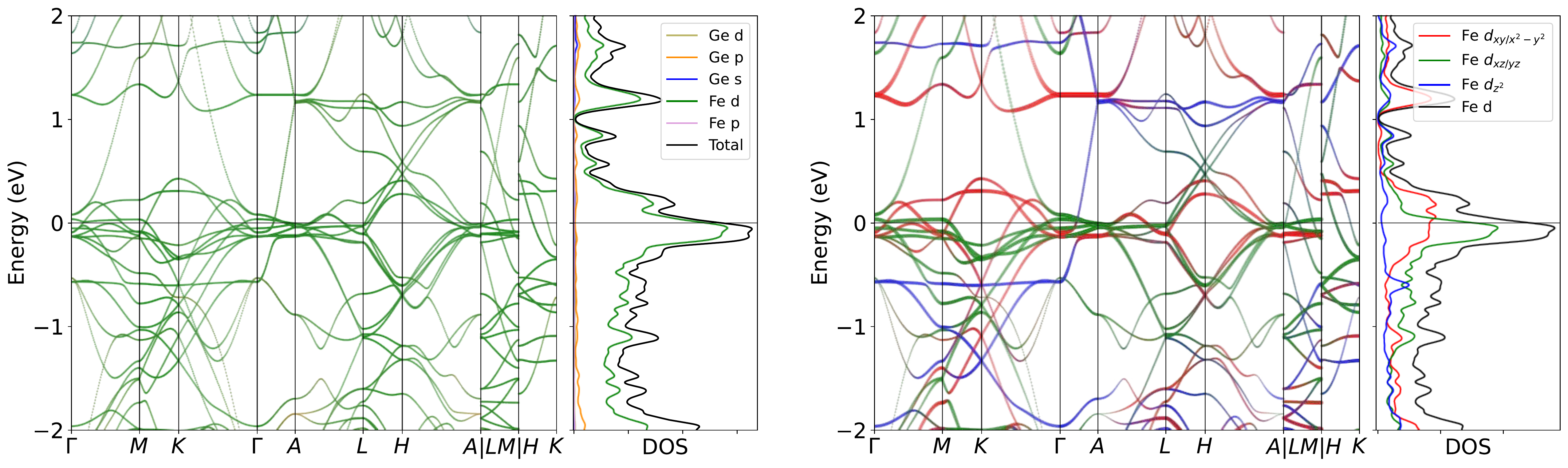}
\caption{Band structure for LiFe$_6$Ge$_6$ without SOC in paramagnetic phase. The projection of Fe $3d$ orbitals is also presented. The band structures clearly reveal the dominating of Fe $3d$ orbitals  near the Fermi level, as well as the pronounced peak.}
\label{fig:LiFe6Ge6band}
\end{figure}

\begin{figure}[H]
\centering
\subfloat[Relaxed phonon at low-T.]{
\label{fig:LiFe6Ge6_relaxed_Low}
\includegraphics[width=0.45\textwidth]{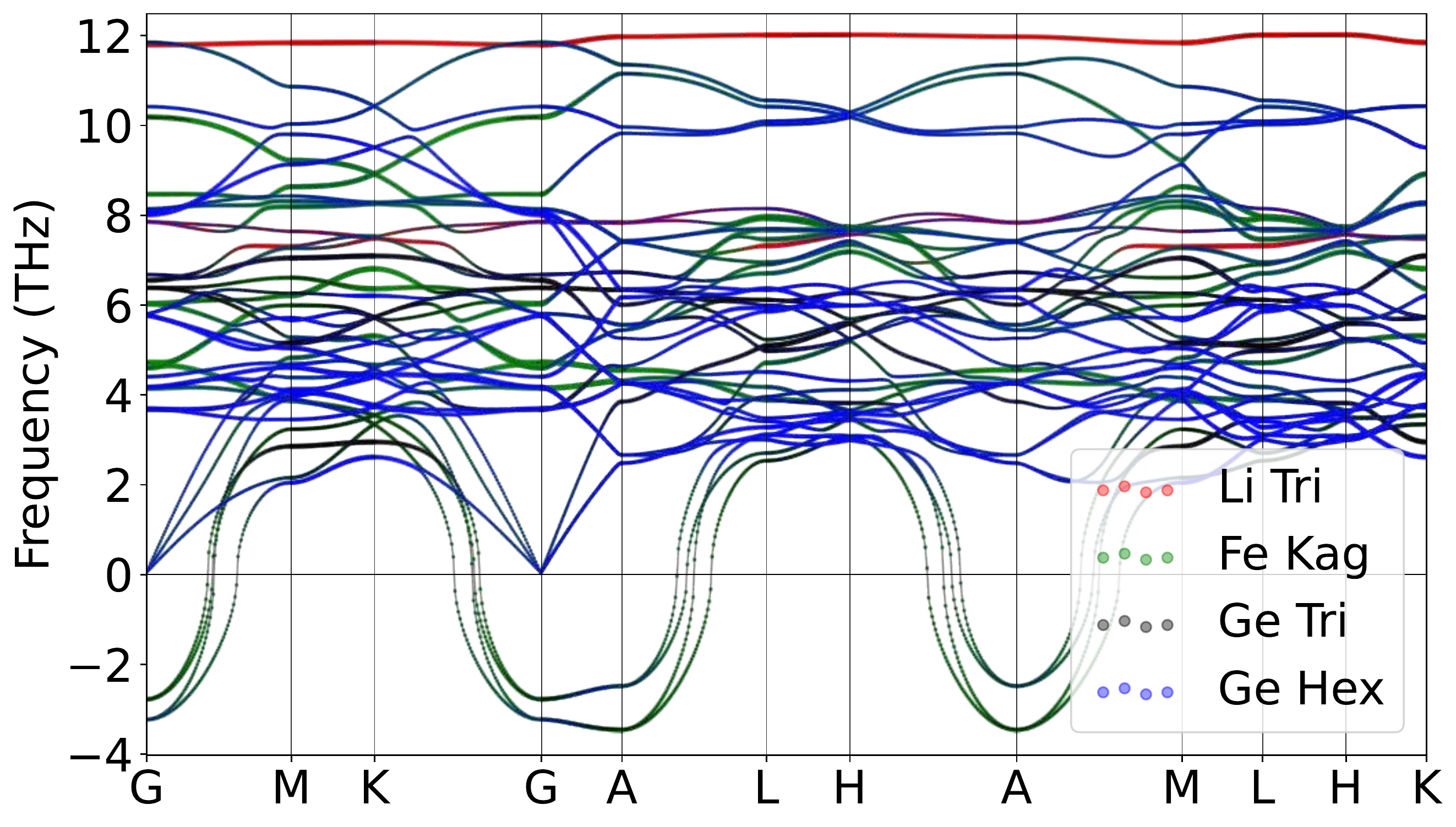}
}
\hspace{5pt}\subfloat[Relaxed phonon at high-T.]{
\label{fig:LiFe6Ge6_relaxed_High}
\includegraphics[width=0.45\textwidth]{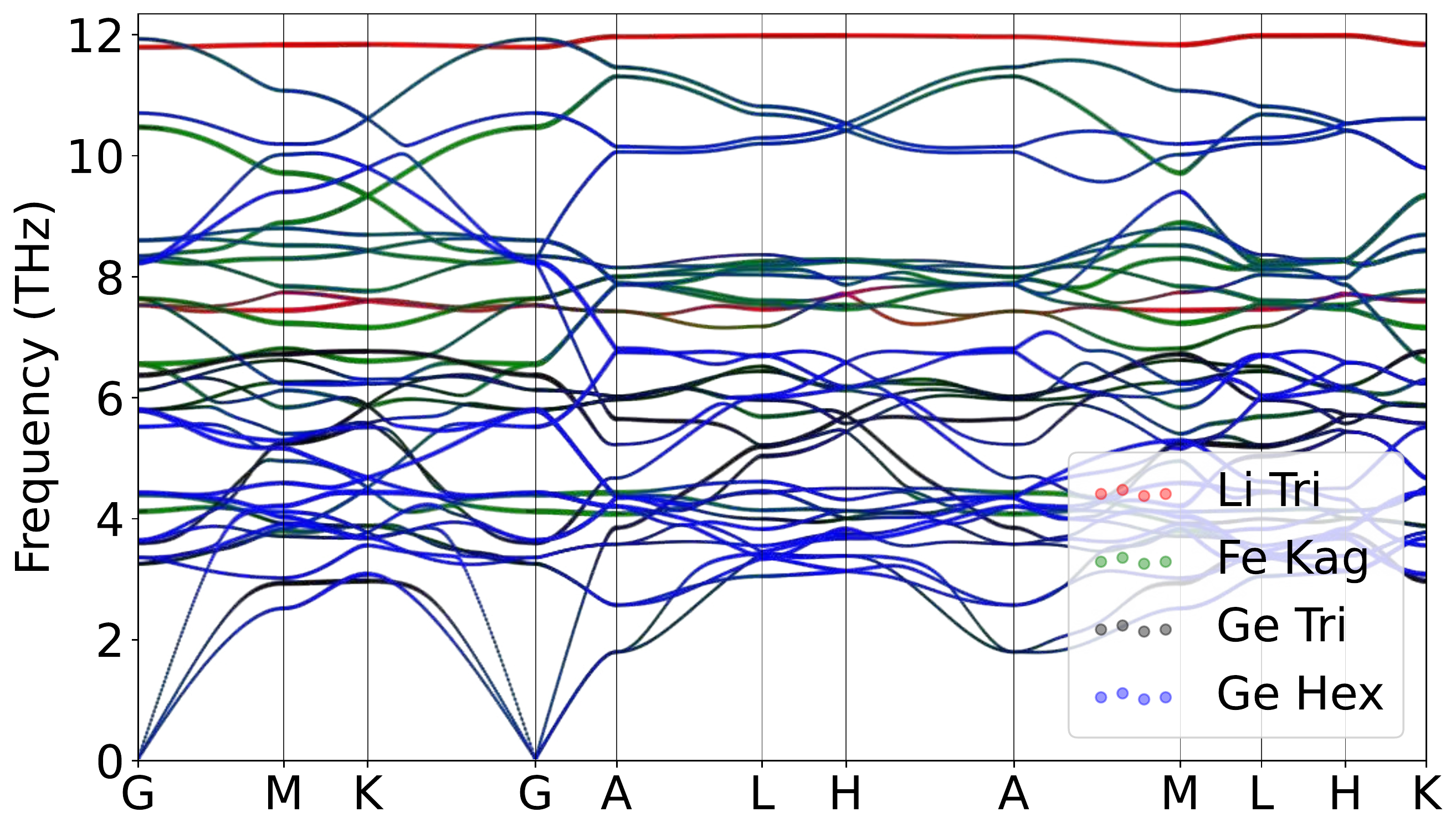}
}
\caption{Atomically projected phonon spectrum for LiFe$_6$Ge$_6$ in paramagnetic phase.}
\label{fig:LiFe6Ge6pho}
\end{figure}

\begin{figure}[htbp]
\centering
\includegraphics[width=0.45\textwidth]{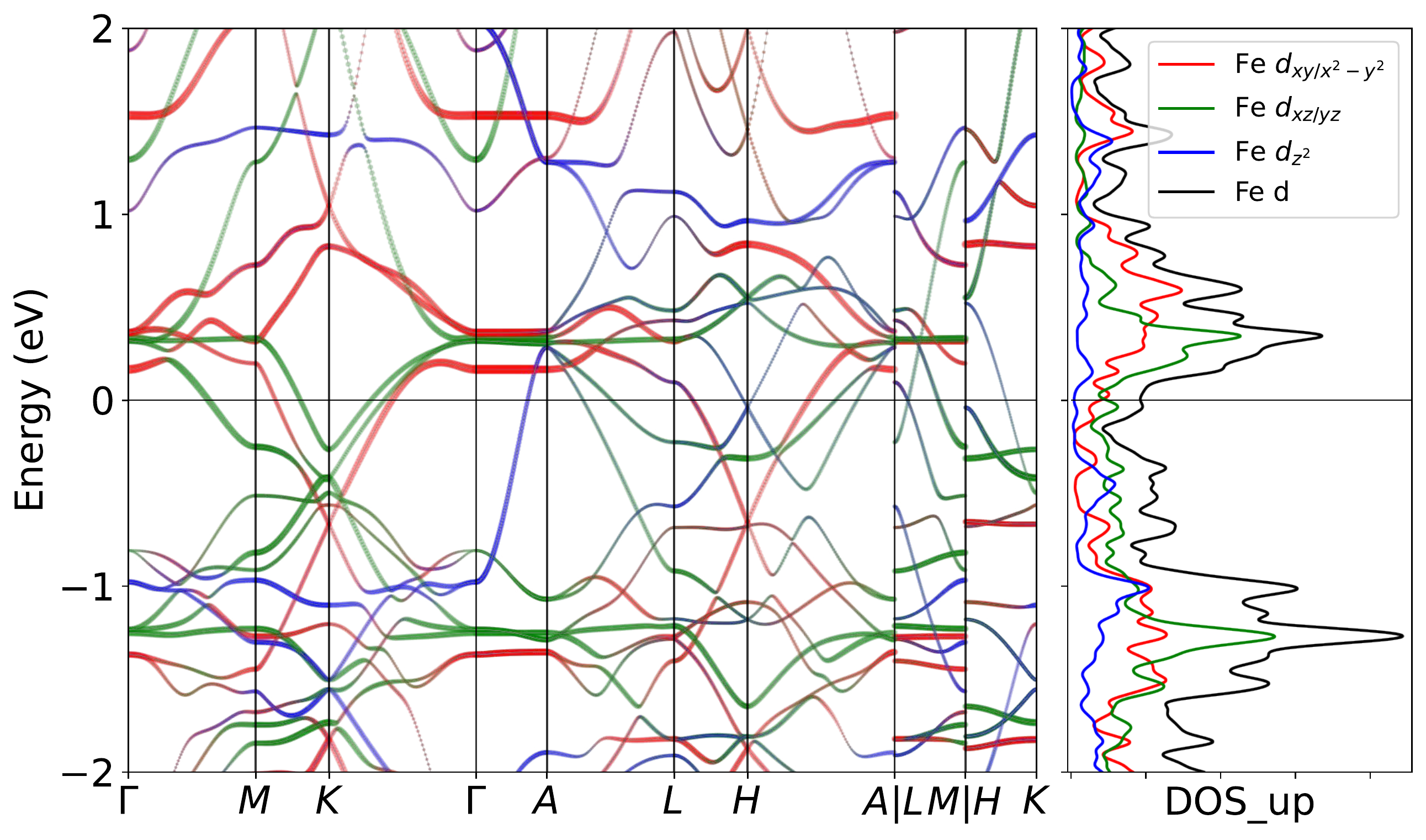}
\includegraphics[width=0.45\textwidth]{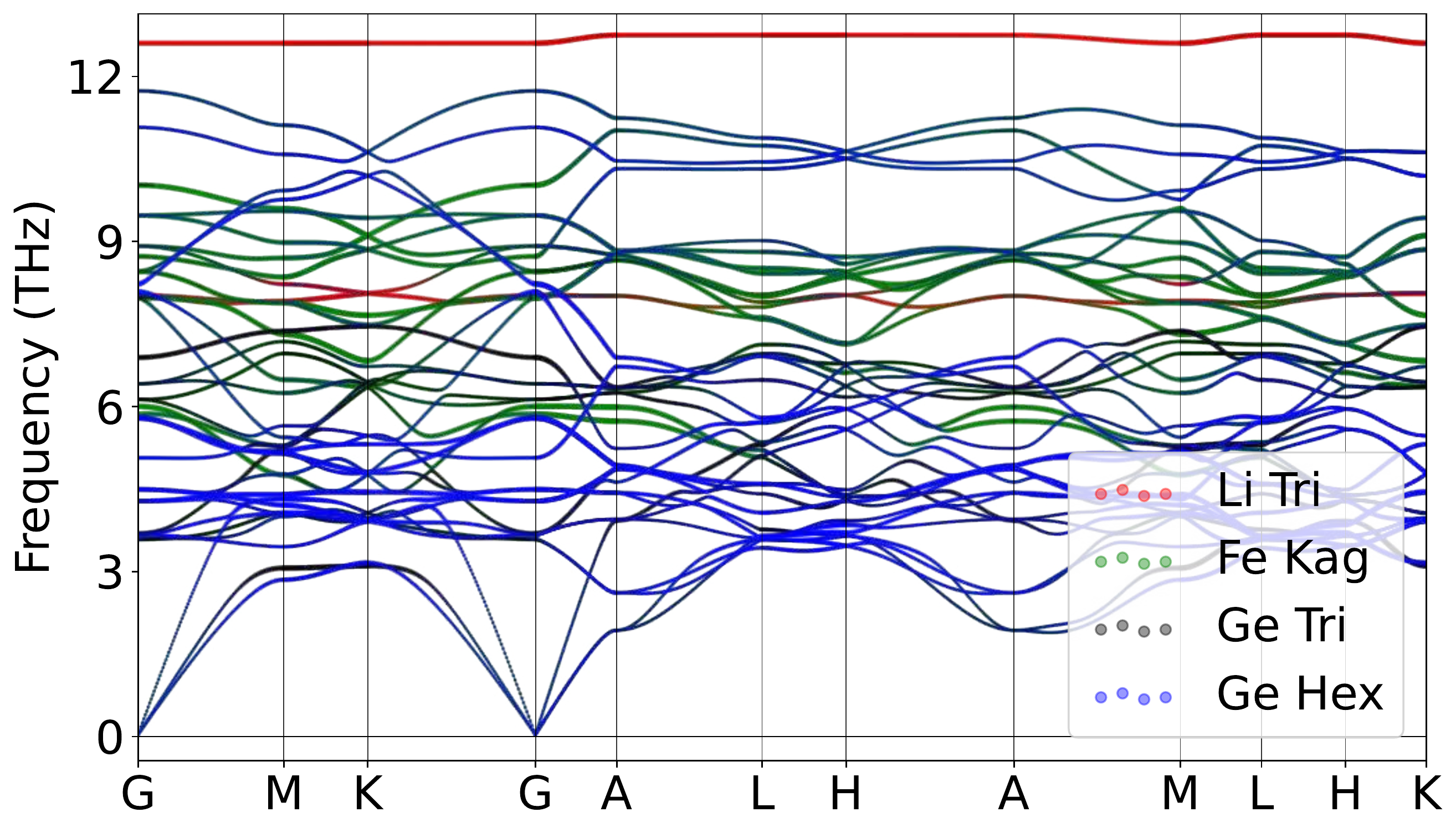}
\caption{Band structure for LiFe$_6$Ge$_6$ with AFM order without SOC for spin (Left)up channel. The projection of Fe $3d$ orbitals is also presented. (Right)Correspondingly, a stable phonon was demonstrated with the collinear magnetic order without Hubbard U at lower temperature regime, weighted by atomic projections.}
\label{fig:LiFe6Ge6mag}
\end{figure}

\begin{figure}[H]
\centering
\label{fig:LiFe6Ge6_relaxed_Low_xz}
\includegraphics[width=0.85\textwidth]{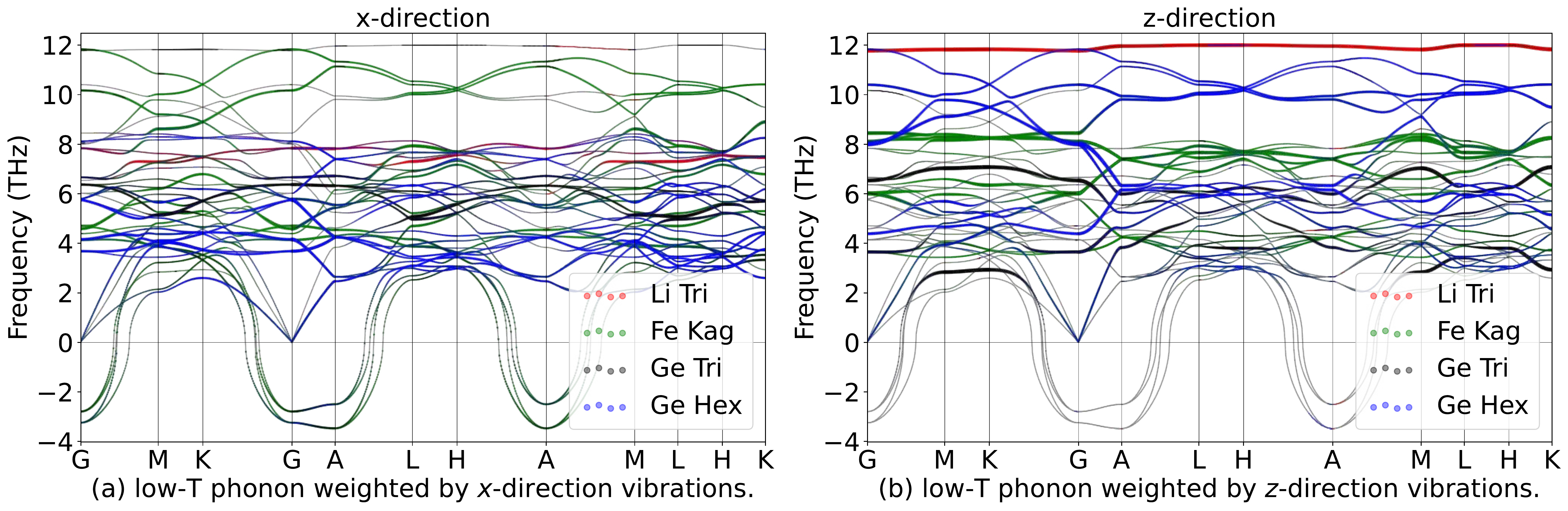}
\hspace{5pt}\label{fig:LiFe6Ge6_relaxed_High_xz}
\includegraphics[width=0.85\textwidth]{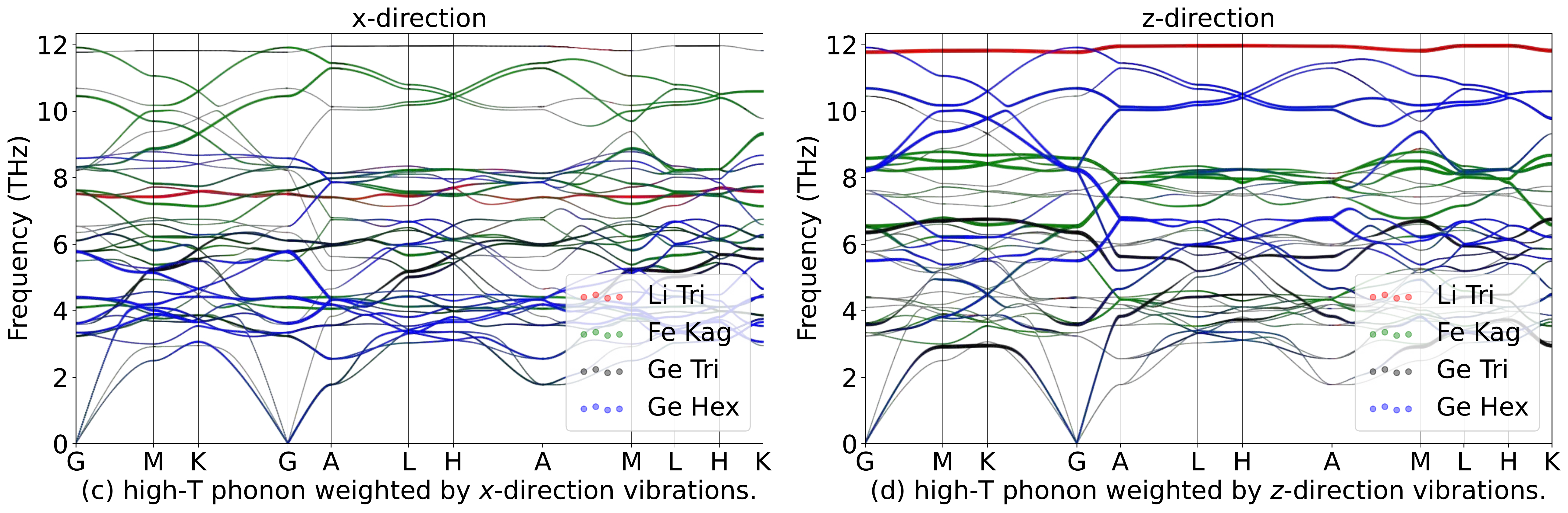}
\caption{Phonon spectrum for LiFe$_6$Ge$_6$in in-plane ($x$) and out-of-plane ($z$) directions in paramagnetic phase.}
\label{fig:LiFe6Ge6phoxyz}
\end{figure}

\begin{table}[htbp]
\global\long\def\arraystretch{1.12}
\captionsetup{width=.95\textwidth}
\caption{IRREPs at high-symmetry points for LiFe$_6$Ge$_6$ phonon spectrum at Low-T.}
\label{tab:191_LiFe6Ge6_Low}
\setlength{\tabcolsep}{1.mm}{
}
\end{table}

\begin{table}[htbp]
\global\long\def\arraystretch{1.12}
\captionsetup{width=.95\textwidth}
\caption{IRREPs at high-symmetry points for LiFe$_6$Ge$_6$ phonon spectrum at High-T.}
\label{tab:191_LiFe6Ge6_High}
\setlength{\tabcolsep}{1.mm}{
%
}
\end{table}
\subsubsection{\label{sms2sec:MgFe6Ge6}\hyperref[smsec:col]{\texorpdfstring{MgFe$_6$Ge$_6$}{}}~{\color{blue}  (High-T stable, Low-T unstable) }}

\begin{table}[htbp]
\global\long\def\arraystretch{1.12}
\captionsetup{width=.85\textwidth}
\caption{Structure parameters of MgFe$_6$Ge$_6$ after relaxation. It contains 360 electrons. }
\label{tab:MgFe6Ge6}
\setlength{\tabcolsep}{4mm}{
%
}
\end{table}

\begin{figure}[htbp]
\centering
\includegraphics[width=0.85\textwidth]{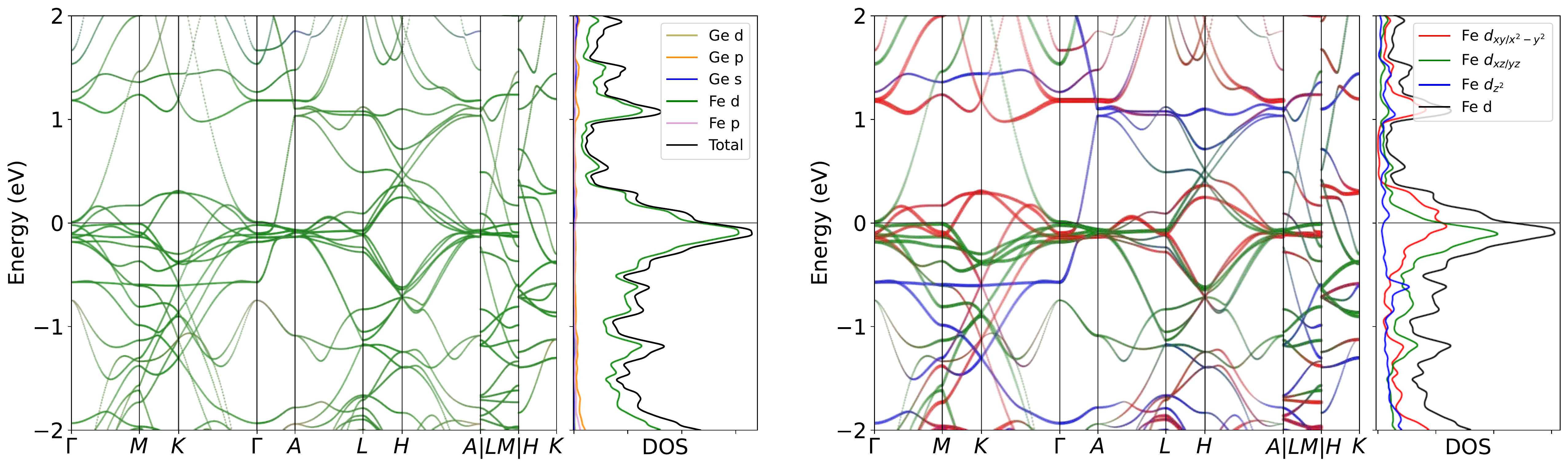}
\caption{Band structure for MgFe$_6$Ge$_6$ without SOC in paramagnetic phase. The projection of Fe $3d$ orbitals is also presented. The band structures clearly reveal the dominating of Fe $3d$ orbitals  near the Fermi level, as well as the pronounced peak.}
\label{fig:MgFe6Ge6band}
\end{figure}

\begin{figure}[H]
\centering
\subfloat[Relaxed phonon at low-T.]{
\label{fig:MgFe6Ge6_relaxed_Low}
\includegraphics[width=0.45\textwidth]{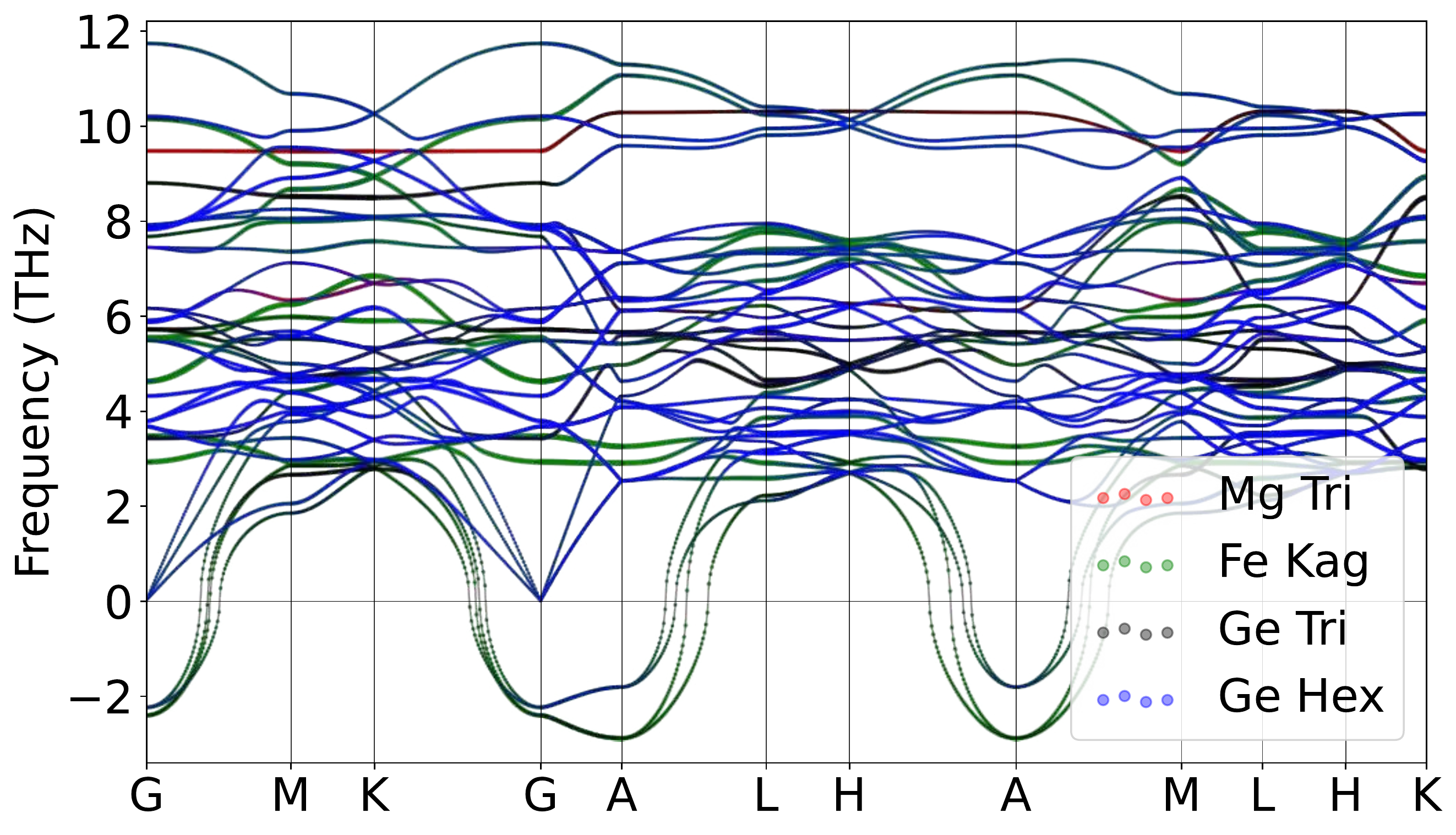}
}
\hspace{5pt}\subfloat[Relaxed phonon at high-T.]{
\label{fig:MgFe6Ge6_relaxed_High}
\includegraphics[width=0.45\textwidth]{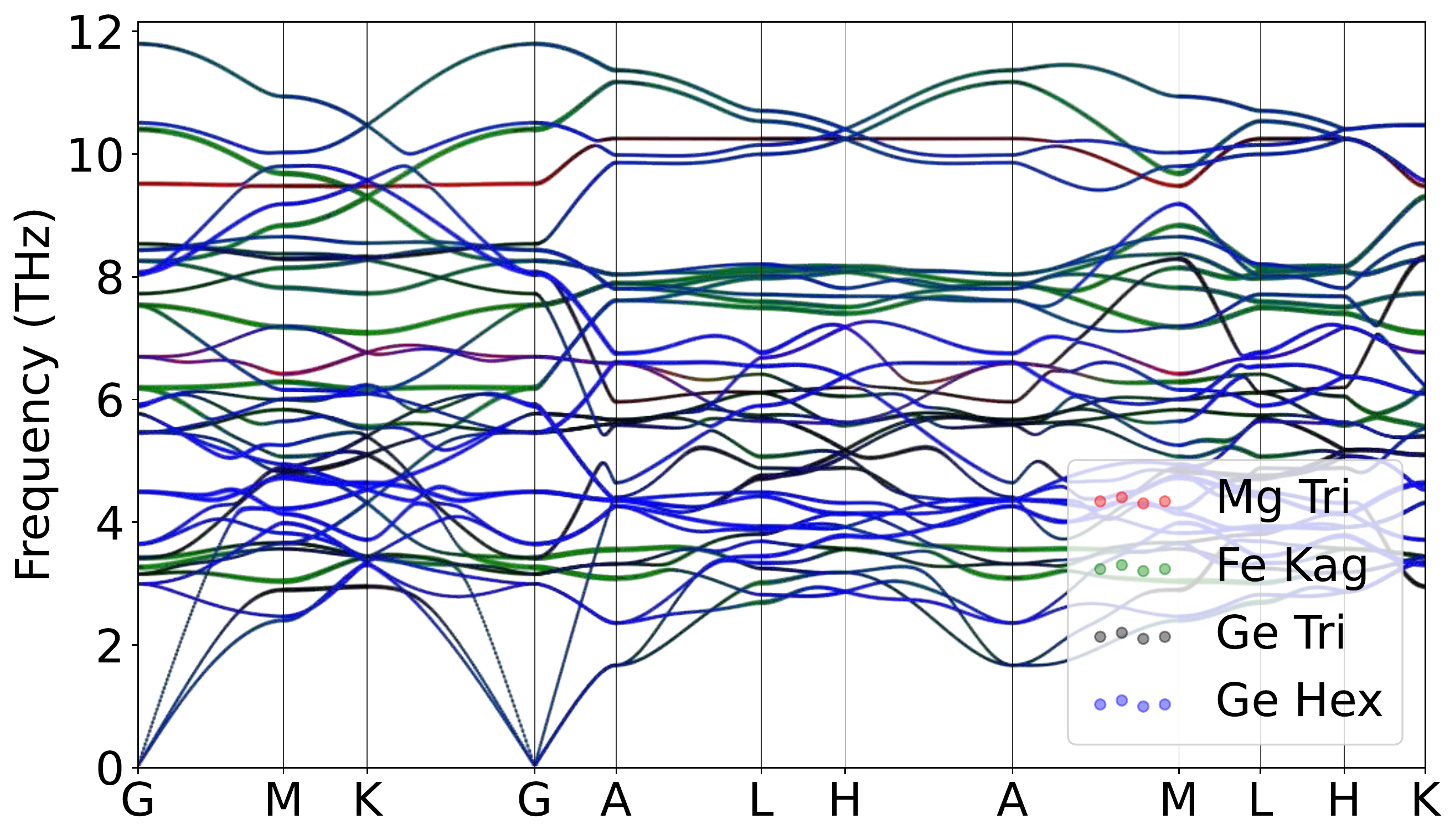}
}
\caption{Atomically projected phonon spectrum for MgFe$_6$Ge$_6$ in paramagnetic phase.}
\label{fig:MgFe6Ge6pho}
\end{figure}

\begin{figure}[htbp]
\centering
\includegraphics[width=0.45\textwidth]{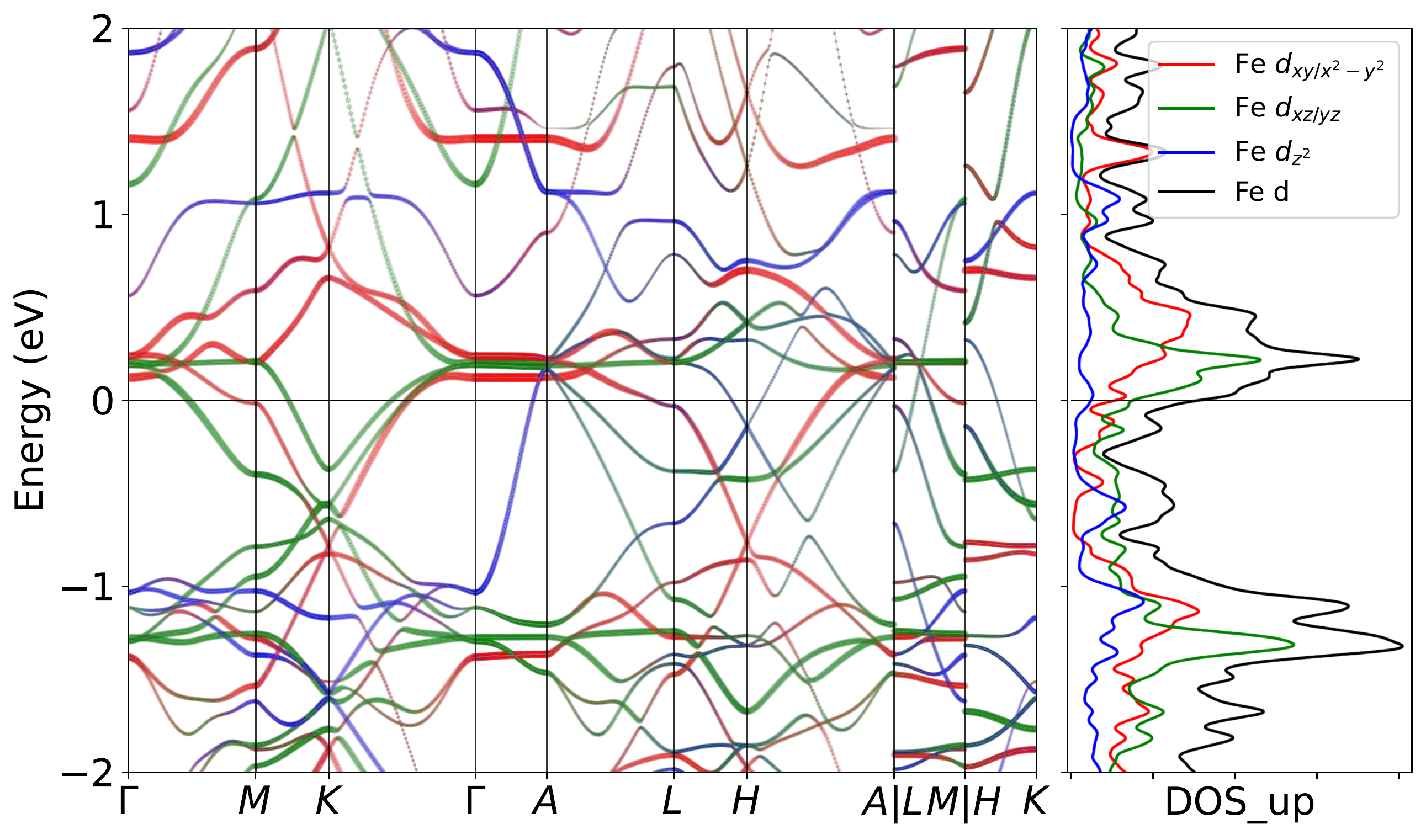}
\includegraphics[width=0.45\textwidth]{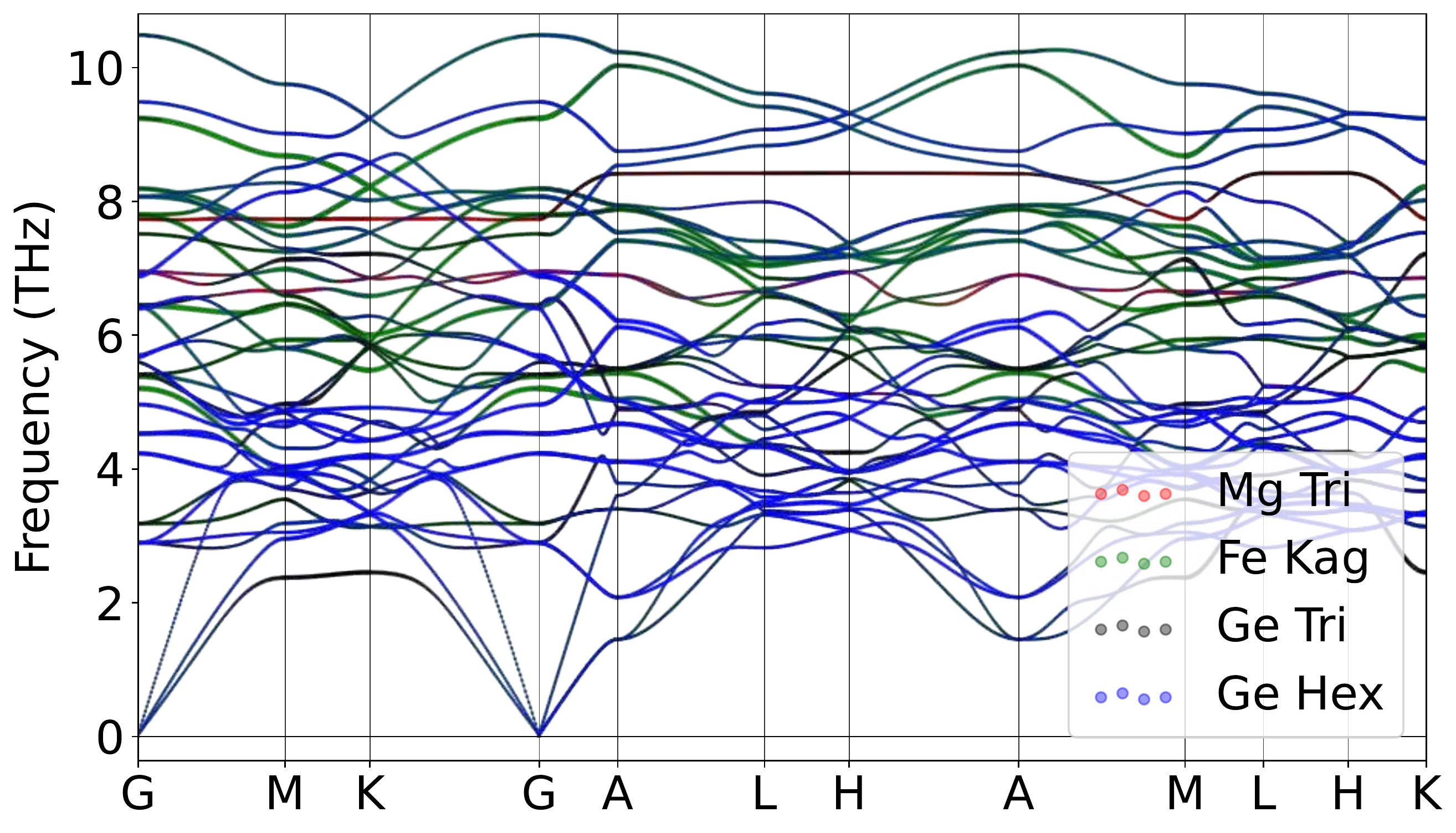}
\caption{Band structure for MgFe$_6$Ge$_6$ with AFM order without SOC for spin (Left)up channel. The projection of Fe $3d$ orbitals is also presented. (Right)Correspondingly, a stable phonon was demonstrated with the collinear magnetic order without Hubbard U at lower temperature regime, weighted by atomic projections.}
\label{fig:MgFe6Ge6mag}
\end{figure}

\begin{figure}[H]
\centering
\label{fig:MgFe6Ge6_relaxed_Low_xz}
\includegraphics[width=0.85\textwidth]{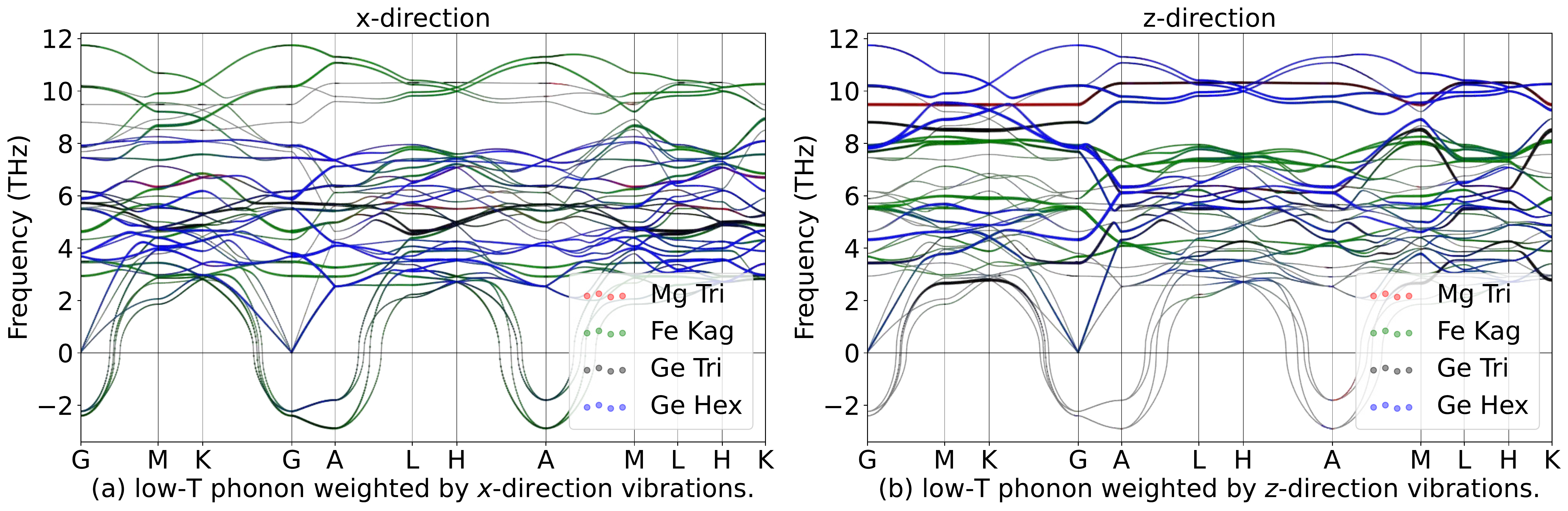}
\hspace{5pt}\label{fig:MgFe6Ge6_relaxed_High_xz}
\includegraphics[width=0.85\textwidth]{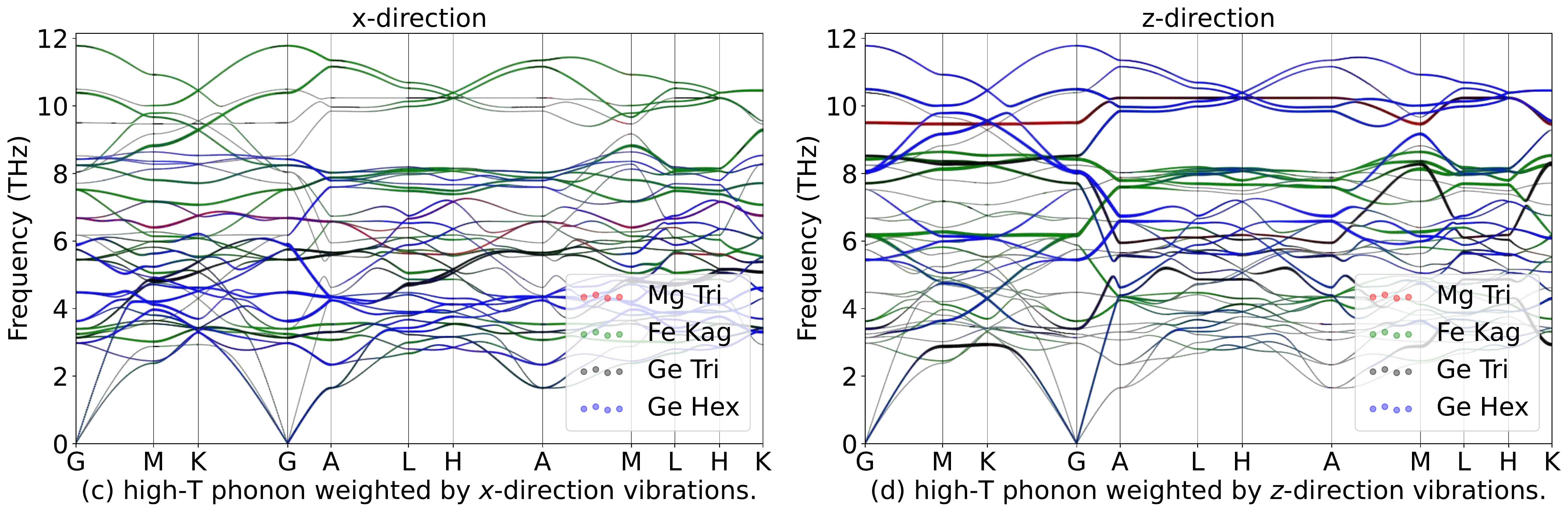}
\caption{Phonon spectrum for MgFe$_6$Ge$_6$in in-plane ($x$) and out-of-plane ($z$) directions in paramagnetic phase.}
\label{fig:MgFe6Ge6phoxyz}
\end{figure}

\begin{table}[htbp]
\global\long\def\arraystretch{1.12}
\captionsetup{width=.95\textwidth}
\caption{IRREPs at high-symmetry points for MgFe$_6$Ge$_6$ phonon spectrum at Low-T.}
\label{tab:191_MgFe6Ge6_Low}
\setlength{\tabcolsep}{1.mm}{
}
\end{table}

\begin{table}[htbp]
\global\long\def\arraystretch{1.12}
\captionsetup{width=.95\textwidth}
\caption{IRREPs at high-symmetry points for MgFe$_6$Ge$_6$ phonon spectrum at High-T.}
\label{tab:191_MgFe6Ge6_High}
\setlength{\tabcolsep}{1.mm}{
%
}
\end{table}
\subsubsection{\label{sms2sec:CaFe6Ge6}\hyperref[smsec:col]{\texorpdfstring{CaFe$_6$Ge$_6$}{}}~{\color{blue}  (High-T stable, Low-T unstable) }}

\begin{table}[htbp]
\global\long\def\arraystretch{1.12}
\captionsetup{width=.85\textwidth}
\caption{Structure parameters of CaFe$_6$Ge$_6$ after relaxation. It contains 368 electrons. }
\label{tab:CaFe6Ge6}
\setlength{\tabcolsep}{4mm}{
%
}
\end{table}

\begin{figure}[htbp]
\centering
\includegraphics[width=0.85\textwidth]{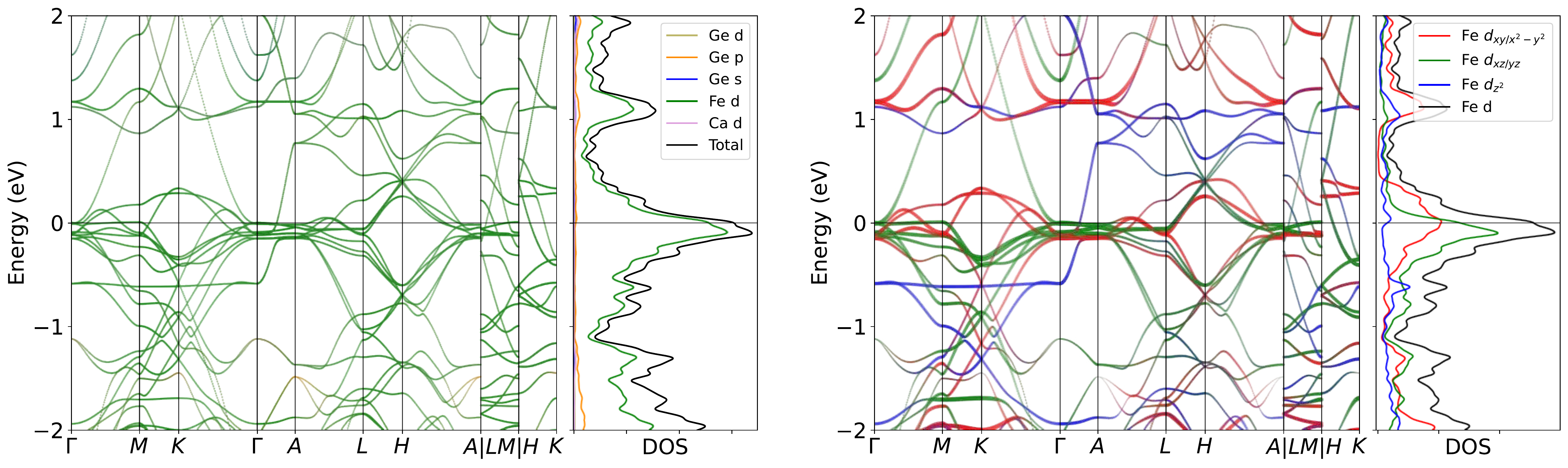}
\caption{Band structure for CaFe$_6$Ge$_6$ without SOC in paramagnetic phase. The projection of Fe $3d$ orbitals is also presented. The band structures clearly reveal the dominating of Fe $3d$ orbitals  near the Fermi level, as well as the pronounced peak.}
\label{fig:CaFe6Ge6band}
\end{figure}

\begin{figure}[H]
\centering
\subfloat[Relaxed phonon at low-T.]{
\label{fig:CaFe6Ge6_relaxed_Low}
\includegraphics[width=0.45\textwidth]{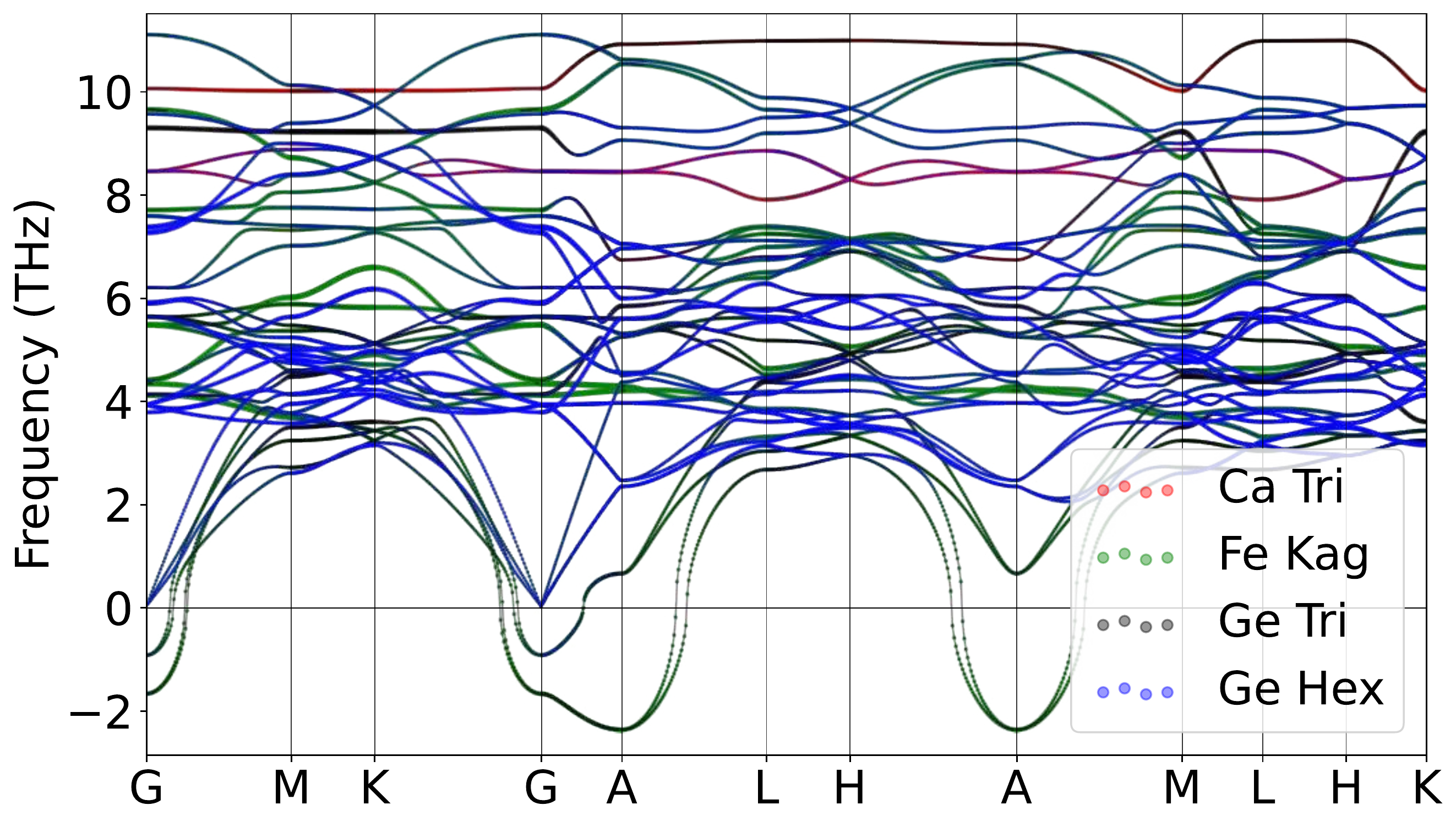}
}
\hspace{5pt}\subfloat[Relaxed phonon at high-T.]{
\label{fig:CaFe6Ge6_relaxed_High}
\includegraphics[width=0.45\textwidth]{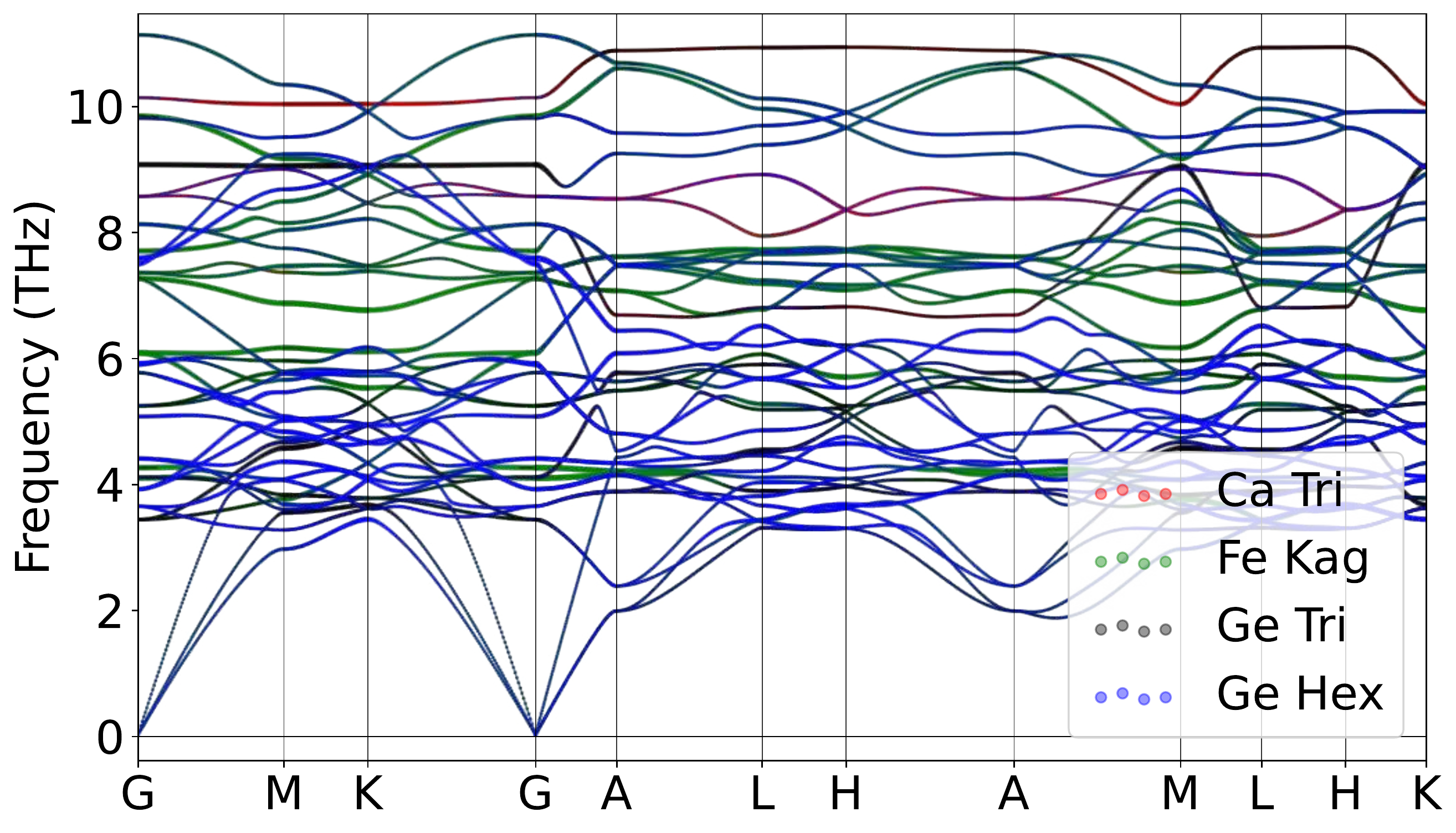}
}
\caption{Atomically projected phonon spectrum for CaFe$_6$Ge$_6$ in paramagnetic phase.}
\label{fig:CaFe6Ge6pho}
\end{figure}

\begin{figure}[htbp]
\centering
\includegraphics[width=0.45\textwidth]{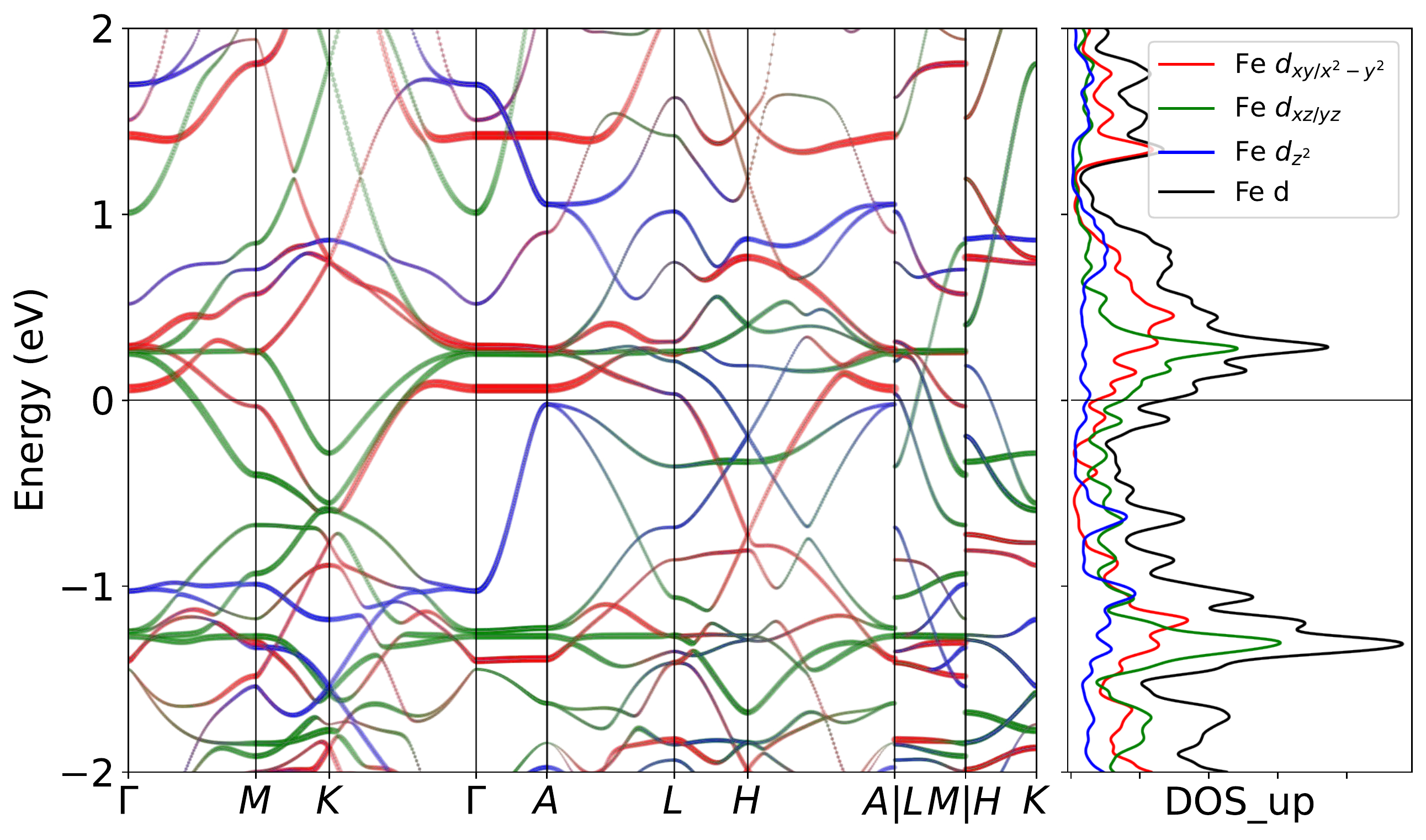}
\includegraphics[width=0.45\textwidth]{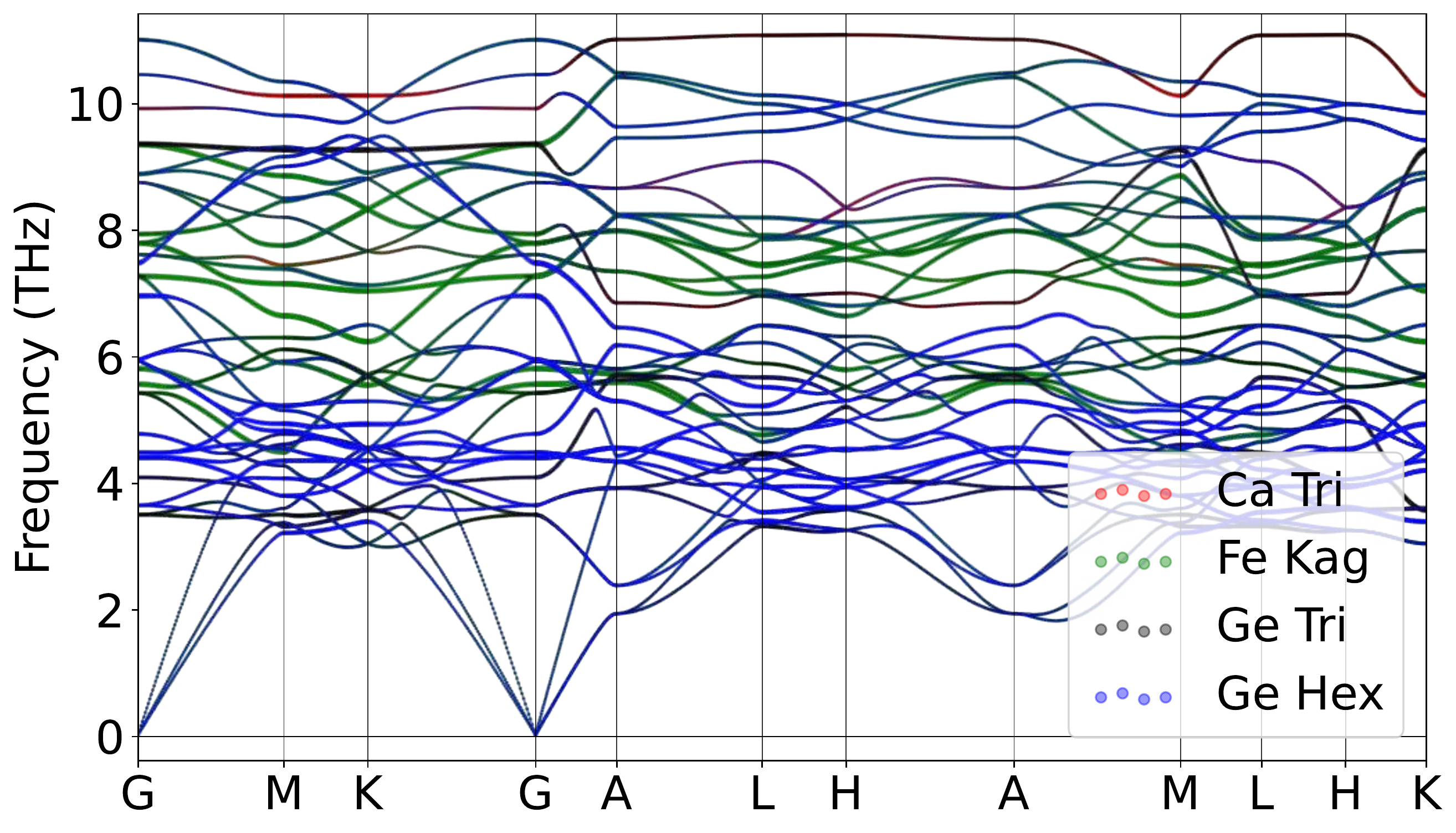}
\caption{Band structure for CaFe$_6$Ge$_6$ with AFM order without SOC for spin (Left)up channel. The projection of Fe $3d$ orbitals is also presented. (Right)Correspondingly, a stable phonon was demonstrated with the collinear magnetic order without Hubbard U at lower temperature regime, weighted by atomic projections.}
\label{fig:CaFe6Ge6mag}
\end{figure}

\begin{figure}[H]
\centering
\label{fig:CaFe6Ge6_relaxed_Low_xz}
\includegraphics[width=0.85\textwidth]{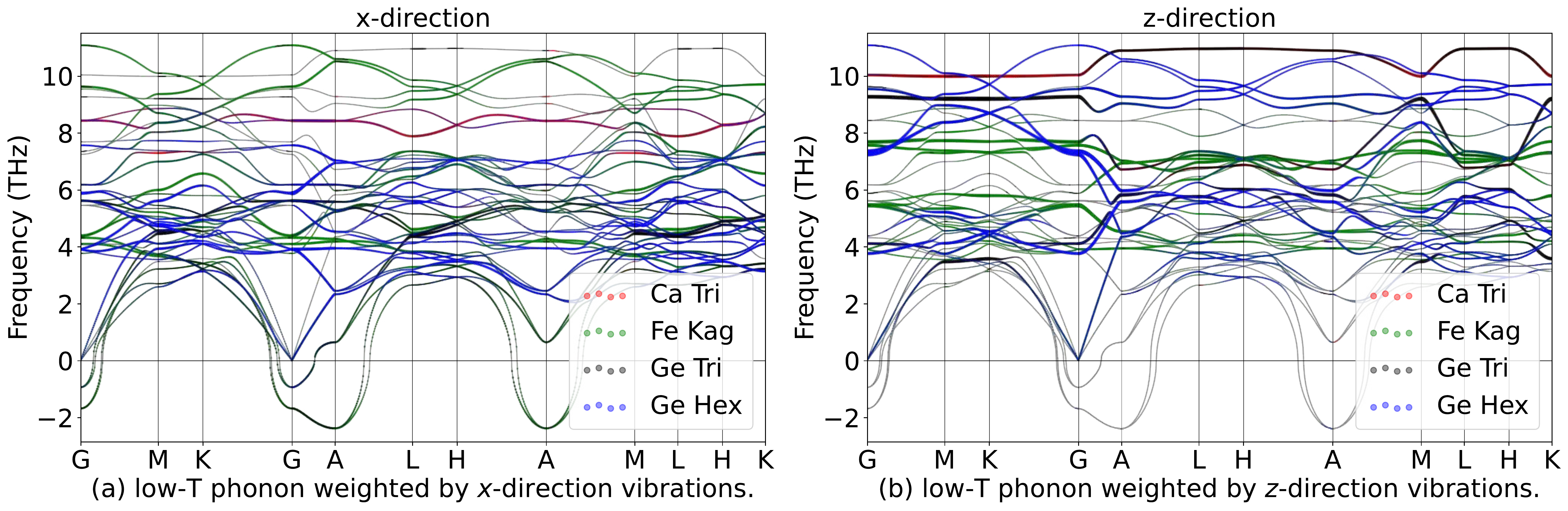}
\hspace{5pt}\label{fig:CaFe6Ge6_relaxed_High_xz}
\includegraphics[width=0.85\textwidth]{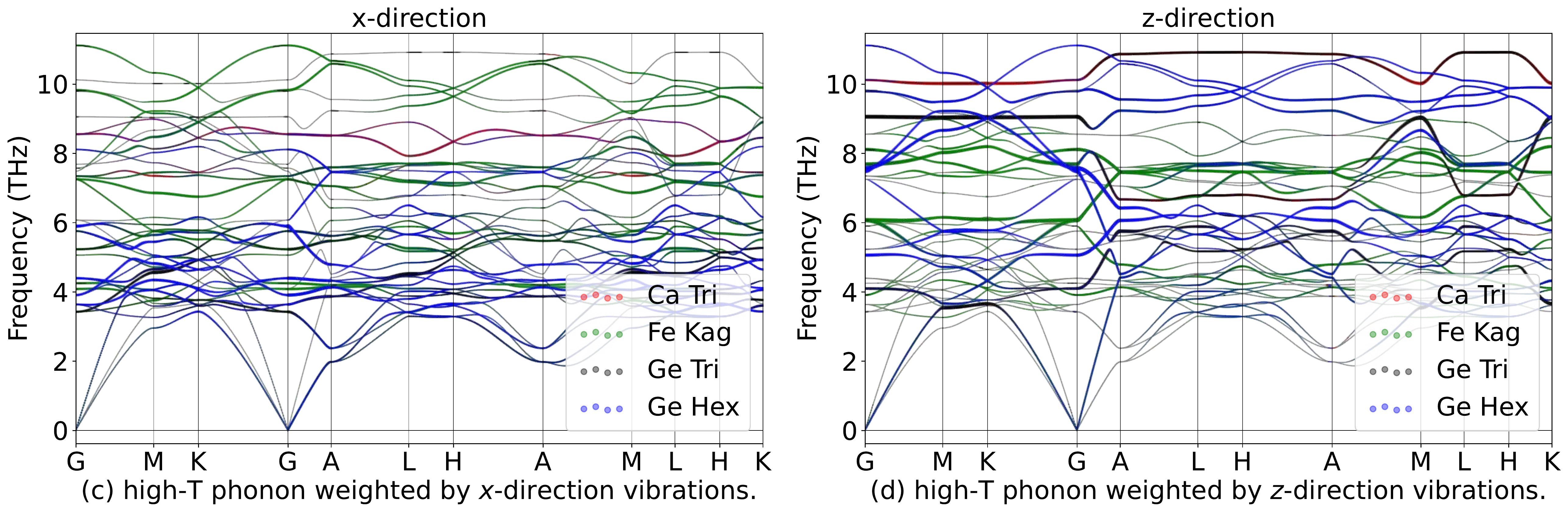}
\caption{Phonon spectrum for CaFe$_6$Ge$_6$in in-plane ($x$) and out-of-plane ($z$) directions in paramagnetic phase.}
\label{fig:CaFe6Ge6phoxyz}
\end{figure}

\begin{table}[htbp]
\global\long\def\arraystretch{1.12}
\captionsetup{width=.95\textwidth}
\caption{IRREPs at high-symmetry points for CaFe$_6$Ge$_6$ phonon spectrum at Low-T.}
\label{tab:191_CaFe6Ge6_Low}
\setlength{\tabcolsep}{1.mm}{
}
\end{table}

\begin{table}[htbp]
\global\long\def\arraystretch{1.12}
\captionsetup{width=.95\textwidth}
\caption{IRREPs at high-symmetry points for CaFe$_6$Ge$_6$ phonon spectrum at High-T.}
\label{tab:191_CaFe6Ge6_High}
\setlength{\tabcolsep}{1.mm}{
%
}
\end{table}
\subsubsection{\label{sms2sec:ScFe6Ge6}\hyperref[smsec:col]{\texorpdfstring{ScFe$_6$Ge$_6$}{}}~{\color{blue}  (High-T stable, Low-T unstable) }}

\begin{table}[htbp]
\global\long\def\arraystretch{1.12}
\captionsetup{width=.85\textwidth}
\caption{Structure parameters of ScFe$_6$Ge$_6$ after relaxation. It contains 369 electrons. }
\label{tab:ScFe6Ge6}
\setlength{\tabcolsep}{4mm}{
%
}
\end{table}

\begin{figure}[htbp]
\centering
\includegraphics[width=0.85\textwidth]{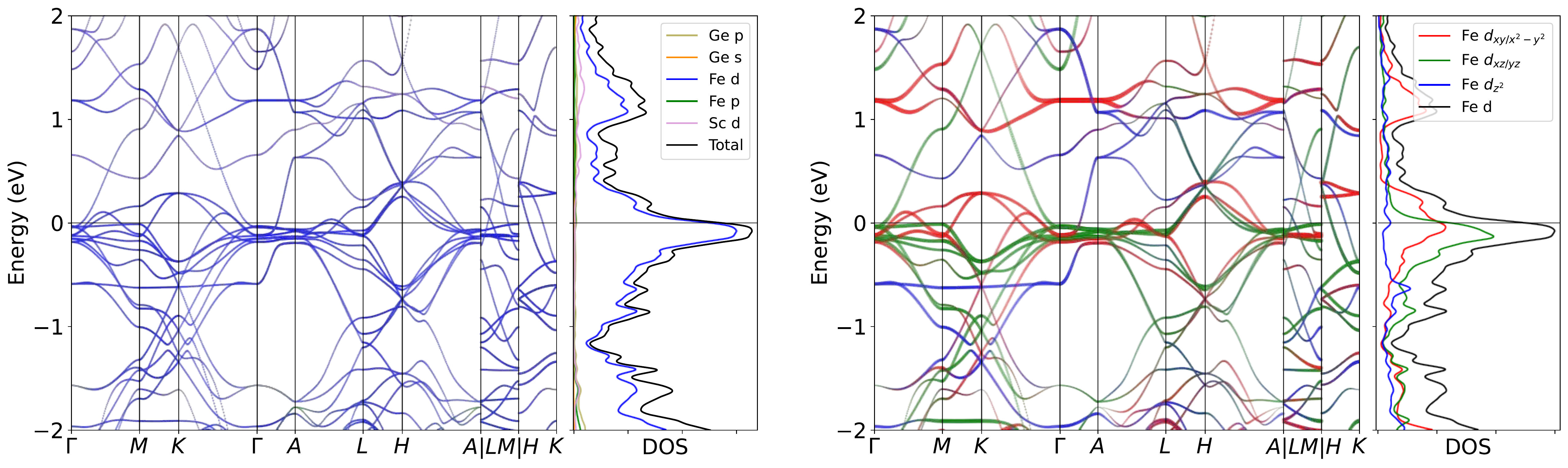}
\caption{Band structure for ScFe$_6$Ge$_6$ without SOC in paramagnetic phase. The projection of Fe $3d$ orbitals is also presented. The band structures clearly reveal the dominating of Fe $3d$ orbitals  near the Fermi level, as well as the pronounced peak.}
\label{fig:ScFe6Ge6band}
\end{figure}

\begin{figure}[H]
\centering
\subfloat[Relaxed phonon at low-T.]{
\label{fig:ScFe6Ge6_relaxed_Low}
\includegraphics[width=0.45\textwidth]{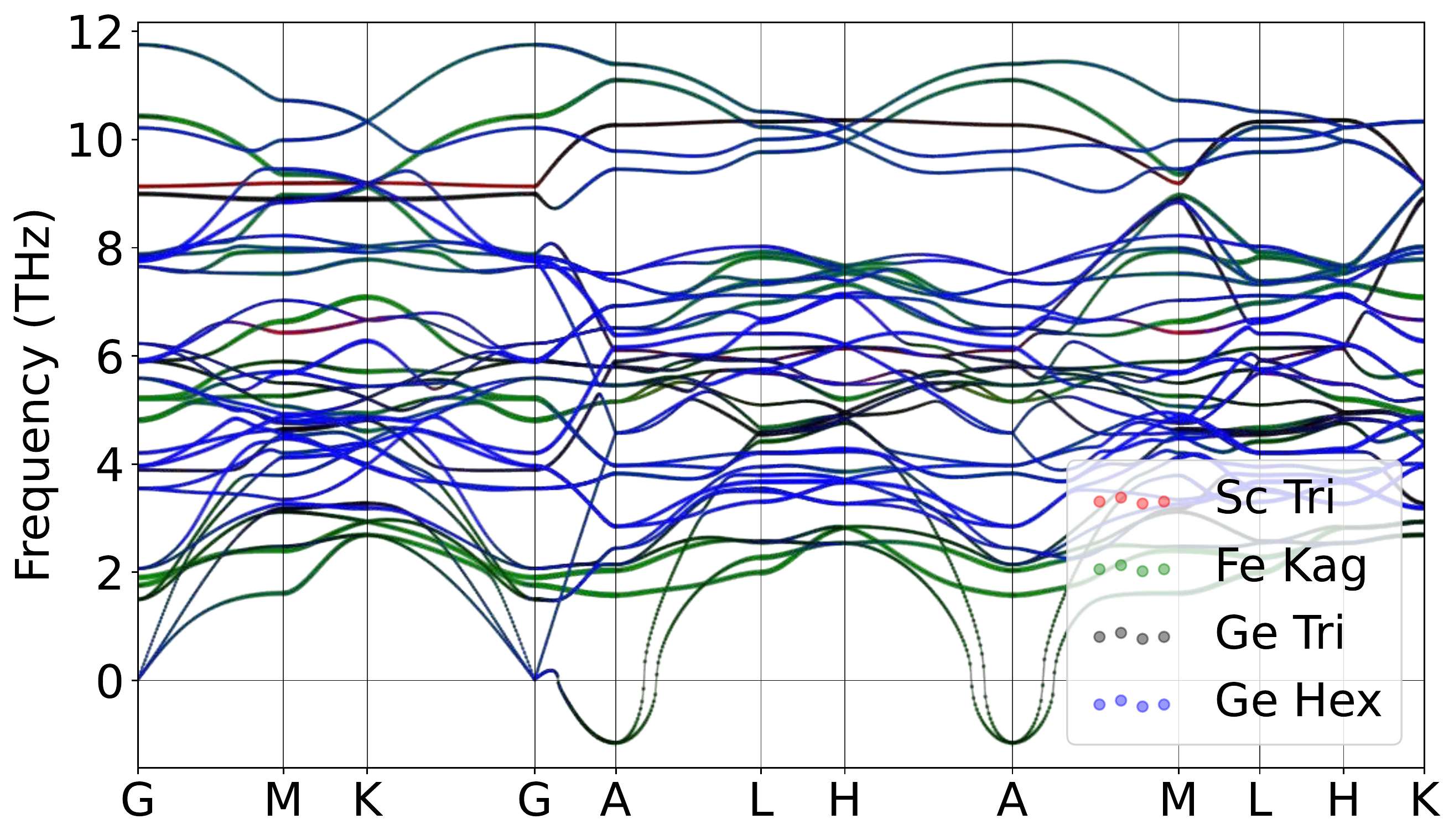}
}
\hspace{5pt}\subfloat[Relaxed phonon at high-T.]{
\label{fig:ScFe6Ge6_relaxed_High}
\includegraphics[width=0.45\textwidth]{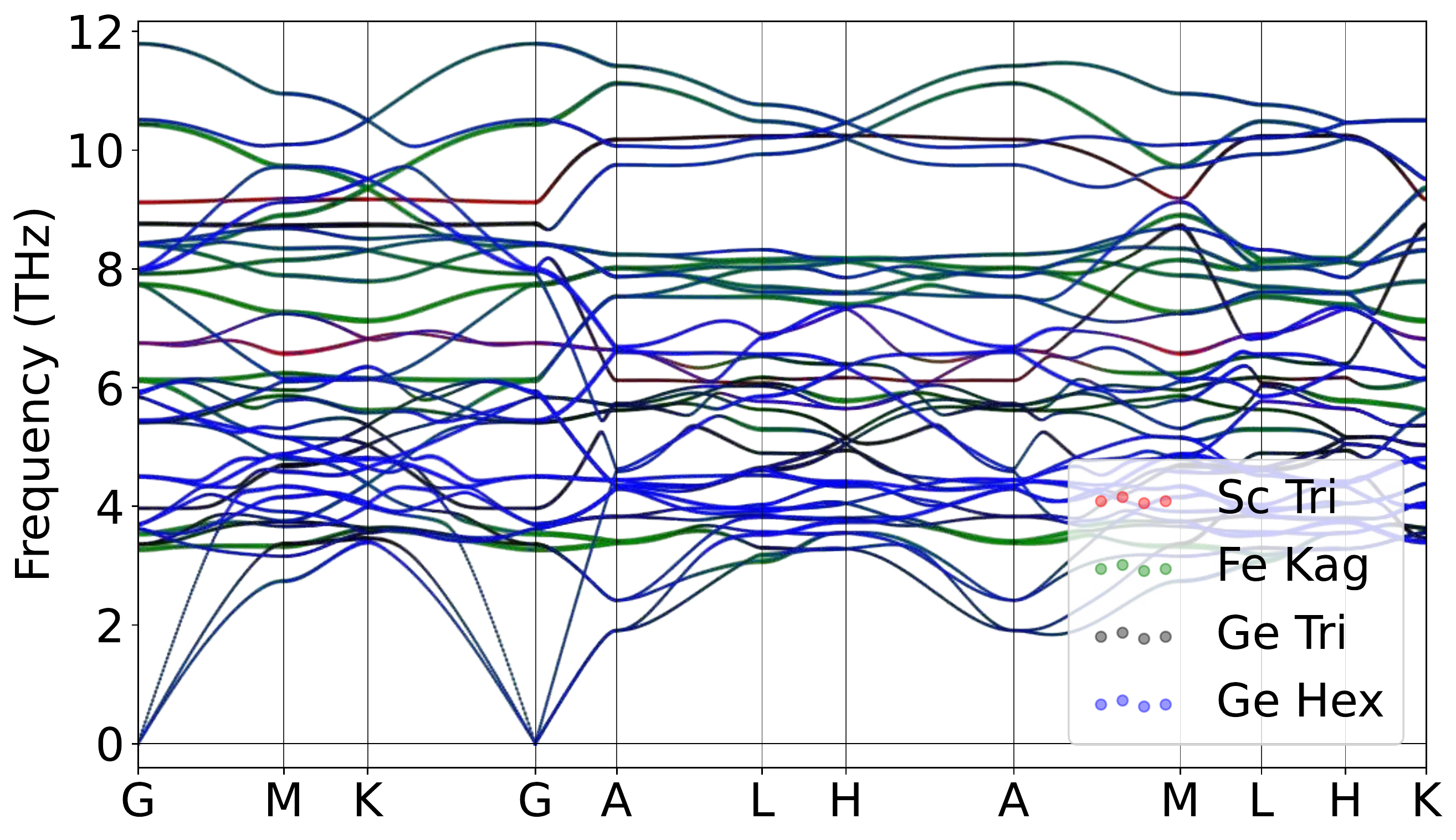}
}
\caption{Atomically projected phonon spectrum for ScFe$_6$Ge$_6$ in paramagnetic phase.}
\label{fig:ScFe6Ge6pho}
\end{figure}

\begin{figure}[htbp]
\centering
\includegraphics[width=0.45\textwidth]{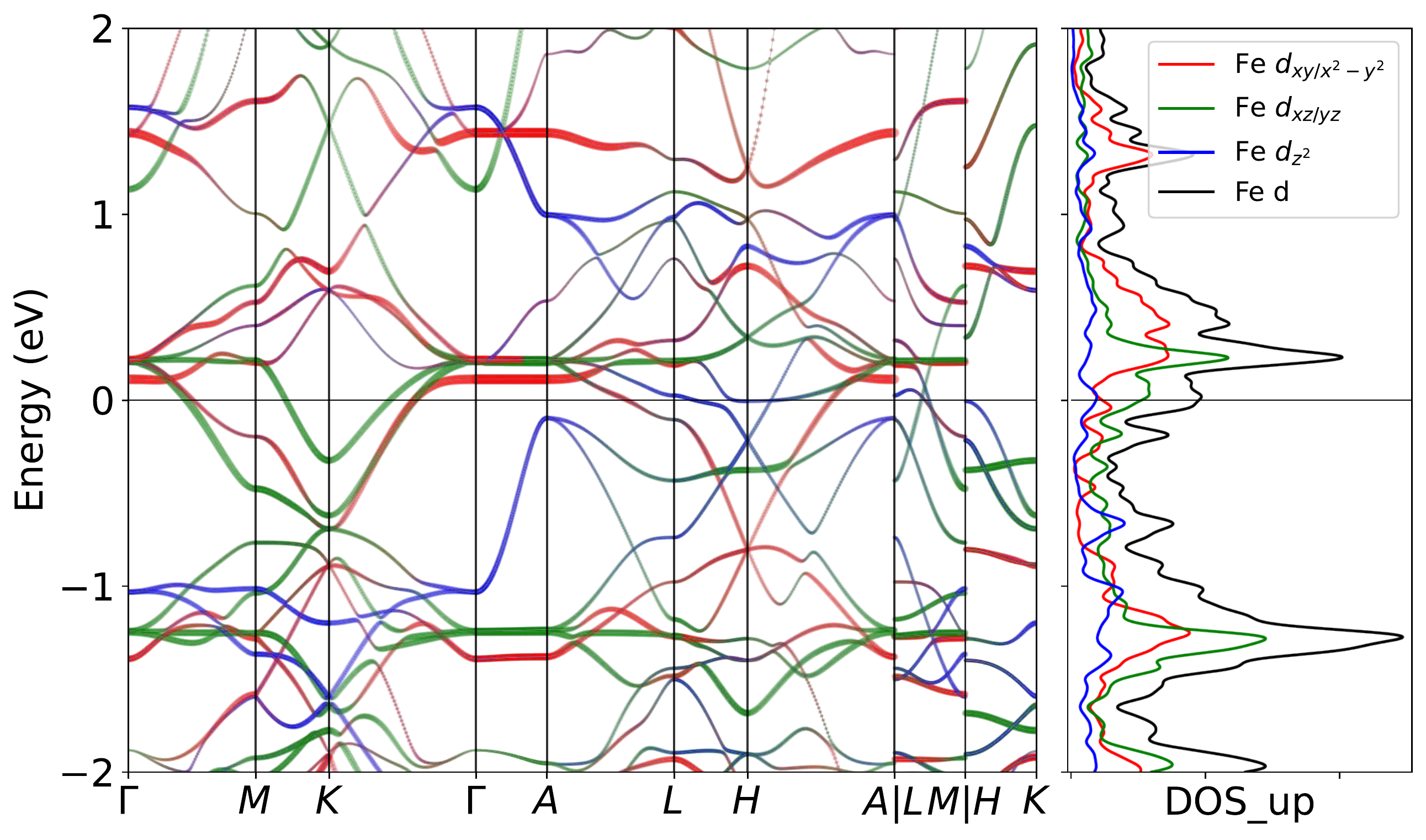}
\includegraphics[width=0.45\textwidth]{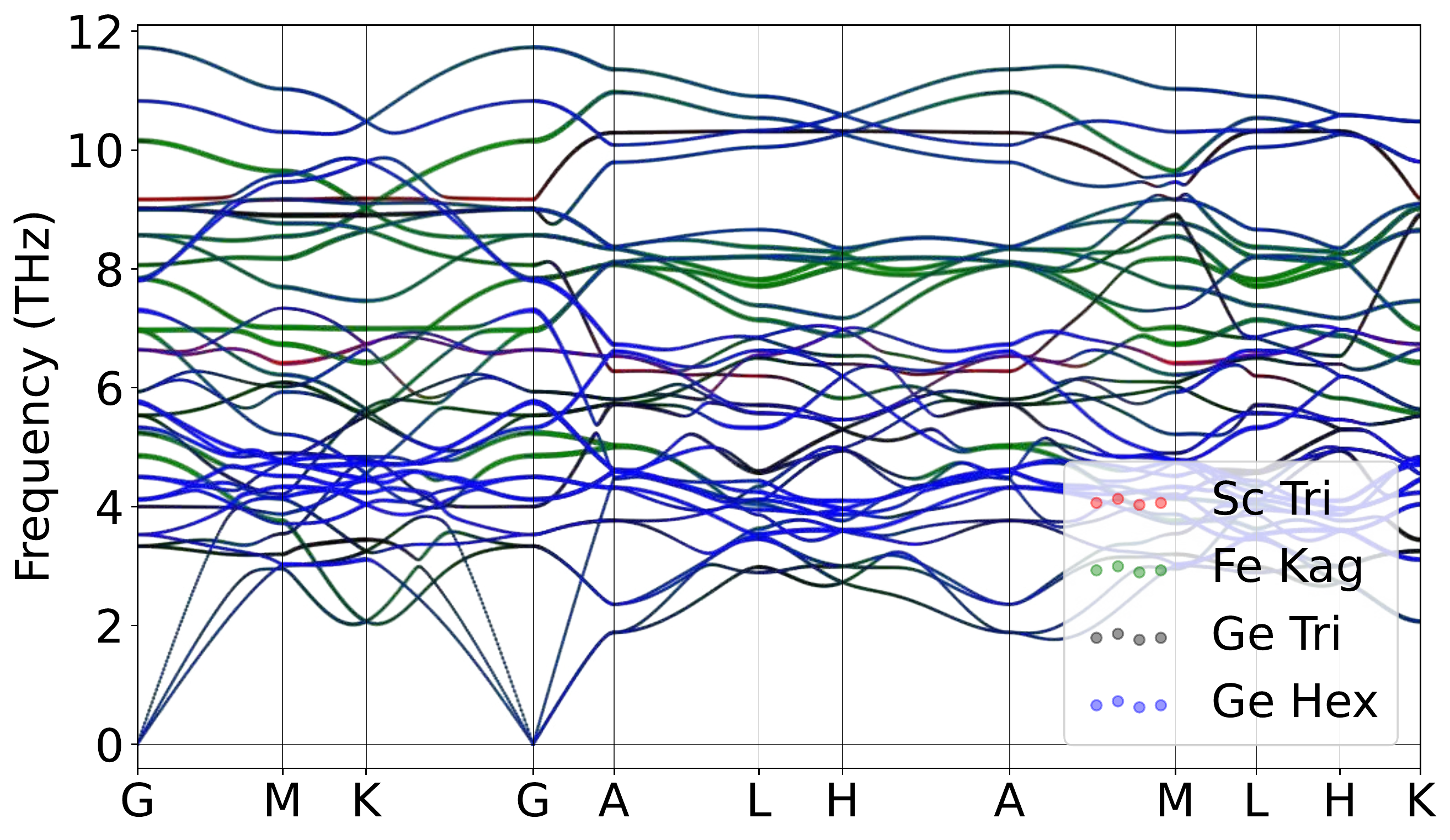}
\caption{Band structure for ScFe$_6$Ge$_6$ with AFM order without SOC for spin (Left)up channel. The projection of Fe $3d$ orbitals is also presented. (Right)Correspondingly, a stable phonon was demonstrated with the collinear magnetic order without Hubbard U at lower temperature regime, weighted by atomic projections.}
\label{fig:ScFe6Ge6mag}
\end{figure}

\begin{figure}[H]
\centering
\label{fig:ScFe6Ge6_relaxed_Low_xz}
\includegraphics[width=0.85\textwidth]{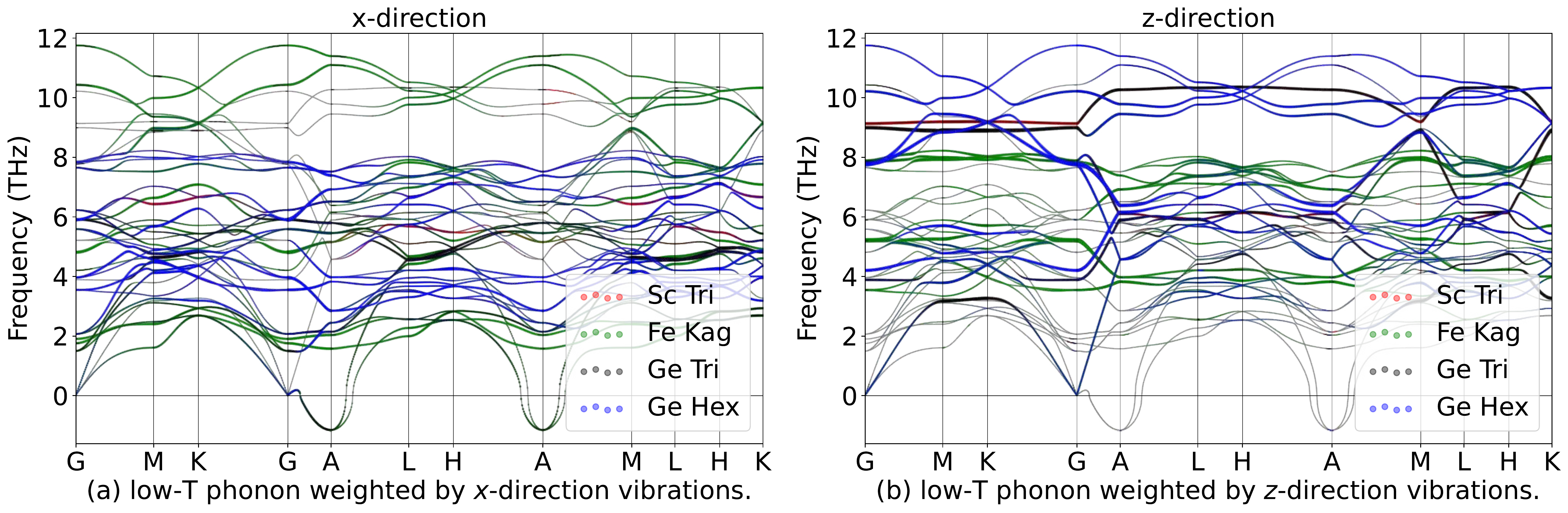}
\hspace{5pt}\label{fig:ScFe6Ge6_relaxed_High_xz}
\includegraphics[width=0.85\textwidth]{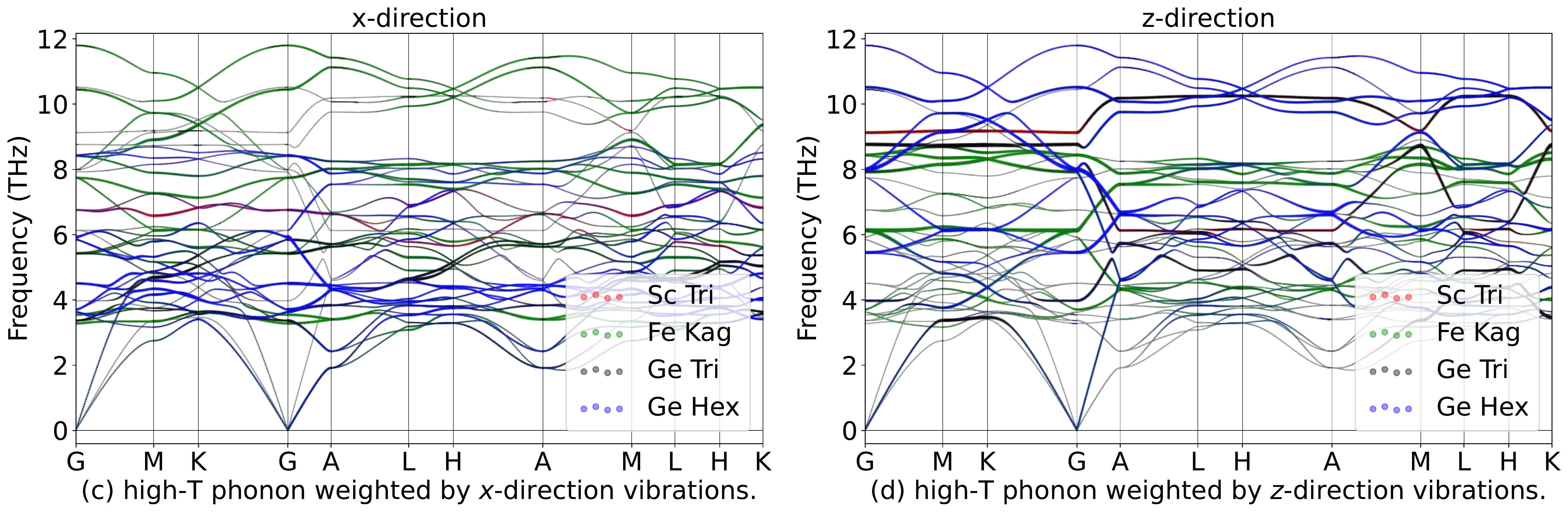}
\caption{Phonon spectrum for ScFe$_6$Ge$_6$in in-plane ($x$) and out-of-plane ($z$) directions in paramagnetic phase.}
\label{fig:ScFe6Ge6phoxyz}
\end{figure}

\begin{table}[htbp]
\global\long\def\arraystretch{1.12}
\captionsetup{width=.95\textwidth}
\caption{IRREPs at high-symmetry points for ScFe$_6$Ge$_6$ phonon spectrum at Low-T.}
\label{tab:191_ScFe6Ge6_Low}
\setlength{\tabcolsep}{1.mm}{
}
\end{table}

\begin{table}[htbp]
\global\long\def\arraystretch{1.12}
\captionsetup{width=.95\textwidth}
\caption{IRREPs at high-symmetry points for ScFe$_6$Ge$_6$ phonon spectrum at High-T.}
\label{tab:191_ScFe6Ge6_High}
\setlength{\tabcolsep}{1.mm}{
%
}
\end{table}
\subsubsection{\label{sms2sec:TiFe6Ge6}\hyperref[smsec:col]{\texorpdfstring{TiFe$_6$Ge$_6$}{}}~{\color{blue}  (High-T stable, Low-T unstable) }}

\begin{table}[htbp]
\global\long\def\arraystretch{1.12}
\captionsetup{width=.85\textwidth}
\caption{Structure parameters of TiFe$_6$Ge$_6$ after relaxation. It contains 370 electrons. }
\label{tab:TiFe6Ge6}
\setlength{\tabcolsep}{4mm}{
%
}
\end{table}

\begin{figure}[htbp]
\centering
\includegraphics[width=0.85\textwidth]{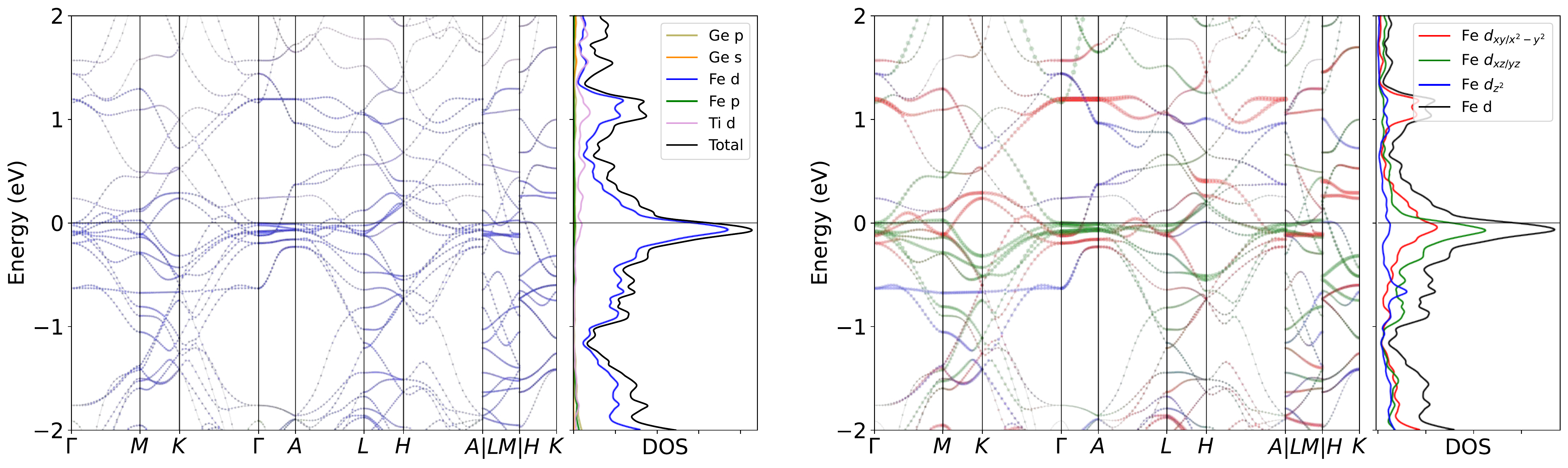}
\caption{Band structure for TiFe$_6$Ge$_6$ without SOC in paramagnetic phase. The projection of Fe $3d$ orbitals is also presented. The band structures clearly reveal the dominating of Fe $3d$ orbitals  near the Fermi level, as well as the pronounced peak.}
\label{fig:TiFe6Ge6band}
\end{figure}

\begin{figure}[H]
\centering
\subfloat[Relaxed phonon at low-T.]{
\label{fig:TiFe6Ge6_relaxed_Low}
\includegraphics[width=0.45\textwidth]{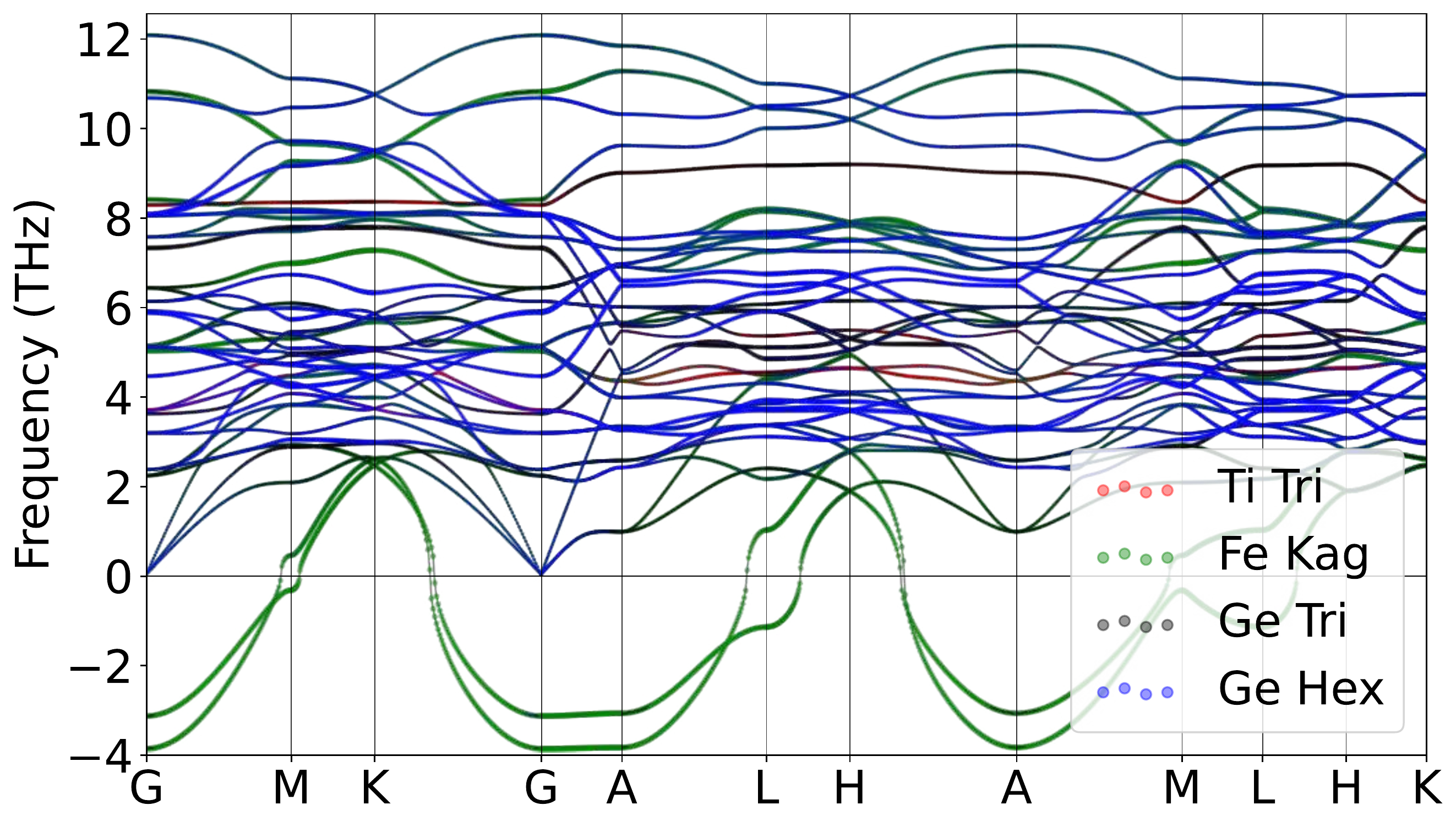}
}
\hspace{5pt}\subfloat[Relaxed phonon at high-T.]{
\label{fig:TiFe6Ge6_relaxed_High}
\includegraphics[width=0.45\textwidth]{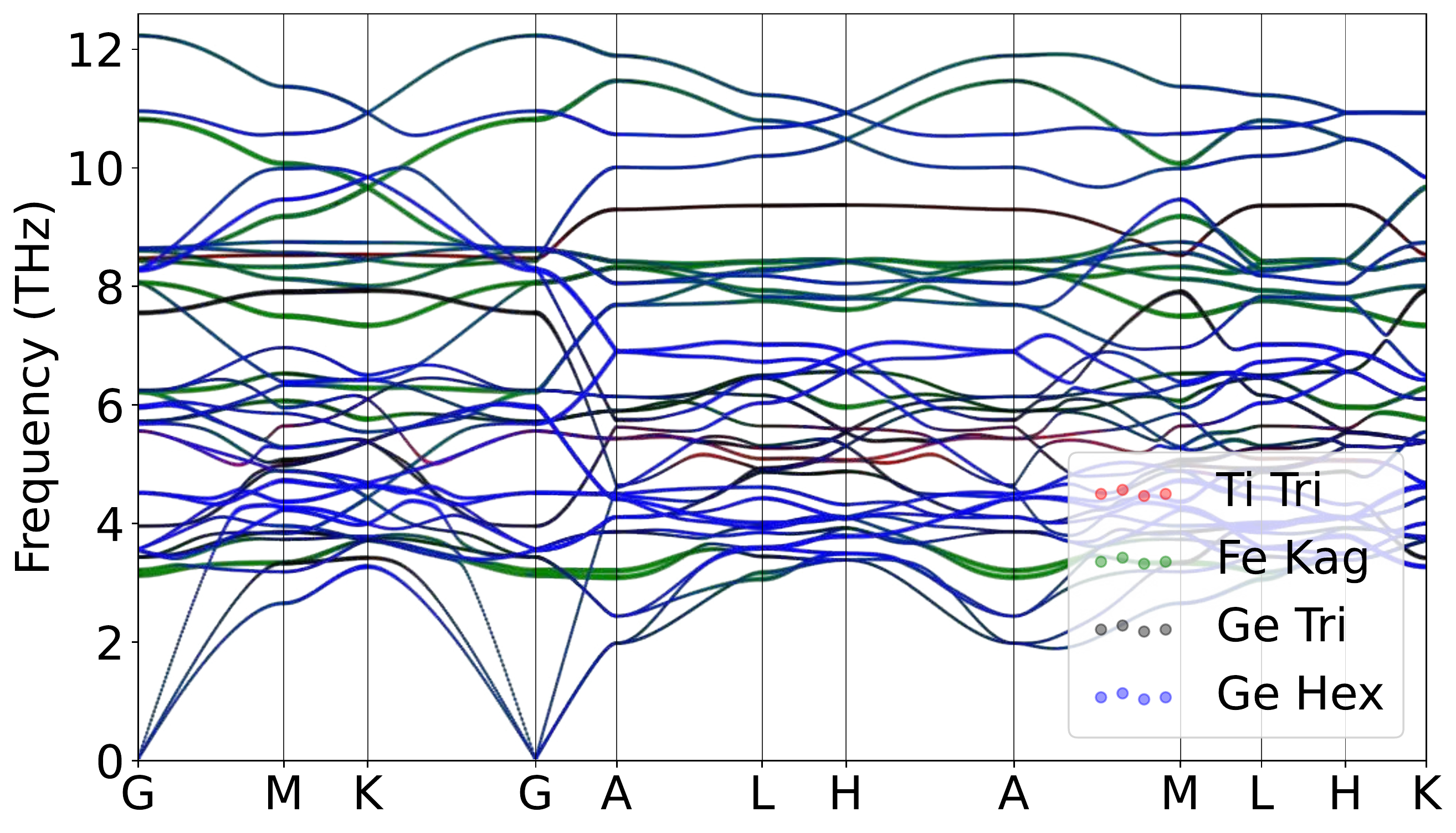}
}
\caption{Atomically projected phonon spectrum for TiFe$_6$Ge$_6$ in paramagnetic phase.}
\label{fig:TiFe6Ge6pho}
\end{figure}

\begin{figure}[htbp]
\centering
\includegraphics[width=0.45\textwidth]{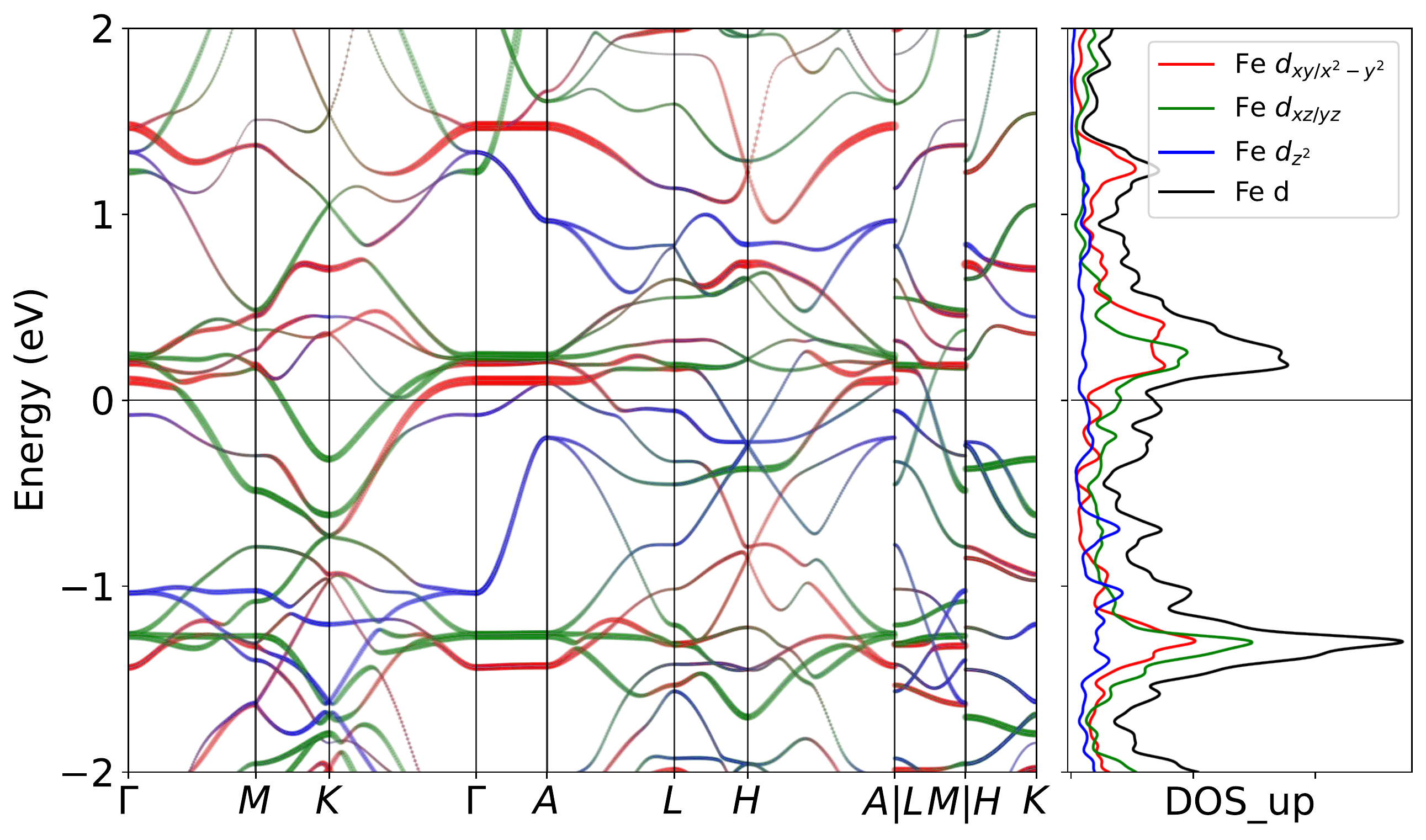}
\includegraphics[width=0.45\textwidth]{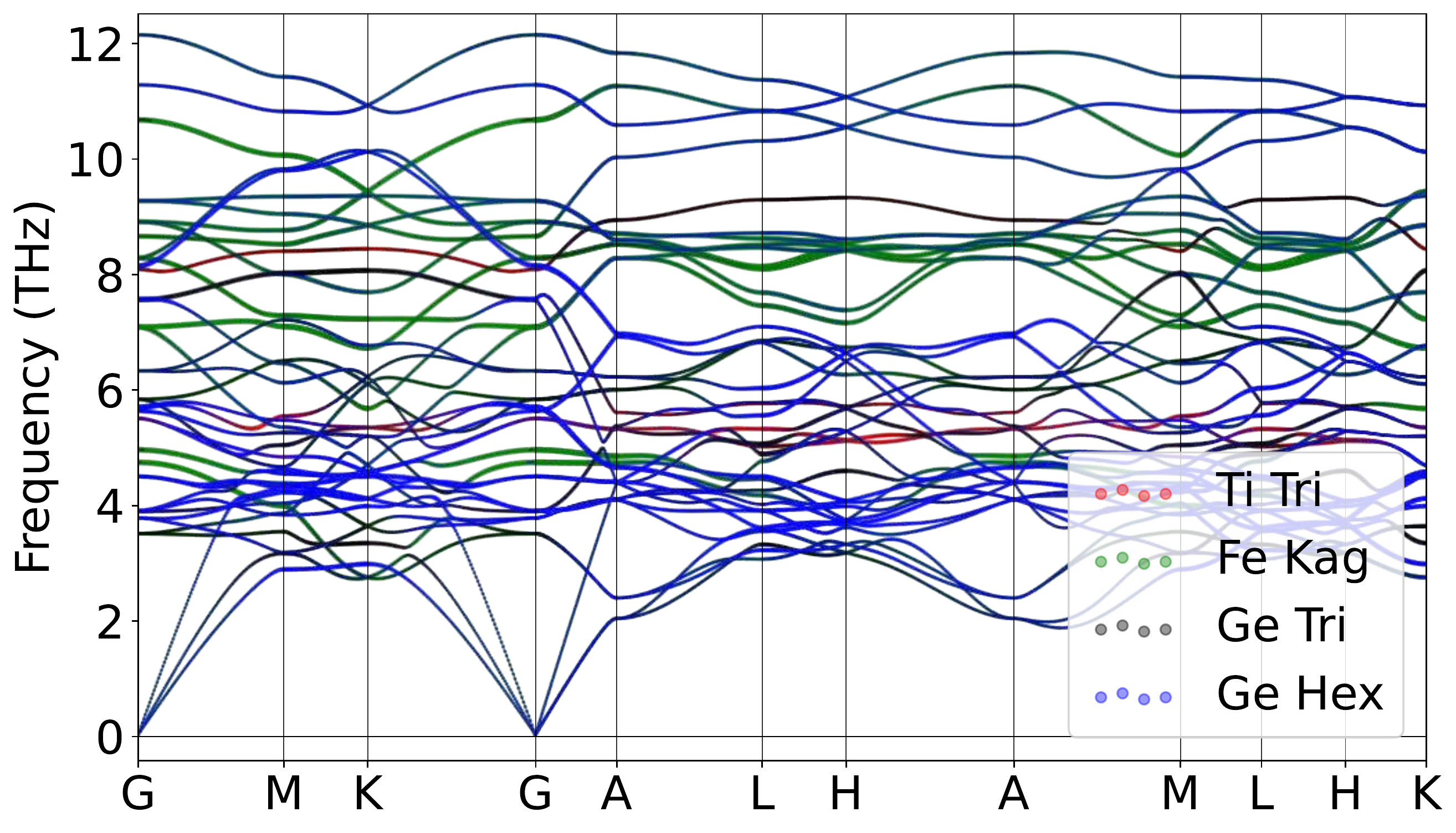}
\caption{Band structure for TiFe$_6$Ge$_6$ with AFM order without SOC for spin (Left)up channel. The projection of Fe $3d$ orbitals is also presented. (Right)Correspondingly, a stable phonon was demonstrated with the collinear magnetic order without Hubbard U at lower temperature regime, weighted by atomic projections.}
\label{fig:TiFe6Ge6mag}
\end{figure}

\begin{figure}[H]
\centering
\label{fig:TiFe6Ge6_relaxed_Low_xz}
\includegraphics[width=0.85\textwidth]{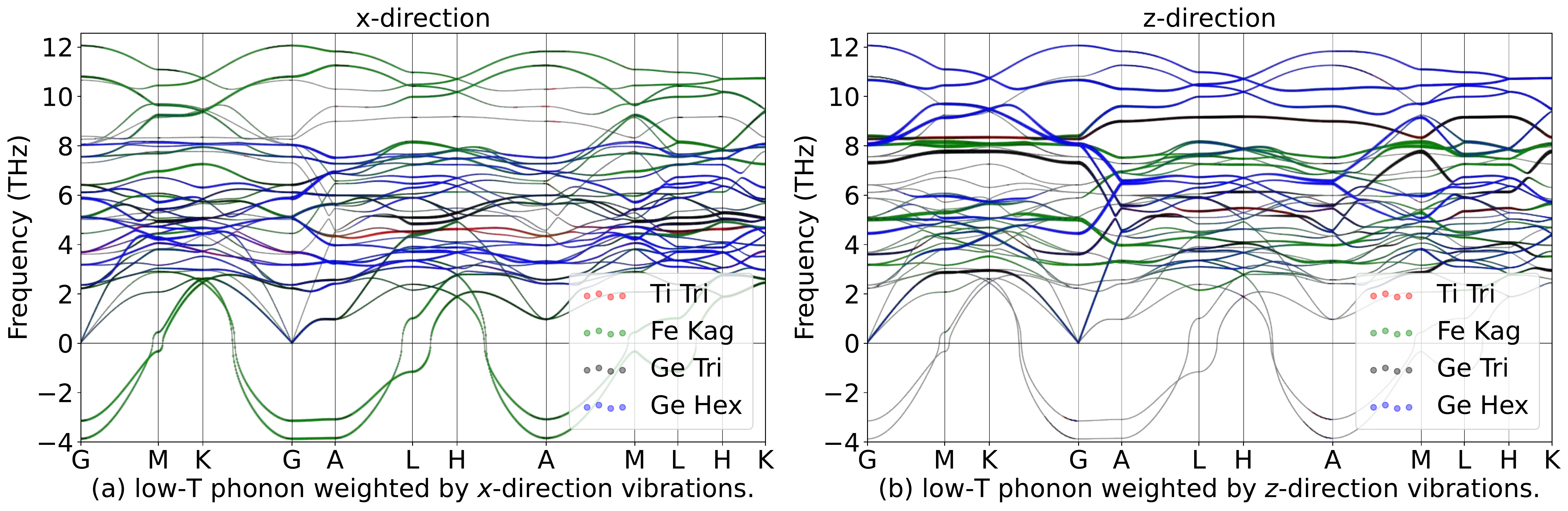}
\hspace{5pt}\label{fig:TiFe6Ge6_relaxed_High_xz}
\includegraphics[width=0.85\textwidth]{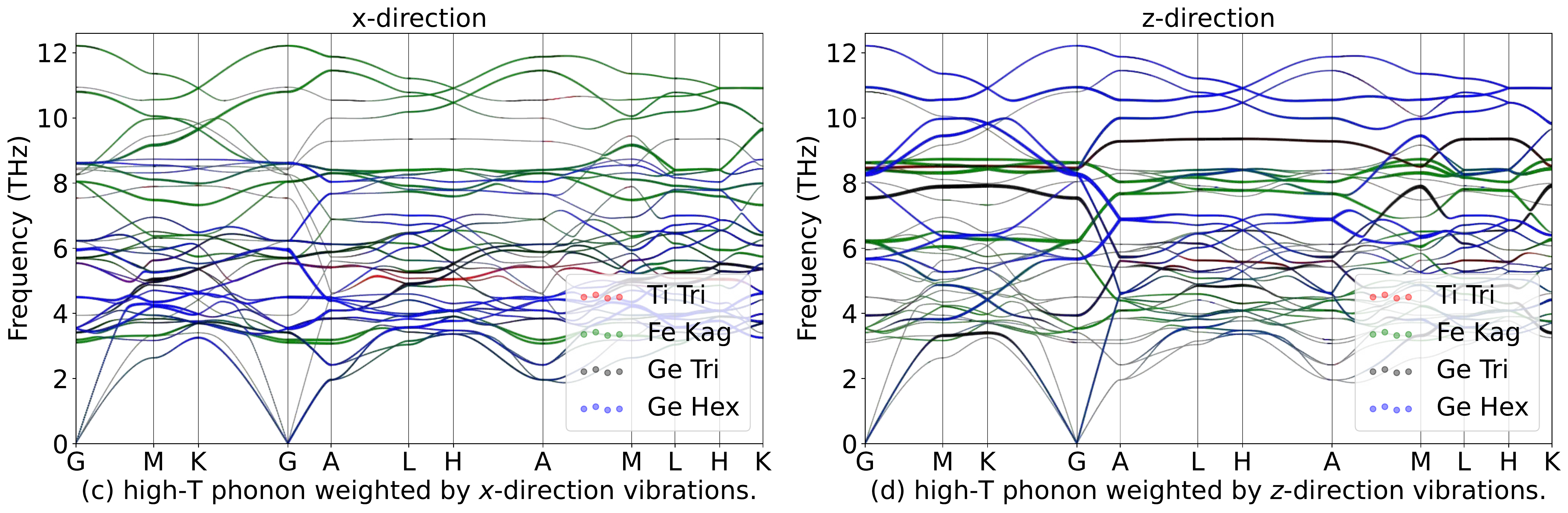}
\caption{Phonon spectrum for TiFe$_6$Ge$_6$in in-plane ($x$) and out-of-plane ($z$) directions in paramagnetic phase.}
\label{fig:TiFe6Ge6phoxyz}
\end{figure}

\begin{table}[htbp]
\global\long\def\arraystretch{1.12}
\captionsetup{width=.95\textwidth}
\caption{IRREPs at high-symmetry points for TiFe$_6$Ge$_6$ phonon spectrum at Low-T.}
\label{tab:191_TiFe6Ge6_Low}
\setlength{\tabcolsep}{1.mm}{
}
\end{table}

\begin{table}[htbp]
\global\long\def\arraystretch{1.12}
\captionsetup{width=.95\textwidth}
\caption{IRREPs at high-symmetry points for TiFe$_6$Ge$_6$ phonon spectrum at High-T.}
\label{tab:191_TiFe6Ge6_High}
\setlength{\tabcolsep}{1.mm}{
%
}
\end{table}
\subsubsection{\label{sms2sec:YFe6Ge6}\hyperref[smsec:col]{\texorpdfstring{YFe$_6$Ge$_6$}{}}~{\color{blue}  (High-T stable, Low-T unstable) }}

\begin{table}[htbp]
\global\long\def\arraystretch{1.12}
\captionsetup{width=.85\textwidth}
\caption{Structure parameters of YFe$_6$Ge$_6$ after relaxation. It contains 387 electrons. }
\label{tab:YFe6Ge6}
\setlength{\tabcolsep}{4mm}{
%
}
\end{table}

\begin{figure}[htbp]
\centering
\includegraphics[width=0.85\textwidth]{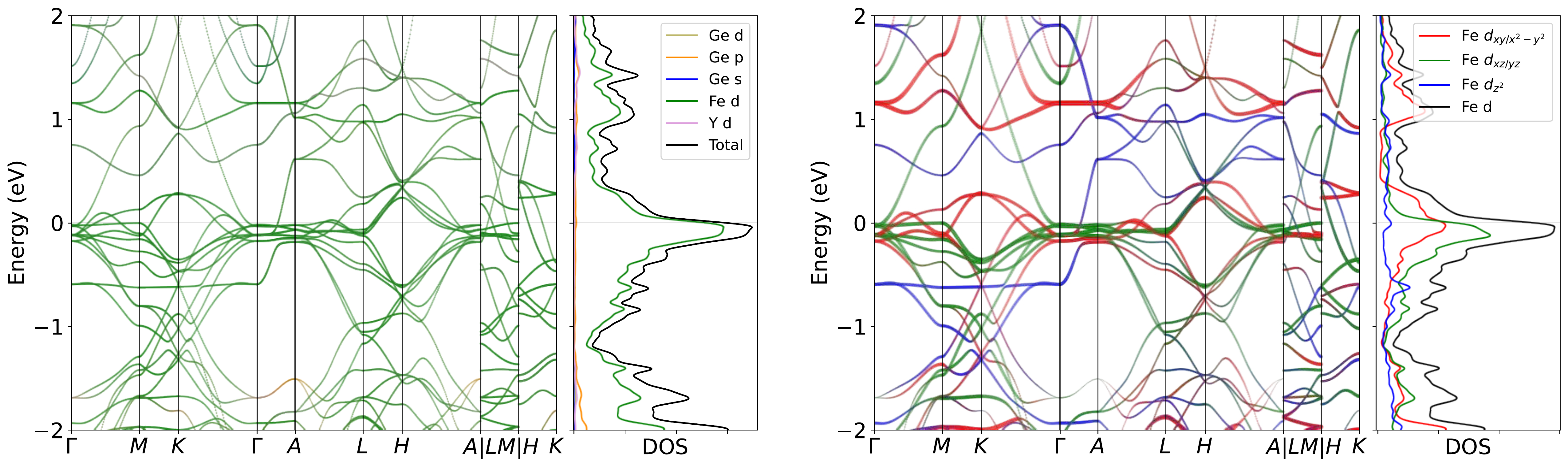}
\caption{Band structure for YFe$_6$Ge$_6$ without SOC in paramagnetic phase. The projection of Fe $3d$ orbitals is also presented. The band structures clearly reveal the dominating of Fe $3d$ orbitals  near the Fermi level, as well as the pronounced peak.}
\label{fig:YFe6Ge6band}
\end{figure}

\begin{figure}[H]
\centering
\subfloat[Relaxed phonon at low-T.]{
\label{fig:YFe6Ge6_relaxed_Low}
\includegraphics[width=0.45\textwidth]{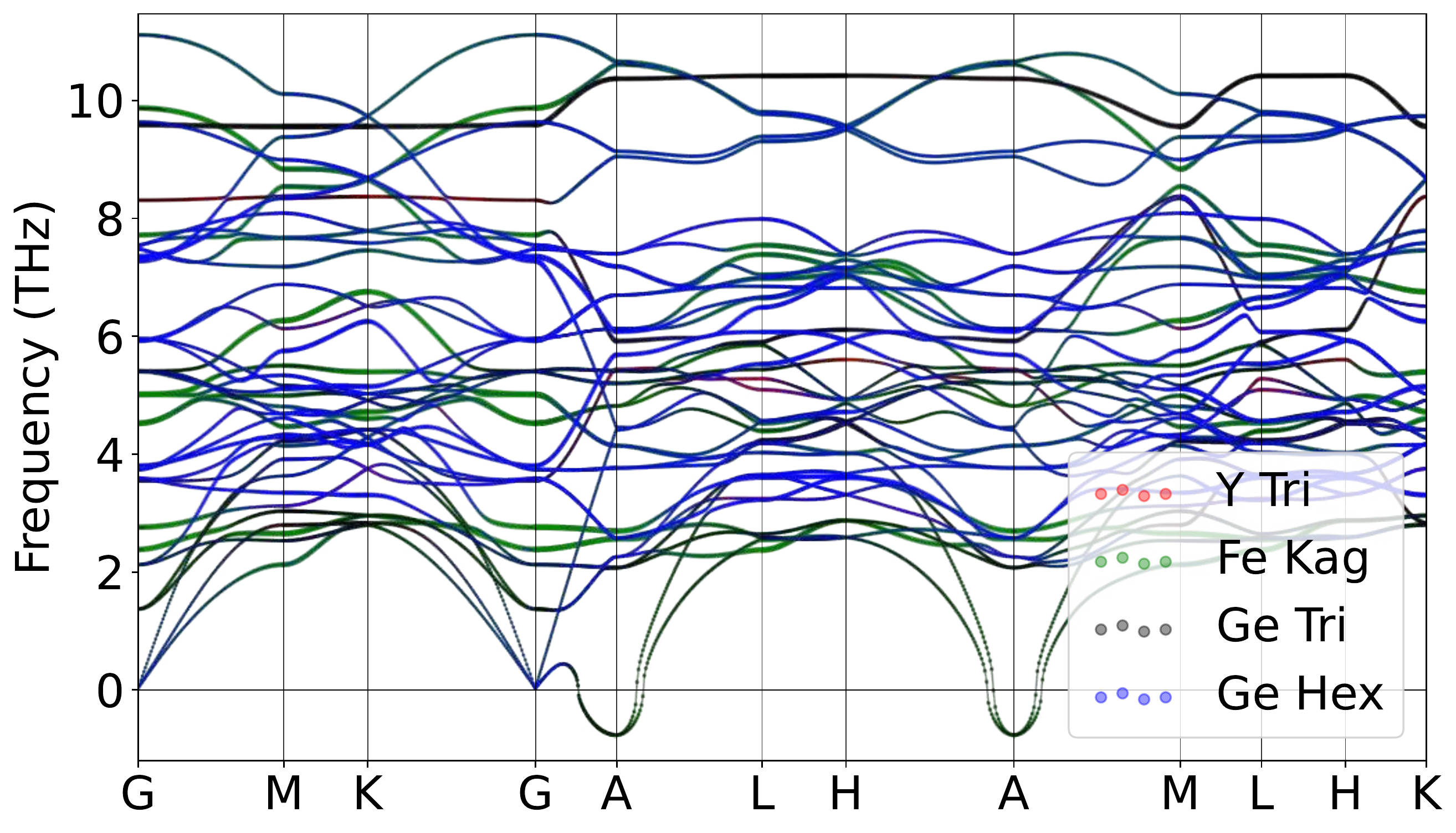}
}
\hspace{5pt}\subfloat[Relaxed phonon at high-T.]{
\label{fig:YFe6Ge6_relaxed_High}
\includegraphics[width=0.45\textwidth]{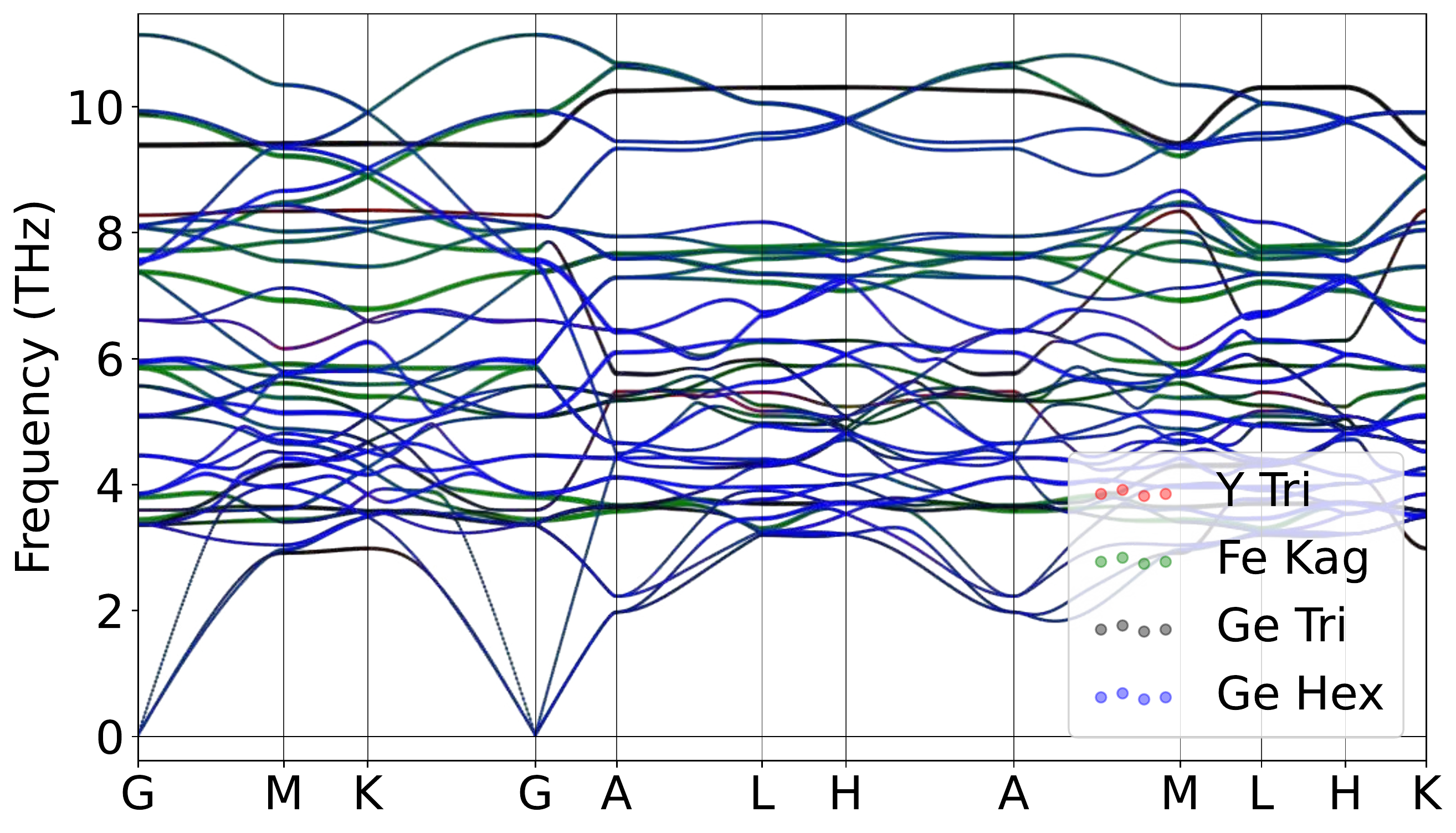}
}
\caption{Atomically projected phonon spectrum for YFe$_6$Ge$_6$ in paramagnetic phase.}
\label{fig:YFe6Ge6pho}
\end{figure}

\begin{figure}[htbp]
\centering
\includegraphics[width=0.45\textwidth]{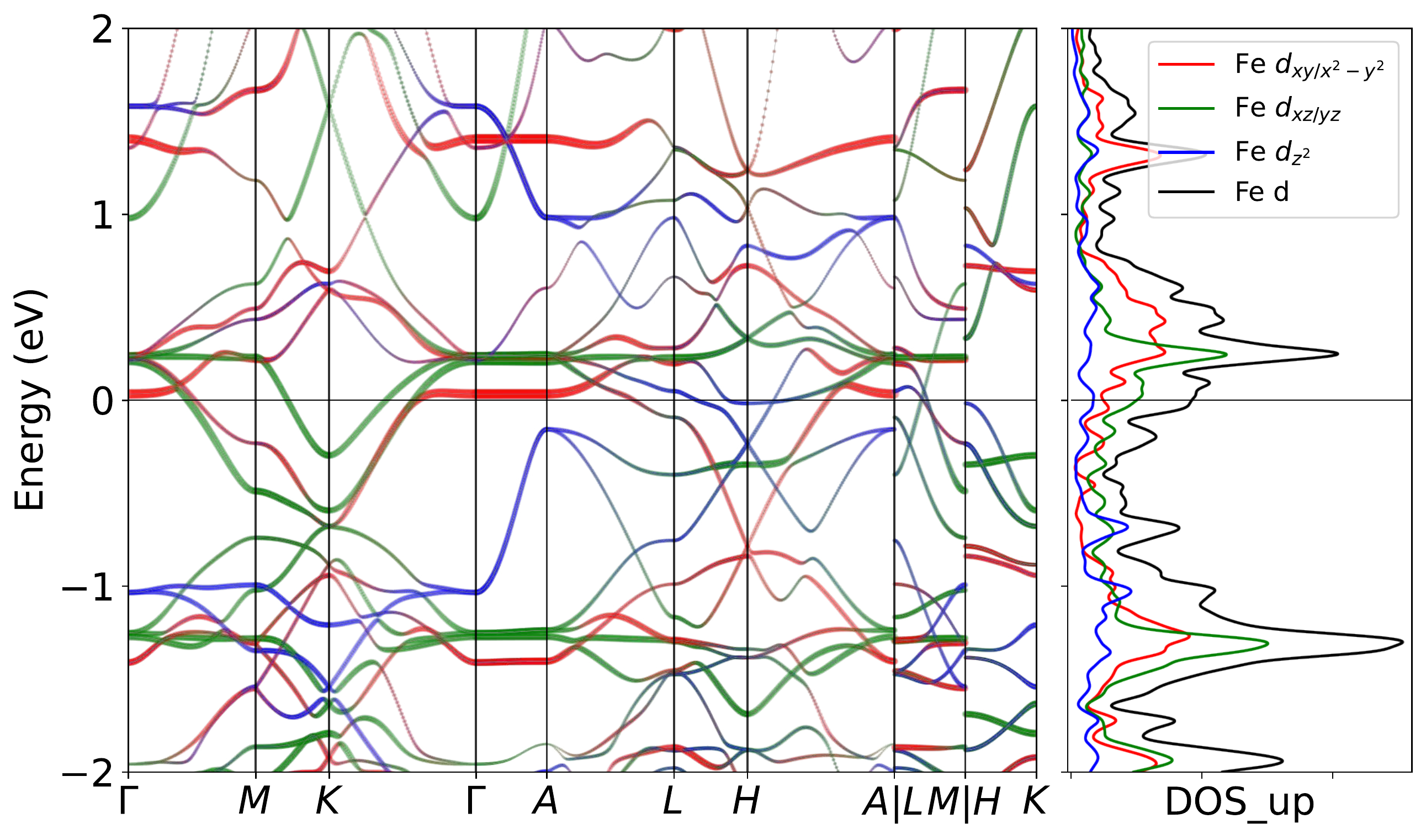}
\includegraphics[width=0.45\textwidth]{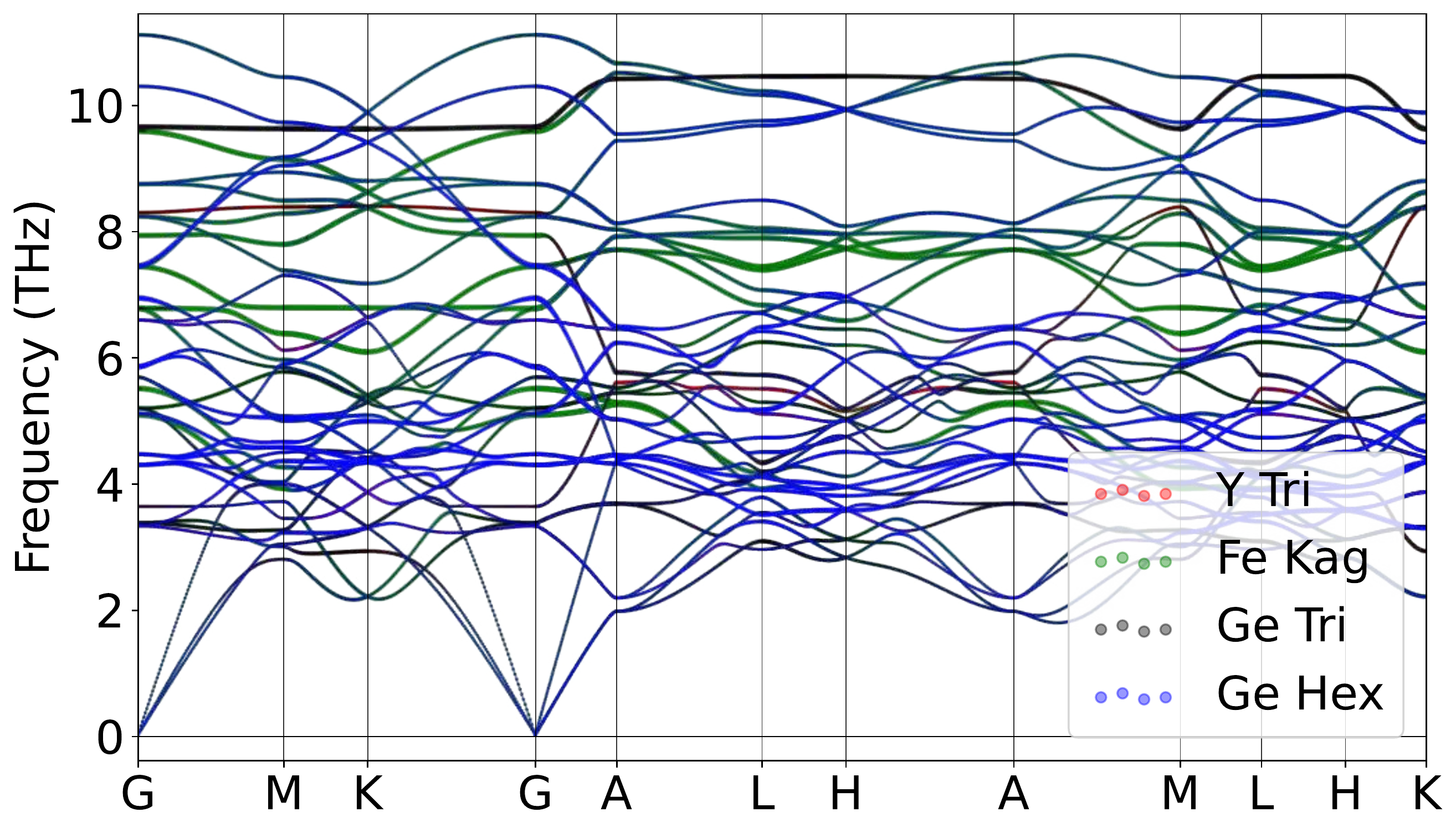}
\caption{Band structure for YFe$_6$Ge$_6$ with AFM order without SOC for spin (Left)up channel. The projection of Fe $3d$ orbitals is also presented. (Right)Correspondingly, a stable phonon was demonstrated with the collinear magnetic order without Hubbard U at lower temperature regime, weighted by atomic projections.}
\label{fig:YFe6Ge6mag}
\end{figure}

\begin{figure}[H]
\centering
\label{fig:YFe6Ge6_relaxed_Low_xz}
\includegraphics[width=0.85\textwidth]{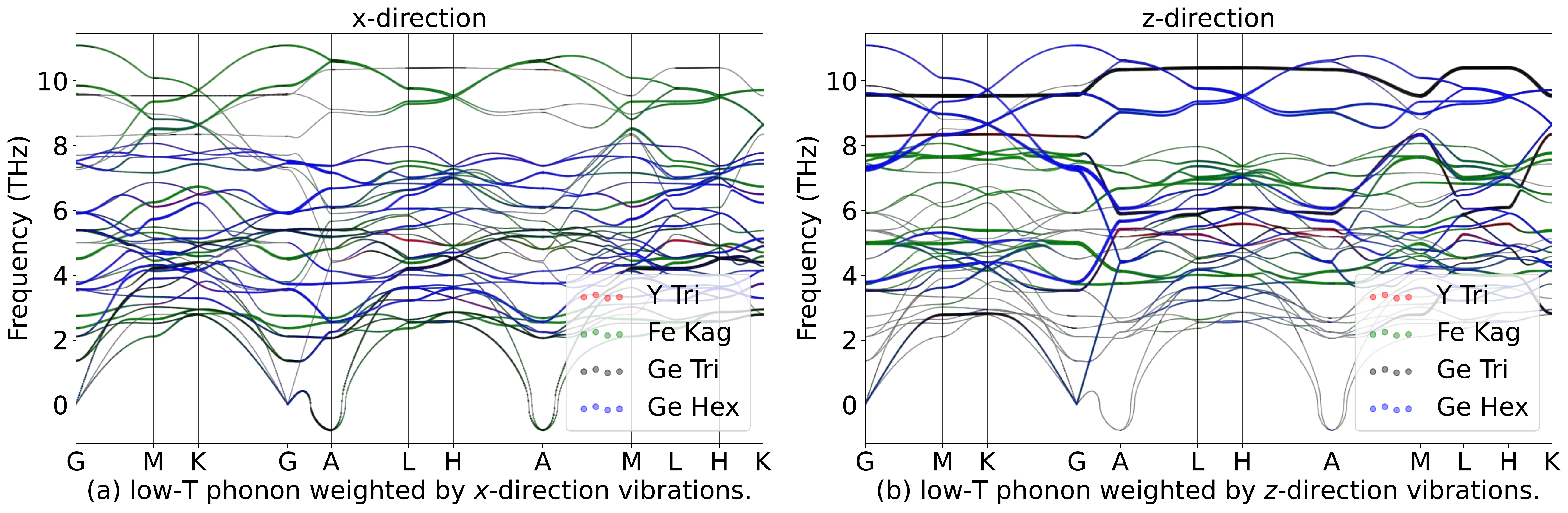}
\hspace{5pt}\label{fig:YFe6Ge6_relaxed_High_xz}
\includegraphics[width=0.85\textwidth]{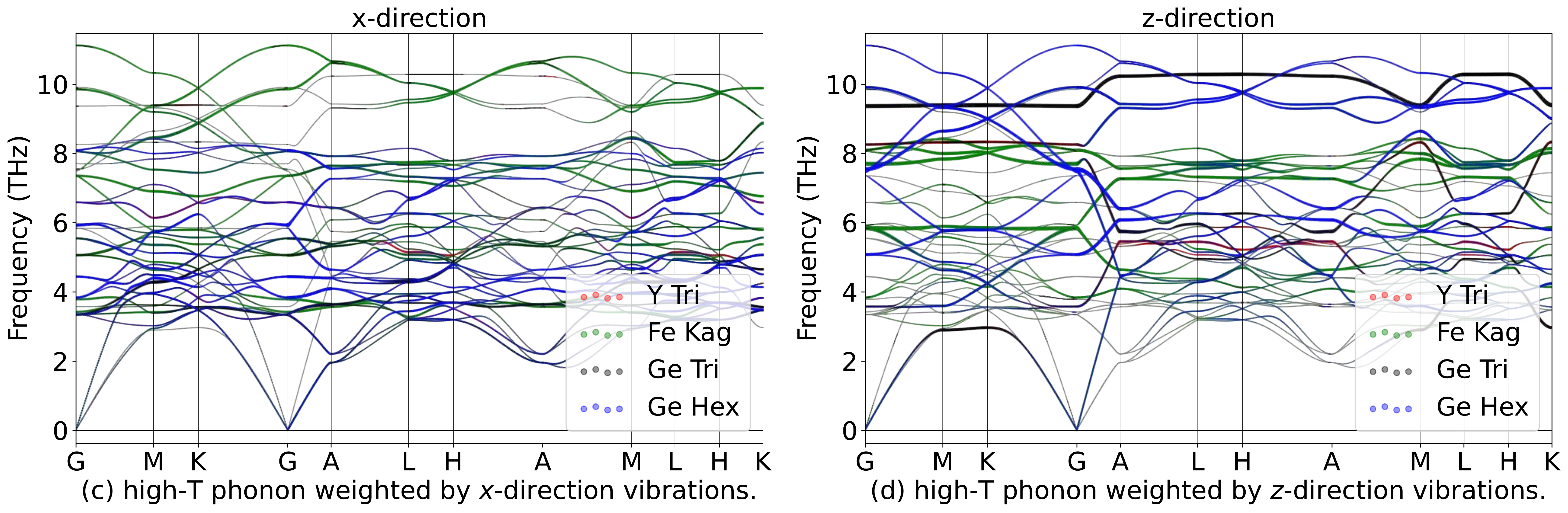}
\caption{Phonon spectrum for YFe$_6$Ge$_6$in in-plane ($x$) and out-of-plane ($z$) directions in paramagnetic phase.}
\label{fig:YFe6Ge6phoxyz}
\end{figure}

\begin{table}[htbp]
\global\long\def\arraystretch{1.12}
\captionsetup{width=.95\textwidth}
\caption{IRREPs at high-symmetry points for YFe$_6$Ge$_6$ phonon spectrum at Low-T.}
\label{tab:191_YFe6Ge6_Low}
\setlength{\tabcolsep}{1.mm}{
}
\end{table}

\begin{table}[htbp]
\global\long\def\arraystretch{1.12}
\captionsetup{width=.95\textwidth}
\caption{IRREPs at high-symmetry points for YFe$_6$Ge$_6$ phonon spectrum at High-T.}
\label{tab:191_YFe6Ge6_High}
\setlength{\tabcolsep}{1.mm}{
%
}
\end{table}
\subsubsection{\label{sms2sec:ZrFe6Ge6}\hyperref[smsec:col]{\texorpdfstring{ZrFe$_6$Ge$_6$}{}}~{\color{blue}  (High-T stable, Low-T unstable) }}

\begin{table}[htbp]
\global\long\def\arraystretch{1.12}
\captionsetup{width=.85\textwidth}
\caption{Structure parameters of ZrFe$_6$Ge$_6$ after relaxation. It contains 388 electrons. }
\label{tab:ZrFe6Ge6}
\setlength{\tabcolsep}{4mm}{
%
}
\end{table}

\begin{figure}[htbp]
\centering
\includegraphics[width=0.85\textwidth]{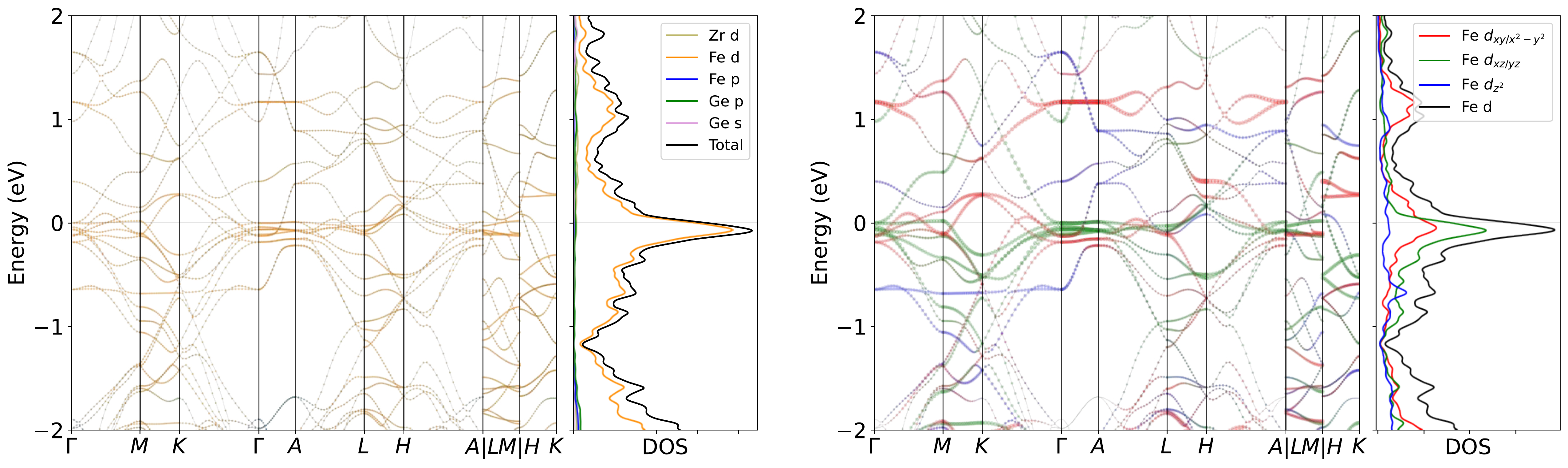}
\caption{Band structure for ZrFe$_6$Ge$_6$ without SOC in paramagnetic phase. The projection of Fe $3d$ orbitals is also presented. The band structures clearly reveal the dominating of Fe $3d$ orbitals  near the Fermi level, as well as the pronounced peak.}
\label{fig:ZrFe6Ge6band}
\end{figure}

\begin{figure}[H]
\centering
\subfloat[Relaxed phonon at low-T.]{
\label{fig:ZrFe6Ge6_relaxed_Low}
\includegraphics[width=0.45\textwidth]{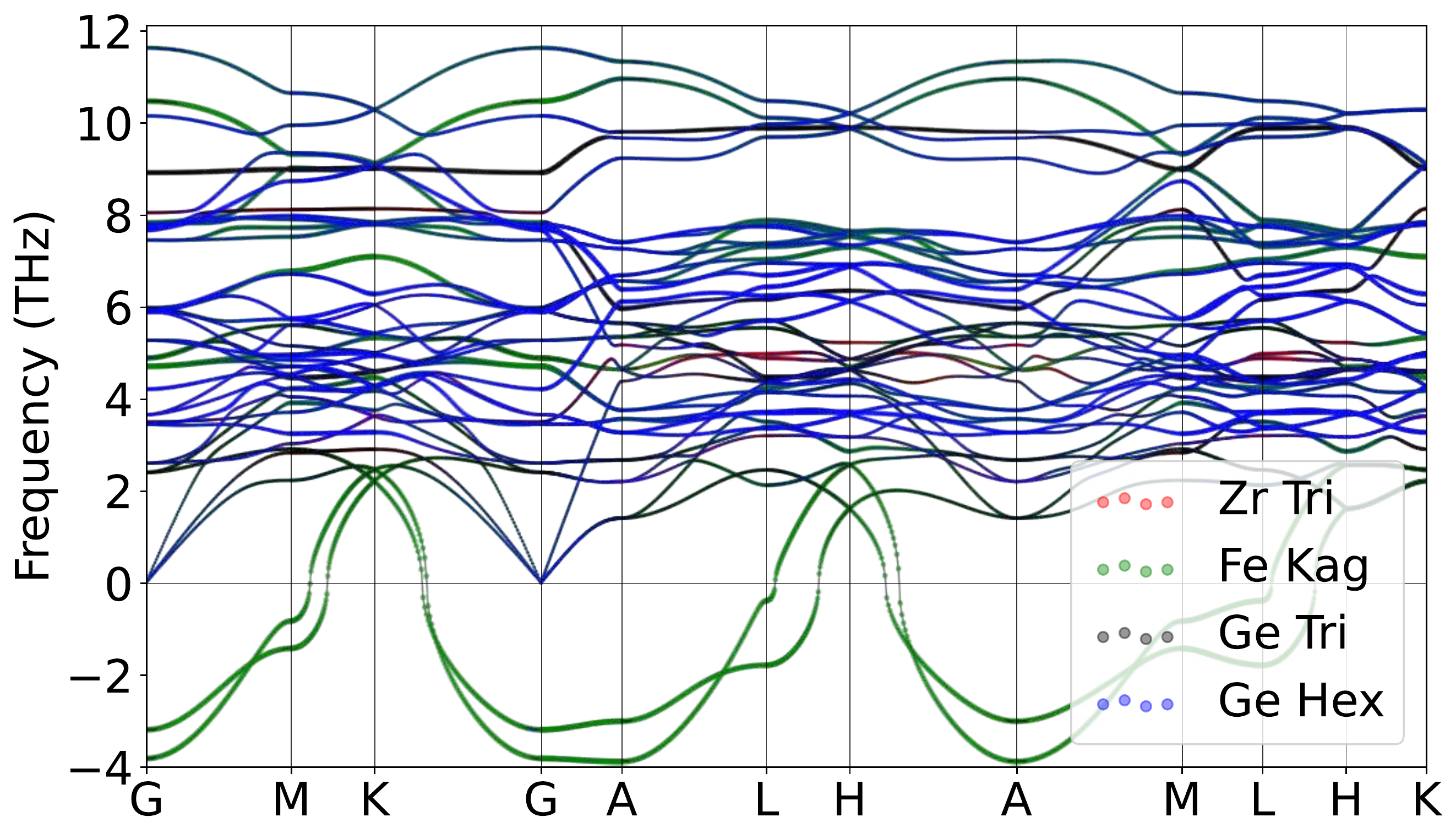}
}
\hspace{5pt}\subfloat[Relaxed phonon at high-T.]{
\label{fig:ZrFe6Ge6_relaxed_High}
\includegraphics[width=0.45\textwidth]{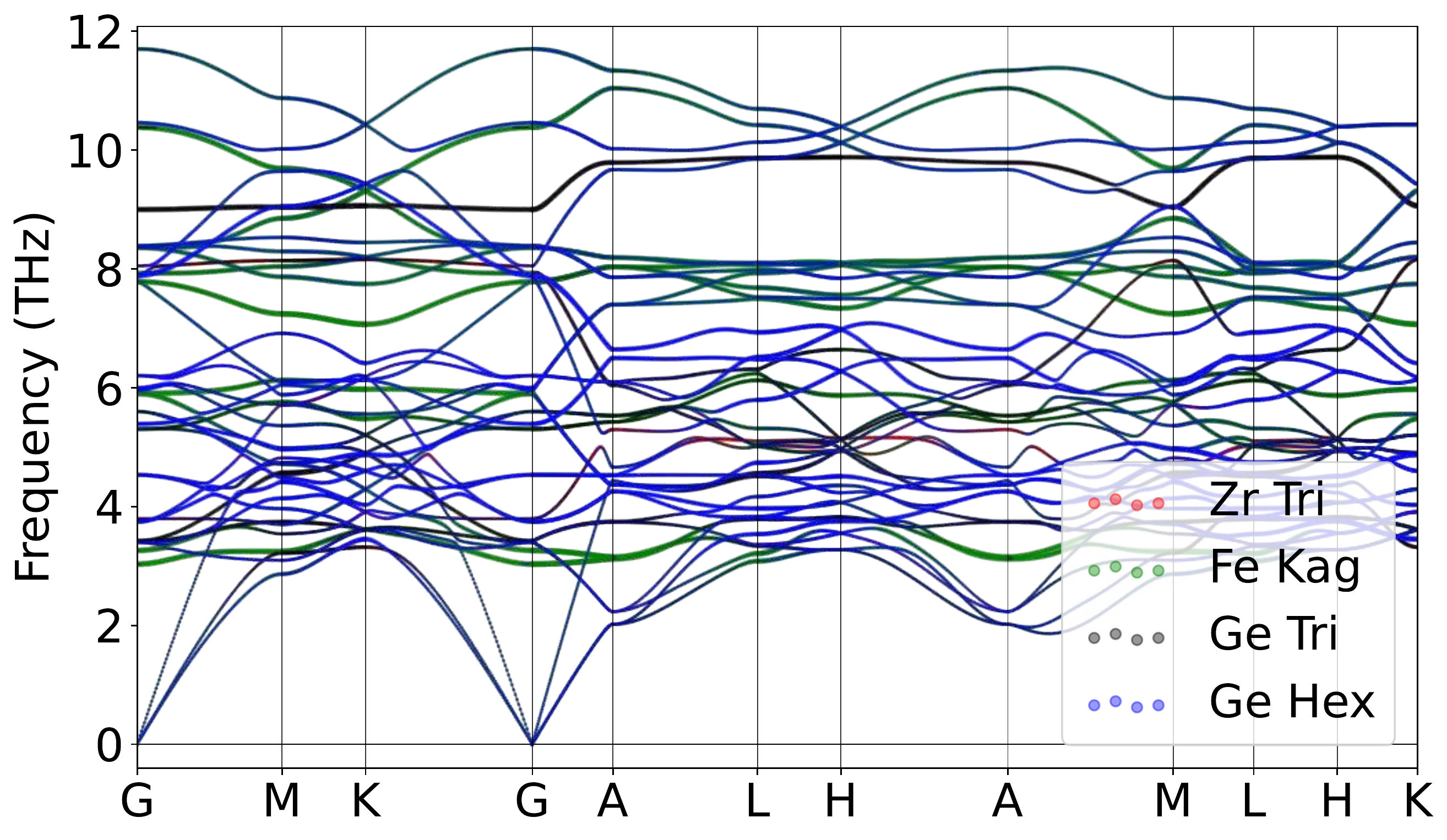}
}
\caption{Atomically projected phonon spectrum for ZrFe$_6$Ge$_6$ in paramagnetic phase.}
\label{fig:ZrFe6Ge6pho}
\end{figure}

\begin{figure}[htbp]
\centering
\includegraphics[width=0.45\textwidth]{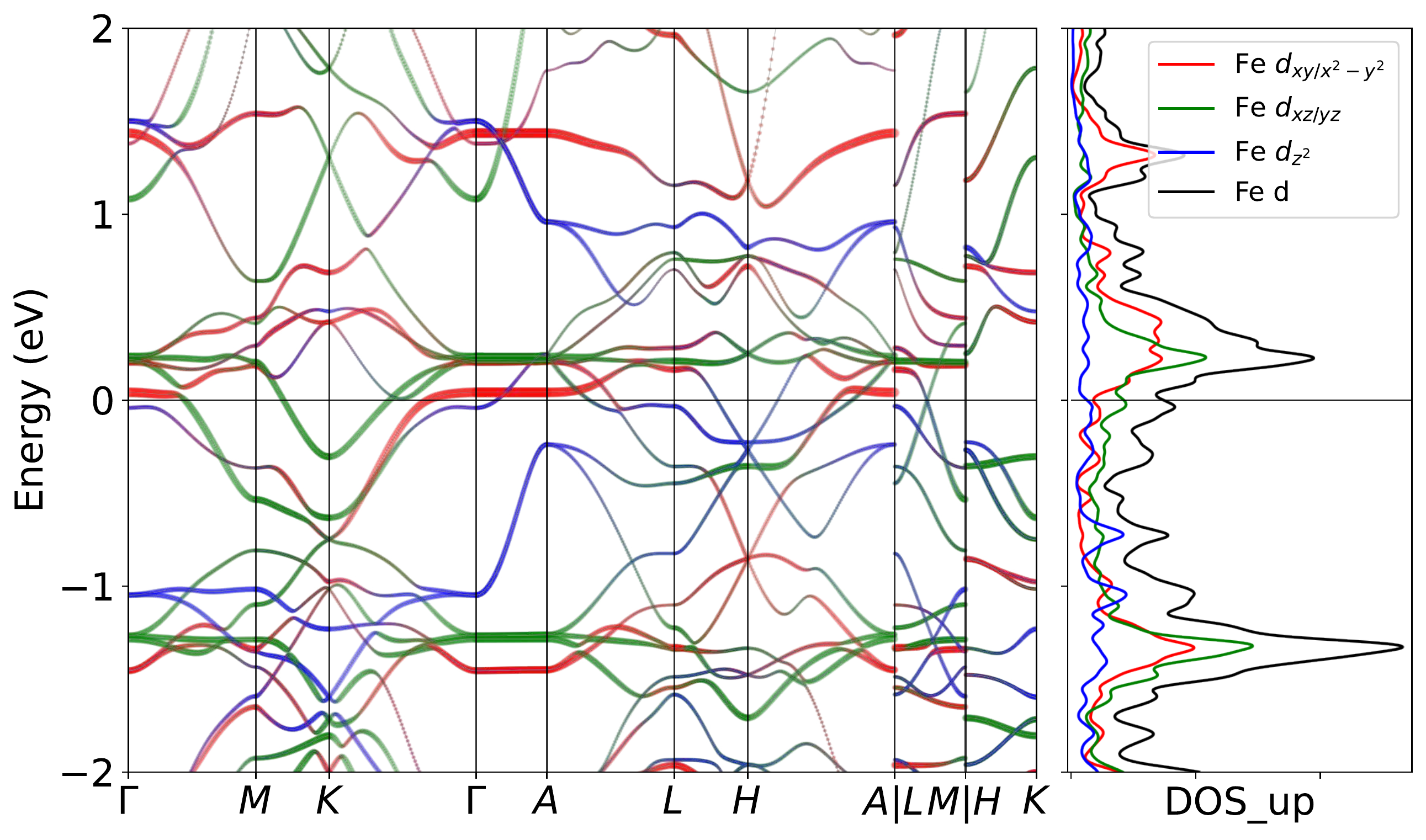}
\includegraphics[width=0.45\textwidth]{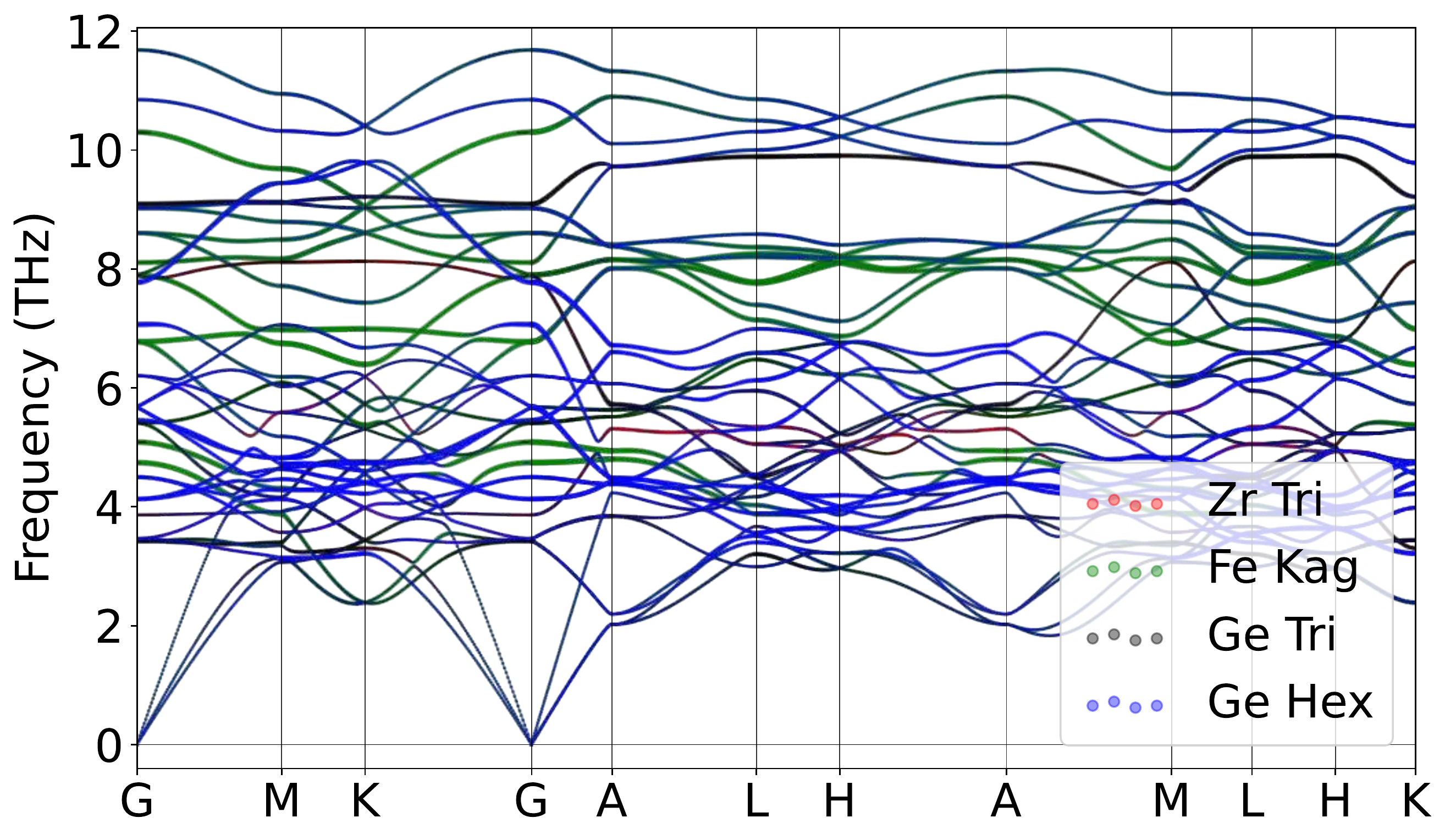}
\caption{Band structure for ZrFe$_6$Ge$_6$ with AFM order without SOC for spin (Left)up channel. The projection of Fe $3d$ orbitals is also presented. (Right)Correspondingly, a stable phonon was demonstrated with the collinear magnetic order without Hubbard U at lower temperature regime, weighted by atomic projections.}
\label{fig:ZrFe6Ge6mag}
\end{figure}

\begin{figure}[H]
\centering
\label{fig:ZrFe6Ge6_relaxed_Low_xz}
\includegraphics[width=0.85\textwidth]{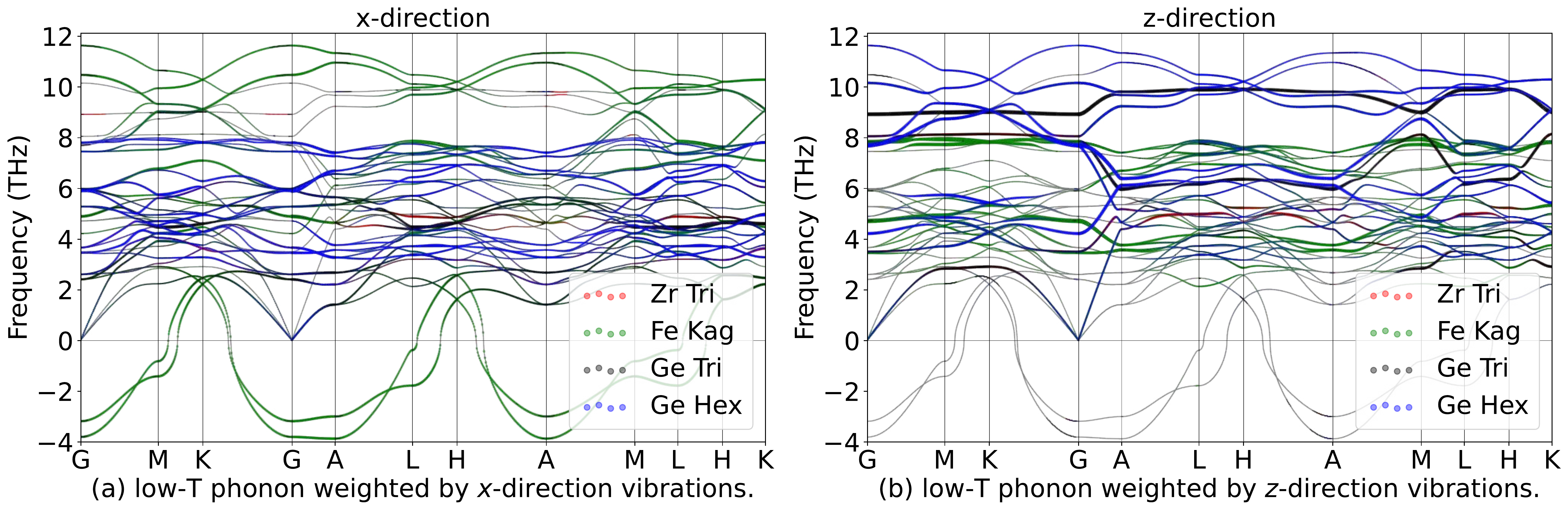}
\hspace{5pt}\label{fig:ZrFe6Ge6_relaxed_High_xz}
\includegraphics[width=0.85\textwidth]{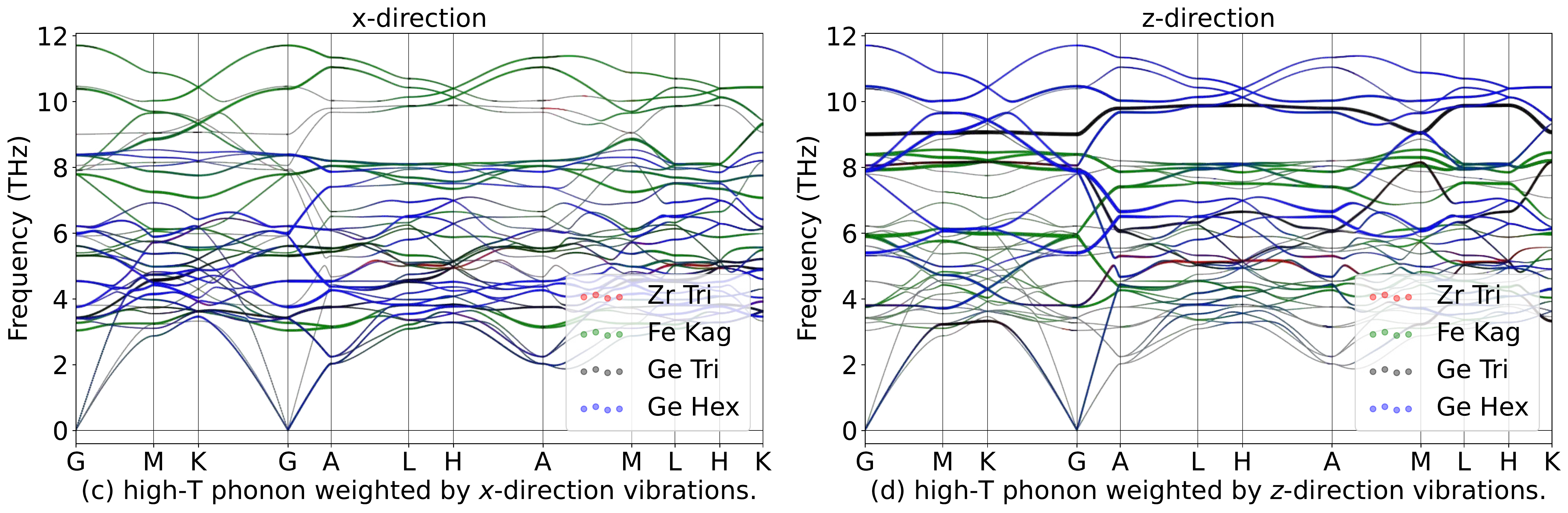}
\caption{Phonon spectrum for ZrFe$_6$Ge$_6$in in-plane ($x$) and out-of-plane ($z$) directions in paramagnetic phase.}
\label{fig:ZrFe6Ge6phoxyz}
\end{figure}

\begin{table}[htbp]
\global\long\def\arraystretch{1.12}
\captionsetup{width=.95\textwidth}
\caption{IRREPs at high-symmetry points for ZrFe$_6$Ge$_6$ phonon spectrum at Low-T.}
\label{tab:191_ZrFe6Ge6_Low}
\setlength{\tabcolsep}{1.mm}{
}
\end{table}

\begin{table}[htbp]
\global\long\def\arraystretch{1.12}
\captionsetup{width=.95\textwidth}
\caption{IRREPs at high-symmetry points for ZrFe$_6$Ge$_6$ phonon spectrum at High-T.}
\label{tab:191_ZrFe6Ge6_High}
\setlength{\tabcolsep}{1.mm}{
%
}
\end{table}
\subsubsection{\label{sms2sec:NbFe6Ge6}\hyperref[smsec:col]{\texorpdfstring{NbFe$_6$Ge$_6$}{}}~{\color{blue}  (High-T stable, Low-T unstable) }}

\begin{table}[htbp]
\global\long\def\arraystretch{1.12}
\captionsetup{width=.85\textwidth}
\caption{Structure parameters of NbFe$_6$Ge$_6$ after relaxation. It contains 389 electrons. }
\label{tab:NbFe6Ge6}
\setlength{\tabcolsep}{4mm}{
%
}
\end{table}

\begin{figure}[htbp]
\centering
\includegraphics[width=0.85\textwidth]{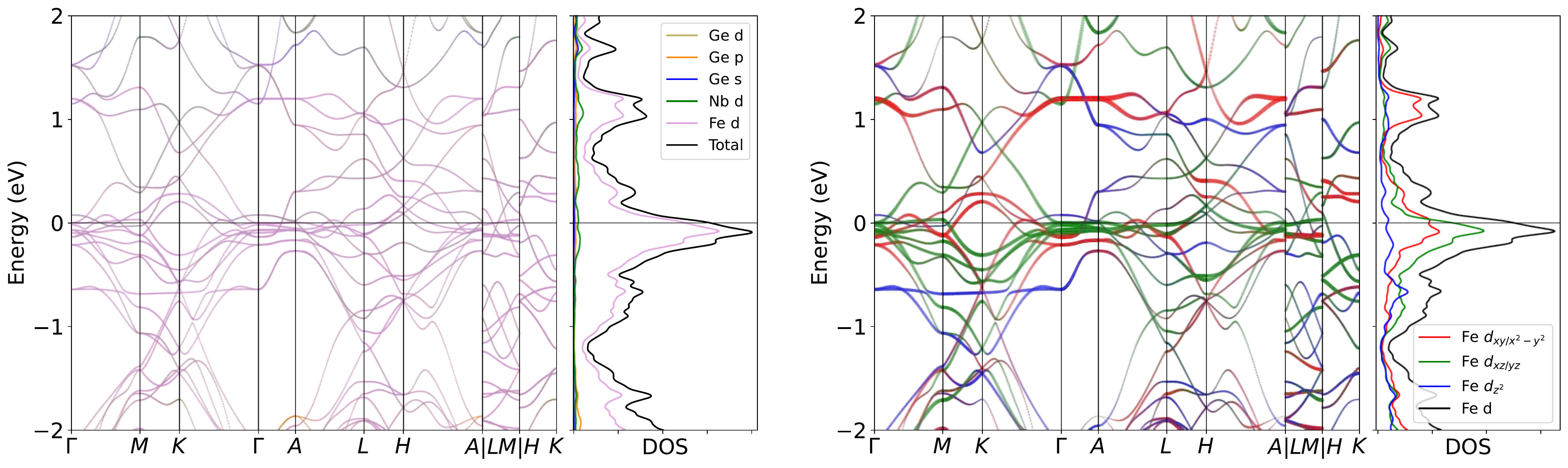}
\caption{Band structure for NbFe$_6$Ge$_6$ without SOC in paramagnetic phase. The projection of Fe $3d$ orbitals is also presented. The band structures clearly reveal the dominating of Fe $3d$ orbitals  near the Fermi level, as well as the pronounced peak.}
\label{fig:NbFe6Ge6band}
\end{figure}

\begin{figure}[H]
\centering
\subfloat[Relaxed phonon at low-T.]{
\label{fig:NbFe6Ge6_relaxed_Low}
\includegraphics[width=0.45\textwidth]{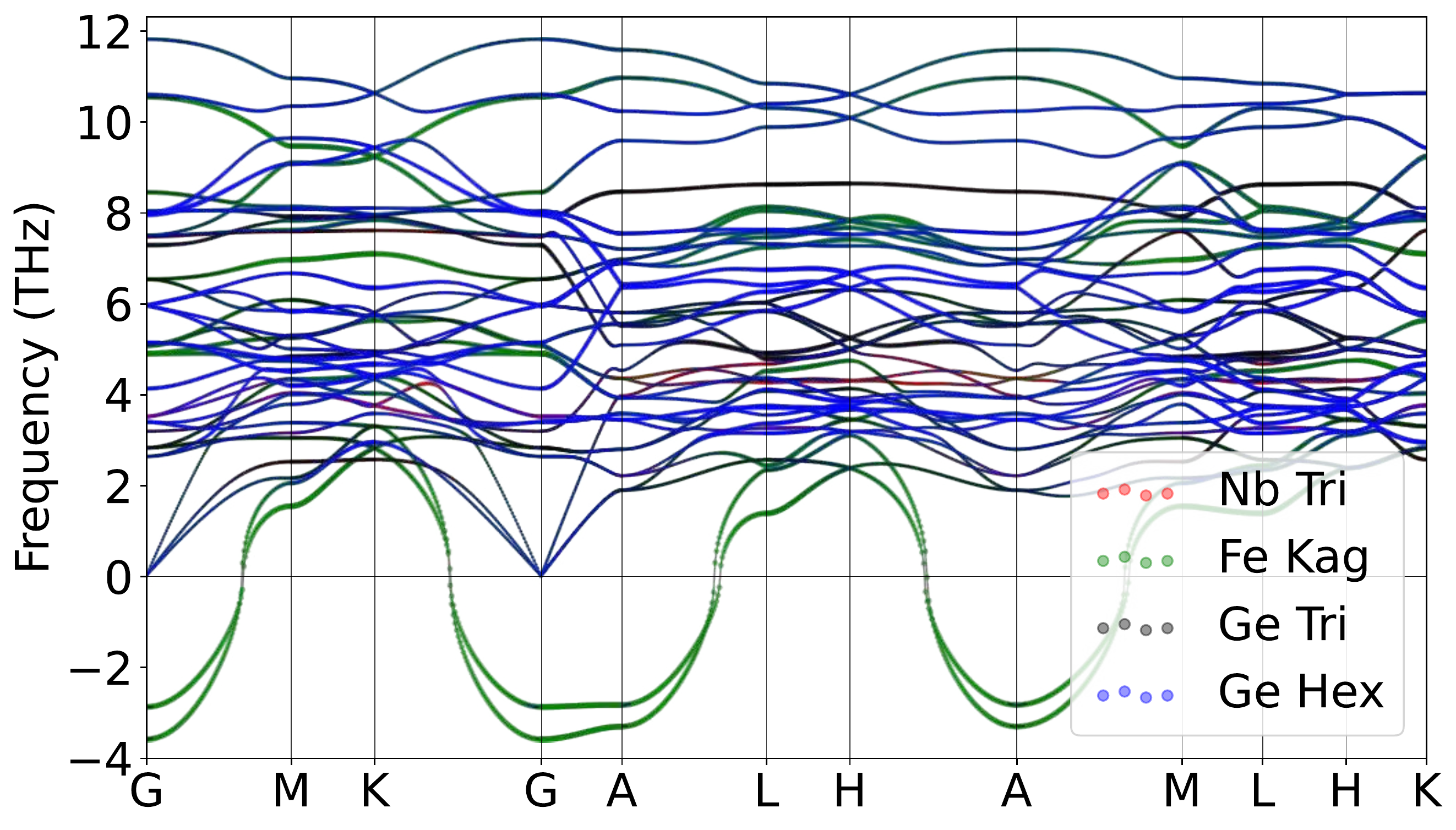}
}
\hspace{5pt}\subfloat[Relaxed phonon at high-T.]{
\label{fig:NbFe6Ge6_relaxed_High}
\includegraphics[width=0.45\textwidth]{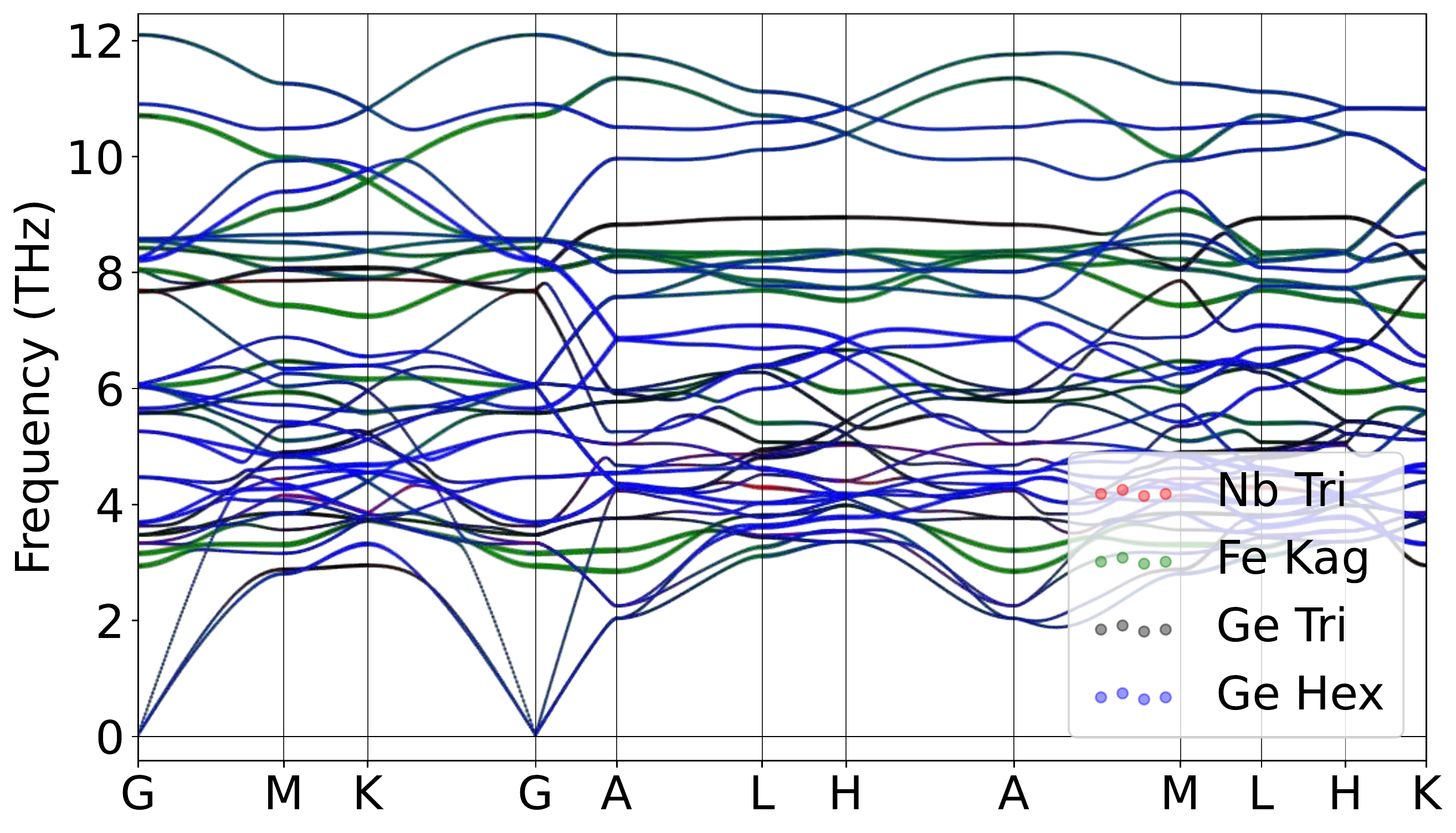}
}
\caption{Atomically projected phonon spectrum for NbFe$_6$Ge$_6$ in paramagnetic phase.}
\label{fig:NbFe6Ge6pho}
\end{figure}

\begin{figure}[htbp]
\centering
\includegraphics[width=0.45\textwidth]{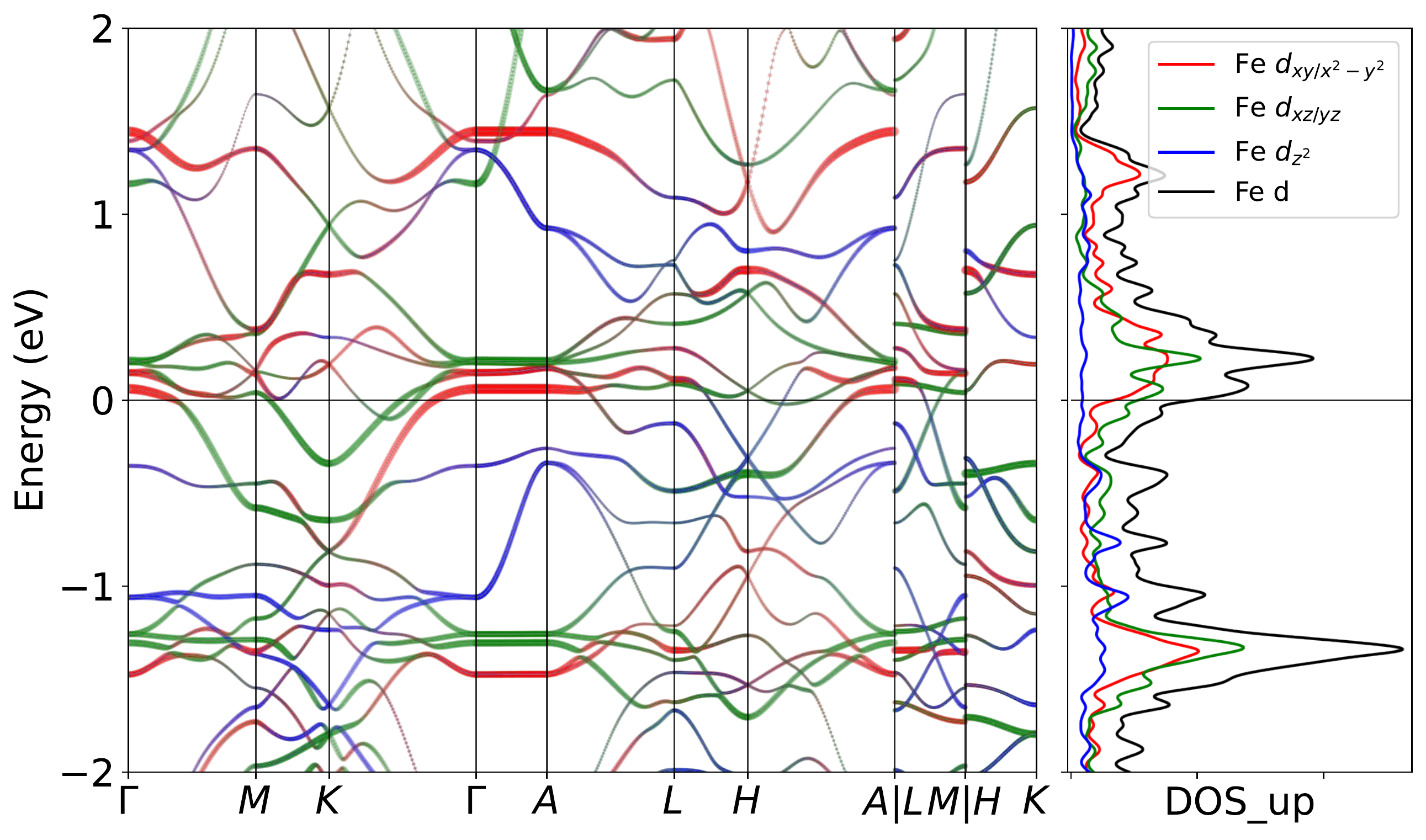}
\includegraphics[width=0.45\textwidth]{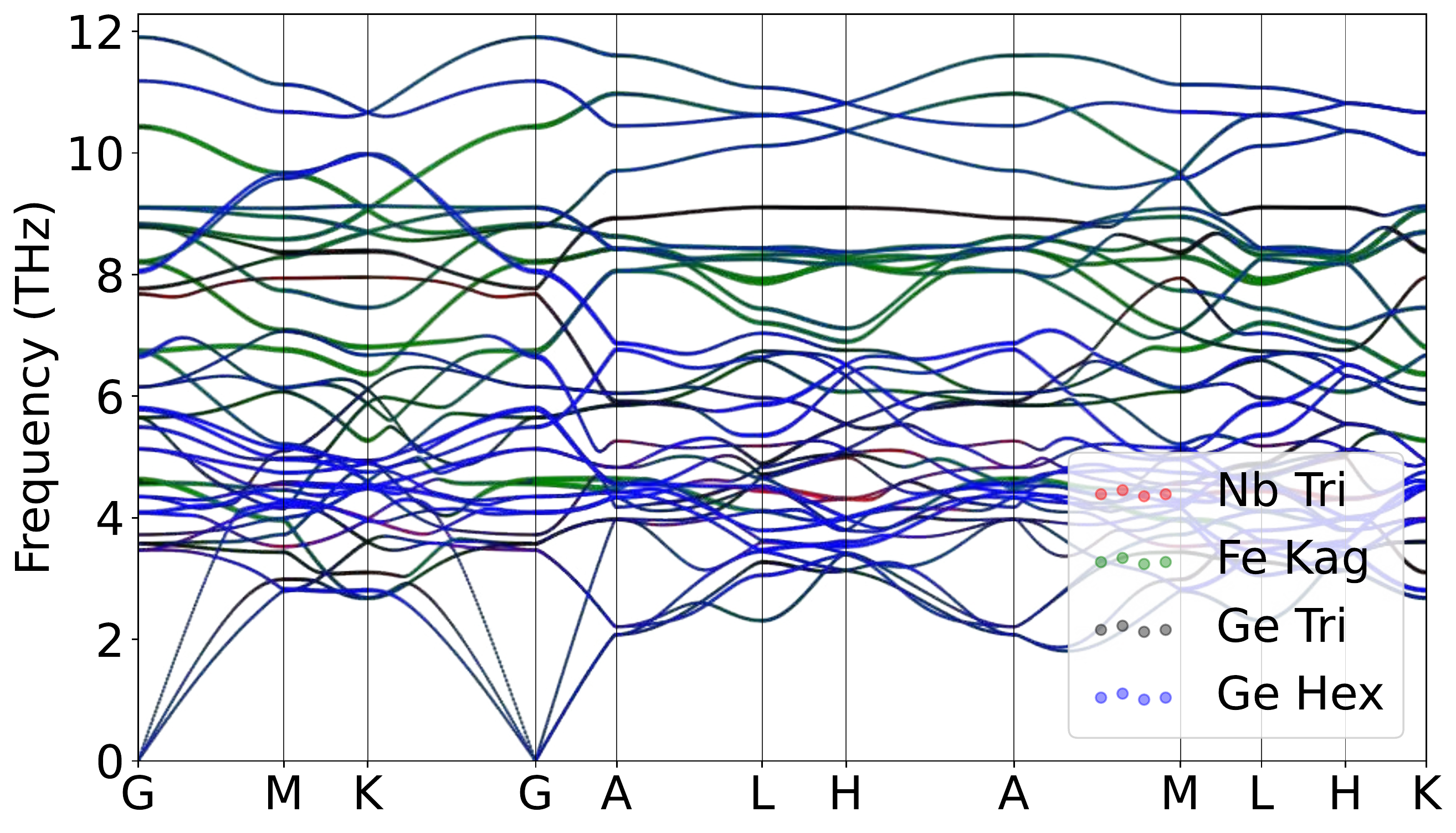}
\caption{Band structure for NbFe$_6$Ge$_6$ with AFM order without SOC for spin (Left)up channel. The projection of Fe $3d$ orbitals is also presented. (Right)Correspondingly, a stable phonon was demonstrated with the collinear magnetic order without Hubbard U at lower temperature regime, weighted by atomic projections.}
\label{fig:NbFe6Ge6mag}
\end{figure}

\begin{figure}[H]
\centering
\label{fig:NbFe6Ge6_relaxed_Low_xz}
\includegraphics[width=0.85\textwidth]{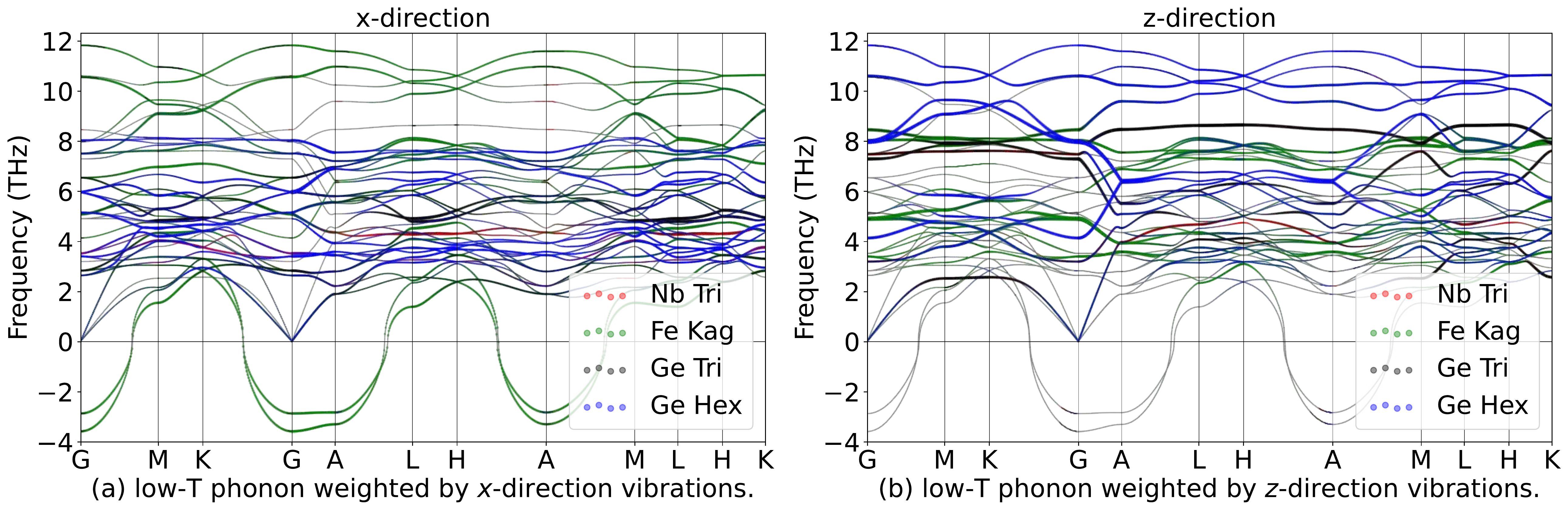}
\hspace{5pt}\label{fig:NbFe6Ge6_relaxed_High_xz}
\includegraphics[width=0.85\textwidth]{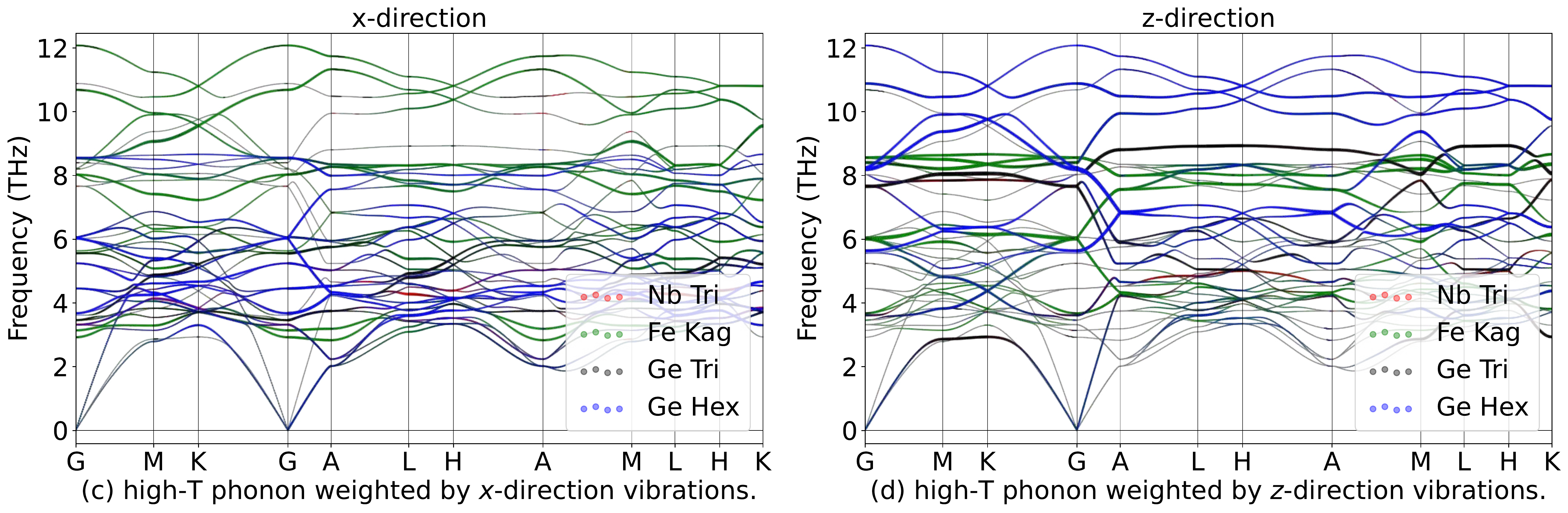}
\caption{Phonon spectrum for NbFe$_6$Ge$_6$in in-plane ($x$) and out-of-plane ($z$) directions in paramagnetic phase.}
\label{fig:NbFe6Ge6phoxyz}
\end{figure}

\begin{table}[htbp]
\global\long\def\arraystretch{1.12}
\captionsetup{width=.95\textwidth}
\caption{IRREPs at high-symmetry points for NbFe$_6$Ge$_6$ phonon spectrum at Low-T.}
\label{tab:191_NbFe6Ge6_Low}
\setlength{\tabcolsep}{1.mm}{
}
\end{table}

\begin{table}[htbp]
\global\long\def\arraystretch{1.12}
\captionsetup{width=.95\textwidth}
\caption{IRREPs at high-symmetry points for NbFe$_6$Ge$_6$ phonon spectrum at High-T.}
\label{tab:191_NbFe6Ge6_High}
\setlength{\tabcolsep}{1.mm}{
%
}
\end{table}
\subsubsection{\label{sms2sec:PrFe6Ge6}\hyperref[smsec:col]{\texorpdfstring{PrFe$_6$Ge$_6$}{}}~{\color{blue}  (High-T stable, Low-T unstable) }}

\begin{table}[htbp]
\global\long\def\arraystretch{1.12}
\captionsetup{width=.85\textwidth}
\caption{Structure parameters of PrFe$_6$Ge$_6$ after relaxation. It contains 407 electrons. }
\label{tab:PrFe6Ge6}
\setlength{\tabcolsep}{4mm}{
%
}
\end{table}

\begin{figure}[htbp]
\centering
\includegraphics[width=0.85\textwidth]{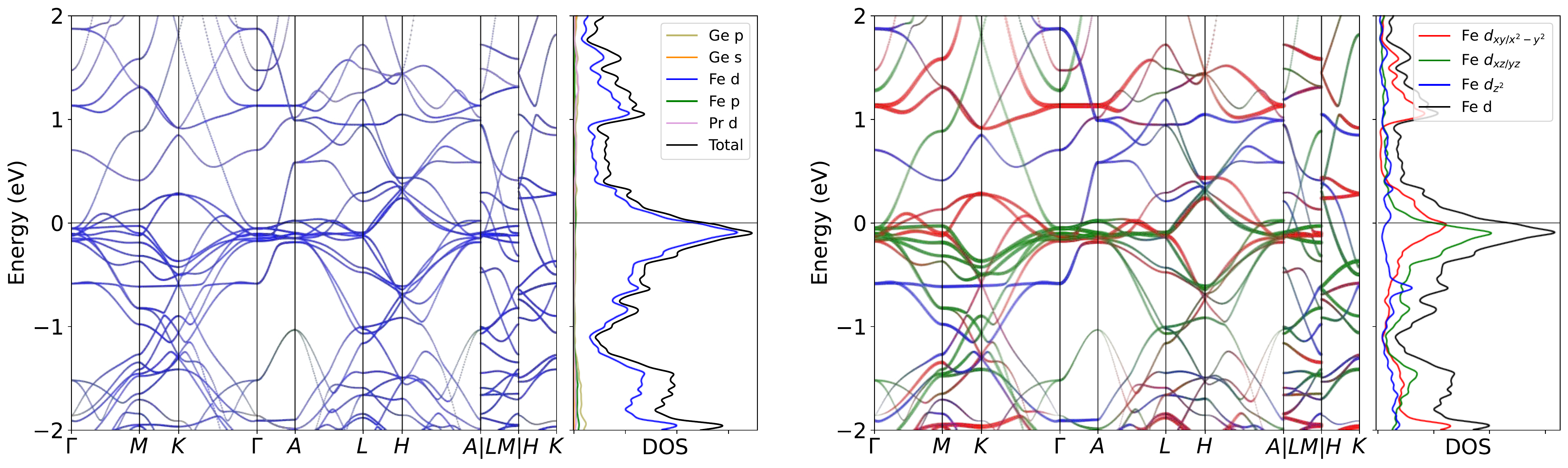}
\caption{Band structure for PrFe$_6$Ge$_6$ without SOC in paramagnetic phase. The projection of Fe $3d$ orbitals is also presented. The band structures clearly reveal the dominating of Fe $3d$ orbitals  near the Fermi level, as well as the pronounced peak.}
\label{fig:PrFe6Ge6band}
\end{figure}

\begin{figure}[H]
\centering
\subfloat[Relaxed phonon at low-T.]{
\label{fig:PrFe6Ge6_relaxed_Low}
\includegraphics[width=0.45\textwidth]{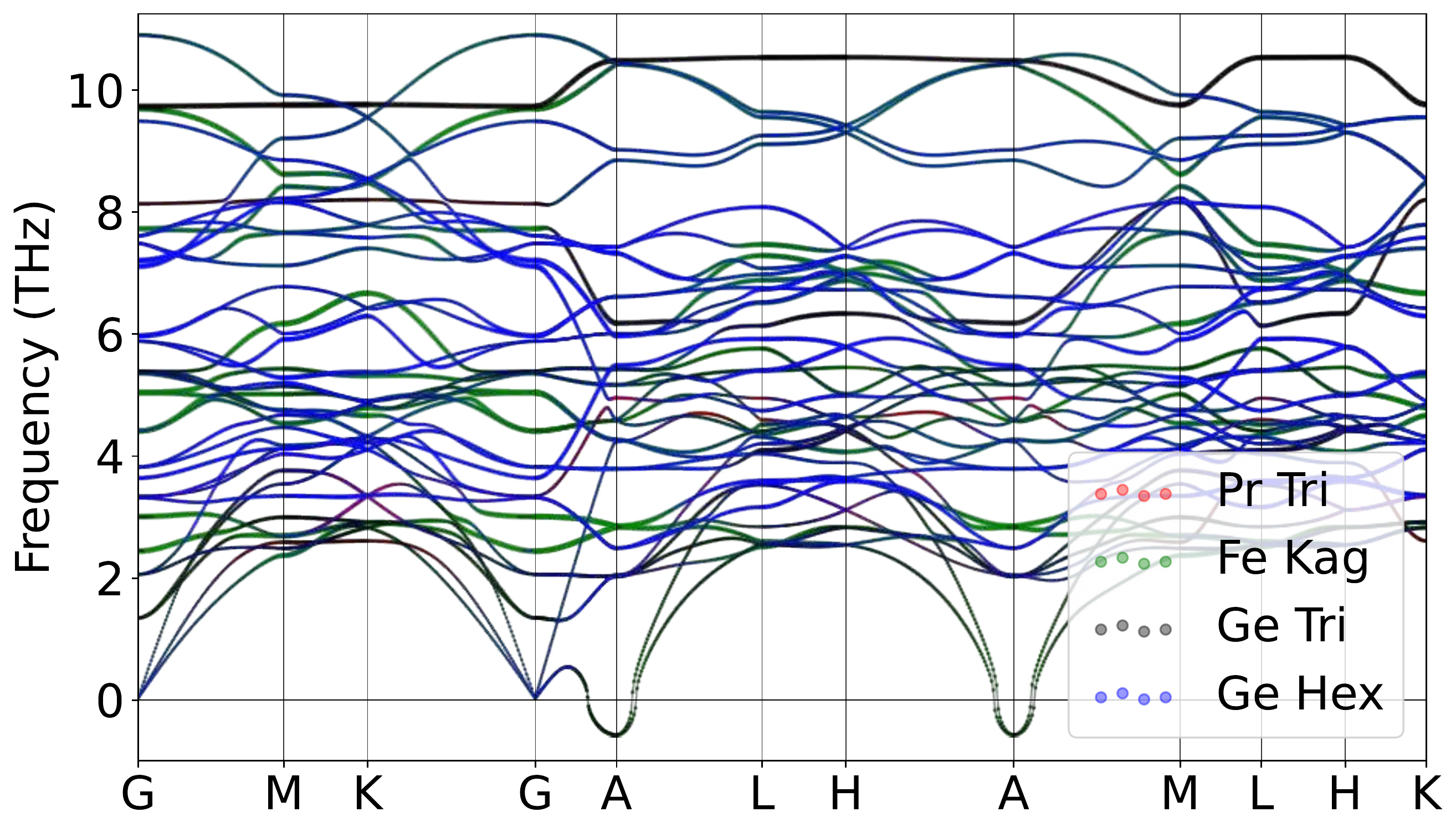}
}
\hspace{5pt}\subfloat[Relaxed phonon at high-T.]{
\label{fig:PrFe6Ge6_relaxed_High}
\includegraphics[width=0.45\textwidth]{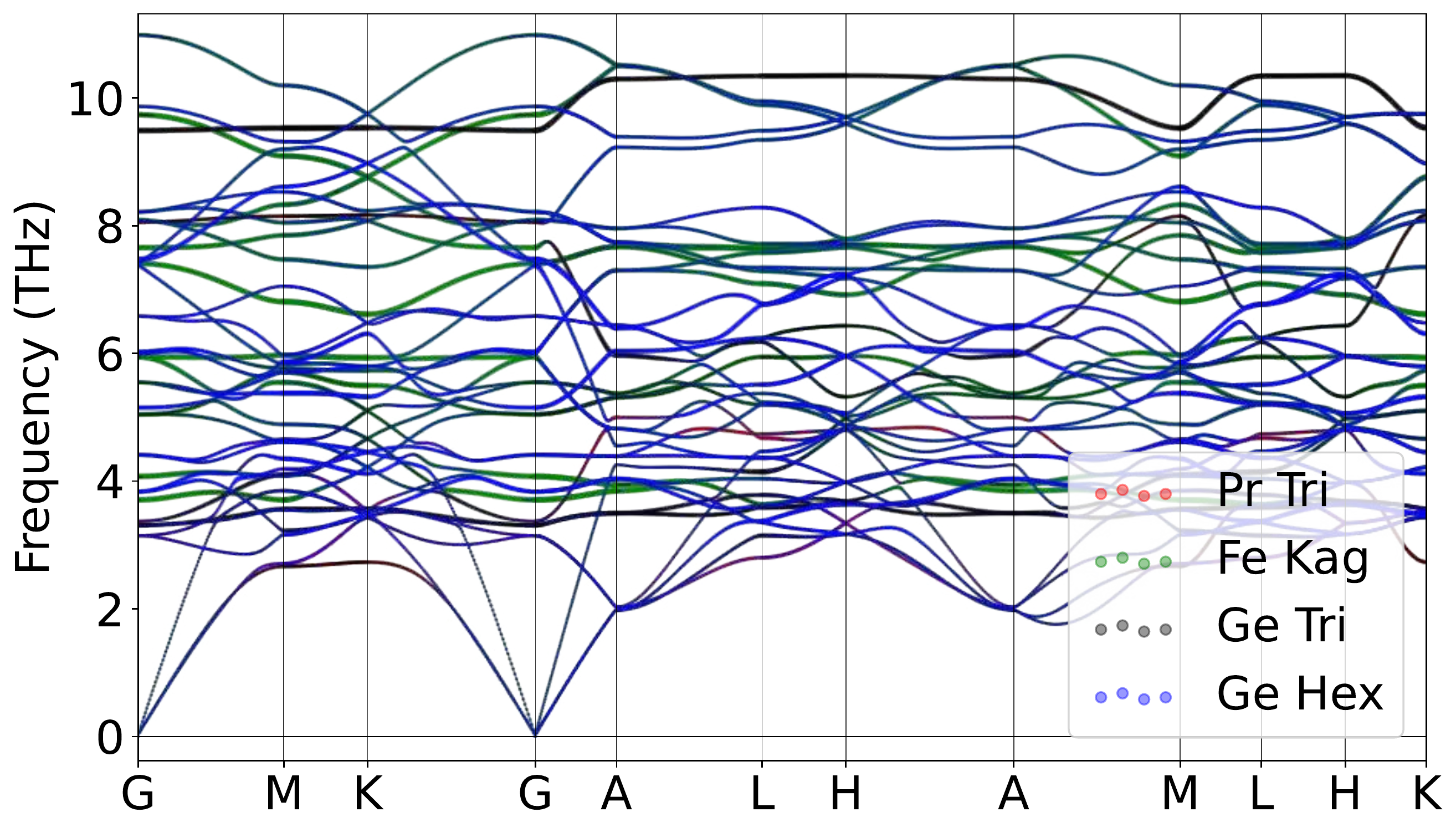}
}
\caption{Atomically projected phonon spectrum for PrFe$_6$Ge$_6$ in paramagnetic phase.}
\label{fig:PrFe6Ge6pho}
\end{figure}

\begin{figure}[htbp]
\centering
\includegraphics[width=0.45\textwidth]{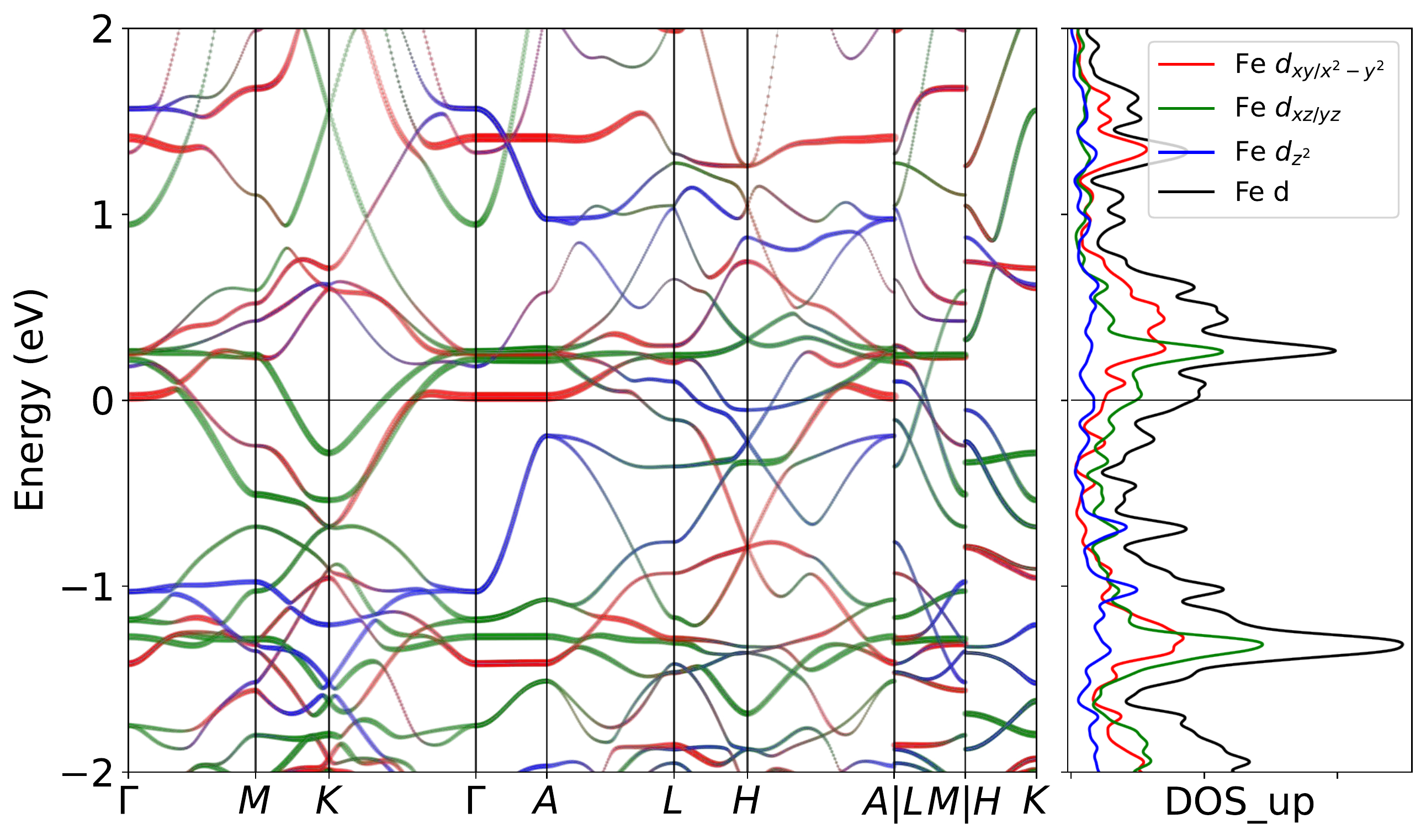}
\includegraphics[width=0.45\textwidth]{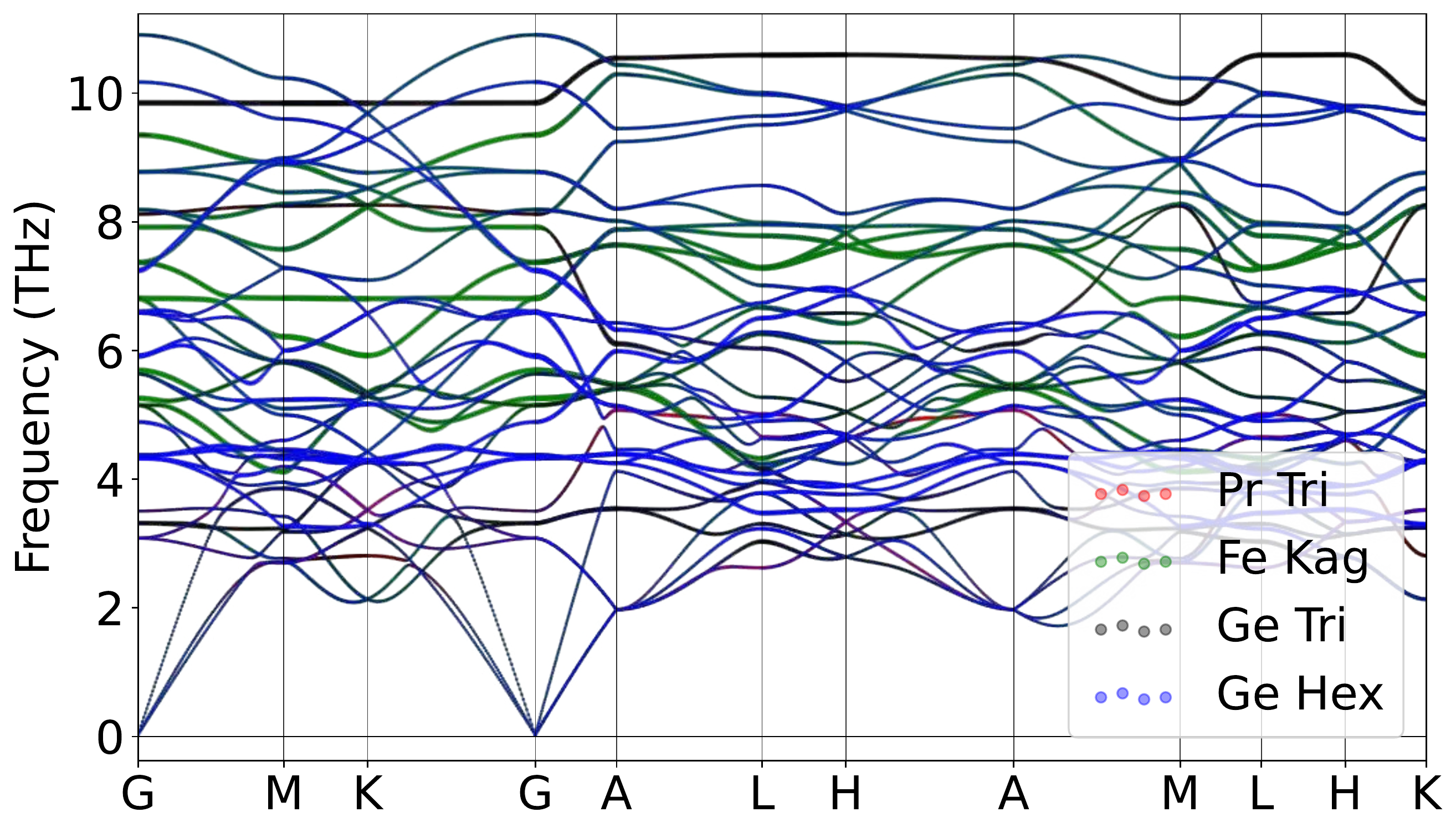}
\caption{Band structure for PrFe$_6$Ge$_6$ with AFM order without SOC for spin (Left)up channel. The projection of Fe $3d$ orbitals is also presented. (Right)Correspondingly, a stable phonon was demonstrated with the collinear magnetic order without Hubbard U at lower temperature regime, weighted by atomic projections.}
\label{fig:PrFe6Ge6mag}
\end{figure}

\begin{figure}[H]
\centering
\label{fig:PrFe6Ge6_relaxed_Low_xz}
\includegraphics[width=0.85\textwidth]{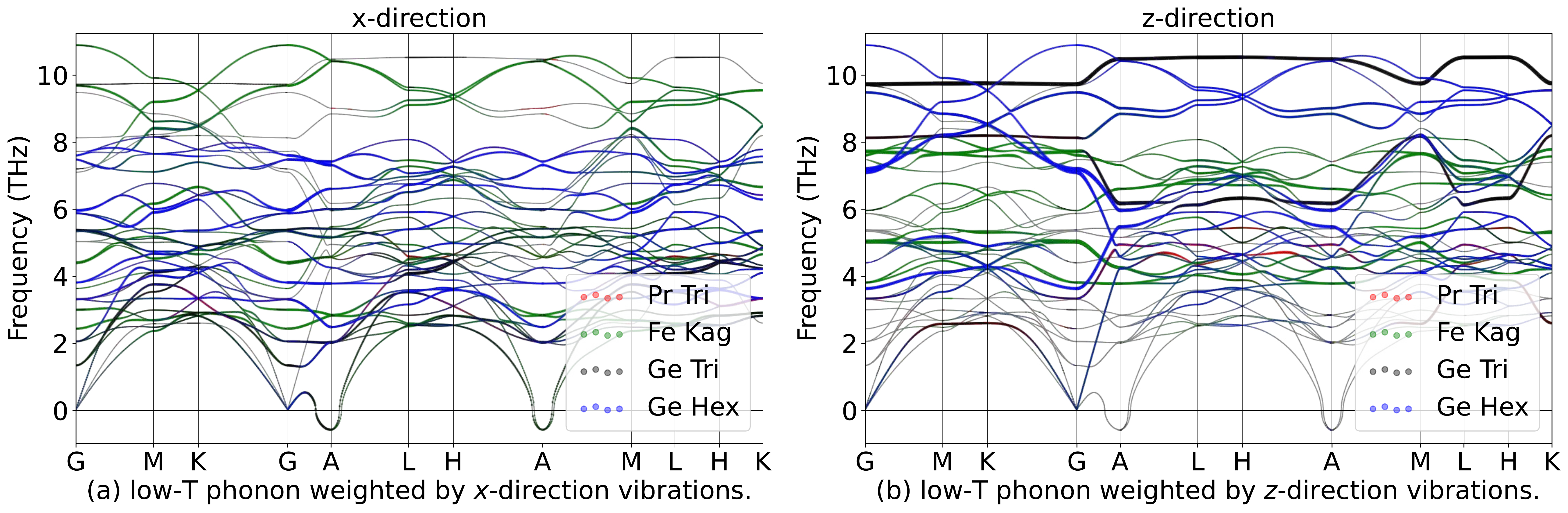}
\hspace{5pt}\label{fig:PrFe6Ge6_relaxed_High_xz}
\includegraphics[width=0.85\textwidth]{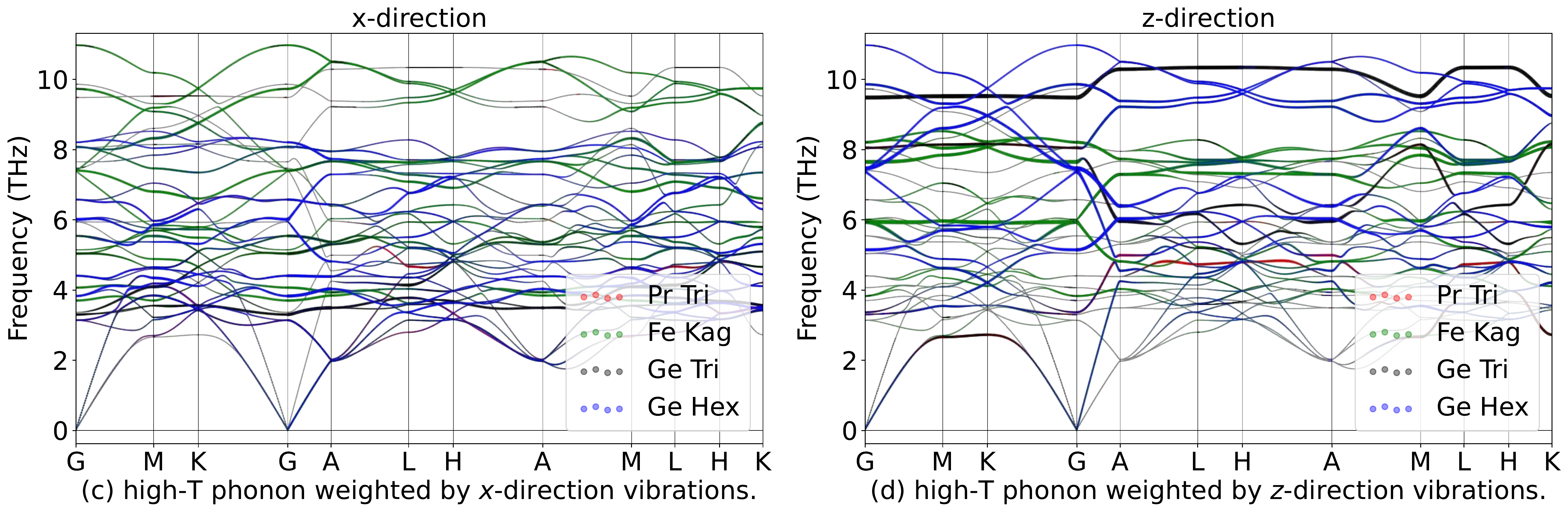}
\caption{Phonon spectrum for PrFe$_6$Ge$_6$in in-plane ($x$) and out-of-plane ($z$) directions in paramagnetic phase.}
\label{fig:PrFe6Ge6phoxyz}
\end{figure}

\begin{table}[htbp]
\global\long\def\arraystretch{1.12}
\captionsetup{width=.95\textwidth}
\caption{IRREPs at high-symmetry points for PrFe$_6$Ge$_6$ phonon spectrum at Low-T.}
\label{tab:191_PrFe6Ge6_Low}
\setlength{\tabcolsep}{1.mm}{
}
\end{table}

\begin{table}[htbp]
\global\long\def\arraystretch{1.12}
\captionsetup{width=.95\textwidth}
\caption{IRREPs at high-symmetry points for PrFe$_6$Ge$_6$ phonon spectrum at High-T.}
\label{tab:191_PrFe6Ge6_High}
\setlength{\tabcolsep}{1.mm}{
%
}
\end{table}
\subsubsection{\label{sms2sec:NdFe6Ge6}\hyperref[smsec:col]{\texorpdfstring{NdFe$_6$Ge$_6$}{}}~{\color{blue}  (High-T stable, Low-T unstable) }}

\begin{table}[htbp]
\global\long\def\arraystretch{1.12}
\captionsetup{width=.85\textwidth}
\caption{Structure parameters of NdFe$_6$Ge$_6$ after relaxation. It contains 408 electrons. }
\label{tab:NdFe6Ge6}
\setlength{\tabcolsep}{4mm}{
%
}
\end{table}

\begin{figure}[htbp]
\centering
\includegraphics[width=0.85\textwidth]{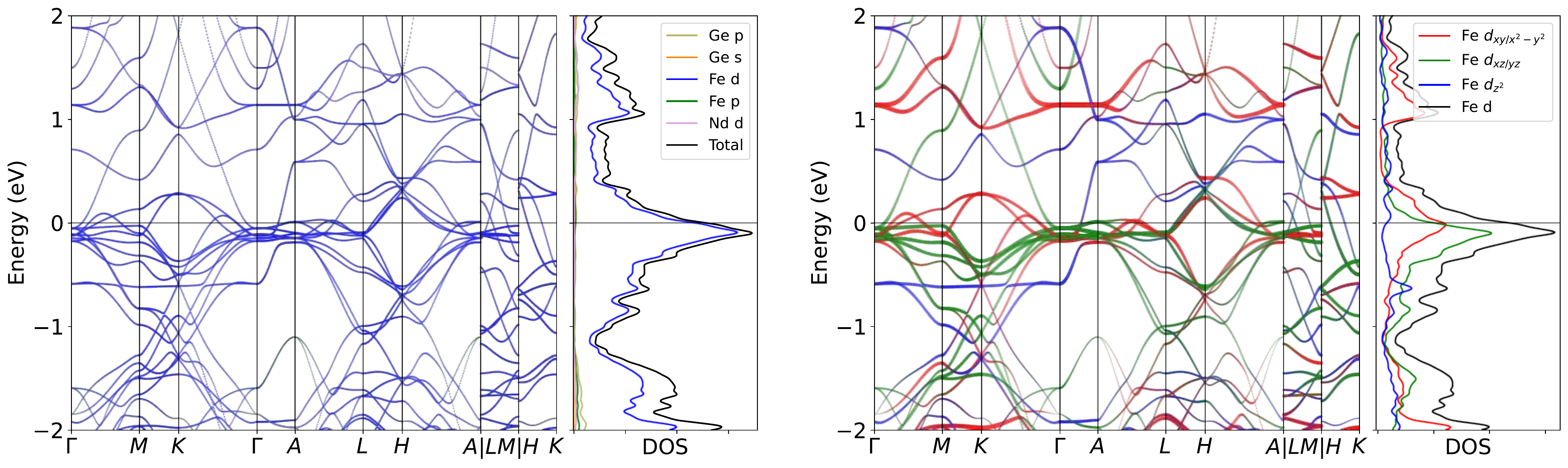}
\caption{Band structure for NdFe$_6$Ge$_6$ without SOC in paramagnetic phase. The projection of Fe $3d$ orbitals is also presented. The band structures clearly reveal the dominating of Fe $3d$ orbitals  near the Fermi level, as well as the pronounced peak.}
\label{fig:NdFe6Ge6band}
\end{figure}

\begin{figure}[H]
\centering
\subfloat[Relaxed phonon at low-T.]{
\label{fig:NdFe6Ge6_relaxed_Low}
\includegraphics[width=0.45\textwidth]{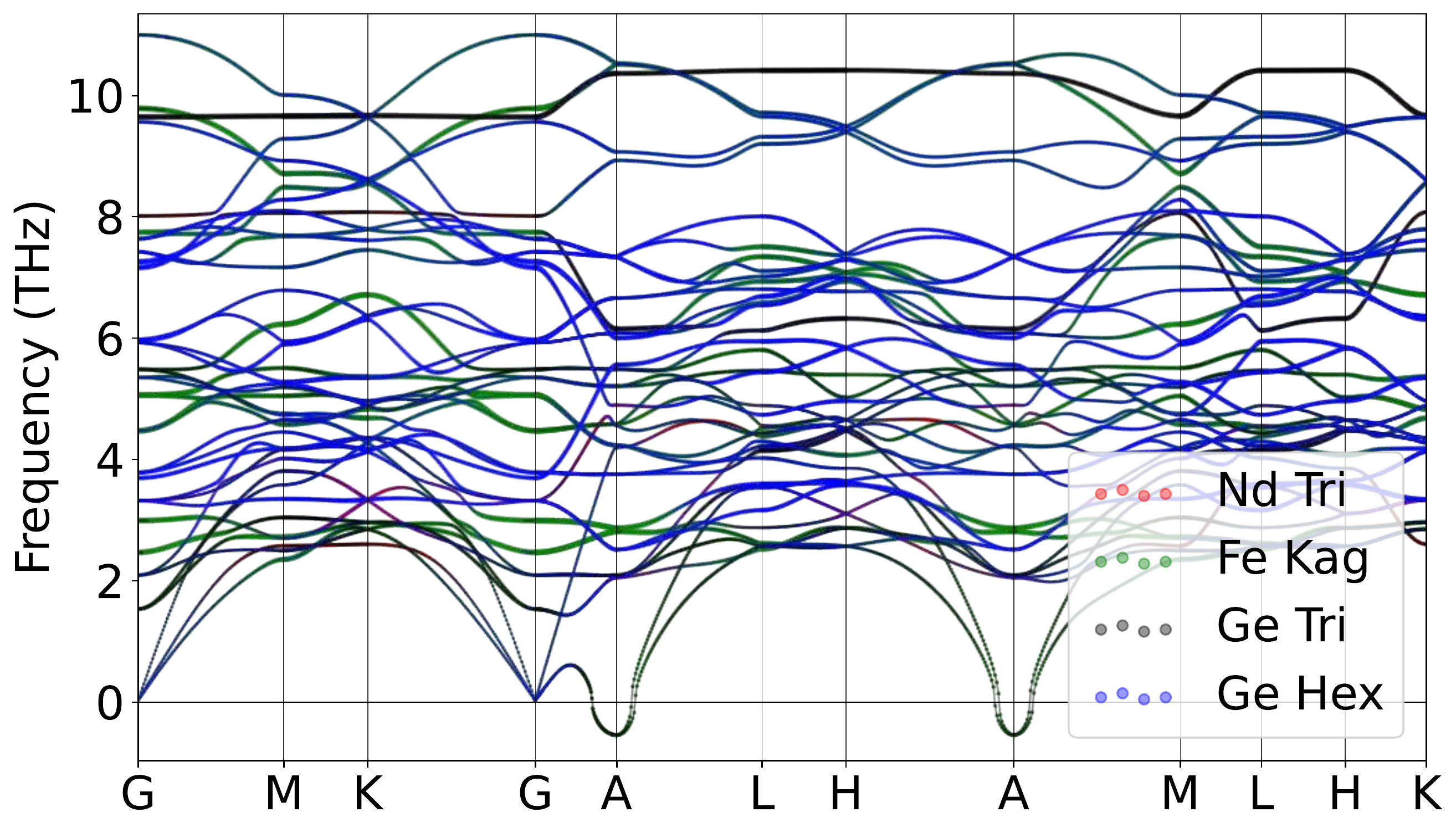}
}
\hspace{5pt}\subfloat[Relaxed phonon at high-T.]{
\label{fig:NdFe6Ge6_relaxed_High}
\includegraphics[width=0.45\textwidth]{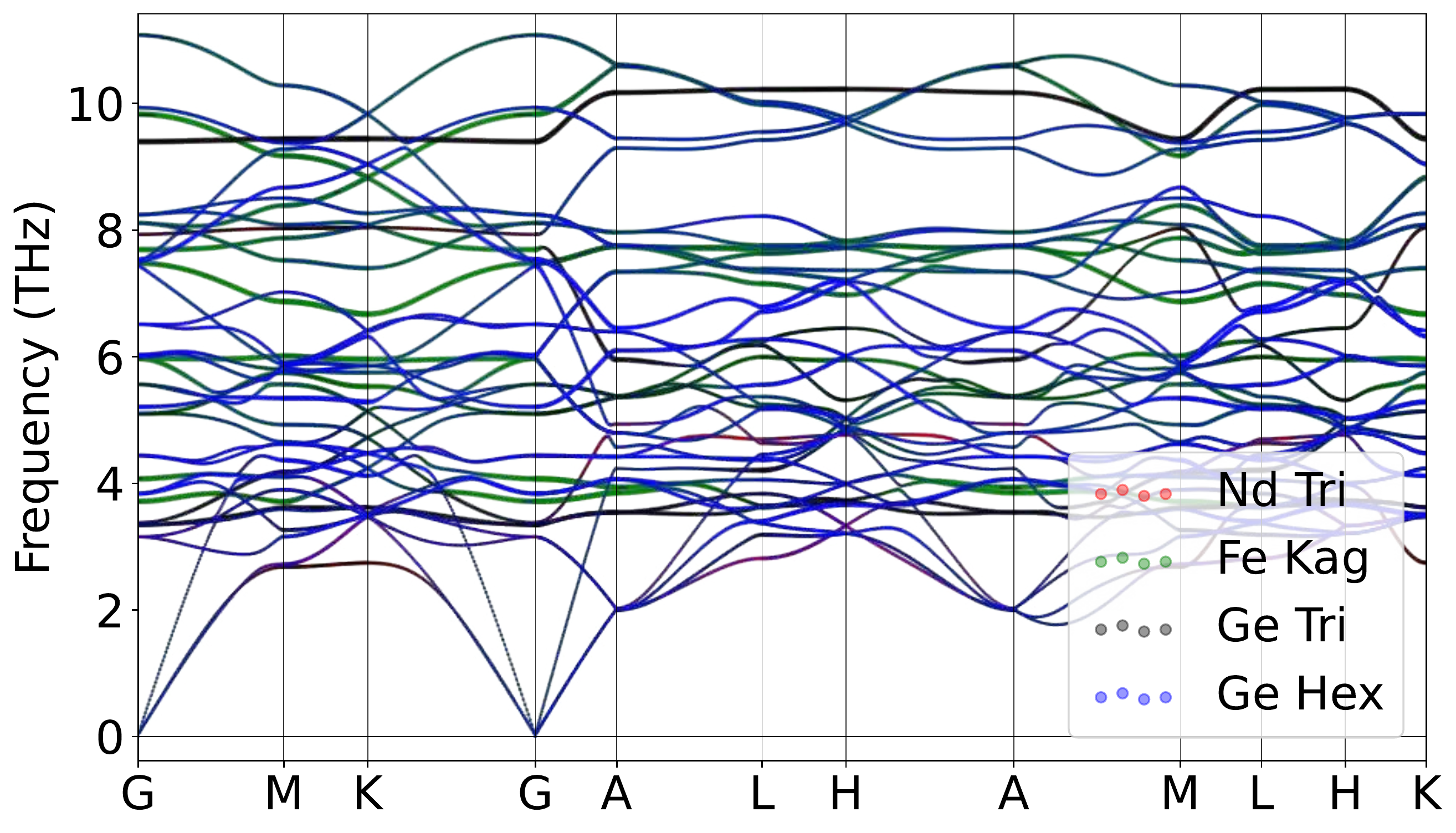}
}
\caption{Atomically projected phonon spectrum for NdFe$_6$Ge$_6$ in paramagnetic phase.}
\label{fig:NdFe6Ge6pho}
\end{figure}

\begin{figure}[htbp]
\centering
\includegraphics[width=0.45\textwidth]{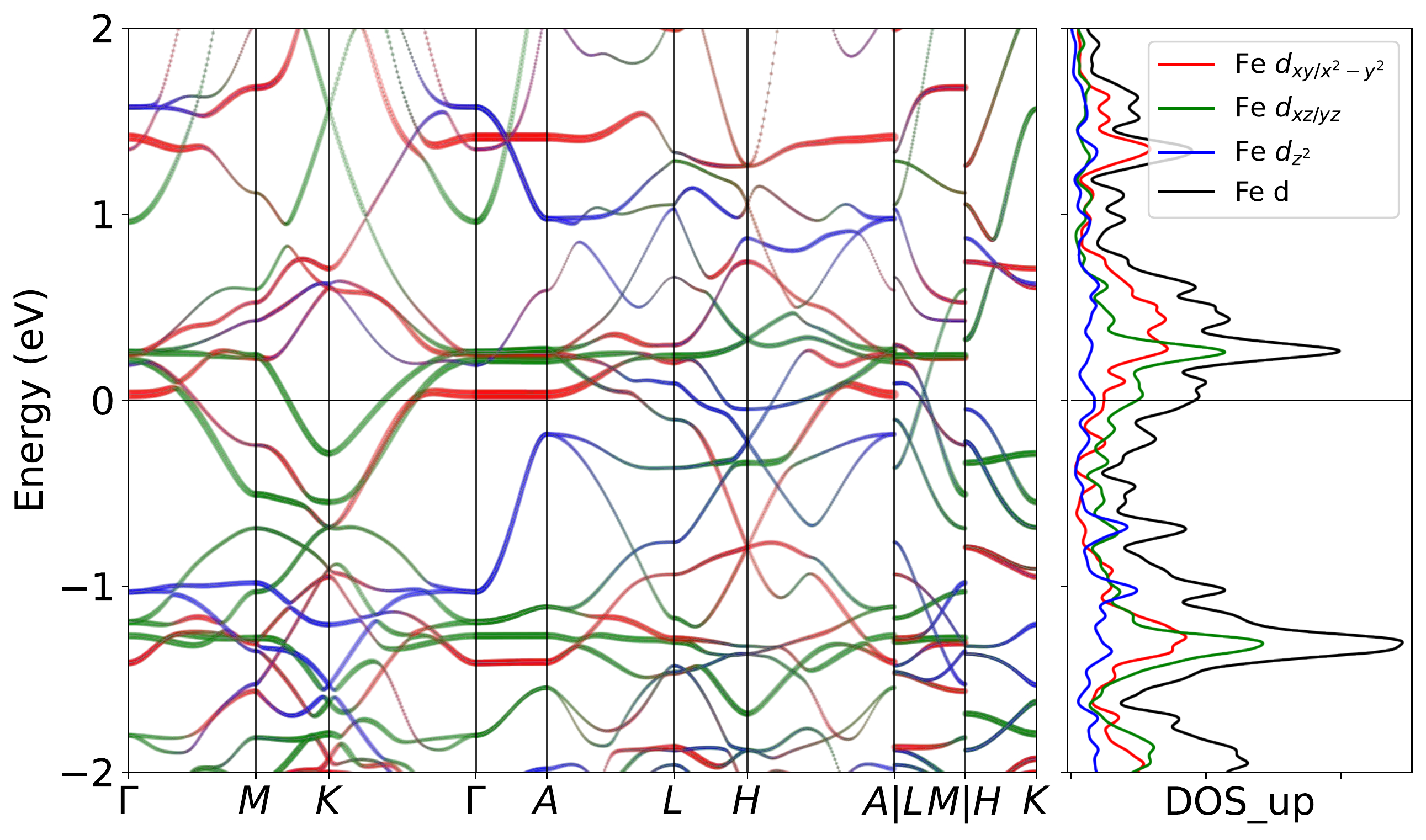}
\includegraphics[width=0.45\textwidth]{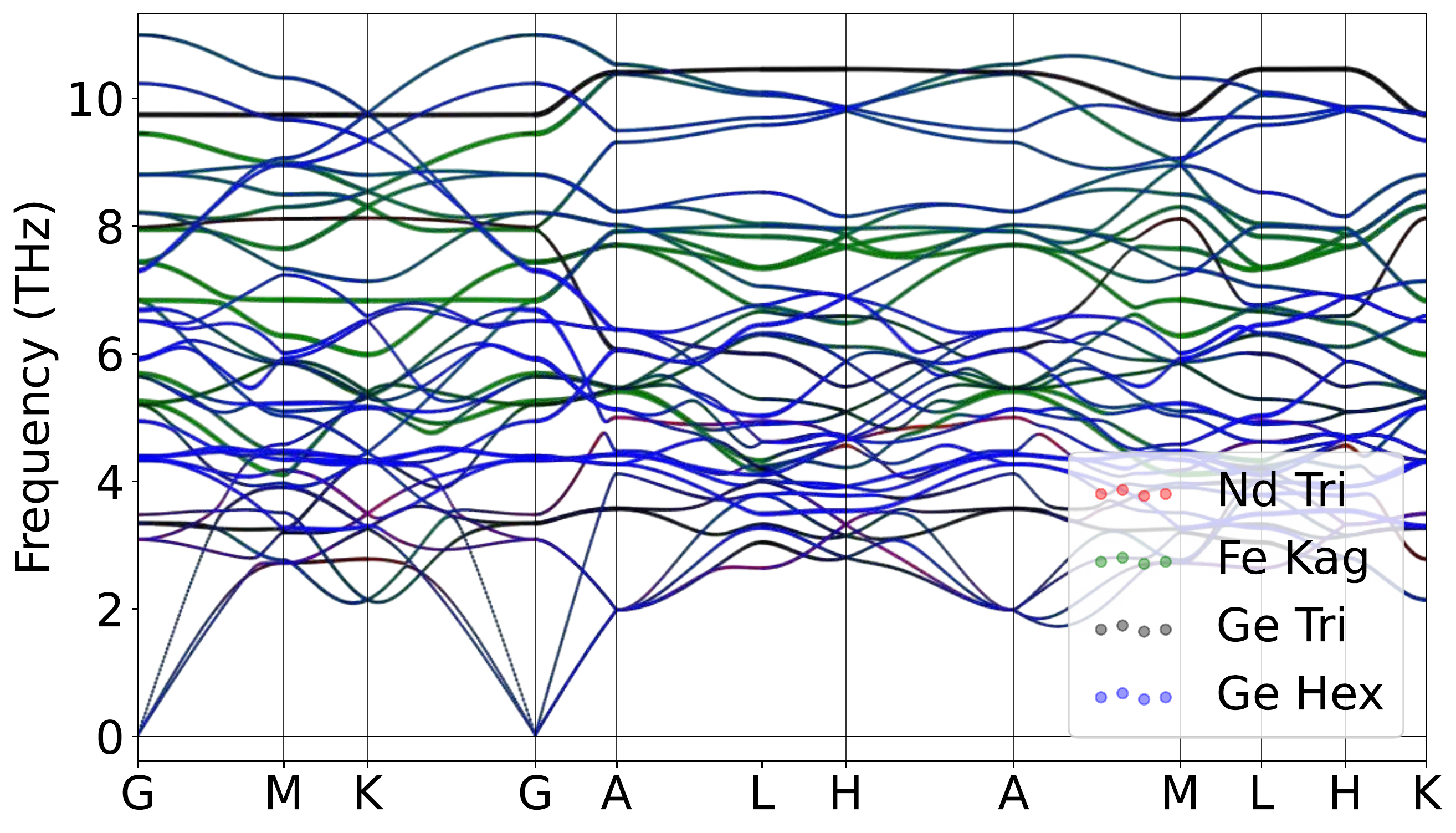}
\caption{Band structure for NdFe$_6$Ge$_6$ with AFM order without SOC for spin (Left)up channel. The projection of Fe $3d$ orbitals is also presented. (Right)Correspondingly, a stable phonon was demonstrated with the collinear magnetic order without Hubbard U at lower temperature regime, weighted by atomic projections.}
\label{fig:NdFe6Ge6mag}
\end{figure}

\begin{figure}[H]
\centering
\label{fig:NdFe6Ge6_relaxed_Low_xz}
\includegraphics[width=0.85\textwidth]{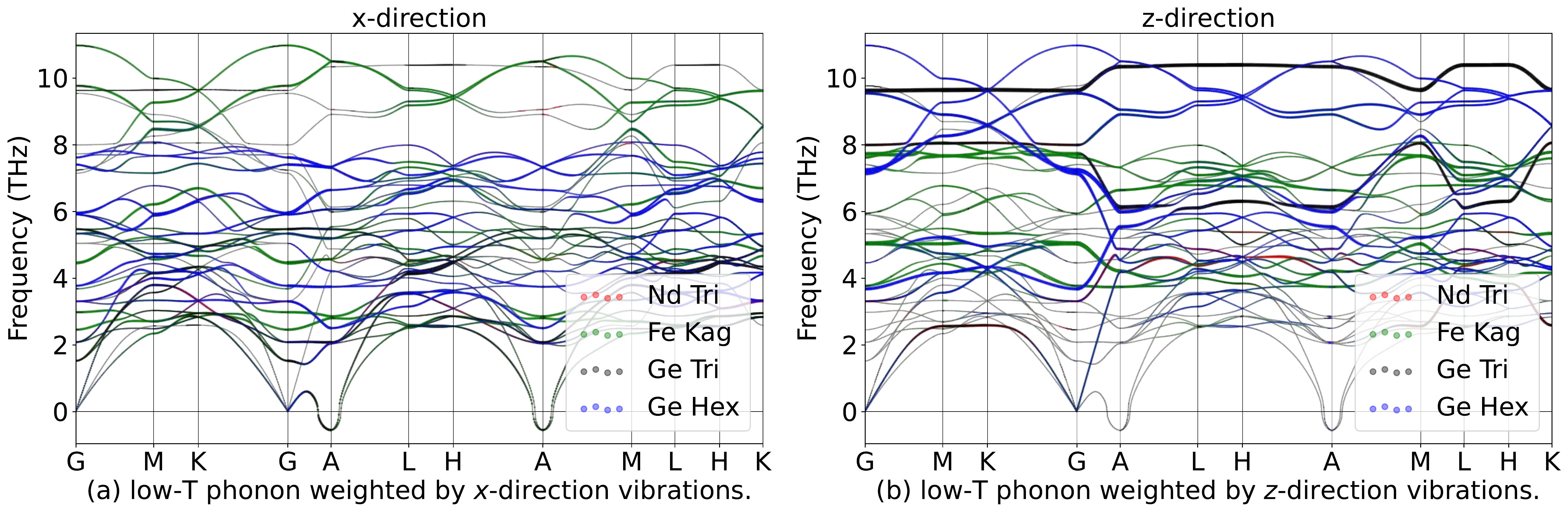}
\hspace{5pt}\label{fig:NdFe6Ge6_relaxed_High_xz}
\includegraphics[width=0.85\textwidth]{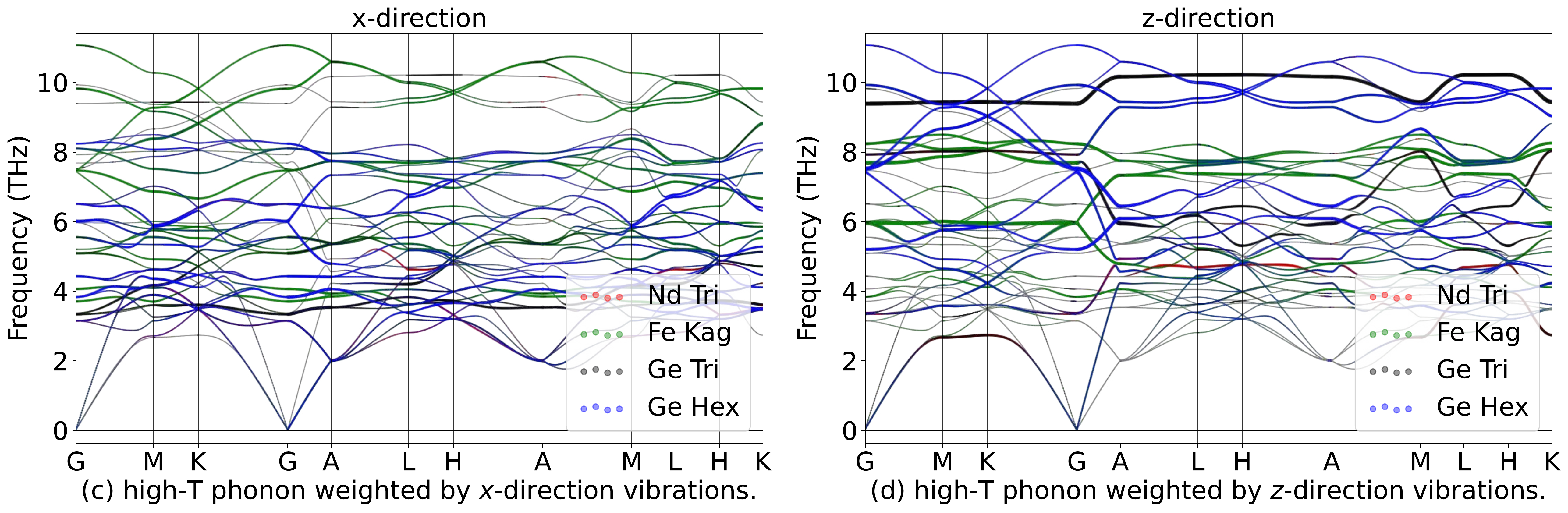}
\caption{Phonon spectrum for NdFe$_6$Ge$_6$in in-plane ($x$) and out-of-plane ($z$) directions in paramagnetic phase.}
\label{fig:NdFe6Ge6phoxyz}
\end{figure}

\begin{table}[htbp]
\global\long\def\arraystretch{1.12}
\captionsetup{width=.95\textwidth}
\caption{IRREPs at high-symmetry points for NdFe$_6$Ge$_6$ phonon spectrum at Low-T.}
\label{tab:191_NdFe6Ge6_Low}
\setlength{\tabcolsep}{1.mm}{
}
\end{table}

\begin{table}[htbp]
\global\long\def\arraystretch{1.12}
\captionsetup{width=.95\textwidth}
\caption{IRREPs at high-symmetry points for NdFe$_6$Ge$_6$ phonon spectrum at High-T.}
\label{tab:191_NdFe6Ge6_High}
\setlength{\tabcolsep}{1.mm}{
%
}
\end{table}
\subsubsection{\label{sms2sec:SmFe6Ge6}\hyperref[smsec:col]{\texorpdfstring{SmFe$_6$Ge$_6$}{}}~{\color{blue}  (High-T stable, Low-T unstable) }}

\begin{table}[htbp]
\global\long\def\arraystretch{1.12}
\captionsetup{width=.85\textwidth}
\caption{Structure parameters of SmFe$_6$Ge$_6$ after relaxation. It contains 410 electrons. }
\label{tab:SmFe6Ge6}
\setlength{\tabcolsep}{4mm}{
%
}
\end{table}

\begin{figure}[htbp]
\centering
\includegraphics[width=0.85\textwidth]{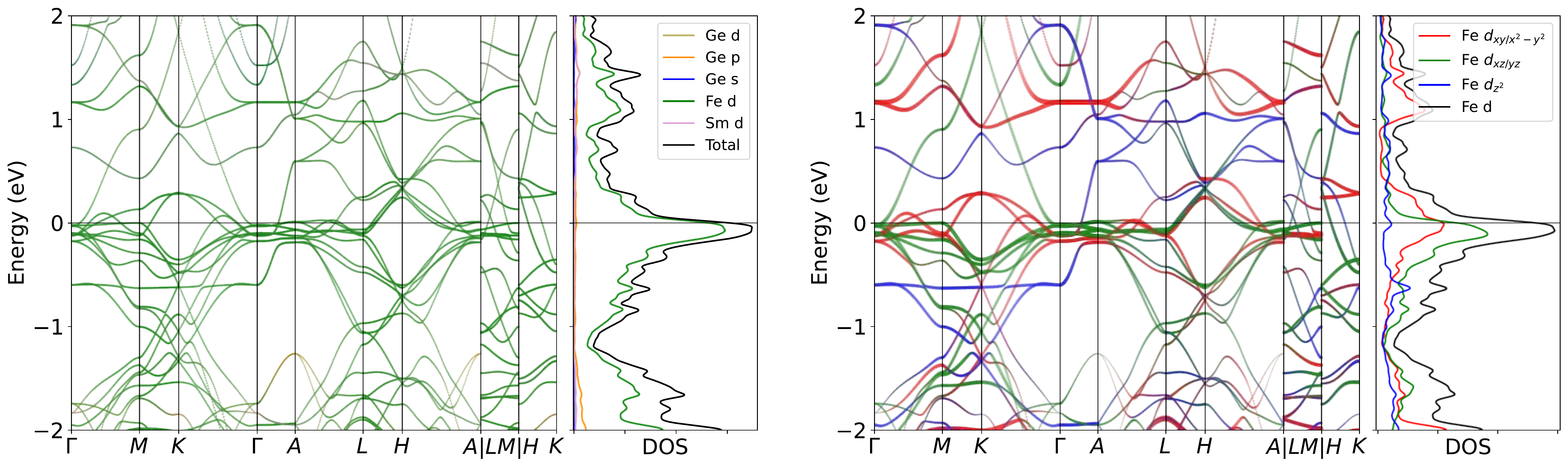}
\caption{Band structure for SmFe$_6$Ge$_6$ without SOC in paramagnetic phase. The projection of Fe $3d$ orbitals is also presented. The band structures clearly reveal the dominating of Fe $3d$ orbitals  near the Fermi level, as well as the pronounced peak.}
\label{fig:SmFe6Ge6band}
\end{figure}

\begin{figure}[H]
\centering
\subfloat[Relaxed phonon at low-T.]{
\label{fig:SmFe6Ge6_relaxed_Low}
\includegraphics[width=0.45\textwidth]{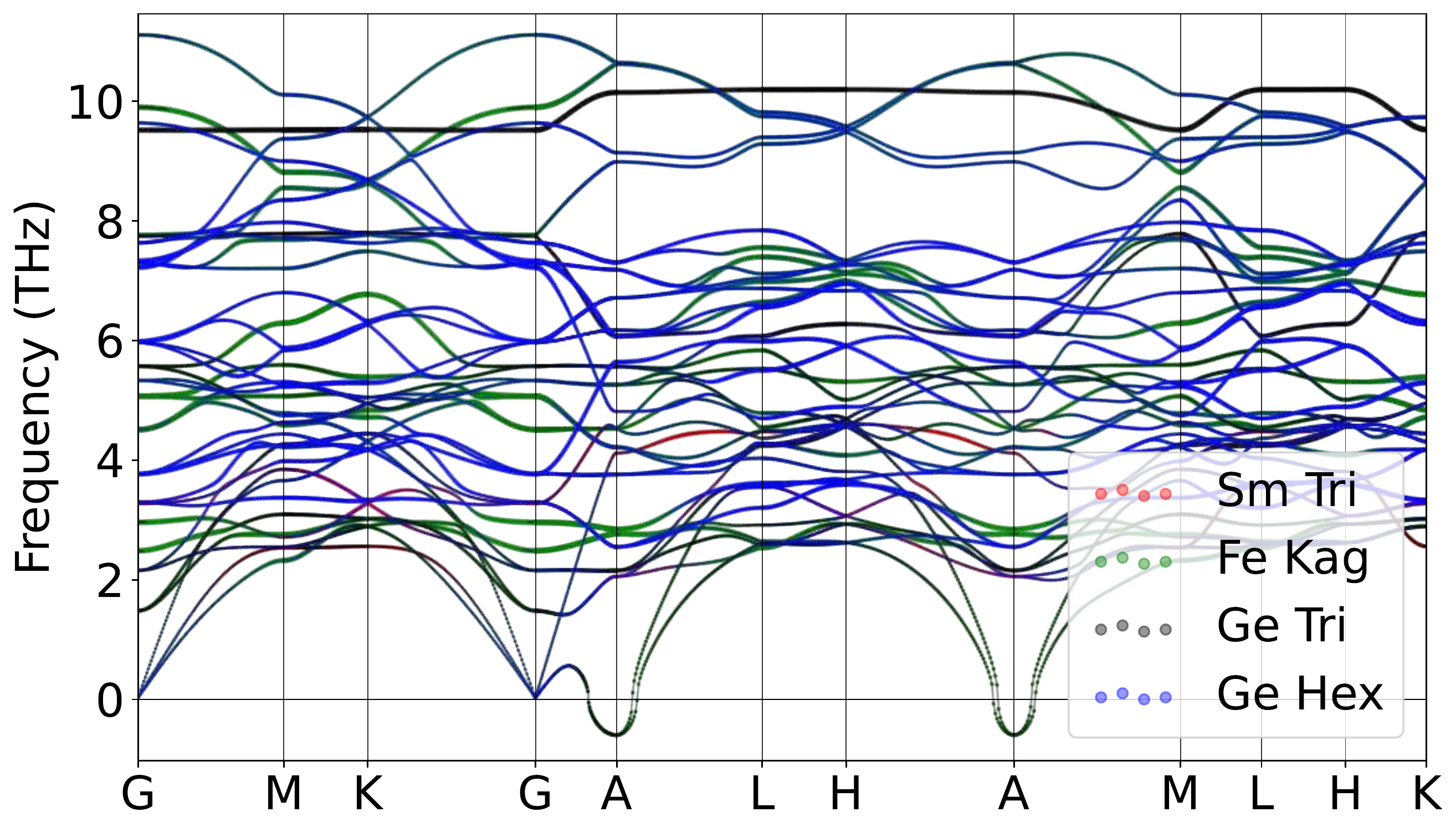}
}
\hspace{5pt}\subfloat[Relaxed phonon at high-T.]{
\label{fig:SmFe6Ge6_relaxed_High}
\includegraphics[width=0.45\textwidth]{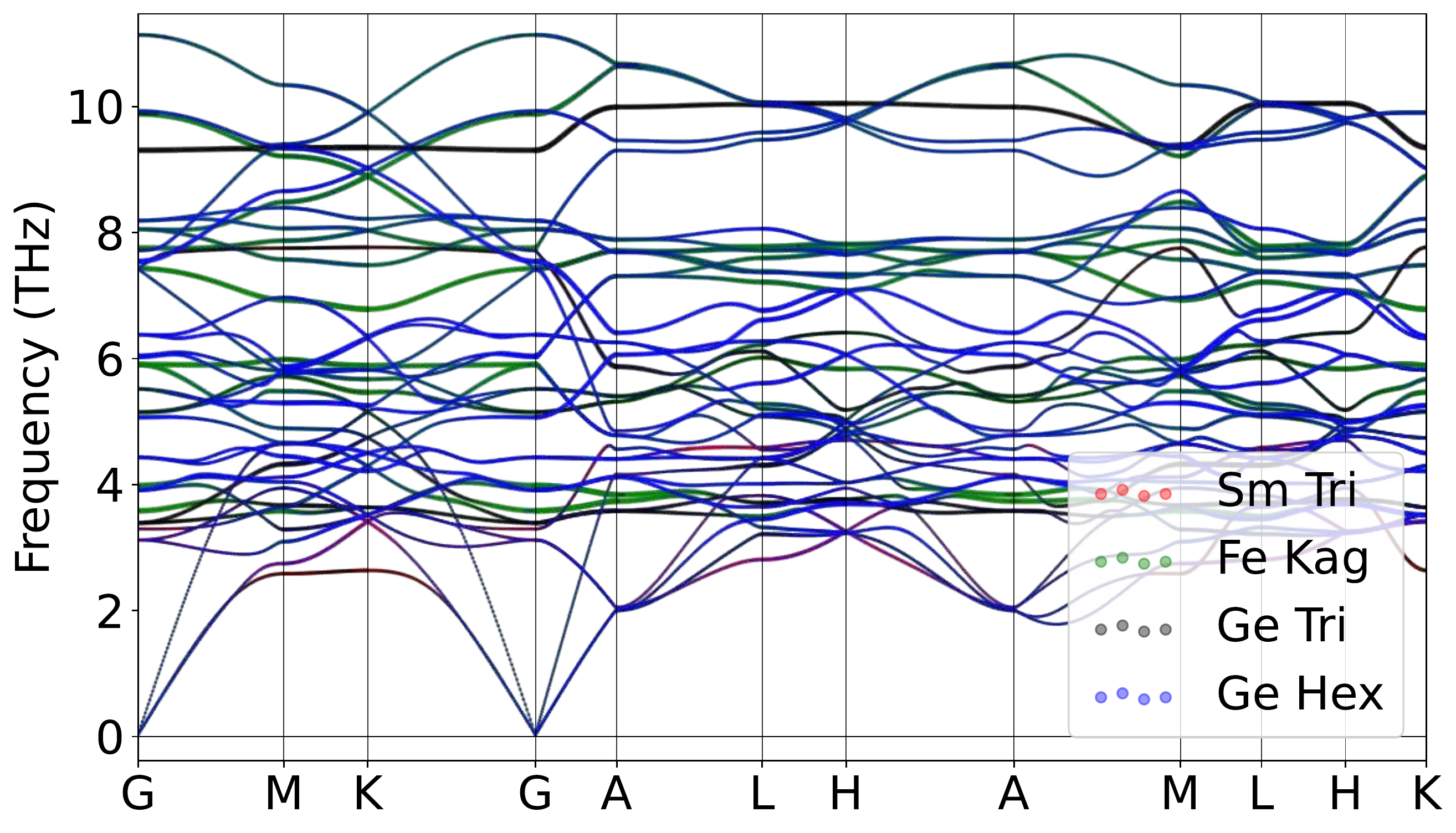}
}
\caption{Atomically projected phonon spectrum for SmFe$_6$Ge$_6$ in paramagnetic phase.}
\label{fig:SmFe6Ge6pho}
\end{figure}

\begin{figure}[htbp]
\centering
\includegraphics[width=0.45\textwidth]{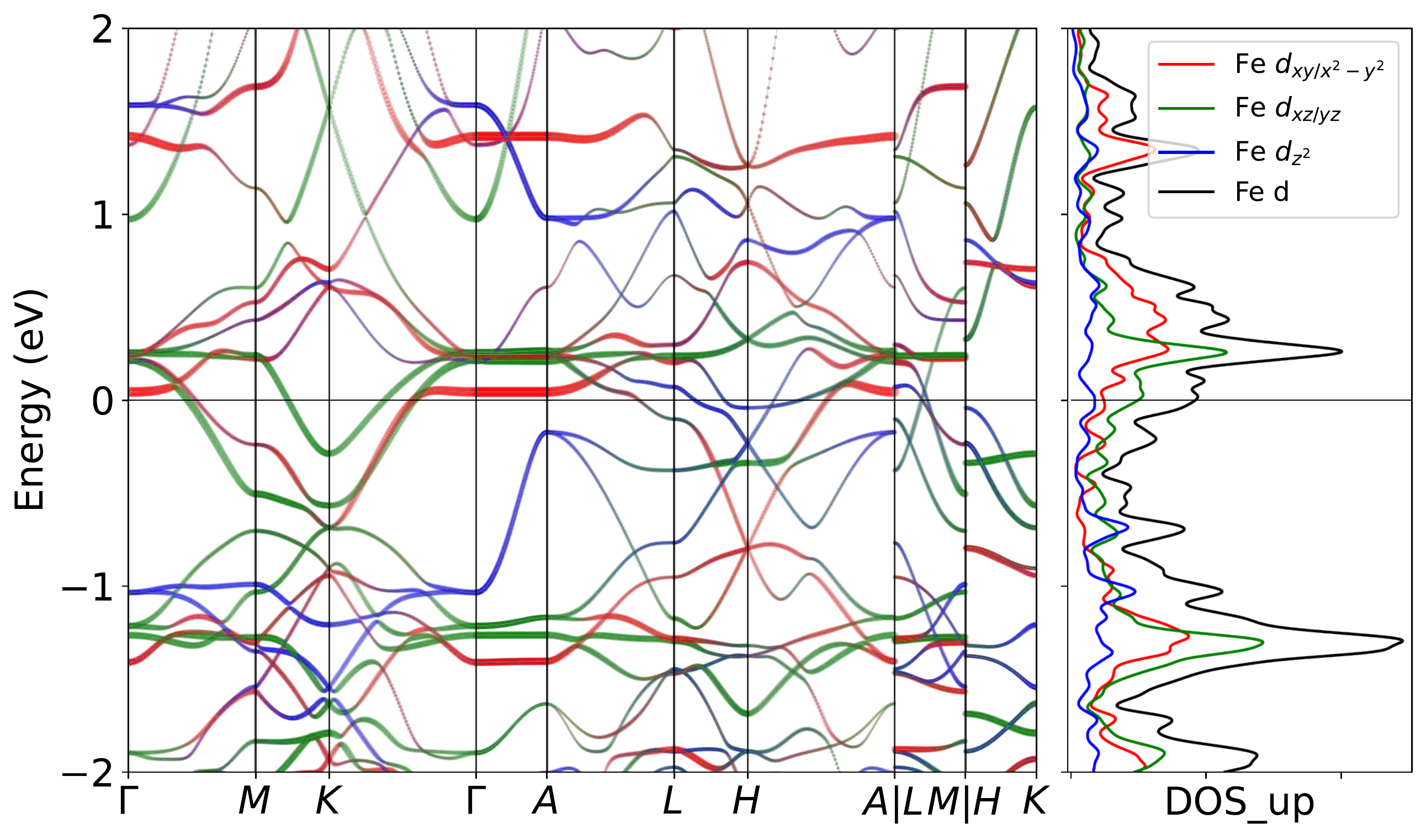}
\includegraphics[width=0.45\textwidth]{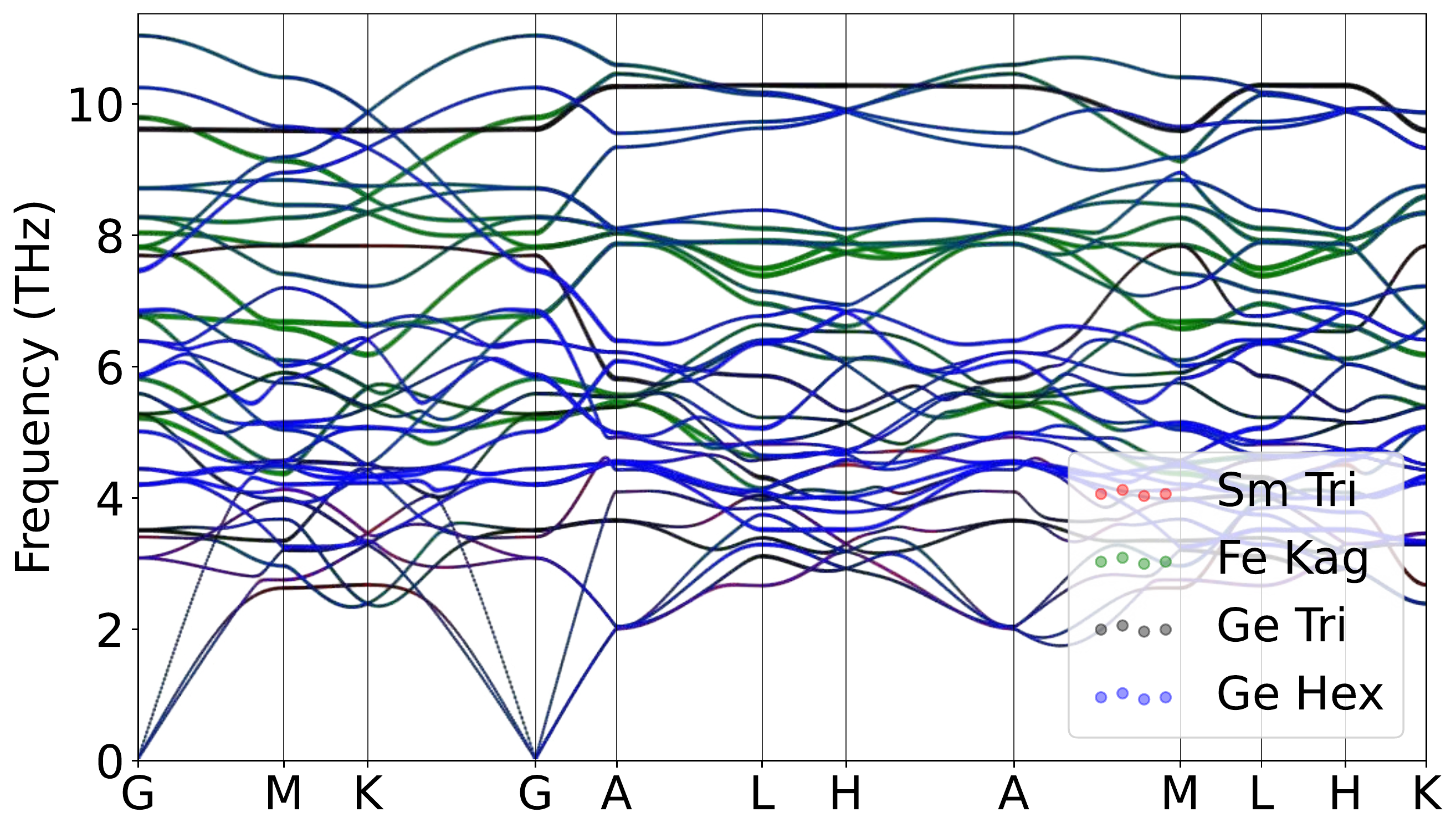}
\caption{Band structure for SmFe$_6$Ge$_6$ with AFM order without SOC for spin (Left)up channel. The projection of Fe $3d$ orbitals is also presented. (Right)Correspondingly, a stable phonon was demonstrated with the collinear magnetic order without Hubbard U at lower temperature regime, weighted by atomic projections.}
\label{fig:SmFe6Ge6mag}
\end{figure}

\begin{figure}[H]
\centering
\label{fig:SmFe6Ge6_relaxed_Low_xz}
\includegraphics[width=0.85\textwidth]{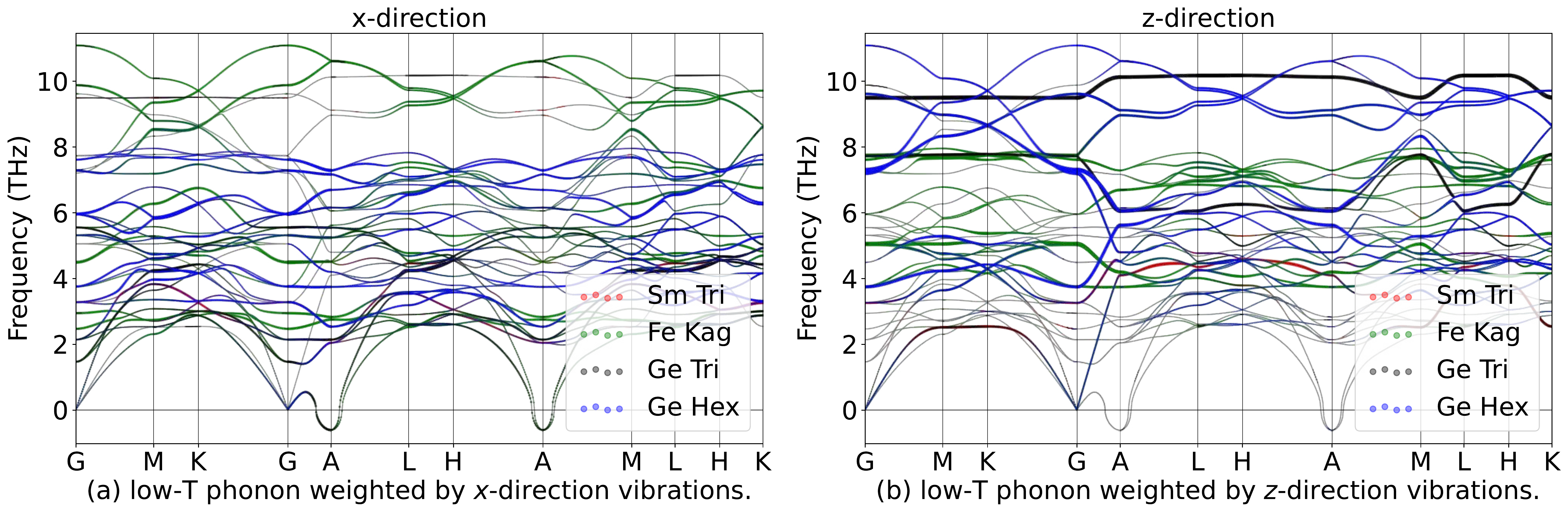}
\hspace{5pt}\label{fig:SmFe6Ge6_relaxed_High_xz}
\includegraphics[width=0.85\textwidth]{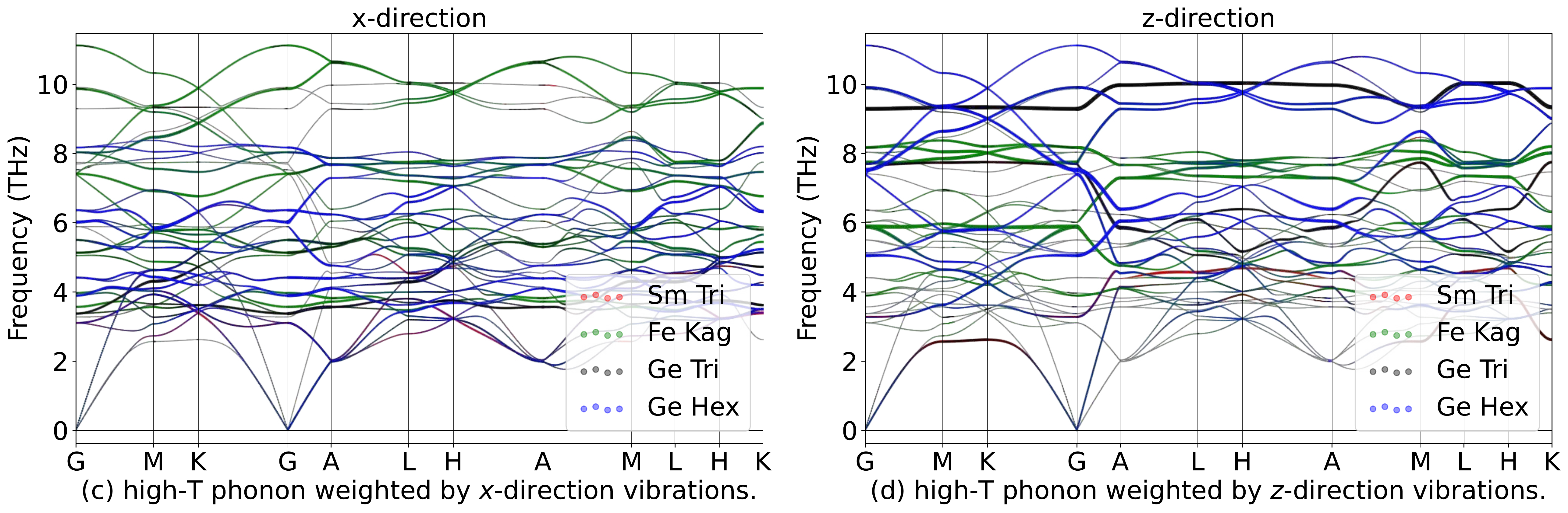}
\caption{Phonon spectrum for SmFe$_6$Ge$_6$in in-plane ($x$) and out-of-plane ($z$) directions in paramagnetic phase.}
\label{fig:SmFe6Ge6phoxyz}
\end{figure}

\begin{table}[htbp]
\global\long\def\arraystretch{1.12}
\captionsetup{width=.95\textwidth}
\caption{IRREPs at high-symmetry points for SmFe$_6$Ge$_6$ phonon spectrum at Low-T.}
\label{tab:191_SmFe6Ge6_Low}
\setlength{\tabcolsep}{1.mm}{
}
\end{table}

\begin{table}[htbp]
\global\long\def\arraystretch{1.12}
\captionsetup{width=.95\textwidth}
\caption{IRREPs at high-symmetry points for SmFe$_6$Ge$_6$ phonon spectrum at High-T.}
\label{tab:191_SmFe6Ge6_High}
\setlength{\tabcolsep}{1.mm}{
%
}
\end{table}
\subsubsection{\label{sms2sec:GdFe6Ge6}\hyperref[smsec:col]{\texorpdfstring{GdFe$_6$Ge$_6$}{}}~{\color{blue}  (High-T stable, Low-T unstable) }}

\begin{table}[htbp]
\global\long\def\arraystretch{1.12}
\captionsetup{width=.85\textwidth}
\caption{Structure parameters of GdFe$_6$Ge$_6$ after relaxation. It contains 412 electrons. }
\label{tab:GdFe6Ge6}
\setlength{\tabcolsep}{4mm}{
%
}
\end{table}

\begin{figure}[htbp]
\centering
\includegraphics[width=0.85\textwidth]{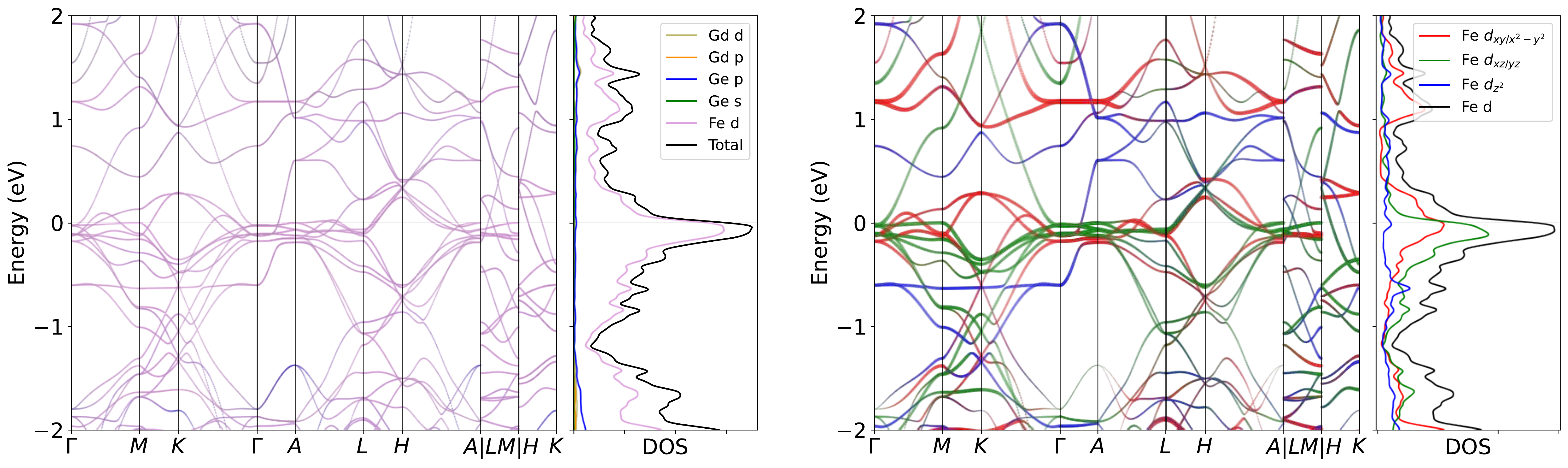}
\caption{Band structure for GdFe$_6$Ge$_6$ without SOC in paramagnetic phase. The projection of Fe $3d$ orbitals is also presented. The band structures clearly reveal the dominating of Fe $3d$ orbitals  near the Fermi level, as well as the pronounced peak.}
\label{fig:GdFe6Ge6band}
\end{figure}

\begin{figure}[H]
\centering
\subfloat[Relaxed phonon at low-T.]{
\label{fig:GdFe6Ge6_relaxed_Low}
\includegraphics[width=0.45\textwidth]{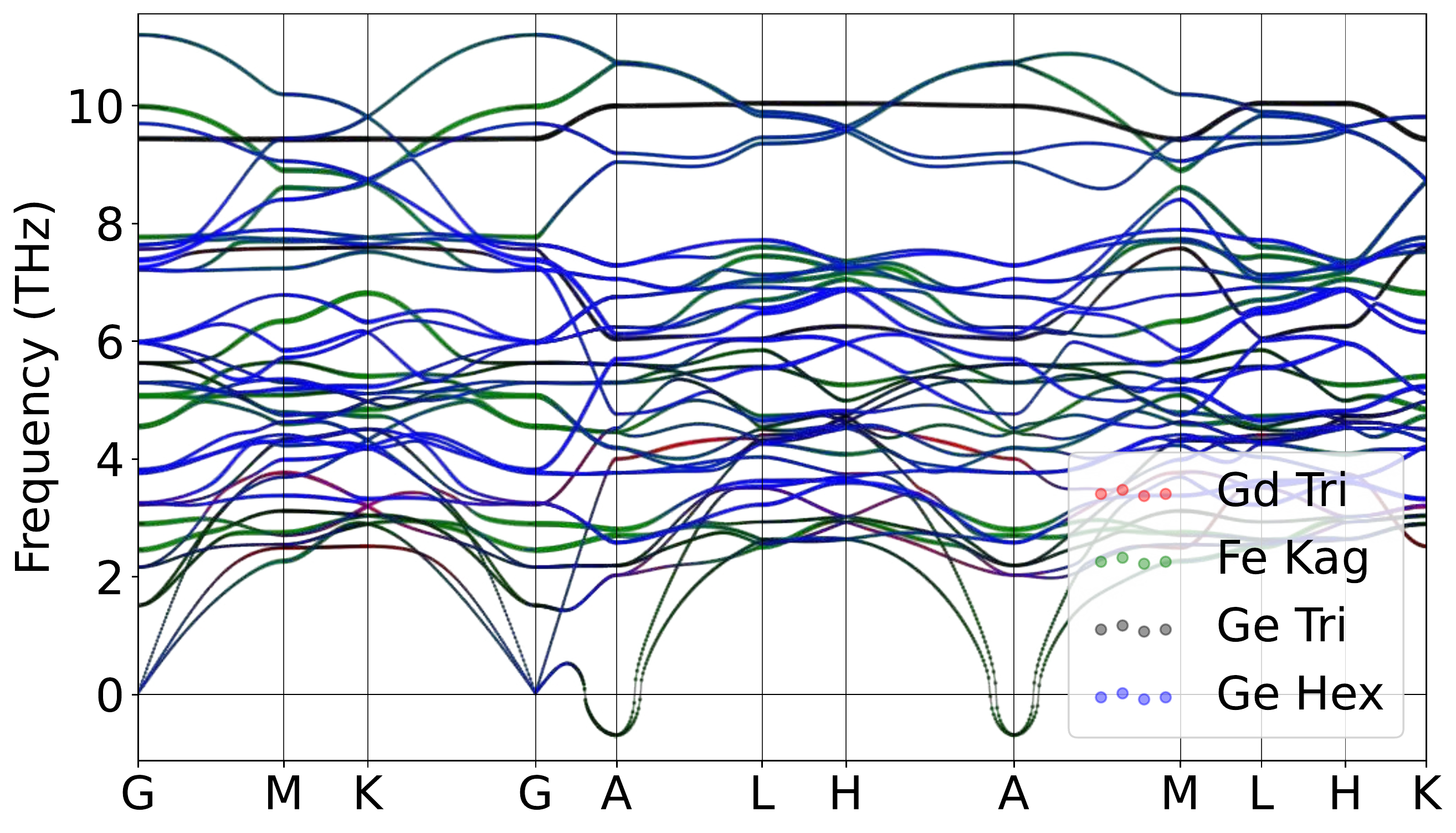}
}
\hspace{5pt}\subfloat[Relaxed phonon at high-T.]{
\label{fig:GdFe6Ge6_relaxed_High}
\includegraphics[width=0.45\textwidth]{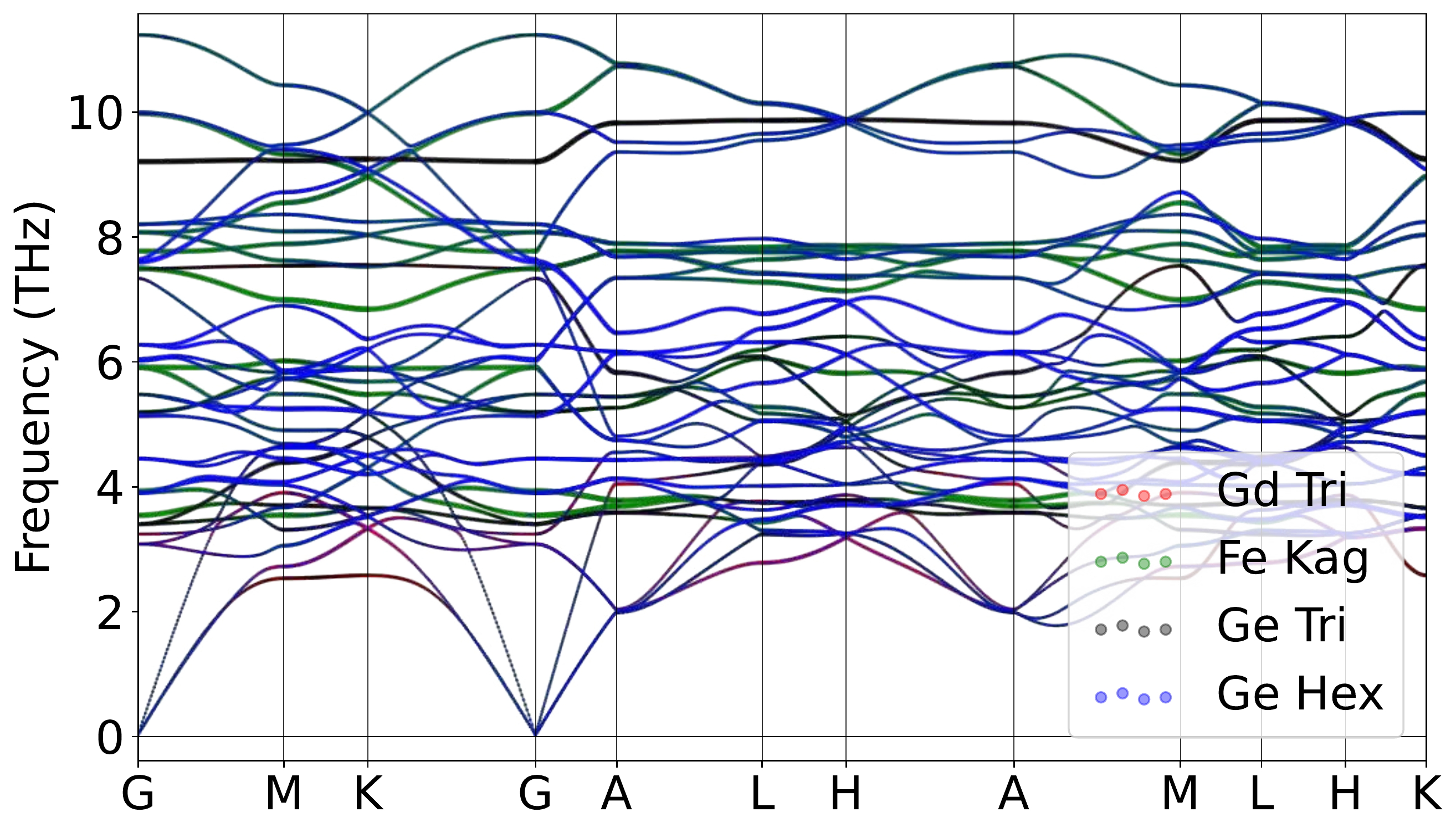}
}
\caption{Atomically projected phonon spectrum for GdFe$_6$Ge$_6$ in paramagnetic phase.}
\label{fig:GdFe6Ge6pho}
\end{figure}

\begin{figure}[htbp]
\centering
\includegraphics[width=0.45\textwidth]{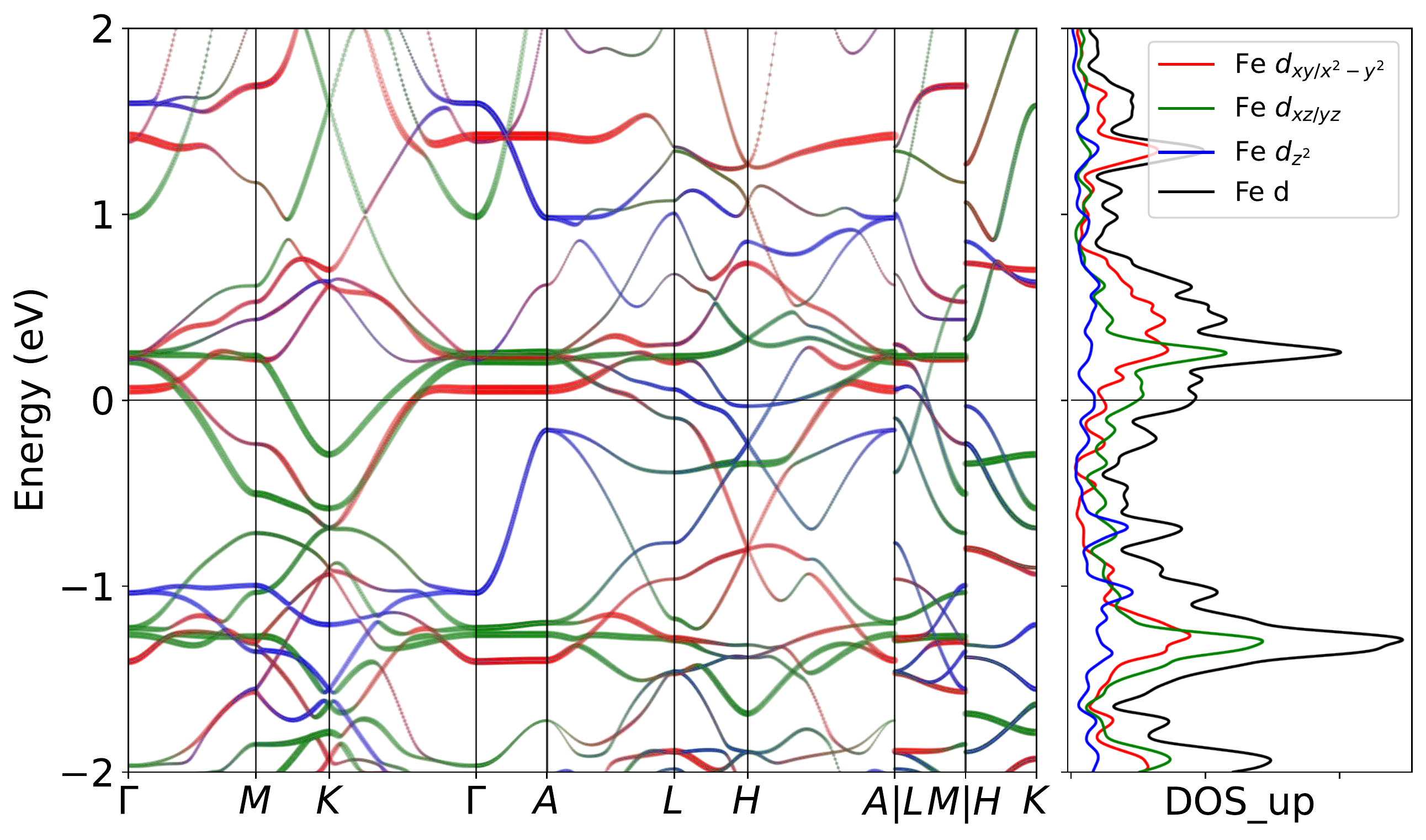}
\includegraphics[width=0.45\textwidth]{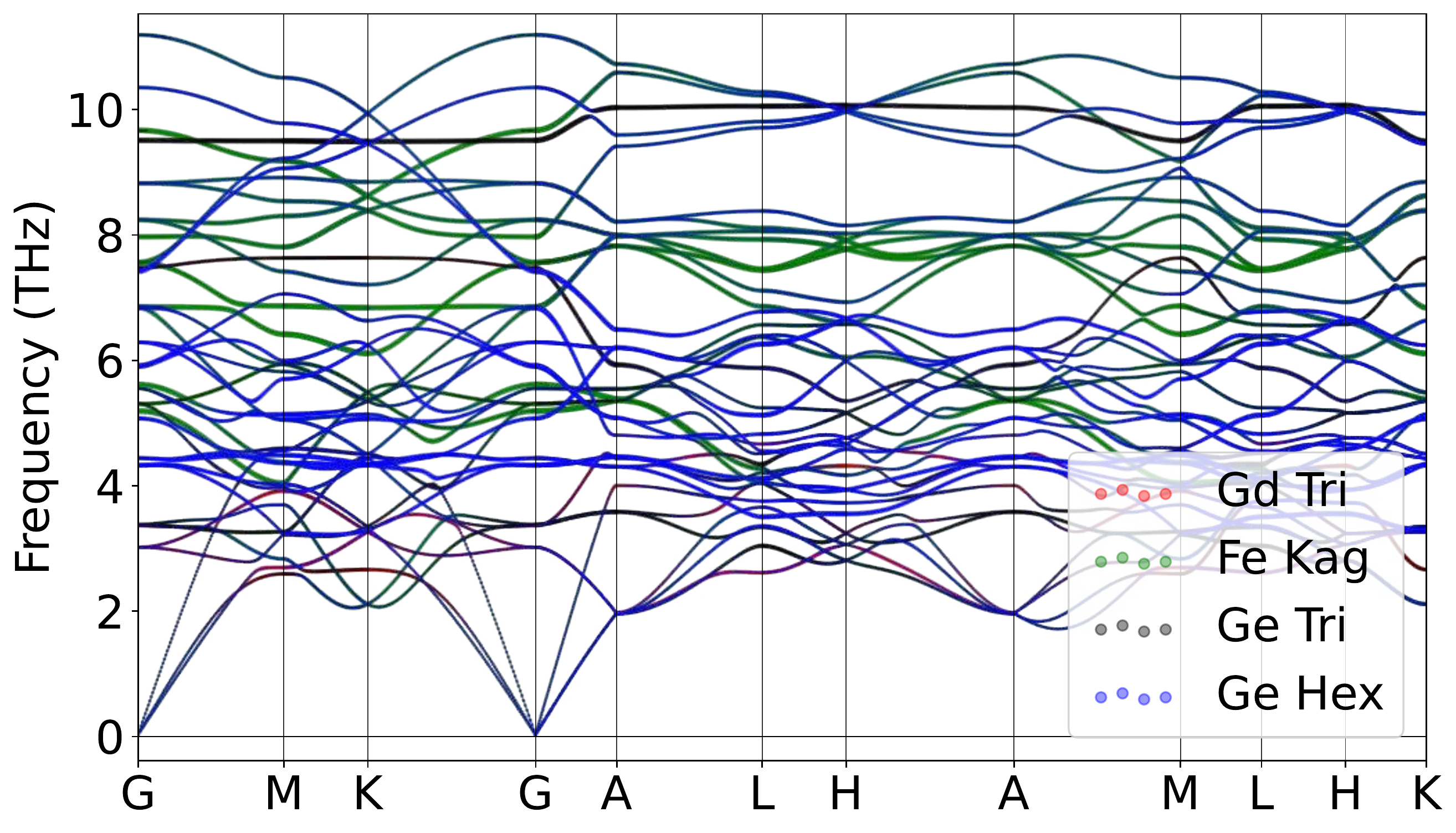}
\caption{Band structure for GdFe$_6$Ge$_6$ with AFM order without SOC for spin (Left)up channel. The projection of Fe $3d$ orbitals is also presented. (Right)Correspondingly, a stable phonon was demonstrated with the collinear magnetic order without Hubbard U at lower temperature regime, weighted by atomic projections.}
\label{fig:GdFe6Ge6mag}
\end{figure}

\begin{figure}[H]
\centering
\label{fig:GdFe6Ge6_relaxed_Low_xz}
\includegraphics[width=0.85\textwidth]{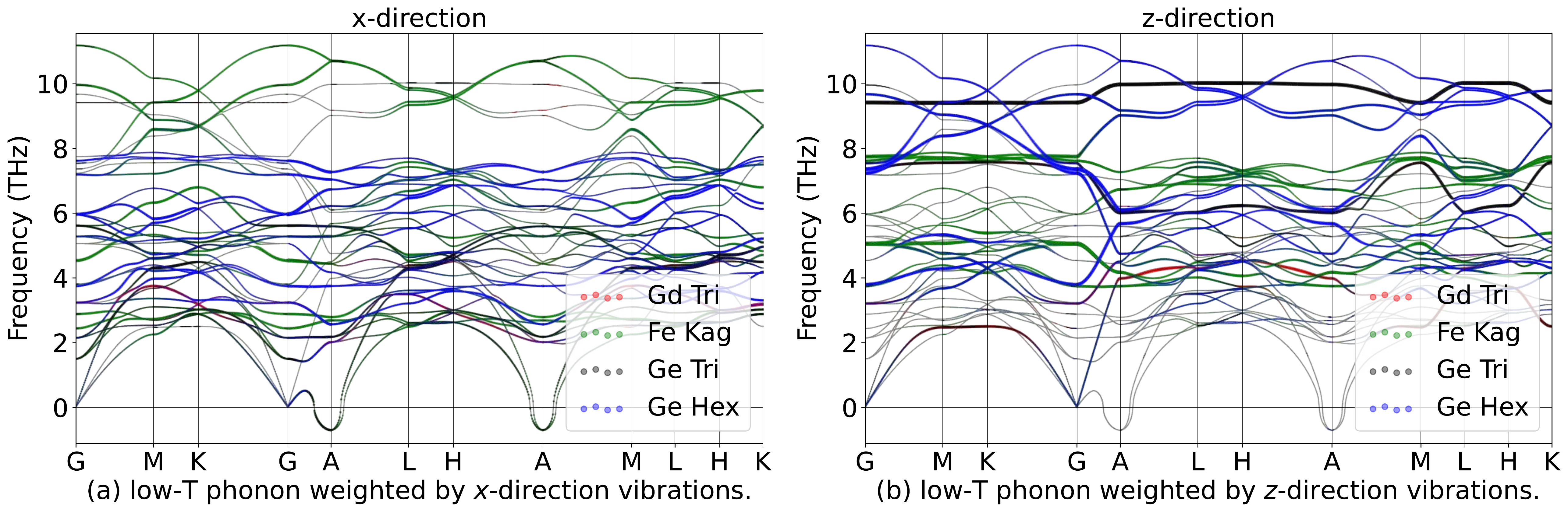}
\hspace{5pt}\label{fig:GdFe6Ge6_relaxed_High_xz}
\includegraphics[width=0.85\textwidth]{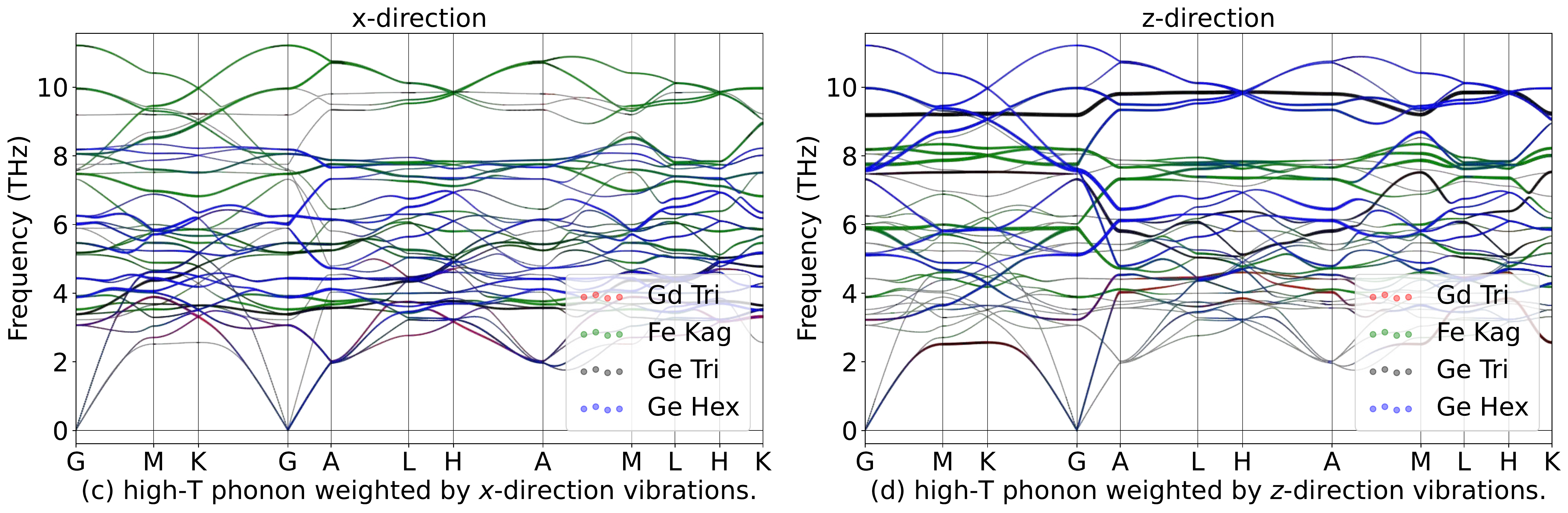}
\caption{Phonon spectrum for GdFe$_6$Ge$_6$in in-plane ($x$) and out-of-plane ($z$) directions in paramagnetic phase.}
\label{fig:GdFe6Ge6phoxyz}
\end{figure}

\begin{table}[htbp]
\global\long\def\arraystretch{1.12}
\captionsetup{width=.95\textwidth}
\caption{IRREPs at high-symmetry points for GdFe$_6$Ge$_6$ phonon spectrum at Low-T.}
\label{tab:191_GdFe6Ge6_Low}
\setlength{\tabcolsep}{1.mm}{
}
\end{table}

\begin{table}[htbp]
\global\long\def\arraystretch{1.12}
\captionsetup{width=.95\textwidth}
\caption{IRREPs at high-symmetry points for GdFe$_6$Ge$_6$ phonon spectrum at High-T.}
\label{tab:191_GdFe6Ge6_High}
\setlength{\tabcolsep}{1.mm}{
%
}
\end{table}
\subsubsection{\label{sms2sec:TbFe6Ge6}\hyperref[smsec:col]{\texorpdfstring{TbFe$_6$Ge$_6$}{}}~{\color{blue}  (High-T stable, Low-T unstable) }}

\begin{table}[htbp]
\global\long\def\arraystretch{1.12}
\captionsetup{width=.85\textwidth}
\caption{Structure parameters of TbFe$_6$Ge$_6$ after relaxation. It contains 413 electrons. }
\label{tab:TbFe6Ge6}
\setlength{\tabcolsep}{4mm}{
%
}
\end{table}

\begin{figure}[htbp]
\centering
\includegraphics[width=0.85\textwidth]{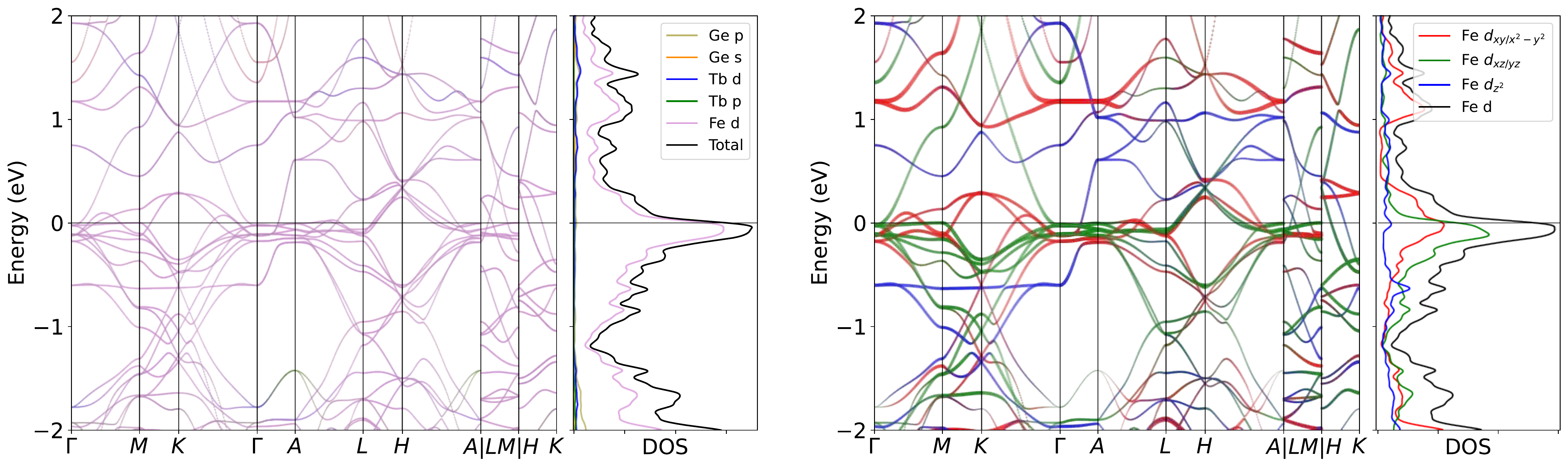}
\caption{Band structure for TbFe$_6$Ge$_6$ without SOC in paramagnetic phase. The projection of Fe $3d$ orbitals is also presented. The band structures clearly reveal the dominating of Fe $3d$ orbitals  near the Fermi level, as well as the pronounced peak.}
\label{fig:TbFe6Ge6band}
\end{figure}

\begin{figure}[H]
\centering
\subfloat[Relaxed phonon at low-T.]{
\label{fig:TbFe6Ge6_relaxed_Low}
\includegraphics[width=0.45\textwidth]{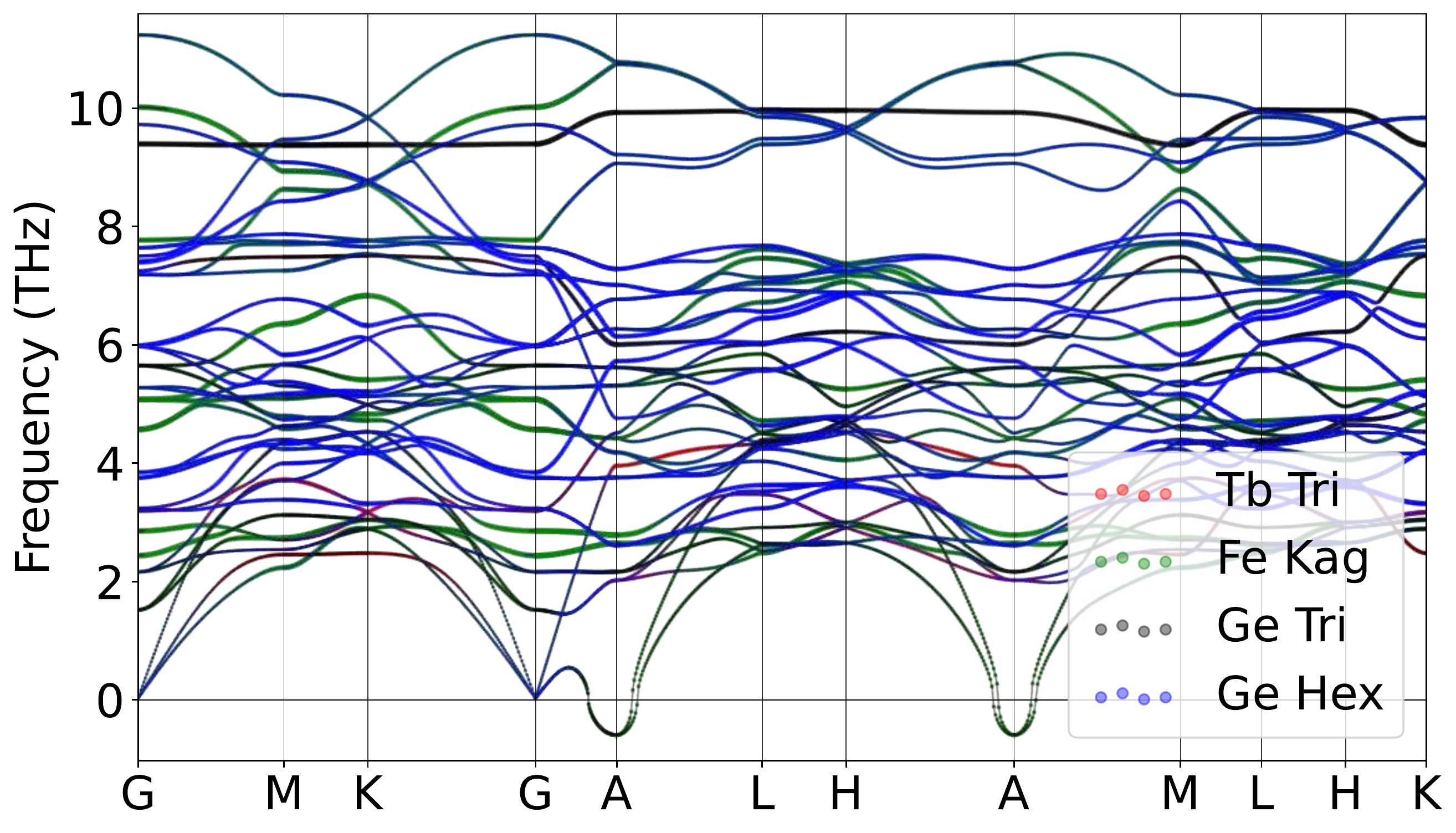}
}
\hspace{5pt}\subfloat[Relaxed phonon at high-T.]{
\label{fig:TbFe6Ge6_relaxed_High}
\includegraphics[width=0.45\textwidth]{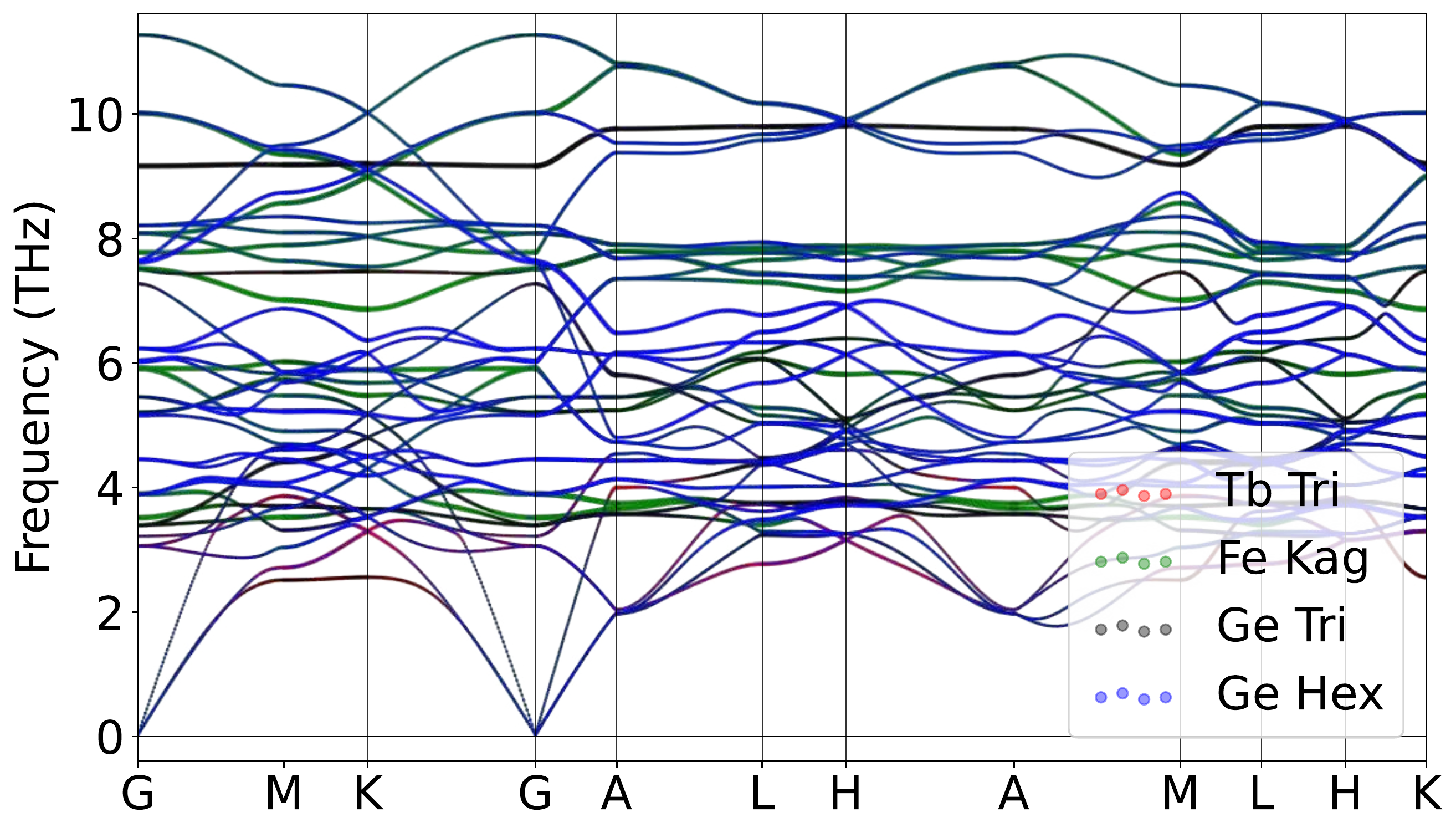}
}
\caption{Atomically projected phonon spectrum for TbFe$_6$Ge$_6$ in paramagnetic phase.}
\label{fig:TbFe6Ge6pho}
\end{figure}

\begin{figure}[htbp]
\centering
\includegraphics[width=0.45\textwidth]{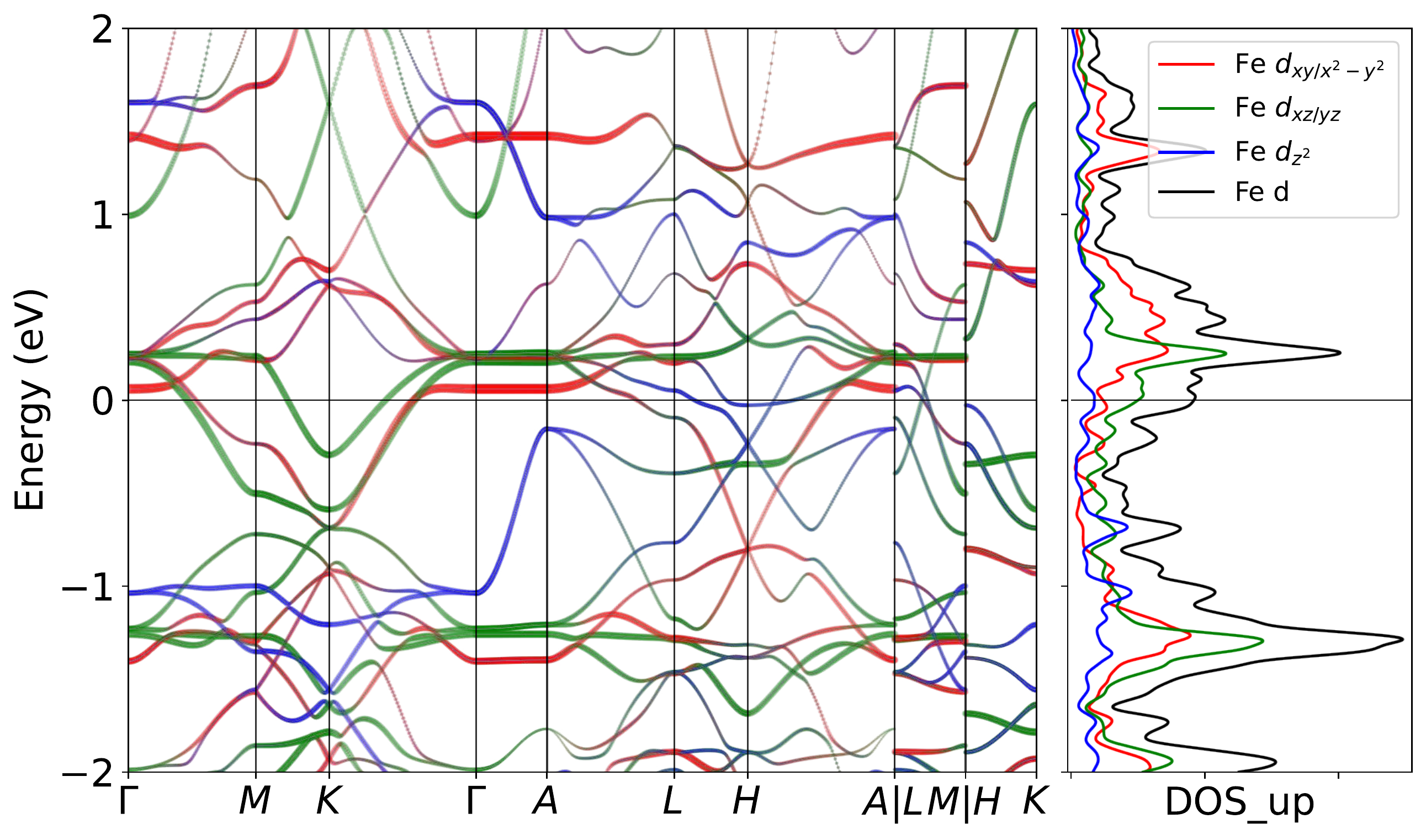}
\includegraphics[width=0.45\textwidth]{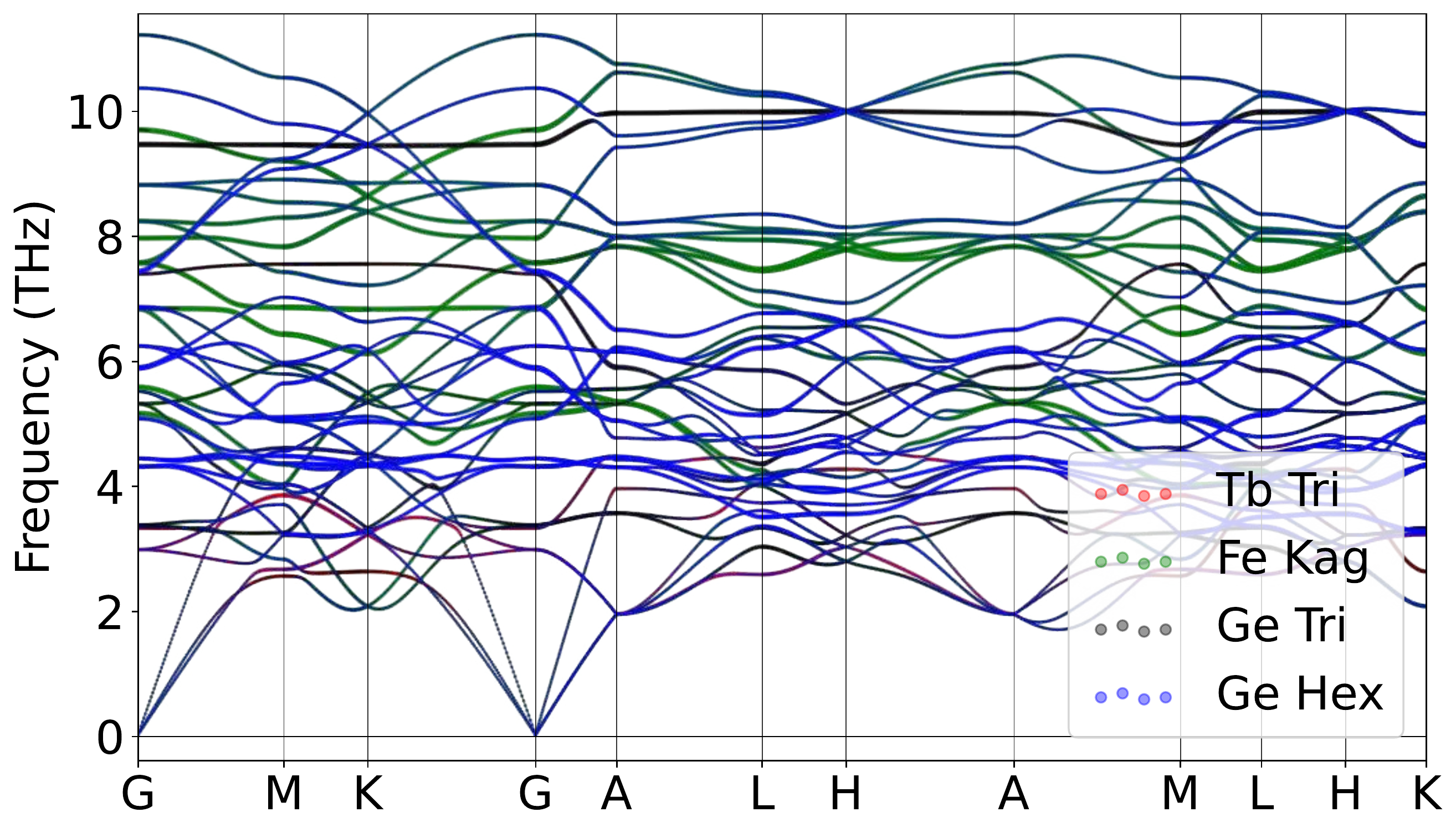}
\caption{Band structure for TbFe$_6$Ge$_6$ with AFM order without SOC for spin (Left)up channel. The projection of Fe $3d$ orbitals is also presented. (Right)Correspondingly, a stable phonon was demonstrated with the collinear magnetic order without Hubbard U at lower temperature regime, weighted by atomic projections.}
\label{fig:TbFe6Ge6mag}
\end{figure}

\begin{figure}[H]
\centering
\label{fig:TbFe6Ge6_relaxed_Low_xz}
\includegraphics[width=0.85\textwidth]{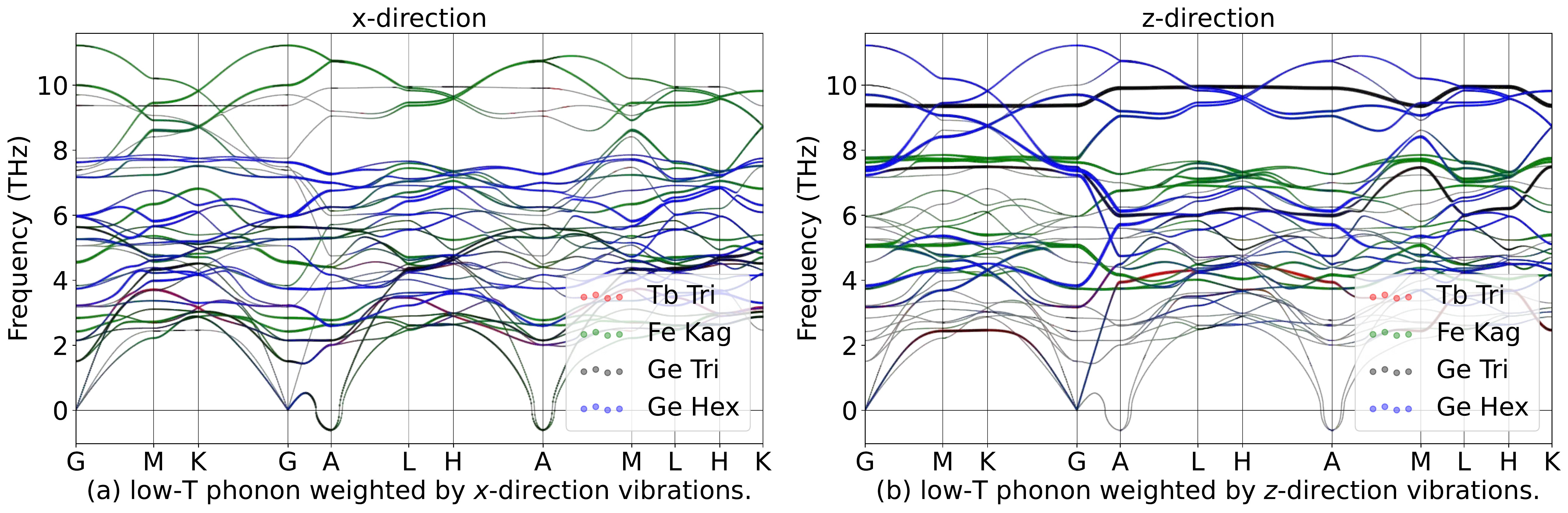}
\hspace{5pt}\label{fig:TbFe6Ge6_relaxed_High_xz}
\includegraphics[width=0.85\textwidth]{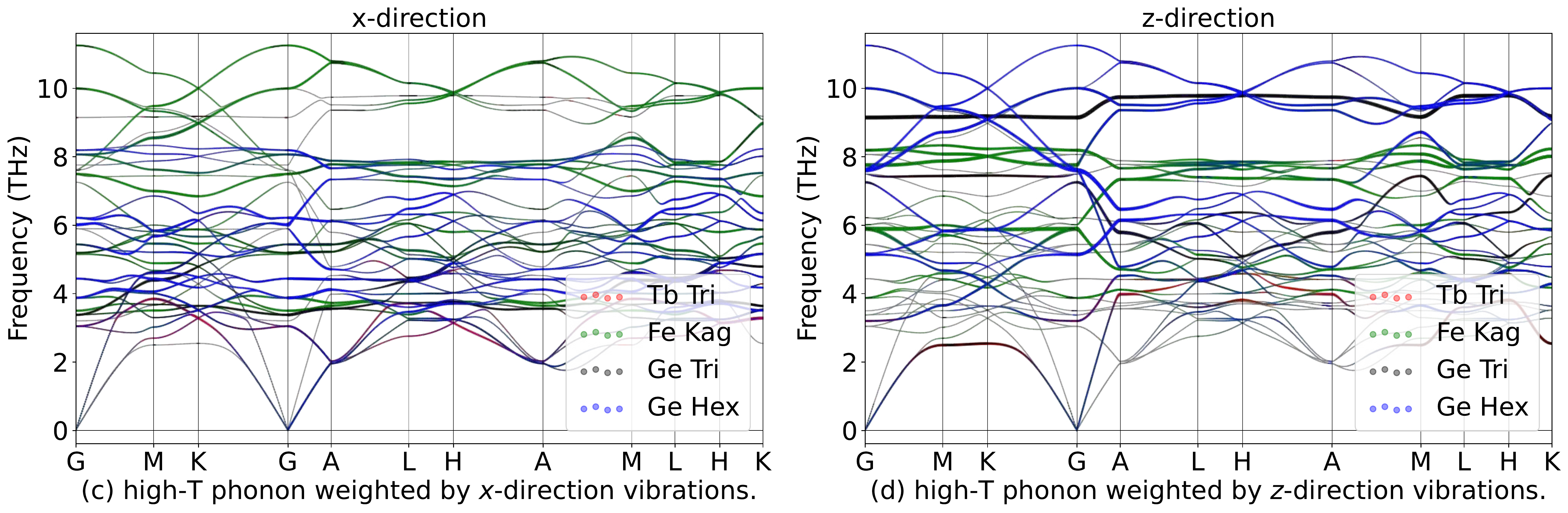}
\caption{Phonon spectrum for TbFe$_6$Ge$_6$in in-plane ($x$) and out-of-plane ($z$) directions in paramagnetic phase.}
\label{fig:TbFe6Ge6phoxyz}
\end{figure}

\begin{table}[htbp]
\global\long\def\arraystretch{1.12}
\captionsetup{width=.95\textwidth}
\caption{IRREPs at high-symmetry points for TbFe$_6$Ge$_6$ phonon spectrum at Low-T.}
\label{tab:191_TbFe6Ge6_Low}
\setlength{\tabcolsep}{1.mm}{
}
\end{table}

\begin{table}[htbp]
\global\long\def\arraystretch{1.12}
\captionsetup{width=.95\textwidth}
\caption{IRREPs at high-symmetry points for TbFe$_6$Ge$_6$ phonon spectrum at High-T.}
\label{tab:191_TbFe6Ge6_High}
\setlength{\tabcolsep}{1.mm}{
%
}
\end{table}
\subsubsection{\label{sms2sec:DyFe6Ge6}\hyperref[smsec:col]{\texorpdfstring{DyFe$_6$Ge$_6$}{}}~{\color{blue}  (High-T stable, Low-T unstable) }}

\begin{table}[htbp]
\global\long\def\arraystretch{1.12}
\captionsetup{width=.85\textwidth}
\caption{Structure parameters of DyFe$_6$Ge$_6$ after relaxation. It contains 414 electrons. }
\label{tab:DyFe6Ge6}
\setlength{\tabcolsep}{4mm}{
%
}
\end{table}

\begin{figure}[htbp]
\centering
\includegraphics[width=0.85\textwidth]{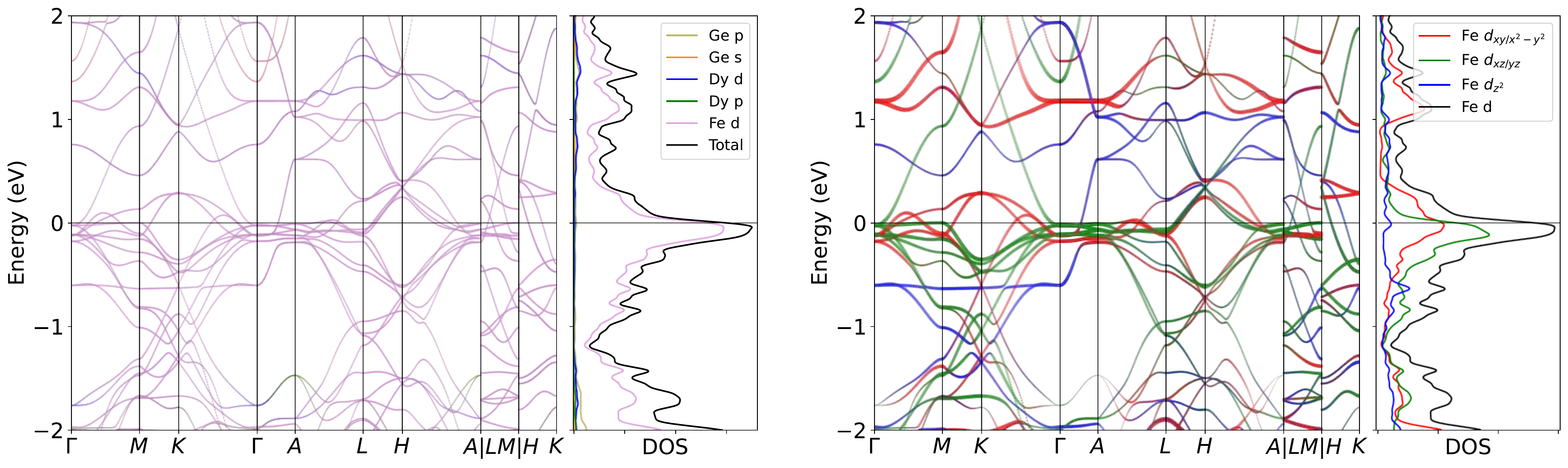}
\caption{Band structure for DyFe$_6$Ge$_6$ without SOC in paramagnetic phase. The projection of Fe $3d$ orbitals is also presented. The band structures clearly reveal the dominating of Fe $3d$ orbitals  near the Fermi level, as well as the pronounced peak.}
\label{fig:DyFe6Ge6band}
\end{figure}

\begin{figure}[H]
\centering
\subfloat[Relaxed phonon at low-T.]{
\label{fig:DyFe6Ge6_relaxed_Low}
\includegraphics[width=0.45\textwidth]{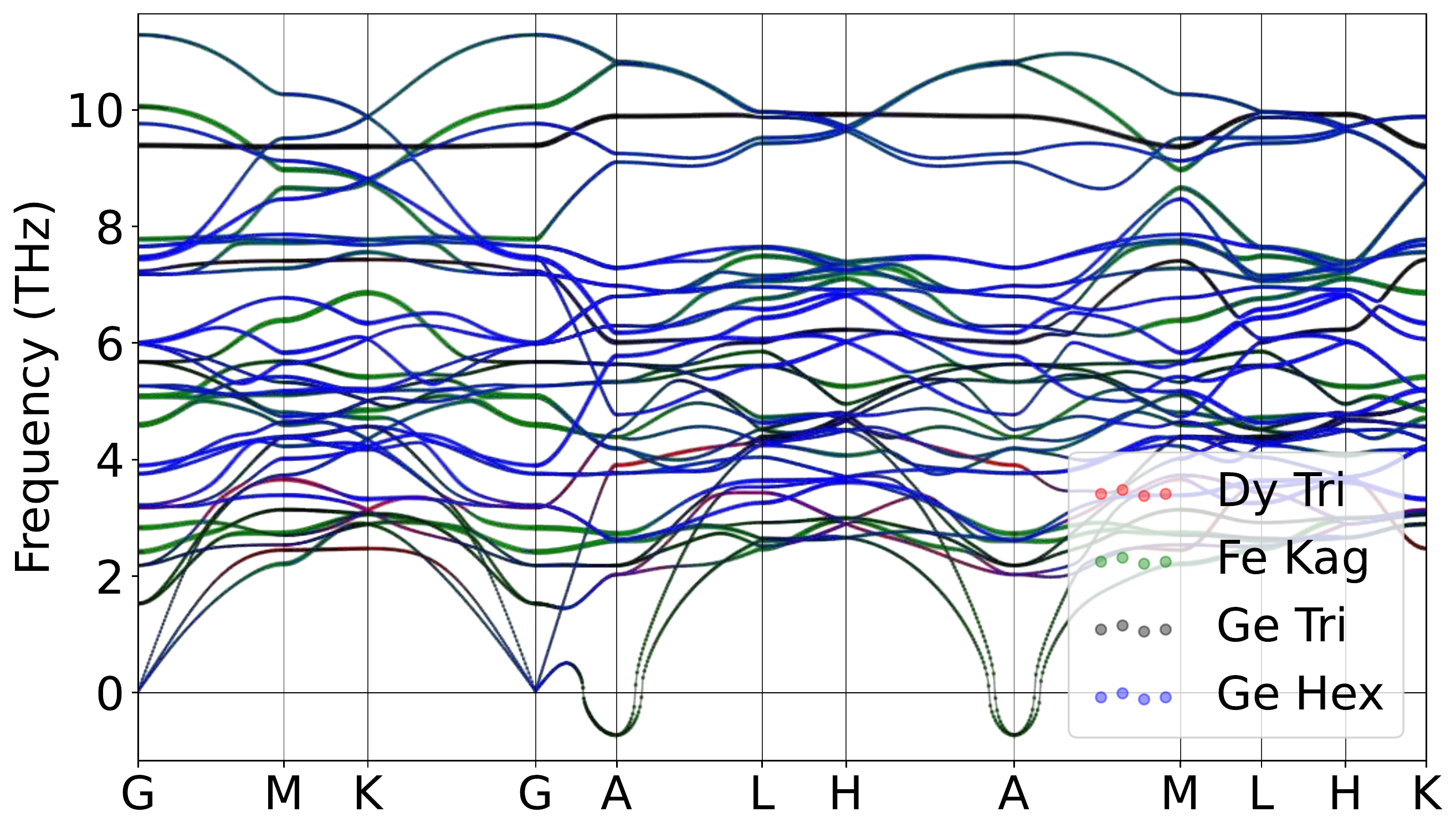}
}
\hspace{5pt}\subfloat[Relaxed phonon at high-T.]{
\label{fig:DyFe6Ge6_relaxed_High}
\includegraphics[width=0.45\textwidth]{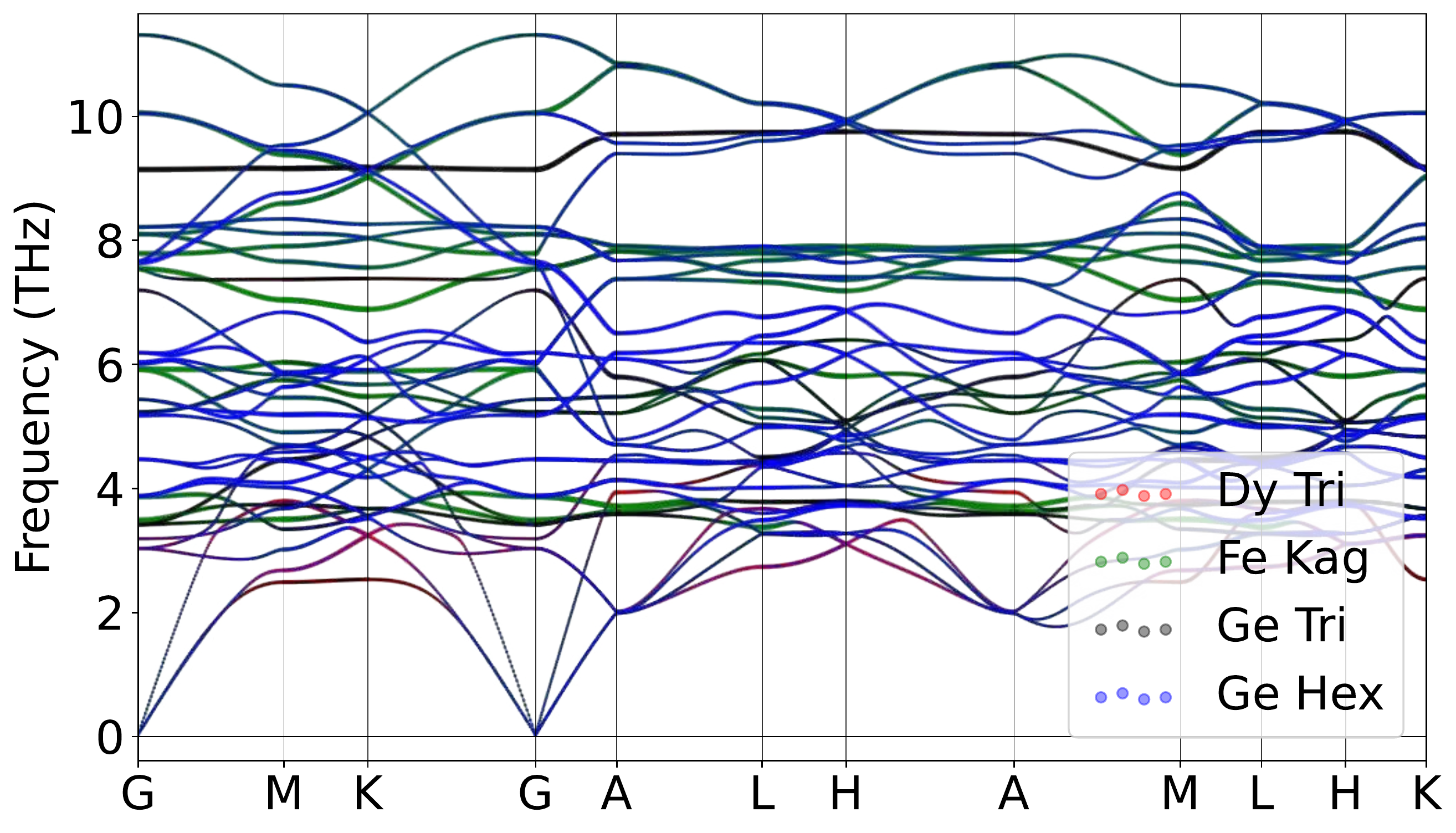}
}
\caption{Atomically projected phonon spectrum for DyFe$_6$Ge$_6$ in paramagnetic phase.}
\label{fig:DyFe6Ge6pho}
\end{figure}

\begin{figure}[htbp]
\centering
\includegraphics[width=0.45\textwidth]{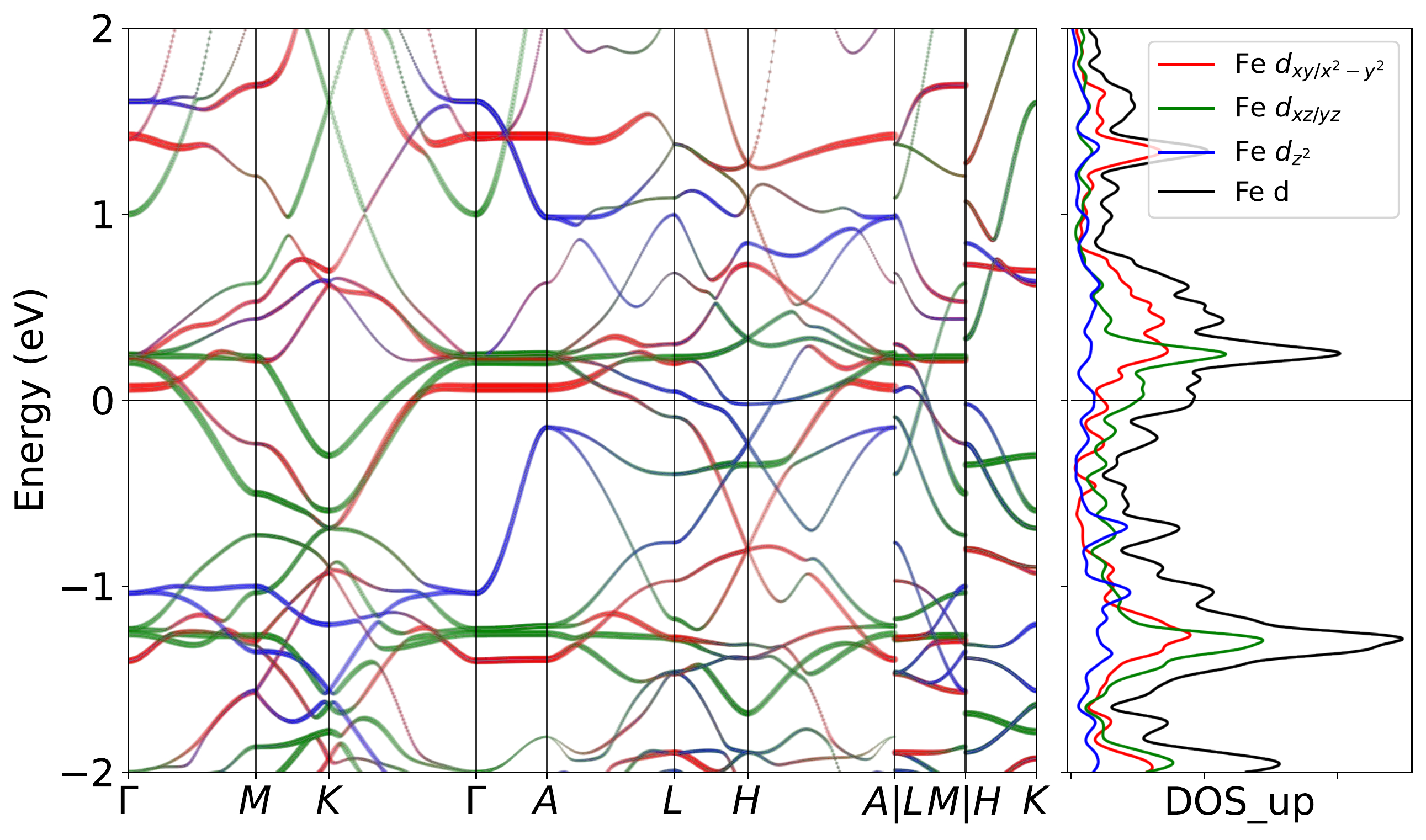}
\includegraphics[width=0.45\textwidth]{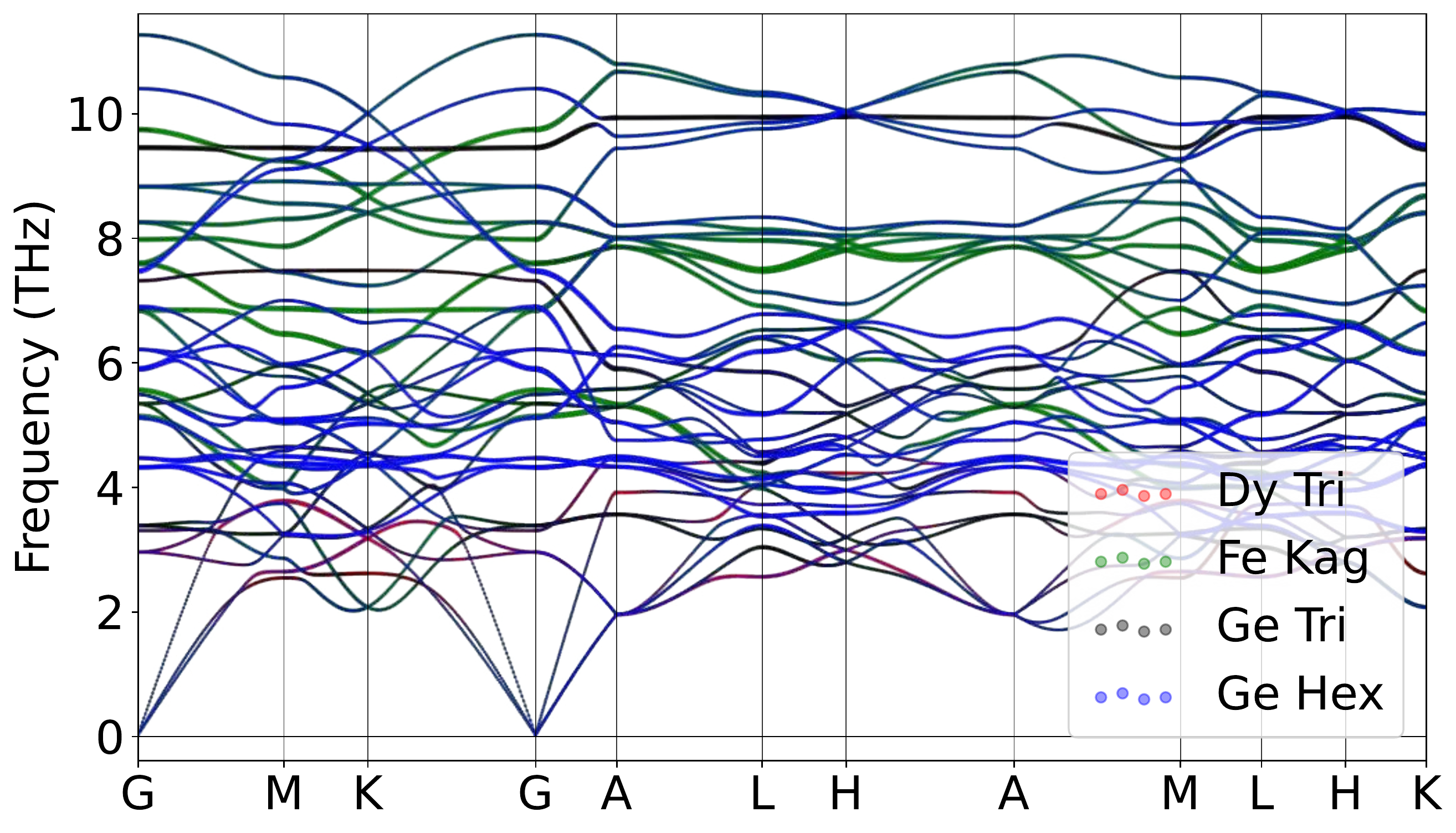}
\caption{Band structure for DyFe$_6$Ge$_6$ with AFM order without SOC for spin (Left)up channel. The projection of Fe $3d$ orbitals is also presented. (Right)Correspondingly, a stable phonon was demonstrated with the collinear magnetic order without Hubbard U at lower temperature regime, weighted by atomic projections.}
\label{fig:DyFe6Ge6mag}
\end{figure}

\begin{figure}[H]
\centering
\label{fig:DyFe6Ge6_relaxed_Low_xz}
\includegraphics[width=0.85\textwidth]{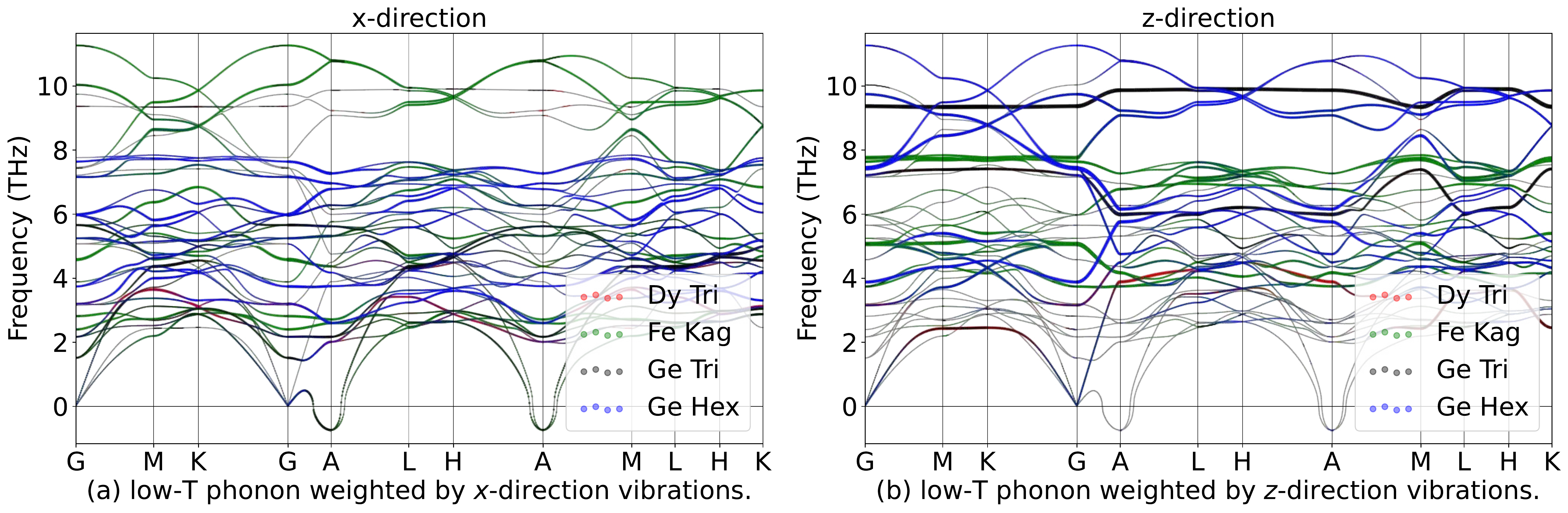}
\hspace{5pt}\label{fig:DyFe6Ge6_relaxed_High_xz}
\includegraphics[width=0.85\textwidth]{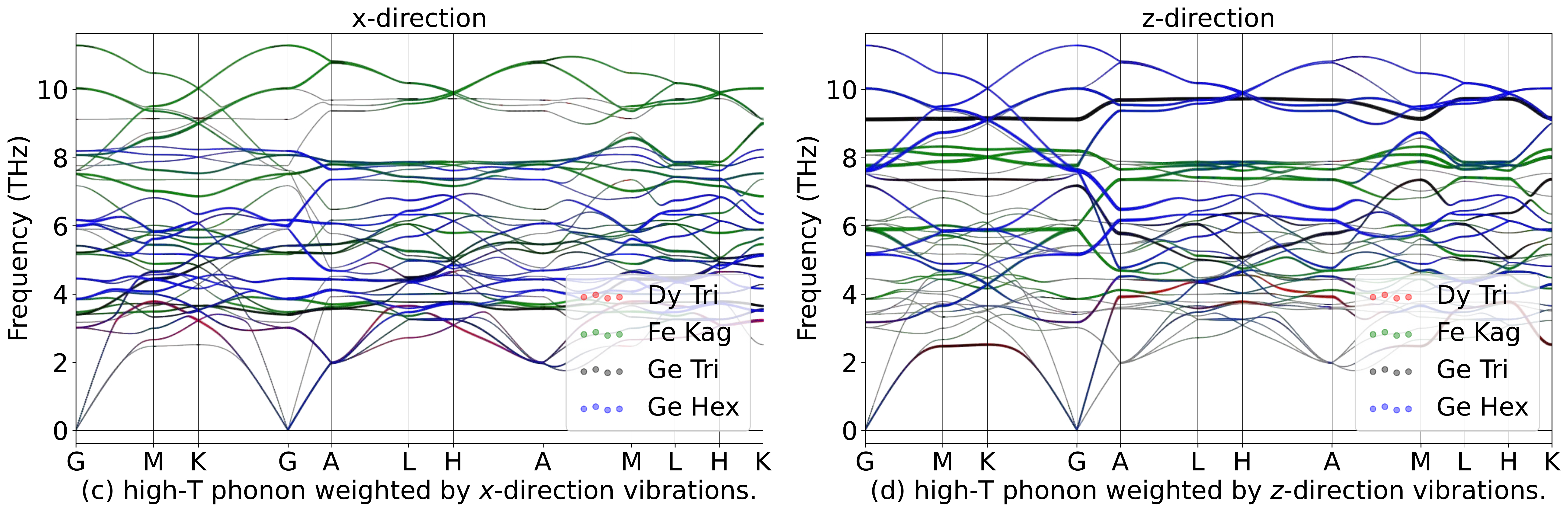}
\caption{Phonon spectrum for DyFe$_6$Ge$_6$in in-plane ($x$) and out-of-plane ($z$) directions in paramagnetic phase.}
\label{fig:DyFe6Ge6phoxyz}
\end{figure}

\begin{table}[htbp]
\global\long\def\arraystretch{1.12}
\captionsetup{width=.95\textwidth}
\caption{IRREPs at high-symmetry points for DyFe$_6$Ge$_6$ phonon spectrum at Low-T.}
\label{tab:191_DyFe6Ge6_Low}
\setlength{\tabcolsep}{1.mm}{
}
\end{table}

\begin{table}[htbp]
\global\long\def\arraystretch{1.12}
\captionsetup{width=.95\textwidth}
\caption{IRREPs at high-symmetry points for DyFe$_6$Ge$_6$ phonon spectrum at High-T.}
\label{tab:191_DyFe6Ge6_High}
\setlength{\tabcolsep}{1.mm}{
%
}
\end{table}
\subsubsection{\label{sms2sec:HoFe6Ge6}\hyperref[smsec:col]{\texorpdfstring{HoFe$_6$Ge$_6$}{}}~{\color{blue}  (High-T stable, Low-T unstable) }}

\begin{table}[htbp]
\global\long\def\arraystretch{1.12}
\captionsetup{width=.85\textwidth}
\caption{Structure parameters of HoFe$_6$Ge$_6$ after relaxation. It contains 415 electrons. }
\label{tab:HoFe6Ge6}
\setlength{\tabcolsep}{4mm}{
%
}
\end{table}

\begin{figure}[htbp]
\centering
\includegraphics[width=0.85\textwidth]{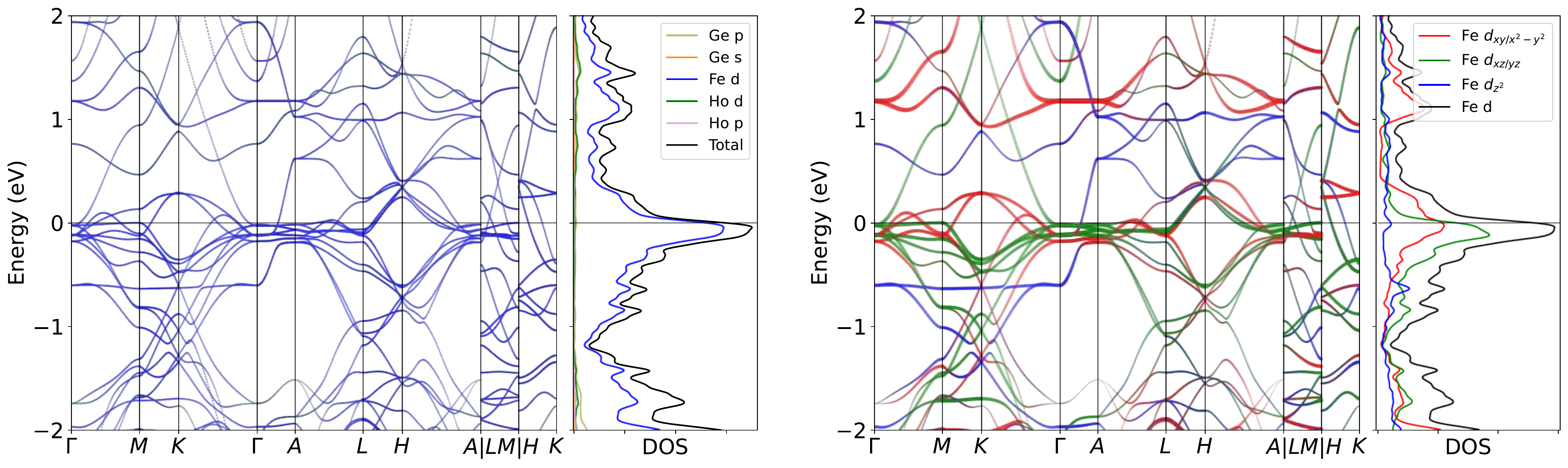}
\caption{Band structure for HoFe$_6$Ge$_6$ without SOC in paramagnetic phase. The projection of Fe $3d$ orbitals is also presented. The band structures clearly reveal the dominating of Fe $3d$ orbitals  near the Fermi level, as well as the pronounced peak.}
\label{fig:HoFe6Ge6band}
\end{figure}

\begin{figure}[H]
\centering
\subfloat[Relaxed phonon at low-T.]{
\label{fig:HoFe6Ge6_relaxed_Low}
\includegraphics[width=0.45\textwidth]{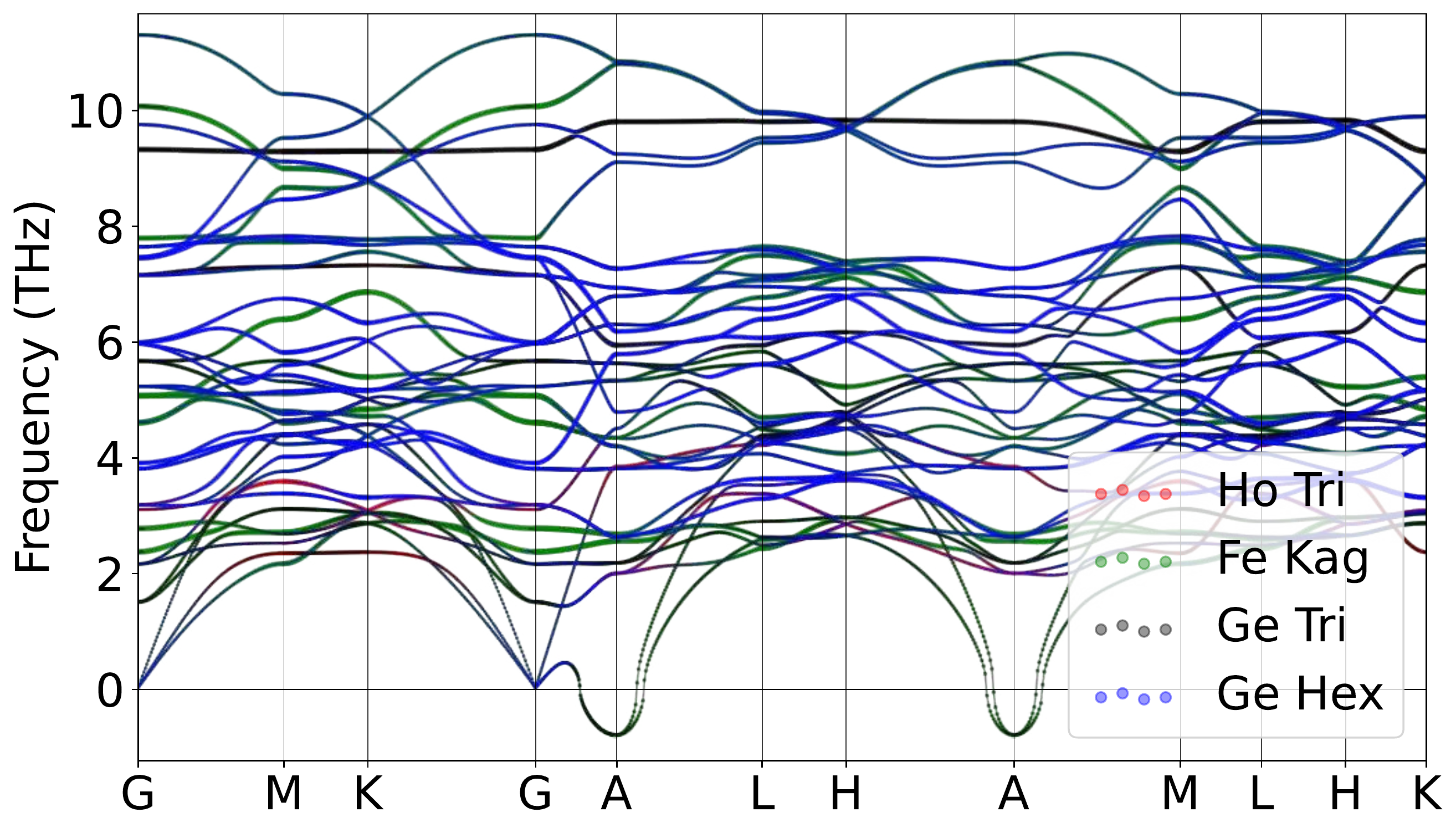}
}
\hspace{5pt}\subfloat[Relaxed phonon at high-T.]{
\label{fig:HoFe6Ge6_relaxed_High}
\includegraphics[width=0.45\textwidth]{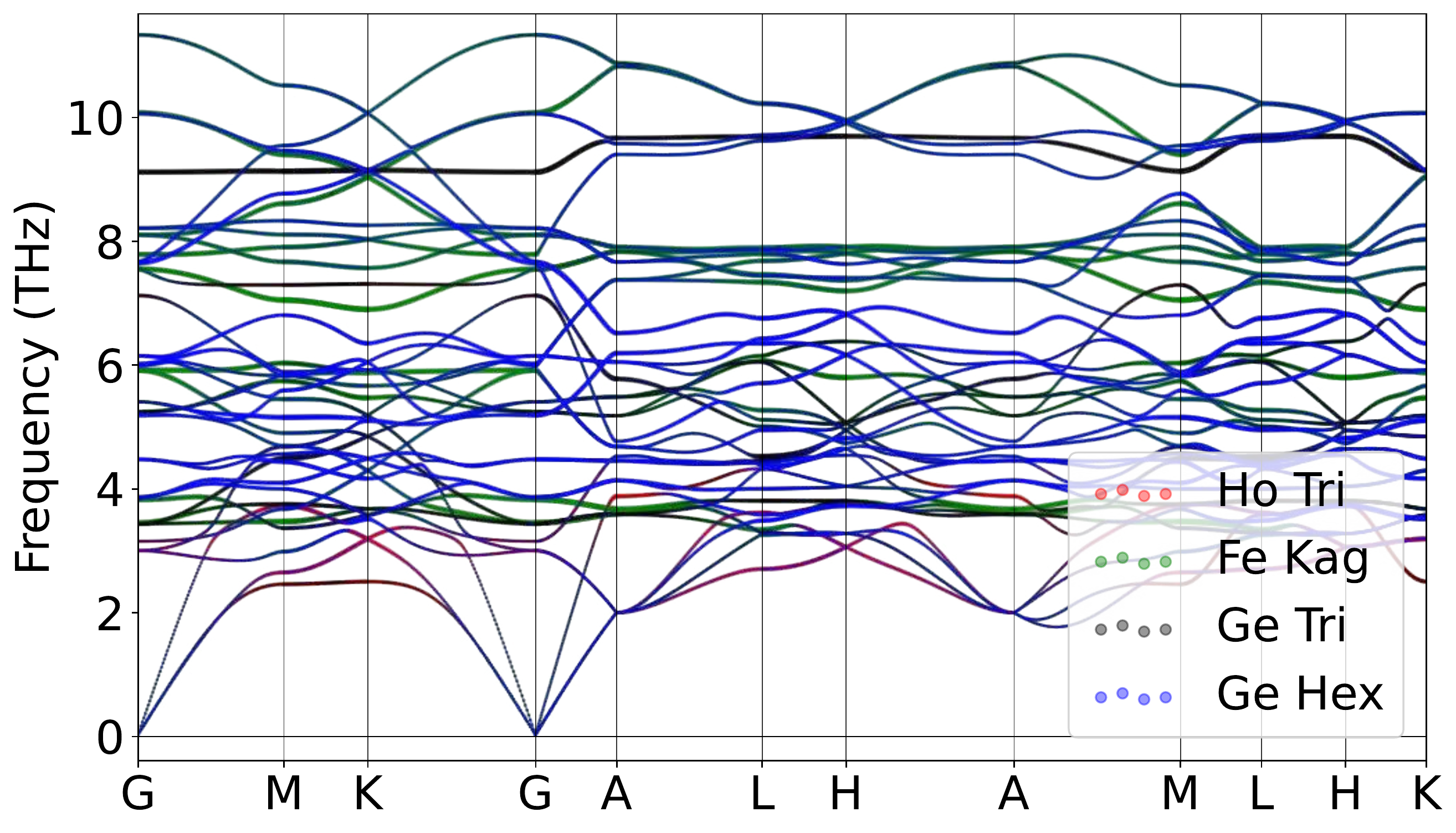}
}
\caption{Atomically projected phonon spectrum for HoFe$_6$Ge$_6$ in paramagnetic phase.}
\label{fig:HoFe6Ge6pho}
\end{figure}

\begin{figure}[htbp]
\centering
\includegraphics[width=0.45\textwidth]{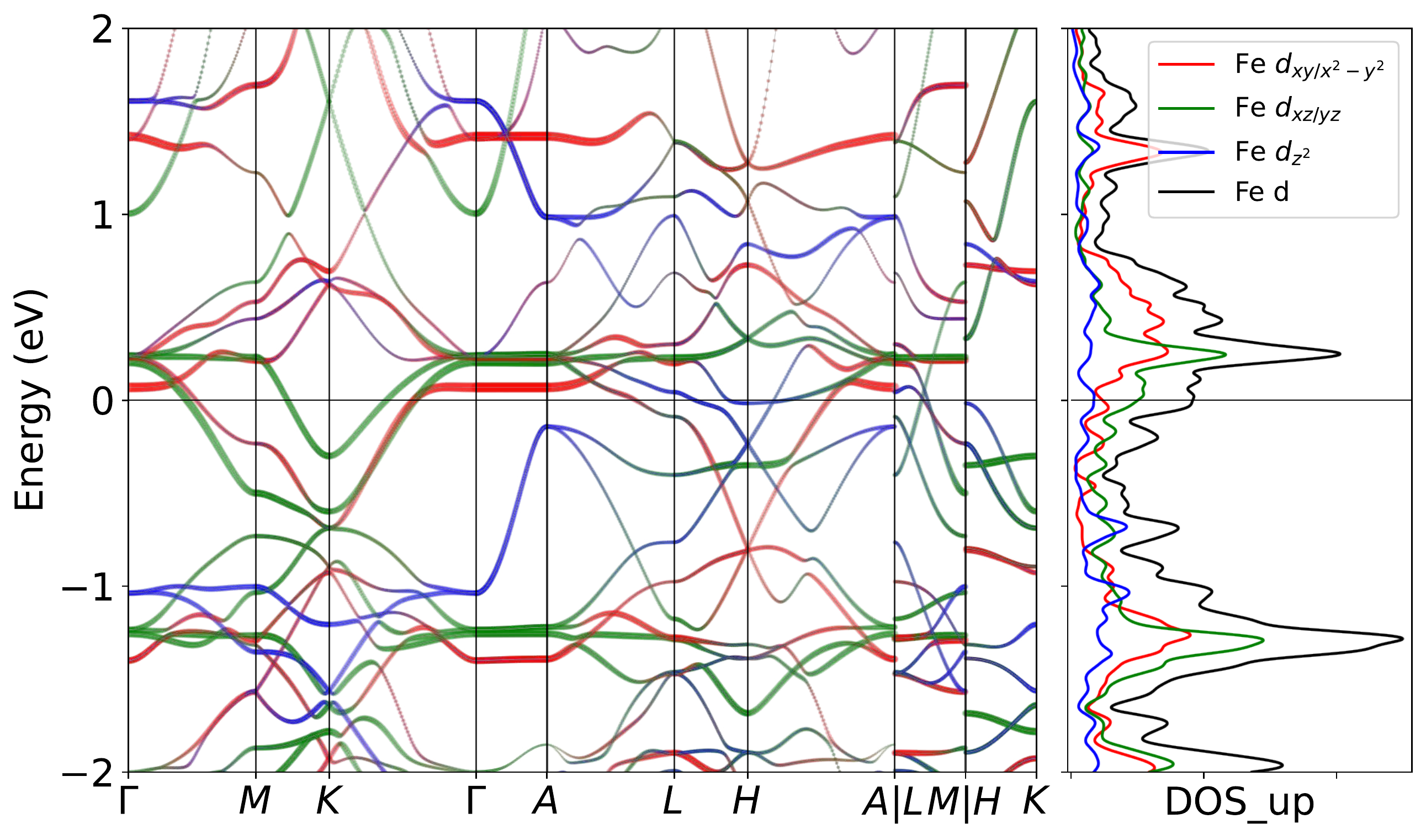}
\includegraphics[width=0.45\textwidth]{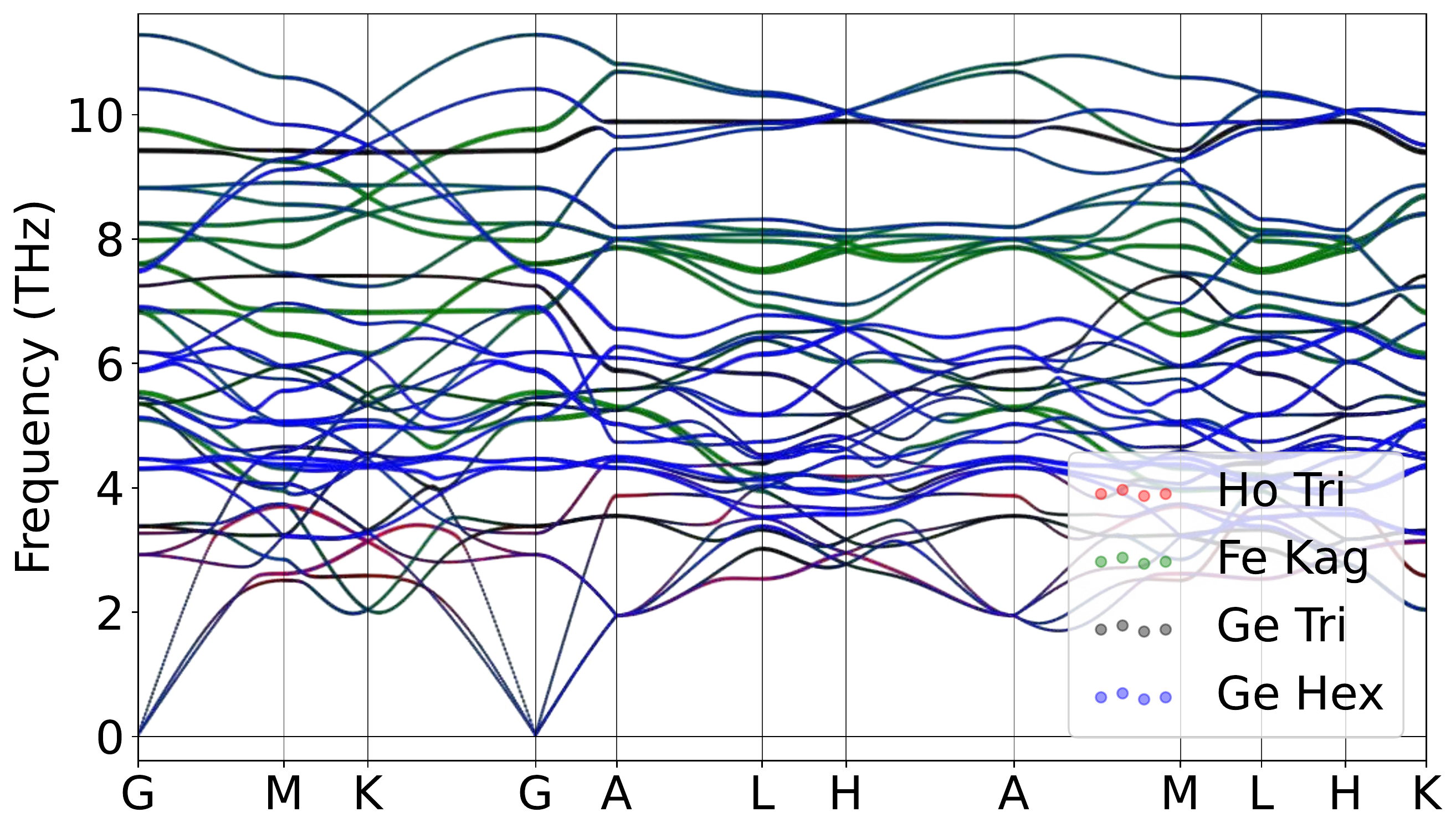}
\caption{Band structure for HoFe$_6$Ge$_6$ with AFM order without SOC for spin (Left)up channel. The projection of Fe $3d$ orbitals is also presented. (Right)Correspondingly, a stable phonon was demonstrated with the collinear magnetic order without Hubbard U at lower temperature regime, weighted by atomic projections.}
\label{fig:HoFe6Ge6mag}
\end{figure}

\begin{figure}[H]
\centering
\label{fig:HoFe6Ge6_relaxed_Low_xz}
\includegraphics[width=0.85\textwidth]{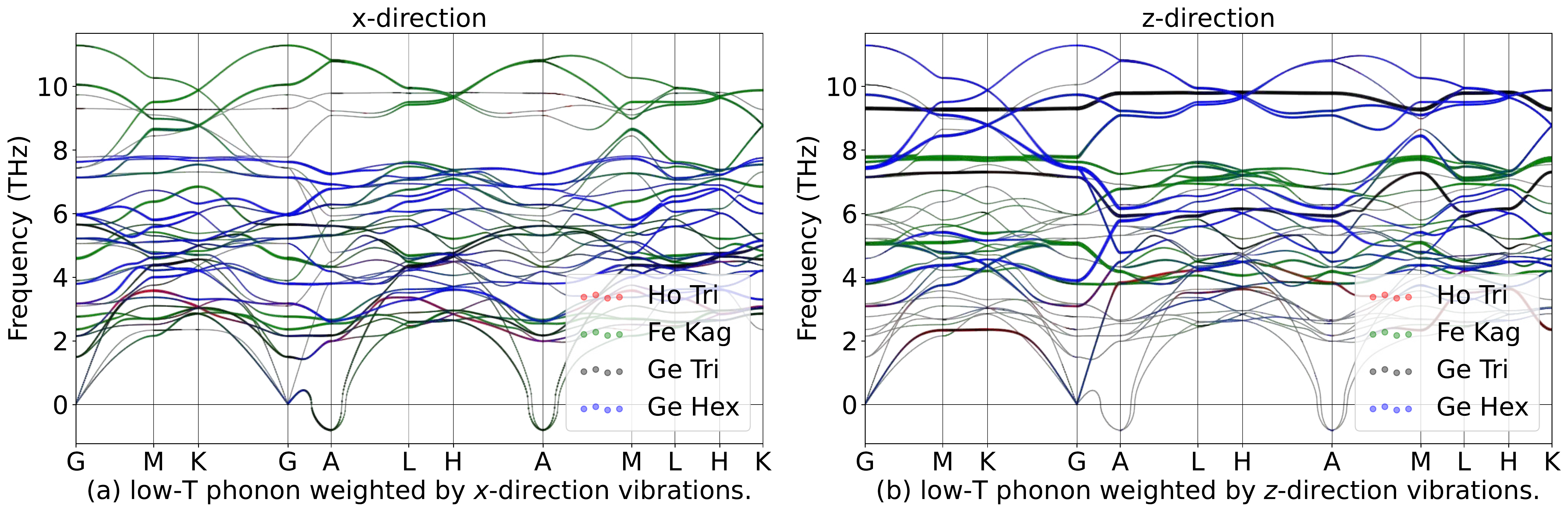}
\hspace{5pt}\label{fig:HoFe6Ge6_relaxed_High_xz}
\includegraphics[width=0.85\textwidth]{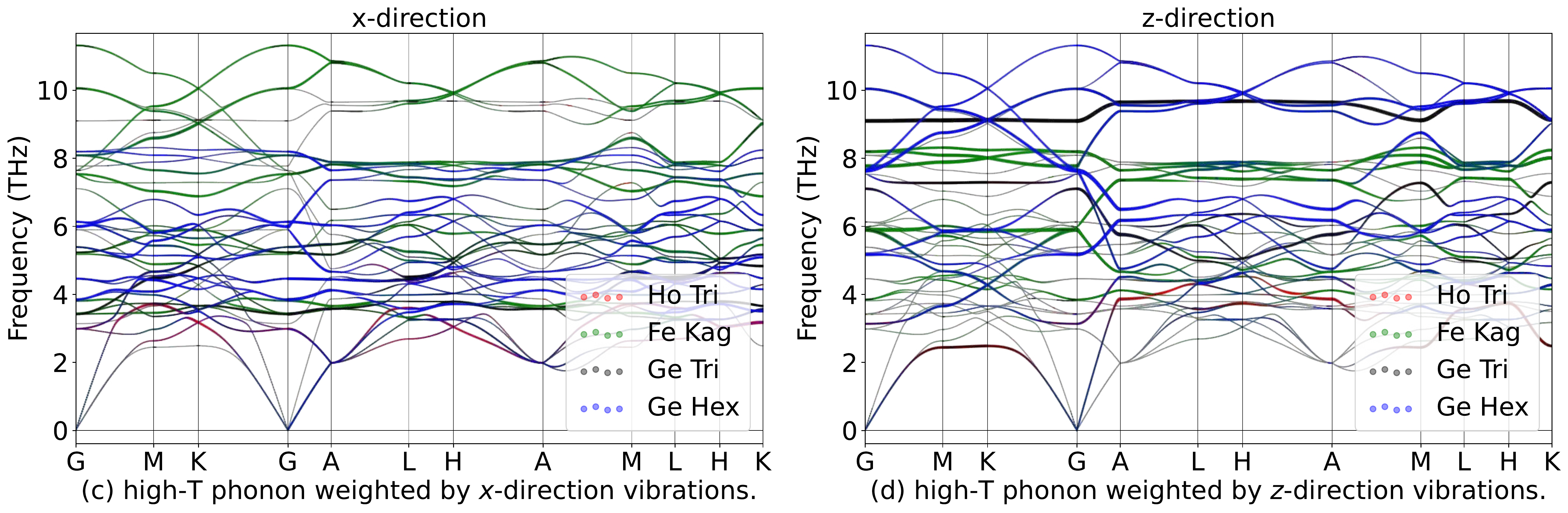}
\caption{Phonon spectrum for HoFe$_6$Ge$_6$in in-plane ($x$) and out-of-plane ($z$) directions in paramagnetic phase.}
\label{fig:HoFe6Ge6phoxyz}
\end{figure}

\begin{table}[htbp]
\global\long\def\arraystretch{1.12}
\captionsetup{width=.95\textwidth}
\caption{IRREPs at high-symmetry points for HoFe$_6$Ge$_6$ phonon spectrum at Low-T.}
\label{tab:191_HoFe6Ge6_Low}
\setlength{\tabcolsep}{1.mm}{
}
\end{table}

\begin{table}[htbp]
\global\long\def\arraystretch{1.12}
\captionsetup{width=.95\textwidth}
\caption{IRREPs at high-symmetry points for HoFe$_6$Ge$_6$ phonon spectrum at High-T.}
\label{tab:191_HoFe6Ge6_High}
\setlength{\tabcolsep}{1.mm}{
%
}
\end{table}
\subsubsection{\label{sms2sec:ErFe6Ge6}\hyperref[smsec:col]{\texorpdfstring{ErFe$_6$Ge$_6$}{}}~{\color{blue}  (High-T stable, Low-T unstable) }}

\begin{table}[htbp]
\global\long\def\arraystretch{1.12}
\captionsetup{width=.85\textwidth}
\caption{Structure parameters of ErFe$_6$Ge$_6$ after relaxation. It contains 416 electrons. }
\label{tab:ErFe6Ge6}
\setlength{\tabcolsep}{4mm}{
%
}
\end{table}

\begin{figure}[htbp]
\centering
\includegraphics[width=0.85\textwidth]{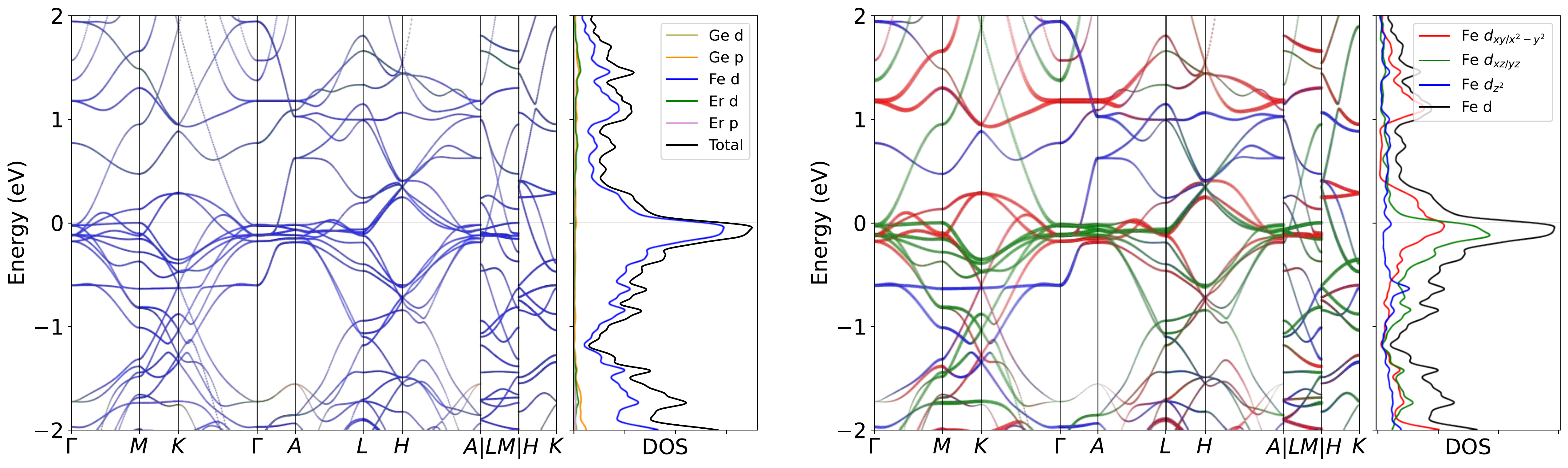}
\caption{Band structure for ErFe$_6$Ge$_6$ without SOC in paramagnetic phase. The projection of Fe $3d$ orbitals is also presented. The band structures clearly reveal the dominating of Fe $3d$ orbitals  near the Fermi level, as well as the pronounced peak.}
\label{fig:ErFe6Ge6band}
\end{figure}

\begin{figure}[H]
\centering
\subfloat[Relaxed phonon at low-T.]{
\label{fig:ErFe6Ge6_relaxed_Low}
\includegraphics[width=0.45\textwidth]{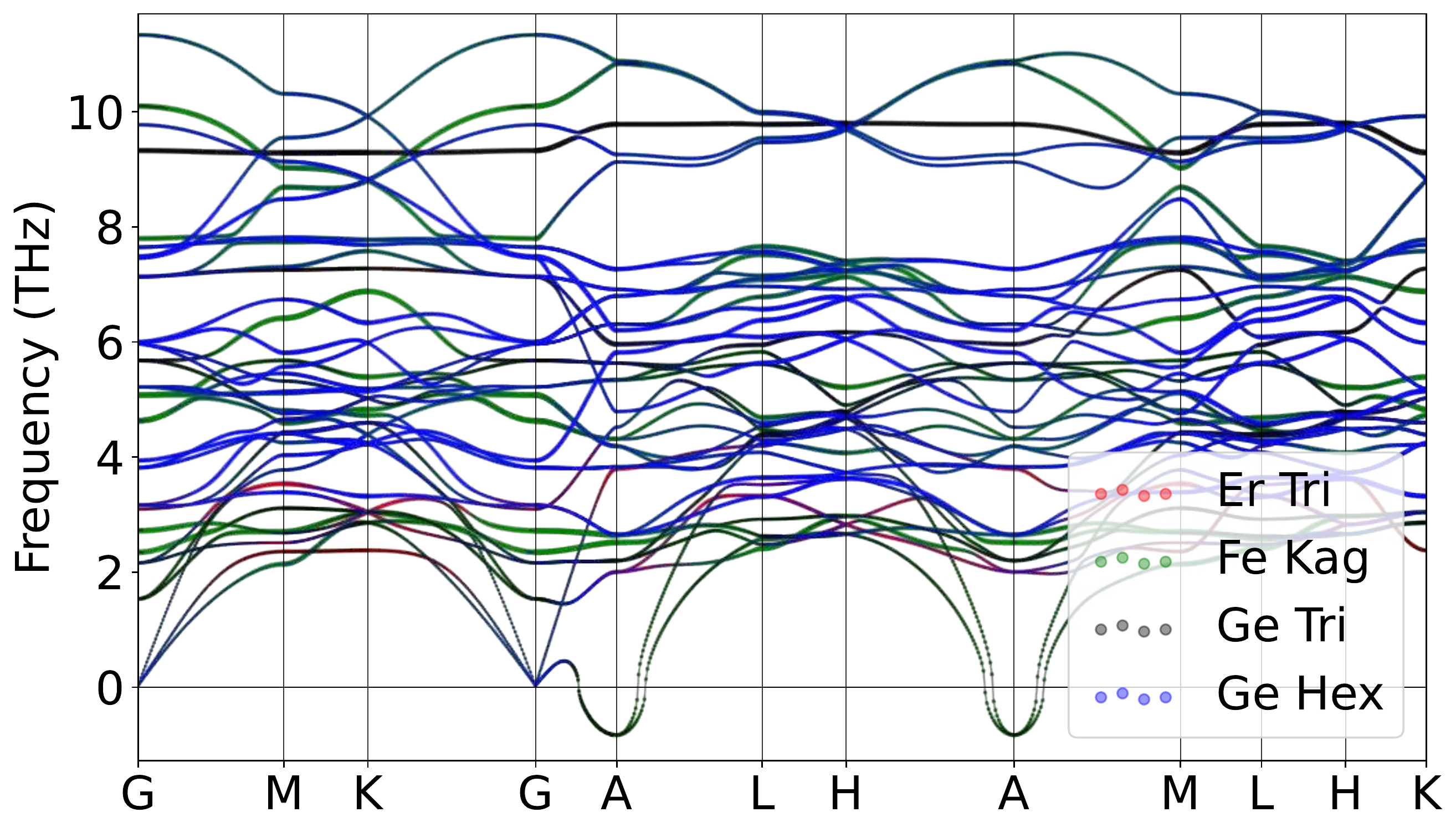}
}
\hspace{5pt}\subfloat[Relaxed phonon at high-T.]{
\label{fig:ErFe6Ge6_relaxed_High}
\includegraphics[width=0.45\textwidth]{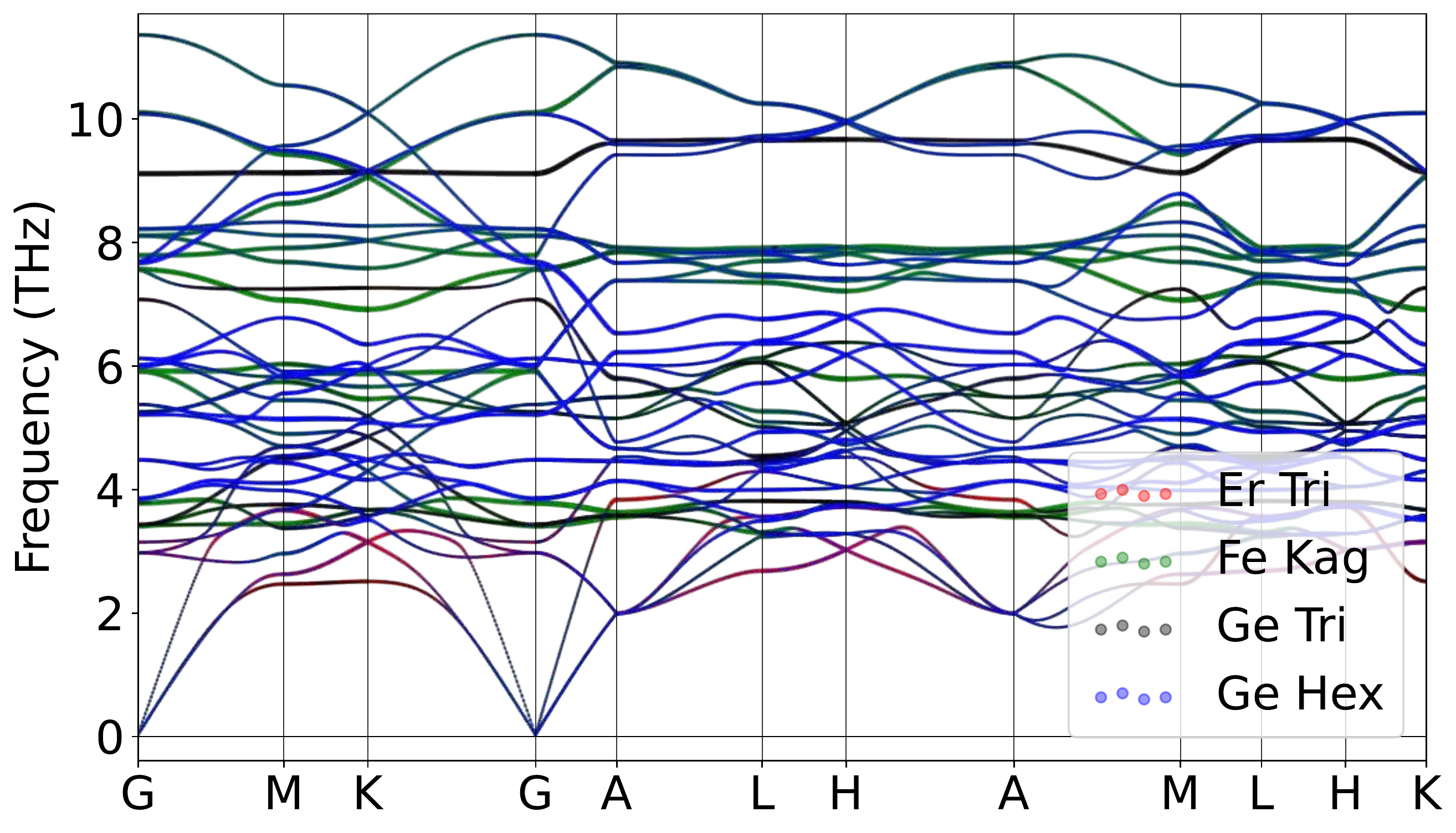}
}
\caption{Atomically projected phonon spectrum for ErFe$_6$Ge$_6$ in paramagnetic phase.}
\label{fig:ErFe6Ge6pho}
\end{figure}

\begin{figure}[htbp]
\centering
\includegraphics[width=0.45\textwidth]{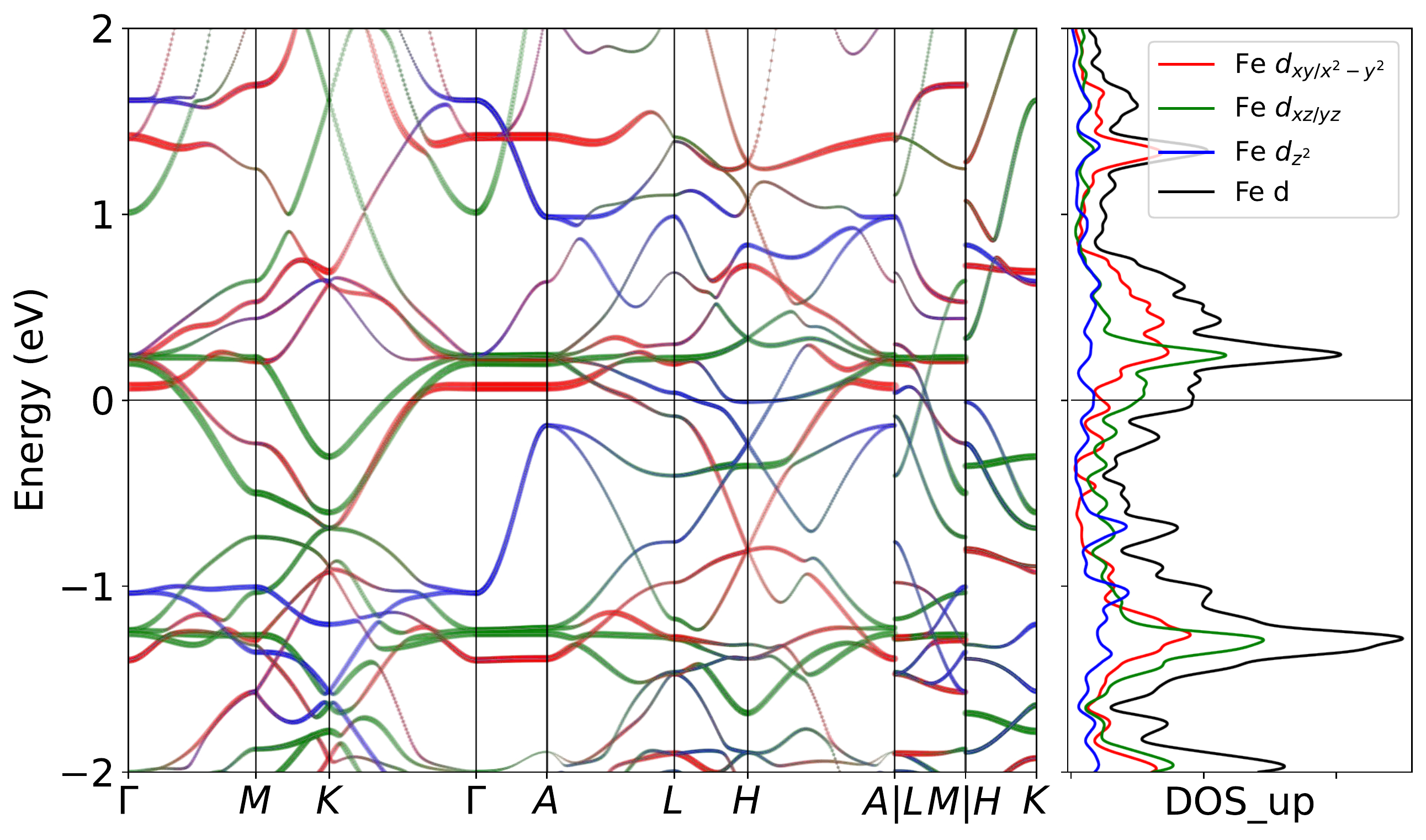}
\includegraphics[width=0.45\textwidth]{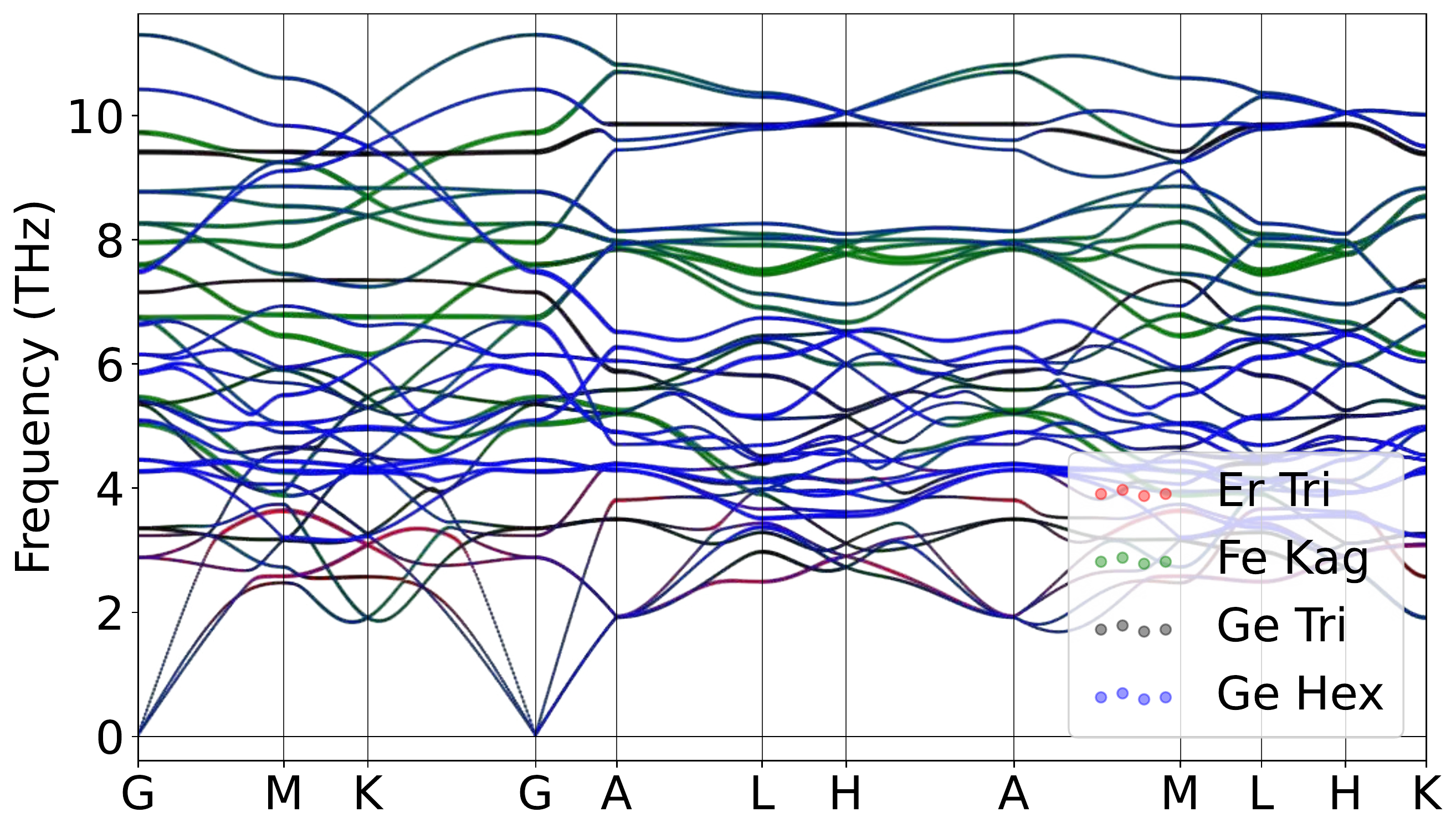}
\caption{Band structure for ErFe$_6$Ge$_6$ with AFM order without SOC for spin (Left)up channel. The projection of Fe $3d$ orbitals is also presented. (Right)Correspondingly, a stable phonon was demonstrated with the collinear magnetic order without Hubbard U at lower temperature regime, weighted by atomic projections.}
\label{fig:ErFe6Ge6mag}
\end{figure}

\begin{figure}[H]
\centering
\label{fig:ErFe6Ge6_relaxed_Low_xz}
\includegraphics[width=0.85\textwidth]{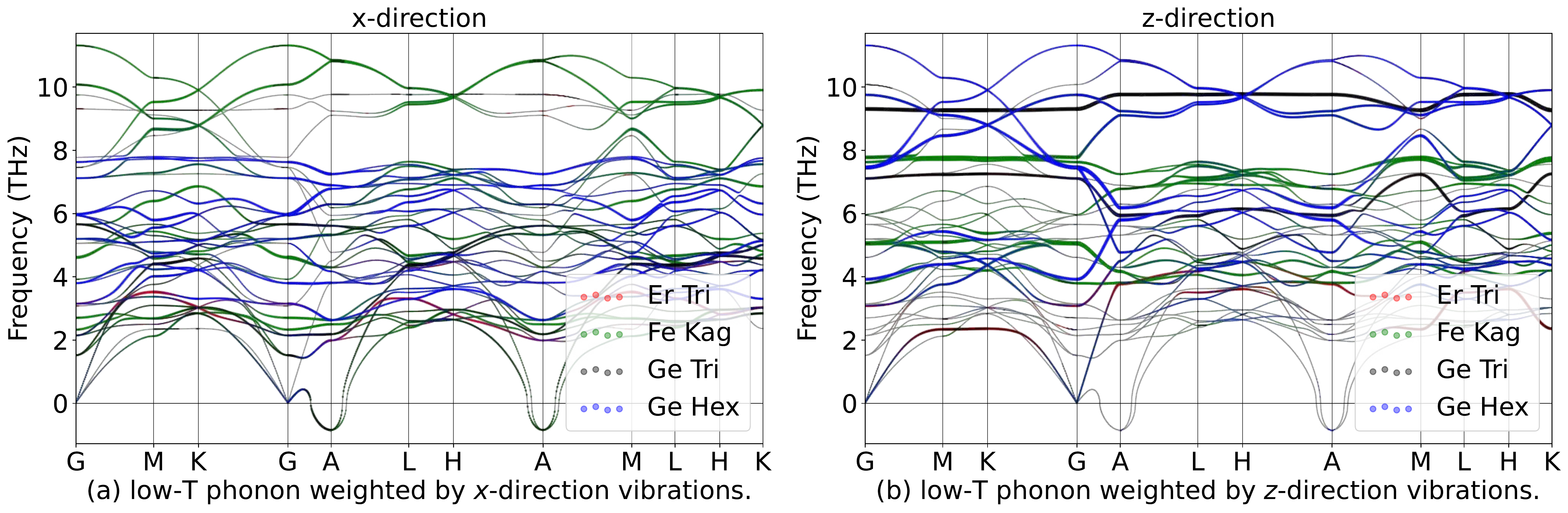}
\hspace{5pt}\label{fig:ErFe6Ge6_relaxed_High_xz}
\includegraphics[width=0.85\textwidth]{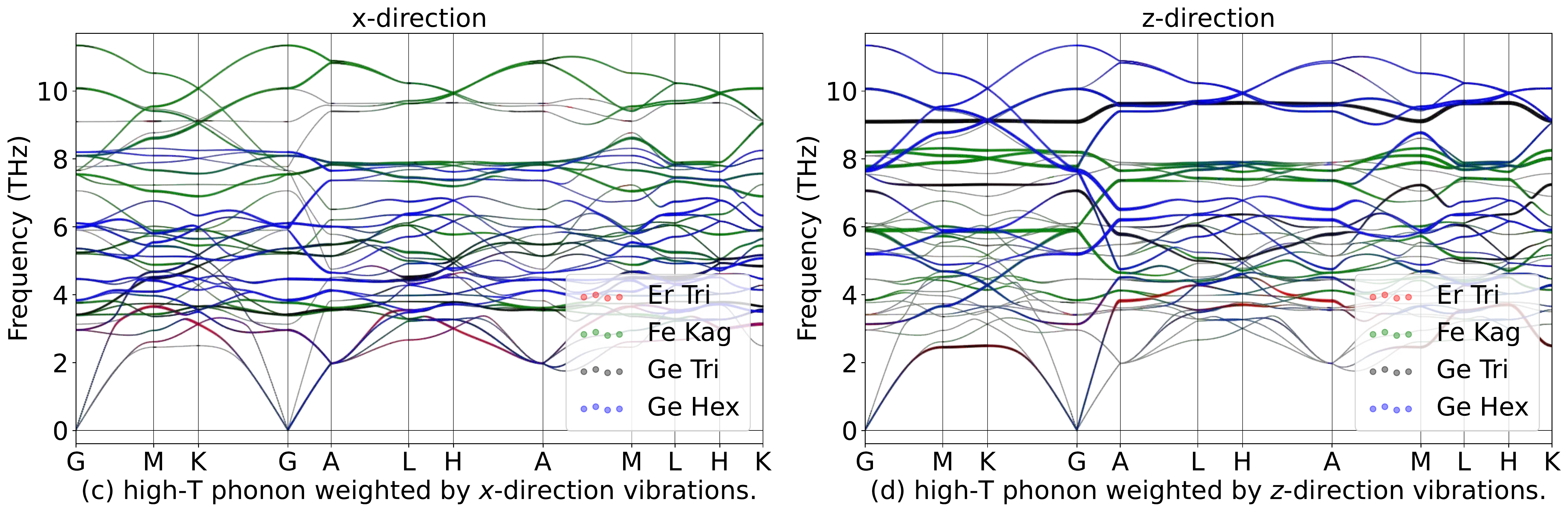}
\caption{Phonon spectrum for ErFe$_6$Ge$_6$in in-plane ($x$) and out-of-plane ($z$) directions in paramagnetic phase.}
\label{fig:ErFe6Ge6phoxyz}
\end{figure}

\begin{table}[htbp]
\global\long\def\arraystretch{1.12}
\captionsetup{width=.95\textwidth}
\caption{IRREPs at high-symmetry points for ErFe$_6$Ge$_6$ phonon spectrum at Low-T.}
\label{tab:191_ErFe6Ge6_Low}
\setlength{\tabcolsep}{1.mm}{
}
\end{table}

\begin{table}[htbp]
\global\long\def\arraystretch{1.12}
\captionsetup{width=.95\textwidth}
\caption{IRREPs at high-symmetry points for ErFe$_6$Ge$_6$ phonon spectrum at High-T.}
\label{tab:191_ErFe6Ge6_High}
\setlength{\tabcolsep}{1.mm}{
%
}
\end{table}
\subsubsection{\label{sms2sec:TmFe6Ge6}\hyperref[smsec:col]{\texorpdfstring{TmFe$_6$Ge$_6$}{}}~{\color{blue}  (High-T stable, Low-T unstable) }}

\begin{table}[htbp]
\global\long\def\arraystretch{1.12}
\captionsetup{width=.85\textwidth}
\caption{Structure parameters of TmFe$_6$Ge$_6$ after relaxation. It contains 417 electrons. }
\label{tab:TmFe6Ge6}
\setlength{\tabcolsep}{4mm}{
%
}
\end{table}

\begin{figure}[htbp]
\centering
\includegraphics[width=0.85\textwidth]{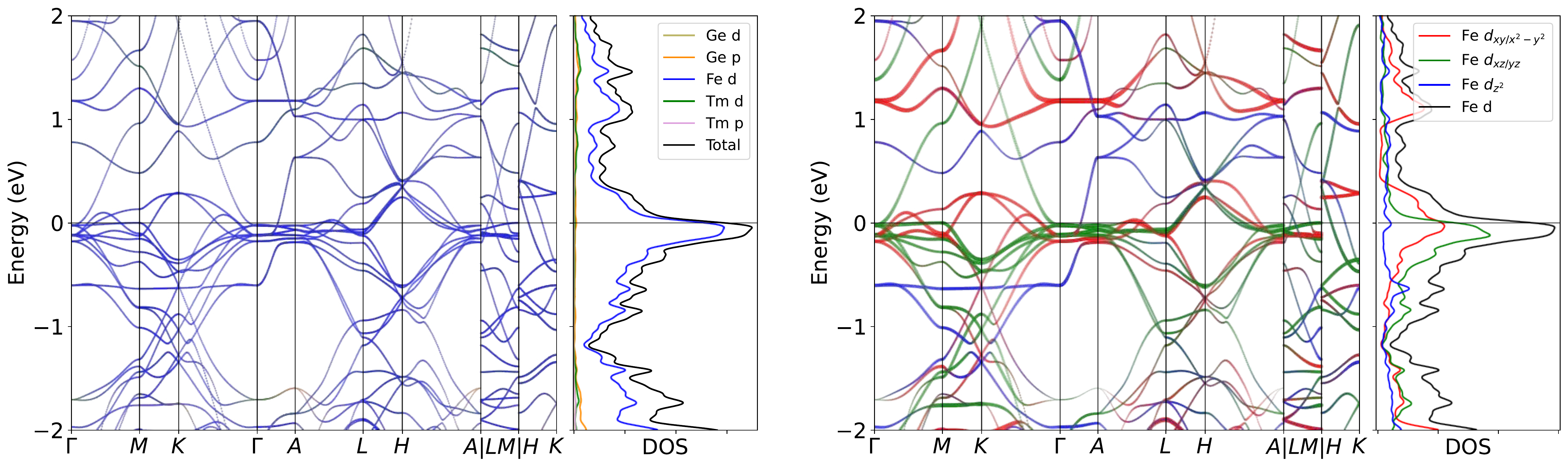}
\caption{Band structure for TmFe$_6$Ge$_6$ without SOC in paramagnetic phase. The projection of Fe $3d$ orbitals is also presented. The band structures clearly reveal the dominating of Fe $3d$ orbitals  near the Fermi level, as well as the pronounced peak.}
\label{fig:TmFe6Ge6band}
\end{figure}

\begin{figure}[H]
\centering
\subfloat[Relaxed phonon at low-T.]{
\label{fig:TmFe6Ge6_relaxed_Low}
\includegraphics[width=0.45\textwidth]{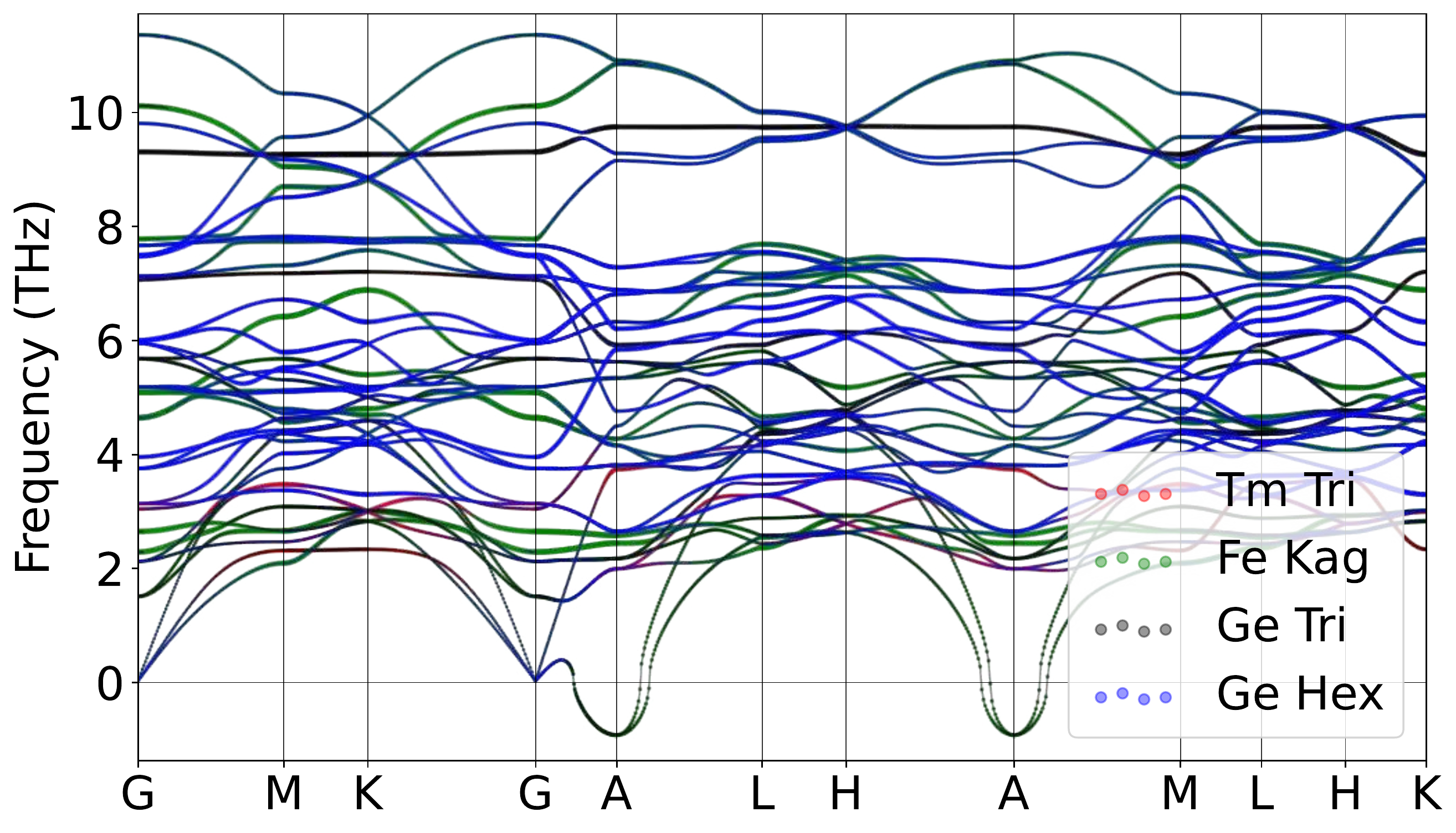}
}
\hspace{5pt}\subfloat[Relaxed phonon at high-T.]{
\label{fig:TmFe6Ge6_relaxed_High}
\includegraphics[width=0.45\textwidth]{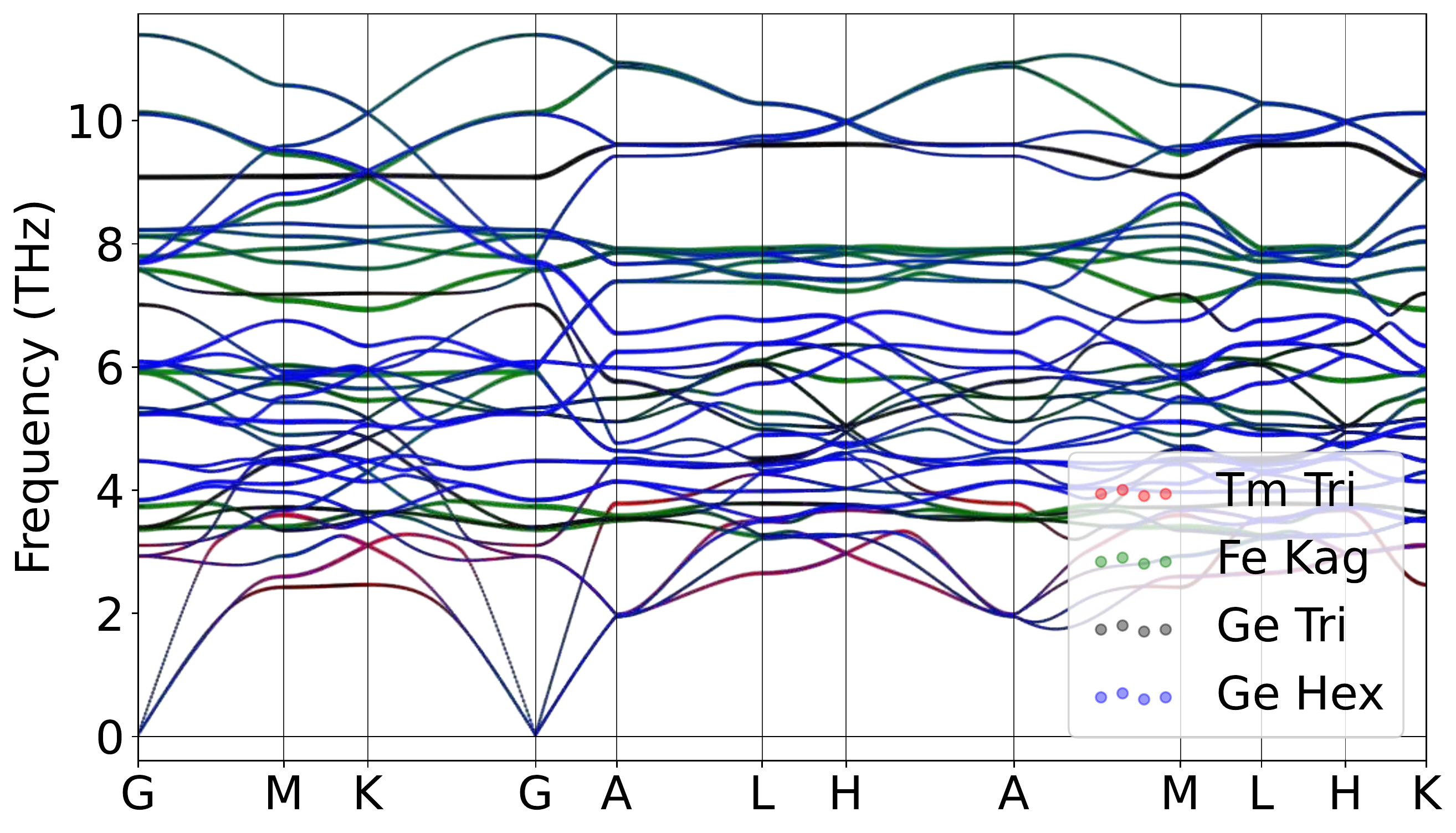}
}
\caption{Atomically projected phonon spectrum for TmFe$_6$Ge$_6$ in paramagnetic phase.}
\label{fig:TmFe6Ge6pho}
\end{figure}

\begin{figure}[htbp]
\centering
\includegraphics[width=0.45\textwidth]{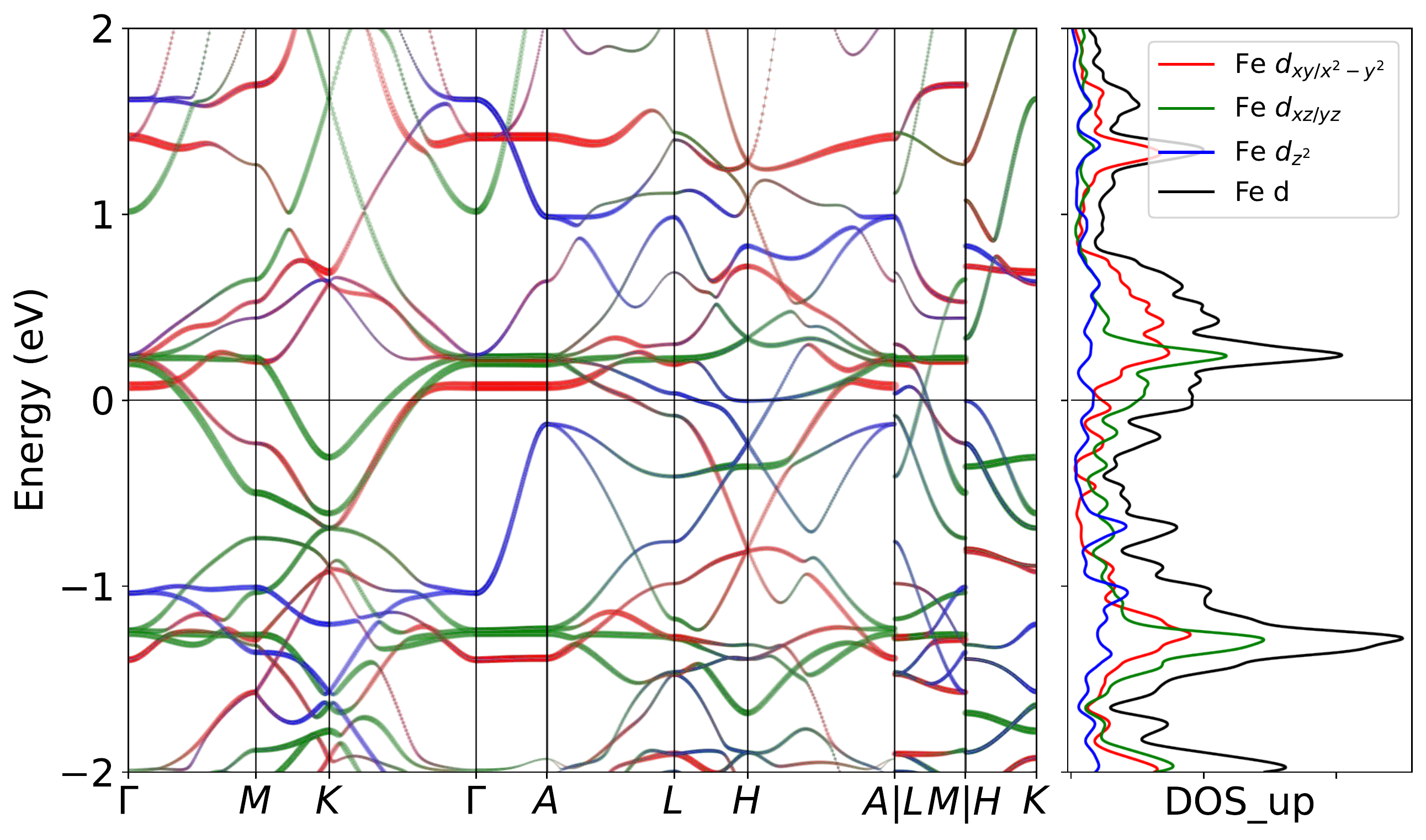}
\includegraphics[width=0.45\textwidth]{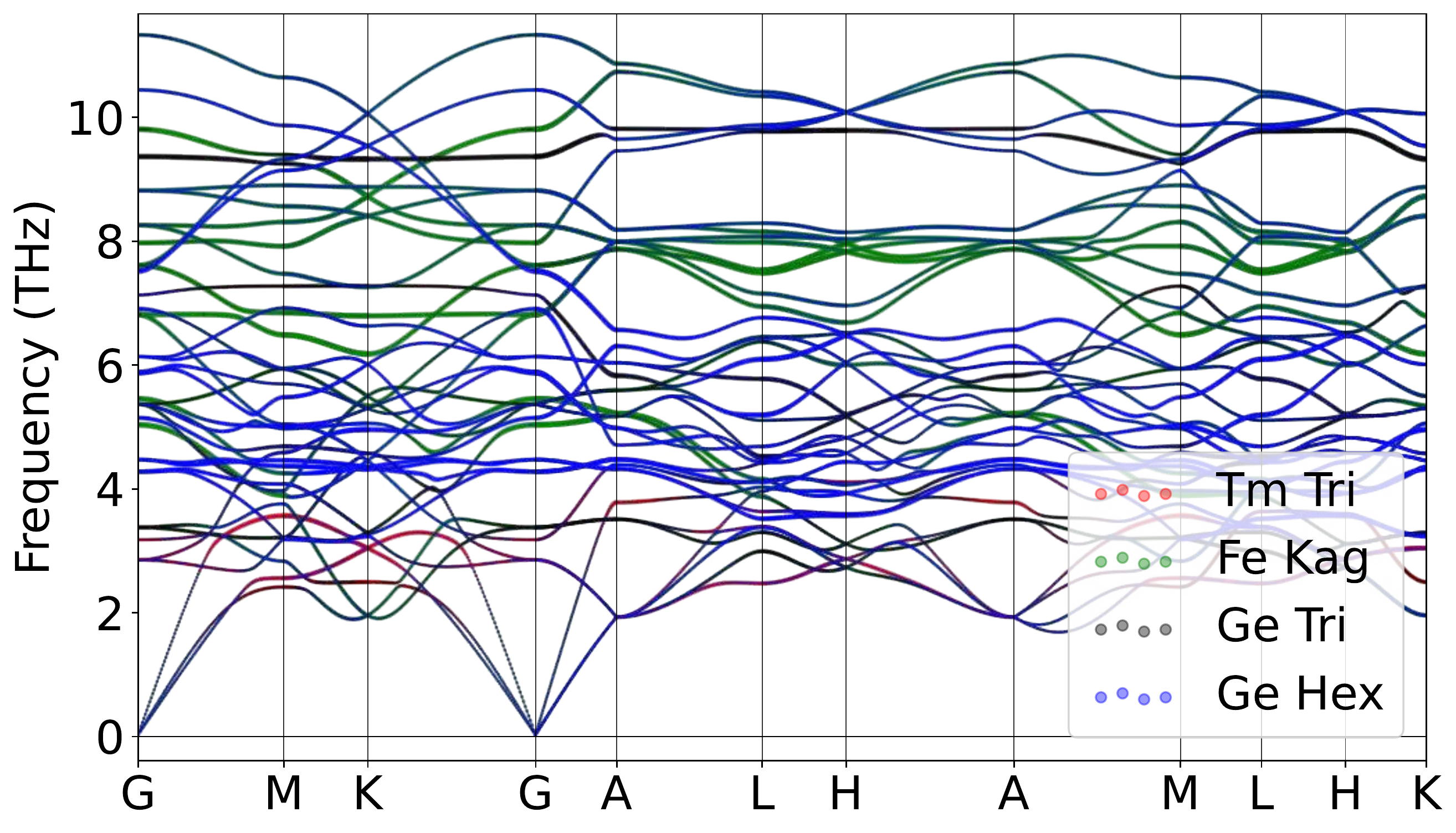}
\caption{Band structure for TmFe$_6$Ge$_6$ with AFM order without SOC for spin (Left)up channel. The projection of Fe $3d$ orbitals is also presented. (Right)Correspondingly, a stable phonon was demonstrated with the collinear magnetic order without Hubbard U at lower temperature regime, weighted by atomic projections.}
\label{fig:TmFe6Ge6mag}
\end{figure}

\begin{figure}[H]
\centering
\label{fig:TmFe6Ge6_relaxed_Low_xz}
\includegraphics[width=0.85\textwidth]{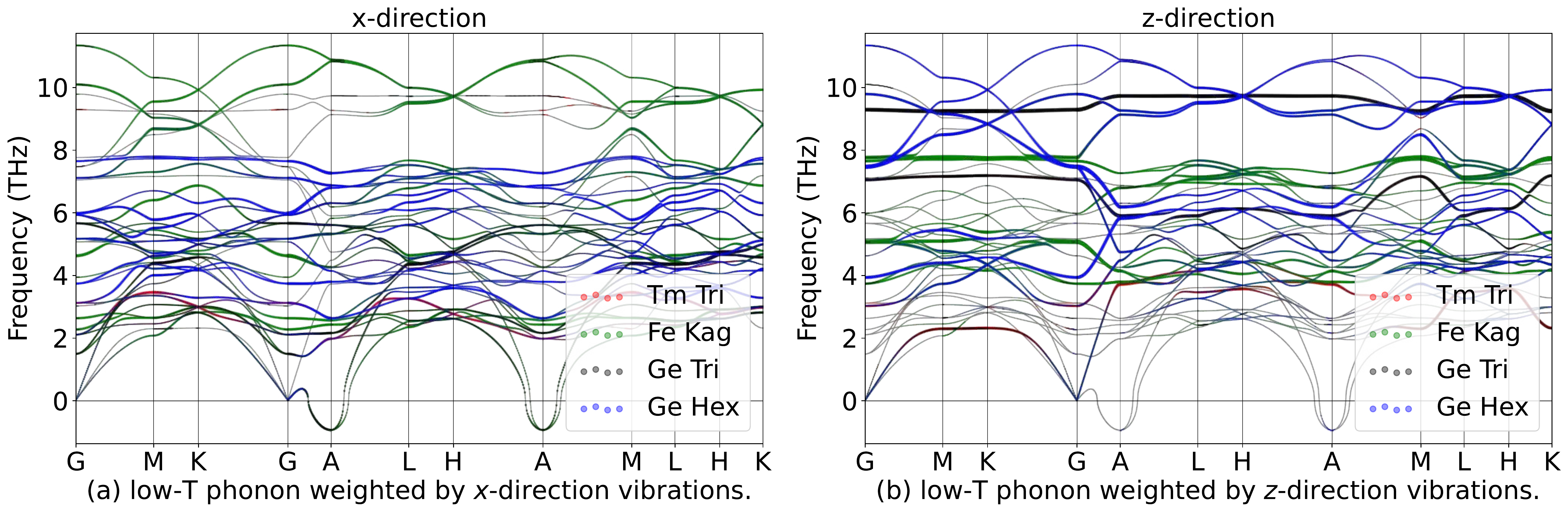}
\hspace{5pt}\label{fig:TmFe6Ge6_relaxed_High_xz}
\includegraphics[width=0.85\textwidth]{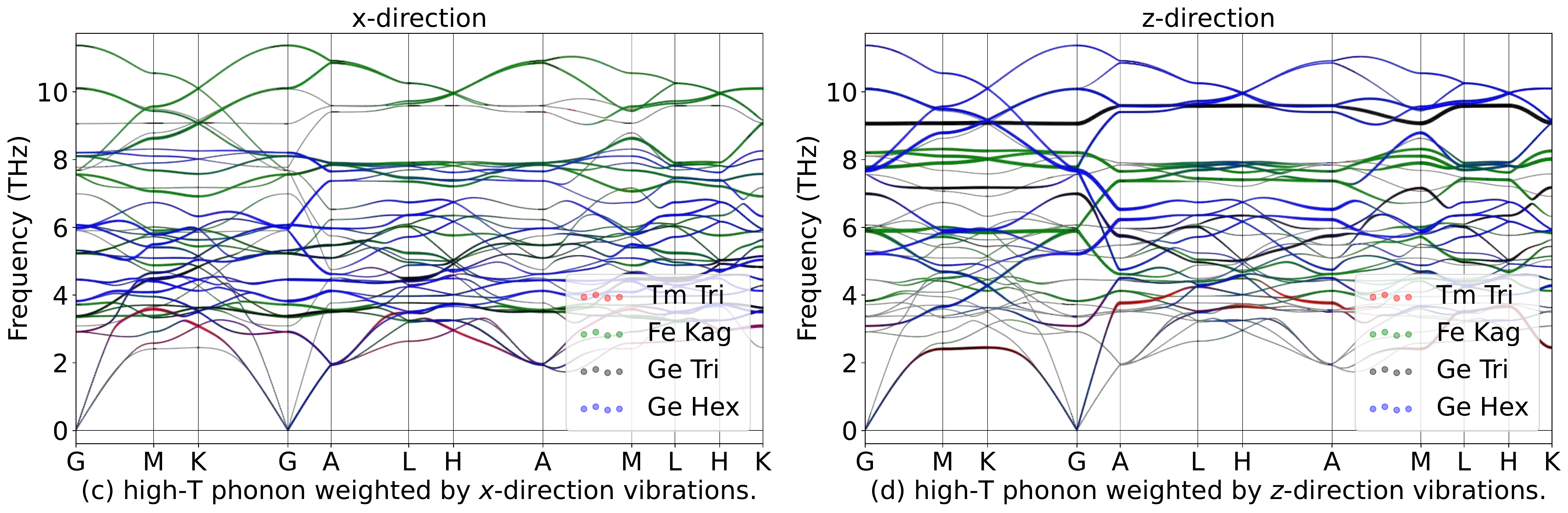}
\caption{Phonon spectrum for TmFe$_6$Ge$_6$in in-plane ($x$) and out-of-plane ($z$) directions in paramagnetic phase.}
\label{fig:TmFe6Ge6phoxyz}
\end{figure}

\begin{table}[htbp]
\global\long\def\arraystretch{1.12}
\captionsetup{width=.95\textwidth}
\caption{IRREPs at high-symmetry points for TmFe$_6$Ge$_6$ phonon spectrum at Low-T.}
\label{tab:191_TmFe6Ge6_Low}
\setlength{\tabcolsep}{1.mm}{
}
\end{table}

\begin{table}[htbp]
\global\long\def\arraystretch{1.12}
\captionsetup{width=.95\textwidth}
\caption{IRREPs at high-symmetry points for TmFe$_6$Ge$_6$ phonon spectrum at High-T.}
\label{tab:191_TmFe6Ge6_High}
\setlength{\tabcolsep}{1.mm}{
%
}
\end{table}
\subsubsection{\label{sms2sec:YbFe6Ge6}\hyperref[smsec:col]{\texorpdfstring{YbFe$_6$Ge$_6$}{}}~{\color{blue}  (High-T stable, Low-T unstable) }}

\begin{table}[htbp]
\global\long\def\arraystretch{1.12}
\captionsetup{width=.85\textwidth}
\caption{Structure parameters of YbFe$_6$Ge$_6$ after relaxation. It contains 418 electrons. }
\label{tab:YbFe6Ge6}
\setlength{\tabcolsep}{4mm}{
%
}
\end{table}

\begin{figure}[htbp]
\centering
\includegraphics[width=0.85\textwidth]{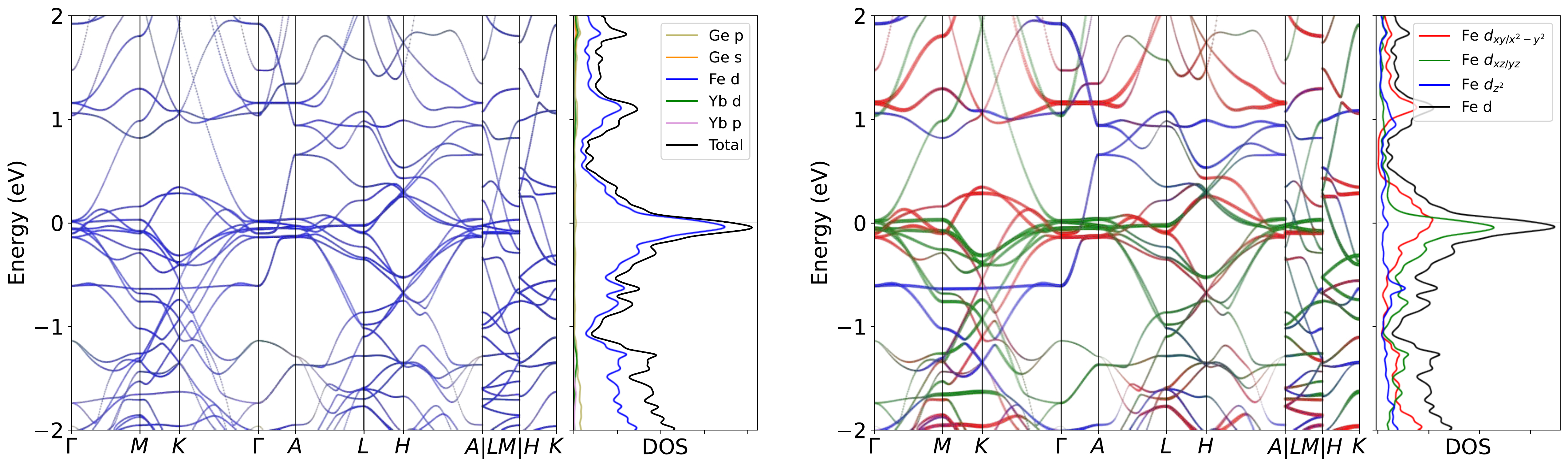}
\caption{Band structure for YbFe$_6$Ge$_6$ without SOC in paramagnetic phase. The projection of Fe $3d$ orbitals is also presented. The band structures clearly reveal the dominating of Fe $3d$ orbitals  near the Fermi level, as well as the pronounced peak.}
\label{fig:YbFe6Ge6band}
\end{figure}

\begin{figure}[H]
\centering
\subfloat[Relaxed phonon at low-T.]{
\label{fig:YbFe6Ge6_relaxed_Low}
\includegraphics[width=0.45\textwidth]{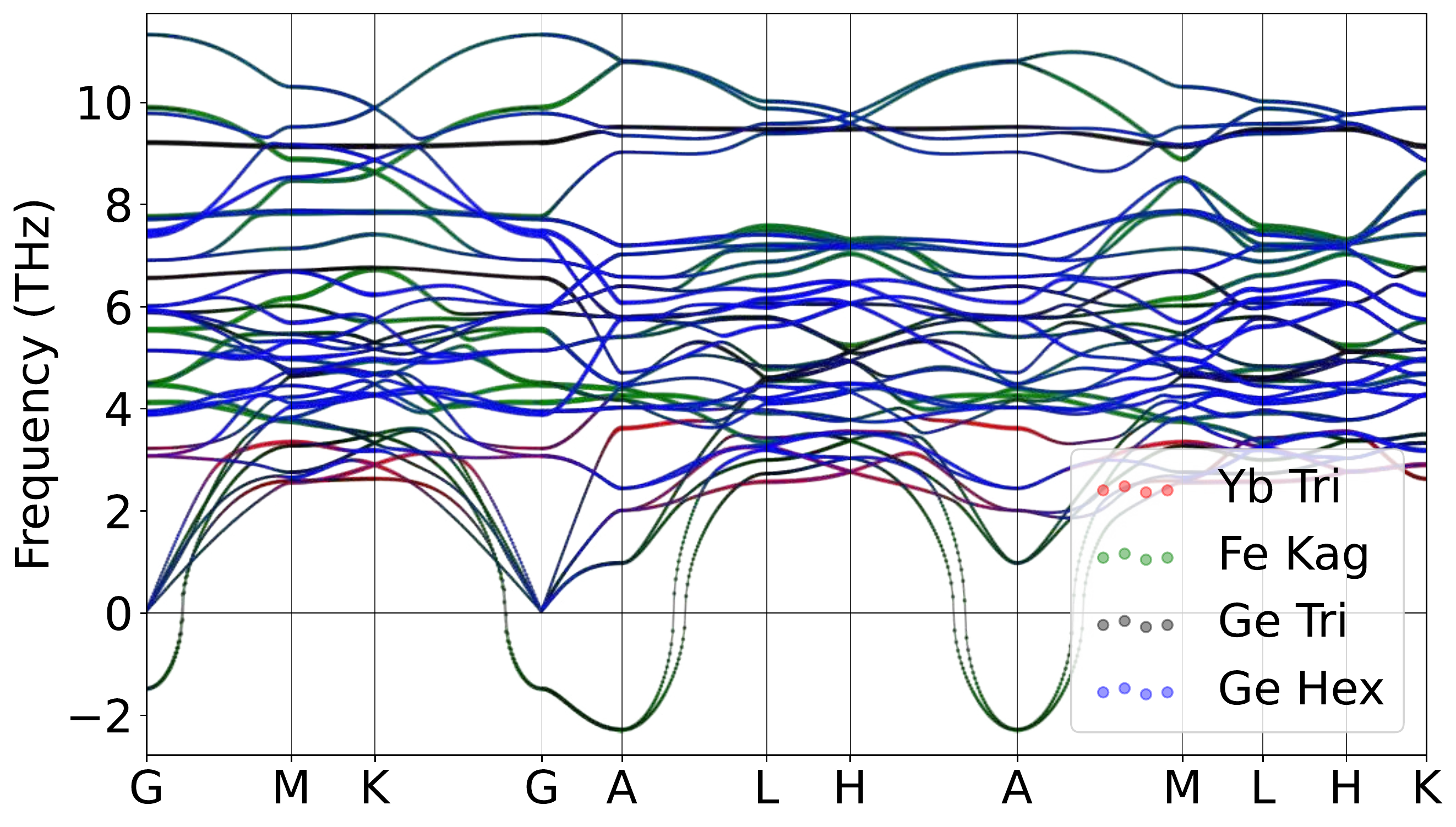}
}
\hspace{5pt}\subfloat[Relaxed phonon at high-T.]{
\label{fig:YbFe6Ge6_relaxed_High}
\includegraphics[width=0.45\textwidth]{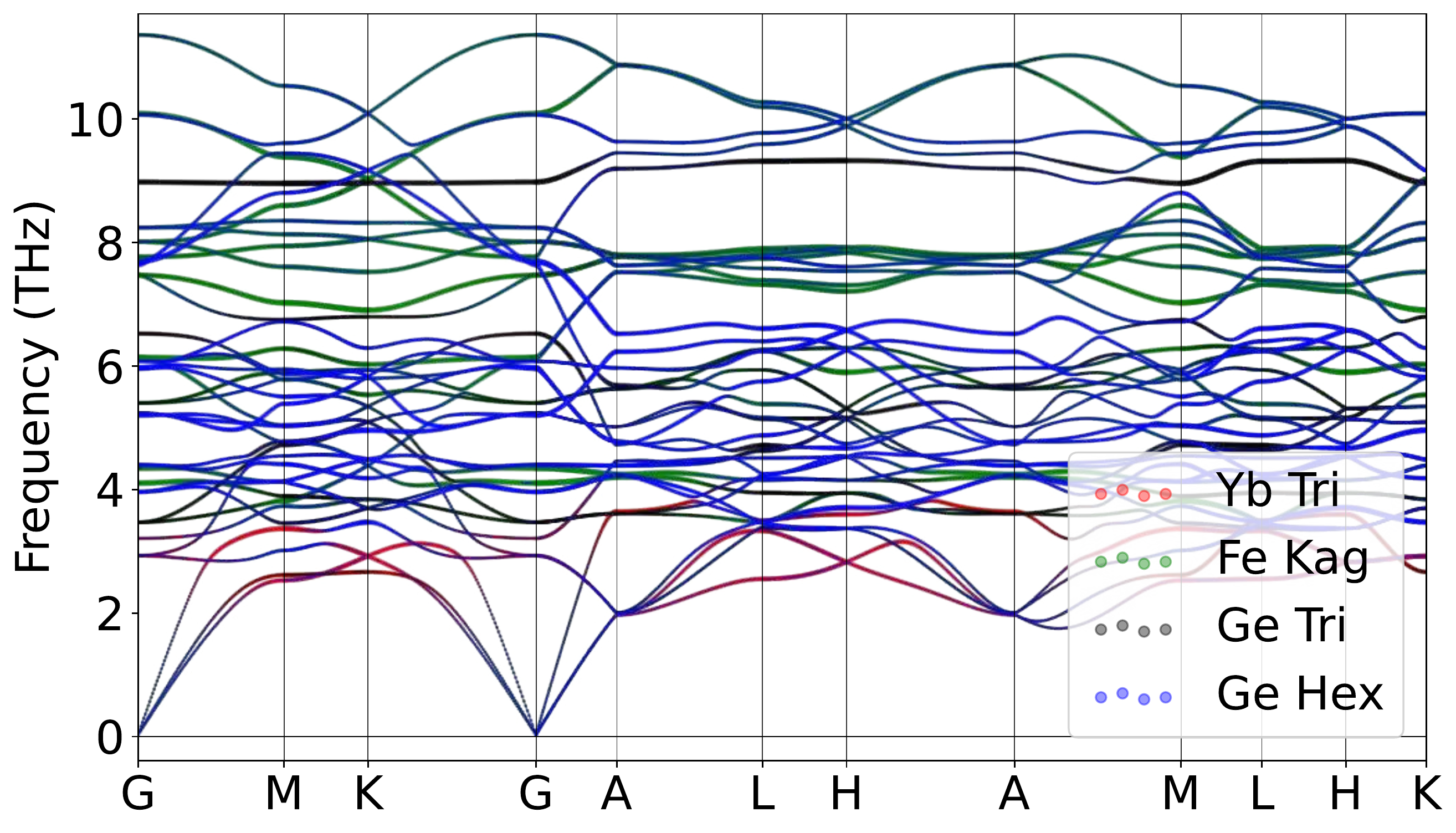}
}
\caption{Atomically projected phonon spectrum for YbFe$_6$Ge$_6$ in paramagnetic phase.}
\label{fig:YbFe6Ge6pho}
\end{figure}

\begin{figure}[htbp]
\centering
\includegraphics[width=0.45\textwidth]{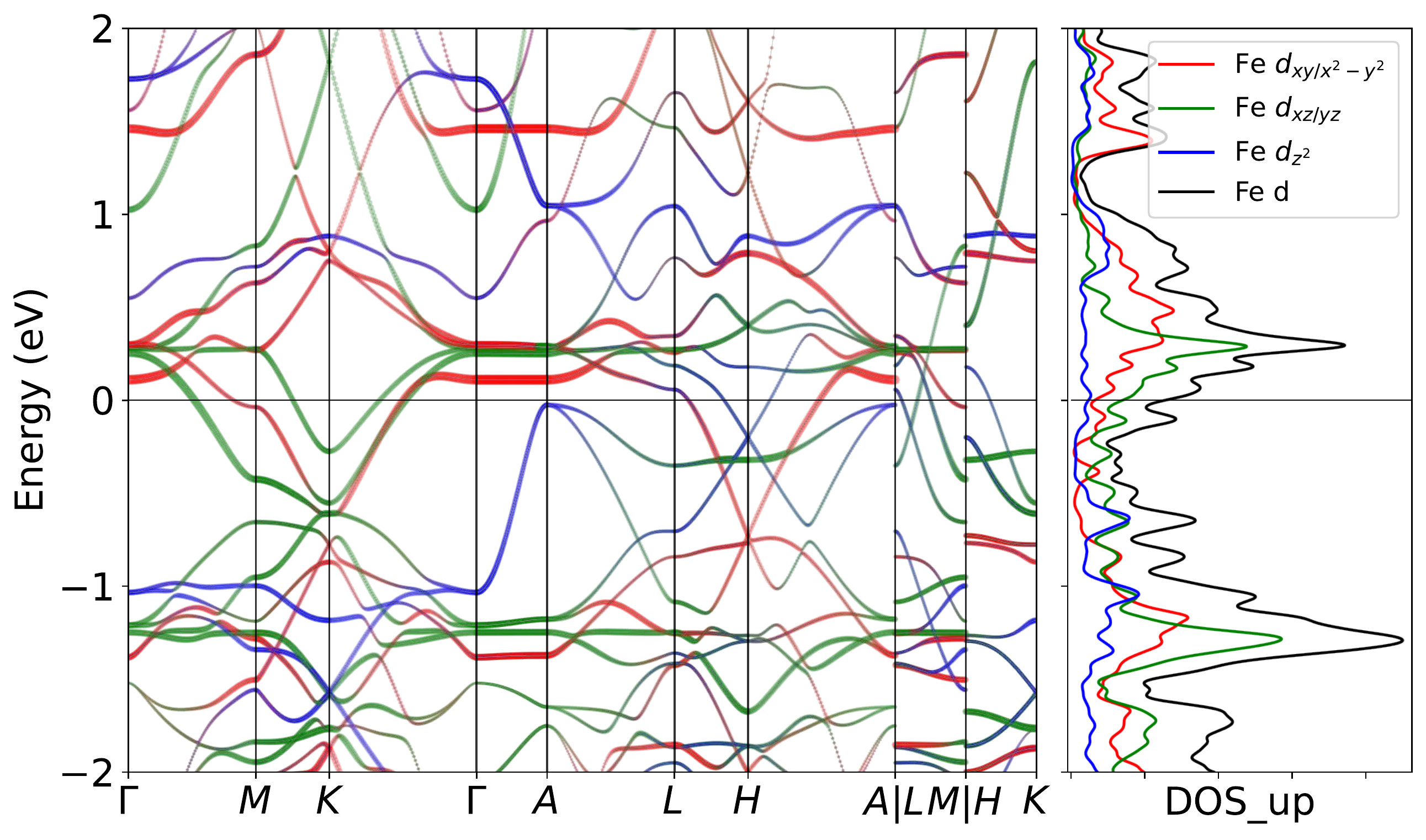}
\includegraphics[width=0.45\textwidth]{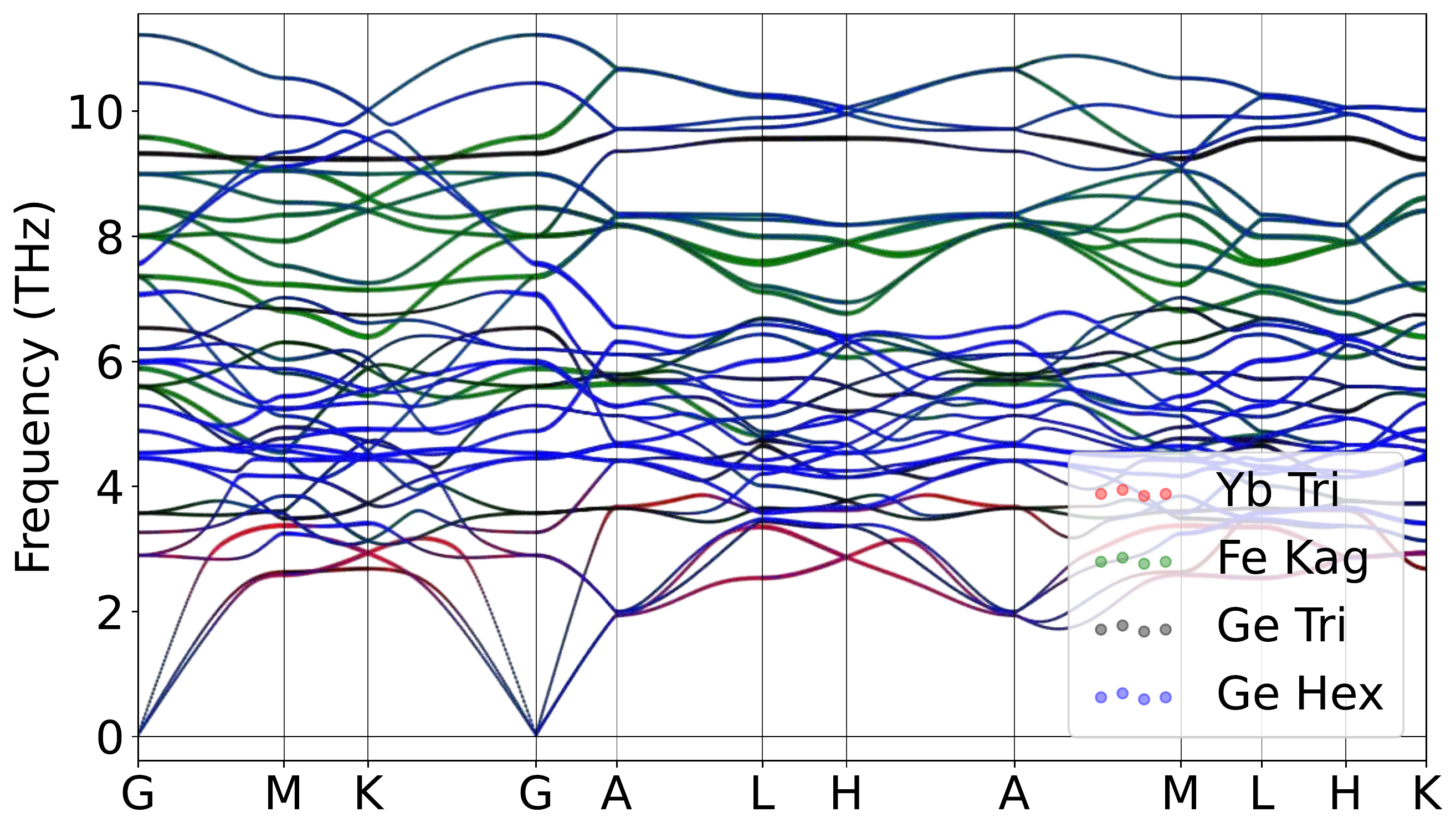}
\caption{Band structure for YbFe$_6$Ge$_6$ with AFM order without SOC for spin (Left)up channel. The projection of Fe $3d$ orbitals is also presented. (Right)Correspondingly, a stable phonon was demonstrated with the collinear magnetic order without Hubbard U at lower temperature regime, weighted by atomic projections.}
\label{fig:YbFe6Ge6mag}
\end{figure}

\begin{figure}[H]
\centering
\label{fig:YbFe6Ge6_relaxed_Low_xz}
\includegraphics[width=0.85\textwidth]{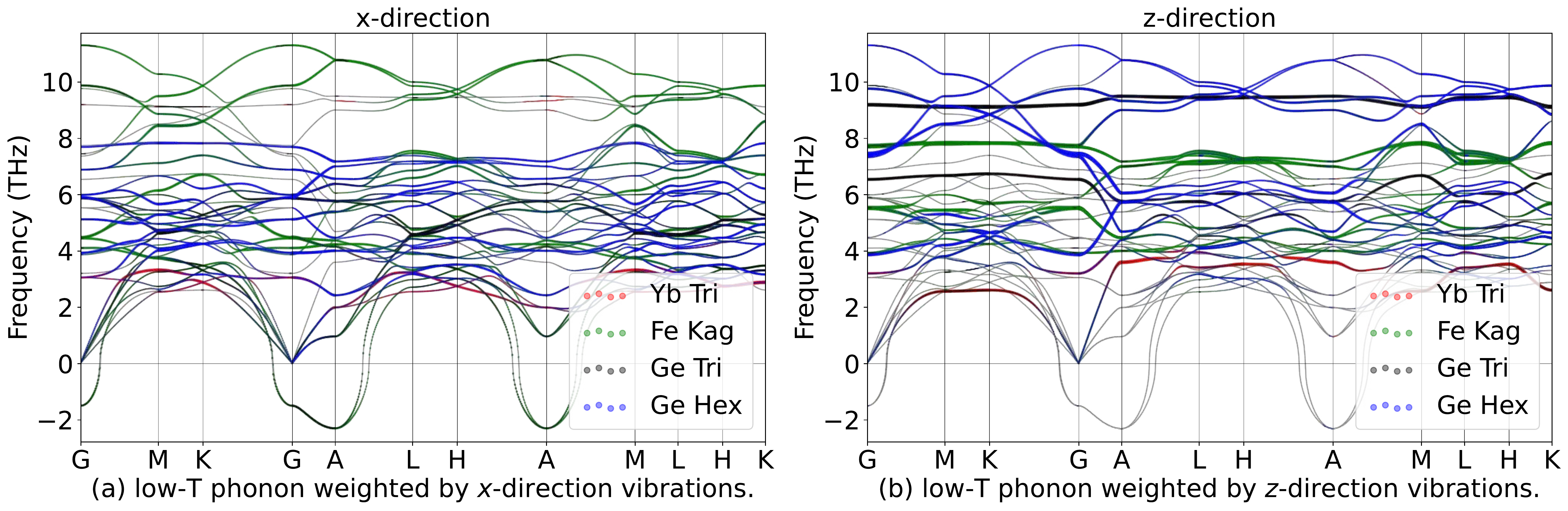}
\hspace{5pt}\label{fig:YbFe6Ge6_relaxed_High_xz}
\includegraphics[width=0.85\textwidth]{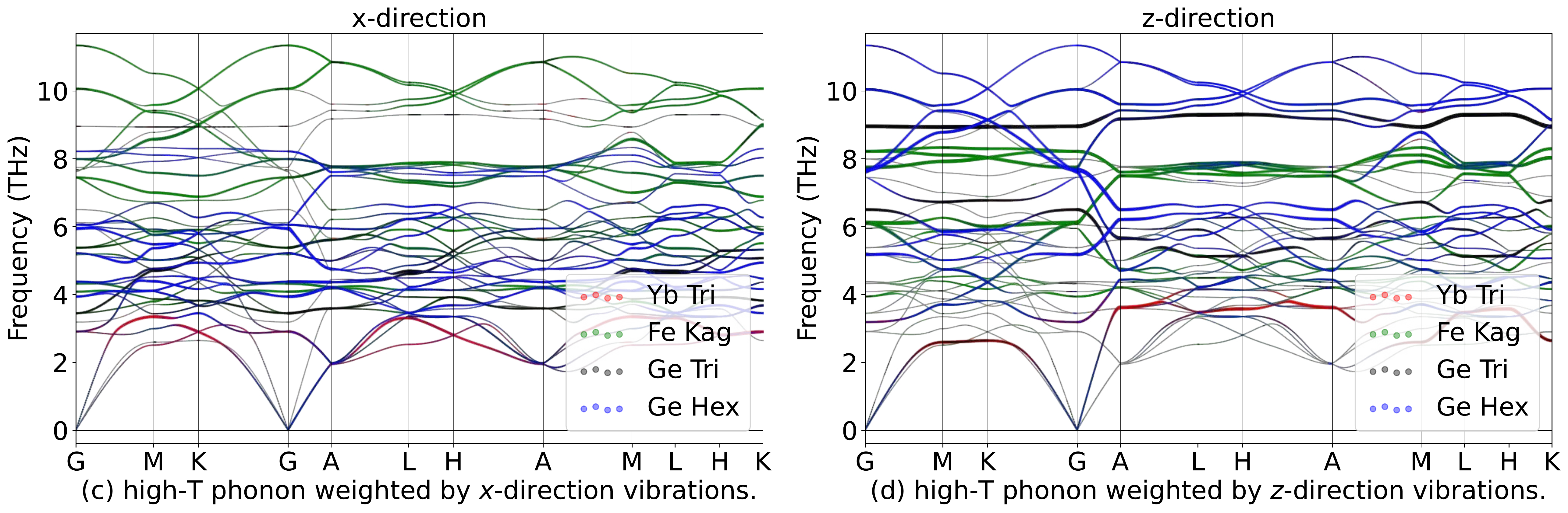}
\caption{Phonon spectrum for YbFe$_6$Ge$_6$in in-plane ($x$) and out-of-plane ($z$) directions in paramagnetic phase.}
\label{fig:YbFe6Ge6phoxyz}
\end{figure}

\begin{table}[htbp]
\global\long\def\arraystretch{1.12}
\captionsetup{width=.95\textwidth}
\caption{IRREPs at high-symmetry points for YbFe$_6$Ge$_6$ phonon spectrum at Low-T.}
\label{tab:191_YbFe6Ge6_Low}
\setlength{\tabcolsep}{1.mm}{
}
\end{table}

\begin{table}[htbp]
\global\long\def\arraystretch{1.12}
\captionsetup{width=.95\textwidth}
\caption{IRREPs at high-symmetry points for YbFe$_6$Ge$_6$ phonon spectrum at High-T.}
\label{tab:191_YbFe6Ge6_High}
\setlength{\tabcolsep}{1.mm}{
%
}
\end{table}
\subsubsection{\label{sms2sec:LuFe6Ge6}\hyperref[smsec:col]{\texorpdfstring{LuFe$_6$Ge$_6$}{}}~{\color{blue}  (High-T stable, Low-T unstable) }}

\begin{table}[htbp]
\global\long\def\arraystretch{1.12}
\captionsetup{width=.85\textwidth}
\caption{Structure parameters of LuFe$_6$Ge$_6$ after relaxation. It contains 419 electrons. }
\label{tab:LuFe6Ge6}
\setlength{\tabcolsep}{4mm}{
%
}
\end{table}

\begin{figure}[htbp]
\centering
\includegraphics[width=0.85\textwidth]{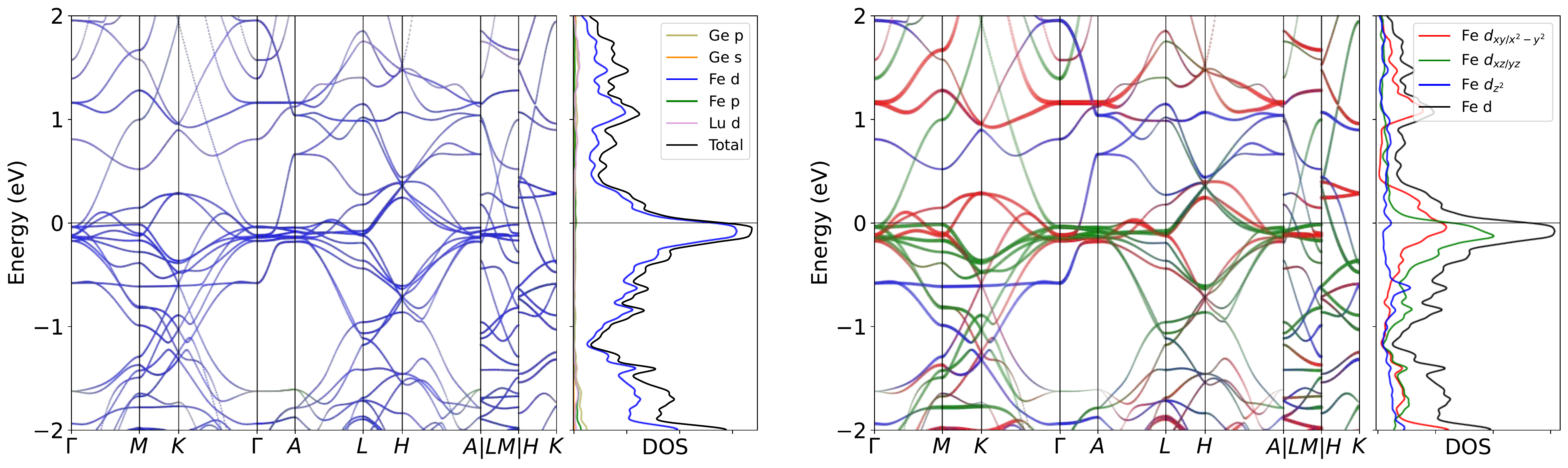}
\caption{Band structure for LuFe$_6$Ge$_6$ without SOC in paramagnetic phase. The projection of Fe $3d$ orbitals is also presented. The band structures clearly reveal the dominating of Fe $3d$ orbitals  near the Fermi level, as well as the pronounced peak.}
\label{fig:LuFe6Ge6band}
\end{figure}

\begin{figure}[H]
\centering
\subfloat[Relaxed phonon at low-T.]{
\label{fig:LuFe6Ge6_relaxed_Low}
\includegraphics[width=0.45\textwidth]{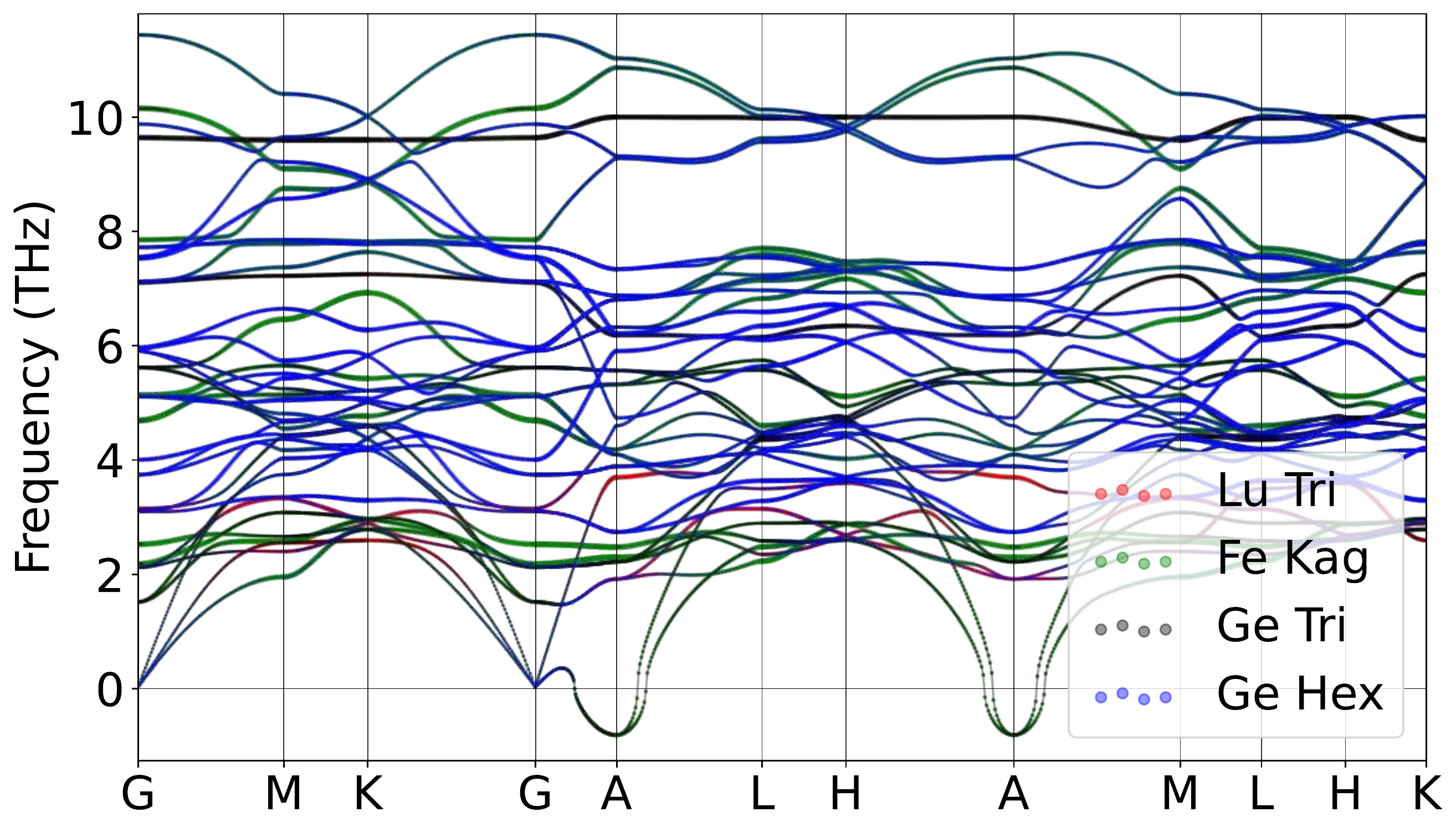}
}
\hspace{5pt}\subfloat[Relaxed phonon at high-T.]{
\label{fig:LuFe6Ge6_relaxed_High}
\includegraphics[width=0.45\textwidth]{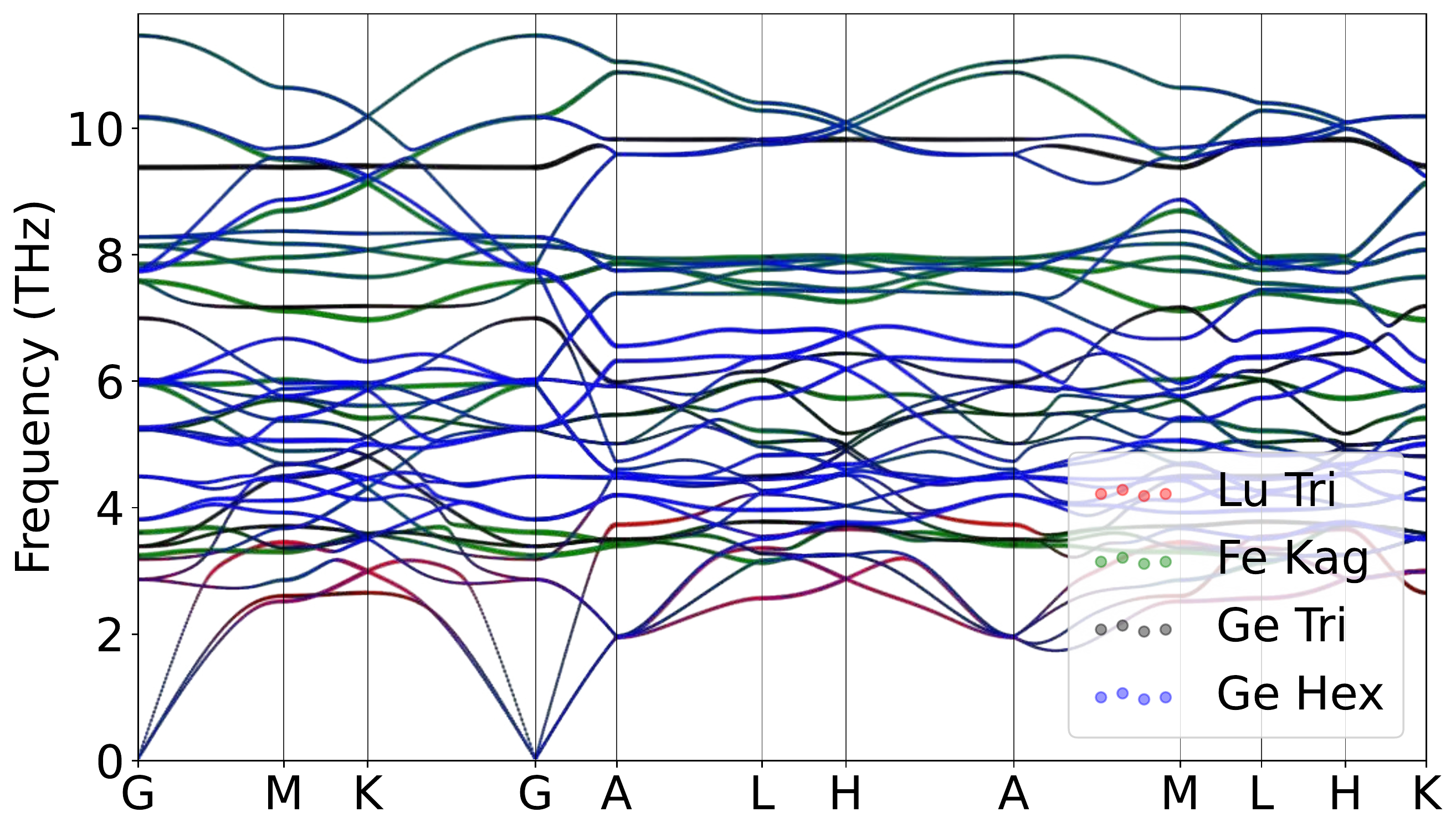}
}
\caption{Atomically projected phonon spectrum for LuFe$_6$Ge$_6$ in paramagnetic phase.}
\label{fig:LuFe6Ge6pho}
\end{figure}

\begin{figure}[htbp]
\centering
\includegraphics[width=0.45\textwidth]{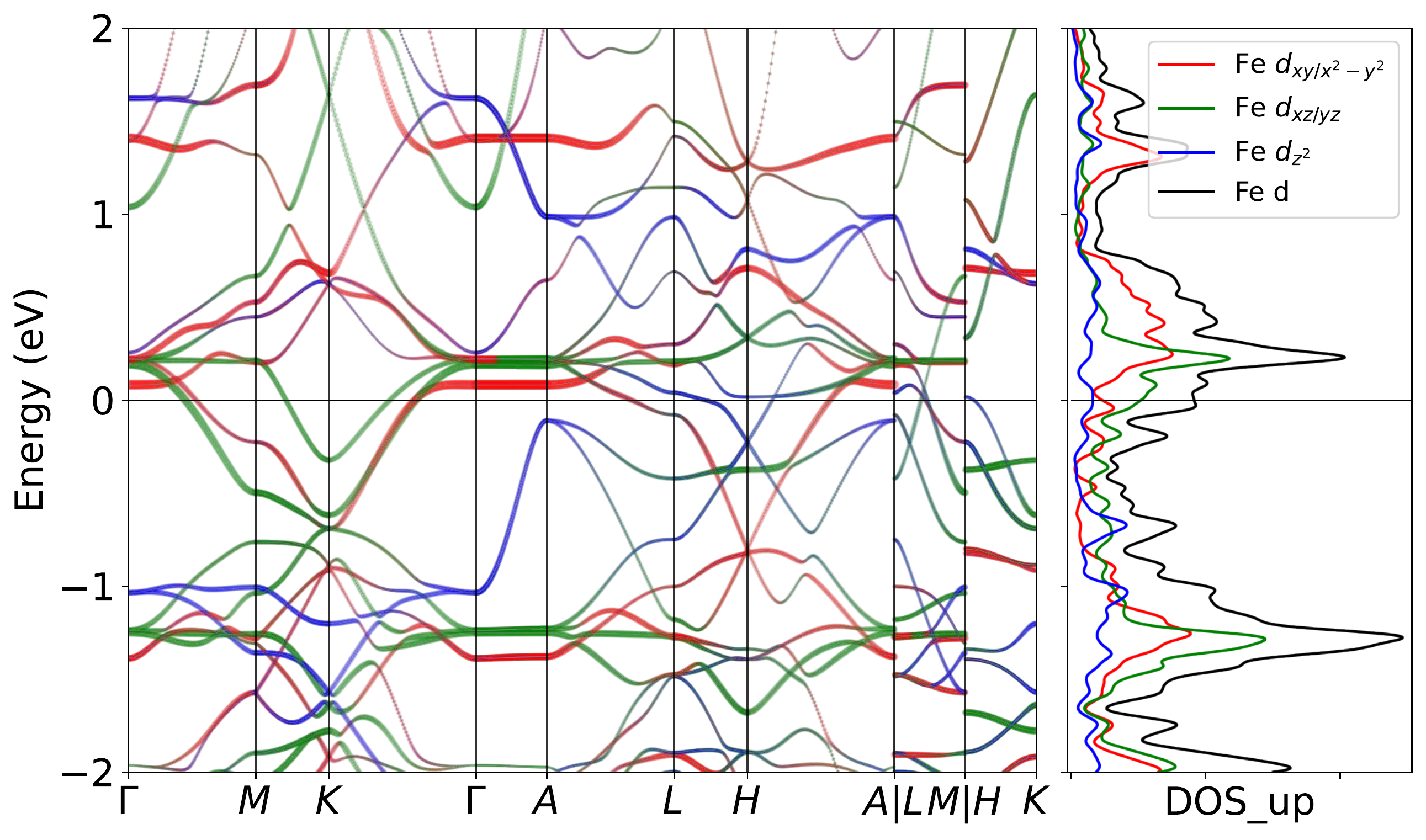}
\includegraphics[width=0.45\textwidth]{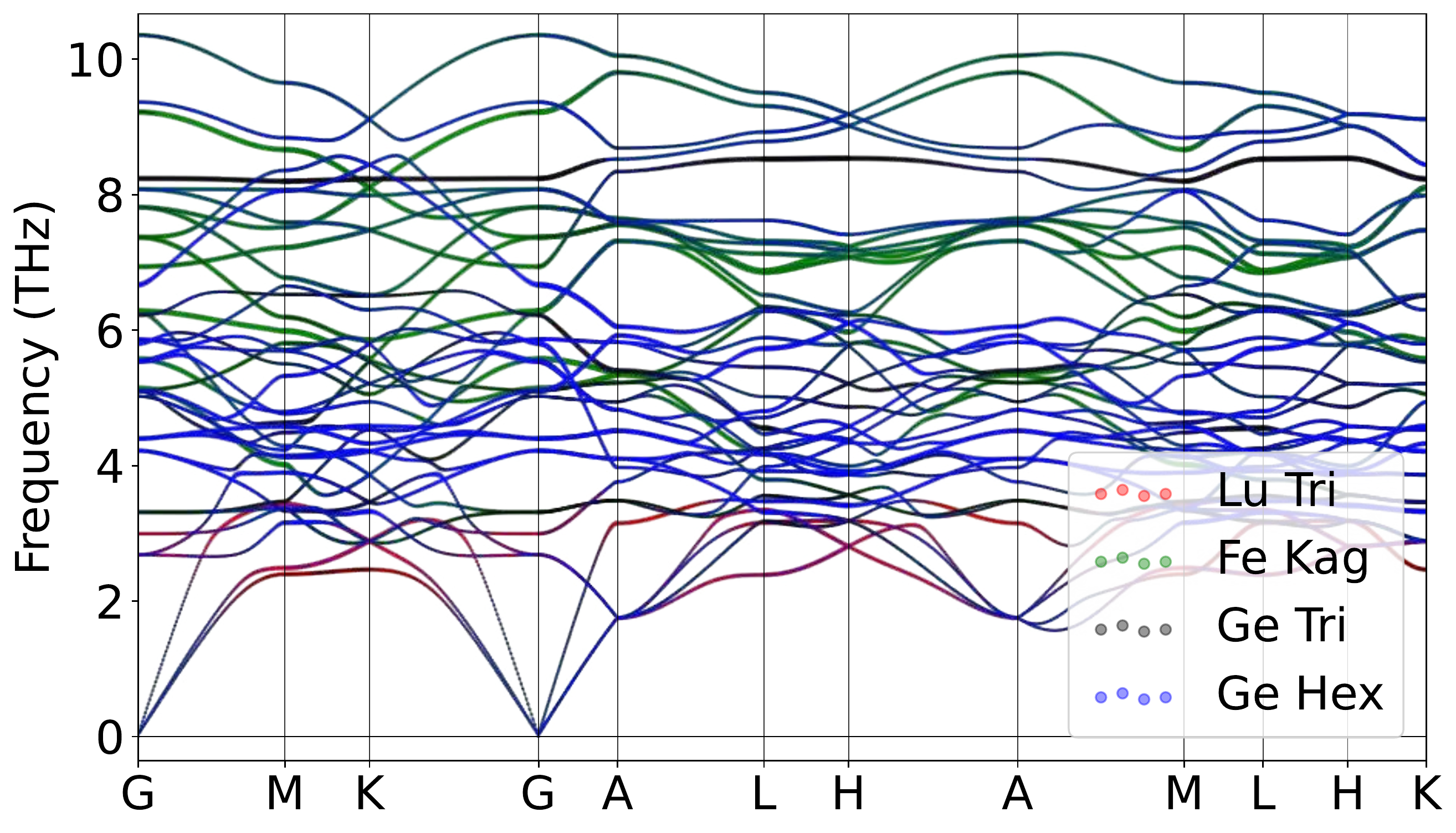}
\caption{Band structure for LuFe$_6$Ge$_6$ with AFM order without SOC for spin (Left)up channel. The projection of Fe $3d$ orbitals is also presented. (Right)Correspondingly, a stable phonon was demonstrated with the collinear magnetic order without Hubbard U at lower temperature regime, weighted by atomic projections.}
\label{fig:LuFe6Ge6mag}
\end{figure}

\begin{figure}[H]
\centering
\label{fig:LuFe6Ge6_relaxed_Low_xz}
\includegraphics[width=0.85\textwidth]{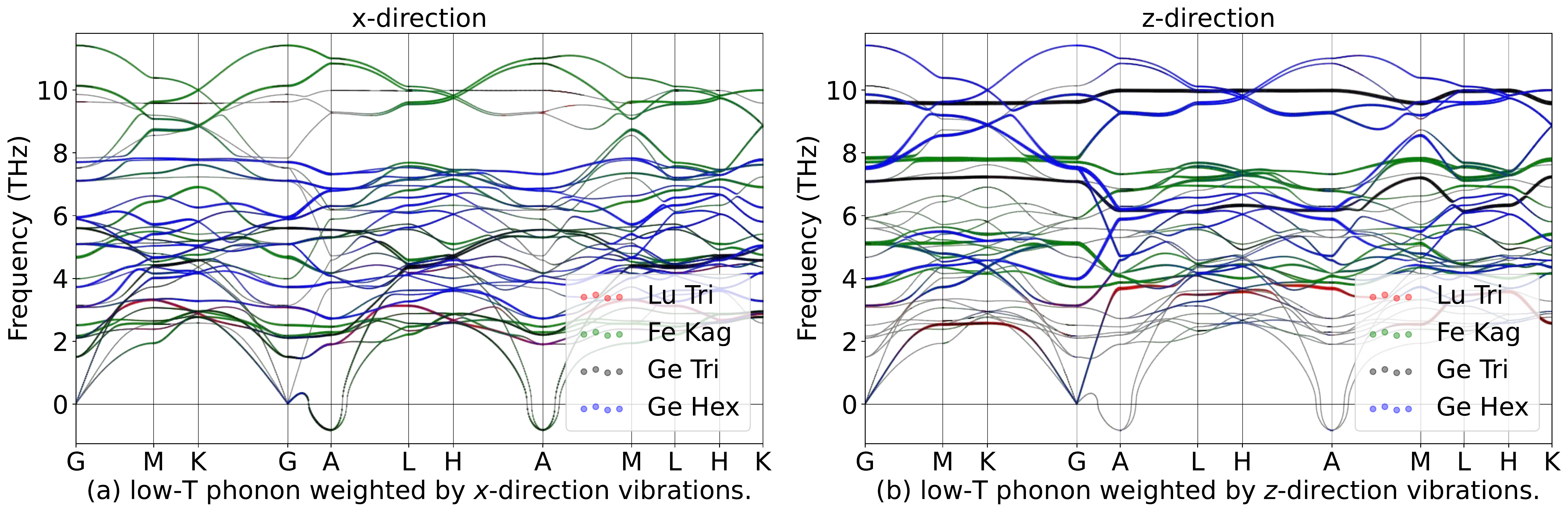}
\hspace{5pt}\label{fig:LuFe6Ge6_relaxed_High_xz}
\includegraphics[width=0.85\textwidth]{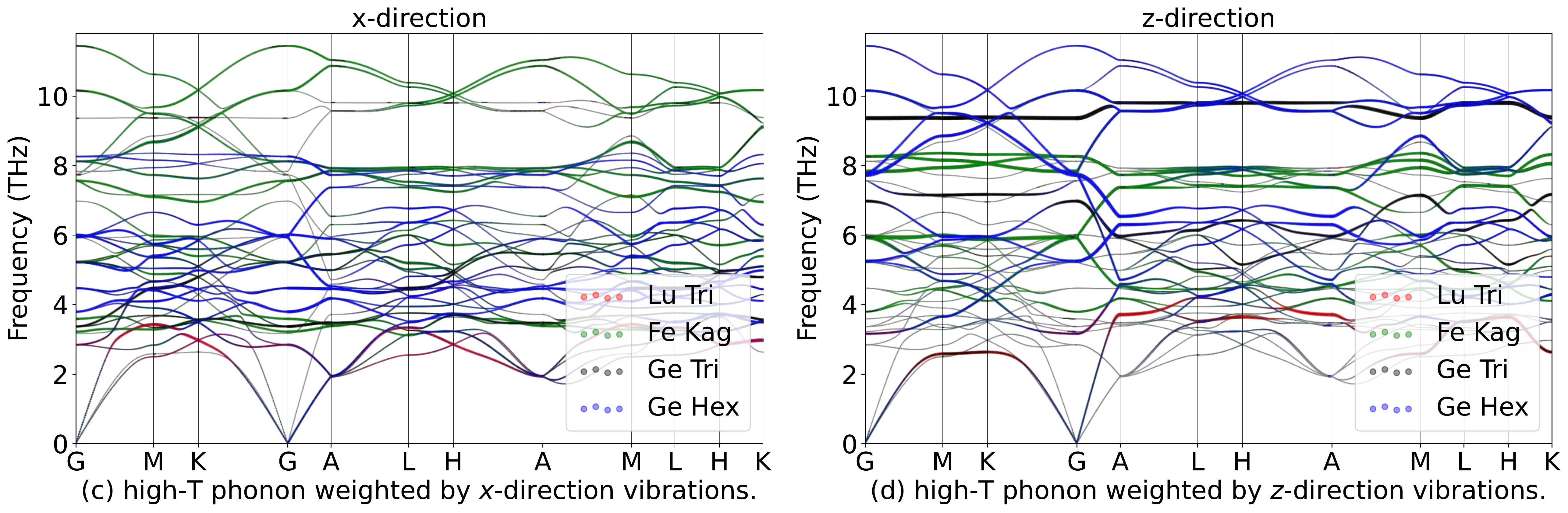}
\caption{Phonon spectrum for LuFe$_6$Ge$_6$in in-plane ($x$) and out-of-plane ($z$) directions in paramagnetic phase.}
\label{fig:LuFe6Ge6phoxyz}
\end{figure}

\begin{table}[htbp]
\global\long\def\arraystretch{1.12}
\captionsetup{width=.95\textwidth}
\caption{IRREPs at high-symmetry points for LuFe$_6$Ge$_6$ phonon spectrum at Low-T.}
\label{tab:191_LuFe6Ge6_Low}
\setlength{\tabcolsep}{1.mm}{
}
\end{table}

\begin{table}[htbp]
\global\long\def\arraystretch{1.12}
\captionsetup{width=.95\textwidth}
\caption{IRREPs at high-symmetry points for LuFe$_6$Ge$_6$ phonon spectrum at High-T.}
\label{tab:191_LuFe6Ge6_High}
\setlength{\tabcolsep}{1.mm}{
%
}
\end{table}
\subsubsection{\label{sms2sec:HfFe6Ge6}\hyperref[smsec:col]{\texorpdfstring{HfFe$_6$Ge$_6$}{}}~{\color{blue}  (High-T stable, Low-T unstable) }}

\begin{table}[htbp]
\global\long\def\arraystretch{1.12}
\captionsetup{width=.85\textwidth}
\caption{Structure parameters of HfFe$_6$Ge$_6$ after relaxation. It contains 420 electrons. }
\label{tab:HfFe6Ge6}
\setlength{\tabcolsep}{4mm}{
%
}
\end{table}

\begin{figure}[htbp]
\centering
\includegraphics[width=0.85\textwidth]{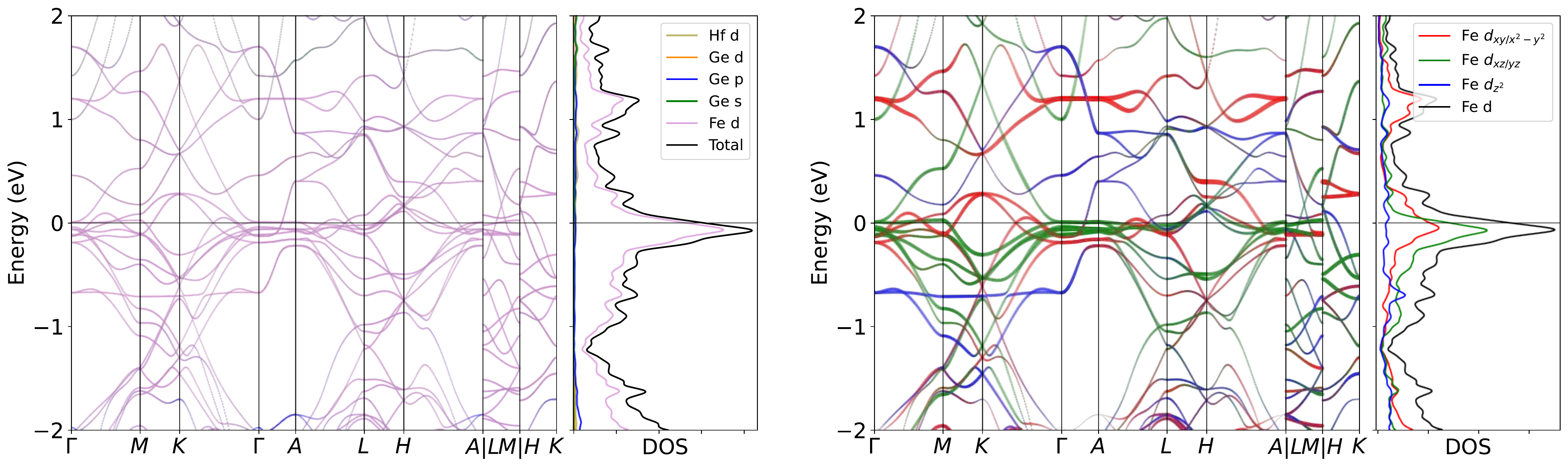}
\caption{Band structure for HfFe$_6$Ge$_6$ without SOC in paramagnetic phase. The projection of Fe $3d$ orbitals is also presented. The band structures clearly reveal the dominating of Fe $3d$ orbitals  near the Fermi level, as well as the pronounced peak.}
\label{fig:HfFe6Ge6band}
\end{figure}

\begin{figure}[H]
\centering
\subfloat[Relaxed phonon at low-T.]{
\label{fig:HfFe6Ge6_relaxed_Low}
\includegraphics[width=0.45\textwidth]{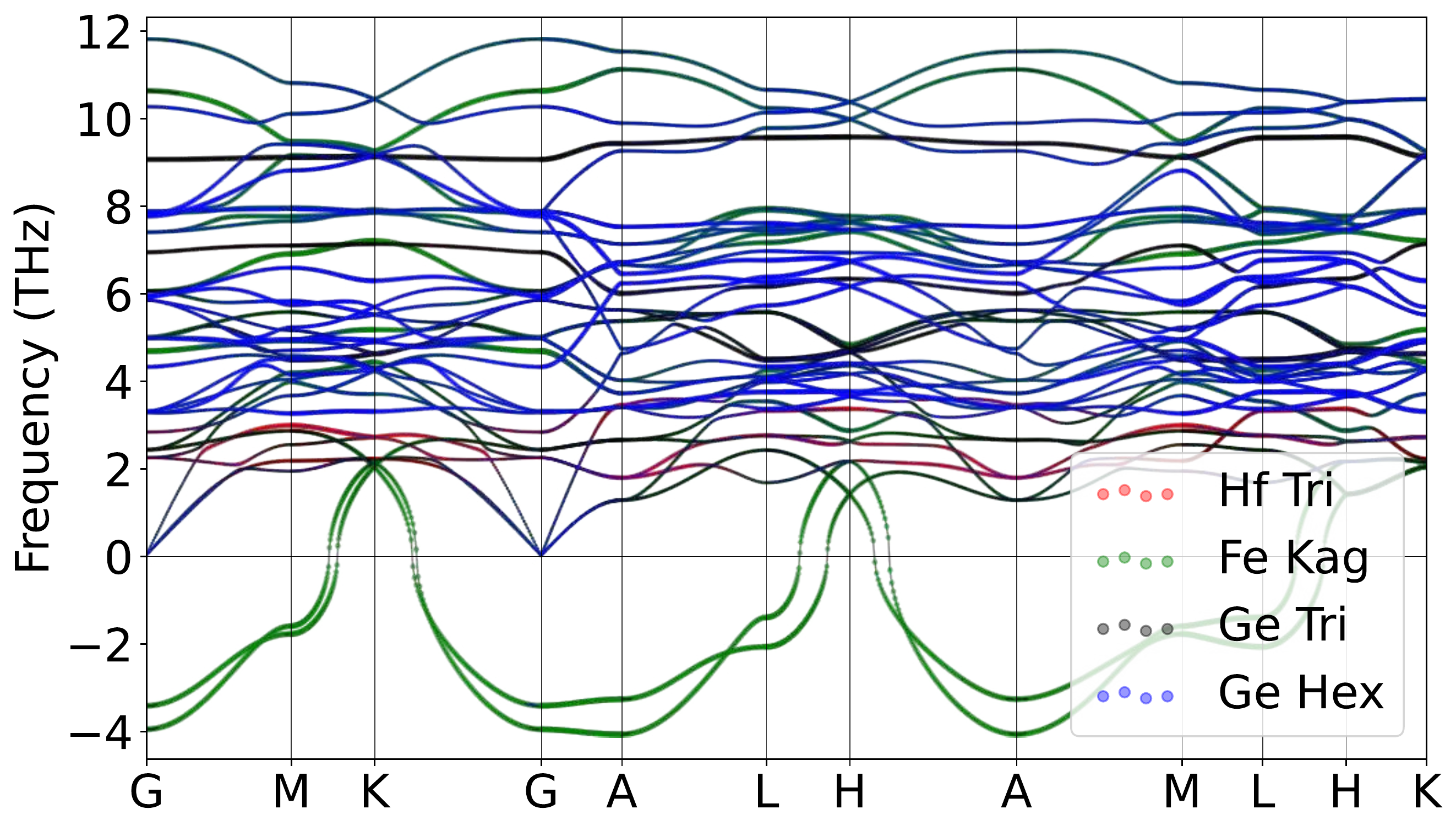}
}
\hspace{5pt}\subfloat[Relaxed phonon at high-T.]{
\label{fig:HfFe6Ge6_relaxed_High}
\includegraphics[width=0.45\textwidth]{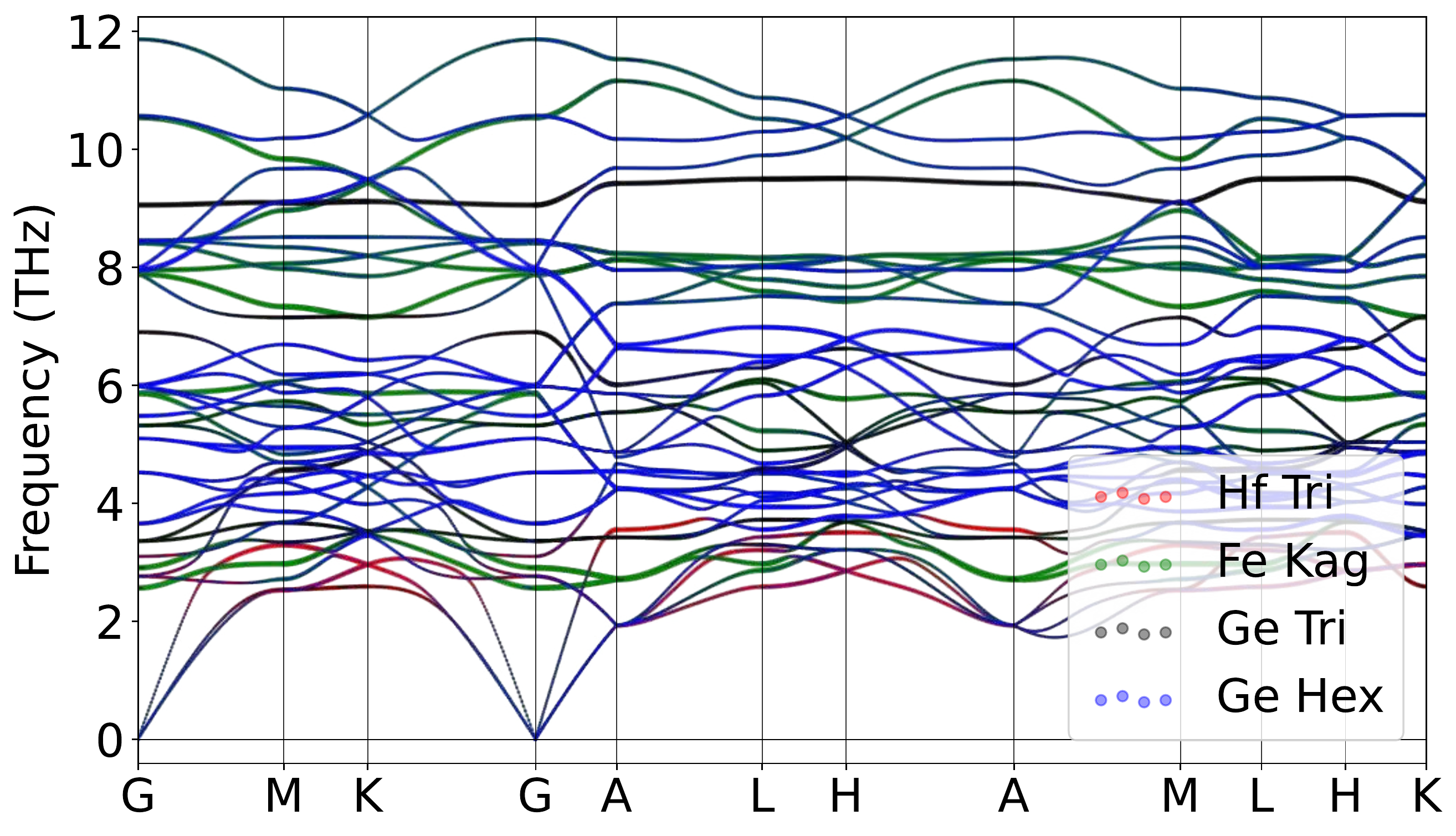}
}
\caption{Atomically projected phonon spectrum for HfFe$_6$Ge$_6$ in paramagnetic phase.}
\label{fig:HfFe6Ge6pho}
\end{figure}

\begin{figure}[htbp]
\centering
\includegraphics[width=0.45\textwidth]{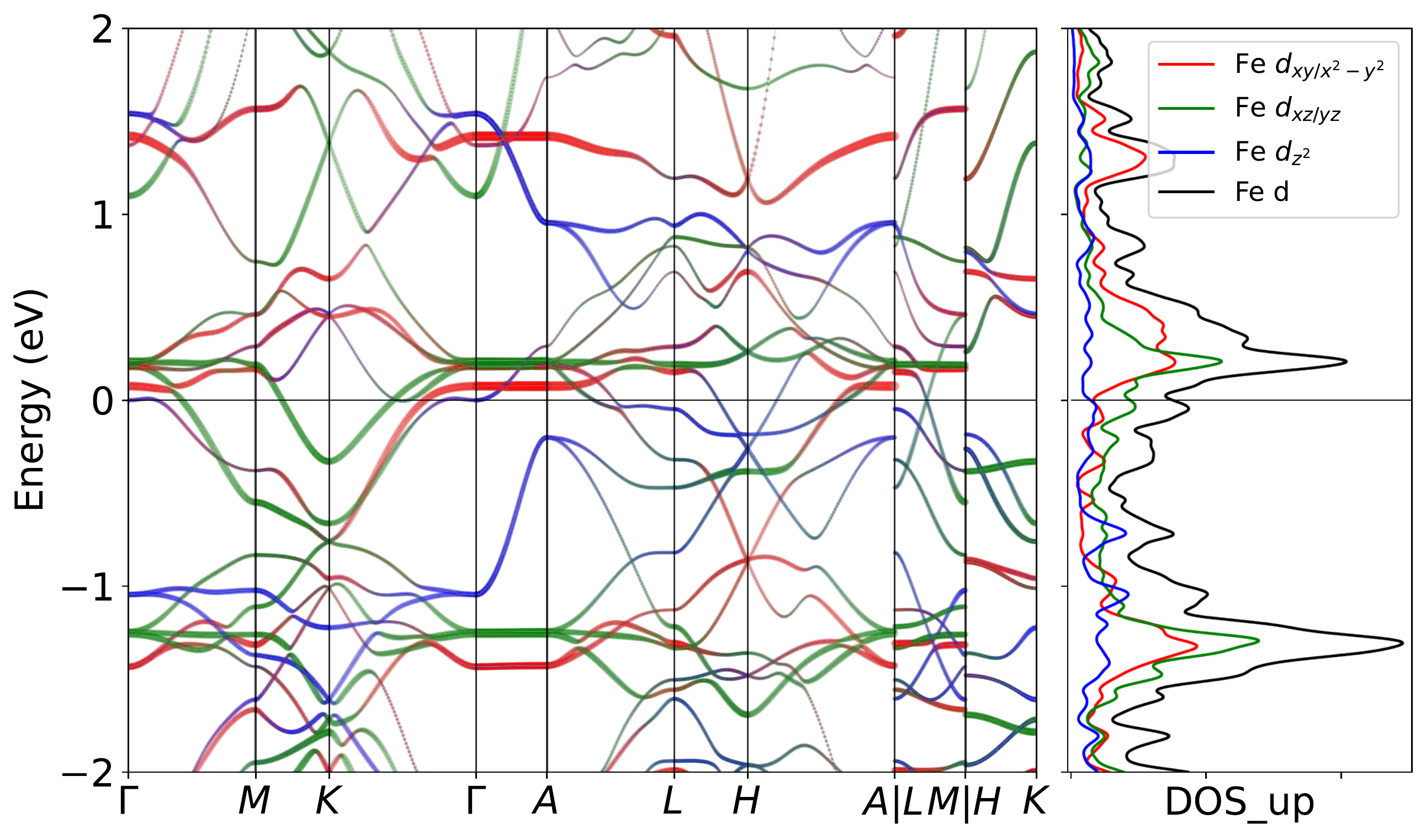}
\includegraphics[width=0.45\textwidth]{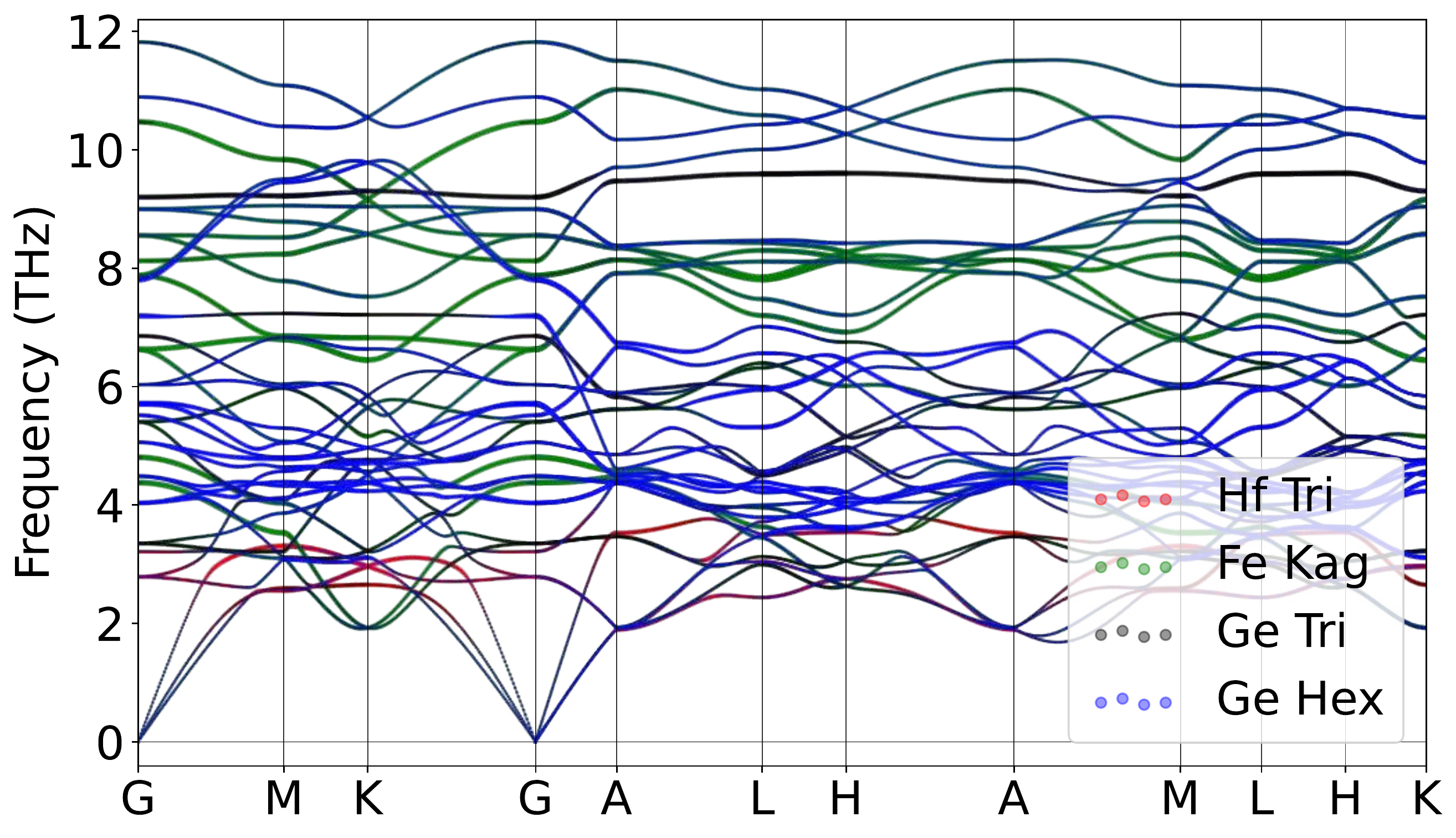}
\caption{Band structure for HfFe$_6$Ge$_6$ with AFM order without SOC for spin (Left)up channel. The projection of Fe $3d$ orbitals is also presented. (Right)Correspondingly, a stable phonon was demonstrated with the collinear magnetic order without Hubbard U at lower temperature regime, weighted by atomic projections.}
\label{fig:HfFe6Ge6mag}
\end{figure}

\begin{figure}[H]
\centering
\label{fig:HfFe6Ge6_relaxed_Low_xz}
\includegraphics[width=0.85\textwidth]{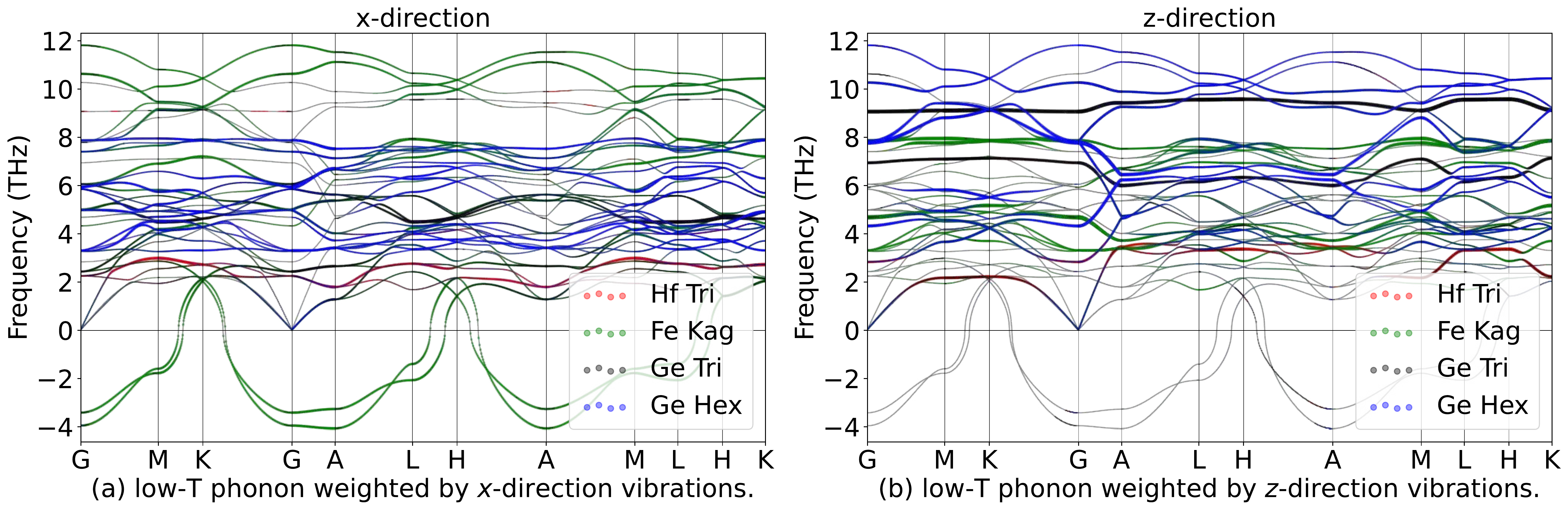}
\hspace{5pt}\label{fig:HfFe6Ge6_relaxed_High_xz}
\includegraphics[width=0.85\textwidth]{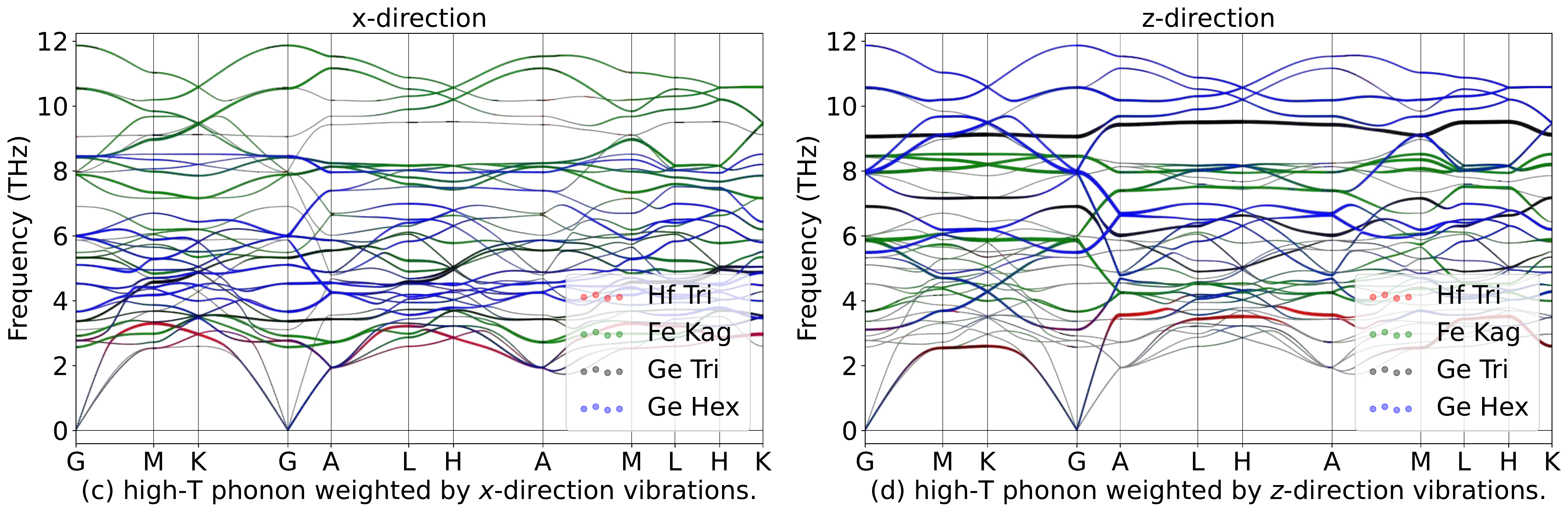}
\caption{Phonon spectrum for HfFe$_6$Ge$_6$in in-plane ($x$) and out-of-plane ($z$) directions in paramagnetic phase.}
\label{fig:HfFe6Ge6phoxyz}
\end{figure}

\begin{table}[htbp]
\global\long\def\arraystretch{1.12}
\captionsetup{width=.95\textwidth}
\caption{IRREPs at high-symmetry points for HfFe$_6$Ge$_6$ phonon spectrum at Low-T.}
\label{tab:191_HfFe6Ge6_Low}
\setlength{\tabcolsep}{1.mm}{
}
\end{table}

\begin{table}[htbp]
\global\long\def\arraystretch{1.12}
\captionsetup{width=.95\textwidth}
\caption{IRREPs at high-symmetry points for HfFe$_6$Ge$_6$ phonon spectrum at High-T.}
\label{tab:191_HfFe6Ge6_High}
\setlength{\tabcolsep}{1.mm}{
%
}
\end{table}
\subsubsection{\label{sms2sec:TaFe6Ge6}\hyperref[smsec:col]{\texorpdfstring{TaFe$_6$Ge$_6$}{}}~{\color{blue}  (High-T stable, Low-T unstable) }}

\begin{table}[htbp]
\global\long\def\arraystretch{1.12}
\captionsetup{width=.85\textwidth}
\caption{Structure parameters of TaFe$_6$Ge$_6$ after relaxation. It contains 421 electrons. }
\label{tab:TaFe6Ge6}
\setlength{\tabcolsep}{4mm}{
%
}
\end{table}

\begin{figure}[htbp]
\centering
\includegraphics[width=0.85\textwidth]{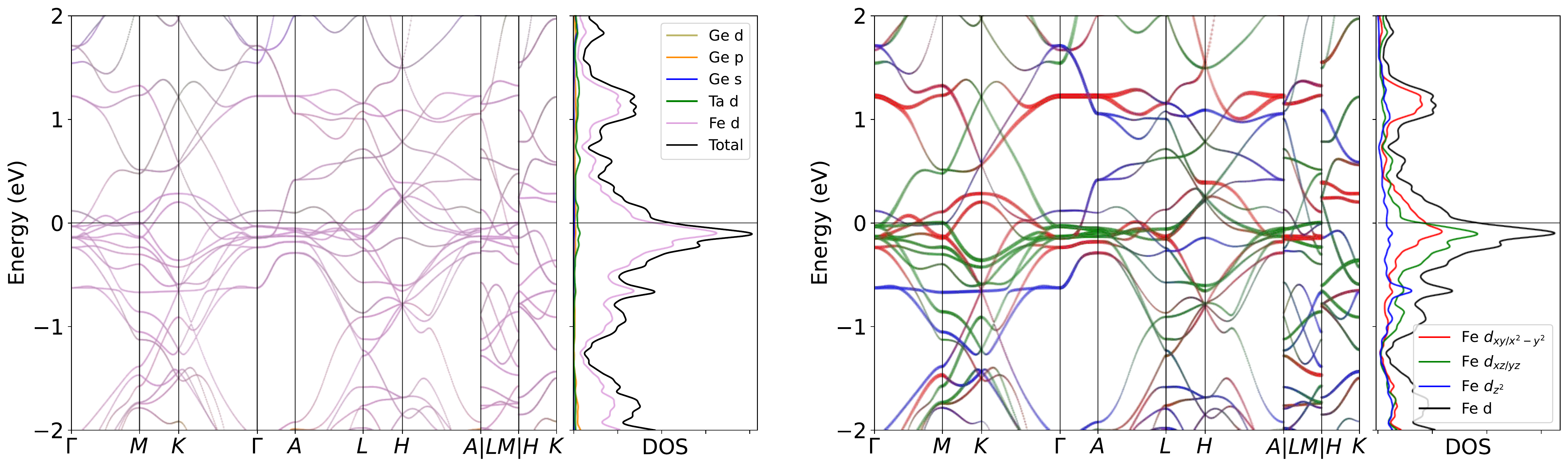}
\caption{Band structure for TaFe$_6$Ge$_6$ without SOC in paramagnetic phase. The projection of Fe $3d$ orbitals is also presented. The band structures clearly reveal the dominating of Fe $3d$ orbitals  near the Fermi level, as well as the pronounced peak.}
\label{fig:TaFe6Ge6band}
\end{figure}

\begin{figure}[H]
\centering
\subfloat[Relaxed phonon at low-T.]{
\label{fig:TaFe6Ge6_relaxed_Low}
\includegraphics[width=0.45\textwidth]{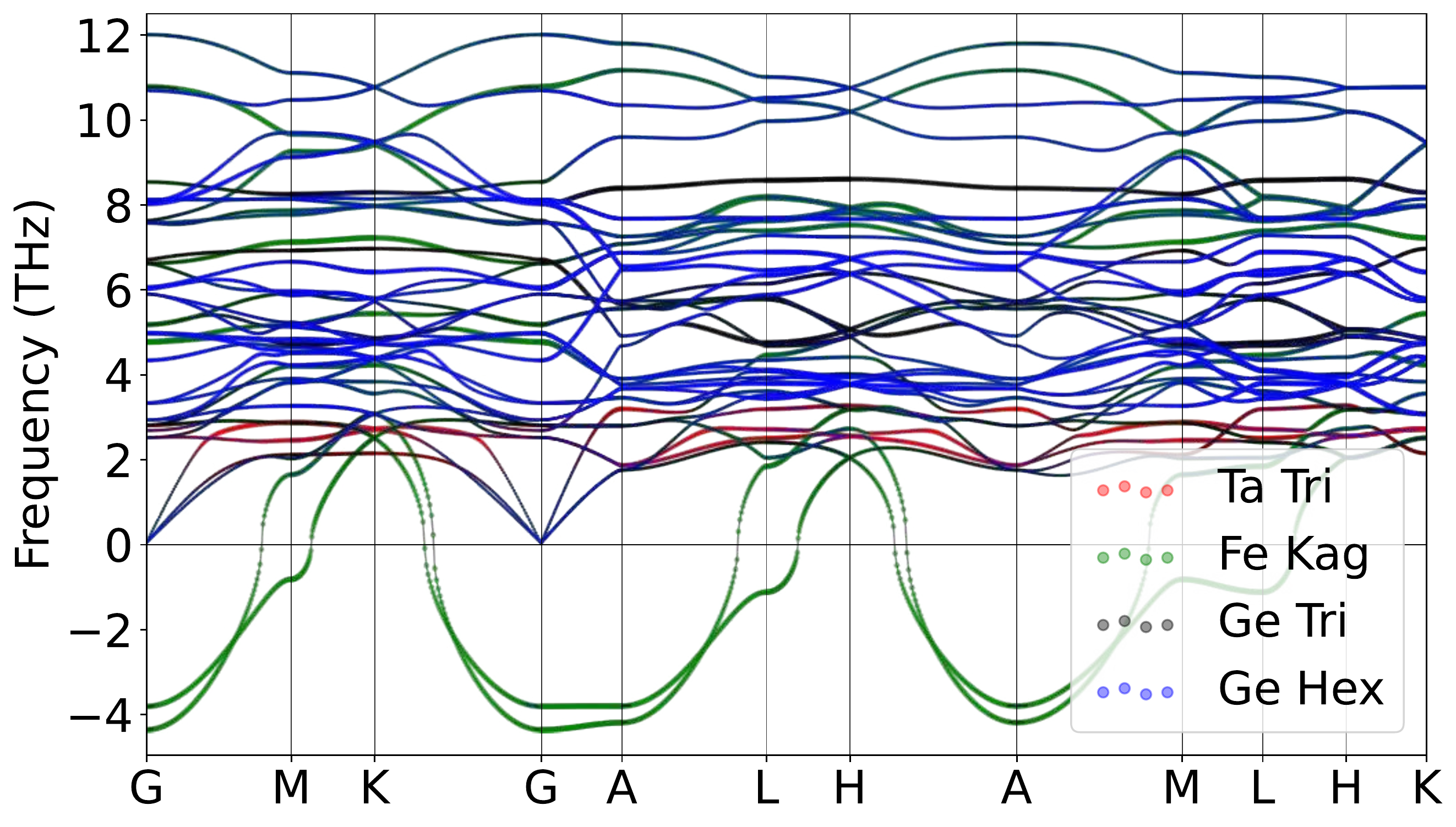}
}
\hspace{5pt}\subfloat[Relaxed phonon at high-T.]{
\label{fig:TaFe6Ge6_relaxed_High}
\includegraphics[width=0.45\textwidth]{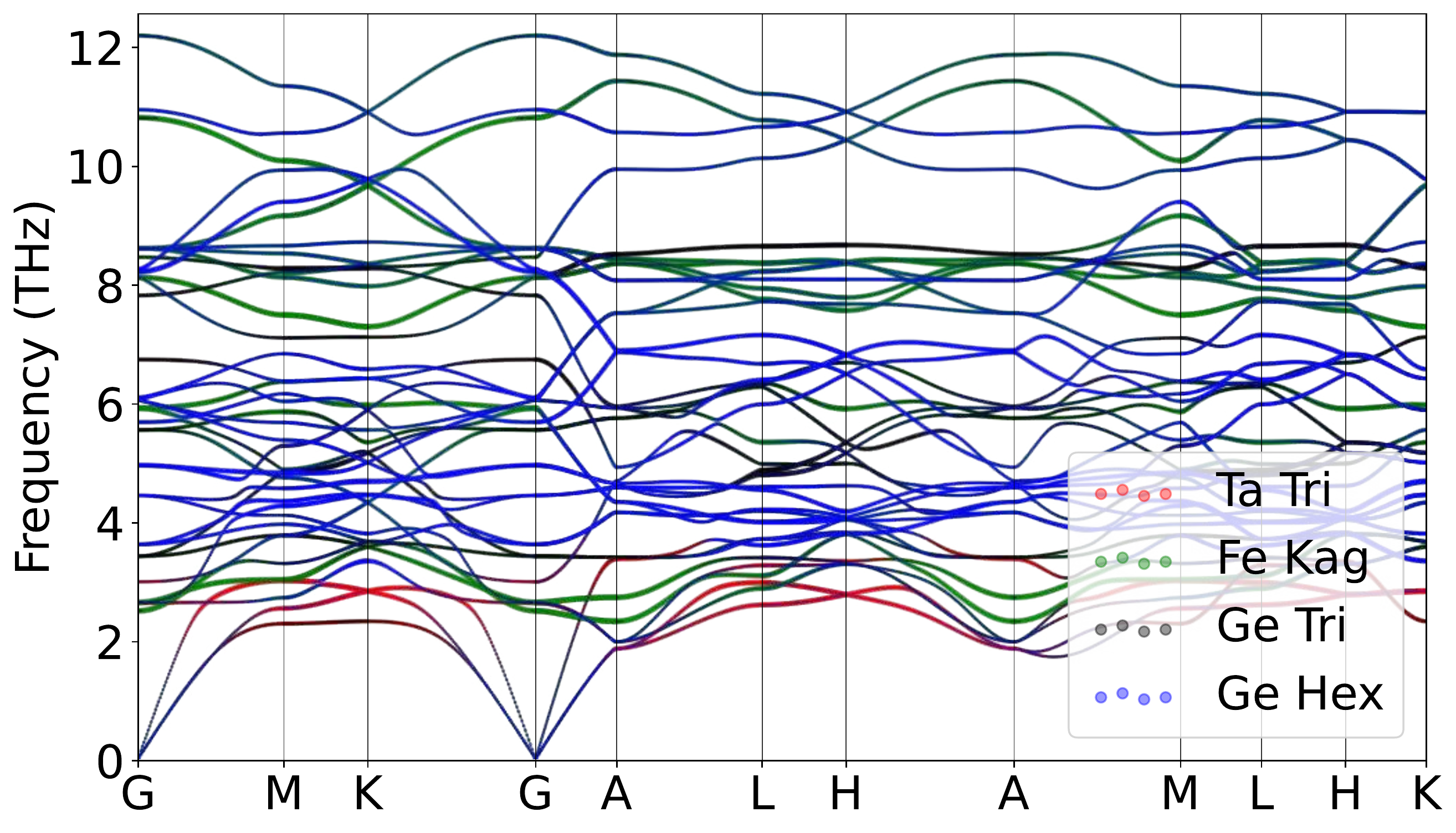}
}
\caption{Atomically projected phonon spectrum for TaFe$_6$Ge$_6$ in paramagnetic phase.}
\label{fig:TaFe6Ge6pho}
\end{figure}

\begin{figure}[htbp]
\centering
\includegraphics[width=0.45\textwidth]{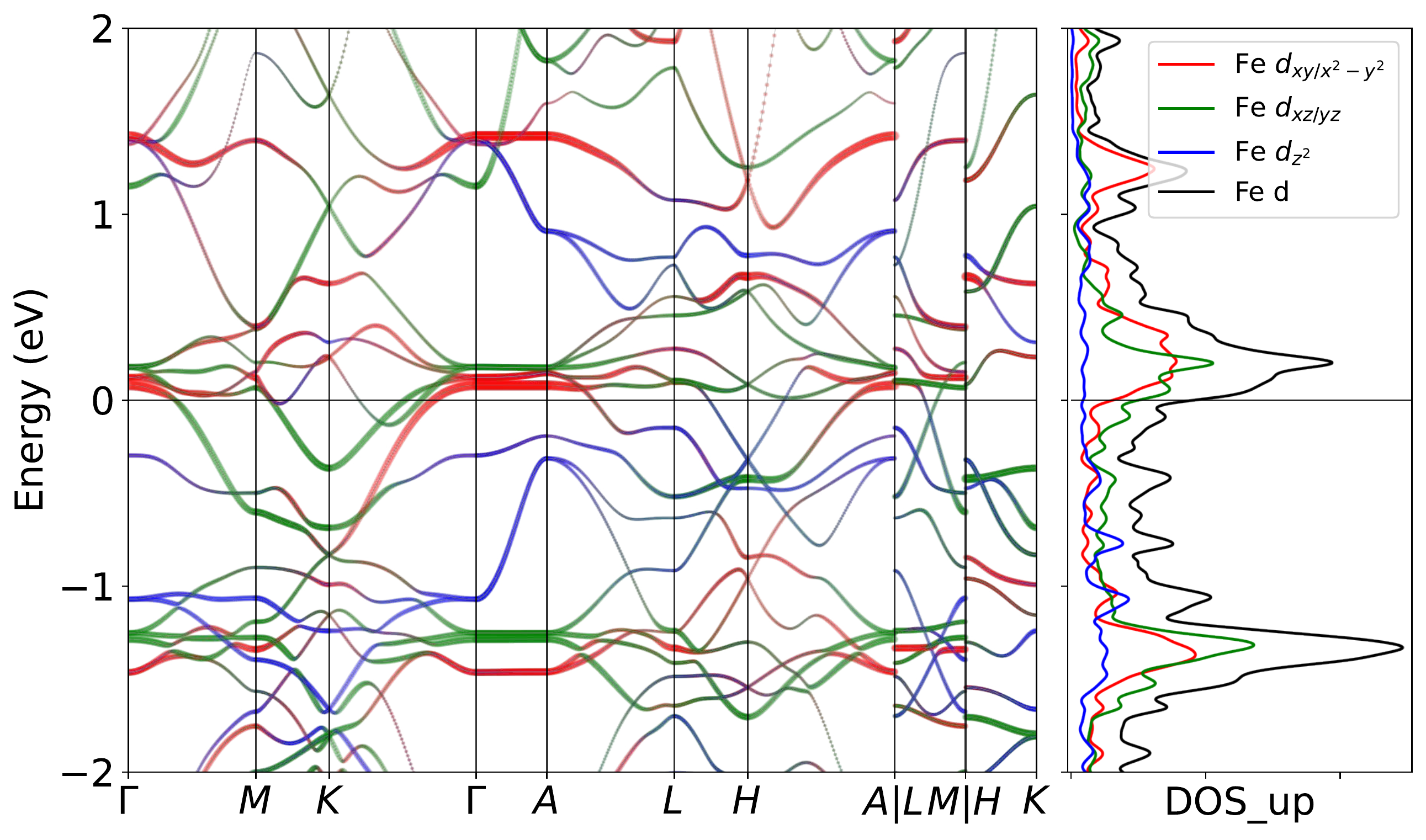}
\includegraphics[width=0.45\textwidth]{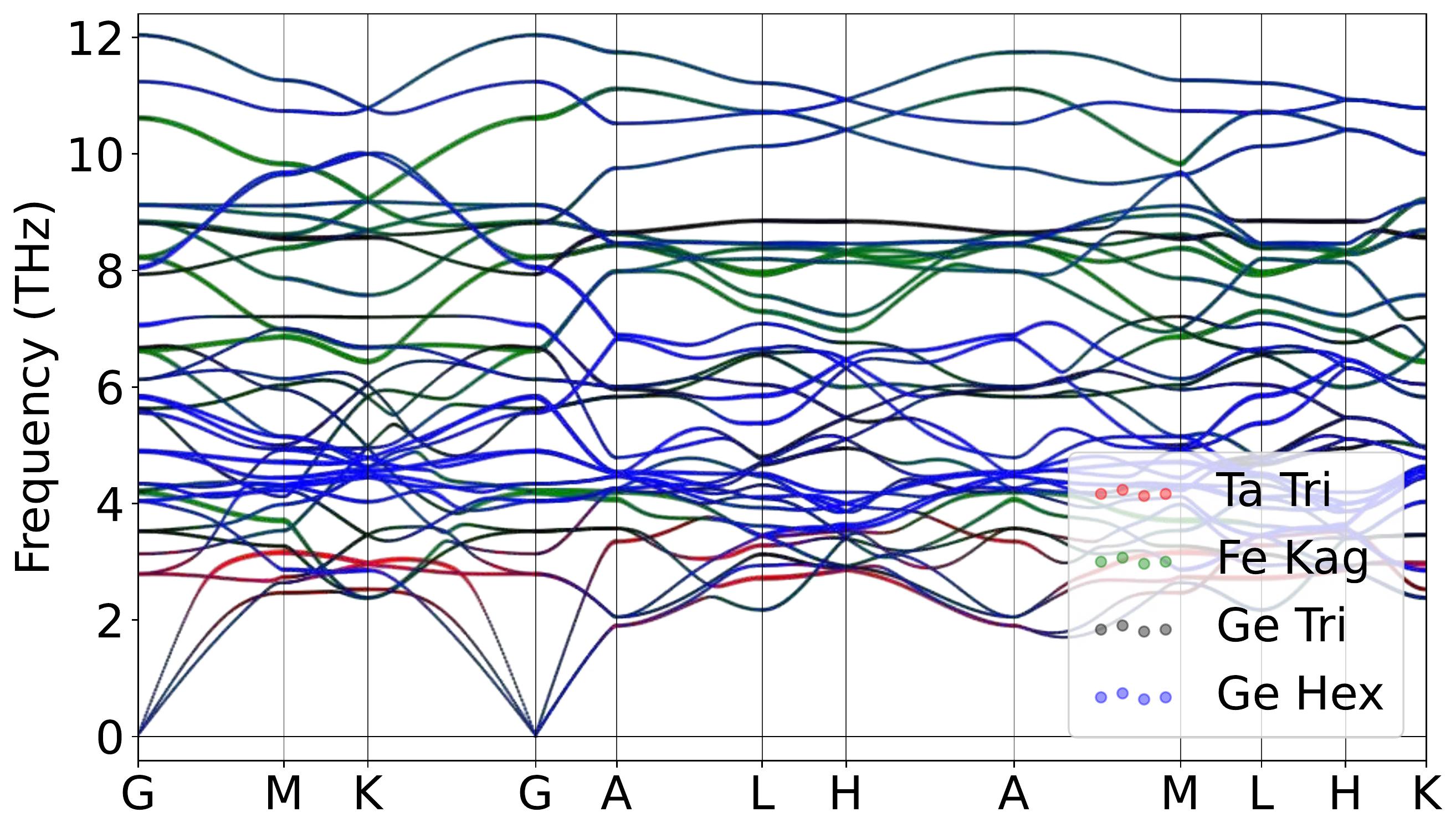}
\caption{Band structure for TaFe$_6$Ge$_6$ with AFM order without SOC for spin (Left)up channel. The projection of Fe $3d$ orbitals is also presented. (Right)Correspondingly, a stable phonon was demonstrated with the collinear magnetic order without Hubbard U at lower temperature regime, weighted by atomic projections.}
\label{fig:TaFe6Ge6mag}
\end{figure}

\begin{figure}[H]
\centering
\label{fig:TaFe6Ge6_relaxed_Low_xz}
\includegraphics[width=0.85\textwidth]{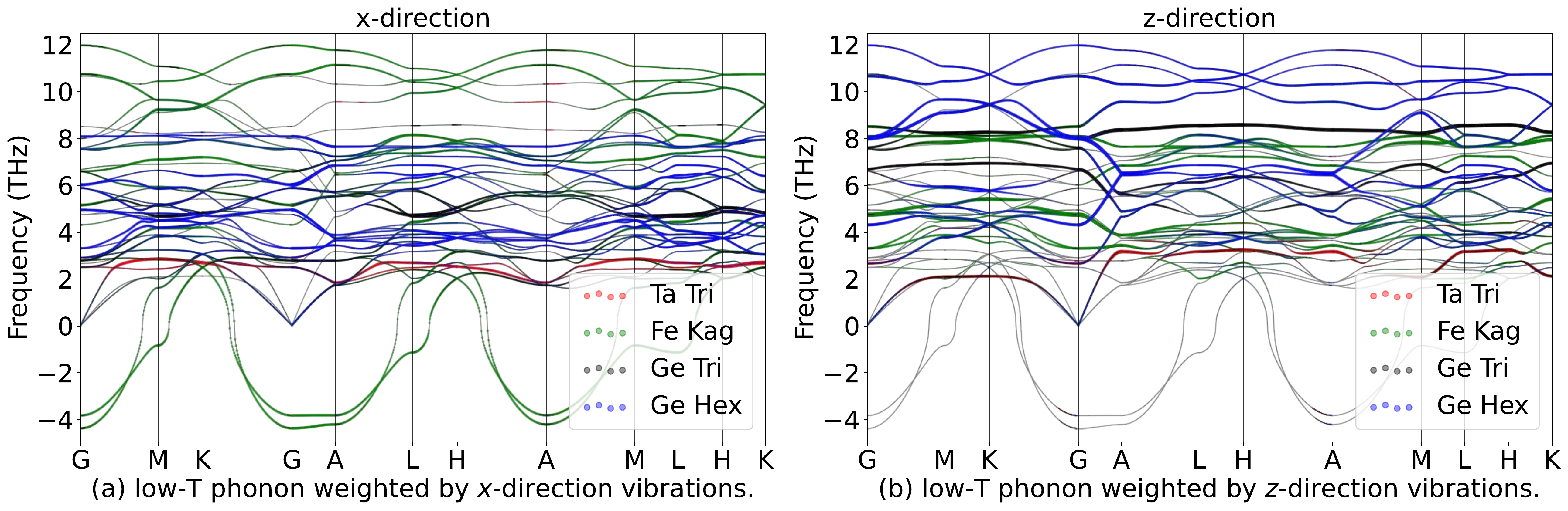}
\hspace{5pt}\label{fig:TaFe6Ge6_relaxed_High_xz}
\includegraphics[width=0.85\textwidth]{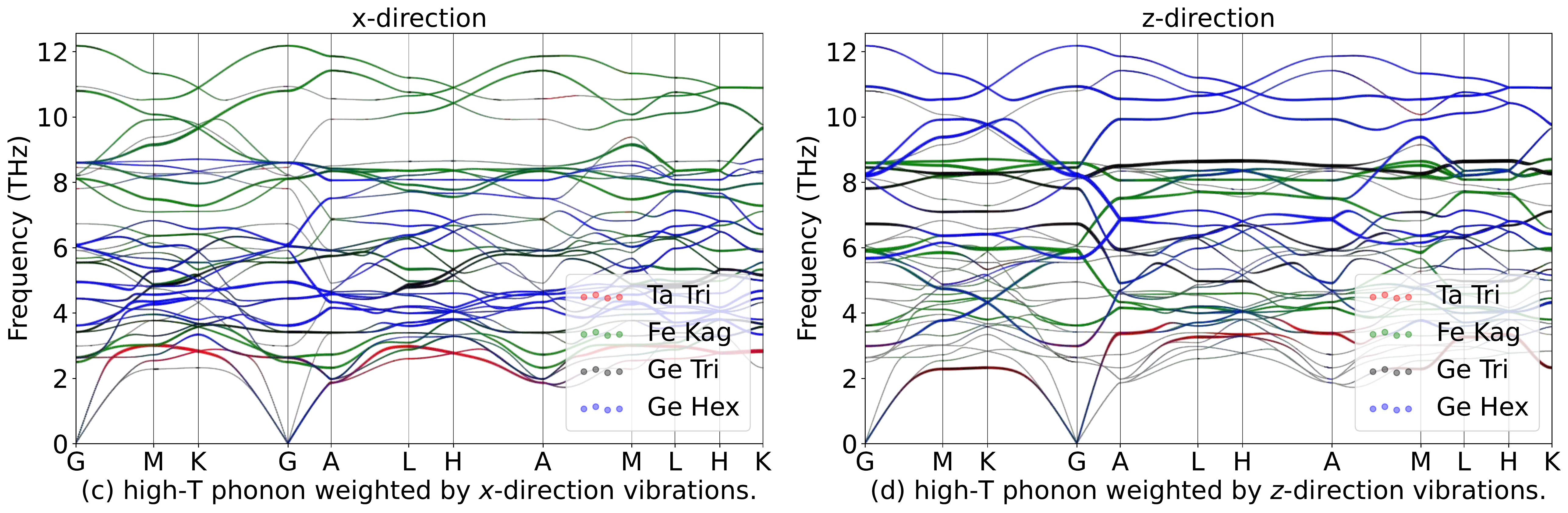}
\caption{Phonon spectrum for TaFe$_6$Ge$_6$in in-plane ($x$) and out-of-plane ($z$) directions in paramagnetic phase.}
\label{fig:TaFe6Ge6phoxyz}
\end{figure}

\begin{table}[htbp]
\global\long\def\arraystretch{1.12}
\captionsetup{width=.95\textwidth}
\caption{IRREPs at high-symmetry points for TaFe$_6$Ge$_6$ phonon spectrum at Low-T.}
\label{tab:191_TaFe6Ge6_Low}
\setlength{\tabcolsep}{1.mm}{
}
\end{table}

\begin{table}[htbp]
\global\long\def\arraystretch{1.12}
\captionsetup{width=.95\textwidth}
\caption{IRREPs at high-symmetry points for TaFe$_6$Ge$_6$ phonon spectrum at High-T.}
\label{tab:191_TaFe6Ge6_High}
\setlength{\tabcolsep}{1.mm}{
%
}
\end{table}
\subsubsection{\label{sms2sec:UFe6Ge6}\hyperref[smsec:col]{\texorpdfstring{UFe$_6$Ge$_6$}{}}~{\color{blue}  (High-T stable, Low-T unstable) }}

\begin{table}[htbp]
\global\long\def\arraystretch{1.12}
\captionsetup{width=.85\textwidth}
\caption{Structure parameters of UFe$_6$Ge$_6$ after relaxation. It contains 440 electrons. }
\label{tab:UFe6Ge6}
\setlength{\tabcolsep}{4mm}{
%
}
\end{table}

\begin{figure}[htbp]
\centering
\includegraphics[width=0.85\textwidth]{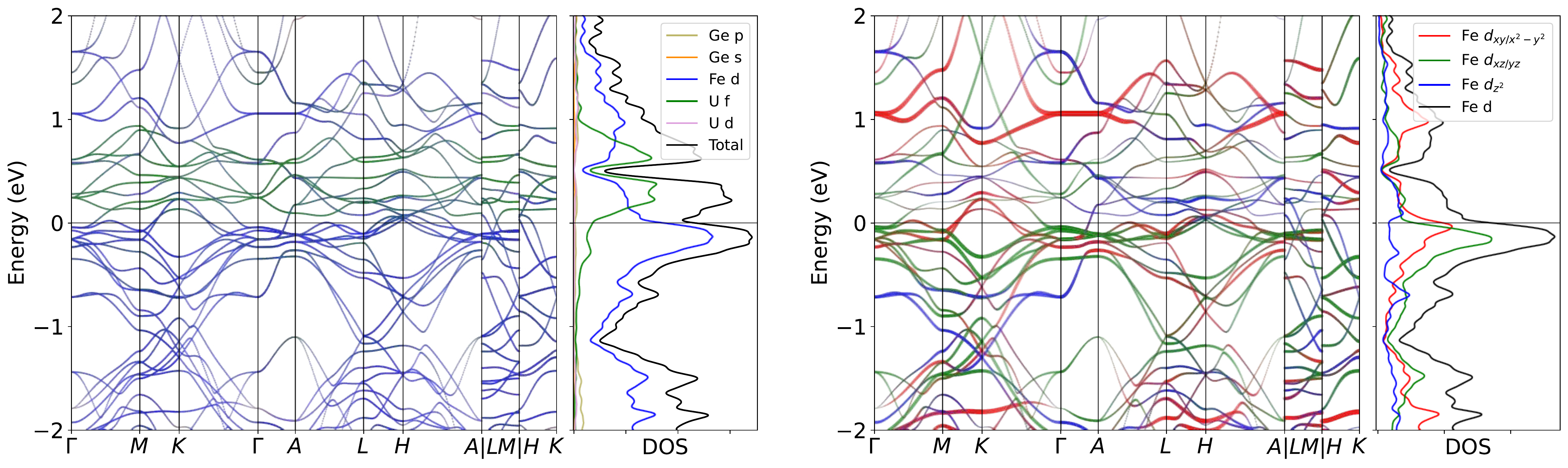}
\caption{Band structure for UFe$_6$Ge$_6$ without SOC in paramagnetic phase. The projection of Fe $3d$ orbitals is also presented. The band structures clearly reveal the dominating of Fe $3d$ orbitals  near the Fermi level, as well as the pronounced peak.}
\label{fig:UFe6Ge6band}
\end{figure}

\begin{figure}[H]
\centering
\subfloat[Relaxed phonon at low-T.]{
\label{fig:UFe6Ge6_relaxed_Low}
\includegraphics[width=0.45\textwidth]{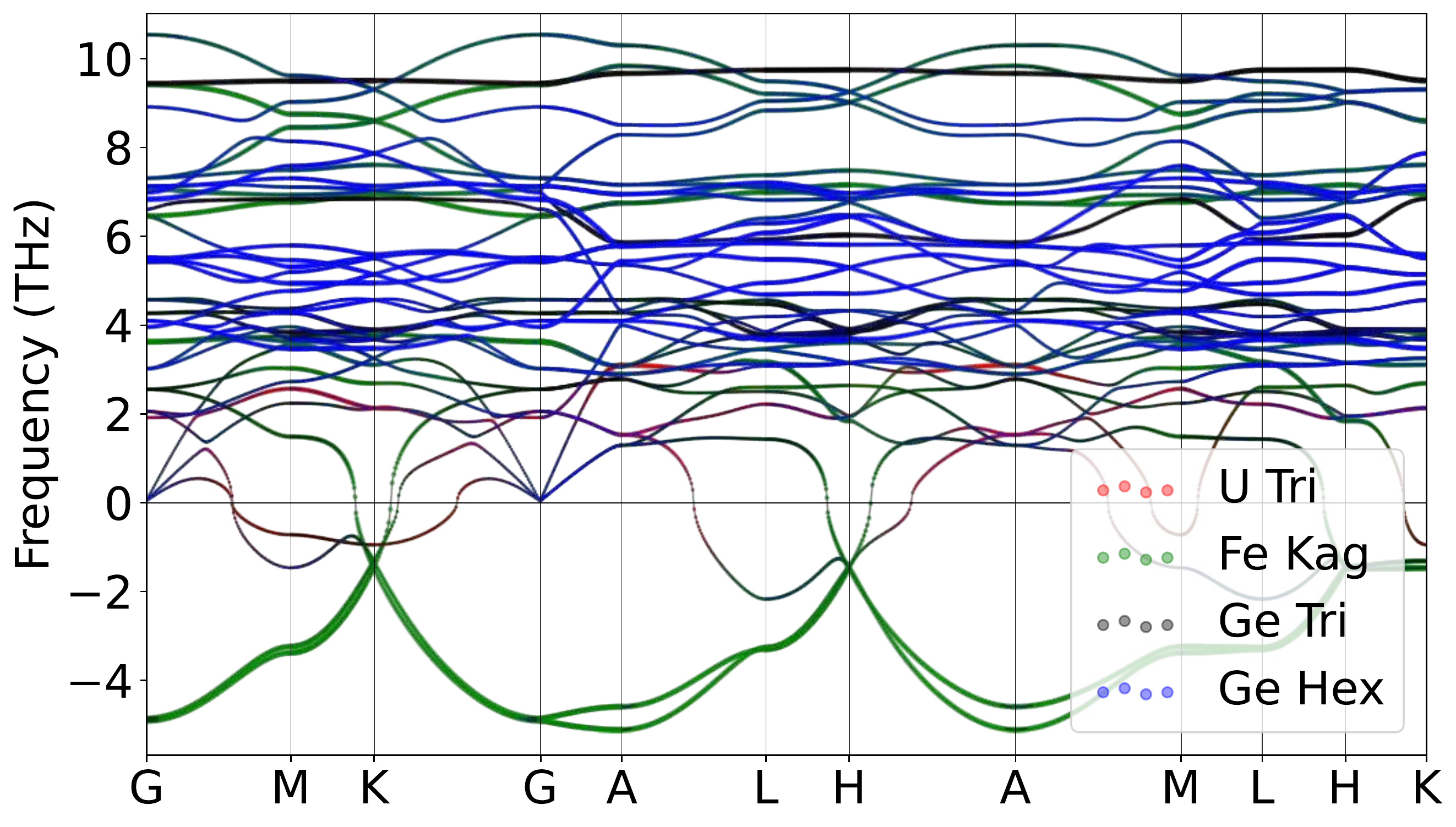}
}
\hspace{5pt}\subfloat[Relaxed phonon at high-T.]{
\label{fig:UFe6Ge6_relaxed_High}
\includegraphics[width=0.45\textwidth]{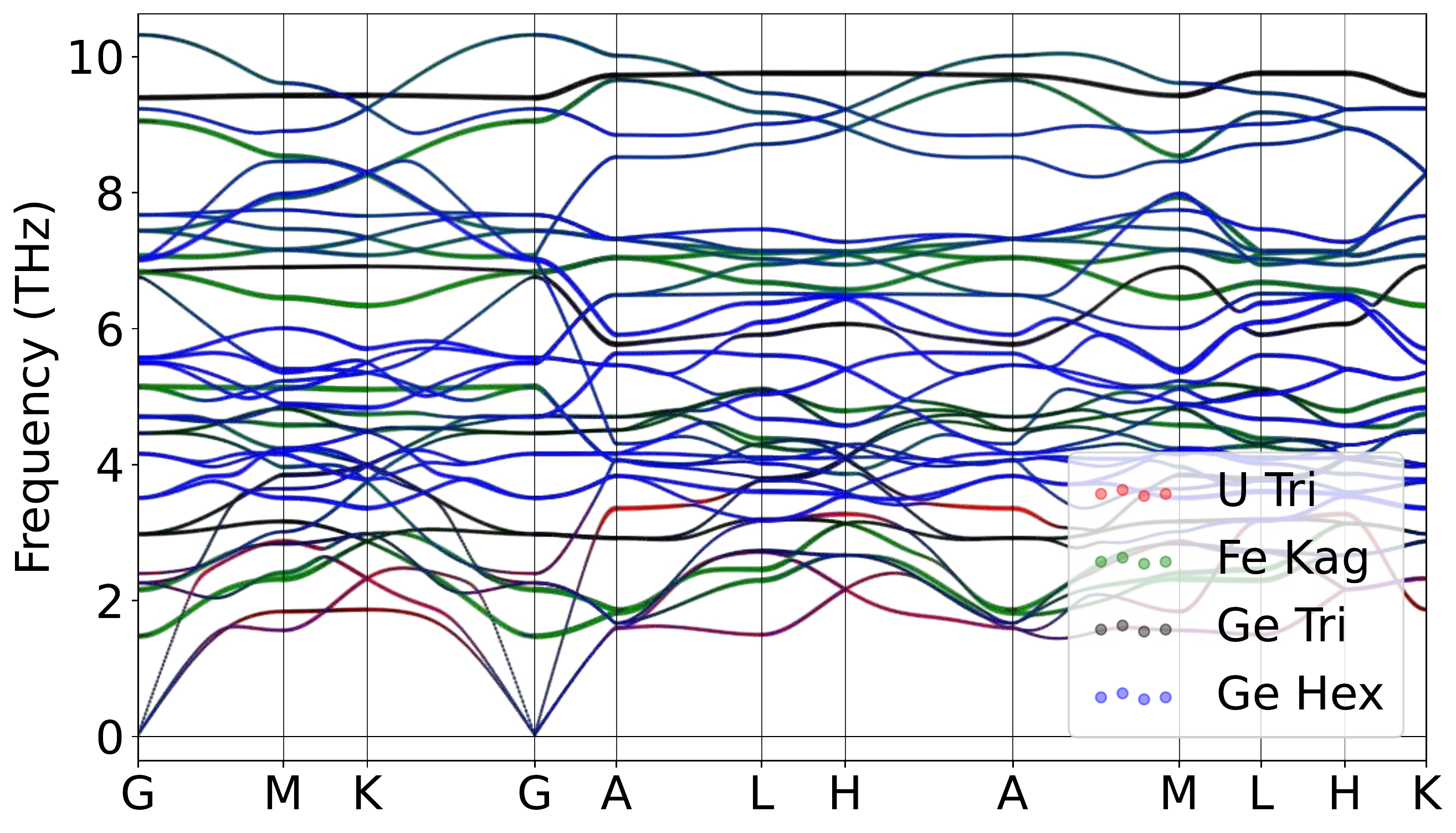}
}
\caption{Atomically projected phonon spectrum for UFe$_6$Ge$_6$ in paramagnetic phase.}
\label{fig:UFe6Ge6pho}
\end{figure}

\begin{figure}[htbp]
\centering
\includegraphics[width=0.45\textwidth]{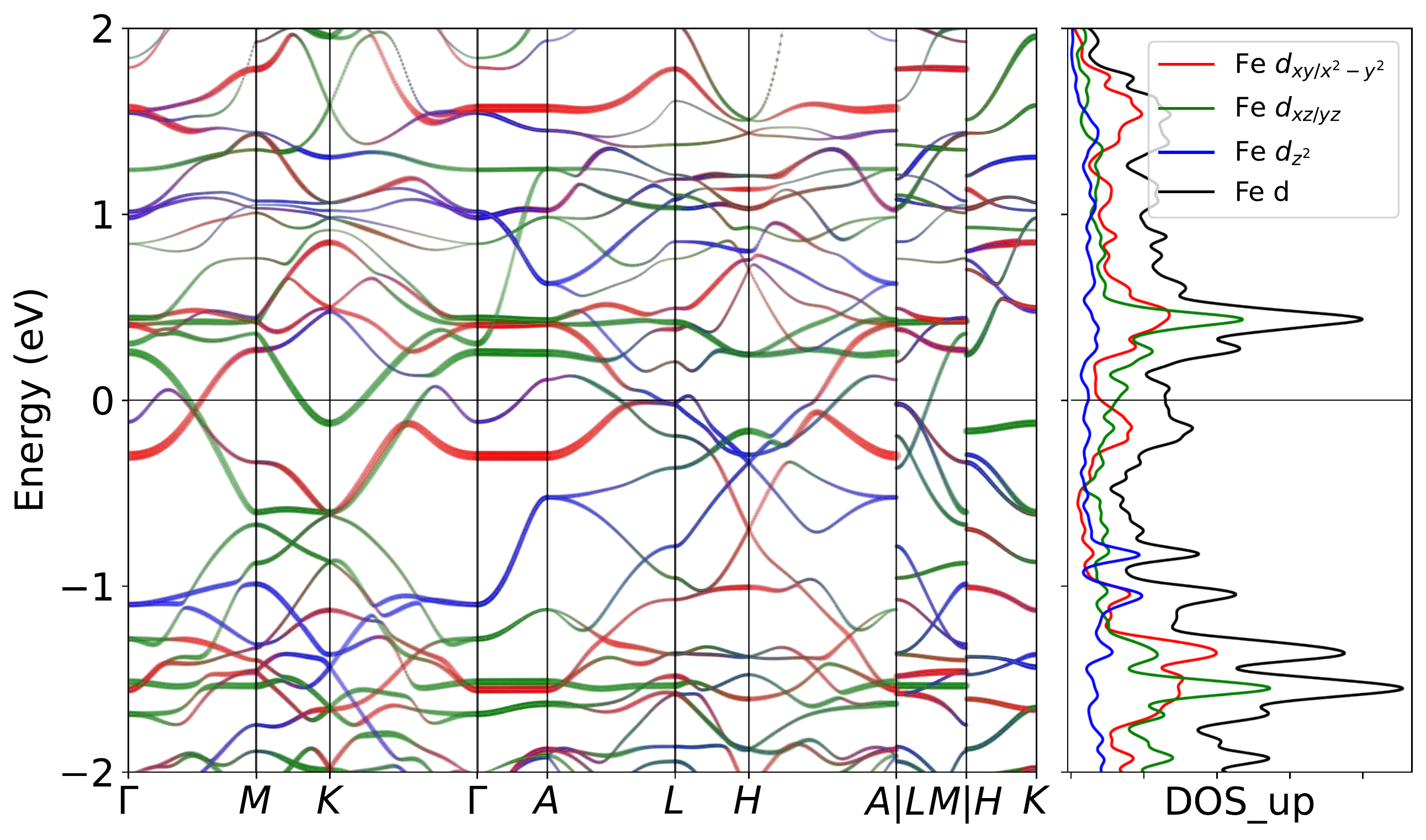}
\caption{Band structure for UFe$_6$Ge$_6$ with AFM order without SOC for spin (Left)up channel. The projection of Fe $3d$ orbitals is also presented.}
\label{fig:UFe6Ge6mag}
\end{figure}

\begin{figure}[H]
\centering
\label{fig:UFe6Ge6_relaxed_Low_xz}
\includegraphics[width=0.85\textwidth]{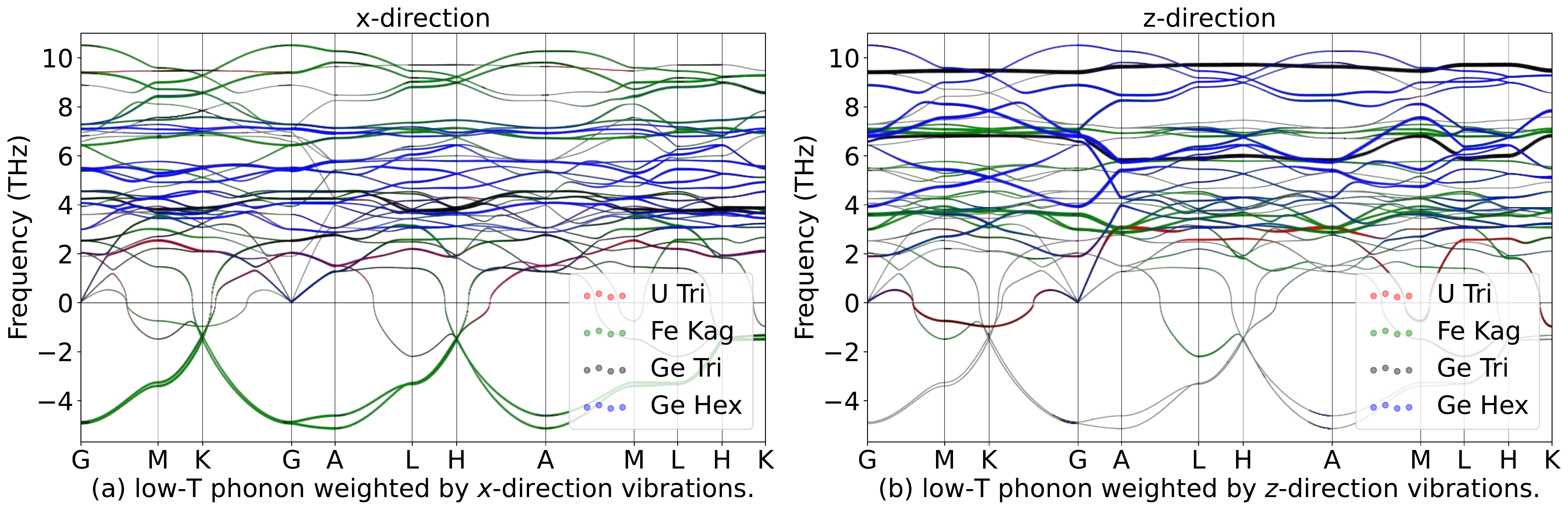}
\hspace{5pt}\label{fig:UFe6Ge6_relaxed_High_xz}
\includegraphics[width=0.85\textwidth]{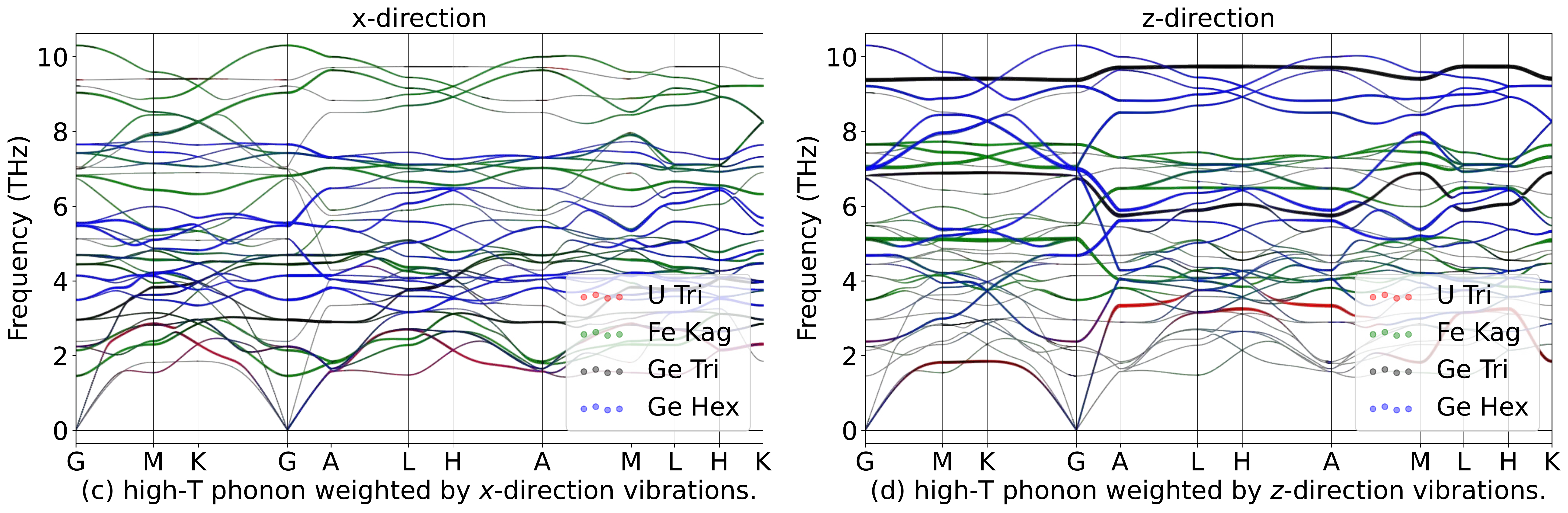}
\caption{Phonon spectrum for UFe$_6$Ge$_6$in in-plane ($x$) and out-of-plane ($z$) directions in paramagnetic phase.}
\label{fig:UFe6Ge6phoxyz}
\end{figure}

\begin{table}[htbp]
\global\long\def\arraystretch{1.12}
\captionsetup{width=.95\textwidth}
\caption{IRREPs at high-symmetry points for UFe$_6$Ge$_6$ phonon spectrum at Low-T.}
\label{tab:191_UFe6Ge6_Low}
\setlength{\tabcolsep}{1.mm}{
}
\end{table}

\begin{table}[htbp]
\global\long\def\arraystretch{1.12}
\captionsetup{width=.95\textwidth}
\caption{IRREPs at high-symmetry points for UFe$_6$Ge$_6$ phonon spectrum at High-T.}
\label{tab:191_UFe6Ge6_High}
\setlength{\tabcolsep}{1.mm}{
%
}
\end{table}
\subsection{\hyperref[smsec:col]{FeSn}}~\label{smssec:FeSn}

\subsubsection{\label{sms2sec:LiFe6Sn6}\hyperref[smsec:col]{\texorpdfstring{LiFe$_6$Sn$_6$}{}}~{\color{blue}  (High-T stable, Low-T unstable) }}

\begin{table}[htbp]
\global\long\def\arraystretch{1.12}
\captionsetup{width=.85\textwidth}
\caption{Structure parameters of LiFe$_6$Sn$_6$ after relaxation. It contains 459 electrons. }
\label{tab:LiFe6Sn6}
\setlength{\tabcolsep}{4mm}{
%
}
\end{table}

\begin{figure}[htbp]
\centering
\includegraphics[width=0.85\textwidth]{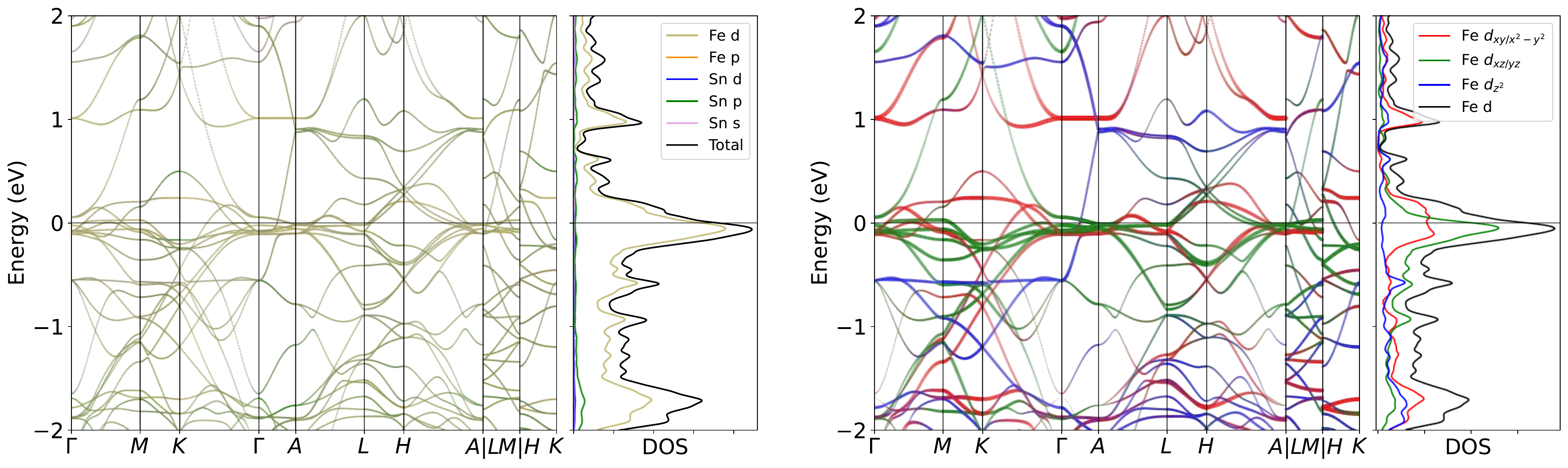}
\caption{Band structure for LiFe$_6$Sn$_6$ without SOC in paramagnetic phase. The projection of Fe $3d$ orbitals is also presented. The band structures clearly reveal the dominating of Fe $3d$ orbitals  near the Fermi level, as well as the pronounced peak.}
\label{fig:LiFe6Sn6band}
\end{figure}

\begin{figure}[H]
\centering
\subfloat[Relaxed phonon at low-T.]{
\label{fig:LiFe6Sn6_relaxed_Low}
\includegraphics[width=0.45\textwidth]{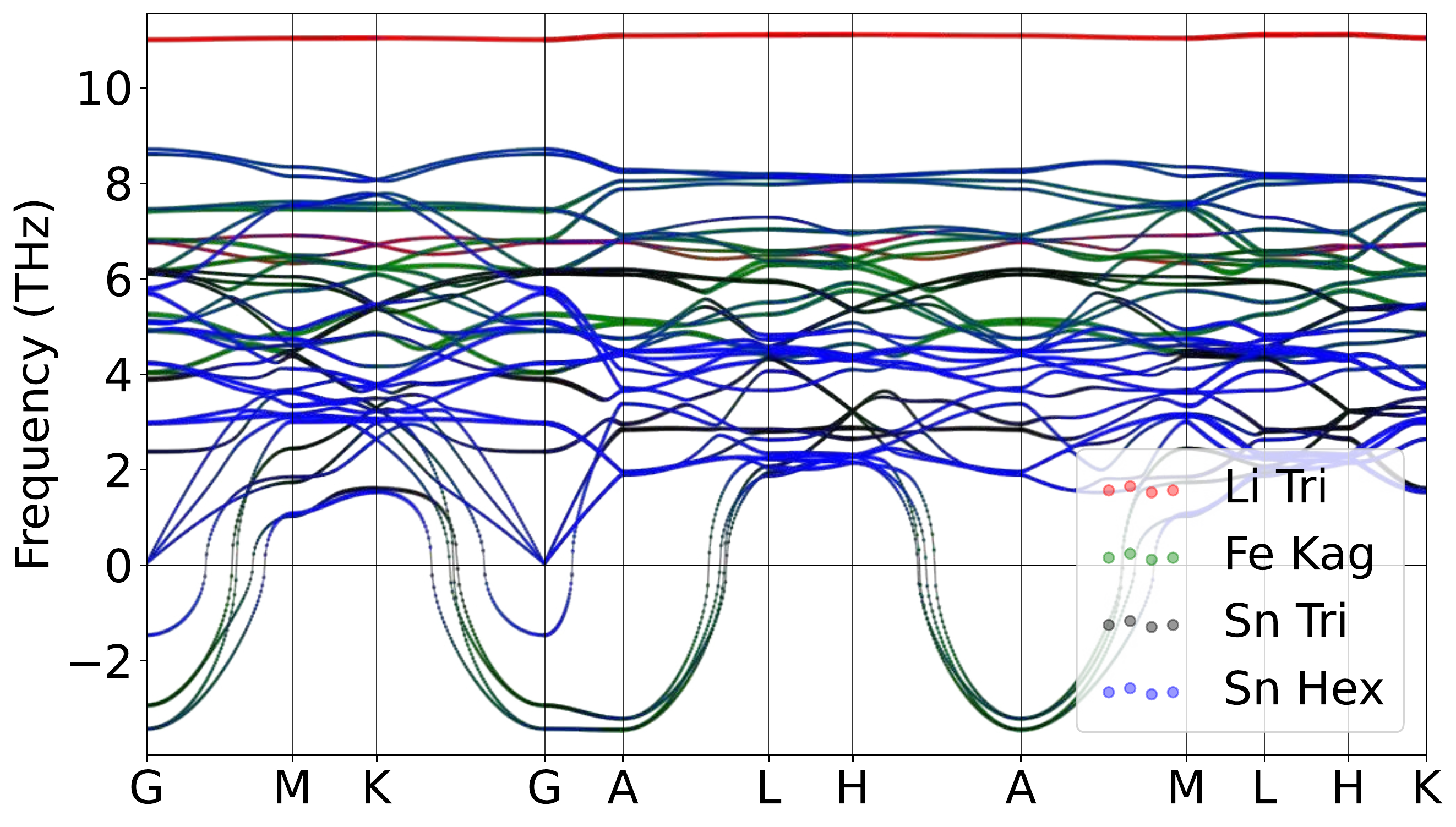}
}
\hspace{5pt}\subfloat[Relaxed phonon at high-T.]{
\label{fig:LiFe6Sn6_relaxed_High}
\includegraphics[width=0.45\textwidth]{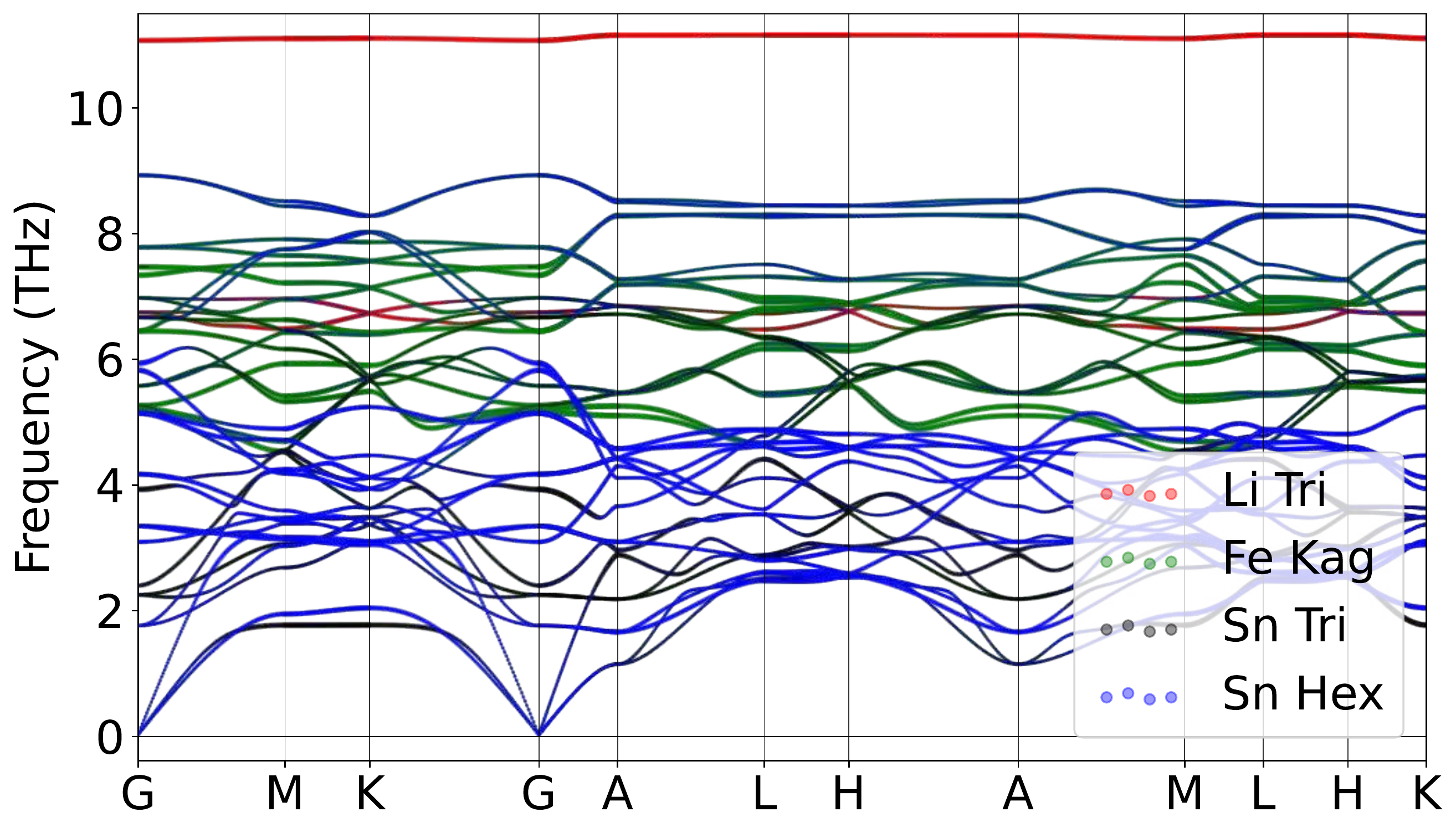}
}
\caption{Atomically projected phonon spectrum for LiFe$_6$Sn$_6$ in paramagnetic phase.}
\label{fig:LiFe6Sn6pho}
\end{figure}

\begin{figure}[htbp]
\centering
\includegraphics[width=0.45\textwidth]{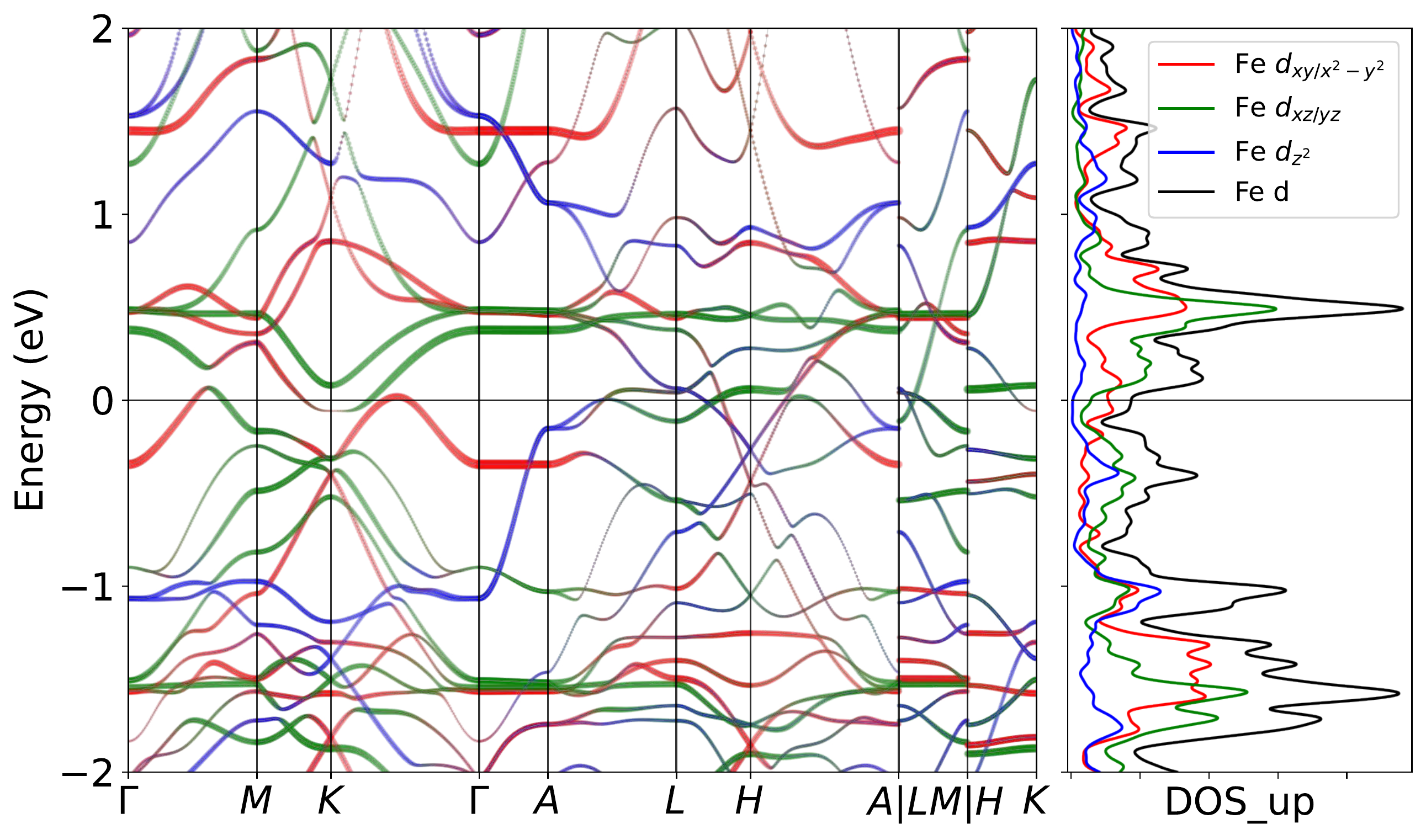}
\includegraphics[width=0.45\textwidth]{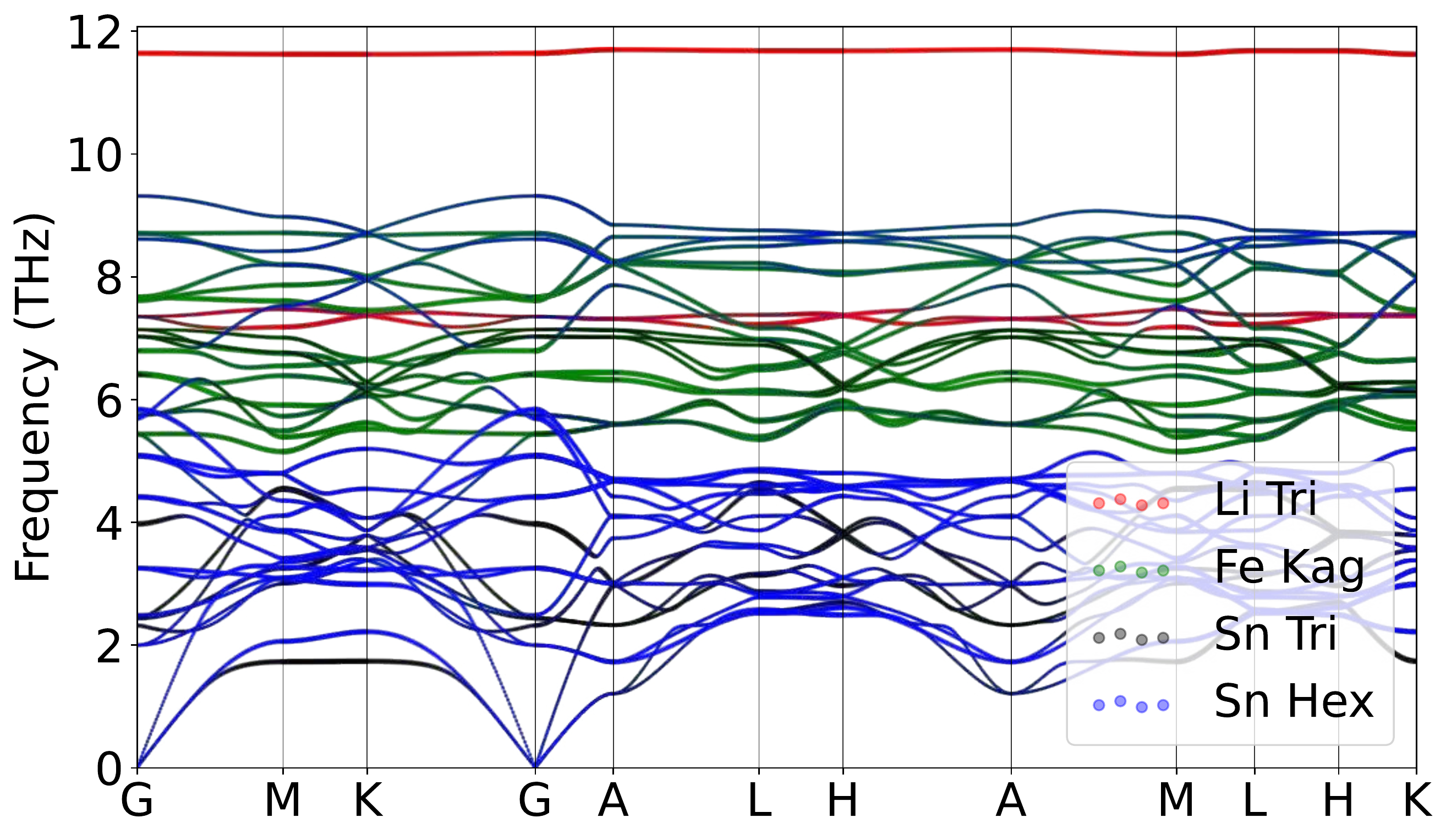}
\caption{Band structure for LiFe$_6$Sn$_6$ with AFM order without SOC for spin (Left)up channel. The projection of Fe $3d$ orbitals is also presented. (Right)Correspondingly, a stable phonon was demonstrated with the collinear magnetic order without Hubbard U at lower temperature regime, weighted by atomic projections.}
\label{fig:LiFe6Sn6mag}
\end{figure}

\begin{figure}[H]
\centering
\label{fig:LiFe6Sn6_relaxed_Low_xz}
\includegraphics[width=0.85\textwidth]{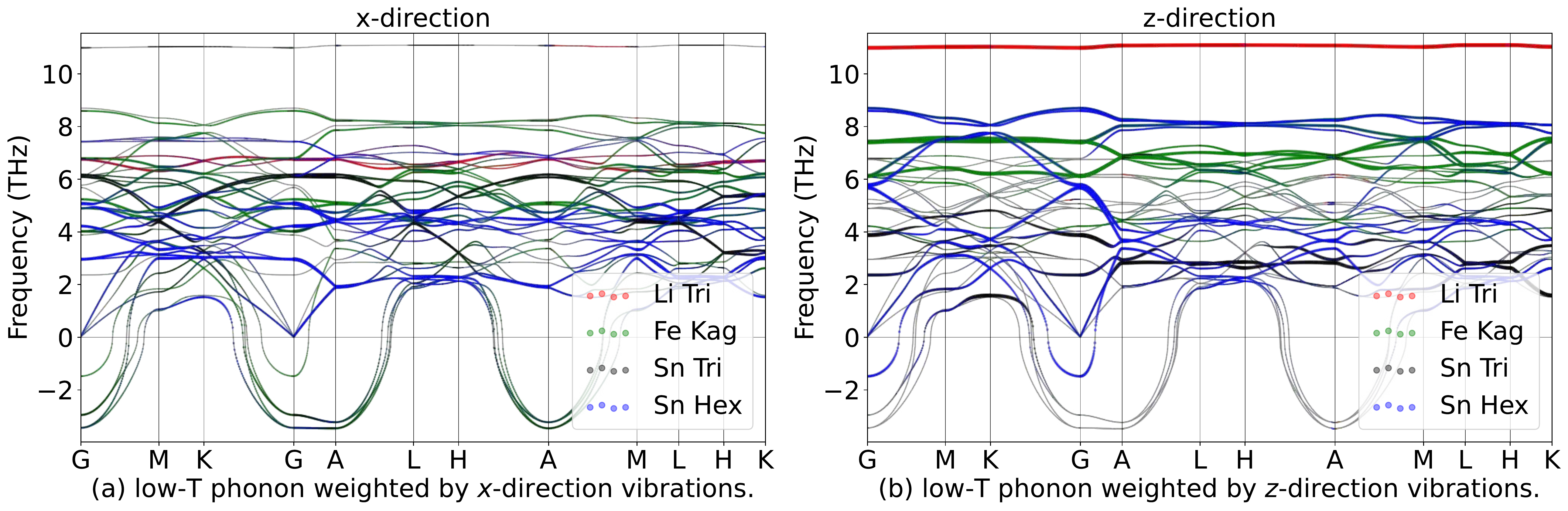}
\hspace{5pt}\label{fig:LiFe6Sn6_relaxed_High_xz}
\includegraphics[width=0.85\textwidth]{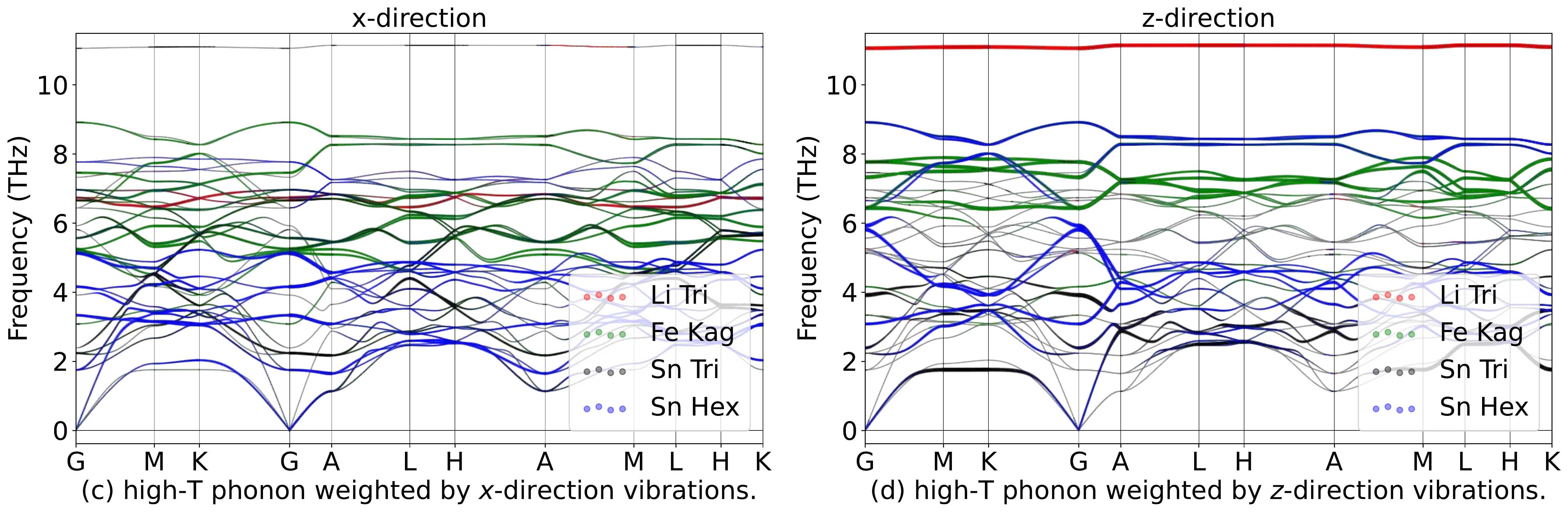}
\caption{Phonon spectrum for LiFe$_6$Sn$_6$in in-plane ($x$) and out-of-plane ($z$) directions in paramagnetic phase.}
\label{fig:LiFe6Sn6phoxyz}
\end{figure}

\begin{table}[htbp]
\global\long\def\arraystretch{1.12}
\captionsetup{width=.95\textwidth}
\caption{IRREPs at high-symmetry points for LiFe$_6$Sn$_6$ phonon spectrum at Low-T.}
\label{tab:191_LiFe6Sn6_Low}
\setlength{\tabcolsep}{1.mm}{
}
\end{table}

\begin{table}[htbp]
\global\long\def\arraystretch{1.12}
\captionsetup{width=.95\textwidth}
\caption{IRREPs at high-symmetry points for LiFe$_6$Sn$_6$ phonon spectrum at High-T.}
\label{tab:191_LiFe6Sn6_High}
\setlength{\tabcolsep}{1.mm}{
%
}
\end{table}
\subsubsection{\label{sms2sec:MgFe6Sn6}\hyperref[smsec:col]{\texorpdfstring{MgFe$_6$Sn$_6$}{}}~{\color{blue}  (High-T stable, Low-T unstable) }}

\begin{table}[htbp]
\global\long\def\arraystretch{1.12}
\captionsetup{width=.85\textwidth}
\caption{Structure parameters of MgFe$_6$Sn$_6$ after relaxation. It contains 468 electrons. }
\label{tab:MgFe6Sn6}
\setlength{\tabcolsep}{4mm}{
%
}
\end{table}

\begin{figure}[htbp]
\centering
\includegraphics[width=0.85\textwidth]{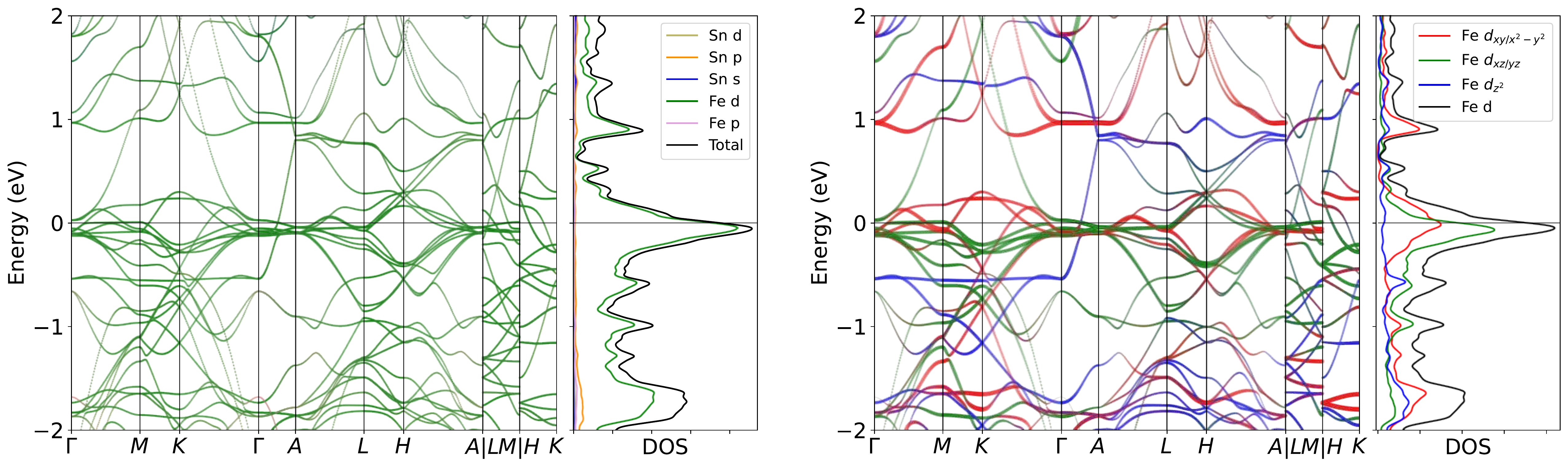}
\caption{Band structure for MgFe$_6$Sn$_6$ without SOC in paramagnetic phase. The projection of Fe $3d$ orbitals is also presented. The band structures clearly reveal the dominating of Fe $3d$ orbitals  near the Fermi level, as well as the pronounced peak.}
\label{fig:MgFe6Sn6band}
\end{figure}

\begin{figure}[H]
\centering
\subfloat[Relaxed phonon at low-T.]{
\label{fig:MgFe6Sn6_relaxed_Low}
\includegraphics[width=0.45\textwidth]{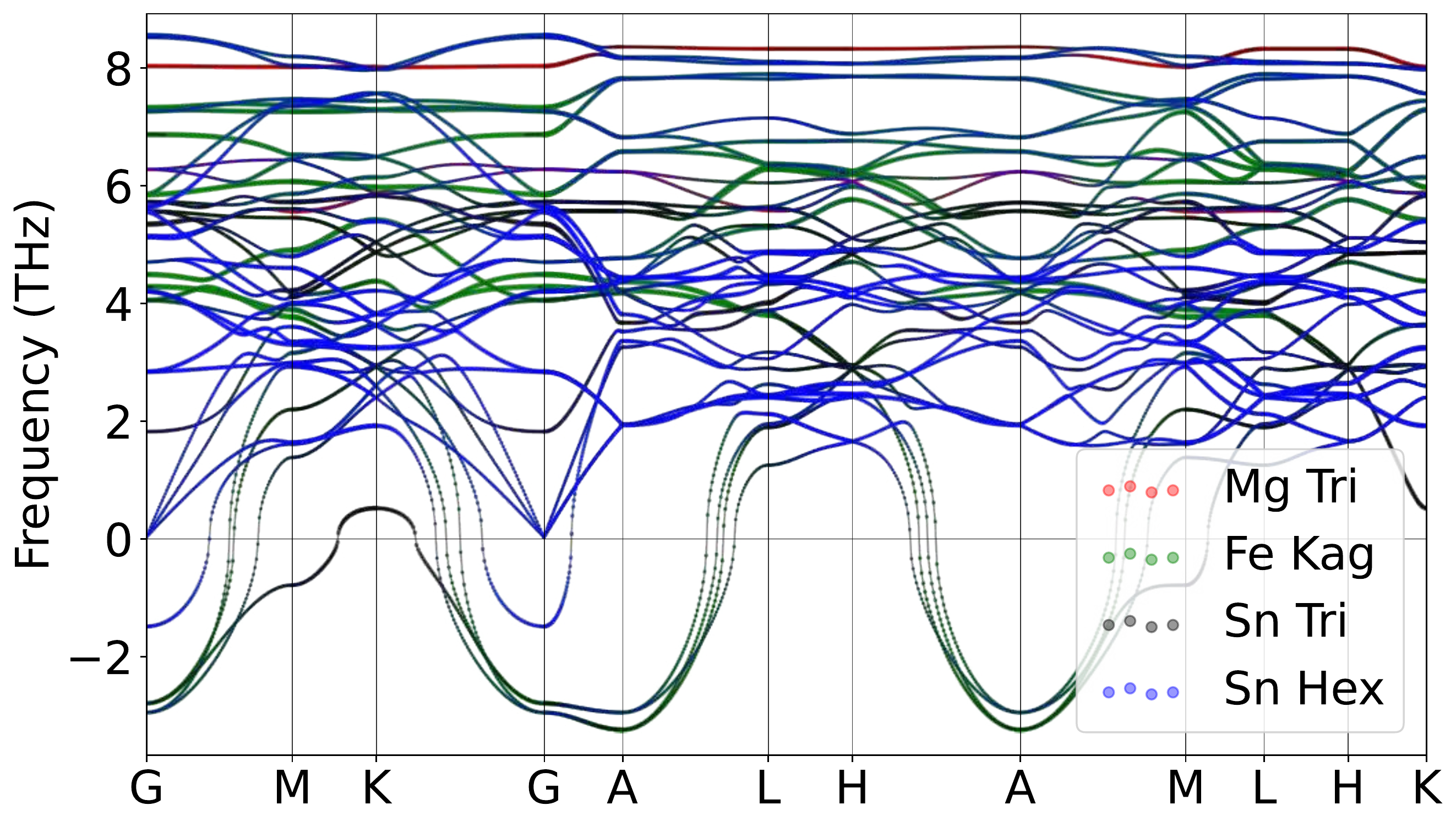}
}
\hspace{5pt}\subfloat[Relaxed phonon at high-T.]{
\label{fig:MgFe6Sn6_relaxed_High}
\includegraphics[width=0.45\textwidth]{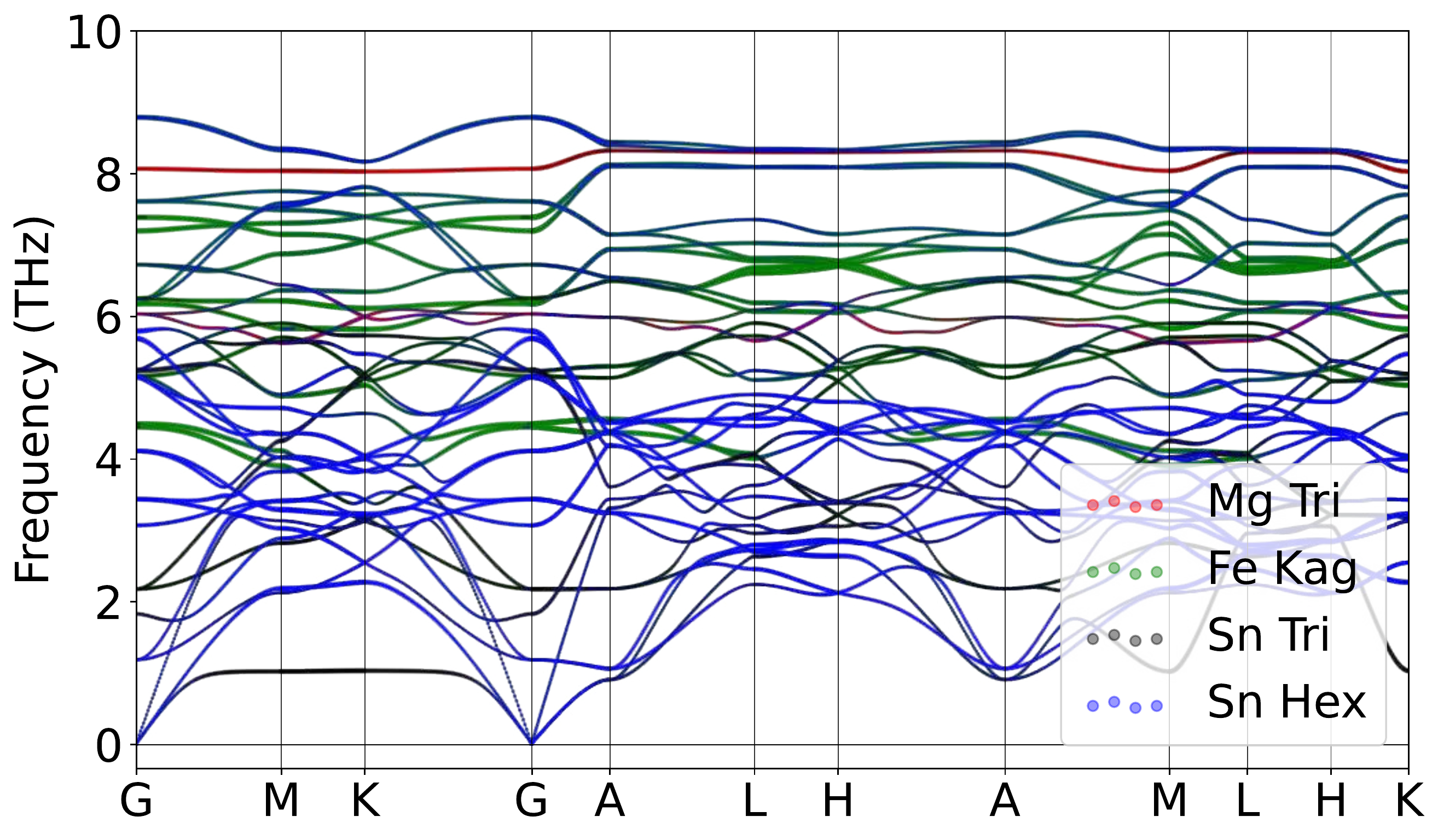}
}
\caption{Atomically projected phonon spectrum for MgFe$_6$Sn$_6$ in paramagnetic phase.}
\label{fig:MgFe6Sn6pho}
\end{figure}

\begin{figure}[htbp]
\centering
\includegraphics[width=0.45\textwidth]{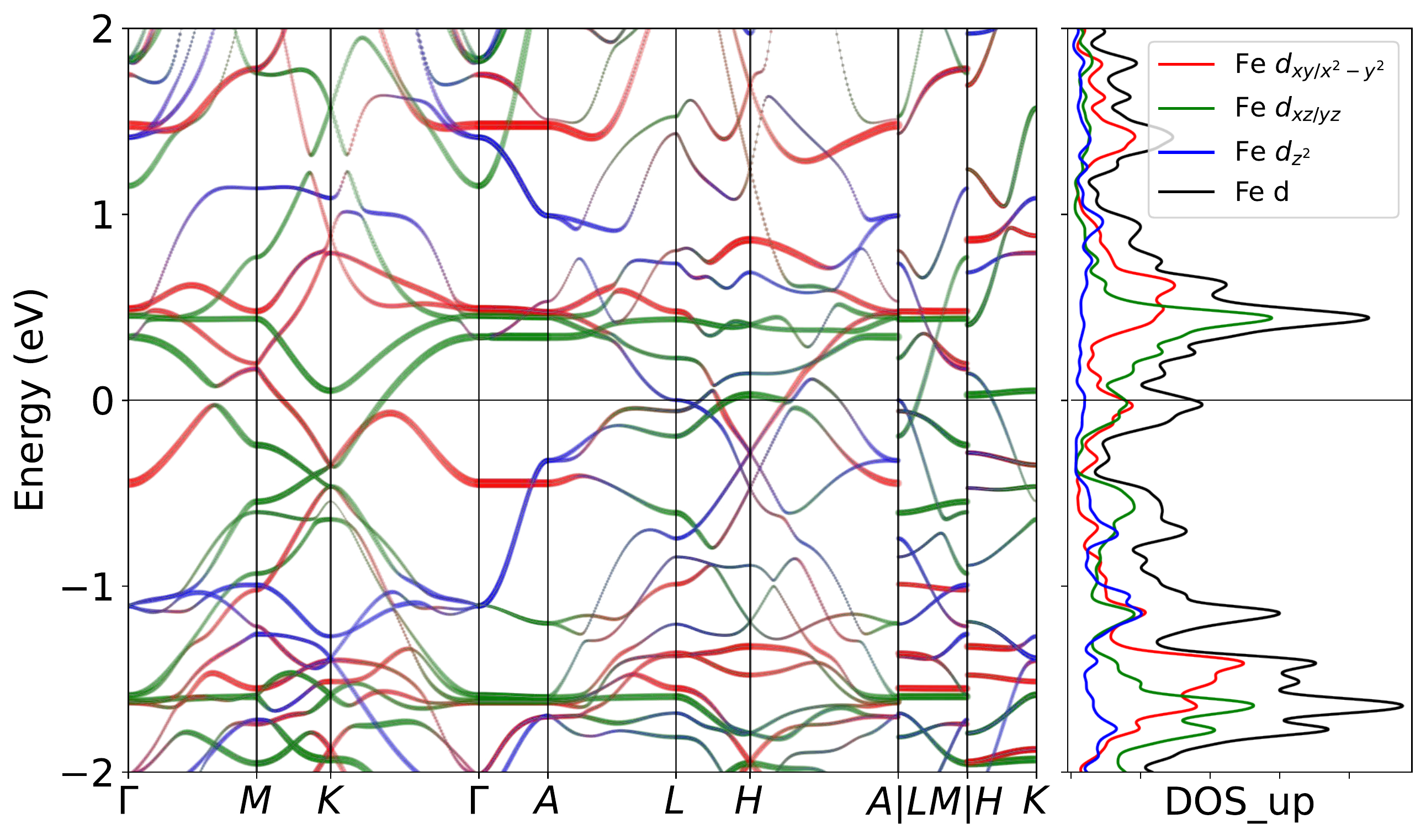}
\includegraphics[width=0.45\textwidth]{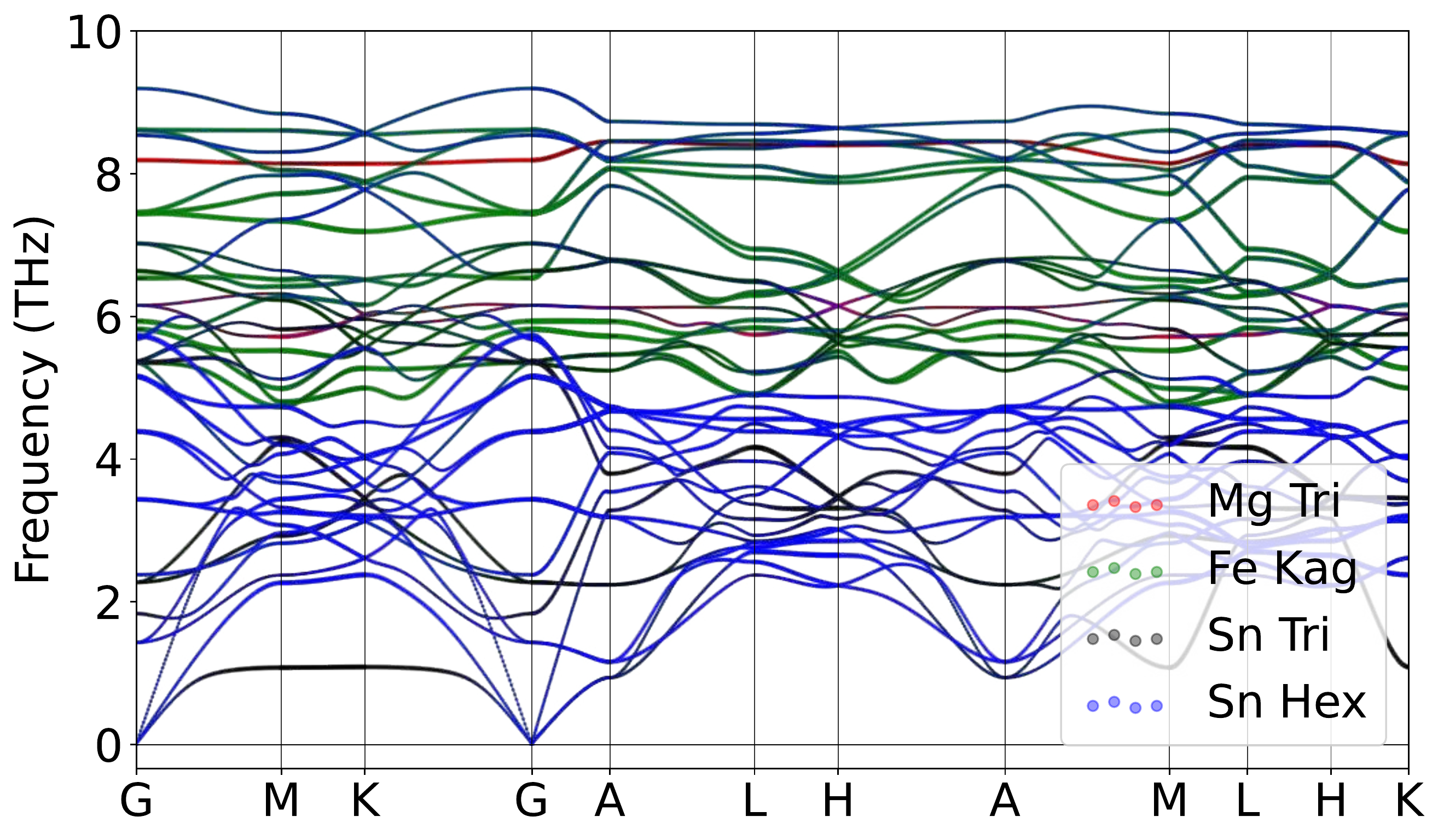}
\caption{Band structure for MgFe$_6$Sn$_6$ with AFM order without SOC for spin (Left)up channel. The projection of Fe $3d$ orbitals is also presented. (Right)Correspondingly, a stable phonon was demonstrated with the collinear magnetic order without Hubbard U at lower temperature regime, weighted by atomic projections.}
\label{fig:MgFe6Sn6mag}
\end{figure}

\begin{figure}[H]
\centering
\label{fig:MgFe6Sn6_relaxed_Low_xz}
\includegraphics[width=0.85\textwidth]{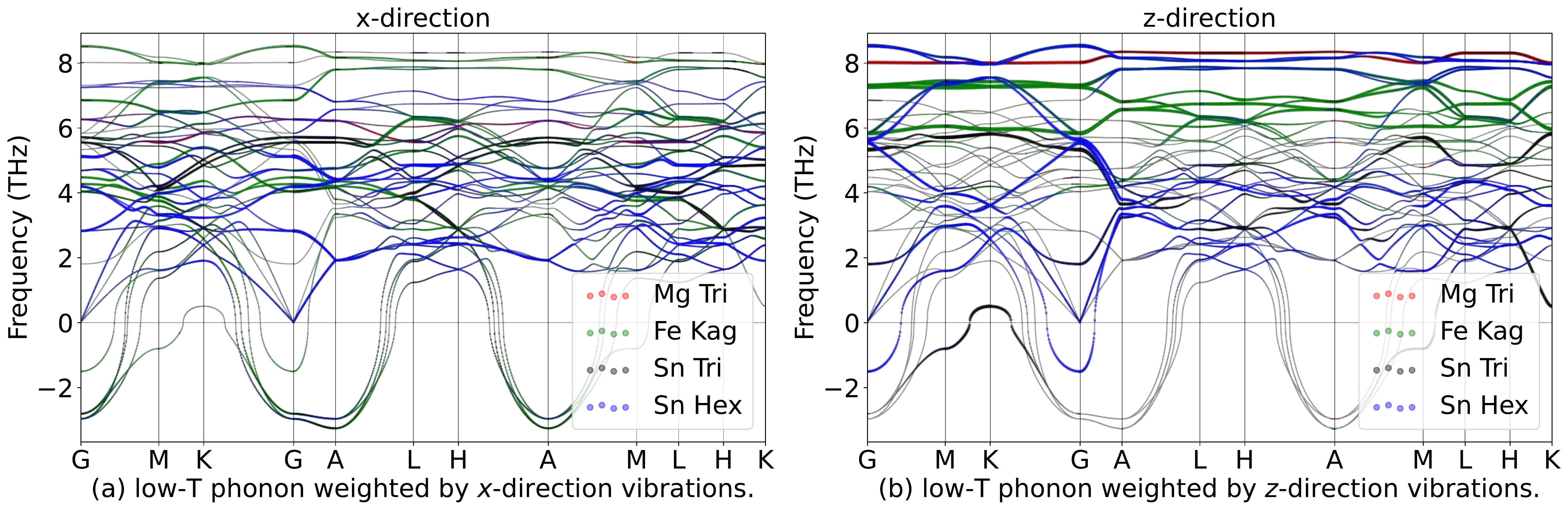}
\hspace{5pt}\label{fig:MgFe6Sn6_relaxed_High_xz}
\includegraphics[width=0.85\textwidth]{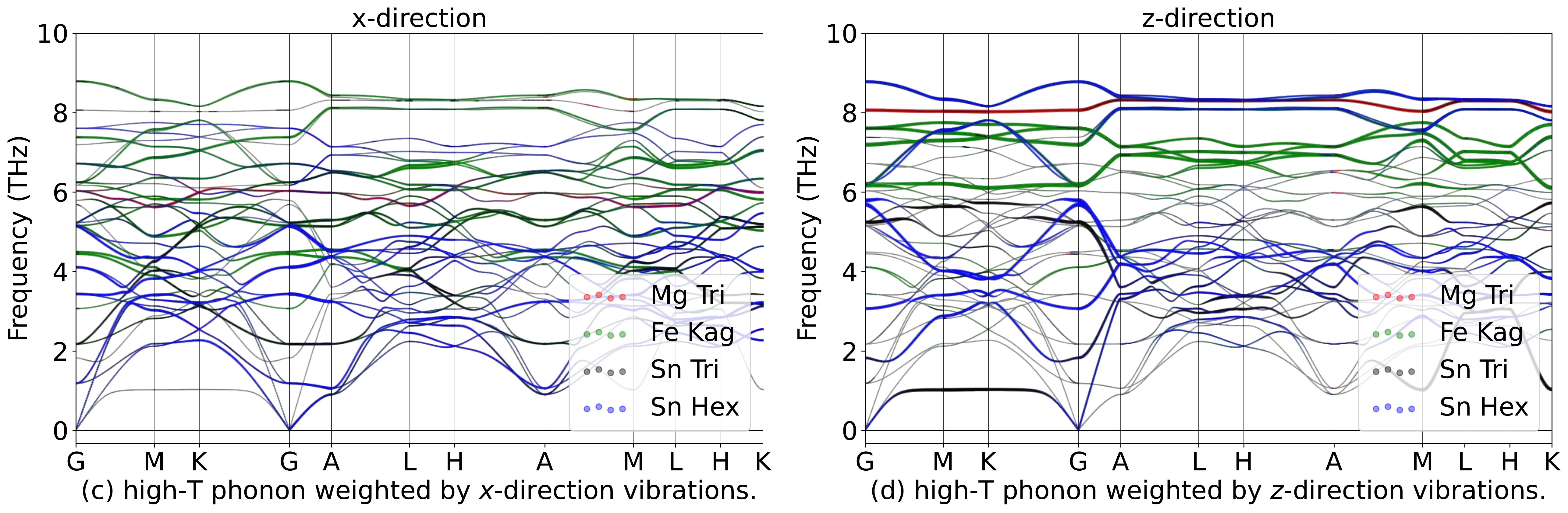}
\caption{Phonon spectrum for MgFe$_6$Sn$_6$in in-plane ($x$) and out-of-plane ($z$) directions in paramagnetic phase.}
\label{fig:MgFe6Sn6phoxyz}
\end{figure}

\begin{table}[htbp]
\global\long\def\arraystretch{1.12}
\captionsetup{width=.95\textwidth}
\caption{IRREPs at high-symmetry points for MgFe$_6$Sn$_6$ phonon spectrum at Low-T.}
\label{tab:191_MgFe6Sn6_Low}
\setlength{\tabcolsep}{1.mm}{
}
\end{table}

\begin{table}[htbp]
\global\long\def\arraystretch{1.12}
\captionsetup{width=.95\textwidth}
\caption{IRREPs at high-symmetry points for MgFe$_6$Sn$_6$ phonon spectrum at High-T.}
\label{tab:191_MgFe6Sn6_High}
\setlength{\tabcolsep}{1.mm}{
%
}
\end{table}
\subsubsection{\label{sms2sec:CaFe6Sn6}\hyperref[smsec:col]{\texorpdfstring{CaFe$_6$Sn$_6$}{}}~{\color{blue}  (High-T stable, Low-T unstable) }}

\begin{table}[htbp]
\global\long\def\arraystretch{1.12}
\captionsetup{width=.85\textwidth}
\caption{Structure parameters of CaFe$_6$Sn$_6$ after relaxation. It contains 476 electrons. }
\label{tab:CaFe6Sn6}
\setlength{\tabcolsep}{4mm}{
%
}
\end{table}

\begin{figure}[htbp]
\centering
\includegraphics[width=0.85\textwidth]{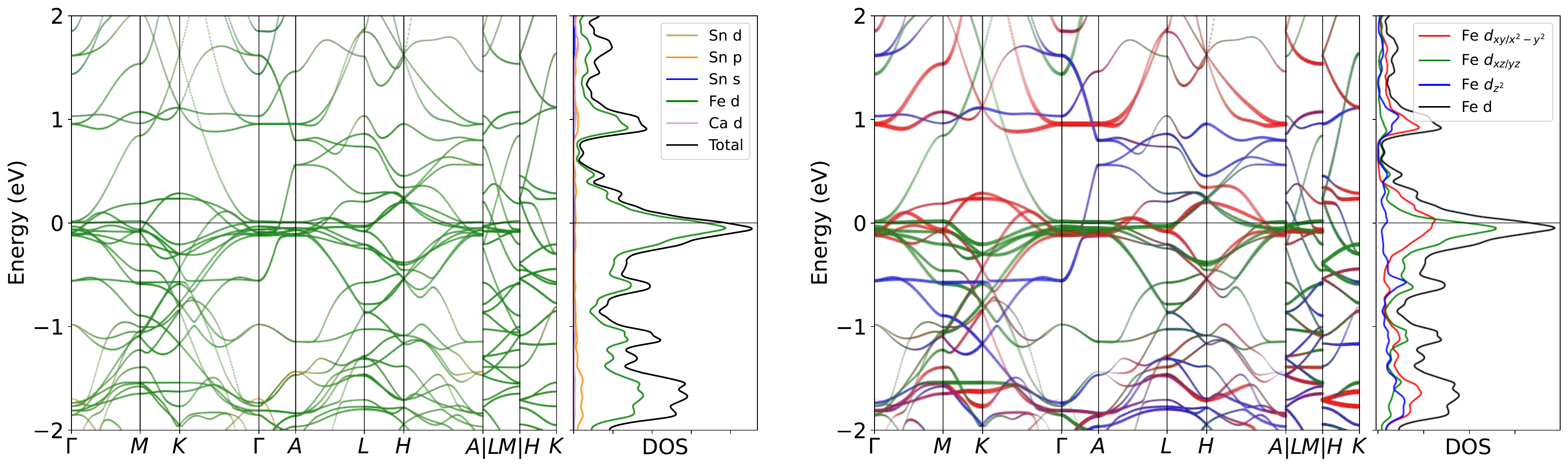}
\caption{Band structure for CaFe$_6$Sn$_6$ without SOC in paramagnetic phase. The projection of Fe $3d$ orbitals is also presented. The band structures clearly reveal the dominating of Fe $3d$ orbitals  near the Fermi level, as well as the pronounced peak.}
\label{fig:CaFe6Sn6band}
\end{figure}

\begin{figure}[H]
\centering
\subfloat[Relaxed phonon at low-T.]{
\label{fig:CaFe6Sn6_relaxed_Low}
\includegraphics[width=0.45\textwidth]{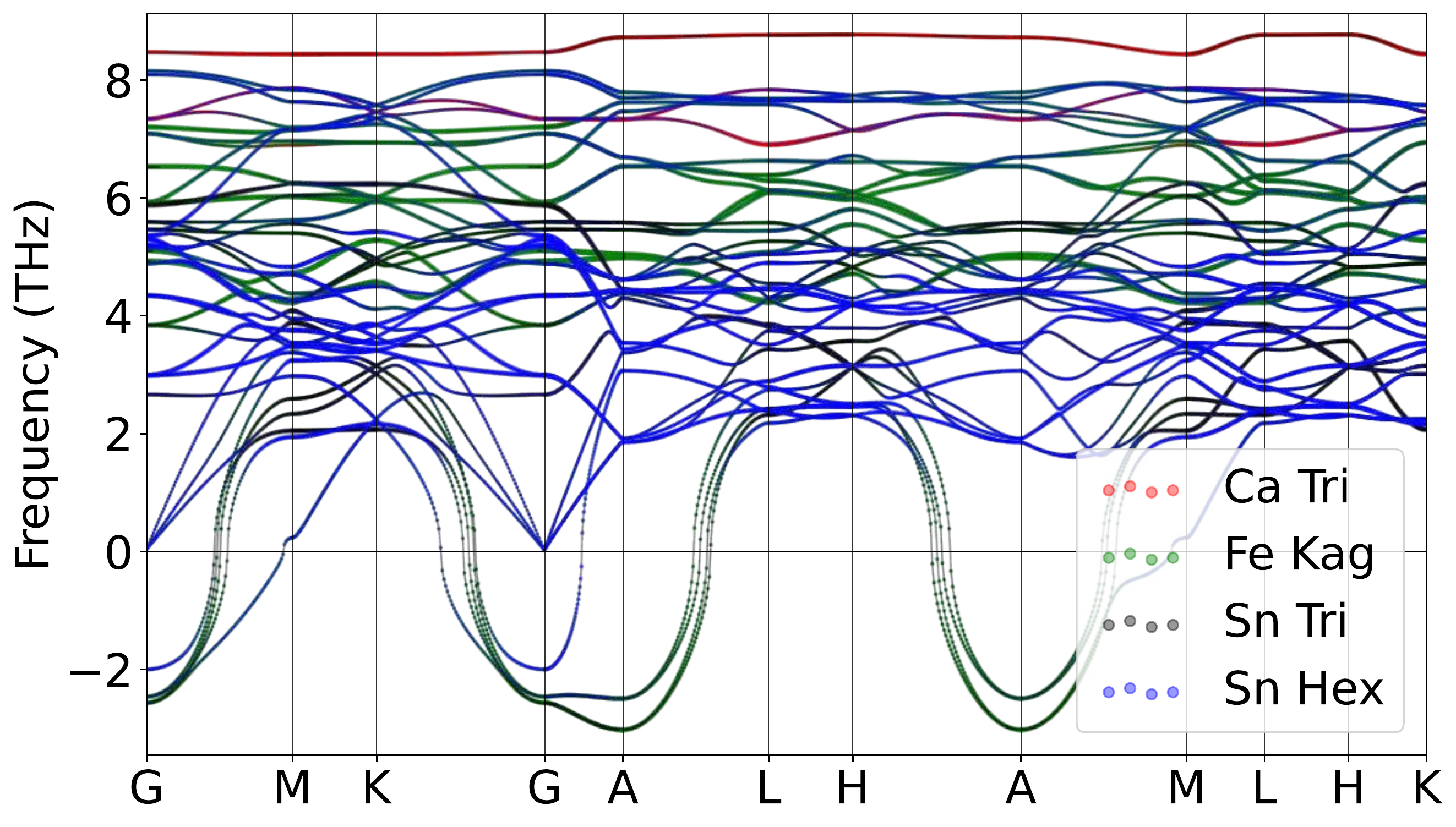}
}
\hspace{5pt}\subfloat[Relaxed phonon at high-T.]{
\label{fig:CaFe6Sn6_relaxed_High}
\includegraphics[width=0.45\textwidth]{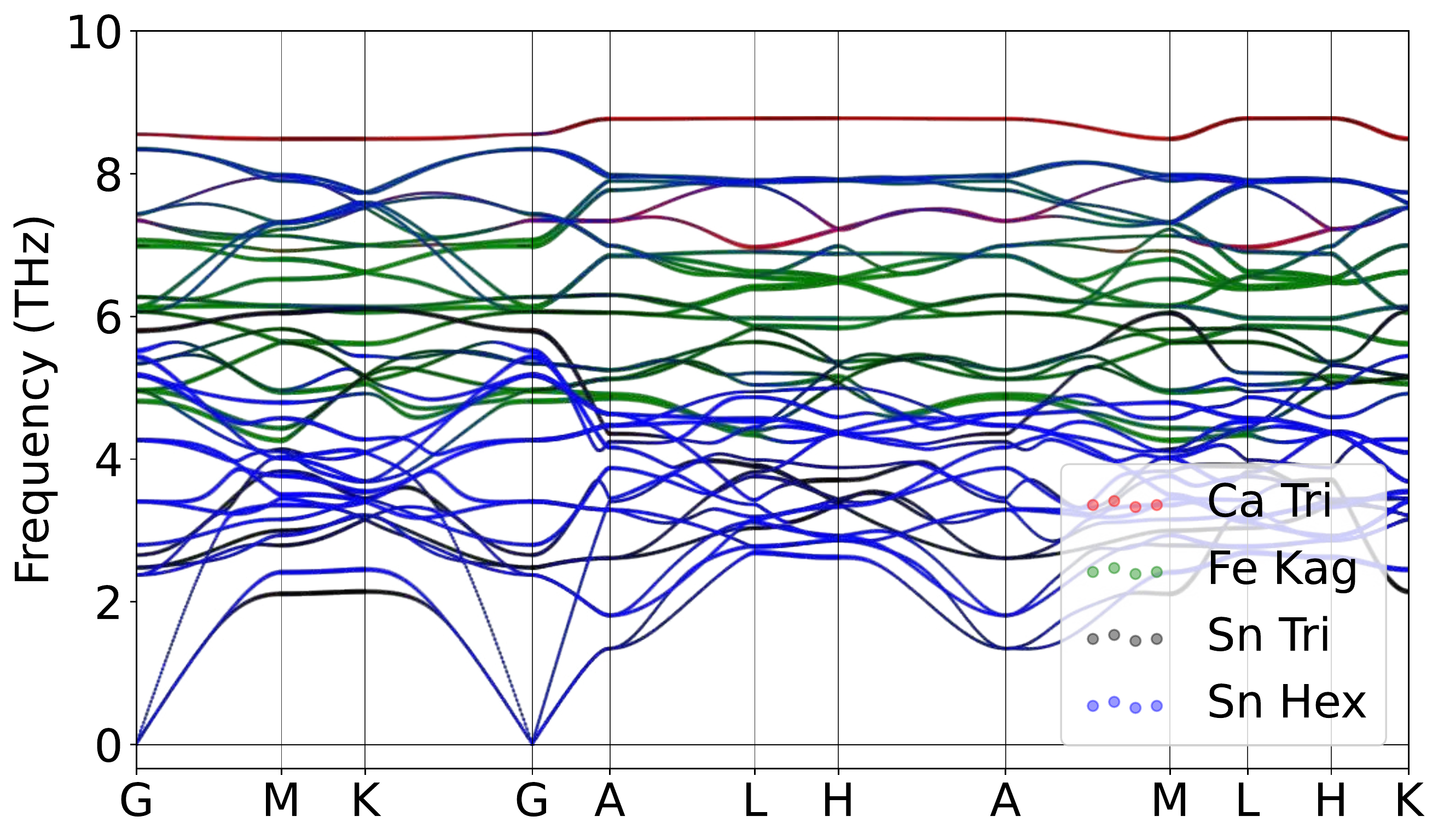}
}
\caption{Atomically projected phonon spectrum for CaFe$_6$Sn$_6$ in paramagnetic phase.}
\label{fig:CaFe6Sn6pho}
\end{figure}

\begin{figure}[htbp]
\centering
\includegraphics[width=0.45\textwidth]{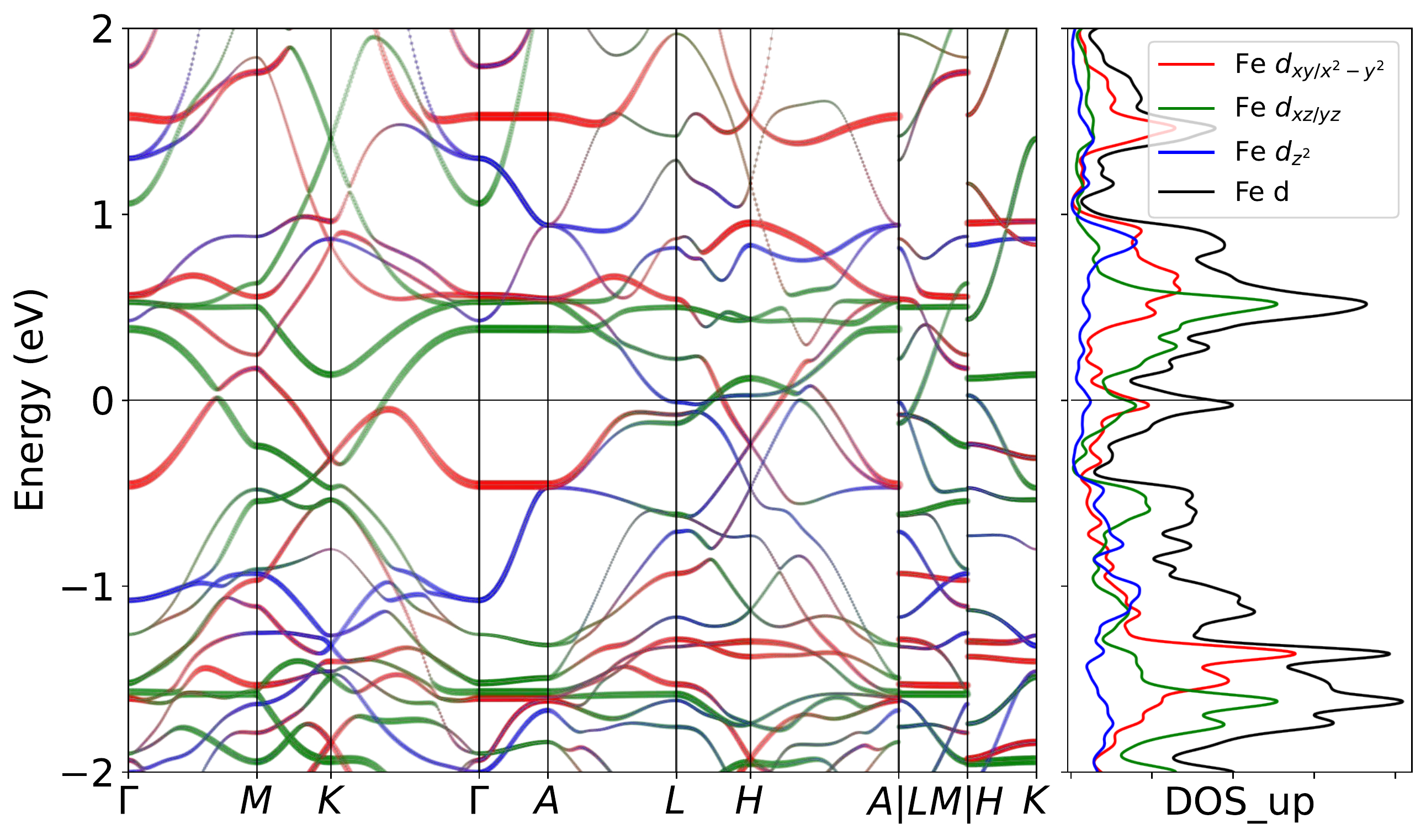}
\includegraphics[width=0.45\textwidth]{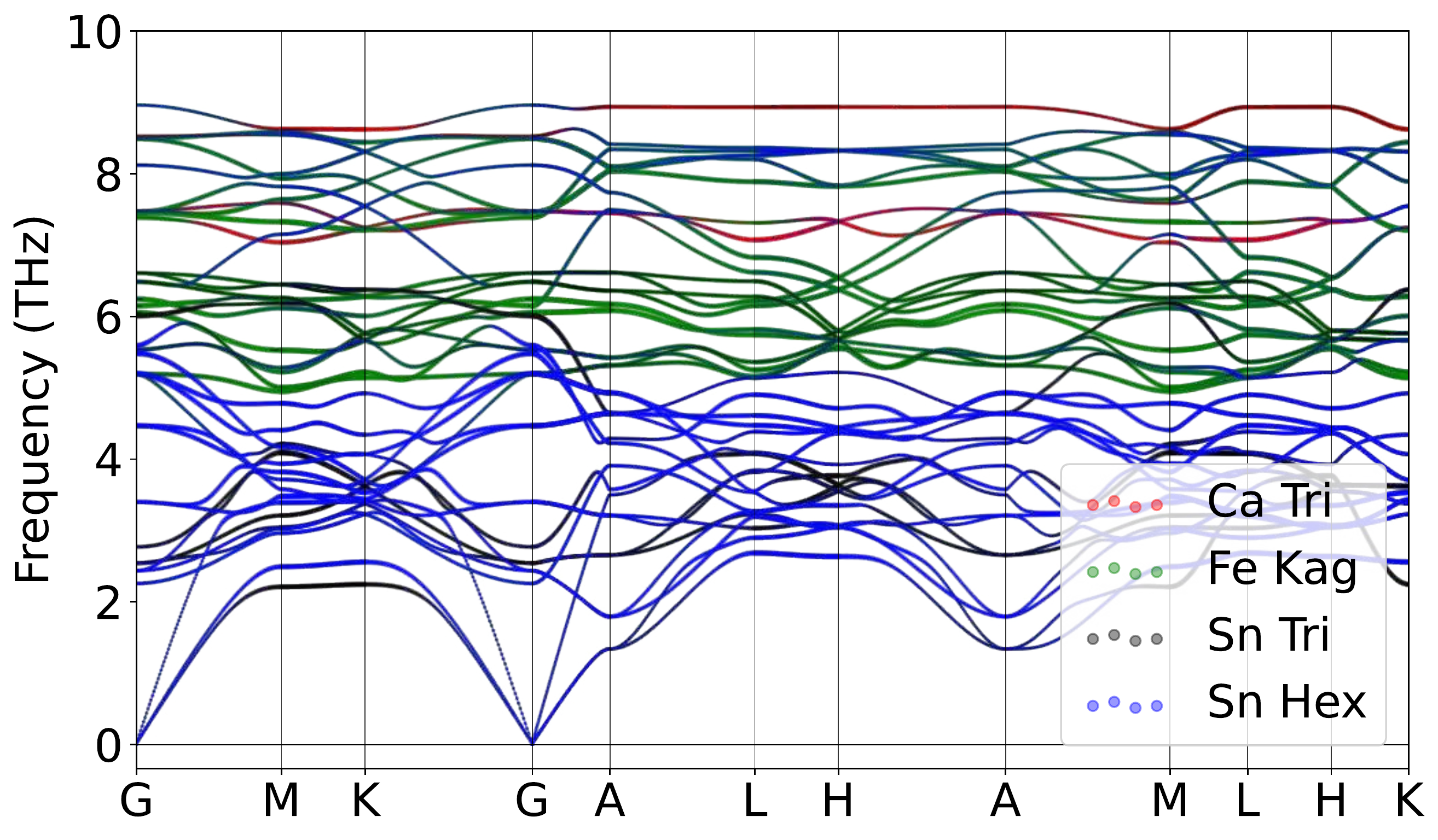}
\caption{Band structure for CaFe$_6$Sn$_6$ with AFM order without SOC for spin (Left)up channel. The projection of Fe $3d$ orbitals is also presented. (Right)Correspondingly, a stable phonon was demonstrated with the collinear magnetic order without Hubbard U at lower temperature regime, weighted by atomic projections.}
\label{fig:CaFe6Sn6mag}
\end{figure}

\begin{figure}[H]
\centering
\label{fig:CaFe6Sn6_relaxed_Low_xz}
\includegraphics[width=0.85\textwidth]{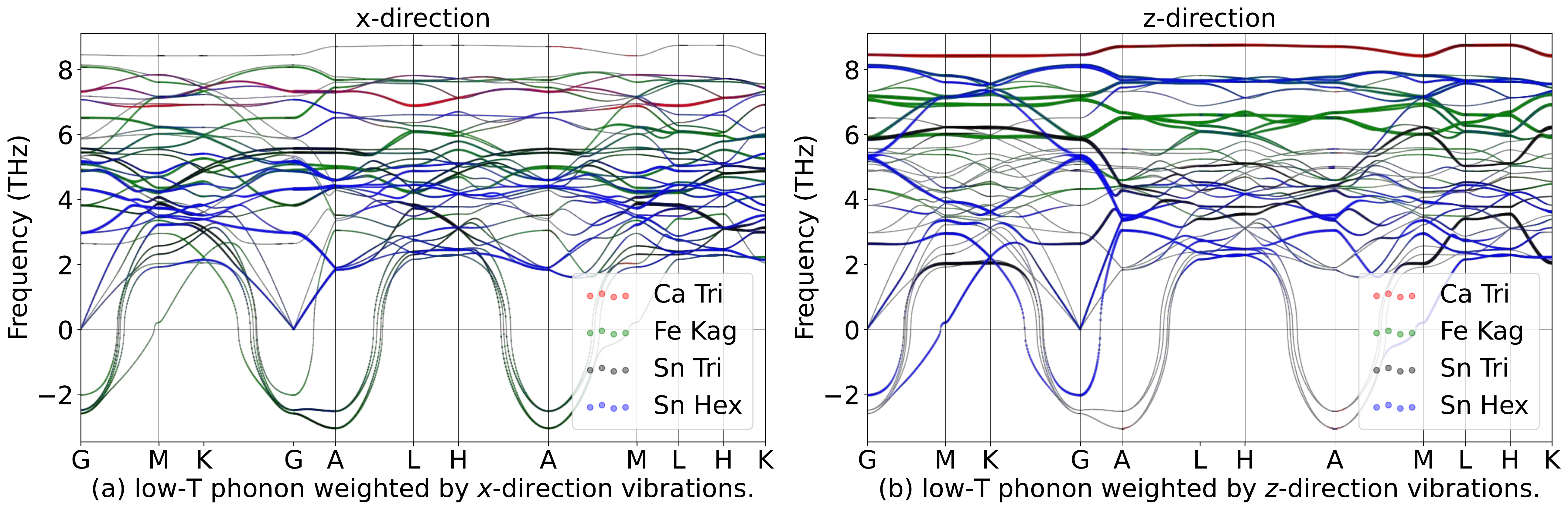}
\hspace{5pt}\label{fig:CaFe6Sn6_relaxed_High_xz}
\includegraphics[width=0.85\textwidth]{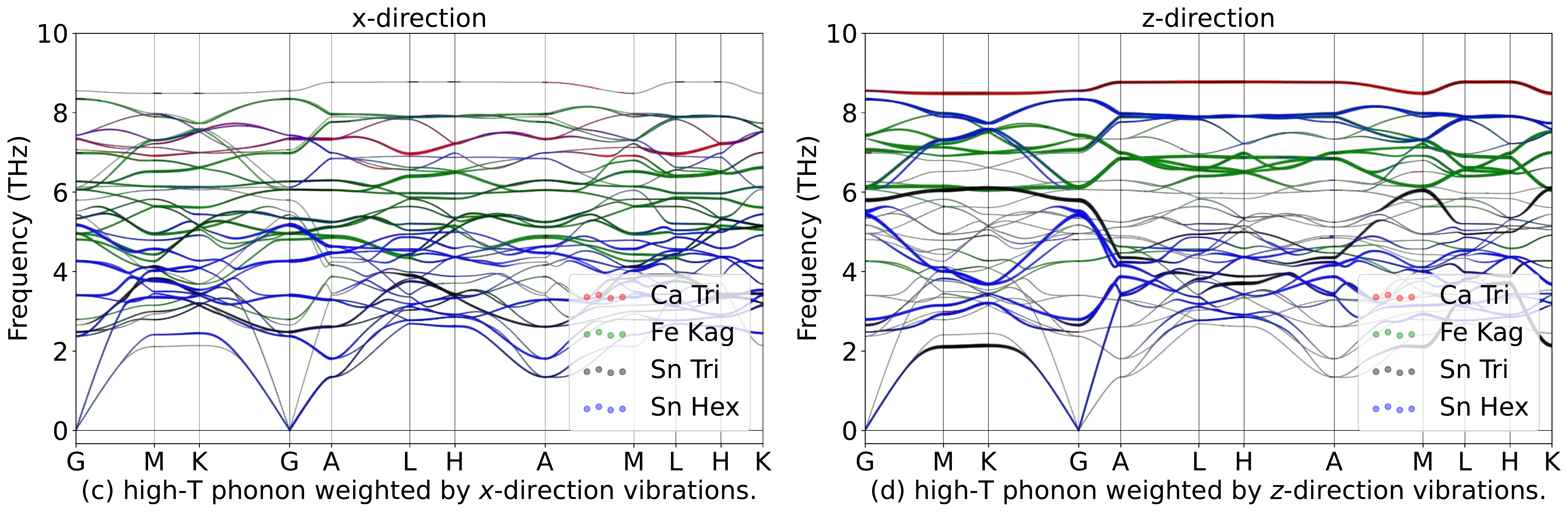}
\caption{Phonon spectrum for CaFe$_6$Sn$_6$in in-plane ($x$) and out-of-plane ($z$) directions in paramagnetic phase.}
\label{fig:CaFe6Sn6phoxyz}
\end{figure}

\begin{table}[htbp]
\global\long\def\arraystretch{1.12}
\captionsetup{width=.95\textwidth}
\caption{IRREPs at high-symmetry points for CaFe$_6$Sn$_6$ phonon spectrum at Low-T.}
\label{tab:191_CaFe6Sn6_Low}
\setlength{\tabcolsep}{1.mm}{
}
\end{table}

\begin{table}[htbp]
\global\long\def\arraystretch{1.12}
\captionsetup{width=.95\textwidth}
\caption{IRREPs at high-symmetry points for CaFe$_6$Sn$_6$ phonon spectrum at High-T.}
\label{tab:191_CaFe6Sn6_High}
\setlength{\tabcolsep}{1.mm}{
%
}
\end{table}
\subsubsection{\label{sms2sec:ScFe6Sn6}\hyperref[smsec:col]{\texorpdfstring{ScFe$_6$Sn$_6$}{}}~{\color{blue}  (High-T stable, Low-T unstable) }}

\begin{table}[htbp]
\global\long\def\arraystretch{1.12}
\captionsetup{width=.85\textwidth}
\caption{Structure parameters of ScFe$_6$Sn$_6$ after relaxation. It contains 477 electrons. }
\label{tab:ScFe6Sn6}
\setlength{\tabcolsep}{4mm}{
%
}
\end{table}

\begin{figure}[htbp]
\centering
\includegraphics[width=0.85\textwidth]{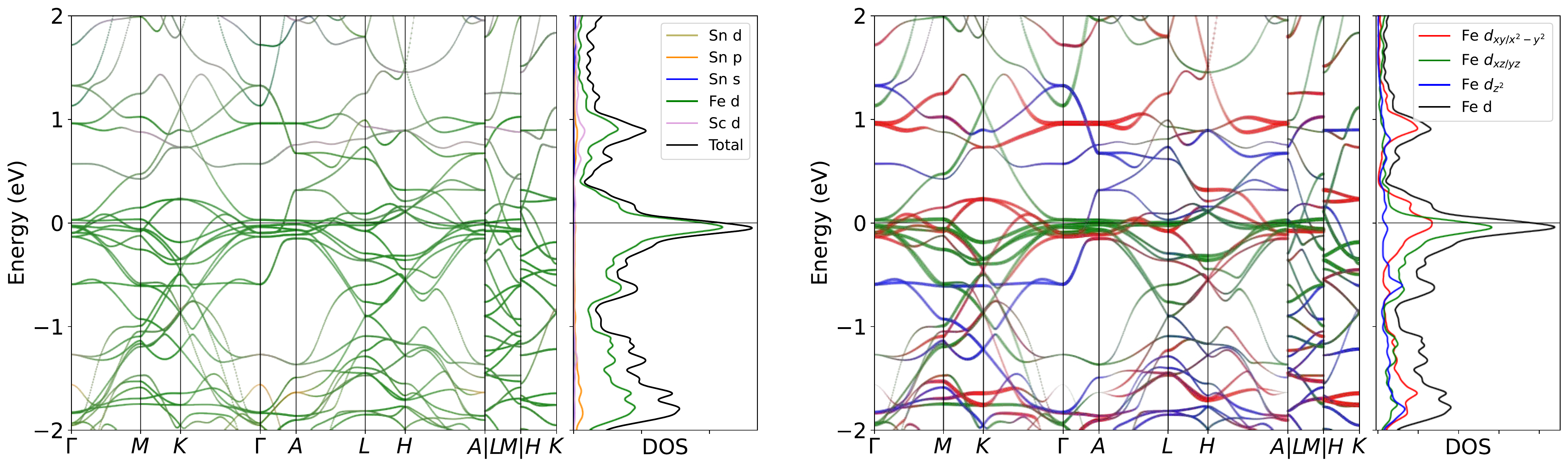}
\caption{Band structure for ScFe$_6$Sn$_6$ without SOC in paramagnetic phase. The projection of Fe $3d$ orbitals is also presented. The band structures clearly reveal the dominating of Fe $3d$ orbitals  near the Fermi level, as well as the pronounced peak.}
\label{fig:ScFe6Sn6band}
\end{figure}

\begin{figure}[H]
\centering
\subfloat[Relaxed phonon at low-T.]{
\label{fig:ScFe6Sn6_relaxed_Low}
\includegraphics[width=0.45\textwidth]{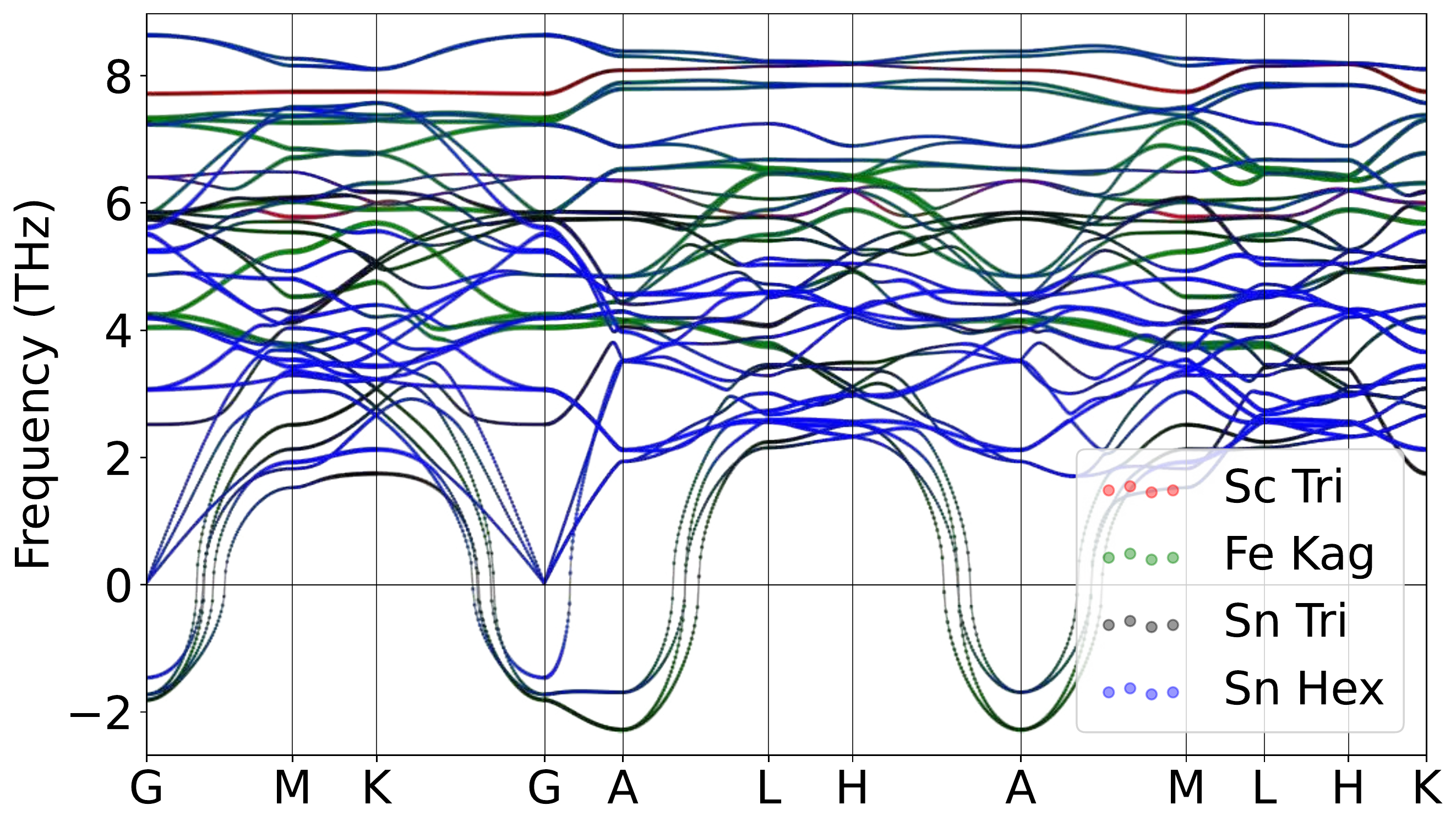}
}
\hspace{5pt}\subfloat[Relaxed phonon at high-T.]{
\label{fig:ScFe6Sn6_relaxed_High}
\includegraphics[width=0.45\textwidth]{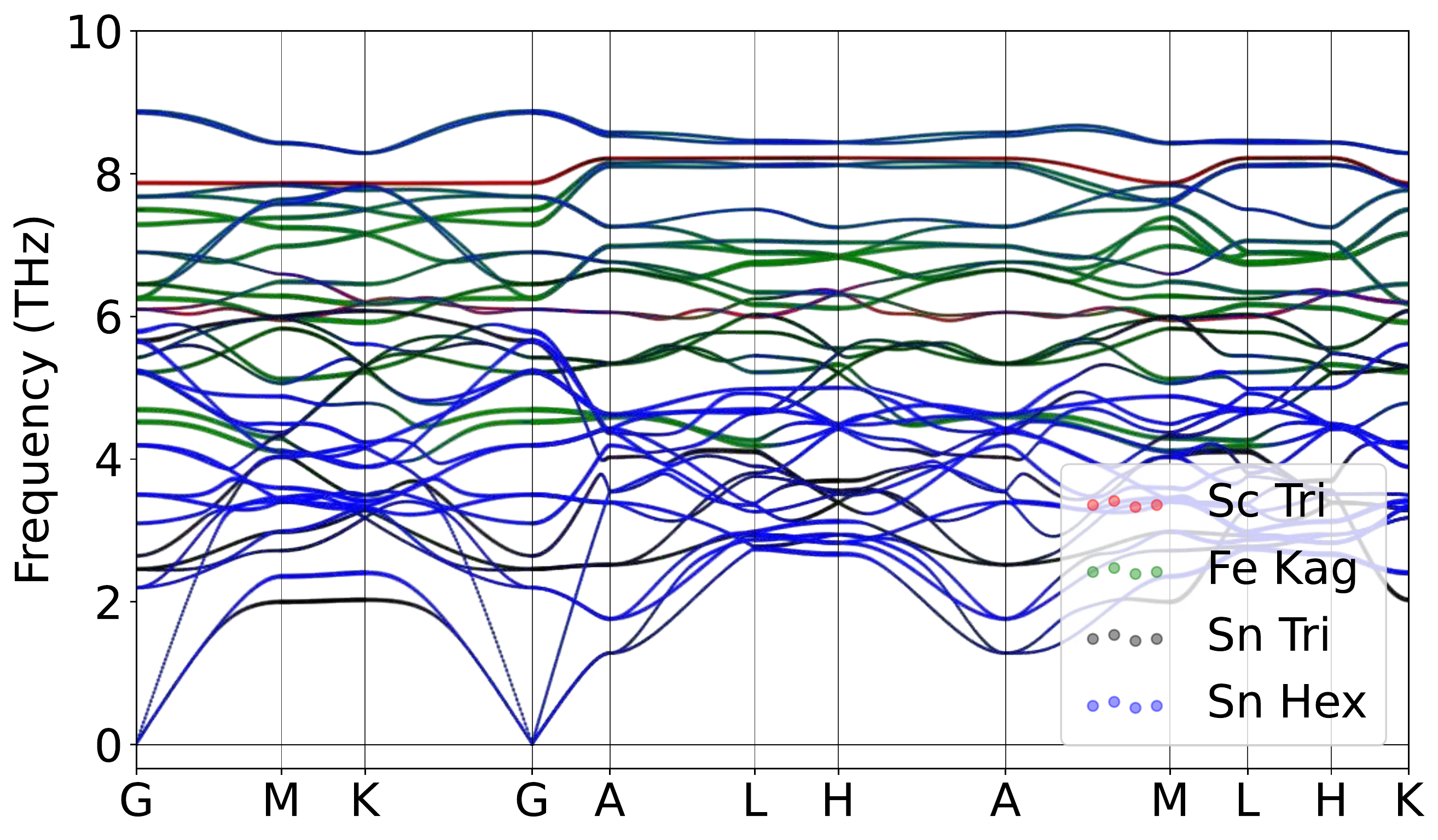}
}
\caption{Atomically projected phonon spectrum for ScFe$_6$Sn$_6$ in paramagnetic phase.}
\label{fig:ScFe6Sn6pho}
\end{figure}

\begin{figure}[htbp]
\centering
\includegraphics[width=0.45\textwidth]{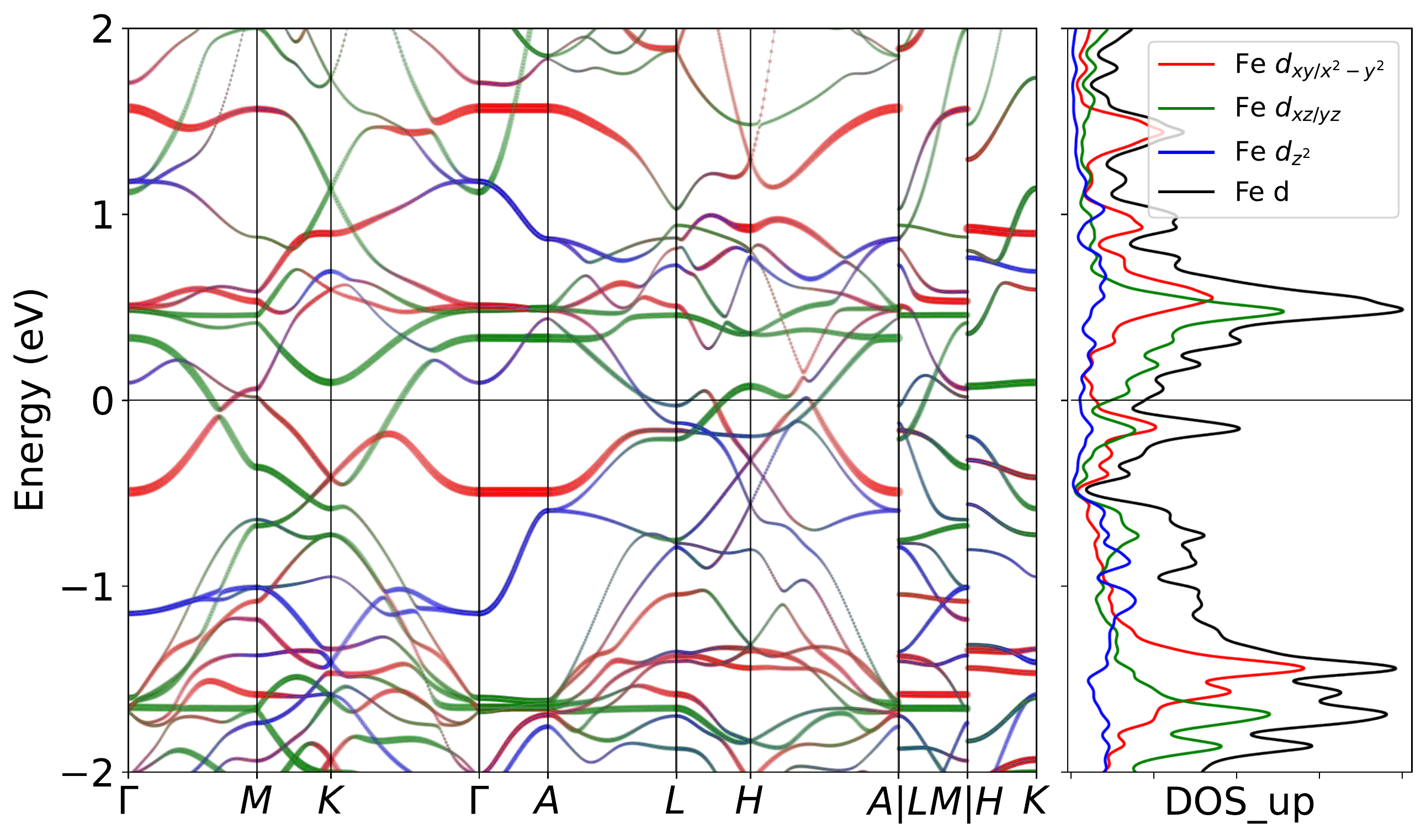}
\includegraphics[width=0.45\textwidth]{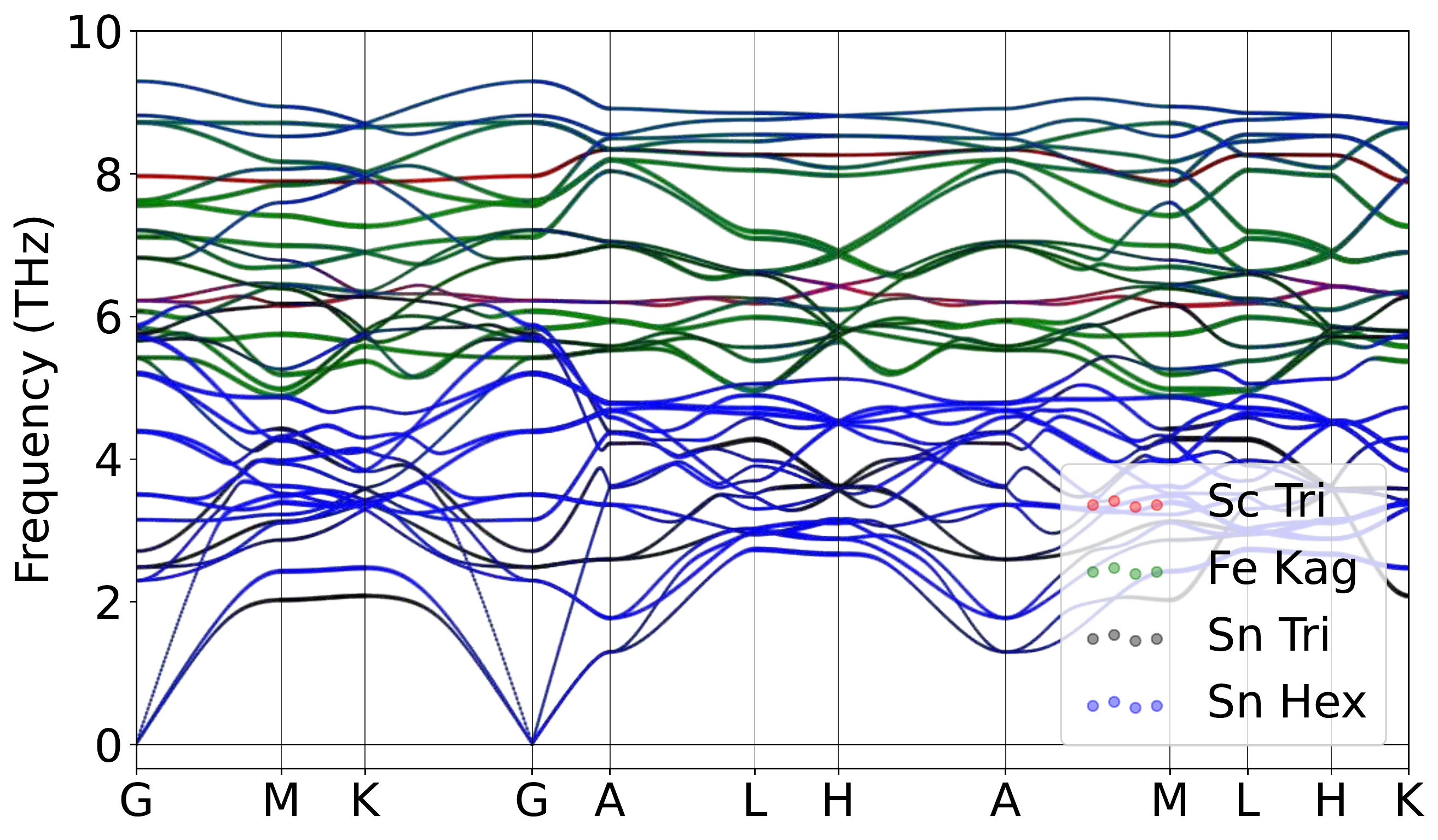}
\caption{Band structure for ScFe$_6$Sn$_6$ with AFM order without SOC for spin (Left)up channel. The projection of Fe $3d$ orbitals is also presented. (Right)Correspondingly, a stable phonon was demonstrated with the collinear magnetic order without Hubbard U at lower temperature regime, weighted by atomic projections.}
\label{fig:ScFe6Sn6mag}
\end{figure}

\begin{figure}[H]
\centering
\label{fig:ScFe6Sn6_relaxed_Low_xz}
\includegraphics[width=0.85\textwidth]{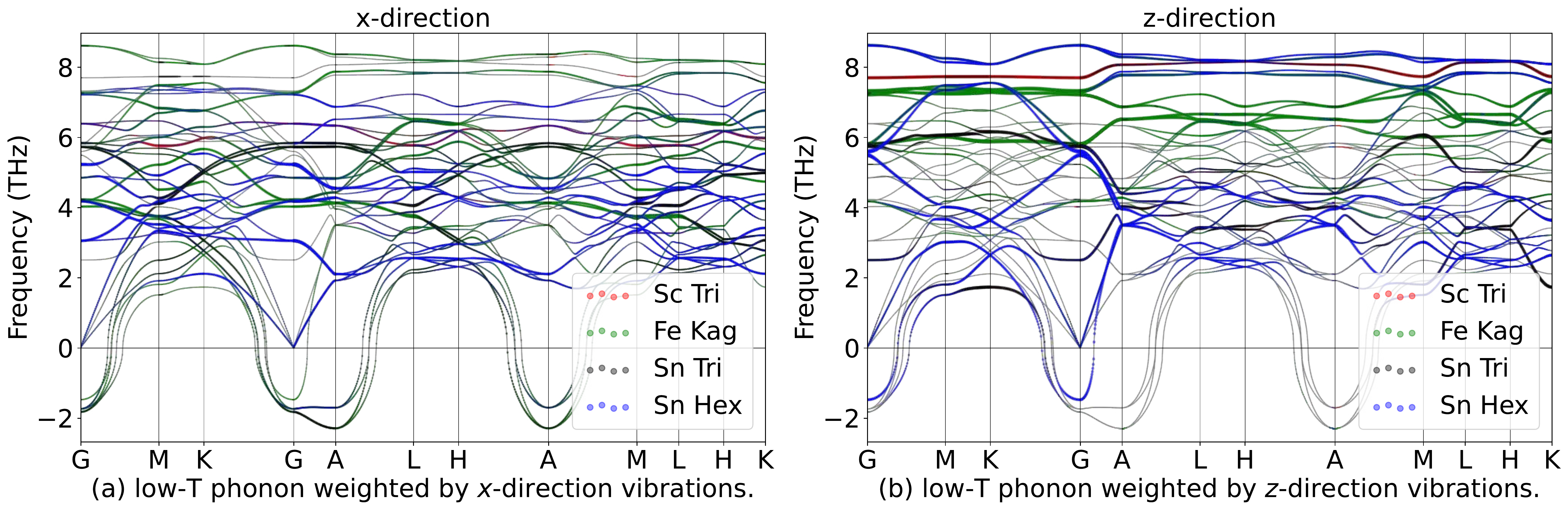}
\hspace{5pt}\label{fig:ScFe6Sn6_relaxed_High_xz}
\includegraphics[width=0.85\textwidth]{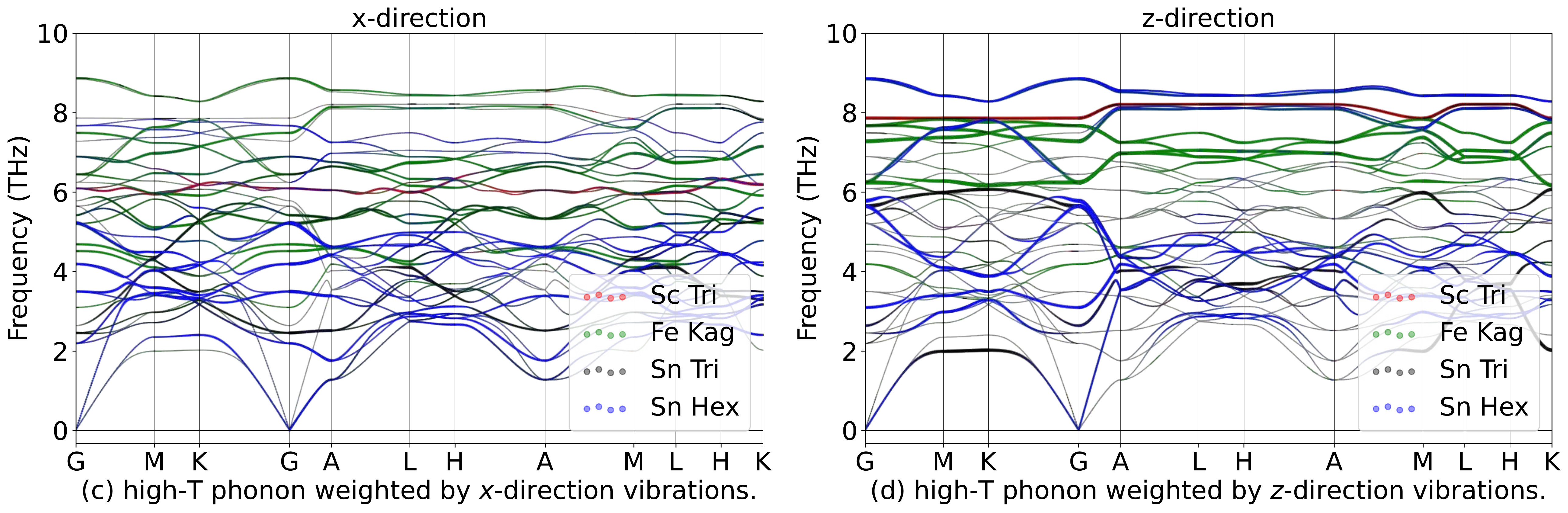}
\caption{Phonon spectrum for ScFe$_6$Sn$_6$in in-plane ($x$) and out-of-plane ($z$) directions in paramagnetic phase.}
\label{fig:ScFe6Sn6phoxyz}
\end{figure}

\begin{table}[htbp]
\global\long\def\arraystretch{1.12}
\captionsetup{width=.95\textwidth}
\caption{IRREPs at high-symmetry points for ScFe$_6$Sn$_6$ phonon spectrum at Low-T.}
\label{tab:191_ScFe6Sn6_Low}
\setlength{\tabcolsep}{1.mm}{
}
\end{table}

\begin{table}[htbp]
\global\long\def\arraystretch{1.12}
\captionsetup{width=.95\textwidth}
\caption{IRREPs at high-symmetry points for ScFe$_6$Sn$_6$ phonon spectrum at High-T.}
\label{tab:191_ScFe6Sn6_High}
\setlength{\tabcolsep}{1.mm}{
%
}
\end{table}
\subsubsection{\label{sms2sec:TiFe6Sn6}\hyperref[smsec:col]{\texorpdfstring{TiFe$_6$Sn$_6$}{}}~{\color{blue}  (High-T stable, Low-T unstable) }}

\begin{table}[htbp]
\global\long\def\arraystretch{1.12}
\captionsetup{width=.85\textwidth}
\caption{Structure parameters of TiFe$_6$Sn$_6$ after relaxation. It contains 478 electrons. }
\label{tab:TiFe6Sn6}
\setlength{\tabcolsep}{4mm}{
%
}
\end{table}

\begin{figure}[htbp]
\centering
\includegraphics[width=0.85\textwidth]{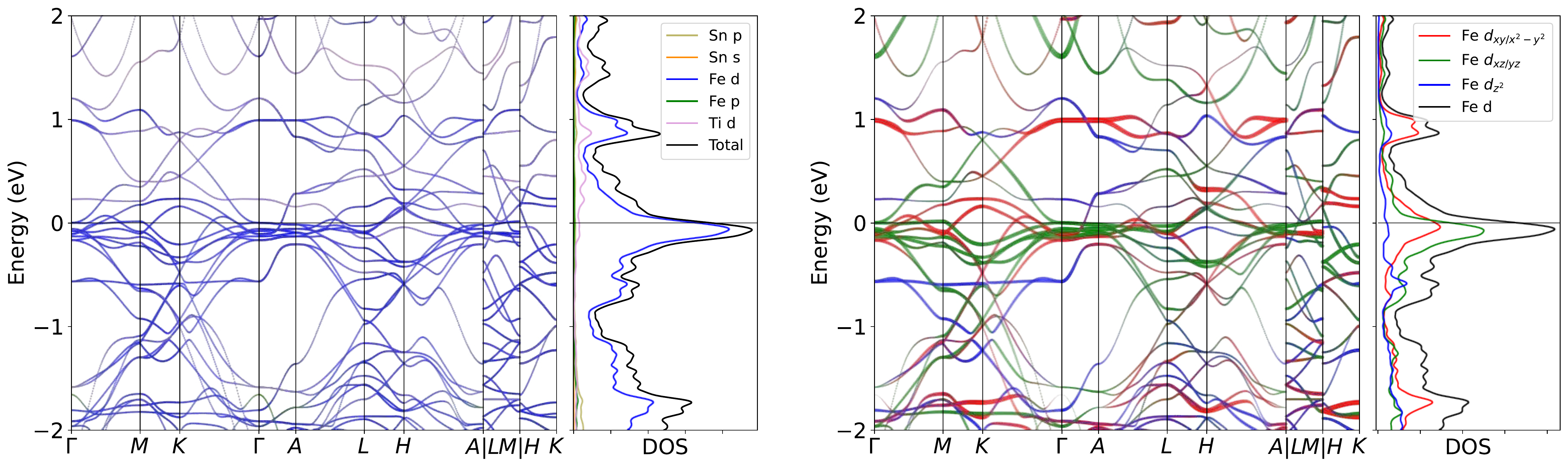}
\caption{Band structure for TiFe$_6$Sn$_6$ without SOC in paramagnetic phase. The projection of Fe $3d$ orbitals is also presented. The band structures clearly reveal the dominating of Fe $3d$ orbitals  near the Fermi level, as well as the pronounced peak.}
\label{fig:TiFe6Sn6band}
\end{figure}

\begin{figure}[H]
\centering
\subfloat[Relaxed phonon at low-T.]{
\label{fig:TiFe6Sn6_relaxed_Low}
\includegraphics[width=0.45\textwidth]{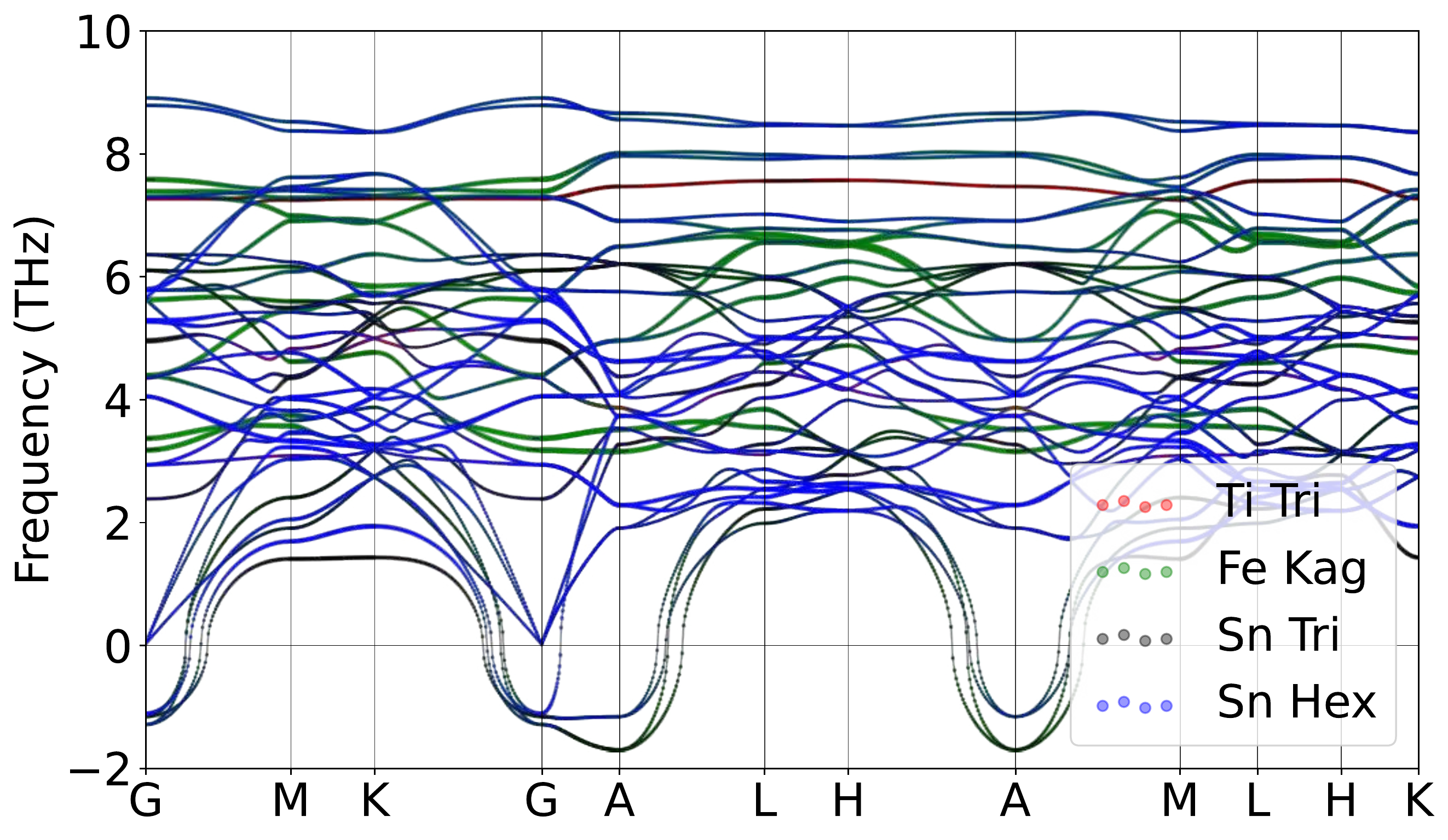}
}
\hspace{5pt}\subfloat[Relaxed phonon at high-T.]{
\label{fig:TiFe6Sn6_relaxed_High}
\includegraphics[width=0.45\textwidth]{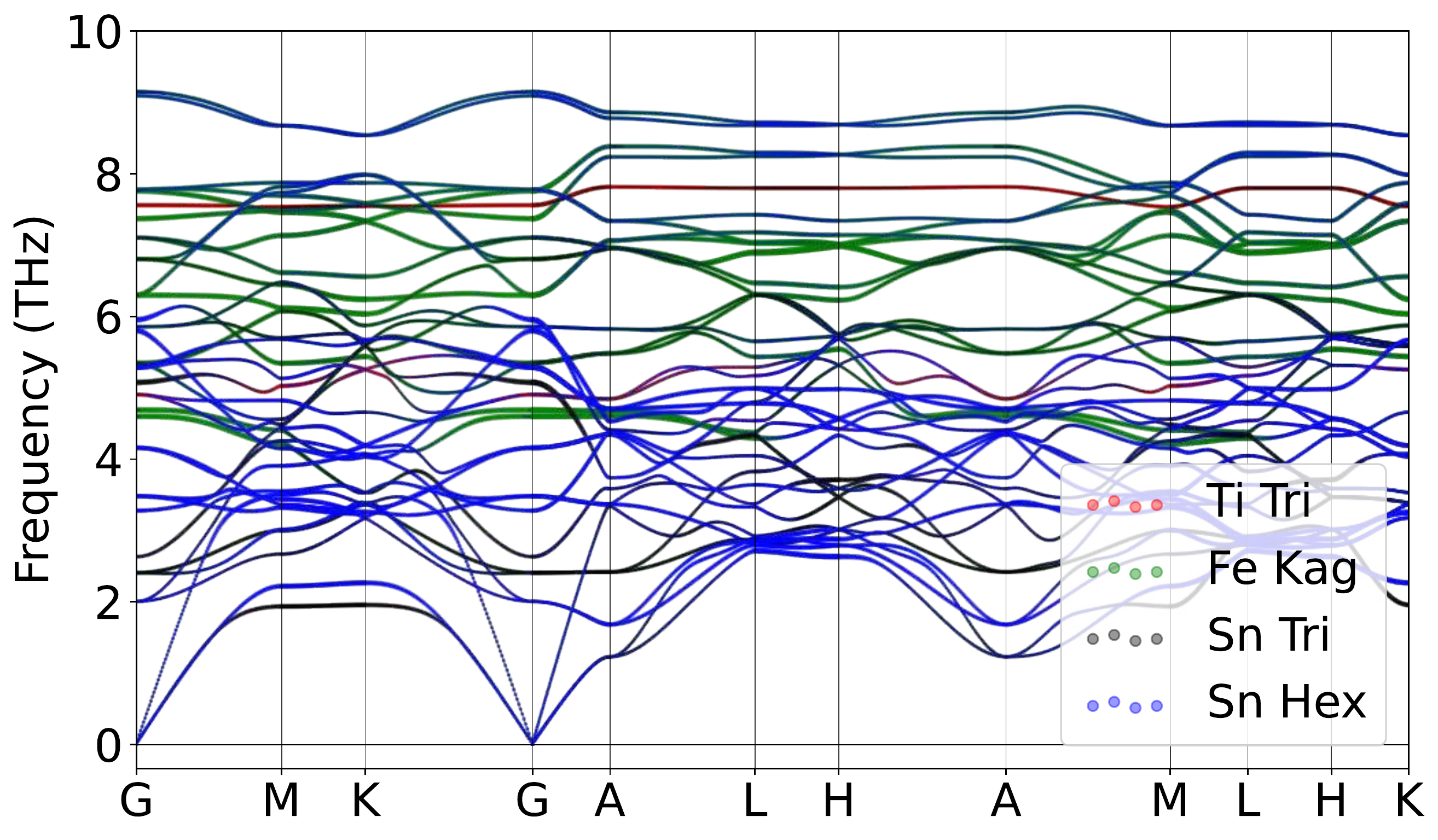}
}
\caption{Atomically projected phonon spectrum for TiFe$_6$Sn$_6$ in paramagnetic phase.}
\label{fig:TiFe6Sn6pho}
\end{figure}

\begin{figure}[htbp]
\centering
\includegraphics[width=0.45\textwidth]{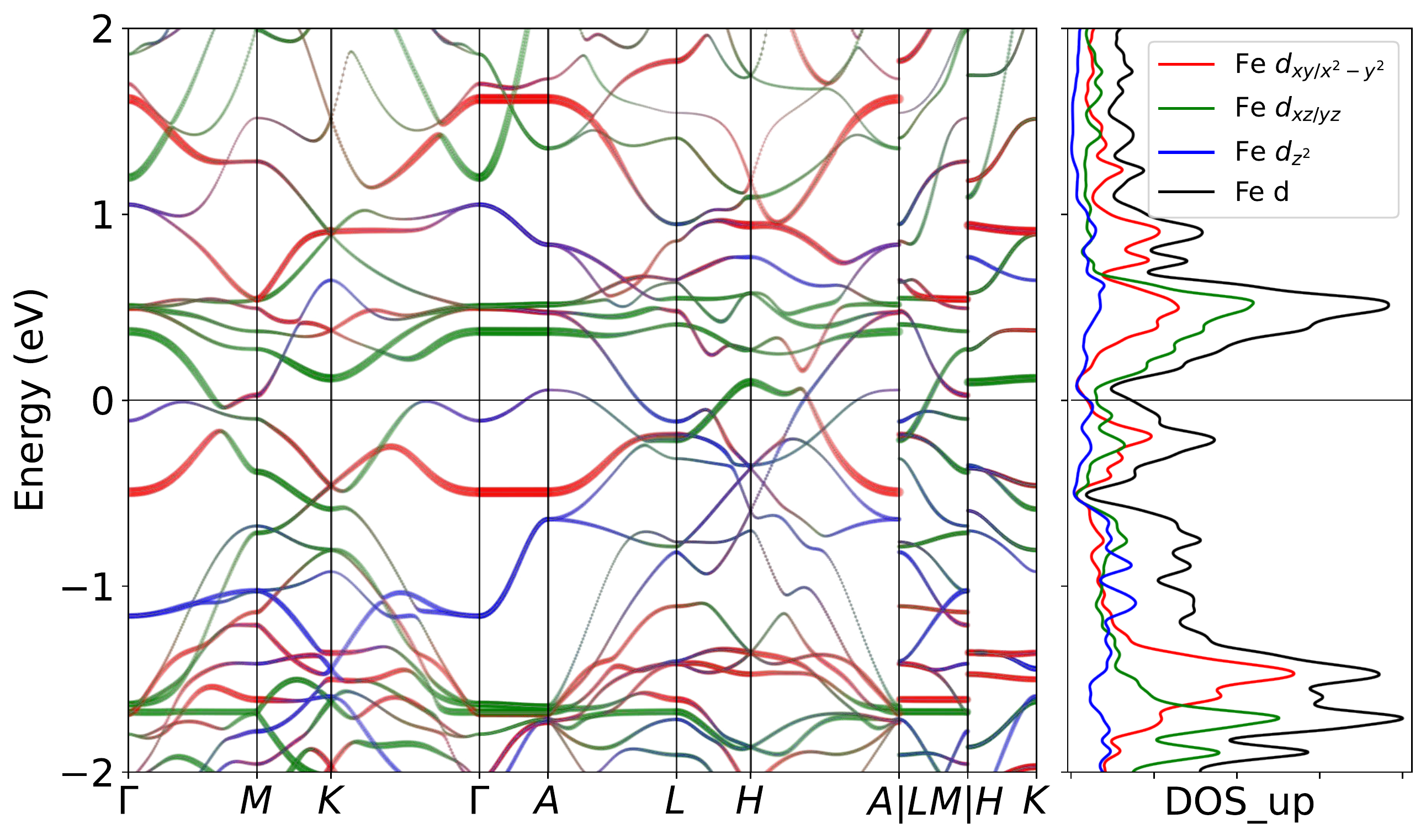}
\includegraphics[width=0.45\textwidth]{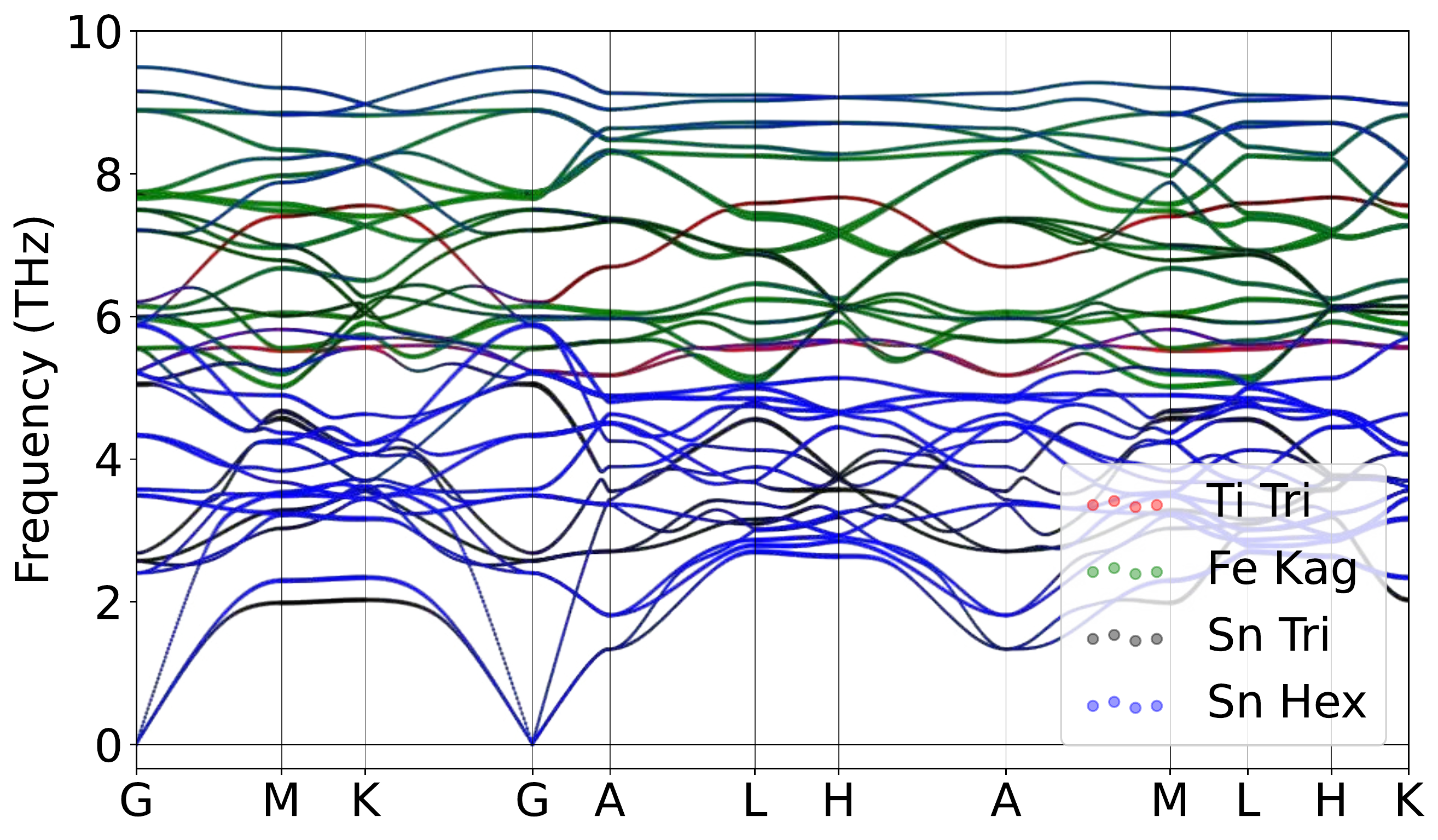}
\caption{Band structure for TiFe$_6$Sn$_6$ with AFM order without SOC for spin (Left)up channel. The projection of Fe $3d$ orbitals is also presented. (Right)Correspondingly, a stable phonon was demonstrated with the collinear magnetic order without Hubbard U at lower temperature regime, weighted by atomic projections.}
\label{fig:TiFe6Sn6mag}
\end{figure}

\begin{figure}[H]
\centering
\label{fig:TiFe6Sn6_relaxed_Low_xz}
\includegraphics[width=0.85\textwidth]{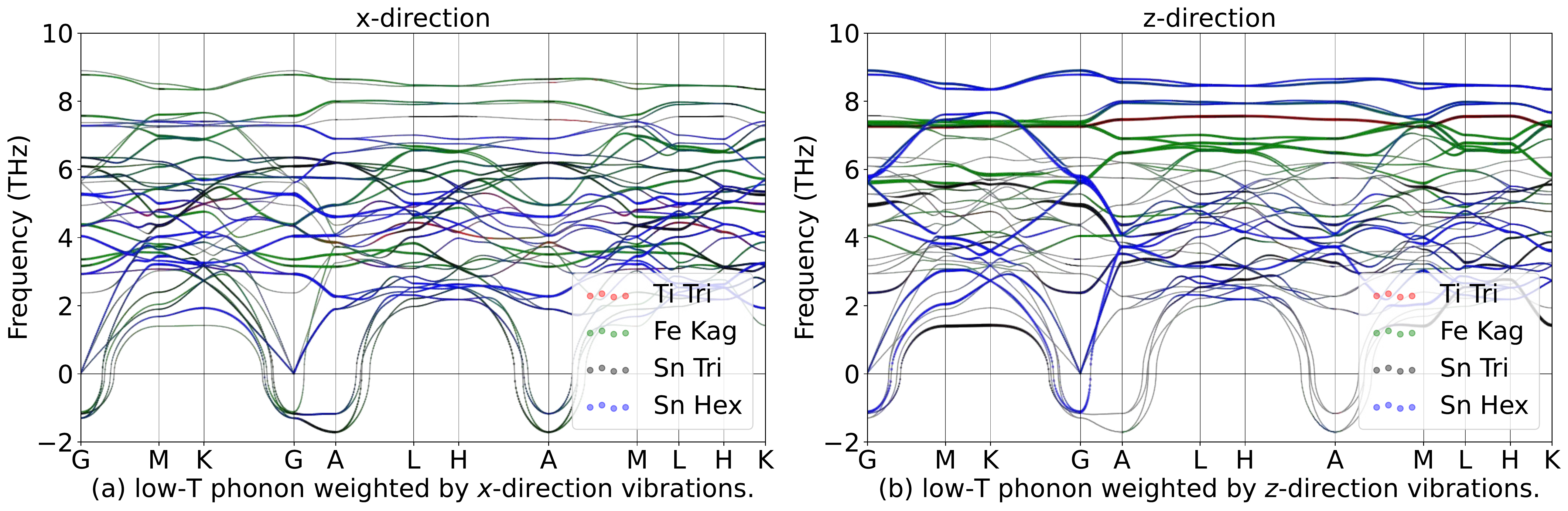}
\hspace{5pt}\label{fig:TiFe6Sn6_relaxed_High_xz}
\includegraphics[width=0.85\textwidth]{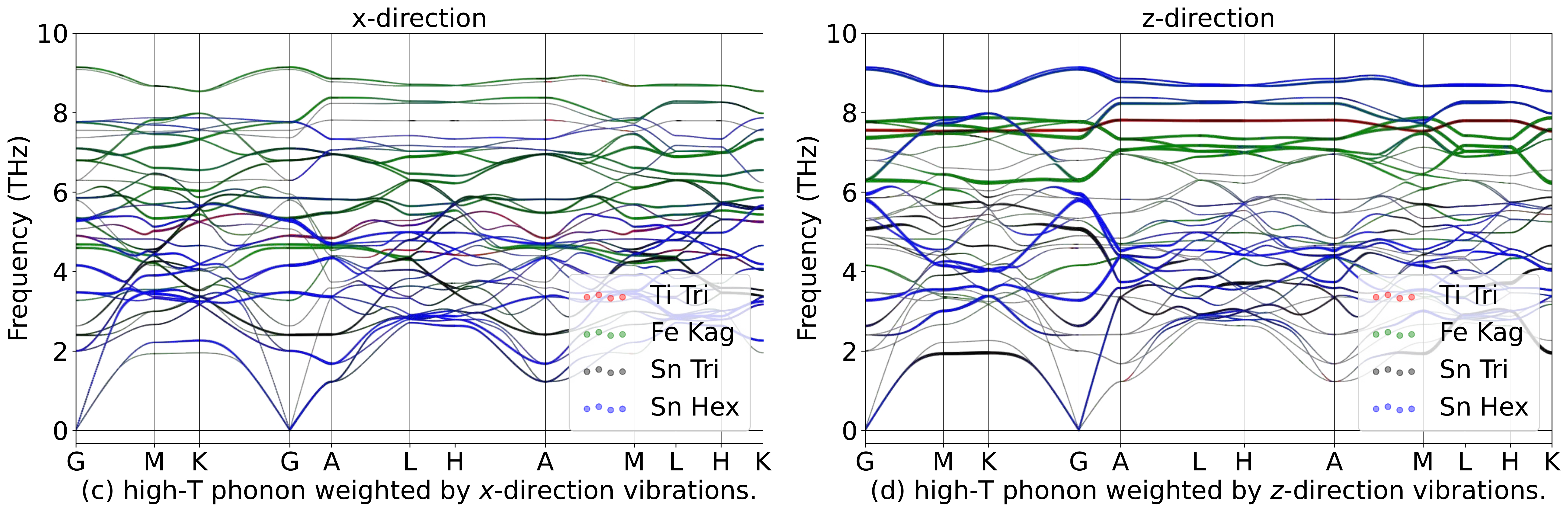}
\caption{Phonon spectrum for TiFe$_6$Sn$_6$in in-plane ($x$) and out-of-plane ($z$) directions in paramagnetic phase.}
\label{fig:TiFe6Sn6phoxyz}
\end{figure}

\begin{table}[htbp]
\global\long\def\arraystretch{1.12}
\captionsetup{width=.95\textwidth}
\caption{IRREPs at high-symmetry points for TiFe$_6$Sn$_6$ phonon spectrum at Low-T.}
\label{tab:191_TiFe6Sn6_Low}
\setlength{\tabcolsep}{1.mm}{
}
\end{table}

\begin{table}[htbp]
\global\long\def\arraystretch{1.12}
\captionsetup{width=.95\textwidth}
\caption{IRREPs at high-symmetry points for TiFe$_6$Sn$_6$ phonon spectrum at High-T.}
\label{tab:191_TiFe6Sn6_High}
\setlength{\tabcolsep}{1.mm}{
%
}
\end{table}
\subsubsection{\label{sms2sec:YFe6Sn6}\hyperref[smsec:col]{\texorpdfstring{YFe$_6$Sn$_6$}{}}~{\color{blue}  (High-T stable, Low-T unstable) }}

\begin{table}[htbp]
\global\long\def\arraystretch{1.12}
\captionsetup{width=.85\textwidth}
\caption{Structure parameters of YFe$_6$Sn$_6$ after relaxation. It contains 495 electrons. }
\label{tab:YFe6Sn6}
\setlength{\tabcolsep}{4mm}{
%
}
\end{table}

\begin{figure}[htbp]
\centering
\includegraphics[width=0.85\textwidth]{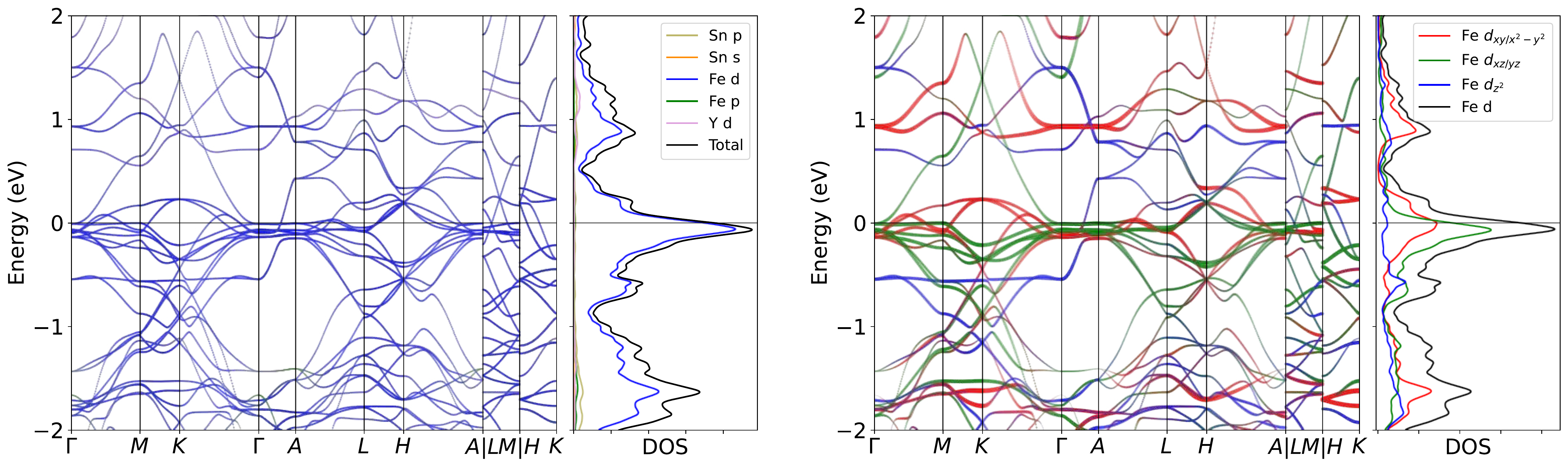}
\caption{Band structure for YFe$_6$Sn$_6$ without SOC in paramagnetic phase. The projection of Fe $3d$ orbitals is also presented. The band structures clearly reveal the dominating of Fe $3d$ orbitals  near the Fermi level, as well as the pronounced peak.}
\label{fig:YFe6Sn6band}
\end{figure}

\begin{figure}[H]
\centering
\subfloat[Relaxed phonon at low-T.]{
\label{fig:YFe6Sn6_relaxed_Low}
\includegraphics[width=0.45\textwidth]{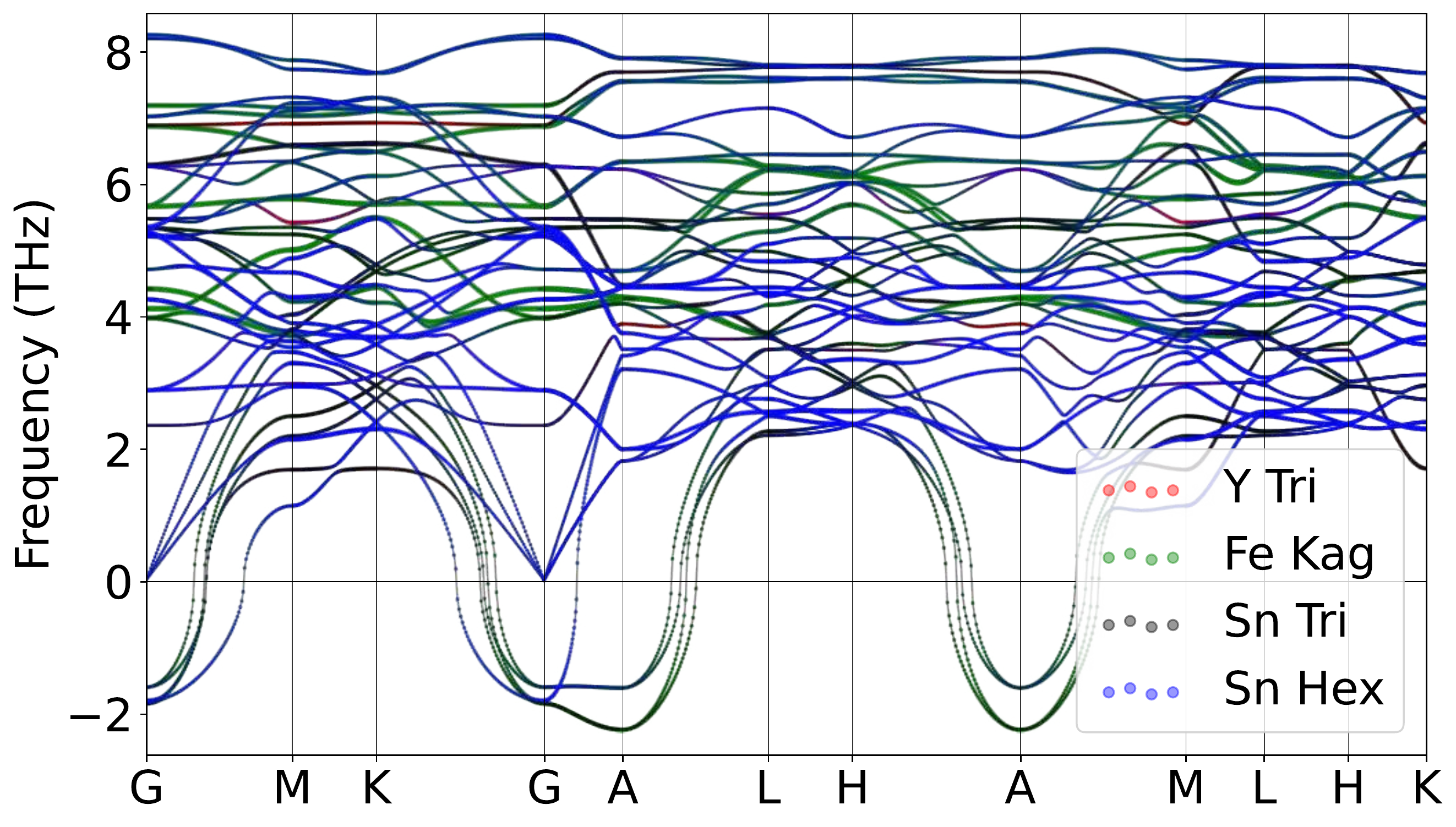}
}
\hspace{5pt}\subfloat[Relaxed phonon at high-T.]{
\label{fig:YFe6Sn6_relaxed_High}
\includegraphics[width=0.45\textwidth]{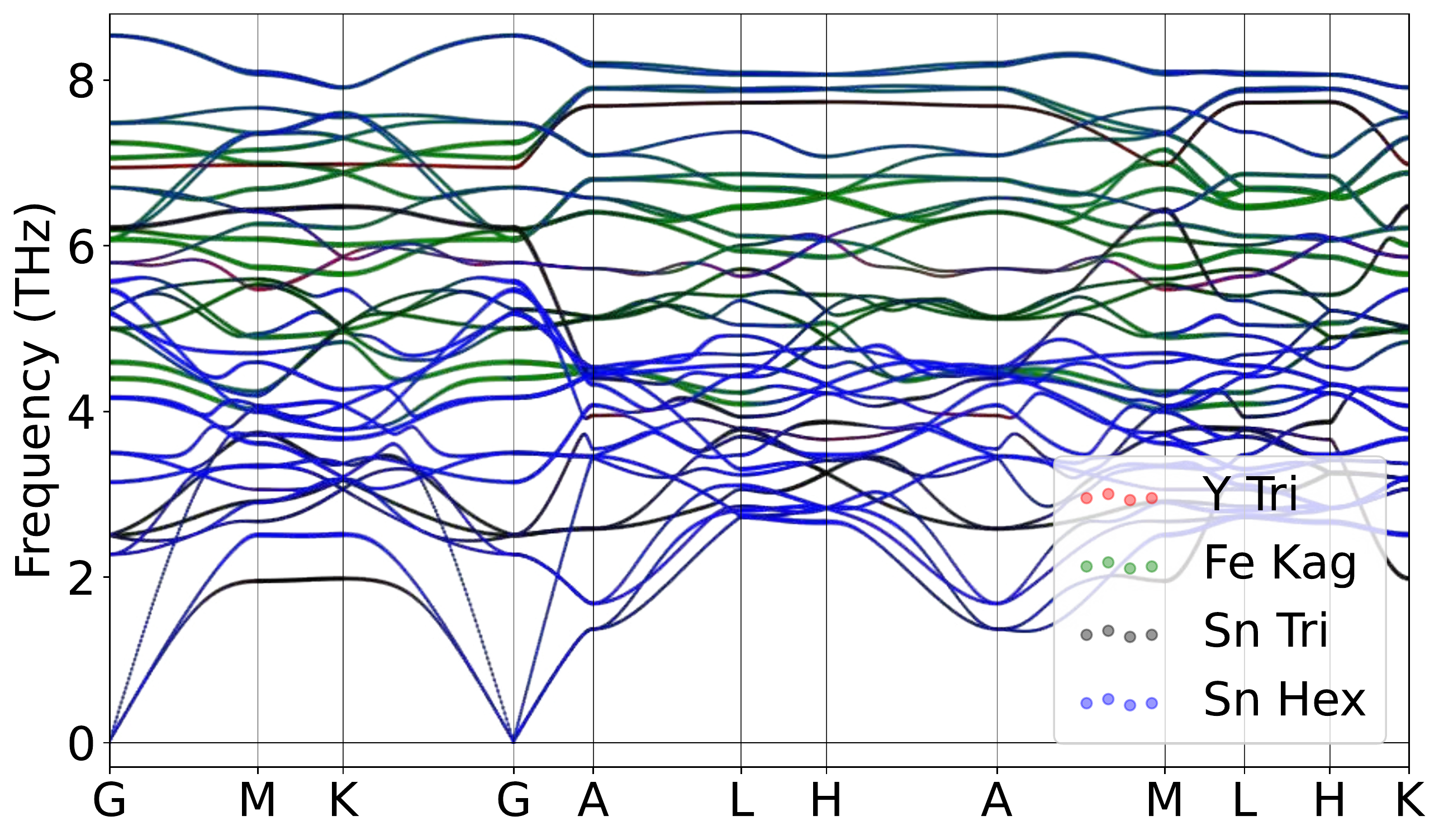}
}
\caption{Atomically projected phonon spectrum for YFe$_6$Sn$_6$ in paramagnetic phase.}
\label{fig:YFe6Sn6pho}
\end{figure}

\begin{figure}[htbp]
\centering
\includegraphics[width=0.45\textwidth]{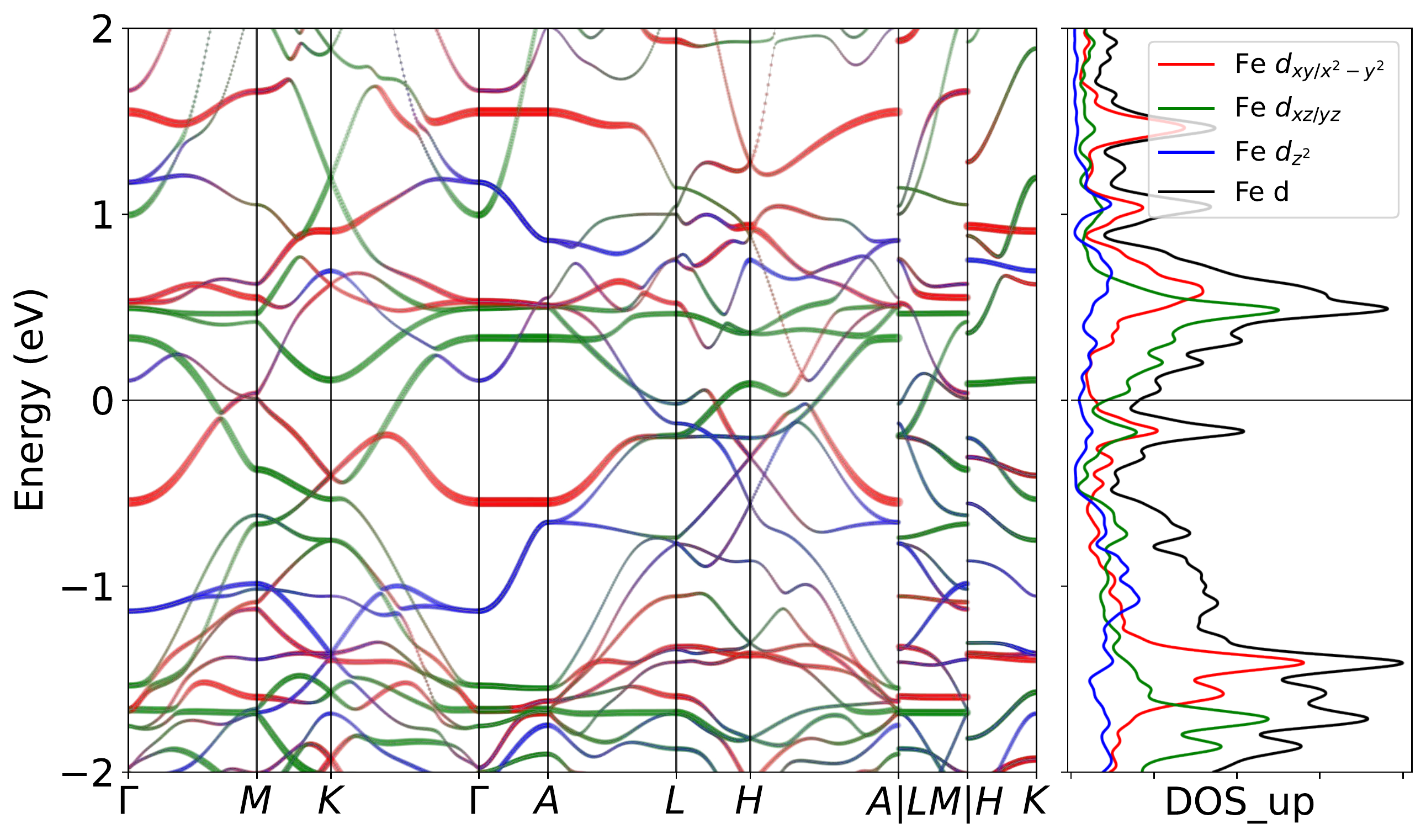}
\includegraphics[width=0.45\textwidth]{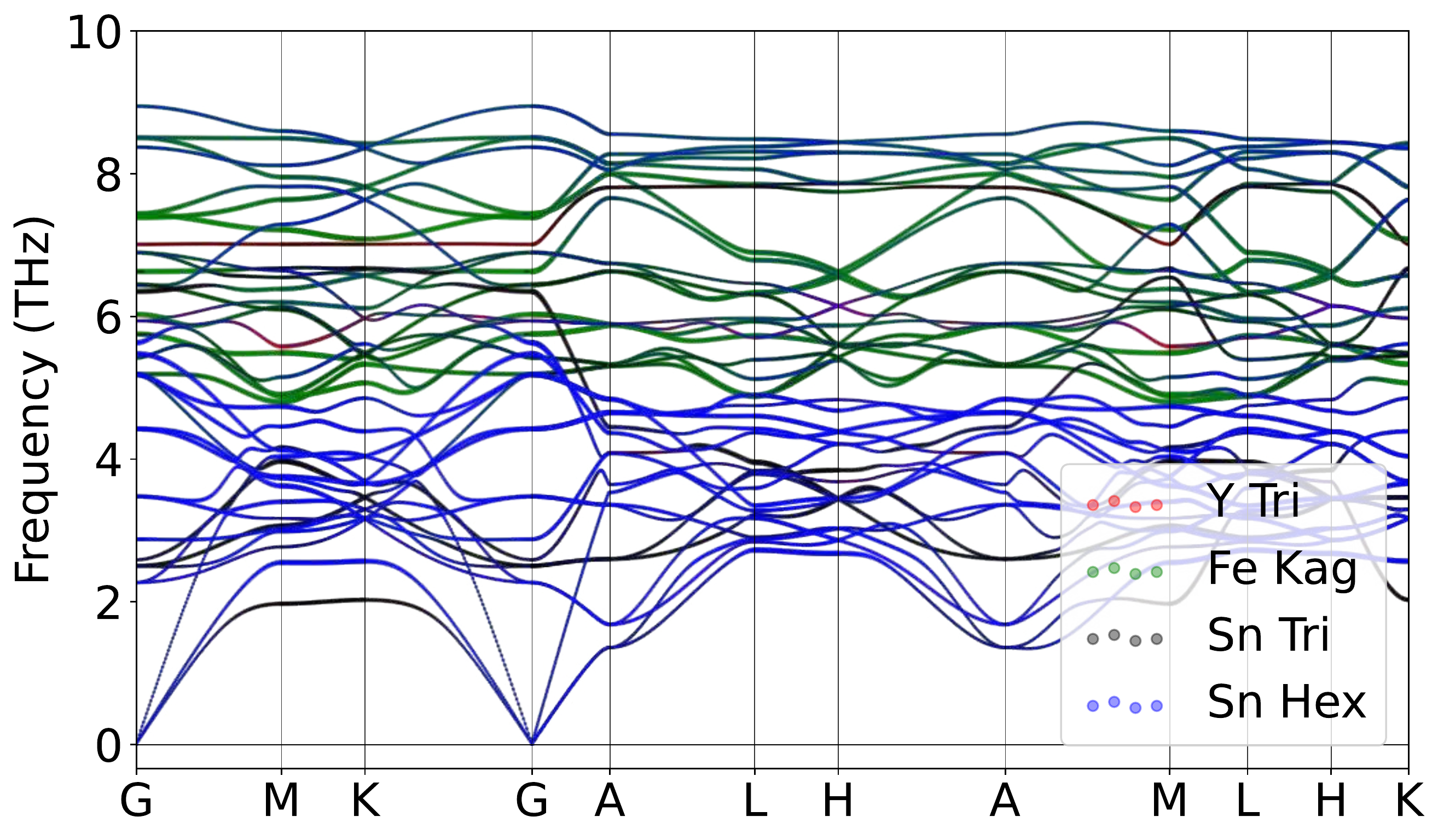}
\caption{Band structure for YFe$_6$Sn$_6$ with AFM order without SOC for spin (Left)up channel. The projection of Fe $3d$ orbitals is also presented. (Right)Correspondingly, a stable phonon was demonstrated with the collinear magnetic order without Hubbard U at lower temperature regime, weighted by atomic projections.}
\label{fig:YFe6Sn6mag}
\end{figure}

\begin{figure}[H]
\centering
\label{fig:YFe6Sn6_relaxed_Low_xz}
\includegraphics[width=0.85\textwidth]{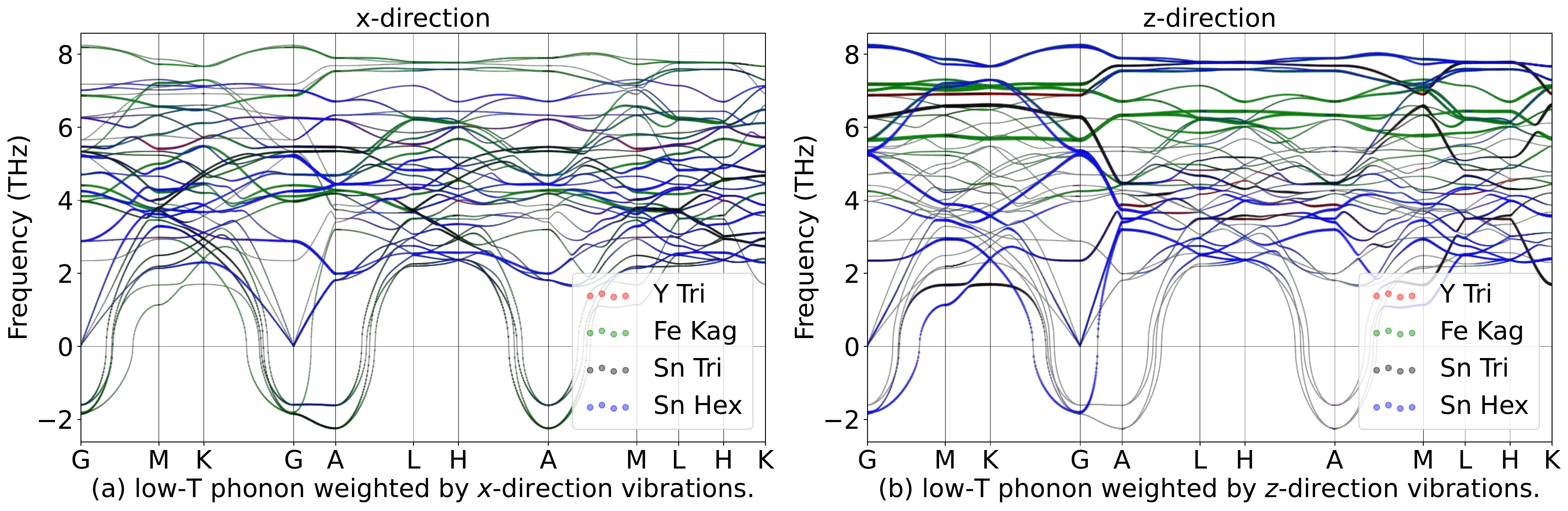}
\hspace{5pt}\label{fig:YFe6Sn6_relaxed_High_xz}
\includegraphics[width=0.85\textwidth]{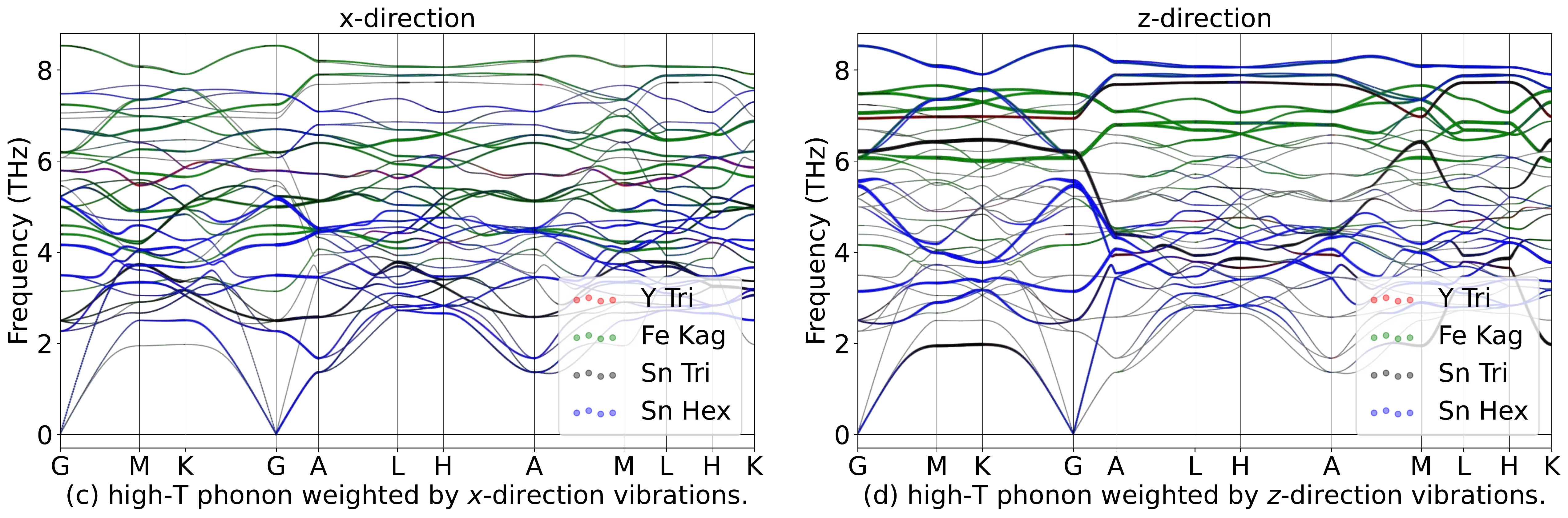}
\caption{Phonon spectrum for YFe$_6$Sn$_6$in in-plane ($x$) and out-of-plane ($z$) directions in paramagnetic phase.}
\label{fig:YFe6Sn6phoxyz}
\end{figure}

\begin{table}[htbp]
\global\long\def\arraystretch{1.12}
\captionsetup{width=.95\textwidth}
\caption{IRREPs at high-symmetry points for YFe$_6$Sn$_6$ phonon spectrum at Low-T.}
\label{tab:191_YFe6Sn6_Low}
\setlength{\tabcolsep}{1.mm}{
}
\end{table}

\begin{table}[htbp]
\global\long\def\arraystretch{1.12}
\captionsetup{width=.95\textwidth}
\caption{IRREPs at high-symmetry points for YFe$_6$Sn$_6$ phonon spectrum at High-T.}
\label{tab:191_YFe6Sn6_High}
\setlength{\tabcolsep}{1.mm}{
%
}
\end{table}
\subsubsection{\label{sms2sec:ZrFe6Sn6}\hyperref[smsec:col]{\texorpdfstring{ZrFe$_6$Sn$_6$}{}}~{\color{blue}  (High-T stable, Low-T unstable) }}

\begin{table}[htbp]
\global\long\def\arraystretch{1.12}
\captionsetup{width=.85\textwidth}
\caption{Structure parameters of ZrFe$_6$Sn$_6$ after relaxation. It contains 496 electrons. }
\label{tab:ZrFe6Sn6}
\setlength{\tabcolsep}{4mm}{
%
}
\end{table}

\begin{figure}[htbp]
\centering
\includegraphics[width=0.85\textwidth]{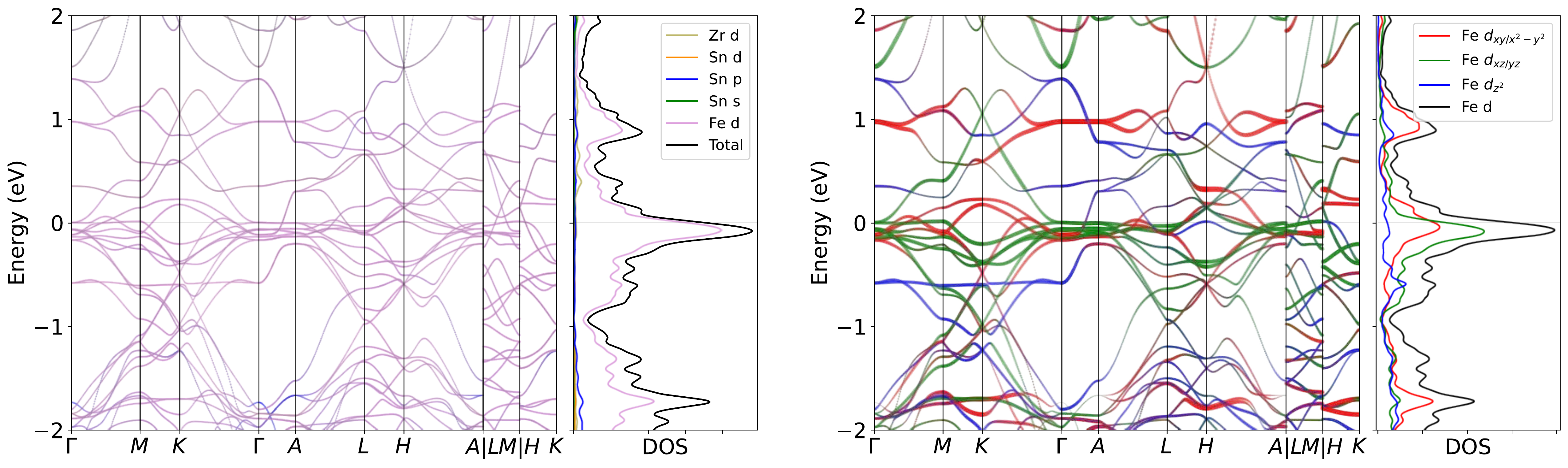}
\caption{Band structure for ZrFe$_6$Sn$_6$ without SOC in paramagnetic phase. The projection of Fe $3d$ orbitals is also presented. The band structures clearly reveal the dominating of Fe $3d$ orbitals  near the Fermi level, as well as the pronounced peak.}
\label{fig:ZrFe6Sn6band}
\end{figure}

\begin{figure}[H]
\centering
\subfloat[Relaxed phonon at low-T.]{
\label{fig:ZrFe6Sn6_relaxed_Low}
\includegraphics[width=0.45\textwidth]{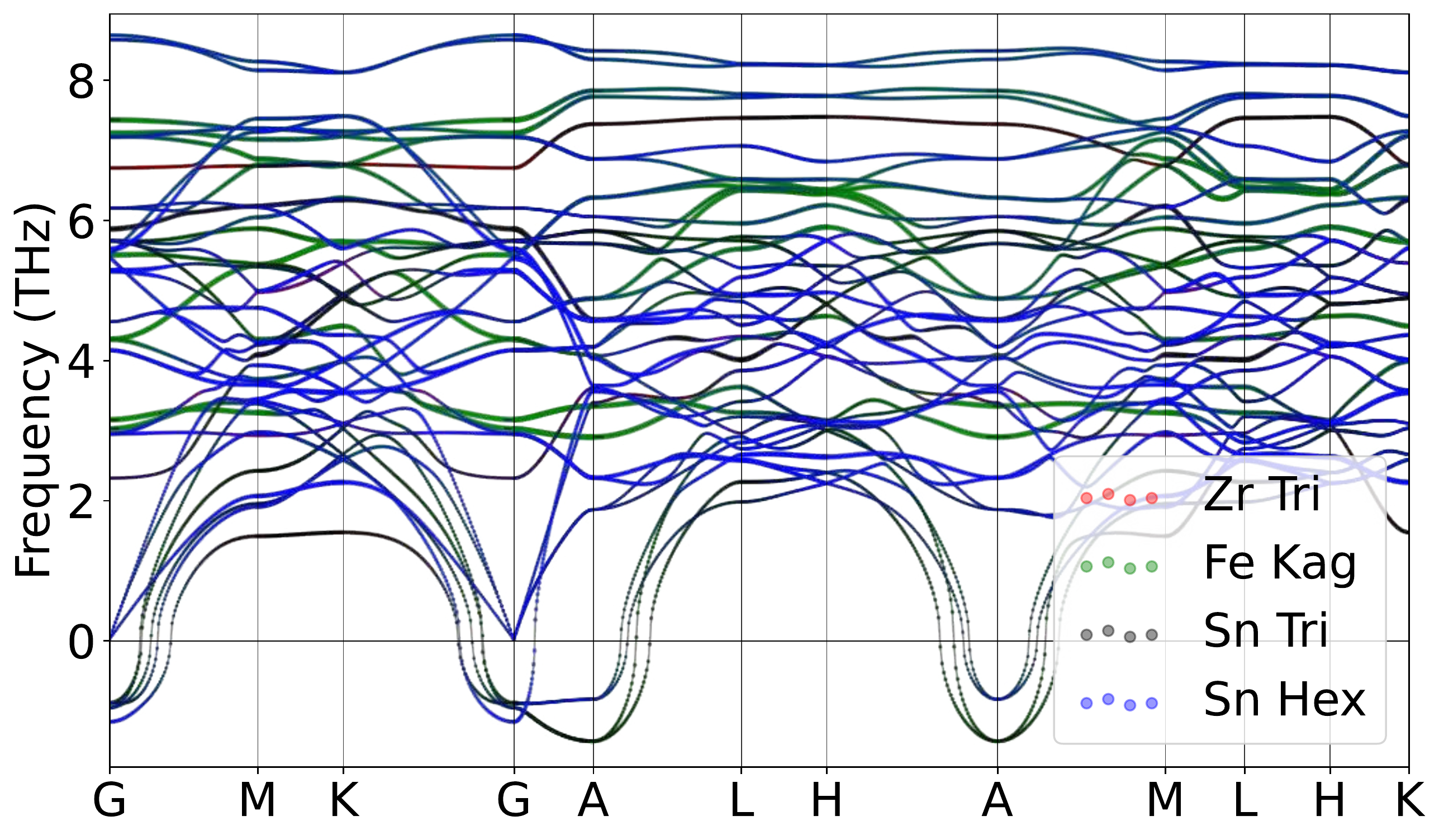}
}
\hspace{5pt}\subfloat[Relaxed phonon at high-T.]{
\label{fig:ZrFe6Sn6_relaxed_High}
\includegraphics[width=0.45\textwidth]{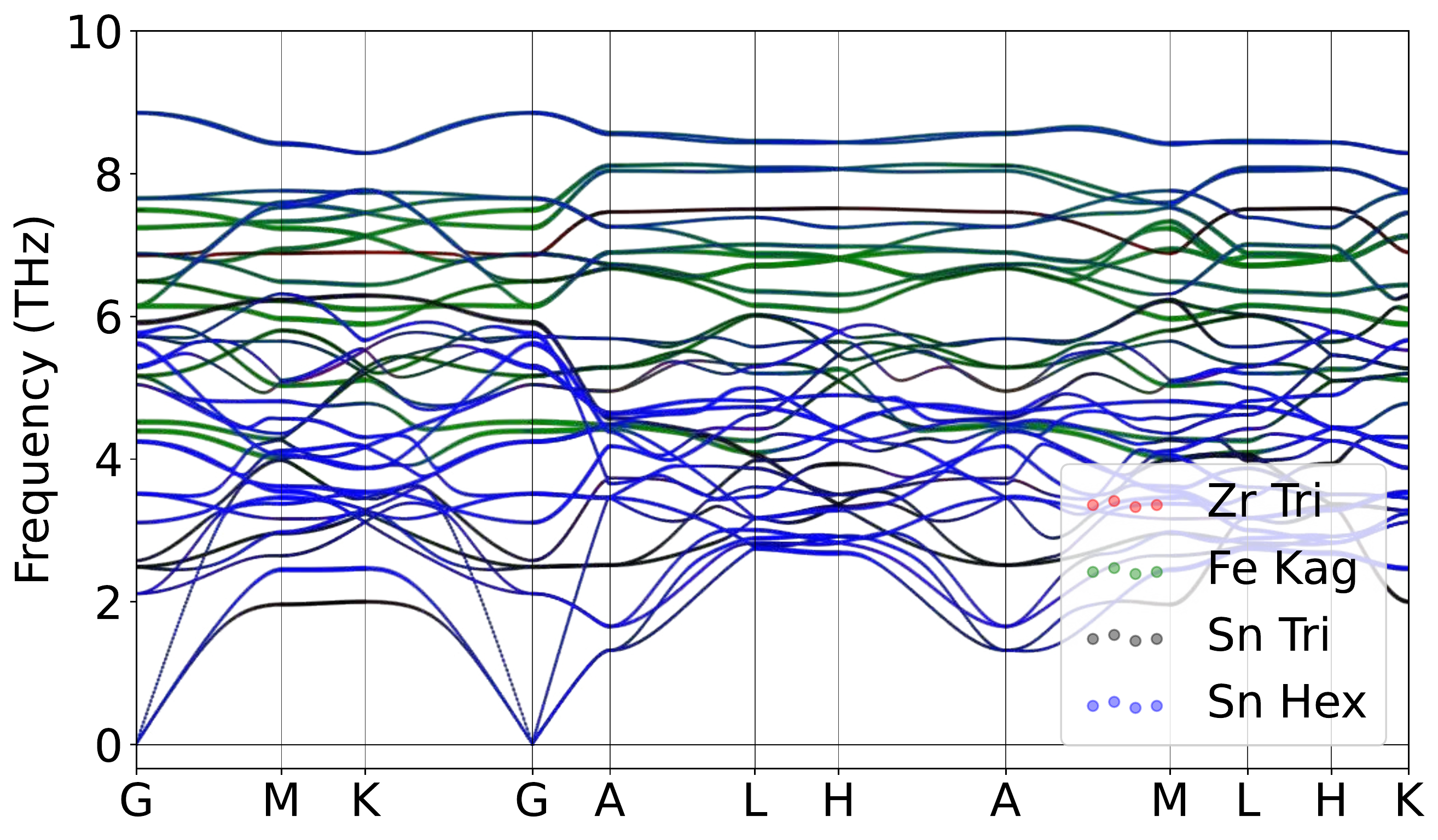}
}
\caption{Atomically projected phonon spectrum for ZrFe$_6$Sn$_6$ in paramagnetic phase.}
\label{fig:ZrFe6Sn6pho}
\end{figure}

\begin{figure}[htbp]
\centering
\includegraphics[width=0.45\textwidth]{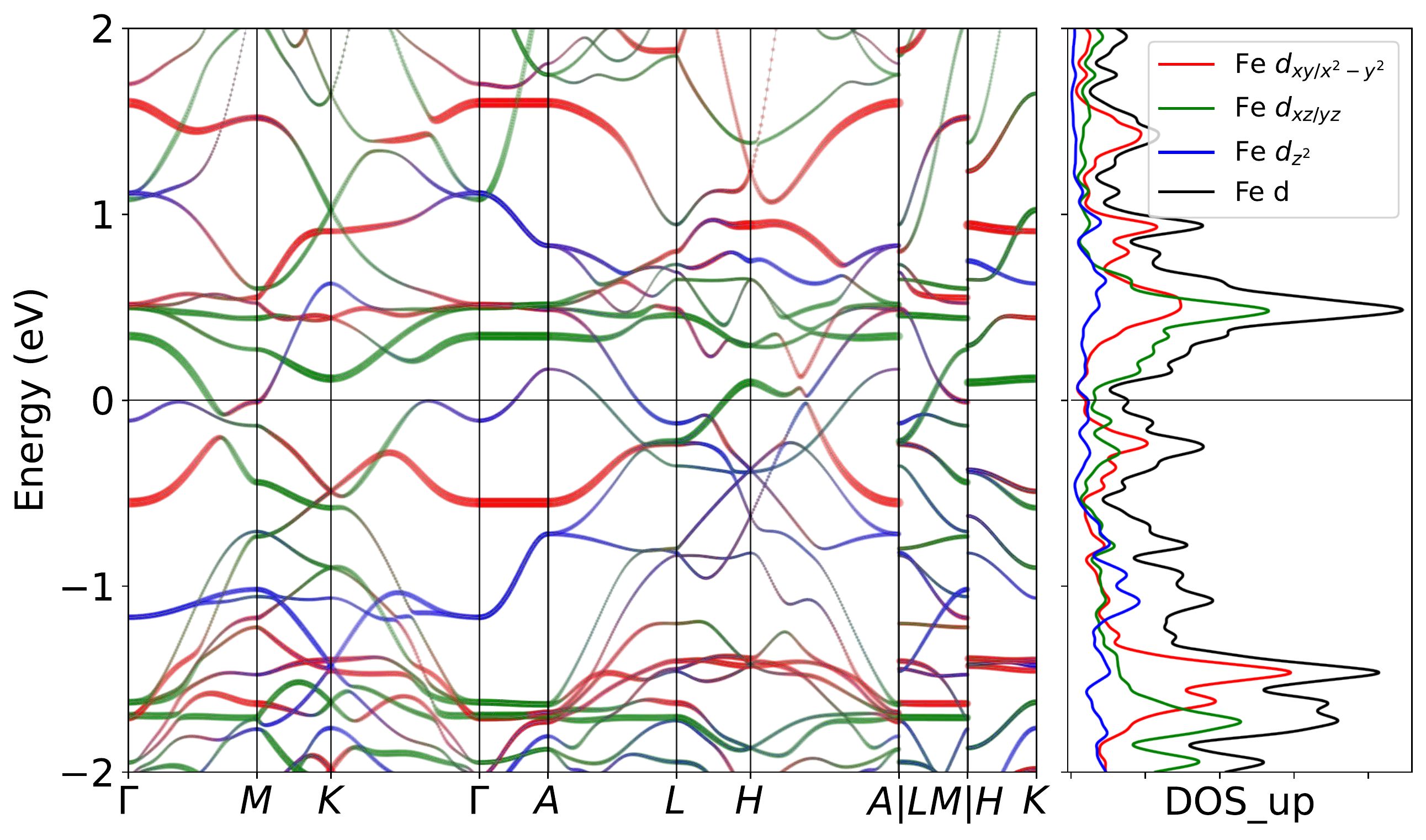}
\includegraphics[width=0.45\textwidth]{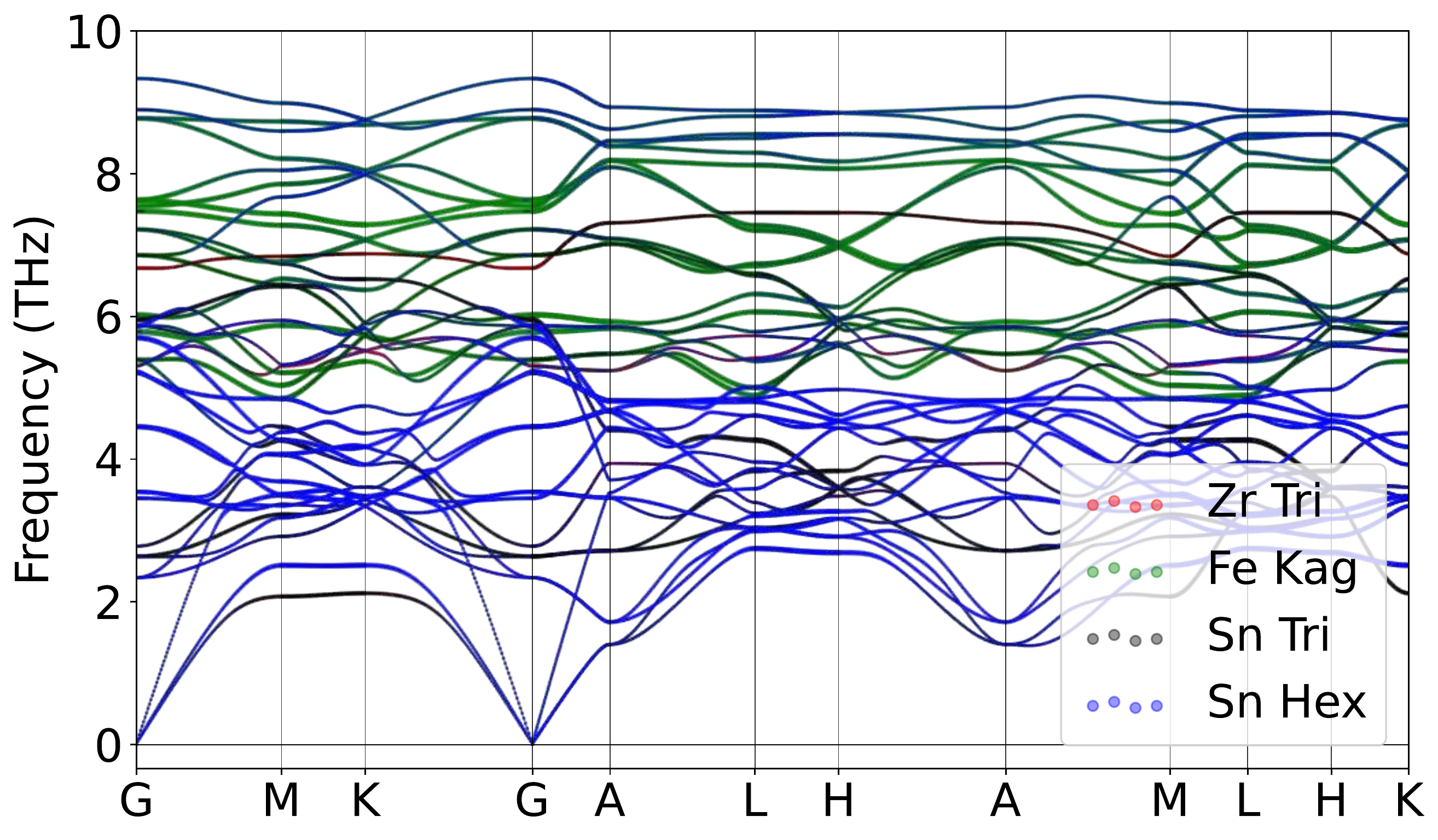}
\caption{Band structure for ZrFe$_6$Sn$_6$ with AFM order without SOC for spin (Left)up channel. The projection of Fe $3d$ orbitals is also presented. (Right)Correspondingly, a stable phonon was demonstrated with the collinear magnetic order without Hubbard U at lower temperature regime, weighted by atomic projections.}
\label{fig:ZrFe6Sn6mag}
\end{figure}

\begin{figure}[H]
\centering
\label{fig:ZrFe6Sn6_relaxed_Low_xz}
\includegraphics[width=0.85\textwidth]{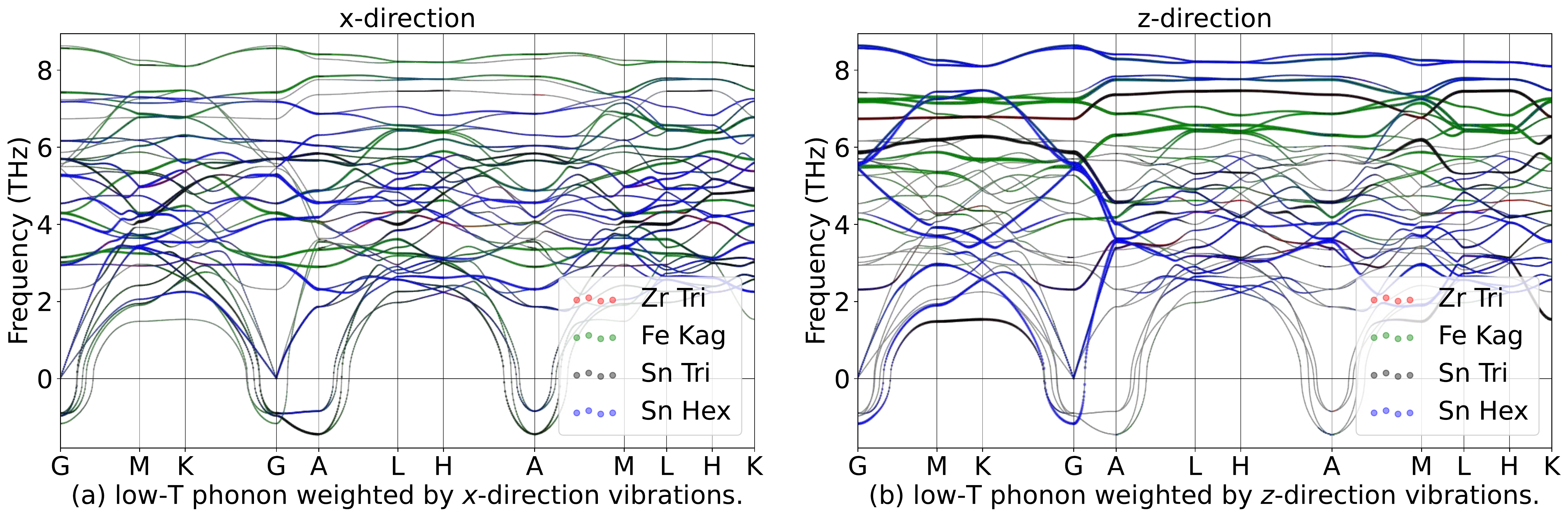}
\hspace{5pt}\label{fig:ZrFe6Sn6_relaxed_High_xz}
\includegraphics[width=0.85\textwidth]{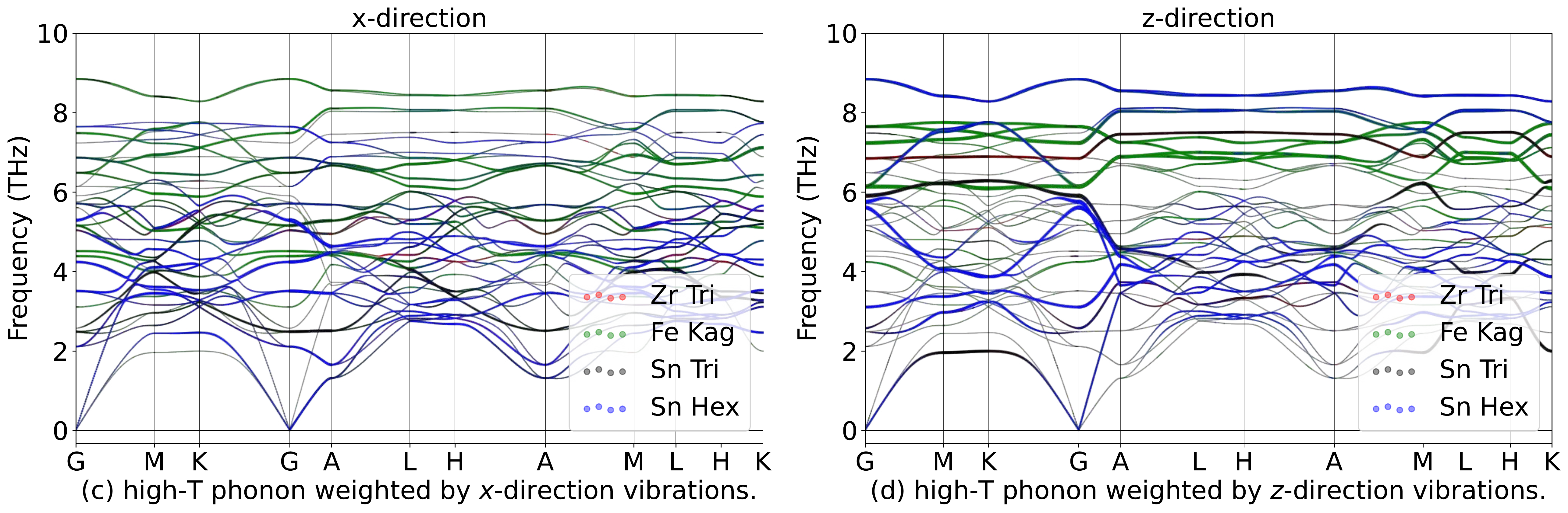}
\caption{Phonon spectrum for ZrFe$_6$Sn$_6$in in-plane ($x$) and out-of-plane ($z$) directions in paramagnetic phase.}
\label{fig:ZrFe6Sn6phoxyz}
\end{figure}

\begin{table}[htbp]
\global\long\def\arraystretch{1.12}
\captionsetup{width=.95\textwidth}
\caption{IRREPs at high-symmetry points for ZrFe$_6$Sn$_6$ phonon spectrum at Low-T.}
\label{tab:191_ZrFe6Sn6_Low}
\setlength{\tabcolsep}{1.mm}{
}
\end{table}

\begin{table}[htbp]
\global\long\def\arraystretch{1.12}
\captionsetup{width=.95\textwidth}
\caption{IRREPs at high-symmetry points for ZrFe$_6$Sn$_6$ phonon spectrum at High-T.}
\label{tab:191_ZrFe6Sn6_High}
\setlength{\tabcolsep}{1.mm}{
%
}
\end{table}
\subsubsection{\label{sms2sec:NbFe6Sn6}\hyperref[smsec:col]{\texorpdfstring{NbFe$_6$Sn$_6$}{}}~{\color{blue}  (High-T stable, Low-T unstable) }}

\begin{table}[htbp]
\global\long\def\arraystretch{1.12}
\captionsetup{width=.85\textwidth}
\caption{Structure parameters of NbFe$_6$Sn$_6$ after relaxation. It contains 497 electrons. }
\label{tab:NbFe6Sn6}
\setlength{\tabcolsep}{4mm}{
%
}
\end{table}

\begin{figure}[htbp]
\centering
\includegraphics[width=0.85\textwidth]{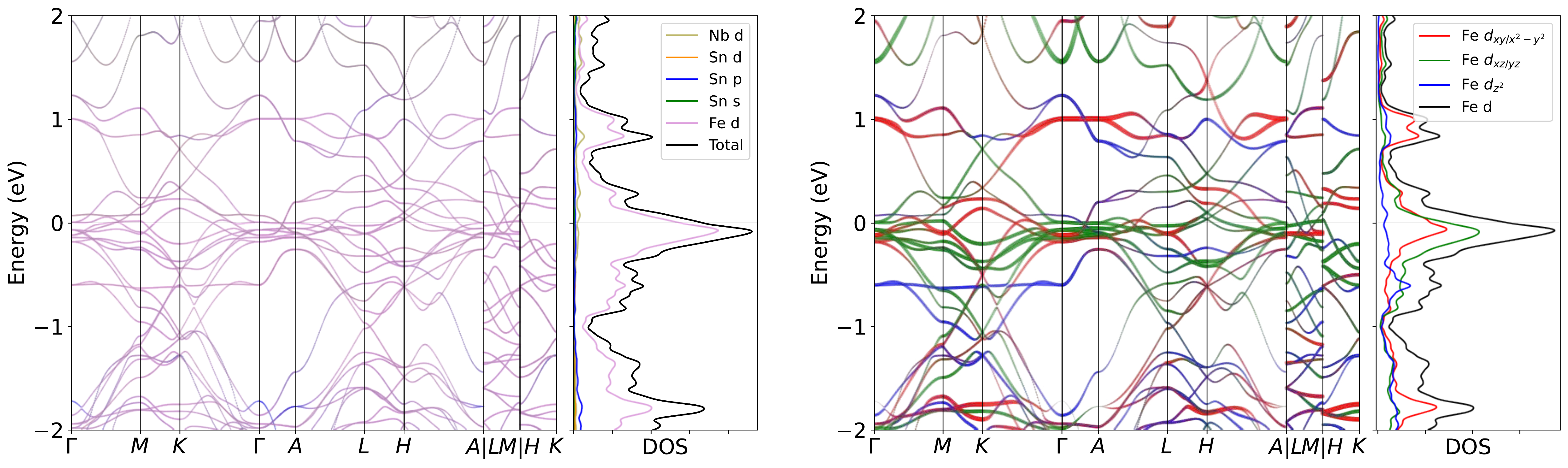}
\caption{Band structure for NbFe$_6$Sn$_6$ without SOC in paramagnetic phase. The projection of Fe $3d$ orbitals is also presented. The band structures clearly reveal the dominating of Fe $3d$ orbitals  near the Fermi level, as well as the pronounced peak.}
\label{fig:NbFe6Sn6band}
\end{figure}

\begin{figure}[H]
\centering
\subfloat[Relaxed phonon at low-T.]{
\label{fig:NbFe6Sn6_relaxed_Low}
\includegraphics[width=0.45\textwidth]{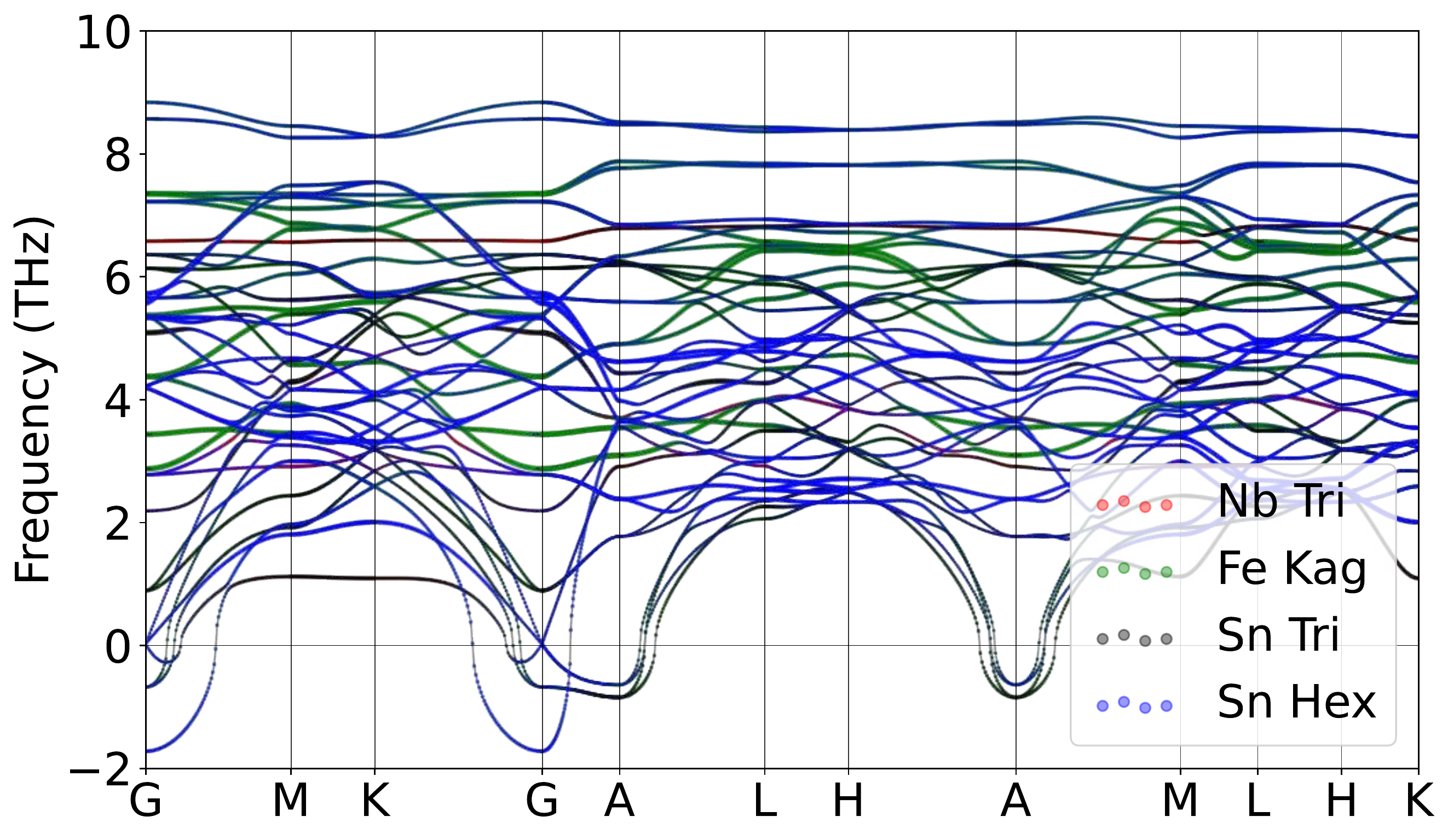}
}
\hspace{5pt}\subfloat[Relaxed phonon at high-T.]{
\label{fig:NbFe6Sn6_relaxed_High}
\includegraphics[width=0.45\textwidth]{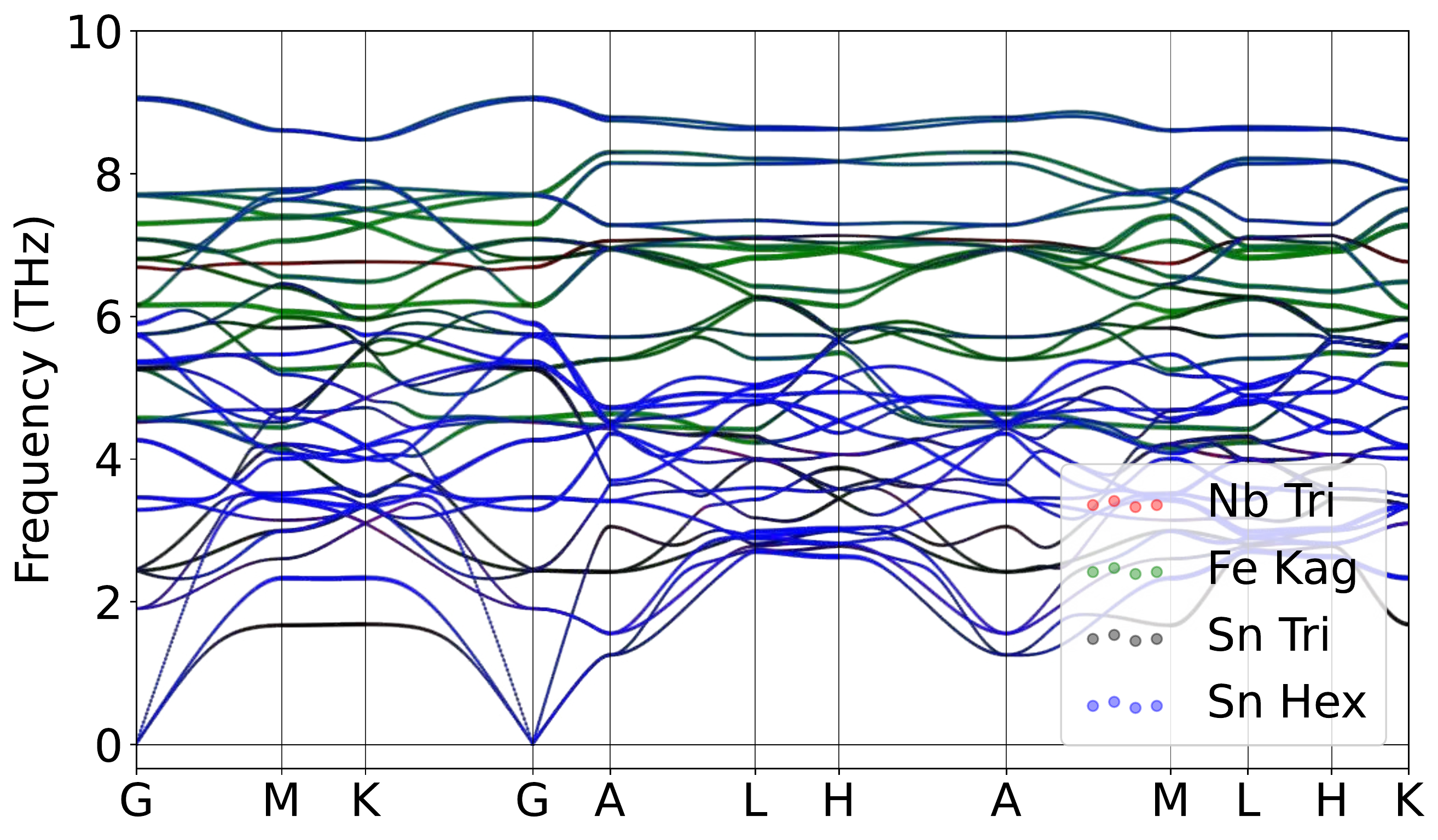}
}
\caption{Atomically projected phonon spectrum for NbFe$_6$Sn$_6$ in paramagnetic phase.}
\label{fig:NbFe6Sn6pho}
\end{figure}

\begin{figure}[htbp]
\centering
\includegraphics[width=0.45\textwidth]{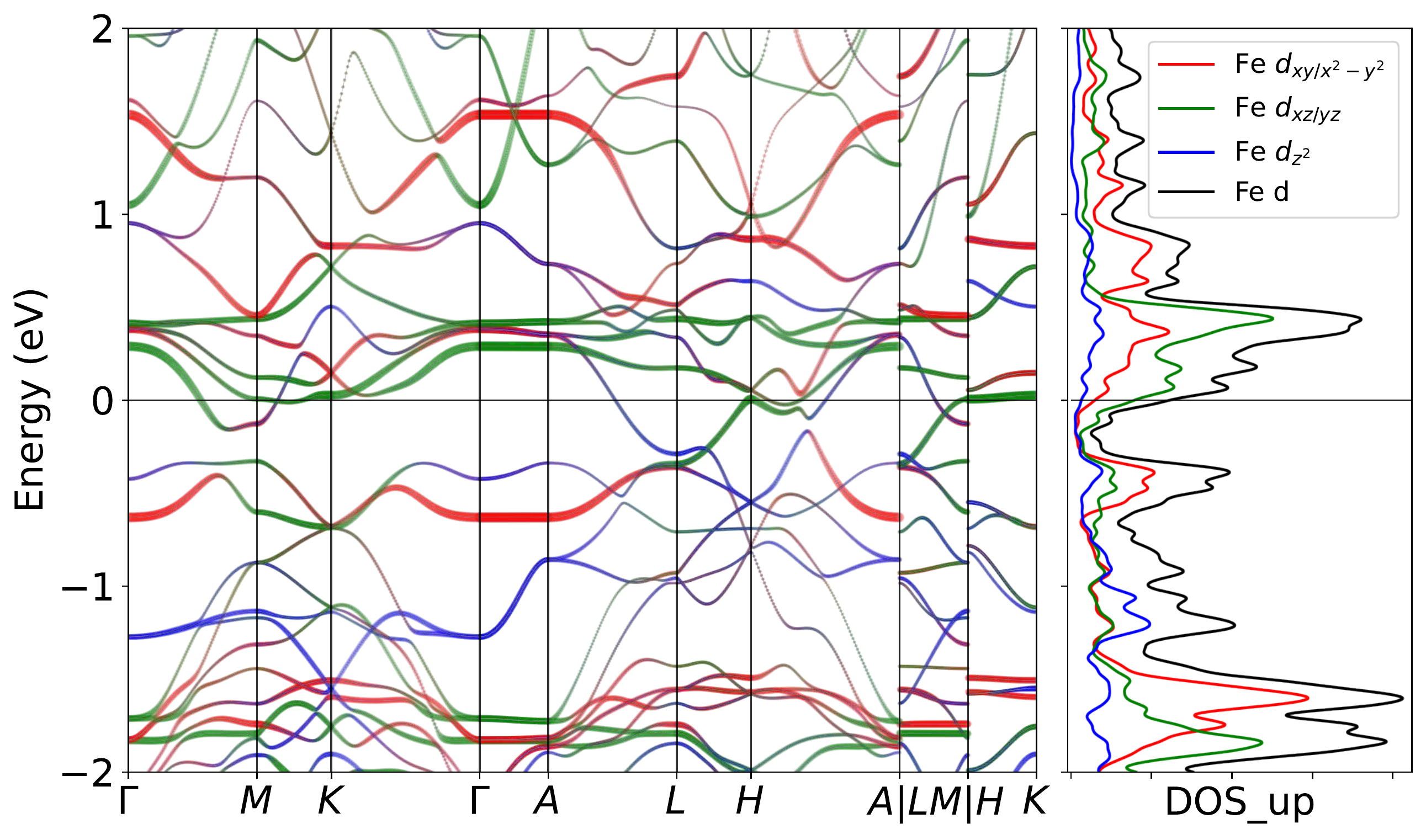}
\includegraphics[width=0.45\textwidth]{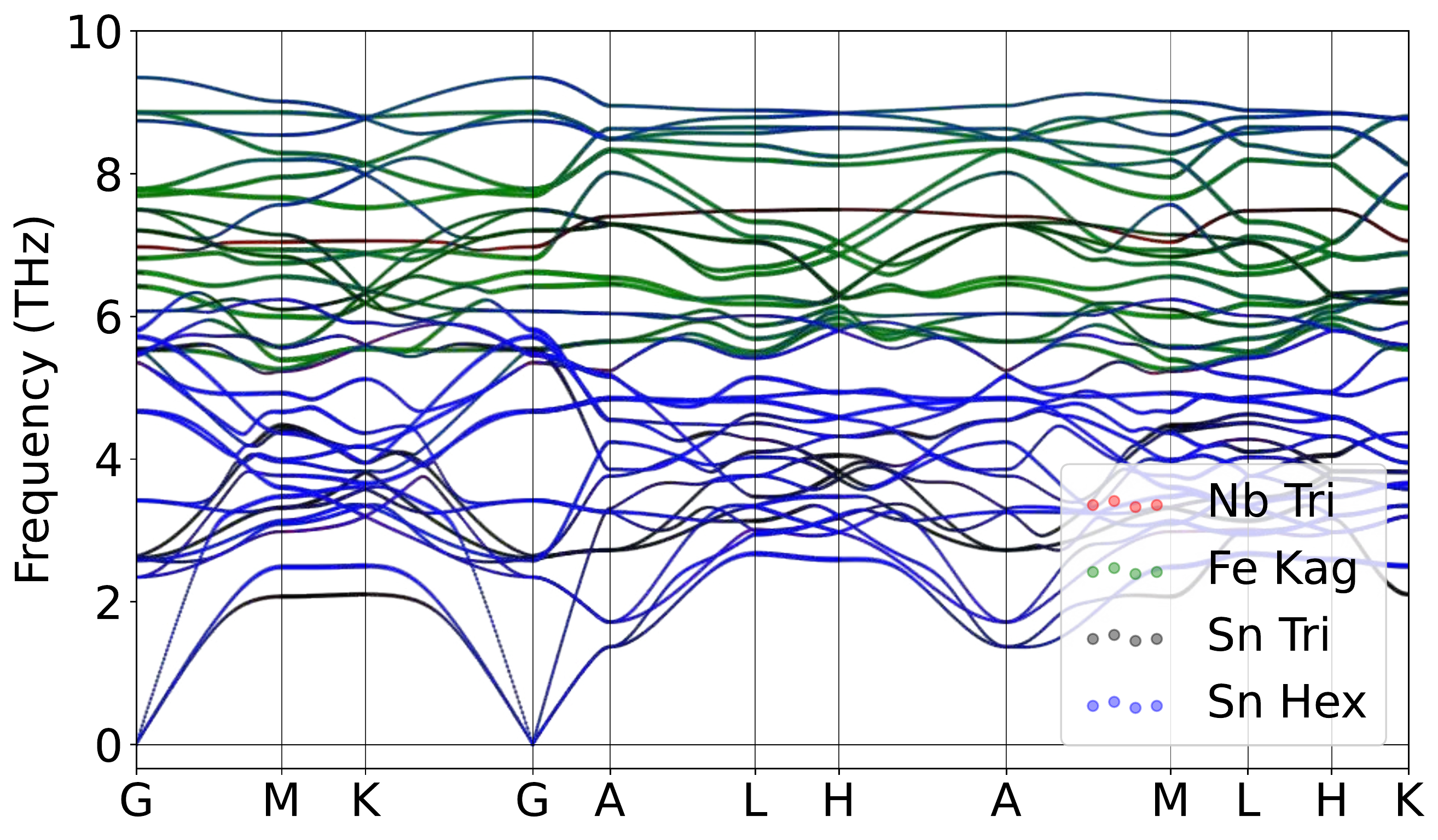}
\caption{Band structure for NbFe$_6$Sn$_6$ with AFM order without SOC for spin (Left)up channel. The projection of Fe $3d$ orbitals is also presented. (Right)Correspondingly, a stable phonon was demonstrated with the collinear magnetic order without Hubbard U at lower temperature regime, weighted by atomic projections.}
\label{fig:NbFe6Sn6mag}
\end{figure}

\begin{figure}[H]
\centering
\label{fig:NbFe6Sn6_relaxed_Low_xz}
\includegraphics[width=0.85\textwidth]{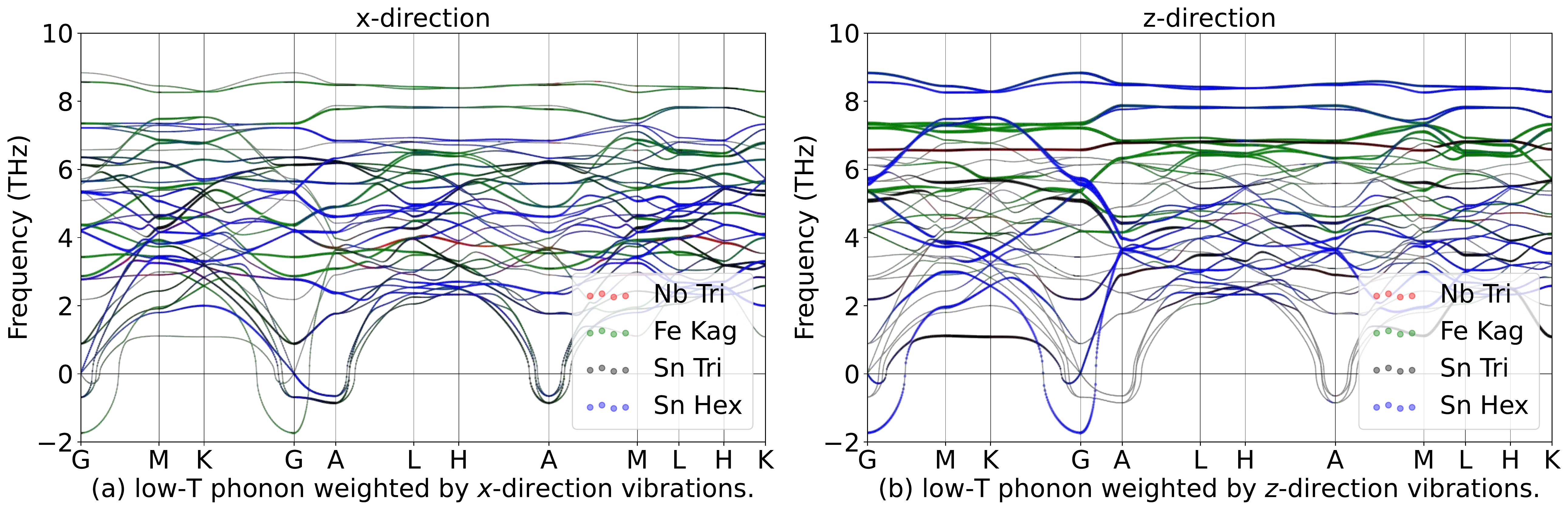}
\hspace{5pt}\label{fig:NbFe6Sn6_relaxed_High_xz}
\includegraphics[width=0.85\textwidth]{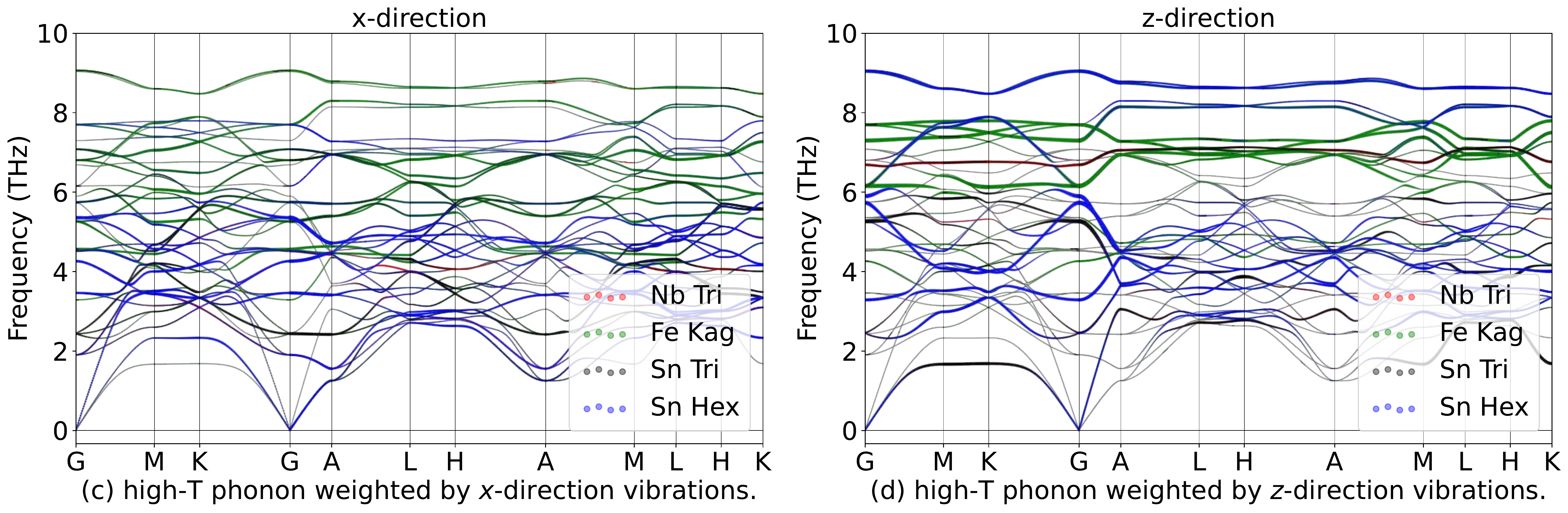}
\caption{Phonon spectrum for NbFe$_6$Sn$_6$in in-plane ($x$) and out-of-plane ($z$) directions in paramagnetic phase.}
\label{fig:NbFe6Sn6phoxyz}
\end{figure}

\begin{table}[htbp]
\global\long\def\arraystretch{1.12}
\captionsetup{width=.95\textwidth}
\caption{IRREPs at high-symmetry points for NbFe$_6$Sn$_6$ phonon spectrum at Low-T.}
\label{tab:191_NbFe6Sn6_Low}
\setlength{\tabcolsep}{1.mm}{
}
\end{table}

\begin{table}[htbp]
\global\long\def\arraystretch{1.12}
\captionsetup{width=.95\textwidth}
\caption{IRREPs at high-symmetry points for NbFe$_6$Sn$_6$ phonon spectrum at High-T.}
\label{tab:191_NbFe6Sn6_High}
\setlength{\tabcolsep}{1.mm}{
%
}
\end{table}
\subsubsection{\label{sms2sec:PrFe6Sn6}\hyperref[smsec:col]{\texorpdfstring{PrFe$_6$Sn$_6$}{}}~{\color{blue}  (High-T stable, Low-T unstable) }}

\begin{table}[htbp]
\global\long\def\arraystretch{1.12}
\captionsetup{width=.85\textwidth}
\caption{Structure parameters of PrFe$_6$Sn$_6$ after relaxation. It contains 515 electrons. }
\label{tab:PrFe6Sn6}
\setlength{\tabcolsep}{4mm}{
%
}
\end{table}

\begin{figure}[htbp]
\centering
\includegraphics[width=0.85\textwidth]{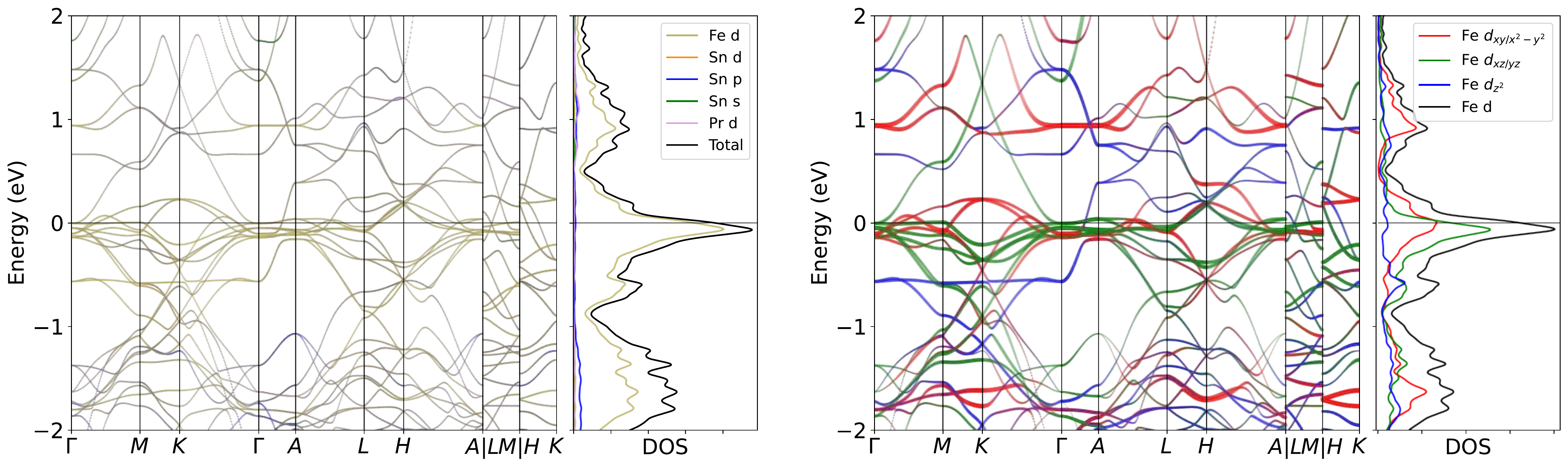}
\caption{Band structure for PrFe$_6$Sn$_6$ without SOC in paramagnetic phase. The projection of Fe $3d$ orbitals is also presented. The band structures clearly reveal the dominating of Fe $3d$ orbitals  near the Fermi level, as well as the pronounced peak.}
\label{fig:PrFe6Sn6band}
\end{figure}

\begin{figure}[H]
\centering
\subfloat[Relaxed phonon at low-T.]{
\label{fig:PrFe6Sn6_relaxed_Low}
\includegraphics[width=0.45\textwidth]{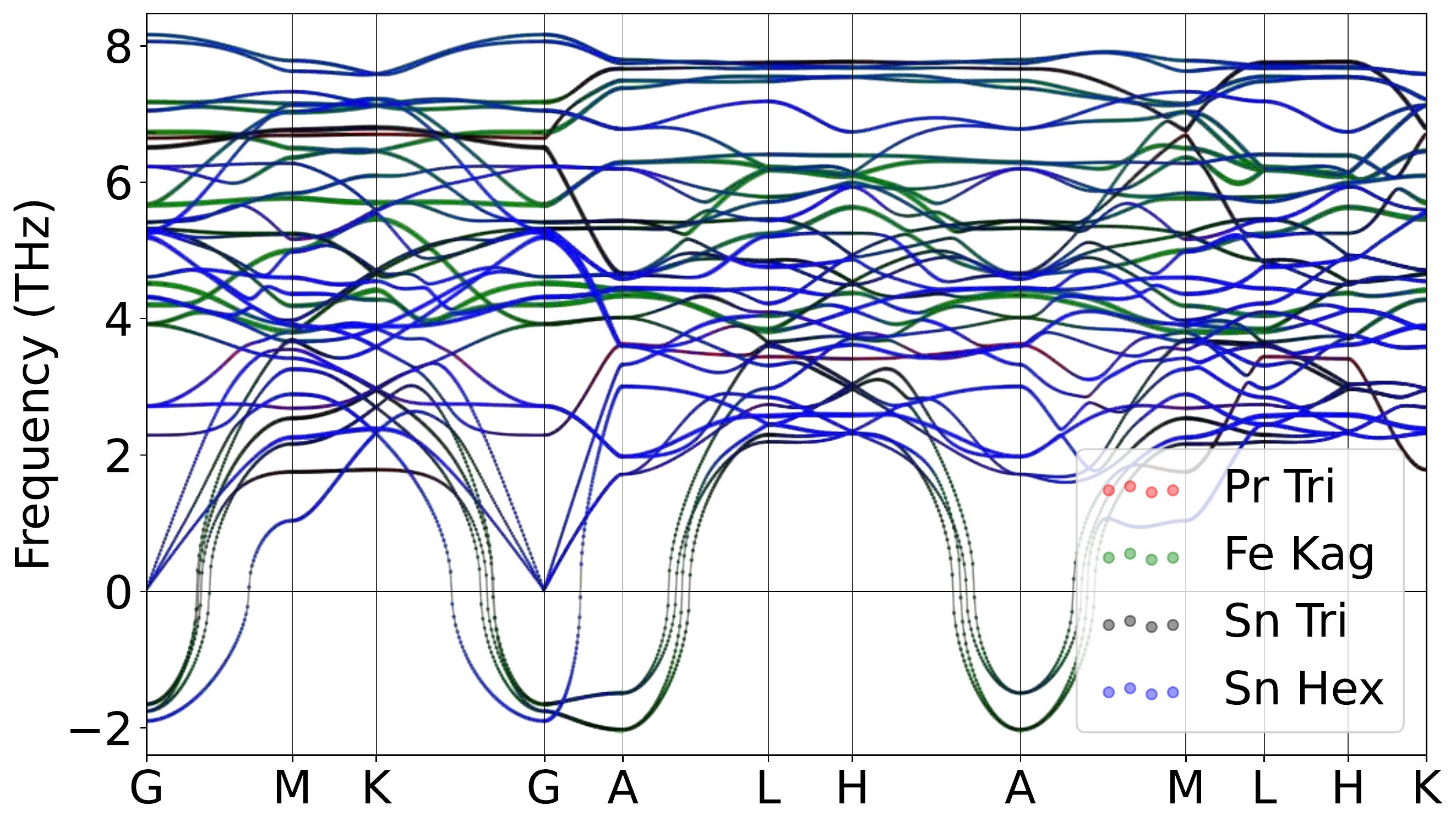}
}
\hspace{5pt}\subfloat[Relaxed phonon at high-T.]{
\label{fig:PrFe6Sn6_relaxed_High}
\includegraphics[width=0.45\textwidth]{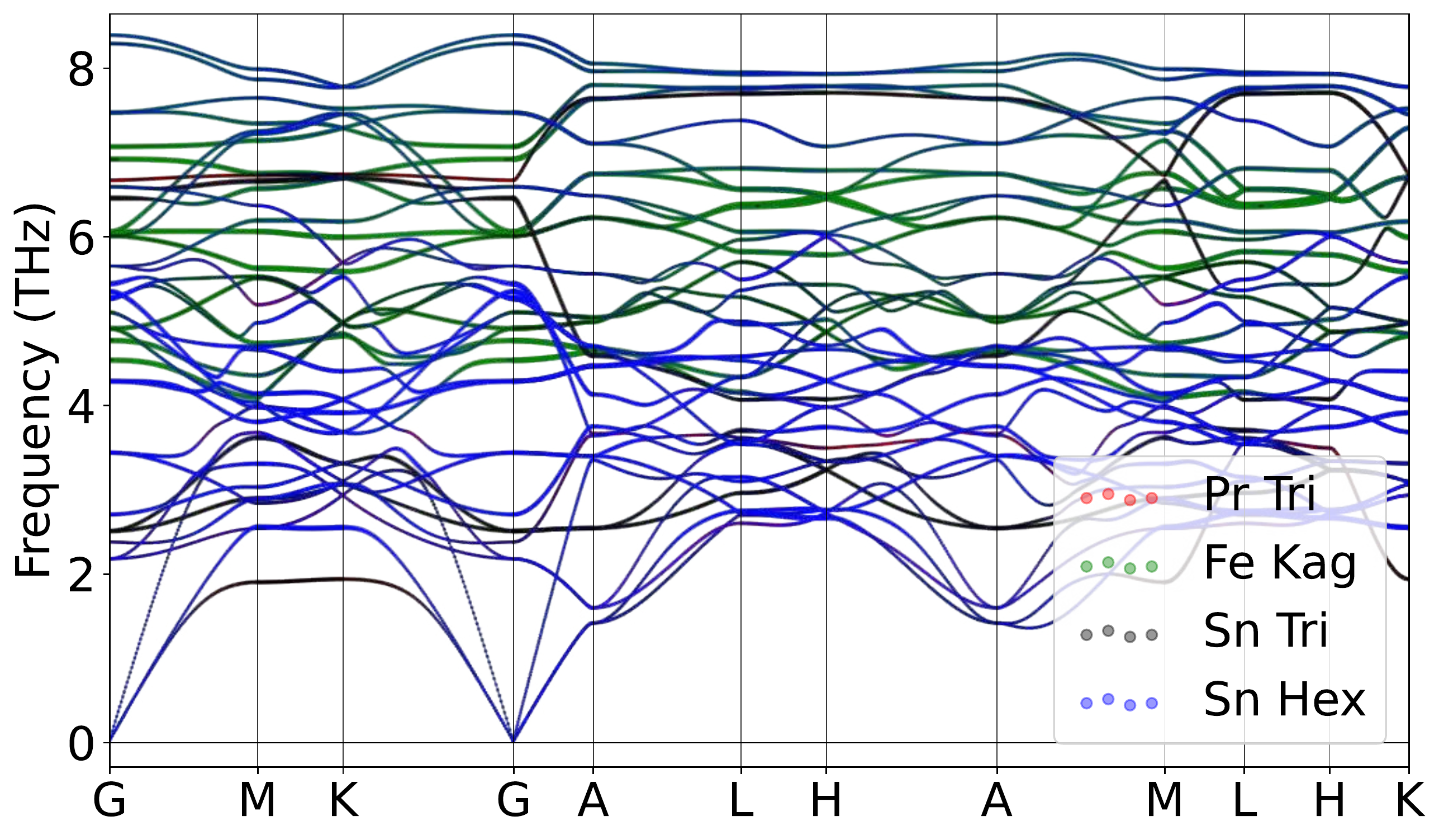}
}
\caption{Atomically projected phonon spectrum for PrFe$_6$Sn$_6$ in paramagnetic phase.}
\label{fig:PrFe6Sn6pho}
\end{figure}

\begin{figure}[H]
\centering
\label{fig:PrFe6Sn6_relaxed_Low_xz}
\includegraphics[width=0.85\textwidth]{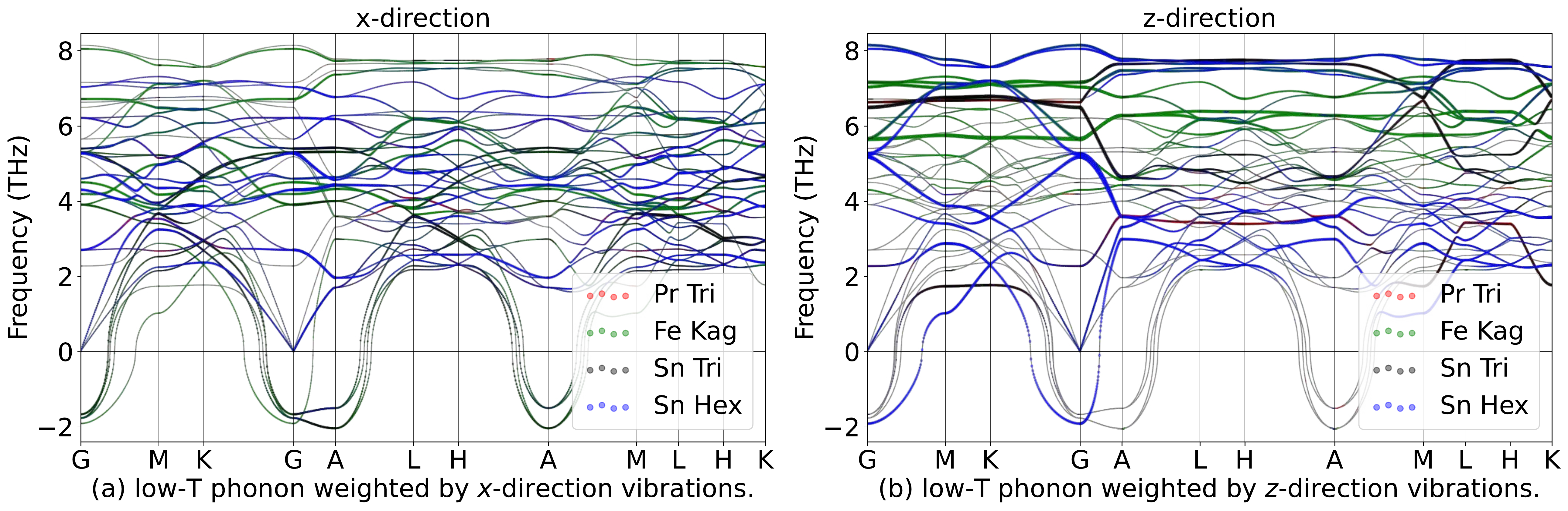}
\hspace{5pt}\label{fig:PrFe6Sn6_relaxed_High_xz}
\includegraphics[width=0.85\textwidth]{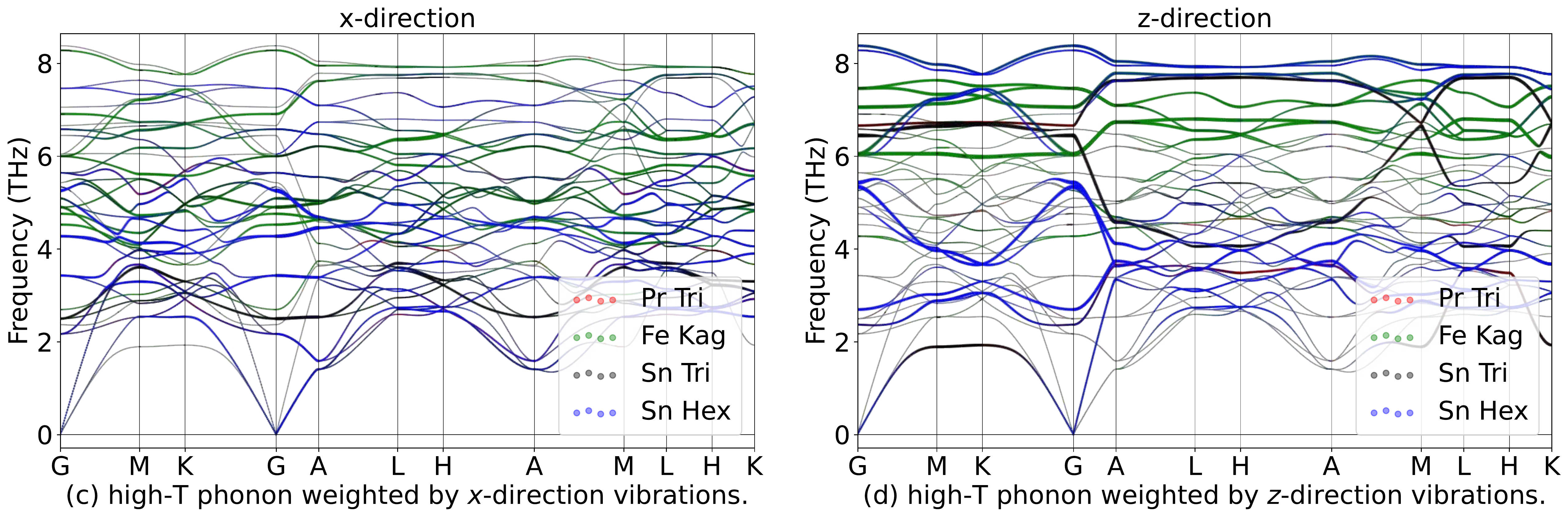}
\caption{Phonon spectrum for PrFe$_6$Sn$_6$in in-plane ($x$) and out-of-plane ($z$) directions in paramagnetic phase.}
\label{fig:PrFe6Sn6phoxyz}
\end{figure}

\begin{table}[htbp]
\global\long\def\arraystretch{1.12}
\captionsetup{width=.95\textwidth}
\caption{IRREPs at high-symmetry points for PrFe$_6$Sn$_6$ phonon spectrum at Low-T.}
\label{tab:191_PrFe6Sn6_Low}
\setlength{\tabcolsep}{1.mm}{
}
\end{table}

\begin{table}[htbp]
\global\long\def\arraystretch{1.12}
\captionsetup{width=.95\textwidth}
\caption{IRREPs at high-symmetry points for PrFe$_6$Sn$_6$ phonon spectrum at High-T.}
\label{tab:191_PrFe6Sn6_High}
\setlength{\tabcolsep}{1.mm}{
%
}
\end{table}
\subsubsection{\label{sms2sec:NdFe6Sn6}\hyperref[smsec:col]{\texorpdfstring{NdFe$_6$Sn$_6$}{}}~{\color{blue}  (High-T stable, Low-T unstable) }}

\begin{table}[htbp]
\global\long\def\arraystretch{1.12}
\captionsetup{width=.85\textwidth}
\caption{Structure parameters of NdFe$_6$Sn$_6$ after relaxation. It contains 516 electrons. }
\label{tab:NdFe6Sn6}
\setlength{\tabcolsep}{4mm}{
%
}
\end{table}

\begin{figure}[htbp]
\centering
\includegraphics[width=0.85\textwidth]{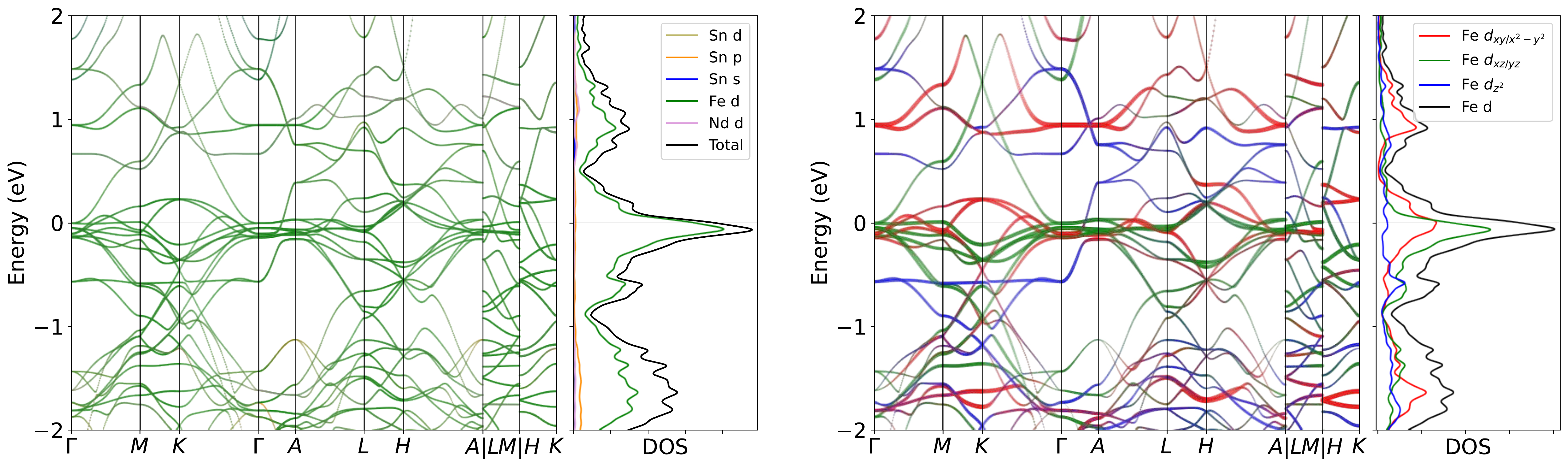}
\caption{Band structure for NdFe$_6$Sn$_6$ without SOC in paramagnetic phase. The projection of Fe $3d$ orbitals is also presented. The band structures clearly reveal the dominating of Fe $3d$ orbitals  near the Fermi level, as well as the pronounced peak.}
\label{fig:NdFe6Sn6band}
\end{figure}

\begin{figure}[H]
\centering
\subfloat[Relaxed phonon at low-T.]{
\label{fig:NdFe6Sn6_relaxed_Low}
\includegraphics[width=0.45\textwidth]{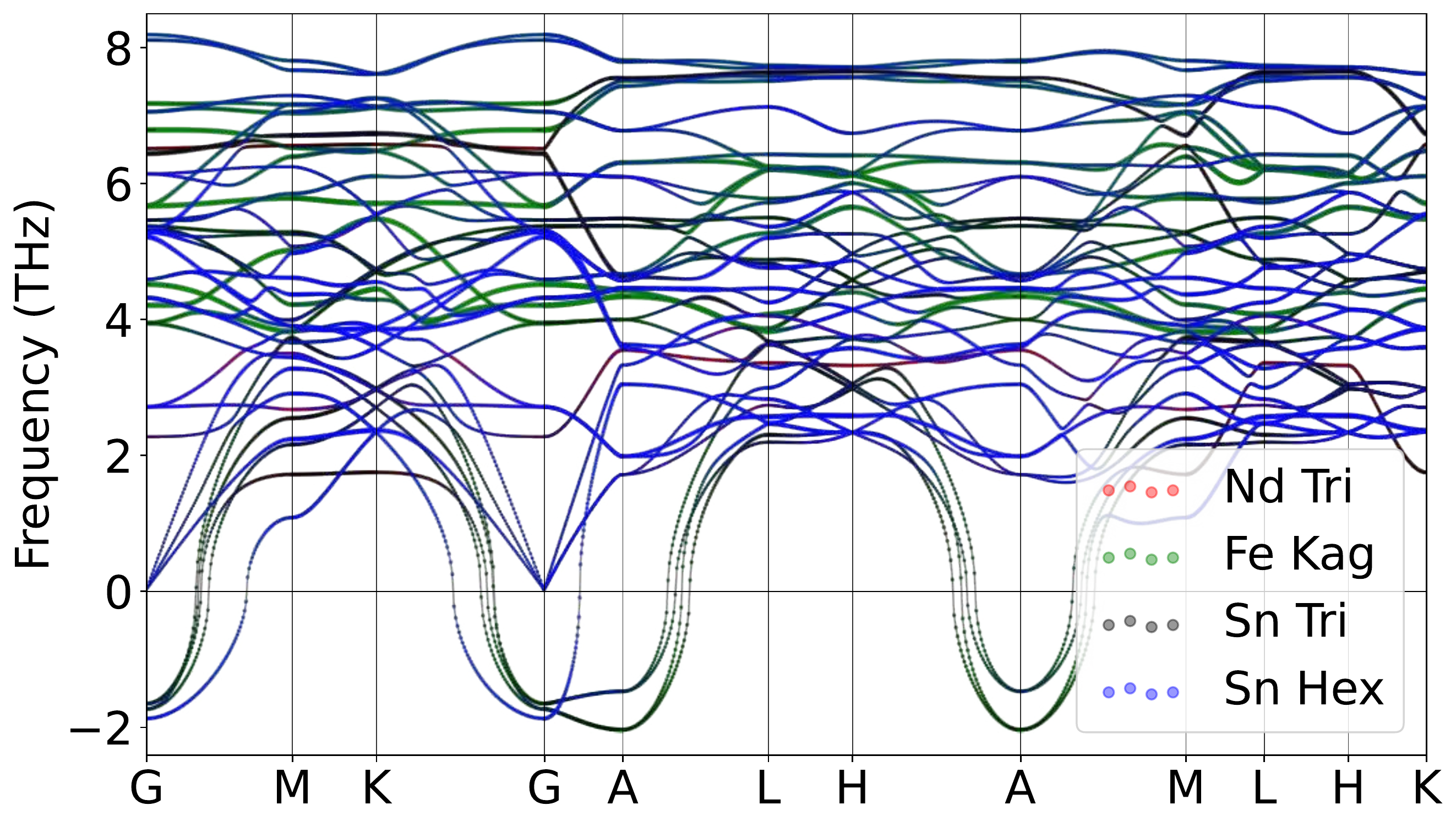}
}
\hspace{5pt}\subfloat[Relaxed phonon at high-T.]{
\label{fig:NdFe6Sn6_relaxed_High}
\includegraphics[width=0.45\textwidth]{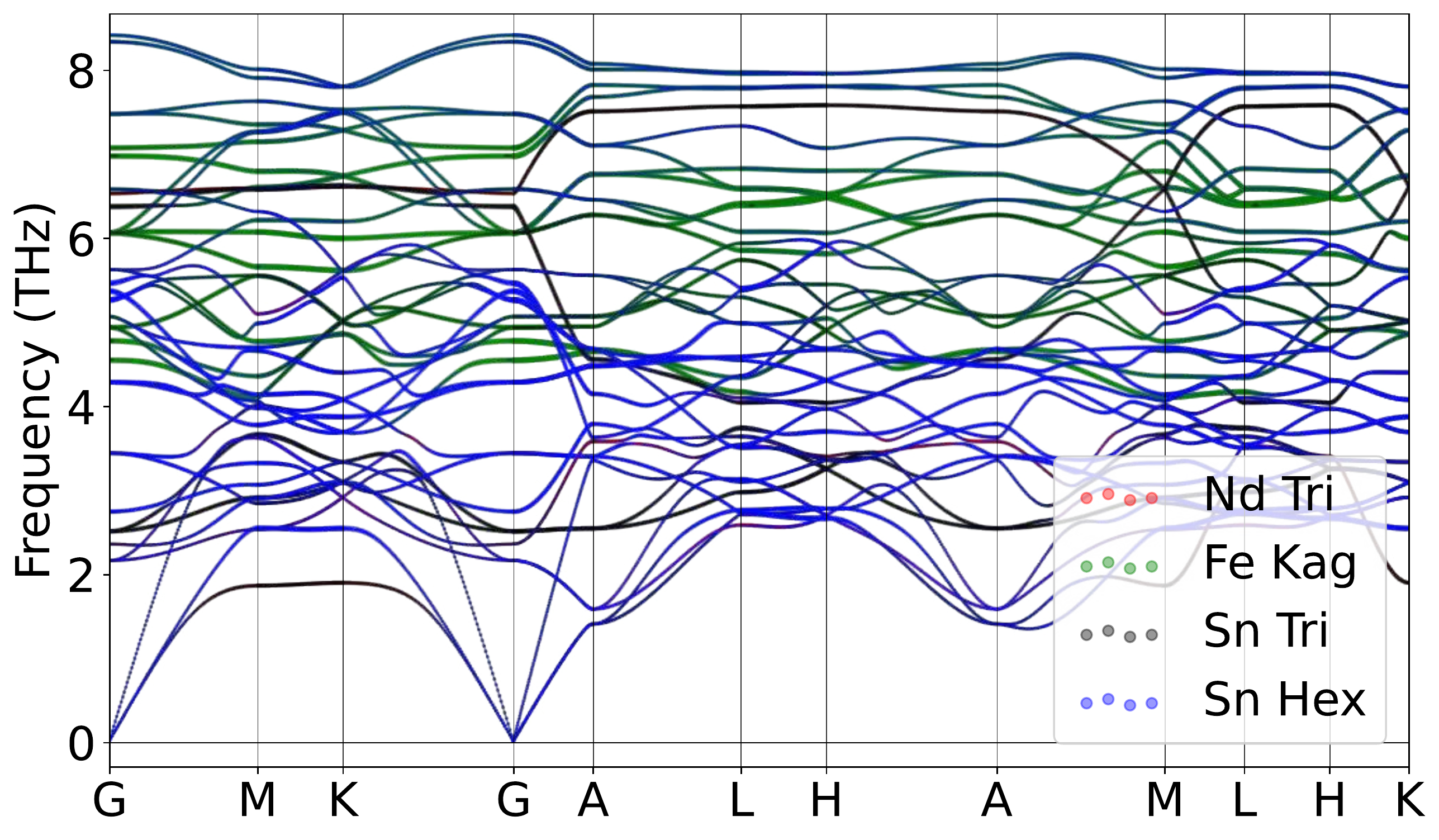}
}
\caption{Atomically projected phonon spectrum for NdFe$_6$Sn$_6$ in paramagnetic phase.}
\label{fig:NdFe6Sn6pho}
\end{figure}

\begin{figure}[H]
\centering
\label{fig:NdFe6Sn6_relaxed_Low_xz}
\includegraphics[width=0.85\textwidth]{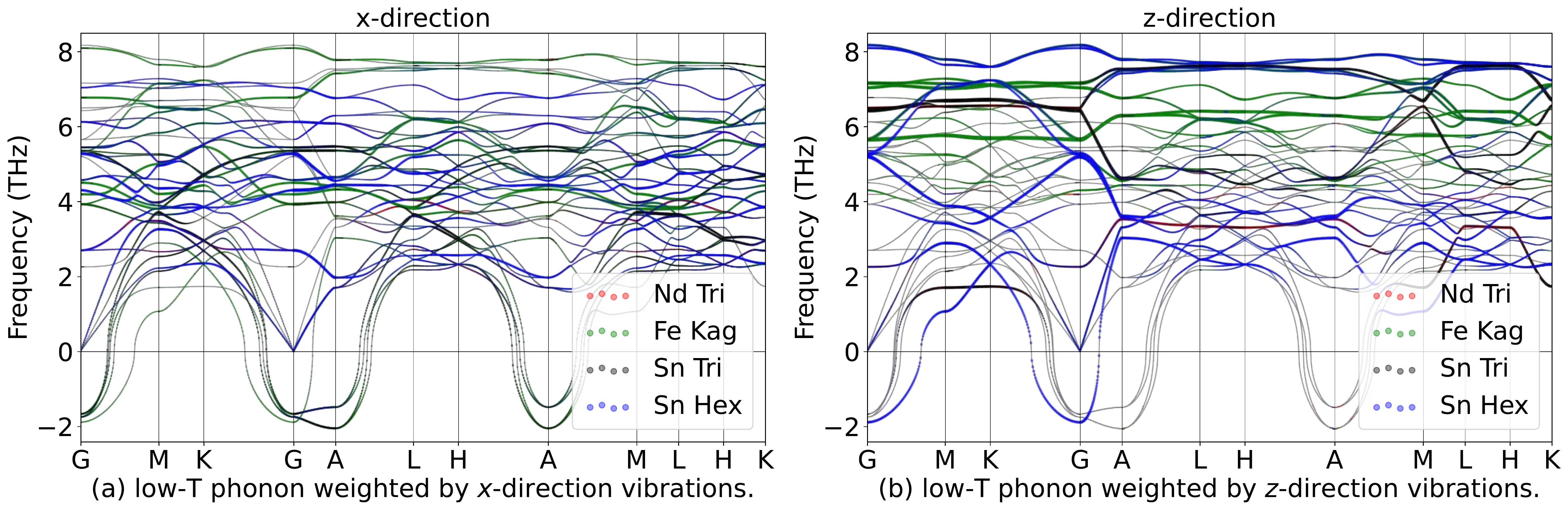}
\hspace{5pt}\label{fig:NdFe6Sn6_relaxed_High_xz}
\includegraphics[width=0.85\textwidth]{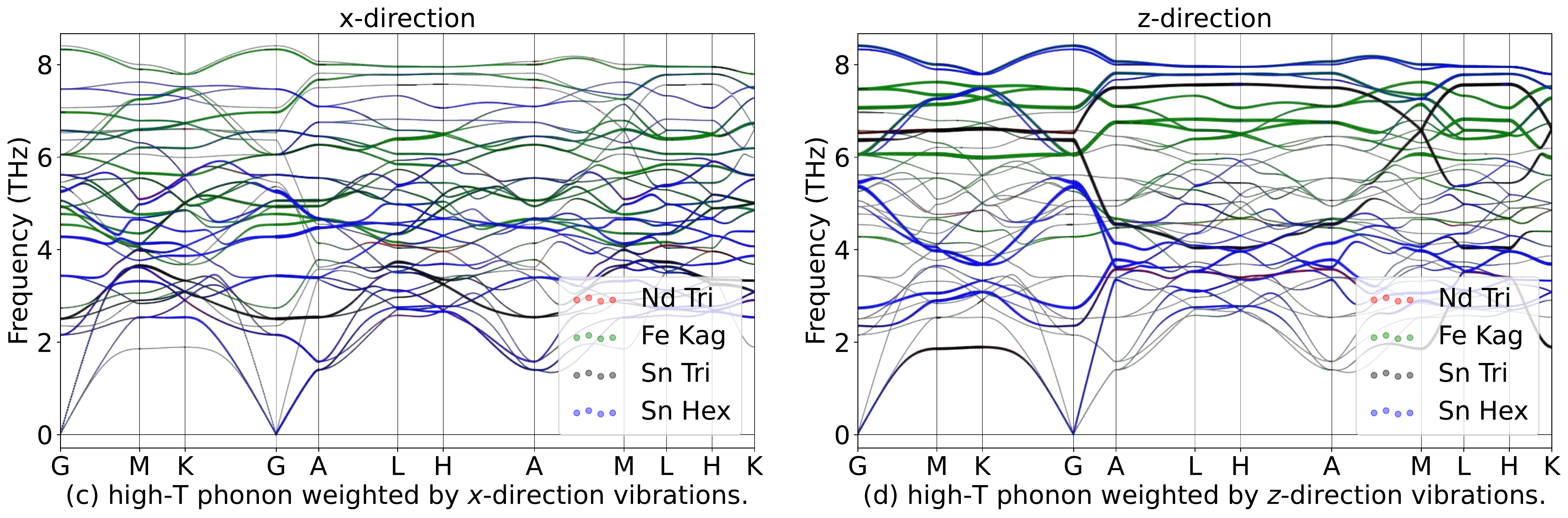}
\caption{Phonon spectrum for NdFe$_6$Sn$_6$in in-plane ($x$) and out-of-plane ($z$) directions in paramagnetic phase.}
\label{fig:NdFe6Sn6phoxyz}
\end{figure}

\begin{table}[htbp]
\global\long\def\arraystretch{1.12}
\captionsetup{width=.95\textwidth}
\caption{IRREPs at high-symmetry points for NdFe$_6$Sn$_6$ phonon spectrum at Low-T.}
\label{tab:191_NdFe6Sn6_Low}
\setlength{\tabcolsep}{1.mm}{
}
\end{table}

\begin{table}[htbp]
\global\long\def\arraystretch{1.12}
\captionsetup{width=.95\textwidth}
\caption{IRREPs at high-symmetry points for NdFe$_6$Sn$_6$ phonon spectrum at High-T.}
\label{tab:191_NdFe6Sn6_High}
\setlength{\tabcolsep}{1.mm}{
%
}
\end{table}
\subsubsection{\label{sms2sec:SmFe6Sn6}\hyperref[smsec:col]{\texorpdfstring{SmFe$_6$Sn$_6$}{}}~{\color{blue}  (High-T stable, Low-T unstable) }}

\begin{table}[htbp]
\global\long\def\arraystretch{1.12}
\captionsetup{width=.85\textwidth}
\caption{Structure parameters of SmFe$_6$Sn$_6$ after relaxation. It contains 518 electrons. }
\label{tab:SmFe6Sn6}
\setlength{\tabcolsep}{4mm}{
%
}
\end{table}

\begin{figure}[htbp]
\centering
\includegraphics[width=0.85\textwidth]{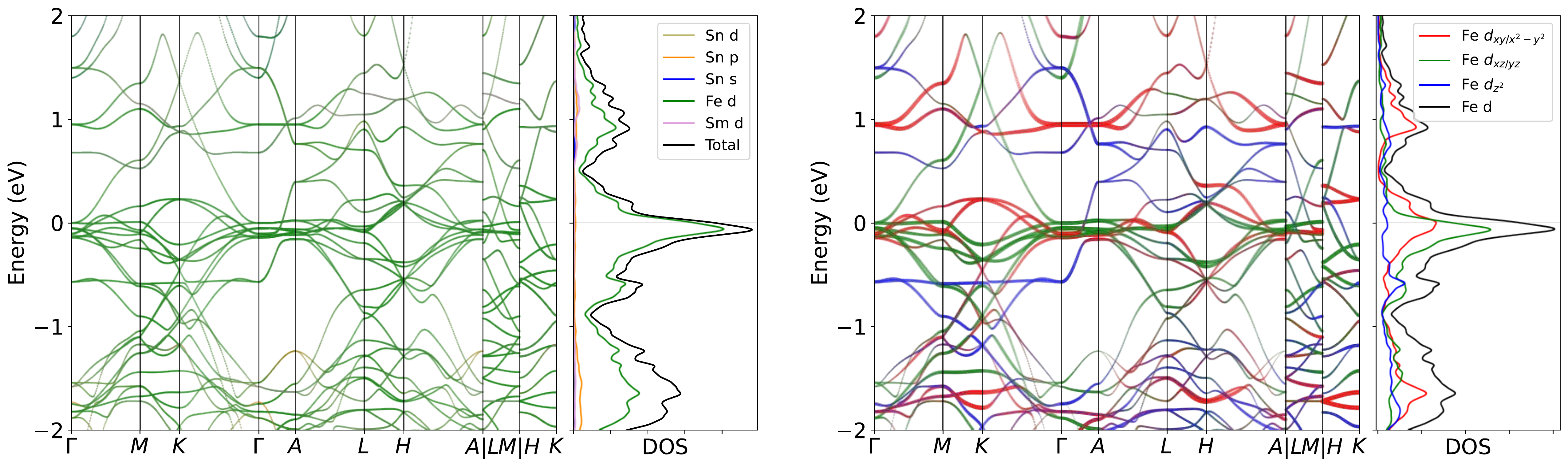}
\caption{Band structure for SmFe$_6$Sn$_6$ without SOC in paramagnetic phase. The projection of Fe $3d$ orbitals is also presented. The band structures clearly reveal the dominating of Fe $3d$ orbitals  near the Fermi level, as well as the pronounced peak.}
\label{fig:SmFe6Sn6band}
\end{figure}

\begin{figure}[H]
\centering
\subfloat[Relaxed phonon at low-T.]{
\label{fig:SmFe6Sn6_relaxed_Low}
\includegraphics[width=0.45\textwidth]{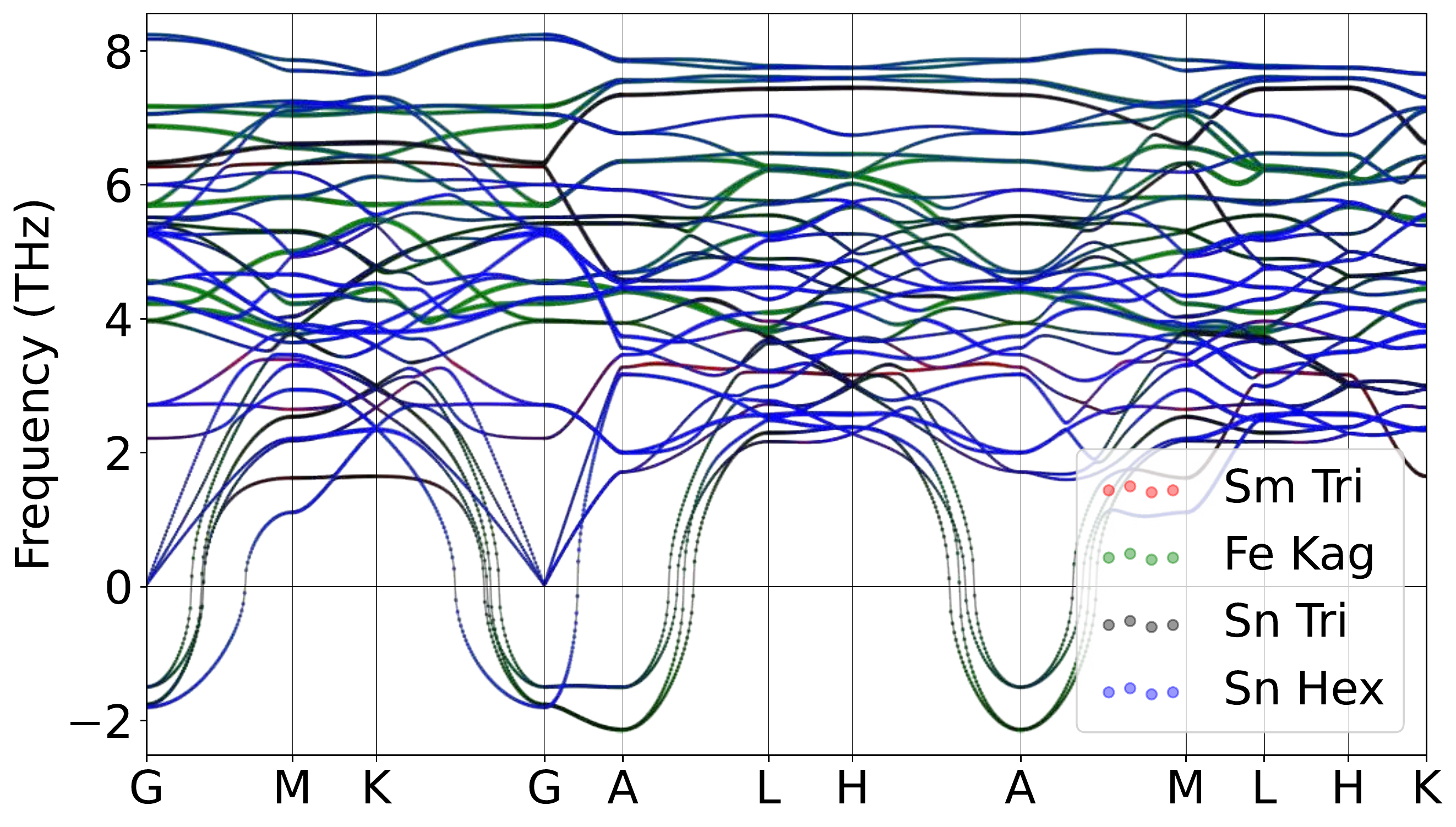}
}
\hspace{5pt}\subfloat[Relaxed phonon at high-T.]{
\label{fig:SmFe6Sn6_relaxed_High}
\includegraphics[width=0.45\textwidth]{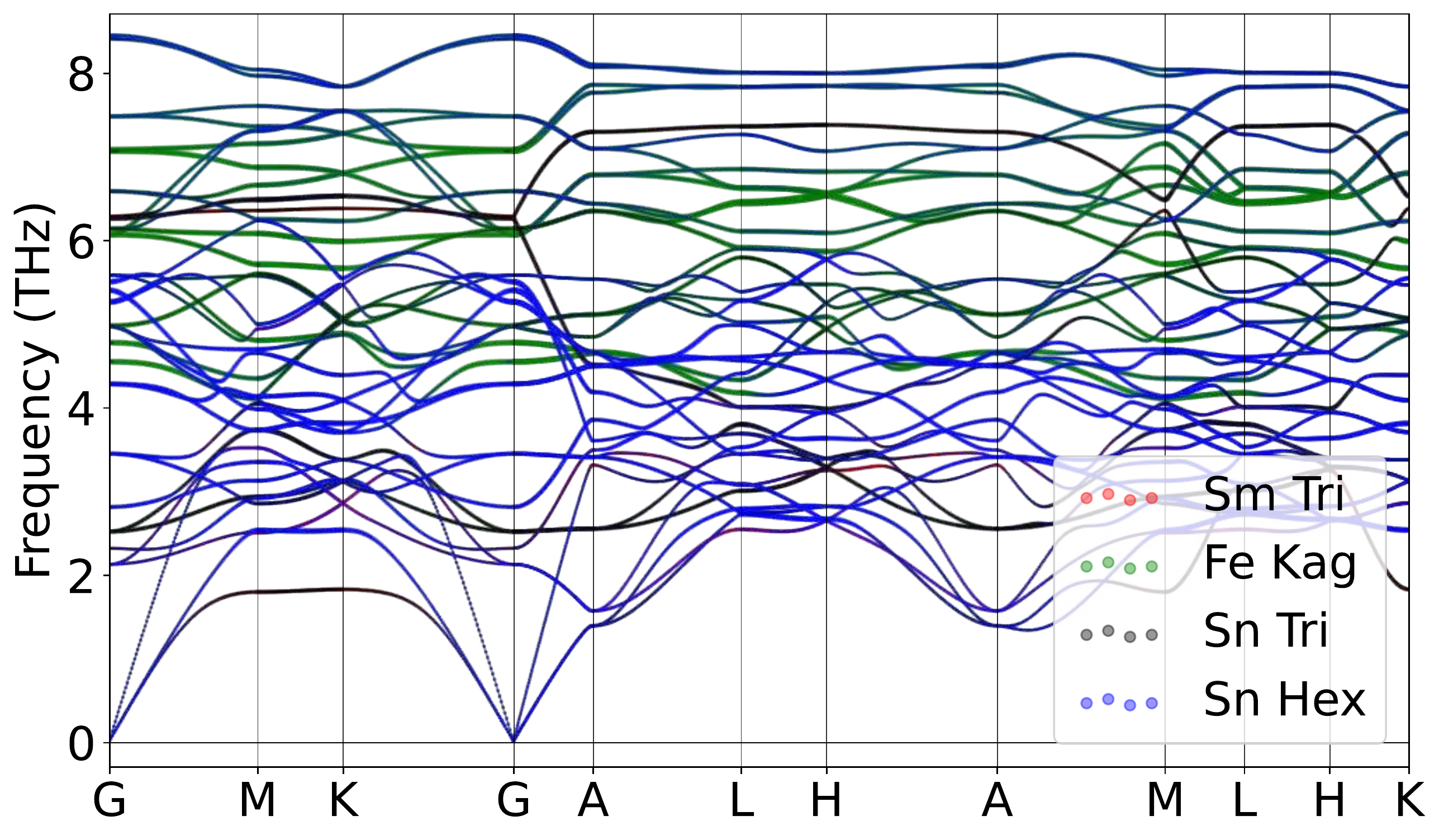}
}
\caption{Atomically projected phonon spectrum for SmFe$_6$Sn$_6$ in paramagnetic phase.}
\label{fig:SmFe6Sn6pho}
\end{figure}

\begin{figure}[htbp]
\centering
\includegraphics[width=0.45\textwidth]{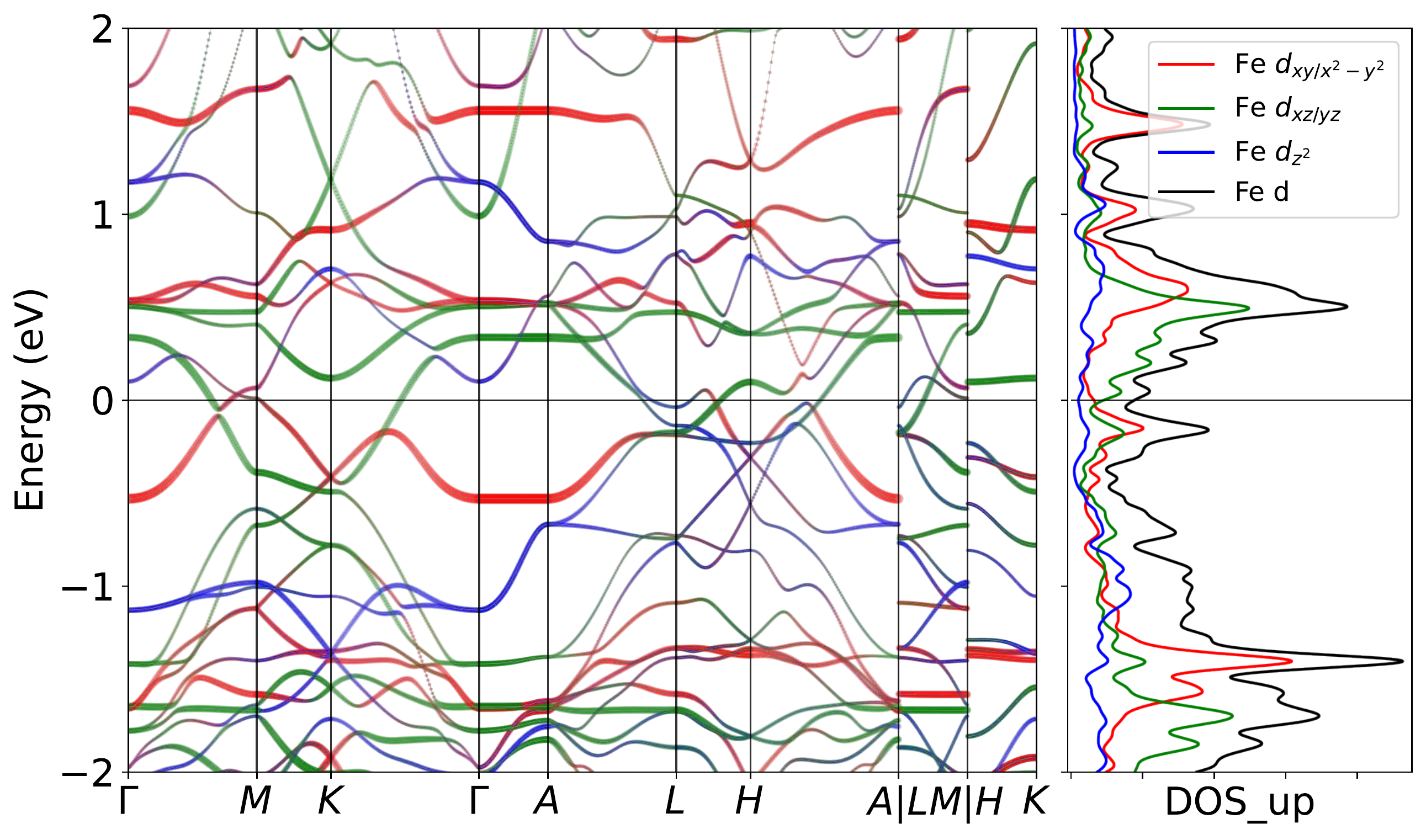}
\includegraphics[width=0.45\textwidth]{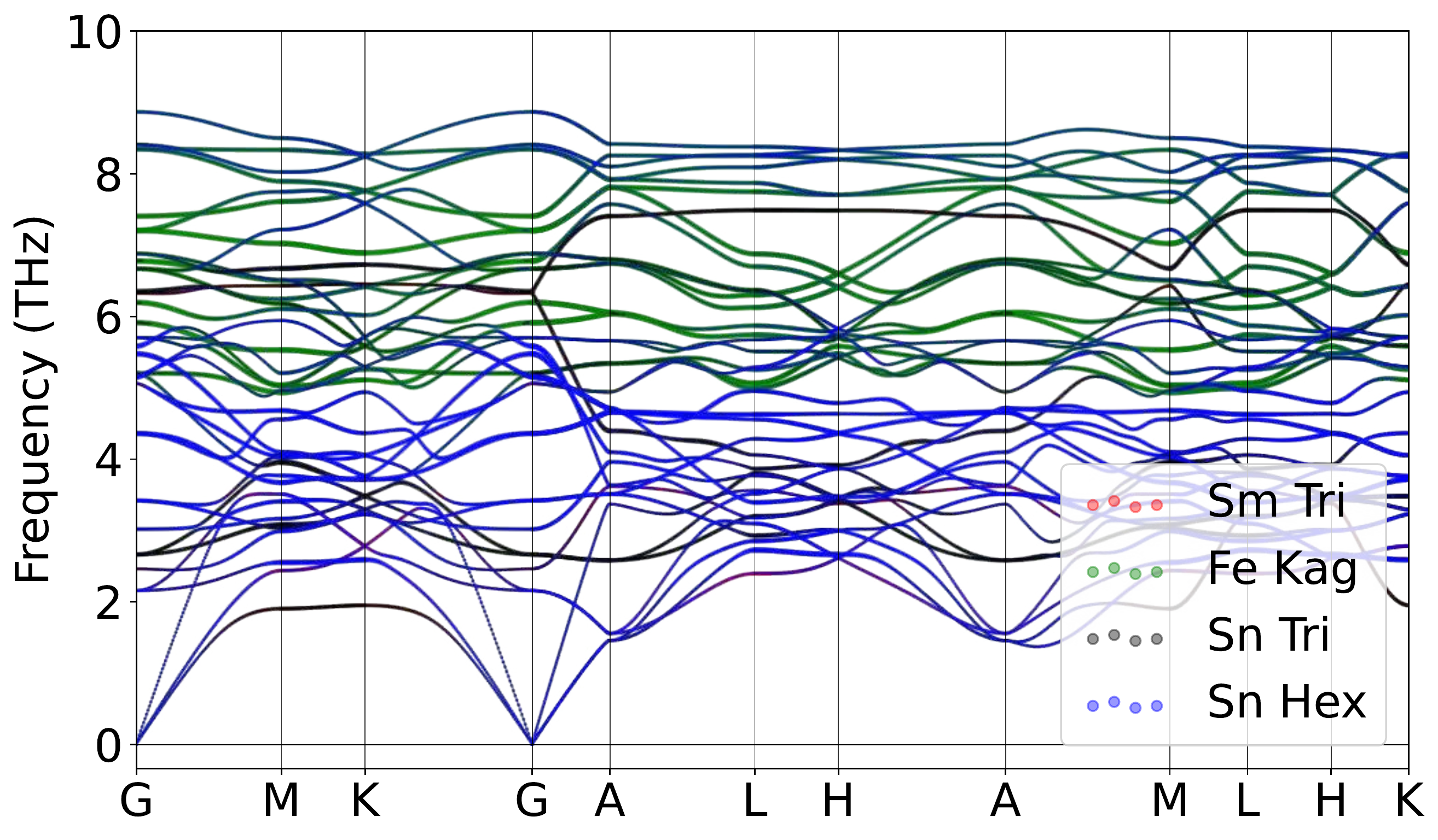}
\caption{Band structure for SmFe$_6$Sn$_6$ with AFM order without SOC for spin (Left)up channel. The projection of Fe $3d$ orbitals is also presented. (Right)Correspondingly, a stable phonon was demonstrated with the collinear magnetic order without Hubbard U at lower temperature regime, weighted by atomic projections.}
\label{fig:SmFe6Sn6mag}
\end{figure}

\begin{figure}[H]
\centering
\label{fig:SmFe6Sn6_relaxed_Low_xz}
\includegraphics[width=0.85\textwidth]{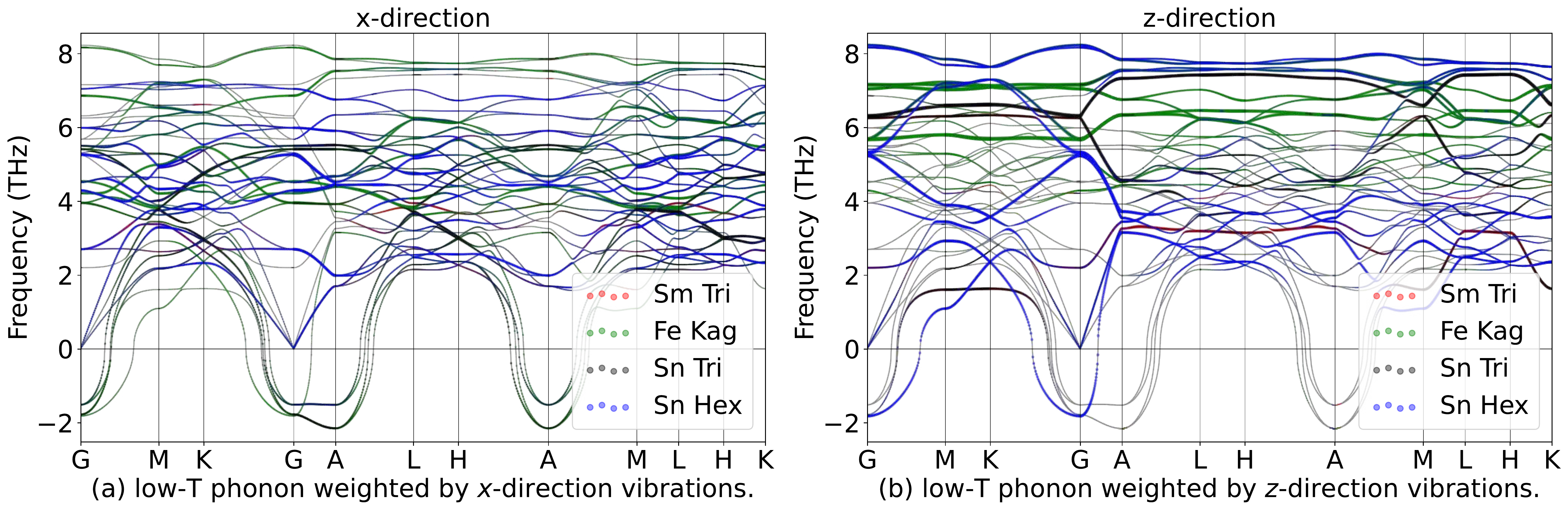}
\hspace{5pt}\label{fig:SmFe6Sn6_relaxed_High_xz}
\includegraphics[width=0.85\textwidth]{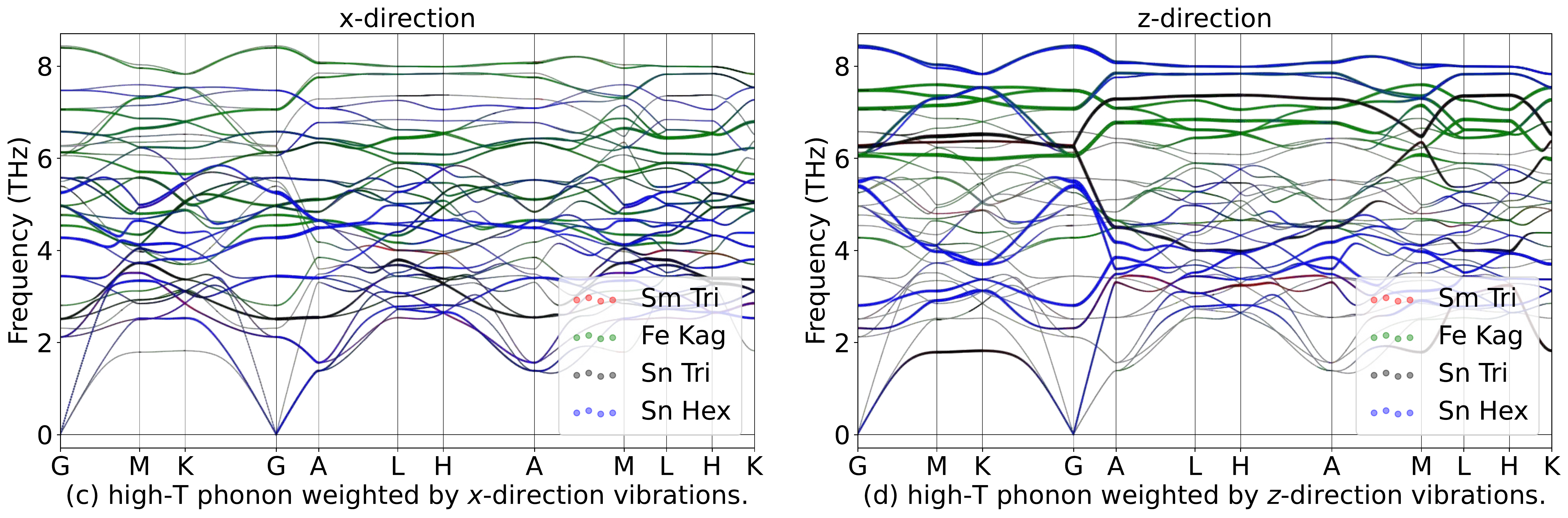}
\caption{Phonon spectrum for SmFe$_6$Sn$_6$in in-plane ($x$) and out-of-plane ($z$) directions in paramagnetic phase.}
\label{fig:SmFe6Sn6phoxyz}
\end{figure}

\begin{table}[htbp]
\global\long\def\arraystretch{1.12}
\captionsetup{width=.95\textwidth}
\caption{IRREPs at high-symmetry points for SmFe$_6$Sn$_6$ phonon spectrum at Low-T.}
\label{tab:191_SmFe6Sn6_Low}
\setlength{\tabcolsep}{1.mm}{
}
\end{table}

\begin{table}[htbp]
\global\long\def\arraystretch{1.12}
\captionsetup{width=.95\textwidth}
\caption{IRREPs at high-symmetry points for SmFe$_6$Sn$_6$ phonon spectrum at High-T.}
\label{tab:191_SmFe6Sn6_High}
\setlength{\tabcolsep}{1.mm}{
%
}
\end{table}
\subsubsection{\label{sms2sec:GdFe6Sn6}\hyperref[smsec:col]{\texorpdfstring{GdFe$_6$Sn$_6$}{}}~{\color{blue}  (High-T stable, Low-T unstable) }}

\begin{table}[htbp]
\global\long\def\arraystretch{1.12}
\captionsetup{width=.85\textwidth}
\caption{Structure parameters of GdFe$_6$Sn$_6$ after relaxation. It contains 520 electrons. }
\label{tab:GdFe6Sn6}
\setlength{\tabcolsep}{4mm}{
%
}
\end{table}

\begin{figure}[htbp]
\centering
\includegraphics[width=0.85\textwidth]{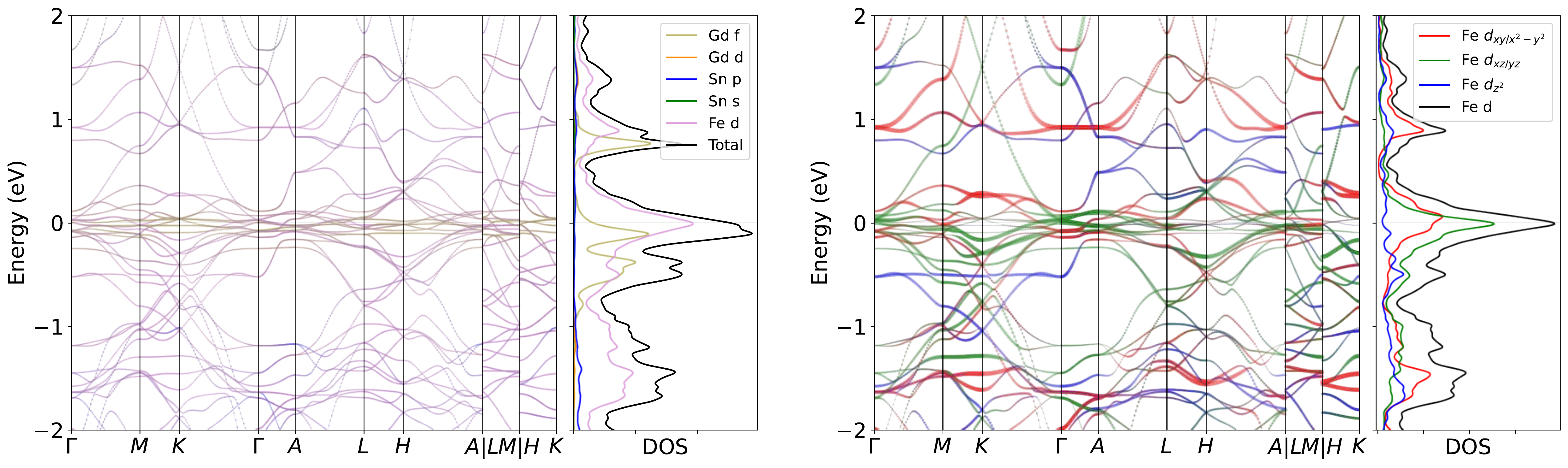}
\caption{Band structure for GdFe$_6$Sn$_6$ without SOC in paramagnetic phase. The projection of Fe $3d$ orbitals is also presented. The band structures clearly reveal the dominating of Fe $3d$ orbitals  near the Fermi level, as well as the pronounced peak.}
\label{fig:GdFe6Sn6band}
\end{figure}

\begin{figure}[H]
\centering
\subfloat[Relaxed phonon at low-T.]{
\label{fig:GdFe6Sn6_relaxed_Low}
\includegraphics[width=0.45\textwidth]{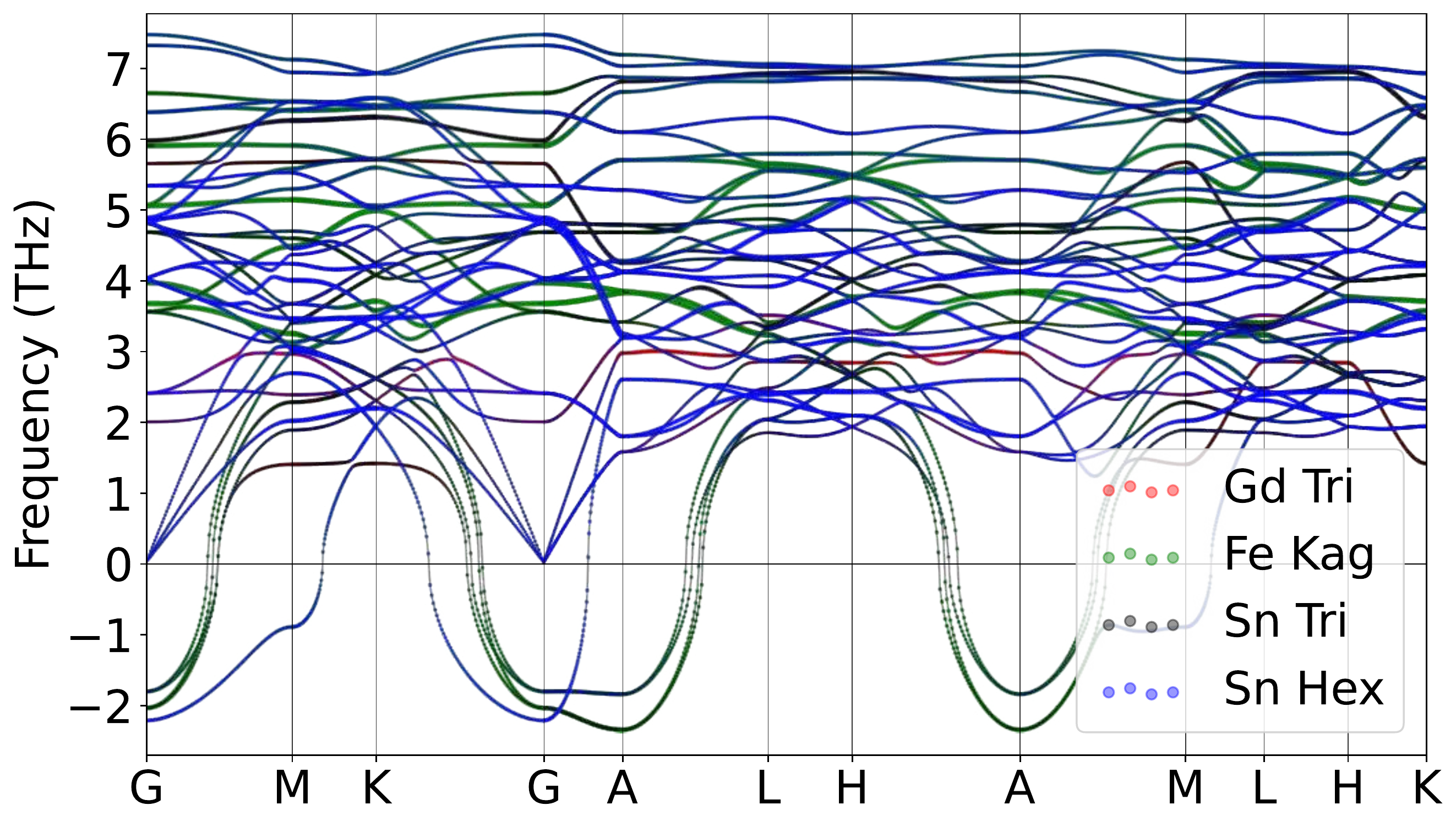}
}
\hspace{5pt}\subfloat[Relaxed phonon at high-T.]{
\label{fig:GdFe6Sn6_relaxed_High}
\includegraphics[width=0.45\textwidth]{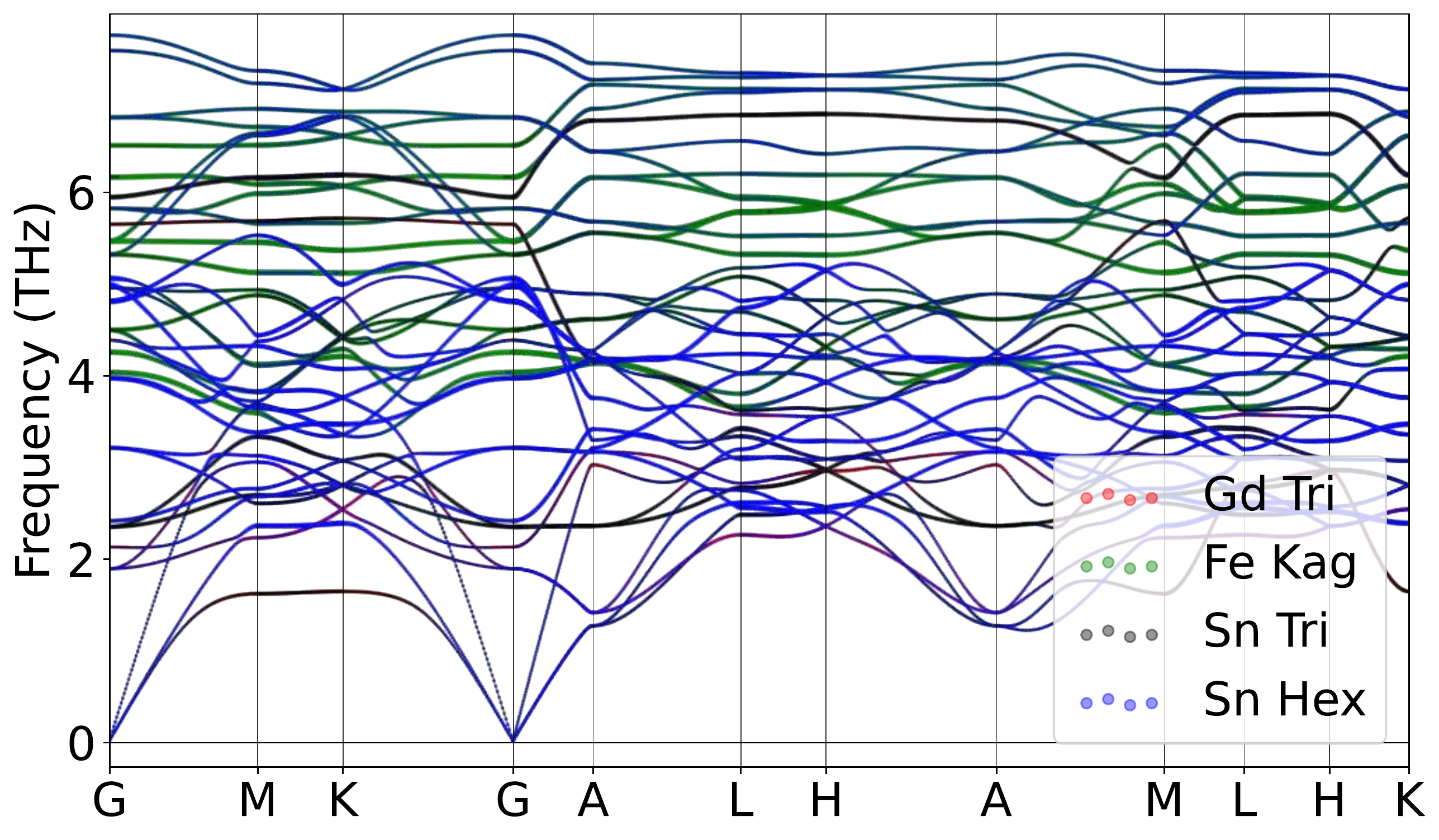}
}
\caption{Atomically projected phonon spectrum for GdFe$_6$Sn$_6$ in paramagnetic phase.}
\label{fig:GdFe6Sn6pho}
\end{figure}

\begin{figure}[htbp]
\centering
\includegraphics[width=0.45\textwidth]{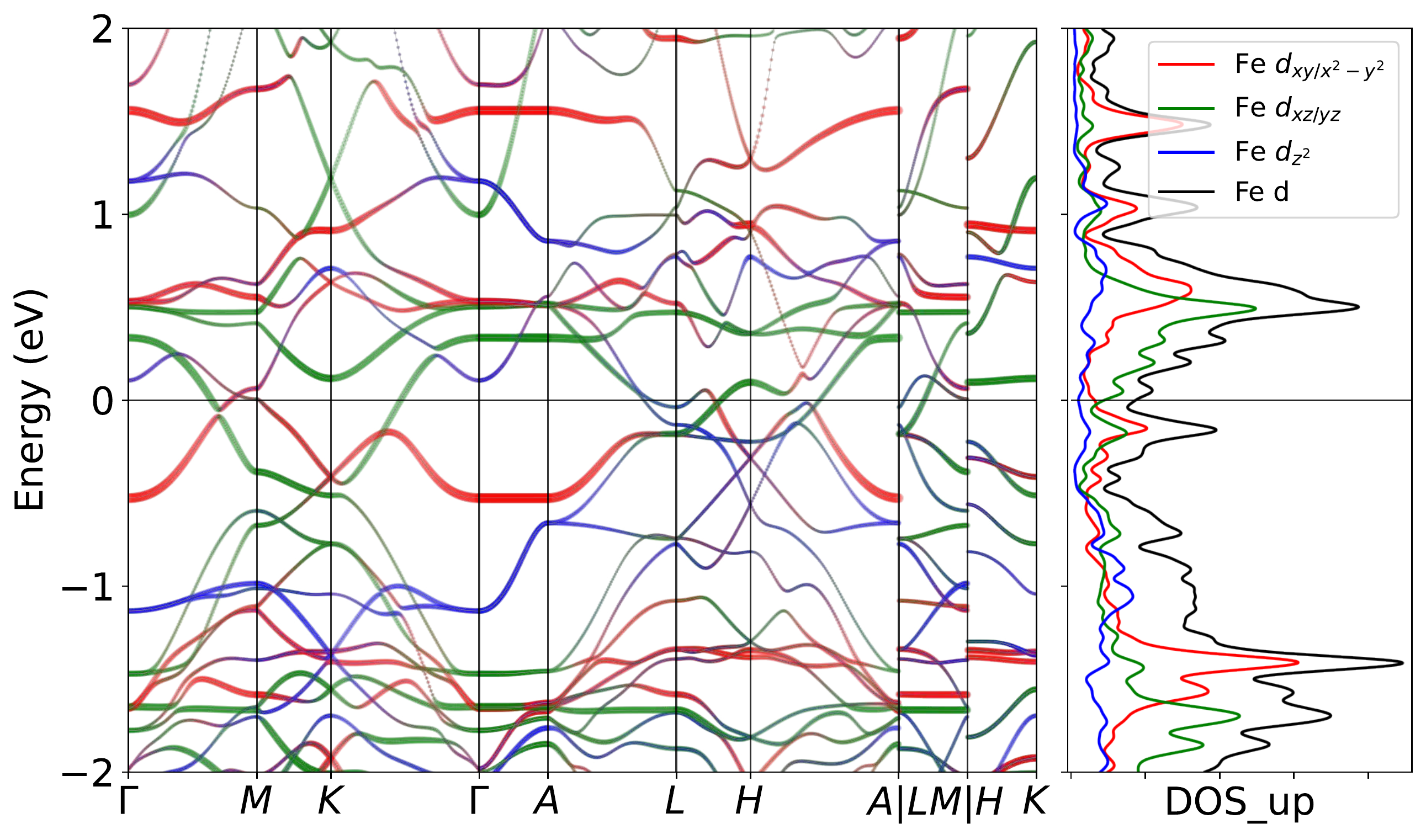}
\includegraphics[width=0.45\textwidth]{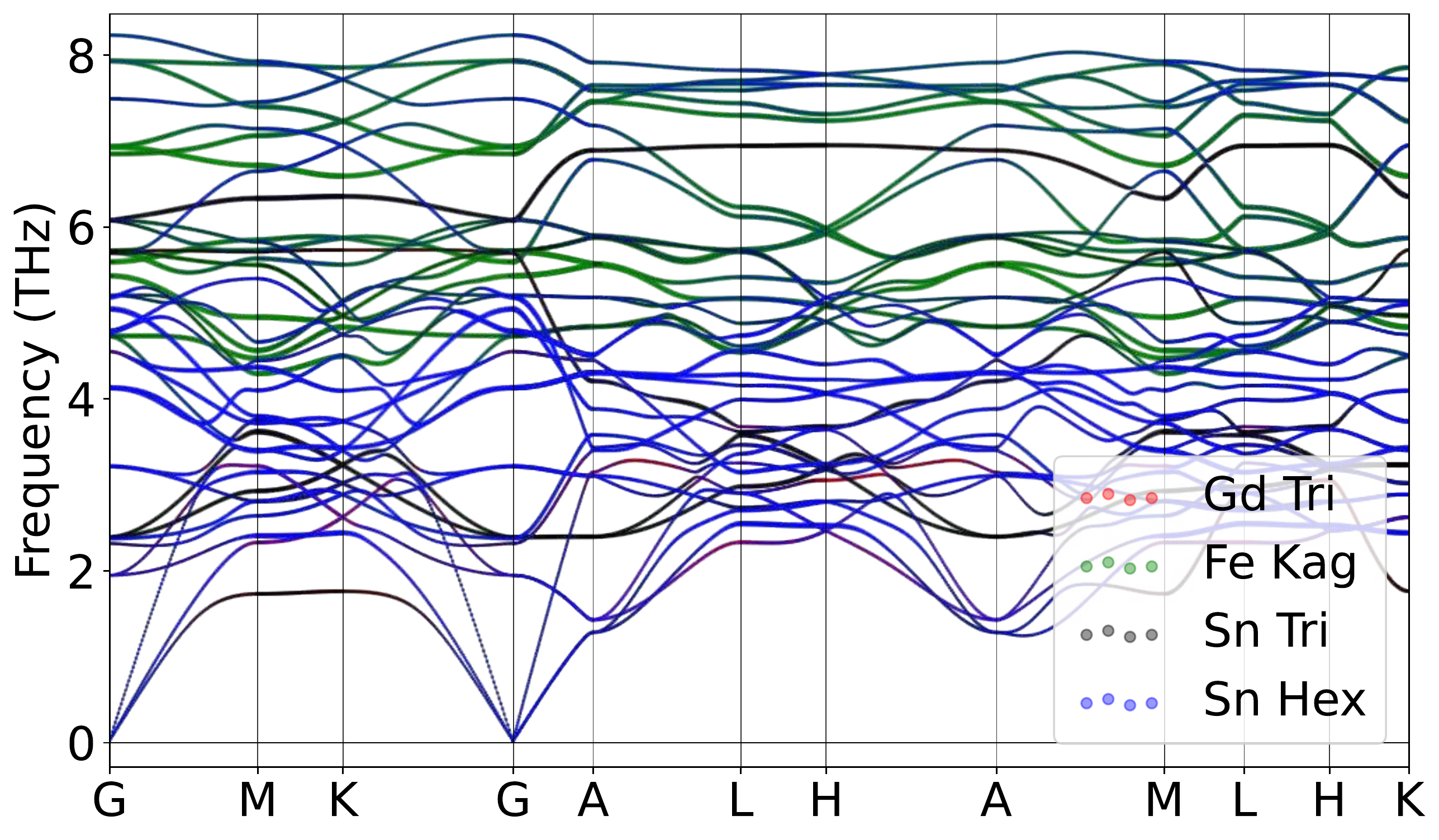}
\caption{Band structure for GdFe$_6$Sn$_6$ with AFM order without SOC for spin (Left)up channel. The projection of Fe $3d$ orbitals is also presented. (Right)Correspondingly, a stable phonon was demonstrated with the collinear magnetic order without Hubbard U at lower temperature regime, weighted by atomic projections.}
\label{fig:GdFe6Sn6mag}
\end{figure}

\begin{figure}[H]
\centering
\label{fig:GdFe6Sn6_relaxed_Low_xz}
\includegraphics[width=0.85\textwidth]{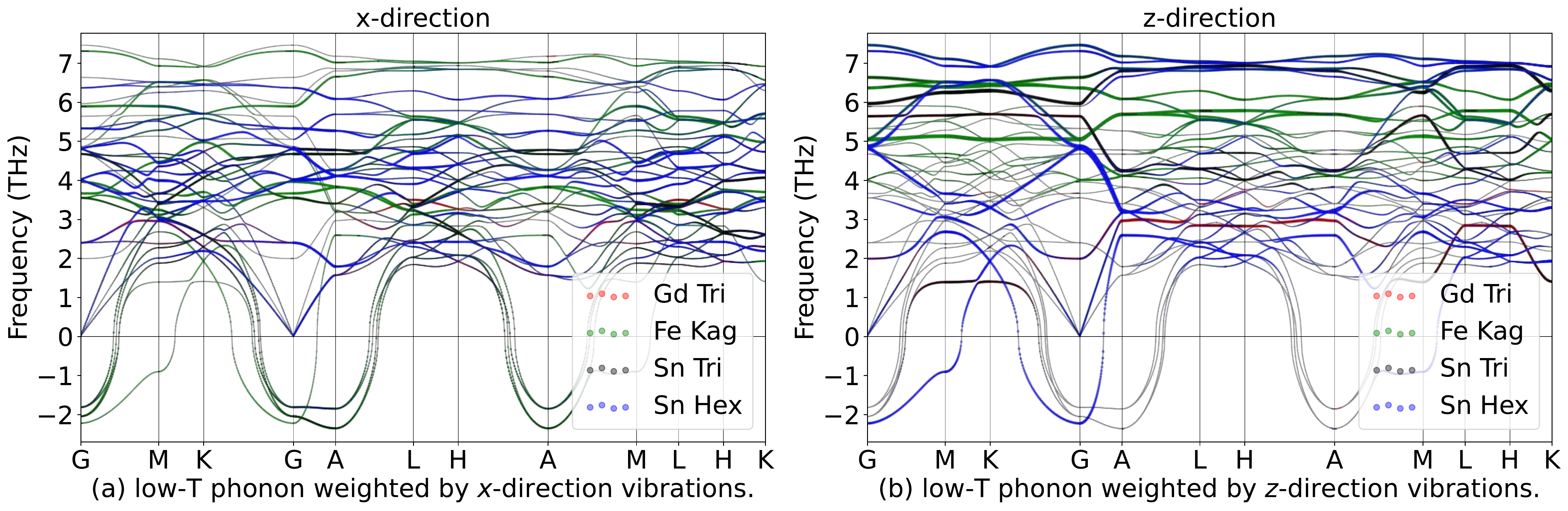}
\hspace{5pt}\label{fig:GdFe6Sn6_relaxed_High_xz}
\includegraphics[width=0.85\textwidth]{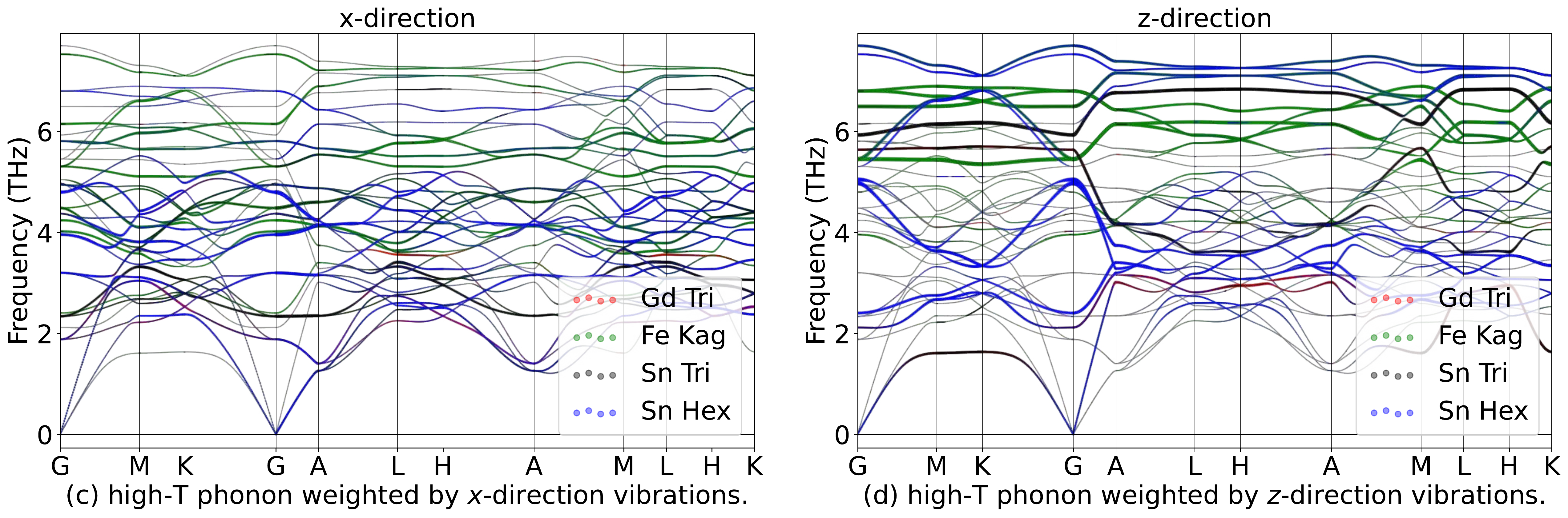}
\caption{Phonon spectrum for GdFe$_6$Sn$_6$in in-plane ($x$) and out-of-plane ($z$) directions in paramagnetic phase.}
\label{fig:GdFe6Sn6phoxyz}
\end{figure}

\begin{table}[htbp]
\global\long\def\arraystretch{1.12}
\captionsetup{width=.95\textwidth}
\caption{IRREPs at high-symmetry points for GdFe$_6$Sn$_6$ phonon spectrum at Low-T.}
\label{tab:191_GdFe6Sn6_Low}
\setlength{\tabcolsep}{1.mm}{
}
\end{table}

\begin{table}[htbp]
\global\long\def\arraystretch{1.12}
\captionsetup{width=.95\textwidth}
\caption{IRREPs at high-symmetry points for GdFe$_6$Sn$_6$ phonon spectrum at High-T.}
\label{tab:191_GdFe6Sn6_High}
\setlength{\tabcolsep}{1.mm}{
%
}
\end{table}
\subsubsection{\label{sms2sec:TbFe6Sn6}\hyperref[smsec:col]{\texorpdfstring{TbFe$_6$Sn$_6$}{}}~{\color{blue}  (High-T stable, Low-T unstable) }}

\begin{table}[htbp]
\global\long\def\arraystretch{1.12}
\captionsetup{width=.85\textwidth}
\caption{Structure parameters of TbFe$_6$Sn$_6$ after relaxation. It contains 521 electrons. }
\label{tab:TbFe6Sn6}
\setlength{\tabcolsep}{4mm}{
%
}
\end{table}

\begin{figure}[htbp]
\centering
\includegraphics[width=0.85\textwidth]{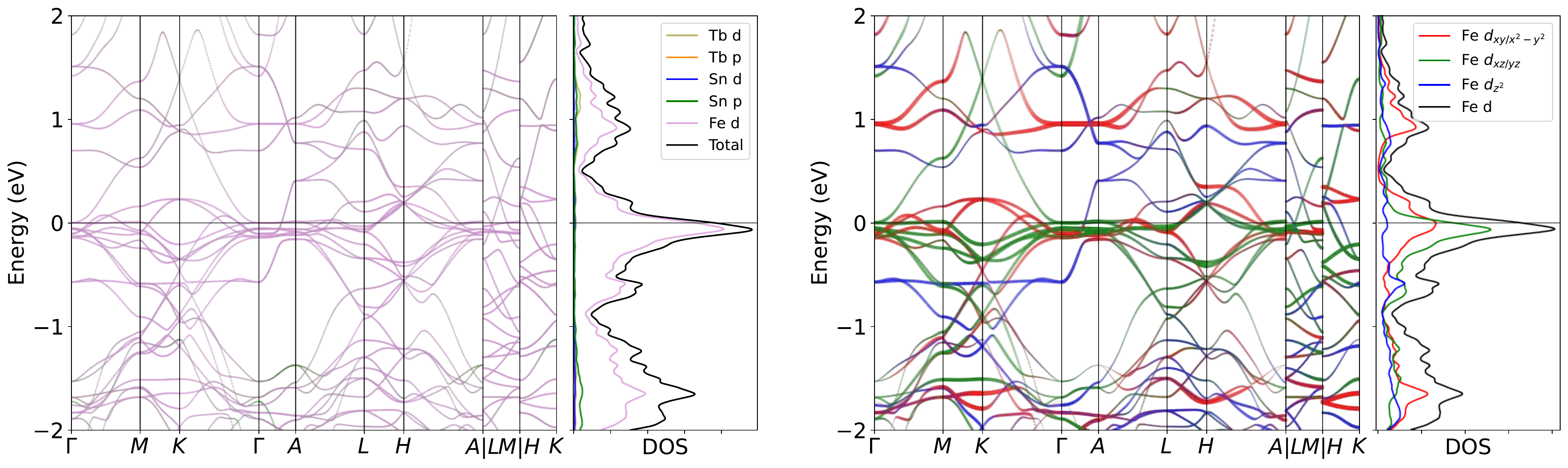}
\caption{Band structure for TbFe$_6$Sn$_6$ without SOC in paramagnetic phase. The projection of Fe $3d$ orbitals is also presented. The band structures clearly reveal the dominating of Fe $3d$ orbitals  near the Fermi level, as well as the pronounced peak.}
\label{fig:TbFe6Sn6band}
\end{figure}

\begin{figure}[H]
\centering
\subfloat[Relaxed phonon at low-T.]{
\label{fig:TbFe6Sn6_relaxed_Low}
\includegraphics[width=0.45\textwidth]{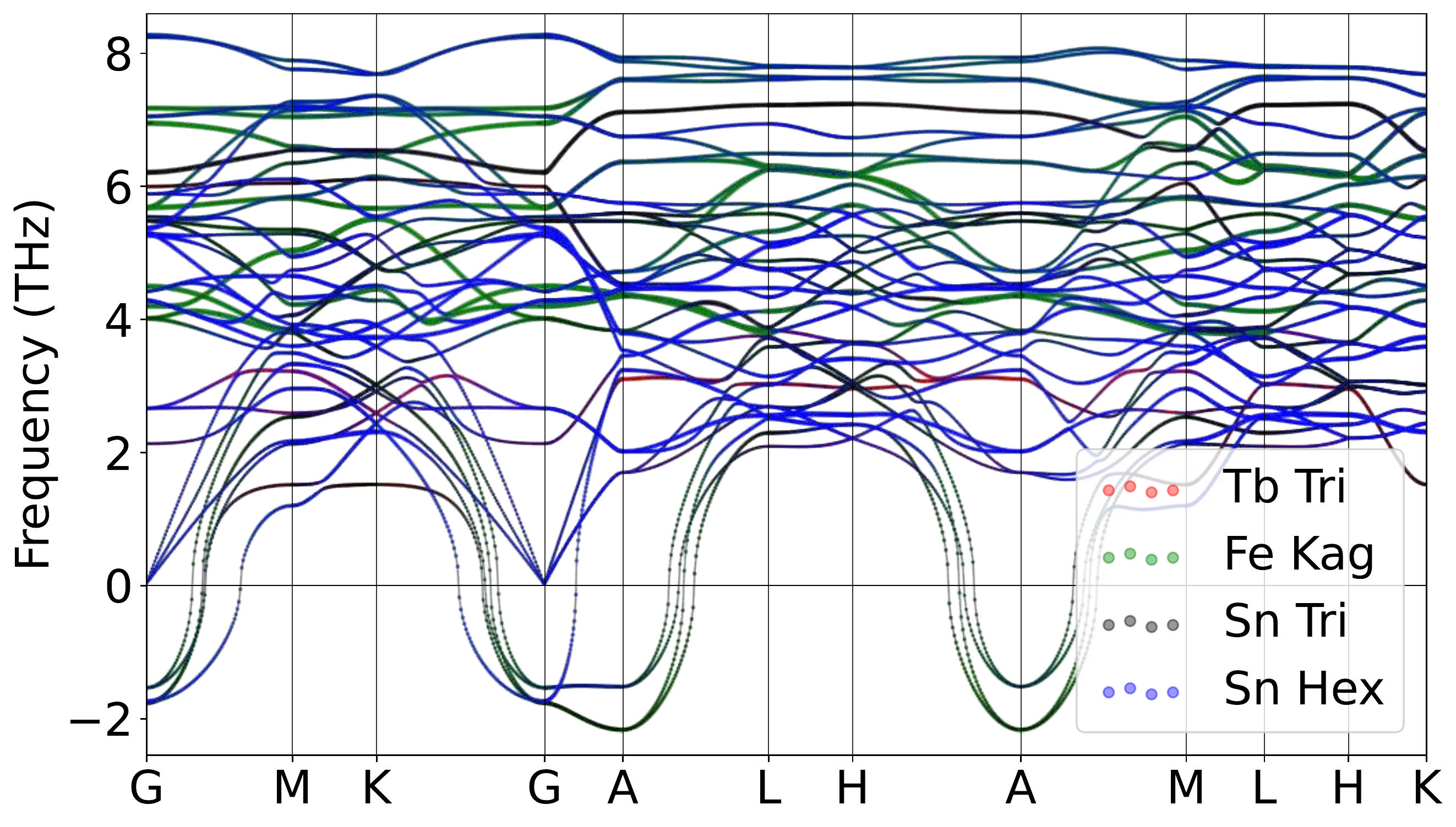}
}
\hspace{5pt}\subfloat[Relaxed phonon at high-T.]{
\label{fig:TbFe6Sn6_relaxed_High}
\includegraphics[width=0.45\textwidth]{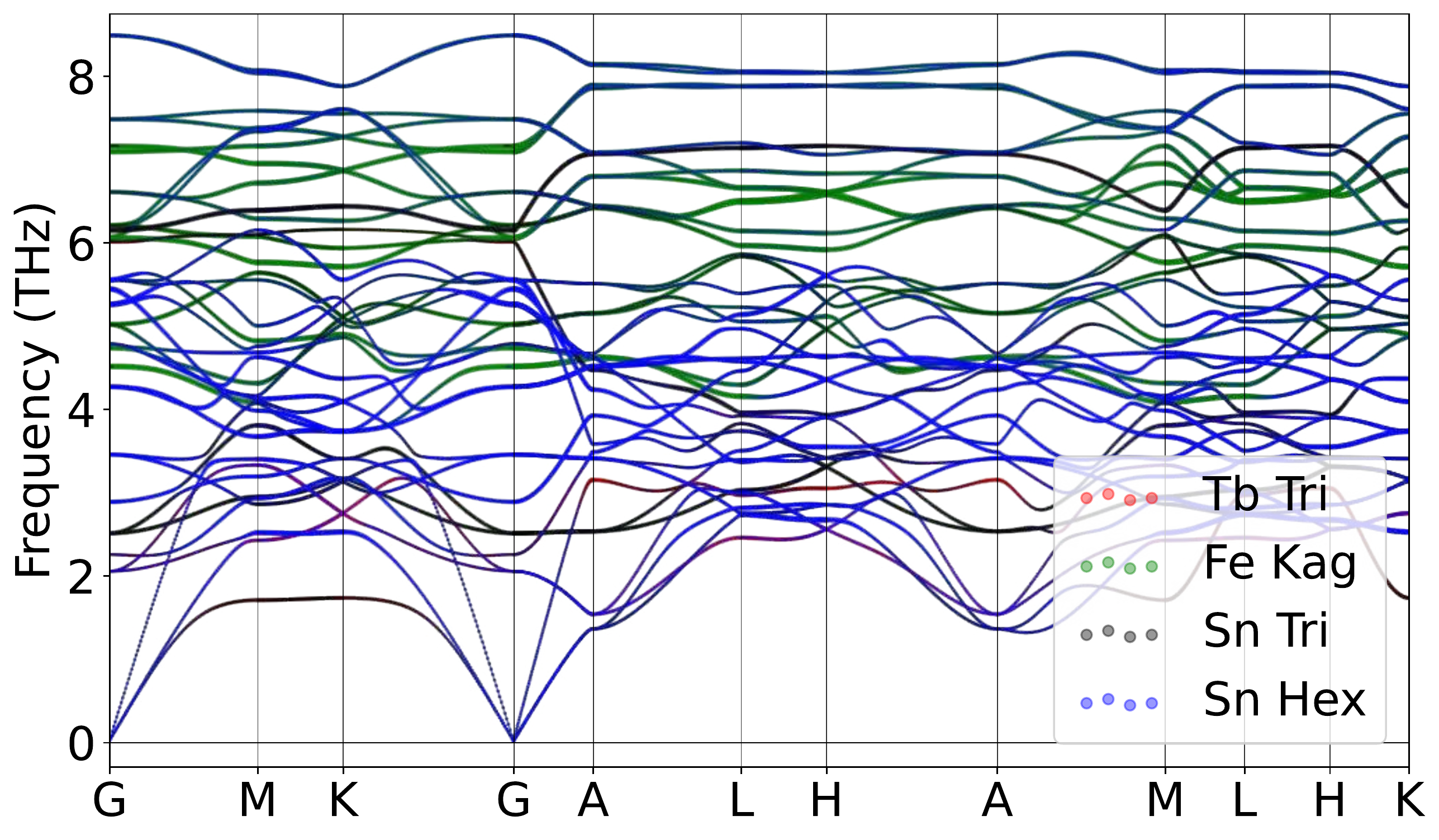}
}
\caption{Atomically projected phonon spectrum for TbFe$_6$Sn$_6$ in paramagnetic phase.}
\label{fig:TbFe6Sn6pho}
\end{figure}

\begin{figure}[htbp]
\centering
\includegraphics[width=0.45\textwidth]{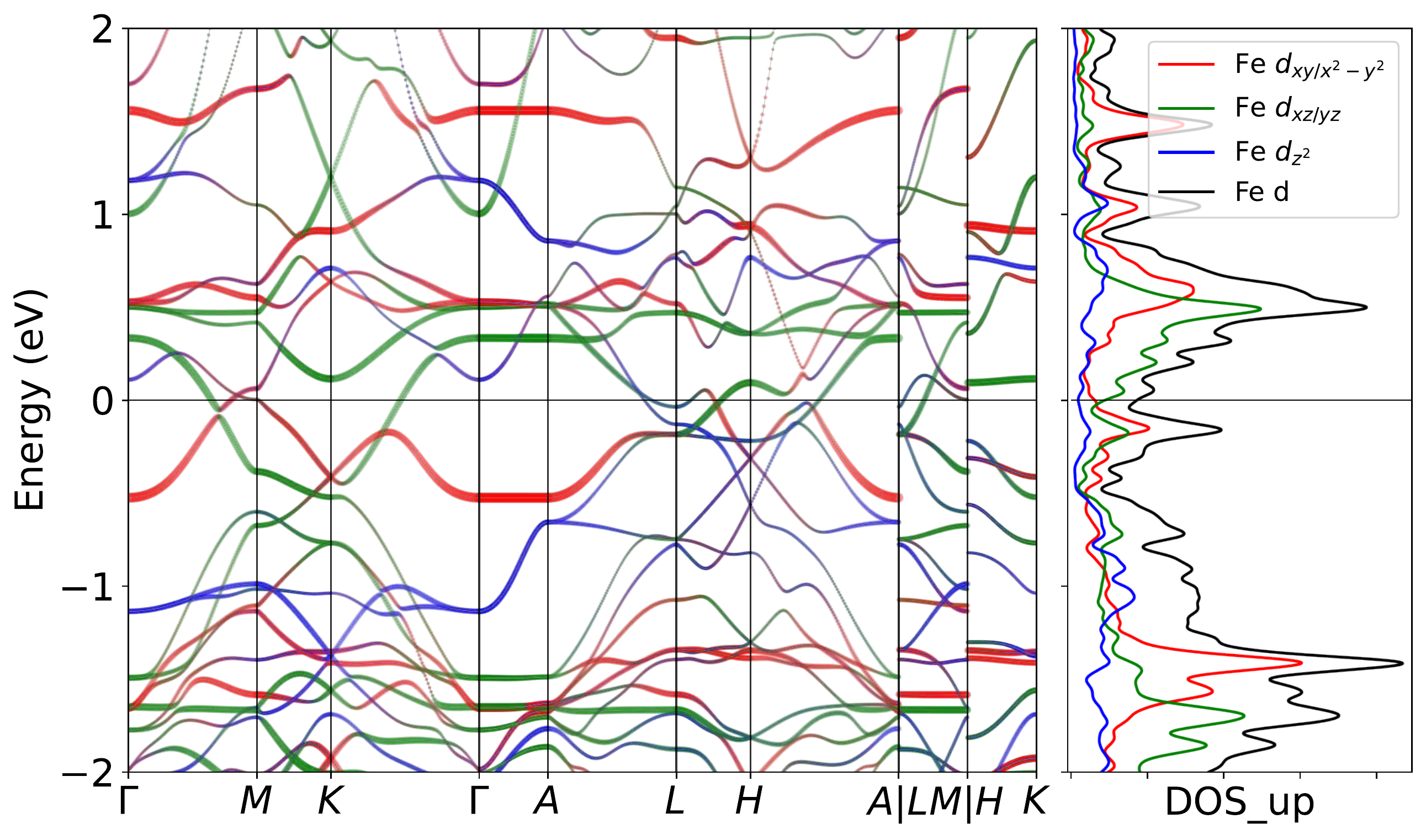}
\includegraphics[width=0.45\textwidth]{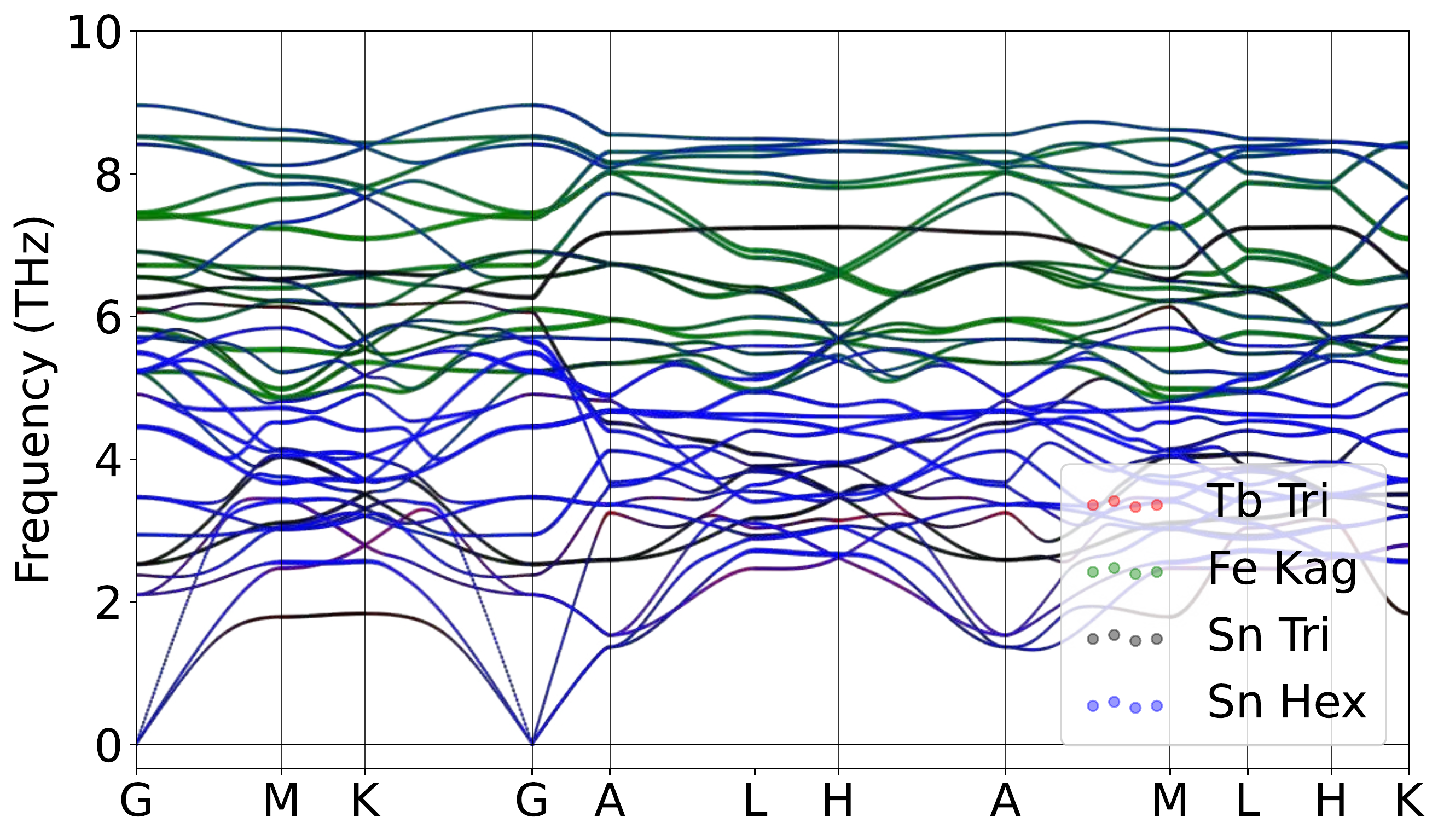}
\caption{Band structure for TbFe$_6$Sn$_6$ with AFM order without SOC for spin (Left)up channel. The projection of Fe $3d$ orbitals is also presented. (Right)Correspondingly, a stable phonon was demonstrated with the collinear magnetic order without Hubbard U at lower temperature regime, weighted by atomic projections.}
\label{fig:TbFe6Sn6mag}
\end{figure}

\begin{figure}[H]
\centering
\label{fig:TbFe6Sn6_relaxed_Low_xz}
\includegraphics[width=0.85\textwidth]{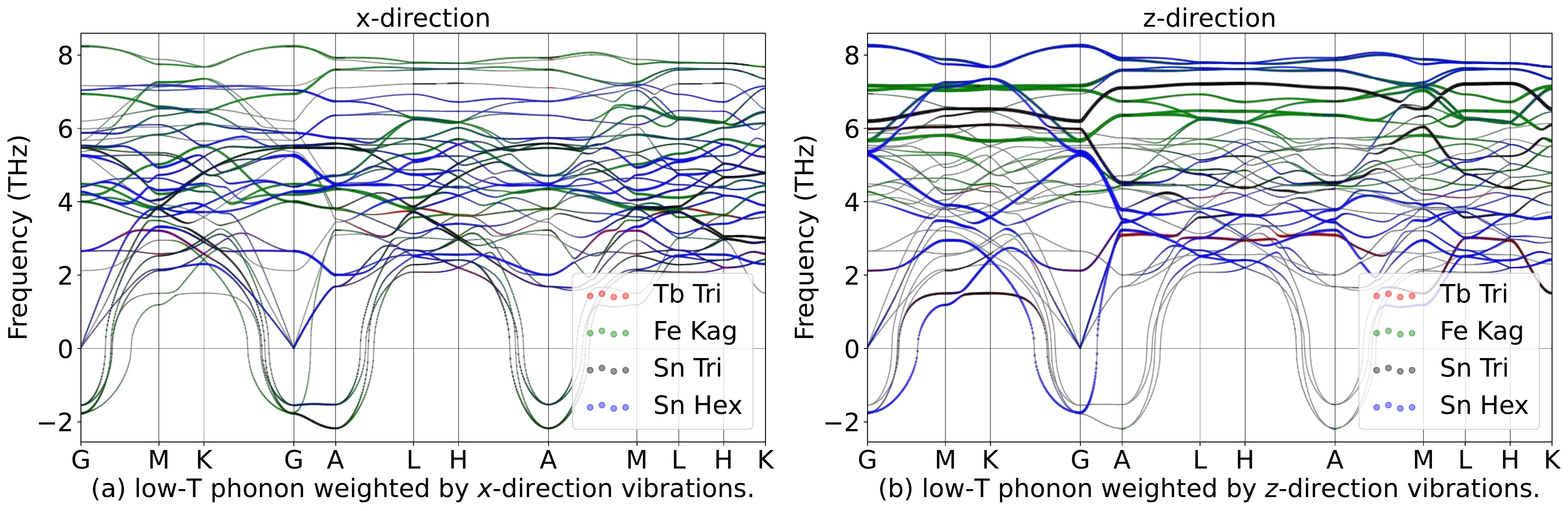}
\hspace{5pt}\label{fig:TbFe6Sn6_relaxed_High_xz}
\includegraphics[width=0.85\textwidth]{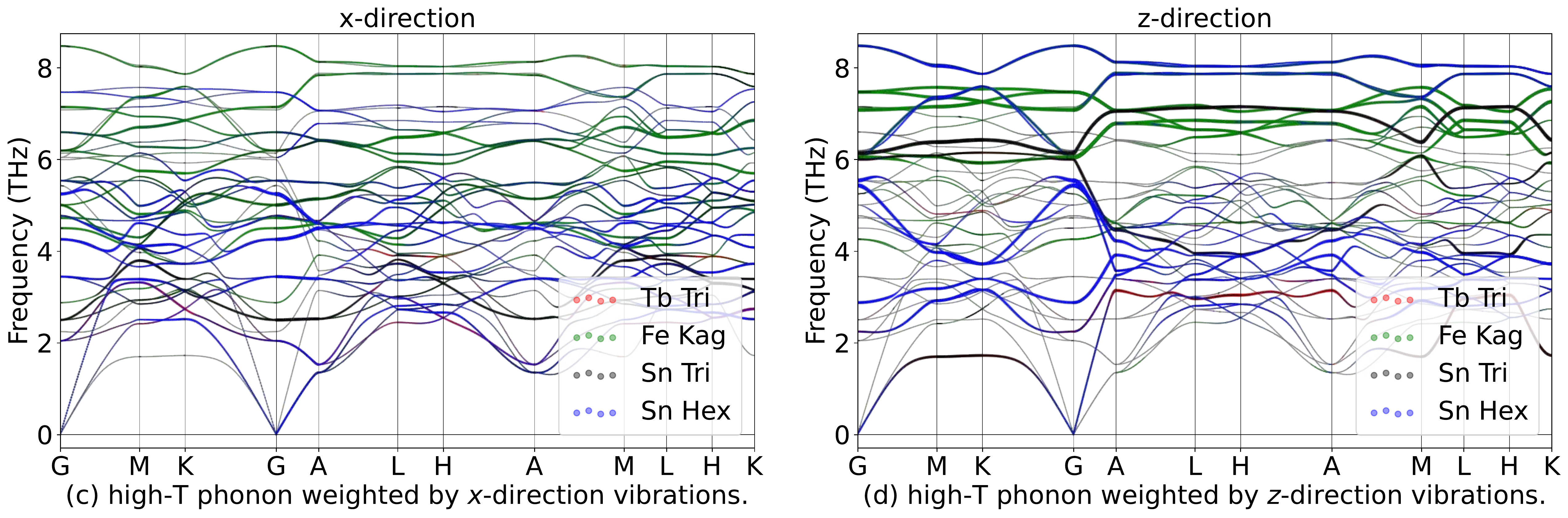}
\caption{Phonon spectrum for TbFe$_6$Sn$_6$in in-plane ($x$) and out-of-plane ($z$) directions in paramagnetic phase.}
\label{fig:TbFe6Sn6phoxyz}
\end{figure}

\begin{table}[htbp]
\global\long\def\arraystretch{1.12}
\captionsetup{width=.95\textwidth}
\caption{IRREPs at high-symmetry points for TbFe$_6$Sn$_6$ phonon spectrum at Low-T.}
\label{tab:191_TbFe6Sn6_Low}
\setlength{\tabcolsep}{1.mm}{
}
\end{table}

\begin{table}[htbp]
\global\long\def\arraystretch{1.12}
\captionsetup{width=.95\textwidth}
\caption{IRREPs at high-symmetry points for TbFe$_6$Sn$_6$ phonon spectrum at High-T.}
\label{tab:191_TbFe6Sn6_High}
\setlength{\tabcolsep}{1.mm}{
%
}
\end{table}
\subsubsection{\label{sms2sec:DyFe6Sn6}\hyperref[smsec:col]{\texorpdfstring{DyFe$_6$Sn$_6$}{}}~{\color{blue}  (High-T stable, Low-T unstable) }}

\begin{table}[htbp]
\global\long\def\arraystretch{1.12}
\captionsetup{width=.85\textwidth}
\caption{Structure parameters of DyFe$_6$Sn$_6$ after relaxation. It contains 522 electrons. }
\label{tab:DyFe6Sn6}
\setlength{\tabcolsep}{4mm}{
%
}
\end{table}

\begin{figure}[htbp]
\centering
\includegraphics[width=0.85\textwidth]{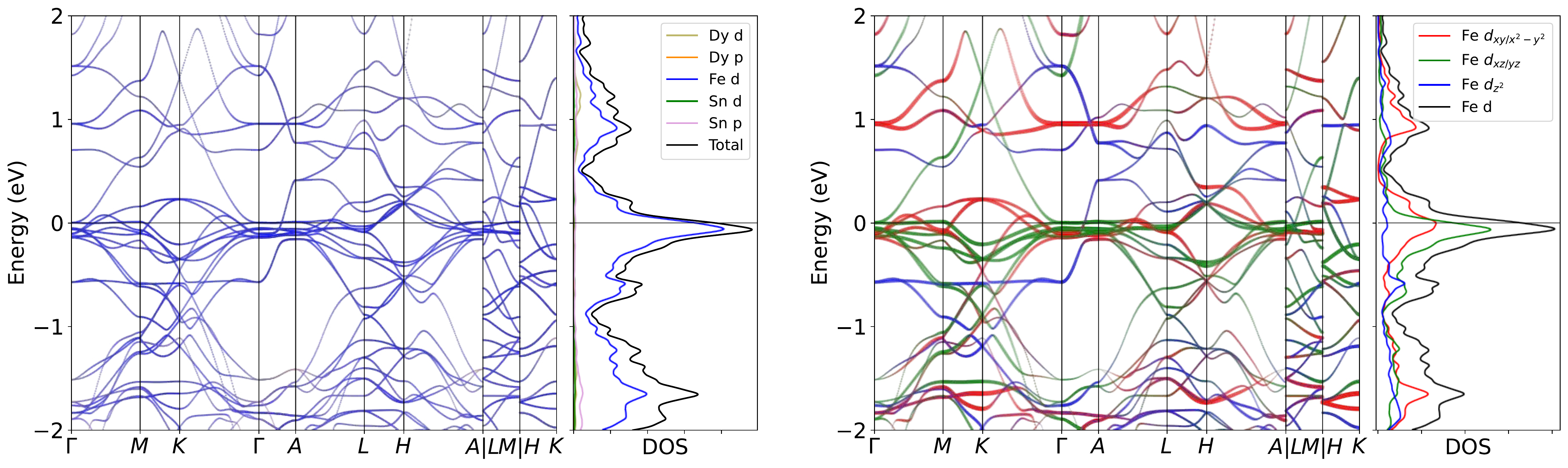}
\caption{Band structure for DyFe$_6$Sn$_6$ without SOC in paramagnetic phase. The projection of Fe $3d$ orbitals is also presented. The band structures clearly reveal the dominating of Fe $3d$ orbitals  near the Fermi level, as well as the pronounced peak.}
\label{fig:DyFe6Sn6band}
\end{figure}

\begin{figure}[H]
\centering
\subfloat[Relaxed phonon at low-T.]{
\label{fig:DyFe6Sn6_relaxed_Low}
\includegraphics[width=0.45\textwidth]{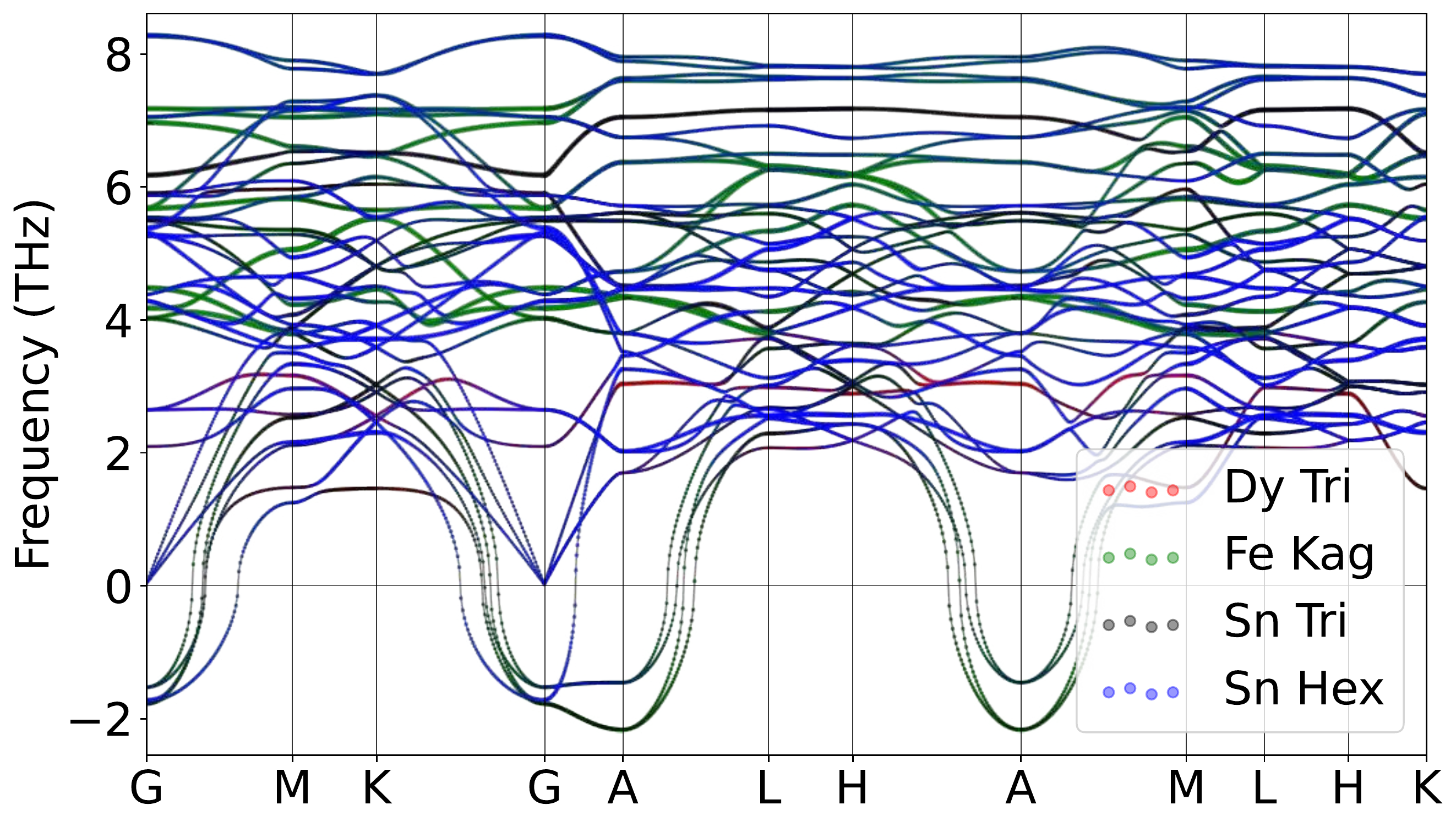}
}
\hspace{5pt}\subfloat[Relaxed phonon at high-T.]{
\label{fig:DyFe6Sn6_relaxed_High}
\includegraphics[width=0.45\textwidth]{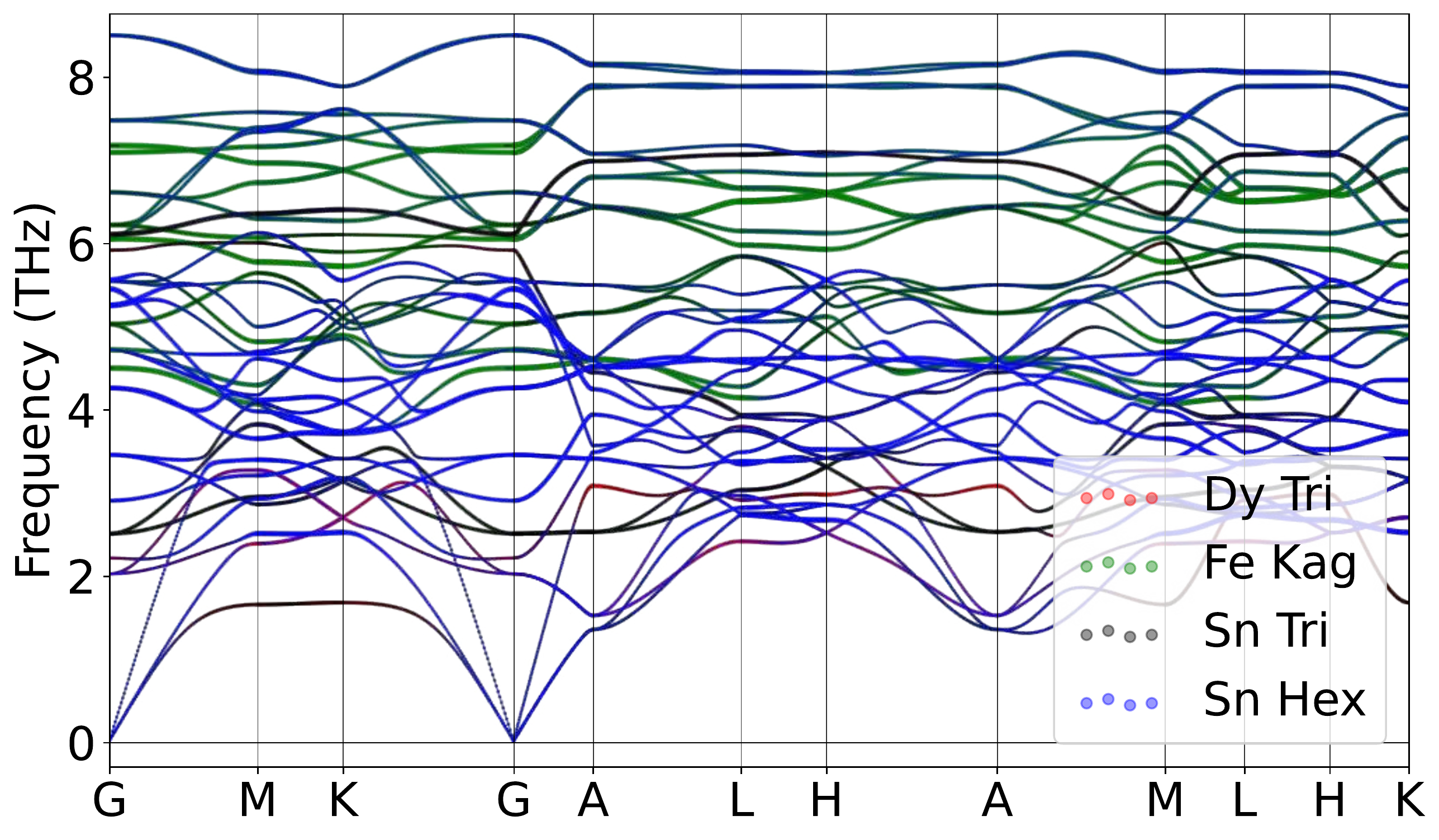}
}
\caption{Atomically projected phonon spectrum for DyFe$_6$Sn$_6$ in paramagnetic phase.}
\label{fig:DyFe6Sn6pho}
\end{figure}

\begin{figure}[htbp]
\centering
\includegraphics[width=0.45\textwidth]{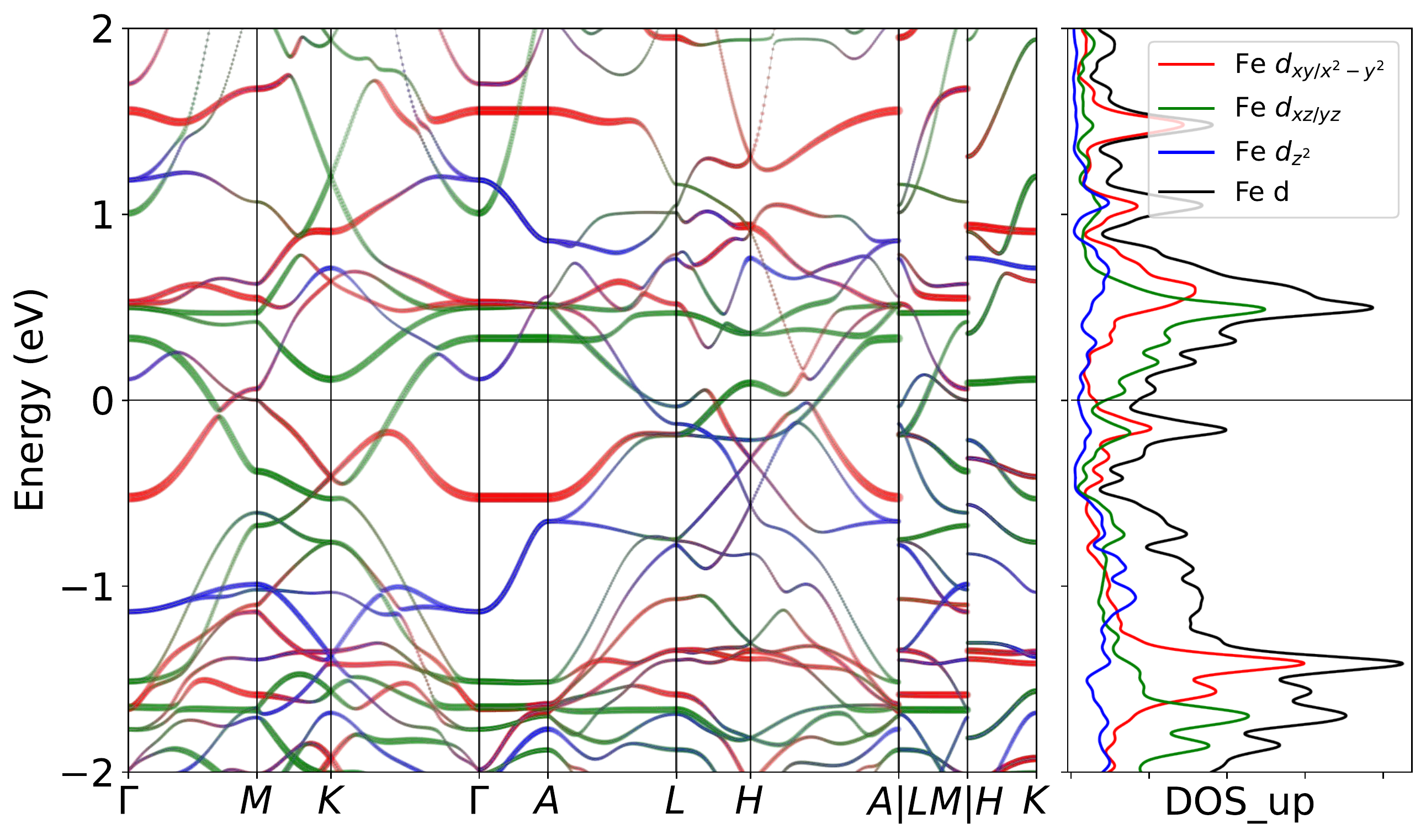}
\includegraphics[width=0.45\textwidth]{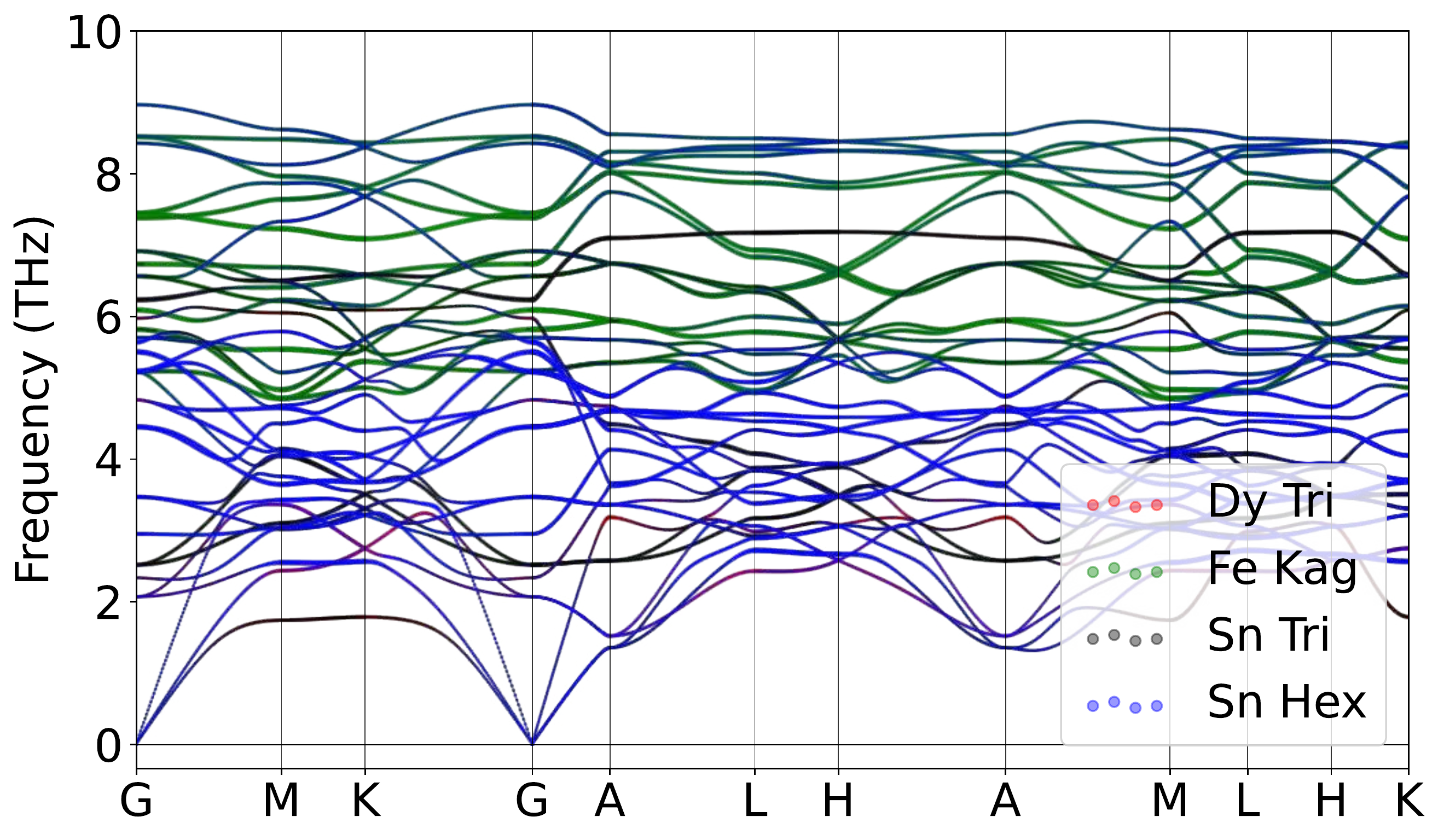}
\caption{Band structure for DyFe$_6$Sn$_6$ with AFM order without SOC for spin (Left)up channel. The projection of Fe $3d$ orbitals is also presented. (Right)Correspondingly, a stable phonon was demonstrated with the collinear magnetic order without Hubbard U at lower temperature regime, weighted by atomic projections.}
\label{fig:DyFe6Sn6mag}
\end{figure}

\begin{figure}[H]
\centering
\label{fig:DyFe6Sn6_relaxed_Low_xz}
\includegraphics[width=0.85\textwidth]{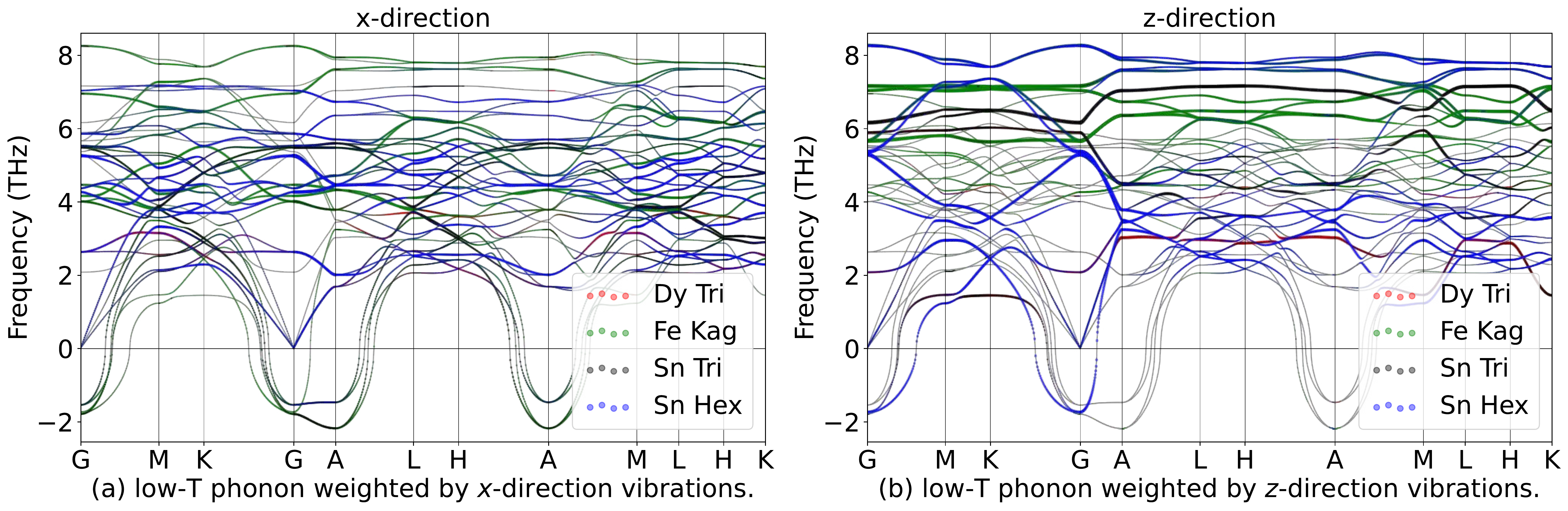}
\hspace{5pt}\label{fig:DyFe6Sn6_relaxed_High_xz}
\includegraphics[width=0.85\textwidth]{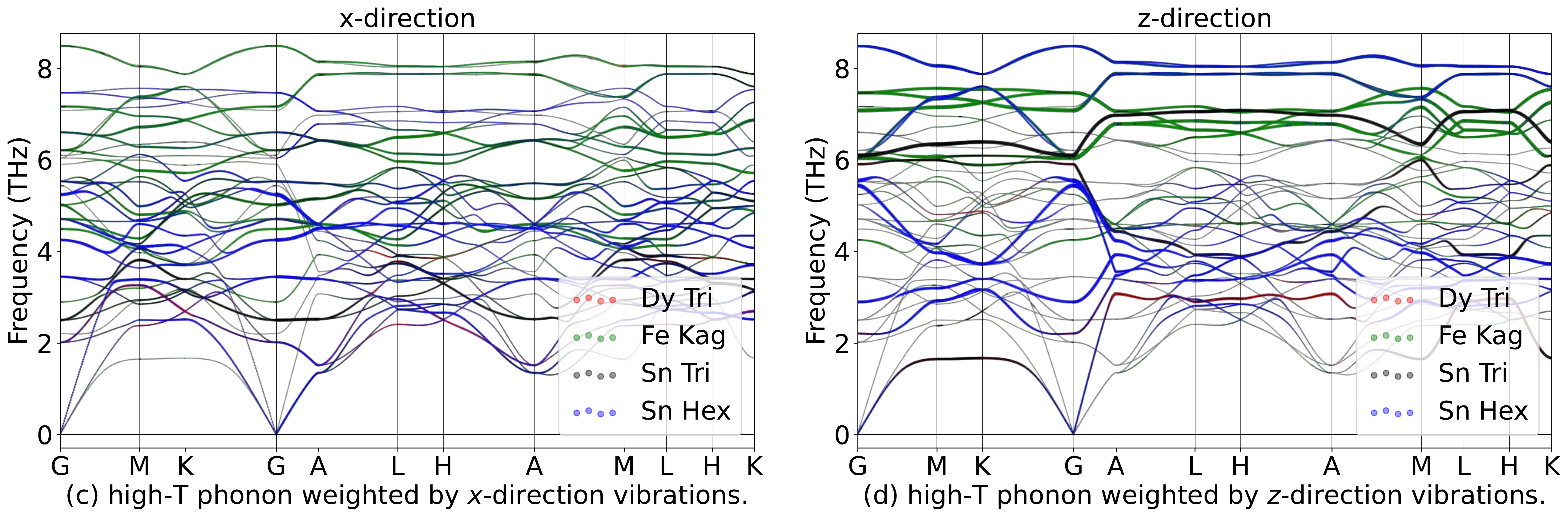}
\caption{Phonon spectrum for DyFe$_6$Sn$_6$in in-plane ($x$) and out-of-plane ($z$) directions in paramagnetic phase.}
\label{fig:DyFe6Sn6phoxyz}
\end{figure}

\begin{table}[htbp]
\global\long\def\arraystretch{1.12}
\captionsetup{width=.95\textwidth}
\caption{IRREPs at high-symmetry points for DyFe$_6$Sn$_6$ phonon spectrum at Low-T.}
\label{tab:191_DyFe6Sn6_Low}
\setlength{\tabcolsep}{1.mm}{
}
\end{table}

\begin{table}[htbp]
\global\long\def\arraystretch{1.12}
\captionsetup{width=.95\textwidth}
\caption{IRREPs at high-symmetry points for DyFe$_6$Sn$_6$ phonon spectrum at High-T.}
\label{tab:191_DyFe6Sn6_High}
\setlength{\tabcolsep}{1.mm}{
%
}
\end{table}
\subsubsection{\label{sms2sec:HoFe6Sn6}\hyperref[smsec:col]{\texorpdfstring{HoFe$_6$Sn$_6$}{}}~{\color{blue}  (High-T stable, Low-T unstable) }}

\begin{table}[htbp]
\global\long\def\arraystretch{1.12}
\captionsetup{width=.85\textwidth}
\caption{Structure parameters of HoFe$_6$Sn$_6$ after relaxation. It contains 523 electrons. }
\label{tab:HoFe6Sn6}
\setlength{\tabcolsep}{4mm}{
%
}
\end{table}

\begin{figure}[htbp]
\centering
\includegraphics[width=0.85\textwidth]{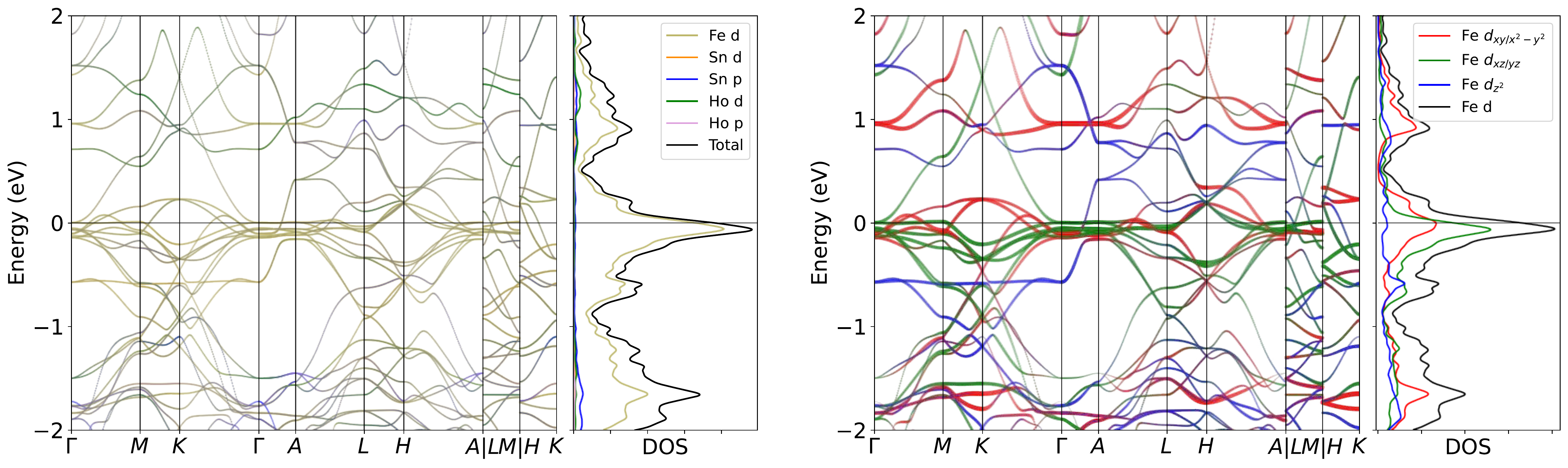}
\caption{Band structure for HoFe$_6$Sn$_6$ without SOC in paramagnetic phase. The projection of Fe $3d$ orbitals is also presented. The band structures clearly reveal the dominating of Fe $3d$ orbitals  near the Fermi level, as well as the pronounced peak.}
\label{fig:HoFe6Sn6band}
\end{figure}

\begin{figure}[H]
\centering
\subfloat[Relaxed phonon at low-T.]{
\label{fig:HoFe6Sn6_relaxed_Low}
\includegraphics[width=0.45\textwidth]{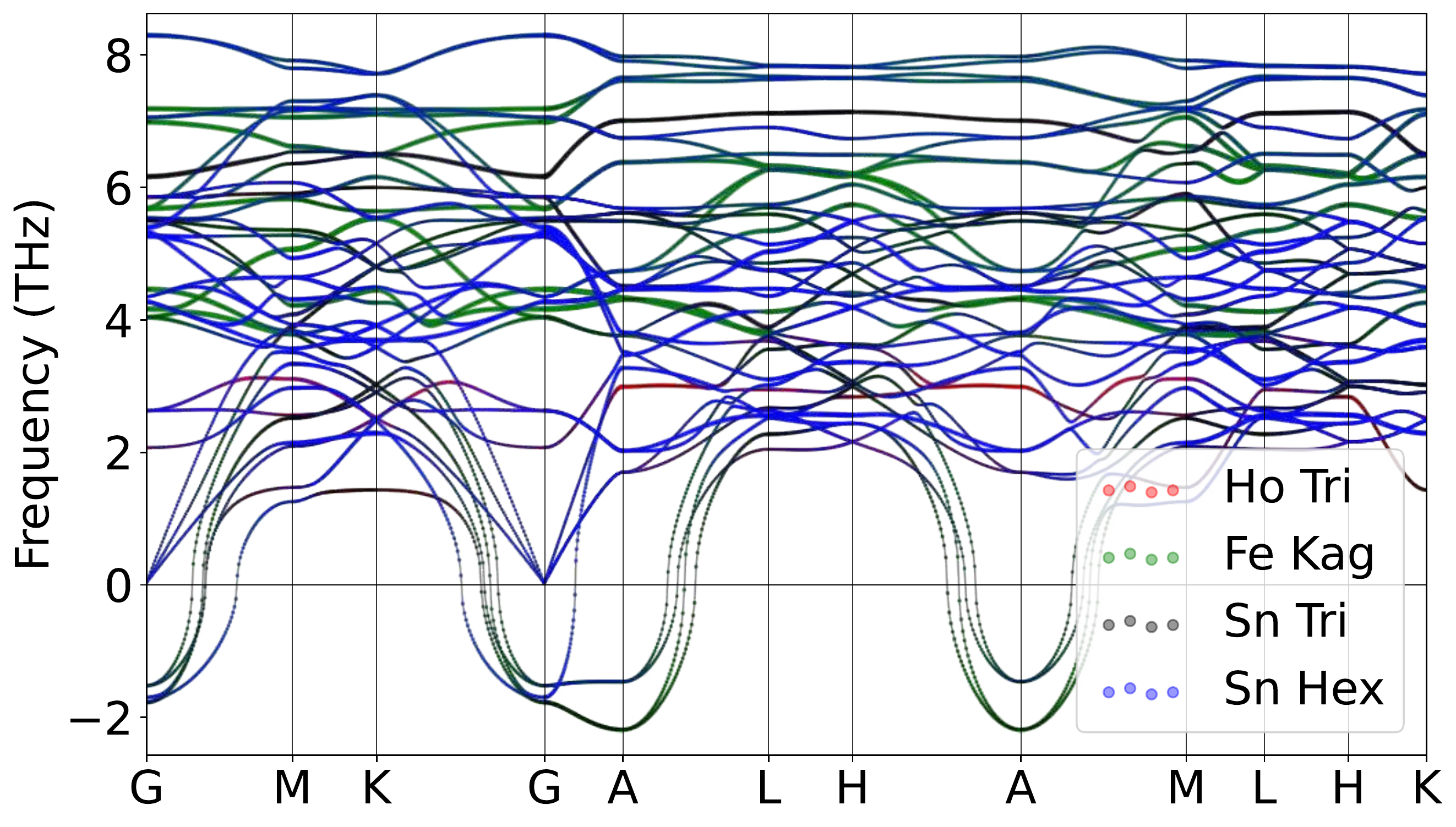}
}
\hspace{5pt}\subfloat[Relaxed phonon at high-T.]{
\label{fig:HoFe6Sn6_relaxed_High}
\includegraphics[width=0.45\textwidth]{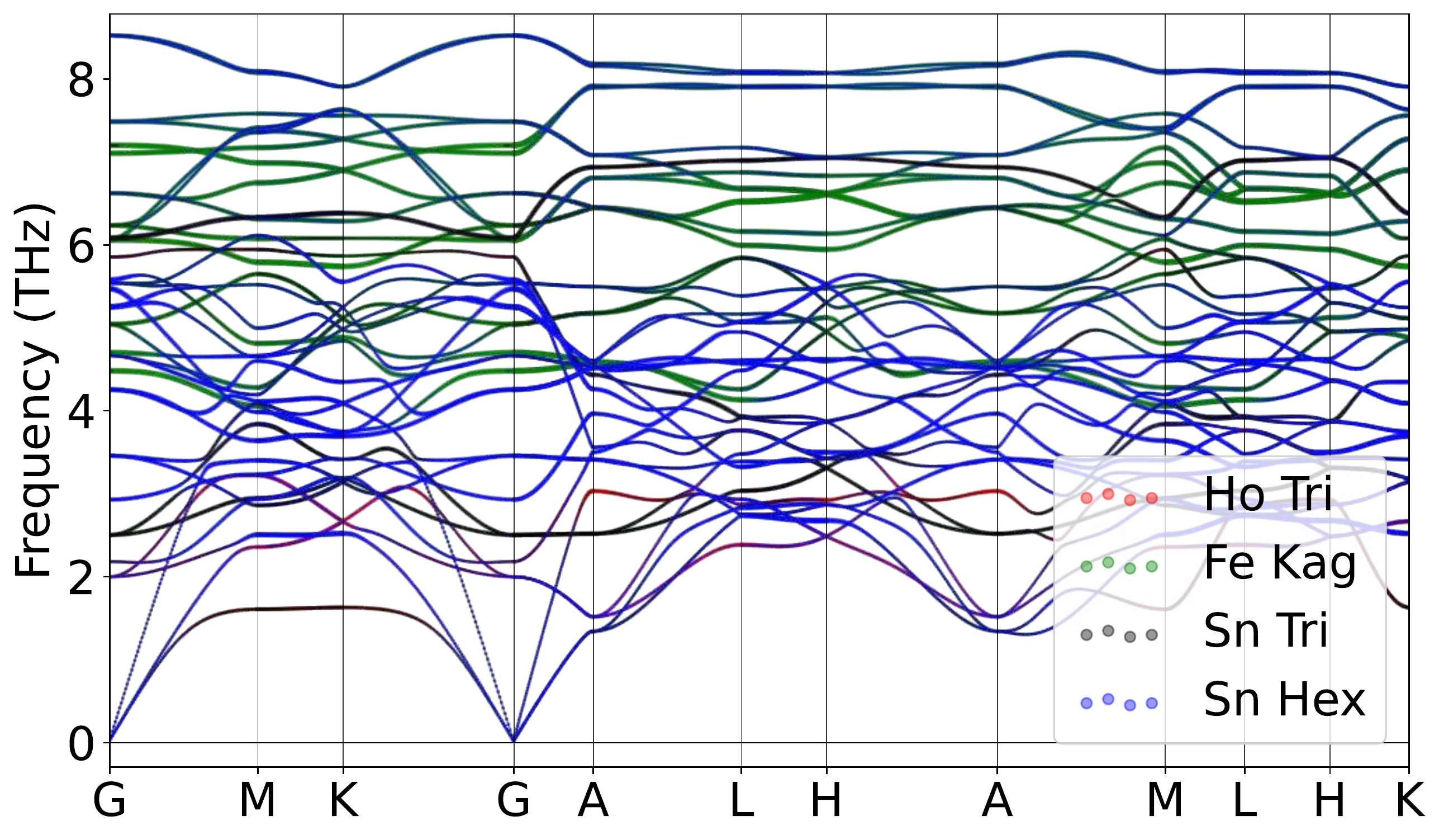}
}
\caption{Atomically projected phonon spectrum for HoFe$_6$Sn$_6$ in paramagnetic phase.}
\label{fig:HoFe6Sn6pho}
\end{figure}

\begin{figure}[htbp]
\centering
\includegraphics[width=0.45\textwidth]{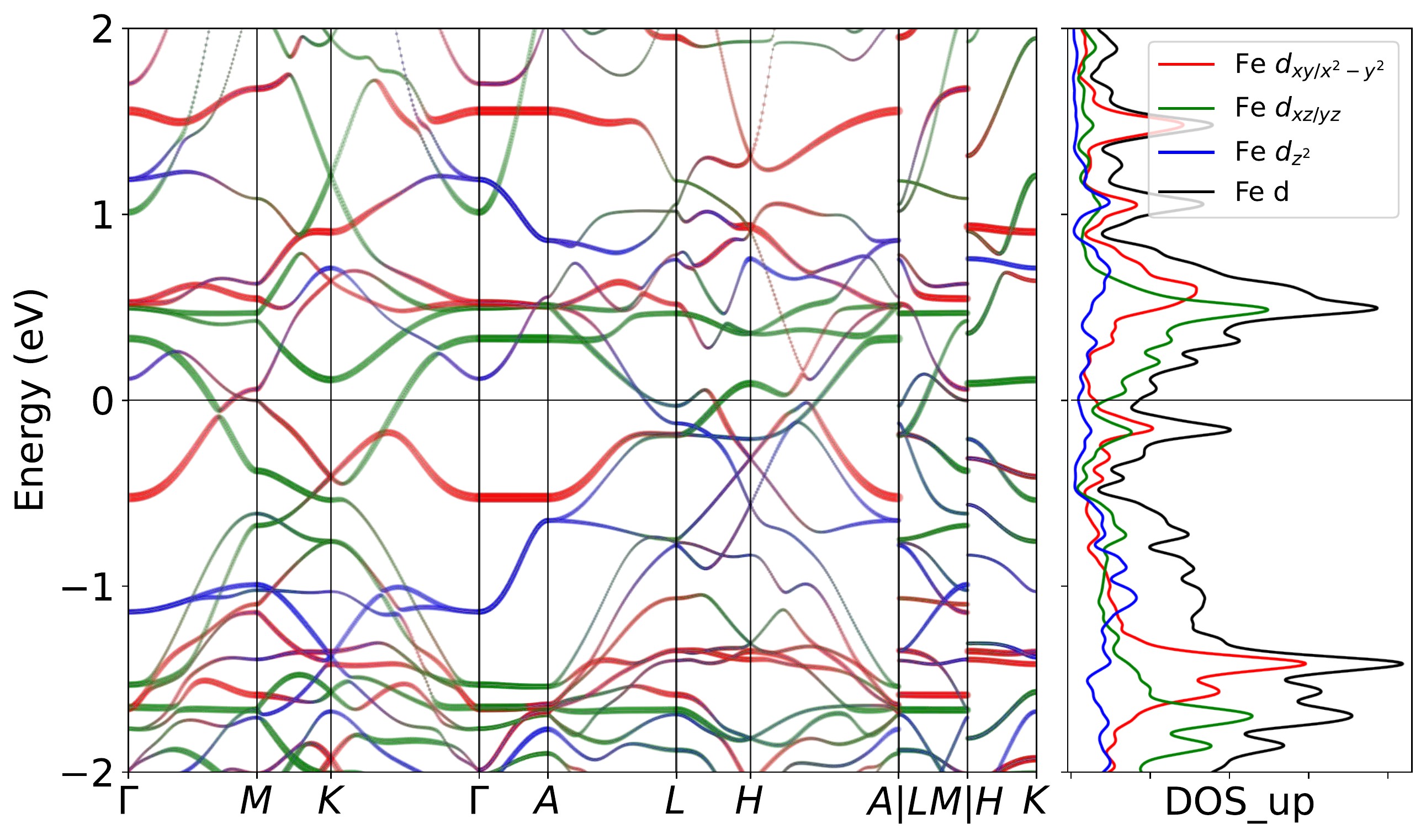}
\includegraphics[width=0.45\textwidth]{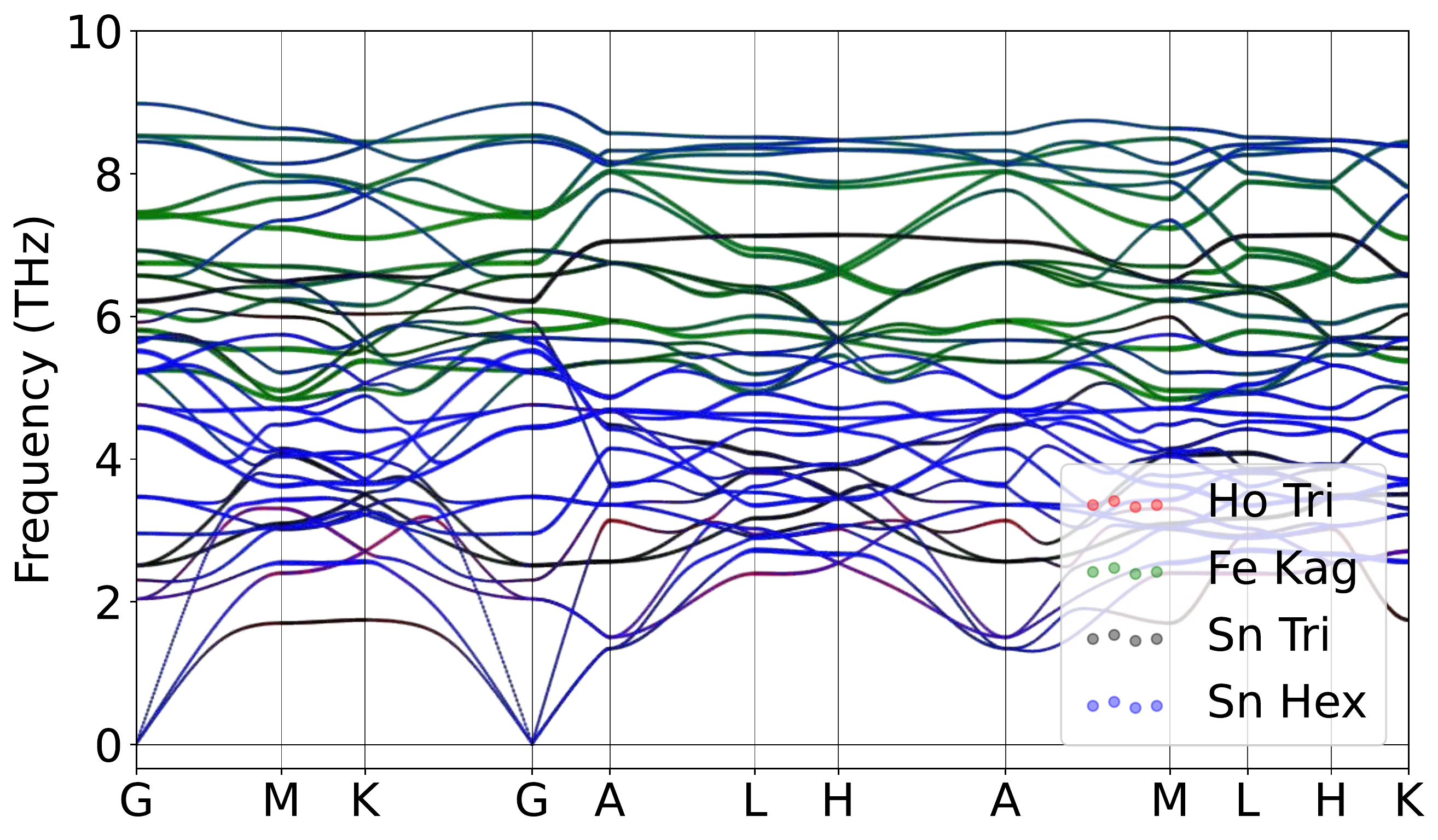}
\caption{Band structure for HoFe$_6$Sn$_6$ with AFM order without SOC for spin (Left)up channel. The projection of Fe $3d$ orbitals is also presented. (Right)Correspondingly, a stable phonon was demonstrated with the collinear magnetic order without Hubbard U at lower temperature regime, weighted by atomic projections.}
\label{fig:HoFe6Sn6mag}
\end{figure}

\begin{figure}[H]
\centering
\label{fig:HoFe6Sn6_relaxed_Low_xz}
\includegraphics[width=0.85\textwidth]{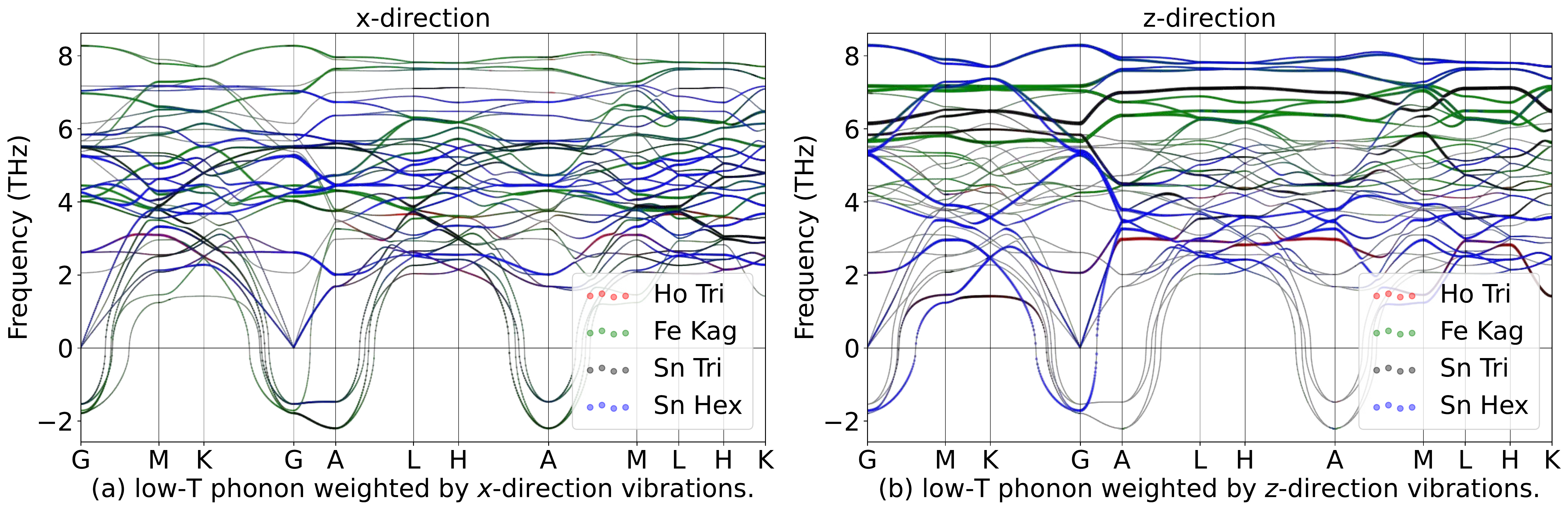}
\hspace{5pt}\label{fig:HoFe6Sn6_relaxed_High_xz}
\includegraphics[width=0.85\textwidth]{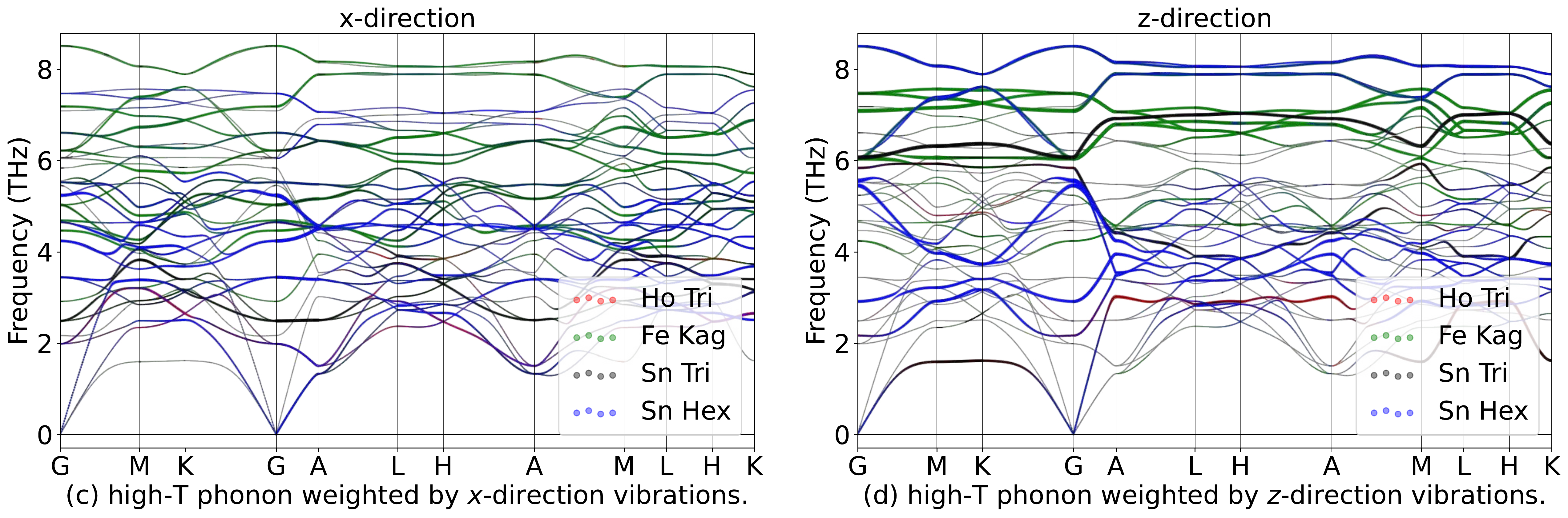}
\caption{Phonon spectrum for HoFe$_6$Sn$_6$in in-plane ($x$) and out-of-plane ($z$) directions in paramagnetic phase.}
\label{fig:HoFe6Sn6phoxyz}
\end{figure}

\begin{table}[htbp]
\global\long\def\arraystretch{1.12}
\captionsetup{width=.95\textwidth}
\caption{IRREPs at high-symmetry points for HoFe$_6$Sn$_6$ phonon spectrum at Low-T.}
\label{tab:191_HoFe6Sn6_Low}
\setlength{\tabcolsep}{1.mm}{
}
\end{table}

\begin{table}[htbp]
\global\long\def\arraystretch{1.12}
\captionsetup{width=.95\textwidth}
\caption{IRREPs at high-symmetry points for HoFe$_6$Sn$_6$ phonon spectrum at High-T.}
\label{tab:191_HoFe6Sn6_High}
\setlength{\tabcolsep}{1.mm}{
%
}
\end{table}
\subsubsection{\label{sms2sec:ErFe6Sn6}\hyperref[smsec:col]{\texorpdfstring{ErFe$_6$Sn$_6$}{}}~{\color{blue}  (High-T stable, Low-T unstable) }}

\begin{table}[htbp]
\global\long\def\arraystretch{1.12}
\captionsetup{width=.85\textwidth}
\caption{Structure parameters of ErFe$_6$Sn$_6$ after relaxation. It contains 524 electrons. }
\label{tab:ErFe6Sn6}
\setlength{\tabcolsep}{4mm}{
%
}
\end{table}

\begin{figure}[htbp]
\centering
\includegraphics[width=0.85\textwidth]{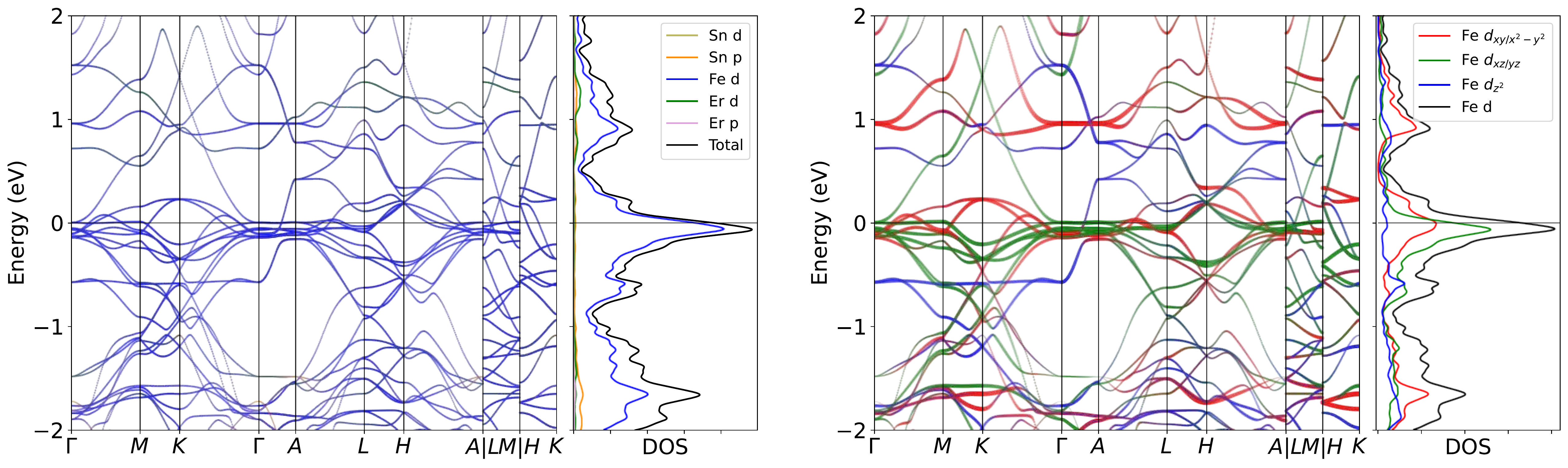}
\caption{Band structure for ErFe$_6$Sn$_6$ without SOC in paramagnetic phase. The projection of Fe $3d$ orbitals is also presented. The band structures clearly reveal the dominating of Fe $3d$ orbitals  near the Fermi level, as well as the pronounced peak.}
\label{fig:ErFe6Sn6band}
\end{figure}

\begin{figure}[H]
\centering
\subfloat[Relaxed phonon at low-T.]{
\label{fig:ErFe6Sn6_relaxed_Low}
\includegraphics[width=0.45\textwidth]{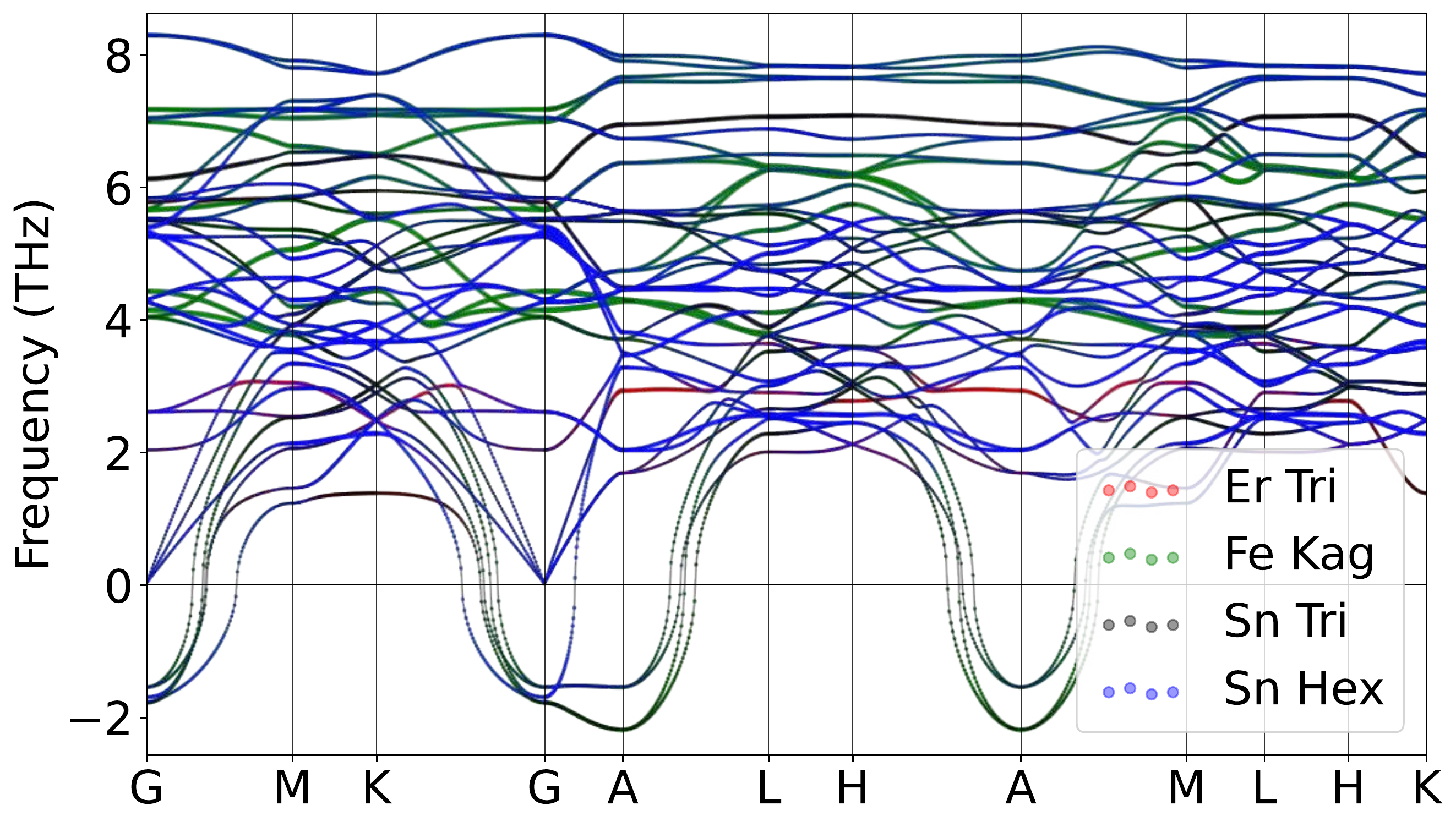}
}
\hspace{5pt}\subfloat[Relaxed phonon at high-T.]{
\label{fig:ErFe6Sn6_relaxed_High}
\includegraphics[width=0.45\textwidth]{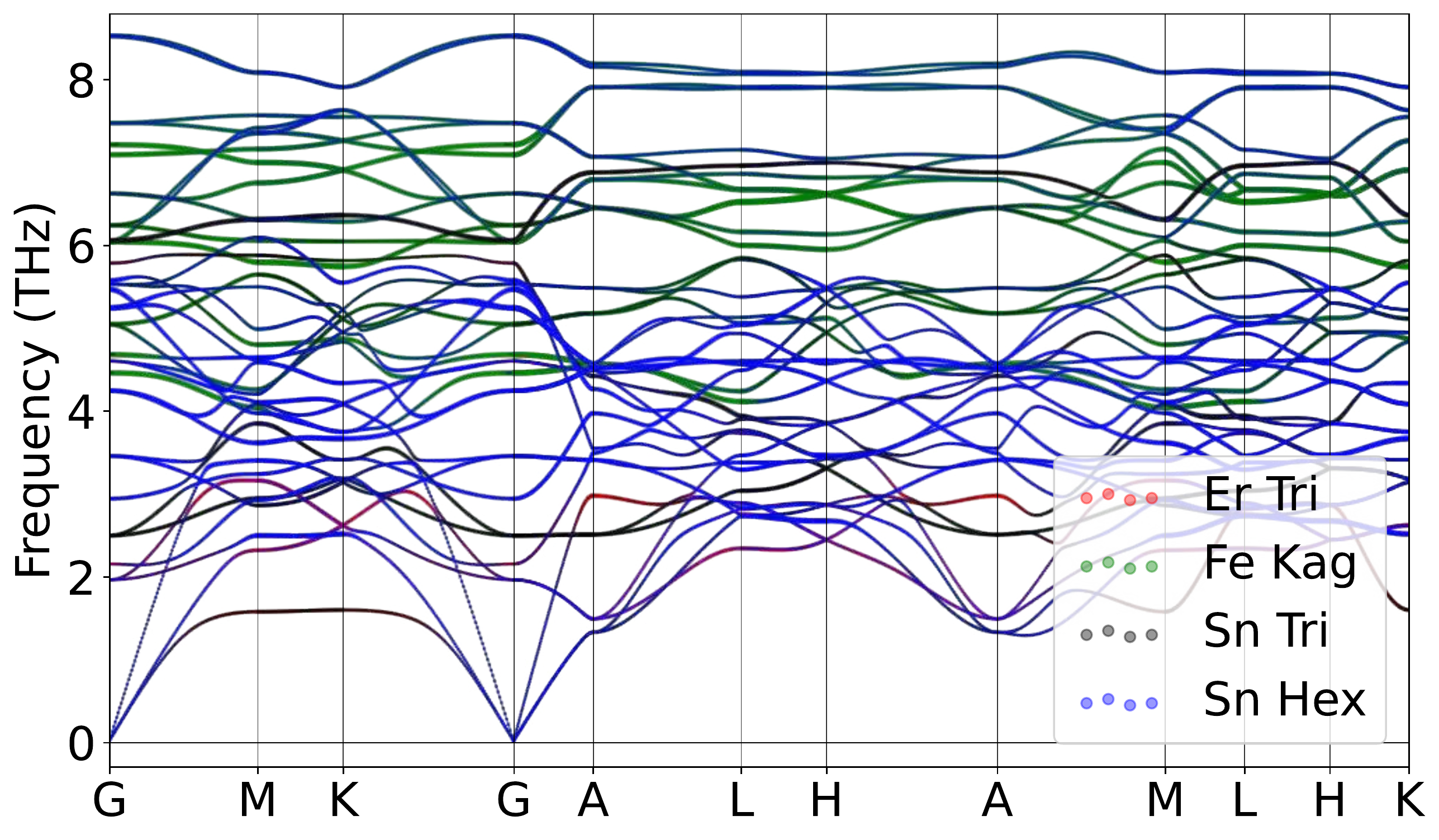}
}
\caption{Atomically projected phonon spectrum for ErFe$_6$Sn$_6$ in paramagnetic phase.}
\label{fig:ErFe6Sn6pho}
\end{figure}

\begin{figure}[htbp]
\centering
\includegraphics[width=0.45\textwidth]{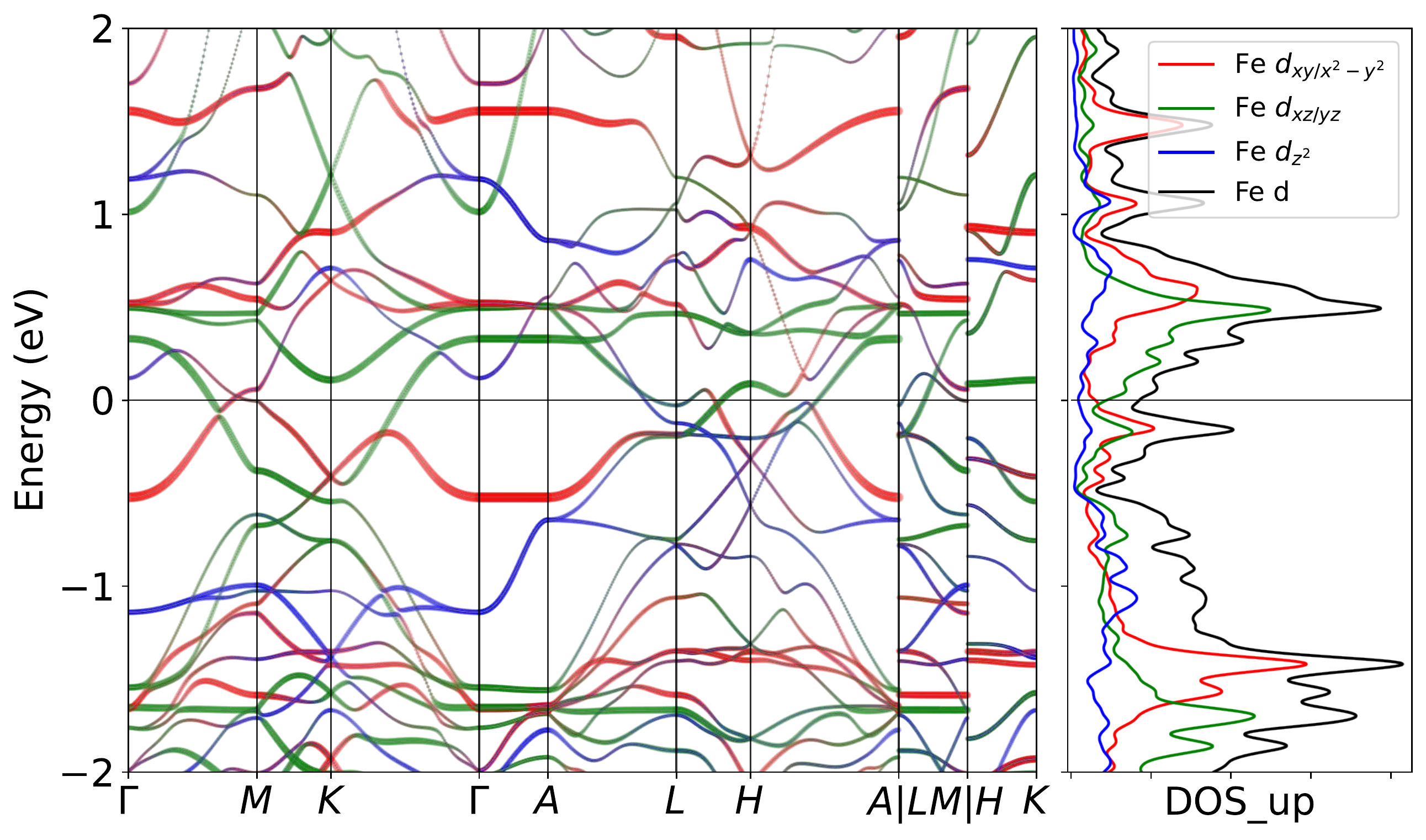}
\includegraphics[width=0.45\textwidth]{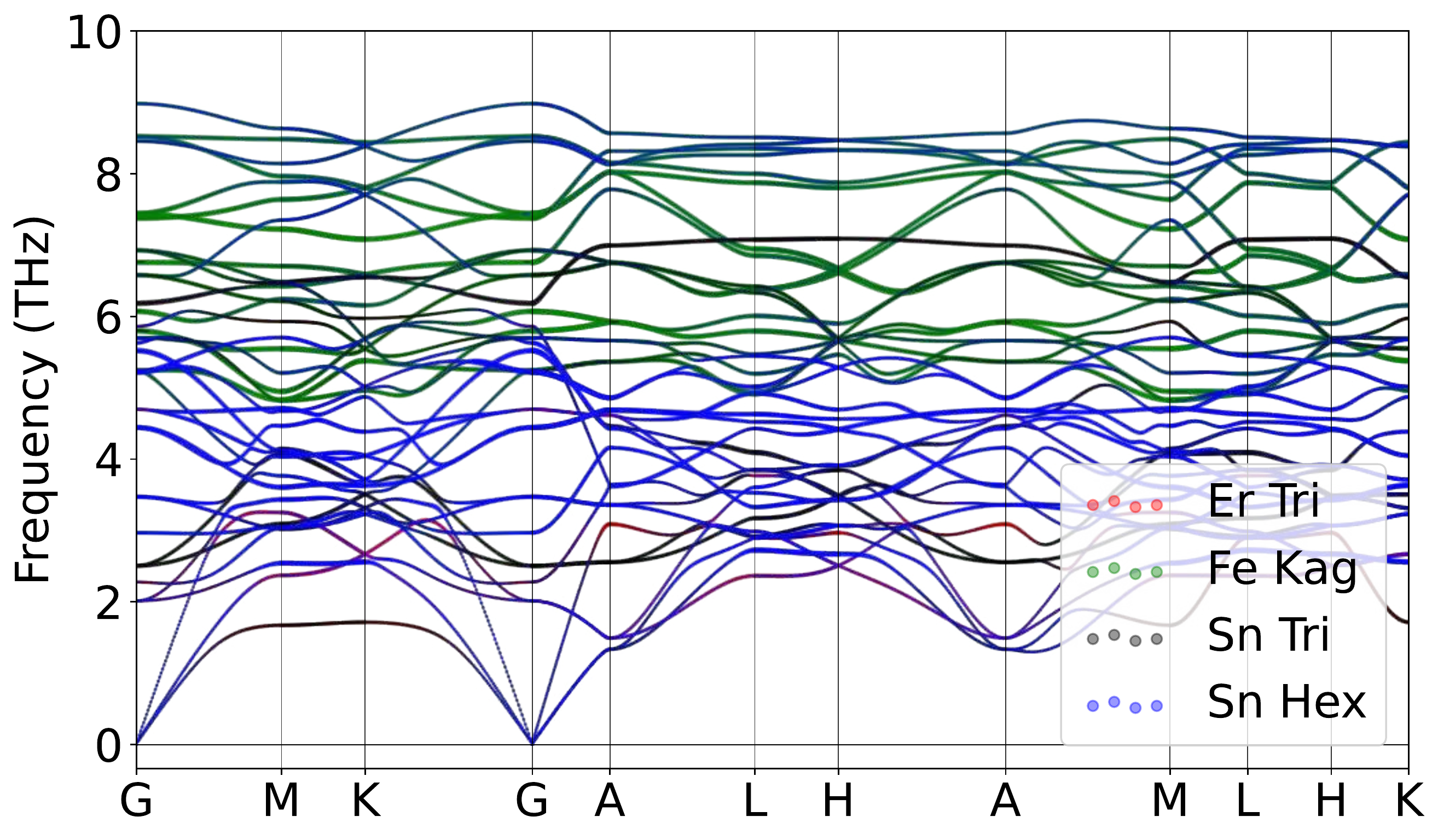}
\caption{Band structure for ErFe$_6$Sn$_6$ with AFM order without SOC for spin (Left)up channel. The projection of Fe $3d$ orbitals is also presented. (Right)Correspondingly, a stable phonon was demonstrated with the collinear magnetic order without Hubbard U at lower temperature regime, weighted by atomic projections.}
\label{fig:ErFe6Sn6mag}
\end{figure}

\begin{figure}[H]
\centering
\label{fig:ErFe6Sn6_relaxed_Low_xz}
\includegraphics[width=0.85\textwidth]{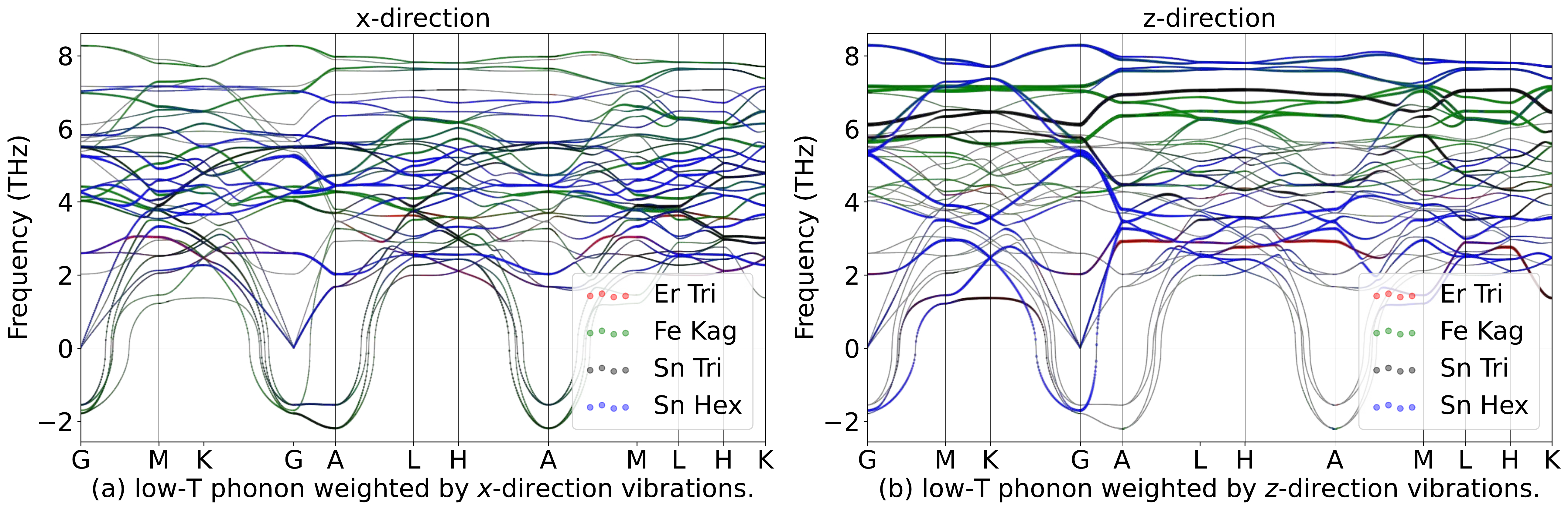}
\hspace{5pt}\label{fig:ErFe6Sn6_relaxed_High_xz}
\includegraphics[width=0.85\textwidth]{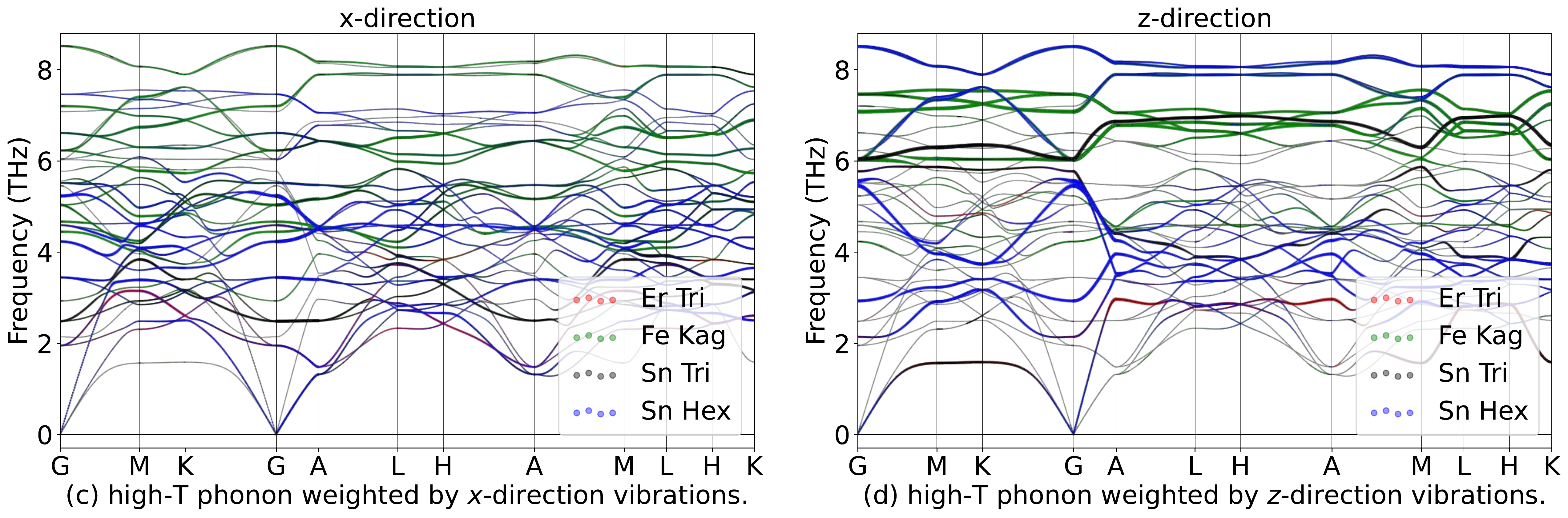}
\caption{Phonon spectrum for ErFe$_6$Sn$_6$in in-plane ($x$) and out-of-plane ($z$) directions in paramagnetic phase.}
\label{fig:ErFe6Sn6phoxyz}
\end{figure}

\begin{table}[htbp]
\global\long\def\arraystretch{1.12}
\captionsetup{width=.95\textwidth}
\caption{IRREPs at high-symmetry points for ErFe$_6$Sn$_6$ phonon spectrum at Low-T.}
\label{tab:191_ErFe6Sn6_Low}
\setlength{\tabcolsep}{1.mm}{
}
\end{table}

\begin{table}[htbp]
\global\long\def\arraystretch{1.12}
\captionsetup{width=.95\textwidth}
\caption{IRREPs at high-symmetry points for ErFe$_6$Sn$_6$ phonon spectrum at High-T.}
\label{tab:191_ErFe6Sn6_High}
\setlength{\tabcolsep}{1.mm}{
%
}
\end{table}
\subsubsection{\label{sms2sec:TmFe6Sn6}\hyperref[smsec:col]{\texorpdfstring{TmFe$_6$Sn$_6$}{}}~{\color{blue}  (High-T stable, Low-T unstable) }}

\begin{table}[htbp]
\global\long\def\arraystretch{1.12}
\captionsetup{width=.85\textwidth}
\caption{Structure parameters of TmFe$_6$Sn$_6$ after relaxation. It contains 525 electrons. }
\label{tab:TmFe6Sn6}
\setlength{\tabcolsep}{4mm}{
%
}
\end{table}

\begin{figure}[htbp]
\centering
\includegraphics[width=0.85\textwidth]{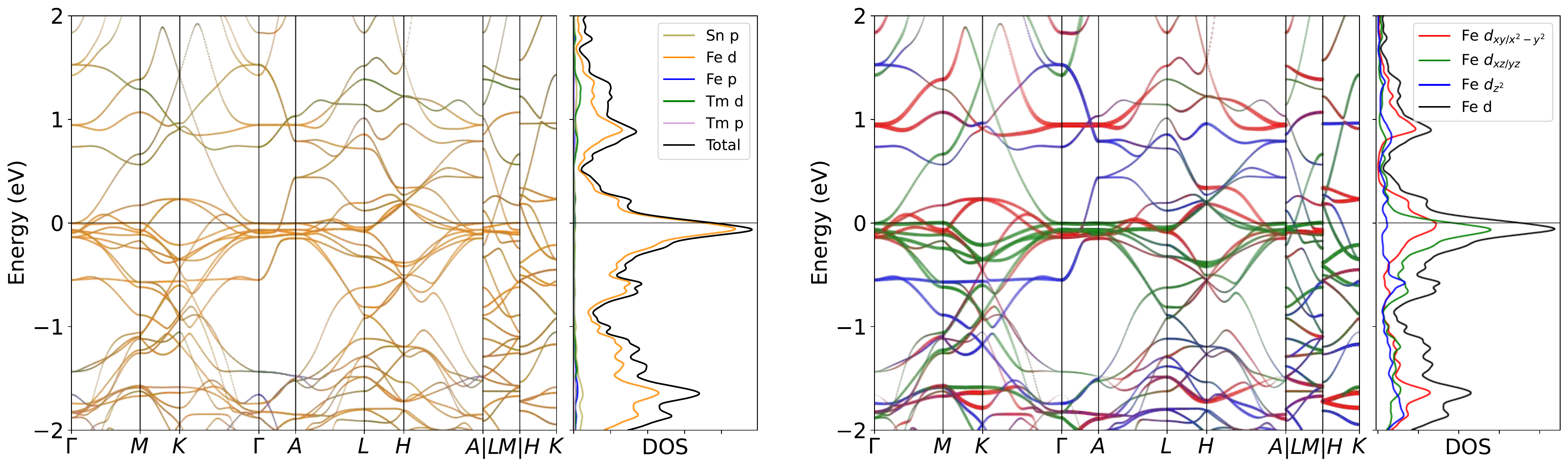}
\caption{Band structure for TmFe$_6$Sn$_6$ without SOC in paramagnetic phase. The projection of Fe $3d$ orbitals is also presented. The band structures clearly reveal the dominating of Fe $3d$ orbitals  near the Fermi level, as well as the pronounced peak.}
\label{fig:TmFe6Sn6band}
\end{figure}

\begin{figure}[H]
\centering
\subfloat[Relaxed phonon at low-T.]{
\label{fig:TmFe6Sn6_relaxed_Low}
\includegraphics[width=0.45\textwidth]{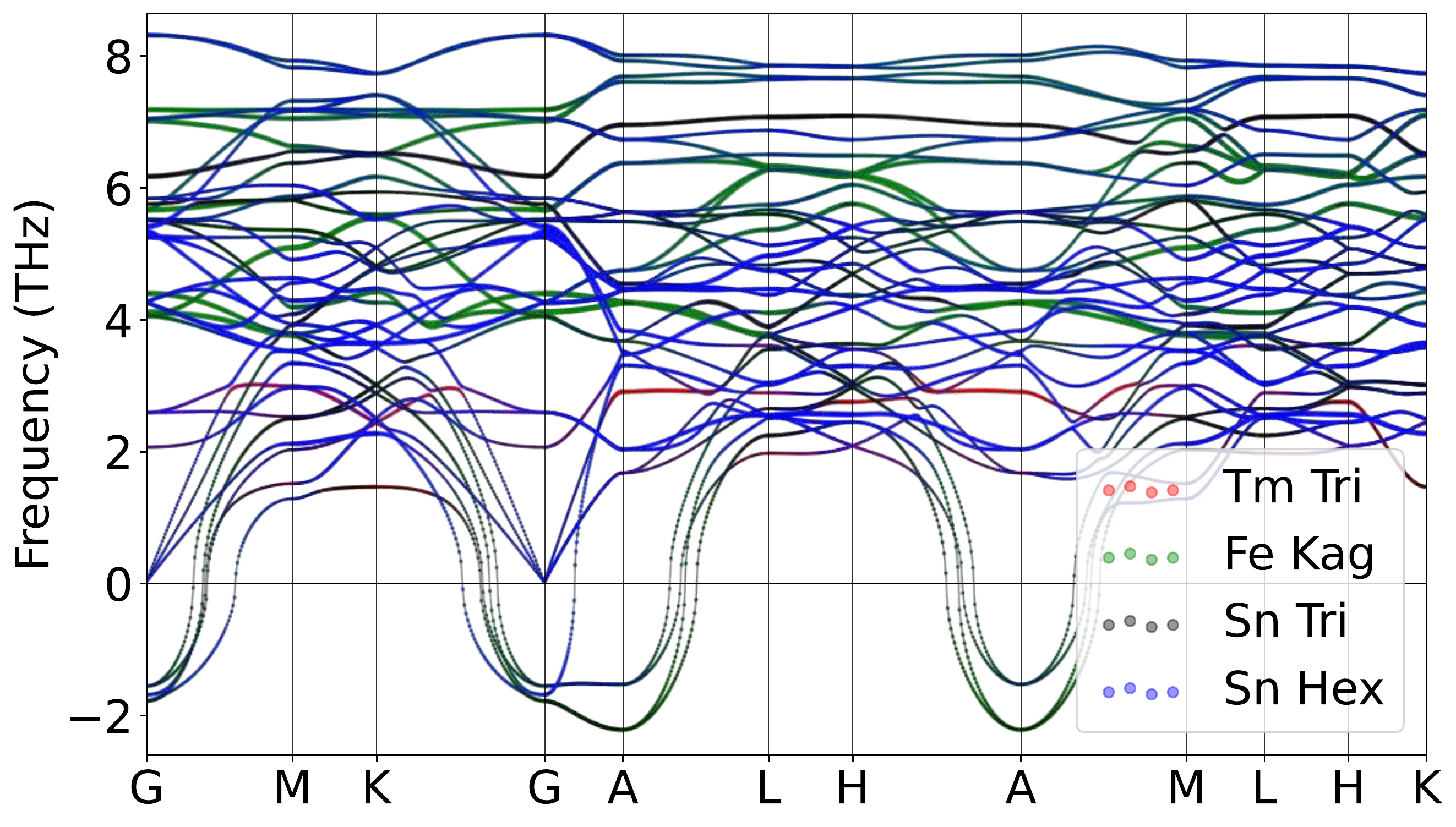}
}
\hspace{5pt}\subfloat[Relaxed phonon at high-T.]{
\label{fig:TmFe6Sn6_relaxed_High}
\includegraphics[width=0.45\textwidth]{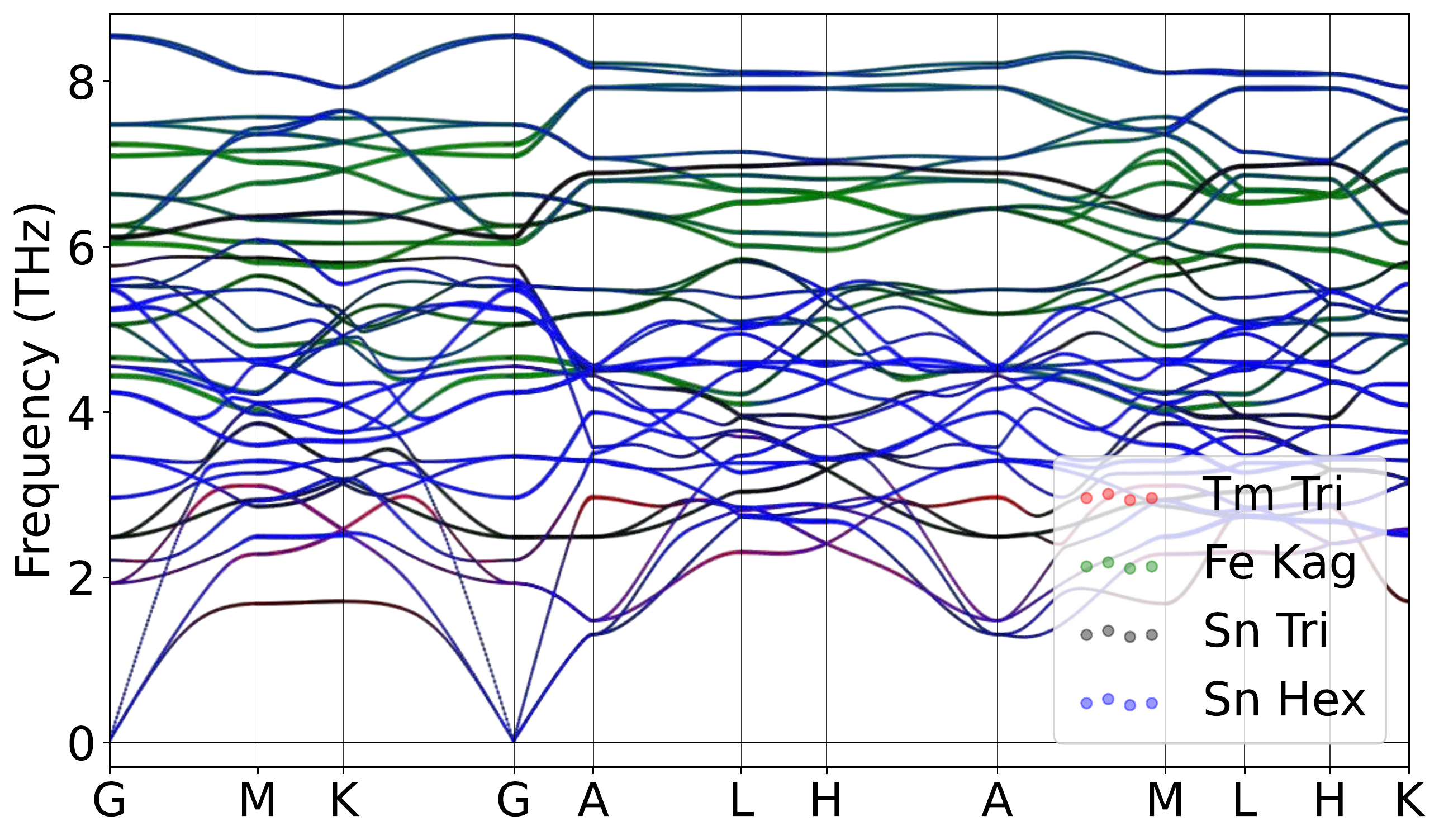}
}
\caption{Atomically projected phonon spectrum for TmFe$_6$Sn$_6$ in paramagnetic phase.}
\label{fig:TmFe6Sn6pho}
\end{figure}

\begin{figure}[htbp]
\centering
\includegraphics[width=0.45\textwidth]{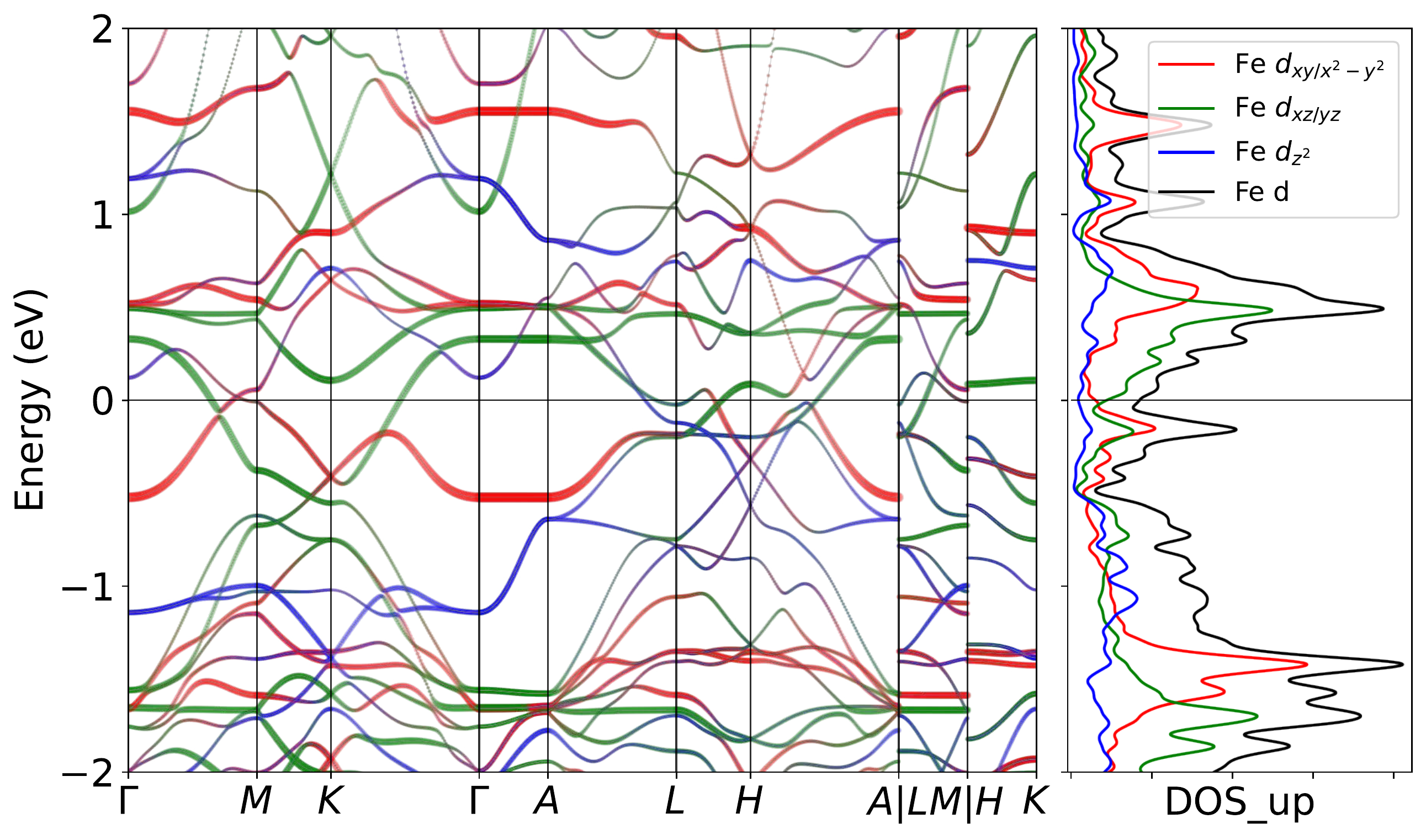}
\includegraphics[width=0.45\textwidth]{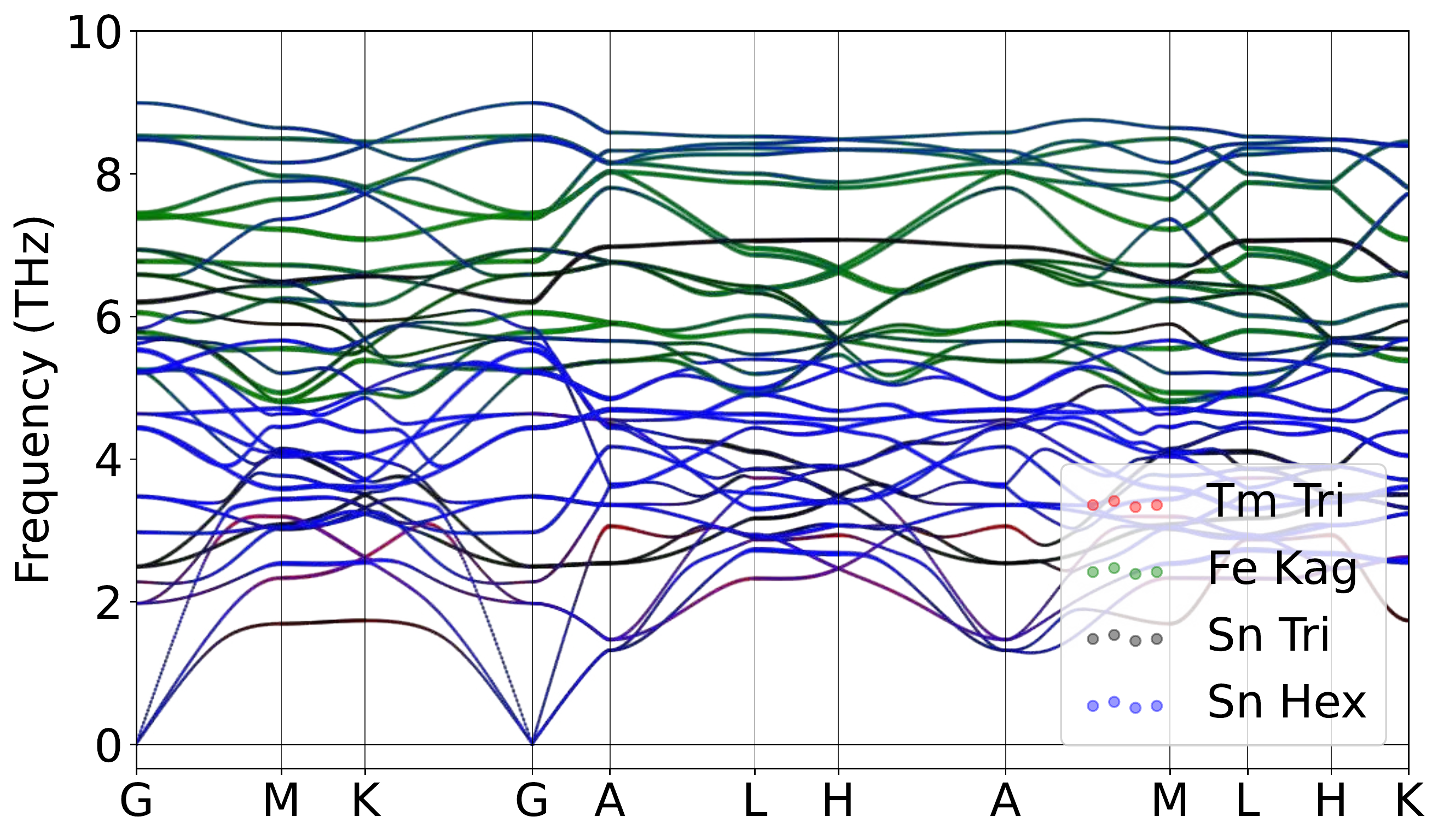}
\caption{Band structure for TmFe$_6$Sn$_6$ with AFM order without SOC for spin (Left)up channel. The projection of Fe $3d$ orbitals is also presented. (Right)Correspondingly, a stable phonon was demonstrated with the collinear magnetic order without Hubbard U at lower temperature regime, weighted by atomic projections.}
\label{fig:TmFe6Sn6mag}
\end{figure}

\begin{figure}[H]
\centering
\label{fig:TmFe6Sn6_relaxed_Low_xz}
\includegraphics[width=0.85\textwidth]{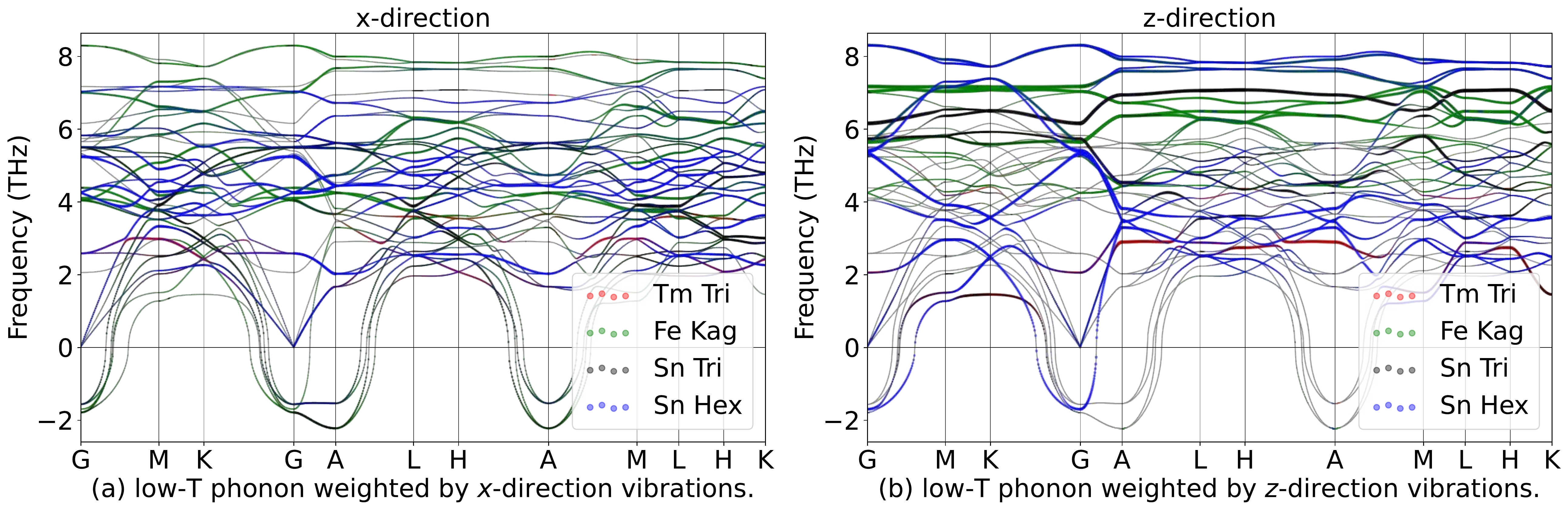}
\hspace{5pt}\label{fig:TmFe6Sn6_relaxed_High_xz}
\includegraphics[width=0.85\textwidth]{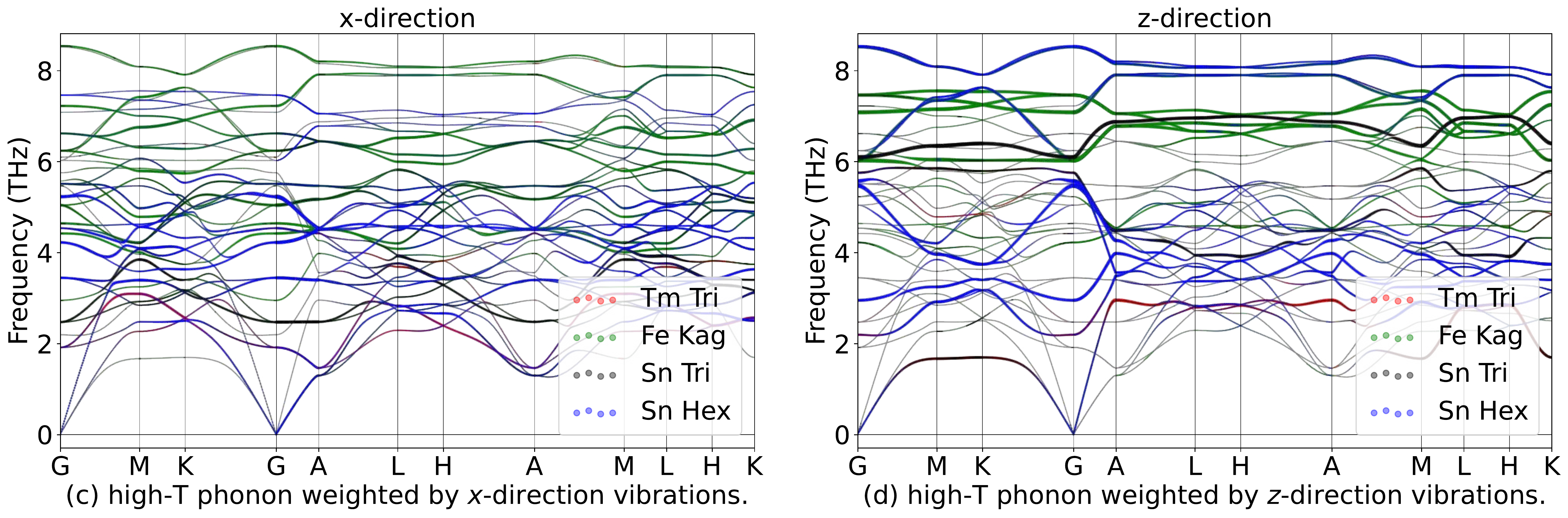}
\caption{Phonon spectrum for TmFe$_6$Sn$_6$in in-plane ($x$) and out-of-plane ($z$) directions in paramagnetic phase.}
\label{fig:TmFe6Sn6phoxyz}
\end{figure}

\begin{table}[htbp]
\global\long\def\arraystretch{1.12}
\captionsetup{width=.95\textwidth}
\caption{IRREPs at high-symmetry points for TmFe$_6$Sn$_6$ phonon spectrum at Low-T.}
\label{tab:191_TmFe6Sn6_Low}
\setlength{\tabcolsep}{1.mm}{
}
\end{table}

\begin{table}[htbp]
\global\long\def\arraystretch{1.12}
\captionsetup{width=.95\textwidth}
\caption{IRREPs at high-symmetry points for TmFe$_6$Sn$_6$ phonon spectrum at High-T.}
\label{tab:191_TmFe6Sn6_High}
\setlength{\tabcolsep}{1.mm}{
%
}
\end{table}
\subsubsection{\label{sms2sec:YbFe6Sn6}\hyperref[smsec:col]{\texorpdfstring{YbFe$_6$Sn$_6$}{}}~{\color{blue}  (High-T stable, Low-T unstable) }}

\begin{table}[htbp]
\global\long\def\arraystretch{1.12}
\captionsetup{width=.85\textwidth}
\caption{Structure parameters of YbFe$_6$Sn$_6$ after relaxation. It contains 526 electrons. }
\label{tab:YbFe6Sn6}
\setlength{\tabcolsep}{4mm}{
%
}
\end{table}

\begin{figure}[htbp]
\centering
\includegraphics[width=0.85\textwidth]{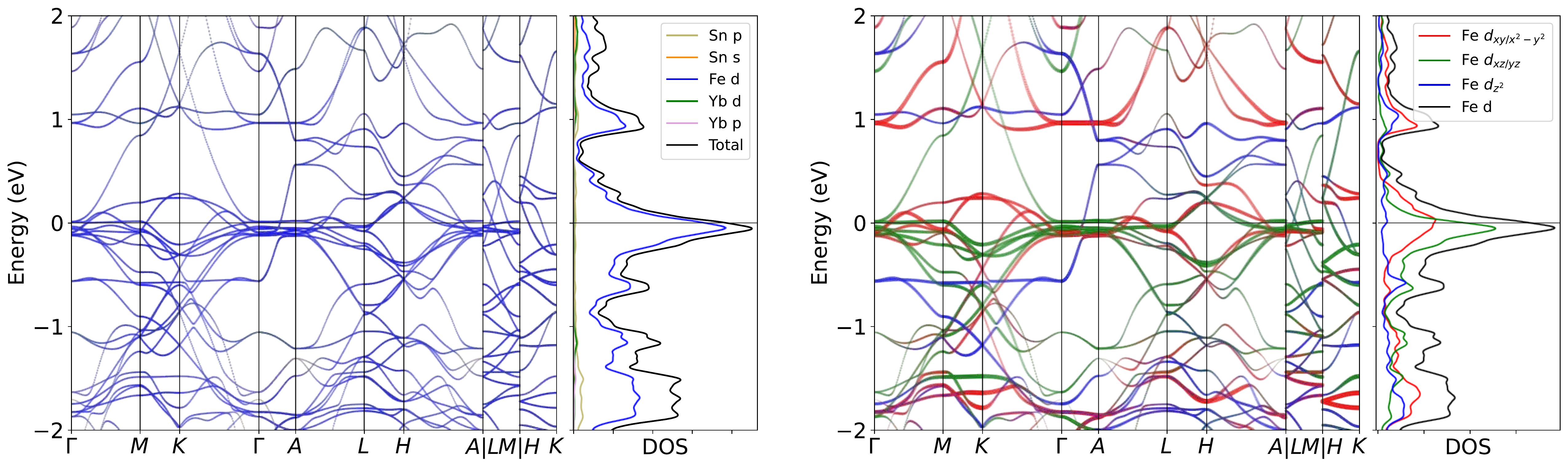}
\caption{Band structure for YbFe$_6$Sn$_6$ without SOC in paramagnetic phase. The projection of Fe $3d$ orbitals is also presented. The band structures clearly reveal the dominating of Fe $3d$ orbitals  near the Fermi level, as well as the pronounced peak.}
\label{fig:YbFe6Sn6band}
\end{figure}

\begin{figure}[H]
\centering
\subfloat[Relaxed phonon at low-T.]{
\label{fig:YbFe6Sn6_relaxed_Low}
\includegraphics[width=0.45\textwidth]{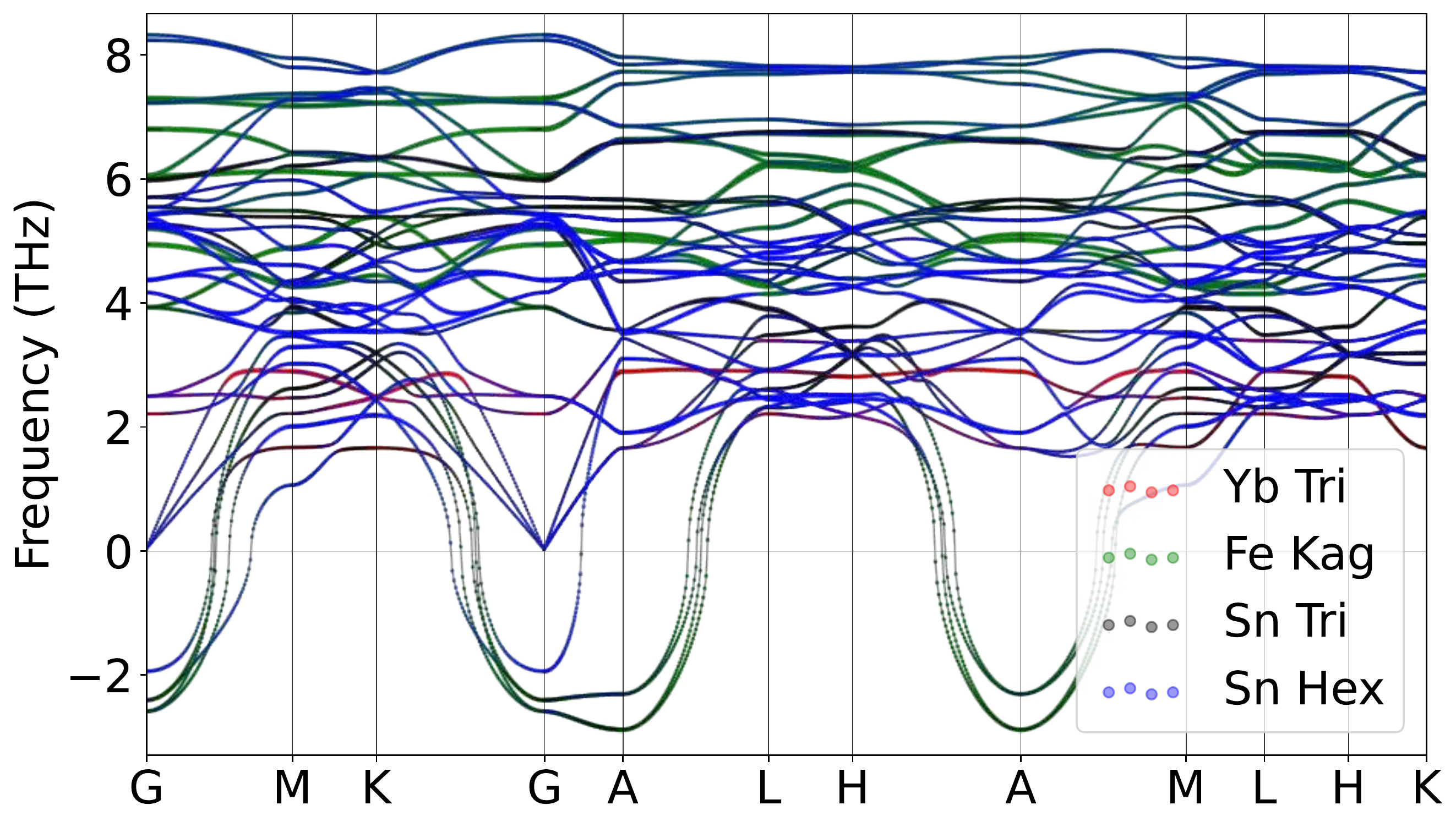}
}
\hspace{5pt}\subfloat[Relaxed phonon at high-T.]{
\label{fig:YbFe6Sn6_relaxed_High}
\includegraphics[width=0.45\textwidth]{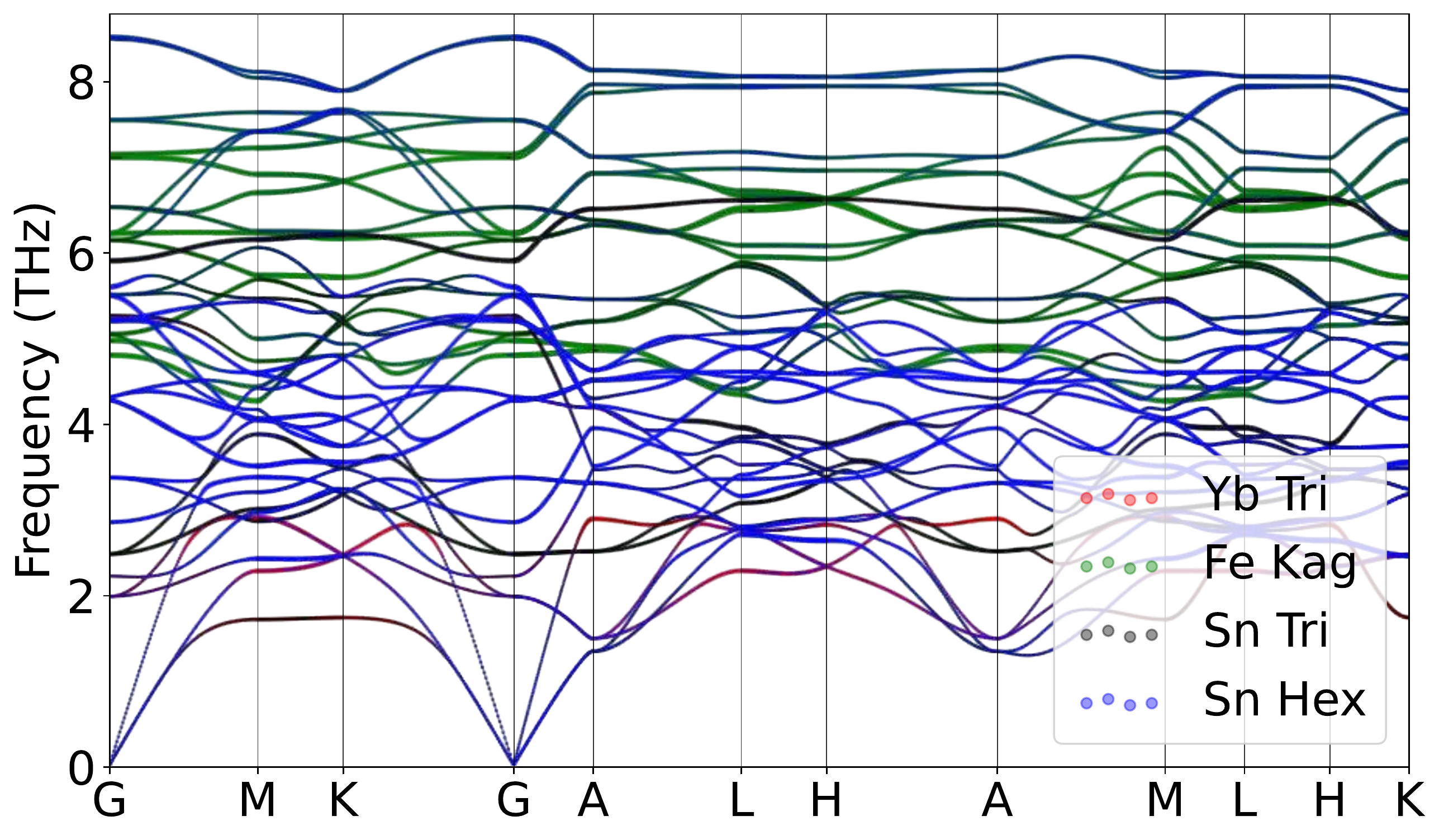}
}
\caption{Atomically projected phonon spectrum for YbFe$_6$Sn$_6$ in paramagnetic phase.}
\label{fig:YbFe6Sn6pho}
\end{figure}

\begin{figure}[htbp]
\centering
\includegraphics[width=0.45\textwidth]{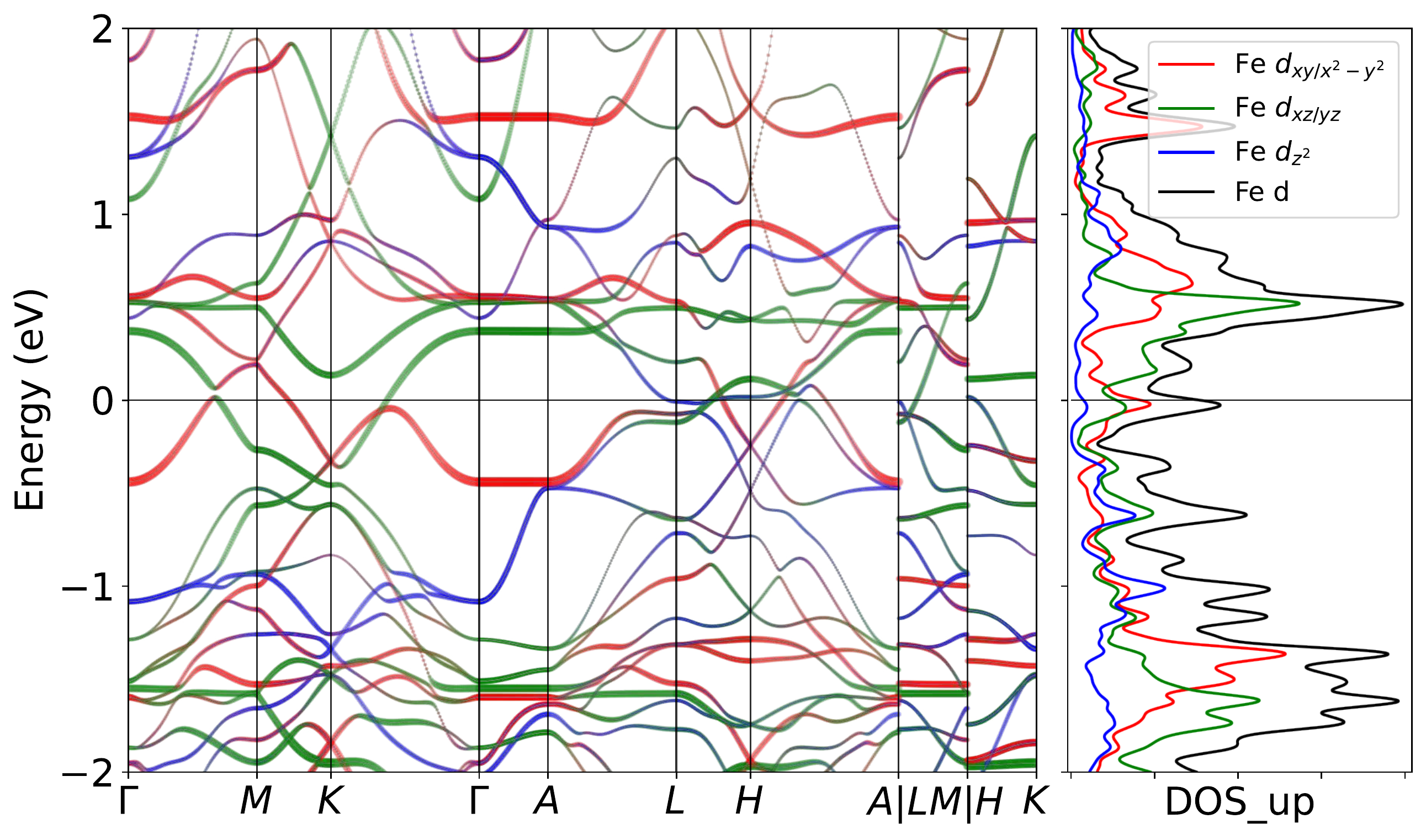}
\includegraphics[width=0.45\textwidth]{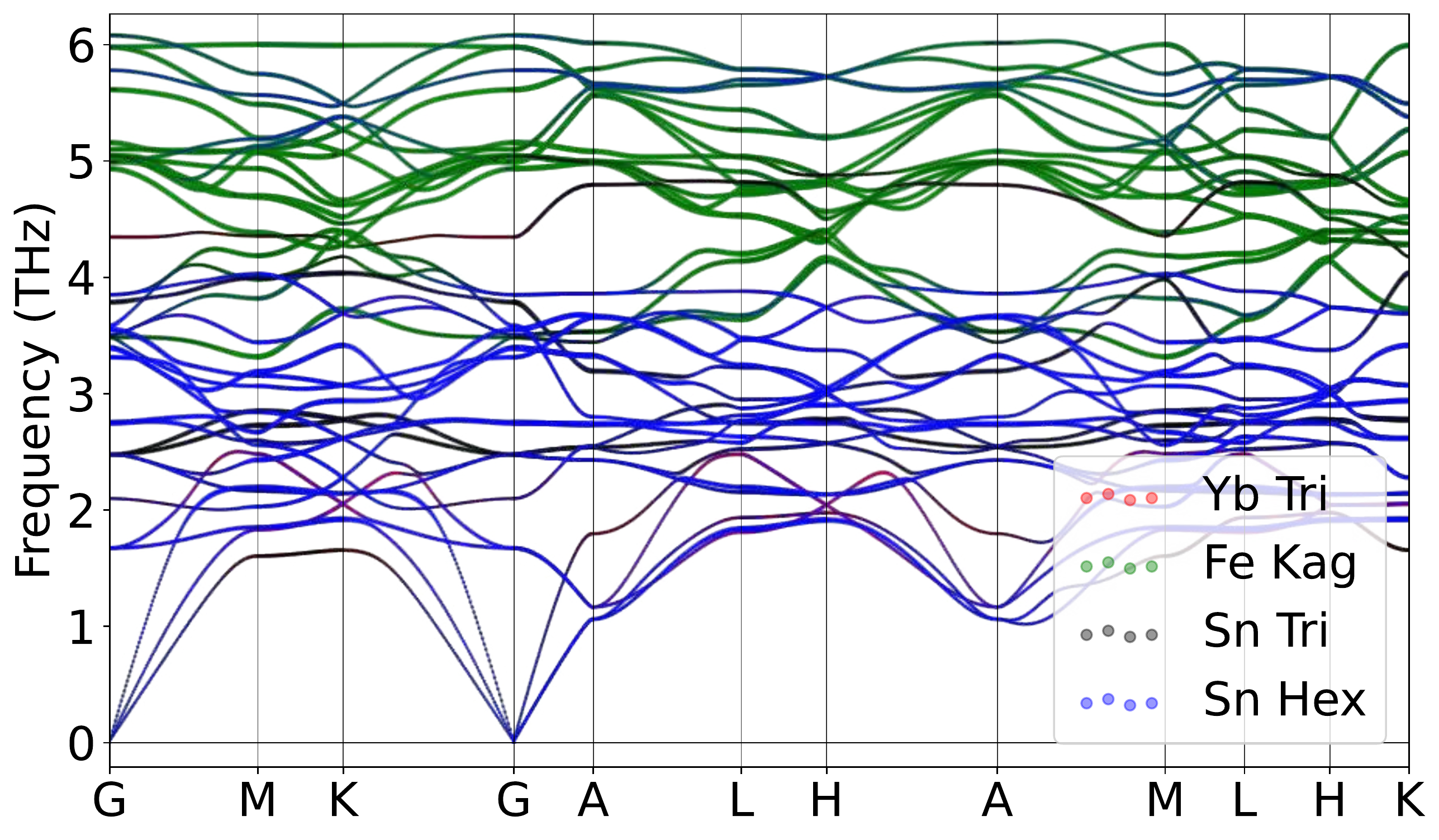}
\caption{Band structure for YbFe$_6$Sn$_6$ with AFM order without SOC for spin (Left)up channel. The projection of Fe $3d$ orbitals is also presented. (Right)Correspondingly, a stable phonon was demonstrated with the collinear magnetic order without Hubbard U at lower temperature regime, weighted by atomic projections.}
\label{fig:YbFe6Sn6mag}
\end{figure}

\begin{figure}[H]
\centering
\label{fig:YbFe6Sn6_relaxed_Low_xz}
\includegraphics[width=0.85\textwidth]{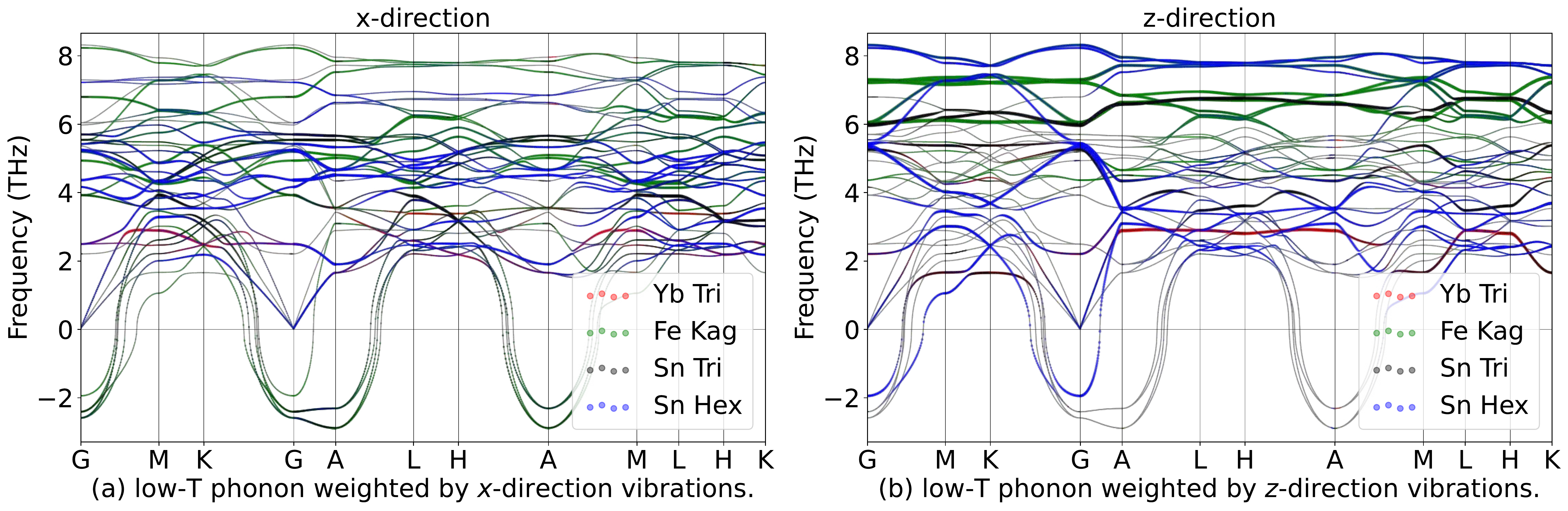}
\hspace{5pt}\label{fig:YbFe6Sn6_relaxed_High_xz}
\includegraphics[width=0.85\textwidth]{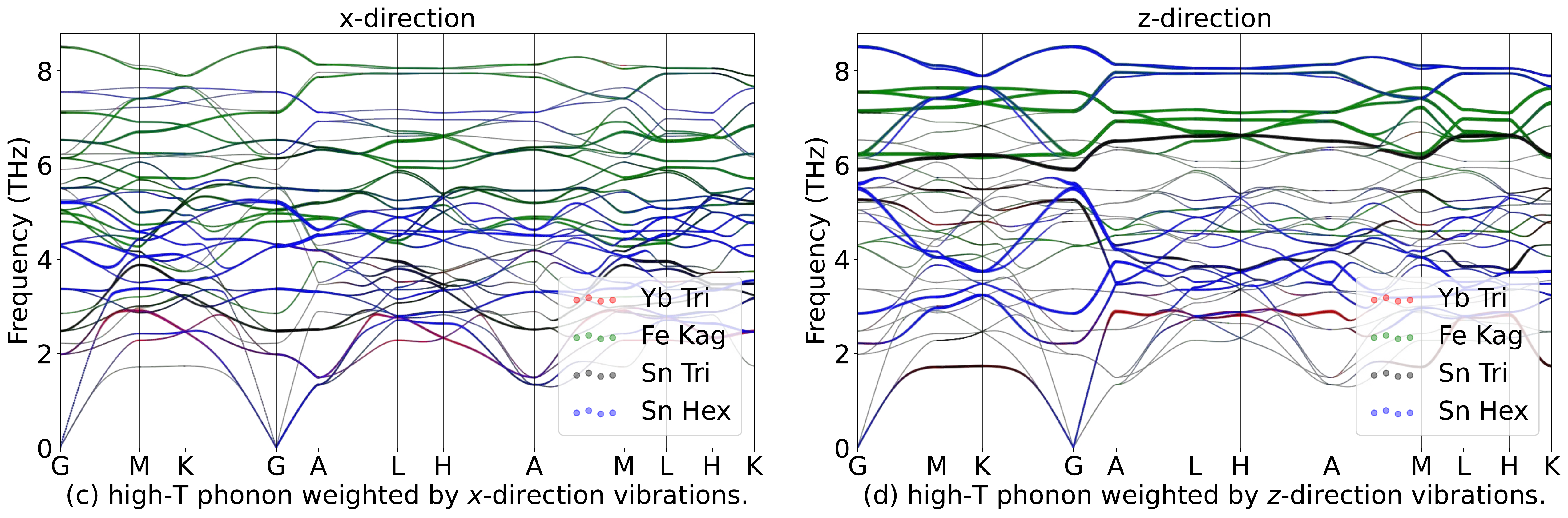}
\caption{Phonon spectrum for YbFe$_6$Sn$_6$in in-plane ($x$) and out-of-plane ($z$) directions in paramagnetic phase.}
\label{fig:YbFe6Sn6phoxyz}
\end{figure}

\begin{table}[htbp]
\global\long\def\arraystretch{1.12}
\captionsetup{width=.95\textwidth}
\caption{IRREPs at high-symmetry points for YbFe$_6$Sn$_6$ phonon spectrum at Low-T.}
\label{tab:191_YbFe6Sn6_Low}
\setlength{\tabcolsep}{1.mm}{
}
\end{table}

\begin{table}[htbp]
\global\long\def\arraystretch{1.12}
\captionsetup{width=.95\textwidth}
\caption{IRREPs at high-symmetry points for YbFe$_6$Sn$_6$ phonon spectrum at High-T.}
\label{tab:191_YbFe6Sn6_High}
\setlength{\tabcolsep}{1.mm}{
%
}
\end{table}
\subsubsection{\label{sms2sec:LuFe6Sn6}\hyperref[smsec:col]{\texorpdfstring{LuFe$_6$Sn$_6$}{}}~{\color{blue}  (High-T stable, Low-T unstable) }}

\begin{table}[htbp]
\global\long\def\arraystretch{1.12}
\captionsetup{width=.85\textwidth}
\caption{Structure parameters of LuFe$_6$Sn$_6$ after relaxation. It contains 527 electrons. }
\label{tab:LuFe6Sn6}
\setlength{\tabcolsep}{4mm}{
%
}
\end{table}

\begin{figure}[htbp]
\centering
\includegraphics[width=0.85\textwidth]{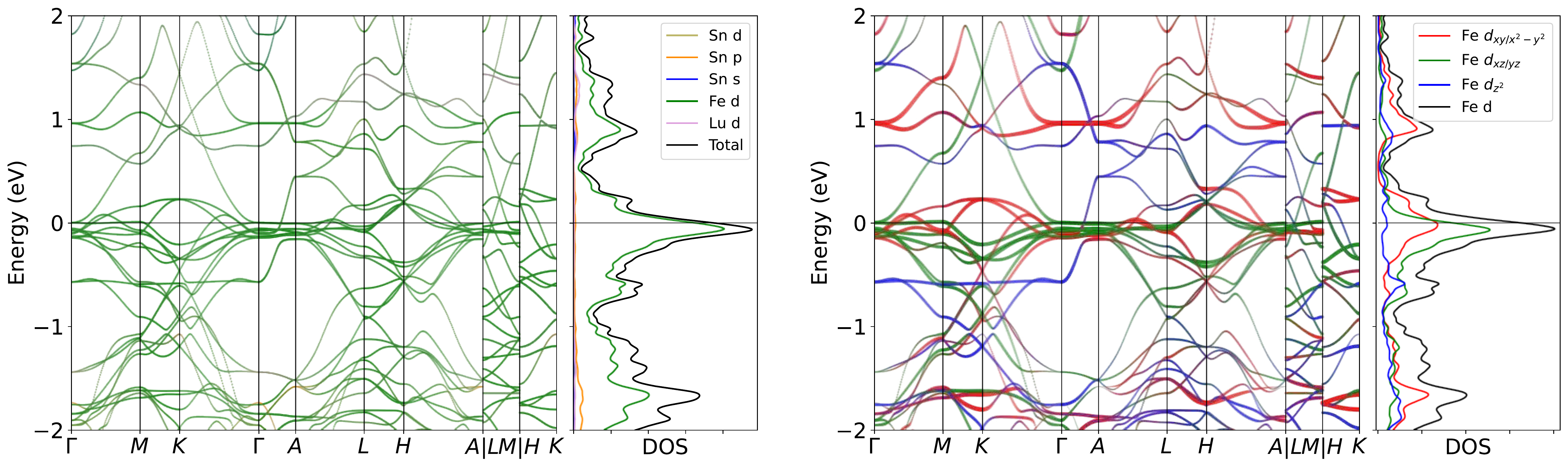}
\caption{Band structure for LuFe$_6$Sn$_6$ without SOC in paramagnetic phase. The projection of Fe $3d$ orbitals is also presented. The band structures clearly reveal the dominating of Fe $3d$ orbitals  near the Fermi level, as well as the pronounced peak.}
\label{fig:LuFe6Sn6band}
\end{figure}

\begin{figure}[H]
\centering
\subfloat[Relaxed phonon at low-T.]{
\label{fig:LuFe6Sn6_relaxed_Low}
\includegraphics[width=0.45\textwidth]{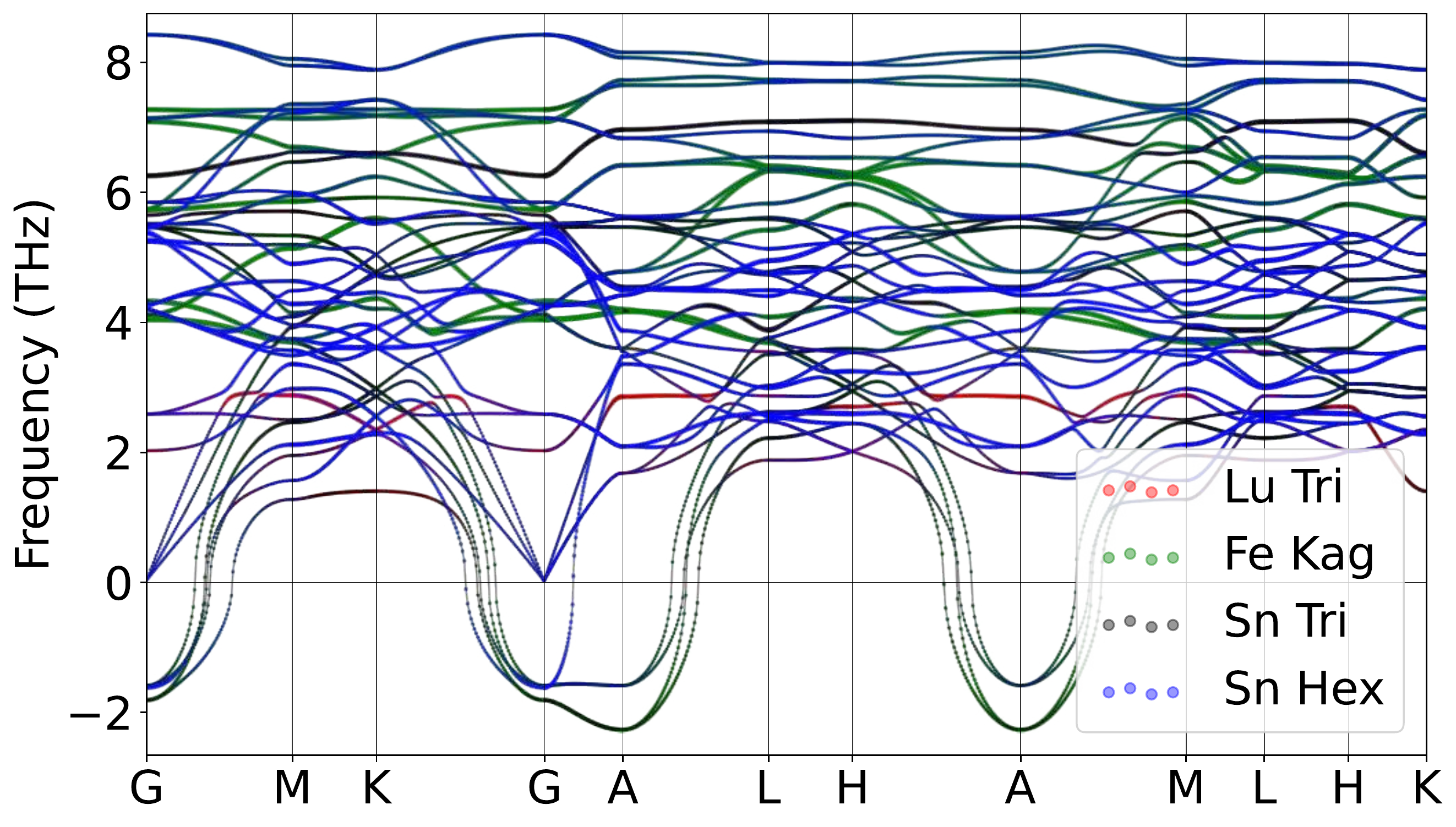}
}
\hspace{5pt}\subfloat[Relaxed phonon at high-T.]{
\label{fig:LuFe6Sn6_relaxed_High}
\includegraphics[width=0.45\textwidth]{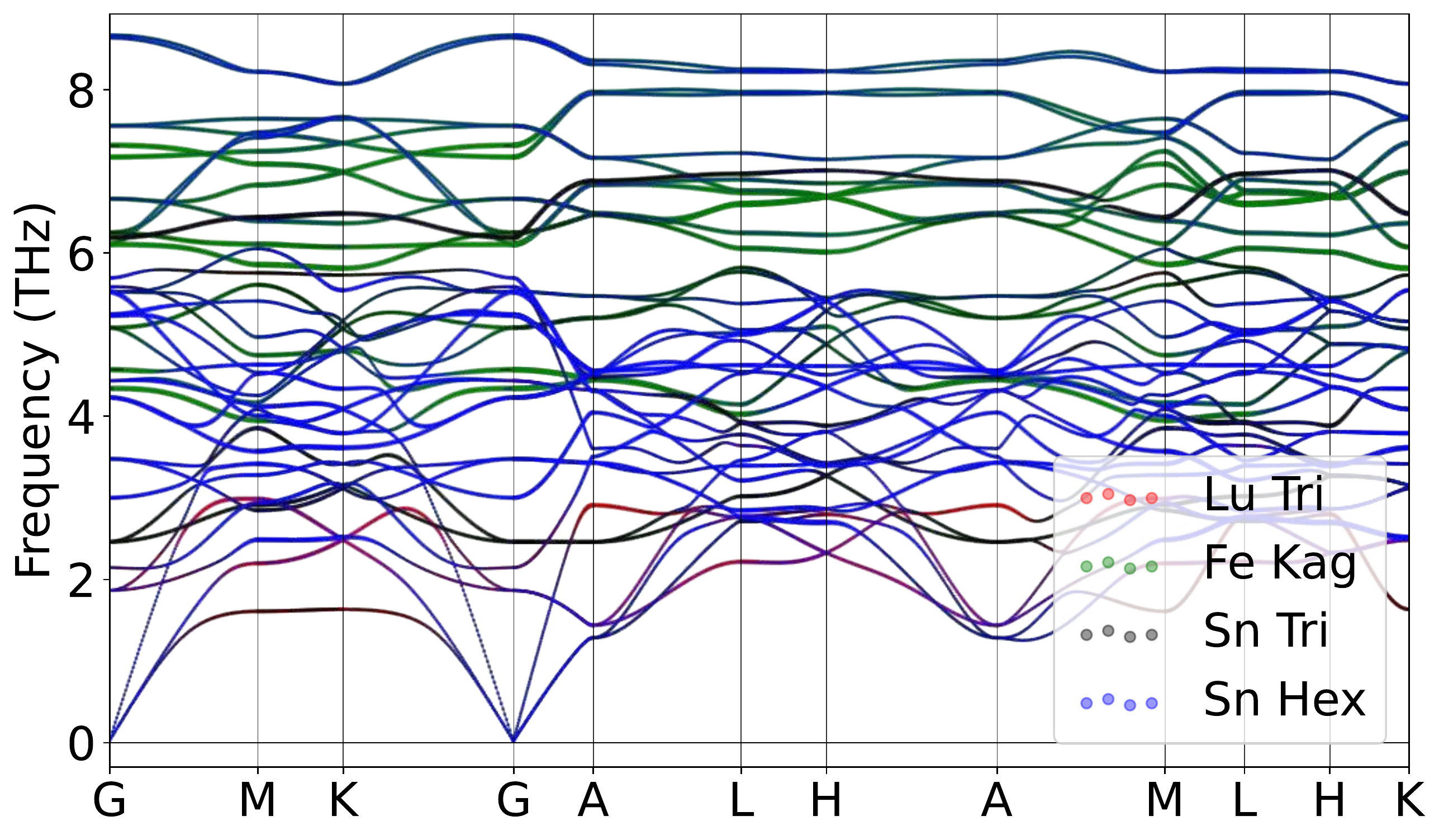}
}
\caption{Atomically projected phonon spectrum for LuFe$_6$Sn$_6$ in paramagnetic phase.}
\label{fig:LuFe6Sn6pho}
\end{figure}

\begin{figure}[htbp]
\centering
\includegraphics[width=0.45\textwidth]{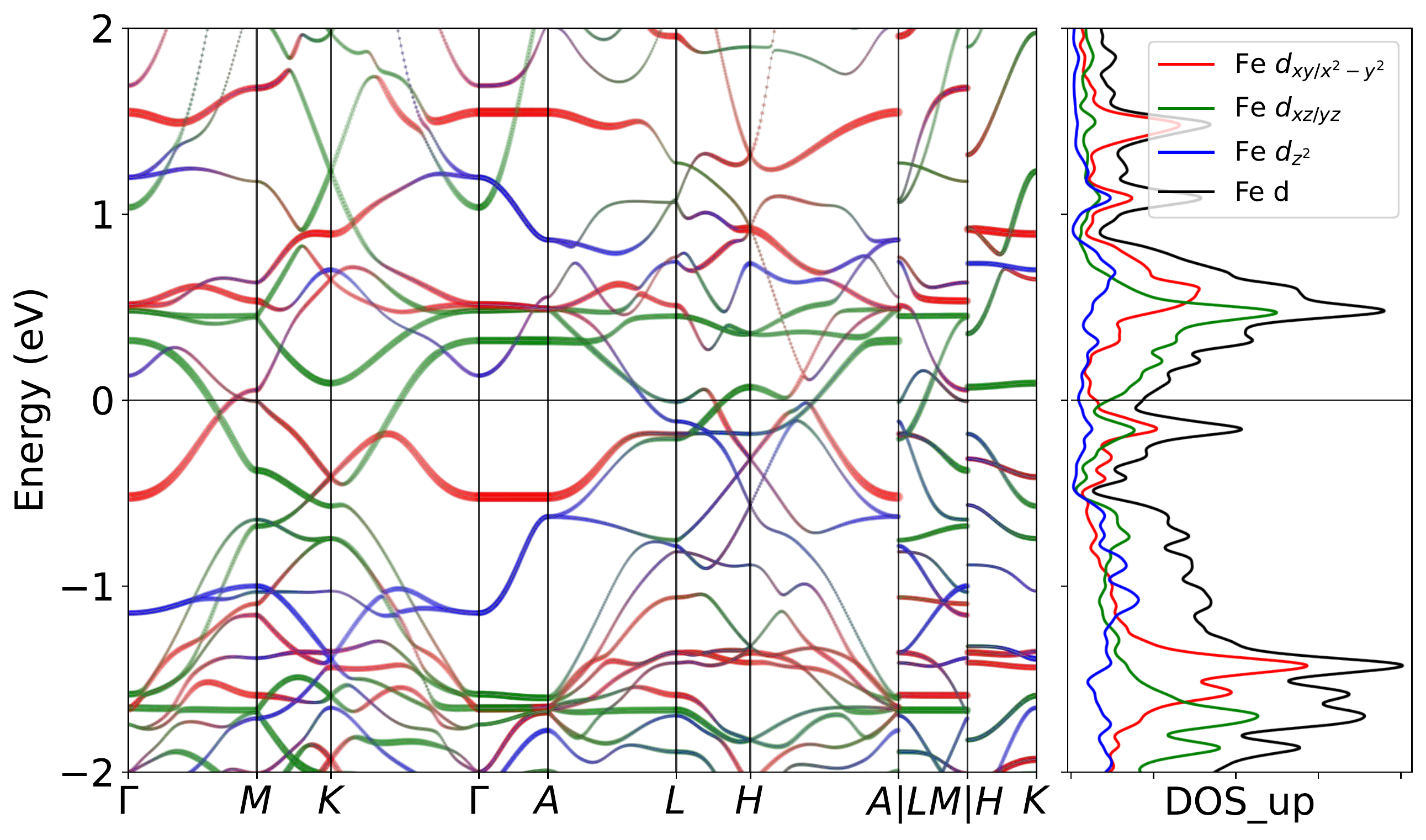}
\includegraphics[width=0.45\textwidth]{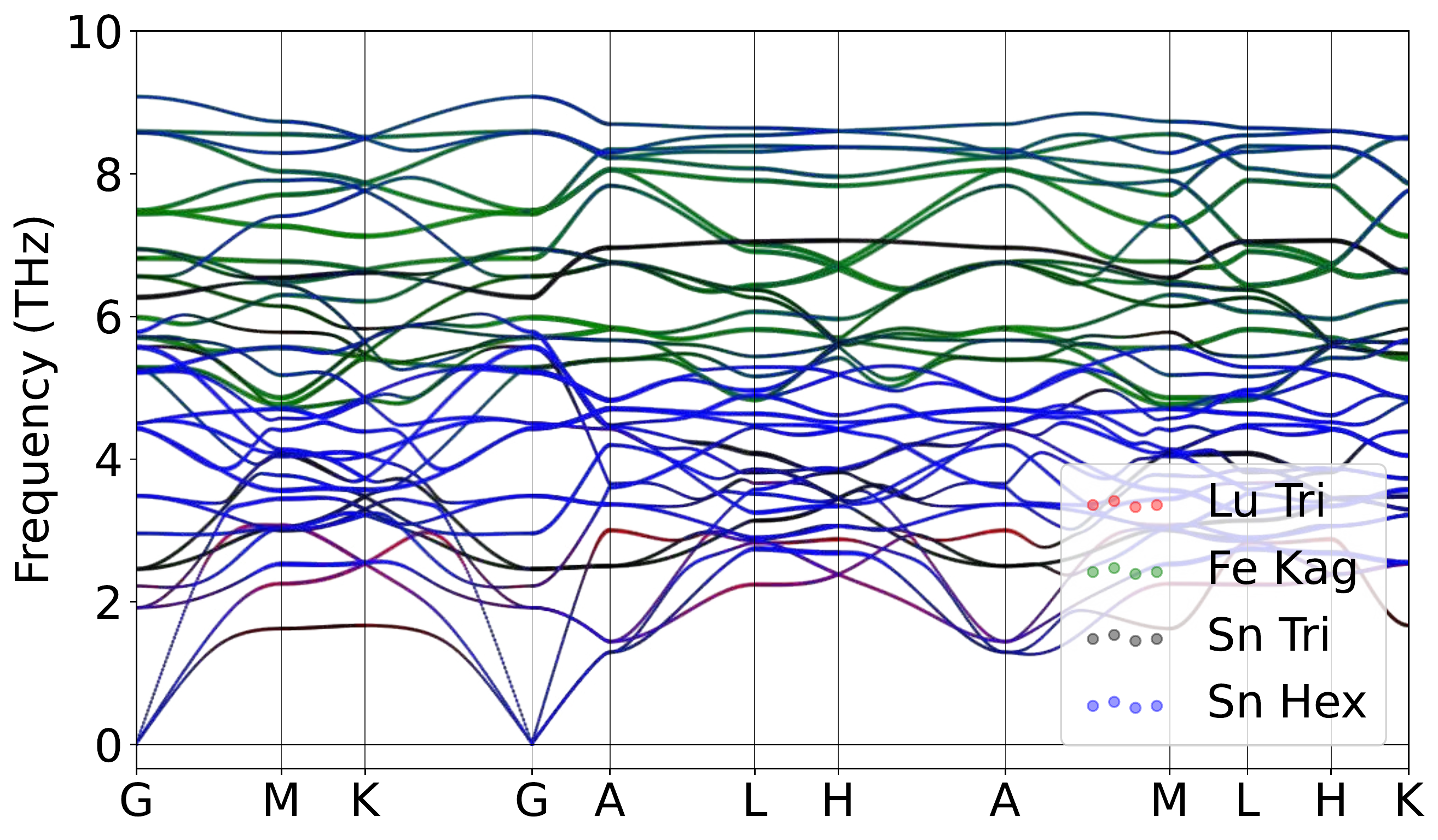}
\caption{Band structure for LuFe$_6$Sn$_6$ with AFM order without SOC for spin (Left)up channel. The projection of Fe $3d$ orbitals is also presented. (Right)Correspondingly, a stable phonon was demonstrated with the collinear magnetic order without Hubbard U at lower temperature regime, weighted by atomic projections.}
\label{fig:LuFe6Sn6mag}
\end{figure}

\begin{figure}[H]
\centering
\label{fig:LuFe6Sn6_relaxed_Low_xz}
\includegraphics[width=0.85\textwidth]{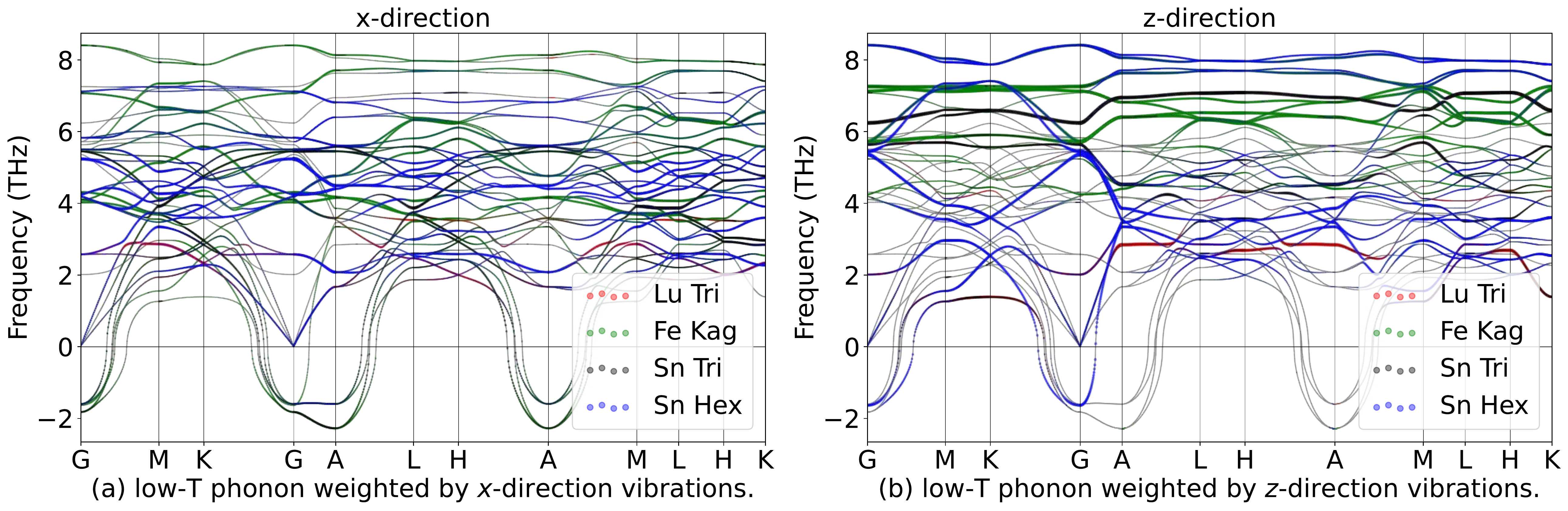}
\hspace{5pt}\label{fig:LuFe6Sn6_relaxed_High_xz}
\includegraphics[width=0.85\textwidth]{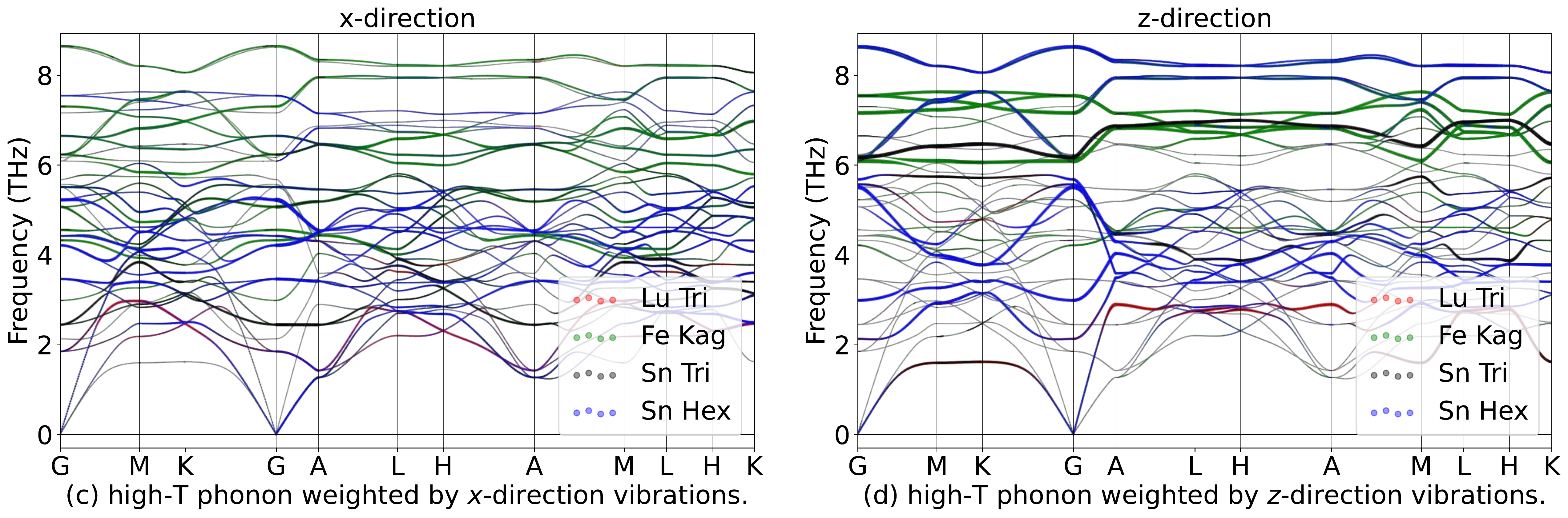}
\caption{Phonon spectrum for LuFe$_6$Sn$_6$in in-plane ($x$) and out-of-plane ($z$) directions in paramagnetic phase.}
\label{fig:LuFe6Sn6phoxyz}
\end{figure}

\begin{table}[htbp]
\global\long\def\arraystretch{1.12}
\captionsetup{width=.95\textwidth}
\caption{IRREPs at high-symmetry points for LuFe$_6$Sn$_6$ phonon spectrum at Low-T.}
\label{tab:191_LuFe6Sn6_Low}
\setlength{\tabcolsep}{1.mm}{
}
\end{table}

\begin{table}[htbp]
\global\long\def\arraystretch{1.12}
\captionsetup{width=.95\textwidth}
\caption{IRREPs at high-symmetry points for LuFe$_6$Sn$_6$ phonon spectrum at High-T.}
\label{tab:191_LuFe6Sn6_High}
\setlength{\tabcolsep}{1.mm}{
%
}
\end{table}
\subsubsection{\label{sms2sec:HfFe6Sn6}\hyperref[smsec:col]{\texorpdfstring{HfFe$_6$Sn$_6$}{}}~{\color{blue}  (High-T stable, Low-T unstable) }}

\begin{table}[htbp]
\global\long\def\arraystretch{1.12}
\captionsetup{width=.85\textwidth}
\caption{Structure parameters of HfFe$_6$Sn$_6$ after relaxation. It contains 528 electrons. }
\label{tab:HfFe6Sn6}
\setlength{\tabcolsep}{4mm}{
%
}
\end{table}

\begin{figure}[htbp]
\centering
\includegraphics[width=0.85\textwidth]{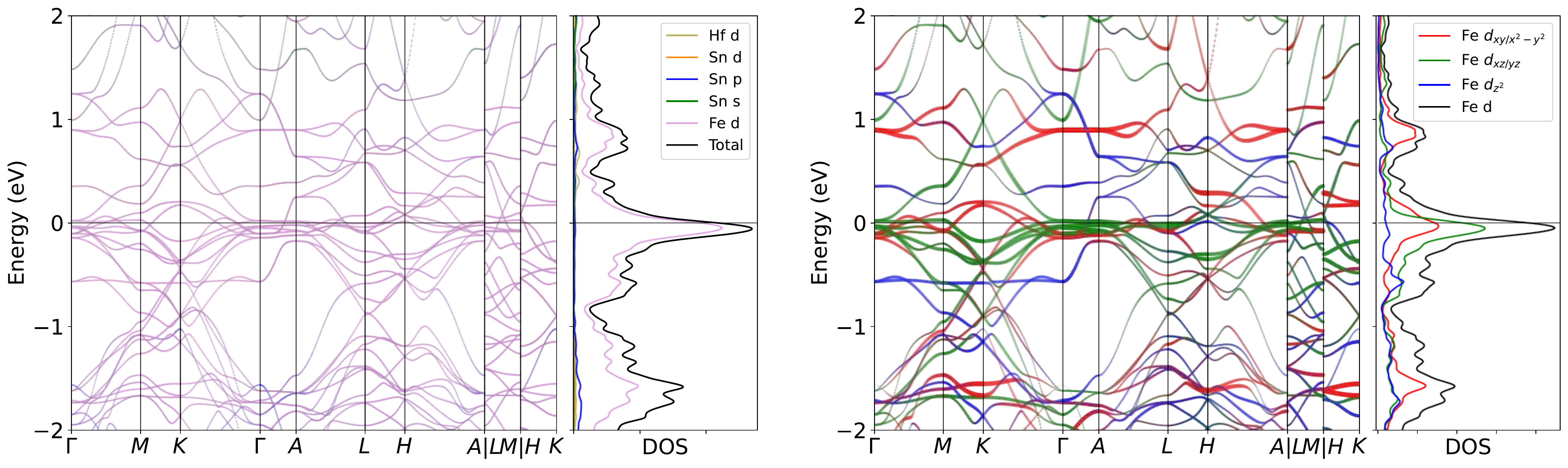}
\caption{Band structure for HfFe$_6$Sn$_6$ without SOC in paramagnetic phase. The projection of Fe $3d$ orbitals is also presented. The band structures clearly reveal the dominating of Fe $3d$ orbitals  near the Fermi level, as well as the pronounced peak.}
\label{fig:HfFe6Sn6band}
\end{figure}

\begin{figure}[H]
\centering
\subfloat[Relaxed phonon at low-T.]{
\label{fig:HfFe6Sn6_relaxed_Low}
\includegraphics[width=0.45\textwidth]{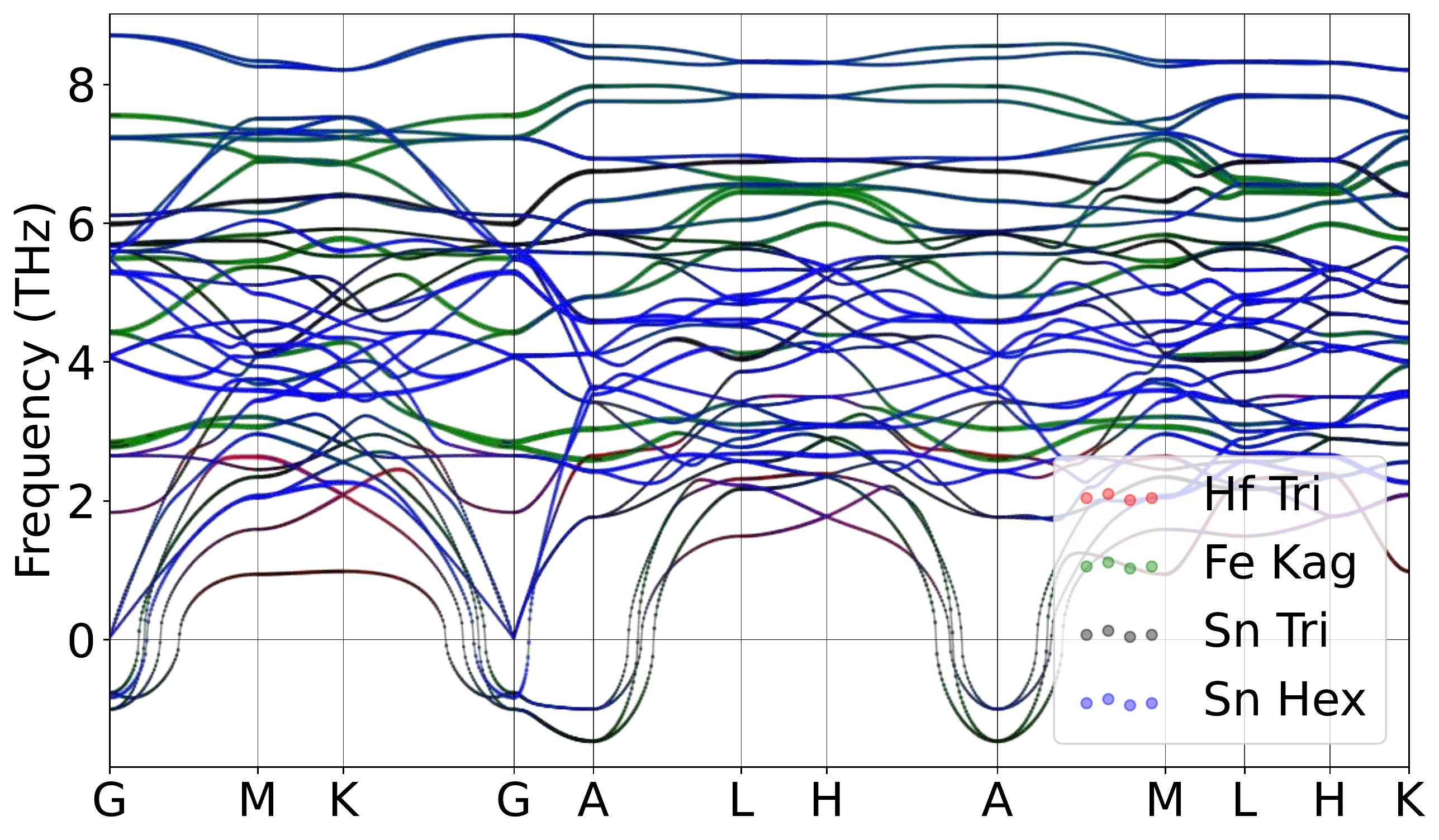}
}
\hspace{5pt}\subfloat[Relaxed phonon at high-T.]{
\label{fig:HfFe6Sn6_relaxed_High}
\includegraphics[width=0.45\textwidth]{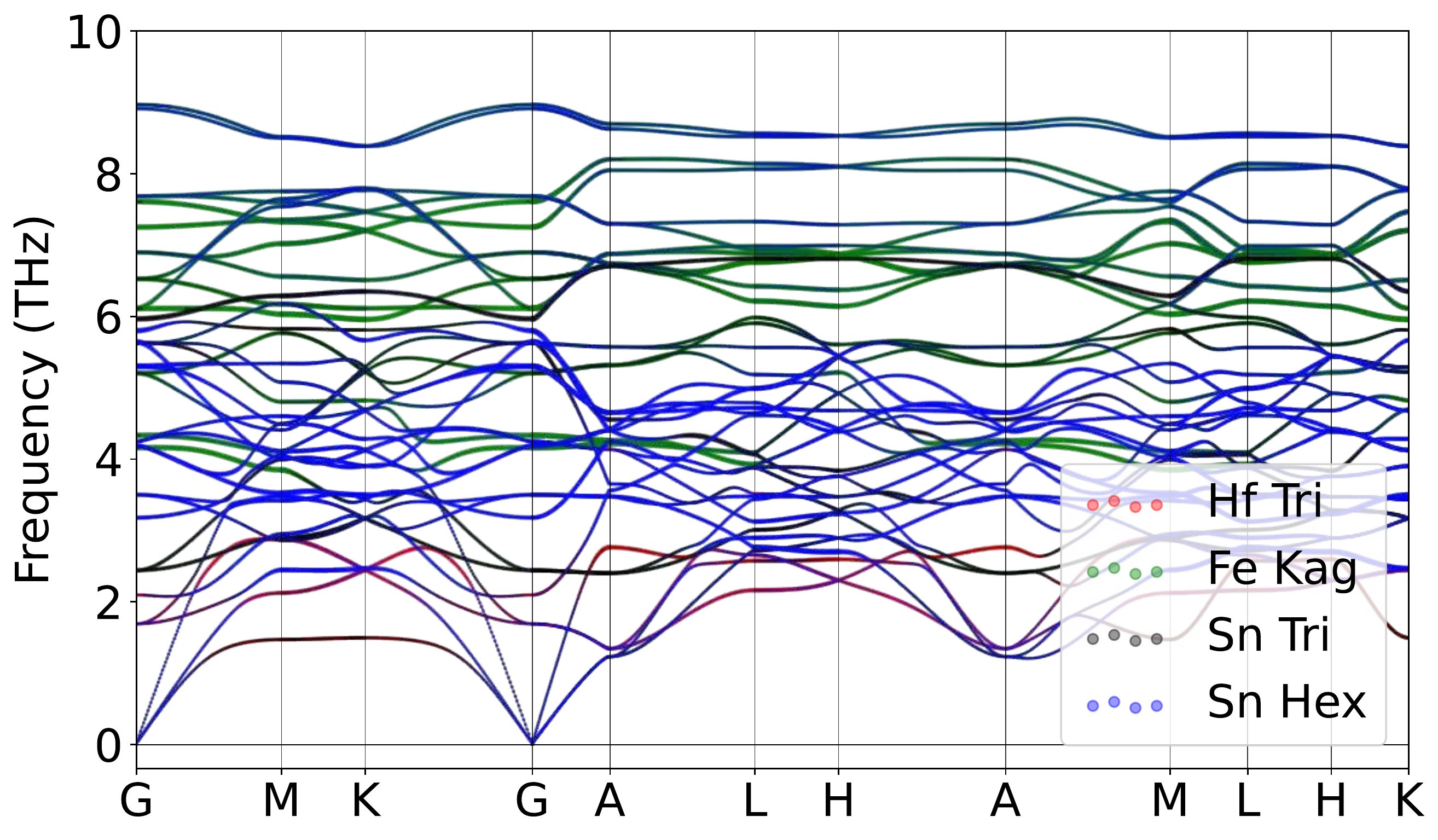}
}
\caption{Atomically projected phonon spectrum for HfFe$_6$Sn$_6$ in paramagnetic phase.}
\label{fig:HfFe6Sn6pho}
\end{figure}

\begin{figure}[htbp]
\centering
\includegraphics[width=0.45\textwidth]{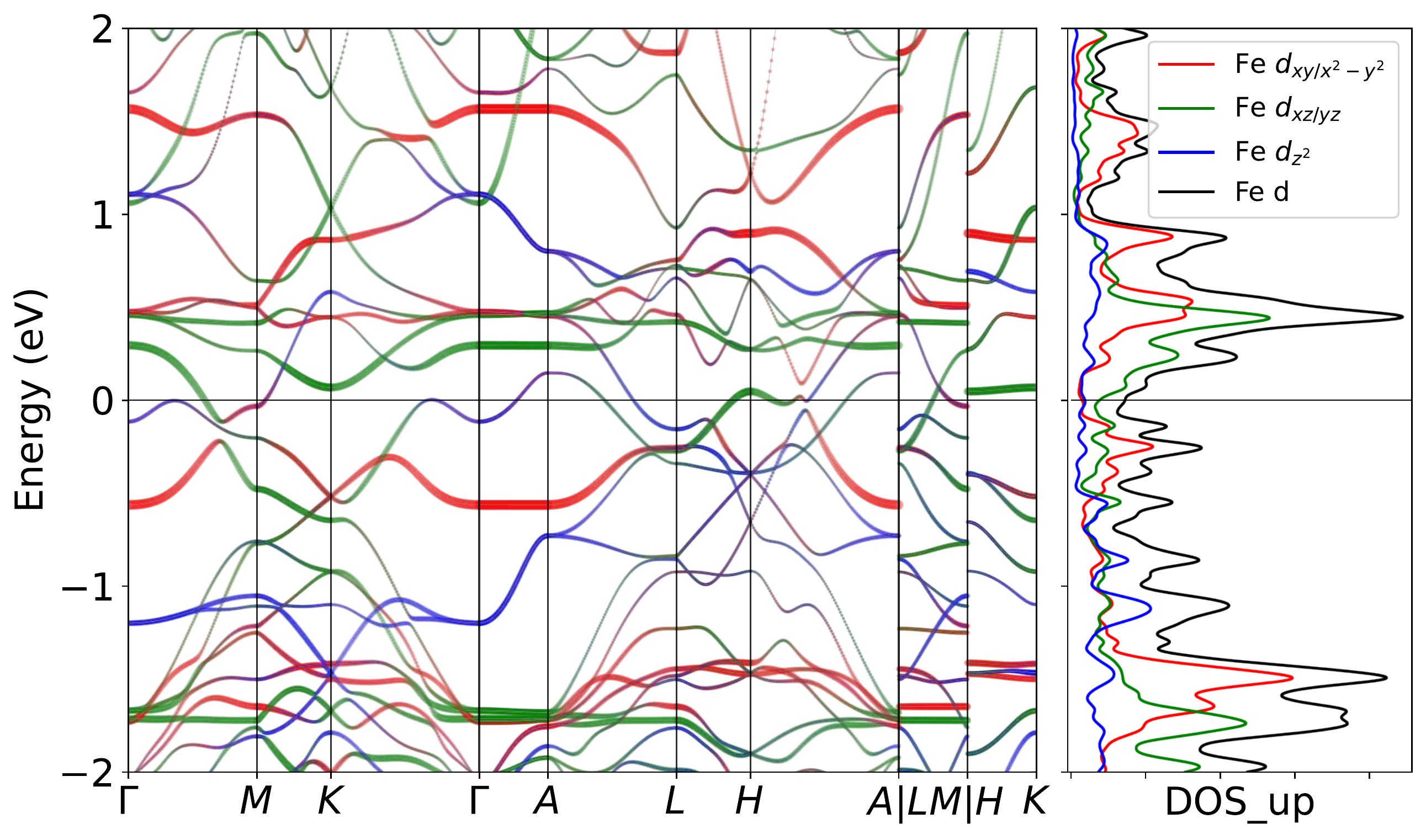}
\includegraphics[width=0.45\textwidth]{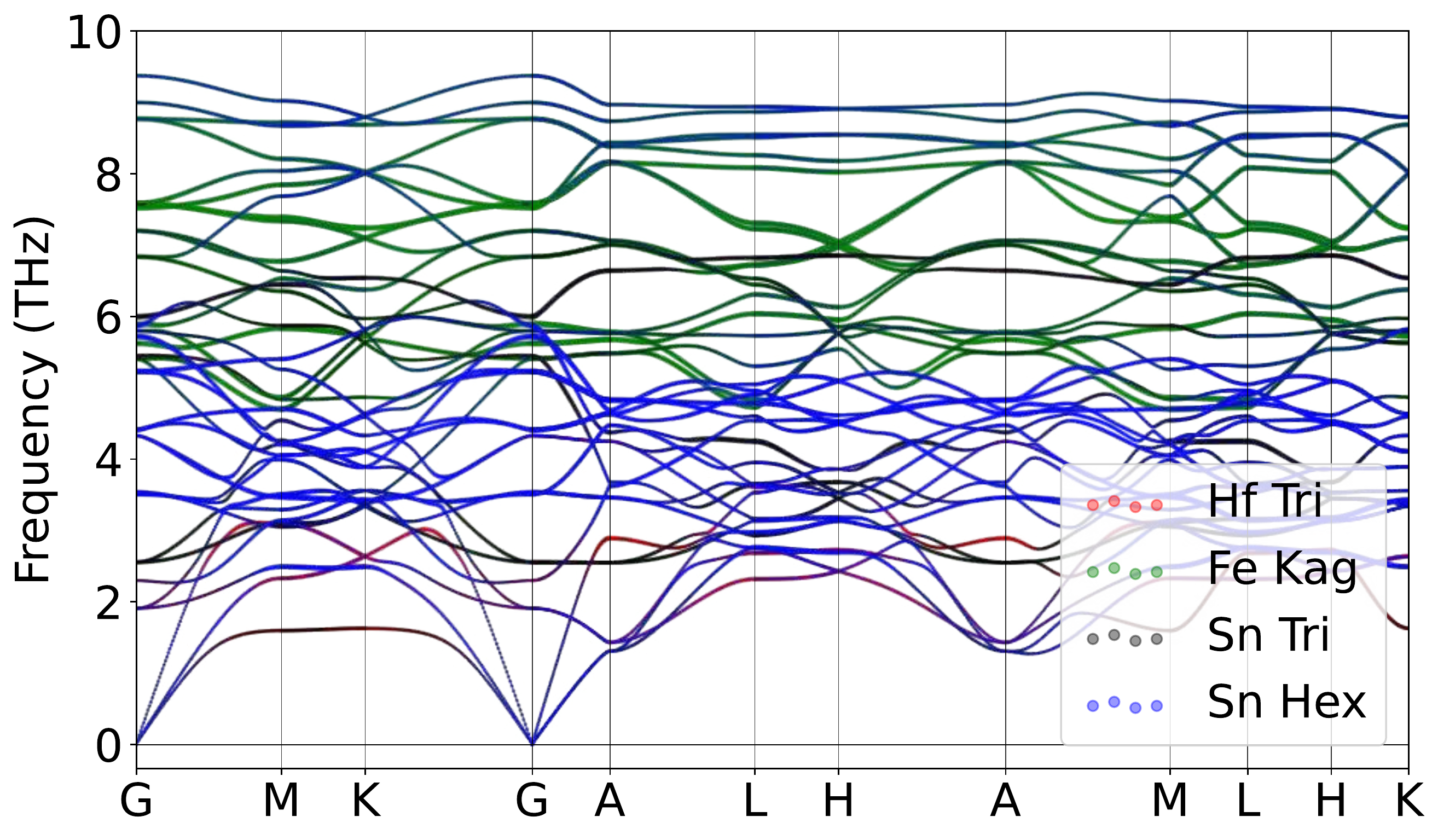}
\caption{Band structure for HfFe$_6$Sn$_6$ with AFM order without SOC for spin (Left)up channel. The projection of Fe $3d$ orbitals is also presented. (Right)Correspondingly, a stable phonon was demonstrated with the collinear magnetic order without Hubbard U at lower temperature regime, weighted by atomic projections.}
\label{fig:HfFe6Sn6mag}
\end{figure}

\begin{figure}[H]
\centering
\label{fig:HfFe6Sn6_relaxed_Low_xz}
\includegraphics[width=0.85\textwidth]{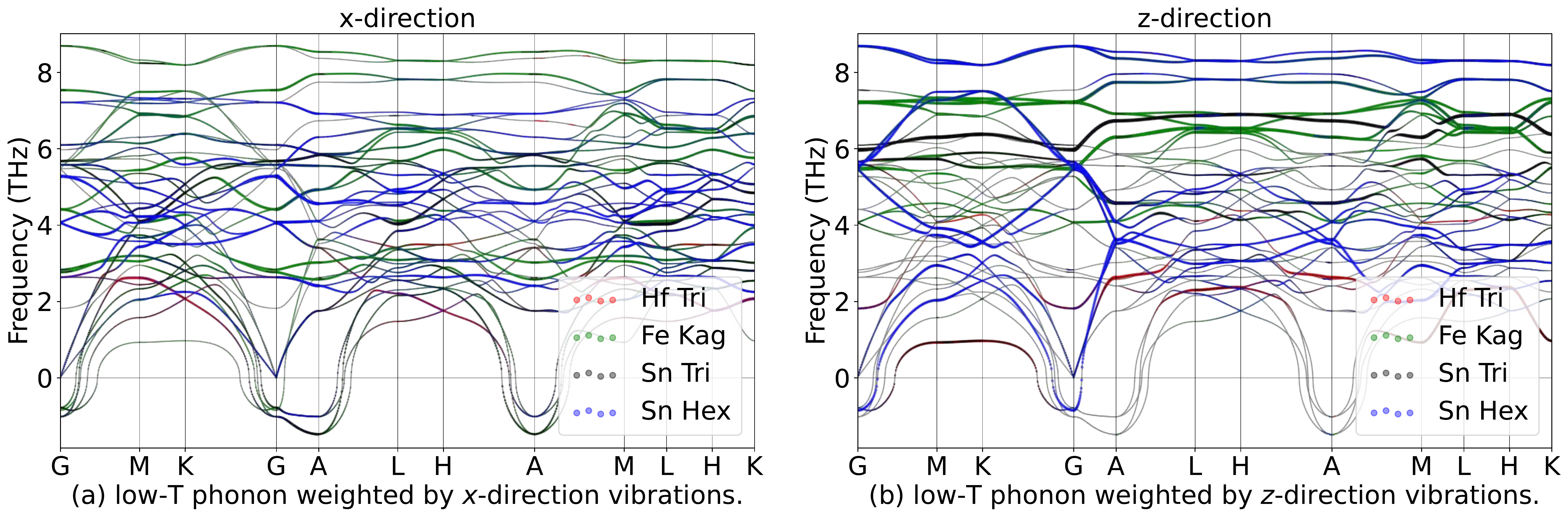}
\hspace{5pt}\label{fig:HfFe6Sn6_relaxed_High_xz}
\includegraphics[width=0.85\textwidth]{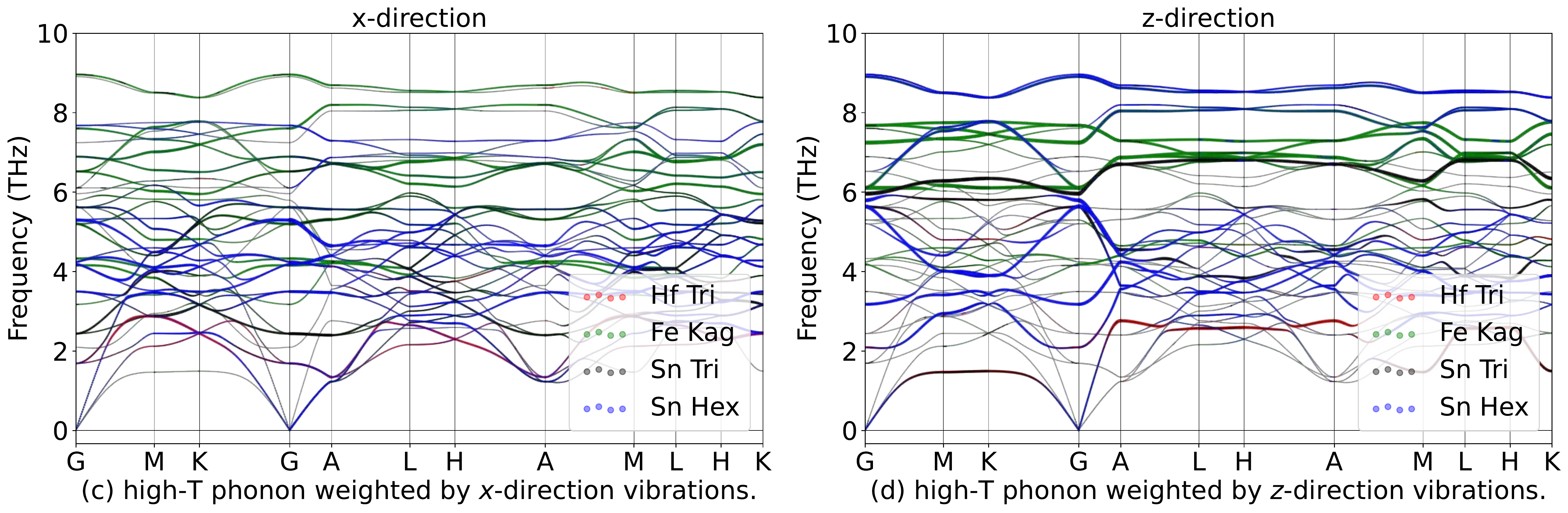}
\caption{Phonon spectrum for HfFe$_6$Sn$_6$in in-plane ($x$) and out-of-plane ($z$) directions in paramagnetic phase.}
\label{fig:HfFe6Sn6phoxyz}
\end{figure}

\begin{table}[htbp]
\global\long\def\arraystretch{1.12}
\captionsetup{width=.95\textwidth}
\caption{IRREPs at high-symmetry points for HfFe$_6$Sn$_6$ phonon spectrum at Low-T.}
\label{tab:191_HfFe6Sn6_Low}
\setlength{\tabcolsep}{1.mm}{
}
\end{table}

\begin{table}[htbp]
\global\long\def\arraystretch{1.12}
\captionsetup{width=.95\textwidth}
\caption{IRREPs at high-symmetry points for HfFe$_6$Sn$_6$ phonon spectrum at High-T.}
\label{tab:191_HfFe6Sn6_High}
\setlength{\tabcolsep}{1.mm}{
%
}
\end{table}
\subsubsection{\label{sms2sec:TaFe6Sn6}\hyperref[smsec:col]{\texorpdfstring{TaFe$_6$Sn$_6$}{}}~{\color{blue}  (High-T stable, Low-T unstable) }}

\begin{table}[htbp]
\global\long\def\arraystretch{1.12}
\captionsetup{width=.85\textwidth}
\caption{Structure parameters of TaFe$_6$Sn$_6$ after relaxation. It contains 529 electrons. }
\label{tab:TaFe6Sn6}
\setlength{\tabcolsep}{4mm}{
%
}
\end{table}

\begin{figure}[htbp]
\centering
\includegraphics[width=0.85\textwidth]{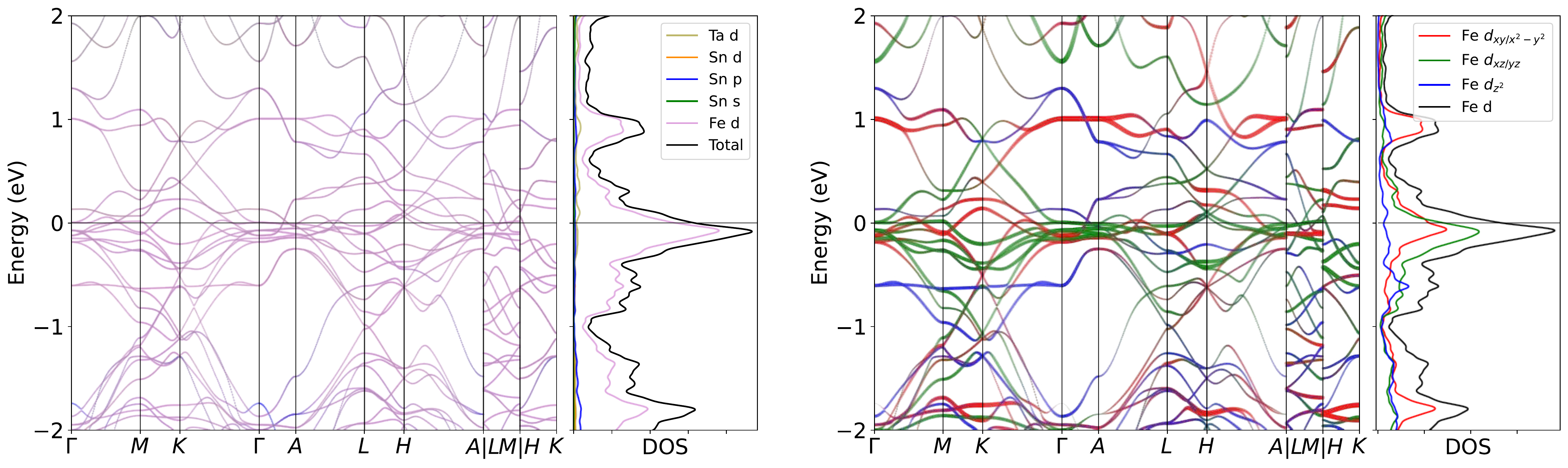}
\caption{Band structure for TaFe$_6$Sn$_6$ without SOC in paramagnetic phase. The projection of Fe $3d$ orbitals is also presented. The band structures clearly reveal the dominating of Fe $3d$ orbitals  near the Fermi level, as well as the pronounced peak.}
\label{fig:TaFe6Sn6band}
\end{figure}

\begin{figure}[H]
\centering
\subfloat[Relaxed phonon at low-T.]{
\label{fig:TaFe6Sn6_relaxed_Low}
\includegraphics[width=0.45\textwidth]{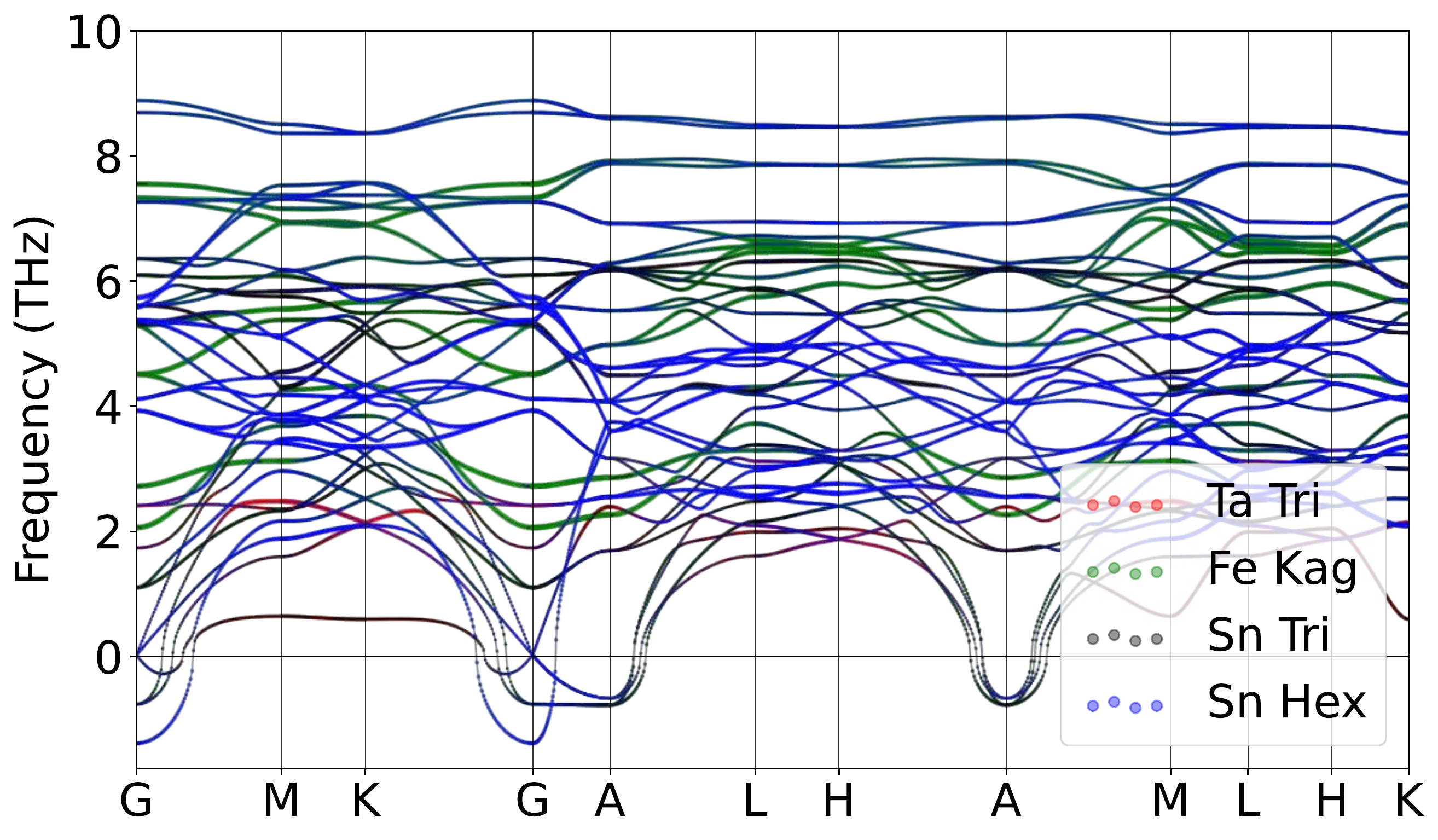}
}
\hspace{5pt}\subfloat[Relaxed phonon at high-T.]{
\label{fig:TaFe6Sn6_relaxed_High}
\includegraphics[width=0.45\textwidth]{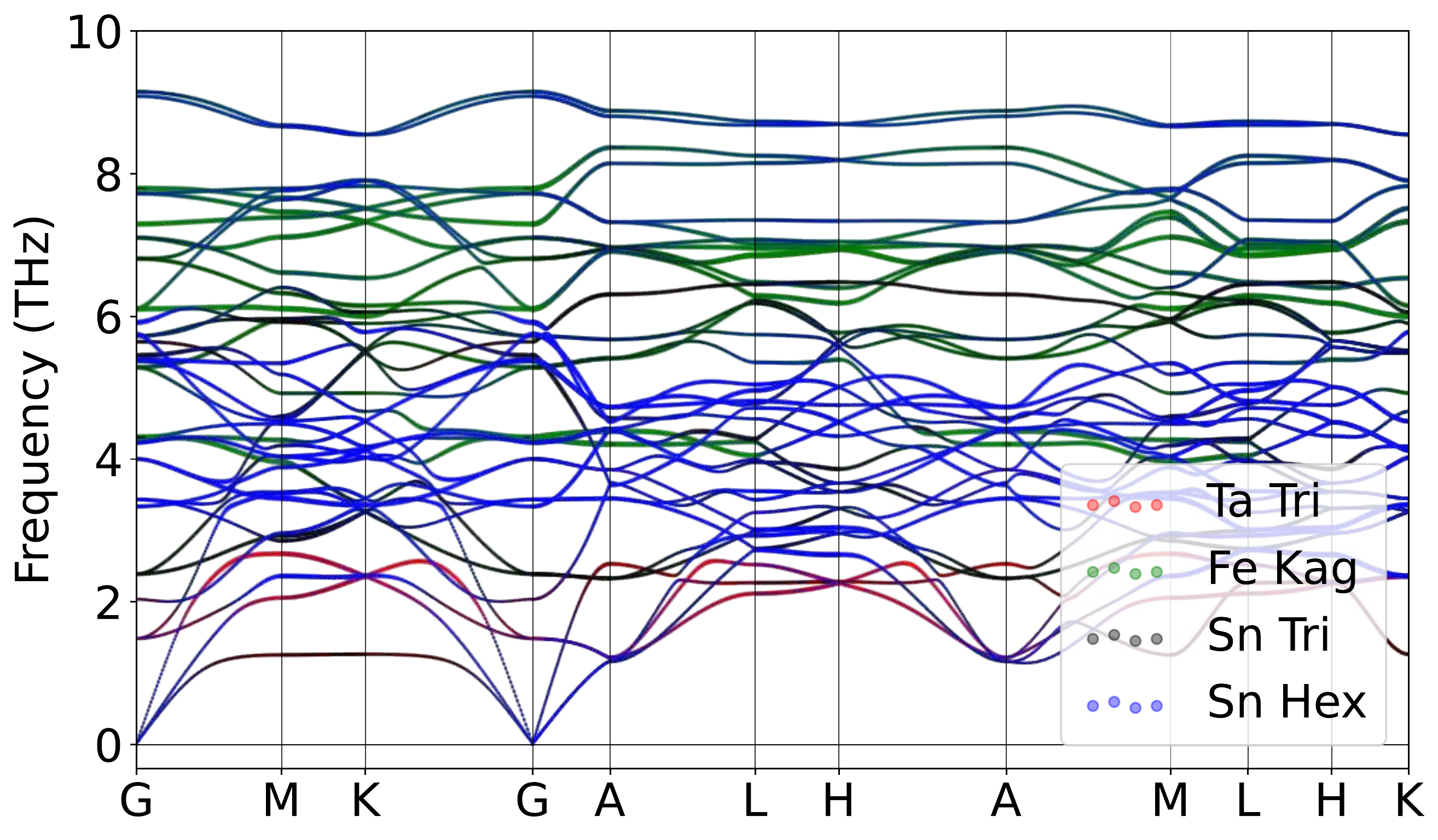}
}
\caption{Atomically projected phonon spectrum for TaFe$_6$Sn$_6$ in paramagnetic phase.}
\label{fig:TaFe6Sn6pho}
\end{figure}

\begin{figure}[htbp]
\centering
\includegraphics[width=0.45\textwidth]{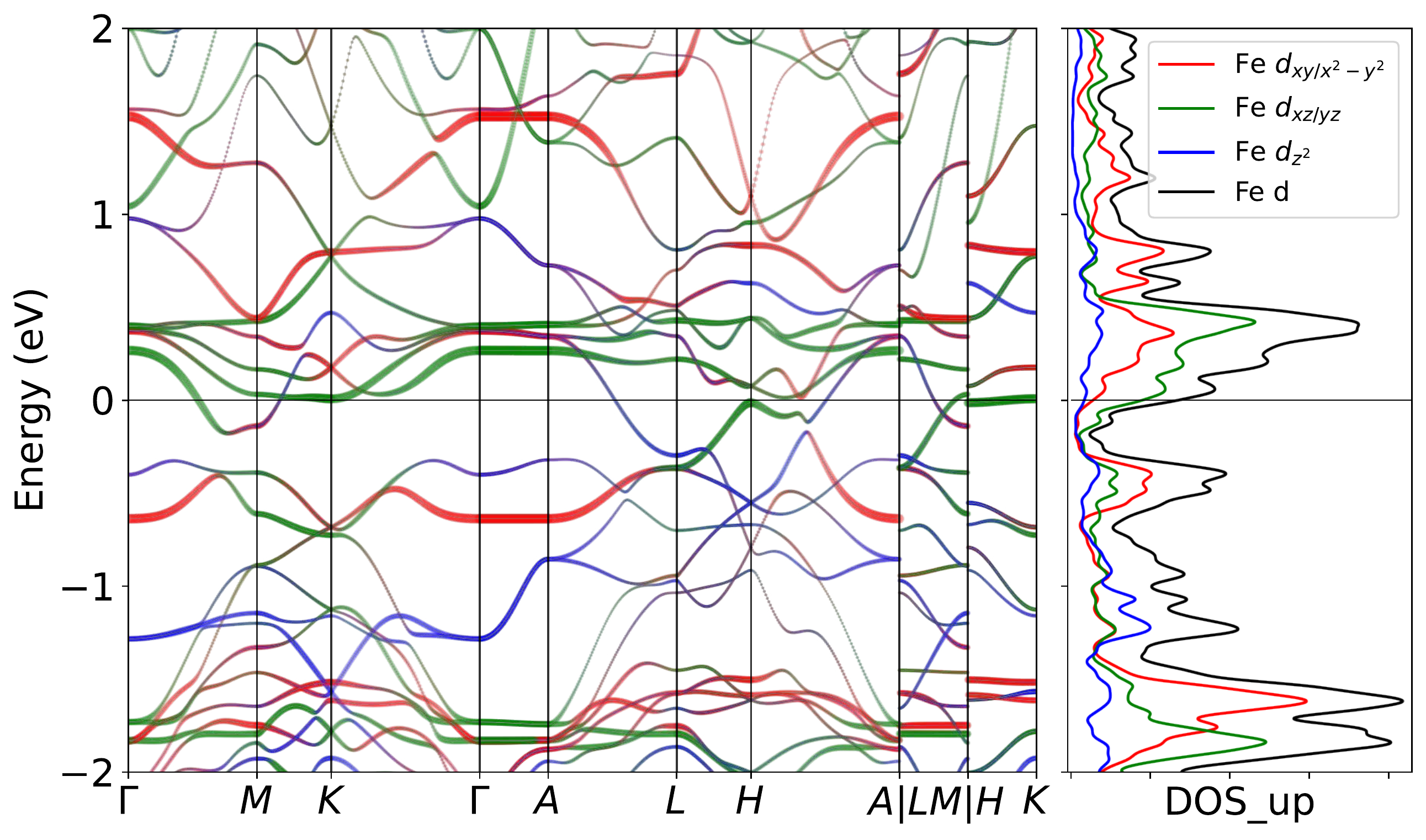}
\includegraphics[width=0.45\textwidth]{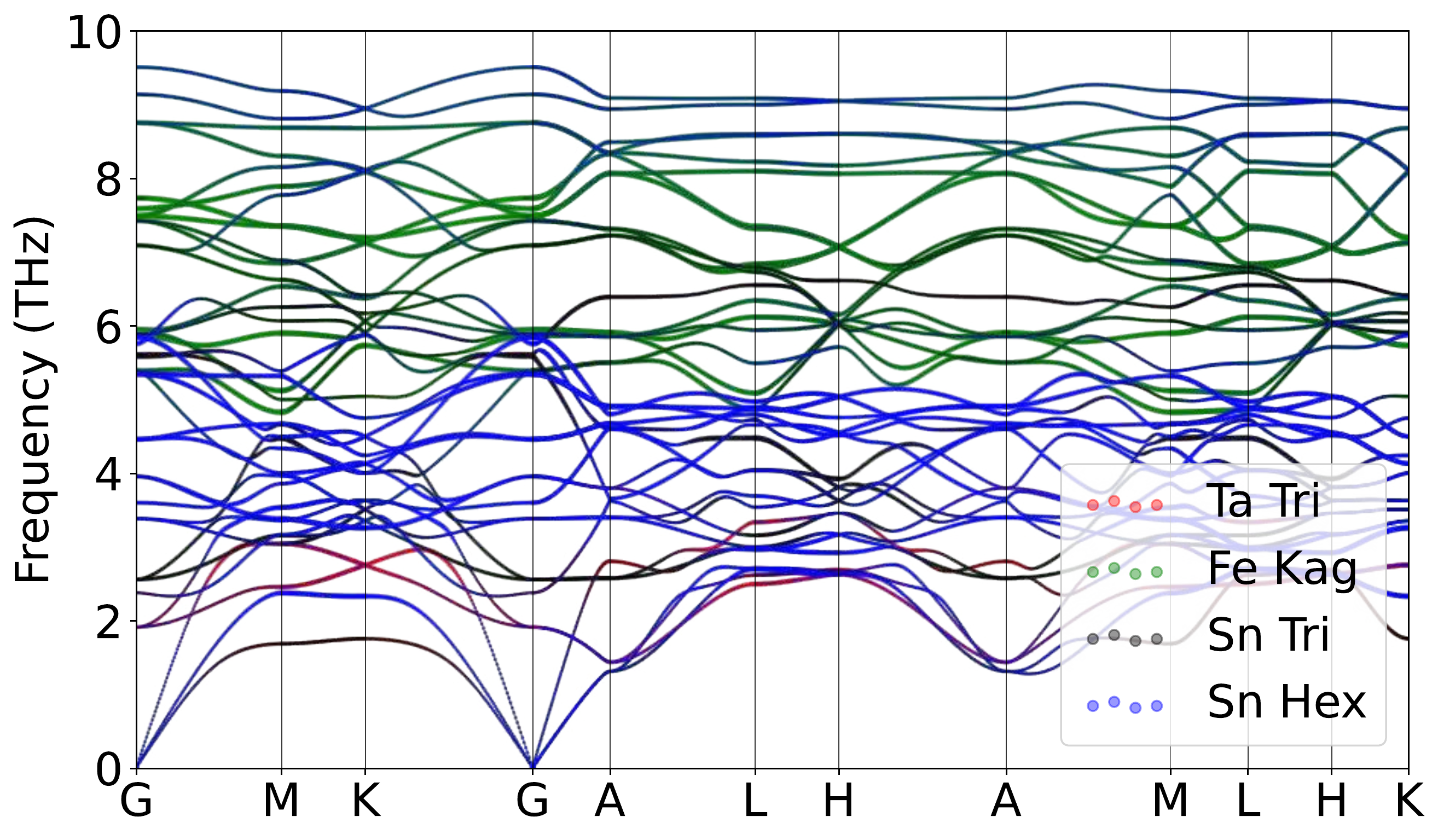}
\caption{Band structure for TaFe$_6$Sn$_6$ with AFM order without SOC for spin (Left)up channel. The projection of Fe $3d$ orbitals is also presented. (Right)Correspondingly, a stable phonon was demonstrated with the collinear magnetic order without Hubbard U at lower temperature regime, weighted by atomic projections.}
\label{fig:TaFe6Sn6mag}
\end{figure}

\begin{figure}[H]
\centering
\label{fig:TaFe6Sn6_relaxed_Low_xz}
\includegraphics[width=0.85\textwidth]{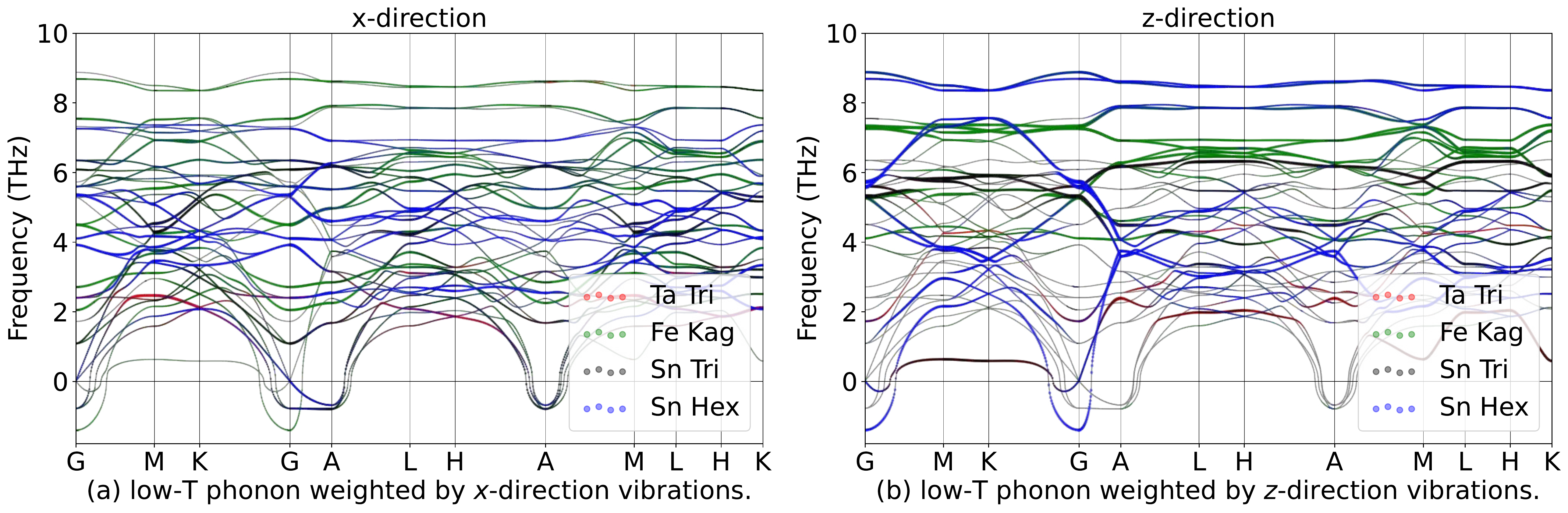}
\hspace{5pt}\label{fig:TaFe6Sn6_relaxed_High_xz}
\includegraphics[width=0.85\textwidth]{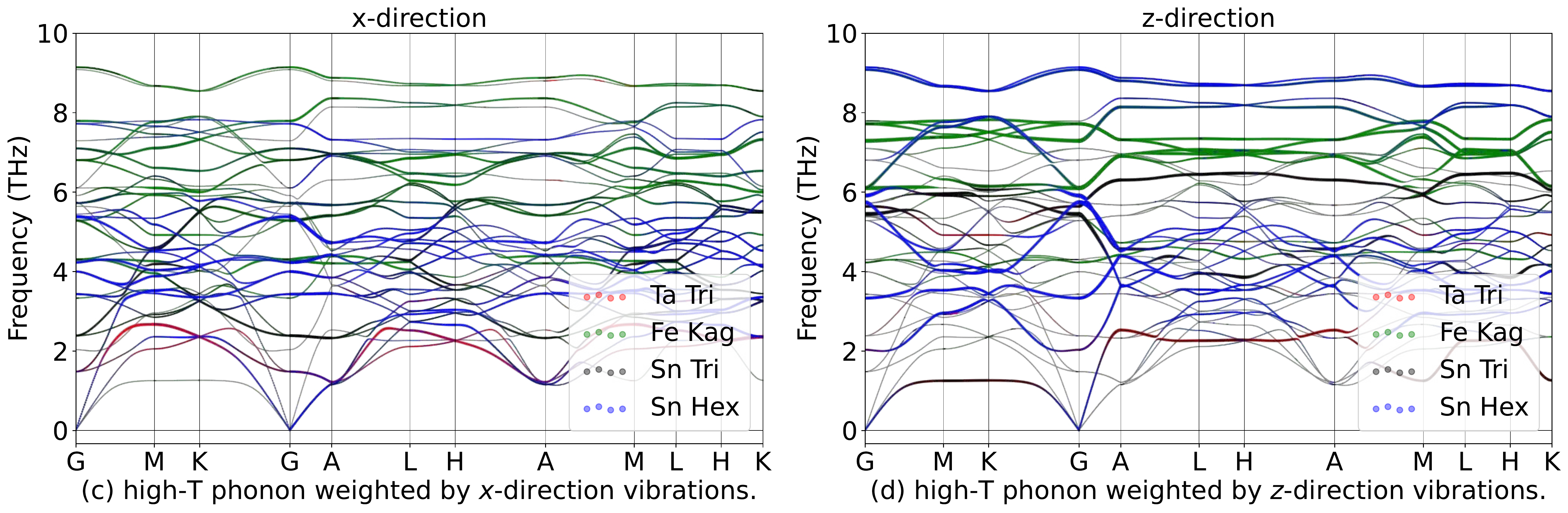}
\caption{Phonon spectrum for TaFe$_6$Sn$_6$in in-plane ($x$) and out-of-plane ($z$) directions in paramagnetic phase.}
\label{fig:TaFe6Sn6phoxyz}
\end{figure}

\begin{table}[htbp]
\global\long\def\arraystretch{1.12}
\captionsetup{width=.95\textwidth}
\caption{IRREPs at high-symmetry points for TaFe$_6$Sn$_6$ phonon spectrum at Low-T.}
\label{tab:191_TaFe6Sn6_Low}
\setlength{\tabcolsep}{1.mm}{
}
\end{table}

\begin{table}[htbp]
\global\long\def\arraystretch{1.12}
\captionsetup{width=.95\textwidth}
\caption{IRREPs at high-symmetry points for TaFe$_6$Sn$_6$ phonon spectrum at High-T.}
\label{tab:191_TaFe6Sn6_High}
\setlength{\tabcolsep}{1.mm}{
%
}
\end{table}
\subsection{\hyperref[smsec:col]{CoGe}}~\label{smssec:CoGe}

\subsubsection{\label{sms2sec:LiCo6Ge6}\hyperref[smsec:col]{\texorpdfstring{LiCo$_6$Ge$_6$}{}}~{\color{blue}  (High-T stable, Low-T unstable) }}

\begin{table}[htbp]
\global\long\def\arraystretch{1.12}
\captionsetup{width=.85\textwidth}
\caption{Structure parameters of LiCo$_6$Ge$_6$ after relaxation. It contains 357 electrons. }
\label{tab:LiCo6Ge6}
\setlength{\tabcolsep}{4mm}{
%
}
\end{table}

\begin{figure}[htbp]
\centering
\includegraphics[width=0.85\textwidth]{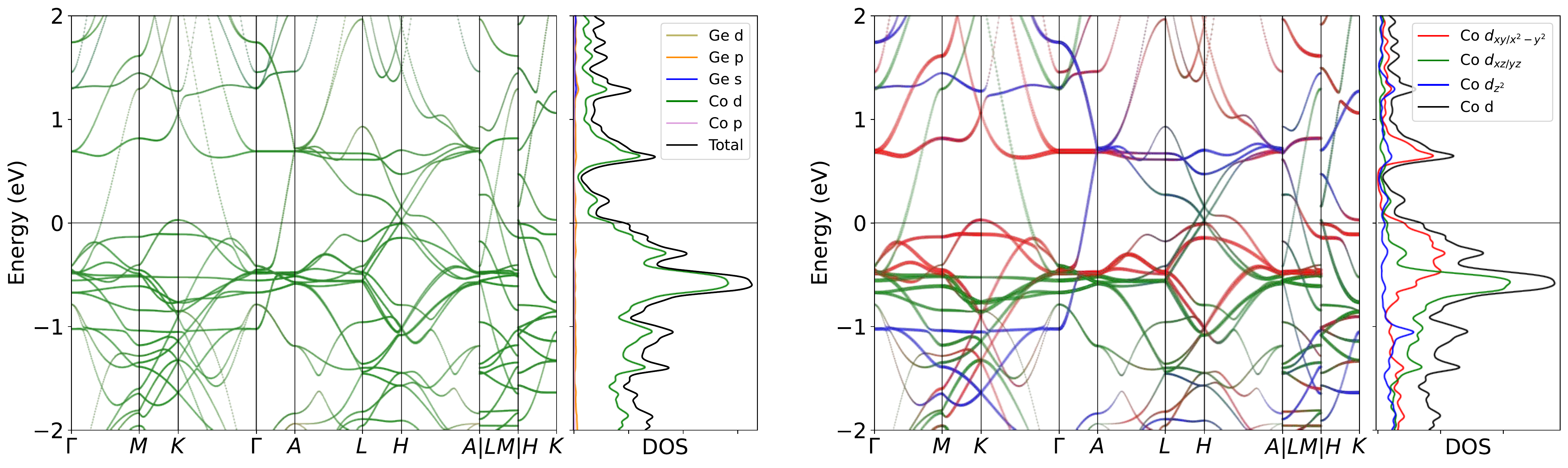}
\caption{Band structure for LiCo$_6$Ge$_6$ without SOC. The projection of Co $3d$ orbitals is also presented. The band structures clearly reveal the dominating of Co $3d$ orbitals  near the Fermi level, as well as the pronounced peak.}
\label{fig:LiCo6Ge6band}
\end{figure}

\begin{figure}[H]
\centering
\subfloat[Relaxed phonon at low-T.]{
\label{fig:LiCo6Ge6_relaxed_Low}
\includegraphics[width=0.45\textwidth]{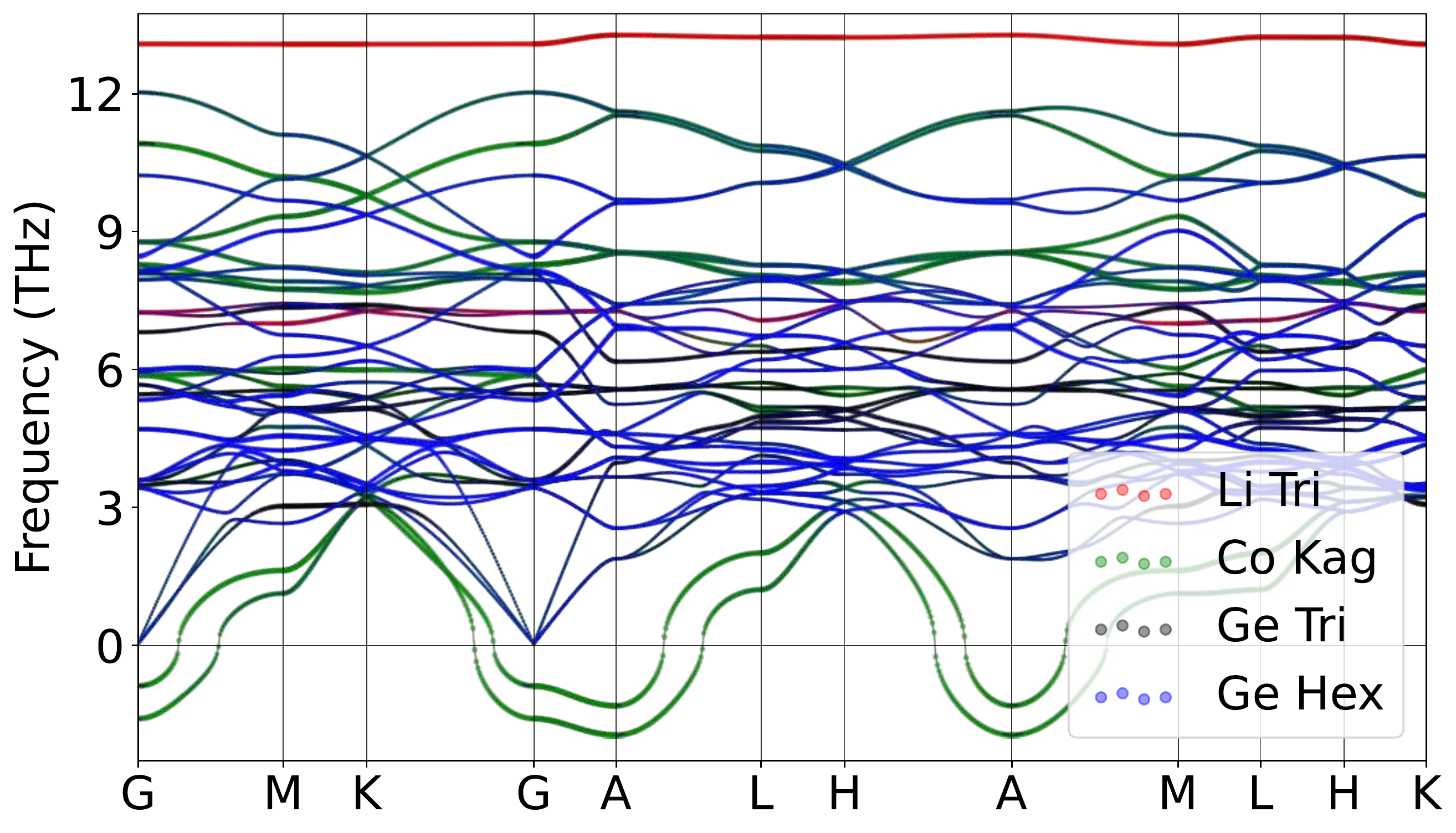}
}
\hspace{5pt}\subfloat[Relaxed phonon at high-T.]{
\label{fig:LiCo6Ge6_relaxed_High}
\includegraphics[width=0.45\textwidth]{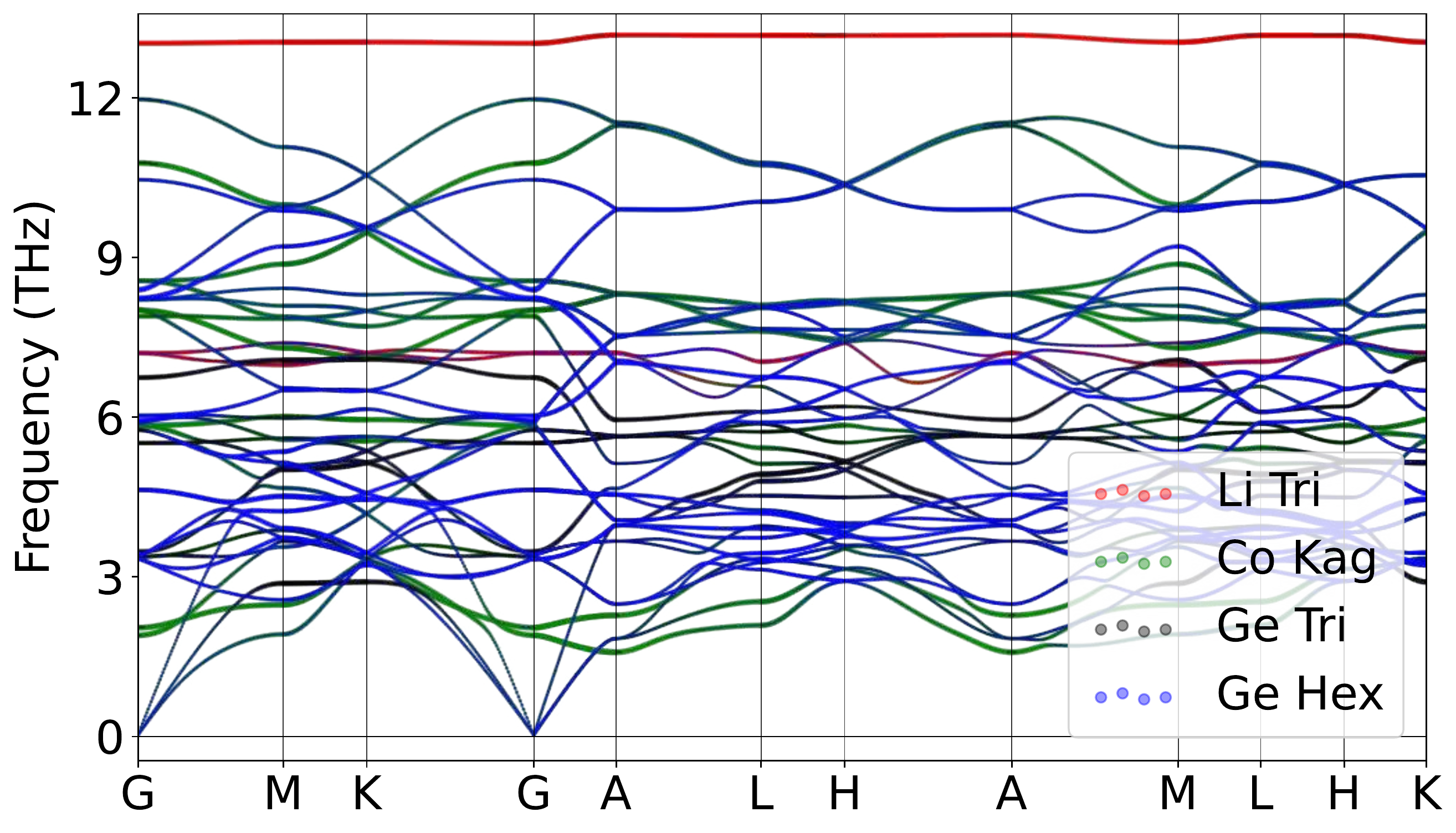}
}
\caption{Atomically projected phonon spectrum for LiCo$_6$Ge$_6$.}
\label{fig:LiCo6Ge6pho}
\end{figure}

\begin{figure}[H]
\centering
\label{fig:LiCo6Ge6_relaxed_Low_xz}
\includegraphics[width=0.85\textwidth]{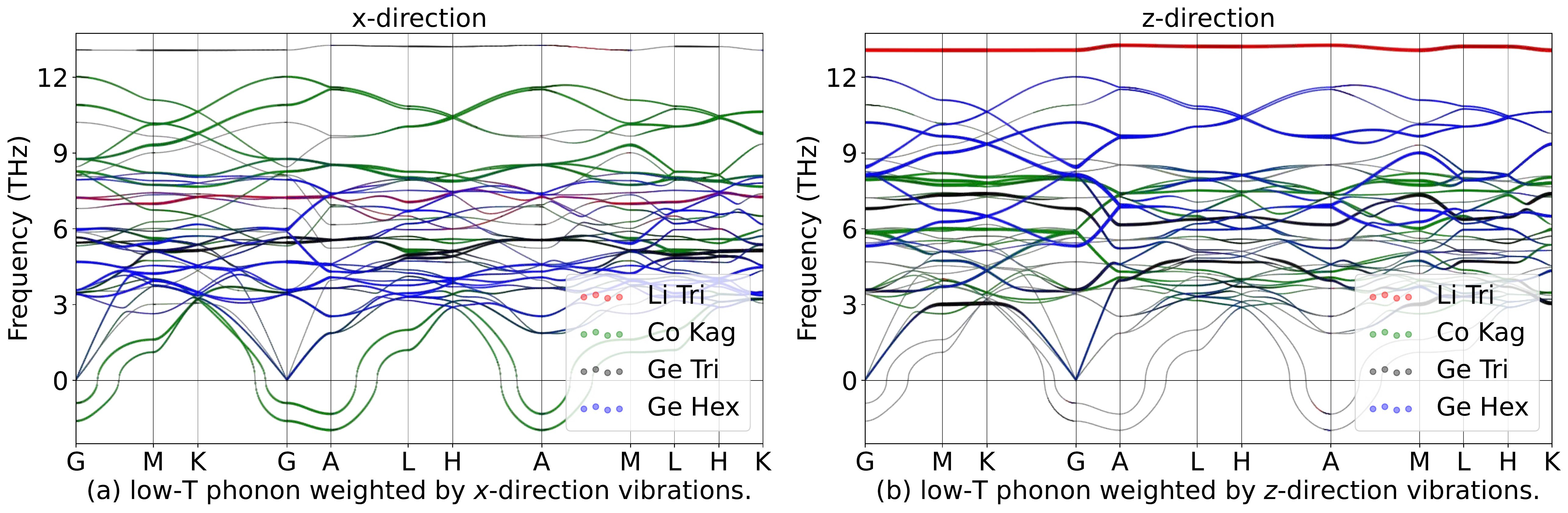}
\hspace{5pt}\label{fig:LiCo6Ge6_relaxed_High_xz}
\includegraphics[width=0.85\textwidth]{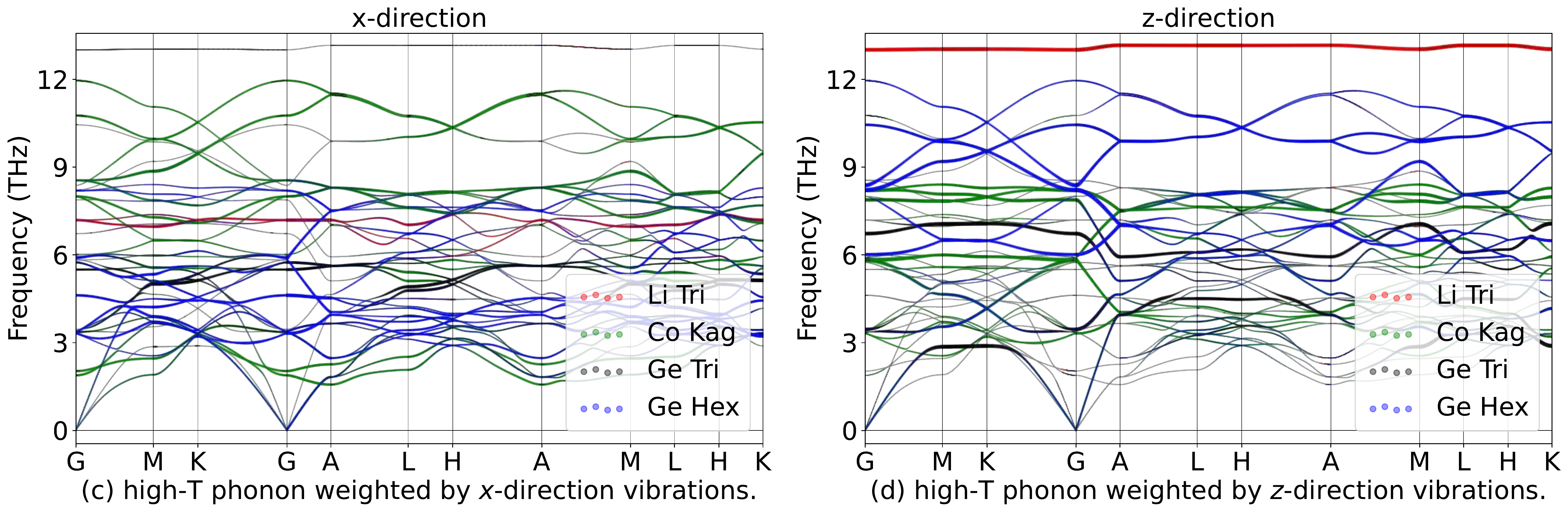}
\caption{Phonon spectrum for LiCo$_6$Ge$_6$in in-plane ($x$) and out-of-plane ($z$) directions.}
\label{fig:LiCo6Ge6phoxyz}
\end{figure}

\begin{table}[htbp]
\global\long\def\arraystretch{1.12}
\captionsetup{width=.95\textwidth}
\caption{IRREPs at high-symmetry points for LiCo$_6$Ge$_6$ phonon spectrum at Low-T.}
\label{tab:191_LiCo6Ge6_Low}
\setlength{\tabcolsep}{1.mm}{
}
\end{table}

\begin{table}[htbp]
\global\long\def\arraystretch{1.12}
\captionsetup{width=.95\textwidth}
\caption{IRREPs at high-symmetry points for LiCo$_6$Ge$_6$ phonon spectrum at High-T.}
\label{tab:191_LiCo6Ge6_High}
\setlength{\tabcolsep}{1.mm}{
%
}
\end{table}
\subsubsection{\label{sms2sec:MgCo6Ge6}\hyperref[smsec:col]{\texorpdfstring{MgCo$_6$Ge$_6$}{}}~{\color{red}  (High-T unstable, Low-T unstable) }}

\begin{table}[htbp]
\global\long\def\arraystretch{1.12}
\captionsetup{width=.85\textwidth}
\caption{Structure parameters of MgCo$_6$Ge$_6$ after relaxation. It contains 366 electrons. }
\label{tab:MgCo6Ge6}
\setlength{\tabcolsep}{4mm}{
%
}
\end{table}

\begin{figure}[htbp]
\centering
\includegraphics[width=0.85\textwidth]{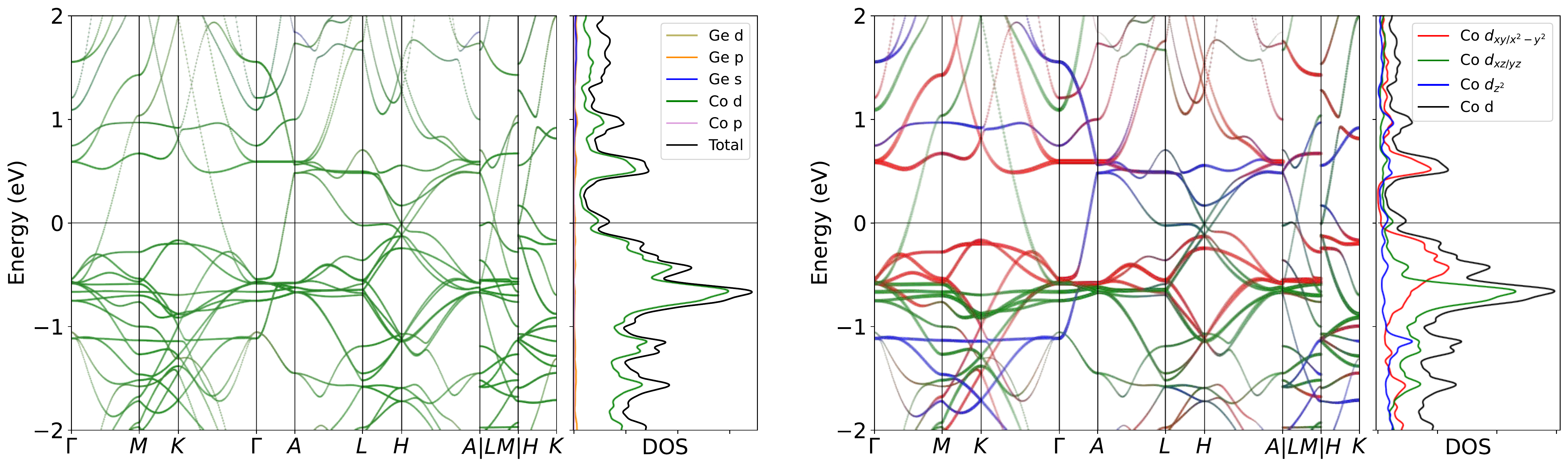}
\caption{Band structure for MgCo$_6$Ge$_6$ without SOC. The projection of Co $3d$ orbitals is also presented. The band structures clearly reveal the dominating of Co $3d$ orbitals  near the Fermi level, as well as the pronounced peak.}
\label{fig:MgCo6Ge6band}
\end{figure}

\begin{figure}[H]
\centering
\subfloat[Relaxed phonon at low-T.]{
\label{fig:MgCo6Ge6_relaxed_Low}
\includegraphics[width=0.45\textwidth]{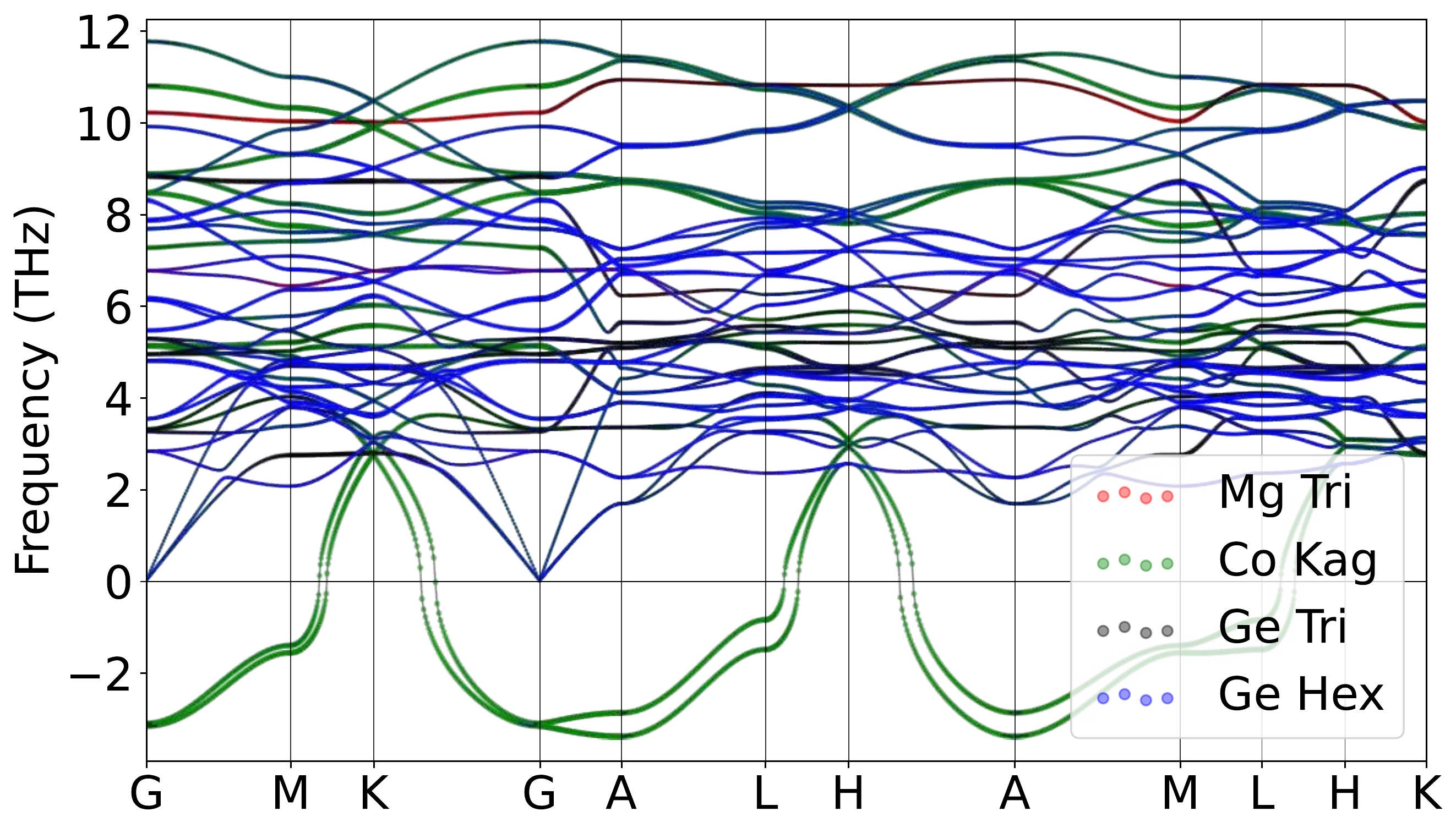}
}
\hspace{5pt}\subfloat[Relaxed phonon at high-T.]{
\label{fig:MgCo6Ge6_relaxed_High}
\includegraphics[width=0.45\textwidth]{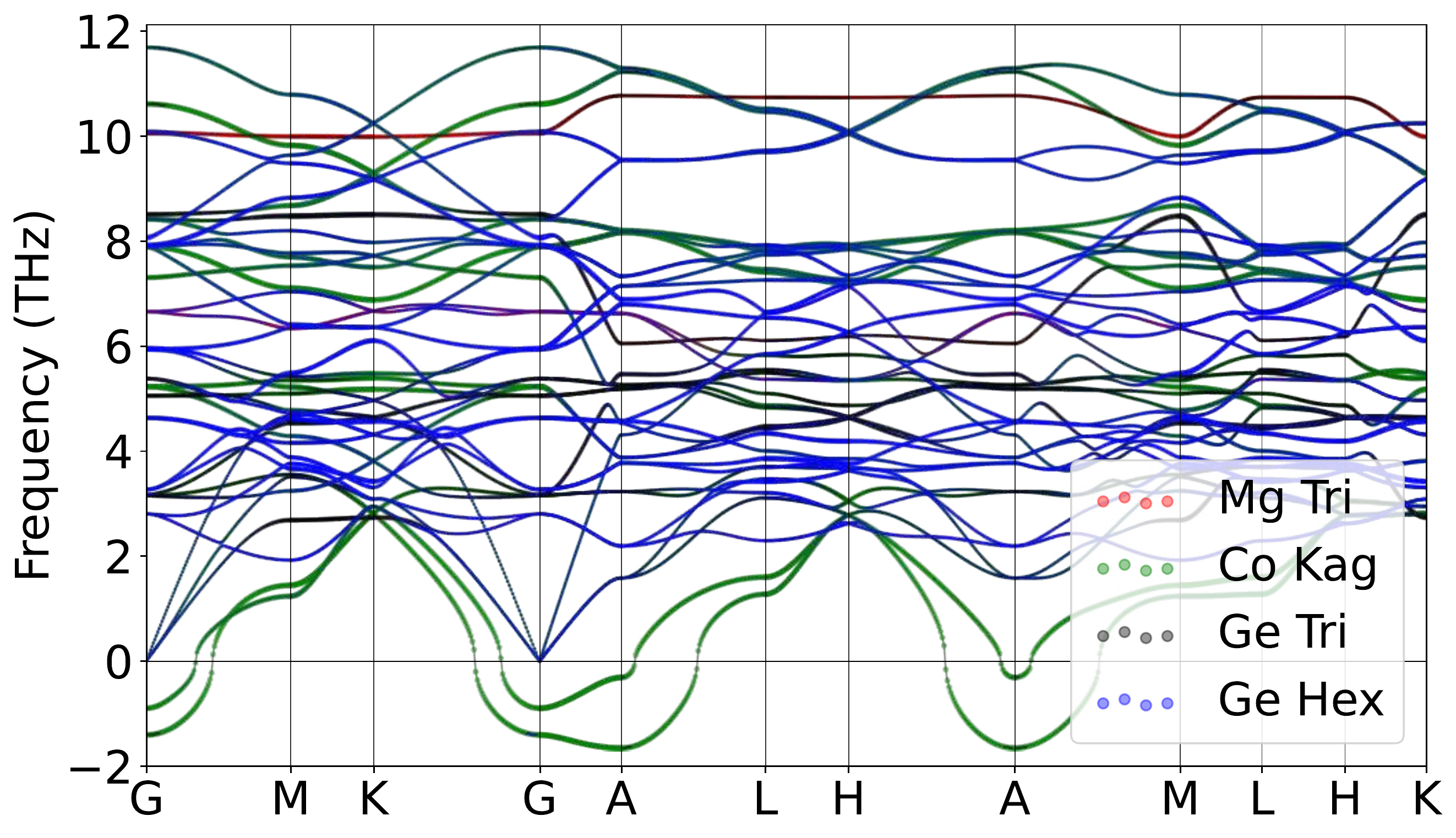}
}
\caption{Atomically projected phonon spectrum for MgCo$_6$Ge$_6$.}
\label{fig:MgCo6Ge6pho}
\end{figure}

\begin{figure}[H]
\centering
\label{fig:MgCo6Ge6_relaxed_Low_xz}
\includegraphics[width=0.85\textwidth]{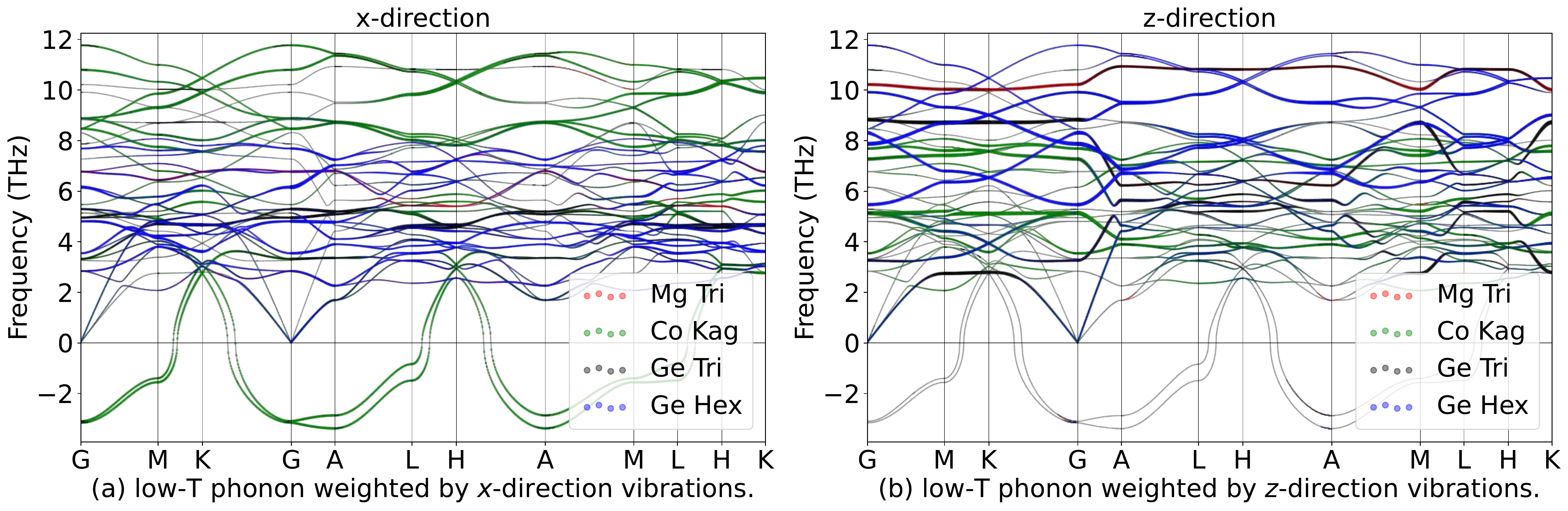}
\hspace{5pt}\label{fig:MgCo6Ge6_relaxed_High_xz}
\includegraphics[width=0.85\textwidth]{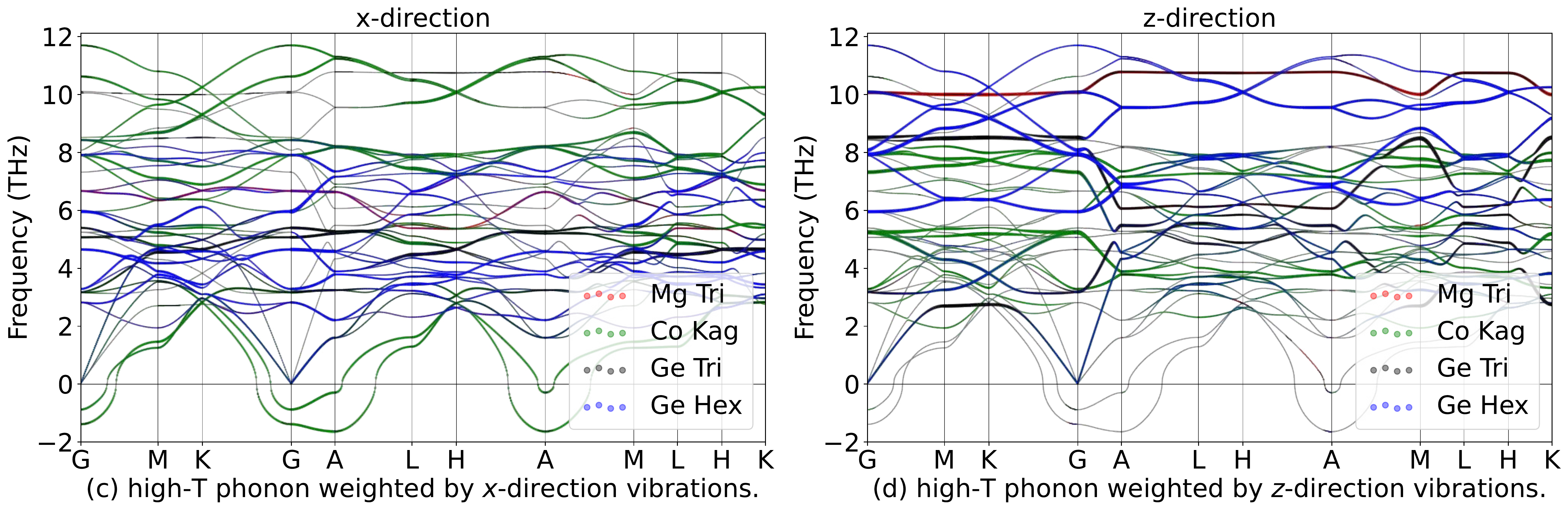}
\caption{Phonon spectrum for MgCo$_6$Ge$_6$in in-plane ($x$) and out-of-plane ($z$) directions.}
\label{fig:MgCo6Ge6phoxyz}
\end{figure}

\begin{table}[htbp]
\global\long\def\arraystretch{1.12}
\captionsetup{width=.95\textwidth}
\caption{IRREPs at high-symmetry points for MgCo$_6$Ge$_6$ phonon spectrum at Low-T.}
\label{tab:191_MgCo6Ge6_Low}
\setlength{\tabcolsep}{1.mm}{
}
\end{table}

\begin{table}[htbp]
\global\long\def\arraystretch{1.12}
\captionsetup{width=.95\textwidth}
\caption{IRREPs at high-symmetry points for MgCo$_6$Ge$_6$ phonon spectrum at High-T.}
\label{tab:191_MgCo6Ge6_High}
\setlength{\tabcolsep}{1.mm}{
%
}
\end{table}
\subsubsection{\label{sms2sec:CaCo6Ge6}\hyperref[smsec:col]{\texorpdfstring{CaCo$_6$Ge$_6$}{}}~{\color{blue}  (High-T stable, Low-T unstable) }}

\begin{table}[htbp]
\global\long\def\arraystretch{1.12}
\captionsetup{width=.85\textwidth}
\caption{Structure parameters of CaCo$_6$Ge$_6$ after relaxation. It contains 374 electrons. }
\label{tab:CaCo6Ge6}
\setlength{\tabcolsep}{4mm}{
%
}
\end{table}

\begin{figure}[htbp]
\centering
\includegraphics[width=0.85\textwidth]{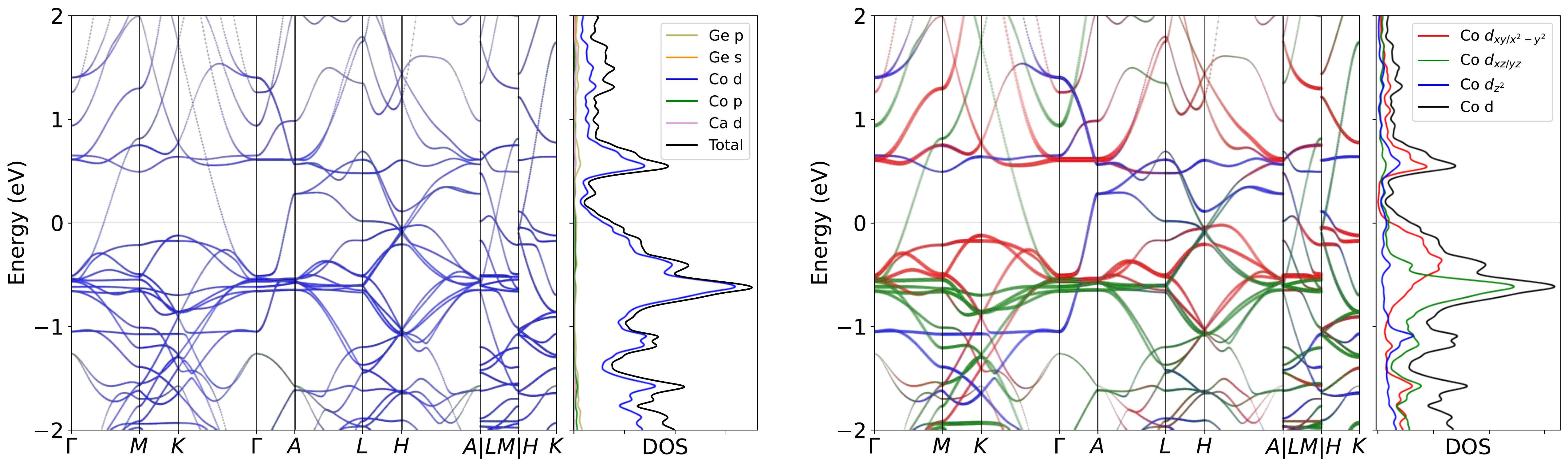}
\caption{Band structure for CaCo$_6$Ge$_6$ without SOC. The projection of Co $3d$ orbitals is also presented. The band structures clearly reveal the dominating of Co $3d$ orbitals  near the Fermi level, as well as the pronounced peak.}
\label{fig:CaCo6Ge6band}
\end{figure}

\begin{figure}[H]
\centering
\subfloat[Relaxed phonon at low-T.]{
\label{fig:CaCo6Ge6_relaxed_Low}
\includegraphics[width=0.45\textwidth]{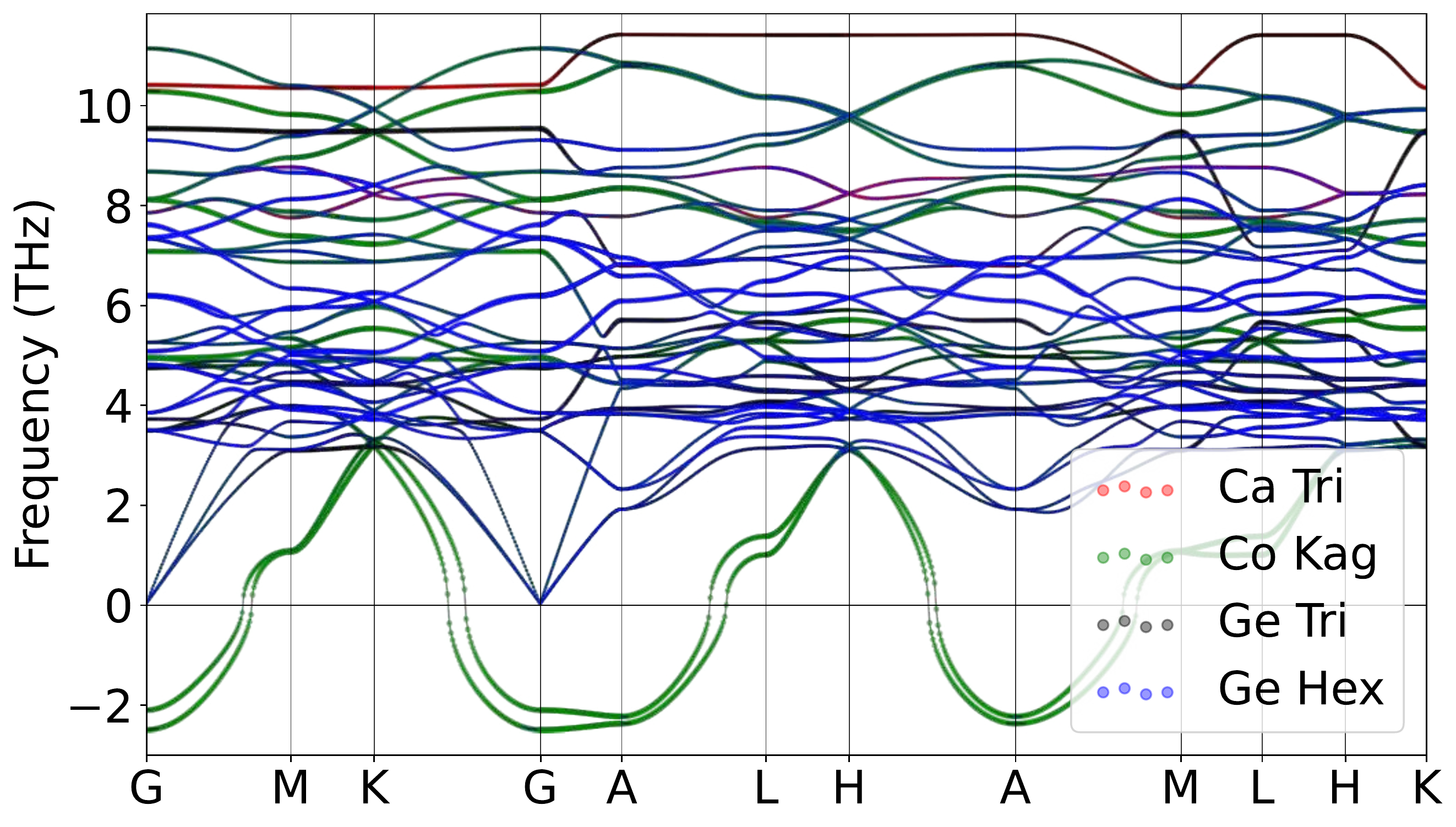}
}
\hspace{5pt}\subfloat[Relaxed phonon at high-T.]{
\label{fig:CaCo6Ge6_relaxed_High}
\includegraphics[width=0.45\textwidth]{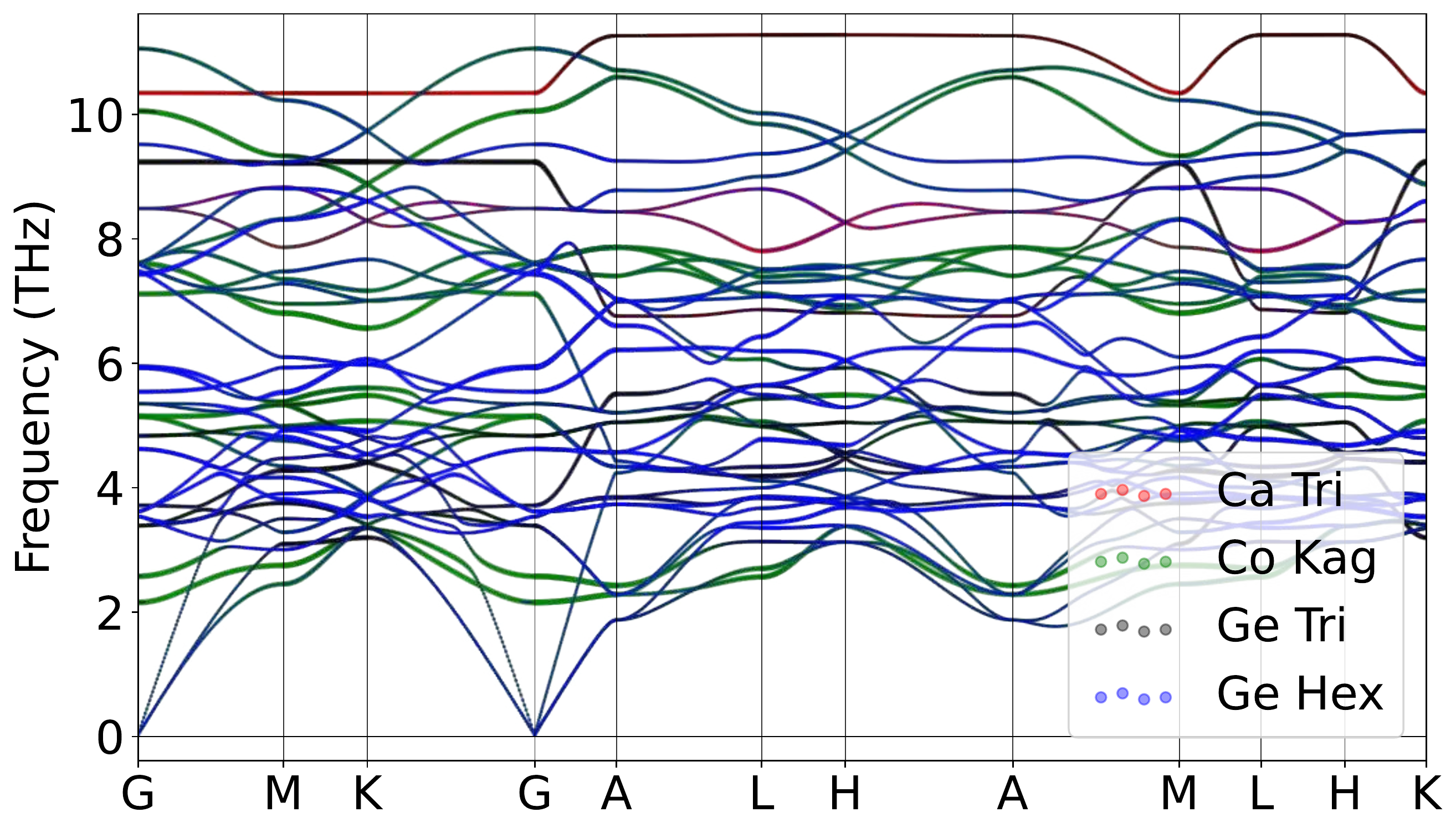}
}
\caption{Atomically projected phonon spectrum for CaCo$_6$Ge$_6$.}
\label{fig:CaCo6Ge6pho}
\end{figure}

\begin{figure}[H]
\centering
\label{fig:CaCo6Ge6_relaxed_Low_xz}
\includegraphics[width=0.85\textwidth]{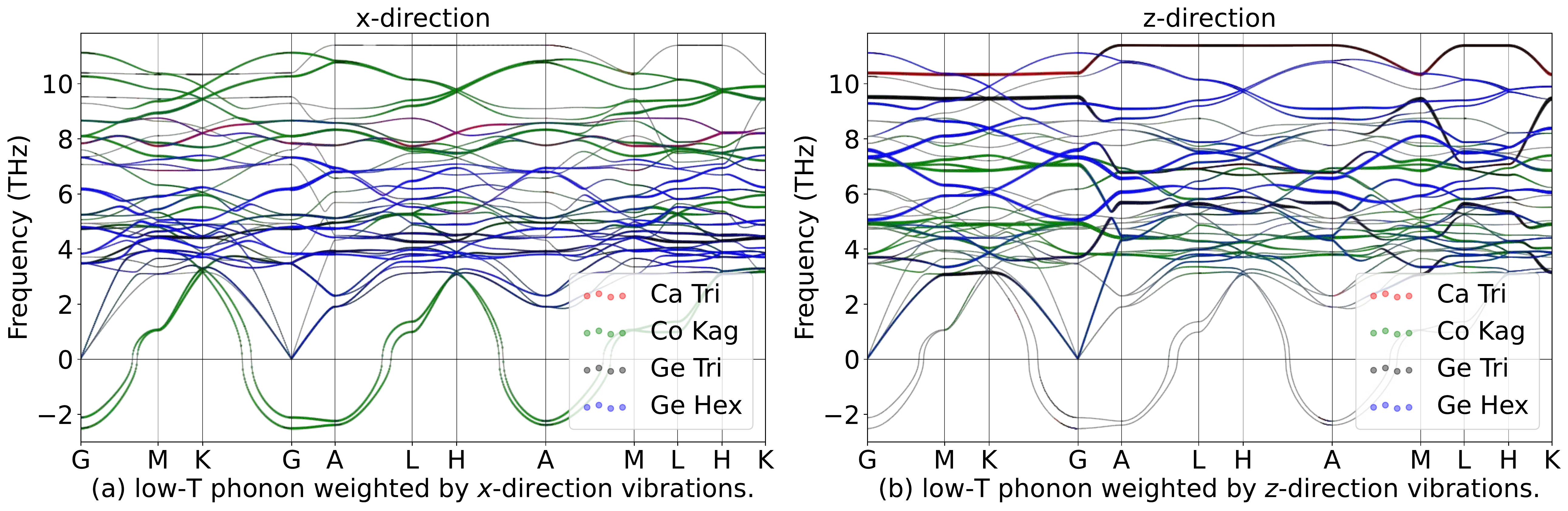}
\hspace{5pt}\label{fig:CaCo6Ge6_relaxed_High_xz}
\includegraphics[width=0.85\textwidth]{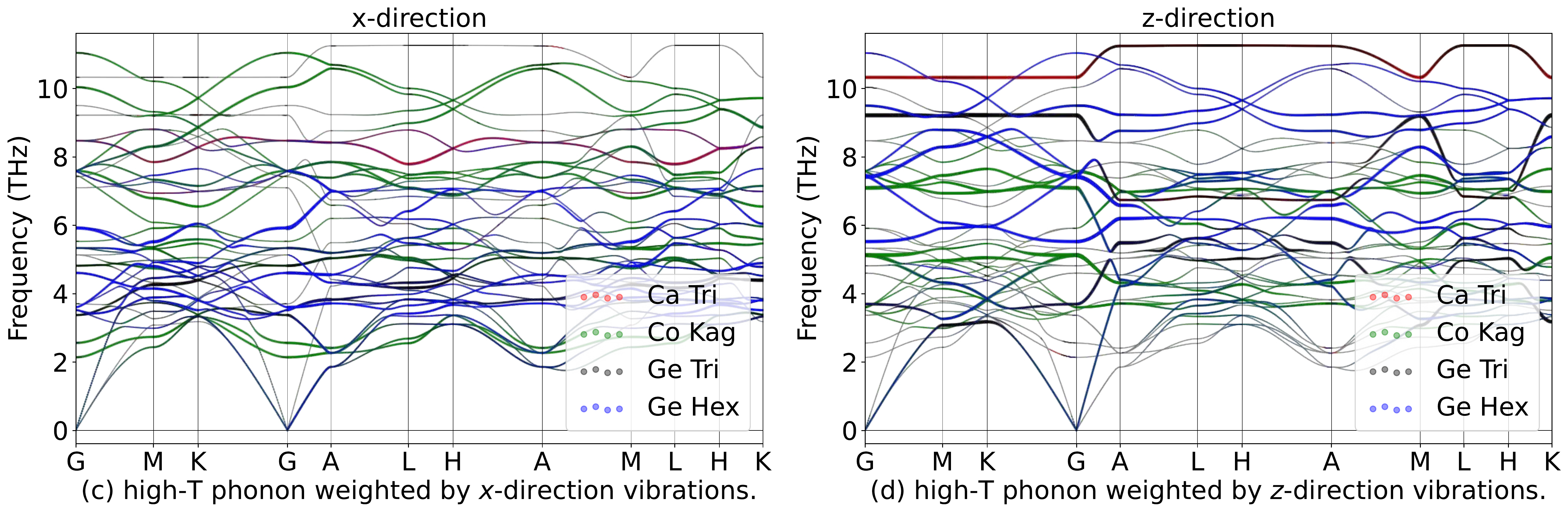}
\caption{Phonon spectrum for CaCo$_6$Ge$_6$in in-plane ($x$) and out-of-plane ($z$) directions.}
\label{fig:CaCo6Ge6phoxyz}
\end{figure}

\begin{table}[htbp]
\global\long\def\arraystretch{1.12}
\captionsetup{width=.95\textwidth}
\caption{IRREPs at high-symmetry points for CaCo$_6$Ge$_6$ phonon spectrum at Low-T.}
\label{tab:191_CaCo6Ge6_Low}
\setlength{\tabcolsep}{1.mm}{
}
\end{table}

\begin{table}[htbp]
\global\long\def\arraystretch{1.12}
\captionsetup{width=.95\textwidth}
\caption{IRREPs at high-symmetry points for CaCo$_6$Ge$_6$ phonon spectrum at High-T.}
\label{tab:191_CaCo6Ge6_High}
\setlength{\tabcolsep}{1.mm}{
%
}
\end{table}
\subsubsection{\label{sms2sec:ScCo6Ge6}\hyperref[smsec:col]{\texorpdfstring{ScCo$_6$Ge$_6$}{}}~{\color{blue}  (High-T stable, Low-T unstable) }}

\begin{table}[htbp]
\global\long\def\arraystretch{1.12}
\captionsetup{width=.85\textwidth}
\caption{Structure parameters of ScCo$_6$Ge$_6$ after relaxation. It contains 375 electrons. }
\label{tab:ScCo6Ge6}
\setlength{\tabcolsep}{4mm}{
%
}
\end{table}

\begin{figure}[htbp]
\centering
\includegraphics[width=0.85\textwidth]{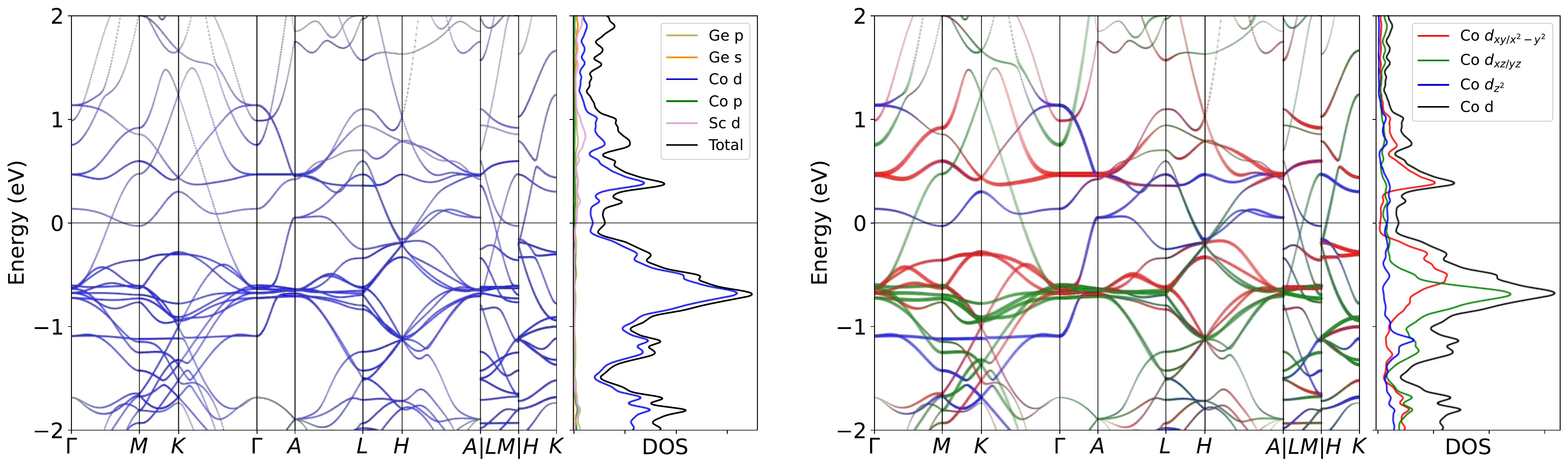}
\caption{Band structure for ScCo$_6$Ge$_6$ without SOC. The projection of Co $3d$ orbitals is also presented. The band structures clearly reveal the dominating of Co $3d$ orbitals  near the Fermi level, as well as the pronounced peak.}
\label{fig:ScCo6Ge6band}
\end{figure}

\begin{figure}[H]
\centering
\subfloat[Relaxed phonon at low-T.]{
\label{fig:ScCo6Ge6_relaxed_Low}
\includegraphics[width=0.45\textwidth]{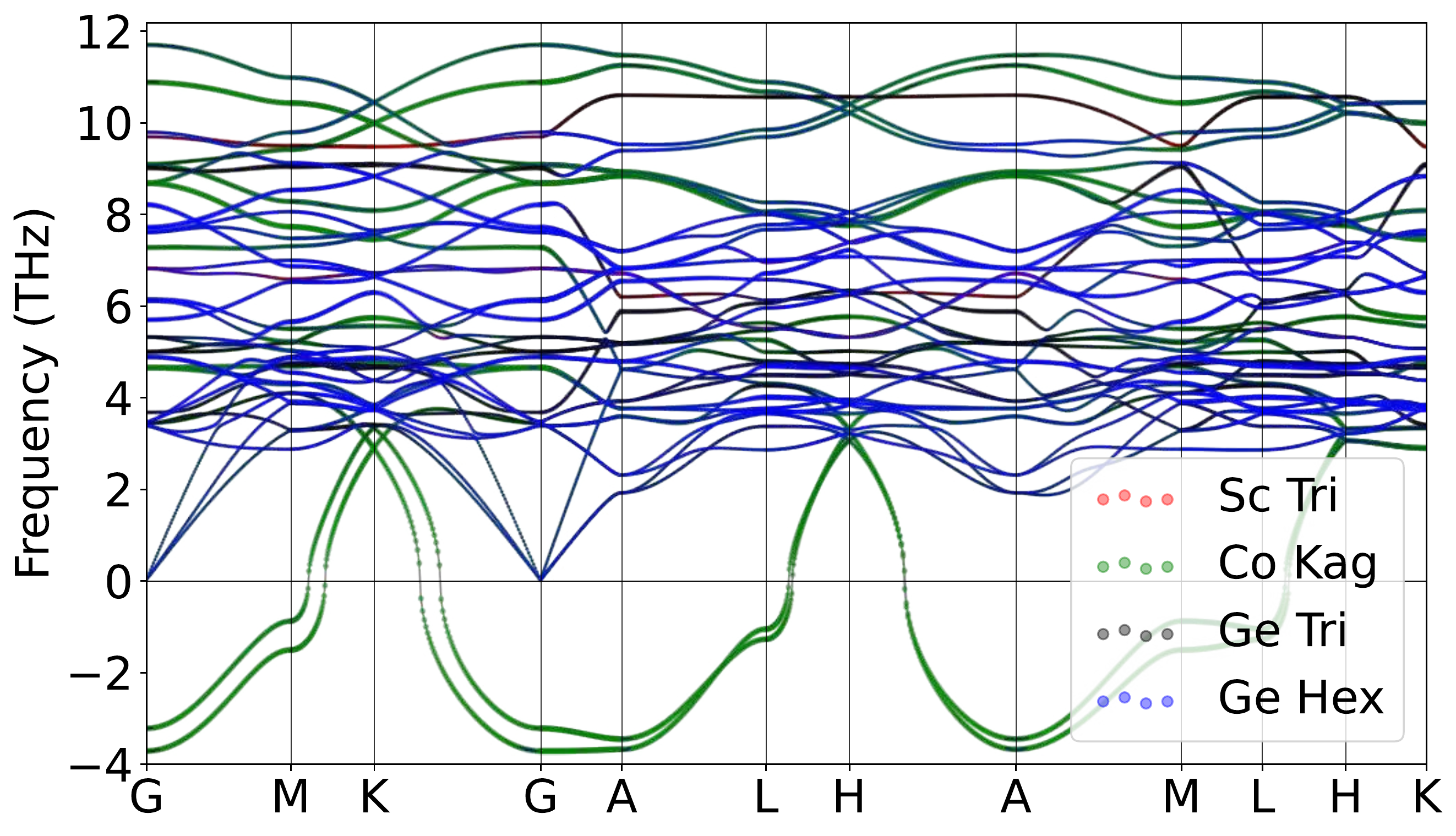}
}
\hspace{5pt}\subfloat[Relaxed phonon at high-T.]{
\label{fig:ScCo6Ge6_relaxed_High}
\includegraphics[width=0.45\textwidth]{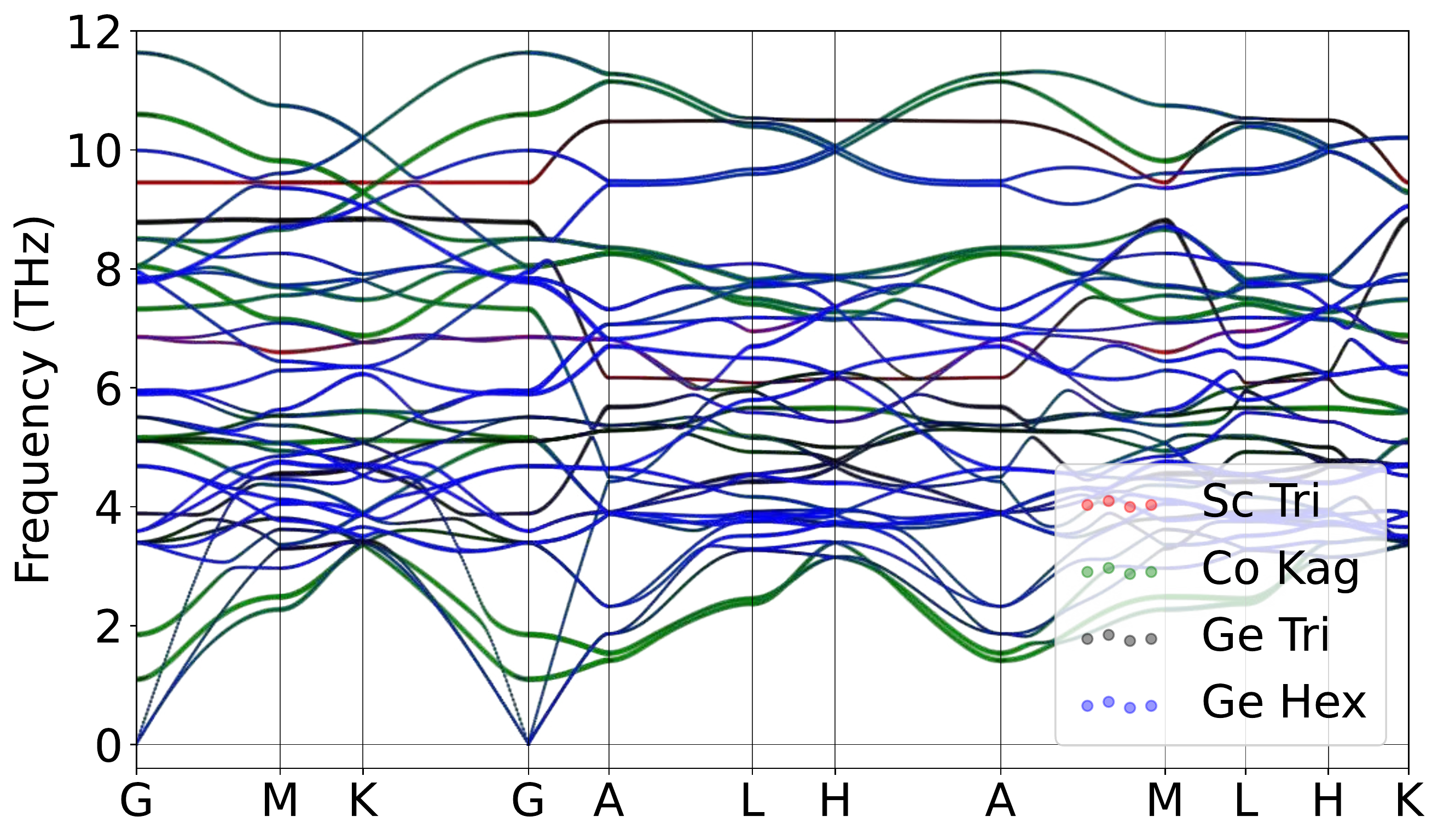}
}
\caption{Atomically projected phonon spectrum for ScCo$_6$Ge$_6$.}
\label{fig:ScCo6Ge6pho}
\end{figure}

\begin{figure}[H]
\centering
\label{fig:ScCo6Ge6_relaxed_Low_xz}
\includegraphics[width=0.85\textwidth]{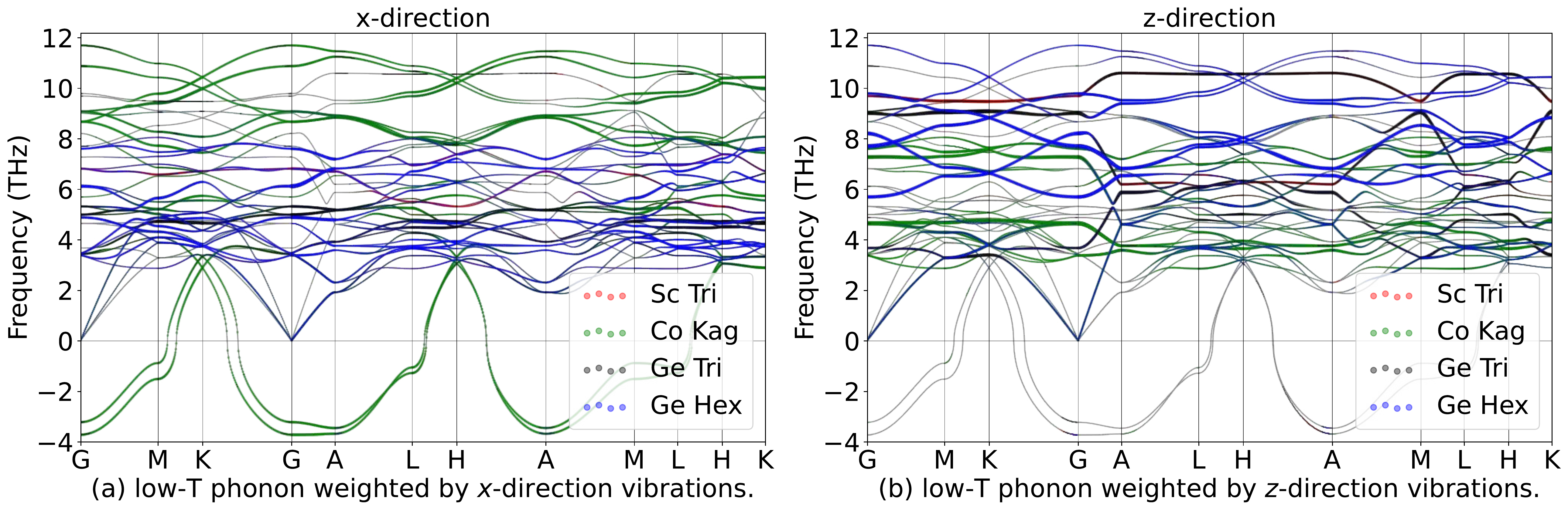}
\hspace{5pt}\label{fig:ScCo6Ge6_relaxed_High_xz}
\includegraphics[width=0.85\textwidth]{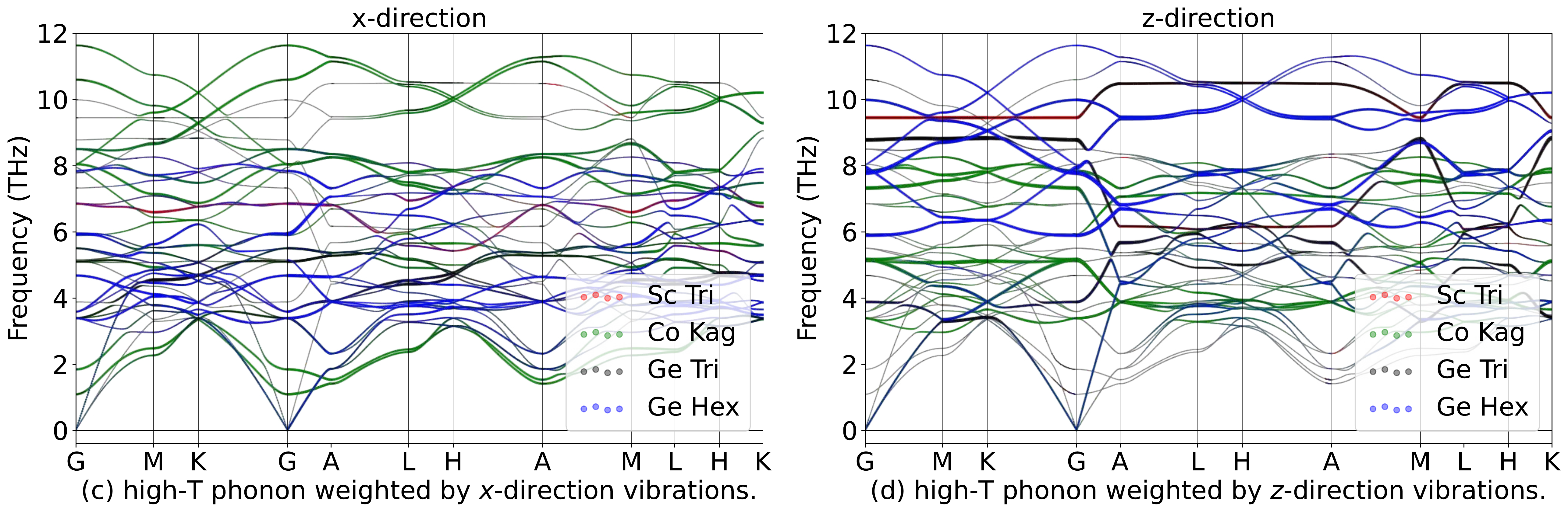}
\caption{Phonon spectrum for ScCo$_6$Ge$_6$in in-plane ($x$) and out-of-plane ($z$) directions.}
\label{fig:ScCo6Ge6phoxyz}
\end{figure}

\begin{table}[htbp]
\global\long\def\arraystretch{1.12}
\captionsetup{width=.95\textwidth}
\caption{IRREPs at high-symmetry points for ScCo$_6$Ge$_6$ phonon spectrum at Low-T.}
\label{tab:191_ScCo6Ge6_Low}
\setlength{\tabcolsep}{1.mm}{
}
\end{table}

\begin{table}[htbp]
\global\long\def\arraystretch{1.12}
\captionsetup{width=.95\textwidth}
\caption{IRREPs at high-symmetry points for ScCo$_6$Ge$_6$ phonon spectrum at High-T.}
\label{tab:191_ScCo6Ge6_High}
\setlength{\tabcolsep}{1.mm}{
%
}
\end{table}
\subsubsection{\label{sms2sec:TiCo6Ge6}\hyperref[smsec:col]{\texorpdfstring{TiCo$_6$Ge$_6$}{}}~{\color{blue}  (High-T stable, Low-T unstable) }}

\begin{table}[htbp]
\global\long\def\arraystretch{1.12}
\captionsetup{width=.85\textwidth}
\caption{Structure parameters of TiCo$_6$Ge$_6$ after relaxation. It contains 376 electrons. }
\label{tab:TiCo6Ge6}
\setlength{\tabcolsep}{4mm}{
%
}
\end{table}

\begin{figure}[htbp]
\centering
\includegraphics[width=0.85\textwidth]{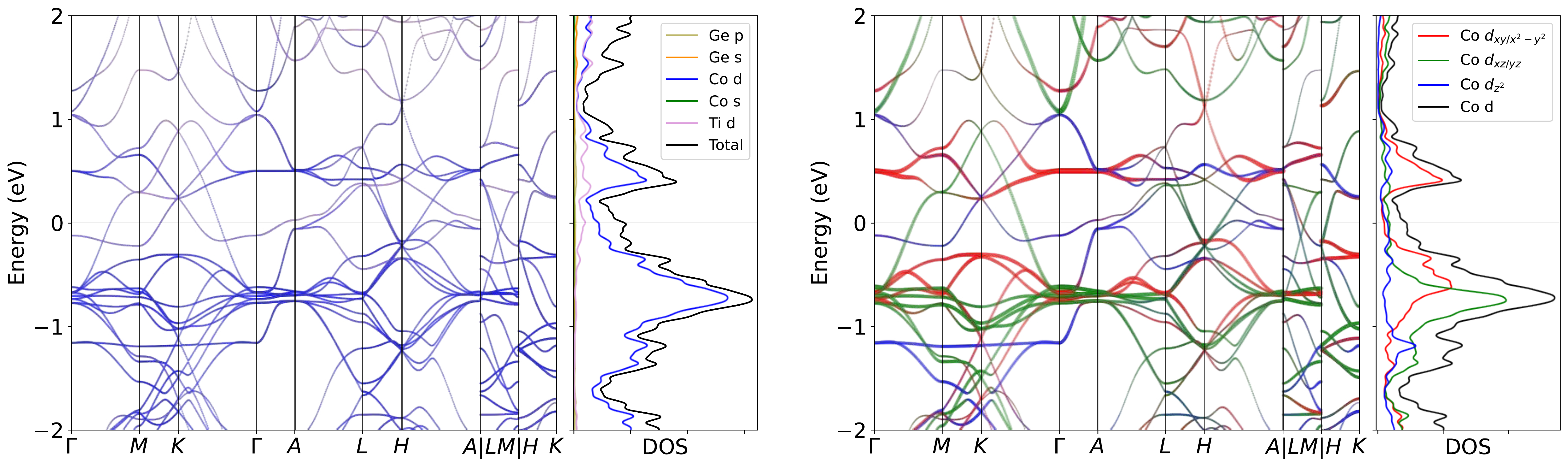}
\caption{Band structure for TiCo$_6$Ge$_6$ without SOC. The projection of Co $3d$ orbitals is also presented. The band structures clearly reveal the dominating of Co $3d$ orbitals  near the Fermi level, as well as the pronounced peak.}
\label{fig:TiCo6Ge6band}
\end{figure}

\begin{figure}[H]
\centering
\subfloat[Relaxed phonon at low-T.]{
\label{fig:TiCo6Ge6_relaxed_Low}
\includegraphics[width=0.45\textwidth]{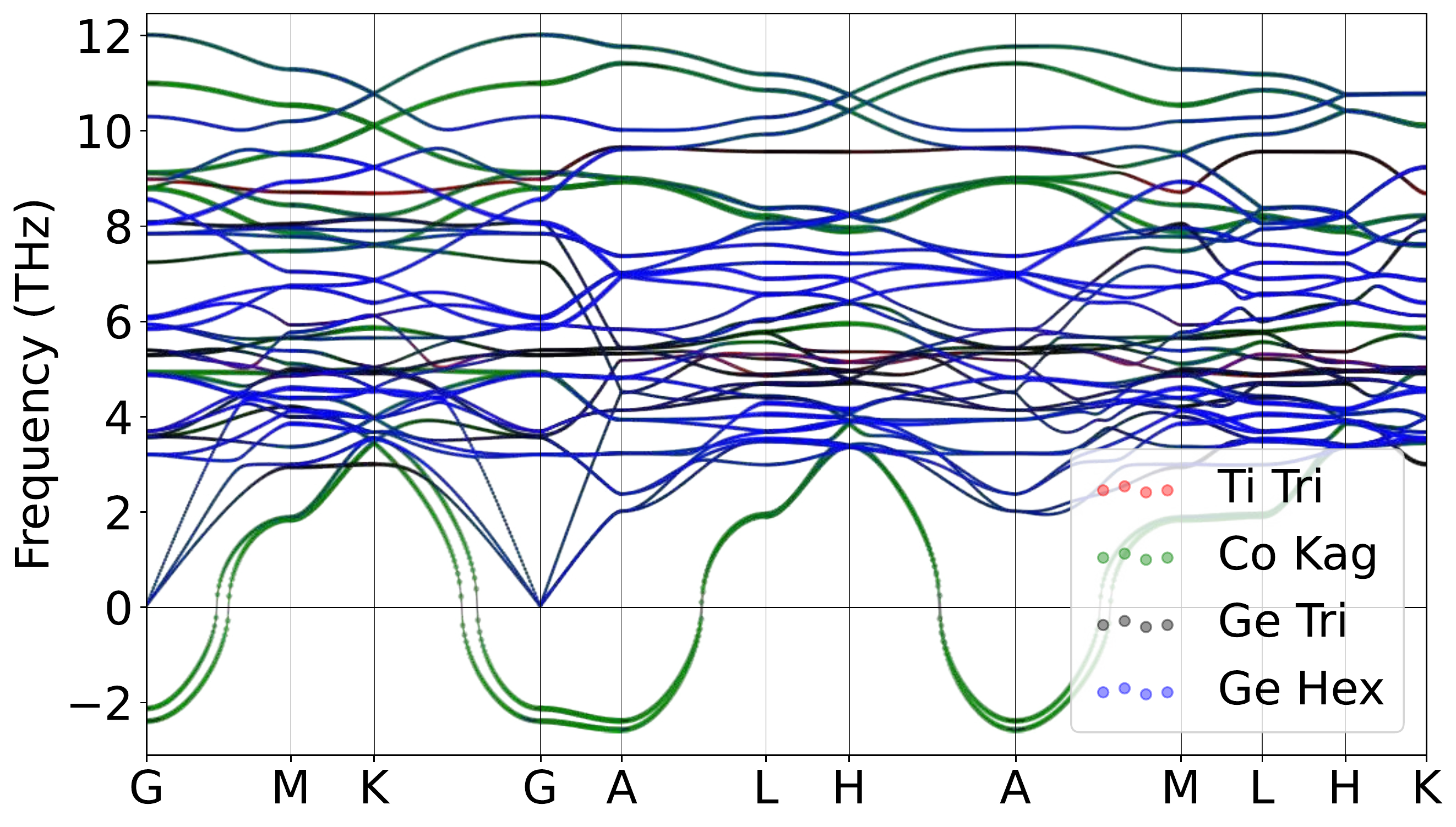}
}
\hspace{5pt}\subfloat[Relaxed phonon at high-T.]{
\label{fig:TiCo6Ge6_relaxed_High}
\includegraphics[width=0.45\textwidth]{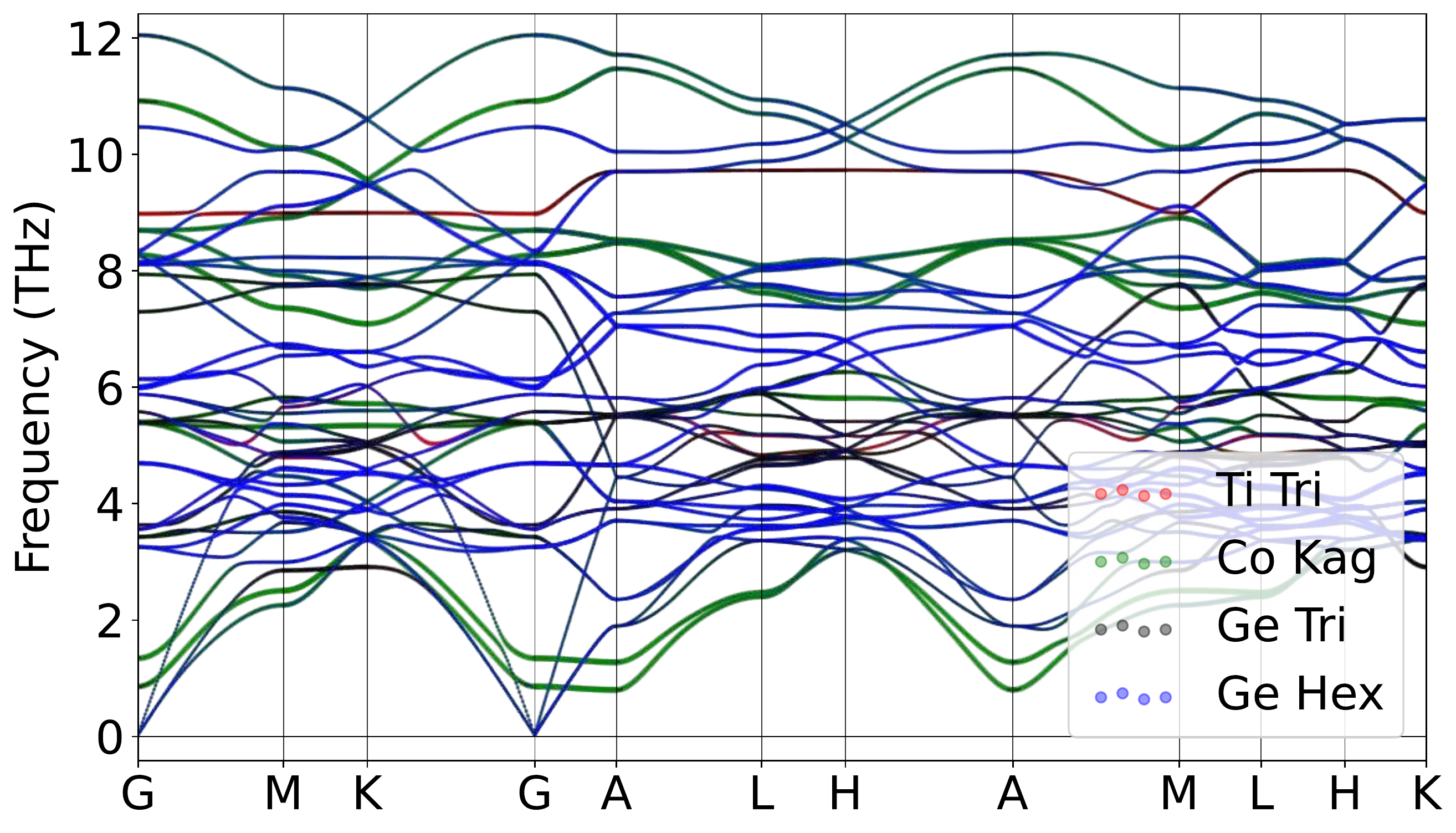}
}
\caption{Atomically projected phonon spectrum for TiCo$_6$Ge$_6$.}
\label{fig:TiCo6Ge6pho}
\end{figure}

\begin{figure}[H]
\centering
\label{fig:TiCo6Ge6_relaxed_Low_xz}
\includegraphics[width=0.85\textwidth]{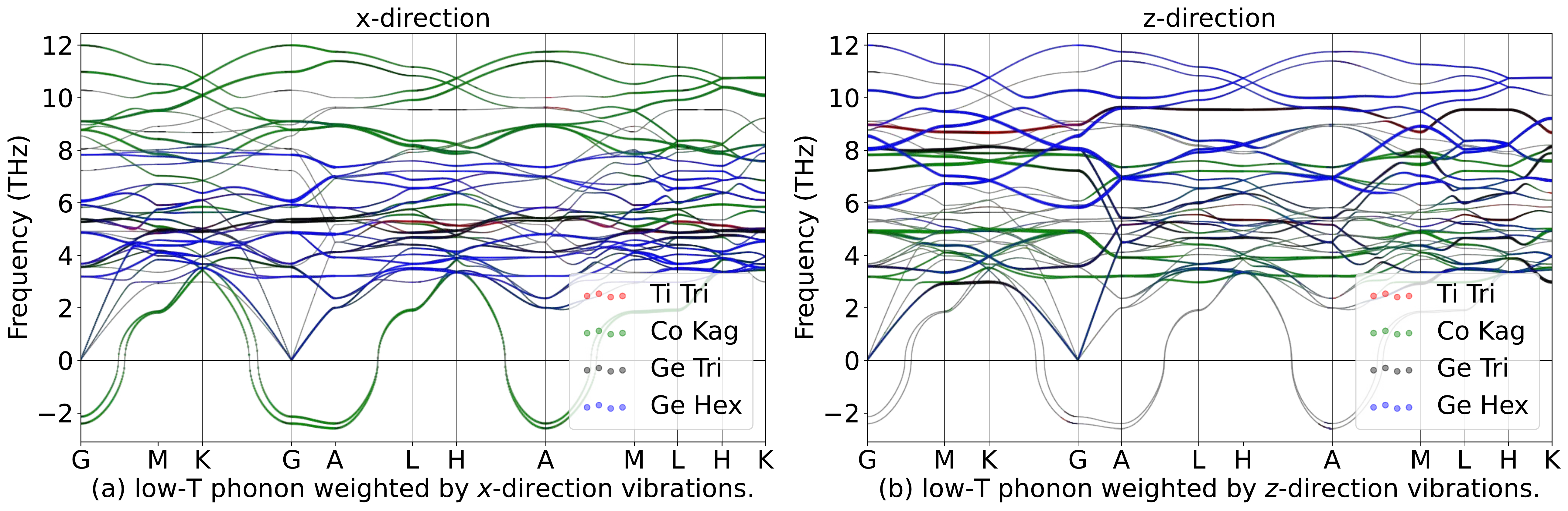}
\hspace{5pt}\label{fig:TiCo6Ge6_relaxed_High_xz}
\includegraphics[width=0.85\textwidth]{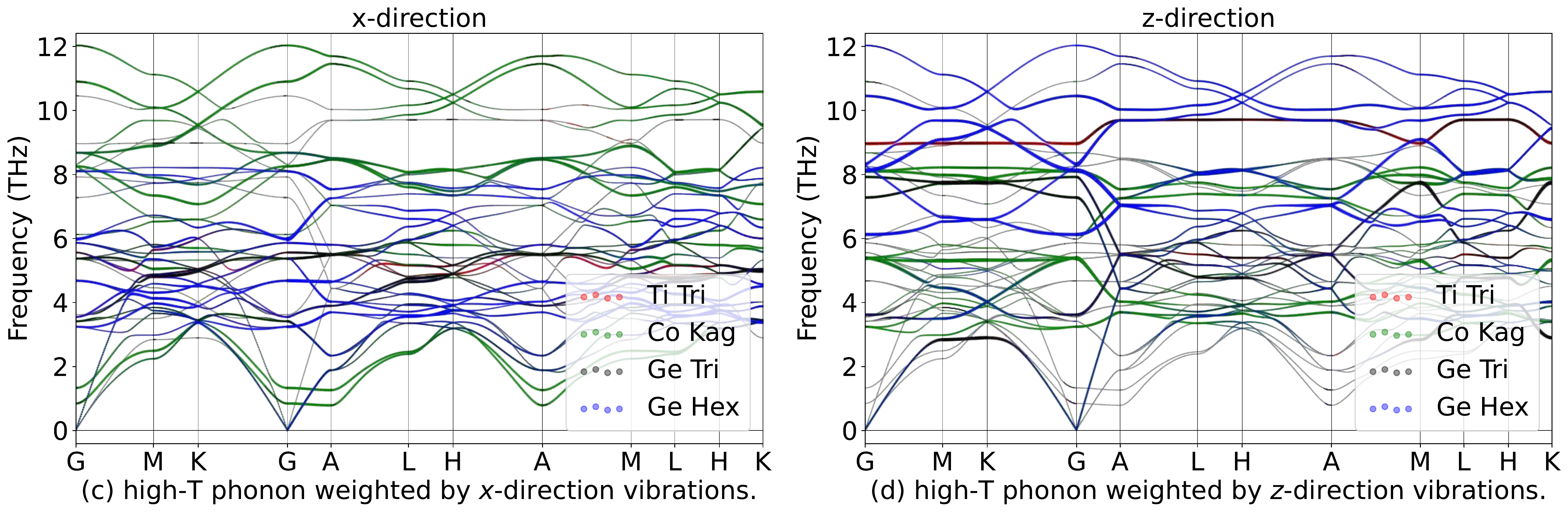}
\caption{Phonon spectrum for TiCo$_6$Ge$_6$in in-plane ($x$) and out-of-plane ($z$) directions.}
\label{fig:TiCo6Ge6phoxyz}
\end{figure}

\begin{table}[htbp]
\global\long\def\arraystretch{1.12}
\captionsetup{width=.95\textwidth}
\caption{IRREPs at high-symmetry points for TiCo$_6$Ge$_6$ phonon spectrum at Low-T.}
\label{tab:191_TiCo6Ge6_Low}
\setlength{\tabcolsep}{1.mm}{
}
\end{table}

\begin{table}[htbp]
\global\long\def\arraystretch{1.12}
\captionsetup{width=.95\textwidth}
\caption{IRREPs at high-symmetry points for TiCo$_6$Ge$_6$ phonon spectrum at High-T.}
\label{tab:191_TiCo6Ge6_High}
\setlength{\tabcolsep}{1.mm}{
%
}
\end{table}
\subsubsection{\label{sms2sec:YCo6Ge6}\hyperref[smsec:col]{\texorpdfstring{YCo$_6$Ge$_6$}{}}~{\color{blue}  (High-T stable, Low-T unstable) }}

\begin{table}[htbp]
\global\long\def\arraystretch{1.12}
\captionsetup{width=.85\textwidth}
\caption{Structure parameters of YCo$_6$Ge$_6$ after relaxation. It contains 393 electrons. }
\label{tab:YCo6Ge6}
\setlength{\tabcolsep}{4mm}{
%
}
\end{table}

\begin{figure}[htbp]
\centering
\includegraphics[width=0.85\textwidth]{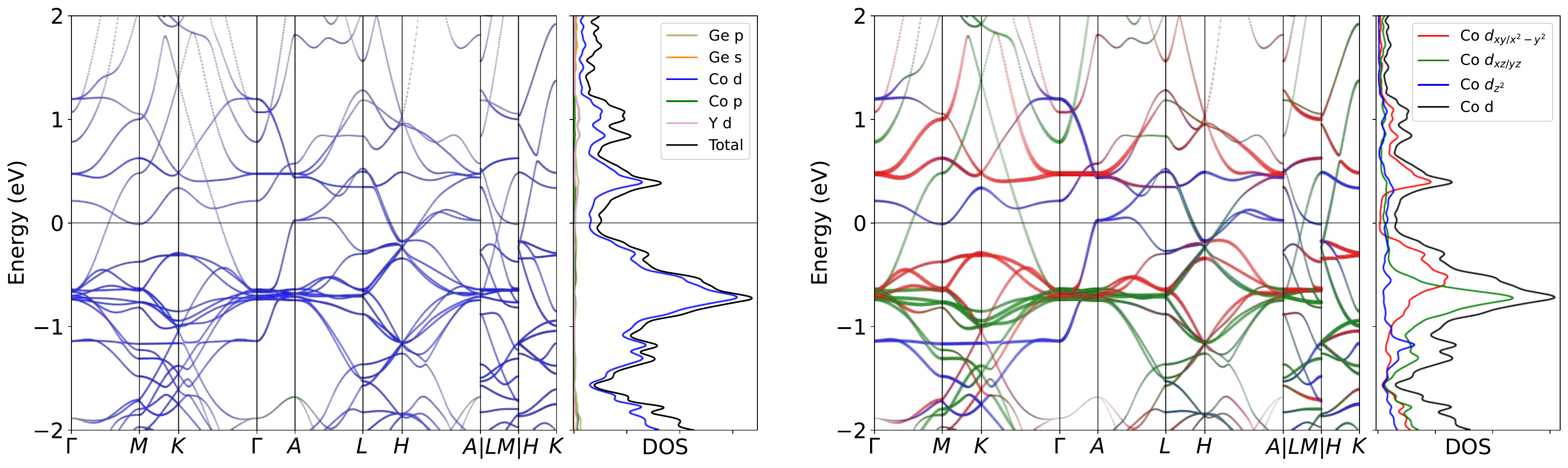}
\caption{Band structure for YCo$_6$Ge$_6$ without SOC. The projection of Co $3d$ orbitals is also presented. The band structures clearly reveal the dominating of Co $3d$ orbitals  near the Fermi level, as well as the pronounced peak.}
\label{fig:YCo6Ge6band}
\end{figure}

\begin{figure}[H]
\centering
\subfloat[Relaxed phonon at low-T.]{
\label{fig:YCo6Ge6_relaxed_Low}
\includegraphics[width=0.45\textwidth]{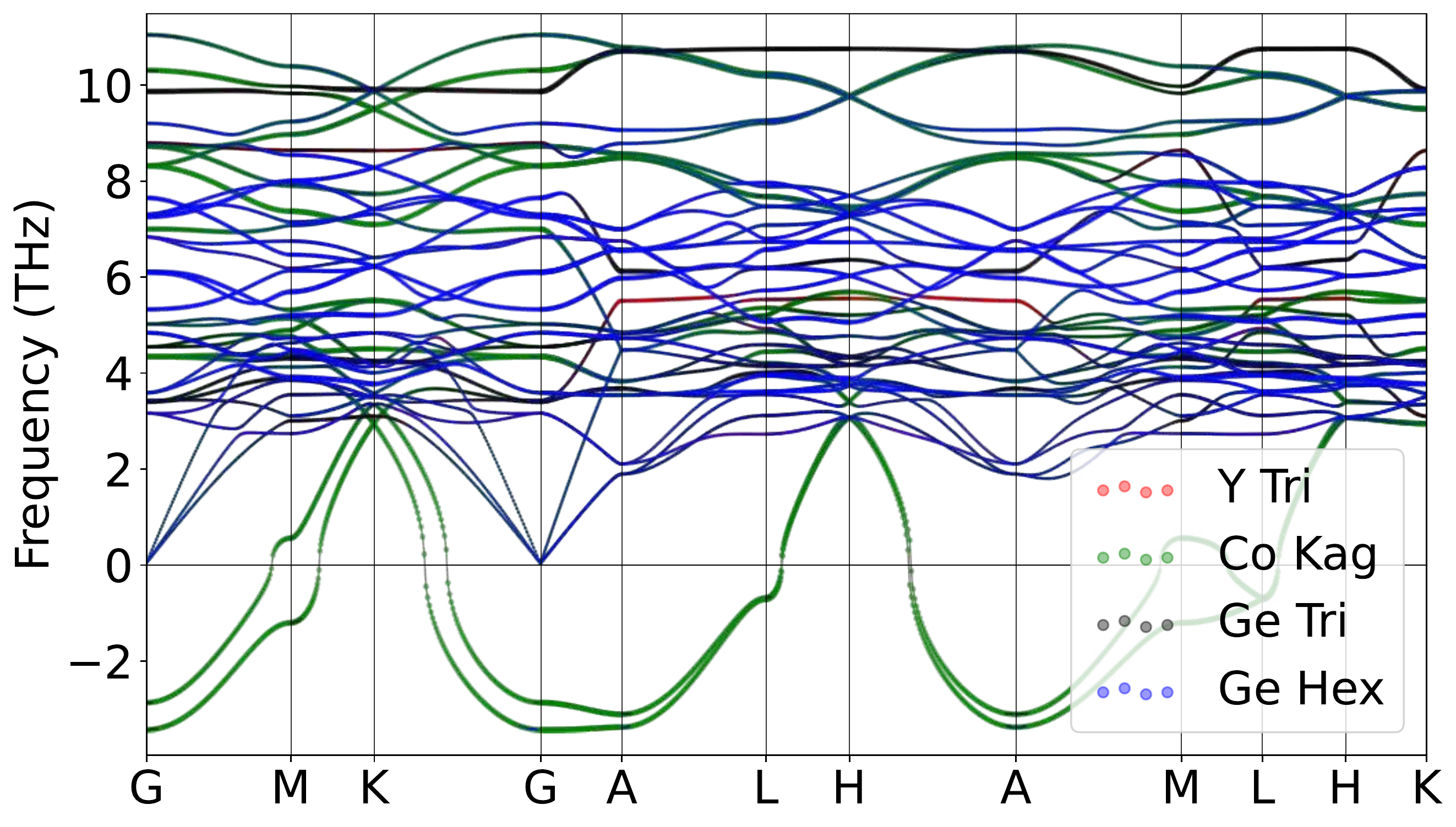}
}
\hspace{5pt}\subfloat[Relaxed phonon at high-T.]{
\label{fig:YCo6Ge6_relaxed_High}
\includegraphics[width=0.45\textwidth]{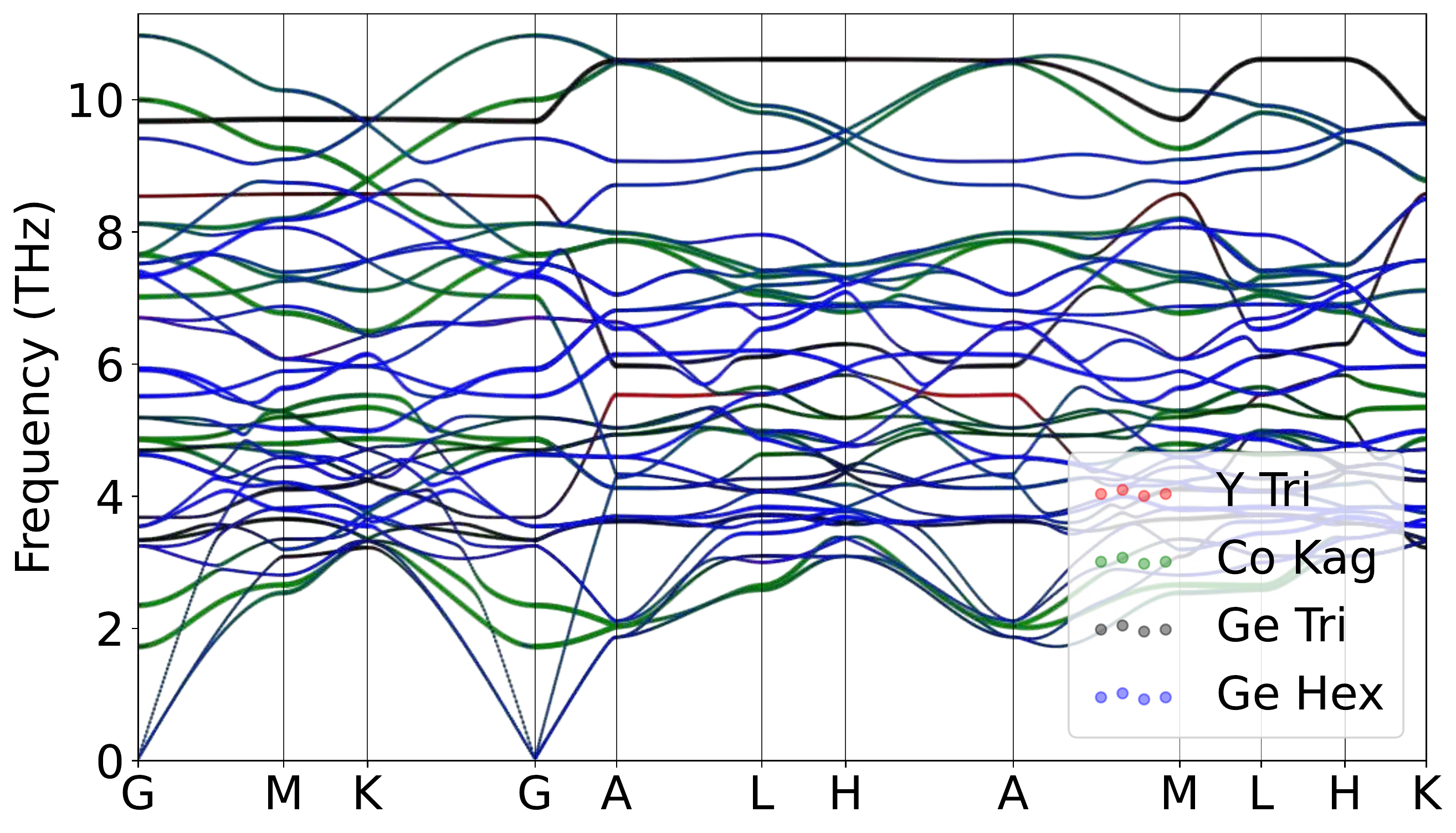}
}
\caption{Atomically projected phonon spectrum for YCo$_6$Ge$_6$.}
\label{fig:YCo6Ge6pho}
\end{figure}

\begin{figure}[H]
\centering
\label{fig:YCo6Ge6_relaxed_Low_xz}
\includegraphics[width=0.85\textwidth]{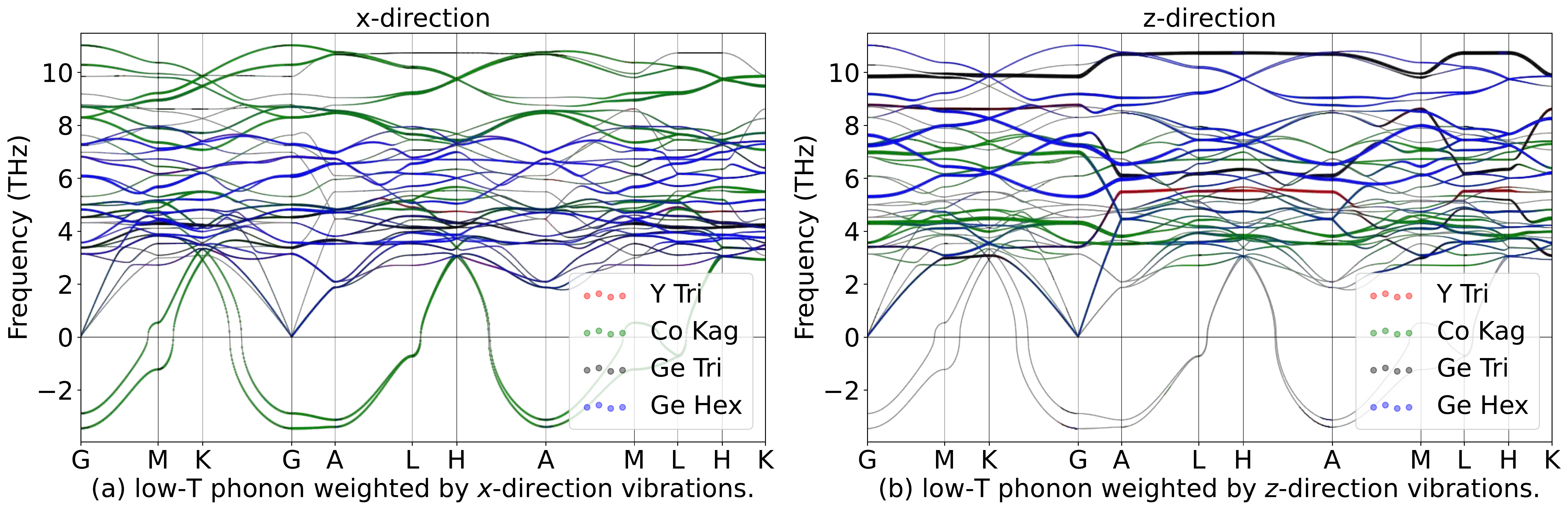}
\hspace{5pt}\label{fig:YCo6Ge6_relaxed_High_xz}
\includegraphics[width=0.85\textwidth]{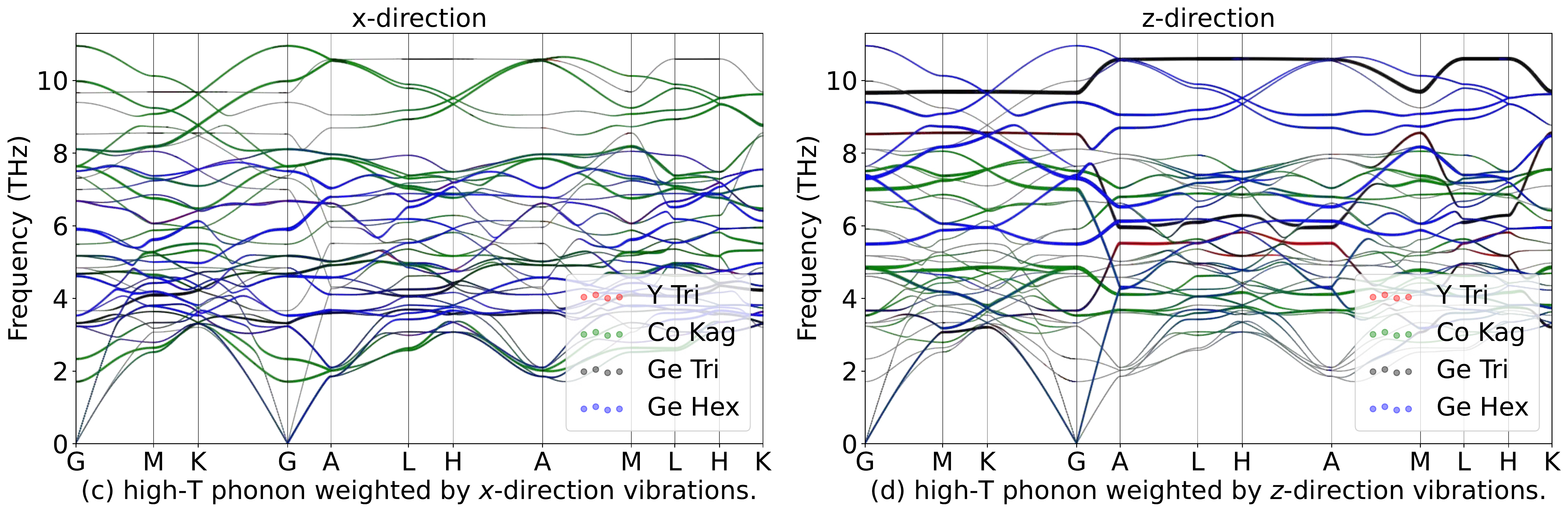}
\caption{Phonon spectrum for YCo$_6$Ge$_6$in in-plane ($x$) and out-of-plane ($z$) directions.}
\label{fig:YCo6Ge6phoxyz}
\end{figure}

\begin{table}[htbp]
\global\long\def\arraystretch{1.12}
\captionsetup{width=.95\textwidth}
\caption{IRREPs at high-symmetry points for YCo$_6$Ge$_6$ phonon spectrum at Low-T.}
\label{tab:191_YCo6Ge6_Low}
\setlength{\tabcolsep}{1.mm}{
}
\end{table}

\begin{table}[htbp]
\global\long\def\arraystretch{1.12}
\captionsetup{width=.95\textwidth}
\caption{IRREPs at high-symmetry points for YCo$_6$Ge$_6$ phonon spectrum at High-T.}
\label{tab:191_YCo6Ge6_High}
\setlength{\tabcolsep}{1.mm}{
%
}
\end{table}
\subsubsection{\label{sms2sec:ZrCo6Ge6}\hyperref[smsec:col]{\texorpdfstring{ZrCo$_6$Ge$_6$}{}}~{\color{blue}  (High-T stable, Low-T unstable) }}

\begin{table}[htbp]
\global\long\def\arraystretch{1.12}
\captionsetup{width=.85\textwidth}
\caption{Structure parameters of ZrCo$_6$Ge$_6$ after relaxation. It contains 394 electrons. }
\label{tab:ZrCo6Ge6}
\setlength{\tabcolsep}{4mm}{
%
}
\end{table}

\begin{figure}[htbp]
\centering
\includegraphics[width=0.85\textwidth]{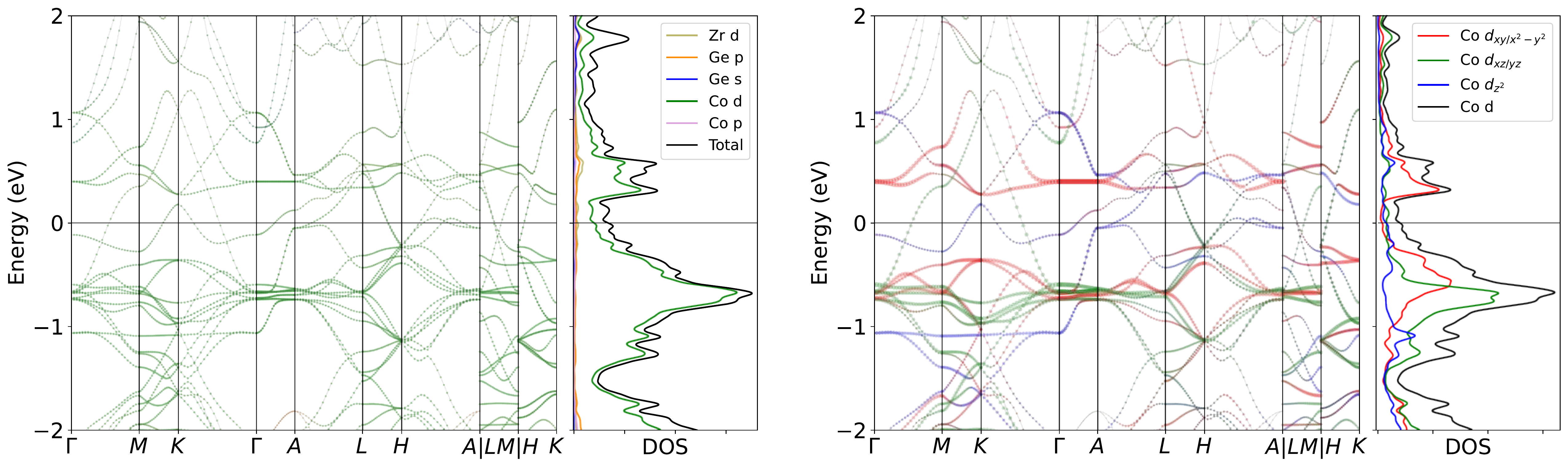}
\caption{Band structure for ZrCo$_6$Ge$_6$ without SOC. The projection of Co $3d$ orbitals is also presented. The band structures clearly reveal the dominating of Co $3d$ orbitals  near the Fermi level, as well as the pronounced peak.}
\label{fig:ZrCo6Ge6band}
\end{figure}

\begin{figure}[H]
\centering
\subfloat[Relaxed phonon at low-T.]{
\label{fig:ZrCo6Ge6_relaxed_Low}
\includegraphics[width=0.45\textwidth]{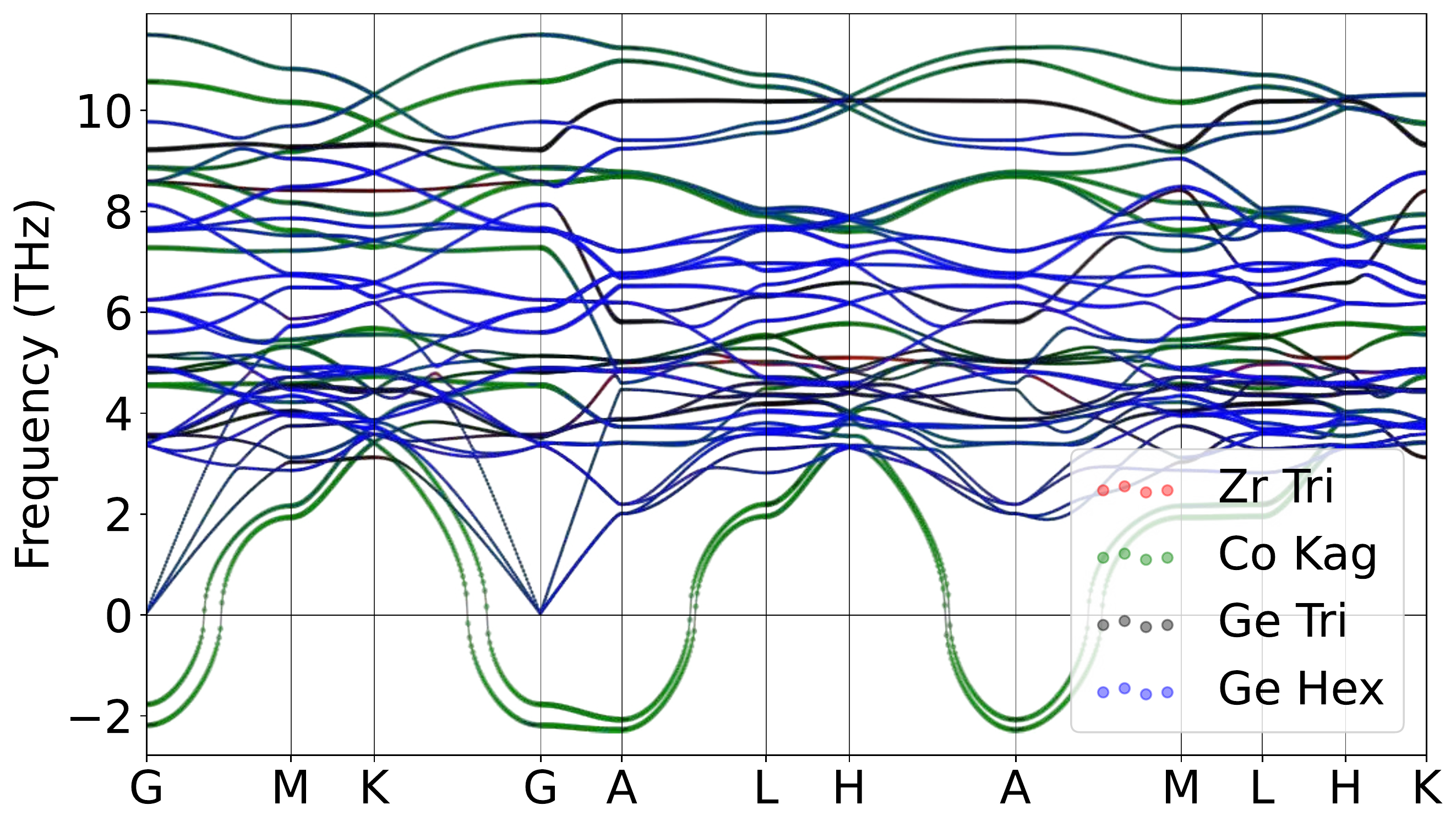}
}
\hspace{5pt}\subfloat[Relaxed phonon at high-T.]{
\label{fig:ZrCo6Ge6_relaxed_High}
\includegraphics[width=0.45\textwidth]{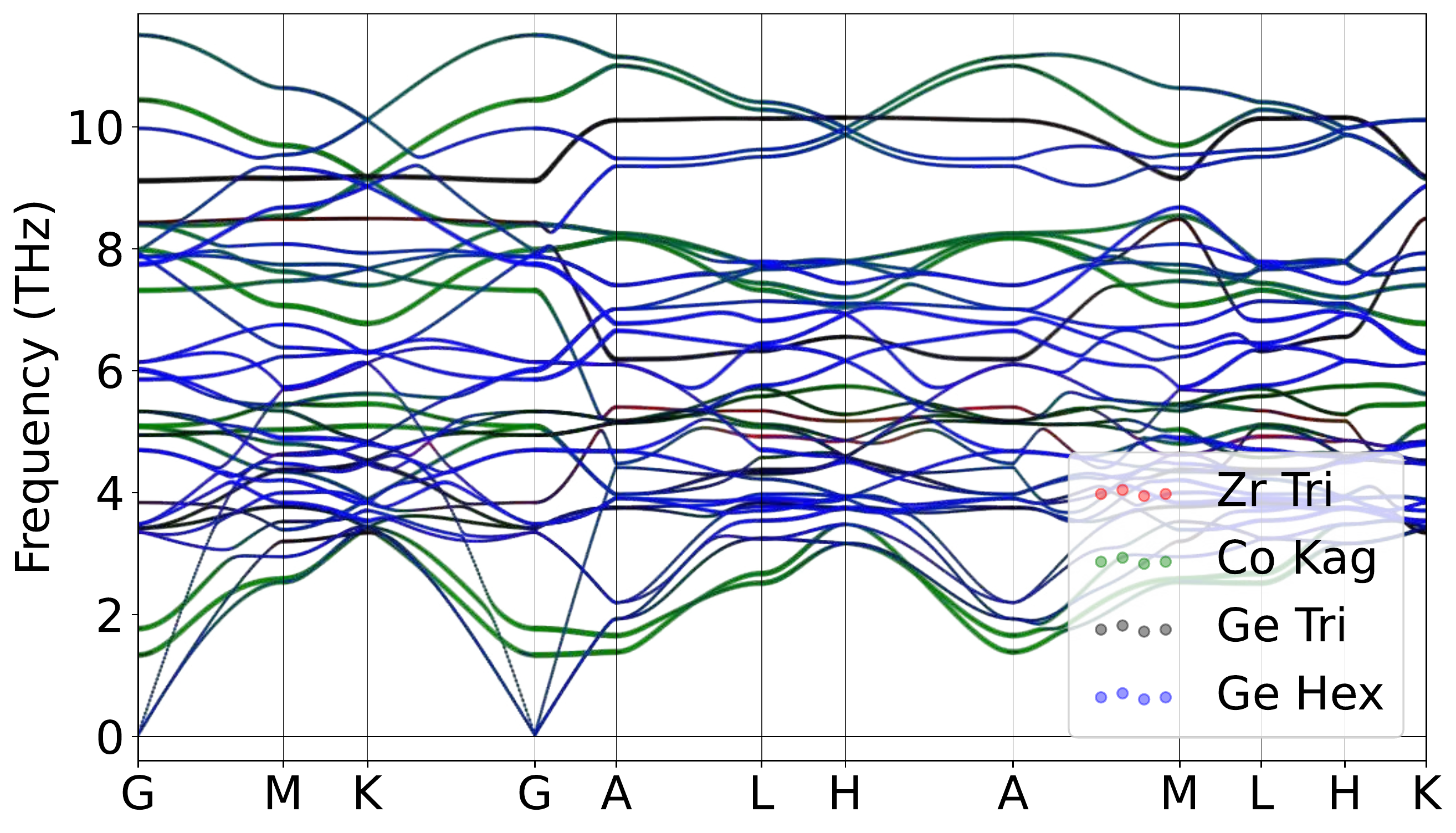}
}
\caption{Atomically projected phonon spectrum for ZrCo$_6$Ge$_6$.}
\label{fig:ZrCo6Ge6pho}
\end{figure}

\begin{figure}[H]
\centering
\label{fig:ZrCo6Ge6_relaxed_Low_xz}
\includegraphics[width=0.85\textwidth]{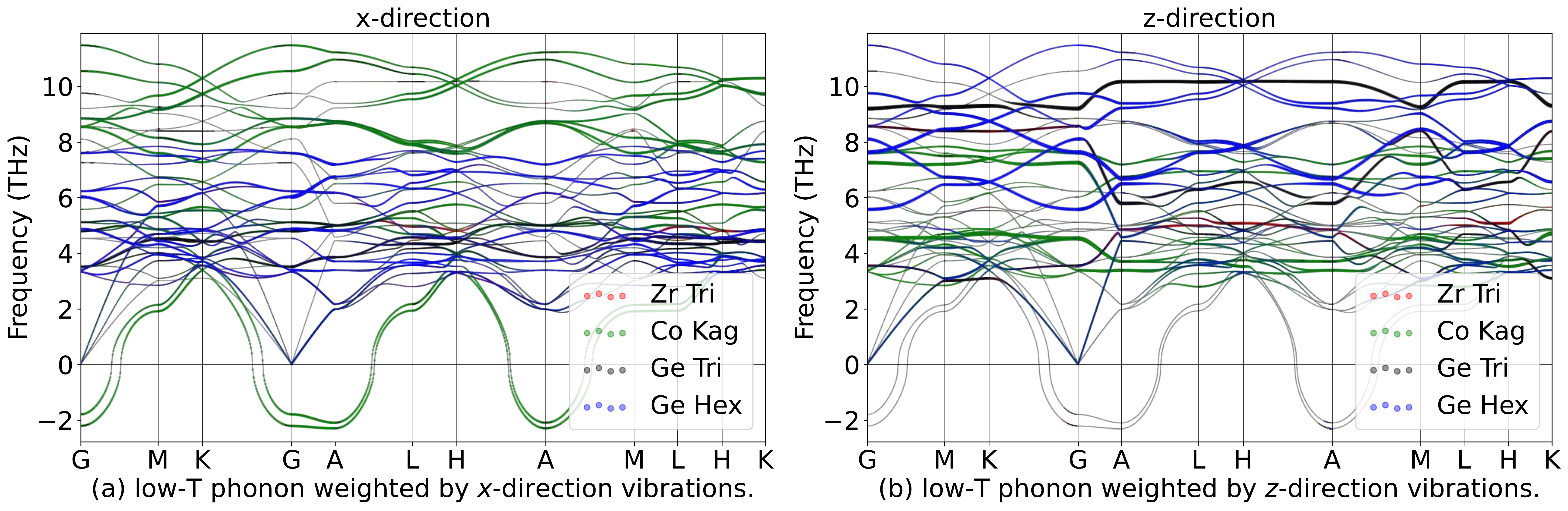}
\hspace{5pt}\label{fig:ZrCo6Ge6_relaxed_High_xz}
\includegraphics[width=0.85\textwidth]{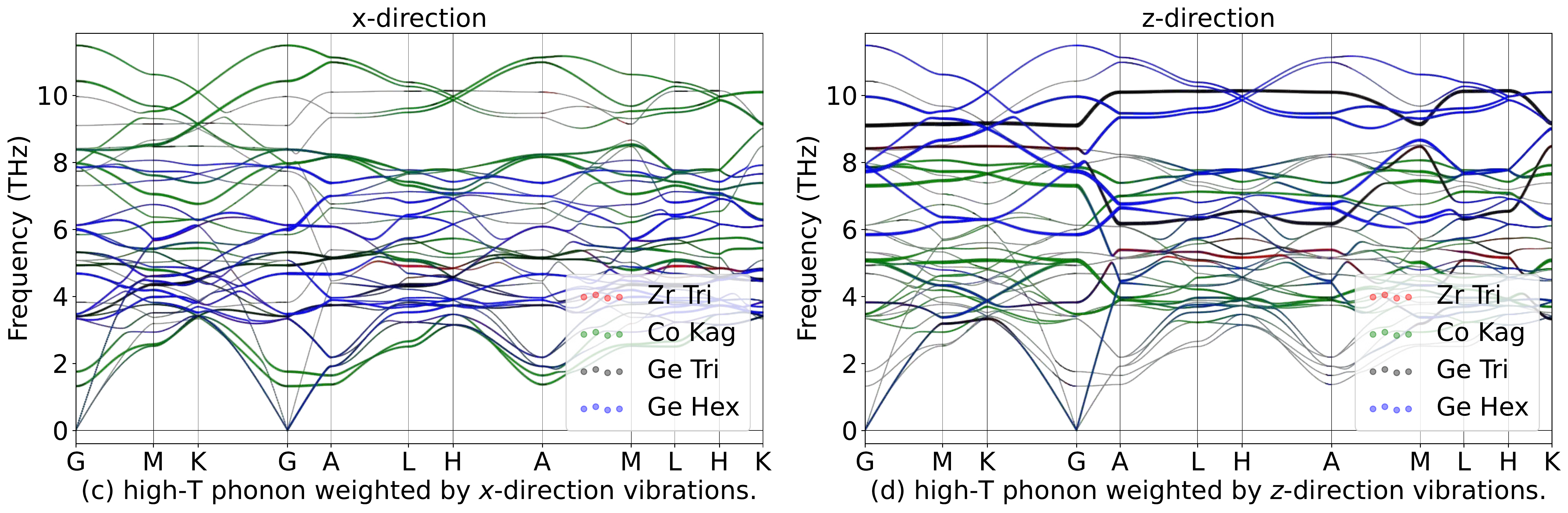}
\caption{Phonon spectrum for ZrCo$_6$Ge$_6$in in-plane ($x$) and out-of-plane ($z$) directions.}
\label{fig:ZrCo6Ge6phoxyz}
\end{figure}

\begin{table}[htbp]
\global\long\def\arraystretch{1.12}
\captionsetup{width=.95\textwidth}
\caption{IRREPs at high-symmetry points for ZrCo$_6$Ge$_6$ phonon spectrum at Low-T.}
\label{tab:191_ZrCo6Ge6_Low}
\setlength{\tabcolsep}{1.mm}{
}
\end{table}

\begin{table}[htbp]
\global\long\def\arraystretch{1.12}
\captionsetup{width=.95\textwidth}
\caption{IRREPs at high-symmetry points for ZrCo$_6$Ge$_6$ phonon spectrum at High-T.}
\label{tab:191_ZrCo6Ge6_High}
\setlength{\tabcolsep}{1.mm}{
%
}
\end{table}
\subsubsection{\label{sms2sec:NbCo6Ge6}\hyperref[smsec:col]{\texorpdfstring{NbCo$_6$Ge$_6$}{}}~{\color{blue}  (High-T stable, Low-T unstable) }}

\begin{table}[htbp]
\global\long\def\arraystretch{1.12}
\captionsetup{width=.85\textwidth}
\caption{Structure parameters of NbCo$_6$Ge$_6$ after relaxation. It contains 395 electrons. }
\label{tab:NbCo6Ge6}
\setlength{\tabcolsep}{4mm}{
%
}
\end{table}

\begin{figure}[htbp]
\centering
\includegraphics[width=0.85\textwidth]{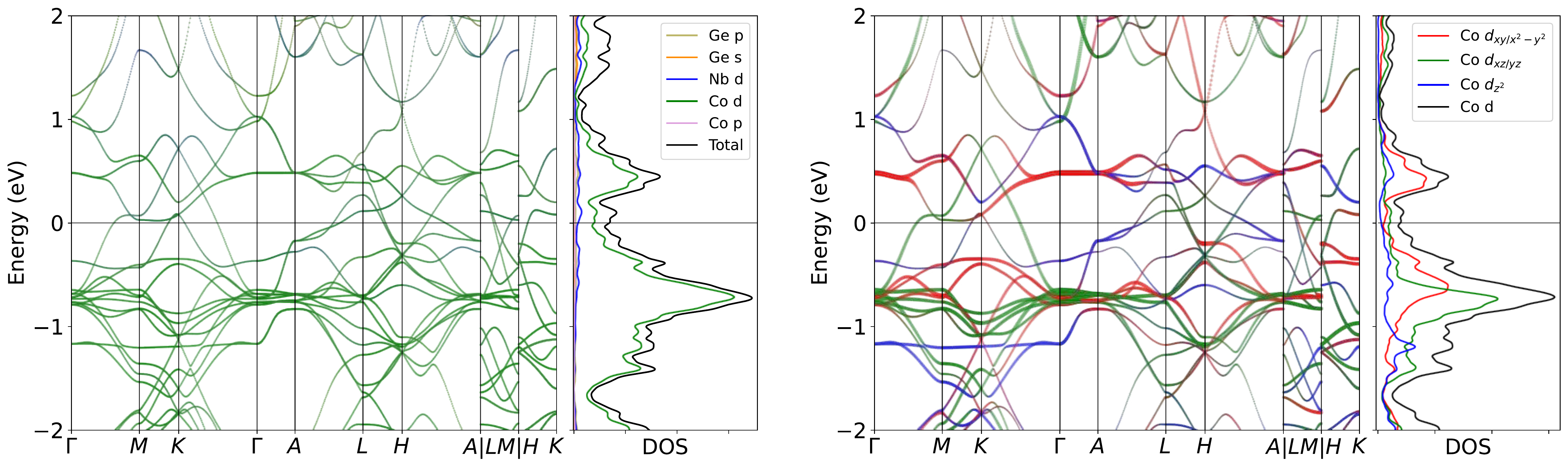}
\caption{Band structure for NbCo$_6$Ge$_6$ without SOC. The projection of Co $3d$ orbitals is also presented. The band structures clearly reveal the dominating of Co $3d$ orbitals  near the Fermi level, as well as the pronounced peak.}
\label{fig:NbCo6Ge6band}
\end{figure}

\begin{figure}[H]
\centering
\subfloat[Relaxed phonon at low-T.]{
\label{fig:NbCo6Ge6_relaxed_Low}
\includegraphics[width=0.45\textwidth]{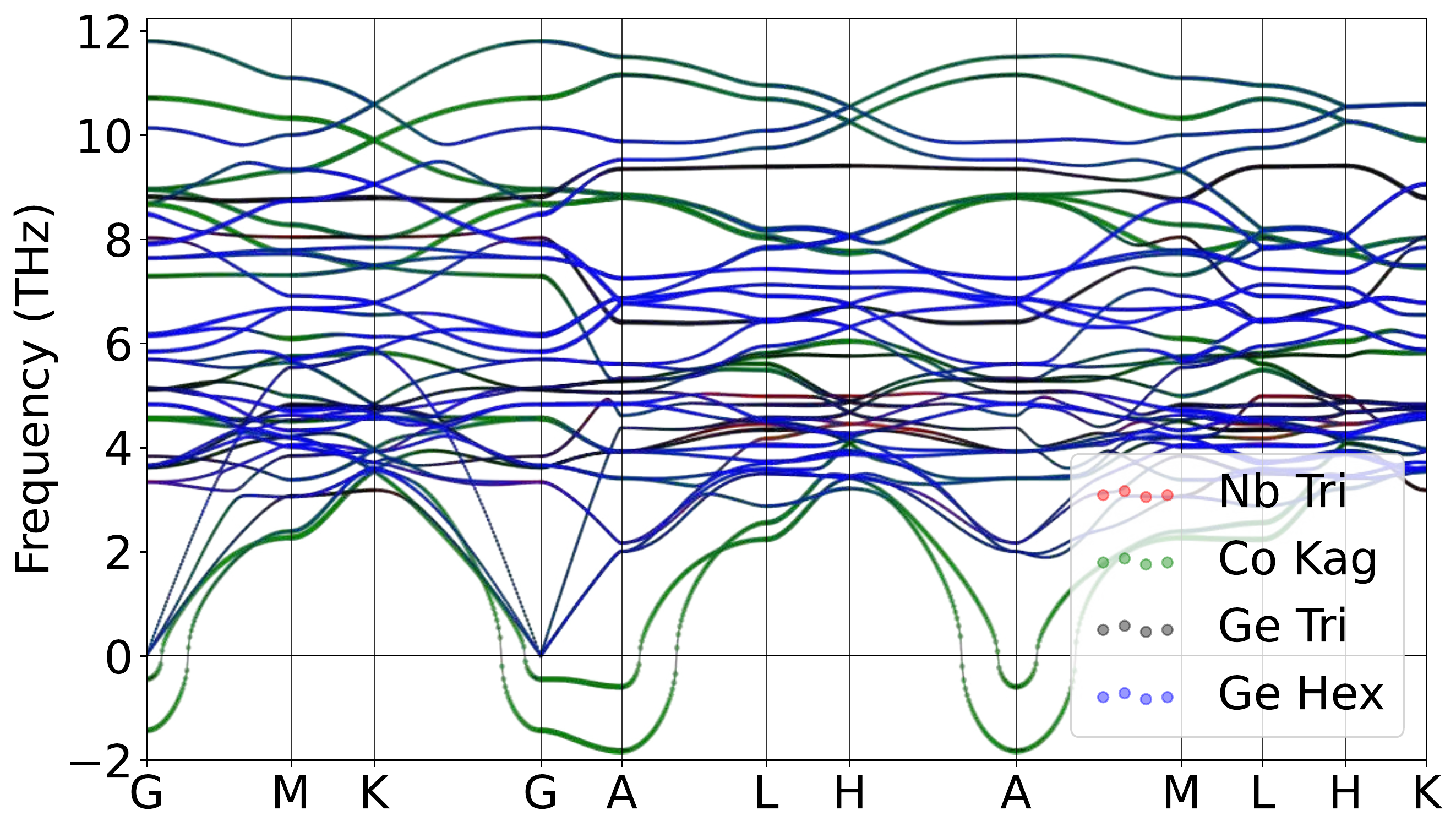}
}
\hspace{5pt}\subfloat[Relaxed phonon at high-T.]{
\label{fig:NbCo6Ge6_relaxed_High}
\includegraphics[width=0.45\textwidth]{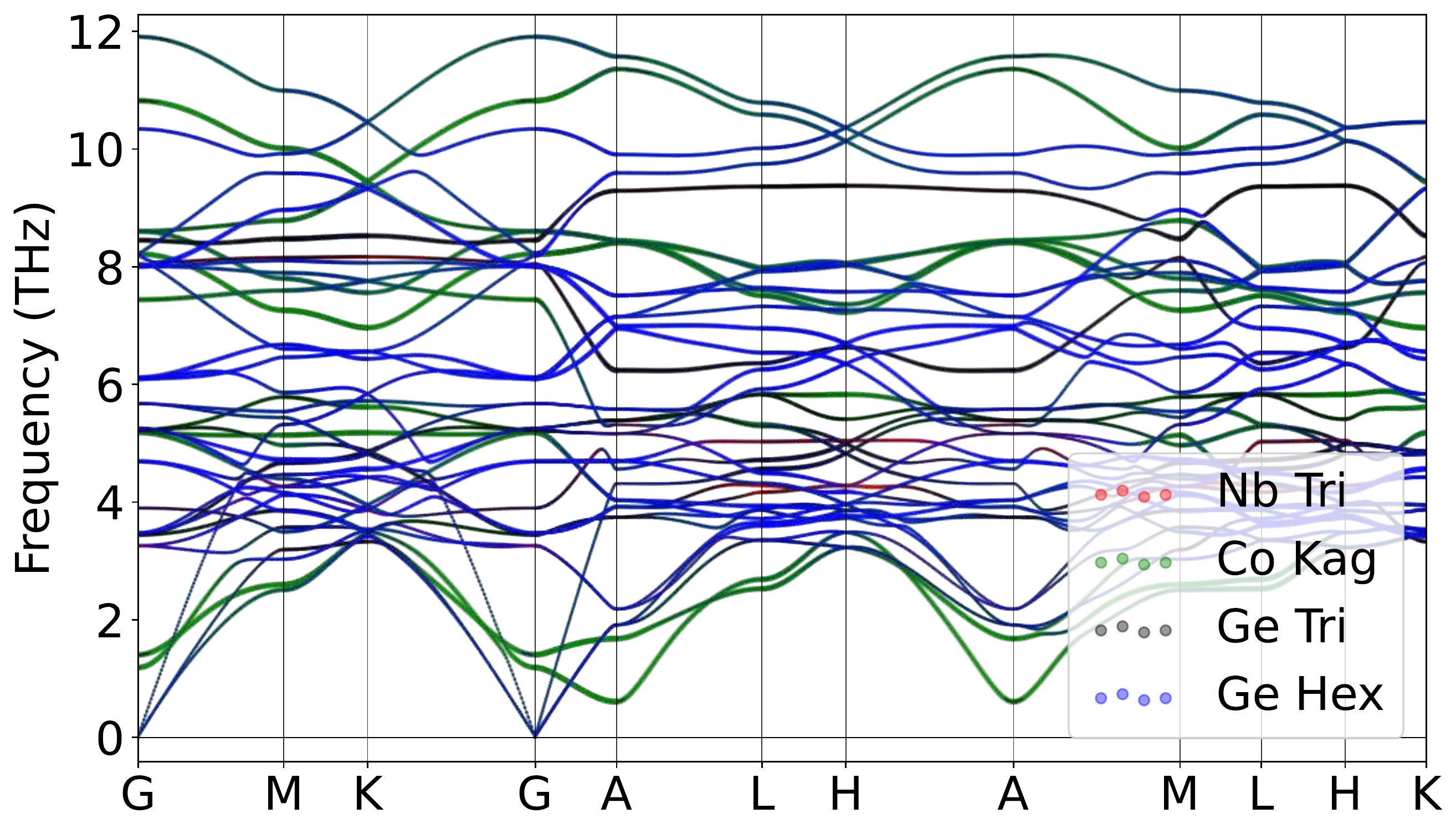}
}
\caption{Atomically projected phonon spectrum for NbCo$_6$Ge$_6$.}
\label{fig:NbCo6Ge6pho}
\end{figure}

\begin{figure}[H]
\centering
\label{fig:NbCo6Ge6_relaxed_Low_xz}
\includegraphics[width=0.85\textwidth]{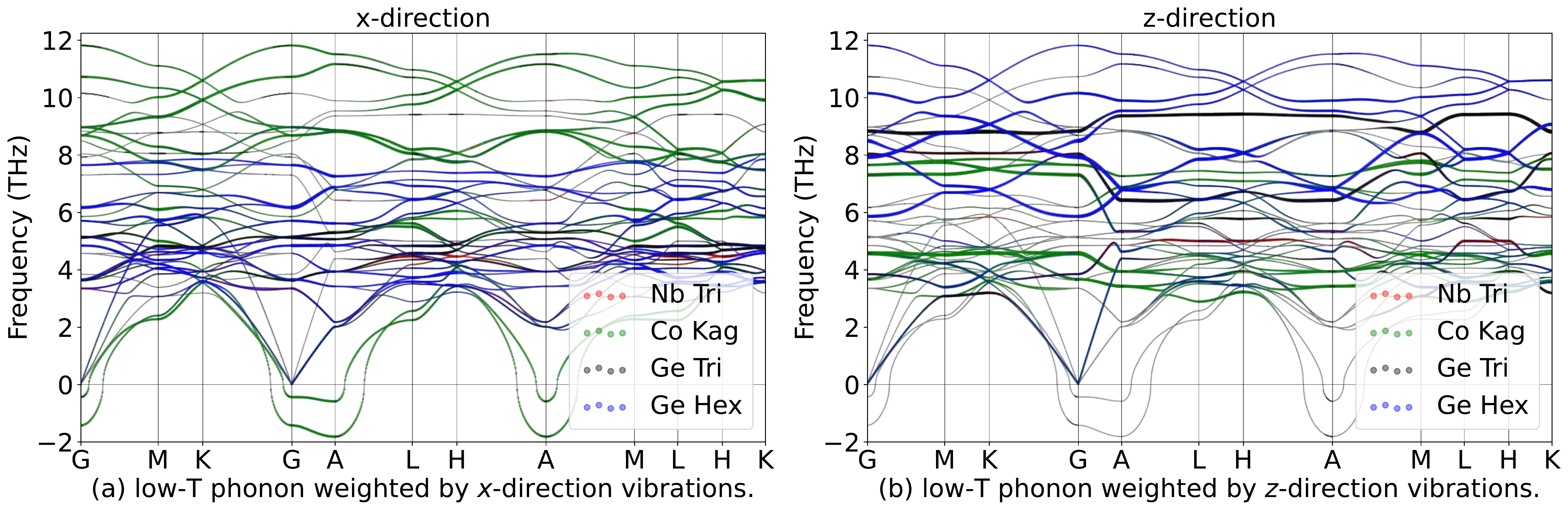}
\hspace{5pt}\label{fig:NbCo6Ge6_relaxed_High_xz}
\includegraphics[width=0.85\textwidth]{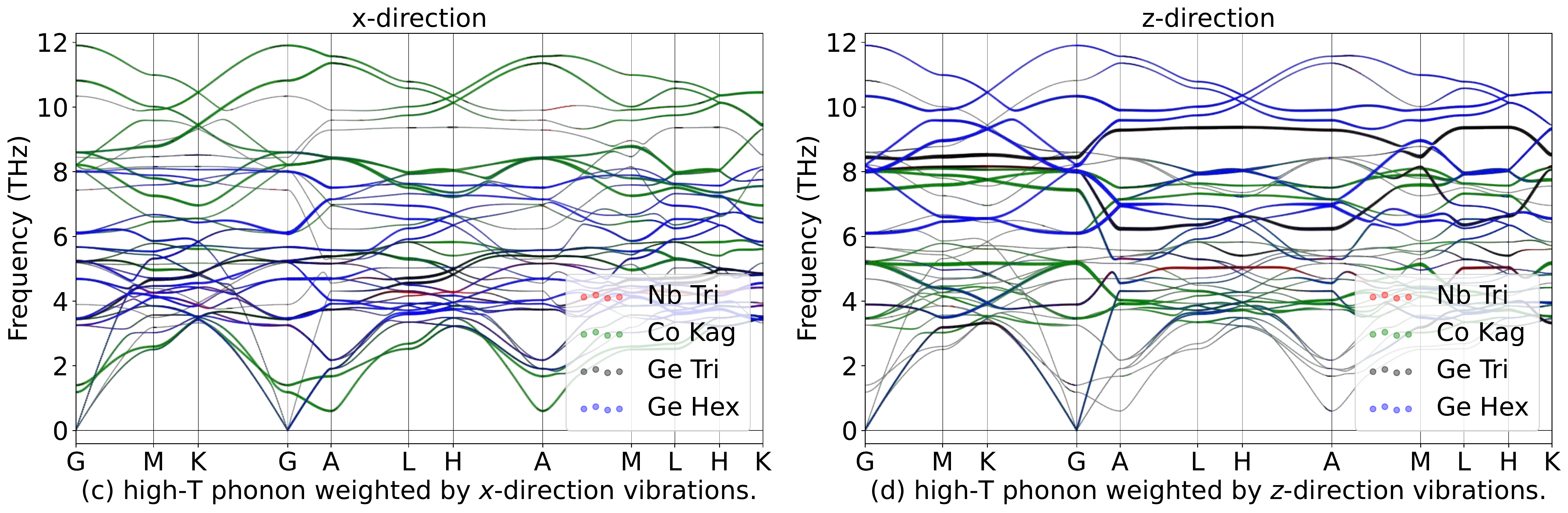}
\caption{Phonon spectrum for NbCo$_6$Ge$_6$in in-plane ($x$) and out-of-plane ($z$) directions.}
\label{fig:NbCo6Ge6phoxyz}
\end{figure}

\begin{table}[htbp]
\global\long\def\arraystretch{1.12}
\captionsetup{width=.95\textwidth}
\caption{IRREPs at high-symmetry points for NbCo$_6$Ge$_6$ phonon spectrum at Low-T.}
\label{tab:191_NbCo6Ge6_Low}
\setlength{\tabcolsep}{1.mm}{
}
\end{table}

\begin{table}[htbp]
\global\long\def\arraystretch{1.12}
\captionsetup{width=.95\textwidth}
\caption{IRREPs at high-symmetry points for NbCo$_6$Ge$_6$ phonon spectrum at High-T.}
\label{tab:191_NbCo6Ge6_High}
\setlength{\tabcolsep}{1.mm}{
%
}
\end{table}
\subsubsection{\label{sms2sec:PrCo6Ge6}\hyperref[smsec:col]{\texorpdfstring{PrCo$_6$Ge$_6$}{}}~{\color{blue}  (High-T stable, Low-T unstable) }}

\begin{table}[htbp]
\global\long\def\arraystretch{1.12}
\captionsetup{width=.85\textwidth}
\caption{Structure parameters of PrCo$_6$Ge$_6$ after relaxation. It contains 413 electrons. }
\label{tab:PrCo6Ge6}
\setlength{\tabcolsep}{4mm}{
%
}
\end{table}

\begin{figure}[htbp]
\centering
\includegraphics[width=0.85\textwidth]{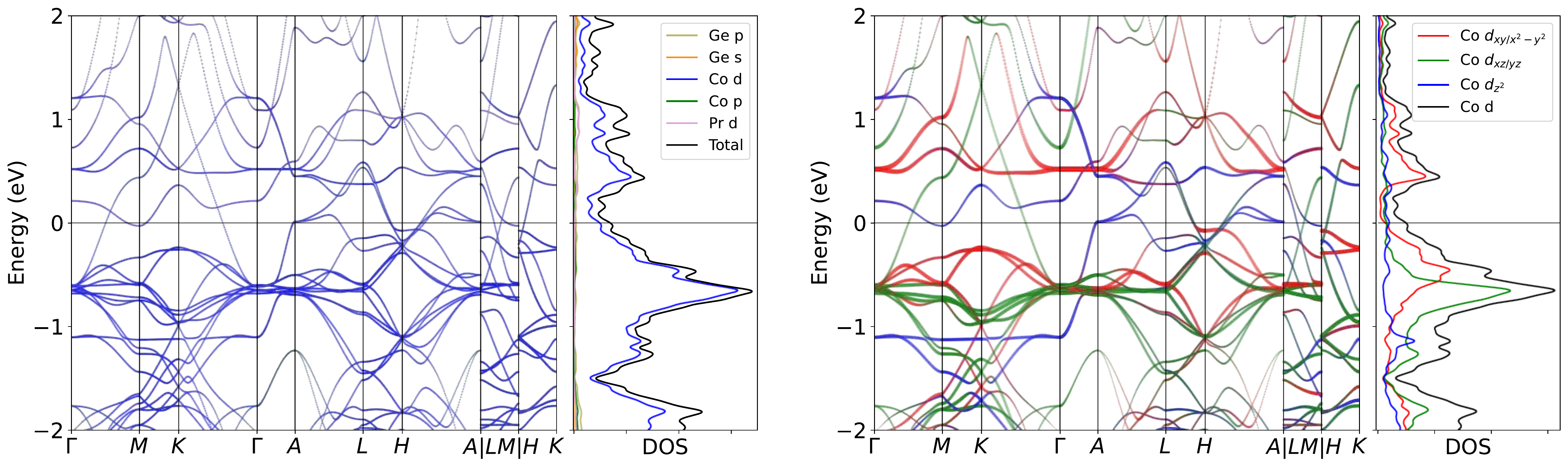}
\caption{Band structure for PrCo$_6$Ge$_6$ without SOC. The projection of Co $3d$ orbitals is also presented. The band structures clearly reveal the dominating of Co $3d$ orbitals  near the Fermi level, as well as the pronounced peak.}
\label{fig:PrCo6Ge6band}
\end{figure}

\begin{figure}[H]
\centering
\subfloat[Relaxed phonon at low-T.]{
\label{fig:PrCo6Ge6_relaxed_Low}
\includegraphics[width=0.45\textwidth]{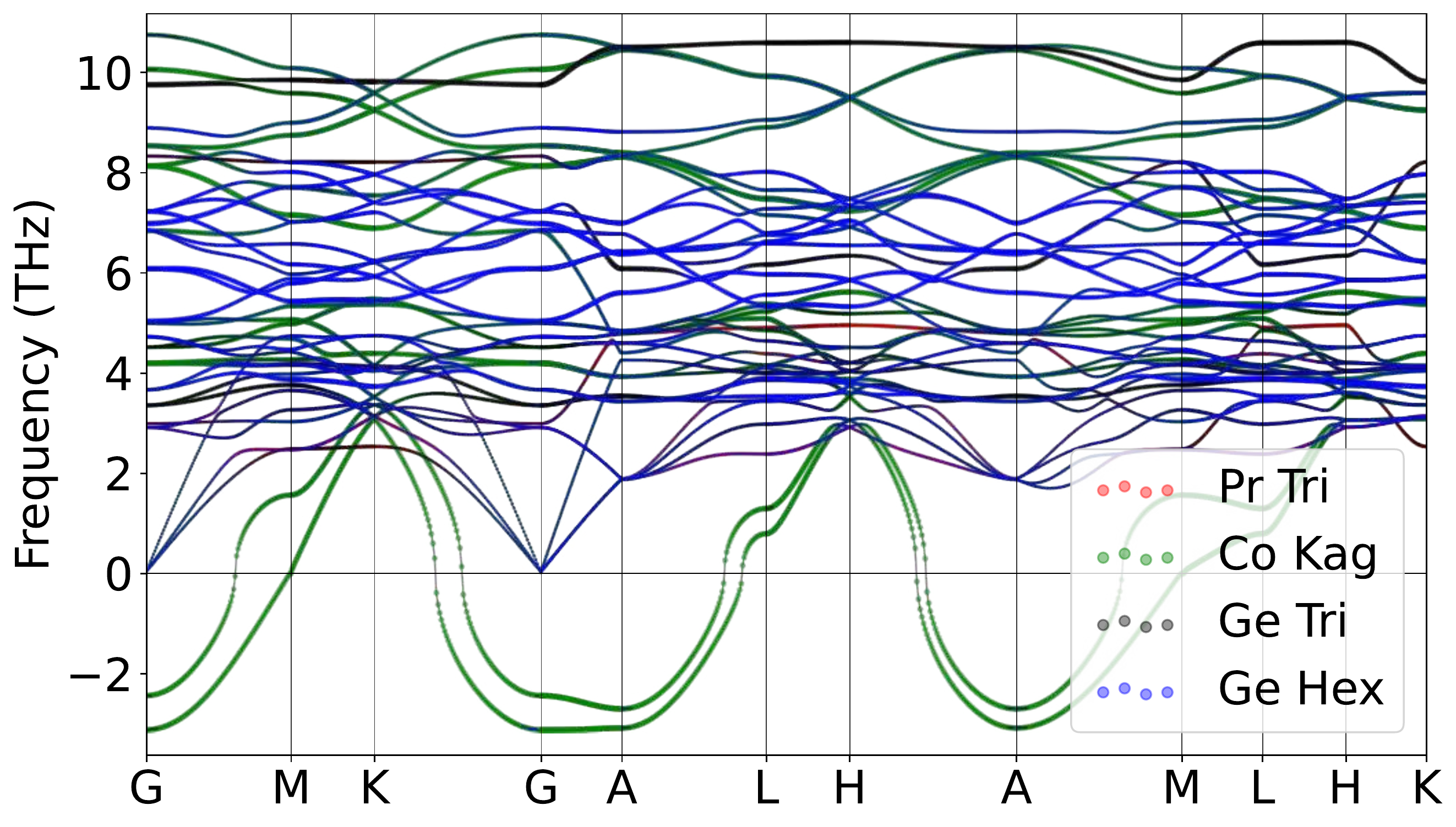}
}
\hspace{5pt}\subfloat[Relaxed phonon at high-T.]{
\label{fig:PrCo6Ge6_relaxed_High}
\includegraphics[width=0.45\textwidth]{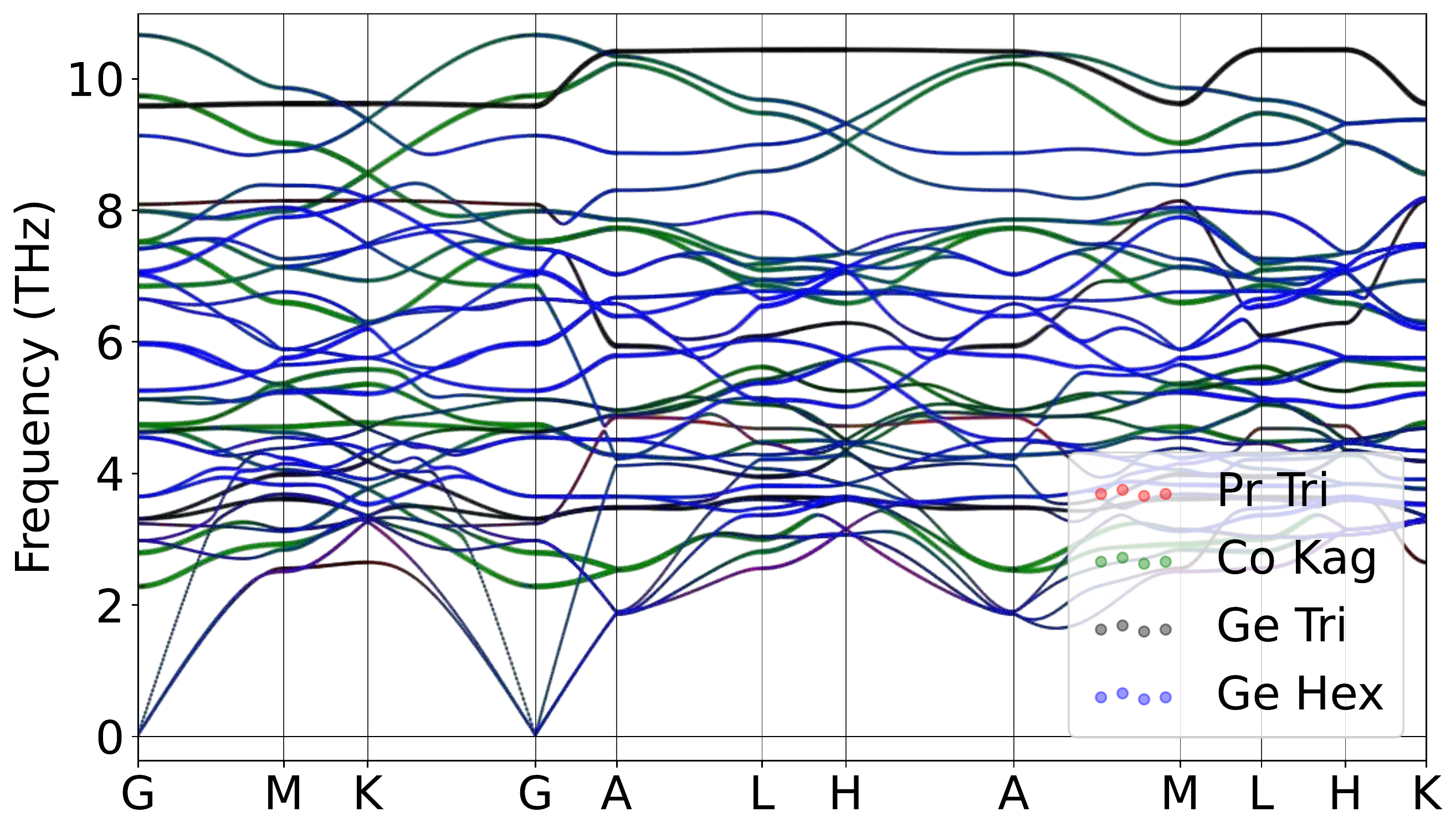}
}
\caption{Atomically projected phonon spectrum for PrCo$_6$Ge$_6$.}
\label{fig:PrCo6Ge6pho}
\end{figure}

\begin{figure}[H]
\centering
\label{fig:PrCo6Ge6_relaxed_Low_xz}
\includegraphics[width=0.85\textwidth]{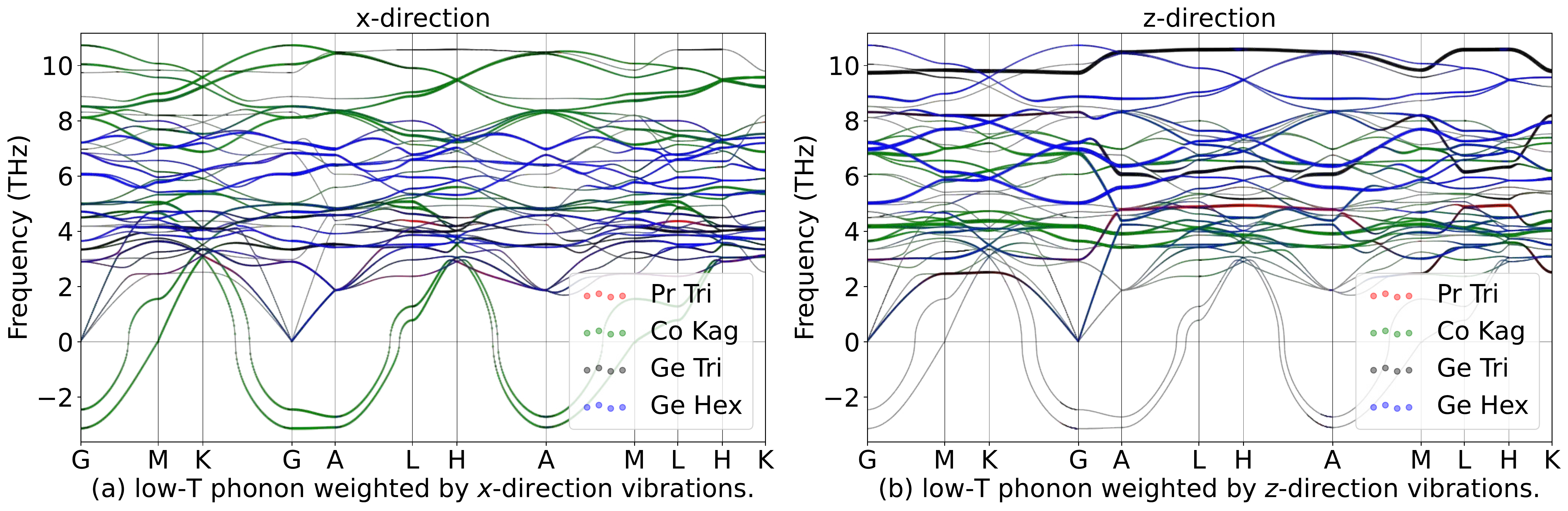}
\hspace{5pt}\label{fig:PrCo6Ge6_relaxed_High_xz}
\includegraphics[width=0.85\textwidth]{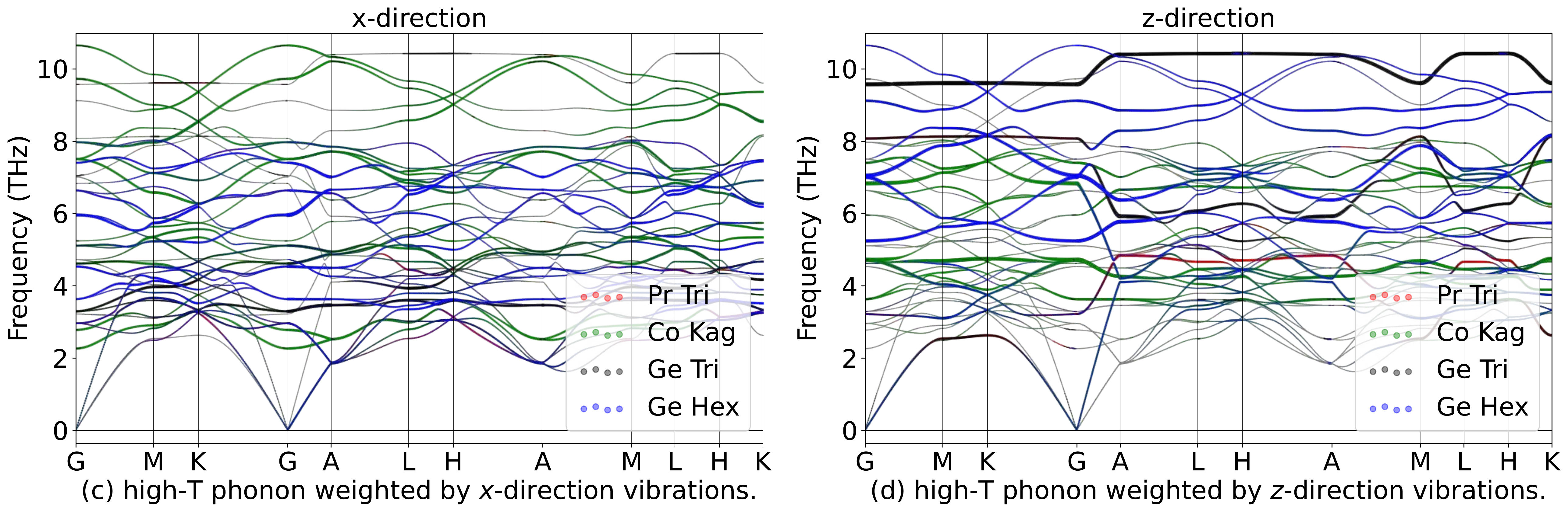}
\caption{Phonon spectrum for PrCo$_6$Ge$_6$in in-plane ($x$) and out-of-plane ($z$) directions.}
\label{fig:PrCo6Ge6phoxyz}
\end{figure}

\begin{table}[htbp]
\global\long\def\arraystretch{1.12}
\captionsetup{width=.95\textwidth}
\caption{IRREPs at high-symmetry points for PrCo$_6$Ge$_6$ phonon spectrum at Low-T.}
\label{tab:191_PrCo6Ge6_Low}
\setlength{\tabcolsep}{1.mm}{
}
\end{table}

\begin{table}[htbp]
\global\long\def\arraystretch{1.12}
\captionsetup{width=.95\textwidth}
\caption{IRREPs at high-symmetry points for PrCo$_6$Ge$_6$ phonon spectrum at High-T.}
\label{tab:191_PrCo6Ge6_High}
\setlength{\tabcolsep}{1.mm}{
%
}
\end{table}
\subsubsection{\label{sms2sec:NdCo6Ge6}\hyperref[smsec:col]{\texorpdfstring{NdCo$_6$Ge$_6$}{}}~{\color{blue}  (High-T stable, Low-T unstable) }}

\begin{table}[htbp]
\global\long\def\arraystretch{1.12}
\captionsetup{width=.85\textwidth}
\caption{Structure parameters of NdCo$_6$Ge$_6$ after relaxation. It contains 414 electrons. }
\label{tab:NdCo6Ge6}
\setlength{\tabcolsep}{4mm}{
%
}
\end{table}

\begin{figure}[htbp]
\centering
\includegraphics[width=0.85\textwidth]{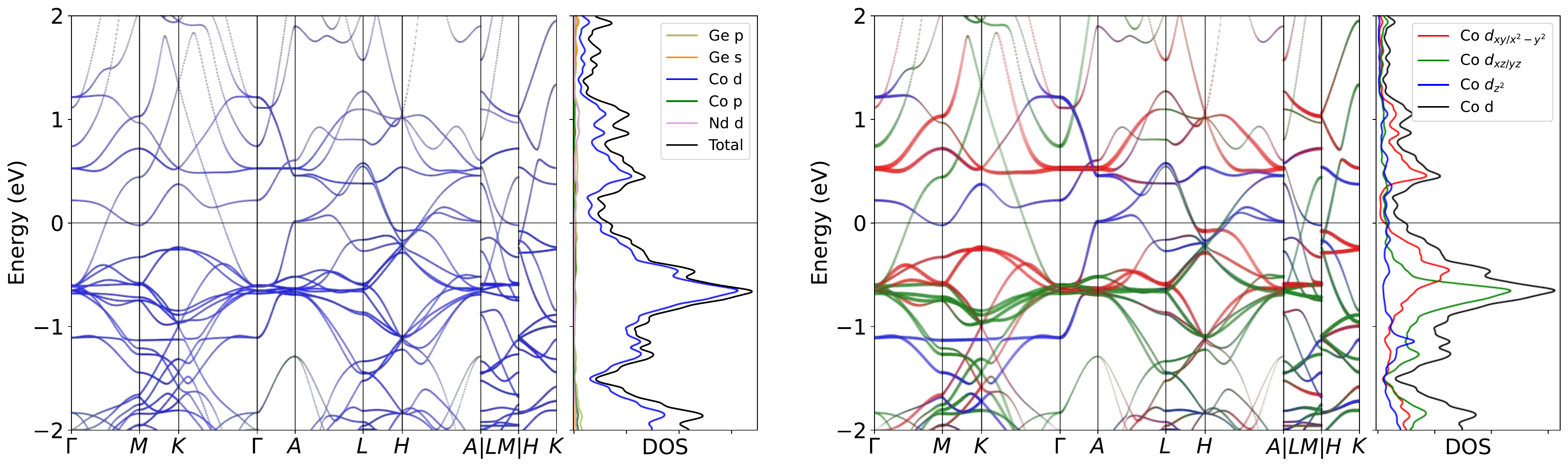}
\caption{Band structure for NdCo$_6$Ge$_6$ without SOC. The projection of Co $3d$ orbitals is also presented. The band structures clearly reveal the dominating of Co $3d$ orbitals  near the Fermi level, as well as the pronounced peak.}
\label{fig:NdCo6Ge6band}
\end{figure}

\begin{figure}[H]
\centering
\subfloat[Relaxed phonon at low-T.]{
\label{fig:NdCo6Ge6_relaxed_Low}
\includegraphics[width=0.45\textwidth]{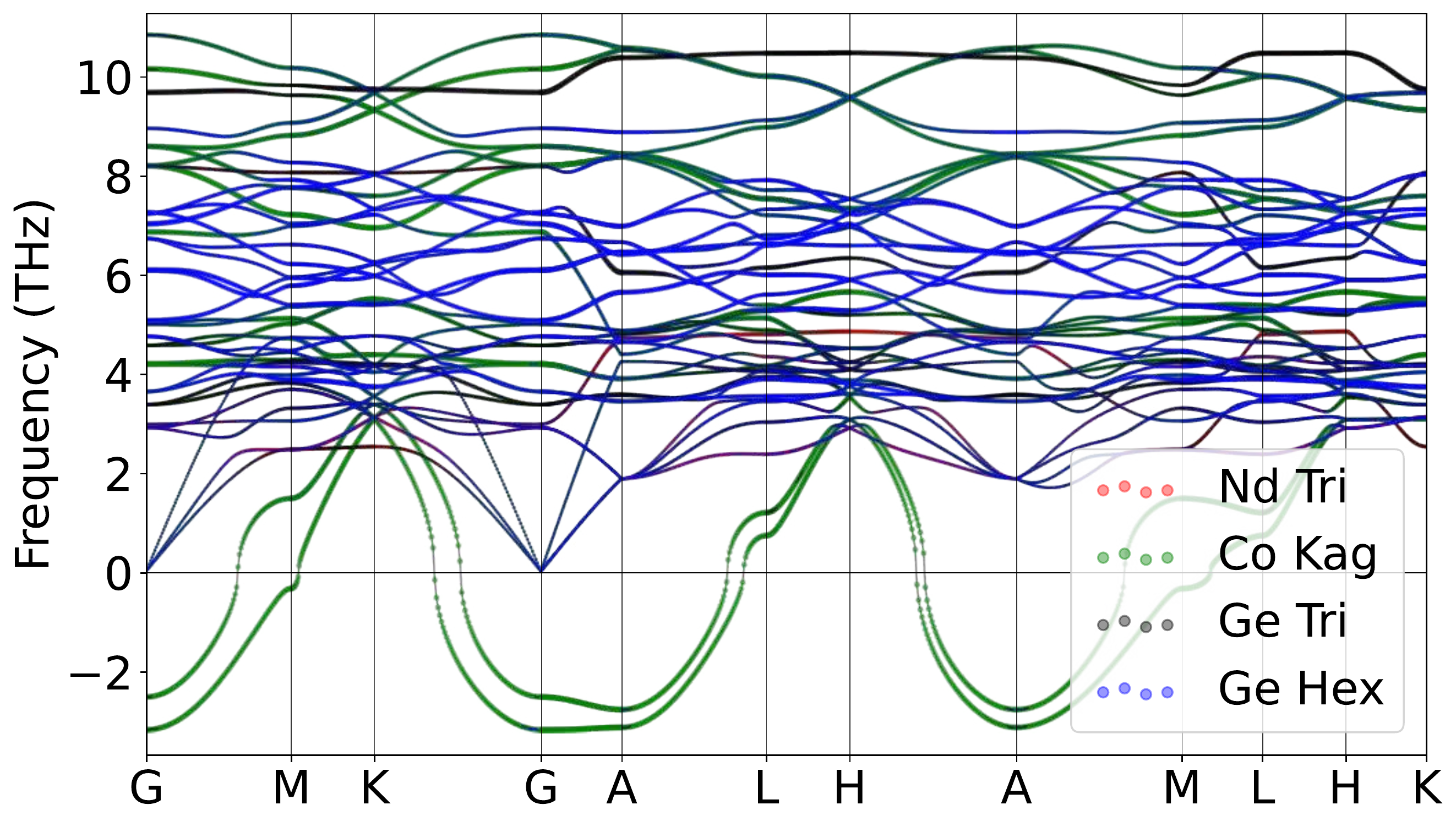}
}
\hspace{5pt}\subfloat[Relaxed phonon at high-T.]{
\label{fig:NdCo6Ge6_relaxed_High}
\includegraphics[width=0.45\textwidth]{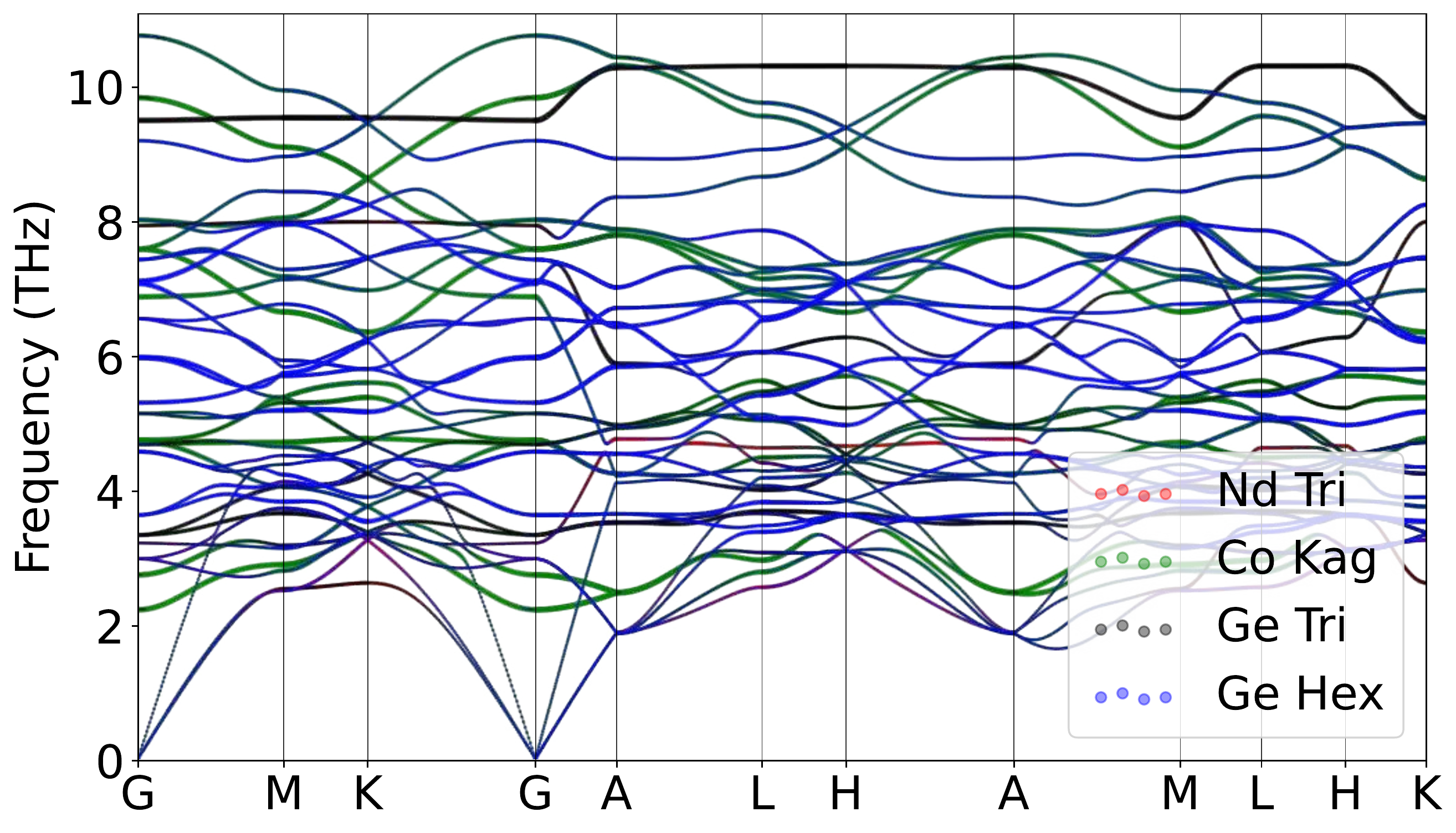}
}
\caption{Atomically projected phonon spectrum for NdCo$_6$Ge$_6$.}
\label{fig:NdCo6Ge6pho}
\end{figure}

\begin{figure}[H]
\centering
\label{fig:NdCo6Ge6_relaxed_Low_xz}
\includegraphics[width=0.85\textwidth]{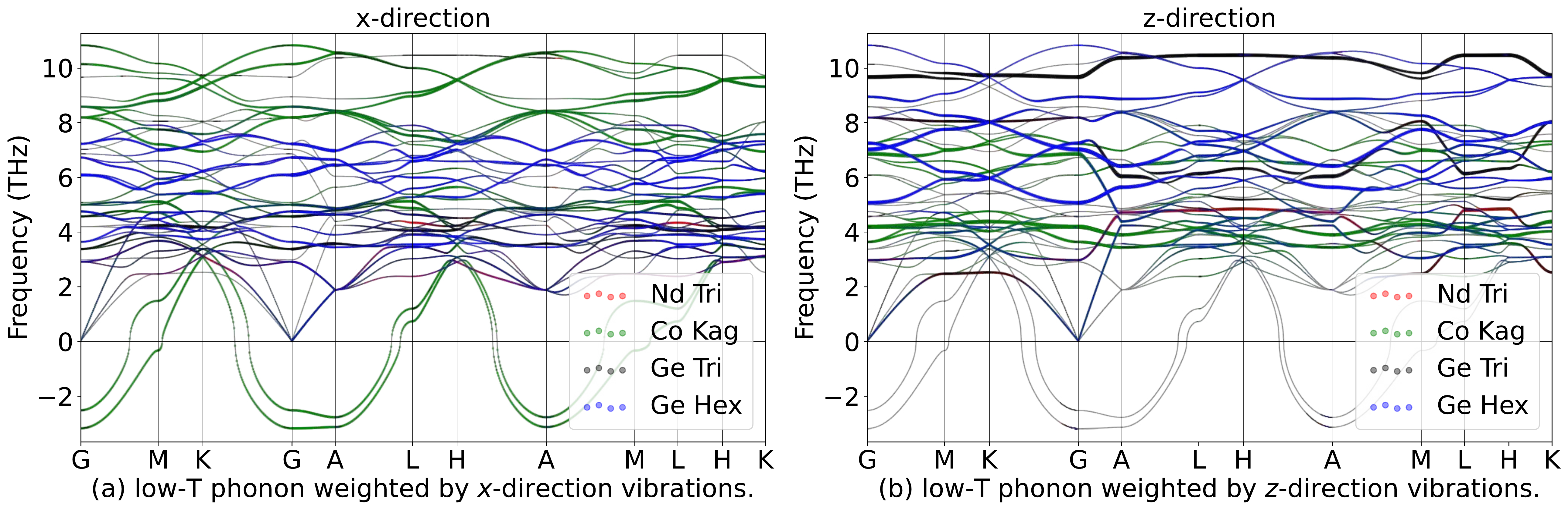}
\hspace{5pt}\label{fig:NdCo6Ge6_relaxed_High_xz}
\includegraphics[width=0.85\textwidth]{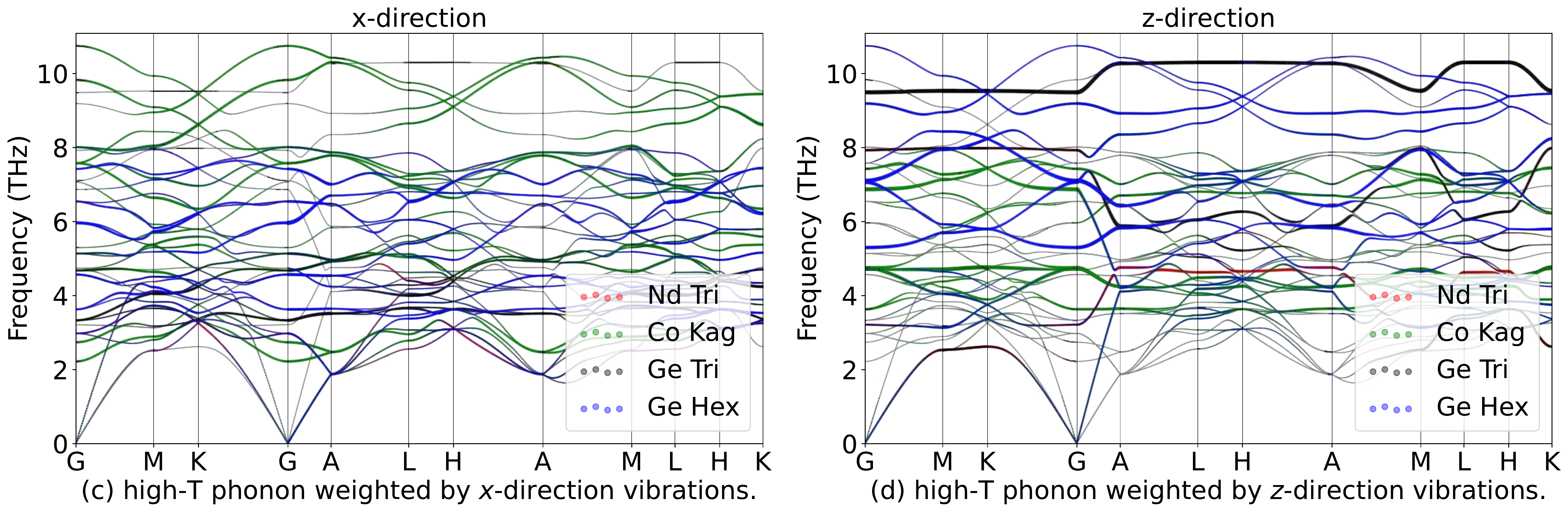}
\caption{Phonon spectrum for NdCo$_6$Ge$_6$in in-plane ($x$) and out-of-plane ($z$) directions.}
\label{fig:NdCo6Ge6phoxyz}
\end{figure}

\begin{table}[htbp]
\global\long\def\arraystretch{1.12}
\captionsetup{width=.95\textwidth}
\caption{IRREPs at high-symmetry points for NdCo$_6$Ge$_6$ phonon spectrum at Low-T.}
\label{tab:191_NdCo6Ge6_Low}
\setlength{\tabcolsep}{1.mm}{
}
\end{table}

\begin{table}[htbp]
\global\long\def\arraystretch{1.12}
\captionsetup{width=.95\textwidth}
\caption{IRREPs at high-symmetry points for NdCo$_6$Ge$_6$ phonon spectrum at High-T.}
\label{tab:191_NdCo6Ge6_High}
\setlength{\tabcolsep}{1.mm}{
%
}
\end{table}
\subsubsection{\label{sms2sec:SmCo6Ge6}\hyperref[smsec:col]{\texorpdfstring{SmCo$_6$Ge$_6$}{}}~{\color{blue}  (High-T stable, Low-T unstable) }}

\begin{table}[htbp]
\global\long\def\arraystretch{1.12}
\captionsetup{width=.85\textwidth}
\caption{Structure parameters of SmCo$_6$Ge$_6$ after relaxation. It contains 416 electrons. }
\label{tab:SmCo6Ge6}
\setlength{\tabcolsep}{4mm}{
%
}
\end{table}

\begin{figure}[htbp]
\centering
\includegraphics[width=0.85\textwidth]{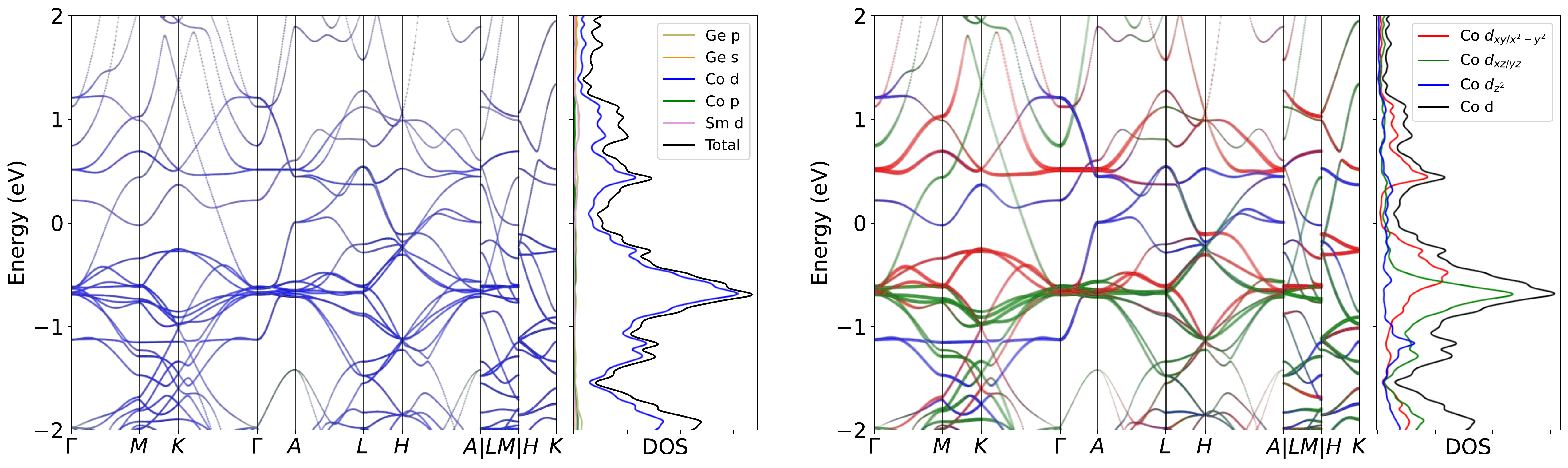}
\caption{Band structure for SmCo$_6$Ge$_6$ without SOC. The projection of Co $3d$ orbitals is also presented. The band structures clearly reveal the dominating of Co $3d$ orbitals  near the Fermi level, as well as the pronounced peak.}
\label{fig:SmCo6Ge6band}
\end{figure}

\begin{figure}[H]
\centering
\subfloat[Relaxed phonon at low-T.]{
\label{fig:SmCo6Ge6_relaxed_Low}
\includegraphics[width=0.45\textwidth]{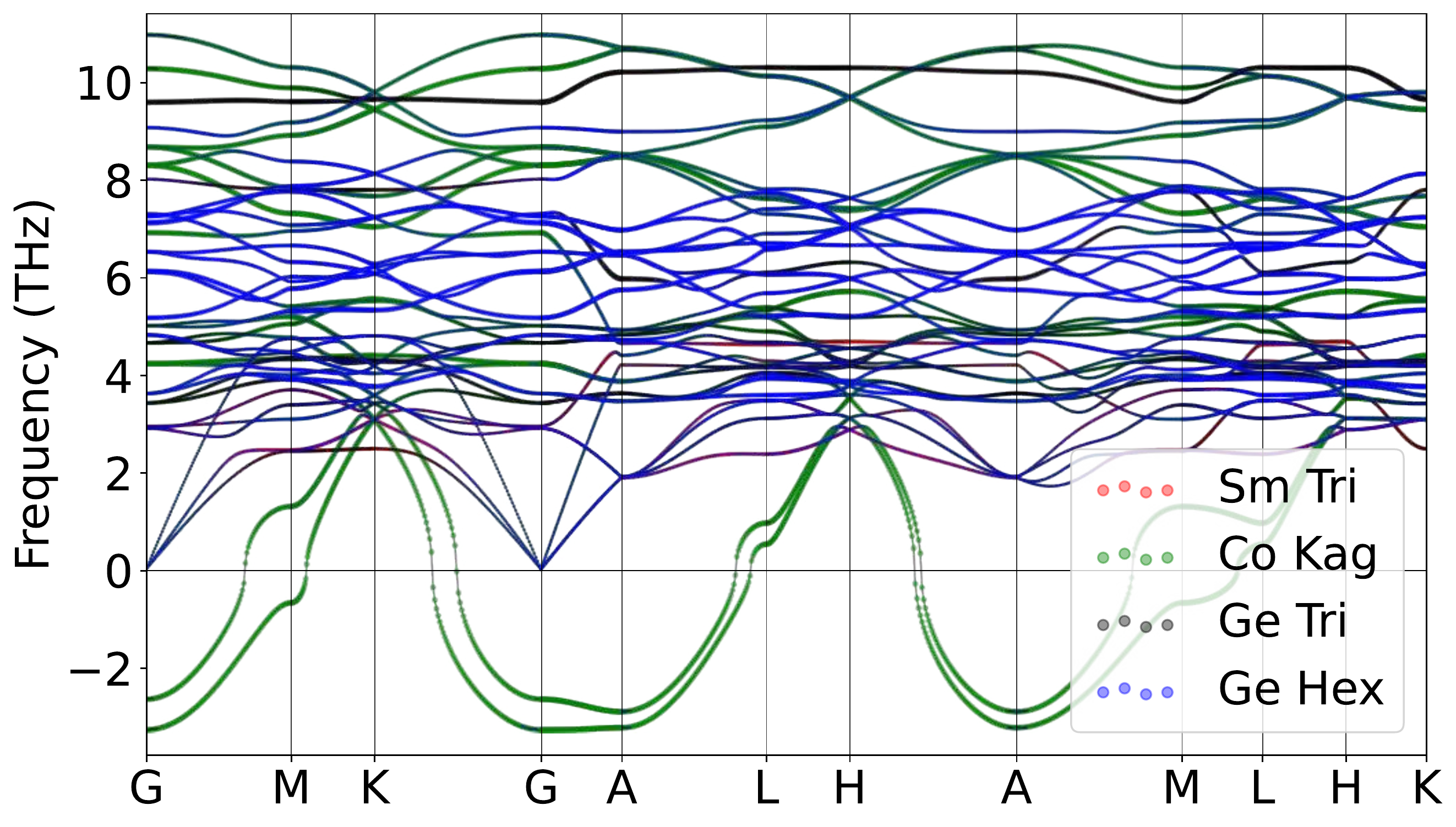}
}
\hspace{5pt}\subfloat[Relaxed phonon at high-T.]{
\label{fig:SmCo6Ge6_relaxed_High}
\includegraphics[width=0.45\textwidth]{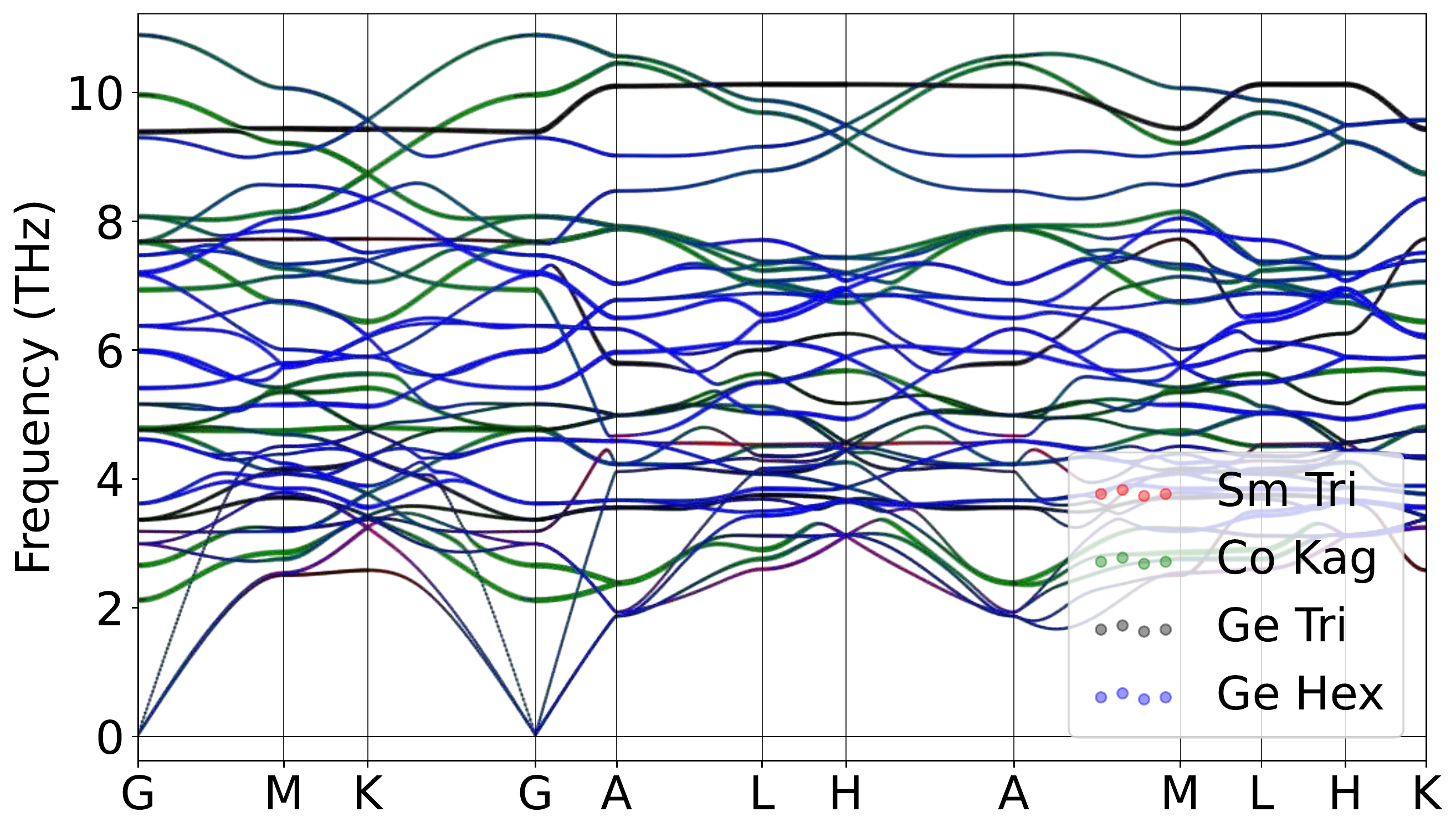}
}
\caption{Atomically projected phonon spectrum for SmCo$_6$Ge$_6$.}
\label{fig:SmCo6Ge6pho}
\end{figure}

\begin{figure}[H]
\centering
\label{fig:SmCo6Ge6_relaxed_Low_xz}
\includegraphics[width=0.85\textwidth]{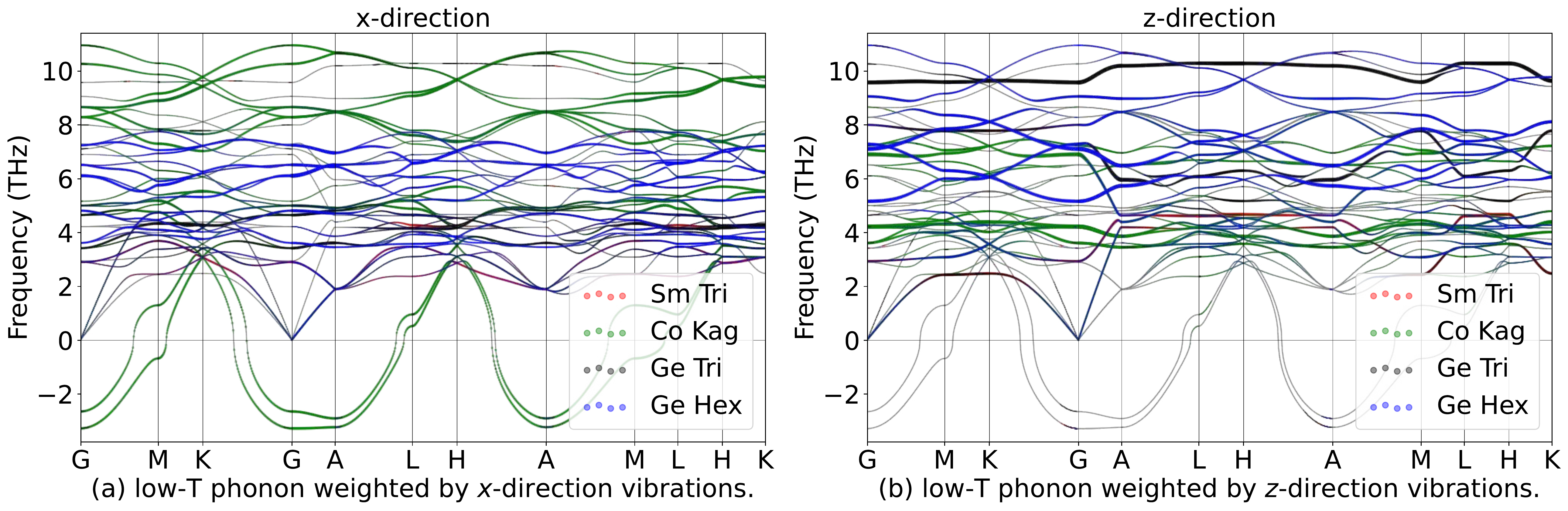}
\hspace{5pt}\label{fig:SmCo6Ge6_relaxed_High_xz}
\includegraphics[width=0.85\textwidth]{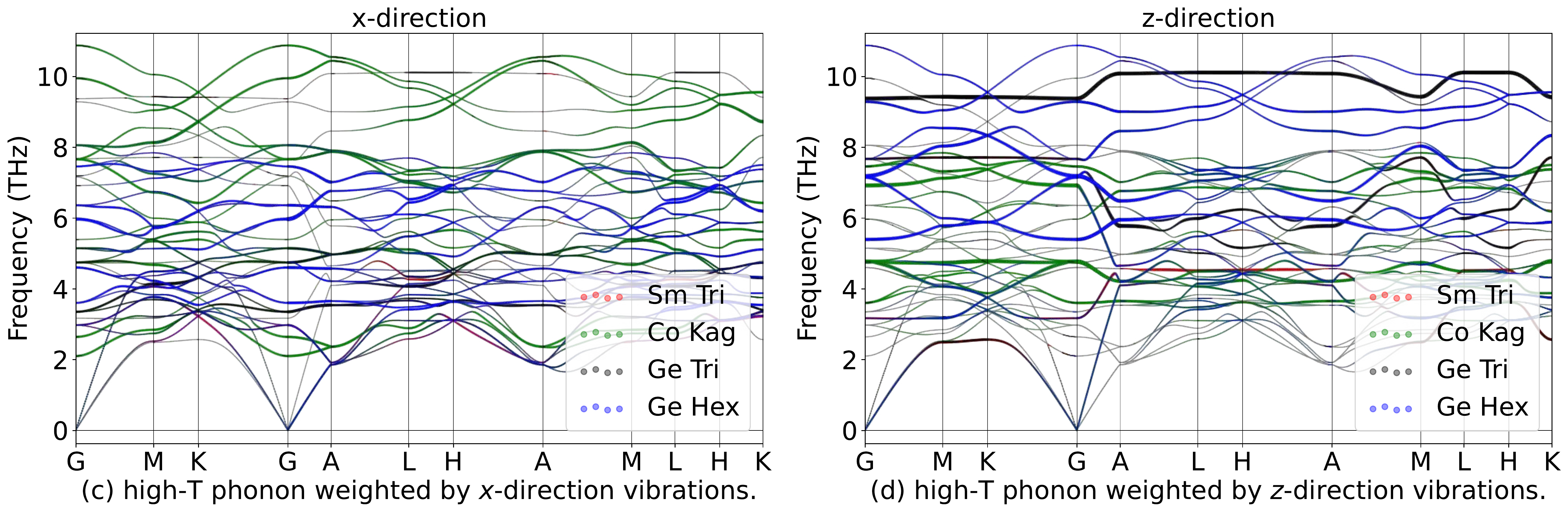}
\caption{Phonon spectrum for SmCo$_6$Ge$_6$in in-plane ($x$) and out-of-plane ($z$) directions.}
\label{fig:SmCo6Ge6phoxyz}
\end{figure}

\begin{table}[htbp]
\global\long\def\arraystretch{1.12}
\captionsetup{width=.95\textwidth}
\caption{IRREPs at high-symmetry points for SmCo$_6$Ge$_6$ phonon spectrum at Low-T.}
\label{tab:191_SmCo6Ge6_Low}
\setlength{\tabcolsep}{1.mm}{
}
\end{table}

\begin{table}[htbp]
\global\long\def\arraystretch{1.12}
\captionsetup{width=.95\textwidth}
\caption{IRREPs at high-symmetry points for SmCo$_6$Ge$_6$ phonon spectrum at High-T.}
\label{tab:191_SmCo6Ge6_High}
\setlength{\tabcolsep}{1.mm}{
%
}
\end{table}
\subsubsection{\label{sms2sec:GdCo6Ge6}\hyperref[smsec:col]{\texorpdfstring{GdCo$_6$Ge$_6$}{}}~{\color{blue}  (High-T stable, Low-T unstable) }}

\begin{table}[htbp]
\global\long\def\arraystretch{1.12}
\captionsetup{width=.85\textwidth}
\caption{Structure parameters of GdCo$_6$Ge$_6$ after relaxation. It contains 418 electrons. }
\label{tab:GdCo6Ge6}
\setlength{\tabcolsep}{4mm}{
%
}
\end{table}

\begin{figure}[htbp]
\centering
\includegraphics[width=0.85\textwidth]{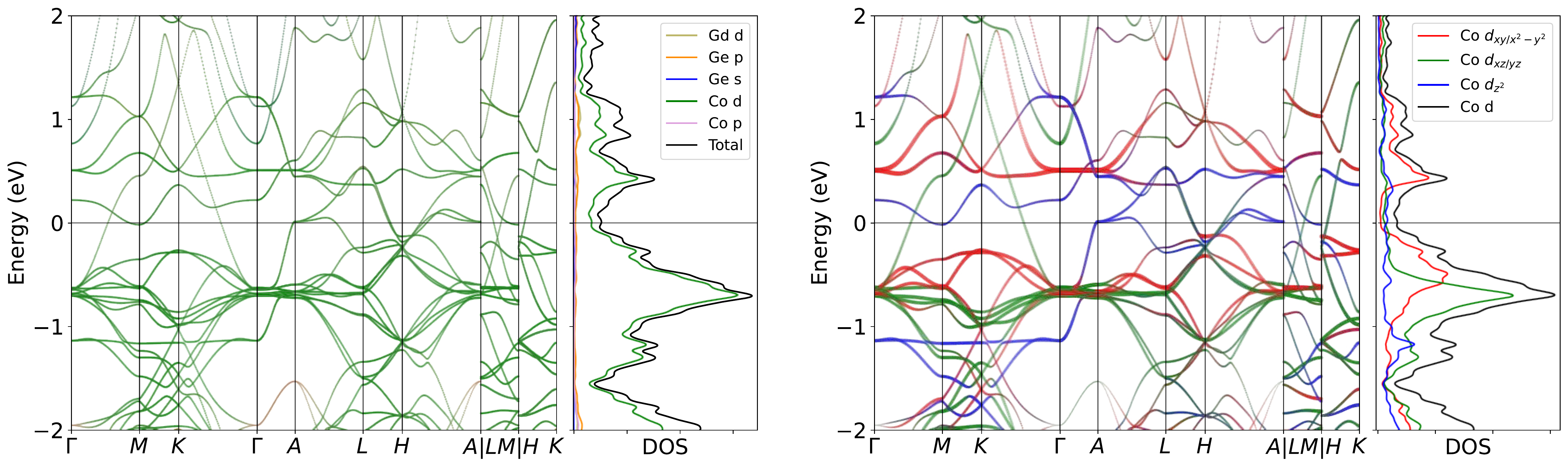}
\caption{Band structure for GdCo$_6$Ge$_6$ without SOC. The projection of Co $3d$ orbitals is also presented. The band structures clearly reveal the dominating of Co $3d$ orbitals  near the Fermi level, as well as the pronounced peak.}
\label{fig:GdCo6Ge6band}
\end{figure}

\begin{figure}[H]
\centering
\subfloat[Relaxed phonon at low-T.]{
\label{fig:GdCo6Ge6_relaxed_Low}
\includegraphics[width=0.45\textwidth]{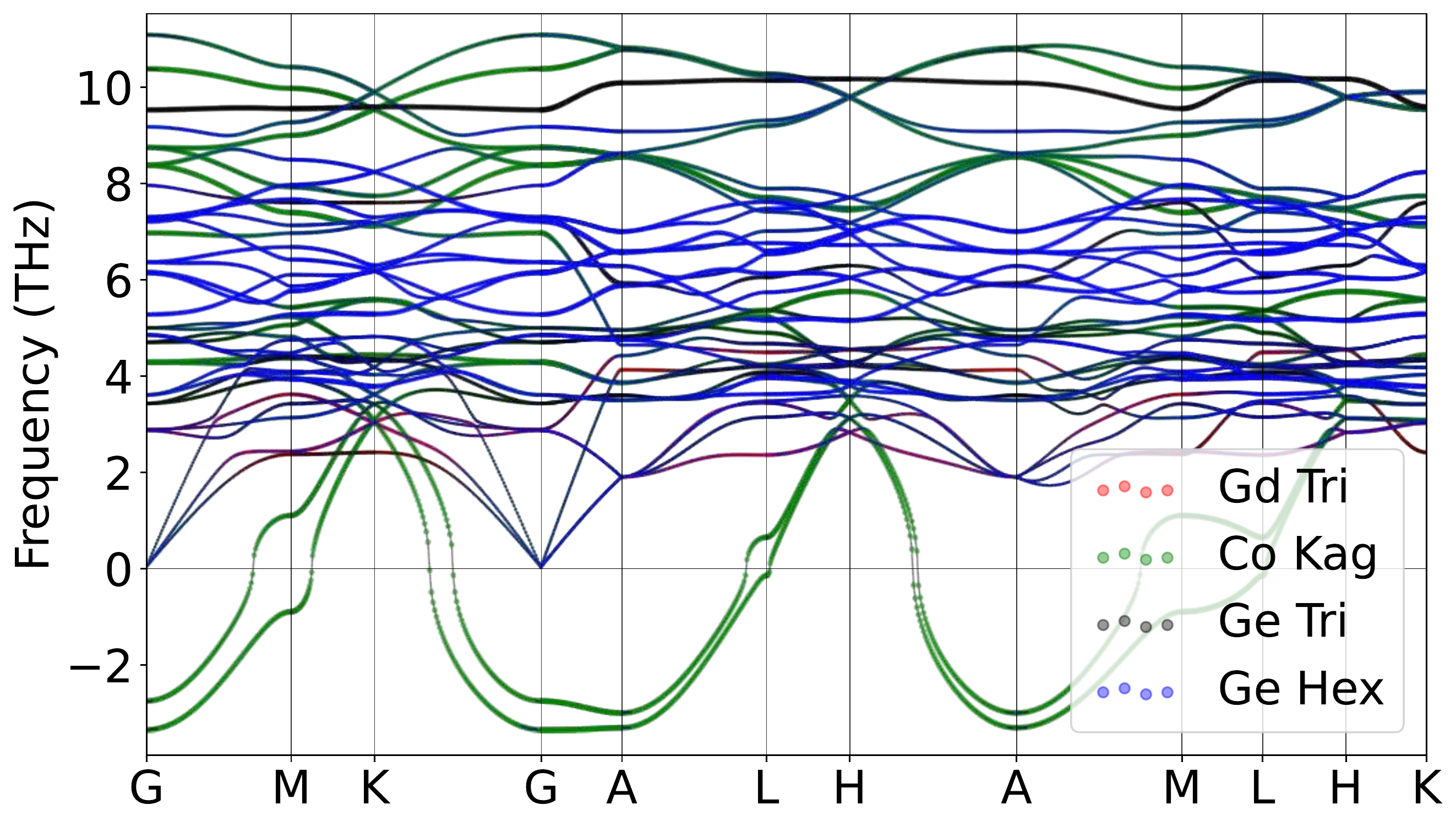}
}
\hspace{5pt}\subfloat[Relaxed phonon at high-T.]{
\label{fig:GdCo6Ge6_relaxed_High}
\includegraphics[width=0.45\textwidth]{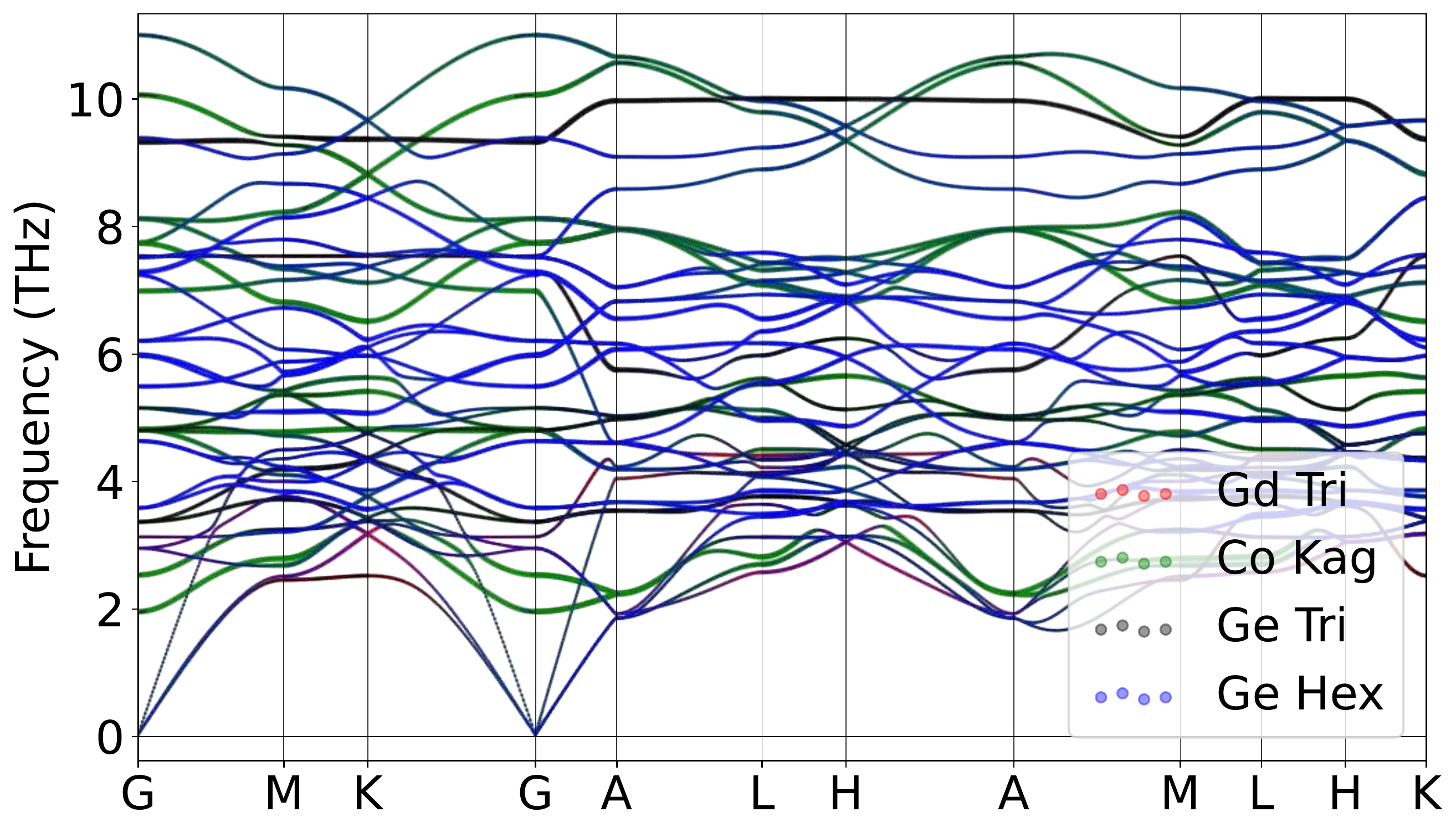}
}
\caption{Atomically projected phonon spectrum for GdCo$_6$Ge$_6$.}
\label{fig:GdCo6Ge6pho}
\end{figure}

\begin{figure}[H]
\centering
\label{fig:GdCo6Ge6_relaxed_Low_xz}
\includegraphics[width=0.85\textwidth]{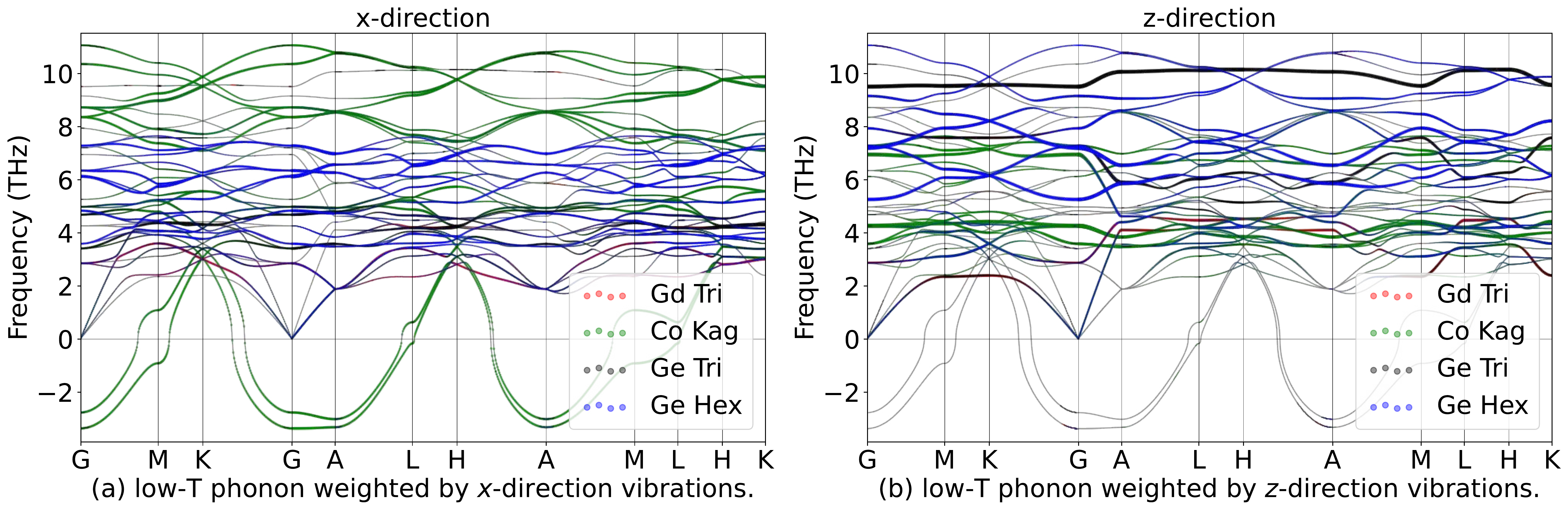}
\hspace{5pt}\label{fig:GdCo6Ge6_relaxed_High_xz}
\includegraphics[width=0.85\textwidth]{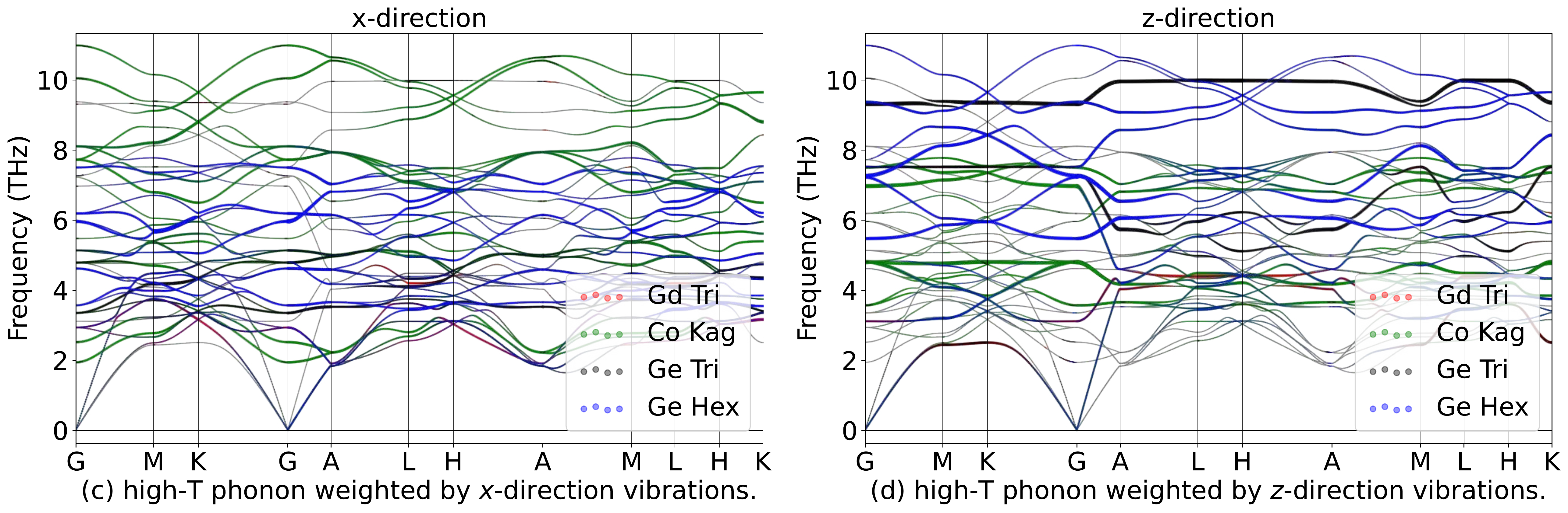}
\caption{Phonon spectrum for GdCo$_6$Ge$_6$in in-plane ($x$) and out-of-plane ($z$) directions.}
\label{fig:GdCo6Ge6phoxyz}
\end{figure}

\begin{table}[htbp]
\global\long\def\arraystretch{1.12}
\captionsetup{width=.95\textwidth}
\caption{IRREPs at high-symmetry points for GdCo$_6$Ge$_6$ phonon spectrum at Low-T.}
\label{tab:191_GdCo6Ge6_Low}
\setlength{\tabcolsep}{1.mm}{
}
\end{table}

\begin{table}[htbp]
\global\long\def\arraystretch{1.12}
\captionsetup{width=.95\textwidth}
\caption{IRREPs at high-symmetry points for GdCo$_6$Ge$_6$ phonon spectrum at High-T.}
\label{tab:191_GdCo6Ge6_High}
\setlength{\tabcolsep}{1.mm}{
%
}
\end{table}
\subsubsection{\label{sms2sec:TbCo6Ge6}\hyperref[smsec:col]{\texorpdfstring{TbCo$_6$Ge$_6$}{}}~{\color{blue}  (High-T stable, Low-T unstable) }}

\begin{table}[htbp]
\global\long\def\arraystretch{1.12}
\captionsetup{width=.85\textwidth}
\caption{Structure parameters of TbCo$_6$Ge$_6$ after relaxation. It contains 419 electrons. }
\label{tab:TbCo6Ge6}
\setlength{\tabcolsep}{4mm}{
%
}
\end{table}

\begin{figure}[htbp]
\centering
\includegraphics[width=0.85\textwidth]{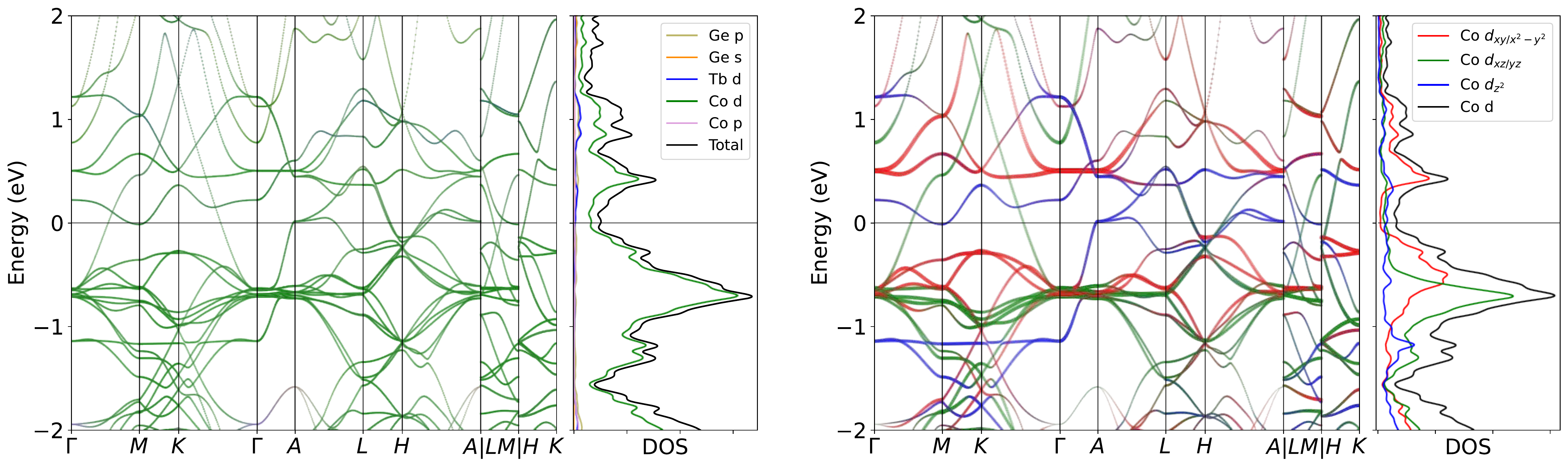}
\caption{Band structure for TbCo$_6$Ge$_6$ without SOC. The projection of Co $3d$ orbitals is also presented. The band structures clearly reveal the dominating of Co $3d$ orbitals  near the Fermi level, as well as the pronounced peak.}
\label{fig:TbCo6Ge6band}
\end{figure}

\begin{figure}[H]
\centering
\subfloat[Relaxed phonon at low-T.]{
\label{fig:TbCo6Ge6_relaxed_Low}
\includegraphics[width=0.45\textwidth]{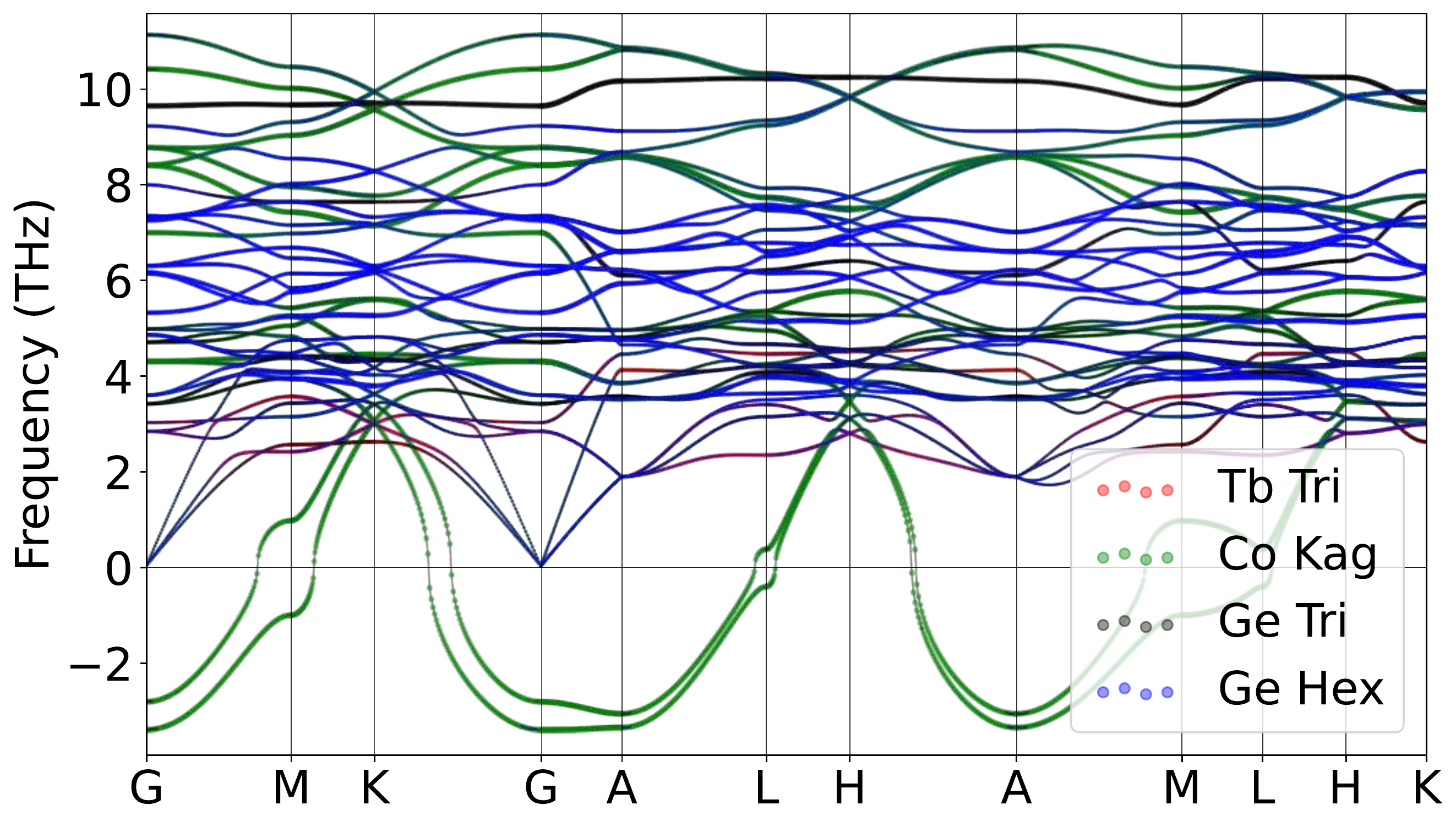}
}
\hspace{5pt}\subfloat[Relaxed phonon at high-T.]{
\label{fig:TbCo6Ge6_relaxed_High}
\includegraphics[width=0.45\textwidth]{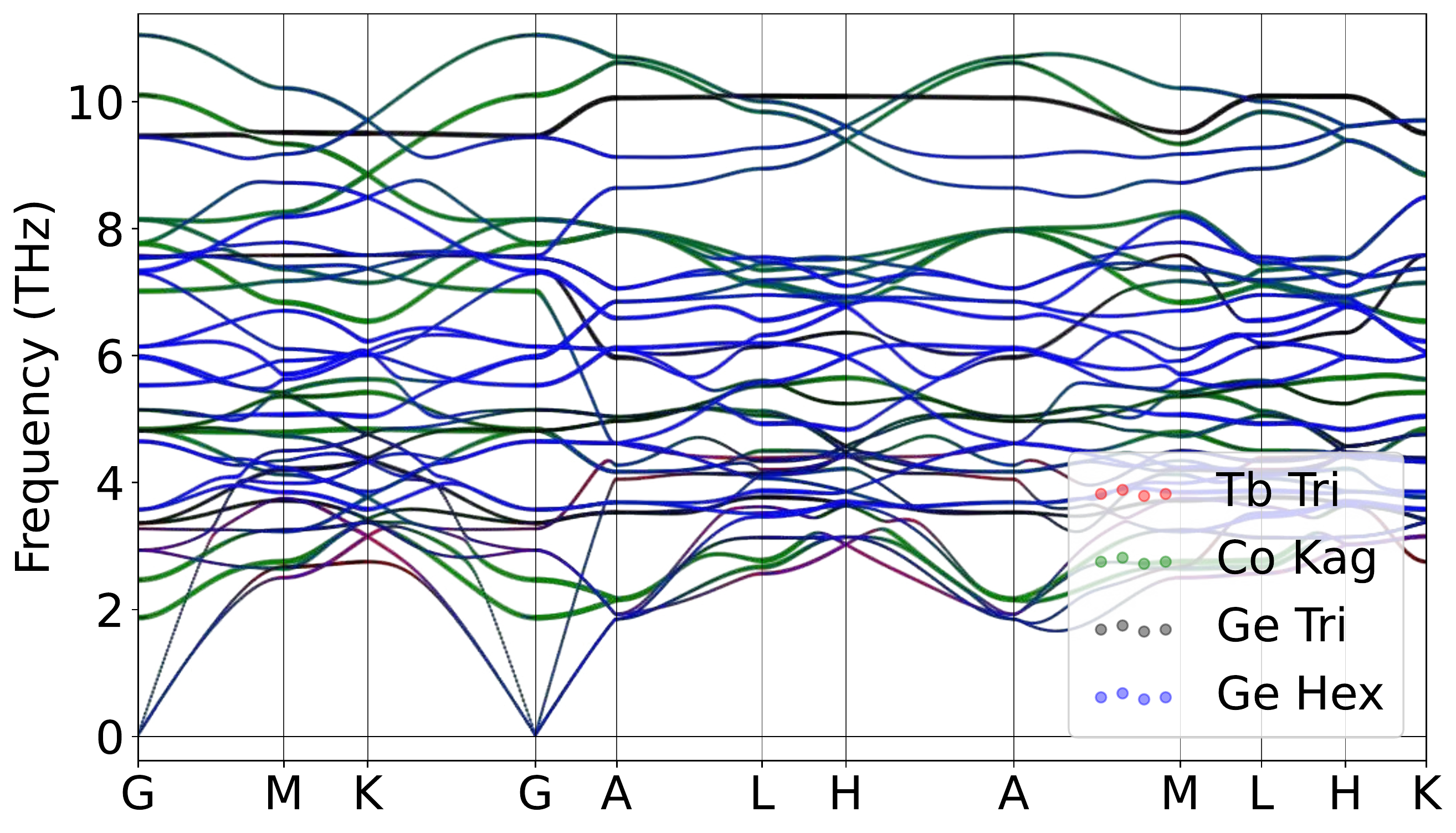}
}
\caption{Atomically projected phonon spectrum for TbCo$_6$Ge$_6$.}
\label{fig:TbCo6Ge6pho}
\end{figure}

\begin{figure}[H]
\centering
\label{fig:TbCo6Ge6_relaxed_Low_xz}
\includegraphics[width=0.85\textwidth]{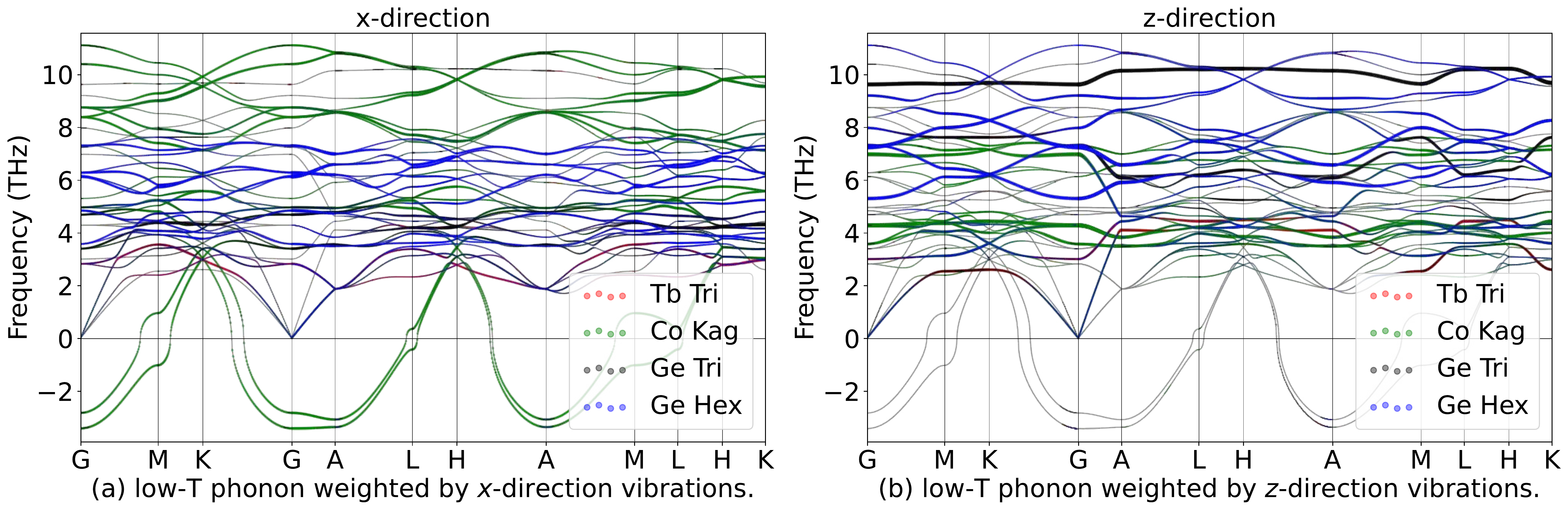}
\hspace{5pt}\label{fig:TbCo6Ge6_relaxed_High_xz}
\includegraphics[width=0.85\textwidth]{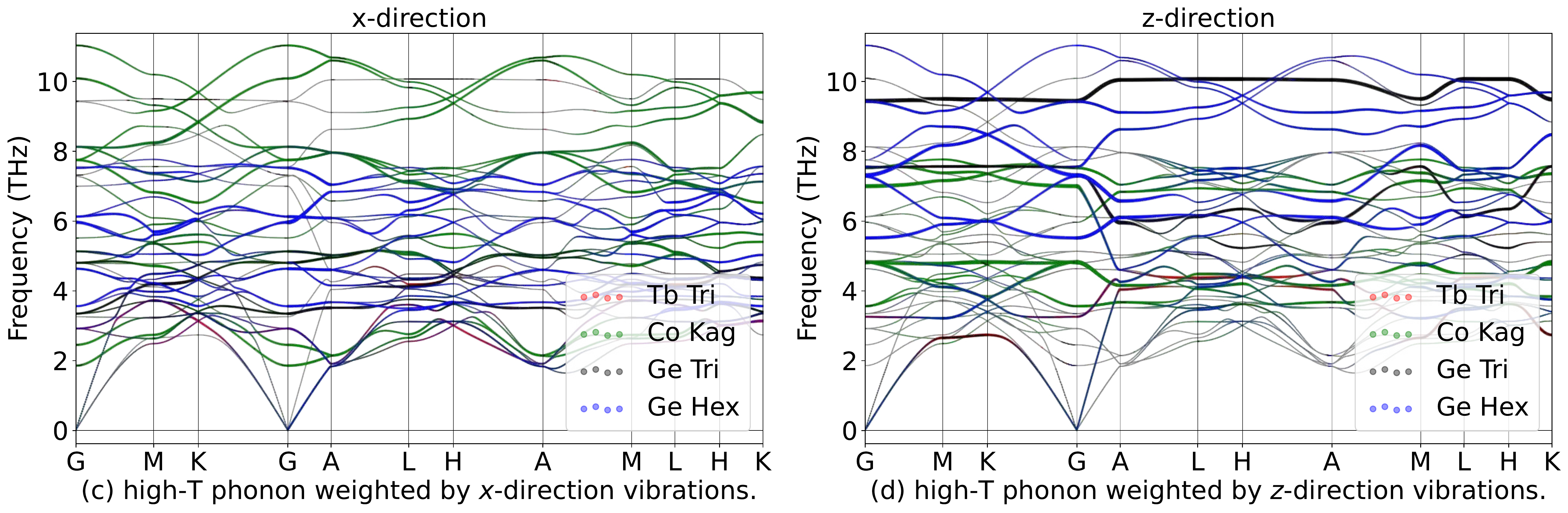}
\caption{Phonon spectrum for TbCo$_6$Ge$_6$in in-plane ($x$) and out-of-plane ($z$) directions.}
\label{fig:TbCo6Ge6phoxyz}
\end{figure}

\begin{table}[htbp]
\global\long\def\arraystretch{1.12}
\captionsetup{width=.95\textwidth}
\caption{IRREPs at high-symmetry points for TbCo$_6$Ge$_6$ phonon spectrum at Low-T.}
\label{tab:191_TbCo6Ge6_Low}
\setlength{\tabcolsep}{1.mm}{
}
\end{table}

\begin{table}[htbp]
\global\long\def\arraystretch{1.12}
\captionsetup{width=.95\textwidth}
\caption{IRREPs at high-symmetry points for TbCo$_6$Ge$_6$ phonon spectrum at High-T.}
\label{tab:191_TbCo6Ge6_High}
\setlength{\tabcolsep}{1.mm}{
%
}
\end{table}
\subsubsection{\label{sms2sec:DyCo6Ge6}\hyperref[smsec:col]{\texorpdfstring{DyCo$_6$Ge$_6$}{}}~{\color{blue}  (High-T stable, Low-T unstable) }}

\begin{table}[htbp]
\global\long\def\arraystretch{1.12}
\captionsetup{width=.85\textwidth}
\caption{Structure parameters of DyCo$_6$Ge$_6$ after relaxation. It contains 420 electrons. }
\label{tab:DyCo6Ge6}
\setlength{\tabcolsep}{4mm}{
%
}
\end{table}

\begin{figure}[htbp]
\centering
\includegraphics[width=0.85\textwidth]{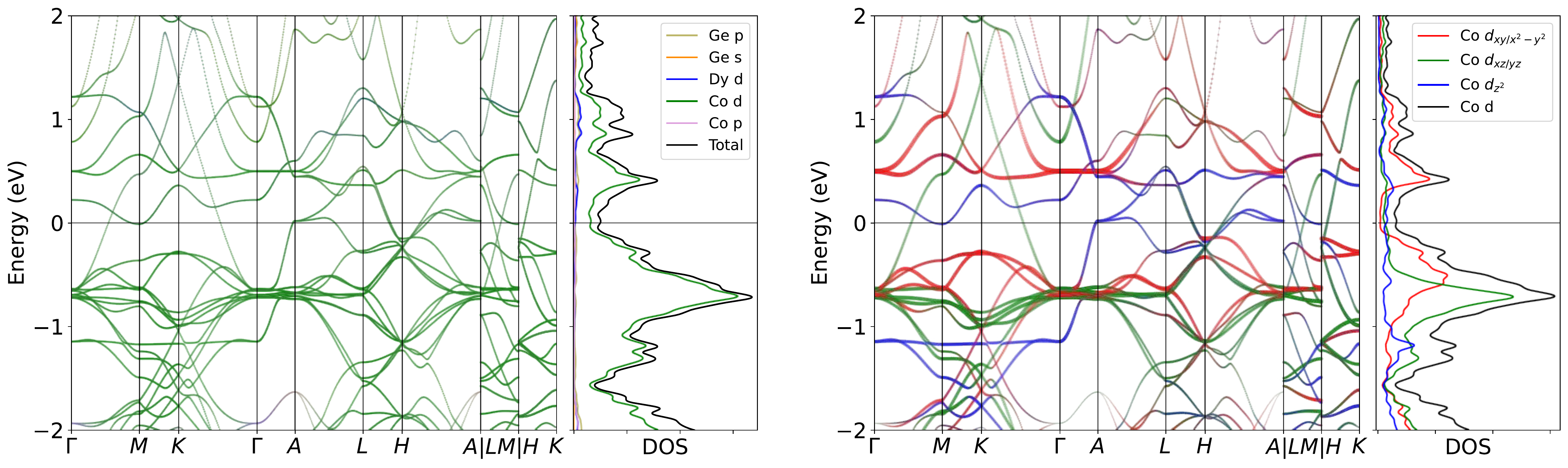}
\caption{Band structure for DyCo$_6$Ge$_6$ without SOC. The projection of Co $3d$ orbitals is also presented. The band structures clearly reveal the dominating of Co $3d$ orbitals  near the Fermi level, as well as the pronounced peak.}
\label{fig:DyCo6Ge6band}
\end{figure}

\begin{figure}[H]
\centering
\subfloat[Relaxed phonon at low-T.]{
\label{fig:DyCo6Ge6_relaxed_Low}
\includegraphics[width=0.45\textwidth]{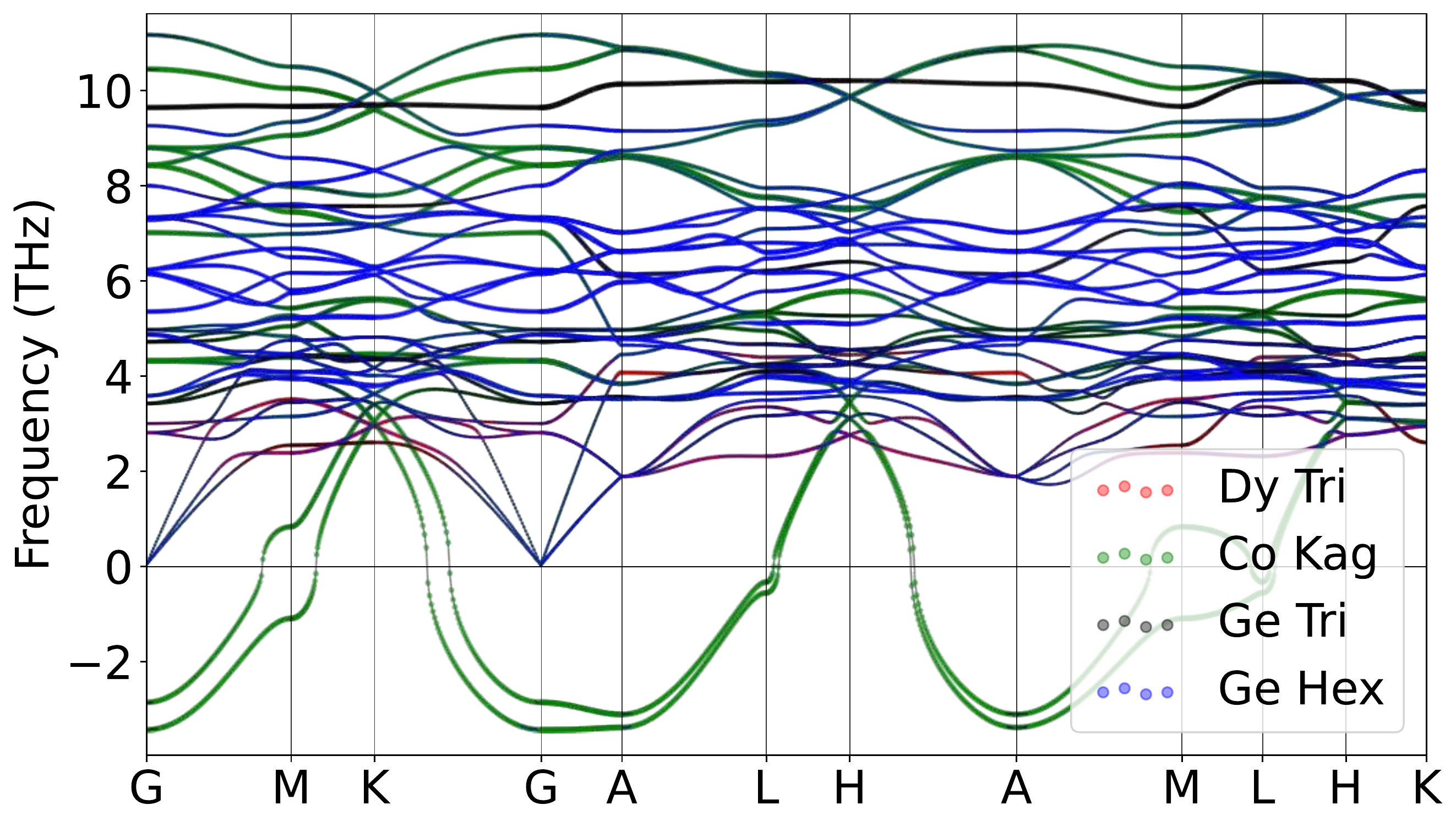}
}
\hspace{5pt}\subfloat[Relaxed phonon at high-T.]{
\label{fig:DyCo6Ge6_relaxed_High}
\includegraphics[width=0.45\textwidth]{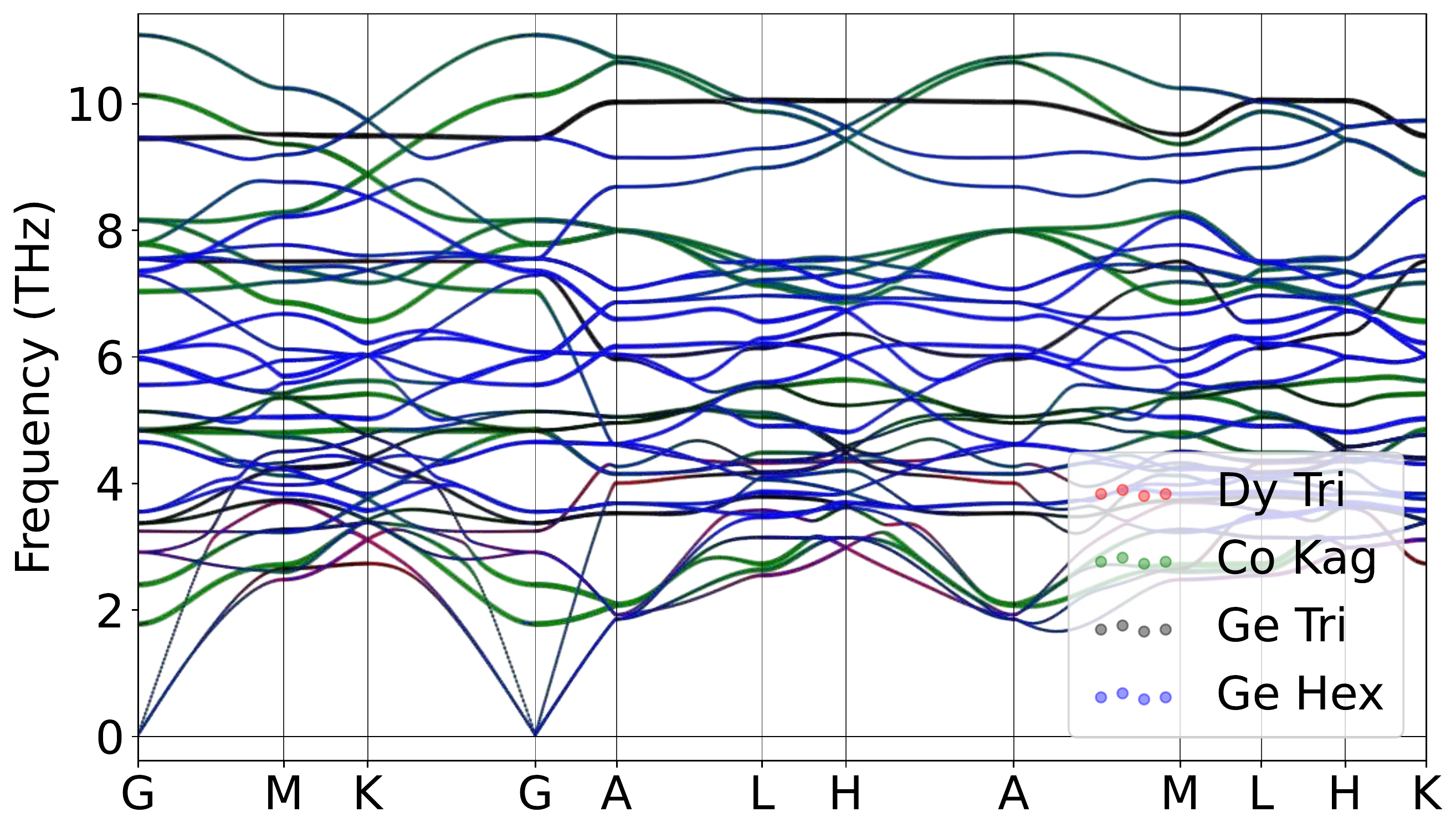}
}
\caption{Atomically projected phonon spectrum for DyCo$_6$Ge$_6$.}
\label{fig:DyCo6Ge6pho}
\end{figure}

\begin{figure}[H]
\centering
\label{fig:DyCo6Ge6_relaxed_Low_xz}
\includegraphics[width=0.85\textwidth]{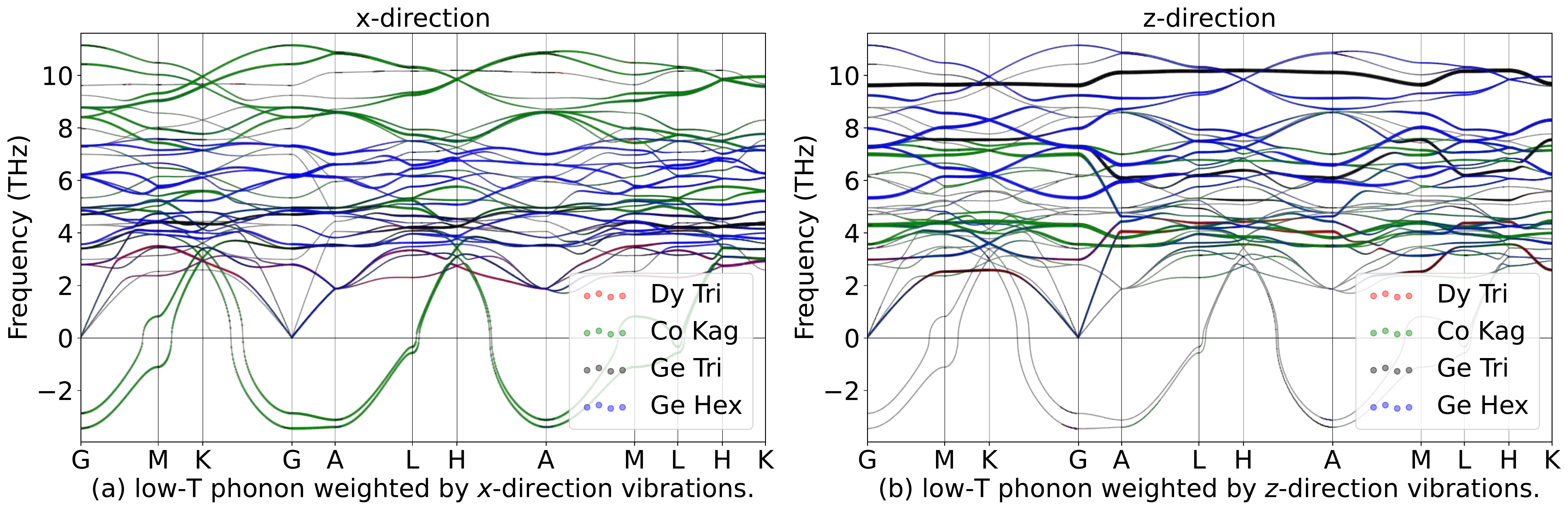}
\hspace{5pt}\label{fig:DyCo6Ge6_relaxed_High_xz}
\includegraphics[width=0.85\textwidth]{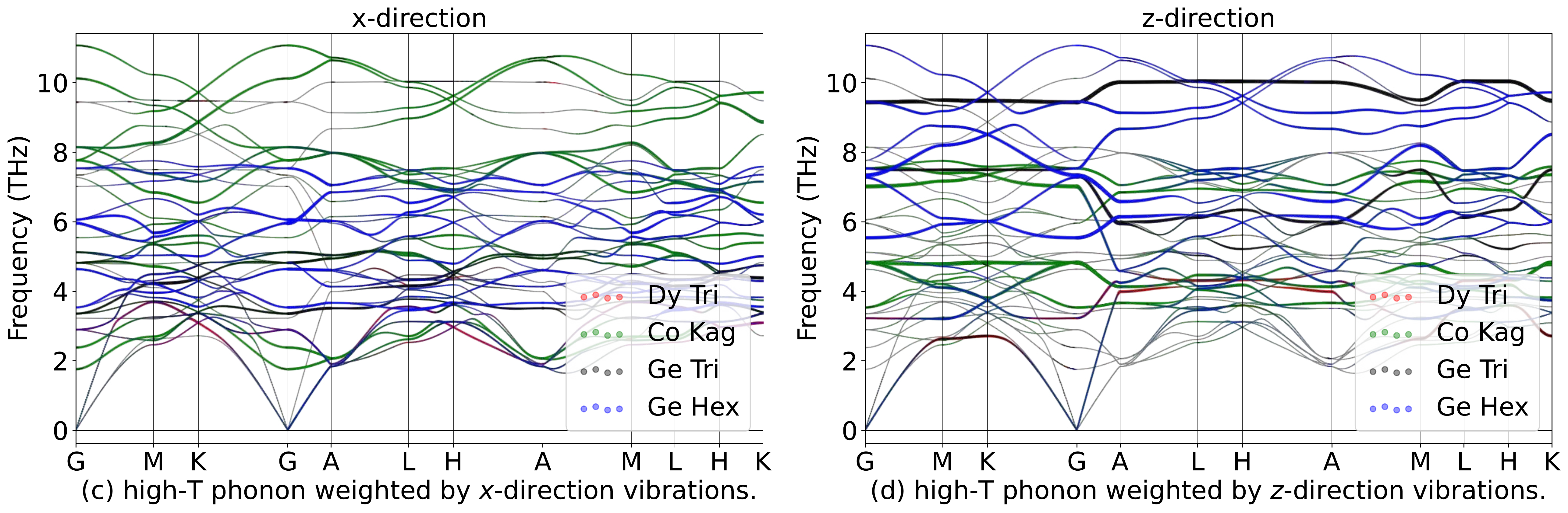}
\caption{Phonon spectrum for DyCo$_6$Ge$_6$in in-plane ($x$) and out-of-plane ($z$) directions.}
\label{fig:DyCo6Ge6phoxyz}
\end{figure}

\begin{table}[htbp]
\global\long\def\arraystretch{1.12}
\captionsetup{width=.95\textwidth}
\caption{IRREPs at high-symmetry points for DyCo$_6$Ge$_6$ phonon spectrum at Low-T.}
\label{tab:191_DyCo6Ge6_Low}
\setlength{\tabcolsep}{1.mm}{
}
\end{table}

\begin{table}[htbp]
\global\long\def\arraystretch{1.12}
\captionsetup{width=.95\textwidth}
\caption{IRREPs at high-symmetry points for DyCo$_6$Ge$_6$ phonon spectrum at High-T.}
\label{tab:191_DyCo6Ge6_High}
\setlength{\tabcolsep}{1.mm}{
%
}
\end{table}
\subsubsection{\label{sms2sec:HoCo6Ge6}\hyperref[smsec:col]{\texorpdfstring{HoCo$_6$Ge$_6$}{}}~{\color{blue}  (High-T stable, Low-T unstable) }}

\begin{table}[htbp]
\global\long\def\arraystretch{1.12}
\captionsetup{width=.85\textwidth}
\caption{Structure parameters of HoCo$_6$Ge$_6$ after relaxation. It contains 421 electrons. }
\label{tab:HoCo6Ge6}
\setlength{\tabcolsep}{4mm}{
%
}
\end{table}

\begin{figure}[htbp]
\centering
\includegraphics[width=0.85\textwidth]{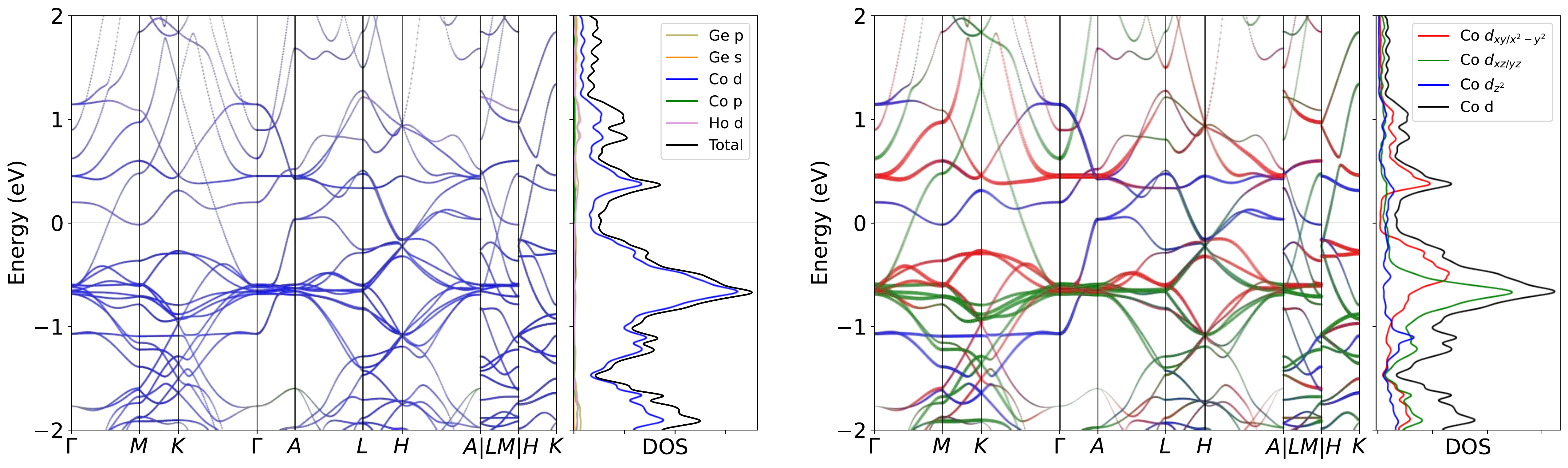}
\caption{Band structure for HoCo$_6$Ge$_6$ without SOC. The projection of Co $3d$ orbitals is also presented. The band structures clearly reveal the dominating of Co $3d$ orbitals  near the Fermi level, as well as the pronounced peak.}
\label{fig:HoCo6Ge6band}
\end{figure}

\begin{figure}[H]
\centering
\subfloat[Relaxed phonon at low-T.]{
\label{fig:HoCo6Ge6_relaxed_Low}
\includegraphics[width=0.45\textwidth]{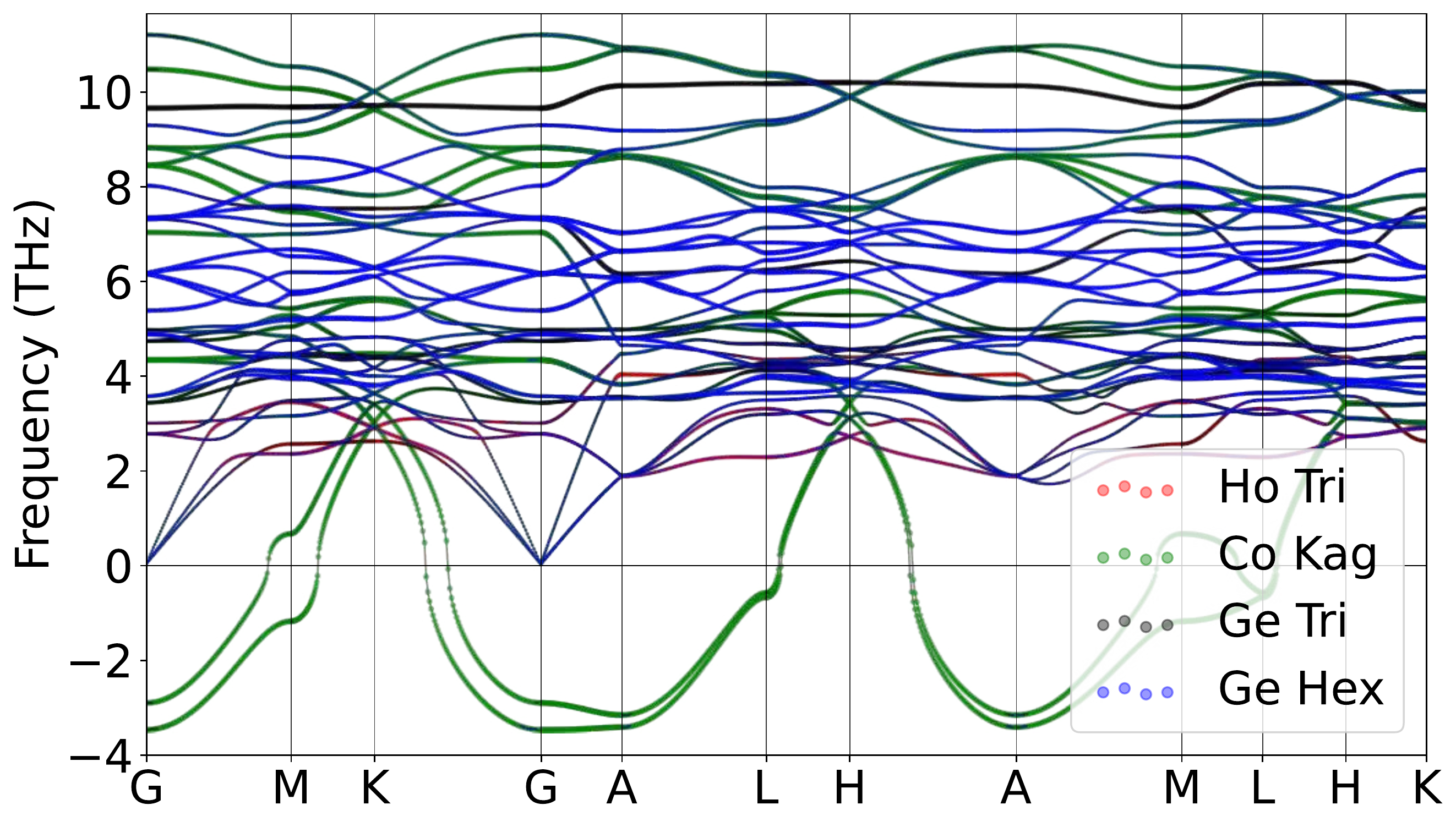}
}
\hspace{5pt}\subfloat[Relaxed phonon at high-T.]{
\label{fig:HoCo6Ge6_relaxed_High}
\includegraphics[width=0.45\textwidth]{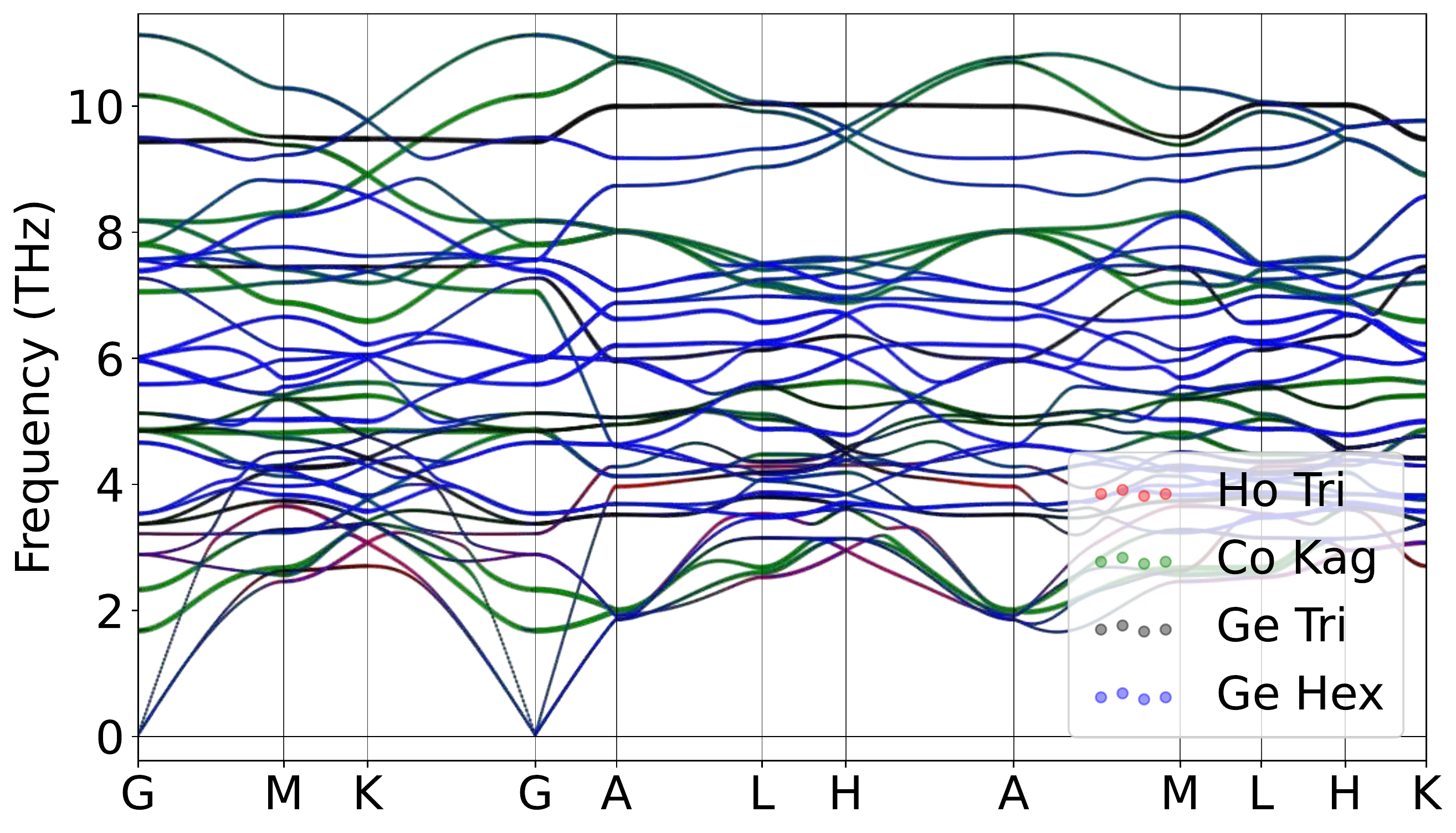}
}
\caption{Atomically projected phonon spectrum for HoCo$_6$Ge$_6$.}
\label{fig:HoCo6Ge6pho}
\end{figure}

\begin{figure}[H]
\centering
\label{fig:HoCo6Ge6_relaxed_Low_xz}
\includegraphics[width=0.85\textwidth]{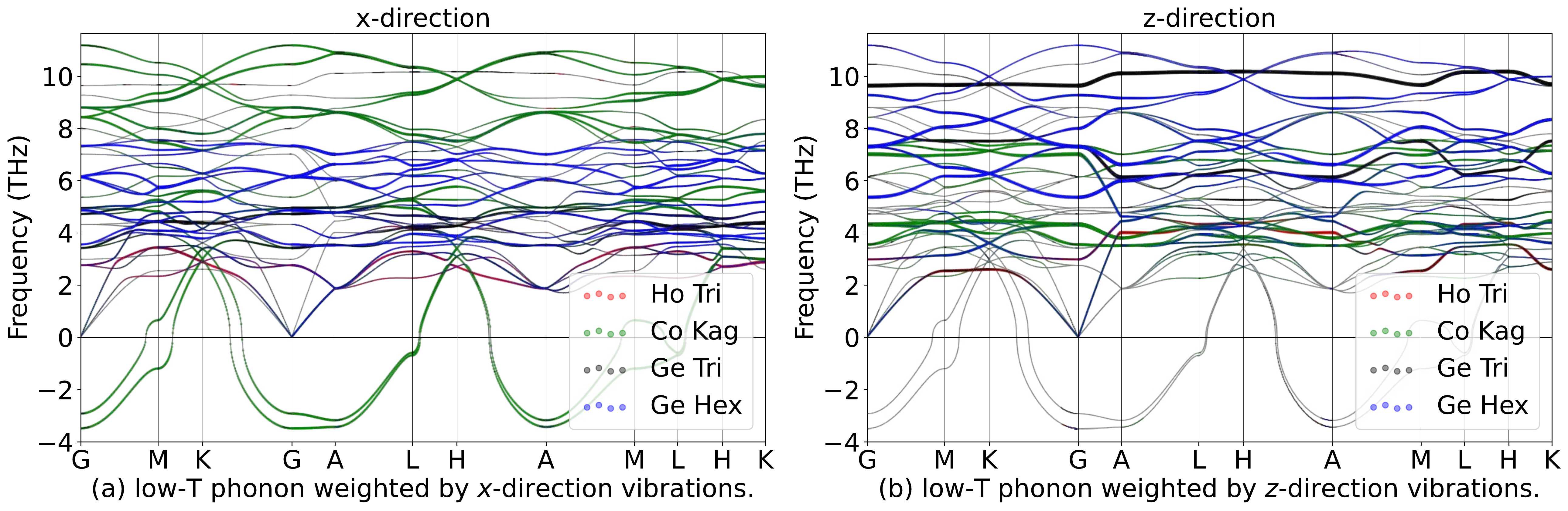}
\hspace{5pt}\label{fig:HoCo6Ge6_relaxed_High_xz}
\includegraphics[width=0.85\textwidth]{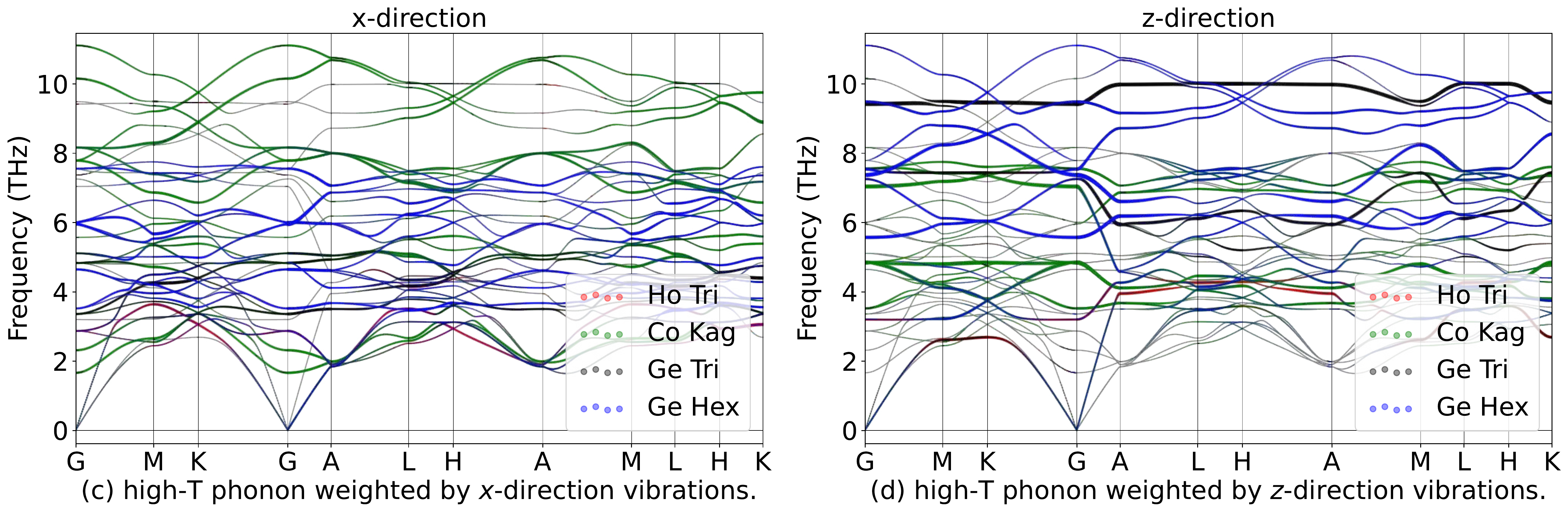}
\caption{Phonon spectrum for HoCo$_6$Ge$_6$in in-plane ($x$) and out-of-plane ($z$) directions.}
\label{fig:HoCo6Ge6phoxyz}
\end{figure}

\begin{table}[htbp]
\global\long\def\arraystretch{1.12}
\captionsetup{width=.95\textwidth}
\caption{IRREPs at high-symmetry points for HoCo$_6$Ge$_6$ phonon spectrum at Low-T.}
\label{tab:191_HoCo6Ge6_Low}
\setlength{\tabcolsep}{1.mm}{
}
\end{table}

\begin{table}[htbp]
\global\long\def\arraystretch{1.12}
\captionsetup{width=.95\textwidth}
\caption{IRREPs at high-symmetry points for HoCo$_6$Ge$_6$ phonon spectrum at High-T.}
\label{tab:191_HoCo6Ge6_High}
\setlength{\tabcolsep}{1.mm}{
%
}
\end{table}
\subsubsection{\label{sms2sec:ErCo6Ge6}\hyperref[smsec:col]{\texorpdfstring{ErCo$_6$Ge$_6$}{}}~{\color{blue}  (High-T stable, Low-T unstable) }}

\begin{table}[htbp]
\global\long\def\arraystretch{1.12}
\captionsetup{width=.85\textwidth}
\caption{Structure parameters of ErCo$_6$Ge$_6$ after relaxation. It contains 422 electrons. }
\label{tab:ErCo6Ge6}
\setlength{\tabcolsep}{4mm}{
%
}
\end{table}

\begin{figure}[htbp]
\centering
\includegraphics[width=0.85\textwidth]{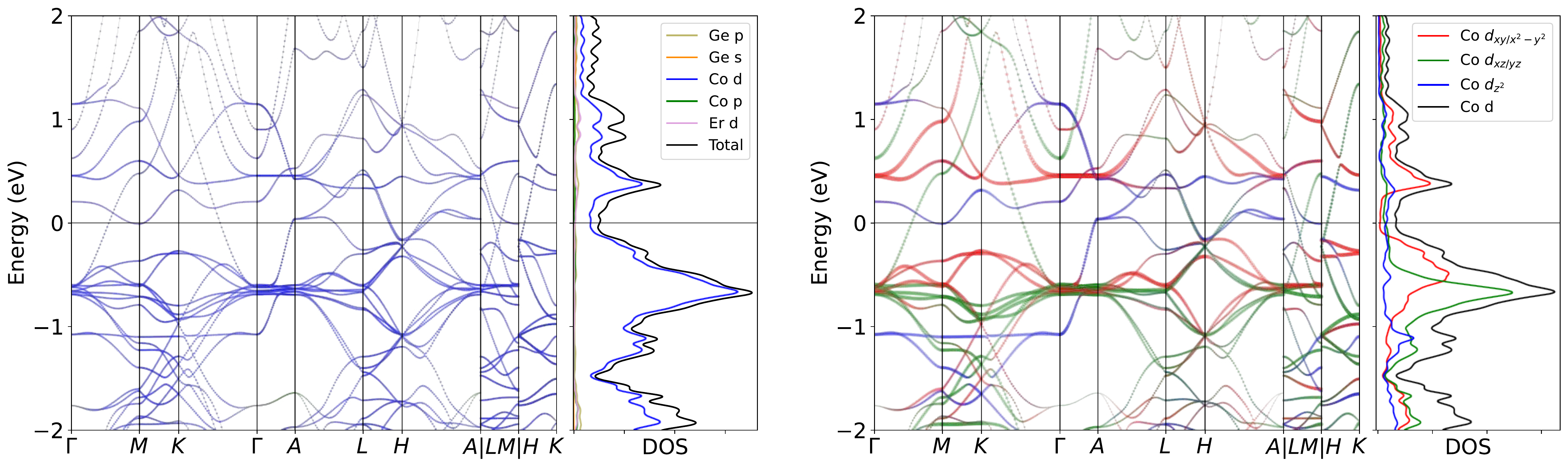}
\caption{Band structure for ErCo$_6$Ge$_6$ without SOC. The projection of Co $3d$ orbitals is also presented. The band structures clearly reveal the dominating of Co $3d$ orbitals  near the Fermi level, as well as the pronounced peak.}
\label{fig:ErCo6Ge6band}
\end{figure}

\begin{figure}[H]
\centering
\subfloat[Relaxed phonon at low-T.]{
\label{fig:ErCo6Ge6_relaxed_Low}
\includegraphics[width=0.45\textwidth]{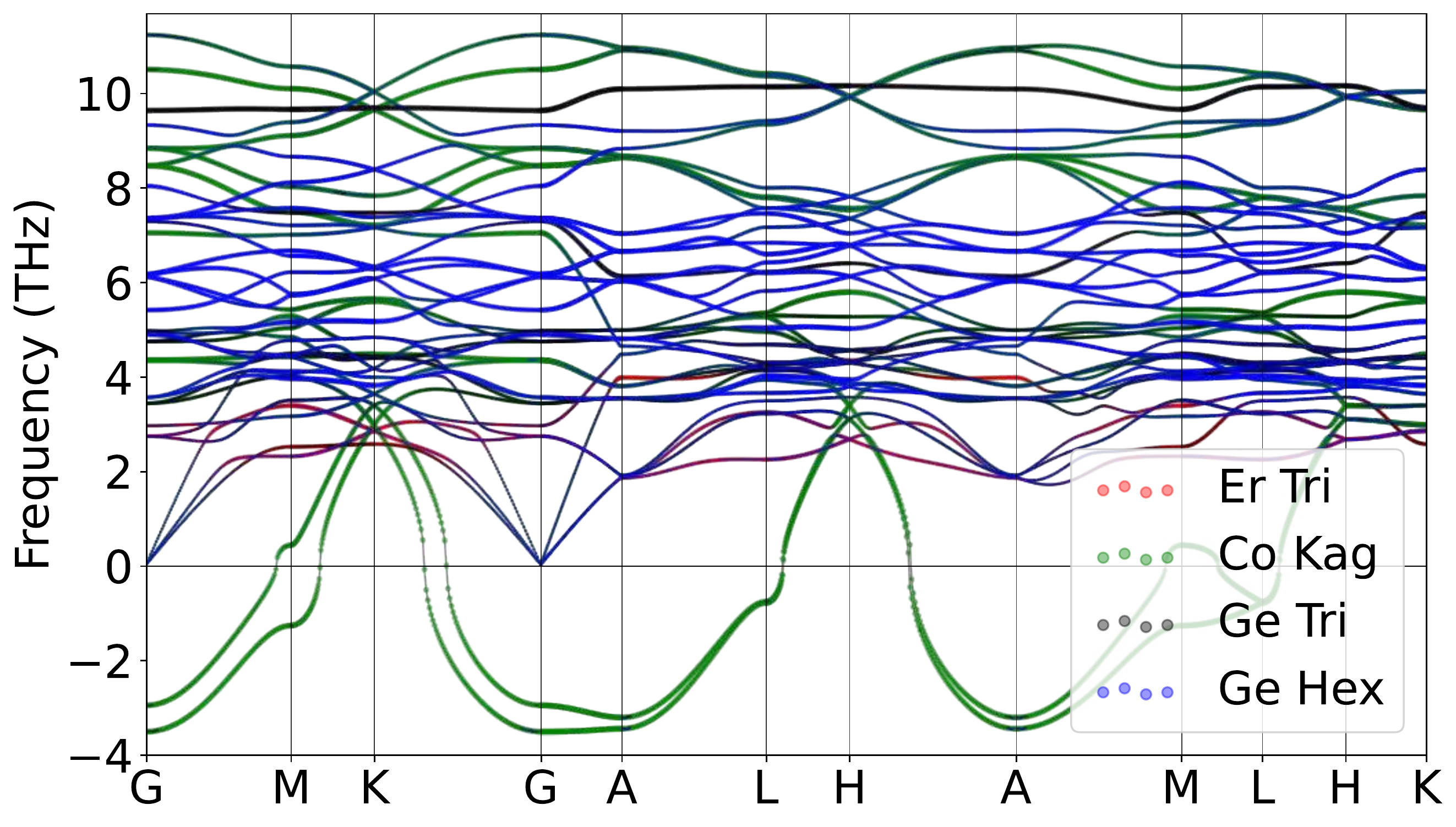}
}
\hspace{5pt}\subfloat[Relaxed phonon at high-T.]{
\label{fig:ErCo6Ge6_relaxed_High}
\includegraphics[width=0.45\textwidth]{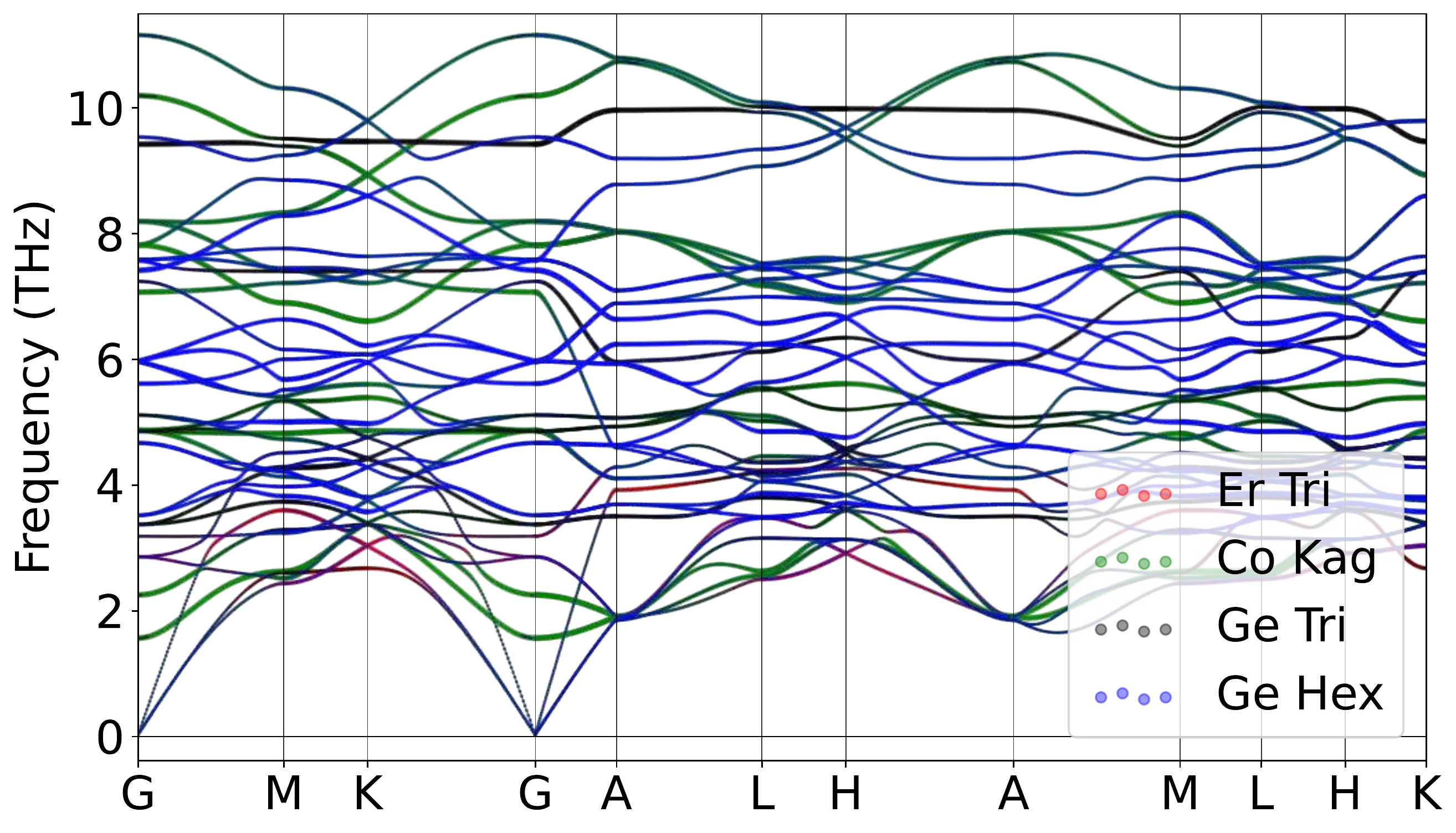}
}
\caption{Atomically projected phonon spectrum for ErCo$_6$Ge$_6$.}
\label{fig:ErCo6Ge6pho}
\end{figure}

\begin{figure}[H]
\centering
\label{fig:ErCo6Ge6_relaxed_Low_xz}
\includegraphics[width=0.85\textwidth]{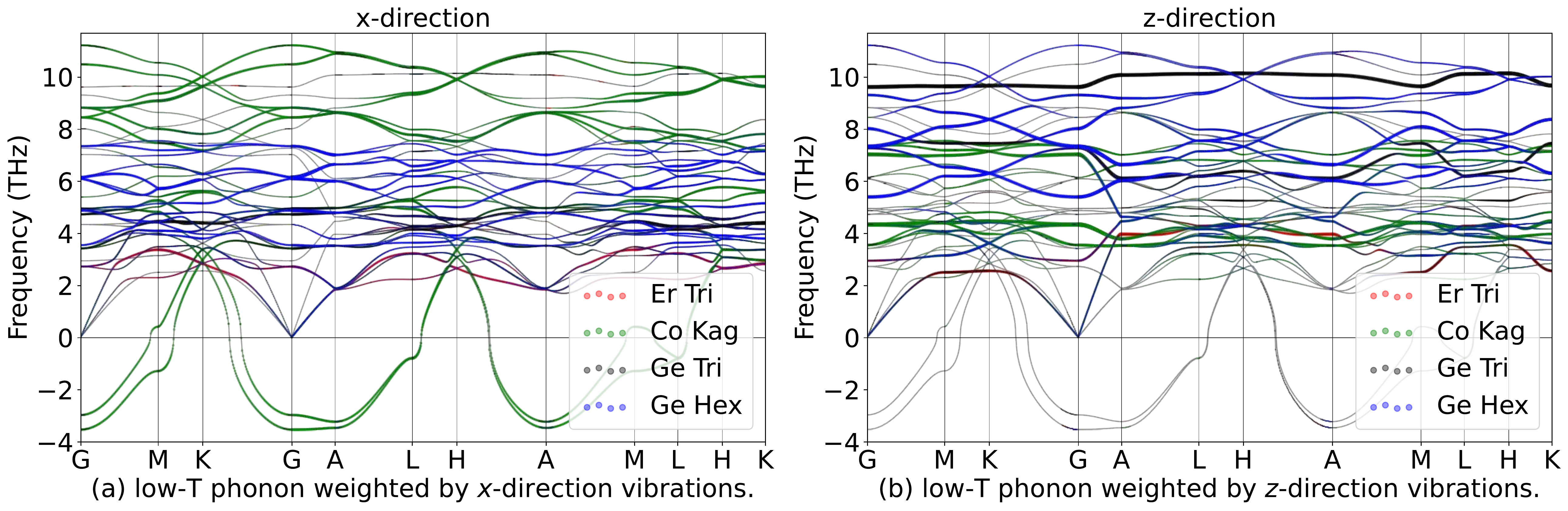}
\hspace{5pt}\label{fig:ErCo6Ge6_relaxed_High_xz}
\includegraphics[width=0.85\textwidth]{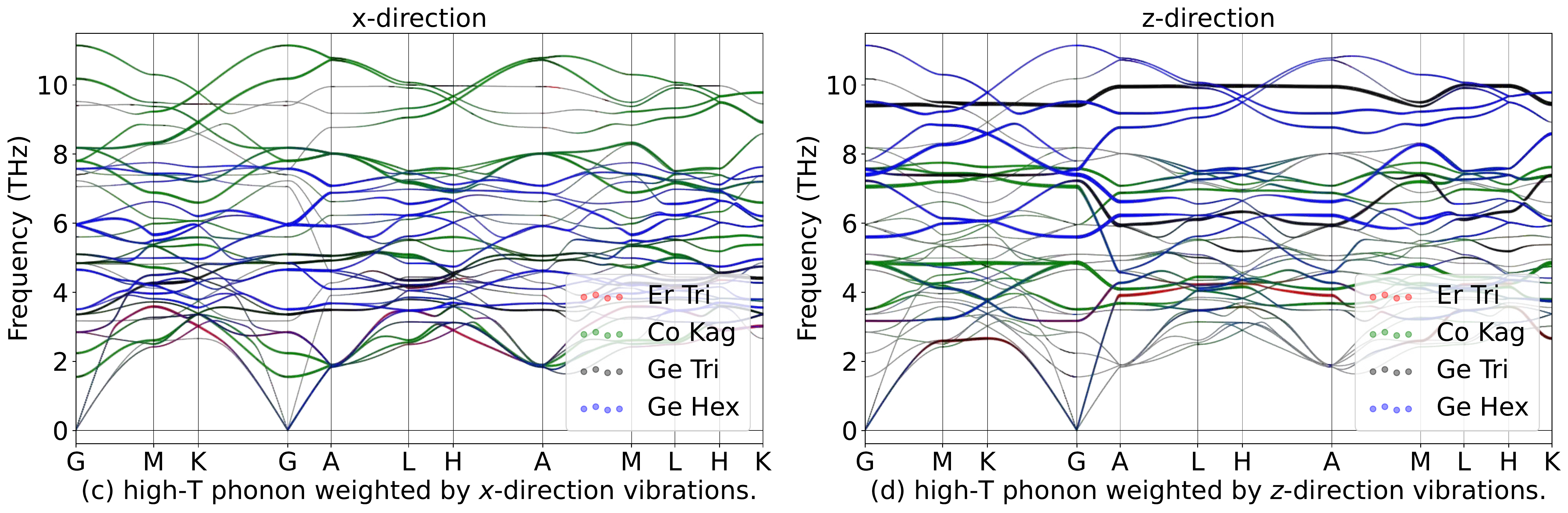}
\caption{Phonon spectrum for ErCo$_6$Ge$_6$in in-plane ($x$) and out-of-plane ($z$) directions.}
\label{fig:ErCo6Ge6phoxyz}
\end{figure}

\begin{table}[htbp]
\global\long\def\arraystretch{1.12}
\captionsetup{width=.95\textwidth}
\caption{IRREPs at high-symmetry points for ErCo$_6$Ge$_6$ phonon spectrum at Low-T.}
\label{tab:191_ErCo6Ge6_Low}
\setlength{\tabcolsep}{1.mm}{
}
\end{table}

\begin{table}[htbp]
\global\long\def\arraystretch{1.12}
\captionsetup{width=.95\textwidth}
\caption{IRREPs at high-symmetry points for ErCo$_6$Ge$_6$ phonon spectrum at High-T.}
\label{tab:191_ErCo6Ge6_High}
\setlength{\tabcolsep}{1.mm}{
%
}
\end{table}
\subsubsection{\label{sms2sec:TmCo6Ge6}\hyperref[smsec:col]{\texorpdfstring{TmCo$_6$Ge$_6$}{}}~{\color{blue}  (High-T stable, Low-T unstable) }}

\begin{table}[htbp]
\global\long\def\arraystretch{1.12}
\captionsetup{width=.85\textwidth}
\caption{Structure parameters of TmCo$_6$Ge$_6$ after relaxation. It contains 423 electrons. }
\label{tab:TmCo6Ge6}
\setlength{\tabcolsep}{4mm}{
%
}
\end{table}

\begin{figure}[htbp]
\centering
\includegraphics[width=0.85\textwidth]{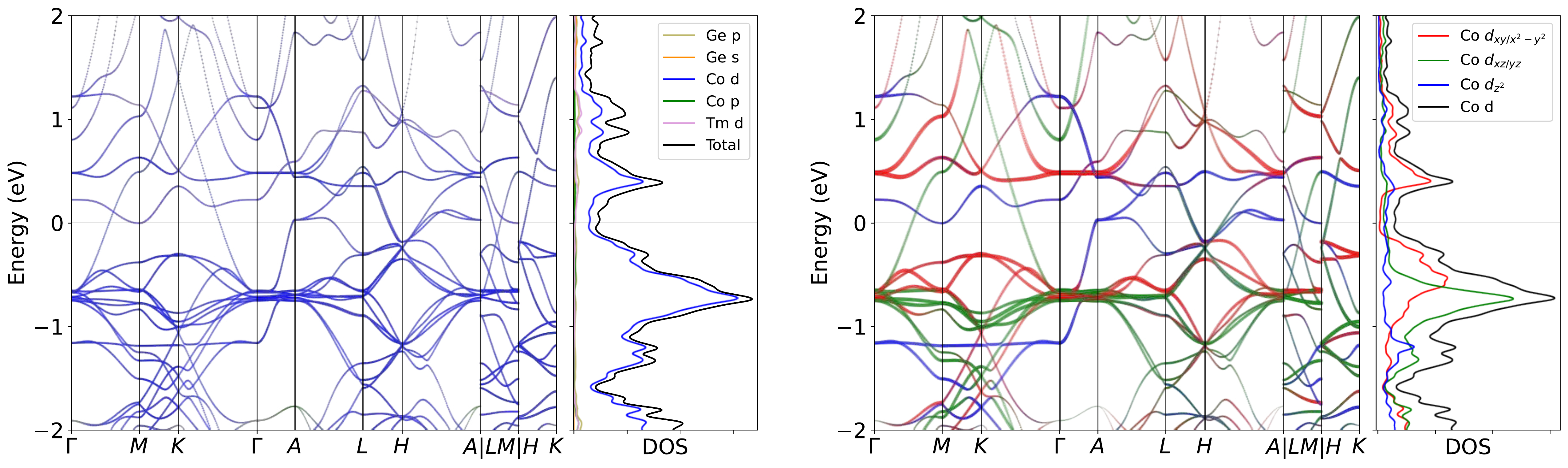}
\caption{Band structure for TmCo$_6$Ge$_6$ without SOC. The projection of Co $3d$ orbitals is also presented. The band structures clearly reveal the dominating of Co $3d$ orbitals  near the Fermi level, as well as the pronounced peak.}
\label{fig:TmCo6Ge6band}
\end{figure}

\begin{figure}[H]
\centering
\subfloat[Relaxed phonon at low-T.]{
\label{fig:TmCo6Ge6_relaxed_Low}
\includegraphics[width=0.45\textwidth]{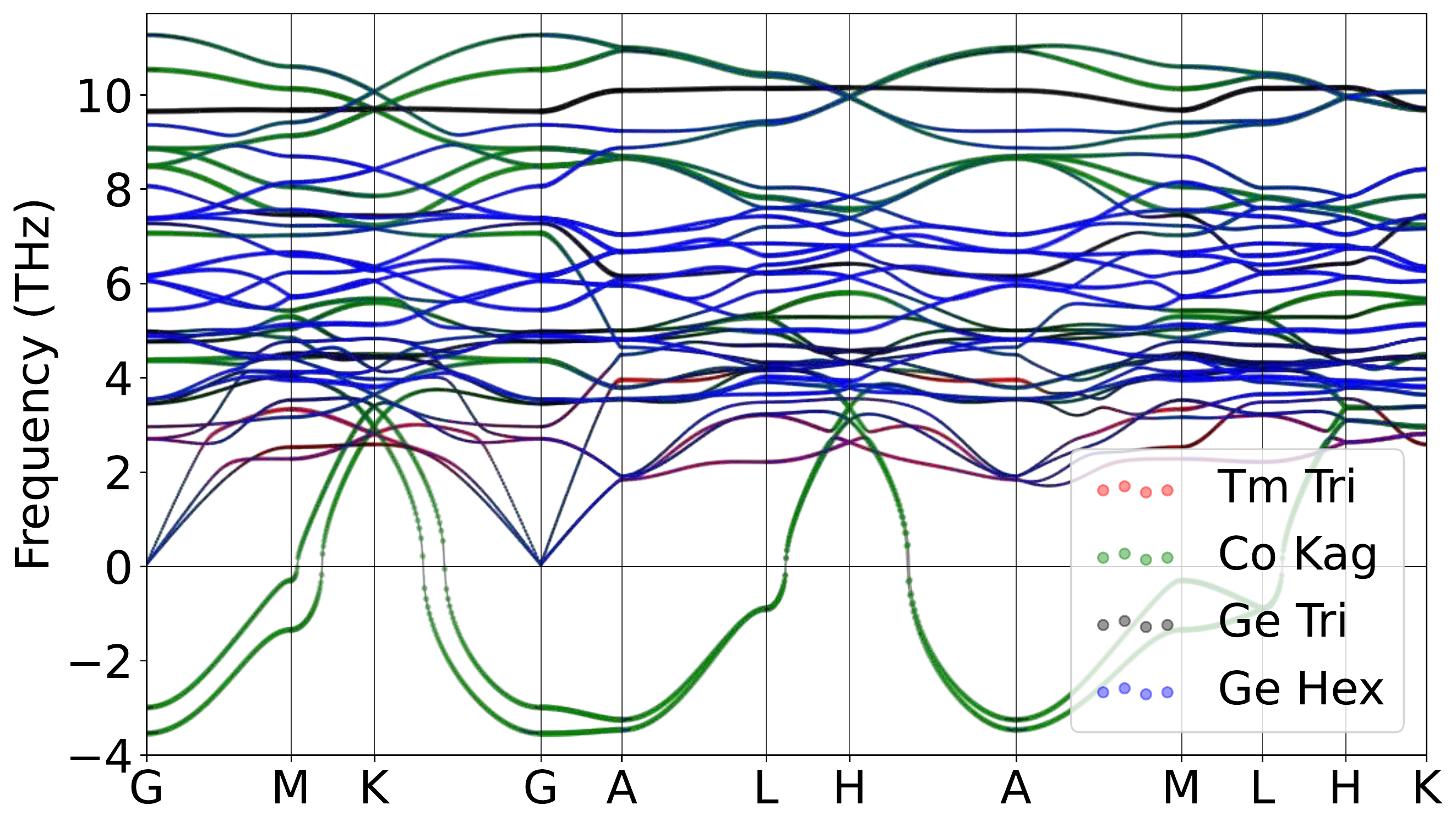}
}
\hspace{5pt}\subfloat[Relaxed phonon at high-T.]{
\label{fig:TmCo6Ge6_relaxed_High}
\includegraphics[width=0.45\textwidth]{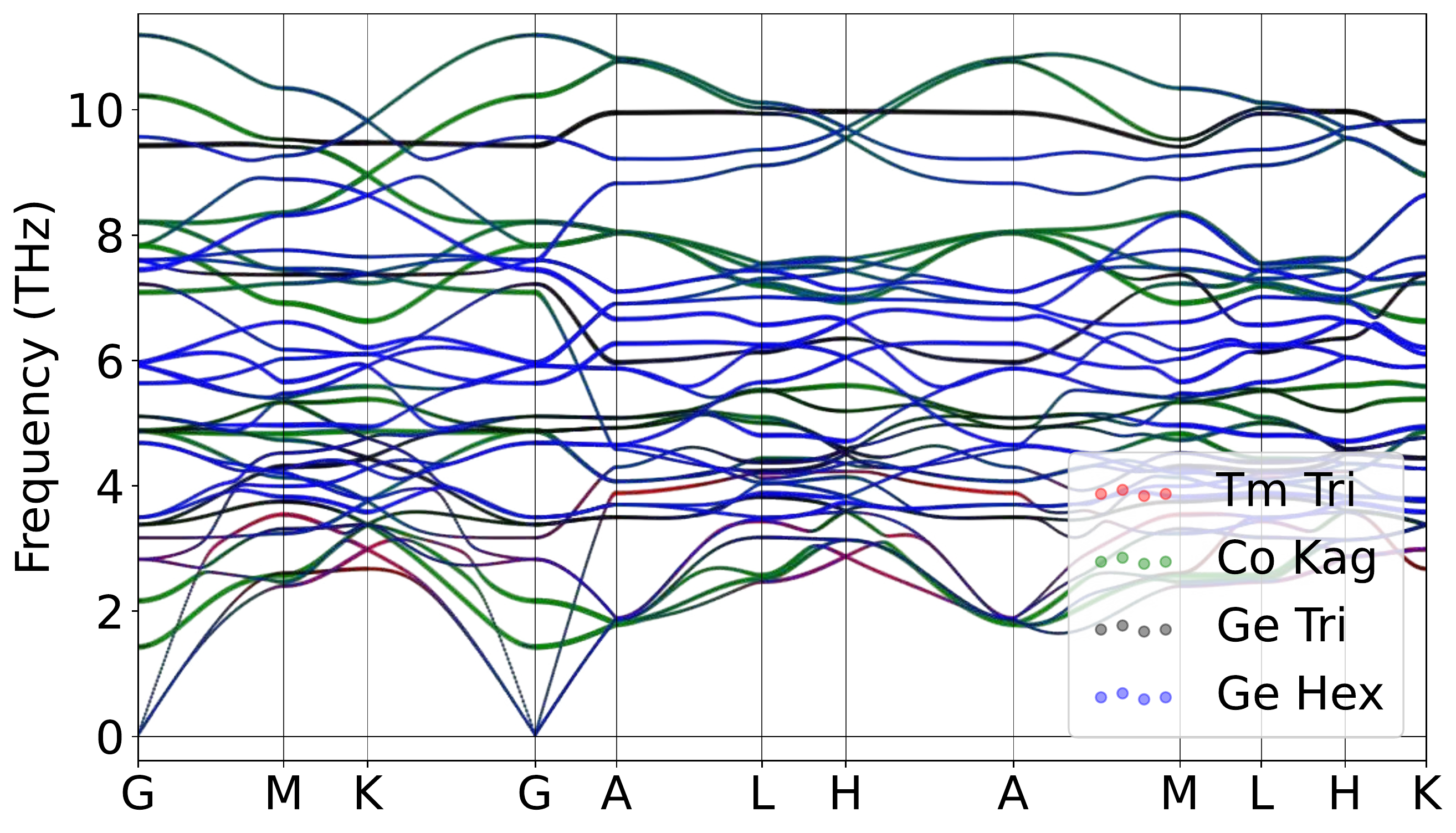}
}
\caption{Atomically projected phonon spectrum for TmCo$_6$Ge$_6$.}
\label{fig:TmCo6Ge6pho}
\end{figure}

\begin{figure}[H]
\centering
\label{fig:TmCo6Ge6_relaxed_Low_xz}
\includegraphics[width=0.85\textwidth]{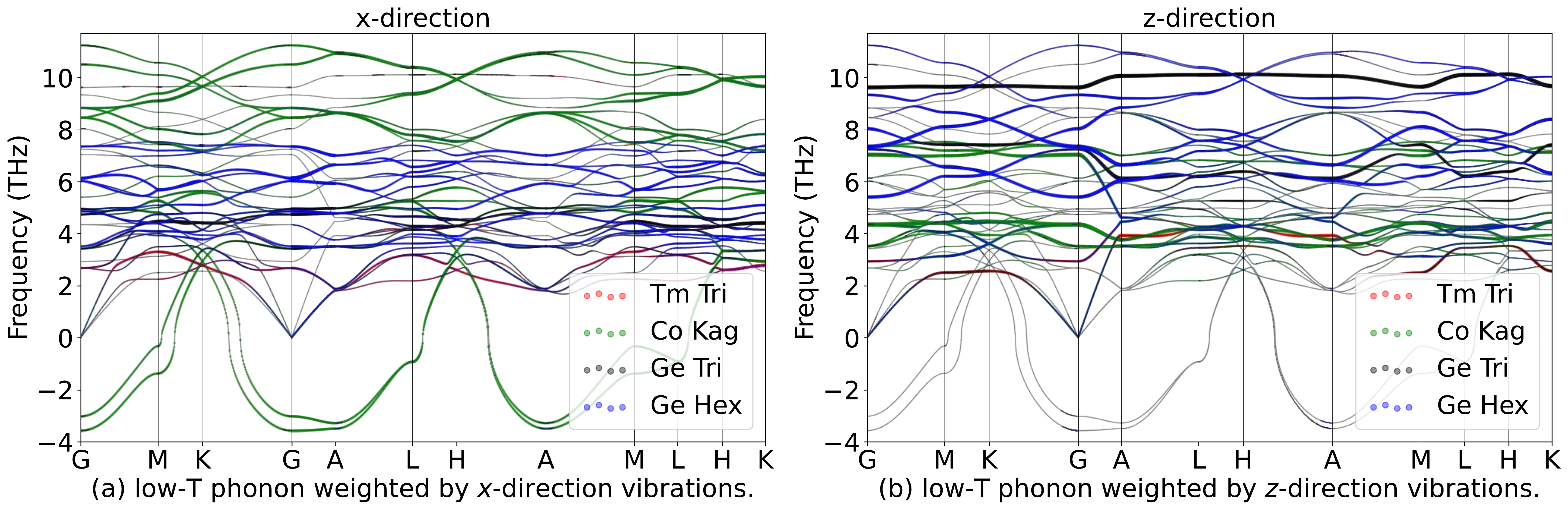}
\hspace{5pt}\label{fig:TmCo6Ge6_relaxed_High_xz}
\includegraphics[width=0.85\textwidth]{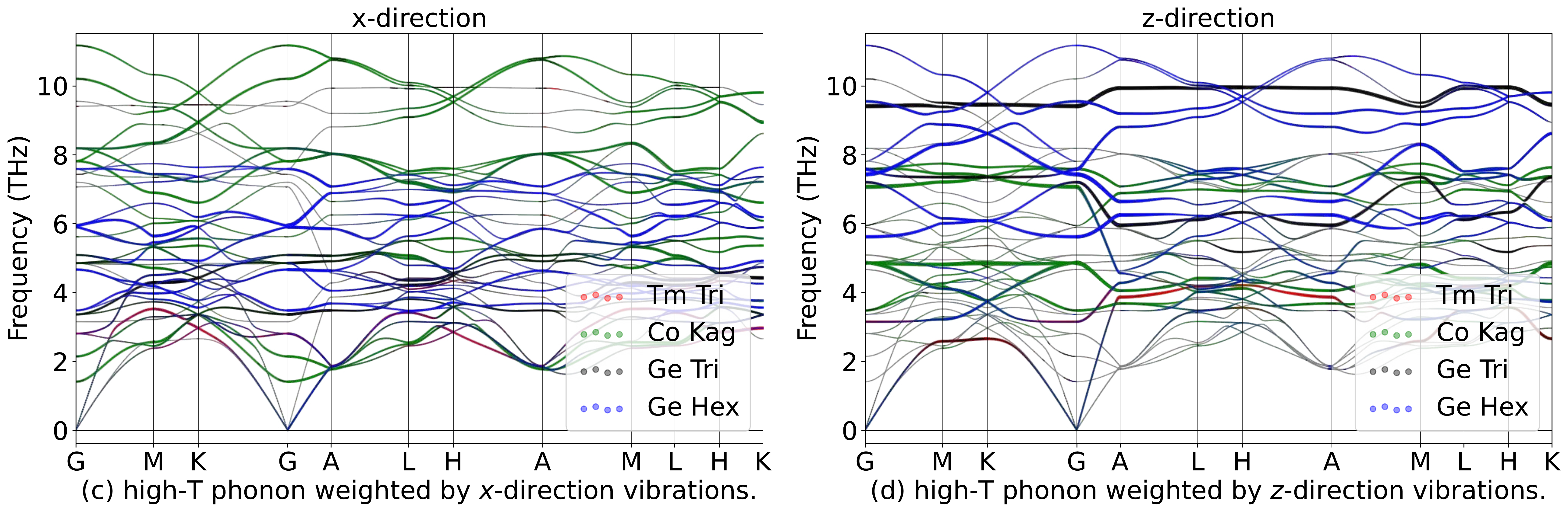}
\caption{Phonon spectrum for TmCo$_6$Ge$_6$in in-plane ($x$) and out-of-plane ($z$) directions.}
\label{fig:TmCo6Ge6phoxyz}
\end{figure}

\begin{table}[htbp]
\global\long\def\arraystretch{1.12}
\captionsetup{width=.95\textwidth}
\caption{IRREPs at high-symmetry points for TmCo$_6$Ge$_6$ phonon spectrum at Low-T.}
\label{tab:191_TmCo6Ge6_Low}
\setlength{\tabcolsep}{1.mm}{
}
\end{table}

\begin{table}[htbp]
\global\long\def\arraystretch{1.12}
\captionsetup{width=.95\textwidth}
\caption{IRREPs at high-symmetry points for TmCo$_6$Ge$_6$ phonon spectrum at High-T.}
\label{tab:191_TmCo6Ge6_High}
\setlength{\tabcolsep}{1.mm}{
%
}
\end{table}
\subsubsection{\label{sms2sec:YbCo6Ge6}\hyperref[smsec:col]{\texorpdfstring{YbCo$_6$Ge$_6$}{}}~{\color{blue}  (High-T stable, Low-T unstable) }}

\begin{table}[htbp]
\global\long\def\arraystretch{1.12}
\captionsetup{width=.85\textwidth}
\caption{Structure parameters of YbCo$_6$Ge$_6$ after relaxation. It contains 424 electrons. }
\label{tab:YbCo6Ge6}
\setlength{\tabcolsep}{4mm}{
%
}
\end{table}

\begin{figure}[htbp]
\centering
\includegraphics[width=0.85\textwidth]{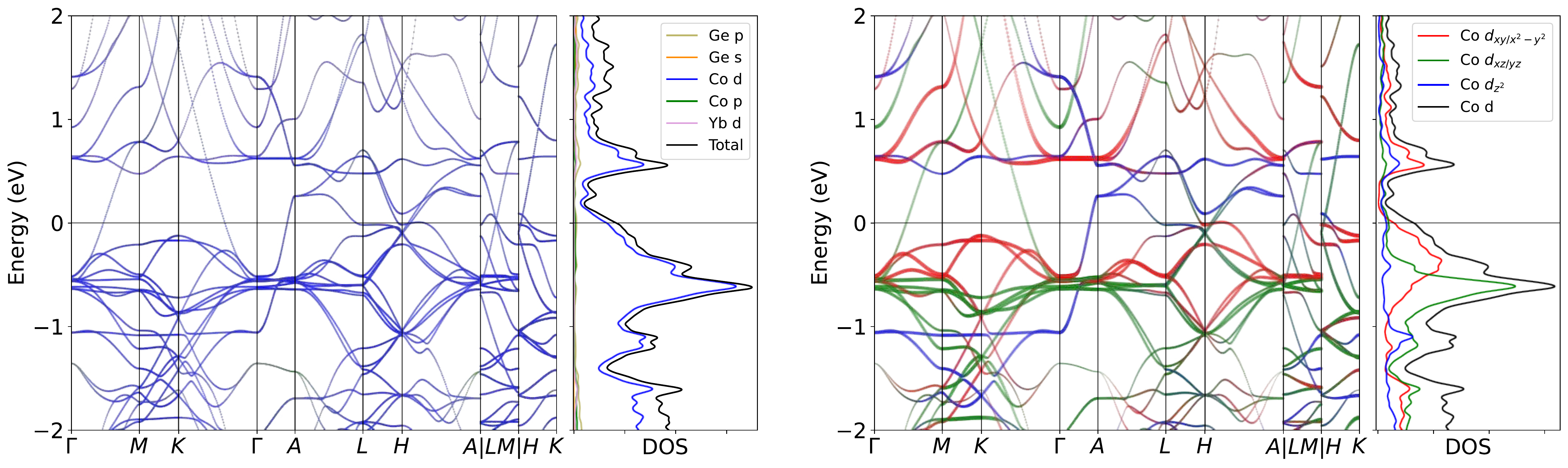}
\caption{Band structure for YbCo$_6$Ge$_6$ without SOC. The projection of Co $3d$ orbitals is also presented. The band structures clearly reveal the dominating of Co $3d$ orbitals  near the Fermi level, as well as the pronounced peak.}
\label{fig:YbCo6Ge6band}
\end{figure}

\begin{figure}[H]
\centering
\subfloat[Relaxed phonon at low-T.]{
\label{fig:YbCo6Ge6_relaxed_Low}
\includegraphics[width=0.45\textwidth]{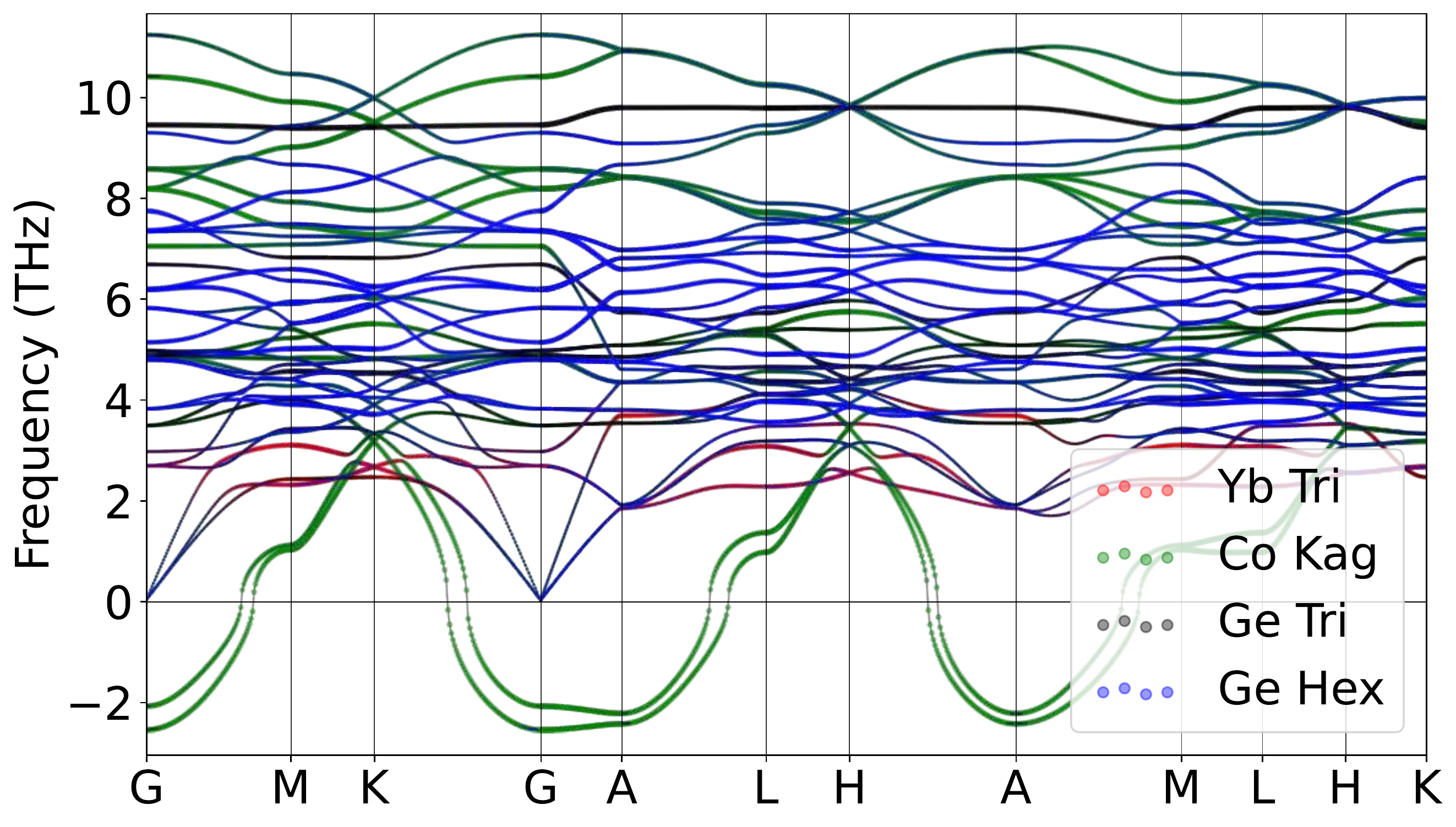}
}
\hspace{5pt}\subfloat[Relaxed phonon at high-T.]{
\label{fig:YbCo6Ge6_relaxed_High}
\includegraphics[width=0.45\textwidth]{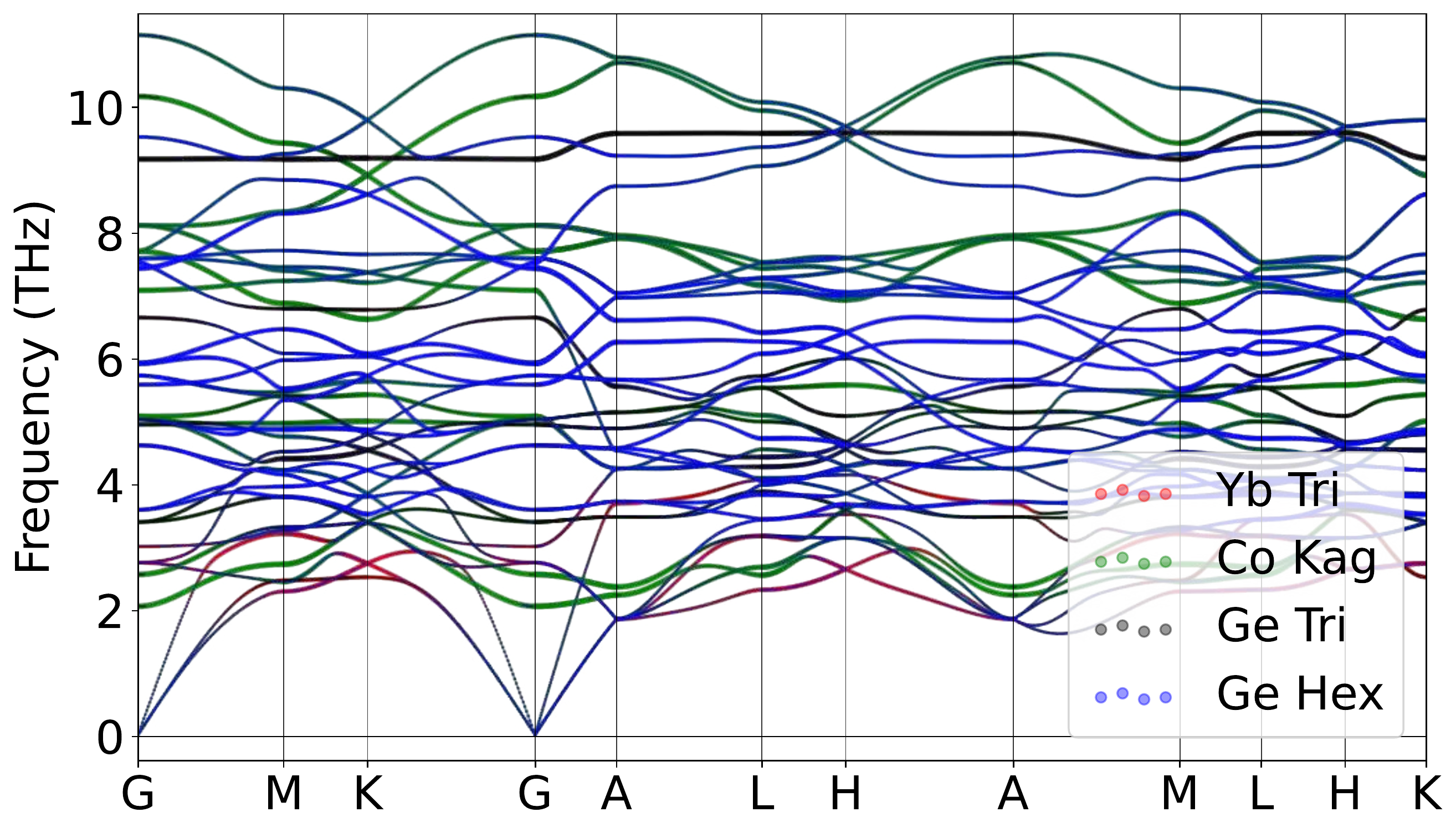}
}
\caption{Atomically projected phonon spectrum for YbCo$_6$Ge$_6$.}
\label{fig:YbCo6Ge6pho}
\end{figure}

\begin{figure}[H]
\centering
\label{fig:YbCo6Ge6_relaxed_Low_xz}
\includegraphics[width=0.85\textwidth]{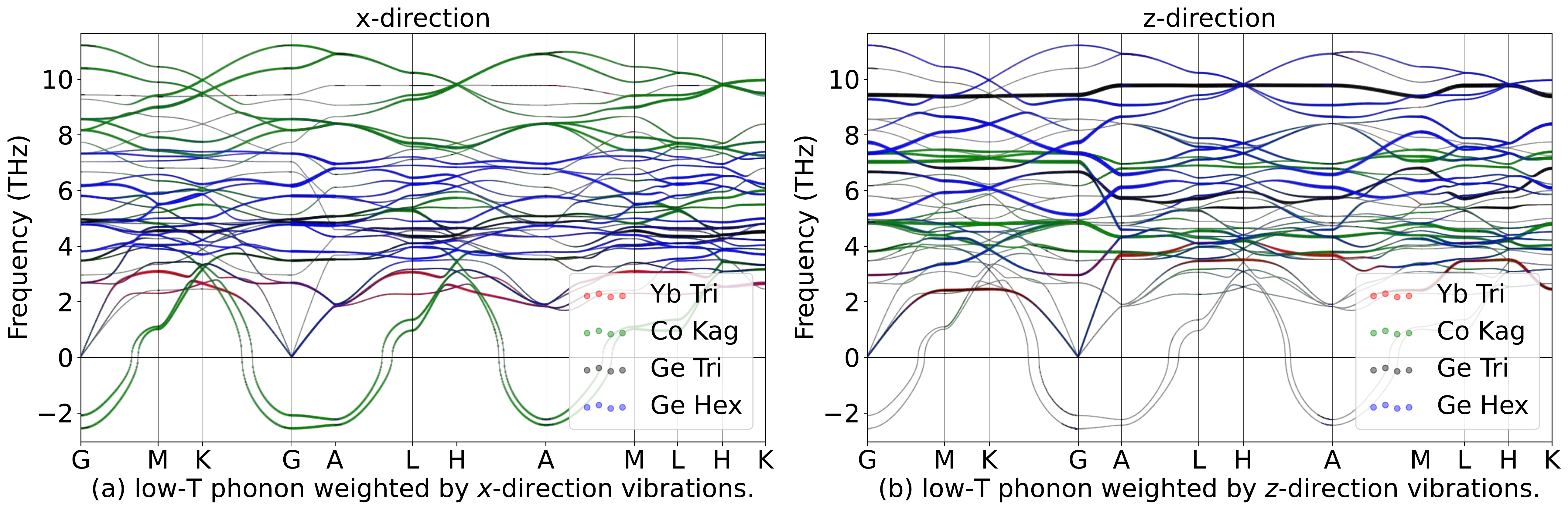}
\hspace{5pt}\label{fig:YbCo6Ge6_relaxed_High_xz}
\includegraphics[width=0.85\textwidth]{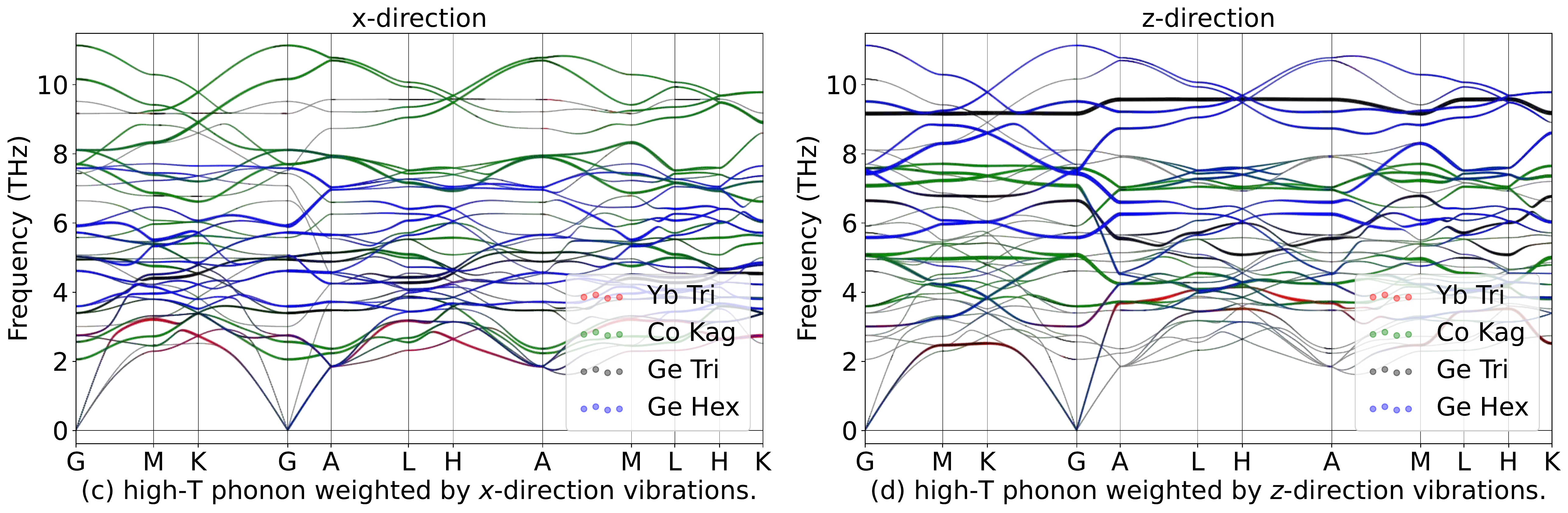}
\caption{Phonon spectrum for YbCo$_6$Ge$_6$in in-plane ($x$) and out-of-plane ($z$) directions.}
\label{fig:YbCo6Ge6phoxyz}
\end{figure}

\begin{table}[htbp]
\global\long\def\arraystretch{1.12}
\captionsetup{width=.95\textwidth}
\caption{IRREPs at high-symmetry points for YbCo$_6$Ge$_6$ phonon spectrum at Low-T.}
\label{tab:191_YbCo6Ge6_Low}
\setlength{\tabcolsep}{1.mm}{
}
\end{table}

\begin{table}[htbp]
\global\long\def\arraystretch{1.12}
\captionsetup{width=.95\textwidth}
\caption{IRREPs at high-symmetry points for YbCo$_6$Ge$_6$ phonon spectrum at High-T.}
\label{tab:191_YbCo6Ge6_High}
\setlength{\tabcolsep}{1.mm}{
%
}
\end{table}
\subsubsection{\label{sms2sec:LuCo6Ge6}\hyperref[smsec:col]{\texorpdfstring{LuCo$_6$Ge$_6$}{}}~{\color{blue}  (High-T stable, Low-T unstable) }}

\begin{table}[htbp]
\global\long\def\arraystretch{1.12}
\captionsetup{width=.85\textwidth}
\caption{Structure parameters of LuCo$_6$Ge$_6$ after relaxation. It contains 425 electrons. }
\label{tab:LuCo6Ge6}
\setlength{\tabcolsep}{4mm}{
%
}
\end{table}

\begin{figure}[htbp]
\centering
\includegraphics[width=0.85\textwidth]{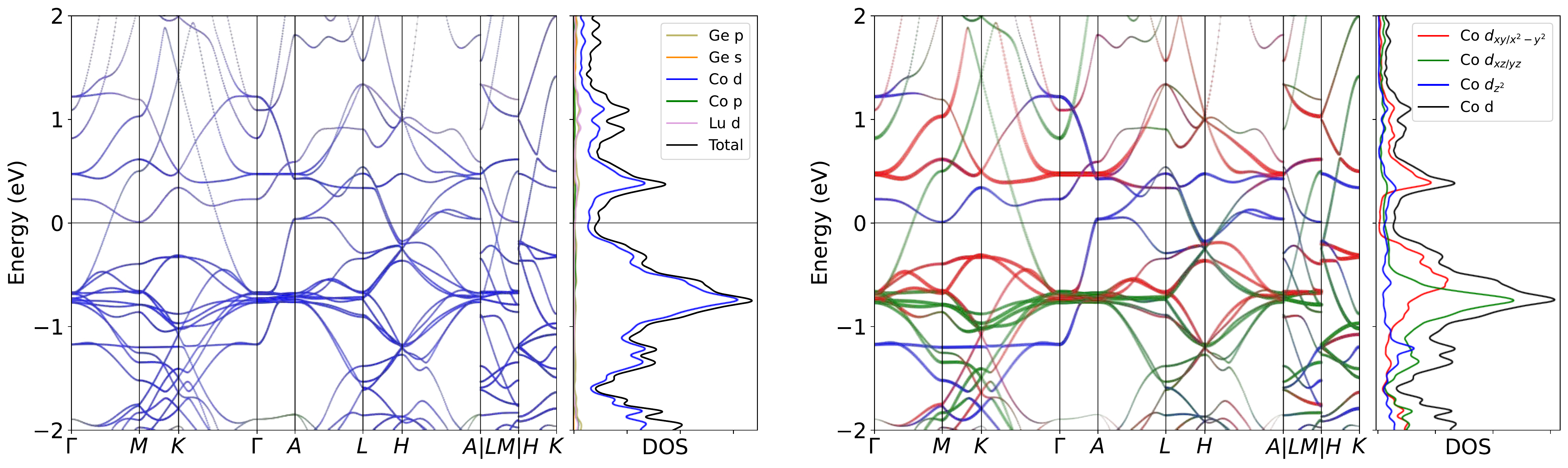}
\caption{Band structure for LuCo$_6$Ge$_6$ without SOC. The projection of Co $3d$ orbitals is also presented. The band structures clearly reveal the dominating of Co $3d$ orbitals  near the Fermi level, as well as the pronounced peak.}
\label{fig:LuCo6Ge6band}
\end{figure}

\begin{figure}[H]
\centering
\subfloat[Relaxed phonon at low-T.]{
\label{fig:LuCo6Ge6_relaxed_Low}
\includegraphics[width=0.45\textwidth]{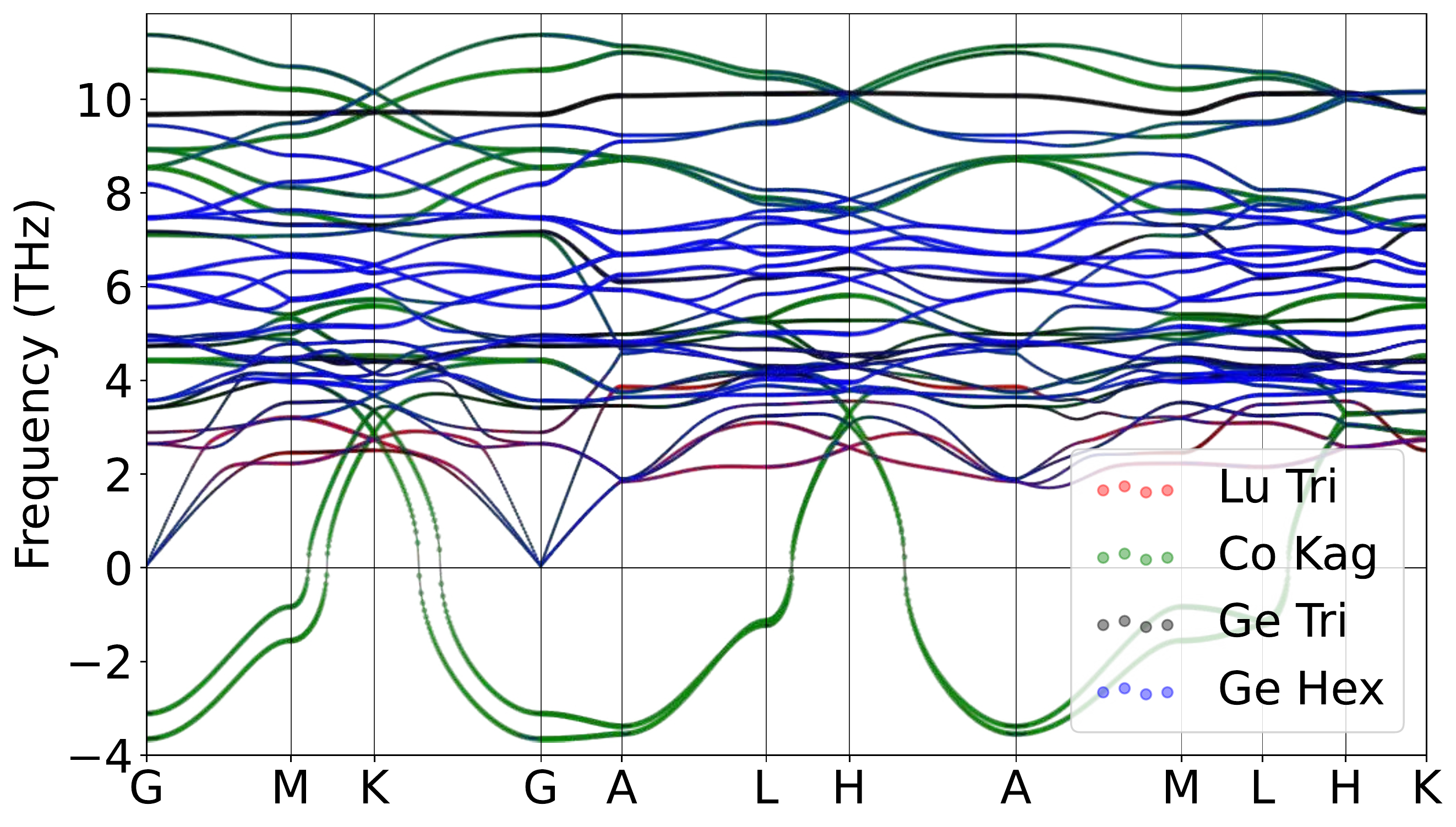}
}
\hspace{5pt}\subfloat[Relaxed phonon at high-T.]{
\label{fig:LuCo6Ge6_relaxed_High}
\includegraphics[width=0.45\textwidth]{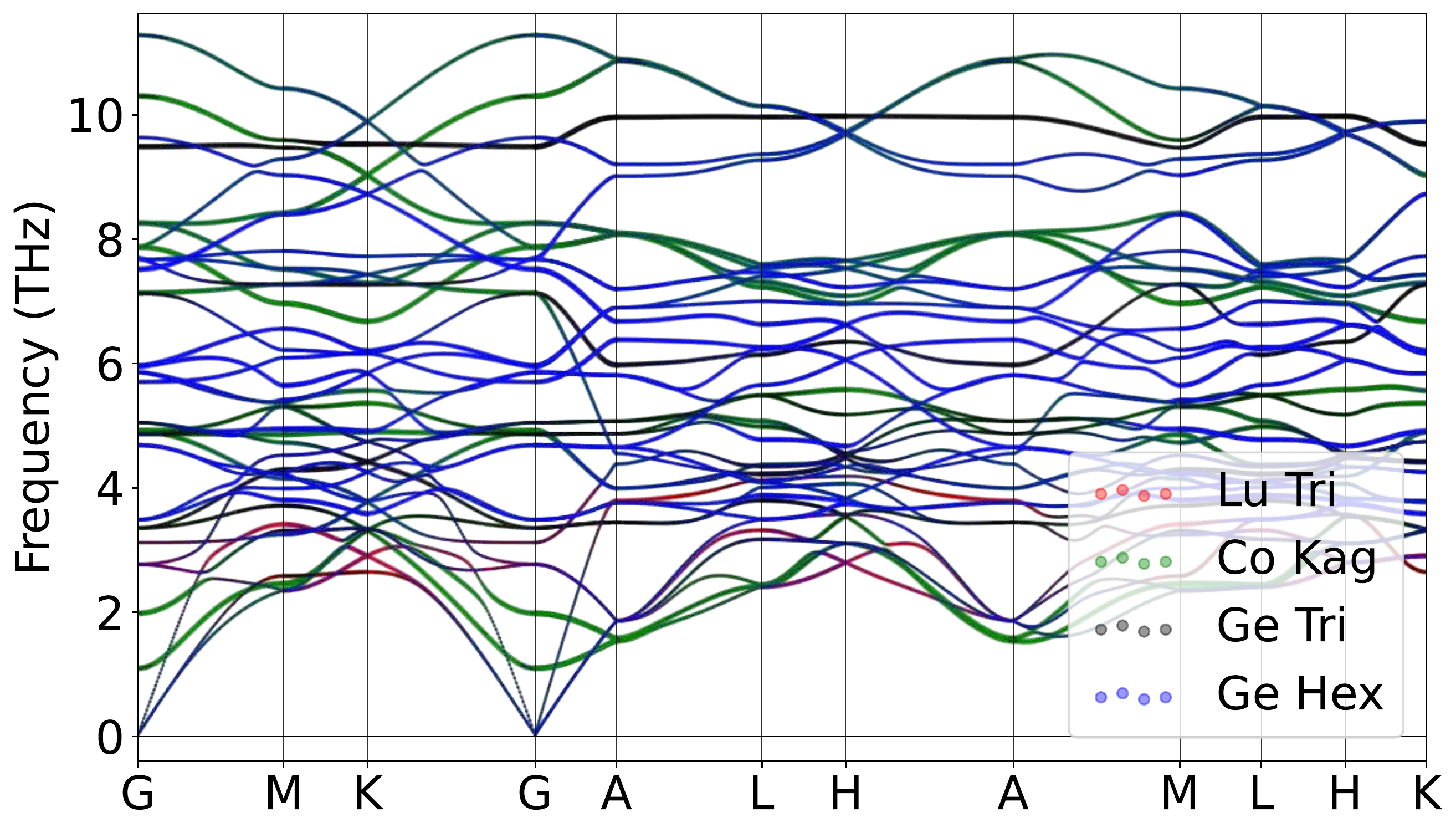}
}
\caption{Atomically projected phonon spectrum for LuCo$_6$Ge$_6$.}
\label{fig:LuCo6Ge6pho}
\end{figure}

\begin{figure}[H]
\centering
\label{fig:LuCo6Ge6_relaxed_Low_xz}
\includegraphics[width=0.85\textwidth]{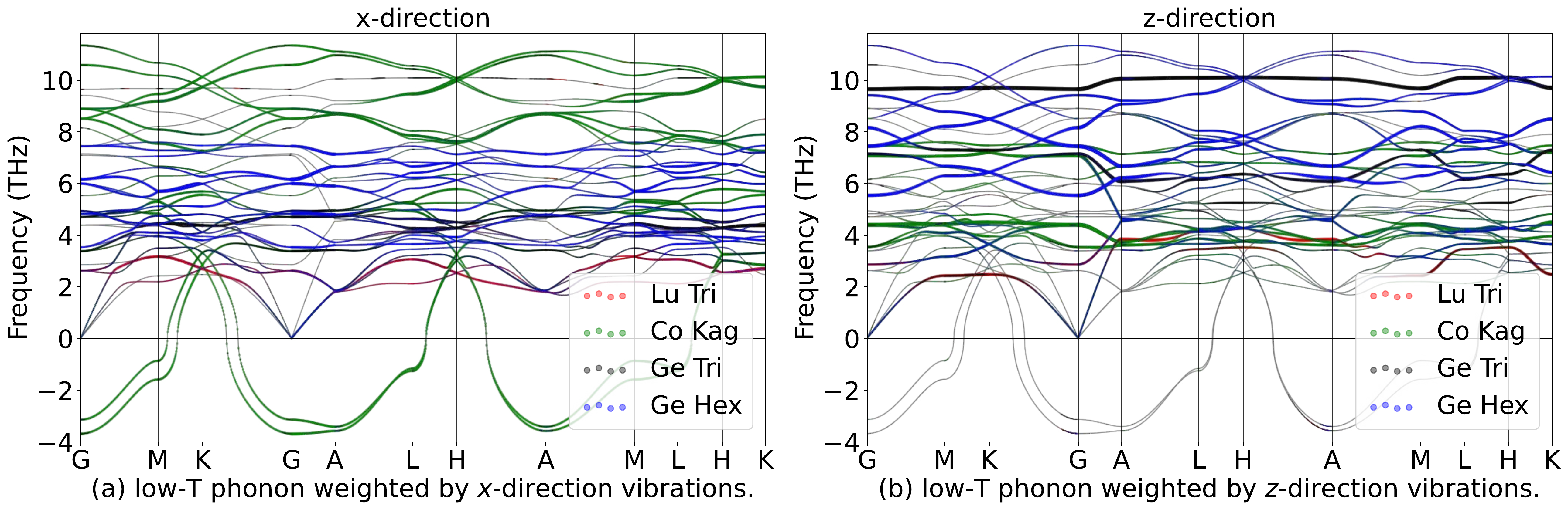}
\hspace{5pt}\label{fig:LuCo6Ge6_relaxed_High_xz}
\includegraphics[width=0.85\textwidth]{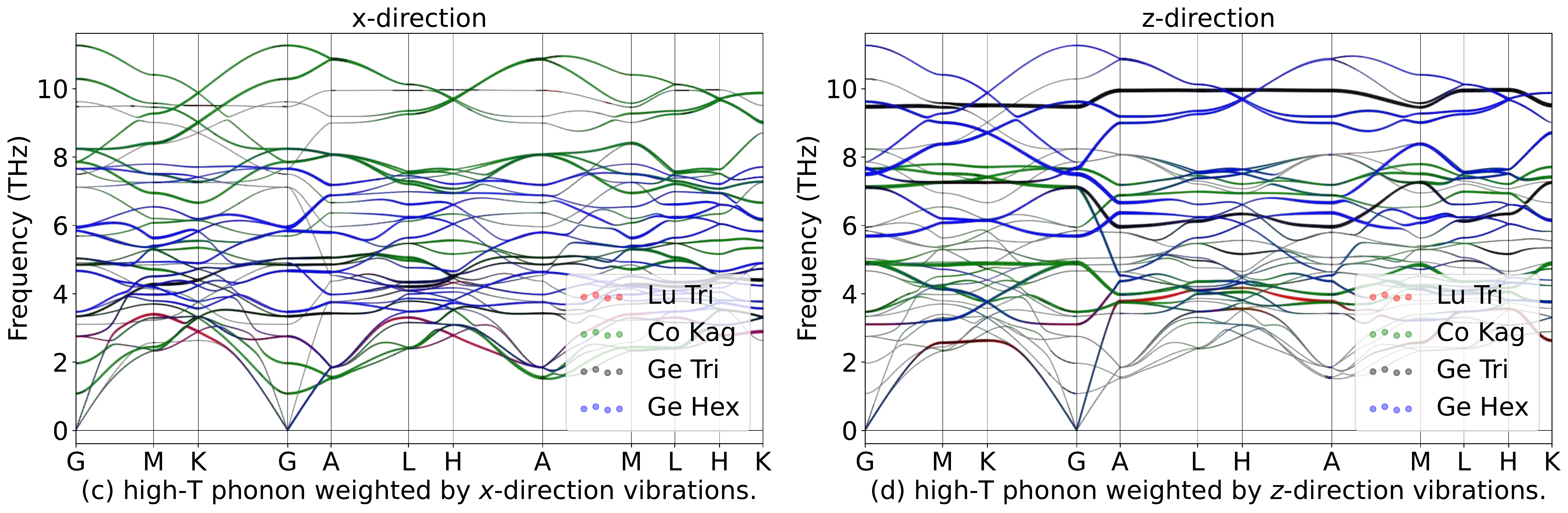}
\caption{Phonon spectrum for LuCo$_6$Ge$_6$in in-plane ($x$) and out-of-plane ($z$) directions.}
\label{fig:LuCo6Ge6phoxyz}
\end{figure}

\begin{table}[htbp]
\global\long\def\arraystretch{1.12}
\captionsetup{width=.95\textwidth}
\caption{IRREPs at high-symmetry points for LuCo$_6$Ge$_6$ phonon spectrum at Low-T.}
\label{tab:191_LuCo6Ge6_Low}
\setlength{\tabcolsep}{1.mm}{
}
\end{table}

\begin{table}[htbp]
\global\long\def\arraystretch{1.12}
\captionsetup{width=.95\textwidth}
\caption{IRREPs at high-symmetry points for LuCo$_6$Ge$_6$ phonon spectrum at High-T.}
\label{tab:191_LuCo6Ge6_High}
\setlength{\tabcolsep}{1.mm}{
%
}
\end{table}
\subsubsection{\label{sms2sec:HfCo6Ge6}\hyperref[smsec:col]{\texorpdfstring{HfCo$_6$Ge$_6$}{}}~{\color{red}  (High-T unstable, Low-T unstable) }}

\begin{table}[htbp]
\global\long\def\arraystretch{1.12}
\captionsetup{width=.85\textwidth}
\caption{Structure parameters of HfCo$_6$Ge$_6$ after relaxation. It contains 426 electrons. }
\label{tab:HfCo6Ge6}
\setlength{\tabcolsep}{4mm}{
%
}
\end{table}

\begin{figure}[htbp]
\centering
\includegraphics[width=0.85\textwidth]{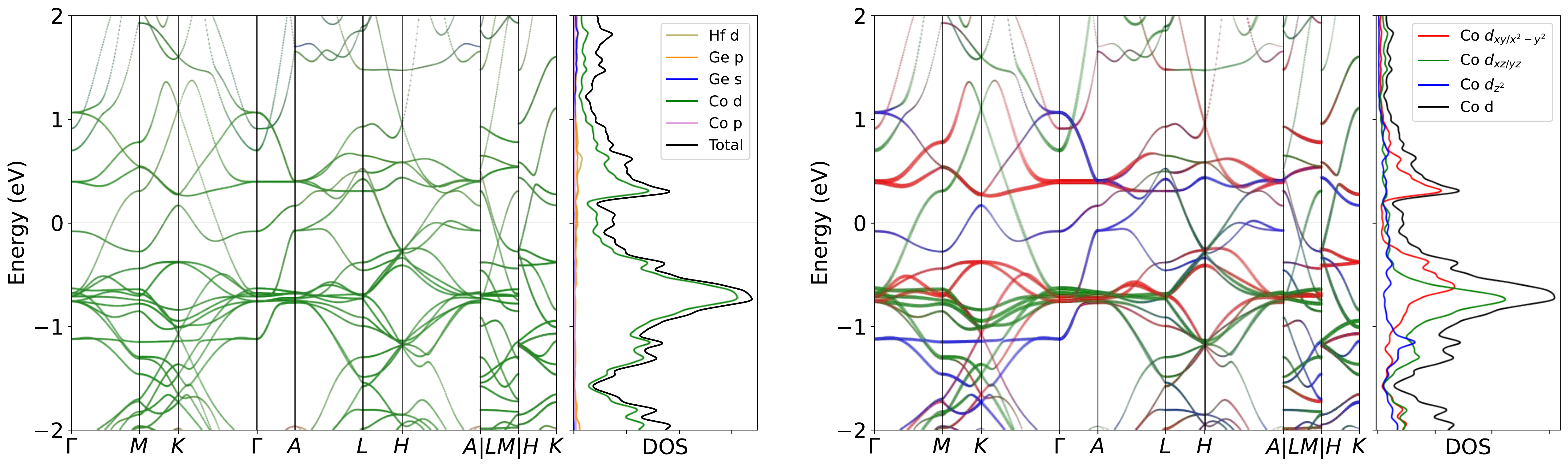}
\caption{Band structure for HfCo$_6$Ge$_6$ without SOC. The projection of Co $3d$ orbitals is also presented. The band structures clearly reveal the dominating of Co $3d$ orbitals  near the Fermi level, as well as the pronounced peak.}
\label{fig:HfCo6Ge6band}
\end{figure}

\begin{figure}[H]
\centering
\subfloat[Relaxed phonon at low-T.]{
\label{fig:HfCo6Ge6_relaxed_Low}
\includegraphics[width=0.45\textwidth]{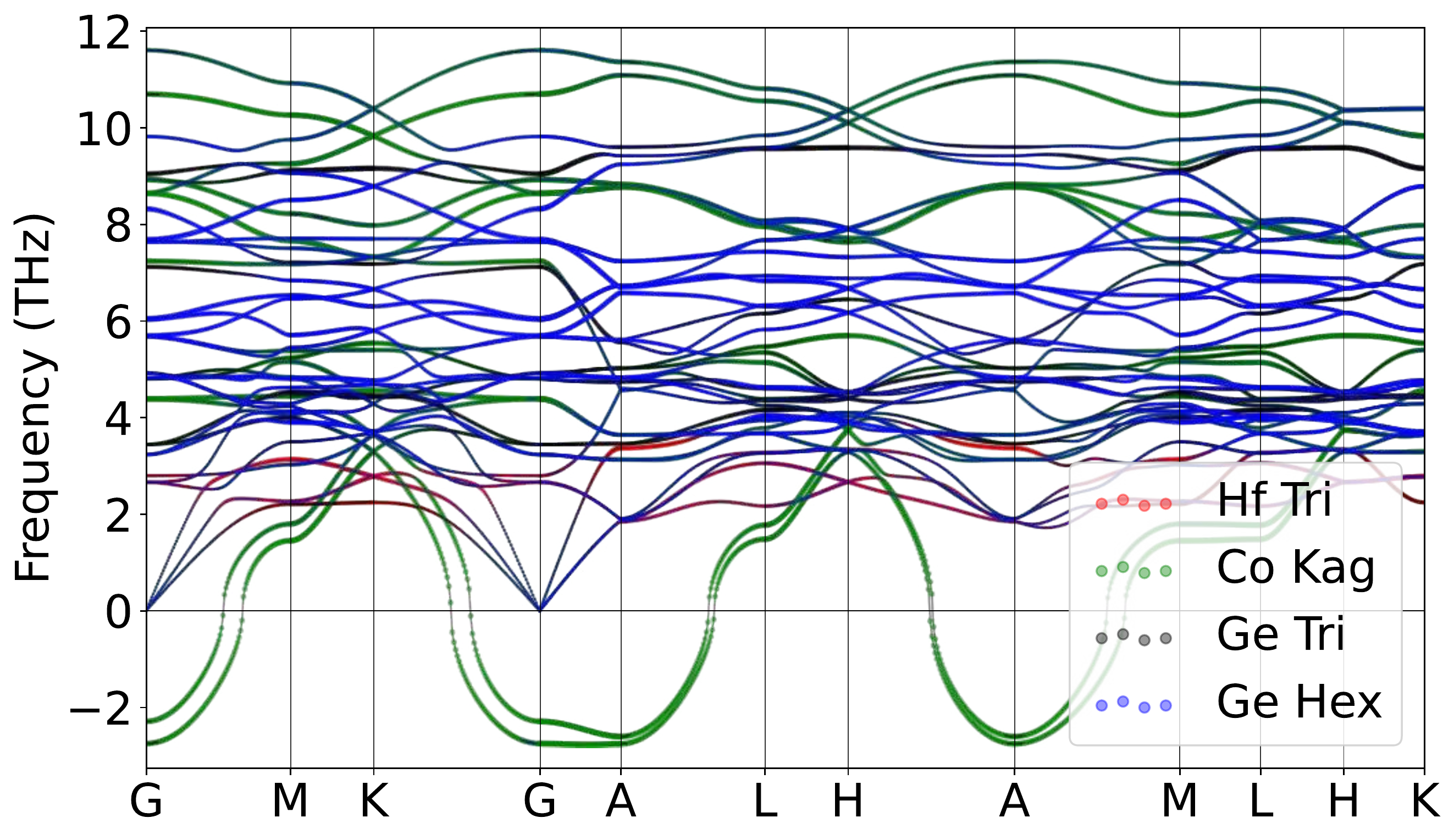}
}
\hspace{5pt}\subfloat[Relaxed phonon at high-T.]{
\label{fig:HfCo6Ge6_relaxed_High}
\includegraphics[width=0.45\textwidth]{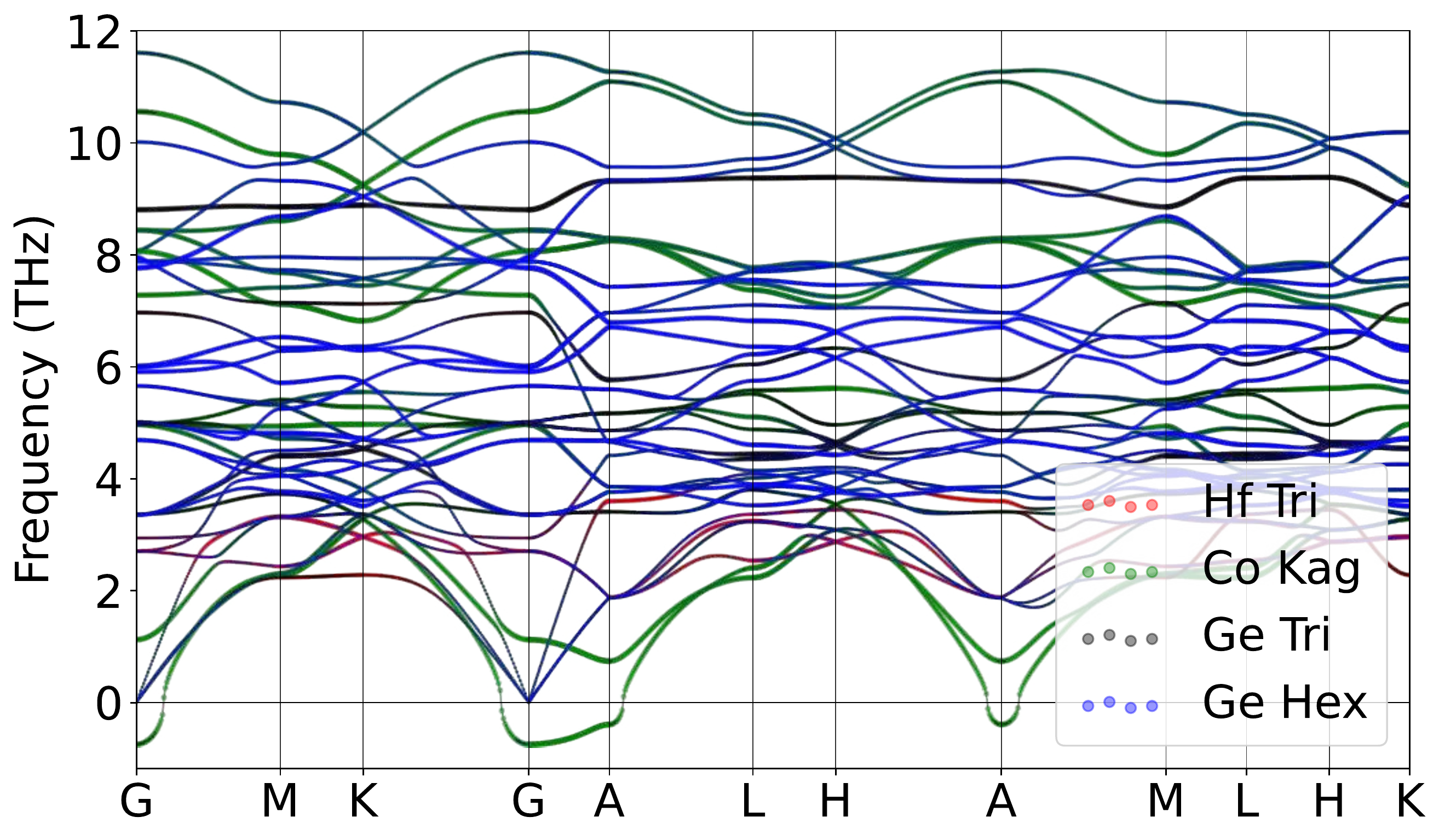}
}
\caption{Atomically projected phonon spectrum for HfCo$_6$Ge$_6$.}
\label{fig:HfCo6Ge6pho}
\end{figure}

\begin{figure}[H]
\centering
\label{fig:HfCo6Ge6_relaxed_Low_xz}
\includegraphics[width=0.85\textwidth]{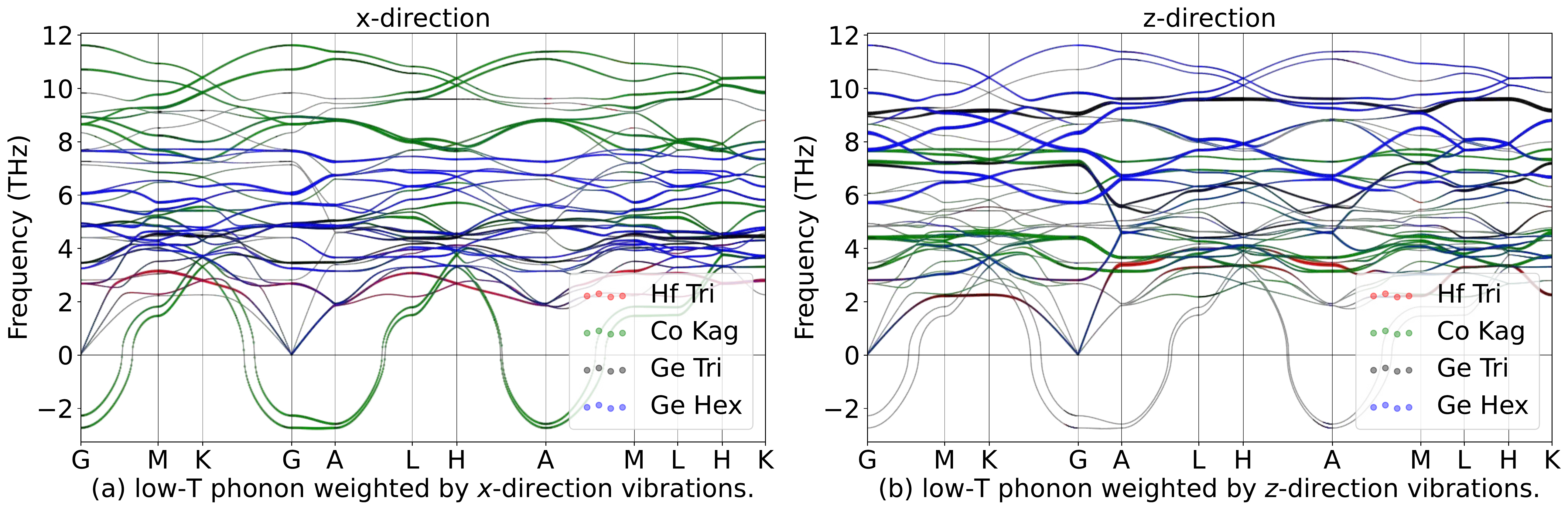}
\hspace{5pt}\label{fig:HfCo6Ge6_relaxed_High_xz}
\includegraphics[width=0.85\textwidth]{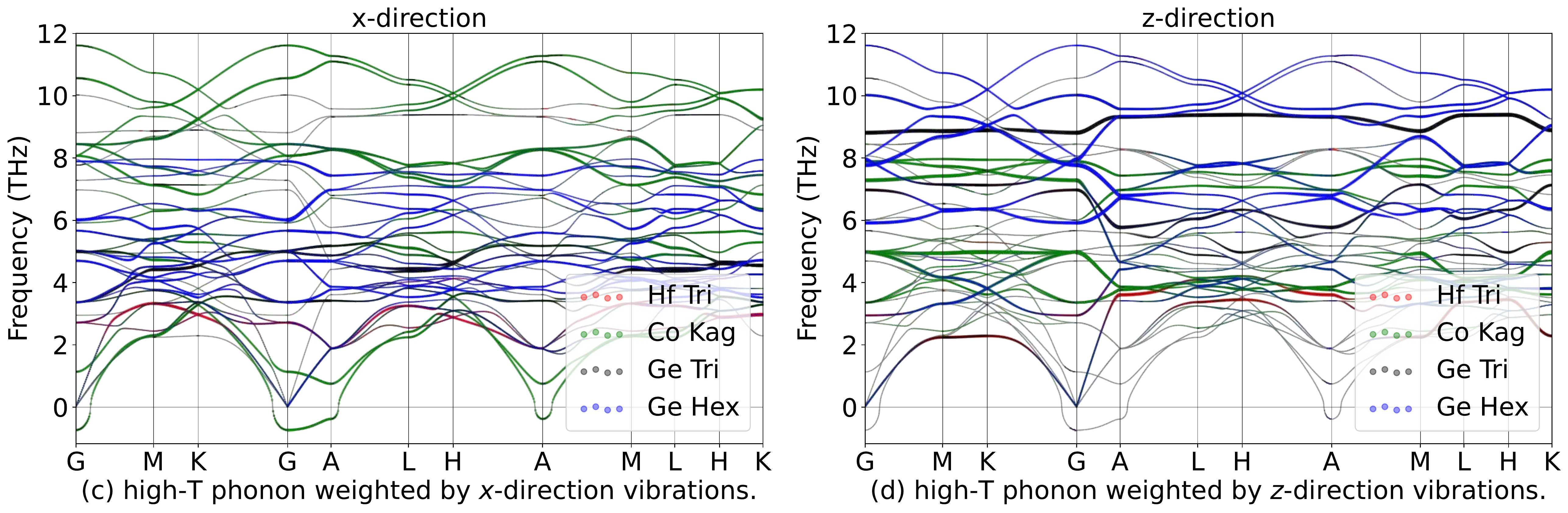}
\caption{Phonon spectrum for HfCo$_6$Ge$_6$in in-plane ($x$) and out-of-plane ($z$) directions.}
\label{fig:HfCo6Ge6phoxyz}
\end{figure}

\begin{table}[htbp]
\global\long\def\arraystretch{1.12}
\captionsetup{width=.95\textwidth}
\caption{IRREPs at high-symmetry points for HfCo$_6$Ge$_6$ phonon spectrum at Low-T.}
\label{tab:191_HfCo6Ge6_Low}
\setlength{\tabcolsep}{1.mm}{
}
\end{table}

\begin{table}[htbp]
\global\long\def\arraystretch{1.12}
\captionsetup{width=.95\textwidth}
\caption{IRREPs at high-symmetry points for HfCo$_6$Ge$_6$ phonon spectrum at High-T.}
\label{tab:191_HfCo6Ge6_High}
\setlength{\tabcolsep}{1.mm}{
%
}
\end{table}
\subsubsection{\label{sms2sec:TaCo6Ge6}\hyperref[smsec:col]{\texorpdfstring{TaCo$_6$Ge$_6$}{}}~{\color{red}  (High-T unstable, Low-T unstable) }}

\begin{table}[htbp]
\global\long\def\arraystretch{1.12}
\captionsetup{width=.85\textwidth}
\caption{Structure parameters of TaCo$_6$Ge$_6$ after relaxation. It contains 427 electrons. }
\label{tab:TaCo6Ge6}
\setlength{\tabcolsep}{4mm}{
%
}
\end{table}

\begin{figure}[htbp]
\centering
\includegraphics[width=0.85\textwidth]{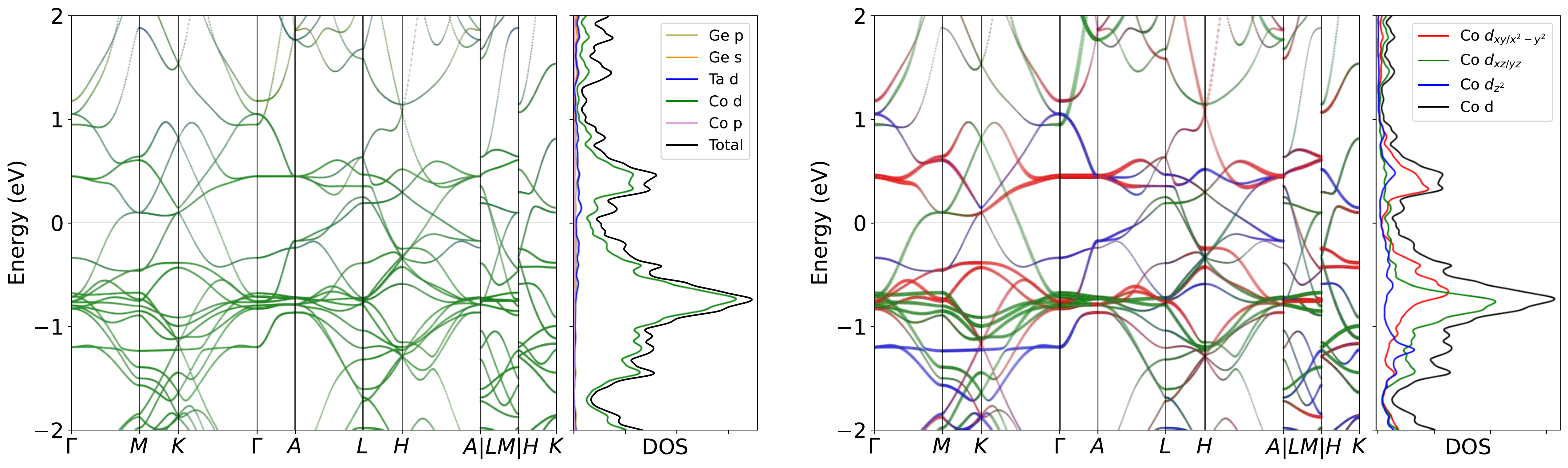}
\caption{Band structure for TaCo$_6$Ge$_6$ without SOC. The projection of Co $3d$ orbitals is also presented. The band structures clearly reveal the dominating of Co $3d$ orbitals  near the Fermi level, as well as the pronounced peak.}
\label{fig:TaCo6Ge6band}
\end{figure}

\begin{figure}[H]
\centering
\subfloat[Relaxed phonon at low-T.]{
\label{fig:TaCo6Ge6_relaxed_Low}
\includegraphics[width=0.45\textwidth]{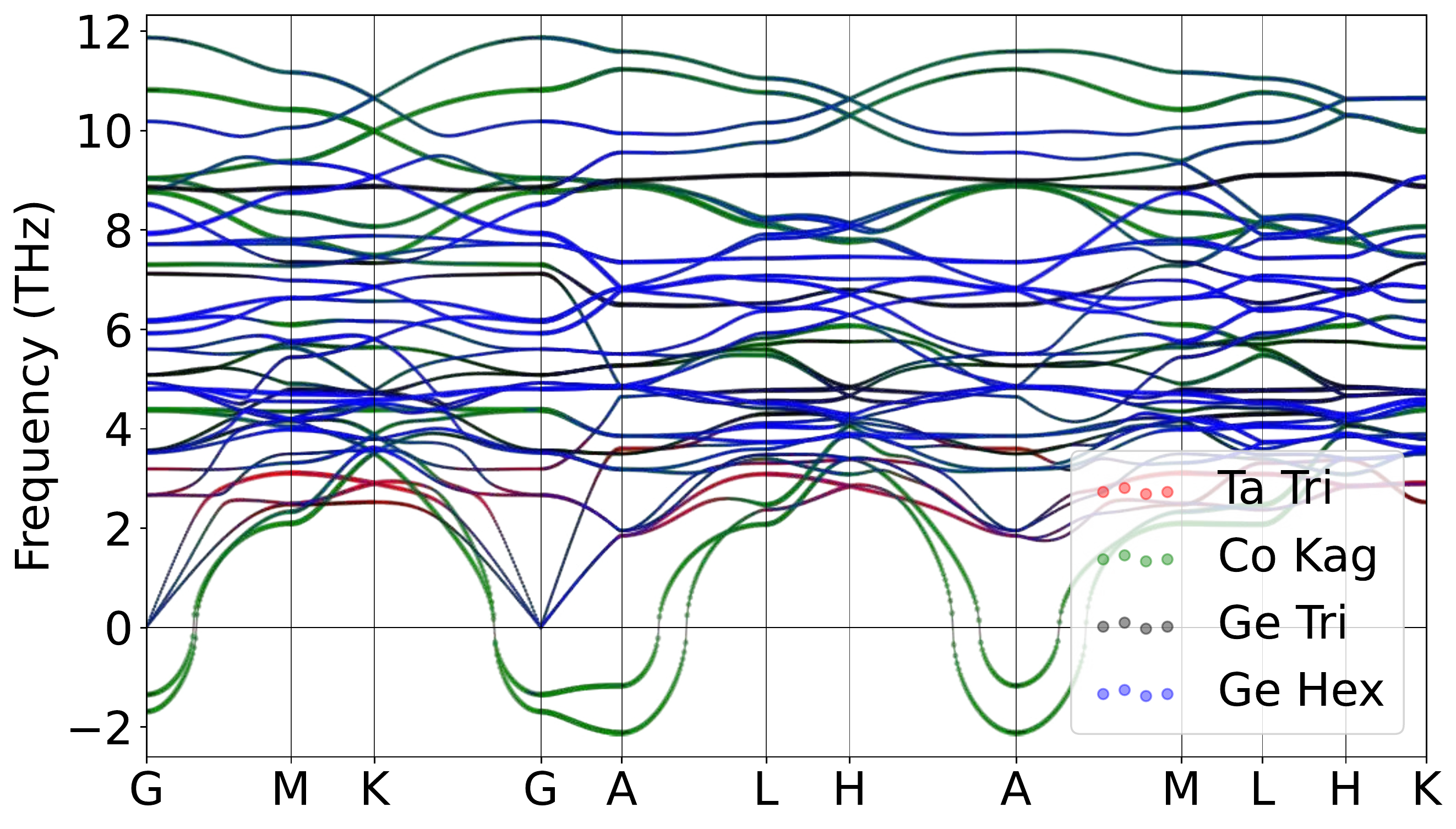}
}
\hspace{5pt}\subfloat[Relaxed phonon at high-T.]{
\label{fig:TaCo6Ge6_relaxed_High}
\includegraphics[width=0.45\textwidth]{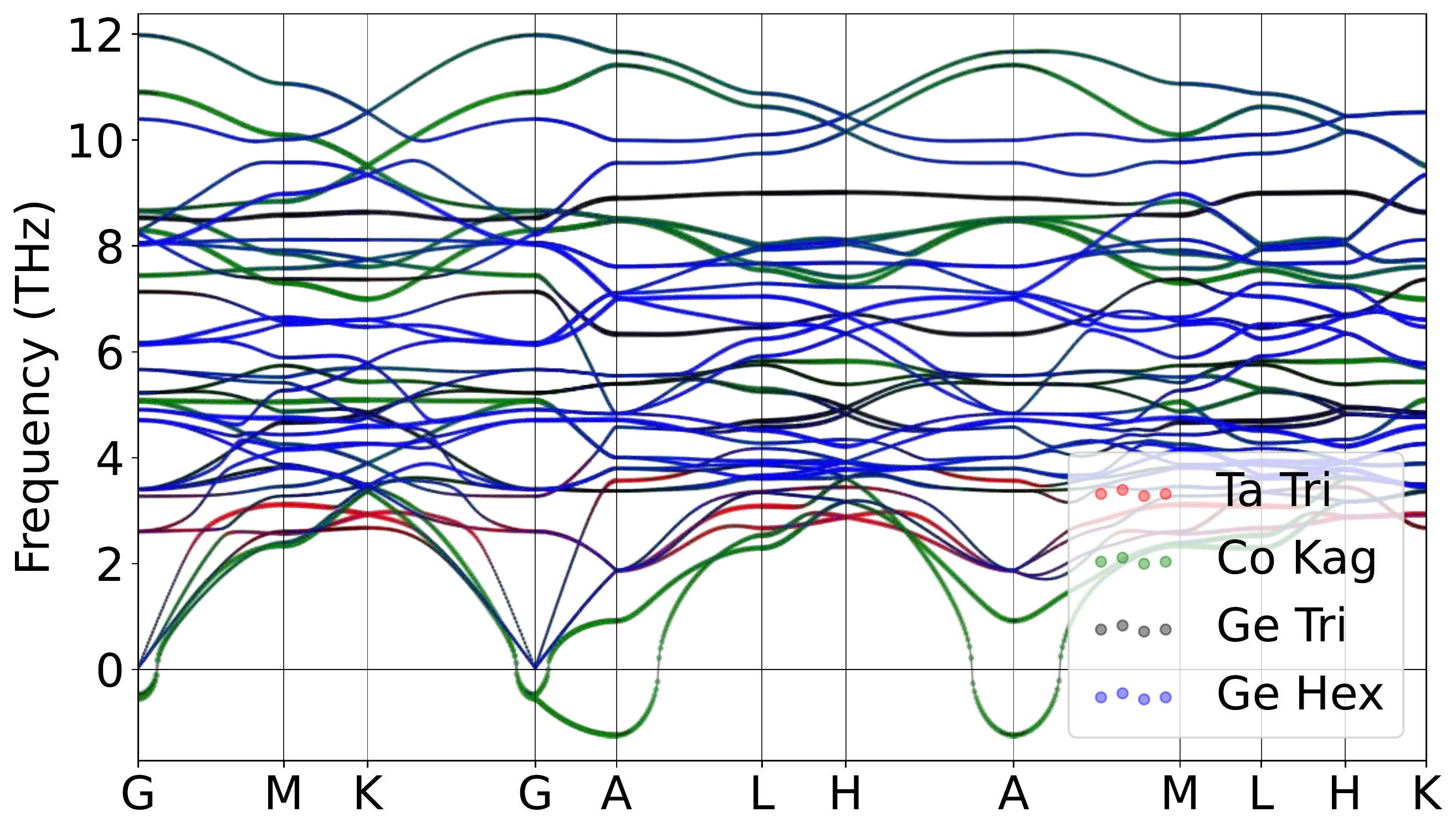}
}
\caption{Atomically projected phonon spectrum for TaCo$_6$Ge$_6$.}
\label{fig:TaCo6Ge6pho}
\end{figure}

\begin{figure}[H]
\centering
\label{fig:TaCo6Ge6_relaxed_Low_xz}
\includegraphics[width=0.85\textwidth]{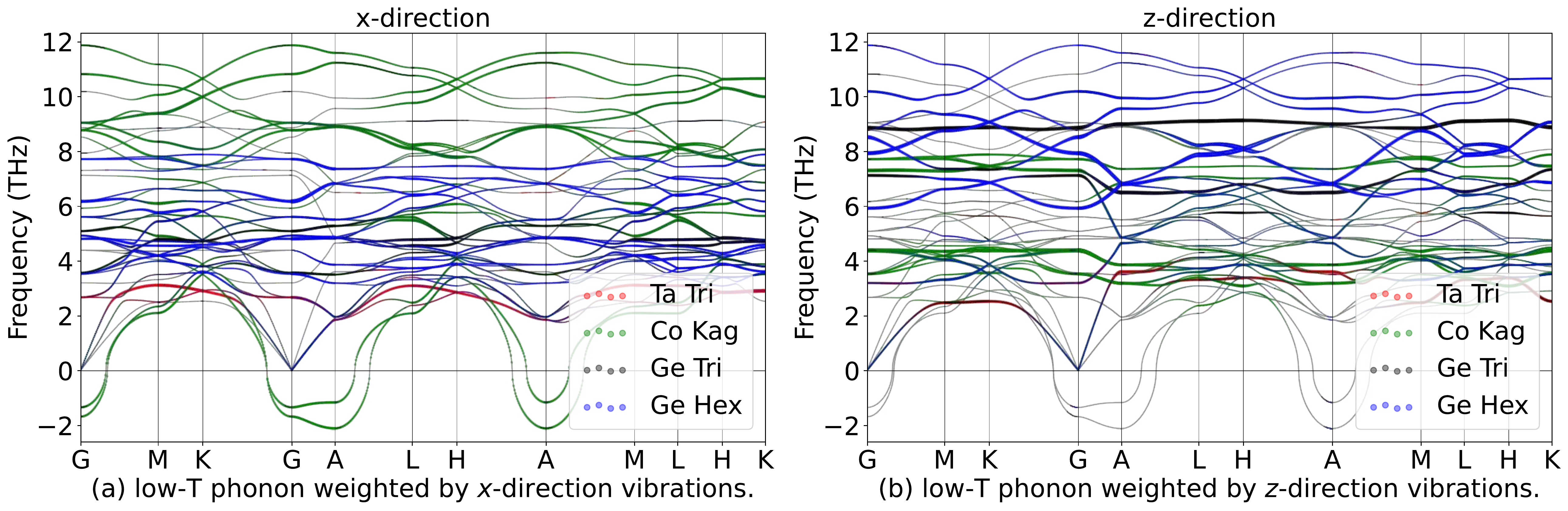}
\hspace{5pt}\label{fig:TaCo6Ge6_relaxed_High_xz}
\includegraphics[width=0.85\textwidth]{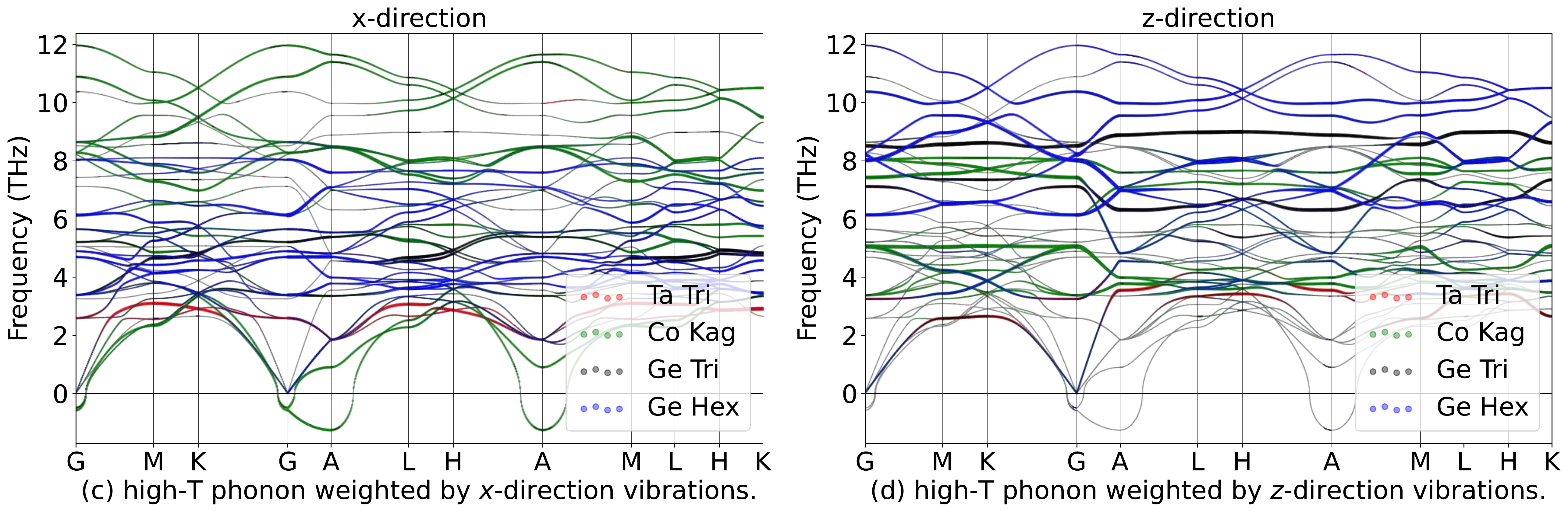}
\caption{Phonon spectrum for TaCo$_6$Ge$_6$in in-plane ($x$) and out-of-plane ($z$) directions.}
\label{fig:TaCo6Ge6phoxyz}
\end{figure}

\begin{table}[htbp]
\global\long\def\arraystretch{1.12}
\captionsetup{width=.95\textwidth}
\caption{IRREPs at high-symmetry points for TaCo$_6$Ge$_6$ phonon spectrum at Low-T.}
\label{tab:191_TaCo6Ge6_Low}
\setlength{\tabcolsep}{1.mm}{
}
\end{table}

\begin{table}[htbp]
\global\long\def\arraystretch{1.12}
\captionsetup{width=.95\textwidth}
\caption{IRREPs at high-symmetry points for TaCo$_6$Ge$_6$ phonon spectrum at High-T.}
\label{tab:191_TaCo6Ge6_High}
\setlength{\tabcolsep}{1.mm}{
%
}
\end{table}
\subsubsection{\label{sms2sec:UCo6Ge6}\hyperref[smsec:col]{\texorpdfstring{UCo$_6$Ge$_6$}{}}~{\color{red}  (High-T unstable, Low-T unstable) }}

\begin{table}[htbp]
\global\long\def\arraystretch{1.12}
\captionsetup{width=.85\textwidth}
\caption{Structure parameters of UCo$_6$Ge$_6$ after relaxation. It contains 446 electrons. }
\label{tab:UCo6Ge6}
\setlength{\tabcolsep}{4mm}{
%
}
\end{table}

\begin{figure}[htbp]
\centering
\includegraphics[width=0.85\textwidth]{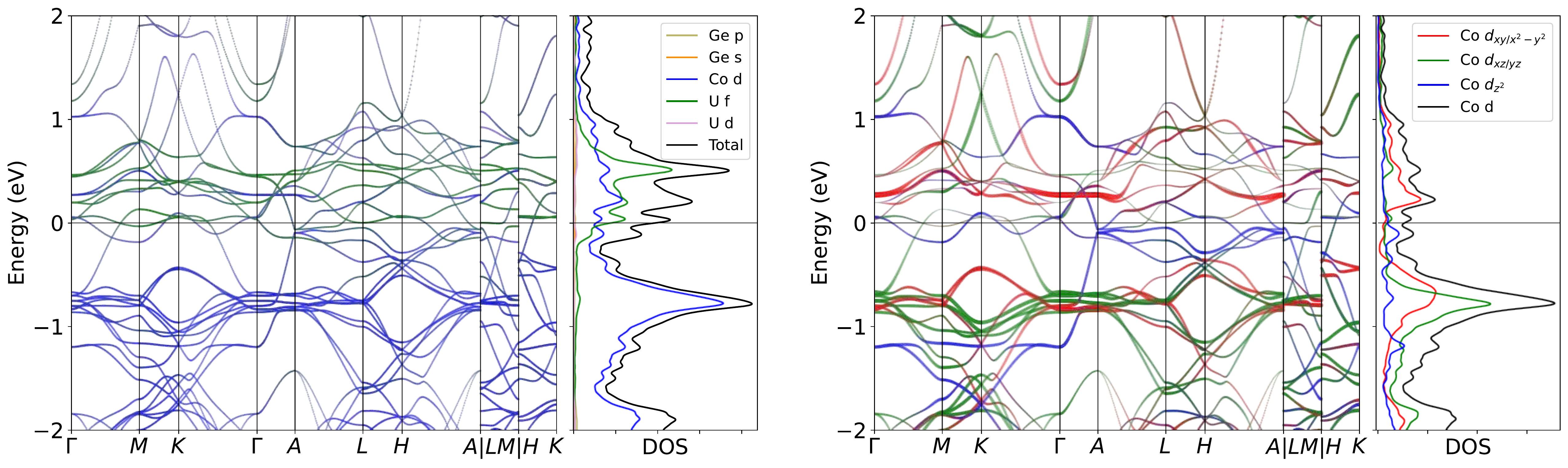}
\caption{Band structure for UCo$_6$Ge$_6$ without SOC. The projection of Co $3d$ orbitals is also presented. The band structures clearly reveal the dominating of Co $3d$ orbitals  near the Fermi level, as well as the pronounced peak.}
\label{fig:UCo6Ge6band}
\end{figure}

\begin{figure}[H]
\centering
\subfloat[Relaxed phonon at low-T.]{
\label{fig:UCo6Ge6_relaxed_Low}
\includegraphics[width=0.45\textwidth]{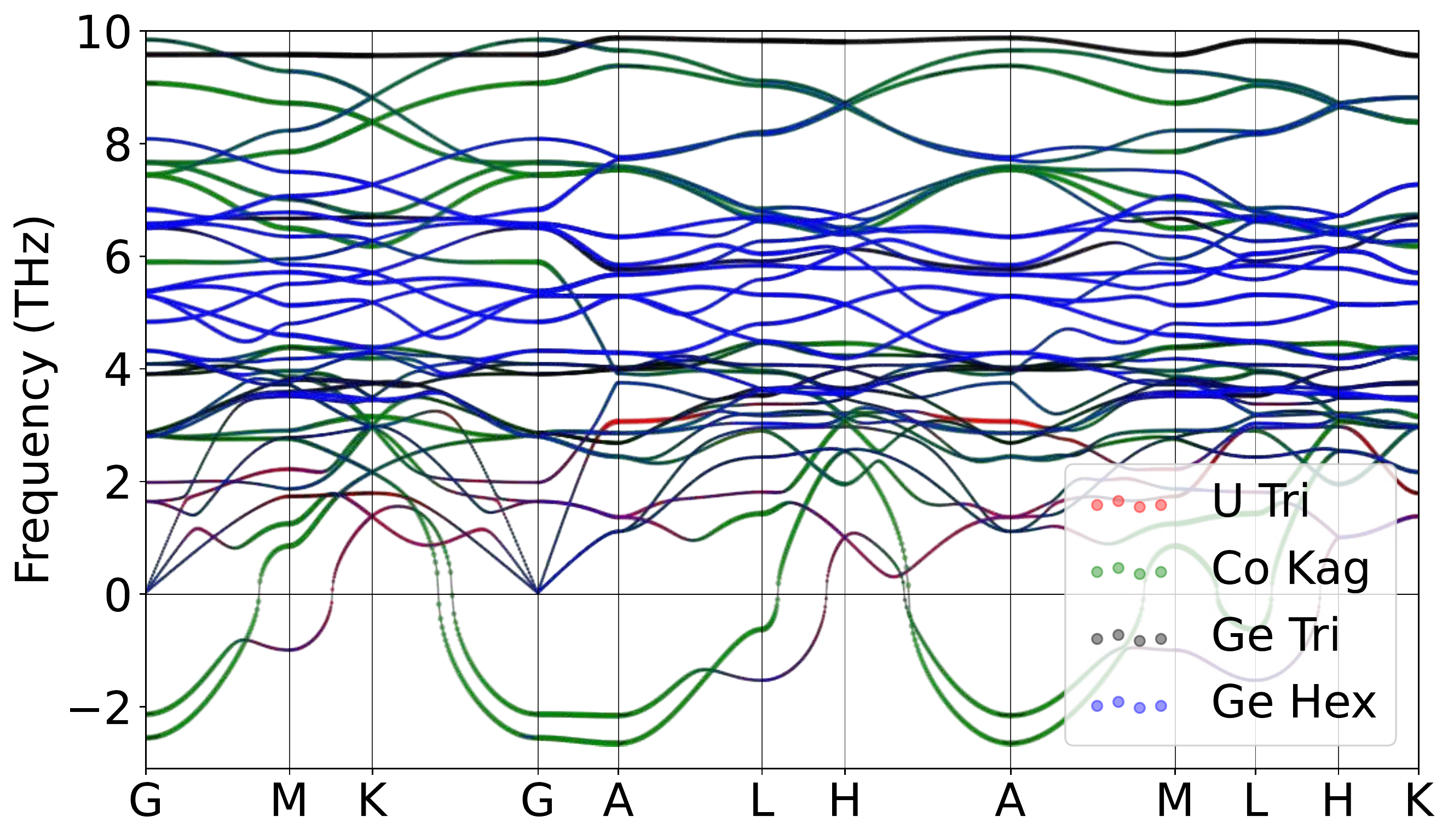}
}
\hspace{5pt}\subfloat[Relaxed phonon at high-T.]{
\label{fig:UCo6Ge6_relaxed_High}
\includegraphics[width=0.45\textwidth]{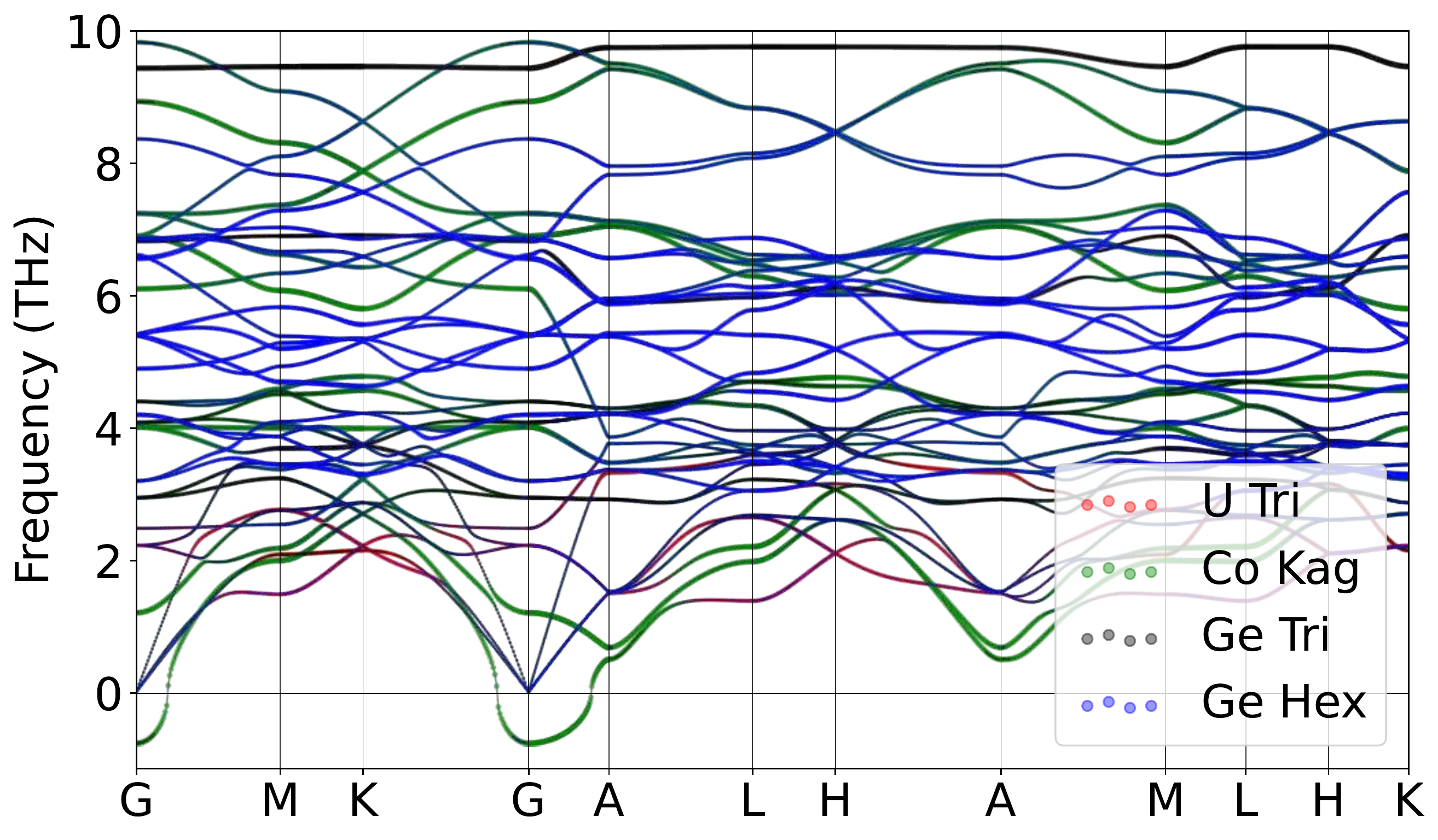}
}
\caption{Atomically projected phonon spectrum for UCo$_6$Ge$_6$.}
\label{fig:UCo6Ge6pho}
\end{figure}

\begin{figure}[H]
\centering
\label{fig:UCo6Ge6_relaxed_Low_xz}
\includegraphics[width=0.85\textwidth]{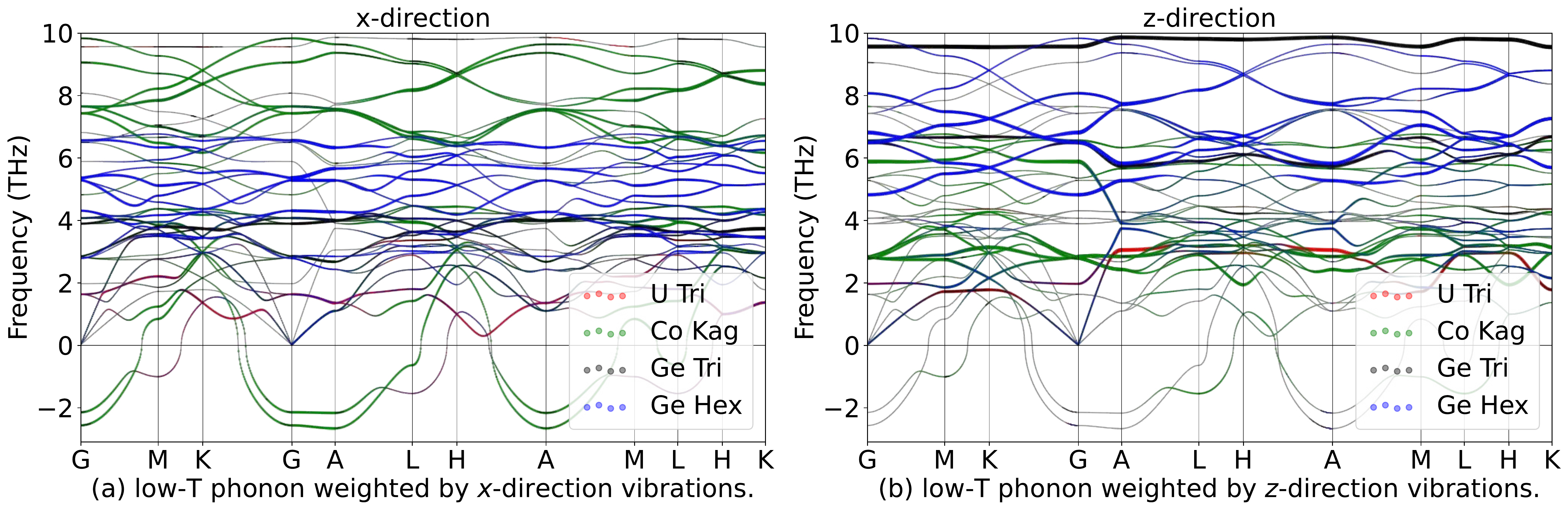}
\hspace{5pt}\label{fig:UCo6Ge6_relaxed_High_xz}
\includegraphics[width=0.85\textwidth]{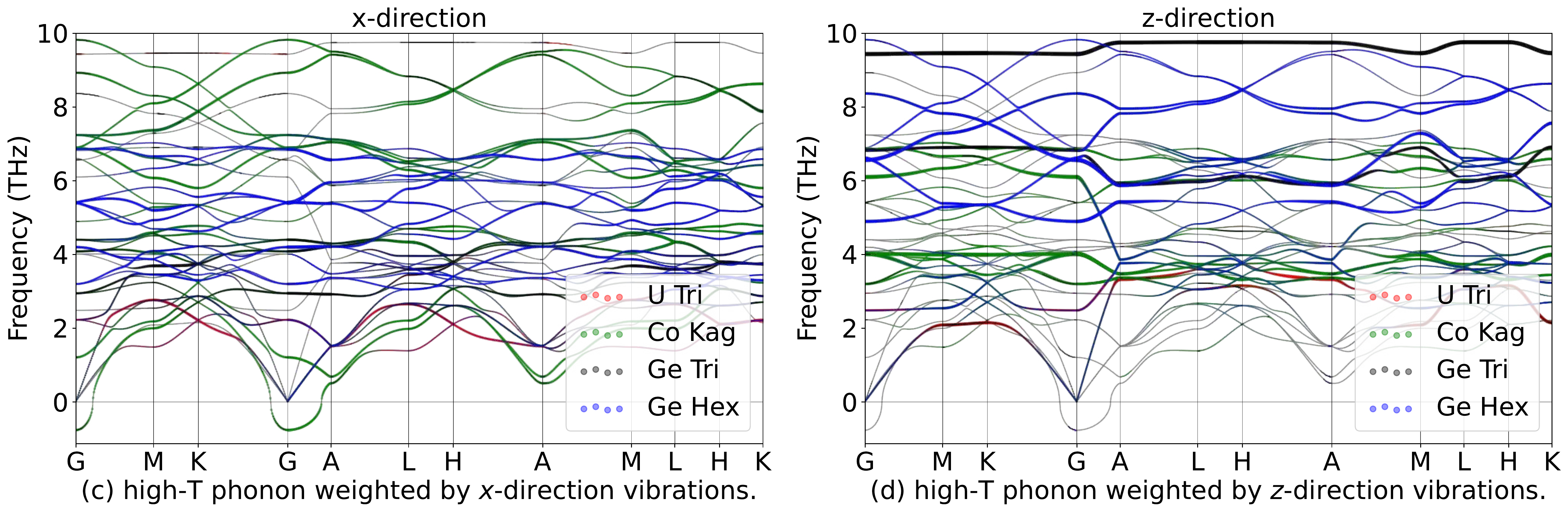}
\caption{Phonon spectrum for UCo$_6$Ge$_6$in in-plane ($x$) and out-of-plane ($z$) directions.}
\label{fig:UCo6Ge6phoxyz}
\end{figure}

\begin{table}[htbp]
\global\long\def\arraystretch{1.12}
\captionsetup{width=.95\textwidth}
\caption{IRREPs at high-symmetry points for UCo$_6$Ge$_6$ phonon spectrum at Low-T.}
\label{tab:191_UCo6Ge6_Low}
\setlength{\tabcolsep}{1.mm}{
}
\end{table}

\begin{table}[htbp]
\global\long\def\arraystretch{1.12}
\captionsetup{width=.95\textwidth}
\caption{IRREPs at high-symmetry points for UCo$_6$Ge$_6$ phonon spectrum at High-T.}
\label{tab:191_UCo6Ge6_High}
\setlength{\tabcolsep}{1.mm}{
%
}
\end{table}
\subsection{\hyperref[smsec:col]{CoSn}}~\label{smssec:CoSn}

\subsubsection{\label{sms2sec:LiCo6Sn6}\hyperref[smsec:col]{\texorpdfstring{LiCo$_6$Sn$_6$}{}}~{\color{green}  (High-T stable, Low-T stable) }}

\begin{table}[htbp]
\global\long\def\arraystretch{1.12}
\captionsetup{width=.85\textwidth}
\caption{Structure parameters of LiCo$_6$Sn$_6$ after relaxation. It contains 465 electrons. }
\label{tab:LiCo6Sn6}
\setlength{\tabcolsep}{4mm}{
%
}
\end{table}

\begin{figure}[htbp]
\centering
\includegraphics[width=0.85\textwidth]{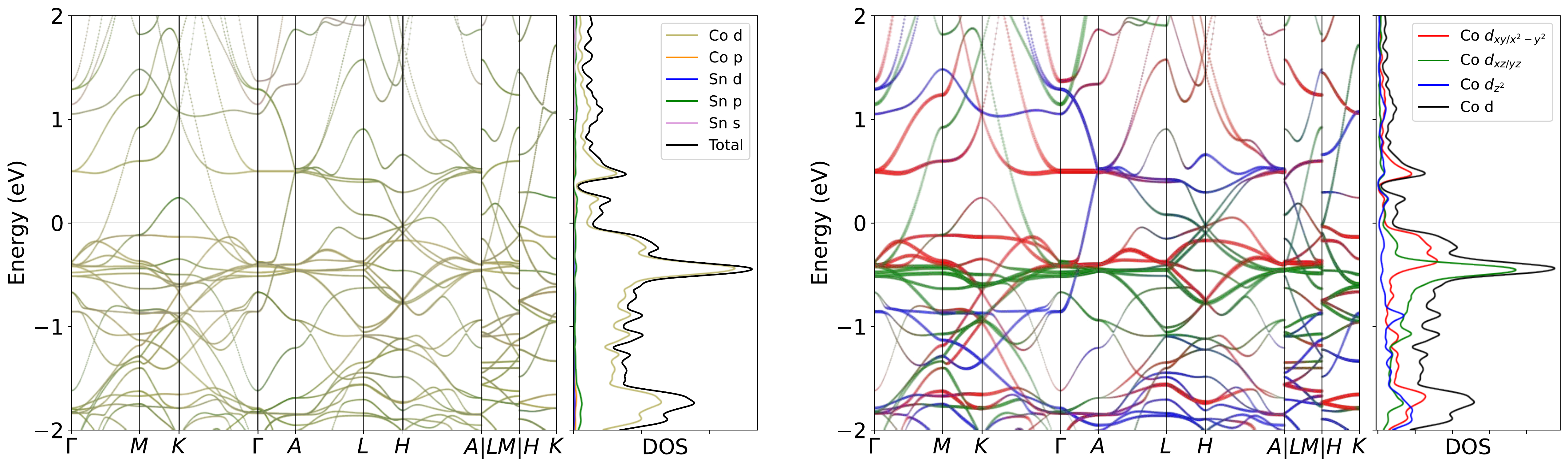}
\caption{Band structure for LiCo$_6$Sn$_6$ without SOC. The projection of Co $3d$ orbitals is also presented. The band structures clearly reveal the dominating of Co $3d$ orbitals  near the Fermi level, as well as the pronounced peak.}
\label{fig:LiCo6Sn6band}
\end{figure}

\begin{figure}[H]
\centering
\subfloat[Relaxed phonon at low-T.]{
\label{fig:LiCo6Sn6_relaxed_Low}
\includegraphics[width=0.45\textwidth]{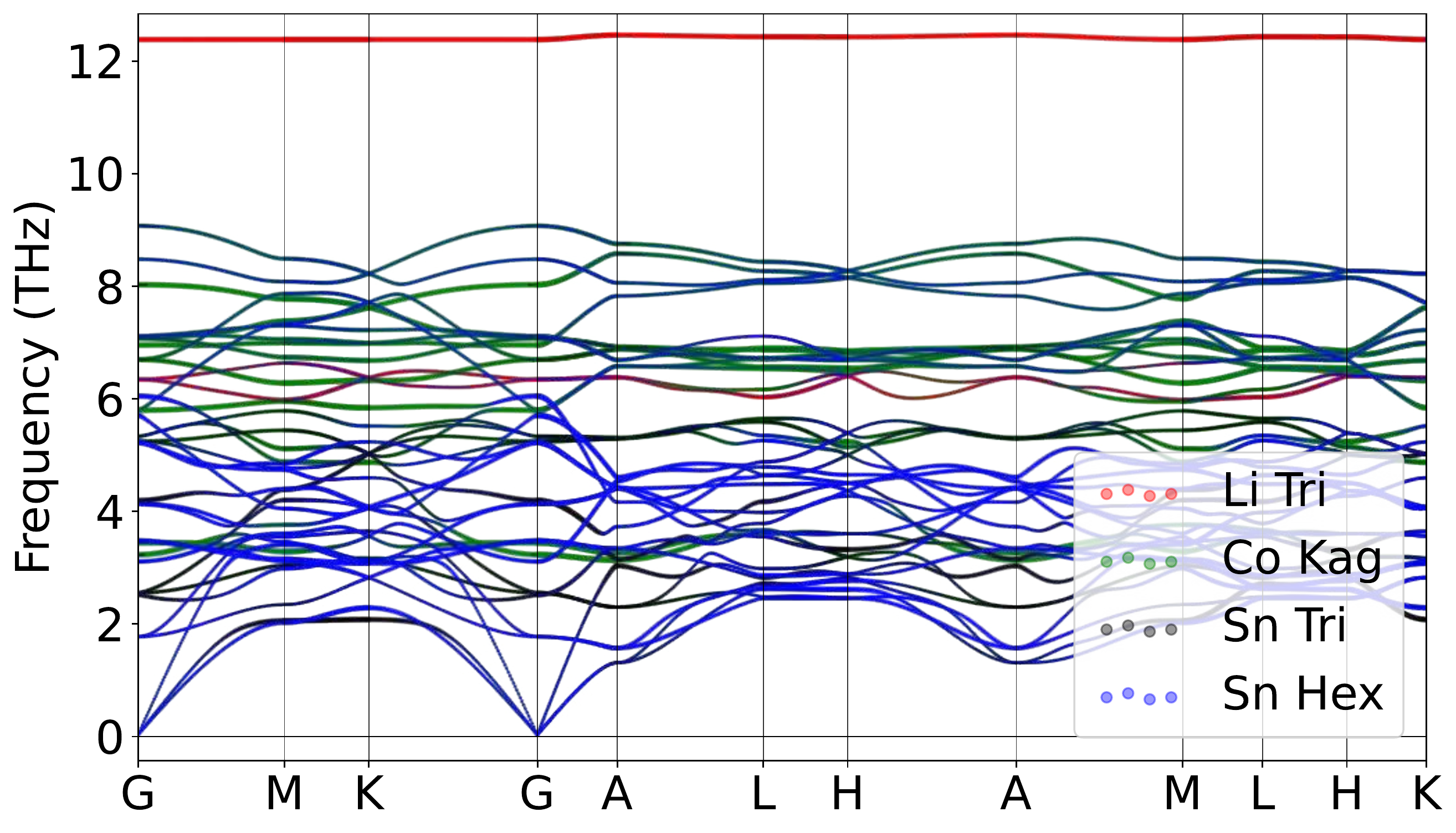}
}
\hspace{5pt}\subfloat[Relaxed phonon at high-T.]{
\label{fig:LiCo6Sn6_relaxed_High}
\includegraphics[width=0.45\textwidth]{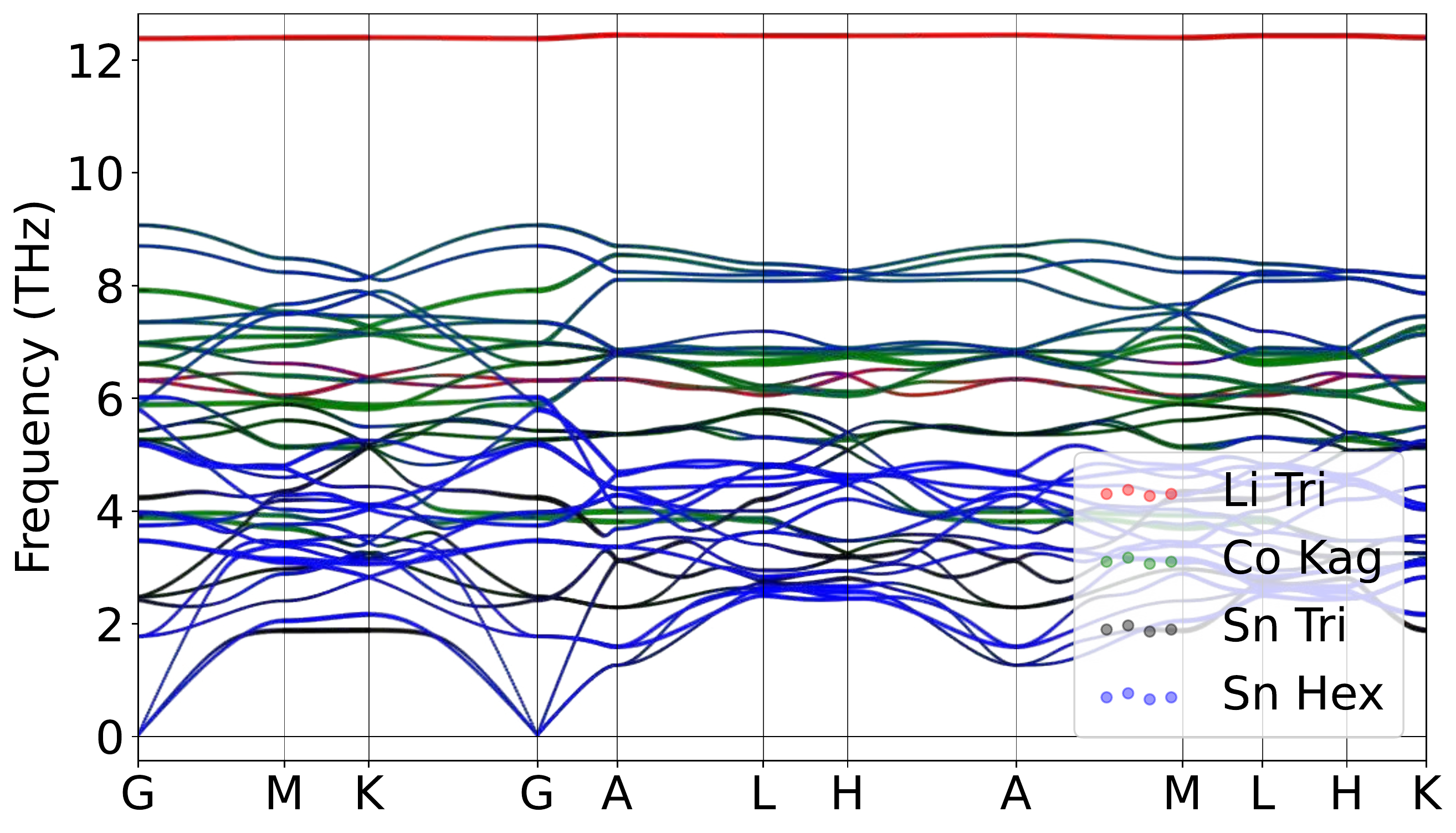}
}
\caption{Atomically projected phonon spectrum for LiCo$_6$Sn$_6$.}
\label{fig:LiCo6Sn6pho}
\end{figure}

\begin{figure}[H]
\centering
\label{fig:LiCo6Sn6_relaxed_Low_xz}
\includegraphics[width=0.85\textwidth]{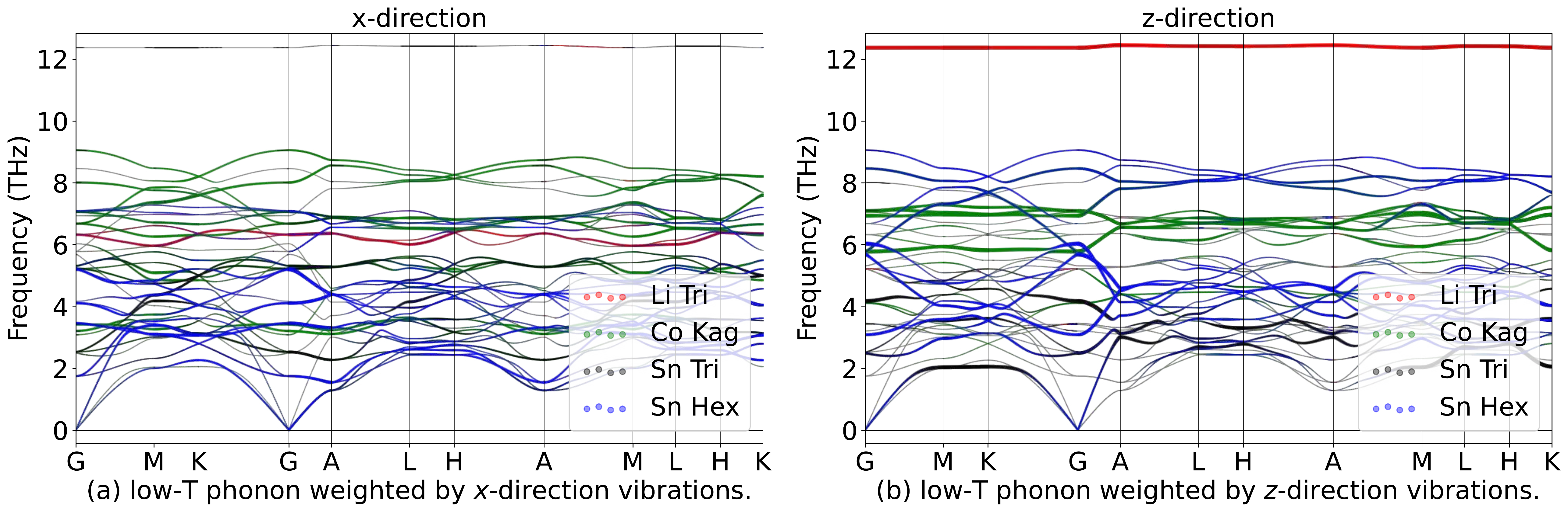}
\hspace{5pt}\label{fig:LiCo6Sn6_relaxed_High_xz}
\includegraphics[width=0.85\textwidth]{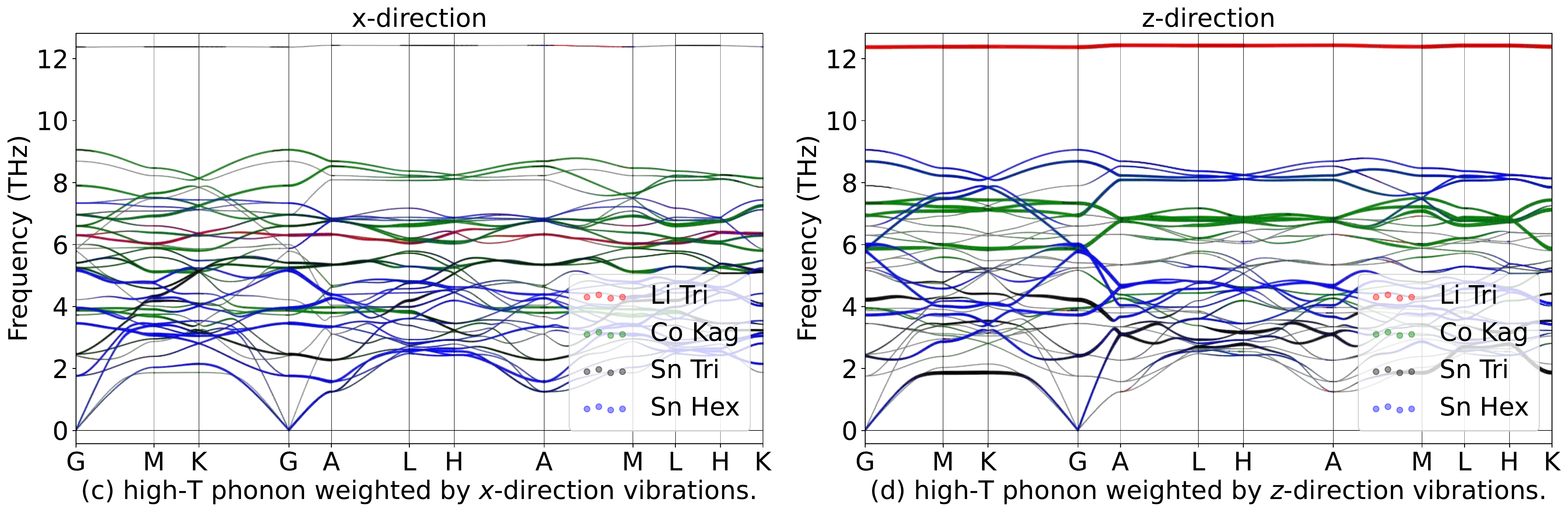}
\caption{Phonon spectrum for LiCo$_6$Sn$_6$in in-plane ($x$) and out-of-plane ($z$) directions.}
\label{fig:LiCo6Sn6phoxyz}
\end{figure}

\begin{table}[htbp]
\global\long\def\arraystretch{1.12}
\captionsetup{width=.95\textwidth}
\caption{IRREPs at high-symmetry points for LiCo$_6$Sn$_6$ phonon spectrum at Low-T.}
\label{tab:191_LiCo6Sn6_Low}
\setlength{\tabcolsep}{1.mm}{
}
\end{table}

\begin{table}[htbp]
\global\long\def\arraystretch{1.12}
\captionsetup{width=.95\textwidth}
\caption{IRREPs at high-symmetry points for LiCo$_6$Sn$_6$ phonon spectrum at High-T.}
\label{tab:191_LiCo6Sn6_High}
\setlength{\tabcolsep}{1.mm}{
%
}
\end{table}
\subsubsection{\label{sms2sec:MgCo6Sn6}\hyperref[smsec:col]{\texorpdfstring{MgCo$_6$Sn$_6$}{}}~{\color{green}  (High-T stable, Low-T stable) }}

\begin{table}[htbp]
\global\long\def\arraystretch{1.12}
\captionsetup{width=.85\textwidth}
\caption{Structure parameters of MgCo$_6$Sn$_6$ after relaxation. It contains 474 electrons. }
\label{tab:MgCo6Sn6}
\setlength{\tabcolsep}{4mm}{
%
}
\end{table}

\begin{figure}[htbp]
\centering
\includegraphics[width=0.85\textwidth]{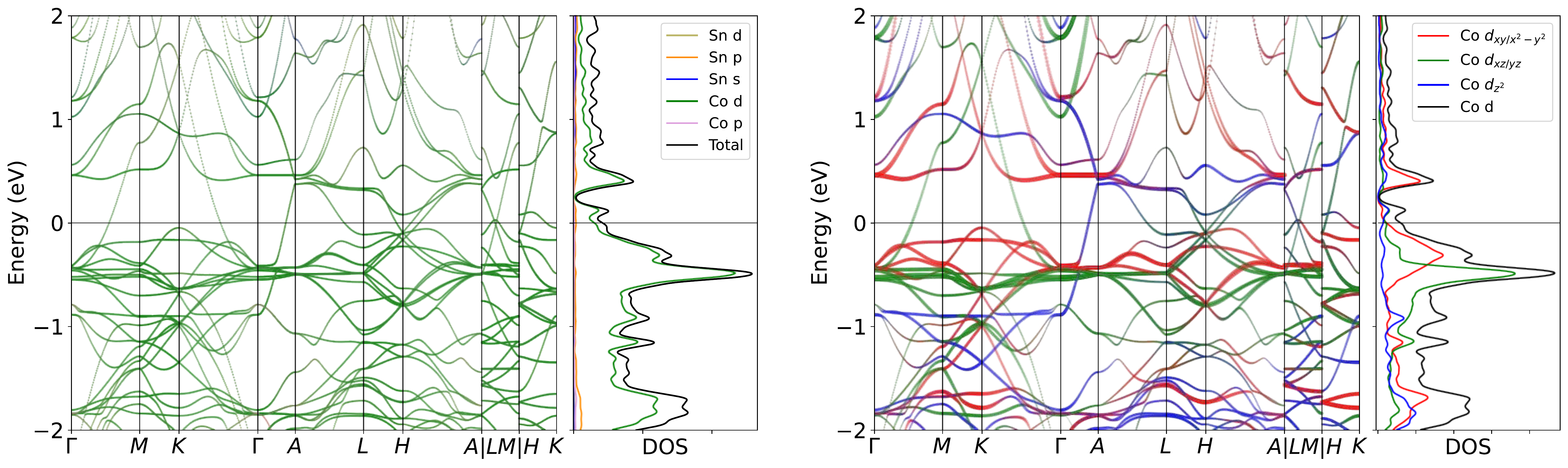}
\caption{Band structure for MgCo$_6$Sn$_6$ without SOC. The projection of Co $3d$ orbitals is also presented. The band structures clearly reveal the dominating of Co $3d$ orbitals  near the Fermi level, as well as the pronounced peak.}
\label{fig:MgCo6Sn6band}
\end{figure}

\begin{figure}[H]
\centering
\subfloat[Relaxed phonon at low-T.]{
\label{fig:MgCo6Sn6_relaxed_Low}
\includegraphics[width=0.45\textwidth]{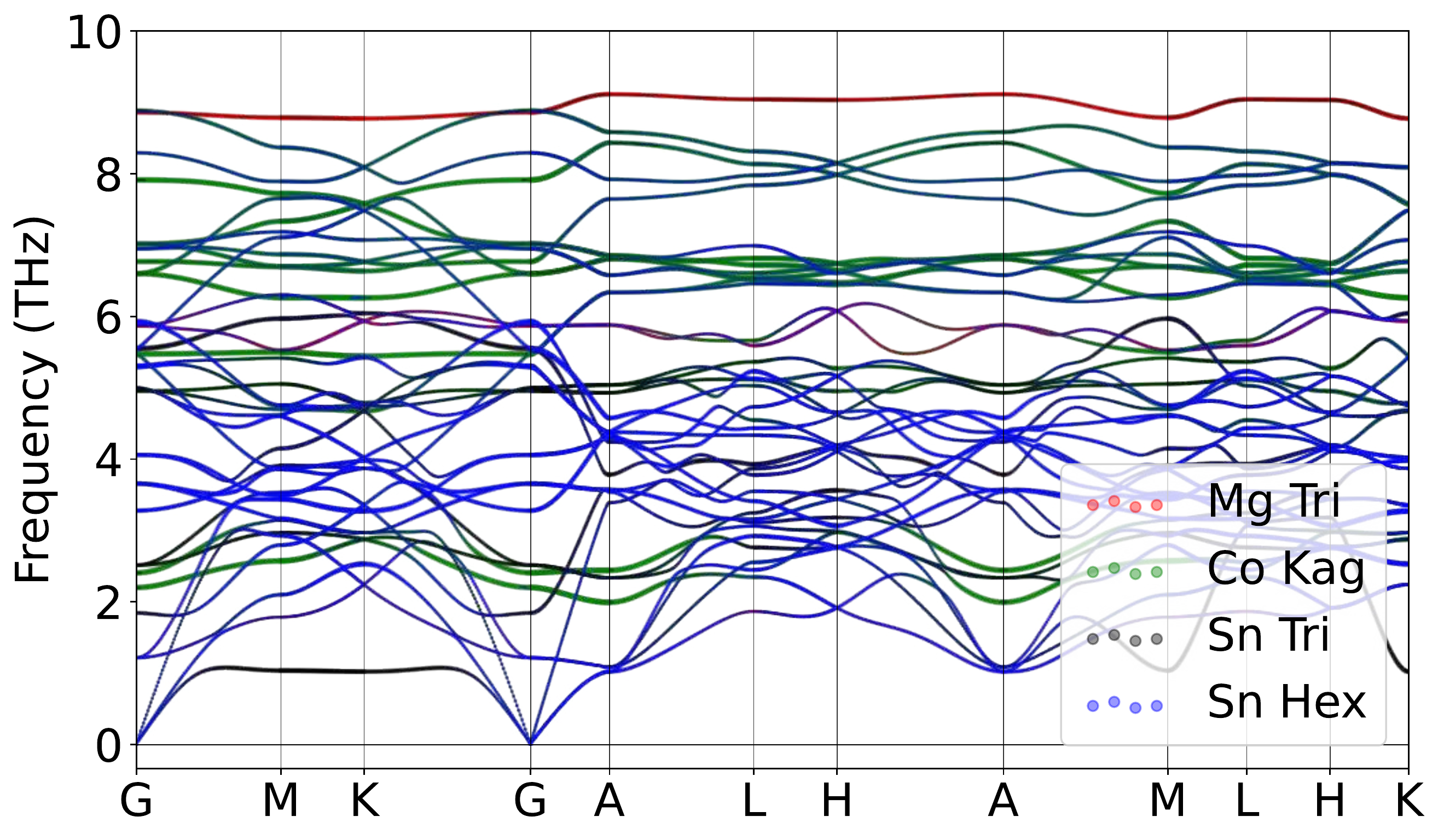}
}
\hspace{5pt}\subfloat[Relaxed phonon at high-T.]{
\label{fig:MgCo6Sn6_relaxed_High}
\includegraphics[width=0.45\textwidth]{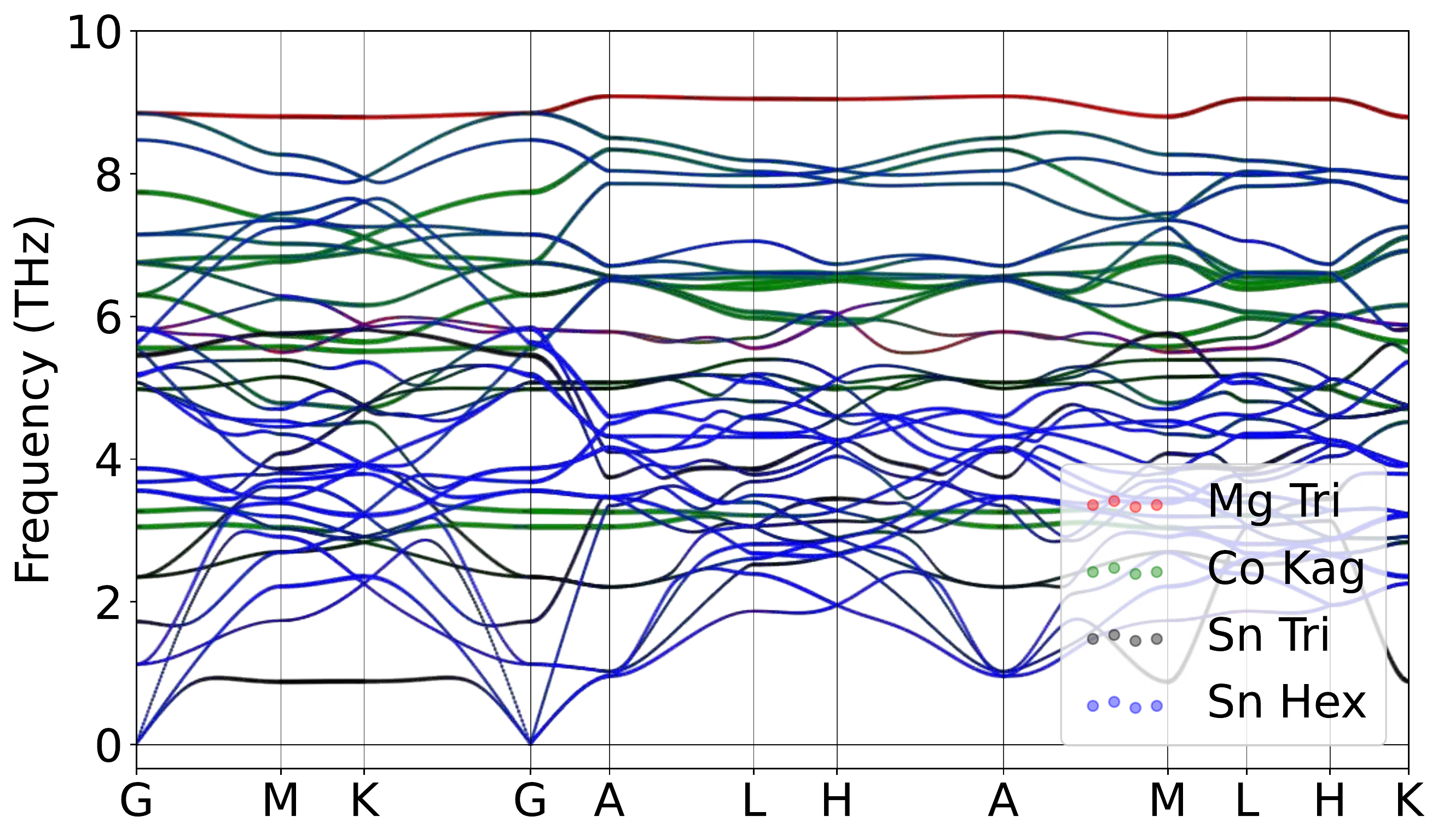}
}
\caption{Atomically projected phonon spectrum for MgCo$_6$Sn$_6$.}
\label{fig:MgCo6Sn6pho}
\end{figure}

\begin{figure}[H]
\centering
\label{fig:MgCo6Sn6_relaxed_Low_xz}
\includegraphics[width=0.85\textwidth]{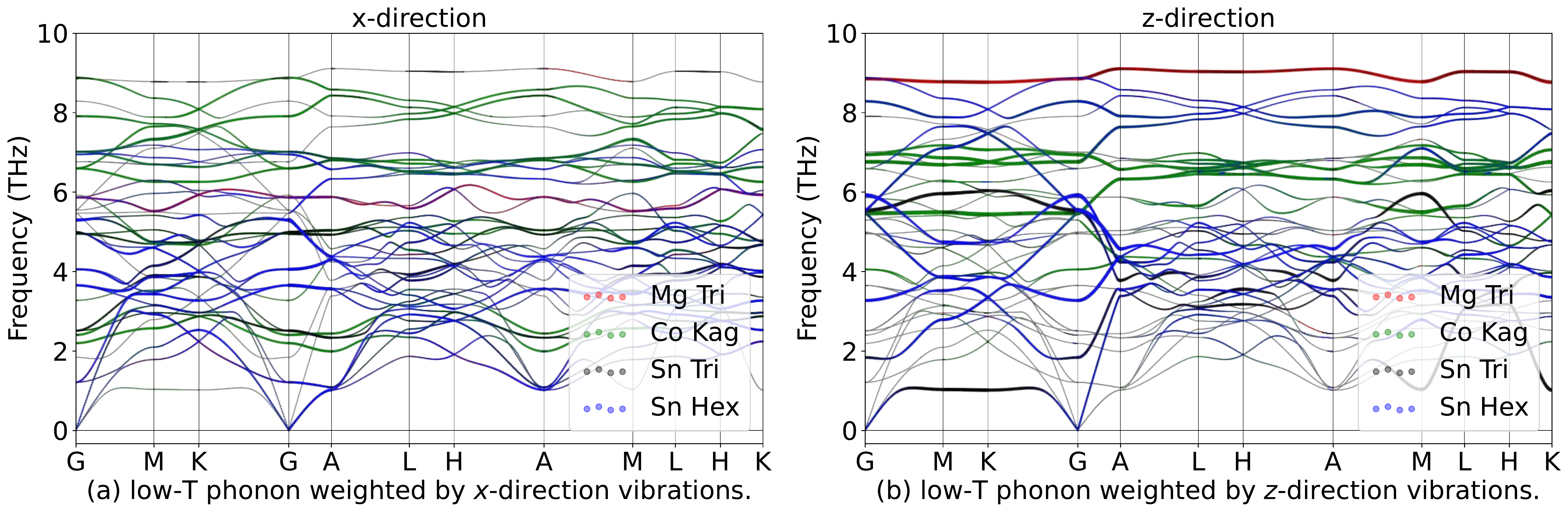}
\hspace{5pt}\label{fig:MgCo6Sn6_relaxed_High_xz}
\includegraphics[width=0.85\textwidth]{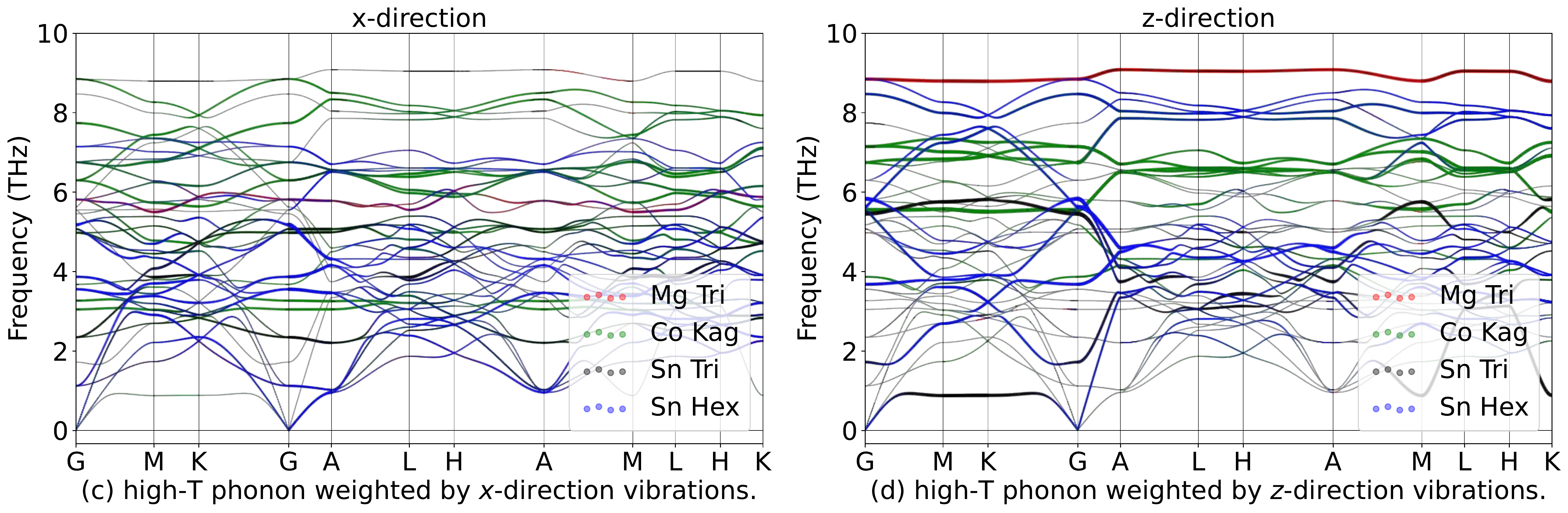}
\caption{Phonon spectrum for MgCo$_6$Sn$_6$in in-plane ($x$) and out-of-plane ($z$) directions.}
\label{fig:MgCo6Sn6phoxyz}
\end{figure}

\begin{table}[htbp]
\global\long\def\arraystretch{1.12}
\captionsetup{width=.95\textwidth}
\caption{IRREPs at high-symmetry points for MgCo$_6$Sn$_6$ phonon spectrum at Low-T.}
\label{tab:191_MgCo6Sn6_Low}
\setlength{\tabcolsep}{1.mm}{
}
\end{table}

\begin{table}[htbp]
\global\long\def\arraystretch{1.12}
\captionsetup{width=.95\textwidth}
\caption{IRREPs at high-symmetry points for MgCo$_6$Sn$_6$ phonon spectrum at High-T.}
\label{tab:191_MgCo6Sn6_High}
\setlength{\tabcolsep}{1.mm}{
%
}
\end{table}
\subsubsection{\label{sms2sec:CaCo6Sn6}\hyperref[smsec:col]{\texorpdfstring{CaCo$_6$Sn$_6$}{}}~{\color{green}  (High-T stable, Low-T stable) }}

\begin{table}[htbp]
\global\long\def\arraystretch{1.12}
\captionsetup{width=.85\textwidth}
\caption{Structure parameters of CaCo$_6$Sn$_6$ after relaxation. It contains 482 electrons. }
\label{tab:CaCo6Sn6}
\setlength{\tabcolsep}{4mm}{
%
}
\end{table}

\begin{figure}[htbp]
\centering
\includegraphics[width=0.85\textwidth]{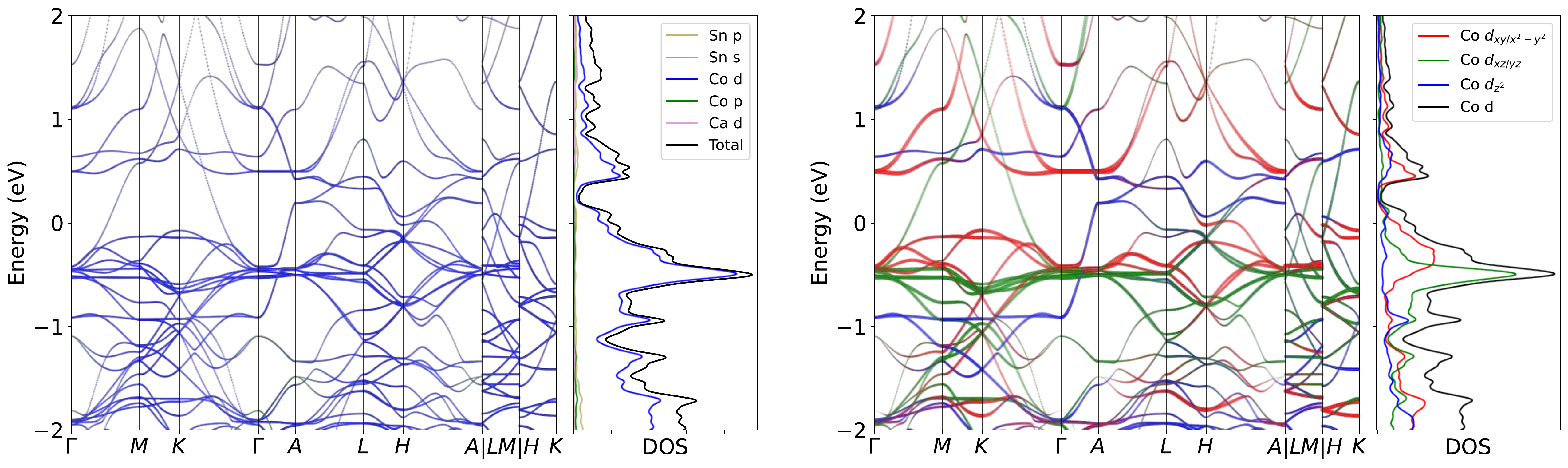}
\caption{Band structure for CaCo$_6$Sn$_6$ without SOC. The projection of Co $3d$ orbitals is also presented. The band structures clearly reveal the dominating of Co $3d$ orbitals  near the Fermi level, as well as the pronounced peak.}
\label{fig:CaCo6Sn6band}
\end{figure}

\begin{figure}[H]
\centering
\subfloat[Relaxed phonon at low-T.]{
\label{fig:CaCo6Sn6_relaxed_Low}
\includegraphics[width=0.45\textwidth]{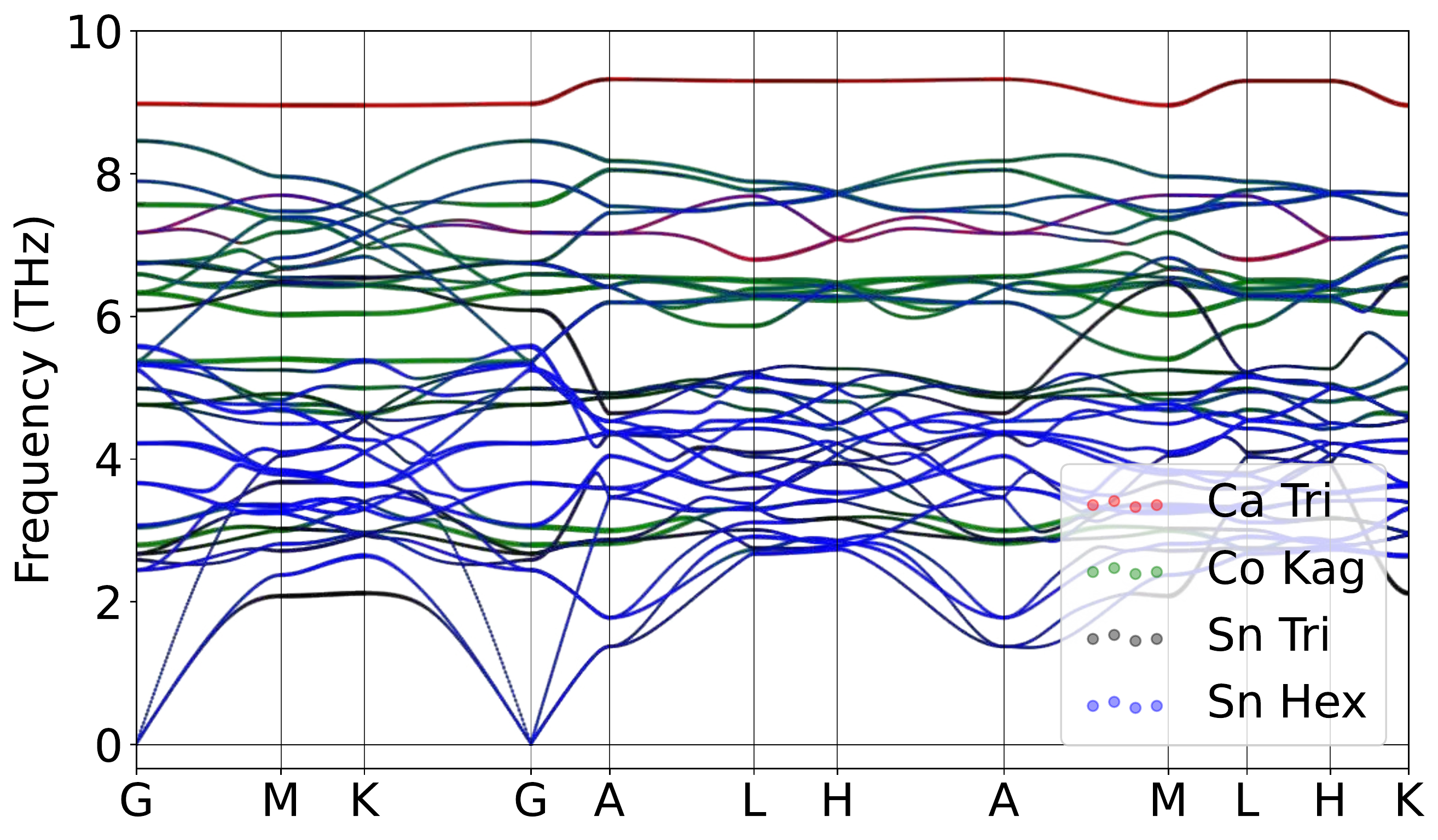}
}
\hspace{5pt}\subfloat[Relaxed phonon at high-T.]{
\label{fig:CaCo6Sn6_relaxed_High}
\includegraphics[width=0.45\textwidth]{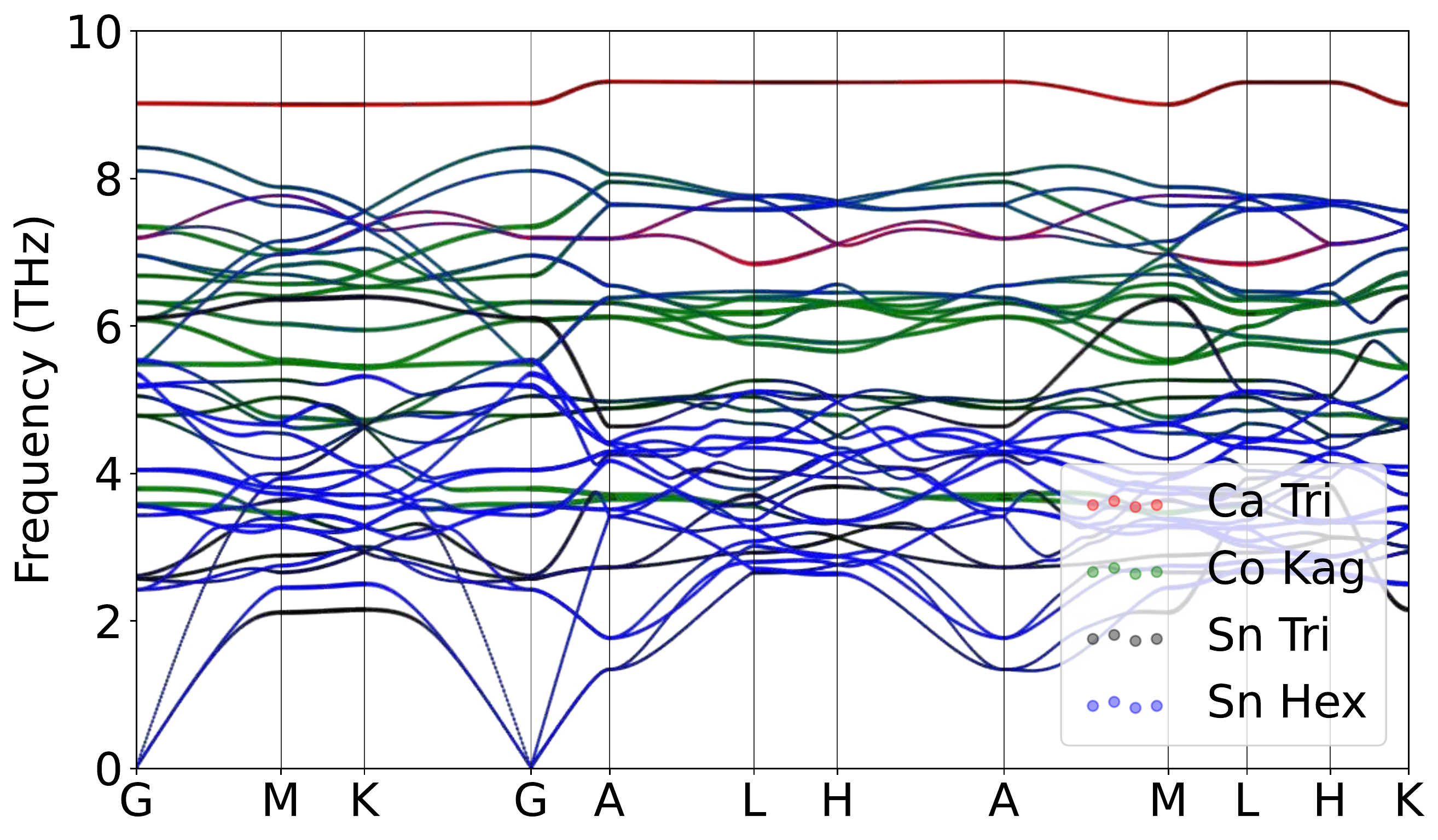}
}
\caption{Atomically projected phonon spectrum for CaCo$_6$Sn$_6$.}
\label{fig:CaCo6Sn6pho}
\end{figure}

\begin{figure}[H]
\centering
\label{fig:CaCo6Sn6_relaxed_Low_xz}
\includegraphics[width=0.85\textwidth]{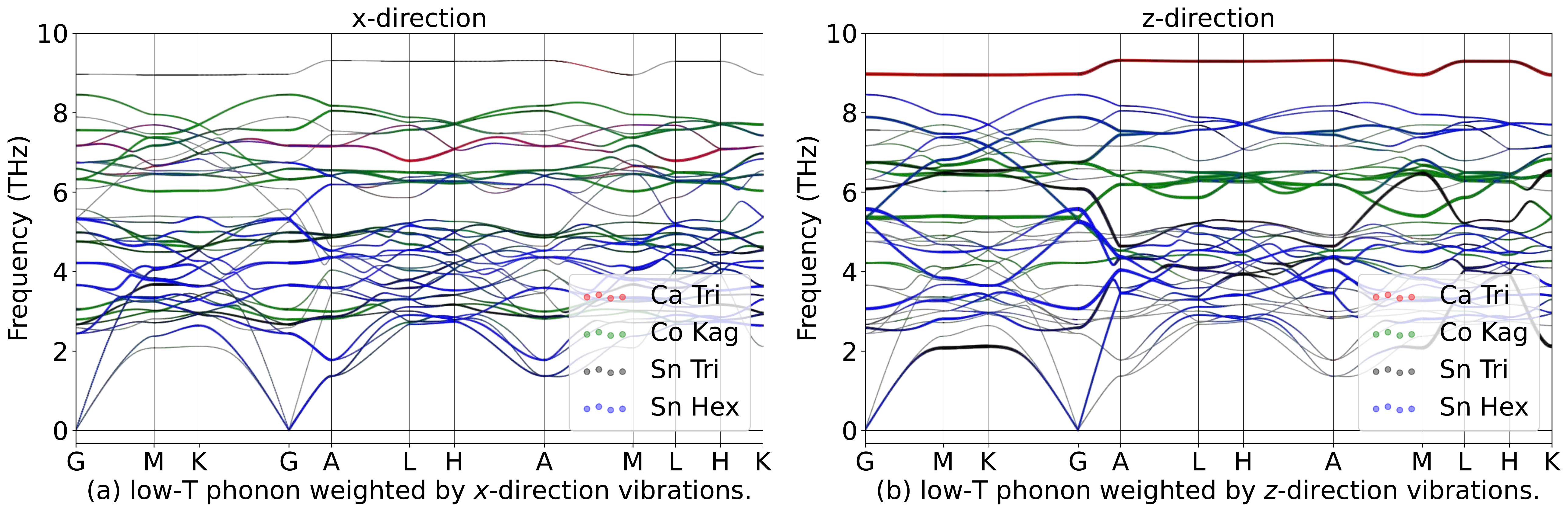}
\hspace{5pt}\label{fig:CaCo6Sn6_relaxed_High_xz}
\includegraphics[width=0.85\textwidth]{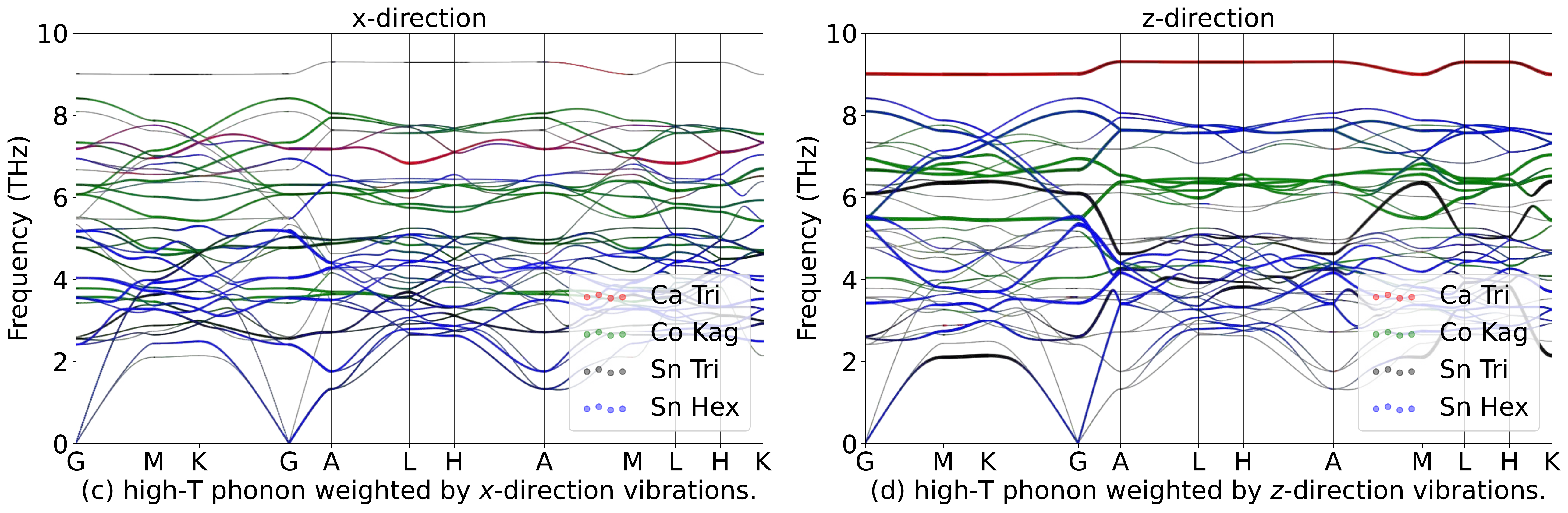}
\caption{Phonon spectrum for CaCo$_6$Sn$_6$in in-plane ($x$) and out-of-plane ($z$) directions.}
\label{fig:CaCo6Sn6phoxyz}
\end{figure}

\begin{table}[htbp]
\global\long\def\arraystretch{1.12}
\captionsetup{width=.95\textwidth}
\caption{IRREPs at high-symmetry points for CaCo$_6$Sn$_6$ phonon spectrum at Low-T.}
\label{tab:191_CaCo6Sn6_Low}
\setlength{\tabcolsep}{1.mm}{
}
\end{table}

\begin{table}[htbp]
\global\long\def\arraystretch{1.12}
\captionsetup{width=.95\textwidth}
\caption{IRREPs at high-symmetry points for CaCo$_6$Sn$_6$ phonon spectrum at High-T.}
\label{tab:191_CaCo6Sn6_High}
\setlength{\tabcolsep}{1.mm}{
%
}
\end{table}
\subsubsection{\label{sms2sec:ScCo6Sn6}\hyperref[smsec:col]{\texorpdfstring{ScCo$_6$Sn$_6$}{}}~{\color{green}  (High-T stable, Low-T stable) }}

\begin{table}[htbp]
\global\long\def\arraystretch{1.12}
\captionsetup{width=.85\textwidth}
\caption{Structure parameters of ScCo$_6$Sn$_6$ after relaxation. It contains 483 electrons. }
\label{tab:ScCo6Sn6}
\setlength{\tabcolsep}{4mm}{
%
}
\end{table}

\begin{figure}[htbp]
\centering
\includegraphics[width=0.85\textwidth]{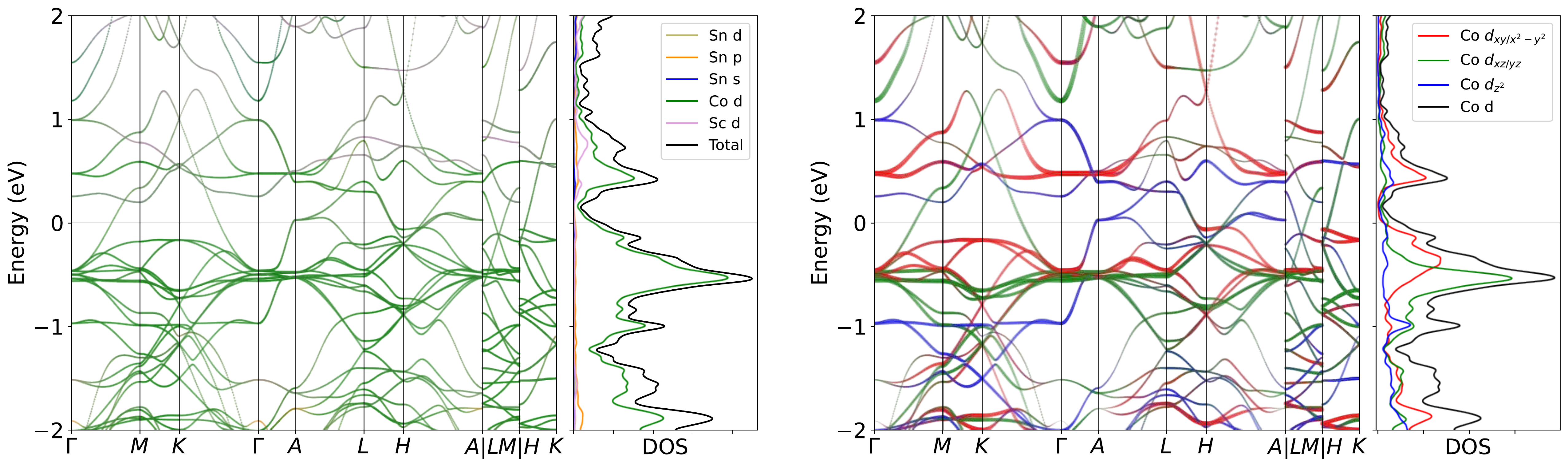}
\caption{Band structure for ScCo$_6$Sn$_6$ without SOC. The projection of Co $3d$ orbitals is also presented. The band structures clearly reveal the dominating of Co $3d$ orbitals  near the Fermi level, as well as the pronounced peak.}
\label{fig:ScCo6Sn6band}
\end{figure}

\begin{figure}[H]
\centering
\subfloat[Relaxed phonon at low-T.]{
\label{fig:ScCo6Sn6_relaxed_Low}
\includegraphics[width=0.45\textwidth]{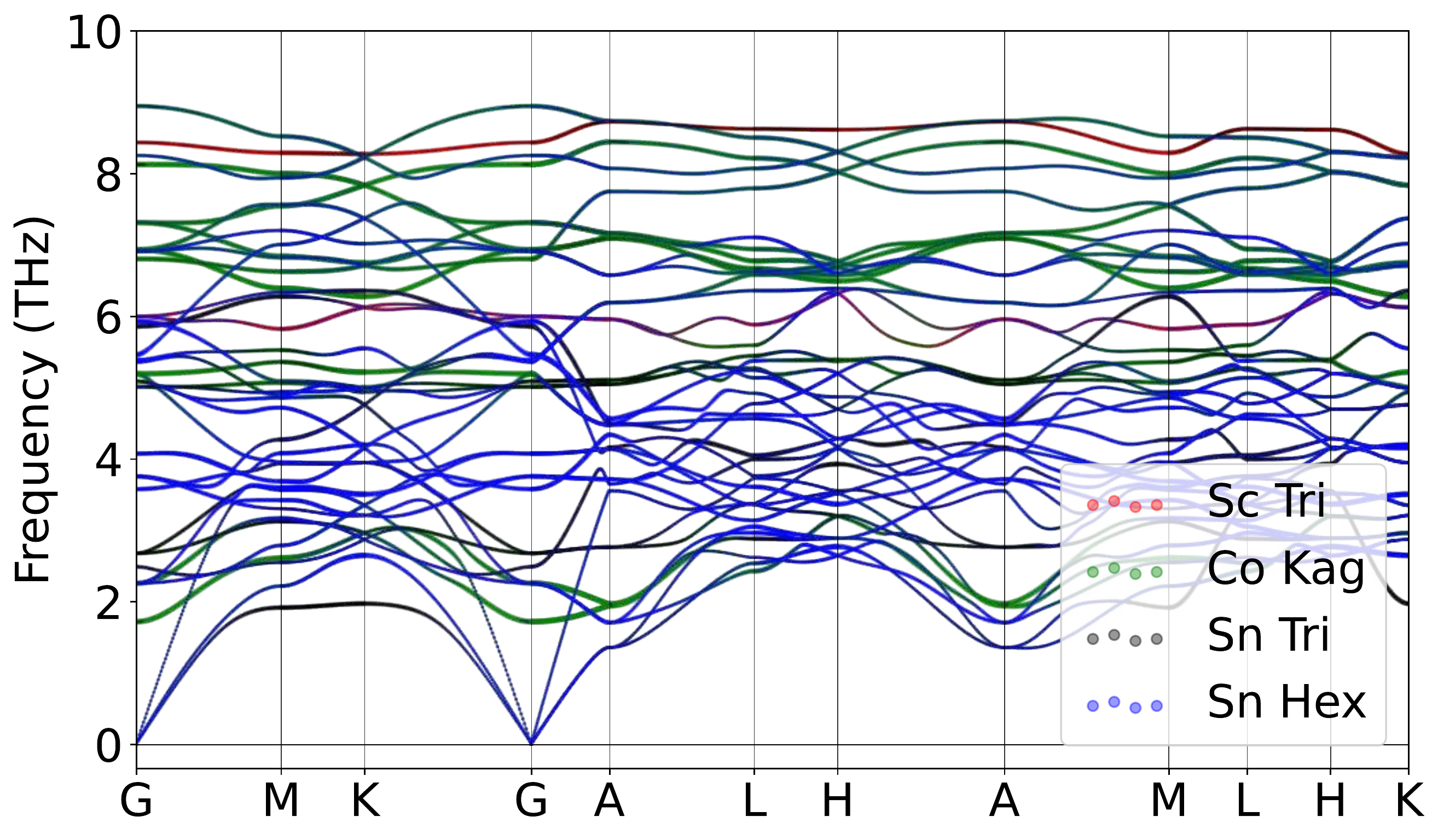}
}
\hspace{5pt}\subfloat[Relaxed phonon at high-T.]{
\label{fig:ScCo6Sn6_relaxed_High}
\includegraphics[width=0.45\textwidth]{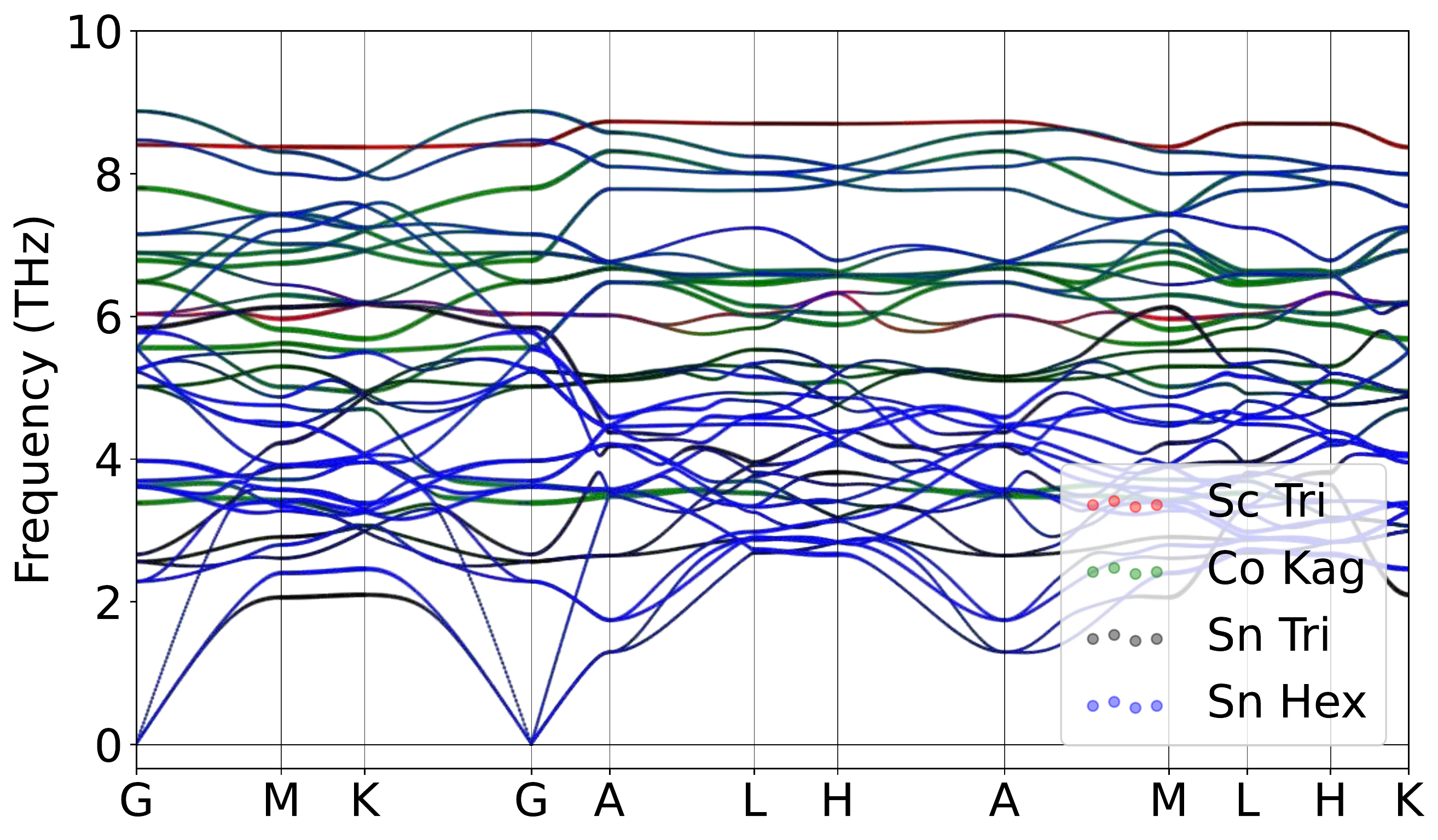}
}
\caption{Atomically projected phonon spectrum for ScCo$_6$Sn$_6$.}
\label{fig:ScCo6Sn6pho}
\end{figure}

\begin{figure}[H]
\centering
\label{fig:ScCo6Sn6_relaxed_Low_xz}
\includegraphics[width=0.85\textwidth]{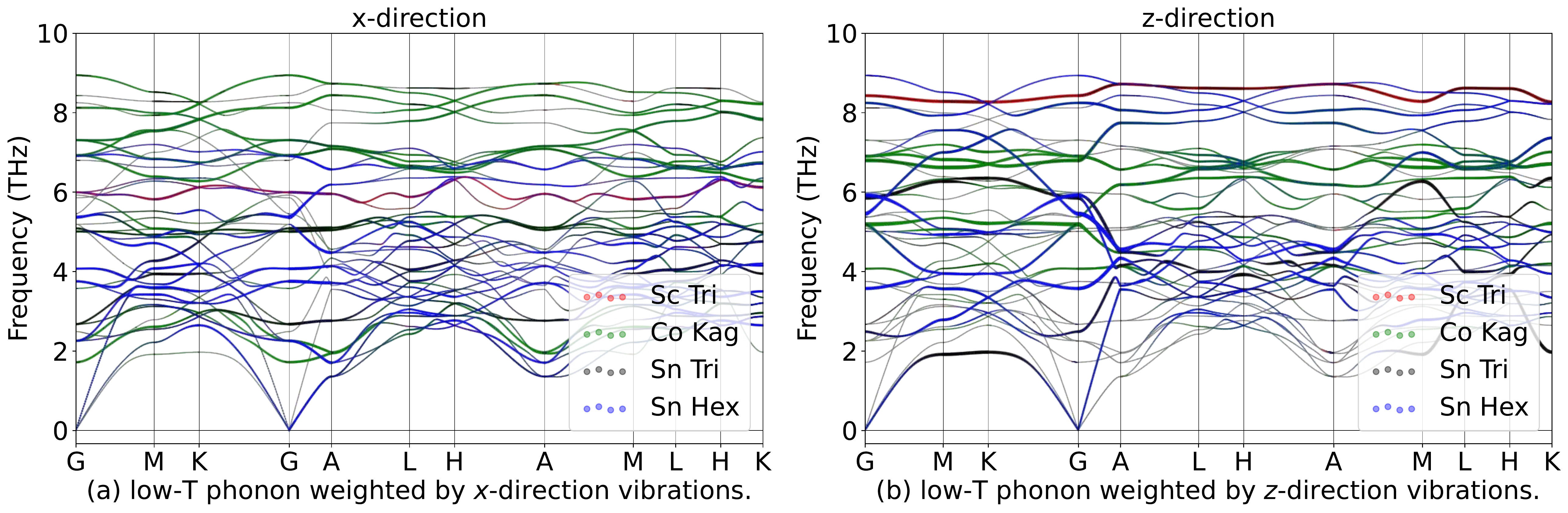}
\hspace{5pt}\label{fig:ScCo6Sn6_relaxed_High_xz}
\includegraphics[width=0.85\textwidth]{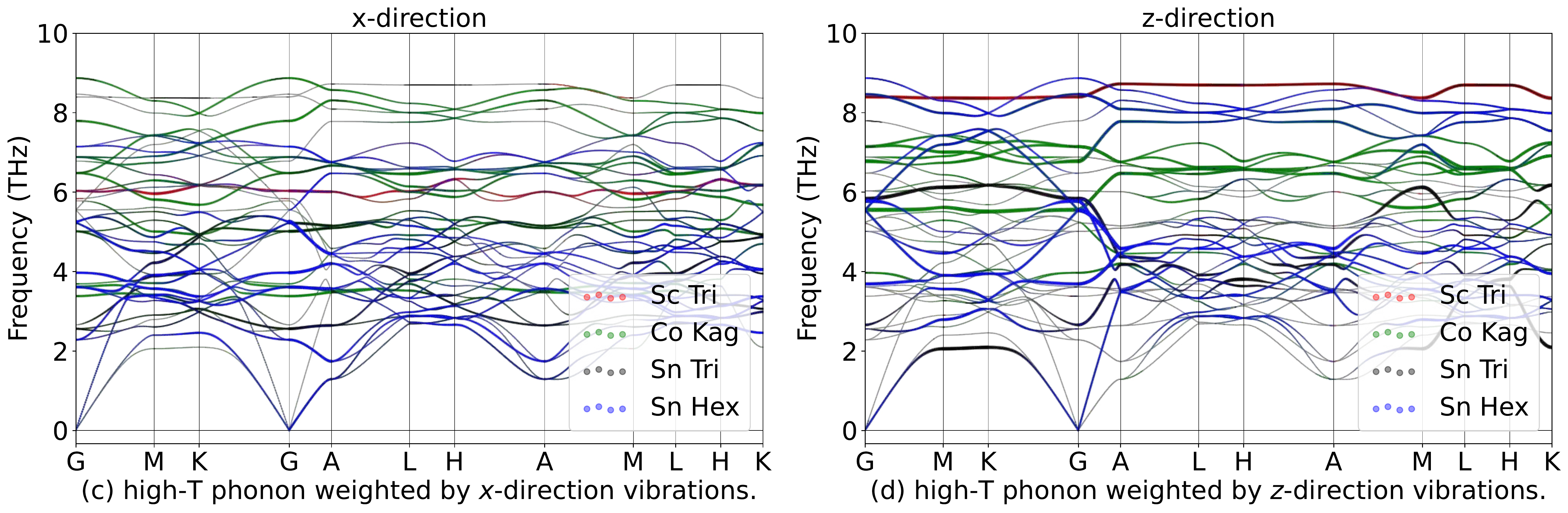}
\caption{Phonon spectrum for ScCo$_6$Sn$_6$in in-plane ($x$) and out-of-plane ($z$) directions.}
\label{fig:ScCo6Sn6phoxyz}
\end{figure}

\begin{table}[htbp]
\global\long\def\arraystretch{1.12}
\captionsetup{width=.95\textwidth}
\caption{IRREPs at high-symmetry points for ScCo$_6$Sn$_6$ phonon spectrum at Low-T.}
\label{tab:191_ScCo6Sn6_Low}
\setlength{\tabcolsep}{1.mm}{
}
\end{table}

\begin{table}[htbp]
\global\long\def\arraystretch{1.12}
\captionsetup{width=.95\textwidth}
\caption{IRREPs at high-symmetry points for ScCo$_6$Sn$_6$ phonon spectrum at High-T.}
\label{tab:191_ScCo6Sn6_High}
\setlength{\tabcolsep}{1.mm}{
%
}
\end{table}
\subsubsection{\label{sms2sec:TiCo6Sn6}\hyperref[smsec:col]{\texorpdfstring{TiCo$_6$Sn$_6$}{}}~{\color{green}  (High-T stable, Low-T stable) }}

\begin{table}[htbp]
\global\long\def\arraystretch{1.12}
\captionsetup{width=.85\textwidth}
\caption{Structure parameters of TiCo$_6$Sn$_6$ after relaxation. It contains 484 electrons. }
\label{tab:TiCo6Sn6}
\setlength{\tabcolsep}{4mm}{
%
}
\end{table}

\begin{figure}[htbp]
\centering
\includegraphics[width=0.85\textwidth]{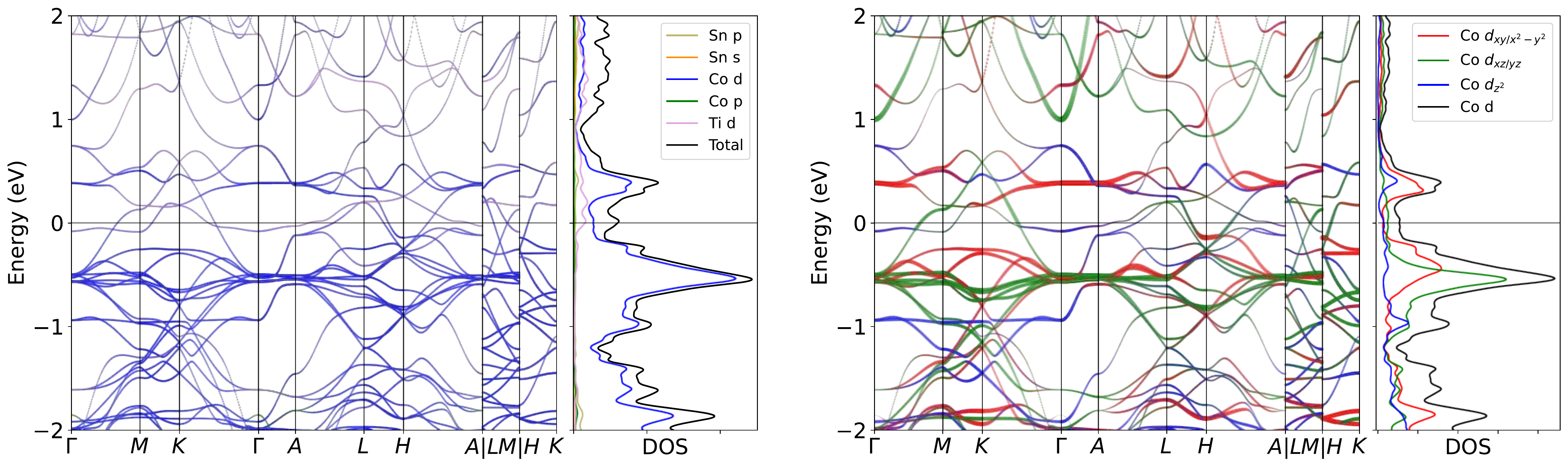}
\caption{Band structure for TiCo$_6$Sn$_6$ without SOC. The projection of Co $3d$ orbitals is also presented. The band structures clearly reveal the dominating of Co $3d$ orbitals  near the Fermi level, as well as the pronounced peak.}
\label{fig:TiCo6Sn6band}
\end{figure}

\begin{figure}[H]
\centering
\subfloat[Relaxed phonon at low-T.]{
\label{fig:TiCo6Sn6_relaxed_Low}
\includegraphics[width=0.45\textwidth]{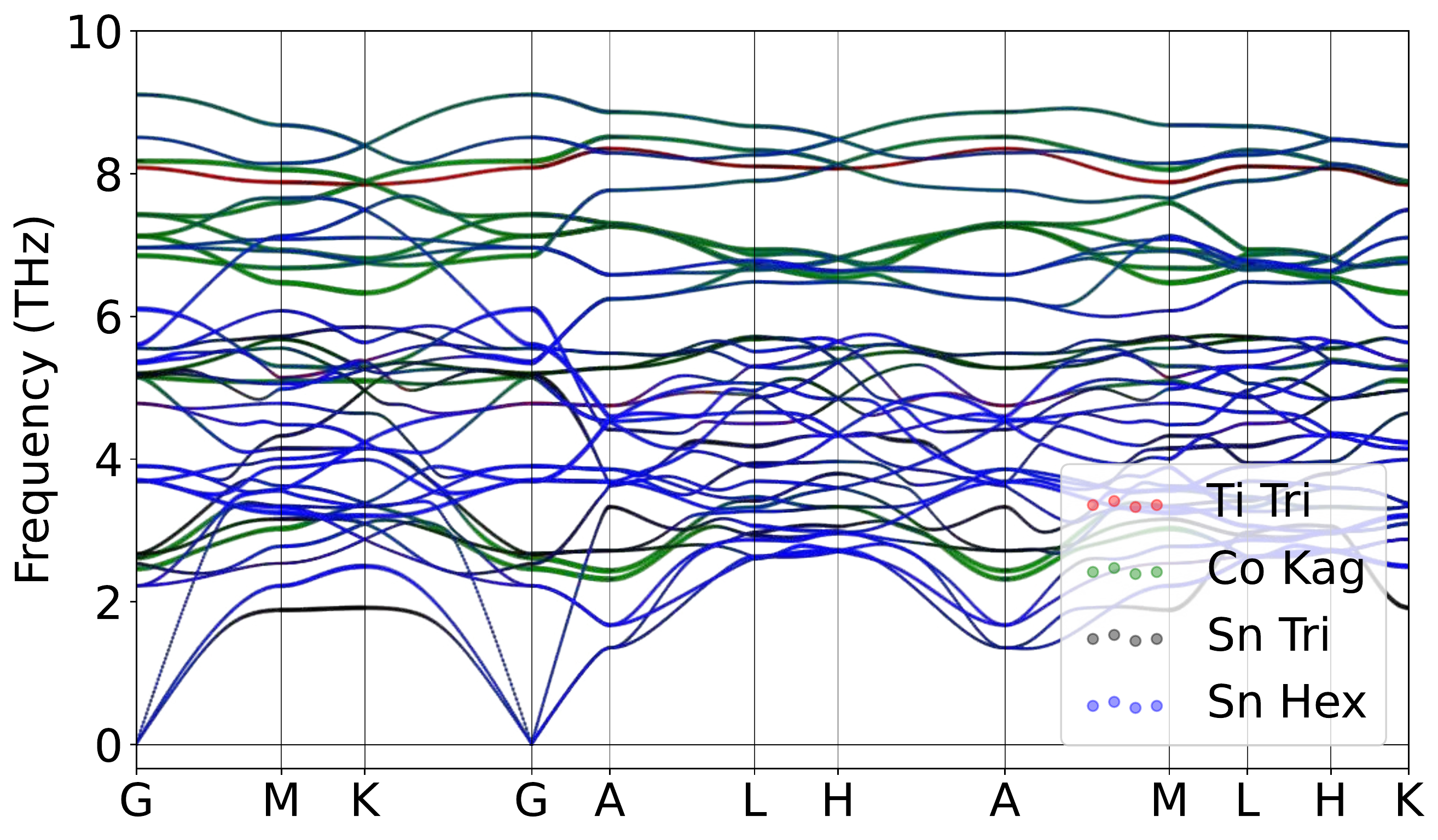}
}
\hspace{5pt}\subfloat[Relaxed phonon at high-T.]{
\label{fig:TiCo6Sn6_relaxed_High}
\includegraphics[width=0.45\textwidth]{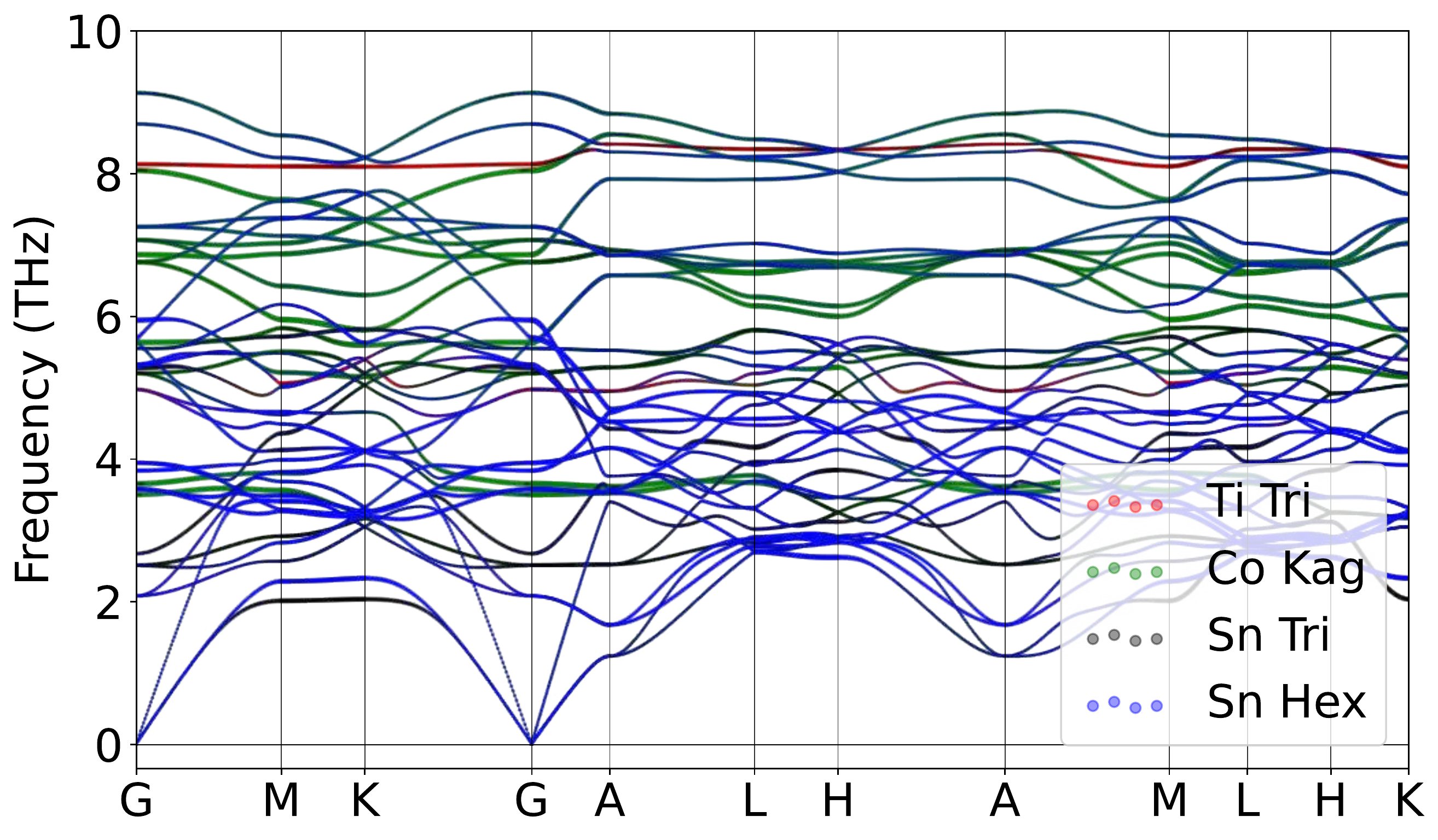}
}
\caption{Atomically projected phonon spectrum for TiCo$_6$Sn$_6$.}
\label{fig:TiCo6Sn6pho}
\end{figure}

\begin{figure}[H]
\centering
\label{fig:TiCo6Sn6_relaxed_Low_xz}
\includegraphics[width=0.85\textwidth]{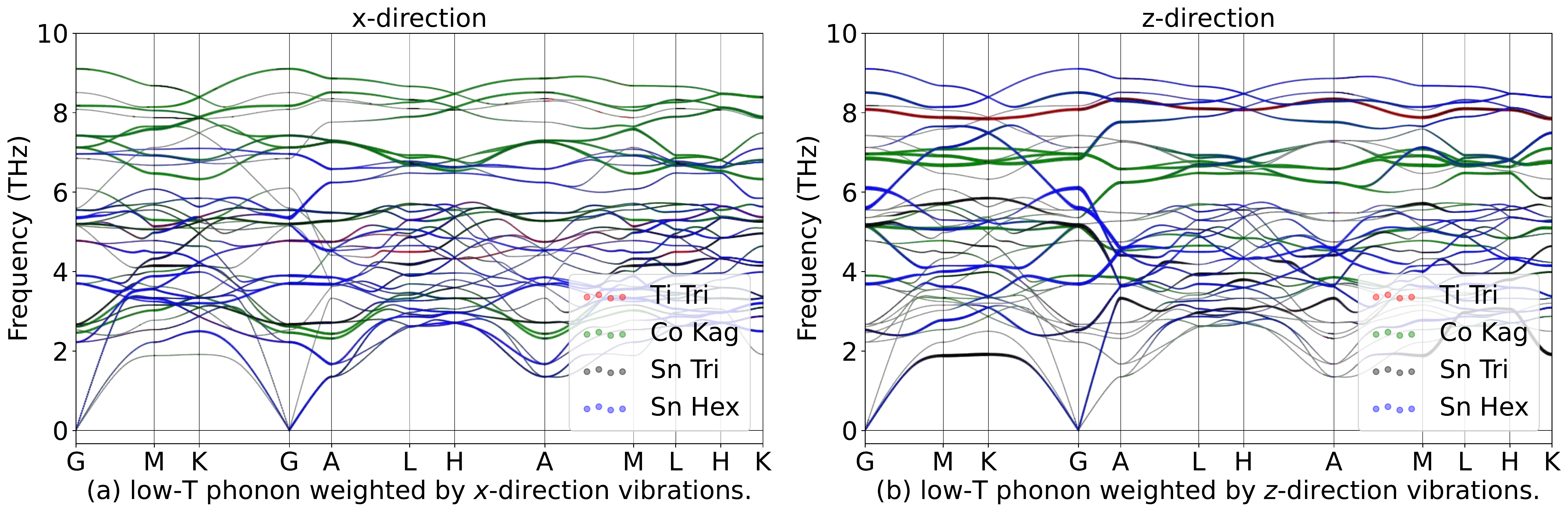}
\hspace{5pt}\label{fig:TiCo6Sn6_relaxed_High_xz}
\includegraphics[width=0.85\textwidth]{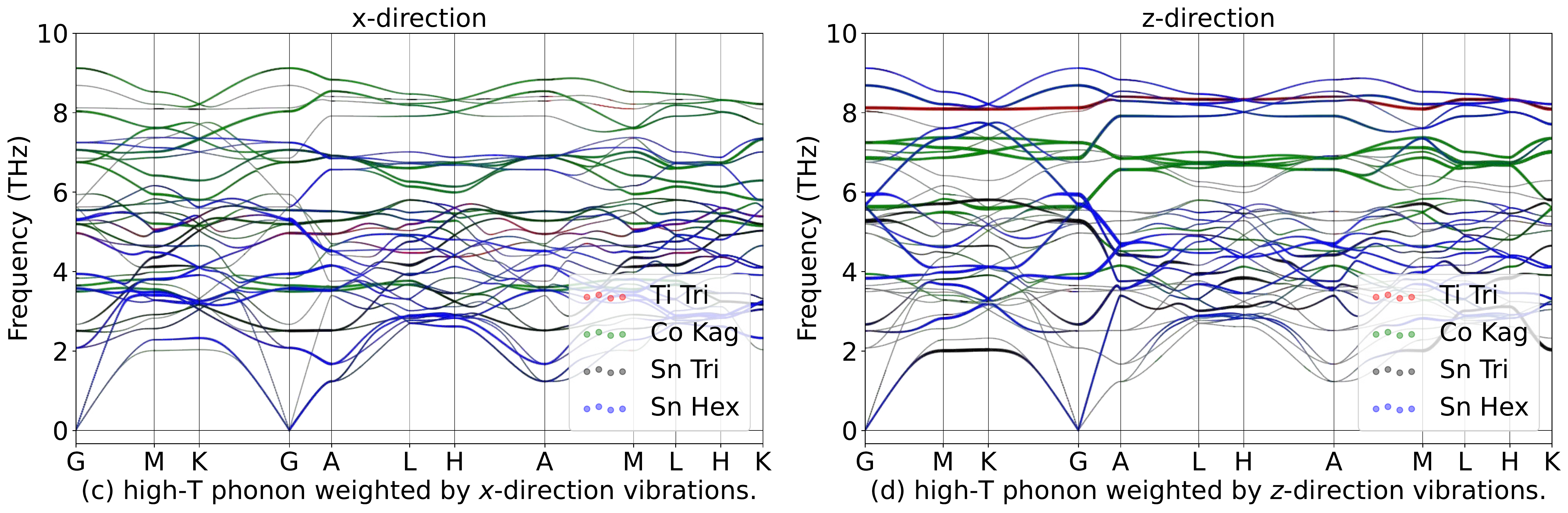}
\caption{Phonon spectrum for TiCo$_6$Sn$_6$in in-plane ($x$) and out-of-plane ($z$) directions.}
\label{fig:TiCo6Sn6phoxyz}
\end{figure}

\begin{table}[htbp]
\global\long\def\arraystretch{1.12}
\captionsetup{width=.95\textwidth}
\caption{IRREPs at high-symmetry points for TiCo$_6$Sn$_6$ phonon spectrum at Low-T.}
\label{tab:191_TiCo6Sn6_Low}
\setlength{\tabcolsep}{1.mm}{
}
\end{table}

\begin{table}[htbp]
\global\long\def\arraystretch{1.12}
\captionsetup{width=.95\textwidth}
\caption{IRREPs at high-symmetry points for TiCo$_6$Sn$_6$ phonon spectrum at High-T.}
\label{tab:191_TiCo6Sn6_High}
\setlength{\tabcolsep}{1.mm}{
%
}
\end{table}
\subsubsection{\label{sms2sec:YCo6Sn6}\hyperref[smsec:col]{\texorpdfstring{YCo$_6$Sn$_6$}{}}~{\color{green}  (High-T stable, Low-T stable) }}

\begin{table}[htbp]
\global\long\def\arraystretch{1.12}
\captionsetup{width=.85\textwidth}
\caption{Structure parameters of YCo$_6$Sn$_6$ after relaxation. It contains 501 electrons. }
\label{tab:YCo6Sn6}
\setlength{\tabcolsep}{4mm}{
%
}
\end{table}

\begin{figure}[htbp]
\centering
\includegraphics[width=0.85\textwidth]{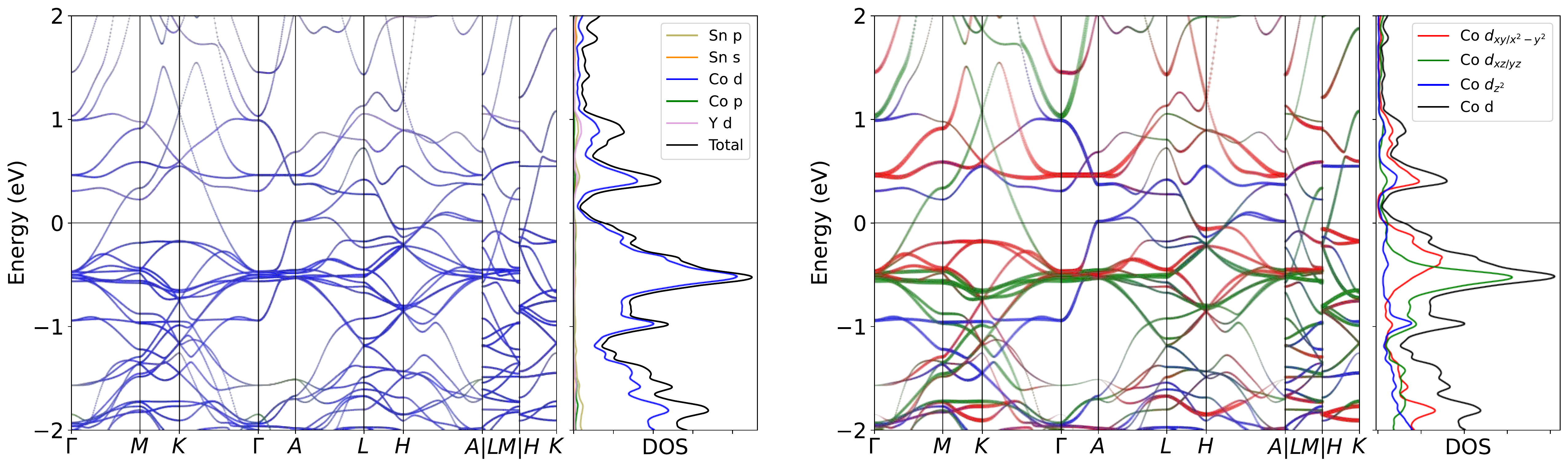}
\caption{Band structure for YCo$_6$Sn$_6$ without SOC. The projection of Co $3d$ orbitals is also presented. The band structures clearly reveal the dominating of Co $3d$ orbitals  near the Fermi level, as well as the pronounced peak.}
\label{fig:YCo6Sn6band}
\end{figure}

\begin{figure}[H]
\centering
\subfloat[Relaxed phonon at low-T.]{
\label{fig:YCo6Sn6_relaxed_Low}
\includegraphics[width=0.45\textwidth]{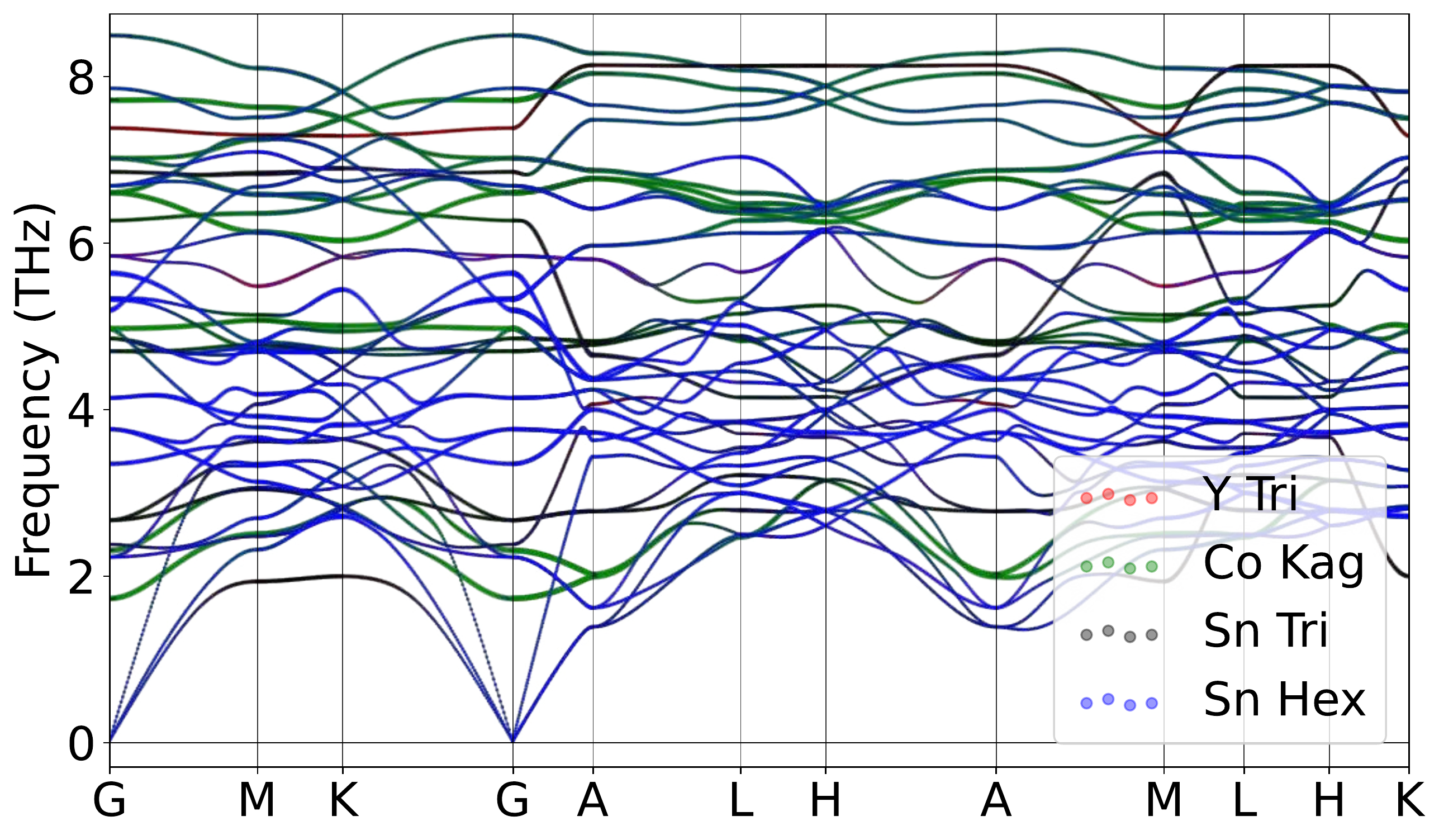}
}
\hspace{5pt}\subfloat[Relaxed phonon at high-T.]{
\label{fig:YCo6Sn6_relaxed_High}
\includegraphics[width=0.45\textwidth]{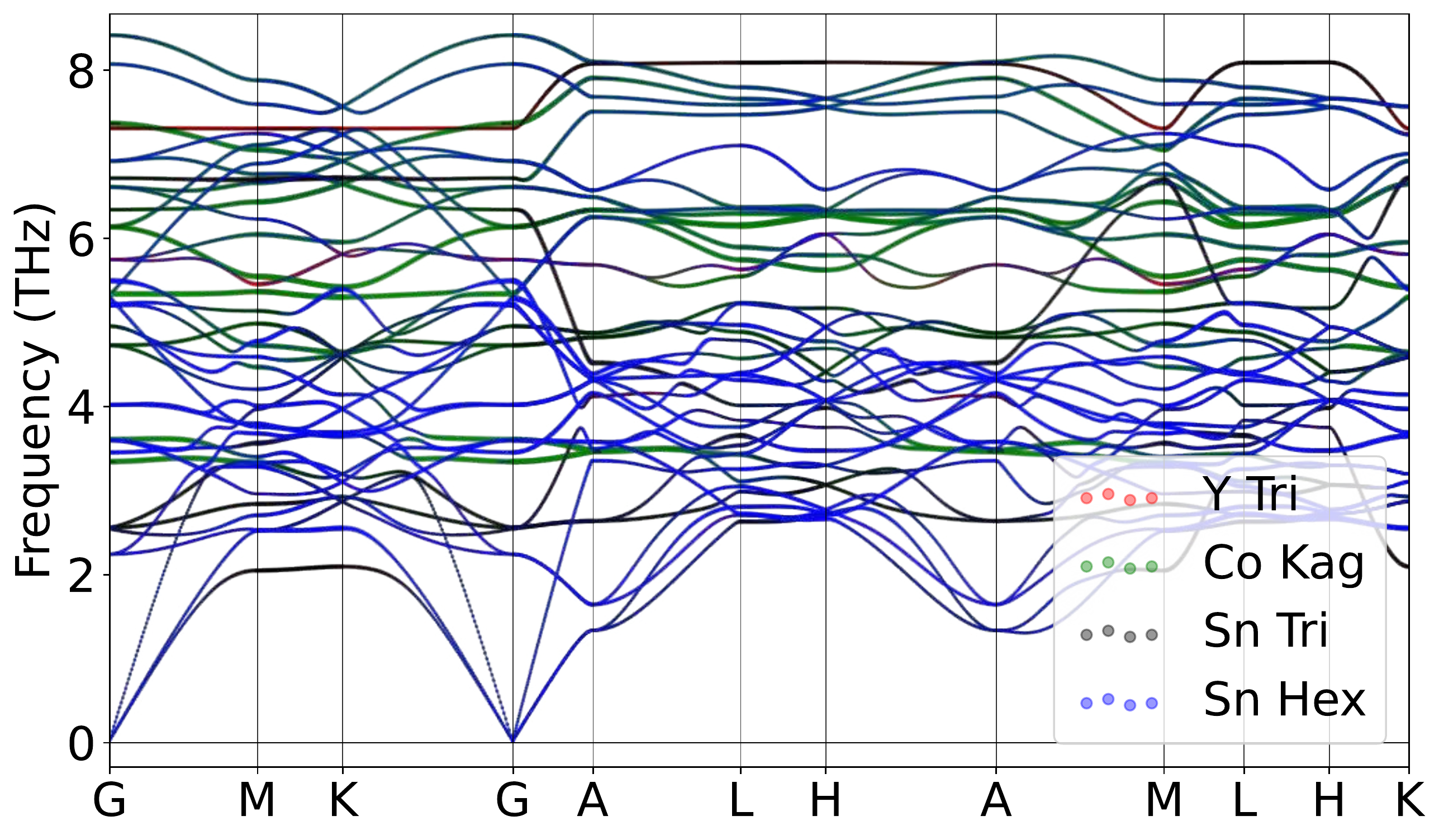}
}
\caption{Atomically projected phonon spectrum for YCo$_6$Sn$_6$.}
\label{fig:YCo6Sn6pho}
\end{figure}

\begin{figure}[H]
\centering
\label{fig:YCo6Sn6_relaxed_Low_xz}
\includegraphics[width=0.85\textwidth]{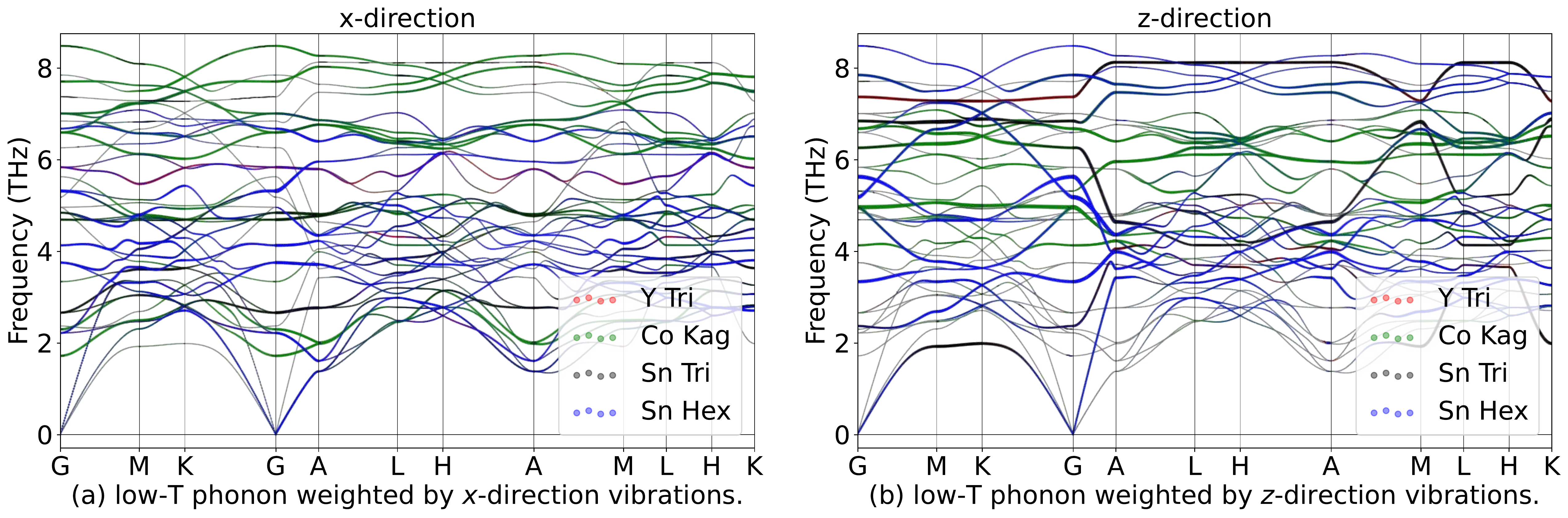}
\hspace{5pt}\label{fig:YCo6Sn6_relaxed_High_xz}
\includegraphics[width=0.85\textwidth]{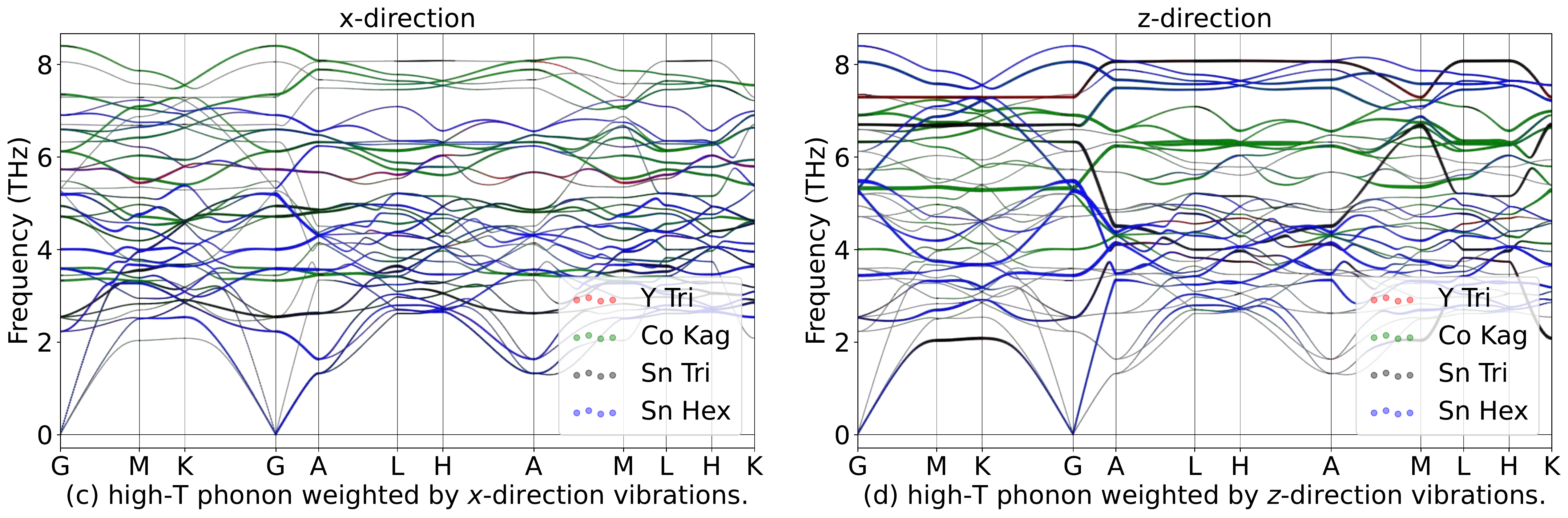}
\caption{Phonon spectrum for YCo$_6$Sn$_6$in in-plane ($x$) and out-of-plane ($z$) directions.}
\label{fig:YCo6Sn6phoxyz}
\end{figure}

\begin{table}[htbp]
\global\long\def\arraystretch{1.12}
\captionsetup{width=.95\textwidth}
\caption{IRREPs at high-symmetry points for YCo$_6$Sn$_6$ phonon spectrum at Low-T.}
\label{tab:191_YCo6Sn6_Low}
\setlength{\tabcolsep}{1.mm}{
}
\end{table}

\begin{table}[htbp]
\global\long\def\arraystretch{1.12}
\captionsetup{width=.95\textwidth}
\caption{IRREPs at high-symmetry points for YCo$_6$Sn$_6$ phonon spectrum at High-T.}
\label{tab:191_YCo6Sn6_High}
\setlength{\tabcolsep}{1.mm}{
%
}
\end{table}
\subsubsection{\label{sms2sec:ZrCo6Sn6}\hyperref[smsec:col]{\texorpdfstring{ZrCo$_6$Sn$_6$}{}}~{\color{green}  (High-T stable, Low-T stable) }}

\begin{table}[htbp]
\global\long\def\arraystretch{1.12}
\captionsetup{width=.85\textwidth}
\caption{Structure parameters of ZrCo$_6$Sn$_6$ after relaxation. It contains 502 electrons. }
\label{tab:ZrCo6Sn6}
\setlength{\tabcolsep}{4mm}{
%
}
\end{table}

\begin{figure}[htbp]
\centering
\includegraphics[width=0.85\textwidth]{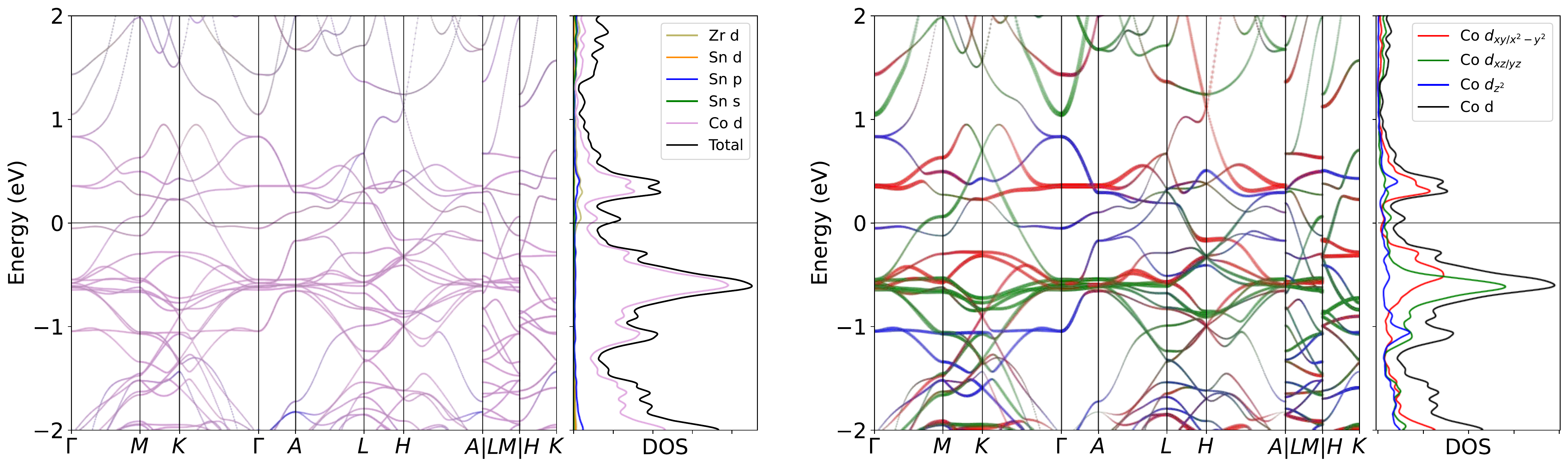}
\caption{Band structure for ZrCo$_6$Sn$_6$ without SOC. The projection of Co $3d$ orbitals is also presented. The band structures clearly reveal the dominating of Co $3d$ orbitals  near the Fermi level, as well as the pronounced peak.}
\label{fig:ZrCo6Sn6band}
\end{figure}

\begin{figure}[H]
\centering
\subfloat[Relaxed phonon at low-T.]{
\label{fig:ZrCo6Sn6_relaxed_Low}
\includegraphics[width=0.45\textwidth]{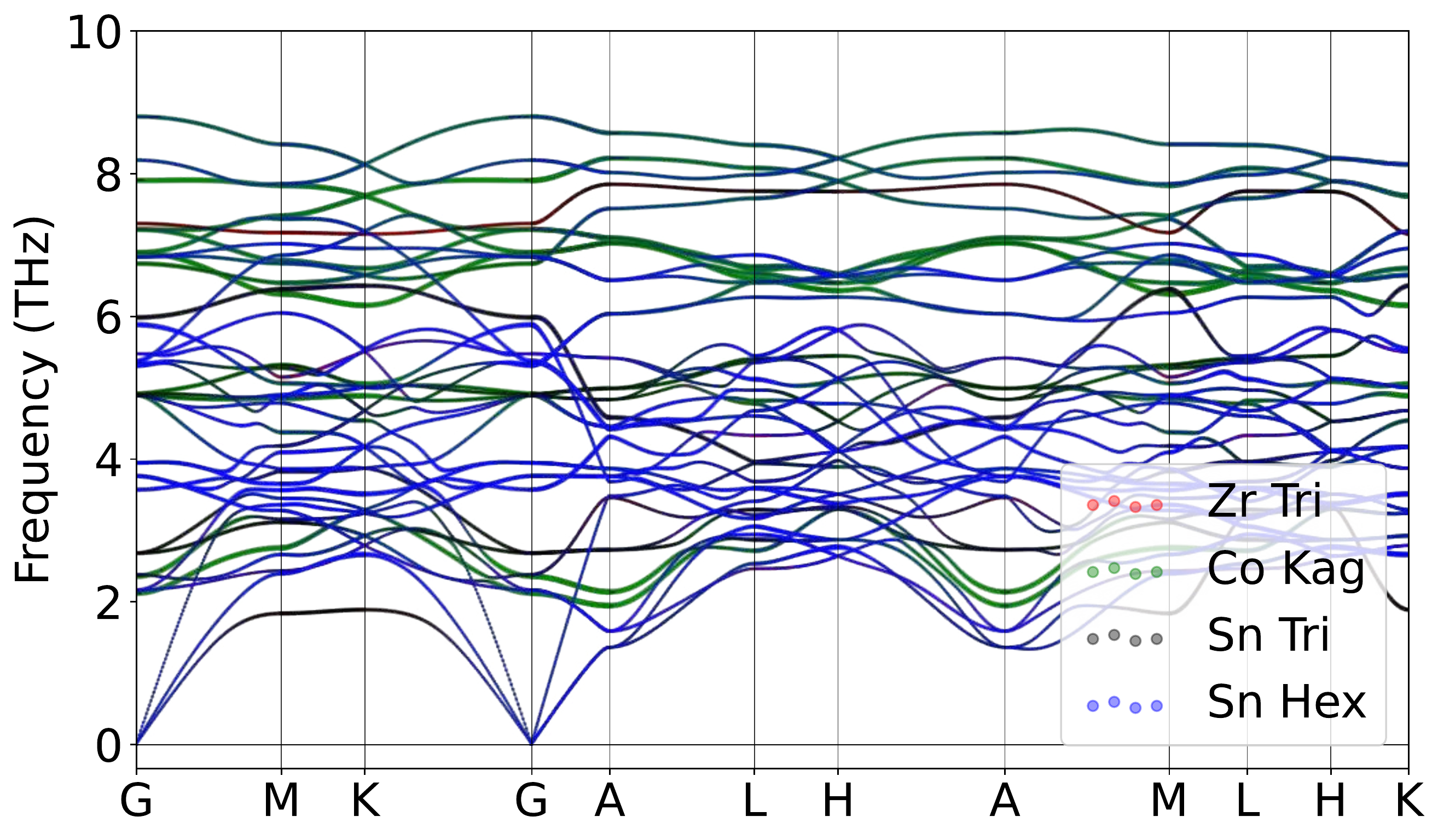}
}
\hspace{5pt}\subfloat[Relaxed phonon at high-T.]{
\label{fig:ZrCo6Sn6_relaxed_High}
\includegraphics[width=0.45\textwidth]{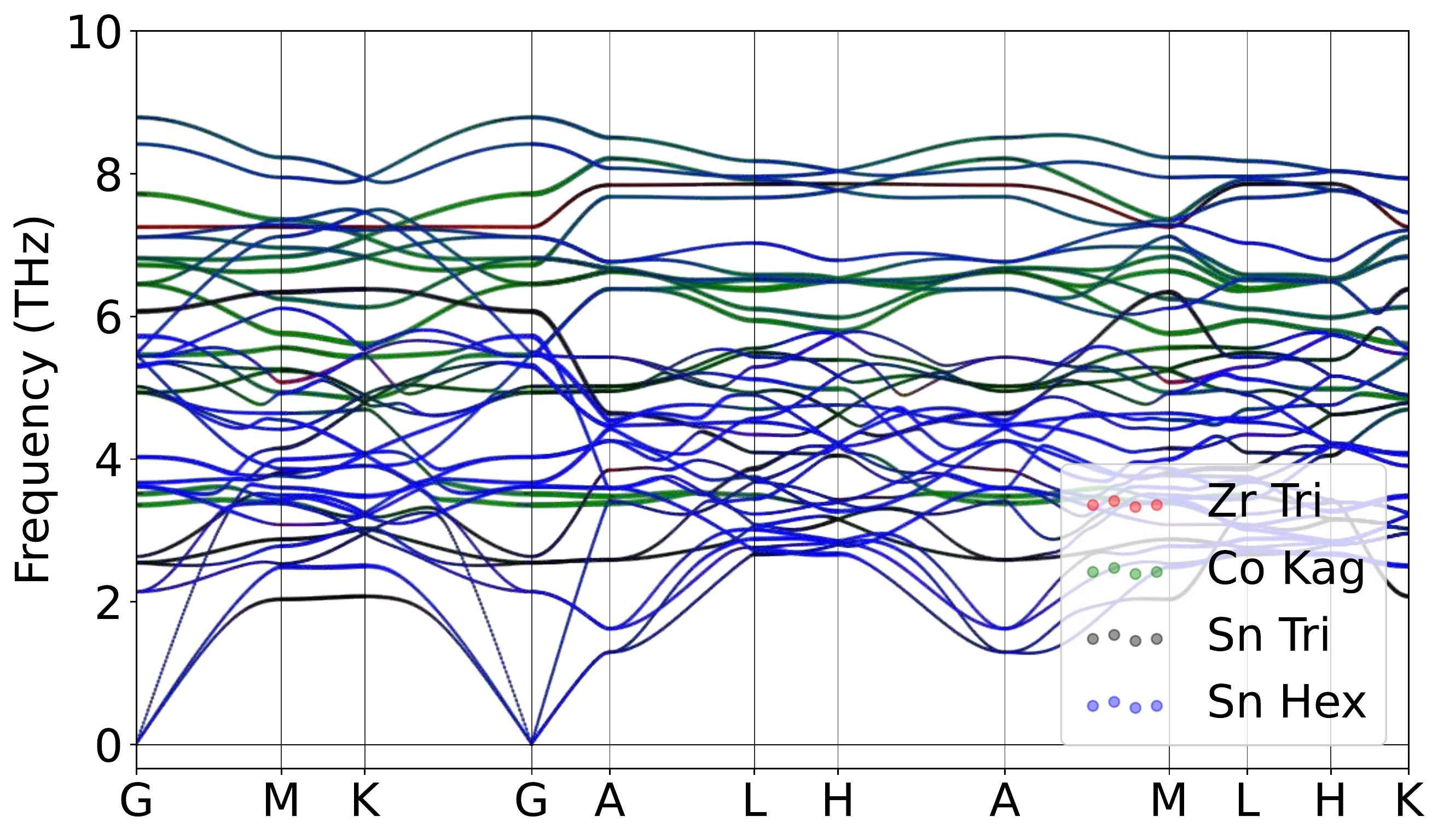}
}
\caption{Atomically projected phonon spectrum for ZrCo$_6$Sn$_6$.}
\label{fig:ZrCo6Sn6pho}
\end{figure}

\begin{figure}[H]
\centering
\label{fig:ZrCo6Sn6_relaxed_Low_xz}
\includegraphics[width=0.85\textwidth]{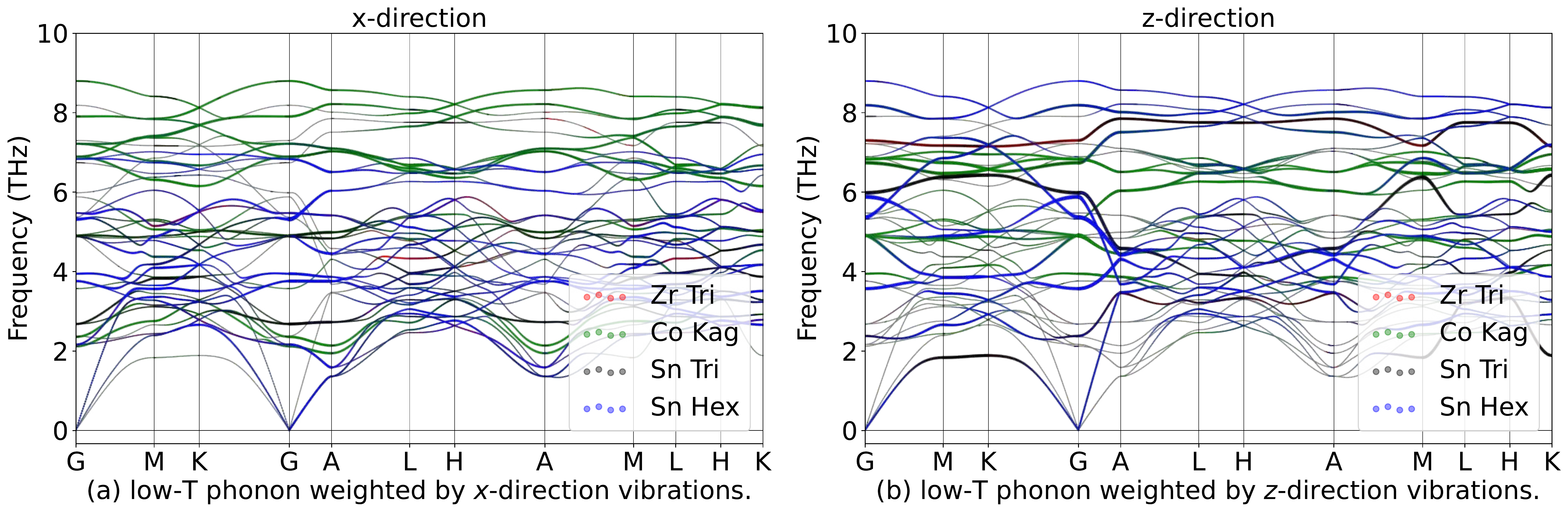}
\hspace{5pt}\label{fig:ZrCo6Sn6_relaxed_High_xz}
\includegraphics[width=0.85\textwidth]{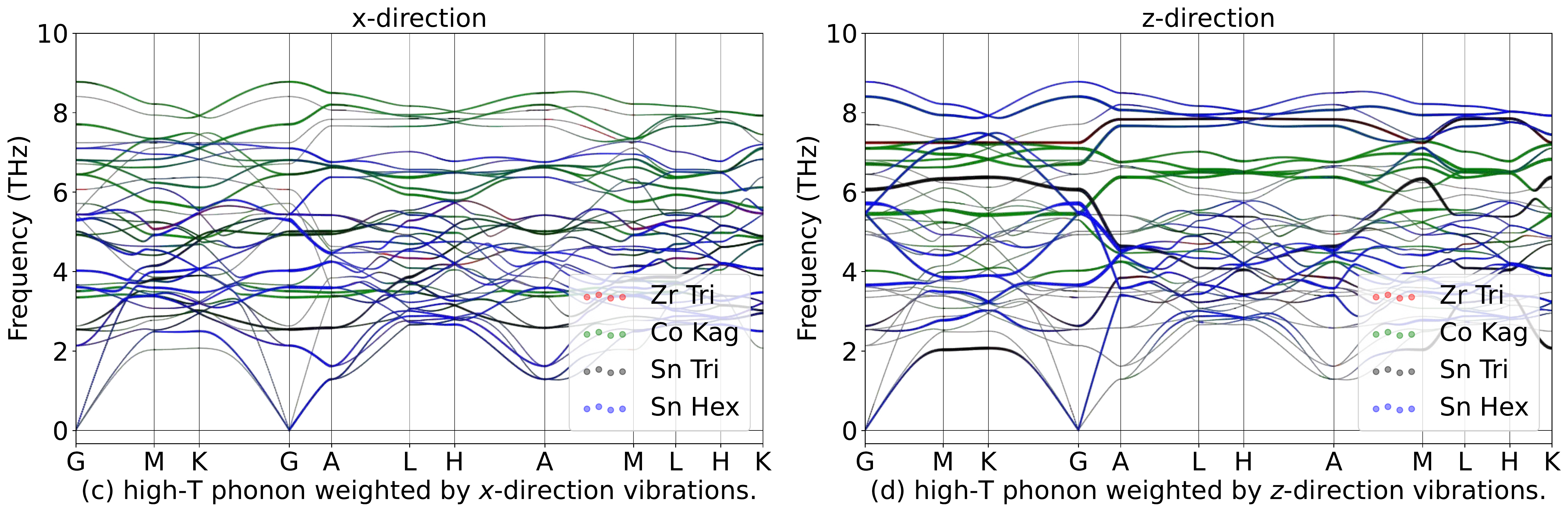}
\caption{Phonon spectrum for ZrCo$_6$Sn$_6$in in-plane ($x$) and out-of-plane ($z$) directions.}
\label{fig:ZrCo6Sn6phoxyz}
\end{figure}

\begin{table}[htbp]
\global\long\def\arraystretch{1.12}
\captionsetup{width=.95\textwidth}
\caption{IRREPs at high-symmetry points for ZrCo$_6$Sn$_6$ phonon spectrum at Low-T.}
\label{tab:191_ZrCo6Sn6_Low}
\setlength{\tabcolsep}{1.mm}{
}
\end{table}

\begin{table}[htbp]
\global\long\def\arraystretch{1.12}
\captionsetup{width=.95\textwidth}
\caption{IRREPs at high-symmetry points for ZrCo$_6$Sn$_6$ phonon spectrum at High-T.}
\label{tab:191_ZrCo6Sn6_High}
\setlength{\tabcolsep}{1.mm}{
%
}
\end{table}
\subsubsection{\label{sms2sec:NbCo6Sn6}\hyperref[smsec:col]{\texorpdfstring{NbCo$_6$Sn$_6$}{}}~{\color{green}  (High-T stable, Low-T stable) }}

\begin{table}[htbp]
\global\long\def\arraystretch{1.12}
\captionsetup{width=.85\textwidth}
\caption{Structure parameters of NbCo$_6$Sn$_6$ after relaxation. It contains 503 electrons. }
\label{tab:NbCo6Sn6}
\setlength{\tabcolsep}{4mm}{
%
}
\end{table}

\begin{figure}[htbp]
\centering
\includegraphics[width=0.85\textwidth]{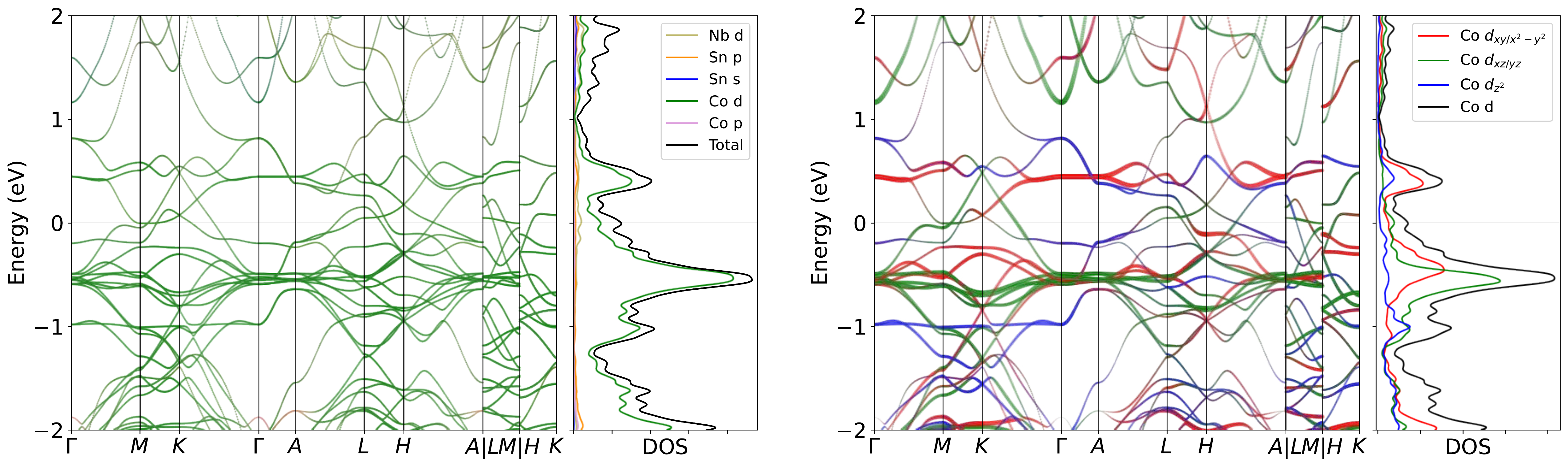}
\caption{Band structure for NbCo$_6$Sn$_6$ without SOC. The projection of Co $3d$ orbitals is also presented. The band structures clearly reveal the dominating of Co $3d$ orbitals  near the Fermi level, as well as the pronounced peak.}
\label{fig:NbCo6Sn6band}
\end{figure}

\begin{figure}[H]
\centering
\subfloat[Relaxed phonon at low-T.]{
\label{fig:NbCo6Sn6_relaxed_Low}
\includegraphics[width=0.45\textwidth]{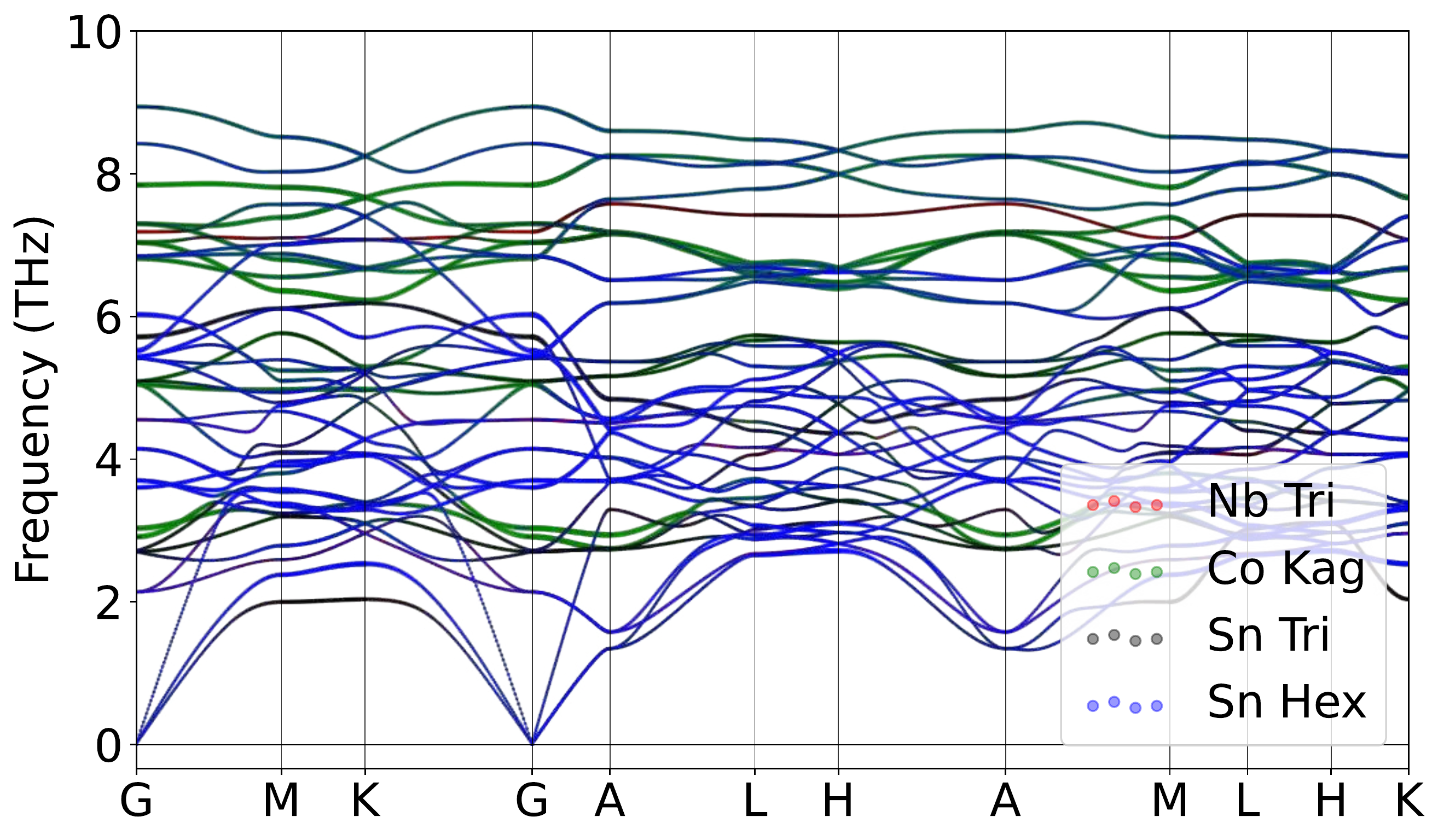}
}
\hspace{5pt}\subfloat[Relaxed phonon at high-T.]{
\label{fig:NbCo6Sn6_relaxed_High}
\includegraphics[width=0.45\textwidth]{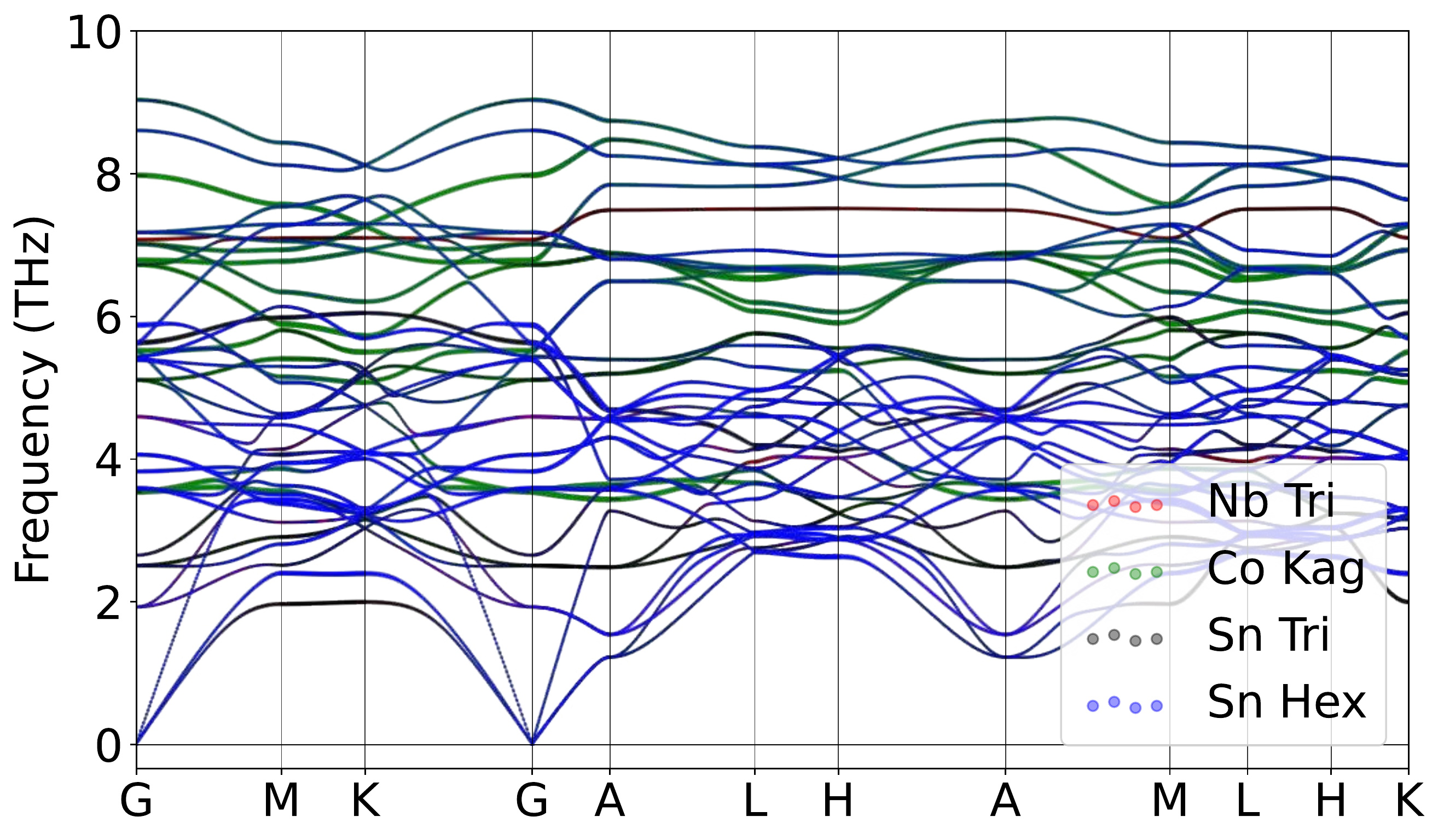}
}
\caption{Atomically projected phonon spectrum for NbCo$_6$Sn$_6$.}
\label{fig:NbCo6Sn6pho}
\end{figure}

\begin{figure}[H]
\centering
\label{fig:NbCo6Sn6_relaxed_Low_xz}
\includegraphics[width=0.85\textwidth]{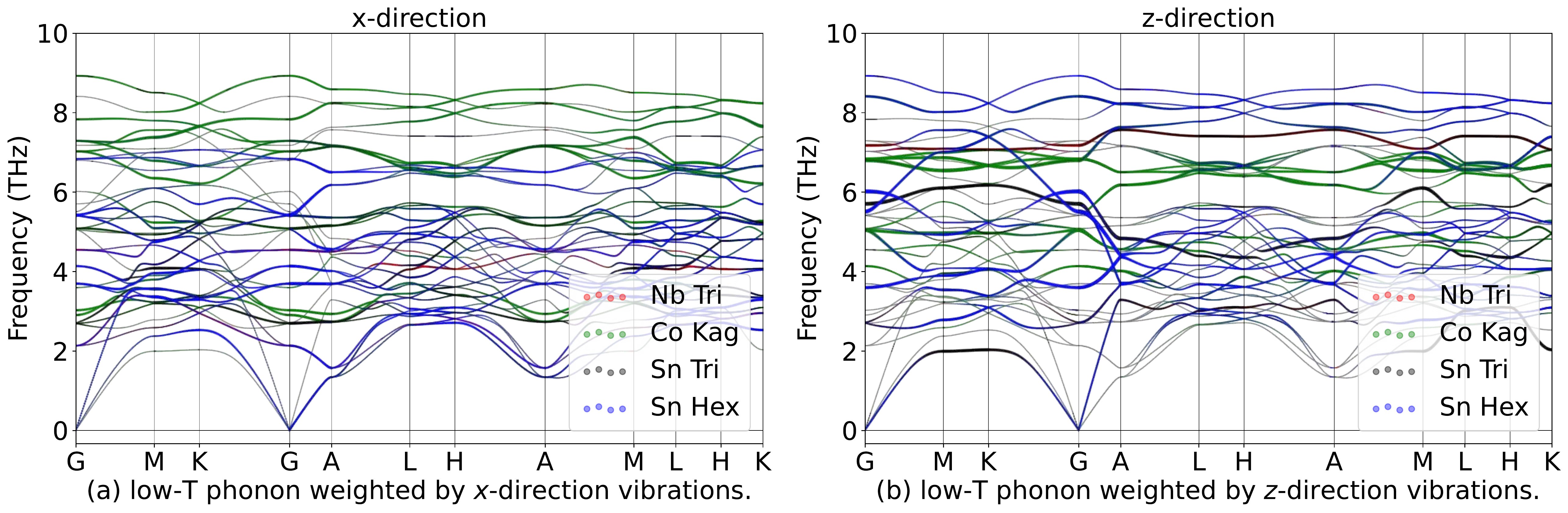}
\hspace{5pt}\label{fig:NbCo6Sn6_relaxed_High_xz}
\includegraphics[width=0.85\textwidth]{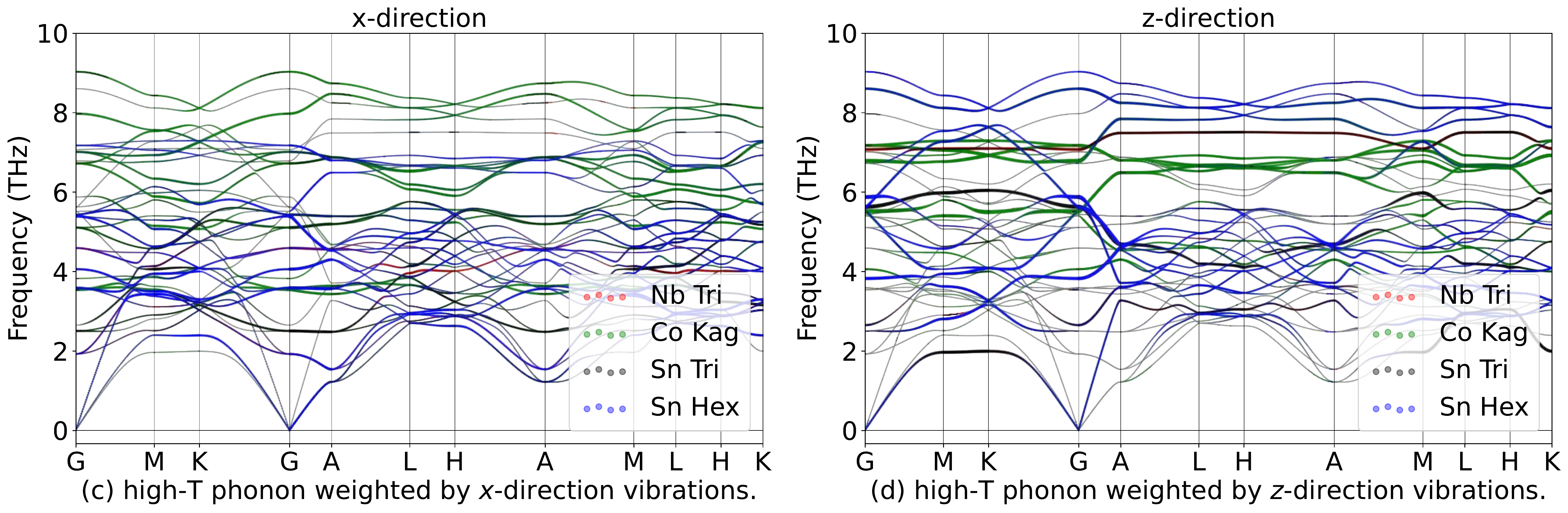}
\caption{Phonon spectrum for NbCo$_6$Sn$_6$in in-plane ($x$) and out-of-plane ($z$) directions.}
\label{fig:NbCo6Sn6phoxyz}
\end{figure}

\begin{table}[htbp]
\global\long\def\arraystretch{1.12}
\captionsetup{width=.95\textwidth}
\caption{IRREPs at high-symmetry points for NbCo$_6$Sn$_6$ phonon spectrum at Low-T.}
\label{tab:191_NbCo6Sn6_Low}
\setlength{\tabcolsep}{1.mm}{
}
\end{table}

\begin{table}[htbp]
\global\long\def\arraystretch{1.12}
\captionsetup{width=.95\textwidth}
\caption{IRREPs at high-symmetry points for NbCo$_6$Sn$_6$ phonon spectrum at High-T.}
\label{tab:191_NbCo6Sn6_High}
\setlength{\tabcolsep}{1.mm}{
%
}
\end{table}
\subsubsection{\label{sms2sec:PrCo6Sn6}\hyperref[smsec:col]{\texorpdfstring{PrCo$_6$Sn$_6$}{}}~{\color{green}  (High-T stable, Low-T stable) }}

\begin{table}[htbp]
\global\long\def\arraystretch{1.12}
\captionsetup{width=.85\textwidth}
\caption{Structure parameters of PrCo$_6$Sn$_6$ after relaxation. It contains 521 electrons. }
\label{tab:PrCo6Sn6}
\setlength{\tabcolsep}{4mm}{
%
}
\end{table}

\begin{figure}[htbp]
\centering
\includegraphics[width=0.85\textwidth]{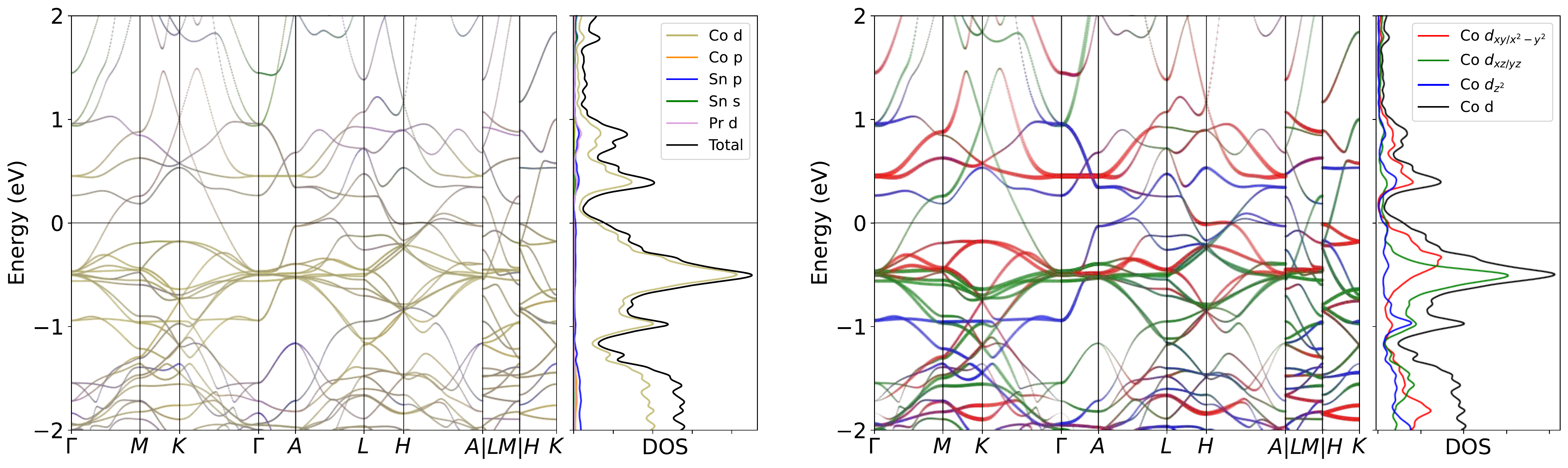}
\caption{Band structure for PrCo$_6$Sn$_6$ without SOC. The projection of Co $3d$ orbitals is also presented. The band structures clearly reveal the dominating of Co $3d$ orbitals  near the Fermi level, as well as the pronounced peak.}
\label{fig:PrCo6Sn6band}
\end{figure}

\begin{figure}[H]
\centering
\subfloat[Relaxed phonon at low-T.]{
\label{fig:PrCo6Sn6_relaxed_Low}
\includegraphics[width=0.45\textwidth]{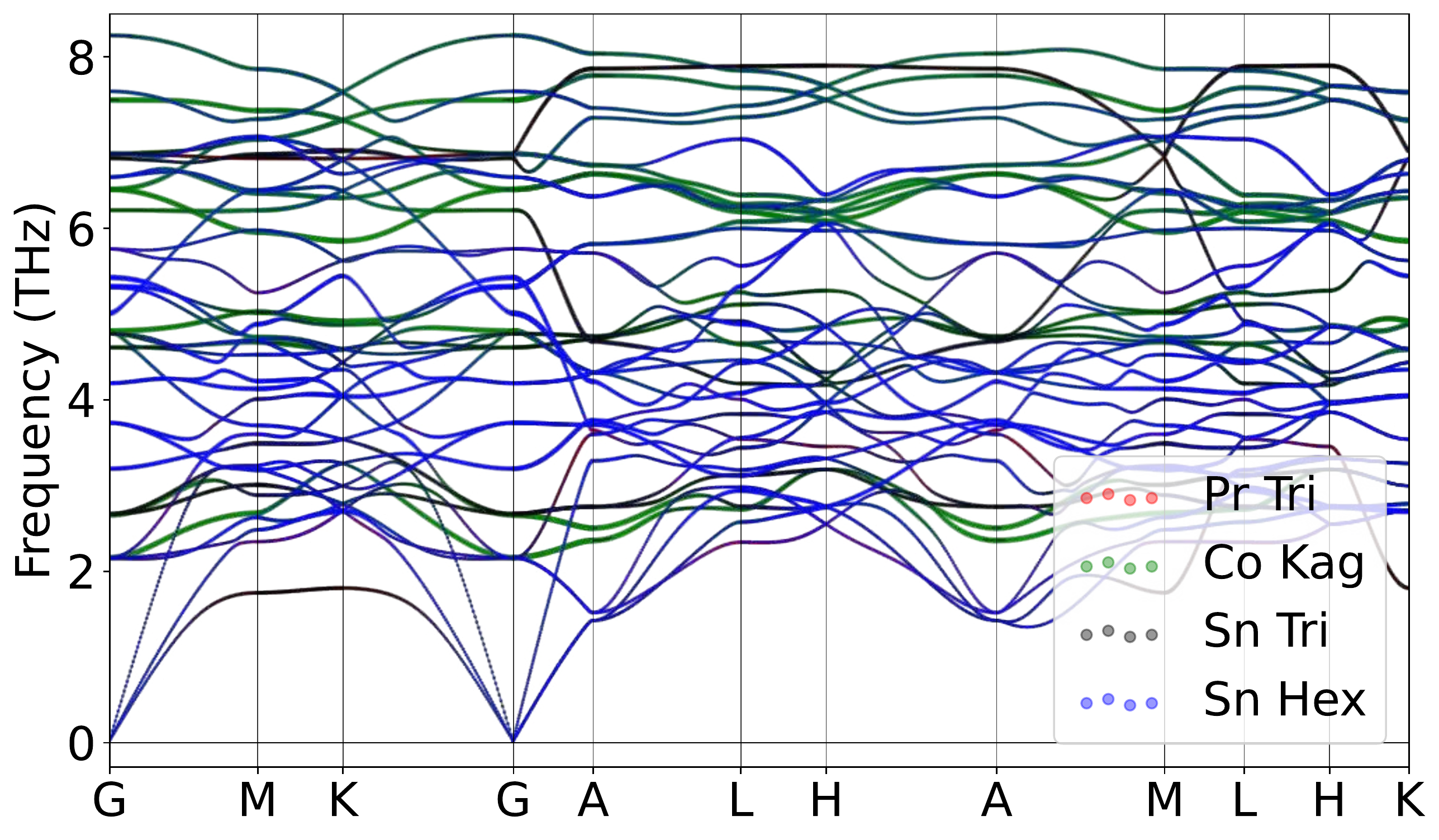}
}
\hspace{5pt}\subfloat[Relaxed phonon at high-T.]{
\label{fig:PrCo6Sn6_relaxed_High}
\includegraphics[width=0.45\textwidth]{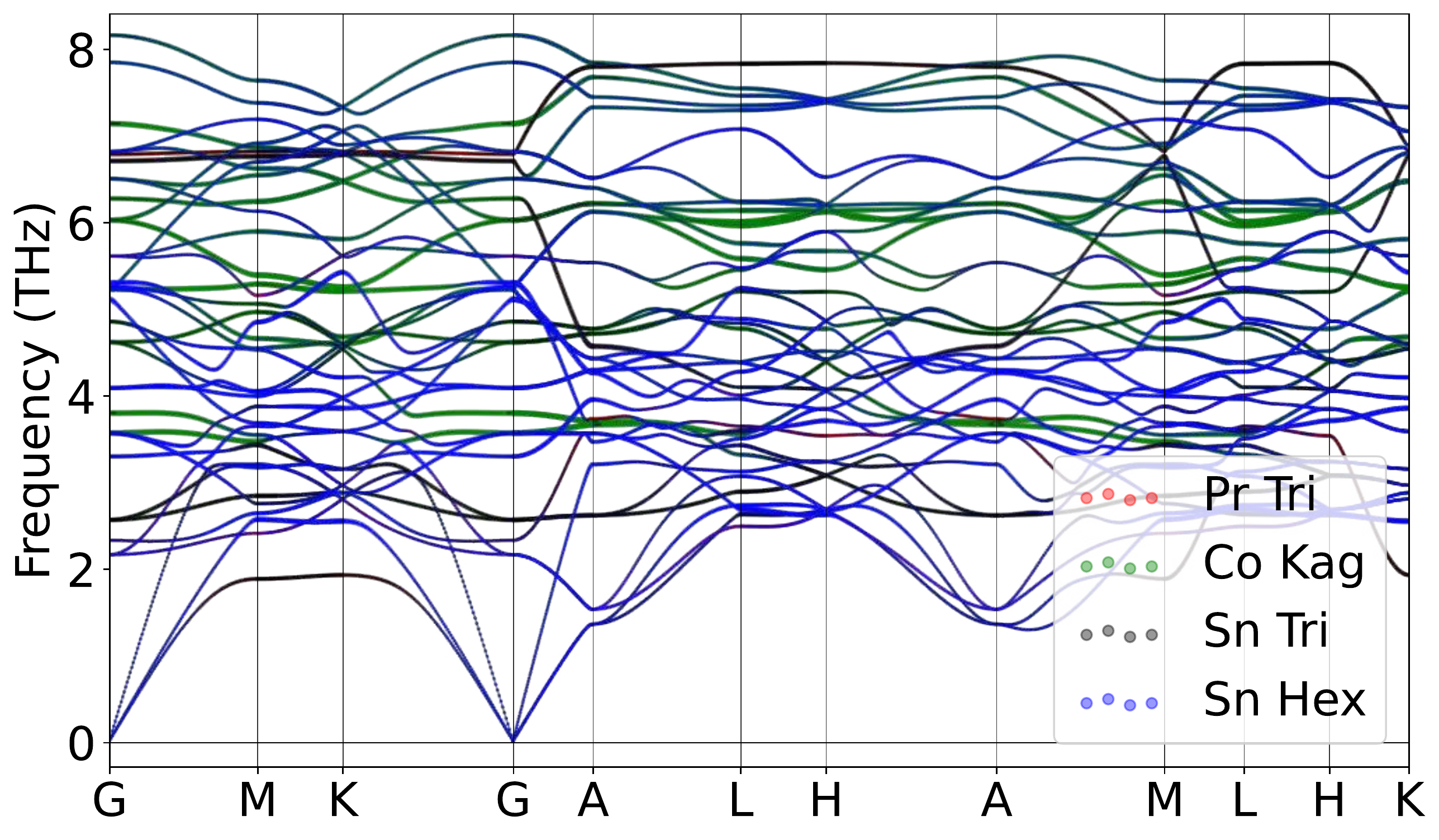}
}
\caption{Atomically projected phonon spectrum for PrCo$_6$Sn$_6$.}
\label{fig:PrCo6Sn6pho}
\end{figure}

\begin{figure}[H]
\centering
\label{fig:PrCo6Sn6_relaxed_Low_xz}
\includegraphics[width=0.85\textwidth]{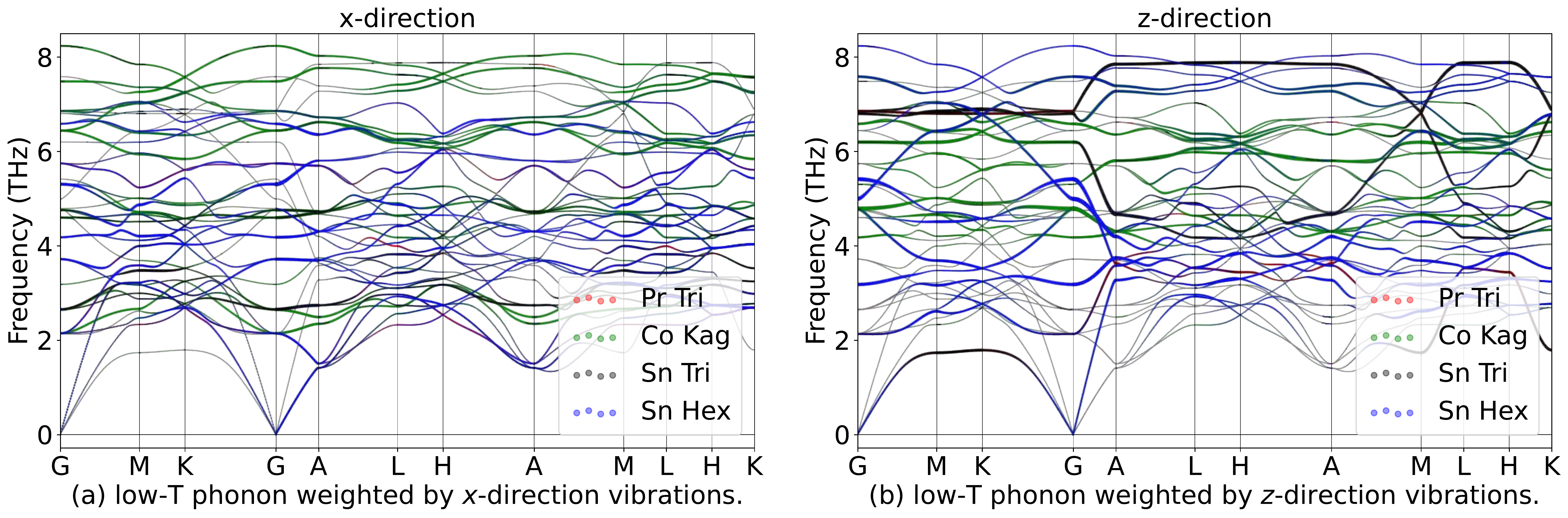}
\hspace{5pt}\label{fig:PrCo6Sn6_relaxed_High_xz}
\includegraphics[width=0.85\textwidth]{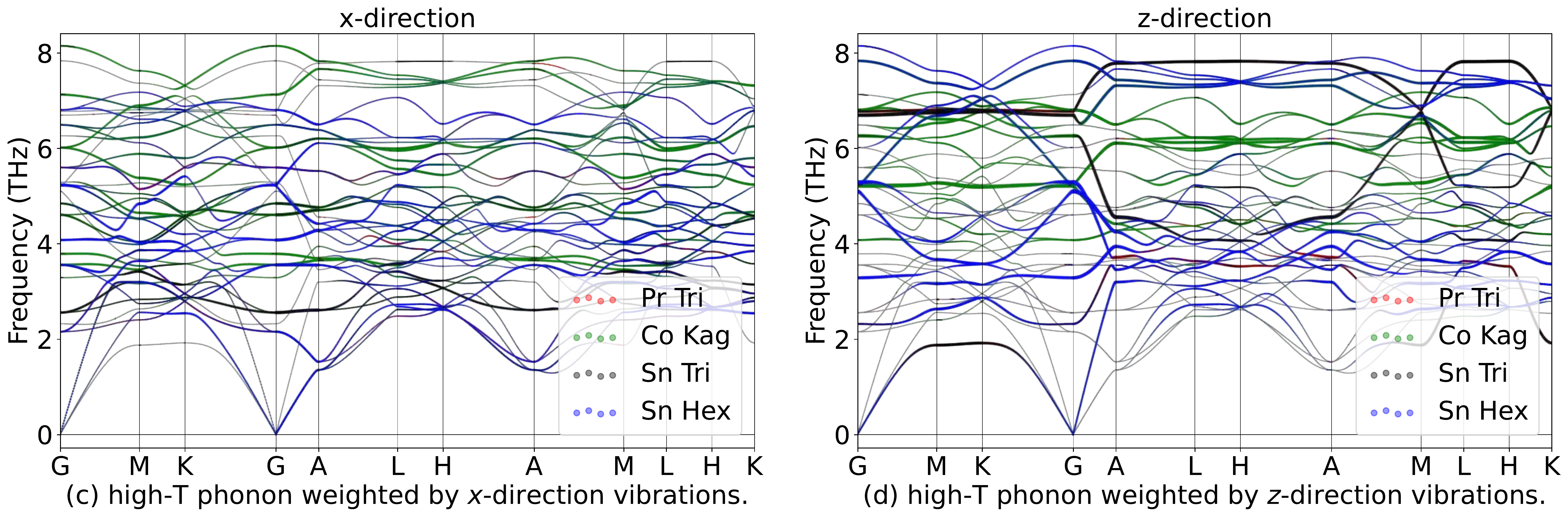}
\caption{Phonon spectrum for PrCo$_6$Sn$_6$in in-plane ($x$) and out-of-plane ($z$) directions.}
\label{fig:PrCo6Sn6phoxyz}
\end{figure}

\begin{table}[htbp]
\global\long\def\arraystretch{1.12}
\captionsetup{width=.95\textwidth}
\caption{IRREPs at high-symmetry points for PrCo$_6$Sn$_6$ phonon spectrum at Low-T.}
\label{tab:191_PrCo6Sn6_Low}
\setlength{\tabcolsep}{1.mm}{
}
\end{table}

\begin{table}[htbp]
\global\long\def\arraystretch{1.12}
\captionsetup{width=.95\textwidth}
\caption{IRREPs at high-symmetry points for PrCo$_6$Sn$_6$ phonon spectrum at High-T.}
\label{tab:191_PrCo6Sn6_High}
\setlength{\tabcolsep}{1.mm}{
%
}
\end{table}
\subsubsection{\label{sms2sec:NdCo6Sn6}\hyperref[smsec:col]{\texorpdfstring{NdCo$_6$Sn$_6$}{}}~{\color{green}  (High-T stable, Low-T stable) }}

\begin{table}[htbp]
\global\long\def\arraystretch{1.12}
\captionsetup{width=.85\textwidth}
\caption{Structure parameters of NdCo$_6$Sn$_6$ after relaxation. It contains 522 electrons. }
\label{tab:NdCo6Sn6}
\setlength{\tabcolsep}{4mm}{
%
}
\end{table}

\begin{figure}[htbp]
\centering
\includegraphics[width=0.85\textwidth]{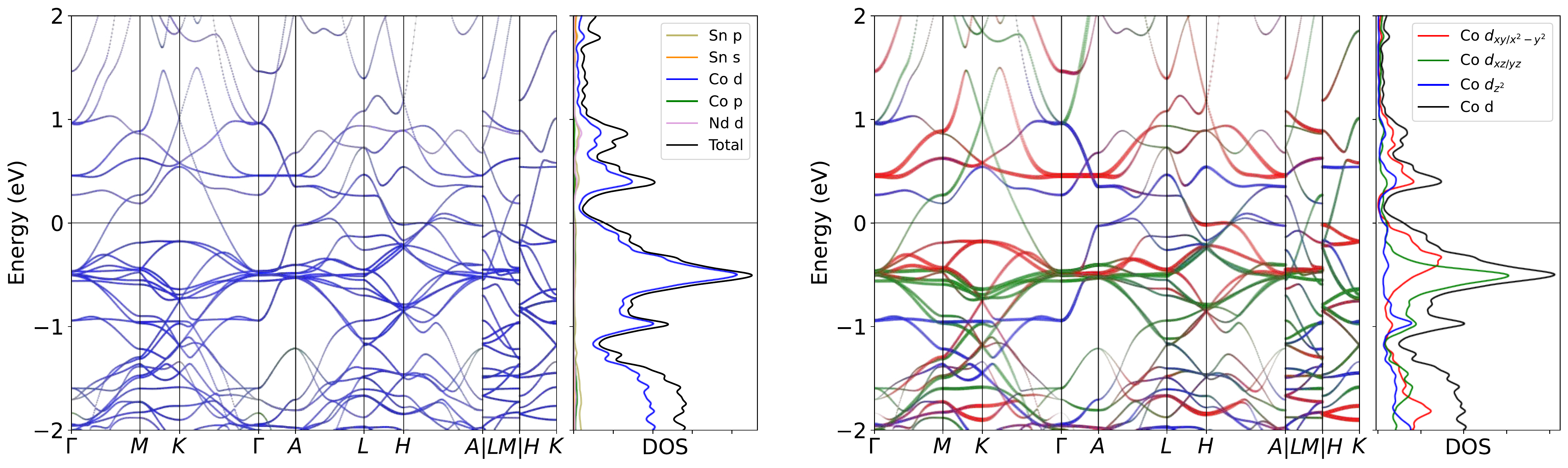}
\caption{Band structure for NdCo$_6$Sn$_6$ without SOC. The projection of Co $3d$ orbitals is also presented. The band structures clearly reveal the dominating of Co $3d$ orbitals  near the Fermi level, as well as the pronounced peak.}
\label{fig:NdCo6Sn6band}
\end{figure}

\begin{figure}[H]
\centering
\subfloat[Relaxed phonon at low-T.]{
\label{fig:NdCo6Sn6_relaxed_Low}
\includegraphics[width=0.45\textwidth]{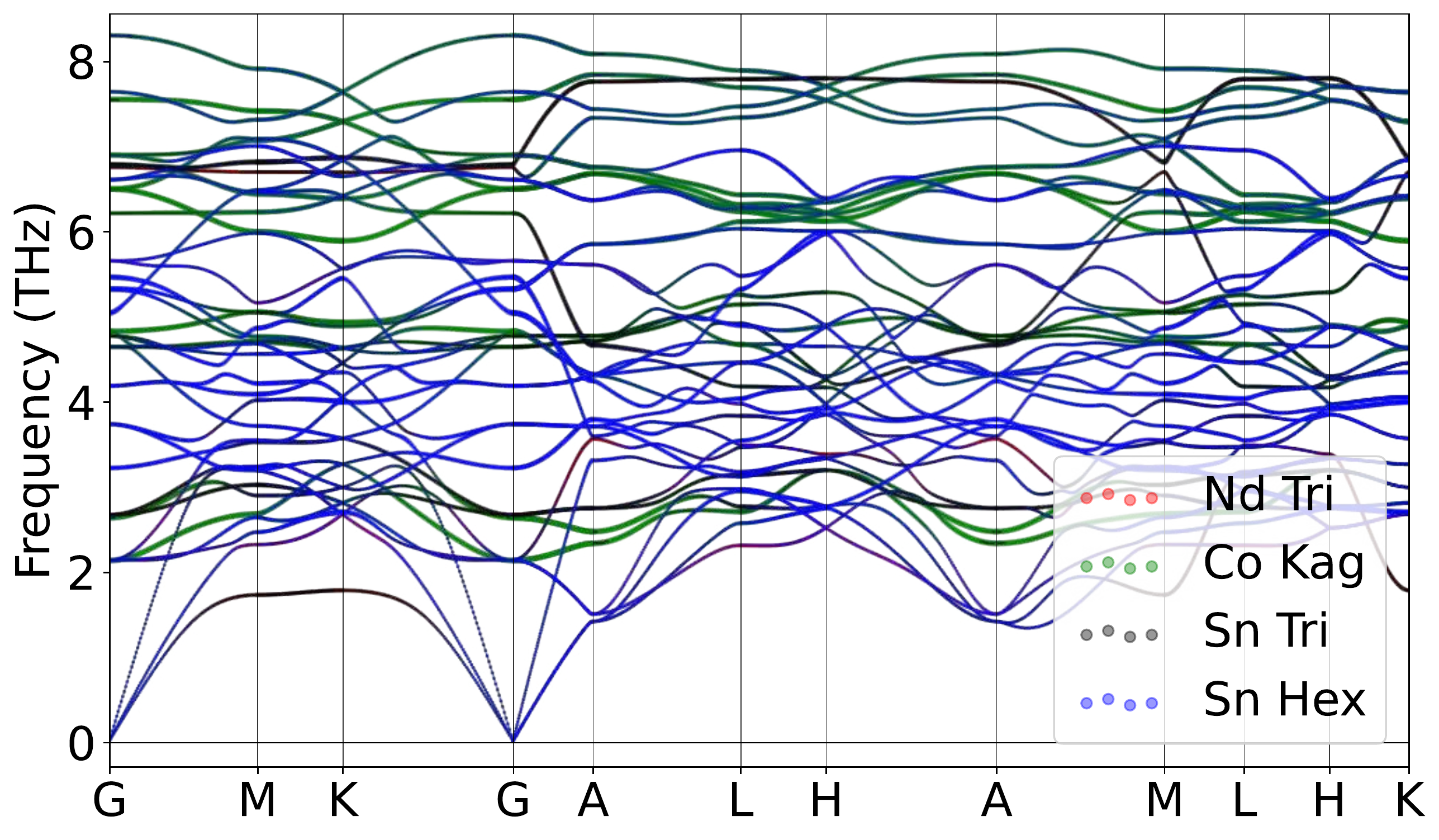}
}
\hspace{5pt}\subfloat[Relaxed phonon at high-T.]{
\label{fig:NdCo6Sn6_relaxed_High}
\includegraphics[width=0.45\textwidth]{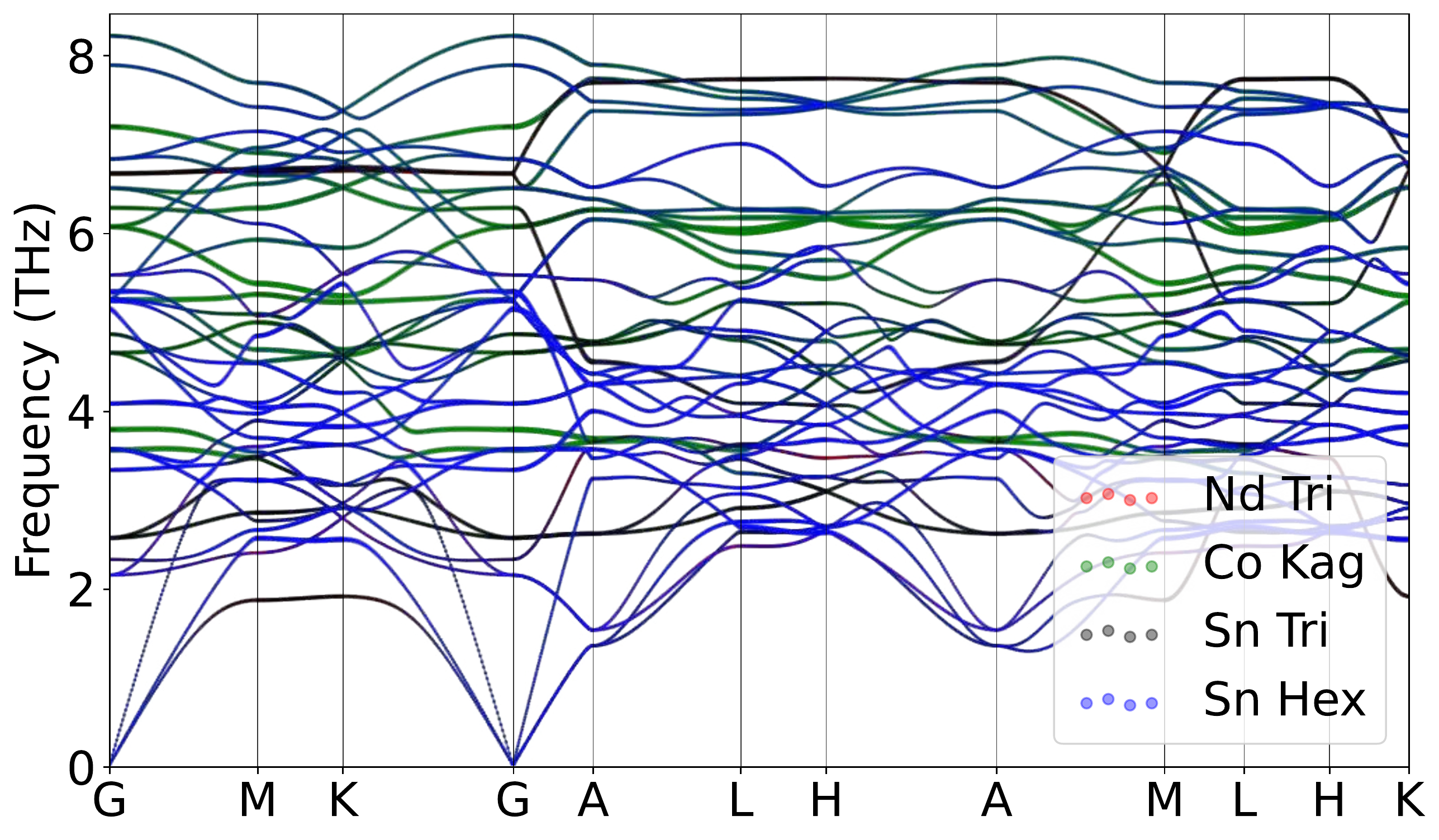}
}
\caption{Atomically projected phonon spectrum for NdCo$_6$Sn$_6$.}
\label{fig:NdCo6Sn6pho}
\end{figure}

\begin{figure}[H]
\centering
\label{fig:NdCo6Sn6_relaxed_Low_xz}
\includegraphics[width=0.85\textwidth]{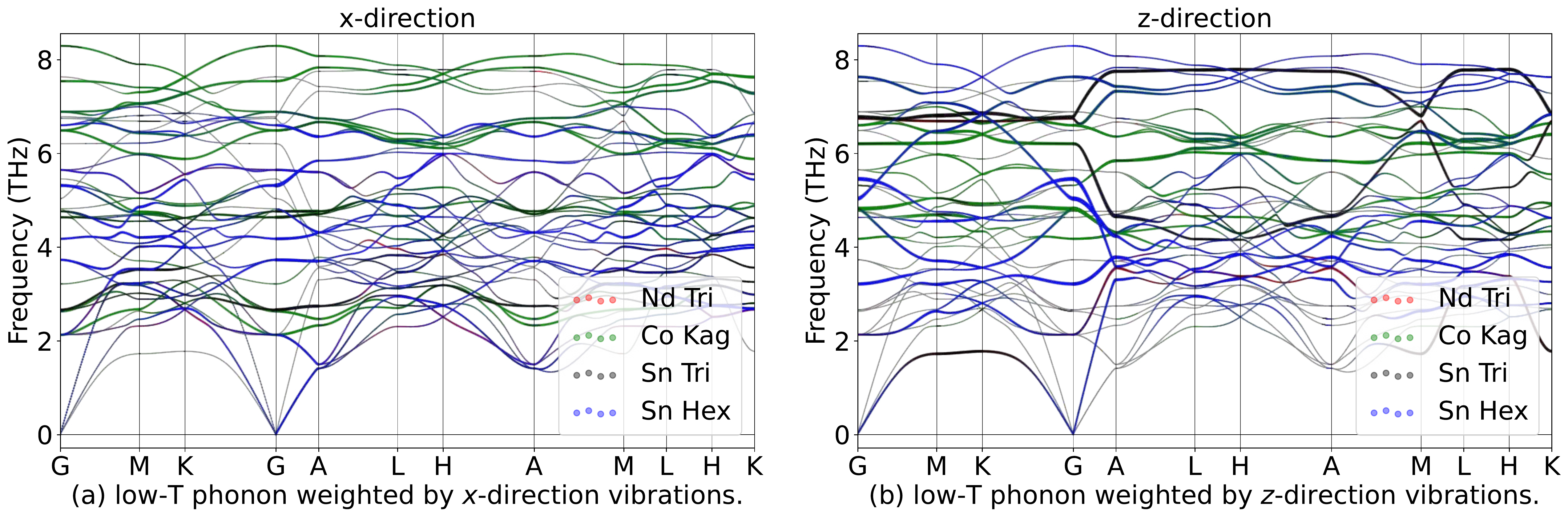}
\hspace{5pt}\label{fig:NdCo6Sn6_relaxed_High_xz}
\includegraphics[width=0.85\textwidth]{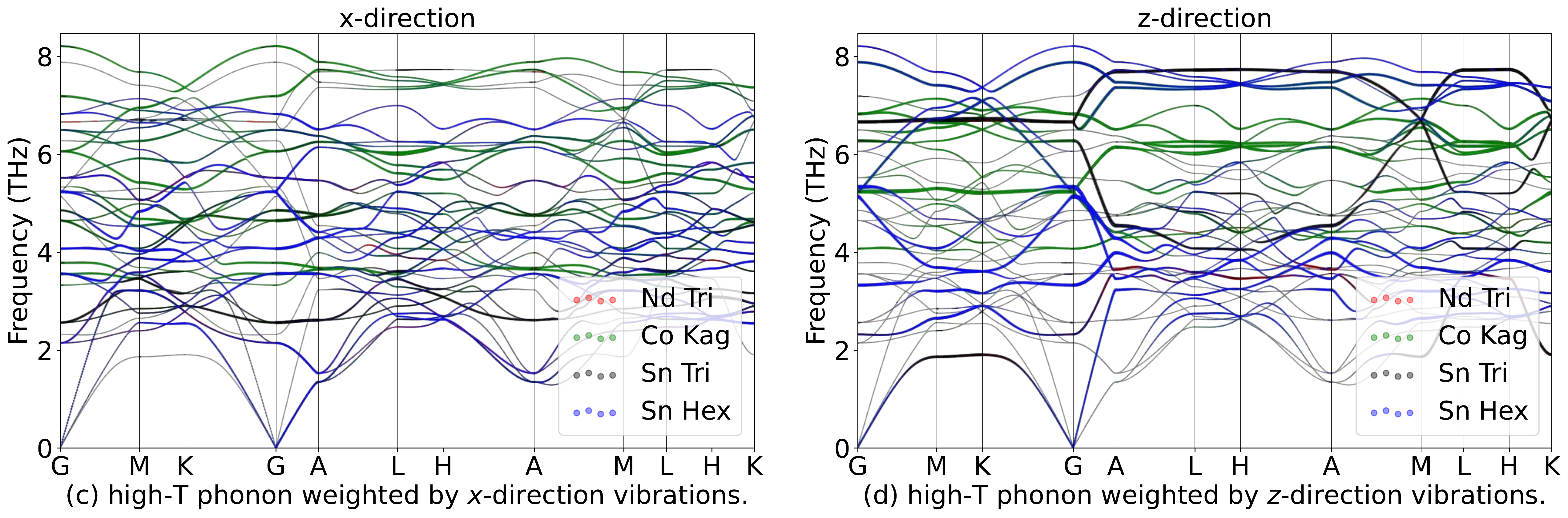}
\caption{Phonon spectrum for NdCo$_6$Sn$_6$in in-plane ($x$) and out-of-plane ($z$) directions.}
\label{fig:NdCo6Sn6phoxyz}
\end{figure}

\begin{table}[htbp]
\global\long\def\arraystretch{1.12}
\captionsetup{width=.95\textwidth}
\caption{IRREPs at high-symmetry points for NdCo$_6$Sn$_6$ phonon spectrum at Low-T.}
\label{tab:191_NdCo6Sn6_Low}
\setlength{\tabcolsep}{1.mm}{
}
\end{table}

\begin{table}[htbp]
\global\long\def\arraystretch{1.12}
\captionsetup{width=.95\textwidth}
\caption{IRREPs at high-symmetry points for NdCo$_6$Sn$_6$ phonon spectrum at High-T.}
\label{tab:191_NdCo6Sn6_High}
\setlength{\tabcolsep}{1.mm}{
%
}
\end{table}
\subsubsection{\label{sms2sec:SmCo6Sn6}\hyperref[smsec:col]{\texorpdfstring{SmCo$_6$Sn$_6$}{}}~{\color{green}  (High-T stable, Low-T stable) }}

\begin{table}[htbp]
\global\long\def\arraystretch{1.12}
\captionsetup{width=.85\textwidth}
\caption{Structure parameters of SmCo$_6$Sn$_6$ after relaxation. It contains 524 electrons. }
\label{tab:SmCo6Sn6}
\setlength{\tabcolsep}{4mm}{
%
}
\end{table}

\begin{figure}[htbp]
\centering
\includegraphics[width=0.85\textwidth]{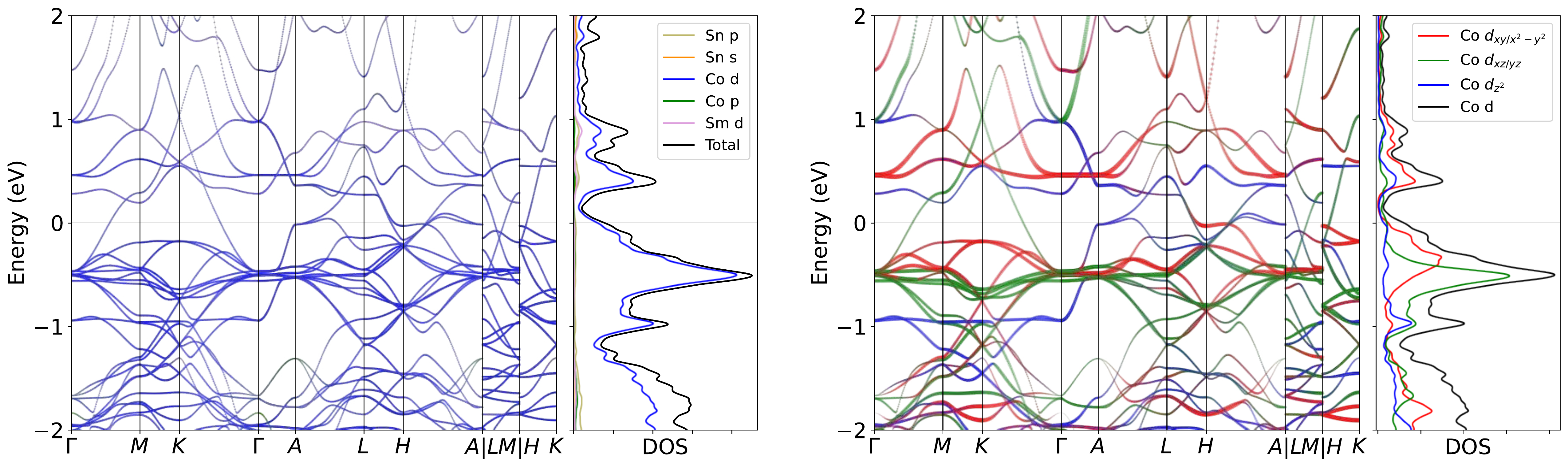}
\caption{Band structure for SmCo$_6$Sn$_6$ without SOC. The projection of Co $3d$ orbitals is also presented. The band structures clearly reveal the dominating of Co $3d$ orbitals  near the Fermi level, as well as the pronounced peak.}
\label{fig:SmCo6Sn6band}
\end{figure}

\begin{figure}[H]
\centering
\subfloat[Relaxed phonon at low-T.]{
\label{fig:SmCo6Sn6_relaxed_Low}
\includegraphics[width=0.45\textwidth]{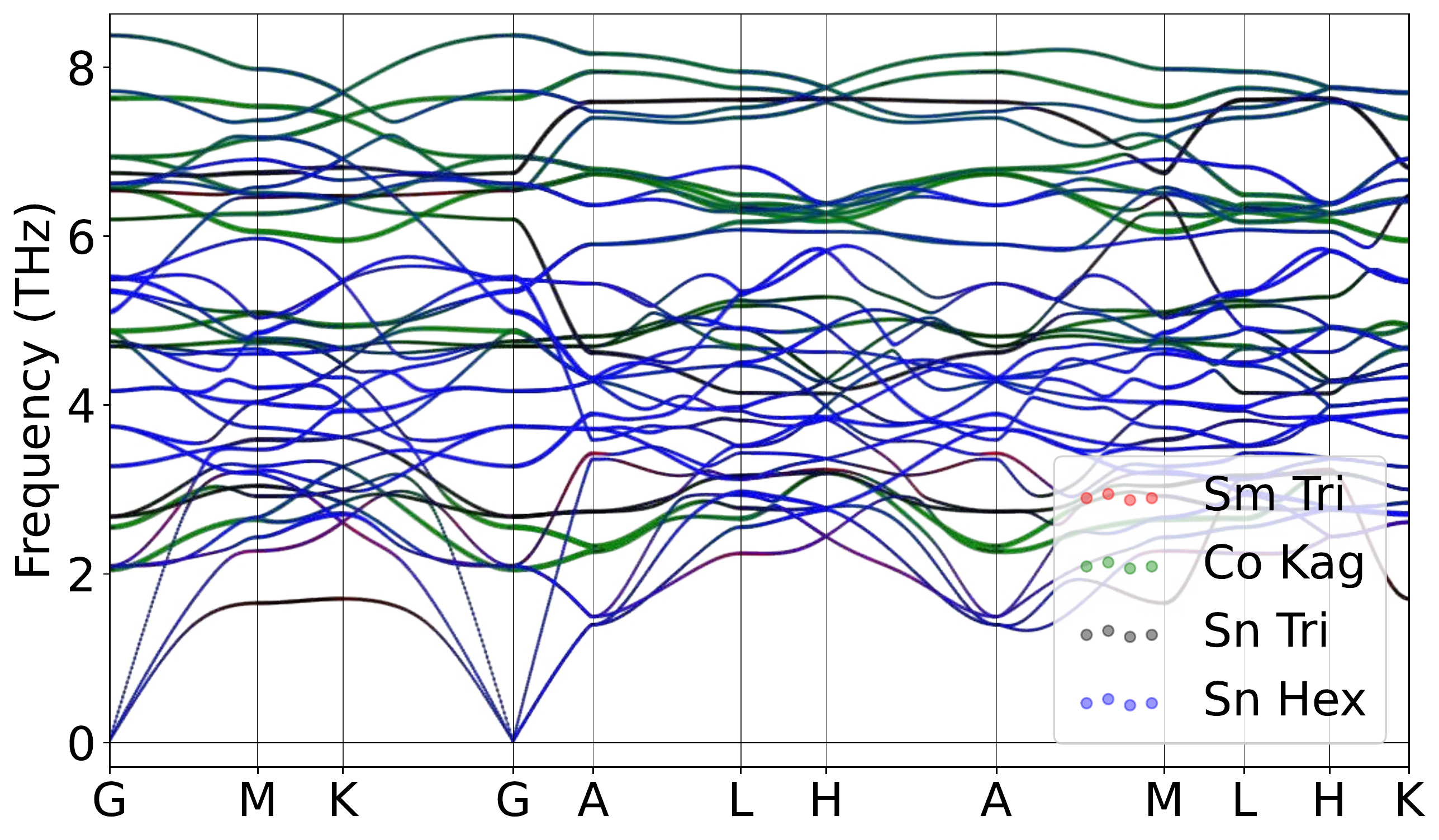}
}
\hspace{5pt}\subfloat[Relaxed phonon at high-T.]{
\label{fig:SmCo6Sn6_relaxed_High}
\includegraphics[width=0.45\textwidth]{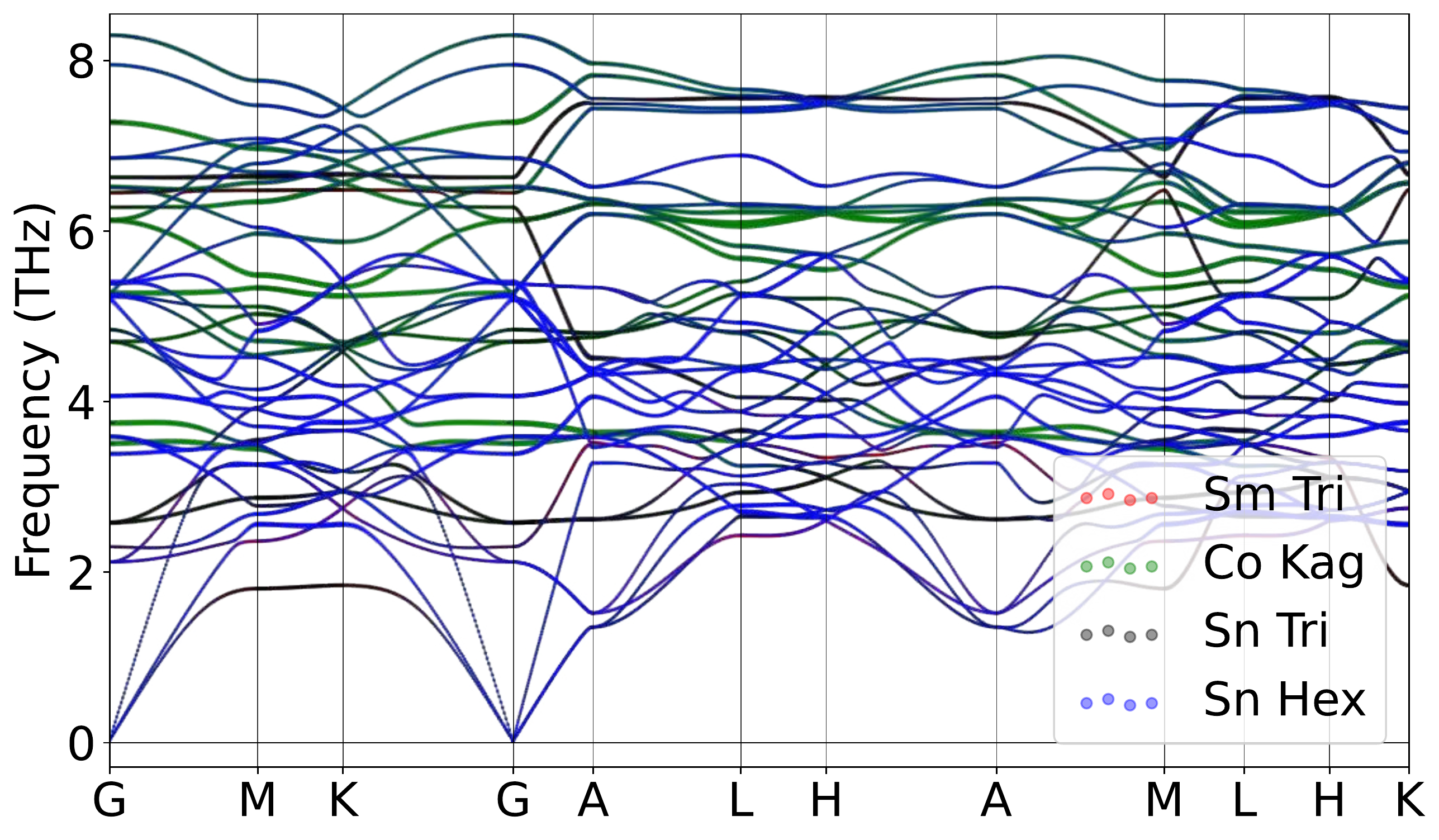}
}
\caption{Atomically projected phonon spectrum for SmCo$_6$Sn$_6$.}
\label{fig:SmCo6Sn6pho}
\end{figure}

\begin{figure}[H]
\centering
\label{fig:SmCo6Sn6_relaxed_Low_xz}
\includegraphics[width=0.85\textwidth]{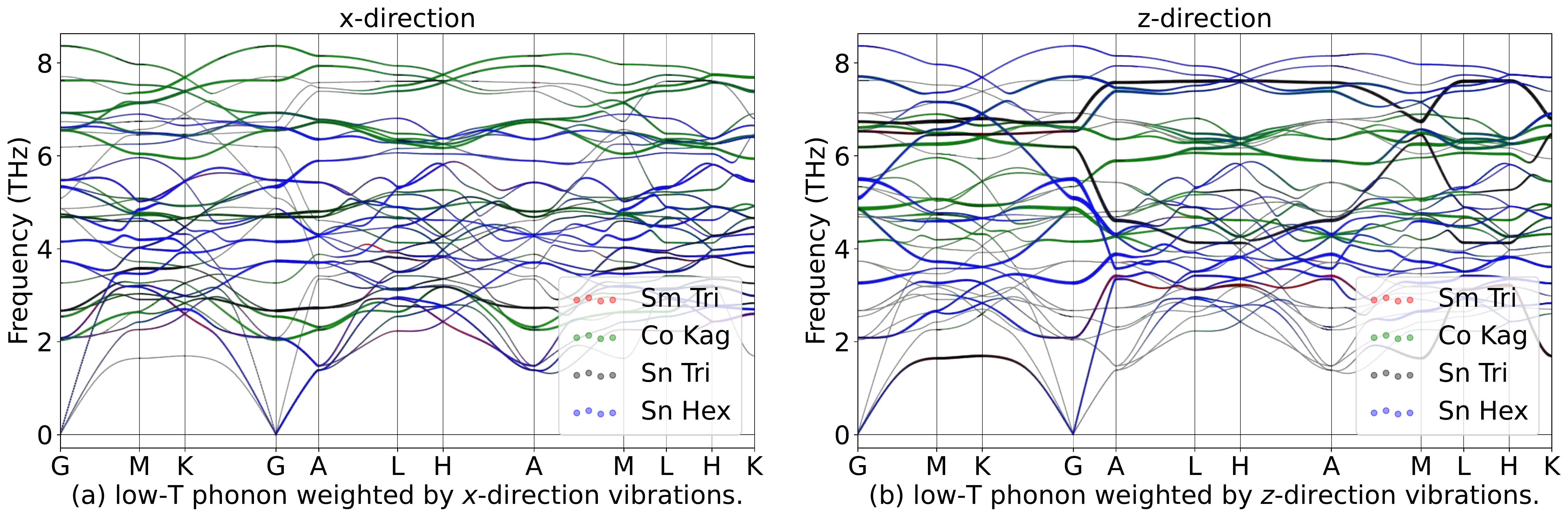}
\hspace{5pt}\label{fig:SmCo6Sn6_relaxed_High_xz}
\includegraphics[width=0.85\textwidth]{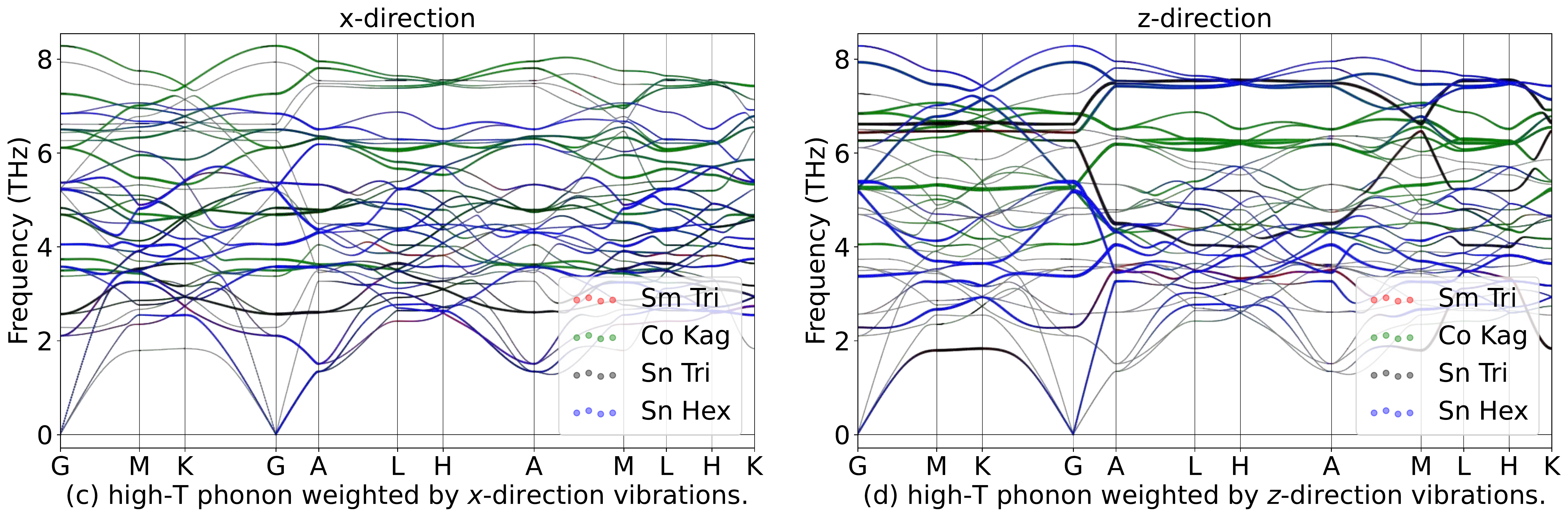}
\caption{Phonon spectrum for SmCo$_6$Sn$_6$in in-plane ($x$) and out-of-plane ($z$) directions.}
\label{fig:SmCo6Sn6phoxyz}
\end{figure}

\begin{table}[htbp]
\global\long\def\arraystretch{1.12}
\captionsetup{width=.95\textwidth}
\caption{IRREPs at high-symmetry points for SmCo$_6$Sn$_6$ phonon spectrum at Low-T.}
\label{tab:191_SmCo6Sn6_Low}
\setlength{\tabcolsep}{1.mm}{
}
\end{table}

\begin{table}[htbp]
\global\long\def\arraystretch{1.12}
\captionsetup{width=.95\textwidth}
\caption{IRREPs at high-symmetry points for SmCo$_6$Sn$_6$ phonon spectrum at High-T.}
\label{tab:191_SmCo6Sn6_High}
\setlength{\tabcolsep}{1.mm}{
%
}
\end{table}
\subsubsection{\label{sms2sec:GdCo6Sn6}\hyperref[smsec:col]{\texorpdfstring{GdCo$_6$Sn$_6$}{}}~{\color{green}  (High-T stable, Low-T stable) }}

\begin{table}[htbp]
\global\long\def\arraystretch{1.12}
\captionsetup{width=.85\textwidth}
\caption{Structure parameters of GdCo$_6$Sn$_6$ after relaxation. It contains 526 electrons. }
\label{tab:GdCo6Sn6}
\setlength{\tabcolsep}{4mm}{
%
}
\end{table}

\begin{figure}[htbp]
\centering
\includegraphics[width=0.85\textwidth]{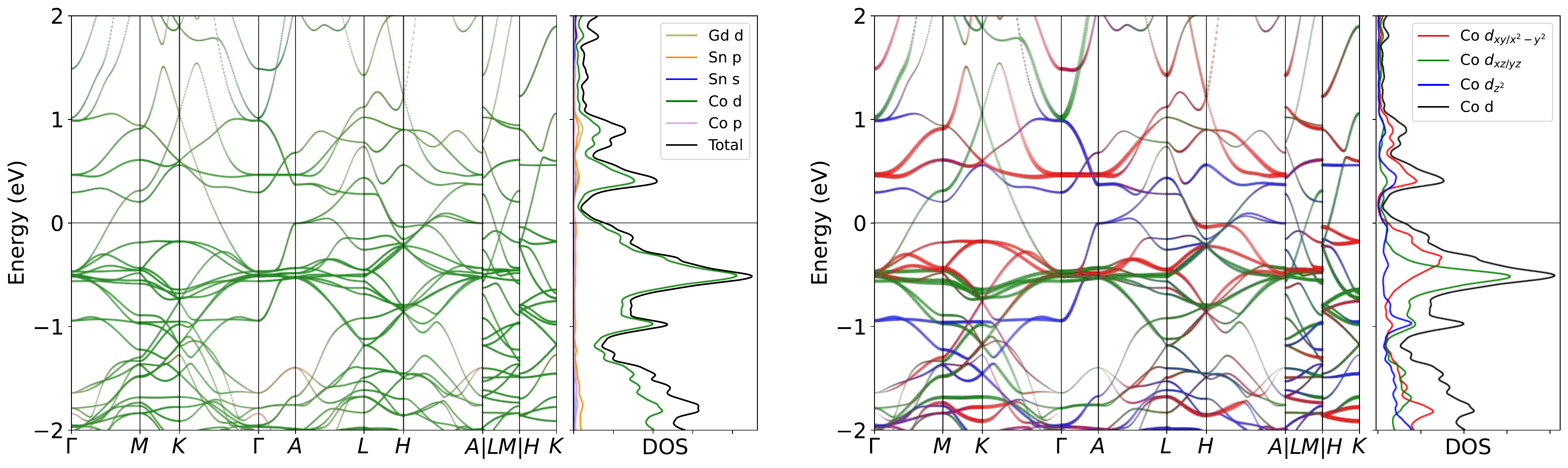}
\caption{Band structure for GdCo$_6$Sn$_6$ without SOC. The projection of Co $3d$ orbitals is also presented. The band structures clearly reveal the dominating of Co $3d$ orbitals  near the Fermi level, as well as the pronounced peak.}
\label{fig:GdCo6Sn6band}
\end{figure}

\begin{figure}[H]
\centering
\subfloat[Relaxed phonon at low-T.]{
\label{fig:GdCo6Sn6_relaxed_Low}
\includegraphics[width=0.45\textwidth]{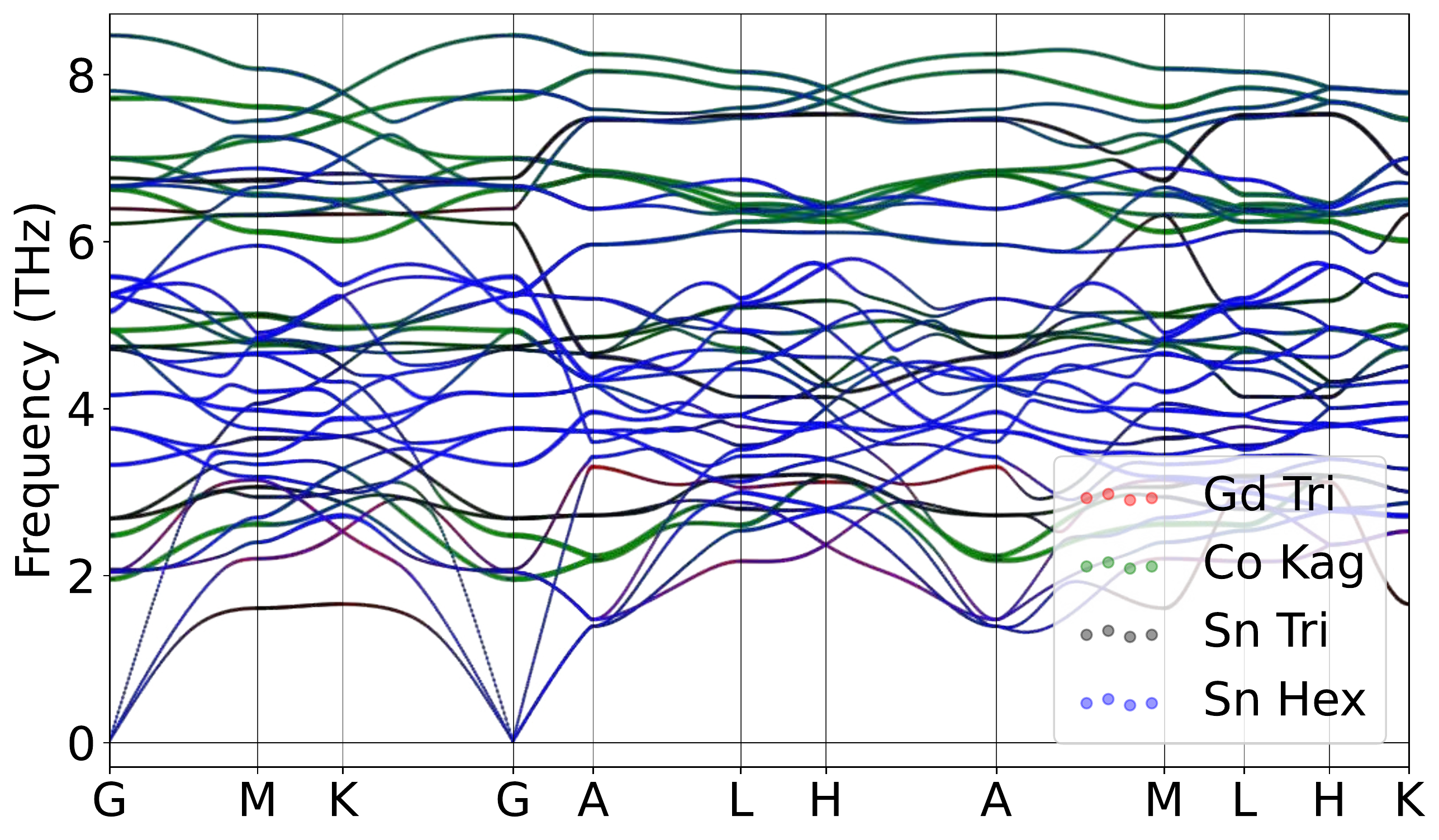}
}
\hspace{5pt}\subfloat[Relaxed phonon at high-T.]{
\label{fig:GdCo6Sn6_relaxed_High}
\includegraphics[width=0.45\textwidth]{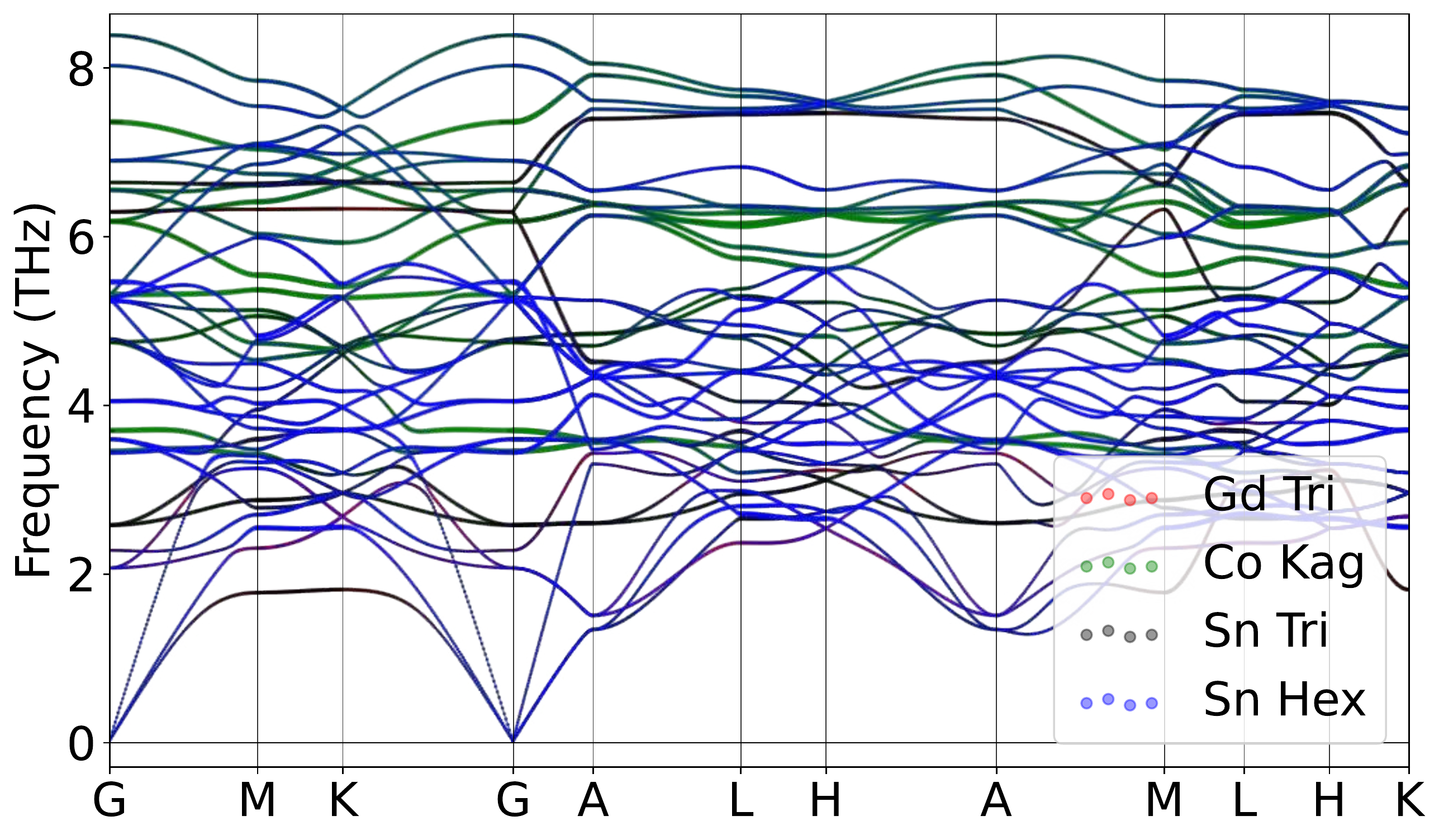}
}
\caption{Atomically projected phonon spectrum for GdCo$_6$Sn$_6$.}
\label{fig:GdCo6Sn6pho}
\end{figure}

\begin{figure}[H]
\centering
\label{fig:GdCo6Sn6_relaxed_Low_xz}
\includegraphics[width=0.85\textwidth]{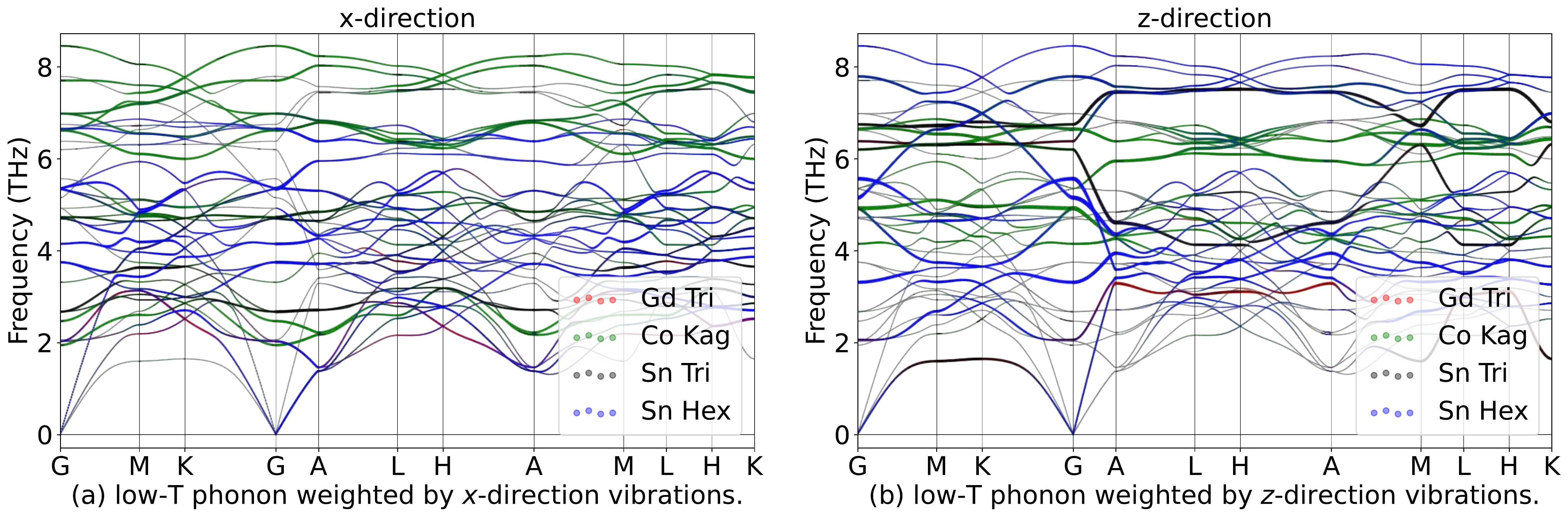}
\hspace{5pt}\label{fig:GdCo6Sn6_relaxed_High_xz}
\includegraphics[width=0.85\textwidth]{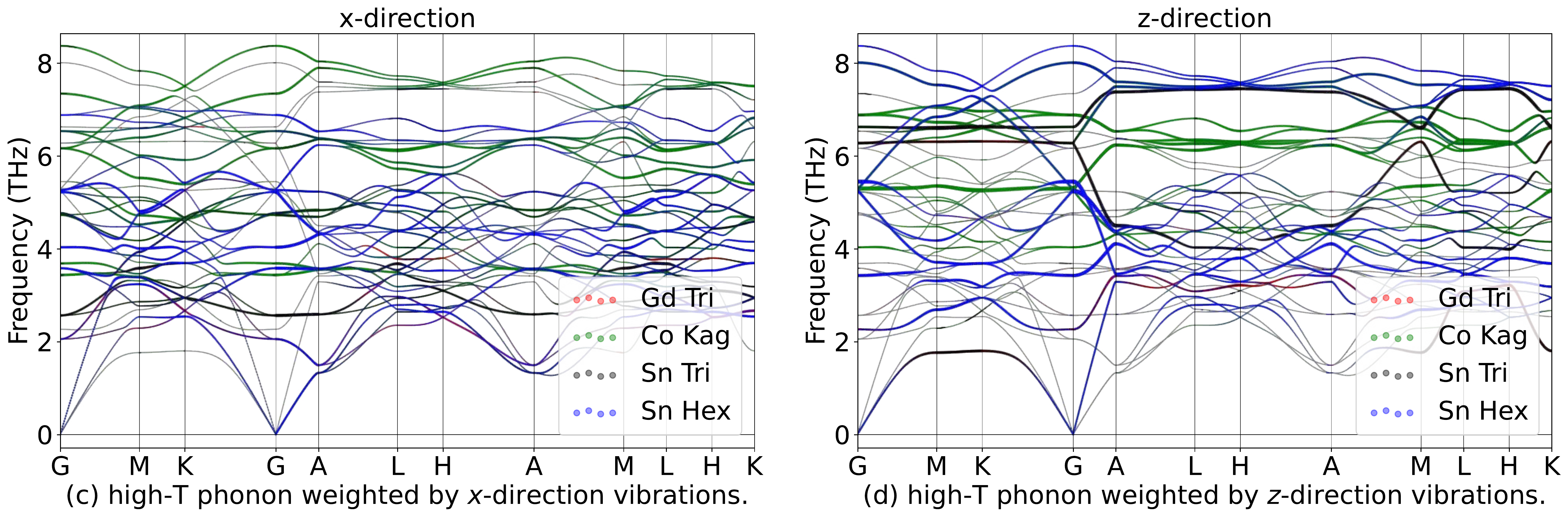}
\caption{Phonon spectrum for GdCo$_6$Sn$_6$in in-plane ($x$) and out-of-plane ($z$) directions.}
\label{fig:GdCo6Sn6phoxyz}
\end{figure}

\begin{table}[htbp]
\global\long\def\arraystretch{1.12}
\captionsetup{width=.95\textwidth}
\caption{IRREPs at high-symmetry points for GdCo$_6$Sn$_6$ phonon spectrum at Low-T.}
\label{tab:191_GdCo6Sn6_Low}
\setlength{\tabcolsep}{1.mm}{
}
\end{table}

\begin{table}[htbp]
\global\long\def\arraystretch{1.12}
\captionsetup{width=.95\textwidth}
\caption{IRREPs at high-symmetry points for GdCo$_6$Sn$_6$ phonon spectrum at High-T.}
\label{tab:191_GdCo6Sn6_High}
\setlength{\tabcolsep}{1.mm}{
%
}
\end{table}
\subsubsection{\label{sms2sec:TbCo6Sn6}\hyperref[smsec:col]{\texorpdfstring{TbCo$_6$Sn$_6$}{}}~{\color{green}  (High-T stable, Low-T stable) }}

\begin{table}[htbp]
\global\long\def\arraystretch{1.12}
\captionsetup{width=.85\textwidth}
\caption{Structure parameters of TbCo$_6$Sn$_6$ after relaxation. It contains 527 electrons. }
\label{tab:TbCo6Sn6}
\setlength{\tabcolsep}{4mm}{
%
}
\end{table}

\begin{figure}[htbp]
\centering
\includegraphics[width=0.85\textwidth]{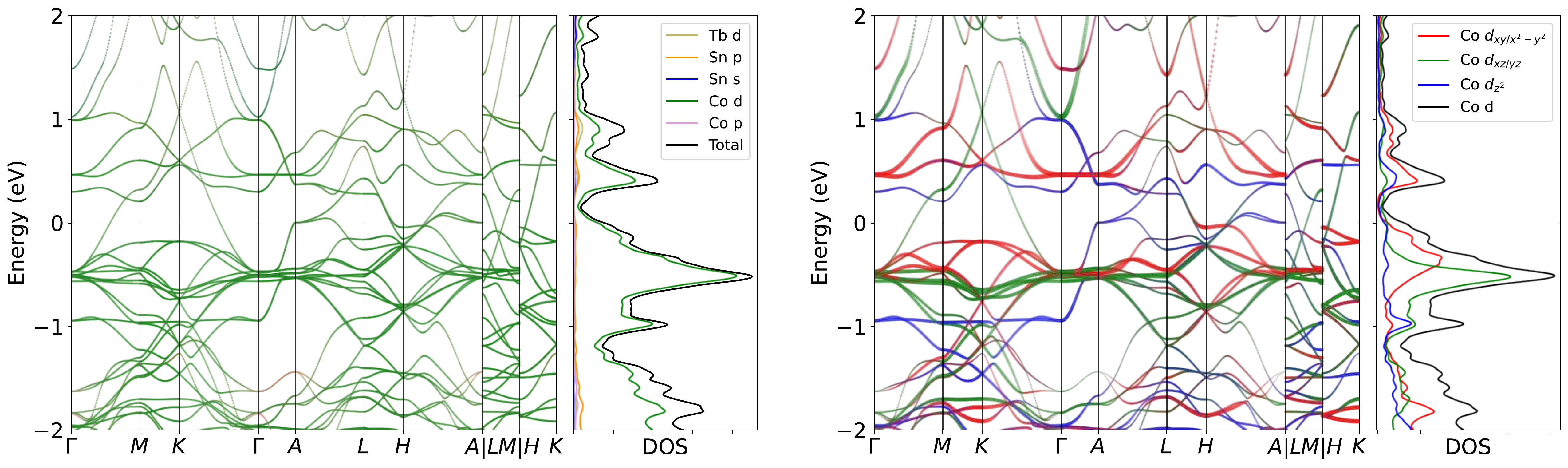}
\caption{Band structure for TbCo$_6$Sn$_6$ without SOC. The projection of Co $3d$ orbitals is also presented. The band structures clearly reveal the dominating of Co $3d$ orbitals  near the Fermi level, as well as the pronounced peak.}
\label{fig:TbCo6Sn6band}
\end{figure}

\begin{figure}[H]
\centering
\subfloat[Relaxed phonon at low-T.]{
\label{fig:TbCo6Sn6_relaxed_Low}
\includegraphics[width=0.45\textwidth]{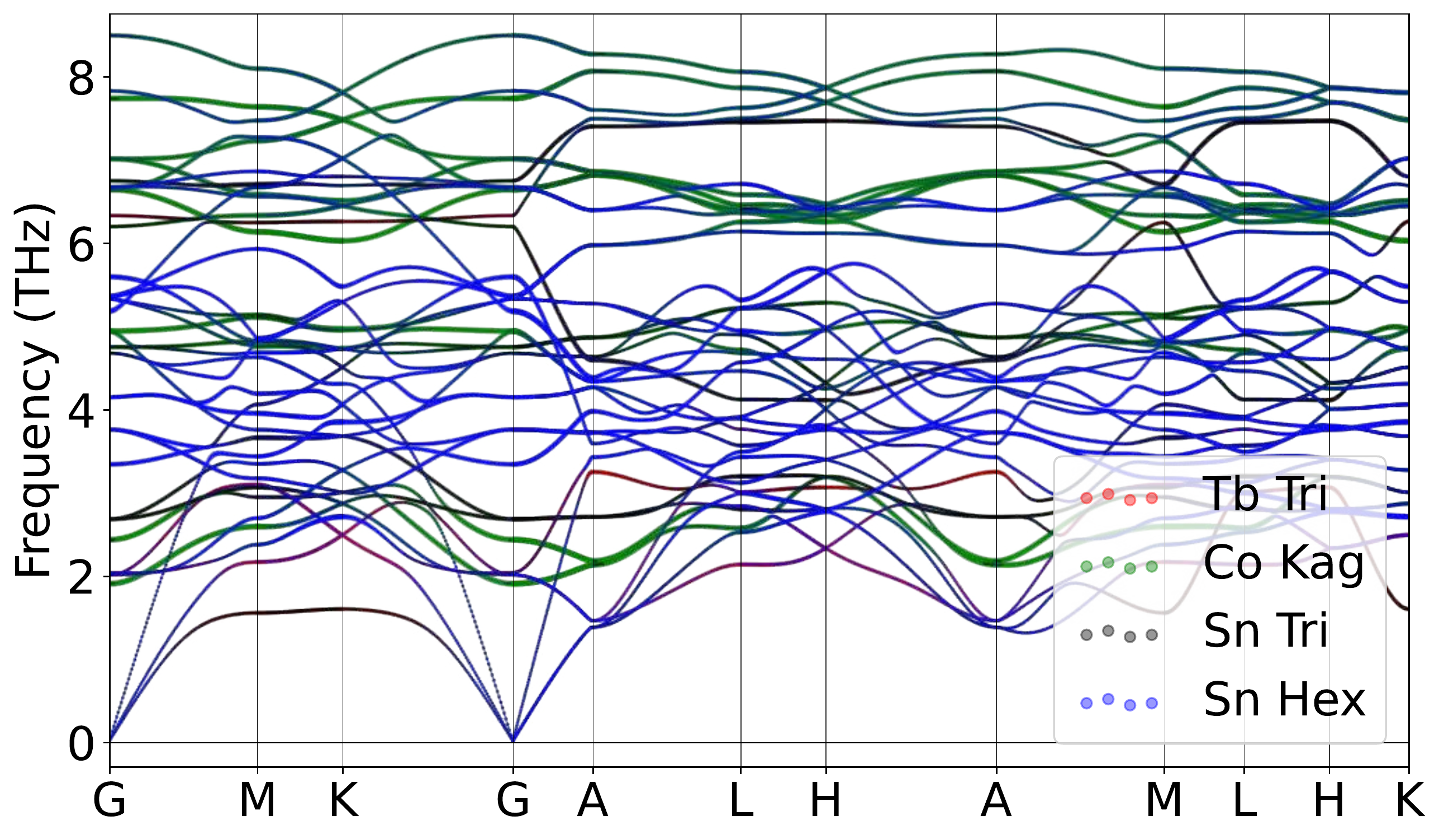}
}
\hspace{5pt}\subfloat[Relaxed phonon at high-T.]{
\label{fig:TbCo6Sn6_relaxed_High}
\includegraphics[width=0.45\textwidth]{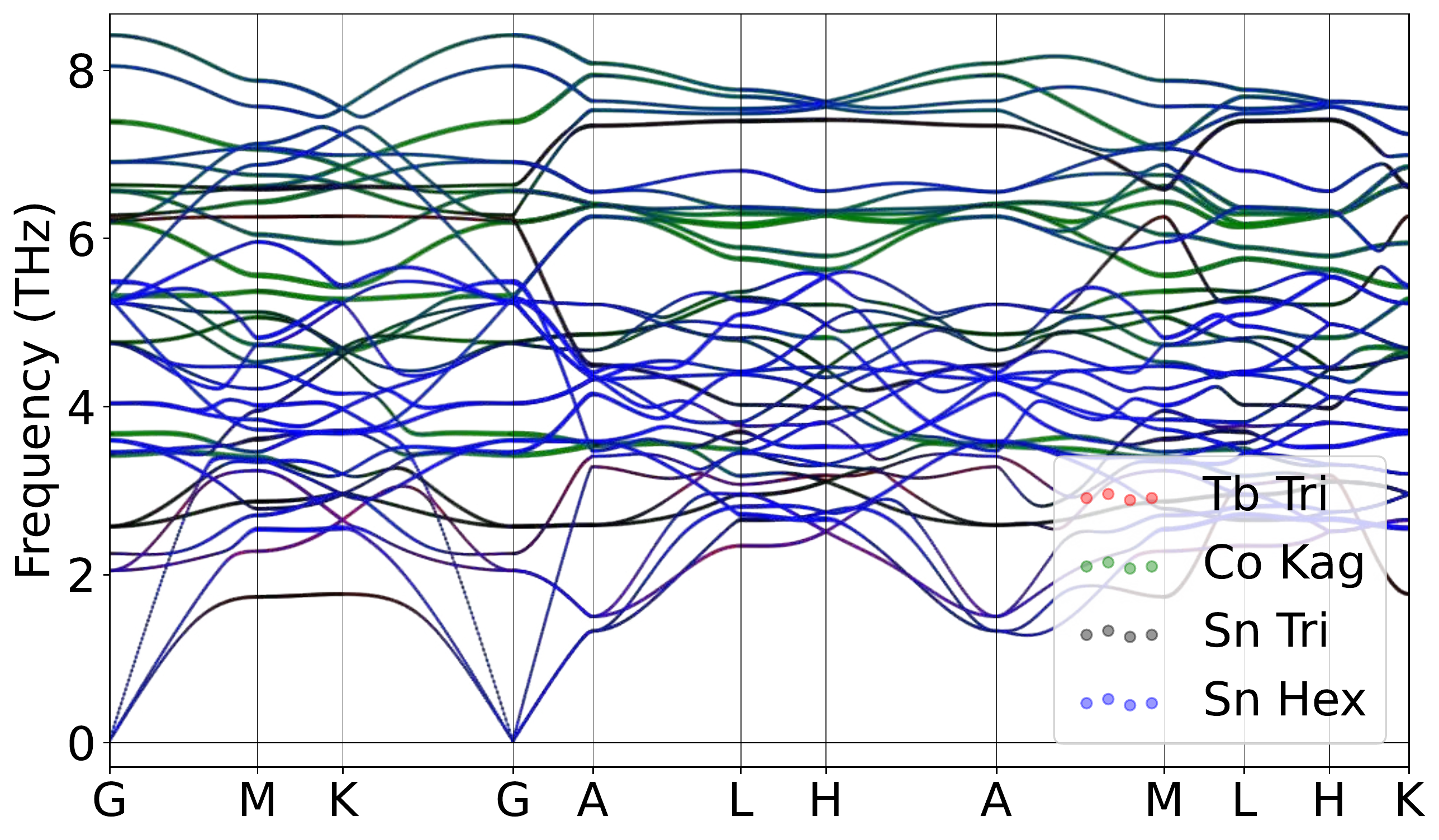}
}
\caption{Atomically projected phonon spectrum for TbCo$_6$Sn$_6$.}
\label{fig:TbCo6Sn6pho}
\end{figure}

\begin{figure}[H]
\centering
\label{fig:TbCo6Sn6_relaxed_Low_xz}
\includegraphics[width=0.85\textwidth]{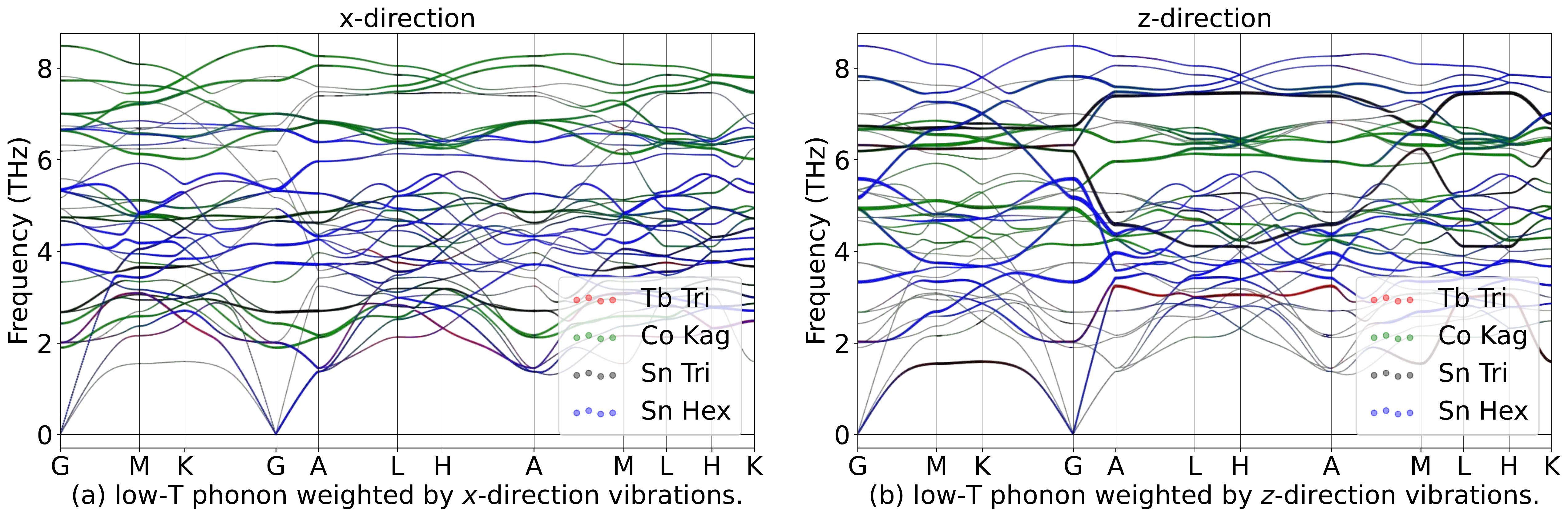}
\hspace{5pt}\label{fig:TbCo6Sn6_relaxed_High_xz}
\includegraphics[width=0.85\textwidth]{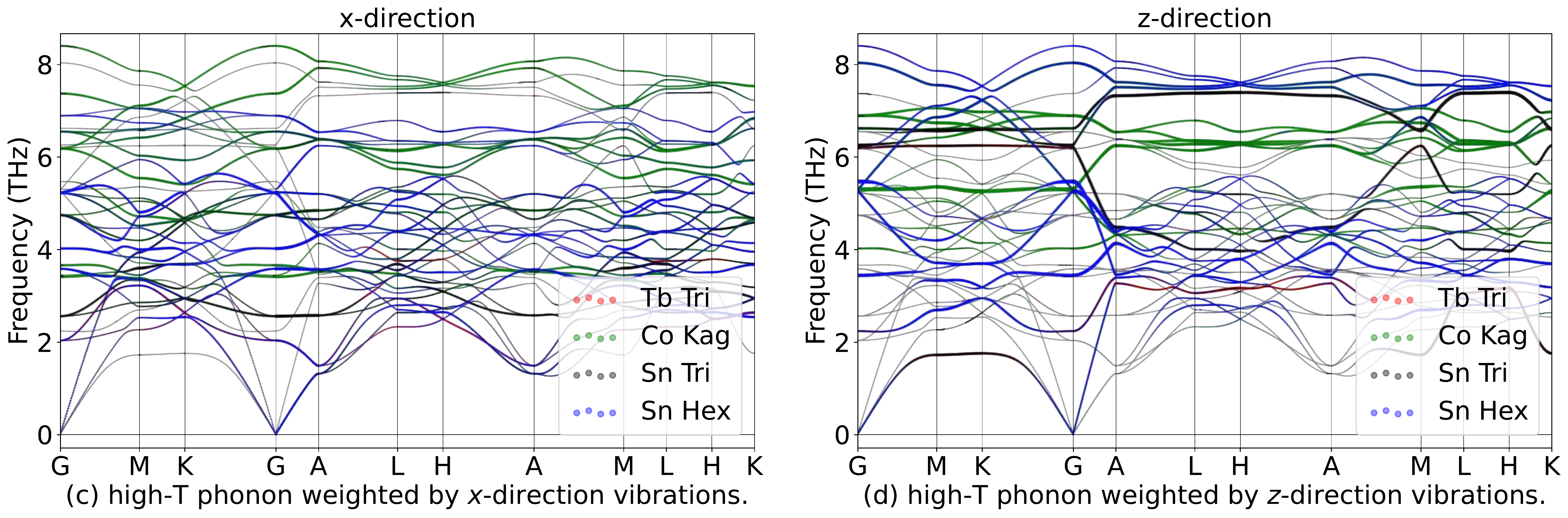}
\caption{Phonon spectrum for TbCo$_6$Sn$_6$in in-plane ($x$) and out-of-plane ($z$) directions.}
\label{fig:TbCo6Sn6phoxyz}
\end{figure}

\begin{table}[htbp]
\global\long\def\arraystretch{1.12}
\captionsetup{width=.95\textwidth}
\caption{IRREPs at high-symmetry points for TbCo$_6$Sn$_6$ phonon spectrum at Low-T.}
\label{tab:191_TbCo6Sn6_Low}
\setlength{\tabcolsep}{1.mm}{
}
\end{table}

\begin{table}[htbp]
\global\long\def\arraystretch{1.12}
\captionsetup{width=.95\textwidth}
\caption{IRREPs at high-symmetry points for TbCo$_6$Sn$_6$ phonon spectrum at High-T.}
\label{tab:191_TbCo6Sn6_High}
\setlength{\tabcolsep}{1.mm}{
%
}
\end{table}
\subsubsection{\label{sms2sec:DyCo6Sn6}\hyperref[smsec:col]{\texorpdfstring{DyCo$_6$Sn$_6$}{}}~{\color{green}  (High-T stable, Low-T stable) }}

\begin{table}[htbp]
\global\long\def\arraystretch{1.12}
\captionsetup{width=.85\textwidth}
\caption{Structure parameters of DyCo$_6$Sn$_6$ after relaxation. It contains 528 electrons. }
\label{tab:DyCo6Sn6}
\setlength{\tabcolsep}{4mm}{
%
}
\end{table}

\begin{figure}[htbp]
\centering
\includegraphics[width=0.85\textwidth]{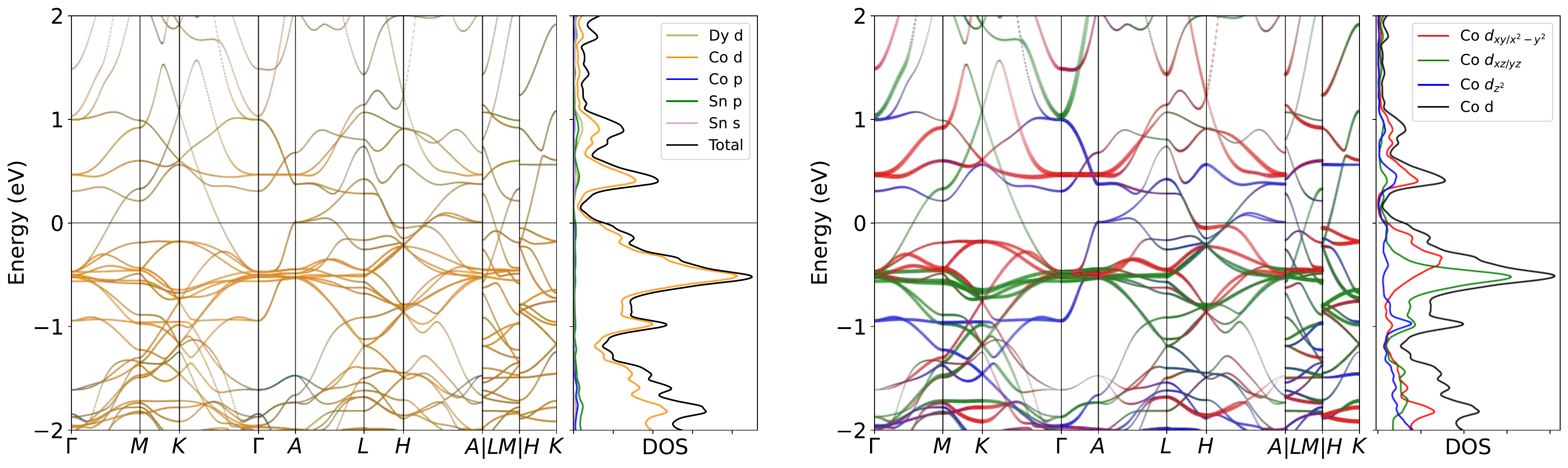}
\caption{Band structure for DyCo$_6$Sn$_6$ without SOC. The projection of Co $3d$ orbitals is also presented. The band structures clearly reveal the dominating of Co $3d$ orbitals  near the Fermi level, as well as the pronounced peak.}
\label{fig:DyCo6Sn6band}
\end{figure}

\begin{figure}[H]
\centering
\subfloat[Relaxed phonon at low-T.]{
\label{fig:DyCo6Sn6_relaxed_Low}
\includegraphics[width=0.45\textwidth]{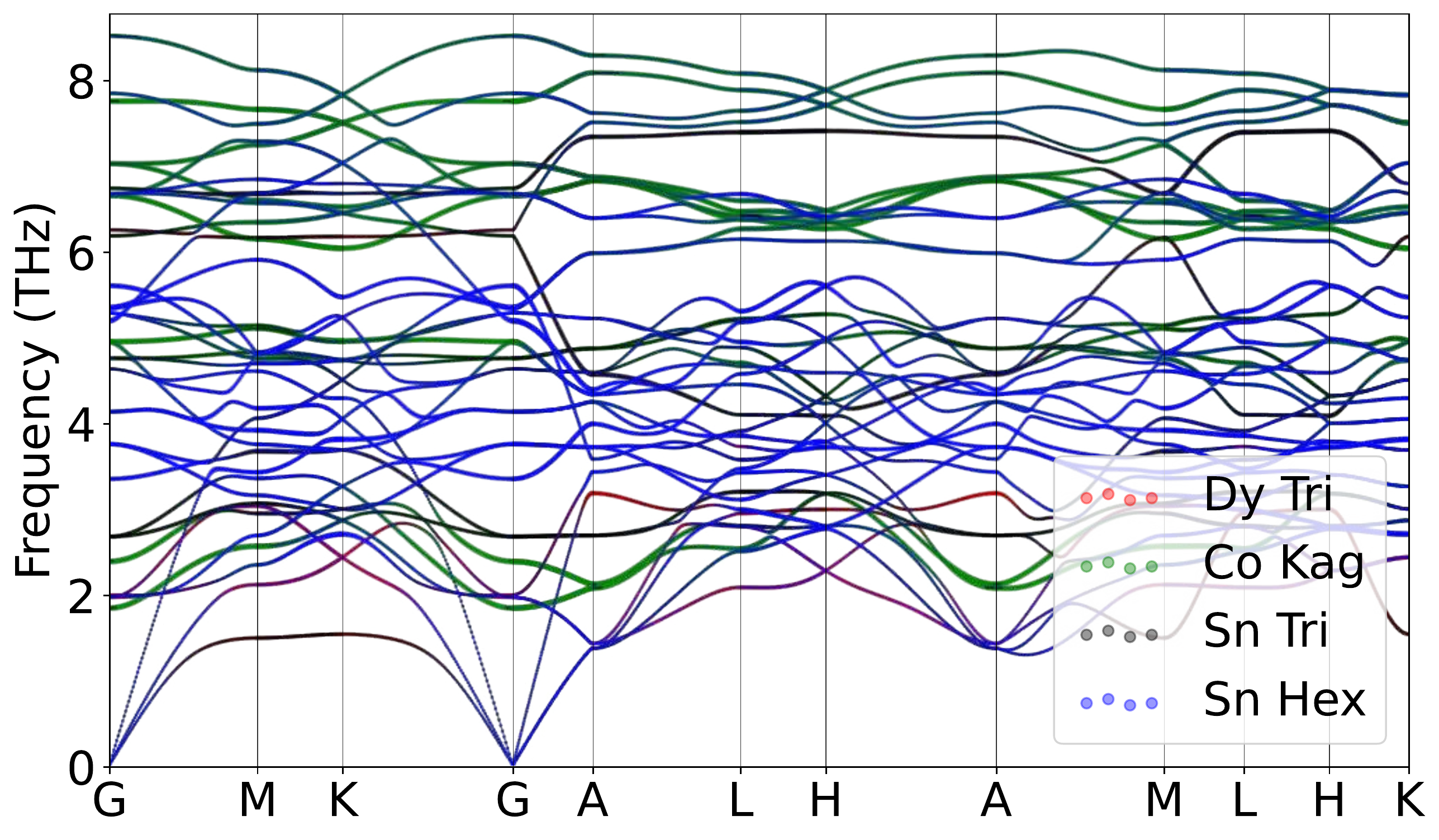}
}
\hspace{5pt}\subfloat[Relaxed phonon at high-T.]{
\label{fig:DyCo6Sn6_relaxed_High}
\includegraphics[width=0.45\textwidth]{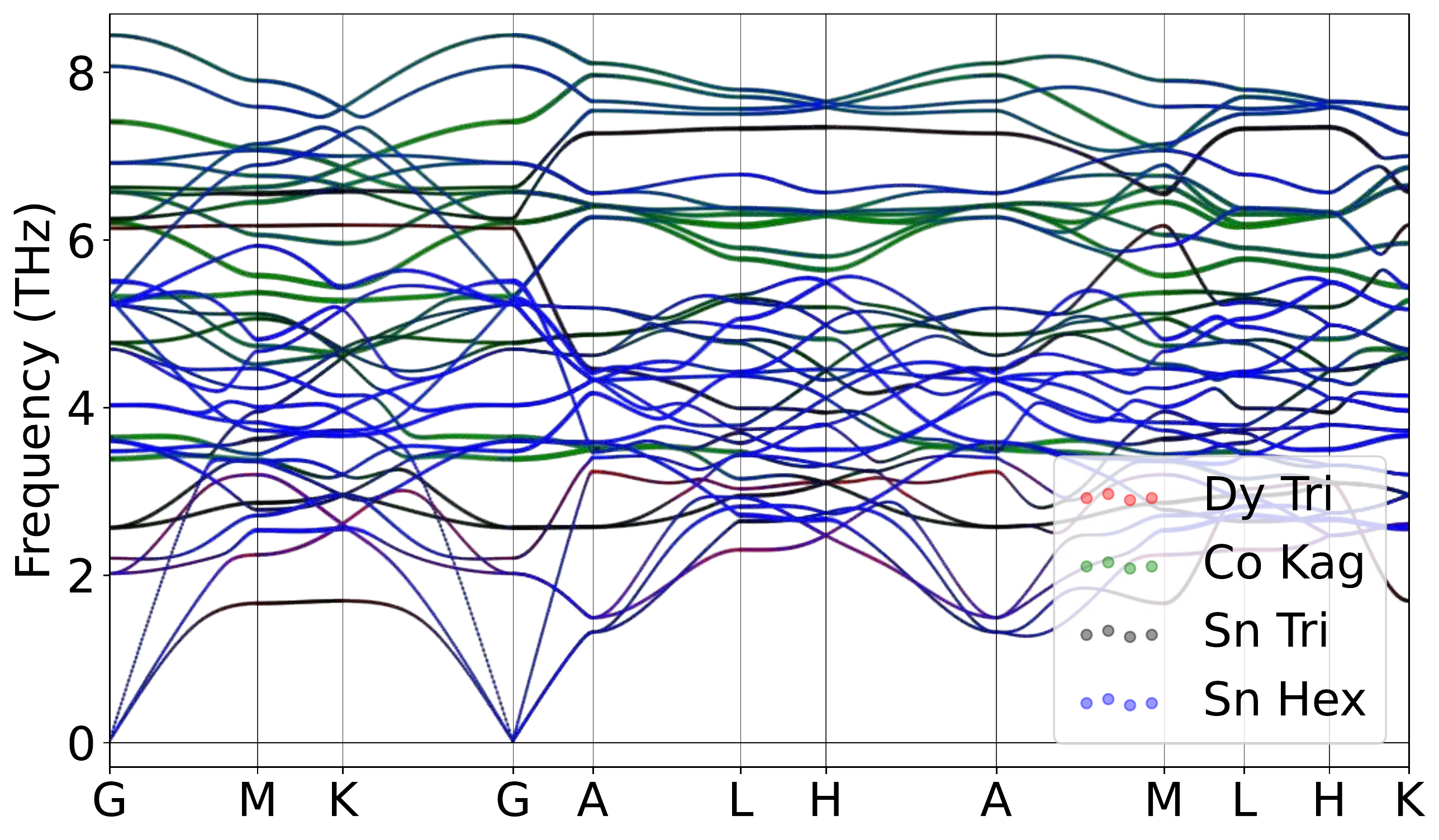}
}
\caption{Atomically projected phonon spectrum for DyCo$_6$Sn$_6$.}
\label{fig:DyCo6Sn6pho}
\end{figure}

\begin{figure}[H]
\centering
\label{fig:DyCo6Sn6_relaxed_Low_xz}
\includegraphics[width=0.85\textwidth]{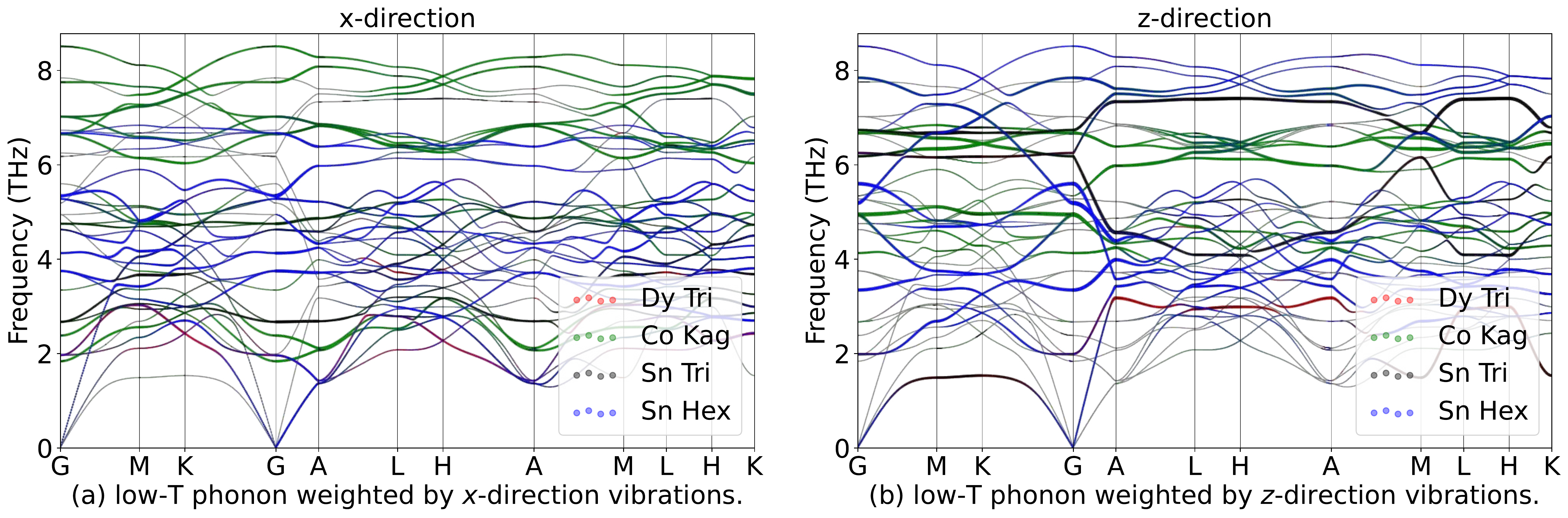}
\hspace{5pt}\label{fig:DyCo6Sn6_relaxed_High_xz}
\includegraphics[width=0.85\textwidth]{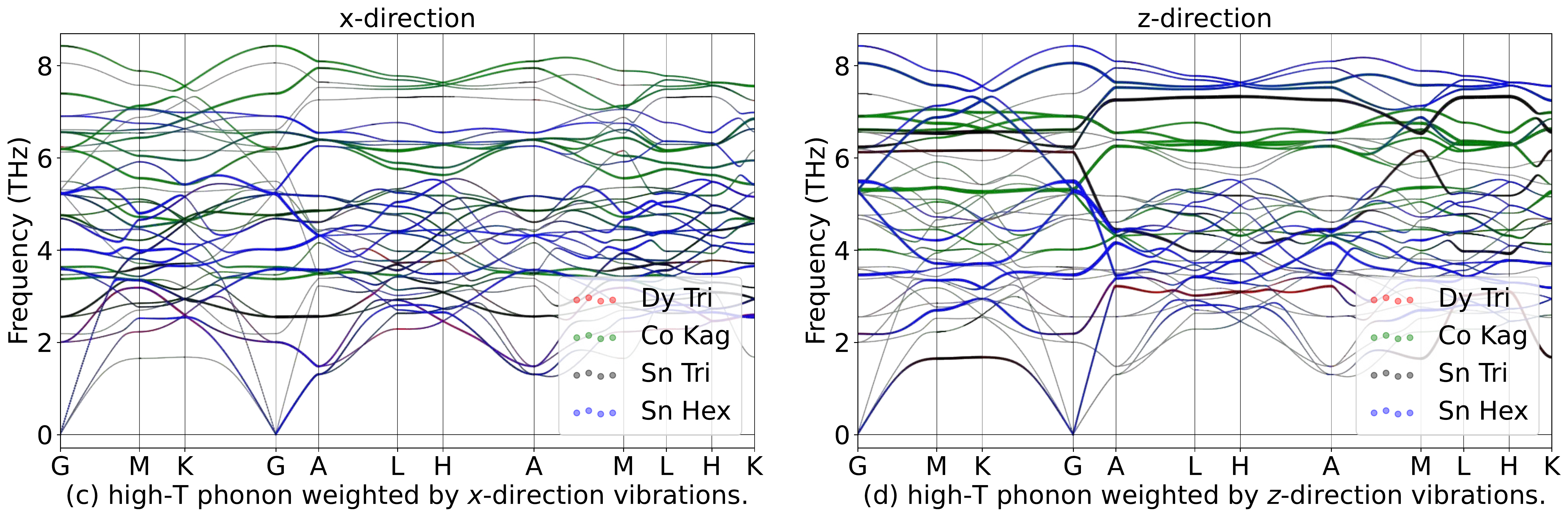}
\caption{Phonon spectrum for DyCo$_6$Sn$_6$in in-plane ($x$) and out-of-plane ($z$) directions.}
\label{fig:DyCo6Sn6phoxyz}
\end{figure}

\begin{table}[htbp]
\global\long\def\arraystretch{1.12}
\captionsetup{width=.95\textwidth}
\caption{IRREPs at high-symmetry points for DyCo$_6$Sn$_6$ phonon spectrum at Low-T.}
\label{tab:191_DyCo6Sn6_Low}
\setlength{\tabcolsep}{1.mm}{
}
\end{table}

\begin{table}[htbp]
\global\long\def\arraystretch{1.12}
\captionsetup{width=.95\textwidth}
\caption{IRREPs at high-symmetry points for DyCo$_6$Sn$_6$ phonon spectrum at High-T.}
\label{tab:191_DyCo6Sn6_High}
\setlength{\tabcolsep}{1.mm}{
%
}
\end{table}
\subsubsection{\label{sms2sec:HoCo6Sn6}\hyperref[smsec:col]{\texorpdfstring{HoCo$_6$Sn$_6$}{}}~{\color{green}  (High-T stable, Low-T stable) }}

\begin{table}[htbp]
\global\long\def\arraystretch{1.12}
\captionsetup{width=.85\textwidth}
\caption{Structure parameters of HoCo$_6$Sn$_6$ after relaxation. It contains 529 electrons. }
\label{tab:HoCo6Sn6}
\setlength{\tabcolsep}{4mm}{
%
}
\end{table}

\begin{figure}[htbp]
\centering
\includegraphics[width=0.85\textwidth]{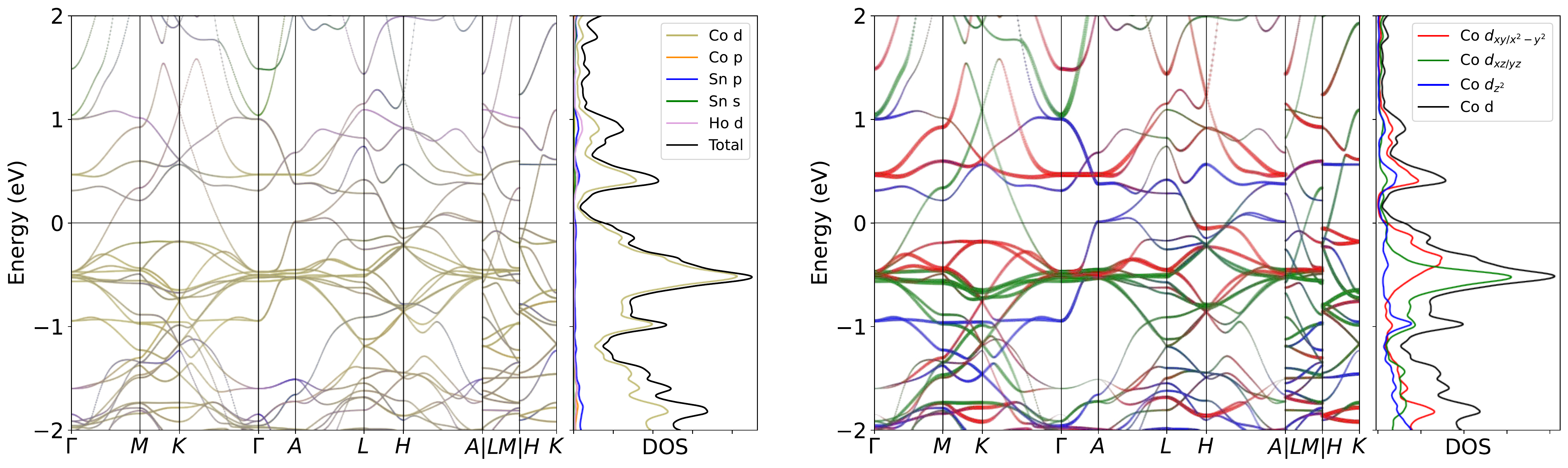}
\caption{Band structure for HoCo$_6$Sn$_6$ without SOC. The projection of Co $3d$ orbitals is also presented. The band structures clearly reveal the dominating of Co $3d$ orbitals  near the Fermi level, as well as the pronounced peak.}
\label{fig:HoCo6Sn6band}
\end{figure}

\begin{figure}[H]
\centering
\subfloat[Relaxed phonon at low-T.]{
\label{fig:HoCo6Sn6_relaxed_Low}
\includegraphics[width=0.45\textwidth]{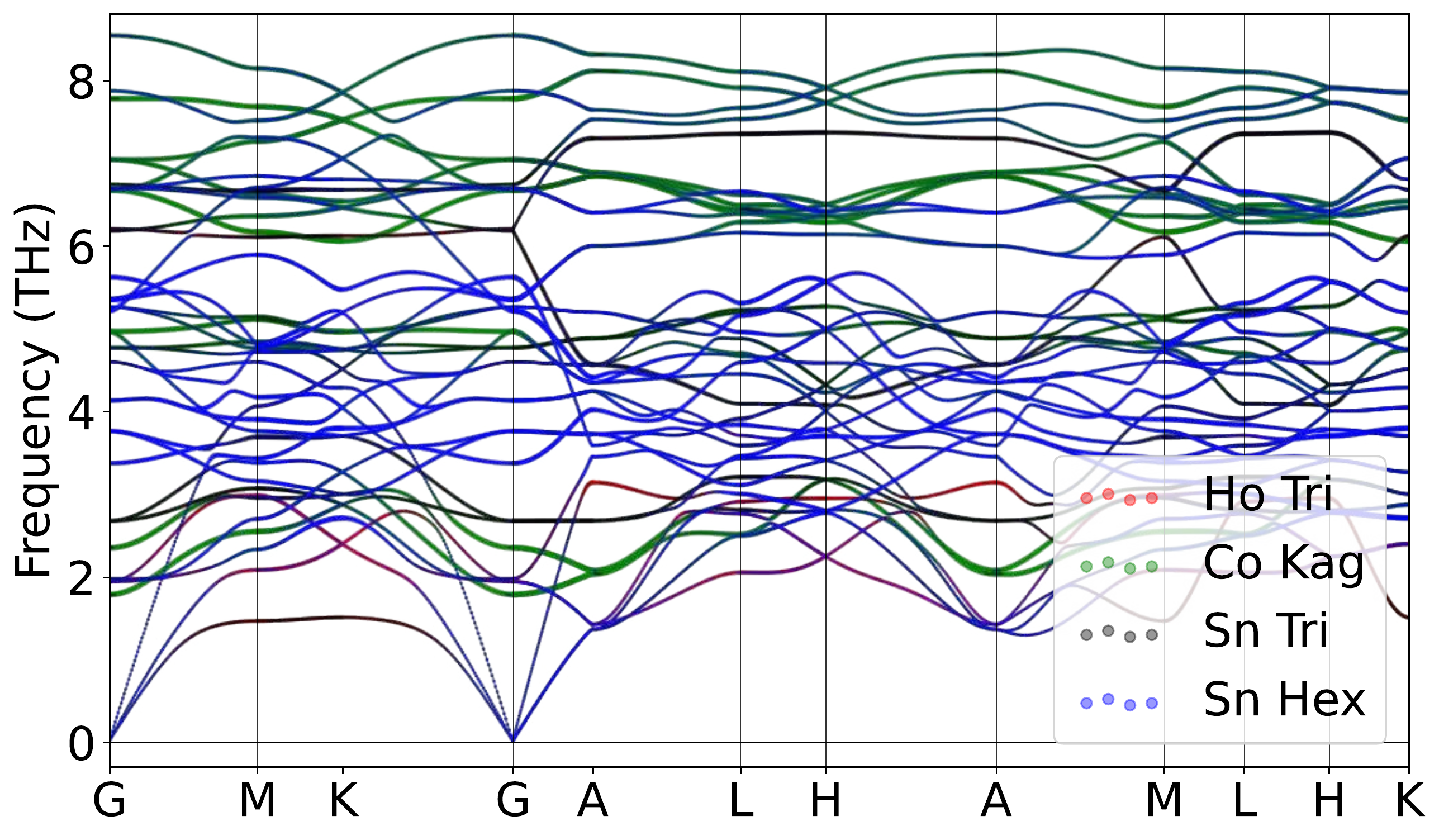}
}
\hspace{5pt}\subfloat[Relaxed phonon at high-T.]{
\label{fig:HoCo6Sn6_relaxed_High}
\includegraphics[width=0.45\textwidth]{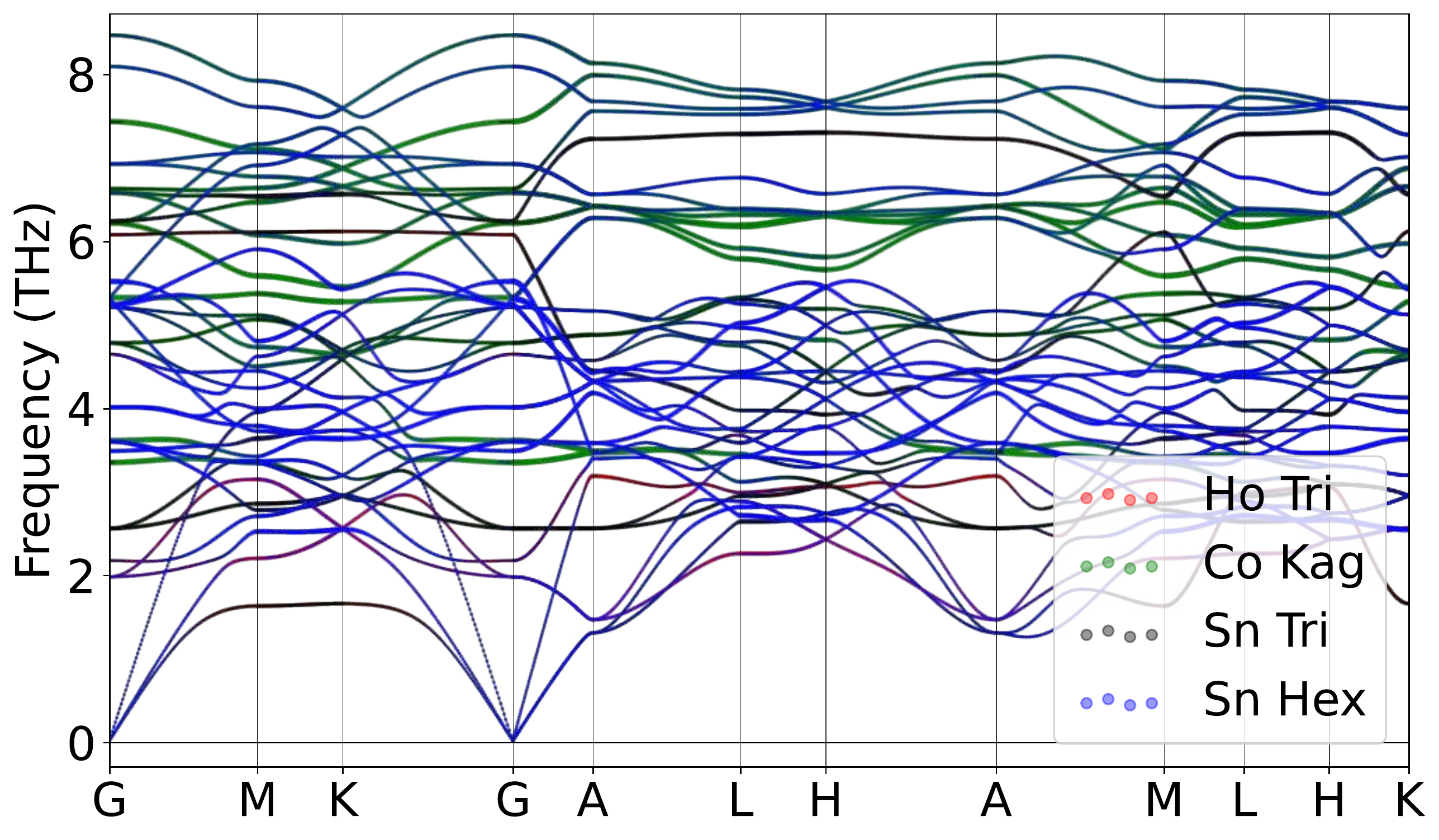}
}
\caption{Atomically projected phonon spectrum for HoCo$_6$Sn$_6$.}
\label{fig:HoCo6Sn6pho}
\end{figure}

\begin{figure}[H]
\centering
\label{fig:HoCo6Sn6_relaxed_Low_xz}
\includegraphics[width=0.85\textwidth]{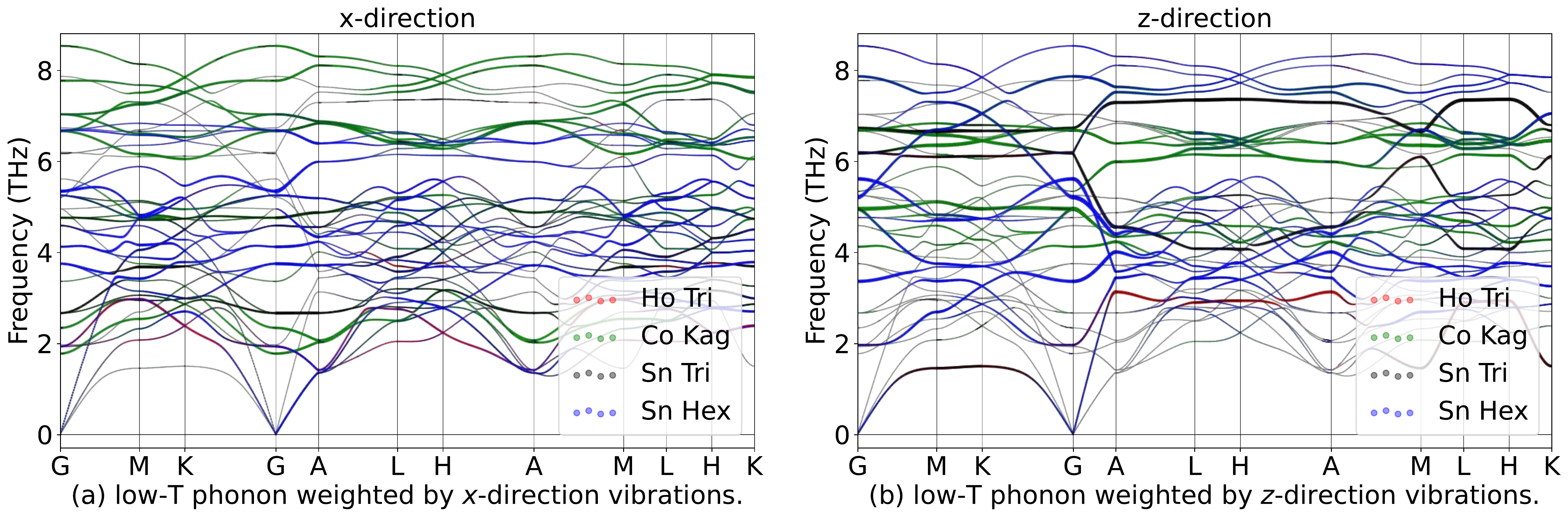}
\hspace{5pt}\label{fig:HoCo6Sn6_relaxed_High_xz}
\includegraphics[width=0.85\textwidth]{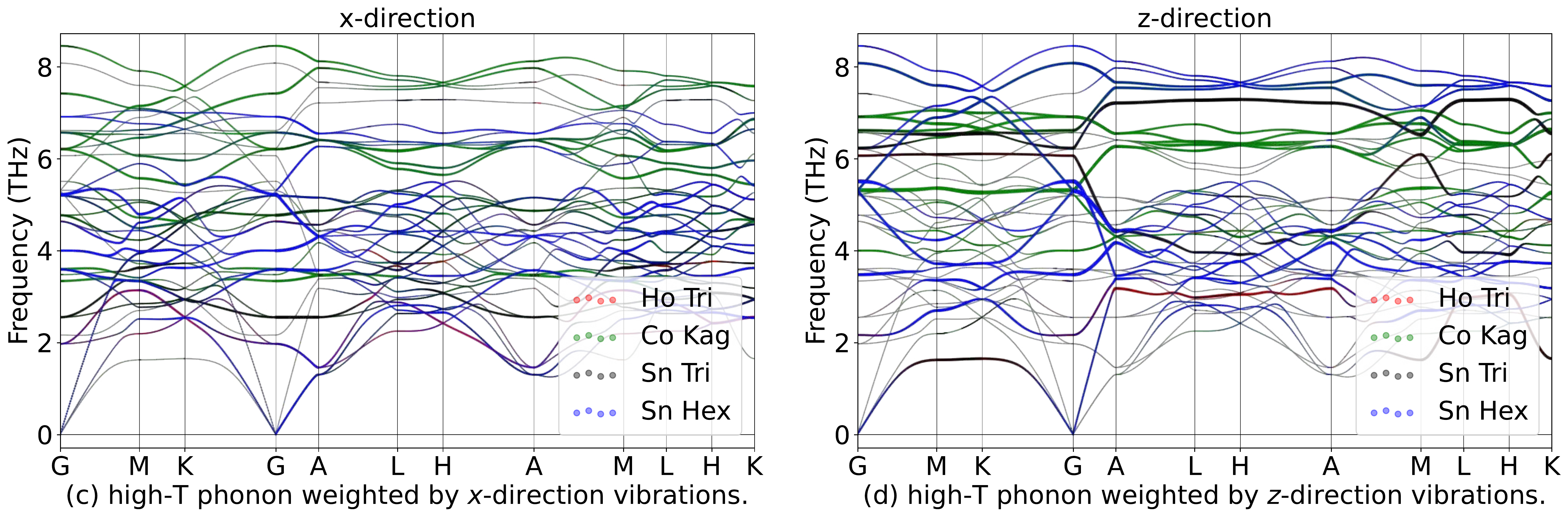}
\caption{Phonon spectrum for HoCo$_6$Sn$_6$in in-plane ($x$) and out-of-plane ($z$) directions.}
\label{fig:HoCo6Sn6phoxyz}
\end{figure}

\begin{table}[htbp]
\global\long\def\arraystretch{1.12}
\captionsetup{width=.95\textwidth}
\caption{IRREPs at high-symmetry points for HoCo$_6$Sn$_6$ phonon spectrum at Low-T.}
\label{tab:191_HoCo6Sn6_Low}
\setlength{\tabcolsep}{1.mm}{
}
\end{table}

\begin{table}[htbp]
\global\long\def\arraystretch{1.12}
\captionsetup{width=.95\textwidth}
\caption{IRREPs at high-symmetry points for HoCo$_6$Sn$_6$ phonon spectrum at High-T.}
\label{tab:191_HoCo6Sn6_High}
\setlength{\tabcolsep}{1.mm}{
%
}
\end{table}
\subsubsection{\label{sms2sec:ErCo6Sn6}\hyperref[smsec:col]{\texorpdfstring{ErCo$_6$Sn$_6$}{}}~{\color{green}  (High-T stable, Low-T stable) }}

\begin{table}[htbp]
\global\long\def\arraystretch{1.12}
\captionsetup{width=.85\textwidth}
\caption{Structure parameters of ErCo$_6$Sn$_6$ after relaxation. It contains 530 electrons. }
\label{tab:ErCo6Sn6}
\setlength{\tabcolsep}{4mm}{
%
}
\end{table}

\begin{figure}[htbp]
\centering
\includegraphics[width=0.85\textwidth]{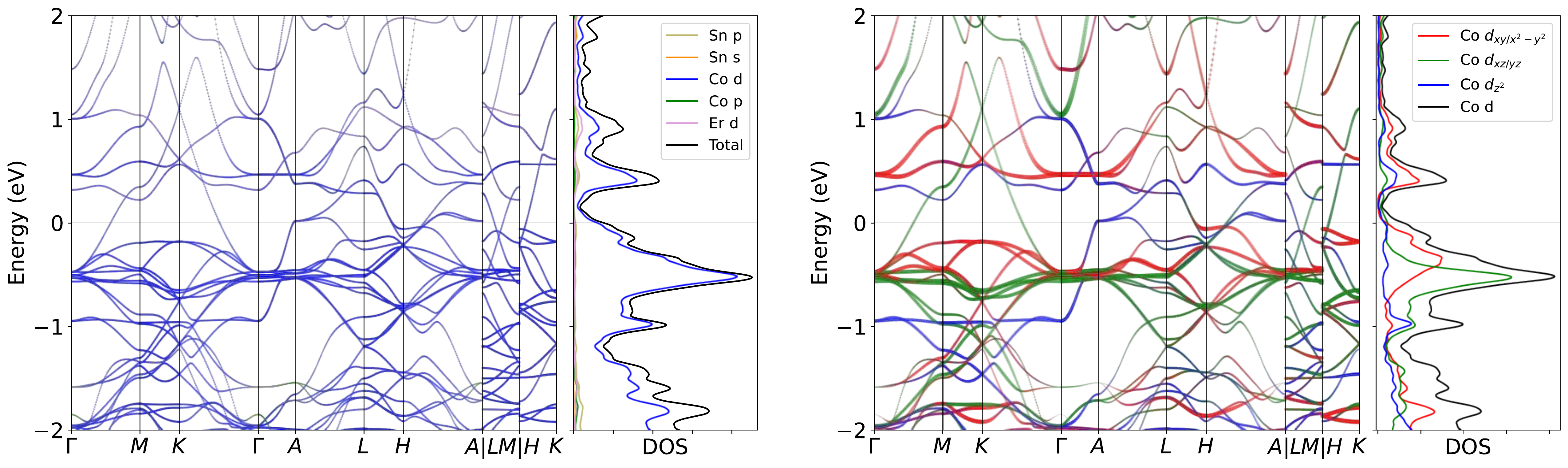}
\caption{Band structure for ErCo$_6$Sn$_6$ without SOC. The projection of Co $3d$ orbitals is also presented. The band structures clearly reveal the dominating of Co $3d$ orbitals  near the Fermi level, as well as the pronounced peak.}
\label{fig:ErCo6Sn6band}
\end{figure}

\begin{figure}[H]
\centering
\subfloat[Relaxed phonon at low-T.]{
\label{fig:ErCo6Sn6_relaxed_Low}
\includegraphics[width=0.45\textwidth]{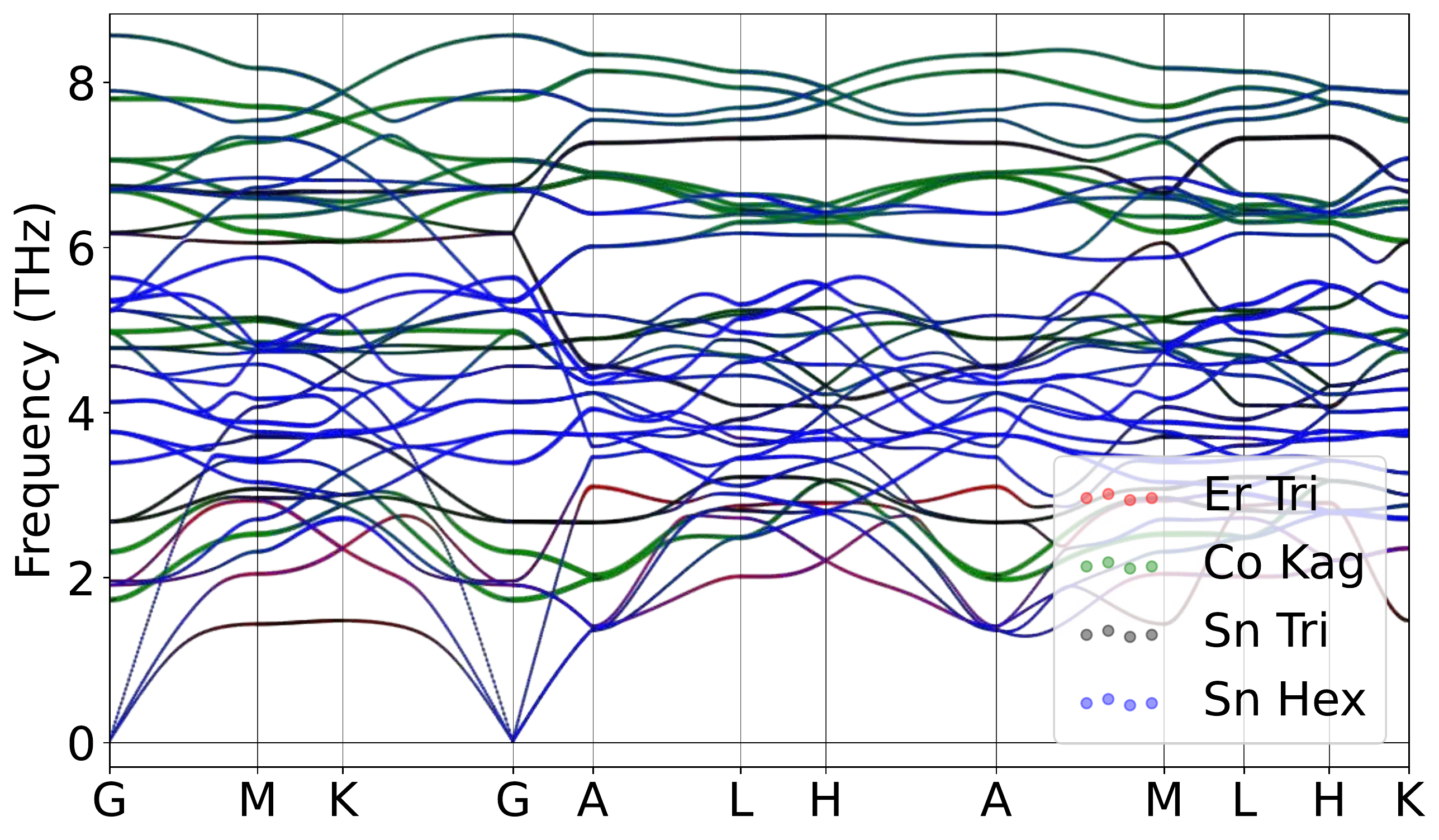}
}
\hspace{5pt}\subfloat[Relaxed phonon at high-T.]{
\label{fig:ErCo6Sn6_relaxed_High}
\includegraphics[width=0.45\textwidth]{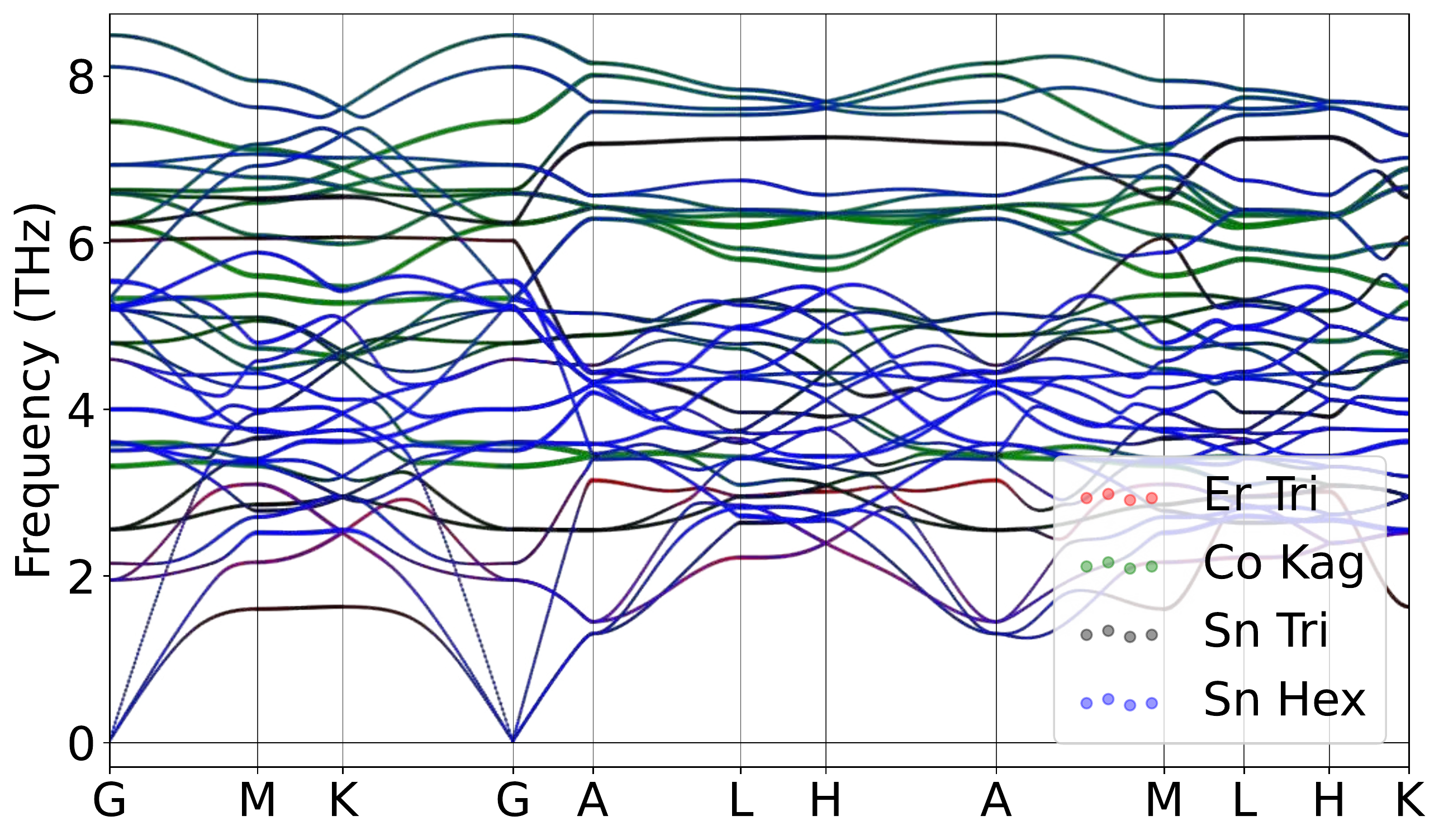}
}
\caption{Atomically projected phonon spectrum for ErCo$_6$Sn$_6$.}
\label{fig:ErCo6Sn6pho}
\end{figure}

\begin{figure}[H]
\centering
\label{fig:ErCo6Sn6_relaxed_Low_xz}
\includegraphics[width=0.85\textwidth]{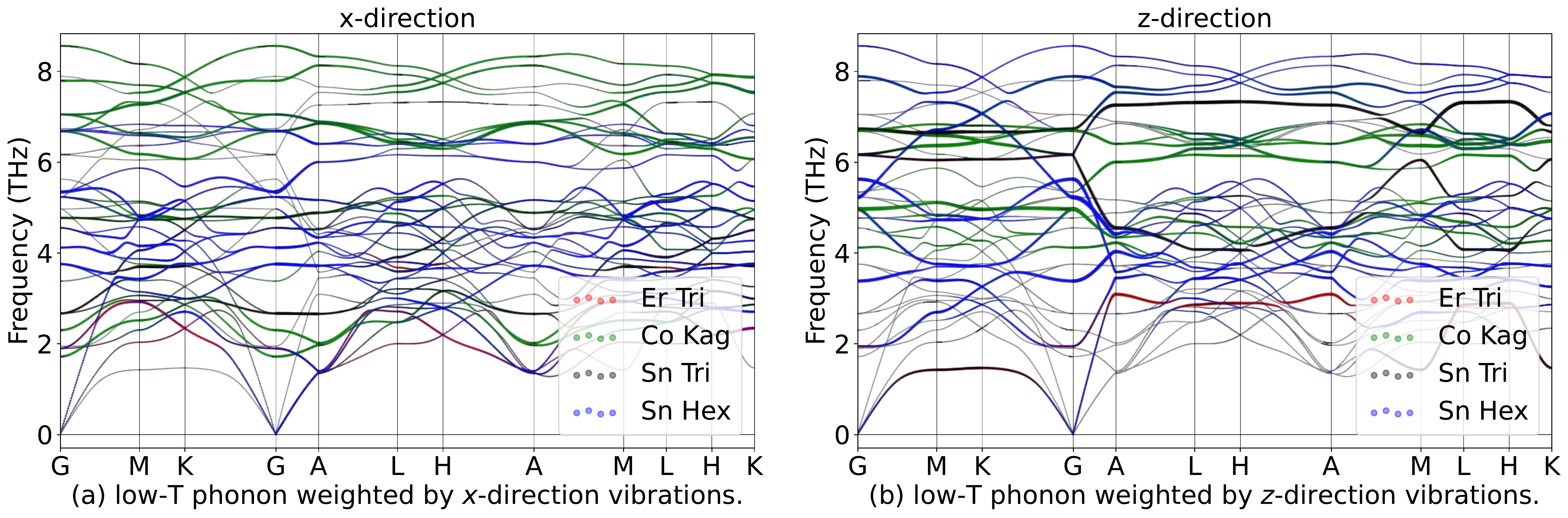}
\hspace{5pt}\label{fig:ErCo6Sn6_relaxed_High_xz}
\includegraphics[width=0.85\textwidth]{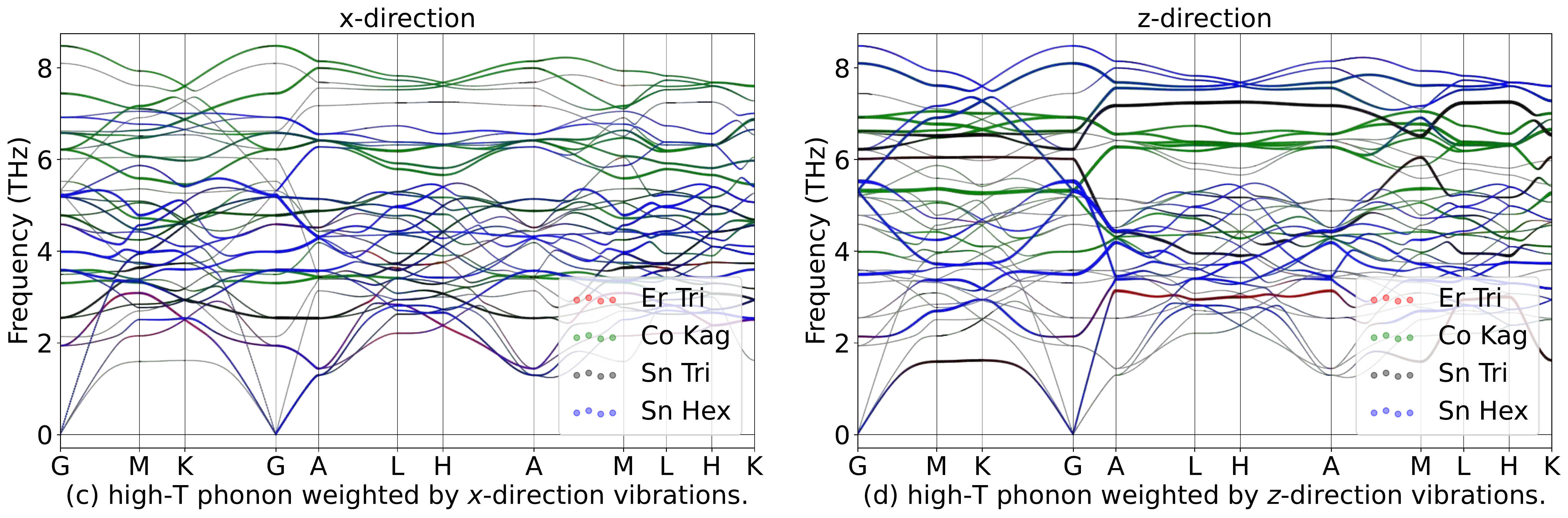}
\caption{Phonon spectrum for ErCo$_6$Sn$_6$in in-plane ($x$) and out-of-plane ($z$) directions.}
\label{fig:ErCo6Sn6phoxyz}
\end{figure}

\begin{table}[htbp]
\global\long\def\arraystretch{1.12}
\captionsetup{width=.95\textwidth}
\caption{IRREPs at high-symmetry points for ErCo$_6$Sn$_6$ phonon spectrum at Low-T.}
\label{tab:191_ErCo6Sn6_Low}
\setlength{\tabcolsep}{1.mm}{
}
\end{table}

\begin{table}[htbp]
\global\long\def\arraystretch{1.12}
\captionsetup{width=.95\textwidth}
\caption{IRREPs at high-symmetry points for ErCo$_6$Sn$_6$ phonon spectrum at High-T.}
\label{tab:191_ErCo6Sn6_High}
\setlength{\tabcolsep}{1.mm}{
%
}
\end{table}
\subsubsection{\label{sms2sec:TmCo6Sn6}\hyperref[smsec:col]{\texorpdfstring{TmCo$_6$Sn$_6$}{}}~{\color{green}  (High-T stable, Low-T stable) }}

\begin{table}[htbp]
\global\long\def\arraystretch{1.12}
\captionsetup{width=.85\textwidth}
\caption{Structure parameters of TmCo$_6$Sn$_6$ after relaxation. It contains 531 electrons. }
\label{tab:TmCo6Sn6}
\setlength{\tabcolsep}{4mm}{
%
}
\end{table}

\begin{figure}[htbp]
\centering
\includegraphics[width=0.85\textwidth]{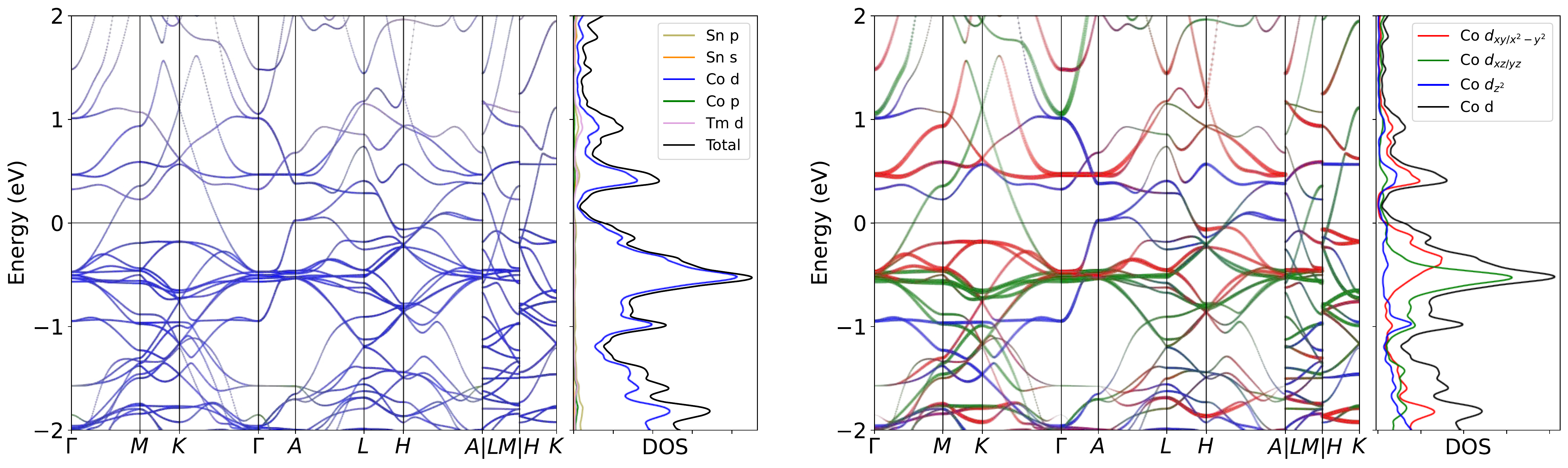}
\caption{Band structure for TmCo$_6$Sn$_6$ without SOC. The projection of Co $3d$ orbitals is also presented. The band structures clearly reveal the dominating of Co $3d$ orbitals  near the Fermi level, as well as the pronounced peak.}
\label{fig:TmCo6Sn6band}
\end{figure}

\begin{figure}[H]
\centering
\subfloat[Relaxed phonon at low-T.]{
\label{fig:TmCo6Sn6_relaxed_Low}
\includegraphics[width=0.45\textwidth]{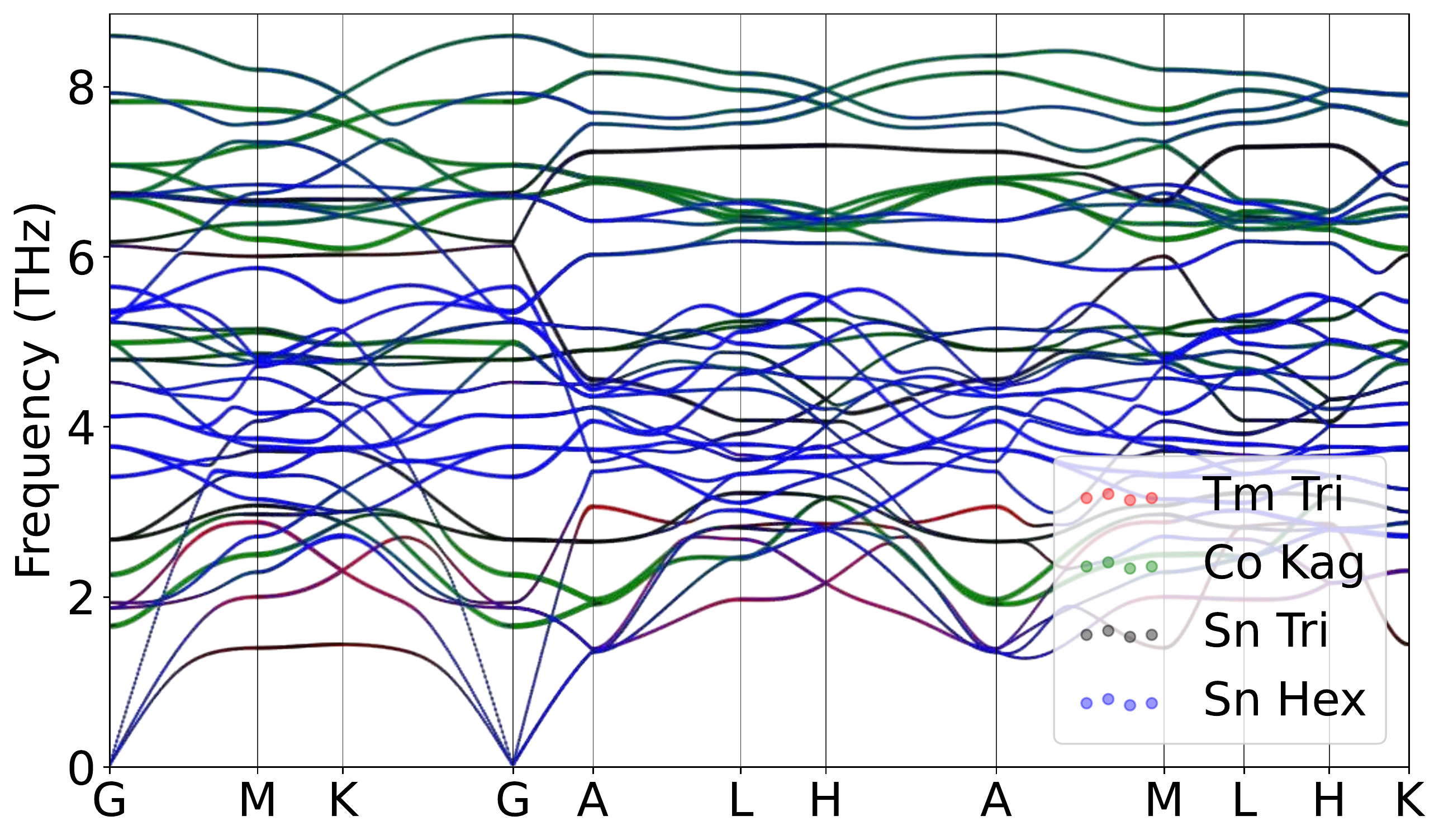}
}
\hspace{5pt}\subfloat[Relaxed phonon at high-T.]{
\label{fig:TmCo6Sn6_relaxed_High}
\includegraphics[width=0.45\textwidth]{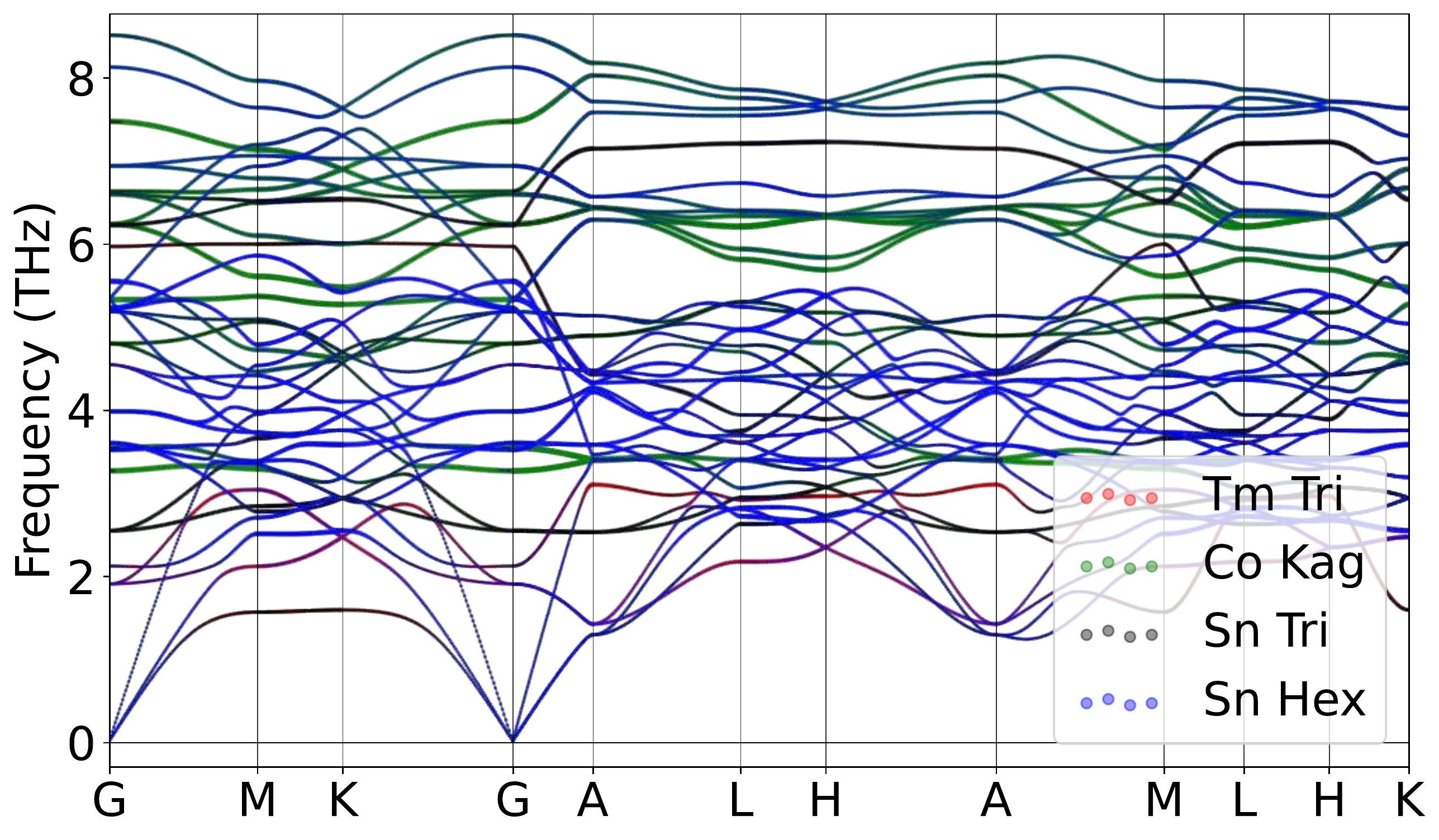}
}
\caption{Atomically projected phonon spectrum for TmCo$_6$Sn$_6$.}
\label{fig:TmCo6Sn6pho}
\end{figure}

\begin{figure}[H]
\centering
\label{fig:TmCo6Sn6_relaxed_Low_xz}
\includegraphics[width=0.85\textwidth]{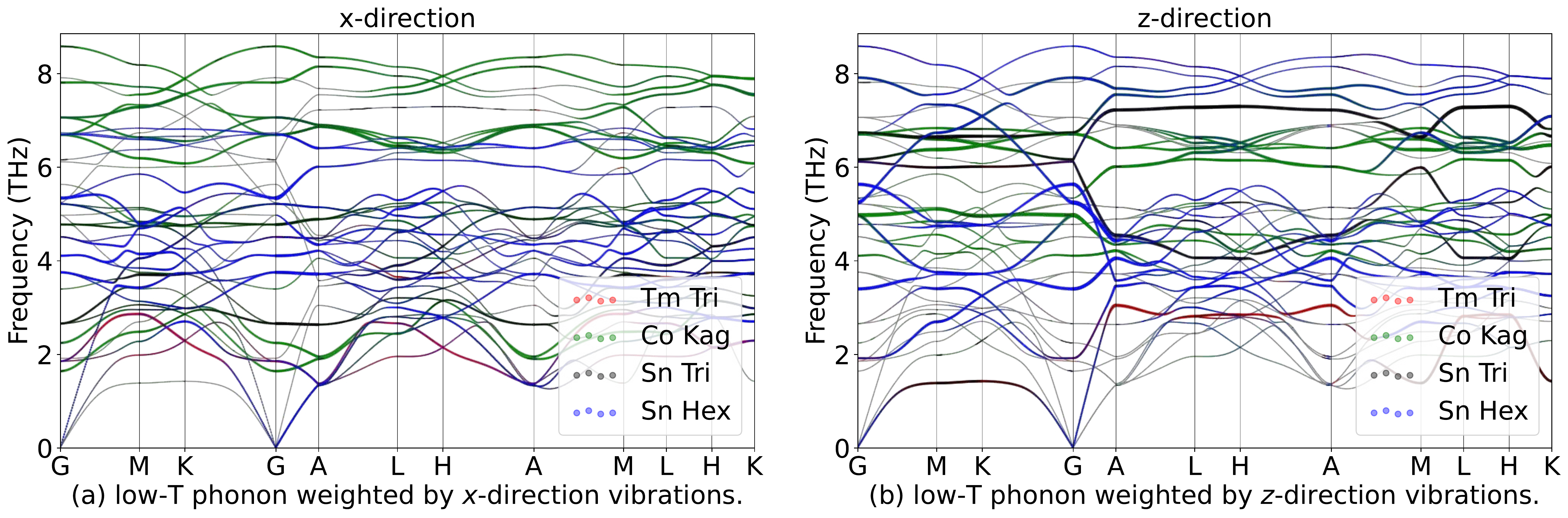}
\hspace{5pt}\label{fig:TmCo6Sn6_relaxed_High_xz}
\includegraphics[width=0.85\textwidth]{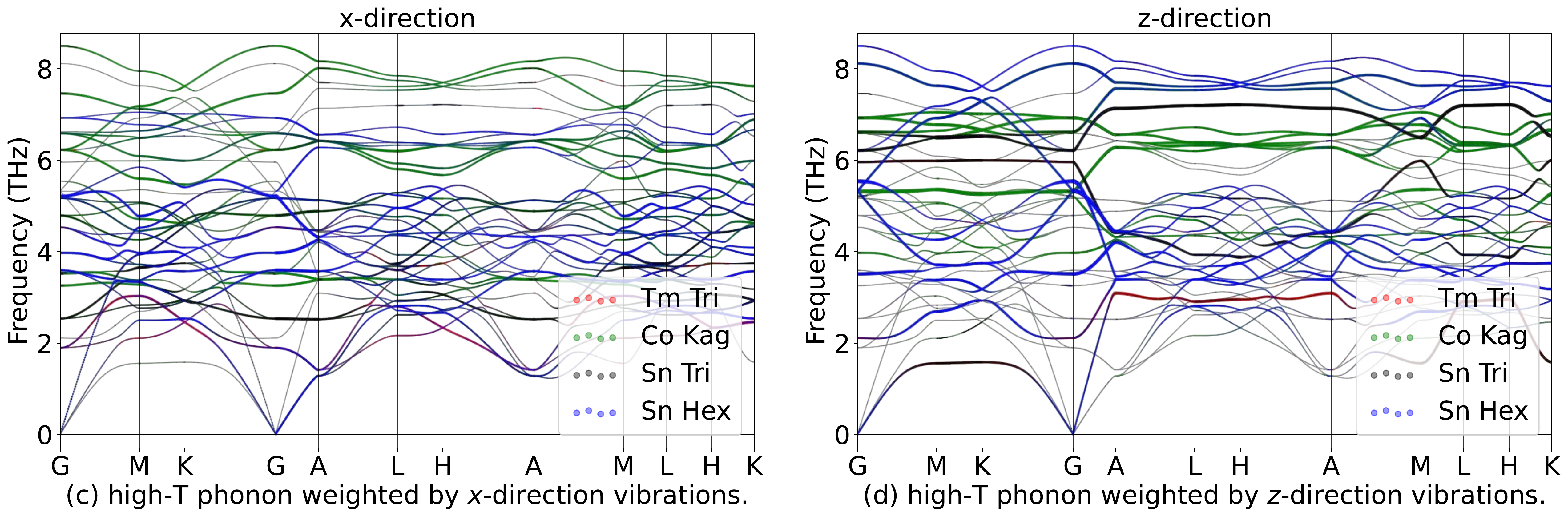}
\caption{Phonon spectrum for TmCo$_6$Sn$_6$in in-plane ($x$) and out-of-plane ($z$) directions.}
\label{fig:TmCo6Sn6phoxyz}
\end{figure}

\begin{table}[htbp]
\global\long\def\arraystretch{1.12}
\captionsetup{width=.95\textwidth}
\caption{IRREPs at high-symmetry points for TmCo$_6$Sn$_6$ phonon spectrum at Low-T.}
\label{tab:191_TmCo6Sn6_Low}
\setlength{\tabcolsep}{1.mm}{
}
\end{table}

\begin{table}[htbp]
\global\long\def\arraystretch{1.12}
\captionsetup{width=.95\textwidth}
\caption{IRREPs at high-symmetry points for TmCo$_6$Sn$_6$ phonon spectrum at High-T.}
\label{tab:191_TmCo6Sn6_High}
\setlength{\tabcolsep}{1.mm}{
%
}
\end{table}
\subsubsection{\label{sms2sec:YbCo6Sn6}\hyperref[smsec:col]{\texorpdfstring{YbCo$_6$Sn$_6$}{}}~{\color{green}  (High-T stable, Low-T stable) }}

\begin{table}[htbp]
\global\long\def\arraystretch{1.12}
\captionsetup{width=.85\textwidth}
\caption{Structure parameters of YbCo$_6$Sn$_6$ after relaxation. It contains 532 electrons. }
\label{tab:YbCo6Sn6}
\setlength{\tabcolsep}{4mm}{
%
}
\end{table}

\begin{figure}[htbp]
\centering
\includegraphics[width=0.85\textwidth]{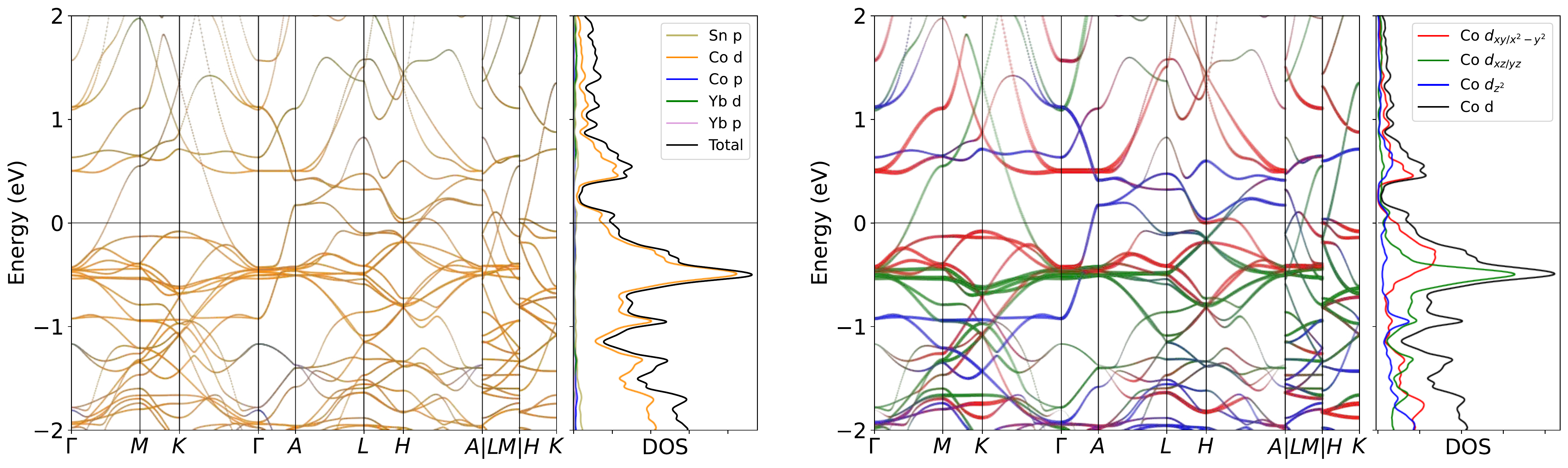}
\caption{Band structure for YbCo$_6$Sn$_6$ without SOC. The projection of Co $3d$ orbitals is also presented. The band structures clearly reveal the dominating of Co $3d$ orbitals  near the Fermi level, as well as the pronounced peak.}
\label{fig:YbCo6Sn6band}
\end{figure}

\begin{figure}[H]
\centering
\subfloat[Relaxed phonon at low-T.]{
\label{fig:YbCo6Sn6_relaxed_Low}
\includegraphics[width=0.45\textwidth]{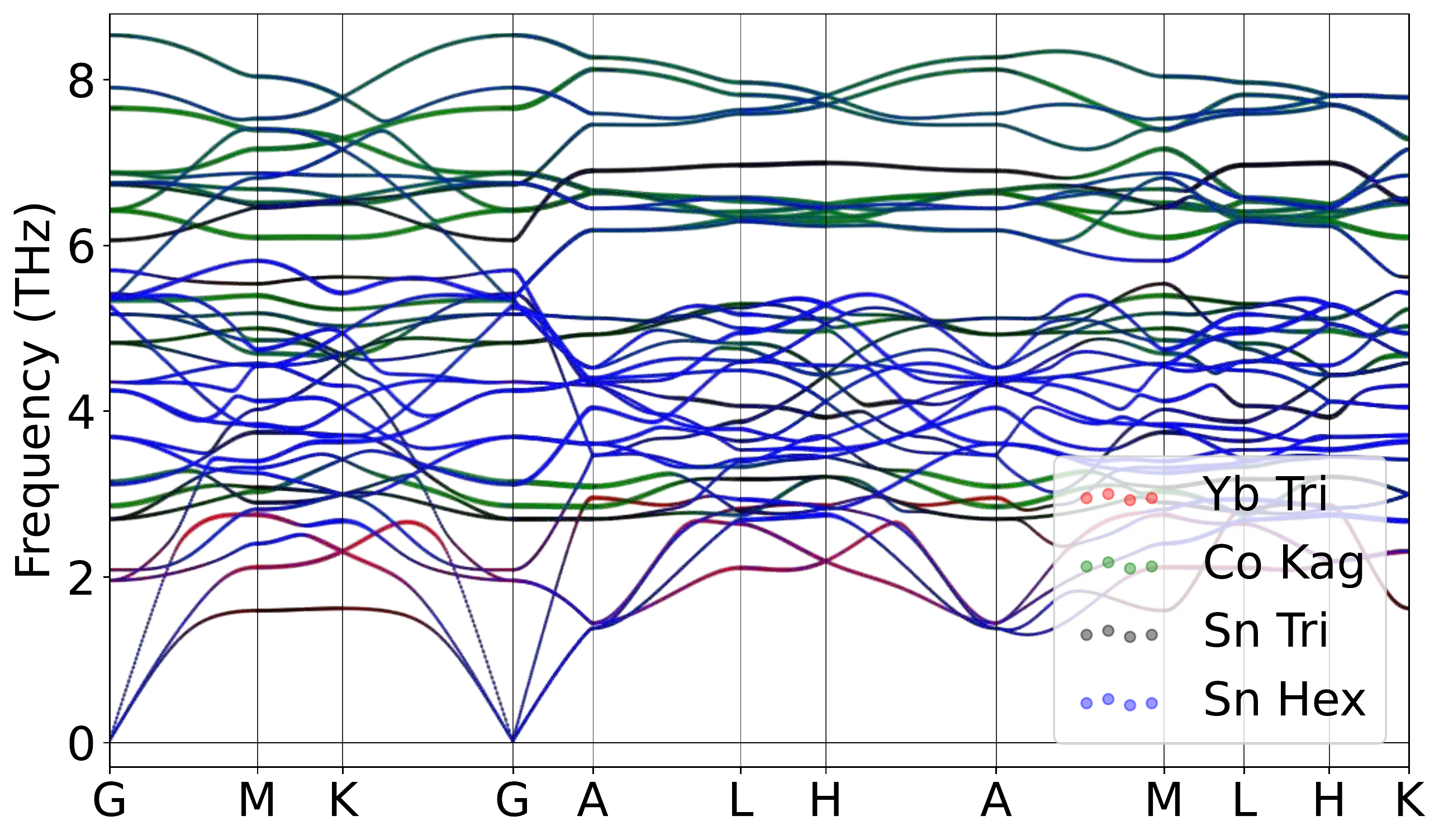}
}
\hspace{5pt}\subfloat[Relaxed phonon at high-T.]{
\label{fig:YbCo6Sn6_relaxed_High}
\includegraphics[width=0.45\textwidth]{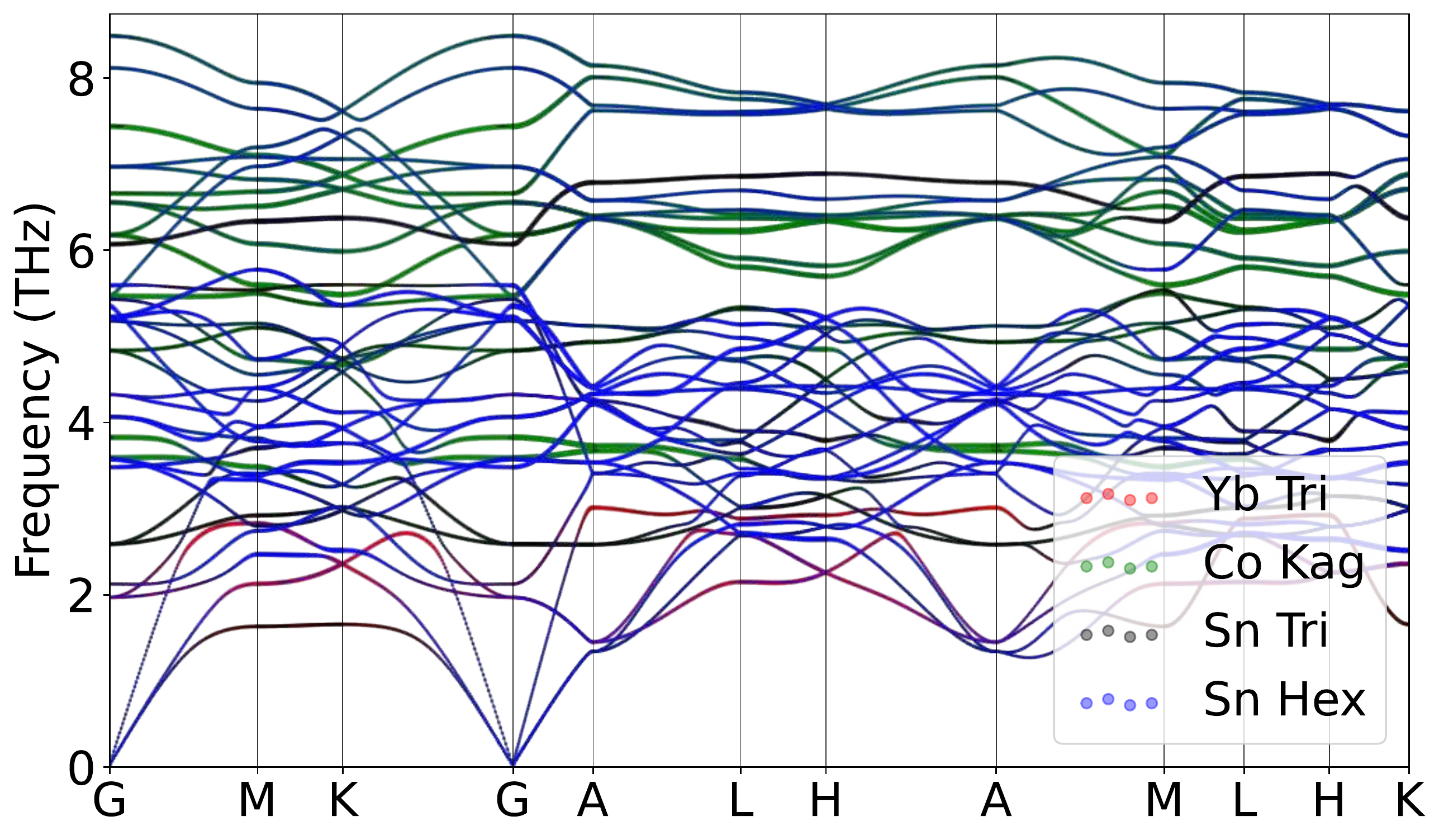}
}
\caption{Atomically projected phonon spectrum for YbCo$_6$Sn$_6$.}
\label{fig:YbCo6Sn6pho}
\end{figure}

\begin{figure}[H]
\centering
\label{fig:YbCo6Sn6_relaxed_Low_xz}
\includegraphics[width=0.85\textwidth]{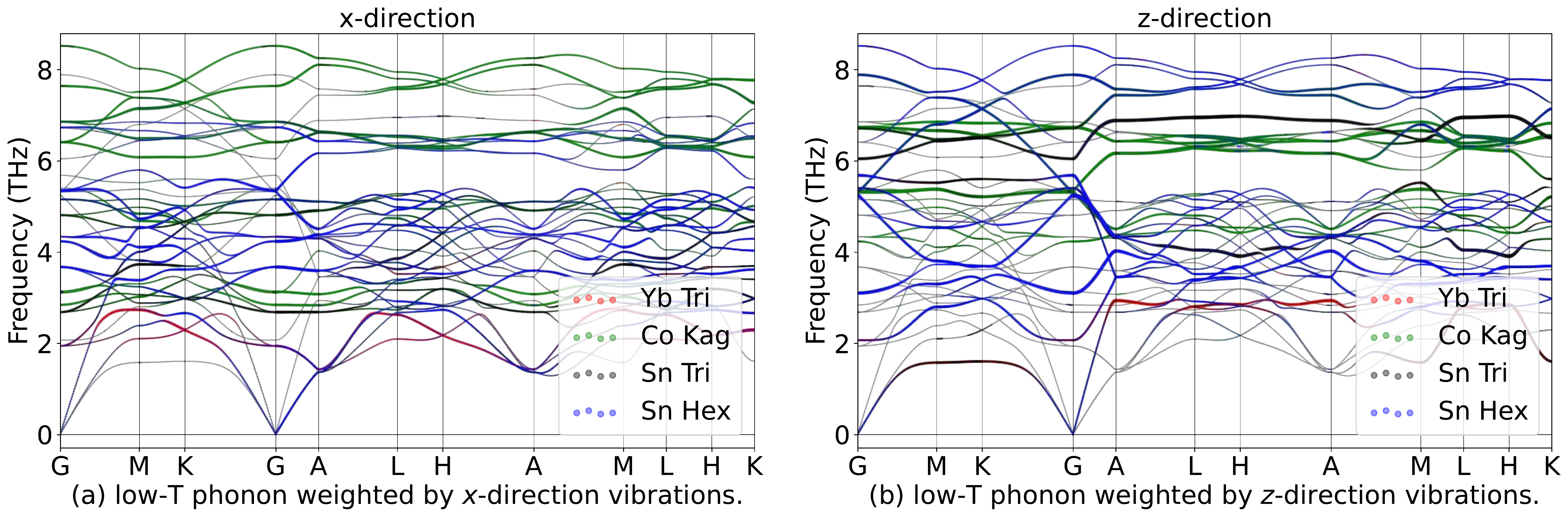}
\hspace{5pt}\label{fig:YbCo6Sn6_relaxed_High_xz}
\includegraphics[width=0.85\textwidth]{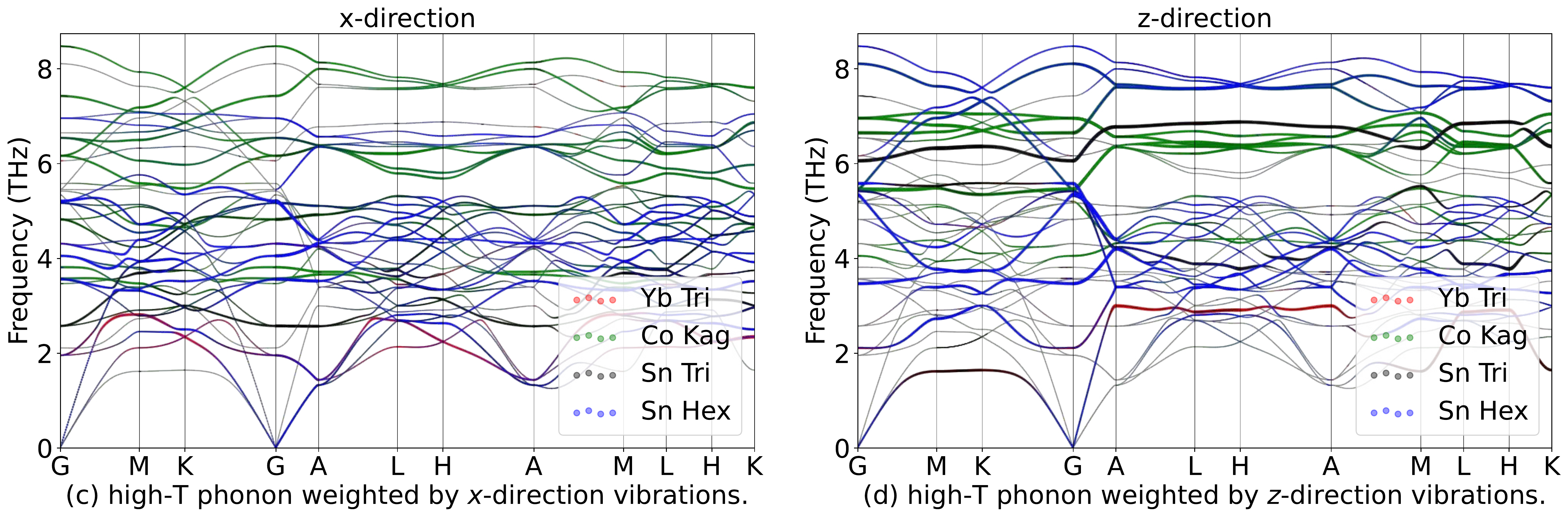}
\caption{Phonon spectrum for YbCo$_6$Sn$_6$in in-plane ($x$) and out-of-plane ($z$) directions.}
\label{fig:YbCo6Sn6phoxyz}
\end{figure}

\begin{table}[htbp]
\global\long\def\arraystretch{1.12}
\captionsetup{width=.95\textwidth}
\caption{IRREPs at high-symmetry points for YbCo$_6$Sn$_6$ phonon spectrum at Low-T.}
\label{tab:191_YbCo6Sn6_Low}
\setlength{\tabcolsep}{1.mm}{
}
\end{table}

\begin{table}[htbp]
\global\long\def\arraystretch{1.12}
\captionsetup{width=.95\textwidth}
\caption{IRREPs at high-symmetry points for YbCo$_6$Sn$_6$ phonon spectrum at High-T.}
\label{tab:191_YbCo6Sn6_High}
\setlength{\tabcolsep}{1.mm}{
%
}
\end{table}
\subsubsection{\label{sms2sec:LuCo6Sn6}\hyperref[smsec:col]{\texorpdfstring{LuCo$_6$Sn$_6$}{}}~{\color{green}  (High-T stable, Low-T stable) }}

\begin{table}[htbp]
\global\long\def\arraystretch{1.12}
\captionsetup{width=.85\textwidth}
\caption{Structure parameters of LuCo$_6$Sn$_6$ after relaxation. It contains 533 electrons. }
\label{tab:LuCo6Sn6}
\setlength{\tabcolsep}{4mm}{
%
}
\end{table}

\begin{figure}[htbp]
\centering
\includegraphics[width=0.85\textwidth]{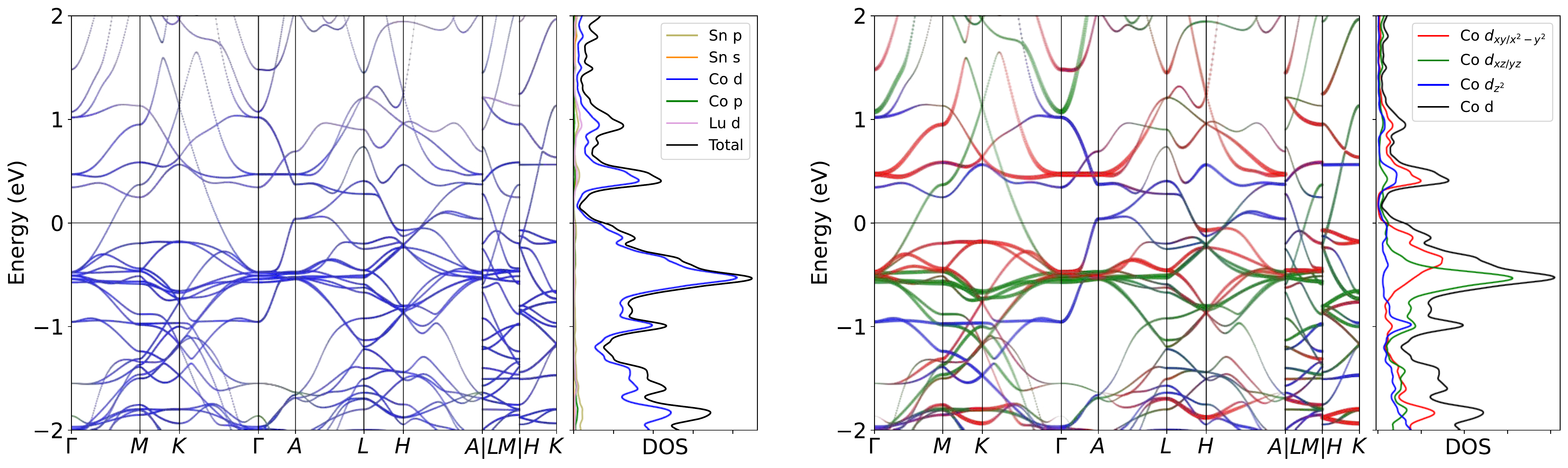}
\caption{Band structure for LuCo$_6$Sn$_6$ without SOC. The projection of Co $3d$ orbitals is also presented. The band structures clearly reveal the dominating of Co $3d$ orbitals  near the Fermi level, as well as the pronounced peak.}
\label{fig:LuCo6Sn6band}
\end{figure}

\begin{figure}[H]
\centering
\subfloat[Relaxed phonon at low-T.]{
\label{fig:LuCo6Sn6_relaxed_Low}
\includegraphics[width=0.45\textwidth]{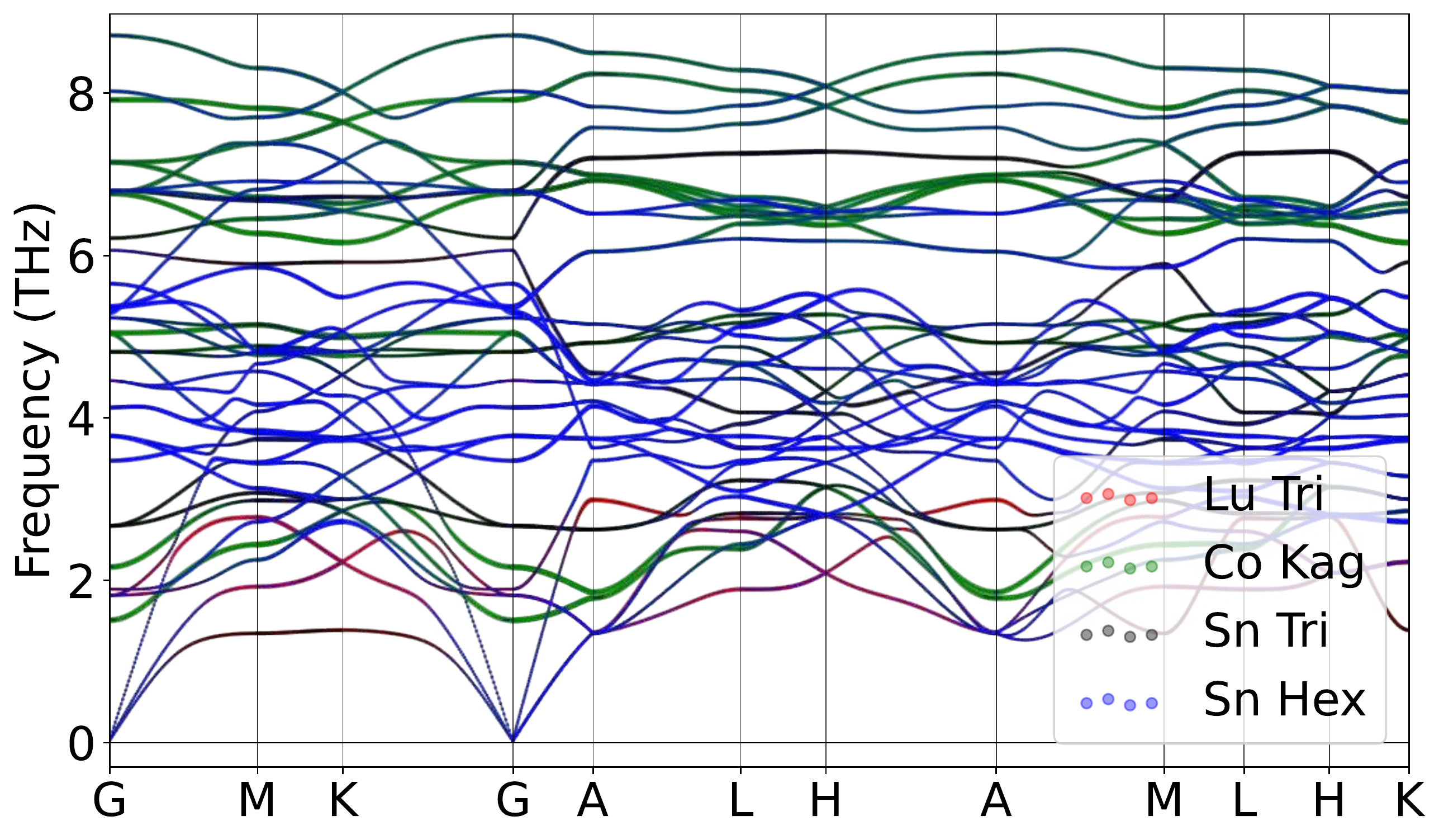}
}
\hspace{5pt}\subfloat[Relaxed phonon at high-T.]{
\label{fig:LuCo6Sn6_relaxed_High}
\includegraphics[width=0.45\textwidth]{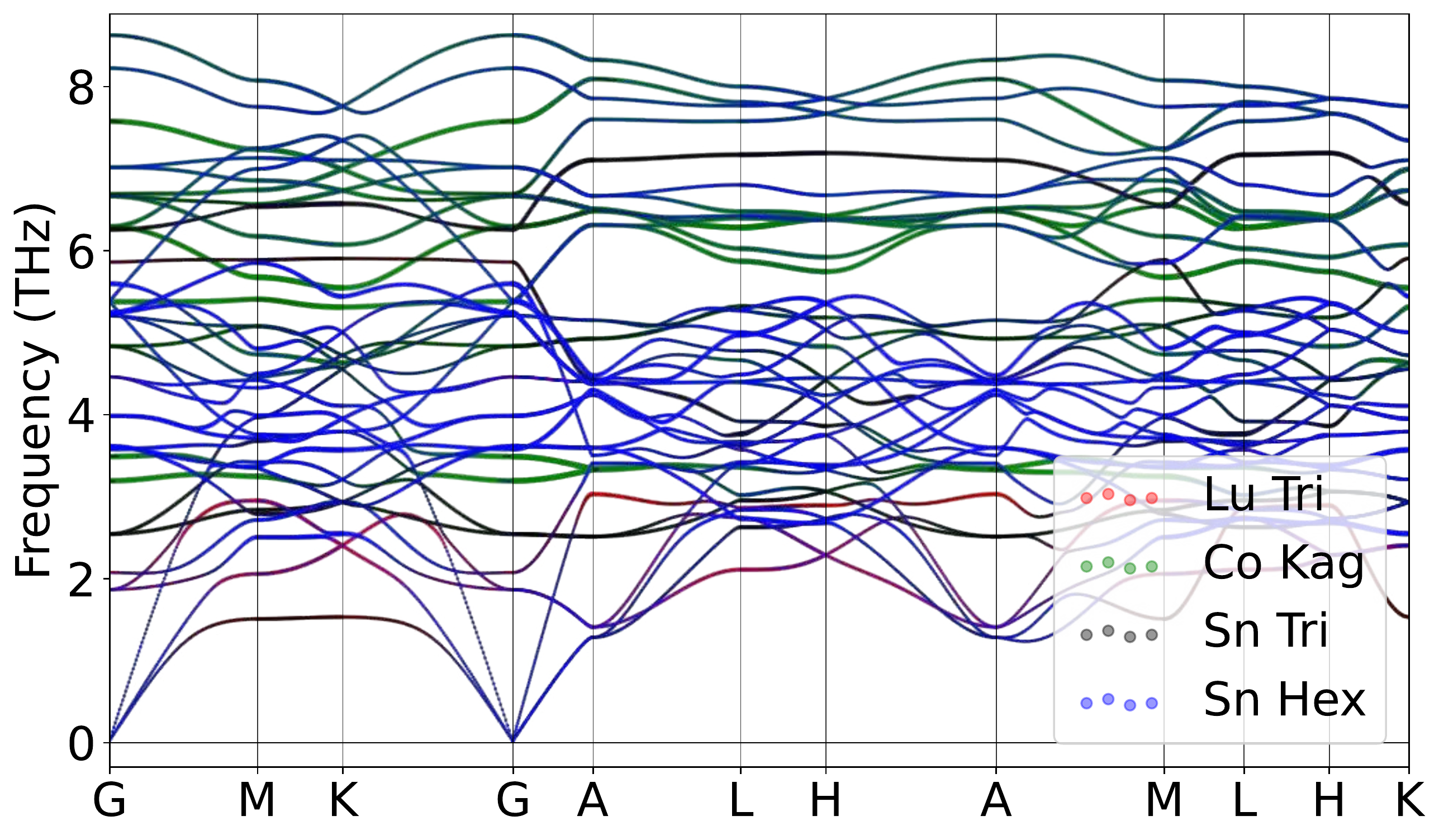}
}
\caption{Atomically projected phonon spectrum for LuCo$_6$Sn$_6$.}
\label{fig:LuCo6Sn6pho}
\end{figure}

\begin{figure}[H]
\centering
\label{fig:LuCo6Sn6_relaxed_Low_xz}
\includegraphics[width=0.85\textwidth]{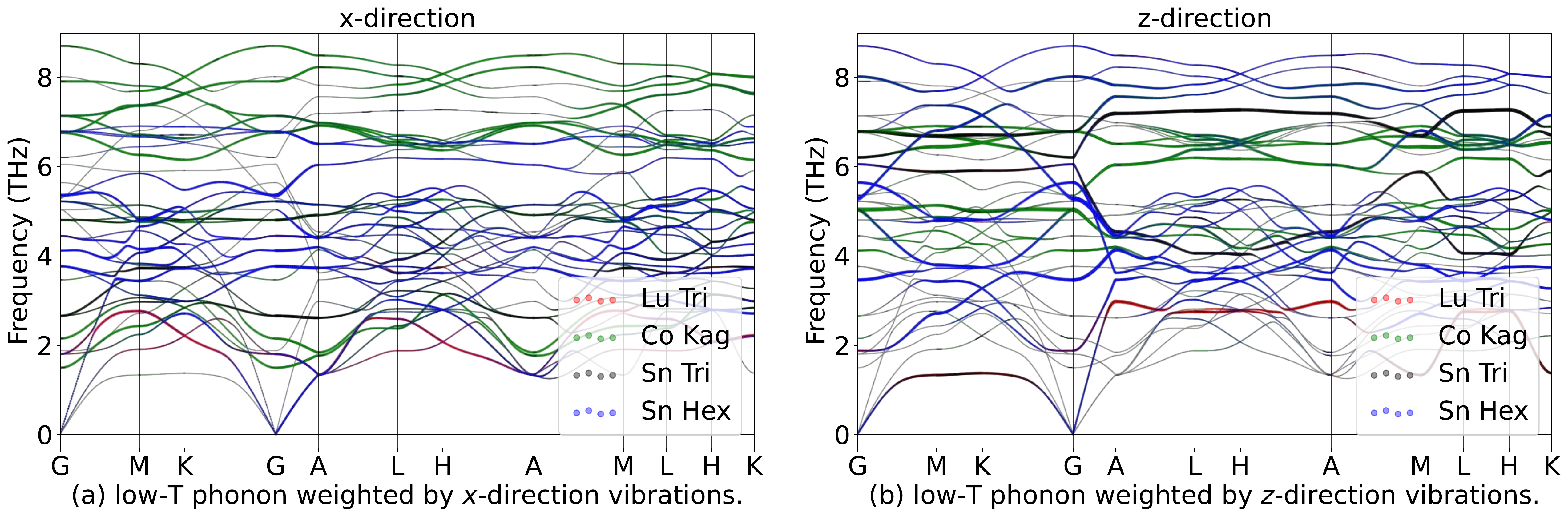}
\hspace{5pt}\label{fig:LuCo6Sn6_relaxed_High_xz}
\includegraphics[width=0.85\textwidth]{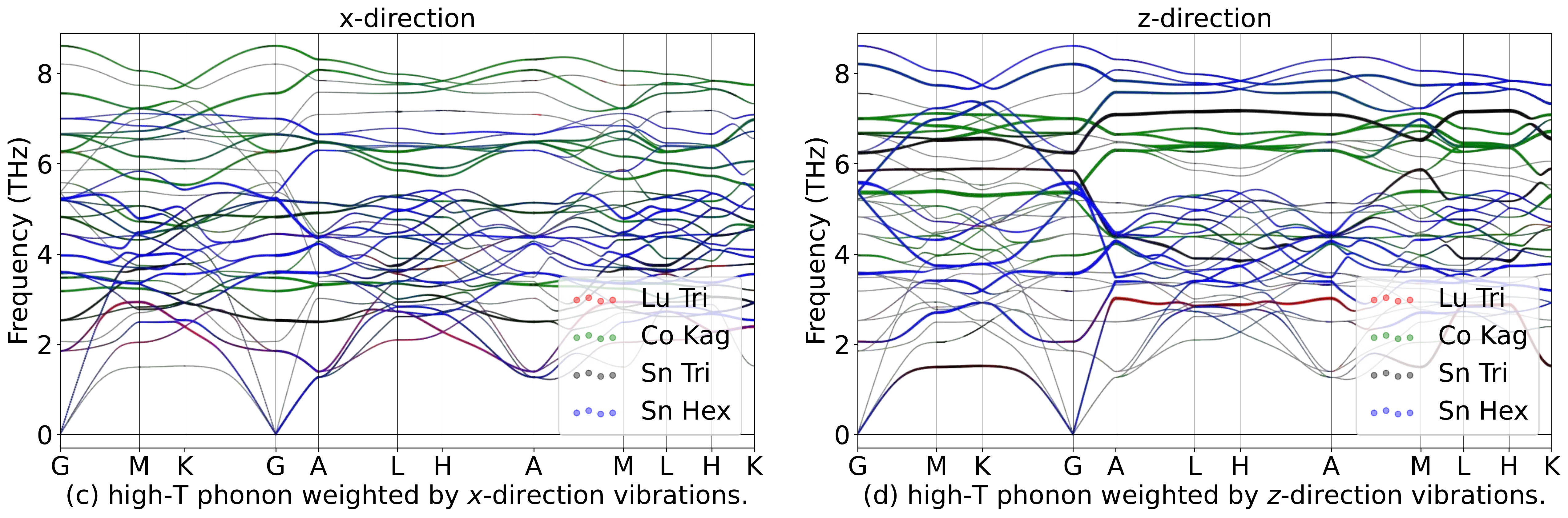}
\caption{Phonon spectrum for LuCo$_6$Sn$_6$in in-plane ($x$) and out-of-plane ($z$) directions.}
\label{fig:LuCo6Sn6phoxyz}
\end{figure}

\begin{table}[htbp]
\global\long\def\arraystretch{1.12}
\captionsetup{width=.95\textwidth}
\caption{IRREPs at high-symmetry points for LuCo$_6$Sn$_6$ phonon spectrum at Low-T.}
\label{tab:191_LuCo6Sn6_Low}
\setlength{\tabcolsep}{1.mm}{
}
\end{table}

\begin{table}[htbp]
\global\long\def\arraystretch{1.12}
\captionsetup{width=.95\textwidth}
\caption{IRREPs at high-symmetry points for LuCo$_6$Sn$_6$ phonon spectrum at High-T.}
\label{tab:191_LuCo6Sn6_High}
\setlength{\tabcolsep}{1.mm}{
%
}
\end{table}
\subsubsection{\label{sms2sec:HfCo6Sn6}\hyperref[smsec:col]{\texorpdfstring{HfCo$_6$Sn$_6$}{}}~{\color{green}  (High-T stable, Low-T stable) }}

\begin{table}[htbp]
\global\long\def\arraystretch{1.12}
\captionsetup{width=.85\textwidth}
\caption{Structure parameters of HfCo$_6$Sn$_6$ after relaxation. It contains 534 electrons. }
\label{tab:HfCo6Sn6}
\setlength{\tabcolsep}{4mm}{
%
}
\end{table}

\begin{figure}[htbp]
\centering
\includegraphics[width=0.85\textwidth]{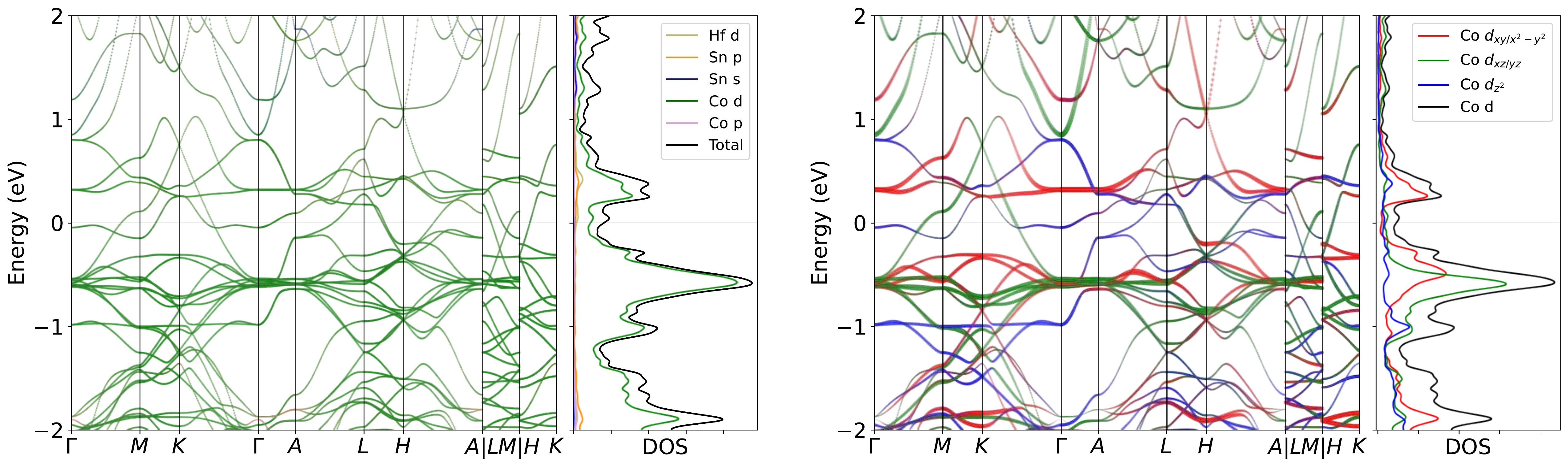}
\caption{Band structure for HfCo$_6$Sn$_6$ without SOC. The projection of Co $3d$ orbitals is also presented. The band structures clearly reveal the dominating of Co $3d$ orbitals  near the Fermi level, as well as the pronounced peak.}
\label{fig:HfCo6Sn6band}
\end{figure}

\begin{figure}[H]
\centering
\subfloat[Relaxed phonon at low-T.]{
\label{fig:HfCo6Sn6_relaxed_Low}
\includegraphics[width=0.45\textwidth]{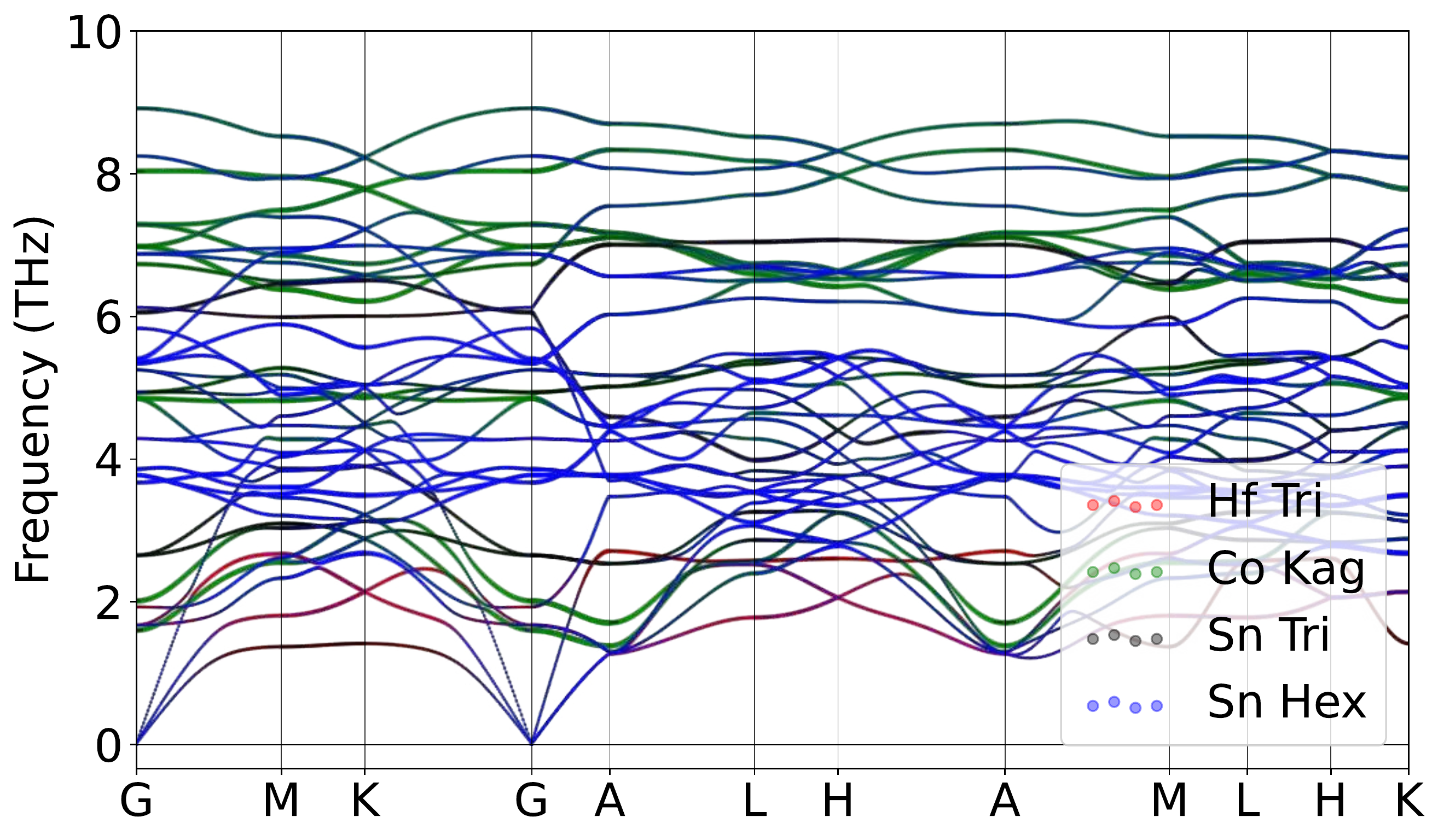}
}
\hspace{5pt}\subfloat[Relaxed phonon at high-T.]{
\label{fig:HfCo6Sn6_relaxed_High}
\includegraphics[width=0.45\textwidth]{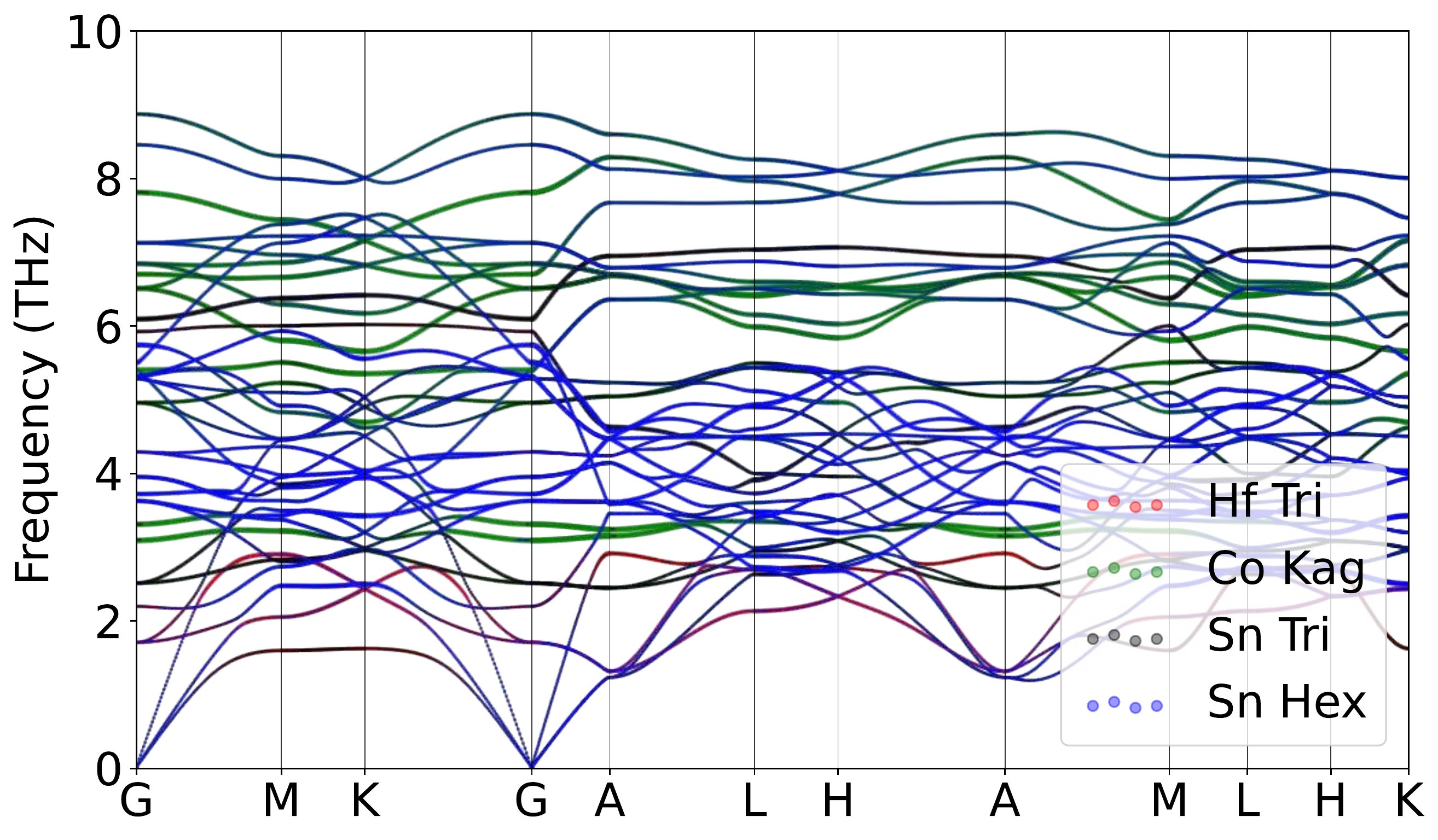}
}
\caption{Atomically projected phonon spectrum for HfCo$_6$Sn$_6$.}
\label{fig:HfCo6Sn6pho}
\end{figure}

\begin{figure}[H]
\centering
\label{fig:HfCo6Sn6_relaxed_Low_xz}
\includegraphics[width=0.85\textwidth]{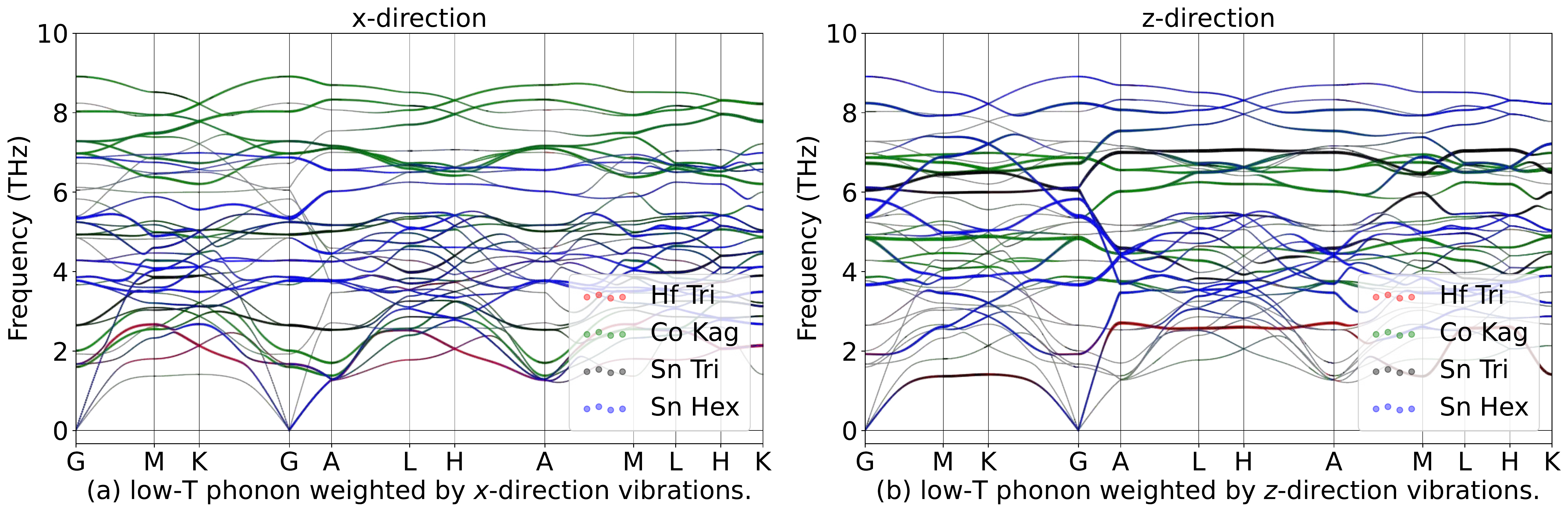}
\hspace{5pt}\label{fig:HfCo6Sn6_relaxed_High_xz}
\includegraphics[width=0.85\textwidth]{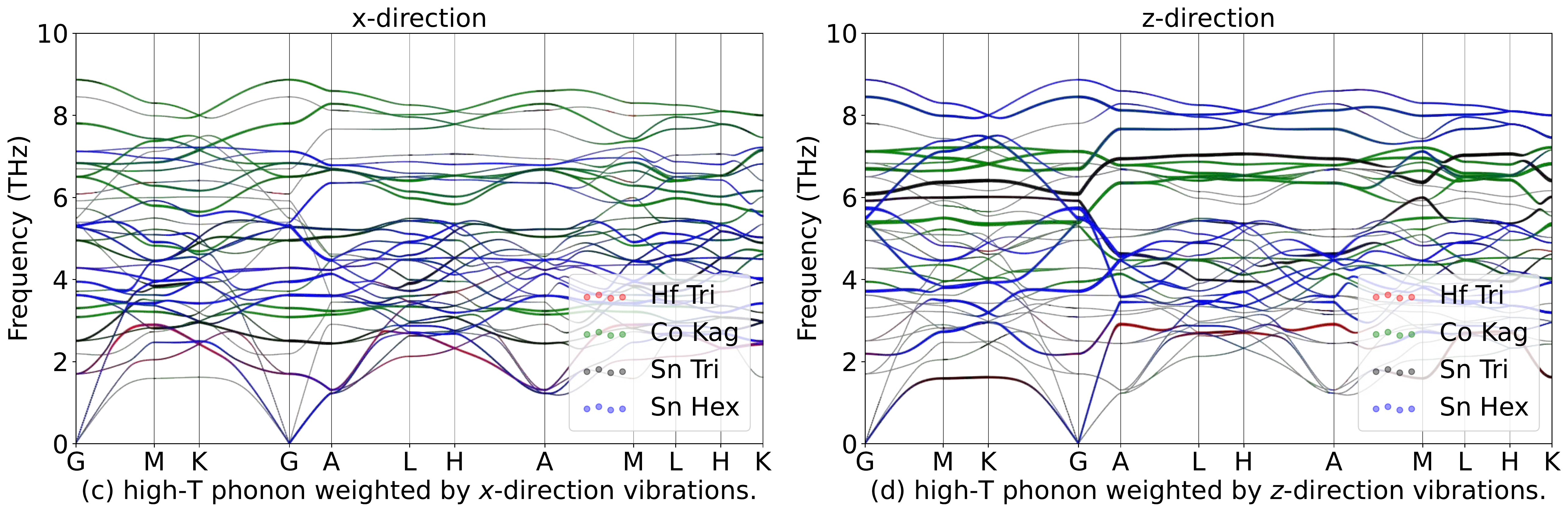}
\caption{Phonon spectrum for HfCo$_6$Sn$_6$in in-plane ($x$) and out-of-plane ($z$) directions.}
\label{fig:HfCo6Sn6phoxyz}
\end{figure}

\begin{table}[htbp]
\global\long\def\arraystretch{1.12}
\captionsetup{width=.95\textwidth}
\caption{IRREPs at high-symmetry points for HfCo$_6$Sn$_6$ phonon spectrum at Low-T.}
\label{tab:191_HfCo6Sn6_Low}
\setlength{\tabcolsep}{1.mm}{
}
\end{table}

\begin{table}[htbp]
\global\long\def\arraystretch{1.12}
\captionsetup{width=.95\textwidth}
\caption{IRREPs at high-symmetry points for HfCo$_6$Sn$_6$ phonon spectrum at High-T.}
\label{tab:191_HfCo6Sn6_High}
\setlength{\tabcolsep}{1.mm}{
%
}
\end{table}
\subsubsection{\label{sms2sec:TaCo6Sn6}\hyperref[smsec:col]{\texorpdfstring{TaCo$_6$Sn$_6$}{}}~{\color{green}  (High-T stable, Low-T stable) }}

\begin{table}[htbp]
\global\long\def\arraystretch{1.12}
\captionsetup{width=.85\textwidth}
\caption{Structure parameters of TaCo$_6$Sn$_6$ after relaxation. It contains 535 electrons. }
\label{tab:TaCo6Sn6}
\setlength{\tabcolsep}{4mm}{
%
}
\end{table}

\begin{figure}[htbp]
\centering
\includegraphics[width=0.85\textwidth]{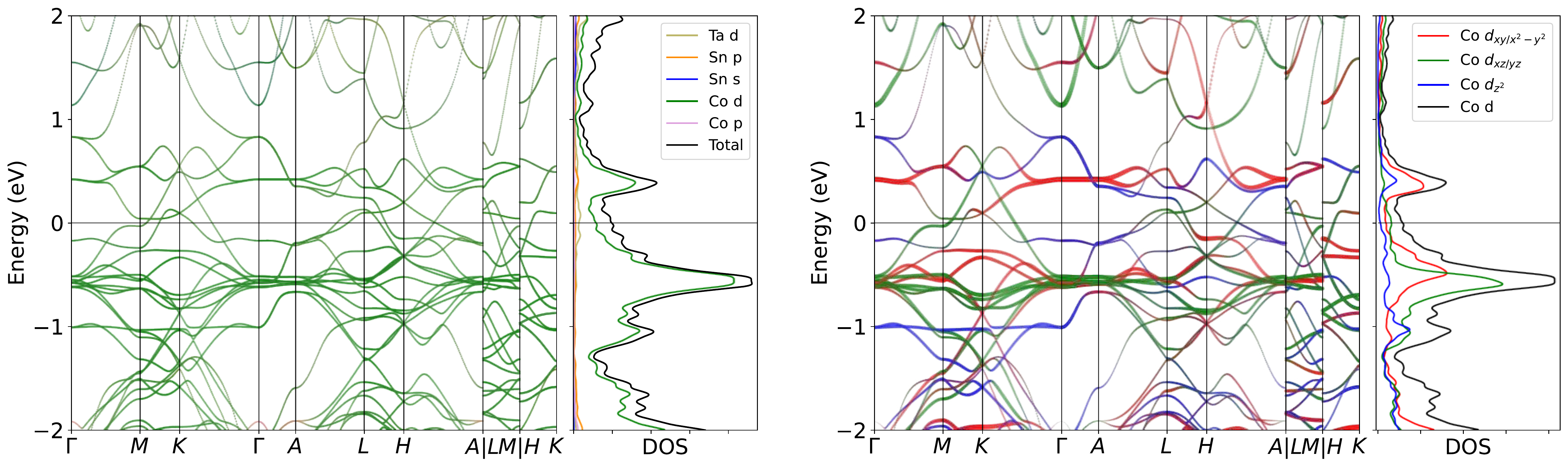}
\caption{Band structure for TaCo$_6$Sn$_6$ without SOC. The projection of Co $3d$ orbitals is also presented. The band structures clearly reveal the dominating of Co $3d$ orbitals  near the Fermi level, as well as the pronounced peak.}
\label{fig:TaCo6Sn6band}
\end{figure}

\begin{figure}[H]
\centering
\subfloat[Relaxed phonon at low-T.]{
\label{fig:TaCo6Sn6_relaxed_Low}
\includegraphics[width=0.45\textwidth]{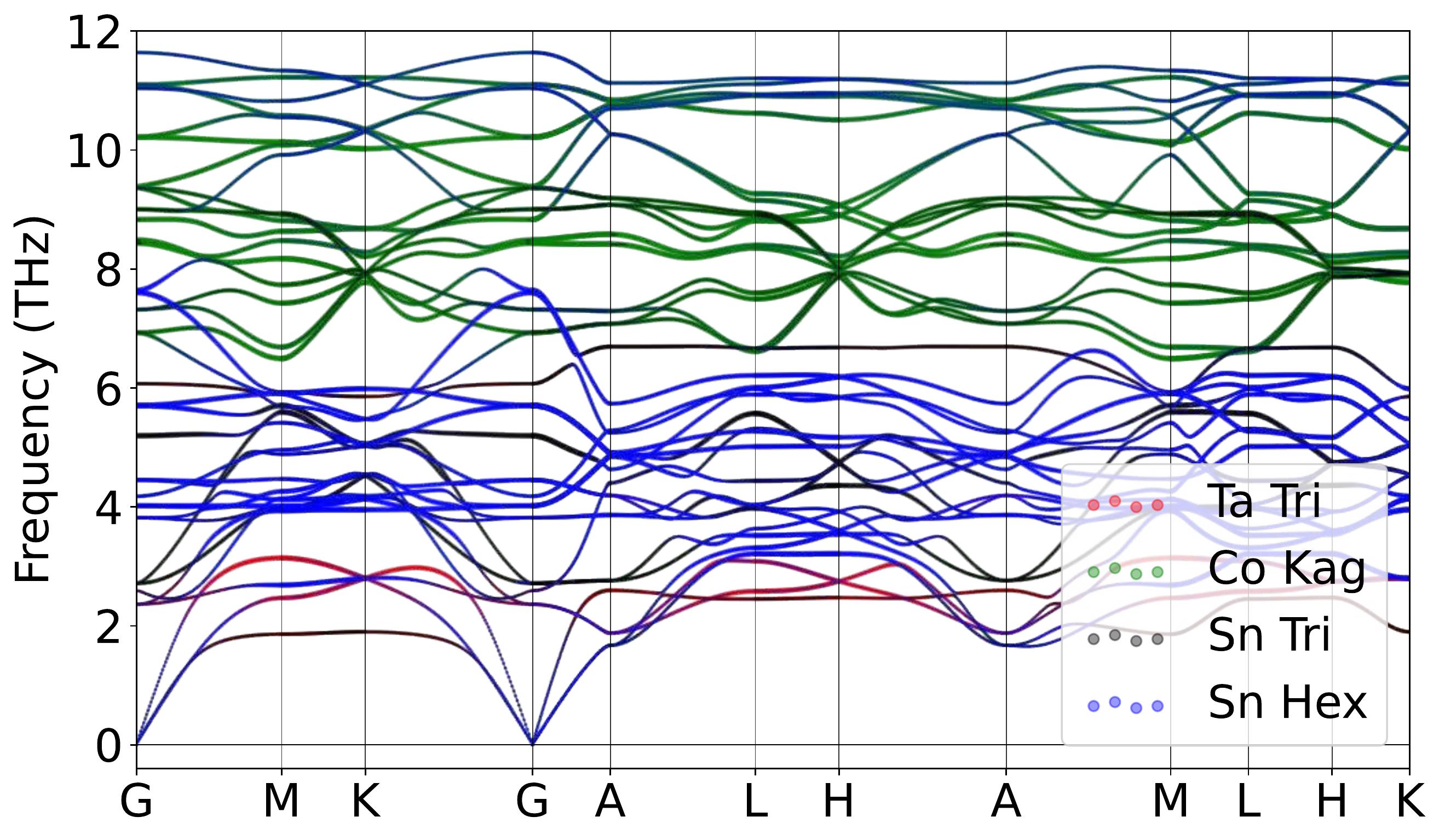}
}
\hspace{5pt}\subfloat[Relaxed phonon at high-T.]{
\label{fig:TaCo6Sn6_relaxed_High}
\includegraphics[width=0.45\textwidth]{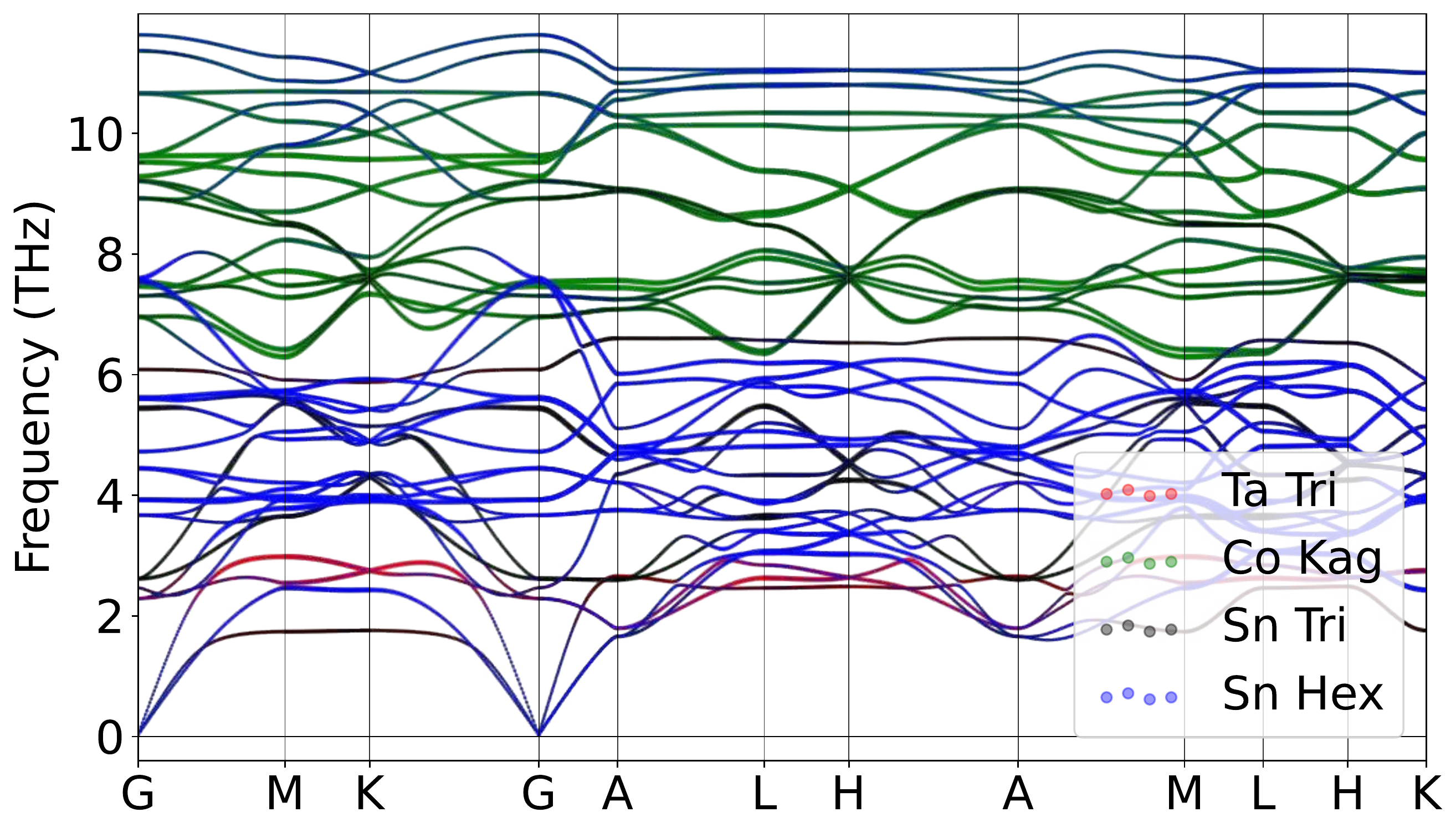}
}
\caption{Atomically projected phonon spectrum for TaCo$_6$Sn$_6$.}
\label{fig:TaCo6Sn6pho}
\end{figure}

\begin{figure}[H]
\centering
\label{fig:TaCo6Sn6_relaxed_Low_xz}
\includegraphics[width=0.85\textwidth]{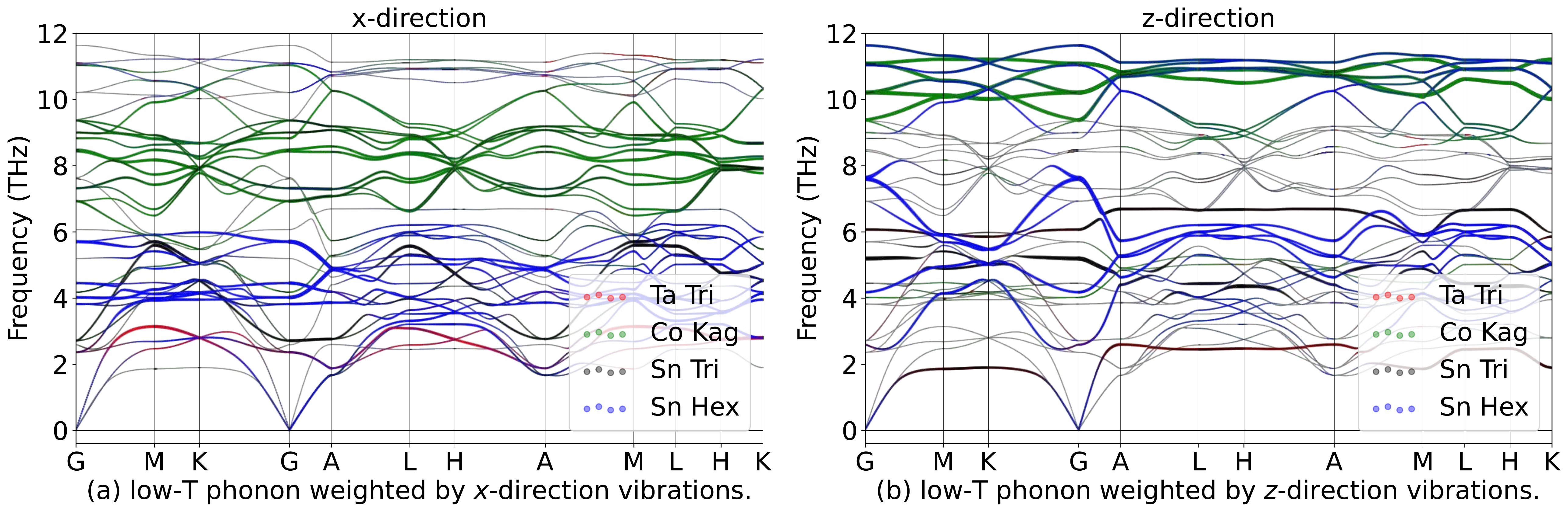}
\hspace{5pt}\label{fig:TaCo6Sn6_relaxed_High_xz}
\includegraphics[width=0.85\textwidth]{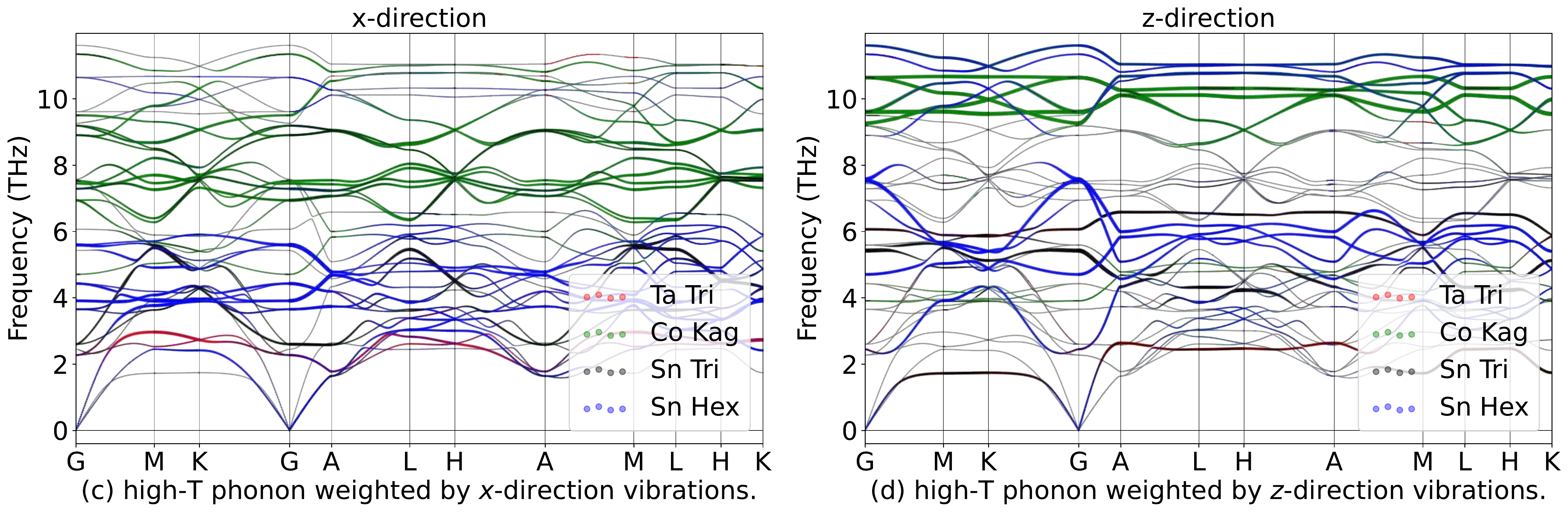}
\caption{Phonon spectrum for TaCo$_6$Sn$_6$in in-plane ($x$) and out-of-plane ($z$) directions.}
\label{fig:TaCo6Sn6phoxyz}
\end{figure}

\begin{table}[htbp]
\global\long\def\arraystretch{1.12}
\captionsetup{width=.95\textwidth}
\caption{IRREPs at high-symmetry points for TaCo$_6$Sn$_6$ phonon spectrum at Low-T.}
\label{tab:191_TaCo6Sn6_Low}
\setlength{\tabcolsep}{1.mm}{
}
\end{table}

\begin{table}[htbp]
\global\long\def\arraystretch{1.12}
\captionsetup{width=.95\textwidth}
\caption{IRREPs at high-symmetry points for TaCo$_6$Sn$_6$ phonon spectrum at High-T.}
\label{tab:191_TaCo6Sn6_High}
\setlength{\tabcolsep}{1.mm}{
%
}
\end{table}
\subsubsection{\label{sms2sec:TlCo6Sn6}\hyperref[smsec:col]{\texorpdfstring{TlCo$_6$Sn$_6$}{}}~{\color{green}  (High-T stable, Low-T stable) }}

\begin{table}[htbp]
\global\long\def\arraystretch{1.12}
\captionsetup{width=.85\textwidth}
\caption{Structure parameters of TlCo$_6$Sn$_6$ after relaxation. It contains 543 electrons. }
\label{tab:TlCo6Sn6}
\setlength{\tabcolsep}{4mm}{
%
}
\end{table}

\begin{figure}[htbp]
\centering
\includegraphics[width=0.85\textwidth]{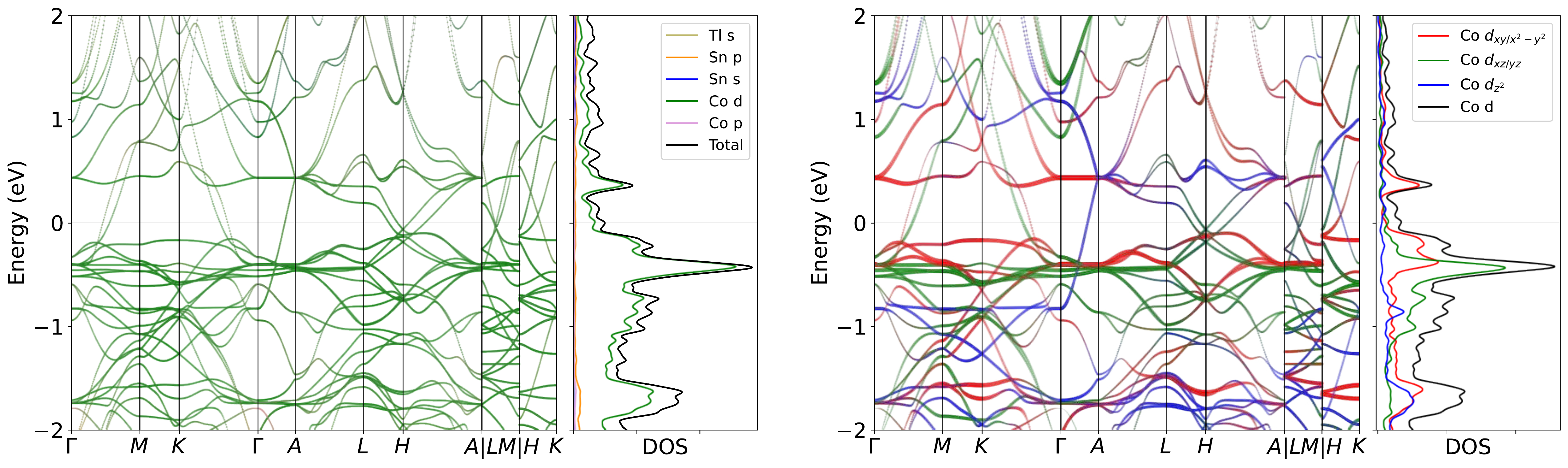}
\caption{Band structure for TlCo$_6$Sn$_6$ without SOC. The projection of Co $3d$ orbitals is also presented. The band structures clearly reveal the dominating of Co $3d$ orbitals  near the Fermi level, as well as the pronounced peak.}
\label{fig:TlCo6Sn6band}
\end{figure}

\begin{figure}[H]
\centering
\subfloat[Relaxed phonon at low-T.]{
\label{fig:TlCo6Sn6_relaxed_Low}
\includegraphics[width=0.45\textwidth]{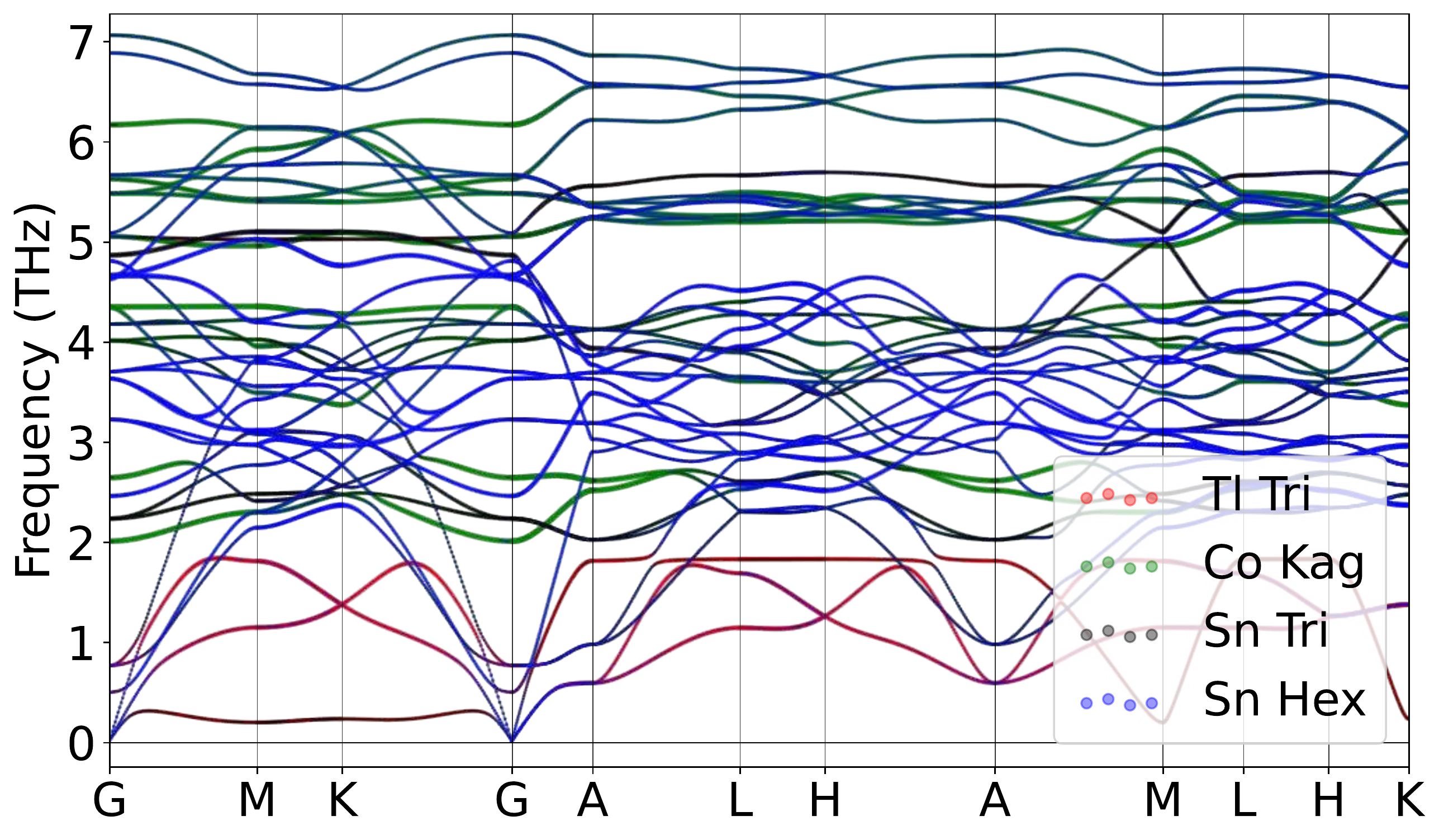}
}
\hspace{5pt}\subfloat[Relaxed phonon at high-T.]{
\label{fig:TlCo6Sn6_relaxed_High}
\includegraphics[width=0.45\textwidth]{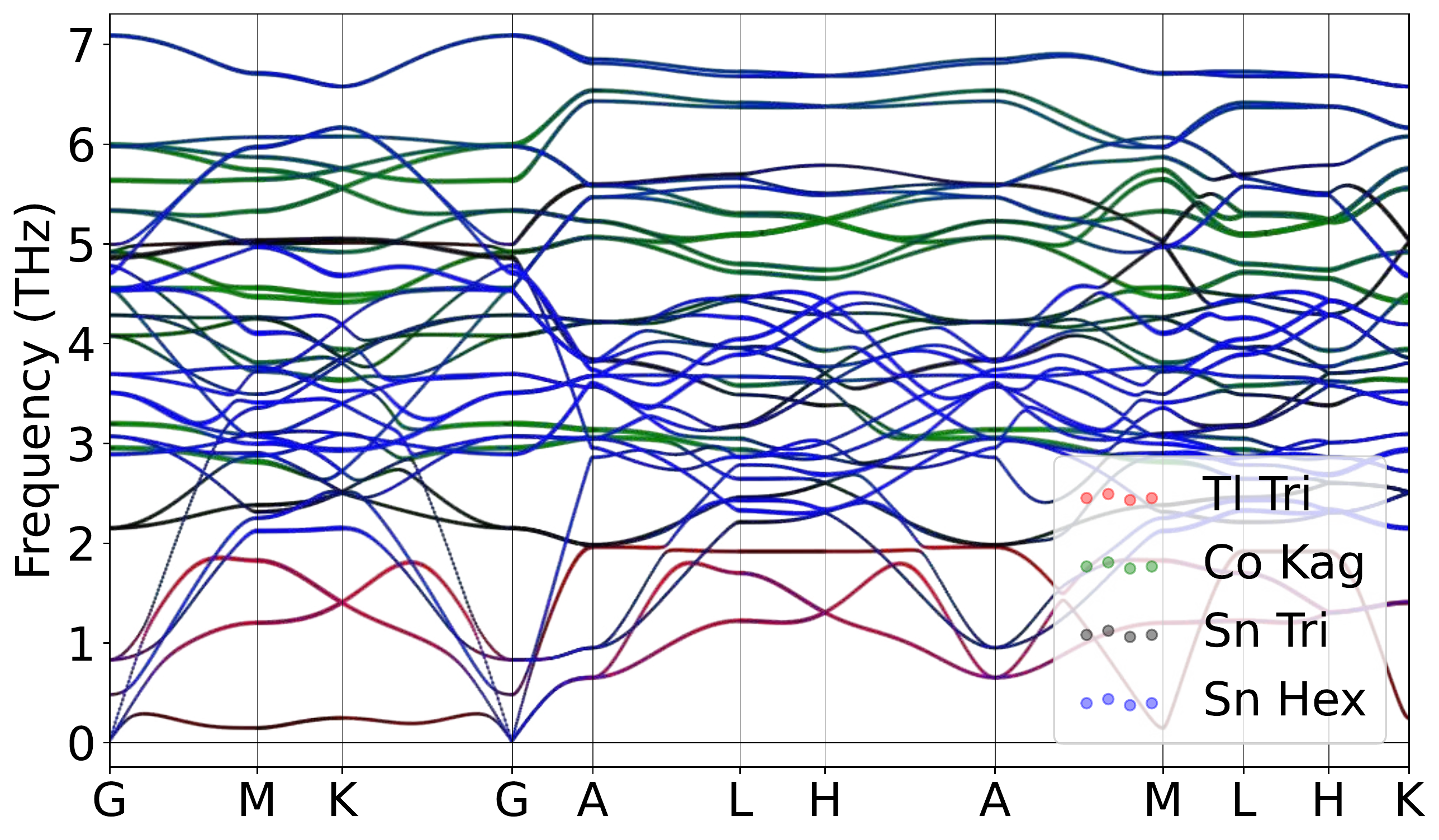}
}
\caption{Atomically projected phonon spectrum for TlCo$_6$Sn$_6$.}
\label{fig:TlCo6Sn6pho}
\end{figure}

\begin{figure}[H]
\centering
\label{fig:TlCo6Sn6_relaxed_Low_xz}
\includegraphics[width=0.85\textwidth]{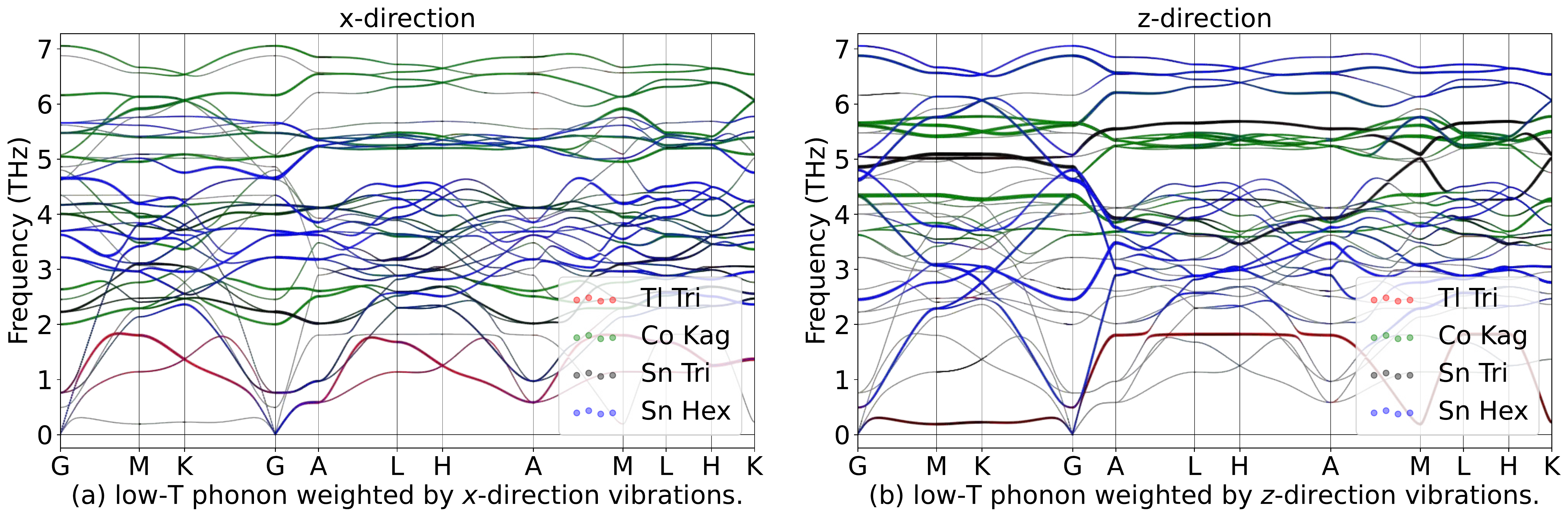}
\hspace{5pt}\label{fig:TlCo6Sn6_relaxed_High_xz}
\includegraphics[width=0.85\textwidth]{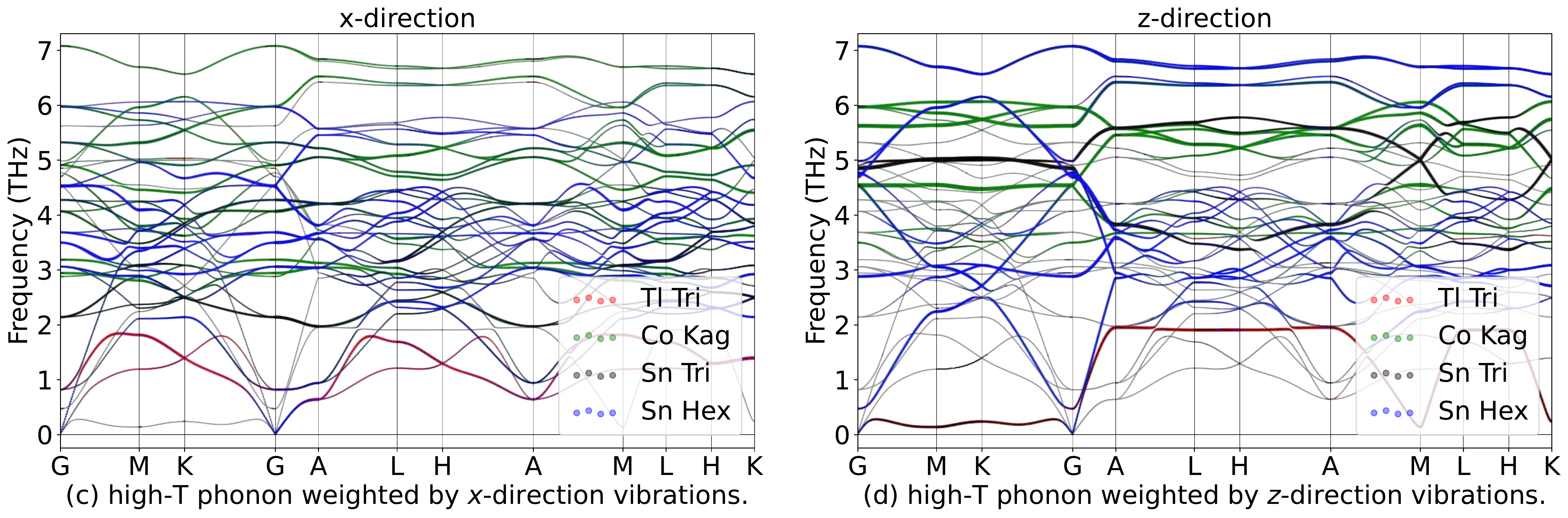}
\caption{Phonon spectrum for TlCo$_6$Sn$_6$in in-plane ($x$) and out-of-plane ($z$) directions.}
\label{fig:TlCo6Sn6phoxyz}
\end{figure}

\begin{table}[htbp]
\global\long\def\arraystretch{1.12}
\captionsetup{width=.95\textwidth}
\caption{IRREPs at high-symmetry points for TlCo$_6$Sn$_6$ phonon spectrum at Low-T.}
\label{tab:191_TlCo6Sn6_Low}
\setlength{\tabcolsep}{1.mm}{
}
\end{table}

\begin{table}[htbp]
\global\long\def\arraystretch{1.12}
\captionsetup{width=.95\textwidth}
\caption{IRREPs at high-symmetry points for TlCo$_6$Sn$_6$ phonon spectrum at High-T.}
\label{tab:191_TlCo6Sn6_High}
\setlength{\tabcolsep}{1.mm}{
%
}
\end{table}
\subsection{\hyperref[smsec:col]{NiSi}}~\label{smssec:NiSi}

\subsubsection{\label{sms2sec:LiNi6Si6}\hyperref[smsec:col]{\texorpdfstring{LiNi$_6$Si$_6$}{}}~{\color{red}  (High-T unstable, Low-T unstable) }}

\begin{table}[htbp]
\global\long\def\arraystretch{1.12}
\captionsetup{width=.85\textwidth}
\caption{Structure parameters of LiNi$_6$Si$_6$ after relaxation. It contains 255 electrons. }
\label{tab:LiNi6Si6}
\setlength{\tabcolsep}{4mm}{
%
}
\end{table}

\begin{figure}[htbp]
\centering
\includegraphics[width=0.85\textwidth]{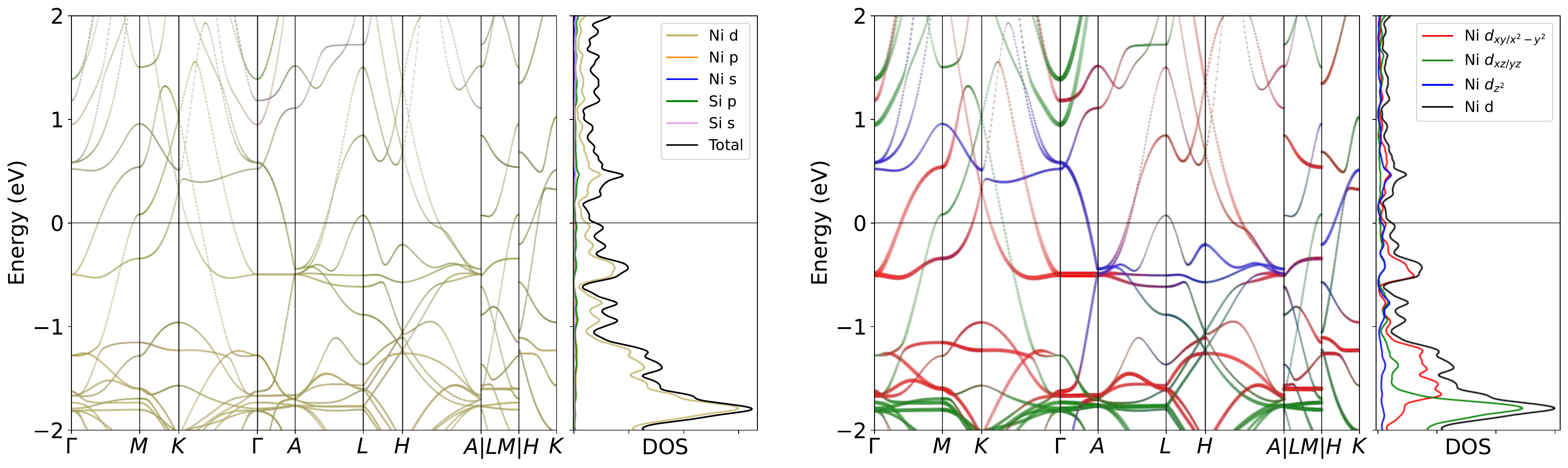}
\caption{Band structure for LiNi$_6$Si$_6$ without SOC. The projection of Ni $3d$ orbitals is also presented. The band structures clearly reveal the dominating of Ni $3d$ orbitals  near the Fermi level, as well as the pronounced peak.}
\label{fig:LiNi6Si6band}
\end{figure}

\begin{figure}[H]
\centering
\subfloat[Relaxed phonon at low-T.]{
\label{fig:LiNi6Si6_relaxed_Low}
\includegraphics[width=0.45\textwidth]{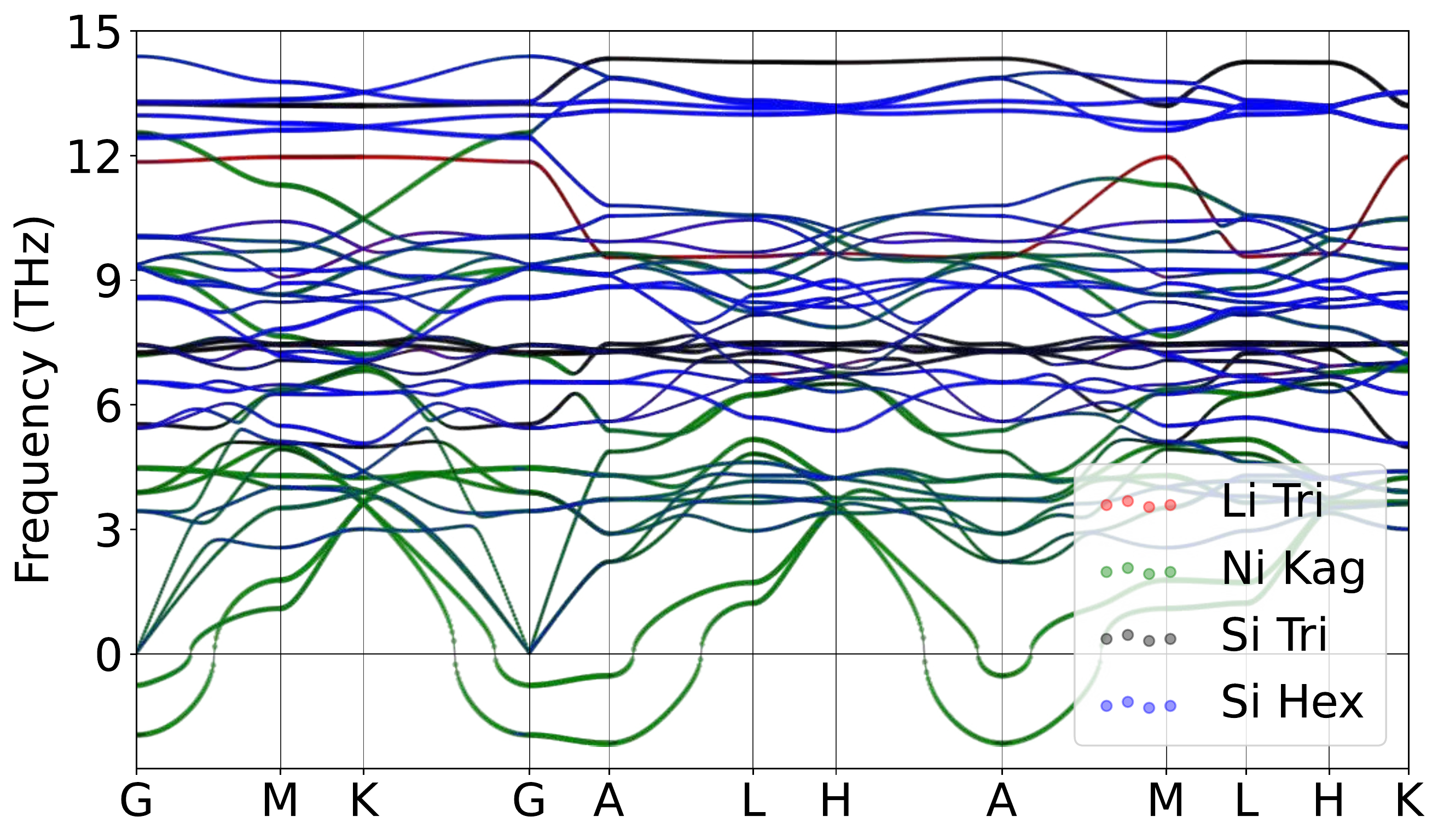}
}
\hspace{5pt}\subfloat[Relaxed phonon at high-T.]{
\label{fig:LiNi6Si6_relaxed_High}
\includegraphics[width=0.45\textwidth]{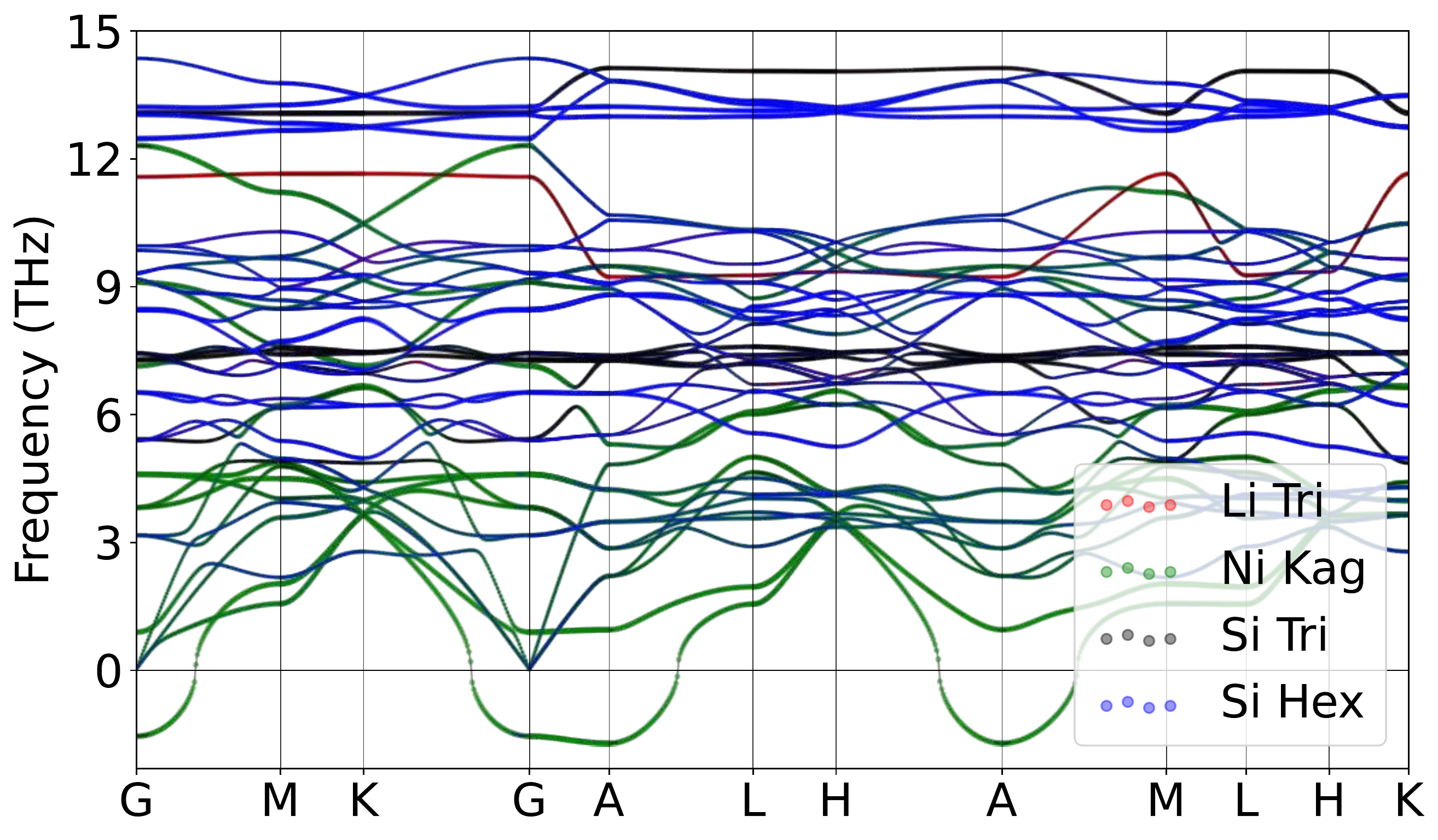}
}
\caption{Atomically projected phonon spectrum for LiNi$_6$Si$_6$.}
\label{fig:LiNi6Si6pho}
\end{figure}

\begin{figure}[H]
\centering
\label{fig:LiNi6Si6_relaxed_Low_xz}
\includegraphics[width=0.85\textwidth]{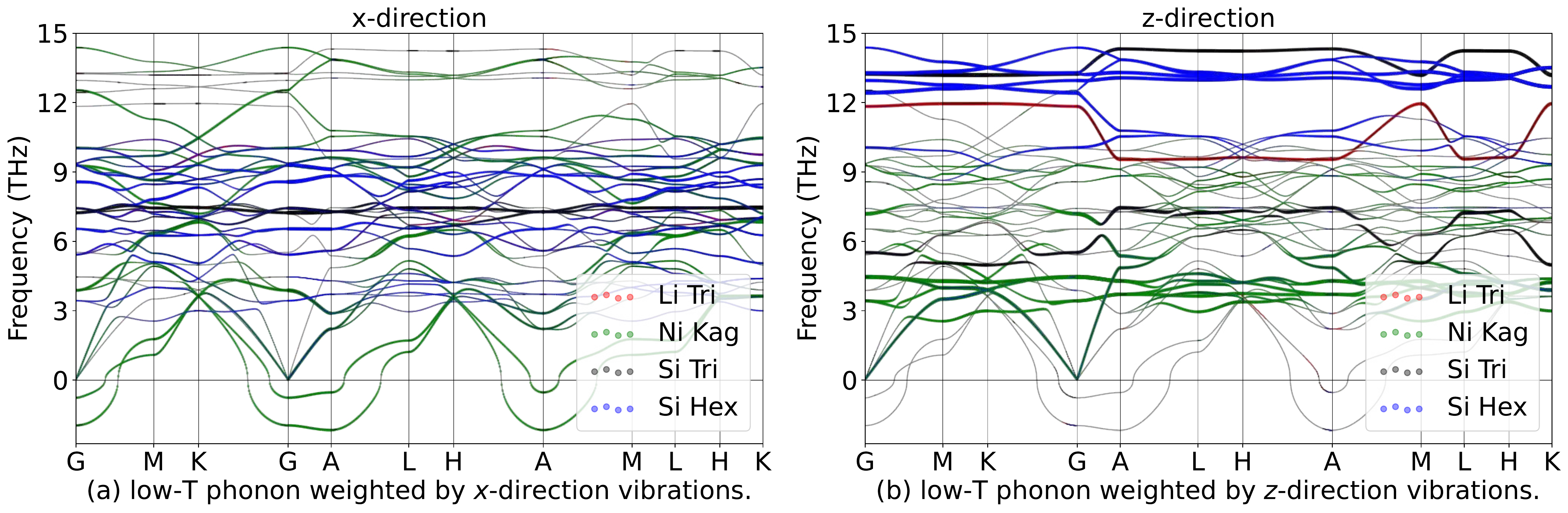}
\hspace{5pt}\label{fig:LiNi6Si6_relaxed_High_xz}
\includegraphics[width=0.85\textwidth]{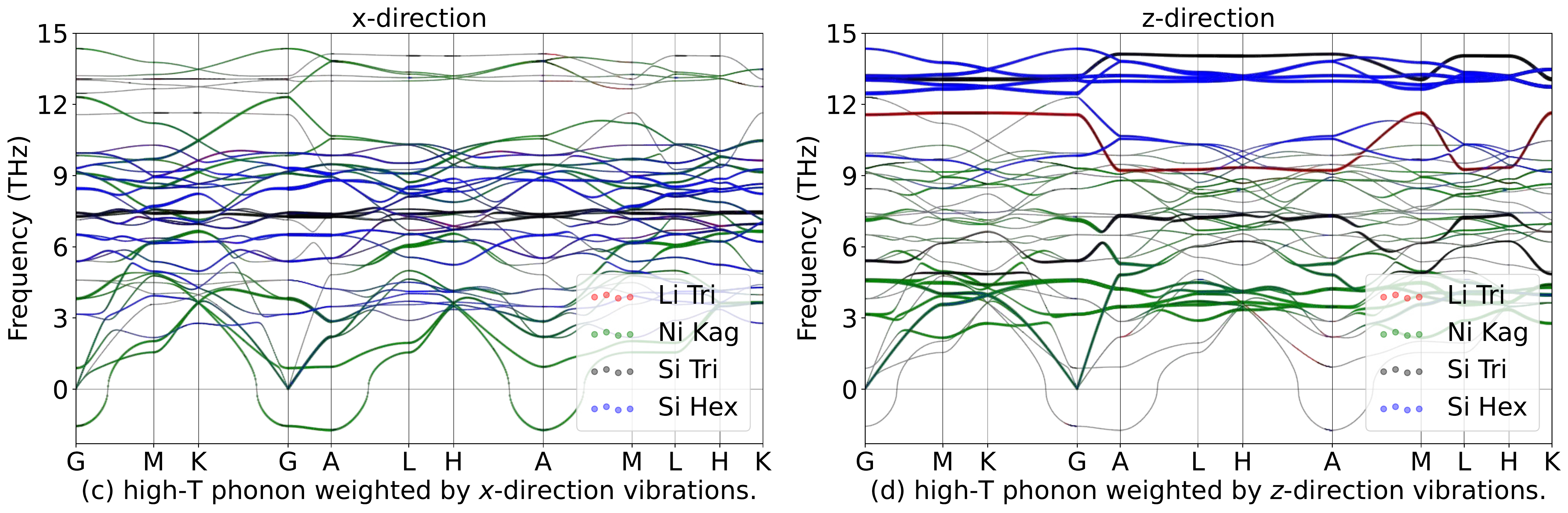}
\caption{Phonon spectrum for LiNi$_6$Si$_6$in in-plane ($x$) and out-of-plane ($z$) directions.}
\label{fig:LiNi6Si6phoxyz}
\end{figure}

\begin{table}[htbp]
\global\long\def\arraystretch{1.12}
\captionsetup{width=.95\textwidth}
\caption{IRREPs at high-symmetry points for LiNi$_6$Si$_6$ phonon spectrum at Low-T.}
\label{tab:191_LiNi6Si6_Low}
\setlength{\tabcolsep}{1.mm}{
}
\end{table}

\begin{table}[htbp]
\global\long\def\arraystretch{1.12}
\captionsetup{width=.95\textwidth}
\caption{IRREPs at high-symmetry points for LiNi$_6$Si$_6$ phonon spectrum at High-T.}
\label{tab:191_LiNi6Si6_High}
\setlength{\tabcolsep}{1.mm}{
%
}
\end{table}
\subsubsection{\label{sms2sec:MgNi6Si6}\hyperref[smsec:col]{\texorpdfstring{MgNi$_6$Si$_6$}{}}~{\color{red}  (High-T unstable, Low-T unstable) }}

\begin{table}[htbp]
\global\long\def\arraystretch{1.12}
\captionsetup{width=.85\textwidth}
\caption{Structure parameters of MgNi$_6$Si$_6$ after relaxation. It contains 264 electrons. }
\label{tab:MgNi6Si6}
\setlength{\tabcolsep}{4mm}{
%
}
\end{table}

\begin{figure}[htbp]
\centering
\includegraphics[width=0.85\textwidth]{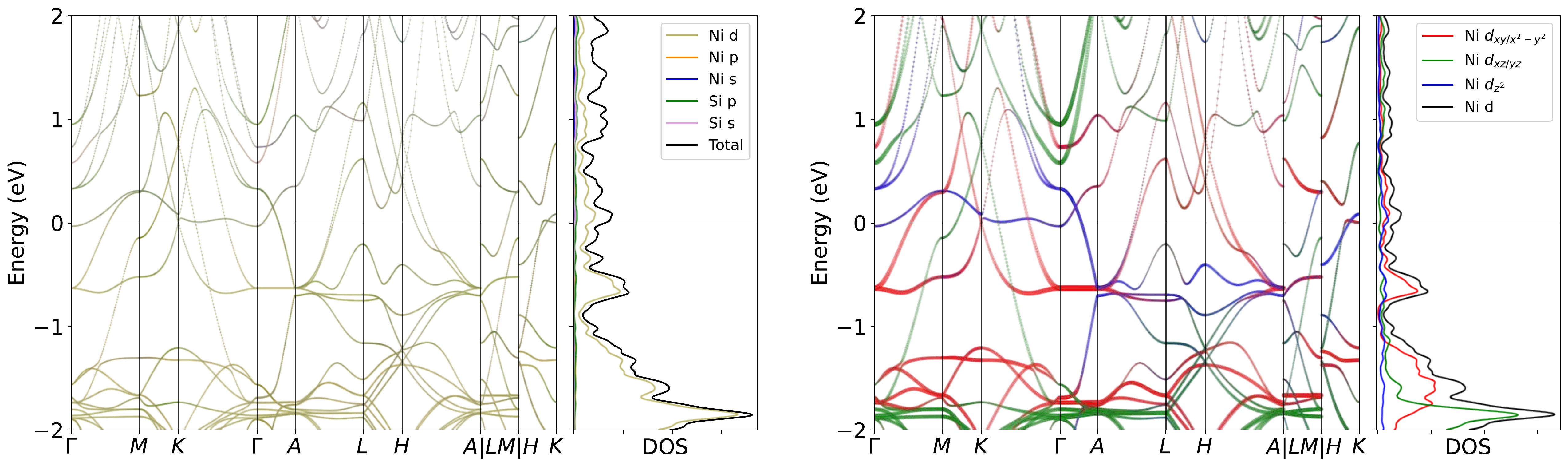}
\caption{Band structure for MgNi$_6$Si$_6$ without SOC. The projection of Ni $3d$ orbitals is also presented. The band structures clearly reveal the dominating of Ni $3d$ orbitals  near the Fermi level, as well as the pronounced peak.}
\label{fig:MgNi6Si6band}
\end{figure}

\begin{figure}[H]
\centering
\subfloat[Relaxed phonon at low-T.]{
\label{fig:MgNi6Si6_relaxed_Low}
\includegraphics[width=0.45\textwidth]{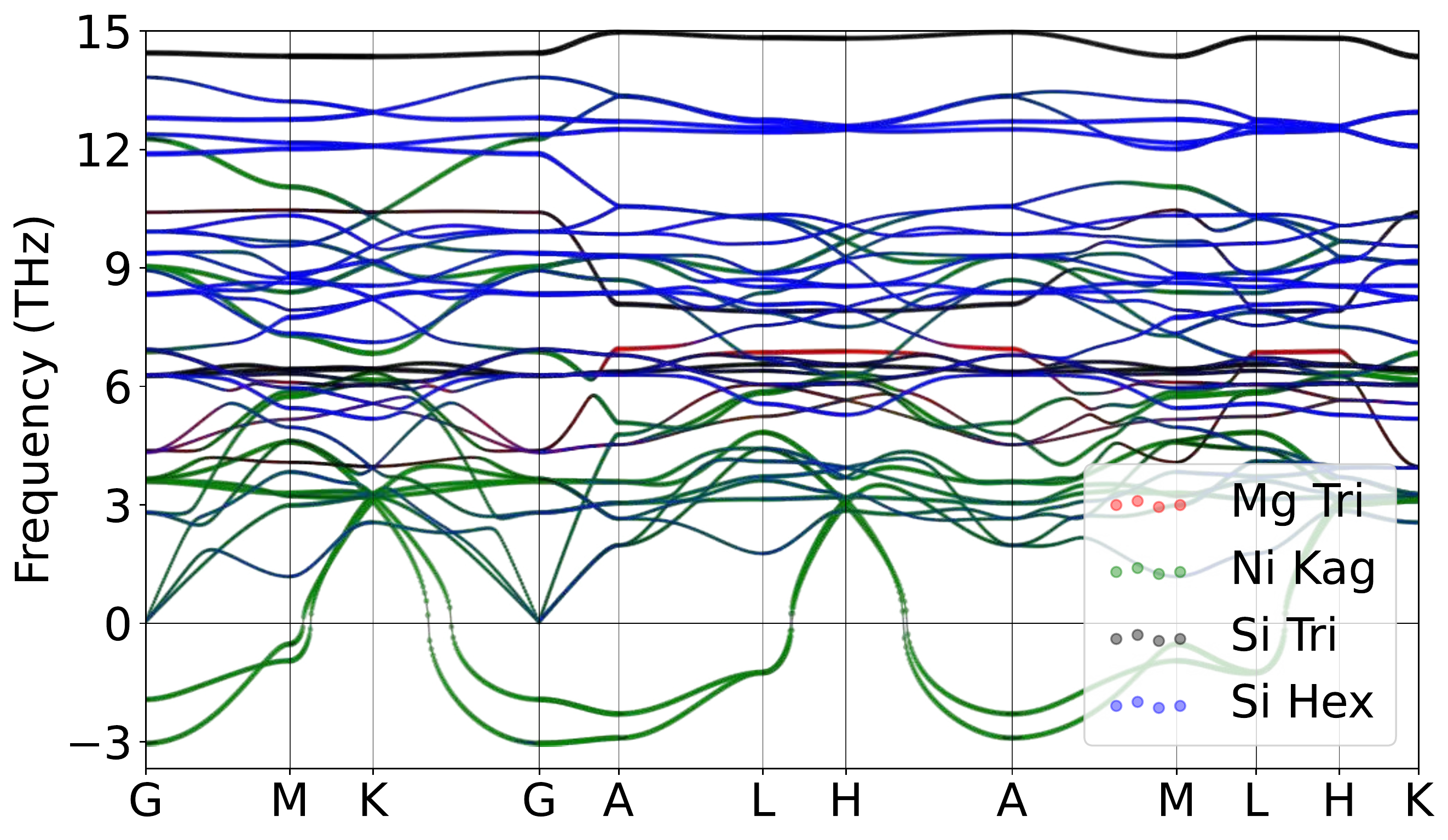}
}
\hspace{5pt}\subfloat[Relaxed phonon at high-T.]{
\label{fig:MgNi6Si6_relaxed_High}
\includegraphics[width=0.45\textwidth]{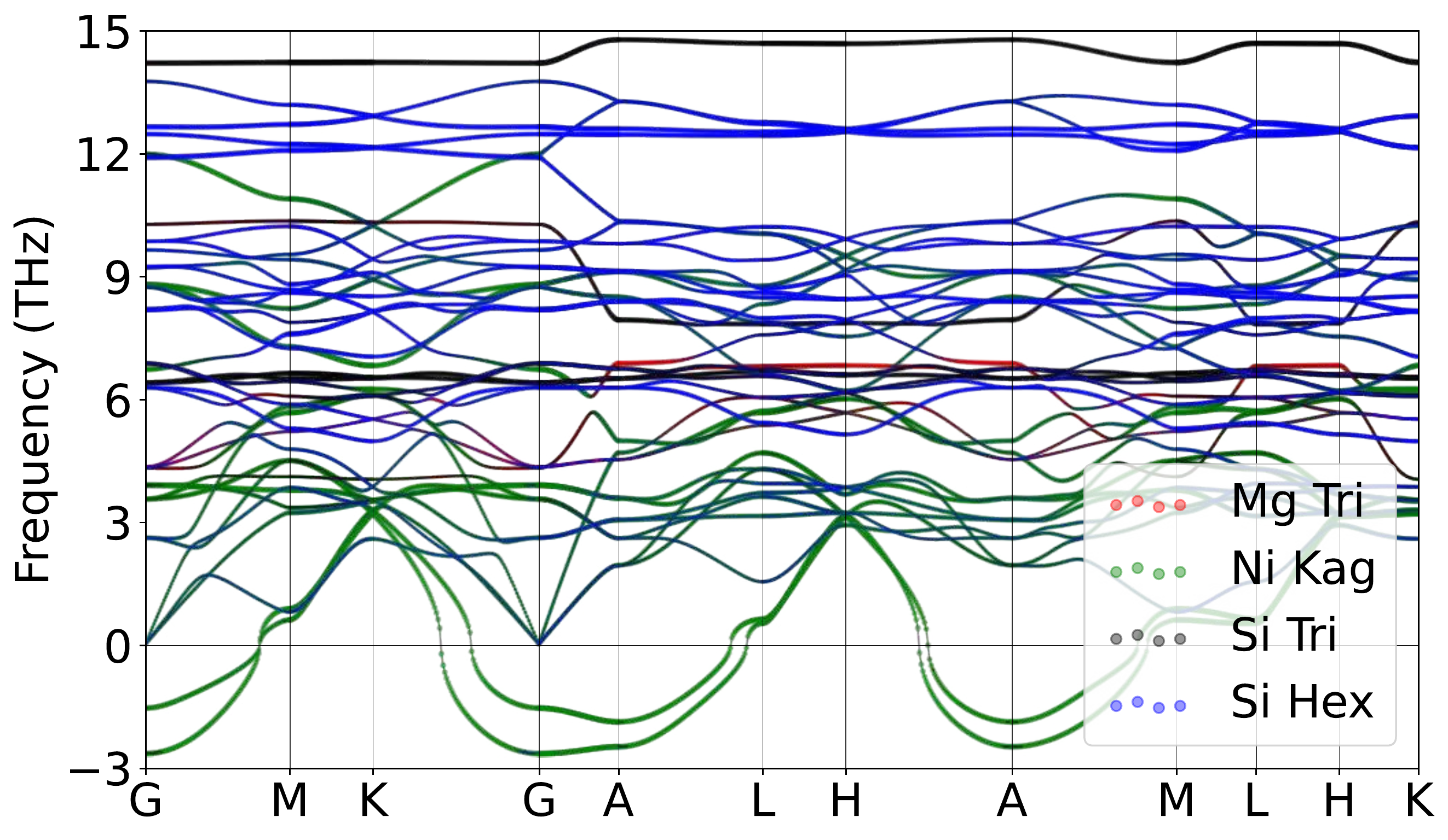}
}
\caption{Atomically projected phonon spectrum for MgNi$_6$Si$_6$.}
\label{fig:MgNi6Si6pho}
\end{figure}

\begin{figure}[H]
\centering
\label{fig:MgNi6Si6_relaxed_Low_xz}
\includegraphics[width=0.85\textwidth]{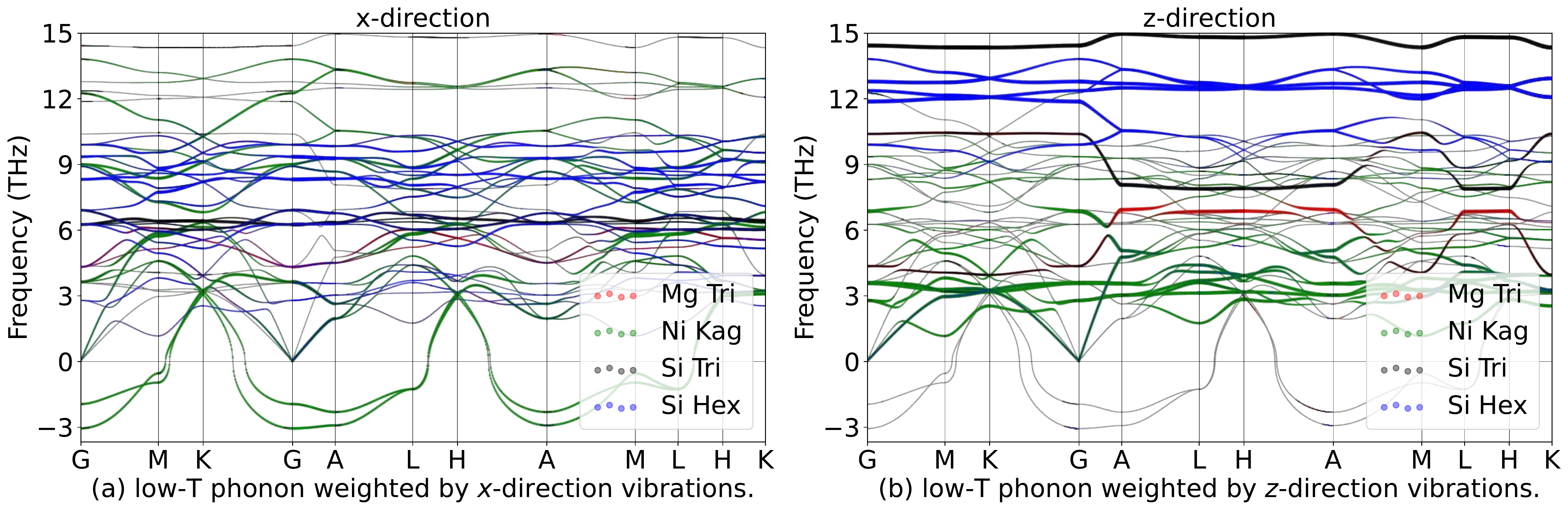}
\hspace{5pt}\label{fig:MgNi6Si6_relaxed_High_xz}
\includegraphics[width=0.85\textwidth]{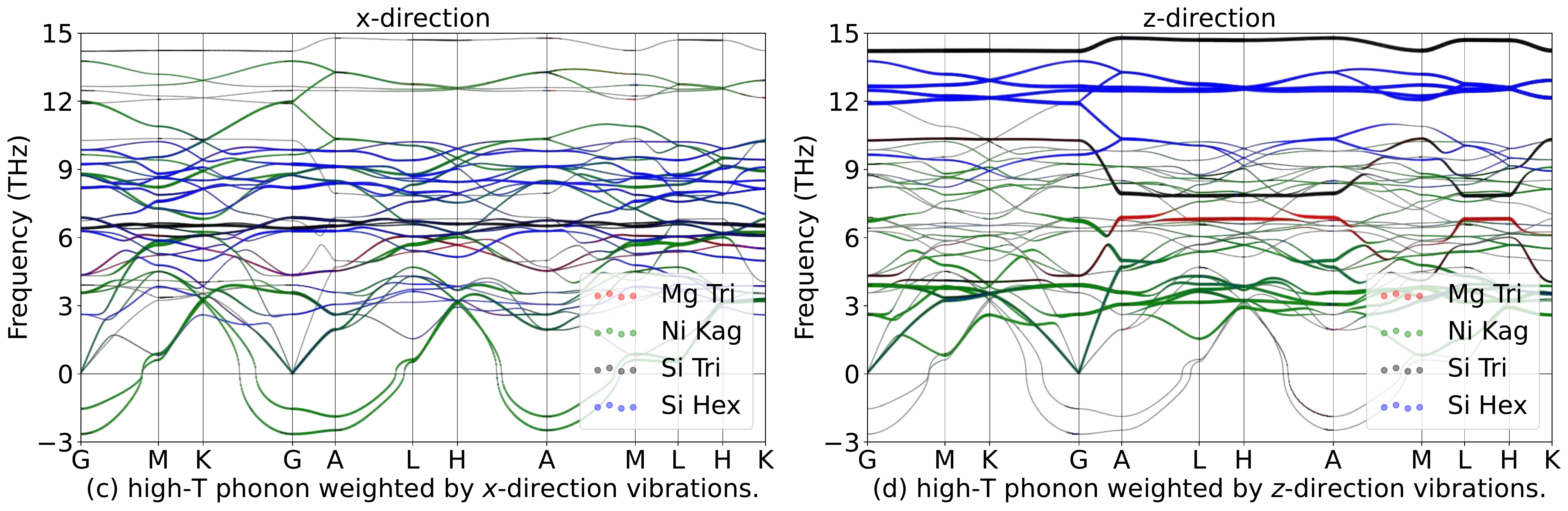}
\caption{Phonon spectrum for MgNi$_6$Si$_6$in in-plane ($x$) and out-of-plane ($z$) directions.}
\label{fig:MgNi6Si6phoxyz}
\end{figure}

\begin{table}[htbp]
\global\long\def\arraystretch{1.12}
\captionsetup{width=.95\textwidth}
\caption{IRREPs at high-symmetry points for MgNi$_6$Si$_6$ phonon spectrum at Low-T.}
\label{tab:191_MgNi6Si6_Low}
\setlength{\tabcolsep}{1.mm}{
}
\end{table}

\begin{table}[htbp]
\global\long\def\arraystretch{1.12}
\captionsetup{width=.95\textwidth}
\caption{IRREPs at high-symmetry points for MgNi$_6$Si$_6$ phonon spectrum at High-T.}
\label{tab:191_MgNi6Si6_High}
\setlength{\tabcolsep}{1.mm}{
%
}
\end{table}
\subsubsection{\label{sms2sec:CaNi6Si6}\hyperref[smsec:col]{\texorpdfstring{CaNi$_6$Si$_6$}{}}~{\color{blue}  (High-T stable, Low-T unstable) }}

\begin{table}[htbp]
\global\long\def\arraystretch{1.12}
\captionsetup{width=.85\textwidth}
\caption{Structure parameters of CaNi$_6$Si$_6$ after relaxation. It contains 272 electrons. }
\label{tab:CaNi6Si6}
\setlength{\tabcolsep}{4mm}{
%
}
\end{table}

\begin{figure}[htbp]
\centering
\includegraphics[width=0.85\textwidth]{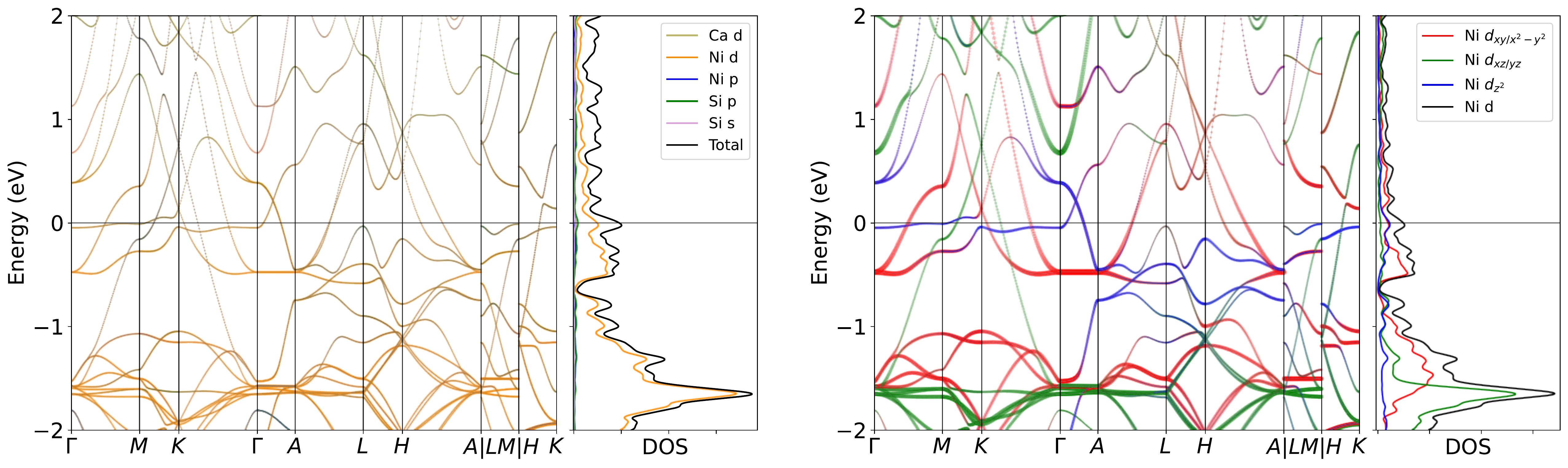}
\caption{Band structure for CaNi$_6$Si$_6$ without SOC. The projection of Ni $3d$ orbitals is also presented. The band structures clearly reveal the dominating of Ni $3d$ orbitals  near the Fermi level, as well as the pronounced peak.}
\label{fig:CaNi6Si6band}
\end{figure}

\begin{figure}[H]
\centering
\subfloat[Relaxed phonon at low-T.]{
\label{fig:CaNi6Si6_relaxed_Low}
\includegraphics[width=0.45\textwidth]{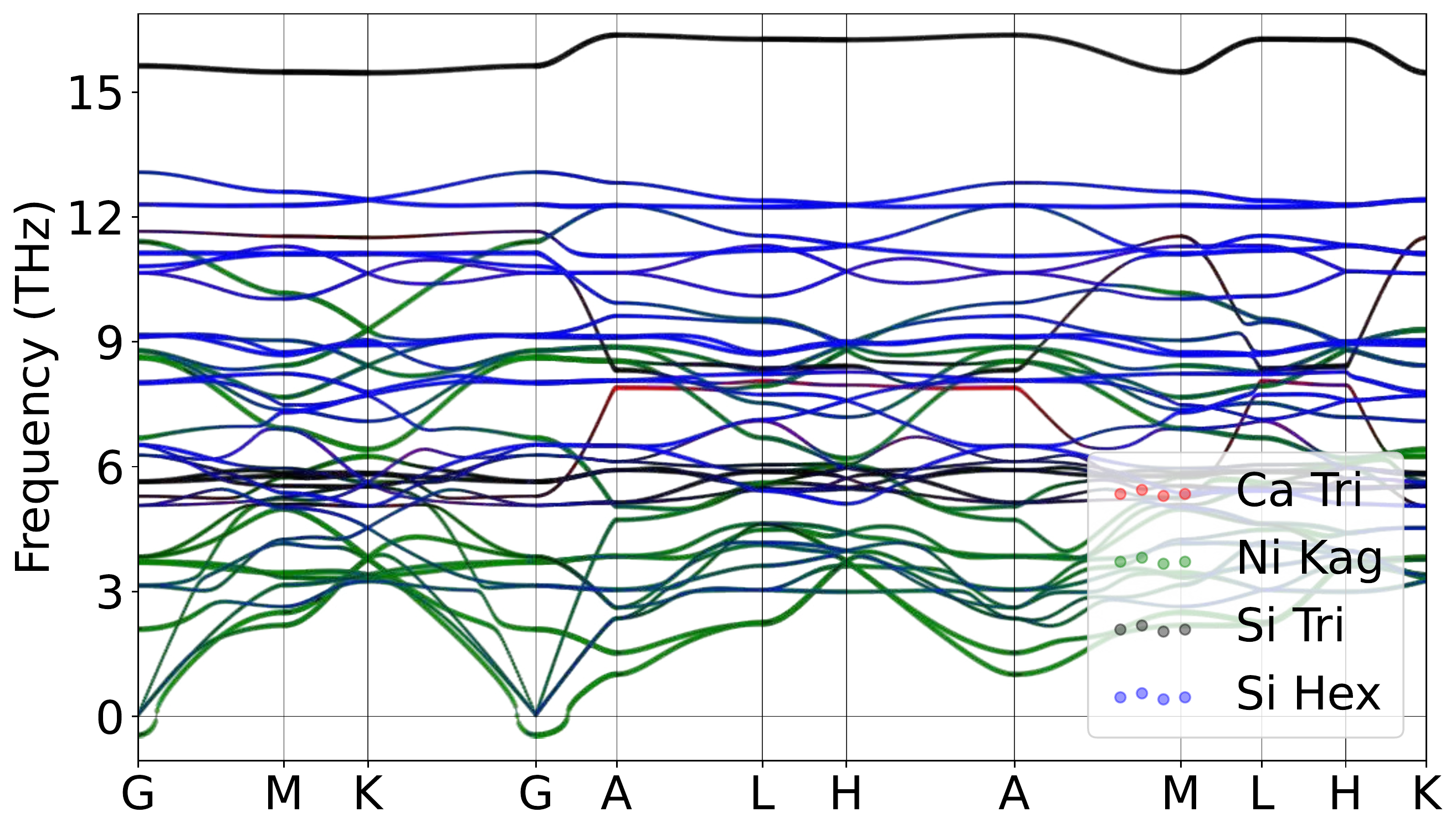}
}
\hspace{5pt}\subfloat[Relaxed phonon at high-T.]{
\label{fig:CaNi6Si6_relaxed_High}
\includegraphics[width=0.45\textwidth]{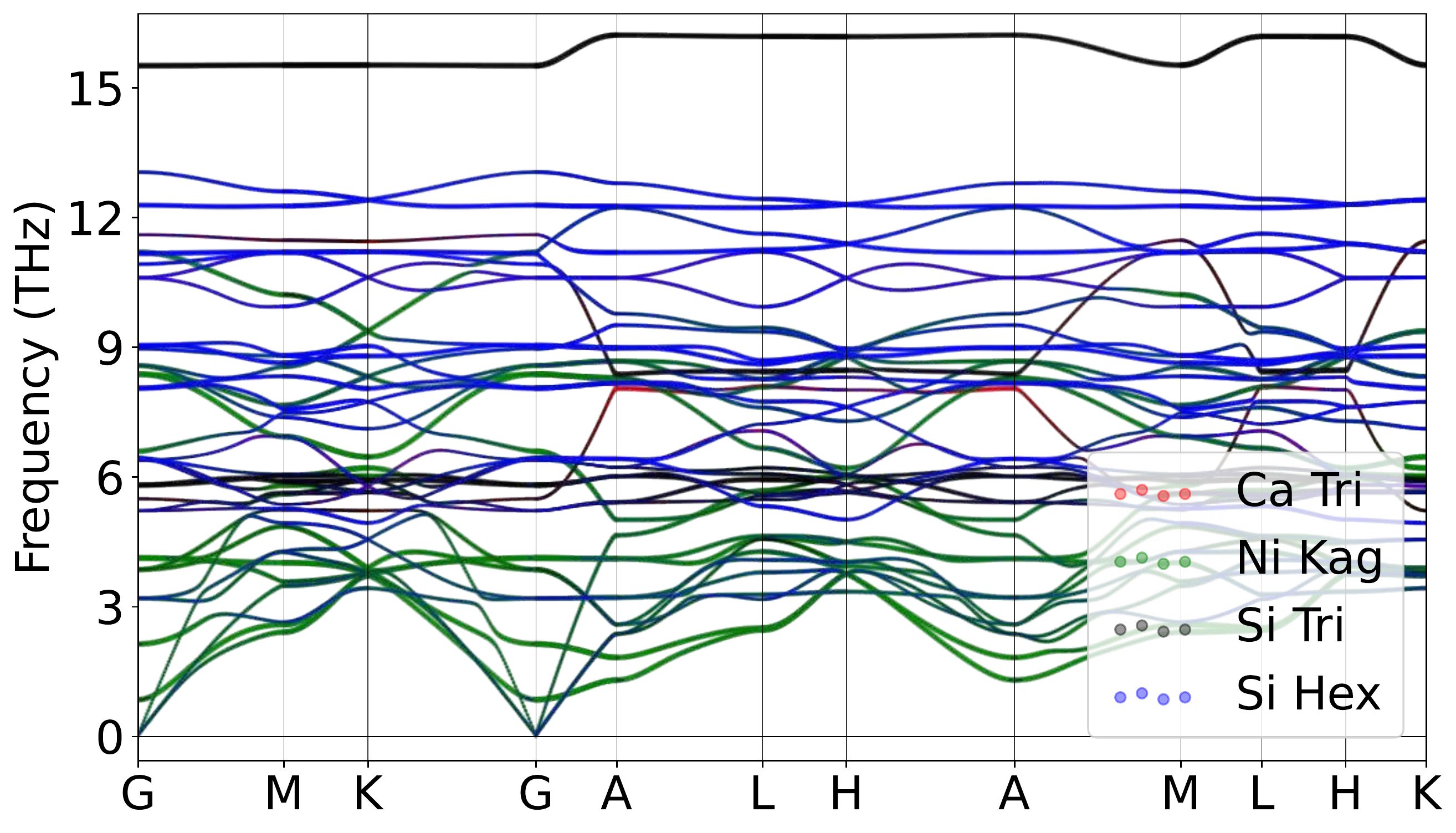}
}
\caption{Atomically projected phonon spectrum for CaNi$_6$Si$_6$.}
\label{fig:CaNi6Si6pho}
\end{figure}

\begin{figure}[H]
\centering
\label{fig:CaNi6Si6_relaxed_Low_xz}
\includegraphics[width=0.85\textwidth]{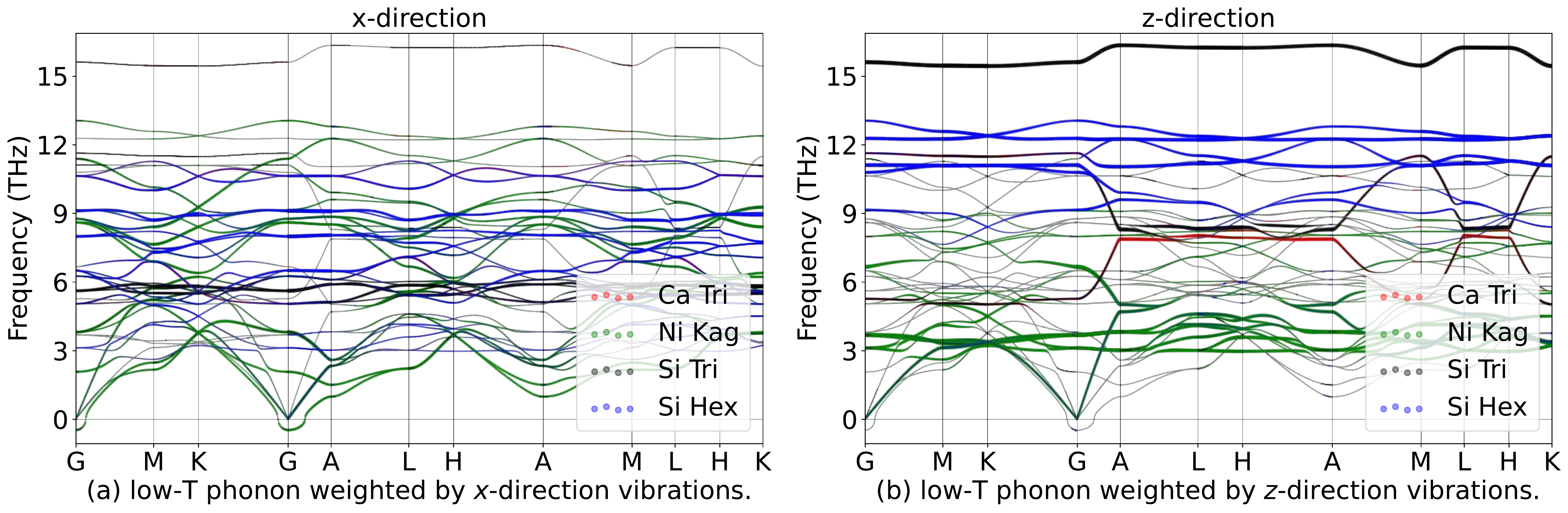}
\hspace{5pt}\label{fig:CaNi6Si6_relaxed_High_xz}
\includegraphics[width=0.85\textwidth]{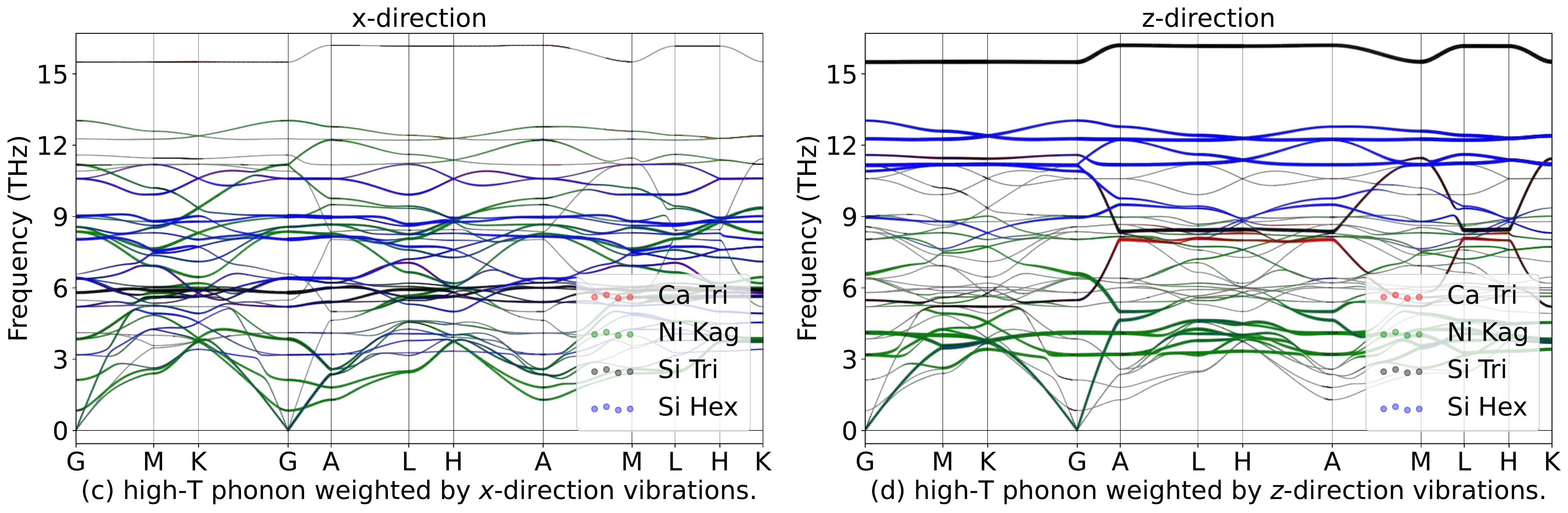}
\caption{Phonon spectrum for CaNi$_6$Si$_6$in in-plane ($x$) and out-of-plane ($z$) directions.}
\label{fig:CaNi6Si6phoxyz}
\end{figure}

\begin{table}[htbp]
\global\long\def\arraystretch{1.12}
\captionsetup{width=.95\textwidth}
\caption{IRREPs at high-symmetry points for CaNi$_6$Si$_6$ phonon spectrum at Low-T.}
\label{tab:191_CaNi6Si6_Low}
\setlength{\tabcolsep}{1.mm}{
}
\end{table}

\begin{table}[htbp]
\global\long\def\arraystretch{1.12}
\captionsetup{width=.95\textwidth}
\caption{IRREPs at high-symmetry points for CaNi$_6$Si$_6$ phonon spectrum at High-T.}
\label{tab:191_CaNi6Si6_High}
\setlength{\tabcolsep}{1.mm}{
%
}
\end{table}
\subsubsection{\label{sms2sec:ScNi6Si6}\hyperref[smsec:col]{\texorpdfstring{ScNi$_6$Si$_6$}{}}~{\color{red}  (High-T unstable, Low-T unstable) }}

\begin{table}[htbp]
\global\long\def\arraystretch{1.12}
\captionsetup{width=.85\textwidth}
\caption{Structure parameters of ScNi$_6$Si$_6$ after relaxation. It contains 273 electrons. }
\label{tab:ScNi6Si6}
\setlength{\tabcolsep}{4mm}{
%
}
\end{table}

\begin{figure}[htbp]
\centering
\includegraphics[width=0.85\textwidth]{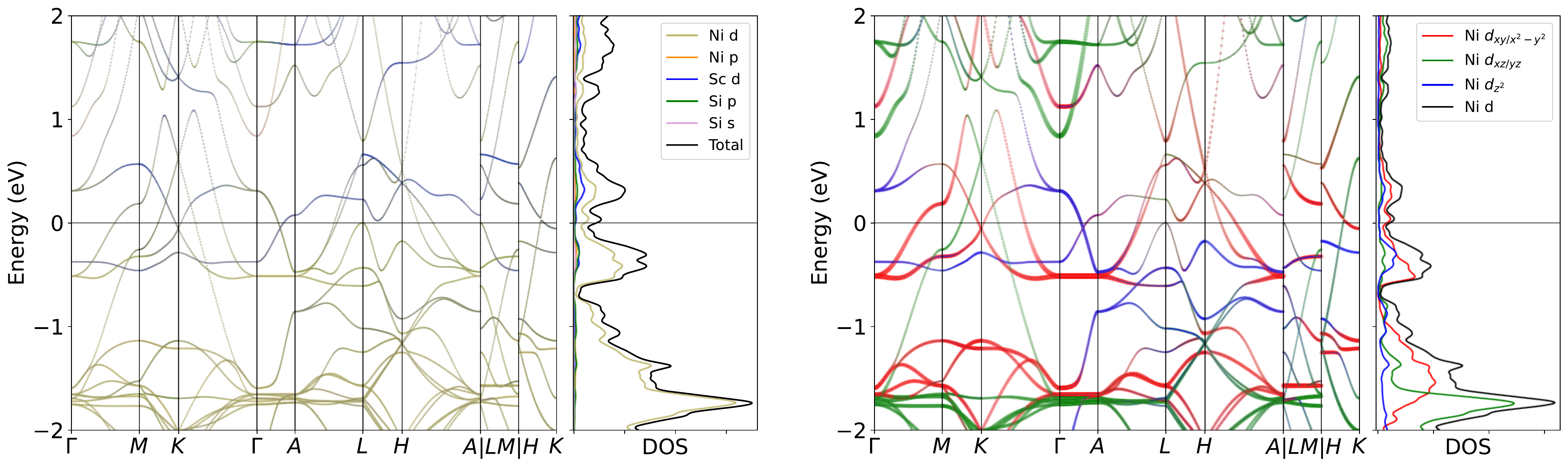}
\caption{Band structure for ScNi$_6$Si$_6$ without SOC. The projection of Ni $3d$ orbitals is also presented. The band structures clearly reveal the dominating of Ni $3d$ orbitals  near the Fermi level, as well as the pronounced peak.}
\label{fig:ScNi6Si6band}
\end{figure}

\begin{figure}[H]
\centering
\subfloat[Relaxed phonon at low-T.]{
\label{fig:ScNi6Si6_relaxed_Low}
\includegraphics[width=0.45\textwidth]{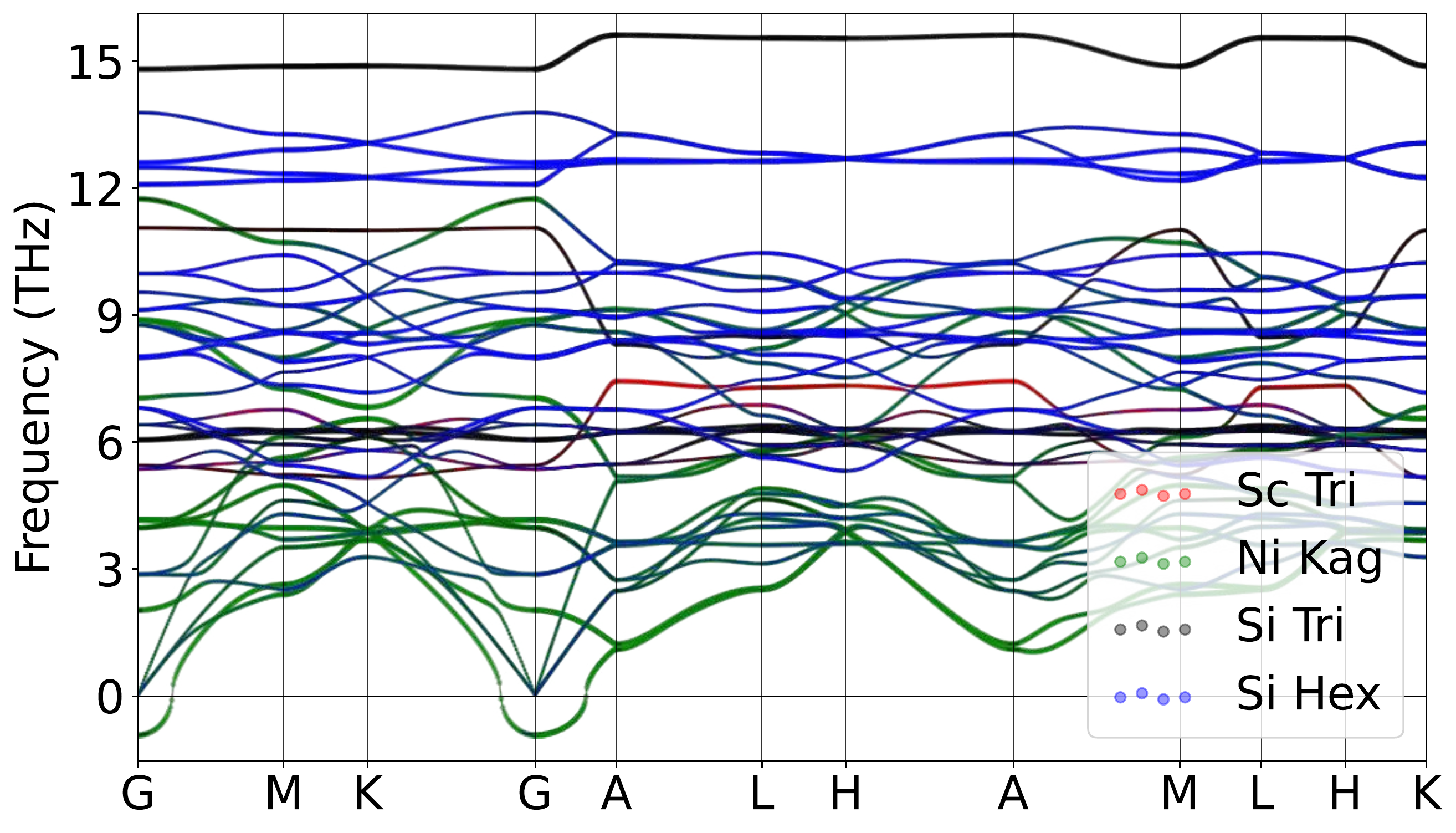}
}
\hspace{5pt}\subfloat[Relaxed phonon at high-T.]{
\label{fig:ScNi6Si6_relaxed_High}
\includegraphics[width=0.45\textwidth]{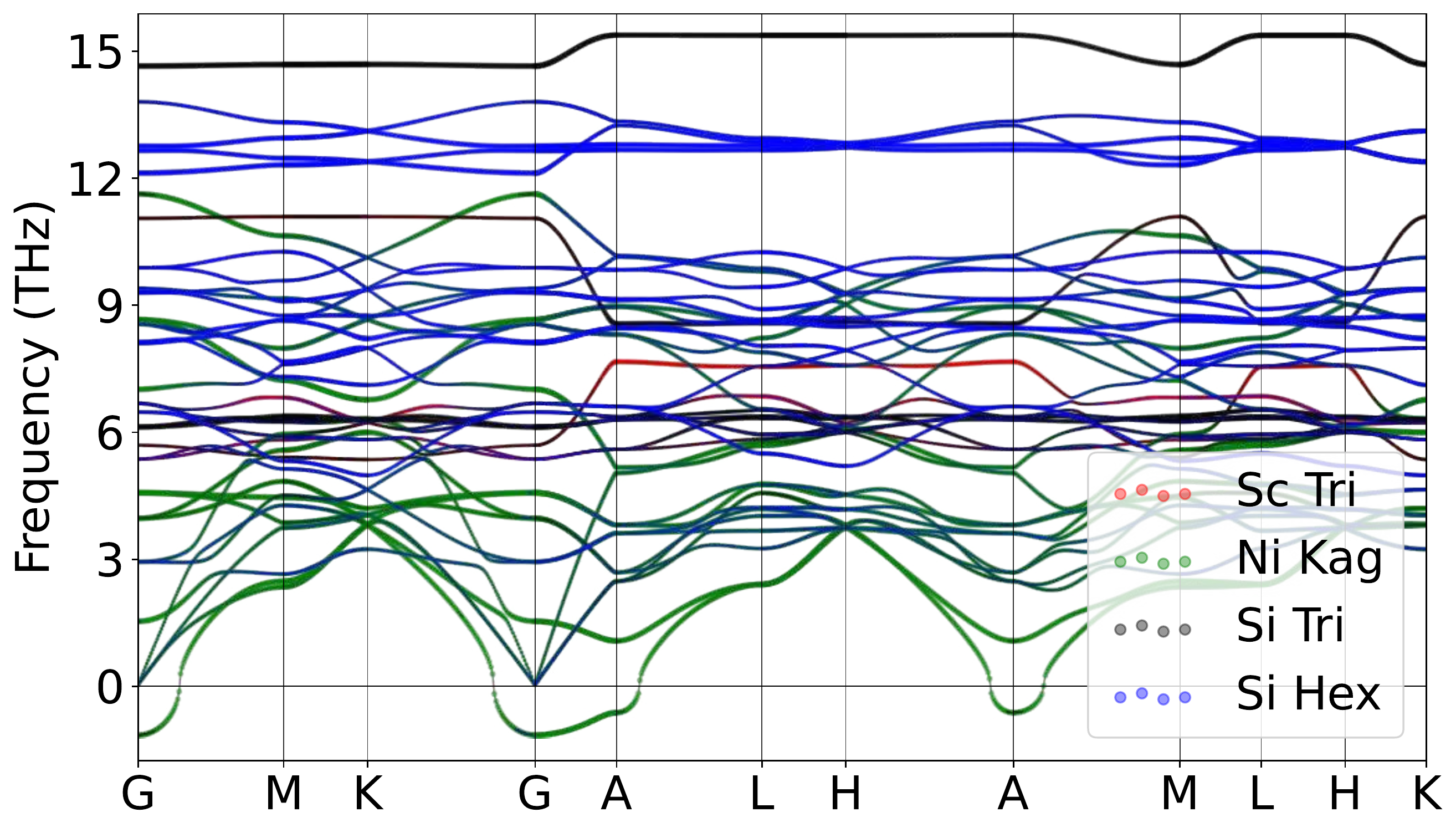}
}
\caption{Atomically projected phonon spectrum for ScNi$_6$Si$_6$.}
\label{fig:ScNi6Si6pho}
\end{figure}

\begin{figure}[H]
\centering
\label{fig:ScNi6Si6_relaxed_Low_xz}
\includegraphics[width=0.85\textwidth]{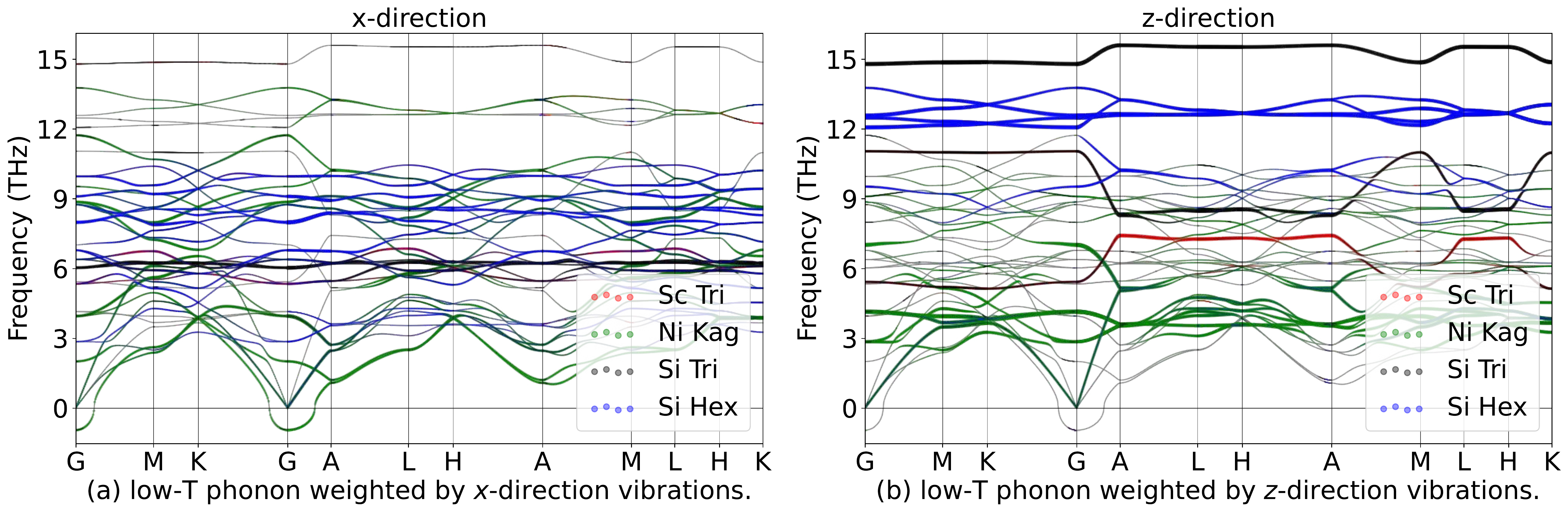}
\hspace{5pt}\label{fig:ScNi6Si6_relaxed_High_xz}
\includegraphics[width=0.85\textwidth]{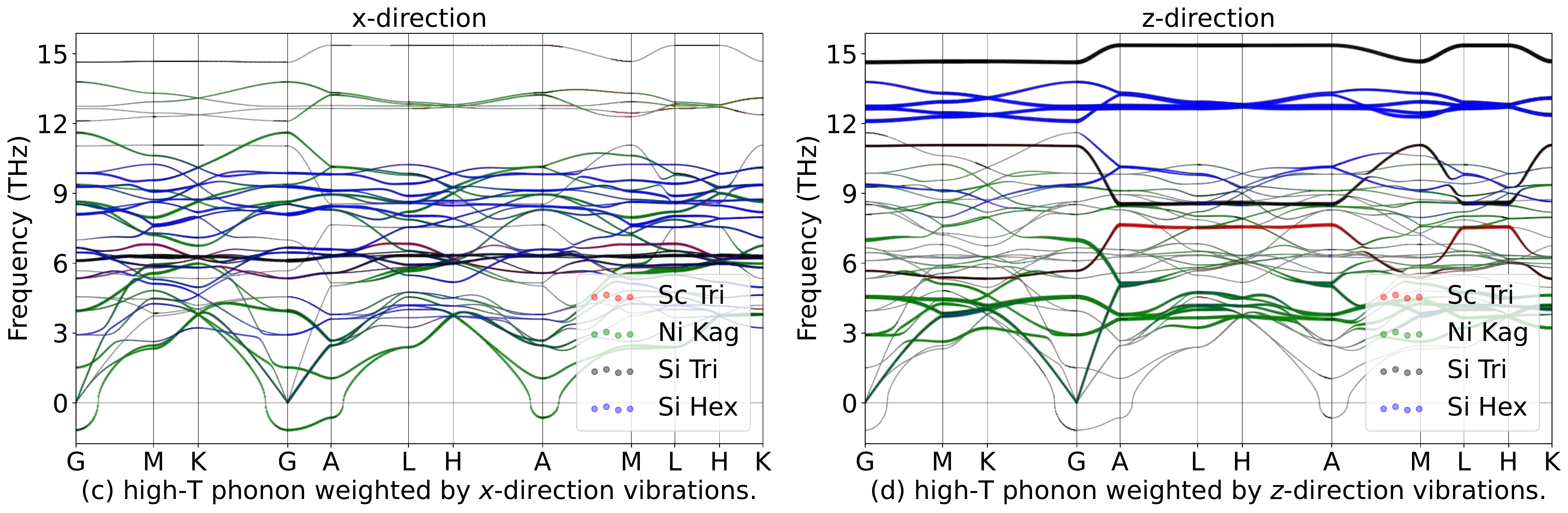}
\caption{Phonon spectrum for ScNi$_6$Si$_6$in in-plane ($x$) and out-of-plane ($z$) directions.}
\label{fig:ScNi6Si6phoxyz}
\end{figure}

\begin{table}[htbp]
\global\long\def\arraystretch{1.12}
\captionsetup{width=.95\textwidth}
\caption{IRREPs at high-symmetry points for ScNi$_6$Si$_6$ phonon spectrum at Low-T.}
\label{tab:191_ScNi6Si6_Low}
\setlength{\tabcolsep}{1.mm}{
}
\end{table}

\begin{table}[htbp]
\global\long\def\arraystretch{1.12}
\captionsetup{width=.95\textwidth}
\caption{IRREPs at high-symmetry points for ScNi$_6$Si$_6$ phonon spectrum at High-T.}
\label{tab:191_ScNi6Si6_High}
\setlength{\tabcolsep}{1.mm}{
%
}
\end{table}
\subsubsection{\label{sms2sec:TiNi6Si6}\hyperref[smsec:col]{\texorpdfstring{TiNi$_6$Si$_6$}{}}~{\color{red}  (High-T unstable, Low-T unstable) }}

\begin{table}[htbp]
\global\long\def\arraystretch{1.12}
\captionsetup{width=.85\textwidth}
\caption{Structure parameters of TiNi$_6$Si$_6$ after relaxation. It contains 274 electrons. }
\label{tab:TiNi6Si6}
\setlength{\tabcolsep}{4mm}{
%
}
\end{table}

\begin{figure}[htbp]
\centering
\includegraphics[width=0.85\textwidth]{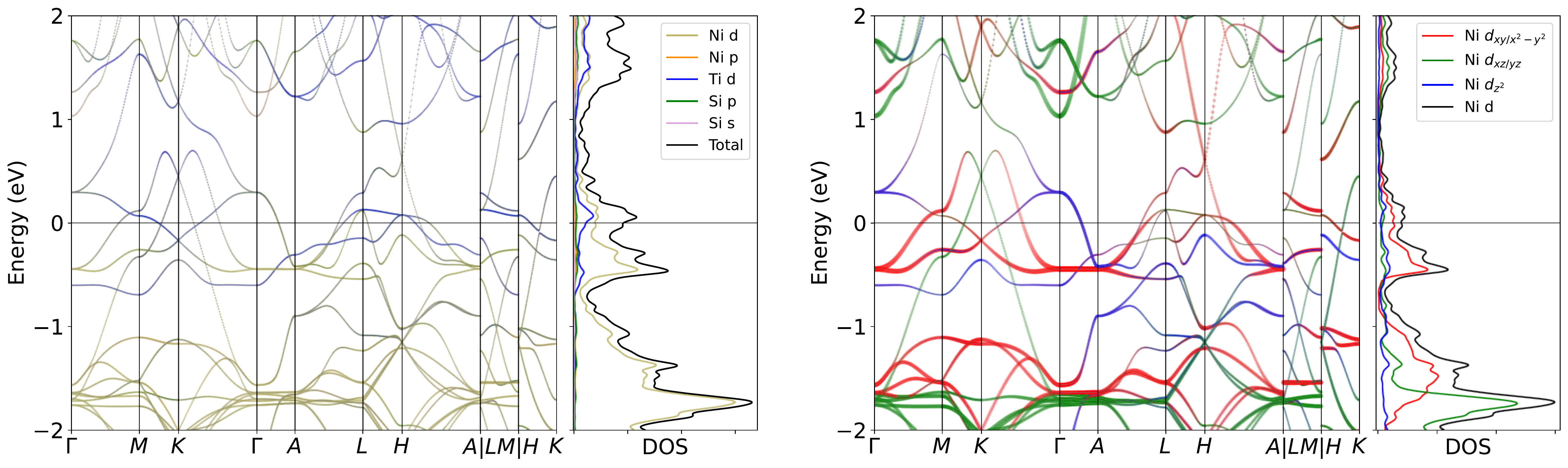}
\caption{Band structure for TiNi$_6$Si$_6$ without SOC. The projection of Ni $3d$ orbitals is also presented. The band structures clearly reveal the dominating of Ni $3d$ orbitals  near the Fermi level, as well as the pronounced peak.}
\label{fig:TiNi6Si6band}
\end{figure}

\begin{figure}[H]
\centering
\subfloat[Relaxed phonon at low-T.]{
\label{fig:TiNi6Si6_relaxed_Low}
\includegraphics[width=0.45\textwidth]{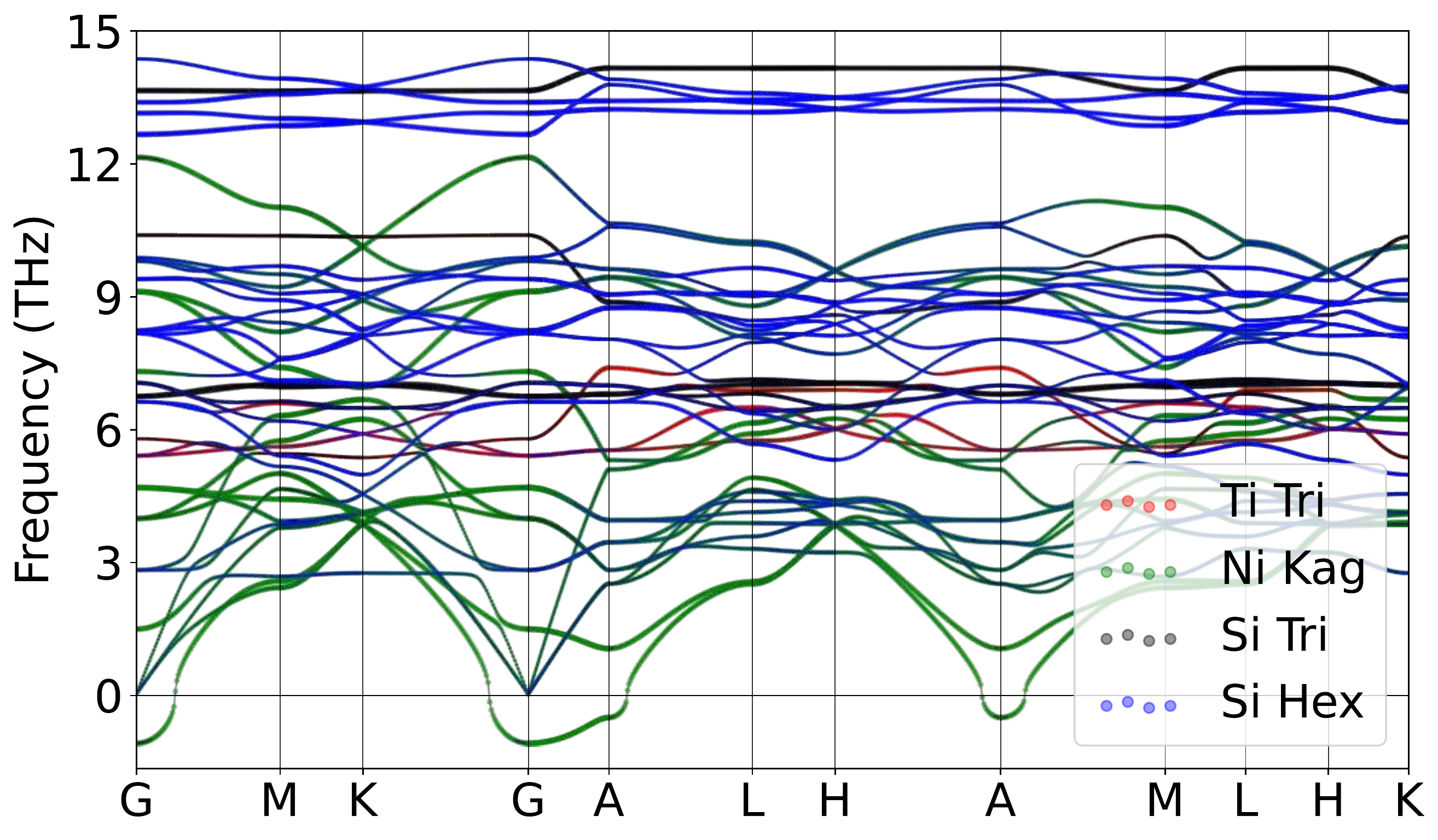}
}
\hspace{5pt}\subfloat[Relaxed phonon at high-T.]{
\label{fig:TiNi6Si6_relaxed_High}
\includegraphics[width=0.45\textwidth]{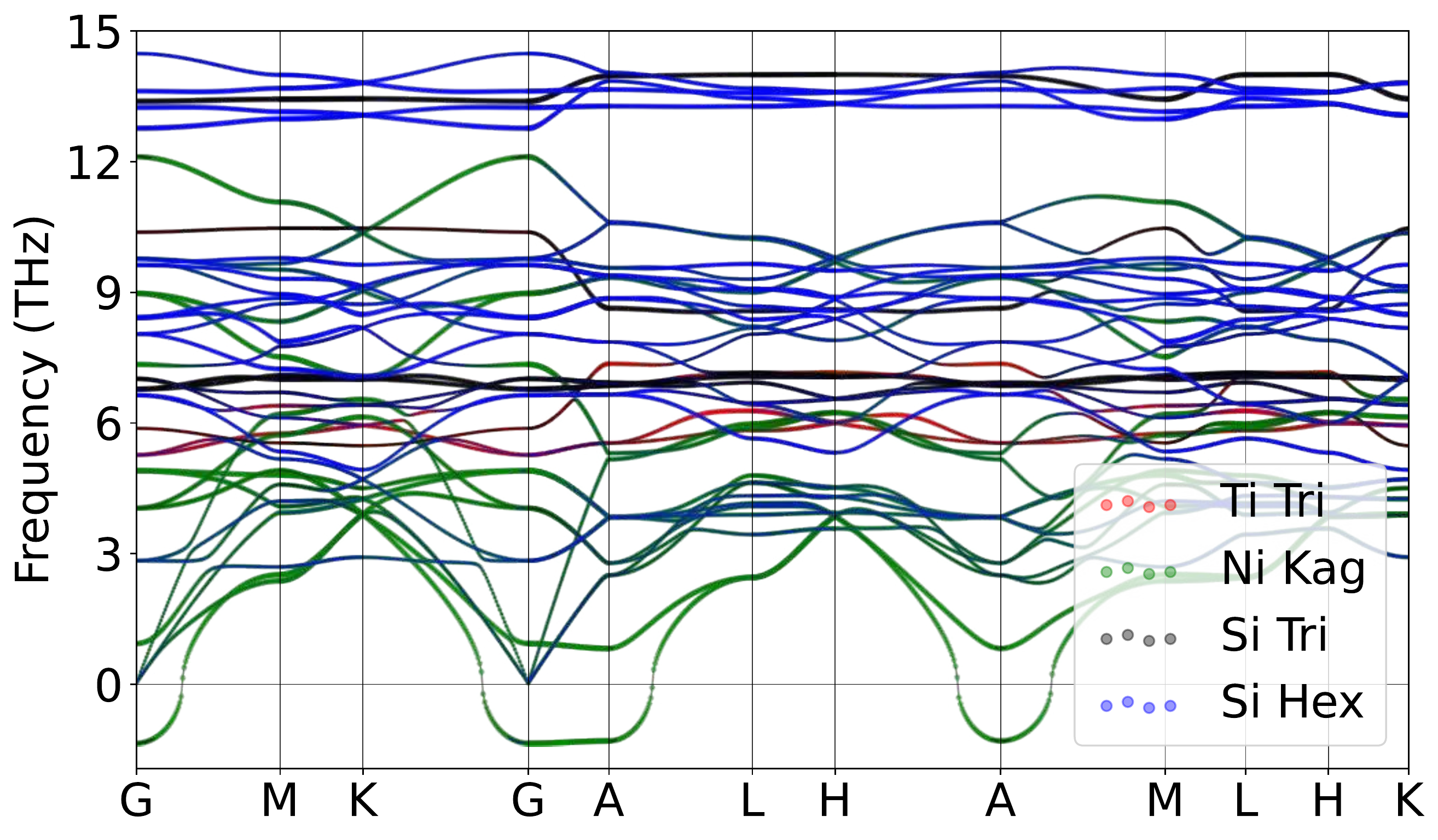}
}
\caption{Atomically projected phonon spectrum for TiNi$_6$Si$_6$.}
\label{fig:TiNi6Si6pho}
\end{figure}

\begin{figure}[H]
\centering
\label{fig:TiNi6Si6_relaxed_Low_xz}
\includegraphics[width=0.85\textwidth]{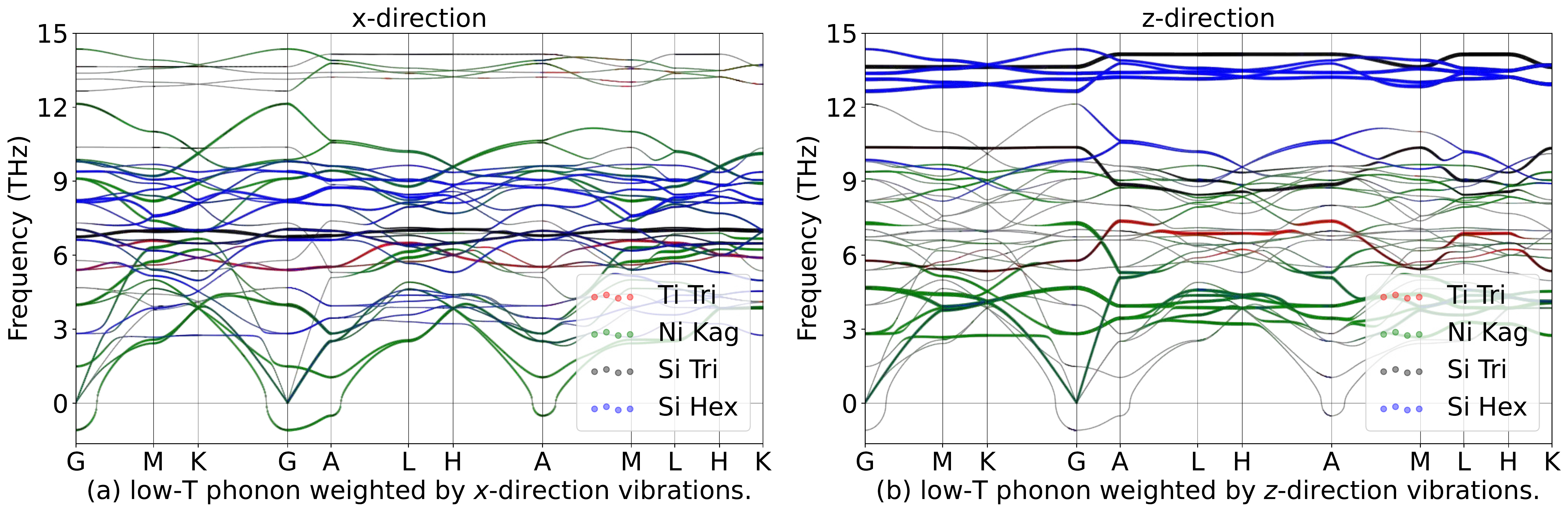}
\hspace{5pt}\label{fig:TiNi6Si6_relaxed_High_xz}
\includegraphics[width=0.85\textwidth]{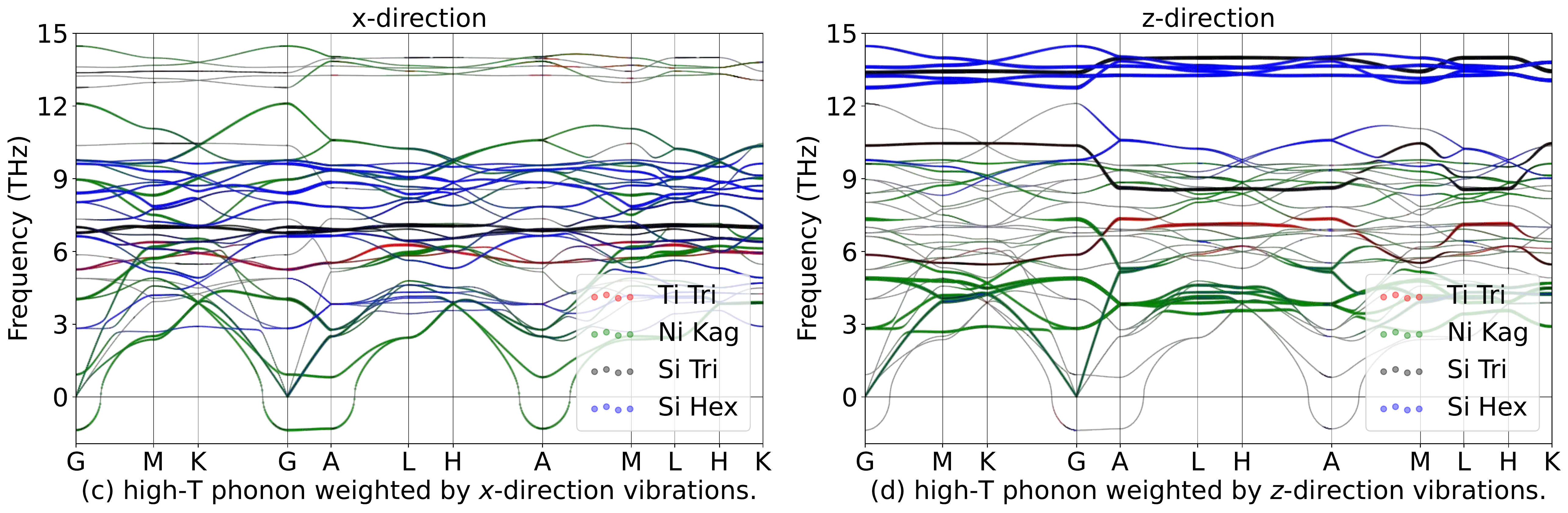}
\caption{Phonon spectrum for TiNi$_6$Si$_6$in in-plane ($x$) and out-of-plane ($z$) directions.}
\label{fig:TiNi6Si6phoxyz}
\end{figure}

\begin{table}[htbp]
\global\long\def\arraystretch{1.12}
\captionsetup{width=.95\textwidth}
\caption{IRREPs at high-symmetry points for TiNi$_6$Si$_6$ phonon spectrum at Low-T.}
\label{tab:191_TiNi6Si6_Low}
\setlength{\tabcolsep}{1.mm}{
}
\end{table}

\begin{table}[htbp]
\global\long\def\arraystretch{1.12}
\captionsetup{width=.95\textwidth}
\caption{IRREPs at high-symmetry points for TiNi$_6$Si$_6$ phonon spectrum at High-T.}
\label{tab:191_TiNi6Si6_High}
\setlength{\tabcolsep}{1.mm}{
%
}
\end{table}
\subsubsection{\label{sms2sec:YNi6Si6}\hyperref[smsec:col]{\texorpdfstring{YNi$_6$Si$_6$}{}}~{\color{green}  (High-T stable, Low-T stable) }}

\begin{table}[htbp]
\global\long\def\arraystretch{1.12}
\captionsetup{width=.85\textwidth}
\caption{Structure parameters of YNi$_6$Si$_6$ after relaxation. It contains 291 electrons. }
\label{tab:YNi6Si6}
\setlength{\tabcolsep}{4mm}{
%
}
\end{table}

\begin{figure}[htbp]
\centering
\includegraphics[width=0.85\textwidth]{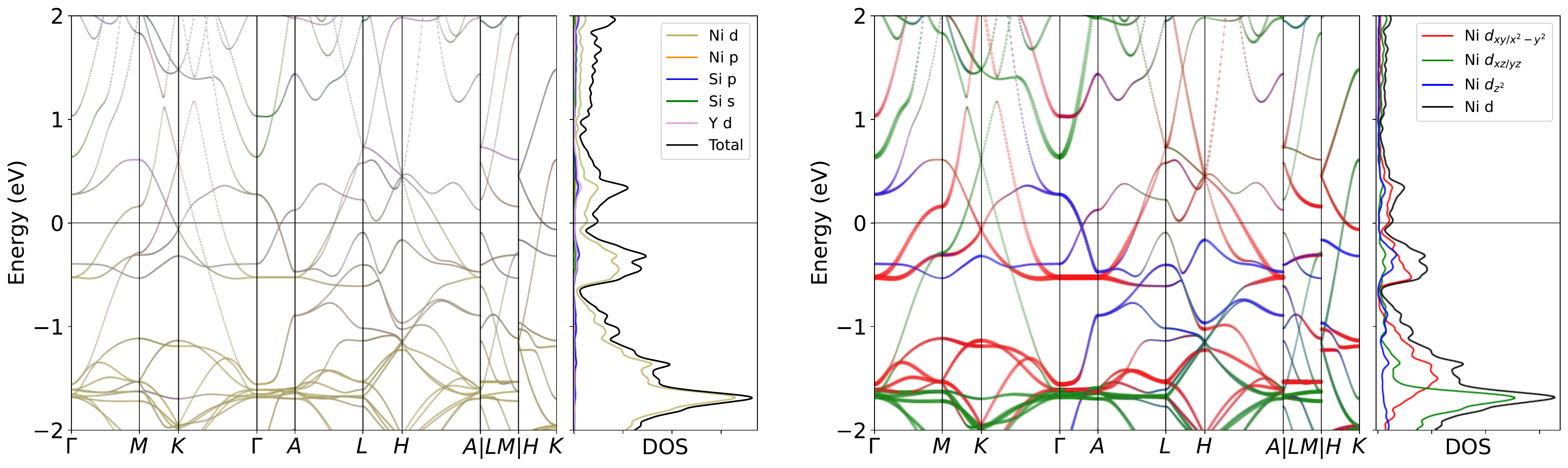}
\caption{Band structure for YNi$_6$Si$_6$ without SOC. The projection of Ni $3d$ orbitals is also presented. The band structures clearly reveal the dominating of Ni $3d$ orbitals  near the Fermi level, as well as the pronounced peak.}
\label{fig:YNi6Si6band}
\end{figure}

\begin{figure}[H]
\centering
\subfloat[Relaxed phonon at low-T.]{
\label{fig:YNi6Si6_relaxed_Low}
\includegraphics[width=0.45\textwidth]{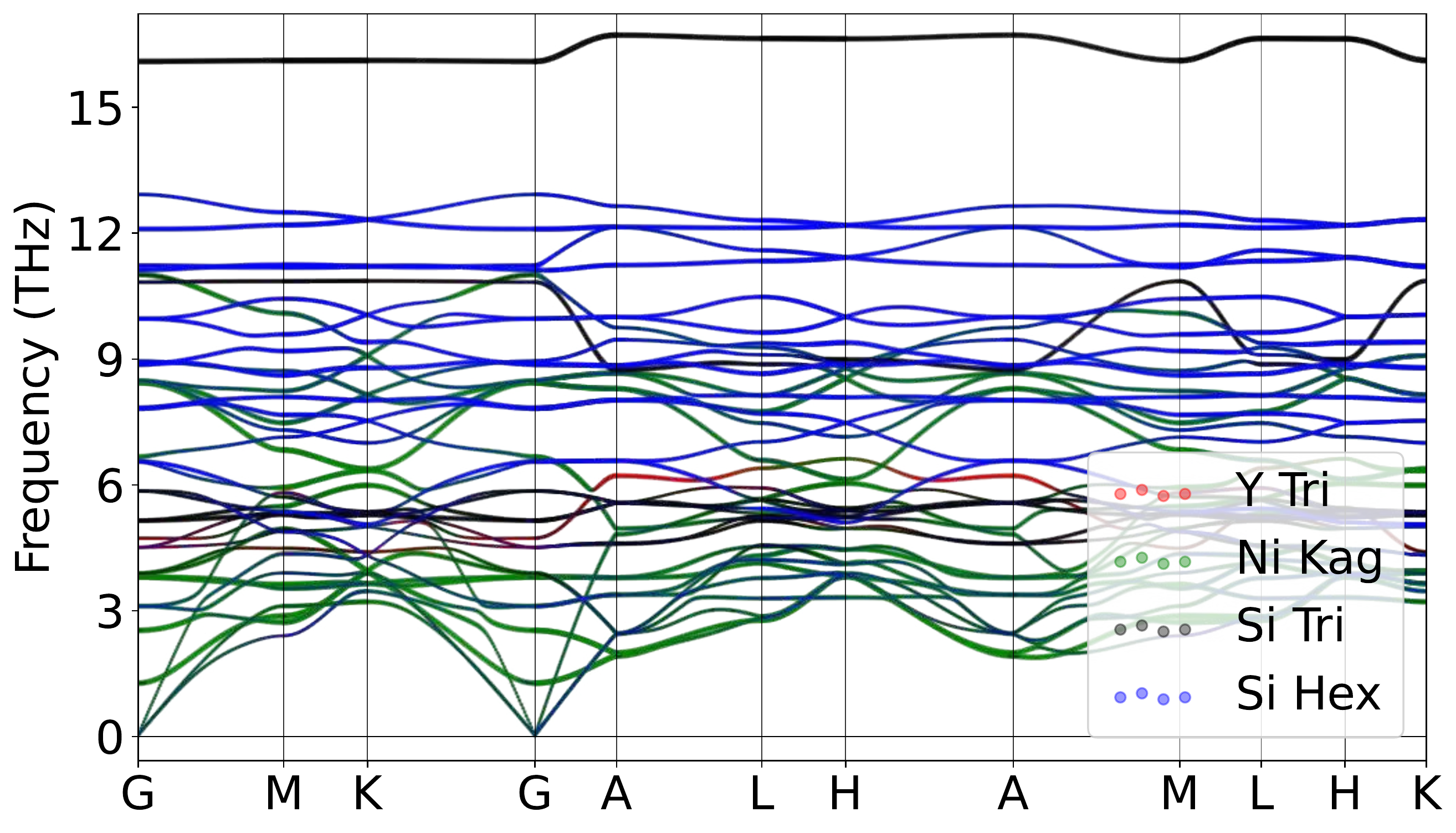}
}
\hspace{5pt}\subfloat[Relaxed phonon at high-T.]{
\label{fig:YNi6Si6_relaxed_High}
\includegraphics[width=0.45\textwidth]{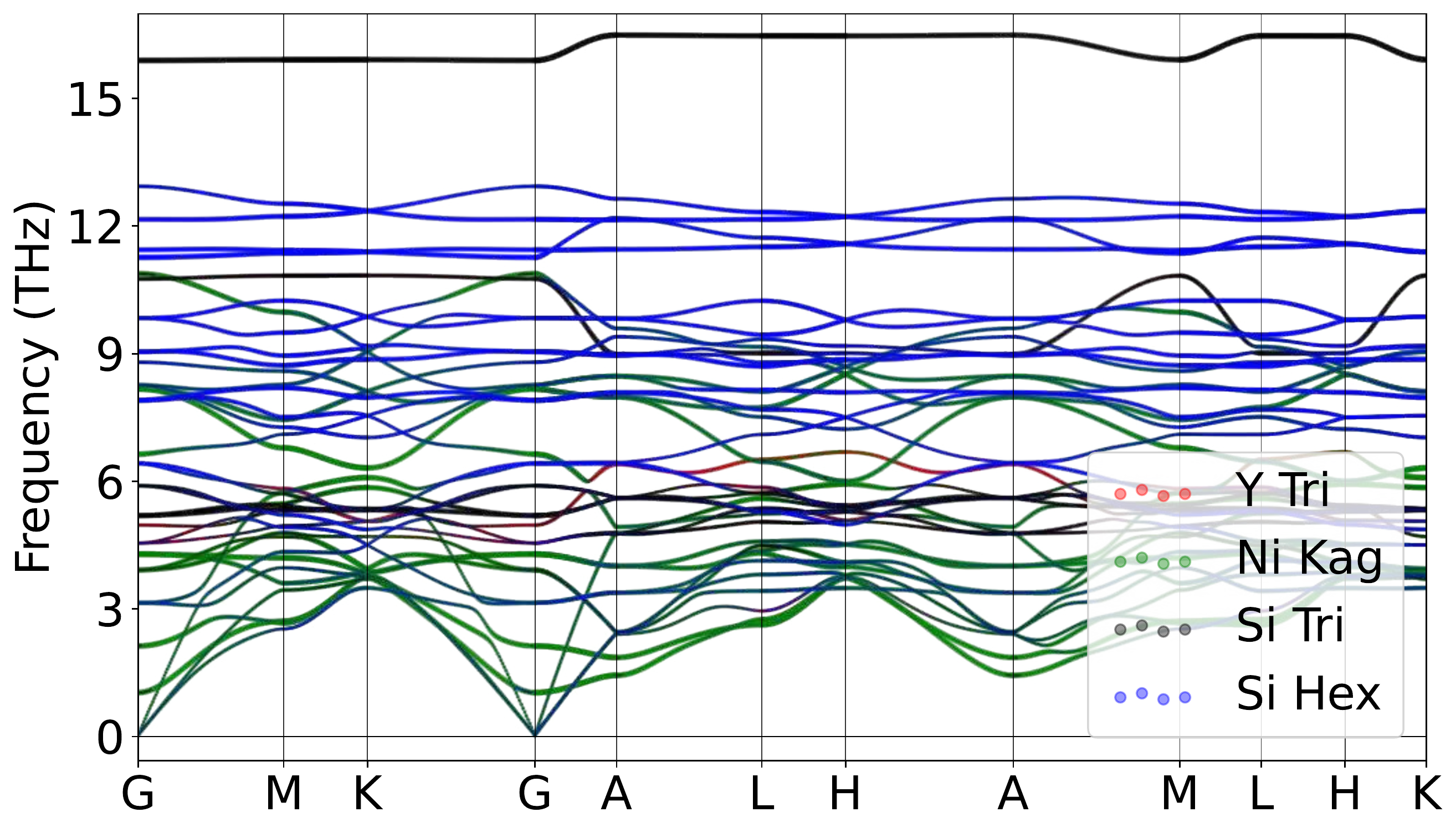}
}
\caption{Atomically projected phonon spectrum for YNi$_6$Si$_6$.}
\label{fig:YNi6Si6pho}
\end{figure}

\begin{figure}[H]
\centering
\label{fig:YNi6Si6_relaxed_Low_xz}
\includegraphics[width=0.85\textwidth]{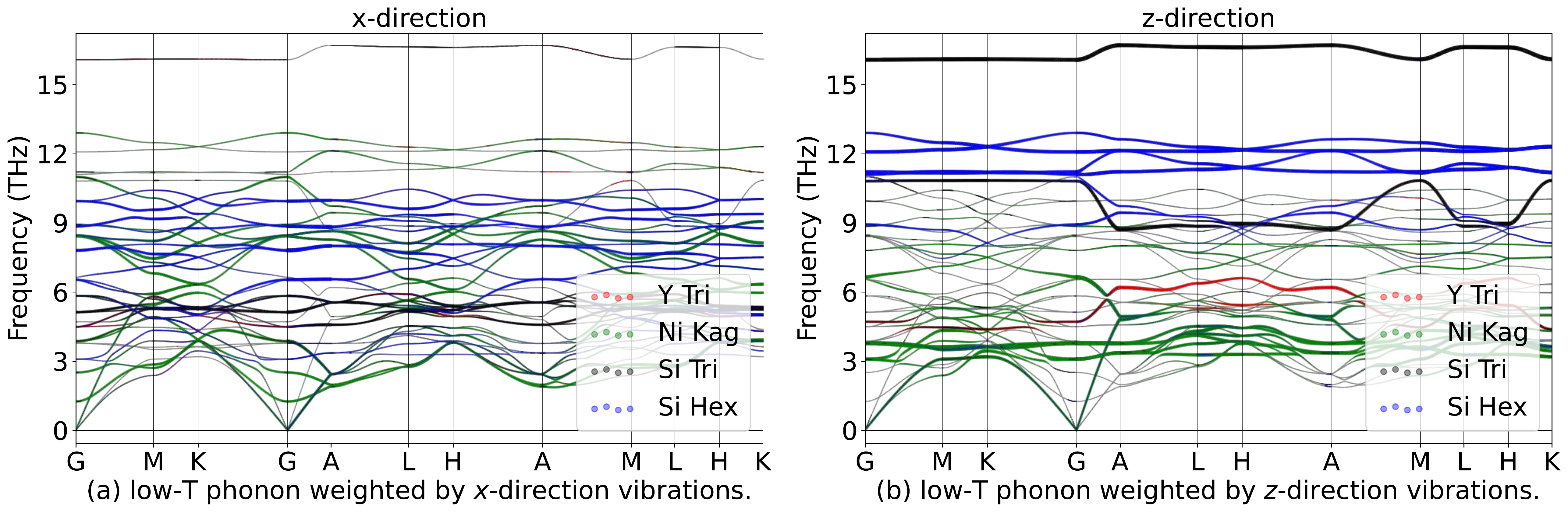}
\hspace{5pt}\label{fig:YNi6Si6_relaxed_High_xz}
\includegraphics[width=0.85\textwidth]{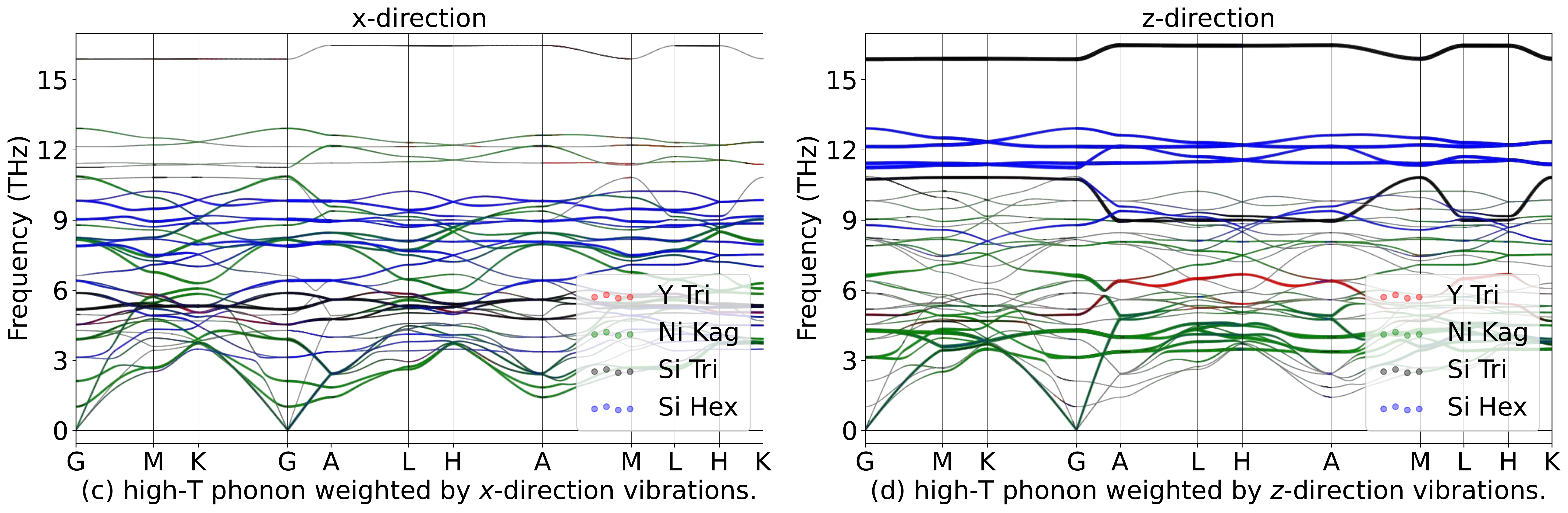}
\caption{Phonon spectrum for YNi$_6$Si$_6$in in-plane ($x$) and out-of-plane ($z$) directions.}
\label{fig:YNi6Si6phoxyz}
\end{figure}

\begin{table}[htbp]
\global\long\def\arraystretch{1.12}
\captionsetup{width=.95\textwidth}
\caption{IRREPs at high-symmetry points for YNi$_6$Si$_6$ phonon spectrum at Low-T.}
\label{tab:191_YNi6Si6_Low}
\setlength{\tabcolsep}{1.mm}{
}
\end{table}

\begin{table}[htbp]
\global\long\def\arraystretch{1.12}
\captionsetup{width=.95\textwidth}
\caption{IRREPs at high-symmetry points for YNi$_6$Si$_6$ phonon spectrum at High-T.}
\label{tab:191_YNi6Si6_High}
\setlength{\tabcolsep}{1.mm}{
%
}
\end{table}
\subsubsection{\label{sms2sec:ZrNi6Si6}\hyperref[smsec:col]{\texorpdfstring{ZrNi$_6$Si$_6$}{}}~{\color{green}  (High-T stable, Low-T stable) }}

\begin{table}[htbp]
\global\long\def\arraystretch{1.12}
\captionsetup{width=.85\textwidth}
\caption{Structure parameters of ZrNi$_6$Si$_6$ after relaxation. It contains 292 electrons. }
\label{tab:ZrNi6Si6}
\setlength{\tabcolsep}{4mm}{
%
}
\end{table}

\begin{figure}[htbp]
\centering
\includegraphics[width=0.85\textwidth]{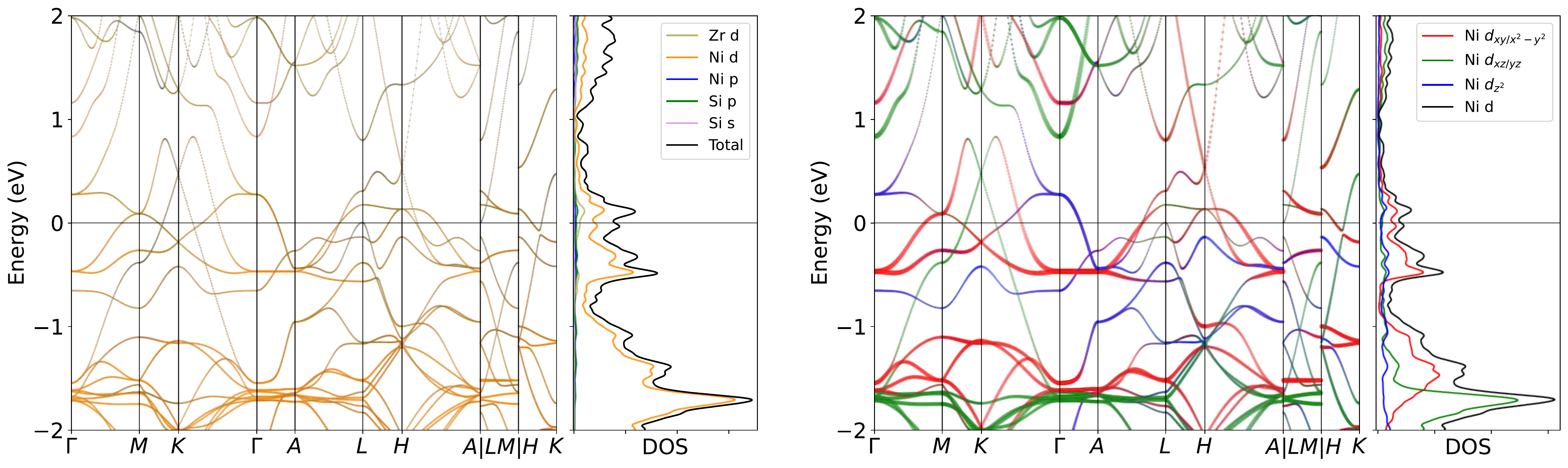}
\caption{Band structure for ZrNi$_6$Si$_6$ without SOC. The projection of Ni $3d$ orbitals is also presented. The band structures clearly reveal the dominating of Ni $3d$ orbitals  near the Fermi level, as well as the pronounced peak.}
\label{fig:ZrNi6Si6band}
\end{figure}

\begin{figure}[H]
\centering
\subfloat[Relaxed phonon at low-T.]{
\label{fig:ZrNi6Si6_relaxed_Low}
\includegraphics[width=0.45\textwidth]{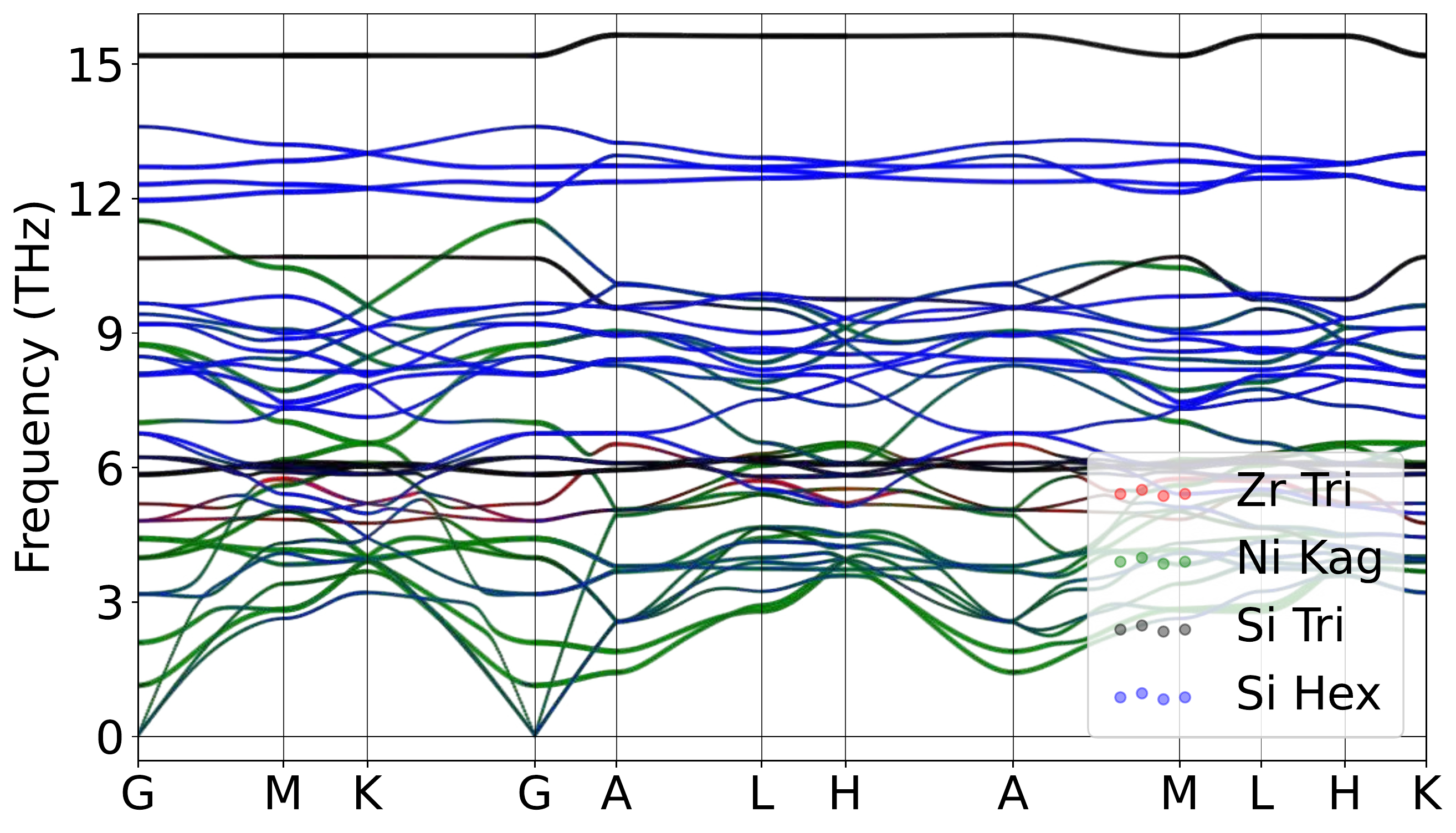}
}
\hspace{5pt}\subfloat[Relaxed phonon at high-T.]{
\label{fig:ZrNi6Si6_relaxed_High}
\includegraphics[width=0.45\textwidth]{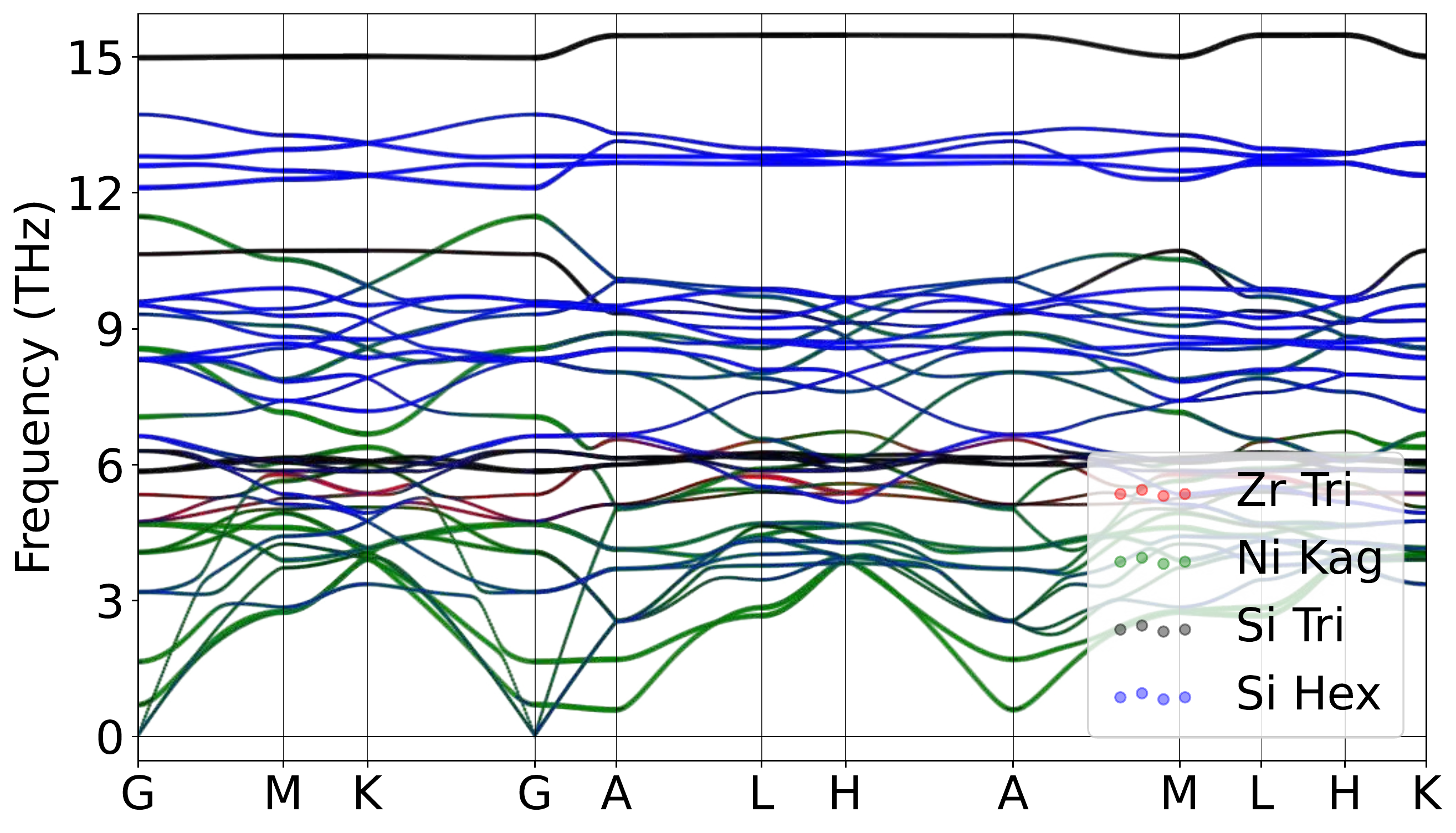}
}
\caption{Atomically projected phonon spectrum for ZrNi$_6$Si$_6$.}
\label{fig:ZrNi6Si6pho}
\end{figure}

\begin{figure}[H]
\centering
\label{fig:ZrNi6Si6_relaxed_Low_xz}
\includegraphics[width=0.85\textwidth]{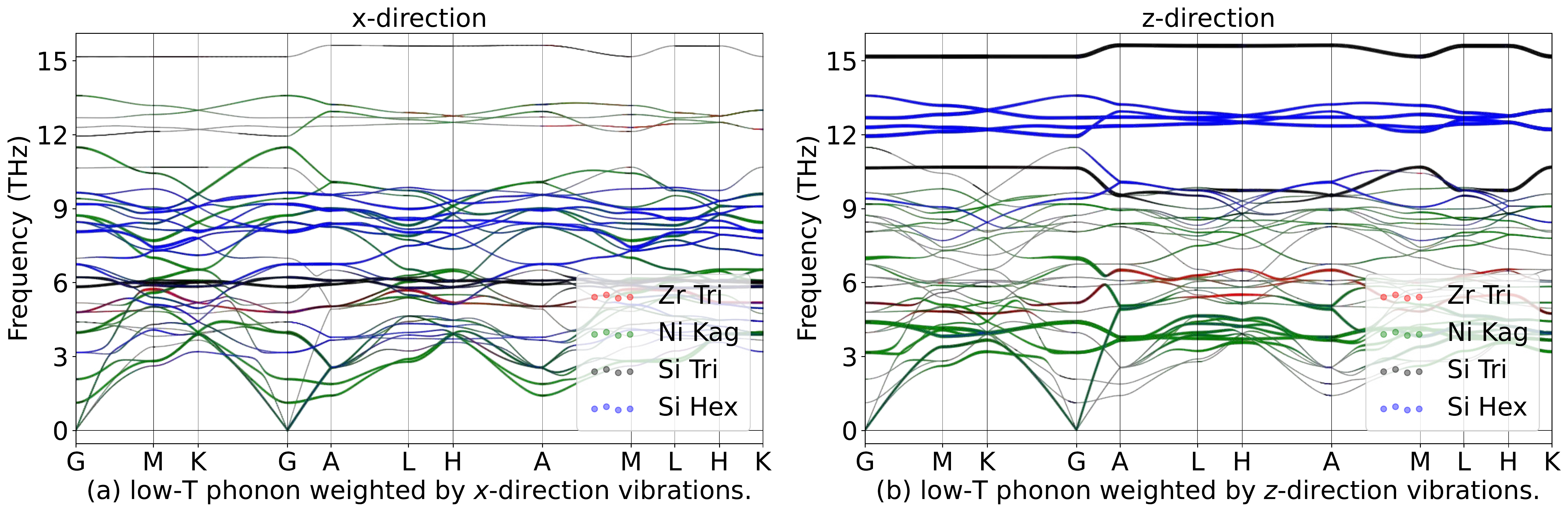}
\hspace{5pt}\label{fig:ZrNi6Si6_relaxed_High_xz}
\includegraphics[width=0.85\textwidth]{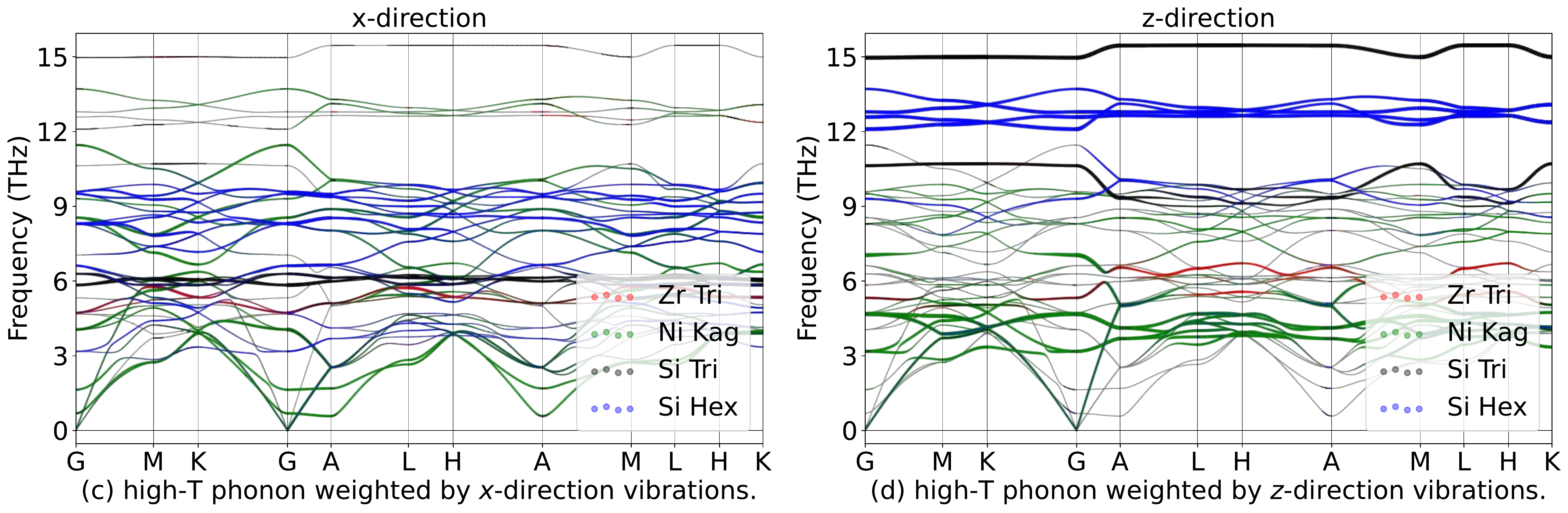}
\caption{Phonon spectrum for ZrNi$_6$Si$_6$in in-plane ($x$) and out-of-plane ($z$) directions.}
\label{fig:ZrNi6Si6phoxyz}
\end{figure}

\begin{table}[htbp]
\global\long\def\arraystretch{1.12}
\captionsetup{width=.95\textwidth}
\caption{IRREPs at high-symmetry points for ZrNi$_6$Si$_6$ phonon spectrum at Low-T.}
\label{tab:191_ZrNi6Si6_Low}
\setlength{\tabcolsep}{1.mm}{
}
\end{table}

\begin{table}[htbp]
\global\long\def\arraystretch{1.12}
\captionsetup{width=.95\textwidth}
\caption{IRREPs at high-symmetry points for ZrNi$_6$Si$_6$ phonon spectrum at High-T.}
\label{tab:191_ZrNi6Si6_High}
\setlength{\tabcolsep}{1.mm}{
%
}
\end{table}
\subsubsection{\label{sms2sec:NbNi6Si6}\hyperref[smsec:col]{\texorpdfstring{NbNi$_6$Si$_6$}{}}~{\color{red}  (High-T unstable, Low-T unstable) }}

\begin{table}[htbp]
\global\long\def\arraystretch{1.12}
\captionsetup{width=.85\textwidth}
\caption{Structure parameters of NbNi$_6$Si$_6$ after relaxation. It contains 293 electrons. }
\label{tab:NbNi6Si6}
\setlength{\tabcolsep}{4mm}{
%
}
\end{table}

\begin{figure}[htbp]
\centering
\includegraphics[width=0.85\textwidth]{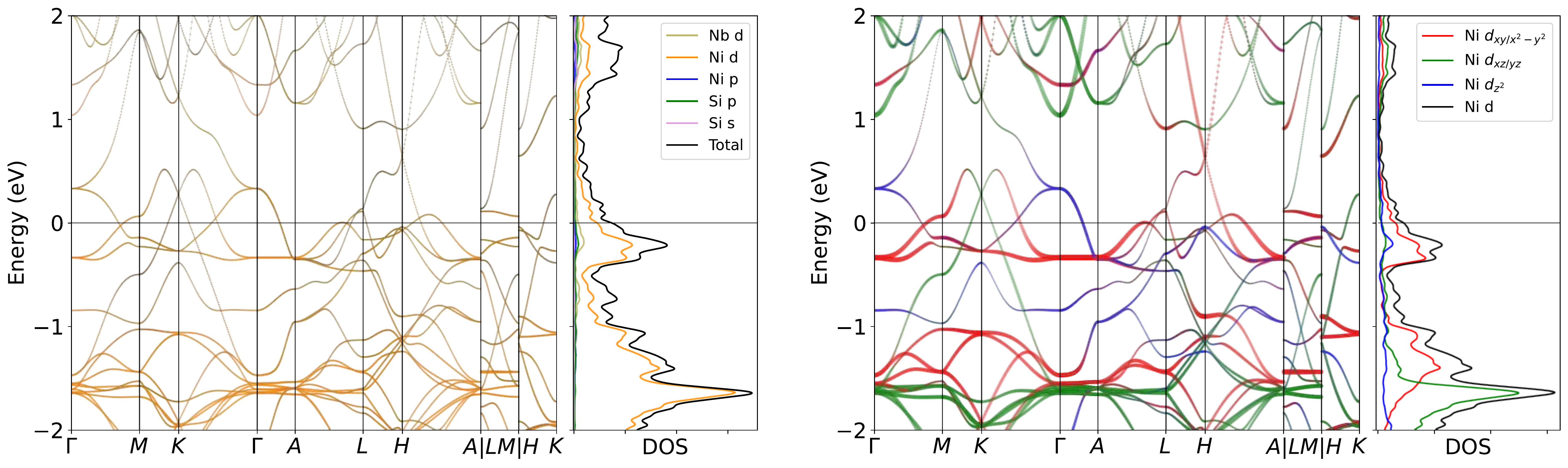}
\caption{Band structure for NbNi$_6$Si$_6$ without SOC. The projection of Ni $3d$ orbitals is also presented. The band structures clearly reveal the dominating of Ni $3d$ orbitals  near the Fermi level, as well as the pronounced peak.}
\label{fig:NbNi6Si6band}
\end{figure}

\begin{figure}[H]
\centering
\subfloat[Relaxed phonon at low-T.]{
\label{fig:NbNi6Si6_relaxed_Low}
\includegraphics[width=0.45\textwidth]{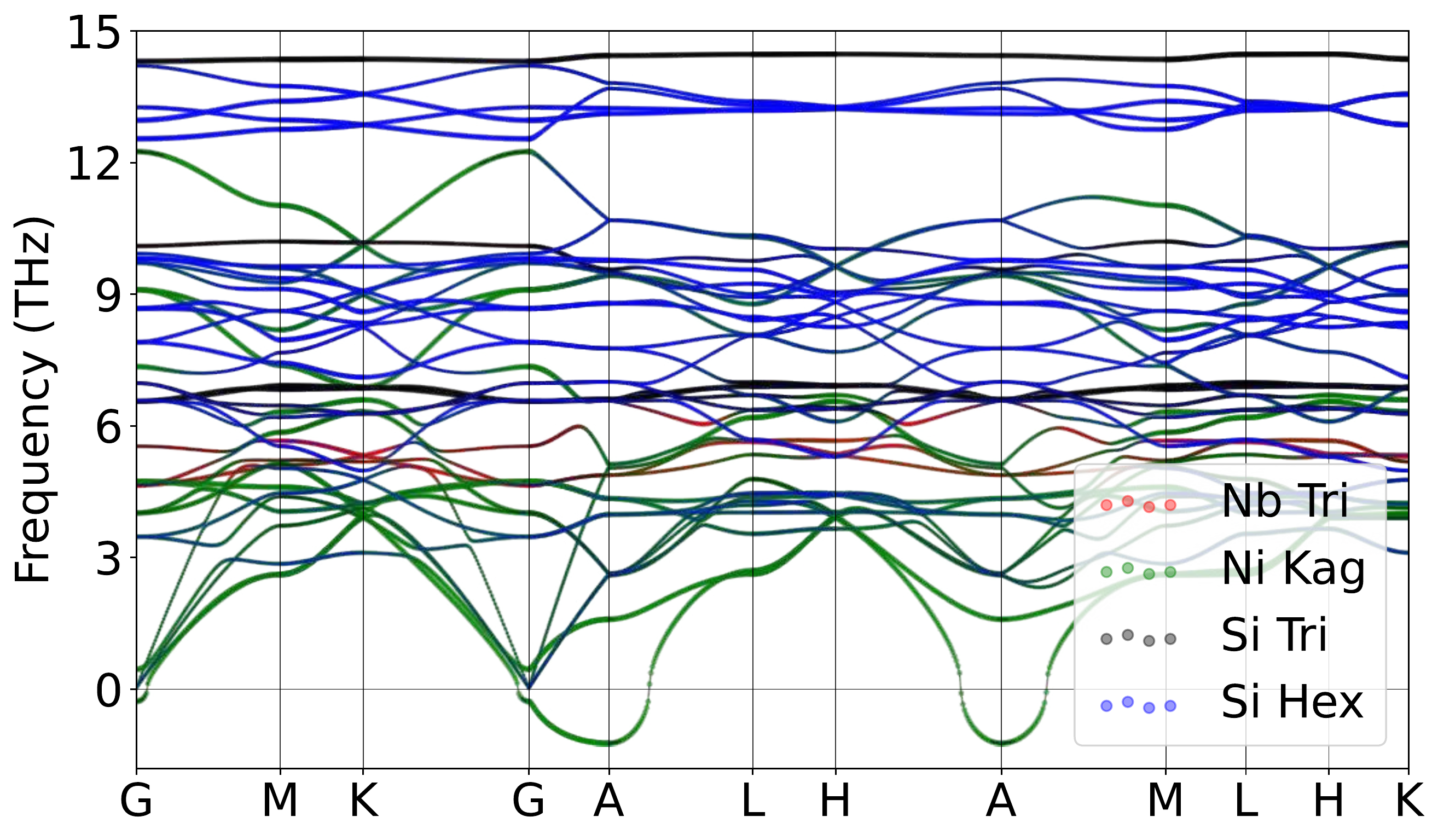}
}
\hspace{5pt}\subfloat[Relaxed phonon at high-T.]{
\label{fig:NbNi6Si6_relaxed_High}
\includegraphics[width=0.45\textwidth]{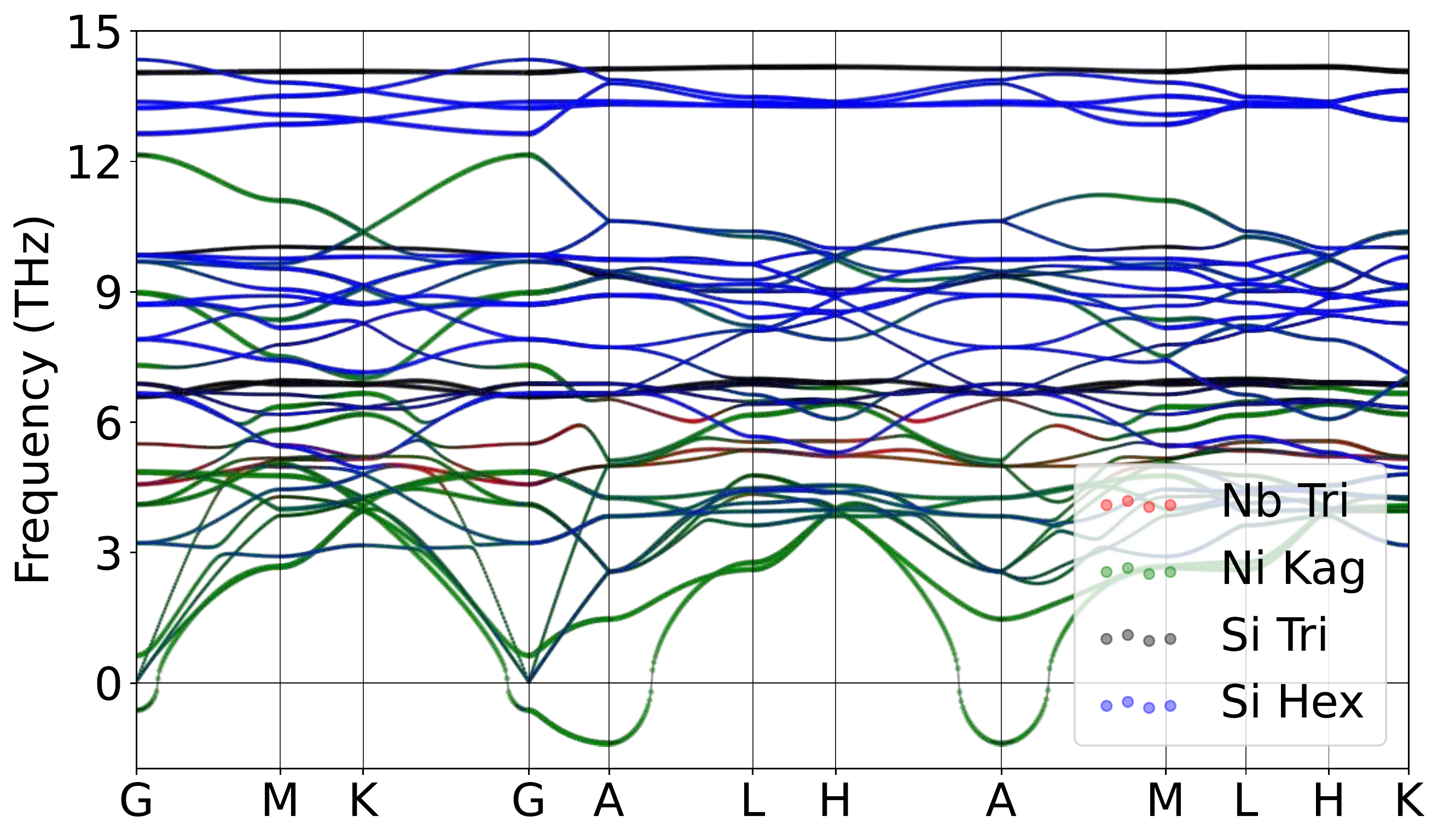}
}
\caption{Atomically projected phonon spectrum for NbNi$_6$Si$_6$.}
\label{fig:NbNi6Si6pho}
\end{figure}

\begin{figure}[H]
\centering
\label{fig:NbNi6Si6_relaxed_Low_xz}
\includegraphics[width=0.85\textwidth]{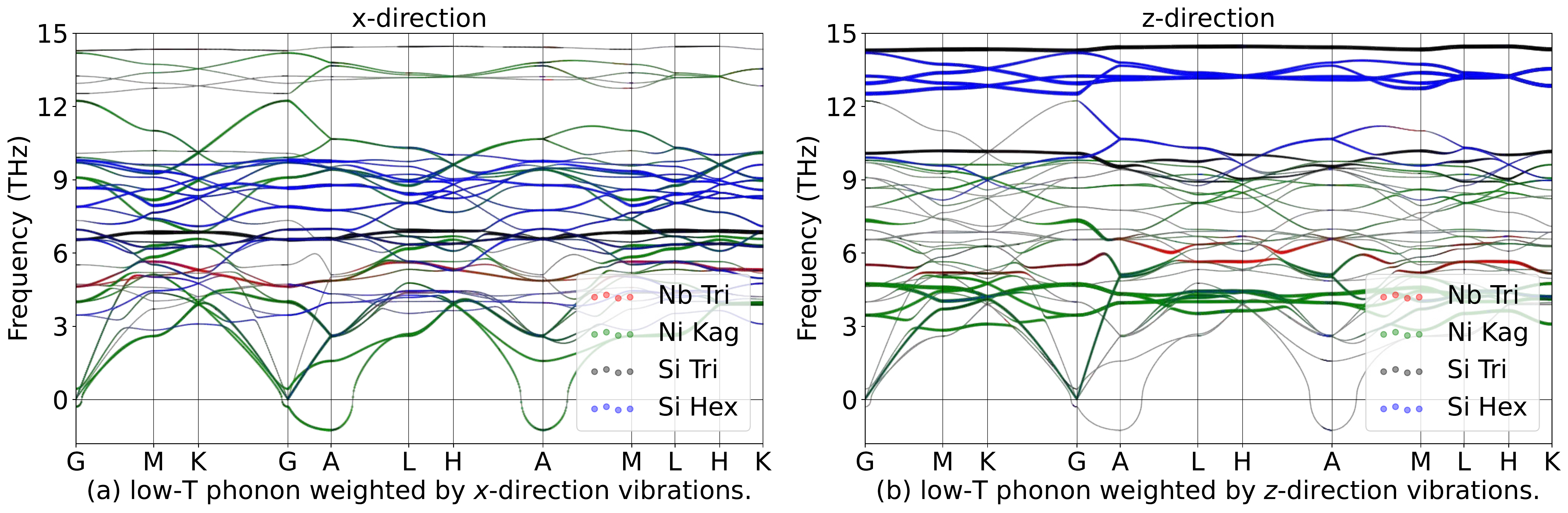}
\hspace{5pt}\label{fig:NbNi6Si6_relaxed_High_xz}
\includegraphics[width=0.85\textwidth]{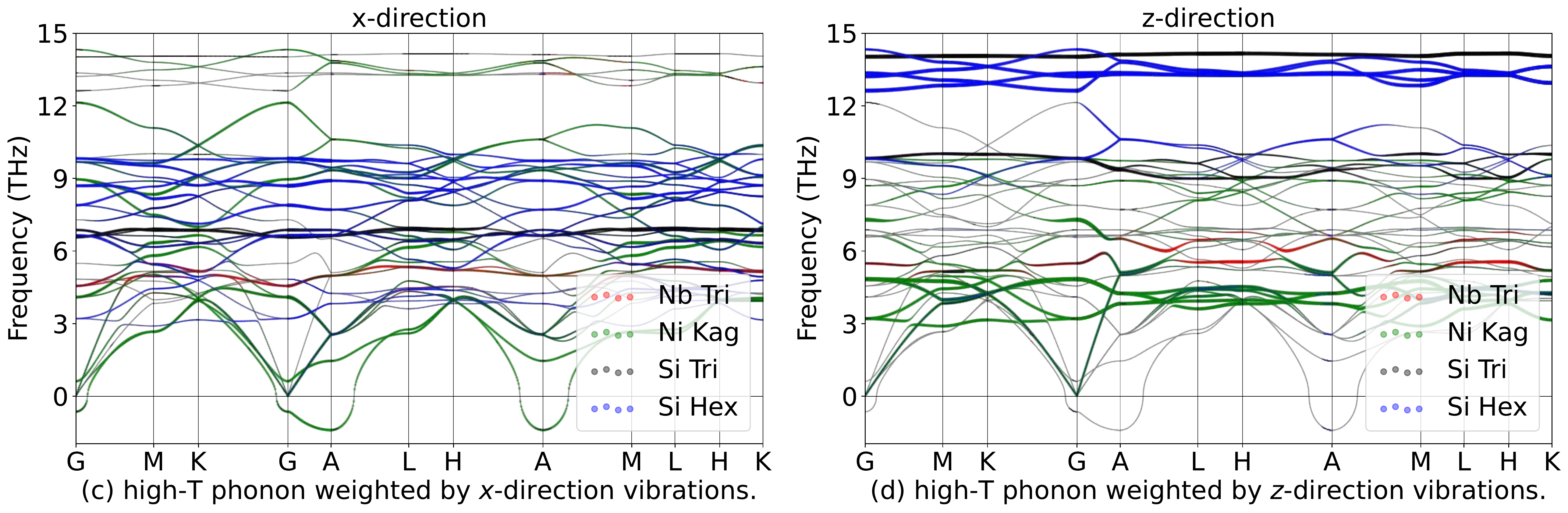}
\caption{Phonon spectrum for NbNi$_6$Si$_6$in in-plane ($x$) and out-of-plane ($z$) directions.}
\label{fig:NbNi6Si6phoxyz}
\end{figure}

\begin{table}[htbp]
\global\long\def\arraystretch{1.12}
\captionsetup{width=.95\textwidth}
\caption{IRREPs at high-symmetry points for NbNi$_6$Si$_6$ phonon spectrum at Low-T.}
\label{tab:191_NbNi6Si6_Low}
\setlength{\tabcolsep}{1.mm}{
}
\end{table}

\begin{table}[htbp]
\global\long\def\arraystretch{1.12}
\captionsetup{width=.95\textwidth}
\caption{IRREPs at high-symmetry points for NbNi$_6$Si$_6$ phonon spectrum at High-T.}
\label{tab:191_NbNi6Si6_High}
\setlength{\tabcolsep}{1.mm}{
%
}
\end{table}
\subsubsection{\label{sms2sec:PrNi6Si6}\hyperref[smsec:col]{\texorpdfstring{PrNi$_6$Si$_6$}{}}~{\color{green}  (High-T stable, Low-T stable) }}

\begin{table}[htbp]
\global\long\def\arraystretch{1.12}
\captionsetup{width=.85\textwidth}
\caption{Structure parameters of PrNi$_6$Si$_6$ after relaxation. It contains 311 electrons. }
\label{tab:PrNi6Si6}
\setlength{\tabcolsep}{4mm}{
%
}
\end{table}

\begin{figure}[htbp]
\centering
\includegraphics[width=0.85\textwidth]{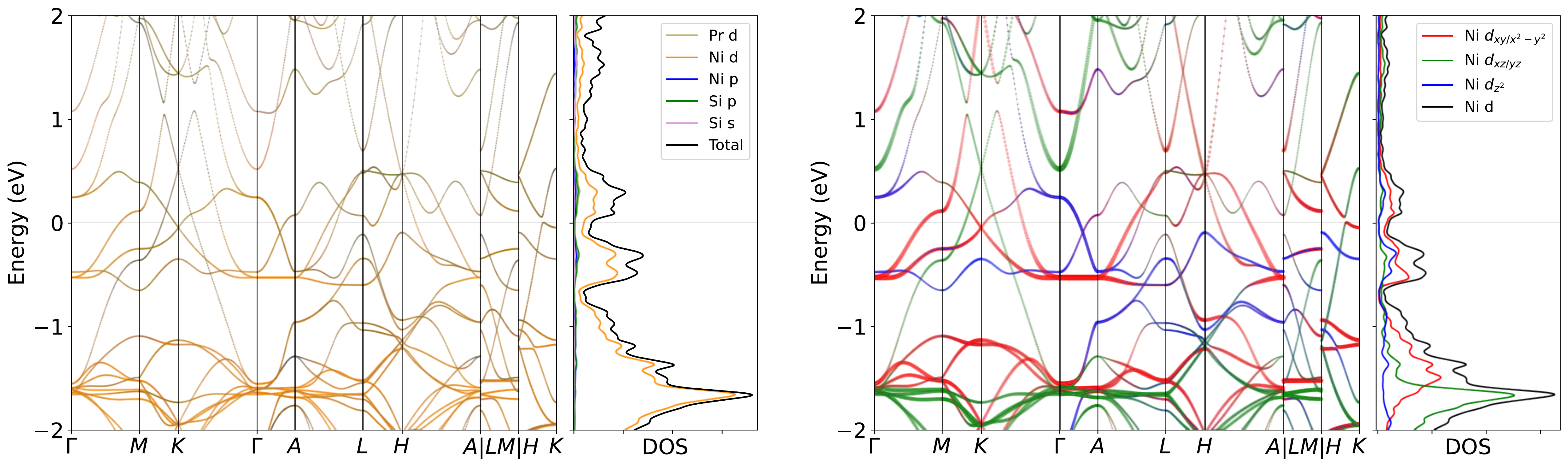}
\caption{Band structure for PrNi$_6$Si$_6$ without SOC. The projection of Ni $3d$ orbitals is also presented. The band structures clearly reveal the dominating of Ni $3d$ orbitals  near the Fermi level, as well as the pronounced peak.}
\label{fig:PrNi6Si6band}
\end{figure}

\begin{figure}[H]
\centering
\subfloat[Relaxed phonon at low-T.]{
\label{fig:PrNi6Si6_relaxed_Low}
\includegraphics[width=0.45\textwidth]{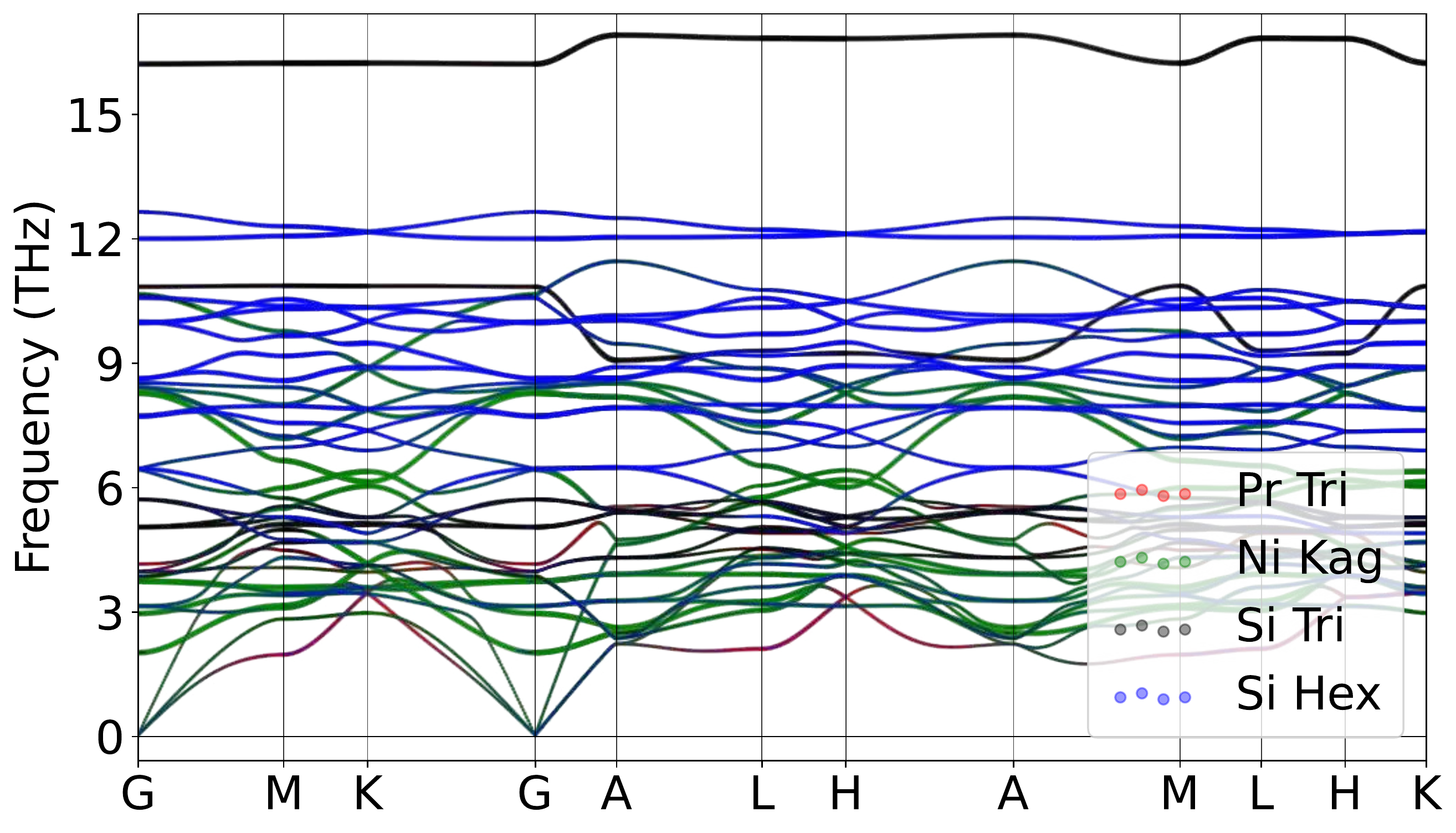}
}
\hspace{5pt}\subfloat[Relaxed phonon at high-T.]{
\label{fig:PrNi6Si6_relaxed_High}
\includegraphics[width=0.45\textwidth]{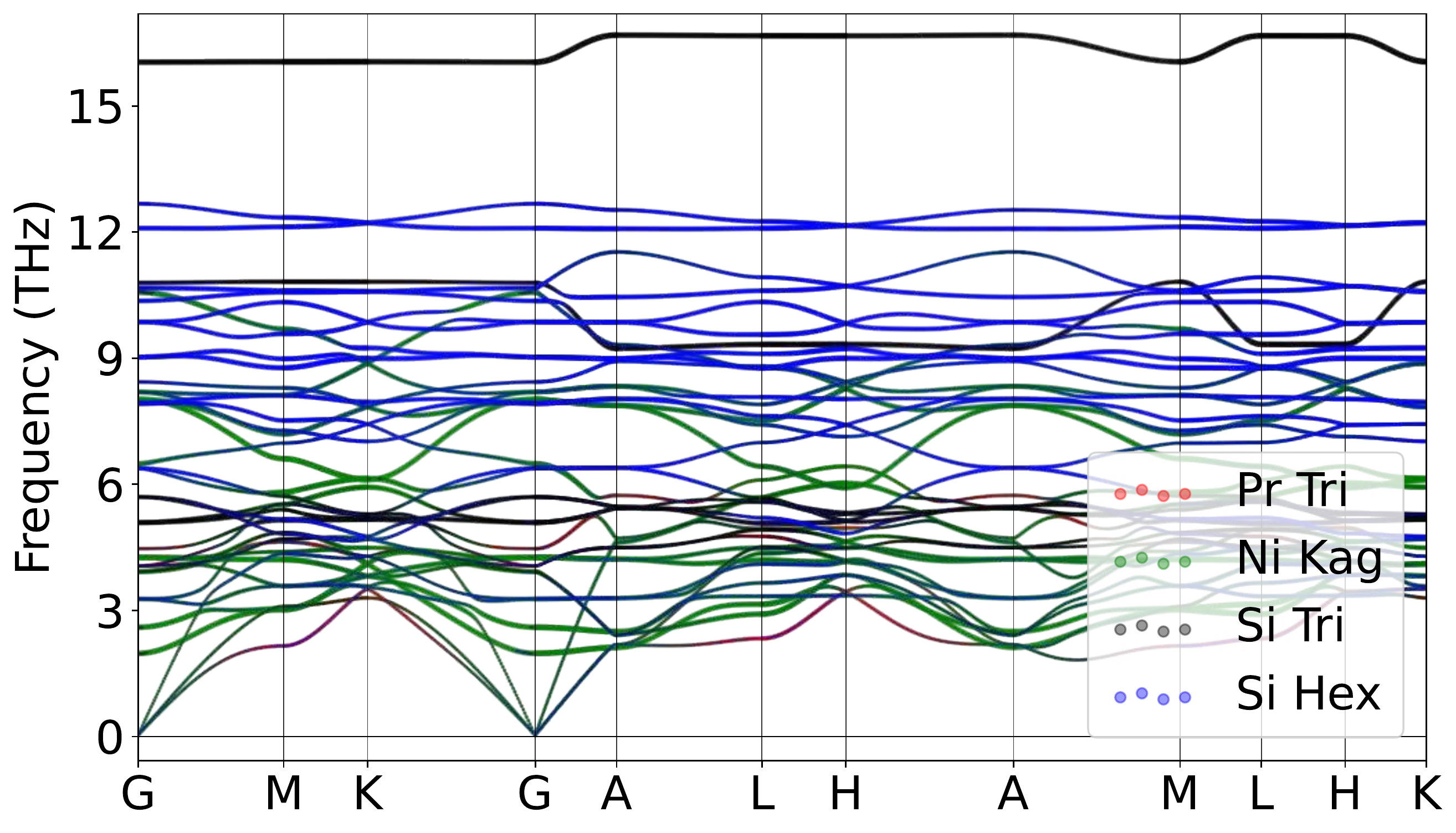}
}
\caption{Atomically projected phonon spectrum for PrNi$_6$Si$_6$.}
\label{fig:PrNi6Si6pho}
\end{figure}

\begin{figure}[H]
\centering
\label{fig:PrNi6Si6_relaxed_Low_xz}
\includegraphics[width=0.85\textwidth]{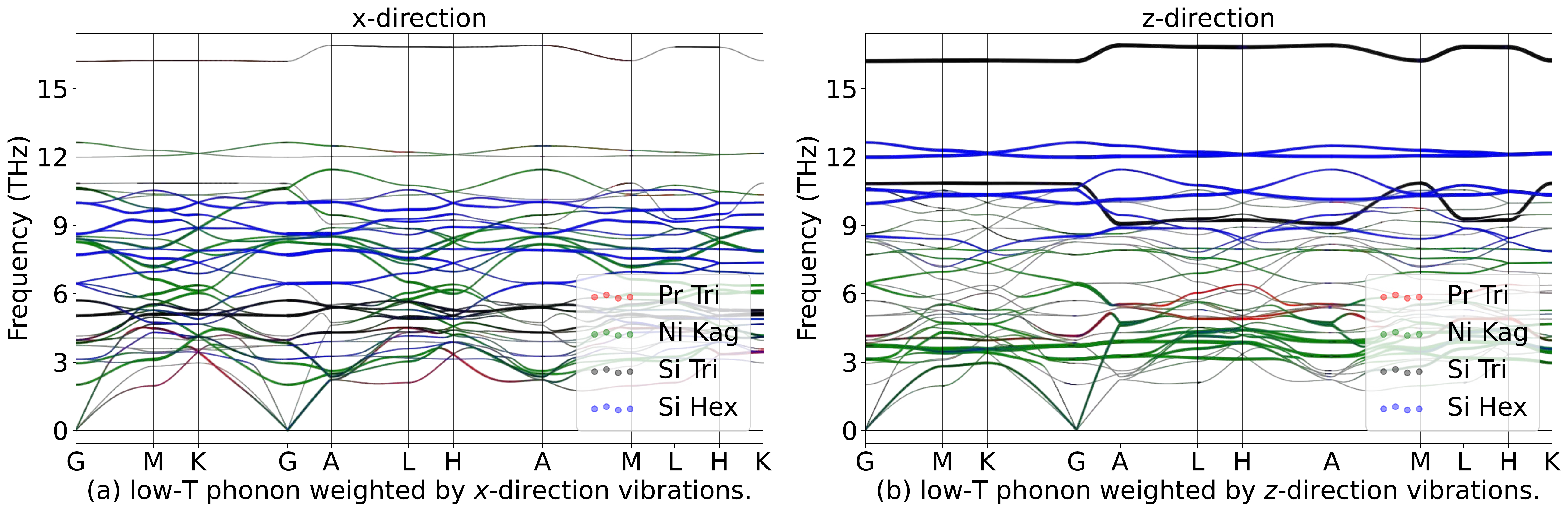}
\hspace{5pt}\label{fig:PrNi6Si6_relaxed_High_xz}
\includegraphics[width=0.85\textwidth]{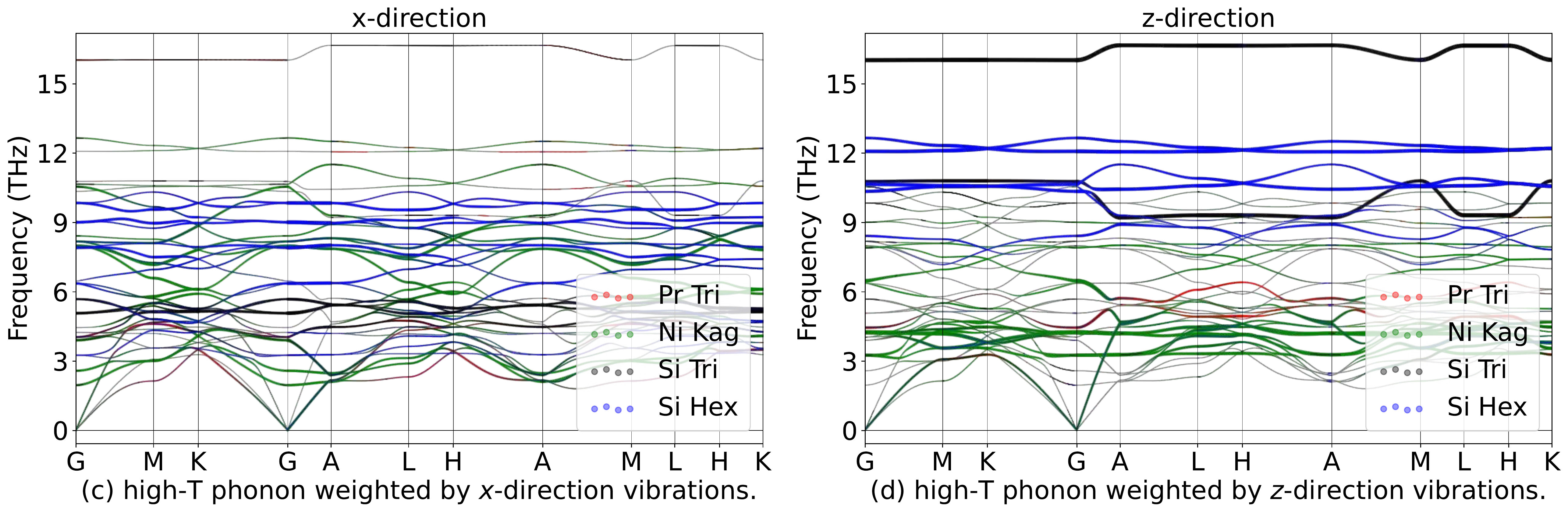}
\caption{Phonon spectrum for PrNi$_6$Si$_6$in in-plane ($x$) and out-of-plane ($z$) directions.}
\label{fig:PrNi6Si6phoxyz}
\end{figure}

\begin{table}[htbp]
\global\long\def\arraystretch{1.12}
\captionsetup{width=.95\textwidth}
\caption{IRREPs at high-symmetry points for PrNi$_6$Si$_6$ phonon spectrum at Low-T.}
\label{tab:191_PrNi6Si6_Low}
\setlength{\tabcolsep}{1.mm}{
}
\end{table}

\begin{table}[htbp]
\global\long\def\arraystretch{1.12}
\captionsetup{width=.95\textwidth}
\caption{IRREPs at high-symmetry points for PrNi$_6$Si$_6$ phonon spectrum at High-T.}
\label{tab:191_PrNi6Si6_High}
\setlength{\tabcolsep}{1.mm}{
%
}
\end{table}
\subsubsection{\label{sms2sec:NdNi6Si6}\hyperref[smsec:col]{\texorpdfstring{NdNi$_6$Si$_6$}{}}~{\color{green}  (High-T stable, Low-T stable) }}

\begin{table}[htbp]
\global\long\def\arraystretch{1.12}
\captionsetup{width=.85\textwidth}
\caption{Structure parameters of NdNi$_6$Si$_6$ after relaxation. It contains 312 electrons. }
\label{tab:NdNi6Si6}
\setlength{\tabcolsep}{4mm}{
%
}
\end{table}

\begin{figure}[htbp]
\centering
\includegraphics[width=0.85\textwidth]{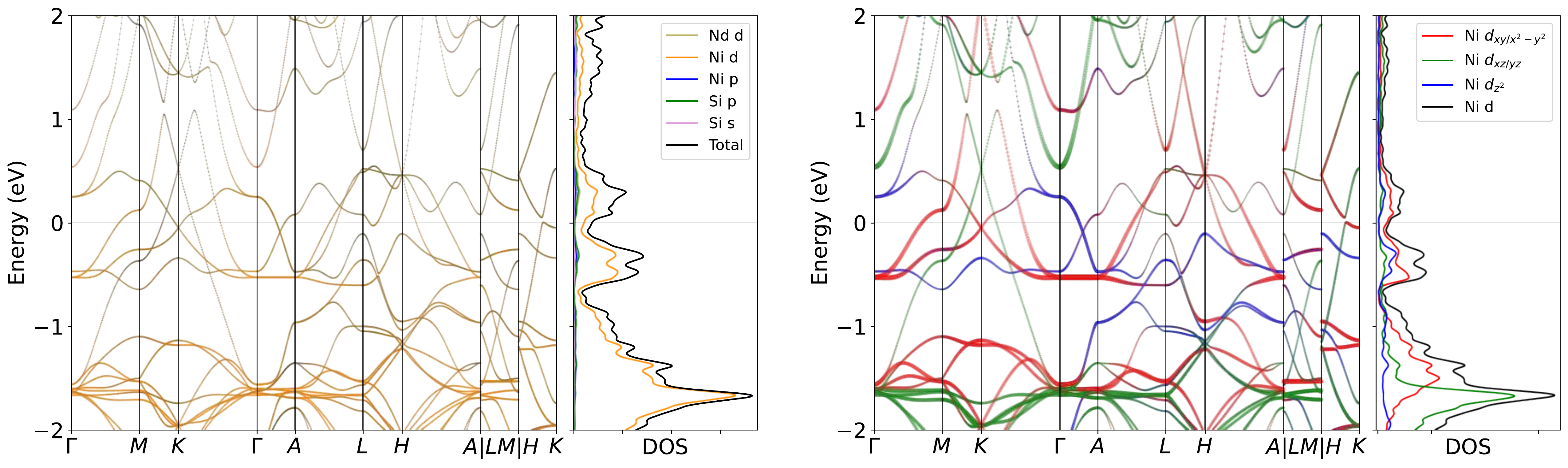}
\caption{Band structure for NdNi$_6$Si$_6$ without SOC. The projection of Ni $3d$ orbitals is also presented. The band structures clearly reveal the dominating of Ni $3d$ orbitals  near the Fermi level, as well as the pronounced peak.}
\label{fig:NdNi6Si6band}
\end{figure}

\begin{figure}[H]
\centering
\subfloat[Relaxed phonon at low-T.]{
\label{fig:NdNi6Si6_relaxed_Low}
\includegraphics[width=0.45\textwidth]{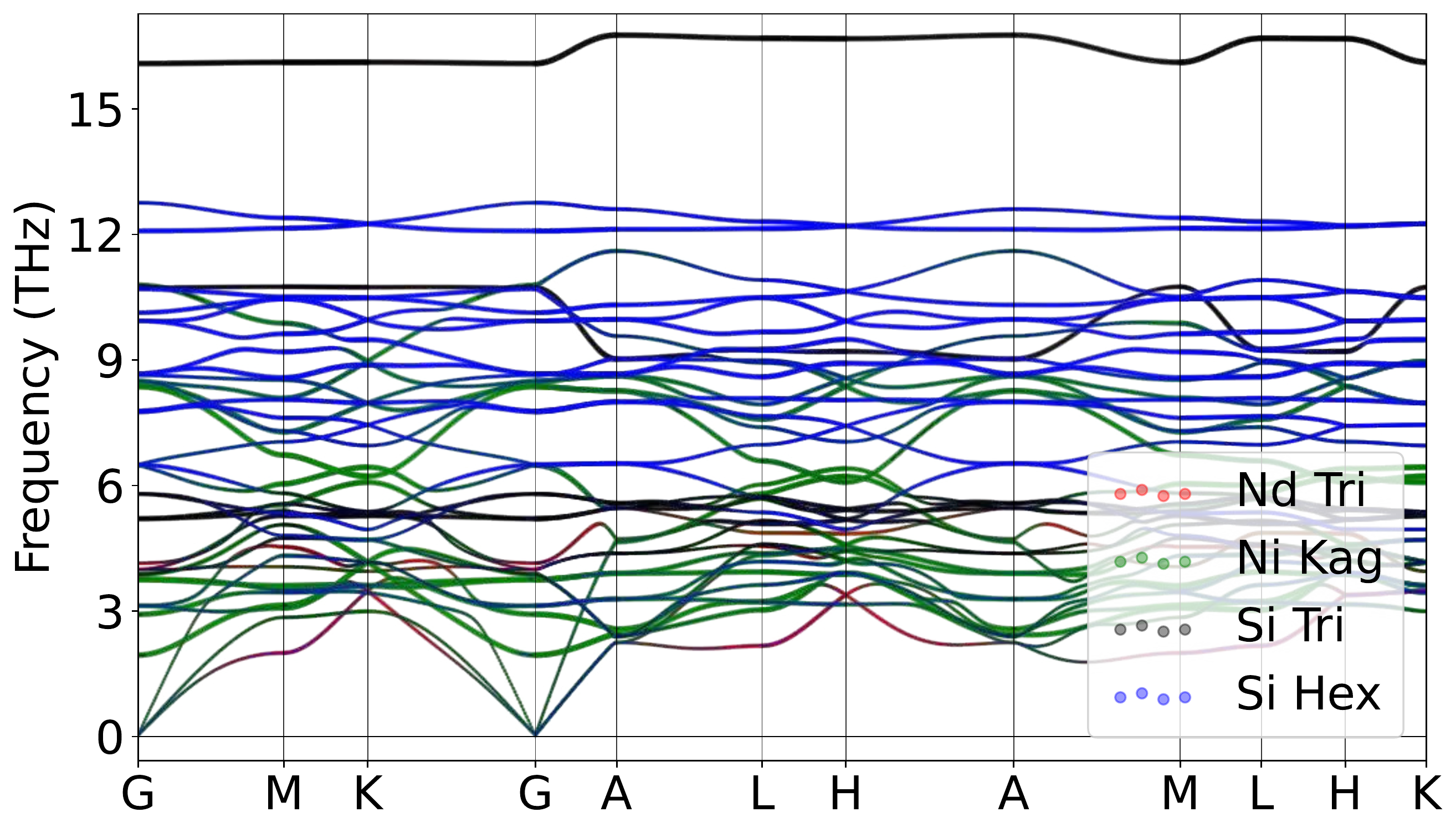}
}
\hspace{5pt}\subfloat[Relaxed phonon at high-T.]{
\label{fig:NdNi6Si6_relaxed_High}
\includegraphics[width=0.45\textwidth]{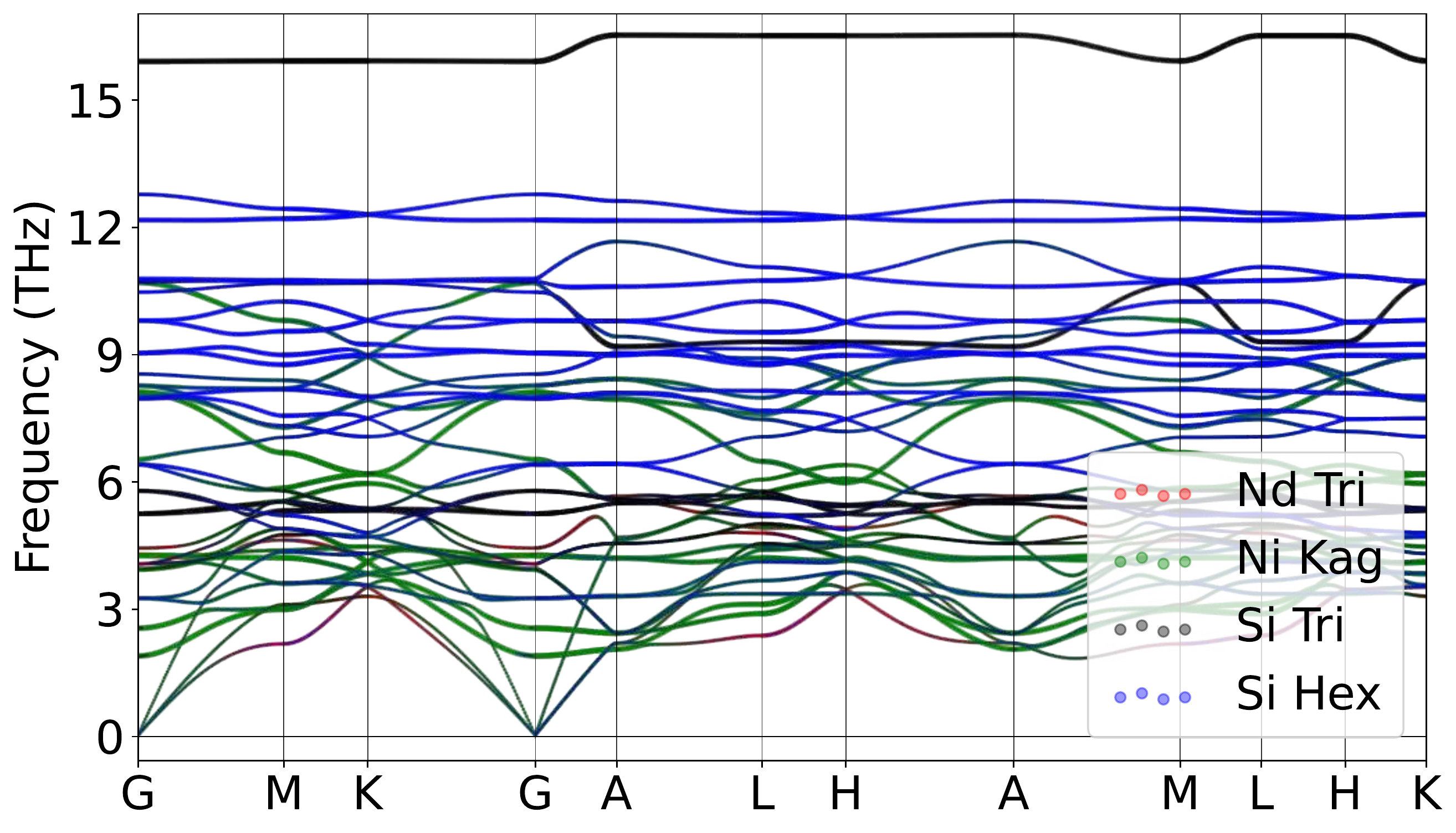}
}
\caption{Atomically projected phonon spectrum for NdNi$_6$Si$_6$.}
\label{fig:NdNi6Si6pho}
\end{figure}

\begin{figure}[H]
\centering
\label{fig:NdNi6Si6_relaxed_Low_xz}
\includegraphics[width=0.85\textwidth]{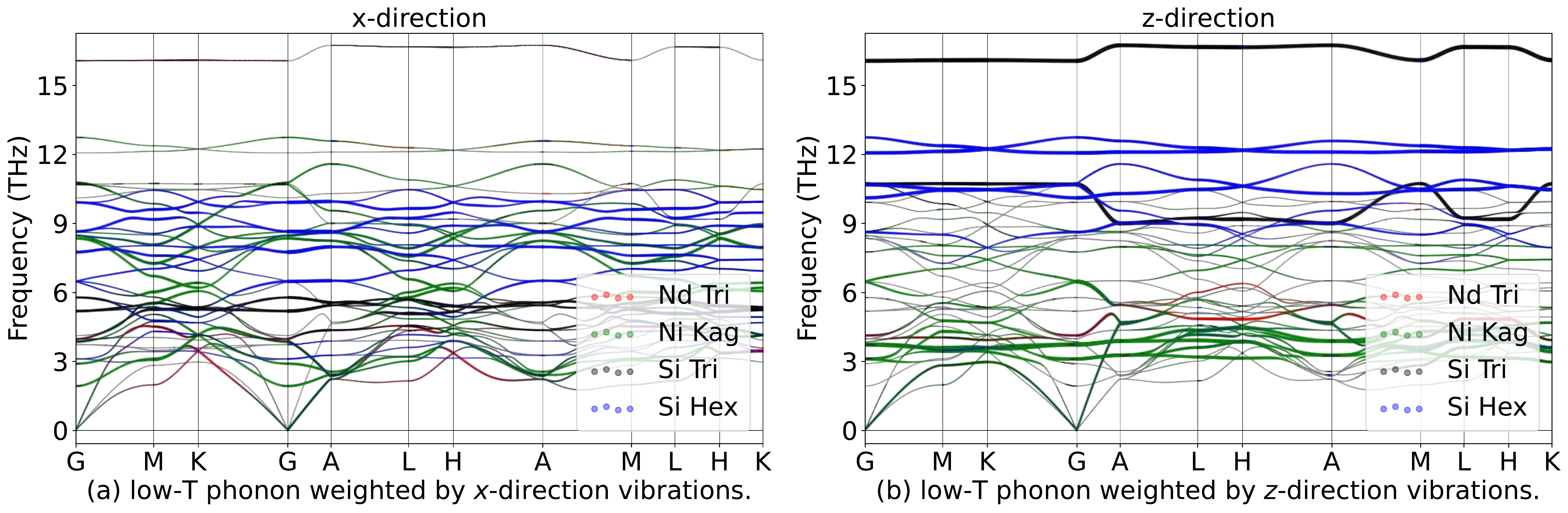}
\hspace{5pt}\label{fig:NdNi6Si6_relaxed_High_xz}
\includegraphics[width=0.85\textwidth]{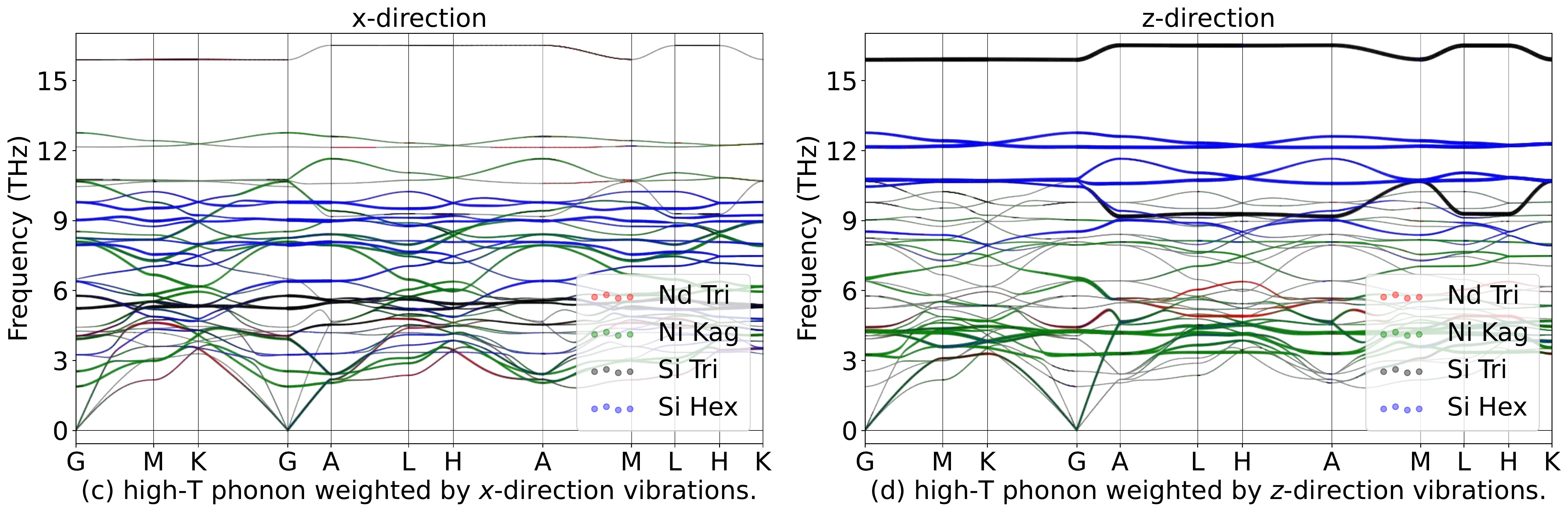}
\caption{Phonon spectrum for NdNi$_6$Si$_6$in in-plane ($x$) and out-of-plane ($z$) directions.}
\label{fig:NdNi6Si6phoxyz}
\end{figure}

\begin{table}[htbp]
\global\long\def\arraystretch{1.12}
\captionsetup{width=.95\textwidth}
\caption{IRREPs at high-symmetry points for NdNi$_6$Si$_6$ phonon spectrum at Low-T.}
\label{tab:191_NdNi6Si6_Low}
\setlength{\tabcolsep}{1.mm}{
}
\end{table}

\begin{table}[htbp]
\global\long\def\arraystretch{1.12}
\captionsetup{width=.95\textwidth}
\caption{IRREPs at high-symmetry points for NdNi$_6$Si$_6$ phonon spectrum at High-T.}
\label{tab:191_NdNi6Si6_High}
\setlength{\tabcolsep}{1.mm}{
%
}
\end{table}
\subsubsection{\label{sms2sec:SmNi6Si6}\hyperref[smsec:col]{\texorpdfstring{SmNi$_6$Si$_6$}{}}~{\color{green}  (High-T stable, Low-T stable) }}

\begin{table}[htbp]
\global\long\def\arraystretch{1.12}
\captionsetup{width=.85\textwidth}
\caption{Structure parameters of SmNi$_6$Si$_6$ after relaxation. It contains 314 electrons. }
\label{tab:SmNi6Si6}
\setlength{\tabcolsep}{4mm}{
%
}
\end{table}

\begin{figure}[htbp]
\centering
\includegraphics[width=0.85\textwidth]{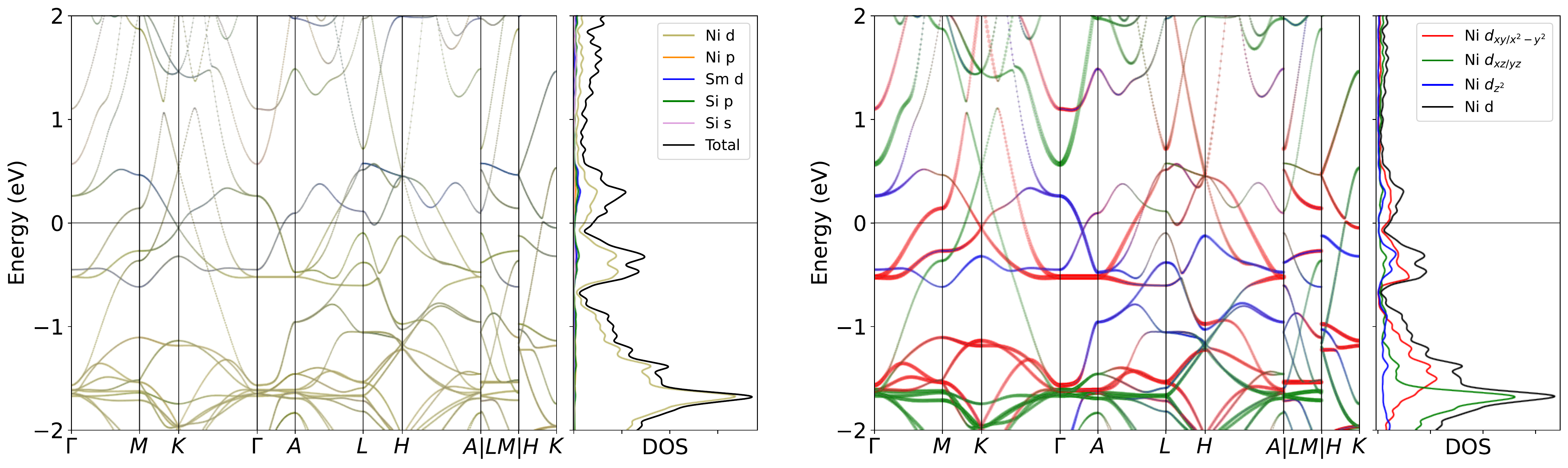}
\caption{Band structure for SmNi$_6$Si$_6$ without SOC. The projection of Ni $3d$ orbitals is also presented. The band structures clearly reveal the dominating of Ni $3d$ orbitals  near the Fermi level, as well as the pronounced peak.}
\label{fig:SmNi6Si6band}
\end{figure}

\begin{figure}[H]
\centering
\subfloat[Relaxed phonon at low-T.]{
\label{fig:SmNi6Si6_relaxed_Low}
\includegraphics[width=0.45\textwidth]{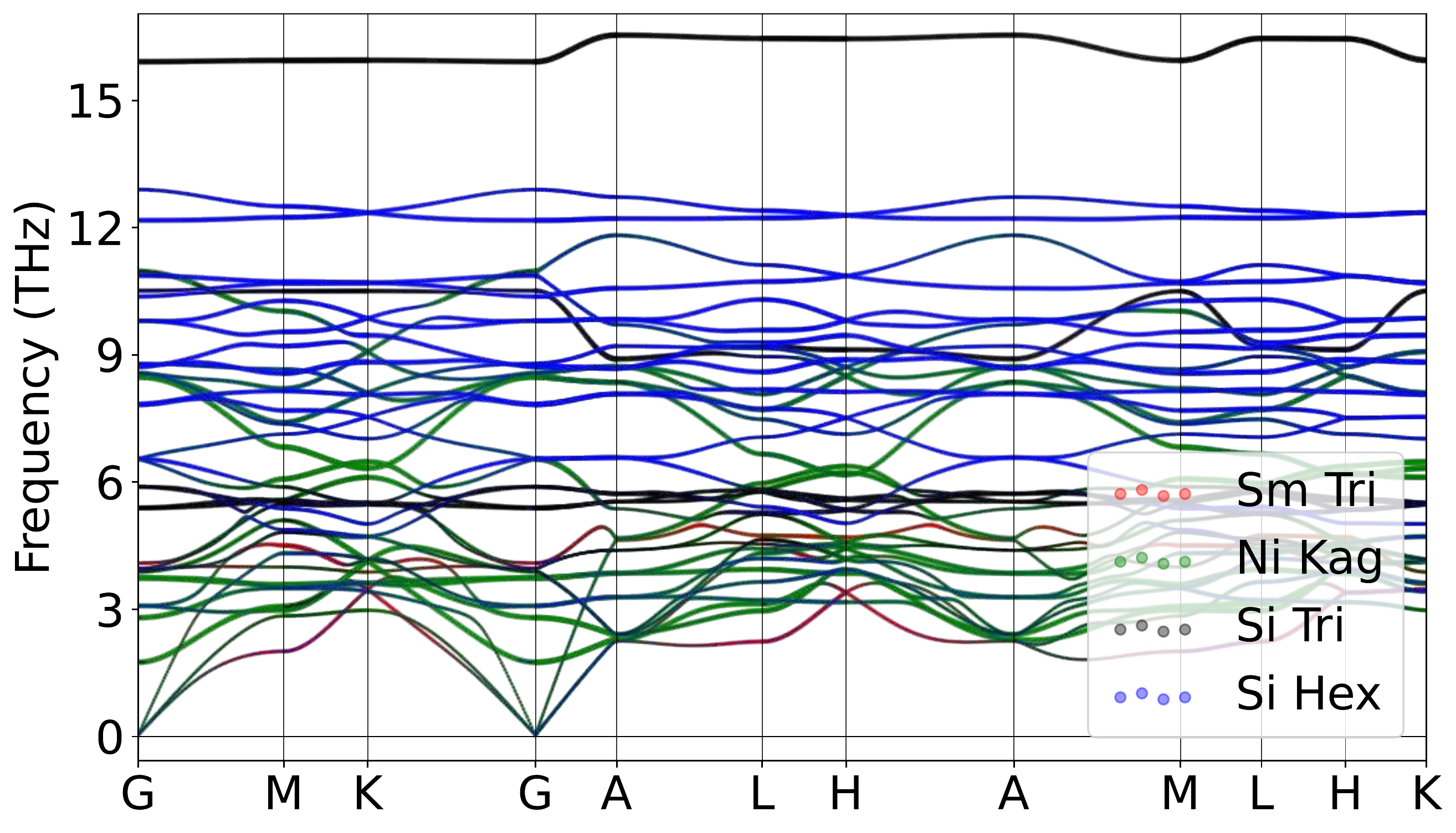}
}
\hspace{5pt}\subfloat[Relaxed phonon at high-T.]{
\label{fig:SmNi6Si6_relaxed_High}
\includegraphics[width=0.45\textwidth]{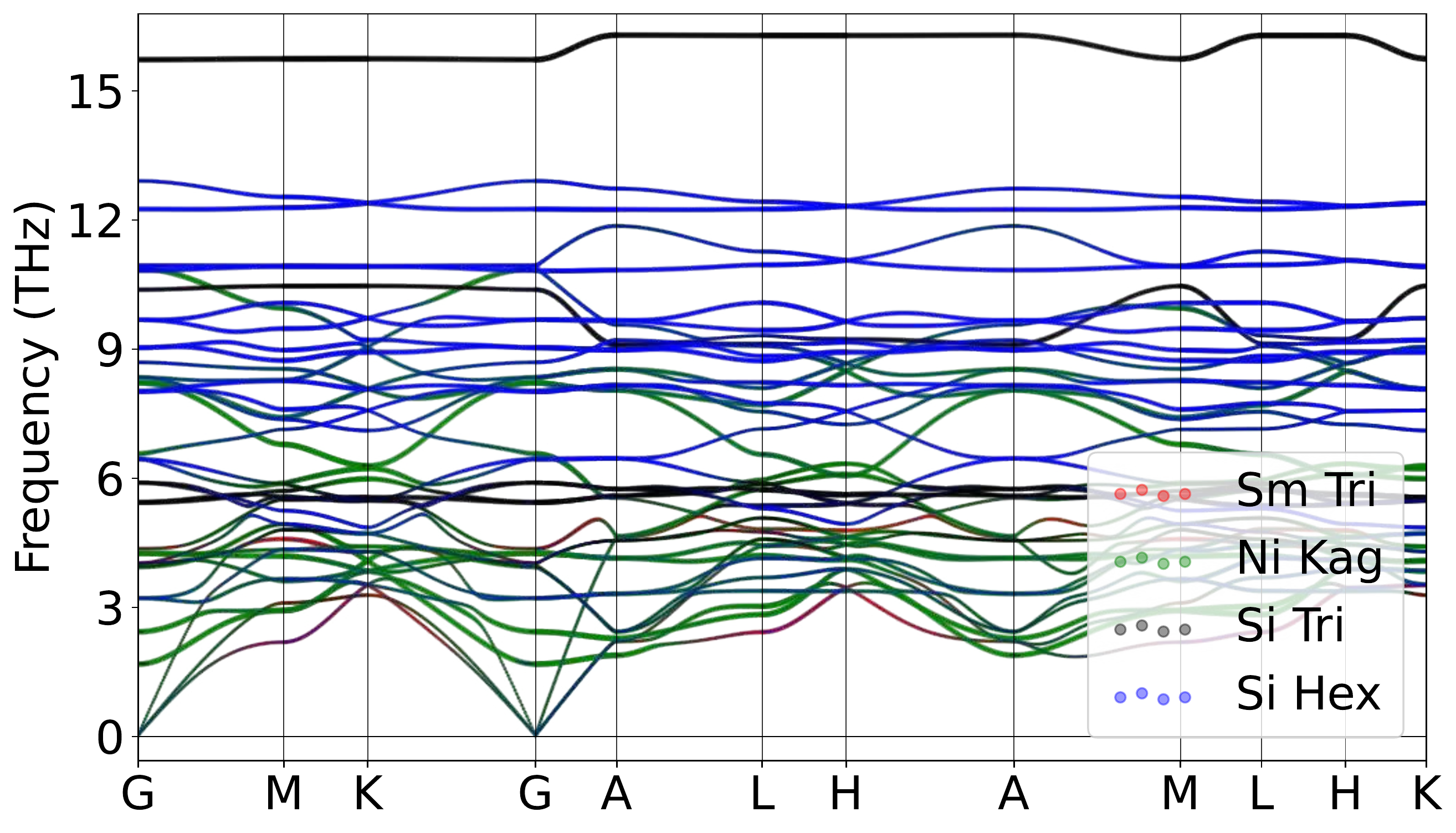}
}
\caption{Atomically projected phonon spectrum for SmNi$_6$Si$_6$.}
\label{fig:SmNi6Si6pho}
\end{figure}

\begin{figure}[H]
\centering
\label{fig:SmNi6Si6_relaxed_Low_xz}
\includegraphics[width=0.85\textwidth]{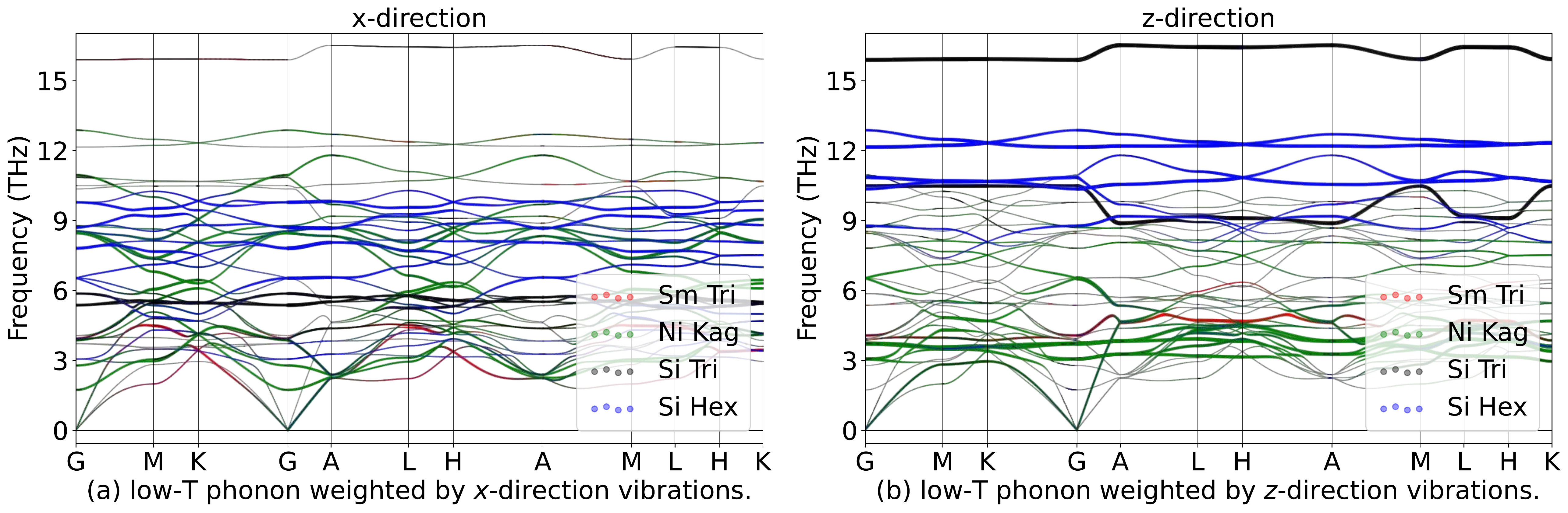}
\hspace{5pt}\label{fig:SmNi6Si6_relaxed_High_xz}
\includegraphics[width=0.85\textwidth]{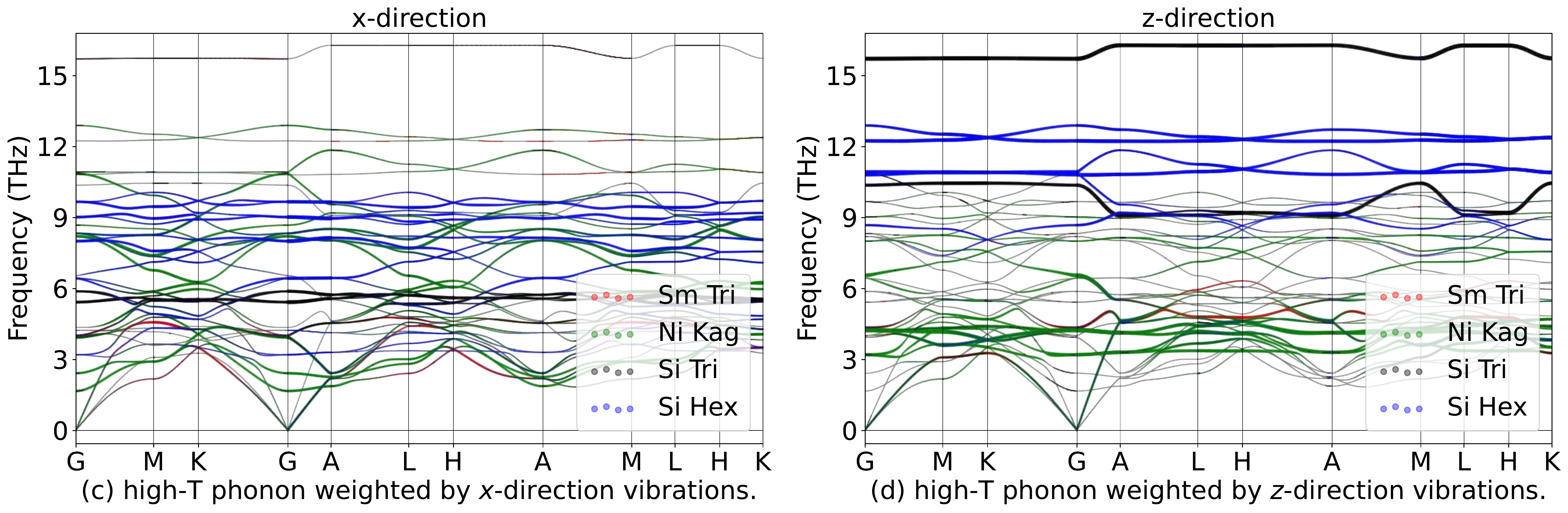}
\caption{Phonon spectrum for SmNi$_6$Si$_6$in in-plane ($x$) and out-of-plane ($z$) directions.}
\label{fig:SmNi6Si6phoxyz}
\end{figure}

\begin{table}[htbp]
\global\long\def\arraystretch{1.12}
\captionsetup{width=.95\textwidth}
\caption{IRREPs at high-symmetry points for SmNi$_6$Si$_6$ phonon spectrum at Low-T.}
\label{tab:191_SmNi6Si6_Low}
\setlength{\tabcolsep}{1.mm}{
}
\end{table}

\begin{table}[htbp]
\global\long\def\arraystretch{1.12}
\captionsetup{width=.95\textwidth}
\caption{IRREPs at high-symmetry points for SmNi$_6$Si$_6$ phonon spectrum at High-T.}
\label{tab:191_SmNi6Si6_High}
\setlength{\tabcolsep}{1.mm}{
%
}
\end{table}
\subsubsection{\label{sms2sec:GdNi6Si6}\hyperref[smsec:col]{\texorpdfstring{GdNi$_6$Si$_6$}{}}~{\color{green}  (High-T stable, Low-T stable) }}

\begin{table}[htbp]
\global\long\def\arraystretch{1.12}
\captionsetup{width=.85\textwidth}
\caption{Structure parameters of GdNi$_6$Si$_6$ after relaxation. It contains 316 electrons. }
\label{tab:GdNi6Si6}
\setlength{\tabcolsep}{4mm}{
%
}
\end{table}

\begin{figure}[htbp]
\centering
\includegraphics[width=0.85\textwidth]{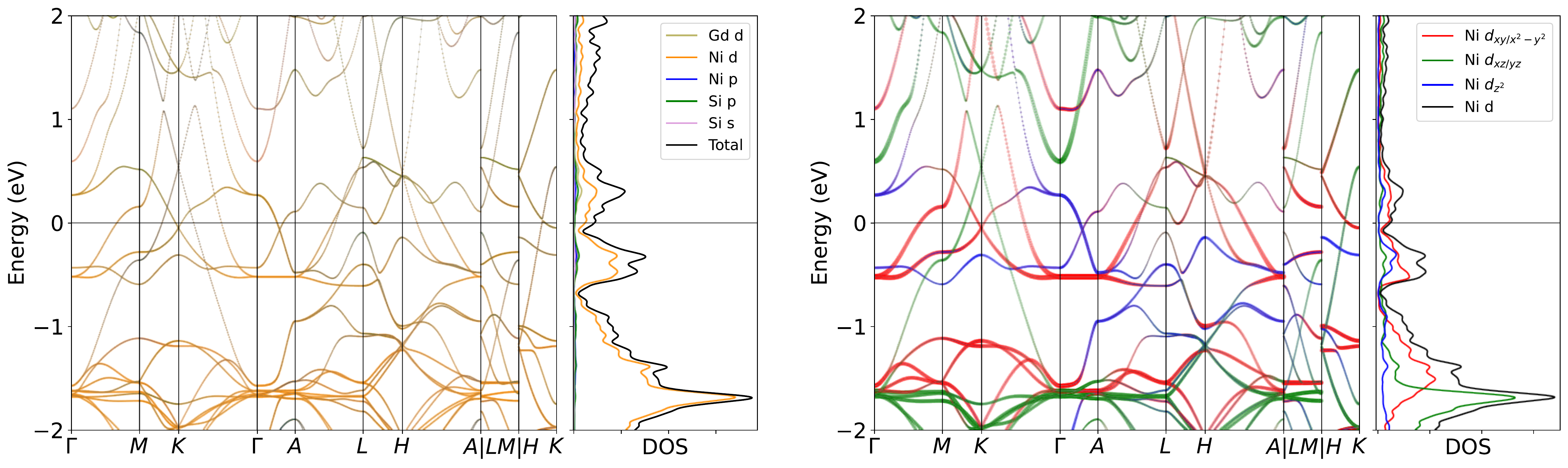}
\caption{Band structure for GdNi$_6$Si$_6$ without SOC. The projection of Ni $3d$ orbitals is also presented. The band structures clearly reveal the dominating of Ni $3d$ orbitals  near the Fermi level, as well as the pronounced peak.}
\label{fig:GdNi6Si6band}
\end{figure}

\begin{figure}[H]
\centering
\subfloat[Relaxed phonon at low-T.]{
\label{fig:GdNi6Si6_relaxed_Low}
\includegraphics[width=0.45\textwidth]{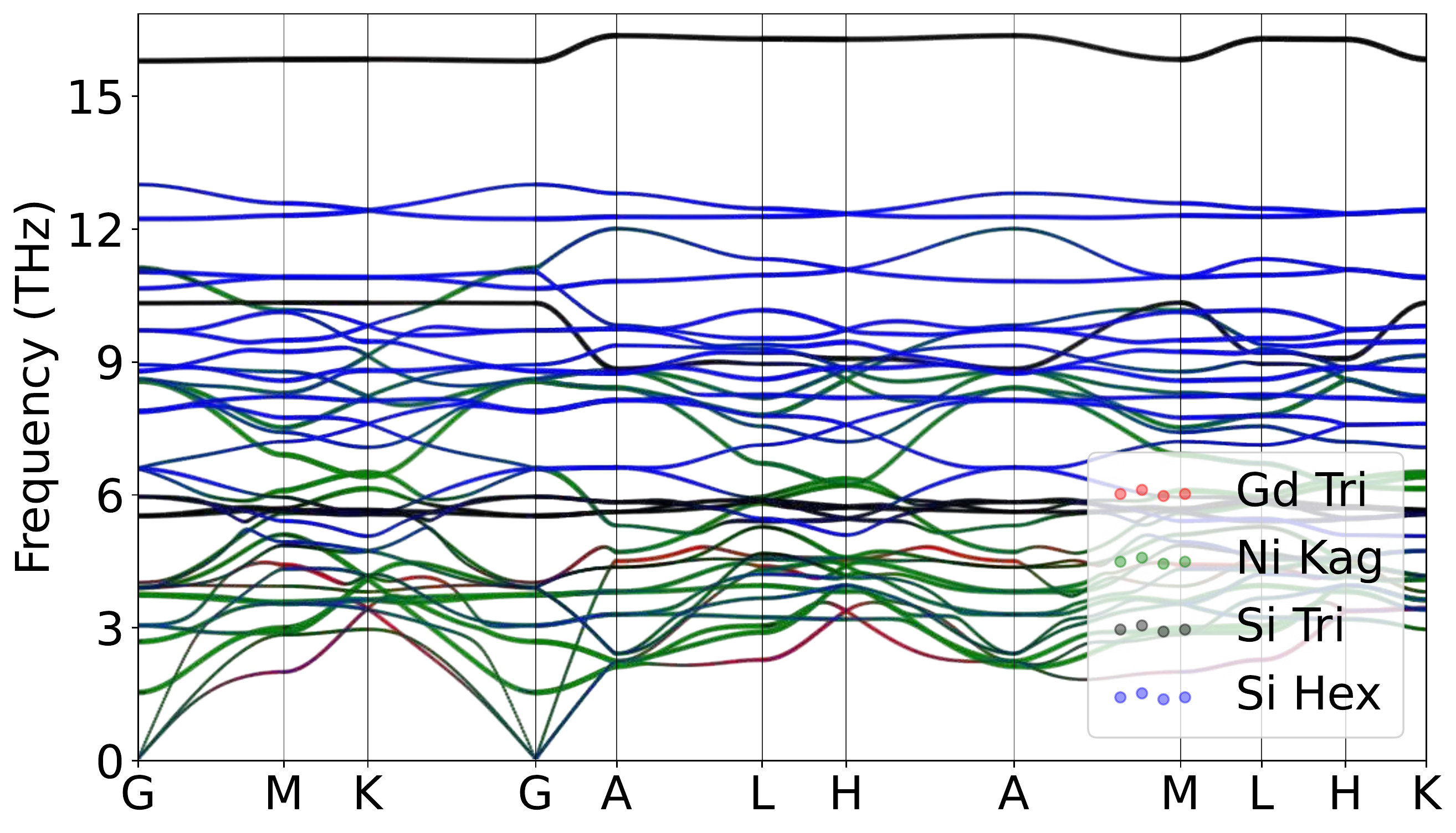}
}
\hspace{5pt}\subfloat[Relaxed phonon at high-T.]{
\label{fig:GdNi6Si6_relaxed_High}
\includegraphics[width=0.45\textwidth]{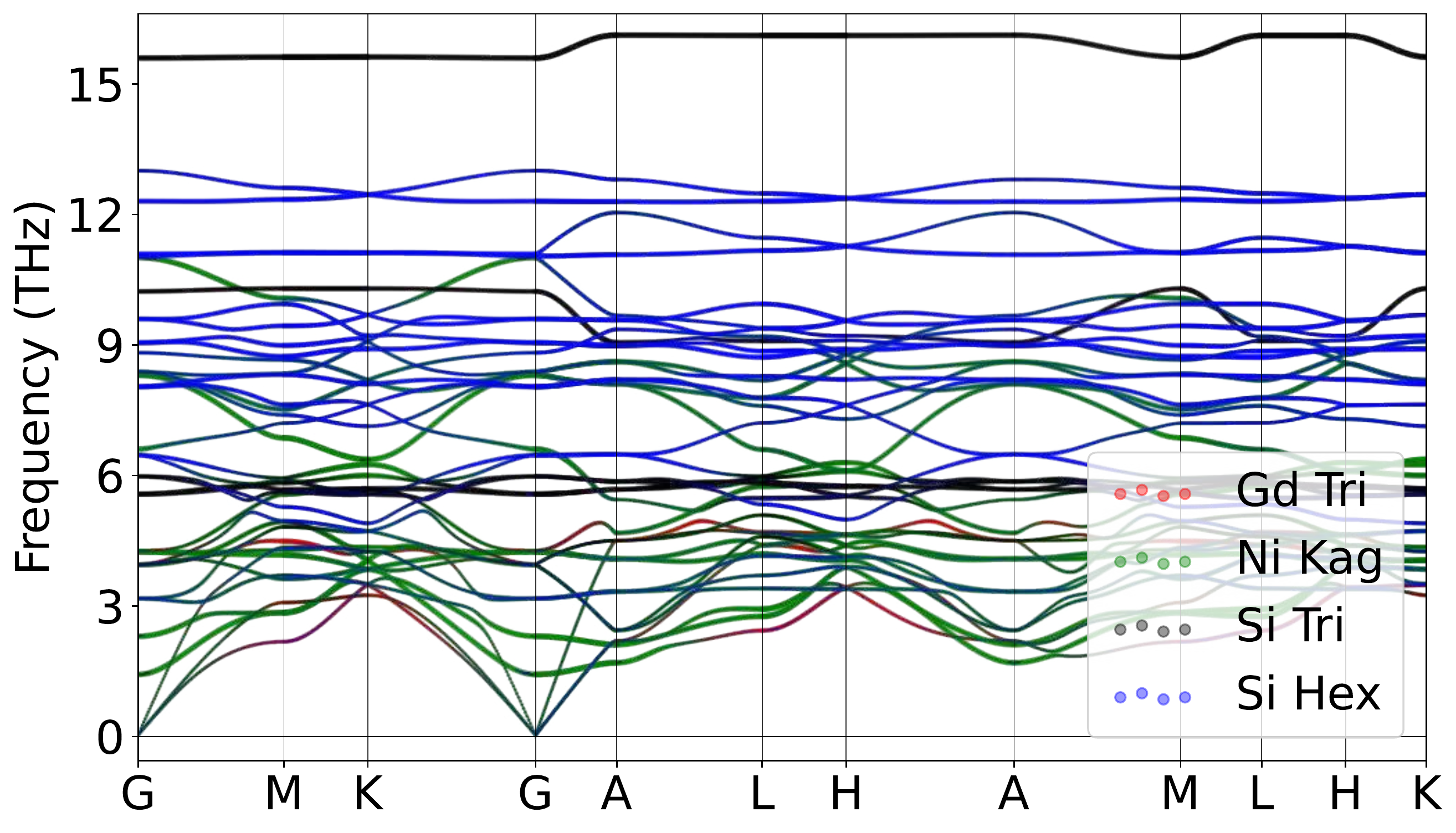}
}
\caption{Atomically projected phonon spectrum for GdNi$_6$Si$_6$.}
\label{fig:GdNi6Si6pho}
\end{figure}

\begin{figure}[H]
\centering
\label{fig:GdNi6Si6_relaxed_Low_xz}
\includegraphics[width=0.85\textwidth]{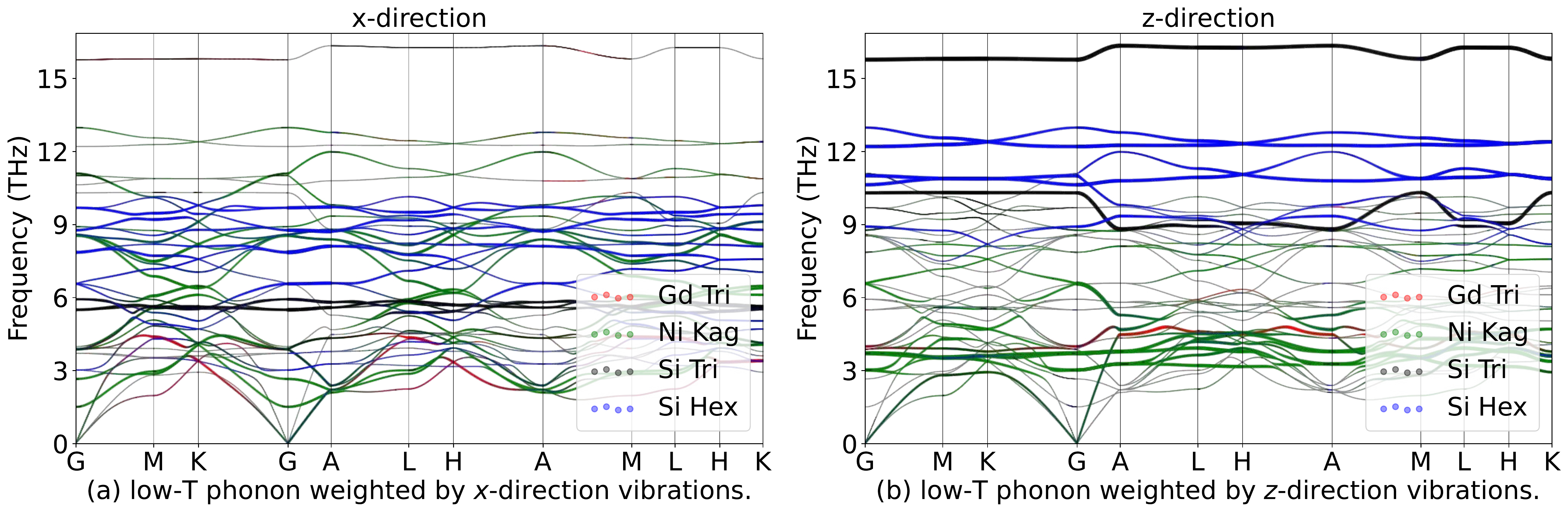}
\hspace{5pt}\label{fig:GdNi6Si6_relaxed_High_xz}
\includegraphics[width=0.85\textwidth]{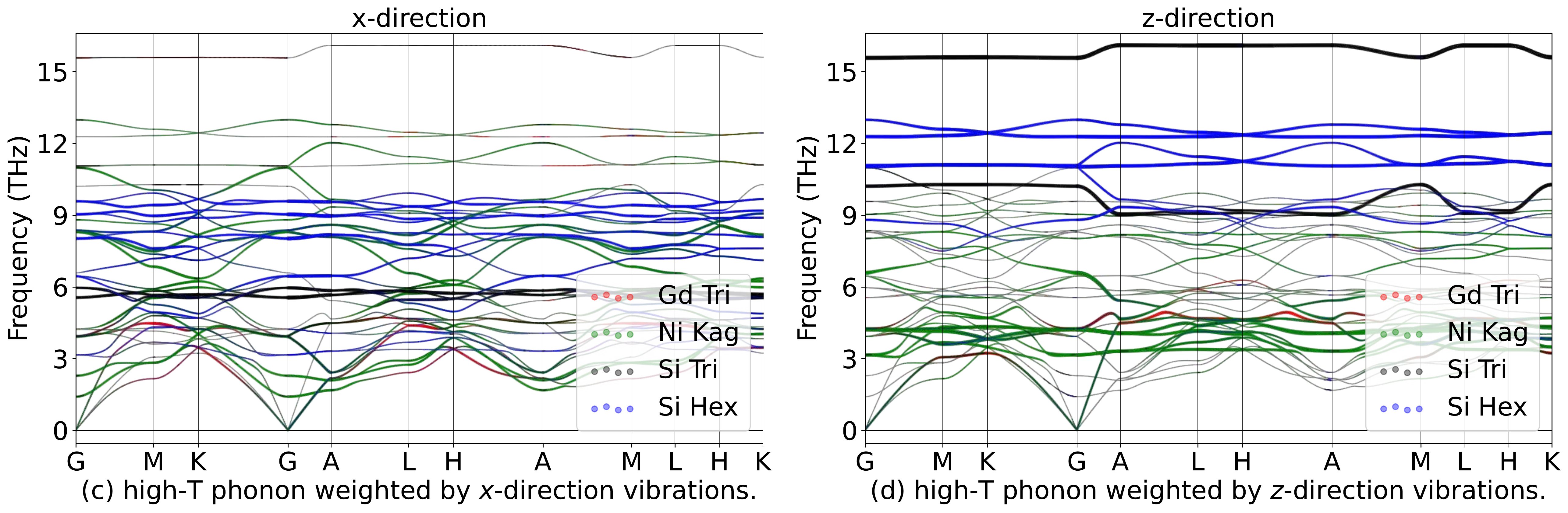}
\caption{Phonon spectrum for GdNi$_6$Si$_6$in in-plane ($x$) and out-of-plane ($z$) directions.}
\label{fig:GdNi6Si6phoxyz}
\end{figure}

\begin{table}[htbp]
\global\long\def\arraystretch{1.12}
\captionsetup{width=.95\textwidth}
\caption{IRREPs at high-symmetry points for GdNi$_6$Si$_6$ phonon spectrum at Low-T.}
\label{tab:191_GdNi6Si6_Low}
\setlength{\tabcolsep}{1.mm}{
}
\end{table}

\begin{table}[htbp]
\global\long\def\arraystretch{1.12}
\captionsetup{width=.95\textwidth}
\caption{IRREPs at high-symmetry points for GdNi$_6$Si$_6$ phonon spectrum at High-T.}
\label{tab:191_GdNi6Si6_High}
\setlength{\tabcolsep}{1.mm}{
%
}
\end{table}
\subsubsection{\label{sms2sec:TbNi6Si6}\hyperref[smsec:col]{\texorpdfstring{TbNi$_6$Si$_6$}{}}~{\color{green}  (High-T stable, Low-T stable) }}

\begin{table}[htbp]
\global\long\def\arraystretch{1.12}
\captionsetup{width=.85\textwidth}
\caption{Structure parameters of TbNi$_6$Si$_6$ after relaxation. It contains 317 electrons. }
\label{tab:TbNi6Si6}
\setlength{\tabcolsep}{4mm}{
%
}
\end{table}

\begin{figure}[htbp]
\centering
\includegraphics[width=0.85\textwidth]{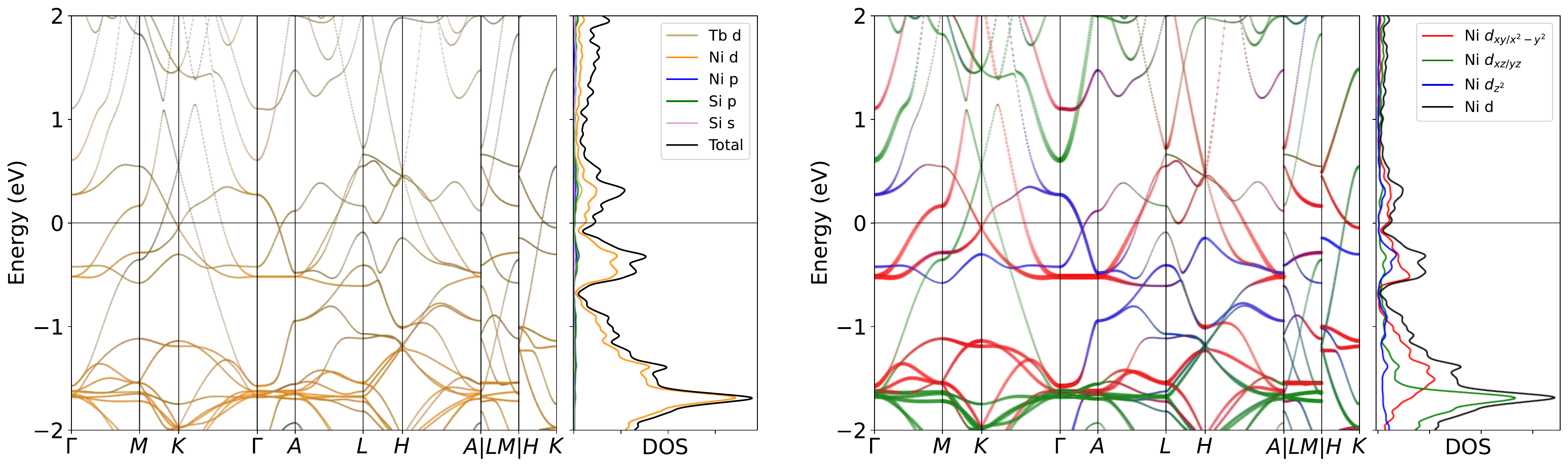}
\caption{Band structure for TbNi$_6$Si$_6$ without SOC. The projection of Ni $3d$ orbitals is also presented. The band structures clearly reveal the dominating of Ni $3d$ orbitals  near the Fermi level, as well as the pronounced peak.}
\label{fig:TbNi6Si6band}
\end{figure}

\begin{figure}[H]
\centering
\subfloat[Relaxed phonon at low-T.]{
\label{fig:TbNi6Si6_relaxed_Low}
\includegraphics[width=0.45\textwidth]{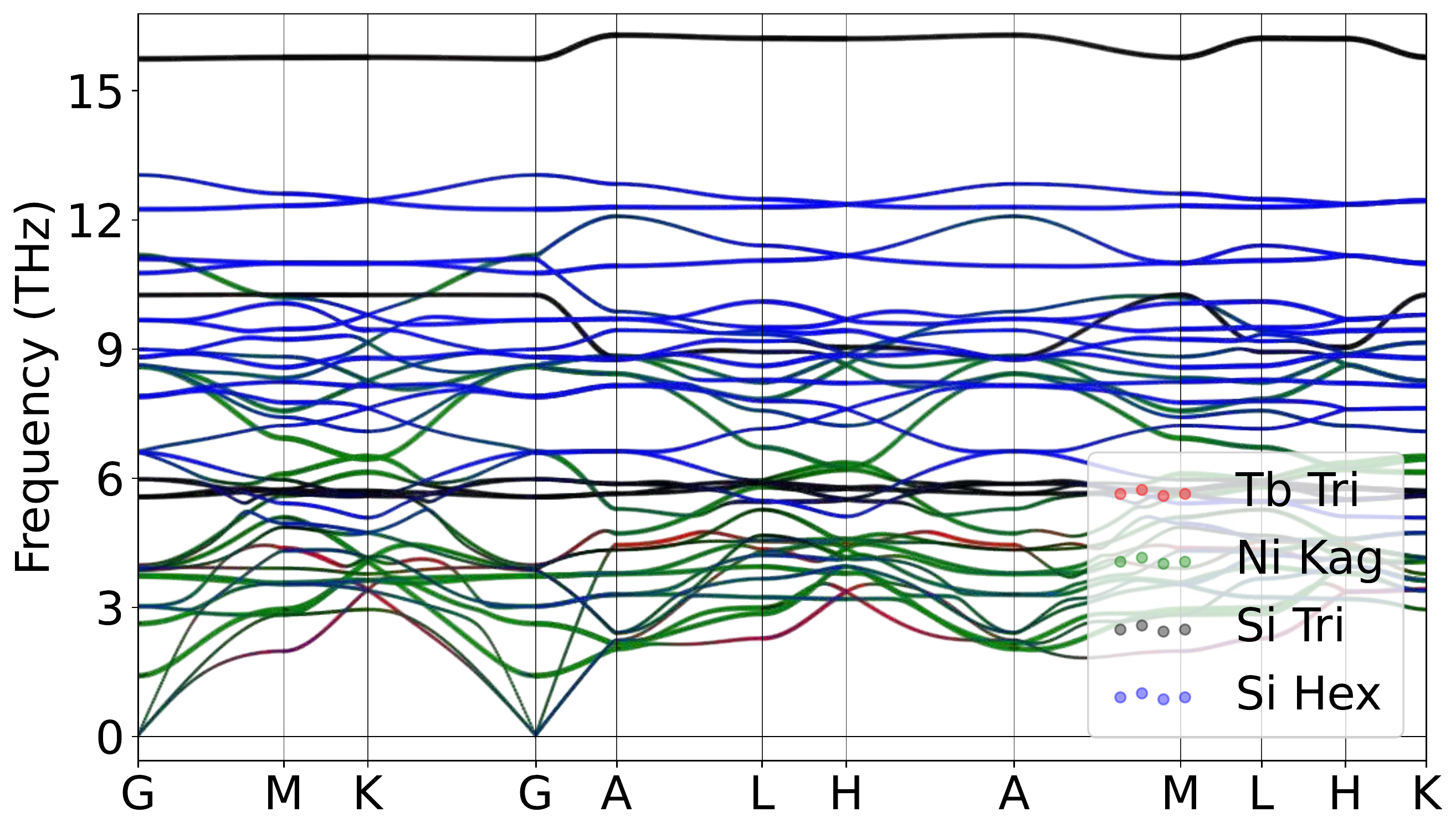}
}
\hspace{5pt}\subfloat[Relaxed phonon at high-T.]{
\label{fig:TbNi6Si6_relaxed_High}
\includegraphics[width=0.45\textwidth]{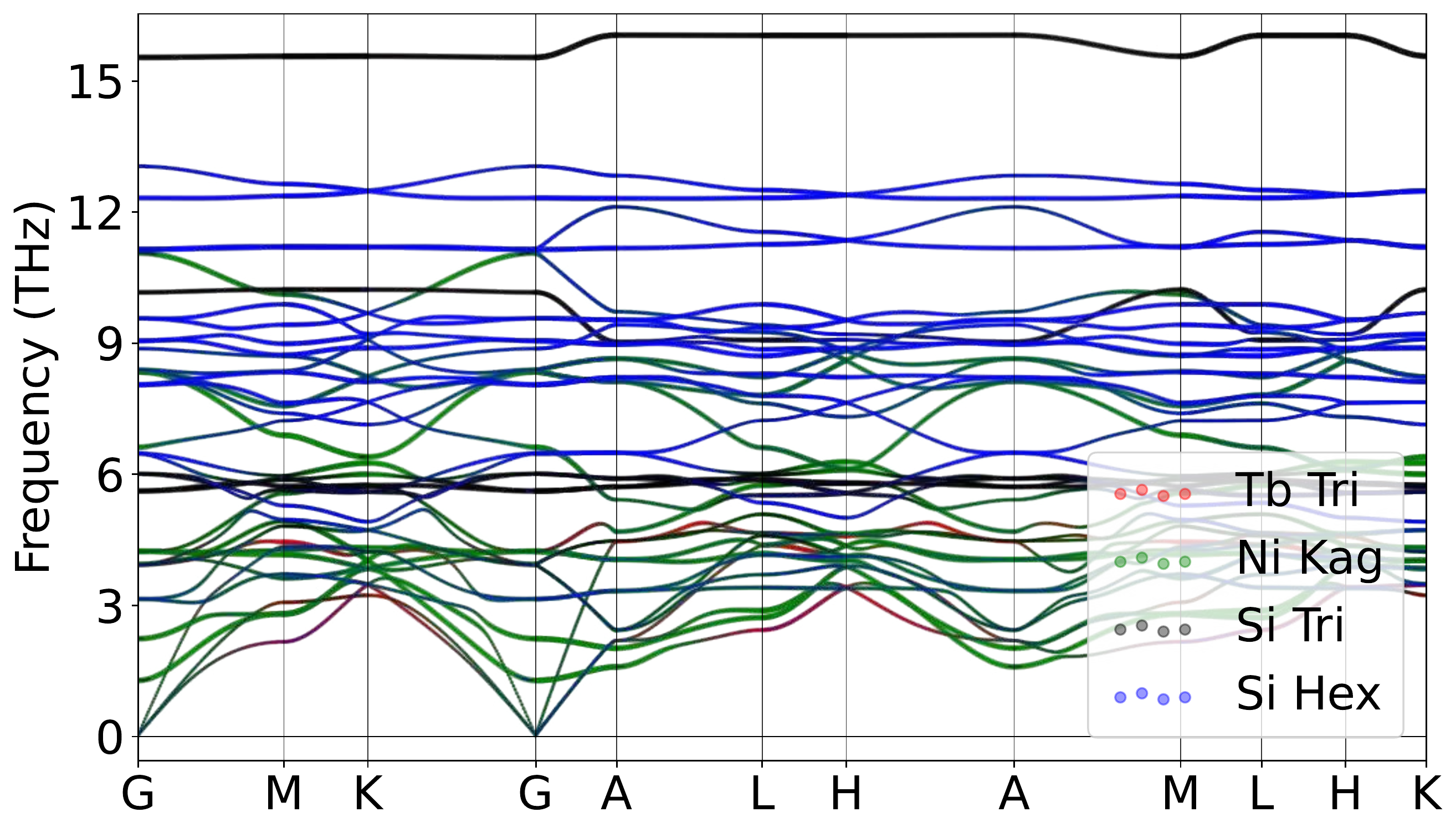}
}
\caption{Atomically projected phonon spectrum for TbNi$_6$Si$_6$.}
\label{fig:TbNi6Si6pho}
\end{figure}

\begin{figure}[H]
\centering
\label{fig:TbNi6Si6_relaxed_Low_xz}
\includegraphics[width=0.85\textwidth]{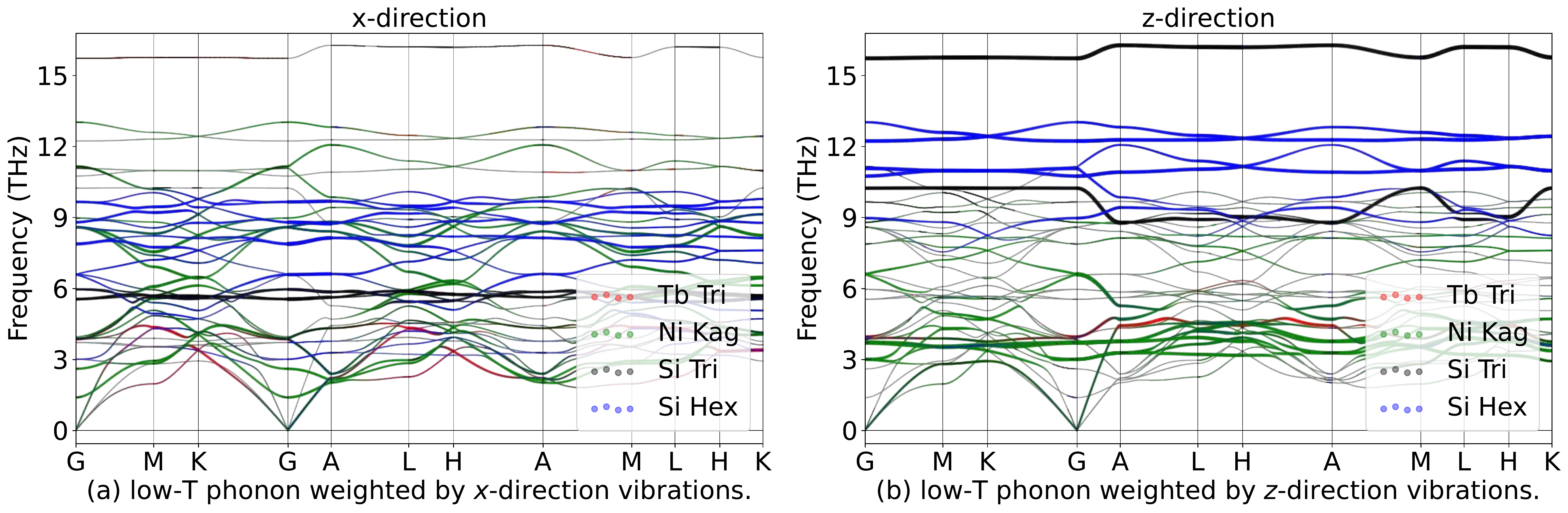}
\hspace{5pt}\label{fig:TbNi6Si6_relaxed_High_xz}
\includegraphics[width=0.85\textwidth]{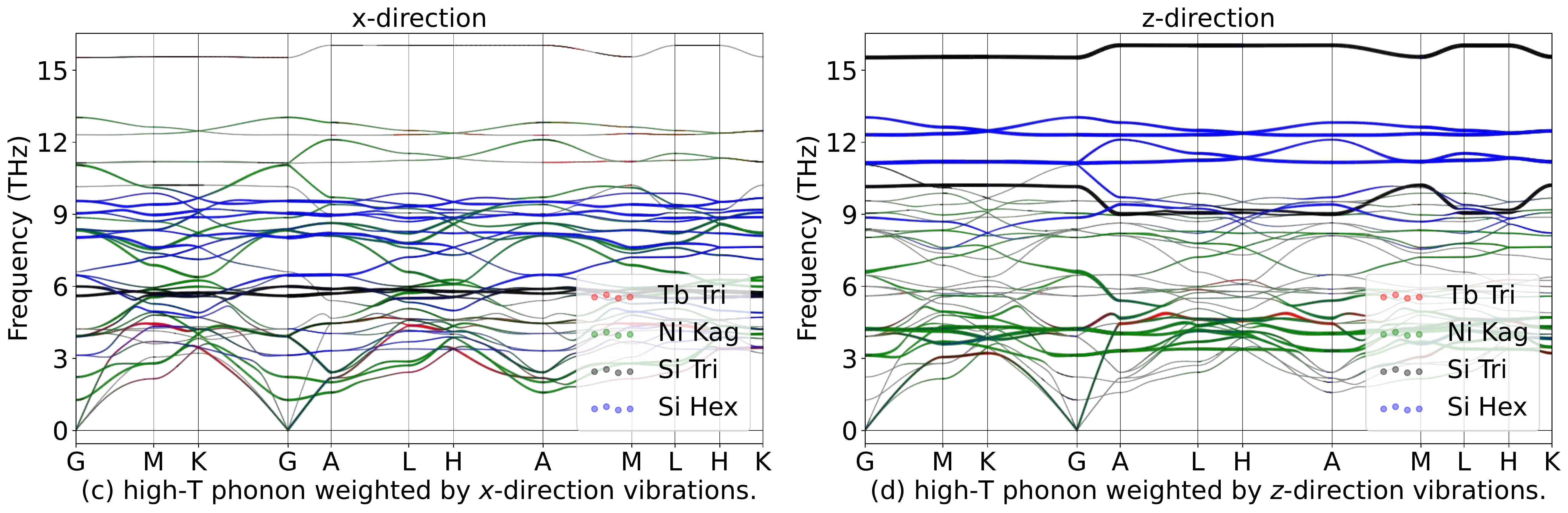}
\caption{Phonon spectrum for TbNi$_6$Si$_6$in in-plane ($x$) and out-of-plane ($z$) directions.}
\label{fig:TbNi6Si6phoxyz}
\end{figure}

\begin{table}[htbp]
\global\long\def\arraystretch{1.12}
\captionsetup{width=.95\textwidth}
\caption{IRREPs at high-symmetry points for TbNi$_6$Si$_6$ phonon spectrum at Low-T.}
\label{tab:191_TbNi6Si6_Low}
\setlength{\tabcolsep}{1.mm}{
}
\end{table}

\begin{table}[htbp]
\global\long\def\arraystretch{1.12}
\captionsetup{width=.95\textwidth}
\caption{IRREPs at high-symmetry points for TbNi$_6$Si$_6$ phonon spectrum at High-T.}
\label{tab:191_TbNi6Si6_High}
\setlength{\tabcolsep}{1.mm}{
%
}
\end{table}
\subsubsection{\label{sms2sec:DyNi6Si6}\hyperref[smsec:col]{\texorpdfstring{DyNi$_6$Si$_6$}{}}~{\color{green}  (High-T stable, Low-T stable) }}

\begin{table}[htbp]
\global\long\def\arraystretch{1.12}
\captionsetup{width=.85\textwidth}
\caption{Structure parameters of DyNi$_6$Si$_6$ after relaxation. It contains 318 electrons. }
\label{tab:DyNi6Si6}
\setlength{\tabcolsep}{4mm}{
%
}
\end{table}

\begin{figure}[htbp]
\centering
\includegraphics[width=0.85\textwidth]{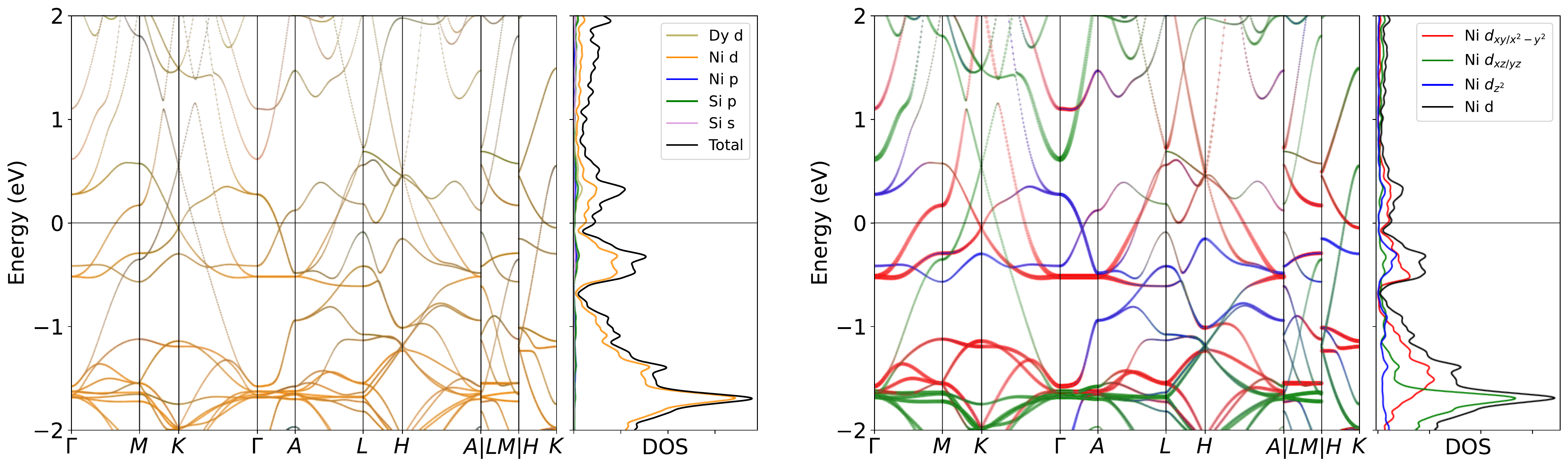}
\caption{Band structure for DyNi$_6$Si$_6$ without SOC. The projection of Ni $3d$ orbitals is also presented. The band structures clearly reveal the dominating of Ni $3d$ orbitals  near the Fermi level, as well as the pronounced peak.}
\label{fig:DyNi6Si6band}
\end{figure}

\begin{figure}[H]
\centering
\subfloat[Relaxed phonon at low-T.]{
\label{fig:DyNi6Si6_relaxed_Low}
\includegraphics[width=0.45\textwidth]{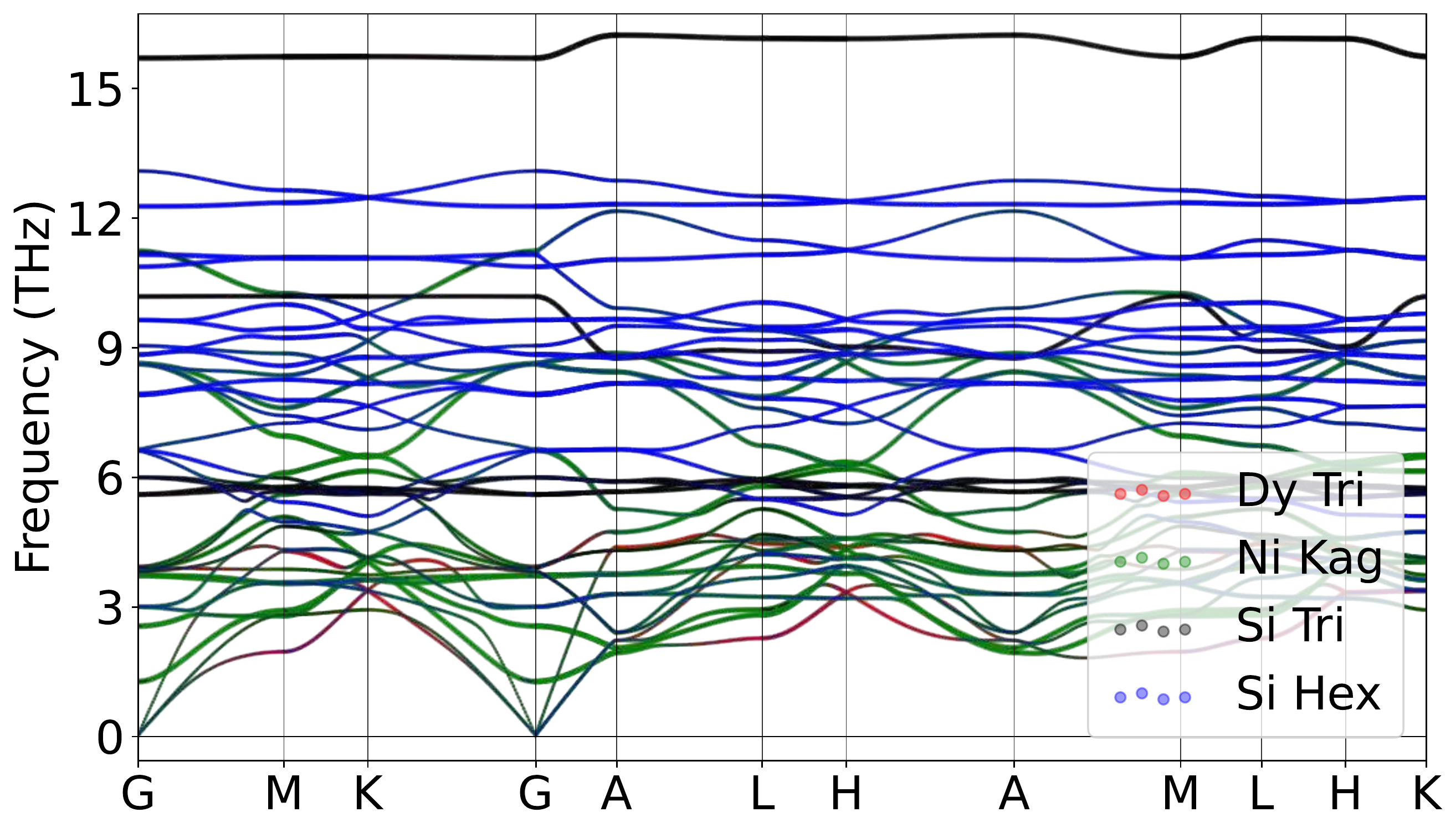}
}
\hspace{5pt}\subfloat[Relaxed phonon at high-T.]{
\label{fig:DyNi6Si6_relaxed_High}
\includegraphics[width=0.45\textwidth]{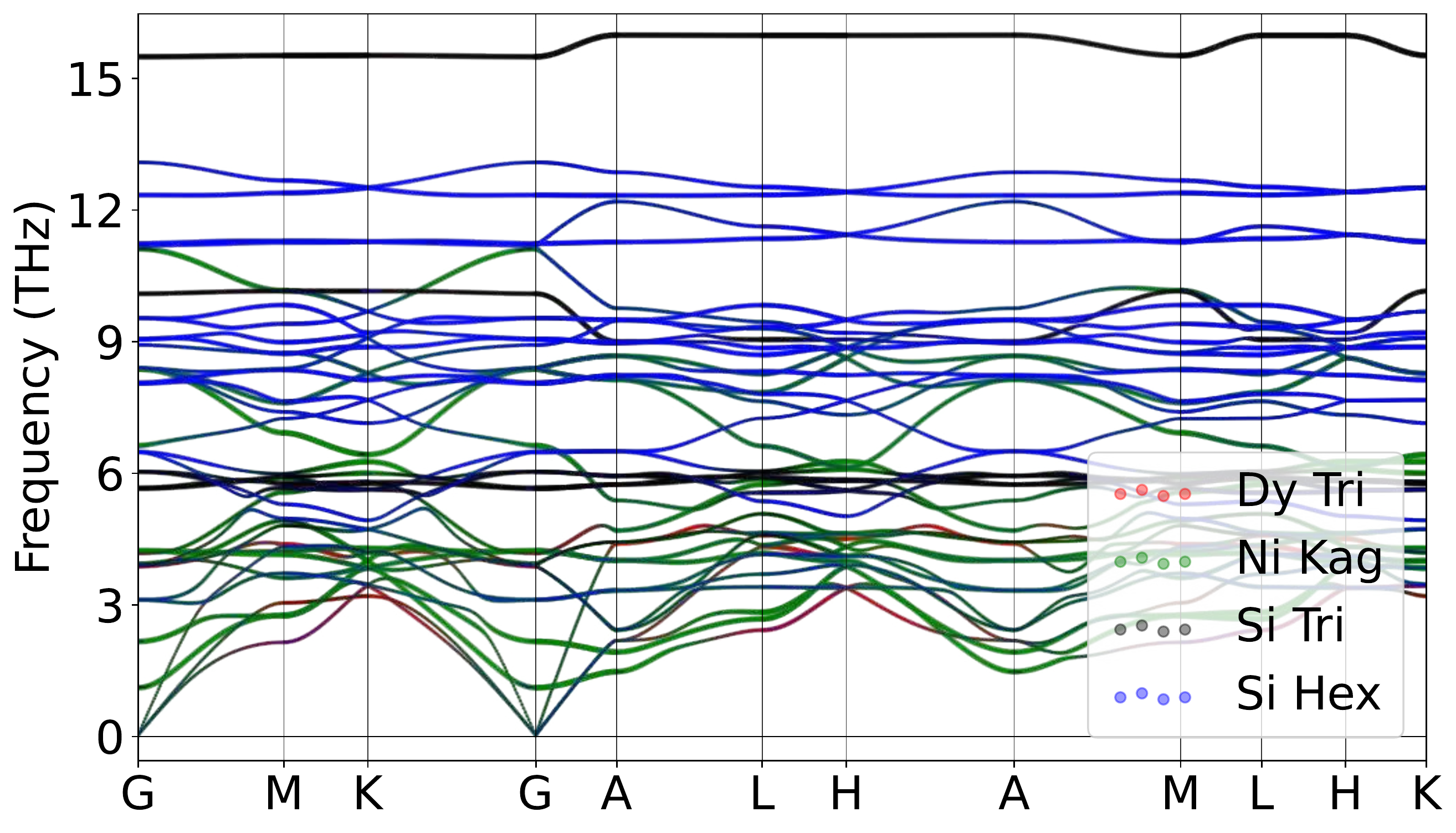}
}
\caption{Atomically projected phonon spectrum for DyNi$_6$Si$_6$.}
\label{fig:DyNi6Si6pho}
\end{figure}

\begin{figure}[H]
\centering
\label{fig:DyNi6Si6_relaxed_Low_xz}
\includegraphics[width=0.85\textwidth]{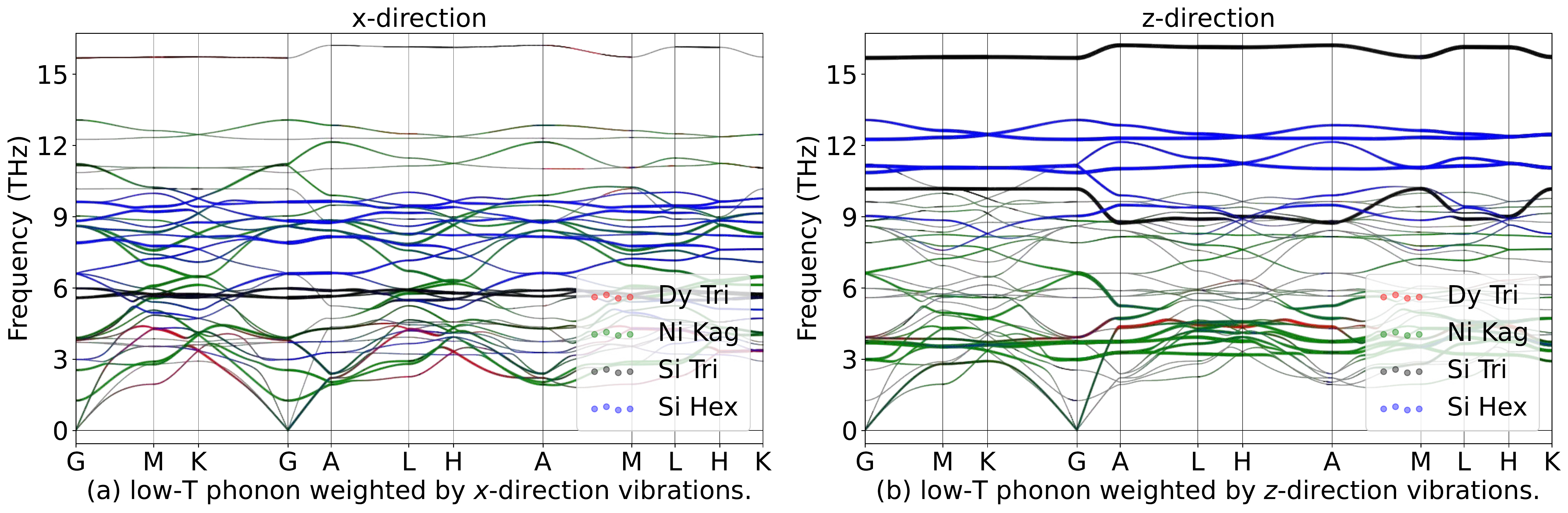}
\hspace{5pt}\label{fig:DyNi6Si6_relaxed_High_xz}
\includegraphics[width=0.85\textwidth]{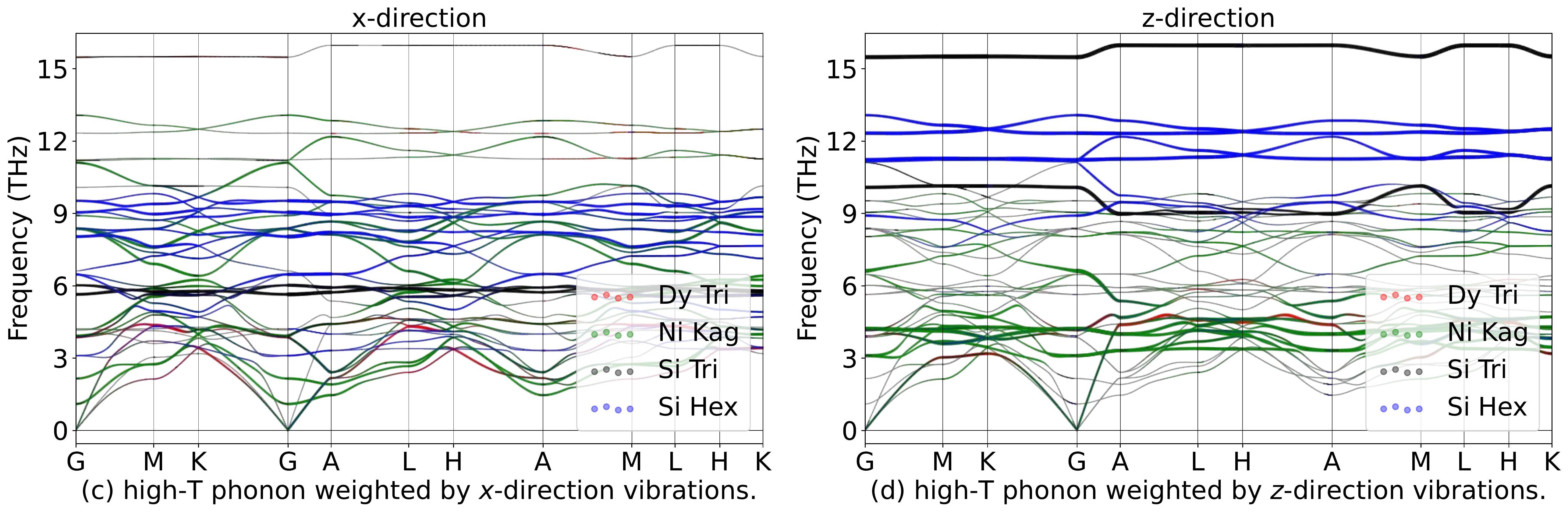}
\caption{Phonon spectrum for DyNi$_6$Si$_6$in in-plane ($x$) and out-of-plane ($z$) directions.}
\label{fig:DyNi6Si6phoxyz}
\end{figure}

\begin{table}[htbp]
\global\long\def\arraystretch{1.12}
\captionsetup{width=.95\textwidth}
\caption{IRREPs at high-symmetry points for DyNi$_6$Si$_6$ phonon spectrum at Low-T.}
\label{tab:191_DyNi6Si6_Low}
\setlength{\tabcolsep}{1.mm}{
}
\end{table}

\begin{table}[htbp]
\global\long\def\arraystretch{1.12}
\captionsetup{width=.95\textwidth}
\caption{IRREPs at high-symmetry points for DyNi$_6$Si$_6$ phonon spectrum at High-T.}
\label{tab:191_DyNi6Si6_High}
\setlength{\tabcolsep}{1.mm}{
%
}
\end{table}
\subsubsection{\label{sms2sec:HoNi6Si6}\hyperref[smsec:col]{\texorpdfstring{HoNi$_6$Si$_6$}{}}~{\color{green}  (High-T stable, Low-T stable) }}

\begin{table}[htbp]
\global\long\def\arraystretch{1.12}
\captionsetup{width=.85\textwidth}
\caption{Structure parameters of HoNi$_6$Si$_6$ after relaxation. It contains 319 electrons. }
\label{tab:HoNi6Si6}
\setlength{\tabcolsep}{4mm}{
%
}
\end{table}

\begin{figure}[htbp]
\centering
\includegraphics[width=0.85\textwidth]{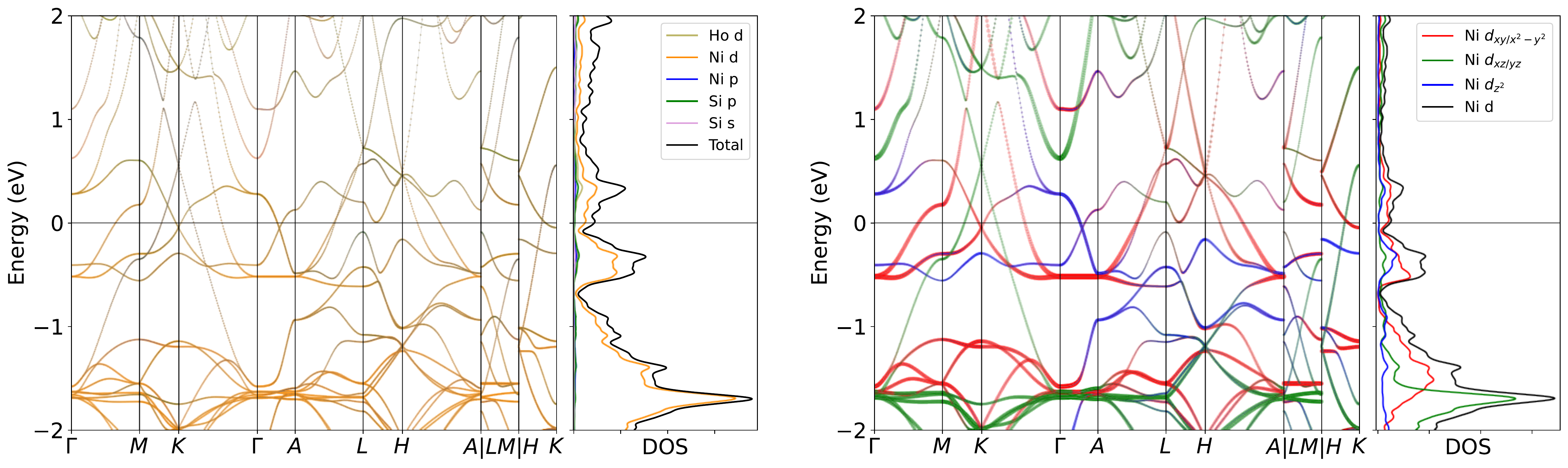}
\caption{Band structure for HoNi$_6$Si$_6$ without SOC. The projection of Ni $3d$ orbitals is also presented. The band structures clearly reveal the dominating of Ni $3d$ orbitals  near the Fermi level, as well as the pronounced peak.}
\label{fig:HoNi6Si6band}
\end{figure}

\begin{figure}[H]
\centering
\subfloat[Relaxed phonon at low-T.]{
\label{fig:HoNi6Si6_relaxed_Low}
\includegraphics[width=0.45\textwidth]{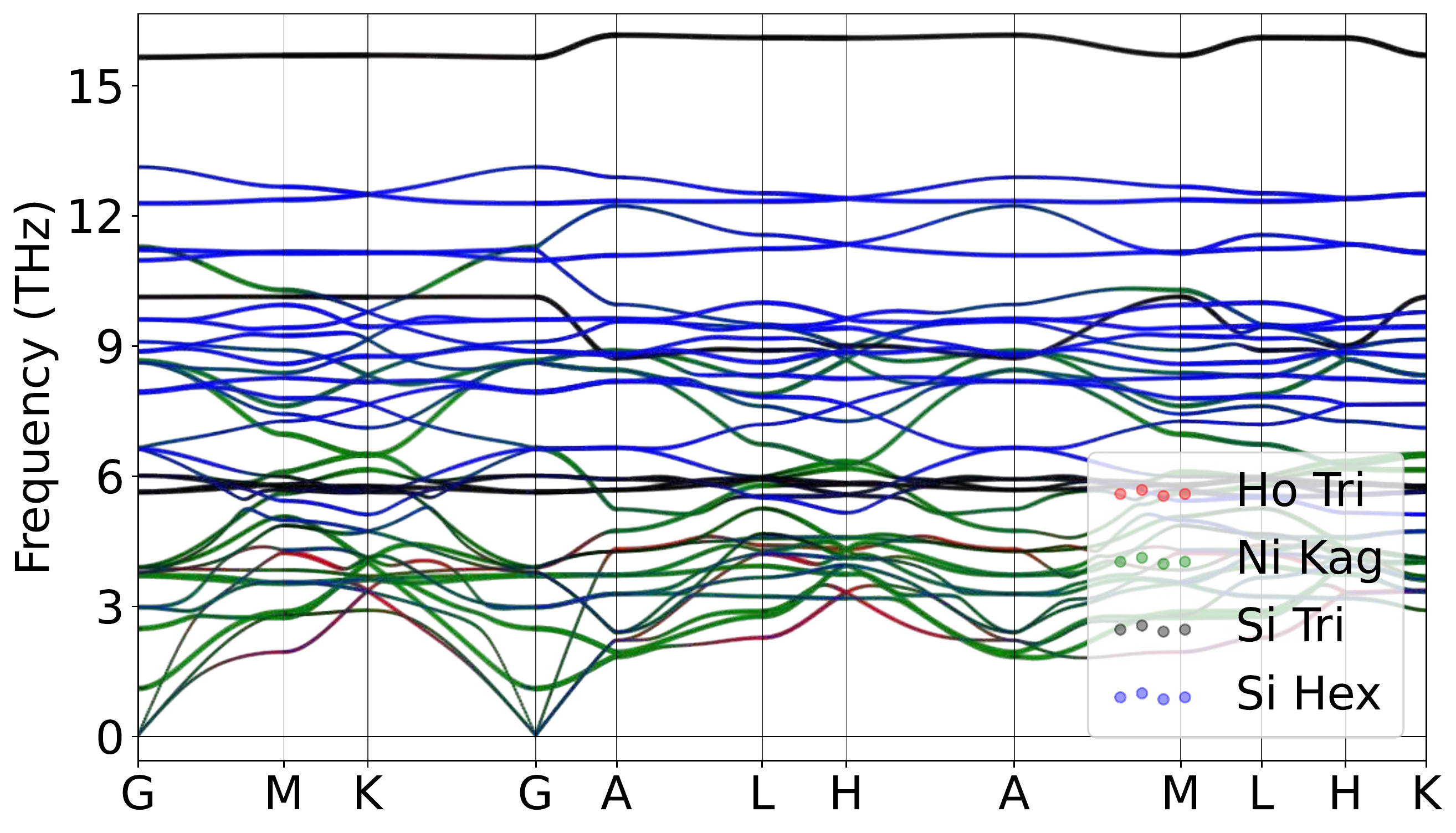}
}
\hspace{5pt}\subfloat[Relaxed phonon at high-T.]{
\label{fig:HoNi6Si6_relaxed_High}
\includegraphics[width=0.45\textwidth]{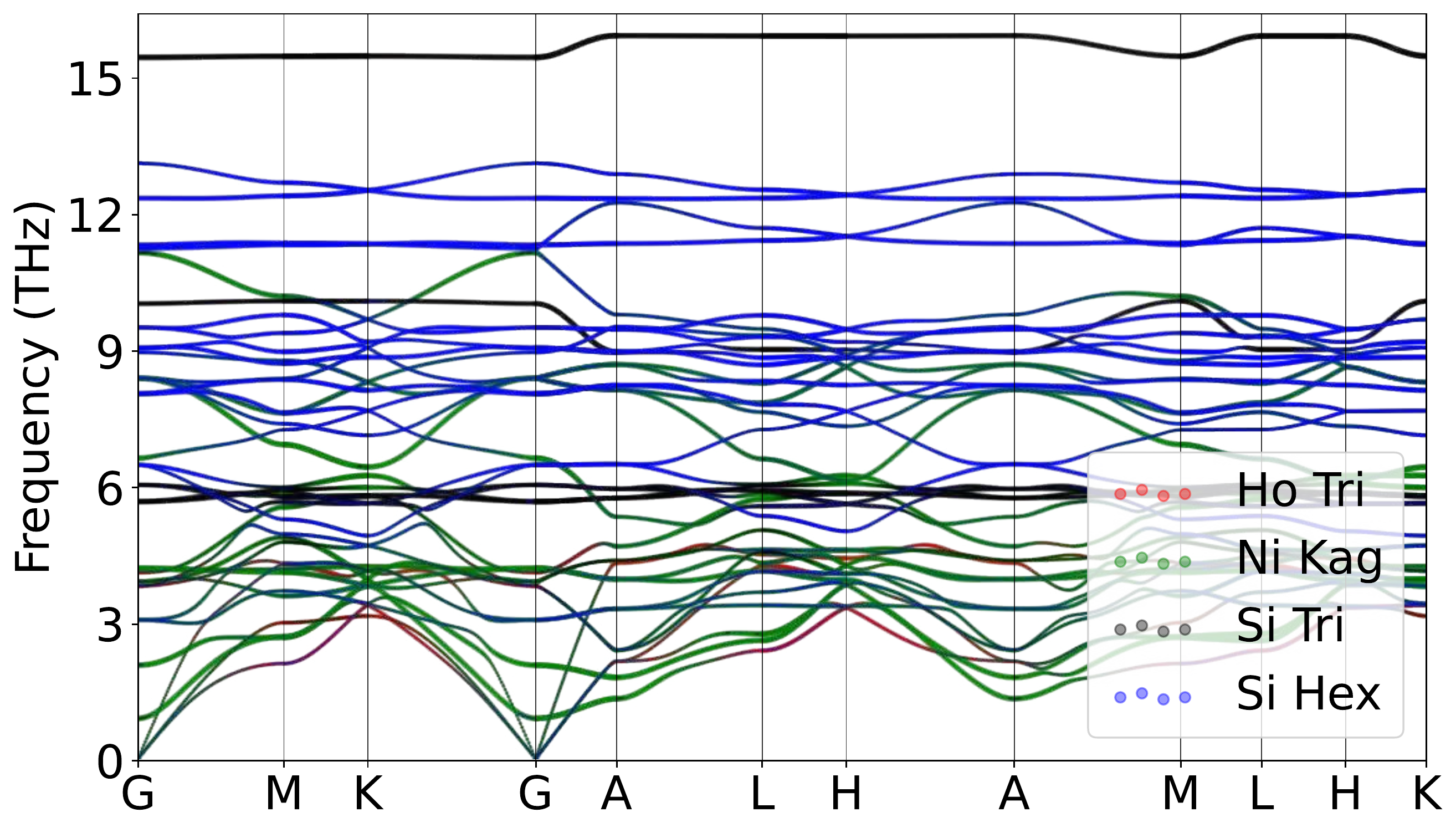}
}
\caption{Atomically projected phonon spectrum for HoNi$_6$Si$_6$.}
\label{fig:HoNi6Si6pho}
\end{figure}

\begin{figure}[H]
\centering
\label{fig:HoNi6Si6_relaxed_Low_xz}
\includegraphics[width=0.85\textwidth]{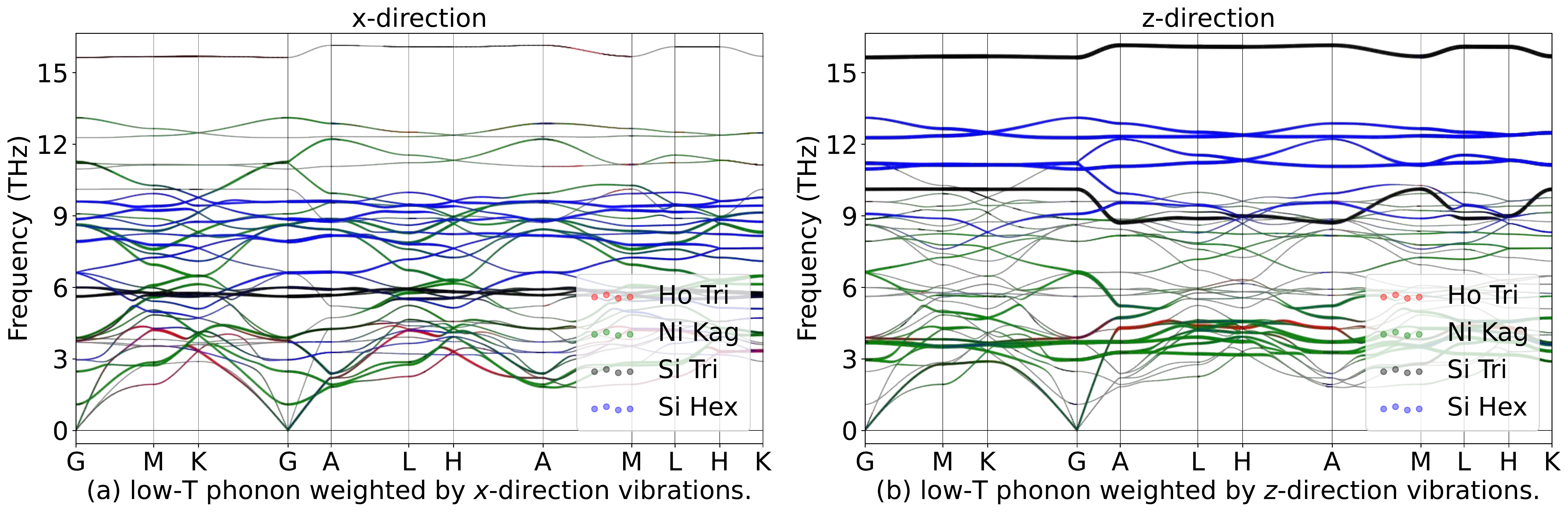}
\hspace{5pt}\label{fig:HoNi6Si6_relaxed_High_xz}
\includegraphics[width=0.85\textwidth]{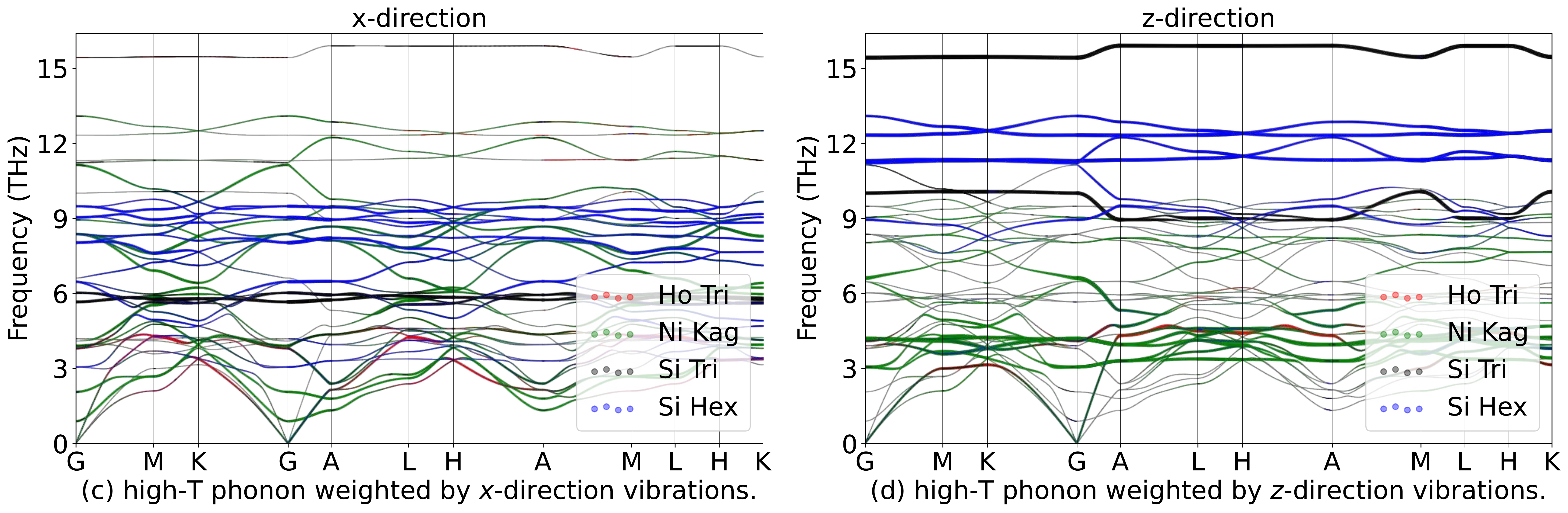}
\caption{Phonon spectrum for HoNi$_6$Si$_6$in in-plane ($x$) and out-of-plane ($z$) directions.}
\label{fig:HoNi6Si6phoxyz}
\end{figure}

\begin{table}[htbp]
\global\long\def\arraystretch{1.12}
\captionsetup{width=.95\textwidth}
\caption{IRREPs at high-symmetry points for HoNi$_6$Si$_6$ phonon spectrum at Low-T.}
\label{tab:191_HoNi6Si6_Low}
\setlength{\tabcolsep}{1.mm}{
}
\end{table}

\begin{table}[htbp]
\global\long\def\arraystretch{1.12}
\captionsetup{width=.95\textwidth}
\caption{IRREPs at high-symmetry points for HoNi$_6$Si$_6$ phonon spectrum at High-T.}
\label{tab:191_HoNi6Si6_High}
\setlength{\tabcolsep}{1.mm}{
%
}
\end{table}
\subsubsection{\label{sms2sec:ErNi6Si6}\hyperref[smsec:col]{\texorpdfstring{ErNi$_6$Si$_6$}{}}~{\color{green}  (High-T stable, Low-T stable) }}

\begin{table}[htbp]
\global\long\def\arraystretch{1.12}
\captionsetup{width=.85\textwidth}
\caption{Structure parameters of ErNi$_6$Si$_6$ after relaxation. It contains 320 electrons. }
\label{tab:ErNi6Si6}
\setlength{\tabcolsep}{4mm}{
%
}
\end{table}

\begin{figure}[htbp]
\centering
\includegraphics[width=0.85\textwidth]{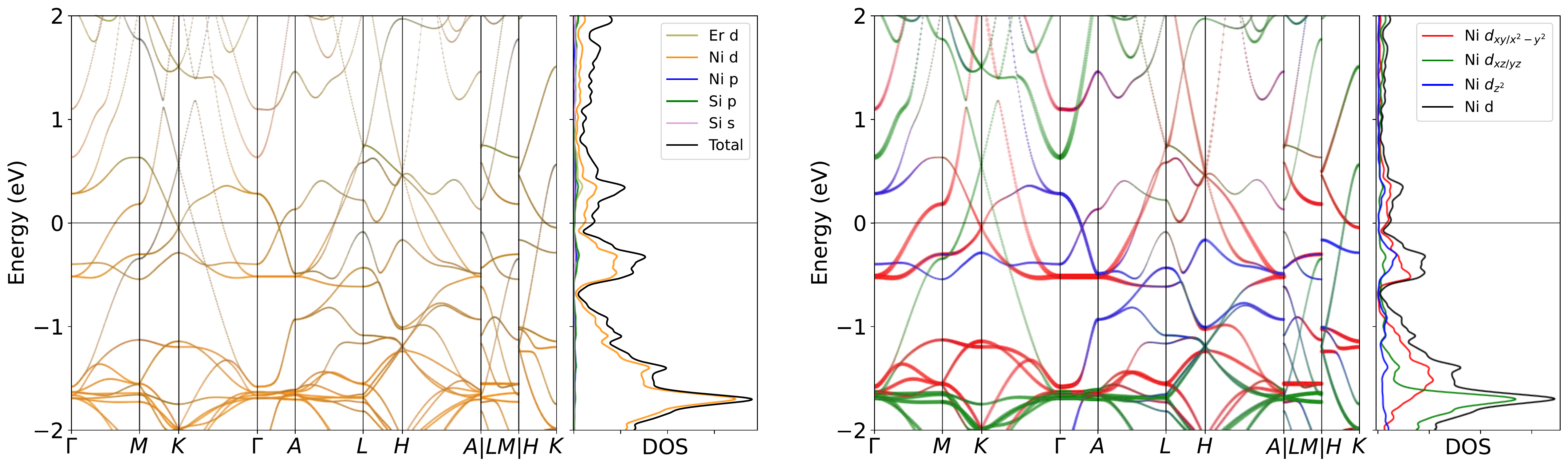}
\caption{Band structure for ErNi$_6$Si$_6$ without SOC. The projection of Ni $3d$ orbitals is also presented. The band structures clearly reveal the dominating of Ni $3d$ orbitals  near the Fermi level, as well as the pronounced peak.}
\label{fig:ErNi6Si6band}
\end{figure}

\begin{figure}[H]
\centering
\subfloat[Relaxed phonon at low-T.]{
\label{fig:ErNi6Si6_relaxed_Low}
\includegraphics[width=0.45\textwidth]{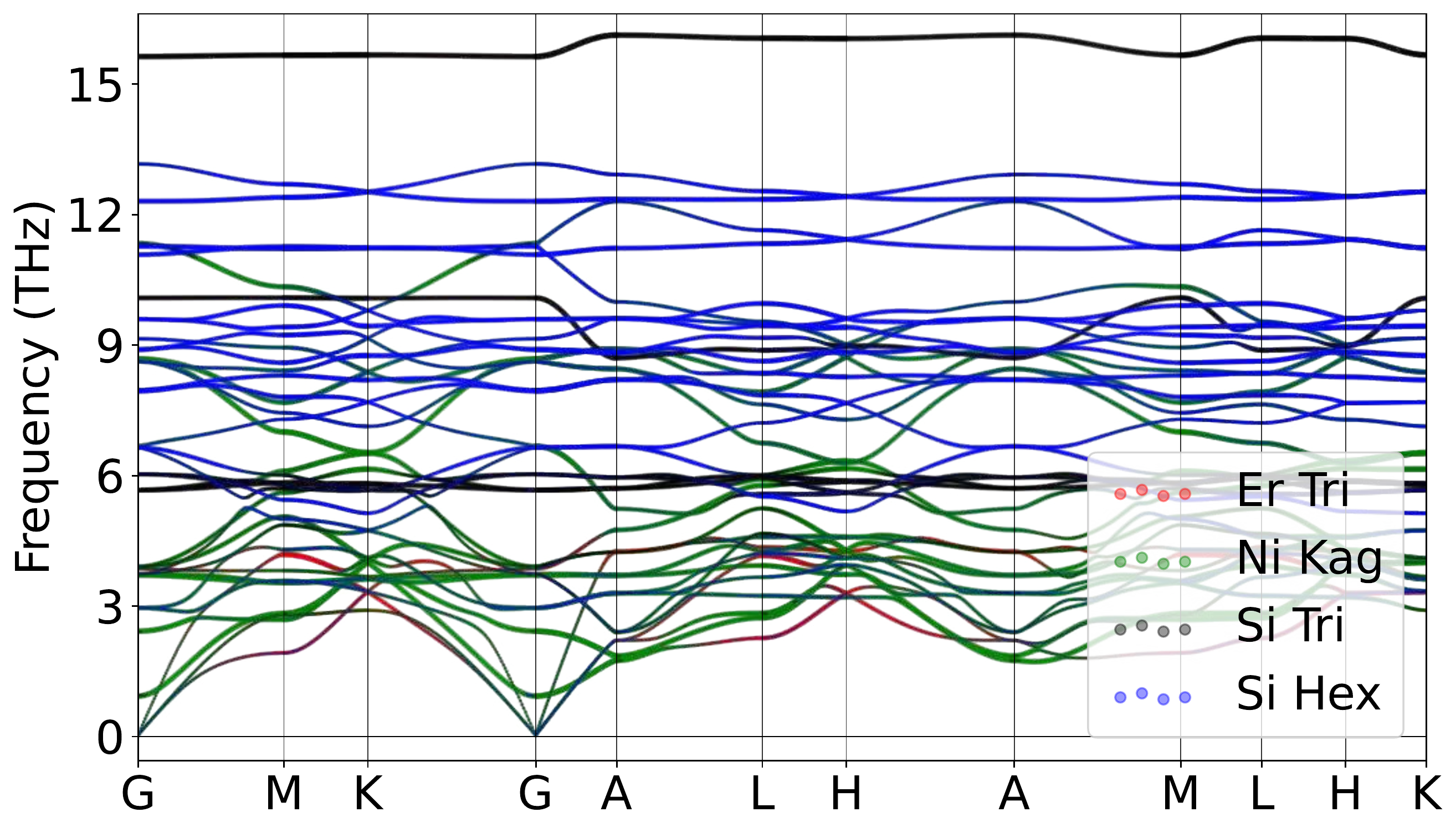}
}
\hspace{5pt}\subfloat[Relaxed phonon at high-T.]{
\label{fig:ErNi6Si6_relaxed_High}
\includegraphics[width=0.45\textwidth]{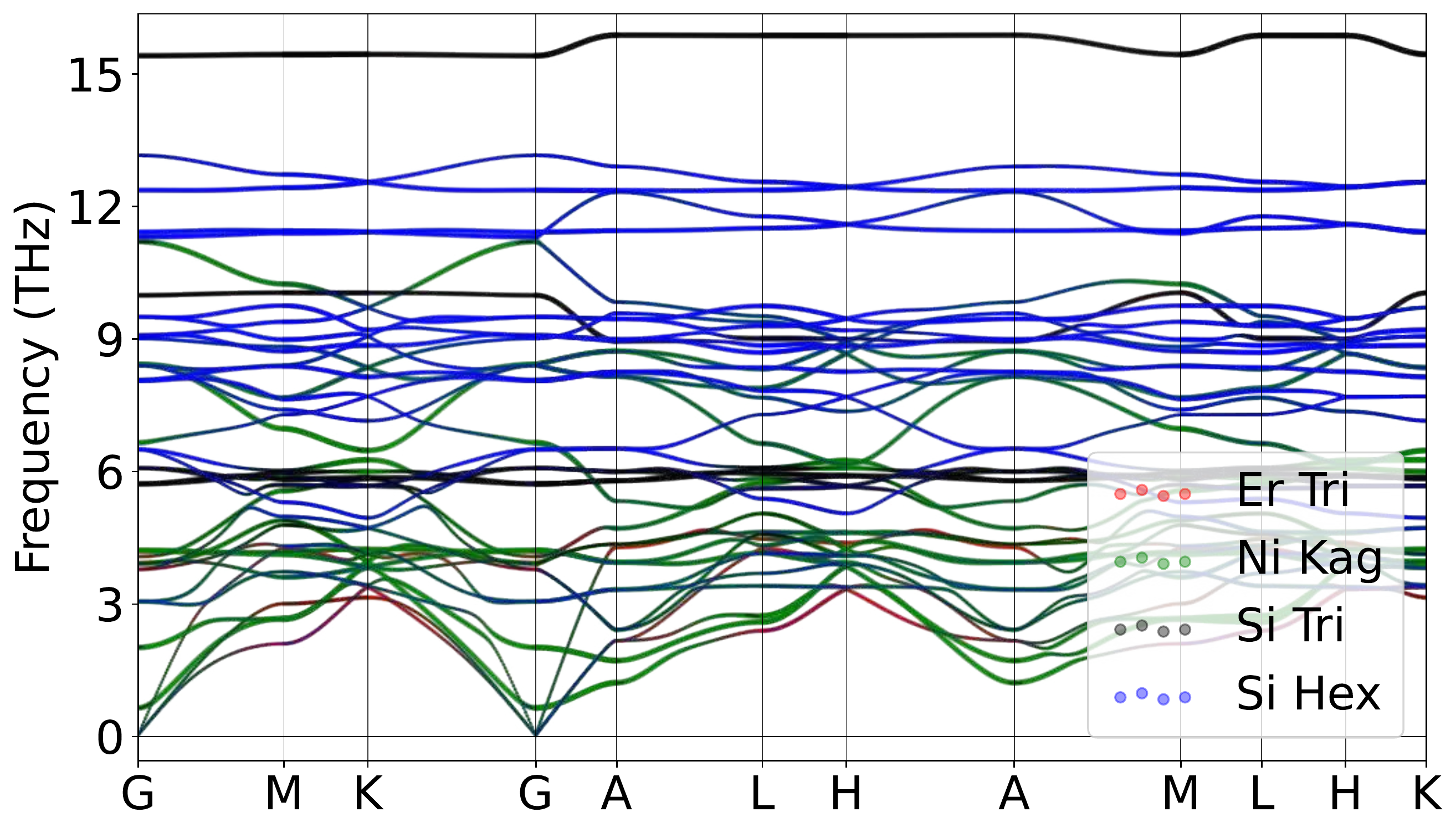}
}
\caption{Atomically projected phonon spectrum for ErNi$_6$Si$_6$.}
\label{fig:ErNi6Si6pho}
\end{figure}

\begin{figure}[H]
\centering
\label{fig:ErNi6Si6_relaxed_Low_xz}
\includegraphics[width=0.85\textwidth]{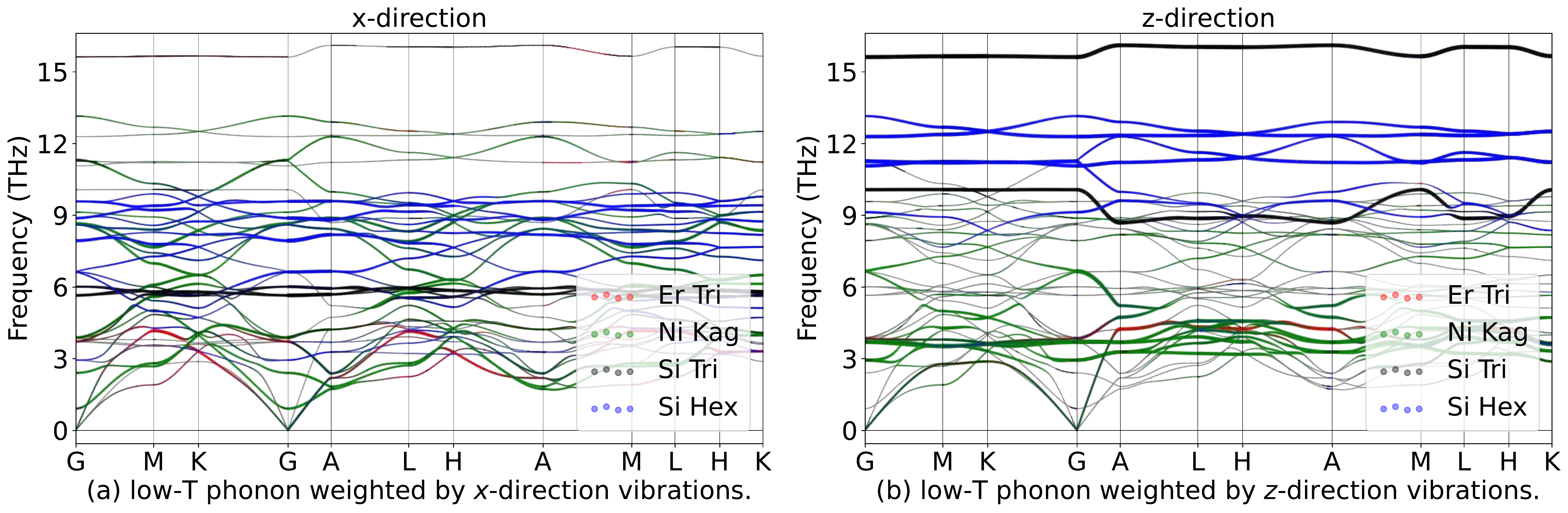}
\hspace{5pt}\label{fig:ErNi6Si6_relaxed_High_xz}
\includegraphics[width=0.85\textwidth]{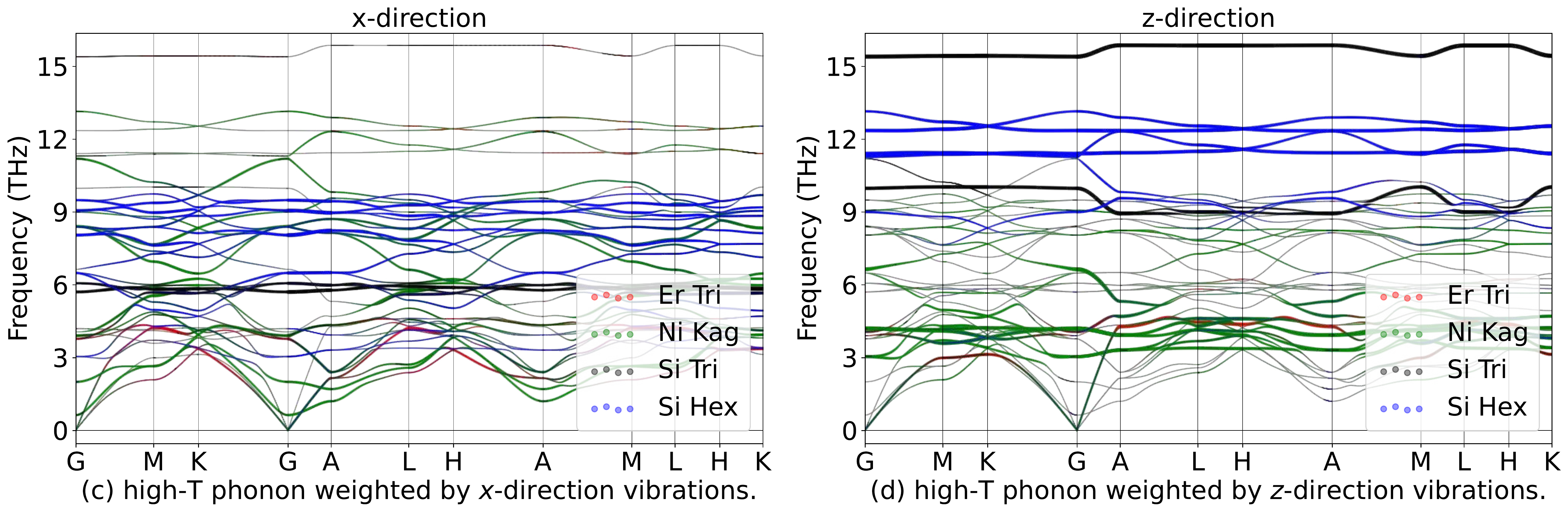}
\caption{Phonon spectrum for ErNi$_6$Si$_6$in in-plane ($x$) and out-of-plane ($z$) directions.}
\label{fig:ErNi6Si6phoxyz}
\end{figure}

\begin{table}[htbp]
\global\long\def\arraystretch{1.12}
\captionsetup{width=.95\textwidth}
\caption{IRREPs at high-symmetry points for ErNi$_6$Si$_6$ phonon spectrum at Low-T.}
\label{tab:191_ErNi6Si6_Low}
\setlength{\tabcolsep}{1.mm}{
}
\end{table}

\begin{table}[htbp]
\global\long\def\arraystretch{1.12}
\captionsetup{width=.95\textwidth}
\caption{IRREPs at high-symmetry points for ErNi$_6$Si$_6$ phonon spectrum at High-T.}
\label{tab:191_ErNi6Si6_High}
\setlength{\tabcolsep}{1.mm}{
%
}
\end{table}
\subsubsection{\label{sms2sec:TmNi6Si6}\hyperref[smsec:col]{\texorpdfstring{TmNi$_6$Si$_6$}{}}~{\color{black} (TBD) }}

\begin{table}[htbp]
\global\long\def\arraystretch{1.12}
\captionsetup{width=.85\textwidth}
\caption{Structure parameters of TmNi$_6$Si$_6$ after relaxation. It contains 321 electrons. }
\label{tab:TmNi6Si6}
\setlength{\tabcolsep}{4mm}{
%
}
\end{table}

\begin{figure}[htbp]
\centering
\includegraphics[width=0.85\textwidth]{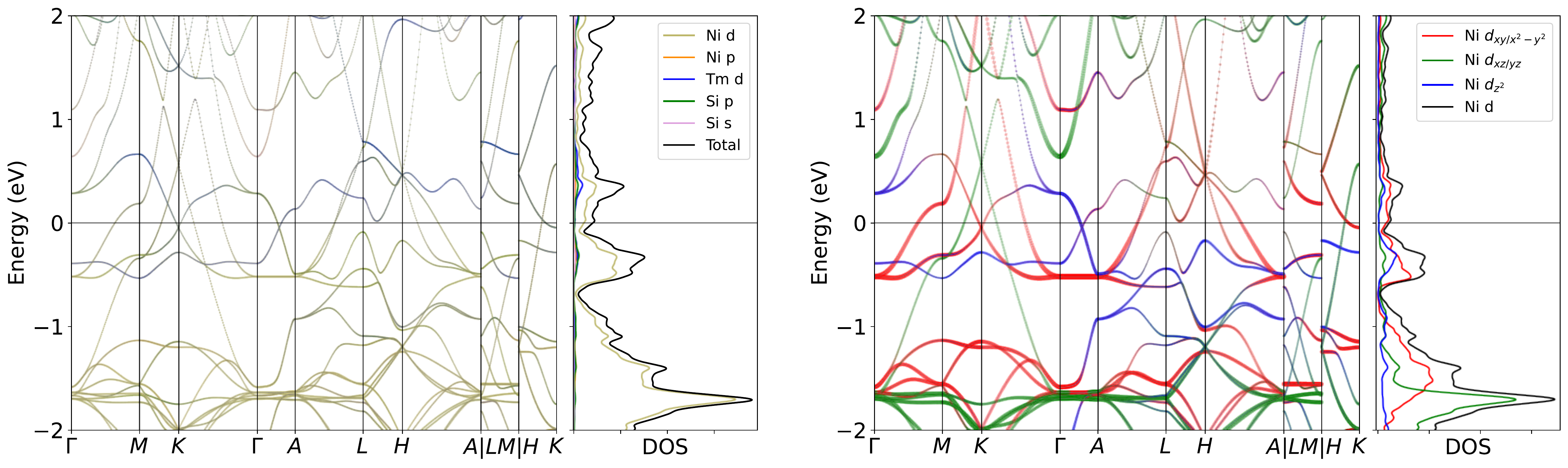}
\caption{Band structure for TmNi$_6$Si$_6$ without SOC. The projection of Ni $3d$ orbitals is also presented. The band structures clearly reveal the dominating of Ni $3d$ orbitals  near the Fermi level, as well as the pronounced peak.}
\label{fig:TmNi6Si6band}
\end{figure}

\begin{figure}[H]
\centering
\subfloat[Relaxed phonon at low-T.]{
\label{fig:TmNi6Si6_relaxed_Low}
\includegraphics[width=0.45\textwidth]{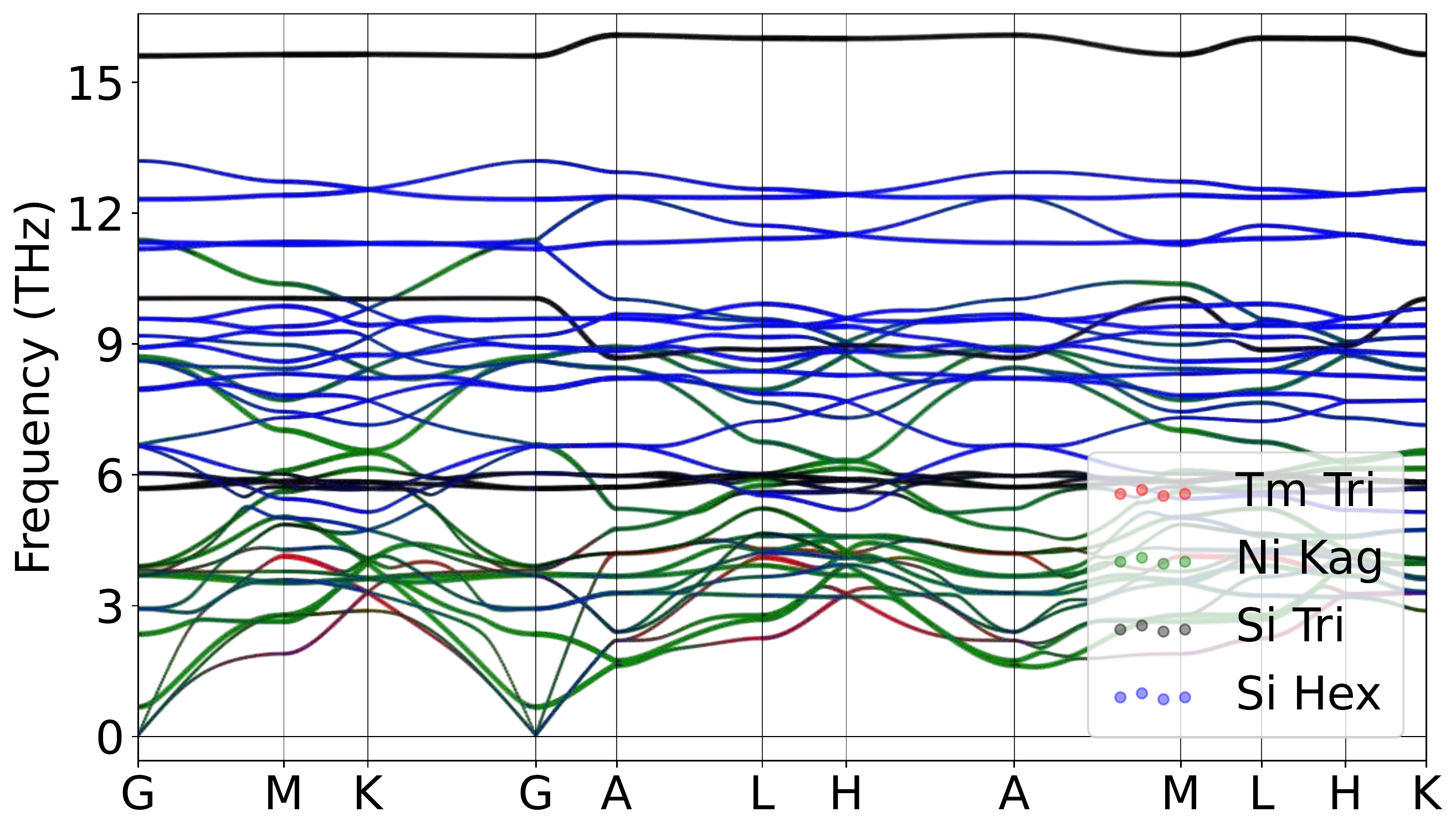}
}
\hspace{5pt}\subfloat[Relaxed phonon at high-T.]{
\label{fig:TmNi6Si6_relaxed_High}
\includegraphics[width=0.45\textwidth]{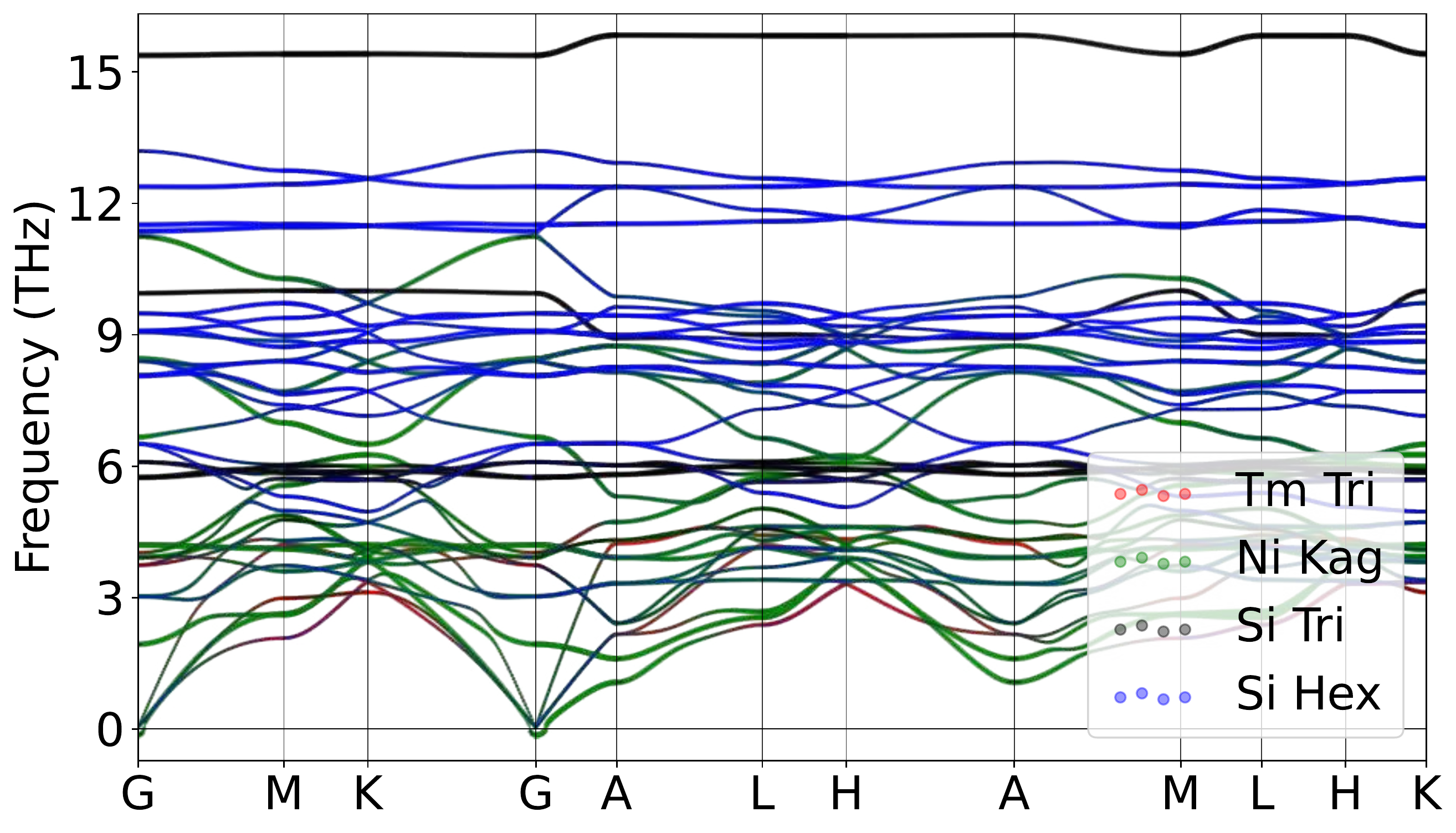}
}
\caption{Atomically projected phonon spectrum for TmNi$_6$Si$_6$.}
\label{fig:TmNi6Si6pho}
\end{figure}

\begin{figure}[H]
\centering
\label{fig:TmNi6Si6_relaxed_Low_xz}
\includegraphics[width=0.85\textwidth]{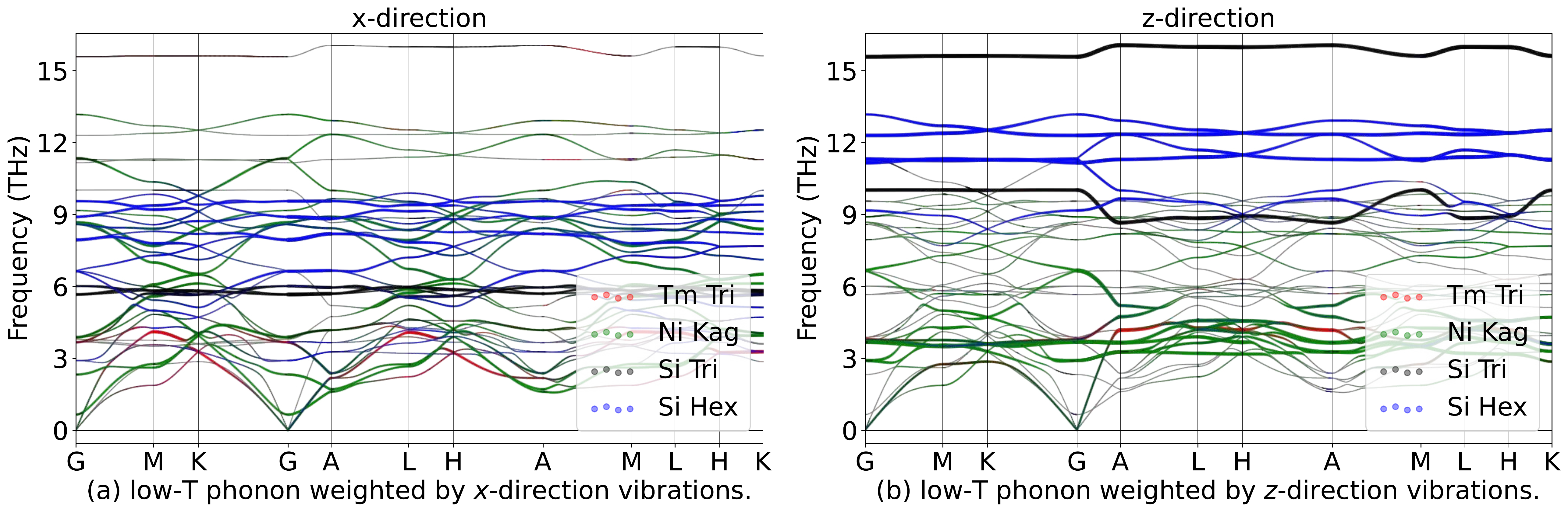}
\hspace{5pt}\label{fig:TmNi6Si6_relaxed_High_xz}
\includegraphics[width=0.85\textwidth]{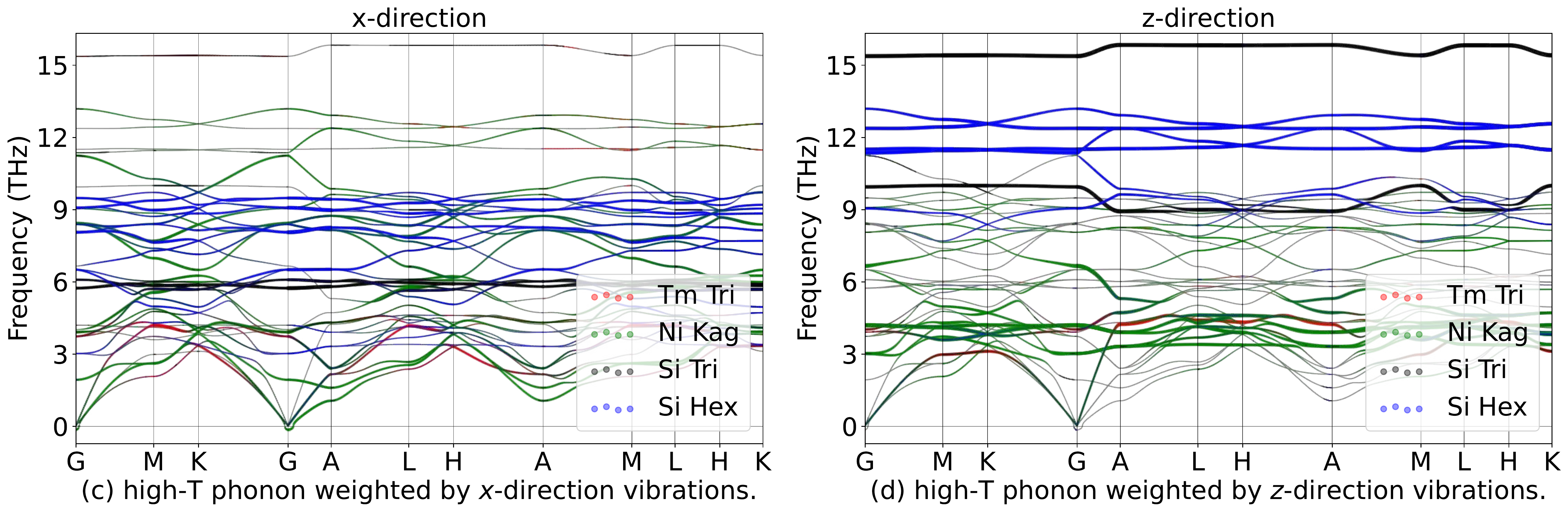}
\caption{Phonon spectrum for TmNi$_6$Si$_6$in in-plane ($x$) and out-of-plane ($z$) directions.}
\label{fig:TmNi6Si6phoxyz}
\end{figure}

\begin{table}[htbp]
\global\long\def\arraystretch{1.12}
\captionsetup{width=.95\textwidth}
\caption{IRREPs at high-symmetry points for TmNi$_6$Si$_6$ phonon spectrum at Low-T.}
\label{tab:191_TmNi6Si6_Low}
\setlength{\tabcolsep}{1.mm}{
}
\end{table}

\begin{table}[htbp]
\global\long\def\arraystretch{1.12}
\captionsetup{width=.95\textwidth}
\caption{IRREPs at high-symmetry points for TmNi$_6$Si$_6$ phonon spectrum at High-T.}
\label{tab:191_TmNi6Si6_High}
\setlength{\tabcolsep}{1.mm}{
%
}
\end{table}
\subsubsection{\label{sms2sec:YbNi6Si6}\hyperref[smsec:col]{\texorpdfstring{YbNi$_6$Si$_6$}{}}~{\color{blue}  (High-T stable, Low-T unstable) }}

\begin{table}[htbp]
\global\long\def\arraystretch{1.12}
\captionsetup{width=.85\textwidth}
\caption{Structure parameters of YbNi$_6$Si$_6$ after relaxation. It contains 322 electrons. }
\label{tab:YbNi6Si6}
\setlength{\tabcolsep}{4mm}{
%
}
\end{table}

\begin{figure}[htbp]
\centering
\includegraphics[width=0.85\textwidth]{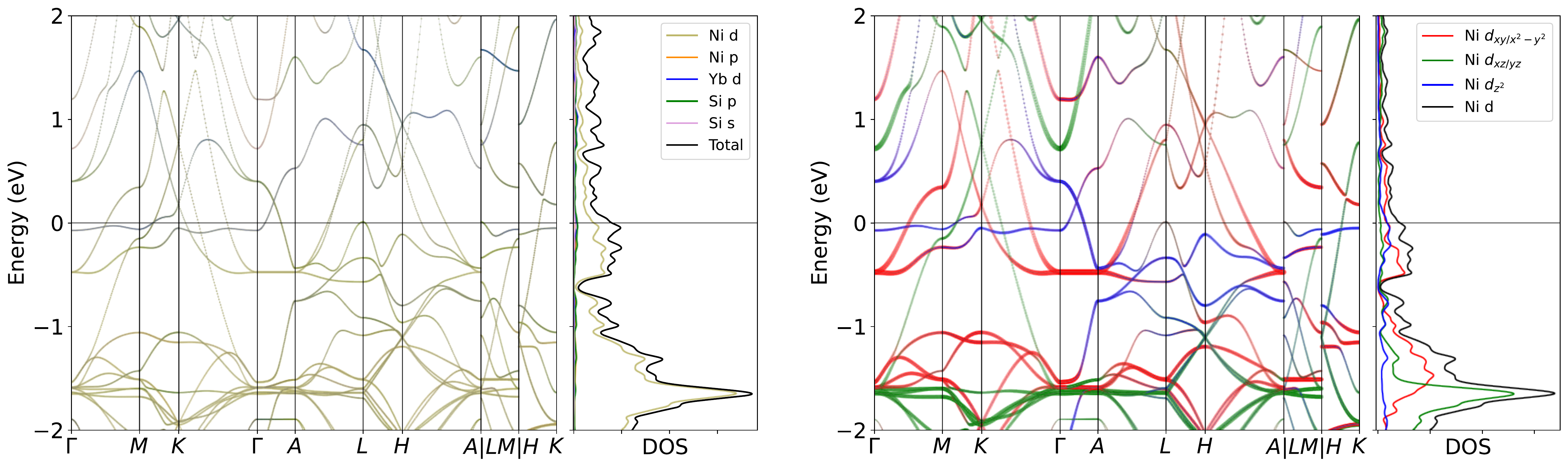}
\caption{Band structure for YbNi$_6$Si$_6$ without SOC. The projection of Ni $3d$ orbitals is also presented. The band structures clearly reveal the dominating of Ni $3d$ orbitals  near the Fermi level, as well as the pronounced peak.}
\label{fig:YbNi6Si6band}
\end{figure}

\begin{figure}[H]
\centering
\subfloat[Relaxed phonon at low-T.]{
\label{fig:YbNi6Si6_relaxed_Low}
\includegraphics[width=0.45\textwidth]{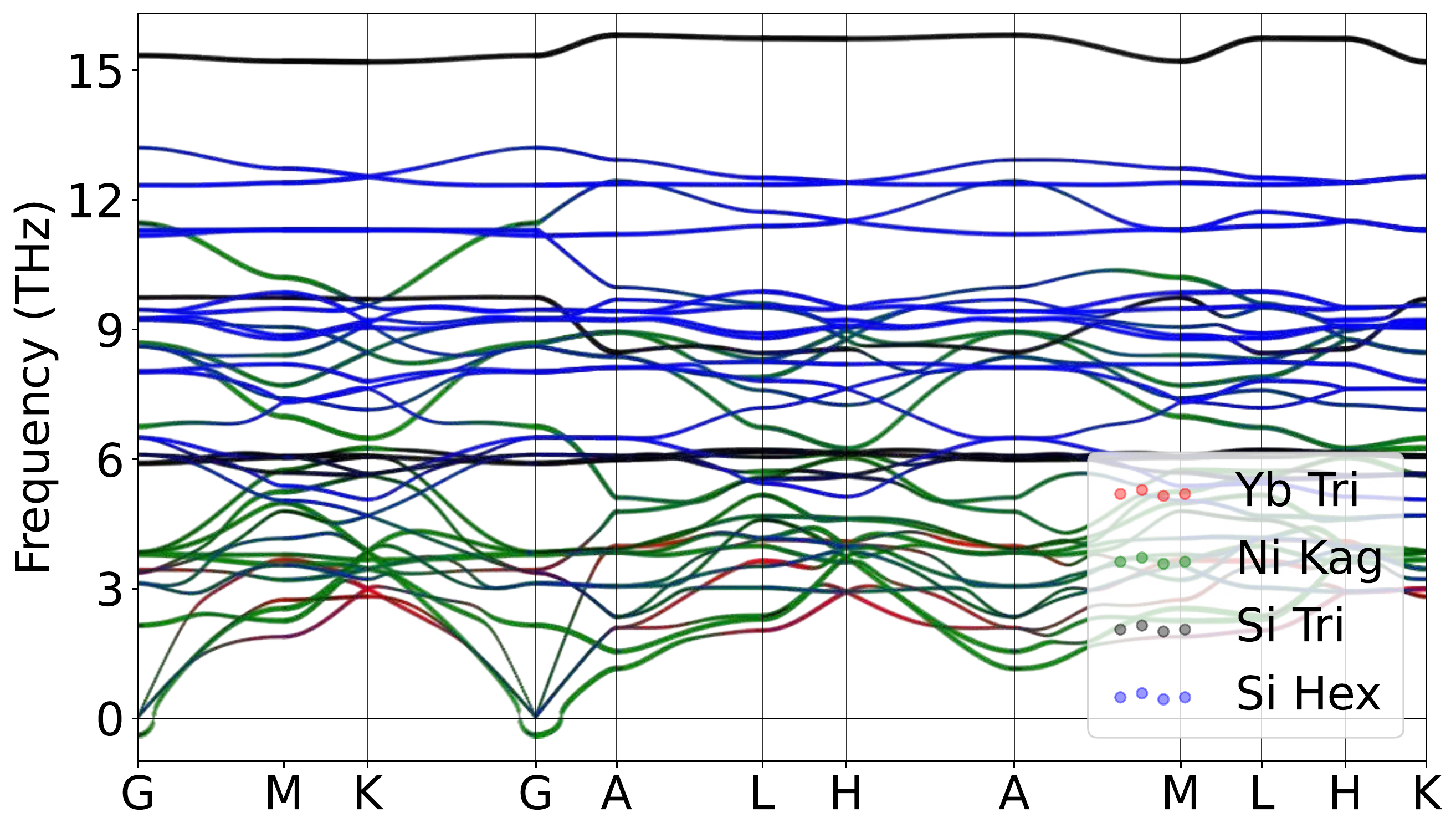}
}
\hspace{5pt}\subfloat[Relaxed phonon at high-T.]{
\label{fig:YbNi6Si6_relaxed_High}
\includegraphics[width=0.45\textwidth]{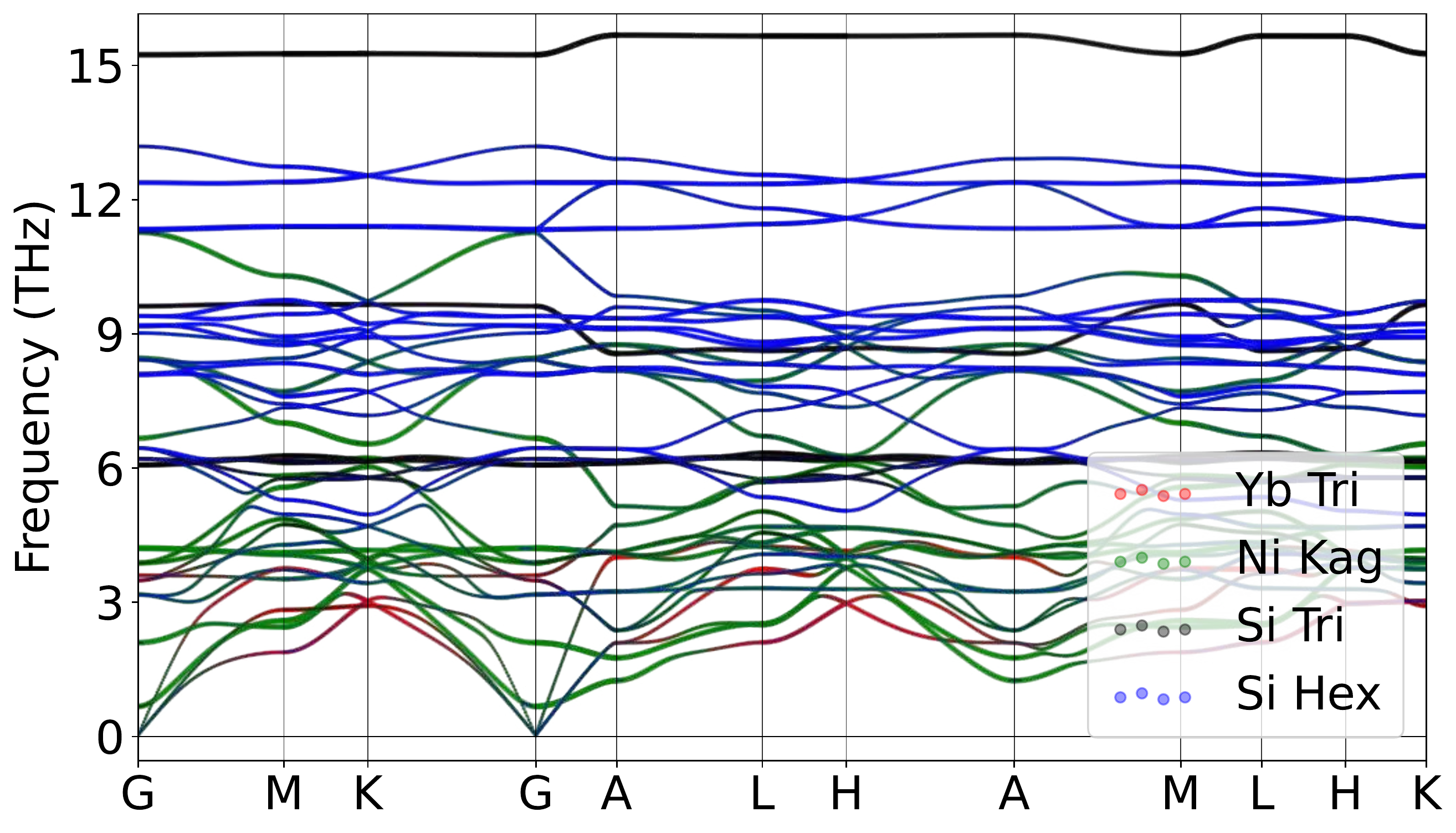}
}
\caption{Atomically projected phonon spectrum for YbNi$_6$Si$_6$.}
\label{fig:YbNi6Si6pho}
\end{figure}

\begin{figure}[H]
\centering
\label{fig:YbNi6Si6_relaxed_Low_xz}
\includegraphics[width=0.85\textwidth]{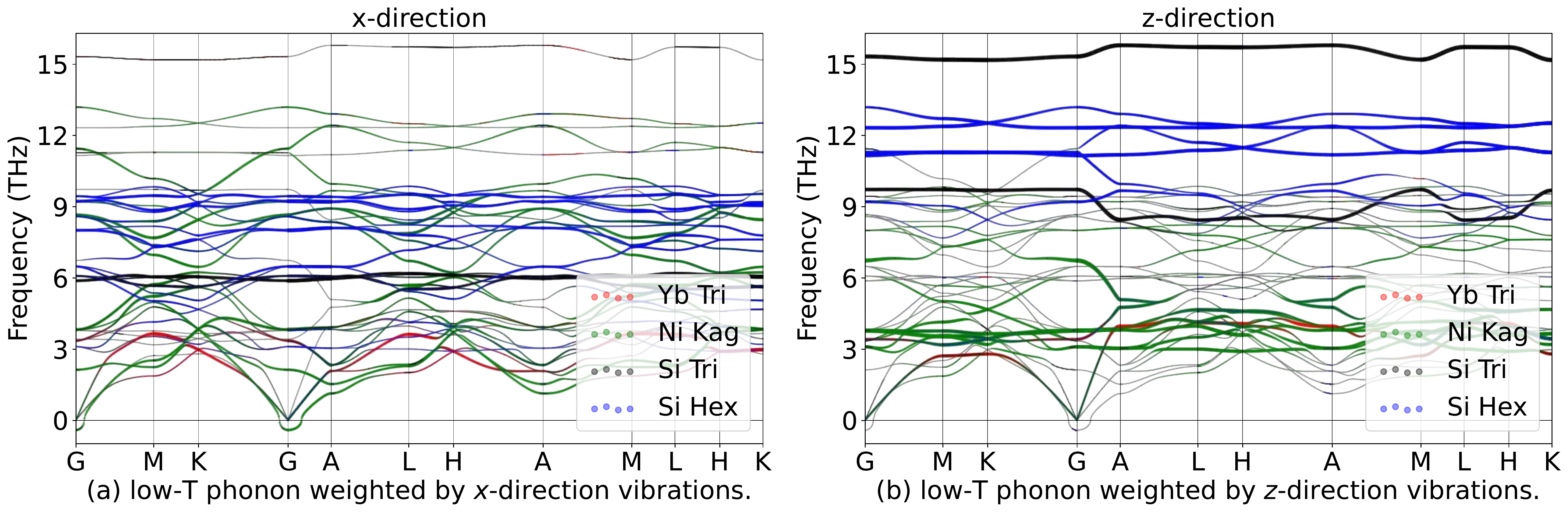}
\hspace{5pt}\label{fig:YbNi6Si6_relaxed_High_xz}
\includegraphics[width=0.85\textwidth]{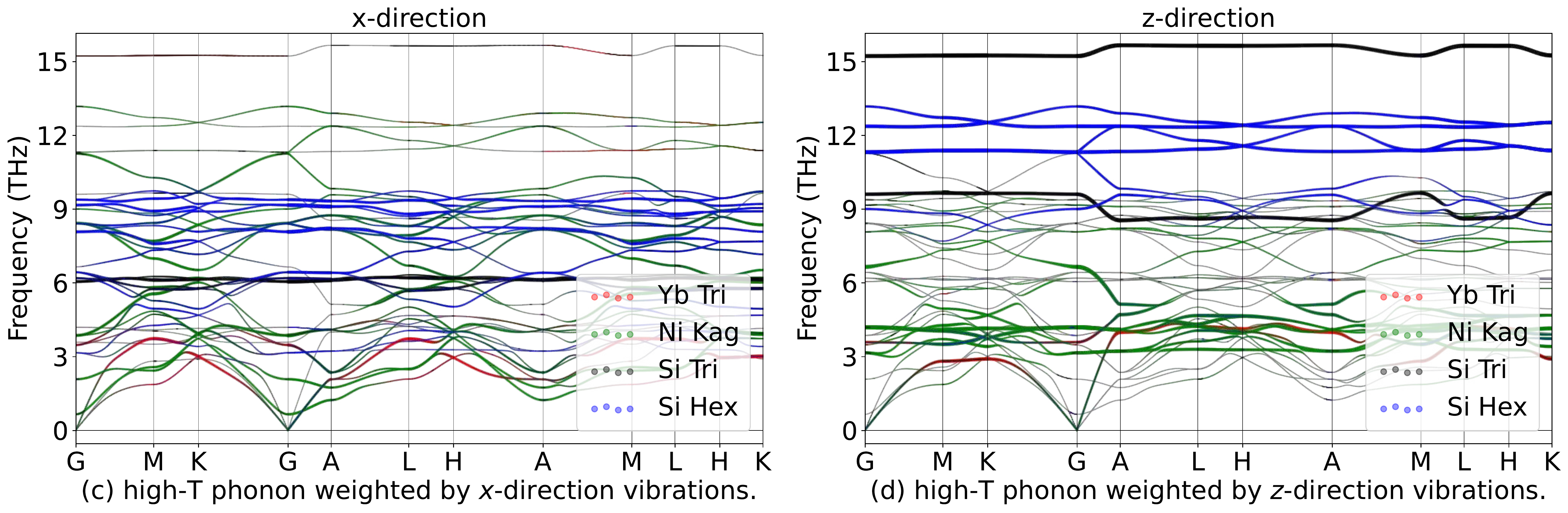}
\caption{Phonon spectrum for YbNi$_6$Si$_6$in in-plane ($x$) and out-of-plane ($z$) directions.}
\label{fig:YbNi6Si6phoxyz}
\end{figure}

\begin{table}[htbp]
\global\long\def\arraystretch{1.12}
\captionsetup{width=.95\textwidth}
\caption{IRREPs at high-symmetry points for YbNi$_6$Si$_6$ phonon spectrum at Low-T.}
\label{tab:191_YbNi6Si6_Low}
\setlength{\tabcolsep}{1.mm}{
}
\end{table}

\begin{table}[htbp]
\global\long\def\arraystretch{1.12}
\captionsetup{width=.95\textwidth}
\caption{IRREPs at high-symmetry points for YbNi$_6$Si$_6$ phonon spectrum at High-T.}
\label{tab:191_YbNi6Si6_High}
\setlength{\tabcolsep}{1.mm}{
%
}
\end{table}
\subsubsection{\label{sms2sec:LuNi6Si6}\hyperref[smsec:col]{\texorpdfstring{LuNi$_6$Si$_6$}{}}~{\color{red}  (High-T unstable, Low-T unstable) }}

\begin{table}[htbp]
\global\long\def\arraystretch{1.12}
\captionsetup{width=.85\textwidth}
\caption{Structure parameters of LuNi$_6$Si$_6$ after relaxation. It contains 323 electrons. }
\label{tab:LuNi6Si6}
\setlength{\tabcolsep}{4mm}{
%
}
\end{table}

\begin{figure}[htbp]
\centering
\includegraphics[width=0.85\textwidth]{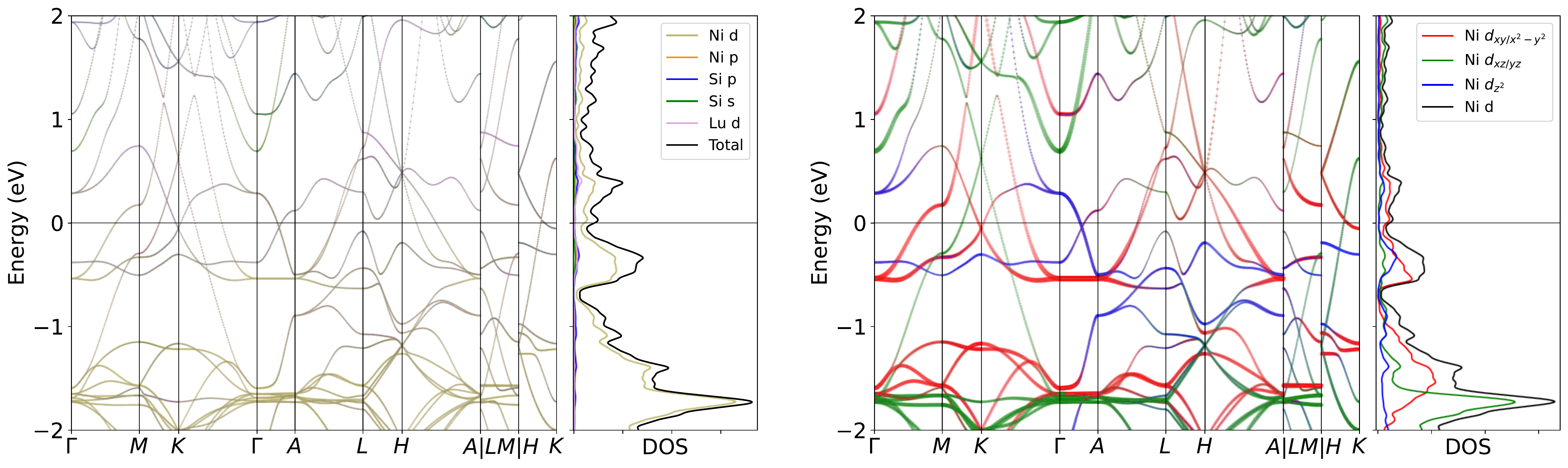}
\caption{Band structure for LuNi$_6$Si$_6$ without SOC. The projection of Ni $3d$ orbitals is also presented. The band structures clearly reveal the dominating of Ni $3d$ orbitals  near the Fermi level, as well as the pronounced peak.}
\label{fig:LuNi6Si6band}
\end{figure}

\begin{figure}[H]
\centering
\subfloat[Relaxed phonon at low-T.]{
\label{fig:LuNi6Si6_relaxed_Low}
\includegraphics[width=0.45\textwidth]{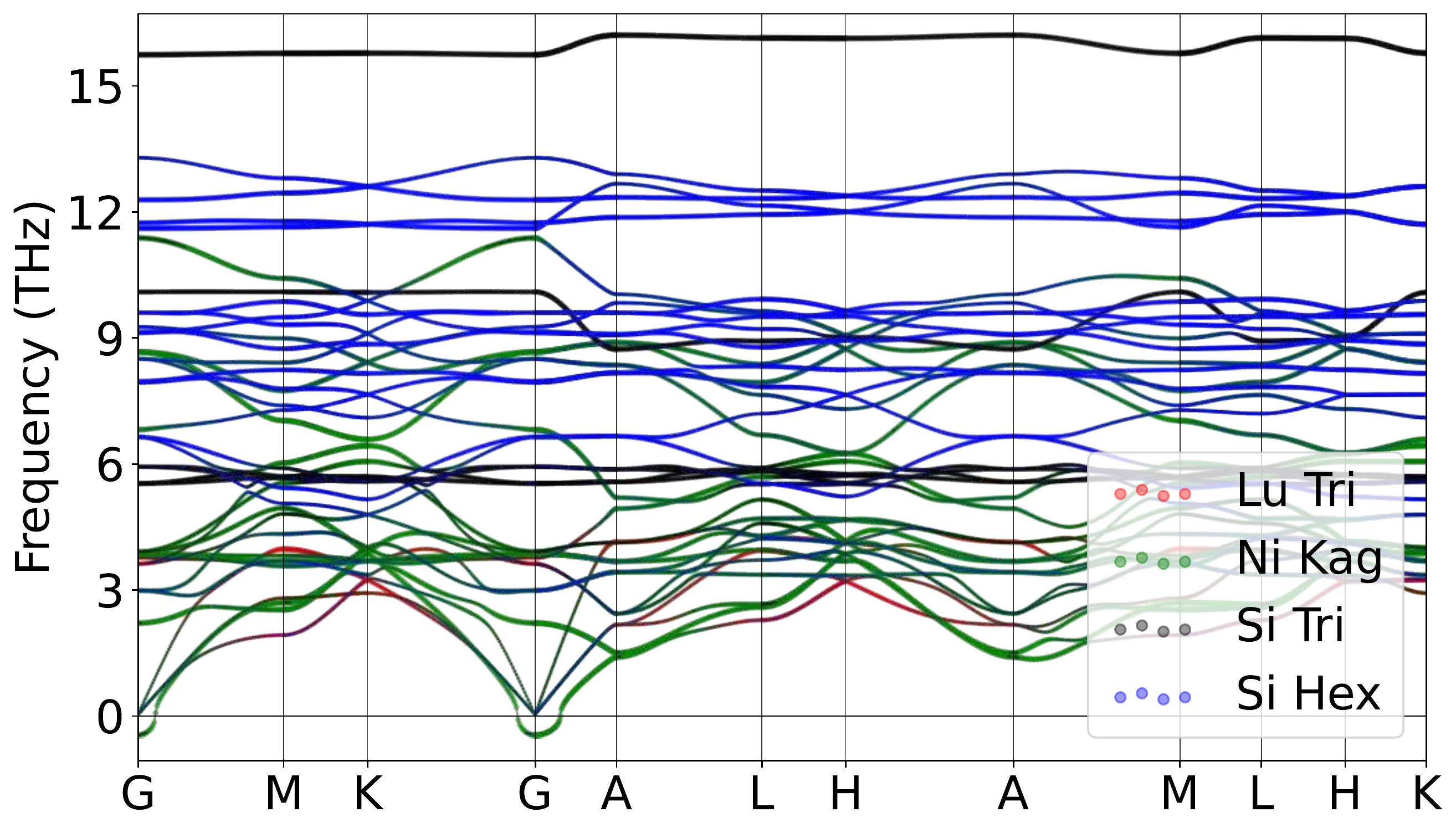}
}
\hspace{5pt}\subfloat[Relaxed phonon at high-T.]{
\label{fig:LuNi6Si6_relaxed_High}
\includegraphics[width=0.45\textwidth]{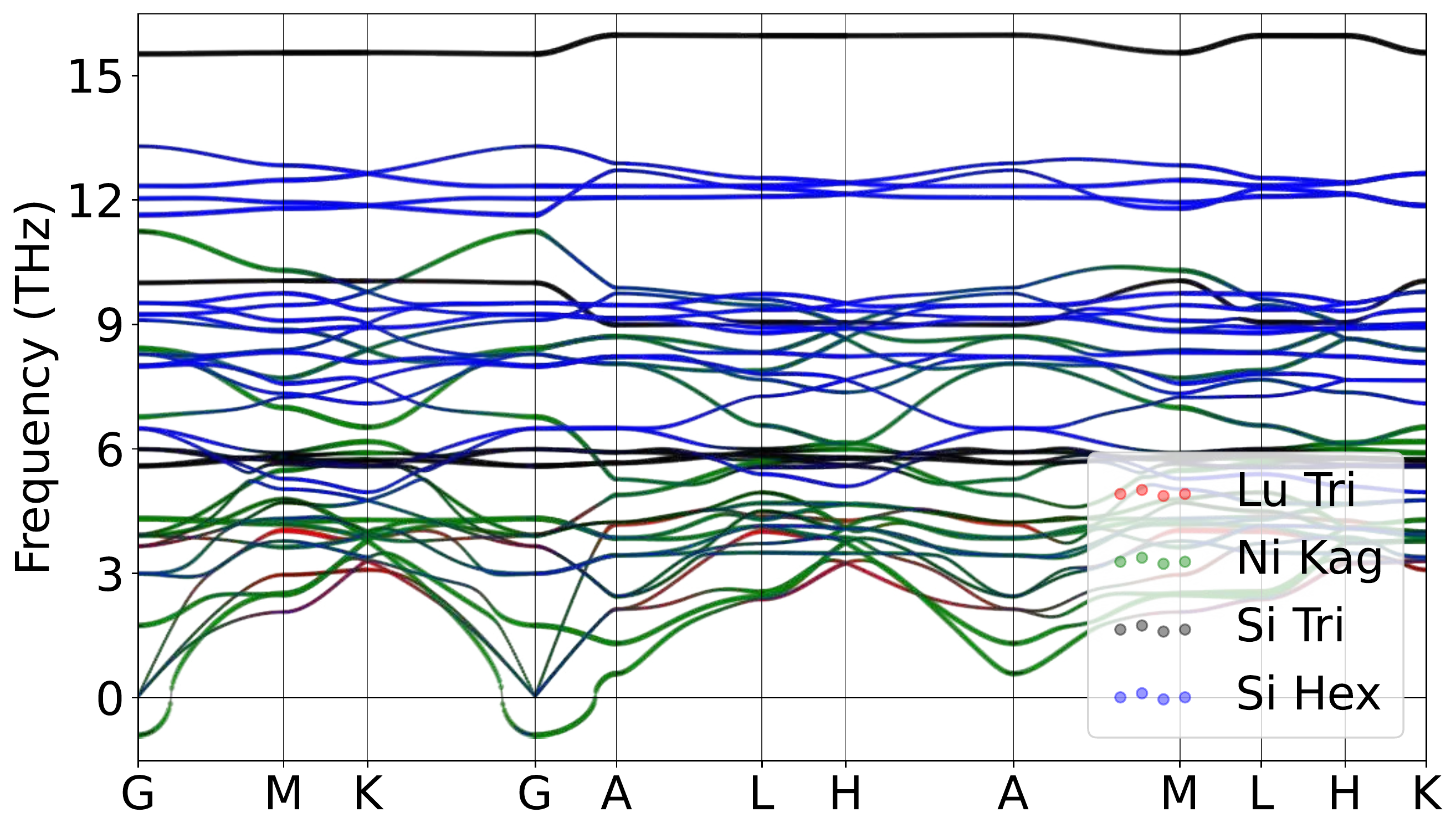}
}
\caption{Atomically projected phonon spectrum for LuNi$_6$Si$_6$.}
\label{fig:LuNi6Si6pho}
\end{figure}

\begin{figure}[H]
\centering
\label{fig:LuNi6Si6_relaxed_Low_xz}
\includegraphics[width=0.85\textwidth]{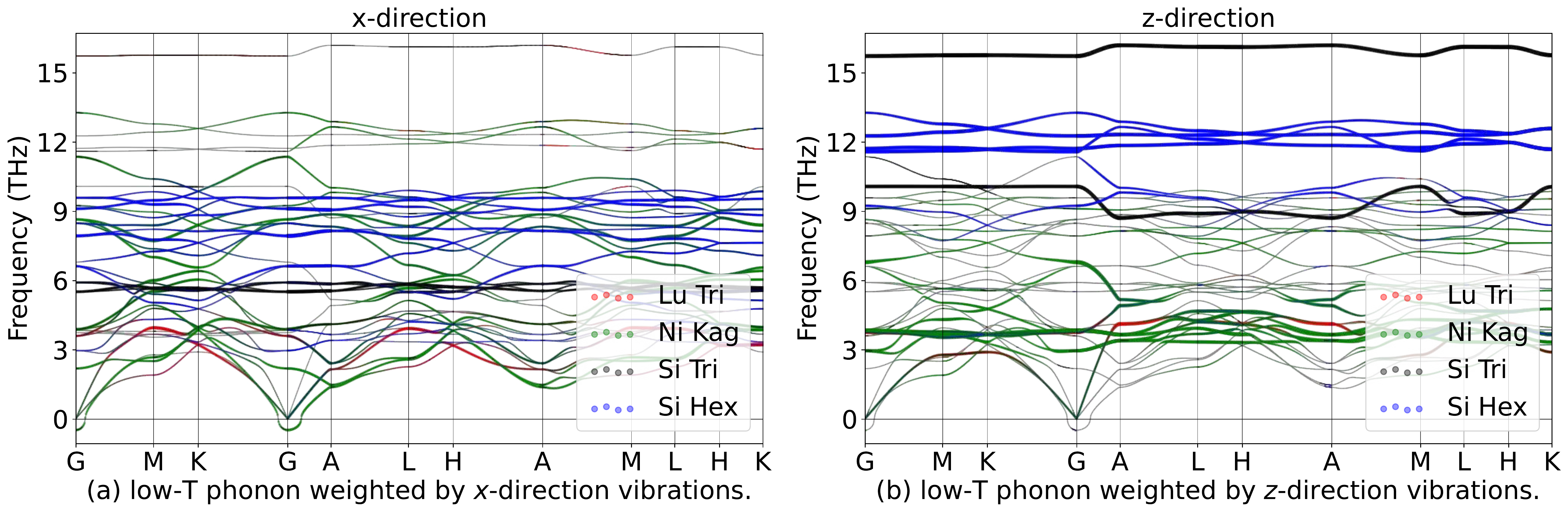}
\hspace{5pt}\label{fig:LuNi6Si6_relaxed_High_xz}
\includegraphics[width=0.85\textwidth]{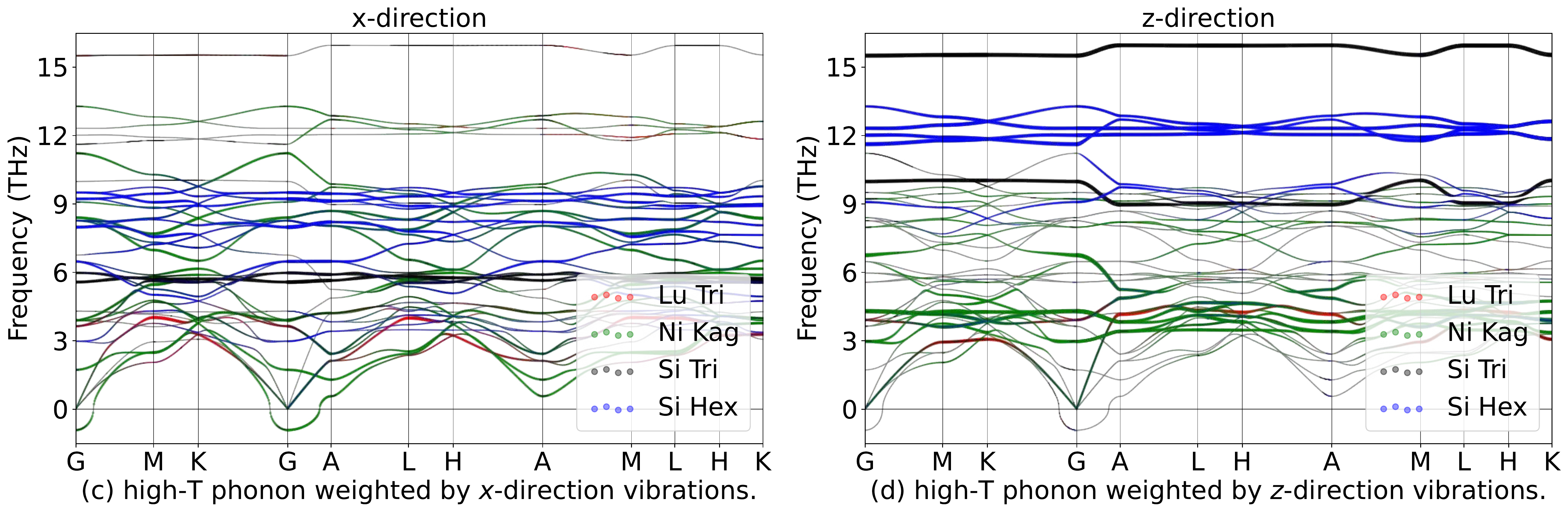}
\caption{Phonon spectrum for LuNi$_6$Si$_6$in in-plane ($x$) and out-of-plane ($z$) directions.}
\label{fig:LuNi6Si6phoxyz}
\end{figure}

\begin{table}[htbp]
\global\long\def\arraystretch{1.12}
\captionsetup{width=.95\textwidth}
\caption{IRREPs at high-symmetry points for LuNi$_6$Si$_6$ phonon spectrum at Low-T.}
\label{tab:191_LuNi6Si6_Low}
\setlength{\tabcolsep}{1.mm}{
}
\end{table}

\begin{table}[htbp]
\global\long\def\arraystretch{1.12}
\captionsetup{width=.95\textwidth}
\caption{IRREPs at high-symmetry points for LuNi$_6$Si$_6$ phonon spectrum at High-T.}
\label{tab:191_LuNi6Si6_High}
\setlength{\tabcolsep}{1.mm}{
%
}
\end{table}
\subsubsection{\label{sms2sec:HfNi6Si6}\hyperref[smsec:col]{\texorpdfstring{HfNi$_6$Si$_6$}{}}~{\color{red}  (High-T unstable, Low-T unstable) }}

\begin{table}[htbp]
\global\long\def\arraystretch{1.12}
\captionsetup{width=.85\textwidth}
\caption{Structure parameters of HfNi$_6$Si$_6$ after relaxation. It contains 324 electrons. }
\label{tab:HfNi6Si6}
\setlength{\tabcolsep}{4mm}{
%
}
\end{table}

\begin{figure}[htbp]
\centering
\includegraphics[width=0.85\textwidth]{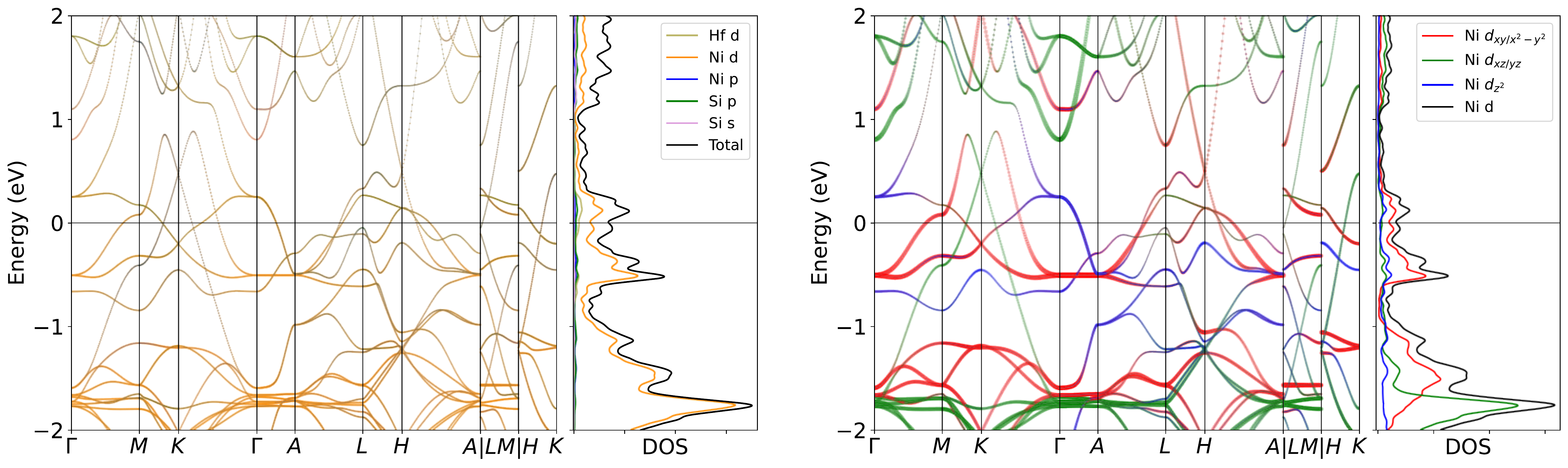}
\caption{Band structure for HfNi$_6$Si$_6$ without SOC. The projection of Ni $3d$ orbitals is also presented. The band structures clearly reveal the dominating of Ni $3d$ orbitals  near the Fermi level, as well as the pronounced peak.}
\label{fig:HfNi6Si6band}
\end{figure}

\begin{figure}[H]
\centering
\subfloat[Relaxed phonon at low-T.]{
\label{fig:HfNi6Si6_relaxed_Low}
\includegraphics[width=0.45\textwidth]{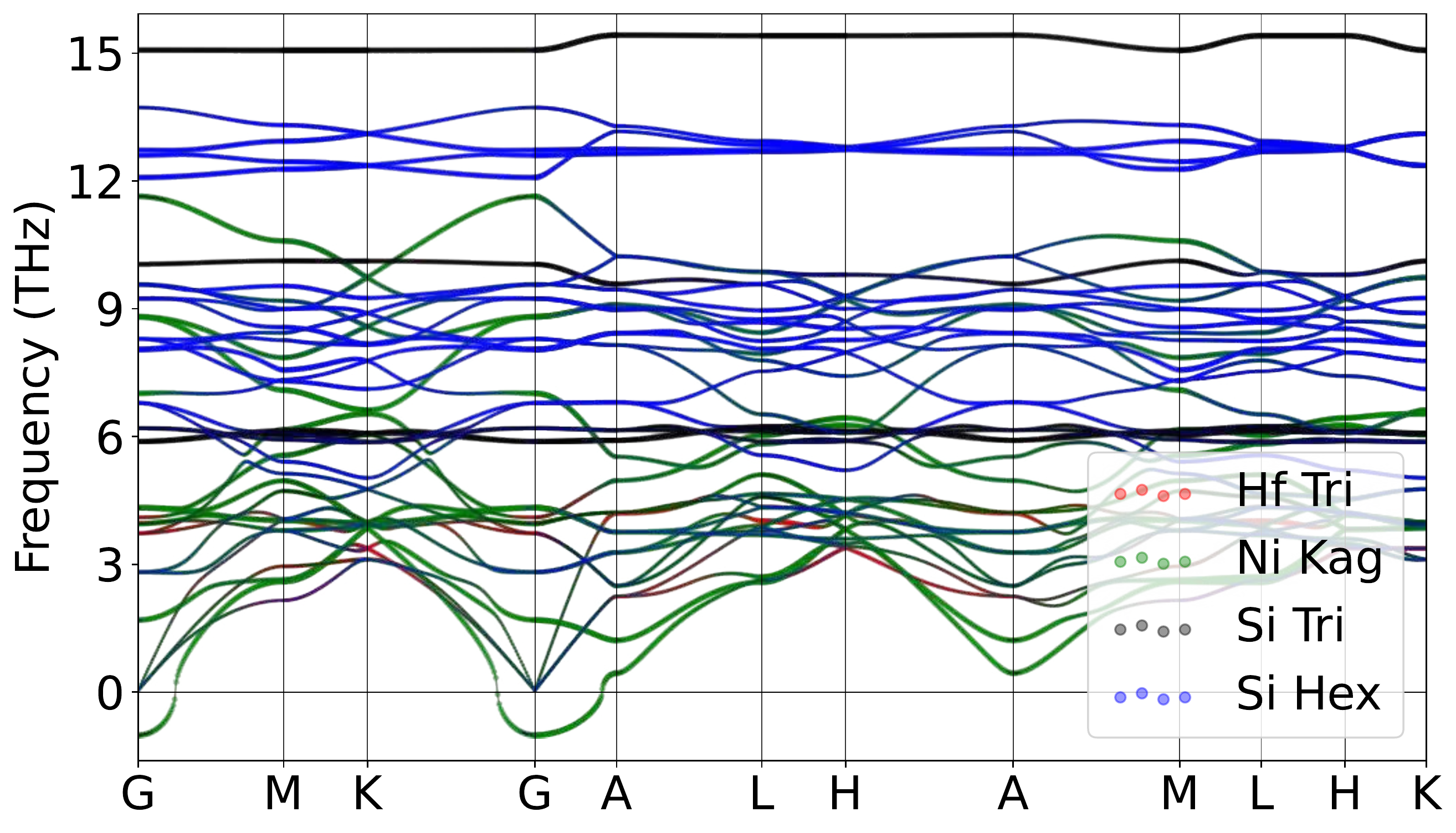}
}
\hspace{5pt}\subfloat[Relaxed phonon at high-T.]{
\label{fig:HfNi6Si6_relaxed_High}
\includegraphics[width=0.45\textwidth]{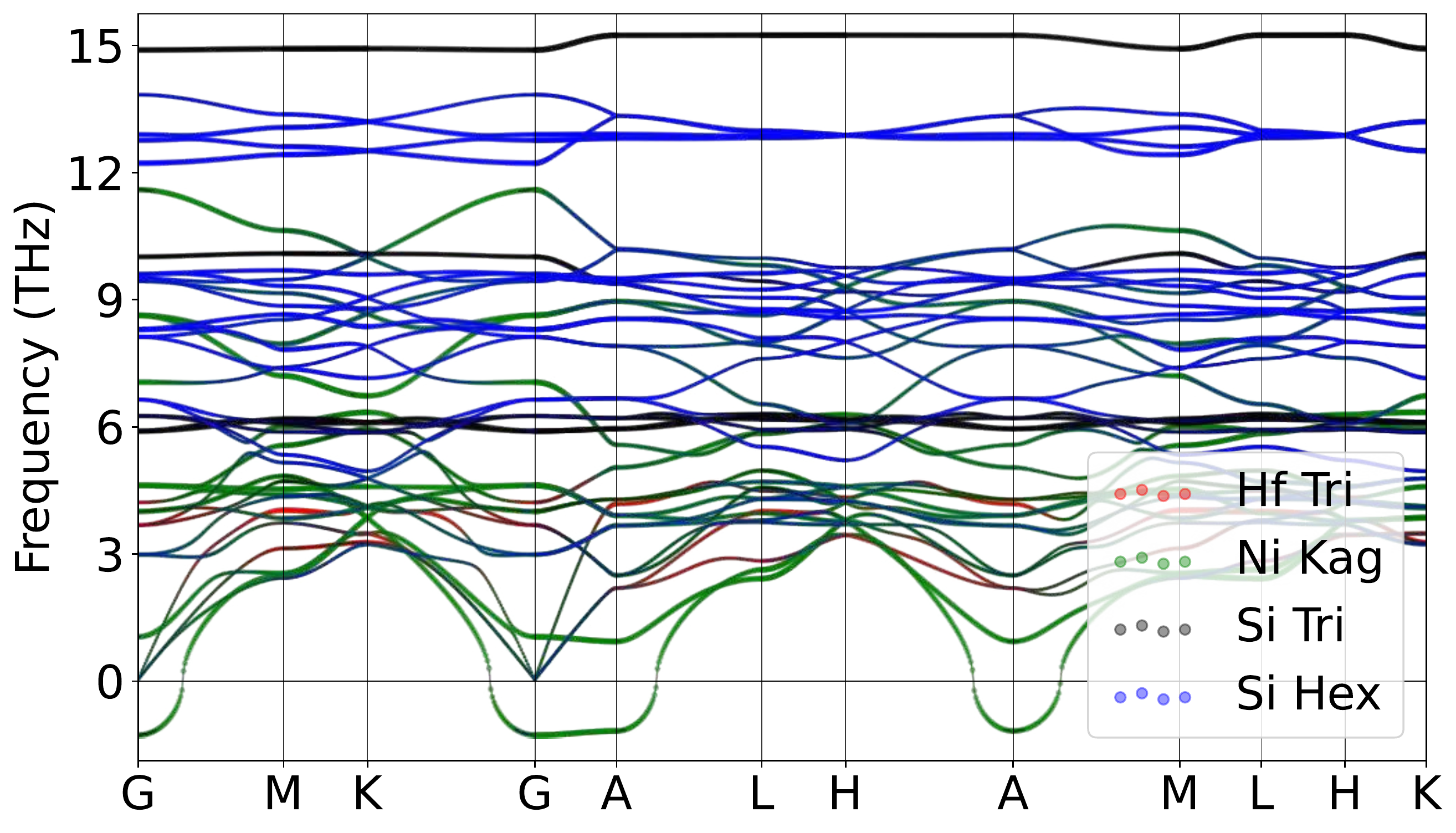}
}
\caption{Atomically projected phonon spectrum for HfNi$_6$Si$_6$.}
\label{fig:HfNi6Si6pho}
\end{figure}

\begin{figure}[H]
\centering
\label{fig:HfNi6Si6_relaxed_Low_xz}
\includegraphics[width=0.85\textwidth]{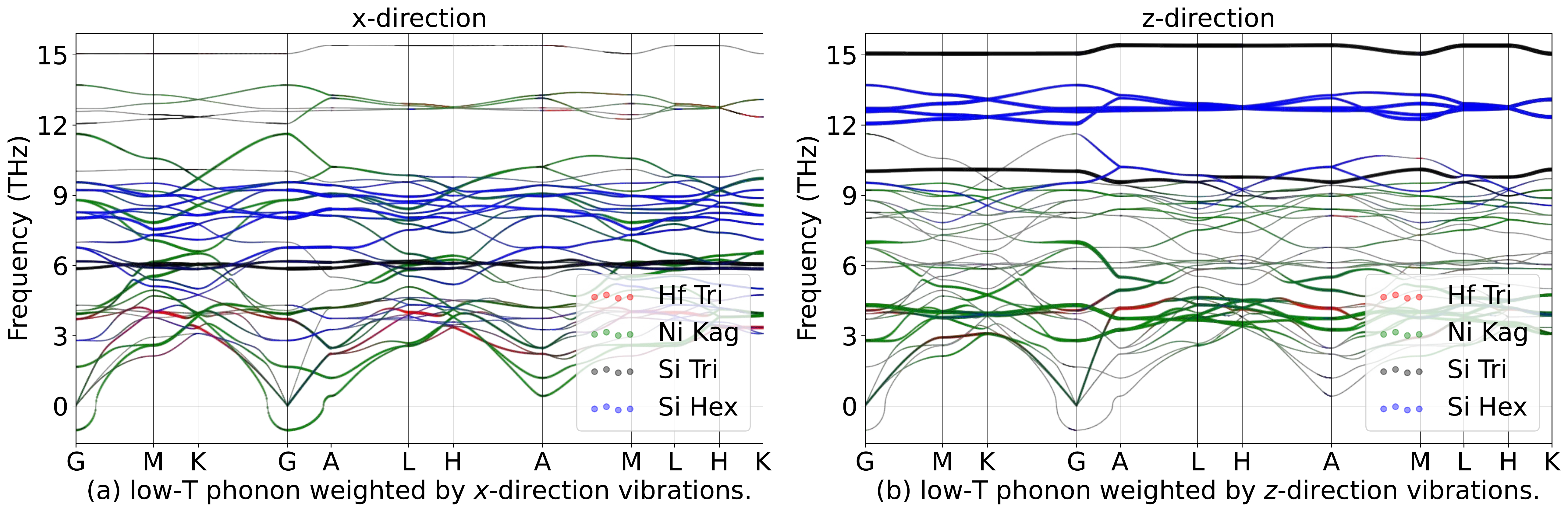}
\hspace{5pt}\label{fig:HfNi6Si6_relaxed_High_xz}
\includegraphics[width=0.85\textwidth]{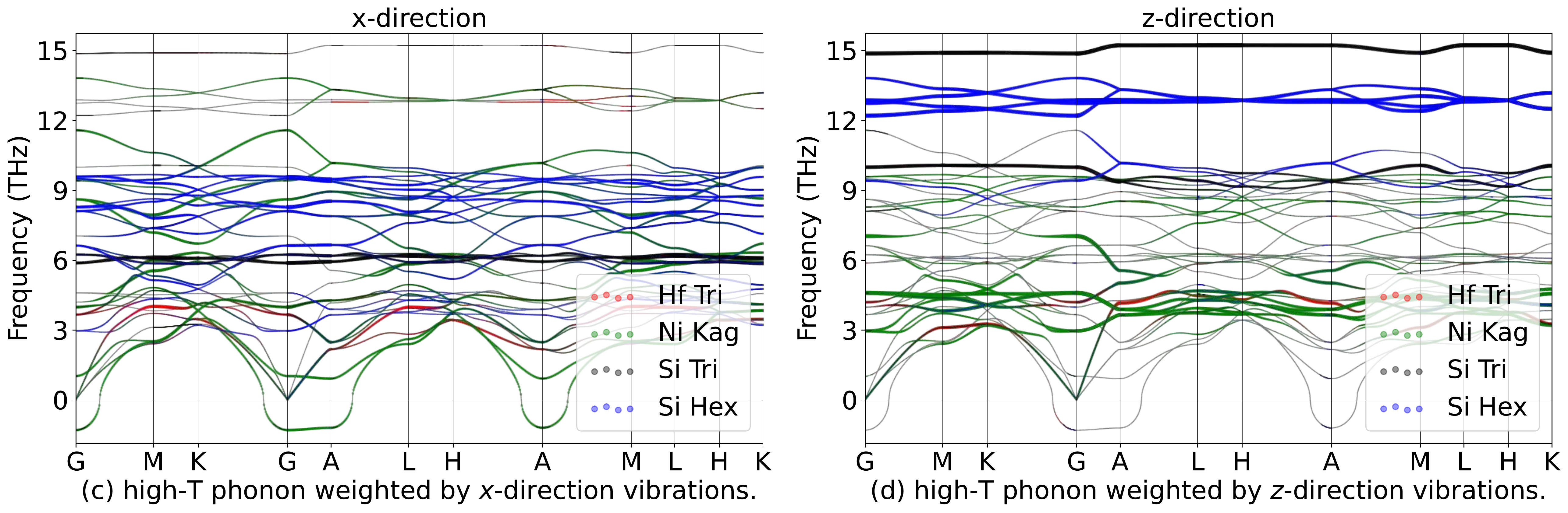}
\caption{Phonon spectrum for HfNi$_6$Si$_6$in in-plane ($x$) and out-of-plane ($z$) directions.}
\label{fig:HfNi6Si6phoxyz}
\end{figure}

\begin{table}[htbp]
\global\long\def\arraystretch{1.12}
\captionsetup{width=.95\textwidth}
\caption{IRREPs at high-symmetry points for HfNi$_6$Si$_6$ phonon spectrum at Low-T.}
\label{tab:191_HfNi6Si6_Low}
\setlength{\tabcolsep}{1.mm}{
}
\end{table}

\begin{table}[htbp]
\global\long\def\arraystretch{1.12}
\captionsetup{width=.95\textwidth}
\caption{IRREPs at high-symmetry points for HfNi$_6$Si$_6$ phonon spectrum at High-T.}
\label{tab:191_HfNi6Si6_High}
\setlength{\tabcolsep}{1.mm}{
%
}
\end{table}
\subsubsection{\label{sms2sec:TaNi6Si6}\hyperref[smsec:col]{\texorpdfstring{TaNi$_6$Si$_6$}{}}~{\color{red}  (High-T unstable, Low-T unstable) }}

\begin{table}[htbp]
\global\long\def\arraystretch{1.12}
\captionsetup{width=.85\textwidth}
\caption{Structure parameters of TaNi$_6$Si$_6$ after relaxation. It contains 325 electrons. }
\label{tab:TaNi6Si6}
\setlength{\tabcolsep}{4mm}{
%
}
\end{table}

\begin{figure}[htbp]
\centering
\includegraphics[width=0.85\textwidth]{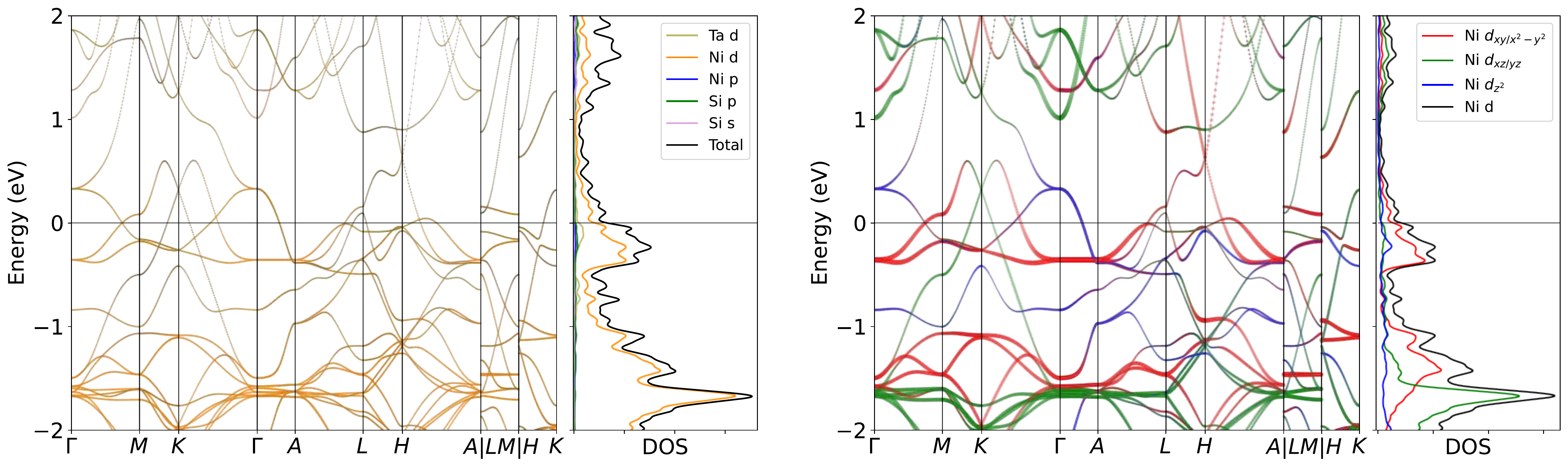}
\caption{Band structure for TaNi$_6$Si$_6$ without SOC. The projection of Ni $3d$ orbitals is also presented. The band structures clearly reveal the dominating of Ni $3d$ orbitals  near the Fermi level, as well as the pronounced peak.}
\label{fig:TaNi6Si6band}
\end{figure}

\begin{figure}[H]
\centering
\subfloat[Relaxed phonon at low-T.]{
\label{fig:TaNi6Si6_relaxed_Low}
\includegraphics[width=0.45\textwidth]{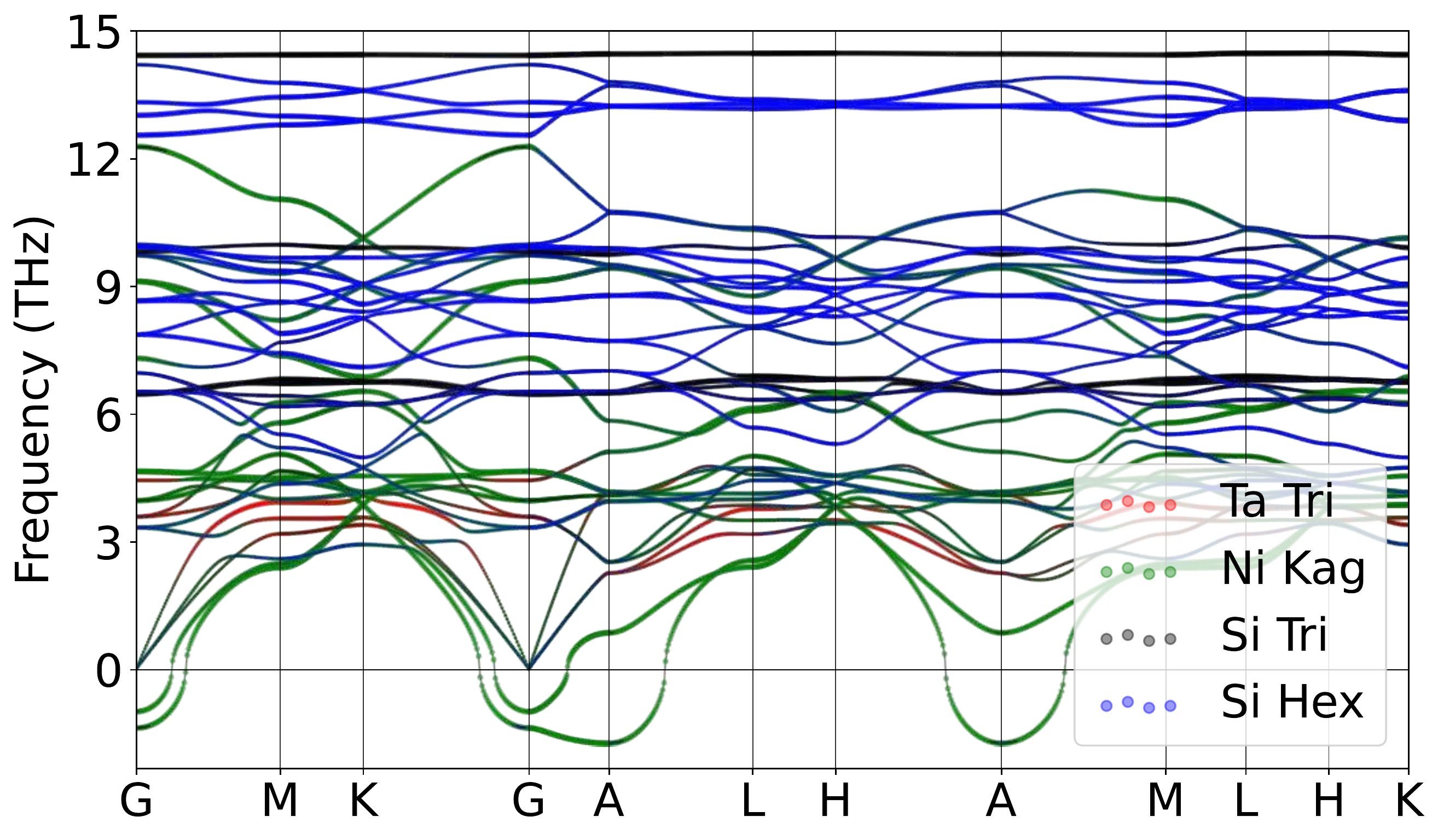}
}
\hspace{5pt}\subfloat[Relaxed phonon at high-T.]{
\label{fig:TaNi6Si6_relaxed_High}
\includegraphics[width=0.45\textwidth]{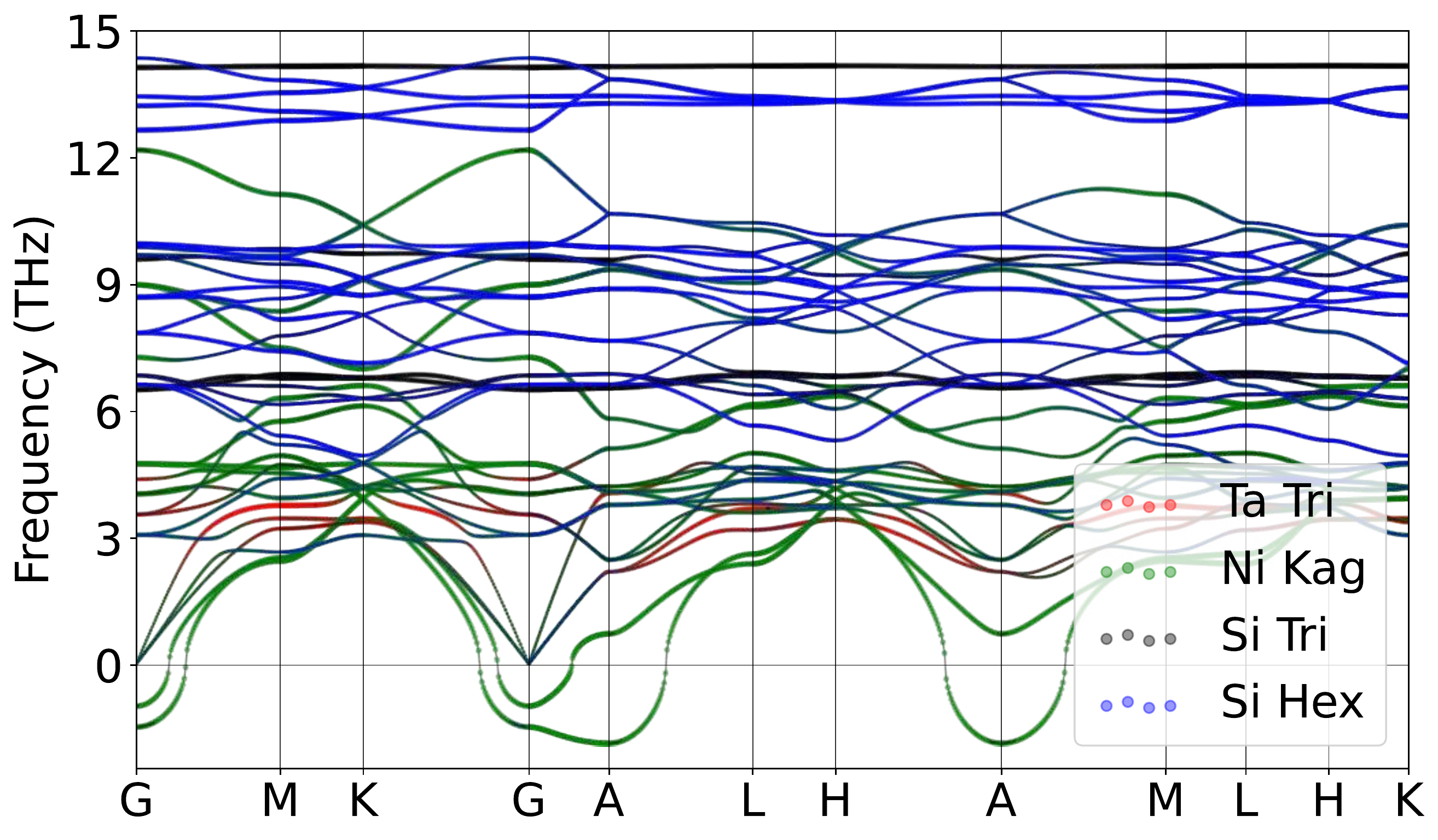}
}
\caption{Atomically projected phonon spectrum for TaNi$_6$Si$_6$.}
\label{fig:TaNi6Si6pho}
\end{figure}

\begin{figure}[H]
\centering
\label{fig:TaNi6Si6_relaxed_Low_xz}
\includegraphics[width=0.85\textwidth]{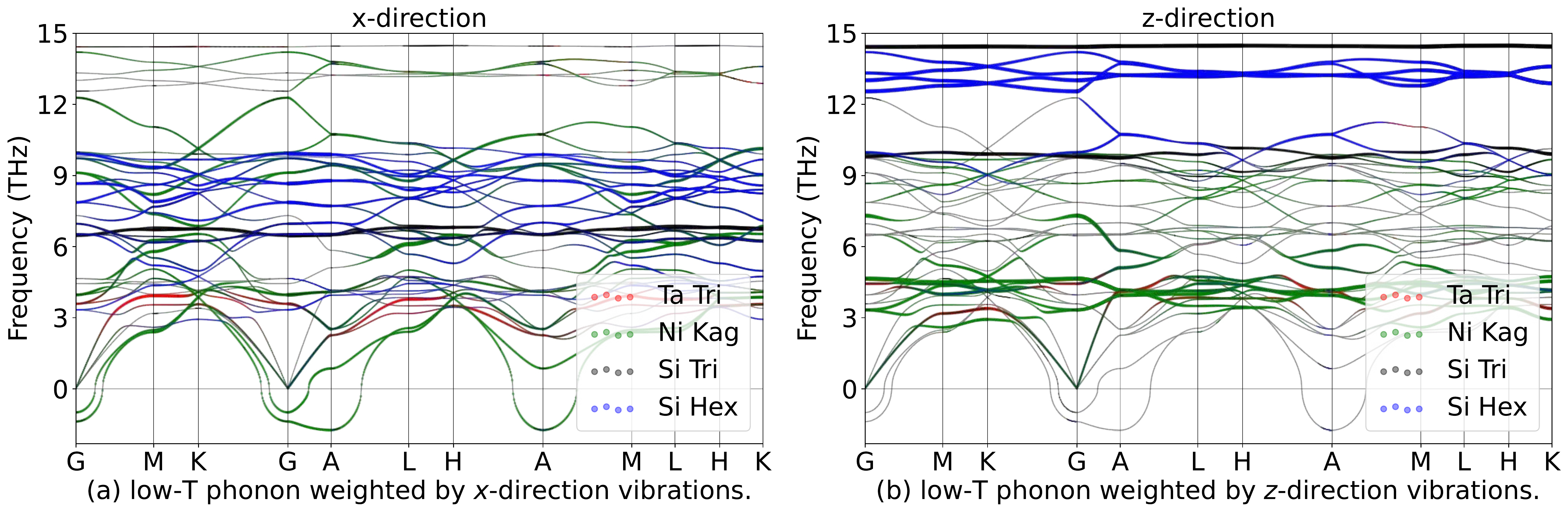}
\hspace{5pt}\label{fig:TaNi6Si6_relaxed_High_xz}
\includegraphics[width=0.85\textwidth]{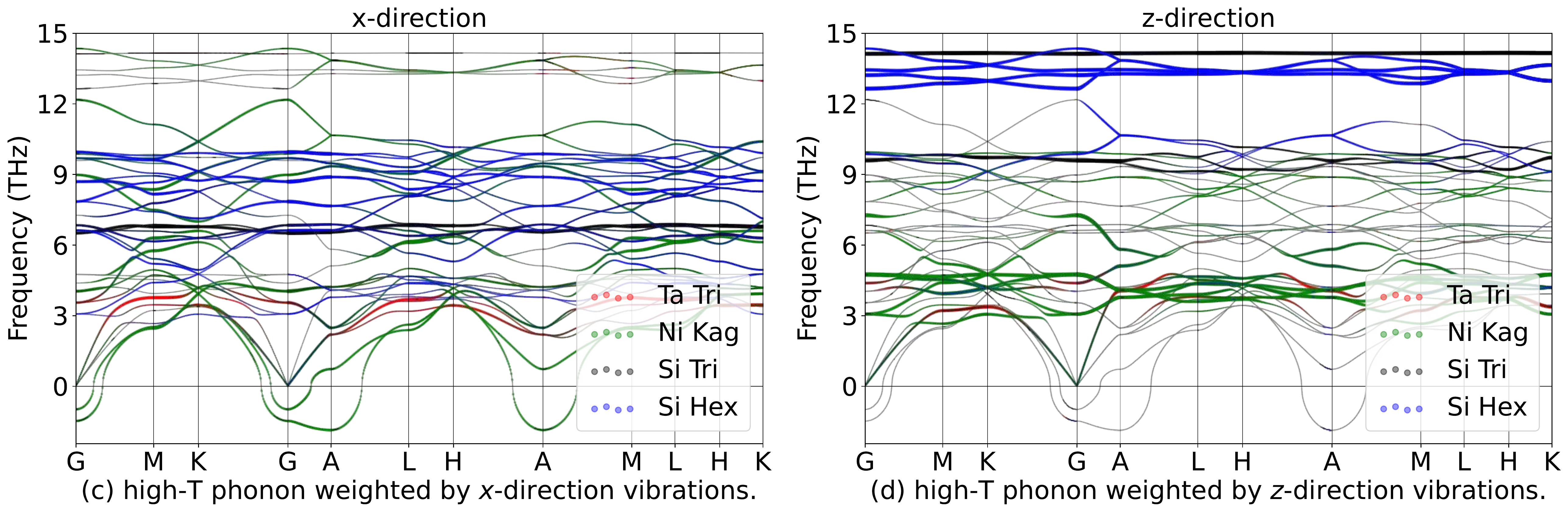}
\caption{Phonon spectrum for TaNi$_6$Si$_6$in in-plane ($x$) and out-of-plane ($z$) directions.}
\label{fig:TaNi6Si6phoxyz}
\end{figure}

\begin{table}[htbp]
\global\long\def\arraystretch{1.12}
\captionsetup{width=.95\textwidth}
\caption{IRREPs at high-symmetry points for TaNi$_6$Si$_6$ phonon spectrum at Low-T.}
\label{tab:191_TaNi6Si6_Low}
\setlength{\tabcolsep}{1.mm}{
}
\end{table}

\begin{table}[htbp]
\global\long\def\arraystretch{1.12}
\captionsetup{width=.95\textwidth}
\caption{IRREPs at high-symmetry points for TaNi$_6$Si$_6$ phonon spectrum at High-T.}
\label{tab:191_TaNi6Si6_High}
\setlength{\tabcolsep}{1.mm}{
%
}
\end{table}
\subsection{\hyperref[smsec:col]{NiGe}}~\label{smssec:NiGe}

\subsubsection{\label{sms2sec:LiNi6Ge6}\hyperref[smsec:col]{\texorpdfstring{LiNi$_6$Ge$_6$}{}}~{\color{red}  (High-T unstable, Low-T unstable) }}

\begin{table}[htbp]
\global\long\def\arraystretch{1.12}
\captionsetup{width=.85\textwidth}
\caption{Structure parameters of LiNi$_6$Ge$_6$ after relaxation. It contains 363 electrons. }
\label{tab:LiNi6Ge6}
\setlength{\tabcolsep}{4mm}{
%
}
\end{table}

\begin{figure}[htbp]
\centering
\includegraphics[width=0.85\textwidth]{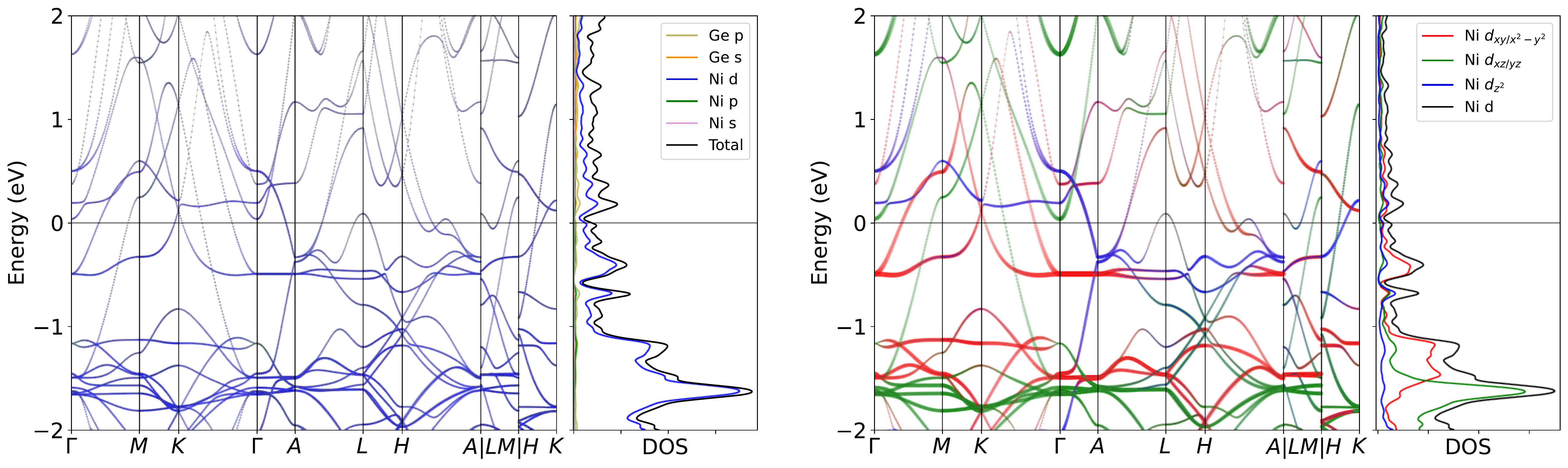}
\caption{Band structure for LiNi$_6$Ge$_6$ without SOC. The projection of Ni $3d$ orbitals is also presented. The band structures clearly reveal the dominating of Ni $3d$ orbitals  near the Fermi level, as well as the pronounced peak.}
\label{fig:LiNi6Ge6band}
\end{figure}

\begin{figure}[H]
\centering
\subfloat[Relaxed phonon at low-T.]{
\label{fig:LiNi6Ge6_relaxed_Low}
\includegraphics[width=0.45\textwidth]{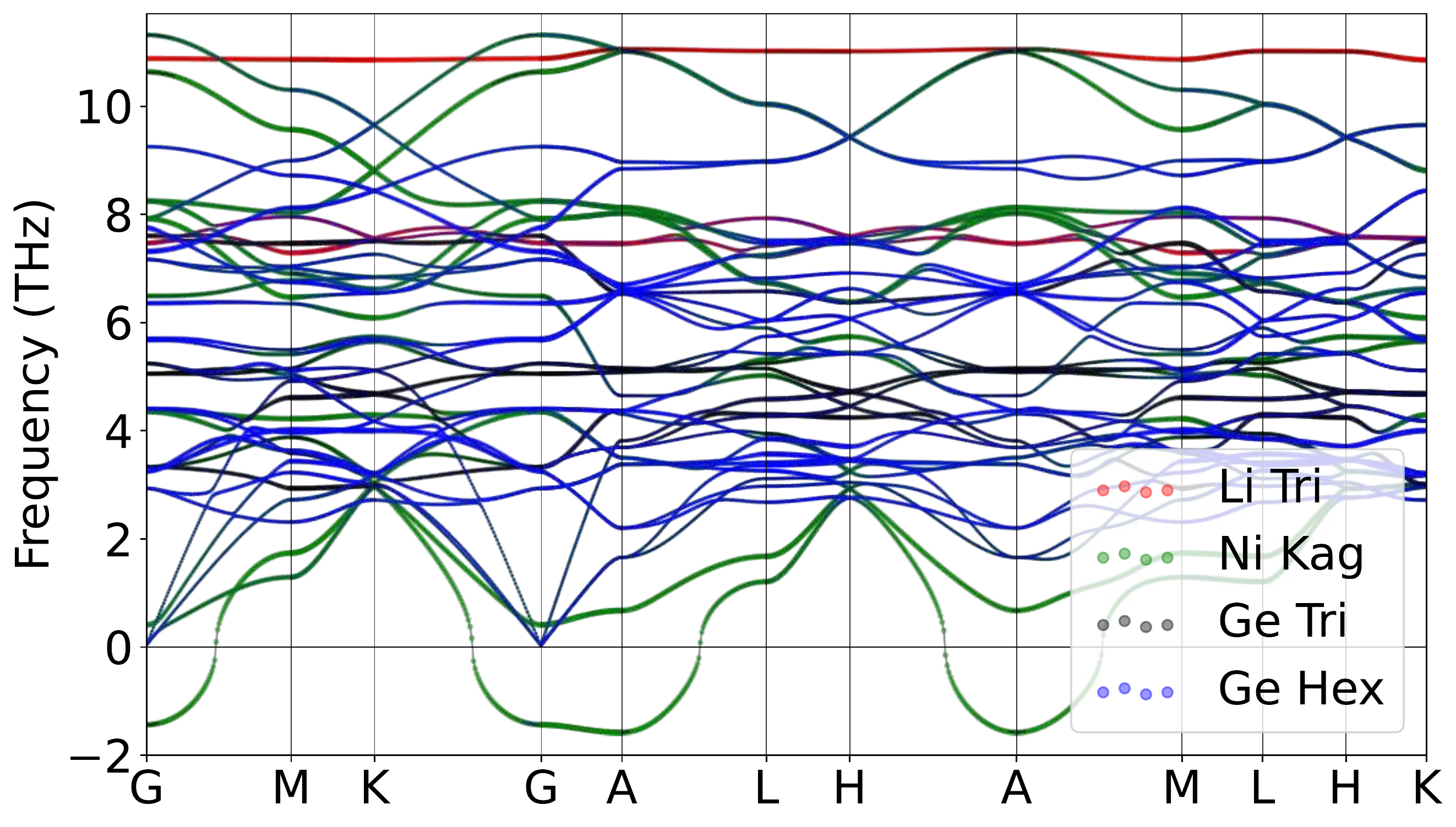}
}
\hspace{5pt}\subfloat[Relaxed phonon at high-T.]{
\label{fig:LiNi6Ge6_relaxed_High}
\includegraphics[width=0.45\textwidth]{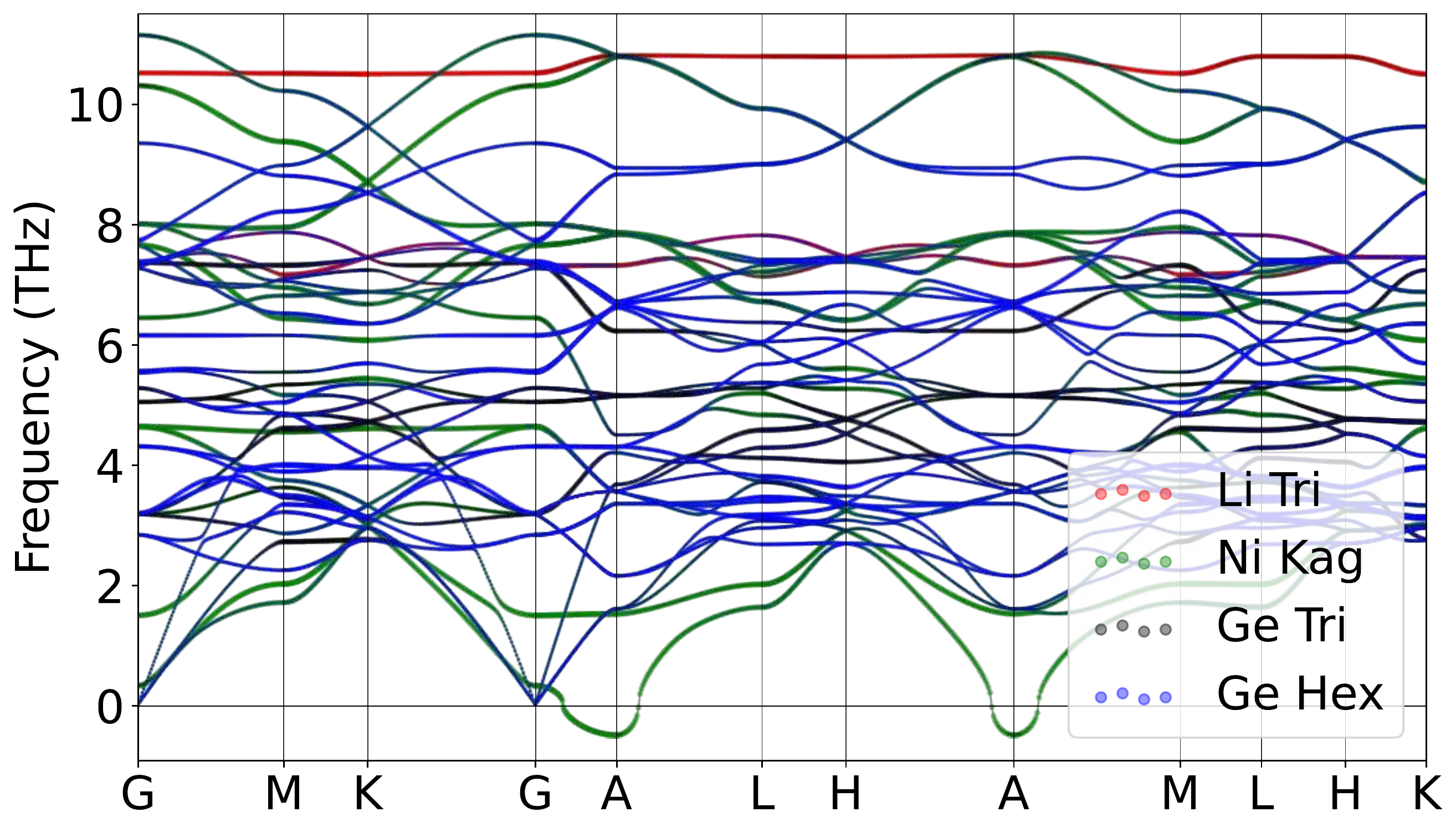}
}
\caption{Atomically projected phonon spectrum for LiNi$_6$Ge$_6$.}
\label{fig:LiNi6Ge6pho}
\end{figure}

\begin{figure}[H]
\centering
\label{fig:LiNi6Ge6_relaxed_Low_xz}
\includegraphics[width=0.85\textwidth]{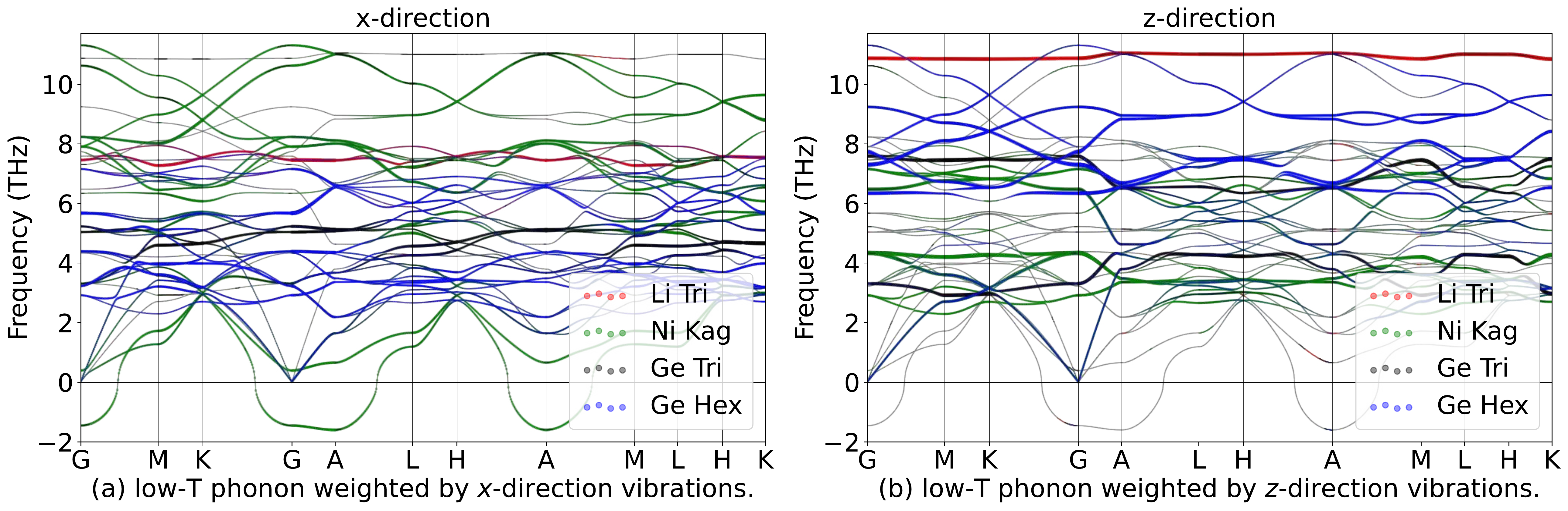}
\hspace{5pt}\label{fig:LiNi6Ge6_relaxed_High_xz}
\includegraphics[width=0.85\textwidth]{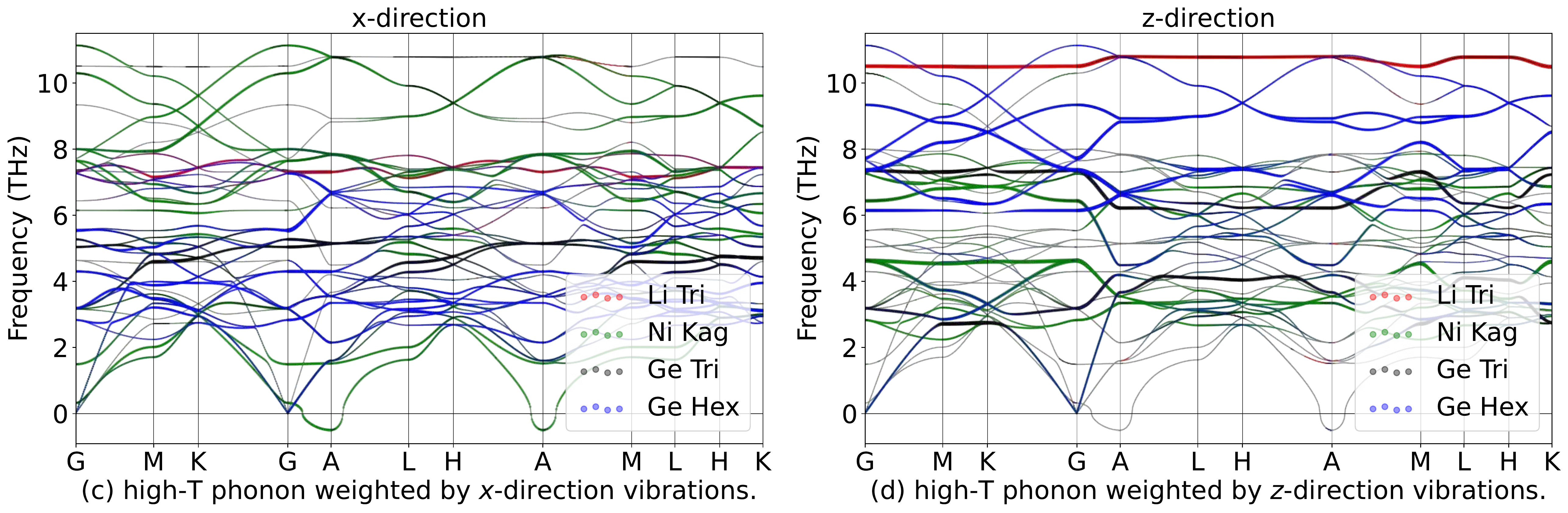}
\caption{Phonon spectrum for LiNi$_6$Ge$_6$in in-plane ($x$) and out-of-plane ($z$) directions.}
\label{fig:LiNi6Ge6phoxyz}
\end{figure}

\begin{table}[htbp]
\global\long\def\arraystretch{1.12}
\captionsetup{width=.95\textwidth}
\caption{IRREPs at high-symmetry points for LiNi$_6$Ge$_6$ phonon spectrum at Low-T.}
\label{tab:191_LiNi6Ge6_Low}
\setlength{\tabcolsep}{1.mm}{
}
\end{table}

\begin{table}[htbp]
\global\long\def\arraystretch{1.12}
\captionsetup{width=.95\textwidth}
\caption{IRREPs at high-symmetry points for LiNi$_6$Ge$_6$ phonon spectrum at High-T.}
\label{tab:191_LiNi6Ge6_High}
\setlength{\tabcolsep}{1.mm}{
%
}
\end{table}
\subsubsection{\label{sms2sec:MgNi6Ge6}\hyperref[smsec:col]{\texorpdfstring{MgNi$_6$Ge$_6$}{}}~{\color{red}  (High-T unstable, Low-T unstable) }}

\begin{table}[htbp]
\global\long\def\arraystretch{1.12}
\captionsetup{width=.85\textwidth}
\caption{Structure parameters of MgNi$_6$Ge$_6$ after relaxation. It contains 372 electrons. }
\label{tab:MgNi6Ge6}
\setlength{\tabcolsep}{4mm}{
%
}
\end{table}

\begin{figure}[htbp]
\centering
\includegraphics[width=0.85\textwidth]{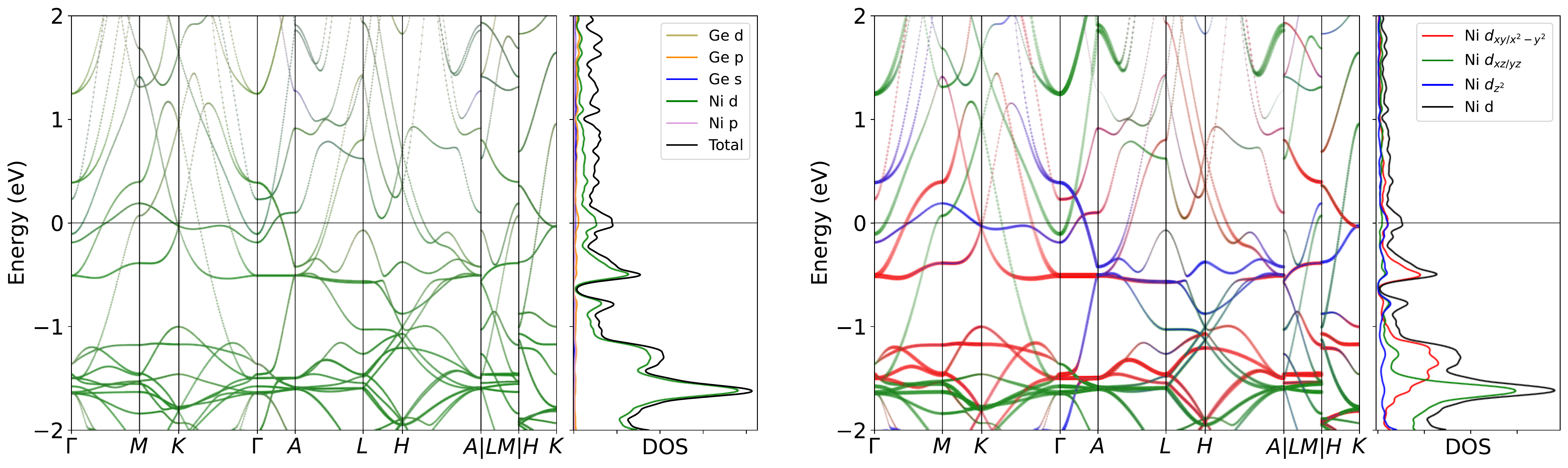}
\caption{Band structure for MgNi$_6$Ge$_6$ without SOC. The projection of Ni $3d$ orbitals is also presented. The band structures clearly reveal the dominating of Ni $3d$ orbitals  near the Fermi level, as well as the pronounced peak.}
\label{fig:MgNi6Ge6band}
\end{figure}

\begin{figure}[H]
\centering
\subfloat[Relaxed phonon at low-T.]{
\label{fig:MgNi6Ge6_relaxed_Low}
\includegraphics[width=0.45\textwidth]{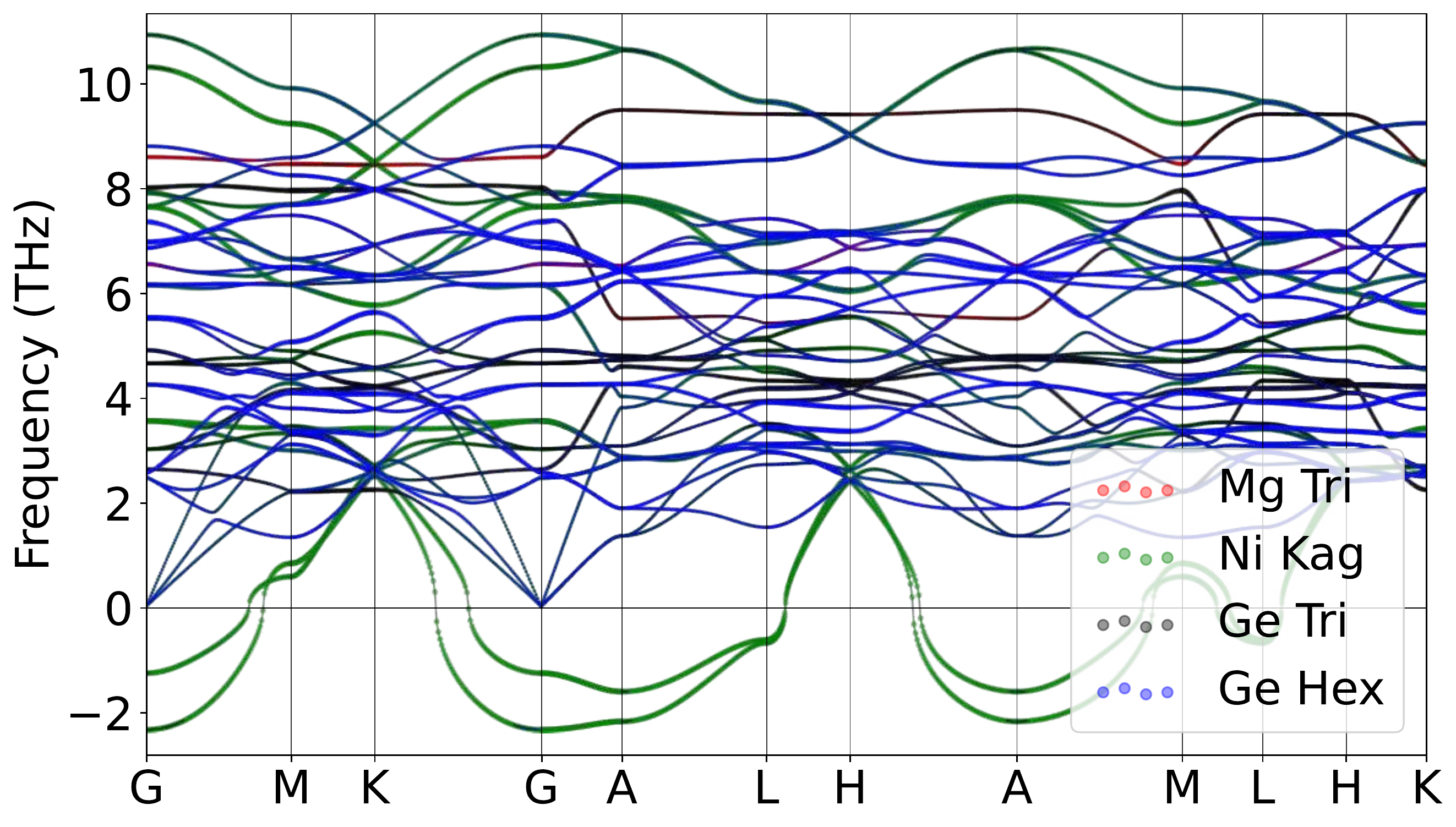}
}
\hspace{5pt}\subfloat[Relaxed phonon at high-T.]{
\label{fig:MgNi6Ge6_relaxed_High}
\includegraphics[width=0.45\textwidth]{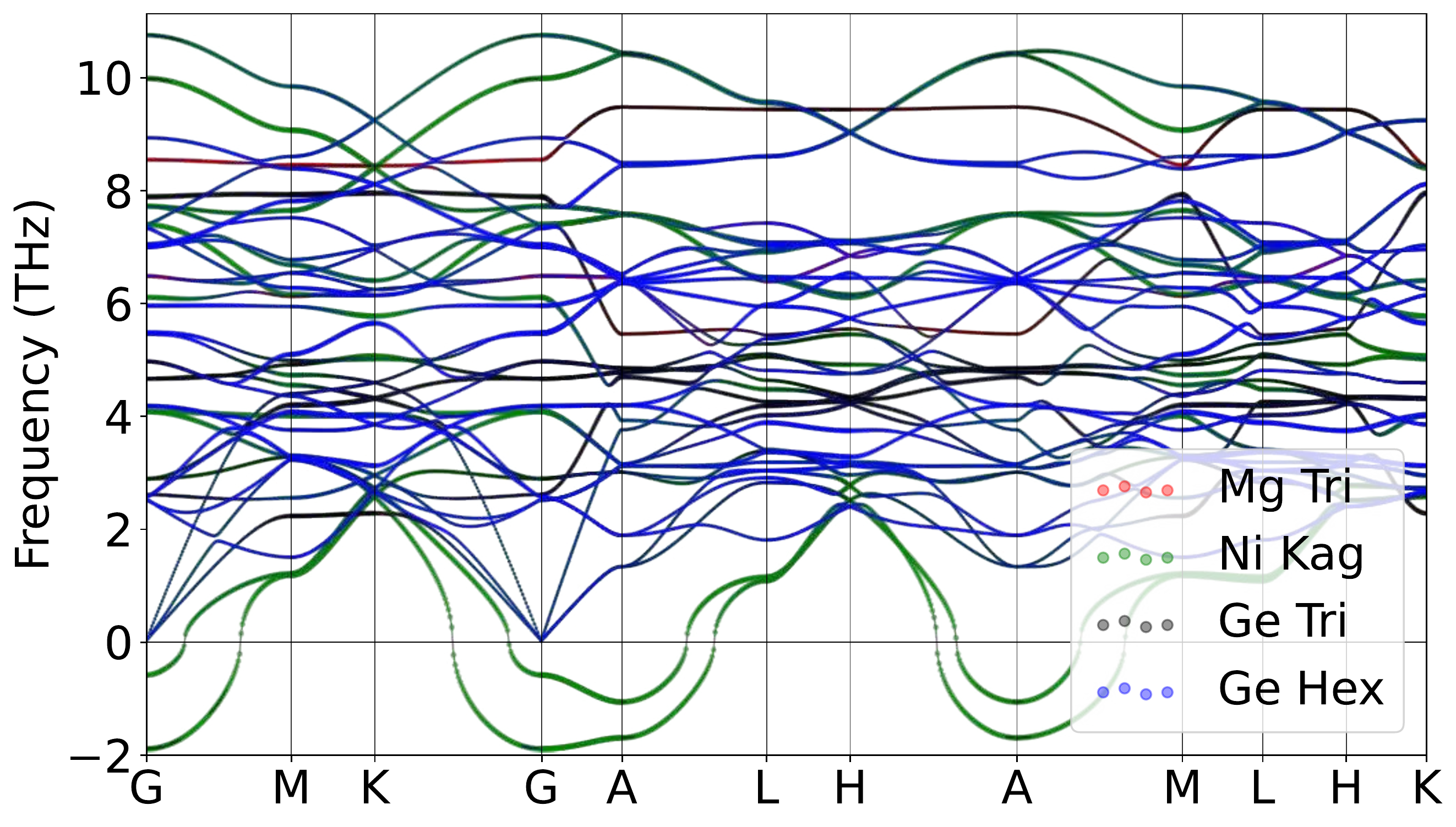}
}
\caption{Atomically projected phonon spectrum for MgNi$_6$Ge$_6$.}
\label{fig:MgNi6Ge6pho}
\end{figure}

\begin{figure}[H]
\centering
\label{fig:MgNi6Ge6_relaxed_Low_xz}
\includegraphics[width=0.85\textwidth]{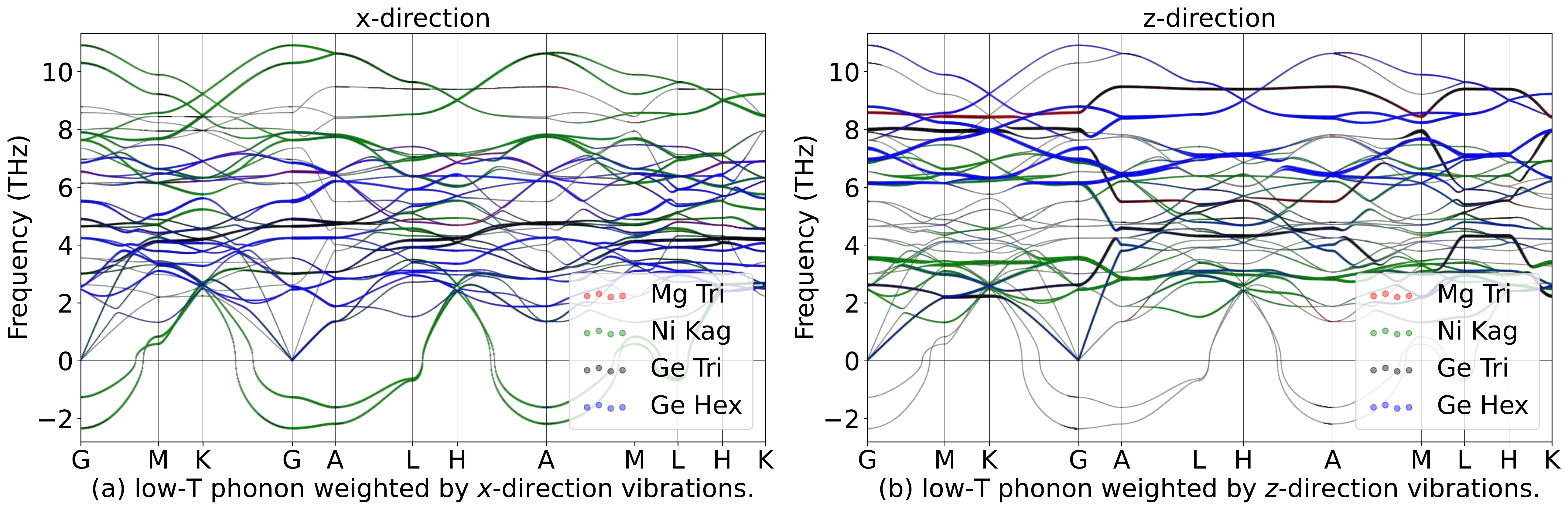}
\hspace{5pt}\label{fig:MgNi6Ge6_relaxed_High_xz}
\includegraphics[width=0.85\textwidth]{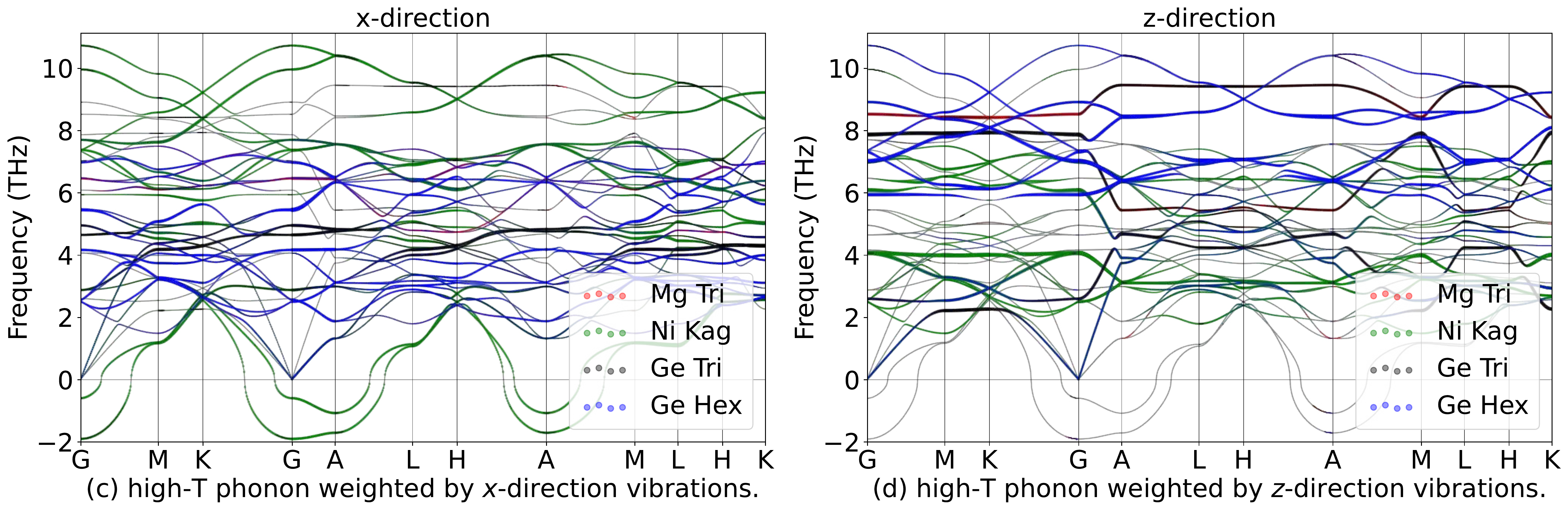}
\caption{Phonon spectrum for MgNi$_6$Ge$_6$in in-plane ($x$) and out-of-plane ($z$) directions.}
\label{fig:MgNi6Ge6phoxyz}
\end{figure}

\begin{table}[htbp]
\global\long\def\arraystretch{1.12}
\captionsetup{width=.95\textwidth}
\caption{IRREPs at high-symmetry points for MgNi$_6$Ge$_6$ phonon spectrum at Low-T.}
\label{tab:191_MgNi6Ge6_Low}
\setlength{\tabcolsep}{1.mm}{
}
\end{table}

\begin{table}[htbp]
\global\long\def\arraystretch{1.12}
\captionsetup{width=.95\textwidth}
\caption{IRREPs at high-symmetry points for MgNi$_6$Ge$_6$ phonon spectrum at High-T.}
\label{tab:191_MgNi6Ge6_High}
\setlength{\tabcolsep}{1.mm}{
%
}
\end{table}
\subsubsection{\label{sms2sec:CaNi6Ge6}\hyperref[smsec:col]{\texorpdfstring{CaNi$_6$Ge$_6$}{}}~{\color{green}  (High-T stable, Low-T stable) }}

\begin{table}[htbp]
\global\long\def\arraystretch{1.12}
\captionsetup{width=.85\textwidth}
\caption{Structure parameters of CaNi$_6$Ge$_6$ after relaxation. It contains 380 electrons. }
\label{tab:CaNi6Ge6}
\setlength{\tabcolsep}{4mm}{
%
}
\end{table}

\begin{figure}[htbp]
\centering
\includegraphics[width=0.85\textwidth]{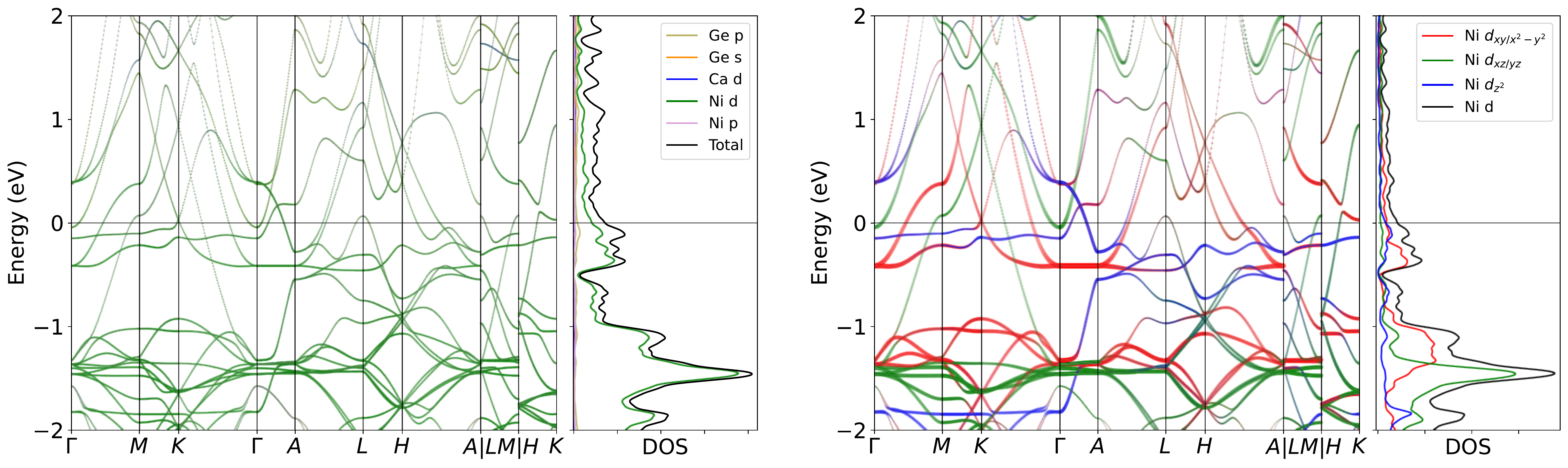}
\caption{Band structure for CaNi$_6$Ge$_6$ without SOC. The projection of Ni $3d$ orbitals is also presented. The band structures clearly reveal the dominating of Ni $3d$ orbitals  near the Fermi level, as well as the pronounced peak.}
\label{fig:CaNi6Ge6band}
\end{figure}

\begin{figure}[H]
\centering
\subfloat[Relaxed phonon at low-T.]{
\label{fig:CaNi6Ge6_relaxed_Low}
\includegraphics[width=0.45\textwidth]{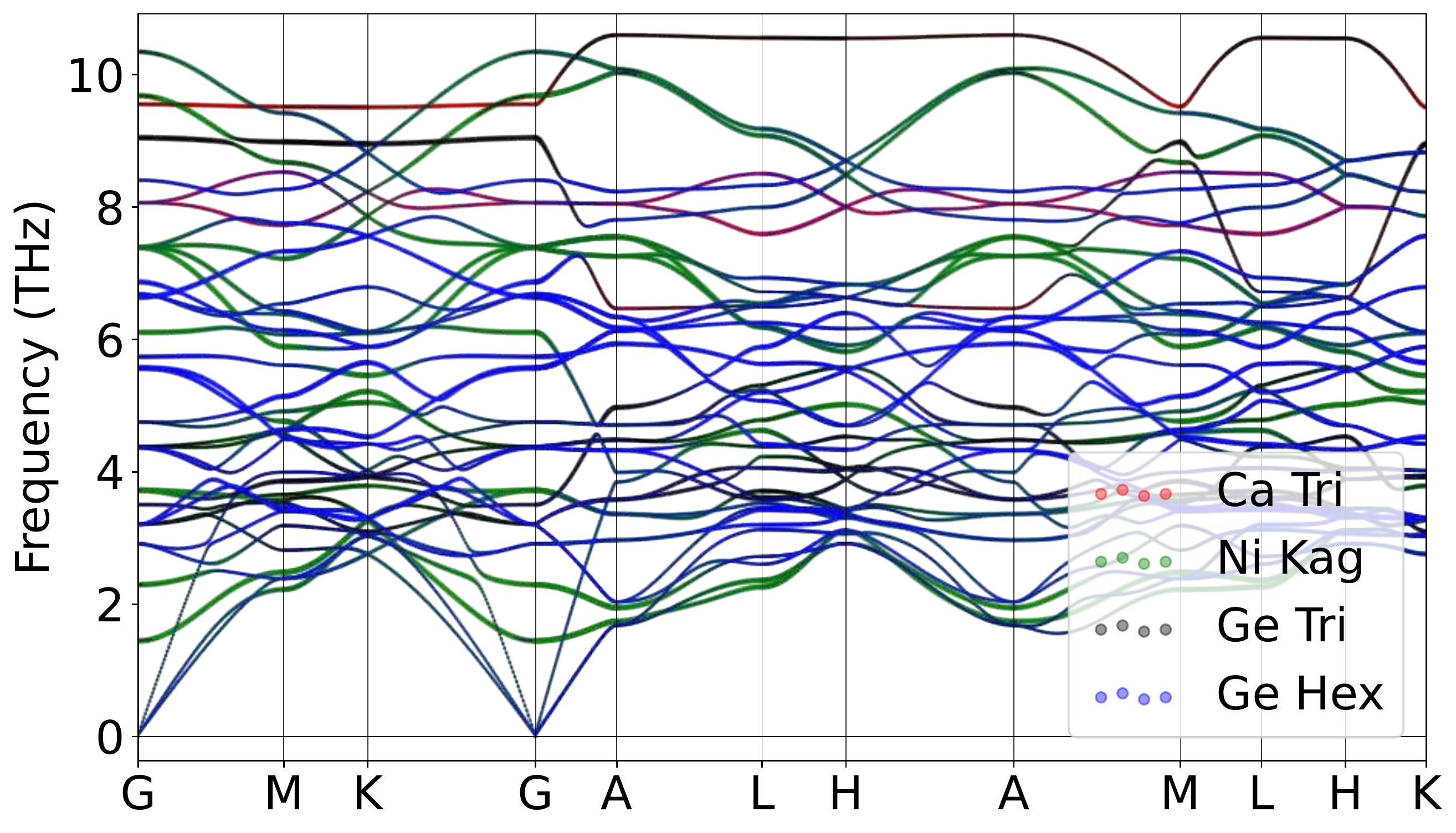}
}
\hspace{5pt}\subfloat[Relaxed phonon at high-T.]{
\label{fig:CaNi6Ge6_relaxed_High}
\includegraphics[width=0.45\textwidth]{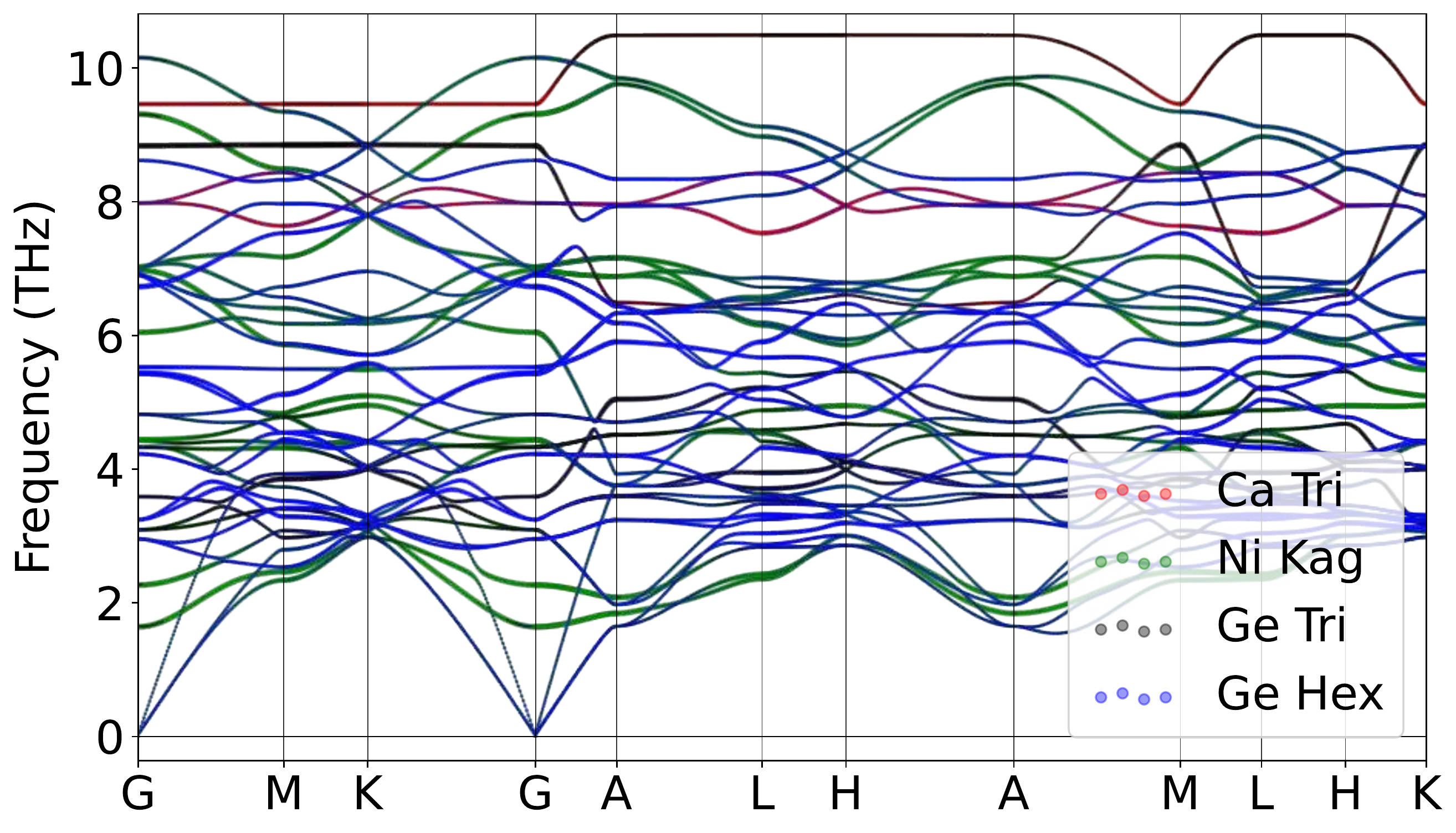}
}
\caption{Atomically projected phonon spectrum for CaNi$_6$Ge$_6$.}
\label{fig:CaNi6Ge6pho}
\end{figure}

\begin{figure}[H]
\centering
\label{fig:CaNi6Ge6_relaxed_Low_xz}
\includegraphics[width=0.85\textwidth]{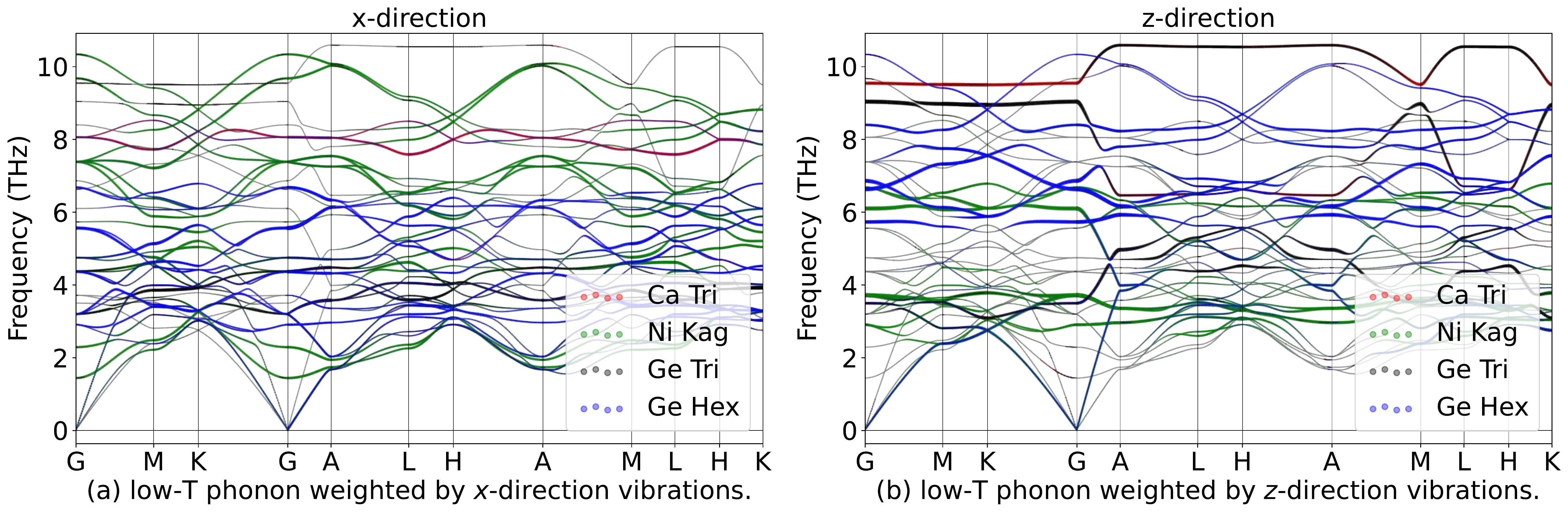}
\hspace{5pt}\label{fig:CaNi6Ge6_relaxed_High_xz}
\includegraphics[width=0.85\textwidth]{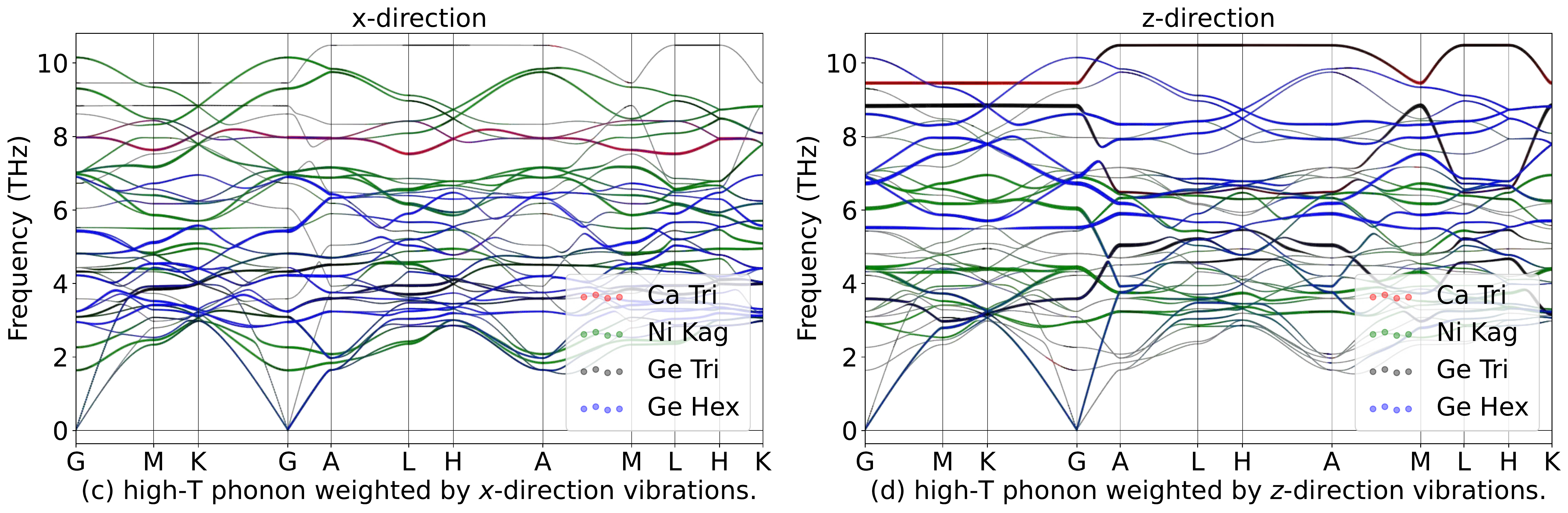}
\caption{Phonon spectrum for CaNi$_6$Ge$_6$in in-plane ($x$) and out-of-plane ($z$) directions.}
\label{fig:CaNi6Ge6phoxyz}
\end{figure}

\begin{table}[htbp]
\global\long\def\arraystretch{1.12}
\captionsetup{width=.95\textwidth}
\caption{IRREPs at high-symmetry points for CaNi$_6$Ge$_6$ phonon spectrum at Low-T.}
\label{tab:191_CaNi6Ge6_Low}
\setlength{\tabcolsep}{1.mm}{
}
\end{table}

\begin{table}[htbp]
\global\long\def\arraystretch{1.12}
\captionsetup{width=.95\textwidth}
\caption{IRREPs at high-symmetry points for CaNi$_6$Ge$_6$ phonon spectrum at High-T.}
\label{tab:191_CaNi6Ge6_High}
\setlength{\tabcolsep}{1.mm}{
%
}
\end{table}
\subsubsection{\label{sms2sec:ScNi6Ge6}\hyperref[smsec:col]{\texorpdfstring{ScNi$_6$Ge$_6$}{}}~{\color{red}  (High-T unstable, Low-T unstable) }}

\begin{table}[htbp]
\global\long\def\arraystretch{1.12}
\captionsetup{width=.85\textwidth}
\caption{Structure parameters of ScNi$_6$Ge$_6$ after relaxation. It contains 381 electrons. }
\label{tab:ScNi6Ge6}
\setlength{\tabcolsep}{4mm}{
%
}
\end{table}

\begin{figure}[htbp]
\centering
\includegraphics[width=0.85\textwidth]{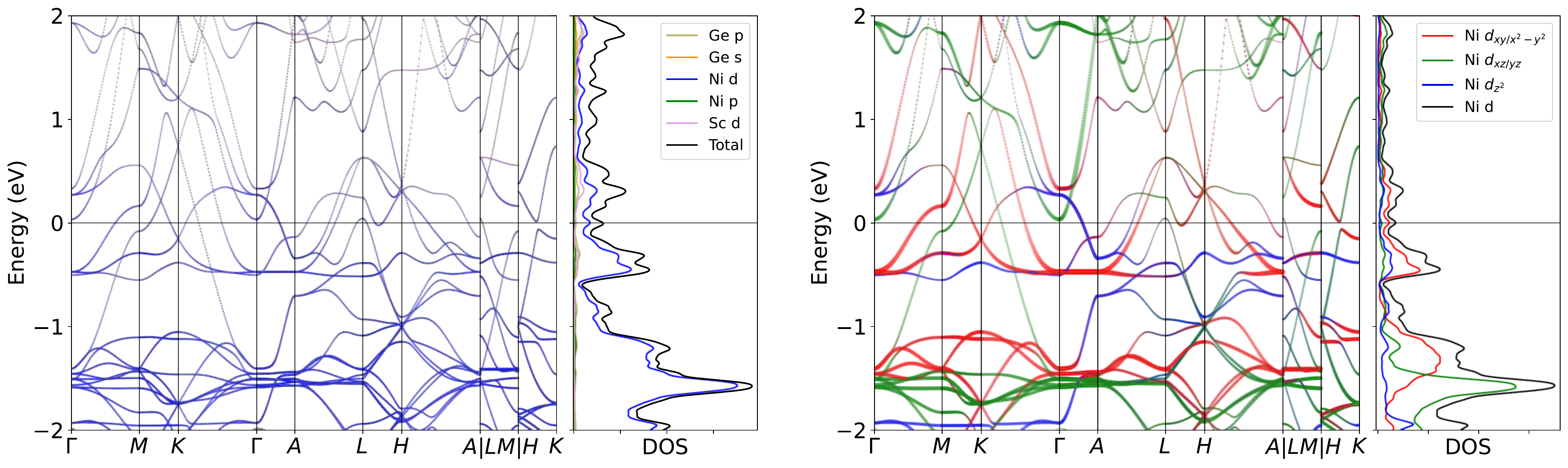}
\caption{Band structure for ScNi$_6$Ge$_6$ without SOC. The projection of Ni $3d$ orbitals is also presented. The band structures clearly reveal the dominating of Ni $3d$ orbitals  near the Fermi level, as well as the pronounced peak.}
\label{fig:ScNi6Ge6band}
\end{figure}

\begin{figure}[H]
\centering
\subfloat[Relaxed phonon at low-T.]{
\label{fig:ScNi6Ge6_relaxed_Low}
\includegraphics[width=0.45\textwidth]{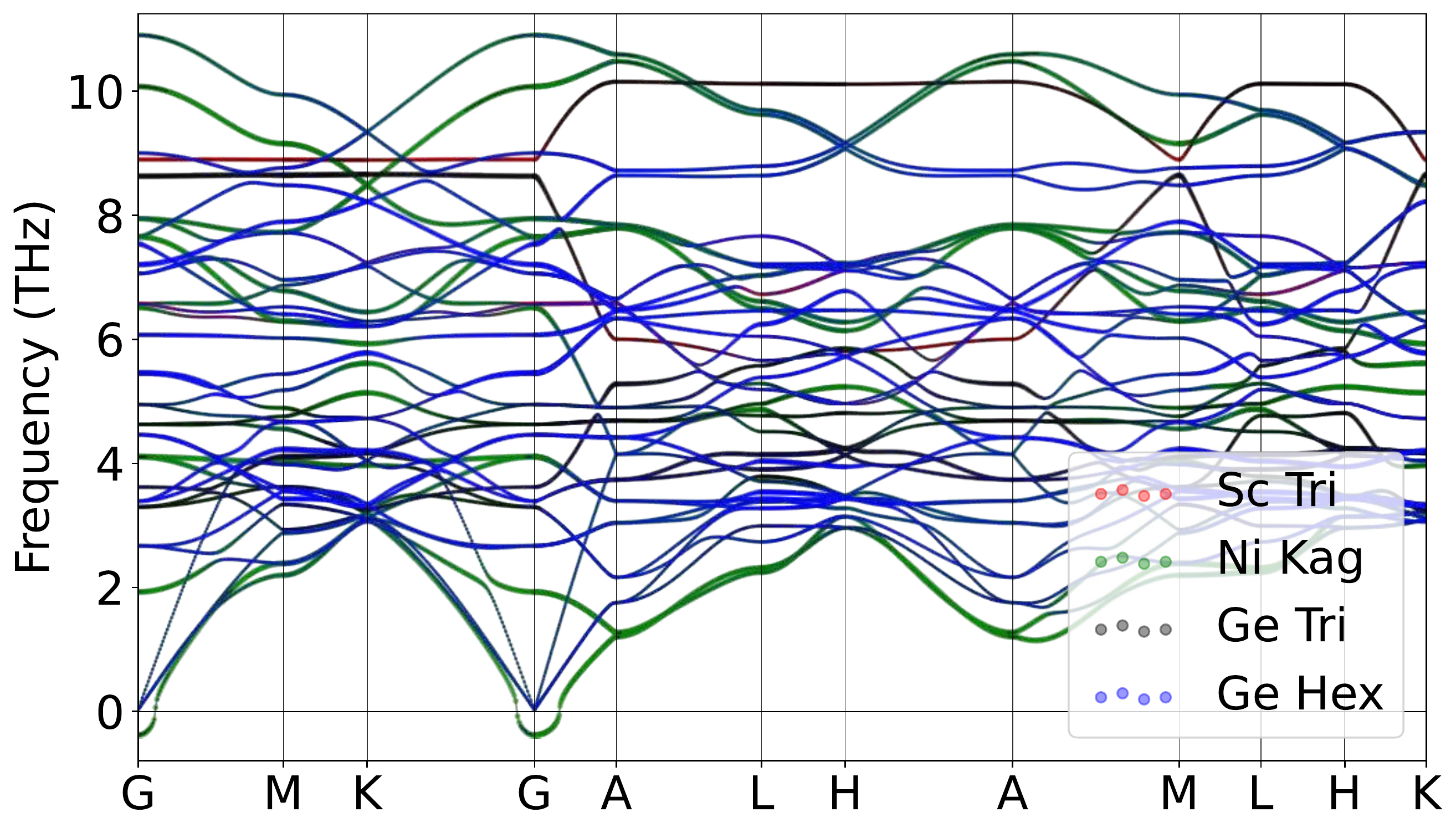}
}
\hspace{5pt}\subfloat[Relaxed phonon at high-T.]{
\label{fig:ScNi6Ge6_relaxed_High}
\includegraphics[width=0.45\textwidth]{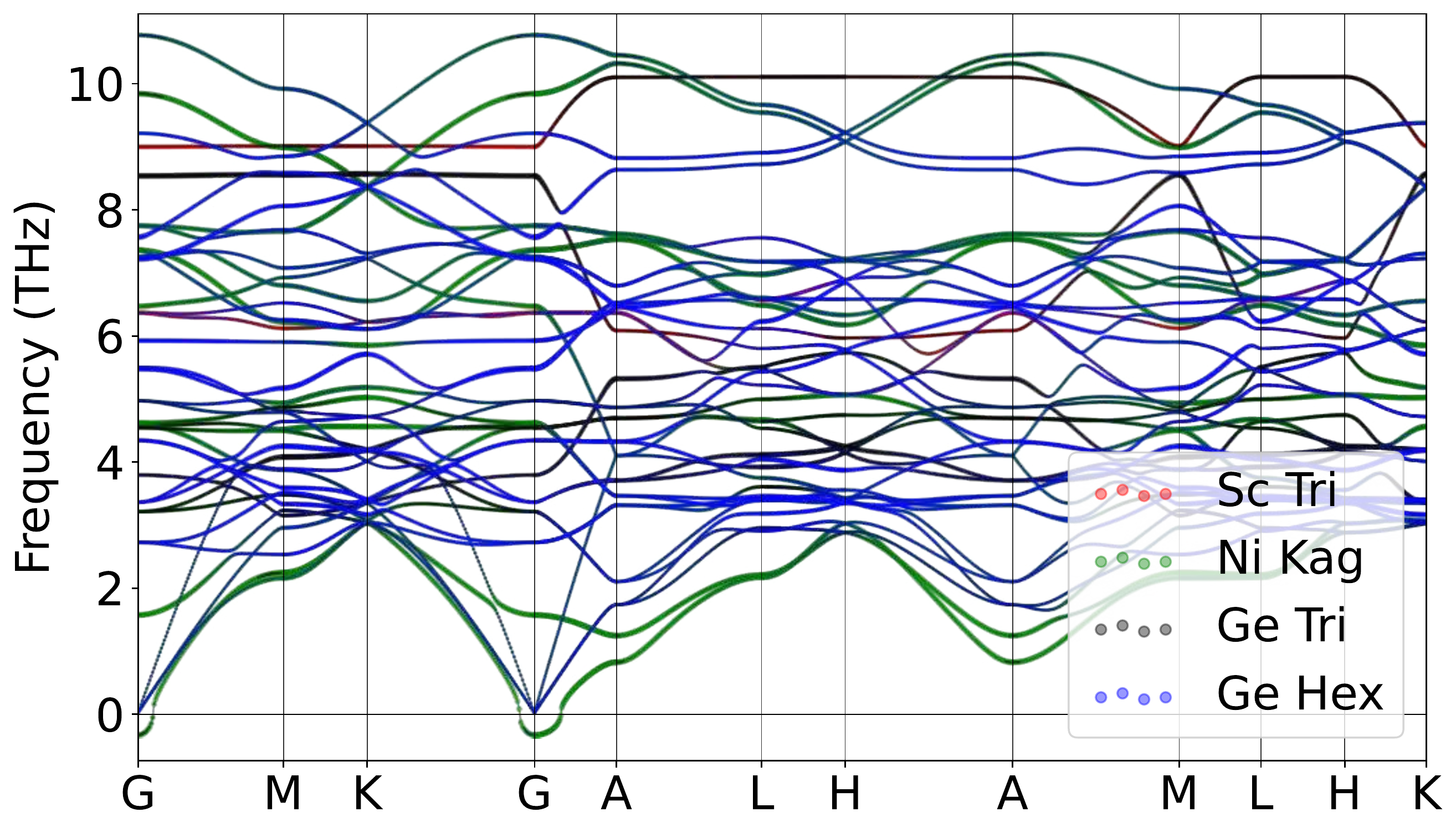}
}
\caption{Atomically projected phonon spectrum for ScNi$_6$Ge$_6$.}
\label{fig:ScNi6Ge6pho}
\end{figure}

\begin{figure}[H]
\centering
\label{fig:ScNi6Ge6_relaxed_Low_xz}
\includegraphics[width=0.85\textwidth]{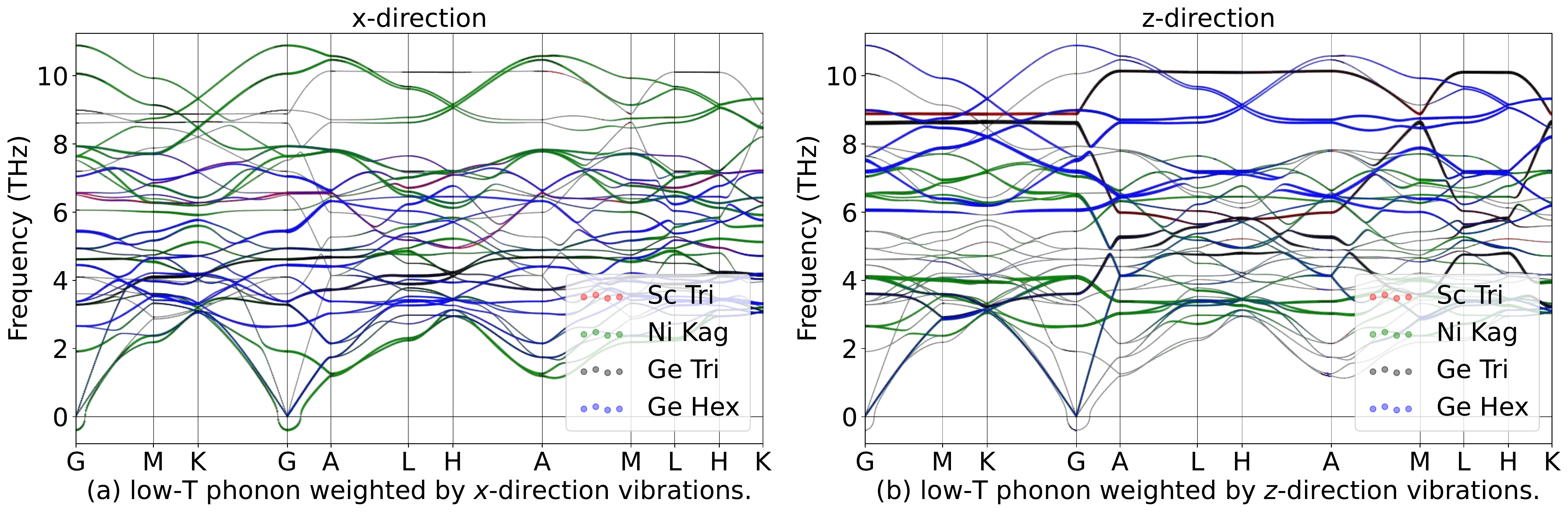}
\hspace{5pt}\label{fig:ScNi6Ge6_relaxed_High_xz}
\includegraphics[width=0.85\textwidth]{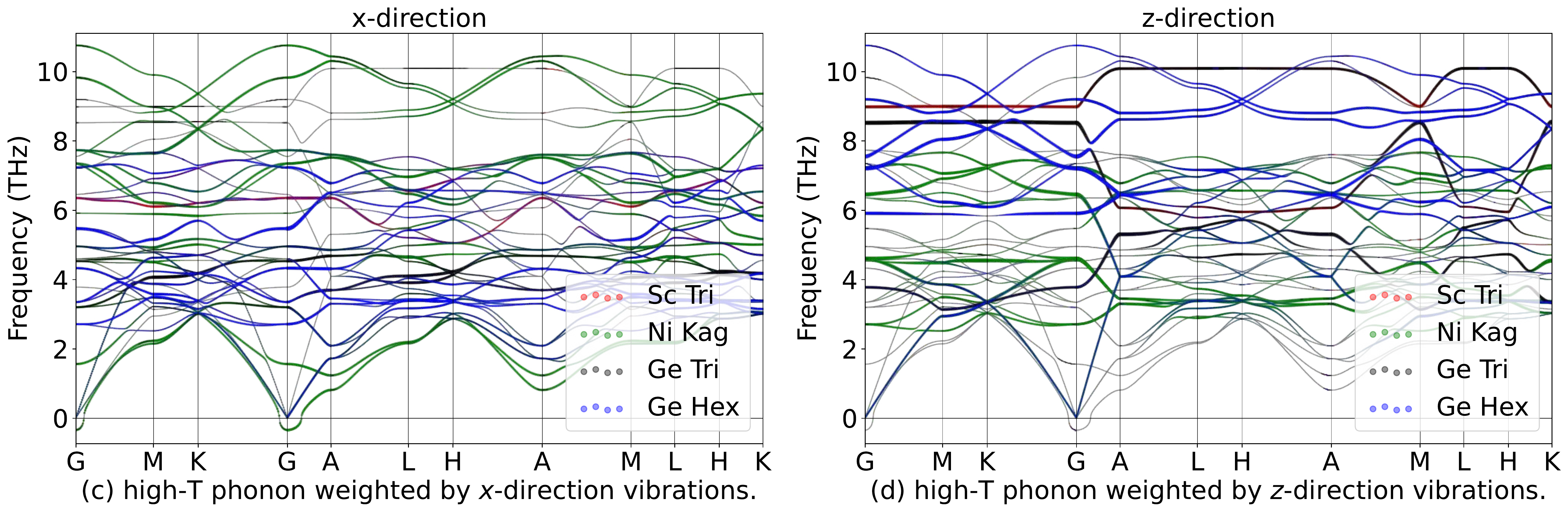}
\caption{Phonon spectrum for ScNi$_6$Ge$_6$in in-plane ($x$) and out-of-plane ($z$) directions.}
\label{fig:ScNi6Ge6phoxyz}
\end{figure}

\begin{table}[htbp]
\global\long\def\arraystretch{1.12}
\captionsetup{width=.95\textwidth}
\caption{IRREPs at high-symmetry points for ScNi$_6$Ge$_6$ phonon spectrum at Low-T.}
\label{tab:191_ScNi6Ge6_Low}
\setlength{\tabcolsep}{1.mm}{
}
\end{table}

\begin{table}[htbp]
\global\long\def\arraystretch{1.12}
\captionsetup{width=.95\textwidth}
\caption{IRREPs at high-symmetry points for ScNi$_6$Ge$_6$ phonon spectrum at High-T.}
\label{tab:191_ScNi6Ge6_High}
\setlength{\tabcolsep}{1.mm}{
%
}
\end{table}
\subsubsection{\label{sms2sec:TiNi6Ge6}\hyperref[smsec:col]{\texorpdfstring{TiNi$_6$Ge$_6$}{}}~{\color{red}  (High-T unstable, Low-T unstable) }}

\begin{table}[htbp]
\global\long\def\arraystretch{1.12}
\captionsetup{width=.85\textwidth}
\caption{Structure parameters of TiNi$_6$Ge$_6$ after relaxation. It contains 382 electrons. }
\label{tab:TiNi6Ge6}
\setlength{\tabcolsep}{4mm}{
%
}
\end{table}

\begin{figure}[htbp]
\centering
\includegraphics[width=0.85\textwidth]{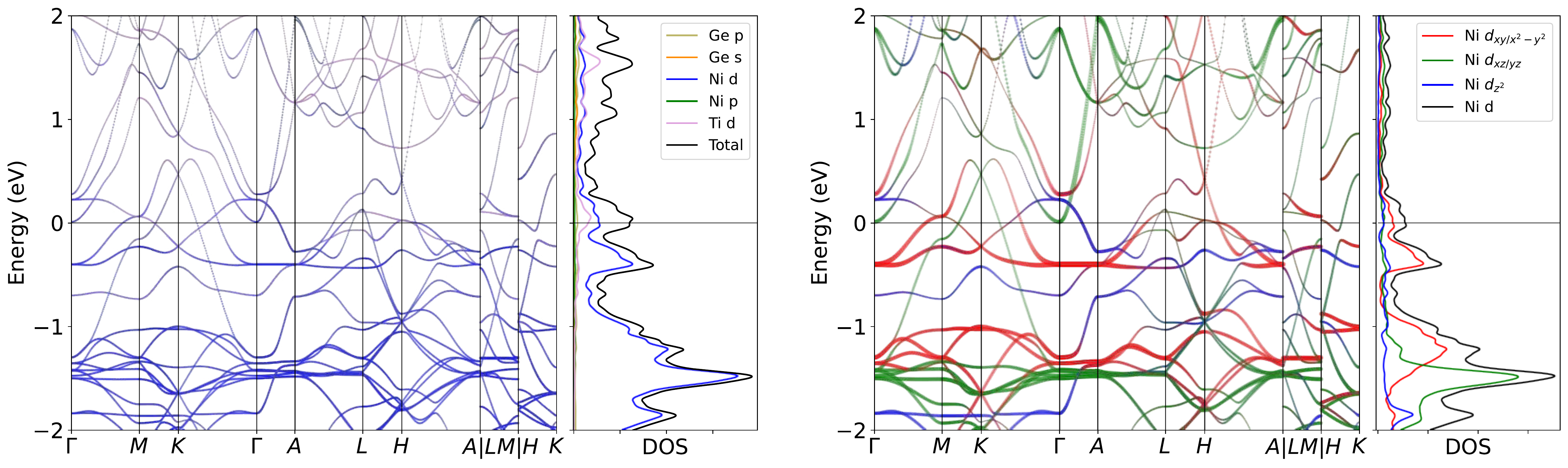}
\caption{Band structure for TiNi$_6$Ge$_6$ without SOC. The projection of Ni $3d$ orbitals is also presented. The band structures clearly reveal the dominating of Ni $3d$ orbitals  near the Fermi level, as well as the pronounced peak.}
\label{fig:TiNi6Ge6band}
\end{figure}

\begin{figure}[H]
\centering
\subfloat[Relaxed phonon at low-T.]{
\label{fig:TiNi6Ge6_relaxed_Low}
\includegraphics[width=0.45\textwidth]{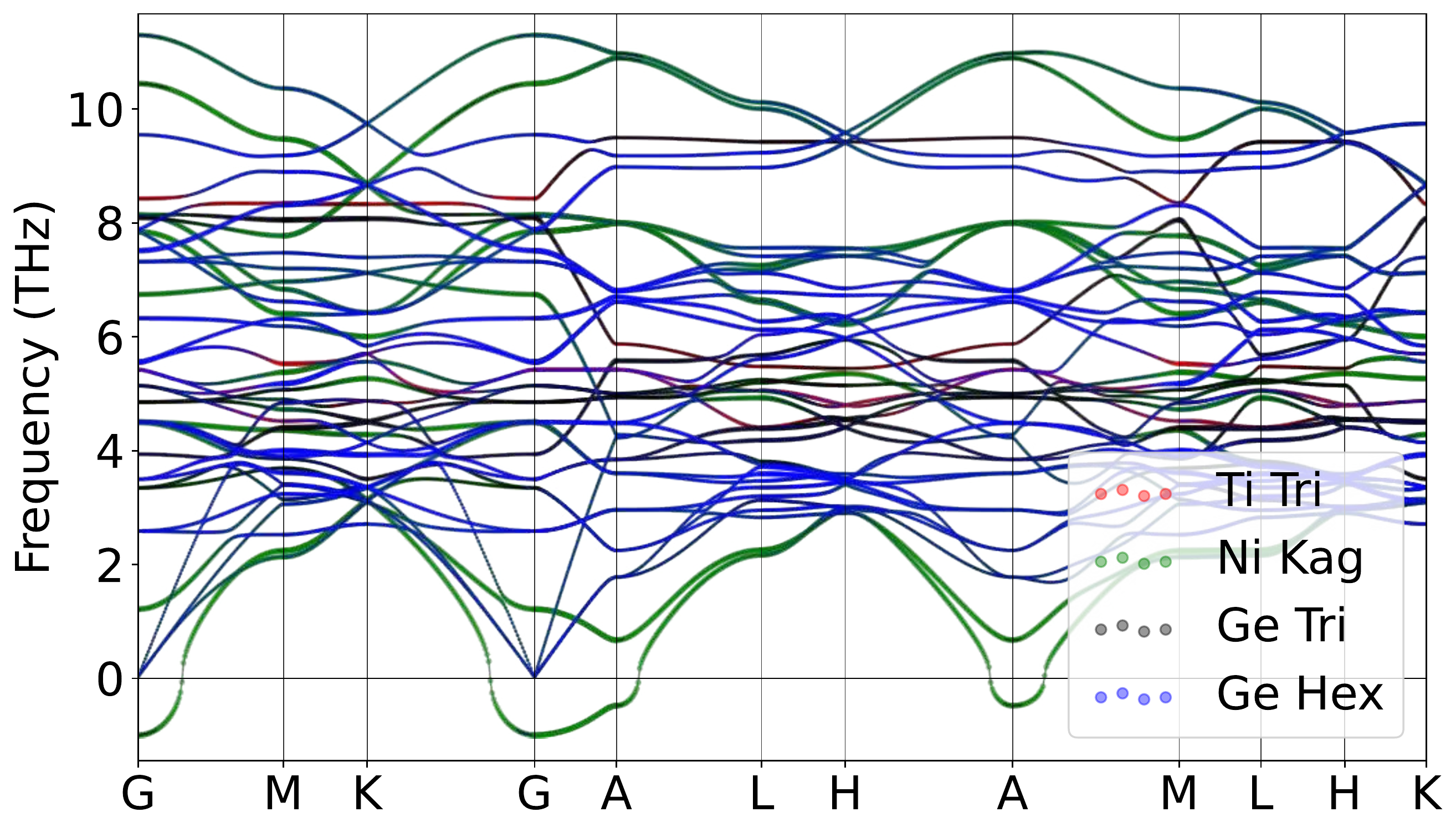}
}
\hspace{5pt}\subfloat[Relaxed phonon at high-T.]{
\label{fig:TiNi6Ge6_relaxed_High}
\includegraphics[width=0.45\textwidth]{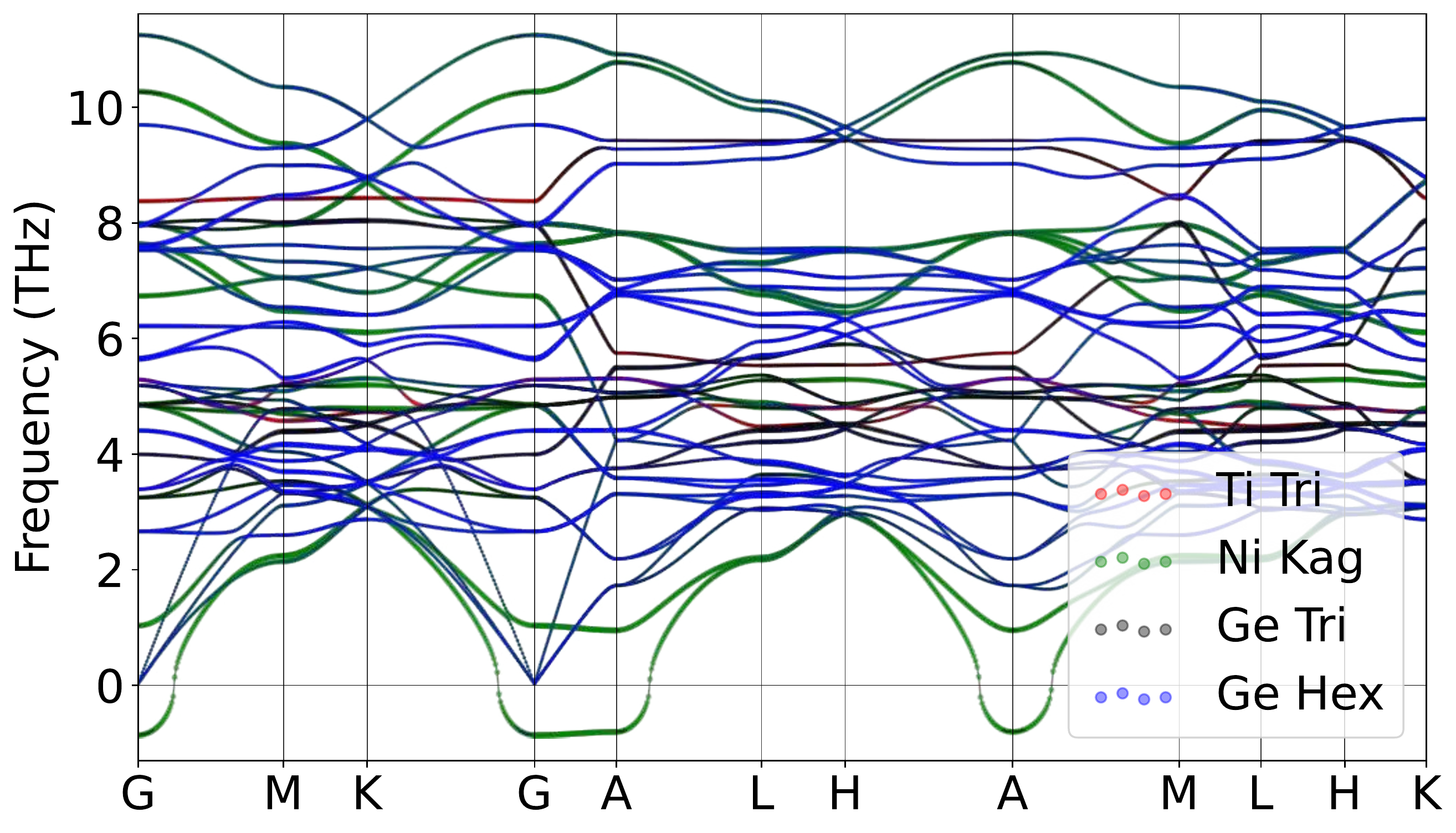}
}
\caption{Atomically projected phonon spectrum for TiNi$_6$Ge$_6$.}
\label{fig:TiNi6Ge6pho}
\end{figure}

\begin{figure}[H]
\centering
\label{fig:TiNi6Ge6_relaxed_Low_xz}
\includegraphics[width=0.85\textwidth]{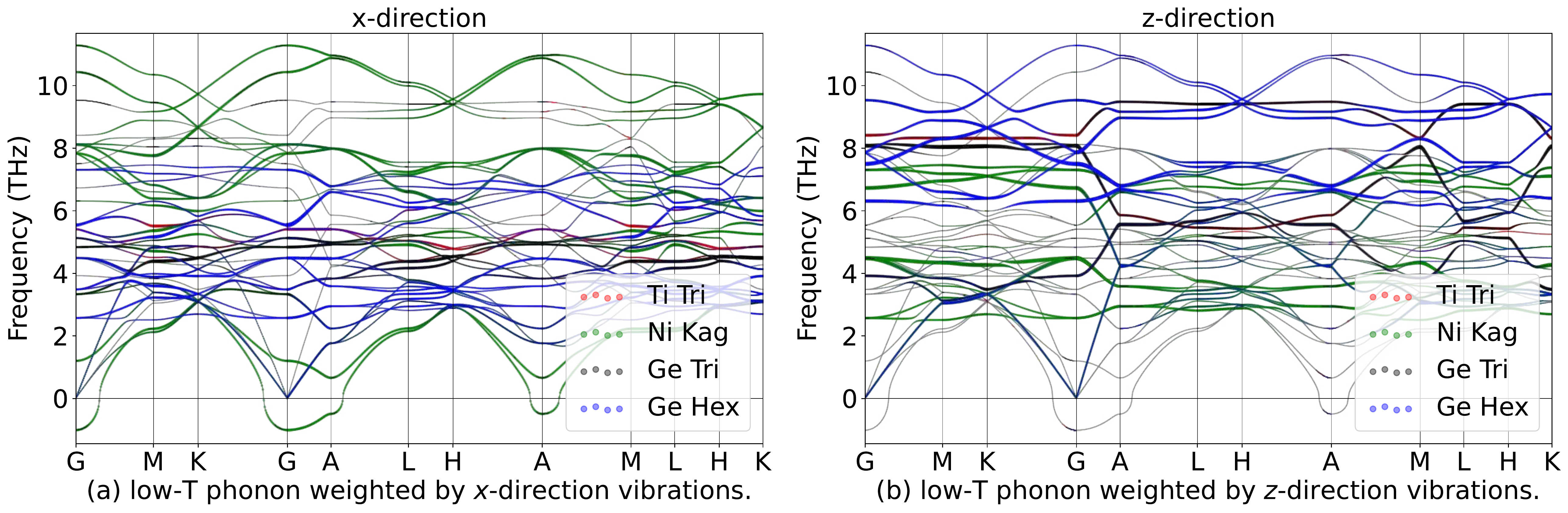}
\hspace{5pt}\label{fig:TiNi6Ge6_relaxed_High_xz}
\includegraphics[width=0.85\textwidth]{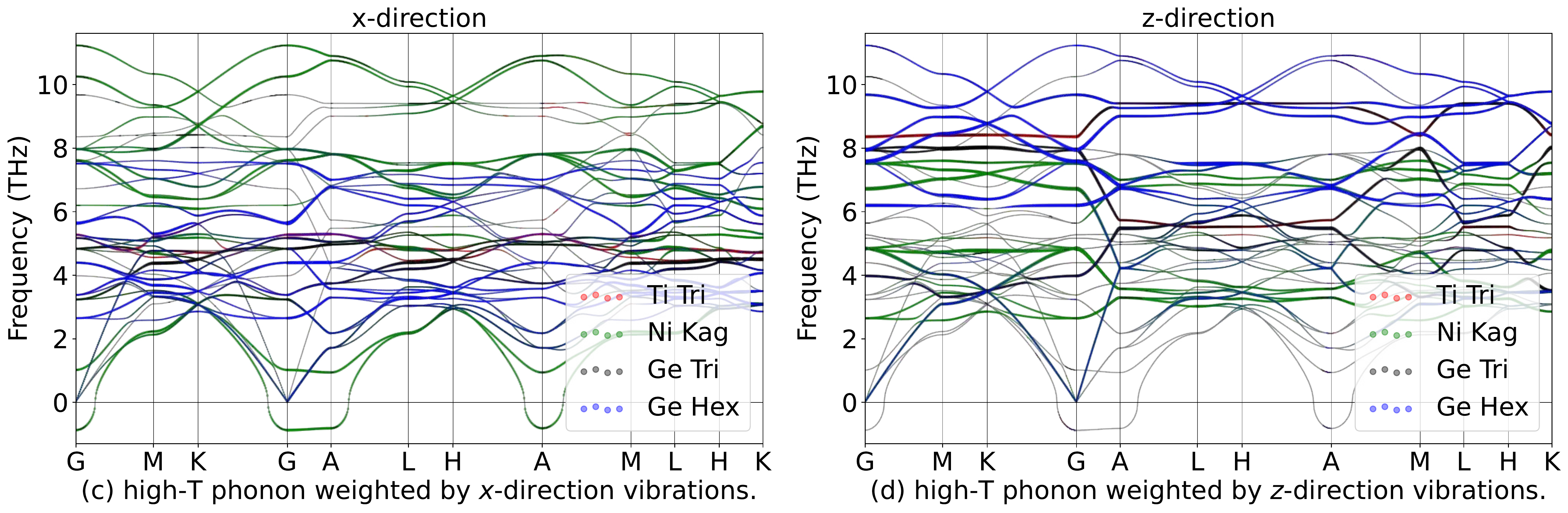}
\caption{Phonon spectrum for TiNi$_6$Ge$_6$in in-plane ($x$) and out-of-plane ($z$) directions.}
\label{fig:TiNi6Ge6phoxyz}
\end{figure}

\begin{table}[htbp]
\global\long\def\arraystretch{1.12}
\captionsetup{width=.95\textwidth}
\caption{IRREPs at high-symmetry points for TiNi$_6$Ge$_6$ phonon spectrum at Low-T.}
\label{tab:191_TiNi6Ge6_Low}
\setlength{\tabcolsep}{1.mm}{
}
\end{table}

\begin{table}[htbp]
\global\long\def\arraystretch{1.12}
\captionsetup{width=.95\textwidth}
\caption{IRREPs at high-symmetry points for TiNi$_6$Ge$_6$ phonon spectrum at High-T.}
\label{tab:191_TiNi6Ge6_High}
\setlength{\tabcolsep}{1.mm}{
%
}
\end{table}
\subsubsection{\label{sms2sec:YNi6Ge6}\hyperref[smsec:col]{\texorpdfstring{YNi$_6$Ge$_6$}{}}~{\color{green}  (High-T stable, Low-T stable) }}

\begin{table}[htbp]
\global\long\def\arraystretch{1.12}
\captionsetup{width=.85\textwidth}
\caption{Structure parameters of YNi$_6$Ge$_6$ after relaxation. It contains 399 electrons. }
\label{tab:YNi6Ge6}
\setlength{\tabcolsep}{4mm}{
%
}
\end{table}

\begin{figure}[htbp]
\centering
\includegraphics[width=0.85\textwidth]{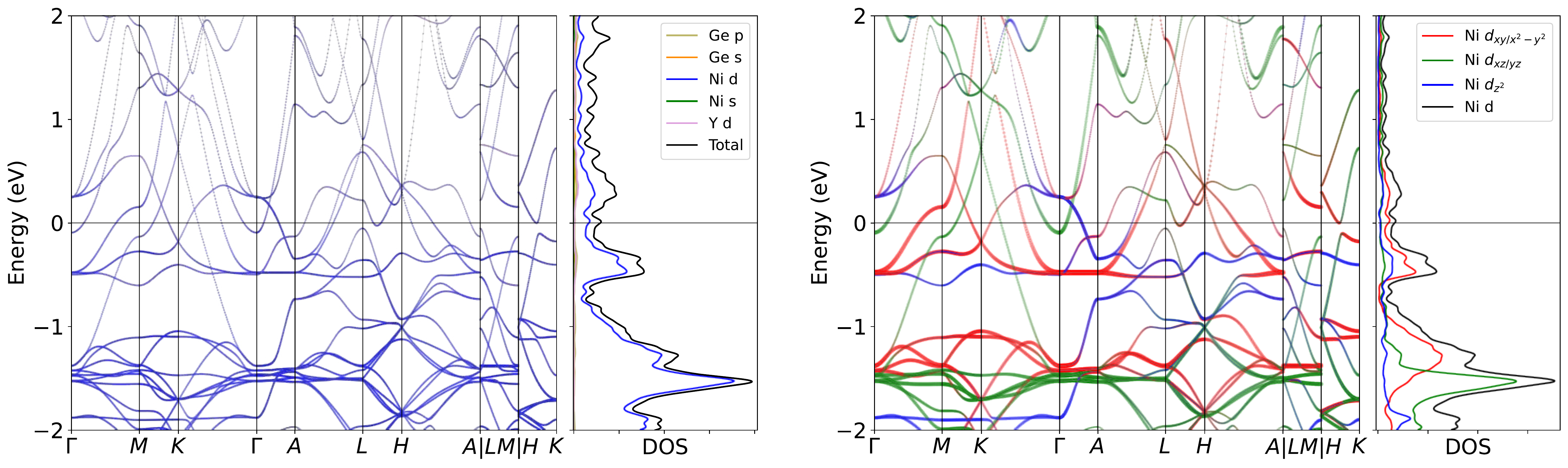}
\caption{Band structure for YNi$_6$Ge$_6$ without SOC. The projection of Ni $3d$ orbitals is also presented. The band structures clearly reveal the dominating of Ni $3d$ orbitals  near the Fermi level, as well as the pronounced peak.}
\label{fig:YNi6Ge6band}
\end{figure}

\begin{figure}[H]
\centering
\subfloat[Relaxed phonon at low-T.]{
\label{fig:YNi6Ge6_relaxed_Low}
\includegraphics[width=0.45\textwidth]{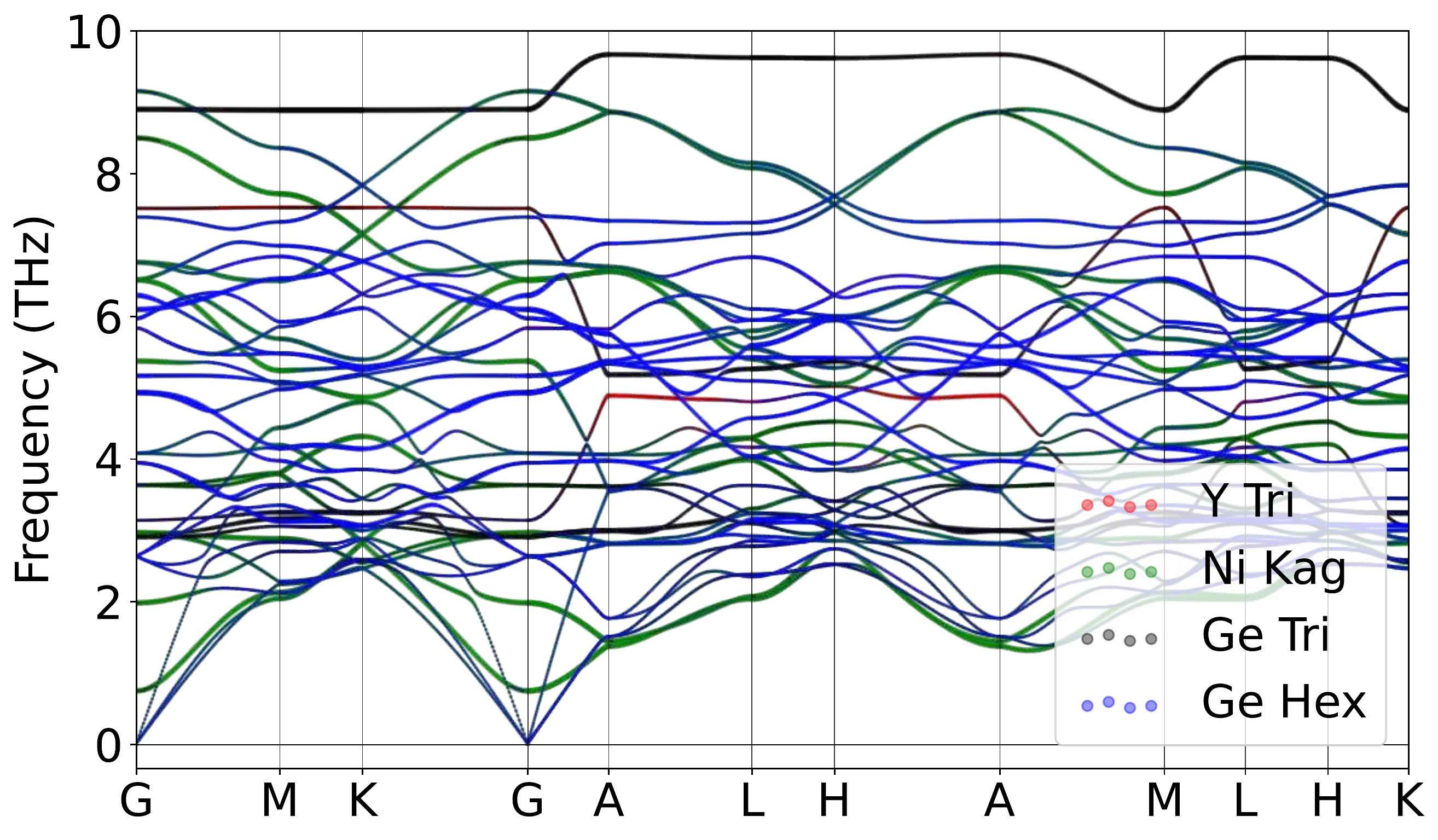}
}
\hspace{5pt}\subfloat[Relaxed phonon at high-T.]{
\label{fig:YNi6Ge6_relaxed_High}
\includegraphics[width=0.45\textwidth]{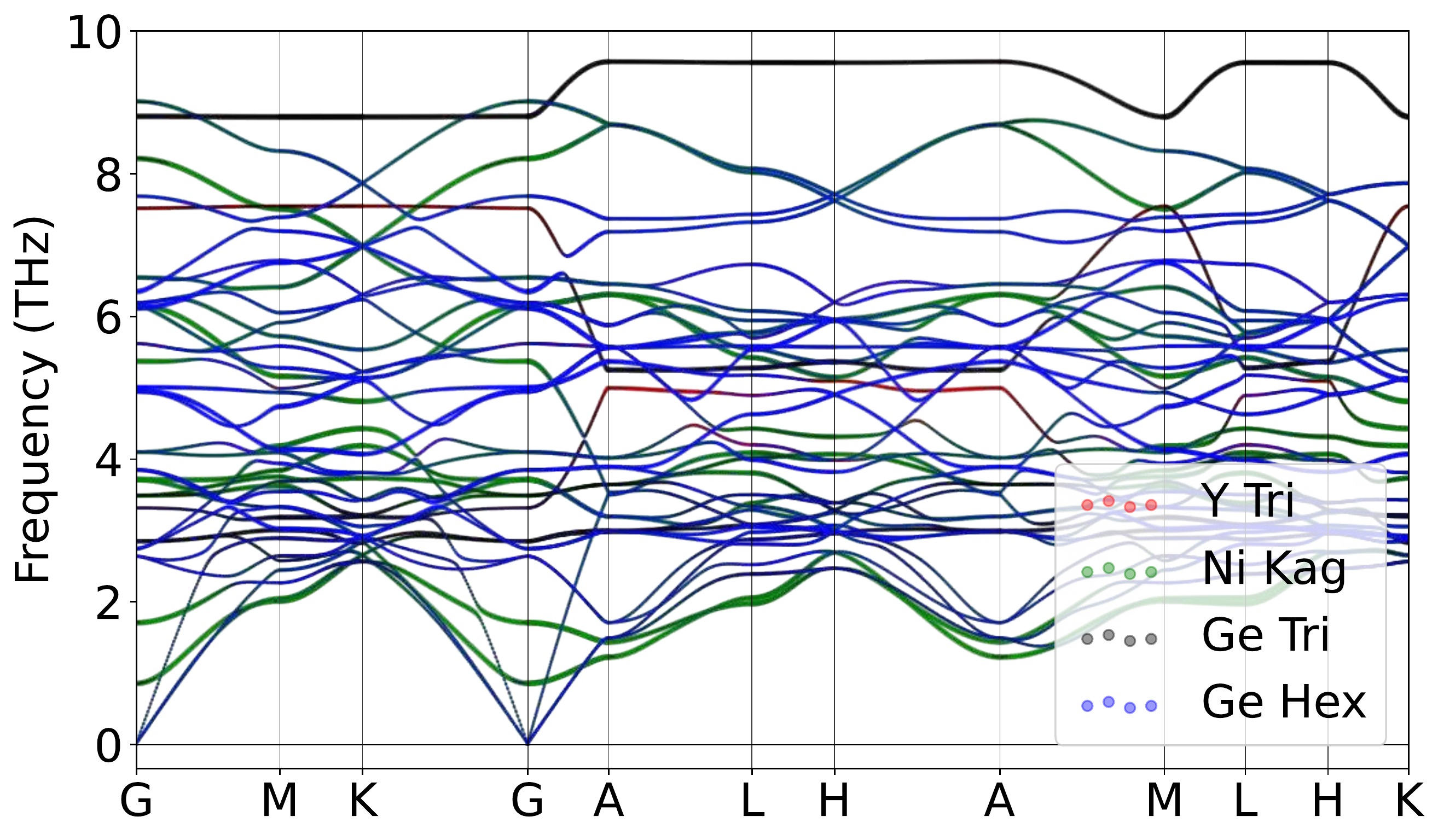}
}
\caption{Atomically projected phonon spectrum for YNi$_6$Ge$_6$.}
\label{fig:YNi6Ge6pho}
\end{figure}

\begin{figure}[H]
\centering
\label{fig:YNi6Ge6_relaxed_Low_xz}
\includegraphics[width=0.85\textwidth]{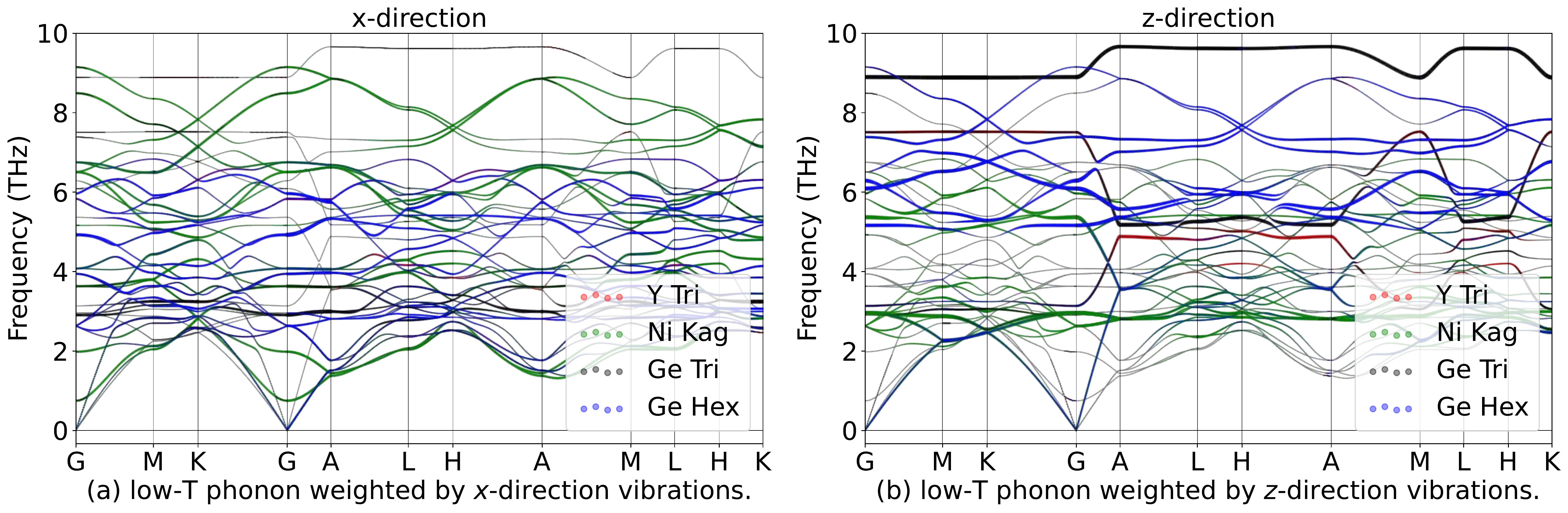}
\hspace{5pt}\label{fig:YNi6Ge6_relaxed_High_xz}
\includegraphics[width=0.85\textwidth]{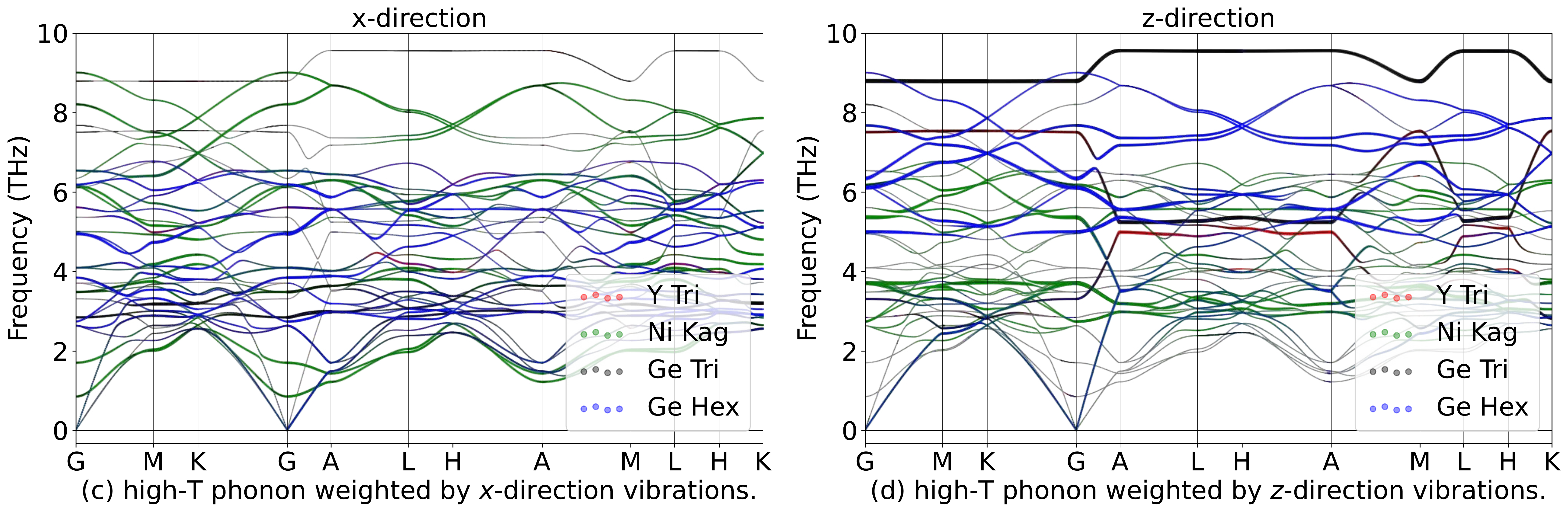}
\caption{Phonon spectrum for YNi$_6$Ge$_6$in in-plane ($x$) and out-of-plane ($z$) directions.}
\label{fig:YNi6Ge6phoxyz}
\end{figure}

\begin{table}[htbp]
\global\long\def\arraystretch{1.12}
\captionsetup{width=.95\textwidth}
\caption{IRREPs at high-symmetry points for YNi$_6$Ge$_6$ phonon spectrum at Low-T.}
\label{tab:191_YNi6Ge6_Low}
\setlength{\tabcolsep}{1.mm}{
}
\end{table}

\begin{table}[htbp]
\global\long\def\arraystretch{1.12}
\captionsetup{width=.95\textwidth}
\caption{IRREPs at high-symmetry points for YNi$_6$Ge$_6$ phonon spectrum at High-T.}
\label{tab:191_YNi6Ge6_High}
\setlength{\tabcolsep}{1.mm}{
%
}
\end{table}
\subsubsection{\label{sms2sec:ZrNi6Ge6}\hyperref[smsec:col]{\texorpdfstring{ZrNi$_6$Ge$_6$}{}}~{\color{green}  (High-T stable, Low-T stable) }}

\begin{table}[htbp]
\global\long\def\arraystretch{1.12}
\captionsetup{width=.85\textwidth}
\caption{Structure parameters of ZrNi$_6$Ge$_6$ after relaxation. It contains 400 electrons. }
\label{tab:ZrNi6Ge6}
\setlength{\tabcolsep}{4mm}{
%
}
\end{table}

\begin{figure}[htbp]
\centering
\includegraphics[width=0.85\textwidth]{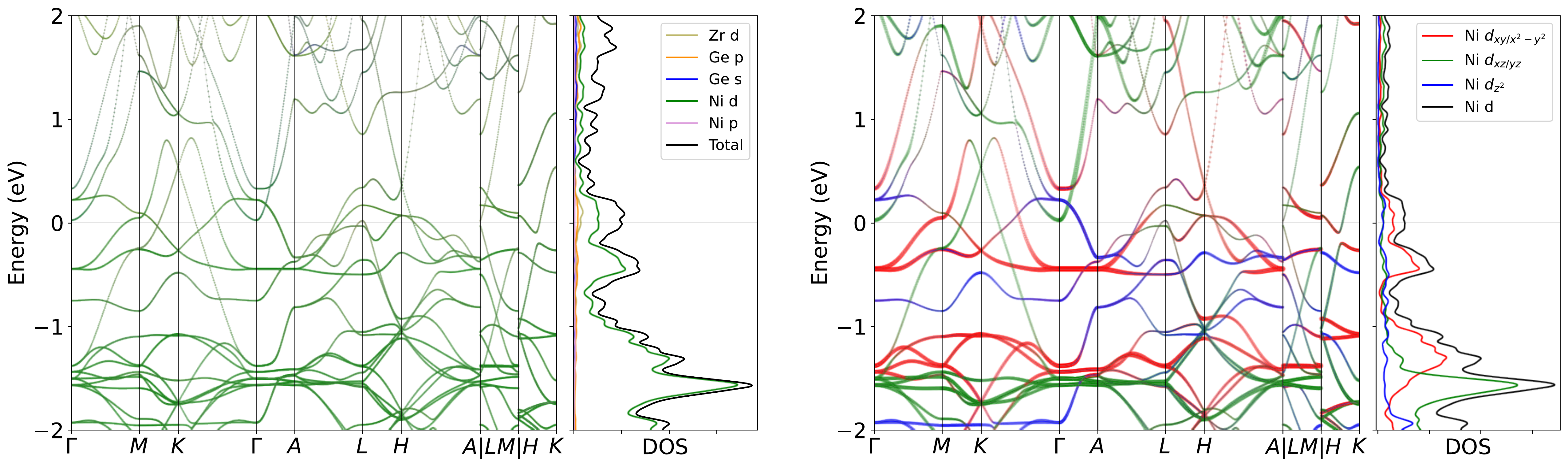}
\caption{Band structure for ZrNi$_6$Ge$_6$ without SOC. The projection of Ni $3d$ orbitals is also presented. The band structures clearly reveal the dominating of Ni $3d$ orbitals  near the Fermi level, as well as the pronounced peak.}
\label{fig:ZrNi6Ge6band}
\end{figure}

\begin{figure}[H]
\centering
\subfloat[Relaxed phonon at low-T.]{
\label{fig:ZrNi6Ge6_relaxed_Low}
\includegraphics[width=0.45\textwidth]{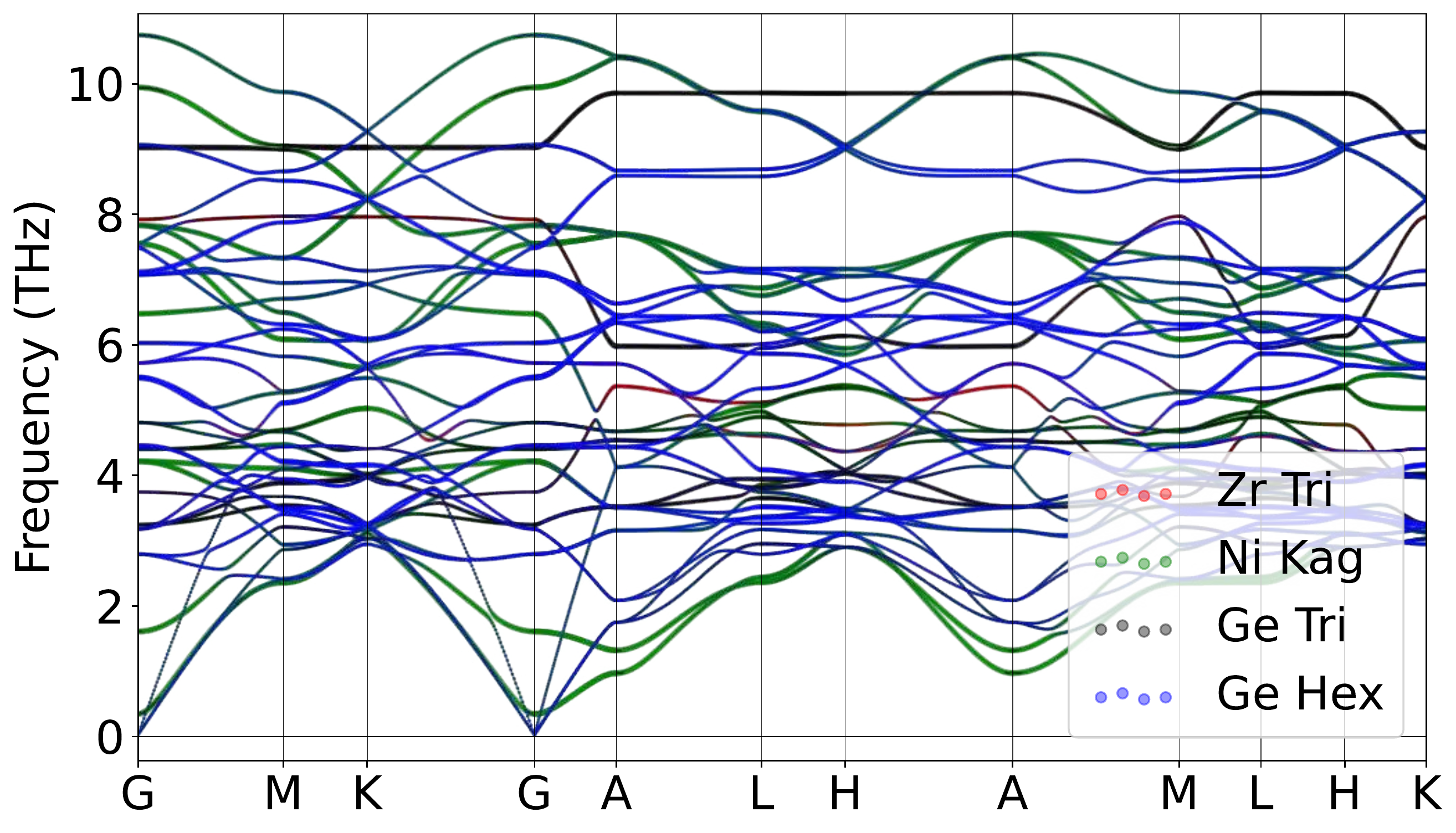}
}
\hspace{5pt}\subfloat[Relaxed phonon at high-T.]{
\label{fig:ZrNi6Ge6_relaxed_High}
\includegraphics[width=0.45\textwidth]{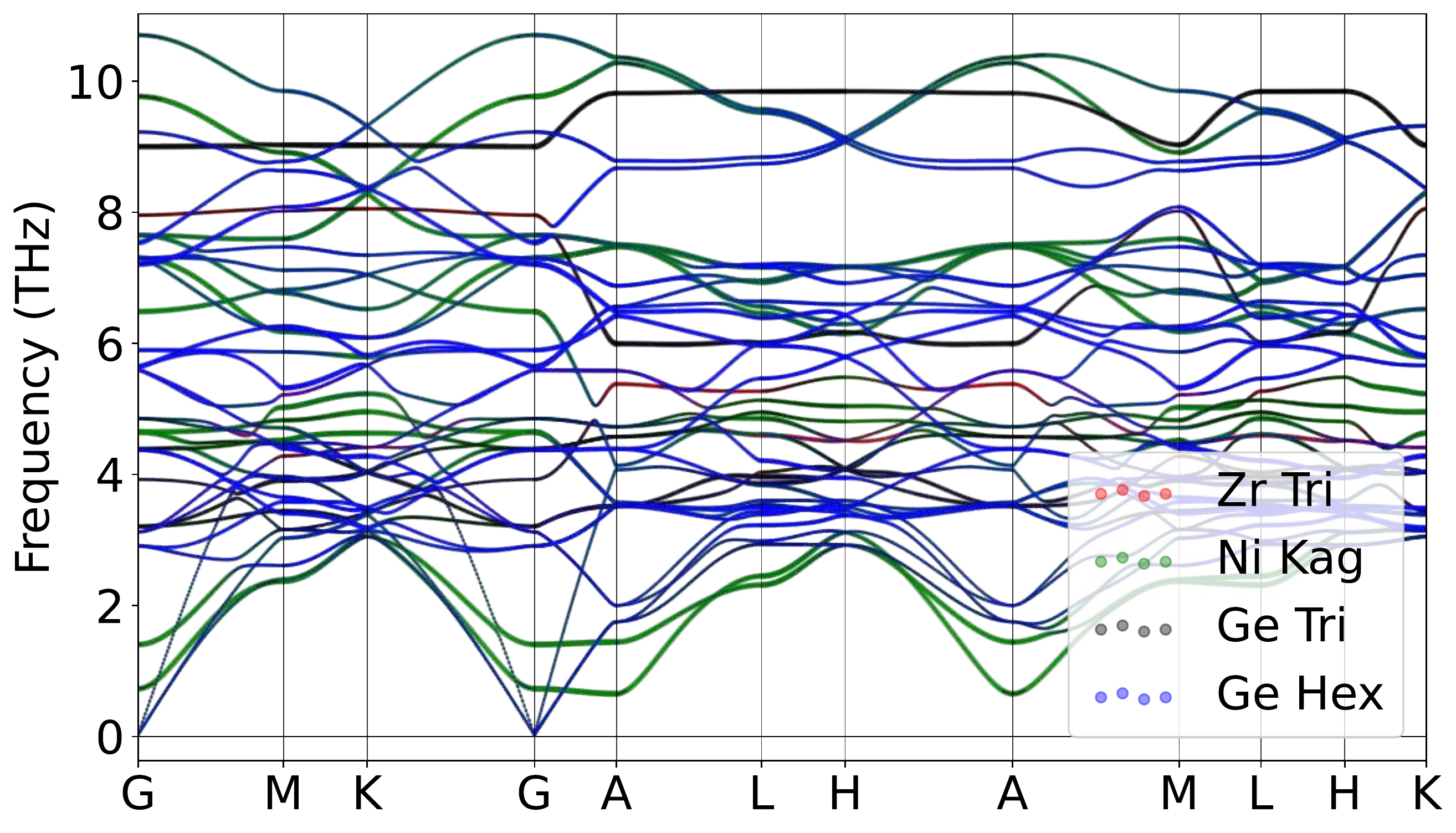}
}
\caption{Atomically projected phonon spectrum for ZrNi$_6$Ge$_6$.}
\label{fig:ZrNi6Ge6pho}
\end{figure}

\begin{figure}[H]
\centering
\label{fig:ZrNi6Ge6_relaxed_Low_xz}
\includegraphics[width=0.85\textwidth]{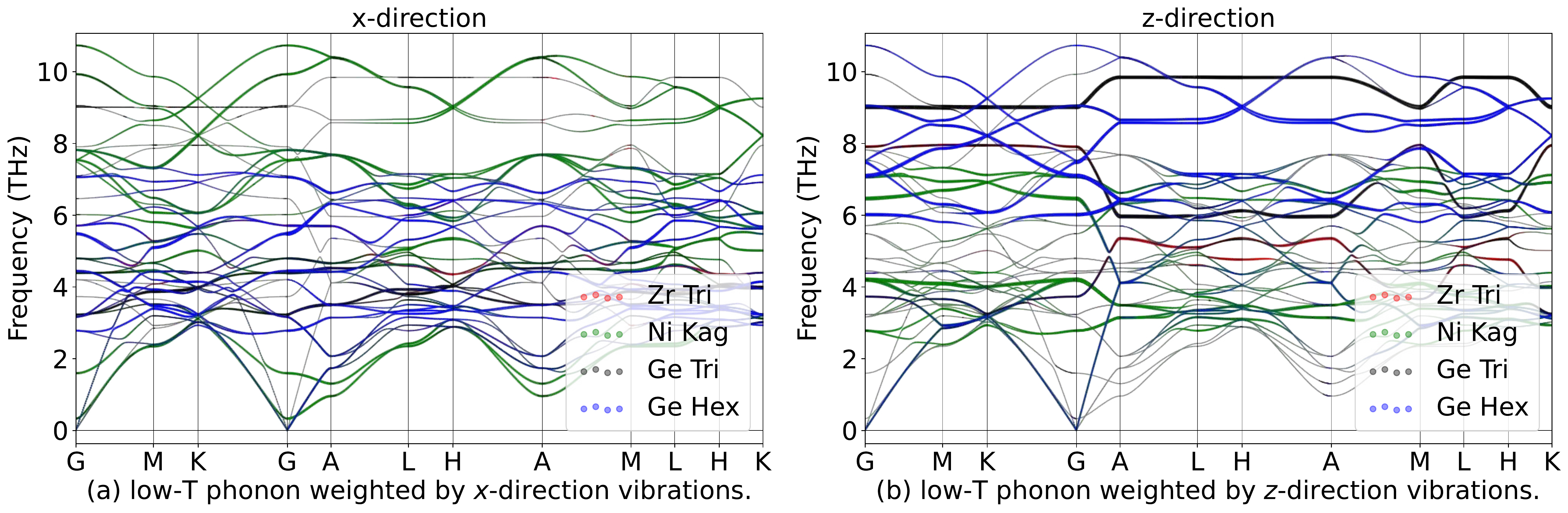}
\hspace{5pt}\label{fig:ZrNi6Ge6_relaxed_High_xz}
\includegraphics[width=0.85\textwidth]{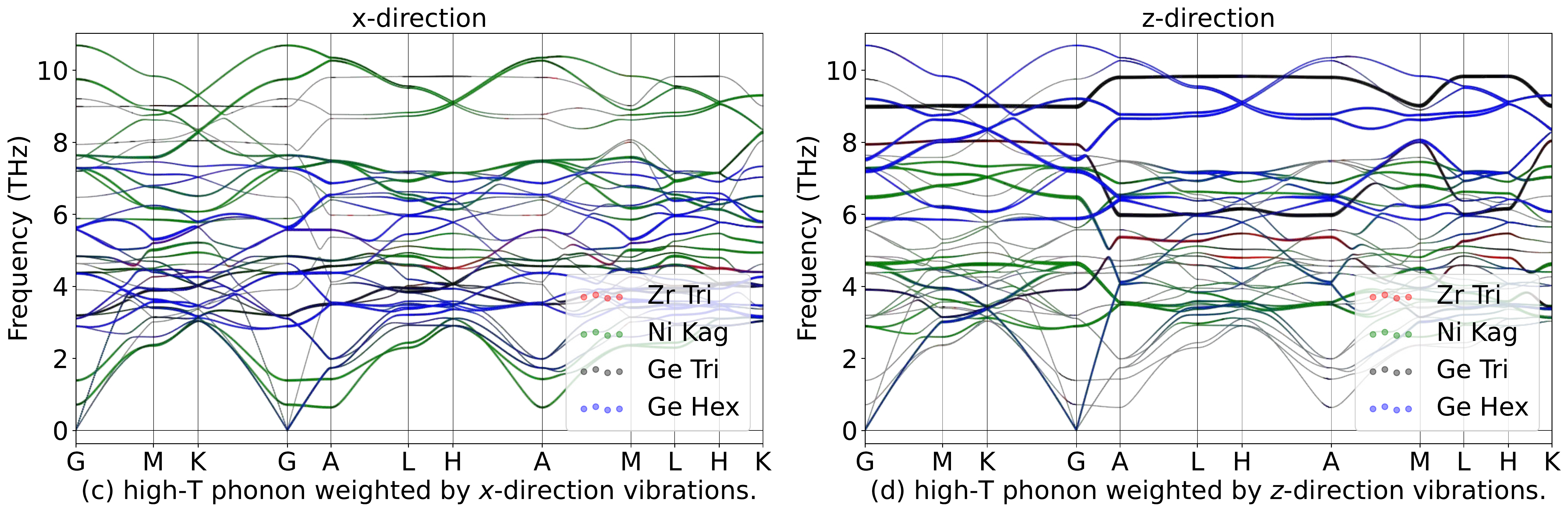}
\caption{Phonon spectrum for ZrNi$_6$Ge$_6$in in-plane ($x$) and out-of-plane ($z$) directions.}
\label{fig:ZrNi6Ge6phoxyz}
\end{figure}

\begin{table}[htbp]
\global\long\def\arraystretch{1.12}
\captionsetup{width=.95\textwidth}
\caption{IRREPs at high-symmetry points for ZrNi$_6$Ge$_6$ phonon spectrum at Low-T.}
\label{tab:191_ZrNi6Ge6_Low}
\setlength{\tabcolsep}{1.mm}{
}
\end{table}

\begin{table}[htbp]
\global\long\def\arraystretch{1.12}
\captionsetup{width=.95\textwidth}
\caption{IRREPs at high-symmetry points for ZrNi$_6$Ge$_6$ phonon spectrum at High-T.}
\label{tab:191_ZrNi6Ge6_High}
\setlength{\tabcolsep}{1.mm}{
%
}
\end{table}
\subsubsection{\label{sms2sec:NbNi6Ge6}\hyperref[smsec:col]{\texorpdfstring{NbNi$_6$Ge$_6$}{}}~{\color{red}  (High-T unstable, Low-T unstable) }}

\begin{table}[htbp]
\global\long\def\arraystretch{1.12}
\captionsetup{width=.85\textwidth}
\caption{Structure parameters of NbNi$_6$Ge$_6$ after relaxation. It contains 401 electrons. }
\label{tab:NbNi6Ge6}
\setlength{\tabcolsep}{4mm}{
%
}
\end{table}

\begin{figure}[htbp]
\centering
\includegraphics[width=0.85\textwidth]{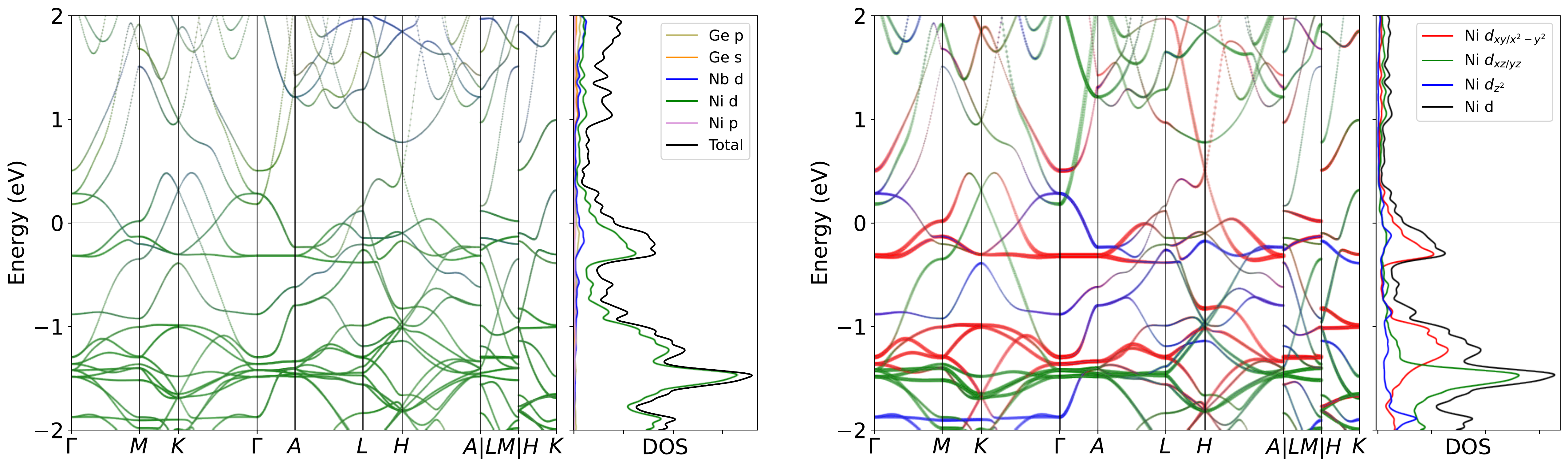}
\caption{Band structure for NbNi$_6$Ge$_6$ without SOC. The projection of Ni $3d$ orbitals is also presented. The band structures clearly reveal the dominating of Ni $3d$ orbitals  near the Fermi level, as well as the pronounced peak.}
\label{fig:NbNi6Ge6band}
\end{figure}

\begin{figure}[H]
\centering
\subfloat[Relaxed phonon at low-T.]{
\label{fig:NbNi6Ge6_relaxed_Low}
\includegraphics[width=0.45\textwidth]{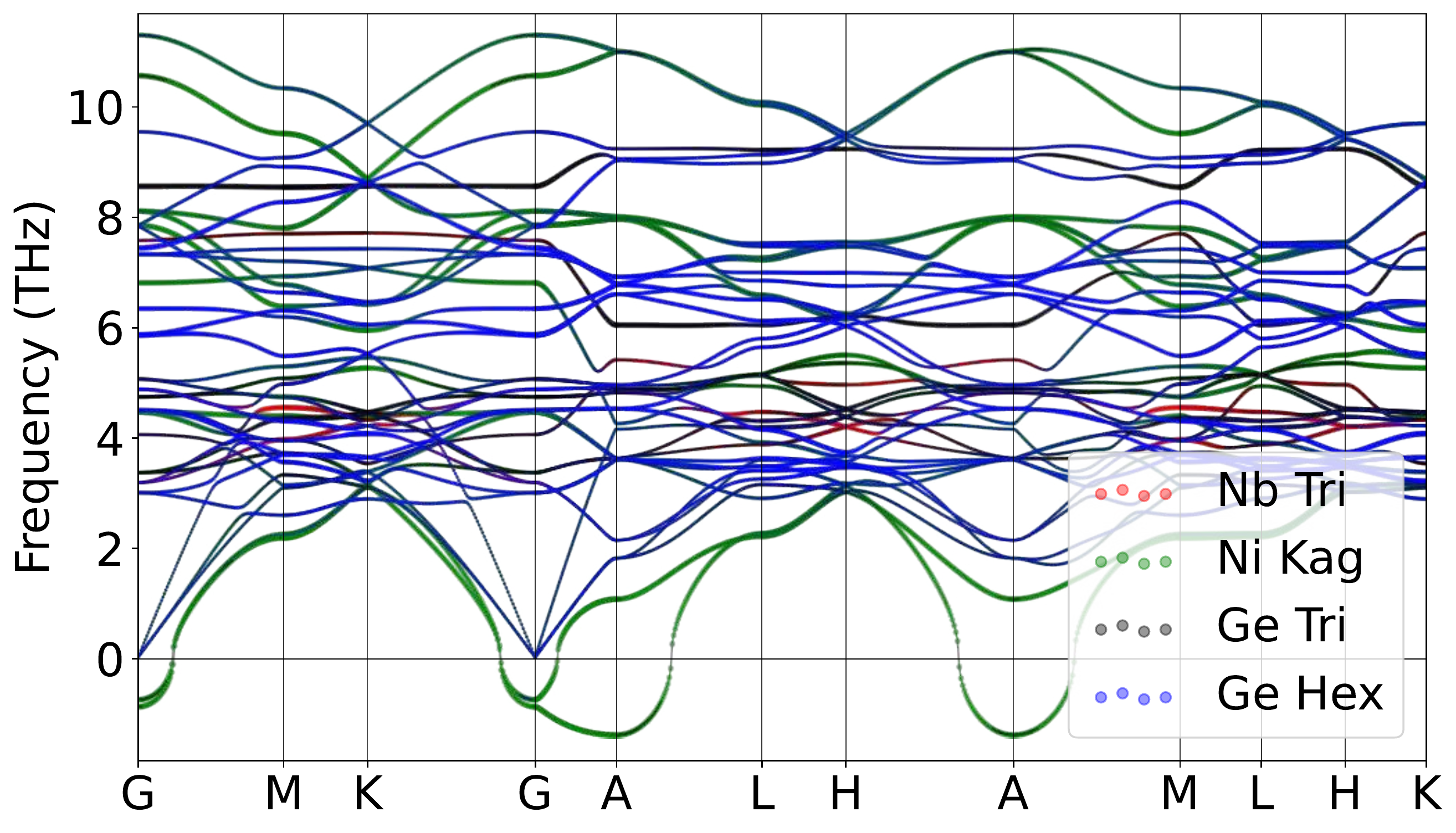}
}
\hspace{5pt}\subfloat[Relaxed phonon at high-T.]{
\label{fig:NbNi6Ge6_relaxed_High}
\includegraphics[width=0.45\textwidth]{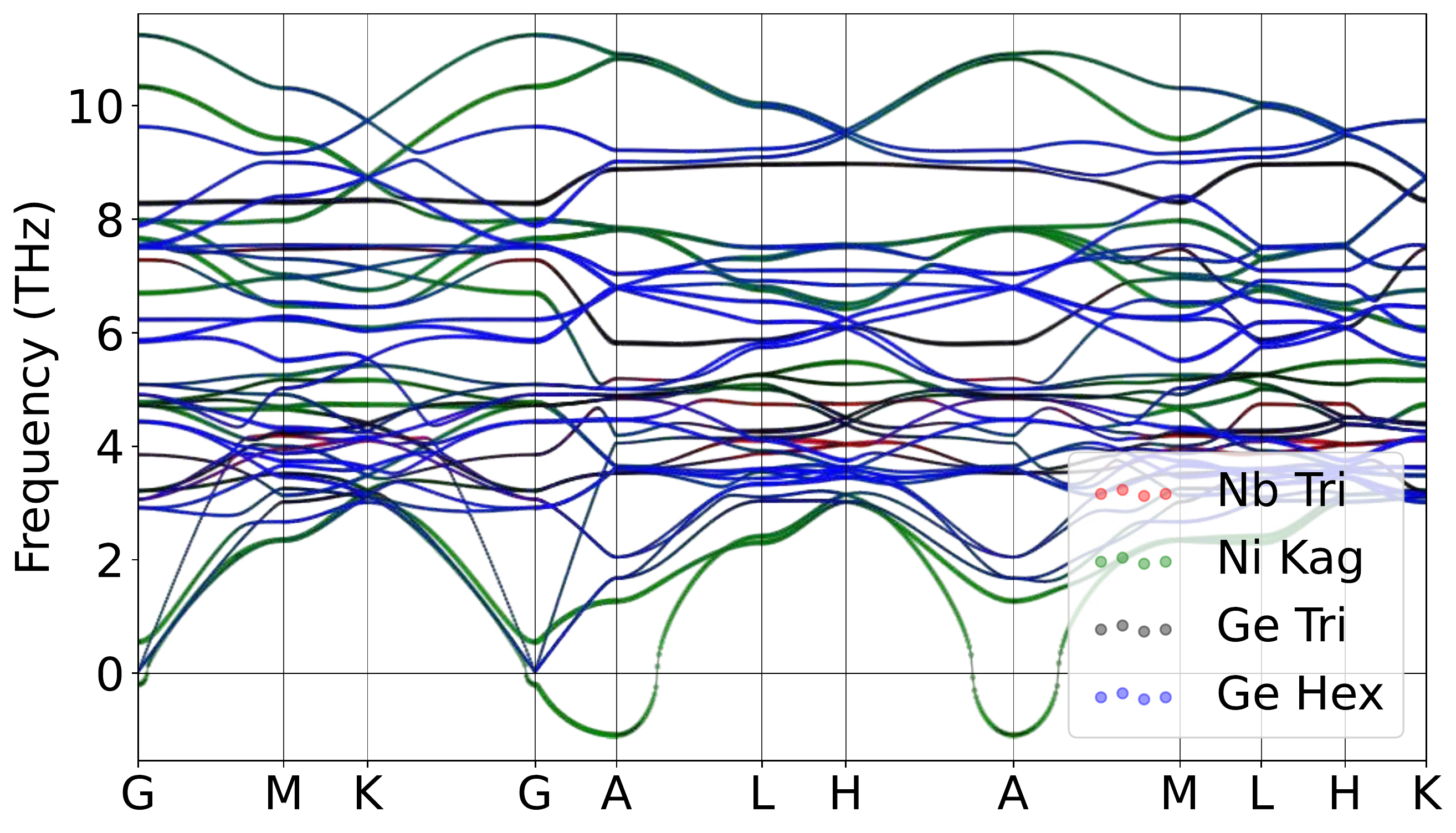}
}
\caption{Atomically projected phonon spectrum for NbNi$_6$Ge$_6$.}
\label{fig:NbNi6Ge6pho}
\end{figure}

\begin{figure}[H]
\centering
\label{fig:NbNi6Ge6_relaxed_Low_xz}
\includegraphics[width=0.85\textwidth]{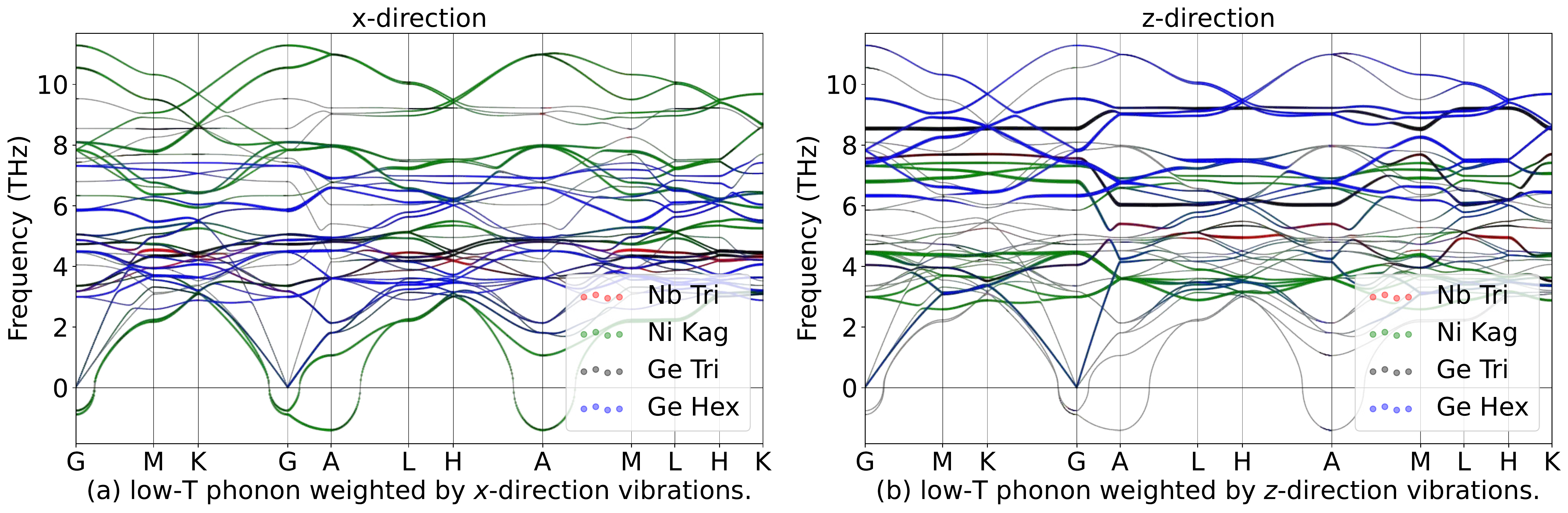}
\hspace{5pt}\label{fig:NbNi6Ge6_relaxed_High_xz}
\includegraphics[width=0.85\textwidth]{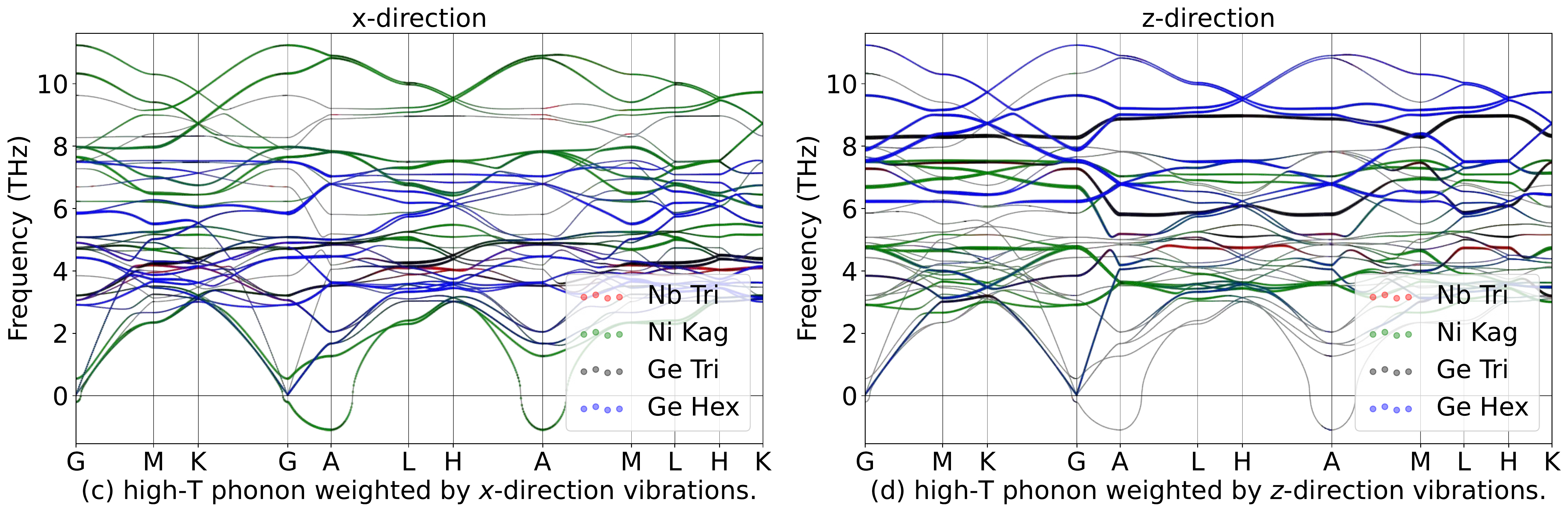}
\caption{Phonon spectrum for NbNi$_6$Ge$_6$in in-plane ($x$) and out-of-plane ($z$) directions.}
\label{fig:NbNi6Ge6phoxyz}
\end{figure}

\begin{table}[htbp]
\global\long\def\arraystretch{1.12}
\captionsetup{width=.95\textwidth}
\caption{IRREPs at high-symmetry points for NbNi$_6$Ge$_6$ phonon spectrum at Low-T.}
\label{tab:191_NbNi6Ge6_Low}
\setlength{\tabcolsep}{1.mm}{
}
\end{table}

\begin{table}[htbp]
\global\long\def\arraystretch{1.12}
\captionsetup{width=.95\textwidth}
\caption{IRREPs at high-symmetry points for NbNi$_6$Ge$_6$ phonon spectrum at High-T.}
\label{tab:191_NbNi6Ge6_High}
\setlength{\tabcolsep}{1.mm}{
%
}
\end{table}
\subsubsection{\label{sms2sec:PrNi6Ge6}\hyperref[smsec:col]{\texorpdfstring{PrNi$_6$Ge$_6$}{}}~{\color{green}  (High-T stable, Low-T stable) }}

\begin{table}[htbp]
\global\long\def\arraystretch{1.12}
\captionsetup{width=.85\textwidth}
\caption{Structure parameters of PrNi$_6$Ge$_6$ after relaxation. It contains 419 electrons. }
\label{tab:PrNi6Ge6}
\setlength{\tabcolsep}{4mm}{
%
}
\end{table}

\begin{figure}[htbp]
\centering
\includegraphics[width=0.85\textwidth]{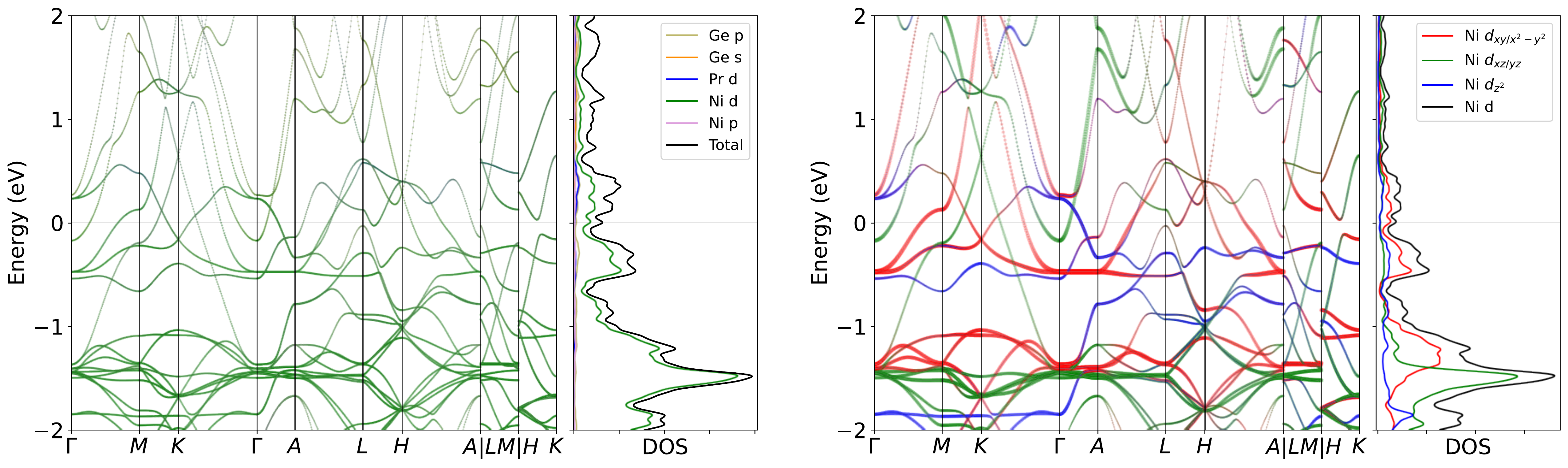}
\caption{Band structure for PrNi$_6$Ge$_6$ without SOC. The projection of Ni $3d$ orbitals is also presented. The band structures clearly reveal the dominating of Ni $3d$ orbitals  near the Fermi level, as well as the pronounced peak.}
\label{fig:PrNi6Ge6band}
\end{figure}

\begin{figure}[H]
\centering
\subfloat[Relaxed phonon at low-T.]{
\label{fig:PrNi6Ge6_relaxed_Low}
\includegraphics[width=0.45\textwidth]{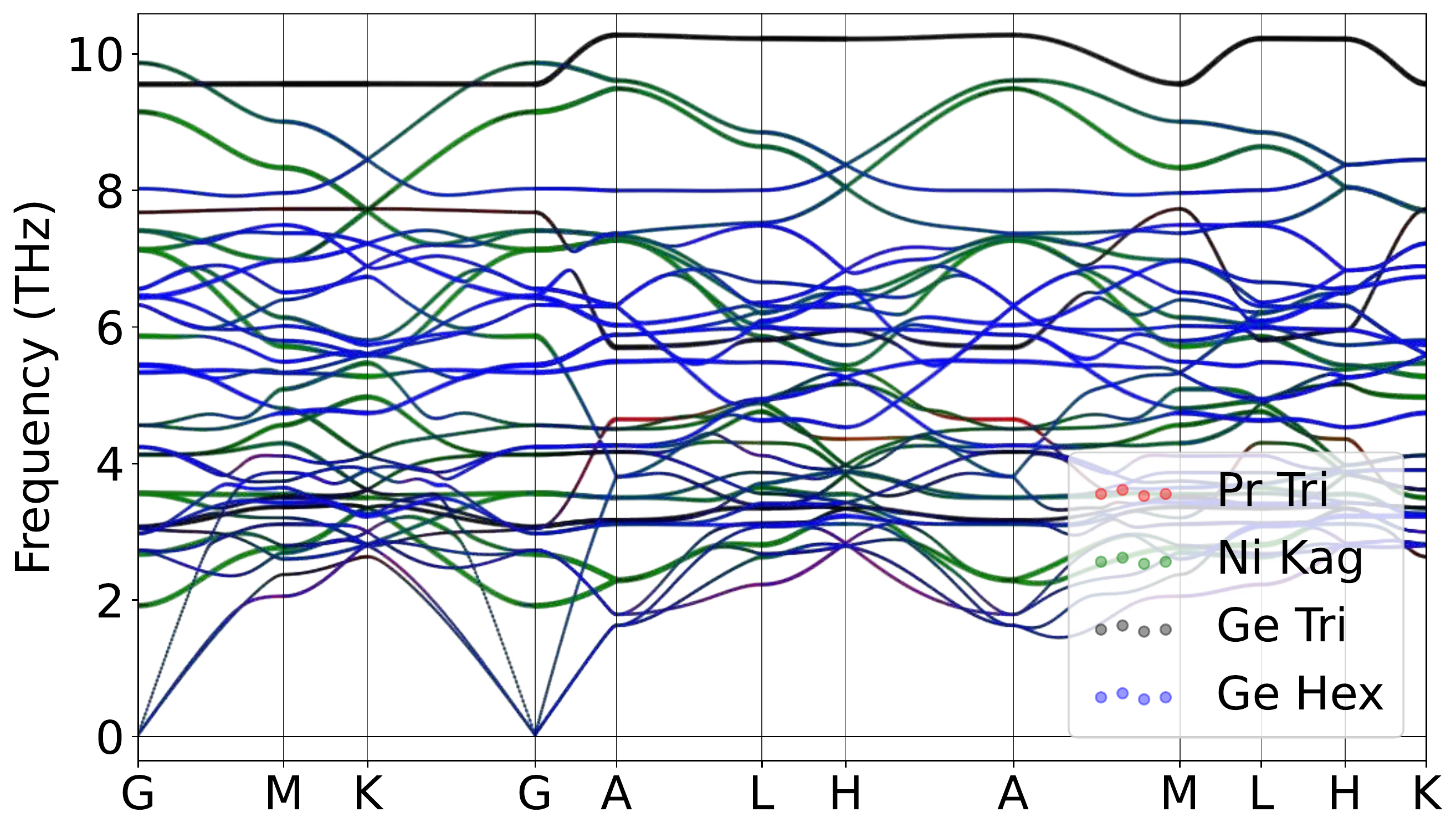}
}
\hspace{5pt}\subfloat[Relaxed phonon at high-T.]{
\label{fig:PrNi6Ge6_relaxed_High}
\includegraphics[width=0.45\textwidth]{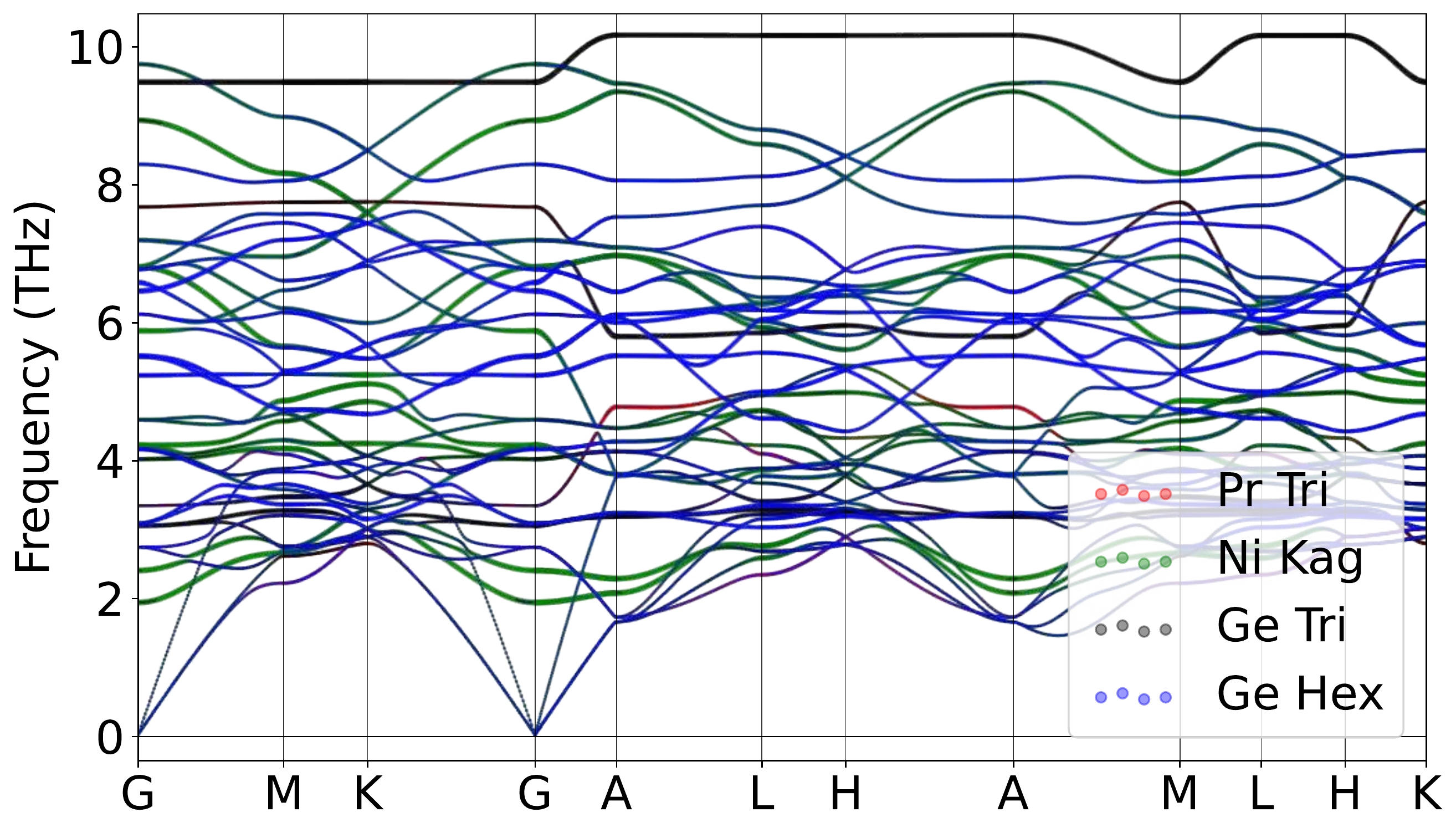}
}
\caption{Atomically projected phonon spectrum for PrNi$_6$Ge$_6$.}
\label{fig:PrNi6Ge6pho}
\end{figure}

\begin{figure}[H]
\centering
\label{fig:PrNi6Ge6_relaxed_Low_xz}
\includegraphics[width=0.85\textwidth]{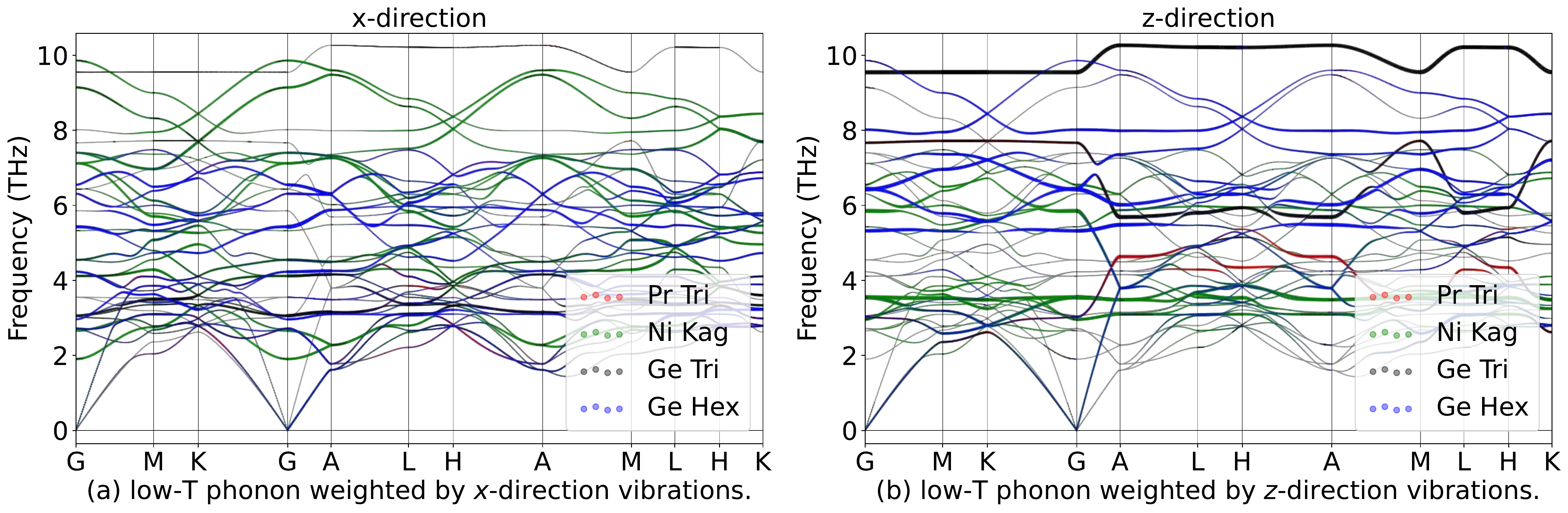}
\hspace{5pt}\label{fig:PrNi6Ge6_relaxed_High_xz}
\includegraphics[width=0.85\textwidth]{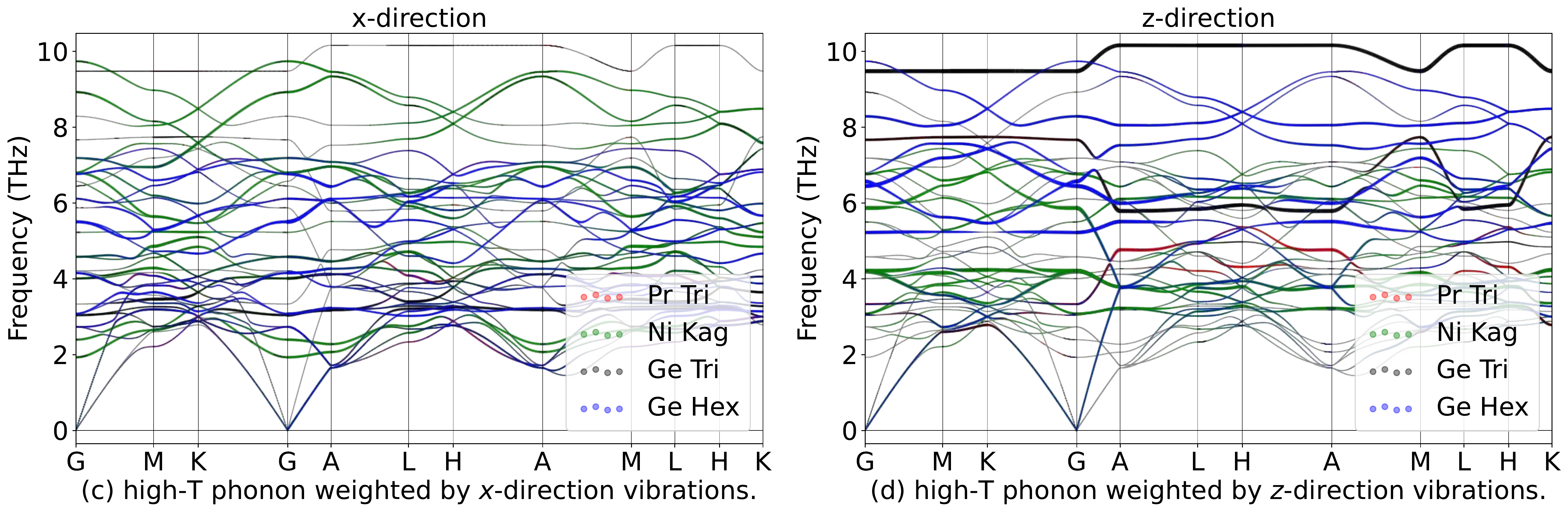}
\caption{Phonon spectrum for PrNi$_6$Ge$_6$in in-plane ($x$) and out-of-plane ($z$) directions.}
\label{fig:PrNi6Ge6phoxyz}
\end{figure}

\begin{table}[htbp]
\global\long\def\arraystretch{1.12}
\captionsetup{width=.95\textwidth}
\caption{IRREPs at high-symmetry points for PrNi$_6$Ge$_6$ phonon spectrum at Low-T.}
\label{tab:191_PrNi6Ge6_Low}
\setlength{\tabcolsep}{1.mm}{
}
\end{table}

\begin{table}[htbp]
\global\long\def\arraystretch{1.12}
\captionsetup{width=.95\textwidth}
\caption{IRREPs at high-symmetry points for PrNi$_6$Ge$_6$ phonon spectrum at High-T.}
\label{tab:191_PrNi6Ge6_High}
\setlength{\tabcolsep}{1.mm}{
%
}
\end{table}
\subsubsection{\label{sms2sec:NdNi6Ge6}\hyperref[smsec:col]{\texorpdfstring{NdNi$_6$Ge$_6$}{}}~{\color{green}  (High-T stable, Low-T stable) }}

\begin{table}[htbp]
\global\long\def\arraystretch{1.12}
\captionsetup{width=.85\textwidth}
\caption{Structure parameters of NdNi$_6$Ge$_6$ after relaxation. It contains 420 electrons. }
\label{tab:NdNi6Ge6}
\setlength{\tabcolsep}{4mm}{
%
}
\end{table}

\begin{figure}[htbp]
\centering
\includegraphics[width=0.85\textwidth]{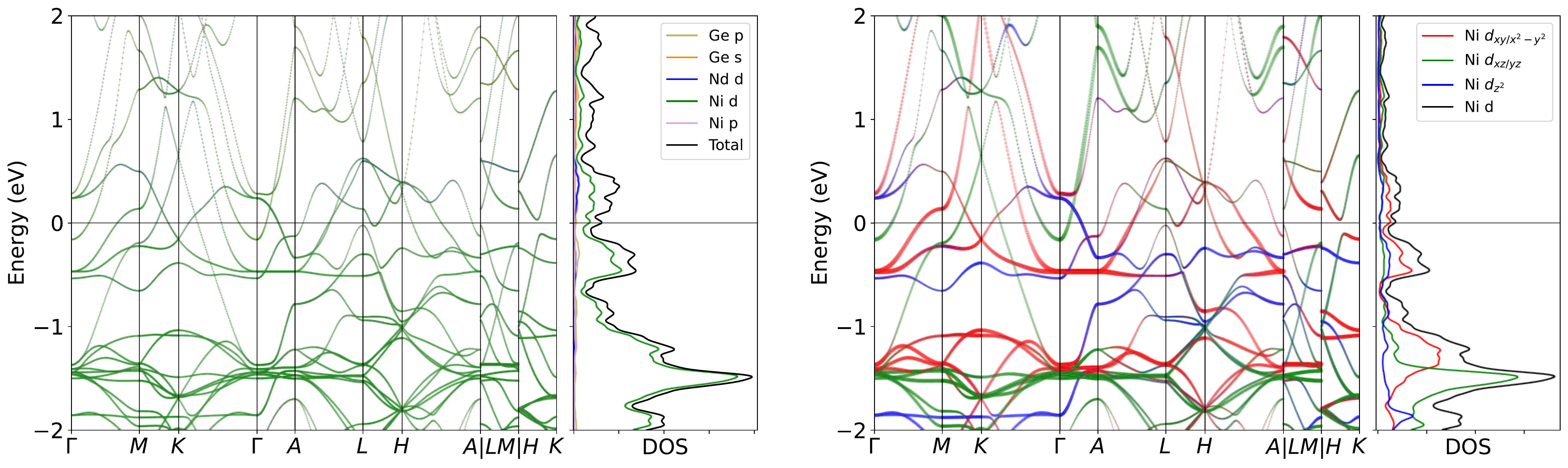}
\caption{Band structure for NdNi$_6$Ge$_6$ without SOC. The projection of Ni $3d$ orbitals is also presented. The band structures clearly reveal the dominating of Ni $3d$ orbitals  near the Fermi level, as well as the pronounced peak.}
\label{fig:NdNi6Ge6band}
\end{figure}

\begin{figure}[H]
\centering
\subfloat[Relaxed phonon at low-T.]{
\label{fig:NdNi6Ge6_relaxed_Low}
\includegraphics[width=0.45\textwidth]{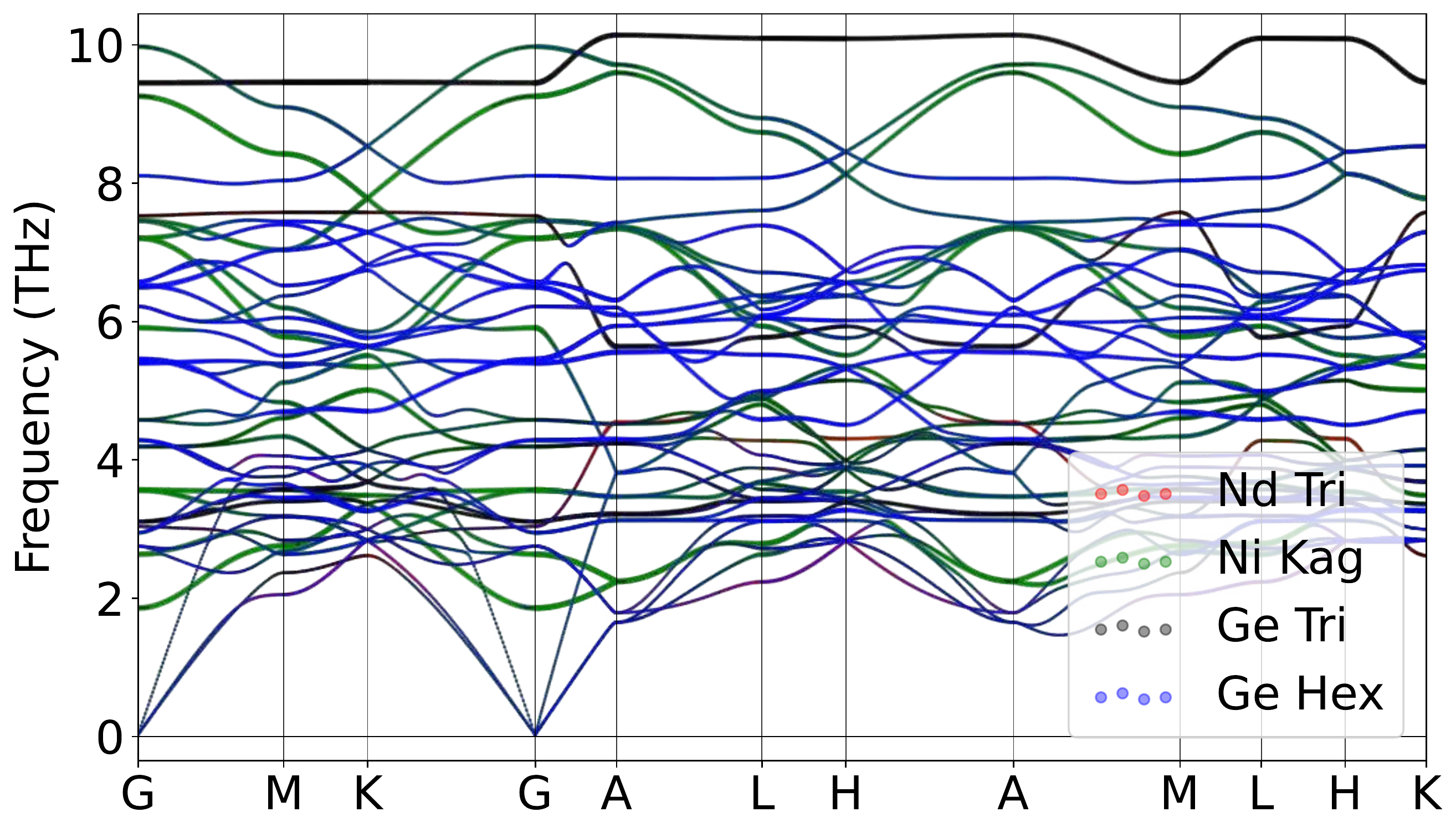}
}
\hspace{5pt}\subfloat[Relaxed phonon at high-T.]{
\label{fig:NdNi6Ge6_relaxed_High}
\includegraphics[width=0.45\textwidth]{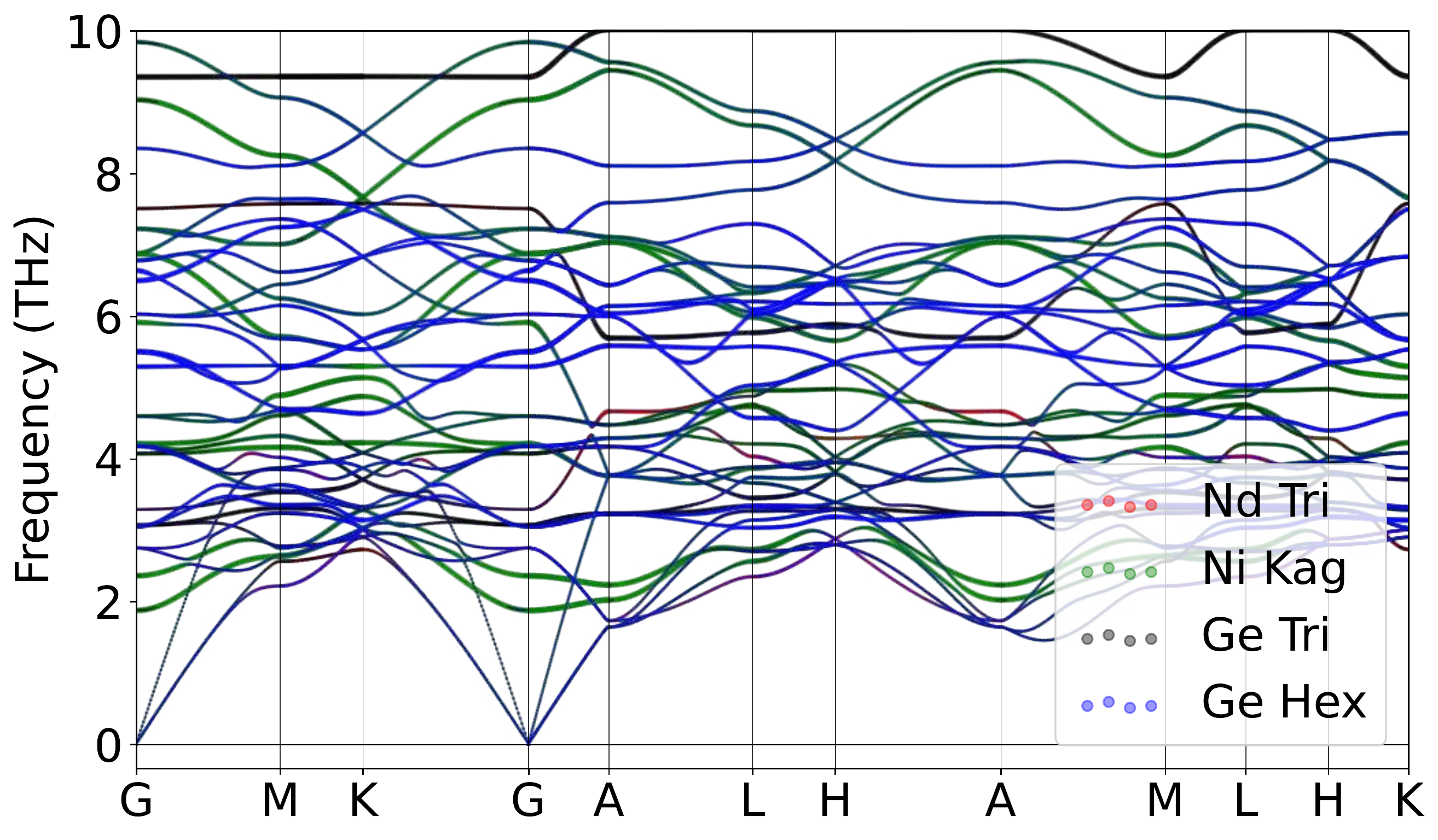}
}
\caption{Atomically projected phonon spectrum for NdNi$_6$Ge$_6$.}
\label{fig:NdNi6Ge6pho}
\end{figure}

\begin{figure}[H]
\centering
\label{fig:NdNi6Ge6_relaxed_Low_xz}
\includegraphics[width=0.85\textwidth]{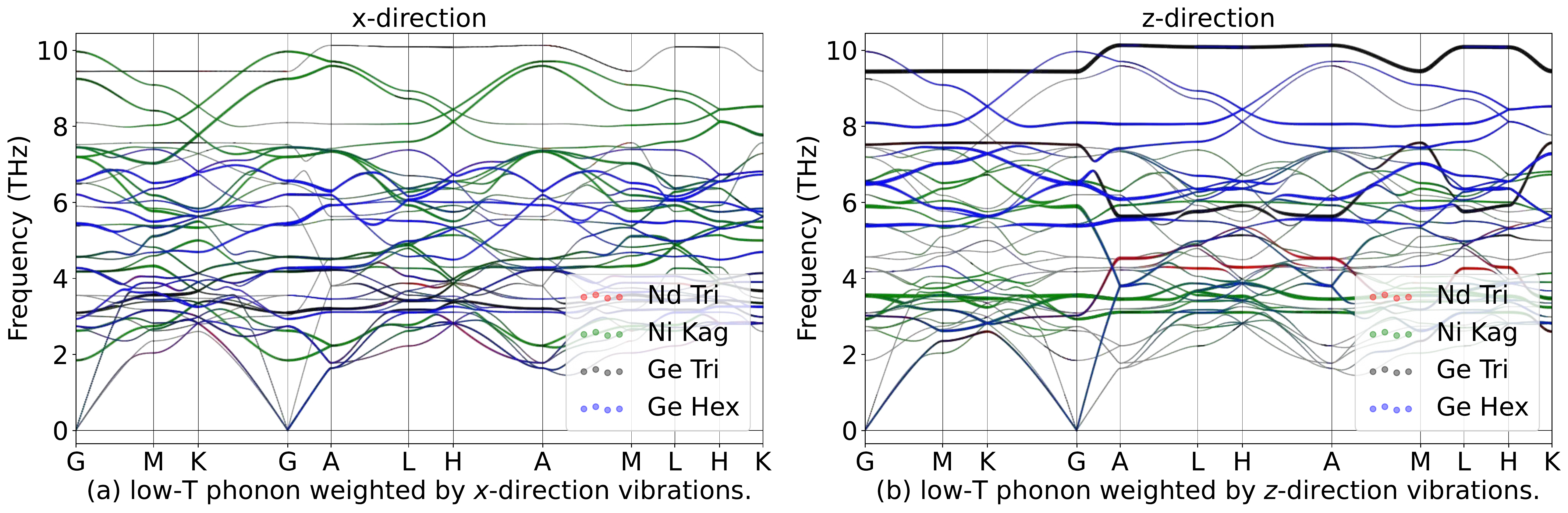}
\hspace{5pt}\label{fig:NdNi6Ge6_relaxed_High_xz}
\includegraphics[width=0.85\textwidth]{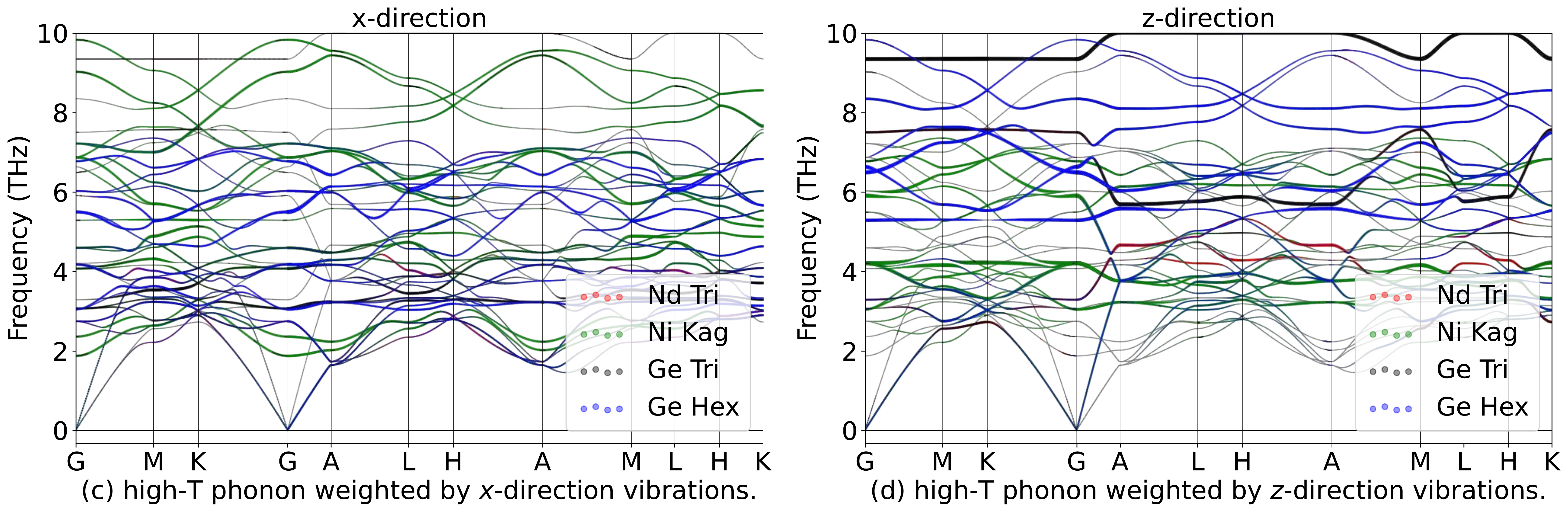}
\caption{Phonon spectrum for NdNi$_6$Ge$_6$in in-plane ($x$) and out-of-plane ($z$) directions.}
\label{fig:NdNi6Ge6phoxyz}
\end{figure}

\begin{table}[htbp]
\global\long\def\arraystretch{1.12}
\captionsetup{width=.95\textwidth}
\caption{IRREPs at high-symmetry points for NdNi$_6$Ge$_6$ phonon spectrum at Low-T.}
\label{tab:191_NdNi6Ge6_Low}
\setlength{\tabcolsep}{1.mm}{
}
\end{table}

\begin{table}[htbp]
\global\long\def\arraystretch{1.12}
\captionsetup{width=.95\textwidth}
\caption{IRREPs at high-symmetry points for NdNi$_6$Ge$_6$ phonon spectrum at High-T.}
\label{tab:191_NdNi6Ge6_High}
\setlength{\tabcolsep}{1.mm}{
%
}
\end{table}
\subsubsection{\label{sms2sec:SmNi6Ge6}\hyperref[smsec:col]{\texorpdfstring{SmNi$_6$Ge$_6$}{}}~{\color{green}  (High-T stable, Low-T stable) }}

\begin{table}[htbp]
\global\long\def\arraystretch{1.12}
\captionsetup{width=.85\textwidth}
\caption{Structure parameters of SmNi$_6$Ge$_6$ after relaxation. It contains 422 electrons. }
\label{tab:SmNi6Ge6}
\setlength{\tabcolsep}{4mm}{
%
}
\end{table}

\begin{figure}[htbp]
\centering
\includegraphics[width=0.85\textwidth]{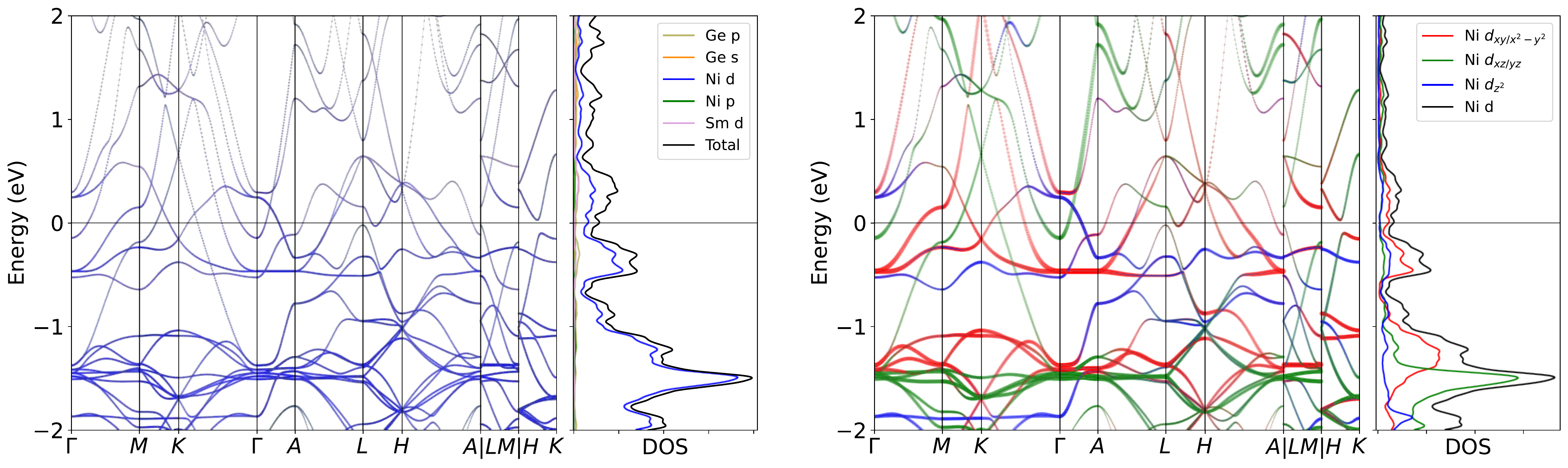}
\caption{Band structure for SmNi$_6$Ge$_6$ without SOC. The projection of Ni $3d$ orbitals is also presented. The band structures clearly reveal the dominating of Ni $3d$ orbitals  near the Fermi level, as well as the pronounced peak.}
\label{fig:SmNi6Ge6band}
\end{figure}

\begin{figure}[H]
\centering
\subfloat[Relaxed phonon at low-T.]{
\label{fig:SmNi6Ge6_relaxed_Low}
\includegraphics[width=0.45\textwidth]{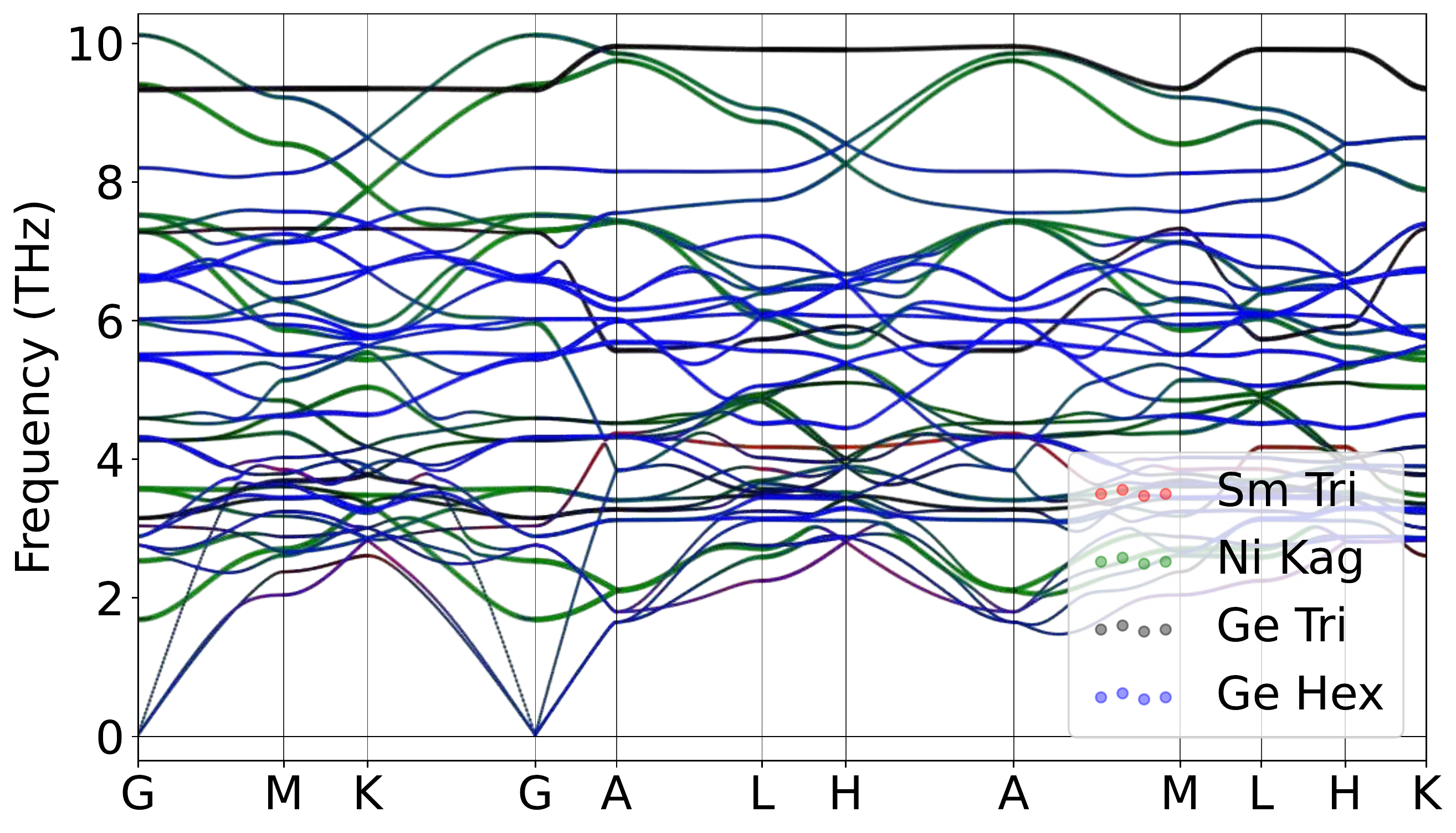}
}
\hspace{5pt}\subfloat[Relaxed phonon at high-T.]{
\label{fig:SmNi6Ge6_relaxed_High}
\includegraphics[width=0.45\textwidth]{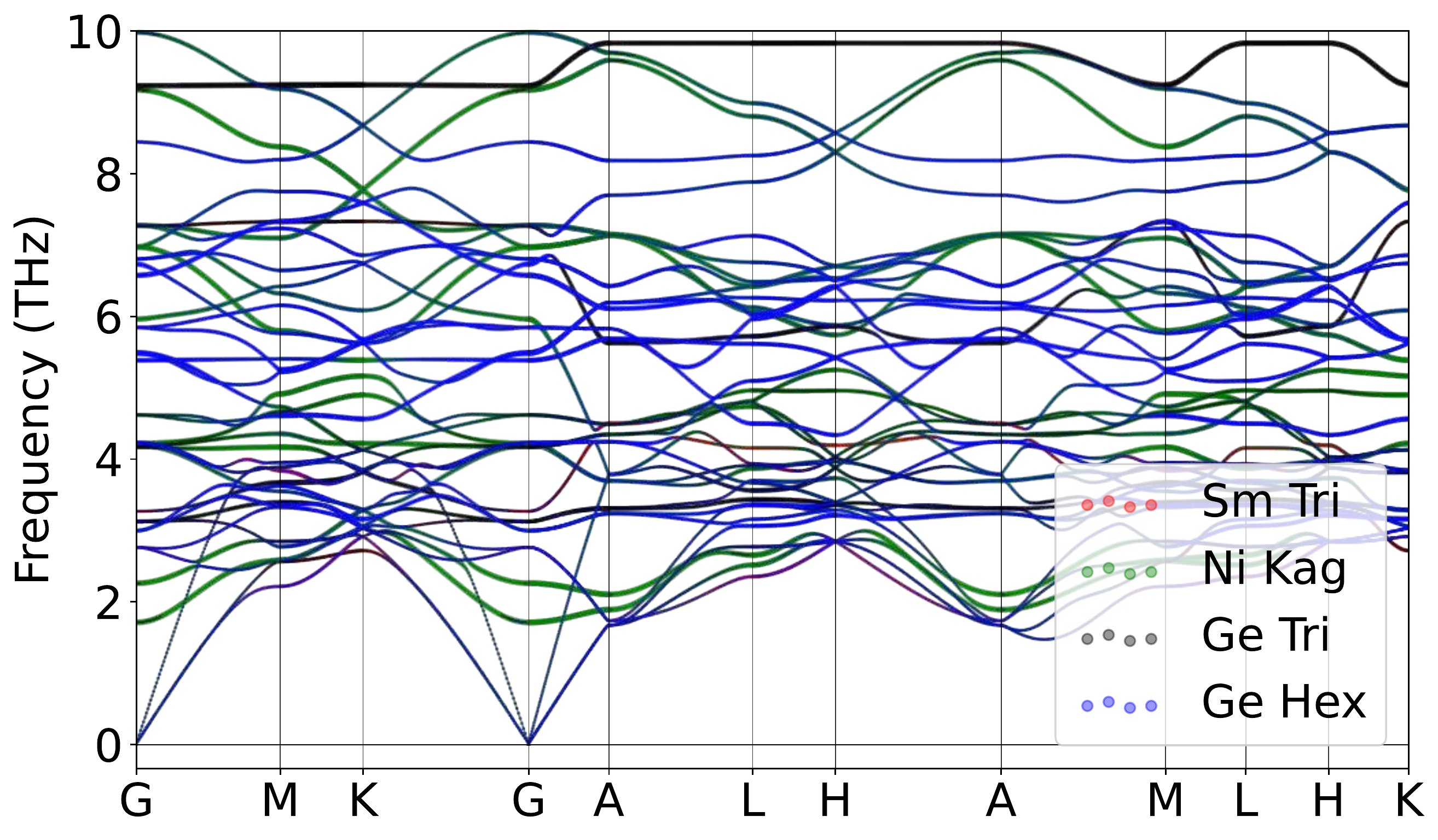}
}
\caption{Atomically projected phonon spectrum for SmNi$_6$Ge$_6$.}
\label{fig:SmNi6Ge6pho}
\end{figure}

\begin{figure}[H]
\centering
\label{fig:SmNi6Ge6_relaxed_Low_xz}
\includegraphics[width=0.85\textwidth]{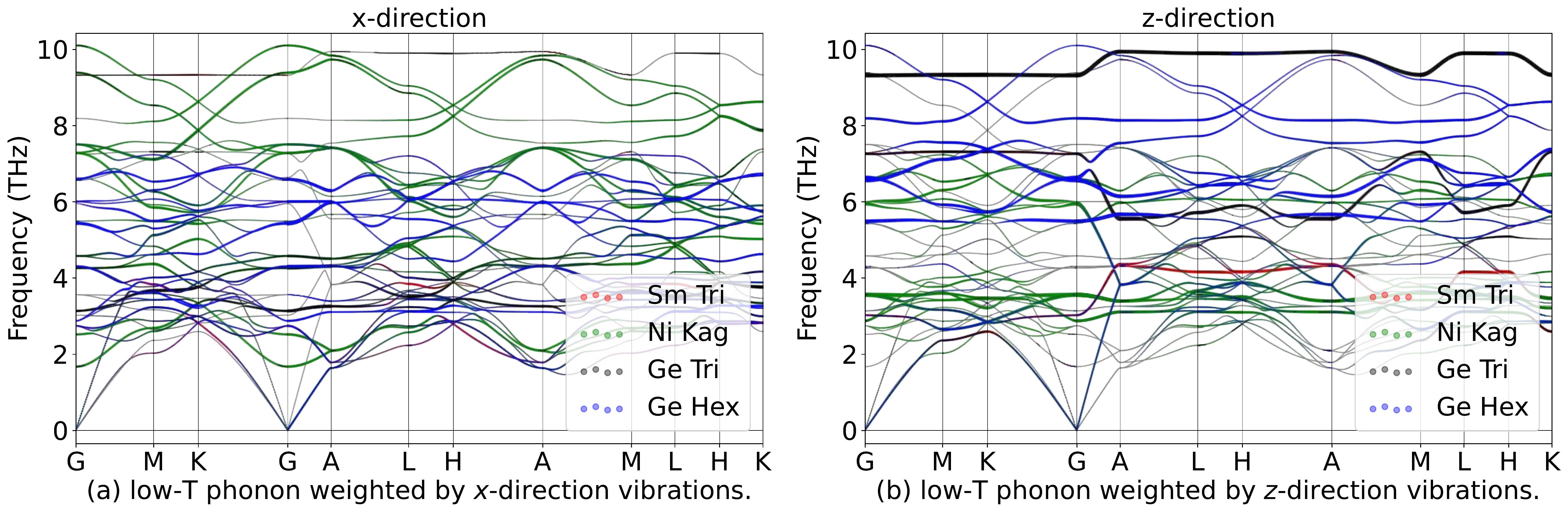}
\hspace{5pt}\label{fig:SmNi6Ge6_relaxed_High_xz}
\includegraphics[width=0.85\textwidth]{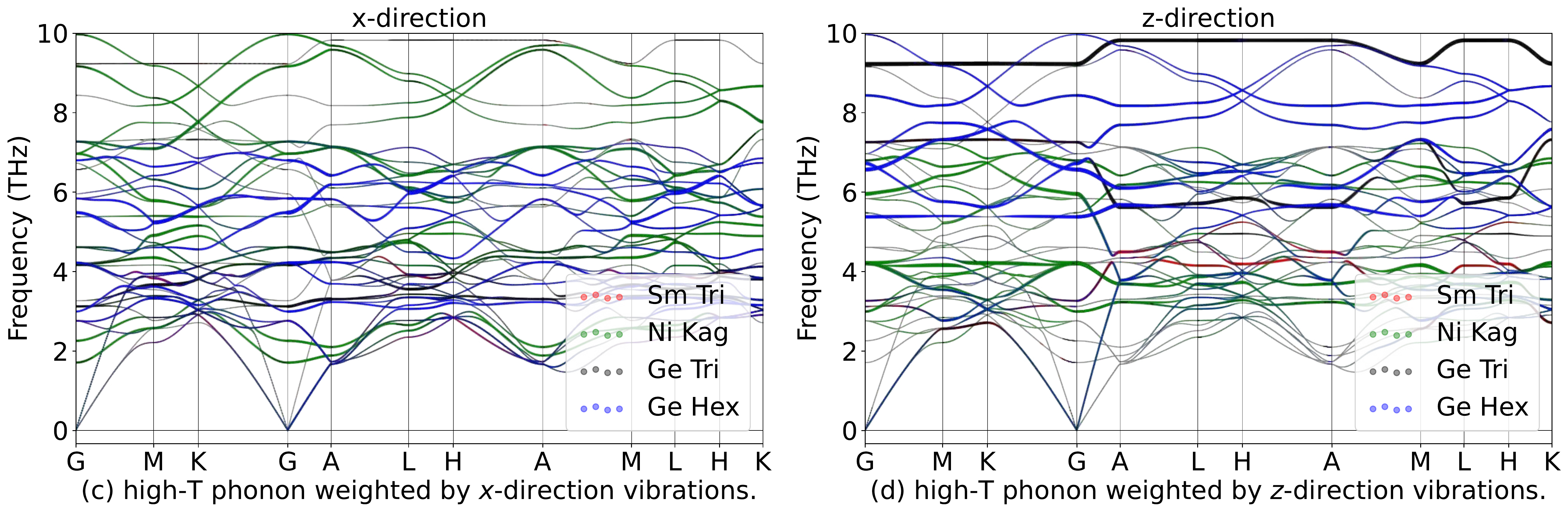}
\caption{Phonon spectrum for SmNi$_6$Ge$_6$in in-plane ($x$) and out-of-plane ($z$) directions.}
\label{fig:SmNi6Ge6phoxyz}
\end{figure}

\begin{table}[htbp]
\global\long\def\arraystretch{1.12}
\captionsetup{width=.95\textwidth}
\caption{IRREPs at high-symmetry points for SmNi$_6$Ge$_6$ phonon spectrum at Low-T.}
\label{tab:191_SmNi6Ge6_Low}
\setlength{\tabcolsep}{1.mm}{
}
\end{table}

\begin{table}[htbp]
\global\long\def\arraystretch{1.12}
\captionsetup{width=.95\textwidth}
\caption{IRREPs at high-symmetry points for SmNi$_6$Ge$_6$ phonon spectrum at High-T.}
\label{tab:191_SmNi6Ge6_High}
\setlength{\tabcolsep}{1.mm}{
%
}
\end{table}
\subsubsection{\label{sms2sec:GdNi6Ge6}\hyperref[smsec:col]{\texorpdfstring{GdNi$_6$Ge$_6$}{}}~{\color{green}  (High-T stable, Low-T stable) }}

\begin{table}[htbp]
\global\long\def\arraystretch{1.12}
\captionsetup{width=.85\textwidth}
\caption{Structure parameters of GdNi$_6$Ge$_6$ after relaxation. It contains 424 electrons. }
\label{tab:GdNi6Ge6}
\setlength{\tabcolsep}{4mm}{
%
}
\end{table}

\begin{figure}[htbp]
\centering
\includegraphics[width=0.85\textwidth]{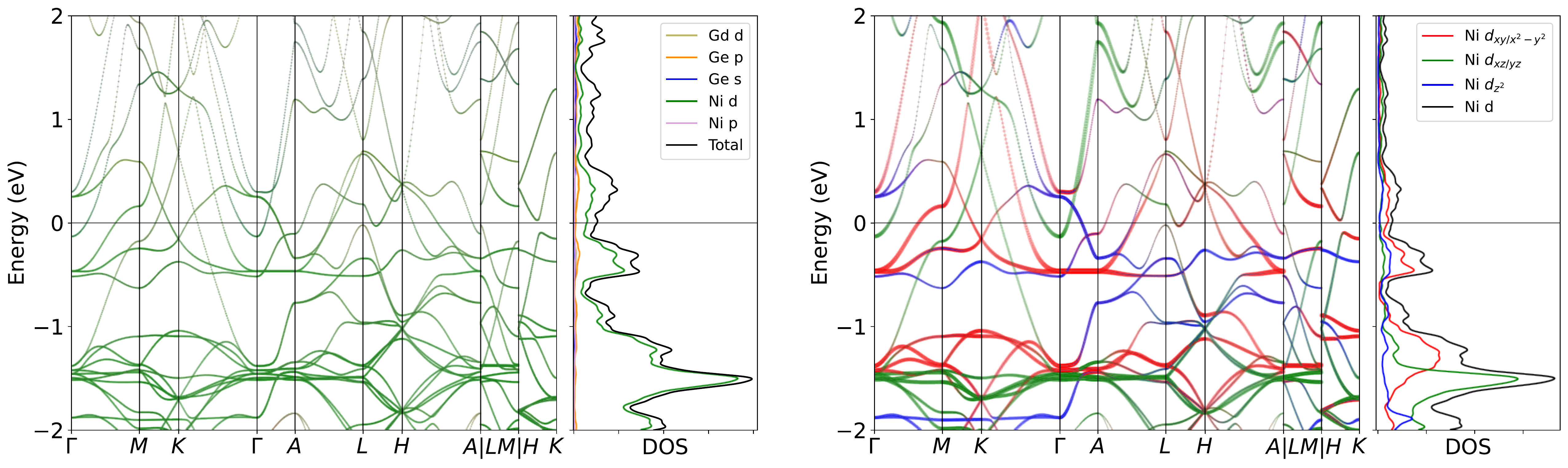}
\caption{Band structure for GdNi$_6$Ge$_6$ without SOC. The projection of Ni $3d$ orbitals is also presented. The band structures clearly reveal the dominating of Ni $3d$ orbitals  near the Fermi level, as well as the pronounced peak.}
\label{fig:GdNi6Ge6band}
\end{figure}

\begin{figure}[H]
\centering
\subfloat[Relaxed phonon at low-T.]{
\label{fig:GdNi6Ge6_relaxed_Low}
\includegraphics[width=0.45\textwidth]{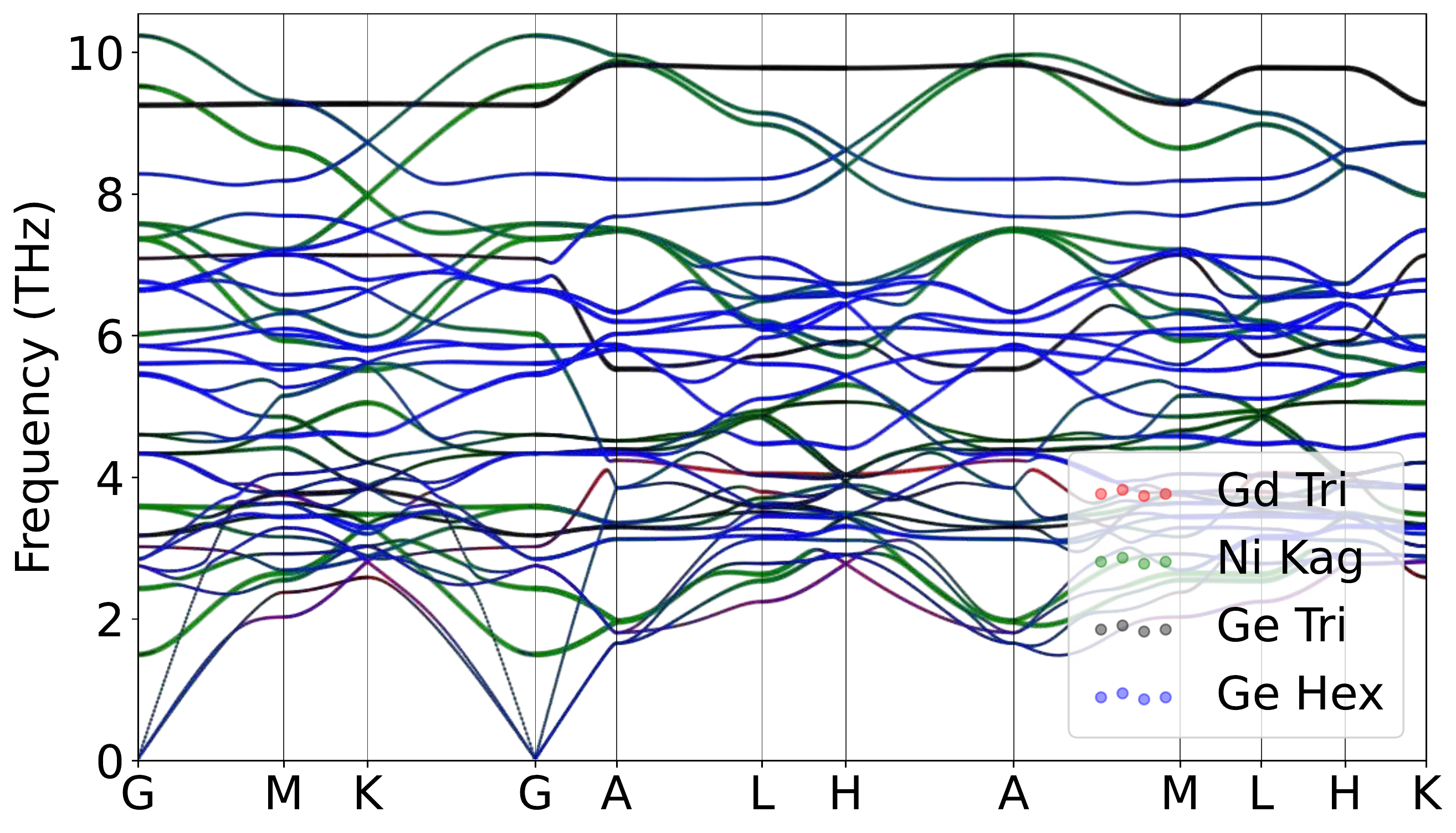}
}
\hspace{5pt}\subfloat[Relaxed phonon at high-T.]{
\label{fig:GdNi6Ge6_relaxed_High}
\includegraphics[width=0.45\textwidth]{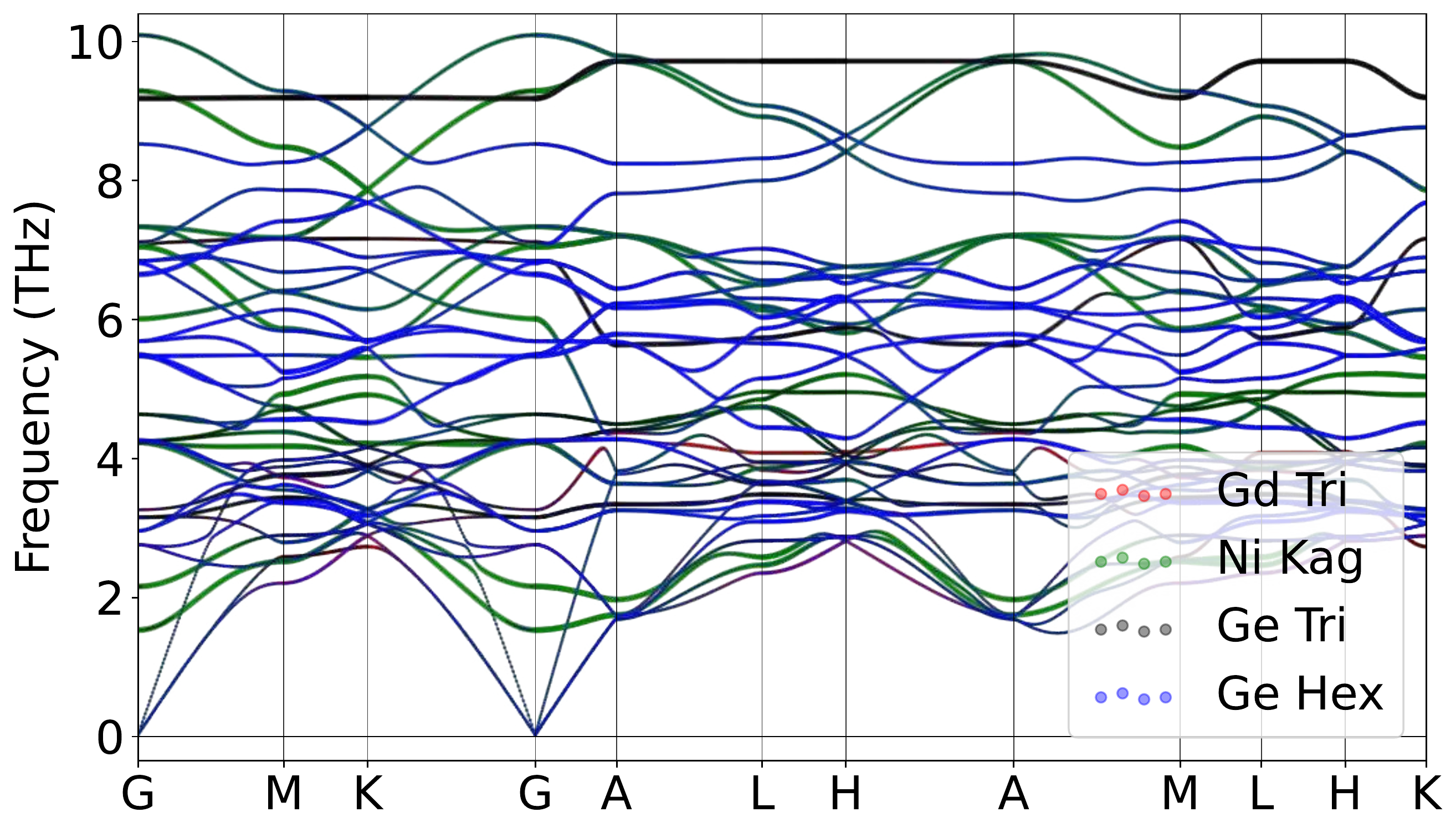}
}
\caption{Atomically projected phonon spectrum for GdNi$_6$Ge$_6$.}
\label{fig:GdNi6Ge6pho}
\end{figure}

\begin{figure}[H]
\centering
\label{fig:GdNi6Ge6_relaxed_Low_xz}
\includegraphics[width=0.85\textwidth]{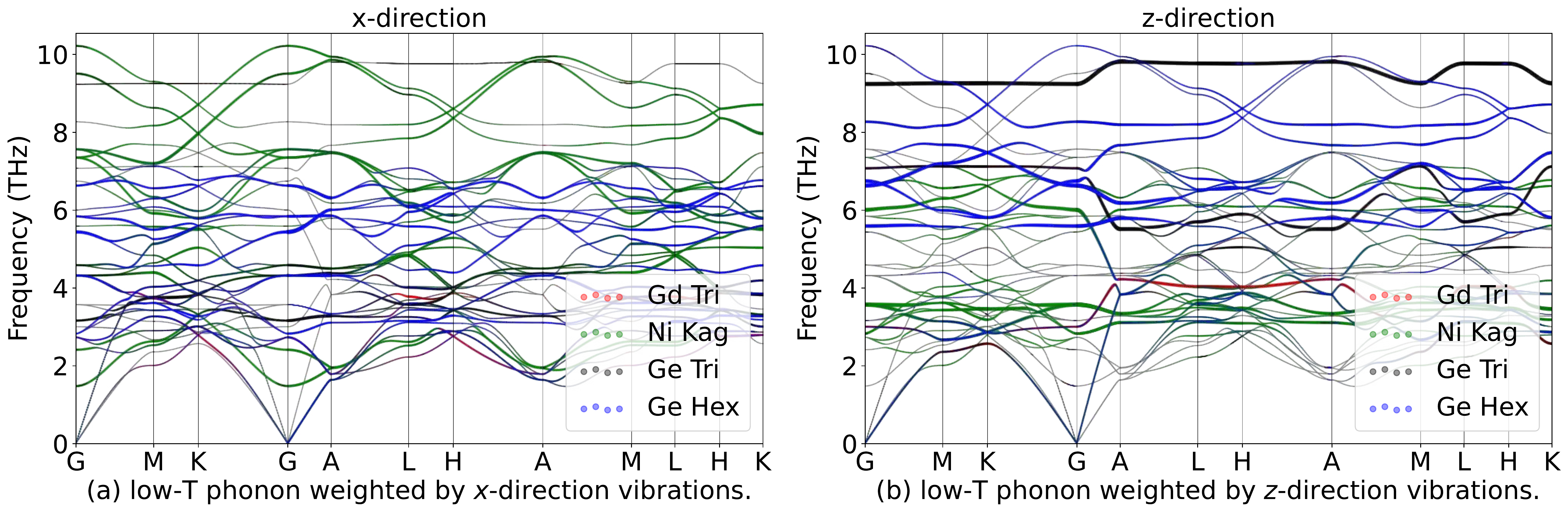}
\hspace{5pt}\label{fig:GdNi6Ge6_relaxed_High_xz}
\includegraphics[width=0.85\textwidth]{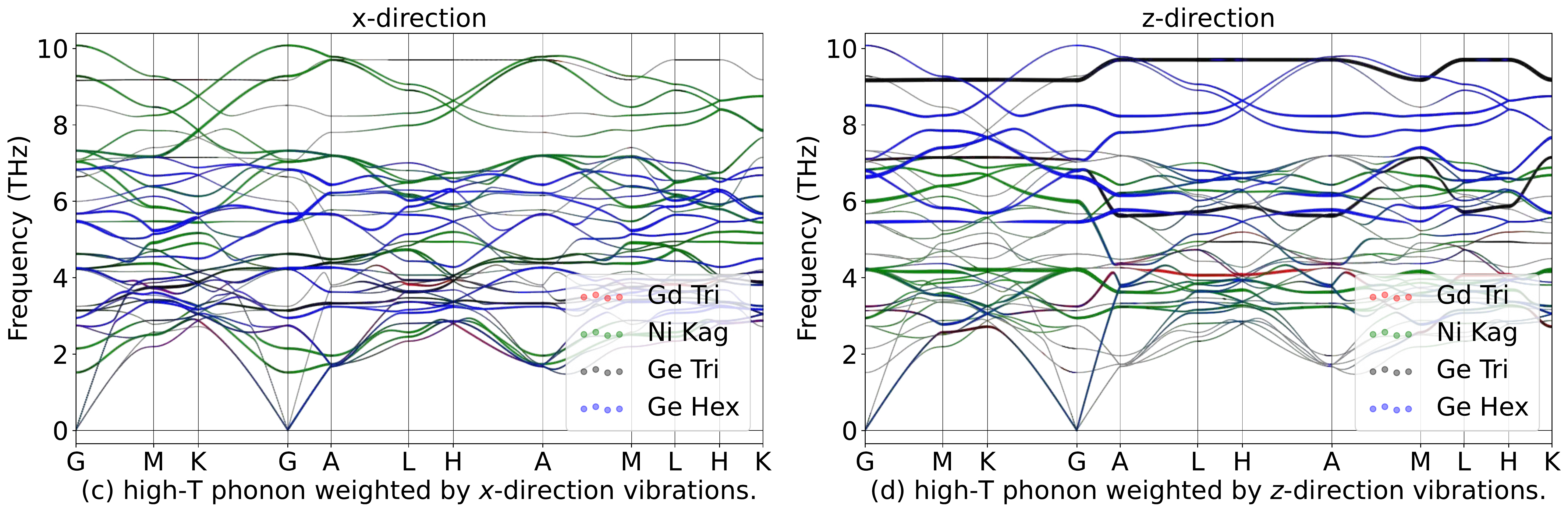}
\caption{Phonon spectrum for GdNi$_6$Ge$_6$in in-plane ($x$) and out-of-plane ($z$) directions.}
\label{fig:GdNi6Ge6phoxyz}
\end{figure}

\begin{table}[htbp]
\global\long\def\arraystretch{1.12}
\captionsetup{width=.95\textwidth}
\caption{IRREPs at high-symmetry points for GdNi$_6$Ge$_6$ phonon spectrum at Low-T.}
\label{tab:191_GdNi6Ge6_Low}
\setlength{\tabcolsep}{1.mm}{
}
\end{table}

\begin{table}[htbp]
\global\long\def\arraystretch{1.12}
\captionsetup{width=.95\textwidth}
\caption{IRREPs at high-symmetry points for GdNi$_6$Ge$_6$ phonon spectrum at High-T.}
\label{tab:191_GdNi6Ge6_High}
\setlength{\tabcolsep}{1.mm}{
%
}
\end{table}
\subsubsection{\label{sms2sec:TbNi6Ge6}\hyperref[smsec:col]{\texorpdfstring{TbNi$_6$Ge$_6$}{}}~{\color{green}  (High-T stable, Low-T stable) }}

\begin{table}[htbp]
\global\long\def\arraystretch{1.12}
\captionsetup{width=.85\textwidth}
\caption{Structure parameters of TbNi$_6$Ge$_6$ after relaxation. It contains 425 electrons. }
\label{tab:TbNi6Ge6}
\setlength{\tabcolsep}{4mm}{
%
}
\end{table}

\begin{figure}[htbp]
\centering
\includegraphics[width=0.85\textwidth]{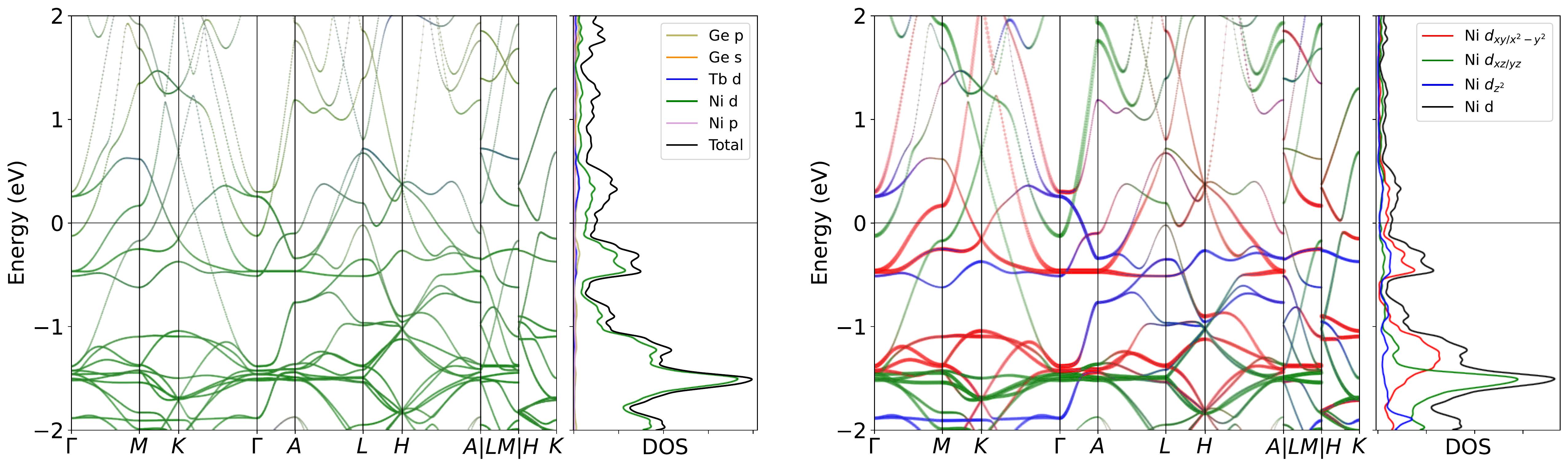}
\caption{Band structure for TbNi$_6$Ge$_6$ without SOC. The projection of Ni $3d$ orbitals is also presented. The band structures clearly reveal the dominating of Ni $3d$ orbitals  near the Fermi level, as well as the pronounced peak.}
\label{fig:TbNi6Ge6band}
\end{figure}

\begin{figure}[H]
\centering
\subfloat[Relaxed phonon at low-T.]{
\label{fig:TbNi6Ge6_relaxed_Low}
\includegraphics[width=0.45\textwidth]{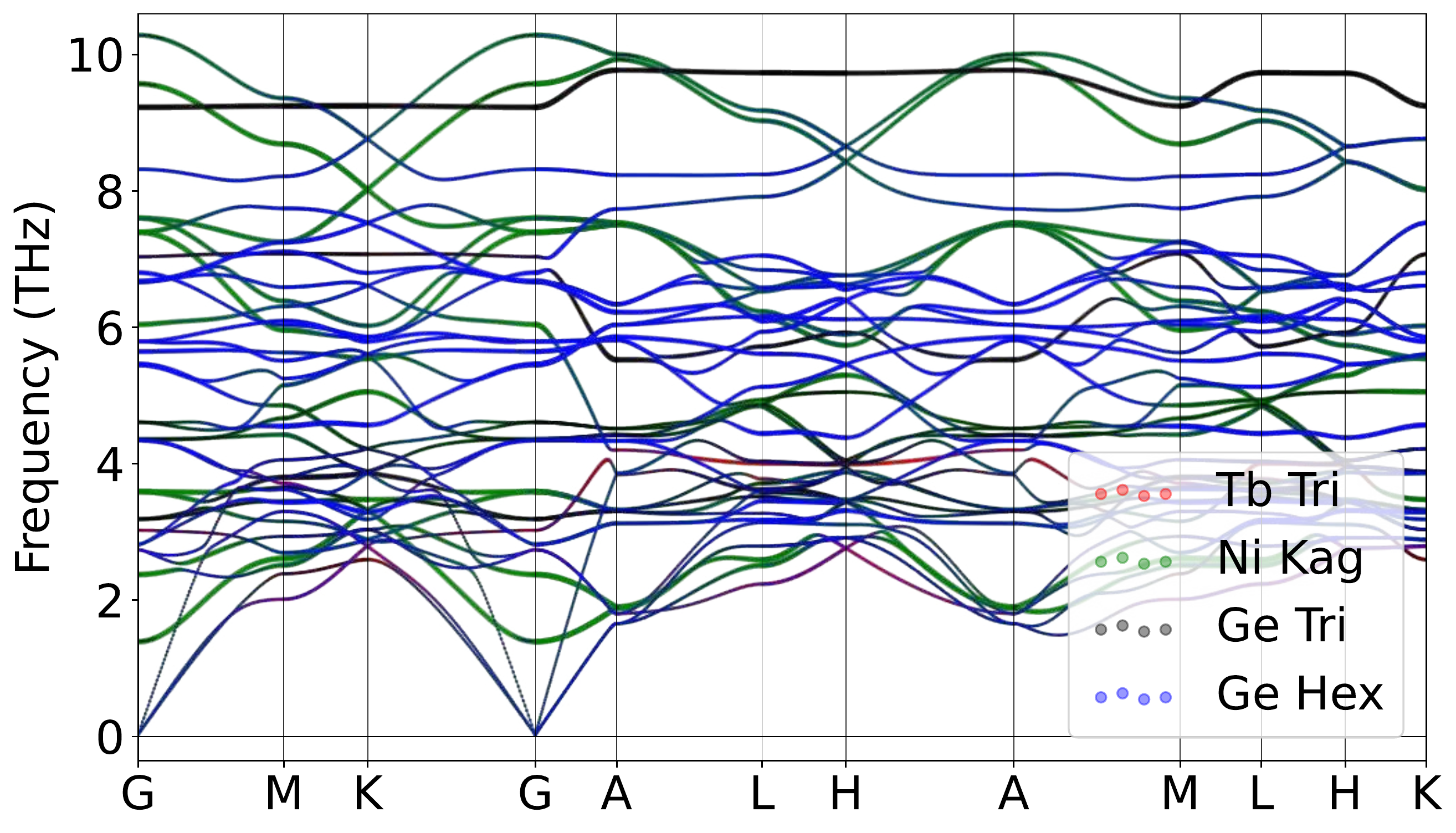}
}
\hspace{5pt}\subfloat[Relaxed phonon at high-T.]{
\label{fig:TbNi6Ge6_relaxed_High}
\includegraphics[width=0.45\textwidth]{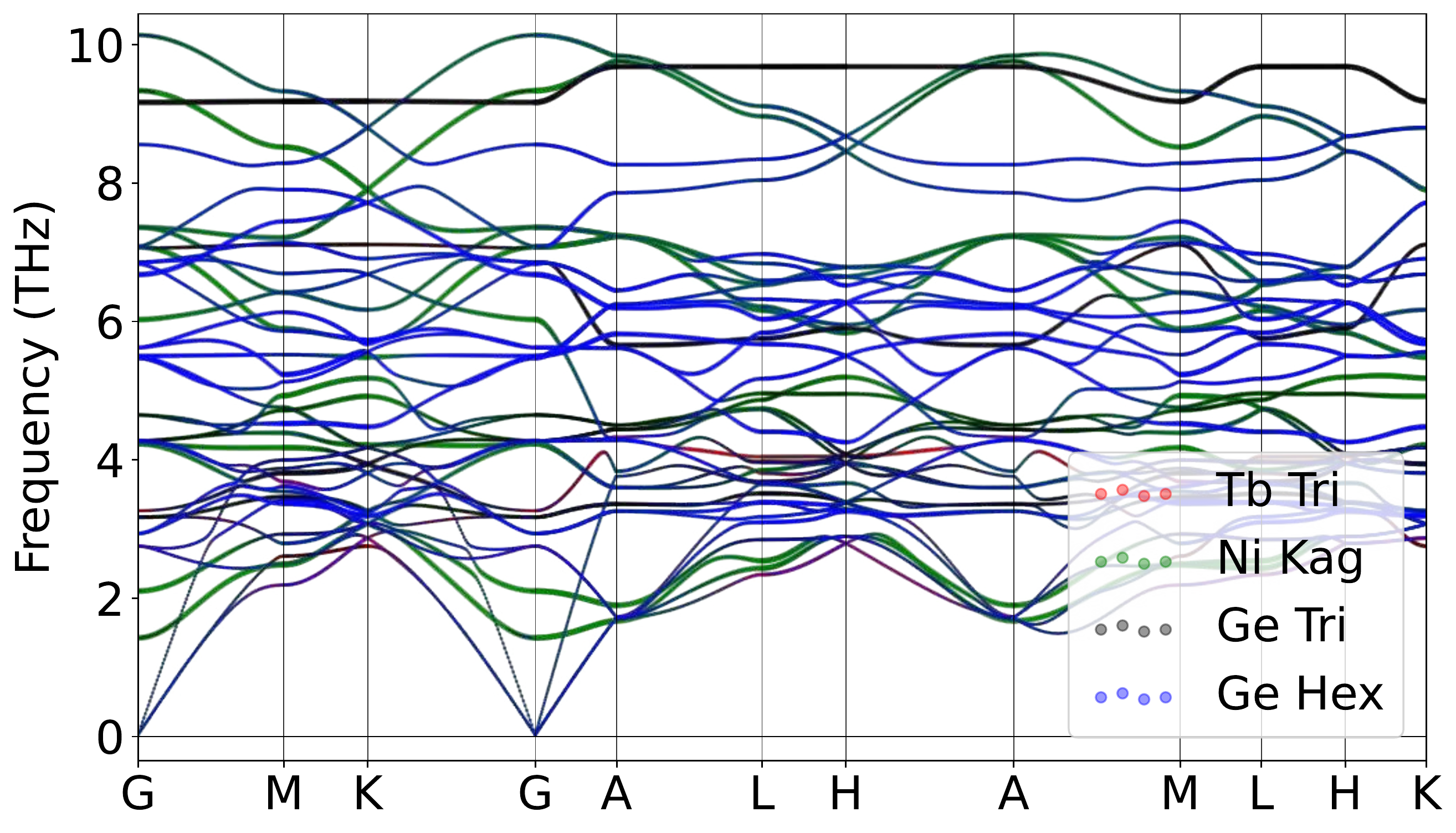}
}
\caption{Atomically projected phonon spectrum for TbNi$_6$Ge$_6$.}
\label{fig:TbNi6Ge6pho}
\end{figure}

\begin{figure}[H]
\centering
\label{fig:TbNi6Ge6_relaxed_Low_xz}
\includegraphics[width=0.85\textwidth]{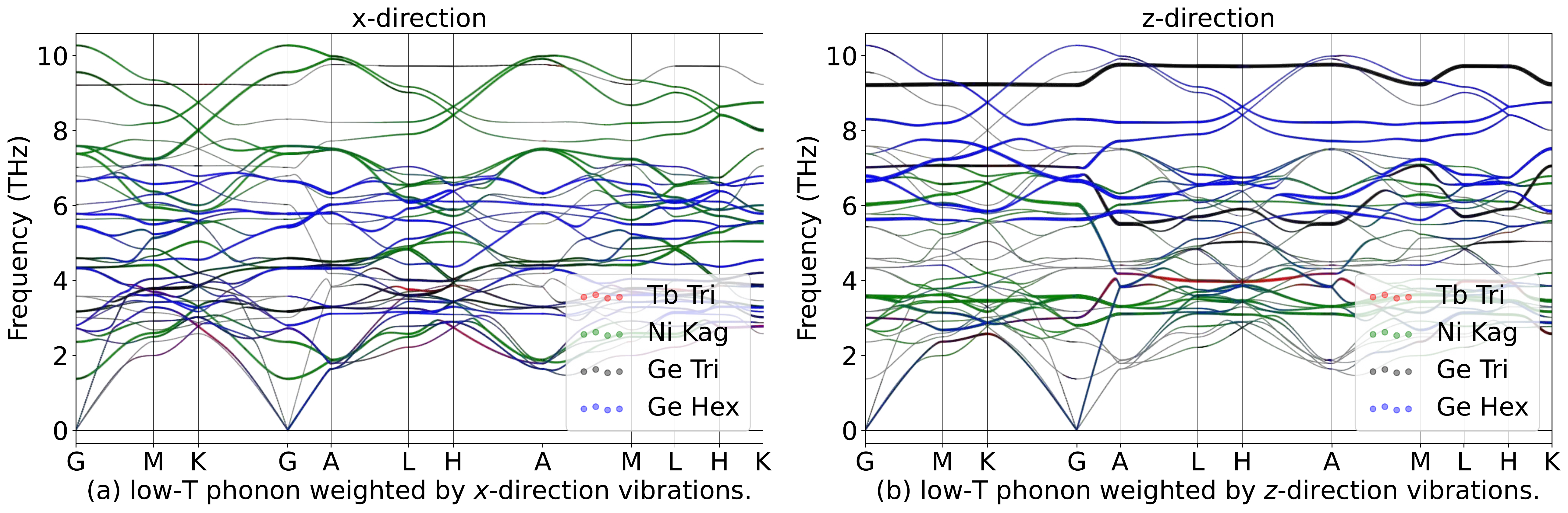}
\hspace{5pt}\label{fig:TbNi6Ge6_relaxed_High_xz}
\includegraphics[width=0.85\textwidth]{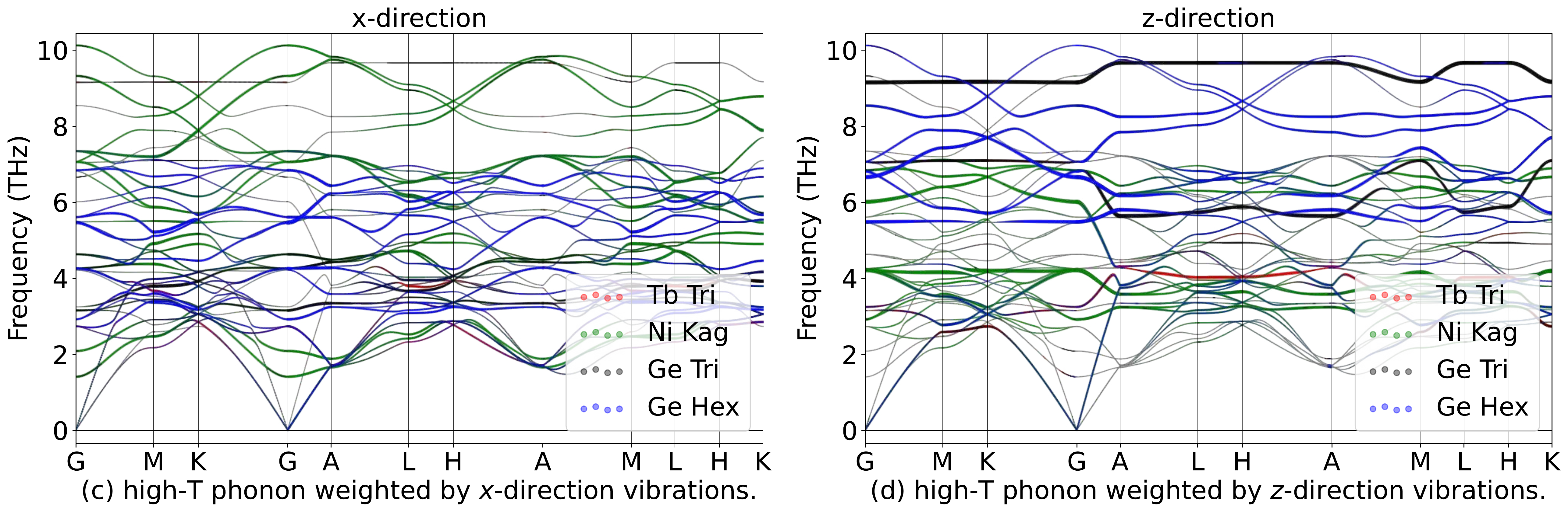}
\caption{Phonon spectrum for TbNi$_6$Ge$_6$in in-plane ($x$) and out-of-plane ($z$) directions.}
\label{fig:TbNi6Ge6phoxyz}
\end{figure}

\begin{table}[htbp]
\global\long\def\arraystretch{1.12}
\captionsetup{width=.95\textwidth}
\caption{IRREPs at high-symmetry points for TbNi$_6$Ge$_6$ phonon spectrum at Low-T.}
\label{tab:191_TbNi6Ge6_Low}
\setlength{\tabcolsep}{1.mm}{
}
\end{table}

\begin{table}[htbp]
\global\long\def\arraystretch{1.12}
\captionsetup{width=.95\textwidth}
\caption{IRREPs at high-symmetry points for TbNi$_6$Ge$_6$ phonon spectrum at High-T.}
\label{tab:191_TbNi6Ge6_High}
\setlength{\tabcolsep}{1.mm}{
%
}
\end{table}
\subsubsection{\label{sms2sec:DyNi6Ge6}\hyperref[smsec:col]{\texorpdfstring{DyNi$_6$Ge$_6$}{}}~{\color{green}  (High-T stable, Low-T stable) }}

\begin{table}[htbp]
\global\long\def\arraystretch{1.12}
\captionsetup{width=.85\textwidth}
\caption{Structure parameters of DyNi$_6$Ge$_6$ after relaxation. It contains 426 electrons. }
\label{tab:DyNi6Ge6}
\setlength{\tabcolsep}{4mm}{
%
}
\end{table}

\begin{figure}[htbp]
\centering
\includegraphics[width=0.85\textwidth]{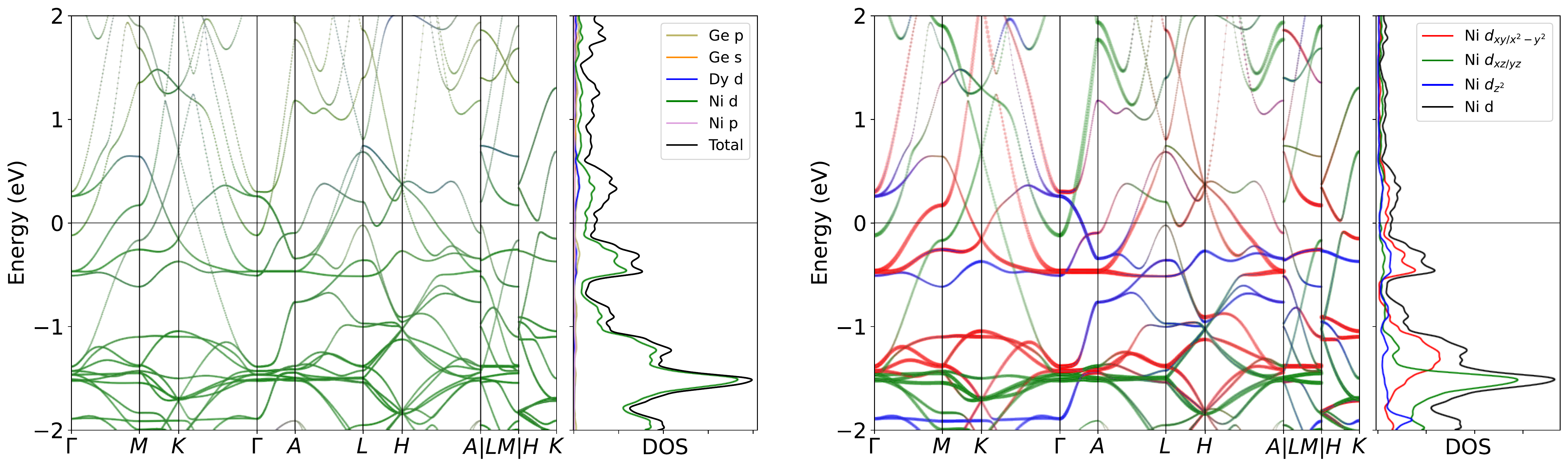}
\caption{Band structure for DyNi$_6$Ge$_6$ without SOC. The projection of Ni $3d$ orbitals is also presented. The band structures clearly reveal the dominating of Ni $3d$ orbitals  near the Fermi level, as well as the pronounced peak.}
\label{fig:DyNi6Ge6band}
\end{figure}

\begin{figure}[H]
\centering
\subfloat[Relaxed phonon at low-T.]{
\label{fig:DyNi6Ge6_relaxed_Low}
\includegraphics[width=0.45\textwidth]{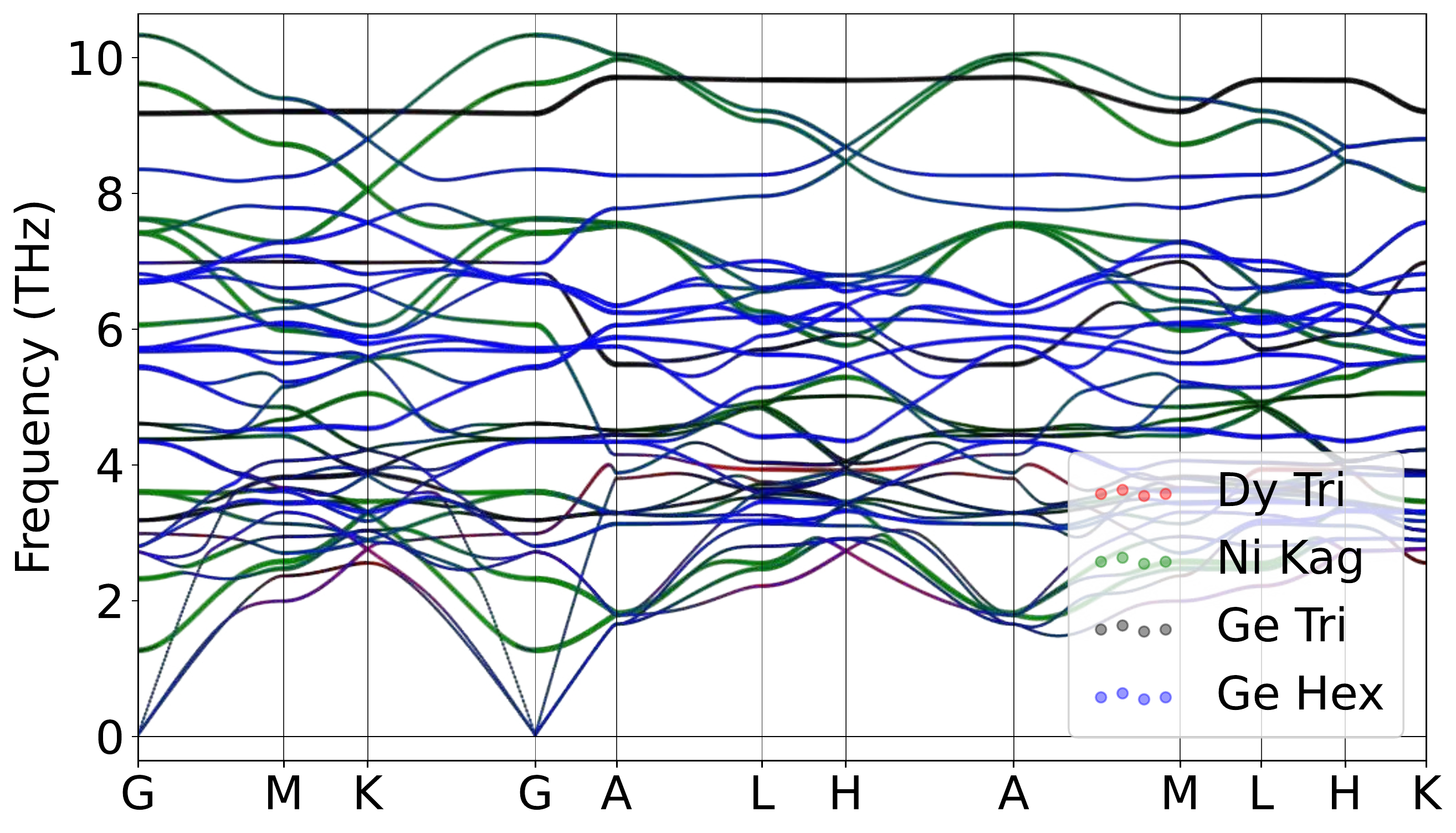}
}
\hspace{5pt}\subfloat[Relaxed phonon at high-T.]{
\label{fig:DyNi6Ge6_relaxed_High}
\includegraphics[width=0.45\textwidth]{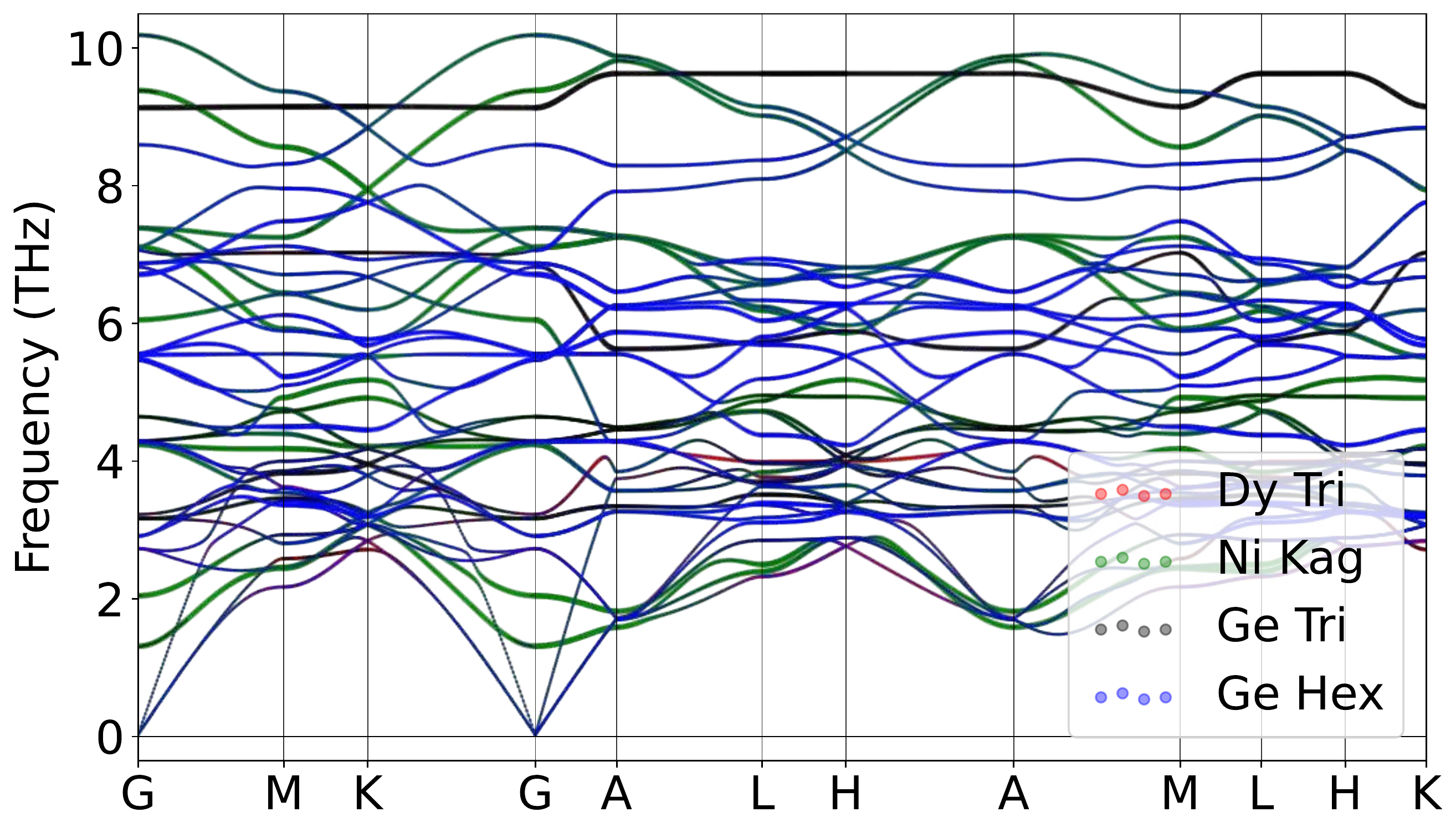}
}
\caption{Atomically projected phonon spectrum for DyNi$_6$Ge$_6$.}
\label{fig:DyNi6Ge6pho}
\end{figure}

\begin{figure}[H]
\centering
\label{fig:DyNi6Ge6_relaxed_Low_xz}
\includegraphics[width=0.85\textwidth]{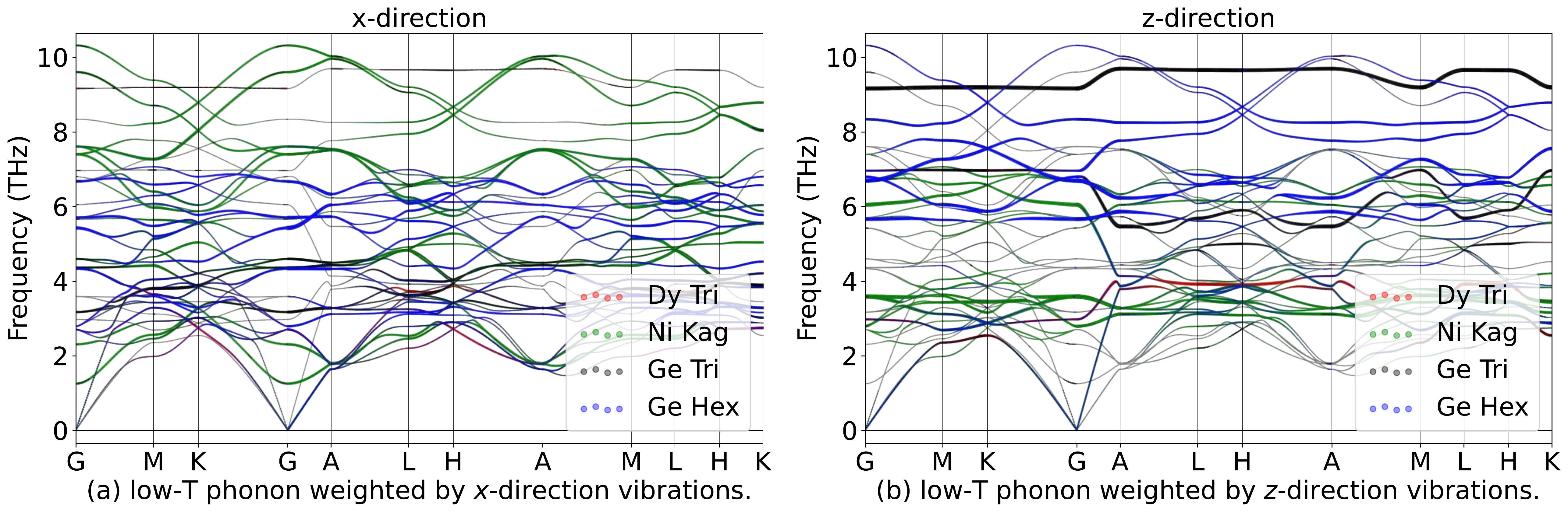}
\hspace{5pt}\label{fig:DyNi6Ge6_relaxed_High_xz}
\includegraphics[width=0.85\textwidth]{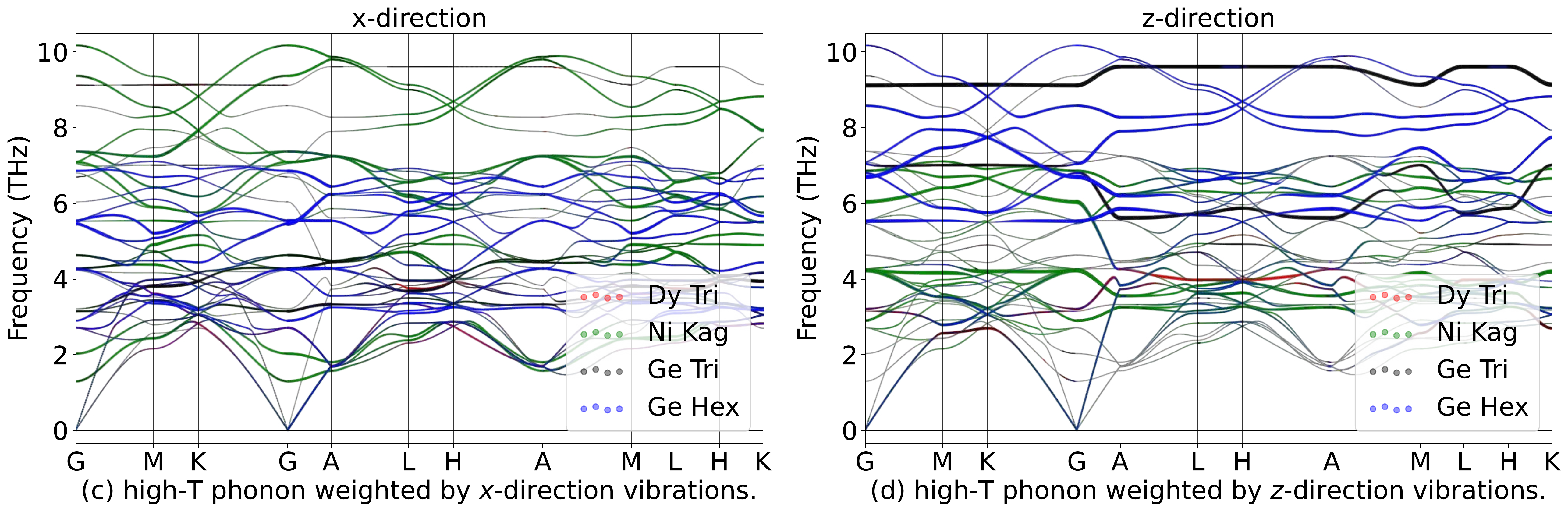}
\caption{Phonon spectrum for DyNi$_6$Ge$_6$in in-plane ($x$) and out-of-plane ($z$) directions.}
\label{fig:DyNi6Ge6phoxyz}
\end{figure}

\begin{table}[htbp]
\global\long\def\arraystretch{1.12}
\captionsetup{width=.95\textwidth}
\caption{IRREPs at high-symmetry points for DyNi$_6$Ge$_6$ phonon spectrum at Low-T.}
\label{tab:191_DyNi6Ge6_Low}
\setlength{\tabcolsep}{1.mm}{
}
\end{table}

\begin{table}[htbp]
\global\long\def\arraystretch{1.12}
\captionsetup{width=.95\textwidth}
\caption{IRREPs at high-symmetry points for DyNi$_6$Ge$_6$ phonon spectrum at High-T.}
\label{tab:191_DyNi6Ge6_High}
\setlength{\tabcolsep}{1.mm}{
%
}
\end{table}
\subsubsection{\label{sms2sec:HoNi6Ge6}\hyperref[smsec:col]{\texorpdfstring{HoNi$_6$Ge$_6$}{}}~{\color{green}  (High-T stable, Low-T stable) }}

\begin{table}[htbp]
\global\long\def\arraystretch{1.12}
\captionsetup{width=.85\textwidth}
\caption{Structure parameters of HoNi$_6$Ge$_6$ after relaxation. It contains 427 electrons. }
\label{tab:HoNi6Ge6}
\setlength{\tabcolsep}{4mm}{
%
}
\end{table}

\begin{figure}[htbp]
\centering
\includegraphics[width=0.85\textwidth]{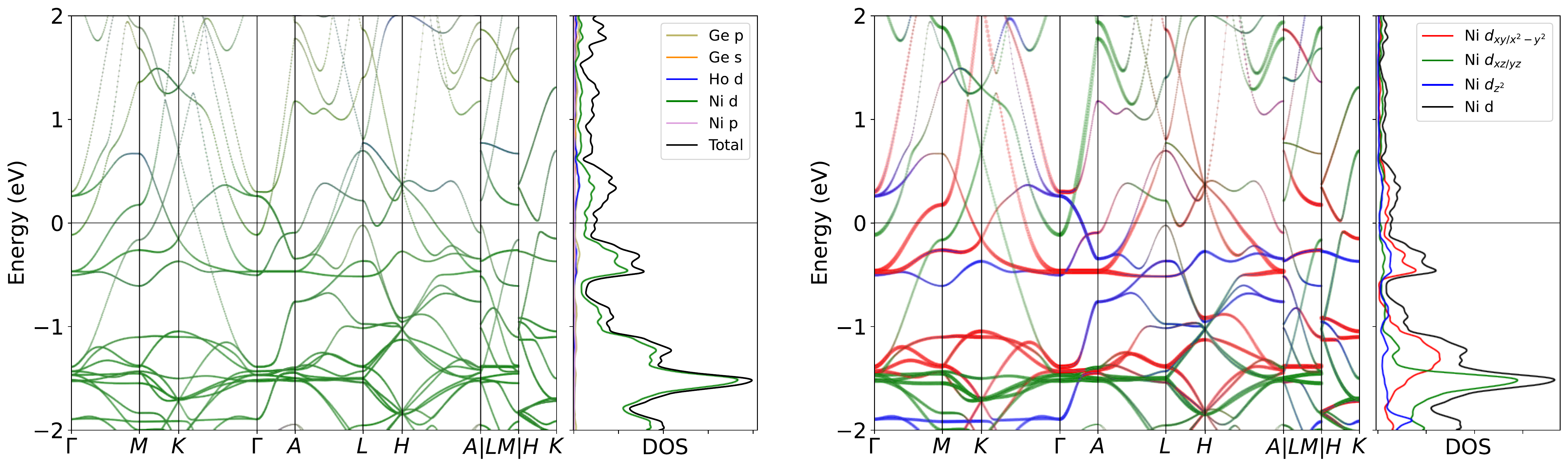}
\caption{Band structure for HoNi$_6$Ge$_6$ without SOC. The projection of Ni $3d$ orbitals is also presented. The band structures clearly reveal the dominating of Ni $3d$ orbitals  near the Fermi level, as well as the pronounced peak.}
\label{fig:HoNi6Ge6band}
\end{figure}

\begin{figure}[H]
\centering
\subfloat[Relaxed phonon at low-T.]{
\label{fig:HoNi6Ge6_relaxed_Low}
\includegraphics[width=0.45\textwidth]{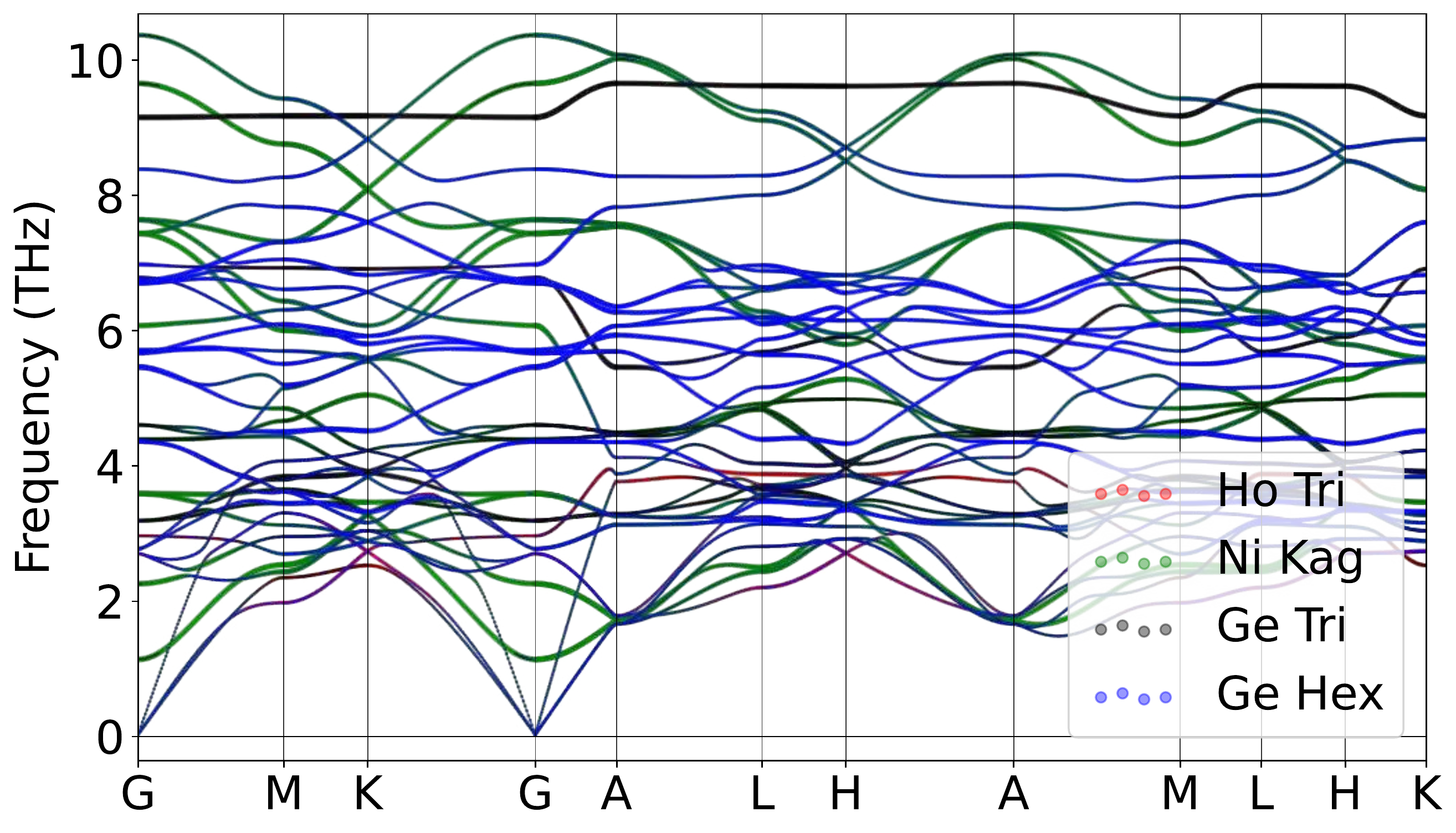}
}
\hspace{5pt}\subfloat[Relaxed phonon at high-T.]{
\label{fig:HoNi6Ge6_relaxed_High}
\includegraphics[width=0.45\textwidth]{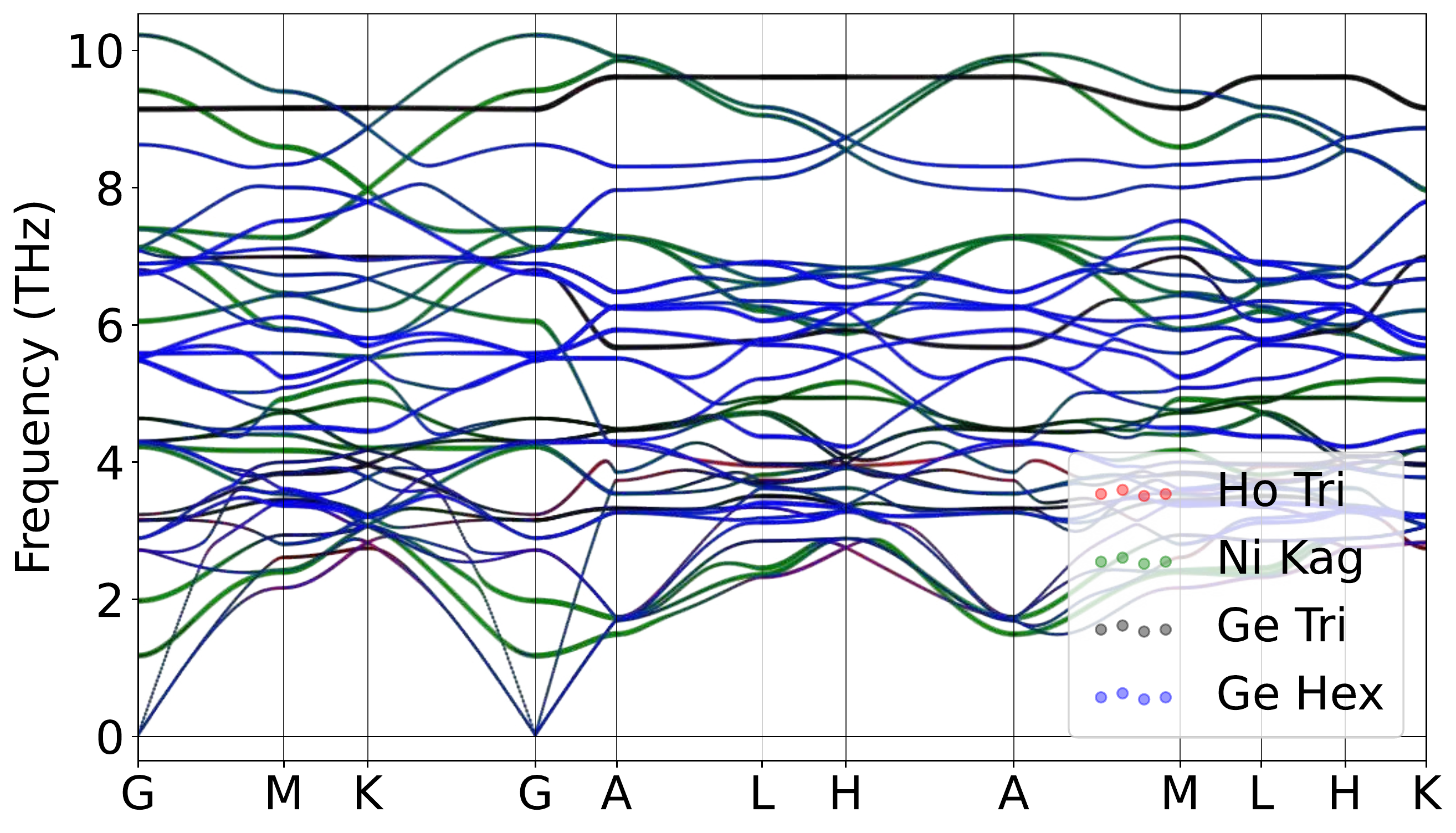}
}
\caption{Atomically projected phonon spectrum for HoNi$_6$Ge$_6$.}
\label{fig:HoNi6Ge6pho}
\end{figure}

\begin{figure}[H]
\centering
\label{fig:HoNi6Ge6_relaxed_Low_xz}
\includegraphics[width=0.85\textwidth]{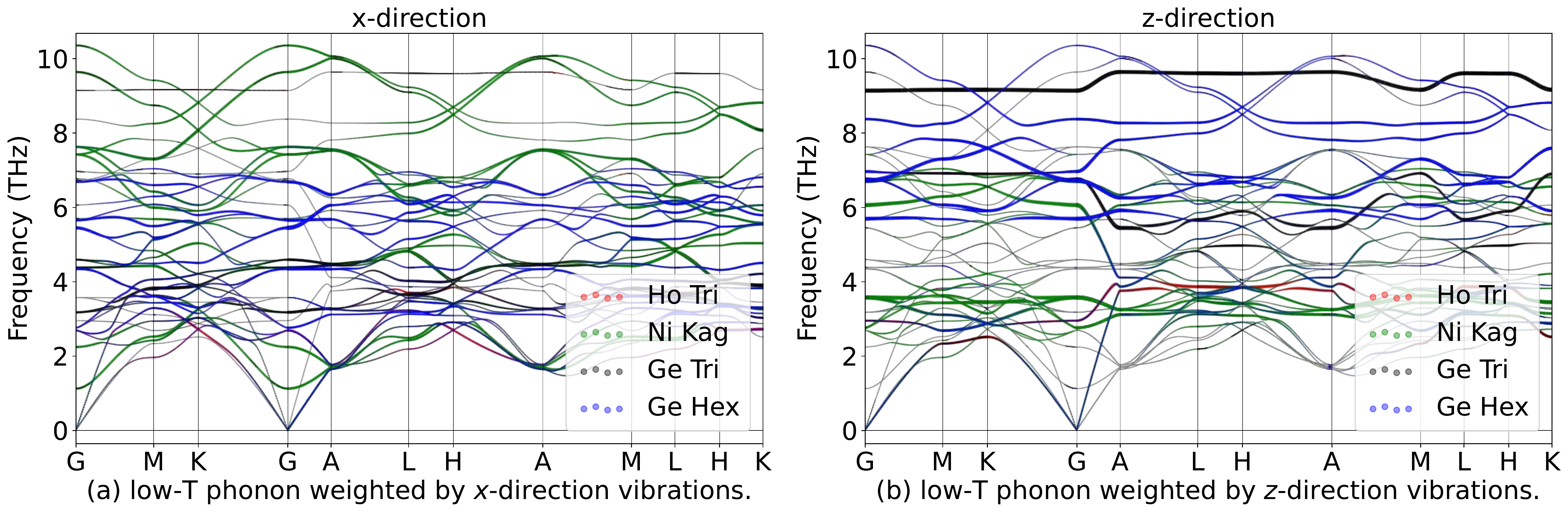}
\hspace{5pt}\label{fig:HoNi6Ge6_relaxed_High_xz}
\includegraphics[width=0.85\textwidth]{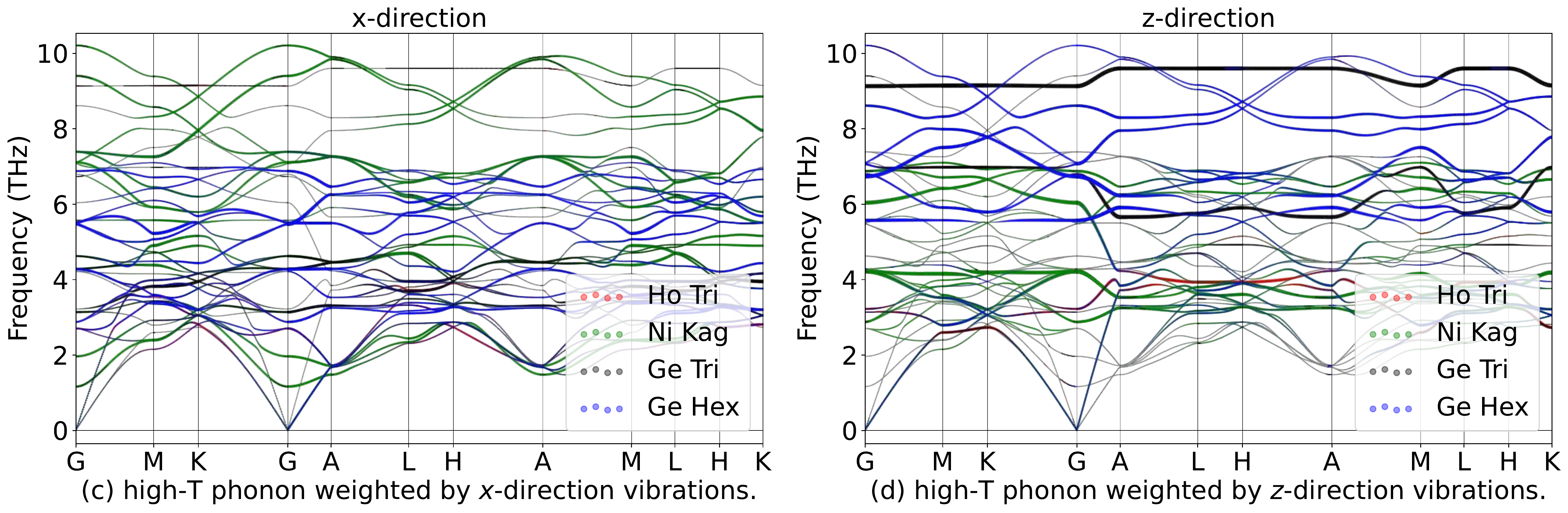}
\caption{Phonon spectrum for HoNi$_6$Ge$_6$in in-plane ($x$) and out-of-plane ($z$) directions.}
\label{fig:HoNi6Ge6phoxyz}
\end{figure}

\begin{table}[htbp]
\global\long\def\arraystretch{1.12}
\captionsetup{width=.95\textwidth}
\caption{IRREPs at high-symmetry points for HoNi$_6$Ge$_6$ phonon spectrum at Low-T.}
\label{tab:191_HoNi6Ge6_Low}
\setlength{\tabcolsep}{1.mm}{
}
\end{table}

\begin{table}[htbp]
\global\long\def\arraystretch{1.12}
\captionsetup{width=.95\textwidth}
\caption{IRREPs at high-symmetry points for HoNi$_6$Ge$_6$ phonon spectrum at High-T.}
\label{tab:191_HoNi6Ge6_High}
\setlength{\tabcolsep}{1.mm}{
%
}
\end{table}
\subsubsection{\label{sms2sec:ErNi6Ge6}\hyperref[smsec:col]{\texorpdfstring{ErNi$_6$Ge$_6$}{}}~{\color{green}  (High-T stable, Low-T stable) }}

\begin{table}[htbp]
\global\long\def\arraystretch{1.12}
\captionsetup{width=.85\textwidth}
\caption{Structure parameters of ErNi$_6$Ge$_6$ after relaxation. It contains 428 electrons. }
\label{tab:ErNi6Ge6}
\setlength{\tabcolsep}{4mm}{
%
}
\end{table}

\begin{figure}[htbp]
\centering
\includegraphics[width=0.85\textwidth]{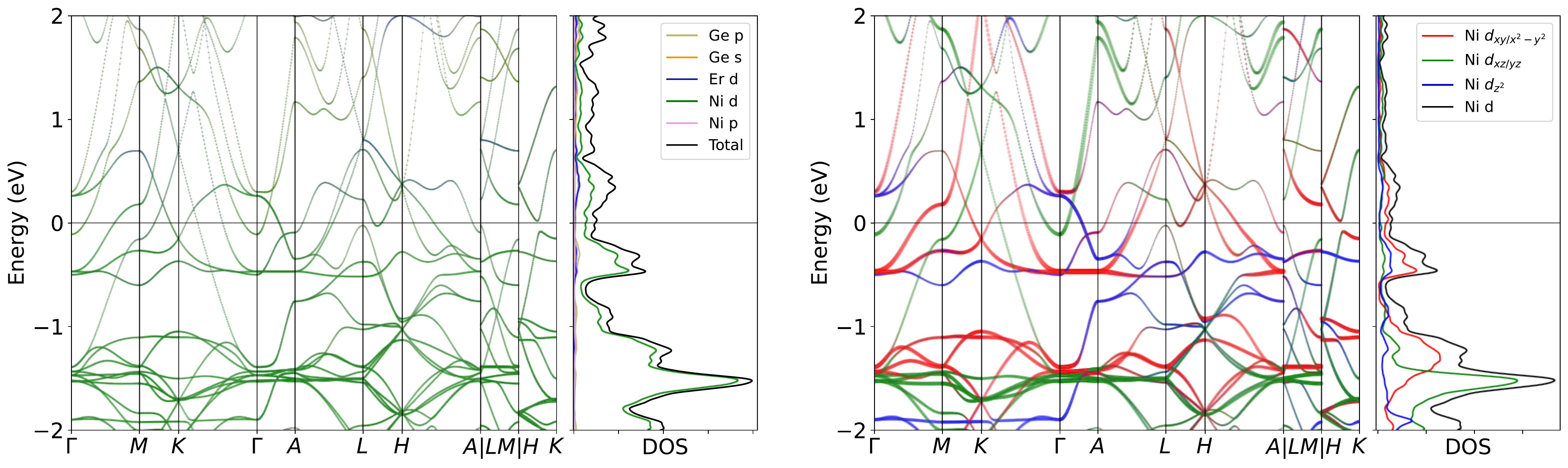}
\caption{Band structure for ErNi$_6$Ge$_6$ without SOC. The projection of Ni $3d$ orbitals is also presented. The band structures clearly reveal the dominating of Ni $3d$ orbitals  near the Fermi level, as well as the pronounced peak.}
\label{fig:ErNi6Ge6band}
\end{figure}

\begin{figure}[H]
\centering
\subfloat[Relaxed phonon at low-T.]{
\label{fig:ErNi6Ge6_relaxed_Low}
\includegraphics[width=0.45\textwidth]{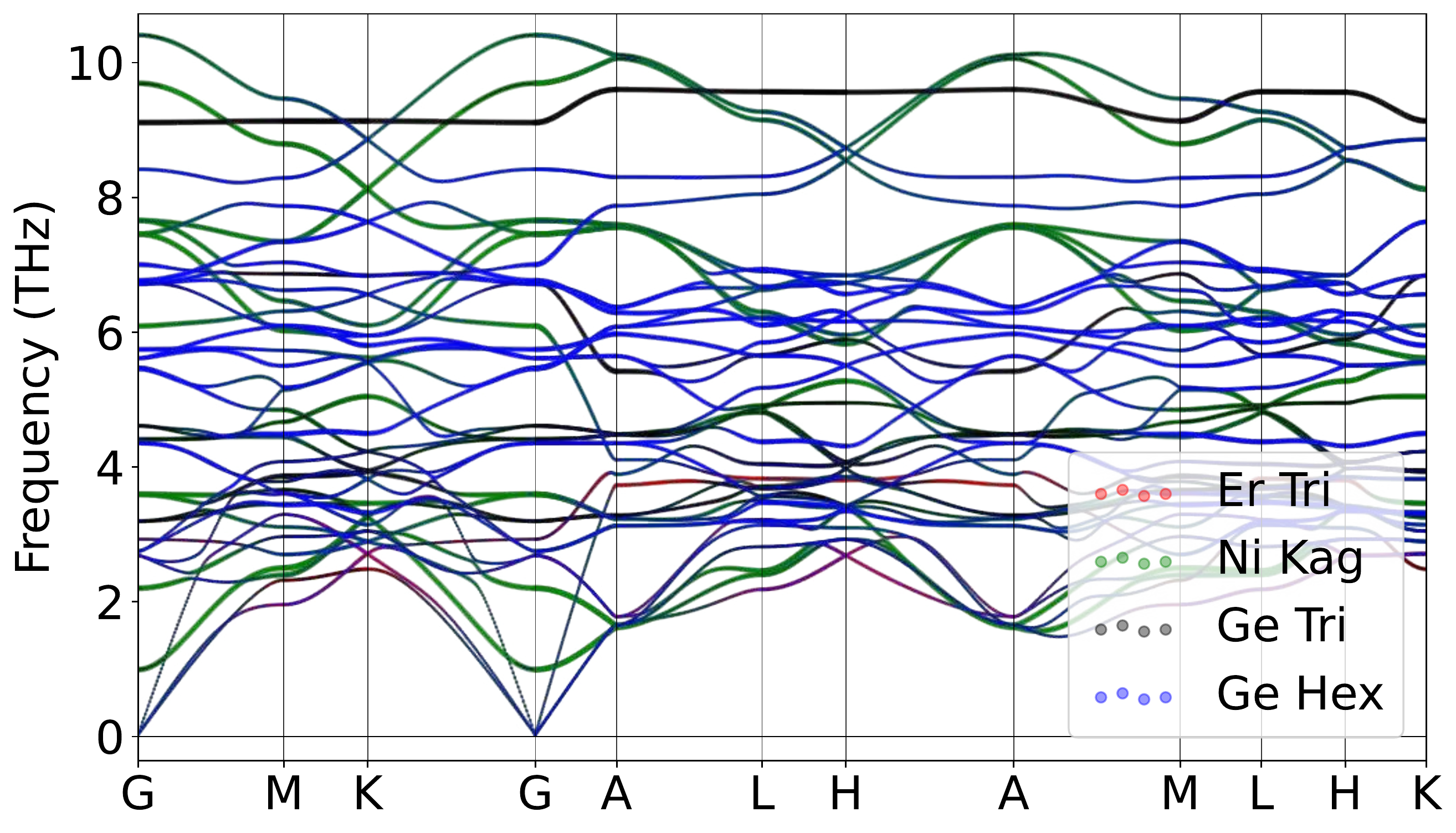}
}
\hspace{5pt}\subfloat[Relaxed phonon at high-T.]{
\label{fig:ErNi6Ge6_relaxed_High}
\includegraphics[width=0.45\textwidth]{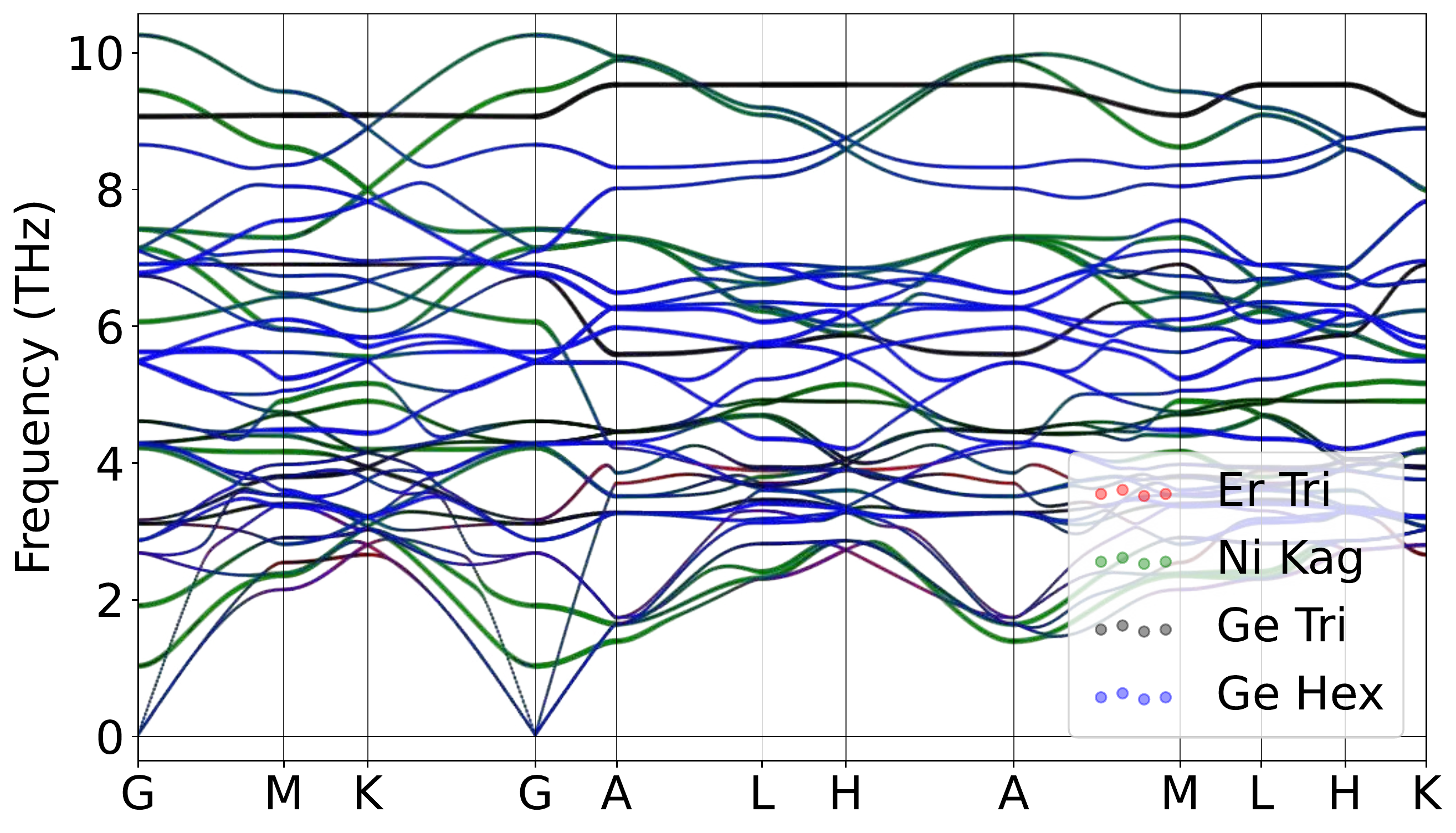}
}
\caption{Atomically projected phonon spectrum for ErNi$_6$Ge$_6$.}
\label{fig:ErNi6Ge6pho}
\end{figure}

\begin{figure}[H]
\centering
\label{fig:ErNi6Ge6_relaxed_Low_xz}
\includegraphics[width=0.85\textwidth]{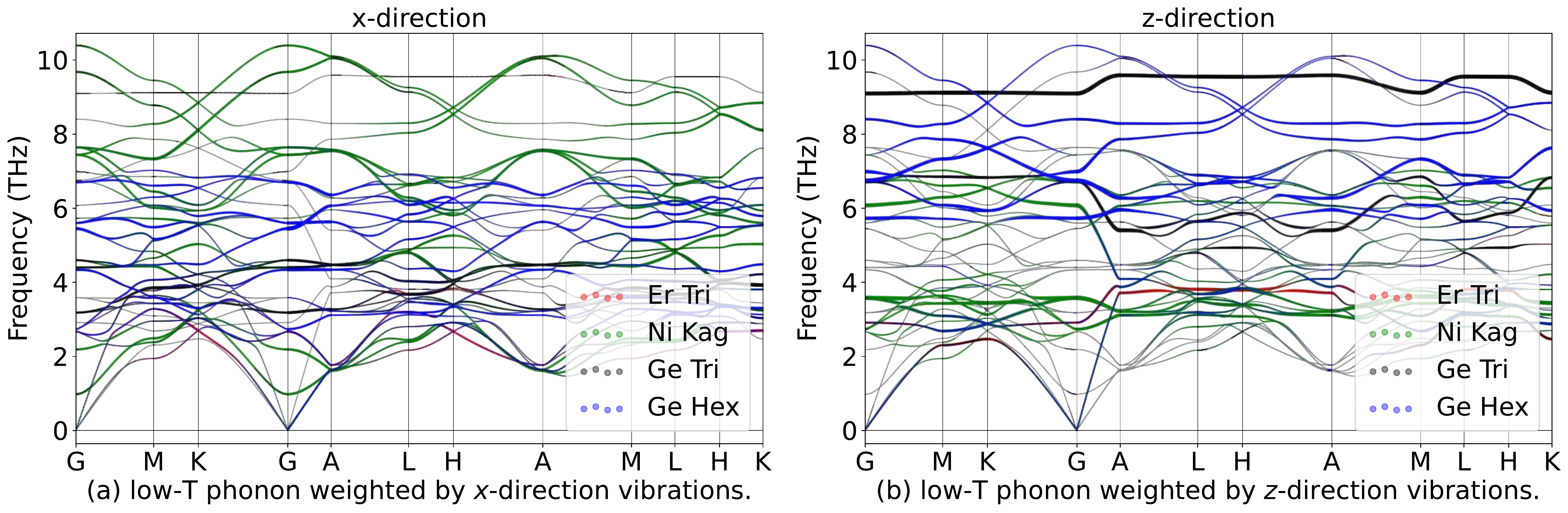}
\hspace{5pt}\label{fig:ErNi6Ge6_relaxed_High_xz}
\includegraphics[width=0.85\textwidth]{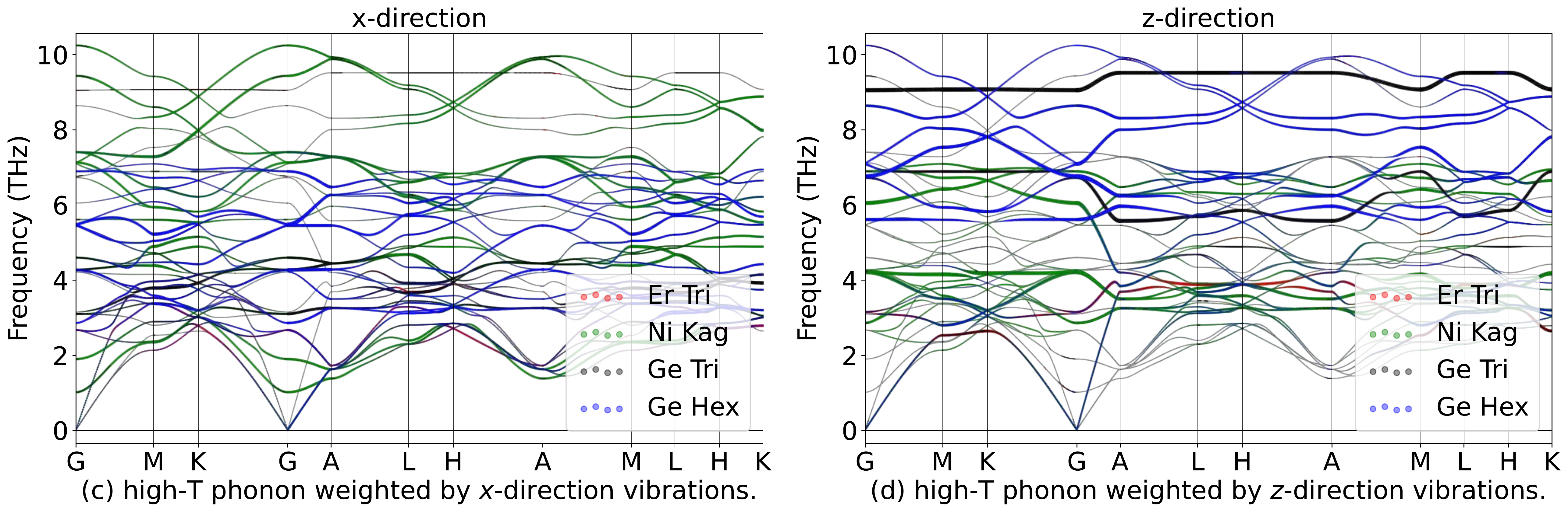}
\caption{Phonon spectrum for ErNi$_6$Ge$_6$in in-plane ($x$) and out-of-plane ($z$) directions.}
\label{fig:ErNi6Ge6phoxyz}
\end{figure}

\begin{table}[htbp]
\global\long\def\arraystretch{1.12}
\captionsetup{width=.95\textwidth}
\caption{IRREPs at high-symmetry points for ErNi$_6$Ge$_6$ phonon spectrum at Low-T.}
\label{tab:191_ErNi6Ge6_Low}
\setlength{\tabcolsep}{1.mm}{
}
\end{table}

\begin{table}[htbp]
\global\long\def\arraystretch{1.12}
\captionsetup{width=.95\textwidth}
\caption{IRREPs at high-symmetry points for ErNi$_6$Ge$_6$ phonon spectrum at High-T.}
\label{tab:191_ErNi6Ge6_High}
\setlength{\tabcolsep}{1.mm}{
%
}
\end{table}
\subsubsection{\label{sms2sec:TmNi6Ge6}\hyperref[smsec:col]{\texorpdfstring{TmNi$_6$Ge$_6$}{}}~{\color{green}  (High-T stable, Low-T stable) }}

\begin{table}[htbp]
\global\long\def\arraystretch{1.12}
\captionsetup{width=.85\textwidth}
\caption{Structure parameters of TmNi$_6$Ge$_6$ after relaxation. It contains 429 electrons. }
\label{tab:TmNi6Ge6}
\setlength{\tabcolsep}{4mm}{
%
}
\end{table}

\begin{figure}[htbp]
\centering
\includegraphics[width=0.85\textwidth]{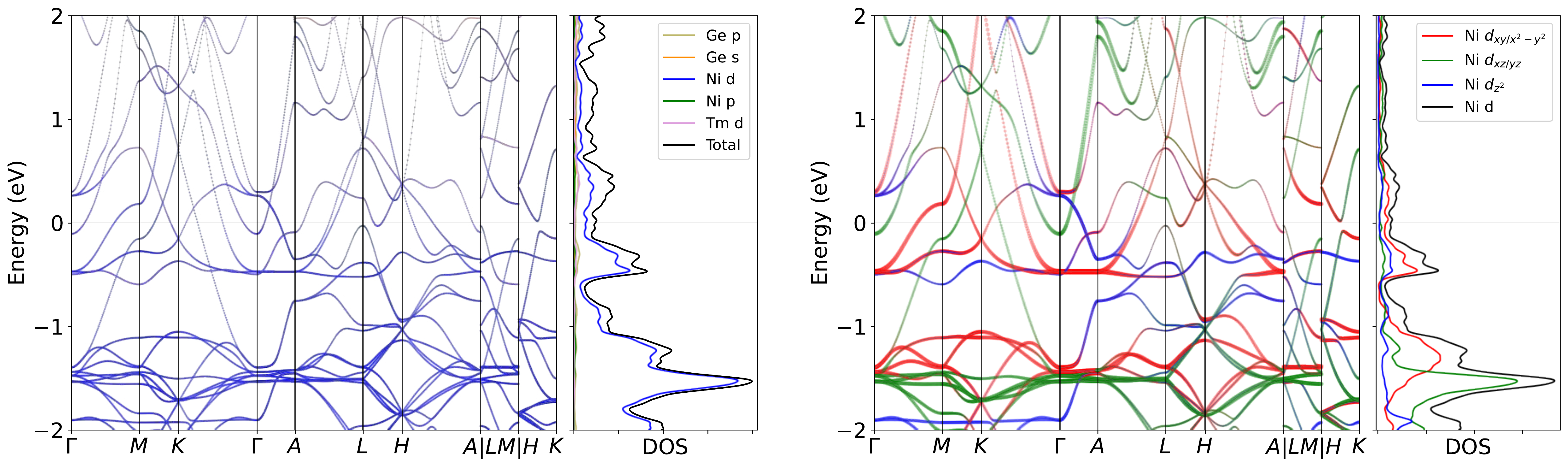}
\caption{Band structure for TmNi$_6$Ge$_6$ without SOC. The projection of Ni $3d$ orbitals is also presented. The band structures clearly reveal the dominating of Ni $3d$ orbitals  near the Fermi level, as well as the pronounced peak.}
\label{fig:TmNi6Ge6band}
\end{figure}

\begin{figure}[H]
\centering
\subfloat[Relaxed phonon at low-T.]{
\label{fig:TmNi6Ge6_relaxed_Low}
\includegraphics[width=0.45\textwidth]{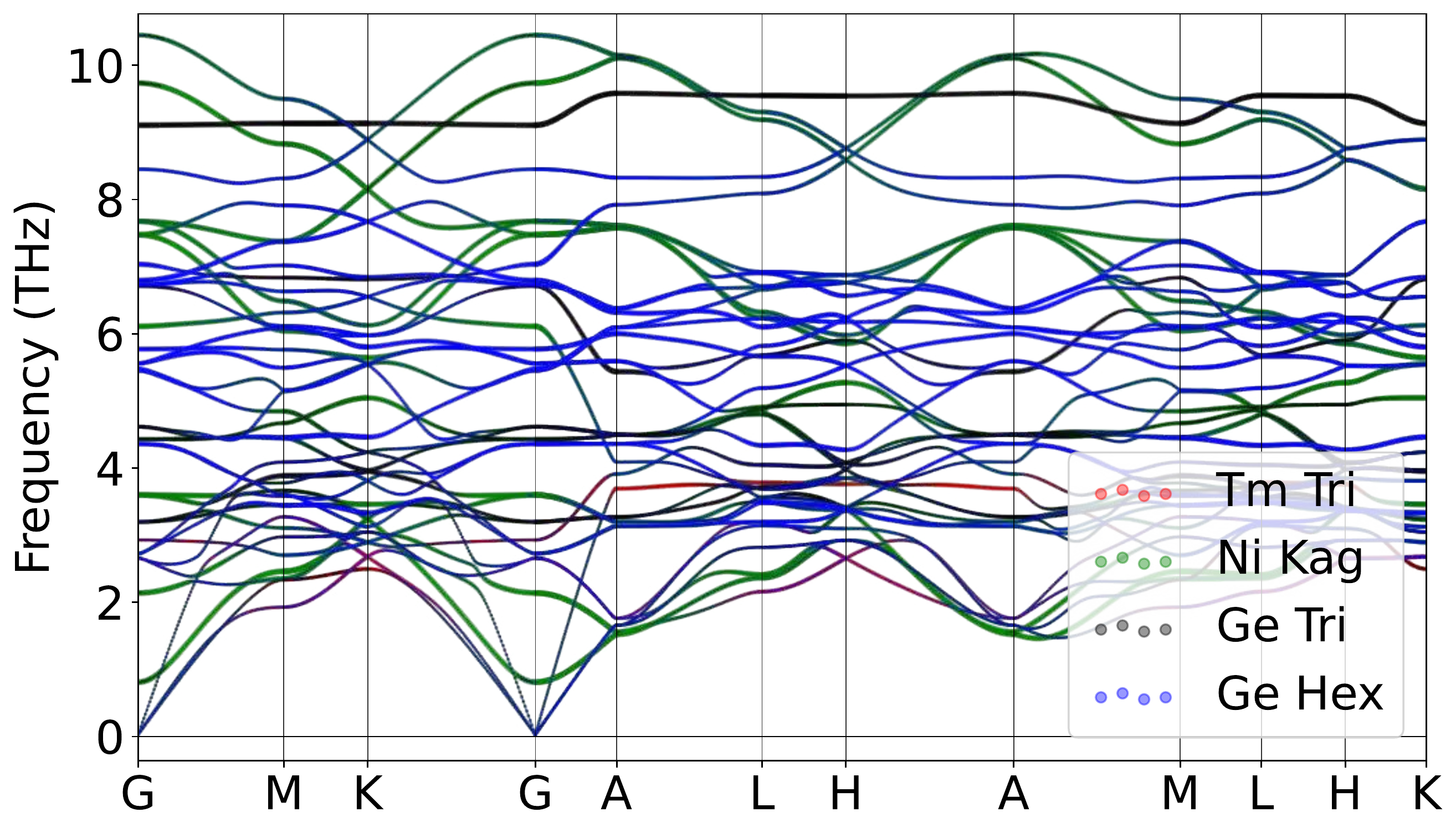}
}
\hspace{5pt}\subfloat[Relaxed phonon at high-T.]{
\label{fig:TmNi6Ge6_relaxed_High}
\includegraphics[width=0.45\textwidth]{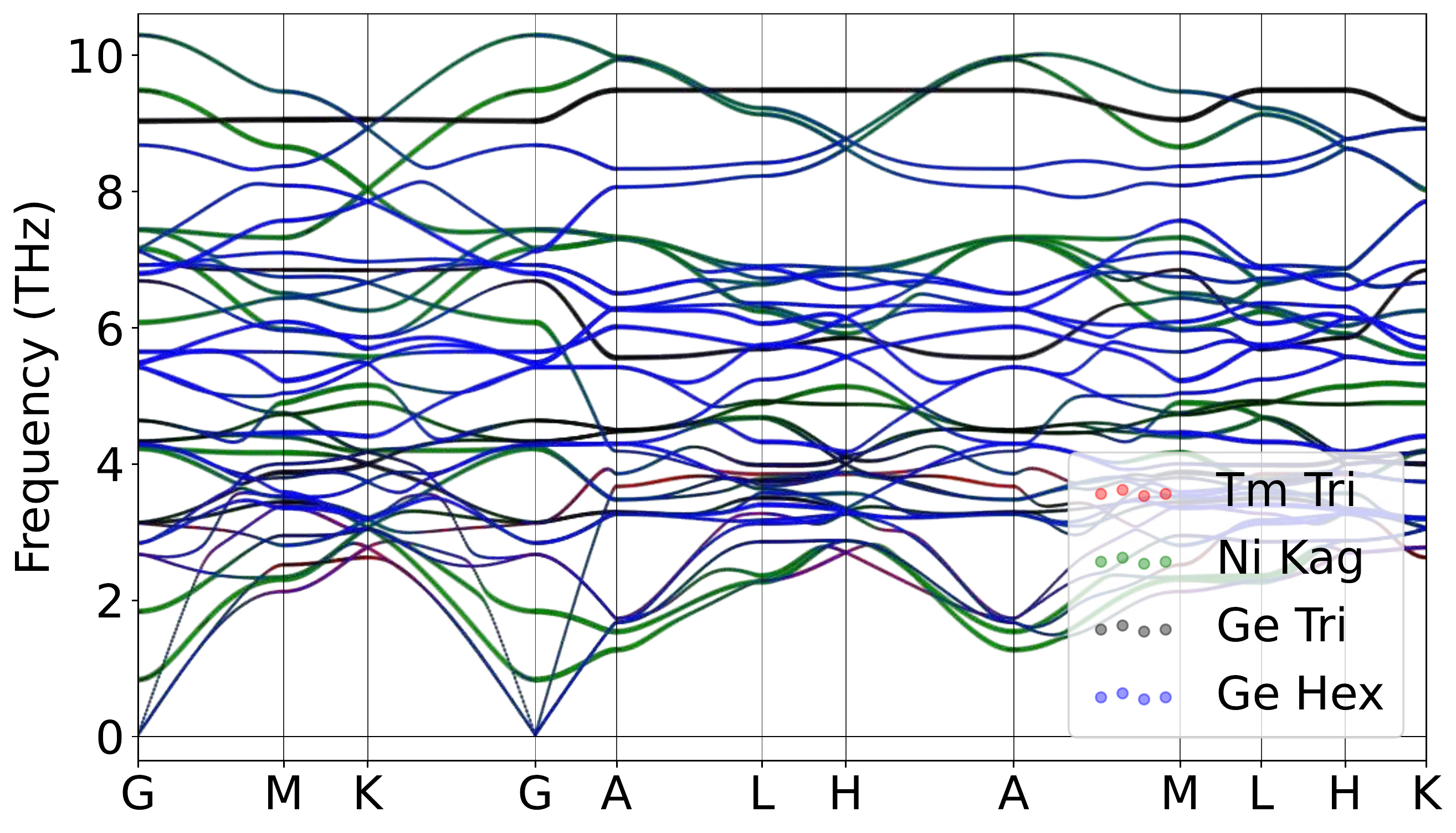}
}
\caption{Atomically projected phonon spectrum for TmNi$_6$Ge$_6$.}
\label{fig:TmNi6Ge6pho}
\end{figure}

\begin{figure}[H]
\centering
\label{fig:TmNi6Ge6_relaxed_Low_xz}
\includegraphics[width=0.85\textwidth]{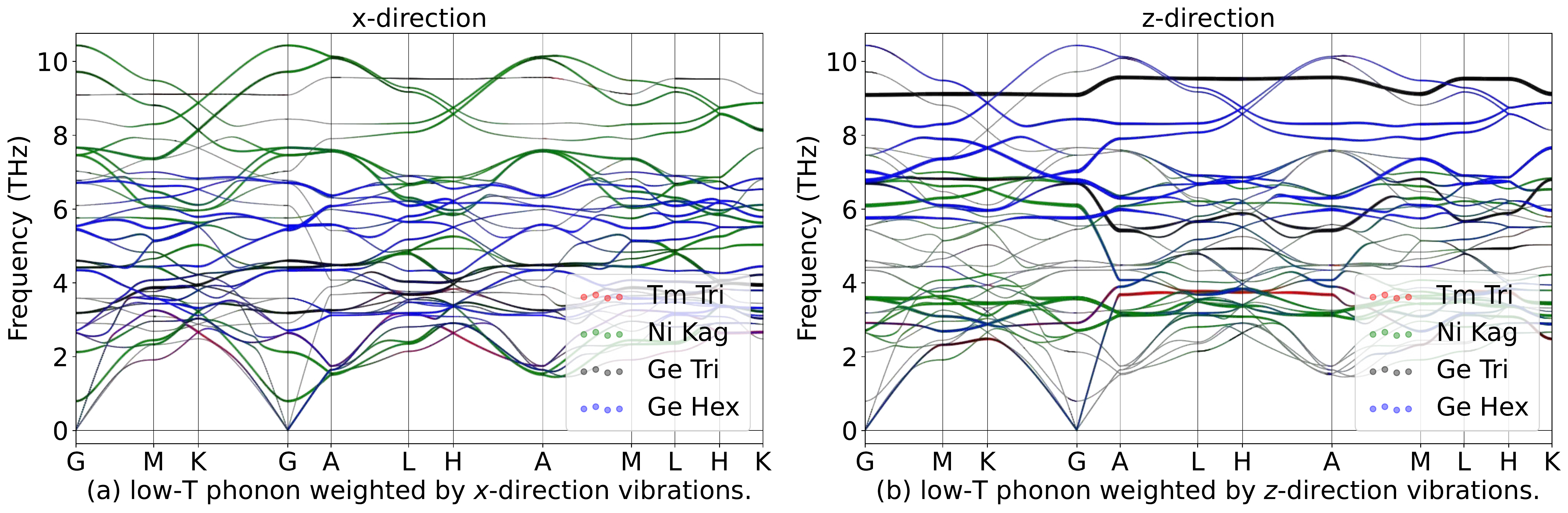}
\hspace{5pt}\label{fig:TmNi6Ge6_relaxed_High_xz}
\includegraphics[width=0.85\textwidth]{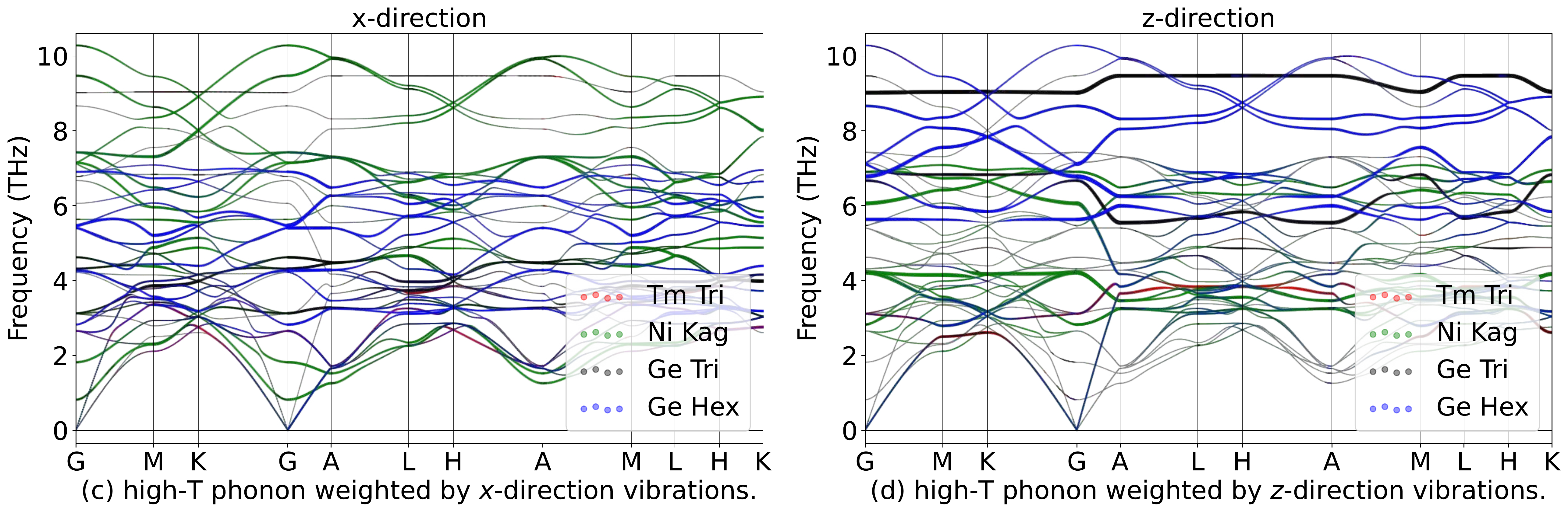}
\caption{Phonon spectrum for TmNi$_6$Ge$_6$in in-plane ($x$) and out-of-plane ($z$) directions.}
\label{fig:TmNi6Ge6phoxyz}
\end{figure}

\begin{table}[htbp]
\global\long\def\arraystretch{1.12}
\captionsetup{width=.95\textwidth}
\caption{IRREPs at high-symmetry points for TmNi$_6$Ge$_6$ phonon spectrum at Low-T.}
\label{tab:191_TmNi6Ge6_Low}
\setlength{\tabcolsep}{1.mm}{
}
\end{table}

\begin{table}[htbp]
\global\long\def\arraystretch{1.12}
\captionsetup{width=.95\textwidth}
\caption{IRREPs at high-symmetry points for TmNi$_6$Ge$_6$ phonon spectrum at High-T.}
\label{tab:191_TmNi6Ge6_High}
\setlength{\tabcolsep}{1.mm}{
%
}
\end{table}
\subsubsection{\label{sms2sec:YbNi6Ge6}\hyperref[smsec:col]{\texorpdfstring{YbNi$_6$Ge$_6$}{}}~{\color{green}  (High-T stable, Low-T stable) }}

\begin{table}[htbp]
\global\long\def\arraystretch{1.12}
\captionsetup{width=.85\textwidth}
\caption{Structure parameters of YbNi$_6$Ge$_6$ after relaxation. It contains 430 electrons. }
\label{tab:YbNi6Ge6}
\setlength{\tabcolsep}{4mm}{
%
}
\end{table}

\begin{figure}[htbp]
\centering
\includegraphics[width=0.85\textwidth]{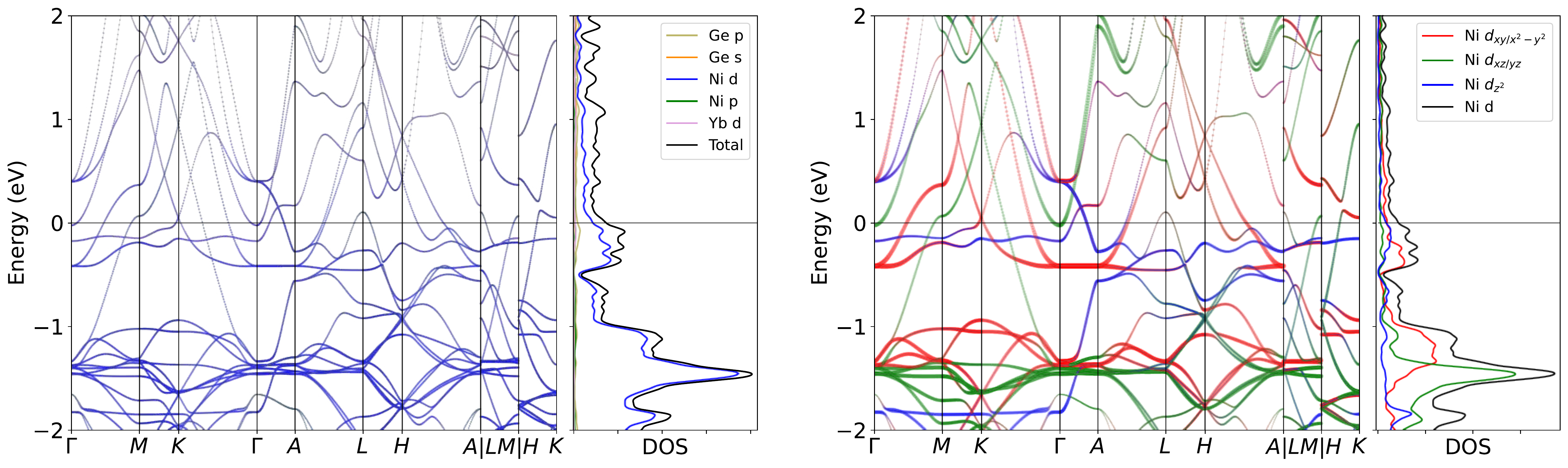}
\caption{Band structure for YbNi$_6$Ge$_6$ without SOC. The projection of Ni $3d$ orbitals is also presented. The band structures clearly reveal the dominating of Ni $3d$ orbitals  near the Fermi level, as well as the pronounced peak.}
\label{fig:YbNi6Ge6band}
\end{figure}

\begin{figure}[H]
\centering
\subfloat[Relaxed phonon at low-T.]{
\label{fig:YbNi6Ge6_relaxed_Low}
\includegraphics[width=0.45\textwidth]{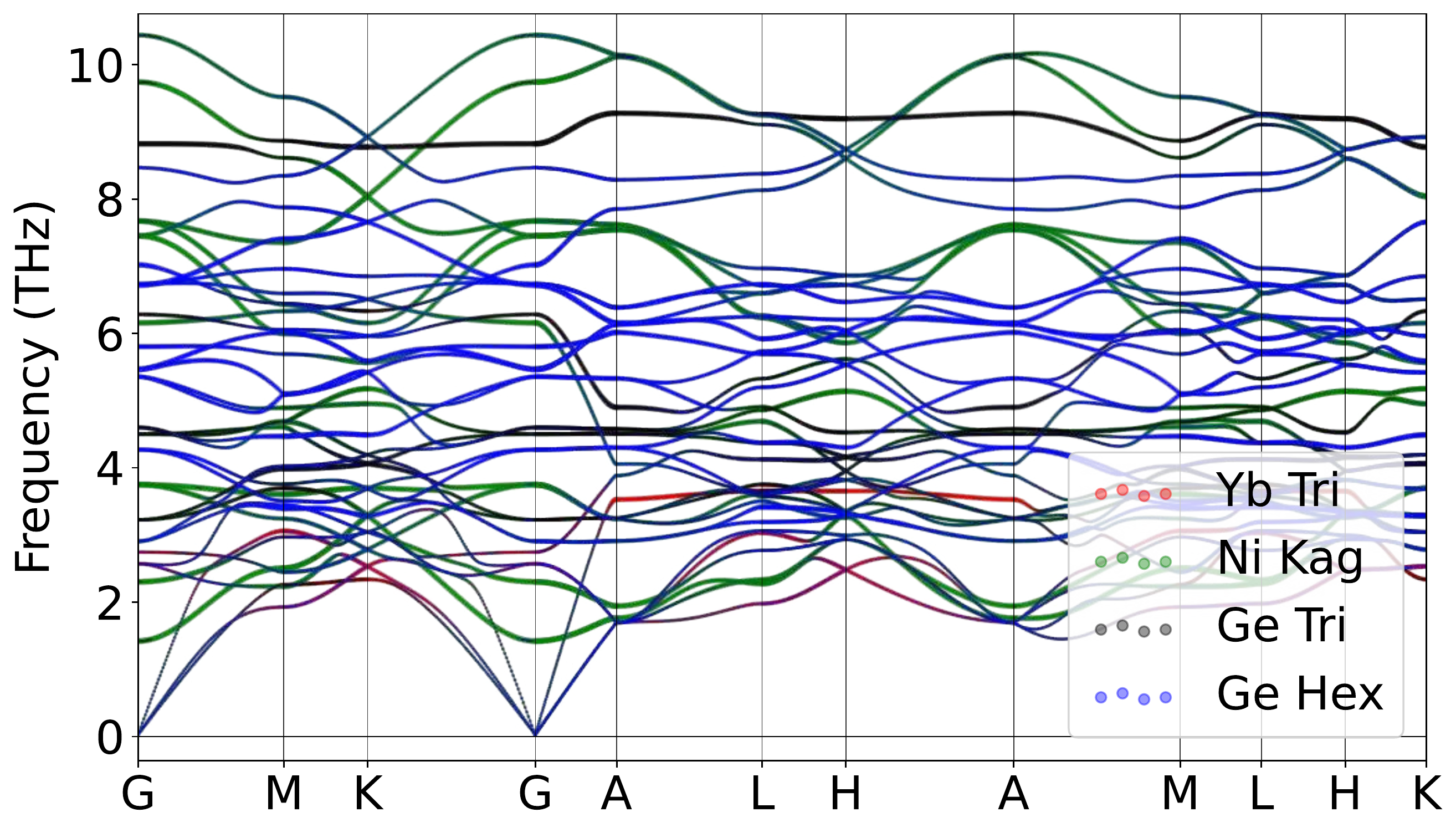}
}
\hspace{5pt}\subfloat[Relaxed phonon at high-T.]{
\label{fig:YbNi6Ge6_relaxed_High}
\includegraphics[width=0.45\textwidth]{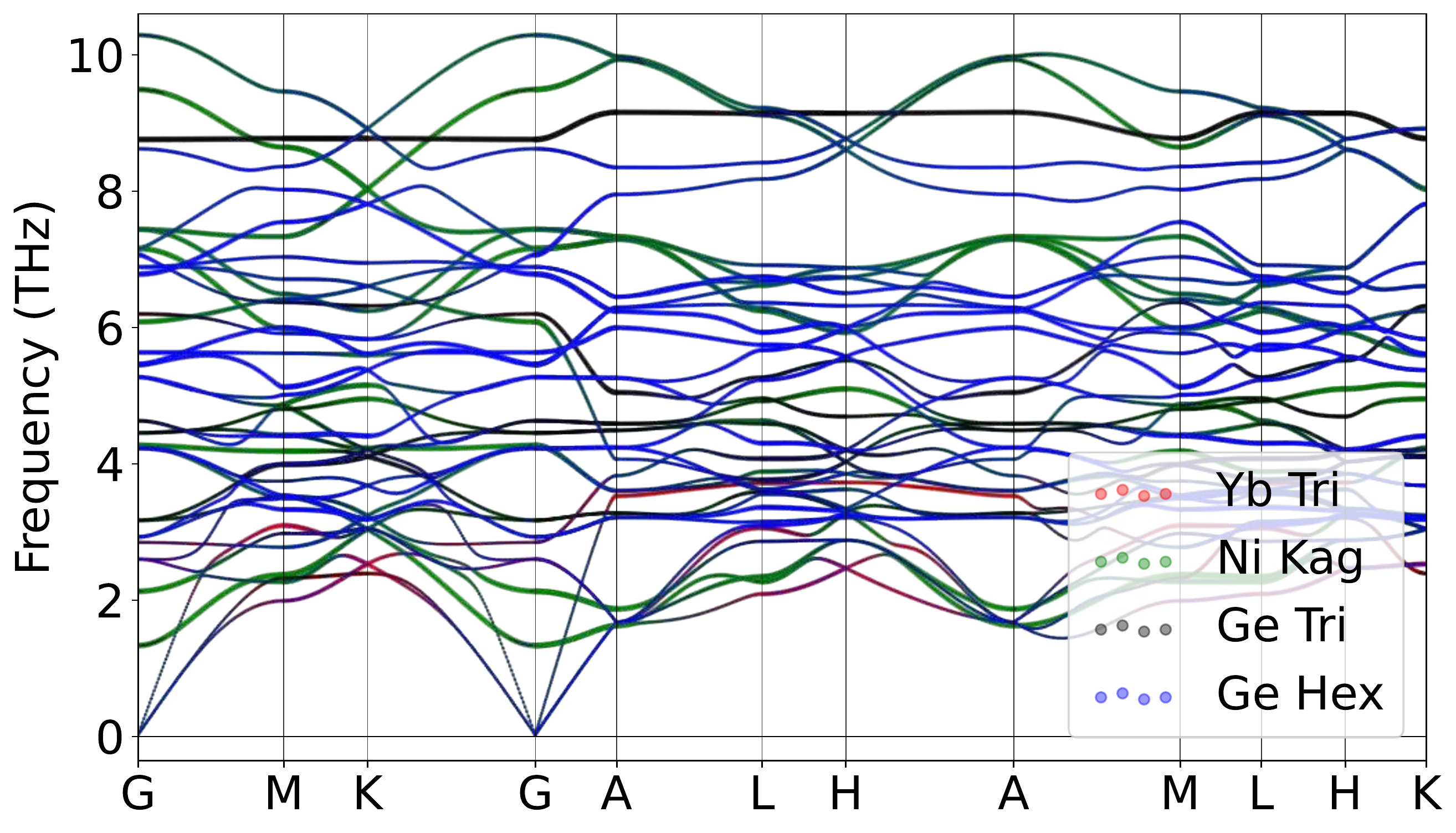}
}
\caption{Atomically projected phonon spectrum for YbNi$_6$Ge$_6$.}
\label{fig:YbNi6Ge6pho}
\end{figure}

\begin{figure}[H]
\centering
\label{fig:YbNi6Ge6_relaxed_Low_xz}
\includegraphics[width=0.85\textwidth]{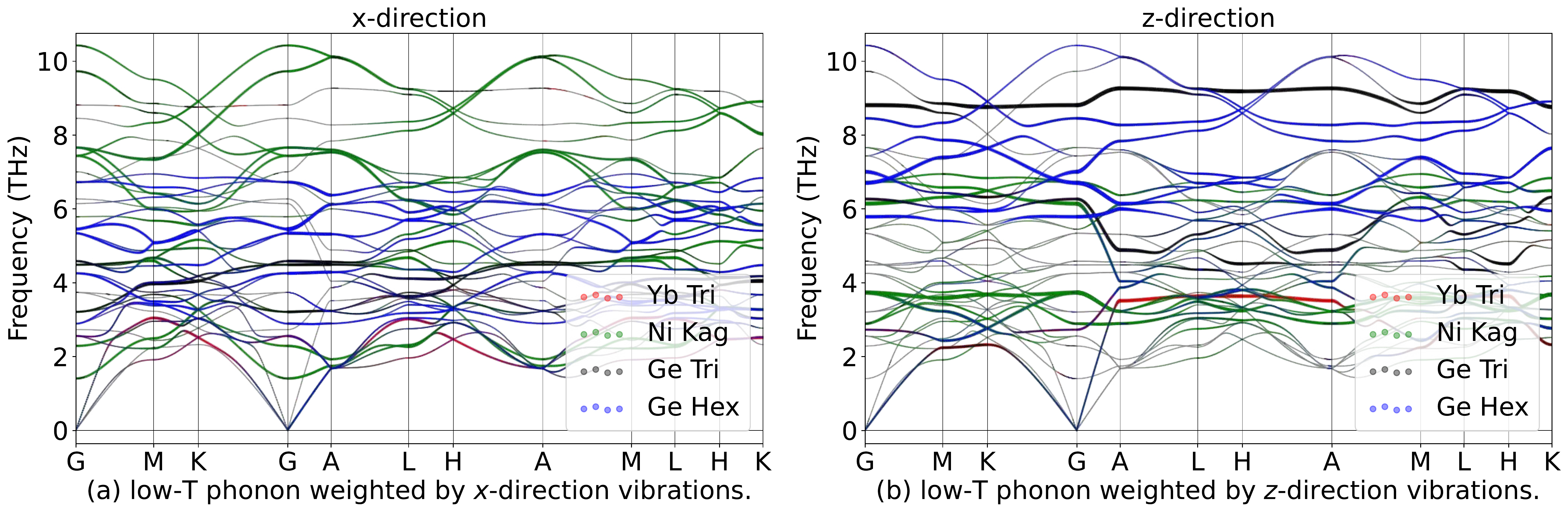}
\hspace{5pt}\label{fig:YbNi6Ge6_relaxed_High_xz}
\includegraphics[width=0.85\textwidth]{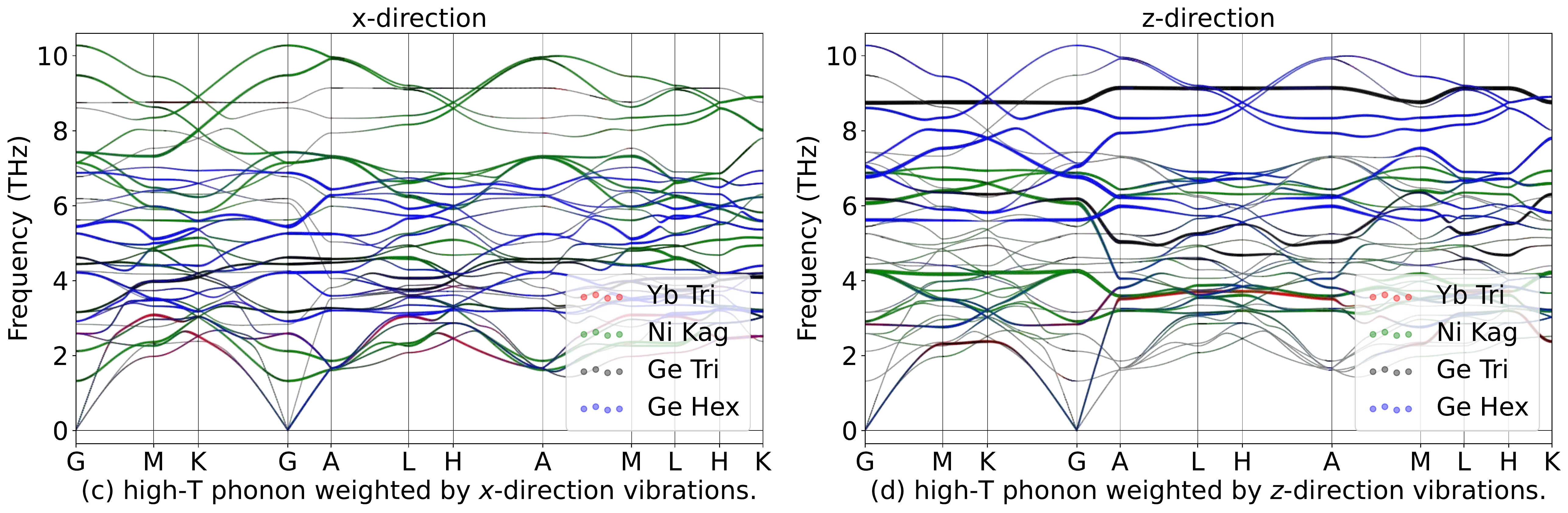}
\caption{Phonon spectrum for YbNi$_6$Ge$_6$in in-plane ($x$) and out-of-plane ($z$) directions.}
\label{fig:YbNi6Ge6phoxyz}
\end{figure}

\begin{table}[htbp]
\global\long\def\arraystretch{1.12}
\captionsetup{width=.95\textwidth}
\caption{IRREPs at high-symmetry points for YbNi$_6$Ge$_6$ phonon spectrum at Low-T.}
\label{tab:191_YbNi6Ge6_Low}
\setlength{\tabcolsep}{1.mm}{
}
\end{table}

\begin{table}[htbp]
\global\long\def\arraystretch{1.12}
\captionsetup{width=.95\textwidth}
\caption{IRREPs at high-symmetry points for YbNi$_6$Ge$_6$ phonon spectrum at High-T.}
\label{tab:191_YbNi6Ge6_High}
\setlength{\tabcolsep}{1.mm}{
%
}
\end{table}
\subsubsection{\label{sms2sec:LuNi6Ge6}\hyperref[smsec:col]{\texorpdfstring{LuNi$_6$Ge$_6$}{}}~{\color{green}  (High-T stable, Low-T stable) }}

\begin{table}[htbp]
\global\long\def\arraystretch{1.12}
\captionsetup{width=.85\textwidth}
\caption{Structure parameters of LuNi$_6$Ge$_6$ after relaxation. It contains 431 electrons. }
\label{tab:LuNi6Ge6}
\setlength{\tabcolsep}{4mm}{
%
}
\end{table}

\begin{figure}[htbp]
\centering
\includegraphics[width=0.85\textwidth]{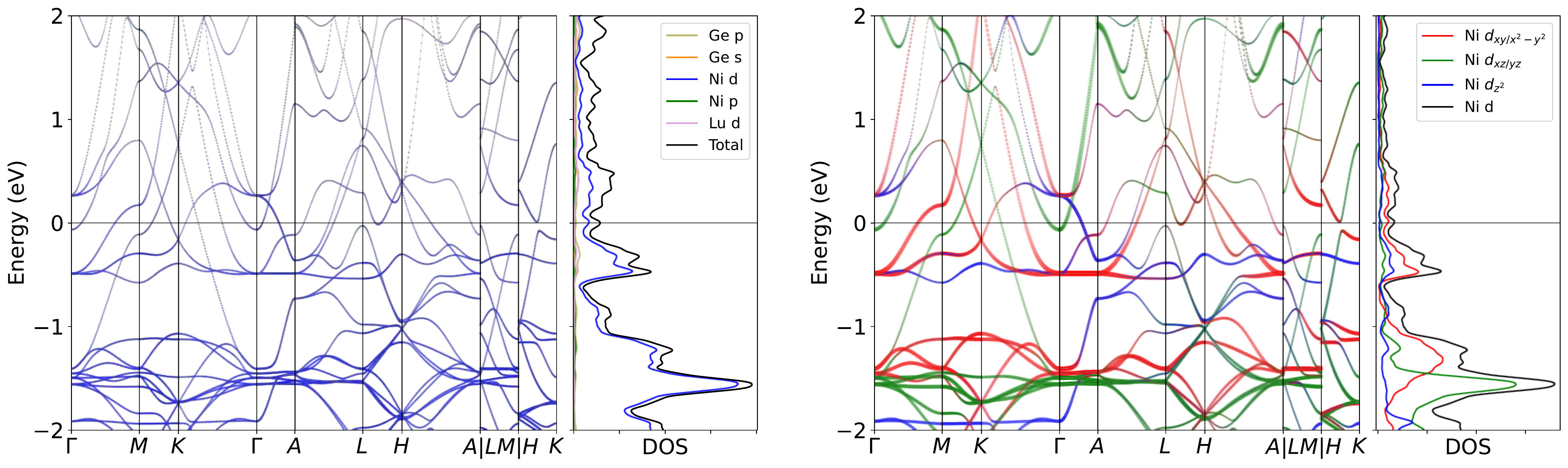}
\caption{Band structure for LuNi$_6$Ge$_6$ without SOC. The projection of Ni $3d$ orbitals is also presented. The band structures clearly reveal the dominating of Ni $3d$ orbitals  near the Fermi level, as well as the pronounced peak.}
\label{fig:LuNi6Ge6band}
\end{figure}

\begin{figure}[H]
\centering
\subfloat[Relaxed phonon at low-T.]{
\label{fig:LuNi6Ge6_relaxed_Low}
\includegraphics[width=0.45\textwidth]{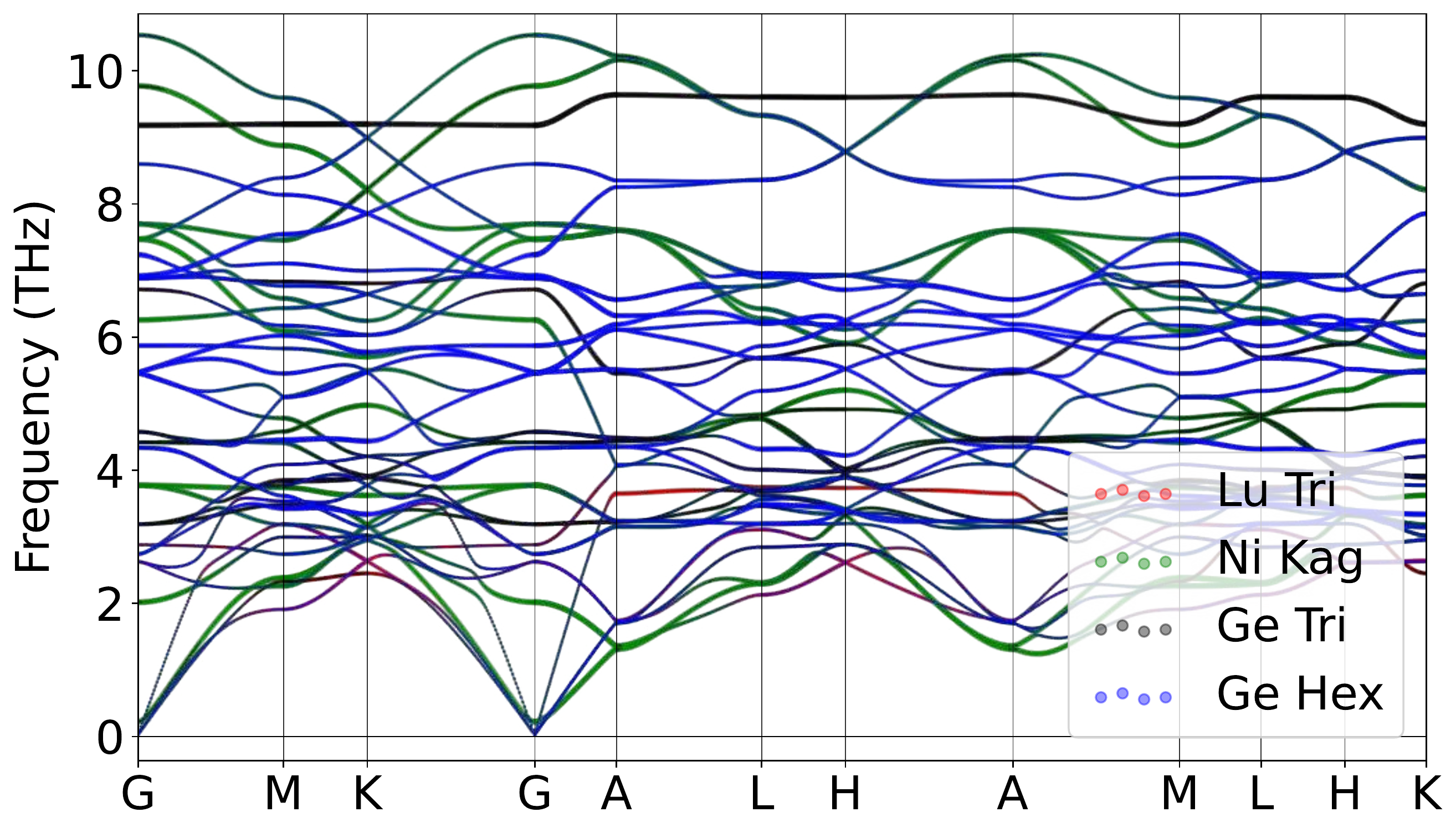}
}
\hspace{5pt}\subfloat[Relaxed phonon at high-T.]{
\label{fig:LuNi6Ge6_relaxed_High}
\includegraphics[width=0.45\textwidth]{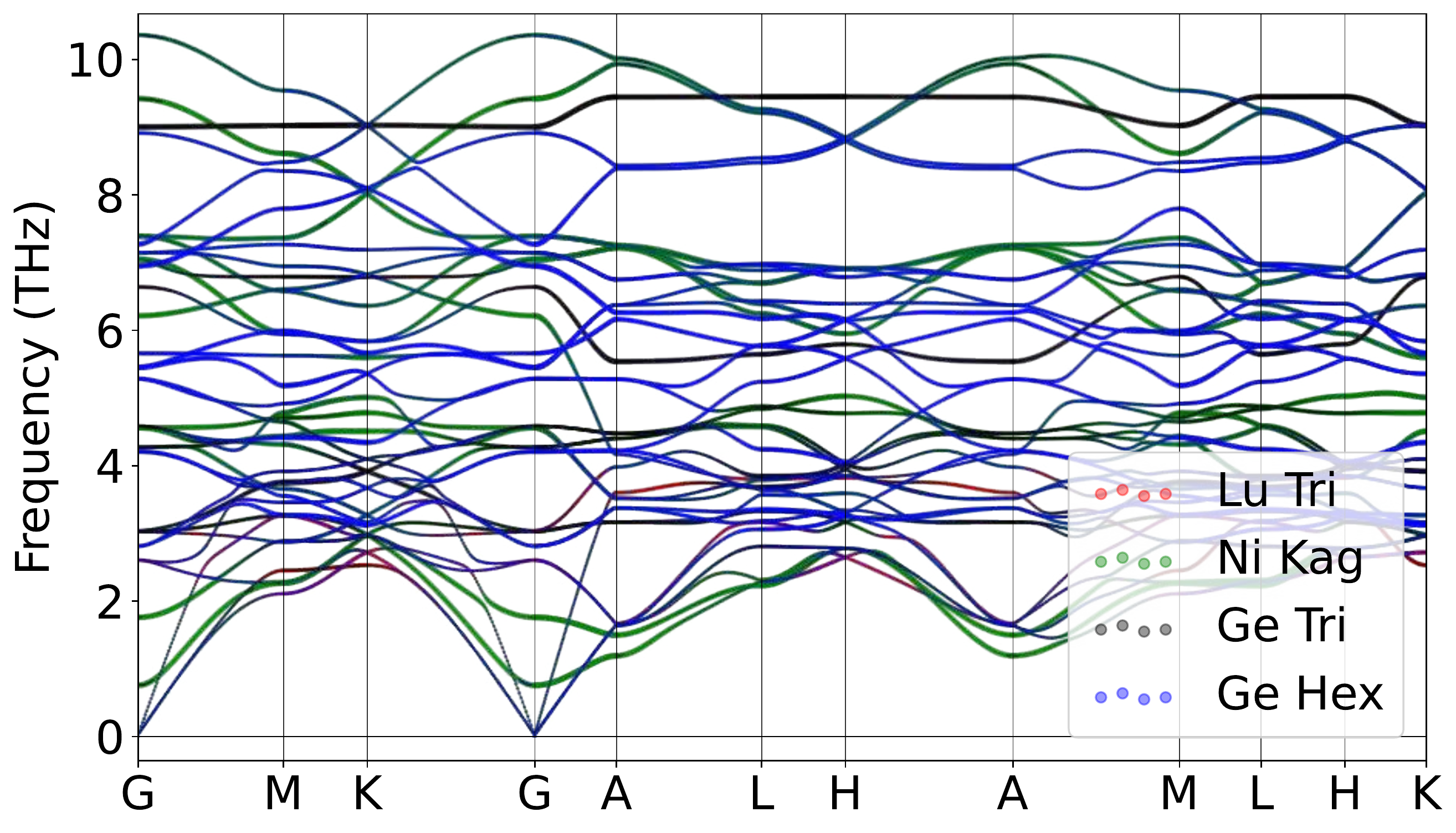}
}
\caption{Atomically projected phonon spectrum for LuNi$_6$Ge$_6$.}
\label{fig:LuNi6Ge6pho}
\end{figure}

\begin{figure}[H]
\centering
\label{fig:LuNi6Ge6_relaxed_Low_xz}
\includegraphics[width=0.85\textwidth]{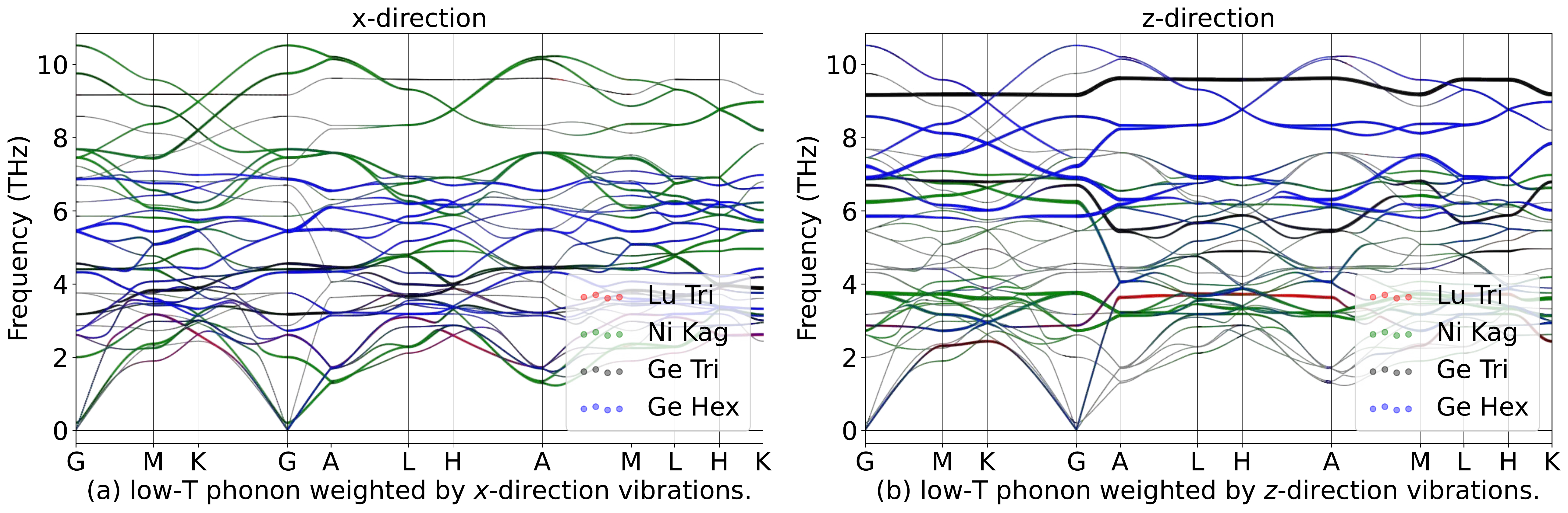}
\hspace{5pt}\label{fig:LuNi6Ge6_relaxed_High_xz}
\includegraphics[width=0.85\textwidth]{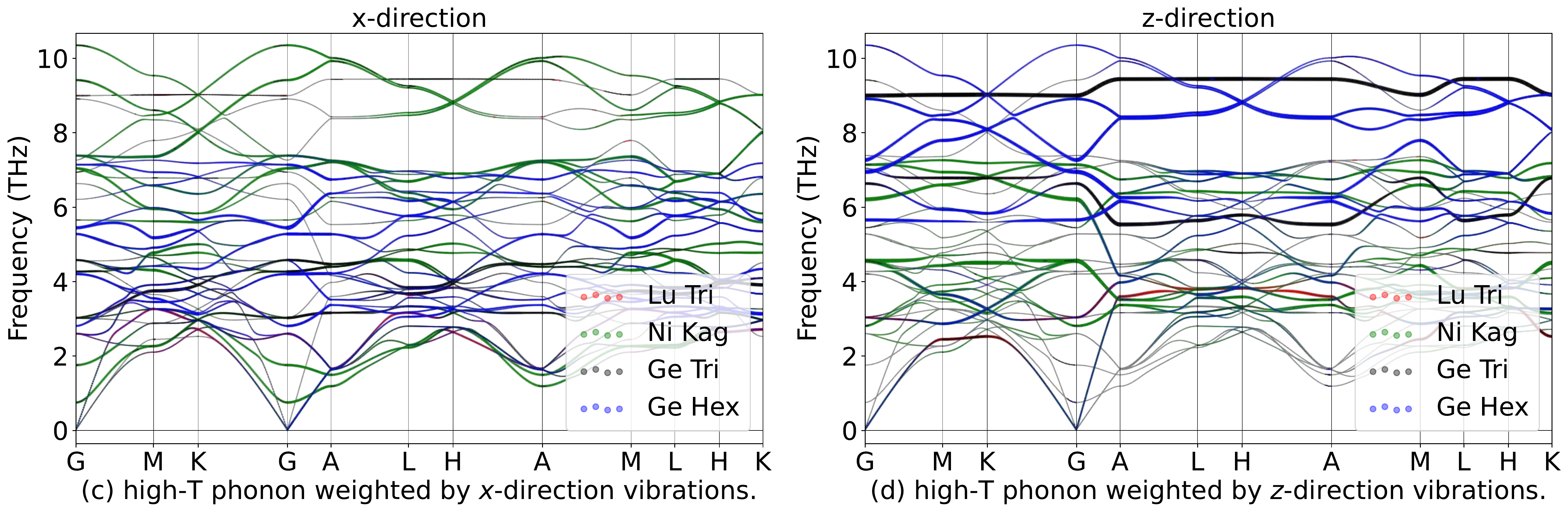}
\caption{Phonon spectrum for LuNi$_6$Ge$_6$in in-plane ($x$) and out-of-plane ($z$) directions.}
\label{fig:LuNi6Ge6phoxyz}
\end{figure}

\begin{table}[htbp]
\global\long\def\arraystretch{1.12}
\captionsetup{width=.95\textwidth}
\caption{IRREPs at high-symmetry points for LuNi$_6$Ge$_6$ phonon spectrum at Low-T.}
\label{tab:191_LuNi6Ge6_Low}
\setlength{\tabcolsep}{1.mm}{
}
\end{table}

\begin{table}[htbp]
\global\long\def\arraystretch{1.12}
\captionsetup{width=.95\textwidth}
\caption{IRREPs at high-symmetry points for LuNi$_6$Ge$_6$ phonon spectrum at High-T.}
\label{tab:191_LuNi6Ge6_High}
\setlength{\tabcolsep}{1.mm}{
%
}
\end{table}
\subsubsection{\label{sms2sec:HfNi6Ge6}\hyperref[smsec:col]{\texorpdfstring{HfNi$_6$Ge$_6$}{}}~{\color{red}  (High-T unstable, Low-T unstable) }}

\begin{table}[htbp]
\global\long\def\arraystretch{1.12}
\captionsetup{width=.85\textwidth}
\caption{Structure parameters of HfNi$_6$Ge$_6$ after relaxation. It contains 432 electrons. }
\label{tab:HfNi6Ge6}
\setlength{\tabcolsep}{4mm}{
%
}
\end{table}

\begin{figure}[htbp]
\centering
\includegraphics[width=0.85\textwidth]{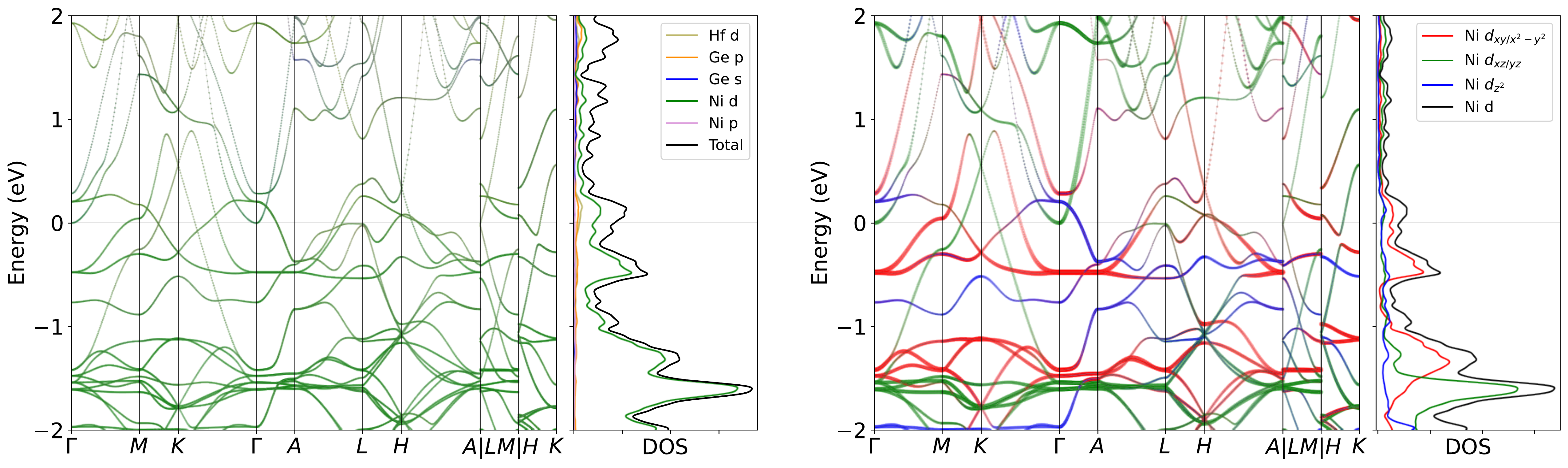}
\caption{Band structure for HfNi$_6$Ge$_6$ without SOC. The projection of Ni $3d$ orbitals is also presented. The band structures clearly reveal the dominating of Ni $3d$ orbitals  near the Fermi level, as well as the pronounced peak.}
\label{fig:HfNi6Ge6band}
\end{figure}

\begin{figure}[H]
\centering
\subfloat[Relaxed phonon at low-T.]{
\label{fig:HfNi6Ge6_relaxed_Low}
\includegraphics[width=0.45\textwidth]{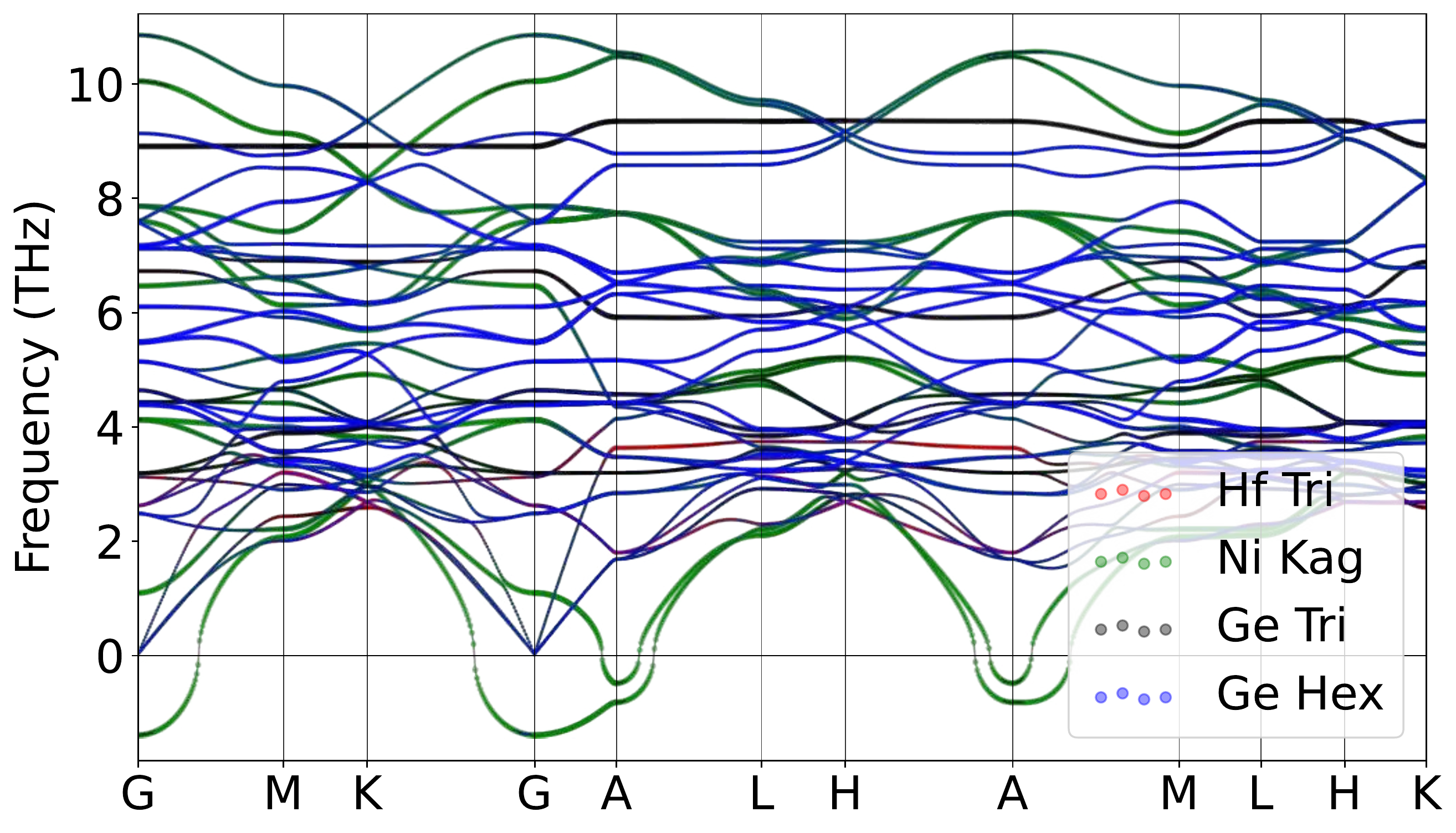}
}
\hspace{5pt}\subfloat[Relaxed phonon at high-T.]{
\label{fig:HfNi6Ge6_relaxed_High}
\includegraphics[width=0.45\textwidth]{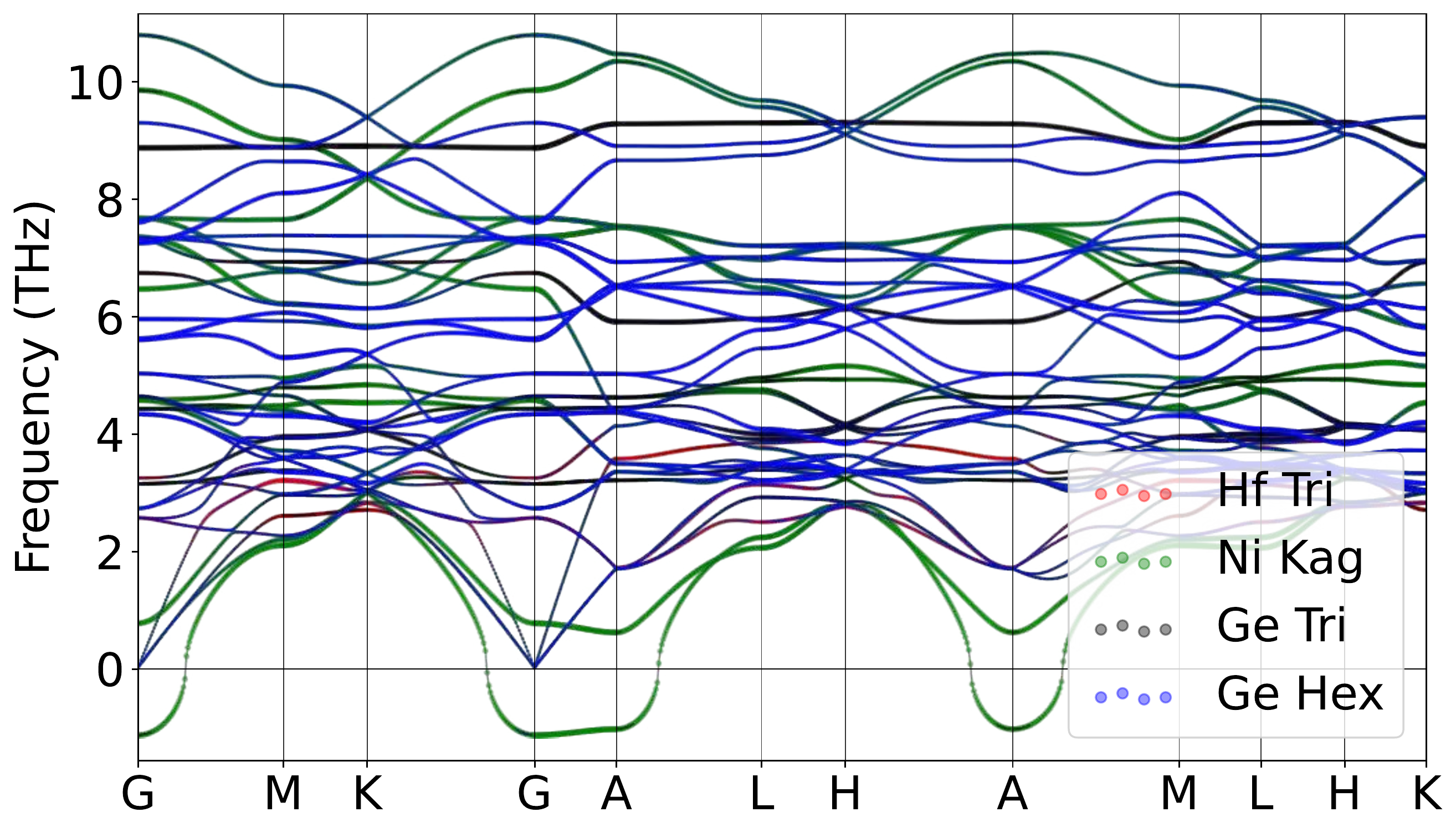}
}
\caption{Atomically projected phonon spectrum for HfNi$_6$Ge$_6$.}
\label{fig:HfNi6Ge6pho}
\end{figure}

\begin{figure}[H]
\centering
\label{fig:HfNi6Ge6_relaxed_Low_xz}
\includegraphics[width=0.85\textwidth]{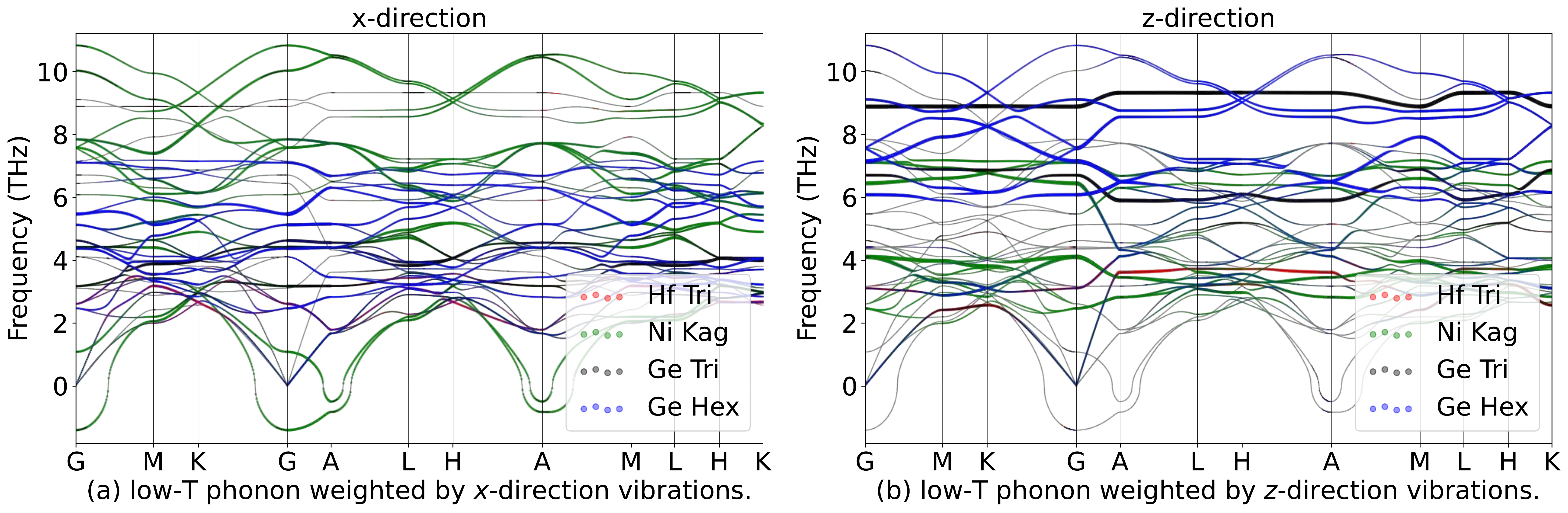}
\hspace{5pt}\label{fig:HfNi6Ge6_relaxed_High_xz}
\includegraphics[width=0.85\textwidth]{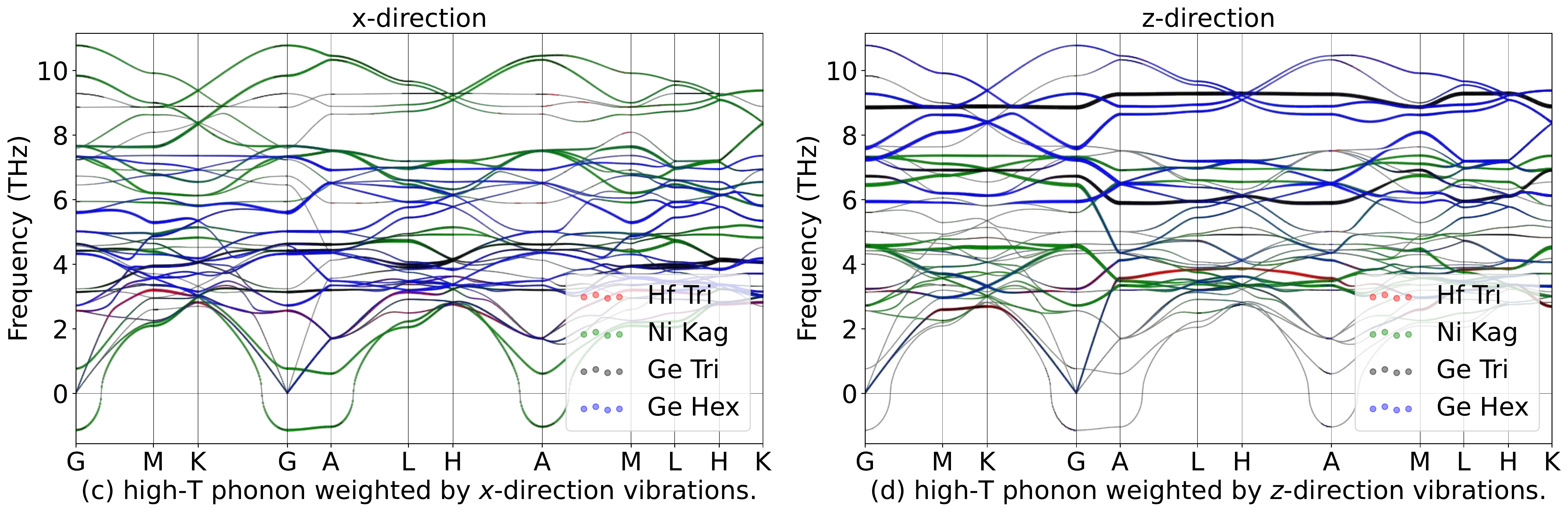}
\caption{Phonon spectrum for HfNi$_6$Ge$_6$in in-plane ($x$) and out-of-plane ($z$) directions.}
\label{fig:HfNi6Ge6phoxyz}
\end{figure}

\begin{table}[htbp]
\global\long\def\arraystretch{1.12}
\captionsetup{width=.95\textwidth}
\caption{IRREPs at high-symmetry points for HfNi$_6$Ge$_6$ phonon spectrum at Low-T.}
\label{tab:191_HfNi6Ge6_Low}
\setlength{\tabcolsep}{1.mm}{
}
\end{table}

\begin{table}[htbp]
\global\long\def\arraystretch{1.12}
\captionsetup{width=.95\textwidth}
\caption{IRREPs at high-symmetry points for HfNi$_6$Ge$_6$ phonon spectrum at High-T.}
\label{tab:191_HfNi6Ge6_High}
\setlength{\tabcolsep}{1.mm}{
%
}
\end{table}
\subsubsection{\label{sms2sec:TaNi6Ge6}\hyperref[smsec:col]{\texorpdfstring{TaNi$_6$Ge$_6$}{}}~{\color{red}  (High-T unstable, Low-T unstable) }}

\begin{table}[htbp]
\global\long\def\arraystretch{1.12}
\captionsetup{width=.85\textwidth}
\caption{Structure parameters of TaNi$_6$Ge$_6$ after relaxation. It contains 433 electrons. }
\label{tab:TaNi6Ge6}
\setlength{\tabcolsep}{4mm}{
%
}
\end{table}

\begin{figure}[htbp]
\centering
\includegraphics[width=0.85\textwidth]{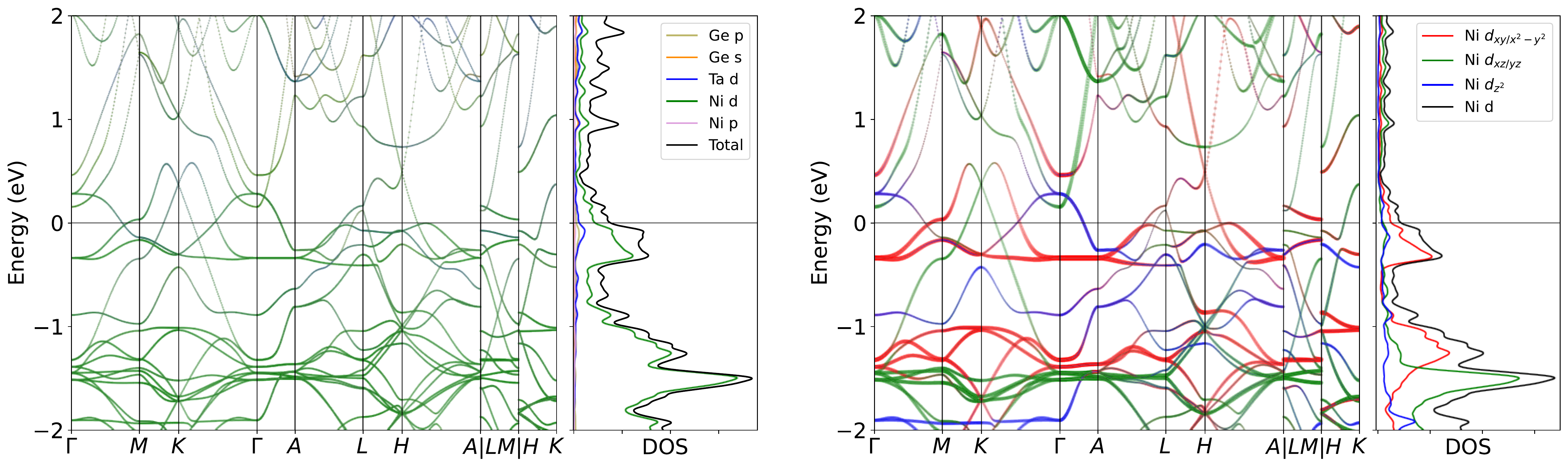}
\caption{Band structure for TaNi$_6$Ge$_6$ without SOC. The projection of Ni $3d$ orbitals is also presented. The band structures clearly reveal the dominating of Ni $3d$ orbitals  near the Fermi level, as well as the pronounced peak.}
\label{fig:TaNi6Ge6band}
\end{figure}

\begin{figure}[H]
\centering
\subfloat[Relaxed phonon at low-T.]{
\label{fig:TaNi6Ge6_relaxed_Low}
\includegraphics[width=0.45\textwidth]{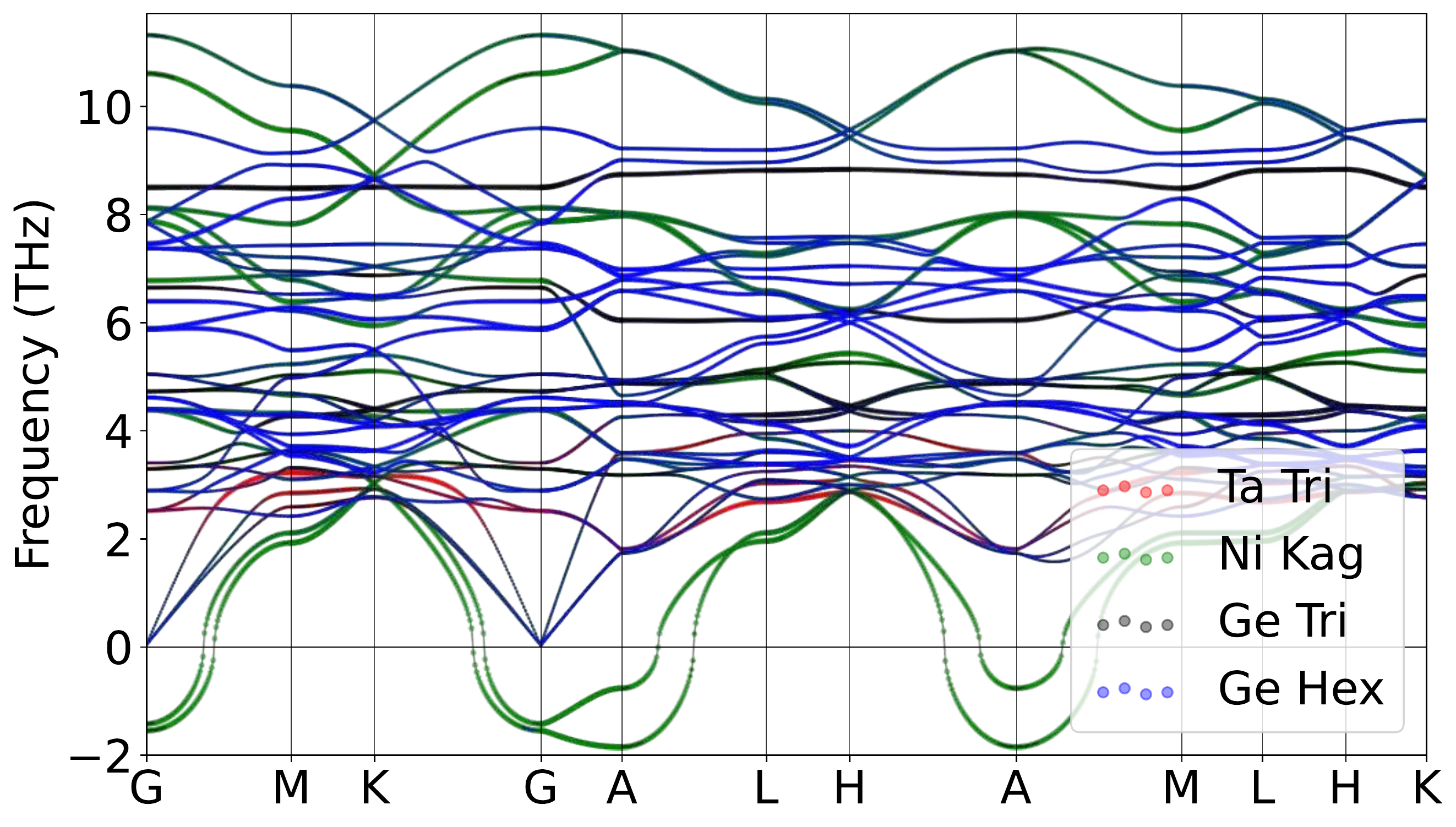}
}
\hspace{5pt}\subfloat[Relaxed phonon at high-T.]{
\label{fig:TaNi6Ge6_relaxed_High}
\includegraphics[width=0.45\textwidth]{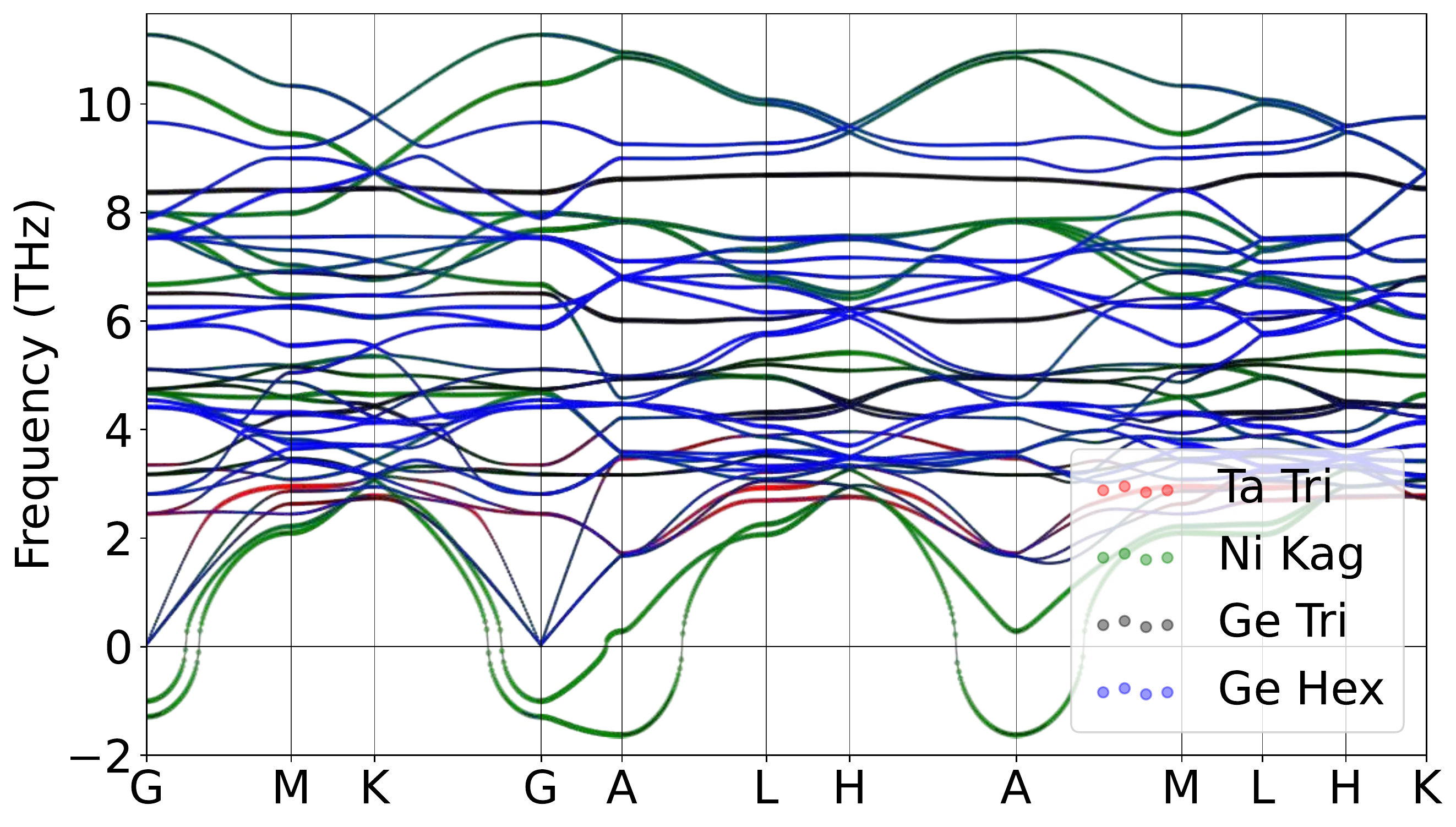}
}
\caption{Atomically projected phonon spectrum for TaNi$_6$Ge$_6$.}
\label{fig:TaNi6Ge6pho}
\end{figure}

\begin{figure}[H]
\centering
\label{fig:TaNi6Ge6_relaxed_Low_xz}
\includegraphics[width=0.85\textwidth]{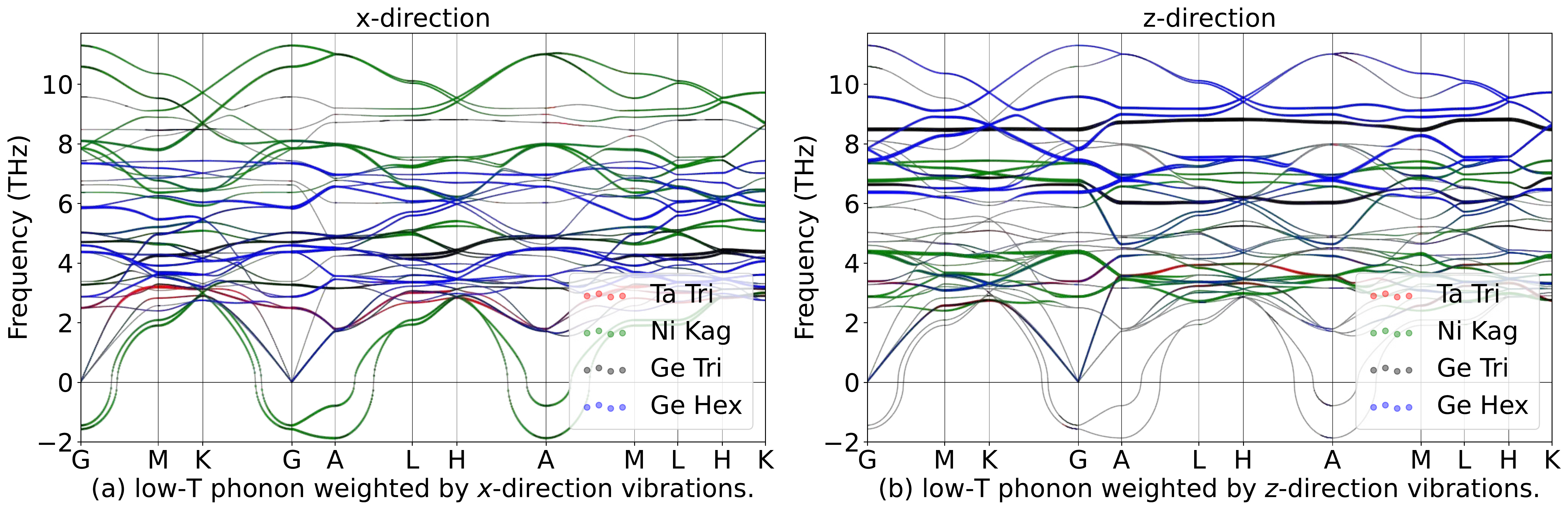}
\hspace{5pt}\label{fig:TaNi6Ge6_relaxed_High_xz}
\includegraphics[width=0.85\textwidth]{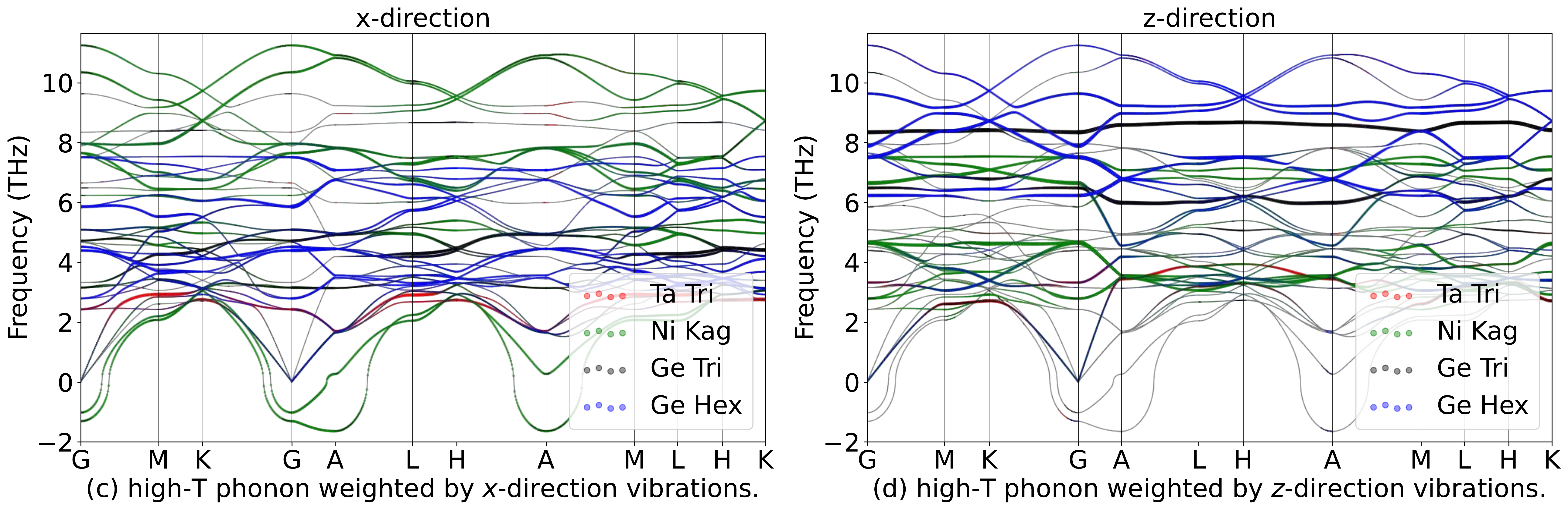}
\caption{Phonon spectrum for TaNi$_6$Ge$_6$in in-plane ($x$) and out-of-plane ($z$) directions.}
\label{fig:TaNi6Ge6phoxyz}
\end{figure}

\begin{table}[htbp]
\global\long\def\arraystretch{1.12}
\captionsetup{width=.95\textwidth}
\caption{IRREPs at high-symmetry points for TaNi$_6$Ge$_6$ phonon spectrum at Low-T.}
\label{tab:191_TaNi6Ge6_Low}
\setlength{\tabcolsep}{1.mm}{
}
\end{table}

\begin{table}[htbp]
\global\long\def\arraystretch{1.12}
\captionsetup{width=.95\textwidth}
\caption{IRREPs at high-symmetry points for TaNi$_6$Ge$_6$ phonon spectrum at High-T.}
\label{tab:191_TaNi6Ge6_High}
\setlength{\tabcolsep}{1.mm}{
%
}
\end{table}
\subsection{\hyperref[smsec:col]{NiIn}}~\label{smssec:NiIn}

\subsubsection{\label{sms2sec:LiNi6In6}\hyperref[smsec:col]{\texorpdfstring{LiNi$_6$In$_6$}{}}~{\color{green}  (High-T stable, Low-T stable) }}

\begin{table}[htbp]
\global\long\def\arraystretch{1.12}
\captionsetup{width=.85\textwidth}
\caption{Structure parameters of LiNi$_6$In$_6$ after relaxation. It contains 465 electrons. }
\label{tab:LiNi6In6}
\setlength{\tabcolsep}{4mm}{
%
}
\end{table}

\begin{figure}[htbp]
\centering
\includegraphics[width=0.85\textwidth]{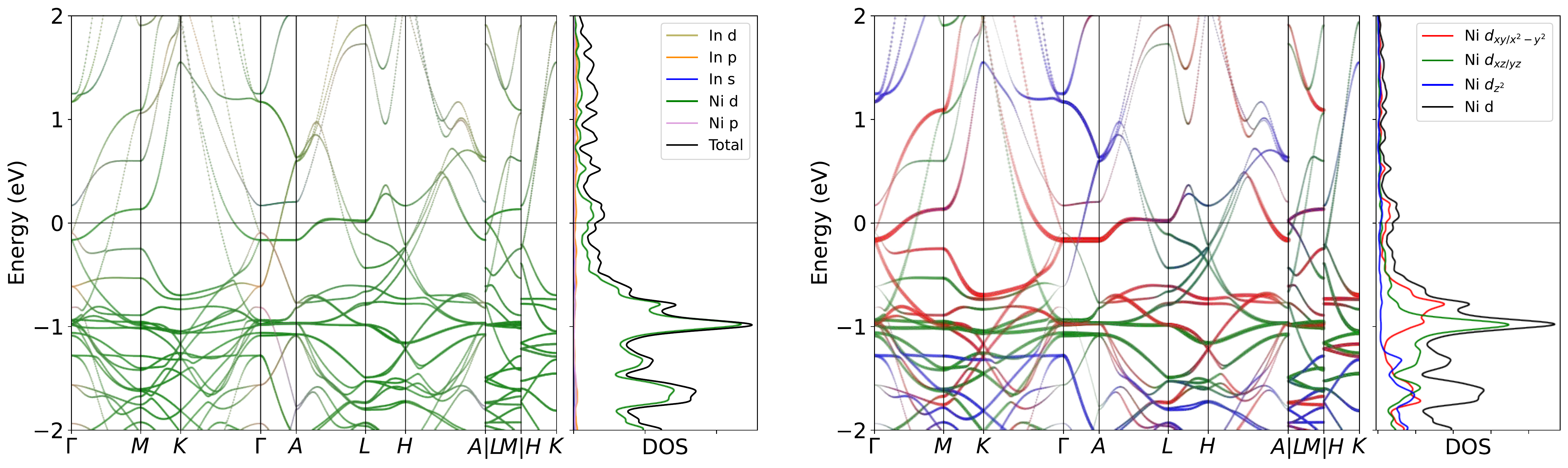}
\caption{Band structure for LiNi$_6$In$_6$ without SOC. The projection of Ni $3d$ orbitals is also presented. The band structures clearly reveal the dominating of Ni $3d$ orbitals  near the Fermi level, as well as the pronounced peak.}
\label{fig:LiNi6In6band}
\end{figure}

\begin{figure}[H]
\centering
\subfloat[Relaxed phonon at low-T.]{
\label{fig:LiNi6In6_relaxed_Low}
\includegraphics[width=0.45\textwidth]{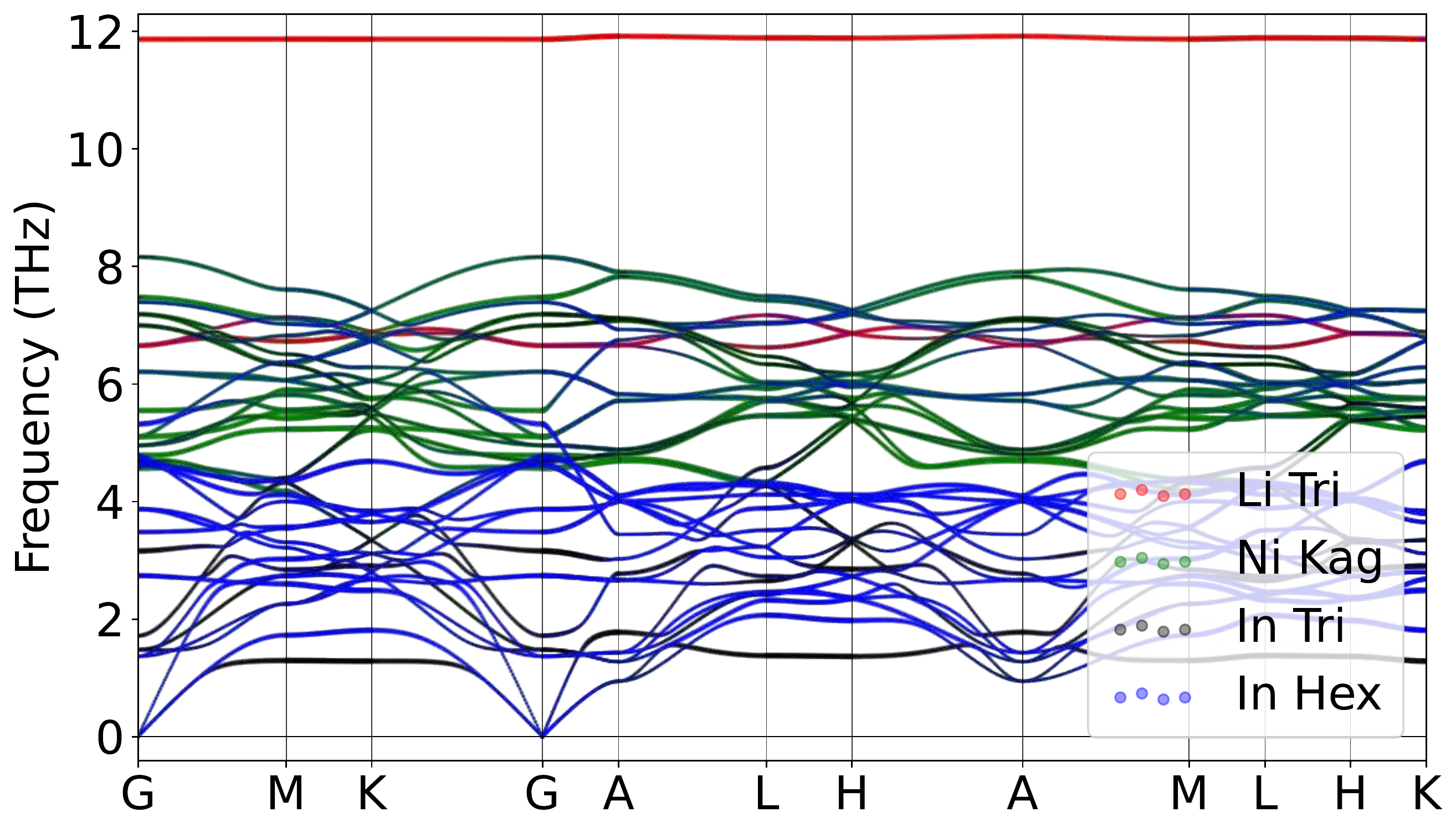}
}
\hspace{5pt}\subfloat[Relaxed phonon at high-T.]{
\label{fig:LiNi6In6_relaxed_High}
\includegraphics[width=0.45\textwidth]{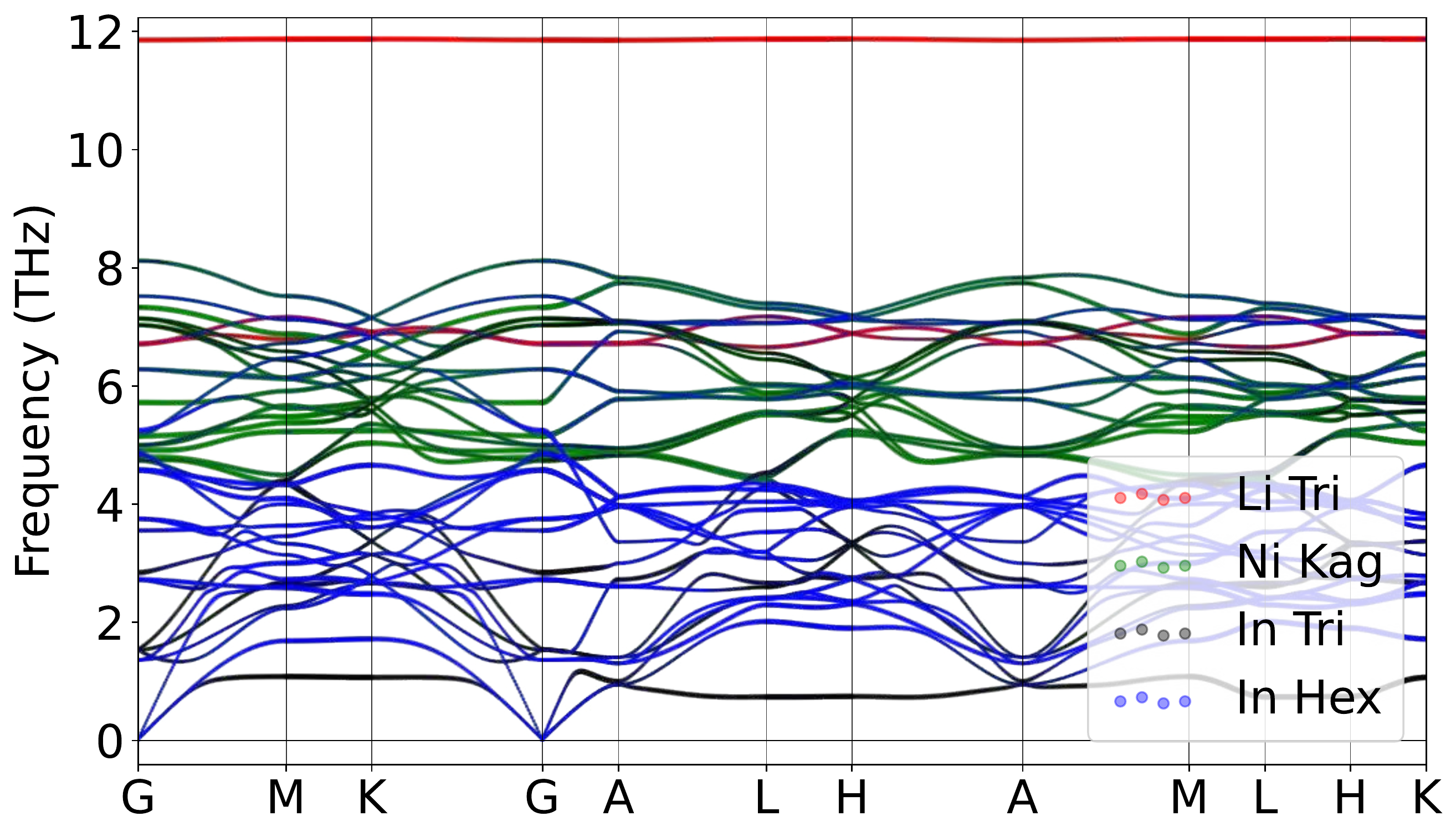}
}
\caption{Atomically projected phonon spectrum for LiNi$_6$In$_6$.}
\label{fig:LiNi6In6pho}
\end{figure}

\begin{figure}[H]
\centering
\label{fig:LiNi6In6_relaxed_Low_xz}
\includegraphics[width=0.85\textwidth]{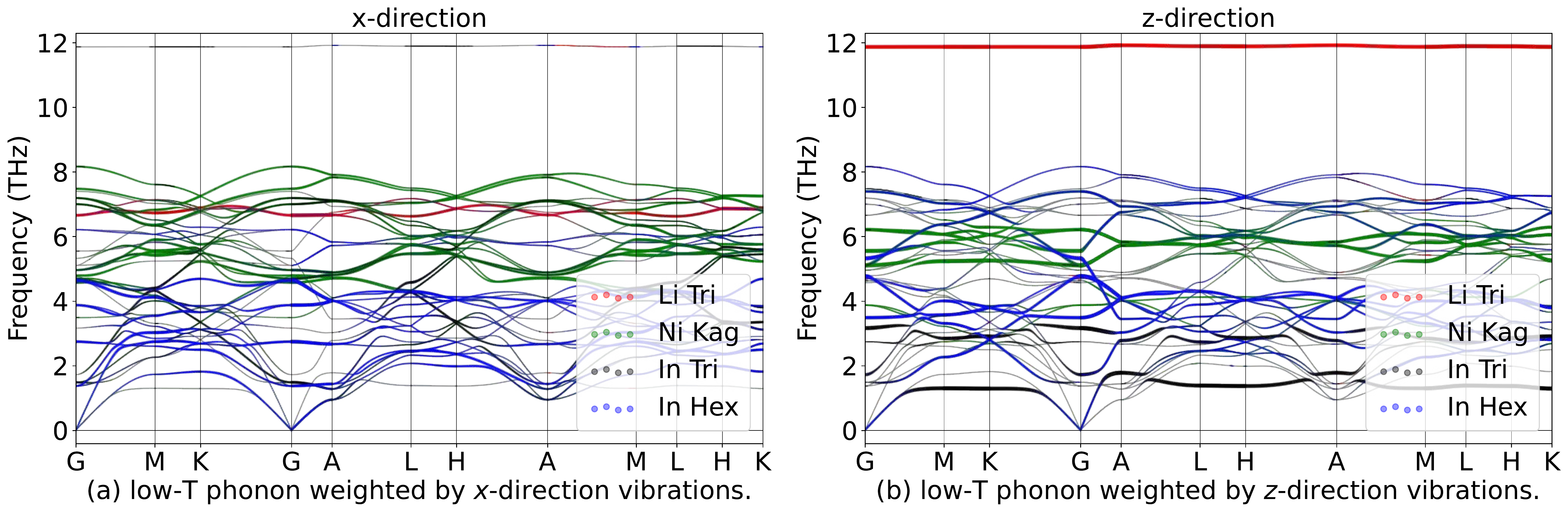}
\hspace{5pt}\label{fig:LiNi6In6_relaxed_High_xz}
\includegraphics[width=0.85\textwidth]{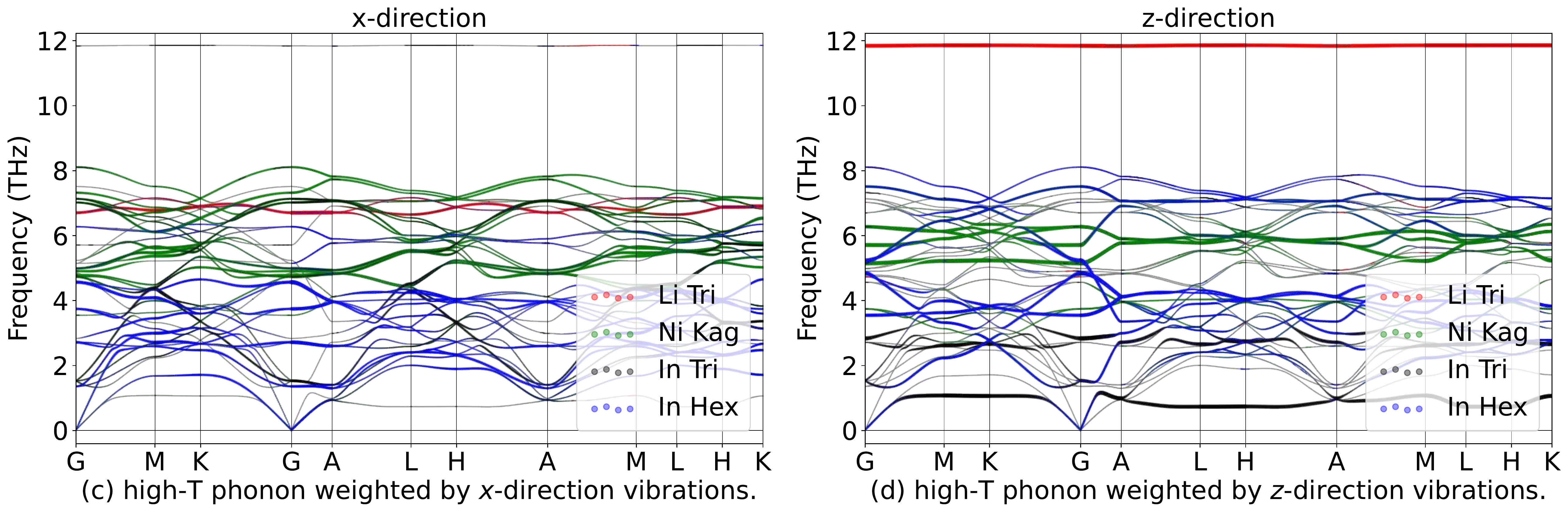}
\caption{Phonon spectrum for LiNi$_6$In$_6$in in-plane ($x$) and out-of-plane ($z$) directions.}
\label{fig:LiNi6In6phoxyz}
\end{figure}

\begin{table}[htbp]
\global\long\def\arraystretch{1.12}
\captionsetup{width=.95\textwidth}
\caption{IRREPs at high-symmetry points for LiNi$_6$In$_6$ phonon spectrum at Low-T.}
\label{tab:191_LiNi6In6_Low}
\setlength{\tabcolsep}{1.mm}{
}
\end{table}

\begin{table}[htbp]
\global\long\def\arraystretch{1.12}
\captionsetup{width=.95\textwidth}
\caption{IRREPs at high-symmetry points for LiNi$_6$In$_6$ phonon spectrum at High-T.}
\label{tab:191_LiNi6In6_High}
\setlength{\tabcolsep}{1.mm}{
%
}
\end{table}
\subsubsection{\label{sms2sec:MgNi6In6}\hyperref[smsec:col]{\texorpdfstring{MgNi$_6$In$_6$}{}}~{\color{red}  (High-T unstable, Low-T unstable) }}

\begin{table}[htbp]
\global\long\def\arraystretch{1.12}
\captionsetup{width=.85\textwidth}
\caption{Structure parameters of MgNi$_6$In$_6$ after relaxation. It contains 474 electrons. }
\label{tab:MgNi6In6}
\setlength{\tabcolsep}{4mm}{
%
}
\end{table}

\begin{figure}[htbp]
\centering
\includegraphics[width=0.85\textwidth]{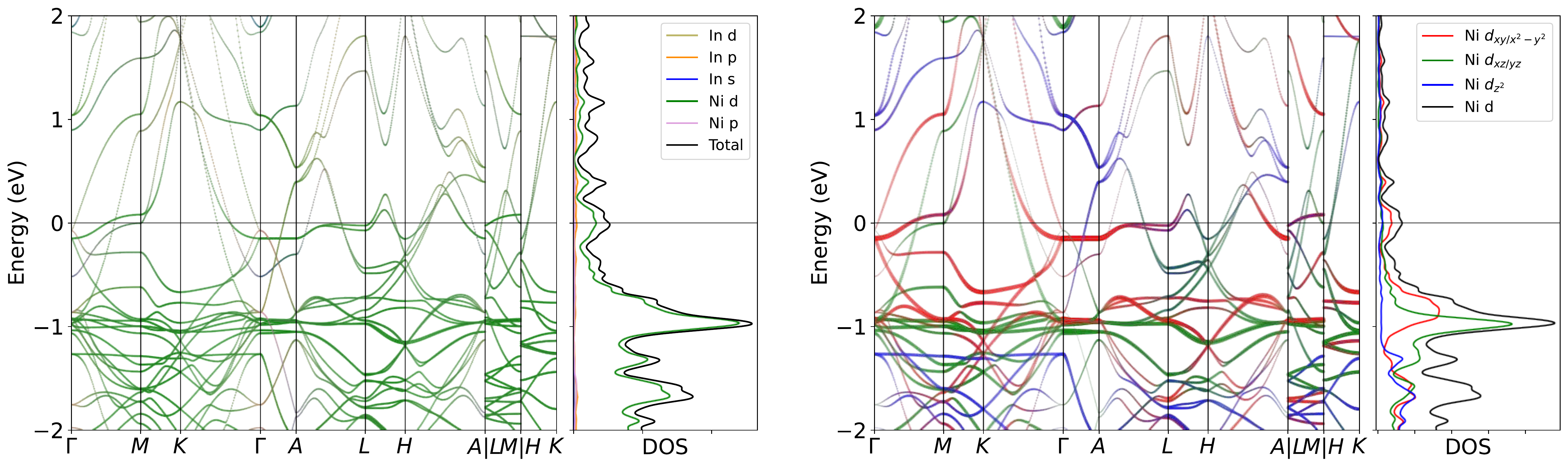}
\caption{Band structure for MgNi$_6$In$_6$ without SOC. The projection of Ni $3d$ orbitals is also presented. The band structures clearly reveal the dominating of Ni $3d$ orbitals  near the Fermi level, as well as the pronounced peak.}
\label{fig:MgNi6In6band}
\end{figure}

\begin{figure}[H]
\centering
\subfloat[Relaxed phonon at low-T.]{
\label{fig:MgNi6In6_relaxed_Low}
\includegraphics[width=0.45\textwidth]{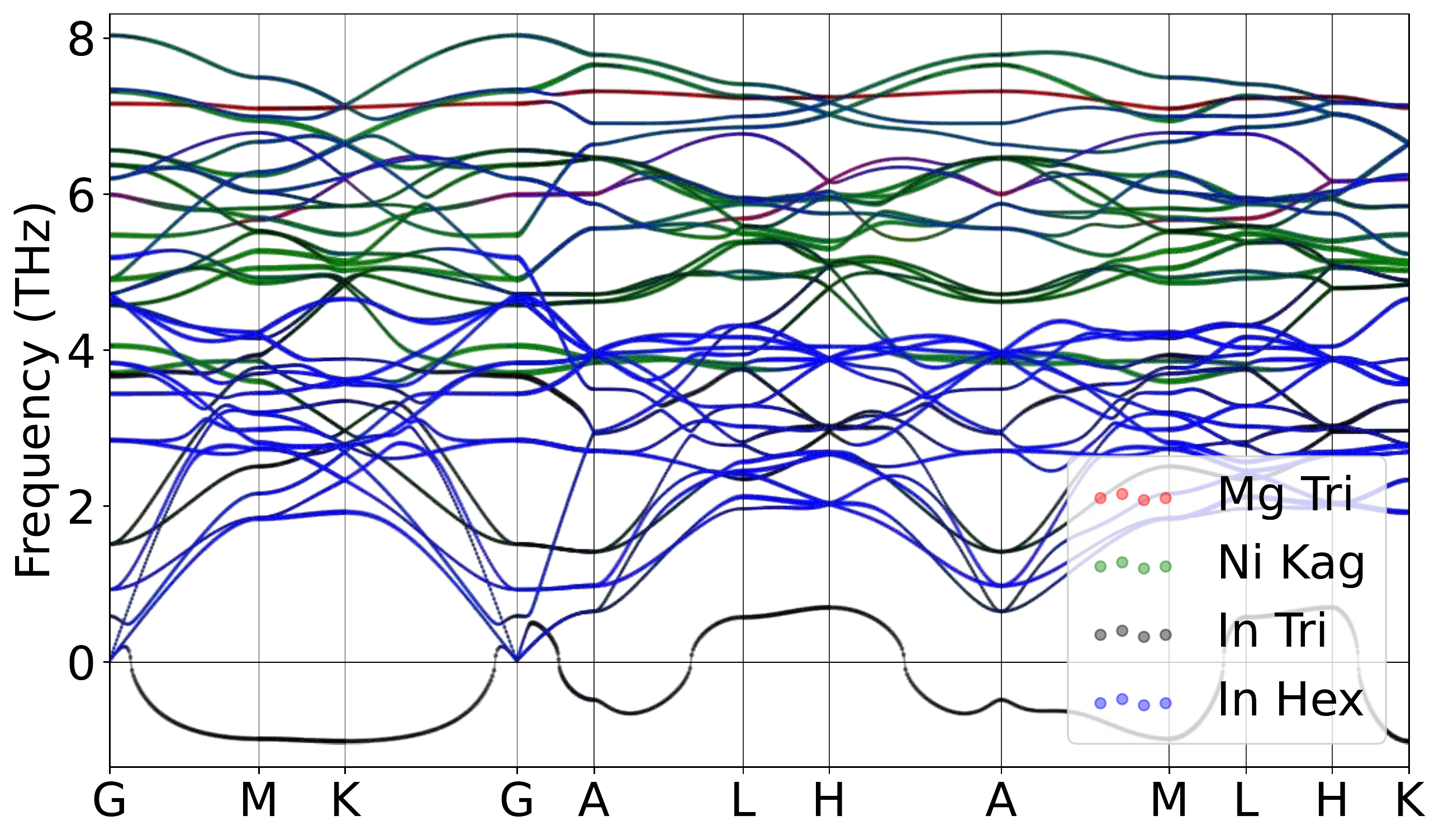}
}
\hspace{5pt}\subfloat[Relaxed phonon at high-T.]{
\label{fig:MgNi6In6_relaxed_High}
\includegraphics[width=0.45\textwidth]{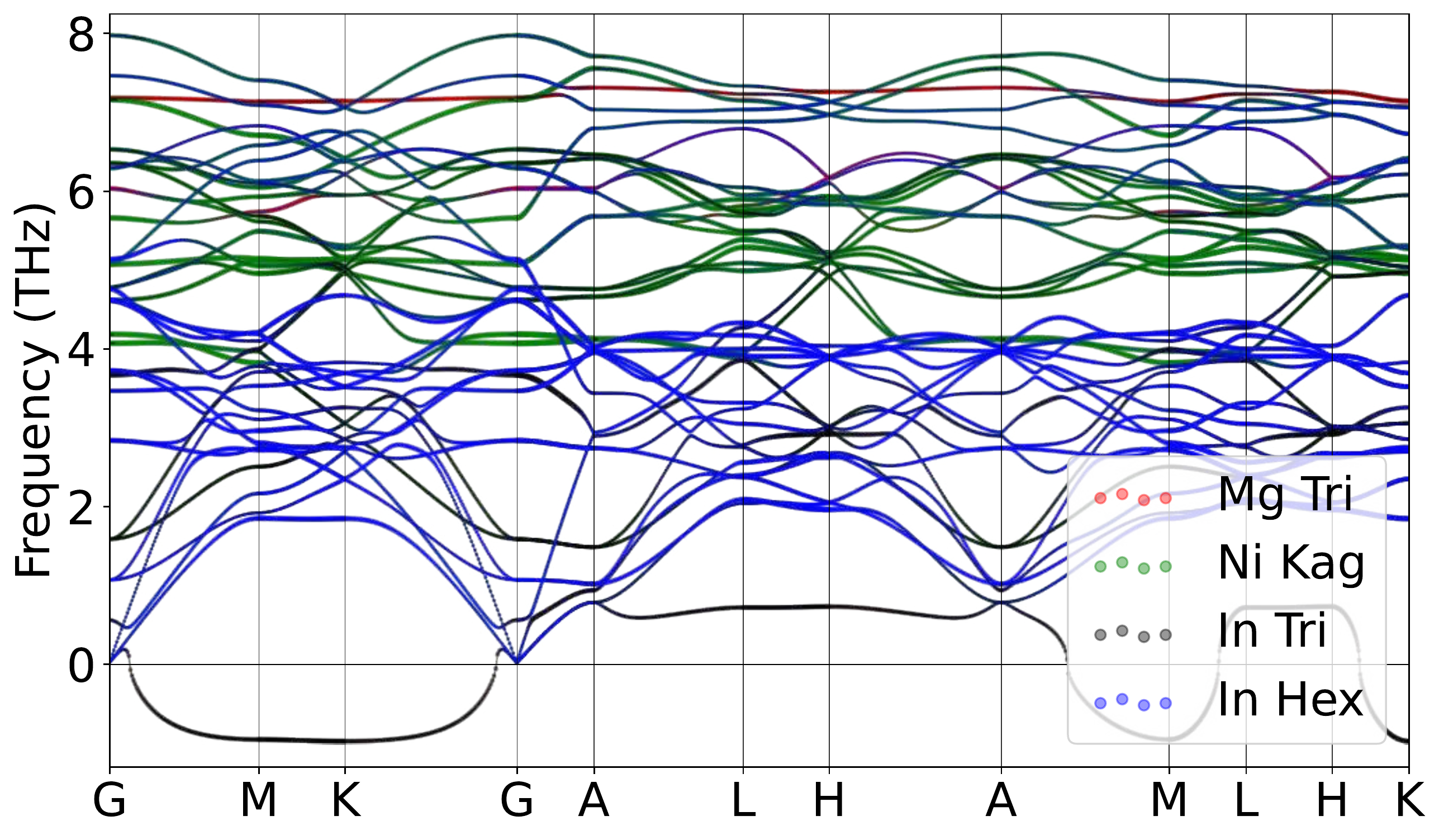}
}
\caption{Atomically projected phonon spectrum for MgNi$_6$In$_6$.}
\label{fig:MgNi6In6pho}
\end{figure}

\begin{figure}[H]
\centering
\label{fig:MgNi6In6_relaxed_Low_xz}
\includegraphics[width=0.85\textwidth]{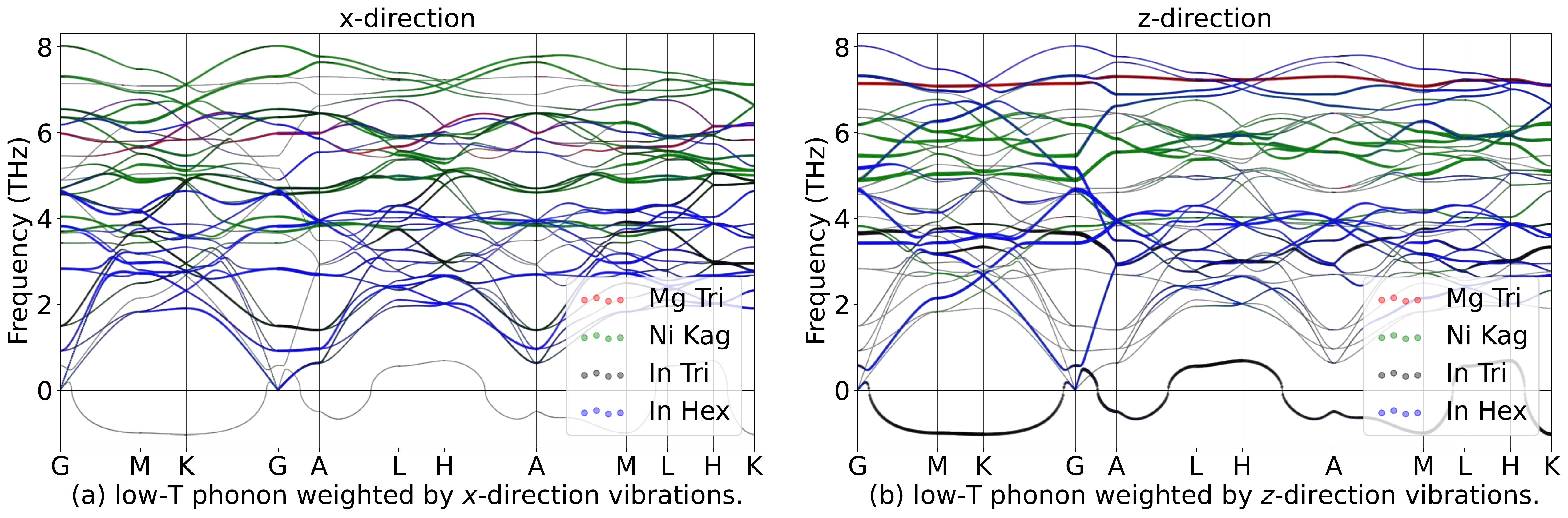}
\hspace{5pt}\label{fig:MgNi6In6_relaxed_High_xz}
\includegraphics[width=0.85\textwidth]{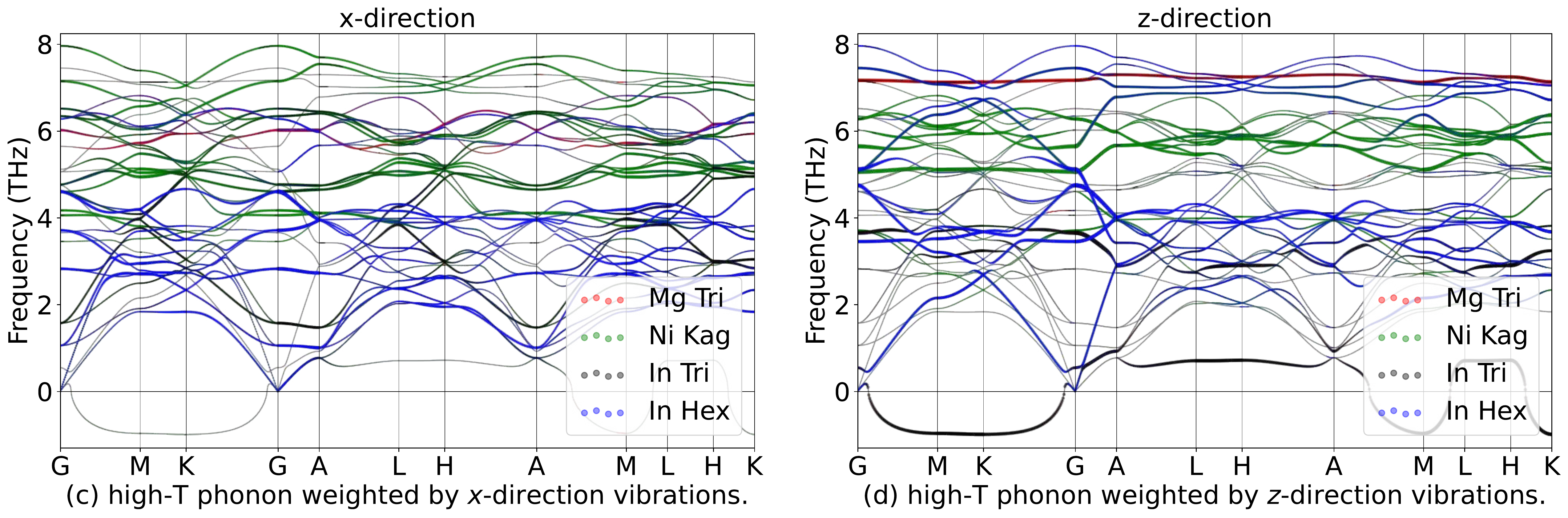}
\caption{Phonon spectrum for MgNi$_6$In$_6$in in-plane ($x$) and out-of-plane ($z$) directions.}
\label{fig:MgNi6In6phoxyz}
\end{figure}

\begin{table}[htbp]
\global\long\def\arraystretch{1.12}
\captionsetup{width=.95\textwidth}
\caption{IRREPs at high-symmetry points for MgNi$_6$In$_6$ phonon spectrum at Low-T.}
\label{tab:191_MgNi6In6_Low}
\setlength{\tabcolsep}{1.mm}{
}
\end{table}

\begin{table}[htbp]
\global\long\def\arraystretch{1.12}
\captionsetup{width=.95\textwidth}
\caption{IRREPs at high-symmetry points for MgNi$_6$In$_6$ phonon spectrum at High-T.}
\label{tab:191_MgNi6In6_High}
\setlength{\tabcolsep}{1.mm}{
%
}
\end{table}
\subsubsection{\label{sms2sec:CaNi6In6}\hyperref[smsec:col]{\texorpdfstring{CaNi$_6$In$_6$}{}}~{\color{green}  (High-T stable, Low-T stable) }}

\begin{table}[htbp]
\global\long\def\arraystretch{1.12}
\captionsetup{width=.85\textwidth}
\caption{Structure parameters of CaNi$_6$In$_6$ after relaxation. It contains 482 electrons. }
\label{tab:CaNi6In6}
\setlength{\tabcolsep}{4mm}{
%
}
\end{table}

\begin{figure}[htbp]
\centering
\includegraphics[width=0.85\textwidth]{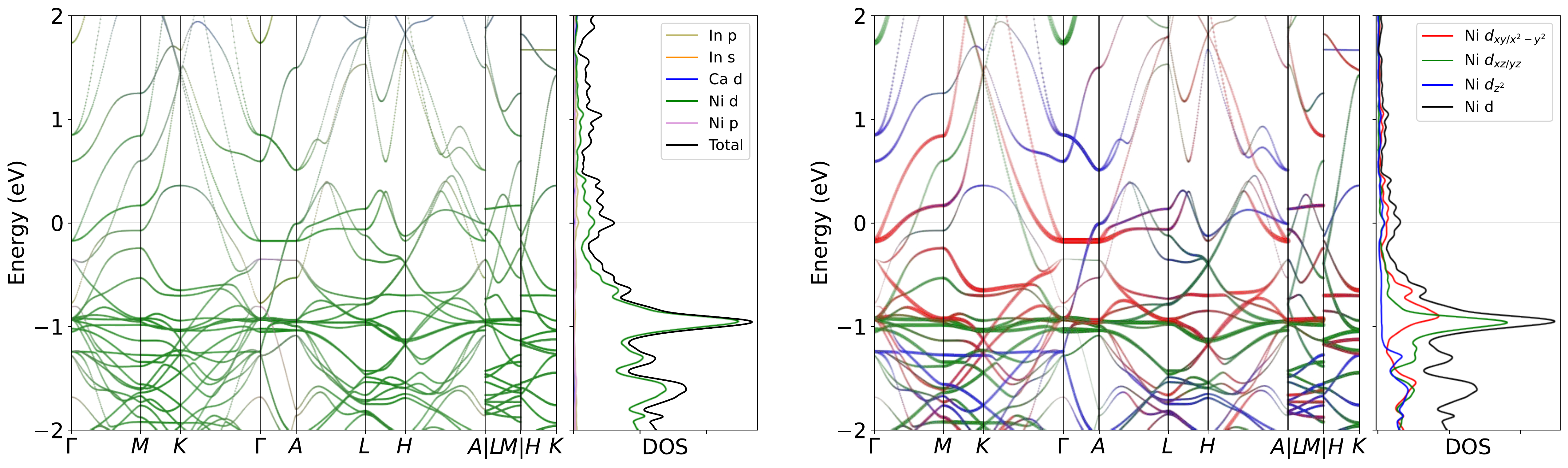}
\caption{Band structure for CaNi$_6$In$_6$ without SOC. The projection of Ni $3d$ orbitals is also presented. The band structures clearly reveal the dominating of Ni $3d$ orbitals  near the Fermi level, as well as the pronounced peak.}
\label{fig:CaNi6In6band}
\end{figure}

\begin{figure}[H]
\centering
\subfloat[Relaxed phonon at low-T.]{
\label{fig:CaNi6In6_relaxed_Low}
\includegraphics[width=0.45\textwidth]{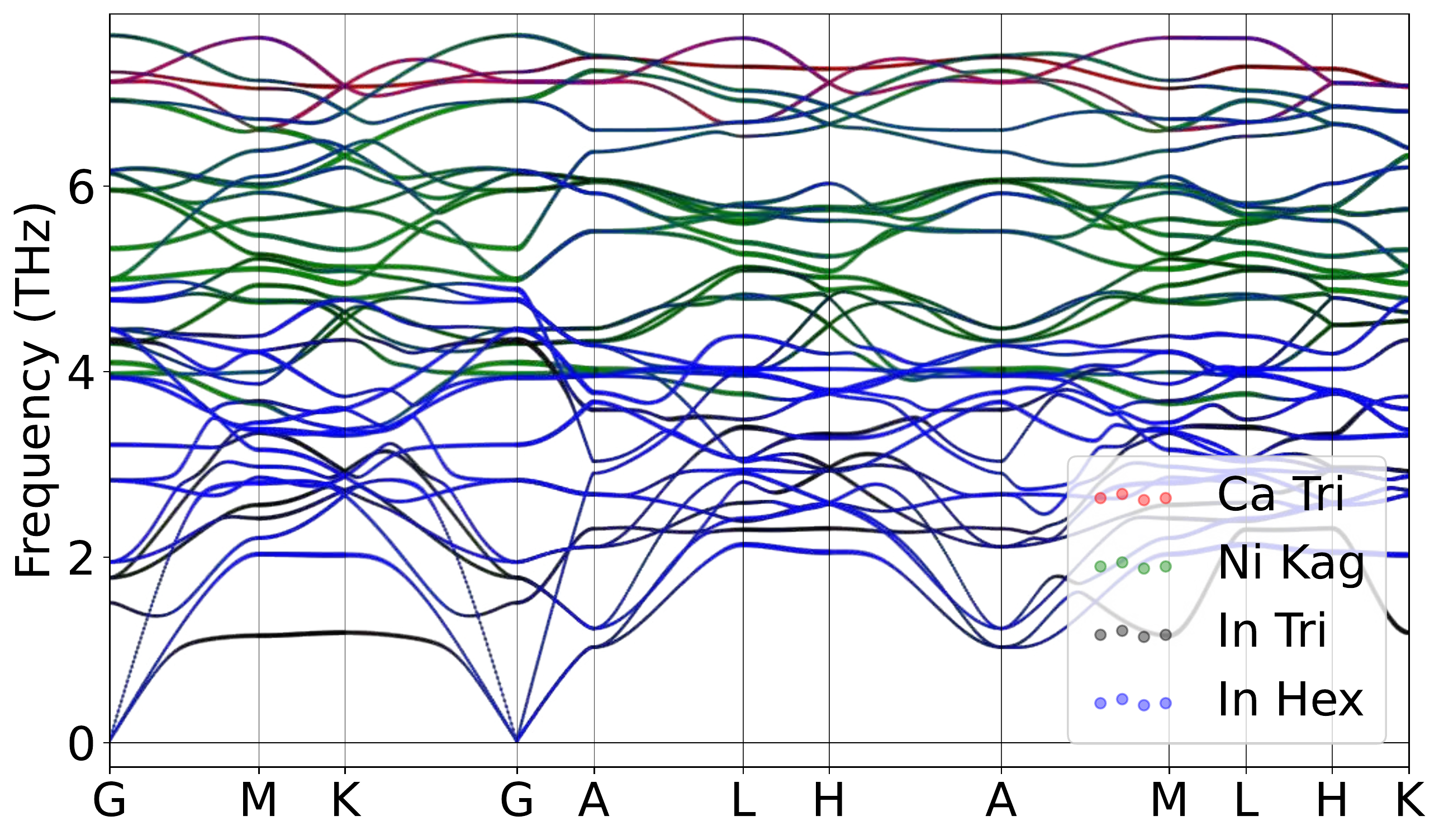}
}
\hspace{5pt}\subfloat[Relaxed phonon at high-T.]{
\label{fig:CaNi6In6_relaxed_High}
\includegraphics[width=0.45\textwidth]{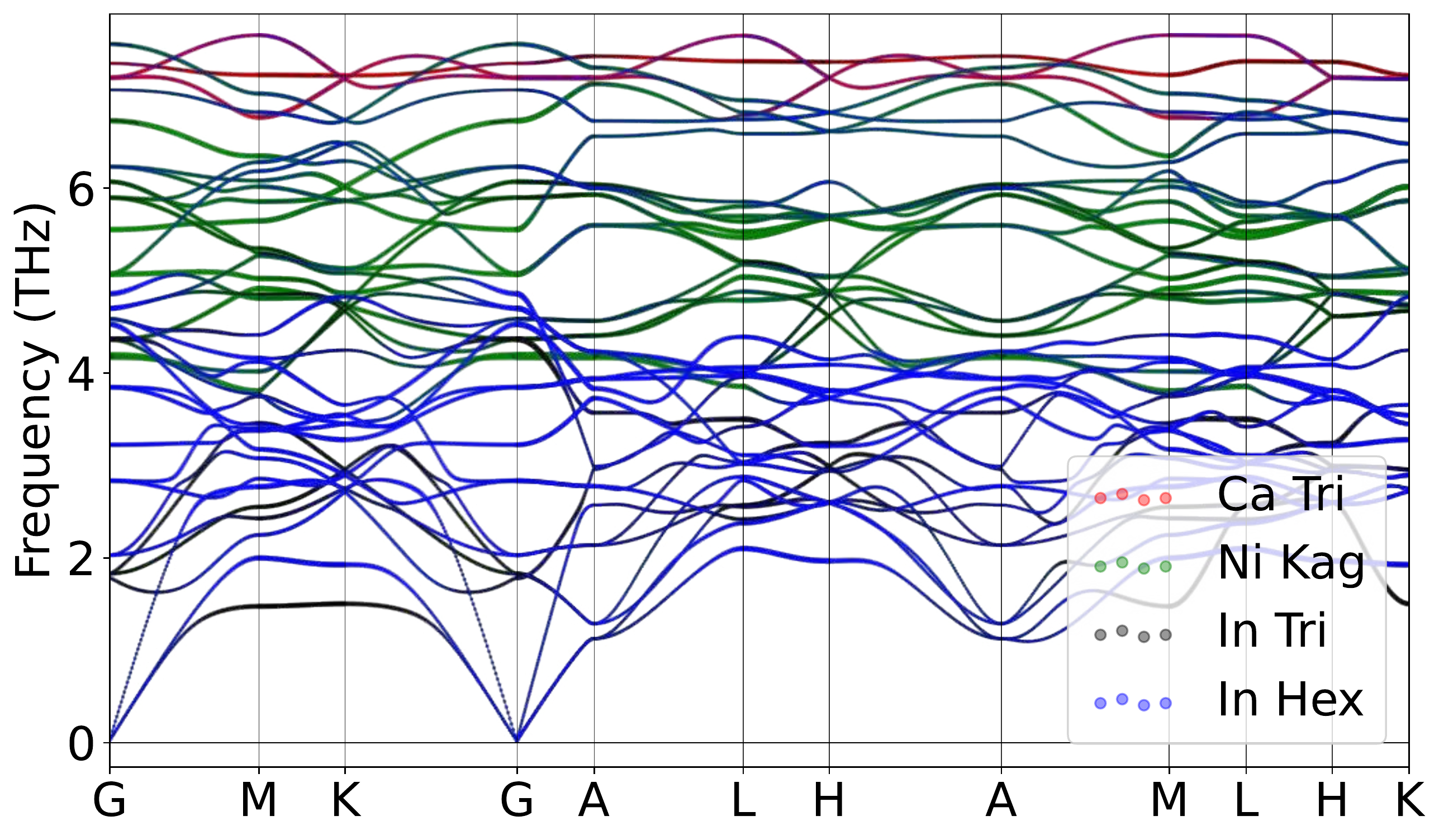}
}
\caption{Atomically projected phonon spectrum for CaNi$_6$In$_6$.}
\label{fig:CaNi6In6pho}
\end{figure}

\begin{figure}[H]
\centering
\label{fig:CaNi6In6_relaxed_Low_xz}
\includegraphics[width=0.85\textwidth]{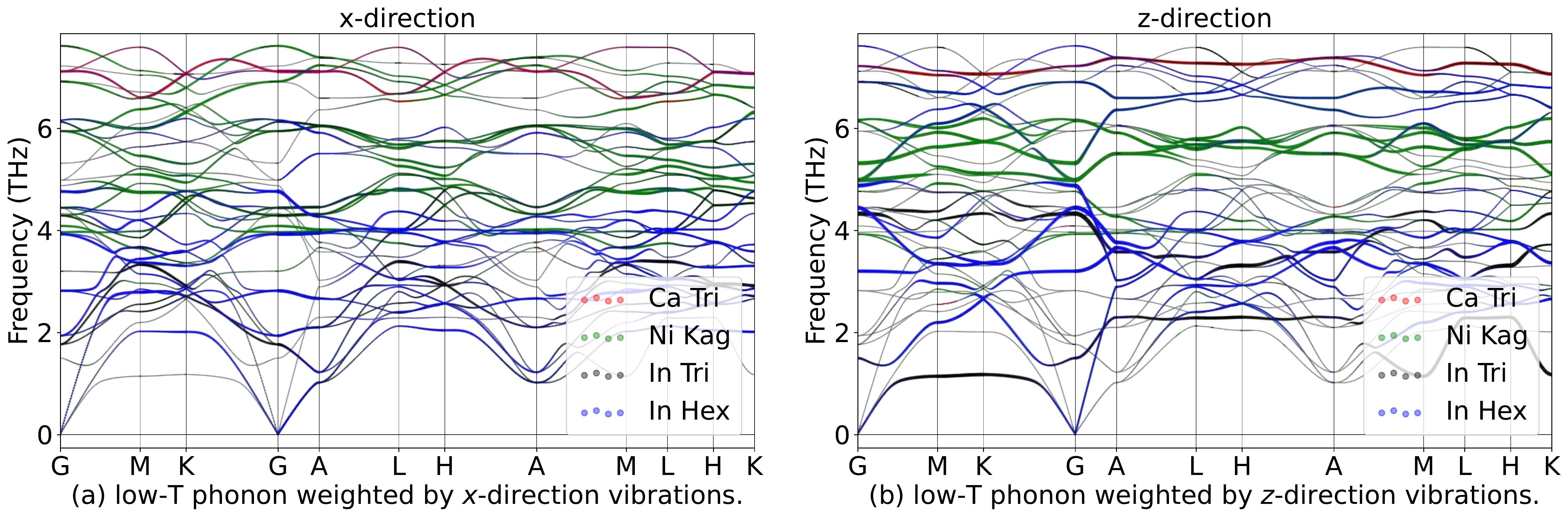}
\hspace{5pt}\label{fig:CaNi6In6_relaxed_High_xz}
\includegraphics[width=0.85\textwidth]{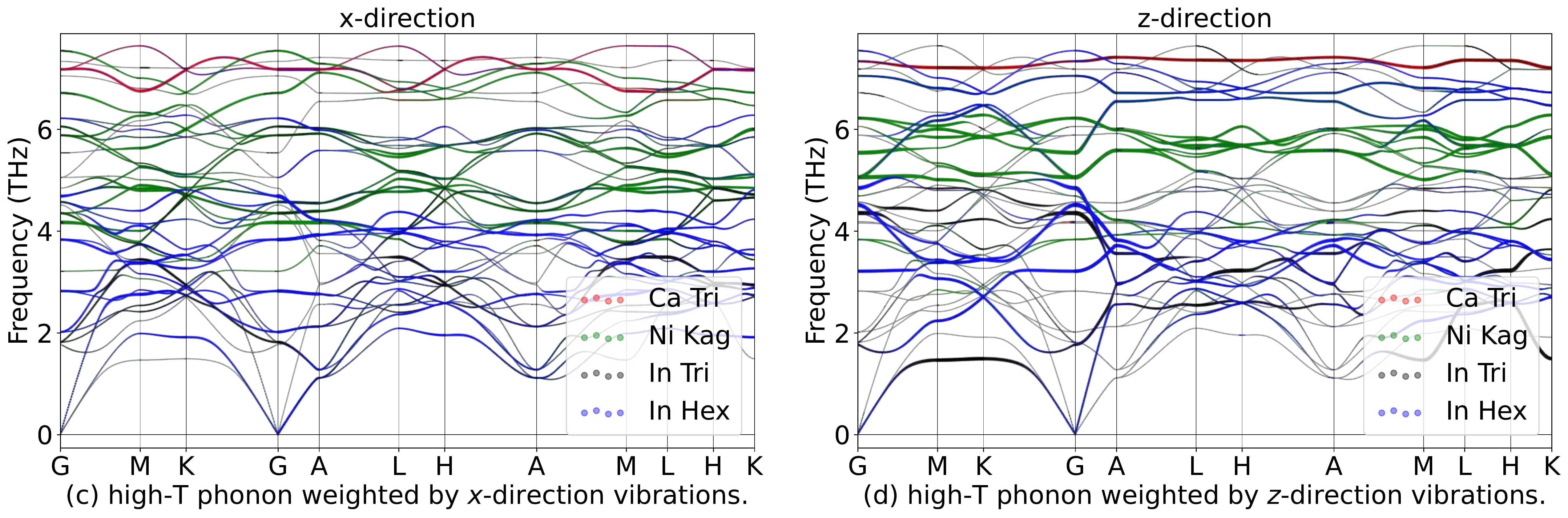}
\caption{Phonon spectrum for CaNi$_6$In$_6$in in-plane ($x$) and out-of-plane ($z$) directions.}
\label{fig:CaNi6In6phoxyz}
\end{figure}

\begin{table}[htbp]
\global\long\def\arraystretch{1.12}
\captionsetup{width=.95\textwidth}
\caption{IRREPs at high-symmetry points for CaNi$_6$In$_6$ phonon spectrum at Low-T.}
\label{tab:191_CaNi6In6_Low}
\setlength{\tabcolsep}{1.mm}{
}
\end{table}

\begin{table}[htbp]
\global\long\def\arraystretch{1.12}
\captionsetup{width=.95\textwidth}
\caption{IRREPs at high-symmetry points for CaNi$_6$In$_6$ phonon spectrum at High-T.}
\label{tab:191_CaNi6In6_High}
\setlength{\tabcolsep}{1.mm}{
%
}
\end{table}
\subsubsection{\label{sms2sec:ScNi6In6}\hyperref[smsec:col]{\texorpdfstring{ScNi$_6$In$_6$}{}}~{\color{green}  (High-T stable, Low-T stable) }}

\begin{table}[htbp]
\global\long\def\arraystretch{1.12}
\captionsetup{width=.85\textwidth}
\caption{Structure parameters of ScNi$_6$In$_6$ after relaxation. It contains 483 electrons. }
\label{tab:ScNi6In6}
\setlength{\tabcolsep}{4mm}{
%
}
\end{table}

\begin{figure}[htbp]
\centering
\includegraphics[width=0.85\textwidth]{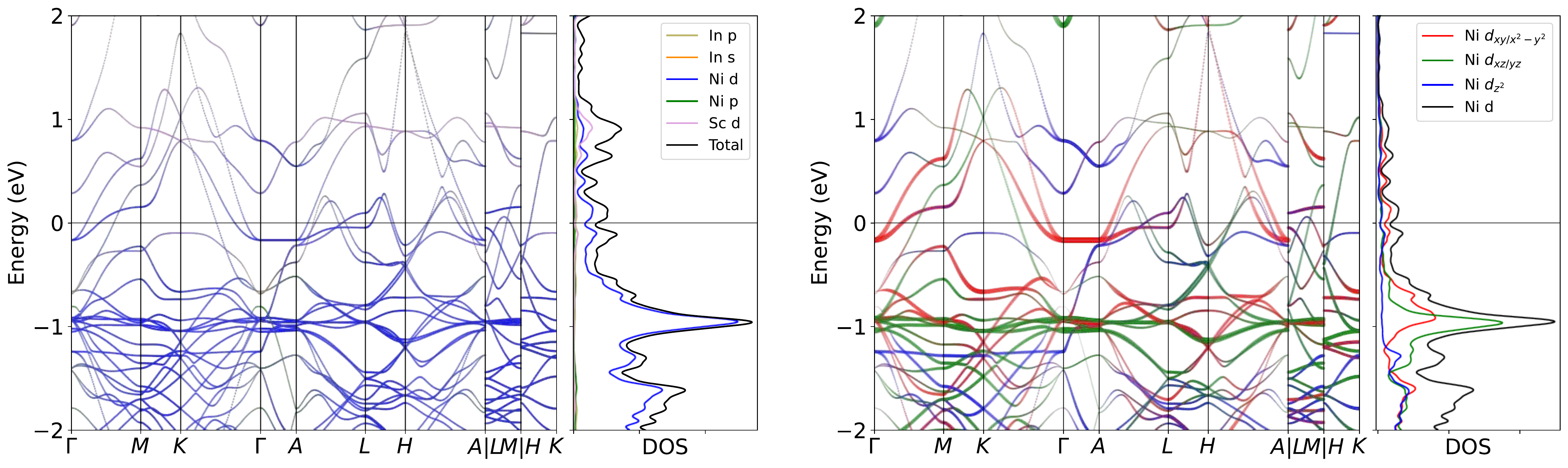}
\caption{Band structure for ScNi$_6$In$_6$ without SOC. The projection of Ni $3d$ orbitals is also presented. The band structures clearly reveal the dominating of Ni $3d$ orbitals  near the Fermi level, as well as the pronounced peak.}
\label{fig:ScNi6In6band}
\end{figure}

\begin{figure}[H]
\centering
\subfloat[Relaxed phonon at low-T.]{
\label{fig:ScNi6In6_relaxed_Low}
\includegraphics[width=0.45\textwidth]{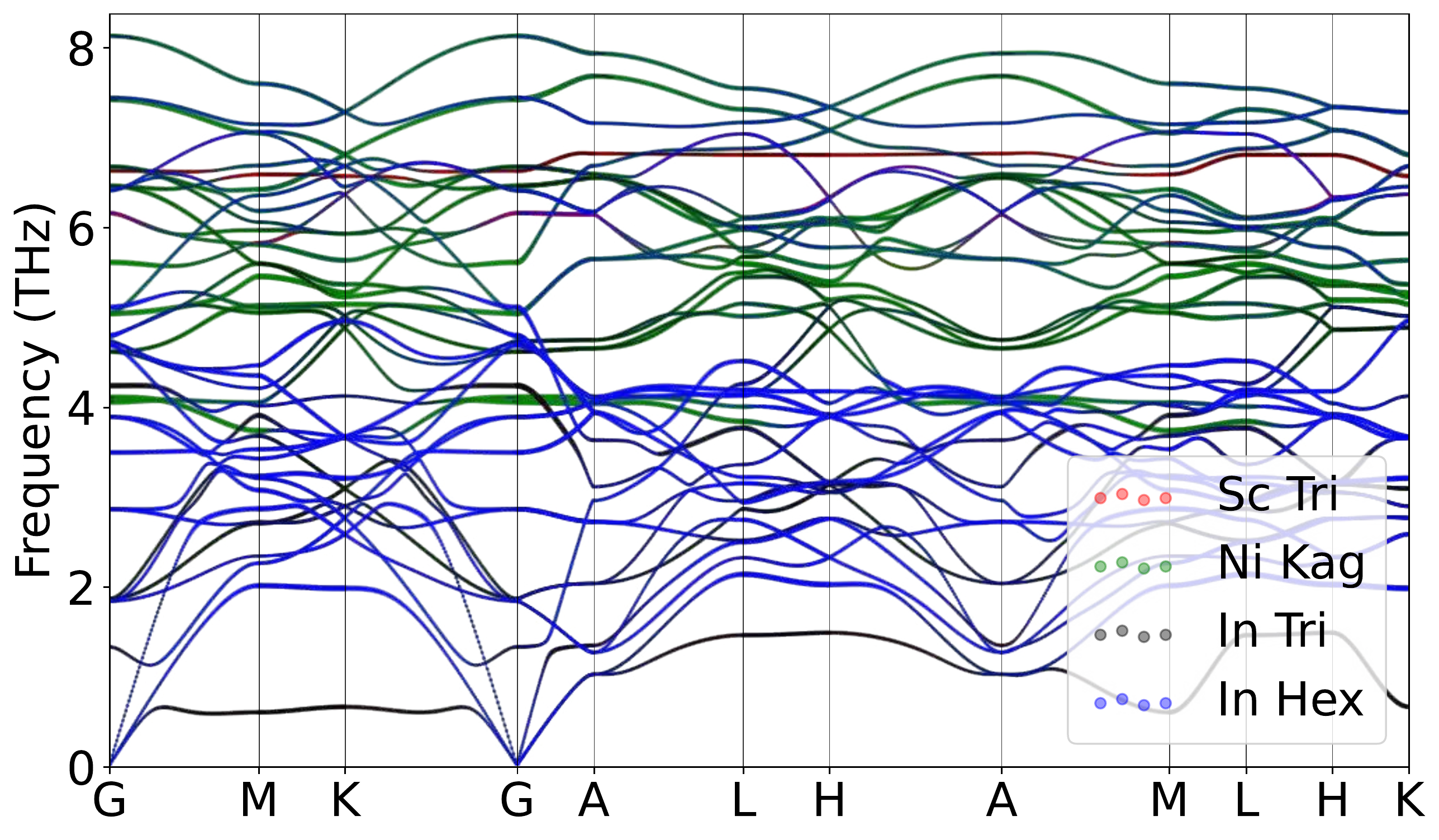}
}
\hspace{5pt}\subfloat[Relaxed phonon at high-T.]{
\label{fig:ScNi6In6_relaxed_High}
\includegraphics[width=0.45\textwidth]{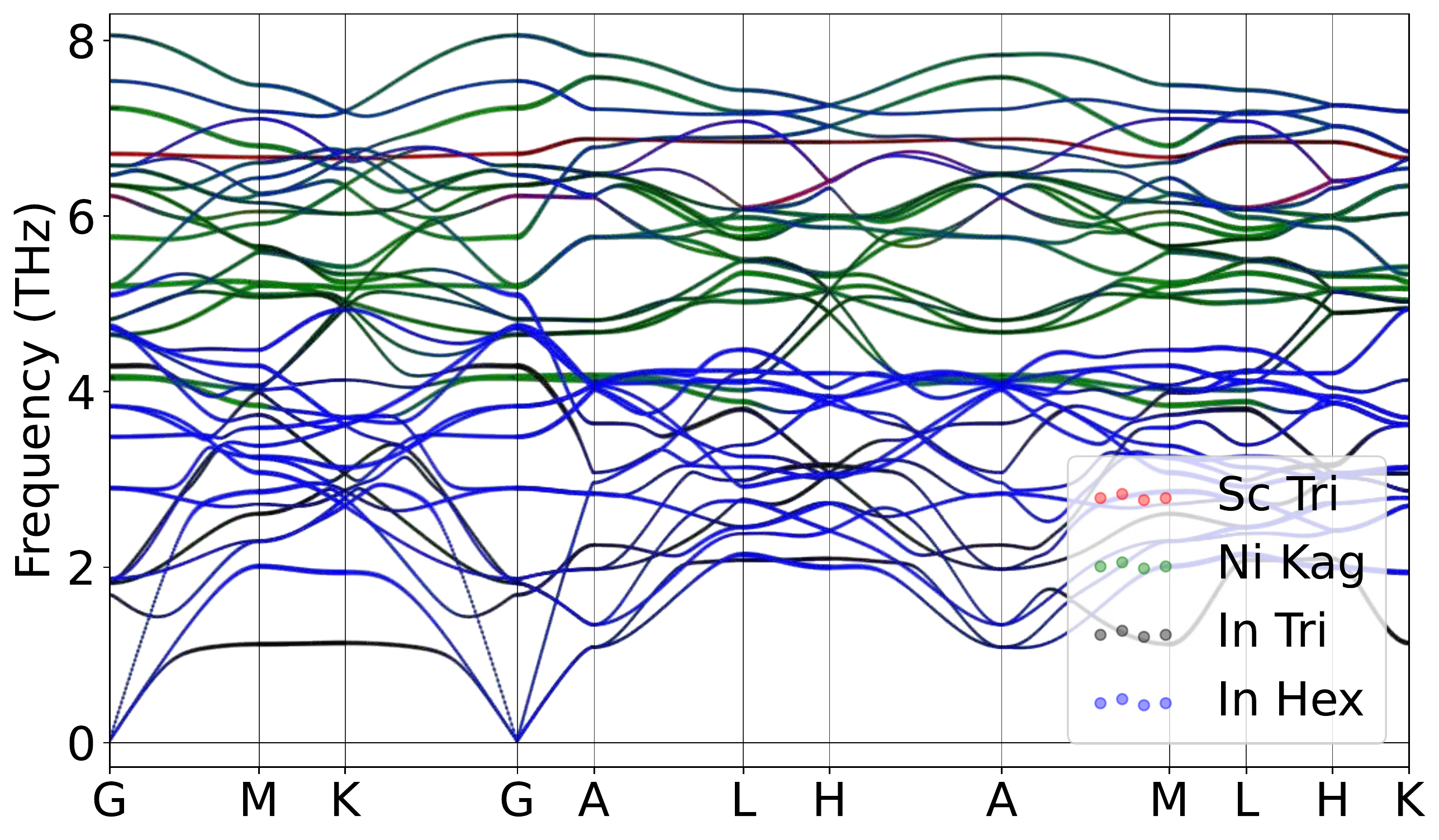}
}
\caption{Atomically projected phonon spectrum for ScNi$_6$In$_6$.}
\label{fig:ScNi6In6pho}
\end{figure}

\begin{figure}[H]
\centering
\label{fig:ScNi6In6_relaxed_Low_xz}
\includegraphics[width=0.85\textwidth]{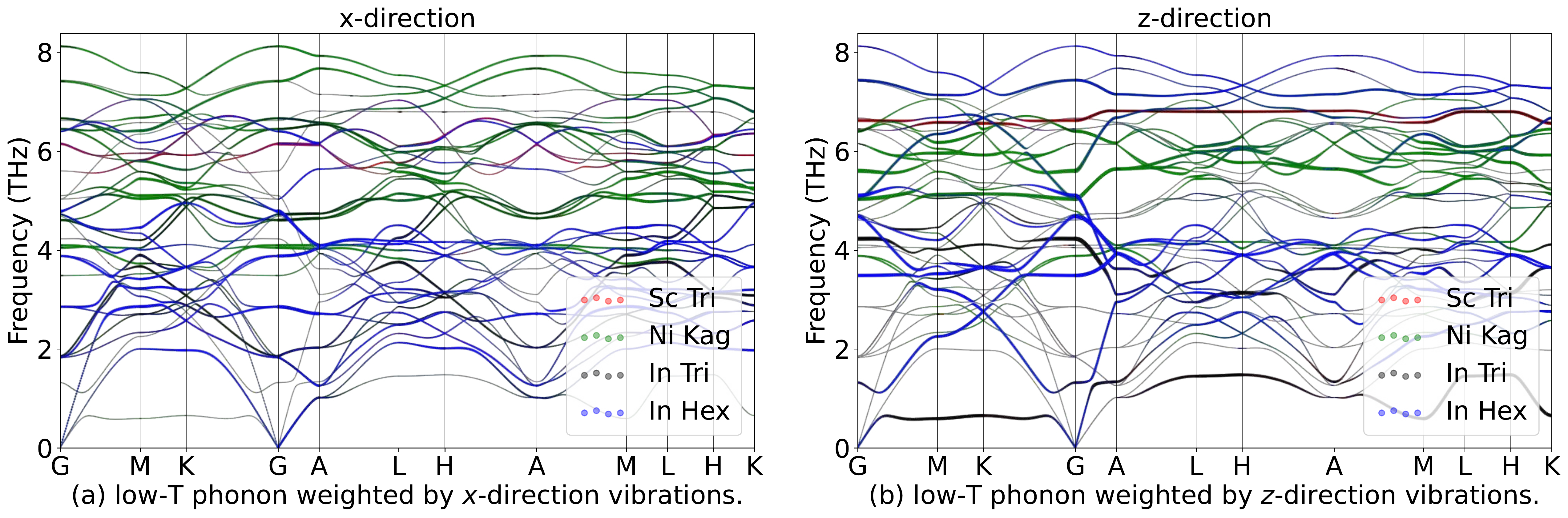}
\hspace{5pt}\label{fig:ScNi6In6_relaxed_High_xz}
\includegraphics[width=0.85\textwidth]{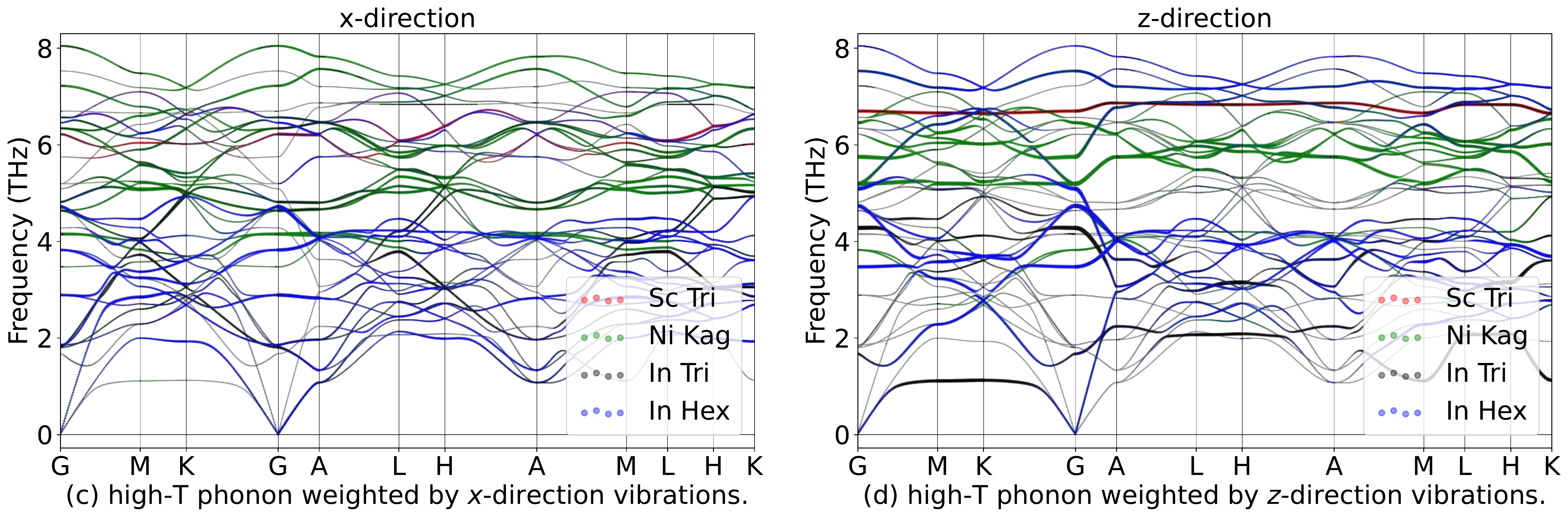}
\caption{Phonon spectrum for ScNi$_6$In$_6$in in-plane ($x$) and out-of-plane ($z$) directions.}
\label{fig:ScNi6In6phoxyz}
\end{figure}

\begin{table}[htbp]
\global\long\def\arraystretch{1.12}
\captionsetup{width=.95\textwidth}
\caption{IRREPs at high-symmetry points for ScNi$_6$In$_6$ phonon spectrum at Low-T.}
\label{tab:191_ScNi6In6_Low}
\setlength{\tabcolsep}{1.mm}{
}
\end{table}

\begin{table}[htbp]
\global\long\def\arraystretch{1.12}
\captionsetup{width=.95\textwidth}
\caption{IRREPs at high-symmetry points for ScNi$_6$In$_6$ phonon spectrum at High-T.}
\label{tab:191_ScNi6In6_High}
\setlength{\tabcolsep}{1.mm}{
%
}
\end{table}
\subsubsection{\label{sms2sec:TiNi6In6}\hyperref[smsec:col]{\texorpdfstring{TiNi$_6$In$_6$}{}}~{\color{blue}  (High-T stable, Low-T unstable) }}

\begin{table}[htbp]
\global\long\def\arraystretch{1.12}
\captionsetup{width=.85\textwidth}
\caption{Structure parameters of TiNi$_6$In$_6$ after relaxation. It contains 484 electrons. }
\label{tab:TiNi6In6}
\setlength{\tabcolsep}{4mm}{
%
}
\end{table}

\begin{figure}[htbp]
\centering
\includegraphics[width=0.85\textwidth]{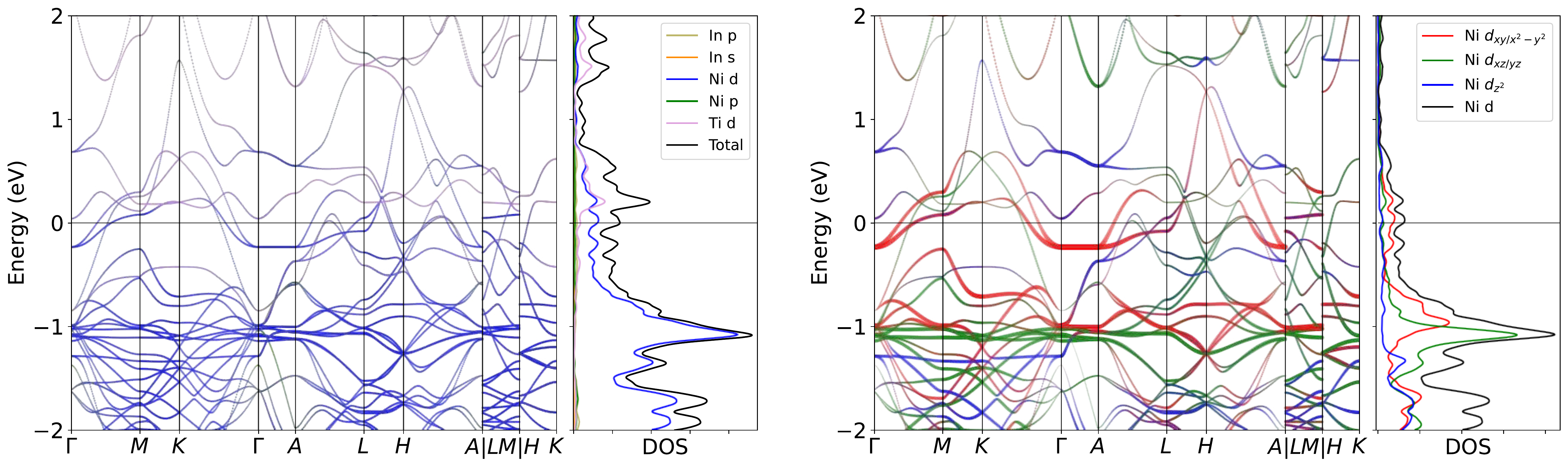}
\caption{Band structure for TiNi$_6$In$_6$ without SOC. The projection of Ni $3d$ orbitals is also presented. The band structures clearly reveal the dominating of Ni $3d$ orbitals  near the Fermi level, as well as the pronounced peak.}
\label{fig:TiNi6In6band}
\end{figure}

\begin{figure}[H]
\centering
\subfloat[Relaxed phonon at low-T.]{
\label{fig:TiNi6In6_relaxed_Low}
\includegraphics[width=0.45\textwidth]{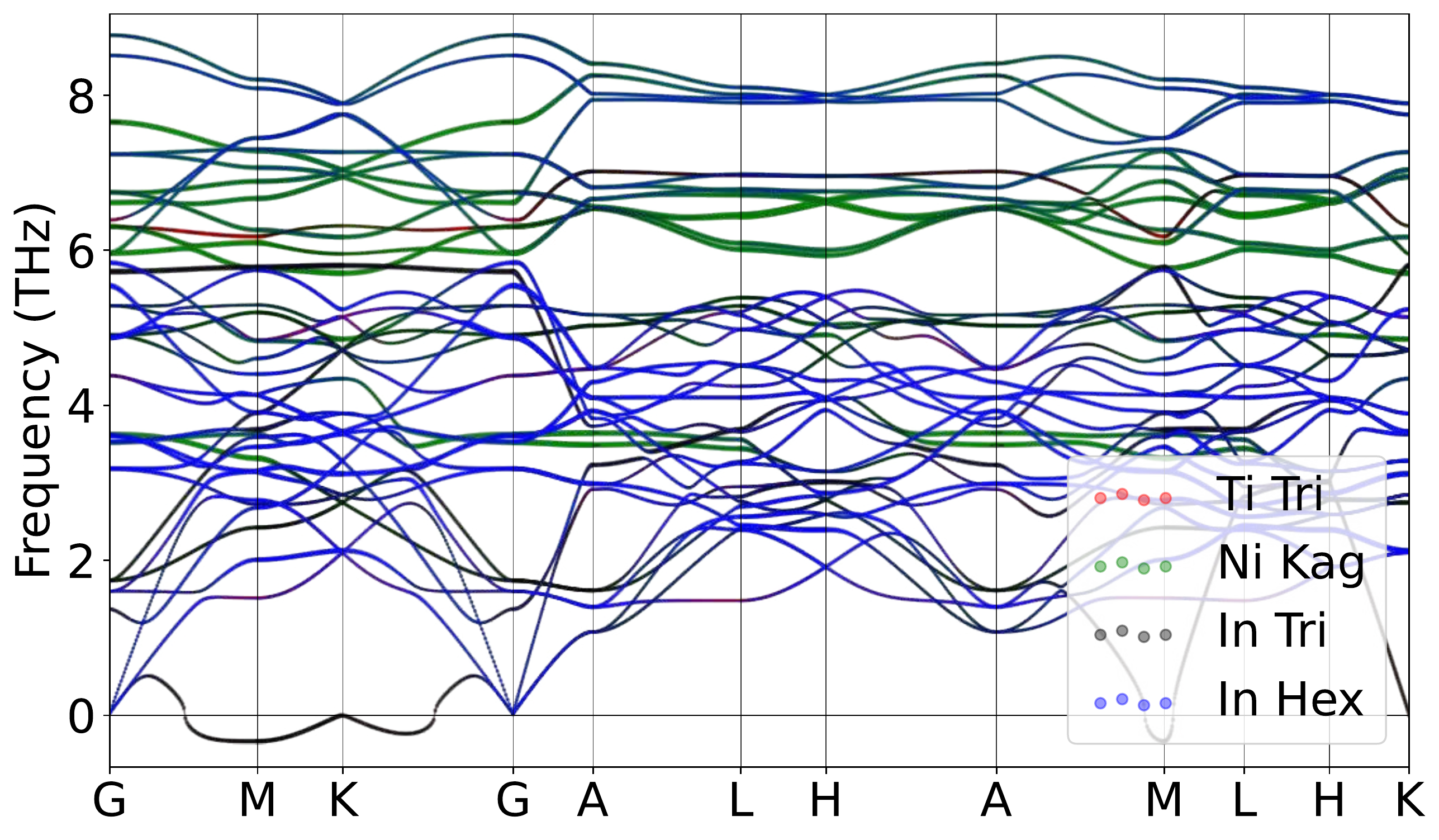}
}
\hspace{5pt}\subfloat[Relaxed phonon at high-T.]{
\label{fig:TiNi6In6_relaxed_High}
\includegraphics[width=0.45\textwidth]{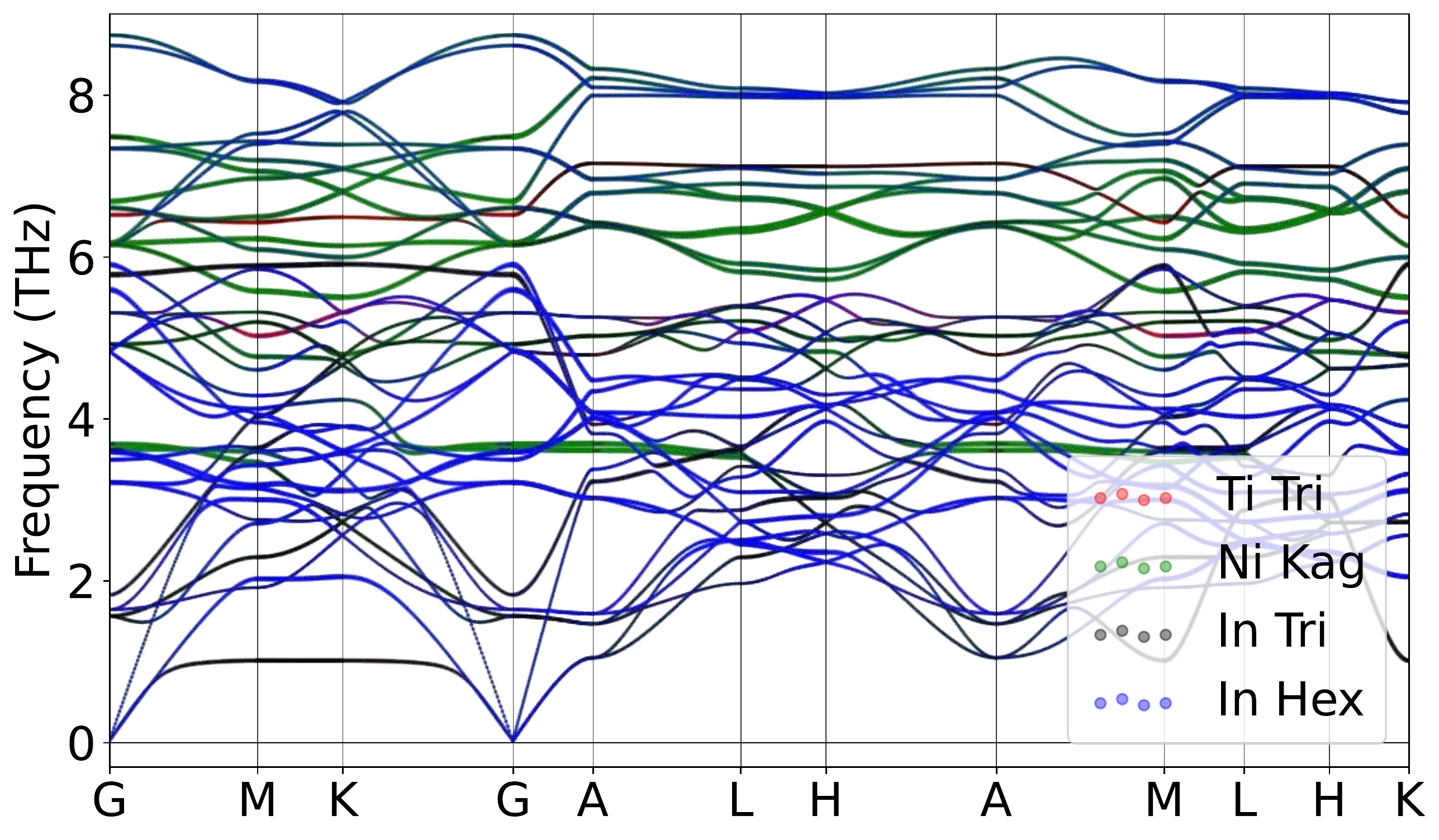}
}
\caption{Atomically projected phonon spectrum for TiNi$_6$In$_6$.}
\label{fig:TiNi6In6pho}
\end{figure}

\begin{figure}[H]
\centering
\label{fig:TiNi6In6_relaxed_Low_xz}
\includegraphics[width=0.85\textwidth]{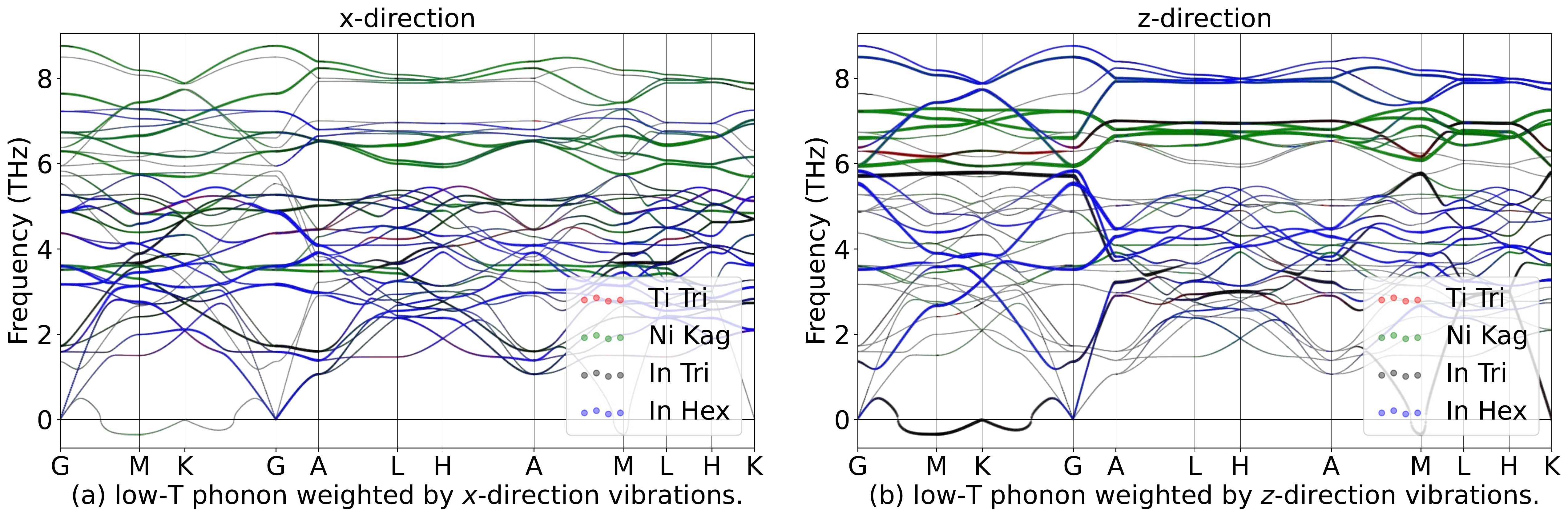}
\hspace{5pt}\label{fig:TiNi6In6_relaxed_High_xz}
\includegraphics[width=0.85\textwidth]{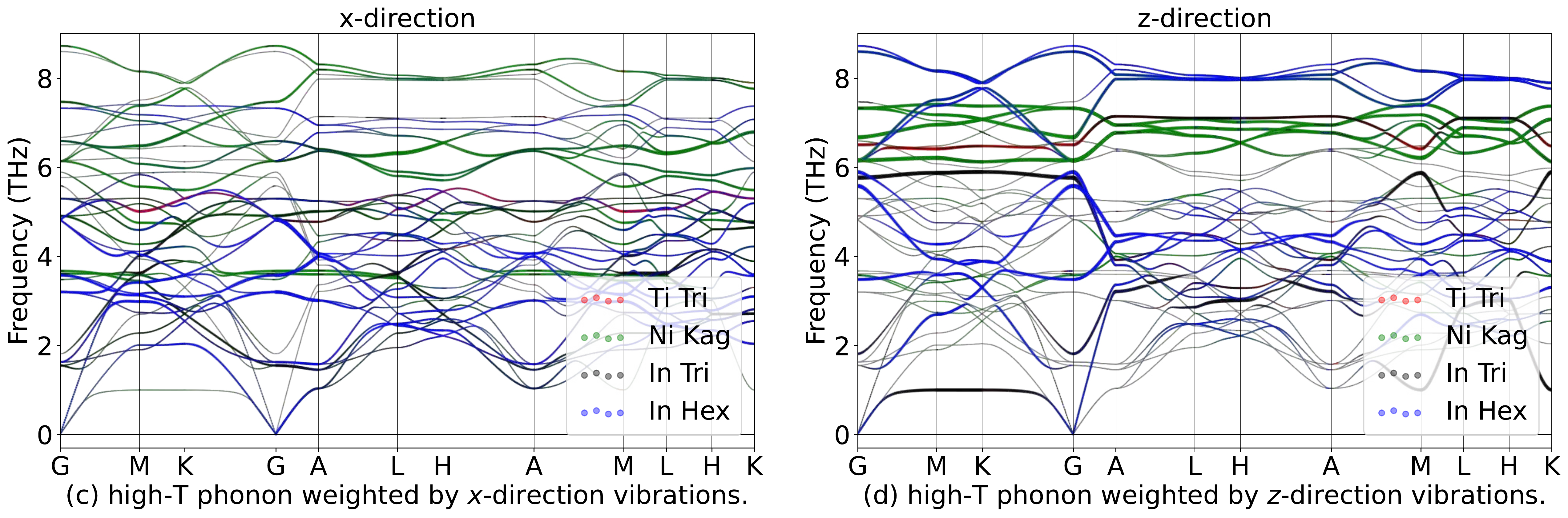}
\caption{Phonon spectrum for TiNi$_6$In$_6$in in-plane ($x$) and out-of-plane ($z$) directions.}
\label{fig:TiNi6In6phoxyz}
\end{figure}

\begin{table}[htbp]
\global\long\def\arraystretch{1.12}
\captionsetup{width=.95\textwidth}
\caption{IRREPs at high-symmetry points for TiNi$_6$In$_6$ phonon spectrum at Low-T.}
\label{tab:191_TiNi6In6_Low}
\setlength{\tabcolsep}{1.mm}{
}
\end{table}

\begin{table}[htbp]
\global\long\def\arraystretch{1.12}
\captionsetup{width=.95\textwidth}
\caption{IRREPs at high-symmetry points for TiNi$_6$In$_6$ phonon spectrum at High-T.}
\label{tab:191_TiNi6In6_High}
\setlength{\tabcolsep}{1.mm}{
%
}
\end{table}
\subsubsection{\label{sms2sec:YNi6In6}\hyperref[smsec:col]{\texorpdfstring{YNi$_6$In$_6$}{}}~{\color{green}  (High-T stable, Low-T stable) }}

\begin{table}[htbp]
\global\long\def\arraystretch{1.12}
\captionsetup{width=.85\textwidth}
\caption{Structure parameters of YNi$_6$In$_6$ after relaxation. It contains 501 electrons. }
\label{tab:YNi6In6}
\setlength{\tabcolsep}{4mm}{
%
}
\end{table}

\begin{figure}[htbp]
\centering
\includegraphics[width=0.85\textwidth]{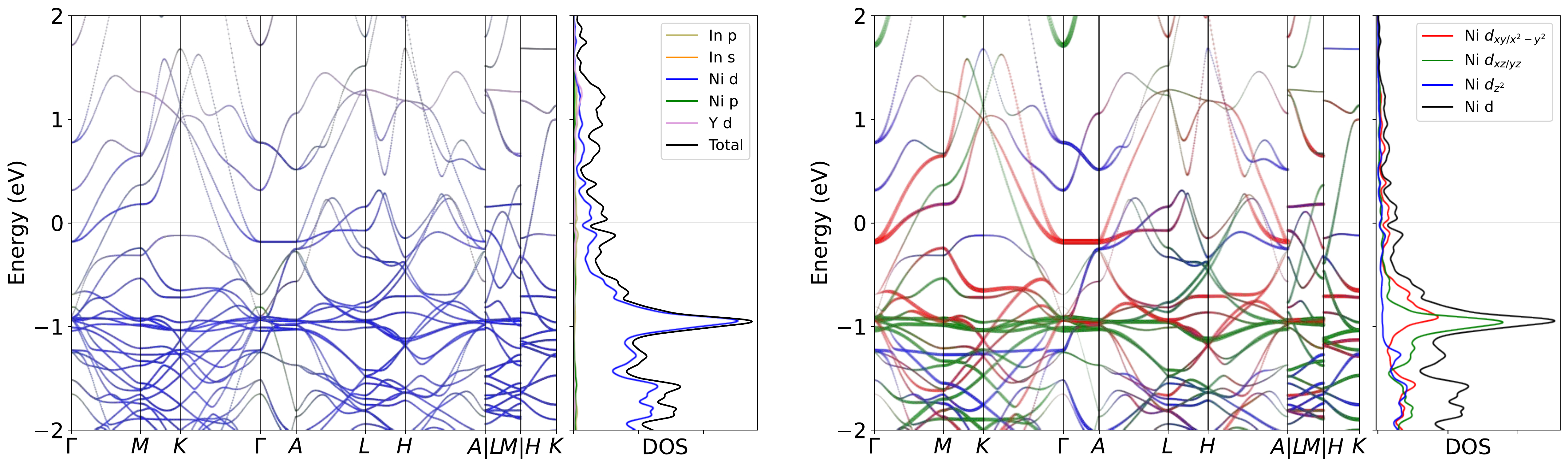}
\caption{Band structure for YNi$_6$In$_6$ without SOC. The projection of Ni $3d$ orbitals is also presented. The band structures clearly reveal the dominating of Ni $3d$ orbitals  near the Fermi level, as well as the pronounced peak.}
\label{fig:YNi6In6band}
\end{figure}

\begin{figure}[H]
\centering
\subfloat[Relaxed phonon at low-T.]{
\label{fig:YNi6In6_relaxed_Low}
\includegraphics[width=0.45\textwidth]{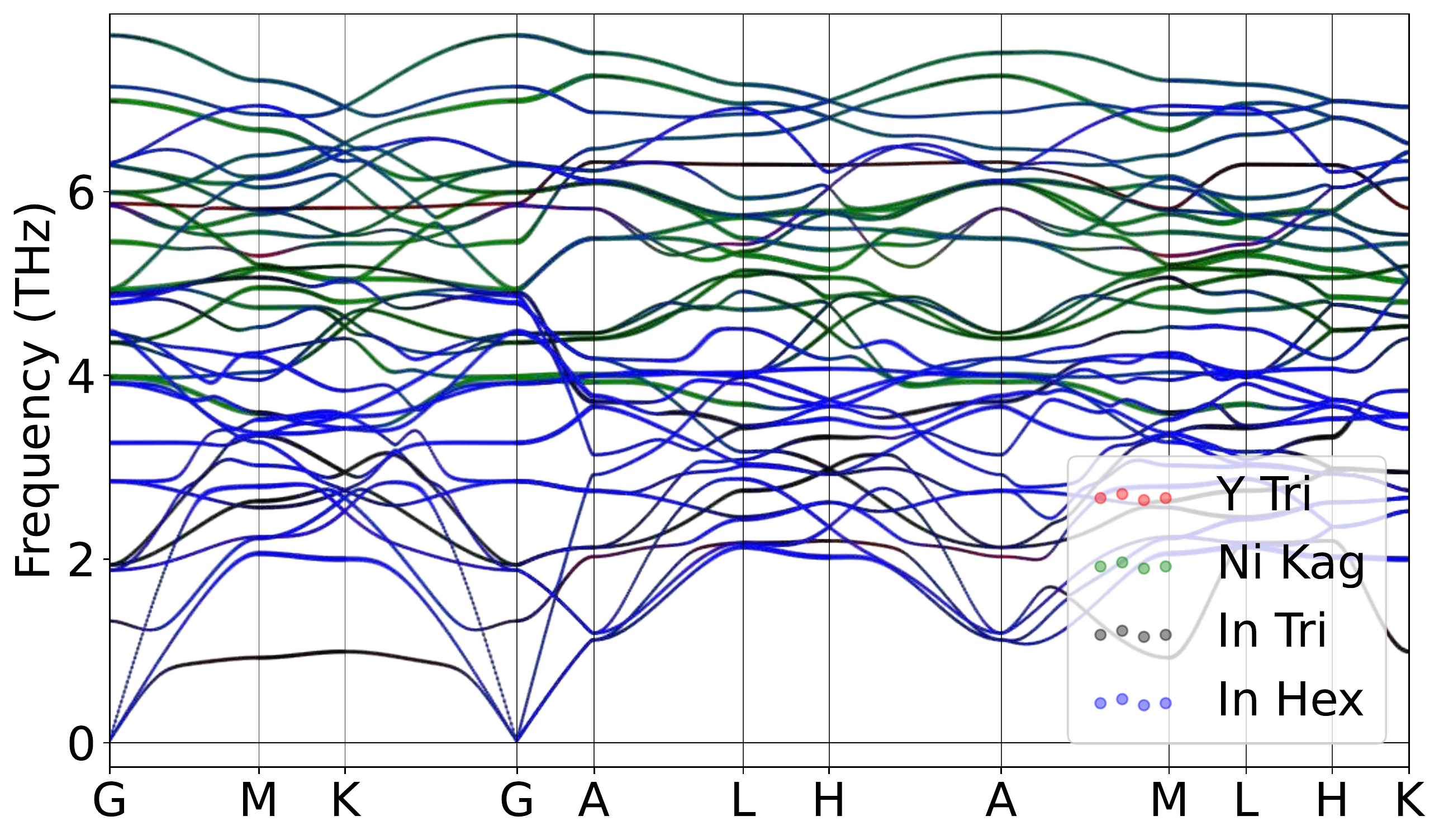}
}
\hspace{5pt}\subfloat[Relaxed phonon at high-T.]{
\label{fig:YNi6In6_relaxed_High}
\includegraphics[width=0.45\textwidth]{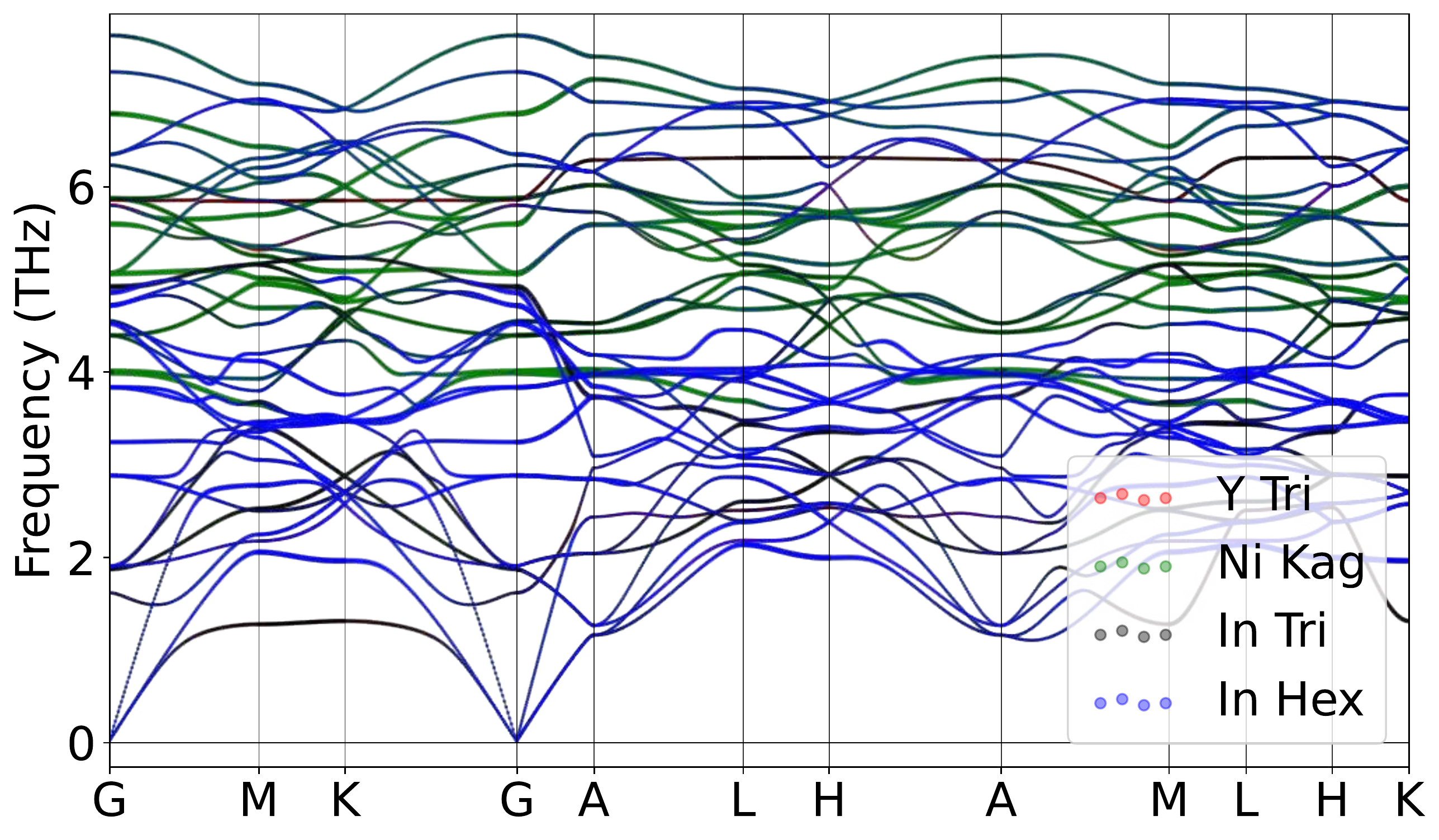}
}
\caption{Atomically projected phonon spectrum for YNi$_6$In$_6$.}
\label{fig:YNi6In6pho}
\end{figure}

\begin{figure}[H]
\centering
\label{fig:YNi6In6_relaxed_Low_xz}
\includegraphics[width=0.85\textwidth]{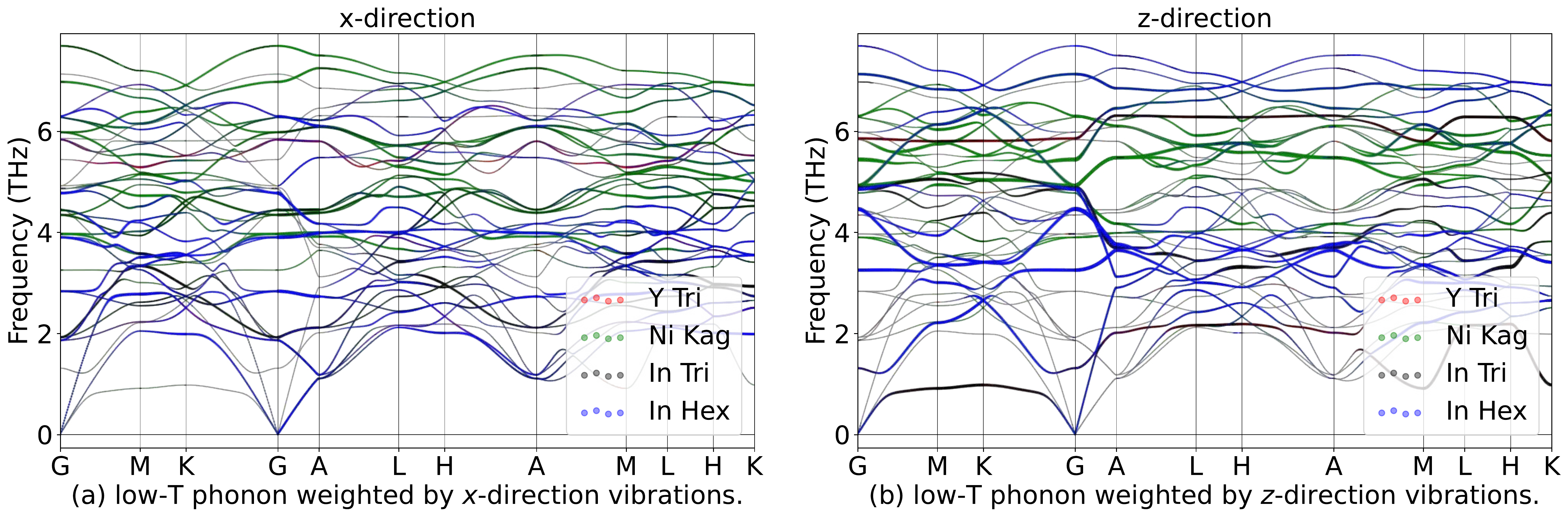}
\hspace{5pt}\label{fig:YNi6In6_relaxed_High_xz}
\includegraphics[width=0.85\textwidth]{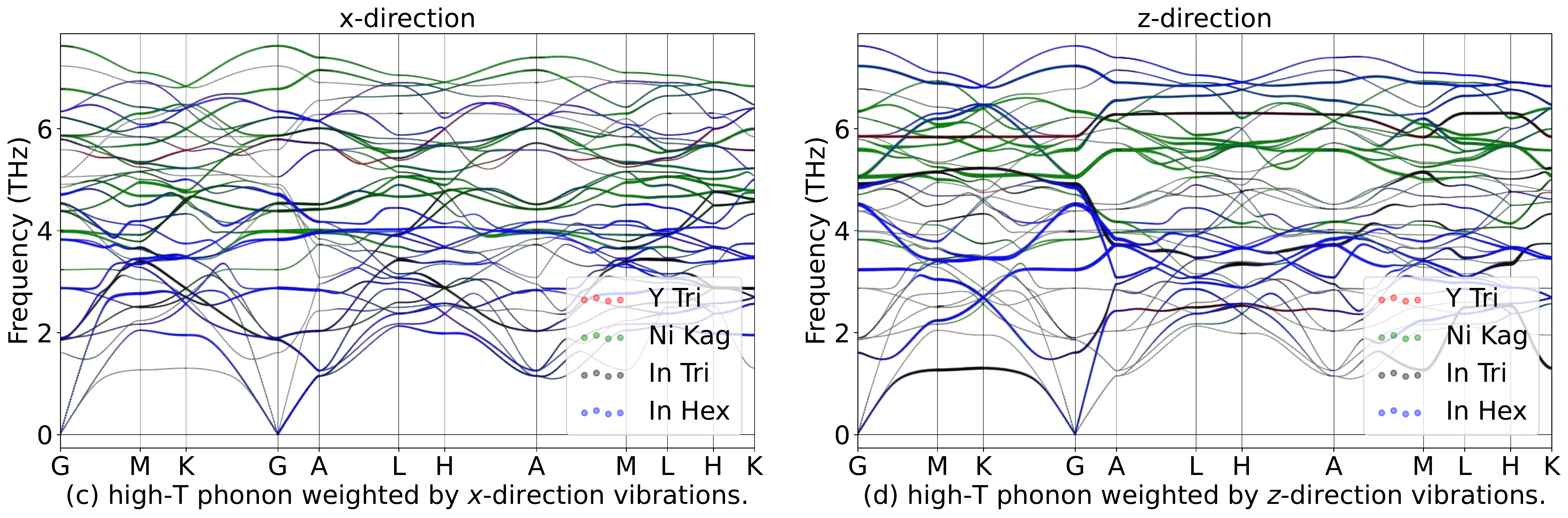}
\caption{Phonon spectrum for YNi$_6$In$_6$in in-plane ($x$) and out-of-plane ($z$) directions.}
\label{fig:YNi6In6phoxyz}
\end{figure}

\begin{table}[htbp]
\global\long\def\arraystretch{1.12}
\captionsetup{width=.95\textwidth}
\caption{IRREPs at high-symmetry points for YNi$_6$In$_6$ phonon spectrum at Low-T.}
\label{tab:191_YNi6In6_Low}
\setlength{\tabcolsep}{1.mm}{
}
\end{table}

\begin{table}[htbp]
\global\long\def\arraystretch{1.12}
\captionsetup{width=.95\textwidth}
\caption{IRREPs at high-symmetry points for YNi$_6$In$_6$ phonon spectrum at High-T.}
\label{tab:191_YNi6In6_High}
\setlength{\tabcolsep}{1.mm}{
%
}
\end{table}
\subsubsection{\label{sms2sec:ZrNi6In6}\hyperref[smsec:col]{\texorpdfstring{ZrNi$_6$In$_6$}{}}~{\color{green}  (High-T stable, Low-T stable) }}

\begin{table}[htbp]
\global\long\def\arraystretch{1.12}
\captionsetup{width=.85\textwidth}
\caption{Structure parameters of ZrNi$_6$In$_6$ after relaxation. It contains 502 electrons. }
\label{tab:ZrNi6In6}
\setlength{\tabcolsep}{4mm}{
%
}
\end{table}

\begin{figure}[htbp]
\centering
\includegraphics[width=0.85\textwidth]{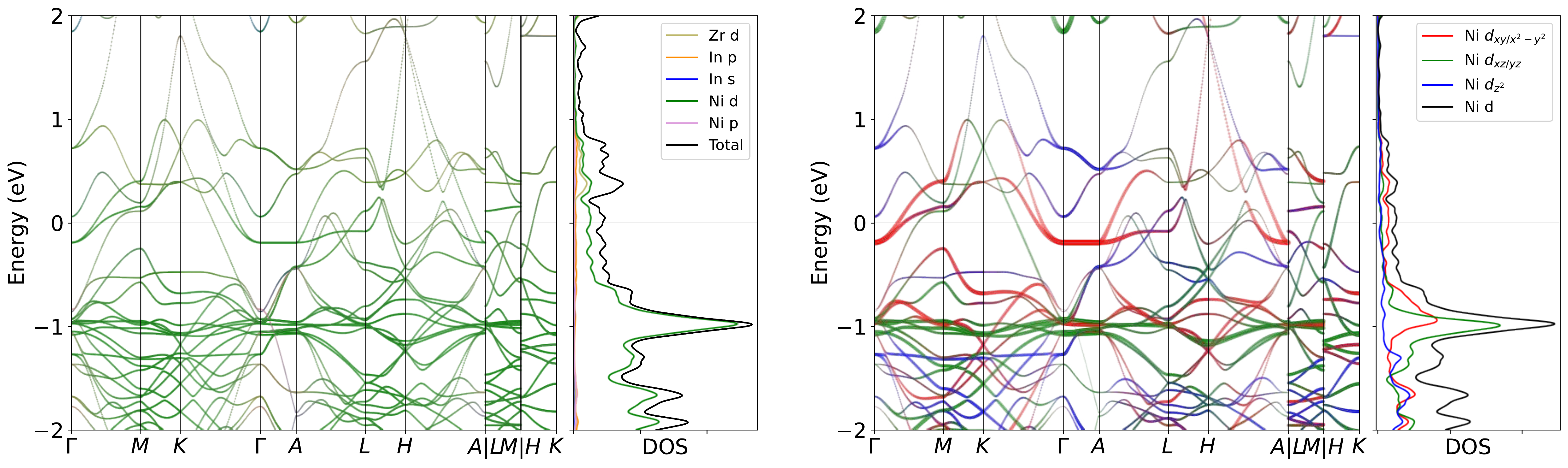}
\caption{Band structure for ZrNi$_6$In$_6$ without SOC. The projection of Ni $3d$ orbitals is also presented. The band structures clearly reveal the dominating of Ni $3d$ orbitals  near the Fermi level, as well as the pronounced peak.}
\label{fig:ZrNi6In6band}
\end{figure}

\begin{figure}[H]
\centering
\subfloat[Relaxed phonon at low-T.]{
\label{fig:ZrNi6In6_relaxed_Low}
\includegraphics[width=0.45\textwidth]{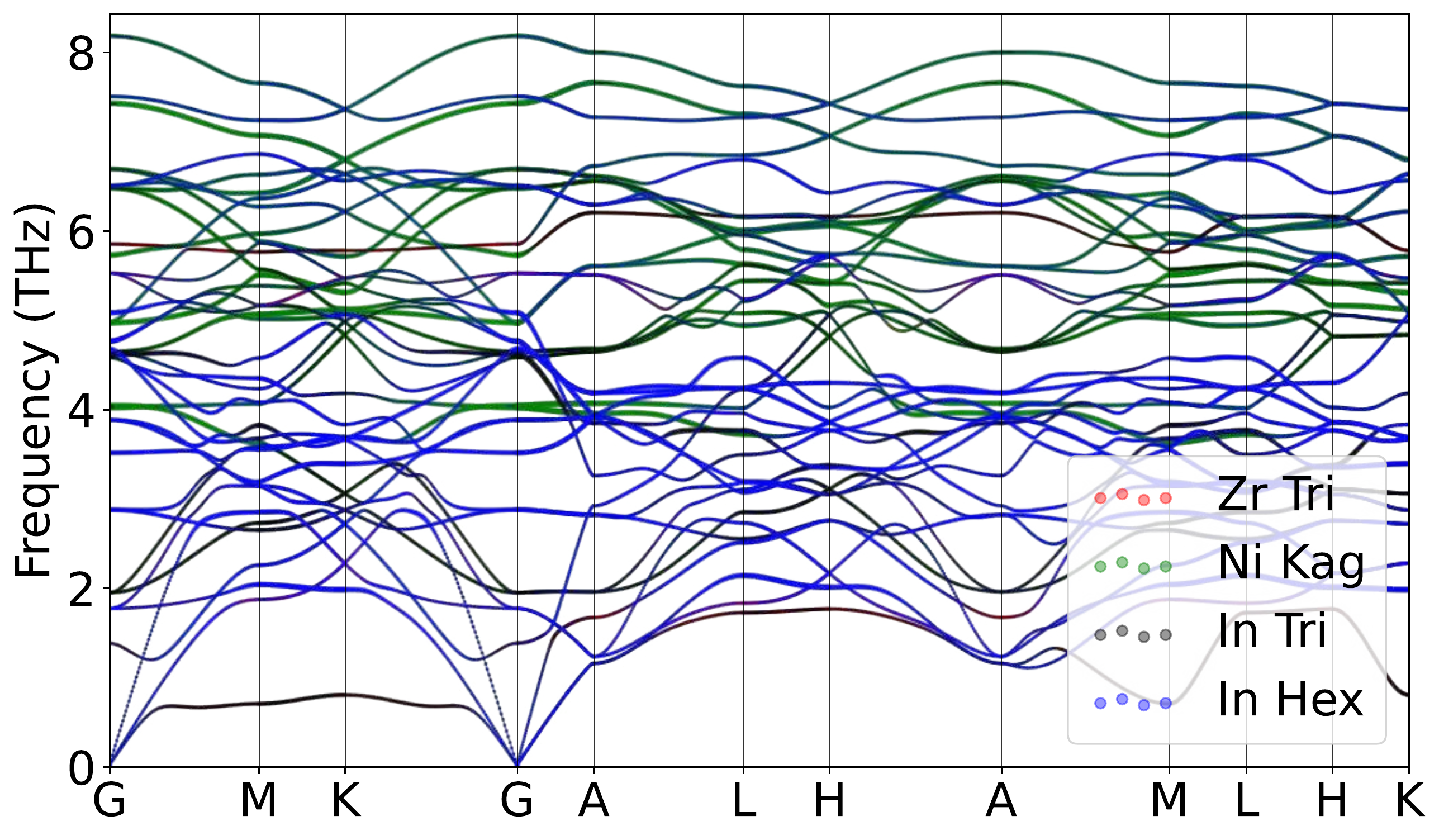}
}
\hspace{5pt}\subfloat[Relaxed phonon at high-T.]{
\label{fig:ZrNi6In6_relaxed_High}
\includegraphics[width=0.45\textwidth]{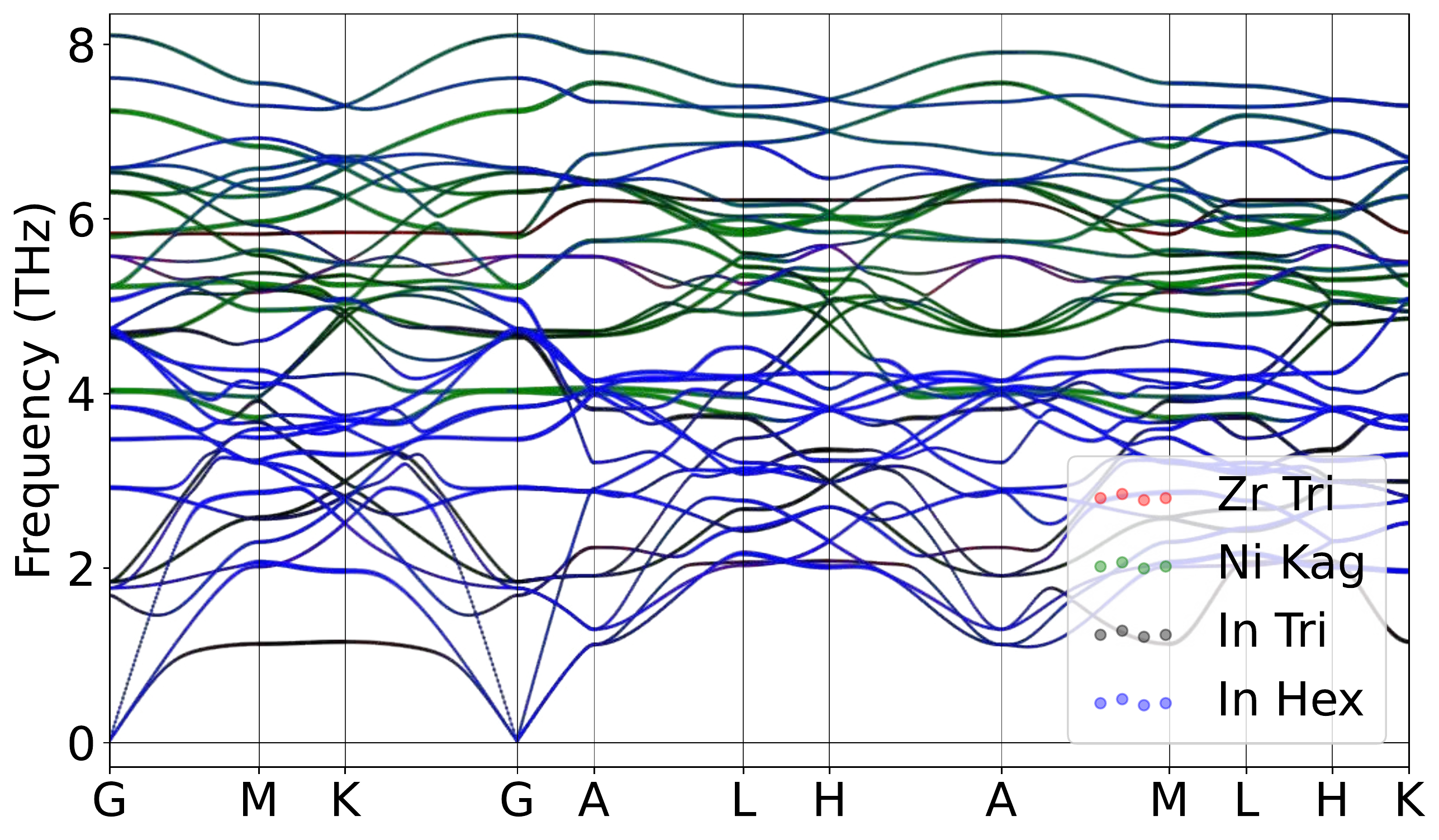}
}
\caption{Atomically projected phonon spectrum for ZrNi$_6$In$_6$.}
\label{fig:ZrNi6In6pho}
\end{figure}

\begin{figure}[H]
\centering
\label{fig:ZrNi6In6_relaxed_Low_xz}
\includegraphics[width=0.85\textwidth]{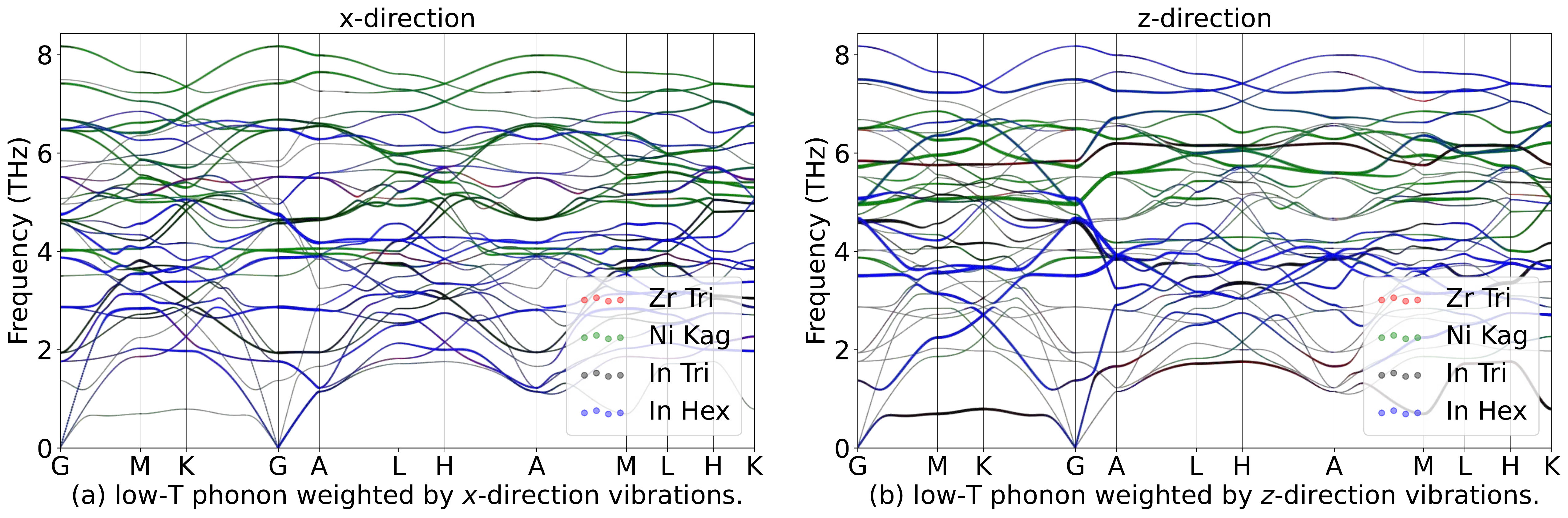}
\hspace{5pt}\label{fig:ZrNi6In6_relaxed_High_xz}
\includegraphics[width=0.85\textwidth]{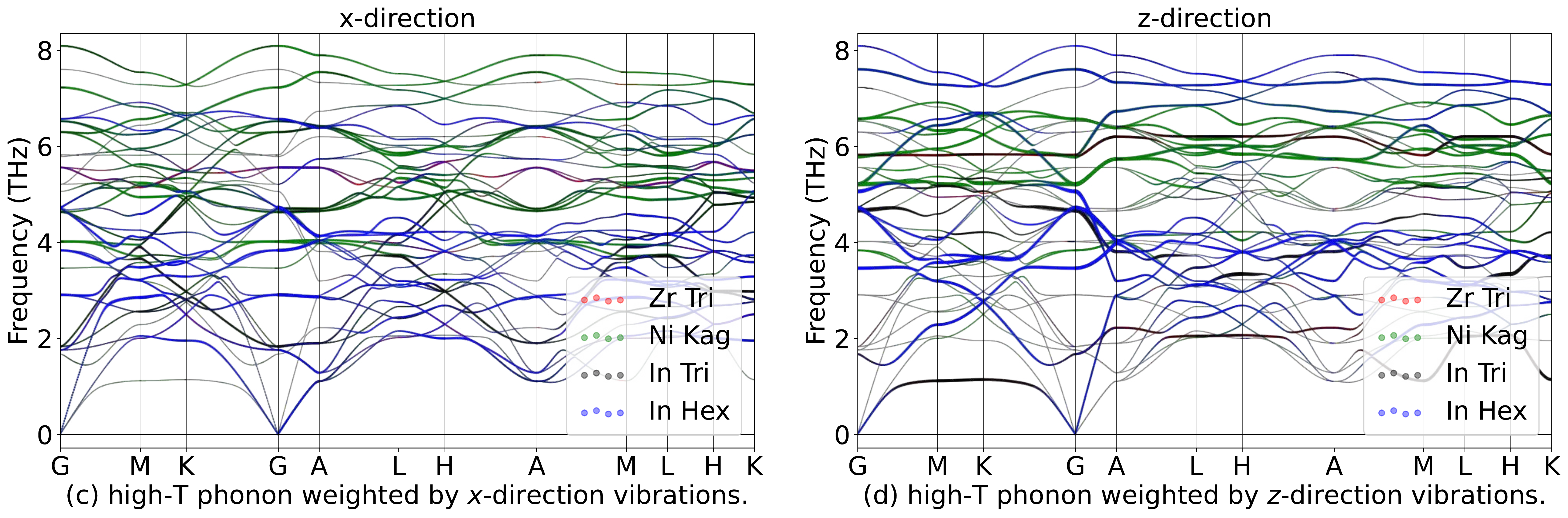}
\caption{Phonon spectrum for ZrNi$_6$In$_6$in in-plane ($x$) and out-of-plane ($z$) directions.}
\label{fig:ZrNi6In6phoxyz}
\end{figure}

\begin{table}[htbp]
\global\long\def\arraystretch{1.12}
\captionsetup{width=.95\textwidth}
\caption{IRREPs at high-symmetry points for ZrNi$_6$In$_6$ phonon spectrum at Low-T.}
\label{tab:191_ZrNi6In6_Low}
\setlength{\tabcolsep}{1.mm}{
}
\end{table}

\begin{table}[htbp]
\global\long\def\arraystretch{1.12}
\captionsetup{width=.95\textwidth}
\caption{IRREPs at high-symmetry points for ZrNi$_6$In$_6$ phonon spectrum at High-T.}
\label{tab:191_ZrNi6In6_High}
\setlength{\tabcolsep}{1.mm}{
%
}
\end{table}
\subsubsection{\label{sms2sec:NbNi6In6}\hyperref[smsec:col]{\texorpdfstring{NbNi$_6$In$_6$}{}}~{\color{green}  (High-T stable, Low-T stable) }}

\begin{table}[htbp]
\global\long\def\arraystretch{1.12}
\captionsetup{width=.85\textwidth}
\caption{Structure parameters of NbNi$_6$In$_6$ after relaxation. It contains 503 electrons. }
\label{tab:NbNi6In6}
\setlength{\tabcolsep}{4mm}{
%
}
\end{table}

\begin{figure}[htbp]
\centering
\includegraphics[width=0.85\textwidth]{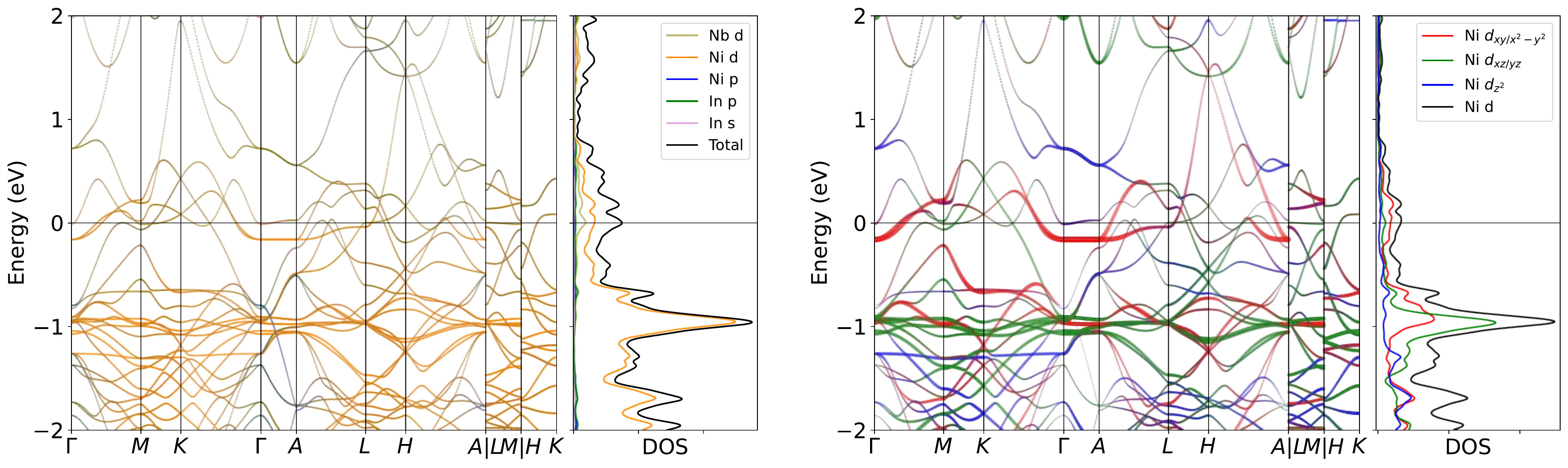}
\caption{Band structure for NbNi$_6$In$_6$ without SOC. The projection of Ni $3d$ orbitals is also presented. The band structures clearly reveal the dominating of Ni $3d$ orbitals  near the Fermi level, as well as the pronounced peak.}
\label{fig:NbNi6In6band}
\end{figure}

\begin{figure}[H]
\centering
\subfloat[Relaxed phonon at low-T.]{
\label{fig:NbNi6In6_relaxed_Low}
\includegraphics[width=0.45\textwidth]{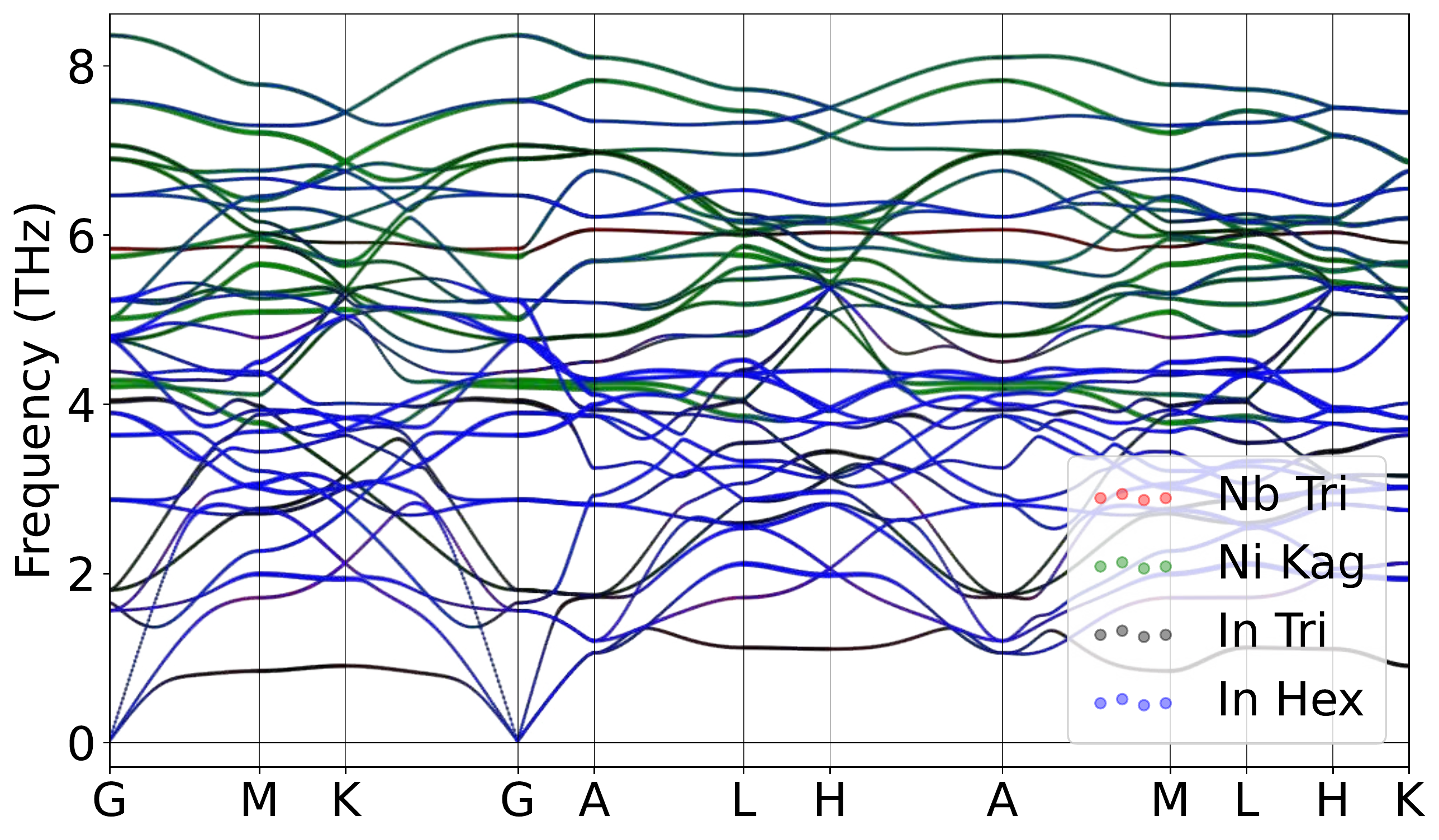}
}
\hspace{5pt}\subfloat[Relaxed phonon at high-T.]{
\label{fig:NbNi6In6_relaxed_High}
\includegraphics[width=0.45\textwidth]{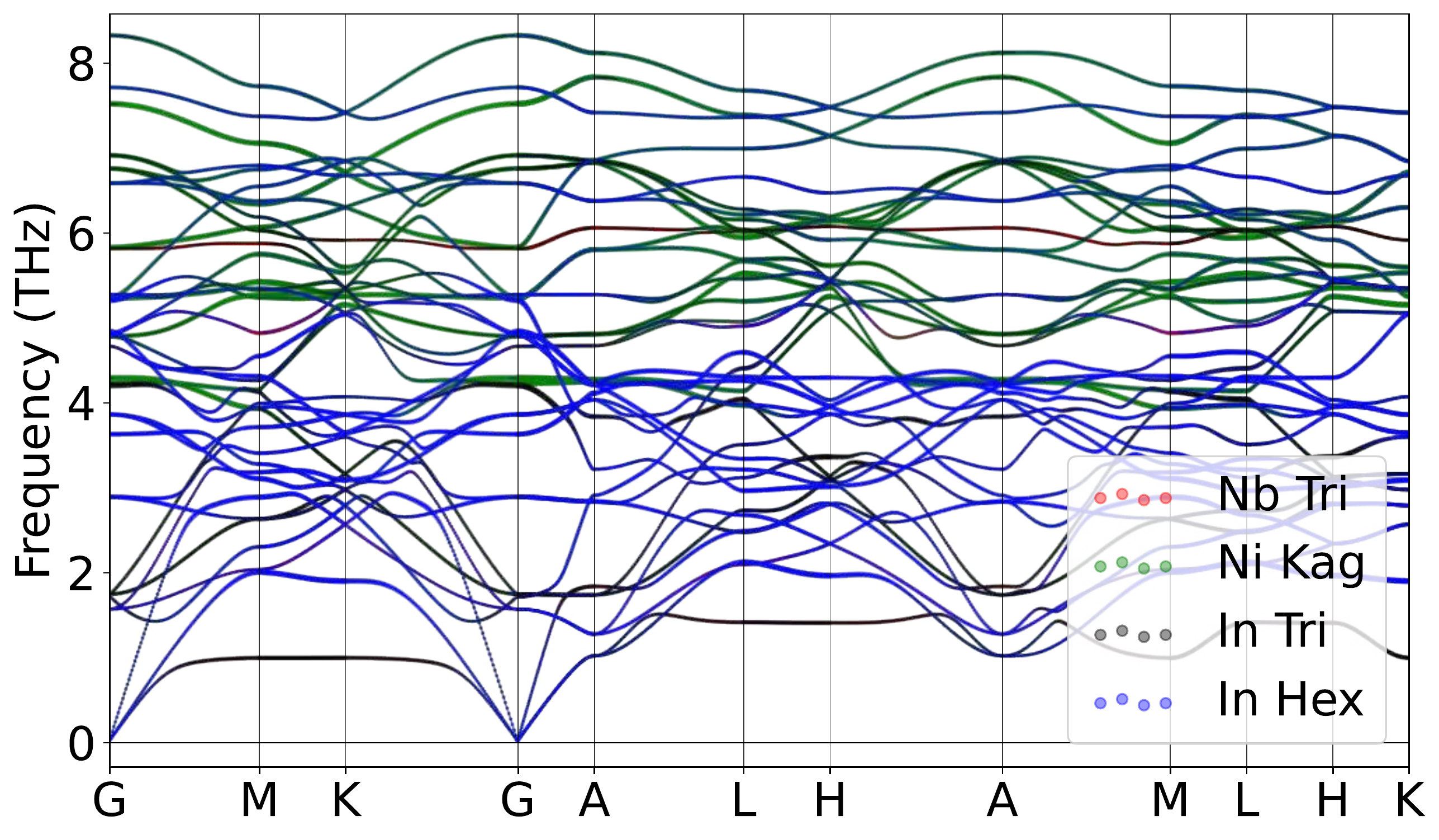}
}
\caption{Atomically projected phonon spectrum for NbNi$_6$In$_6$.}
\label{fig:NbNi6In6pho}
\end{figure}

\begin{figure}[H]
\centering
\label{fig:NbNi6In6_relaxed_Low_xz}
\includegraphics[width=0.85\textwidth]{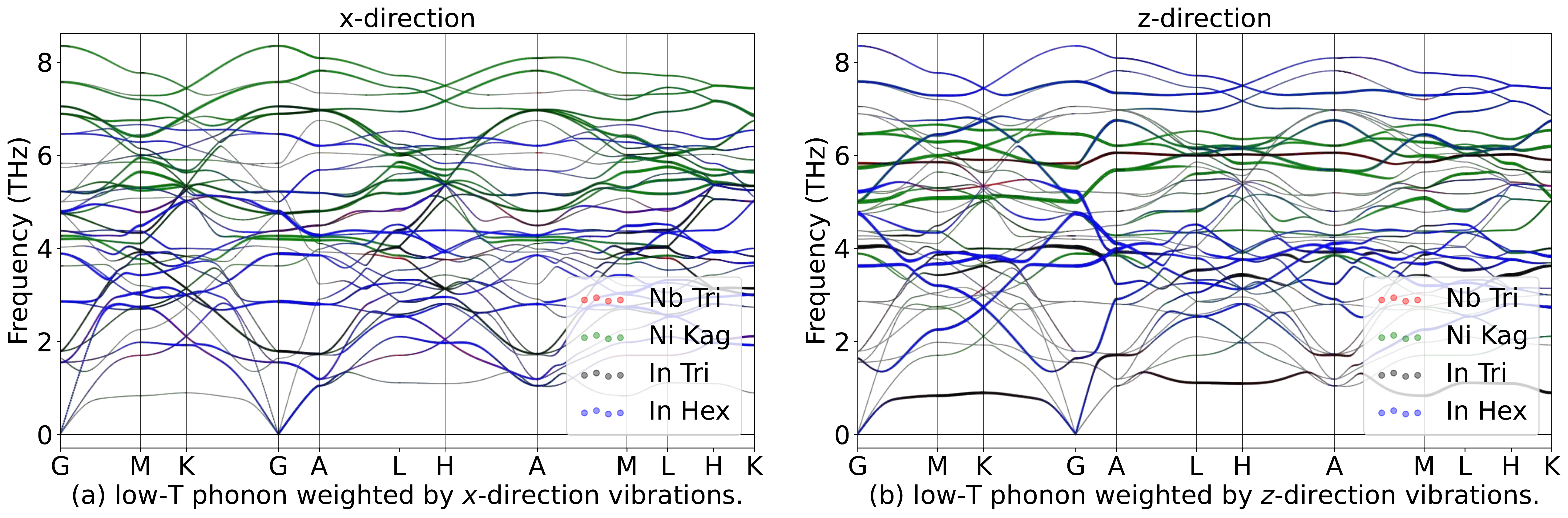}
\hspace{5pt}\label{fig:NbNi6In6_relaxed_High_xz}
\includegraphics[width=0.85\textwidth]{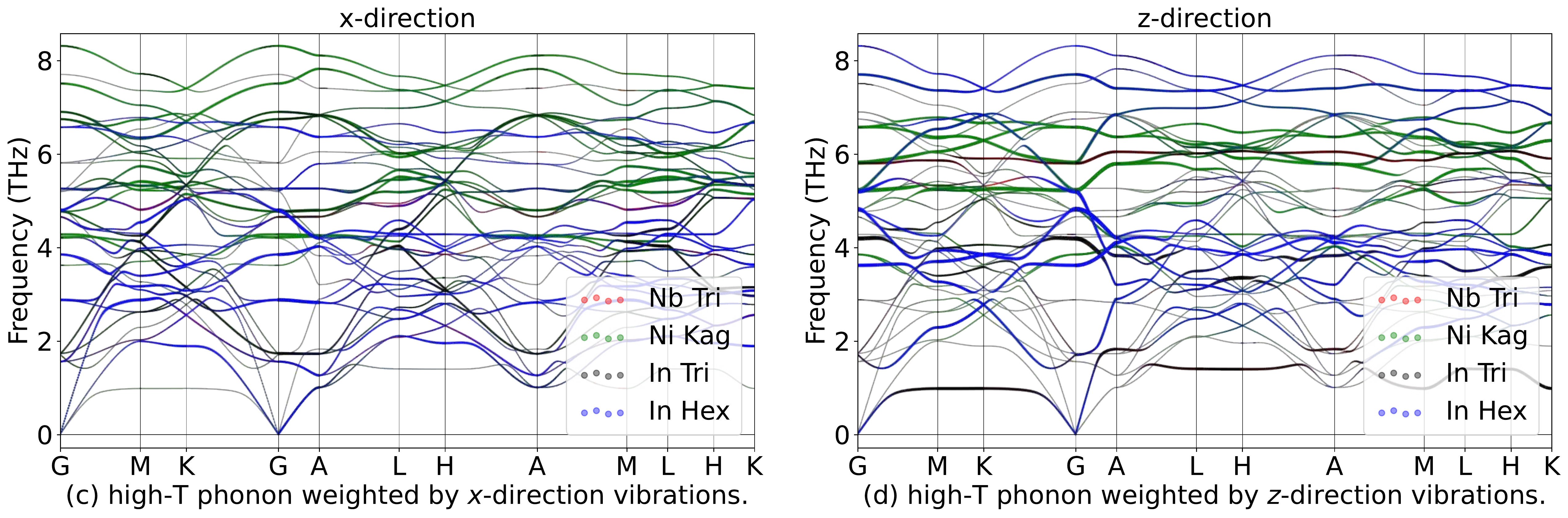}
\caption{Phonon spectrum for NbNi$_6$In$_6$in in-plane ($x$) and out-of-plane ($z$) directions.}
\label{fig:NbNi6In6phoxyz}
\end{figure}

\begin{table}[htbp]
\global\long\def\arraystretch{1.12}
\captionsetup{width=.95\textwidth}
\caption{IRREPs at high-symmetry points for NbNi$_6$In$_6$ phonon spectrum at Low-T.}
\label{tab:191_NbNi6In6_Low}
\setlength{\tabcolsep}{1.mm}{
}
\end{table}

\begin{table}[htbp]
\global\long\def\arraystretch{1.12}
\captionsetup{width=.95\textwidth}
\caption{IRREPs at high-symmetry points for NbNi$_6$In$_6$ phonon spectrum at High-T.}
\label{tab:191_NbNi6In6_High}
\setlength{\tabcolsep}{1.mm}{
%
}
\end{table}
\subsubsection{\label{sms2sec:PrNi6In6}\hyperref[smsec:col]{\texorpdfstring{PrNi$_6$In$_6$}{}}~{\color{green}  (High-T stable, Low-T stable) }}

\begin{table}[htbp]
\global\long\def\arraystretch{1.12}
\captionsetup{width=.85\textwidth}
\caption{Structure parameters of PrNi$_6$In$_6$ after relaxation. It contains 521 electrons. }
\label{tab:PrNi6In6}
\setlength{\tabcolsep}{4mm}{
%
}
\end{table}

\begin{figure}[htbp]
\centering
\includegraphics[width=0.85\textwidth]{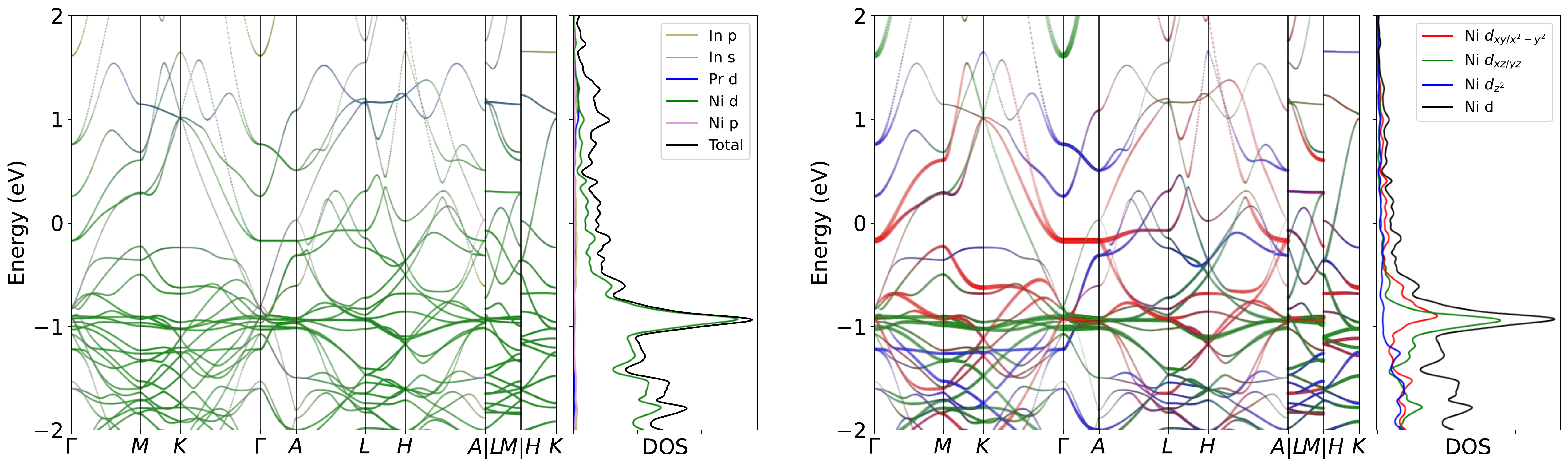}
\caption{Band structure for PrNi$_6$In$_6$ without SOC. The projection of Ni $3d$ orbitals is also presented. The band structures clearly reveal the dominating of Ni $3d$ orbitals  near the Fermi level, as well as the pronounced peak.}
\label{fig:PrNi6In6band}
\end{figure}

\begin{figure}[H]
\centering
\subfloat[Relaxed phonon at low-T.]{
\label{fig:PrNi6In6_relaxed_Low}
\includegraphics[width=0.45\textwidth]{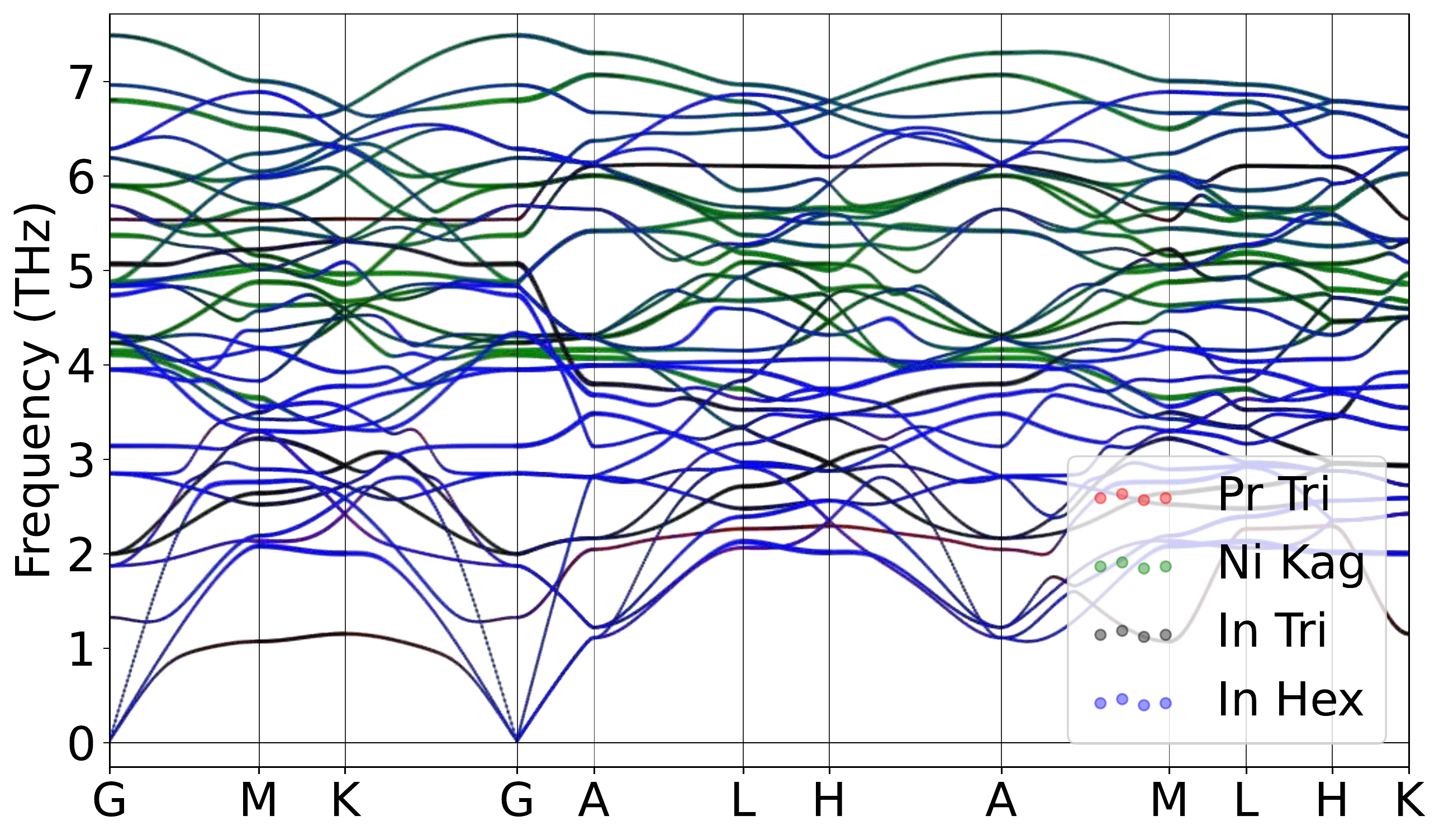}
}
\hspace{5pt}\subfloat[Relaxed phonon at high-T.]{
\label{fig:PrNi6In6_relaxed_High}
\includegraphics[width=0.45\textwidth]{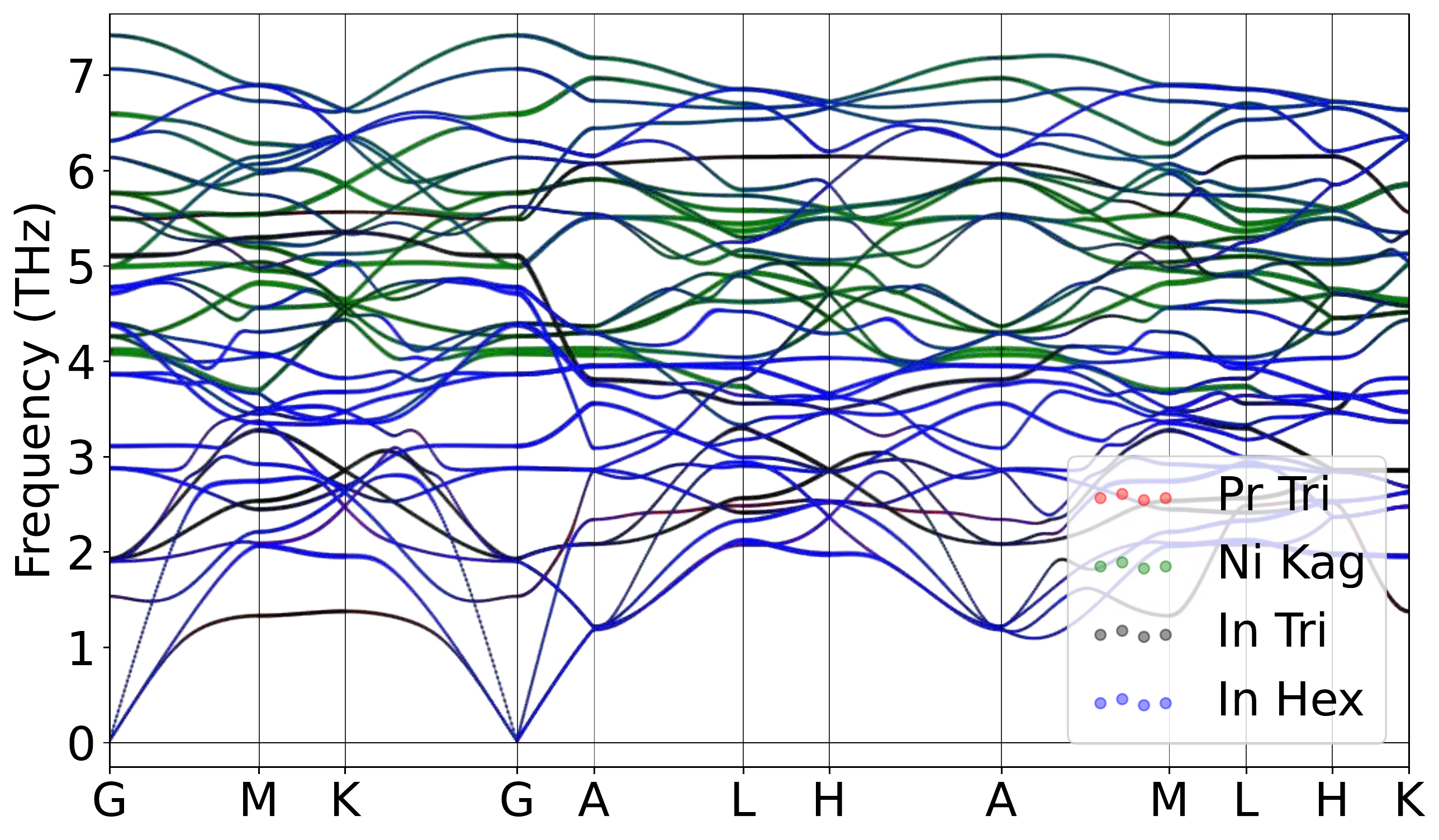}
}
\caption{Atomically projected phonon spectrum for PrNi$_6$In$_6$.}
\label{fig:PrNi6In6pho}
\end{figure}

\begin{figure}[H]
\centering
\label{fig:PrNi6In6_relaxed_Low_xz}
\includegraphics[width=0.85\textwidth]{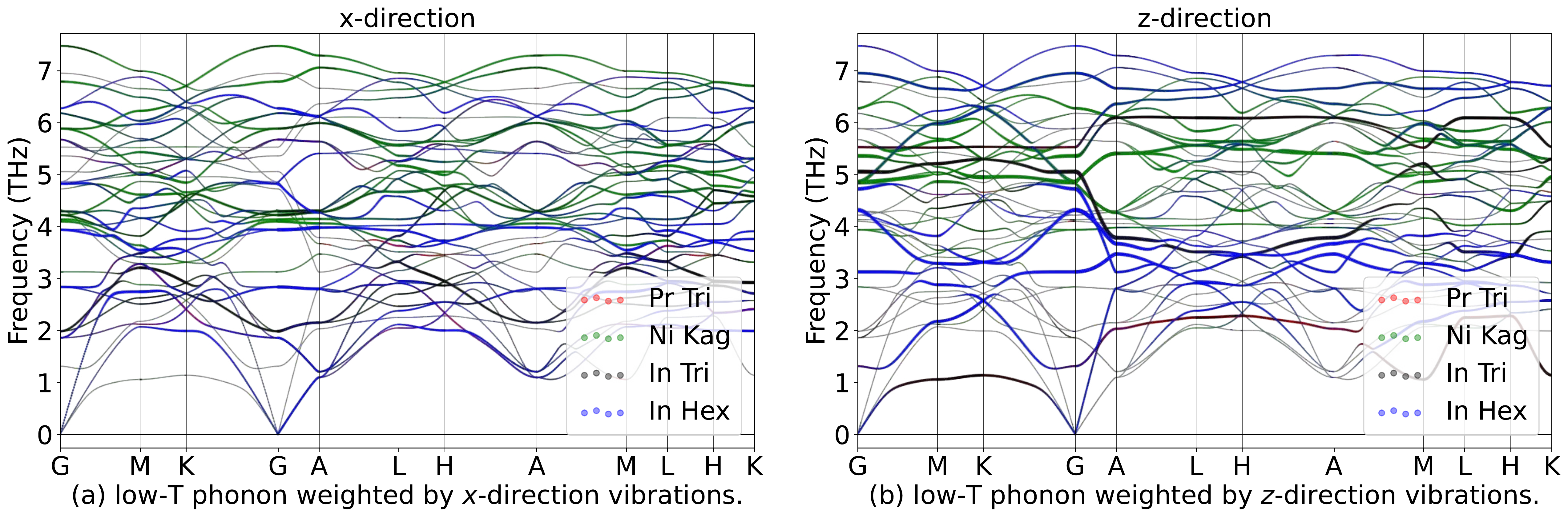}
\hspace{5pt}\label{fig:PrNi6In6_relaxed_High_xz}
\includegraphics[width=0.85\textwidth]{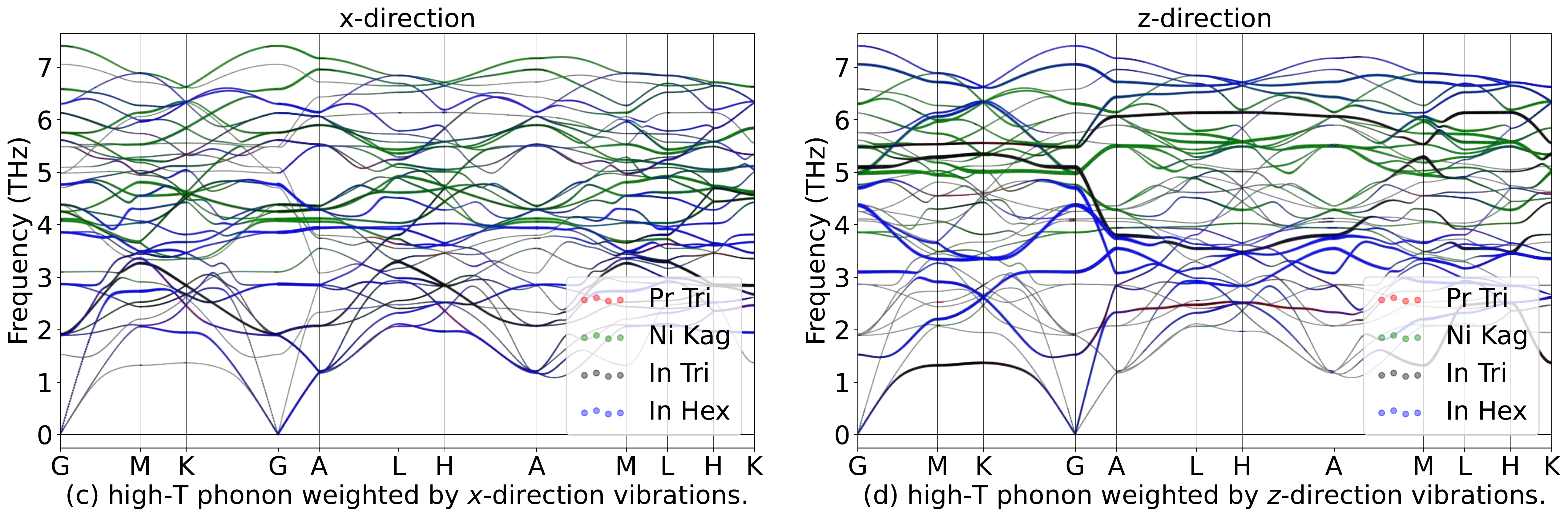}
\caption{Phonon spectrum for PrNi$_6$In$_6$in in-plane ($x$) and out-of-plane ($z$) directions.}
\label{fig:PrNi6In6phoxyz}
\end{figure}

\begin{table}[htbp]
\global\long\def\arraystretch{1.12}
\captionsetup{width=.95\textwidth}
\caption{IRREPs at high-symmetry points for PrNi$_6$In$_6$ phonon spectrum at Low-T.}
\label{tab:191_PrNi6In6_Low}
\setlength{\tabcolsep}{1.mm}{
}
\end{table}

\begin{table}[htbp]
\global\long\def\arraystretch{1.12}
\captionsetup{width=.95\textwidth}
\caption{IRREPs at high-symmetry points for PrNi$_6$In$_6$ phonon spectrum at High-T.}
\label{tab:191_PrNi6In6_High}
\setlength{\tabcolsep}{1.mm}{
%
}
\end{table}
\subsubsection{\label{sms2sec:NdNi6In6}\hyperref[smsec:col]{\texorpdfstring{NdNi$_6$In$_6$}{}}~{\color{green}  (High-T stable, Low-T stable) }}

\begin{table}[htbp]
\global\long\def\arraystretch{1.12}
\captionsetup{width=.85\textwidth}
\caption{Structure parameters of NdNi$_6$In$_6$ after relaxation. It contains 522 electrons. }
\label{tab:NdNi6In6}
\setlength{\tabcolsep}{4mm}{
%
}
\end{table}

\begin{figure}[htbp]
\centering
\includegraphics[width=0.85\textwidth]{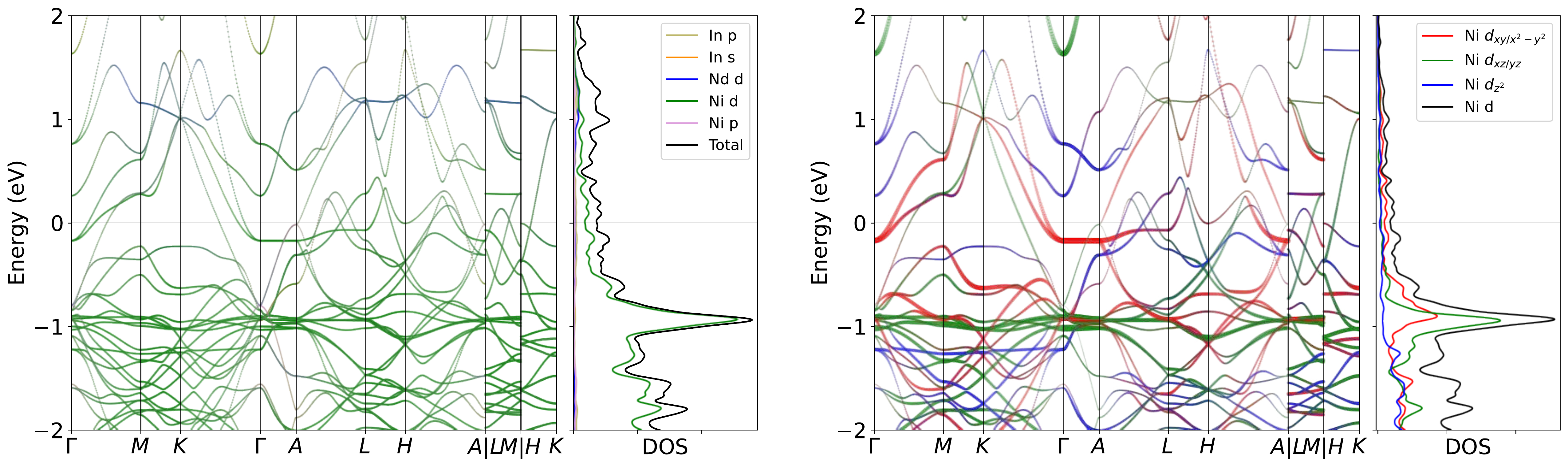}
\caption{Band structure for NdNi$_6$In$_6$ without SOC. The projection of Ni $3d$ orbitals is also presented. The band structures clearly reveal the dominating of Ni $3d$ orbitals  near the Fermi level, as well as the pronounced peak.}
\label{fig:NdNi6In6band}
\end{figure}

\begin{figure}[H]
\centering
\subfloat[Relaxed phonon at low-T.]{
\label{fig:NdNi6In6_relaxed_Low}
\includegraphics[width=0.45\textwidth]{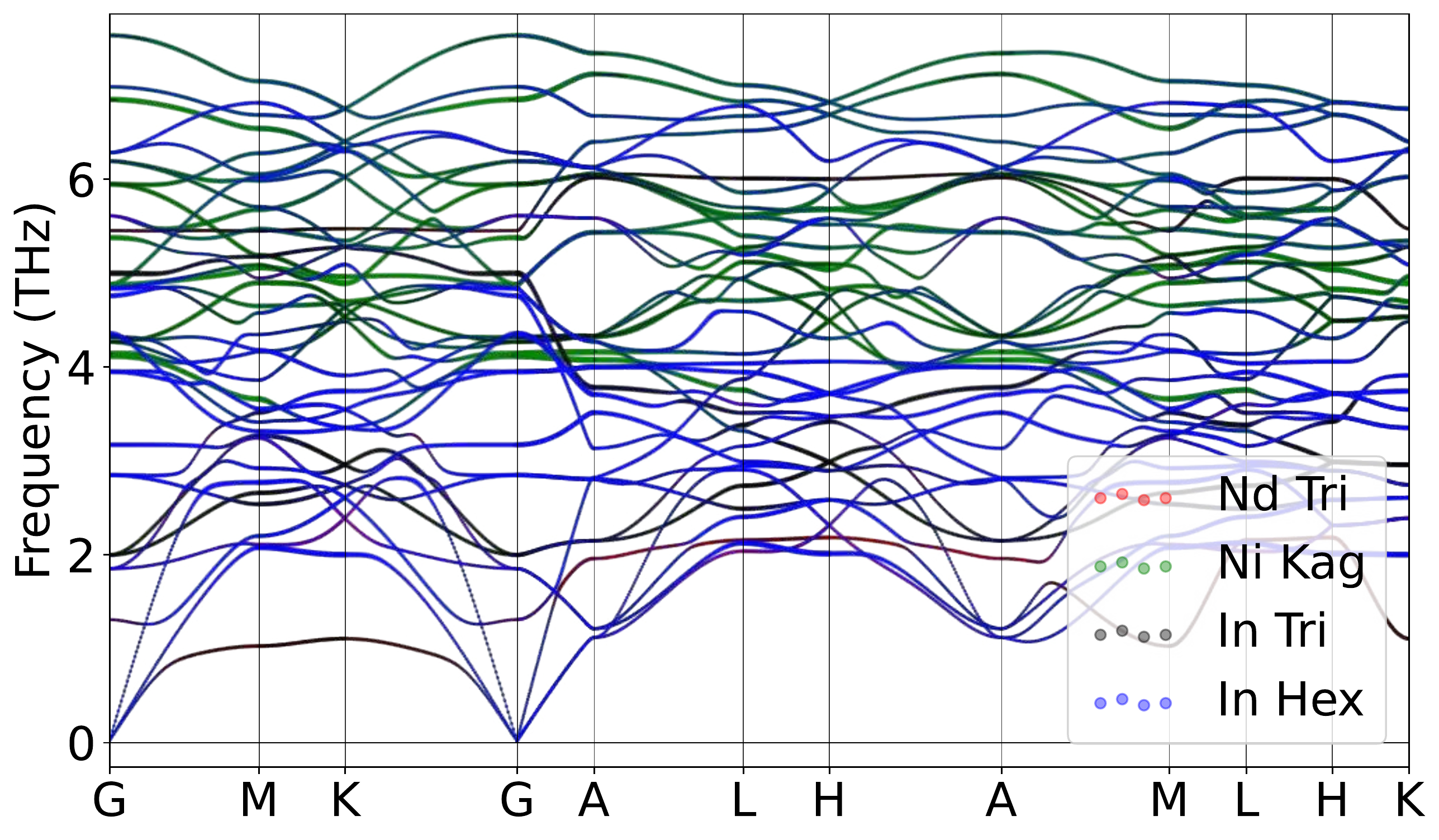}
}
\hspace{5pt}\subfloat[Relaxed phonon at high-T.]{
\label{fig:NdNi6In6_relaxed_High}
\includegraphics[width=0.45\textwidth]{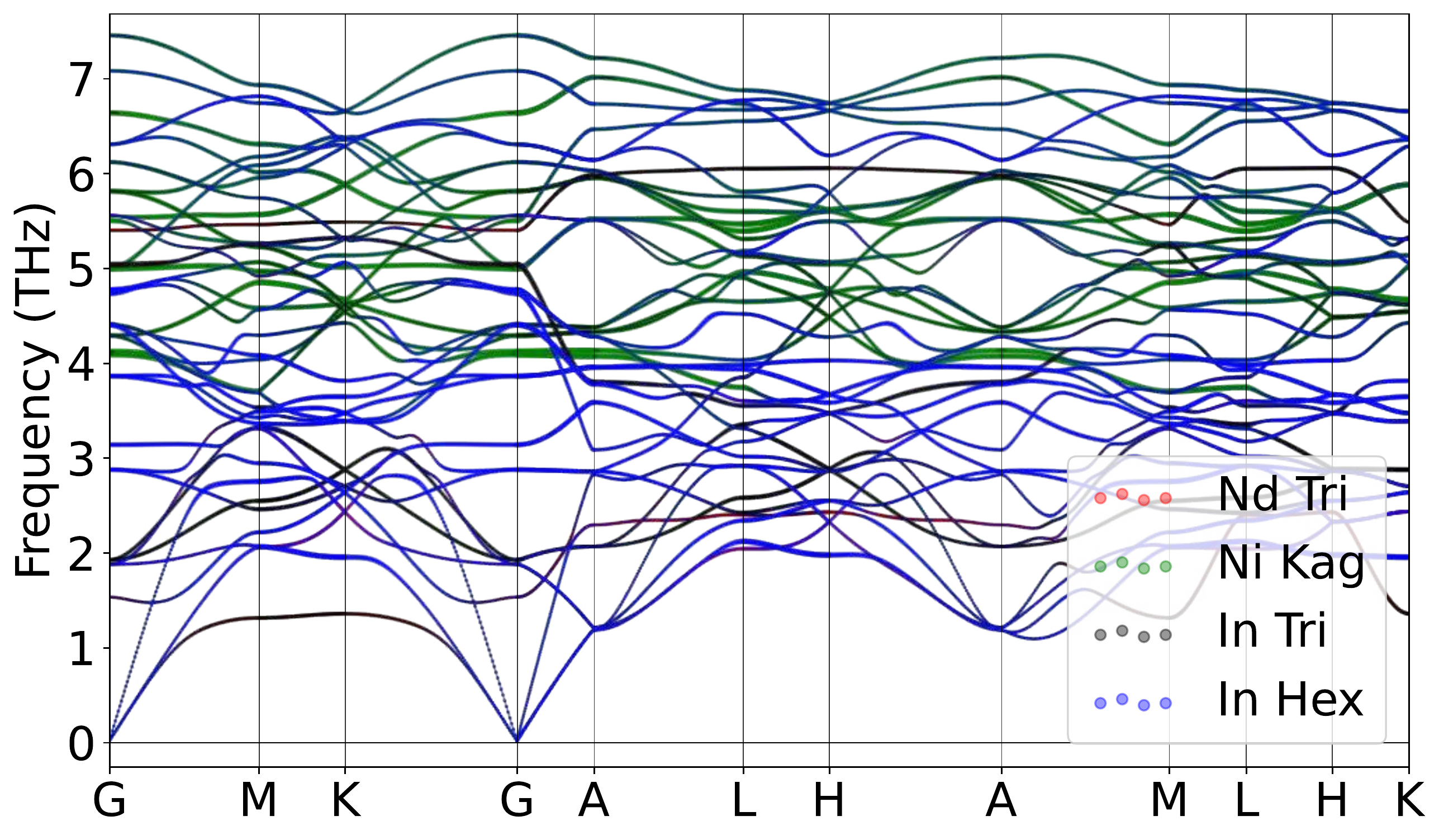}
}
\caption{Atomically projected phonon spectrum for NdNi$_6$In$_6$.}
\label{fig:NdNi6In6pho}
\end{figure}

\begin{figure}[H]
\centering
\label{fig:NdNi6In6_relaxed_Low_xz}
\includegraphics[width=0.85\textwidth]{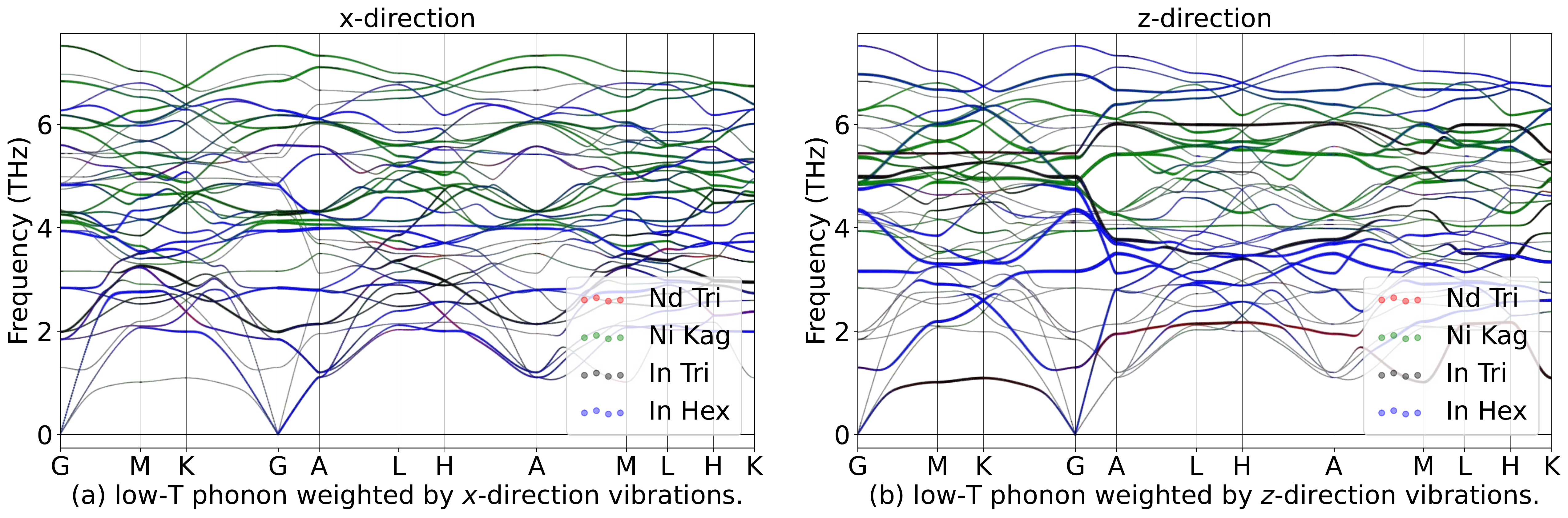}
\hspace{5pt}\label{fig:NdNi6In6_relaxed_High_xz}
\includegraphics[width=0.85\textwidth]{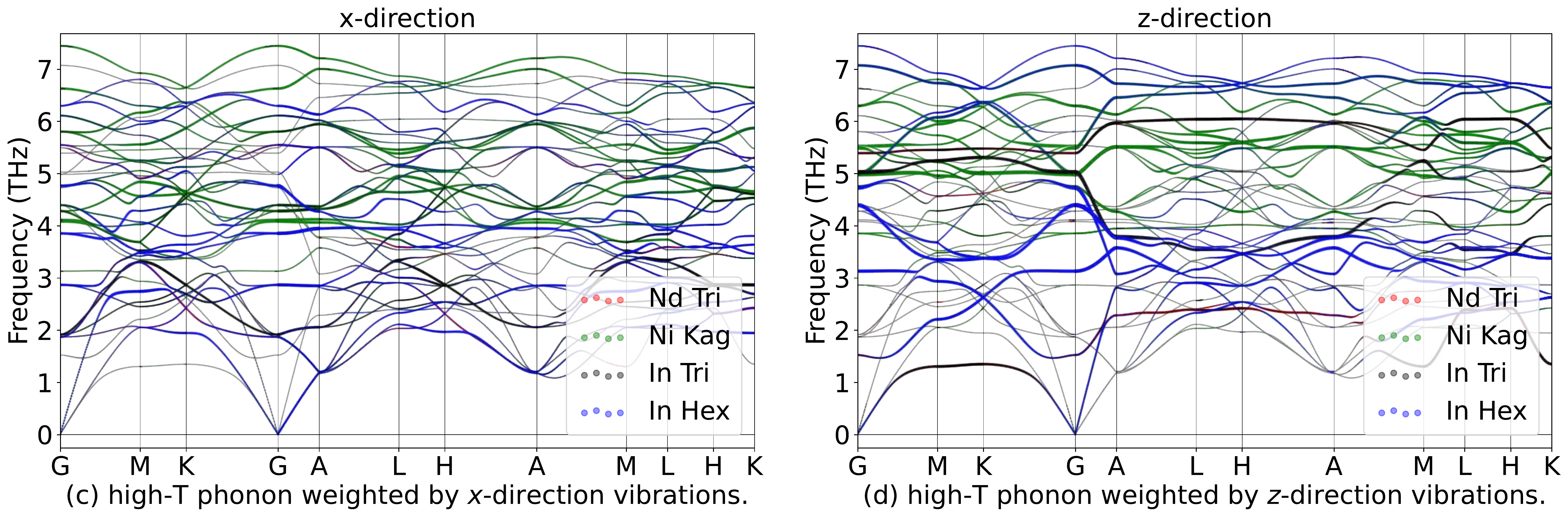}
\caption{Phonon spectrum for NdNi$_6$In$_6$in in-plane ($x$) and out-of-plane ($z$) directions.}
\label{fig:NdNi6In6phoxyz}
\end{figure}

\begin{table}[htbp]
\global\long\def\arraystretch{1.12}
\captionsetup{width=.95\textwidth}
\caption{IRREPs at high-symmetry points for NdNi$_6$In$_6$ phonon spectrum at Low-T.}
\label{tab:191_NdNi6In6_Low}
\setlength{\tabcolsep}{1.mm}{
}
\end{table}

\begin{table}[htbp]
\global\long\def\arraystretch{1.12}
\captionsetup{width=.95\textwidth}
\caption{IRREPs at high-symmetry points for NdNi$_6$In$_6$ phonon spectrum at High-T.}
\label{tab:191_NdNi6In6_High}
\setlength{\tabcolsep}{1.mm}{
%
}
\end{table}
\subsubsection{\label{sms2sec:SmNi6In6}\hyperref[smsec:col]{\texorpdfstring{SmNi$_6$In$_6$}{}}~{\color{green}  (High-T stable, Low-T stable) }}

\begin{table}[htbp]
\global\long\def\arraystretch{1.12}
\captionsetup{width=.85\textwidth}
\caption{Structure parameters of SmNi$_6$In$_6$ after relaxation. It contains 524 electrons. }
\label{tab:SmNi6In6}
\setlength{\tabcolsep}{4mm}{
%
}
\end{table}

\begin{figure}[htbp]
\centering
\includegraphics[width=0.85\textwidth]{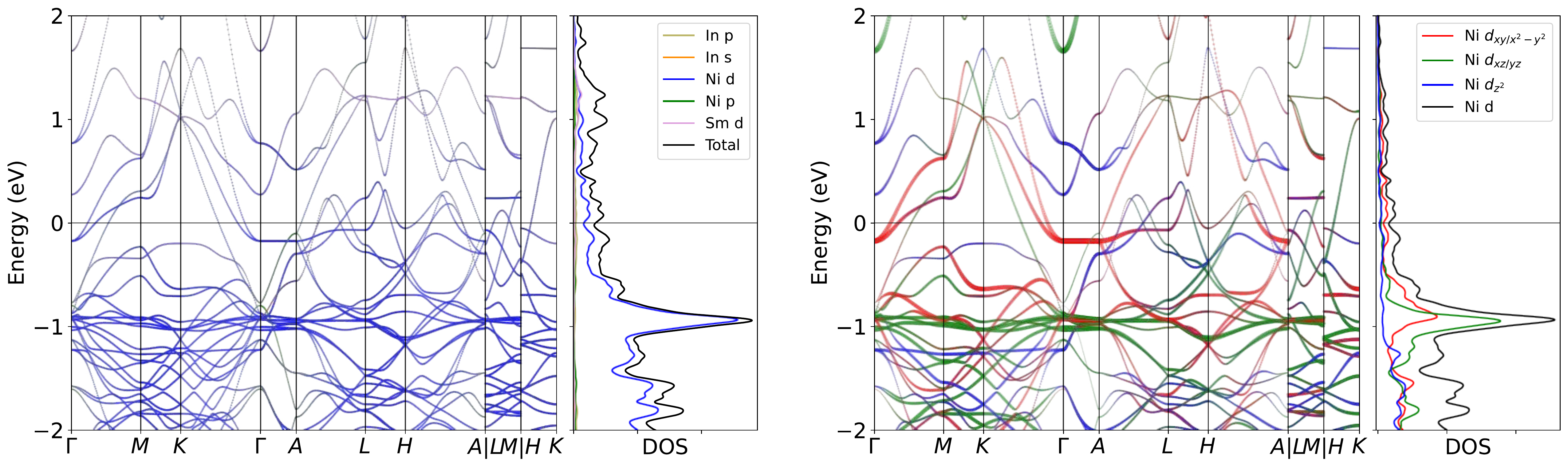}
\caption{Band structure for SmNi$_6$In$_6$ without SOC. The projection of Ni $3d$ orbitals is also presented. The band structures clearly reveal the dominating of Ni $3d$ orbitals  near the Fermi level, as well as the pronounced peak.}
\label{fig:SmNi6In6band}
\end{figure}

\begin{figure}[H]
\centering
\subfloat[Relaxed phonon at low-T.]{
\label{fig:SmNi6In6_relaxed_Low}
\includegraphics[width=0.45\textwidth]{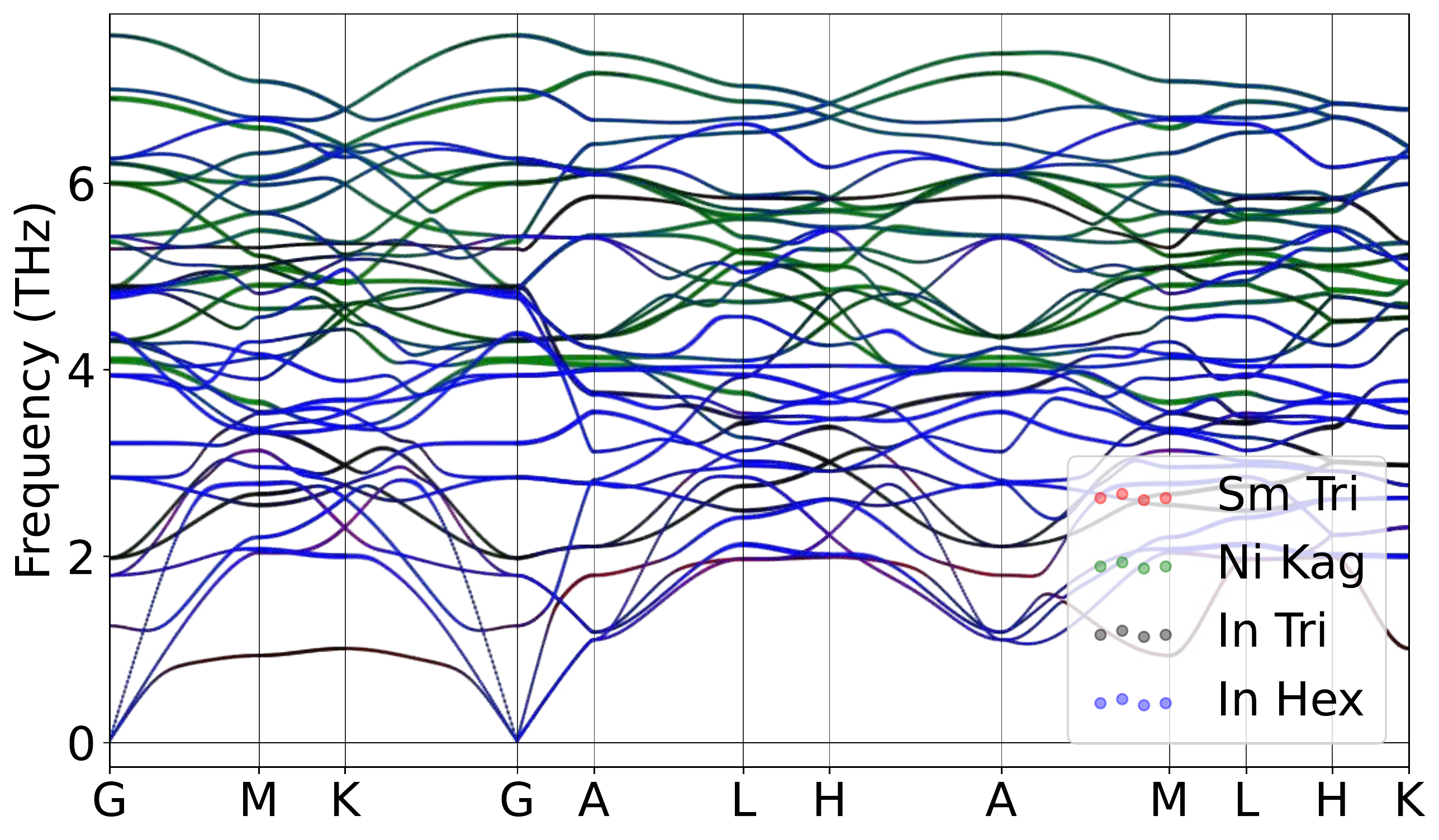}
}
\hspace{5pt}\subfloat[Relaxed phonon at high-T.]{
\label{fig:SmNi6In6_relaxed_High}
\includegraphics[width=0.45\textwidth]{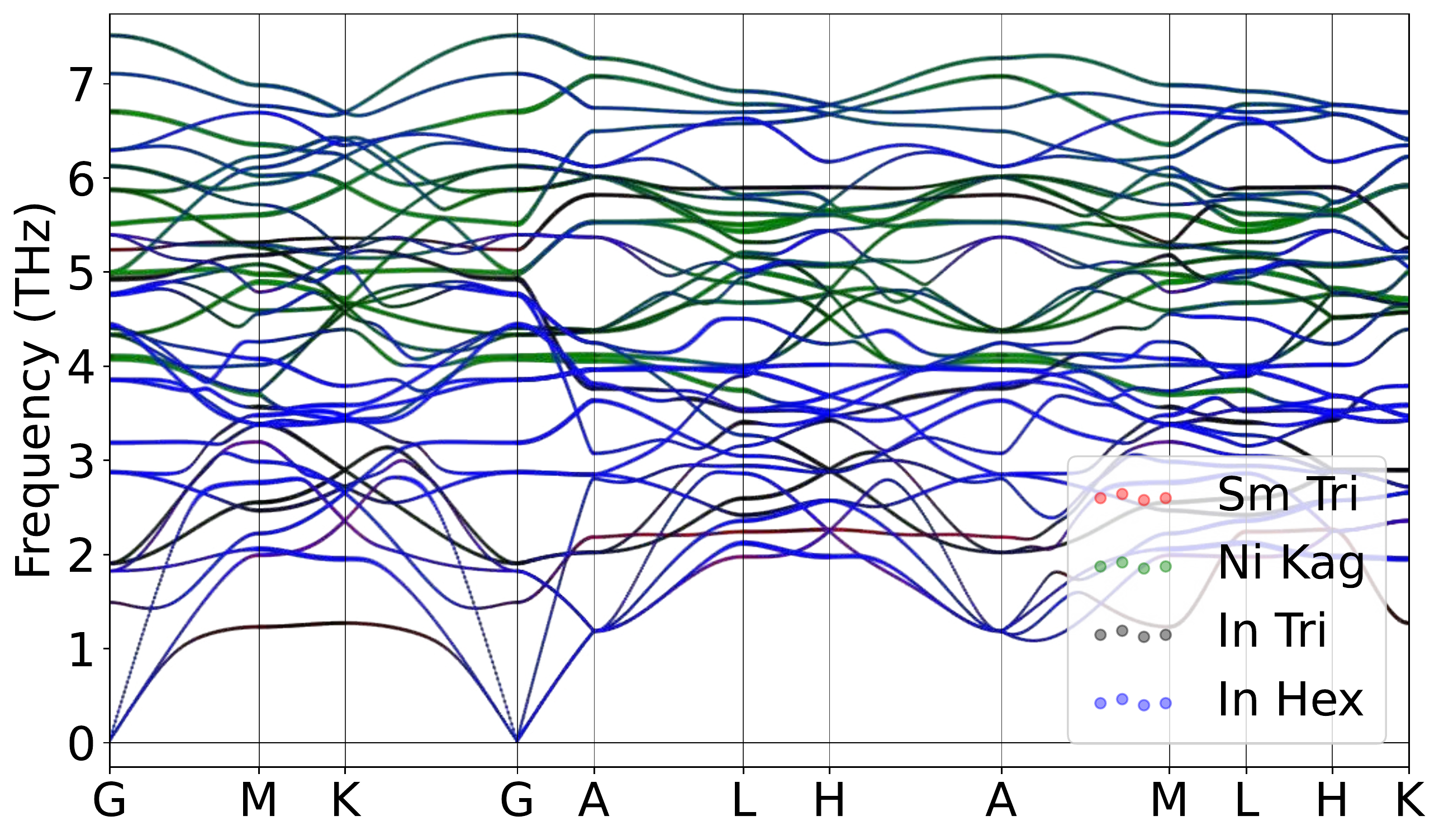}
}
\caption{Atomically projected phonon spectrum for SmNi$_6$In$_6$.}
\label{fig:SmNi6In6pho}
\end{figure}

\begin{figure}[H]
\centering
\label{fig:SmNi6In6_relaxed_Low_xz}
\includegraphics[width=0.85\textwidth]{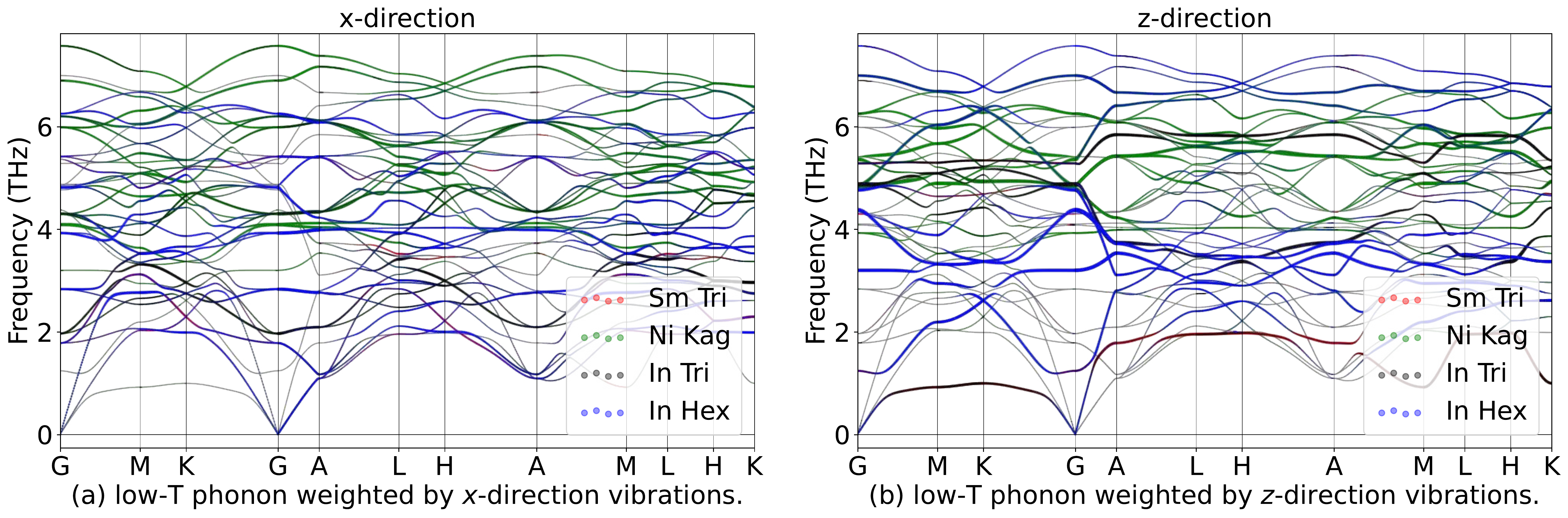}
\hspace{5pt}\label{fig:SmNi6In6_relaxed_High_xz}
\includegraphics[width=0.85\textwidth]{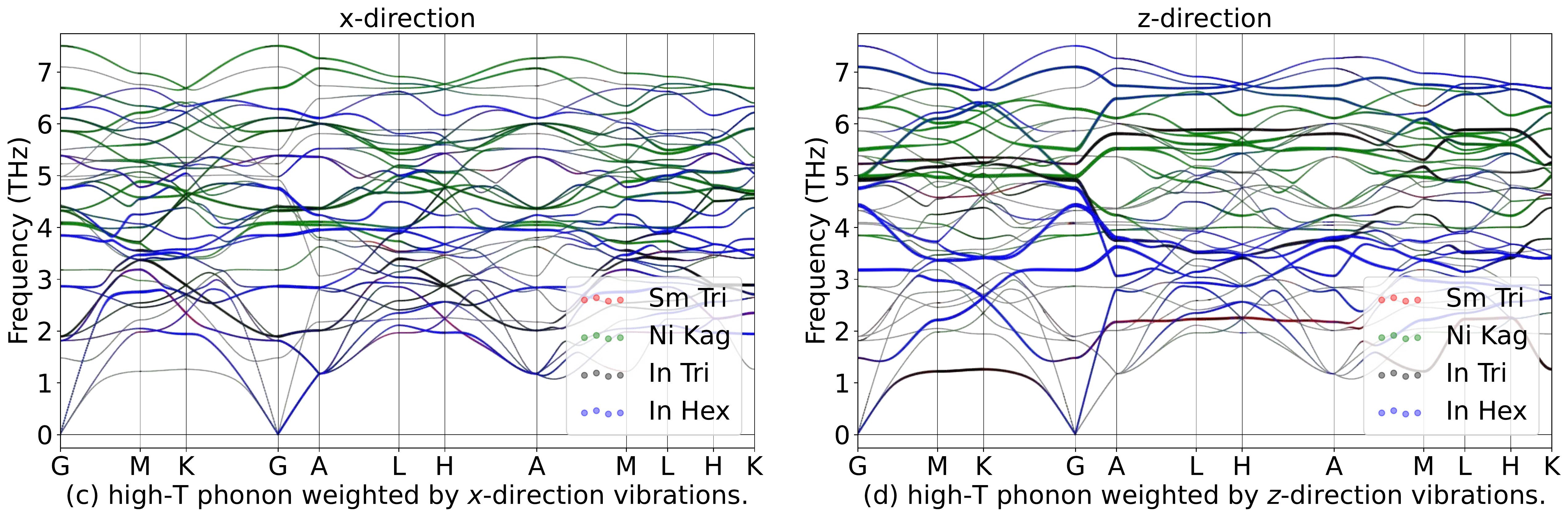}
\caption{Phonon spectrum for SmNi$_6$In$_6$in in-plane ($x$) and out-of-plane ($z$) directions.}
\label{fig:SmNi6In6phoxyz}
\end{figure}

\begin{table}[htbp]
\global\long\def\arraystretch{1.12}
\captionsetup{width=.95\textwidth}
\caption{IRREPs at high-symmetry points for SmNi$_6$In$_6$ phonon spectrum at Low-T.}
\label{tab:191_SmNi6In6_Low}
\setlength{\tabcolsep}{1.mm}{
}
\end{table}

\begin{table}[htbp]
\global\long\def\arraystretch{1.12}
\captionsetup{width=.95\textwidth}
\caption{IRREPs at high-symmetry points for SmNi$_6$In$_6$ phonon spectrum at High-T.}
\label{tab:191_SmNi6In6_High}
\setlength{\tabcolsep}{1.mm}{
%
}
\end{table}
\subsubsection{\label{sms2sec:GdNi6In6}\hyperref[smsec:col]{\texorpdfstring{GdNi$_6$In$_6$}{}}~{\color{green}  (High-T stable, Low-T stable) }}

\begin{table}[htbp]
\global\long\def\arraystretch{1.12}
\captionsetup{width=.85\textwidth}
\caption{Structure parameters of GdNi$_6$In$_6$ after relaxation. It contains 526 electrons. }
\label{tab:GdNi6In6}
\setlength{\tabcolsep}{4mm}{
%
}
\end{table}

\begin{figure}[htbp]
\centering
\includegraphics[width=0.85\textwidth]{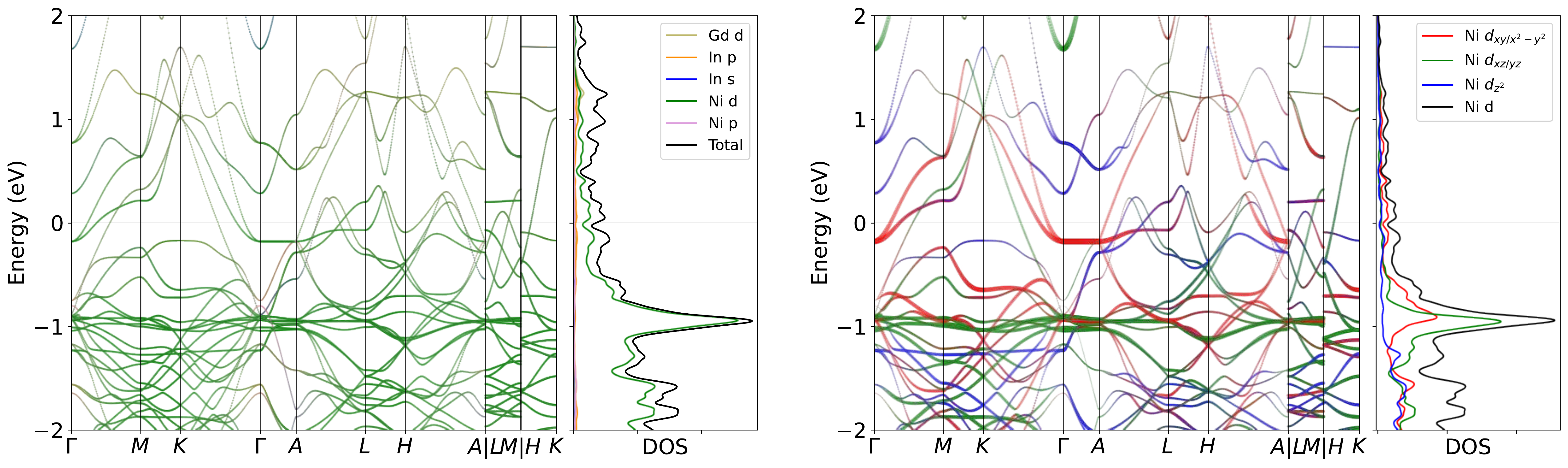}
\caption{Band structure for GdNi$_6$In$_6$ without SOC. The projection of Ni $3d$ orbitals is also presented. The band structures clearly reveal the dominating of Ni $3d$ orbitals  near the Fermi level, as well as the pronounced peak.}
\label{fig:GdNi6In6band}
\end{figure}

\begin{figure}[H]
\centering
\subfloat[Relaxed phonon at low-T.]{
\label{fig:GdNi6In6_relaxed_Low}
\includegraphics[width=0.45\textwidth]{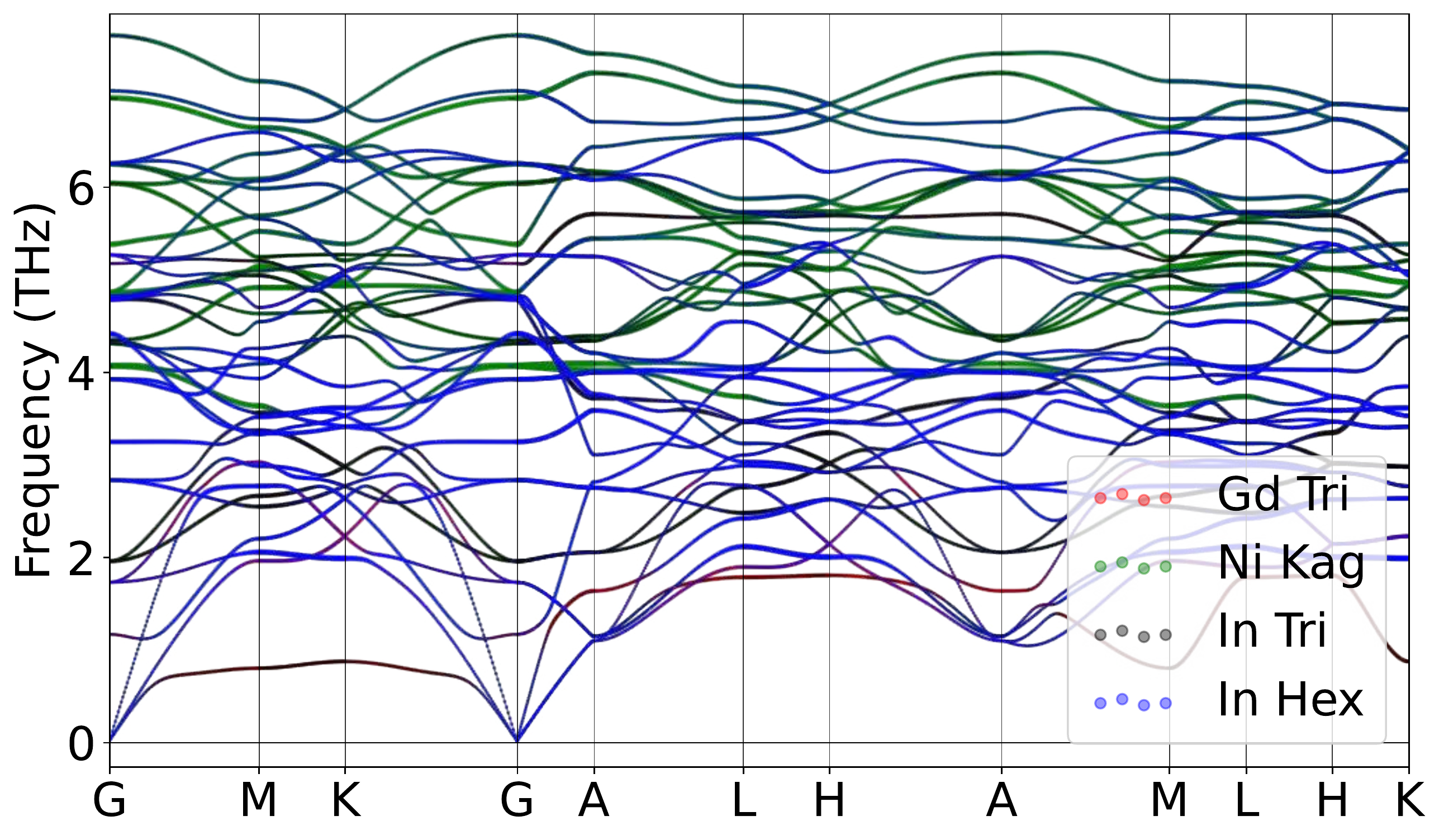}
}
\hspace{5pt}\subfloat[Relaxed phonon at high-T.]{
\label{fig:GdNi6In6_relaxed_High}
\includegraphics[width=0.45\textwidth]{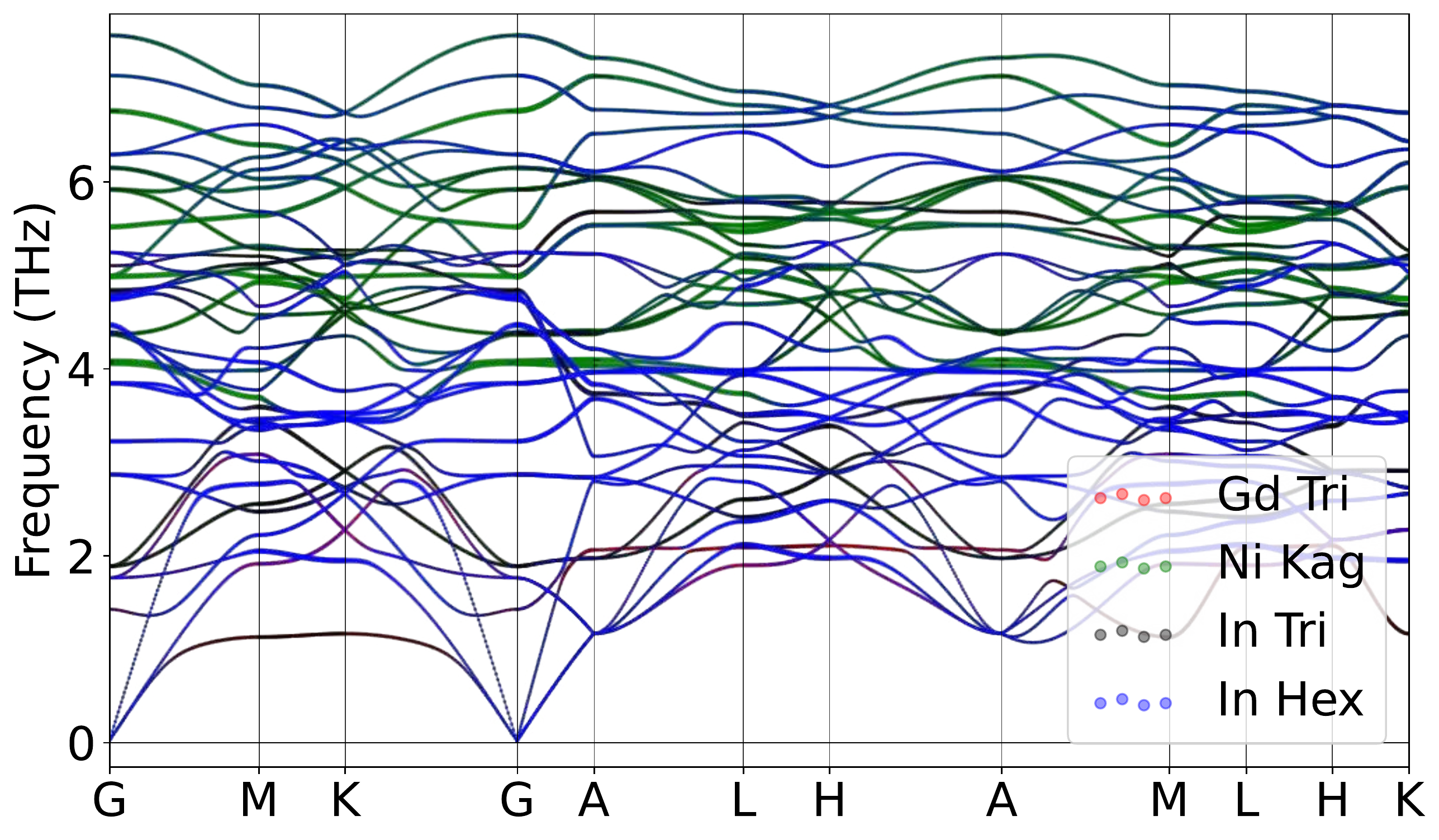}
}
\caption{Atomically projected phonon spectrum for GdNi$_6$In$_6$.}
\label{fig:GdNi6In6pho}
\end{figure}

\begin{figure}[H]
\centering
\label{fig:GdNi6In6_relaxed_Low_xz}
\includegraphics[width=0.85\textwidth]{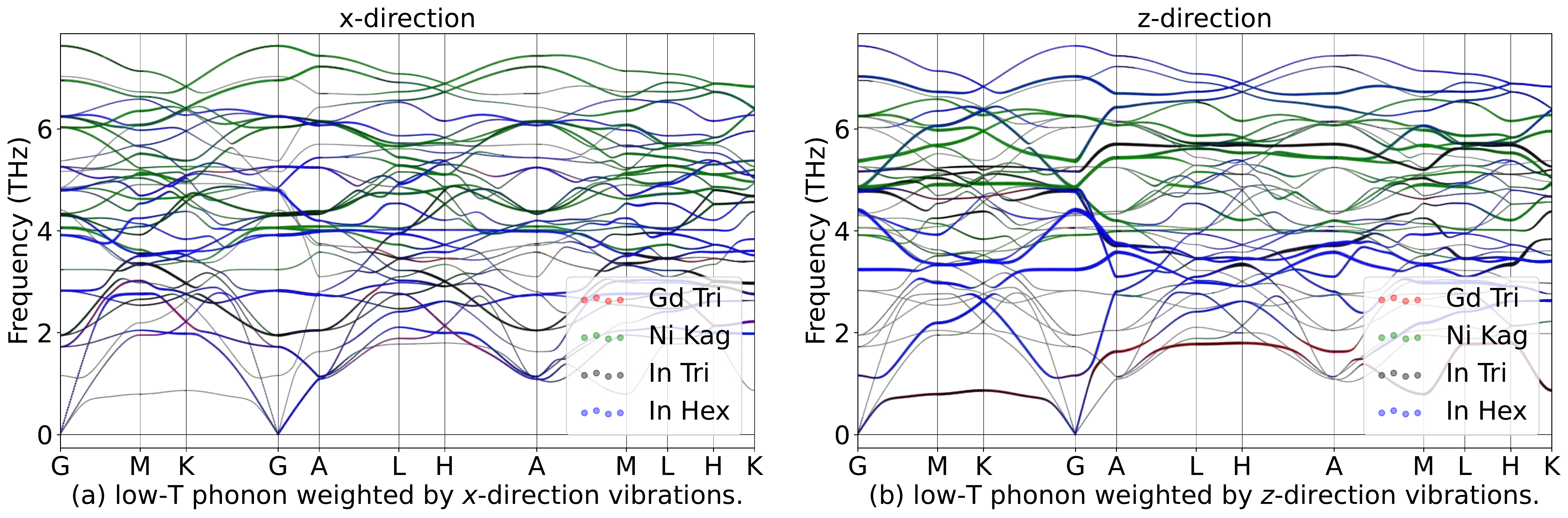}
\hspace{5pt}\label{fig:GdNi6In6_relaxed_High_xz}
\includegraphics[width=0.85\textwidth]{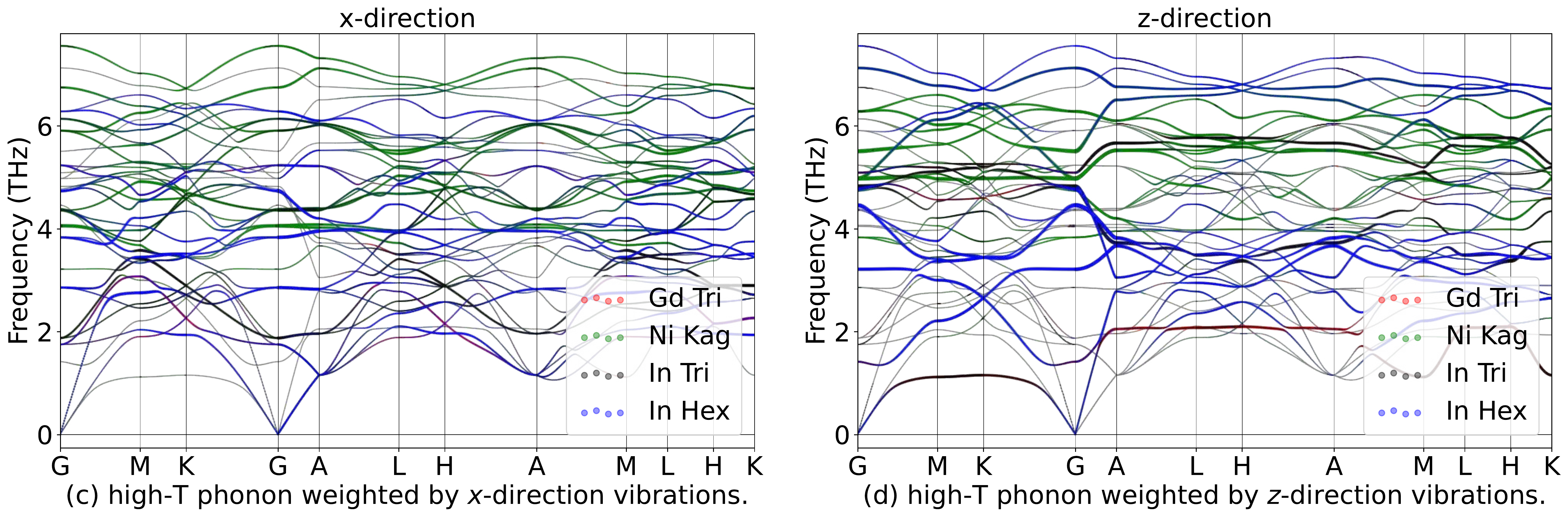}
\caption{Phonon spectrum for GdNi$_6$In$_6$in in-plane ($x$) and out-of-plane ($z$) directions.}
\label{fig:GdNi6In6phoxyz}
\end{figure}

\begin{table}[htbp]
\global\long\def\arraystretch{1.12}
\captionsetup{width=.95\textwidth}
\caption{IRREPs at high-symmetry points for GdNi$_6$In$_6$ phonon spectrum at Low-T.}
\label{tab:191_GdNi6In6_Low}
\setlength{\tabcolsep}{1.mm}{
}
\end{table}

\begin{table}[htbp]
\global\long\def\arraystretch{1.12}
\captionsetup{width=.95\textwidth}
\caption{IRREPs at high-symmetry points for GdNi$_6$In$_6$ phonon spectrum at High-T.}
\label{tab:191_GdNi6In6_High}
\setlength{\tabcolsep}{1.mm}{
%
}
\end{table}
\subsubsection{\label{sms2sec:TbNi6In6}\hyperref[smsec:col]{\texorpdfstring{TbNi$_6$In$_6$}{}}~{\color{green}  (High-T stable, Low-T stable) }}

\begin{table}[htbp]
\global\long\def\arraystretch{1.12}
\captionsetup{width=.85\textwidth}
\caption{Structure parameters of TbNi$_6$In$_6$ after relaxation. It contains 527 electrons. }
\label{tab:TbNi6In6}
\setlength{\tabcolsep}{4mm}{
%
}
\end{table}

\begin{figure}[htbp]
\centering
\includegraphics[width=0.85\textwidth]{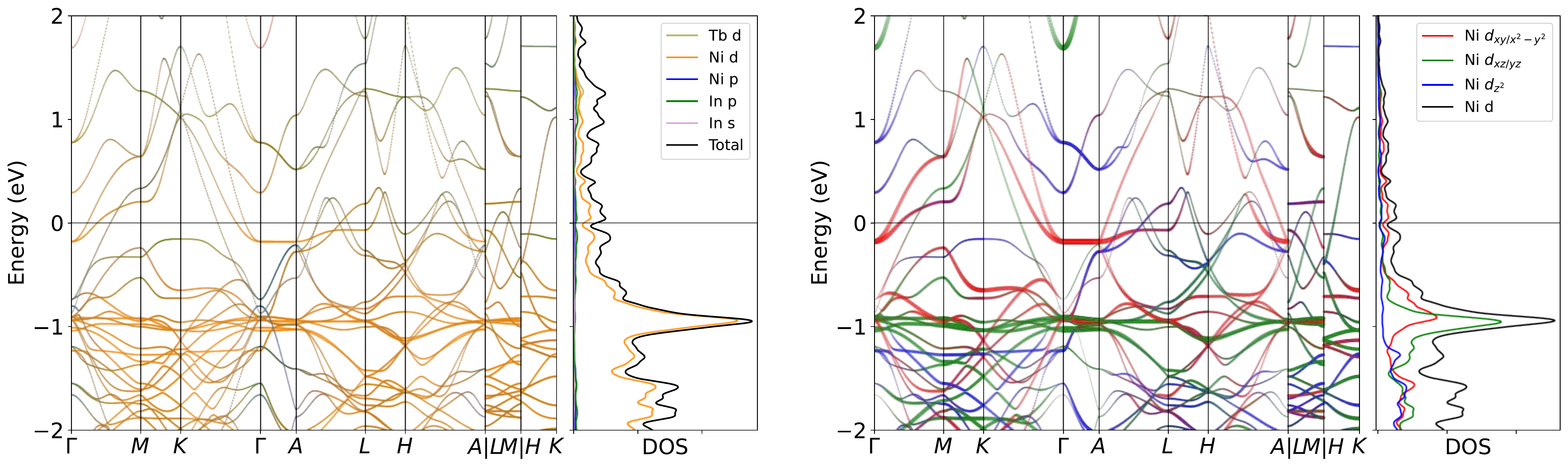}
\caption{Band structure for TbNi$_6$In$_6$ without SOC. The projection of Ni $3d$ orbitals is also presented. The band structures clearly reveal the dominating of Ni $3d$ orbitals  near the Fermi level, as well as the pronounced peak.}
\label{fig:TbNi6In6band}
\end{figure}

\begin{figure}[H]
\centering
\subfloat[Relaxed phonon at low-T.]{
\label{fig:TbNi6In6_relaxed_Low}
\includegraphics[width=0.45\textwidth]{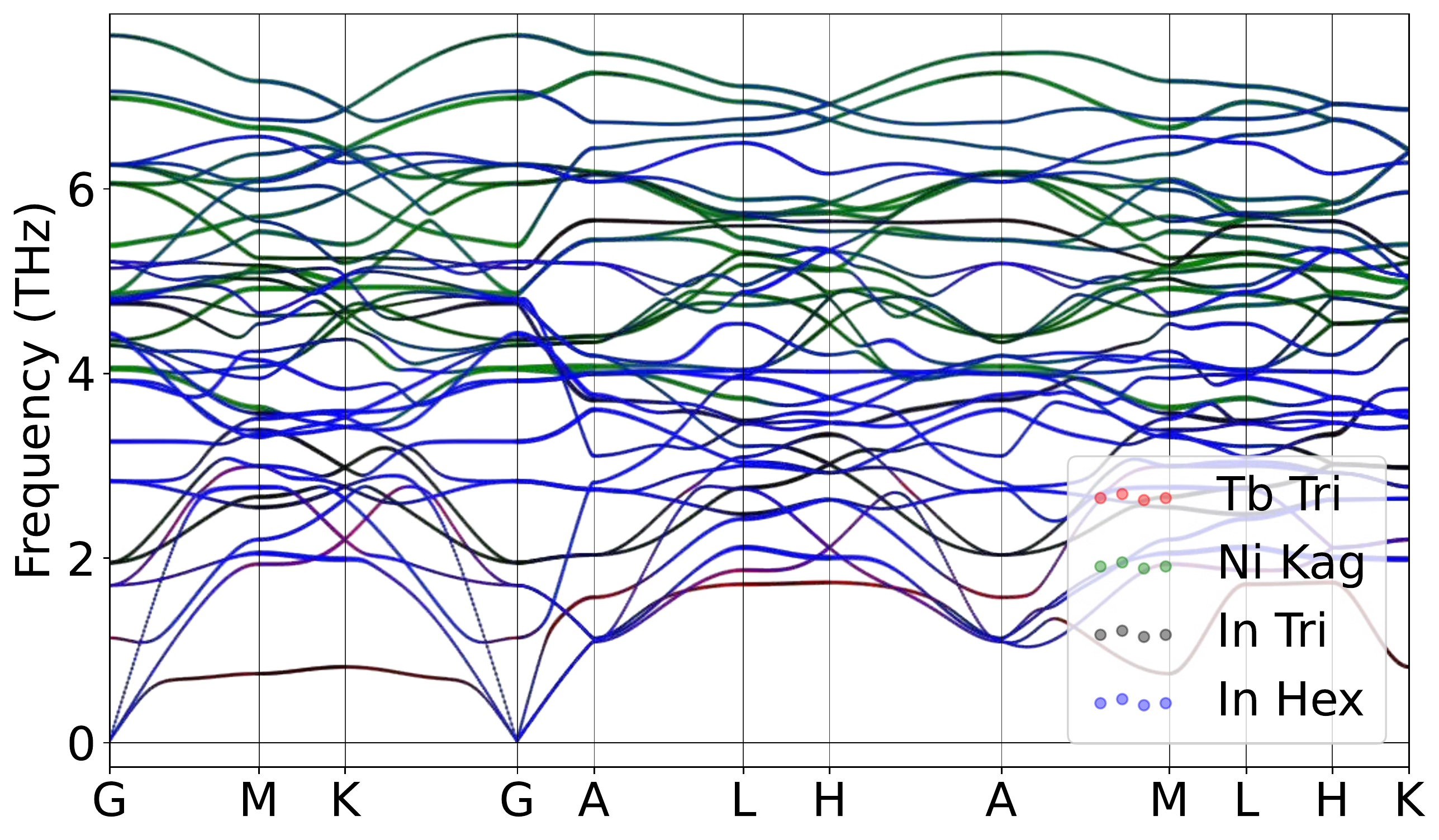}
}
\hspace{5pt}\subfloat[Relaxed phonon at high-T.]{
\label{fig:TbNi6In6_relaxed_High}
\includegraphics[width=0.45\textwidth]{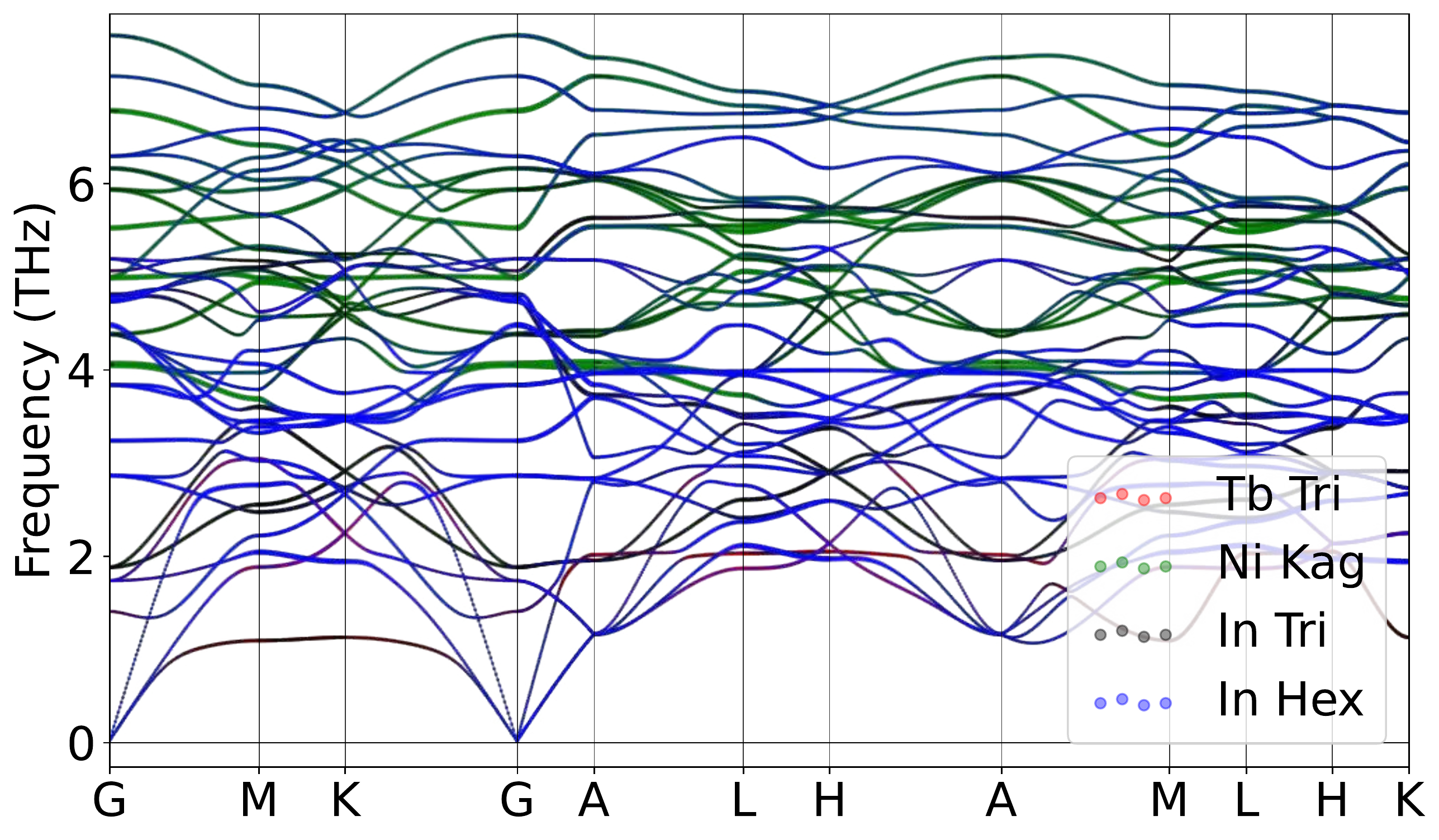}
}
\caption{Atomically projected phonon spectrum for TbNi$_6$In$_6$.}
\label{fig:TbNi6In6pho}
\end{figure}

\begin{figure}[H]
\centering
\label{fig:TbNi6In6_relaxed_Low_xz}
\includegraphics[width=0.85\textwidth]{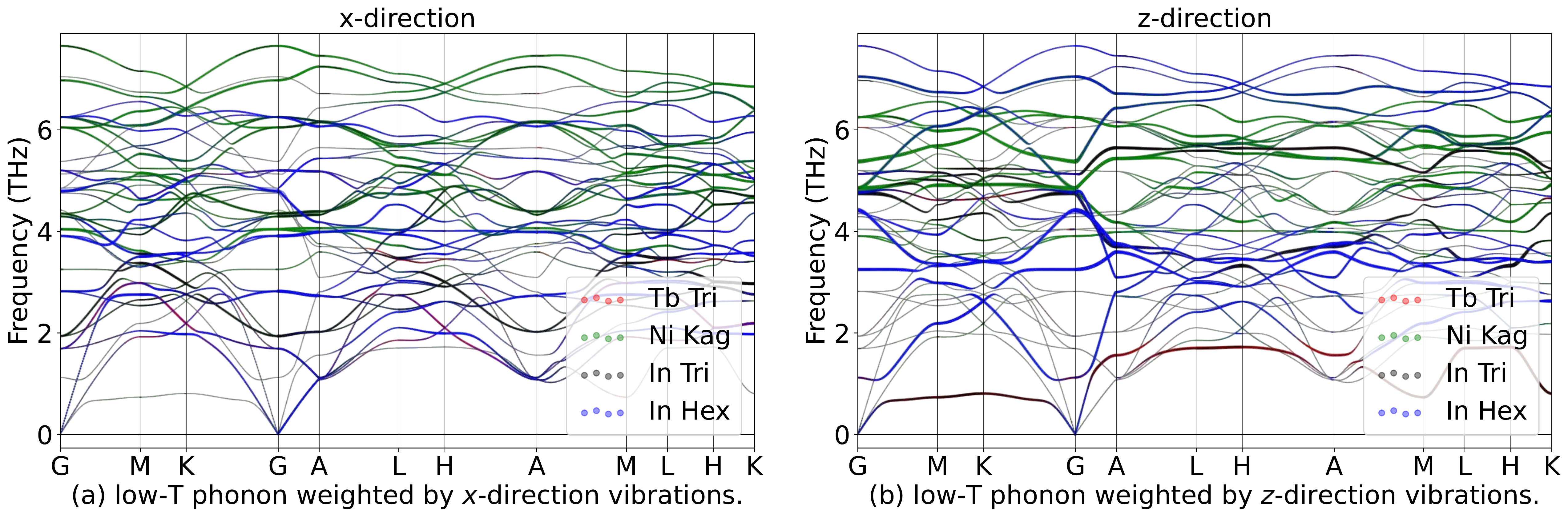}
\hspace{5pt}\label{fig:TbNi6In6_relaxed_High_xz}
\includegraphics[width=0.85\textwidth]{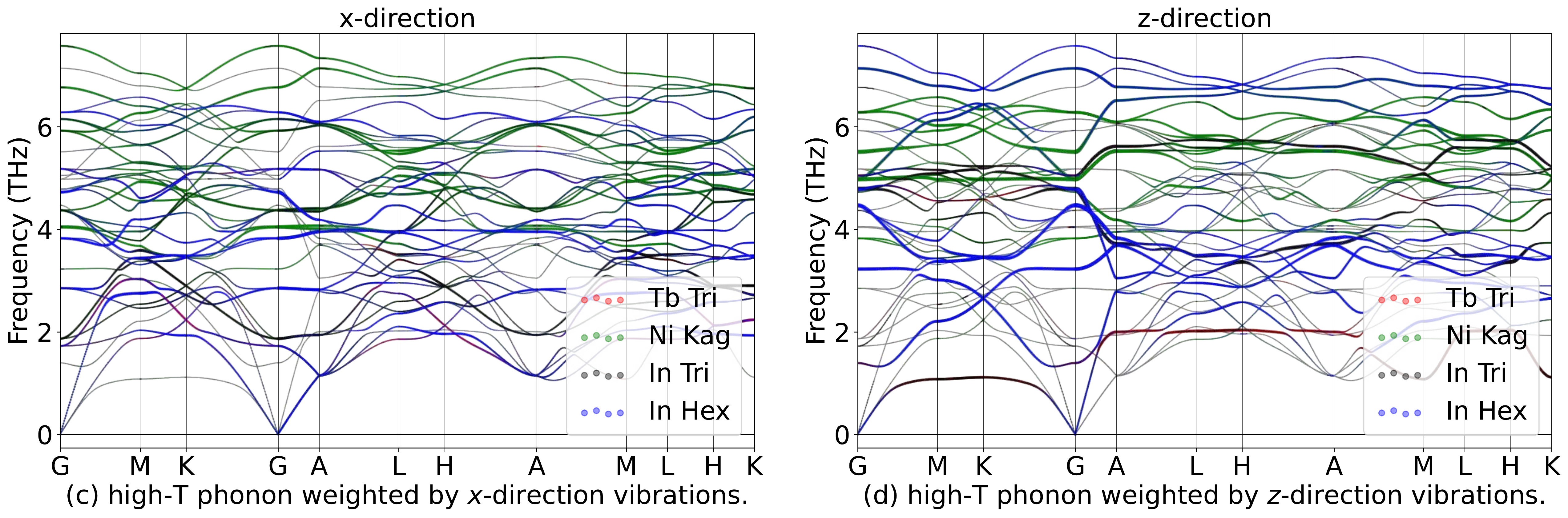}
\caption{Phonon spectrum for TbNi$_6$In$_6$in in-plane ($x$) and out-of-plane ($z$) directions.}
\label{fig:TbNi6In6phoxyz}
\end{figure}

\begin{table}[htbp]
\global\long\def\arraystretch{1.12}
\captionsetup{width=.95\textwidth}
\caption{IRREPs at high-symmetry points for TbNi$_6$In$_6$ phonon spectrum at Low-T.}
\label{tab:191_TbNi6In6_Low}
\setlength{\tabcolsep}{1.mm}{
}
\end{table}

\begin{table}[htbp]
\global\long\def\arraystretch{1.12}
\captionsetup{width=.95\textwidth}
\caption{IRREPs at high-symmetry points for TbNi$_6$In$_6$ phonon spectrum at High-T.}
\label{tab:191_TbNi6In6_High}
\setlength{\tabcolsep}{1.mm}{
%
}
\end{table}
\subsubsection{\label{sms2sec:DyNi6In6}\hyperref[smsec:col]{\texorpdfstring{DyNi$_6$In$_6$}{}}~{\color{green}  (High-T stable, Low-T stable) }}

\begin{table}[htbp]
\global\long\def\arraystretch{1.12}
\captionsetup{width=.85\textwidth}
\caption{Structure parameters of DyNi$_6$In$_6$ after relaxation. It contains 528 electrons. }
\label{tab:DyNi6In6}
\setlength{\tabcolsep}{4mm}{
%
}
\end{table}

\begin{figure}[htbp]
\centering
\includegraphics[width=0.85\textwidth]{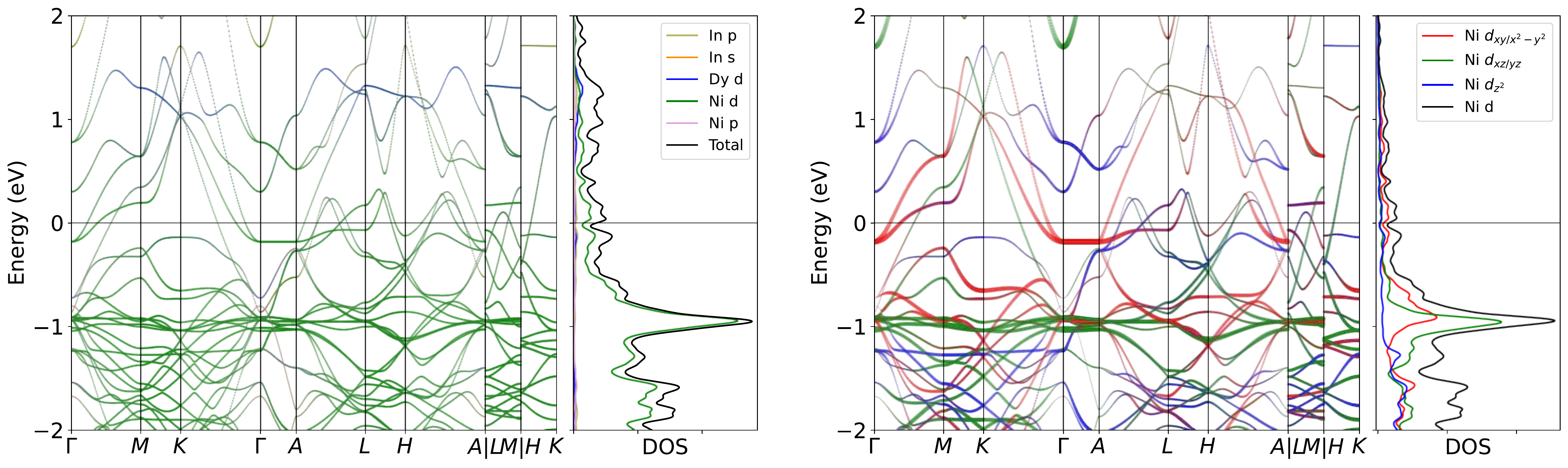}
\caption{Band structure for DyNi$_6$In$_6$ without SOC. The projection of Ni $3d$ orbitals is also presented. The band structures clearly reveal the dominating of Ni $3d$ orbitals  near the Fermi level, as well as the pronounced peak.}
\label{fig:DyNi6In6band}
\end{figure}

\begin{figure}[H]
\centering
\subfloat[Relaxed phonon at low-T.]{
\label{fig:DyNi6In6_relaxed_Low}
\includegraphics[width=0.45\textwidth]{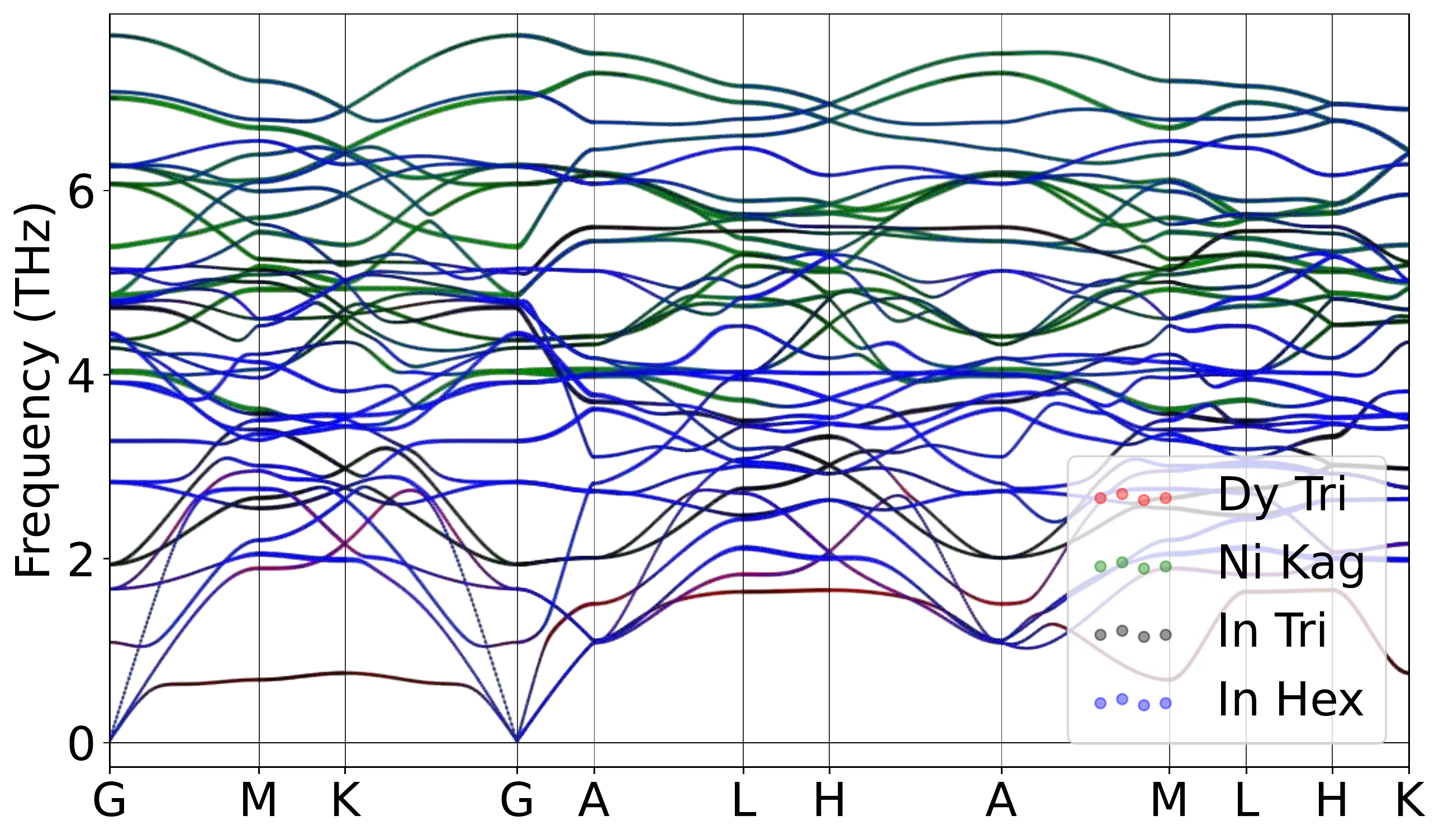}
}
\hspace{5pt}\subfloat[Relaxed phonon at high-T.]{
\label{fig:DyNi6In6_relaxed_High}
\includegraphics[width=0.45\textwidth]{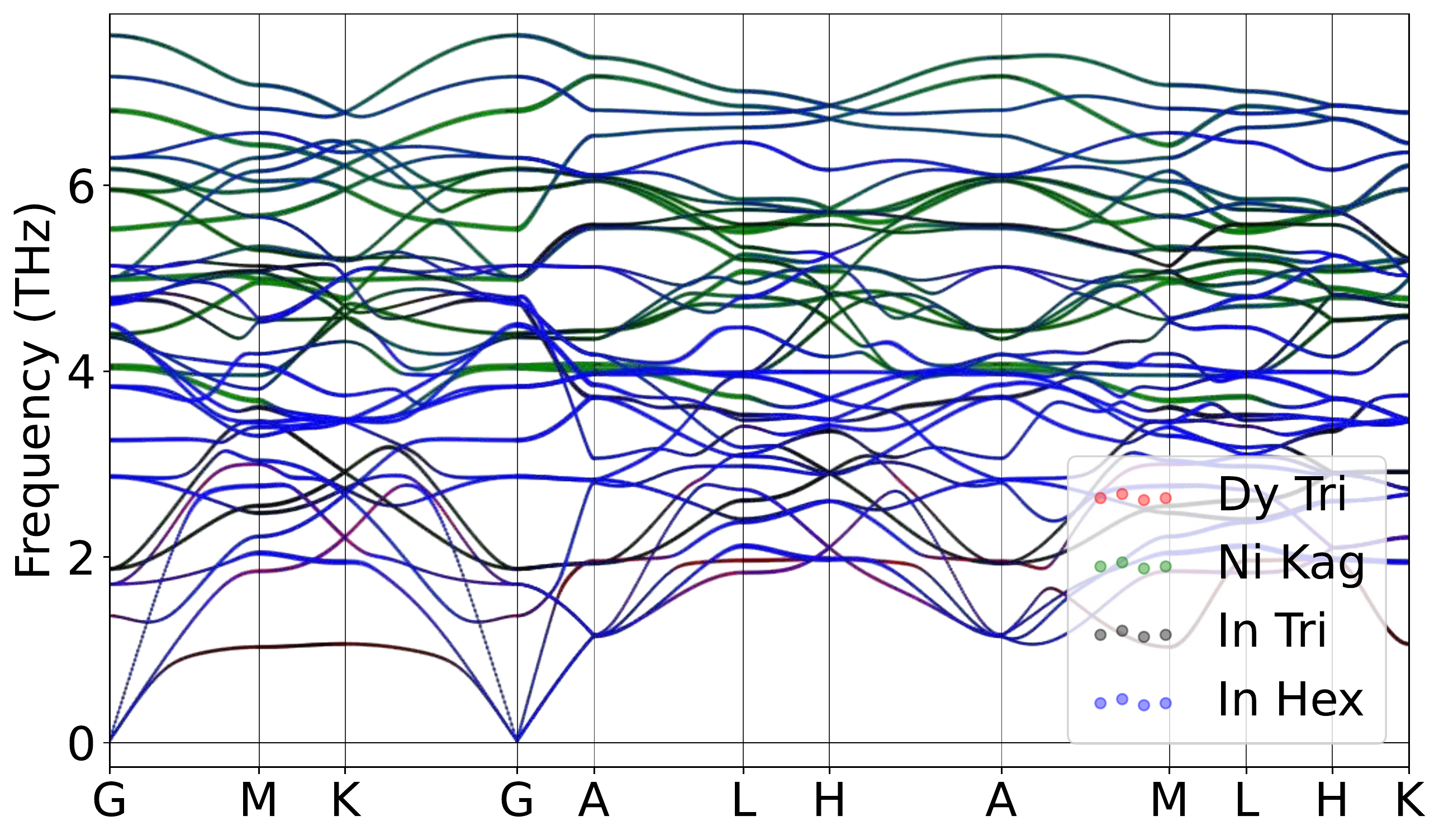}
}
\caption{Atomically projected phonon spectrum for DyNi$_6$In$_6$.}
\label{fig:DyNi6In6pho}
\end{figure}

\begin{figure}[H]
\centering
\label{fig:DyNi6In6_relaxed_Low_xz}
\includegraphics[width=0.85\textwidth]{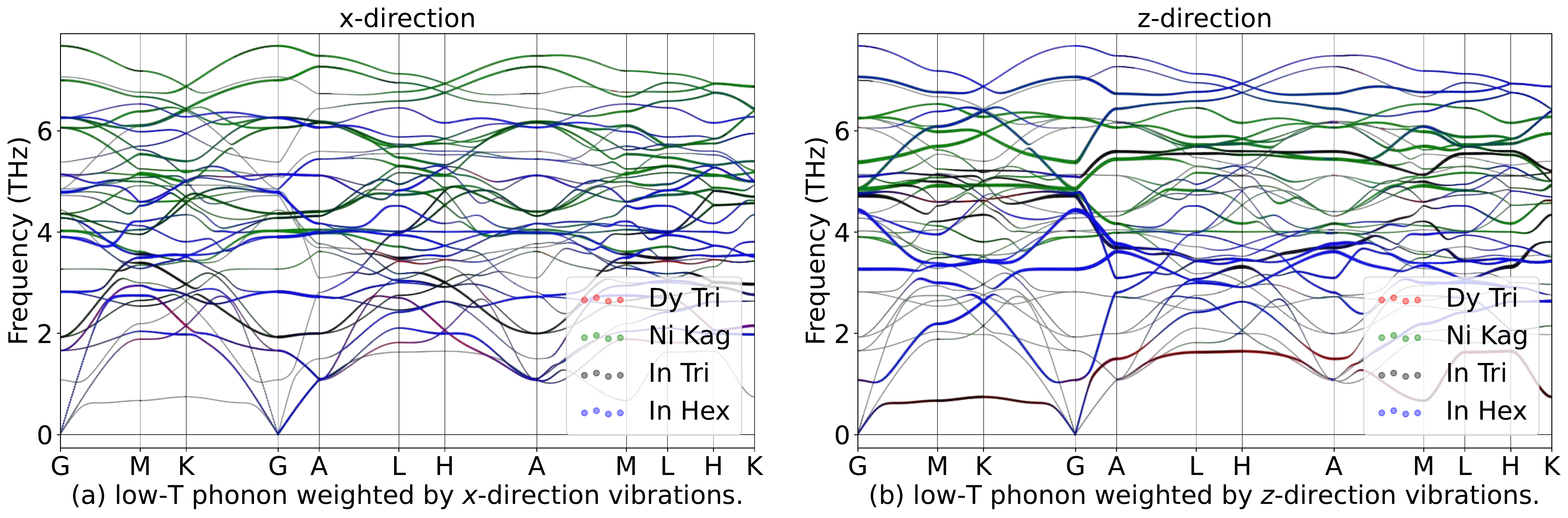}
\hspace{5pt}\label{fig:DyNi6In6_relaxed_High_xz}
\includegraphics[width=0.85\textwidth]{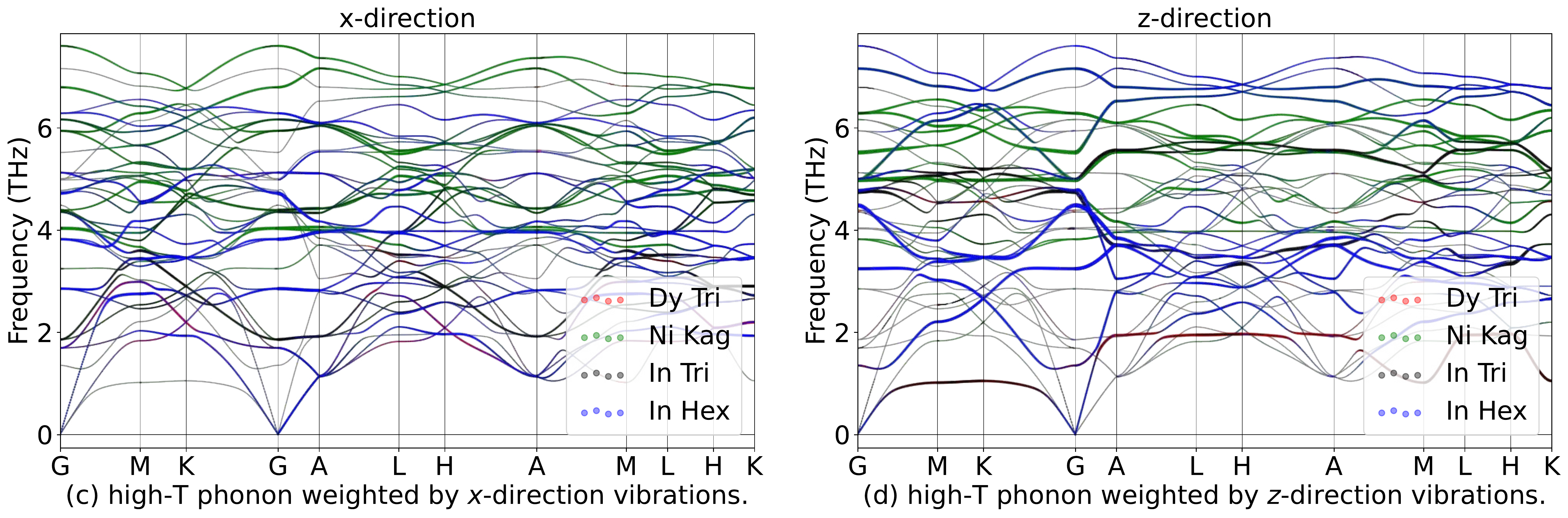}
\caption{Phonon spectrum for DyNi$_6$In$_6$in in-plane ($x$) and out-of-plane ($z$) directions.}
\label{fig:DyNi6In6phoxyz}
\end{figure}

\begin{table}[htbp]
\global\long\def\arraystretch{1.12}
\captionsetup{width=.95\textwidth}
\caption{IRREPs at high-symmetry points for DyNi$_6$In$_6$ phonon spectrum at Low-T.}
\label{tab:191_DyNi6In6_Low}
\setlength{\tabcolsep}{1.mm}{
}
\end{table}

\begin{table}[htbp]
\global\long\def\arraystretch{1.12}
\captionsetup{width=.95\textwidth}
\caption{IRREPs at high-symmetry points for DyNi$_6$In$_6$ phonon spectrum at High-T.}
\label{tab:191_DyNi6In6_High}
\setlength{\tabcolsep}{1.mm}{
%
}
\end{table}
\subsubsection{\label{sms2sec:HoNi6In6}\hyperref[smsec:col]{\texorpdfstring{HoNi$_6$In$_6$}{}}~{\color{green}  (High-T stable, Low-T stable) }}

\begin{table}[htbp]
\global\long\def\arraystretch{1.12}
\captionsetup{width=.85\textwidth}
\caption{Structure parameters of HoNi$_6$In$_6$ after relaxation. It contains 529 electrons. }
\label{tab:HoNi6In6}
\setlength{\tabcolsep}{4mm}{
%
}
\end{table}

\begin{figure}[htbp]
\centering
\includegraphics[width=0.85\textwidth]{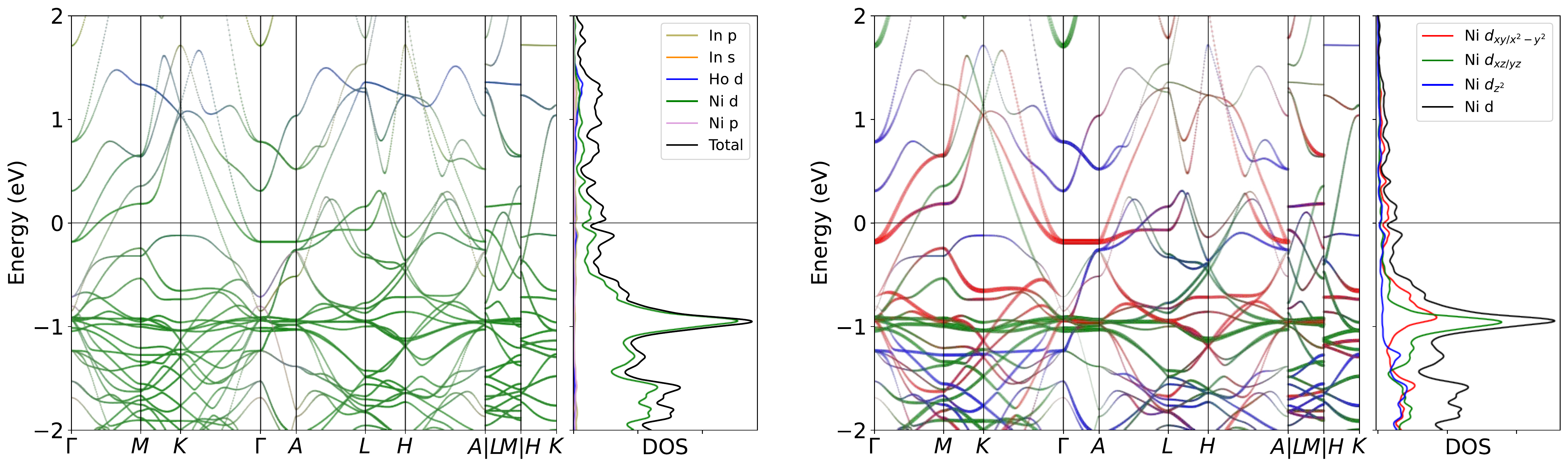}
\caption{Band structure for HoNi$_6$In$_6$ without SOC. The projection of Ni $3d$ orbitals is also presented. The band structures clearly reveal the dominating of Ni $3d$ orbitals  near the Fermi level, as well as the pronounced peak.}
\label{fig:HoNi6In6band}
\end{figure}

\begin{figure}[H]
\centering
\subfloat[Relaxed phonon at low-T.]{
\label{fig:HoNi6In6_relaxed_Low}
\includegraphics[width=0.45\textwidth]{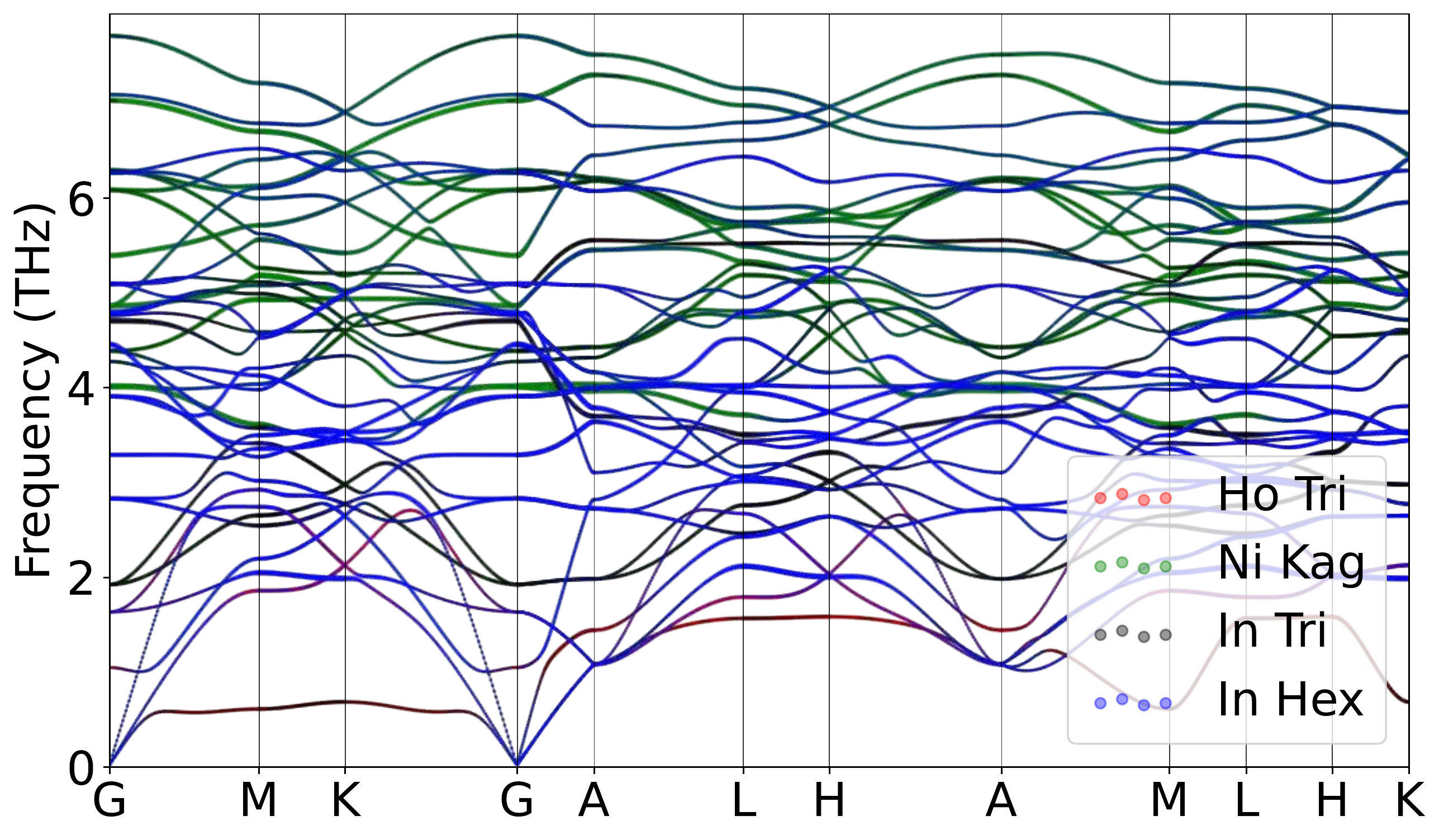}
}
\hspace{5pt}\subfloat[Relaxed phonon at high-T.]{
\label{fig:HoNi6In6_relaxed_High}
\includegraphics[width=0.45\textwidth]{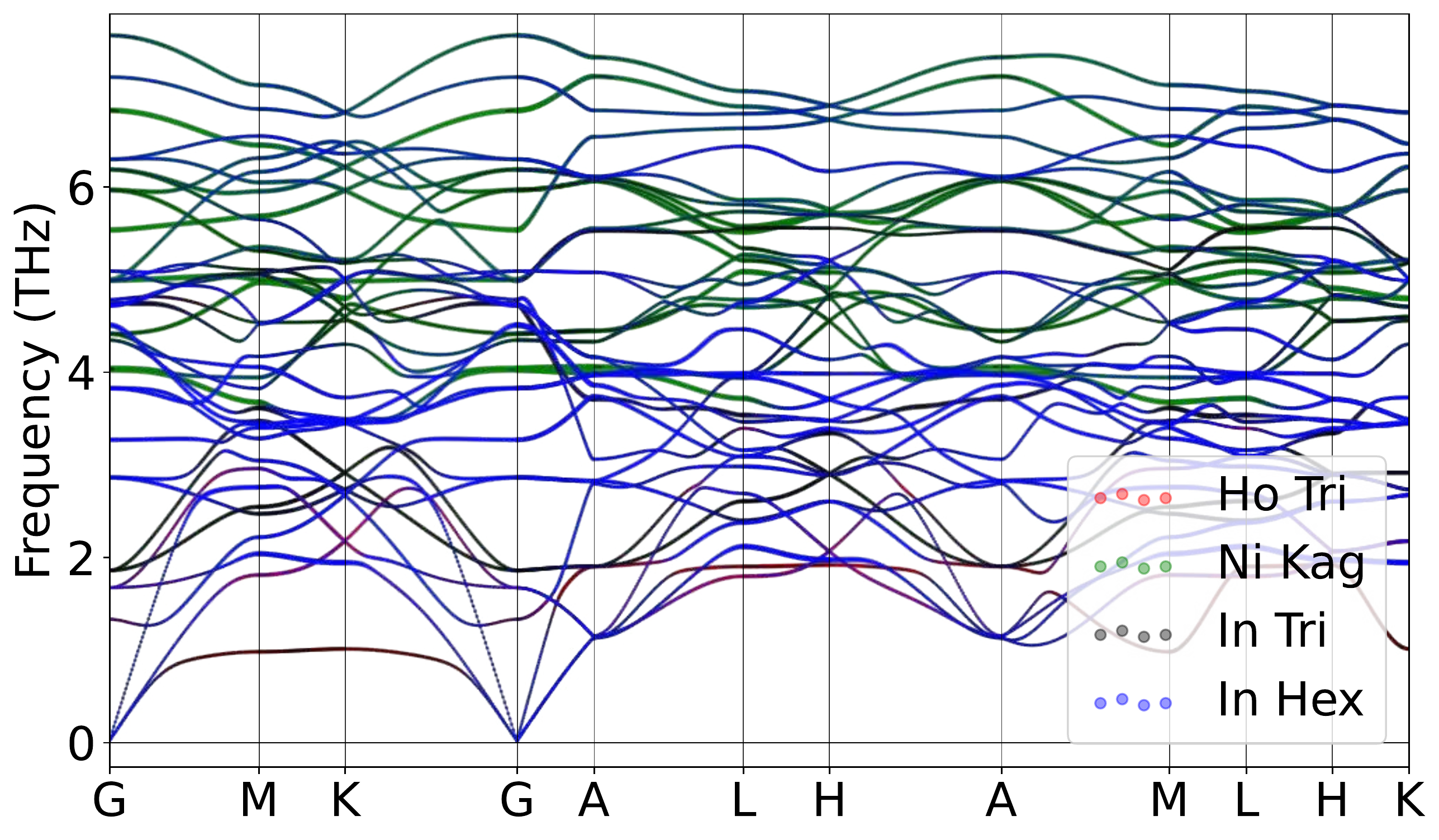}
}
\caption{Atomically projected phonon spectrum for HoNi$_6$In$_6$.}
\label{fig:HoNi6In6pho}
\end{figure}

\begin{figure}[H]
\centering
\label{fig:HoNi6In6_relaxed_Low_xz}
\includegraphics[width=0.85\textwidth]{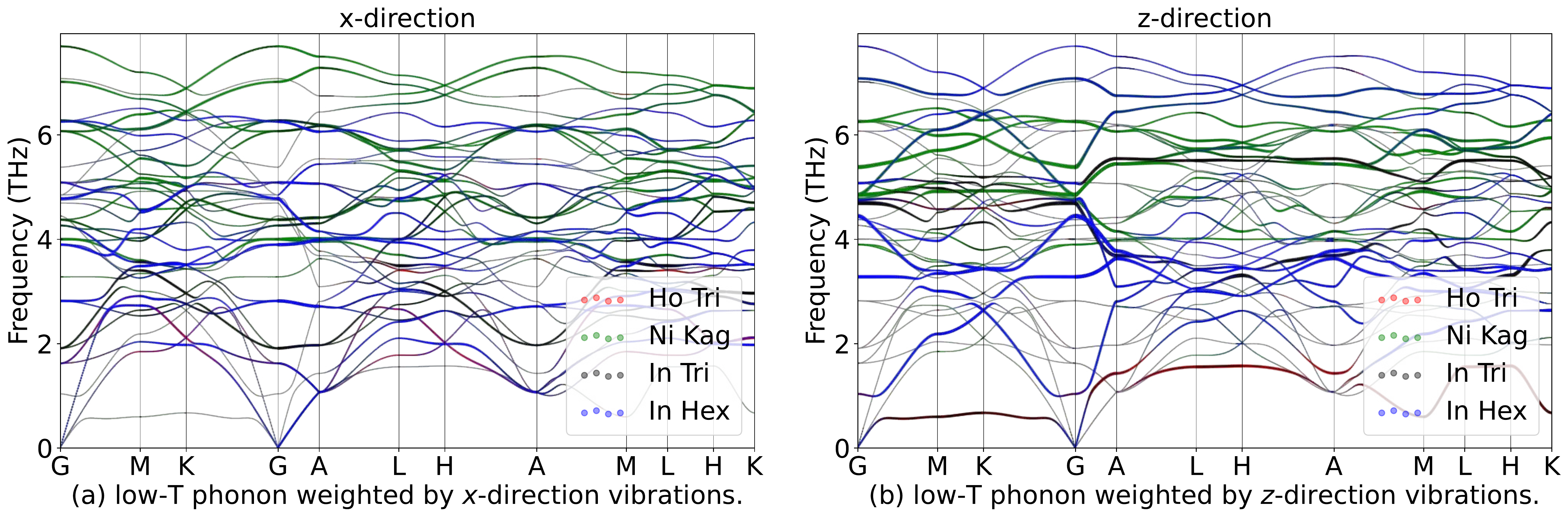}
\hspace{5pt}\label{fig:HoNi6In6_relaxed_High_xz}
\includegraphics[width=0.85\textwidth]{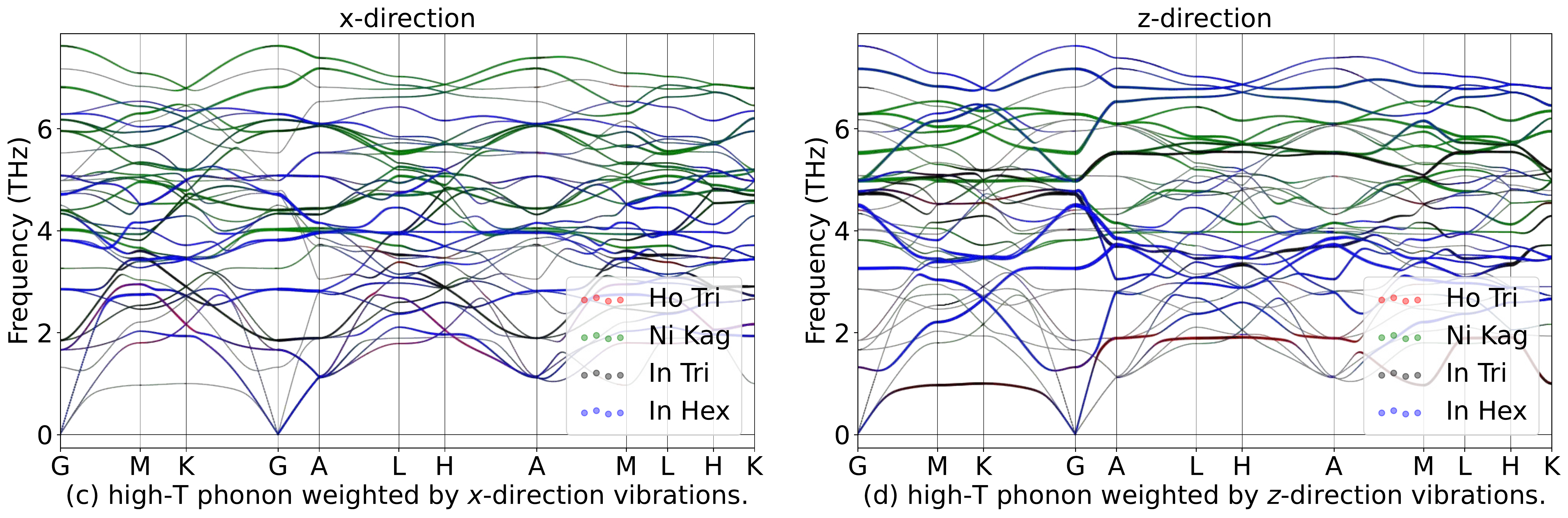}
\caption{Phonon spectrum for HoNi$_6$In$_6$in in-plane ($x$) and out-of-plane ($z$) directions.}
\label{fig:HoNi6In6phoxyz}
\end{figure}

\begin{table}[htbp]
\global\long\def\arraystretch{1.12}
\captionsetup{width=.95\textwidth}
\caption{IRREPs at high-symmetry points for HoNi$_6$In$_6$ phonon spectrum at Low-T.}
\label{tab:191_HoNi6In6_Low}
\setlength{\tabcolsep}{1.mm}{
}
\end{table}

\begin{table}[htbp]
\global\long\def\arraystretch{1.12}
\captionsetup{width=.95\textwidth}
\caption{IRREPs at high-symmetry points for HoNi$_6$In$_6$ phonon spectrum at High-T.}
\label{tab:191_HoNi6In6_High}
\setlength{\tabcolsep}{1.mm}{
%
}
\end{table}
\subsubsection{\label{sms2sec:ErNi6In6}\hyperref[smsec:col]{\texorpdfstring{ErNi$_6$In$_6$}{}}~{\color{green}  (High-T stable, Low-T stable) }}

\begin{table}[htbp]
\global\long\def\arraystretch{1.12}
\captionsetup{width=.85\textwidth}
\caption{Structure parameters of ErNi$_6$In$_6$ after relaxation. It contains 530 electrons. }
\label{tab:ErNi6In6}
\setlength{\tabcolsep}{4mm}{
%
}
\end{table}

\begin{figure}[htbp]
\centering
\includegraphics[width=0.85\textwidth]{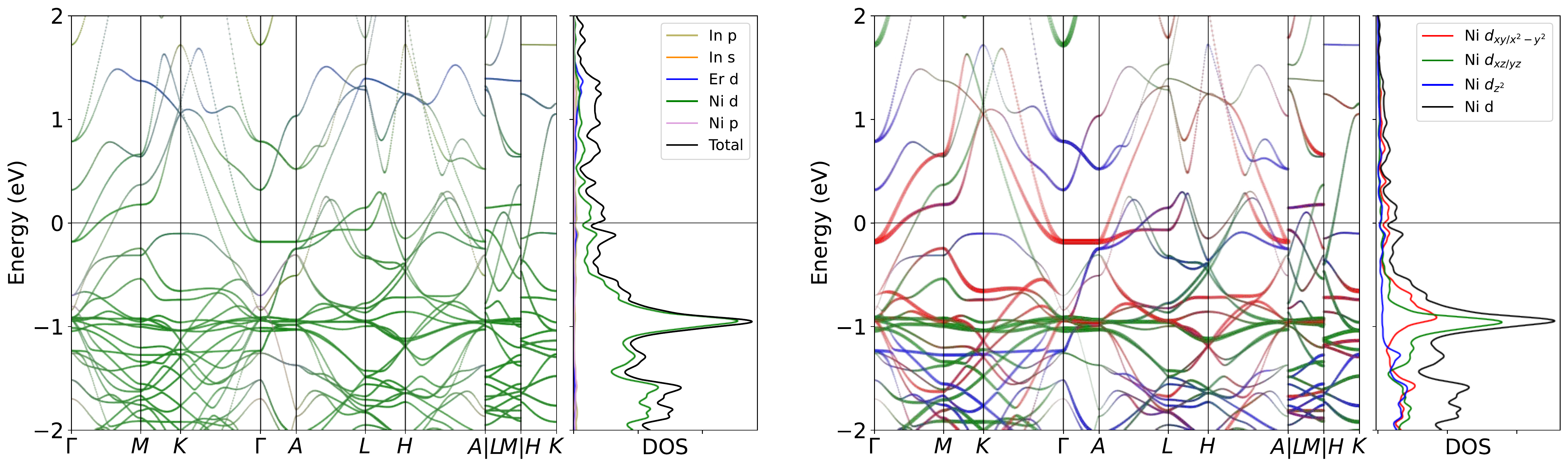}
\caption{Band structure for ErNi$_6$In$_6$ without SOC. The projection of Ni $3d$ orbitals is also presented. The band structures clearly reveal the dominating of Ni $3d$ orbitals  near the Fermi level, as well as the pronounced peak.}
\label{fig:ErNi6In6band}
\end{figure}

\begin{figure}[H]
\centering
\subfloat[Relaxed phonon at low-T.]{
\label{fig:ErNi6In6_relaxed_Low}
\includegraphics[width=0.45\textwidth]{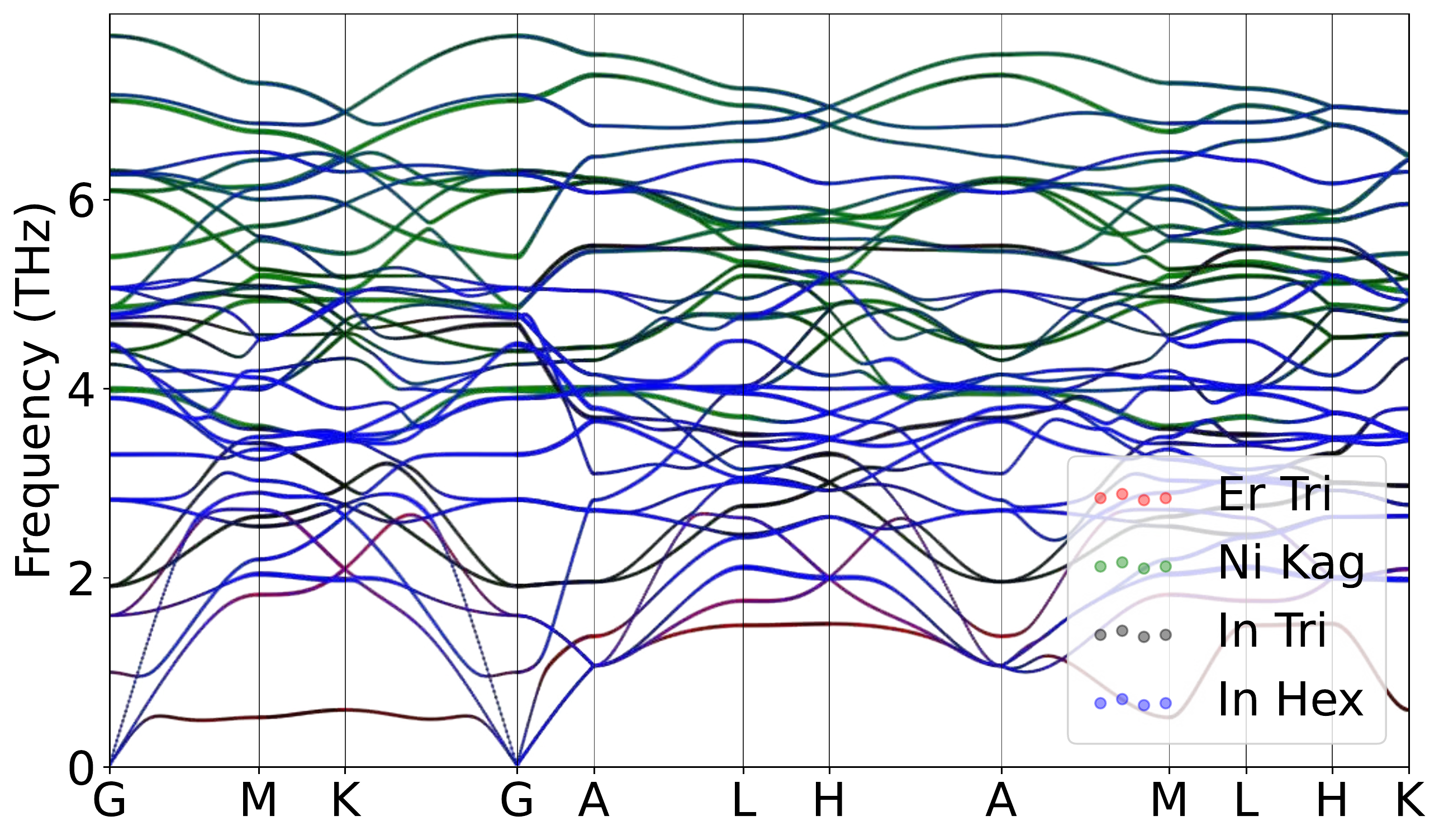}
}
\hspace{5pt}\subfloat[Relaxed phonon at high-T.]{
\label{fig:ErNi6In6_relaxed_High}
\includegraphics[width=0.45\textwidth]{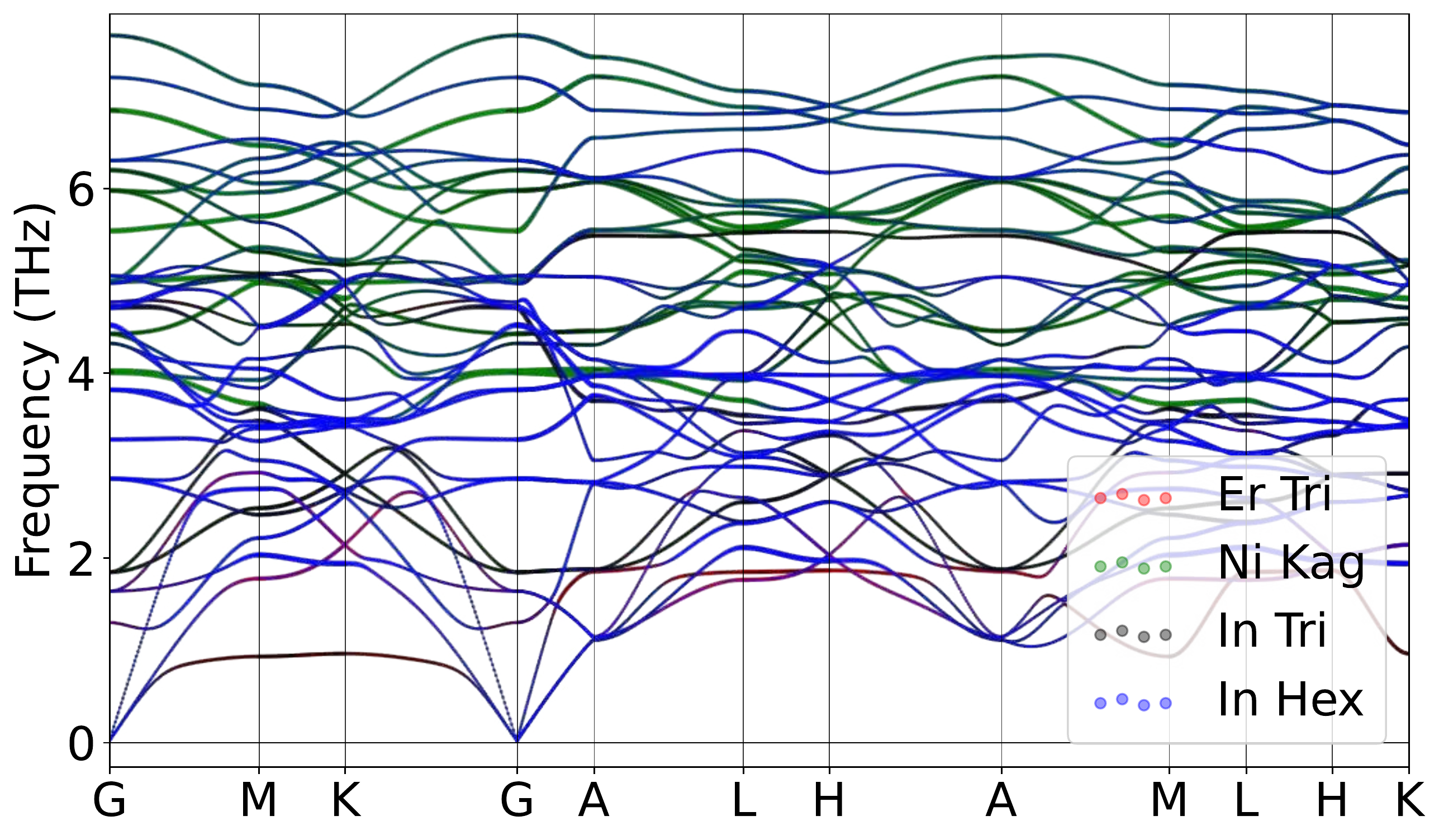}
}
\caption{Atomically projected phonon spectrum for ErNi$_6$In$_6$.}
\label{fig:ErNi6In6pho}
\end{figure}

\begin{figure}[H]
\centering
\label{fig:ErNi6In6_relaxed_Low_xz}
\includegraphics[width=0.85\textwidth]{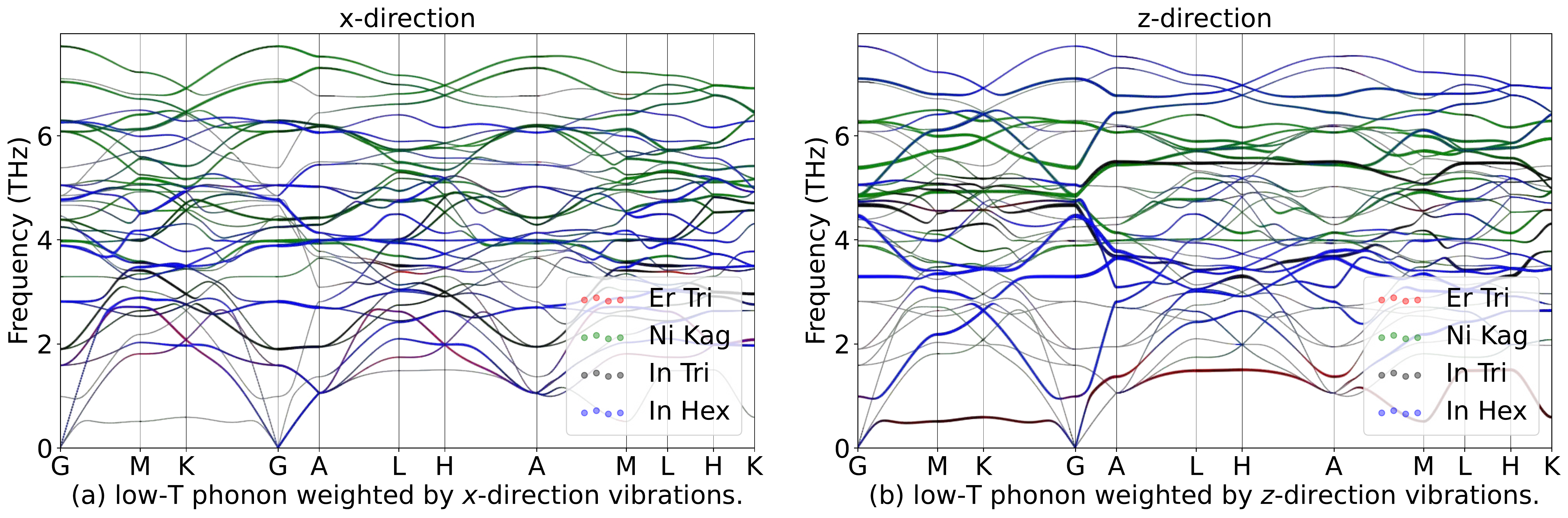}
\hspace{5pt}\label{fig:ErNi6In6_relaxed_High_xz}
\includegraphics[width=0.85\textwidth]{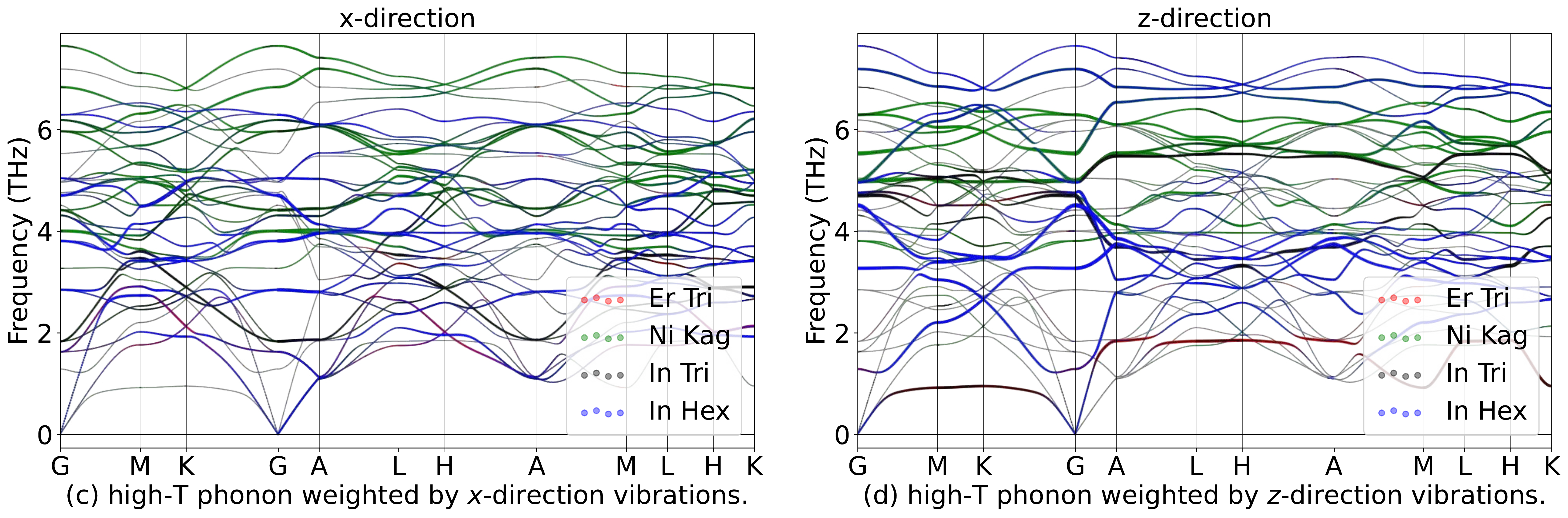}
\caption{Phonon spectrum for ErNi$_6$In$_6$in in-plane ($x$) and out-of-plane ($z$) directions.}
\label{fig:ErNi6In6phoxyz}
\end{figure}

\begin{table}[htbp]
\global\long\def\arraystretch{1.12}
\captionsetup{width=.95\textwidth}
\caption{IRREPs at high-symmetry points for ErNi$_6$In$_6$ phonon spectrum at Low-T.}
\label{tab:191_ErNi6In6_Low}
\setlength{\tabcolsep}{1.mm}{
}
\end{table}

\begin{table}[htbp]
\global\long\def\arraystretch{1.12}
\captionsetup{width=.95\textwidth}
\caption{IRREPs at high-symmetry points for ErNi$_6$In$_6$ phonon spectrum at High-T.}
\label{tab:191_ErNi6In6_High}
\setlength{\tabcolsep}{1.mm}{
%
}
\end{table}
\subsubsection{\label{sms2sec:TmNi6In6}\hyperref[smsec:col]{\texorpdfstring{TmNi$_6$In$_6$}{}}~{\color{blue}  (High-T stable, Low-T unstable) }}

\begin{table}[htbp]
\global\long\def\arraystretch{1.12}
\captionsetup{width=.85\textwidth}
\caption{Structure parameters of TmNi$_6$In$_6$ after relaxation. It contains 531 electrons. }
\label{tab:TmNi6In6}
\setlength{\tabcolsep}{4mm}{
%
}
\end{table}

\begin{figure}[htbp]
\centering
\includegraphics[width=0.85\textwidth]{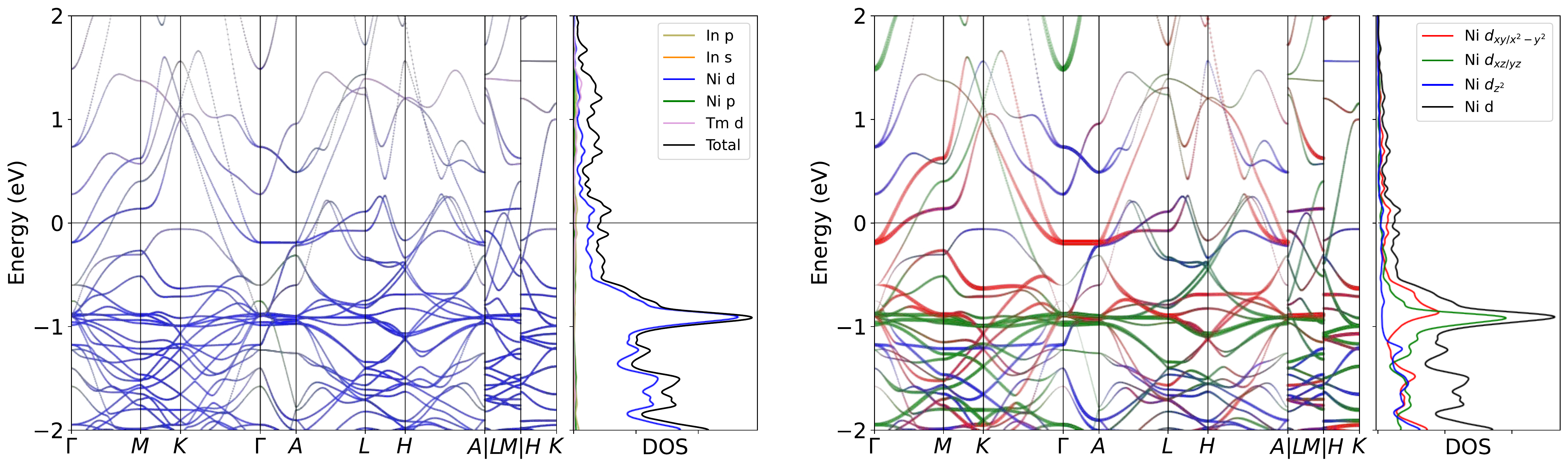}
\caption{Band structure for TmNi$_6$In$_6$ without SOC. The projection of Ni $3d$ orbitals is also presented. The band structures clearly reveal the dominating of Ni $3d$ orbitals  near the Fermi level, as well as the pronounced peak.}
\label{fig:TmNi6In6band}
\end{figure}

\begin{figure}[H]
\centering
\subfloat[Relaxed phonon at low-T.]{
\label{fig:TmNi6In6_relaxed_Low}
\includegraphics[width=0.45\textwidth]{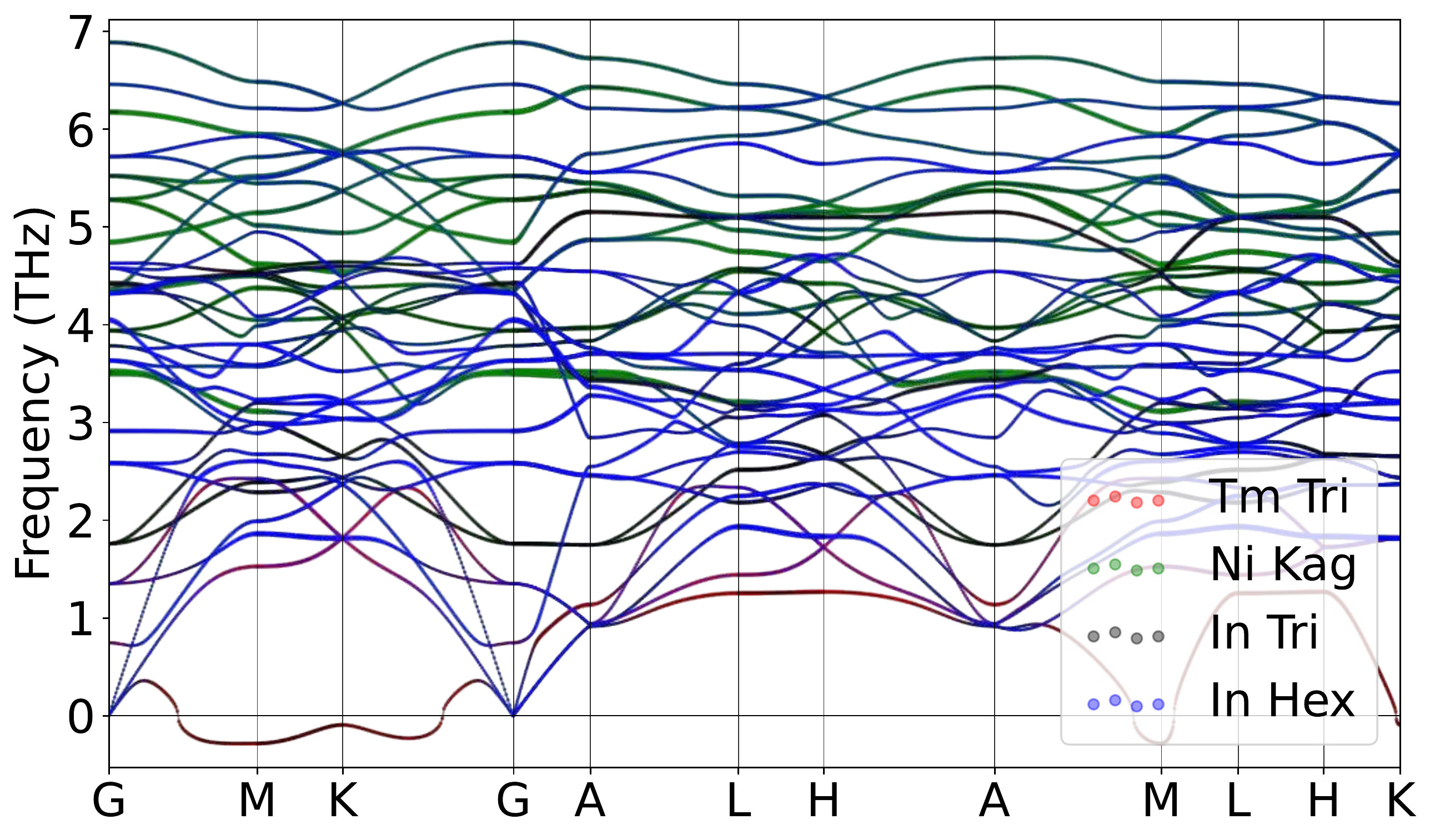}
}
\hspace{5pt}\subfloat[Relaxed phonon at high-T.]{
\label{fig:TmNi6In6_relaxed_High}
\includegraphics[width=0.45\textwidth]{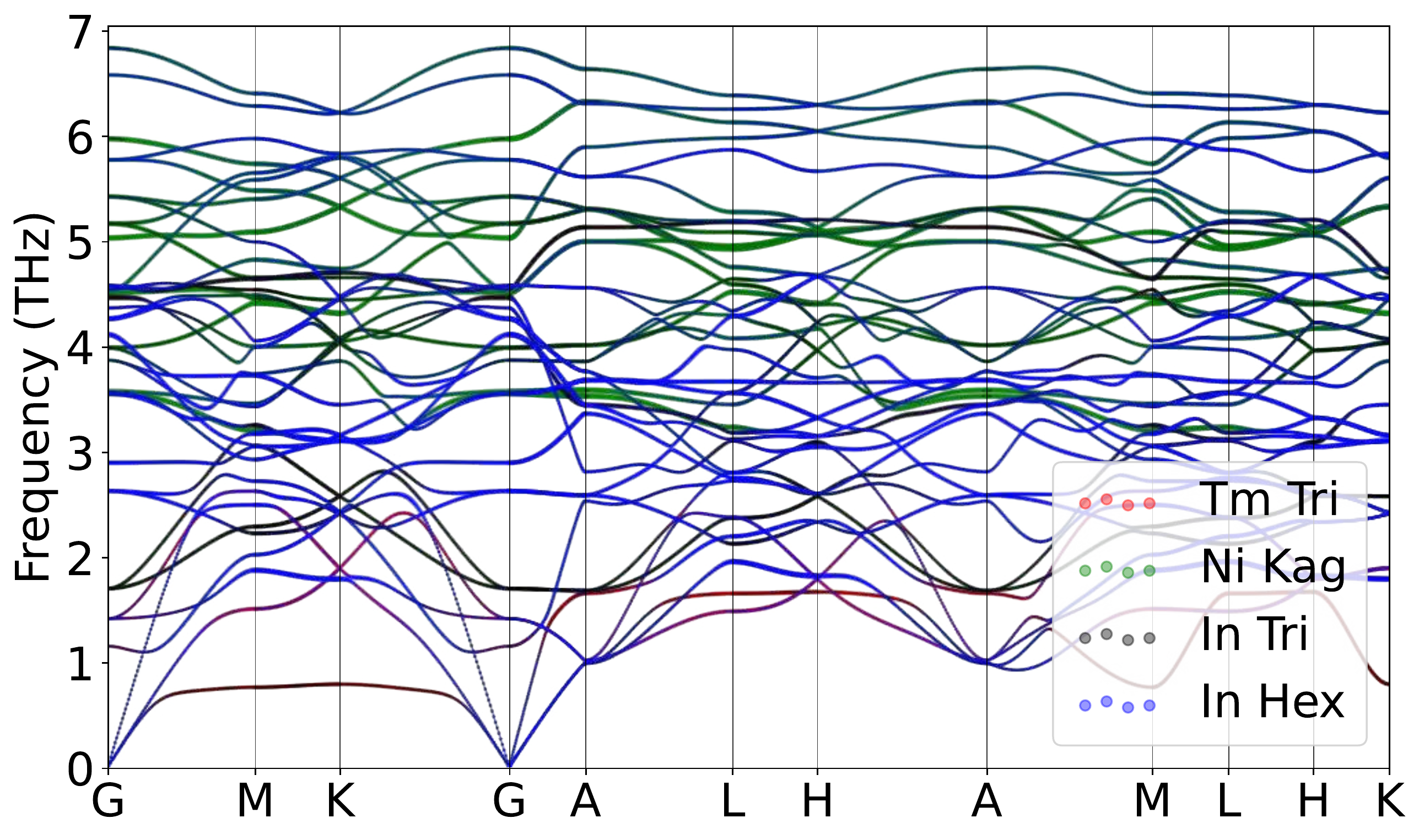}
}
\caption{Atomically projected phonon spectrum for TmNi$_6$In$_6$.}
\label{fig:TmNi6In6pho}
\end{figure}

\begin{figure}[H]
\centering
\label{fig:TmNi6In6_relaxed_Low_xz}
\includegraphics[width=0.85\textwidth]{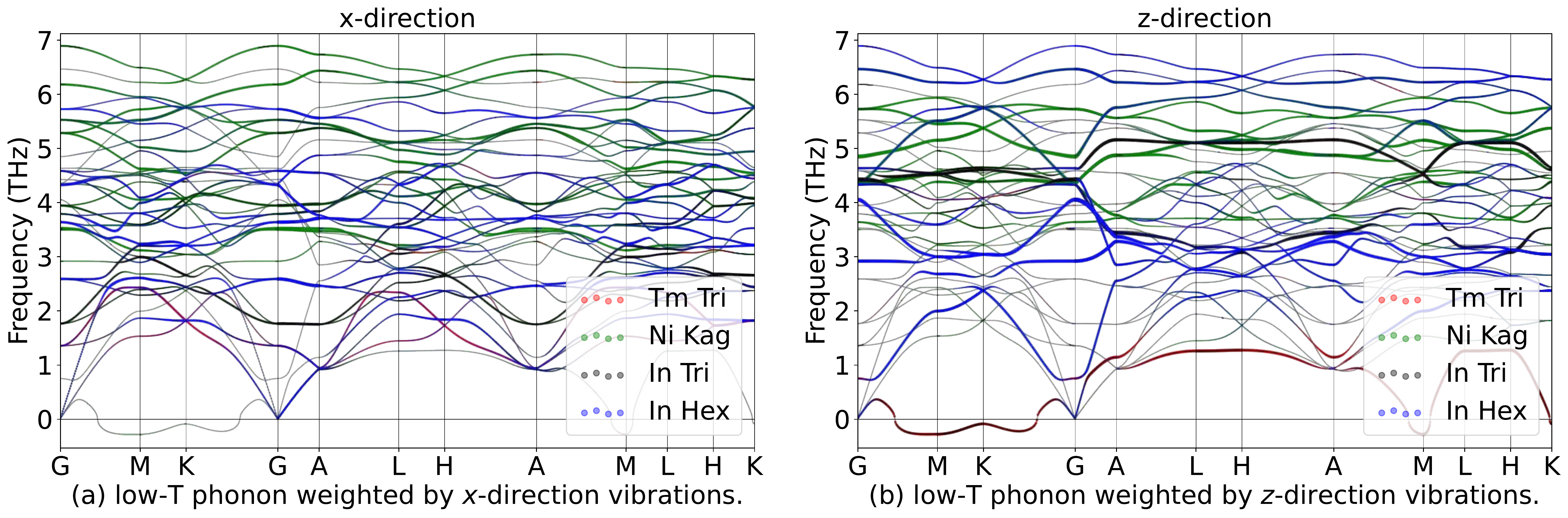}
\hspace{5pt}\label{fig:TmNi6In6_relaxed_High_xz}
\includegraphics[width=0.85\textwidth]{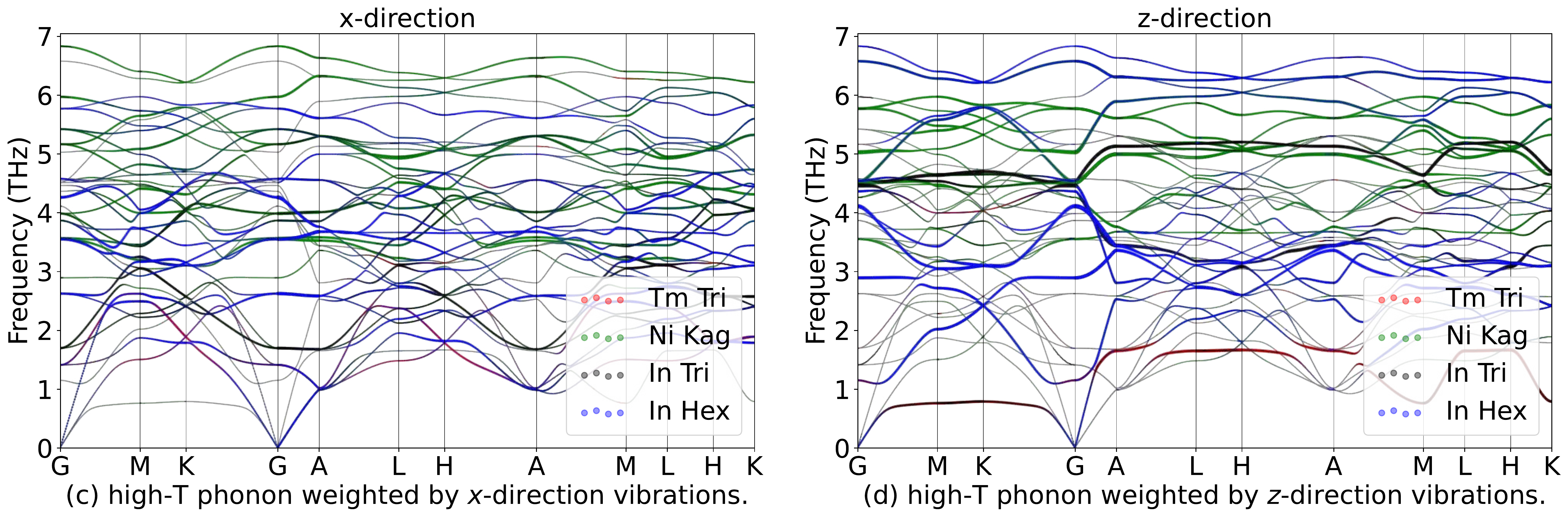}
\caption{Phonon spectrum for TmNi$_6$In$_6$in in-plane ($x$) and out-of-plane ($z$) directions.}
\label{fig:TmNi6In6phoxyz}
\end{figure}

\begin{table}[htbp]
\global\long\def\arraystretch{1.12}
\captionsetup{width=.95\textwidth}
\caption{IRREPs at high-symmetry points for TmNi$_6$In$_6$ phonon spectrum at Low-T.}
\label{tab:191_TmNi6In6_Low}
\setlength{\tabcolsep}{1.mm}{
}
\end{table}

\begin{table}[htbp]
\global\long\def\arraystretch{1.12}
\captionsetup{width=.95\textwidth}
\caption{IRREPs at high-symmetry points for TmNi$_6$In$_6$ phonon spectrum at High-T.}
\label{tab:191_TmNi6In6_High}
\setlength{\tabcolsep}{1.mm}{
%
}
\end{table}
\subsubsection{\label{sms2sec:YbNi6In6}\hyperref[smsec:col]{\texorpdfstring{YbNi$_6$In$_6$}{}}~{\color{green}  (High-T stable, Low-T stable) }}

\begin{table}[htbp]
\global\long\def\arraystretch{1.12}
\captionsetup{width=.85\textwidth}
\caption{Structure parameters of YbNi$_6$In$_6$ after relaxation. It contains 532 electrons. }
\label{tab:YbNi6In6}
\setlength{\tabcolsep}{4mm}{
%
}
\end{table}

\begin{figure}[htbp]
\centering
\includegraphics[width=0.85\textwidth]{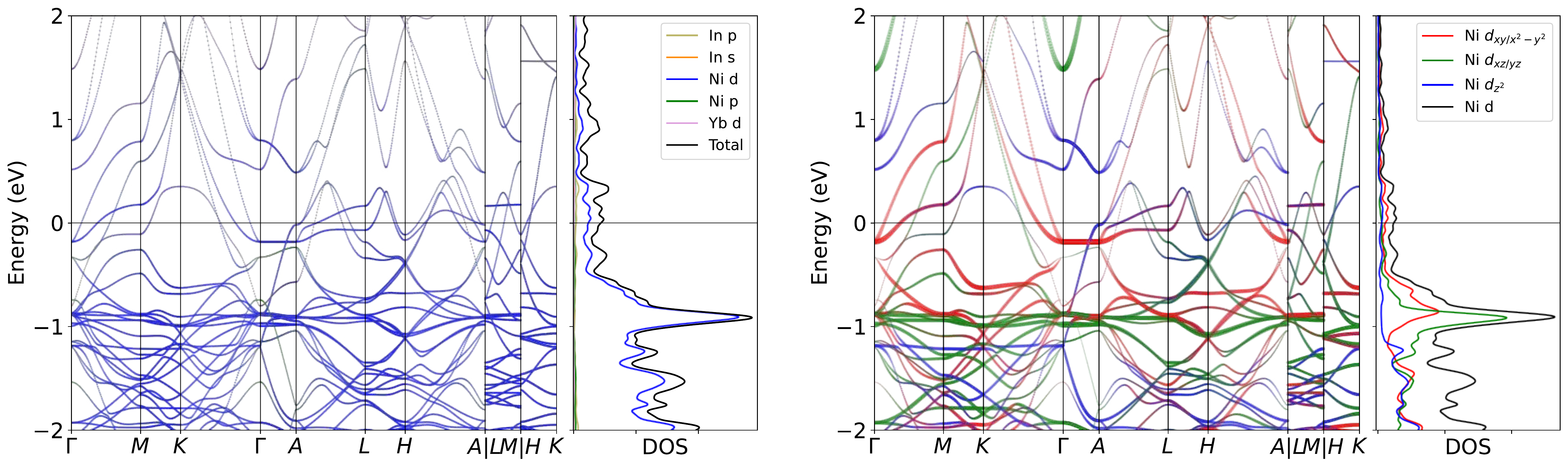}
\caption{Band structure for YbNi$_6$In$_6$ without SOC. The projection of Ni $3d$ orbitals is also presented. The band structures clearly reveal the dominating of Ni $3d$ orbitals  near the Fermi level, as well as the pronounced peak.}
\label{fig:YbNi6In6band}
\end{figure}

\begin{figure}[H]
\centering
\subfloat[Relaxed phonon at low-T.]{
\label{fig:YbNi6In6_relaxed_Low}
\includegraphics[width=0.45\textwidth]{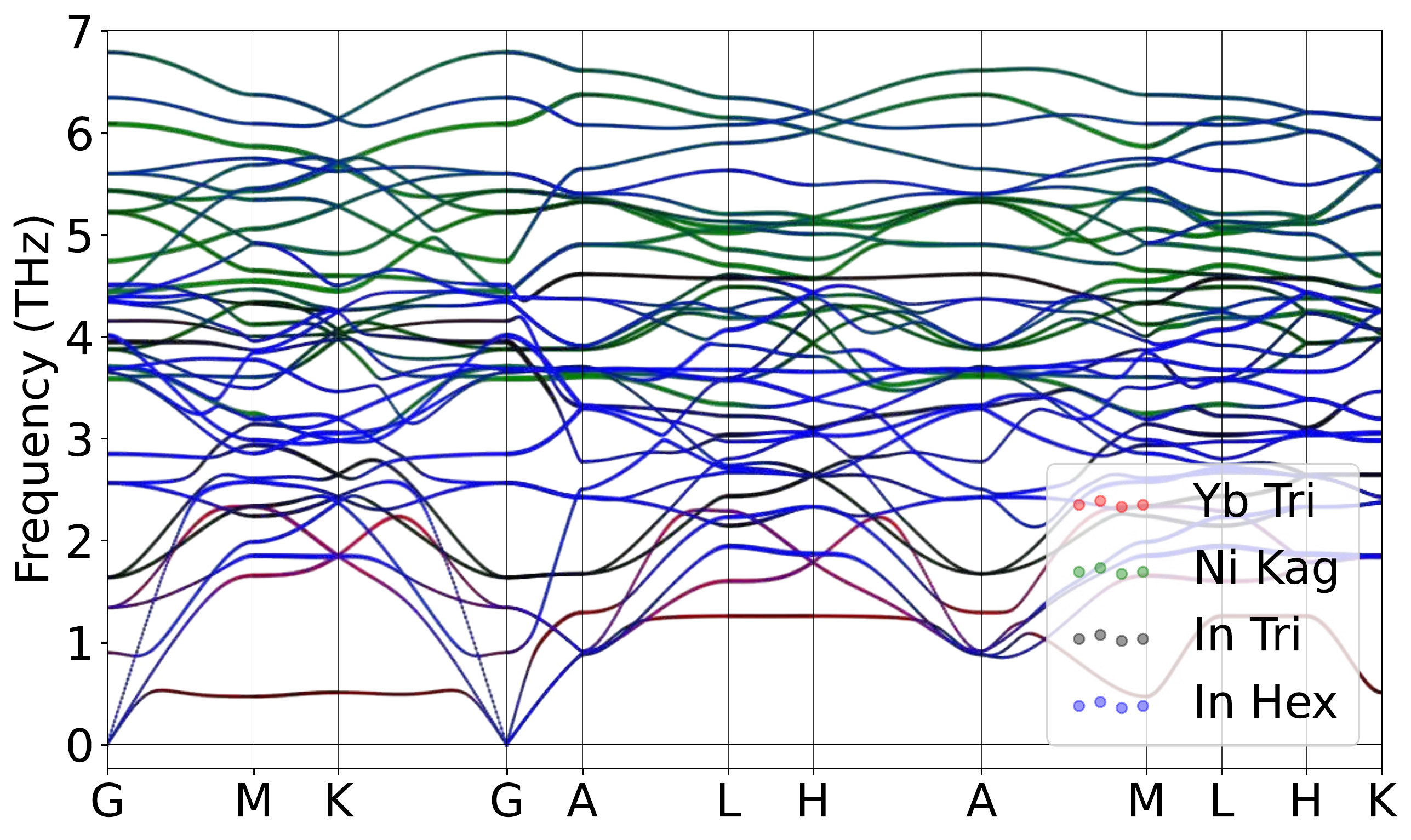}
}
\hspace{5pt}\subfloat[Relaxed phonon at high-T.]{
\label{fig:YbNi6In6_relaxed_High}
\includegraphics[width=0.45\textwidth]{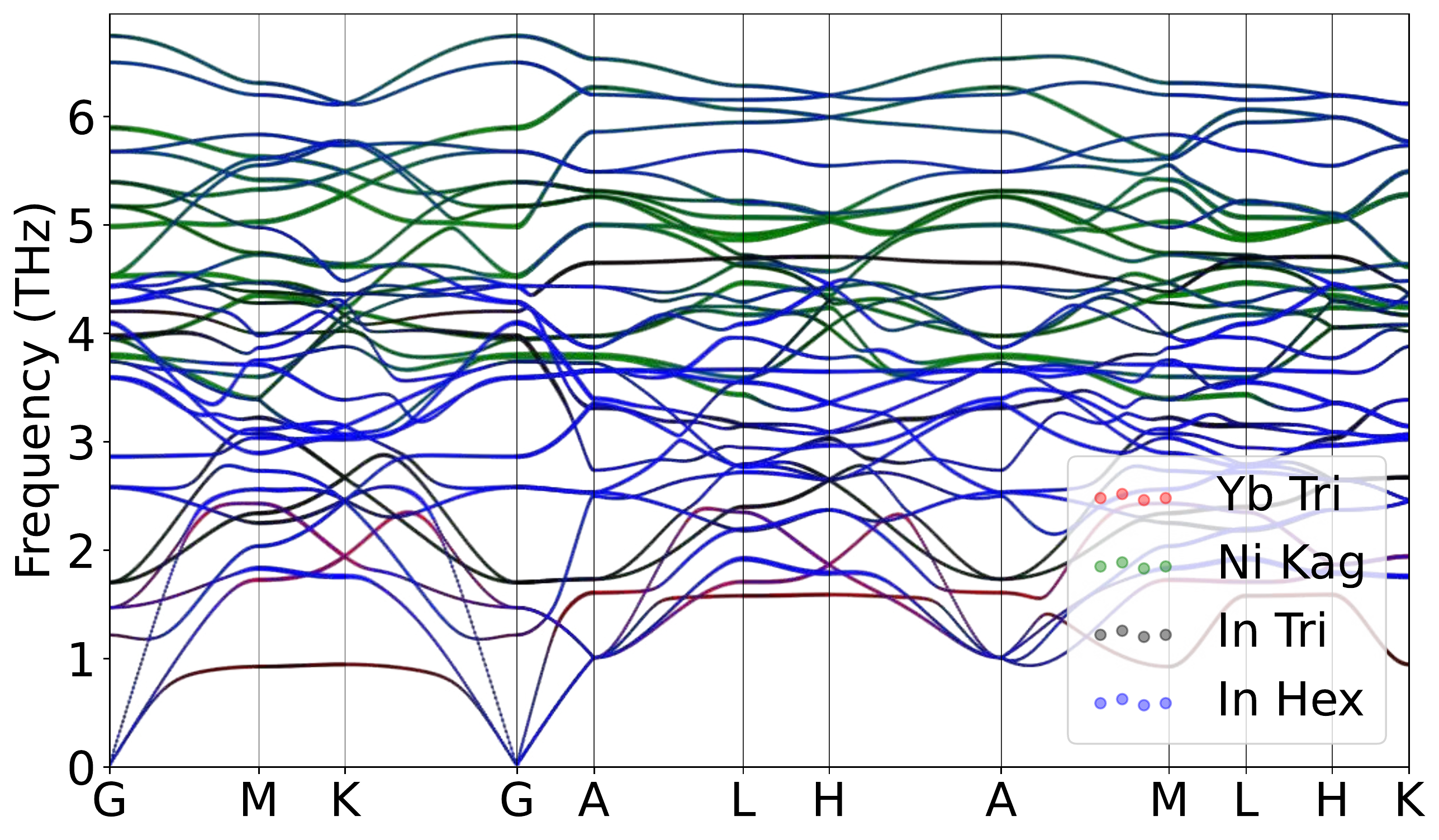}
}
\caption{Atomically projected phonon spectrum for YbNi$_6$In$_6$.}
\label{fig:YbNi6In6pho}
\end{figure}

\begin{figure}[H]
\centering
\label{fig:YbNi6In6_relaxed_Low_xz}
\includegraphics[width=0.85\textwidth]{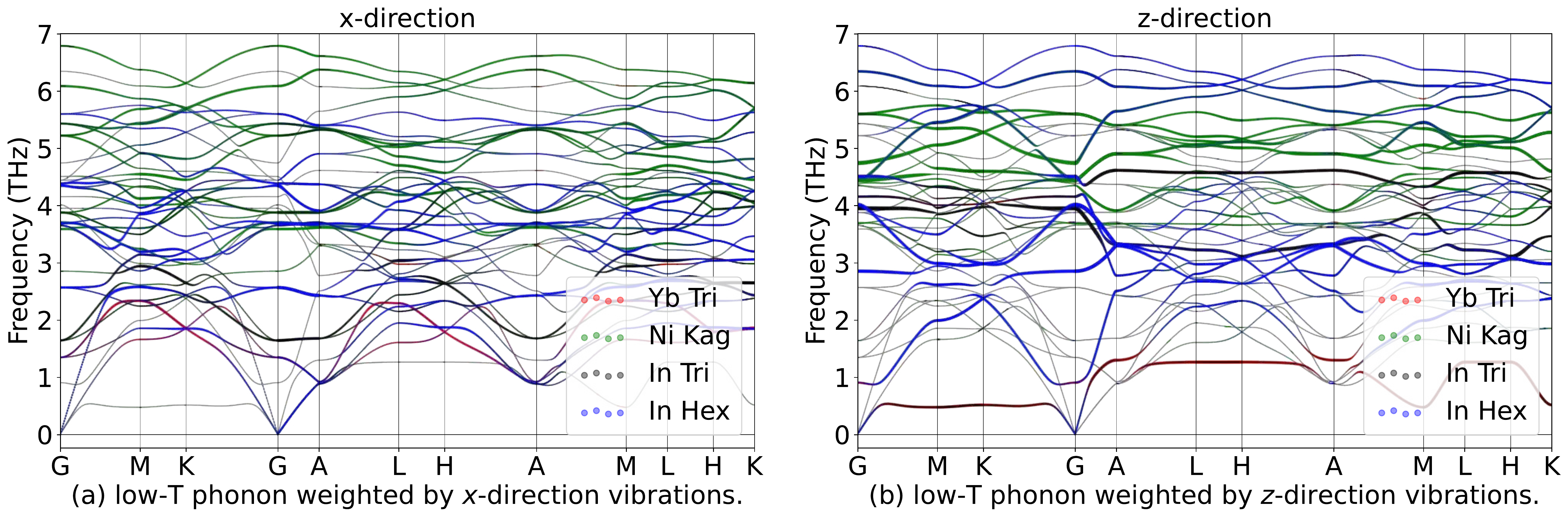}
\hspace{5pt}\label{fig:YbNi6In6_relaxed_High_xz}
\includegraphics[width=0.85\textwidth]{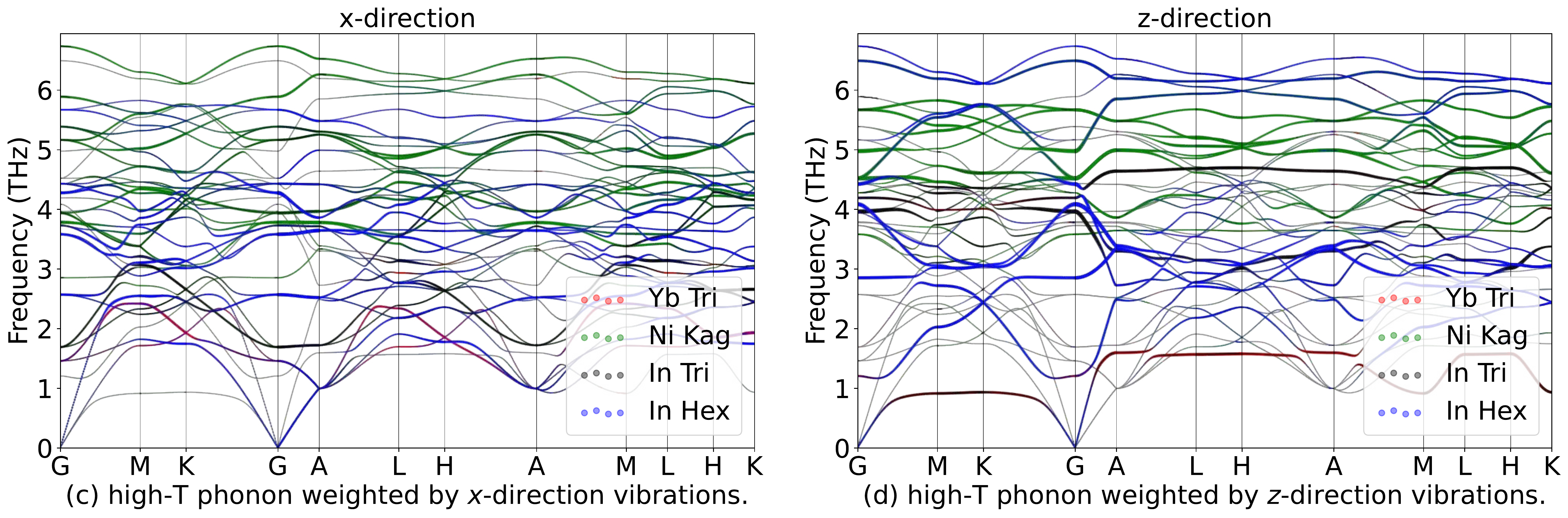}
\caption{Phonon spectrum for YbNi$_6$In$_6$in in-plane ($x$) and out-of-plane ($z$) directions.}
\label{fig:YbNi6In6phoxyz}
\end{figure}

\begin{table}[htbp]
\global\long\def\arraystretch{1.12}
\captionsetup{width=.95\textwidth}
\caption{IRREPs at high-symmetry points for YbNi$_6$In$_6$ phonon spectrum at Low-T.}
\label{tab:191_YbNi6In6_Low}
\setlength{\tabcolsep}{1.mm}{
}
\end{table}

\begin{table}[htbp]
\global\long\def\arraystretch{1.12}
\captionsetup{width=.95\textwidth}
\caption{IRREPs at high-symmetry points for YbNi$_6$In$_6$ phonon spectrum at High-T.}
\label{tab:191_YbNi6In6_High}
\setlength{\tabcolsep}{1.mm}{
%
}
\end{table}
\subsubsection{\label{sms2sec:LuNi6In6}\hyperref[smsec:col]{\texorpdfstring{LuNi$_6$In$_6$}{}}~{\color{blue}  (High-T stable, Low-T unstable) }}

\begin{table}[htbp]
\global\long\def\arraystretch{1.12}
\captionsetup{width=.85\textwidth}
\caption{Structure parameters of LuNi$_6$In$_6$ after relaxation. It contains 533 electrons. }
\label{tab:LuNi6In6}
\setlength{\tabcolsep}{4mm}{
%
}
\end{table}

\begin{figure}[htbp]
\centering
\includegraphics[width=0.85\textwidth]{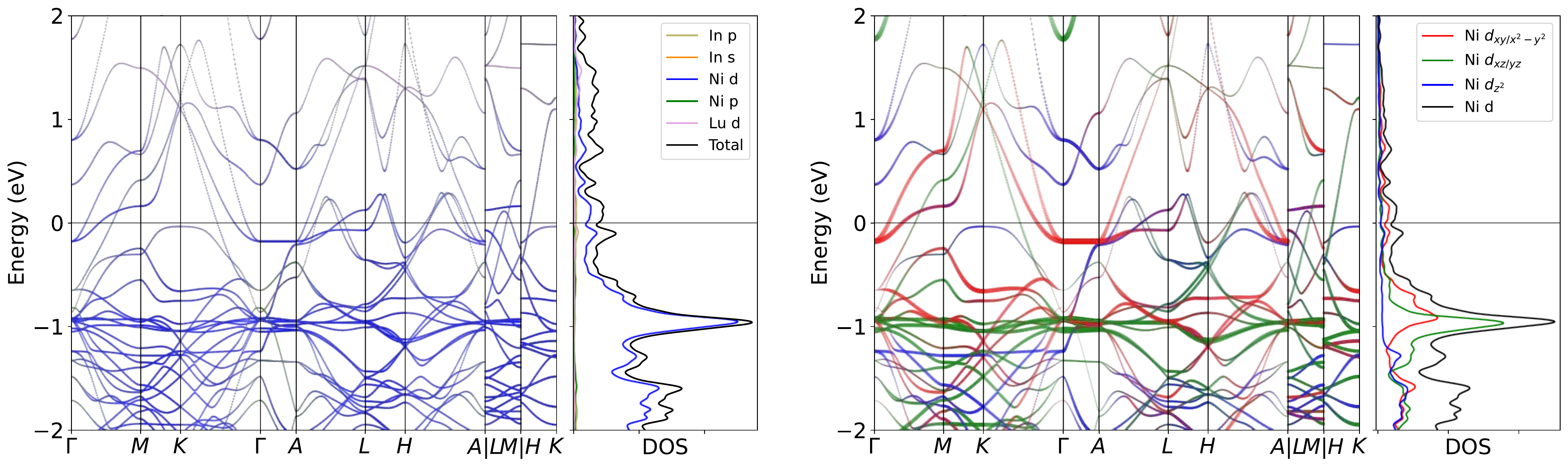}
\caption{Band structure for LuNi$_6$In$_6$ without SOC. The projection of Ni $3d$ orbitals is also presented. The band structures clearly reveal the dominating of Ni $3d$ orbitals  near the Fermi level, as well as the pronounced peak.}
\label{fig:LuNi6In6band}
\end{figure}

\begin{figure}[H]
\centering
\subfloat[Relaxed phonon at low-T.]{
\label{fig:LuNi6In6_relaxed_Low}
\includegraphics[width=0.45\textwidth]{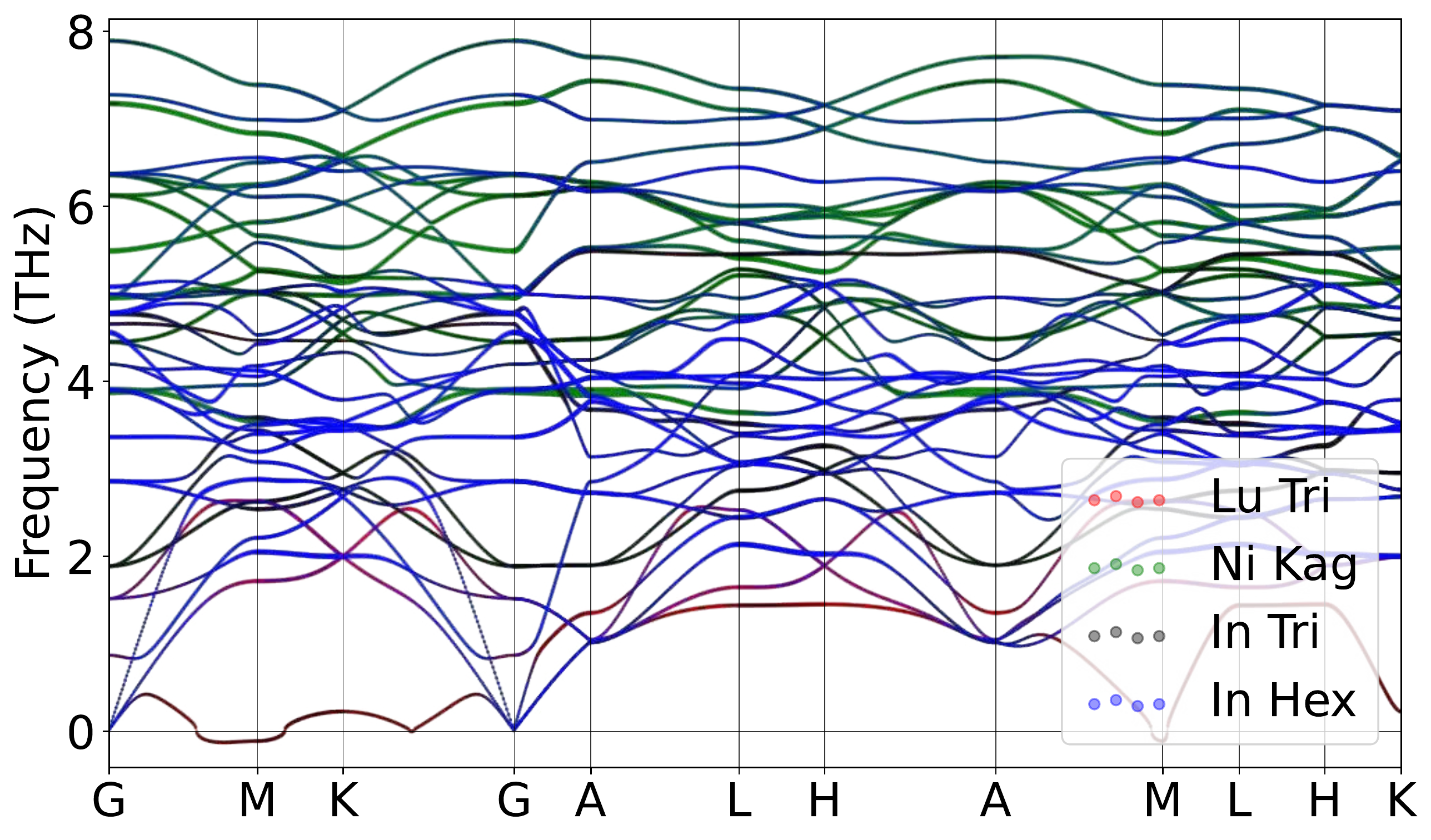}
}
\hspace{5pt}\subfloat[Relaxed phonon at high-T.]{
\label{fig:LuNi6In6_relaxed_High}
\includegraphics[width=0.45\textwidth]{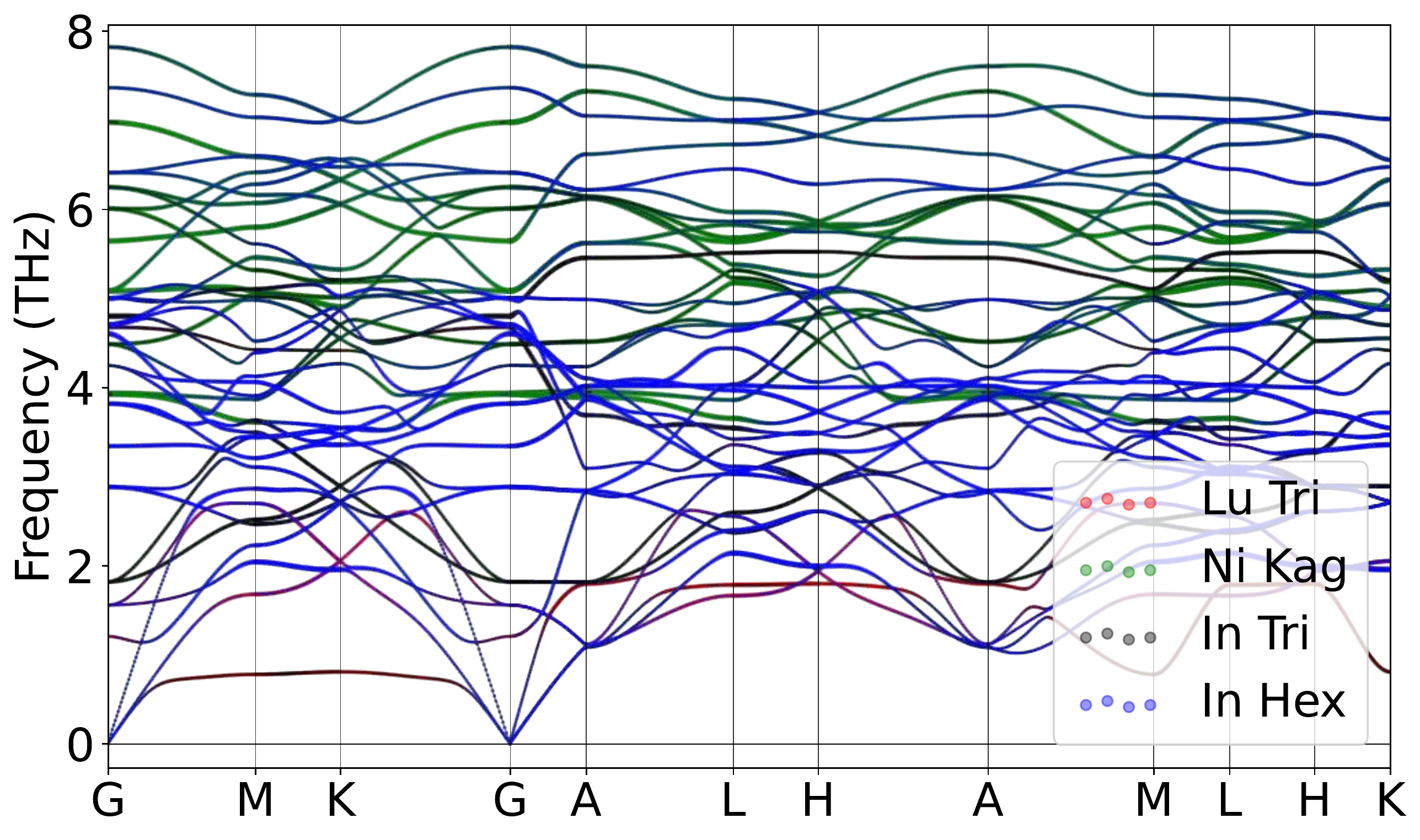}
}
\caption{Atomically projected phonon spectrum for LuNi$_6$In$_6$.}
\label{fig:LuNi6In6pho}
\end{figure}

\begin{figure}[H]
\centering
\label{fig:LuNi6In6_relaxed_Low_xz}
\includegraphics[width=0.85\textwidth]{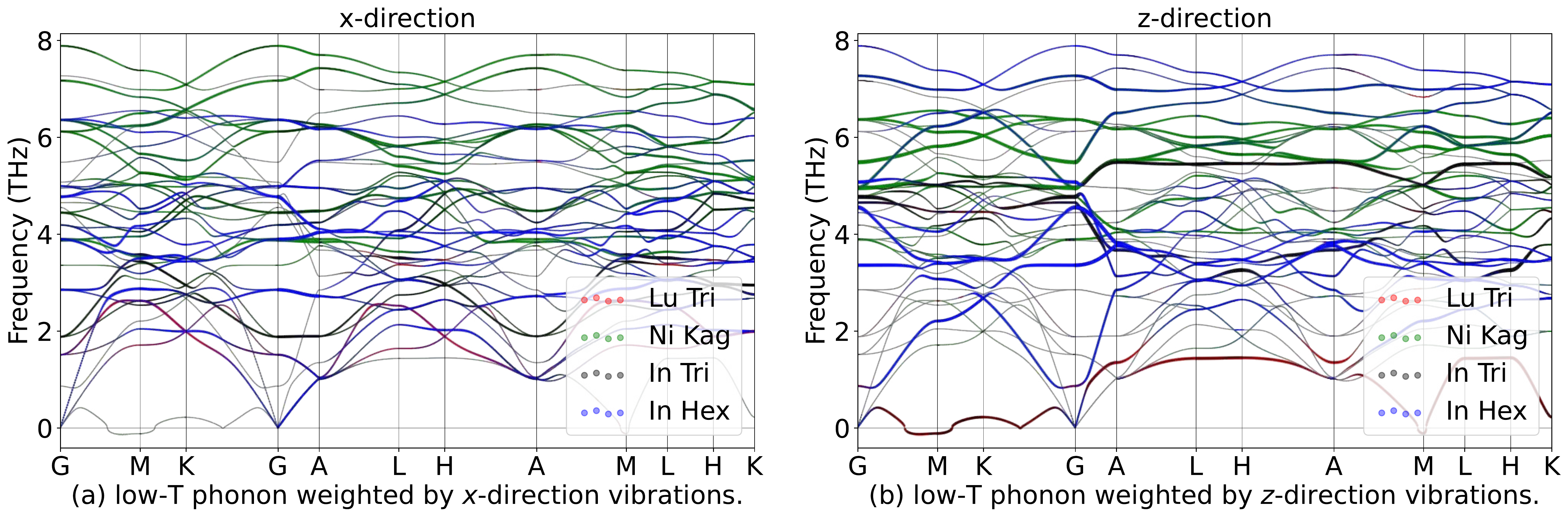}
\hspace{5pt}\label{fig:LuNi6In6_relaxed_High_xz}
\includegraphics[width=0.85\textwidth]{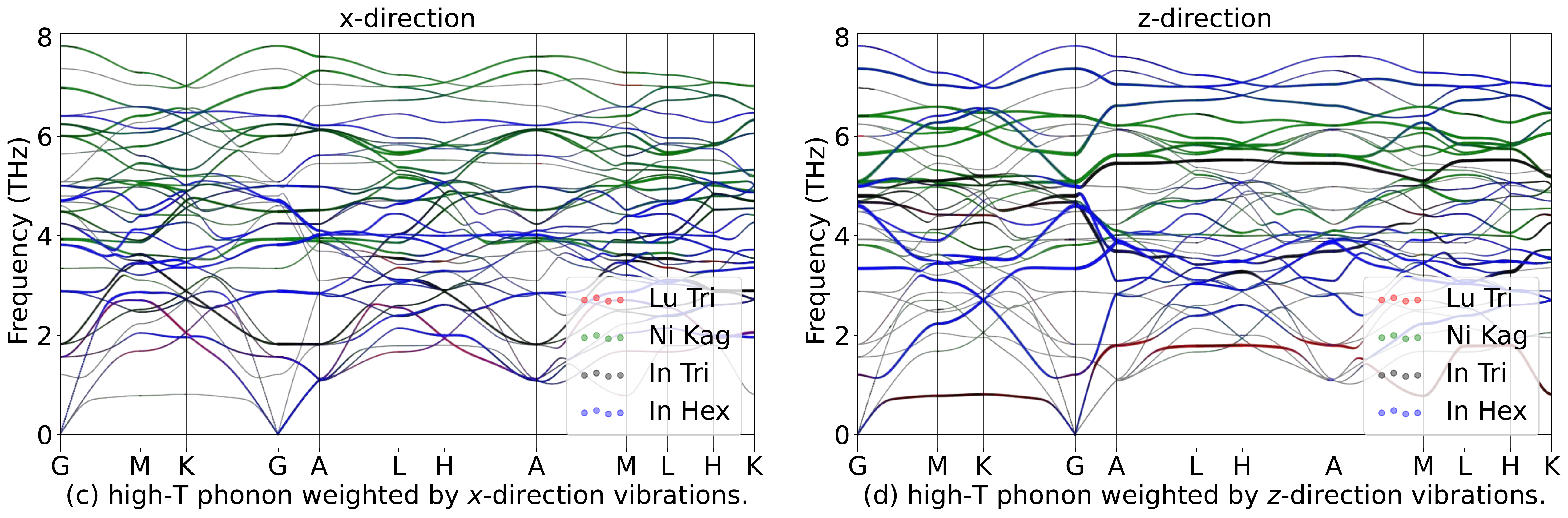}
\caption{Phonon spectrum for LuNi$_6$In$_6$in in-plane ($x$) and out-of-plane ($z$) directions.}
\label{fig:LuNi6In6phoxyz}
\end{figure}

\begin{table}[htbp]
\global\long\def\arraystretch{1.12}
\captionsetup{width=.95\textwidth}
\caption{IRREPs at high-symmetry points for LuNi$_6$In$_6$ phonon spectrum at Low-T.}
\label{tab:191_LuNi6In6_Low}
\setlength{\tabcolsep}{1.mm}{
}
\end{table}

\begin{table}[htbp]
\global\long\def\arraystretch{1.12}
\captionsetup{width=.95\textwidth}
\caption{IRREPs at high-symmetry points for LuNi$_6$In$_6$ phonon spectrum at High-T.}
\label{tab:191_LuNi6In6_High}
\setlength{\tabcolsep}{1.mm}{
%
}
\end{table}
\subsubsection{\label{sms2sec:HfNi6In6}\hyperref[smsec:col]{\texorpdfstring{HfNi$_6$In$_6$}{}}~{\color{blue}  (High-T stable, Low-T unstable) }}

\begin{table}[htbp]
\global\long\def\arraystretch{1.12}
\captionsetup{width=.85\textwidth}
\caption{Structure parameters of HfNi$_6$In$_6$ after relaxation. It contains 534 electrons. }
\label{tab:HfNi6In6}
\setlength{\tabcolsep}{4mm}{
%
}
\end{table}

\begin{figure}[htbp]
\centering
\includegraphics[width=0.85\textwidth]{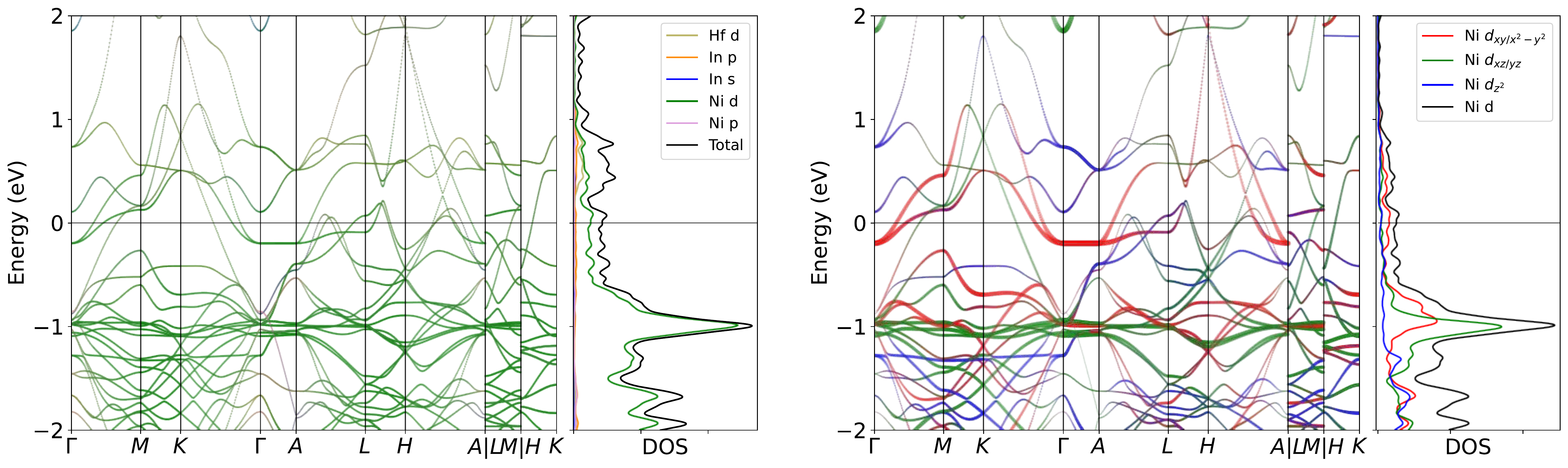}
\caption{Band structure for HfNi$_6$In$_6$ without SOC. The projection of Ni $3d$ orbitals is also presented. The band structures clearly reveal the dominating of Ni $3d$ orbitals  near the Fermi level, as well as the pronounced peak.}
\label{fig:HfNi6In6band}
\end{figure}

\begin{figure}[H]
\centering
\subfloat[Relaxed phonon at low-T.]{
\label{fig:HfNi6In6_relaxed_Low}
\includegraphics[width=0.45\textwidth]{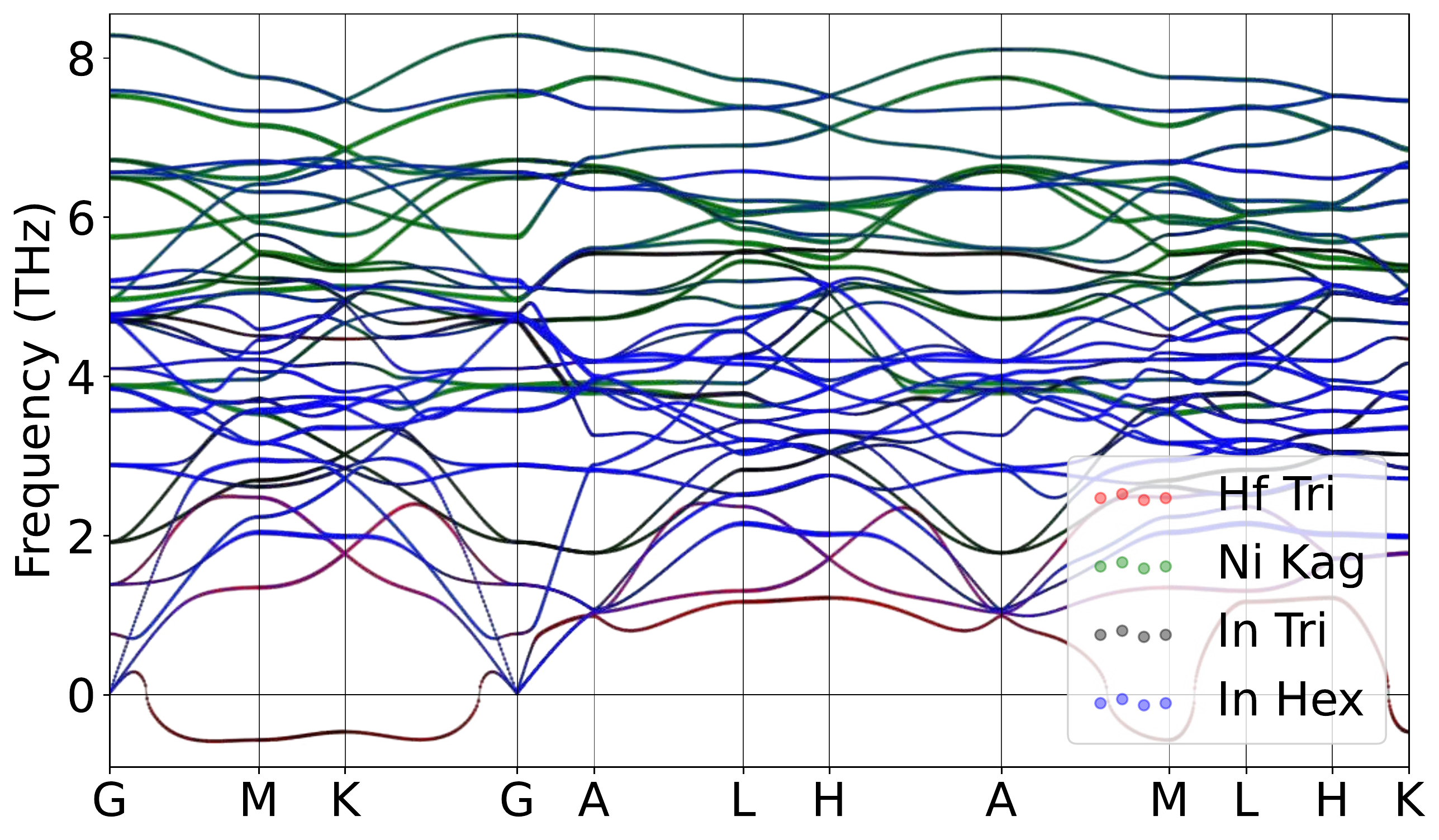}
}
\hspace{5pt}\subfloat[Relaxed phonon at high-T.]{
\label{fig:HfNi6In6_relaxed_High}
\includegraphics[width=0.45\textwidth]{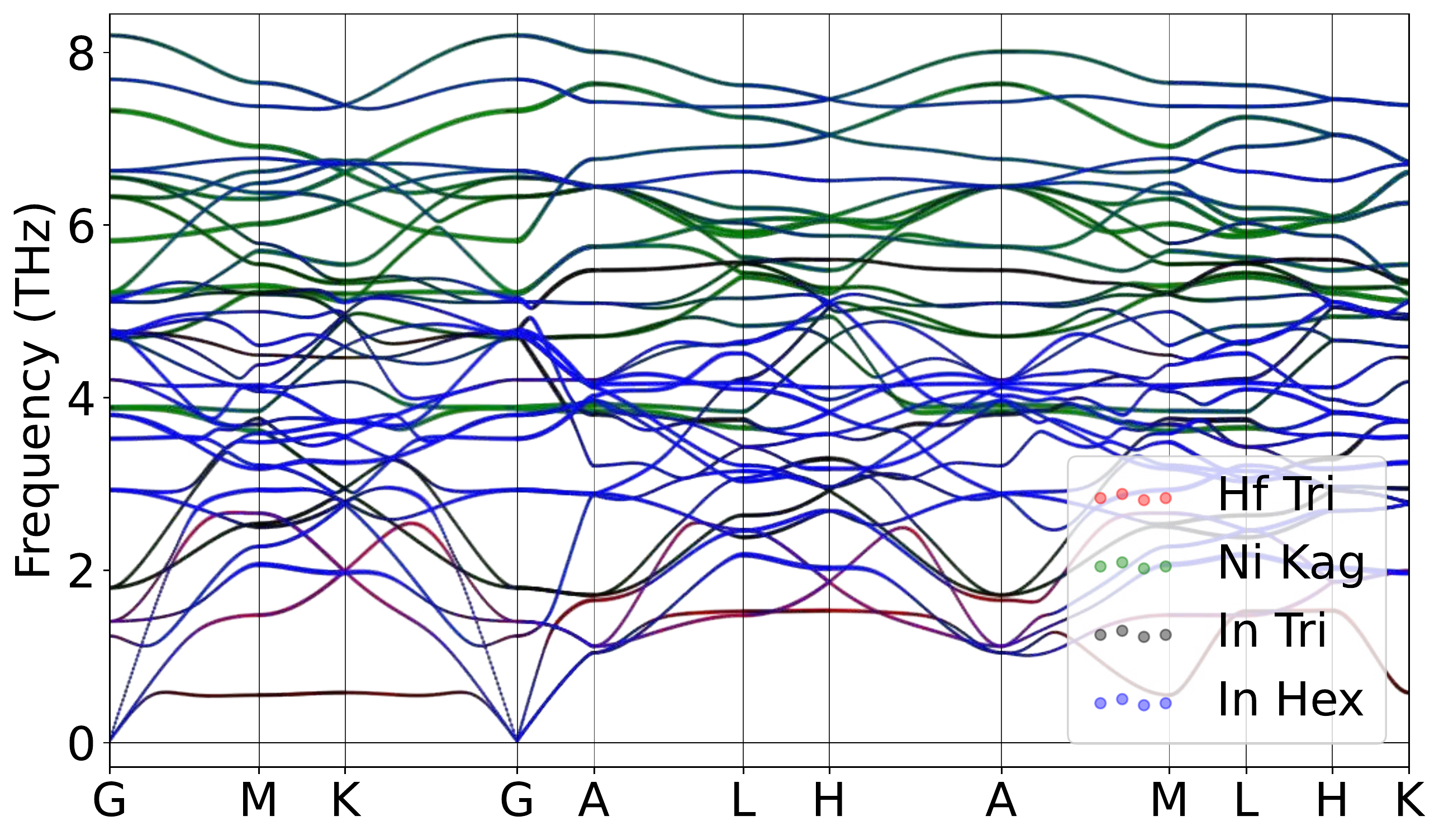}
}
\caption{Atomically projected phonon spectrum for HfNi$_6$In$_6$.}
\label{fig:HfNi6In6pho}
\end{figure}

\begin{figure}[H]
\centering
\label{fig:HfNi6In6_relaxed_Low_xz}
\includegraphics[width=0.85\textwidth]{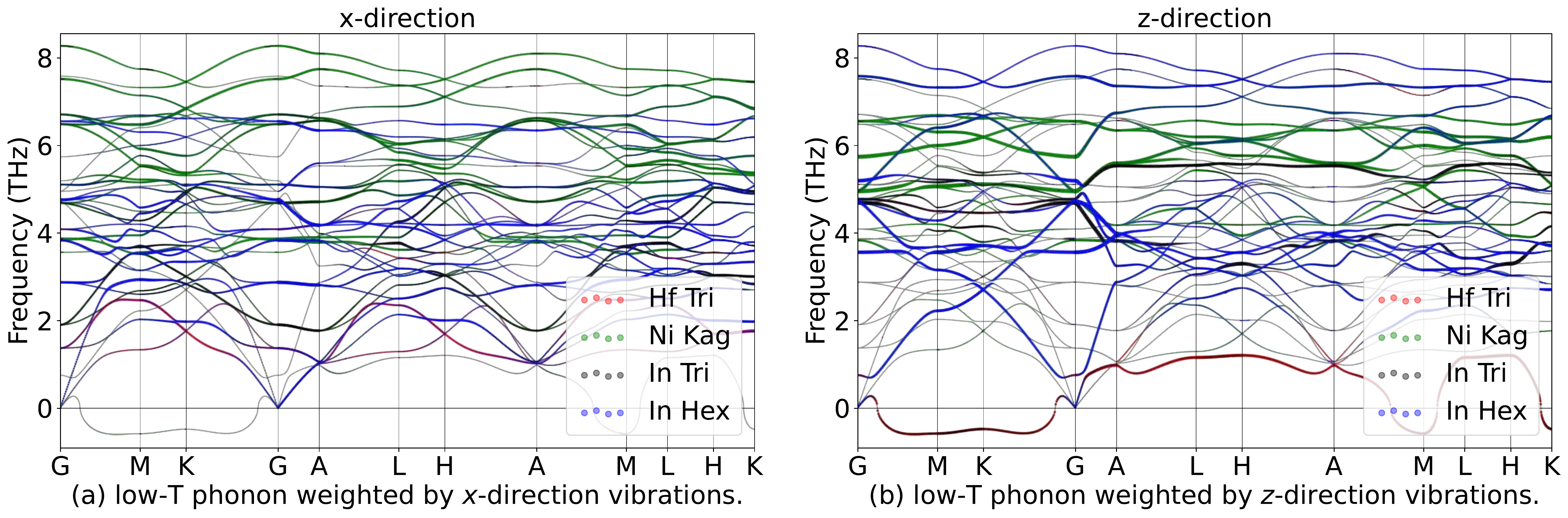}
\hspace{5pt}\label{fig:HfNi6In6_relaxed_High_xz}
\includegraphics[width=0.85\textwidth]{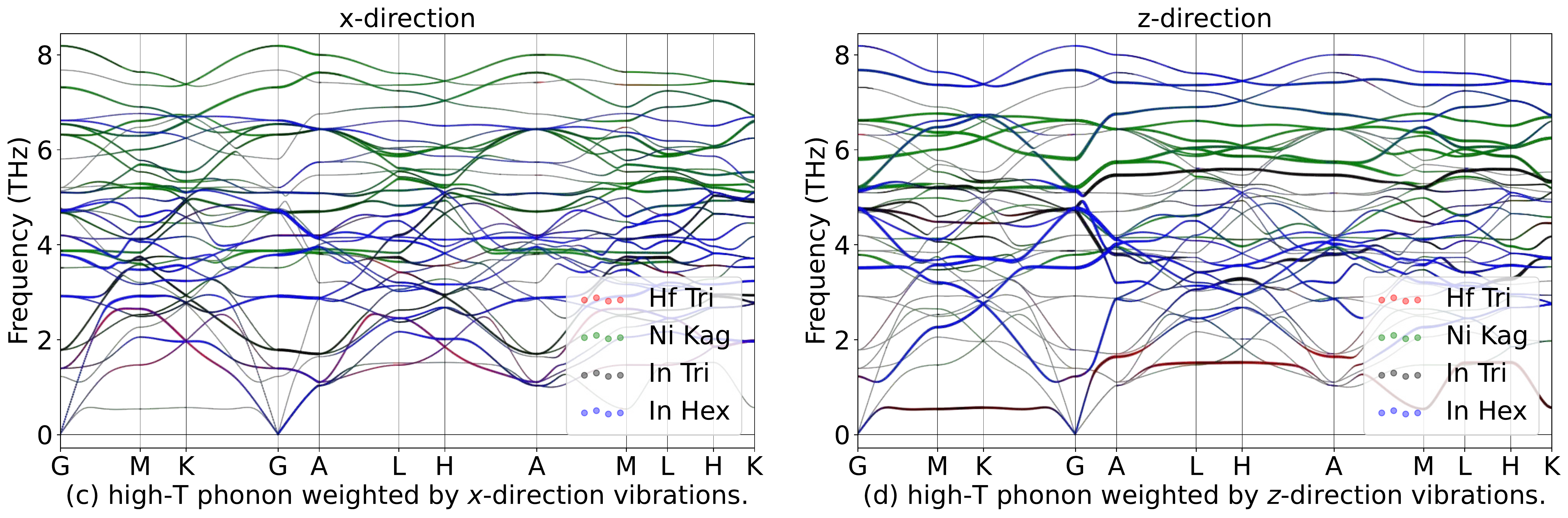}
\caption{Phonon spectrum for HfNi$_6$In$_6$in in-plane ($x$) and out-of-plane ($z$) directions.}
\label{fig:HfNi6In6phoxyz}
\end{figure}

\begin{table}[htbp]
\global\long\def\arraystretch{1.12}
\captionsetup{width=.95\textwidth}
\caption{IRREPs at high-symmetry points for HfNi$_6$In$_6$ phonon spectrum at Low-T.}
\label{tab:191_HfNi6In6_Low}
\setlength{\tabcolsep}{1.mm}{
}
\end{table}

\begin{table}[htbp]
\global\long\def\arraystretch{1.12}
\captionsetup{width=.95\textwidth}
\caption{IRREPs at high-symmetry points for HfNi$_6$In$_6$ phonon spectrum at High-T.}
\label{tab:191_HfNi6In6_High}
\setlength{\tabcolsep}{1.mm}{
%
}
\end{table}
\subsubsection{\label{sms2sec:TaNi6In6}\hyperref[smsec:col]{\texorpdfstring{TaNi$_6$In$_6$}{}}~{\color{blue}  (High-T stable, Low-T unstable) }}

\begin{table}[htbp]
\global\long\def\arraystretch{1.12}
\captionsetup{width=.85\textwidth}
\caption{Structure parameters of TaNi$_6$In$_6$ after relaxation. It contains 535 electrons. }
\label{tab:TaNi6In6}
\setlength{\tabcolsep}{4mm}{
%
}
\end{table}

\begin{figure}[htbp]
\centering
\includegraphics[width=0.85\textwidth]{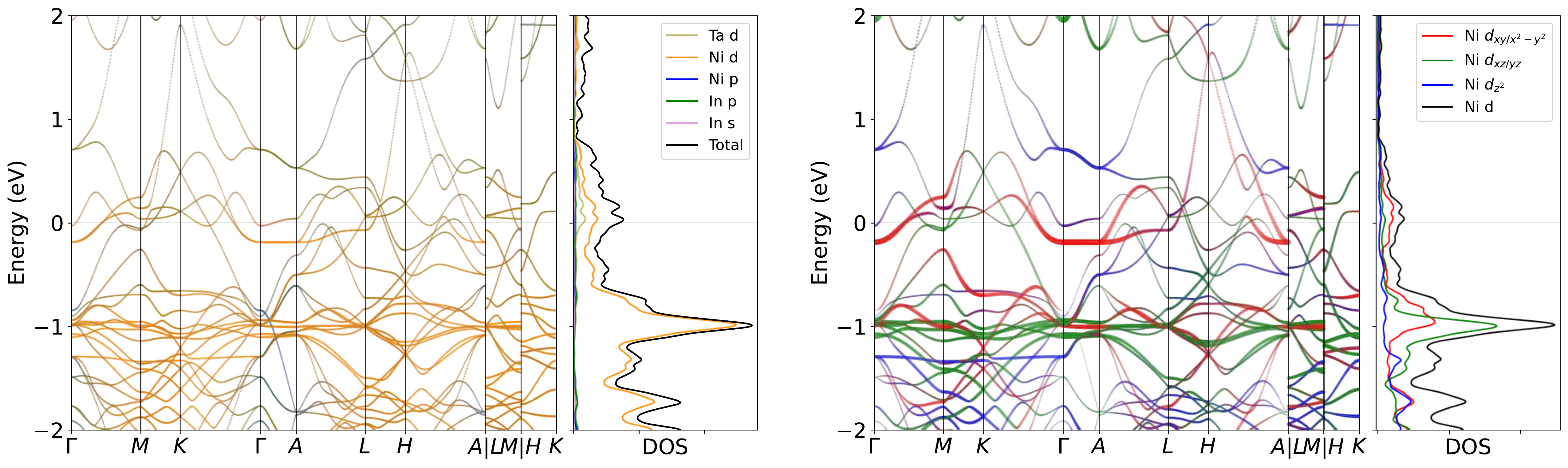}
\caption{Band structure for TaNi$_6$In$_6$ without SOC. The projection of Ni $3d$ orbitals is also presented. The band structures clearly reveal the dominating of Ni $3d$ orbitals  near the Fermi level, as well as the pronounced peak.}
\label{fig:TaNi6In6band}
\end{figure}

\begin{figure}[H]
\centering
\subfloat[Relaxed phonon at low-T.]{
\label{fig:TaNi6In6_relaxed_Low}
\includegraphics[width=0.45\textwidth]{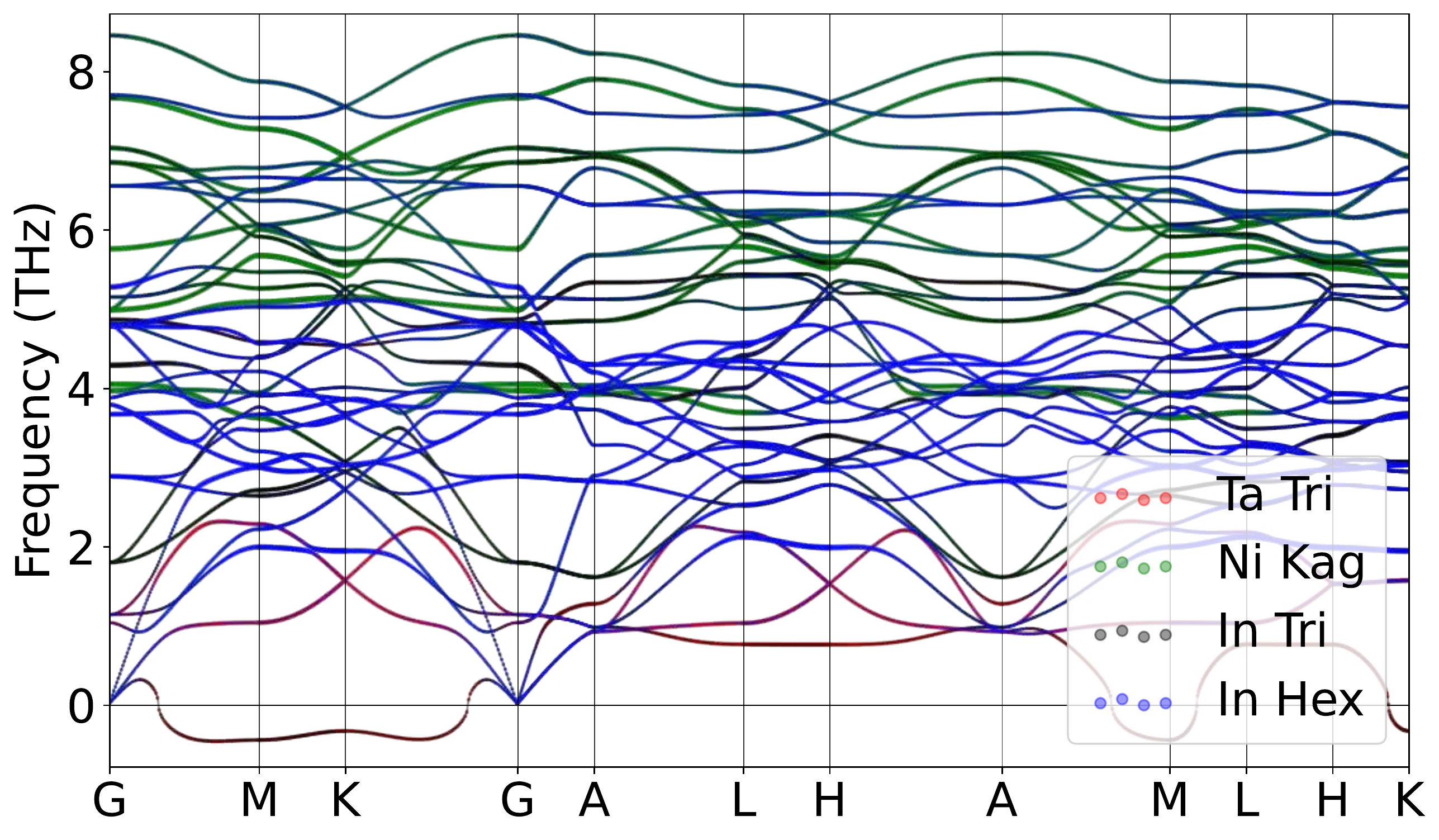}
}
\hspace{5pt}\subfloat[Relaxed phonon at high-T.]{
\label{fig:TaNi6In6_relaxed_High}
\includegraphics[width=0.45\textwidth]{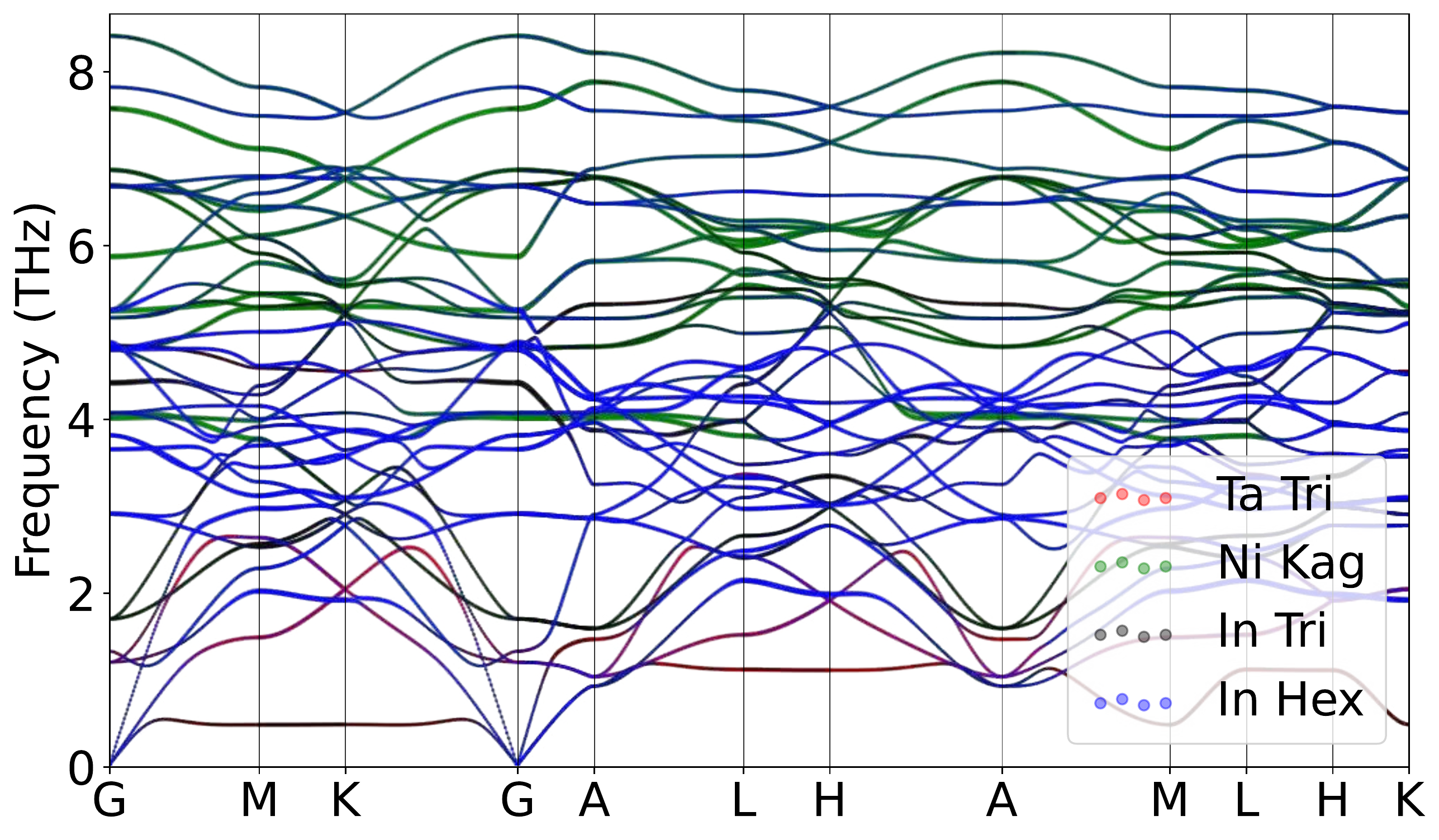}
}
\caption{Atomically projected phonon spectrum for TaNi$_6$In$_6$.}
\label{fig:TaNi6In6pho}
\end{figure}

\begin{figure}[H]
\centering
\label{fig:TaNi6In6_relaxed_Low_xz}
\includegraphics[width=0.85\textwidth]{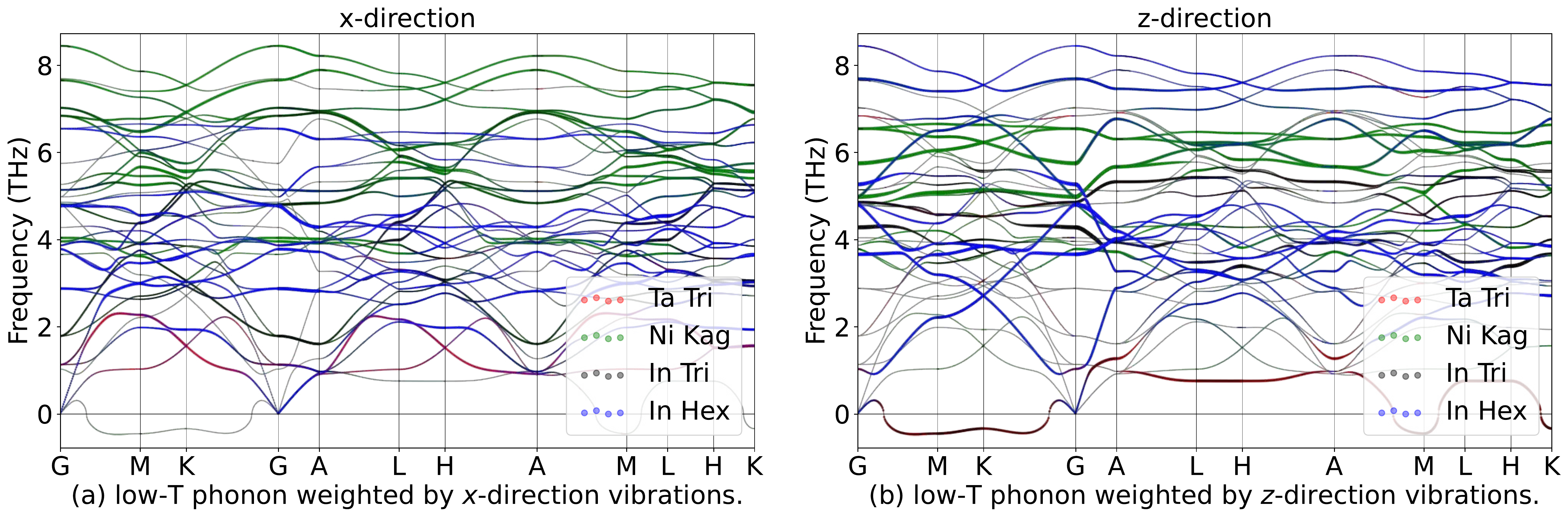}
\hspace{5pt}\label{fig:TaNi6In6_relaxed_High_xz}
\includegraphics[width=0.85\textwidth]{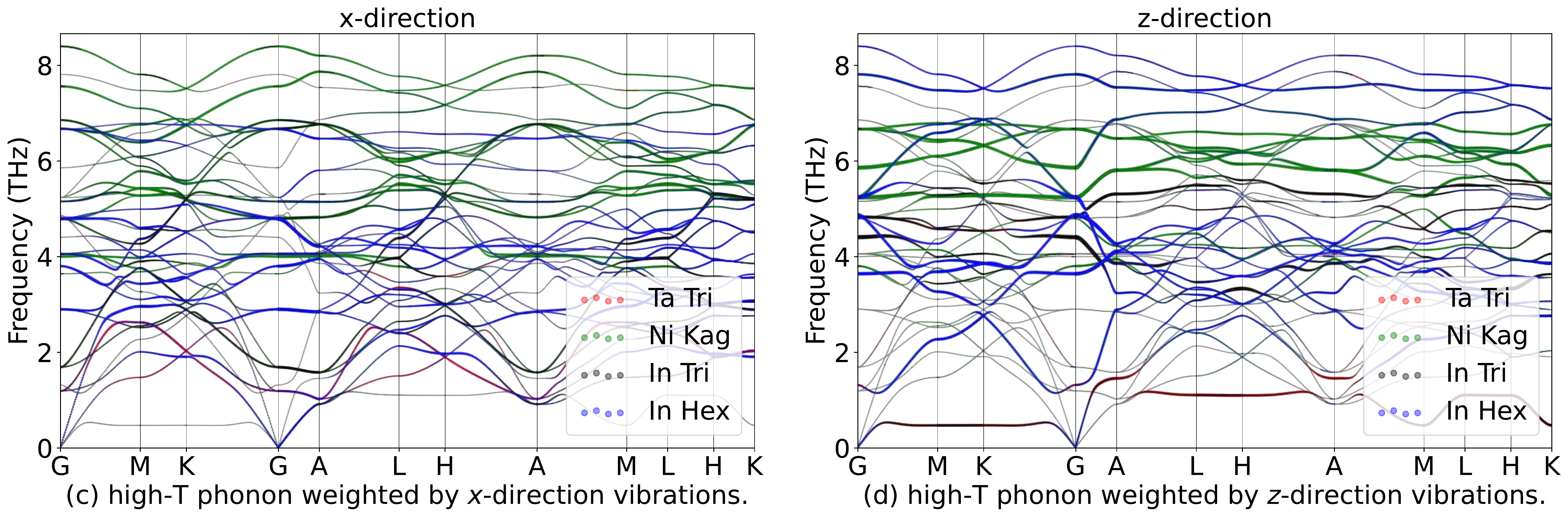}
\caption{Phonon spectrum for TaNi$_6$In$_6$in in-plane ($x$) and out-of-plane ($z$) directions.}
\label{fig:TaNi6In6phoxyz}
\end{figure}

\begin{table}[htbp]
\global\long\def\arraystretch{1.12}
\captionsetup{width=.95\textwidth}
\caption{IRREPs at high-symmetry points for TaNi$_6$In$_6$ phonon spectrum at Low-T.}
\label{tab:191_TaNi6In6_Low}
\setlength{\tabcolsep}{1.mm}{
}
\end{table}

\begin{table}[htbp]
\global\long\def\arraystretch{1.12}
\captionsetup{width=.95\textwidth}
\caption{IRREPs at high-symmetry points for TaNi$_6$In$_6$ phonon spectrum at High-T.}
\label{tab:191_TaNi6In6_High}
\setlength{\tabcolsep}{1.mm}{
%
}
\end{table}
\subsection{\hyperref[smsec:col]{NiSn}}~\label{smssec:NiSn}

\subsubsection{\label{sms2sec:LiNi6Sn6}\hyperref[smsec:col]{\texorpdfstring{LiNi$_6$Sn$_6$}{}}~{\color{green}  (High-T stable, Low-T stable) }}

\begin{table}[htbp]
\global\long\def\arraystretch{1.12}
\captionsetup{width=.85\textwidth}
\caption{Structure parameters of LiNi$_6$Sn$_6$ after relaxation. It contains 471 electrons. }
\label{tab:LiNi6Sn6}
\setlength{\tabcolsep}{4mm}{
%
}
\end{table}

\begin{figure}[htbp]
\centering
\includegraphics[width=0.85\textwidth]{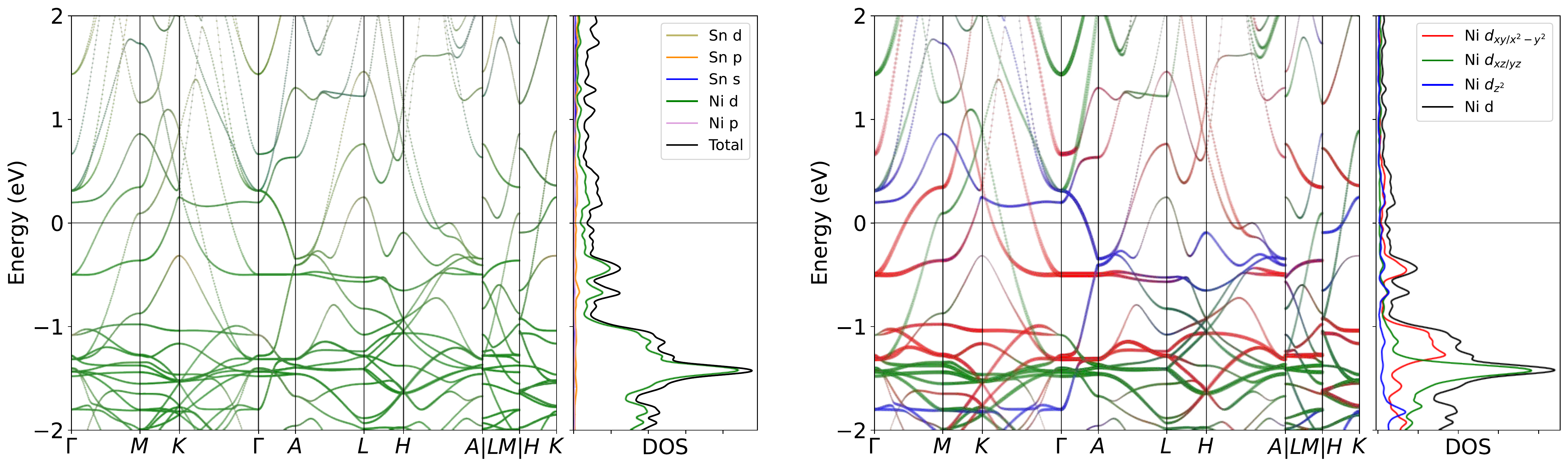}
\caption{Band structure for LiNi$_6$Sn$_6$ without SOC. The projection of Ni $3d$ orbitals is also presented. The band structures clearly reveal the dominating of Ni $3d$ orbitals  near the Fermi level, as well as the pronounced peak.}
\label{fig:LiNi6Sn6band}
\end{figure}

\begin{figure}[H]
\centering
\subfloat[Relaxed phonon at low-T.]{
\label{fig:LiNi6Sn6_relaxed_Low}
\includegraphics[width=0.45\textwidth]{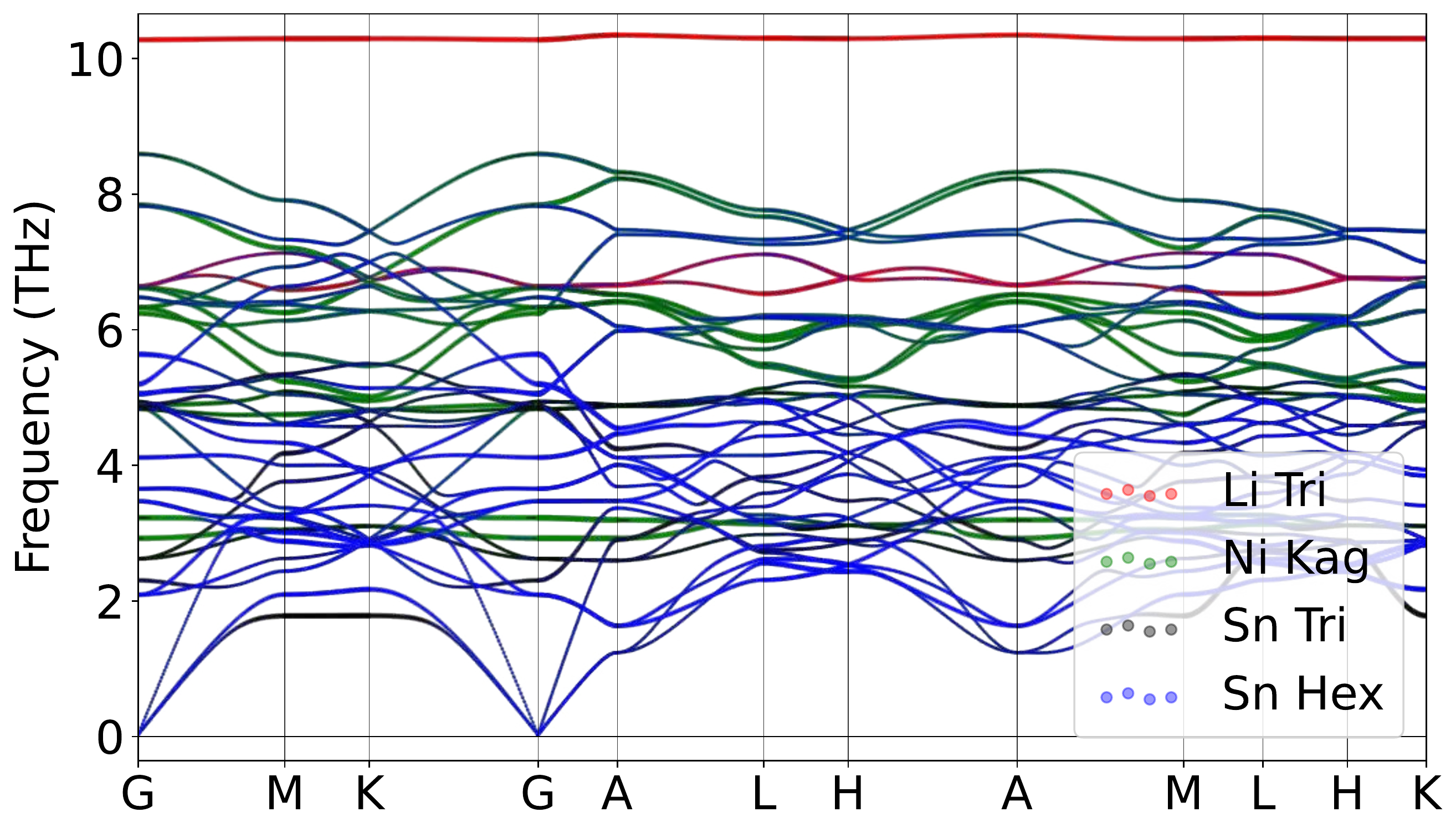}
}
\hspace{5pt}\subfloat[Relaxed phonon at high-T.]{
\label{fig:LiNi6Sn6_relaxed_High}
\includegraphics[width=0.45\textwidth]{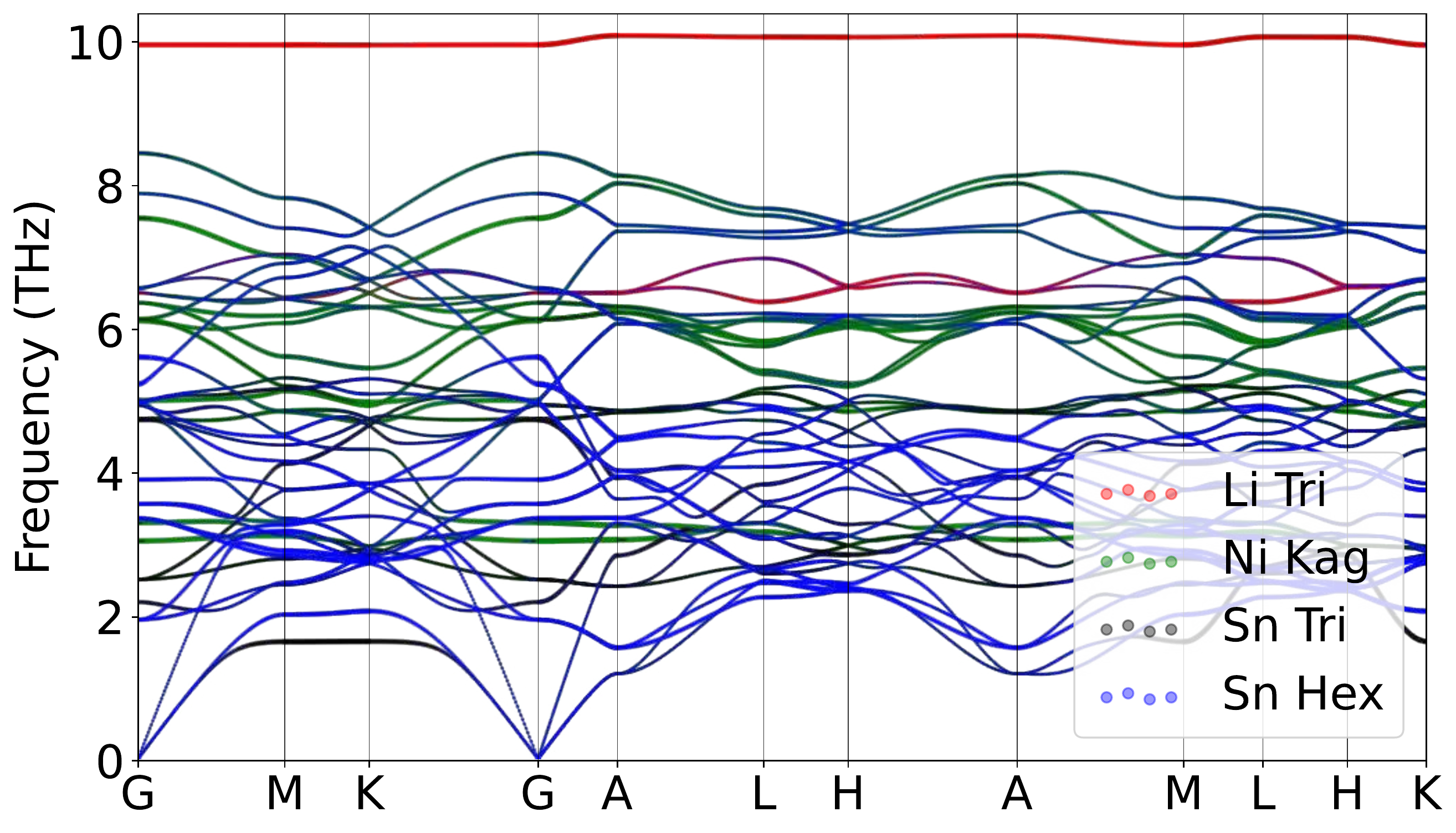}
}
\caption{Atomically projected phonon spectrum for LiNi$_6$Sn$_6$.}
\label{fig:LiNi6Sn6pho}
\end{figure}

\begin{figure}[H]
\centering
\label{fig:LiNi6Sn6_relaxed_Low_xz}
\includegraphics[width=0.85\textwidth]{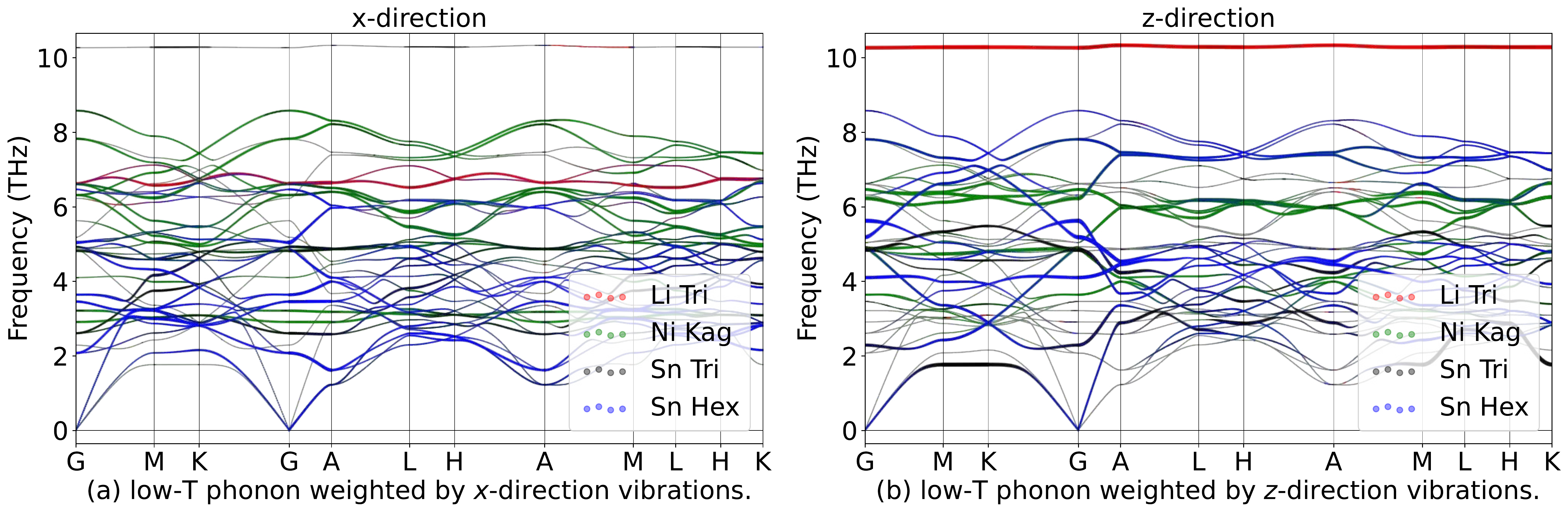}
\hspace{5pt}\label{fig:LiNi6Sn6_relaxed_High_xz}
\includegraphics[width=0.85\textwidth]{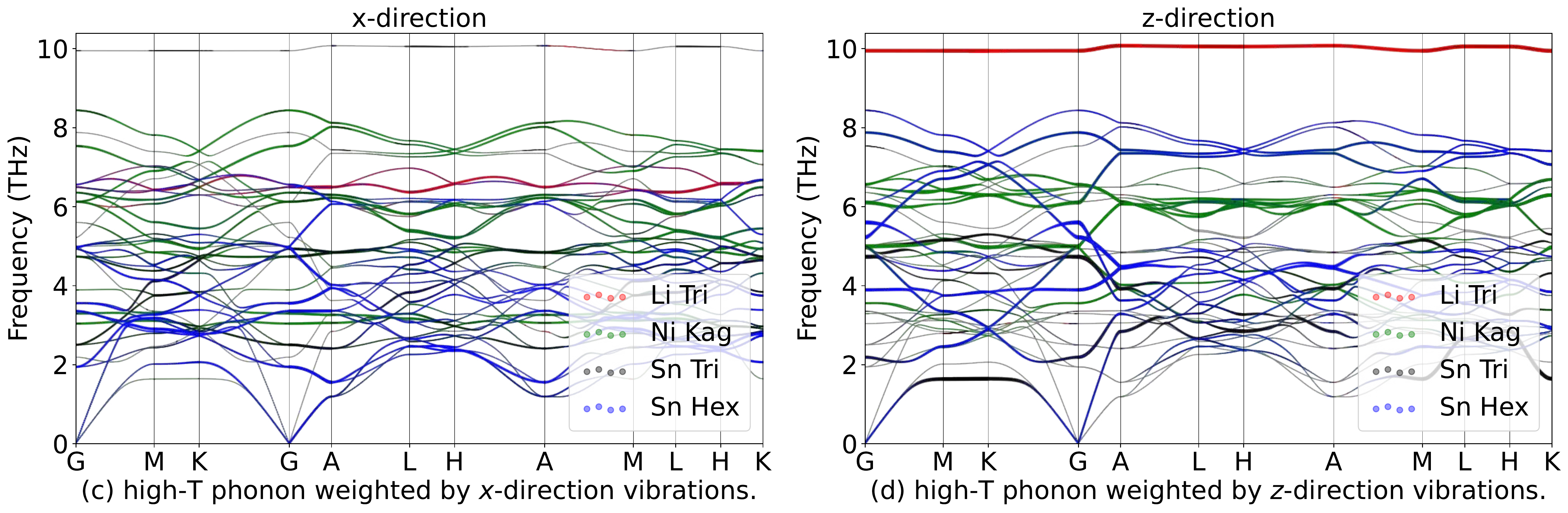}
\caption{Phonon spectrum for LiNi$_6$Sn$_6$in in-plane ($x$) and out-of-plane ($z$) directions.}
\label{fig:LiNi6Sn6phoxyz}
\end{figure}

\begin{table}[htbp]
\global\long\def\arraystretch{1.12}
\captionsetup{width=.95\textwidth}
\caption{IRREPs at high-symmetry points for LiNi$_6$Sn$_6$ phonon spectrum at Low-T.}
\label{tab:191_LiNi6Sn6_Low}
\setlength{\tabcolsep}{1.mm}{
}
\end{table}

\begin{table}[htbp]
\global\long\def\arraystretch{1.12}
\captionsetup{width=.95\textwidth}
\caption{IRREPs at high-symmetry points for LiNi$_6$Sn$_6$ phonon spectrum at High-T.}
\label{tab:191_LiNi6Sn6_High}
\setlength{\tabcolsep}{1.mm}{
%
}
\end{table}
\subsubsection{\label{sms2sec:MgNi6Sn6}\hyperref[smsec:col]{\texorpdfstring{MgNi$_6$Sn$_6$}{}}~{\color{green}  (High-T stable, Low-T stable) }}

\begin{table}[htbp]
\global\long\def\arraystretch{1.12}
\captionsetup{width=.85\textwidth}
\caption{Structure parameters of MgNi$_6$Sn$_6$ after relaxation. It contains 480 electrons. }
\label{tab:MgNi6Sn6}
\setlength{\tabcolsep}{4mm}{
%
}
\end{table}

\begin{figure}[htbp]
\centering
\includegraphics[width=0.85\textwidth]{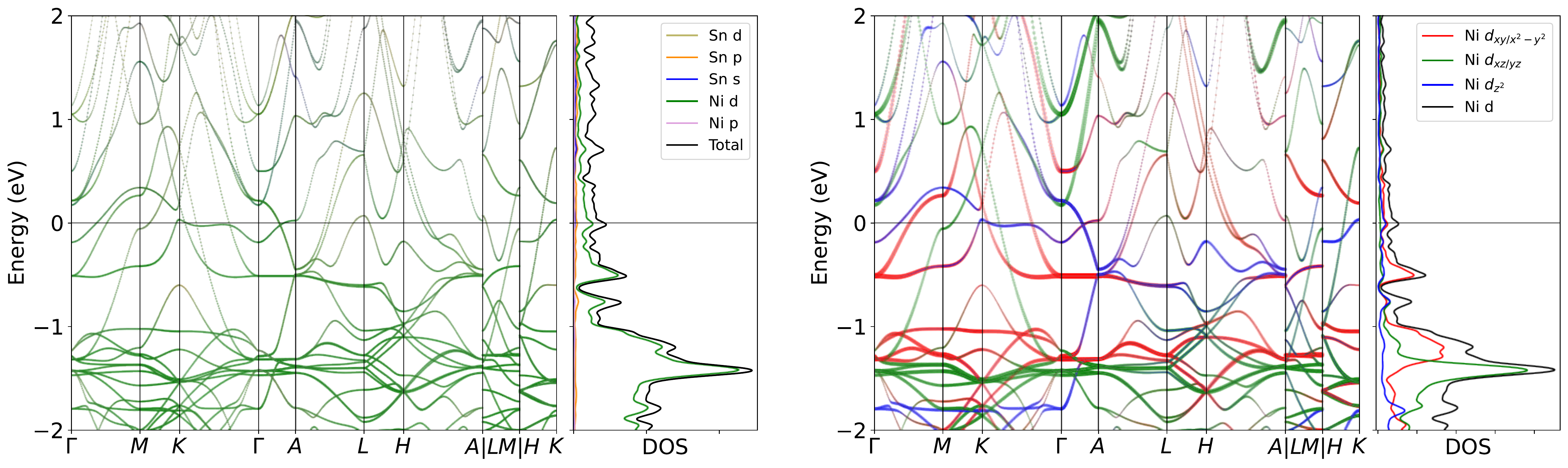}
\caption{Band structure for MgNi$_6$Sn$_6$ without SOC. The projection of Ni $3d$ orbitals is also presented. The band structures clearly reveal the dominating of Ni $3d$ orbitals  near the Fermi level, as well as the pronounced peak.}
\label{fig:MgNi6Sn6band}
\end{figure}

\begin{figure}[H]
\centering
\subfloat[Relaxed phonon at low-T.]{
\label{fig:MgNi6Sn6_relaxed_Low}
\includegraphics[width=0.45\textwidth]{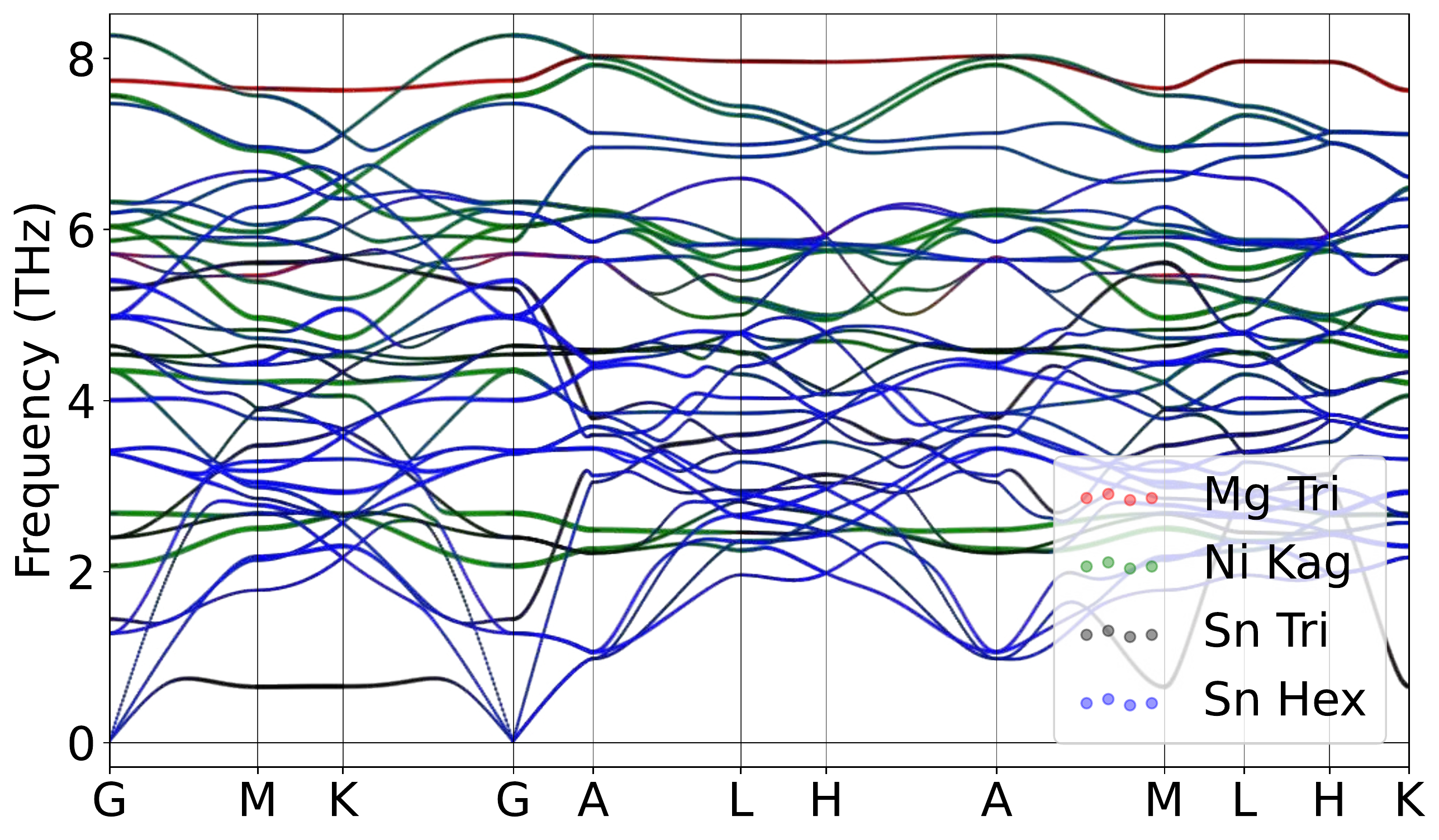}
}
\hspace{5pt}\subfloat[Relaxed phonon at high-T.]{
\label{fig:MgNi6Sn6_relaxed_High}
\includegraphics[width=0.45\textwidth]{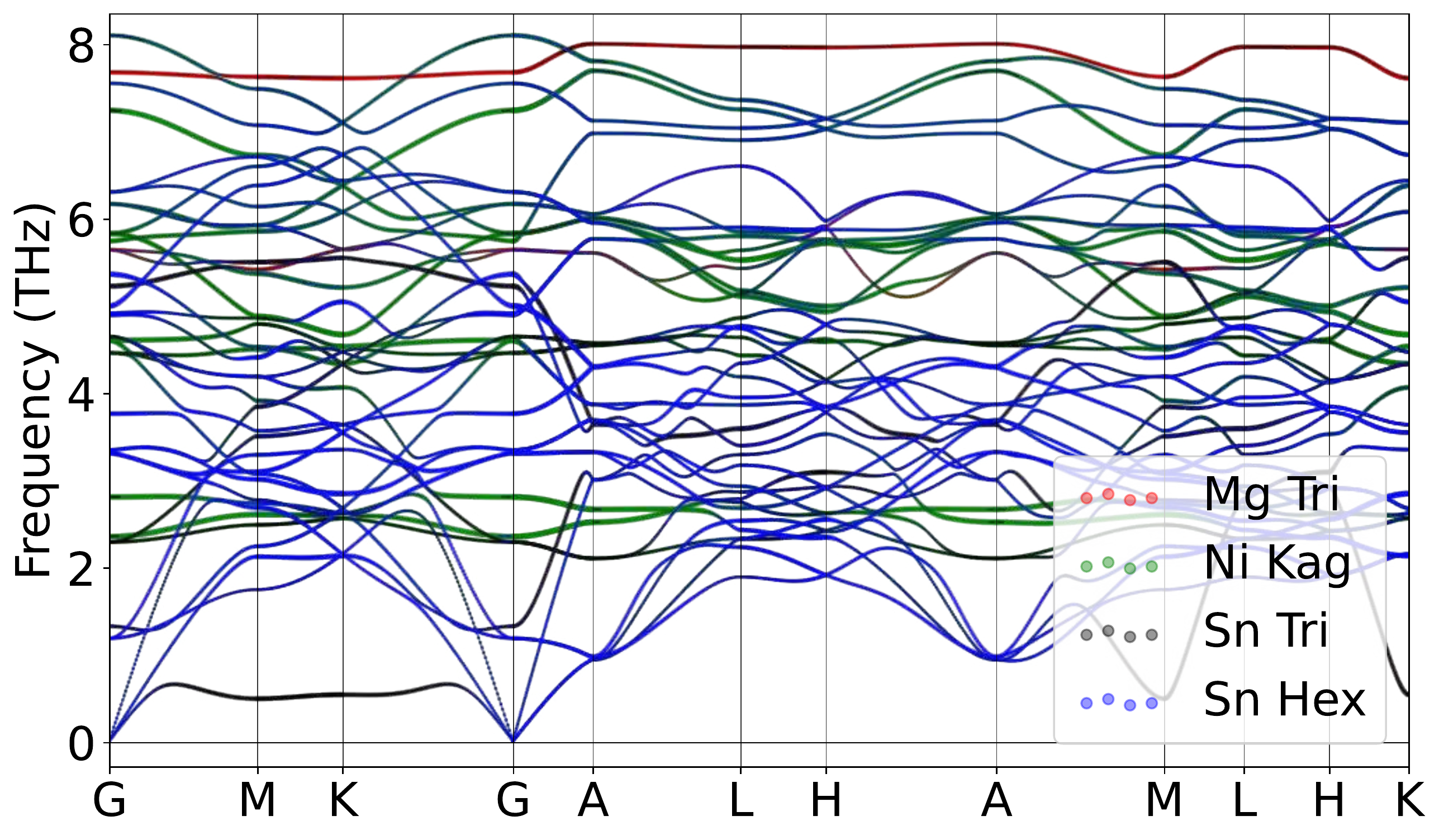}
}
\caption{Atomically projected phonon spectrum for MgNi$_6$Sn$_6$.}
\label{fig:MgNi6Sn6pho}
\end{figure}

\begin{figure}[H]
\centering
\label{fig:MgNi6Sn6_relaxed_Low_xz}
\includegraphics[width=0.85\textwidth]{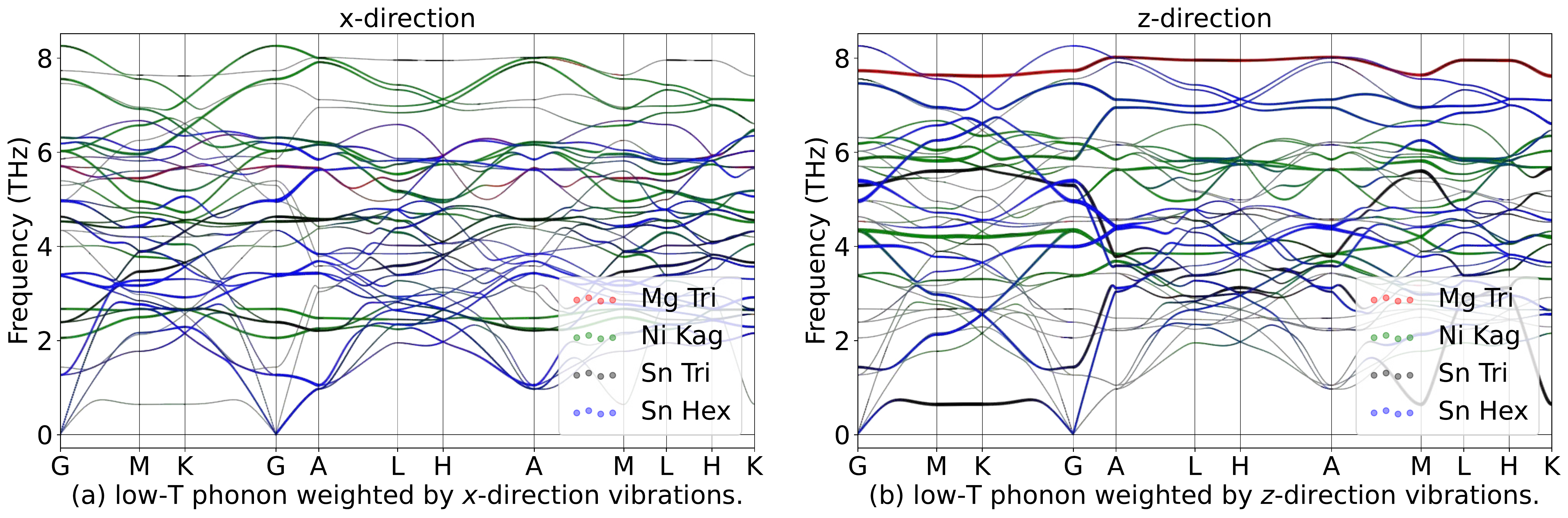}
\hspace{5pt}\label{fig:MgNi6Sn6_relaxed_High_xz}
\includegraphics[width=0.85\textwidth]{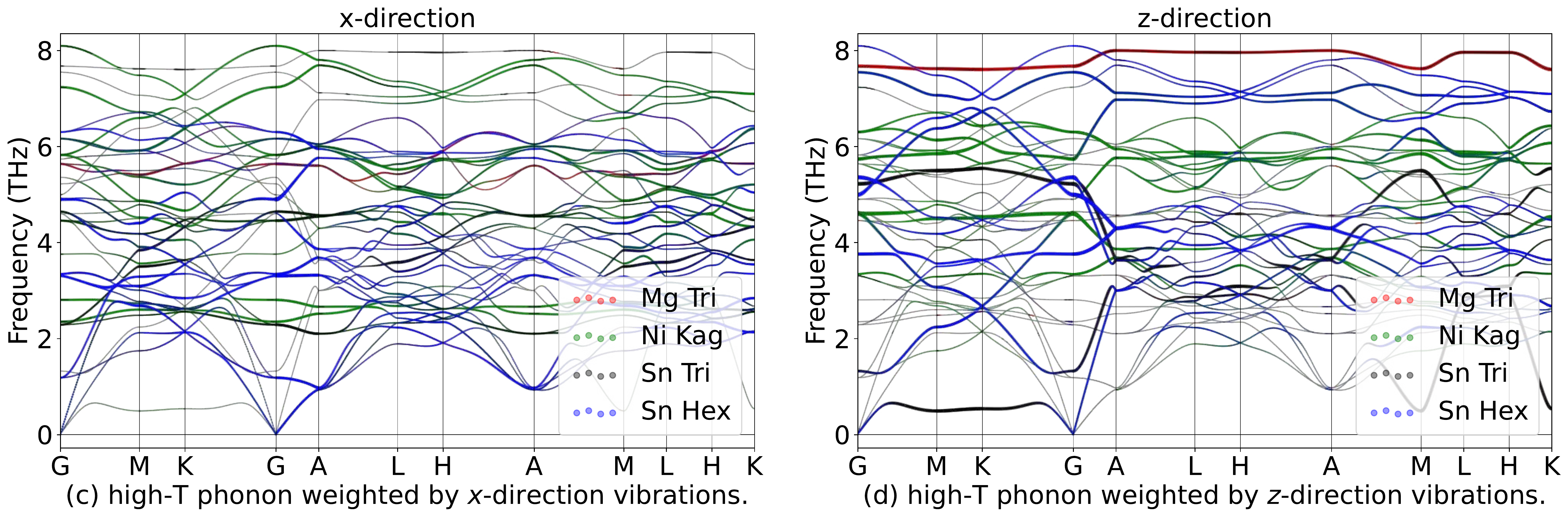}
\caption{Phonon spectrum for MgNi$_6$Sn$_6$in in-plane ($x$) and out-of-plane ($z$) directions.}
\label{fig:MgNi6Sn6phoxyz}
\end{figure}

\begin{table}[htbp]
\global\long\def\arraystretch{1.12}
\captionsetup{width=.95\textwidth}
\caption{IRREPs at high-symmetry points for MgNi$_6$Sn$_6$ phonon spectrum at Low-T.}
\label{tab:191_MgNi6Sn6_Low}
\setlength{\tabcolsep}{1.mm}{
}
\end{table}

\begin{table}[htbp]
\global\long\def\arraystretch{1.12}
\captionsetup{width=.95\textwidth}
\caption{IRREPs at high-symmetry points for MgNi$_6$Sn$_6$ phonon spectrum at High-T.}
\label{tab:191_MgNi6Sn6_High}
\setlength{\tabcolsep}{1.mm}{
%
}
\end{table}
\subsubsection{\label{sms2sec:CaNi6Sn6}\hyperref[smsec:col]{\texorpdfstring{CaNi$_6$Sn$_6$}{}}~{\color{green}  (High-T stable, Low-T stable) }}

\begin{table}[htbp]
\global\long\def\arraystretch{1.12}
\captionsetup{width=.85\textwidth}
\caption{Structure parameters of CaNi$_6$Sn$_6$ after relaxation. It contains 488 electrons. }
\label{tab:CaNi6Sn6}
\setlength{\tabcolsep}{4mm}{
%
}
\end{table}

\begin{figure}[htbp]
\centering
\includegraphics[width=0.85\textwidth]{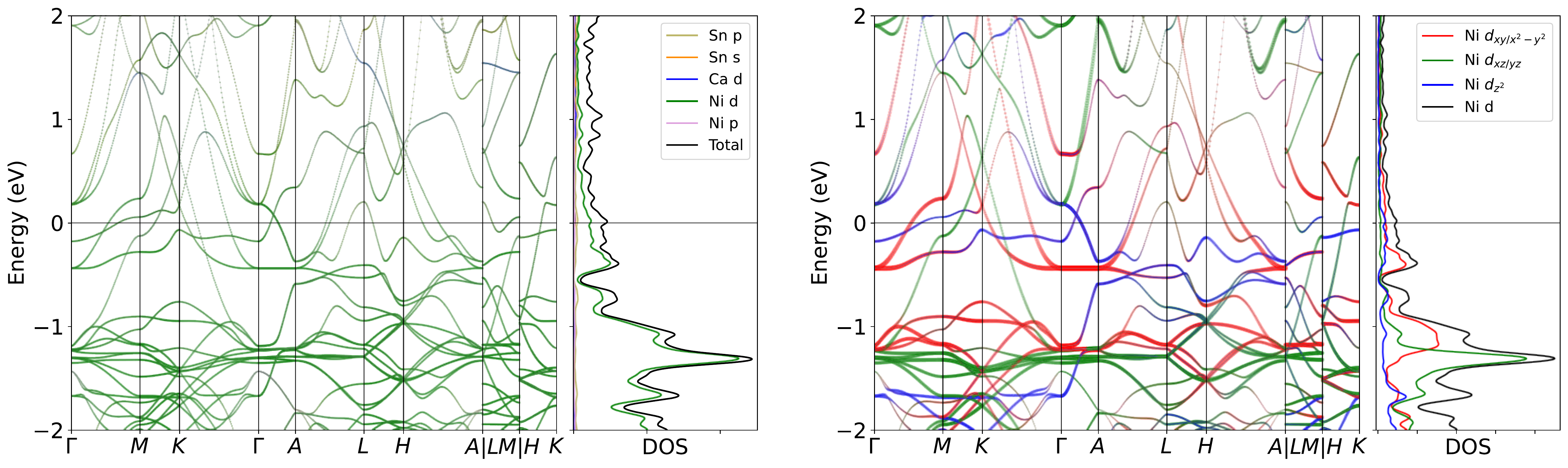}
\caption{Band structure for CaNi$_6$Sn$_6$ without SOC. The projection of Ni $3d$ orbitals is also presented. The band structures clearly reveal the dominating of Ni $3d$ orbitals  near the Fermi level, as well as the pronounced peak.}
\label{fig:CaNi6Sn6band}
\end{figure}

\begin{figure}[H]
\centering
\subfloat[Relaxed phonon at low-T.]{
\label{fig:CaNi6Sn6_relaxed_Low}
\includegraphics[width=0.45\textwidth]{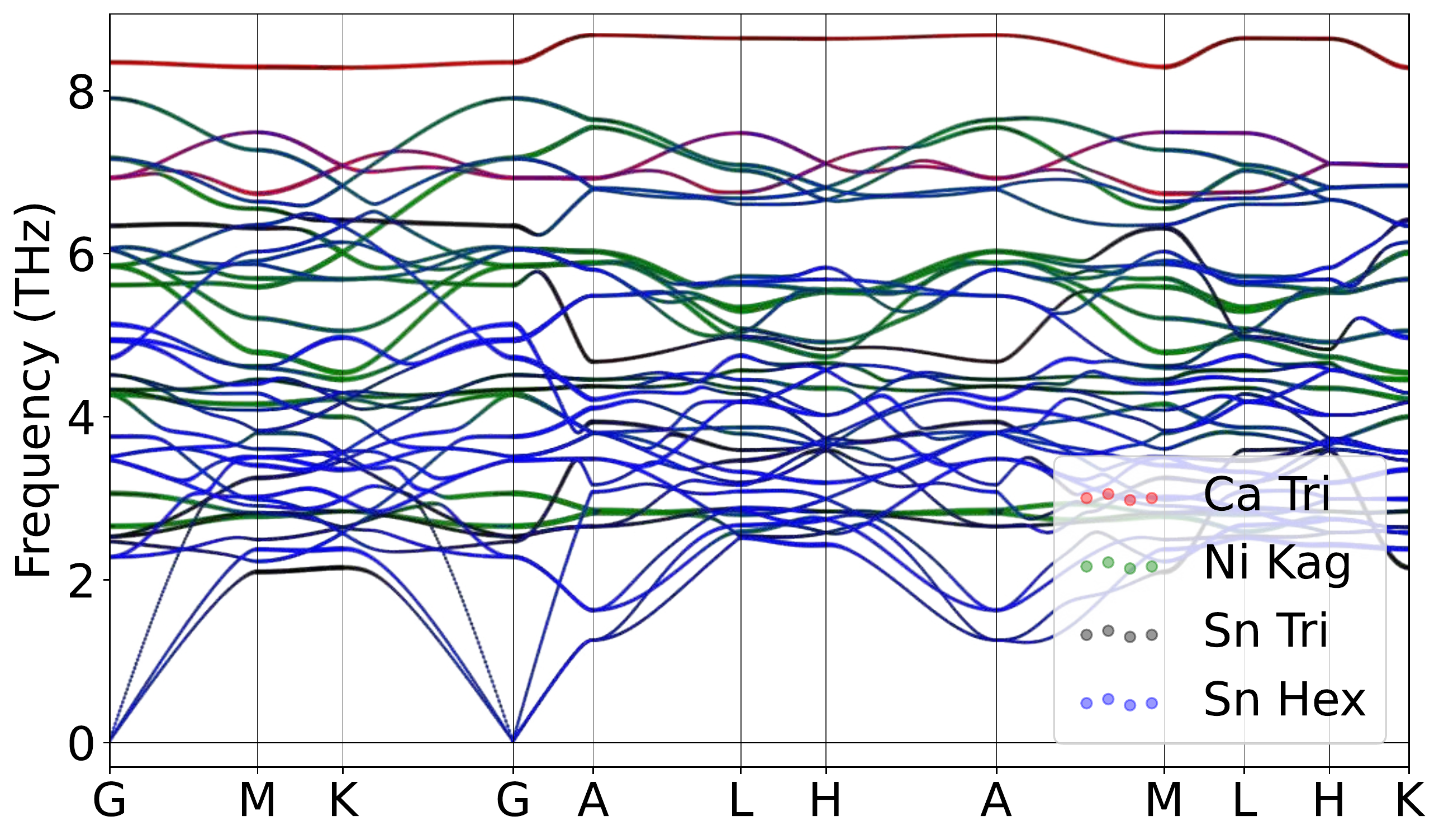}
}
\hspace{5pt}\subfloat[Relaxed phonon at high-T.]{
\label{fig:CaNi6Sn6_relaxed_High}
\includegraphics[width=0.45\textwidth]{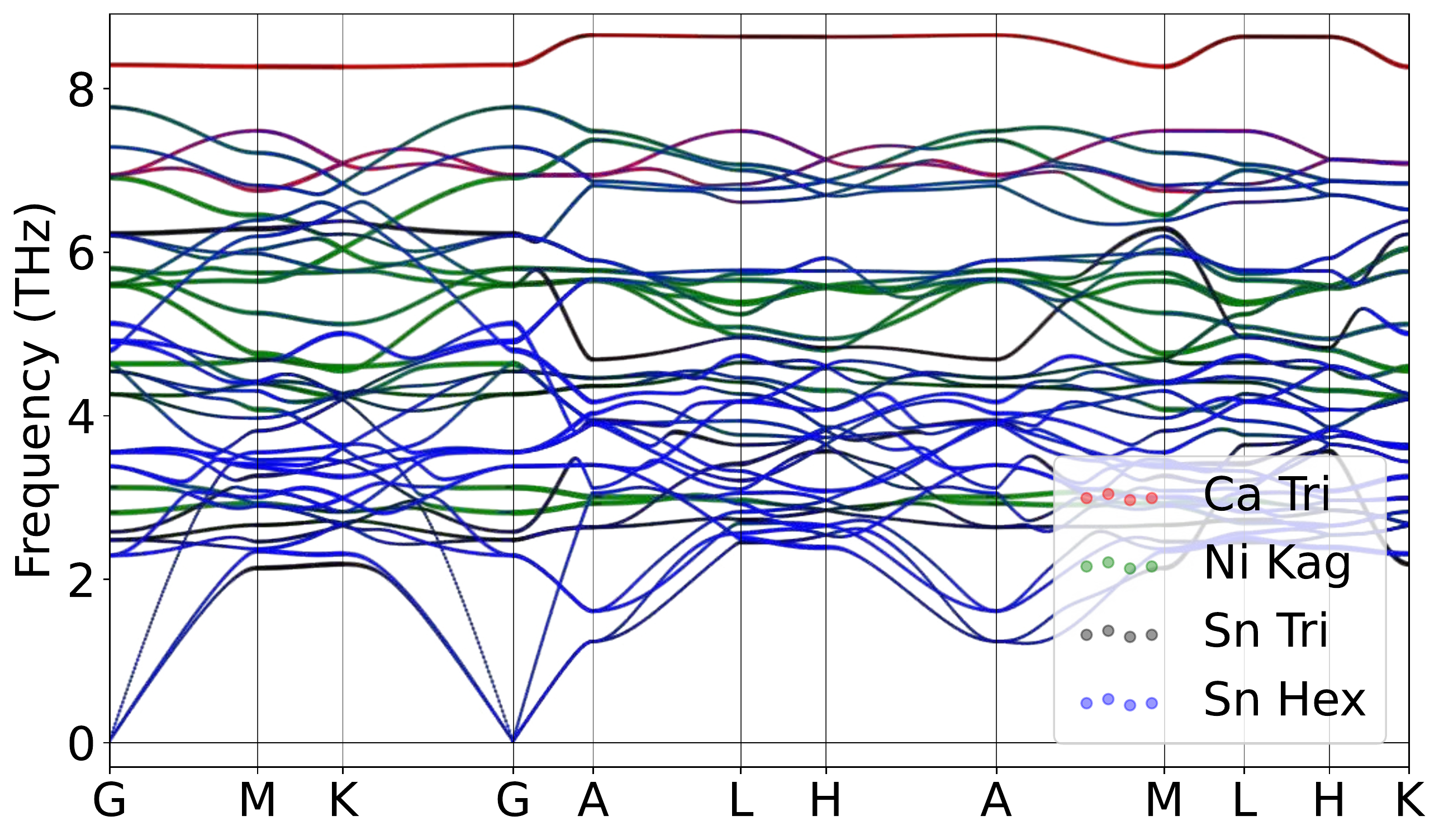}
}
\caption{Atomically projected phonon spectrum for CaNi$_6$Sn$_6$.}
\label{fig:CaNi6Sn6pho}
\end{figure}

\begin{figure}[H]
\centering
\label{fig:CaNi6Sn6_relaxed_Low_xz}
\includegraphics[width=0.85\textwidth]{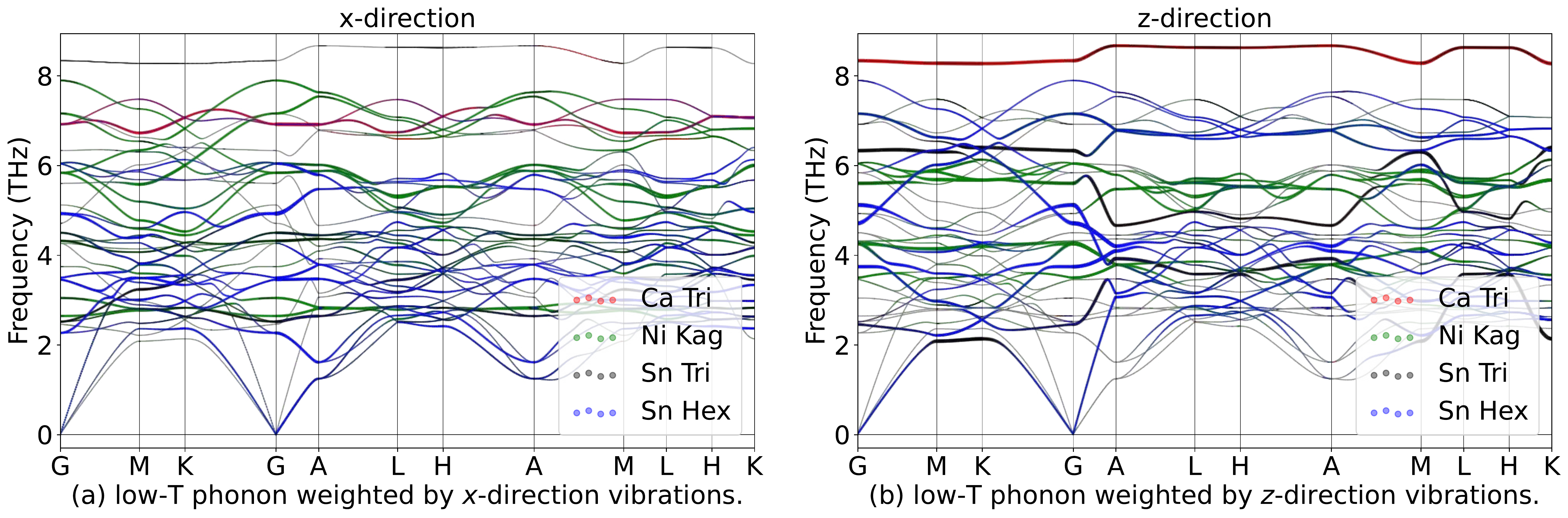}
\hspace{5pt}\label{fig:CaNi6Sn6_relaxed_High_xz}
\includegraphics[width=0.85\textwidth]{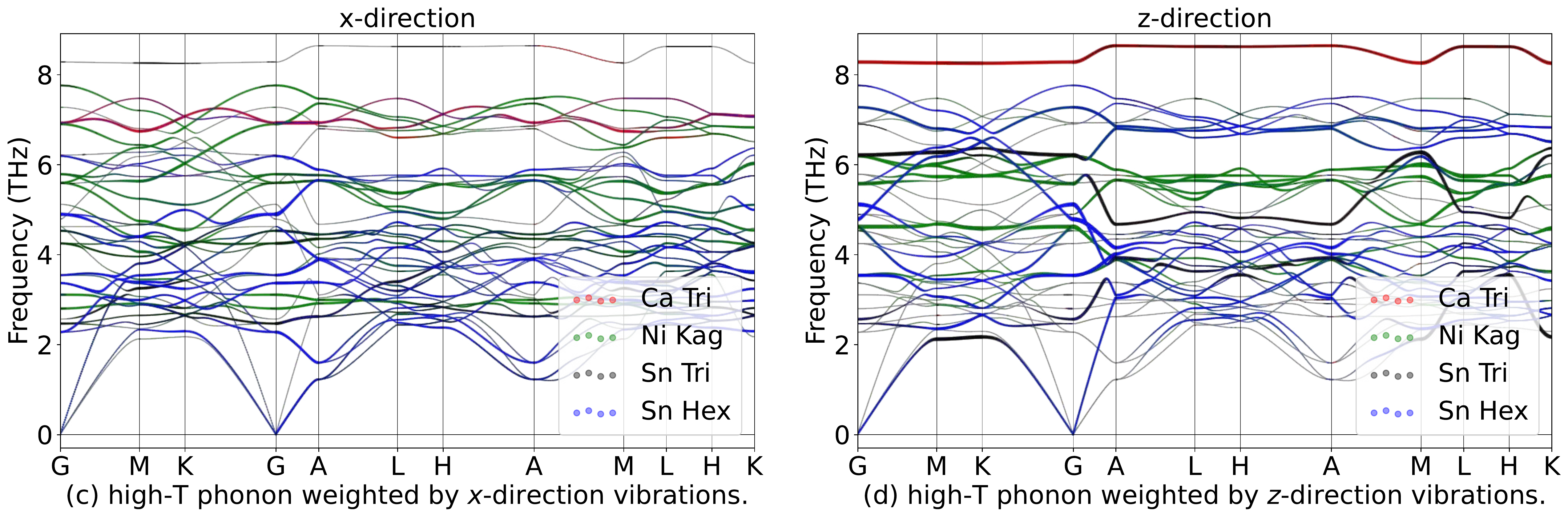}
\caption{Phonon spectrum for CaNi$_6$Sn$_6$in in-plane ($x$) and out-of-plane ($z$) directions.}
\label{fig:CaNi6Sn6phoxyz}
\end{figure}

\begin{table}[htbp]
\global\long\def\arraystretch{1.12}
\captionsetup{width=.95\textwidth}
\caption{IRREPs at high-symmetry points for CaNi$_6$Sn$_6$ phonon spectrum at Low-T.}
\label{tab:191_CaNi6Sn6_Low}
\setlength{\tabcolsep}{1.mm}{
}
\end{table}

\begin{table}[htbp]
\global\long\def\arraystretch{1.12}
\captionsetup{width=.95\textwidth}
\caption{IRREPs at high-symmetry points for CaNi$_6$Sn$_6$ phonon spectrum at High-T.}
\label{tab:191_CaNi6Sn6_High}
\setlength{\tabcolsep}{1.mm}{
%
}
\end{table}
\subsubsection{\label{sms2sec:ScNi6Sn6}\hyperref[smsec:col]{\texorpdfstring{ScNi$_6$Sn$_6$}{}}~{\color{green}  (High-T stable, Low-T stable) }}

\begin{table}[htbp]
\global\long\def\arraystretch{1.12}
\captionsetup{width=.85\textwidth}
\caption{Structure parameters of ScNi$_6$Sn$_6$ after relaxation. It contains 489 electrons. }
\label{tab:ScNi6Sn6}
\setlength{\tabcolsep}{4mm}{
%
}
\end{table}

\begin{figure}[htbp]
\centering
\includegraphics[width=0.85\textwidth]{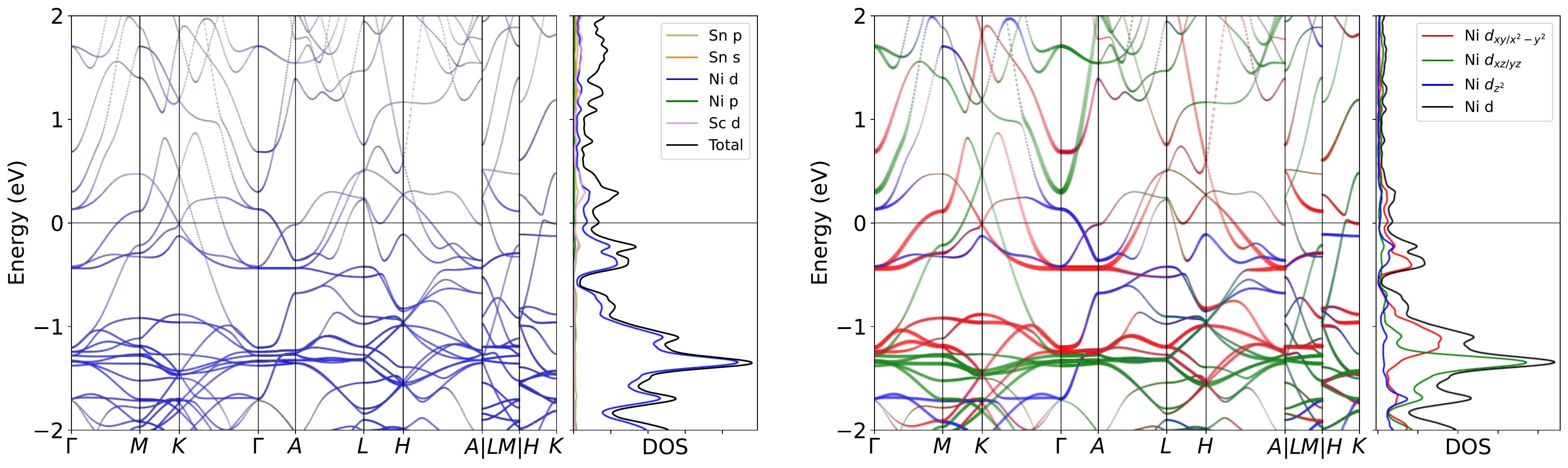}
\caption{Band structure for ScNi$_6$Sn$_6$ without SOC. The projection of Ni $3d$ orbitals is also presented. The band structures clearly reveal the dominating of Ni $3d$ orbitals  near the Fermi level, as well as the pronounced peak.}
\label{fig:ScNi6Sn6band}
\end{figure}

\begin{figure}[H]
\centering
\subfloat[Relaxed phonon at low-T.]{
\label{fig:ScNi6Sn6_relaxed_Low}
\includegraphics[width=0.45\textwidth]{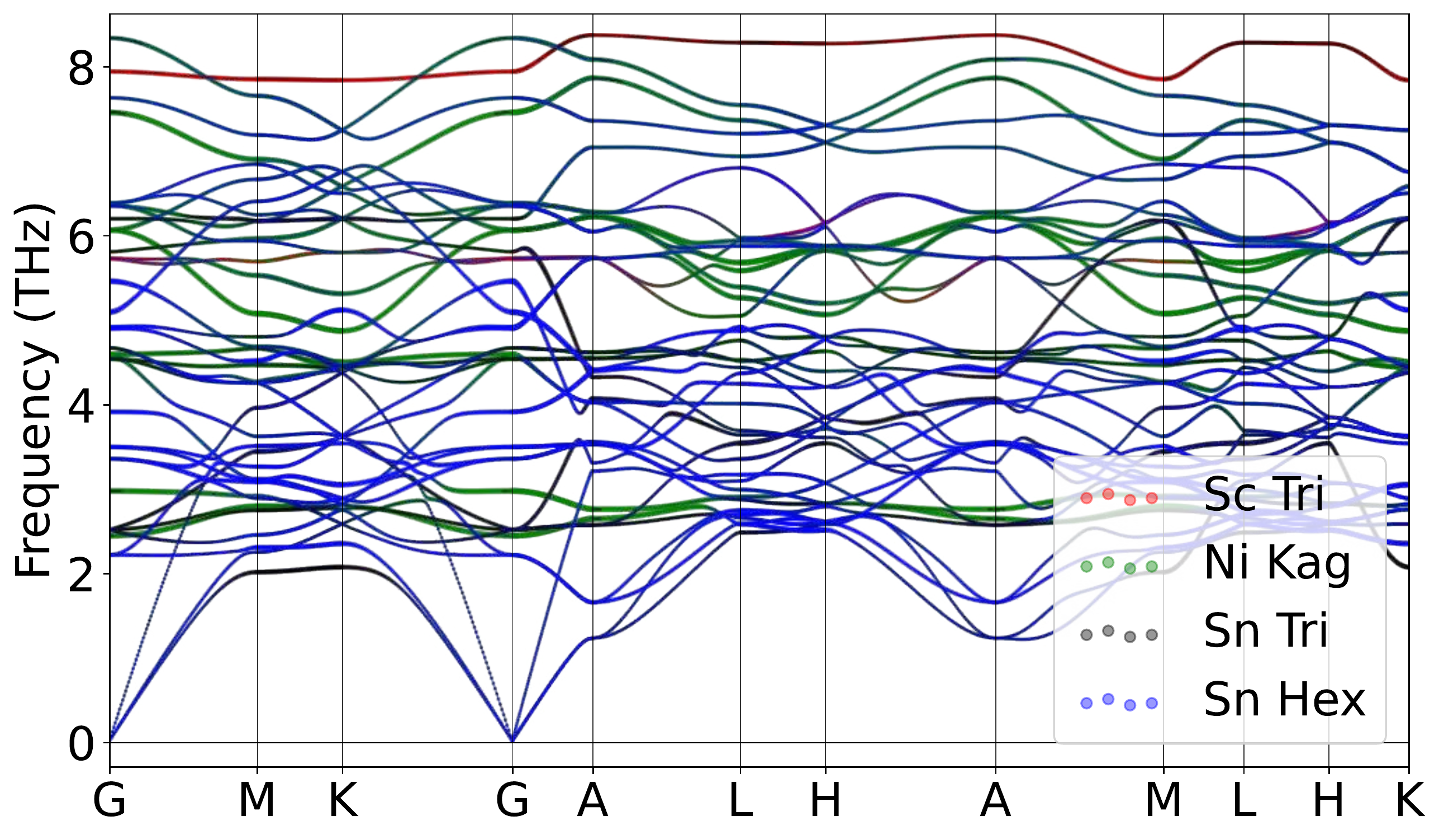}
}
\hspace{5pt}\subfloat[Relaxed phonon at high-T.]{
\label{fig:ScNi6Sn6_relaxed_High}
\includegraphics[width=0.45\textwidth]{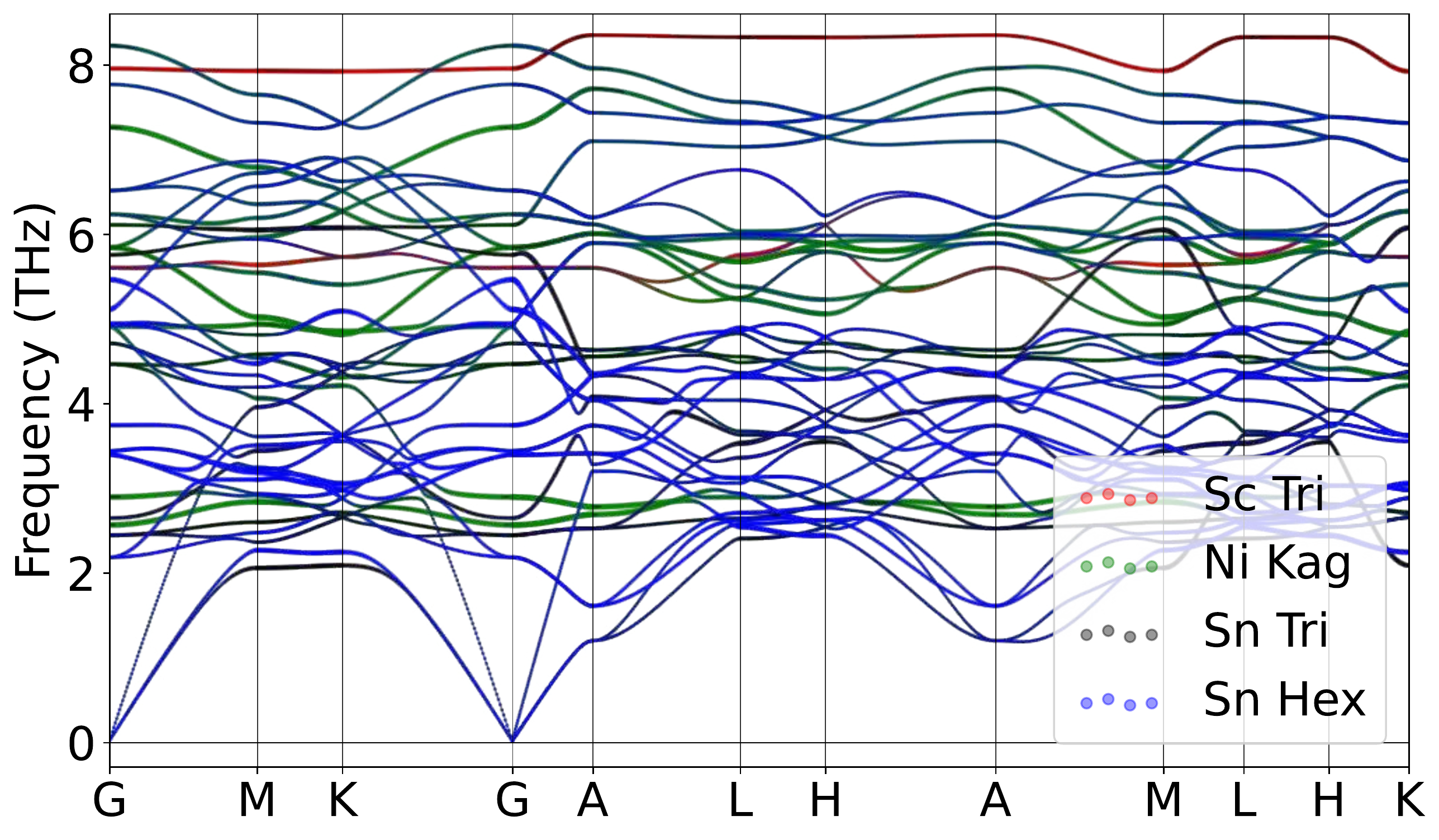}
}
\caption{Atomically projected phonon spectrum for ScNi$_6$Sn$_6$.}
\label{fig:ScNi6Sn6pho}
\end{figure}

\begin{figure}[H]
\centering
\label{fig:ScNi6Sn6_relaxed_Low_xz}
\includegraphics[width=0.85\textwidth]{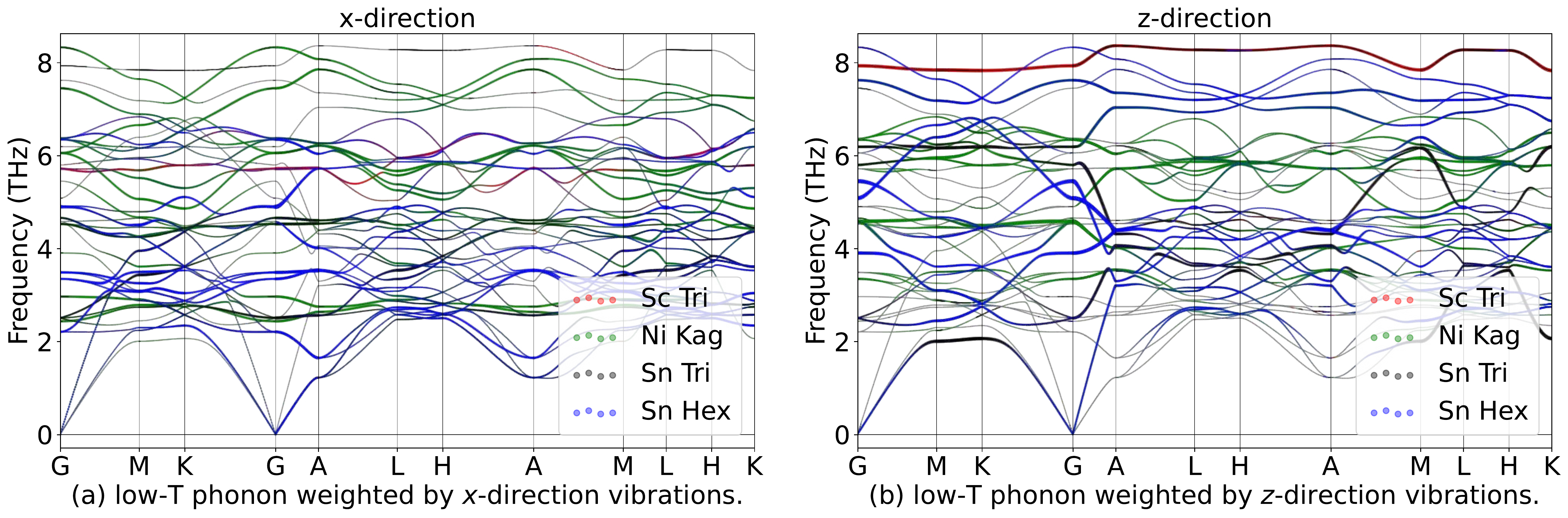}
\hspace{5pt}\label{fig:ScNi6Sn6_relaxed_High_xz}
\includegraphics[width=0.85\textwidth]{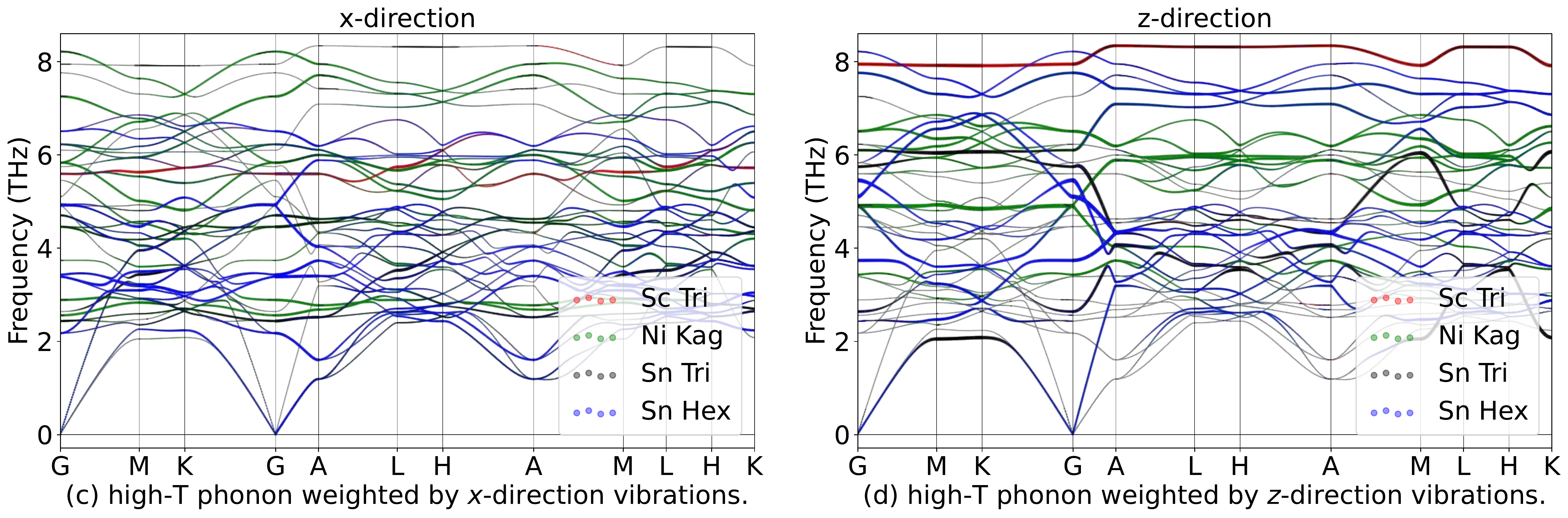}
\caption{Phonon spectrum for ScNi$_6$Sn$_6$in in-plane ($x$) and out-of-plane ($z$) directions.}
\label{fig:ScNi6Sn6phoxyz}
\end{figure}

\begin{table}[htbp]
\global\long\def\arraystretch{1.12}
\captionsetup{width=.95\textwidth}
\caption{IRREPs at high-symmetry points for ScNi$_6$Sn$_6$ phonon spectrum at Low-T.}
\label{tab:191_ScNi6Sn6_Low}
\setlength{\tabcolsep}{1.mm}{
}
\end{table}

\begin{table}[htbp]
\global\long\def\arraystretch{1.12}
\captionsetup{width=.95\textwidth}
\caption{IRREPs at high-symmetry points for ScNi$_6$Sn$_6$ phonon spectrum at High-T.}
\label{tab:191_ScNi6Sn6_High}
\setlength{\tabcolsep}{1.mm}{
%
}
\end{table}
\subsubsection{\label{sms2sec:TiNi6Sn6}\hyperref[smsec:col]{\texorpdfstring{TiNi$_6$Sn$_6$}{}}~{\color{green}  (High-T stable, Low-T stable) }}

\begin{table}[htbp]
\global\long\def\arraystretch{1.12}
\captionsetup{width=.85\textwidth}
\caption{Structure parameters of TiNi$_6$Sn$_6$ after relaxation. It contains 490 electrons. }
\label{tab:TiNi6Sn6}
\setlength{\tabcolsep}{4mm}{
%
}
\end{table}

\begin{figure}[htbp]
\centering
\includegraphics[width=0.85\textwidth]{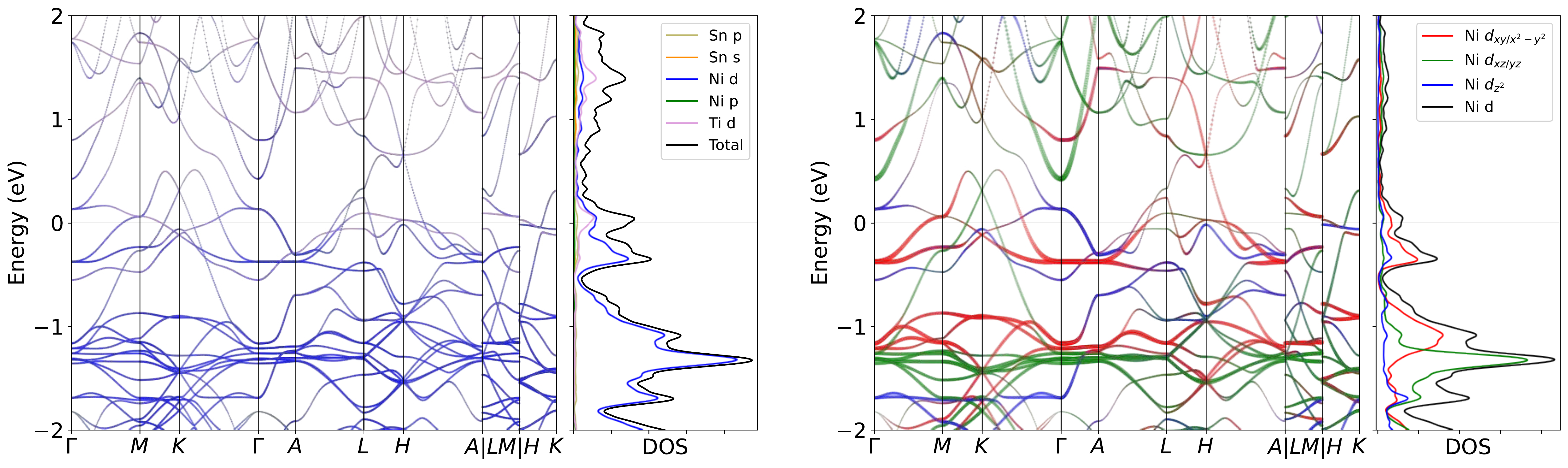}
\caption{Band structure for TiNi$_6$Sn$_6$ without SOC. The projection of Ni $3d$ orbitals is also presented. The band structures clearly reveal the dominating of Ni $3d$ orbitals  near the Fermi level, as well as the pronounced peak.}
\label{fig:TiNi6Sn6band}
\end{figure}

\begin{figure}[H]
\centering
\subfloat[Relaxed phonon at low-T.]{
\label{fig:TiNi6Sn6_relaxed_Low}
\includegraphics[width=0.45\textwidth]{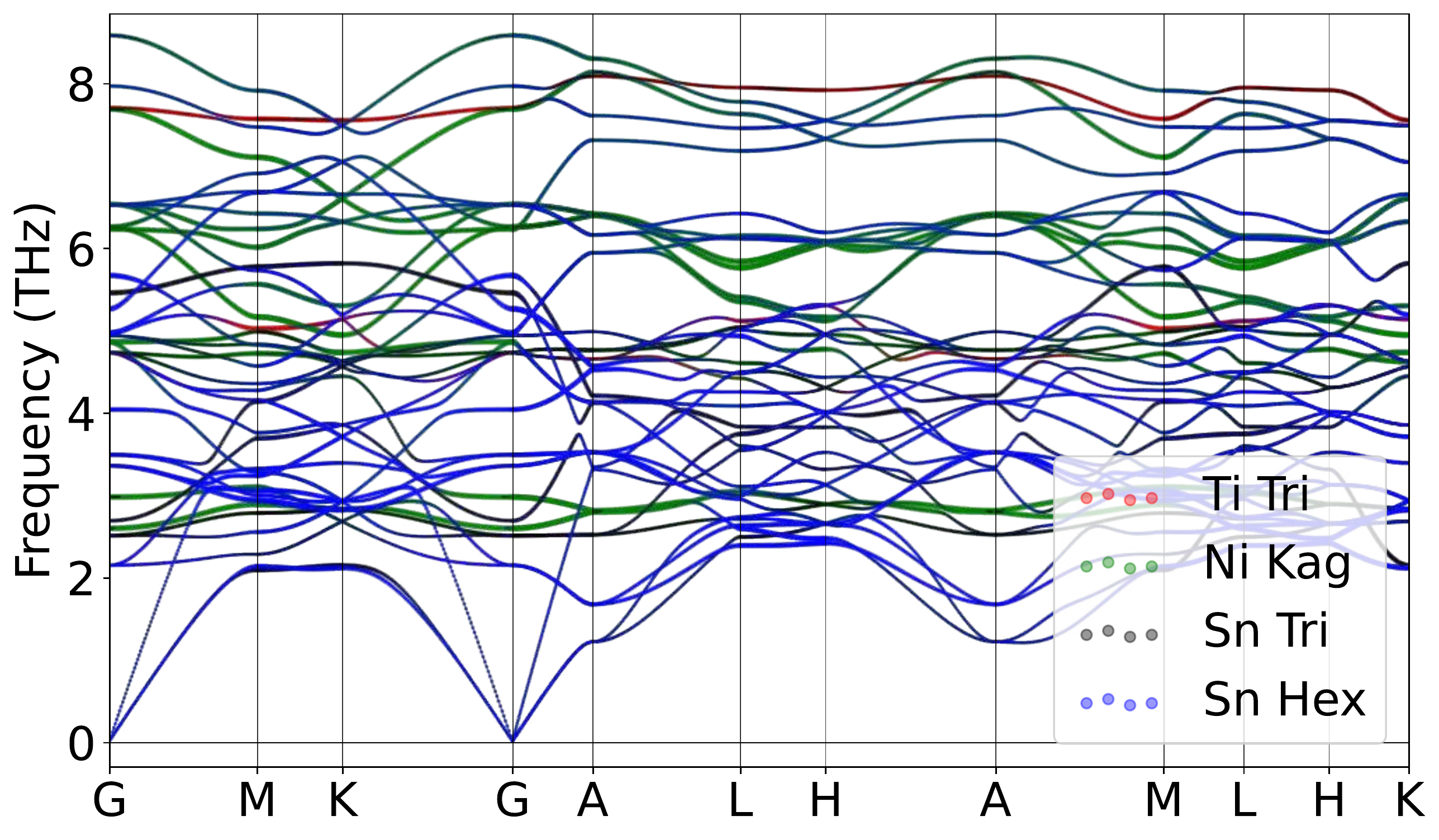}
}
\hspace{5pt}\subfloat[Relaxed phonon at high-T.]{
\label{fig:TiNi6Sn6_relaxed_High}
\includegraphics[width=0.45\textwidth]{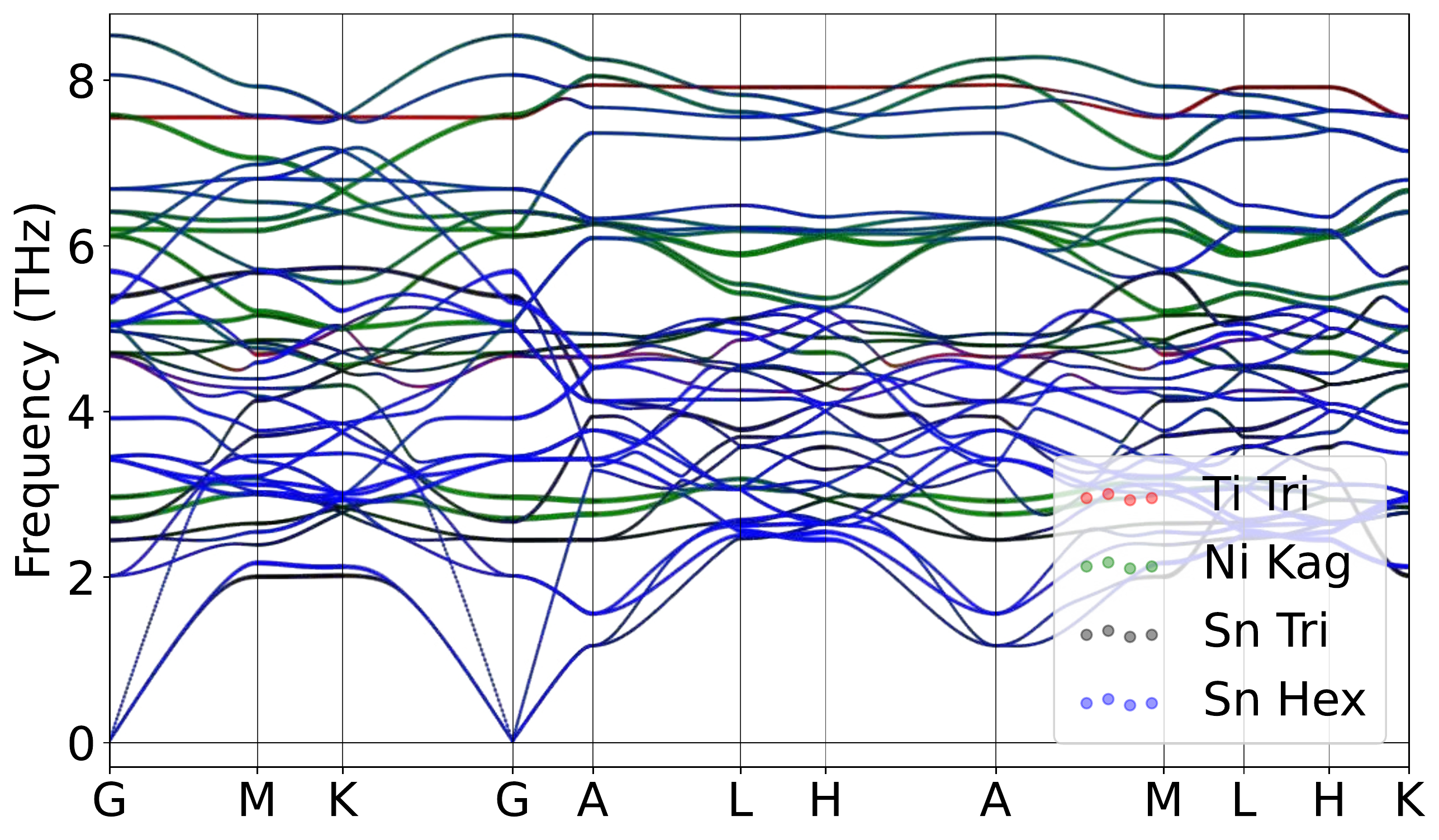}
}
\caption{Atomically projected phonon spectrum for TiNi$_6$Sn$_6$.}
\label{fig:TiNi6Sn6pho}
\end{figure}

\begin{figure}[H]
\centering
\label{fig:TiNi6Sn6_relaxed_Low_xz}
\includegraphics[width=0.85\textwidth]{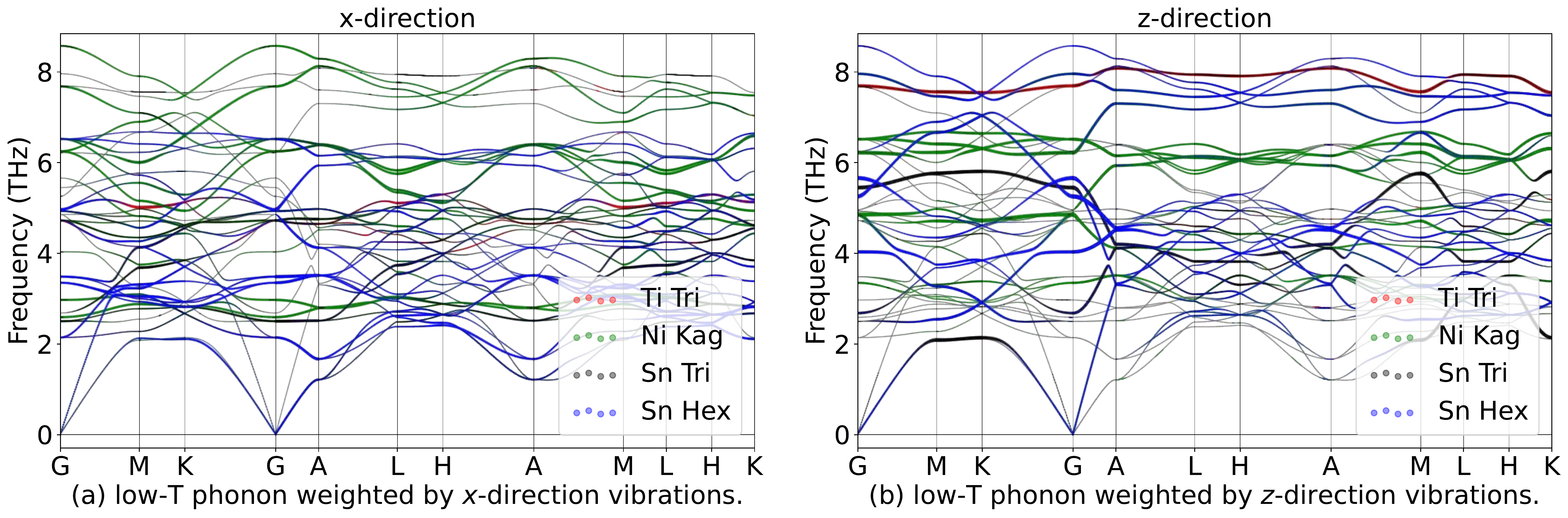}
\hspace{5pt}\label{fig:TiNi6Sn6_relaxed_High_xz}
\includegraphics[width=0.85\textwidth]{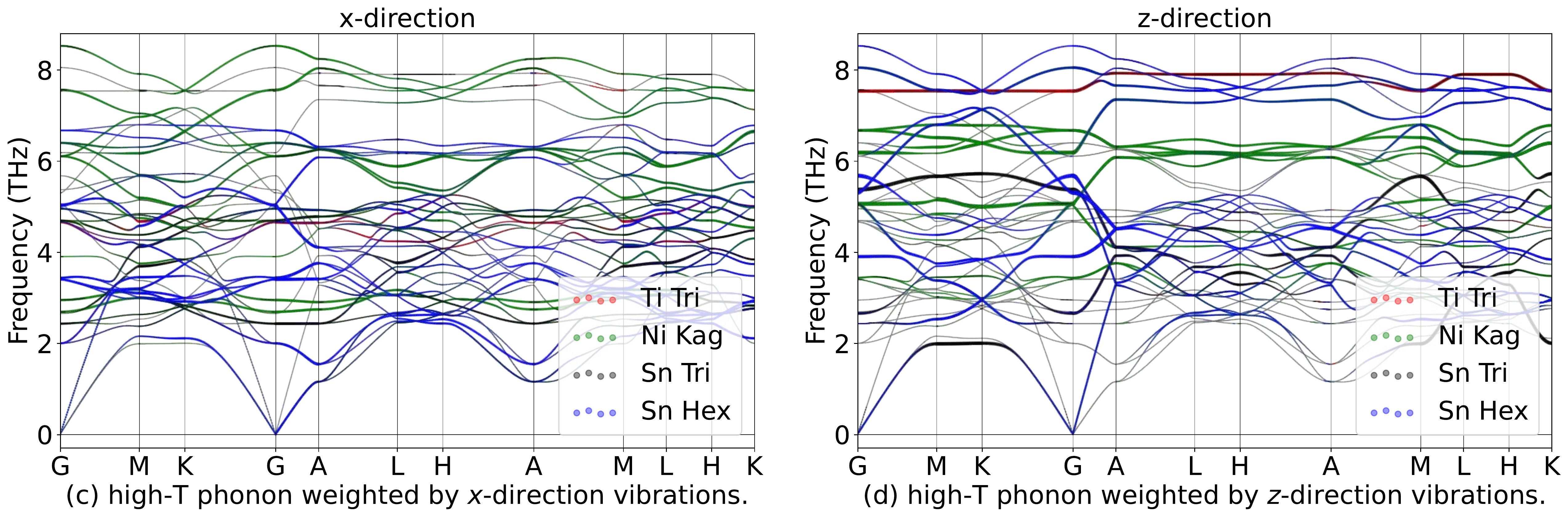}
\caption{Phonon spectrum for TiNi$_6$Sn$_6$in in-plane ($x$) and out-of-plane ($z$) directions.}
\label{fig:TiNi6Sn6phoxyz}
\end{figure}

\begin{table}[htbp]
\global\long\def\arraystretch{1.12}
\captionsetup{width=.95\textwidth}
\caption{IRREPs at high-symmetry points for TiNi$_6$Sn$_6$ phonon spectrum at Low-T.}
\label{tab:191_TiNi6Sn6_Low}
\setlength{\tabcolsep}{1.mm}{
}
\end{table}

\begin{table}[htbp]
\global\long\def\arraystretch{1.12}
\captionsetup{width=.95\textwidth}
\caption{IRREPs at high-symmetry points for TiNi$_6$Sn$_6$ phonon spectrum at High-T.}
\label{tab:191_TiNi6Sn6_High}
\setlength{\tabcolsep}{1.mm}{
%
}
\end{table}
\subsubsection{\label{sms2sec:YNi6Sn6}\hyperref[smsec:col]{\texorpdfstring{YNi$_6$Sn$_6$}{}}~{\color{green}  (High-T stable, Low-T stable) }}

\begin{table}[htbp]
\global\long\def\arraystretch{1.12}
\captionsetup{width=.85\textwidth}
\caption{Structure parameters of YNi$_6$Sn$_6$ after relaxation. It contains 507 electrons. }
\label{tab:YNi6Sn6}
\setlength{\tabcolsep}{4mm}{
%
}
\end{table}

\begin{figure}[htbp]
\centering
\includegraphics[width=0.85\textwidth]{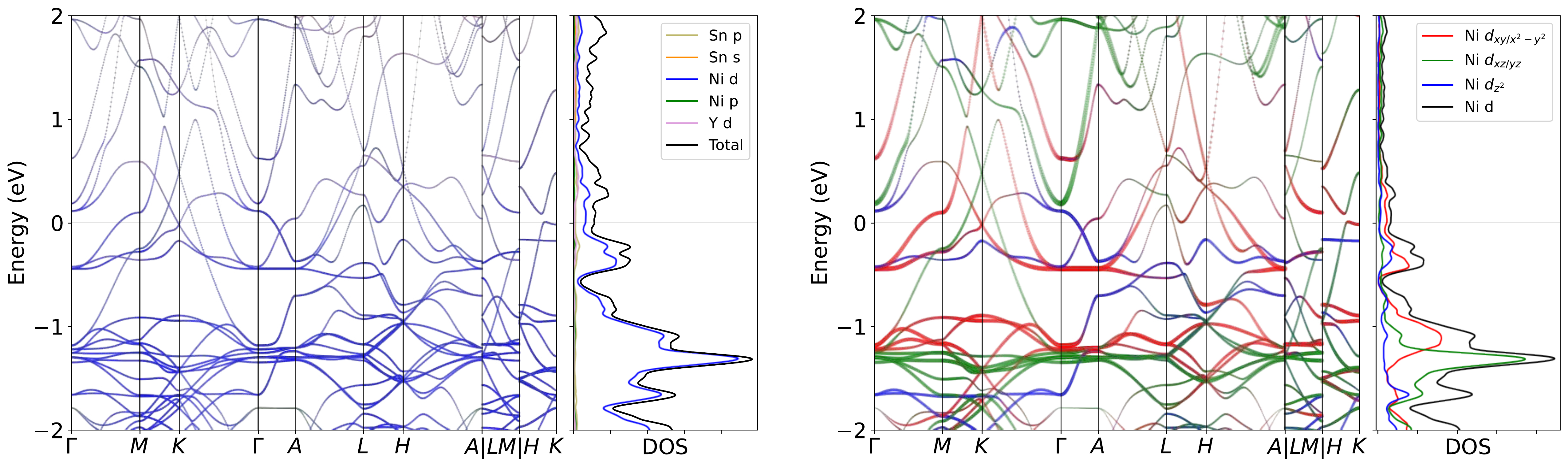}
\caption{Band structure for YNi$_6$Sn$_6$ without SOC. The projection of Ni $3d$ orbitals is also presented. The band structures clearly reveal the dominating of Ni $3d$ orbitals  near the Fermi level, as well as the pronounced peak.}
\label{fig:YNi6Sn6band}
\end{figure}

\begin{figure}[H]
\centering
\subfloat[Relaxed phonon at low-T.]{
\label{fig:YNi6Sn6_relaxed_Low}
\includegraphics[width=0.45\textwidth]{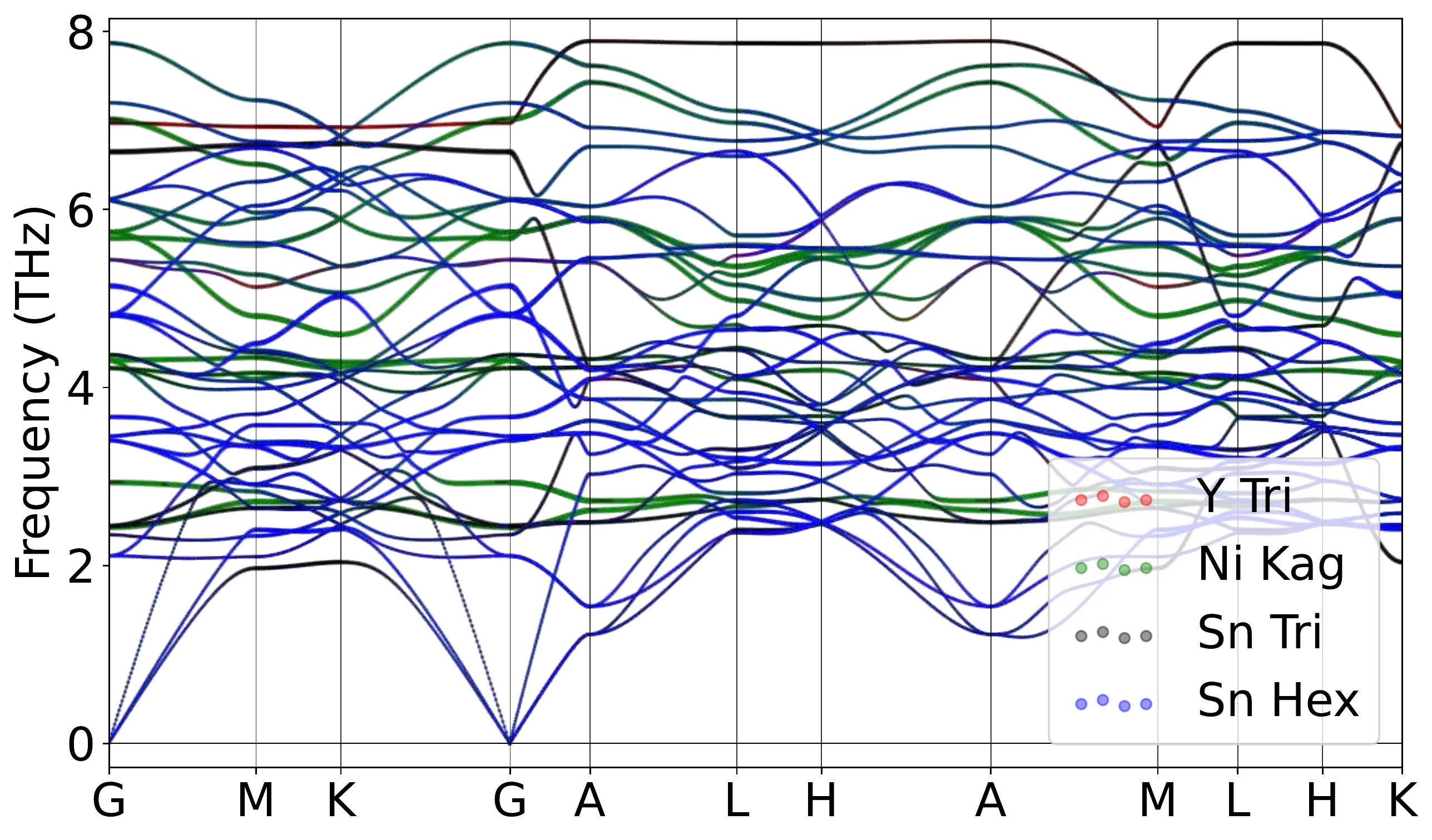}
}
\hspace{5pt}\subfloat[Relaxed phonon at high-T.]{
\label{fig:YNi6Sn6_relaxed_High}
\includegraphics[width=0.45\textwidth]{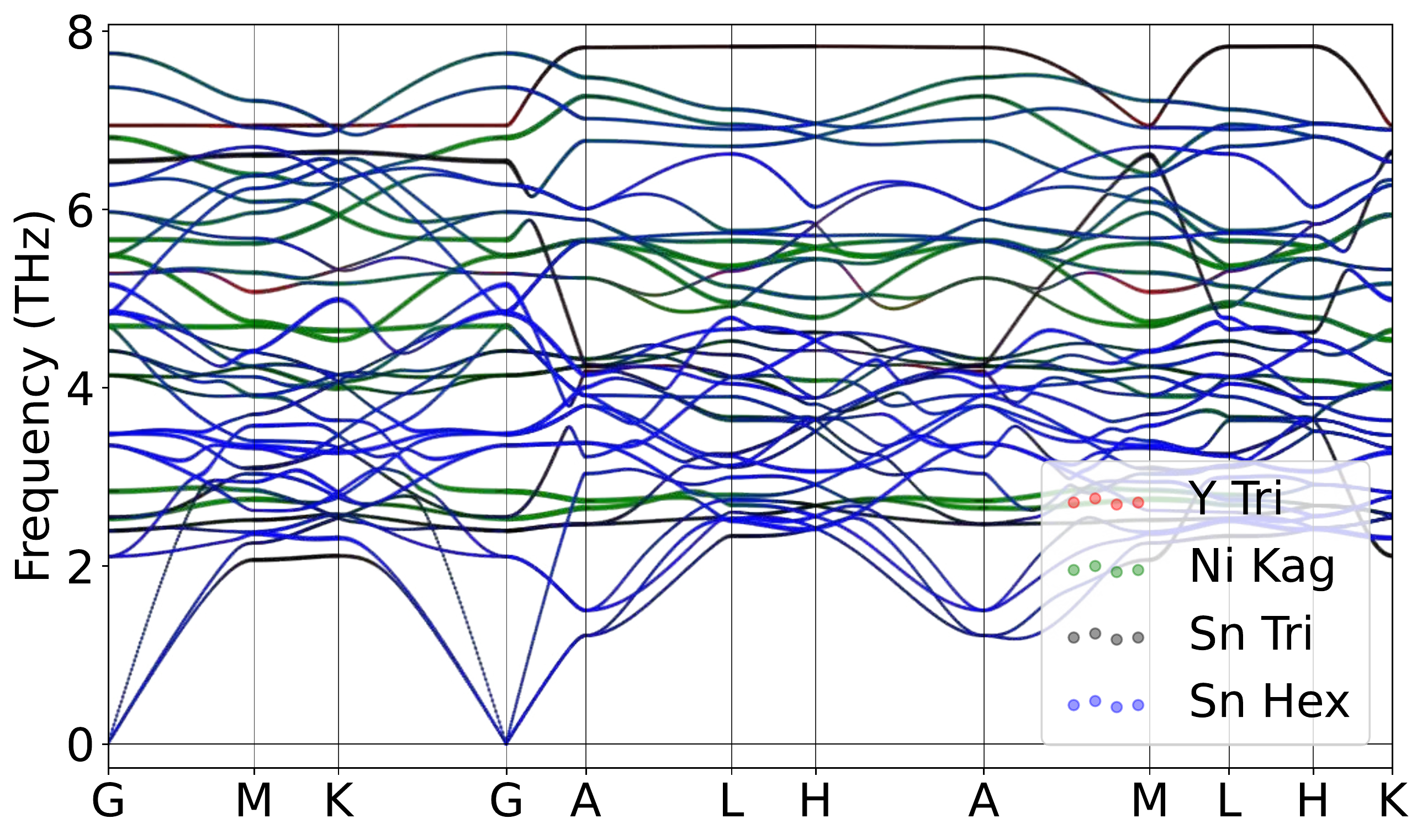}
}
\caption{Atomically projected phonon spectrum for YNi$_6$Sn$_6$.}
\label{fig:YNi6Sn6pho}
\end{figure}

\begin{figure}[H]
\centering
\label{fig:YNi6Sn6_relaxed_Low_xz}
\includegraphics[width=0.85\textwidth]{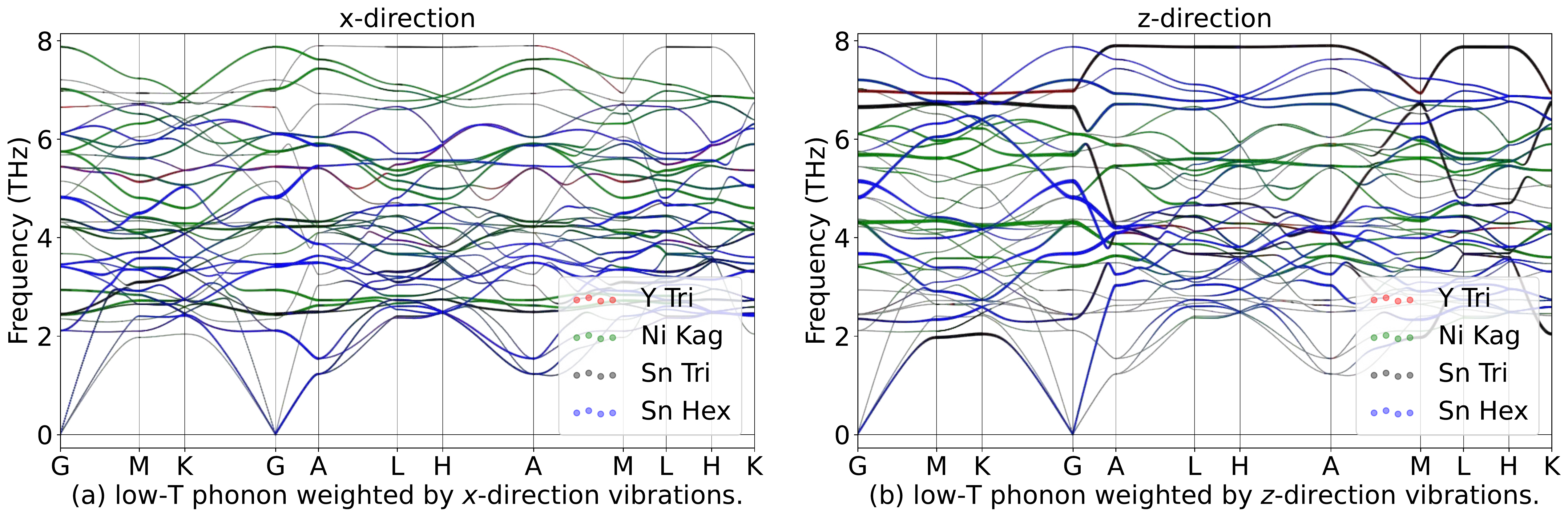}
\hspace{5pt}\label{fig:YNi6Sn6_relaxed_High_xz}
\includegraphics[width=0.85\textwidth]{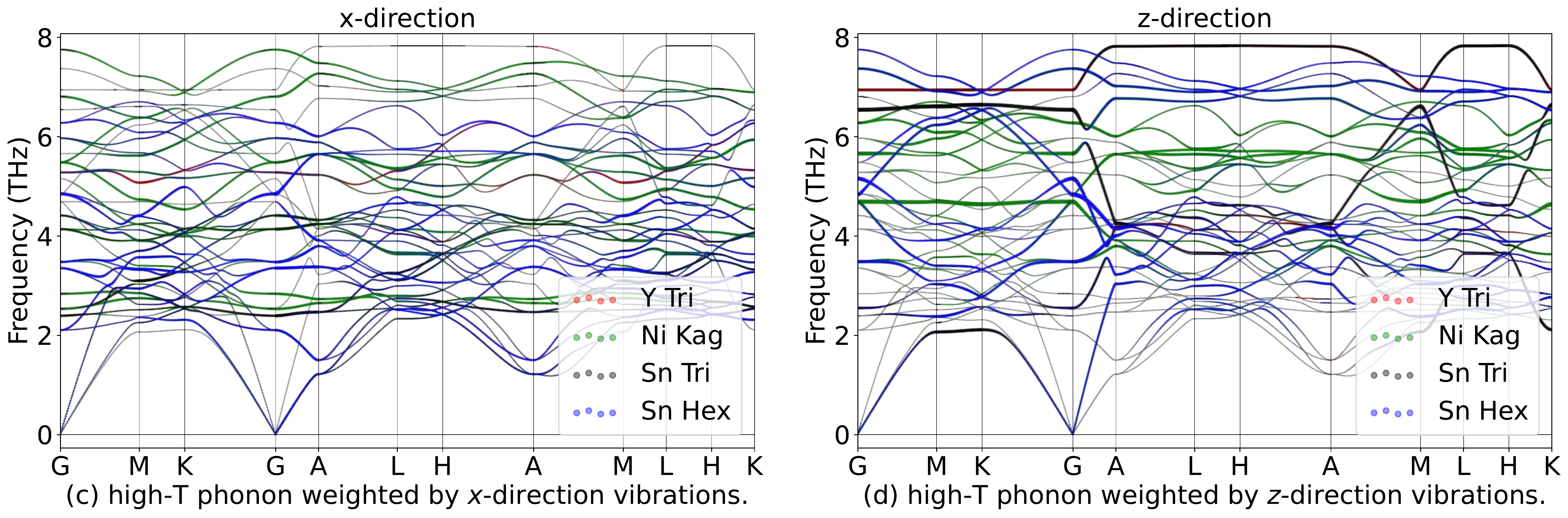}
\caption{Phonon spectrum for YNi$_6$Sn$_6$in in-plane ($x$) and out-of-plane ($z$) directions.}
\label{fig:YNi6Sn6phoxyz}
\end{figure}

\begin{table}[htbp]
\global\long\def\arraystretch{1.12}
\captionsetup{width=.95\textwidth}
\caption{IRREPs at high-symmetry points for YNi$_6$Sn$_6$ phonon spectrum at Low-T.}
\label{tab:191_YNi6Sn6_Low}
\setlength{\tabcolsep}{1.mm}{
}
\end{table}

\begin{table}[htbp]
\global\long\def\arraystretch{1.12}
\captionsetup{width=.95\textwidth}
\caption{IRREPs at high-symmetry points for YNi$_6$Sn$_6$ phonon spectrum at High-T.}
\label{tab:191_YNi6Sn6_High}
\setlength{\tabcolsep}{1.mm}{
%
}
\end{table}
\subsubsection{\label{sms2sec:ZrNi6Sn6}\hyperref[smsec:col]{\texorpdfstring{ZrNi$_6$Sn$_6$}{}}~{\color{green}  (High-T stable, Low-T stable) }}

\begin{table}[htbp]
\global\long\def\arraystretch{1.12}
\captionsetup{width=.85\textwidth}
\caption{Structure parameters of ZrNi$_6$Sn$_6$ after relaxation. It contains 508 electrons. }
\label{tab:ZrNi6Sn6}
\setlength{\tabcolsep}{4mm}{
%
}
\end{table}

\begin{figure}[htbp]
\centering
\includegraphics[width=0.85\textwidth]{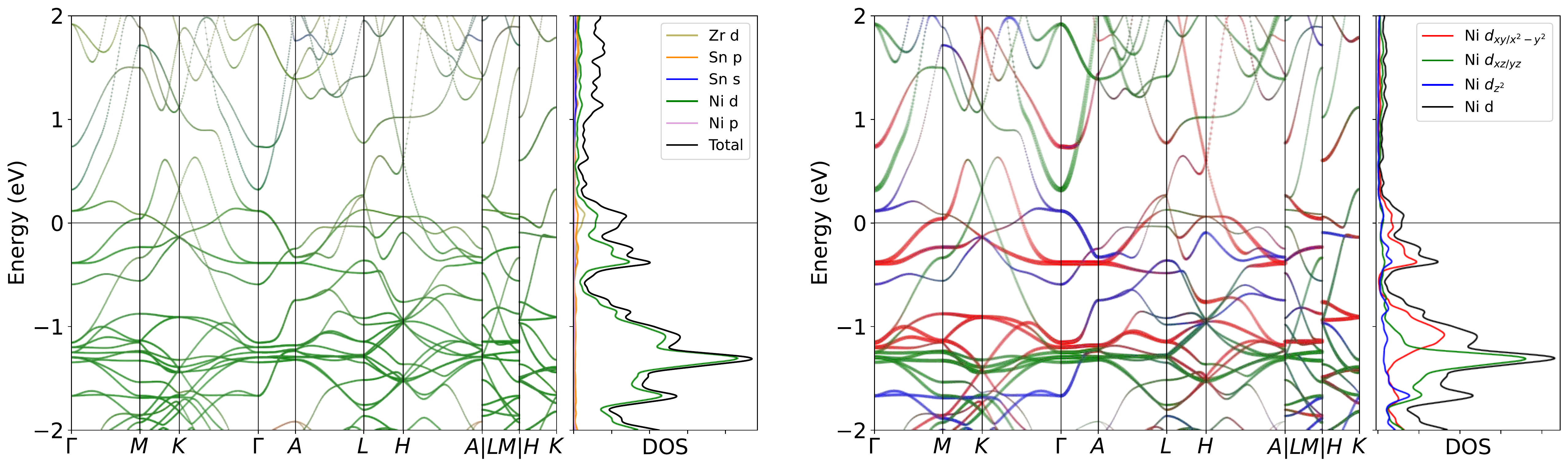}
\caption{Band structure for ZrNi$_6$Sn$_6$ without SOC. The projection of Ni $3d$ orbitals is also presented. The band structures clearly reveal the dominating of Ni $3d$ orbitals  near the Fermi level, as well as the pronounced peak.}
\label{fig:ZrNi6Sn6band}
\end{figure}

\begin{figure}[H]
\centering
\subfloat[Relaxed phonon at low-T.]{
\label{fig:ZrNi6Sn6_relaxed_Low}
\includegraphics[width=0.45\textwidth]{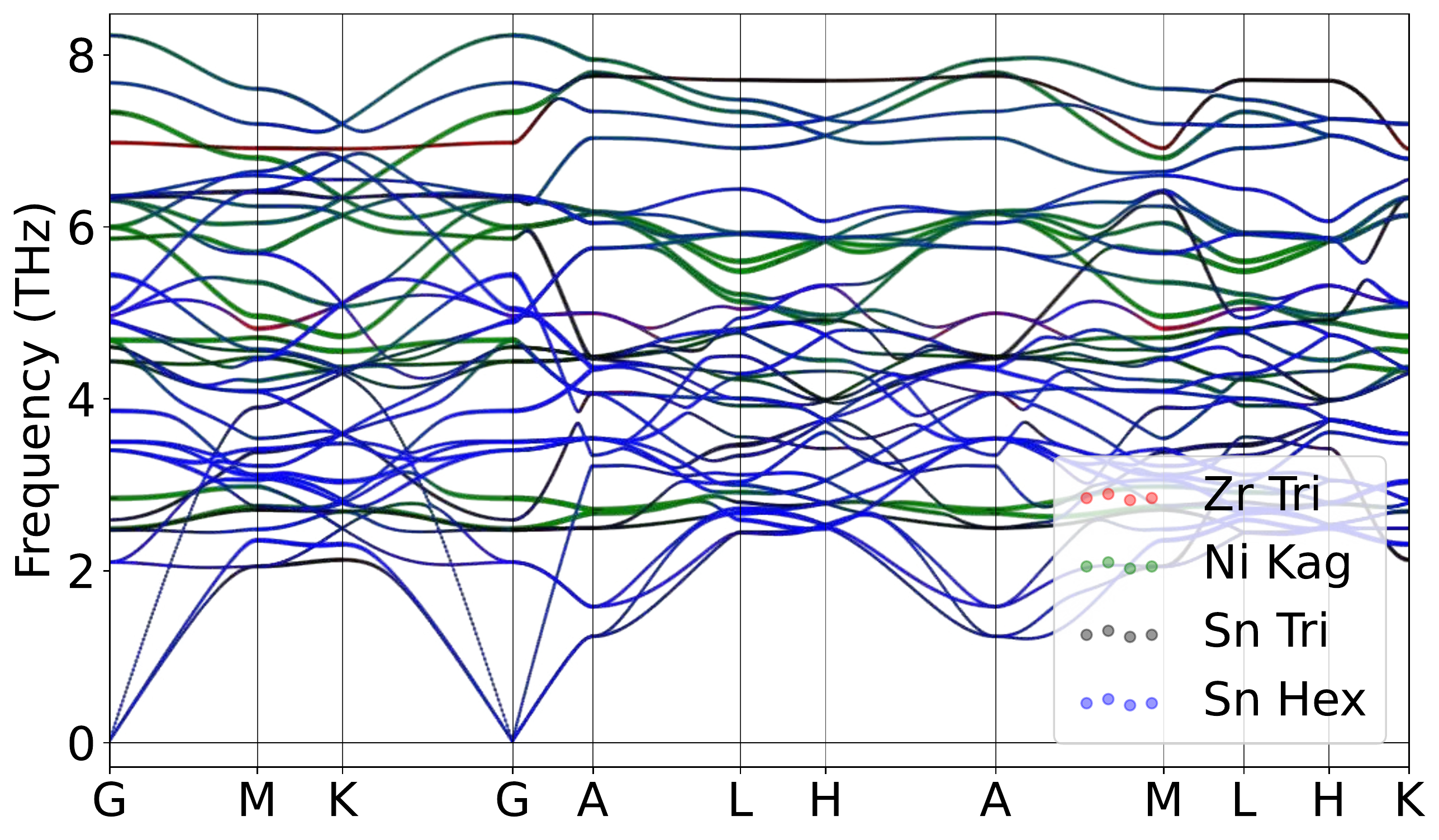}
}
\hspace{5pt}\subfloat[Relaxed phonon at high-T.]{
\label{fig:ZrNi6Sn6_relaxed_High}
\includegraphics[width=0.45\textwidth]{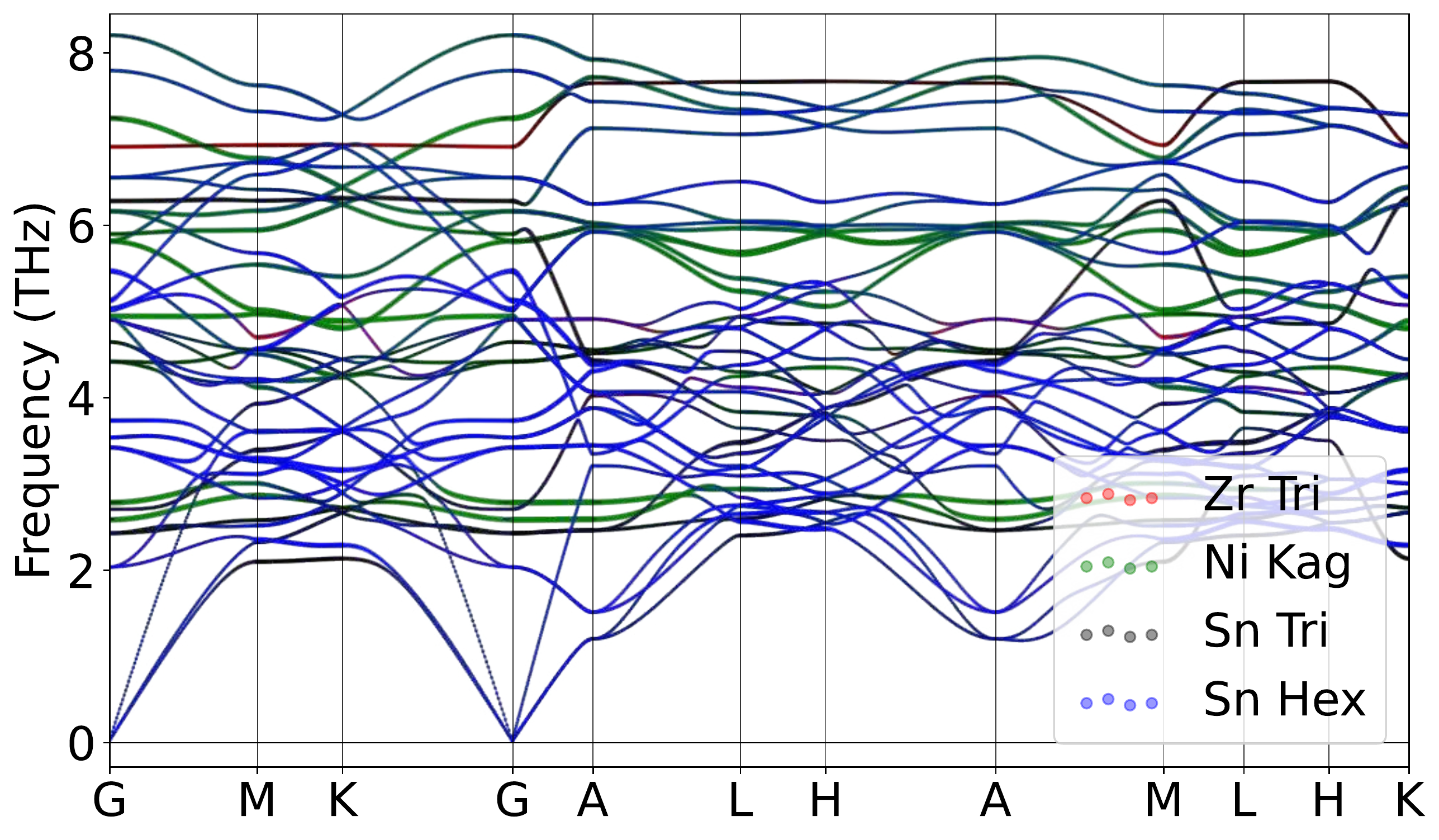}
}
\caption{Atomically projected phonon spectrum for ZrNi$_6$Sn$_6$.}
\label{fig:ZrNi6Sn6pho}
\end{figure}

\begin{figure}[H]
\centering
\label{fig:ZrNi6Sn6_relaxed_Low_xz}
\includegraphics[width=0.85\textwidth]{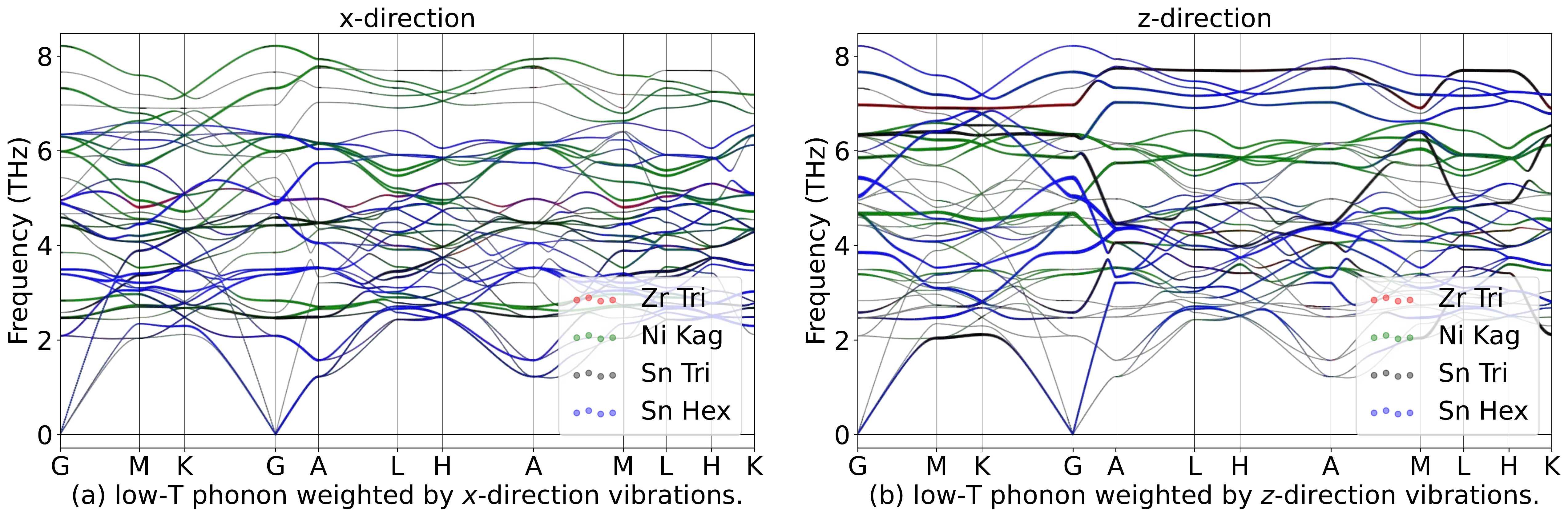}
\hspace{5pt}\label{fig:ZrNi6Sn6_relaxed_High_xz}
\includegraphics[width=0.85\textwidth]{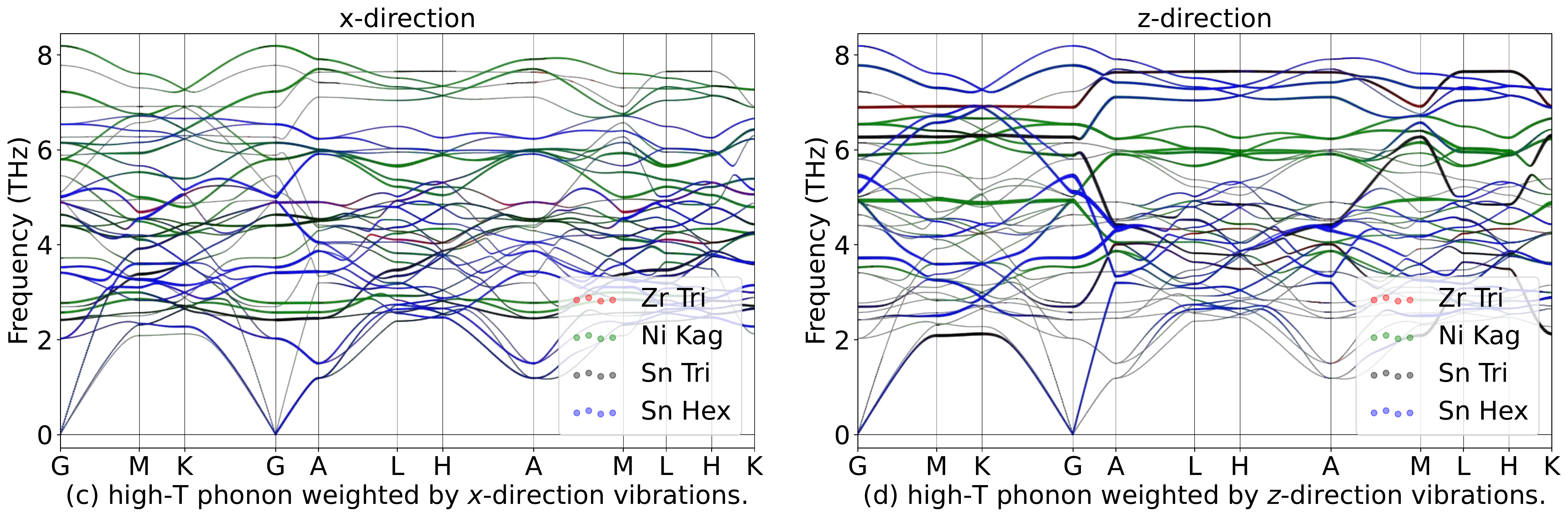}
\caption{Phonon spectrum for ZrNi$_6$Sn$_6$in in-plane ($x$) and out-of-plane ($z$) directions.}
\label{fig:ZrNi6Sn6phoxyz}
\end{figure}

\begin{table}[htbp]
\global\long\def\arraystretch{1.12}
\captionsetup{width=.95\textwidth}
\caption{IRREPs at high-symmetry points for ZrNi$_6$Sn$_6$ phonon spectrum at Low-T.}
\label{tab:191_ZrNi6Sn6_Low}
\setlength{\tabcolsep}{1.mm}{
}
\end{table}

\begin{table}[htbp]
\global\long\def\arraystretch{1.12}
\captionsetup{width=.95\textwidth}
\caption{IRREPs at high-symmetry points for ZrNi$_6$Sn$_6$ phonon spectrum at High-T.}
\label{tab:191_ZrNi6Sn6_High}
\setlength{\tabcolsep}{1.mm}{
%
}
\end{table}
\subsubsection{\label{sms2sec:NbNi6Sn6}\hyperref[smsec:col]{\texorpdfstring{NbNi$_6$Sn$_6$}{}}~{\color{green}  (High-T stable, Low-T stable) }}

\begin{table}[htbp]
\global\long\def\arraystretch{1.12}
\captionsetup{width=.85\textwidth}
\caption{Structure parameters of NbNi$_6$Sn$_6$ after relaxation. It contains 509 electrons. }
\label{tab:NbNi6Sn6}
\setlength{\tabcolsep}{4mm}{
%
}
\end{table}

\begin{figure}[htbp]
\centering
\includegraphics[width=0.85\textwidth]{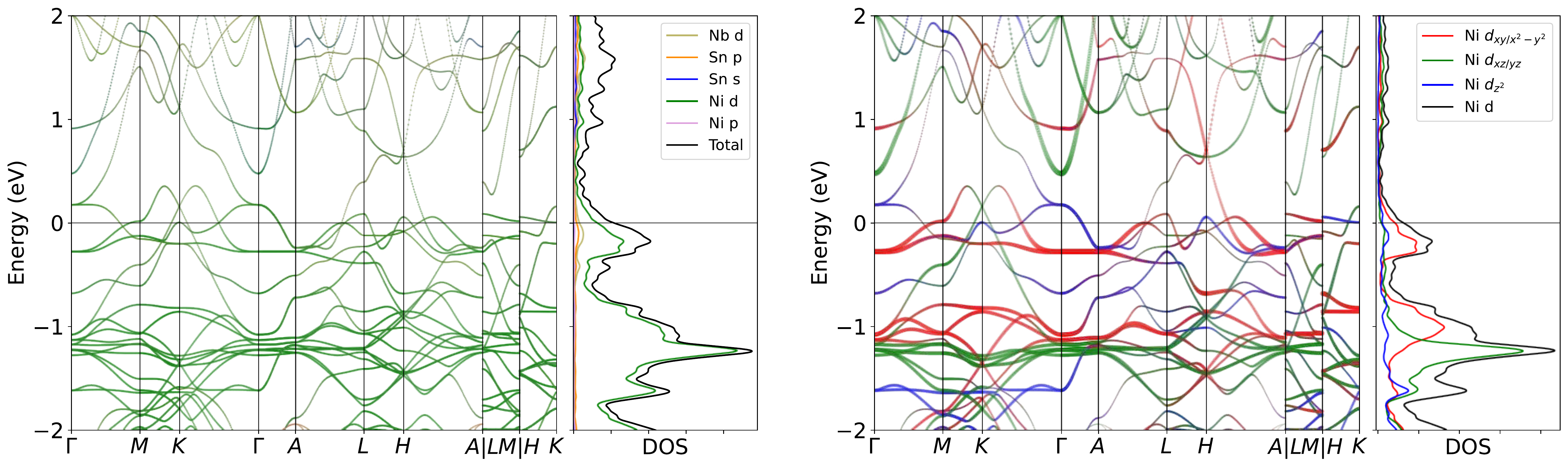}
\caption{Band structure for NbNi$_6$Sn$_6$ without SOC. The projection of Ni $3d$ orbitals is also presented. The band structures clearly reveal the dominating of Ni $3d$ orbitals  near the Fermi level, as well as the pronounced peak.}
\label{fig:NbNi6Sn6band}
\end{figure}

\begin{figure}[H]
\centering
\subfloat[Relaxed phonon at low-T.]{
\label{fig:NbNi6Sn6_relaxed_Low}
\includegraphics[width=0.45\textwidth]{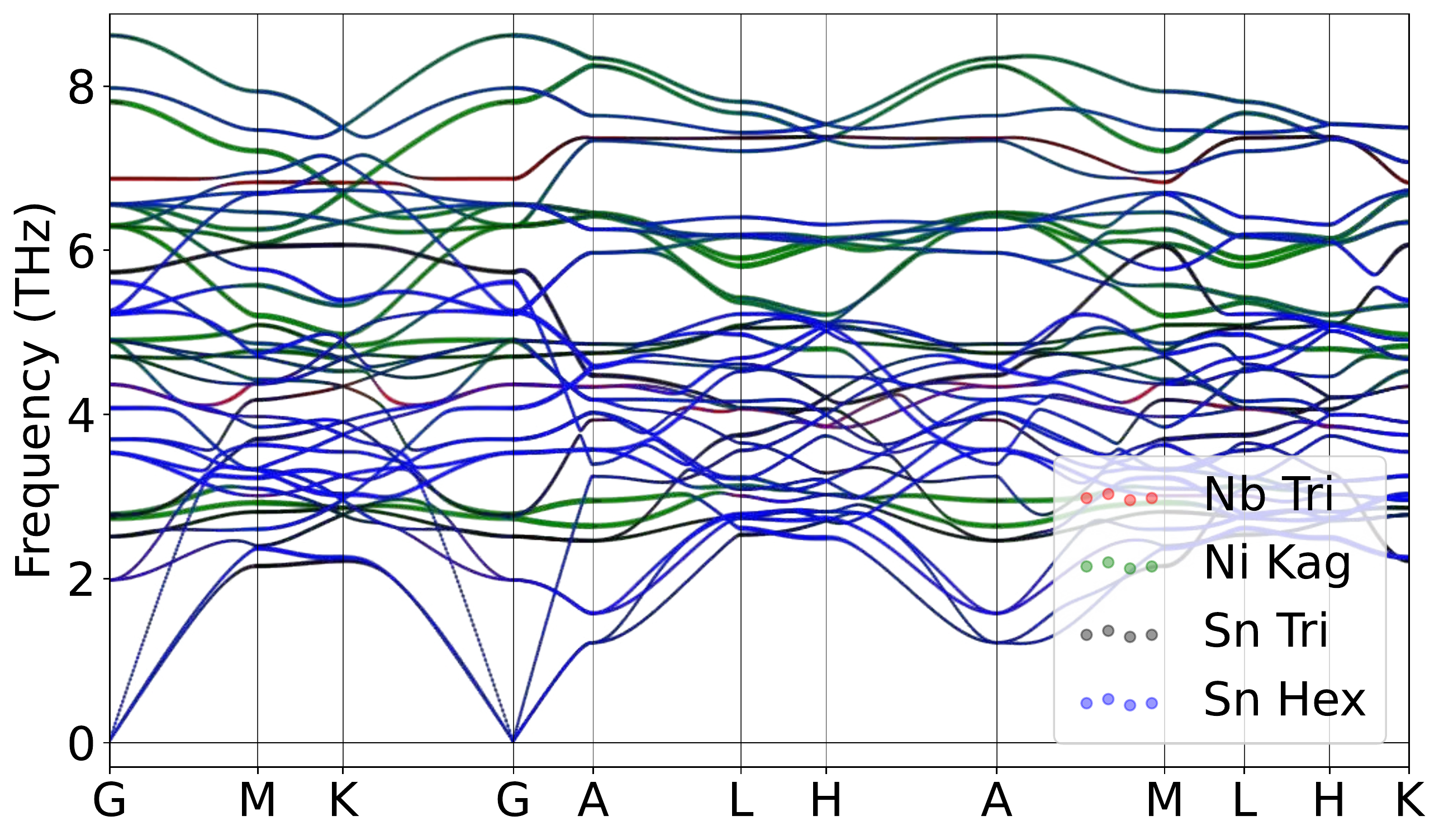}
}
\hspace{5pt}\subfloat[Relaxed phonon at high-T.]{
\label{fig:NbNi6Sn6_relaxed_High}
\includegraphics[width=0.45\textwidth]{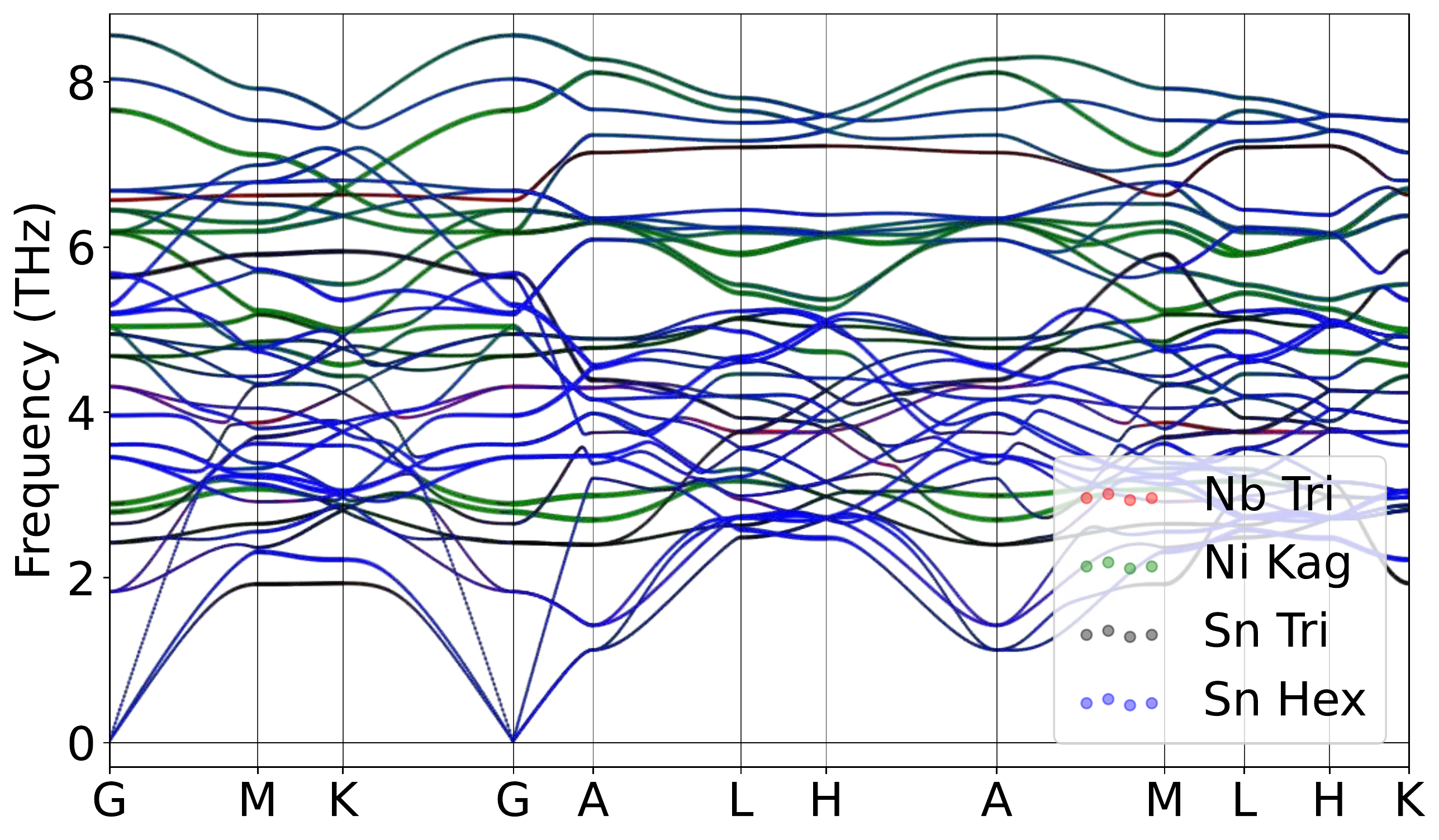}
}
\caption{Atomically projected phonon spectrum for NbNi$_6$Sn$_6$.}
\label{fig:NbNi6Sn6pho}
\end{figure}

\begin{figure}[H]
\centering
\label{fig:NbNi6Sn6_relaxed_Low_xz}
\includegraphics[width=0.85\textwidth]{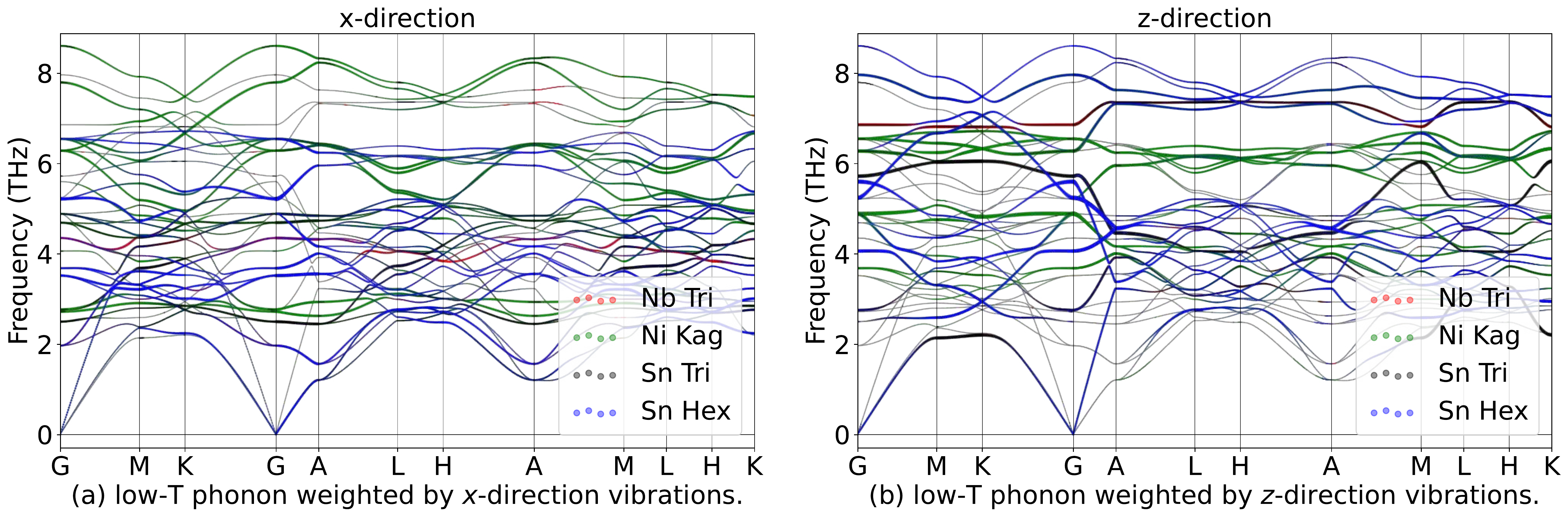}
\hspace{5pt}\label{fig:NbNi6Sn6_relaxed_High_xz}
\includegraphics[width=0.85\textwidth]{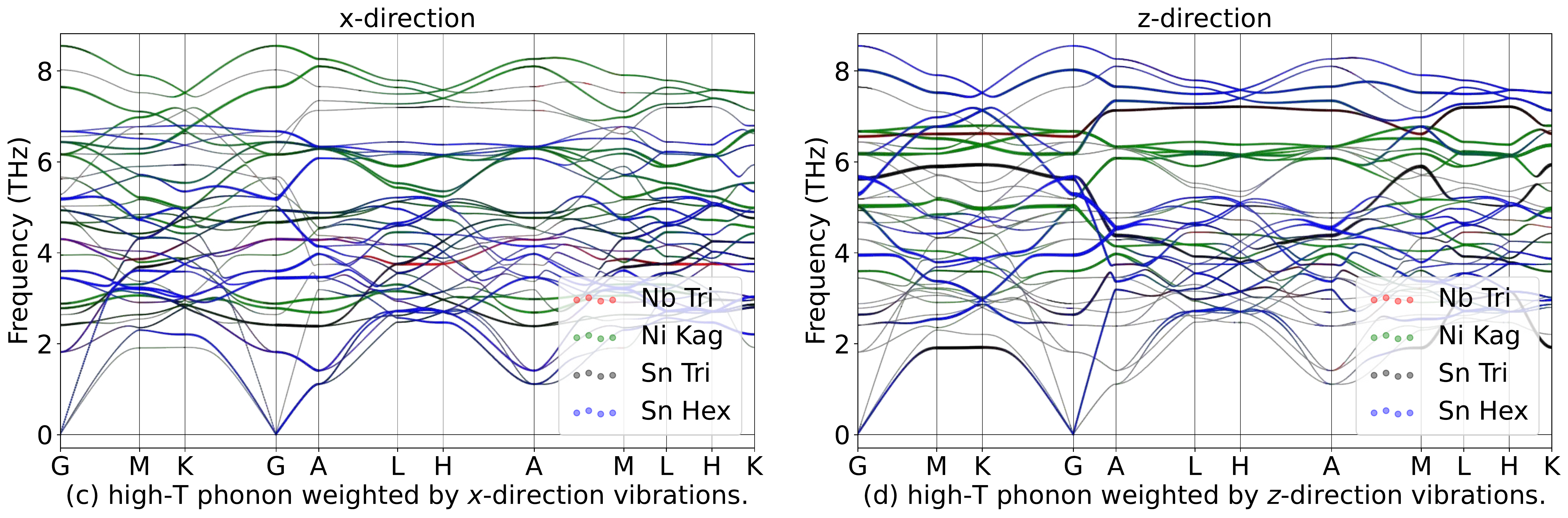}
\caption{Phonon spectrum for NbNi$_6$Sn$_6$in in-plane ($x$) and out-of-plane ($z$) directions.}
\label{fig:NbNi6Sn6phoxyz}
\end{figure}

\begin{table}[htbp]
\global\long\def\arraystretch{1.12}
\captionsetup{width=.95\textwidth}
\caption{IRREPs at high-symmetry points for NbNi$_6$Sn$_6$ phonon spectrum at Low-T.}
\label{tab:191_NbNi6Sn6_Low}
\setlength{\tabcolsep}{1.mm}{
}
\end{table}

\begin{table}[htbp]
\global\long\def\arraystretch{1.12}
\captionsetup{width=.95\textwidth}
\caption{IRREPs at high-symmetry points for NbNi$_6$Sn$_6$ phonon spectrum at High-T.}
\label{tab:191_NbNi6Sn6_High}
\setlength{\tabcolsep}{1.mm}{
%
}
\end{table}
\subsubsection{\label{sms2sec:PrNi6Sn6}\hyperref[smsec:col]{\texorpdfstring{PrNi$_6$Sn$_6$}{}}~{\color{green}  (High-T stable, Low-T stable) }}

\begin{table}[htbp]
\global\long\def\arraystretch{1.12}
\captionsetup{width=.85\textwidth}
\caption{Structure parameters of PrNi$_6$Sn$_6$ after relaxation. It contains 527 electrons. }
\label{tab:PrNi6Sn6}
\setlength{\tabcolsep}{4mm}{
%
}
\end{table}

\begin{figure}[htbp]
\centering
\includegraphics[width=0.85\textwidth]{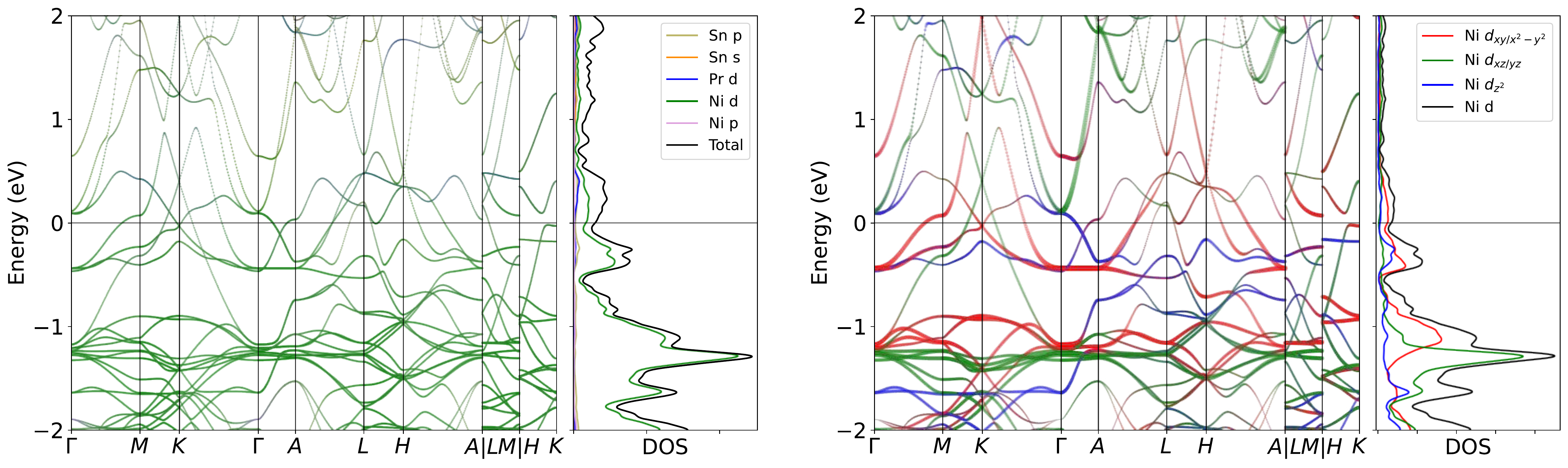}
\caption{Band structure for PrNi$_6$Sn$_6$ without SOC. The projection of Ni $3d$ orbitals is also presented. The band structures clearly reveal the dominating of Ni $3d$ orbitals  near the Fermi level, as well as the pronounced peak.}
\label{fig:PrNi6Sn6band}
\end{figure}

\begin{figure}[H]
\centering
\subfloat[Relaxed phonon at low-T.]{
\label{fig:PrNi6Sn6_relaxed_Low}
\includegraphics[width=0.45\textwidth]{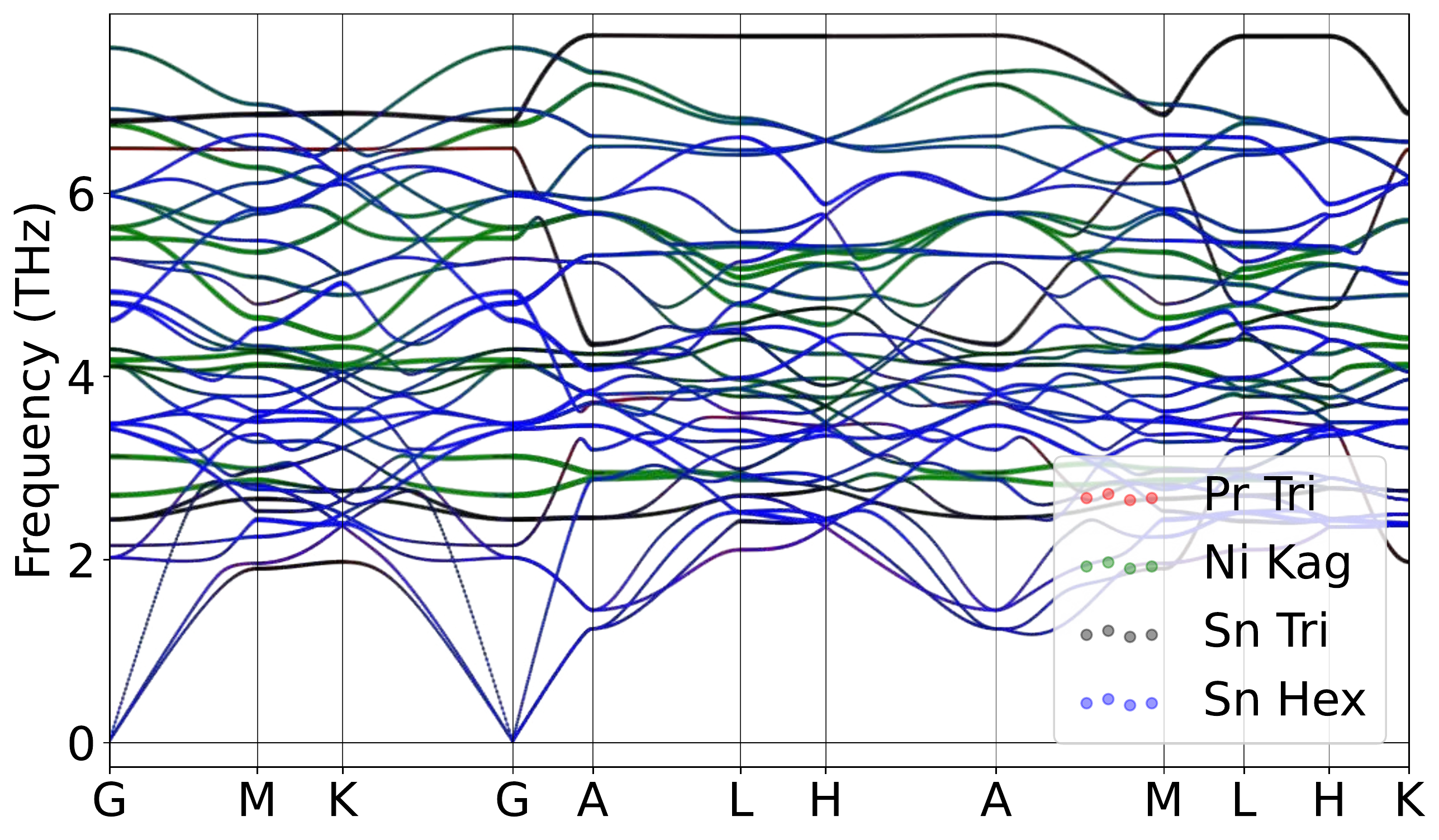}
}
\hspace{5pt}\subfloat[Relaxed phonon at high-T.]{
\label{fig:PrNi6Sn6_relaxed_High}
\includegraphics[width=0.45\textwidth]{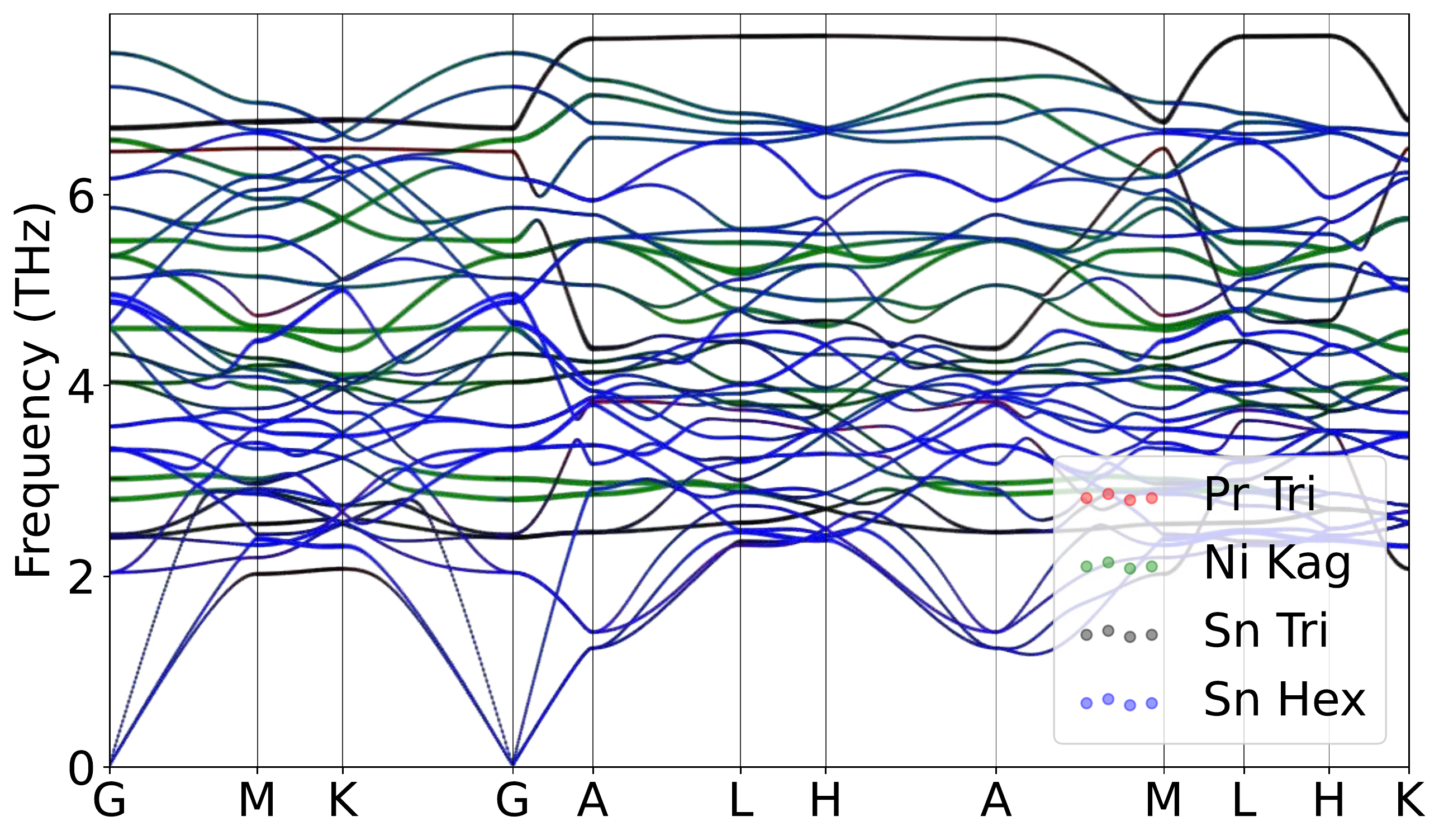}
}
\caption{Atomically projected phonon spectrum for PrNi$_6$Sn$_6$.}
\label{fig:PrNi6Sn6pho}
\end{figure}

\begin{figure}[H]
\centering
\label{fig:PrNi6Sn6_relaxed_Low_xz}
\includegraphics[width=0.85\textwidth]{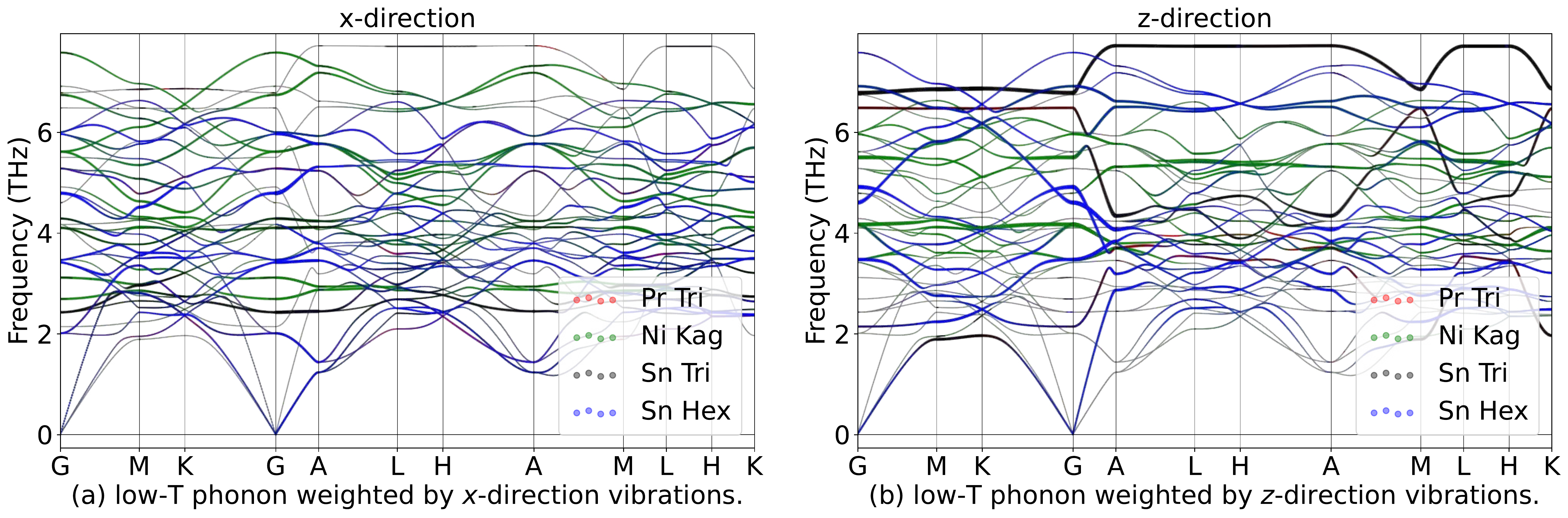}
\hspace{5pt}\label{fig:PrNi6Sn6_relaxed_High_xz}
\includegraphics[width=0.85\textwidth]{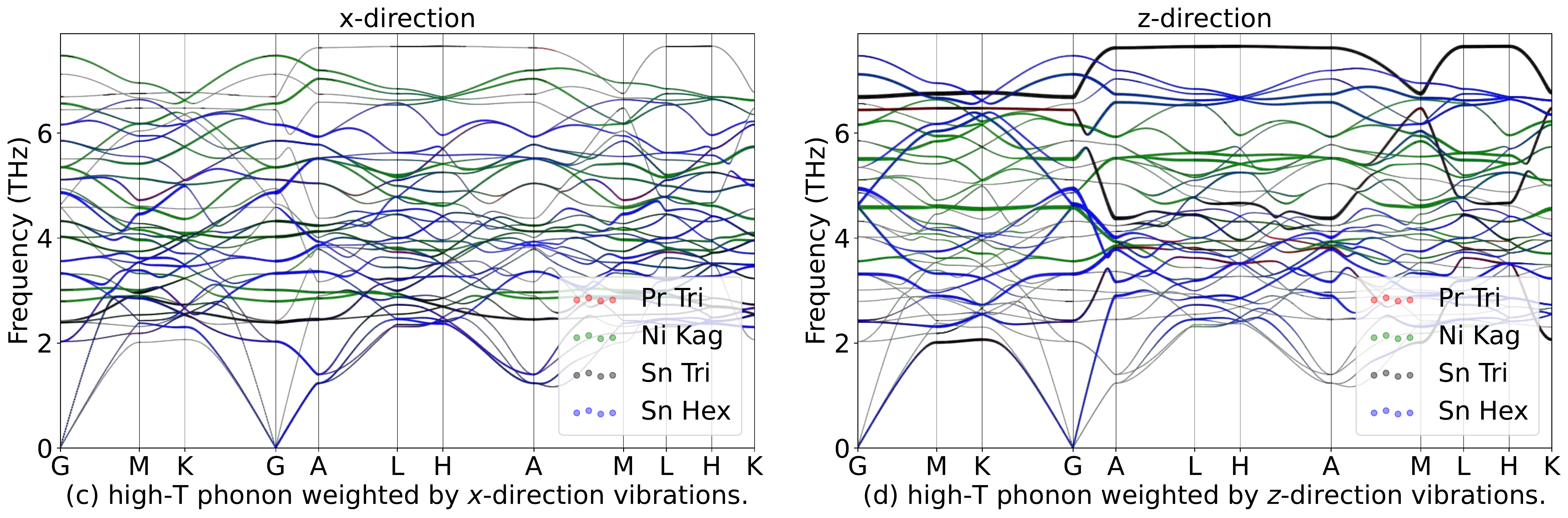}
\caption{Phonon spectrum for PrNi$_6$Sn$_6$in in-plane ($x$) and out-of-plane ($z$) directions.}
\label{fig:PrNi6Sn6phoxyz}
\end{figure}

\begin{table}[htbp]
\global\long\def\arraystretch{1.12}
\captionsetup{width=.95\textwidth}
\caption{IRREPs at high-symmetry points for PrNi$_6$Sn$_6$ phonon spectrum at Low-T.}
\label{tab:191_PrNi6Sn6_Low}
\setlength{\tabcolsep}{1.mm}{
}
\end{table}

\begin{table}[htbp]
\global\long\def\arraystretch{1.12}
\captionsetup{width=.95\textwidth}
\caption{IRREPs at high-symmetry points for PrNi$_6$Sn$_6$ phonon spectrum at High-T.}
\label{tab:191_PrNi6Sn6_High}
\setlength{\tabcolsep}{1.mm}{
%
}
\end{table}
\subsubsection{\label{sms2sec:NdNi6Sn6}\hyperref[smsec:col]{\texorpdfstring{NdNi$_6$Sn$_6$}{}}~{\color{green}  (High-T stable, Low-T stable) }}

\begin{table}[htbp]
\global\long\def\arraystretch{1.12}
\captionsetup{width=.85\textwidth}
\caption{Structure parameters of NdNi$_6$Sn$_6$ after relaxation. It contains 528 electrons. }
\label{tab:NdNi6Sn6}
\setlength{\tabcolsep}{4mm}{
%
}
\end{table}

\begin{figure}[htbp]
\centering
\includegraphics[width=0.85\textwidth]{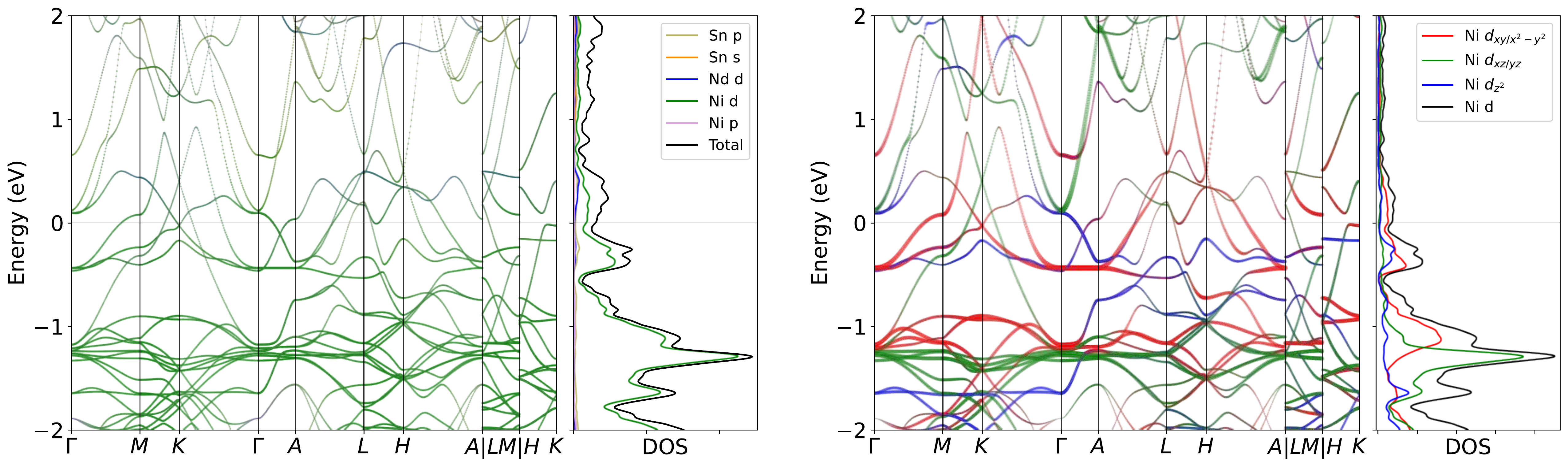}
\caption{Band structure for NdNi$_6$Sn$_6$ without SOC. The projection of Ni $3d$ orbitals is also presented. The band structures clearly reveal the dominating of Ni $3d$ orbitals  near the Fermi level, as well as the pronounced peak.}
\label{fig:NdNi6Sn6band}
\end{figure}

\begin{figure}[H]
\centering
\subfloat[Relaxed phonon at low-T.]{
\label{fig:NdNi6Sn6_relaxed_Low}
\includegraphics[width=0.45\textwidth]{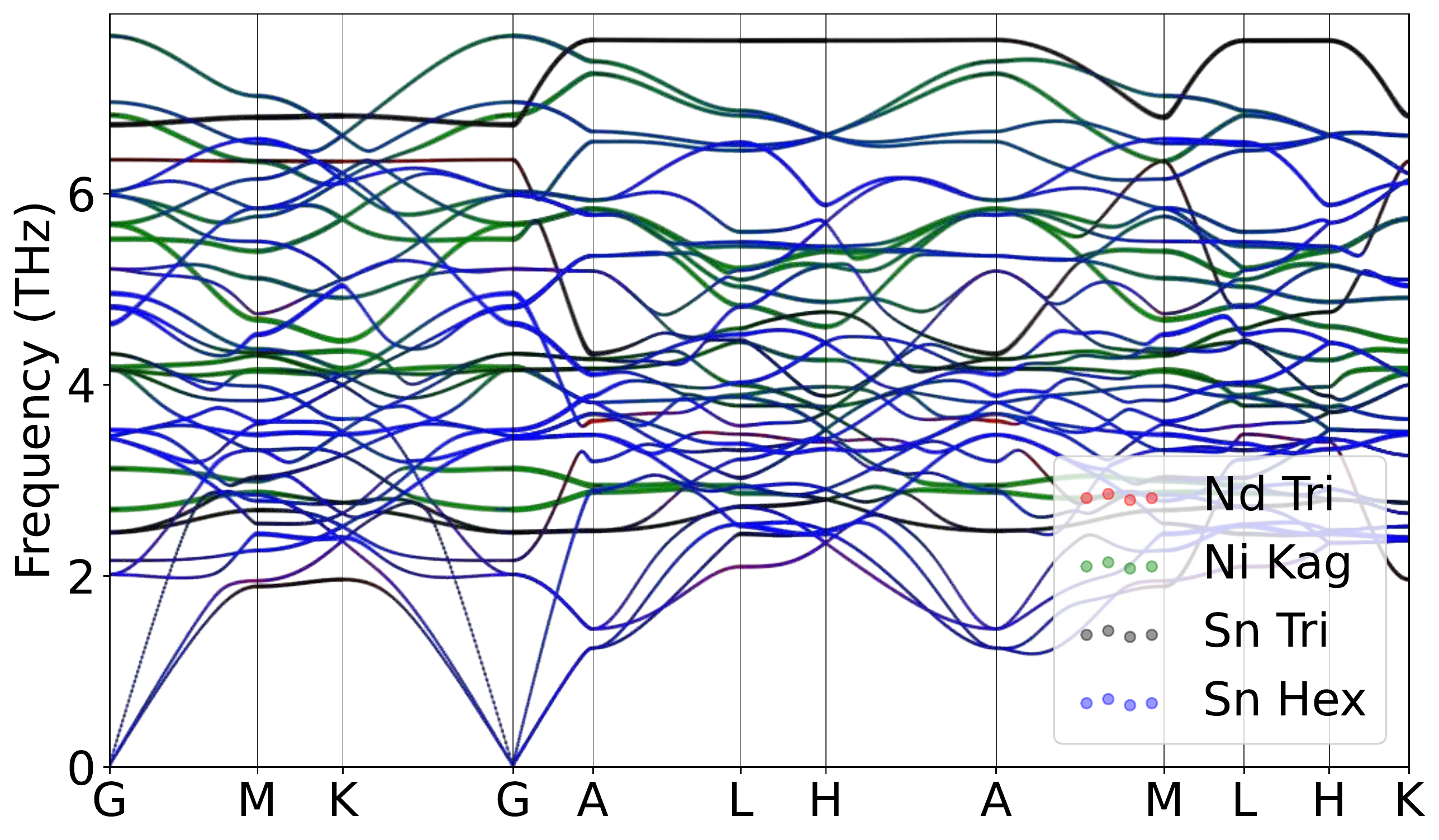}
}
\hspace{5pt}\subfloat[Relaxed phonon at high-T.]{
\label{fig:NdNi6Sn6_relaxed_High}
\includegraphics[width=0.45\textwidth]{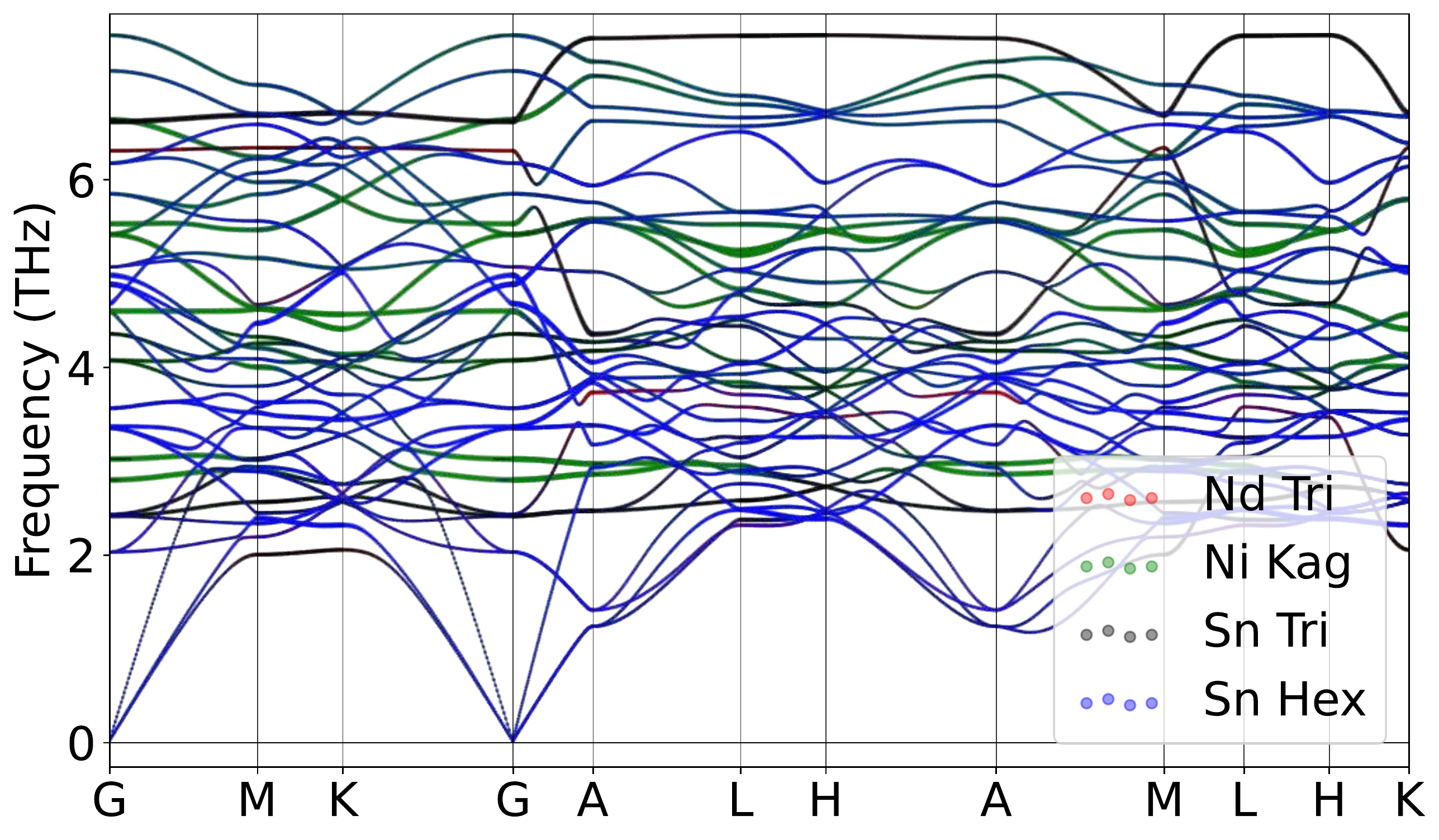}
}
\caption{Atomically projected phonon spectrum for NdNi$_6$Sn$_6$.}
\label{fig:NdNi6Sn6pho}
\end{figure}

\begin{figure}[H]
\centering
\label{fig:NdNi6Sn6_relaxed_Low_xz}
\includegraphics[width=0.85\textwidth]{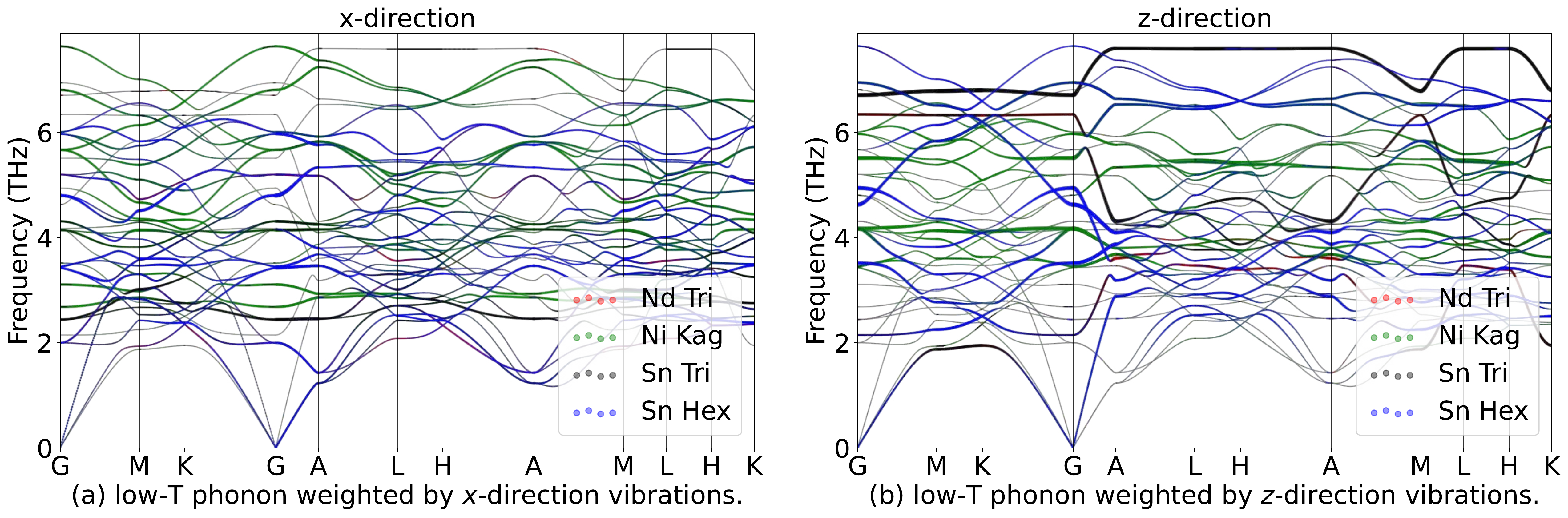}
\hspace{5pt}\label{fig:NdNi6Sn6_relaxed_High_xz}
\includegraphics[width=0.85\textwidth]{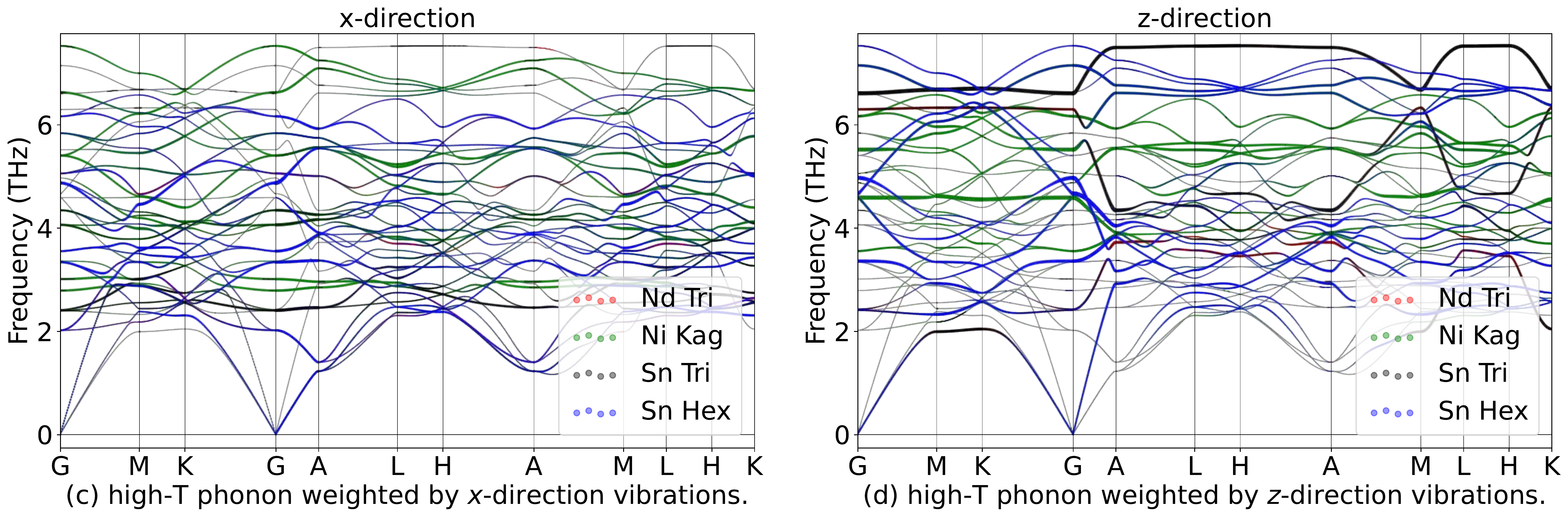}
\caption{Phonon spectrum for NdNi$_6$Sn$_6$in in-plane ($x$) and out-of-plane ($z$) directions.}
\label{fig:NdNi6Sn6phoxyz}
\end{figure}

\begin{table}[htbp]
\global\long\def\arraystretch{1.12}
\captionsetup{width=.95\textwidth}
\caption{IRREPs at high-symmetry points for NdNi$_6$Sn$_6$ phonon spectrum at Low-T.}
\label{tab:191_NdNi6Sn6_Low}
\setlength{\tabcolsep}{1.mm}{
}
\end{table}

\begin{table}[htbp]
\global\long\def\arraystretch{1.12}
\captionsetup{width=.95\textwidth}
\caption{IRREPs at high-symmetry points for NdNi$_6$Sn$_6$ phonon spectrum at High-T.}
\label{tab:191_NdNi6Sn6_High}
\setlength{\tabcolsep}{1.mm}{
%
}
\end{table}
\subsubsection{\label{sms2sec:SmNi6Sn6}\hyperref[smsec:col]{\texorpdfstring{SmNi$_6$Sn$_6$}{}}~{\color{green}  (High-T stable, Low-T stable) }}

\begin{table}[htbp]
\global\long\def\arraystretch{1.12}
\captionsetup{width=.85\textwidth}
\caption{Structure parameters of SmNi$_6$Sn$_6$ after relaxation. It contains 530 electrons. }
\label{tab:SmNi6Sn6}
\setlength{\tabcolsep}{4mm}{
%
}
\end{table}

\begin{figure}[htbp]
\centering
\includegraphics[width=0.85\textwidth]{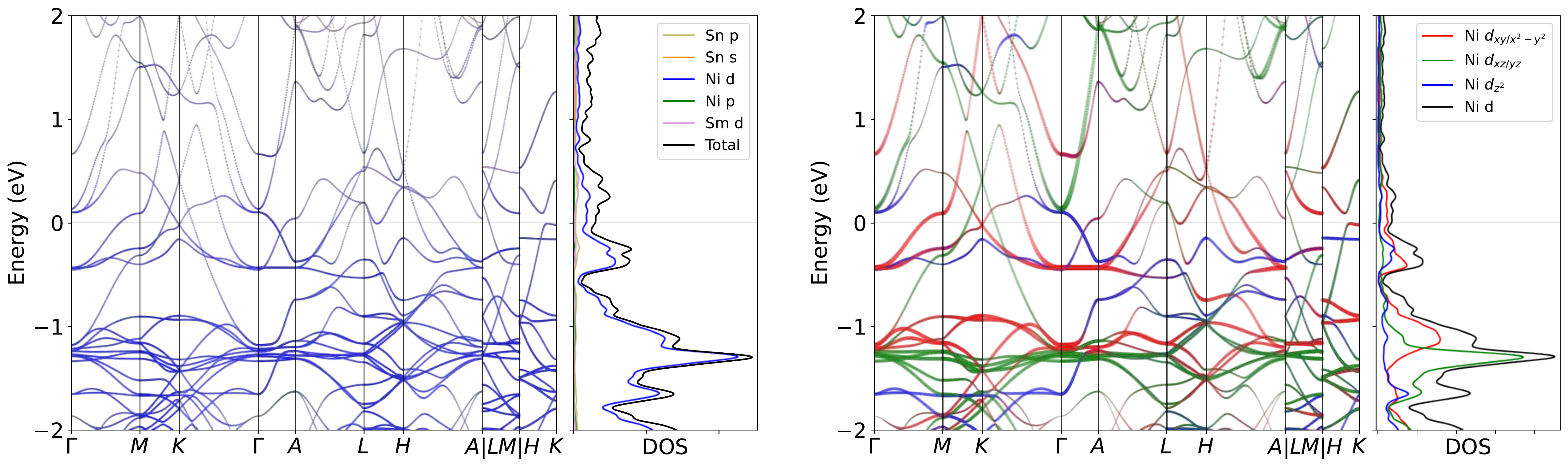}
\caption{Band structure for SmNi$_6$Sn$_6$ without SOC. The projection of Ni $3d$ orbitals is also presented. The band structures clearly reveal the dominating of Ni $3d$ orbitals  near the Fermi level, as well as the pronounced peak.}
\label{fig:SmNi6Sn6band}
\end{figure}

\begin{figure}[H]
\centering
\subfloat[Relaxed phonon at low-T.]{
\label{fig:SmNi6Sn6_relaxed_Low}
\includegraphics[width=0.45\textwidth]{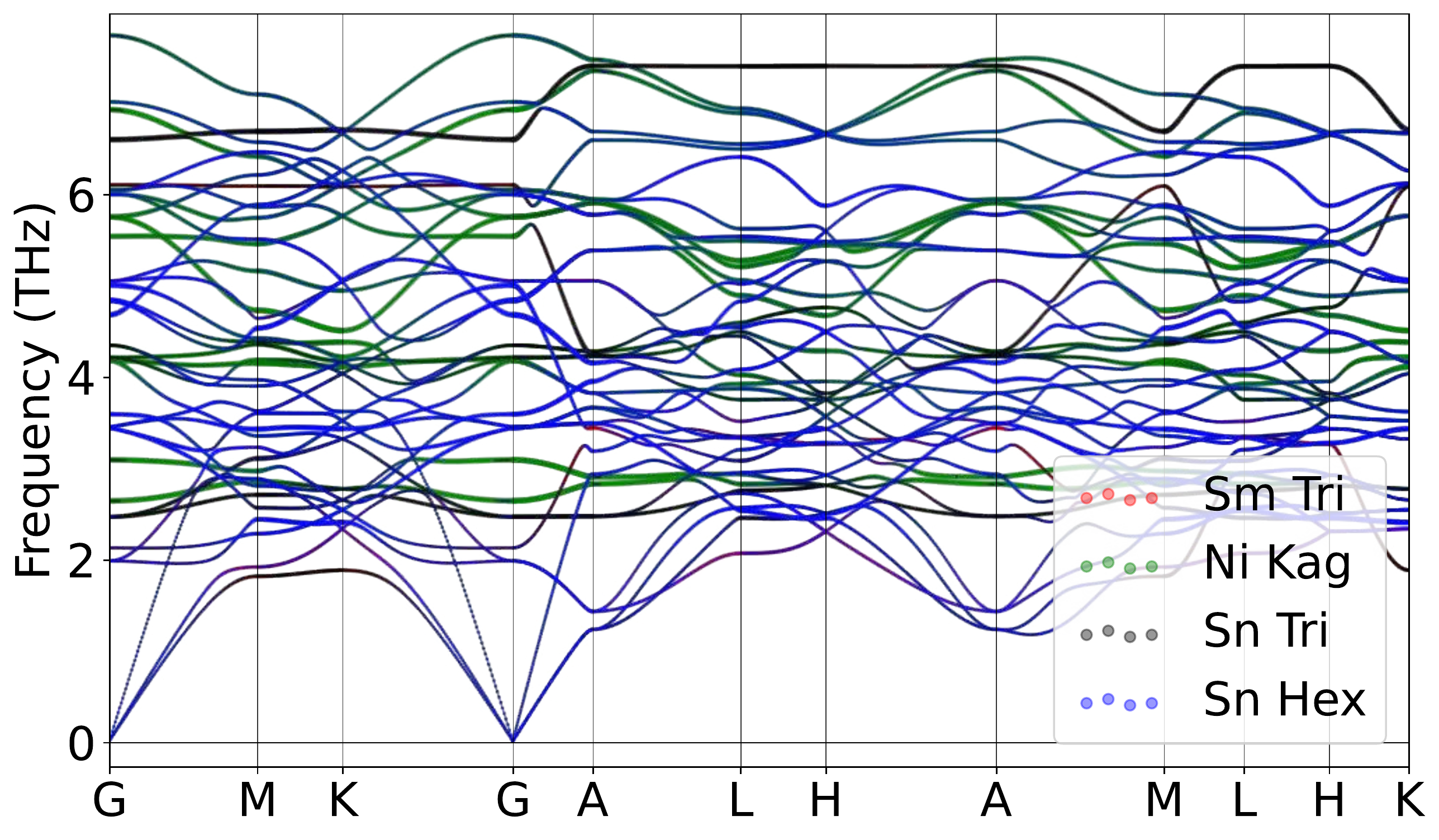}
}
\hspace{5pt}\subfloat[Relaxed phonon at high-T.]{
\label{fig:SmNi6Sn6_relaxed_High}
\includegraphics[width=0.45\textwidth]{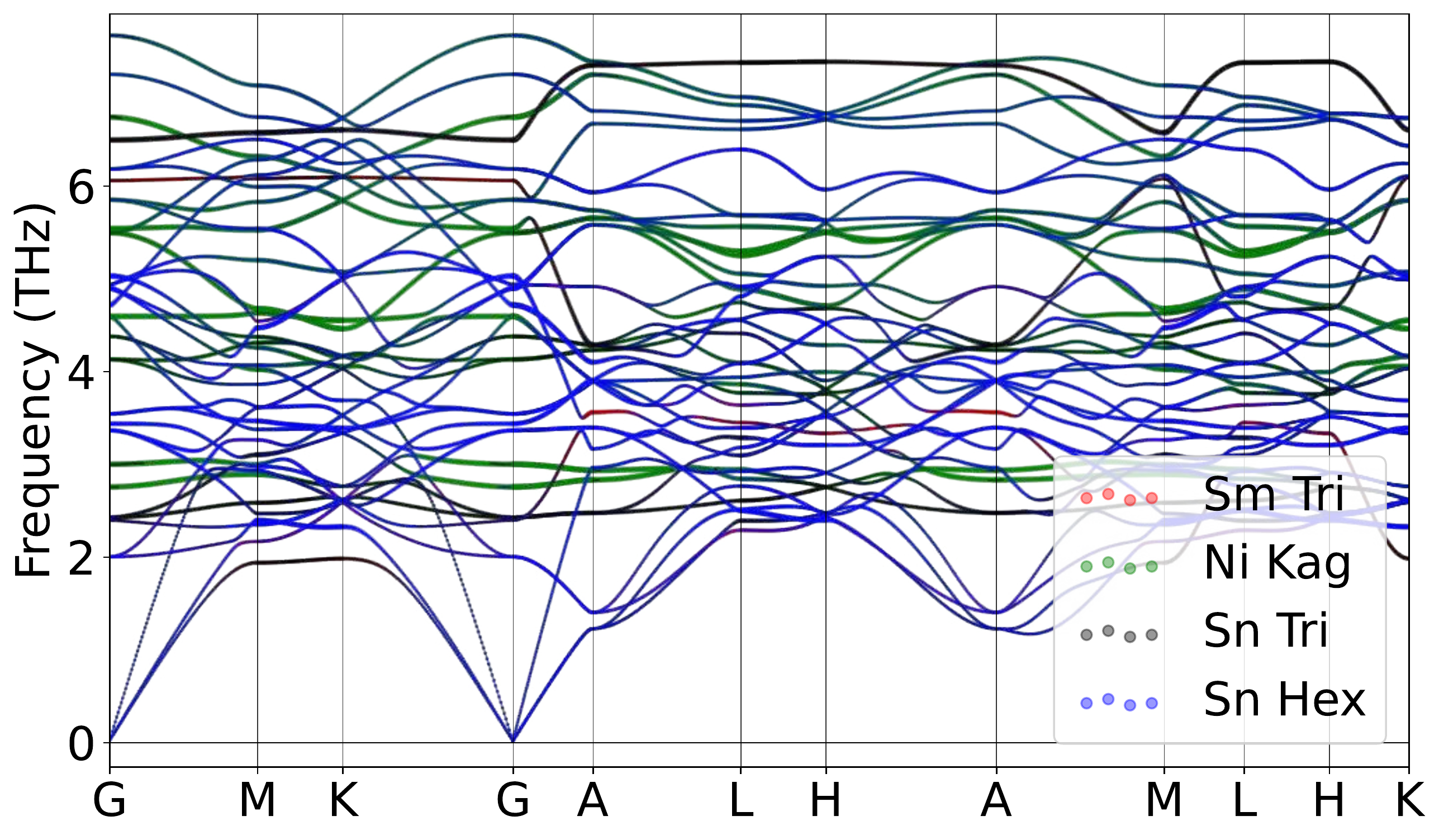}
}
\caption{Atomically projected phonon spectrum for SmNi$_6$Sn$_6$.}
\label{fig:SmNi6Sn6pho}
\end{figure}

\begin{figure}[H]
\centering
\label{fig:SmNi6Sn6_relaxed_Low_xz}
\includegraphics[width=0.85\textwidth]{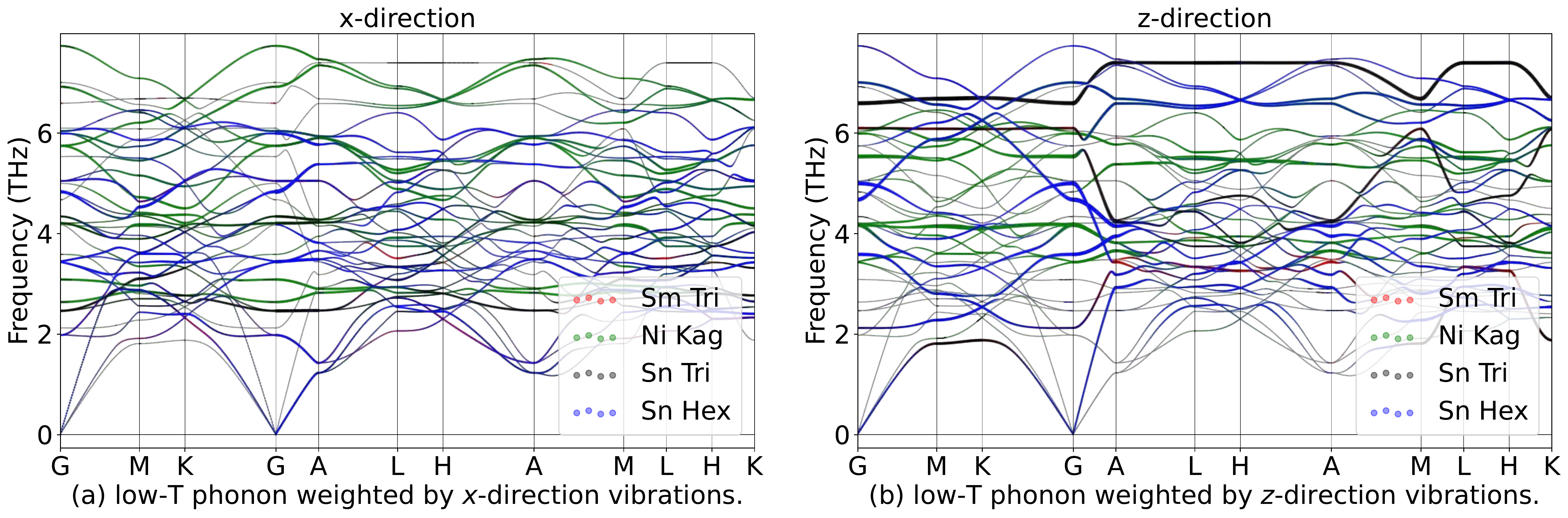}
\hspace{5pt}\label{fig:SmNi6Sn6_relaxed_High_xz}
\includegraphics[width=0.85\textwidth]{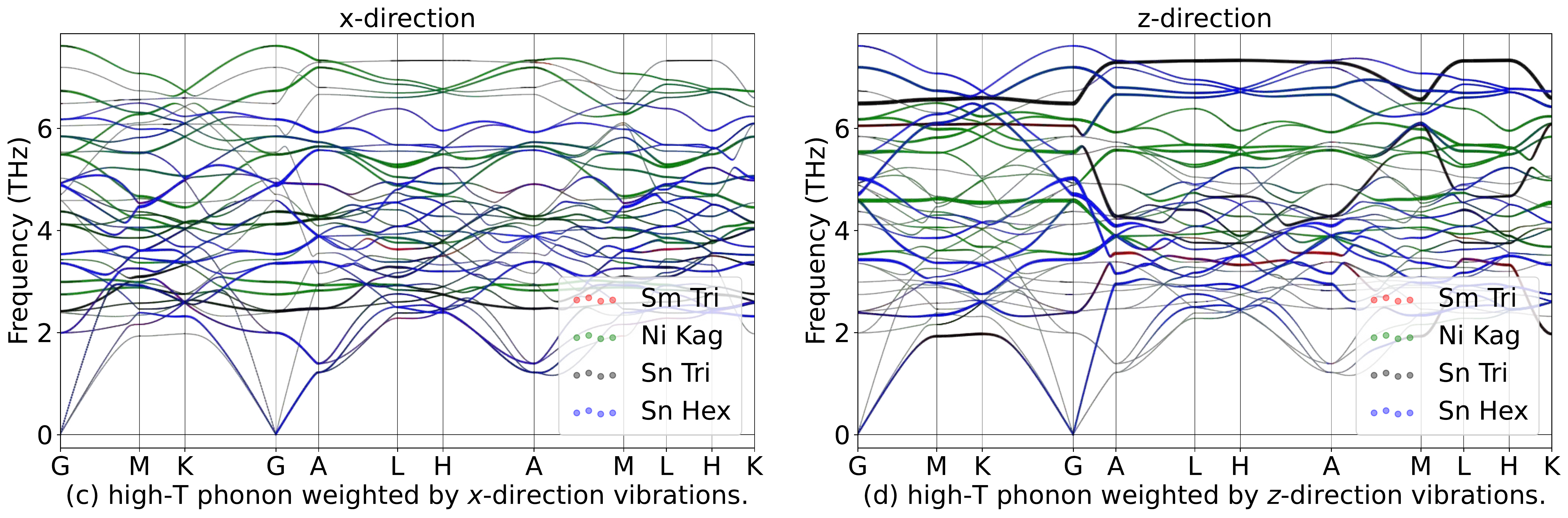}
\caption{Phonon spectrum for SmNi$_6$Sn$_6$in in-plane ($x$) and out-of-plane ($z$) directions.}
\label{fig:SmNi6Sn6phoxyz}
\end{figure}

\begin{table}[htbp]
\global\long\def\arraystretch{1.12}
\captionsetup{width=.95\textwidth}
\caption{IRREPs at high-symmetry points for SmNi$_6$Sn$_6$ phonon spectrum at Low-T.}
\label{tab:191_SmNi6Sn6_Low}
\setlength{\tabcolsep}{1.mm}{
}
\end{table}

\begin{table}[htbp]
\global\long\def\arraystretch{1.12}
\captionsetup{width=.95\textwidth}
\caption{IRREPs at high-symmetry points for SmNi$_6$Sn$_6$ phonon spectrum at High-T.}
\label{tab:191_SmNi6Sn6_High}
\setlength{\tabcolsep}{1.mm}{
%
}
\end{table}
\subsubsection{\label{sms2sec:GdNi6Sn6}\hyperref[smsec:col]{\texorpdfstring{GdNi$_6$Sn$_6$}{}}~{\color{green}  (High-T stable, Low-T stable) }}

\begin{table}[htbp]
\global\long\def\arraystretch{1.12}
\captionsetup{width=.85\textwidth}
\caption{Structure parameters of GdNi$_6$Sn$_6$ after relaxation. It contains 532 electrons. }
\label{tab:GdNi6Sn6}
\setlength{\tabcolsep}{4mm}{
%
}
\end{table}

\begin{figure}[htbp]
\centering
\includegraphics[width=0.85\textwidth]{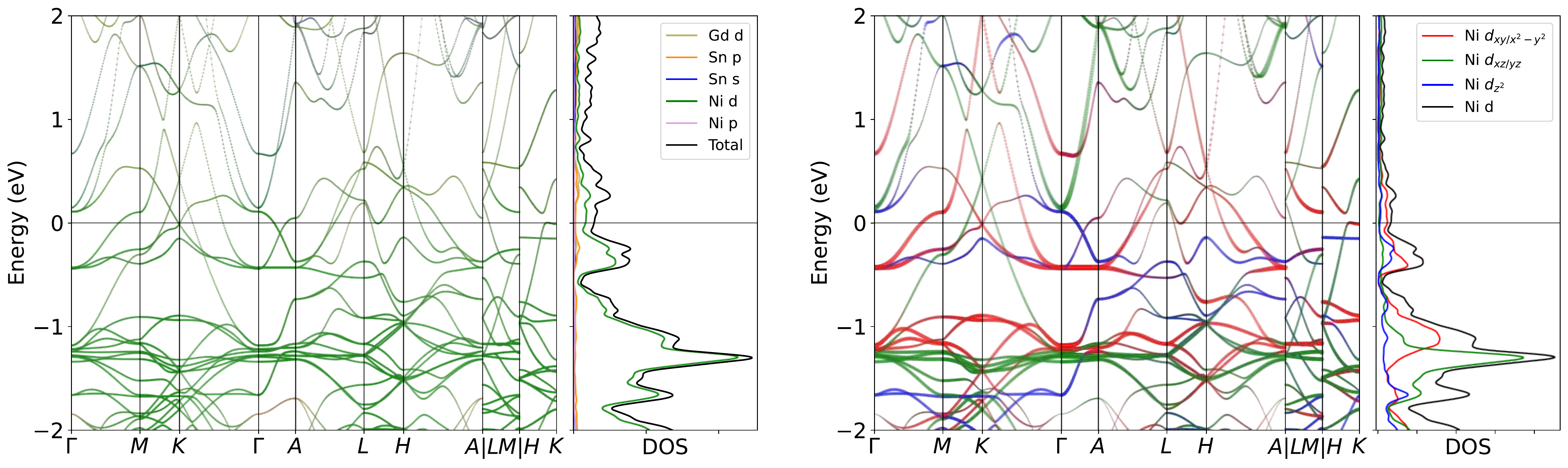}
\caption{Band structure for GdNi$_6$Sn$_6$ without SOC. The projection of Ni $3d$ orbitals is also presented. The band structures clearly reveal the dominating of Ni $3d$ orbitals  near the Fermi level, as well as the pronounced peak.}
\label{fig:GdNi6Sn6band}
\end{figure}

\begin{figure}[H]
\centering
\subfloat[Relaxed phonon at low-T.]{
\label{fig:GdNi6Sn6_relaxed_Low}
\includegraphics[width=0.45\textwidth]{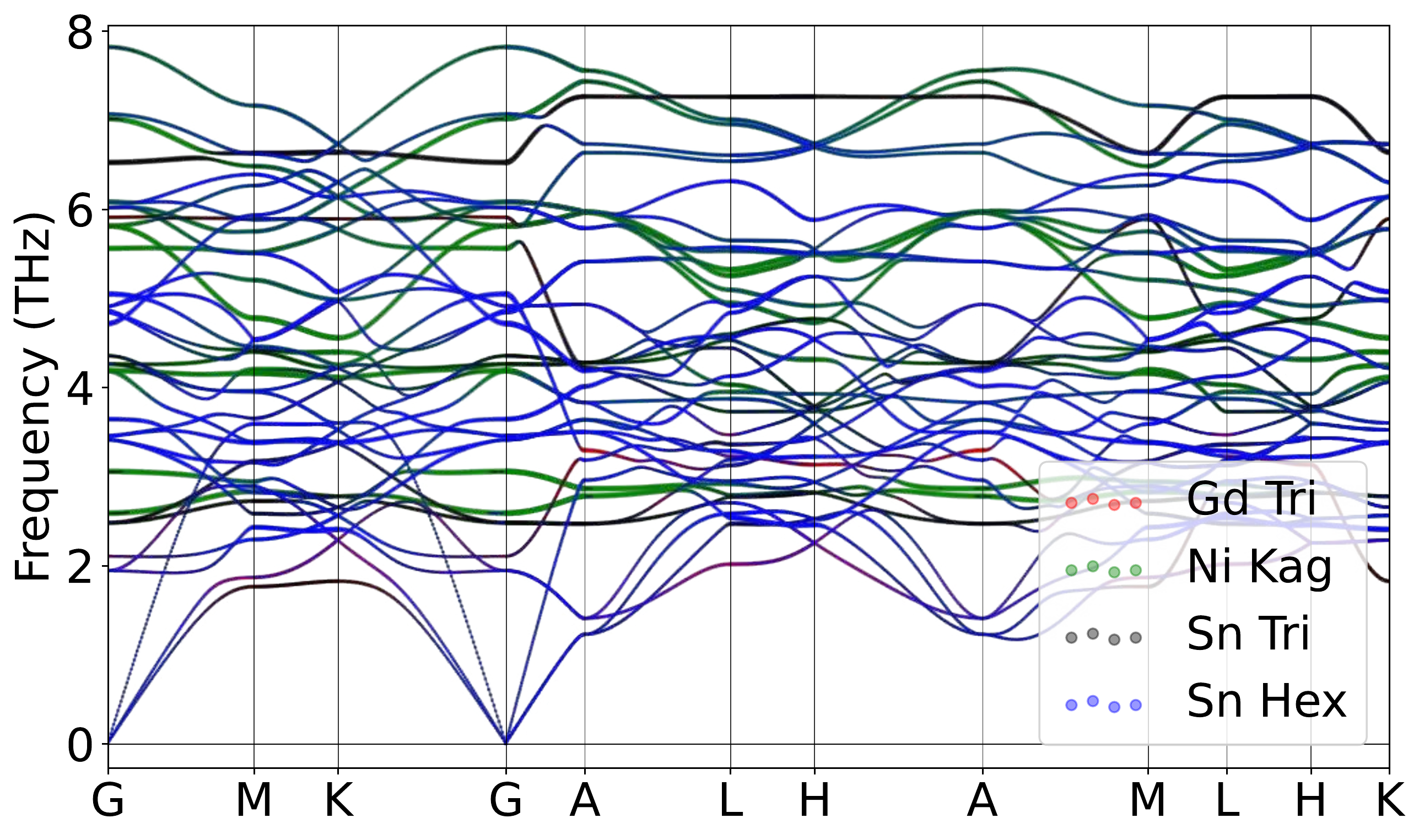}
}
\hspace{5pt}\subfloat[Relaxed phonon at high-T.]{
\label{fig:GdNi6Sn6_relaxed_High}
\includegraphics[width=0.45\textwidth]{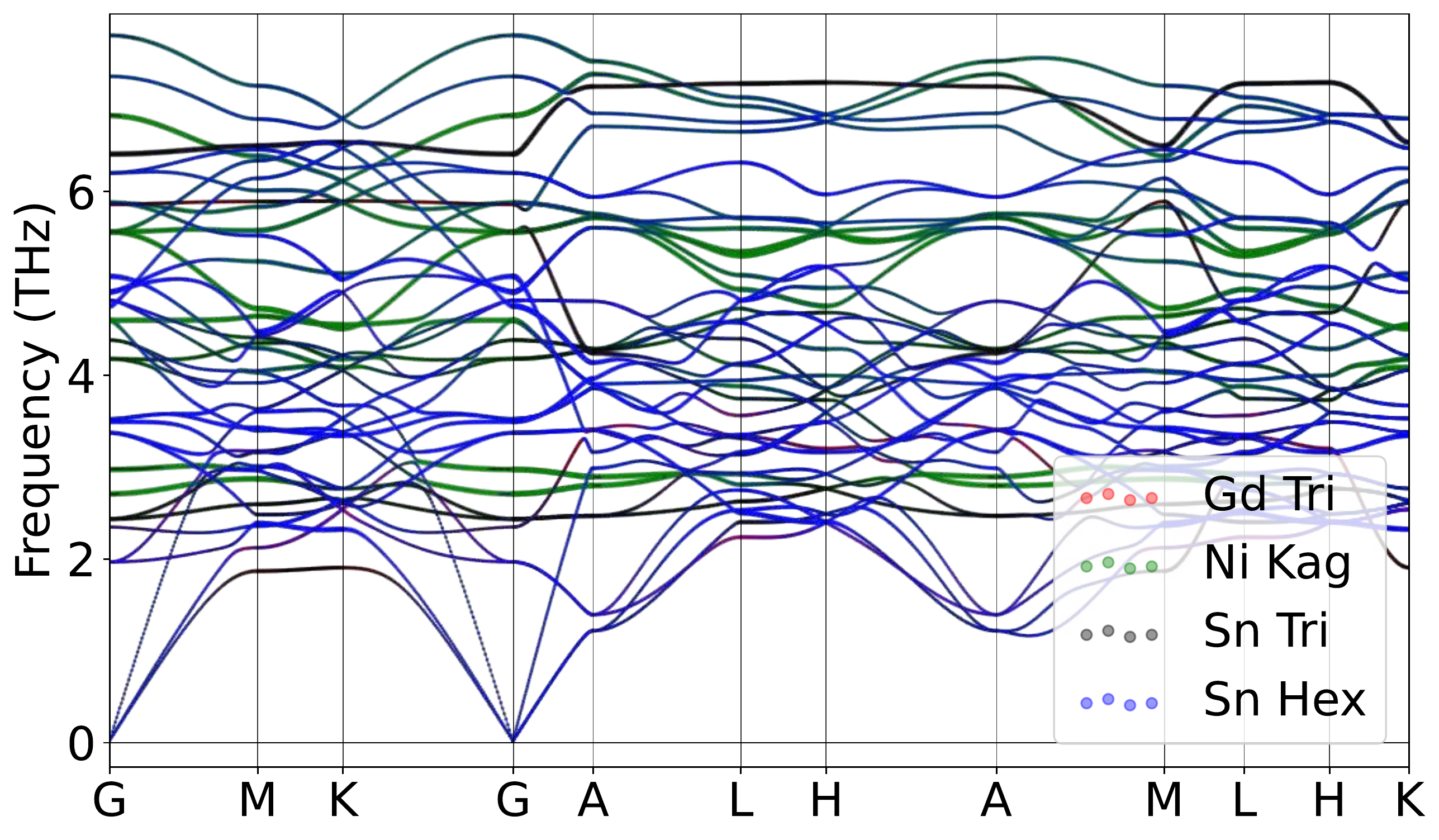}
}
\caption{Atomically projected phonon spectrum for GdNi$_6$Sn$_6$.}
\label{fig:GdNi6Sn6pho}
\end{figure}

\begin{figure}[H]
\centering
\label{fig:GdNi6Sn6_relaxed_Low_xz}
\includegraphics[width=0.85\textwidth]{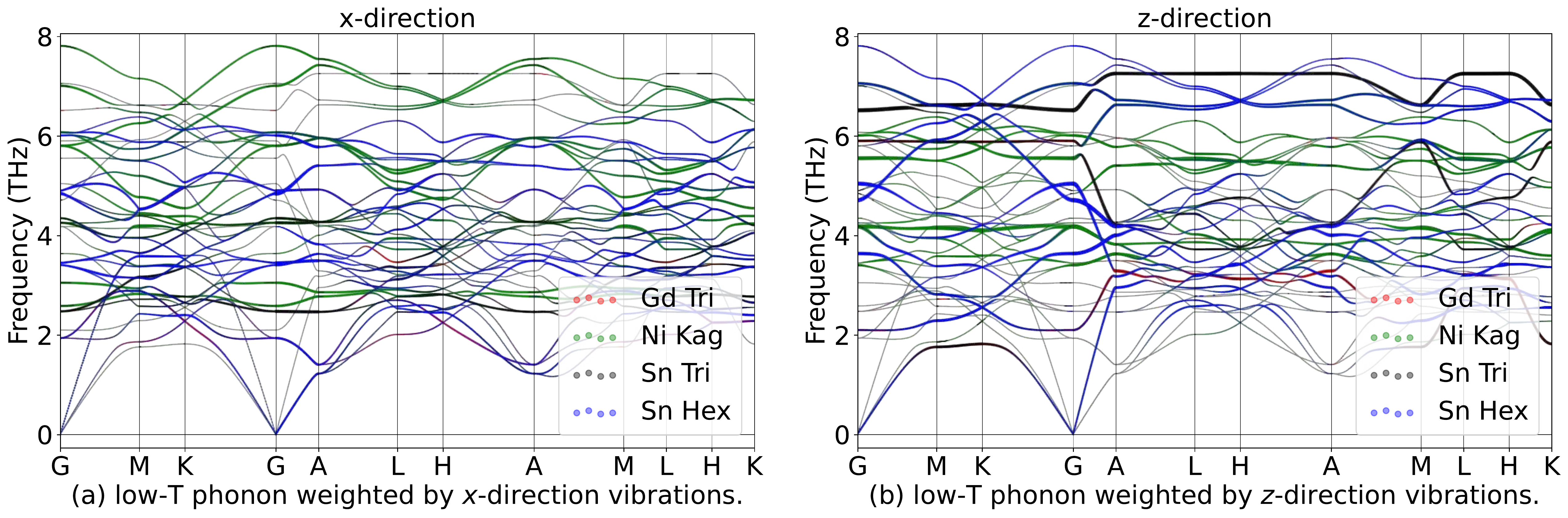}
\hspace{5pt}\label{fig:GdNi6Sn6_relaxed_High_xz}
\includegraphics[width=0.85\textwidth]{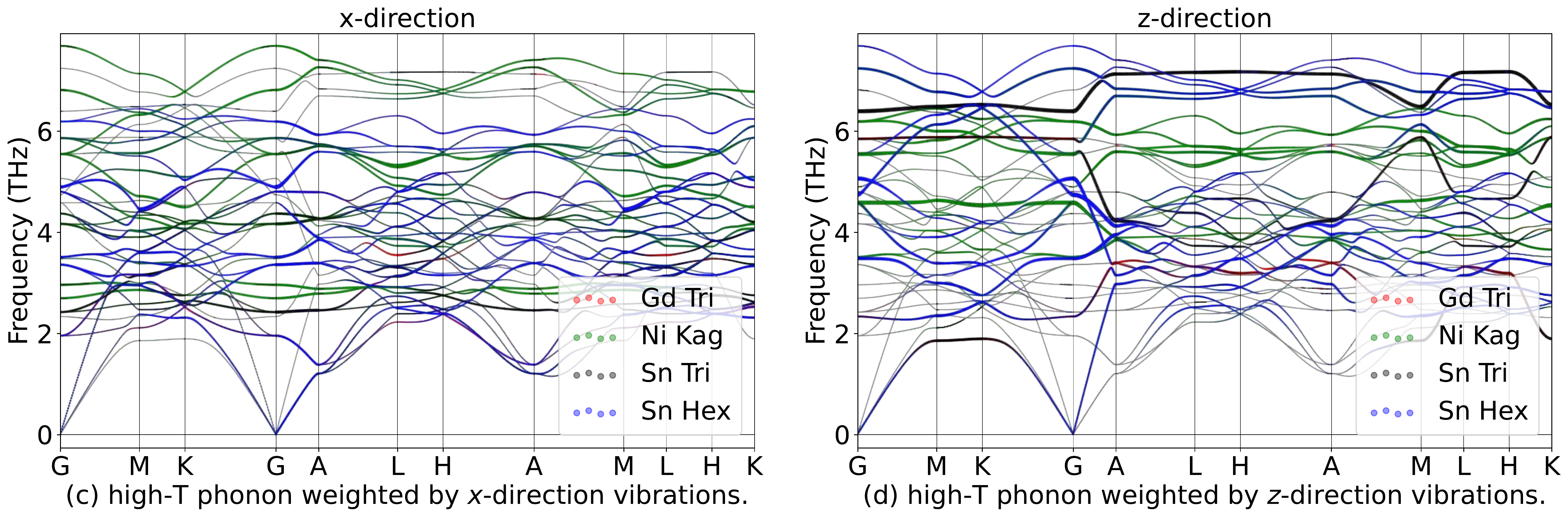}
\caption{Phonon spectrum for GdNi$_6$Sn$_6$in in-plane ($x$) and out-of-plane ($z$) directions.}
\label{fig:GdNi6Sn6phoxyz}
\end{figure}

\begin{table}[htbp]
\global\long\def\arraystretch{1.12}
\captionsetup{width=.95\textwidth}
\caption{IRREPs at high-symmetry points for GdNi$_6$Sn$_6$ phonon spectrum at Low-T.}
\label{tab:191_GdNi6Sn6_Low}
\setlength{\tabcolsep}{1.mm}{
}
\end{table}

\begin{table}[htbp]
\global\long\def\arraystretch{1.12}
\captionsetup{width=.95\textwidth}
\caption{IRREPs at high-symmetry points for GdNi$_6$Sn$_6$ phonon spectrum at High-T.}
\label{tab:191_GdNi6Sn6_High}
\setlength{\tabcolsep}{1.mm}{
%
}
\end{table}
\subsubsection{\label{sms2sec:TbNi6Sn6}\hyperref[smsec:col]{\texorpdfstring{TbNi$_6$Sn$_6$}{}}~{\color{green}  (High-T stable, Low-T stable) }}

\begin{table}[htbp]
\global\long\def\arraystretch{1.12}
\captionsetup{width=.85\textwidth}
\caption{Structure parameters of TbNi$_6$Sn$_6$ after relaxation. It contains 533 electrons. }
\label{tab:TbNi6Sn6}
\setlength{\tabcolsep}{4mm}{
%
}
\end{table}

\begin{figure}[htbp]
\centering
\includegraphics[width=0.85\textwidth]{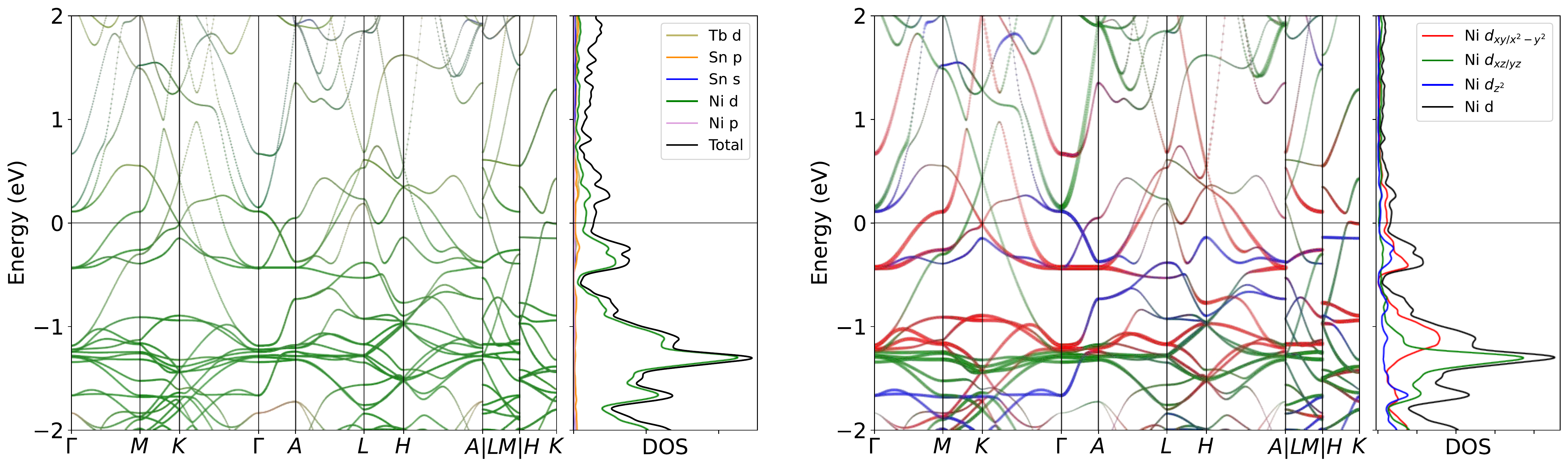}
\caption{Band structure for TbNi$_6$Sn$_6$ without SOC. The projection of Ni $3d$ orbitals is also presented. The band structures clearly reveal the dominating of Ni $3d$ orbitals  near the Fermi level, as well as the pronounced peak.}
\label{fig:TbNi6Sn6band}
\end{figure}

\begin{figure}[H]
\centering
\subfloat[Relaxed phonon at low-T.]{
\label{fig:TbNi6Sn6_relaxed_Low}
\includegraphics[width=0.45\textwidth]{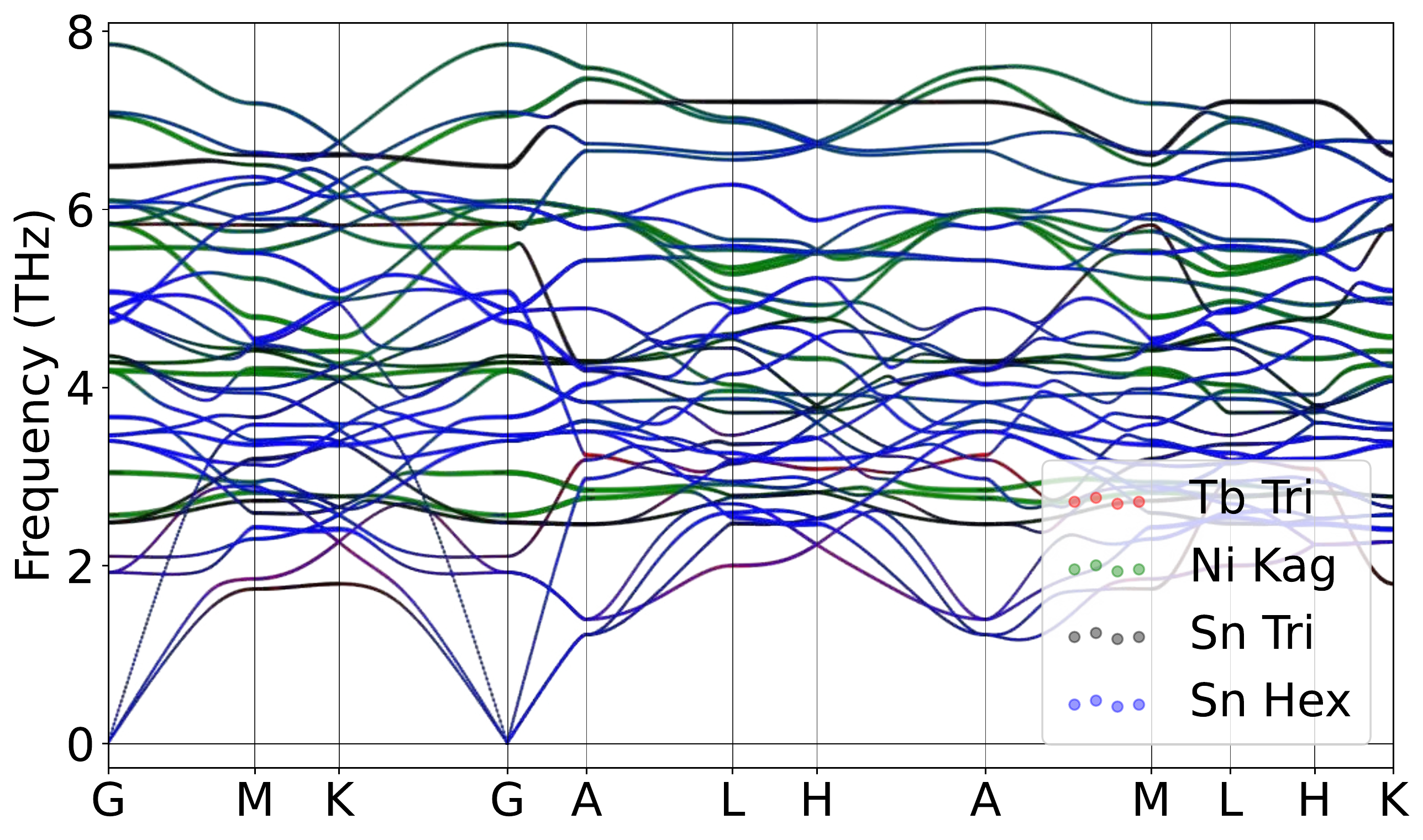}
}
\hspace{5pt}\subfloat[Relaxed phonon at high-T.]{
\label{fig:TbNi6Sn6_relaxed_High}
\includegraphics[width=0.45\textwidth]{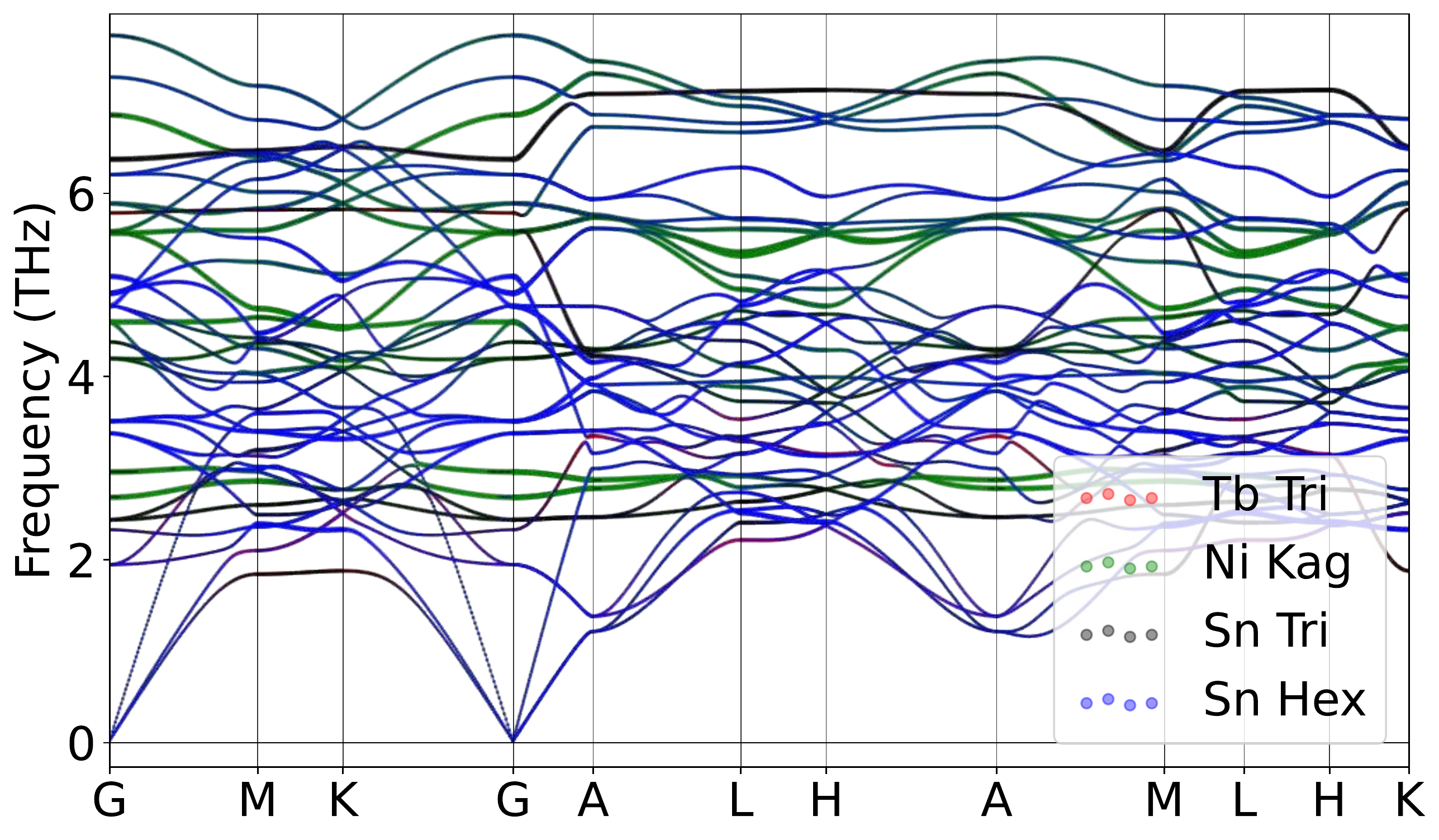}
}
\caption{Atomically projected phonon spectrum for TbNi$_6$Sn$_6$.}
\label{fig:TbNi6Sn6pho}
\end{figure}

\begin{figure}[H]
\centering
\label{fig:TbNi6Sn6_relaxed_Low_xz}
\includegraphics[width=0.85\textwidth]{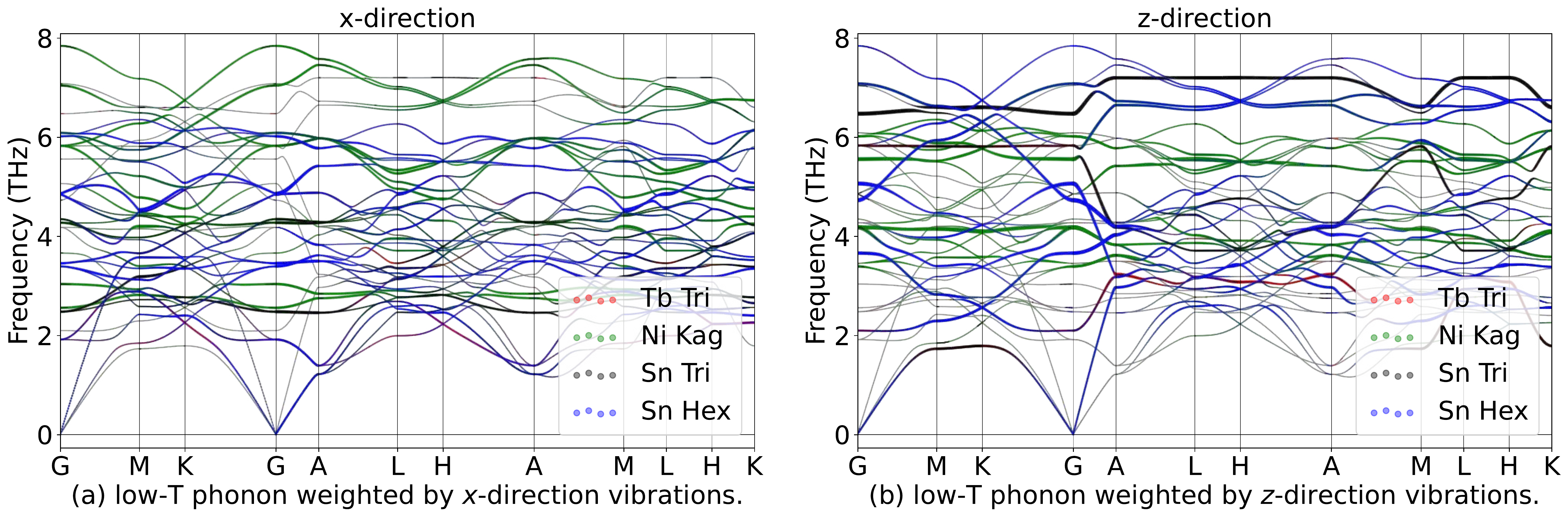}
\hspace{5pt}\label{fig:TbNi6Sn6_relaxed_High_xz}
\includegraphics[width=0.85\textwidth]{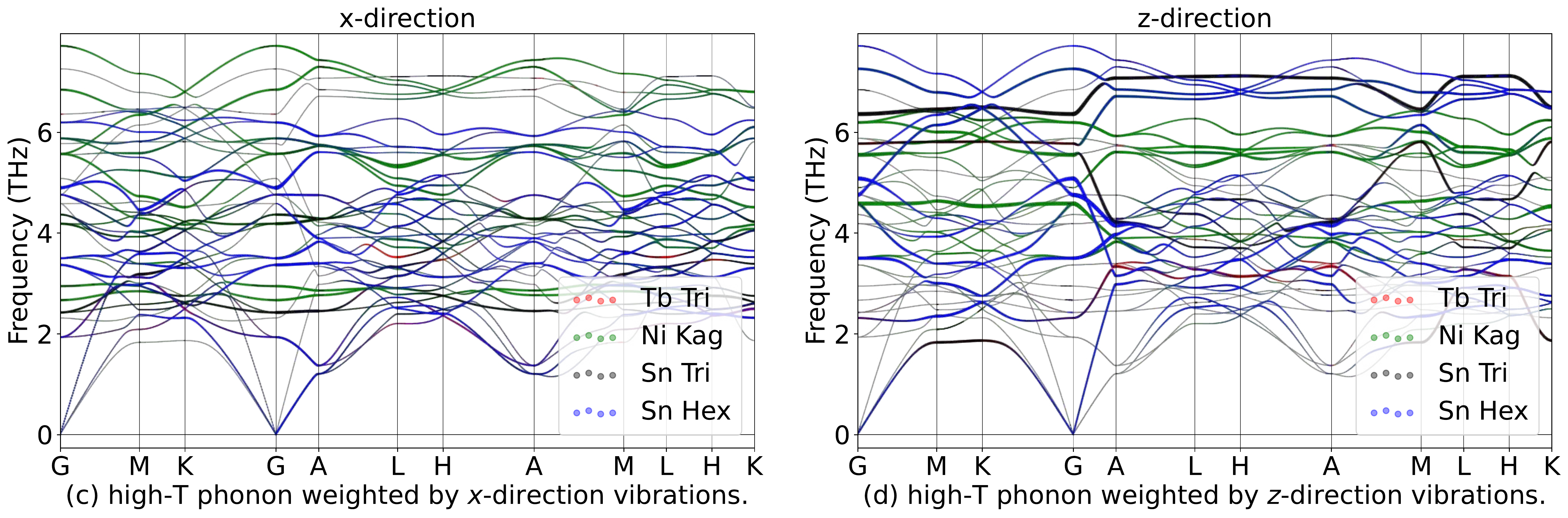}
\caption{Phonon spectrum for TbNi$_6$Sn$_6$in in-plane ($x$) and out-of-plane ($z$) directions.}
\label{fig:TbNi6Sn6phoxyz}
\end{figure}

\begin{table}[htbp]
\global\long\def\arraystretch{1.12}
\captionsetup{width=.95\textwidth}
\caption{IRREPs at high-symmetry points for TbNi$_6$Sn$_6$ phonon spectrum at Low-T.}
\label{tab:191_TbNi6Sn6_Low}
\setlength{\tabcolsep}{1.mm}{
}
\end{table}

\begin{table}[htbp]
\global\long\def\arraystretch{1.12}
\captionsetup{width=.95\textwidth}
\caption{IRREPs at high-symmetry points for TbNi$_6$Sn$_6$ phonon spectrum at High-T.}
\label{tab:191_TbNi6Sn6_High}
\setlength{\tabcolsep}{1.mm}{
%
}
\end{table}
\subsubsection{\label{sms2sec:DyNi6Sn6}\hyperref[smsec:col]{\texorpdfstring{DyNi$_6$Sn$_6$}{}}~{\color{green}  (High-T stable, Low-T stable) }}

\begin{table}[htbp]
\global\long\def\arraystretch{1.12}
\captionsetup{width=.85\textwidth}
\caption{Structure parameters of DyNi$_6$Sn$_6$ after relaxation. It contains 534 electrons. }
\label{tab:DyNi6Sn6}
\setlength{\tabcolsep}{4mm}{
%
}
\end{table}

\begin{figure}[htbp]
\centering
\includegraphics[width=0.85\textwidth]{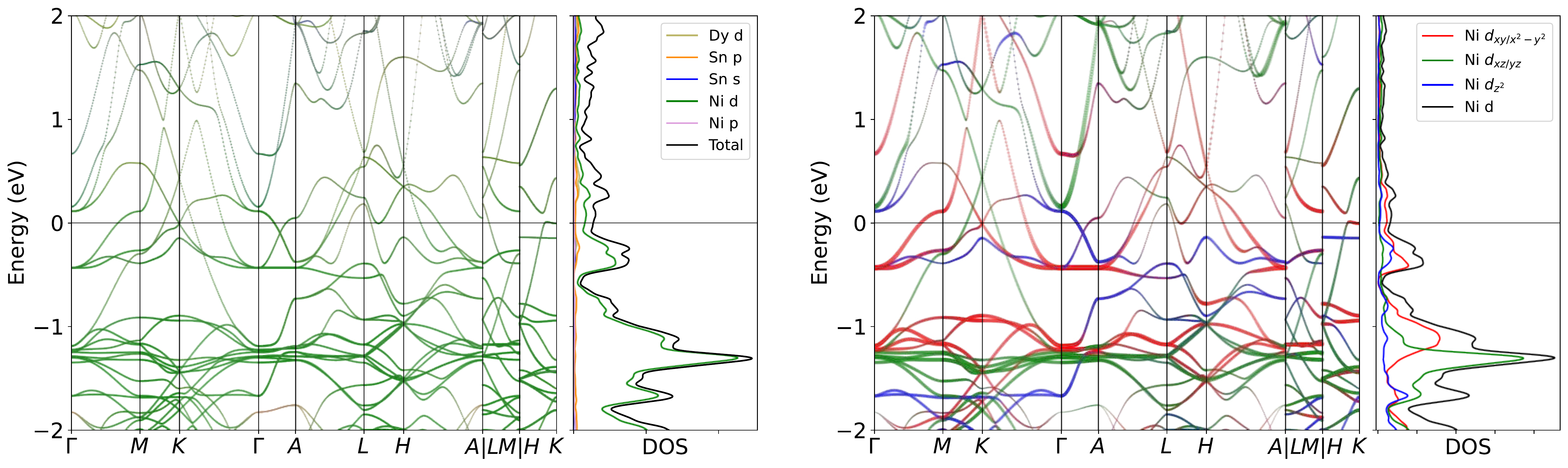}
\caption{Band structure for DyNi$_6$Sn$_6$ without SOC. The projection of Ni $3d$ orbitals is also presented. The band structures clearly reveal the dominating of Ni $3d$ orbitals  near the Fermi level, as well as the pronounced peak.}
\label{fig:DyNi6Sn6band}
\end{figure}

\begin{figure}[H]
\centering
\subfloat[Relaxed phonon at low-T.]{
\label{fig:DyNi6Sn6_relaxed_Low}
\includegraphics[width=0.45\textwidth]{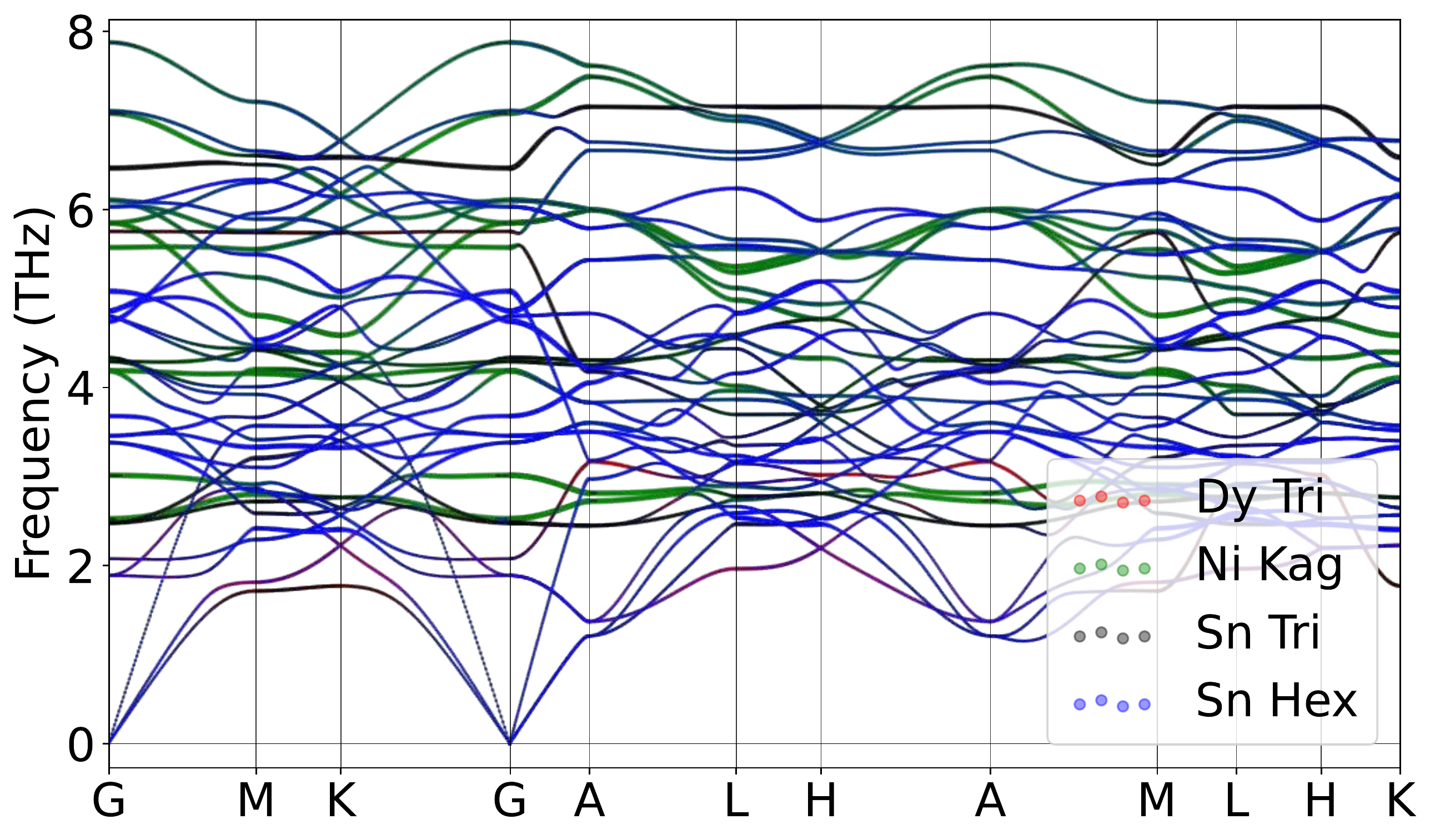}
}
\hspace{5pt}\subfloat[Relaxed phonon at high-T.]{
\label{fig:DyNi6Sn6_relaxed_High}
\includegraphics[width=0.45\textwidth]{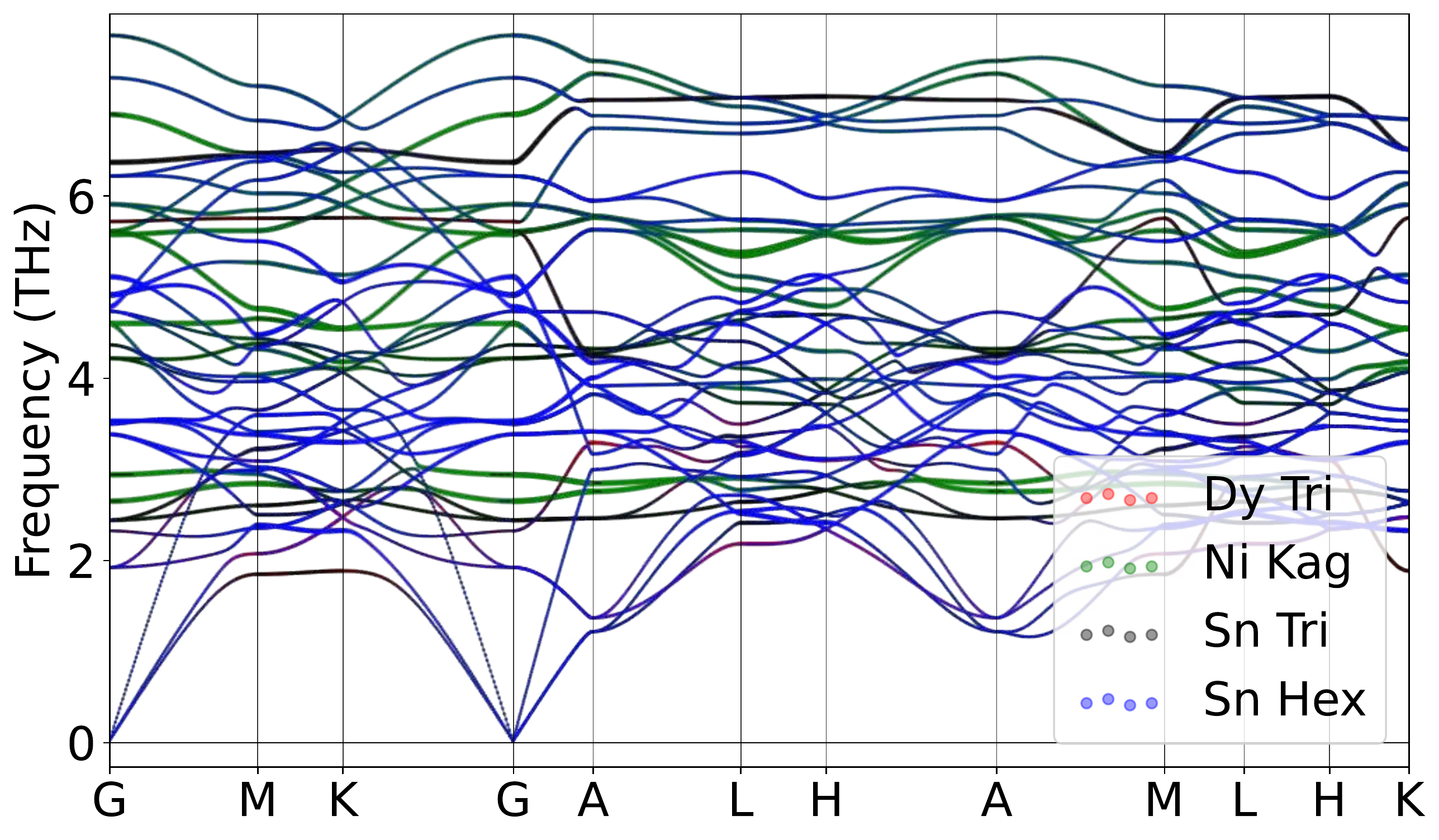}
}
\caption{Atomically projected phonon spectrum for DyNi$_6$Sn$_6$.}
\label{fig:DyNi6Sn6pho}
\end{figure}

\begin{figure}[H]
\centering
\label{fig:DyNi6Sn6_relaxed_Low_xz}
\includegraphics[width=0.85\textwidth]{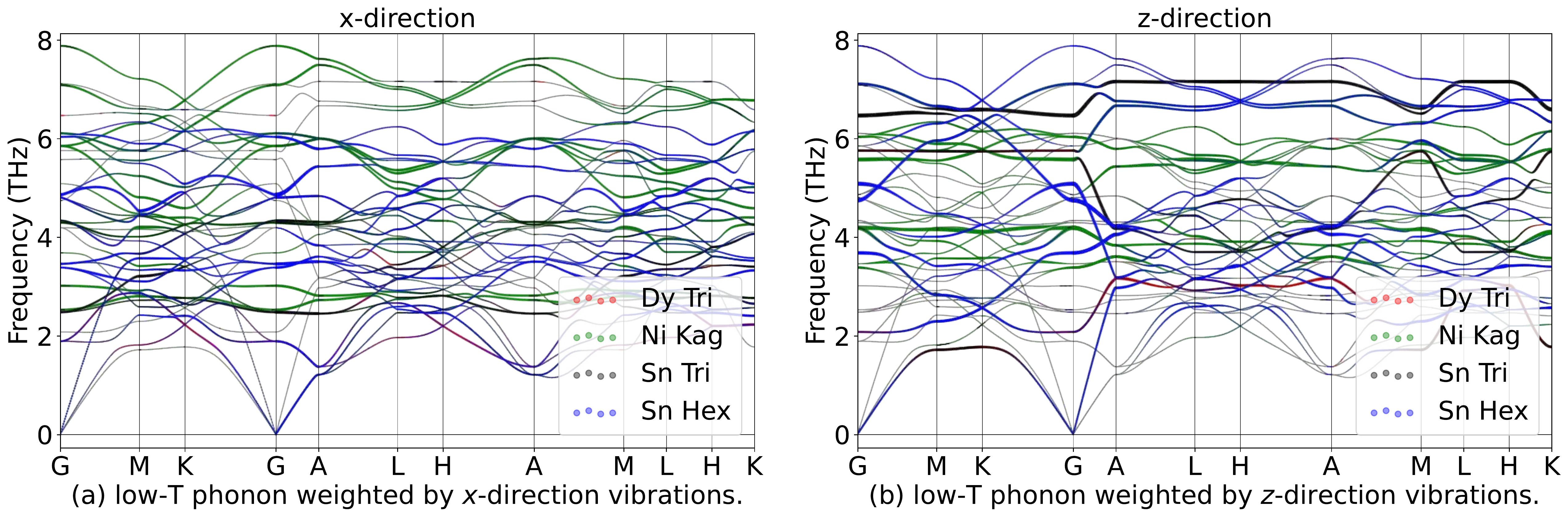}
\hspace{5pt}\label{fig:DyNi6Sn6_relaxed_High_xz}
\includegraphics[width=0.85\textwidth]{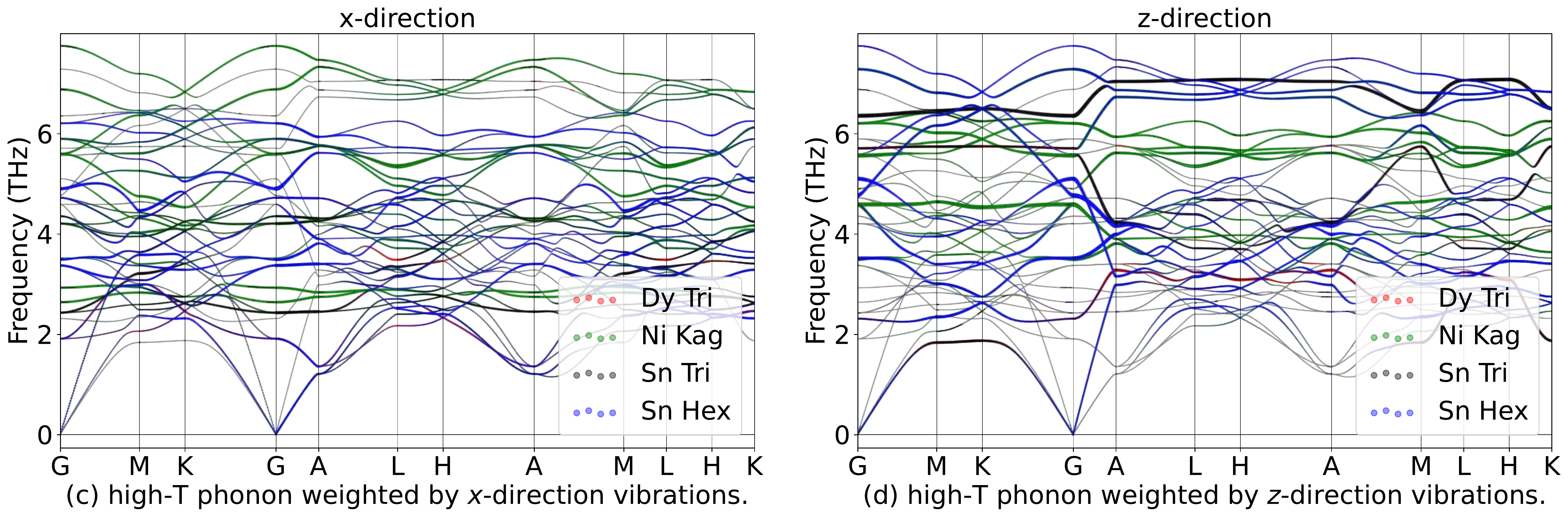}
\caption{Phonon spectrum for DyNi$_6$Sn$_6$in in-plane ($x$) and out-of-plane ($z$) directions.}
\label{fig:DyNi6Sn6phoxyz}
\end{figure}

\begin{table}[htbp]
\global\long\def\arraystretch{1.12}
\captionsetup{width=.95\textwidth}
\caption{IRREPs at high-symmetry points for DyNi$_6$Sn$_6$ phonon spectrum at Low-T.}
\label{tab:191_DyNi6Sn6_Low}
\setlength{\tabcolsep}{1.mm}{
}
\end{table}

\begin{table}[htbp]
\global\long\def\arraystretch{1.12}
\captionsetup{width=.95\textwidth}
\caption{IRREPs at high-symmetry points for DyNi$_6$Sn$_6$ phonon spectrum at High-T.}
\label{tab:191_DyNi6Sn6_High}
\setlength{\tabcolsep}{1.mm}{
%
}
\end{table}
\subsubsection{\label{sms2sec:HoNi6Sn6}\hyperref[smsec:col]{\texorpdfstring{HoNi$_6$Sn$_6$}{}}~{\color{green}  (High-T stable, Low-T stable) }}

\begin{table}[htbp]
\global\long\def\arraystretch{1.12}
\captionsetup{width=.85\textwidth}
\caption{Structure parameters of HoNi$_6$Sn$_6$ after relaxation. It contains 535 electrons. }
\label{tab:HoNi6Sn6}
\setlength{\tabcolsep}{4mm}{
%
}
\end{table}

\begin{figure}[htbp]
\centering
\includegraphics[width=0.85\textwidth]{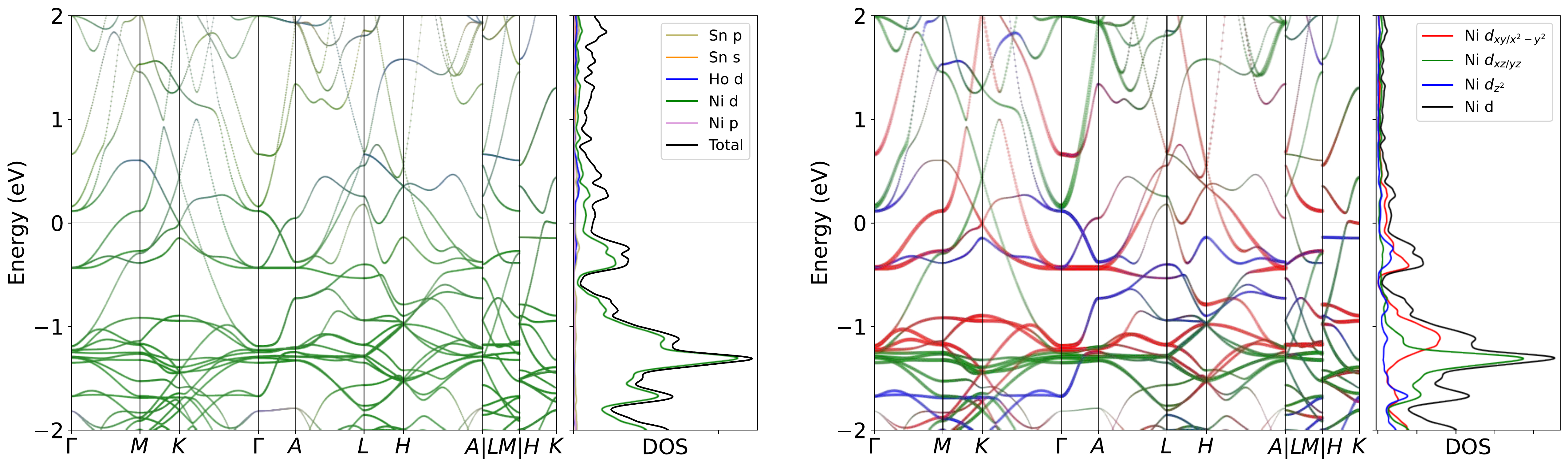}
\caption{Band structure for HoNi$_6$Sn$_6$ without SOC. The projection of Ni $3d$ orbitals is also presented. The band structures clearly reveal the dominating of Ni $3d$ orbitals  near the Fermi level, as well as the pronounced peak.}
\label{fig:HoNi6Sn6band}
\end{figure}

\begin{figure}[H]
\centering
\subfloat[Relaxed phonon at low-T.]{
\label{fig:HoNi6Sn6_relaxed_Low}
\includegraphics[width=0.45\textwidth]{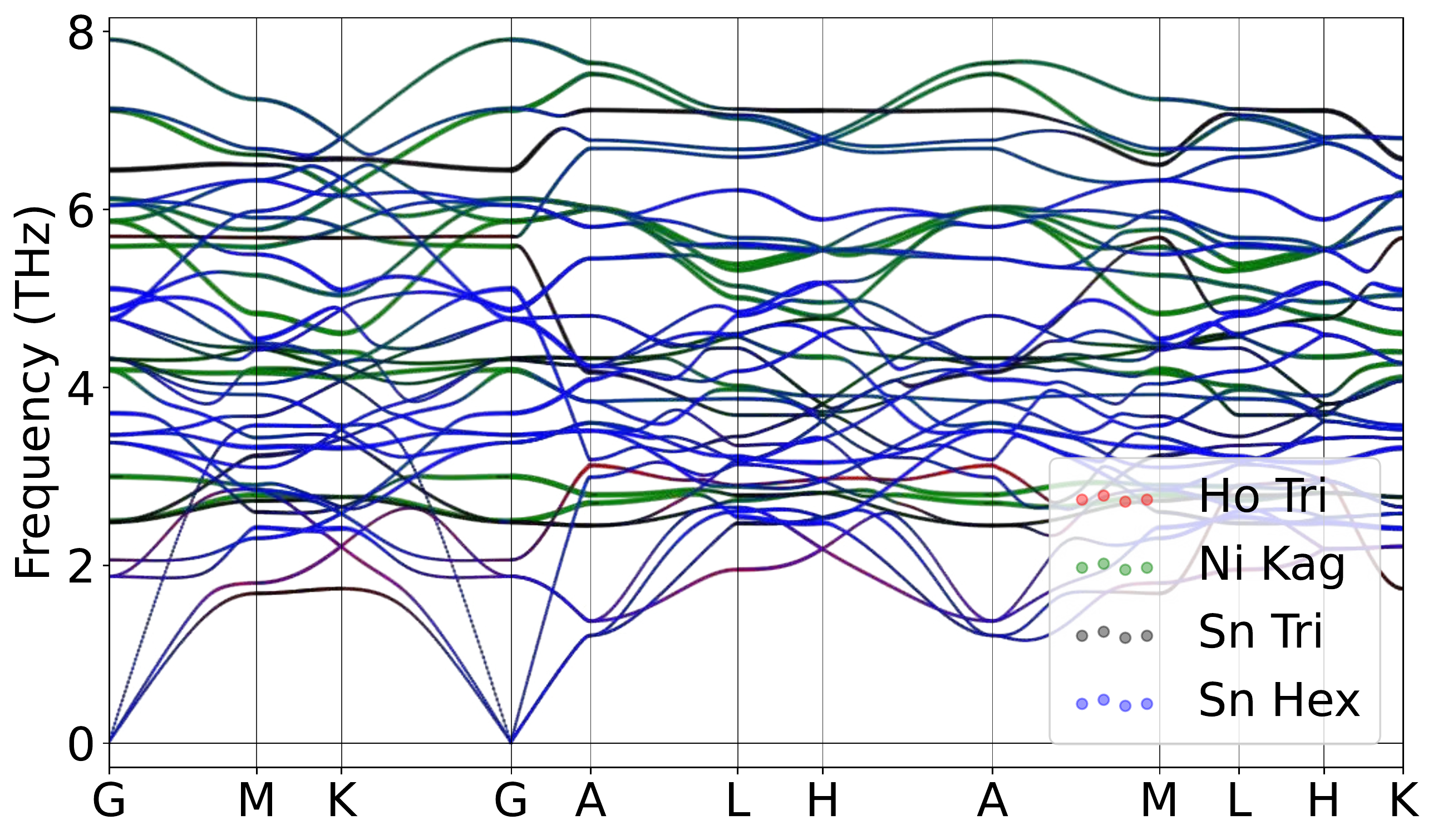}
}
\hspace{5pt}\subfloat[Relaxed phonon at high-T.]{
\label{fig:HoNi6Sn6_relaxed_High}
\includegraphics[width=0.45\textwidth]{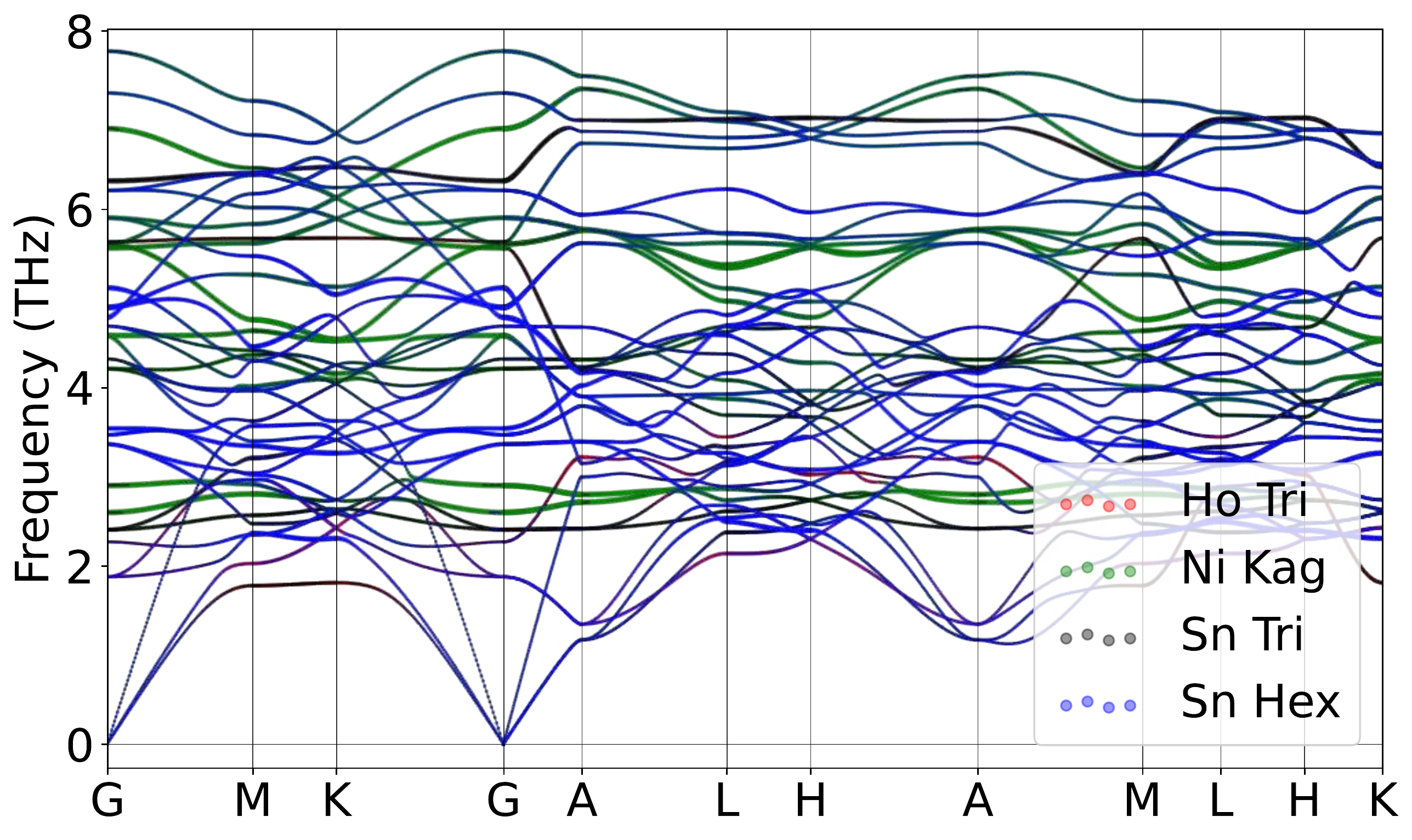}
}
\caption{Atomically projected phonon spectrum for HoNi$_6$Sn$_6$.}
\label{fig:HoNi6Sn6pho}
\end{figure}

\begin{figure}[H]
\centering
\label{fig:HoNi6Sn6_relaxed_Low_xz}
\includegraphics[width=0.85\textwidth]{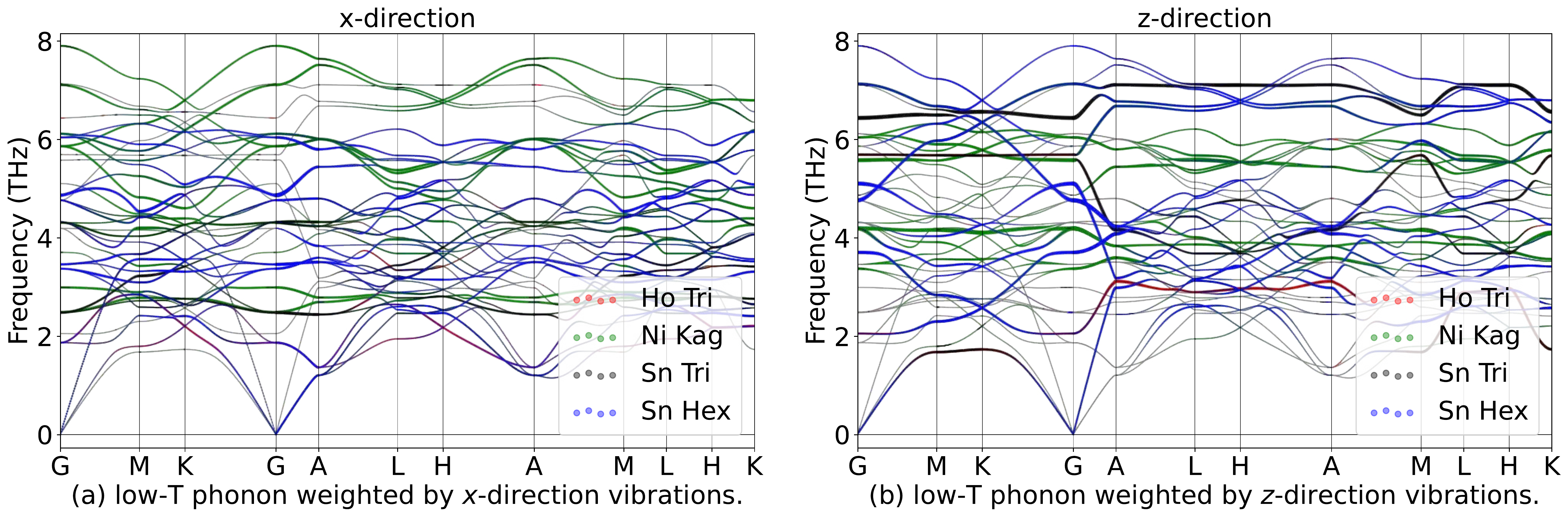}
\hspace{5pt}\label{fig:HoNi6Sn6_relaxed_High_xz}
\includegraphics[width=0.85\textwidth]{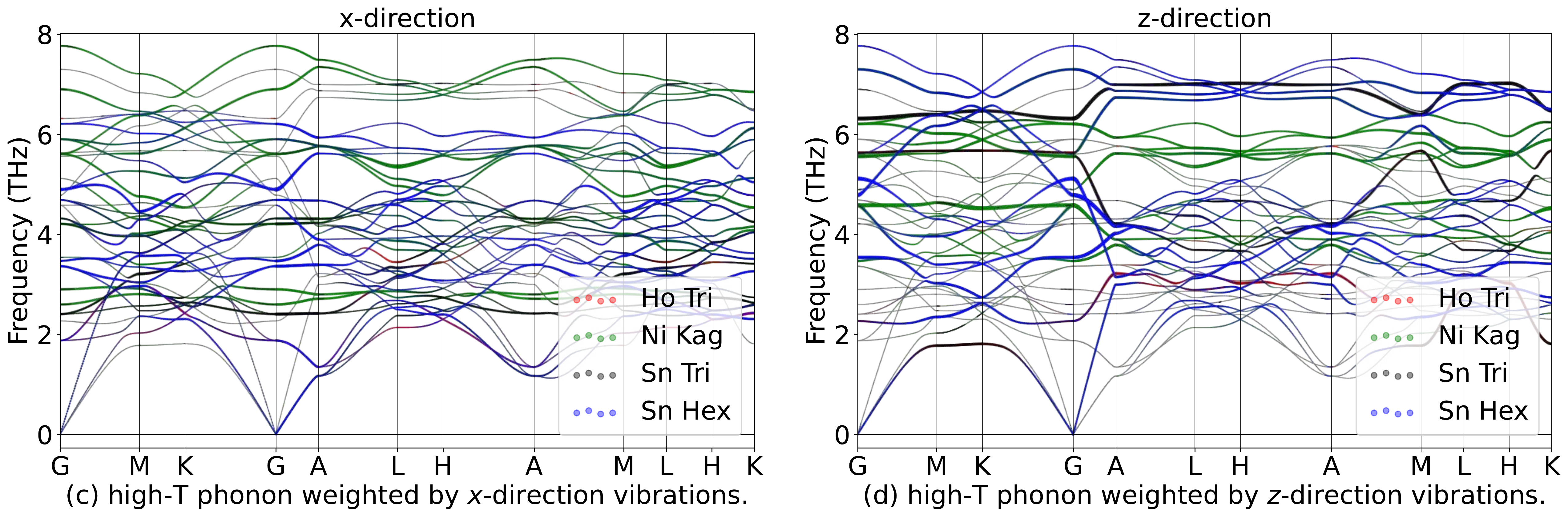}
\caption{Phonon spectrum for HoNi$_6$Sn$_6$in in-plane ($x$) and out-of-plane ($z$) directions.}
\label{fig:HoNi6Sn6phoxyz}
\end{figure}

\begin{table}[htbp]
\global\long\def\arraystretch{1.12}
\captionsetup{width=.95\textwidth}
\caption{IRREPs at high-symmetry points for HoNi$_6$Sn$_6$ phonon spectrum at Low-T.}
\label{tab:191_HoNi6Sn6_Low}
\setlength{\tabcolsep}{1.mm}{
}
\end{table}

\begin{table}[htbp]
\global\long\def\arraystretch{1.12}
\captionsetup{width=.95\textwidth}
\caption{IRREPs at high-symmetry points for HoNi$_6$Sn$_6$ phonon spectrum at High-T.}
\label{tab:191_HoNi6Sn6_High}
\setlength{\tabcolsep}{1.mm}{
%
}
\end{table}
\subsubsection{\label{sms2sec:ErNi6Sn6}\hyperref[smsec:col]{\texorpdfstring{ErNi$_6$Sn$_6$}{}}~{\color{green}  (High-T stable, Low-T stable) }}

\begin{table}[htbp]
\global\long\def\arraystretch{1.12}
\captionsetup{width=.85\textwidth}
\caption{Structure parameters of ErNi$_6$Sn$_6$ after relaxation. It contains 536 electrons. }
\label{tab:ErNi6Sn6}
\setlength{\tabcolsep}{4mm}{
%
}
\end{table}

\begin{figure}[htbp]
\centering
\includegraphics[width=0.85\textwidth]{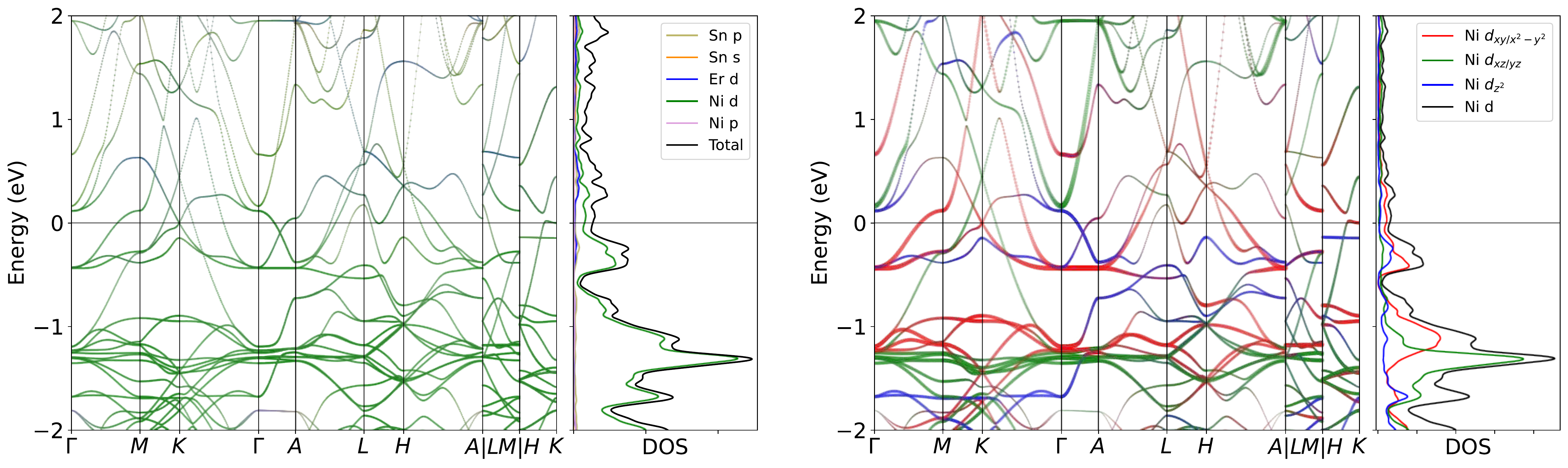}
\caption{Band structure for ErNi$_6$Sn$_6$ without SOC. The projection of Ni $3d$ orbitals is also presented. The band structures clearly reveal the dominating of Ni $3d$ orbitals  near the Fermi level, as well as the pronounced peak.}
\label{fig:ErNi6Sn6band}
\end{figure}

\begin{figure}[H]
\centering
\subfloat[Relaxed phonon at low-T.]{
\label{fig:ErNi6Sn6_relaxed_Low}
\includegraphics[width=0.45\textwidth]{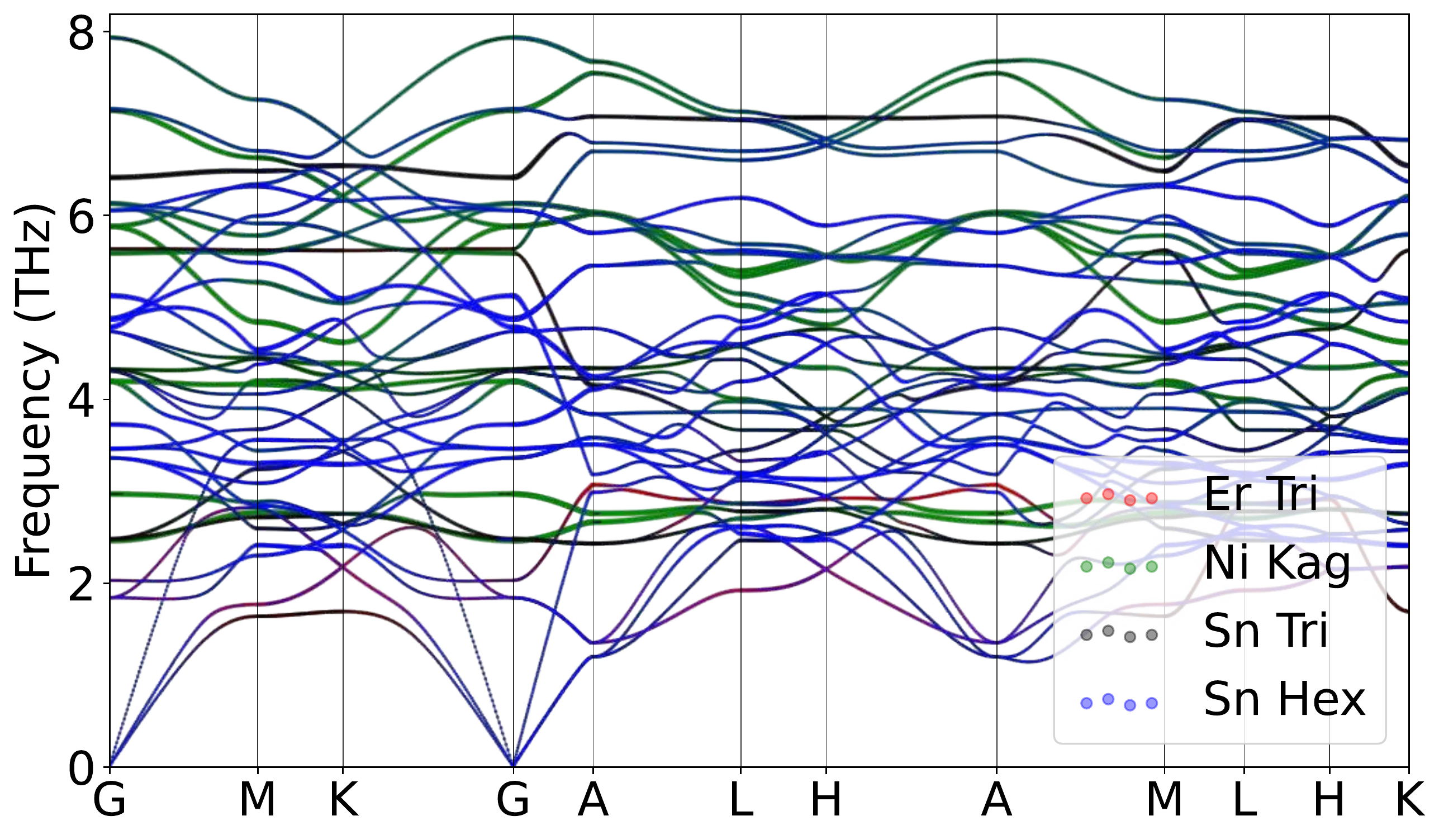}
}
\hspace{5pt}\subfloat[Relaxed phonon at high-T.]{
\label{fig:ErNi6Sn6_relaxed_High}
\includegraphics[width=0.45\textwidth]{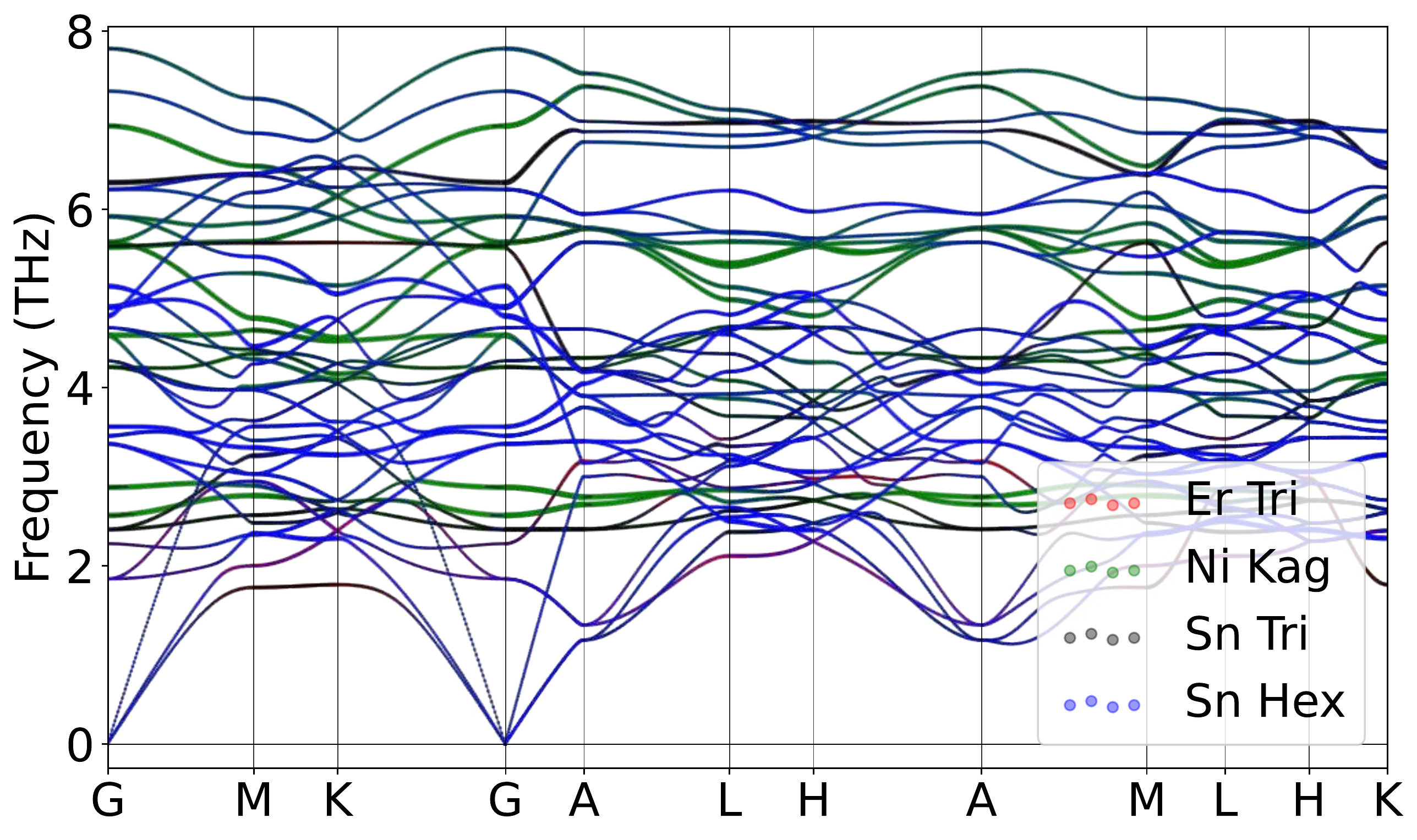}
}
\caption{Atomically projected phonon spectrum for ErNi$_6$Sn$_6$.}
\label{fig:ErNi6Sn6pho}
\end{figure}

\begin{figure}[H]
\centering
\label{fig:ErNi6Sn6_relaxed_Low_xz}
\includegraphics[width=0.85\textwidth]{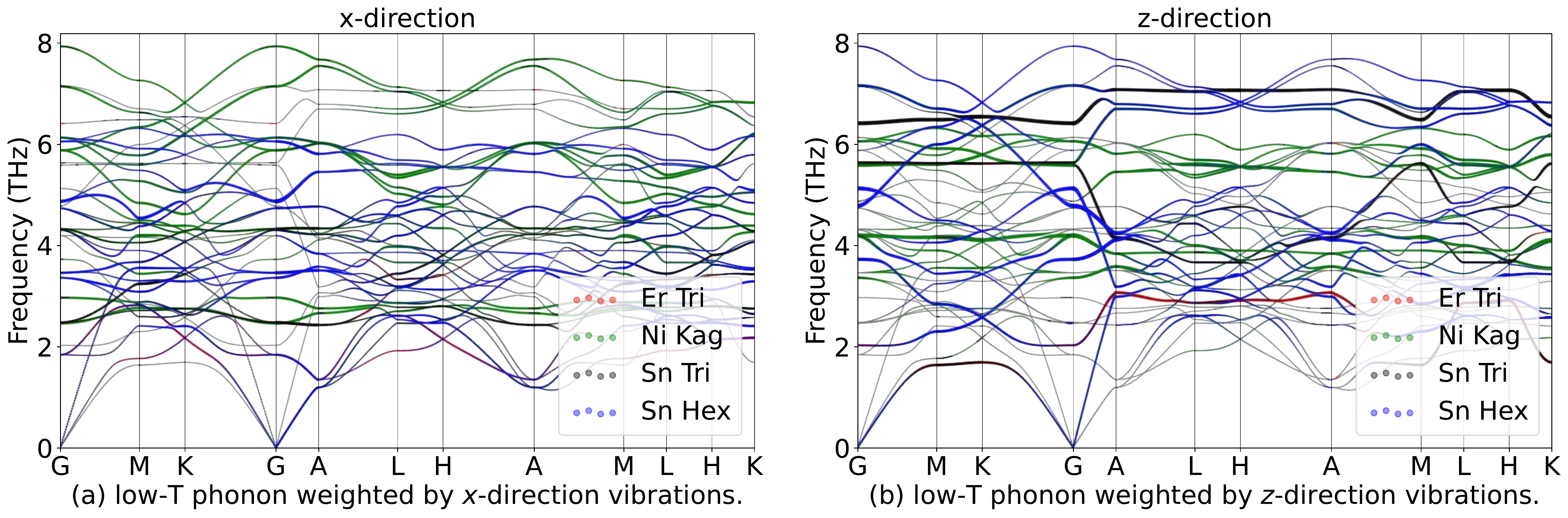}
\hspace{5pt}\label{fig:ErNi6Sn6_relaxed_High_xz}
\includegraphics[width=0.85\textwidth]{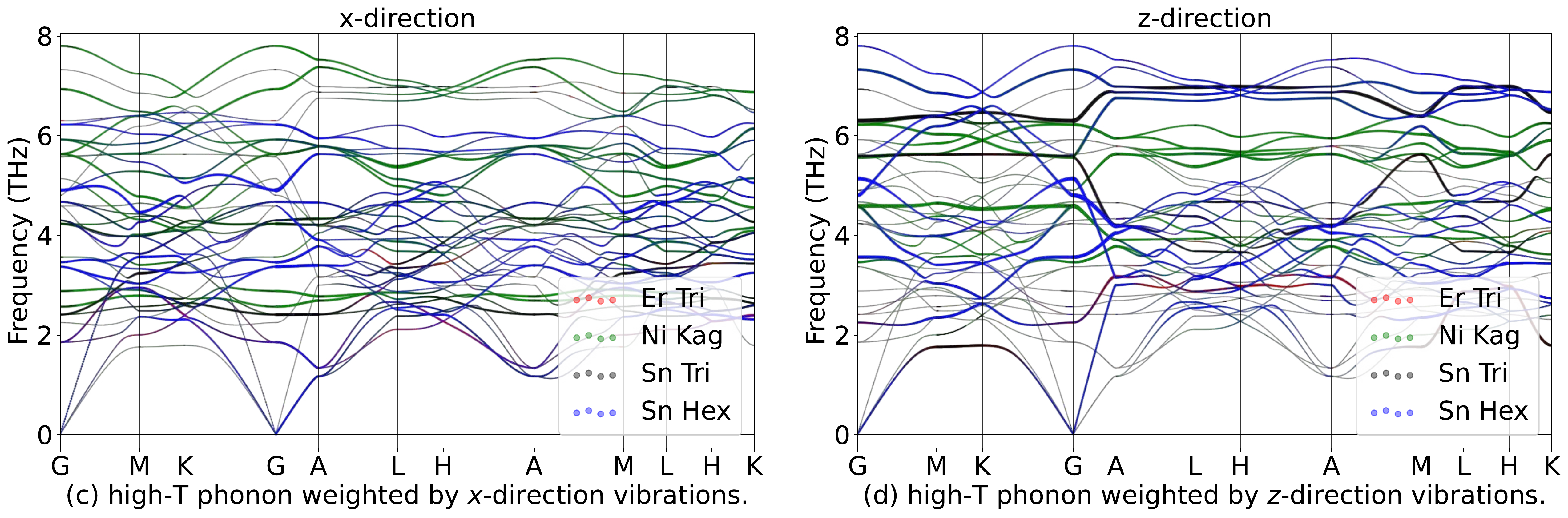}
\caption{Phonon spectrum for ErNi$_6$Sn$_6$in in-plane ($x$) and out-of-plane ($z$) directions.}
\label{fig:ErNi6Sn6phoxyz}
\end{figure}

\begin{table}[htbp]
\global\long\def\arraystretch{1.12}
\captionsetup{width=.95\textwidth}
\caption{IRREPs at high-symmetry points for ErNi$_6$Sn$_6$ phonon spectrum at Low-T.}
\label{tab:191_ErNi6Sn6_Low}
\setlength{\tabcolsep}{1.mm}{
}
\end{table}

\begin{table}[htbp]
\global\long\def\arraystretch{1.12}
\captionsetup{width=.95\textwidth}
\caption{IRREPs at high-symmetry points for ErNi$_6$Sn$_6$ phonon spectrum at High-T.}
\label{tab:191_ErNi6Sn6_High}
\setlength{\tabcolsep}{1.mm}{
%
}
\end{table}
\subsubsection{\label{sms2sec:TmNi6Sn6}\hyperref[smsec:col]{\texorpdfstring{TmNi$_6$Sn$_6$}{}}~{\color{green}  (High-T stable, Low-T stable) }}

\begin{table}[htbp]
\global\long\def\arraystretch{1.12}
\captionsetup{width=.85\textwidth}
\caption{Structure parameters of TmNi$_6$Sn$_6$ after relaxation. It contains 537 electrons. }
\label{tab:TmNi6Sn6}
\setlength{\tabcolsep}{4mm}{
%
}
\end{table}

\begin{figure}[htbp]
\centering
\includegraphics[width=0.85\textwidth]{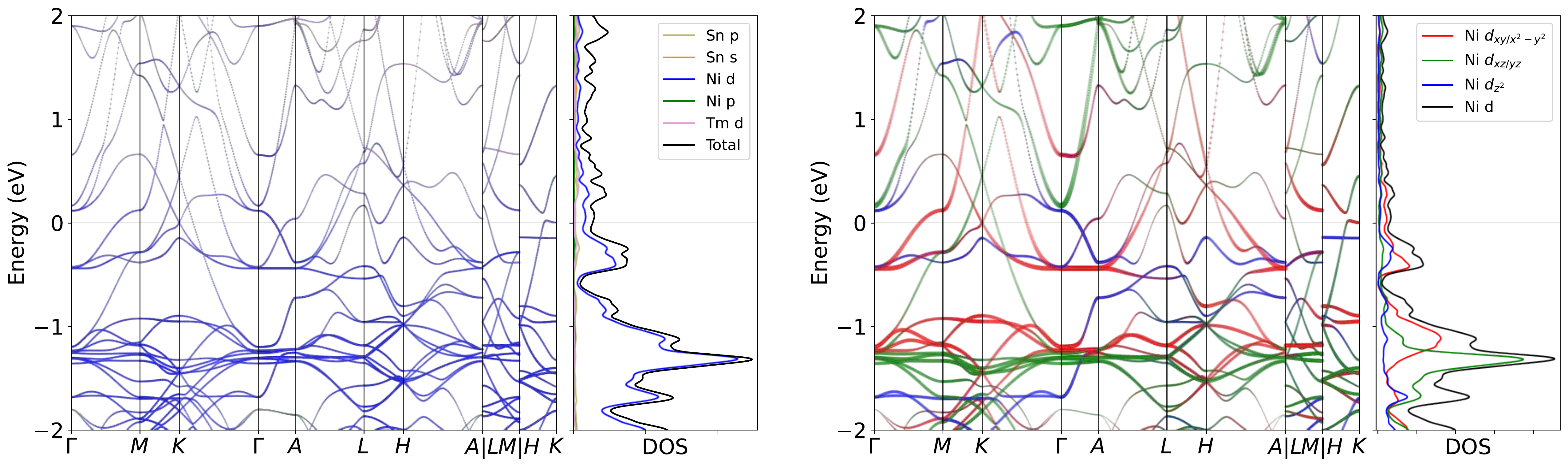}
\caption{Band structure for TmNi$_6$Sn$_6$ without SOC. The projection of Ni $3d$ orbitals is also presented. The band structures clearly reveal the dominating of Ni $3d$ orbitals  near the Fermi level, as well as the pronounced peak.}
\label{fig:TmNi6Sn6band}
\end{figure}

\begin{figure}[H]
\centering
\subfloat[Relaxed phonon at low-T.]{
\label{fig:TmNi6Sn6_relaxed_Low}
\includegraphics[width=0.45\textwidth]{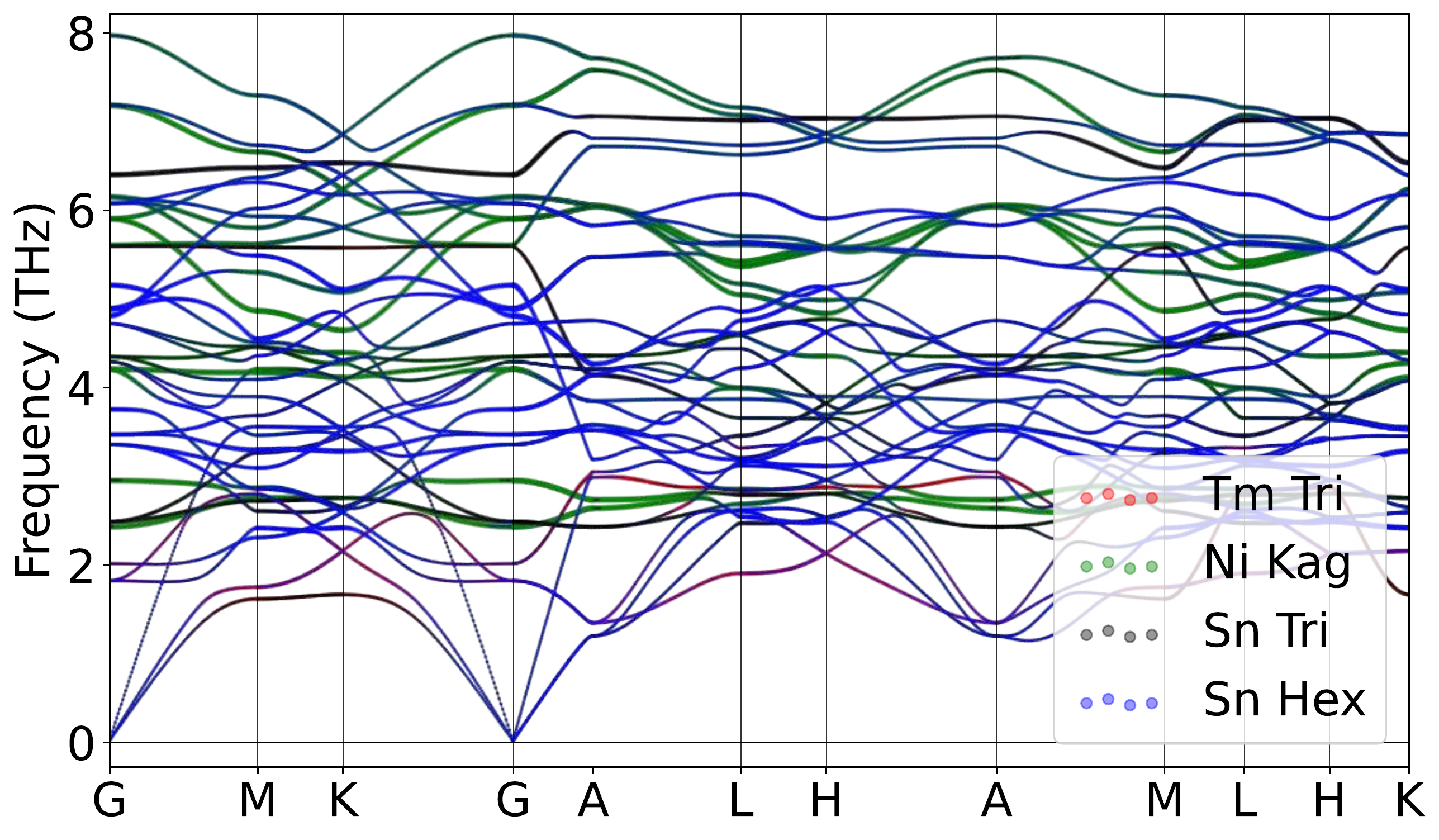}
}
\hspace{5pt}\subfloat[Relaxed phonon at high-T.]{
\label{fig:TmNi6Sn6_relaxed_High}
\includegraphics[width=0.45\textwidth]{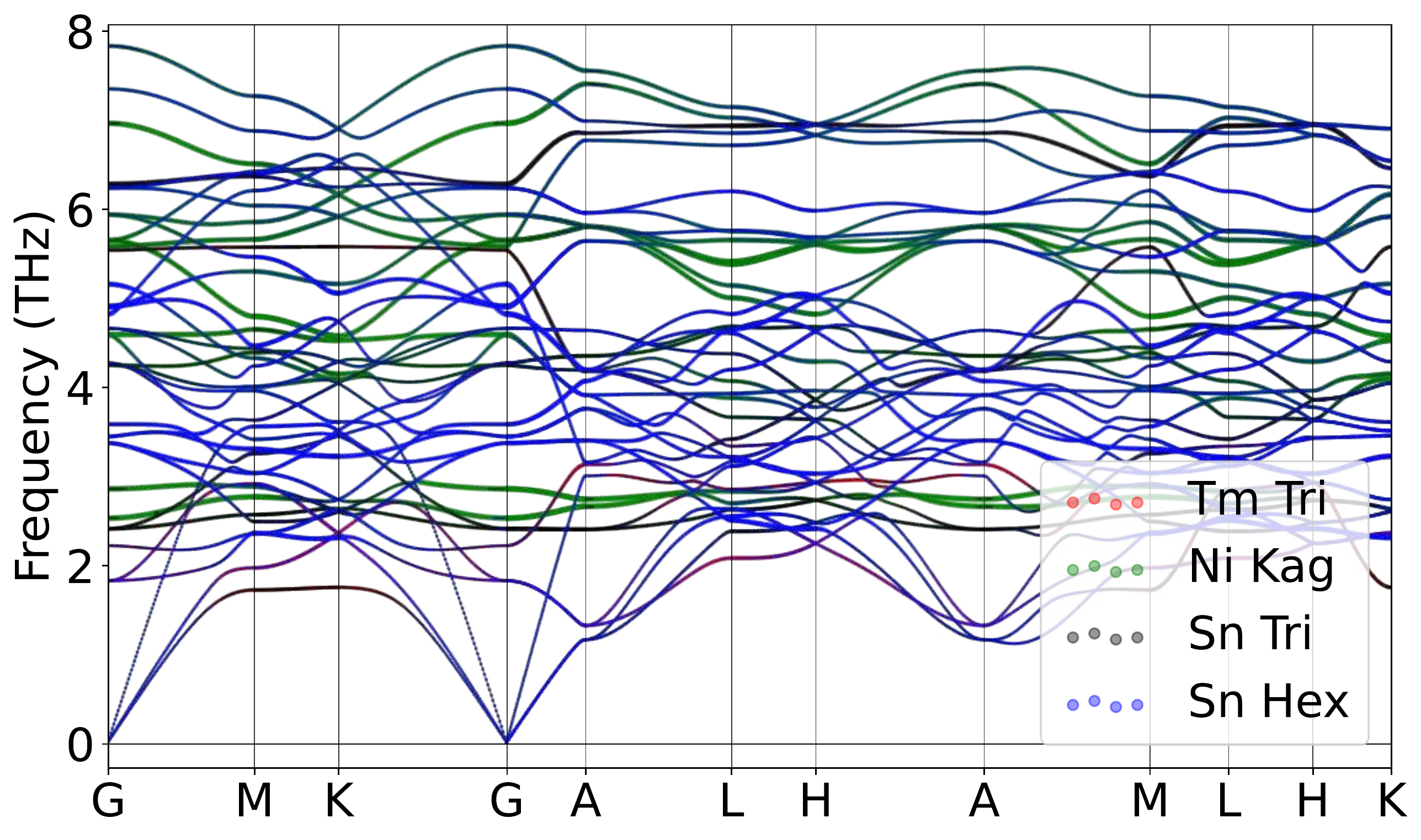}
}
\caption{Atomically projected phonon spectrum for TmNi$_6$Sn$_6$.}
\label{fig:TmNi6Sn6pho}
\end{figure}

\begin{figure}[H]
\centering
\label{fig:TmNi6Sn6_relaxed_Low_xz}
\includegraphics[width=0.85\textwidth]{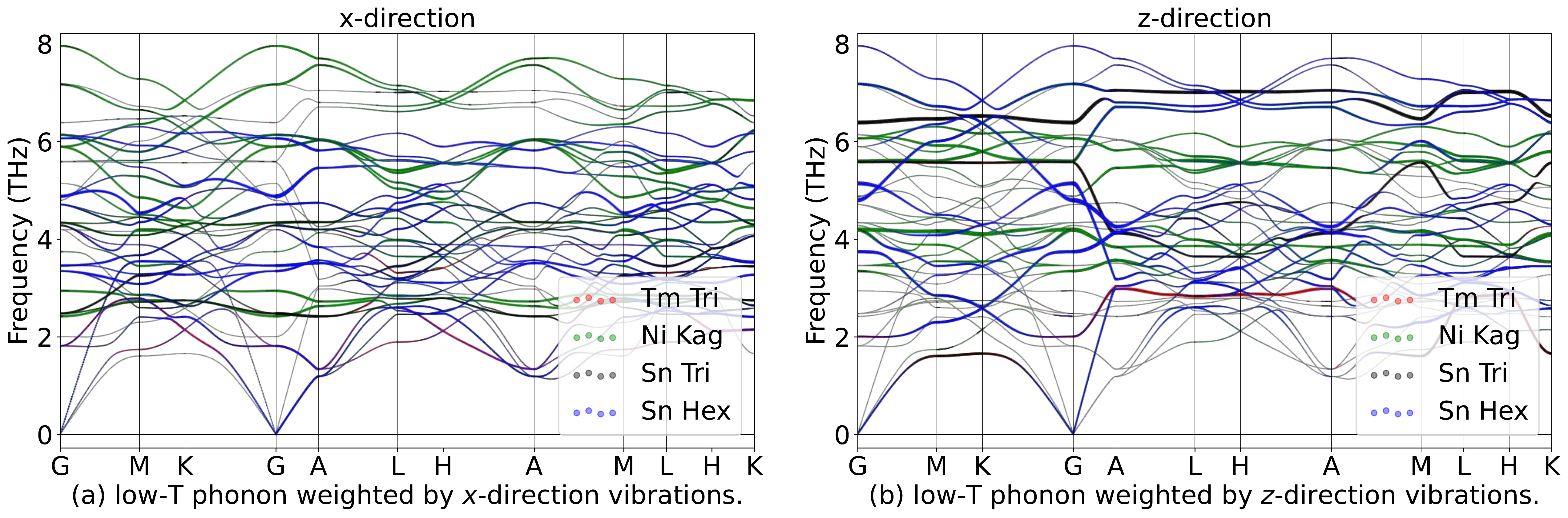}
\hspace{5pt}\label{fig:TmNi6Sn6_relaxed_High_xz}
\includegraphics[width=0.85\textwidth]{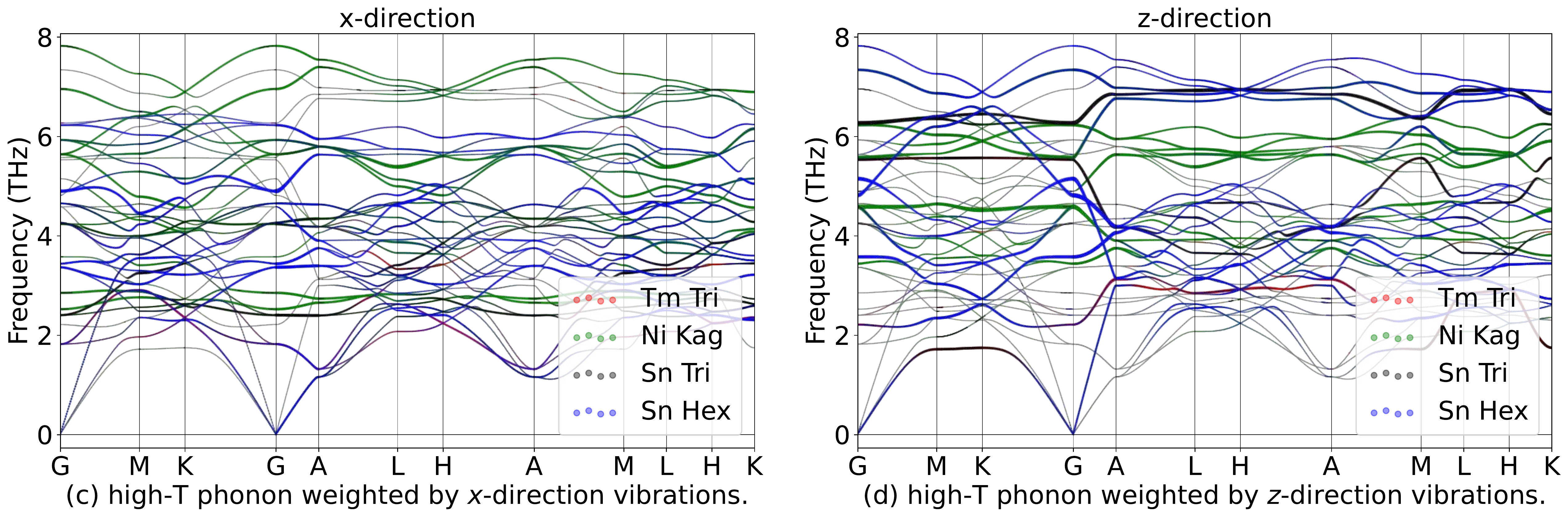}
\caption{Phonon spectrum for TmNi$_6$Sn$_6$in in-plane ($x$) and out-of-plane ($z$) directions.}
\label{fig:TmNi6Sn6phoxyz}
\end{figure}

\begin{table}[htbp]
\global\long\def\arraystretch{1.12}
\captionsetup{width=.95\textwidth}
\caption{IRREPs at high-symmetry points for TmNi$_6$Sn$_6$ phonon spectrum at Low-T.}
\label{tab:191_TmNi6Sn6_Low}
\setlength{\tabcolsep}{1.mm}{
}
\end{table}

\begin{table}[htbp]
\global\long\def\arraystretch{1.12}
\captionsetup{width=.95\textwidth}
\caption{IRREPs at high-symmetry points for TmNi$_6$Sn$_6$ phonon spectrum at High-T.}
\label{tab:191_TmNi6Sn6_High}
\setlength{\tabcolsep}{1.mm}{
%
}
\end{table}
\subsubsection{\label{sms2sec:YbNi6Sn6}\hyperref[smsec:col]{\texorpdfstring{YbNi$_6$Sn$_6$}{}}~{\color{green}  (High-T stable, Low-T stable) }}

\begin{table}[htbp]
\global\long\def\arraystretch{1.12}
\captionsetup{width=.85\textwidth}
\caption{Structure parameters of YbNi$_6$Sn$_6$ after relaxation. It contains 538 electrons. }
\label{tab:YbNi6Sn6}
\setlength{\tabcolsep}{4mm}{
%
}
\end{table}

\begin{figure}[htbp]
\centering
\includegraphics[width=0.85\textwidth]{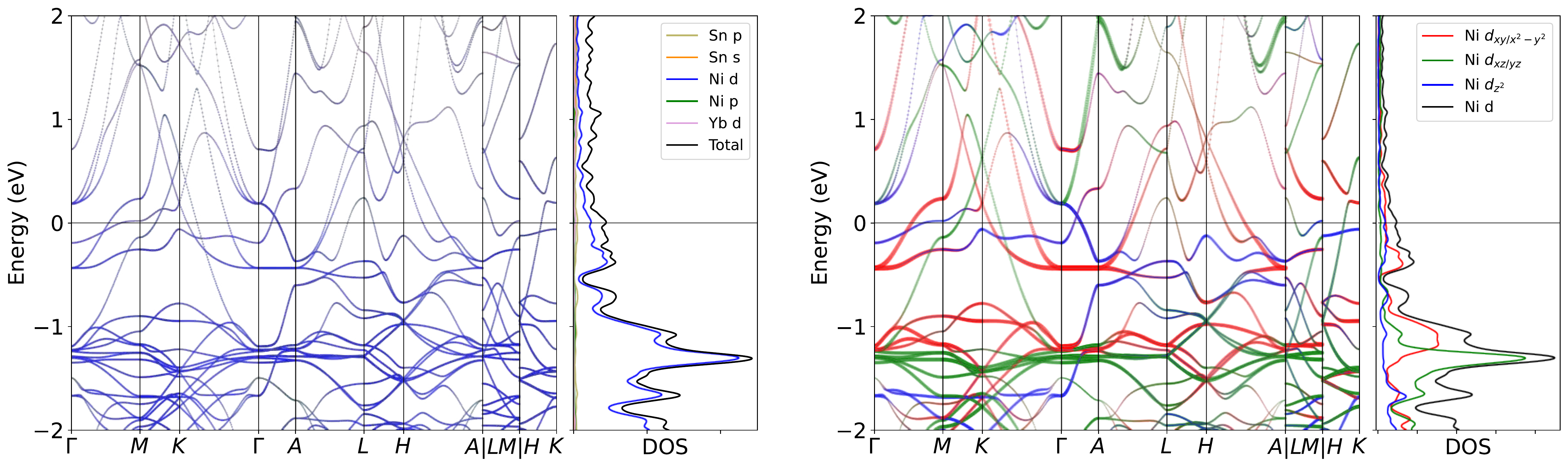}
\caption{Band structure for YbNi$_6$Sn$_6$ without SOC. The projection of Ni $3d$ orbitals is also presented. The band structures clearly reveal the dominating of Ni $3d$ orbitals  near the Fermi level, as well as the pronounced peak.}
\label{fig:YbNi6Sn6band}
\end{figure}

\begin{figure}[H]
\centering
\subfloat[Relaxed phonon at low-T.]{
\label{fig:YbNi6Sn6_relaxed_Low}
\includegraphics[width=0.45\textwidth]{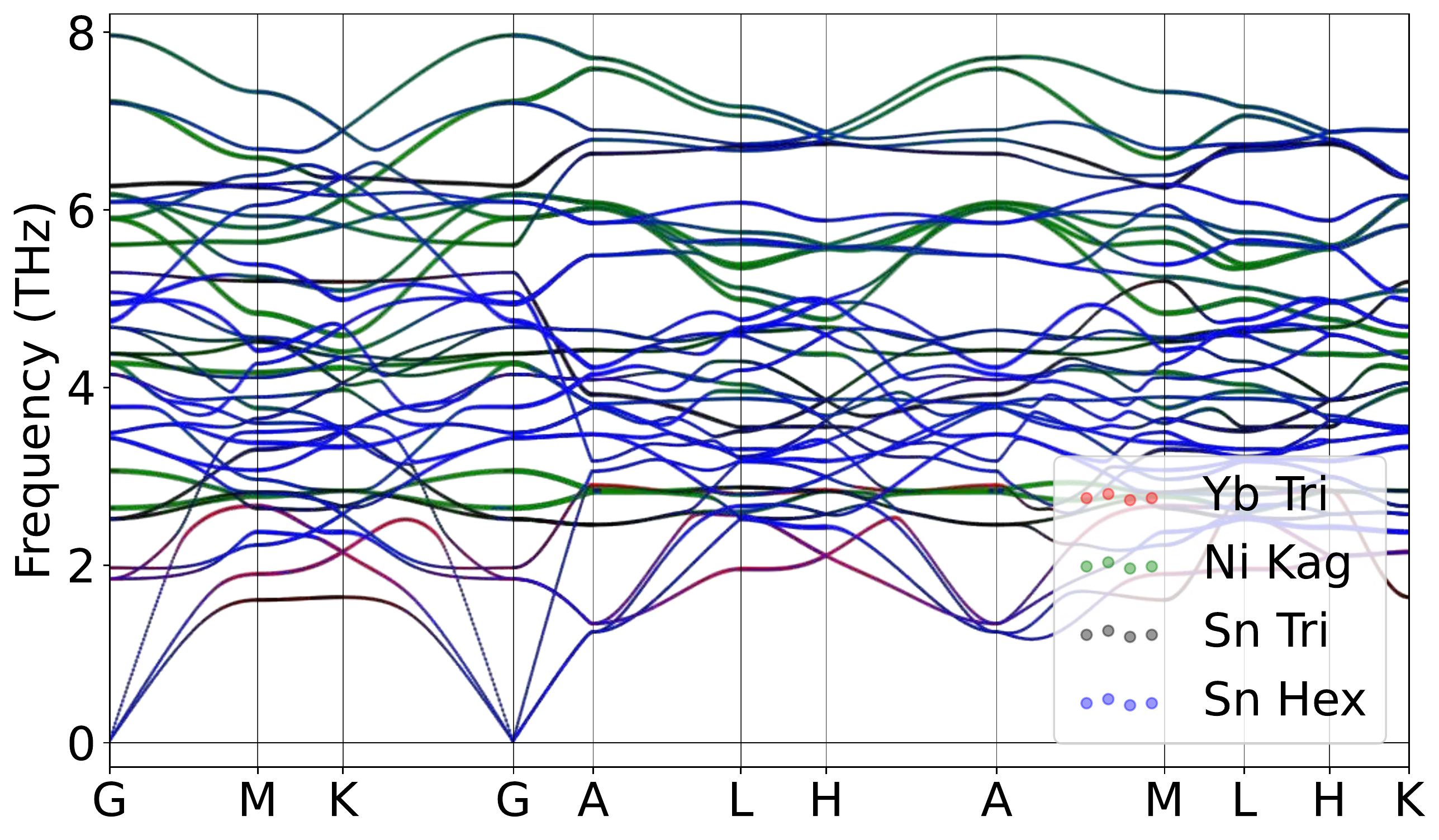}
}
\hspace{5pt}\subfloat[Relaxed phonon at high-T.]{
\label{fig:YbNi6Sn6_relaxed_High}
\includegraphics[width=0.45\textwidth]{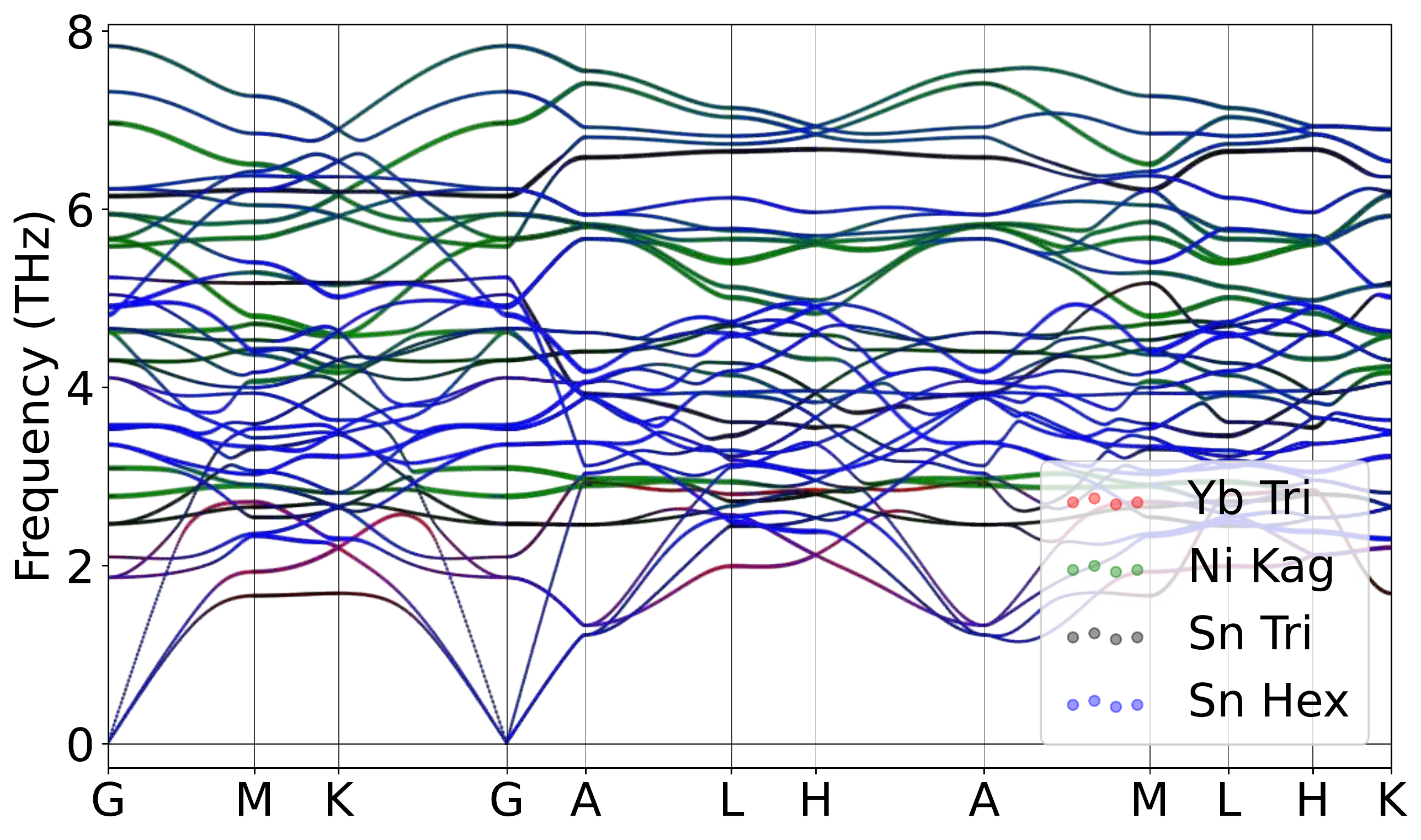}
}
\caption{Atomically projected phonon spectrum for YbNi$_6$Sn$_6$.}
\label{fig:YbNi6Sn6pho}
\end{figure}

\begin{figure}[H]
\centering
\label{fig:YbNi6Sn6_relaxed_Low_xz}
\includegraphics[width=0.85\textwidth]{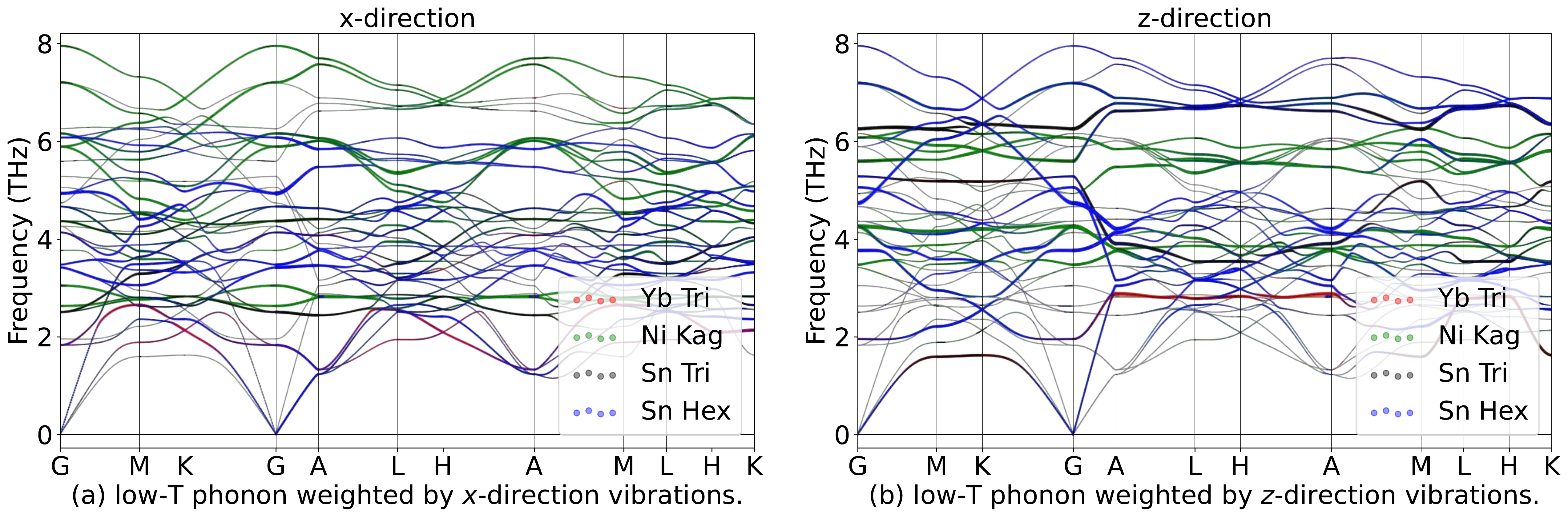}
\hspace{5pt}\label{fig:YbNi6Sn6_relaxed_High_xz}
\includegraphics[width=0.85\textwidth]{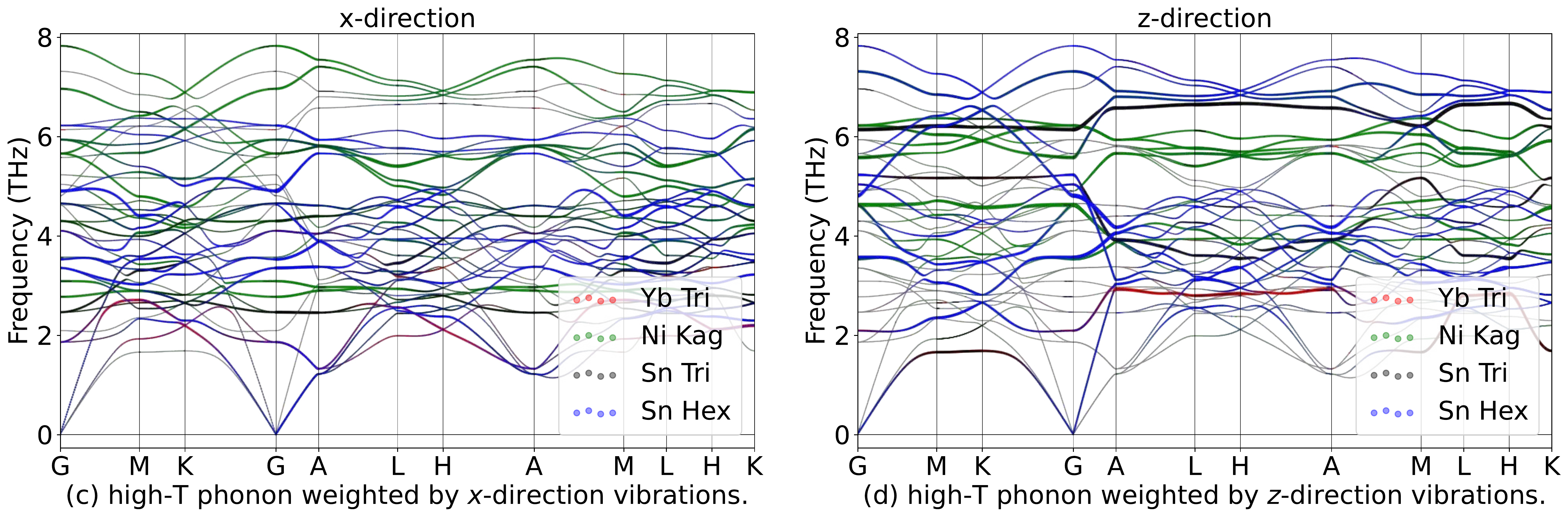}
\caption{Phonon spectrum for YbNi$_6$Sn$_6$in in-plane ($x$) and out-of-plane ($z$) directions.}
\label{fig:YbNi6Sn6phoxyz}
\end{figure}

\begin{table}[htbp]
\global\long\def\arraystretch{1.12}
\captionsetup{width=.95\textwidth}
\caption{IRREPs at high-symmetry points for YbNi$_6$Sn$_6$ phonon spectrum at Low-T.}
\label{tab:191_YbNi6Sn6_Low}
\setlength{\tabcolsep}{1.mm}{
}
\end{table}

\begin{table}[htbp]
\global\long\def\arraystretch{1.12}
\captionsetup{width=.95\textwidth}
\caption{IRREPs at high-symmetry points for YbNi$_6$Sn$_6$ phonon spectrum at High-T.}
\label{tab:191_YbNi6Sn6_High}
\setlength{\tabcolsep}{1.mm}{
%
}
\end{table}
\subsubsection{\label{sms2sec:LuNi6Sn6}\hyperref[smsec:col]{\texorpdfstring{LuNi$_6$Sn$_6$}{}}~{\color{green}  (High-T stable, Low-T stable) }}

\begin{table}[htbp]
\global\long\def\arraystretch{1.12}
\captionsetup{width=.85\textwidth}
\caption{Structure parameters of LuNi$_6$Sn$_6$ after relaxation. It contains 539 electrons. }
\label{tab:LuNi6Sn6}
\setlength{\tabcolsep}{4mm}{
%
}
\end{table}

\begin{figure}[htbp]
\centering
\includegraphics[width=0.85\textwidth]{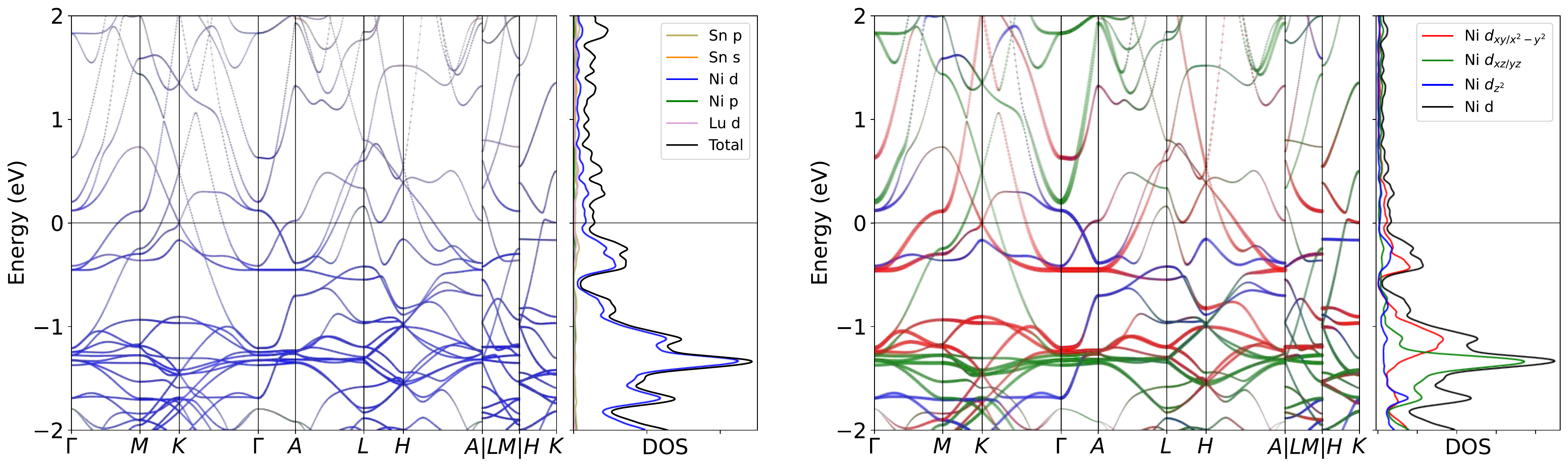}
\caption{Band structure for LuNi$_6$Sn$_6$ without SOC. The projection of Ni $3d$ orbitals is also presented. The band structures clearly reveal the dominating of Ni $3d$ orbitals  near the Fermi level, as well as the pronounced peak.}
\label{fig:LuNi6Sn6band}
\end{figure}

\begin{figure}[H]
\centering
\subfloat[Relaxed phonon at low-T.]{
\label{fig:LuNi6Sn6_relaxed_Low}
\includegraphics[width=0.45\textwidth]{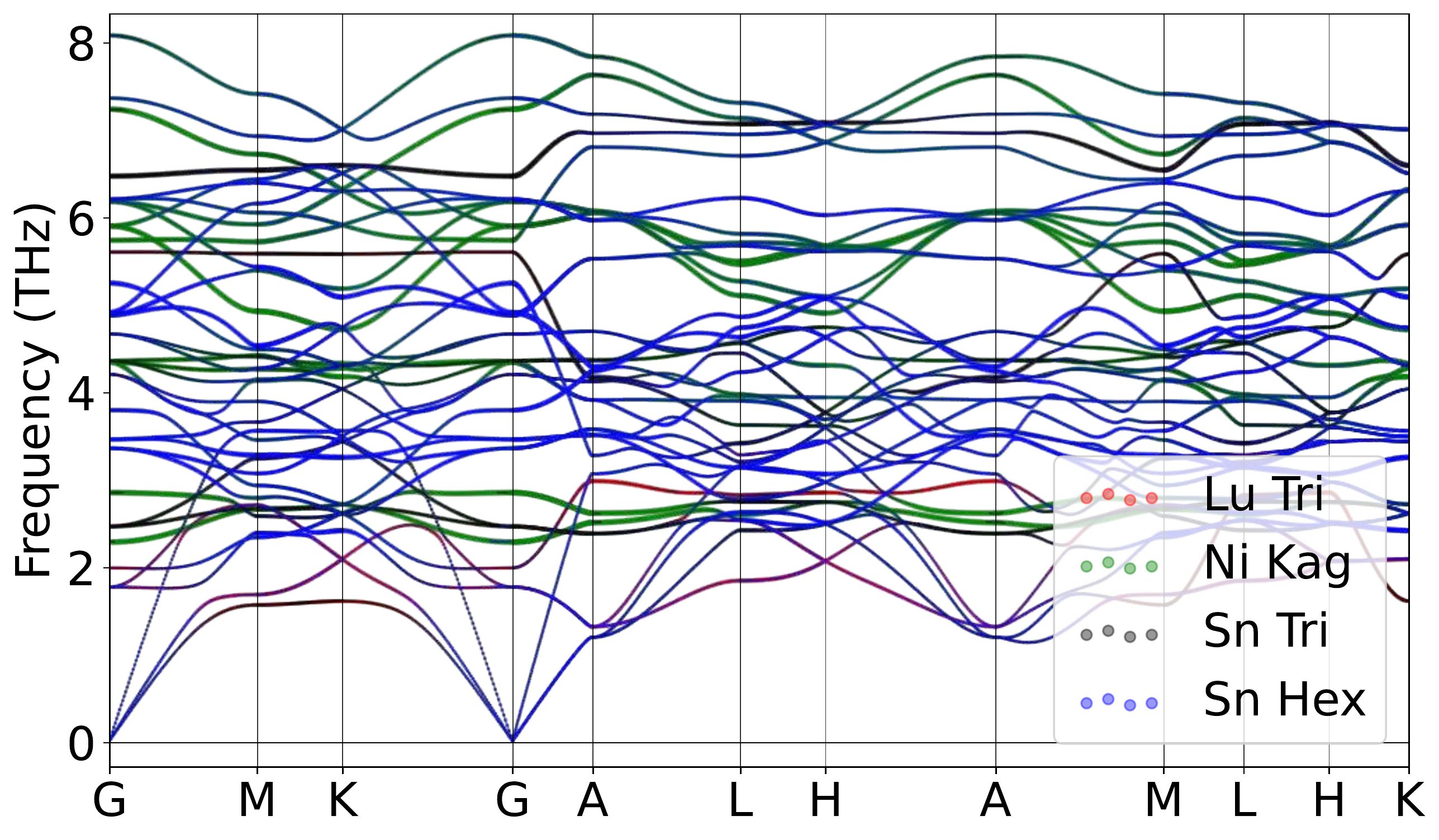}
}
\hspace{5pt}\subfloat[Relaxed phonon at high-T.]{
\label{fig:LuNi6Sn6_relaxed_High}
\includegraphics[width=0.45\textwidth]{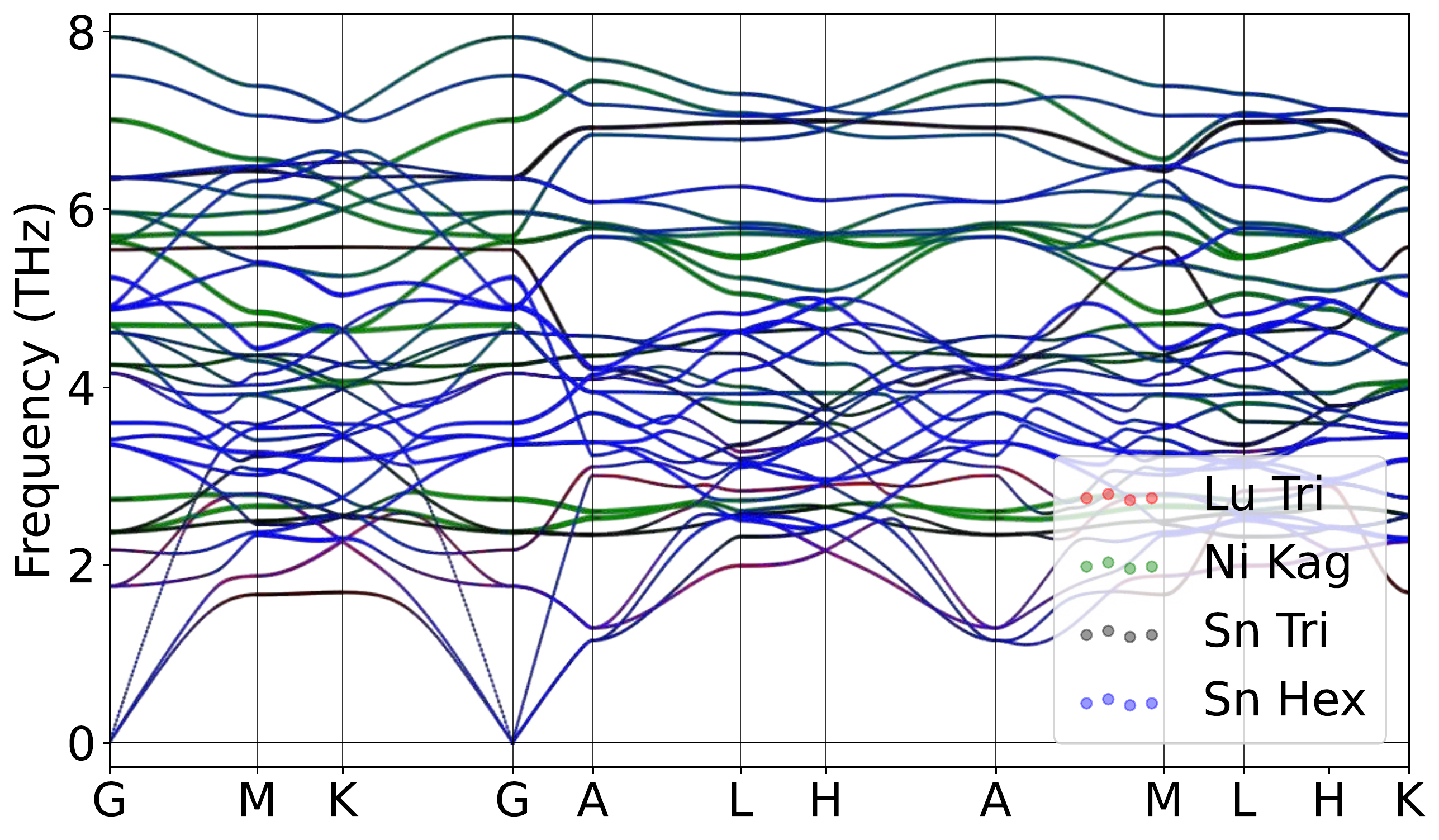}
}
\caption{Atomically projected phonon spectrum for LuNi$_6$Sn$_6$.}
\label{fig:LuNi6Sn6pho}
\end{figure}

\begin{figure}[H]
\centering
\label{fig:LuNi6Sn6_relaxed_Low_xz}
\includegraphics[width=0.85\textwidth]{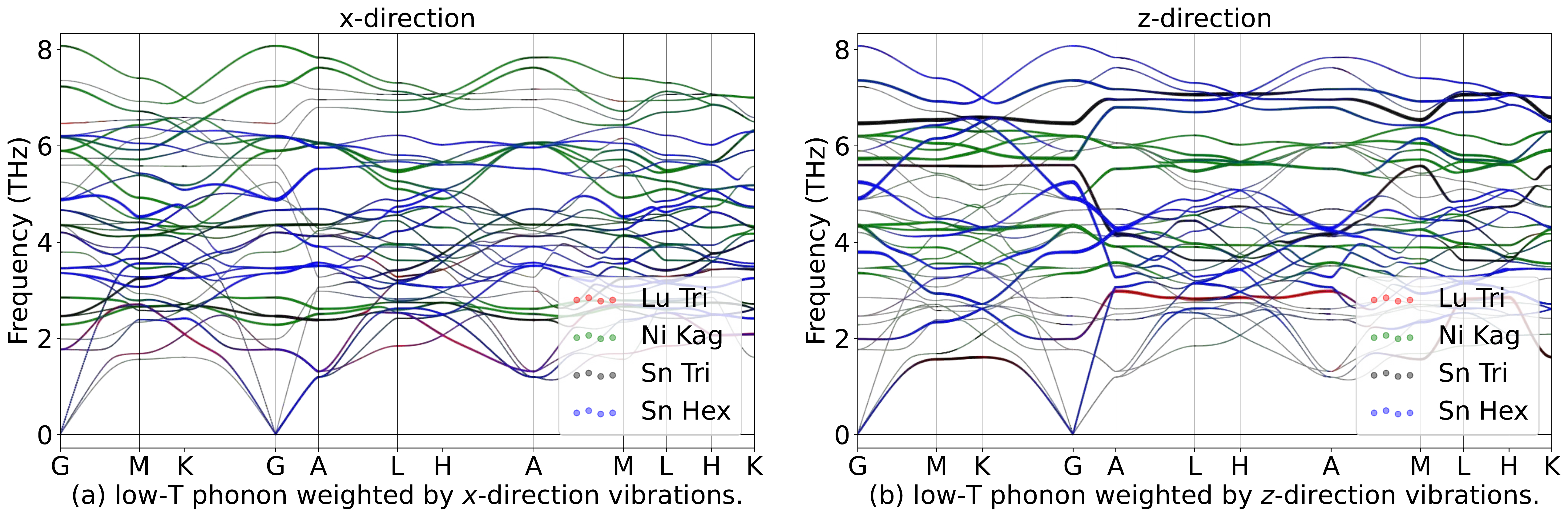}
\hspace{5pt}\label{fig:LuNi6Sn6_relaxed_High_xz}
\includegraphics[width=0.85\textwidth]{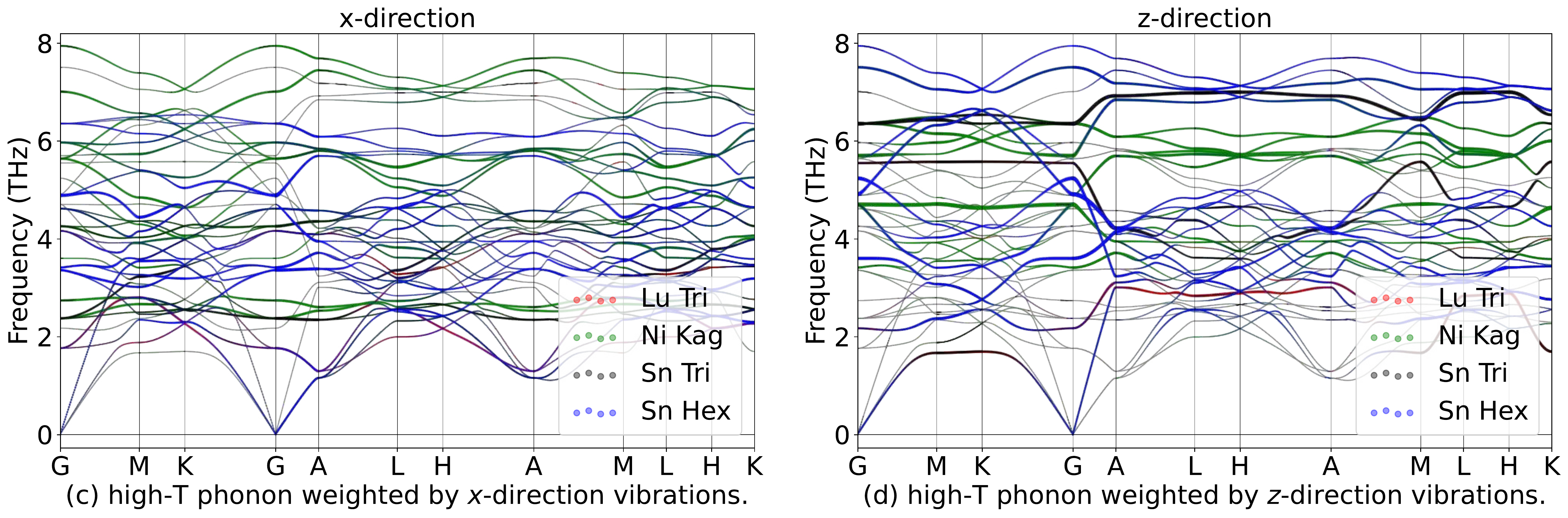}
\caption{Phonon spectrum for LuNi$_6$Sn$_6$in in-plane ($x$) and out-of-plane ($z$) directions.}
\label{fig:LuNi6Sn6phoxyz}
\end{figure}

\begin{table}[htbp]
\global\long\def\arraystretch{1.12}
\captionsetup{width=.95\textwidth}
\caption{IRREPs at high-symmetry points for LuNi$_6$Sn$_6$ phonon spectrum at Low-T.}
\label{tab:191_LuNi6Sn6_Low}
\setlength{\tabcolsep}{1.mm}{
}
\end{table}

\begin{table}[htbp]
\global\long\def\arraystretch{1.12}
\captionsetup{width=.95\textwidth}
\caption{IRREPs at high-symmetry points for LuNi$_6$Sn$_6$ phonon spectrum at High-T.}
\label{tab:191_LuNi6Sn6_High}
\setlength{\tabcolsep}{1.mm}{
%
}
\end{table}
\subsubsection{\label{sms2sec:HfNi6Sn6}\hyperref[smsec:col]{\texorpdfstring{HfNi$_6$Sn$_6$}{}}~{\color{green}  (High-T stable, Low-T stable) }}

\begin{table}[htbp]
\global\long\def\arraystretch{1.12}
\captionsetup{width=.85\textwidth}
\caption{Structure parameters of HfNi$_6$Sn$_6$ after relaxation. It contains 540 electrons. }
\label{tab:HfNi6Sn6}
\setlength{\tabcolsep}{4mm}{
%
}
\end{table}

\begin{figure}[htbp]
\centering
\includegraphics[width=0.85\textwidth]{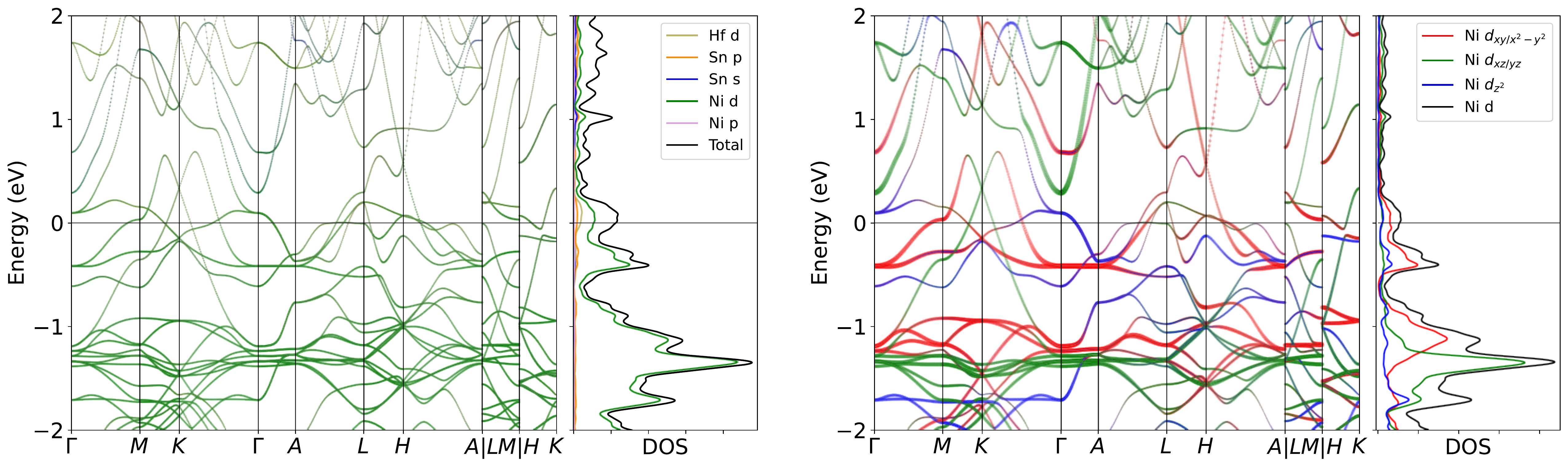}
\caption{Band structure for HfNi$_6$Sn$_6$ without SOC. The projection of Ni $3d$ orbitals is also presented. The band structures clearly reveal the dominating of Ni $3d$ orbitals  near the Fermi level, as well as the pronounced peak.}
\label{fig:HfNi6Sn6band}
\end{figure}

\begin{figure}[H]
\centering
\subfloat[Relaxed phonon at low-T.]{
\label{fig:HfNi6Sn6_relaxed_Low}
\includegraphics[width=0.45\textwidth]{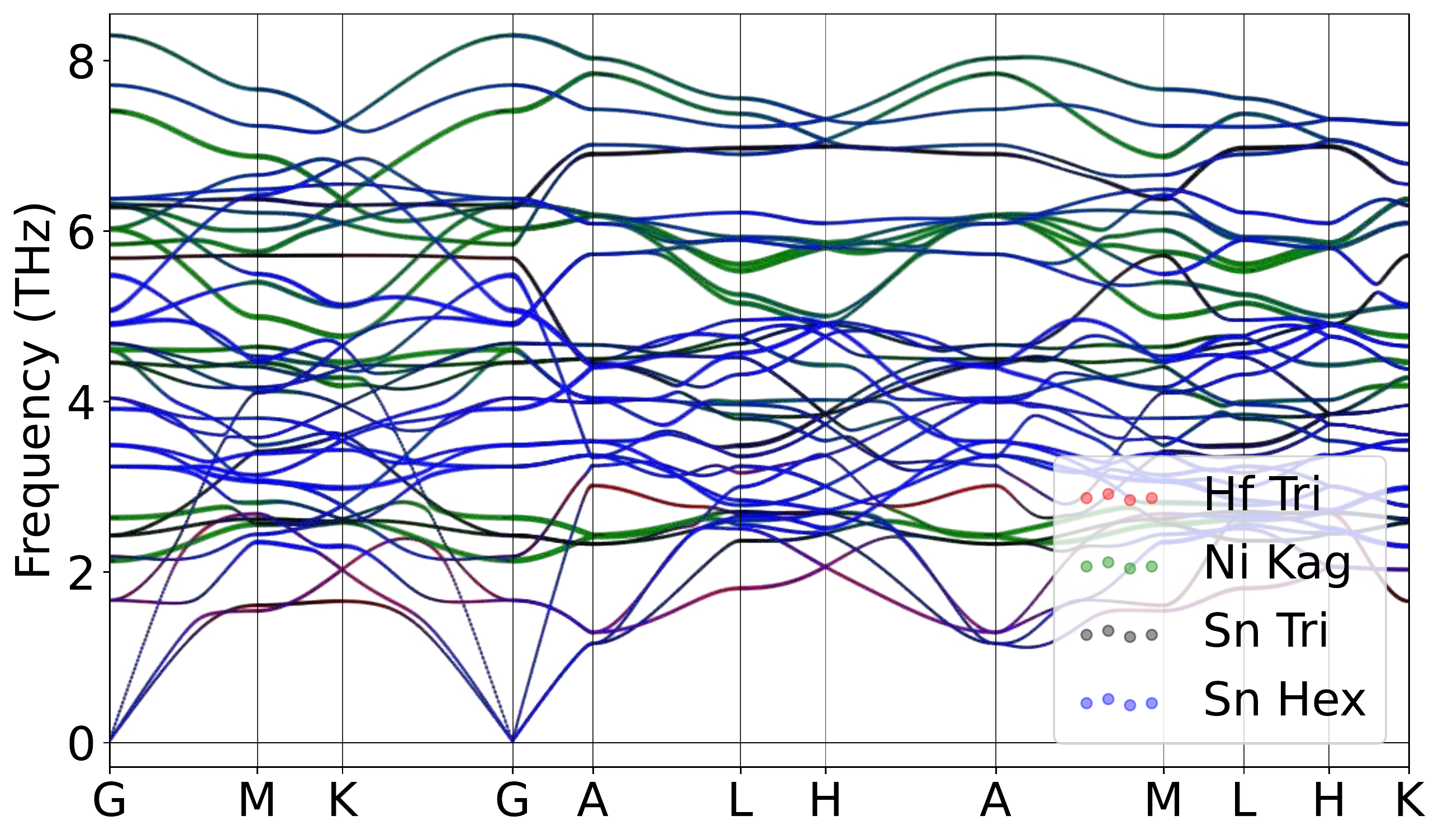}
}
\hspace{5pt}\subfloat[Relaxed phonon at high-T.]{
\label{fig:HfNi6Sn6_relaxed_High}
\includegraphics[width=0.45\textwidth]{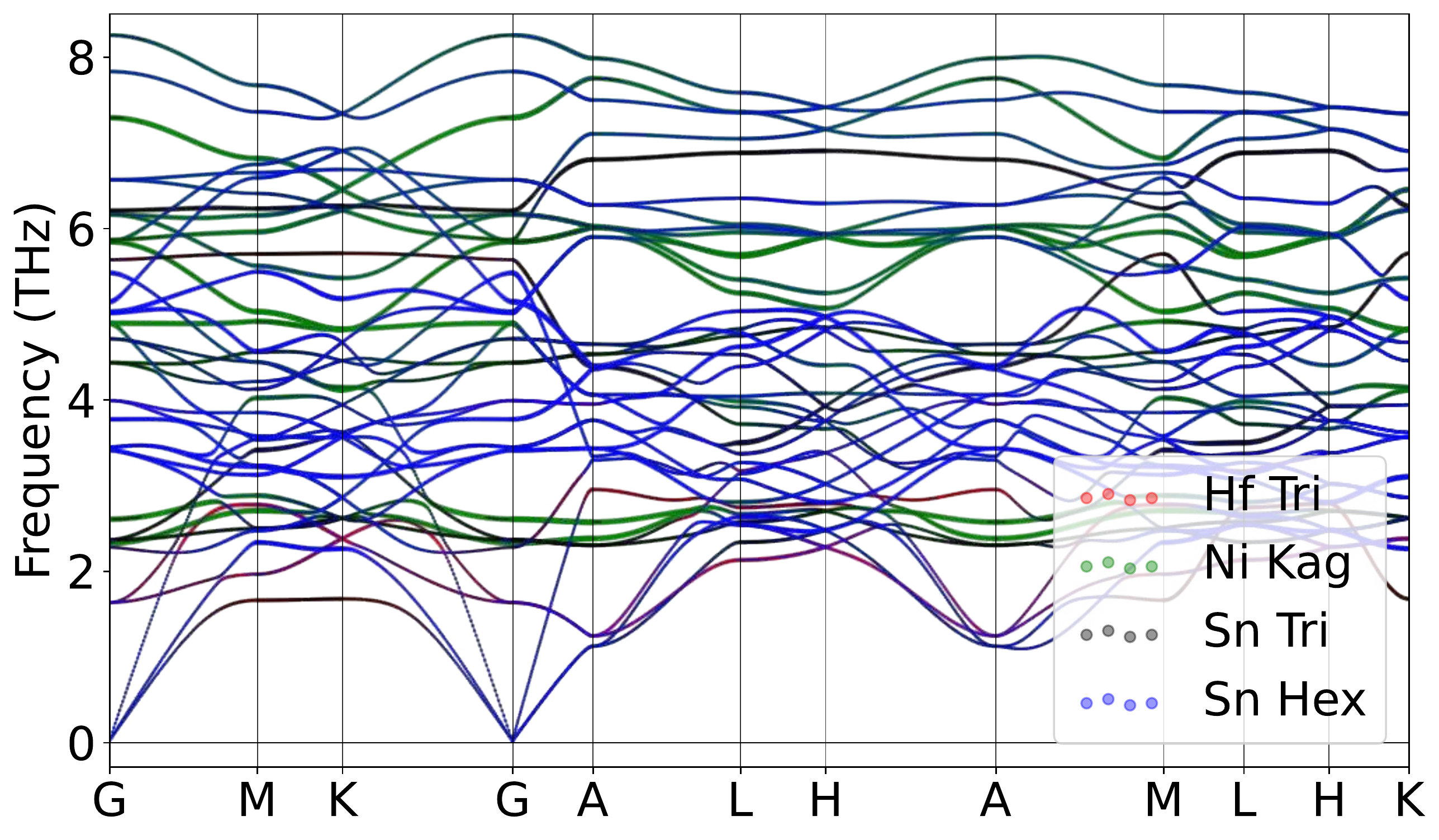}
}
\caption{Atomically projected phonon spectrum for HfNi$_6$Sn$_6$.}
\label{fig:HfNi6Sn6pho}
\end{figure}

\begin{figure}[H]
\centering
\label{fig:HfNi6Sn6_relaxed_Low_xz}
\includegraphics[width=0.85\textwidth]{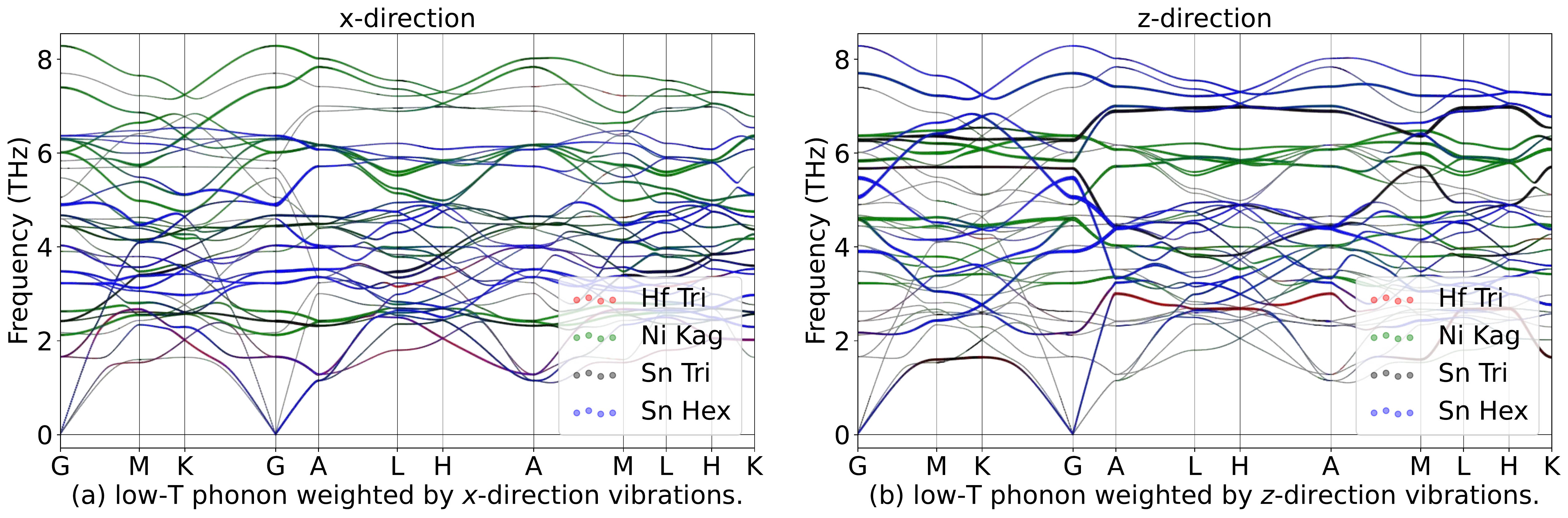}
\hspace{5pt}\label{fig:HfNi6Sn6_relaxed_High_xz}
\includegraphics[width=0.85\textwidth]{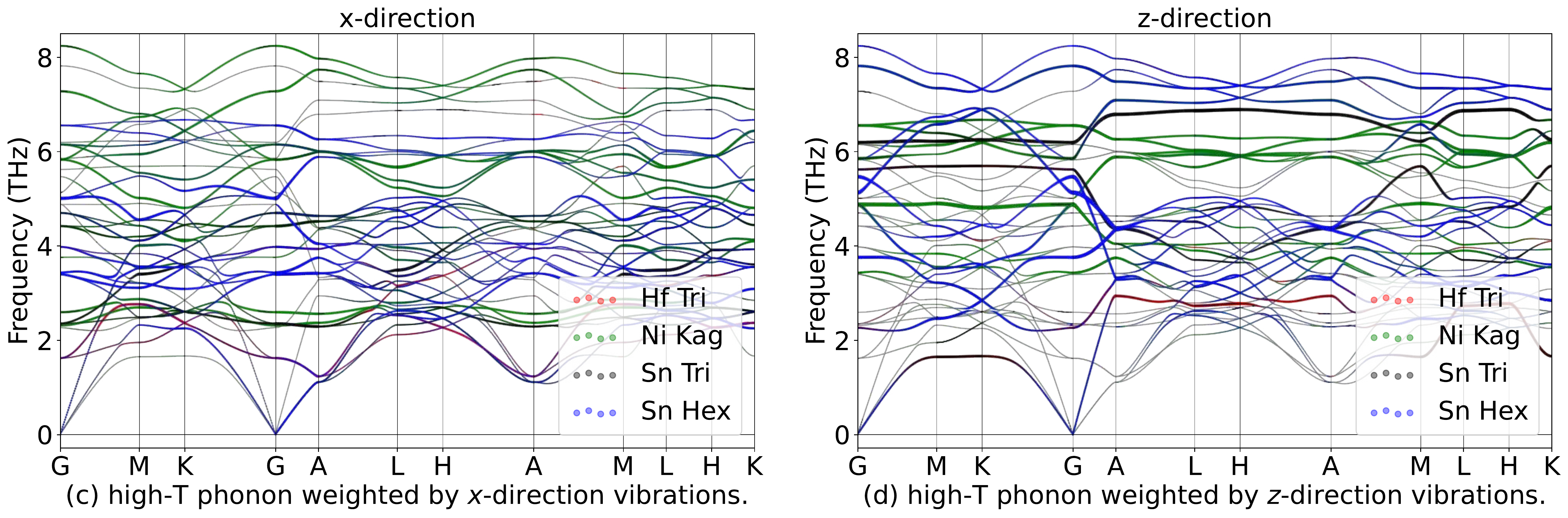}
\caption{Phonon spectrum for HfNi$_6$Sn$_6$in in-plane ($x$) and out-of-plane ($z$) directions.}
\label{fig:HfNi6Sn6phoxyz}
\end{figure}

\begin{table}[htbp]
\global\long\def\arraystretch{1.12}
\captionsetup{width=.95\textwidth}
\caption{IRREPs at high-symmetry points for HfNi$_6$Sn$_6$ phonon spectrum at Low-T.}
\label{tab:191_HfNi6Sn6_Low}
\setlength{\tabcolsep}{1.mm}{
}
\end{table}

\begin{table}[htbp]
\global\long\def\arraystretch{1.12}
\captionsetup{width=.95\textwidth}
\caption{IRREPs at high-symmetry points for HfNi$_6$Sn$_6$ phonon spectrum at High-T.}
\label{tab:191_HfNi6Sn6_High}
\setlength{\tabcolsep}{1.mm}{
%
}
\end{table}
\subsubsection{\label{sms2sec:TaNi6Sn6}\hyperref[smsec:col]{\texorpdfstring{TaNi$_6$Sn$_6$}{}}~{\color{green}  (High-T stable, Low-T stable) }}

\begin{table}[htbp]
\global\long\def\arraystretch{1.12}
\captionsetup{width=.85\textwidth}
\caption{Structure parameters of TaNi$_6$Sn$_6$ after relaxation. It contains 541 electrons. }
\label{tab:TaNi6Sn6}
\setlength{\tabcolsep}{4mm}{
%
}
\end{table}

\begin{figure}[htbp]
\centering
\includegraphics[width=0.85\textwidth]{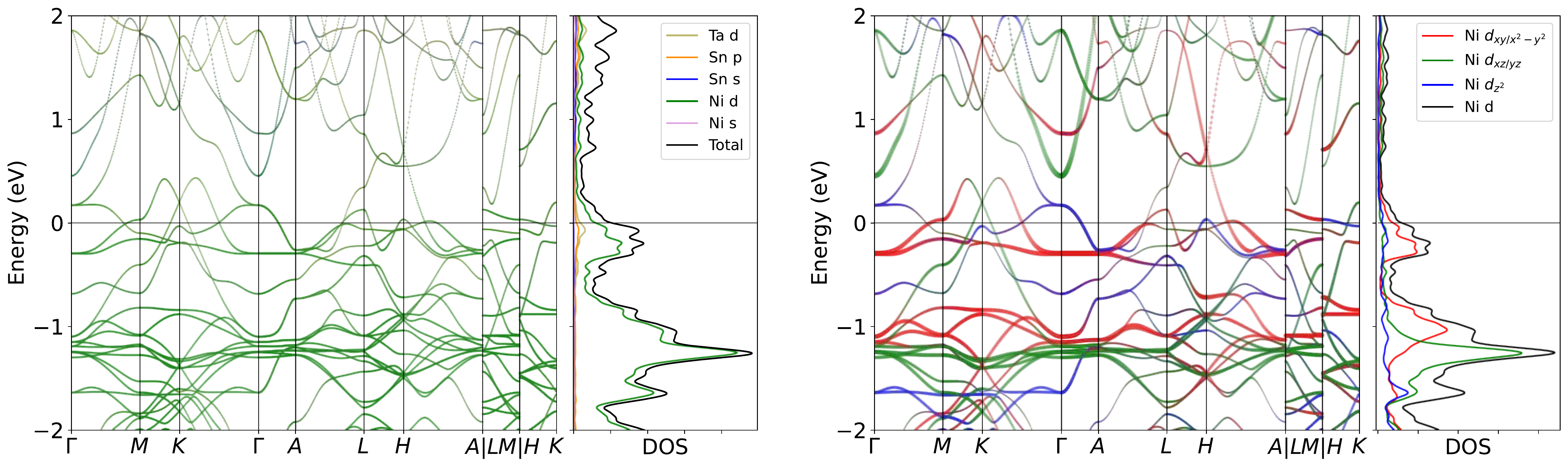}
\caption{Band structure for TaNi$_6$Sn$_6$ without SOC. The projection of Ni $3d$ orbitals is also presented. The band structures clearly reveal the dominating of Ni $3d$ orbitals  near the Fermi level, as well as the pronounced peak.}
\label{fig:TaNi6Sn6band}
\end{figure}

\begin{figure}[H]
\centering
\subfloat[Relaxed phonon at low-T.]{
\label{fig:TaNi6Sn6_relaxed_Low}
\includegraphics[width=0.45\textwidth]{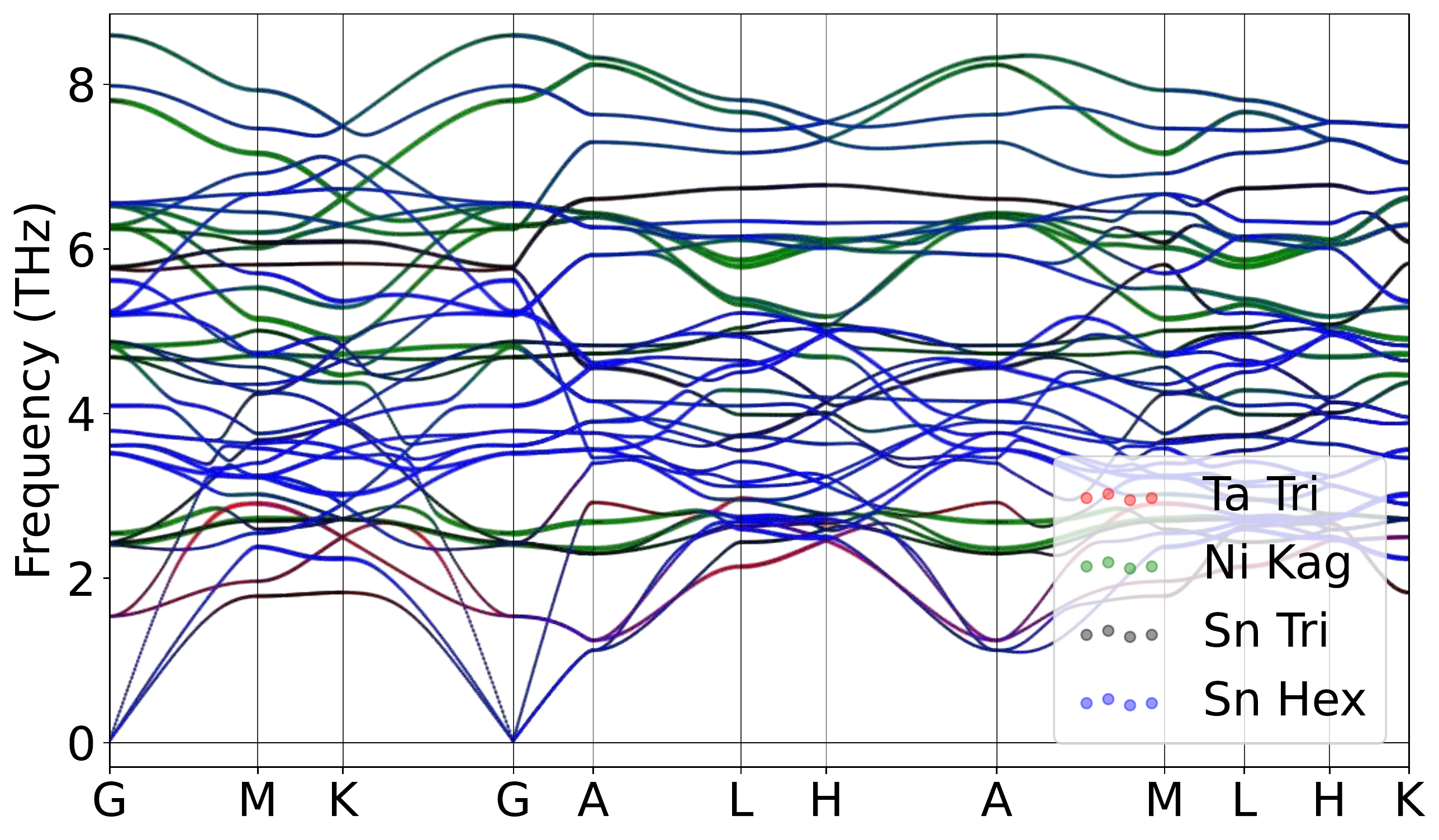}
}
\hspace{5pt}\subfloat[Relaxed phonon at high-T.]{
\label{fig:TaNi6Sn6_relaxed_High}
\includegraphics[width=0.45\textwidth]{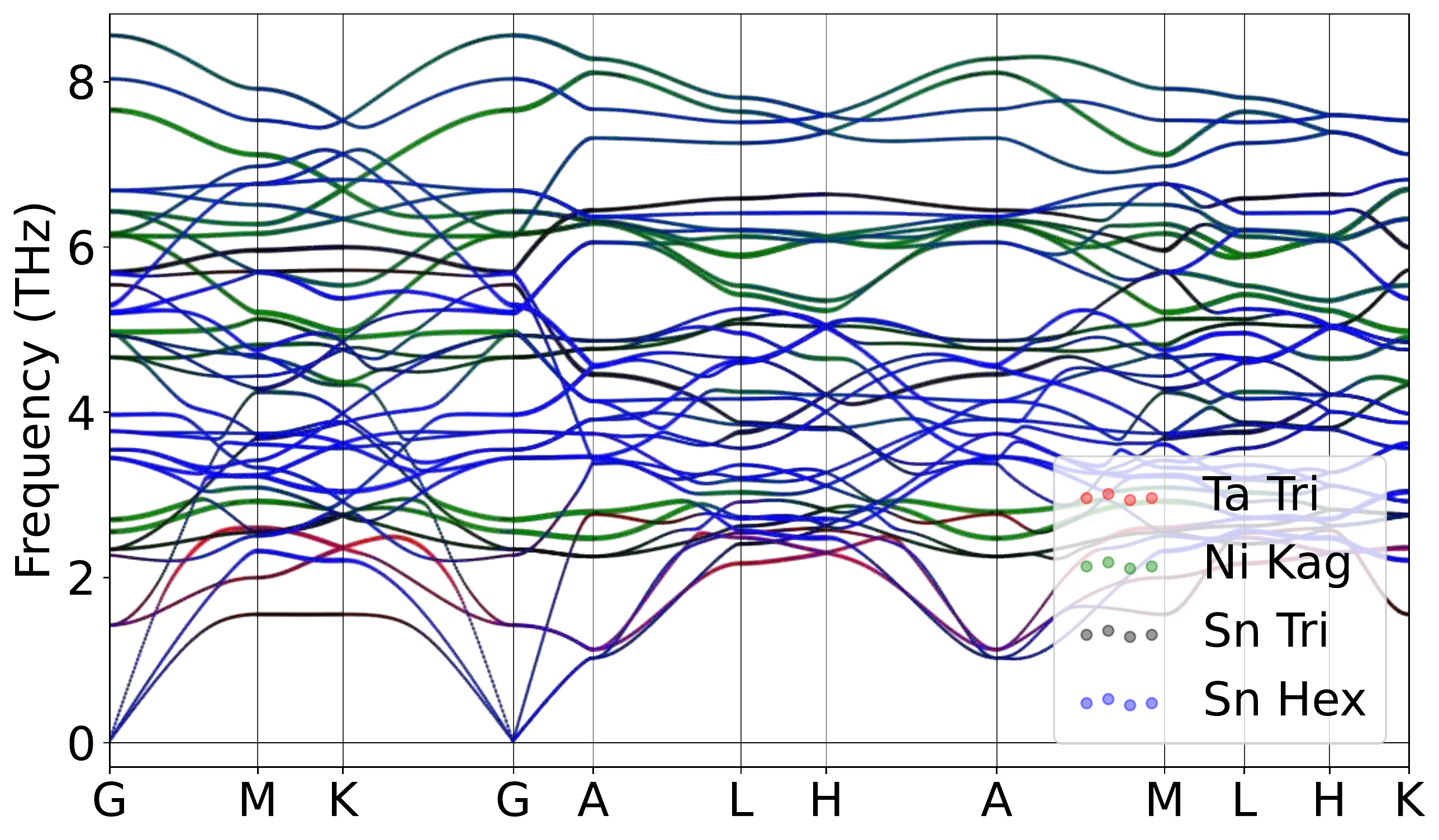}
}
\caption{Atomically projected phonon spectrum for TaNi$_6$Sn$_6$.}
\label{fig:TaNi6Sn6pho}
\end{figure}

\begin{figure}[H]
\centering
\label{fig:TaNi6Sn6_relaxed_Low_xz}
\includegraphics[width=0.85\textwidth]{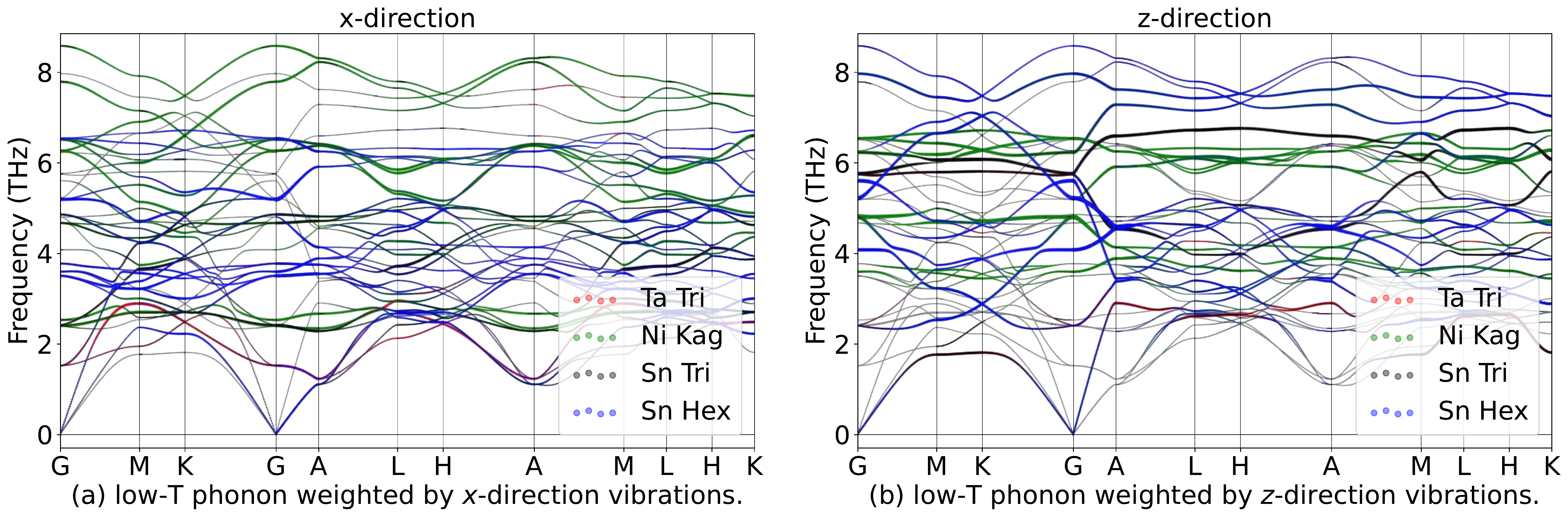}
\hspace{5pt}\label{fig:TaNi6Sn6_relaxed_High_xz}
\includegraphics[width=0.85\textwidth]{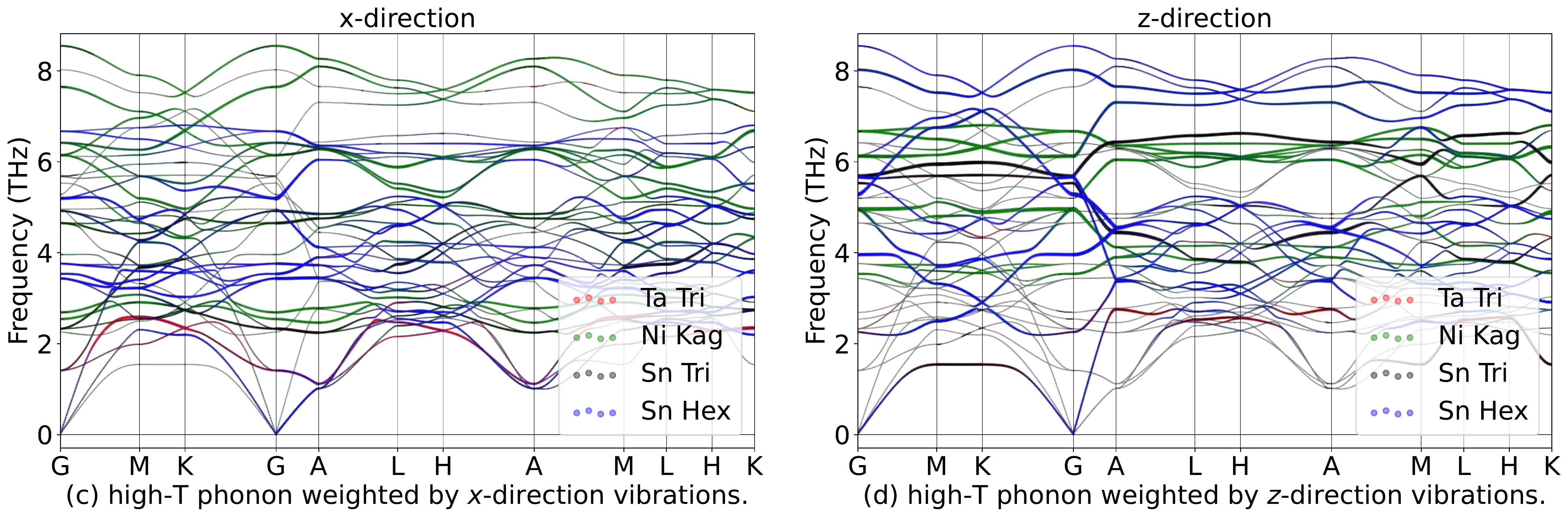}
\caption{Phonon spectrum for TaNi$_6$Sn$_6$in in-plane ($x$) and out-of-plane ($z$) directions.}
\label{fig:TaNi6Sn6phoxyz}
\end{figure}

\begin{table}[htbp]
\global\long\def\arraystretch{1.12}
\captionsetup{width=.95\textwidth}
\caption{IRREPs at high-symmetry points for TaNi$_6$Sn$_6$ phonon spectrum at Low-T.}
\label{tab:191_TaNi6Sn6_Low}
\setlength{\tabcolsep}{1.mm}{
}
\end{table}

\begin{table}[htbp]
\global\long\def\arraystretch{1.12}
\captionsetup{width=.95\textwidth}
\caption{IRREPs at high-symmetry points for TaNi$_6$Sn$_6$ phonon spectrum at High-T.}
\label{tab:191_TaNi6Sn6_High}
\setlength{\tabcolsep}{1.mm}{
%
}
\end{table}
\subsection{\hyperref[smsec:col]{NbSn}}~\label{smssec:NbSn}

\subsubsection{\label{sms2sec:LiNb6Sn6}\hyperref[smsec:col]{\texorpdfstring{LiNb$_6$Sn$_6$}{}}~{\color{red}  (High-T unstable, Low-T unstable) }}

\begin{table}[htbp]
\global\long\def\arraystretch{1.12}
\captionsetup{width=.85\textwidth}
\caption{Structure parameters of LiNb$_6$Sn$_6$ after relaxation. It contains 549 electrons. }
\label{tab:LiNb6Sn6}
\setlength{\tabcolsep}{4mm}{
%
}
\end{table}

\begin{figure}[htbp]
\centering
\includegraphics[width=0.85\textwidth]{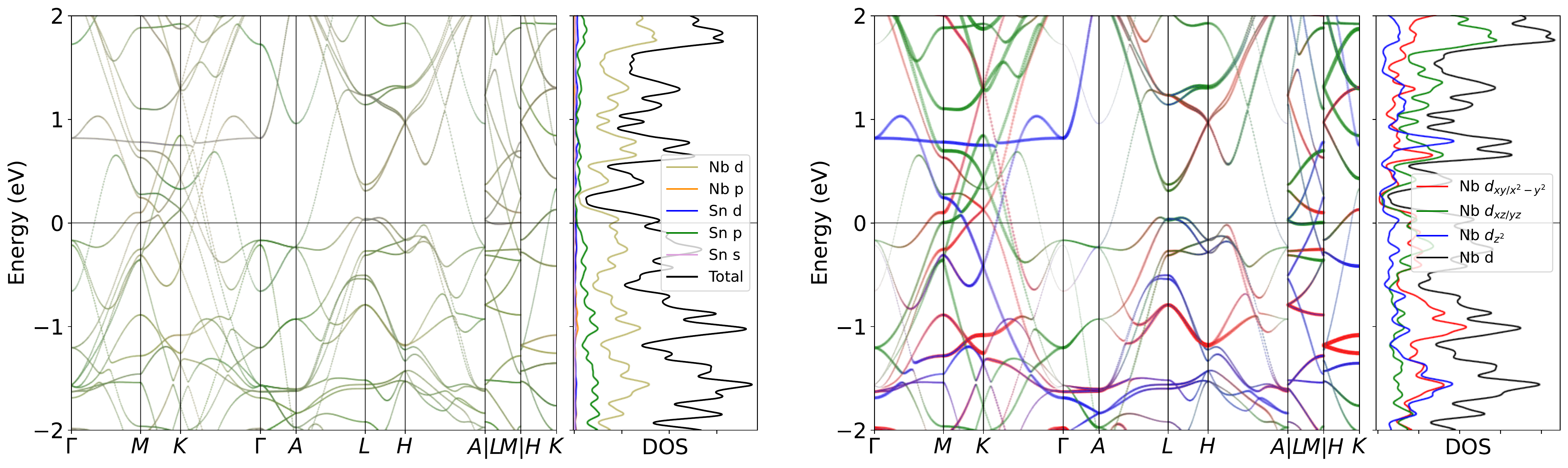}
\caption{Band structure for LiNb$_6$Sn$_6$ without SOC. The projection of Nb $3d$ orbitals is also presented. The band structures clearly reveal the dominating of Nb $3d$ orbitals  near the Fermi level, as well as the pronounced peak.}
\label{fig:LiNb6Sn6band}
\end{figure}

\begin{figure}[H]
\centering
\subfloat[Relaxed phonon at low-T.]{
\label{fig:LiNb6Sn6_relaxed_Low}
\includegraphics[width=0.45\textwidth]{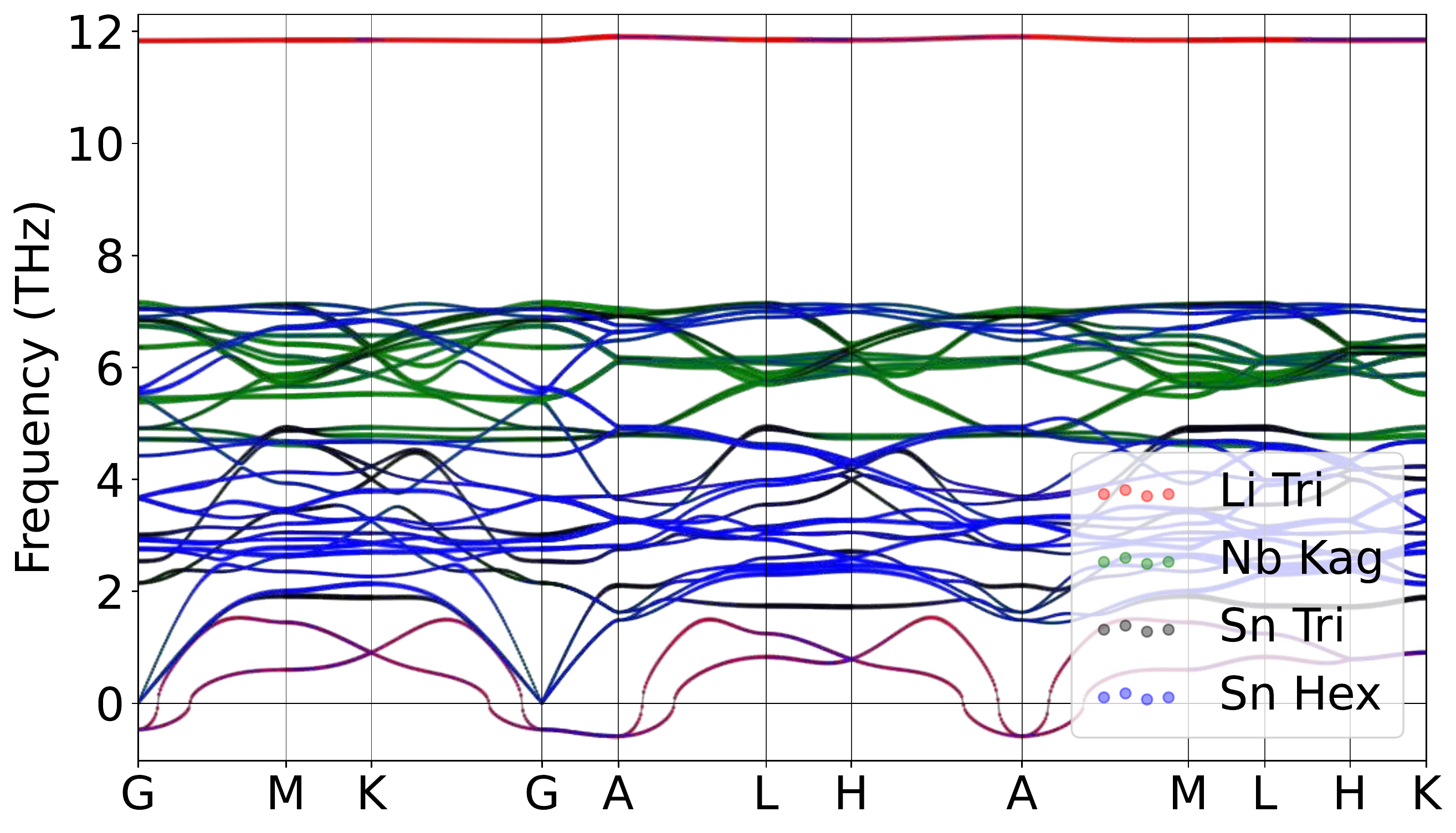}
}
\hspace{5pt}\subfloat[Relaxed phonon at high-T.]{
\label{fig:LiNb6Sn6_relaxed_High}
\includegraphics[width=0.45\textwidth]{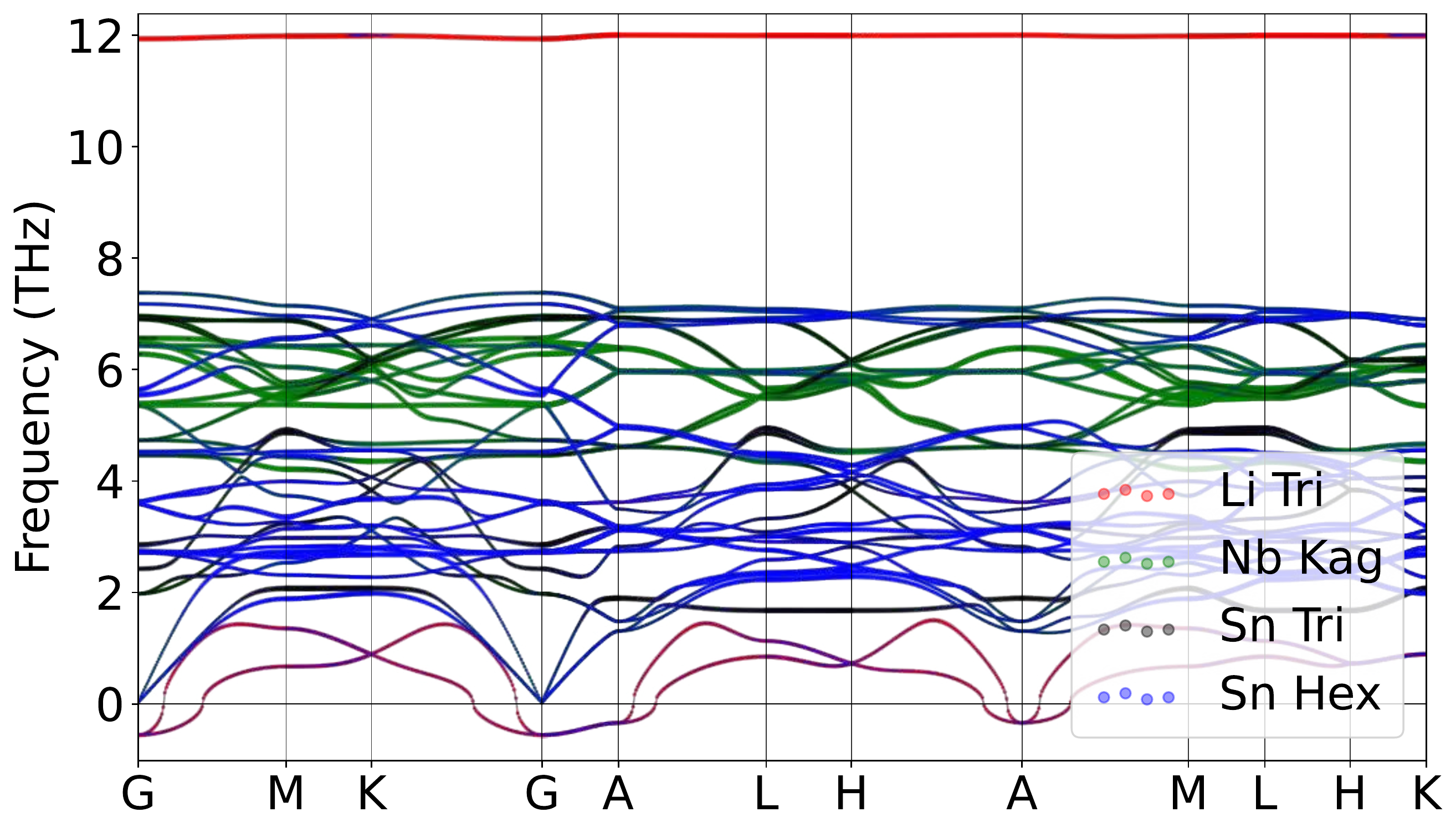}
}
\caption{Atomically projected phonon spectrum for LiNb$_6$Sn$_6$.}
\label{fig:LiNb6Sn6pho}
\end{figure}

\begin{figure}[H]
\centering
\label{fig:LiNb6Sn6_relaxed_Low_xz}
\includegraphics[width=0.85\textwidth]{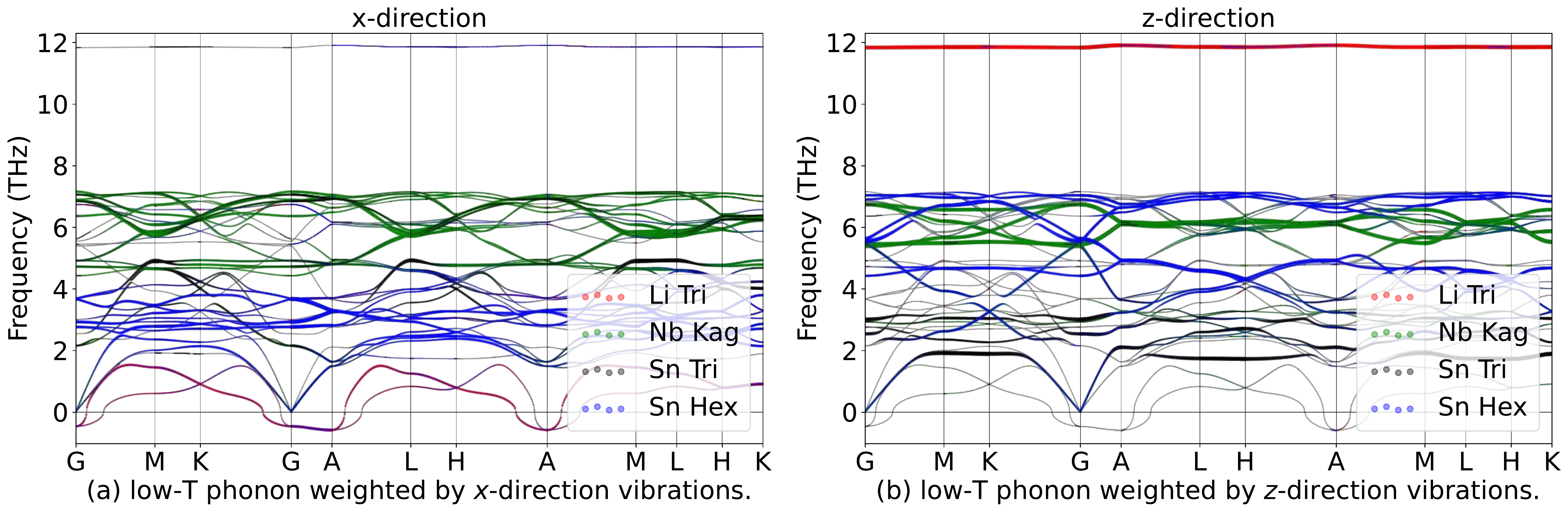}
\hspace{5pt}\label{fig:LiNb6Sn6_relaxed_High_xz}
\includegraphics[width=0.85\textwidth]{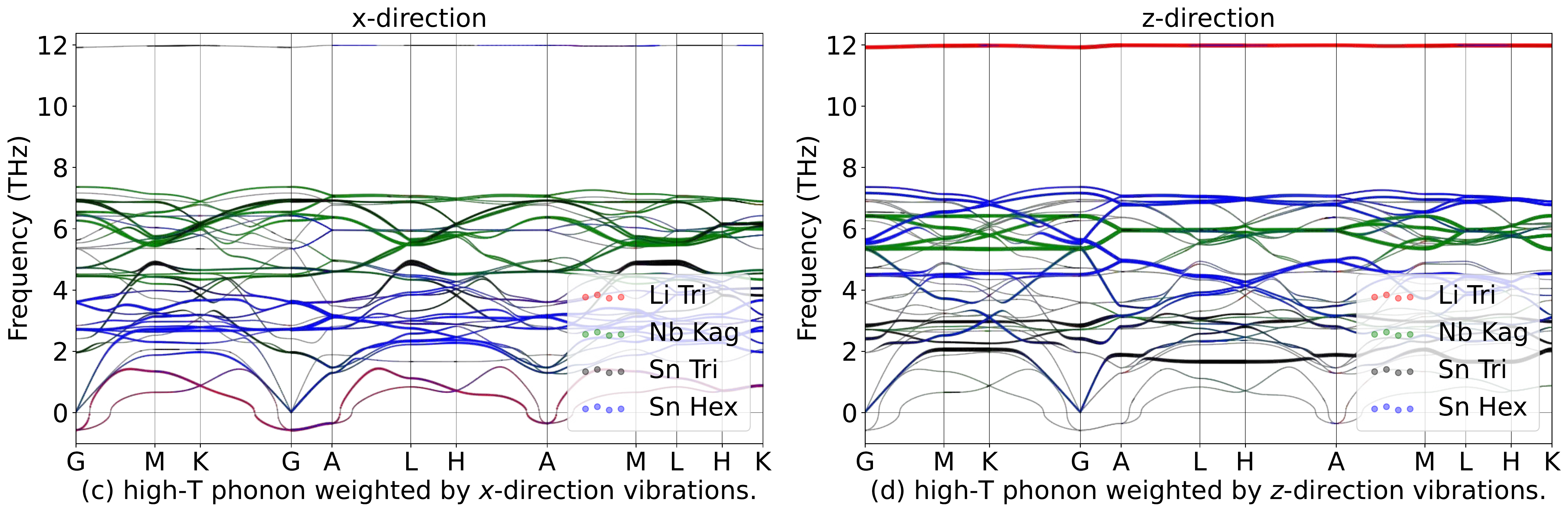}
\caption{Phonon spectrum for LiNb$_6$Sn$_6$in in-plane ($x$) and out-of-plane ($z$) directions.}
\label{fig:LiNb6Sn6phoxyz}
\end{figure}

\begin{table}[htbp]
\global\long\def\arraystretch{1.12}
\captionsetup{width=.95\textwidth}
\caption{IRREPs at high-symmetry points for LiNb$_6$Sn$_6$ phonon spectrum at Low-T.}
\label{tab:191_LiNb6Sn6_Low}
\setlength{\tabcolsep}{1.mm}{
}
\end{table}

\begin{table}[htbp]
\global\long\def\arraystretch{1.12}
\captionsetup{width=.95\textwidth}
\caption{IRREPs at high-symmetry points for LiNb$_6$Sn$_6$ phonon spectrum at High-T.}
\label{tab:191_LiNb6Sn6_High}
\setlength{\tabcolsep}{1.mm}{
%
}
\end{table}
\subsubsection{\label{sms2sec:MgNb6Sn6}\hyperref[smsec:col]{\texorpdfstring{MgNb$_6$Sn$_6$}{}}~{\color{red}  (High-T unstable, Low-T unstable) }}

\begin{table}[htbp]
\global\long\def\arraystretch{1.12}
\captionsetup{width=.85\textwidth}
\caption{Structure parameters of MgNb$_6$Sn$_6$ after relaxation. It contains 558 electrons. }
\label{tab:MgNb6Sn6}
\setlength{\tabcolsep}{4mm}{
%
}
\end{table}

\begin{figure}[htbp]
\centering
\includegraphics[width=0.85\textwidth]{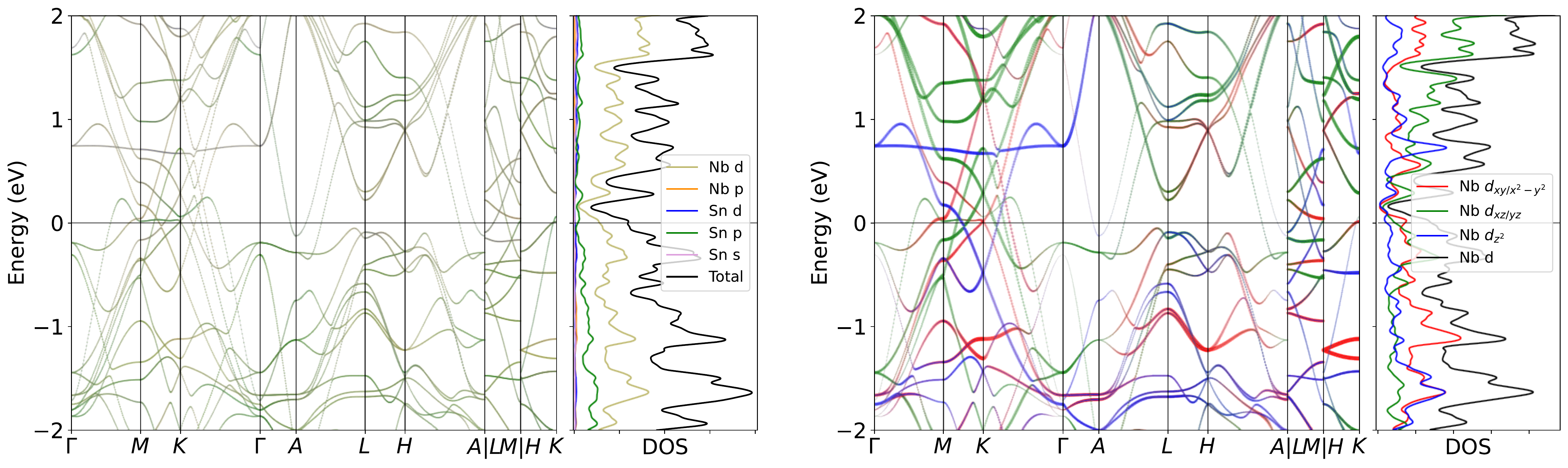}
\caption{Band structure for MgNb$_6$Sn$_6$ without SOC. The projection of Nb $3d$ orbitals is also presented. The band structures clearly reveal the dominating of Nb $3d$ orbitals  near the Fermi level, as well as the pronounced peak.}
\label{fig:MgNb6Sn6band}
\end{figure}

\begin{figure}[H]
\centering
\subfloat[Relaxed phonon at low-T.]{
\label{fig:MgNb6Sn6_relaxed_Low}
\includegraphics[width=0.45\textwidth]{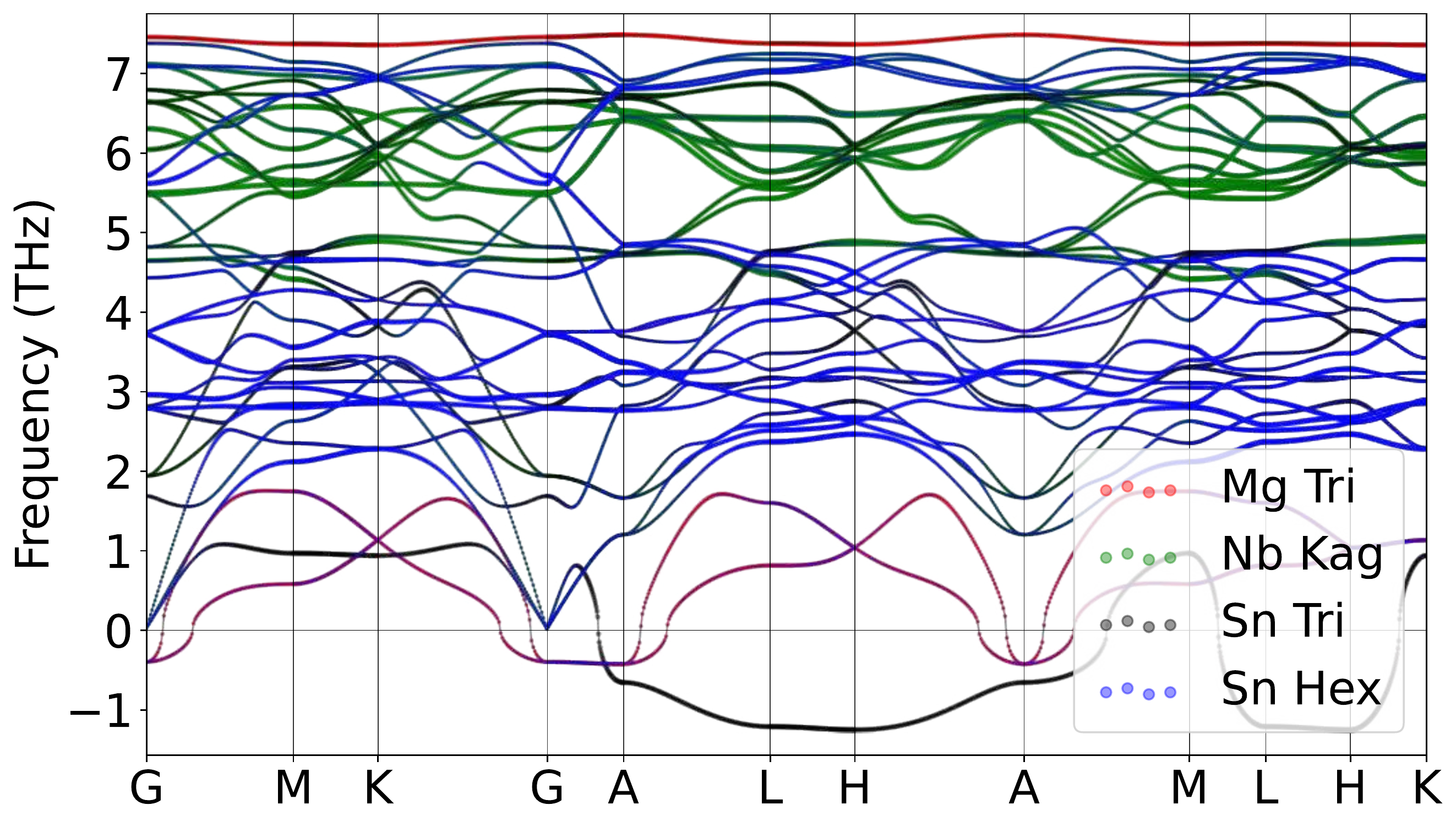}
}
\hspace{5pt}\subfloat[Relaxed phonon at high-T.]{
\label{fig:MgNb6Sn6_relaxed_High}
\includegraphics[width=0.45\textwidth]{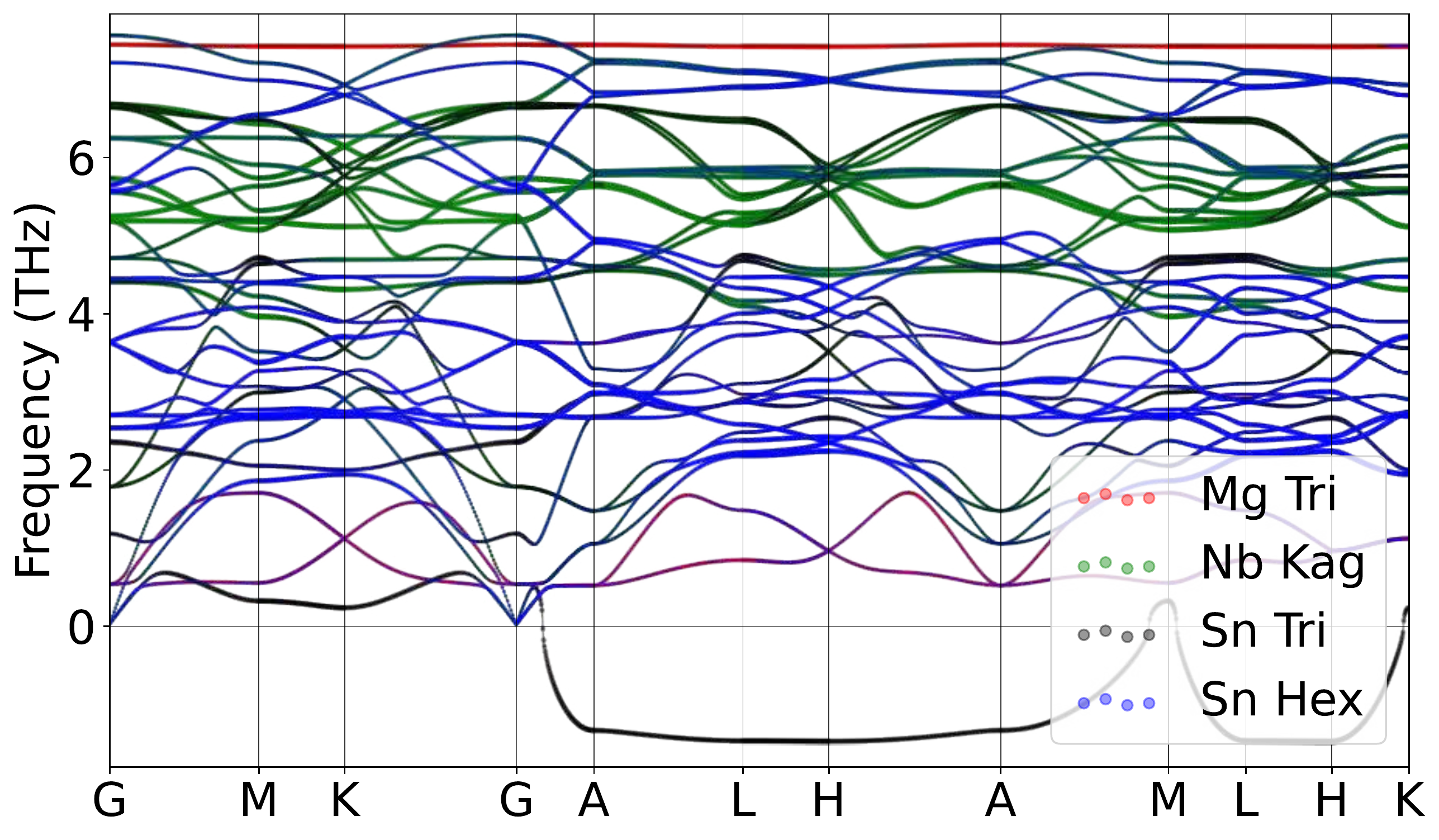}
}
\caption{Atomically projected phonon spectrum for MgNb$_6$Sn$_6$.}
\label{fig:MgNb6Sn6pho}
\end{figure}

\begin{figure}[H]
\centering
\label{fig:MgNb6Sn6_relaxed_Low_xz}
\includegraphics[width=0.85\textwidth]{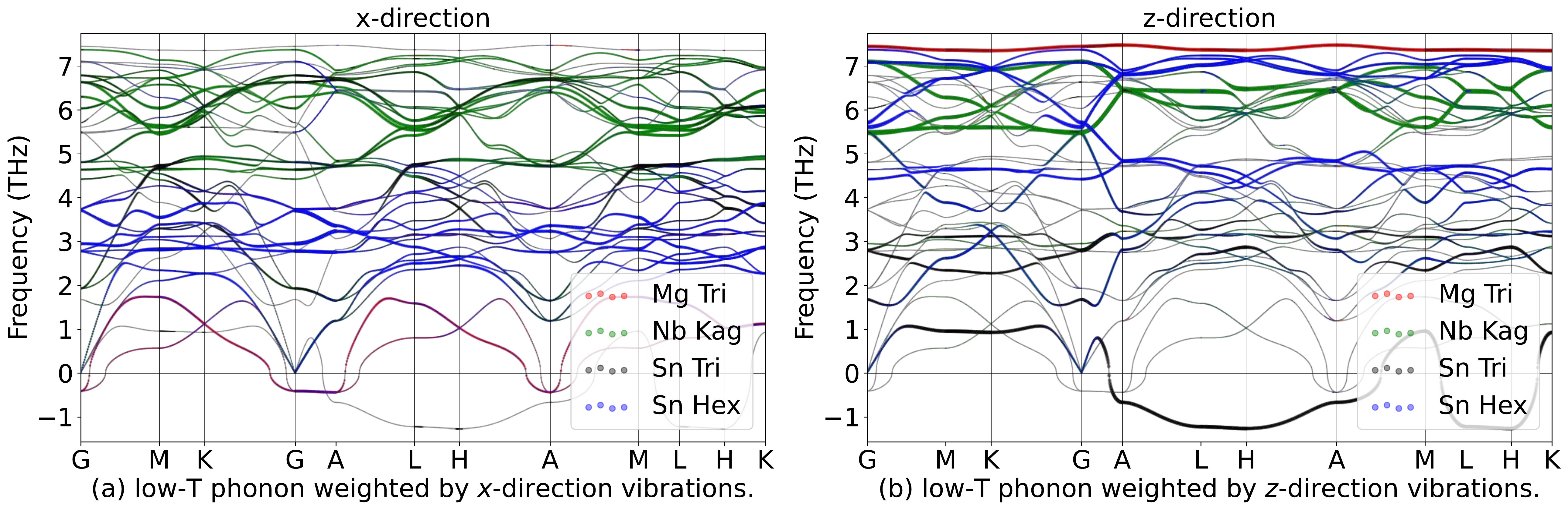}
\hspace{5pt}\label{fig:MgNb6Sn6_relaxed_High_xz}
\includegraphics[width=0.85\textwidth]{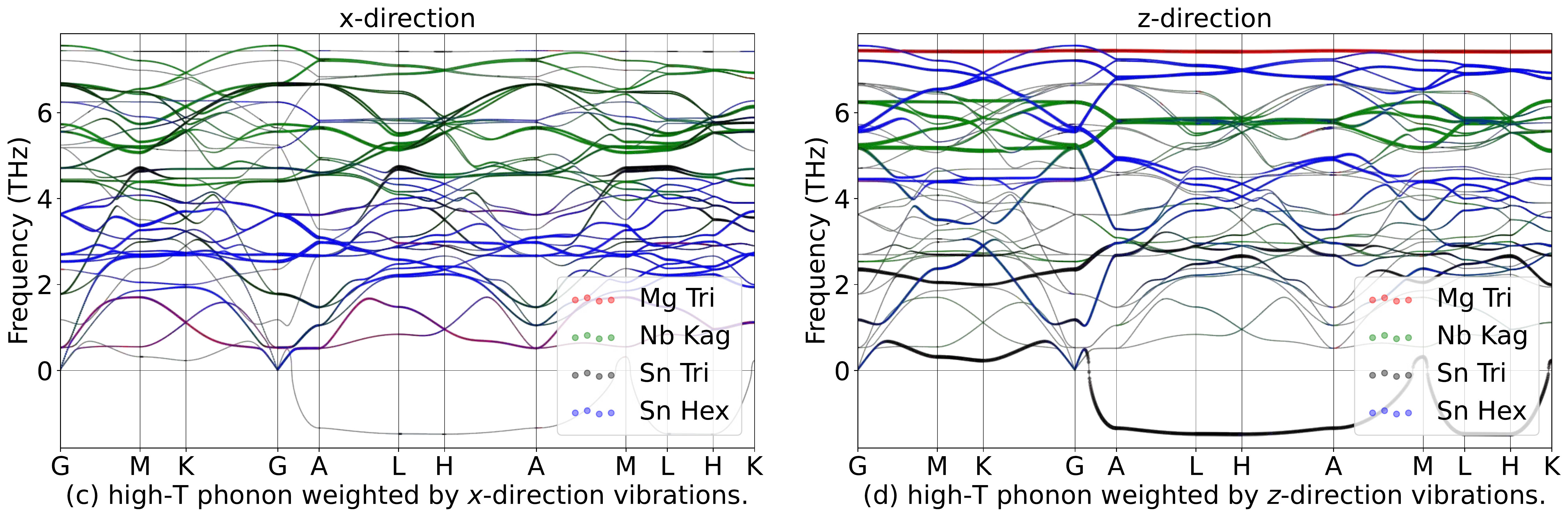}
\caption{Phonon spectrum for MgNb$_6$Sn$_6$in in-plane ($x$) and out-of-plane ($z$) directions.}
\label{fig:MgNb6Sn6phoxyz}
\end{figure}

\begin{table}[htbp]
\global\long\def\arraystretch{1.12}
\captionsetup{width=.95\textwidth}
\caption{IRREPs at high-symmetry points for MgNb$_6$Sn$_6$ phonon spectrum at Low-T.}
\label{tab:191_MgNb6Sn6_Low}
\setlength{\tabcolsep}{1.mm}{
}
\end{table}

\begin{table}[htbp]
\global\long\def\arraystretch{1.12}
\captionsetup{width=.95\textwidth}
\caption{IRREPs at high-symmetry points for MgNb$_6$Sn$_6$ phonon spectrum at High-T.}
\label{tab:191_MgNb6Sn6_High}
\setlength{\tabcolsep}{1.mm}{
%
}
\end{table}
\subsubsection{\label{sms2sec:CaNb6Sn6}\hyperref[smsec:col]{\texorpdfstring{CaNb$_6$Sn$_6$}{}}~{\color{blue}  (High-T stable, Low-T unstable) }}

\begin{table}[htbp]
\global\long\def\arraystretch{1.12}
\captionsetup{width=.85\textwidth}
\caption{Structure parameters of CaNb$_6$Sn$_6$ after relaxation. It contains 566 electrons. }
\label{tab:CaNb6Sn6}
\setlength{\tabcolsep}{4mm}{
%
}
\end{table}

\begin{figure}[htbp]
\centering
\includegraphics[width=0.85\textwidth]{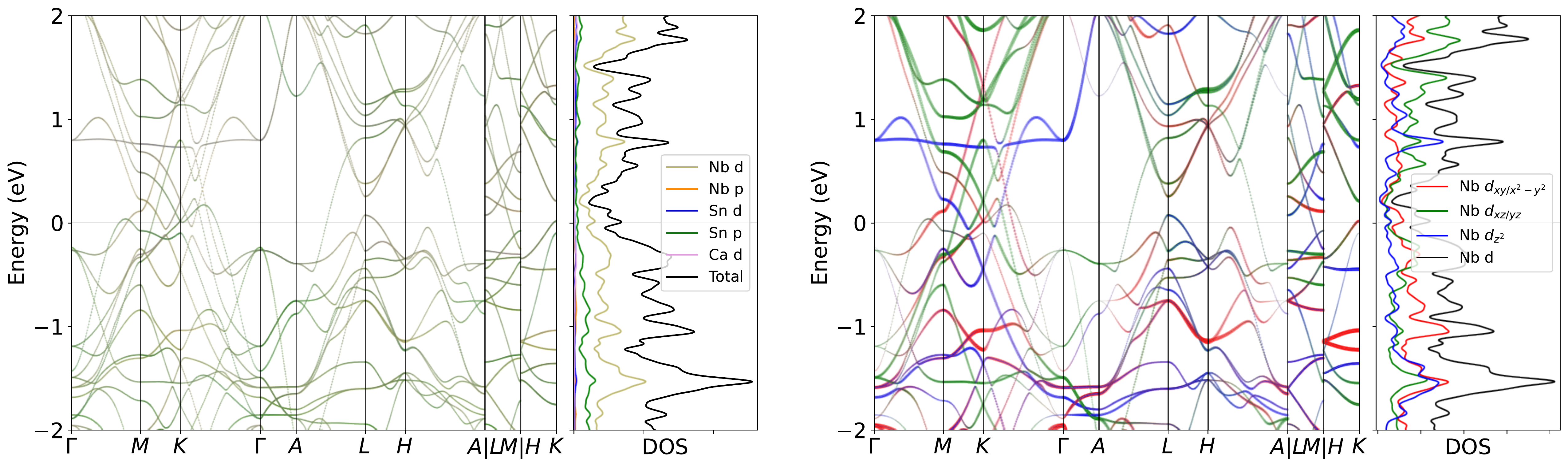}
\caption{Band structure for CaNb$_6$Sn$_6$ without SOC. The projection of Nb $3d$ orbitals is also presented. The band structures clearly reveal the dominating of Nb $3d$ orbitals  near the Fermi level, as well as the pronounced peak.}
\label{fig:CaNb6Sn6band}
\end{figure}

\begin{figure}[H]
\centering
\subfloat[Relaxed phonon at low-T.]{
\label{fig:CaNb6Sn6_relaxed_Low}
\includegraphics[width=0.45\textwidth]{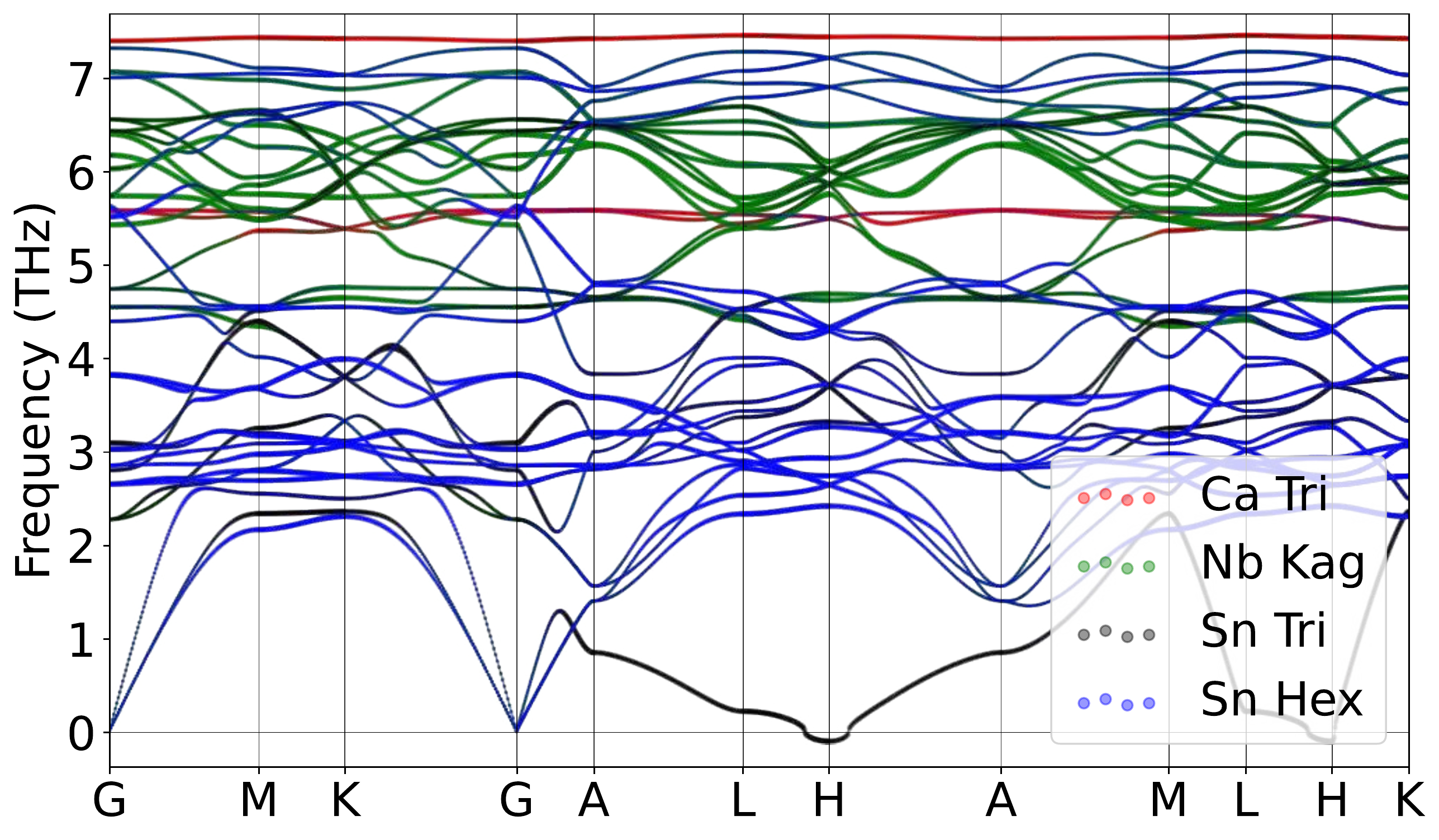}
}
\hspace{5pt}\subfloat[Relaxed phonon at high-T.]{
\label{fig:CaNb6Sn6_relaxed_High}
\includegraphics[width=0.45\textwidth]{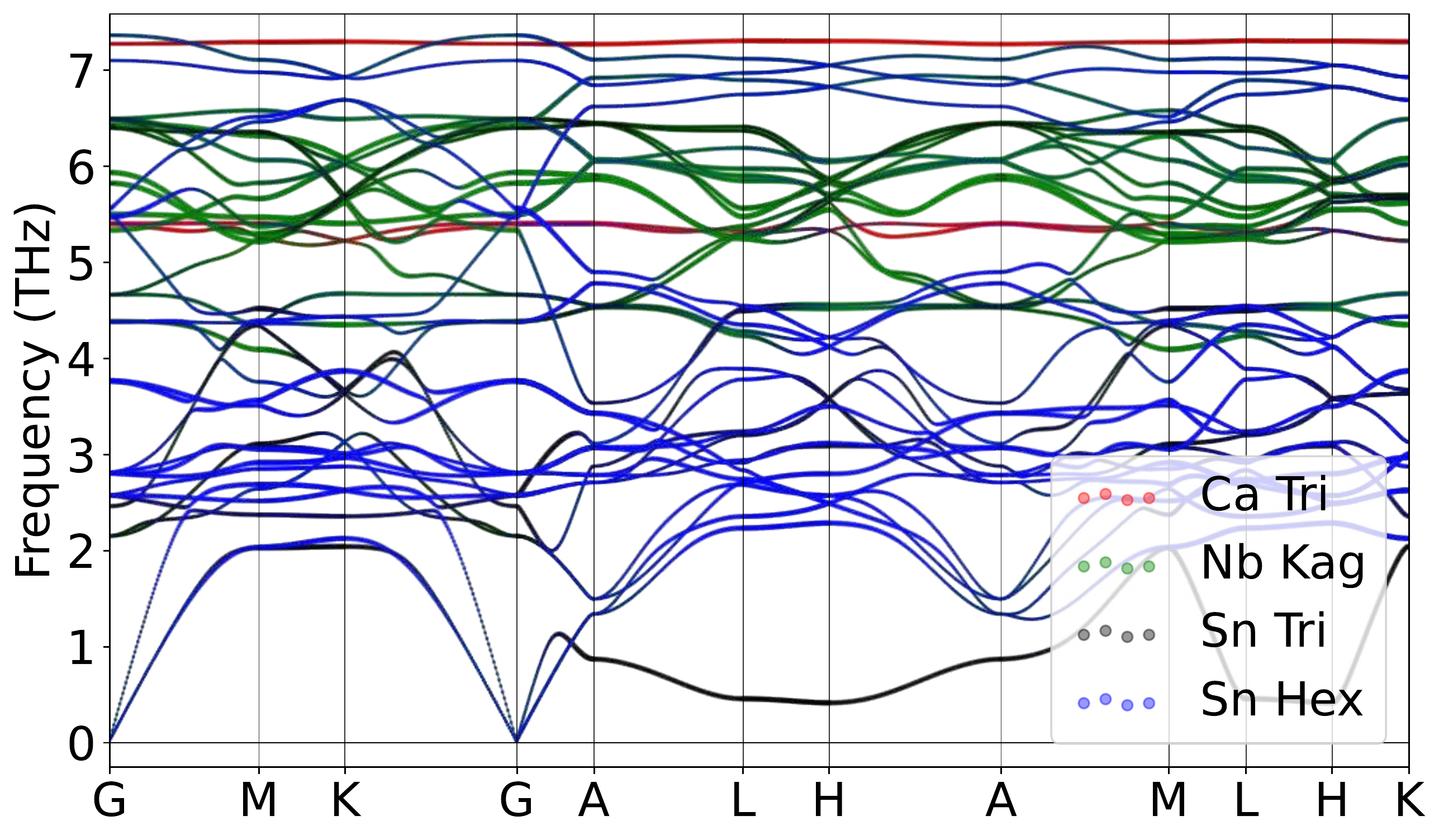}
}
\caption{Atomically projected phonon spectrum for CaNb$_6$Sn$_6$.}
\label{fig:CaNb6Sn6pho}
\end{figure}

\begin{figure}[H]
\centering
\label{fig:CaNb6Sn6_relaxed_Low_xz}
\includegraphics[width=0.85\textwidth]{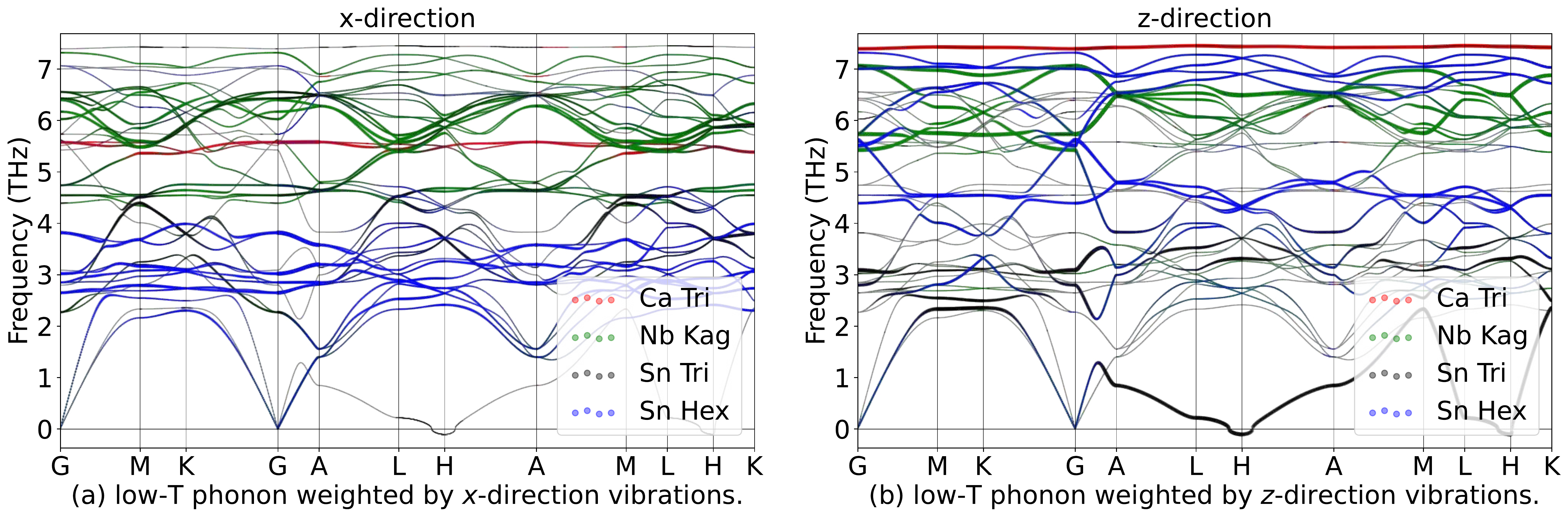}
\hspace{5pt}\label{fig:CaNb6Sn6_relaxed_High_xz}
\includegraphics[width=0.85\textwidth]{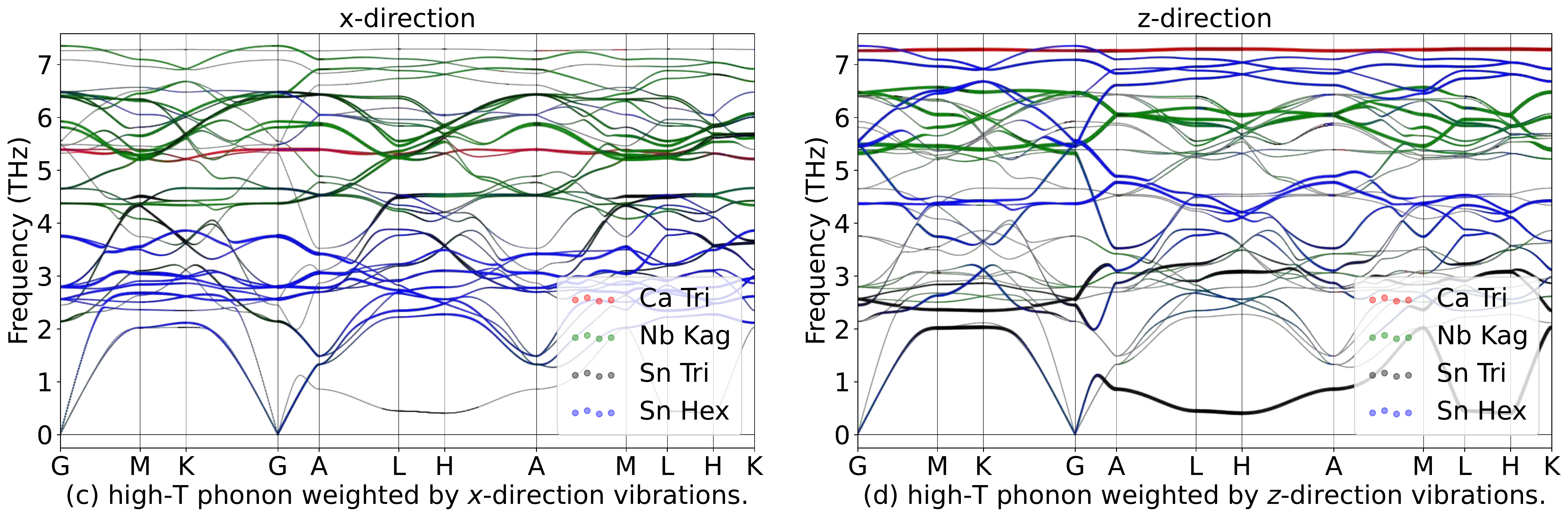}
\caption{Phonon spectrum for CaNb$_6$Sn$_6$in in-plane ($x$) and out-of-plane ($z$) directions.}
\label{fig:CaNb6Sn6phoxyz}
\end{figure}

\begin{table}[htbp]
\global\long\def\arraystretch{1.12}
\captionsetup{width=.95\textwidth}
\caption{IRREPs at high-symmetry points for CaNb$_6$Sn$_6$ phonon spectrum at Low-T.}
\label{tab:191_CaNb6Sn6_Low}
\setlength{\tabcolsep}{1.mm}{
}
\end{table}

\begin{table}[htbp]
\global\long\def\arraystretch{1.12}
\captionsetup{width=.95\textwidth}
\caption{IRREPs at high-symmetry points for CaNb$_6$Sn$_6$ phonon spectrum at High-T.}
\label{tab:191_CaNb6Sn6_High}
\setlength{\tabcolsep}{1.mm}{
%
}
\end{table}
\subsubsection{\label{sms2sec:ScNb6Sn6}\hyperref[smsec:col]{\texorpdfstring{ScNb$_6$Sn$_6$}{}}~{\color{blue}  (High-T stable, Low-T unstable) }}

\begin{table}[htbp]
\global\long\def\arraystretch{1.12}
\captionsetup{width=.85\textwidth}
\caption{Structure parameters of ScNb$_6$Sn$_6$ after relaxation. It contains 567 electrons. }
\label{tab:ScNb6Sn6}
\setlength{\tabcolsep}{4mm}{
%
}
\end{table}

\begin{figure}[htbp]
\centering
\includegraphics[width=0.85\textwidth]{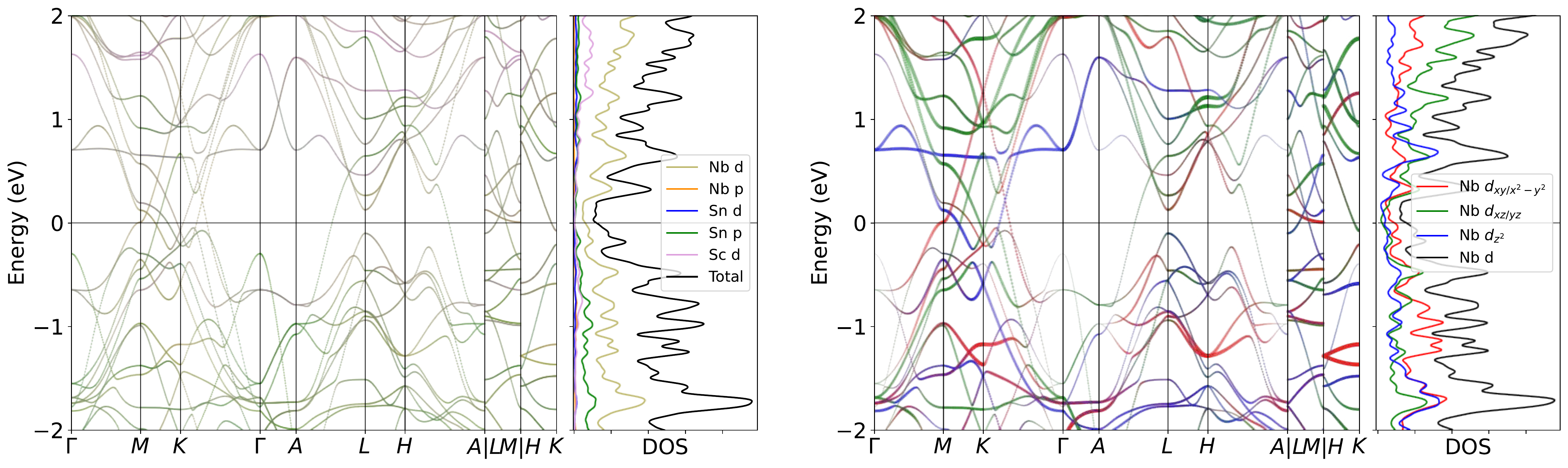}
\caption{Band structure for ScNb$_6$Sn$_6$ without SOC. The projection of Nb $3d$ orbitals is also presented. The band structures clearly reveal the dominating of Nb $3d$ orbitals  near the Fermi level, as well as the pronounced peak.}
\label{fig:ScNb6Sn6band}
\end{figure}

\begin{figure}[H]
\centering
\subfloat[Relaxed phonon at low-T.]{
\label{fig:ScNb6Sn6_relaxed_Low}
\includegraphics[width=0.45\textwidth]{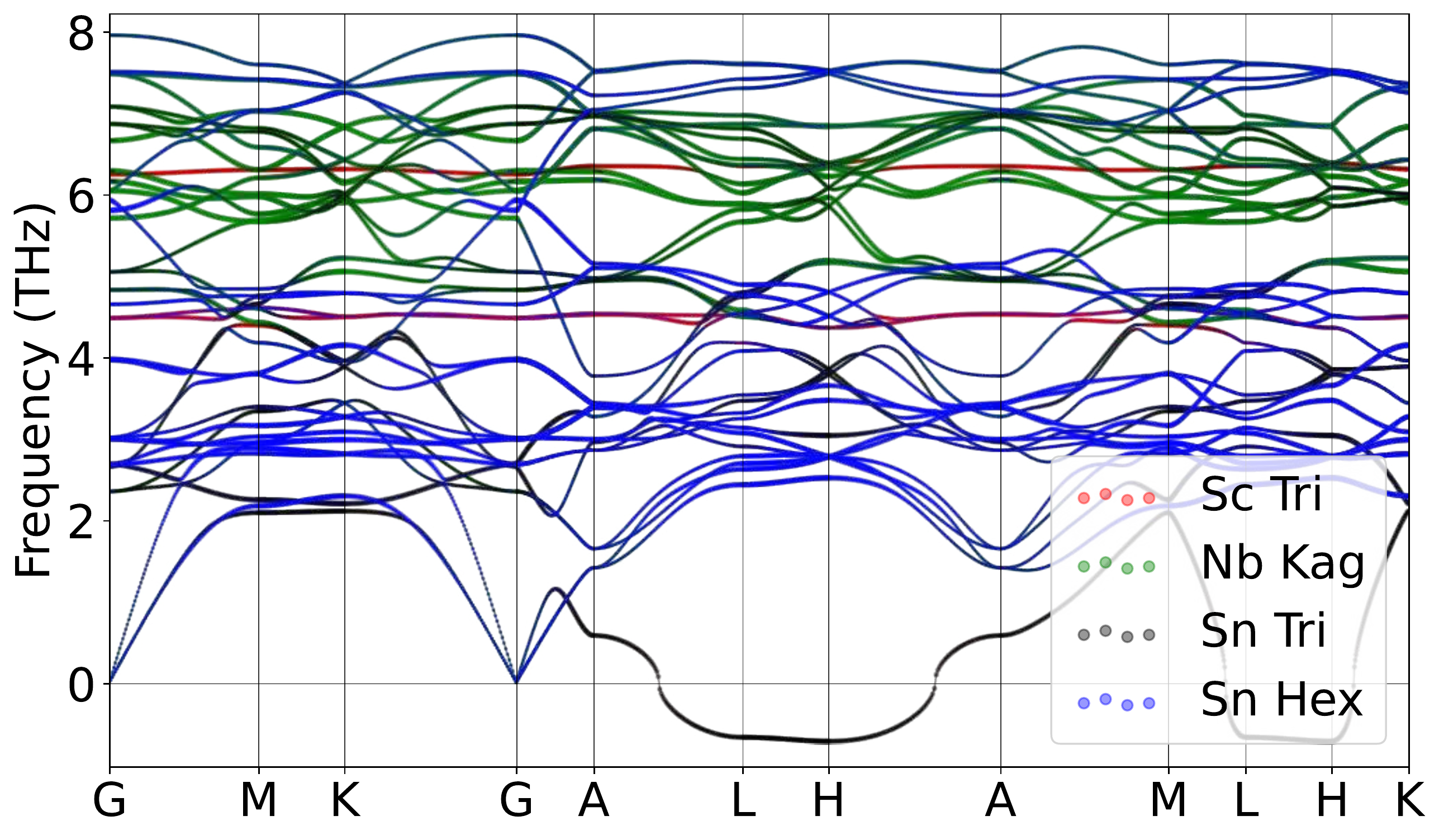}
}
\hspace{5pt}\subfloat[Relaxed phonon at high-T.]{
\label{fig:ScNb6Sn6_relaxed_High}
\includegraphics[width=0.45\textwidth]{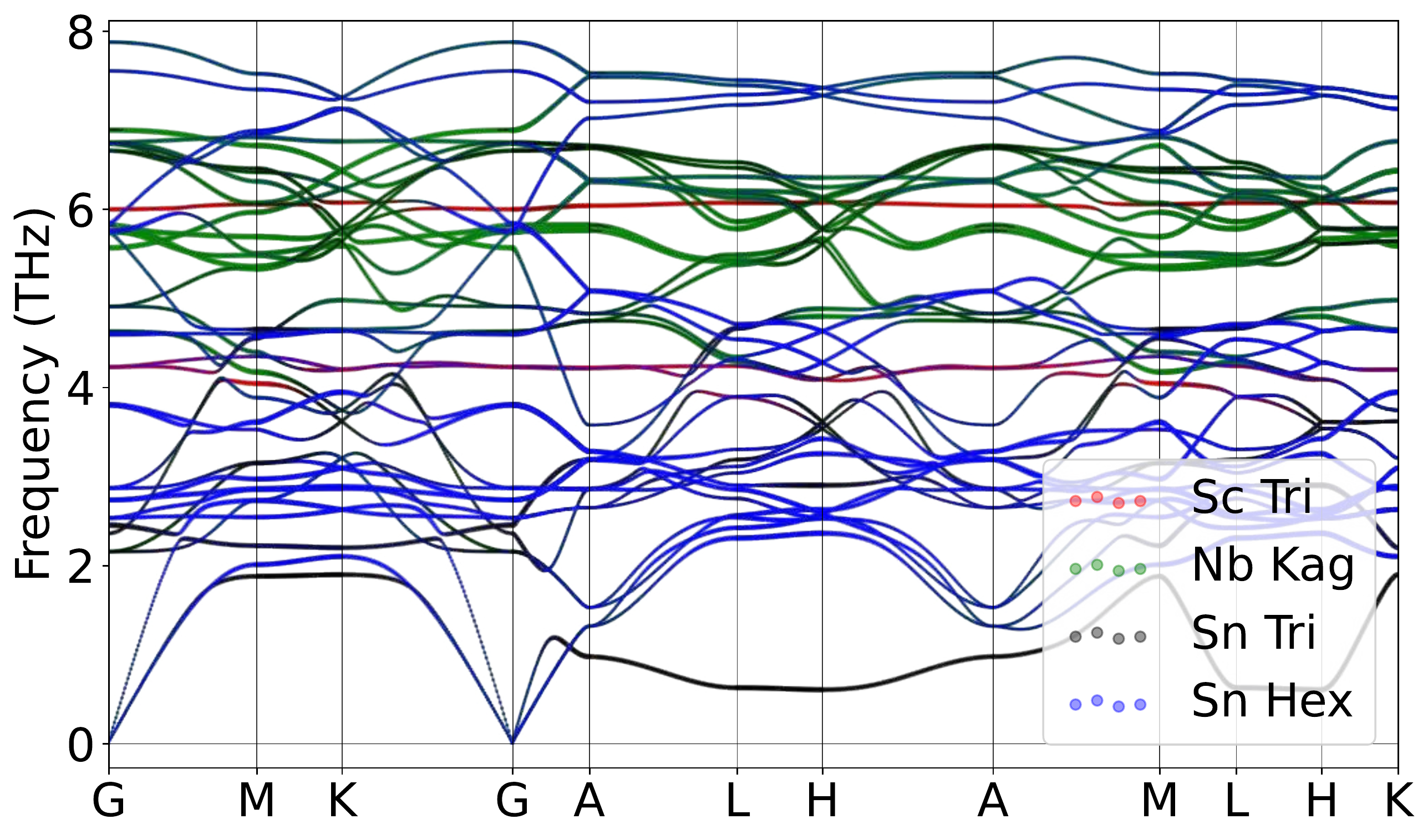}
}
\caption{Atomically projected phonon spectrum for ScNb$_6$Sn$_6$.}
\label{fig:ScNb6Sn6pho}
\end{figure}

\begin{figure}[H]
\centering
\label{fig:ScNb6Sn6_relaxed_Low_xz}
\includegraphics[width=0.85\textwidth]{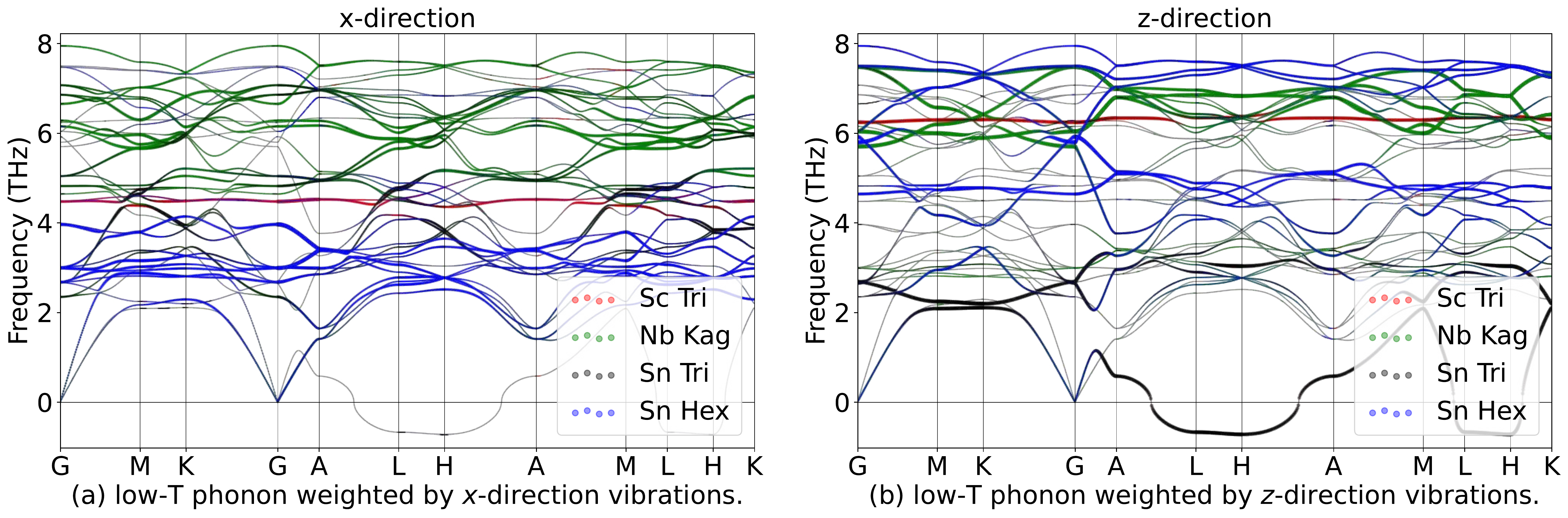}
\hspace{5pt}\label{fig:ScNb6Sn6_relaxed_High_xz}
\includegraphics[width=0.85\textwidth]{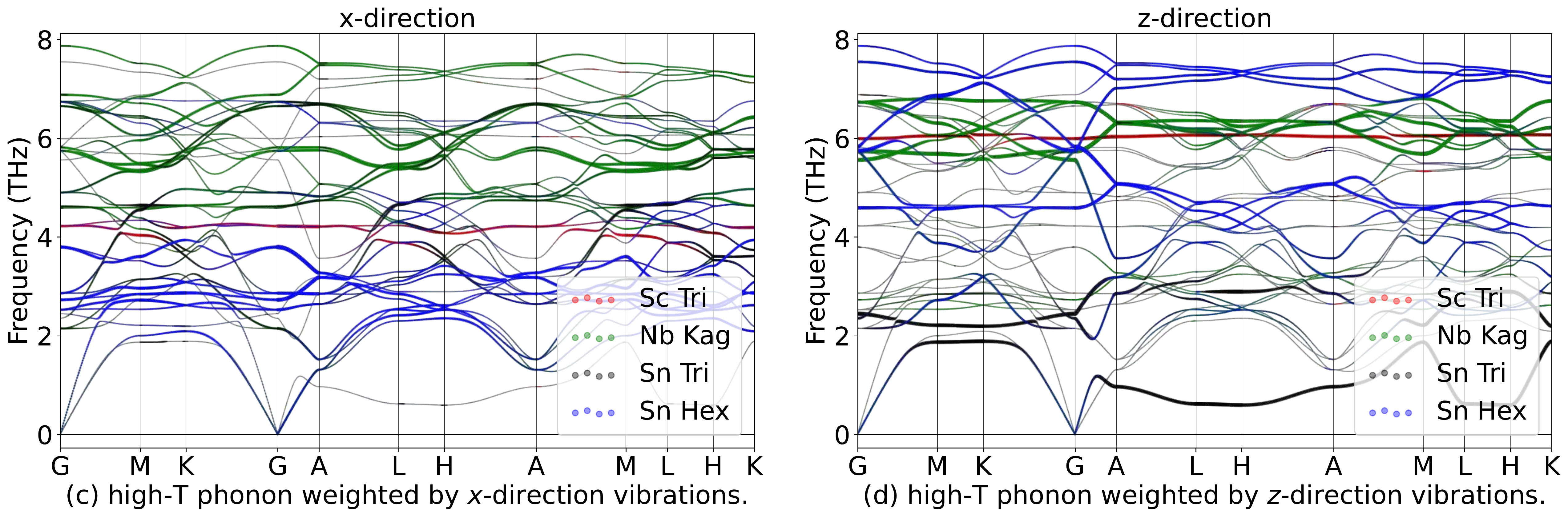}
\caption{Phonon spectrum for ScNb$_6$Sn$_6$in in-plane ($x$) and out-of-plane ($z$) directions.}
\label{fig:ScNb6Sn6phoxyz}
\end{figure}

\begin{table}[htbp]
\global\long\def\arraystretch{1.12}
\captionsetup{width=.95\textwidth}
\caption{IRREPs at high-symmetry points for ScNb$_6$Sn$_6$ phonon spectrum at Low-T.}
\label{tab:191_ScNb6Sn6_Low}
\setlength{\tabcolsep}{1.mm}{
}
\end{table}

\begin{table}[htbp]
\global\long\def\arraystretch{1.12}
\captionsetup{width=.95\textwidth}
\caption{IRREPs at high-symmetry points for ScNb$_6$Sn$_6$ phonon spectrum at High-T.}
\label{tab:191_ScNb6Sn6_High}
\setlength{\tabcolsep}{1.mm}{
%
}
\end{table}
\subsubsection{\label{sms2sec:TiNb6Sn6}\hyperref[smsec:col]{\texorpdfstring{TiNb$_6$Sn$_6$}{}}~{\color{green}  (High-T stable, Low-T stable) }}

\begin{table}[htbp]
\global\long\def\arraystretch{1.12}
\captionsetup{width=.85\textwidth}
\caption{Structure parameters of TiNb$_6$Sn$_6$ after relaxation. It contains 568 electrons. }
\label{tab:TiNb6Sn6}
\setlength{\tabcolsep}{4mm}{
%
}
\end{table}

\begin{figure}[htbp]
\centering
\includegraphics[width=0.85\textwidth]{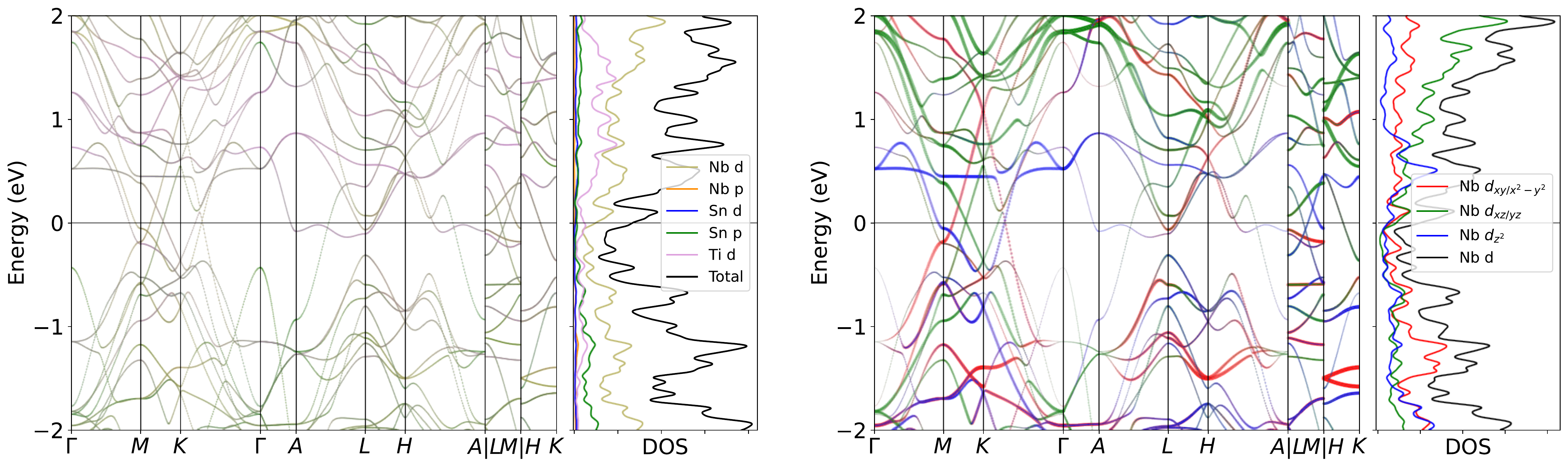}
\caption{Band structure for TiNb$_6$Sn$_6$ without SOC. The projection of Nb $3d$ orbitals is also presented. The band structures clearly reveal the dominating of Nb $3d$ orbitals  near the Fermi level, as well as the pronounced peak.}
\label{fig:TiNb6Sn6band}
\end{figure}

\begin{figure}[H]
\centering
\subfloat[Relaxed phonon at low-T.]{
\label{fig:TiNb6Sn6_relaxed_Low}
\includegraphics[width=0.45\textwidth]{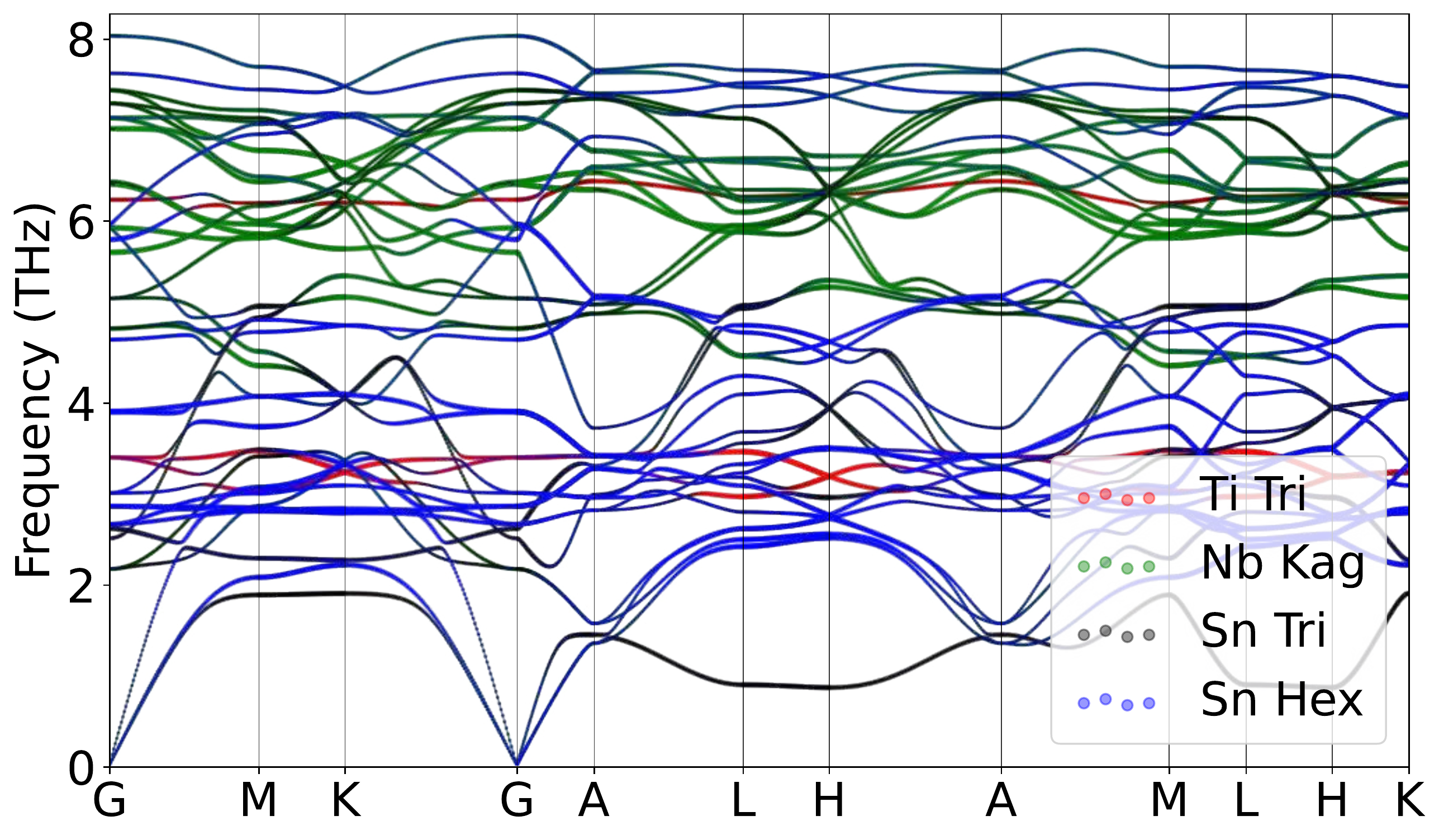}
}
\hspace{5pt}\subfloat[Relaxed phonon at high-T.]{
\label{fig:TiNb6Sn6_relaxed_High}
\includegraphics[width=0.45\textwidth]{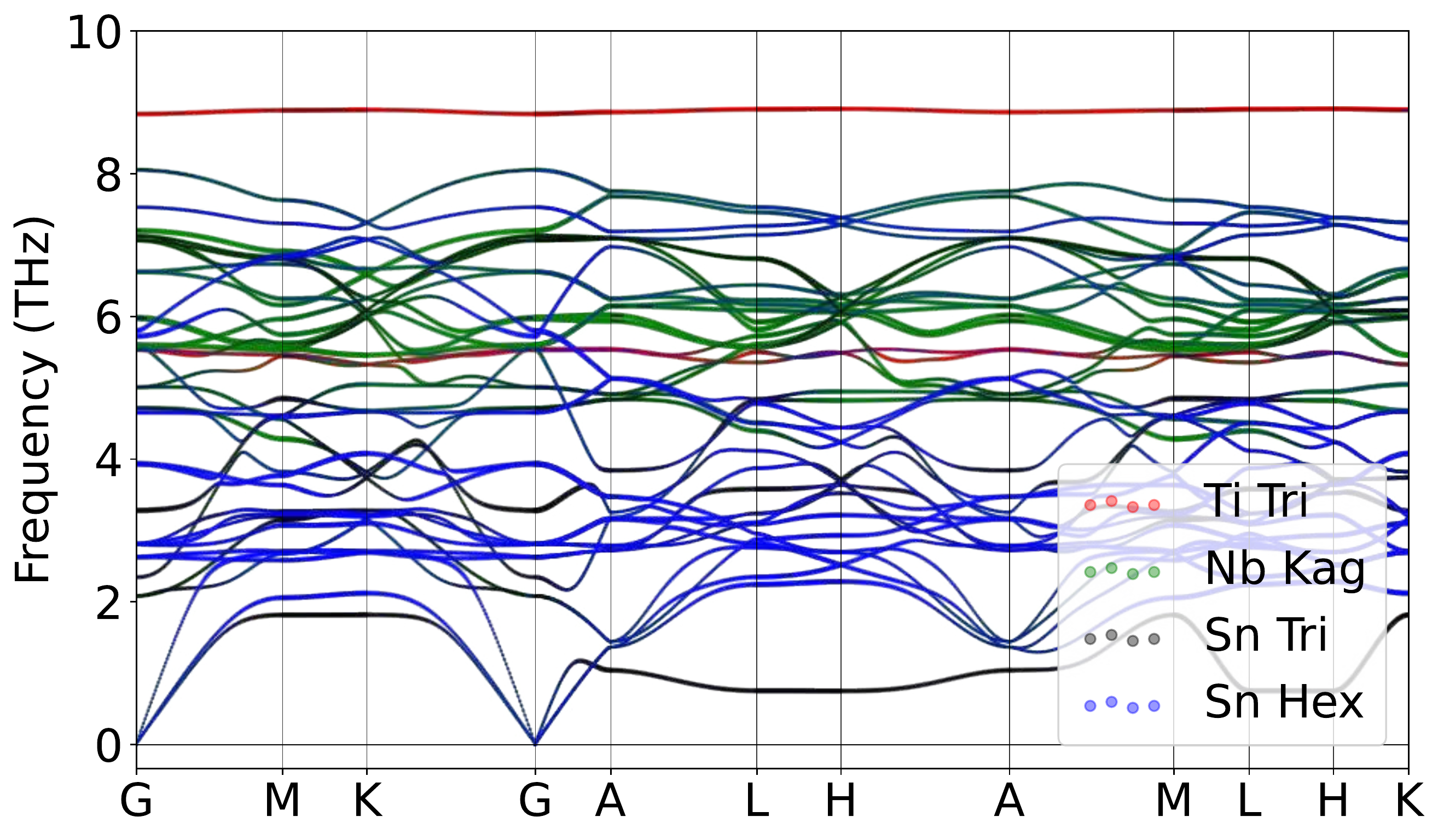}
}
\caption{Atomically projected phonon spectrum for TiNb$_6$Sn$_6$.}
\label{fig:TiNb6Sn6pho}
\end{figure}

\begin{figure}[H]
\centering
\label{fig:TiNb6Sn6_relaxed_Low_xz}
\includegraphics[width=0.85\textwidth]{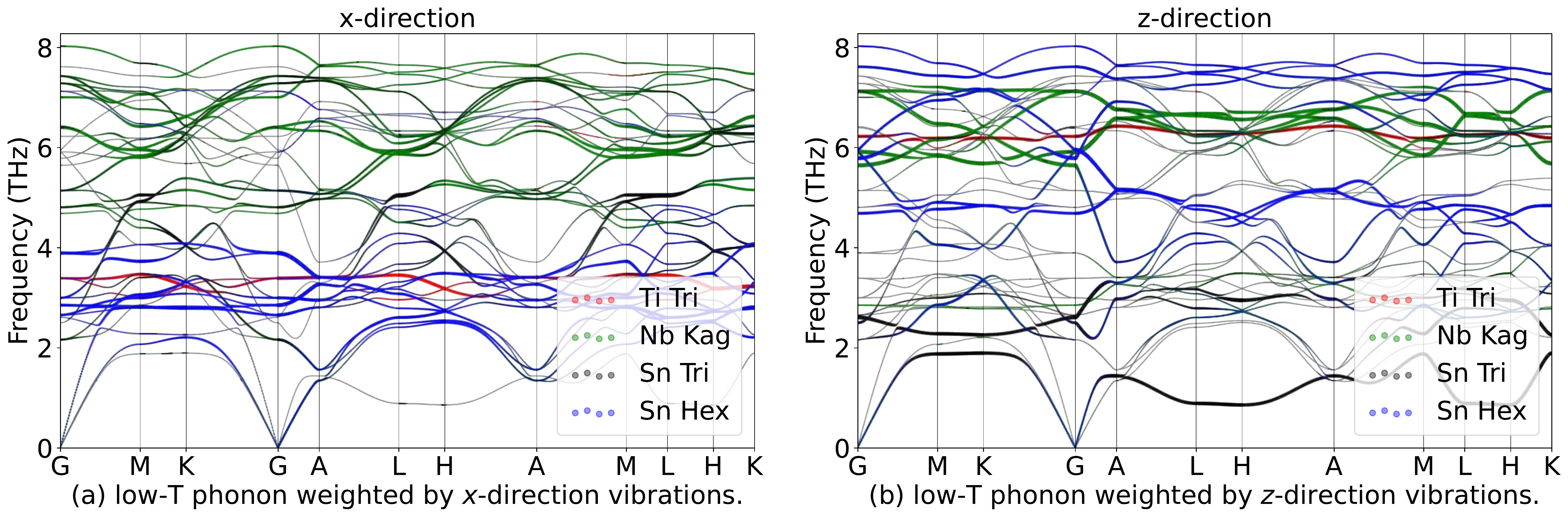}
\hspace{5pt}\label{fig:TiNb6Sn6_relaxed_High_xz}
\includegraphics[width=0.85\textwidth]{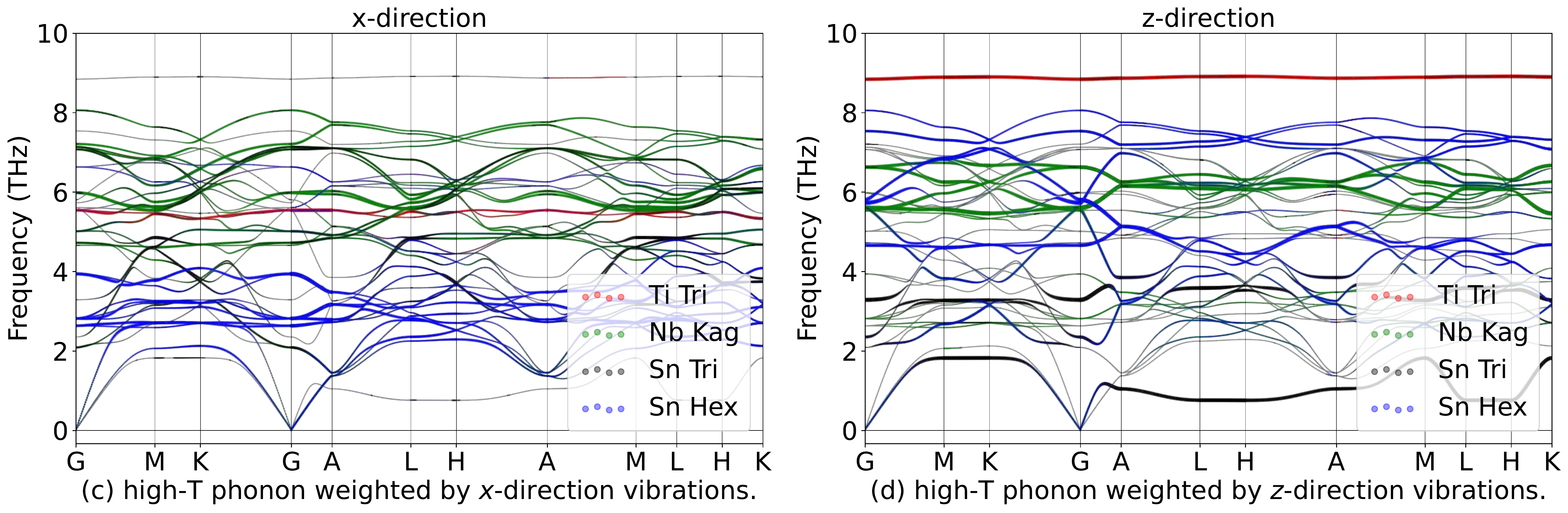}
\caption{Phonon spectrum for TiNb$_6$Sn$_6$in in-plane ($x$) and out-of-plane ($z$) directions.}
\label{fig:TiNb6Sn6phoxyz}
\end{figure}

\begin{table}[htbp]
\global\long\def\arraystretch{1.12}
\captionsetup{width=.95\textwidth}
\caption{IRREPs at high-symmetry points for TiNb$_6$Sn$_6$ phonon spectrum at Low-T.}
\label{tab:191_TiNb6Sn6_Low}
\setlength{\tabcolsep}{1.mm}{
}
\end{table}

\begin{table}[htbp]
\global\long\def\arraystretch{1.12}
\captionsetup{width=.95\textwidth}
\caption{IRREPs at high-symmetry points for TiNb$_6$Sn$_6$ phonon spectrum at High-T.}
\label{tab:191_TiNb6Sn6_High}
\setlength{\tabcolsep}{1.mm}{
%
}
\end{table}
\subsubsection{\label{sms2sec:YNb6Sn6}\hyperref[smsec:col]{\texorpdfstring{YNb$_6$Sn$_6$}{}}~{\color{green}  (High-T stable, Low-T stable) }}

\begin{table}[htbp]
\global\long\def\arraystretch{1.12}
\captionsetup{width=.85\textwidth}
\caption{Structure parameters of YNb$_6$Sn$_6$ after relaxation. It contains 585 electrons. }
\label{tab:YNb6Sn6}
\setlength{\tabcolsep}{4mm}{
%
}
\end{table}

\begin{figure}[htbp]
\centering
\includegraphics[width=0.85\textwidth]{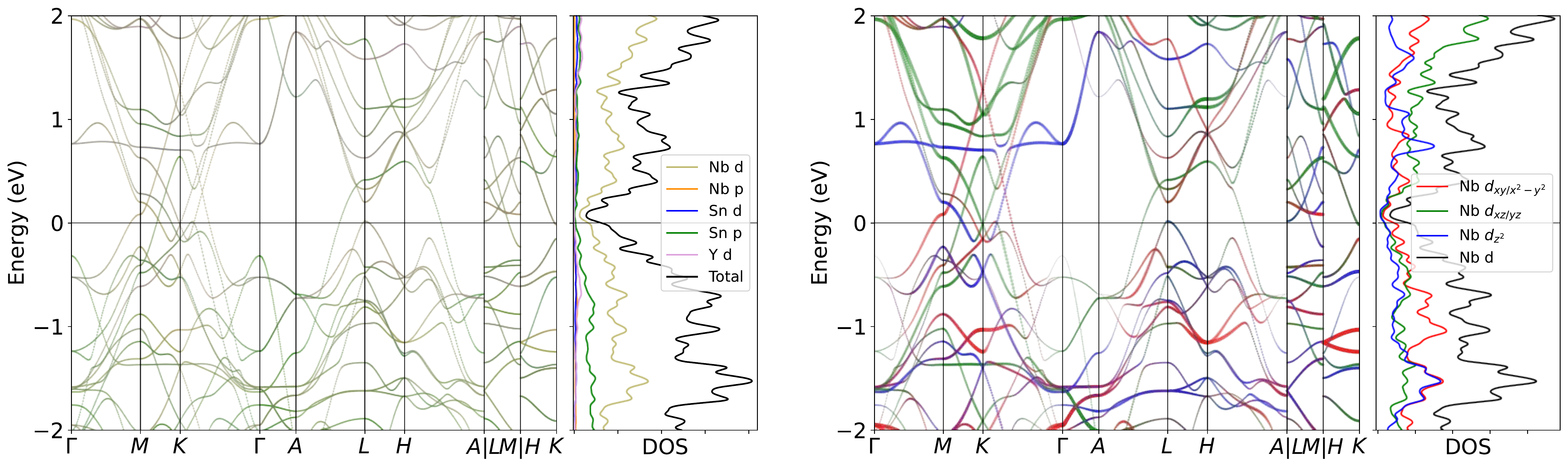}
\caption{Band structure for YNb$_6$Sn$_6$ without SOC. The projection of Nb $3d$ orbitals is also presented. The band structures clearly reveal the dominating of Nb $3d$ orbitals  near the Fermi level, as well as the pronounced peak.}
\label{fig:YNb6Sn6band}
\end{figure}

\begin{figure}[H]
\centering
\subfloat[Relaxed phonon at low-T.]{
\label{fig:YNb6Sn6_relaxed_Low}
\includegraphics[width=0.45\textwidth]{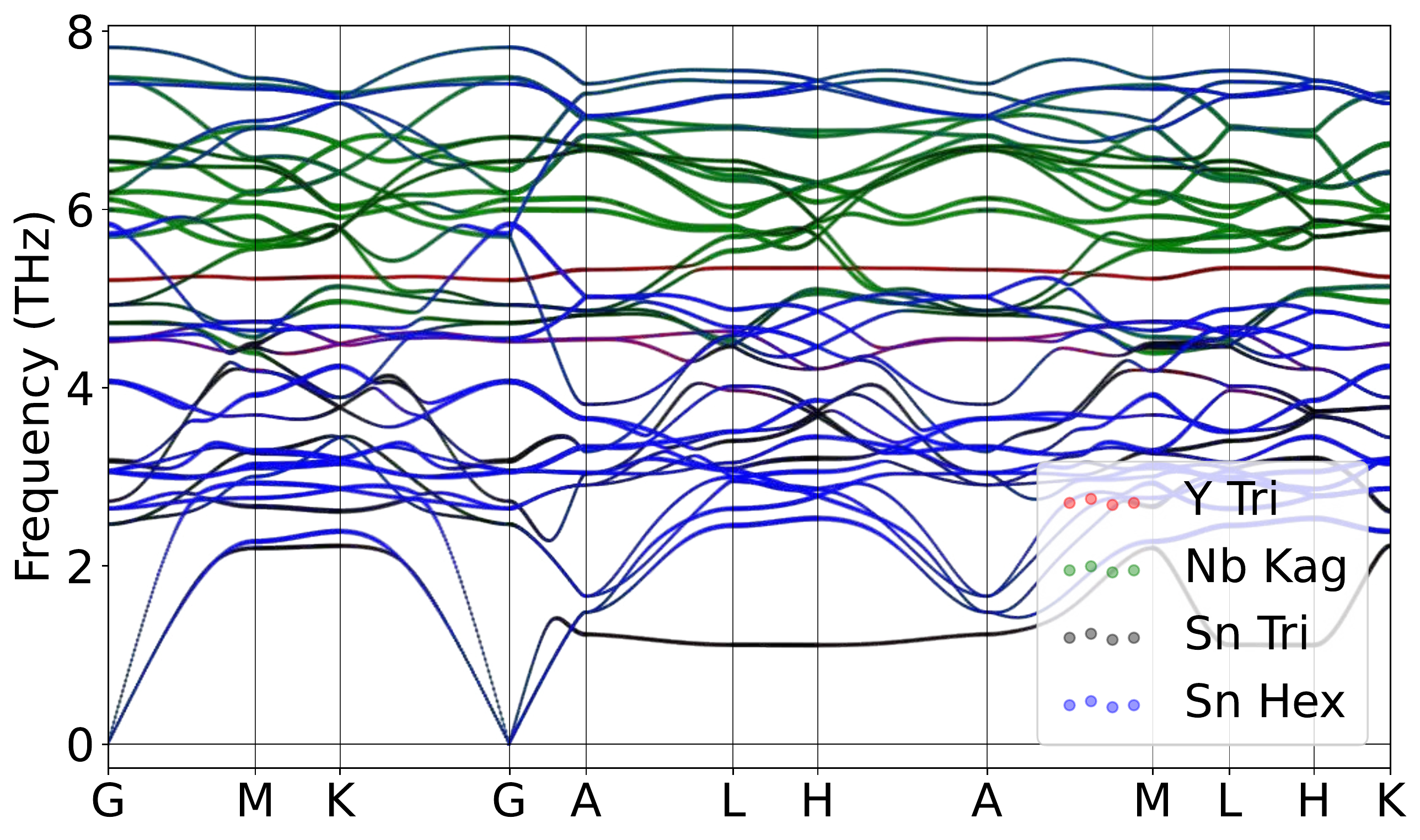}
}
\hspace{5pt}\subfloat[Relaxed phonon at high-T.]{
\label{fig:YNb6Sn6_relaxed_High}
\includegraphics[width=0.45\textwidth]{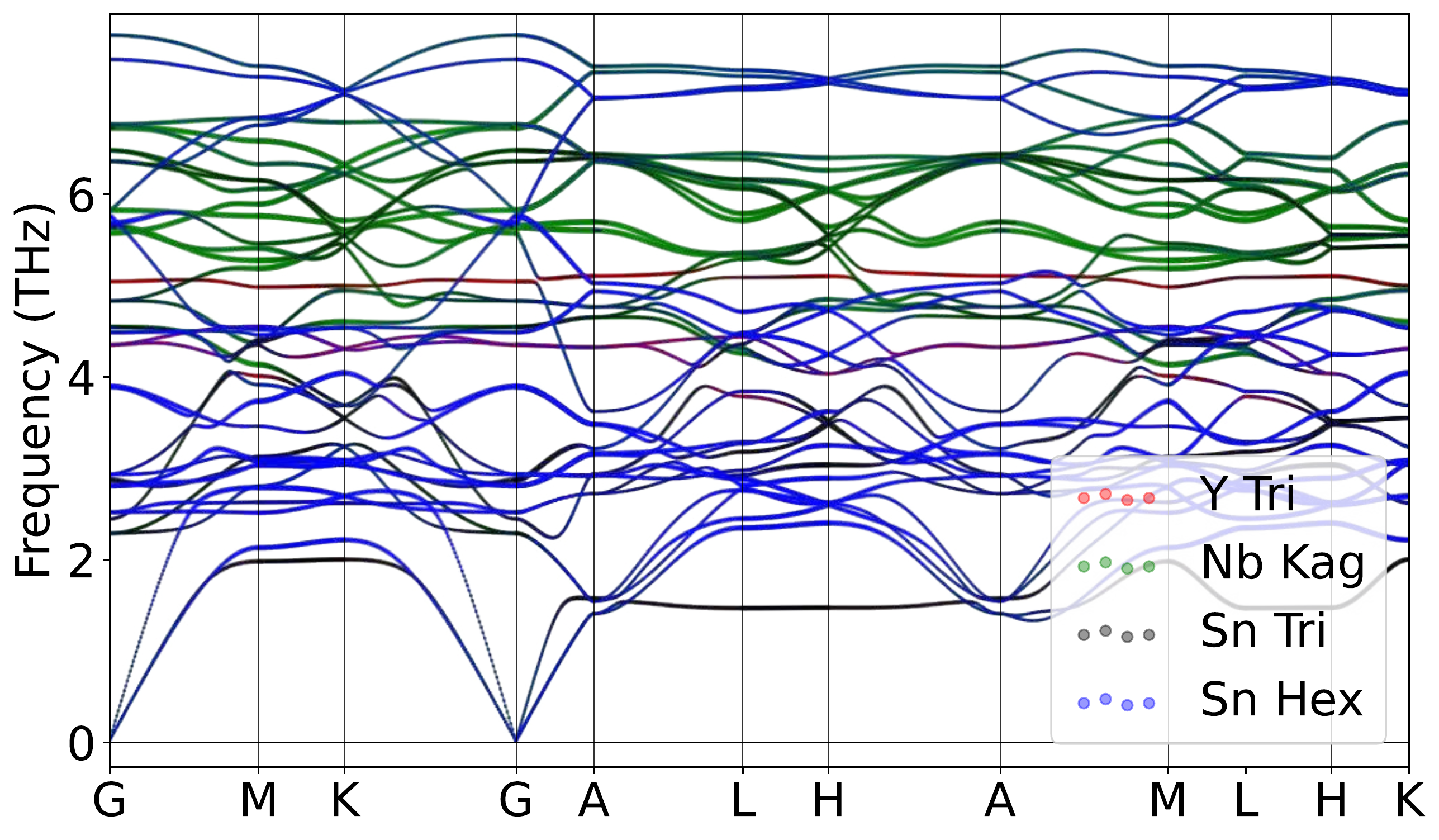}
}
\caption{Atomically projected phonon spectrum for YNb$_6$Sn$_6$.}
\label{fig:YNb6Sn6pho}
\end{figure}

\begin{figure}[H]
\centering
\label{fig:YNb6Sn6_relaxed_Low_xz}
\includegraphics[width=0.85\textwidth]{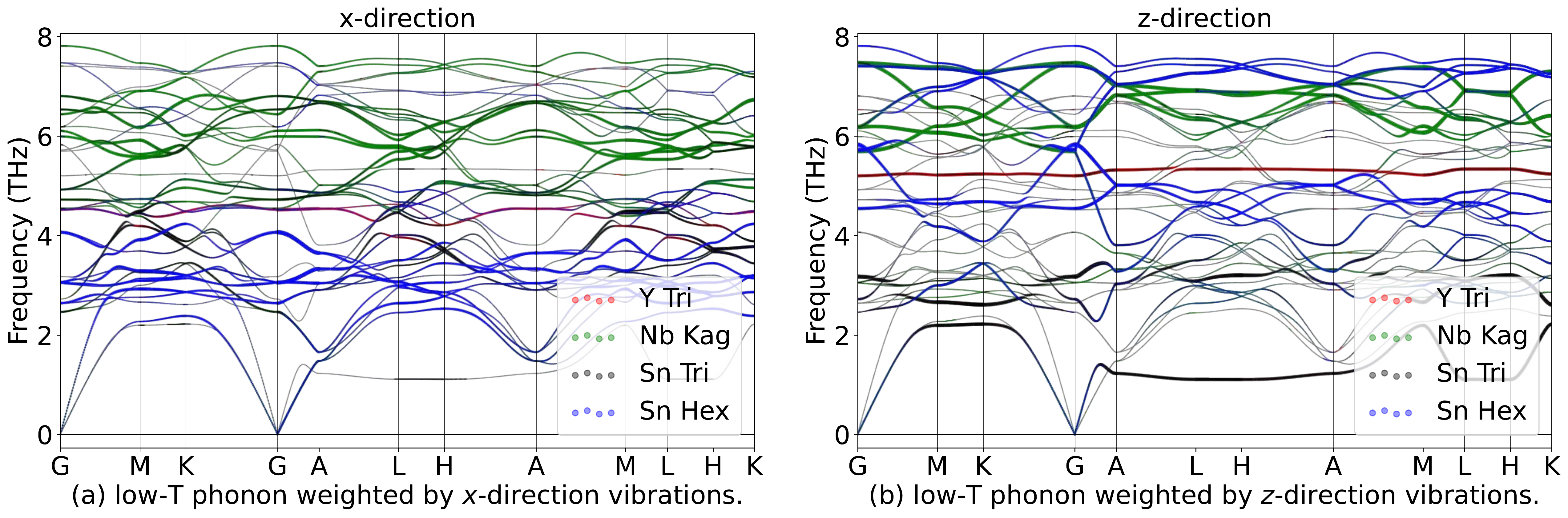}
\hspace{5pt}\label{fig:YNb6Sn6_relaxed_High_xz}
\includegraphics[width=0.85\textwidth]{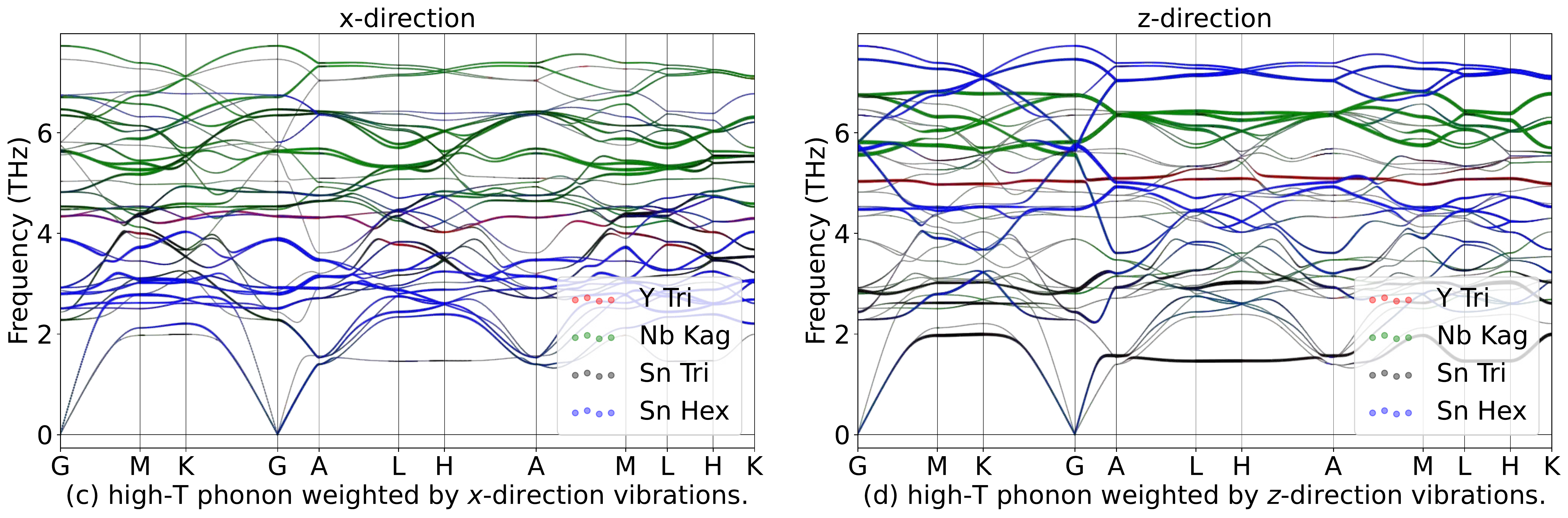}
\caption{Phonon spectrum for YNb$_6$Sn$_6$in in-plane ($x$) and out-of-plane ($z$) directions.}
\label{fig:YNb6Sn6phoxyz}
\end{figure}

\begin{table}[htbp]
\global\long\def\arraystretch{1.12}
\captionsetup{width=.95\textwidth}
\caption{IRREPs at high-symmetry points for YNb$_6$Sn$_6$ phonon spectrum at Low-T.}
\label{tab:191_YNb6Sn6_Low}
\setlength{\tabcolsep}{1.mm}{
}
\end{table}

\begin{table}[htbp]
\global\long\def\arraystretch{1.12}
\captionsetup{width=.95\textwidth}
\caption{IRREPs at high-symmetry points for YNb$_6$Sn$_6$ phonon spectrum at High-T.}
\label{tab:191_YNb6Sn6_High}
\setlength{\tabcolsep}{1.mm}{
%
}
\end{table}
\subsubsection{\label{sms2sec:ZrNb6Sn6}\hyperref[smsec:col]{\texorpdfstring{ZrNb$_6$Sn$_6$}{}}~{\color{green}  (High-T stable, Low-T stable) }}

\begin{table}[htbp]
\global\long\def\arraystretch{1.12}
\captionsetup{width=.85\textwidth}
\caption{Structure parameters of ZrNb$_6$Sn$_6$ after relaxation. It contains 586 electrons. }
\label{tab:ZrNb6Sn6}
\setlength{\tabcolsep}{4mm}{
%
}
\end{table}

\begin{figure}[htbp]
\centering
\includegraphics[width=0.85\textwidth]{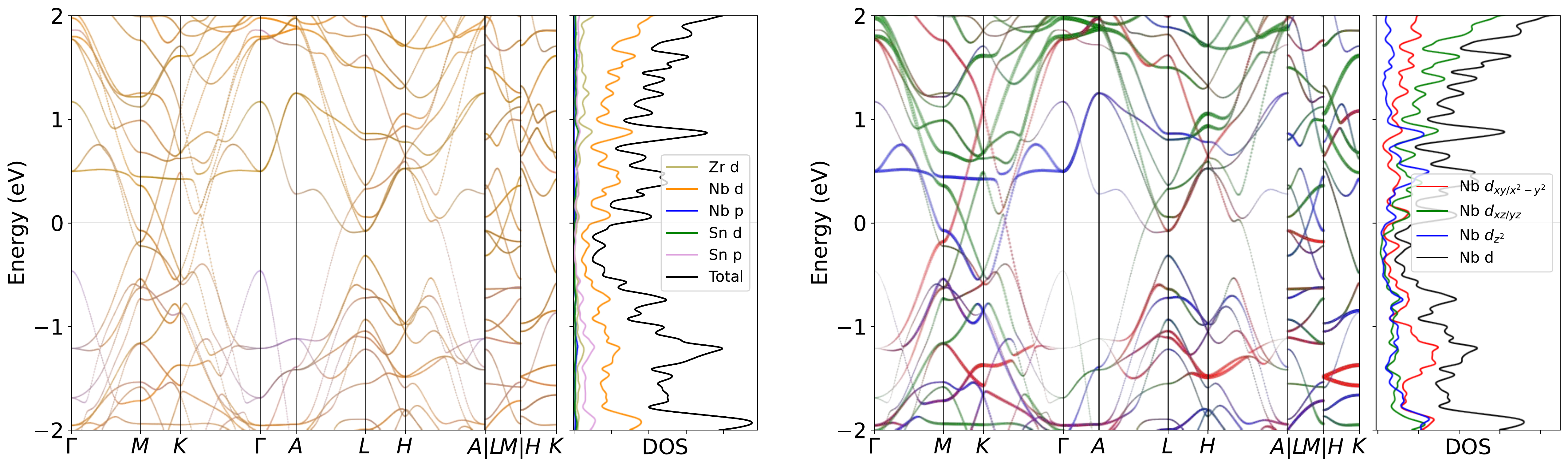}
\caption{Band structure for ZrNb$_6$Sn$_6$ without SOC. The projection of Nb $3d$ orbitals is also presented. The band structures clearly reveal the dominating of Nb $3d$ orbitals  near the Fermi level, as well as the pronounced peak.}
\label{fig:ZrNb6Sn6band}
\end{figure}

\begin{figure}[H]
\centering
\subfloat[Relaxed phonon at low-T.]{
\label{fig:ZrNb6Sn6_relaxed_Low}
\includegraphics[width=0.45\textwidth]{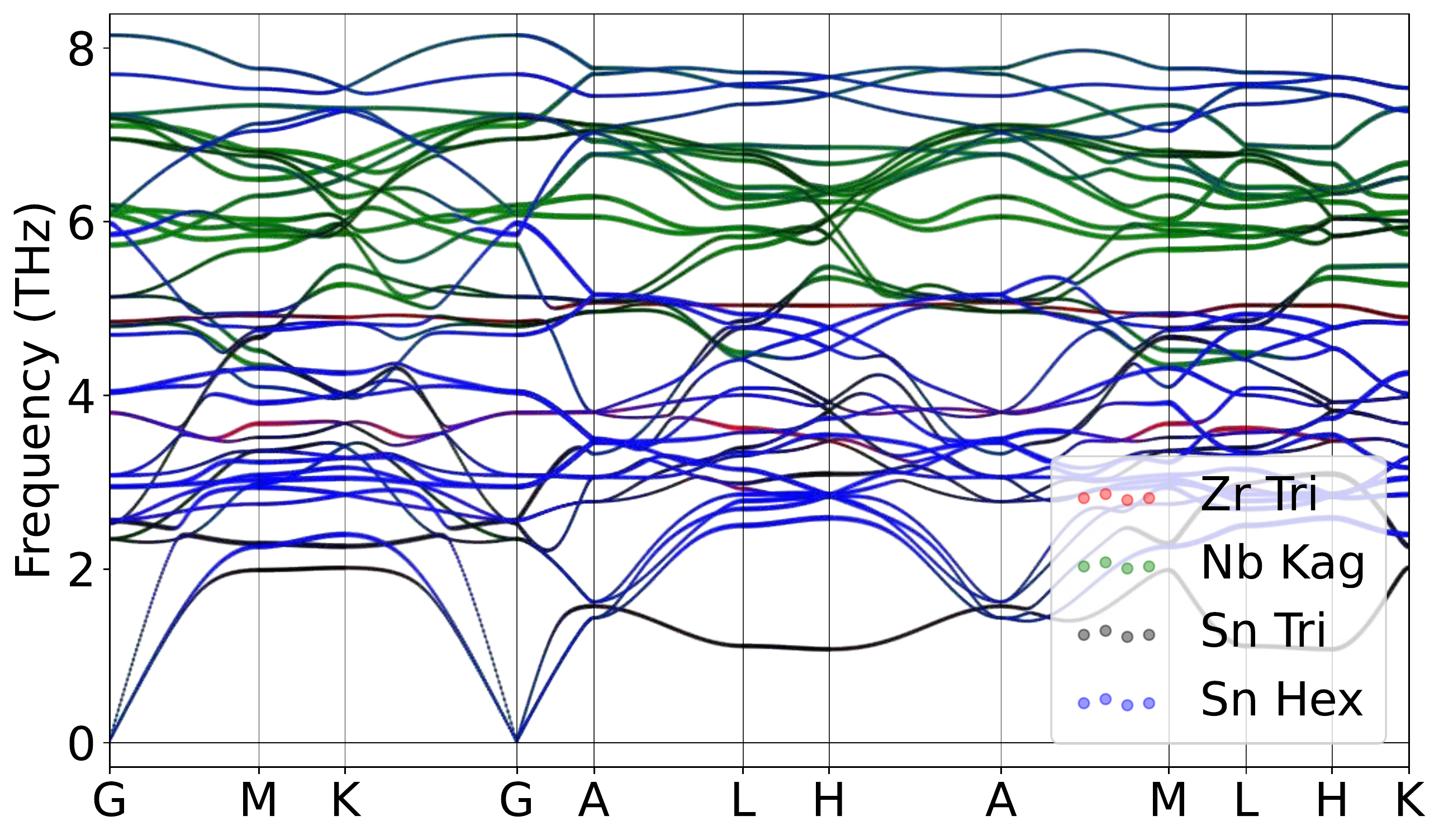}
}
\hspace{5pt}\subfloat[Relaxed phonon at high-T.]{
\label{fig:ZrNb6Sn6_relaxed_High}
\includegraphics[width=0.45\textwidth]{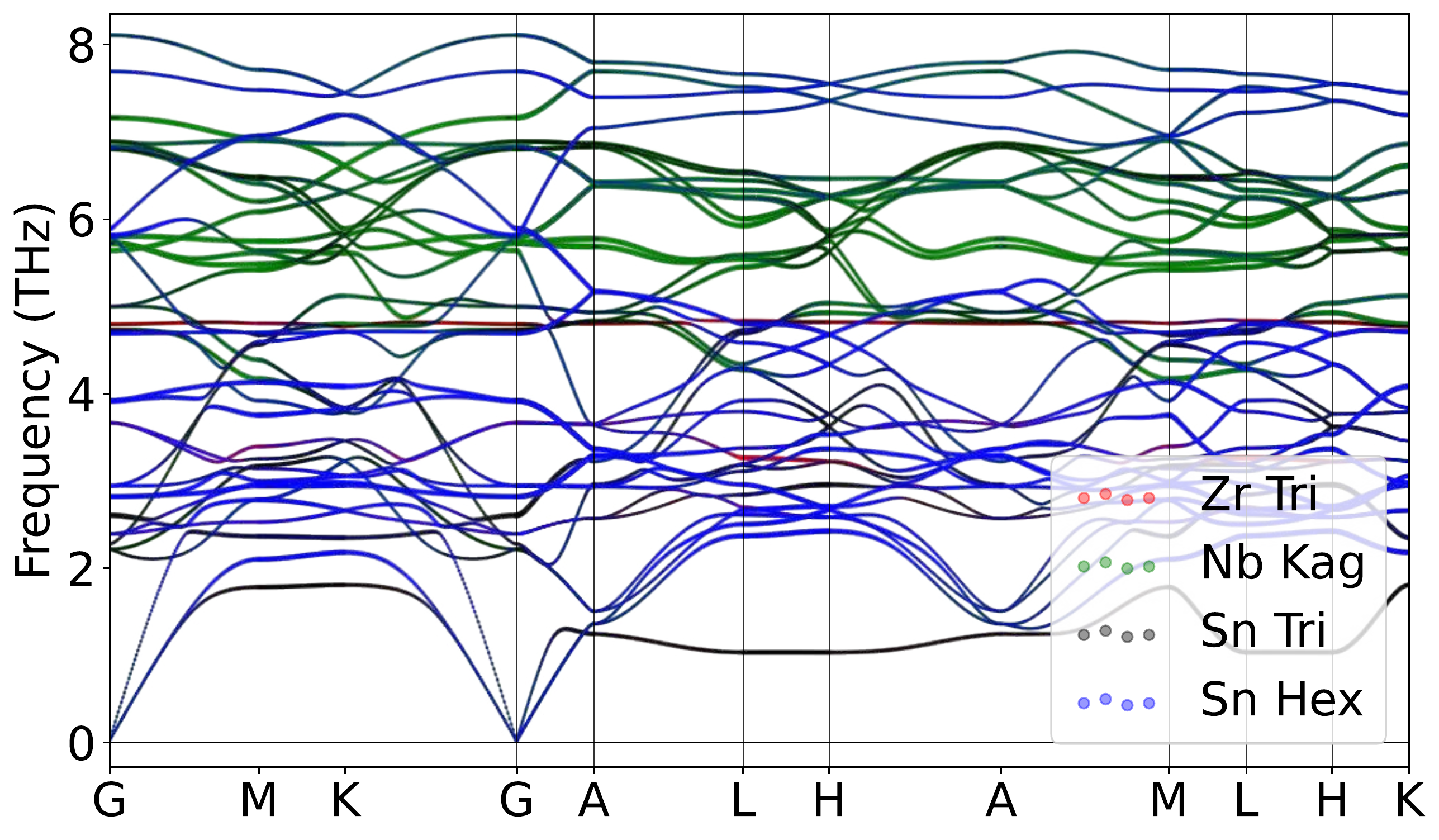}
}
\caption{Atomically projected phonon spectrum for ZrNb$_6$Sn$_6$.}
\label{fig:ZrNb6Sn6pho}
\end{figure}

\begin{figure}[H]
\centering
\label{fig:ZrNb6Sn6_relaxed_Low_xz}
\includegraphics[width=0.85\textwidth]{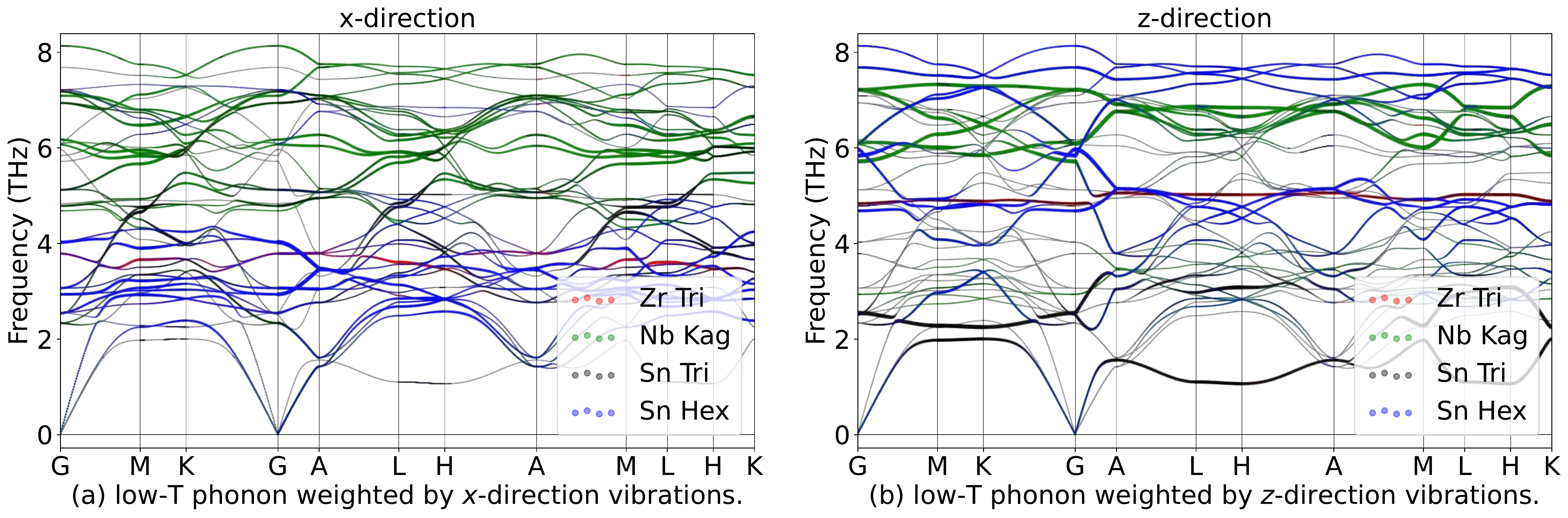}
\hspace{5pt}\label{fig:ZrNb6Sn6_relaxed_High_xz}
\includegraphics[width=0.85\textwidth]{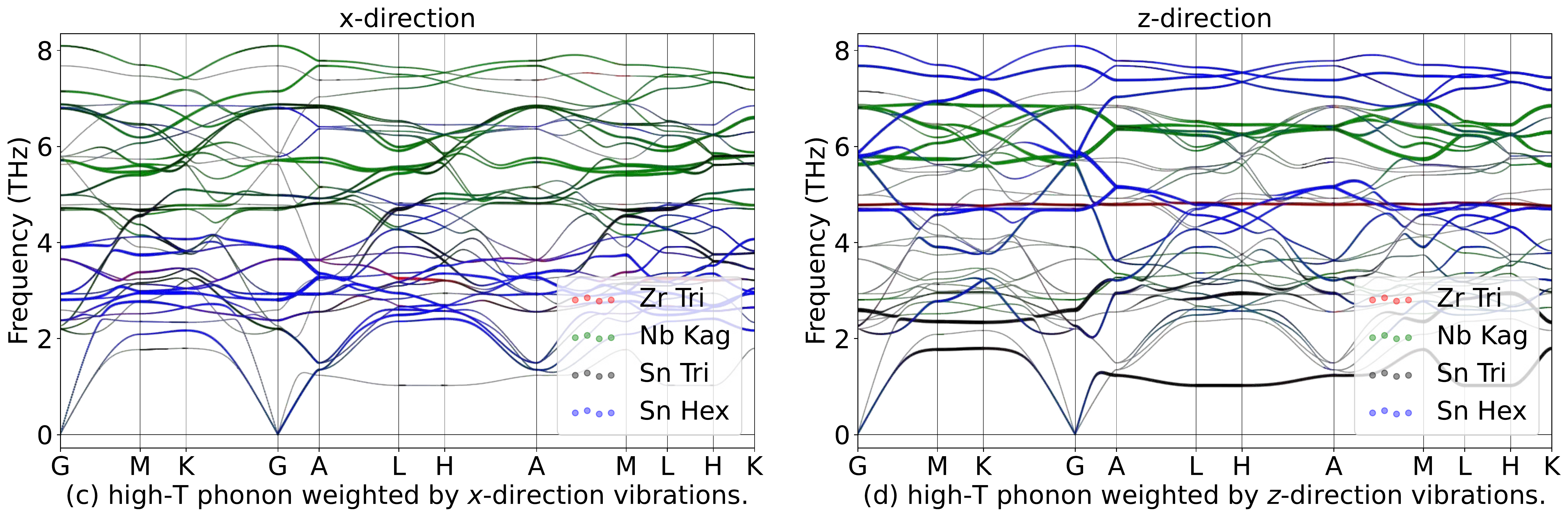}
\caption{Phonon spectrum for ZrNb$_6$Sn$_6$in in-plane ($x$) and out-of-plane ($z$) directions.}
\label{fig:ZrNb6Sn6phoxyz}
\end{figure}

\begin{table}[htbp]
\global\long\def\arraystretch{1.12}
\captionsetup{width=.95\textwidth}
\caption{IRREPs at high-symmetry points for ZrNb$_6$Sn$_6$ phonon spectrum at Low-T.}
\label{tab:191_ZrNb6Sn6_Low}
\setlength{\tabcolsep}{1.mm}{
}
\end{table}

\begin{table}[htbp]
\global\long\def\arraystretch{1.12}
\captionsetup{width=.95\textwidth}
\caption{IRREPs at high-symmetry points for ZrNb$_6$Sn$_6$ phonon spectrum at High-T.}
\label{tab:191_ZrNb6Sn6_High}
\setlength{\tabcolsep}{1.mm}{
%
}
\end{table}
\subsubsection{\label{sms2sec:PrNb6Sn6}\hyperref[smsec:col]{\texorpdfstring{PrNb$_6$Sn$_6$}{}}~{\color{green}  (High-T stable, Low-T stable) }}

\begin{table}[htbp]
\global\long\def\arraystretch{1.12}
\captionsetup{width=.85\textwidth}
\caption{Structure parameters of PrNb$_6$Sn$_6$ after relaxation. It contains 605 electrons. }
\label{tab:PrNb6Sn6}
\setlength{\tabcolsep}{4mm}{
%
}
\end{table}

\begin{figure}[htbp]
\centering
\includegraphics[width=0.85\textwidth]{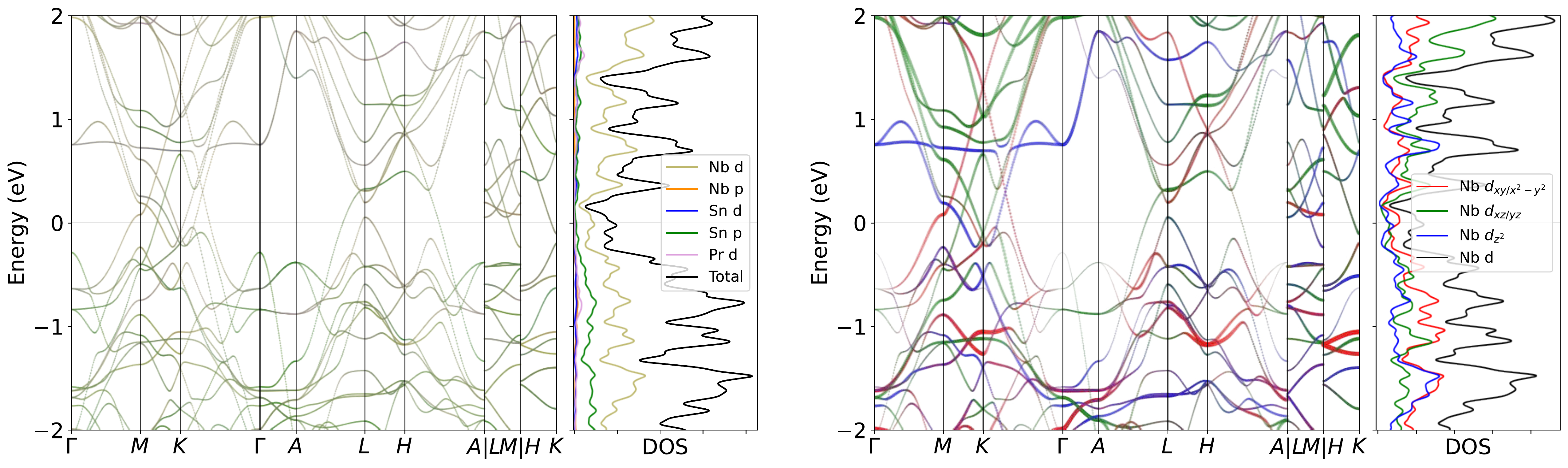}
\caption{Band structure for PrNb$_6$Sn$_6$ without SOC. The projection of Nb $3d$ orbitals is also presented. The band structures clearly reveal the dominating of Nb $3d$ orbitals  near the Fermi level, as well as the pronounced peak.}
\label{fig:PrNb6Sn6band}
\end{figure}

\begin{figure}[H]
\centering
\subfloat[Relaxed phonon at low-T.]{
\label{fig:PrNb6Sn6_relaxed_Low}
\includegraphics[width=0.45\textwidth]{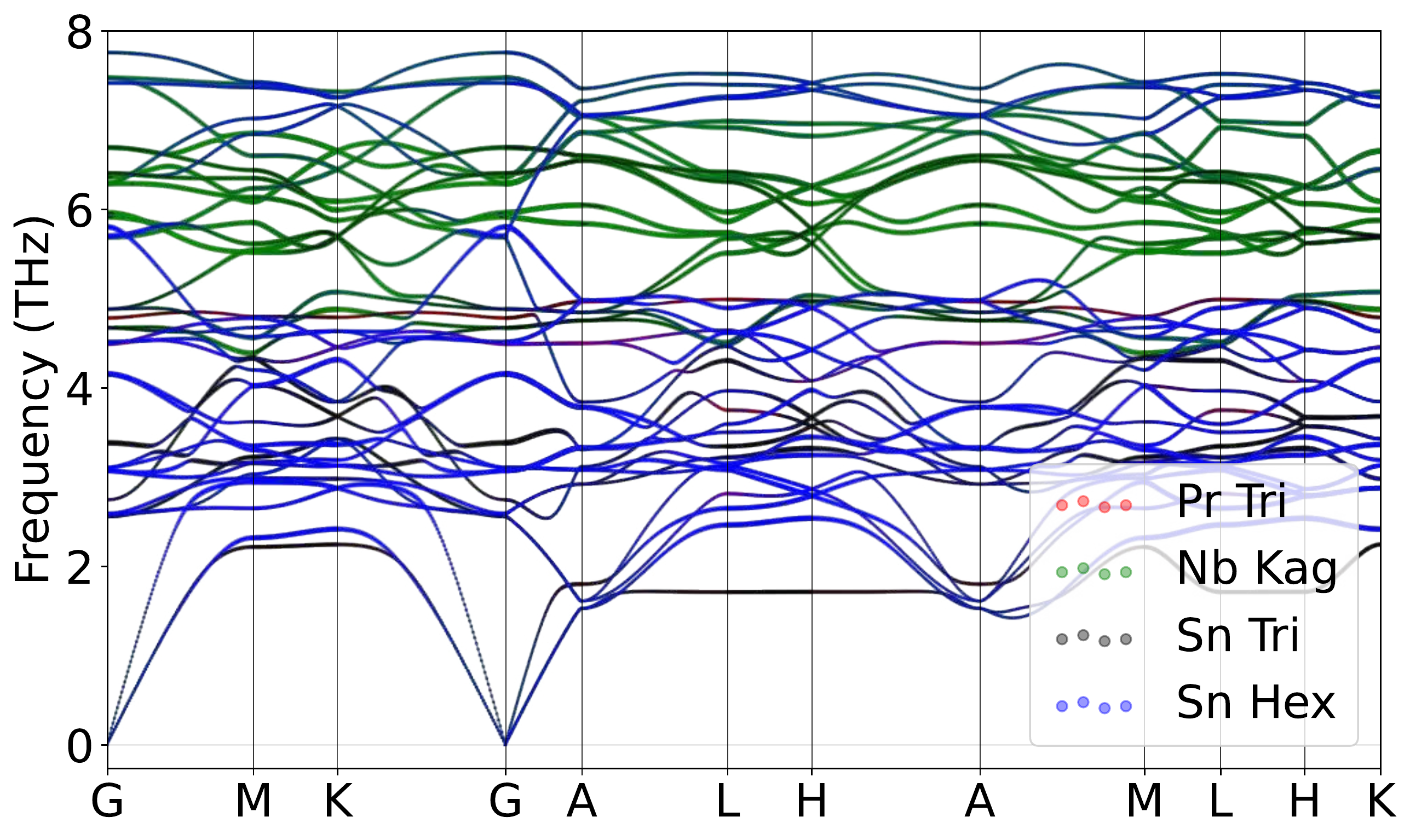}
}
\hspace{5pt}\subfloat[Relaxed phonon at high-T.]{
\label{fig:PrNb6Sn6_relaxed_High}
\includegraphics[width=0.45\textwidth]{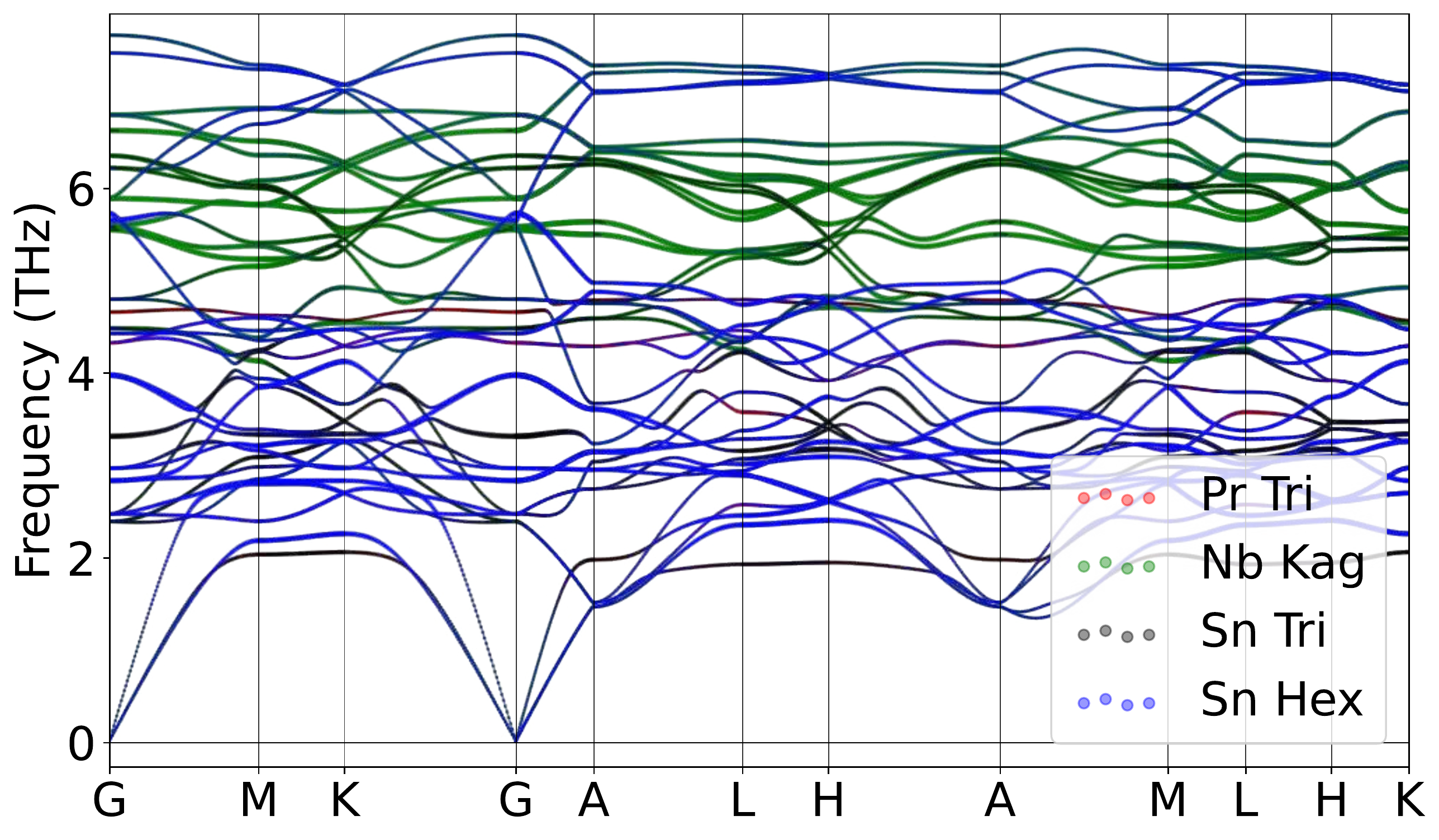}
}
\caption{Atomically projected phonon spectrum for PrNb$_6$Sn$_6$.}
\label{fig:PrNb6Sn6pho}
\end{figure}

\begin{figure}[H]
\centering
\label{fig:PrNb6Sn6_relaxed_Low_xz}
\includegraphics[width=0.85\textwidth]{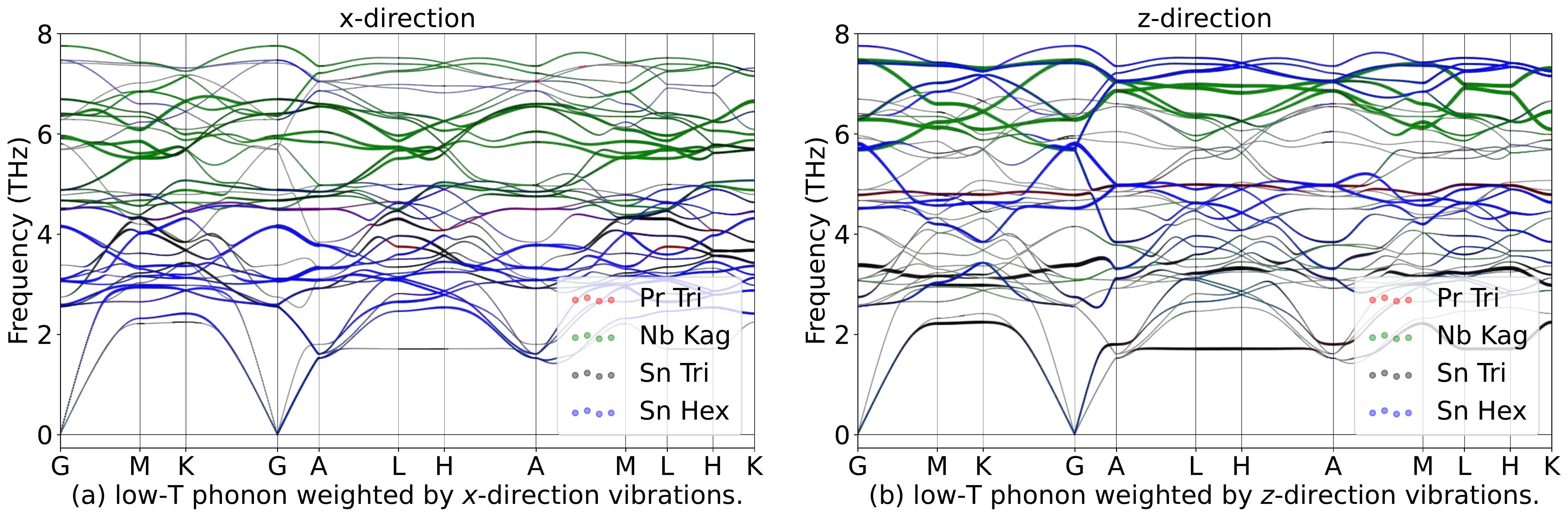}
\hspace{5pt}\label{fig:PrNb6Sn6_relaxed_High_xz}
\includegraphics[width=0.85\textwidth]{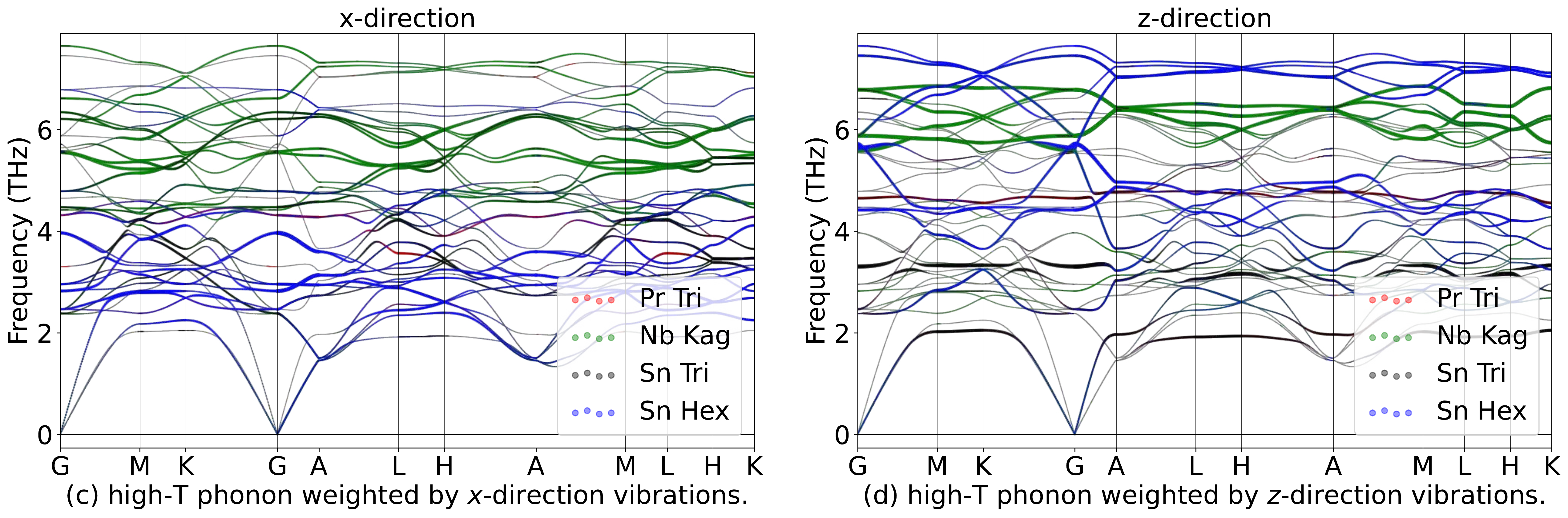}
\caption{Phonon spectrum for PrNb$_6$Sn$_6$in in-plane ($x$) and out-of-plane ($z$) directions.}
\label{fig:PrNb6Sn6phoxyz}
\end{figure}

\begin{table}[htbp]
\global\long\def\arraystretch{1.12}
\captionsetup{width=.95\textwidth}
\caption{IRREPs at high-symmetry points for PrNb$_6$Sn$_6$ phonon spectrum at Low-T.}
\label{tab:191_PrNb6Sn6_Low}
\setlength{\tabcolsep}{1.mm}{
}
\end{table}

\begin{table}[htbp]
\global\long\def\arraystretch{1.12}
\captionsetup{width=.95\textwidth}
\caption{IRREPs at high-symmetry points for PrNb$_6$Sn$_6$ phonon spectrum at High-T.}
\label{tab:191_PrNb6Sn6_High}
\setlength{\tabcolsep}{1.mm}{
%
}
\end{table}
\subsubsection{\label{sms2sec:NdNb6Sn6}\hyperref[smsec:col]{\texorpdfstring{NdNb$_6$Sn$_6$}{}}~{\color{green}  (High-T stable, Low-T stable) }}

\begin{table}[htbp]
\global\long\def\arraystretch{1.12}
\captionsetup{width=.85\textwidth}
\caption{Structure parameters of NdNb$_6$Sn$_6$ after relaxation. It contains 606 electrons. }
\label{tab:NdNb6Sn6}
\setlength{\tabcolsep}{4mm}{
%
}
\end{table}

\begin{figure}[htbp]
\centering
\includegraphics[width=0.85\textwidth]{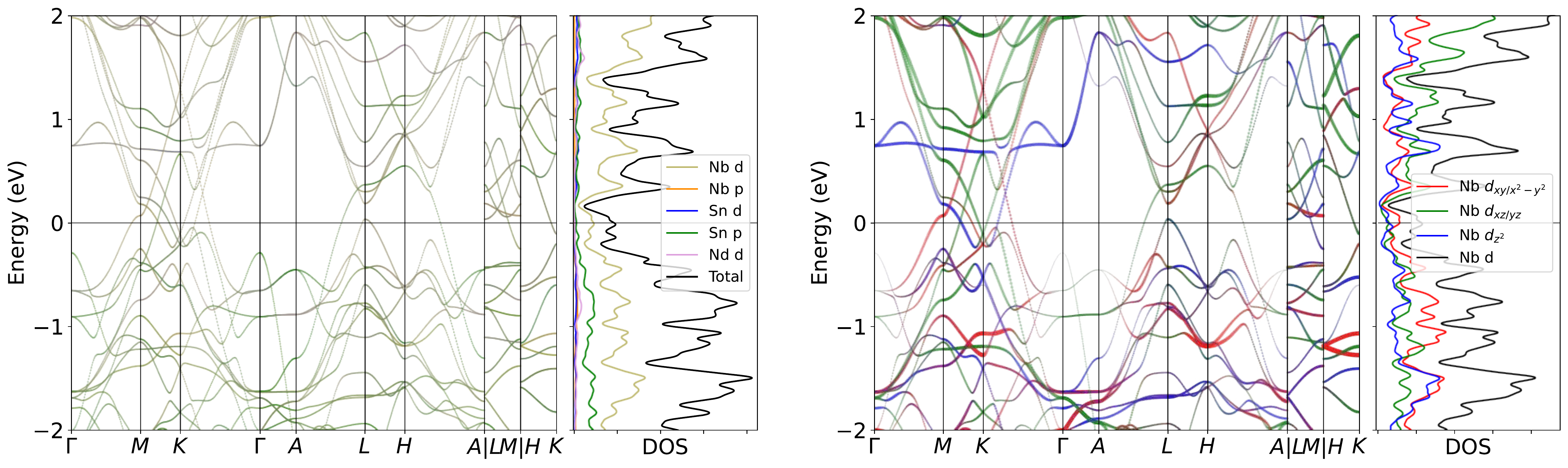}
\caption{Band structure for NdNb$_6$Sn$_6$ without SOC. The projection of Nb $3d$ orbitals is also presented. The band structures clearly reveal the dominating of Nb $3d$ orbitals  near the Fermi level, as well as the pronounced peak.}
\label{fig:NdNb6Sn6band}
\end{figure}

\begin{figure}[H]
\centering
\subfloat[Relaxed phonon at low-T.]{
\label{fig:NdNb6Sn6_relaxed_Low}
\includegraphics[width=0.45\textwidth]{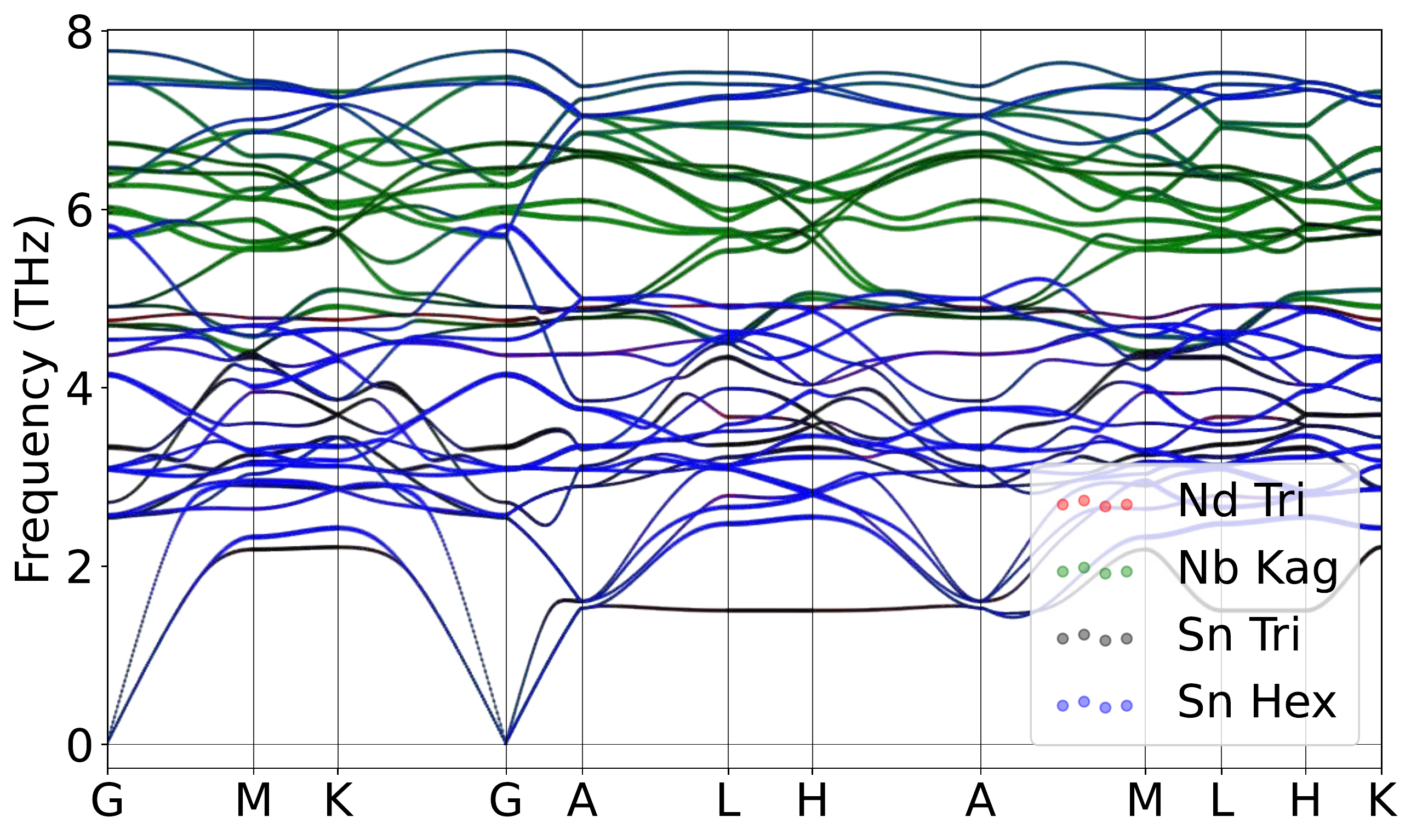}
}
\hspace{5pt}\subfloat[Relaxed phonon at high-T.]{
\label{fig:NdNb6Sn6_relaxed_High}
\includegraphics[width=0.45\textwidth]{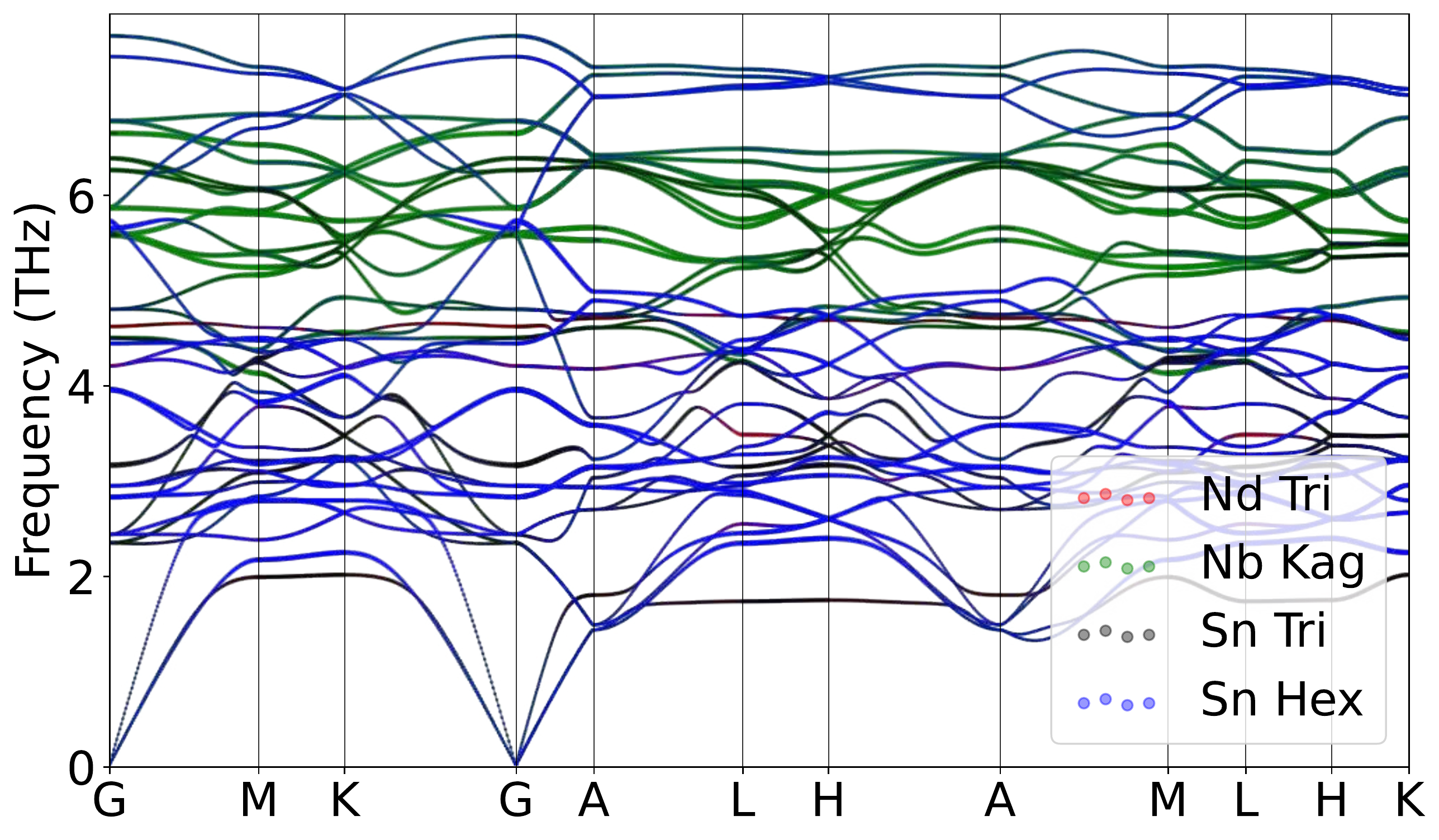}
}
\caption{Atomically projected phonon spectrum for NdNb$_6$Sn$_6$.}
\label{fig:NdNb6Sn6pho}
\end{figure}

\begin{figure}[H]
\centering
\label{fig:NdNb6Sn6_relaxed_Low_xz}
\includegraphics[width=0.85\textwidth]{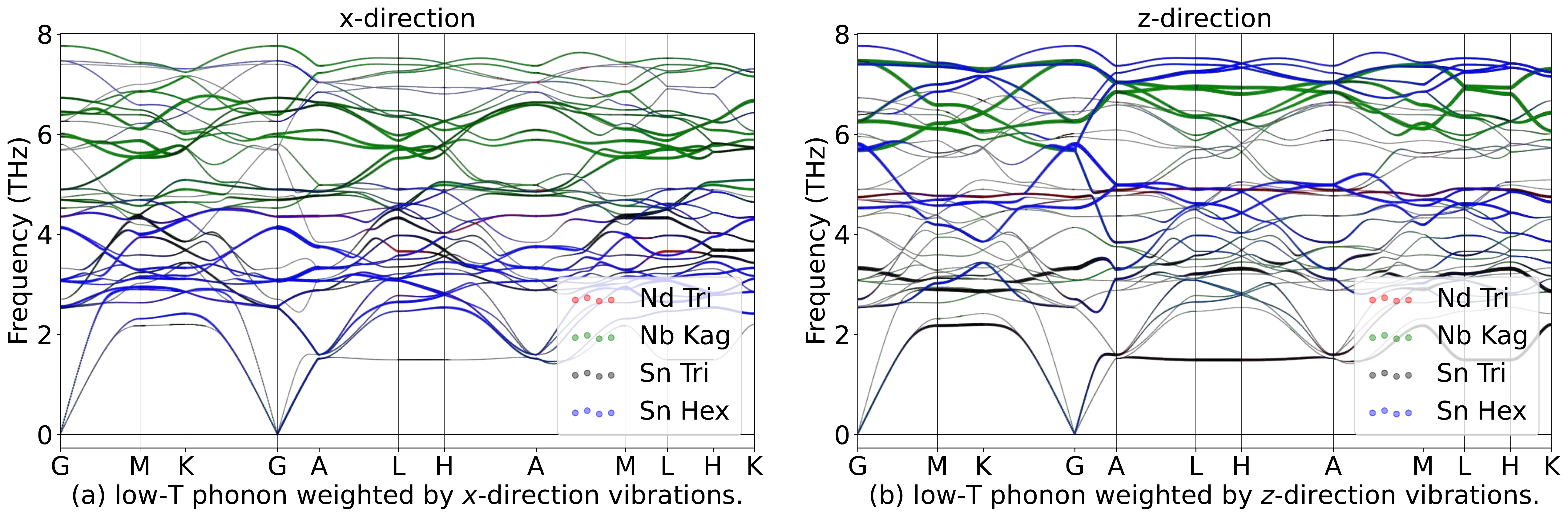}
\hspace{5pt}\label{fig:NdNb6Sn6_relaxed_High_xz}
\includegraphics[width=0.85\textwidth]{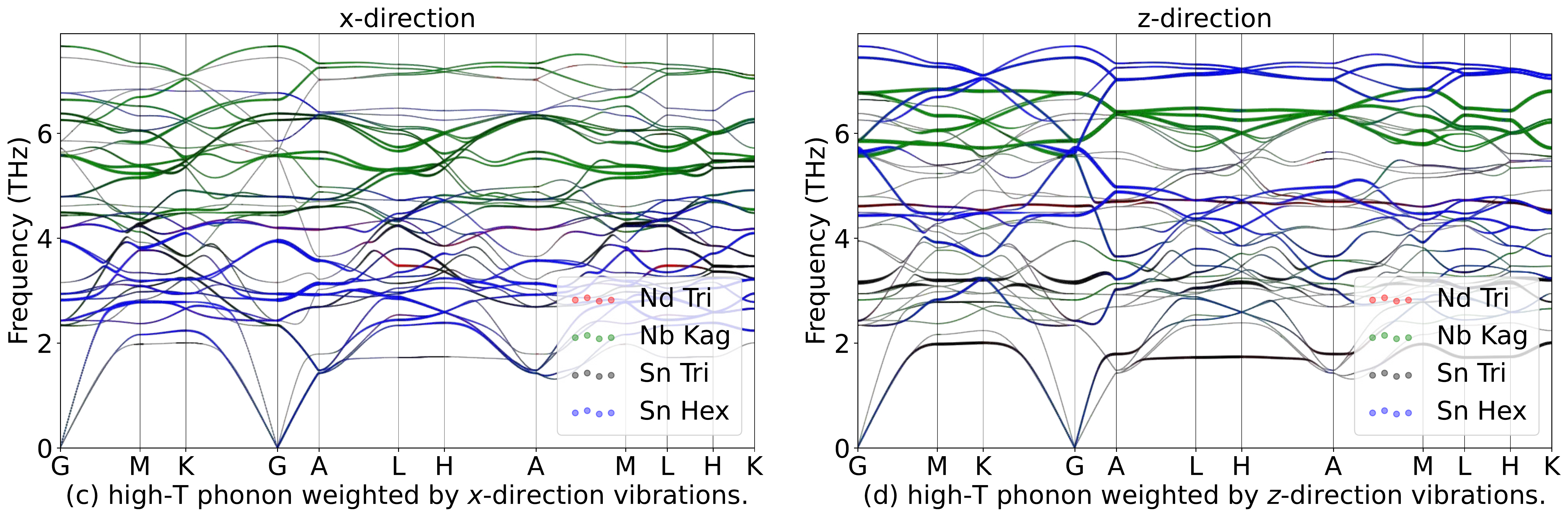}
\caption{Phonon spectrum for NdNb$_6$Sn$_6$in in-plane ($x$) and out-of-plane ($z$) directions.}
\label{fig:NdNb6Sn6phoxyz}
\end{figure}

\begin{table}[htbp]
\global\long\def\arraystretch{1.12}
\captionsetup{width=.95\textwidth}
\caption{IRREPs at high-symmetry points for NdNb$_6$Sn$_6$ phonon spectrum at Low-T.}
\label{tab:191_NdNb6Sn6_Low}
\setlength{\tabcolsep}{1.mm}{
}
\end{table}

\begin{table}[htbp]
\global\long\def\arraystretch{1.12}
\captionsetup{width=.95\textwidth}
\caption{IRREPs at high-symmetry points for NdNb$_6$Sn$_6$ phonon spectrum at High-T.}
\label{tab:191_NdNb6Sn6_High}
\setlength{\tabcolsep}{1.mm}{
%
}
\end{table}
\subsubsection{\label{sms2sec:SmNb6Sn6}\hyperref[smsec:col]{\texorpdfstring{SmNb$_6$Sn$_6$}{}}~{\color{green}  (High-T stable, Low-T stable) }}

\begin{table}[htbp]
\global\long\def\arraystretch{1.12}
\captionsetup{width=.85\textwidth}
\caption{Structure parameters of SmNb$_6$Sn$_6$ after relaxation. It contains 608 electrons. }
\label{tab:SmNb6Sn6}
\setlength{\tabcolsep}{4mm}{
%
}
\end{table}

\begin{figure}[htbp]
\centering
\includegraphics[width=0.85\textwidth]{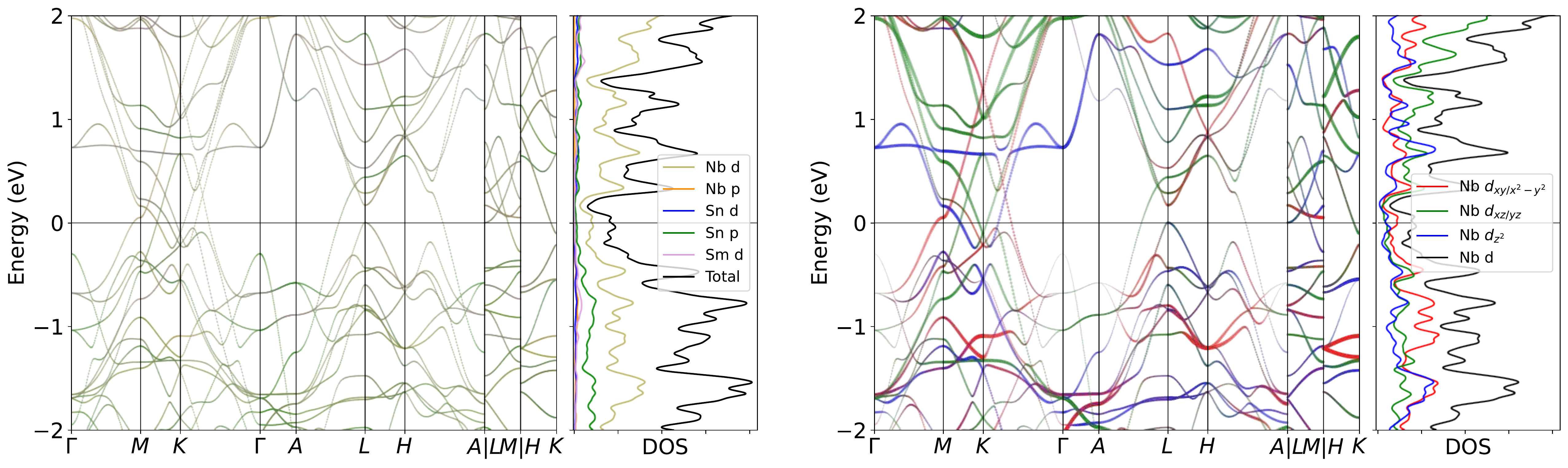}
\caption{Band structure for SmNb$_6$Sn$_6$ without SOC. The projection of Nb $3d$ orbitals is also presented. The band structures clearly reveal the dominating of Nb $3d$ orbitals  near the Fermi level, as well as the pronounced peak.}
\label{fig:SmNb6Sn6band}
\end{figure}

\begin{figure}[H]
\centering
\subfloat[Relaxed phonon at low-T.]{
\label{fig:SmNb6Sn6_relaxed_Low}
\includegraphics[width=0.45\textwidth]{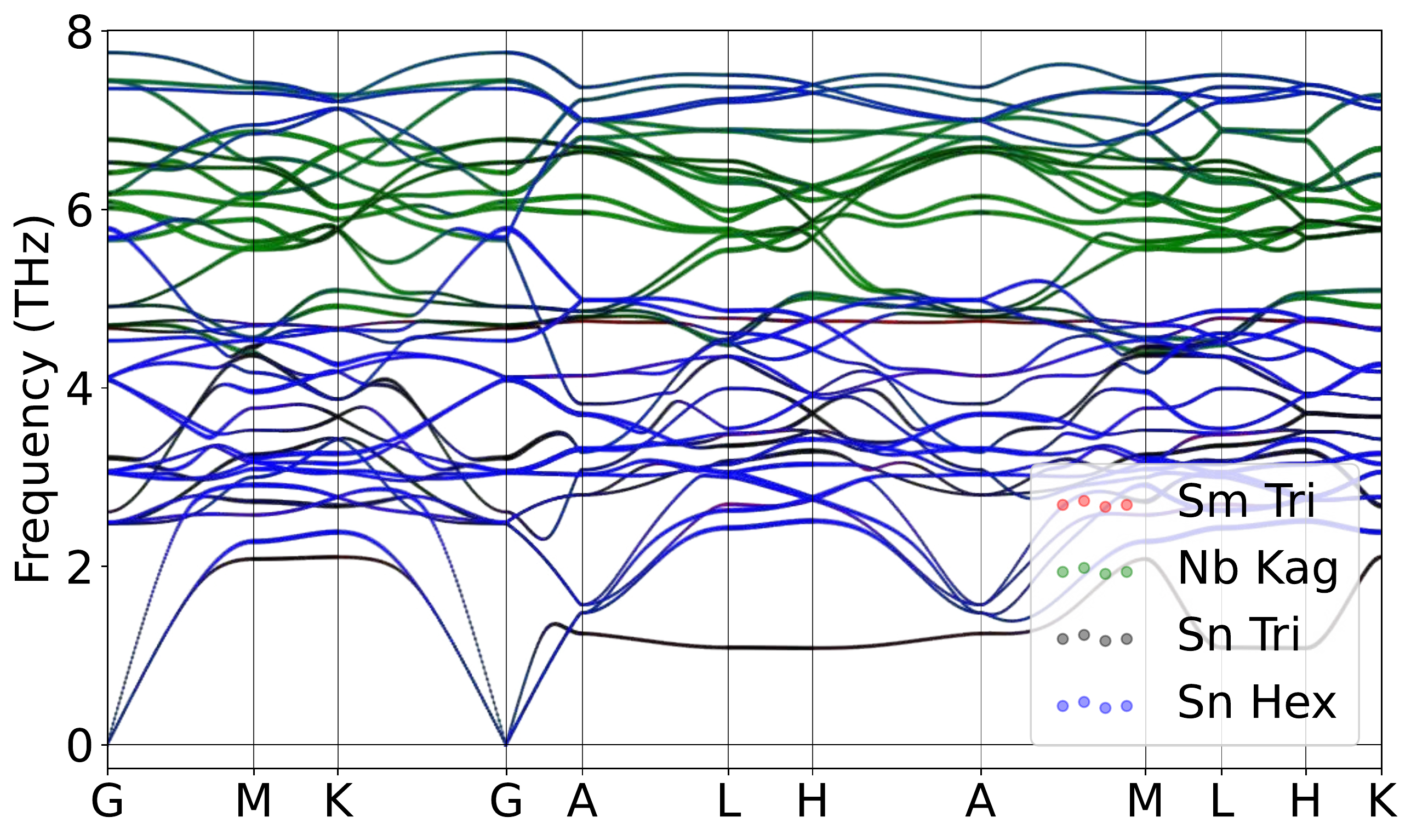}
}
\hspace{5pt}\subfloat[Relaxed phonon at high-T.]{
\label{fig:SmNb6Sn6_relaxed_High}
\includegraphics[width=0.45\textwidth]{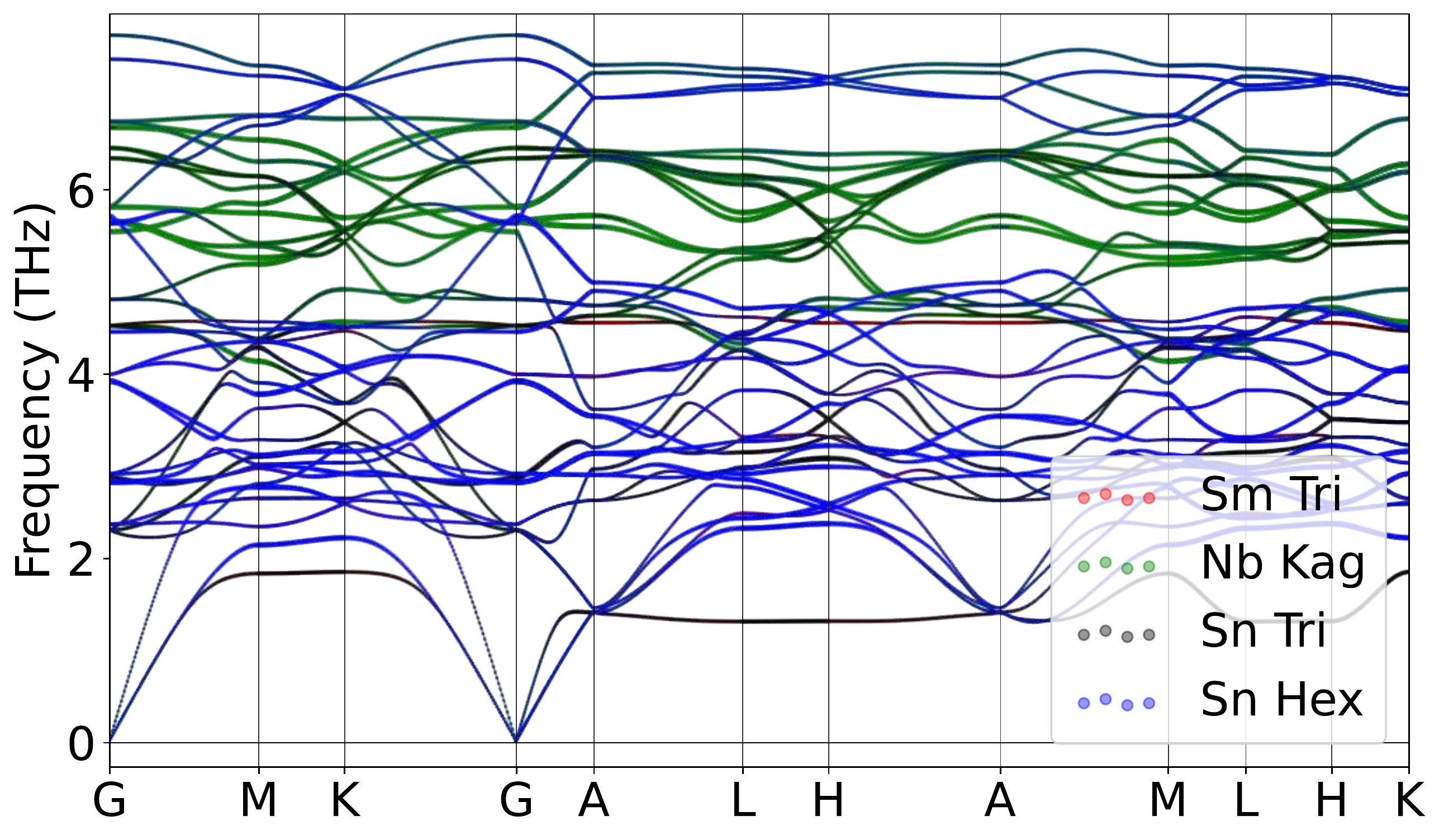}
}
\caption{Atomically projected phonon spectrum for SmNb$_6$Sn$_6$.}
\label{fig:SmNb6Sn6pho}
\end{figure}

\begin{figure}[H]
\centering
\label{fig:SmNb6Sn6_relaxed_Low_xz}
\includegraphics[width=0.85\textwidth]{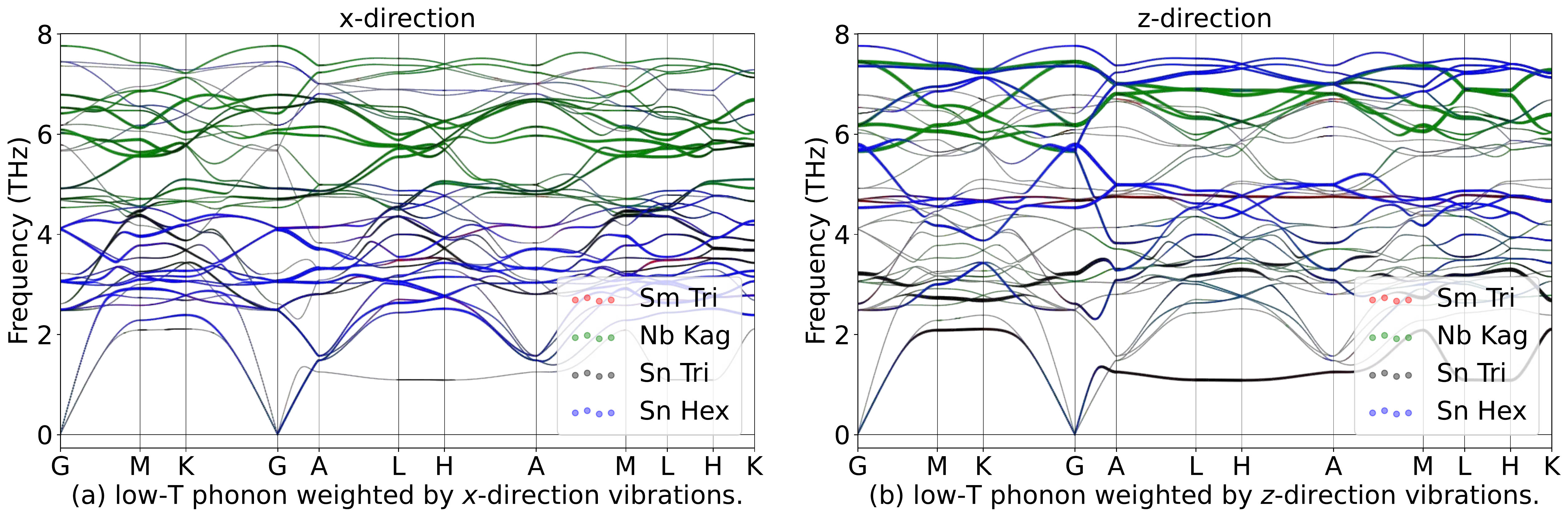}
\hspace{5pt}\label{fig:SmNb6Sn6_relaxed_High_xz}
\includegraphics[width=0.85\textwidth]{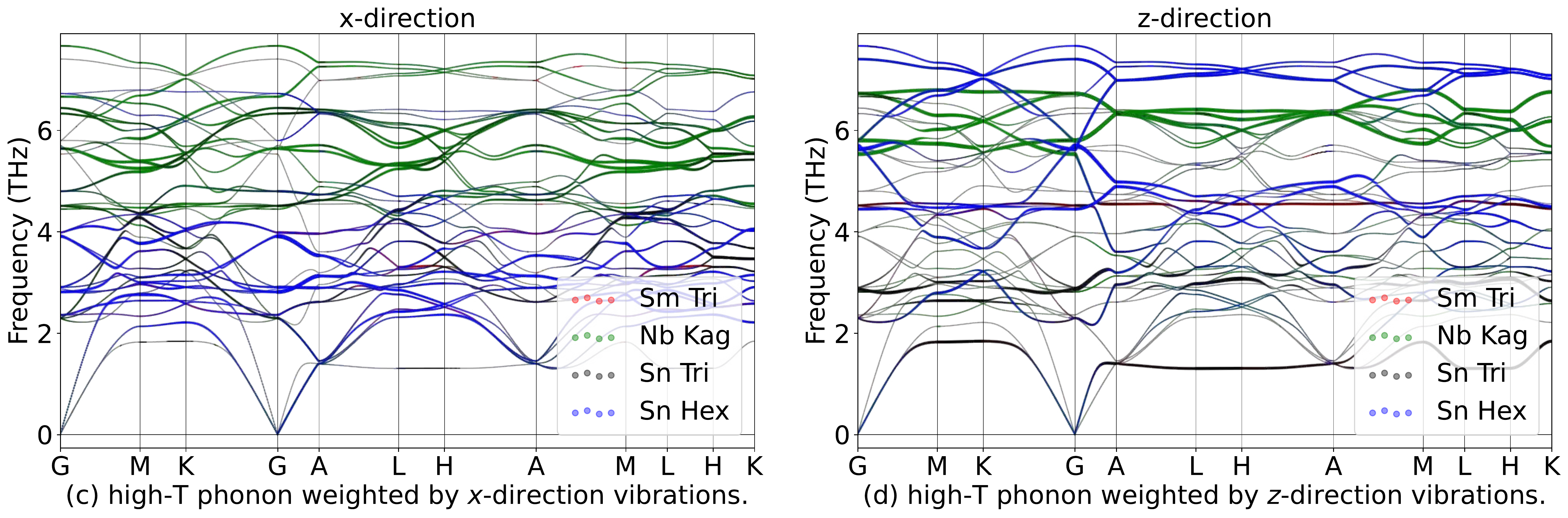}
\caption{Phonon spectrum for SmNb$_6$Sn$_6$in in-plane ($x$) and out-of-plane ($z$) directions.}
\label{fig:SmNb6Sn6phoxyz}
\end{figure}

\begin{table}[htbp]
\global\long\def\arraystretch{1.12}
\captionsetup{width=.95\textwidth}
\caption{IRREPs at high-symmetry points for SmNb$_6$Sn$_6$ phonon spectrum at Low-T.}
\label{tab:191_SmNb6Sn6_Low}
\setlength{\tabcolsep}{1.mm}{
}
\end{table}

\begin{table}[htbp]
\global\long\def\arraystretch{1.12}
\captionsetup{width=.95\textwidth}
\caption{IRREPs at high-symmetry points for SmNb$_6$Sn$_6$ phonon spectrum at High-T.}
\label{tab:191_SmNb6Sn6_High}
\setlength{\tabcolsep}{1.mm}{
%
}
\end{table}
\subsubsection{\label{sms2sec:GdNb6Sn6}\hyperref[smsec:col]{\texorpdfstring{GdNb$_6$Sn$_6$}{}}~{\color{green}  (High-T stable, Low-T stable) }}

\begin{table}[htbp]
\global\long\def\arraystretch{1.12}
\captionsetup{width=.85\textwidth}
\caption{Structure parameters of GdNb$_6$Sn$_6$ after relaxation. It contains 610 electrons. }
\label{tab:GdNb6Sn6}
\setlength{\tabcolsep}{4mm}{
%
}
\end{table}

\begin{figure}[htbp]
\centering
\includegraphics[width=0.85\textwidth]{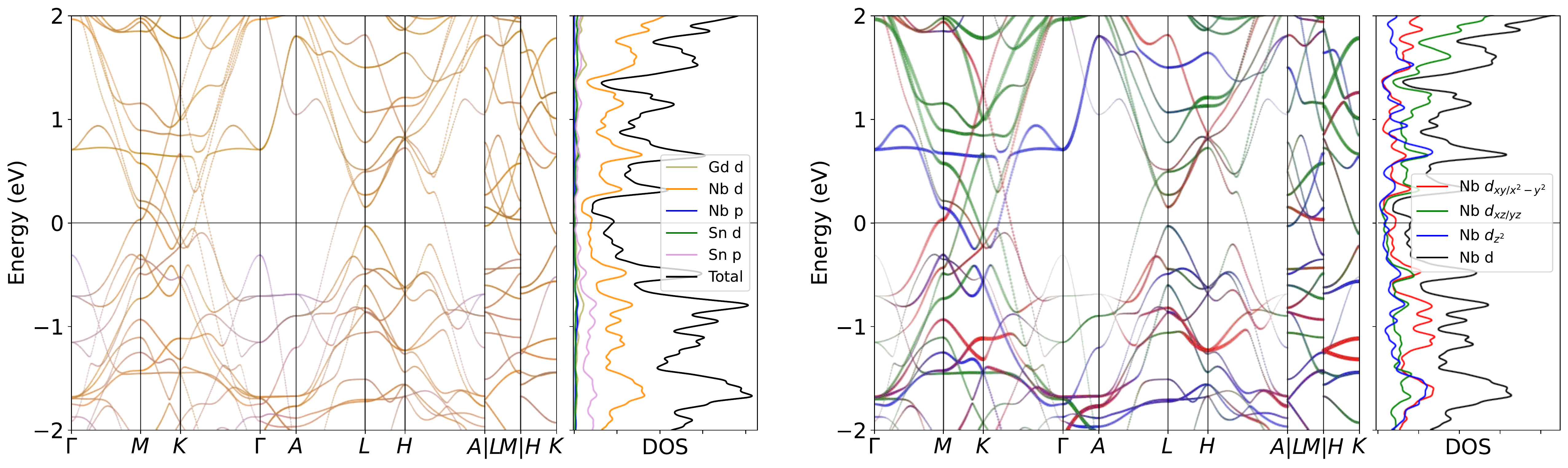}
\caption{Band structure for GdNb$_6$Sn$_6$ without SOC. The projection of Nb $3d$ orbitals is also presented. The band structures clearly reveal the dominating of Nb $3d$ orbitals  near the Fermi level, as well as the pronounced peak.}
\label{fig:GdNb6Sn6band}
\end{figure}

\begin{figure}[H]
\centering
\subfloat[Relaxed phonon at low-T.]{
\label{fig:GdNb6Sn6_relaxed_Low}
\includegraphics[width=0.45\textwidth]{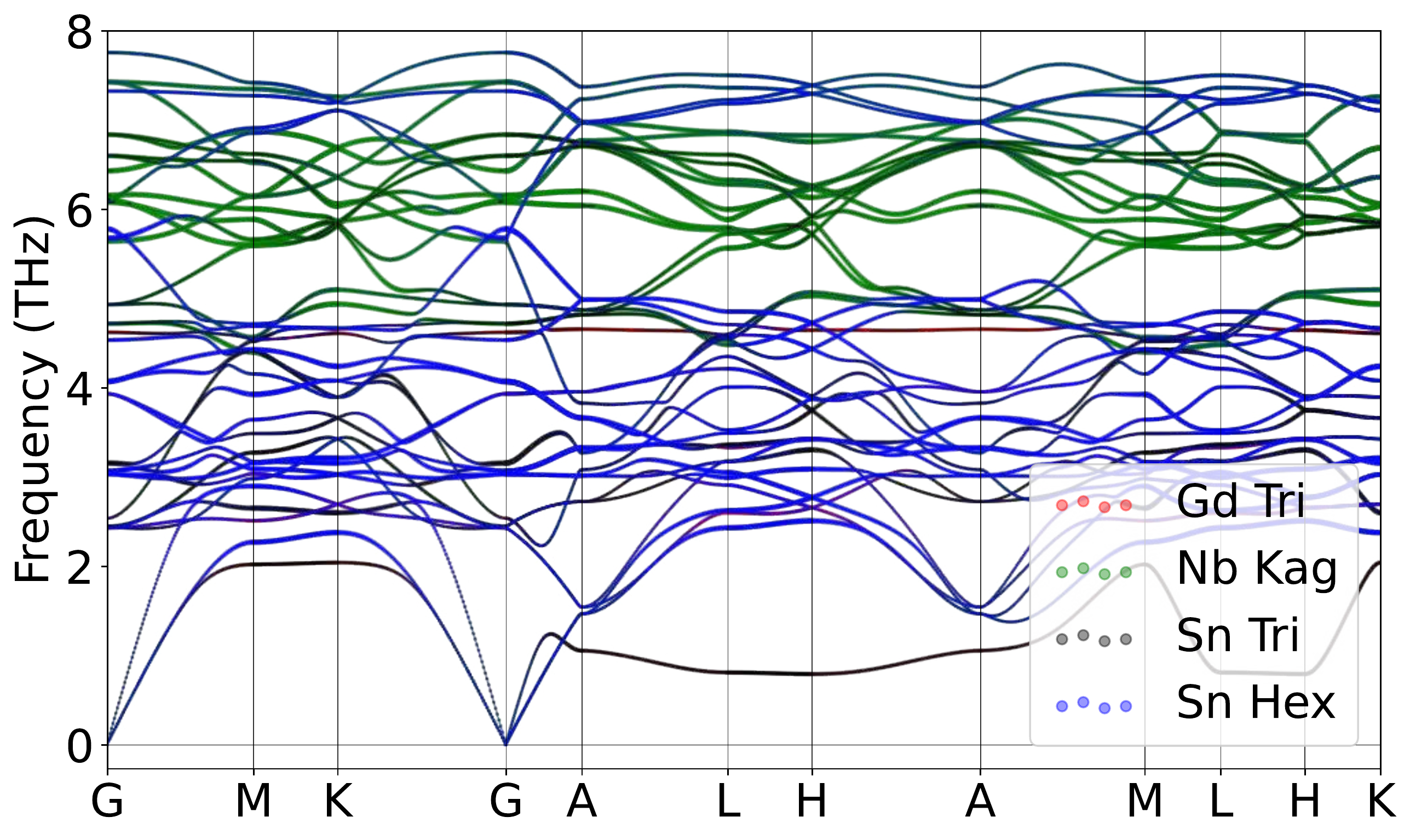}
}
\hspace{5pt}\subfloat[Relaxed phonon at high-T.]{
\label{fig:GdNb6Sn6_relaxed_High}
\includegraphics[width=0.45\textwidth]{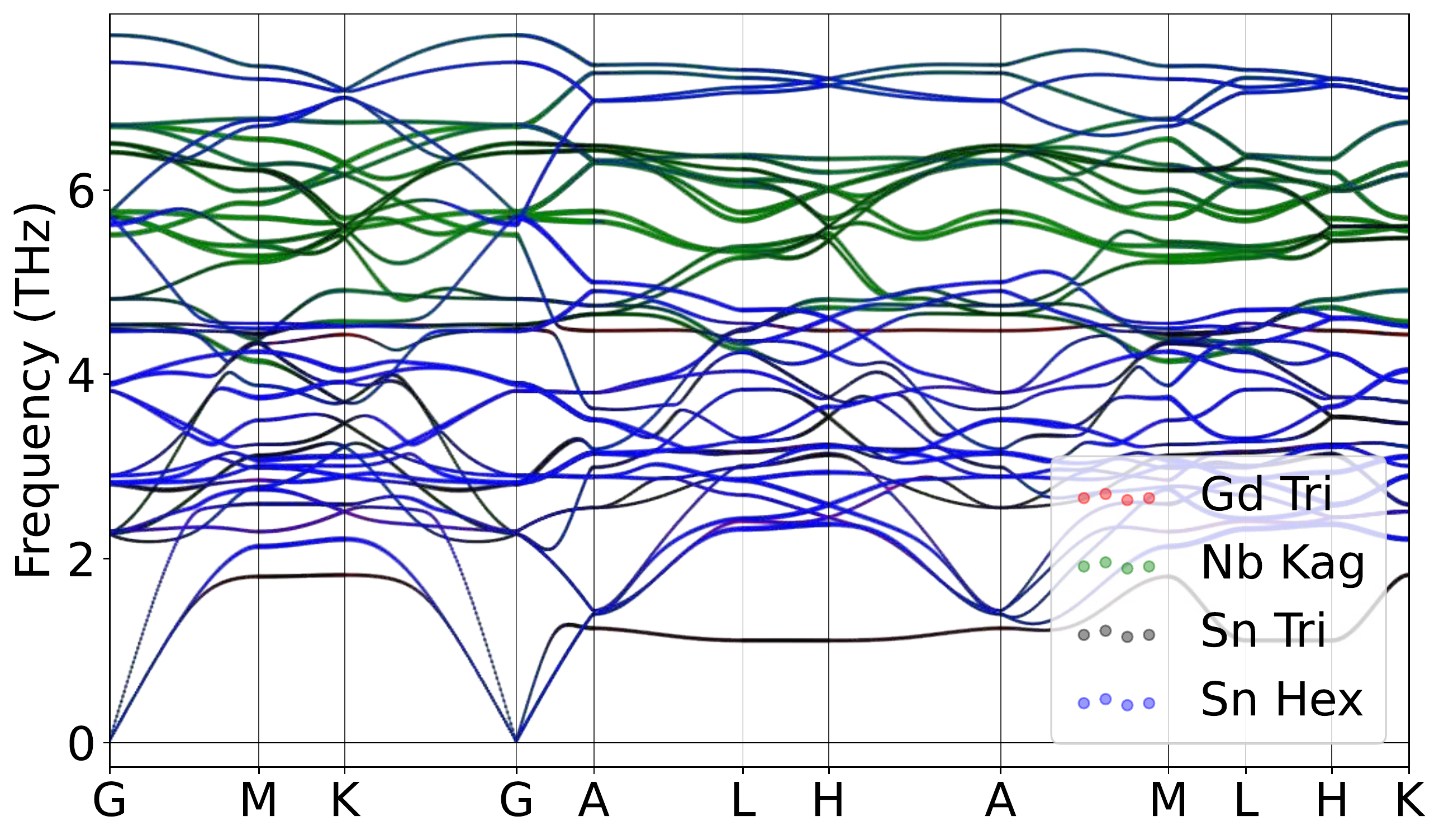}
}
\caption{Atomically projected phonon spectrum for GdNb$_6$Sn$_6$.}
\label{fig:GdNb6Sn6pho}
\end{figure}

\begin{figure}[H]
\centering
\label{fig:GdNb6Sn6_relaxed_Low_xz}
\includegraphics[width=0.85\textwidth]{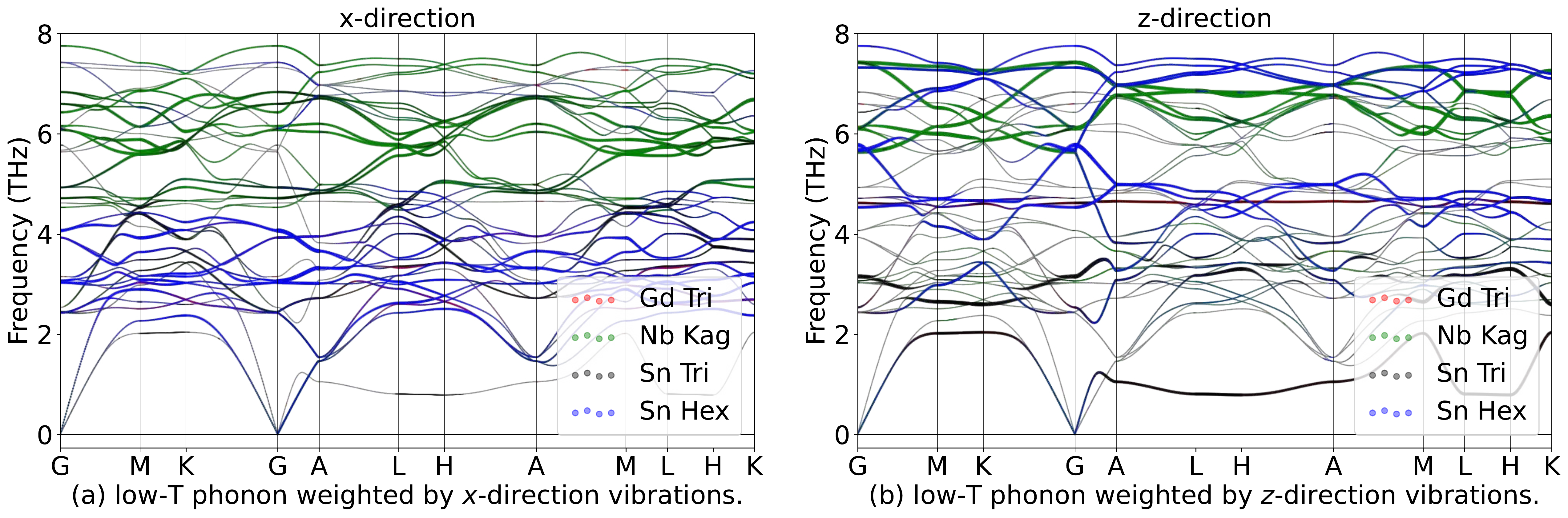}
\hspace{5pt}\label{fig:GdNb6Sn6_relaxed_High_xz}
\includegraphics[width=0.85\textwidth]{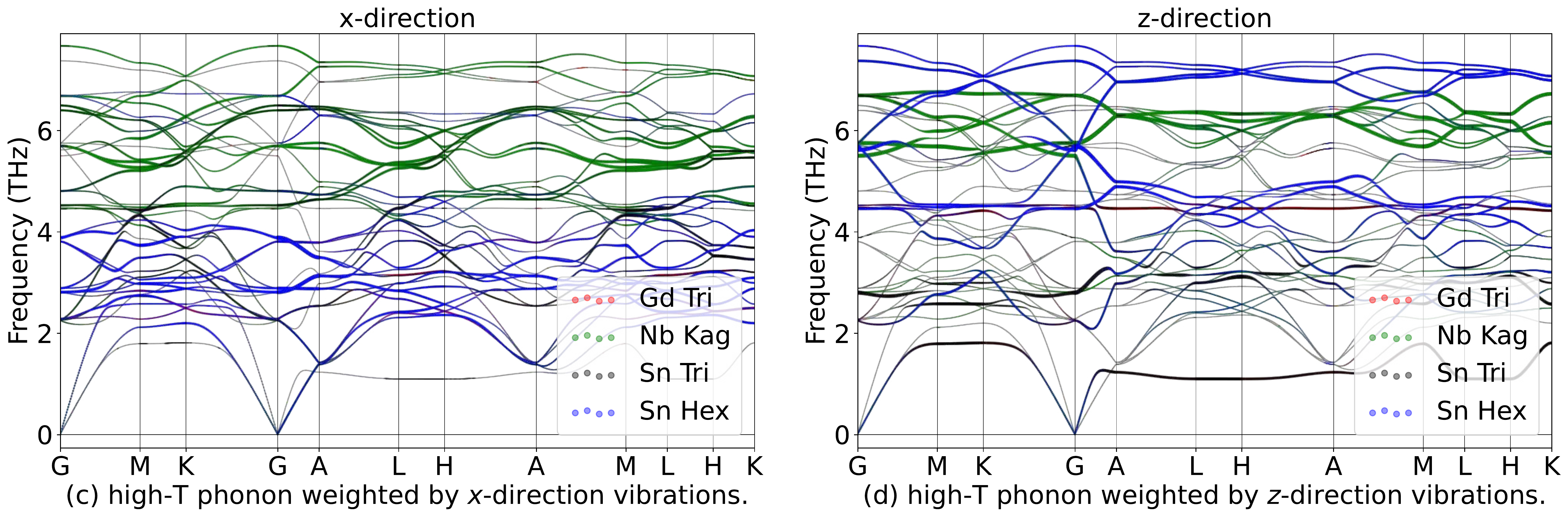}
\caption{Phonon spectrum for GdNb$_6$Sn$_6$in in-plane ($x$) and out-of-plane ($z$) directions.}
\label{fig:GdNb6Sn6phoxyz}
\end{figure}

\begin{table}[htbp]
\global\long\def\arraystretch{1.12}
\captionsetup{width=.95\textwidth}
\caption{IRREPs at high-symmetry points for GdNb$_6$Sn$_6$ phonon spectrum at Low-T.}
\label{tab:191_GdNb6Sn6_Low}
\setlength{\tabcolsep}{1.mm}{
}
\end{table}

\begin{table}[htbp]
\global\long\def\arraystretch{1.12}
\captionsetup{width=.95\textwidth}
\caption{IRREPs at high-symmetry points for GdNb$_6$Sn$_6$ phonon spectrum at High-T.}
\label{tab:191_GdNb6Sn6_High}
\setlength{\tabcolsep}{1.mm}{
%
}
\end{table}
\subsubsection{\label{sms2sec:TbNb6Sn6}\hyperref[smsec:col]{\texorpdfstring{TbNb$_6$Sn$_6$}{}}~{\color{green}  (High-T stable, Low-T stable) }}

\begin{table}[htbp]
\global\long\def\arraystretch{1.12}
\captionsetup{width=.85\textwidth}
\caption{Structure parameters of TbNb$_6$Sn$_6$ after relaxation. It contains 611 electrons. }
\label{tab:TbNb6Sn6}
\setlength{\tabcolsep}{4mm}{
%
}
\end{table}

\begin{figure}[htbp]
\centering
\includegraphics[width=0.85\textwidth]{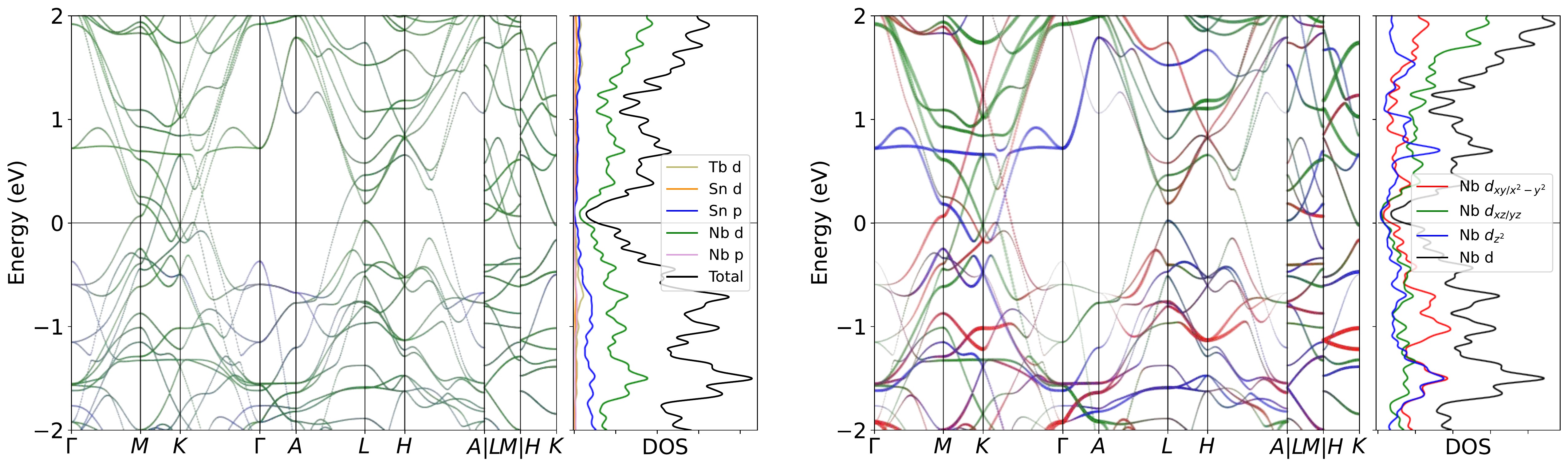}
\caption{Band structure for TbNb$_6$Sn$_6$ without SOC. The projection of Nb $3d$ orbitals is also presented. The band structures clearly reveal the dominating of Nb $3d$ orbitals  near the Fermi level, as well as the pronounced peak.}
\label{fig:TbNb6Sn6band}
\end{figure}

\begin{figure}[H]
\centering
\subfloat[Relaxed phonon at low-T.]{
\label{fig:TbNb6Sn6_relaxed_Low}
\includegraphics[width=0.45\textwidth]{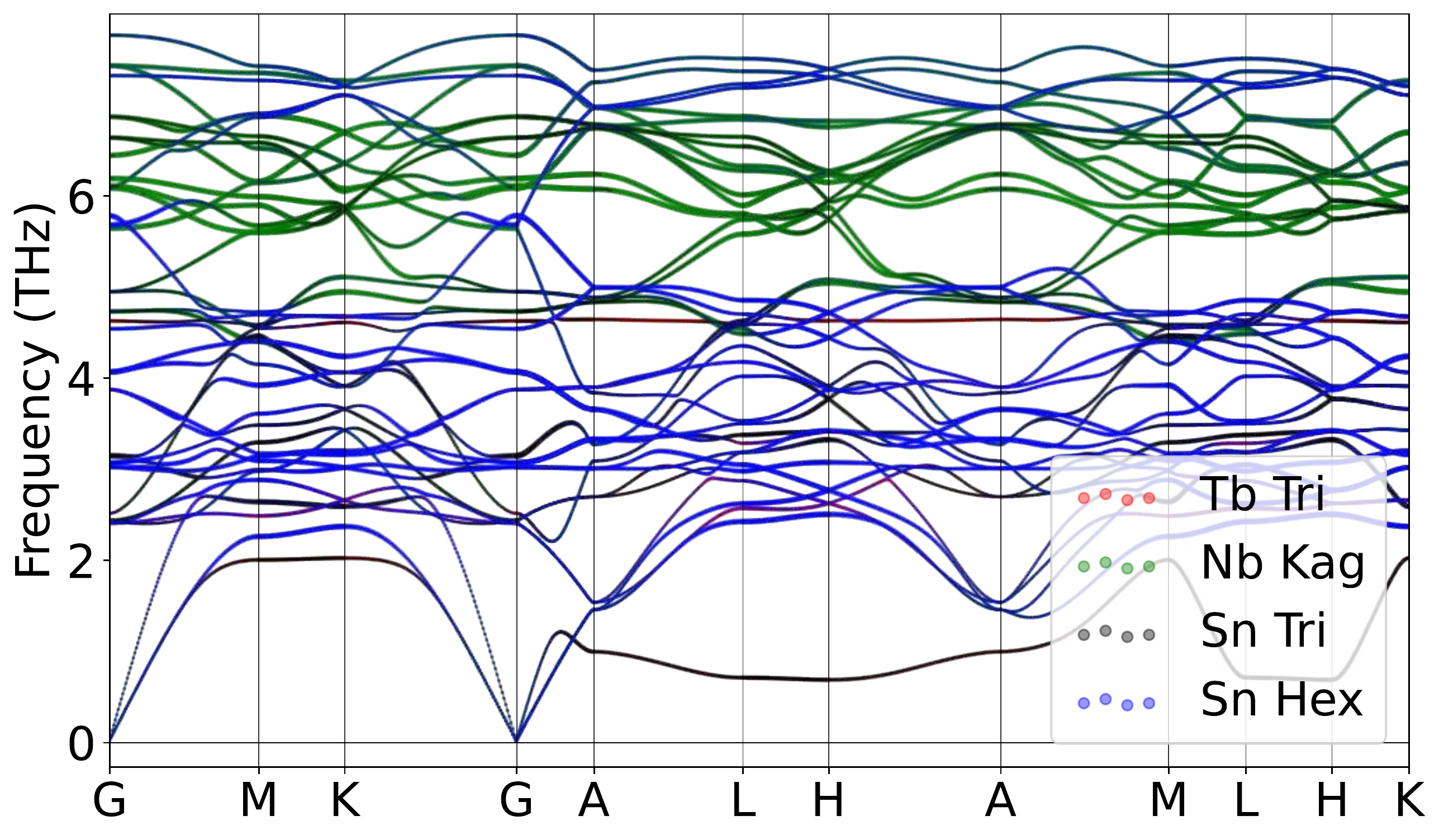}
}
\hspace{5pt}\subfloat[Relaxed phonon at high-T.]{
\label{fig:TbNb6Sn6_relaxed_High}
\includegraphics[width=0.45\textwidth]{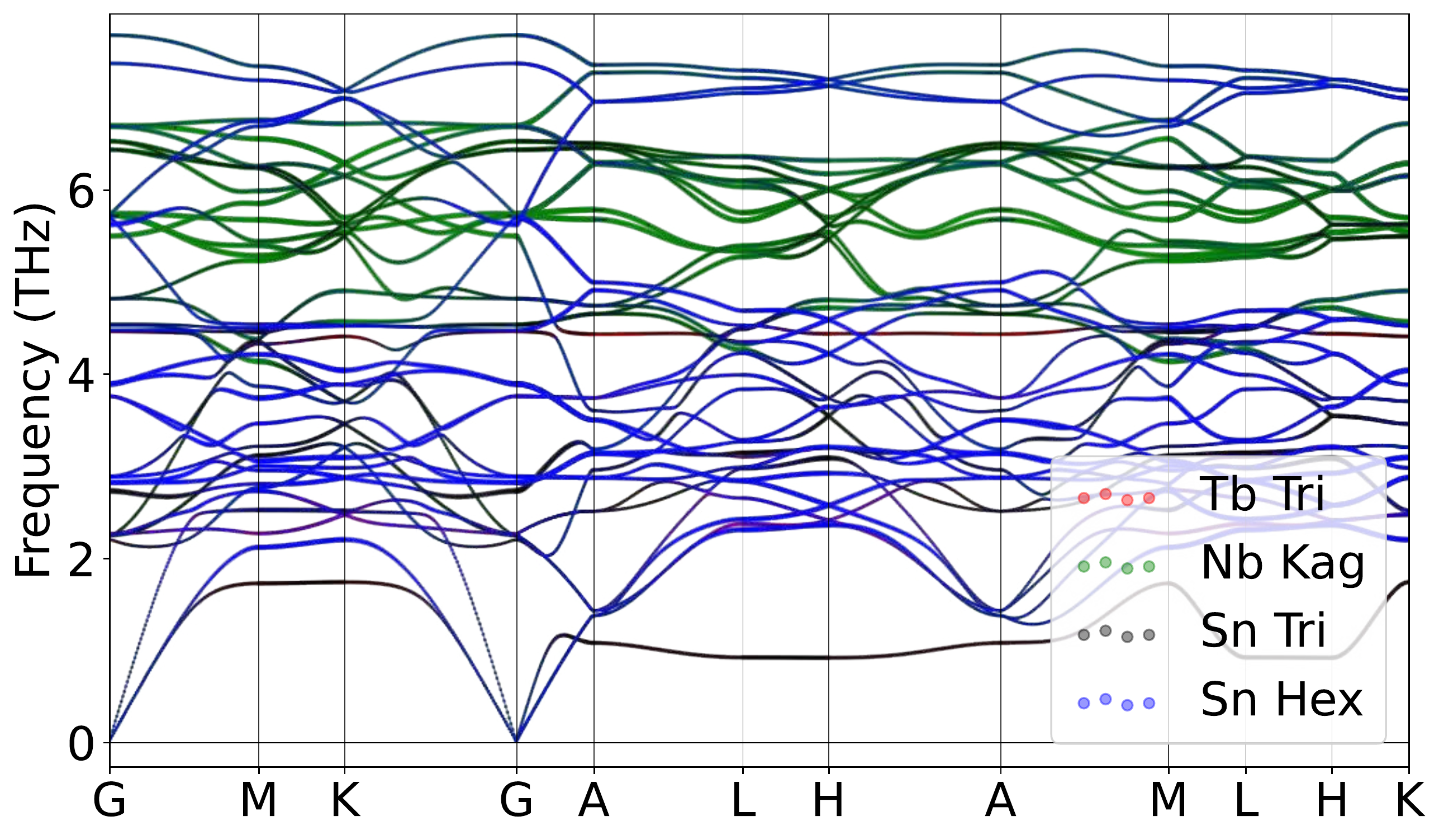}
}
\caption{Atomically projected phonon spectrum for TbNb$_6$Sn$_6$.}
\label{fig:TbNb6Sn6pho}
\end{figure}

\begin{figure}[H]
\centering
\label{fig:TbNb6Sn6_relaxed_Low_xz}
\includegraphics[width=0.85\textwidth]{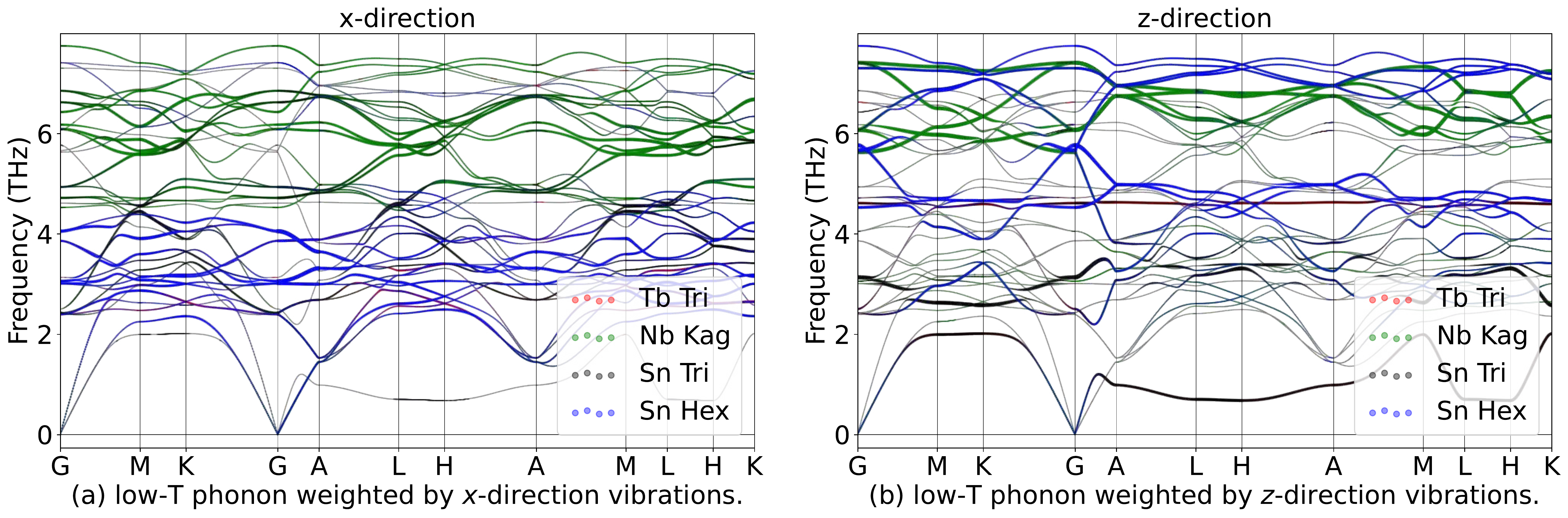}
\hspace{5pt}\label{fig:TbNb6Sn6_relaxed_High_xz}
\includegraphics[width=0.85\textwidth]{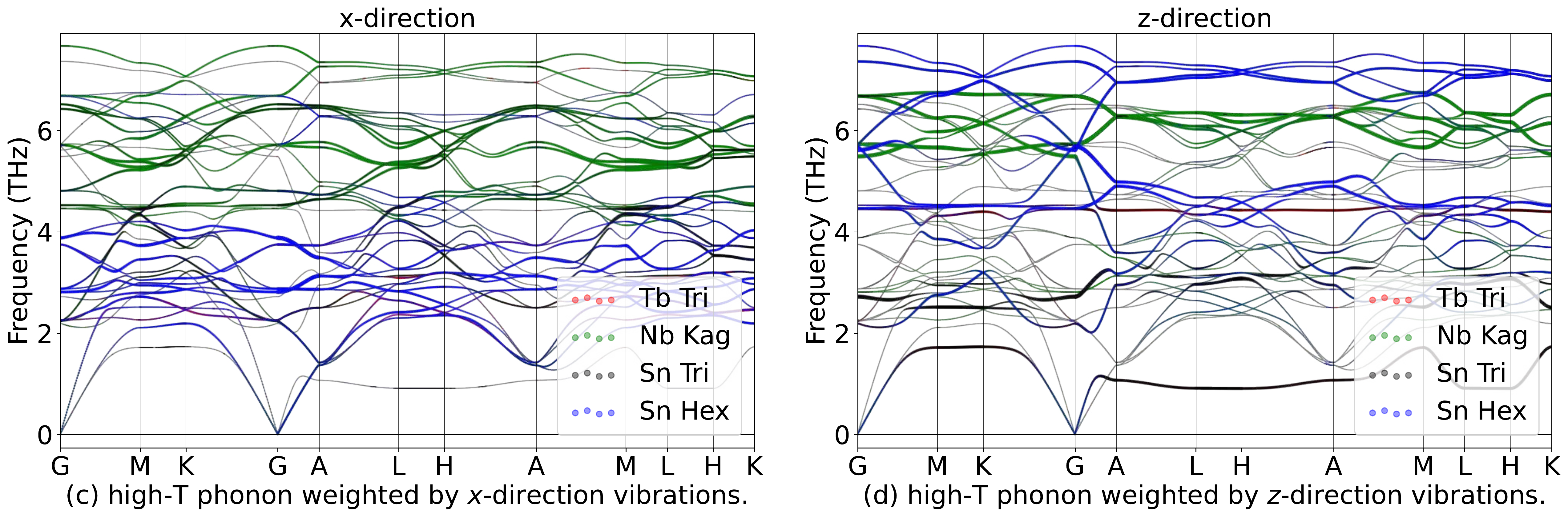}
\caption{Phonon spectrum for TbNb$_6$Sn$_6$in in-plane ($x$) and out-of-plane ($z$) directions.}
\label{fig:TbNb6Sn6phoxyz}
\end{figure}

\begin{table}[htbp]
\global\long\def\arraystretch{1.12}
\captionsetup{width=.95\textwidth}
\caption{IRREPs at high-symmetry points for TbNb$_6$Sn$_6$ phonon spectrum at Low-T.}
\label{tab:191_TbNb6Sn6_Low}
\setlength{\tabcolsep}{1.mm}{
}
\end{table}

\begin{table}[htbp]
\global\long\def\arraystretch{1.12}
\captionsetup{width=.95\textwidth}
\caption{IRREPs at high-symmetry points for TbNb$_6$Sn$_6$ phonon spectrum at High-T.}
\label{tab:191_TbNb6Sn6_High}
\setlength{\tabcolsep}{1.mm}{
%
}
\end{table}
\subsubsection{\label{sms2sec:DyNb6Sn6}\hyperref[smsec:col]{\texorpdfstring{DyNb$_6$Sn$_6$}{}}~{\color{green}  (High-T stable, Low-T stable) }}

\begin{table}[htbp]
\global\long\def\arraystretch{1.12}
\captionsetup{width=.85\textwidth}
\caption{Structure parameters of DyNb$_6$Sn$_6$ after relaxation. It contains 612 electrons. }
\label{tab:DyNb6Sn6}
\setlength{\tabcolsep}{4mm}{
%
}
\end{table}

\begin{figure}[htbp]
\centering
\includegraphics[width=0.85\textwidth]{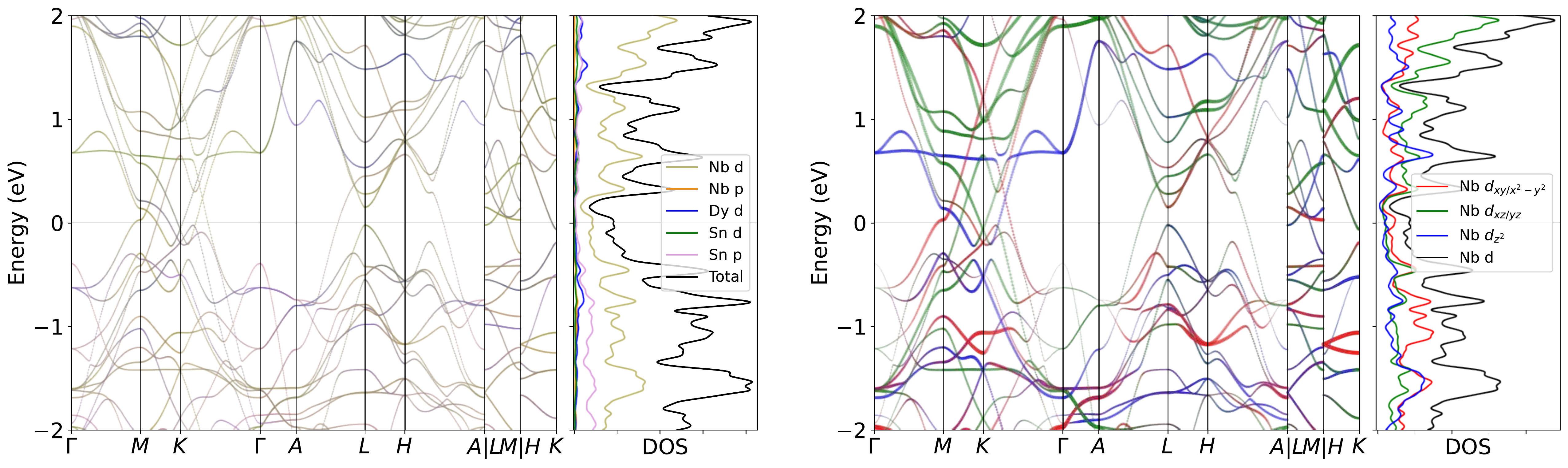}
\caption{Band structure for DyNb$_6$Sn$_6$ without SOC. The projection of Nb $3d$ orbitals is also presented. The band structures clearly reveal the dominating of Nb $3d$ orbitals  near the Fermi level, as well as the pronounced peak.}
\label{fig:DyNb6Sn6band}
\end{figure}

\begin{figure}[H]
\centering
\subfloat[Relaxed phonon at low-T.]{
\label{fig:DyNb6Sn6_relaxed_Low}
\includegraphics[width=0.45\textwidth]{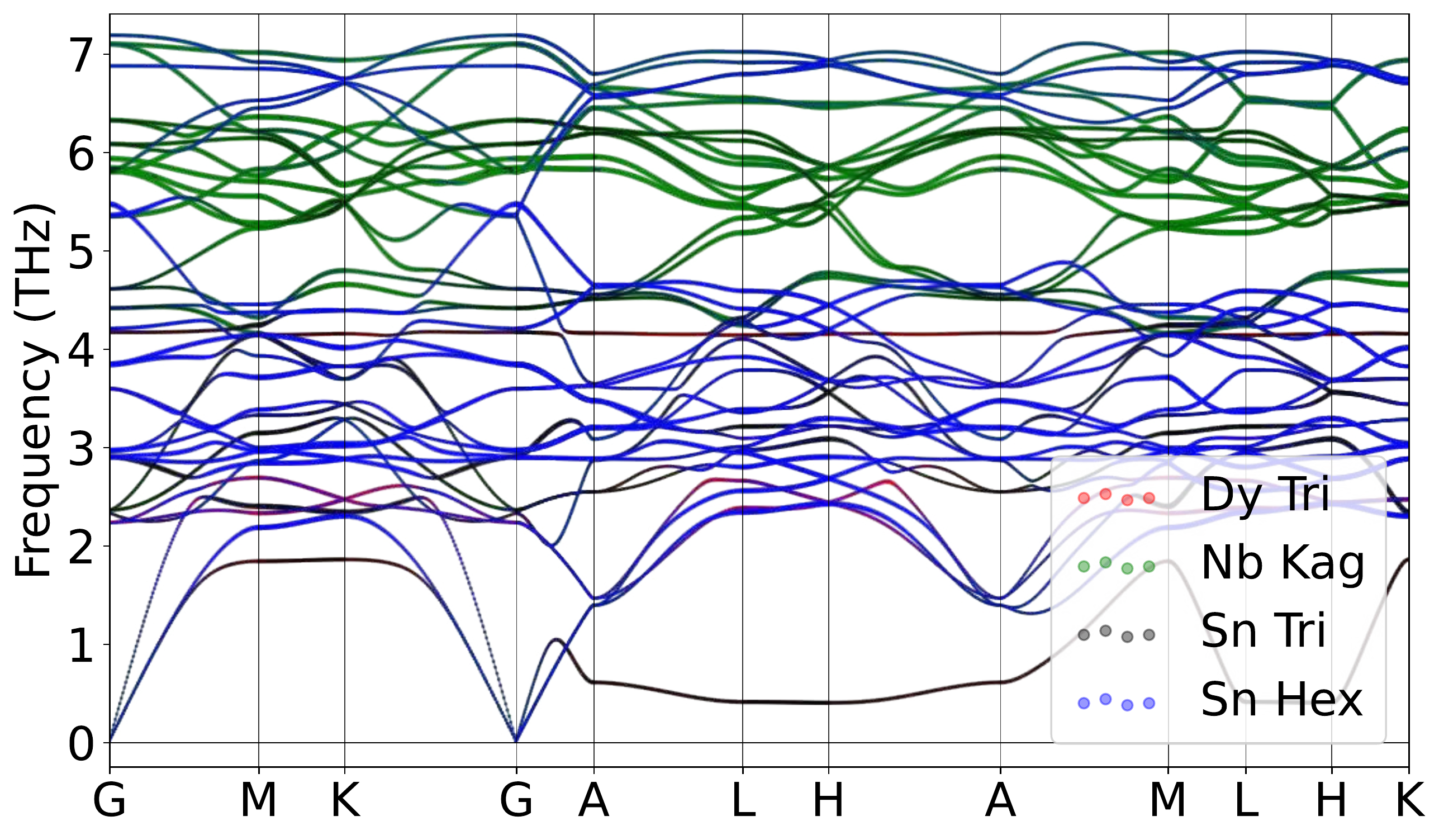}
}
\hspace{5pt}\subfloat[Relaxed phonon at high-T.]{
\label{fig:DyNb6Sn6_relaxed_High}
\includegraphics[width=0.45\textwidth]{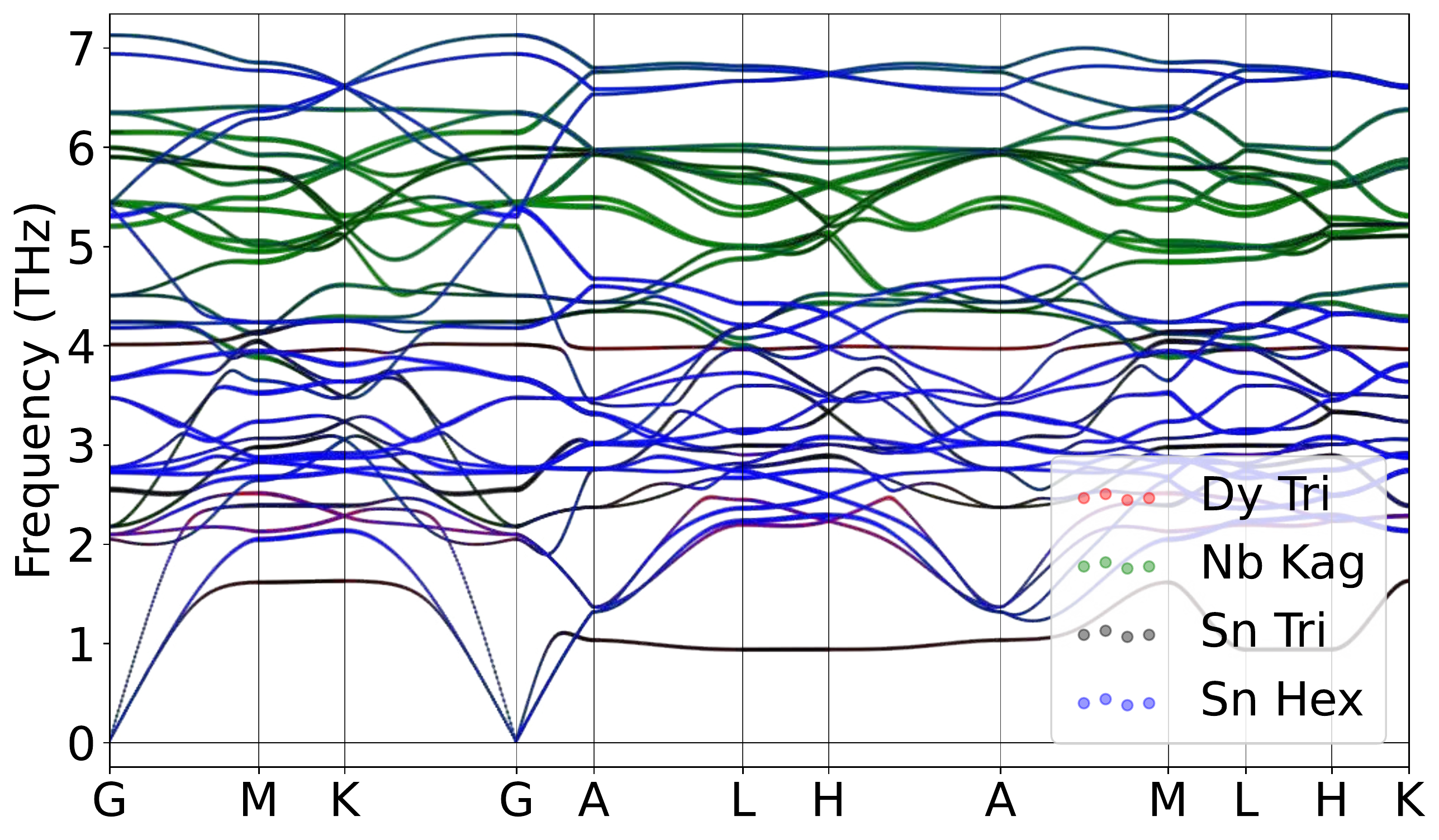}
}
\caption{Atomically projected phonon spectrum for DyNb$_6$Sn$_6$.}
\label{fig:DyNb6Sn6pho}
\end{figure}

\begin{figure}[H]
\centering
\label{fig:DyNb6Sn6_relaxed_Low_xz}
\includegraphics[width=0.85\textwidth]{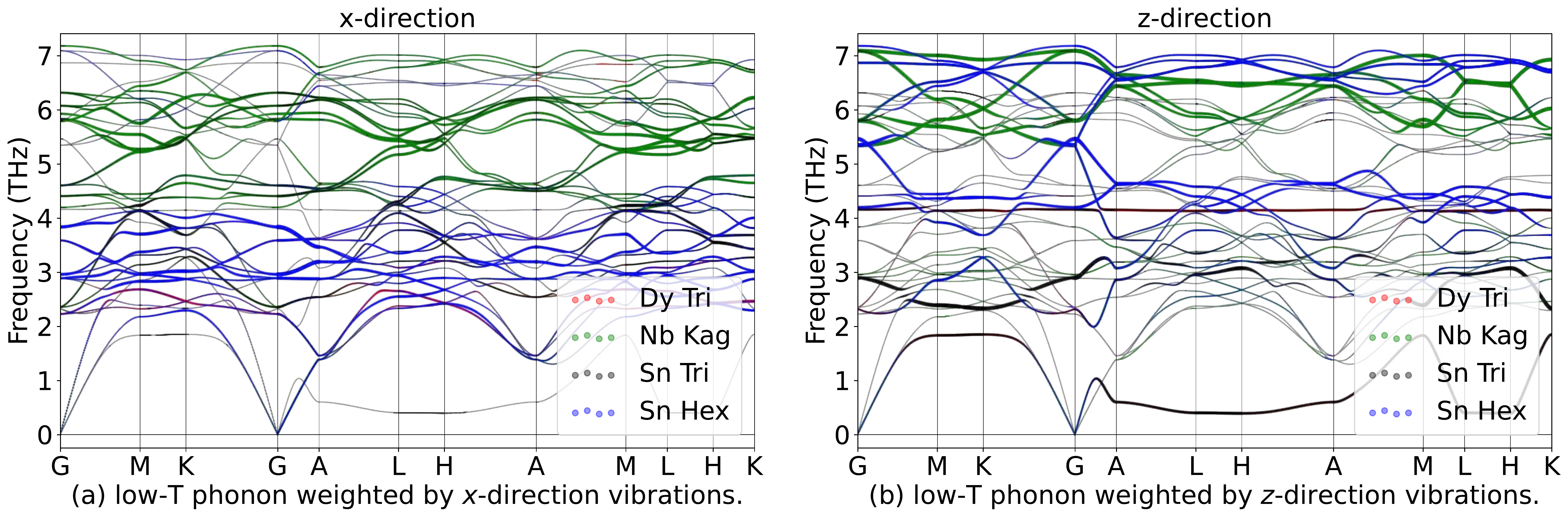}
\hspace{5pt}\label{fig:DyNb6Sn6_relaxed_High_xz}
\includegraphics[width=0.85\textwidth]{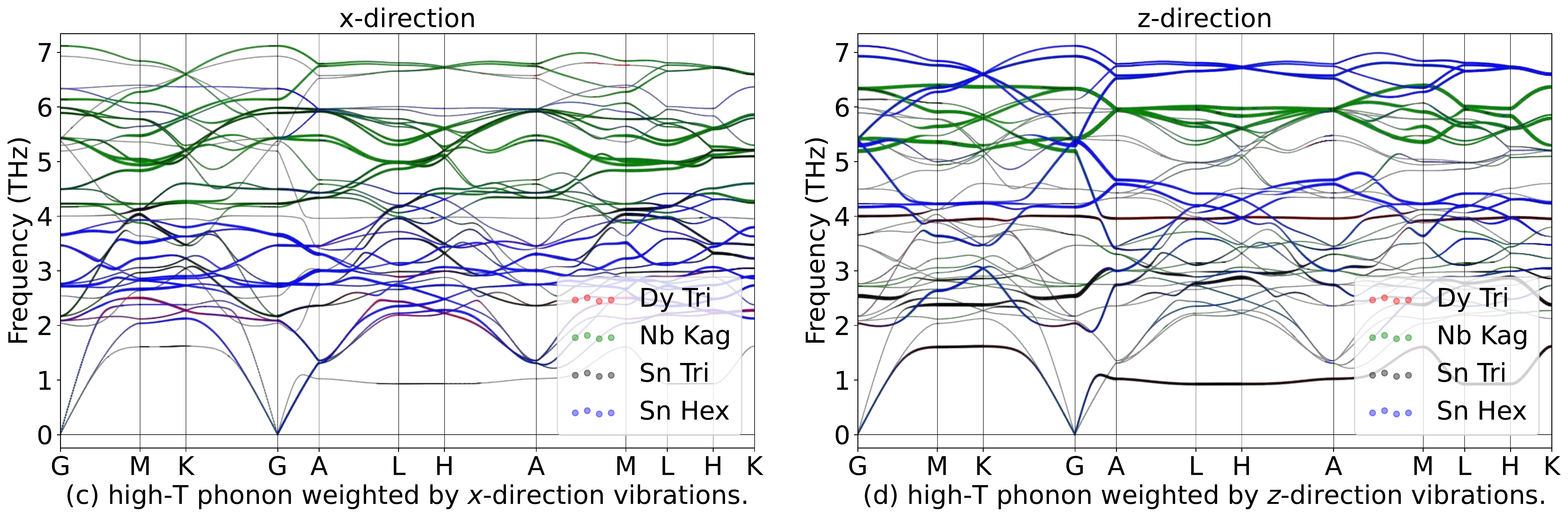}
\caption{Phonon spectrum for DyNb$_6$Sn$_6$in in-plane ($x$) and out-of-plane ($z$) directions.}
\label{fig:DyNb6Sn6phoxyz}
\end{figure}

\begin{table}[htbp]
\global\long\def\arraystretch{1.12}
\captionsetup{width=.95\textwidth}
\caption{IRREPs at high-symmetry points for DyNb$_6$Sn$_6$ phonon spectrum at Low-T.}
\label{tab:191_DyNb6Sn6_Low}
\setlength{\tabcolsep}{1.mm}{
}
\end{table}

\begin{table}[htbp]
\global\long\def\arraystretch{1.12}
\captionsetup{width=.95\textwidth}
\caption{IRREPs at high-symmetry points for DyNb$_6$Sn$_6$ phonon spectrum at High-T.}
\label{tab:191_DyNb6Sn6_High}
\setlength{\tabcolsep}{1.mm}{
%
}
\end{table}
\subsubsection{\label{sms2sec:HoNb6Sn6}\hyperref[smsec:col]{\texorpdfstring{HoNb$_6$Sn$_6$}{}}~{\color{blue}  (High-T stable, Low-T unstable) }}

\begin{table}[htbp]
\global\long\def\arraystretch{1.12}
\captionsetup{width=.85\textwidth}
\caption{Structure parameters of HoNb$_6$Sn$_6$ after relaxation. It contains 613 electrons. }
\label{tab:HoNb6Sn6}
\setlength{\tabcolsep}{4mm}{
%
}
\end{table}

\begin{figure}[htbp]
\centering
\includegraphics[width=0.85\textwidth]{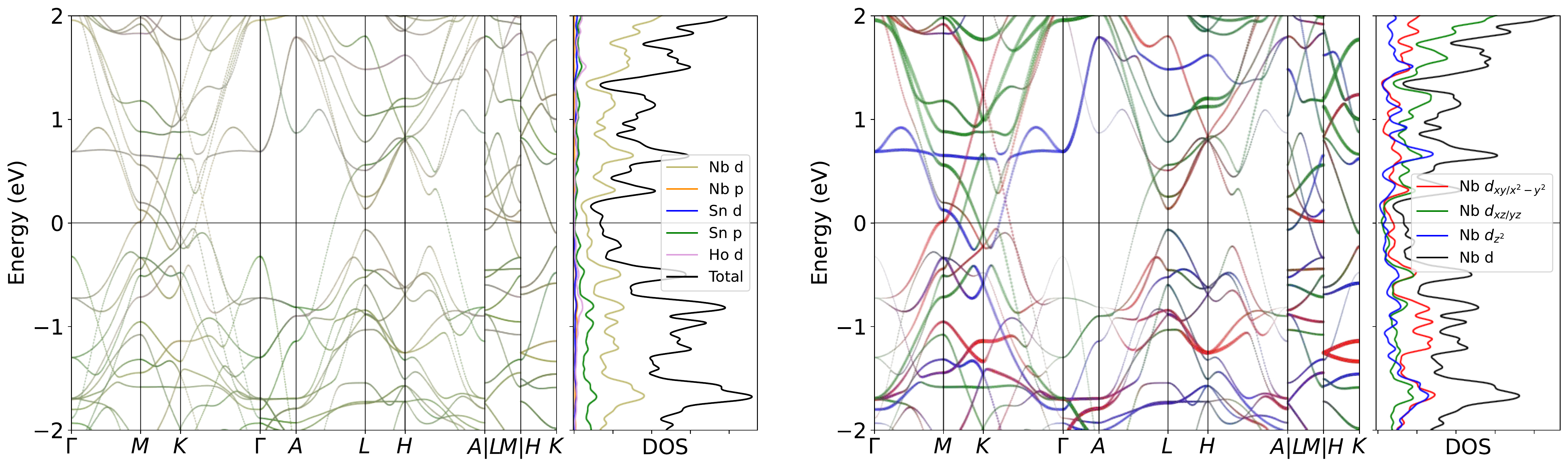}
\caption{Band structure for HoNb$_6$Sn$_6$ without SOC. The projection of Nb $3d$ orbitals is also presented. The band structures clearly reveal the dominating of Nb $3d$ orbitals  near the Fermi level, as well as the pronounced peak.}
\label{fig:HoNb6Sn6band}
\end{figure}

\begin{figure}[H]
\centering
\subfloat[Relaxed phonon at low-T.]{
\label{fig:HoNb6Sn6_relaxed_Low}
\includegraphics[width=0.45\textwidth]{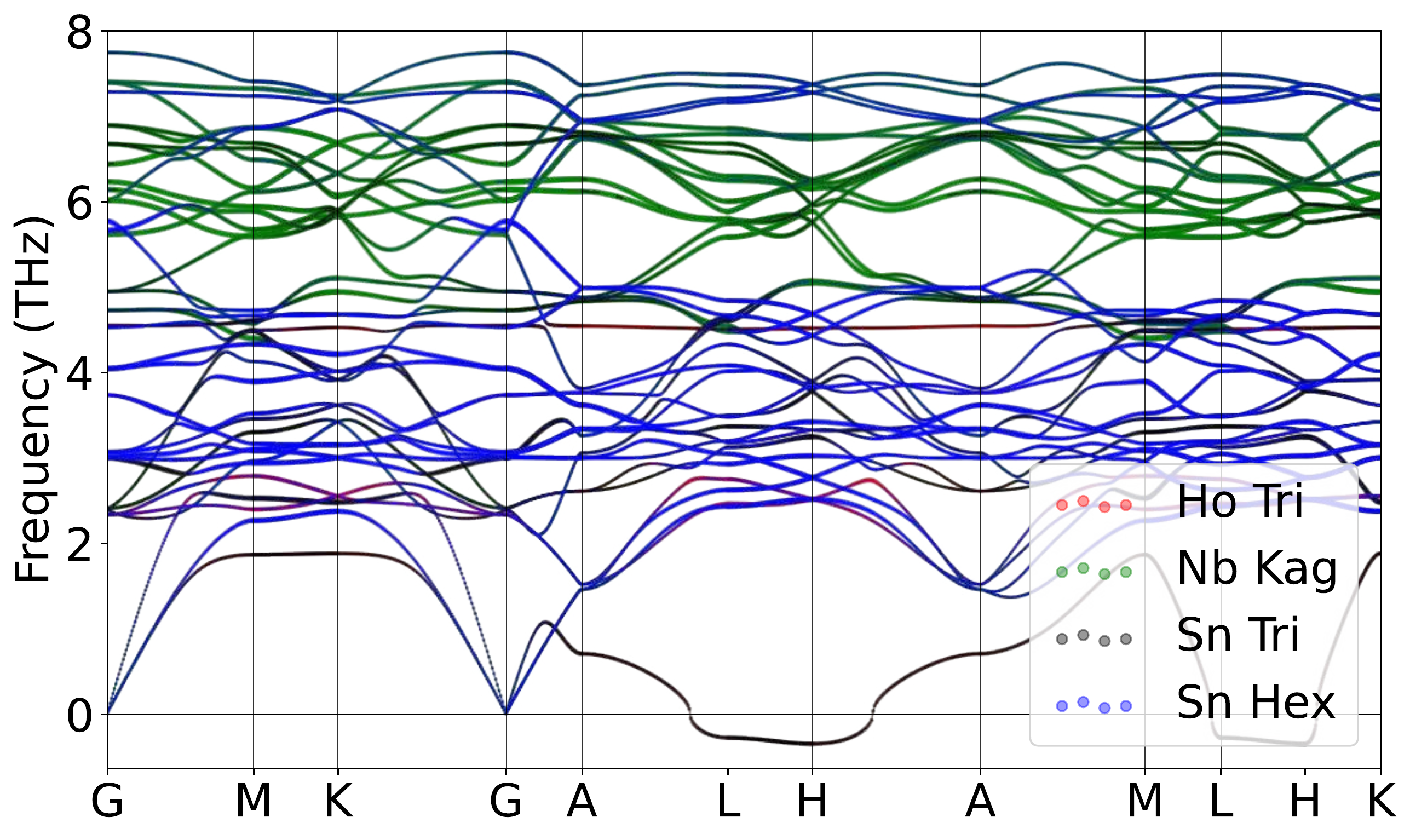}
}
\hspace{5pt}\subfloat[Relaxed phonon at high-T.]{
\label{fig:HoNb6Sn6_relaxed_High}
\includegraphics[width=0.45\textwidth]{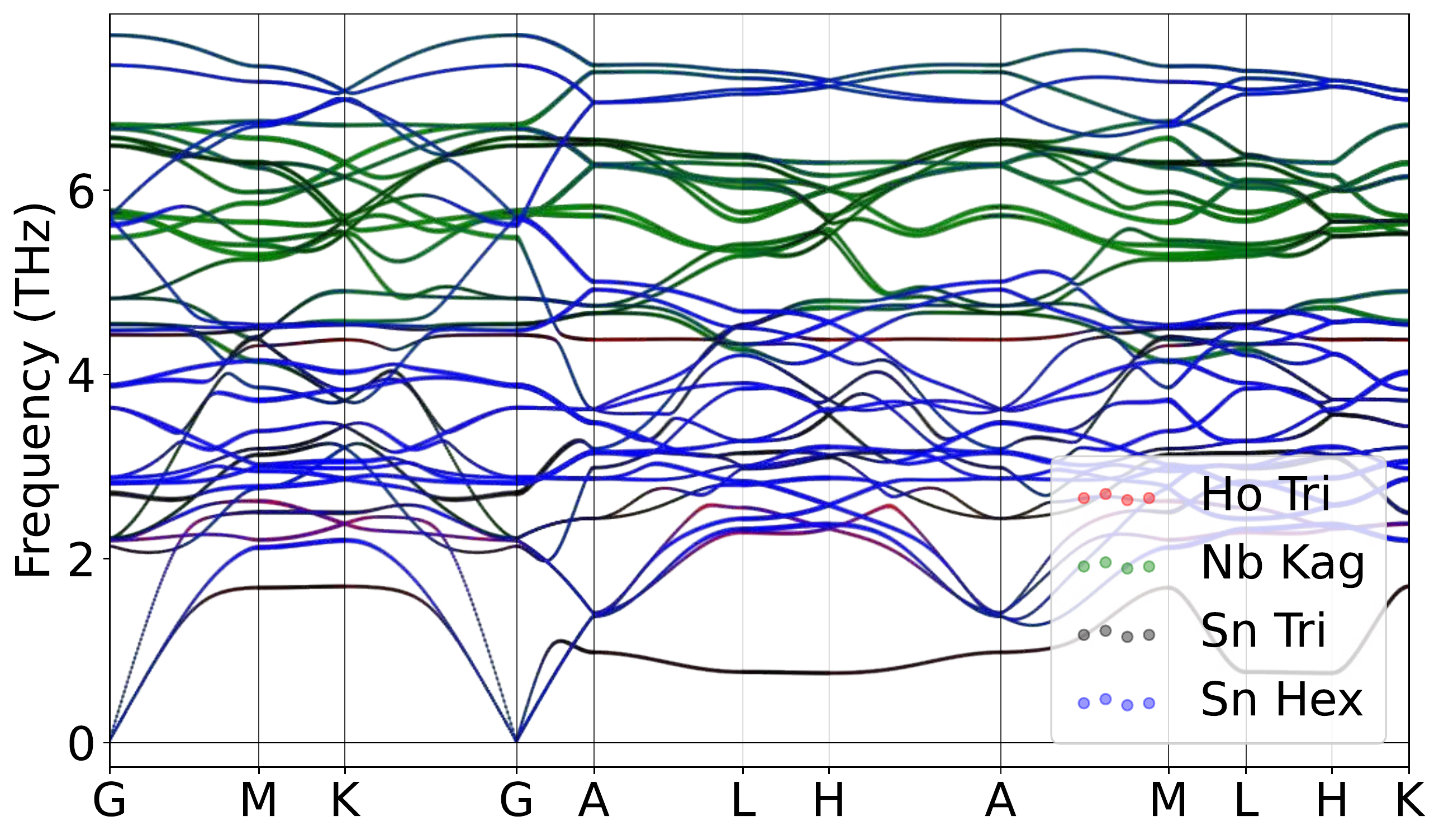}
}
\caption{Atomically projected phonon spectrum for HoNb$_6$Sn$_6$.}
\label{fig:HoNb6Sn6pho}
\end{figure}

\begin{figure}[H]
\centering
\label{fig:HoNb6Sn6_relaxed_Low_xz}
\includegraphics[width=0.85\textwidth]{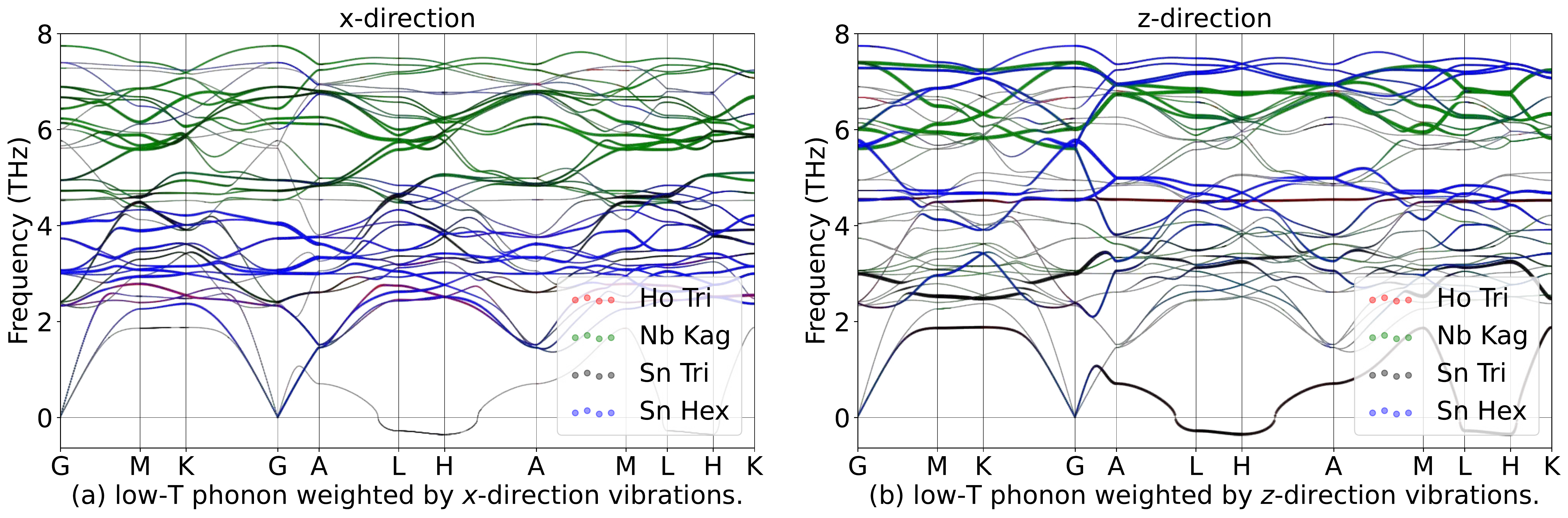}
\hspace{5pt}\label{fig:HoNb6Sn6_relaxed_High_xz}
\includegraphics[width=0.85\textwidth]{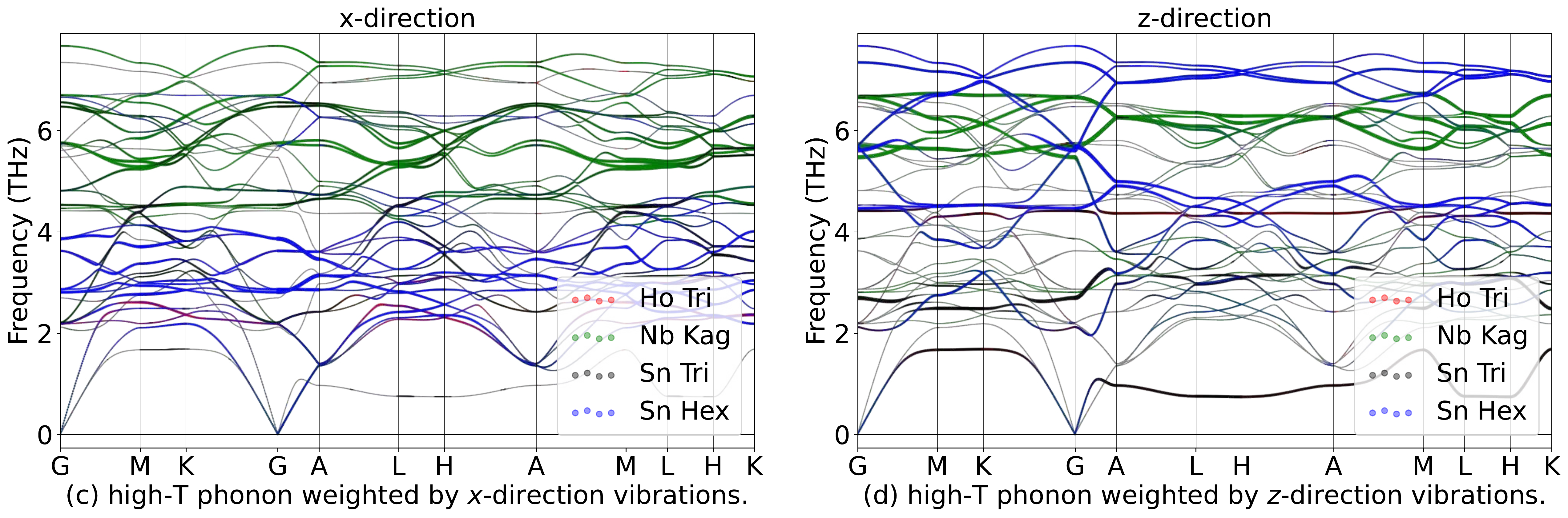}
\caption{Phonon spectrum for HoNb$_6$Sn$_6$in in-plane ($x$) and out-of-plane ($z$) directions.}
\label{fig:HoNb6Sn6phoxyz}
\end{figure}

\begin{table}[htbp]
\global\long\def\arraystretch{1.12}
\captionsetup{width=.95\textwidth}
\caption{IRREPs at high-symmetry points for HoNb$_6$Sn$_6$ phonon spectrum at Low-T.}
\label{tab:191_HoNb6Sn6_Low}
\setlength{\tabcolsep}{1.mm}{
}
\end{table}

\begin{table}[htbp]
\global\long\def\arraystretch{1.12}
\captionsetup{width=.95\textwidth}
\caption{IRREPs at high-symmetry points for HoNb$_6$Sn$_6$ phonon spectrum at High-T.}
\label{tab:191_HoNb6Sn6_High}
\setlength{\tabcolsep}{1.mm}{
%
}
\end{table}
\subsubsection{\label{sms2sec:ErNb6Sn6}\hyperref[smsec:col]{\texorpdfstring{ErNb$_6$Sn$_6$}{}}~{\color{blue}  (High-T stable, Low-T unstable) }}

\begin{table}[htbp]
\global\long\def\arraystretch{1.12}
\captionsetup{width=.85\textwidth}
\caption{Structure parameters of ErNb$_6$Sn$_6$ after relaxation. It contains 614 electrons. }
\label{tab:ErNb6Sn6}
\setlength{\tabcolsep}{4mm}{
%
}
\end{table}

\begin{figure}[htbp]
\centering
\includegraphics[width=0.85\textwidth]{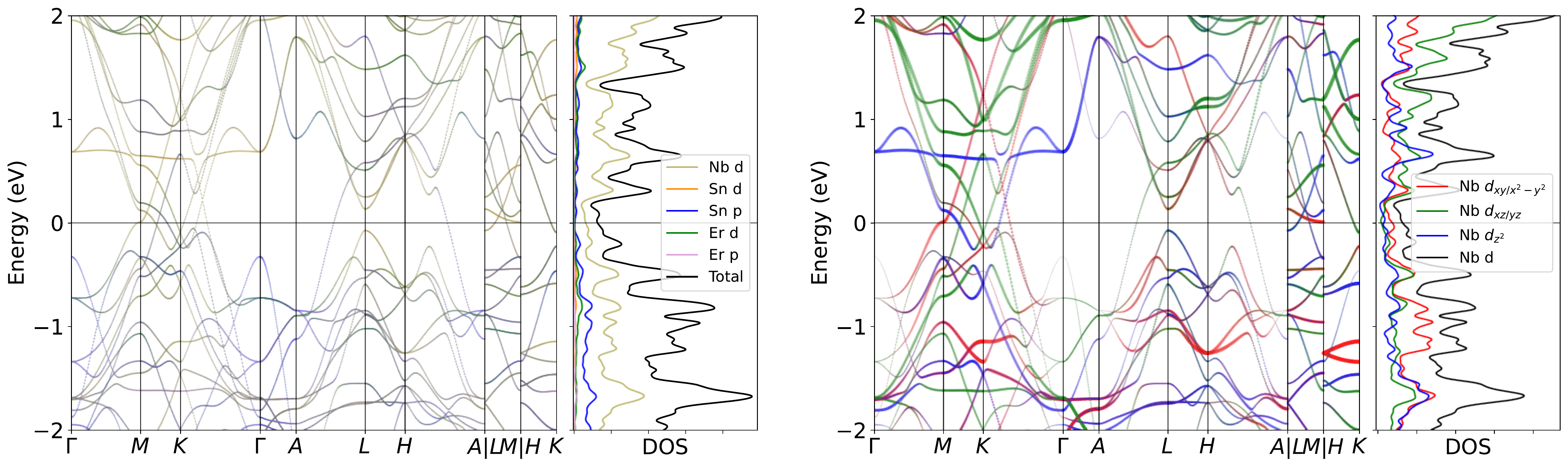}
\caption{Band structure for ErNb$_6$Sn$_6$ without SOC. The projection of Nb $3d$ orbitals is also presented. The band structures clearly reveal the dominating of Nb $3d$ orbitals  near the Fermi level, as well as the pronounced peak.}
\label{fig:ErNb6Sn6band}
\end{figure}

\begin{figure}[H]
\centering
\subfloat[Relaxed phonon at low-T.]{
\label{fig:ErNb6Sn6_relaxed_Low}
\includegraphics[width=0.45\textwidth]{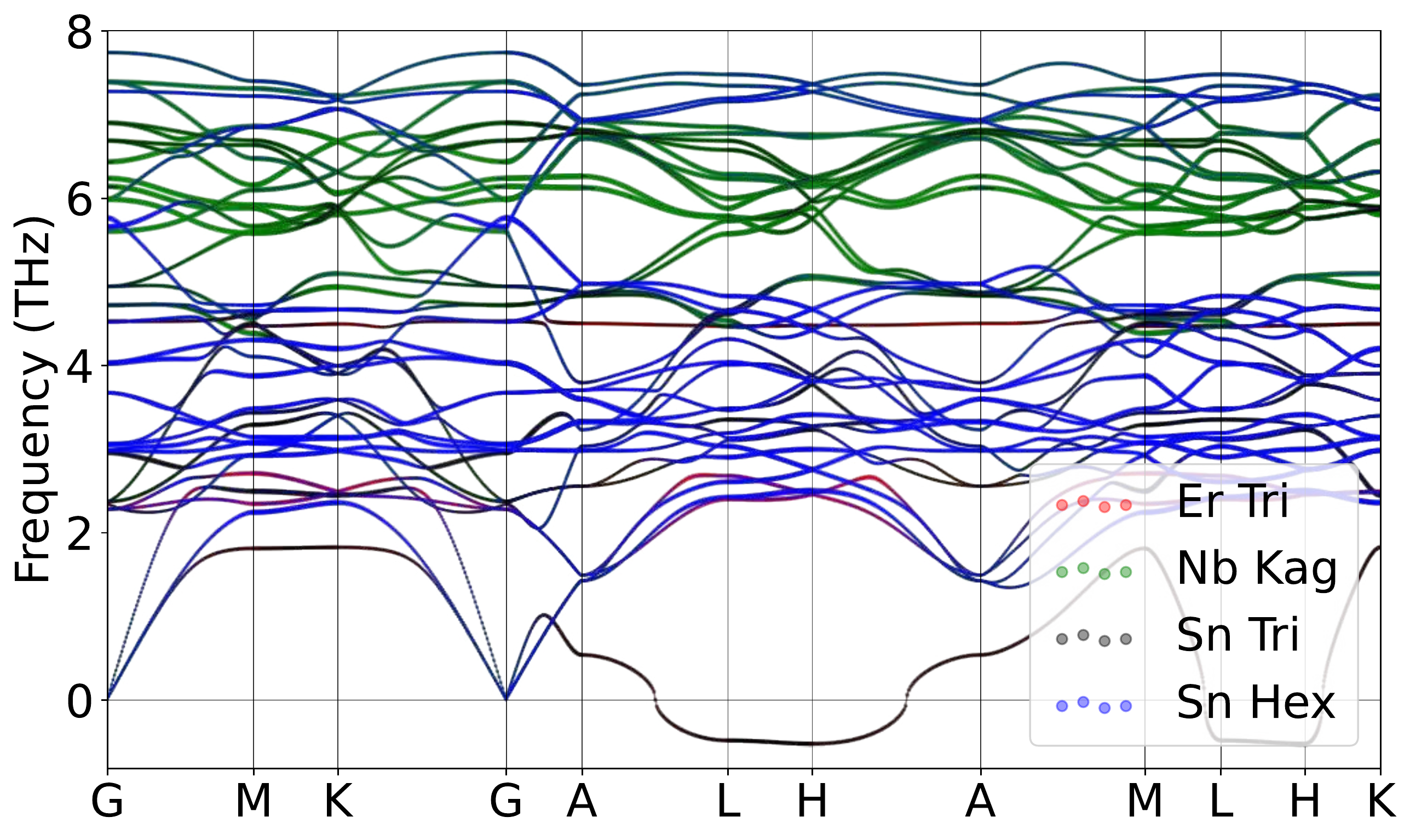}
}
\hspace{5pt}\subfloat[Relaxed phonon at high-T.]{
\label{fig:ErNb6Sn6_relaxed_High}
\includegraphics[width=0.45\textwidth]{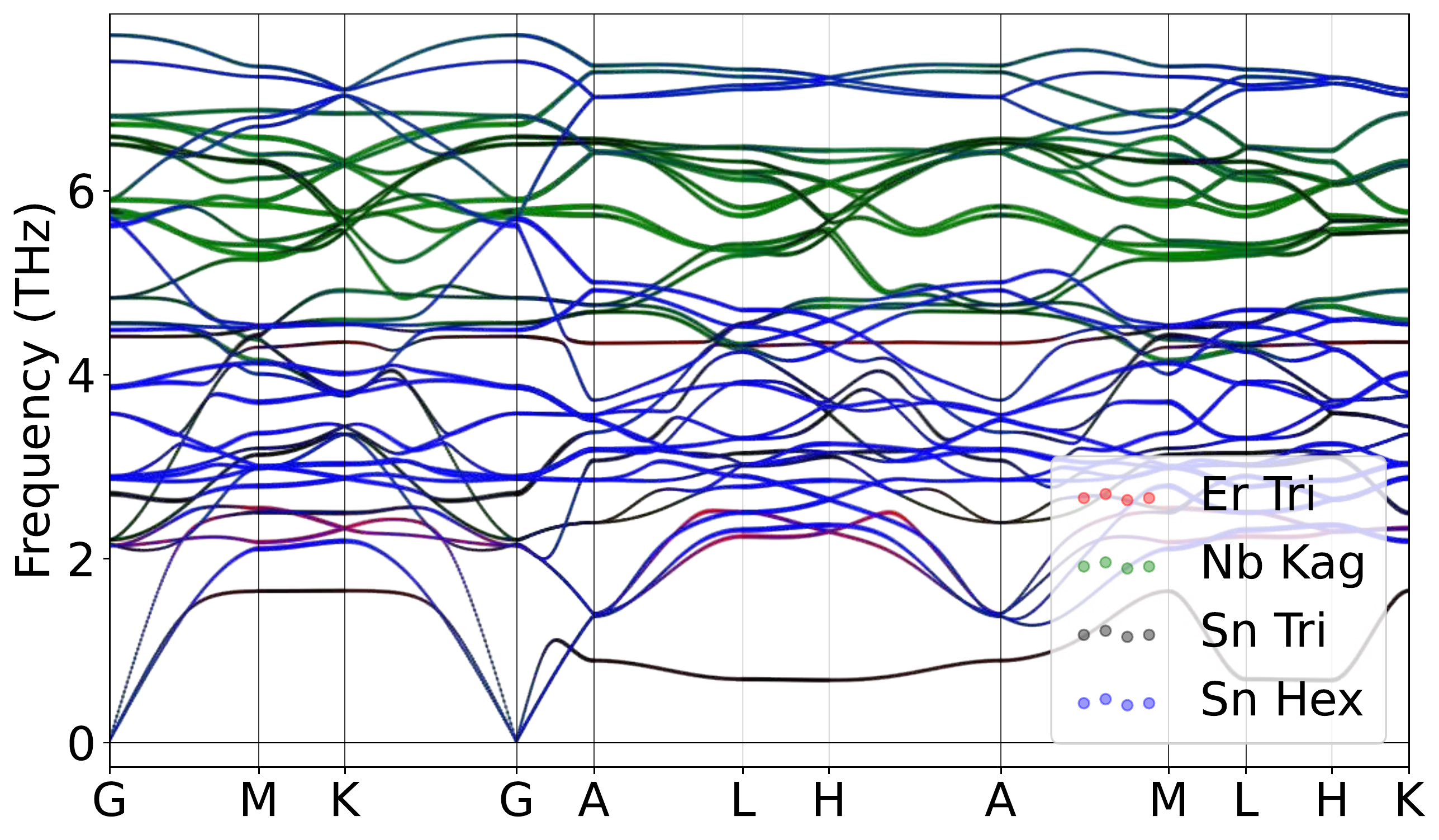}
}
\caption{Atomically projected phonon spectrum for ErNb$_6$Sn$_6$.}
\label{fig:ErNb6Sn6pho}
\end{figure}

\begin{figure}[H]
\centering
\label{fig:ErNb6Sn6_relaxed_Low_xz}
\includegraphics[width=0.85\textwidth]{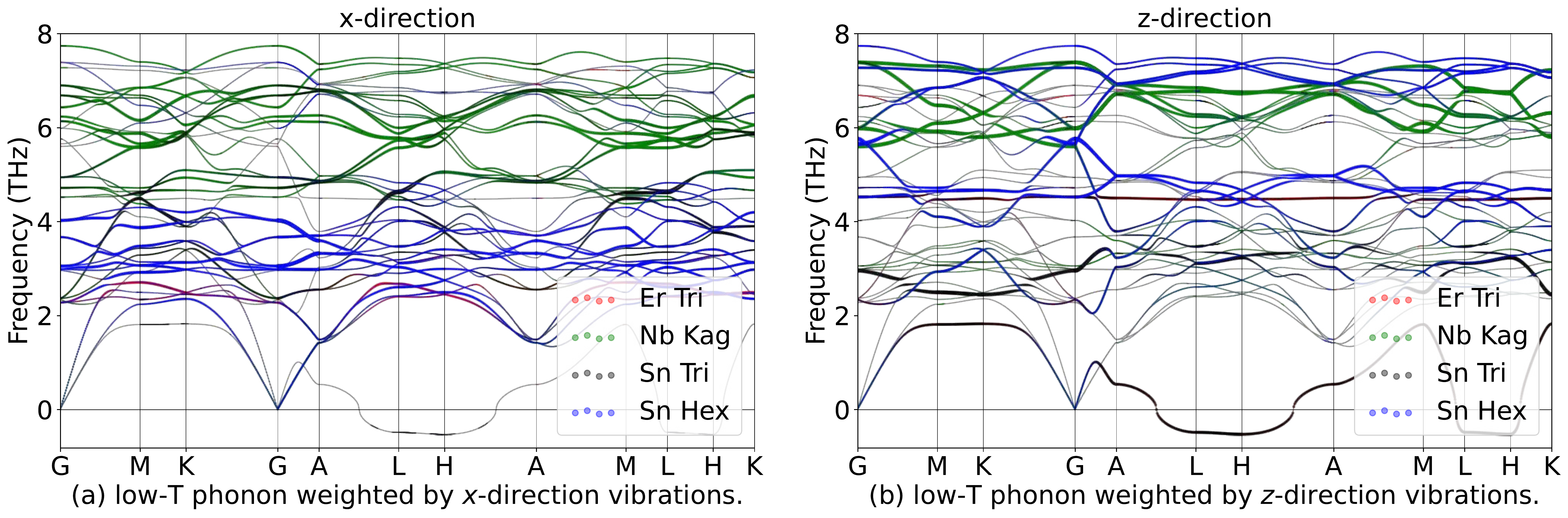}
\hspace{5pt}\label{fig:ErNb6Sn6_relaxed_High_xz}
\includegraphics[width=0.85\textwidth]{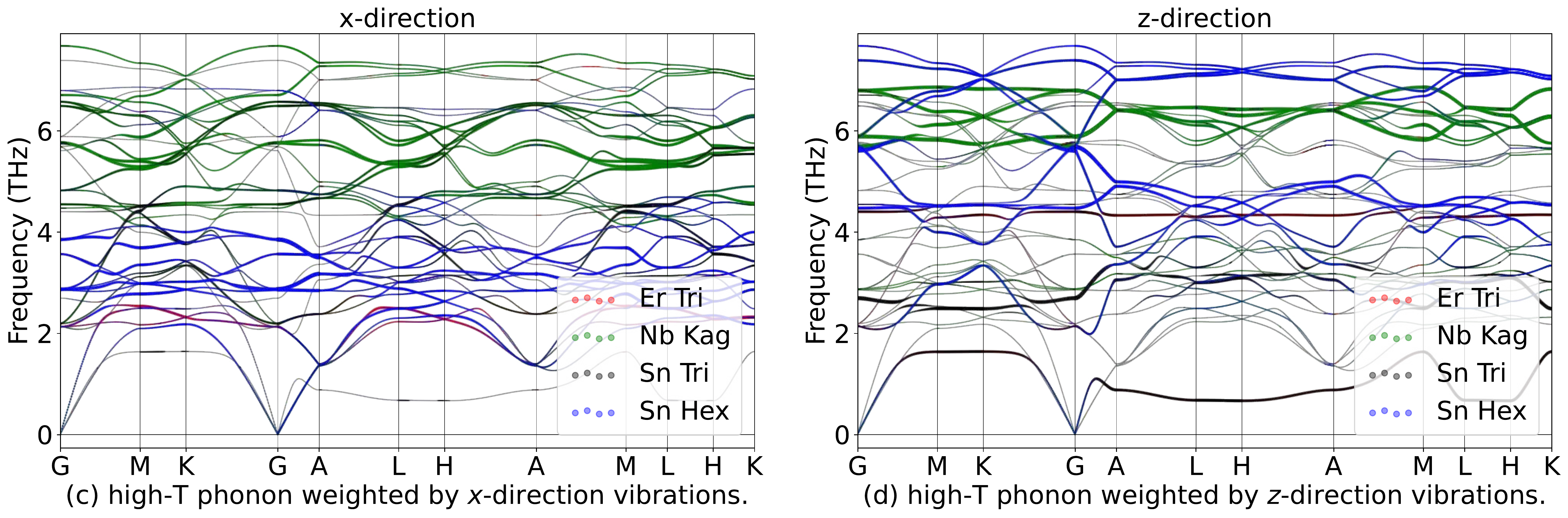}
\caption{Phonon spectrum for ErNb$_6$Sn$_6$in in-plane ($x$) and out-of-plane ($z$) directions.}
\label{fig:ErNb6Sn6phoxyz}
\end{figure}

\begin{table}[htbp]
\global\long\def\arraystretch{1.12}
\captionsetup{width=.95\textwidth}
\caption{IRREPs at high-symmetry points for ErNb$_6$Sn$_6$ phonon spectrum at Low-T.}
\label{tab:191_ErNb6Sn6_Low}
\setlength{\tabcolsep}{1.mm}{
}
\end{table}

\begin{table}[htbp]
\global\long\def\arraystretch{1.12}
\captionsetup{width=.95\textwidth}
\caption{IRREPs at high-symmetry points for ErNb$_6$Sn$_6$ phonon spectrum at High-T.}
\label{tab:191_ErNb6Sn6_High}
\setlength{\tabcolsep}{1.mm}{
%
}
\end{table}
\subsubsection{\label{sms2sec:TmNb6Sn6}\hyperref[smsec:col]{\texorpdfstring{TmNb$_6$Sn$_6$}{}}~{\color{blue}  (High-T stable, Low-T unstable) }}

\begin{table}[htbp]
\global\long\def\arraystretch{1.12}
\captionsetup{width=.85\textwidth}
\caption{Structure parameters of TmNb$_6$Sn$_6$ after relaxation. It contains 615 electrons. }
\label{tab:TmNb6Sn6}
\setlength{\tabcolsep}{4mm}{
%
}
\end{table}

\begin{figure}[htbp]
\centering
\includegraphics[width=0.85\textwidth]{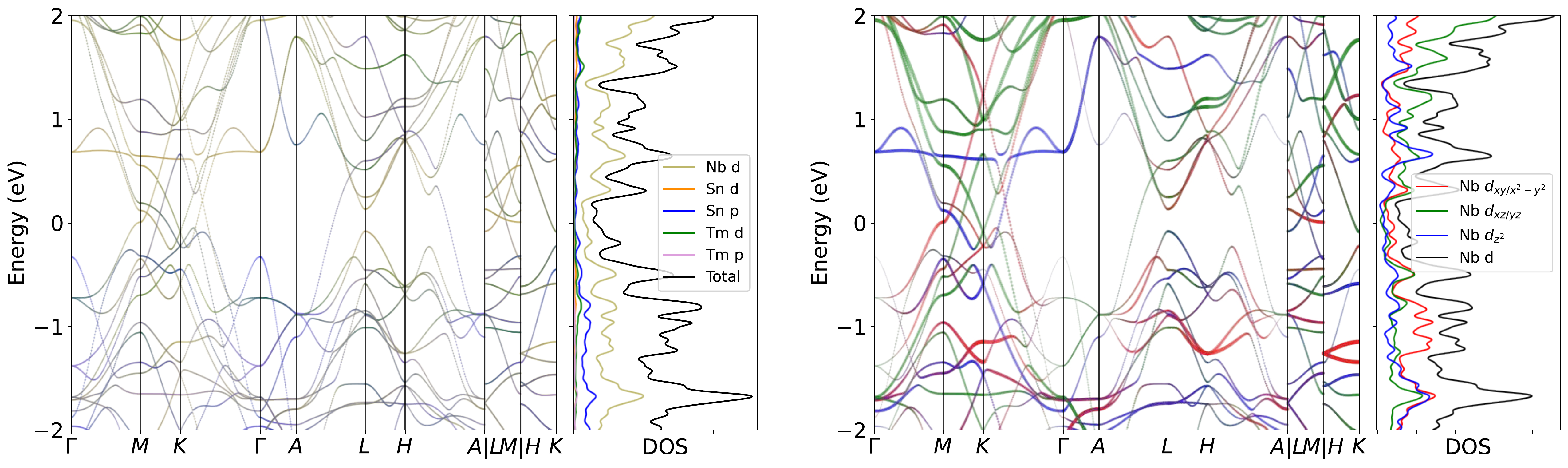}
\caption{Band structure for TmNb$_6$Sn$_6$ without SOC. The projection of Nb $3d$ orbitals is also presented. The band structures clearly reveal the dominating of Nb $3d$ orbitals  near the Fermi level, as well as the pronounced peak.}
\label{fig:TmNb6Sn6band}
\end{figure}

\begin{figure}[H]
\centering
\subfloat[Relaxed phonon at low-T.]{
\label{fig:TmNb6Sn6_relaxed_Low}
\includegraphics[width=0.45\textwidth]{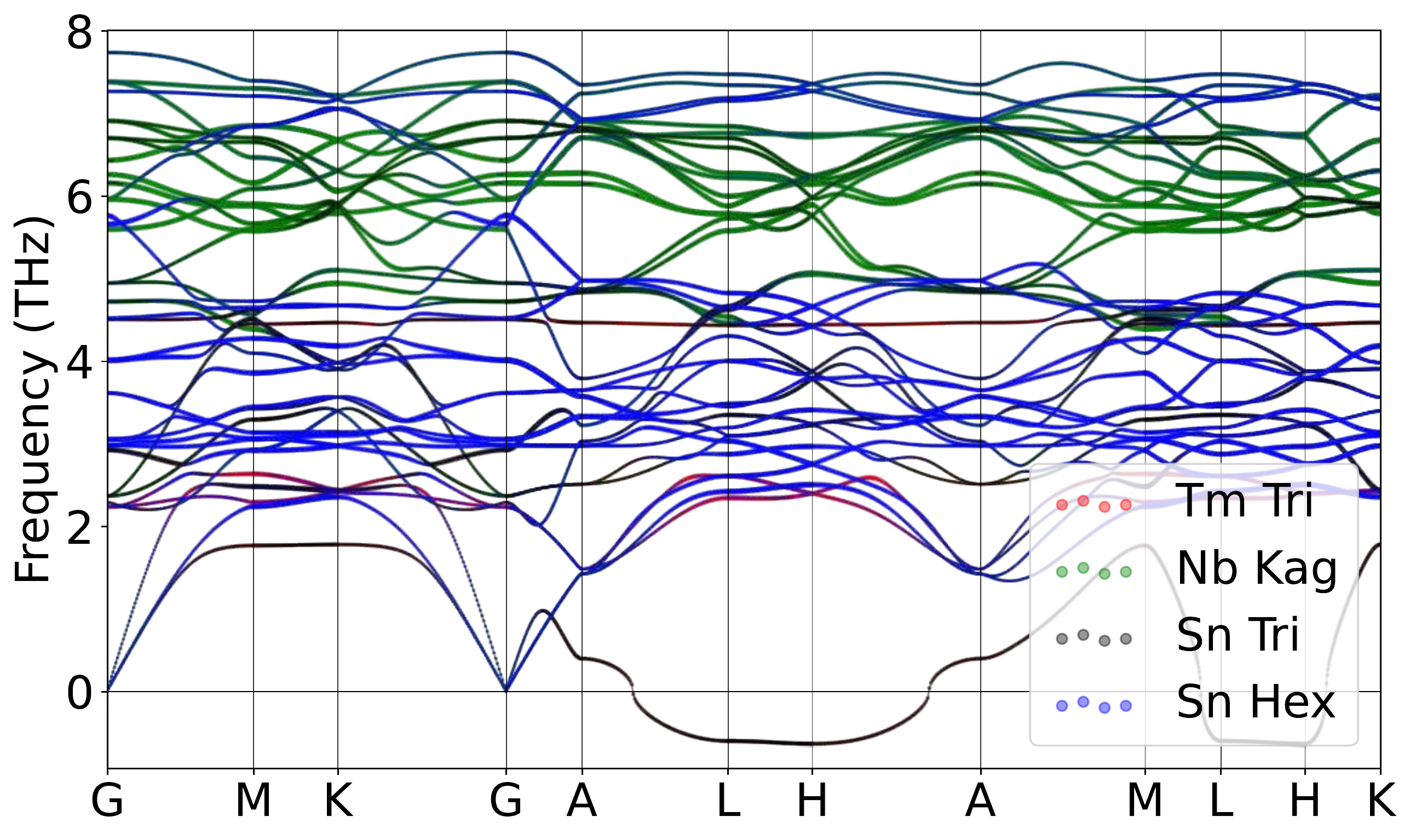}
}
\hspace{5pt}\subfloat[Relaxed phonon at high-T.]{
\label{fig:TmNb6Sn6_relaxed_High}
\includegraphics[width=0.45\textwidth]{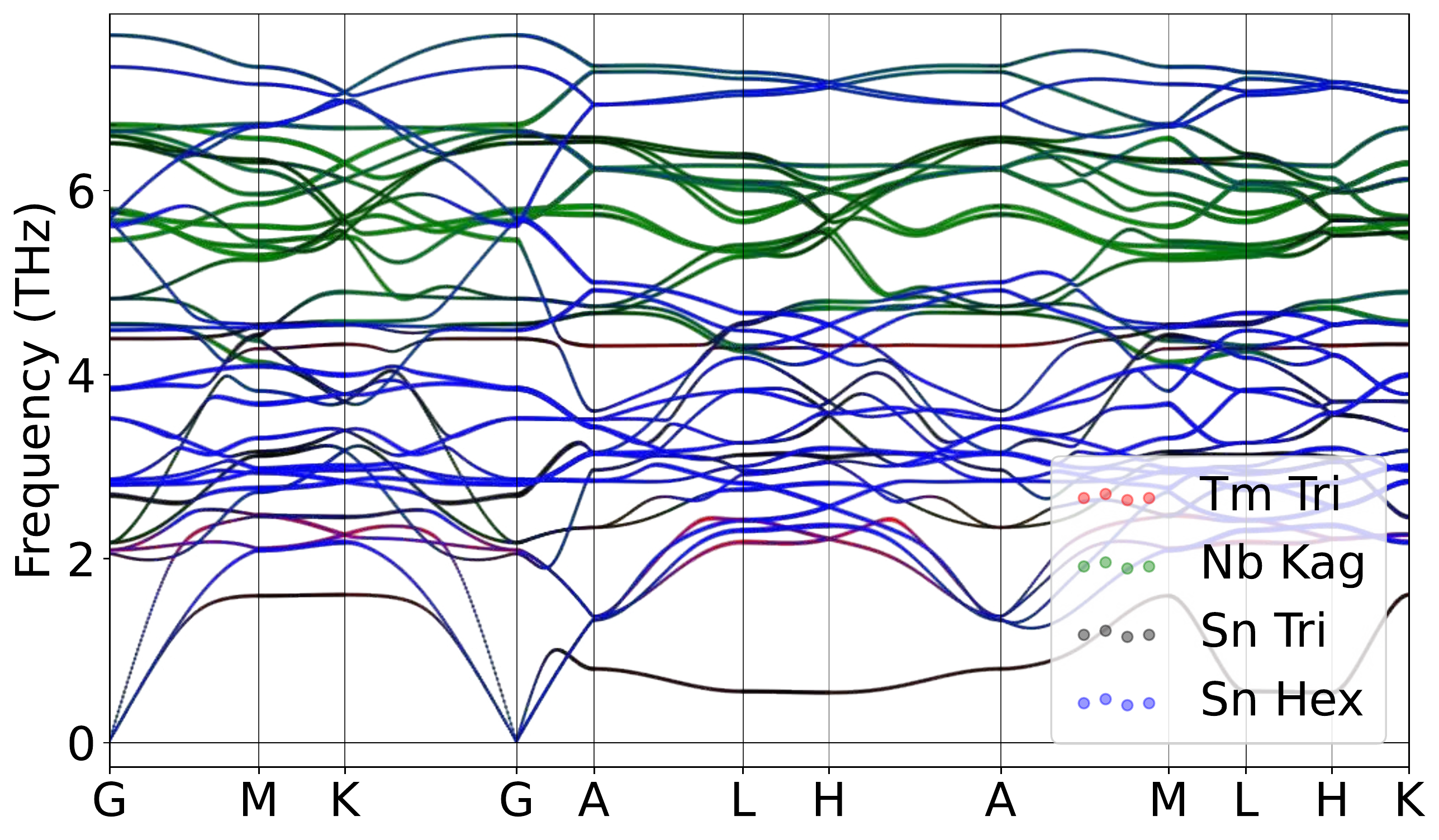}
}
\caption{Atomically projected phonon spectrum for TmNb$_6$Sn$_6$.}
\label{fig:TmNb6Sn6pho}
\end{figure}

\begin{figure}[H]
\centering
\label{fig:TmNb6Sn6_relaxed_Low_xz}
\includegraphics[width=0.85\textwidth]{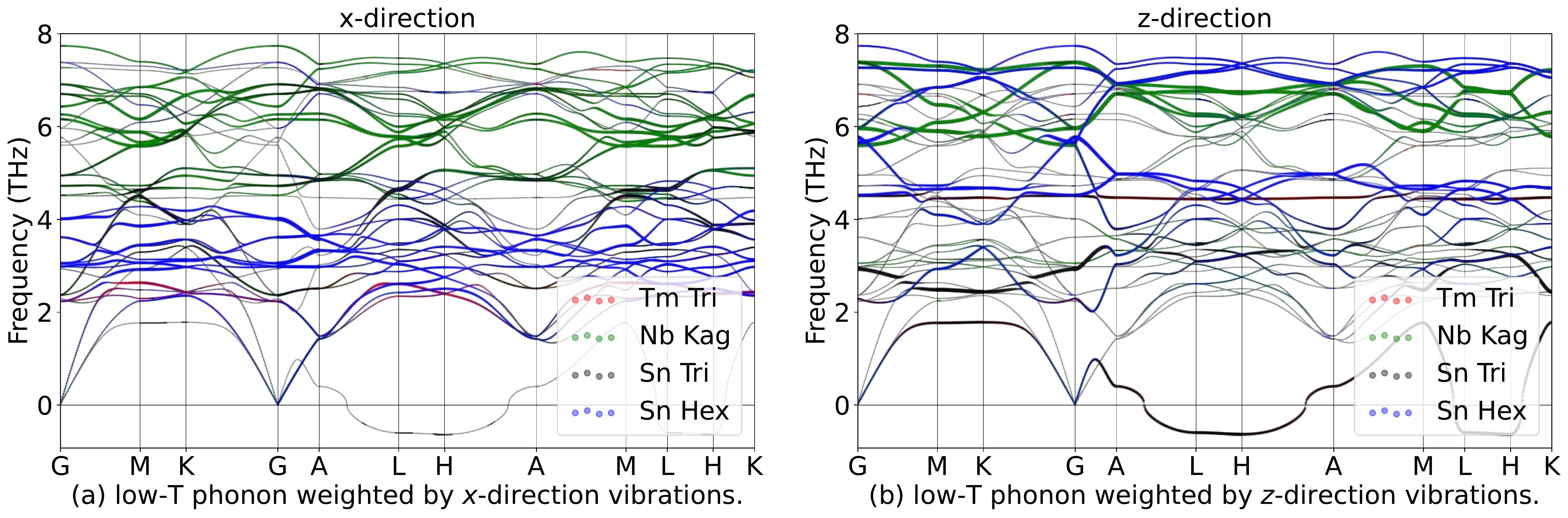}
\hspace{5pt}\label{fig:TmNb6Sn6_relaxed_High_xz}
\includegraphics[width=0.85\textwidth]{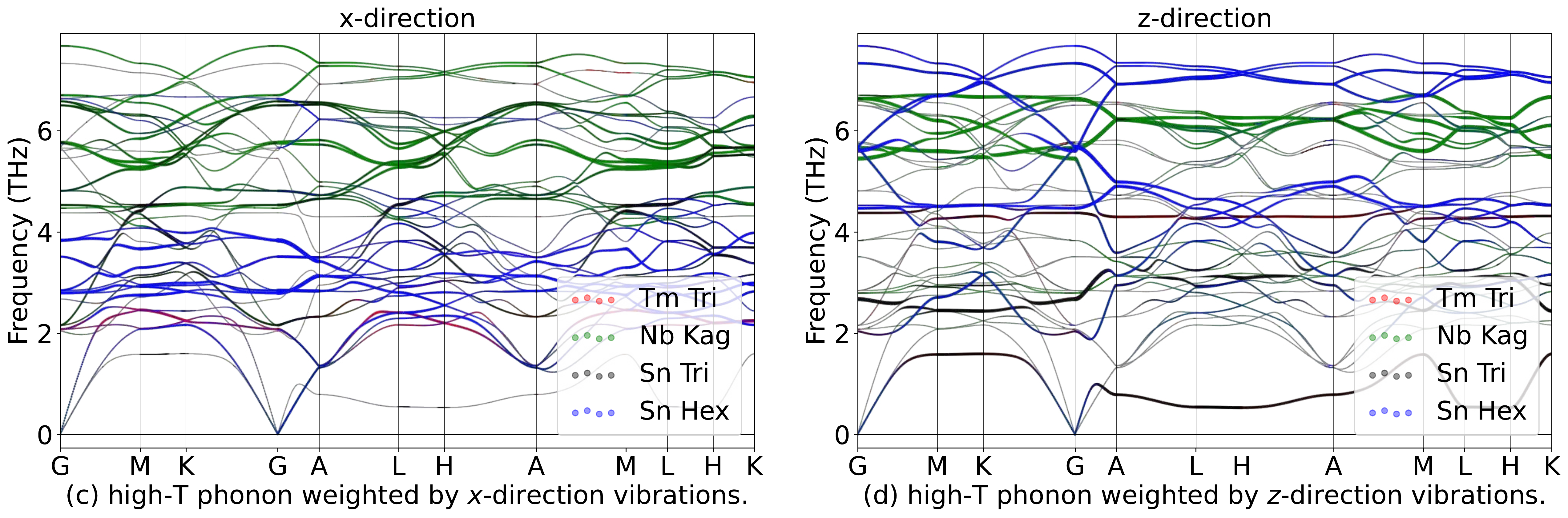}
\caption{Phonon spectrum for TmNb$_6$Sn$_6$in in-plane ($x$) and out-of-plane ($z$) directions.}
\label{fig:TmNb6Sn6phoxyz}
\end{figure}

\begin{table}[htbp]
\global\long\def\arraystretch{1.12}
\captionsetup{width=.95\textwidth}
\caption{IRREPs at high-symmetry points for TmNb$_6$Sn$_6$ phonon spectrum at Low-T.}
\label{tab:191_TmNb6Sn6_Low}
\setlength{\tabcolsep}{1.mm}{
}
\end{table}

\begin{table}[htbp]
\global\long\def\arraystretch{1.12}
\captionsetup{width=.95\textwidth}
\caption{IRREPs at high-symmetry points for TmNb$_6$Sn$_6$ phonon spectrum at High-T.}
\label{tab:191_TmNb6Sn6_High}
\setlength{\tabcolsep}{1.mm}{
%
}
\end{table}
\subsubsection{\label{sms2sec:YbNb6Sn6}\hyperref[smsec:col]{\texorpdfstring{YbNb$_6$Sn$_6$}{}}~{\color{blue}  (High-T stable, Low-T unstable) }}

\begin{table}[htbp]
\global\long\def\arraystretch{1.12}
\captionsetup{width=.85\textwidth}
\caption{Structure parameters of YbNb$_6$Sn$_6$ after relaxation. It contains 616 electrons. }
\label{tab:YbNb6Sn6}
\setlength{\tabcolsep}{4mm}{
%
}
\end{table}

\begin{figure}[htbp]
\centering
\includegraphics[width=0.85\textwidth]{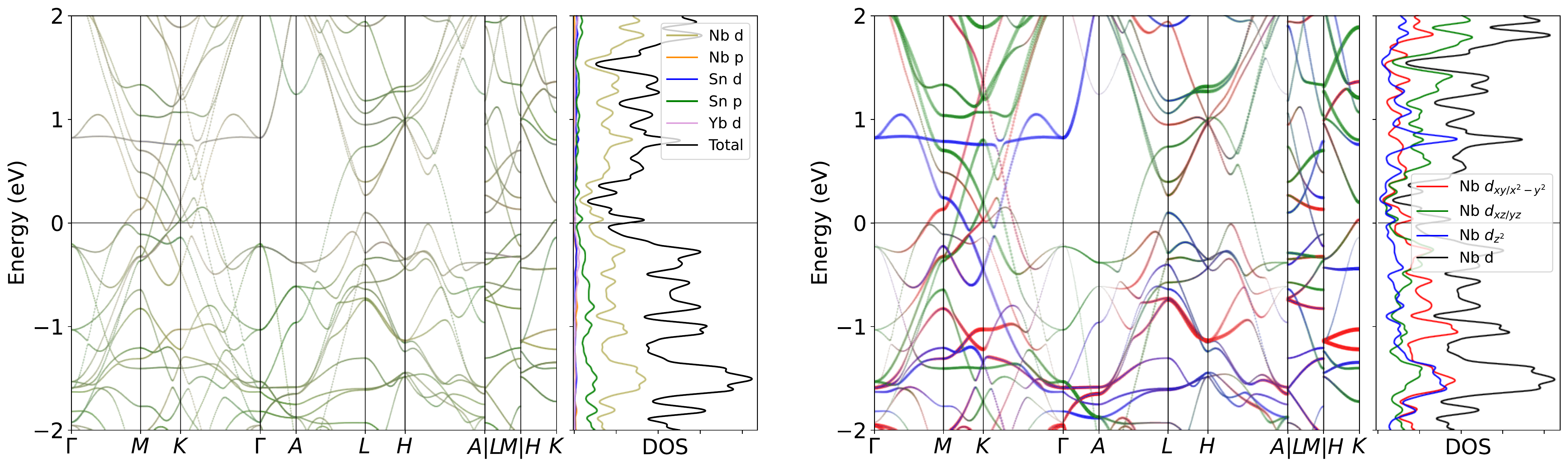}
\caption{Band structure for YbNb$_6$Sn$_6$ without SOC. The projection of Nb $3d$ orbitals is also presented. The band structures clearly reveal the dominating of Nb $3d$ orbitals  near the Fermi level, as well as the pronounced peak.}
\label{fig:YbNb6Sn6band}
\end{figure}

\begin{figure}[H]
\centering
\subfloat[Relaxed phonon at low-T.]{
\label{fig:YbNb6Sn6_relaxed_Low}
\includegraphics[width=0.45\textwidth]{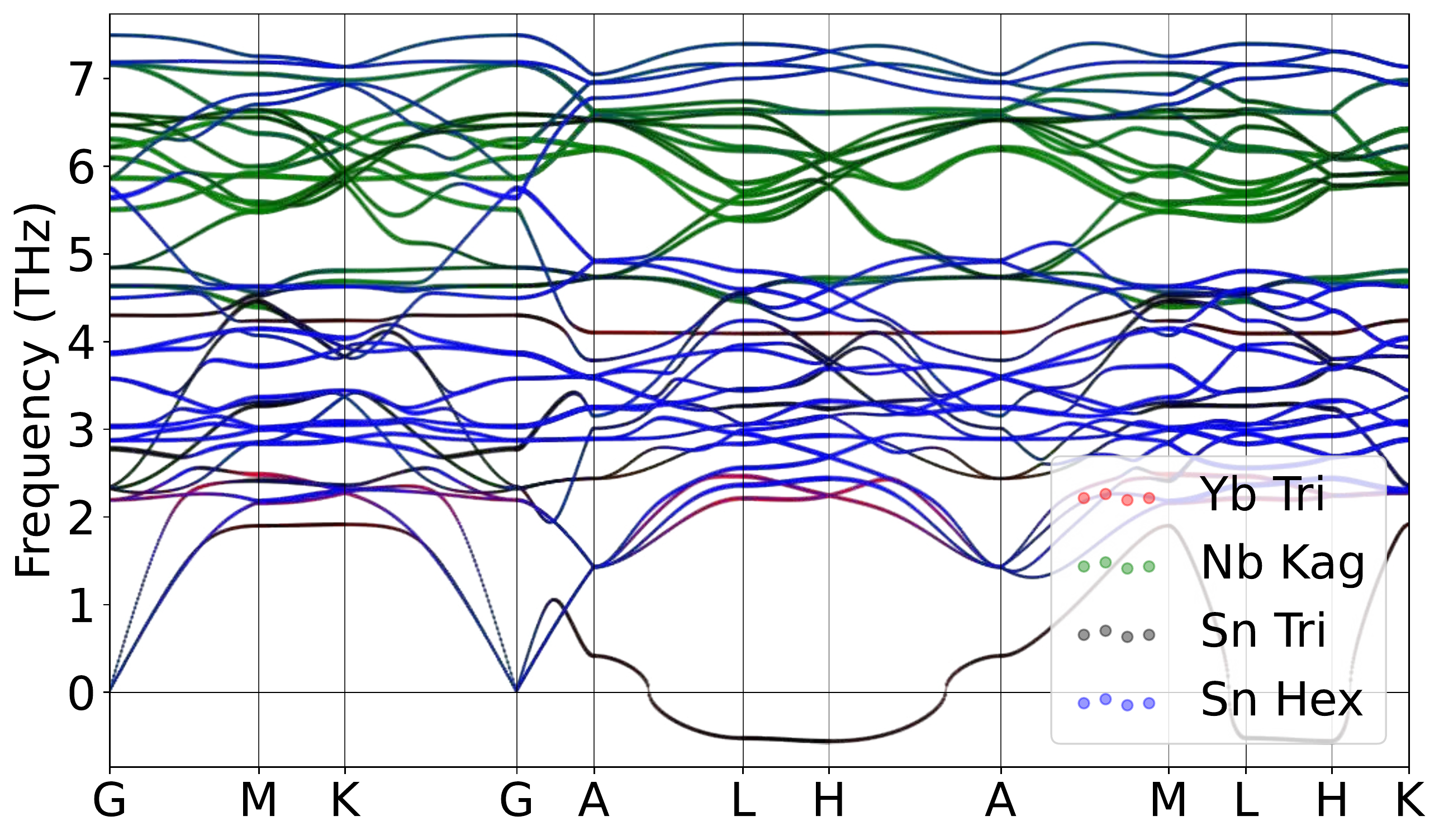}
}
\hspace{5pt}\subfloat[Relaxed phonon at high-T.]{
\label{fig:YbNb6Sn6_relaxed_High}
\includegraphics[width=0.45\textwidth]{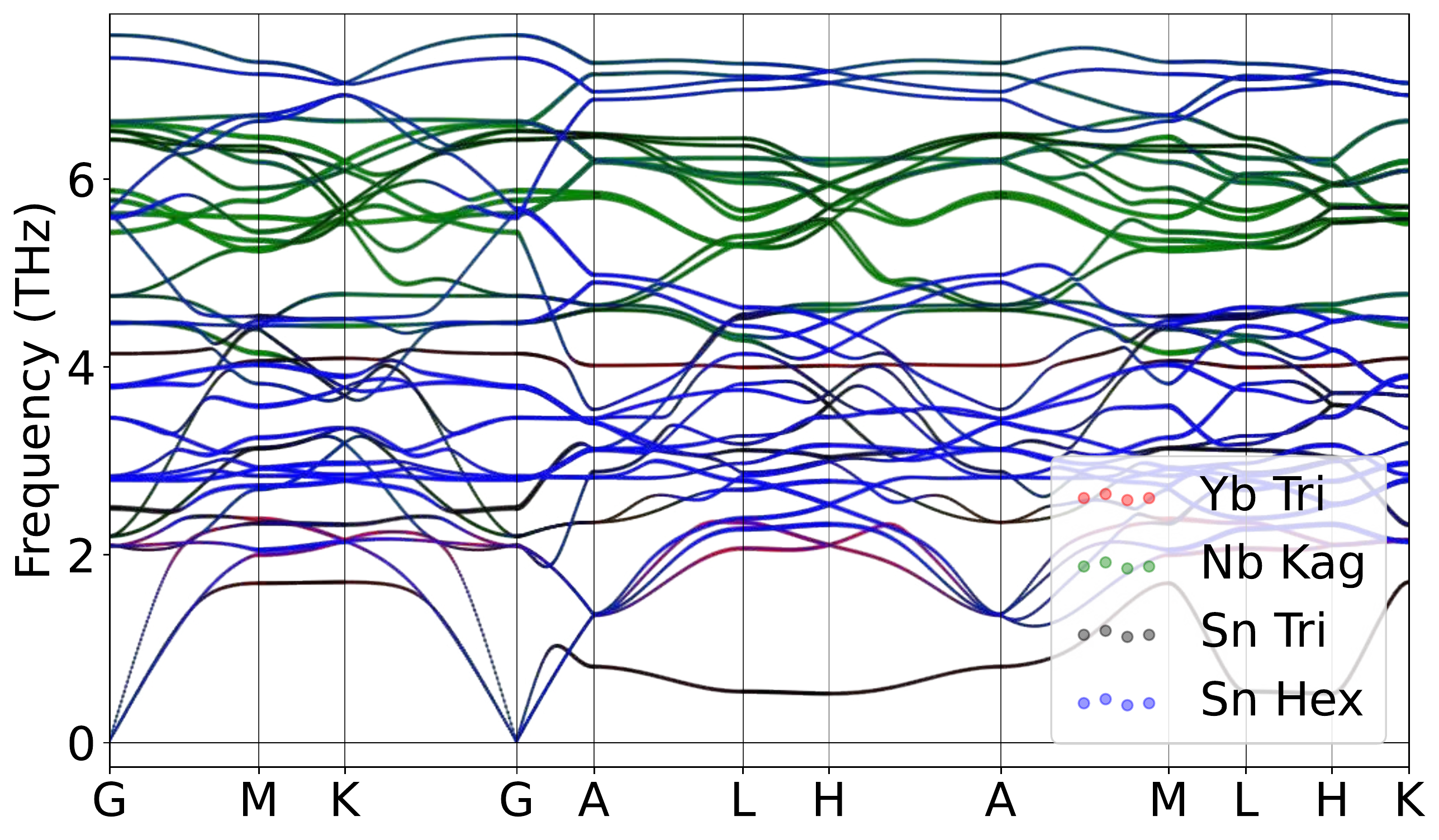}
}
\caption{Atomically projected phonon spectrum for YbNb$_6$Sn$_6$.}
\label{fig:YbNb6Sn6pho}
\end{figure}

\begin{figure}[H]
\centering
\label{fig:YbNb6Sn6_relaxed_Low_xz}
\includegraphics[width=0.85\textwidth]{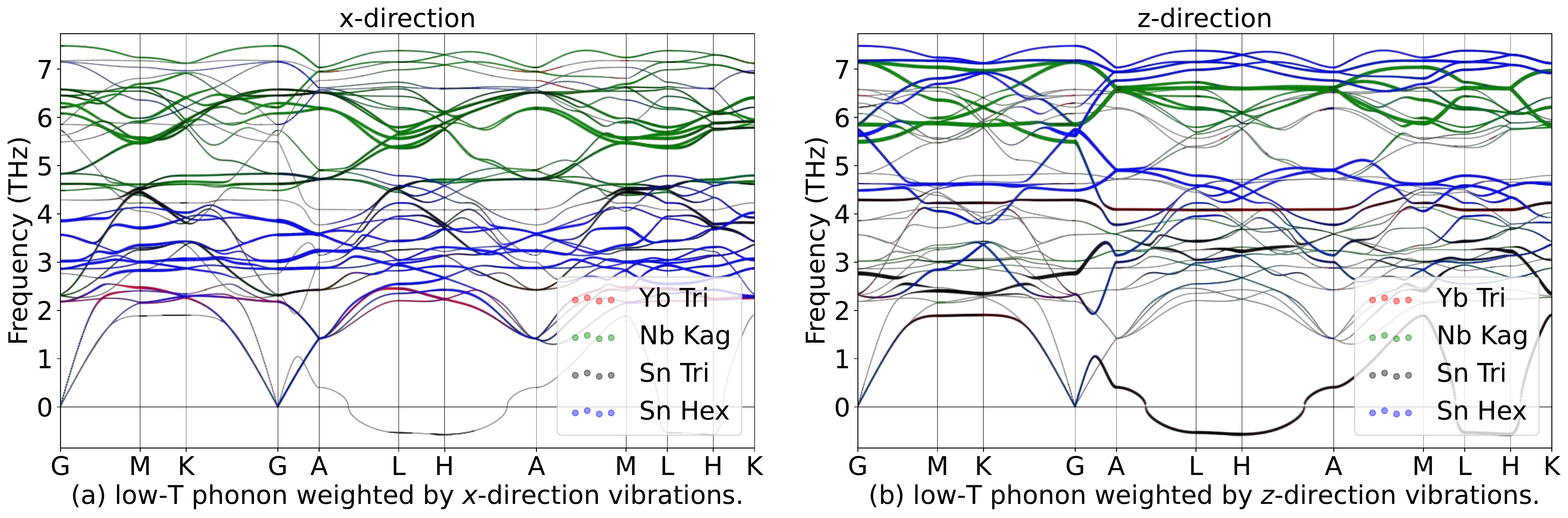}
\hspace{5pt}\label{fig:YbNb6Sn6_relaxed_High_xz}
\includegraphics[width=0.85\textwidth]{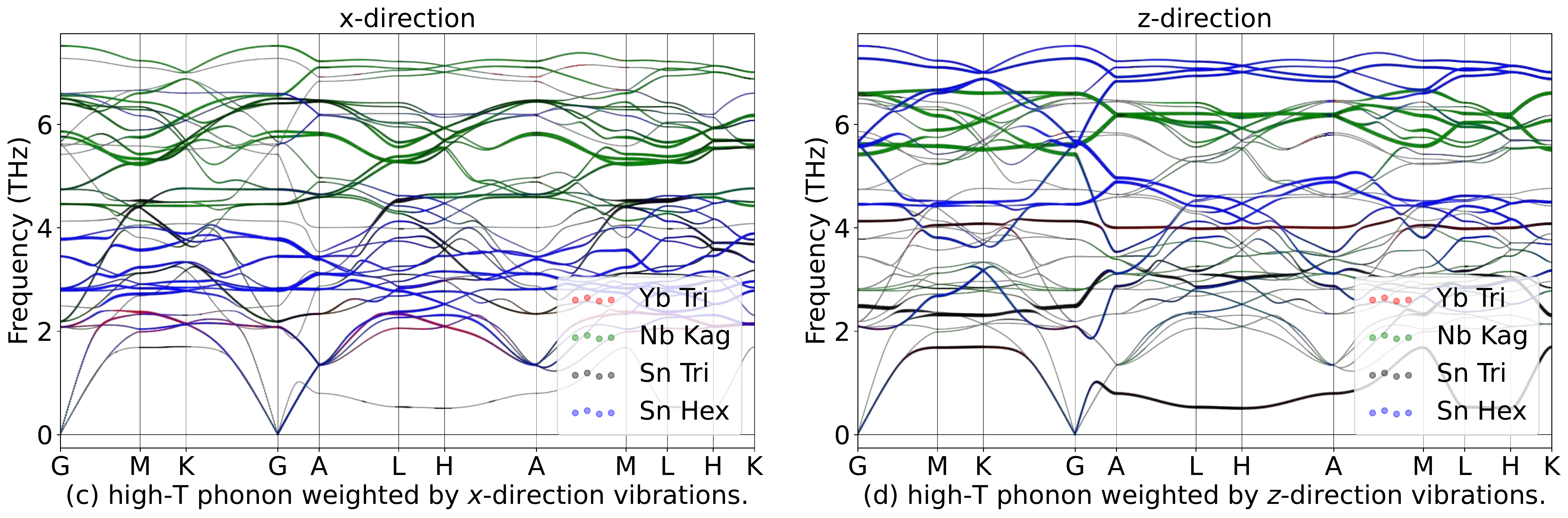}
\caption{Phonon spectrum for YbNb$_6$Sn$_6$in in-plane ($x$) and out-of-plane ($z$) directions.}
\label{fig:YbNb6Sn6phoxyz}
\end{figure}

\begin{table}[htbp]
\global\long\def\arraystretch{1.12}
\captionsetup{width=.95\textwidth}
\caption{IRREPs at high-symmetry points for YbNb$_6$Sn$_6$ phonon spectrum at Low-T.}
\label{tab:191_YbNb6Sn6_Low}
\setlength{\tabcolsep}{1.mm}{
}
\end{table}

\begin{table}[htbp]
\global\long\def\arraystretch{1.12}
\captionsetup{width=.95\textwidth}
\caption{IRREPs at high-symmetry points for YbNb$_6$Sn$_6$ phonon spectrum at High-T.}
\label{tab:191_YbNb6Sn6_High}
\setlength{\tabcolsep}{1.mm}{
%
}
\end{table}
\subsubsection{\label{sms2sec:LuNb6Sn6}\hyperref[smsec:col]{\texorpdfstring{LuNb$_6$Sn$_6$}{}}~{\color{black} (TBD) }}

\begin{table}[htbp]
\global\long\def\arraystretch{1.12}
\captionsetup{width=.85\textwidth}
\caption{Structure parameters of LuNb$_6$Sn$_6$ after relaxation. It contains 617 electrons. }
\label{tab:LuNb6Sn6}
\setlength{\tabcolsep}{4mm}{
%
}
\end{table}

\begin{figure}[htbp]
\centering
\includegraphics[width=0.85\textwidth]{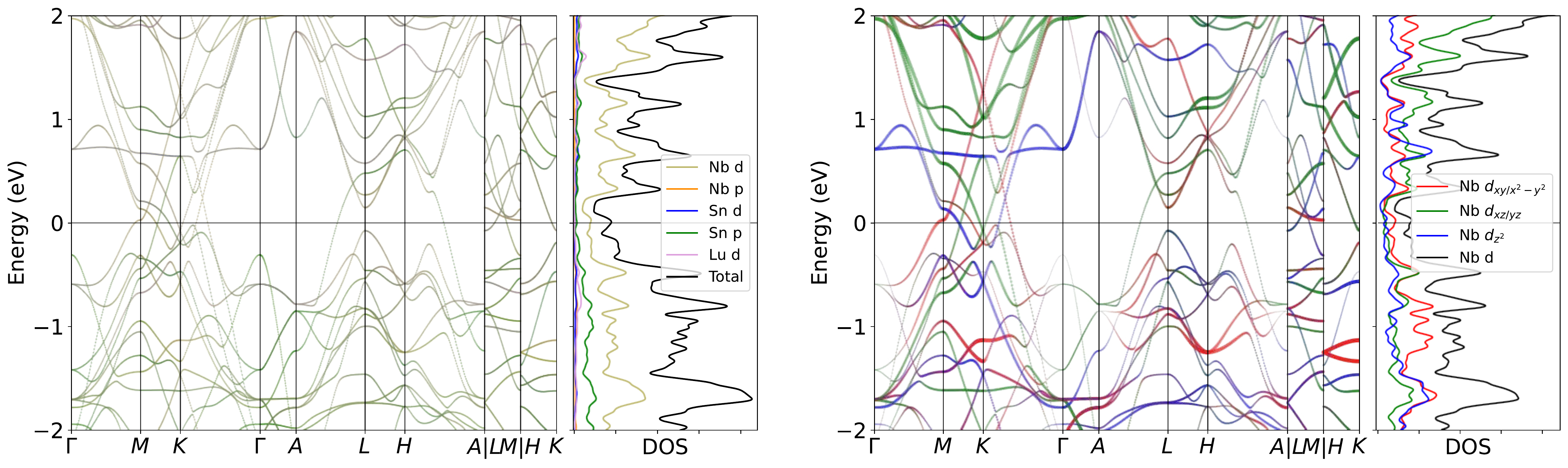}
\caption{Band structure for LuNb$_6$Sn$_6$ without SOC. The projection of Nb $3d$ orbitals is also presented. The band structures clearly reveal the dominating of Nb $3d$ orbitals  near the Fermi level, as well as the pronounced peak.}
\label{fig:LuNb6Sn6band}
\end{figure}

\begin{figure}[H]
\centering
\label{fig:LuNb6Sn6_relaxed_Low}
\includegraphics[width=0.45\textwidth]{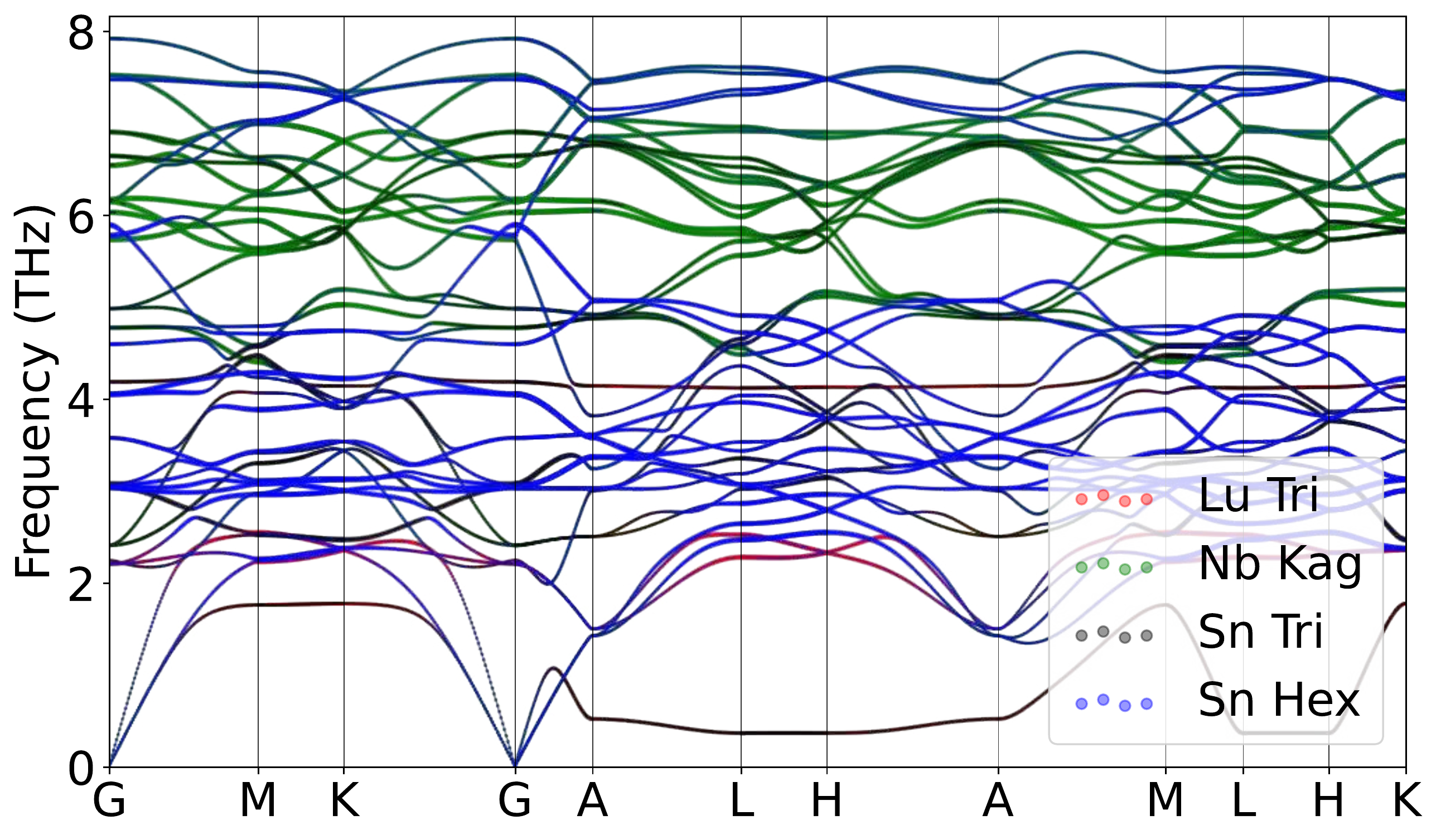}
\caption{Atomically projected phonon spectrum for LuNb$_6$Sn$_6$.}
\label{fig:LuNb6Sn6pho}
\end{figure}

\begin{figure}[H]
\centering
\label{fig:LuNb6Sn6_relaxed_Low_xz}
\includegraphics[width=0.85\textwidth]{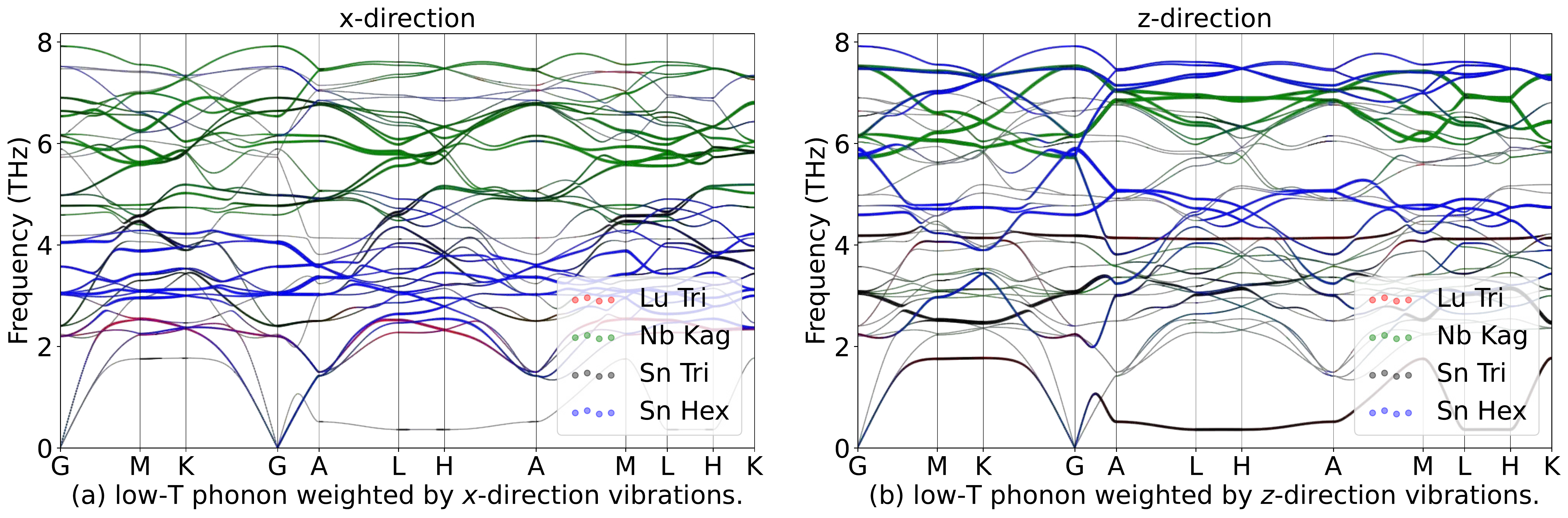}
\caption{Phonon spectrum for LuNb$_6$Sn$_6$in in-plane ($x$) and out-of-plane ($z$) directions.}
\label{fig:LuNb6Sn6phoxyz}
\end{figure}

\subsubsection{\label{sms2sec:HfNb6Sn6}\hyperref[smsec:col]{\texorpdfstring{HfNb$_6$Sn$_6$}{}}~{\color{green}  (High-T stable, Low-T stable) }}

\begin{table}[htbp]
\global\long\def\arraystretch{1.12}
\captionsetup{width=.85\textwidth}
\caption{Structure parameters of HfNb$_6$Sn$_6$ after relaxation. It contains 618 electrons. }
\label{tab:HfNb6Sn6}
\setlength{\tabcolsep}{4mm}{
\begin{tabular}{c|c||c|c}
\hline
\hline
 Atom & Coordinates & \multicolumn{2}{c}{Parameters~($\AA$)} \\
\hline
Hf ~ 1a & ($0,0,0$) & $a$ & $5.6801$ \\
Nb ~ 6i & ($\frac{1}{2},0,0.2476$) & $c$ & $9.4826$ \\
Sn1 ~ 2c & ($\frac{1}{3},\frac{2}{3},0$) & $d^{Tri}_{Z-Z}$ & $3.3484$ \\
Sn2 ~ 2d & ($\frac{1}{3},\frac{2}{3},\frac{1}{2}$) & $d_{T_{1}}$ & $2.8401$ \\
Sn3 ~ 2e & ($0,0,0.3234$) & $d_{T_{2}}$ & $4.6954$ \\
\hline
\hline
\end{tabular}}
\end{table}

\begin{figure}[htbp]
\centering
\includegraphics[width=0.85\textwidth]{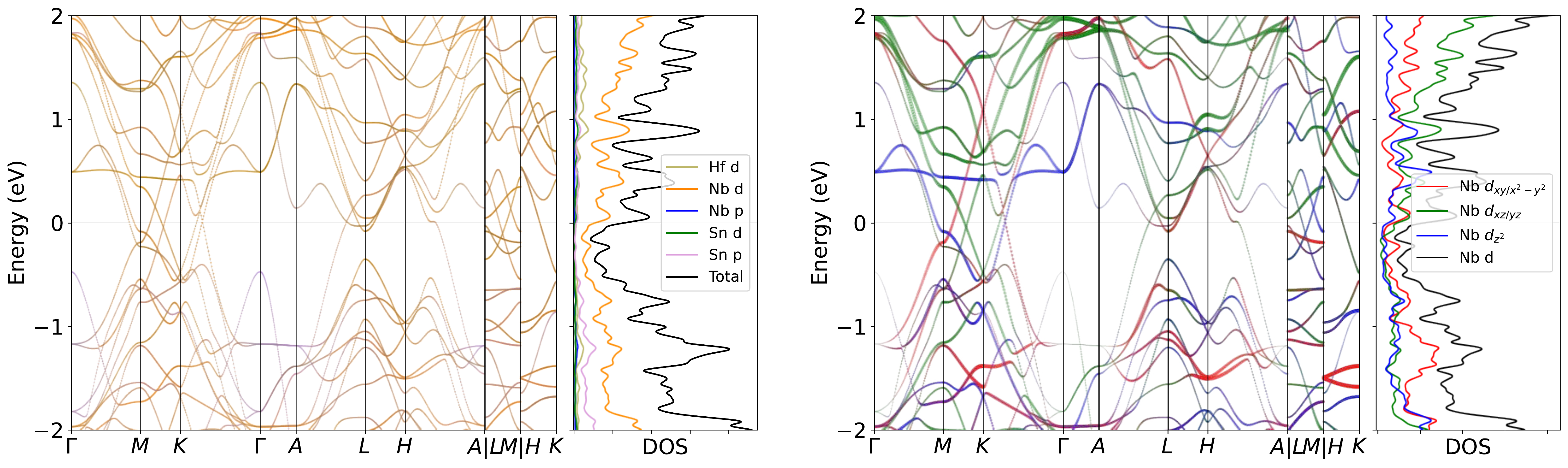}
\caption{Band structure for HfNb$_6$Sn$_6$ without SOC. The projection of Nb $3d$ orbitals is also presented. The band structures clearly reveal the dominating of Nb $3d$ orbitals  near the Fermi level, as well as the pronounced peak.}
\label{fig:HfNb6Sn6band}
\end{figure}

\begin{figure}[H]
\centering
\subfloat[Relaxed phonon at low-T.]{
\label{fig:HfNb6Sn6_relaxed_Low}
\includegraphics[width=0.45\textwidth]{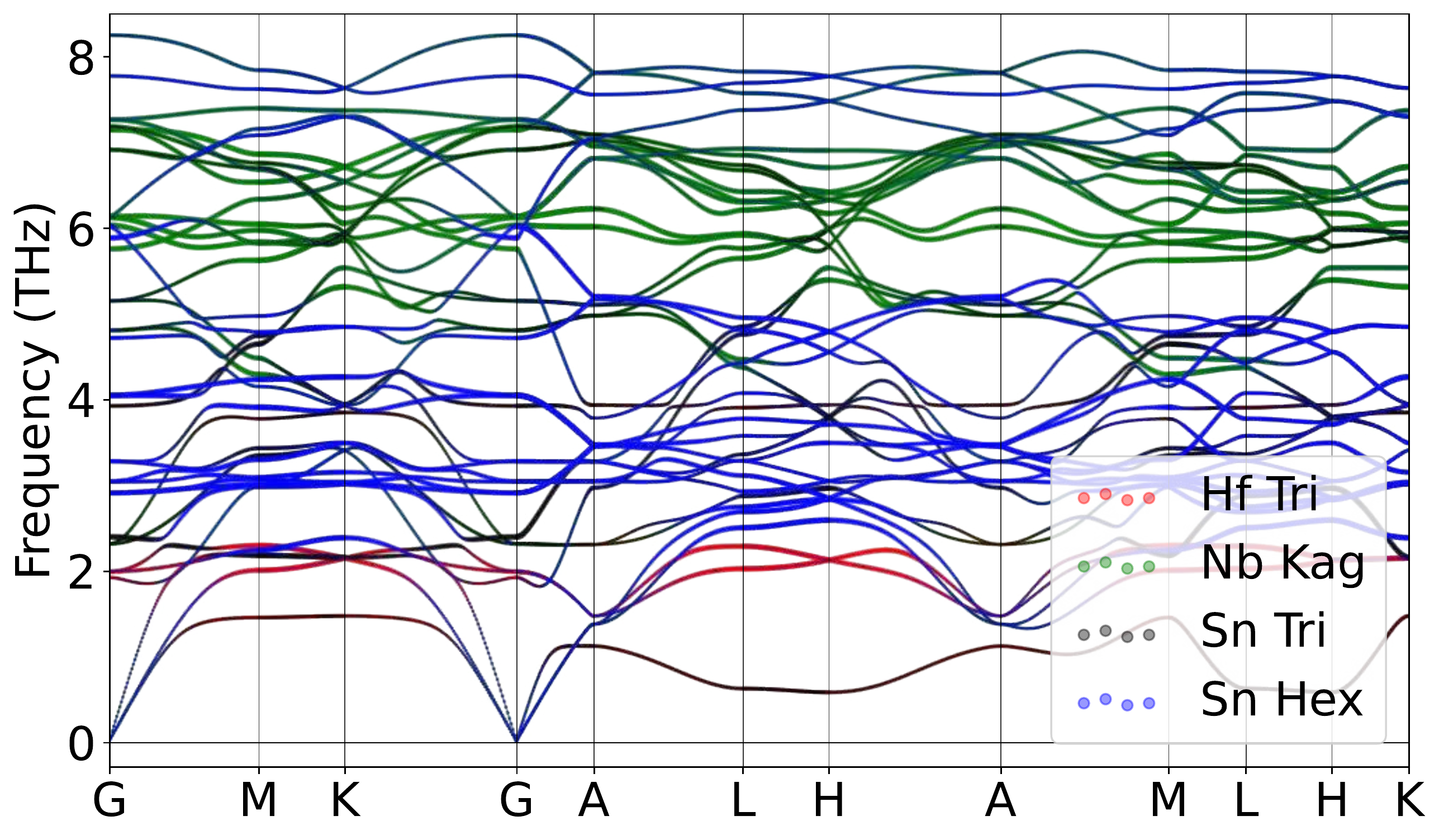}
}
\hspace{5pt}\subfloat[Relaxed phonon at high-T.]{
\label{fig:HfNb6Sn6_relaxed_High}
\includegraphics[width=0.45\textwidth]{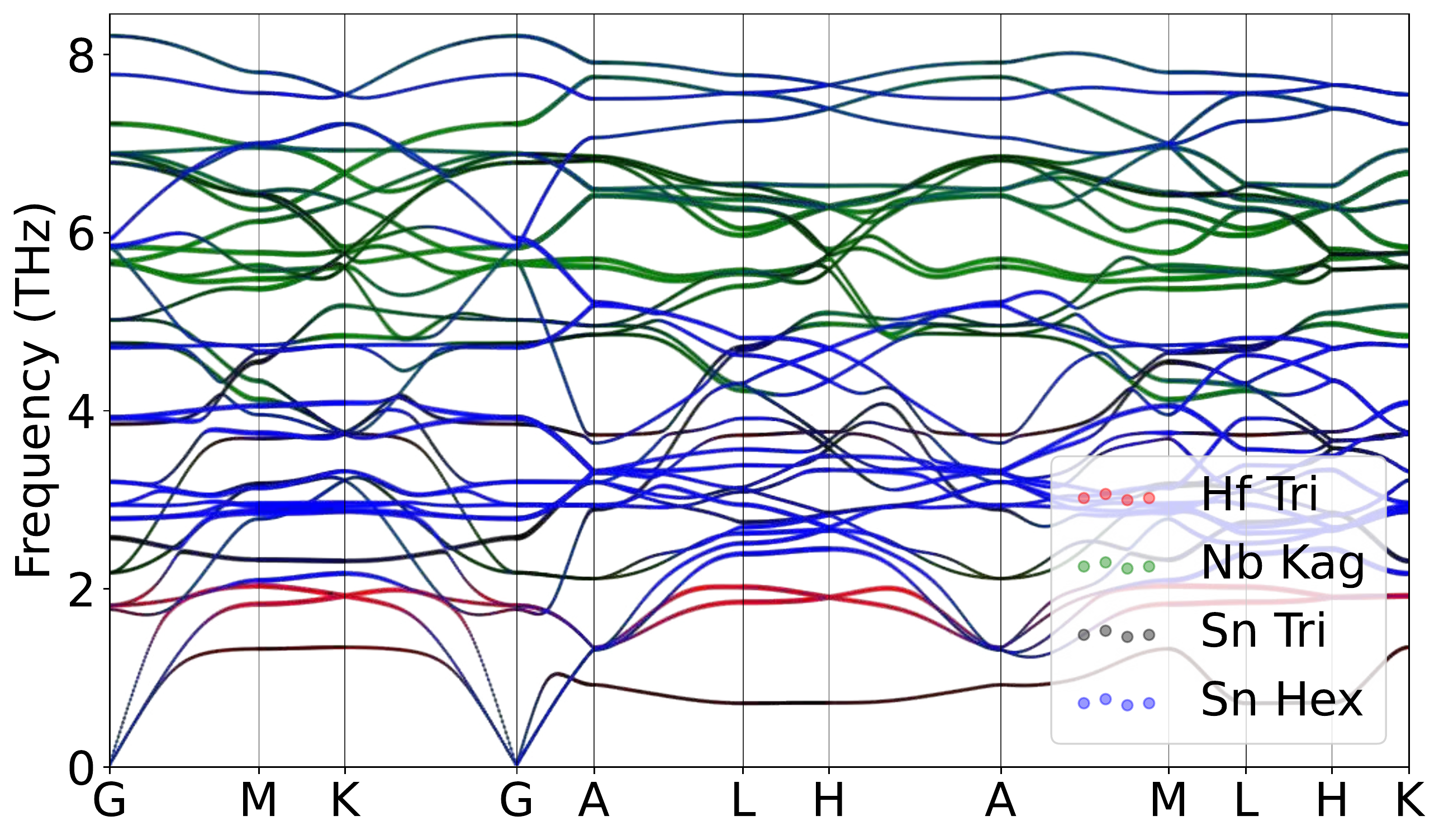}
}
\caption{Atomically projected phonon spectrum for HfNb$_6$Sn$_6$.}
\label{fig:HfNb6Sn6pho}
\end{figure}

\begin{figure}[H]
\centering
\label{fig:HfNb6Sn6_relaxed_Low_xz}
\includegraphics[width=0.85\textwidth]{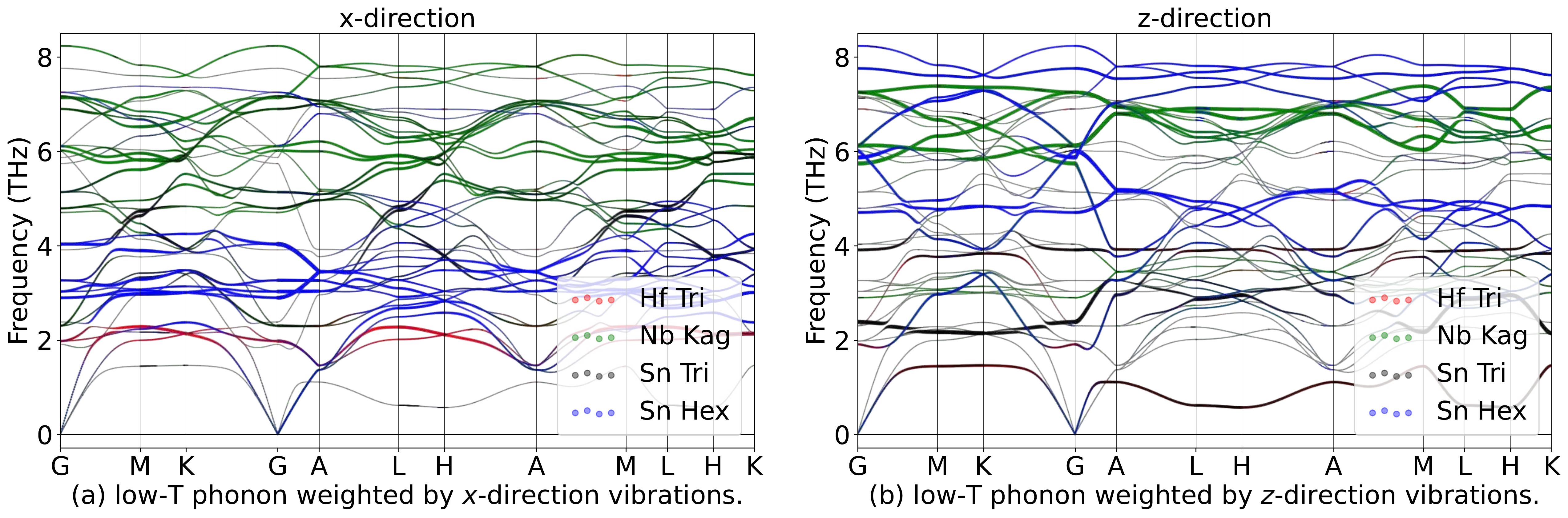}
\hspace{5pt}\label{fig:HfNb6Sn6_relaxed_High_xz}
\includegraphics[width=0.85\textwidth]{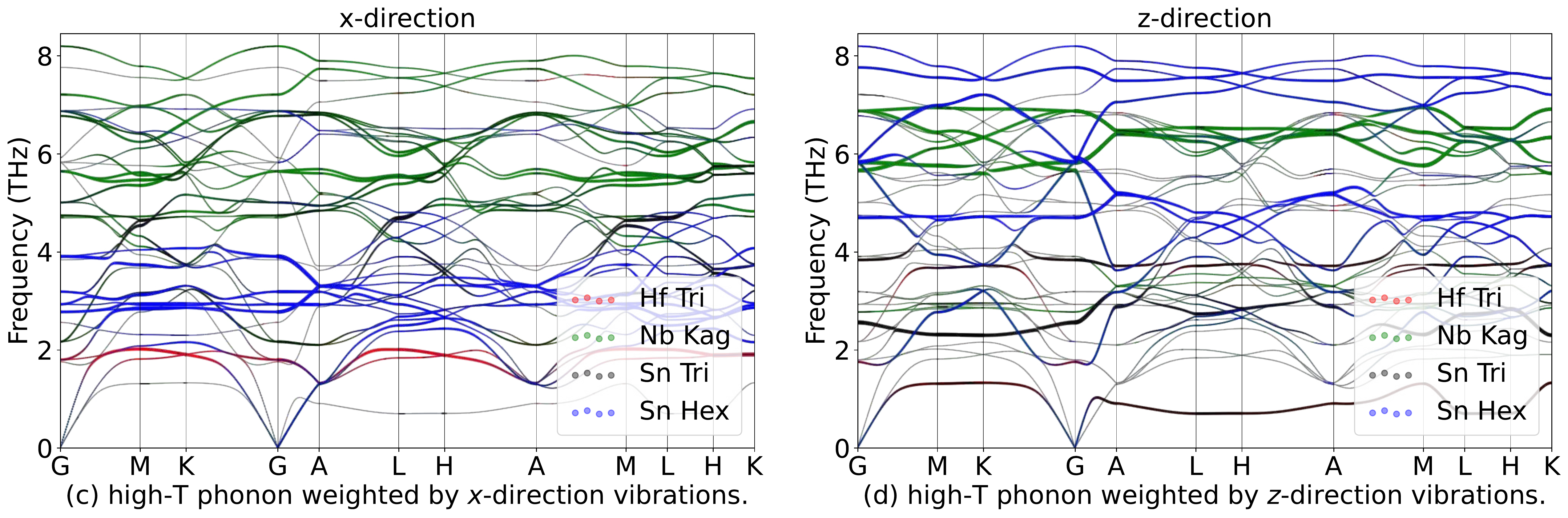}
\caption{Phonon spectrum for HfNb$_6$Sn$_6$in in-plane ($x$) and out-of-plane ($z$) directions.}
\label{fig:HfNb6Sn6phoxyz}
\end{figure}

\begin{table}[htbp]
\global\long\def\arraystretch{1.12}
\captionsetup{width=.95\textwidth}
\caption{IRREPs at high-symmetry points for HfNb$_6$Sn$_6$ phonon spectrum at Low-T.}
\label{tab:191_HfNb6Sn6_Low}
\setlength{\tabcolsep}{1.mm}{
}
\end{table}

\begin{table}[htbp]
\global\long\def\arraystretch{1.12}
\captionsetup{width=.95\textwidth}
\caption{IRREPs at high-symmetry points for HfNb$_6$Sn$_6$ phonon spectrum at High-T.}
\label{tab:191_HfNb6Sn6_High}
\setlength{\tabcolsep}{1.mm}{
%
}
\end{table}
 
\bibliography{ref}